\documentclass[letterpaper,11pt]{article}

\usepackage[margin=1in]{geometry} % full-width

\usepackage{amsmath}
\usepackage{amsthm}
\usepackage{amssymb}
\usepackage[utf8]{inputenc}
\usepackage{hyperref}
\usepackage[english]{babel}
\usepackage{enumitem}
\usepackage{graphicx}
\usepackage{setspace}
\usepackage{comment}
\usepackage{algorithm}
\usepackage{algpseudocode}
\usepackage{multirow}
\usepackage{enumitem}
\usepackage{hyperref}
\usepackage{enumitem}
\usepackage{lineno}
\usepackage{adjustbox}
\newcommand{\floor}[1]{\left \lfloor #1 \right \rfloor}

\newcommand{\polylog}{\mathop{\rm polylog}\limits}

\newcommand{\SA}{\mathsf{SA}}
\newcommand{\ISA}{\mathsf{ISA}}

\newcommand{\lcp}{\mathsf{lcp}}
\newcommand{\lcs}{\mathsf{lcs}}
\newcommand{\reverse}{\mathsf{rev}}
\newcommand{\Occ}{\mathit{Occ}}
\newcommand{\rangesum}{\mathsf{rSum}}
\newcommand{\rangecount}{\mathsf{rCount}}

\newcommand{\val}{\mathsf{val}}
\newcommand{\expr}{\mathsf{expr}}
\newcommand{\prule}{\mathsf{rule}}

\newcommand{\assign}{\mathsf{assign}}
\newcommand{\level}{\mathsf{level}}
\newcommand{\length}{\mathsf{len}}
\newcommand{\vOcc}{\mathsf{vocc}}

\newcommand{\interval}{\mathsf{interval}}

\newcommand{\pathP}{\mathsf{path}}

\newcommand{\RAQ}{\mathbf{RA}}
\newcommand{\LCEQ}{\mathbf{LCE}}
\newcommand{\rLCEQ}{\mathbf{rLCE}}

\newcommand{\RR}{\mathsf{RR}}
\newcommand{\recover}{\mathsf{recover}}
\newcommand{\source}{\mathsf{source}}
\newcommand{\modulo}{\mathsf{mod}}
\newcommand{\succeeding}{\mathsf{succ}}
\newcommand{\preceding}{\mathsf{prec}}
\newcommand{\capture}{\mathsf{capture}}
\newcommand{\run}{\mathsf{run}}
\newcommand{\centerset}{\mathsf{center}}
\newcommand{\str}{\mathsf{str}}
\newcommand{\lex}{\mathsf{lex}}
\newcommand{\leftmost}{\mathsf{leftmost}}
\newcommand{\CCP}{\mathsf{occ}}
\newcommand{\sRecover}{\mathsf{sRecover}}
\newcommand{\nRecover}{\mathsf{nRecover}}
\newcommand{\mRecover}{\mathsf{mRecover}}
\newcommand{\suffix}{\mathsf{suffix}}
\newcommand{\samp}{\mathsf{samp}}
\newcommand{\ofs}{\mathsf{offset}}

\newcommand{\levelQ}{\mathbf{Level}}
\newcommand{\attrQ}{\mathbf{AttrPos}}
\newcommand{\clenQ}{\mathbf{CLen}}
\newcommand{\clcpQ}{\mathbf{CLCP}}
\newcommand{\clcsQ}{\mathbf{CLCS}}
\newcommand{\runQ}{\mathbf{VerRun}}
\newcommand{\sourceQ}{\mathbf{VerSrc}}
\newcommand{\precQ}{\mathbf{VerPrec}}
\newcommand{\rsizeQ}{\mathbf{RSize}}
\newcommand{\sampleQ}{\mathbf{Sample}}
\newcommand{\OVQ}{\mathbf{OV}}
\newcommand{\CAPQ}{\mathbf{CAP}}
\newcommand{\BiSQ}{\mathbf{BiS}}
\newcommand{\BiAQ}{\mathbf{BiA}}
\newcommand{\RSCQ}{\mathbf{RSC}}
\newcommand{\RSSQ}{\mathbf{RSS}}

\newcommand{\RSCQA}{\mathbf{RSC}\text{-}\mathbf{A}}
\newcommand{\RSCQBX}{\mathbf{RSC}\text{-}\mathbf{B1}}
\newcommand{\RSCQBY}{\mathbf{RSC}\text{-}\mathbf{B2}}
\newcommand{\RSCQCX}{\mathbf{RSC}\text{-}\mathbf{C1}}
\newcommand{\RSCQCY}{\mathbf{RSC}\text{-}\mathbf{C2}}
\newcommand{\RSCQDX}{\mathbf{RSC}\text{-}\mathbf{D1}}
\newcommand{\RSCQDY}{\mathbf{RSC}\text{-}\mathbf{D2}}

\newcommand{\RSSQA}{\mathbf{RSS}\text{-}\mathbf{A}}
\newcommand{\RSSQB}{\mathbf{RSS}\text{-}\mathbf{B}}
\newcommand{\RSSQCX}{\mathbf{RSS}\text{-}\mathbf{C1}}
\newcommand{\RSSQCY}{\mathbf{RSS}\text{-}\mathbf{C2}}
\newcommand{\RSSQDX}{\mathbf{RSS}\text{-}\mathbf{D1}}
\newcommand{\RSSQDY}{\mathbf{RSS}\text{-}\mathbf{D2}}
\newcommand{\RSSQEX}{\mathbf{RSS}\text{-}\mathbf{E1}}
\newcommand{\RSSQEY}{\mathbf{RSS}\text{-}\mathbf{E2}}

\newcommand{\answer}{\mathsf{answer}}
\newcommand{\sub}{\mathsf{sub}}

\newcommand{\group}{\mathsf{group}}
\newcommand{\rightRun}{\mathsf{rightRun}}

\newcommand{\LEFT}{\mathsf{left}}
\newcommand{\RIGHT}{\mathsf{right}}
\newcommand{\NEW}{\mathsf{new}}
\newcommand{\OLD}{\mathsf{old}}
\newcommand{\ins}{\mathbf{ins}}
\newcommand{\del}{\mathbf{del}}

\newcommand{\free}{\mathsf{free}}

\newcommand{\symA}{\mathsf{A}}
\newcommand{\symB}{\mathsf{B}}

\newcommand{\bstart}{\mathsf{start}}
\newcommand{\bend}{\mathsf{end}}
\newcommand{\leftLen}{\mathsf{leftLen}}
\newcommand{\rightLen}{\mathsf{rightLen}}
\newcommand{\cover}{\mathsf{cover}}
\newcommand{\SUM}{\mathsf{sum}}

\theoremstyle{plain}
\newtheorem{theorem}{Theorem}[section]
\newtheorem{corollary}[theorem]{Corollary}
\newtheorem{lemma}[theorem]{Lemma}
\newtheorem{proposition}[theorem]{Proposition}

\theoremstyle{definition}
\newtheorem{definition}[theorem]{Definition}

\title{Dynamic Suffix Array in Optimal Compressed Space}

\author{Takaaki Nishimoto \and Yasuo Tabei}
\date{
	RIKEN Center for Advanced Intelligence Project, Tokyo, Japan \\ \textrm{\{takaaki.nishimoto, yasuo.tabei\}@riken.jp}
}

\begin{document}

\maketitle
\begin{abstract}
Big data, encompassing extensive datasets, has seen rapid expansion, notably with a considerable portion being textual data, including strings and texts.
Simple compression methods and standard data structures prove inadequate for processing these datasets, as they require decompression for usage or consume extensive memory resources.
Consequently, this motivation has led to the development of compressed data structures that support various queries for a given string, typically operating in polylogarithmic time and utilizing compressed space proportional to the string's length.
Notably, the suffix array (SA) query is a critical component in implementing a suffix tree, which has a broad spectrum of applications.

A line of research has been conducted on (especially, static) compressed data structures that support the SA query, where the input strings are assumed to be static, meaning they are not subject to alteration. 
A common finding from most of the studies is the suboptimal space efficiency of existing compressed data structures. Kociumaka, Navarro, and Prezza [IEEE Trans. Inf. Theory 2023] have made a significant contribution by introducing an asymptotically minimal space requirement, $O\left(\delta \log\frac{n\log\sigma}{\delta\log n} \log n \right)$ bits, sufficient to represent any string of length $n$, with an alphabet size of $\sigma$, and substring complexity $\delta$, serving as a measure of repetitiveness. The space is referred to as $\delta$-\emph{optimal space}. 
Several compressed data structures supporting random access, count and locate queries in $\delta$-optimal space have also been proposed. 
More recently, Kempa and Kociumaka [FOCS 2023] presented $\delta$-SA, a compressed data structure supporting SA queries in $\delta$-optimal space. 
 However, the data structures introduced thus far are static, unable to accommodate modifications to the input strings, resulting in a limitation on adaptability and utility in dynamic contexts, where an input string evolves over time.

We present the first dynamic compressed data structure that supports the SA query and update in polylogarithmic time and $\delta$-optimal space. More precisely, it can answer SA queries and perform updates in $O(\log^7 n)$ and expected $O(\log^8 n)$ time, respectively, using an expected 
$\delta$-optimal space.
Since our data structure is built by a randomized algorithm, the update operation fails with probability $O(n^{-w})$, where a user-defined constant parameter $w \geq 2$ can be adjusted.  
By selecting a sufficiently large value for $w$, 
we effectively reduce the failure probability to near zero, ensuring robust performance without compromising the efficiency of query support.
In addition, our data structure supports three essential queries for realizing suffix trees - inverse suffix array (ISA), random access (RA) and LCE queries. 
These queries are answered in $O(\log^5{n})$, $O(\log{n})$ and $O(\log n)$ times, respectively, using an expected $\delta$-optimal space.

Our data structure is not a mere extension of the $\delta$-SA, being rooted in an approach of independent interest that is fundamentally different, all aimed at achieving our specified objectives. Among these innovations are interval attractors; efficient queries such as restricted suffix count (RSC) and restricted suffix search (RSS) queries, tailored to answer SA queries with high efficiency; innovative representations of interval attractors through weighted points; and novel algorithms designed for the dynamic updating of our data structures.

\end{abstract}
\clearpage

\addtocontents{toc}{\protect\setcounter{tocdepth}{-1}}
\section{Introduction}\label{sec:introduction}
Big data, referring to massive datasets, has become prevalent in both research and industry.
These datasets are experiencing rapid expansion and predominantly consist of sequential or textual data.
Examples encompass web pages gathered by crawlers~\cite{FerraginaM10}, version-controlled documents such as those on Wikipedia~\cite{DBLP:conf/focs/KempaK20}, and, perhaps most significantly, biological sequences including human genomes~\cite{Przeworski00}.
The volume of these datasets has grown significantly. 
A common property of those textual datasets is their high degree of redundancy. 
For instance, it is observed that 99.9\% of human genomes are identical across different individuals~\cite{Przeworski00}.
These datasets are termed \emph{highly repetitive strings}, prompting researchers to concentrate on devising efficient compression and processing methods tailored for highly repetitive strings. While mere compression methods may not be effective - as one must compress a dataset and then decompress the compressed dataset for use - a truly desirable technique is a compressed data structure that enables us to compress a dataset and process it directly in its compressed format.

To meet this demand, several compressed data structures for highly repetitive strings have been proposed. 
These structures not only maintain data in a compressed format but also support various string queries with a moderate (typically polylogarithmic) increase in time complexity. 
Supported queries include random access~\cite{DBLP:journals/jacm/GagieNP20,9961143,DBLP:journals/jacm/GanardiJL21}, rank and select~\cite{DBLP:conf/cpm/Prezza19,DBLP:conf/isaac/FerraginaMV21,DBLP:journals/jcss/BelazzouguiCGGK21}, longest common extension (LCE)~\cite{DBLP:conf/cpm/I17,DBLP:conf/focs/KempaK23,DBLP:journals/jacm/GagieNP20}, and locate queries~\cite{9961143,DBLP:journals/talg/ChristiansenEKN21,DBLP:conf/icalp/NishimotoT21,ViceVersa}. 
Particularly noteworthy is the suffix array (SA) query (e.g., \cite{DBLP:journals/siamcomp/GrossiV05}), which identifies the starting position of the $i$th lexicographically smallest suffix in a string of length $n$ for any given position $i$ within the range $[1, n]$. The significance of the SA query extends beyond its basic functionality; it is a critical component for implementing a \emph{suffix tree}~\cite{DBLP:conf/focs/Weiner73}, facilitating a wide array of applications. These applications range from finding repeats and computing special subwords to comparing sequences and more, as detailed in a standard textbook~\cite{DBLP:books/cu/Gusfield1997}.

A line of research has been conducted on compressed data structures that support the SA query, addressing various aspects such as linear-time construction~\cite{DBLP:journals/talg/BelazzouguiCKM20}, optimal time or space efficiency~\cite{DBLP:conf/soda/Kempa19,DBLP:conf/soda/KempaK23}, and implementation within BWT-runs bounded space~\cite{DBLP:journals/jacm/GagieNP20,DBLP:conf/spire/BoucherKP022}. 
Notably, the input strings are assumed to be static, meaning they are not subject to alteration.
A common finding from these studies is the suboptimal space efficiency of existing compressed data structures.
Kociumaka, Navarro, and Prezza~\cite{9961143} have made a significant contribution by introducing an asymptotically minimal space requirement, \(O\left(\delta \log\frac{n\log\sigma}{\delta\log n} \log n \right)\) bits, sufficient to represent any string $T$ of length \(n\), with an alphabet size of \(\sigma\), and substring complexity \(\delta\), serving as a measure of repetitiveness. The space is referred to as $\delta$-\emph{optimal space}~\cite{ViceVersa,DBLP:conf/focs/KempaK23}. Furthermore, they also presented a compressed data structure supporting random access and pattern-matching queries in $\delta$-optimal space. 
A compressed data structure supporting locate queries in near-optimal time and $\delta$-optimal space is also proposed~\cite{ViceVersa}.
More recently, Kempa and Kociumaka~\cite{DBLP:journals/corr/abs-2308-03635,DBLP:conf/focs/KempaK23} presented $\delta$-SA, a compressed data structure supporting the SA query in $\delta$-optimal space. 
However, these data structures are static and cannot accommodate modifications to the input strings. 

The predominance of static compressed data structures inherently limits their adaptability and utility in dynamic contexts where an input string evolves over time.
This limitation underscores the critical need for continued research and development of dynamic data structures that can support updates. 
Such structures would not only enhance the versatility of data handling in applications requiring real-time data modification but also 
represent a significant advancement in the field of compressed data structures. 
Although a dynamic data structure supporting the SA query and updates in polylogarithmic time has been proposed~\cite{DBLP:conf/stoc/KempaK22}, it does not operate in compressed space.
State-of-the-art static and dynamic compressed data structures that support the SA query and other queries are summarized in Table~\ref{table:result} in Section~\ref{sec:related_works}.
Therefore, the challenge lies in developing a dynamic compressed data structure that supports both the SA query and updates efficiently in polylogarithmic time while maintaining $\delta$-optimal space.

\emph{Contribution.} 
In this paper, we present the first dynamic compressed data structure that supports the SA query and update in polylogarithmic time and expected $\delta$-optimal space. 
Our data structure not only supports SA queries and updates but also includes essential queries for realizing suffix trees, as follows: 
(i) SA query: For a given position $i$ within the range $[1, n]$, it returns the starting position of the $i$-th lexicographically smallest suffix of the input string in $O(\log^{7} n)$ time;
(ii) ISA query: It returns the number of suffixes of the input string $T$ that are lexicographically smaller than or equal to the suffix starting at a given position in $O(\log^{5} n)$ time;    
(iii) RA query: This query returns the character at a given position of the input string in $O(\log n)$ time; 
(iv) LCE query: It returns the length of the longest common prefix between the two suffixes starting at two given positions of the input string in $O(\log n)$ time;
(v) Update: The data structure supports insertion and deletion of a single character in the input string in expected $O(\log^{8} n)$ time. 
%The update fails with probability $O(n^{-w})$.
%Our compressed data structure supporting these queries uses expected $\delta$-optimal space. 
Since our data structure is built by a randomized algorithm, the update operation fails with probability $O(n^{-w})$ for a user-defined constant parameter $w \geq 2$.  
%By selecting a sufficiently large value for $w$, we effectively reduce the failure probability to near zero, ensuring robust performance without compromising the efficiency of the supported queries.

\section{Related Work and Method Overview}
%To efficiently solve SA and ISA queries, we generally consider a set of positions that are included in the same prefix, which appears at different locations within the suffixes of $T$. Each of these positions has the property that the distance between the position and the starting position of the corresponding prefix is the same. These positions are sampled from the string $T$ during preprocessing, and SA and ISA queries are solved using this set of sampled positions.
To efficiently solve SA and ISA queries, we sample positions within the same prefix that appear at different locations in the suffixes of $T$. Each sampled position has a fixed distance to the start of the corresponding prefix and is used during preprocessing to solve SA and ISA queries.

Kempa and Kociumaka~\cite{DBLP:conf/stoc/KempaK22} introduced string synchronizing sets, which are sets of sampled positions, and a dynamic uncompressed data structure capable of solving SA and ISA queries, as well as handling dynamic updates in $O(\polylog n)$ time, using $\Omega(n \polylog n)$ bits of space. Recently, they reduced the size of string synchronizing sets by leveraging a deterministic version of restricted recompression, and presented a static compressed data structure capable of solving SA and ISA queries in $O(\log^{4+\epsilon} n)$ time and $\delta$-optimal space for any given constant $\epsilon > 0$~\cite{DBLP:conf/focs/KempaK23}.

This paper presents \emph{interval attractors}, defined on derivation trees built via restricted recompression. Using interval attractors, we define restricted suffix count (RSC) and restricted suffix search (RSS) queries and employ a directed acyclic graph (DAG) representation of derivation trees to solve SA and ISA queries. A basic interval attractor representation requires $O(n^2 \log n)$ space, but by leveraging periodicity and mapping them onto two-dimensional grids, we develop compressed structures to store interval attractors in $\delta$-optimal space, achieving $O(\polylog n)$ query time. Additionally, our structures support dynamic updates, such as character insertions and deletions in $T$, with expected $O(\polylog n)$ time.

A similar approach based on string synchronizing sets~\cite{DBLP:conf/stoc/KempaK19} could yield a dynamic structure with comparable complexity. However, our interval attractors provide an advantage, enabling an ISA query solution in $O(\log^{4} n)$ time, improving upon the prior $O(\log^{4+\epsilon} n)$ time complexity~\cite{DBLP:conf/focs/KempaK23}.
The details of our method and the main results are outlined in the subsequent sections.

\section{Preliminaries}
An \emph{interval} $[b, e]$ for two integers $b$ and $e$ represents the set $\{b, b+1, \ldots, e \}$ if $b \geq e$; 
otherwise, the interval $[b, e]$ represents an empty set $\emptyset$ (i.e., $[b, e] = \emptyset$). 
Let $\min \mathcal{S}$~(respectively, $\max \mathcal{S}$) be the smallest integer~(respectively, the largest integer) in a nonempty set $\mathcal{S}$ of integers. 
Let $T$ be a string of length $n \geq 2$ over an alphabet $\Sigma = \{ 1, 2, \ldots, n^{\mathcal{O}(1)} \}$ of size $\sigma$, and 
$|T|$ denotes the length of $T$ (i.e., $|T| = n$). 
We assume a total order over the alphabet $\Sigma$.
Let $T[i]$ be the $i$-th character of $T$ (i.e., $T = T[1], T[2], \ldots, T[n]$) and 
let $T[i..j]$ be the substring of $T$ that begins at position $i$
and ends at position $j$ (i.e., $T[i..j] = T[i], T[i+1], \ldots, T[j]$). 
A \emph{prefix} (respectively, \emph{suffix}) of string $T$ is a substring that begins at position $1$ (respectively, ends at position $n$). 
A prefix of $T$ is called \emph{proper prefix} if it is shorter than $T$. 
Similarly, a suffix of $T$ is called \emph{proper suffix} if it is shorter than $T$. 
$\reverse(T)$ denotes the string obtained by reversing string $T$ (i.e., $\reverse(T) = T[n], T[n-1], \ldots, T[1]$). 
Let $\$$ and $\#$ be the smallest and largest characters in alphabet $\Sigma$, respectively, such that 
neither $\$$ nor $\#$ is contained in string $T$ (i.e., for all $i \in \{1, 2, \ldots, n \}$, both $T[i] \neq \$$ and $T[i] \neq \#$ hold).
We assume $T$ begins with $\$$ and ends with $\$$ 
($T[0] = \$$ and $T[n+1] = \$$) throughout this paper. 

For two characters $c, c^{\prime} \in \Sigma$, 
$c < c^{\prime}$ means that $c$ is smaller than $c^{\prime}$. 
%For a string $P$, $P[i] < P[j]$ means that the $i$-th character of $P$ is smaller than the $j$-th character of $P$. 
For a string $P$, $T \prec P$ holds if and only if either of the following two conditions holds: 
(i) there exists $i \in \{ 1, 2, \ldots, n \}$ such that $T[1..(i-1)] = P[1..(i-1)]$ and $T[i] < P[i]$ hold; (ii) $T = P[1..|T|]$ (i.e., $T$ is a prefix of $P$) and $|T| < |P|$ hold.  
In this case, $T$ is said to be lexicographically smaller than $P$.
$T \preceq P$ means that either $T = P$ or $T \prec P$ holds. 
A substring $T[i..j]$ of string $T$ is called \emph{occurrence} of string $P$ if 
the substring is equal to string $P$ (i.e., $T[i..j] = P$).

A suffix array~\cite{DBLP:journals/siamcomp/ManberM93} ($\SA$) of string $T$ is an integer array of size $n$ such that 
$\SA[i]$ stores the starting position of the $i$-th suffix of $T$ in lexicographical order. 
Formally, $\SA$ is a permutation of $n$ integers $1, 2, \ldots, n$ satisfying $T[\SA[1]..n] \prec \cdots \prec T[\SA[n]..n]$. 
The \emph{suffix array interval}~(\emph{sa-interval}) of string $P$ is an interval $[\eta, \eta^{\prime}] \subseteq \{ 1, 2, \ldots, n \}$ such that 
$\SA[\eta..\eta^{\prime}]$ represents all the occurrence positions of $P$ in string $T$ (i.e., $\forall i \in \{ 1, 2, \ldots, n \}, T[\SA[i]..\SA[i] + |P| - 1] = P \Leftrightarrow \eta \leq i \leq \eta^{\prime}$). 
If $P$ is the concatenation of a suffix $T[s..n]$ and character $\$$, 
then the sa-interval of $P$ is defined as $[s, s]$ for simplicity. 
An inverse suffix array ($\ISA$) of string $T$ is the inverse of suffix array $\SA$. 
That is, for any pair of integers $i, j \in \{ 1, 2, \ldots, n \}$, $\ISA[i] = j$ if and only if $\SA[j] = i$. 

Substring complexity~\cite{DBLP:journals/algorithmica/RaskhodnikovaRRS13,DBLP:journals/talg/ChristiansenEKN21} is a measure of a repetitiveness for strings. 
$\mathcal{S}_{d}$ is defined as a set of substrings of length $d \geq 0$ in string $T$ (i.e., $\mathcal{S}_{d} = \{ T[i..(i+d-1)] \mid i \in [1, n - d + 1] \}$). 
The substring complexity $\delta$ of string $T$ is defined as $\max \{ |\mathcal{S}_{1}| / 1, |\mathcal{S}_{2}| / 2, \ldots, |\mathcal{S}_{n}| / n  \}$. 

We will use base-2 logarithms throughout this paper unless otherwise indicated. 
Our computation model is a unit-cost word RAM~\cite{DBLP:conf/stacs/Hagerup98} with 
multiplication, randomization, and a machine word size of $B = \Theta(\log n)$ bits for input string $T$ of length $n$. 

%Due to space limitations, all proofs, tables, figures, and examples have been moved to Sections \ref{sec:preliminary}-\ref{sec:summary}.

%\section{Queries on Restricted Recompression for Answering SA and ISA queries}
%This section introduces restricted suffix count (RSC) and restricted suffix search (RSS) queries for addressing SA and ISA queries. Initially, we review the concept of restricted recompression, a form of grammar compression that constructs a derivation tree from an input string. Additionally, we present a novel property of restricted recompression related to the height of its derivation tree. Subsequently, we introduce a new concept termed \emph{interval attractors} defined within the derivation tree by restricted recompression. 
%Finally, we define RSC and RSS queries on interval attractors.

\section{Run-length Straight Line Program (RLSLP)}
The \emph{Straight Line Program (SLP)}~\cite{DBLP:journals/njc/KarpinskiRS97} is a form of Chomsky Normal Form for Context-Free Grammar (CFG) that derives a single string $T$. The \emph{Run-length SLP (RLSLP)}~\cite{DBLP:conf/mfcs/TanimuraNBIT17} is an extended version of SLP, which allows a production rule to generate a sequence of nonterminal symbols appearing successively. Formally, RLSLP is defined as a 4-tuple $\mathcal{G}^R=(\mathcal{V}, \Sigma, \mathcal{D}, E)$ where  $\mathcal{V} = \{ X_{1}, X_{2}, \ldots, X_{g} \}$ is a set of \emph{nonterminals}; $\Sigma$ is the alphabet for string $T$; $\mathcal{D} \subseteq \mathcal{V} \times (\Sigma \cup \mathcal{V}^+)$ is a set of production rules. 
Each production rule in $\mathcal{D}$ can take one of the following forms: (i) $X_{i} \rightarrow c$ where $c \in \Sigma$; (ii) $X_{i} \rightarrow X_j, X_k$ $(j, k < i)$ for two distinct nonterminals $X_j, X_k \in \mathcal{V}$, where $X_j \neq X_k$; (iii) $X_{i} \rightarrow X_j$ ($j < i$) where $X_j \in \mathcal{V}$; (iv) $X_{i} \rightarrow (X_{j})^{d}$ ($j < i$) where $(X_{j})^{d}$ represents $d (\geq 2)$ successive appearances of a nonterminal $X_j \in \mathcal{V}$.
$E \in \mathcal{V}$ is the unique nonterminal known as the \emph{start symbol}, which does not appear on the right-hand side of any production rule in $\mathcal{D}$ (i.e., for all $(X_{i} \rightarrow \expr_{i}) \in \mathcal{D}$, $\expr_{i}$ does not contain $E$).

RLSLP is represented as a labeled ordered tree in which each internal node corresponds to a nonterminal symbol in $\mathcal{V}$, each leaf corresponds to a terminal symbol in $\Sigma$, and each edge from a node to its children represents a production rule in $\mathcal{D}$. Furthermore, the root node corresponds to the start symbol $E \in \mathcal{V}$.
Let $H$ represent the height of the derivation tree of the RLSLP $\mathcal{G}^{R}$, and let $\mathcal{U}$ denote the set of all nodes in this derivation tree.

Several algorithms have been proposed to build an RLSLP for a string~(e.g., \emph{signature encoding}~\cite{DBLP:conf/mfcs/TanimuraNBIT17} and \emph{recompression}~\cite{DBLP:journals/talg/Jez15}). 
%Several algorithms have been proposed to build an RLSLP for a string~(e.g., \emph{signature encoding}~\cite{DBLP:conf/mfcs/TanimuraNBIT17}, \emph{recompression}~\cite{DBLP:journals/talg/Jez15}, \emph{signature grammar}~\cite{DBLP:conf/latin/ChristiansenE18} and \emph{restricted block compression}~\cite{DBLP:conf/latin/KociumakaNO22}). 
In the next section, we define the concept of an interval attractor, a common notion applicable to RLSLPs with the construction algorithms listed above.

\section{Interval Attractor}
Each node in the derivation tree of an RLSLP can be associated with a mutually exclusive and collectively exhaustive set of intervals (i.e., a partition) in the input string $T$. An interval attractor is defined for each node based on this partition.

Throughout this paper, let $u, u^{\prime} \in \mathcal{U}$ be nodes at heights $h, h' \in \{0,1, \ldots, H\}$ in the derivation tree of RLSLP $\mathcal{G}^{R}$, 
deriving the substrings that start at positions $\gamma$ and $\gamma^{\prime}$ on $T$, respectively.
Let $\Delta$ denote the set of all possible intervals of length at least 2 on input string $T$ (i.e., $\Delta = \{[s, e] \mid 1 \leq s < e \leq n \}$) throughout this paper.

We can construct the partition $\Delta(u) \subseteq \Delta$ associated with each $u \in \mathcal{U}$ such that any pair of intervals $[s,e] \in \Delta(u)$ and $[s^{\prime}, e^{\prime}] \in \Delta(u^{\prime})$ for any $u^{\prime} \in \mathcal{U}$ $(u \neq u')$ satisfies the following conditions: 
(i) $\gamma \in [s, e-1]$ and $\gamma^{\prime} \in [s^{\prime}, e^{\prime}-1]$; 
%(i) $\gamma \in [s, e]$ and $\gamma^{\prime} \in [s^{\prime}, e^{\prime}]$; 
(ii) If the distance from $s$ to $\gamma$ is unequal to the distance from $s'$ to $\gamma'$ (i.e., $\gamma - s \neq \gamma' - s'$), then substrings $T[s..e]$ and $T[s'..e']$ are different (i.e., $T[s..e] \neq T[s^{\prime}..e^{\prime}]$). (iii)  If $\gamma > \gamma'$ holds, then the substring $T[s..e]$ is not a prefix of $T[s^{\prime}..e^{\prime}]$ (i.e., either $s \neq s^{\prime}$ or $s = s^{\prime} \leq e^{\prime} < e$). (iv) If $h > h'$ holds, then $T[s..e]$ is not a substring of $T[s^{\prime}..e^{\prime}]$ (i.e., $[s, e] \not \subseteq [s', e']$). 
Additionally, if $u = u'$, then for any pair $(s^{\prime\prime}, e^{\prime\prime})$ such that $s^{\prime \prime} \in [\min \{ s, s^{\prime}  \}, s ]$ and $e^{\prime \prime} \in [e, \max \{ e, e^{\prime} \}]$, it holds that $[s^{\prime\prime}, e^{\prime\prime}] \in \Delta(u)$.
Note that for $\Delta(u)$ and $\Delta(u')$ associated with any pair of nodes $u, u' \in \mathcal{U}$, the conditions $\Delta(u) \cap \Delta(u^{\prime}) = \emptyset$ and $\bigcup_{u \in \mathcal{U}} \Delta(u) = \Delta$ hold.

Interval attractor $I(u)$ associated with $u \in \mathcal{U}$ is defined as a pair of intervals $([p,q],[\ell,r])$ such that $p = \min \{ s \}$, $q = \max \{ s \}$, $\ell = \min \{ e \}$, and $r = \max\{ e \}$ for all intervals $[s, e] \in \Delta(u)$.
If $\Delta(u)$ is empty, then the interval attractor $I(u)$ is undefined.

The lexicographical order between suffixes in the sa-interval of a string can be determined by comparing substrings associated with interval attractors as follows.
\begin{theorem}\label{theo:intro_sa_intv_formula}    
    For any two suffixes $T[\SA[i]..n]$ and $T[\SA[i']..n]$ in the sa-interval of a string $P (|P| \geq 2)$ on the suffix array $\SA$ of $T$, 
    there exist nodes $u, u' \in \mathcal{U}$ such that $[\SA[i], \SA[i] + |P|-1] \in \Delta(u)$ and $T[\SA[i']..(\SA[i'] + |P| - 1)] \in \Delta(u')$.
    Also, there exist interval attractors $I(u)=([p,q], [\ell, r])$ and $I(u')=([p',q'], [\ell', r'])$ for such nodes $u, u' \in \mathcal{U}$. 
    If $T[\SA[i]..r+1] \neq T[\SA[i^{\prime}]..r^{\prime}+1]$, 
    then the relation $i < i^{\prime} \Leftrightarrow T[\gamma..r+1] \prec T[\gamma^{\prime}..r^{\prime}+1]$ holds.
\end{theorem}

% Original
\begin{comment}
\color{blue}
The lexicographical order between any suffixes starting from positions $\gamma$ and $\gamma'$ with nodes $u, u' \in \mathcal{U}$ can be determined by comparing substrings associated with interval attractors $I(u)$ and $I(u')$ as follows.
\begin{theorem}
Let $\SA$ be the suffix array of $T$.
For two SA intervals $[\SA[i]..\SA[i] + |P|-1]$ and $[\SA[i^{\prime}]..\SA[i^{\prime}] + |P|-1]$ with the appearance of the same substring $P$ of $T$, 
there exist nodes $u, u^{\prime} \in \mathcal{U}$ in the derivation tree of an RLSLP of $T$ such that $[\SA[i], \SA[i] + |P|-1] \in \Delta(u)$ and $[\SA[i^{\prime}], \SA[i^{\prime}] + |P|-1] \in \Delta(u^{\prime})$, respectively. 
%Let $I(u) = ([p,q], [\ell, r])$ and $I(u') = ([p',q'], [\ell', r'])$ be interval attractors of $u$ and $u'$, respectively. 
If $T[\SA[i]..r+1] \neq T[\SA[i^{\prime}]..r^{\prime}+1]$, 
then the relation $i < i^{\prime} \Leftrightarrow T[\gamma..r+1] \prec T[\gamma^{\prime}..r^{\prime}+1]$ holds.
%If $T[\gamma..r+1] \neq T[\gamma^{\prime}..r^{\prime}+1]$, 
%then the relation $i < i^{\prime} \Leftrightarrow T[\gamma..r+1] \prec T[\gamma^{\prime}..r^{\prime}+1]$ holds.
\end{theorem}
\color{black}
\end{comment}

Based on Theorem~\ref{theo:intro_sa_intv_formula}, we introduce RSC and RSS queries, which can be used to efficiently solve SA and ISA queries in the following section.

\section{RSC and RSS Queries}
%\begin{wrapfigure}{position}[overhang]{width}
% \begin{center}
%		\includegraphics[scale=0.7]{figures/rsc.pdf}
%		\includegraphics[scale=0.7]{figures/rss.pdf}
%
%	  \caption{ 
%   (Left) An illustration of RSC query $\RSCQ(s, e)$. 
%   There exists a node $u \in \mathcal{U}$ such that $[s, e] \in \Delta(u)$, 
%   and interval attractor $I(u)$ is associated with $u$. 
%   $[\eta_{1}, \eta^{\prime}_{1}]$ and $[\eta_{2}, \eta^{\prime}_{2}]$ 
%   are the sa-intervals of two substrings 
%   $T[s..e]$ and $T[s..\min \{ r+1, n \}]$ on the suffix array $\SA$ of $T$, respectively. 
%   This RSC query returns the number of suffixes within yellow interval $[\eta_{1}, \eta_{2}-1]$ on the suffix array. 
%   (Right) An illustration of RSS query $\RSSQ(P, b)$. 
%   The $b$-th suffix on the suffix array appears in the sa-interval of $P$ 
%   and has prefix $P$ as a substring $T[s..e]$.    
%   There exists a node $u \in \mathcal{U}$ such that $[s, e] \in \Delta(u)$, 
%   and interval attractor $I(u)$ is associated with $u$. 
%   This RSS query returns substring $T[s..r+1]$.   
%	  }
%\label{fig:rsc_rss}
% \end{center}
%\end{wrapfigure}

\begin{figure}[ht]
\centering
\begin{minipage}[b]{0.45\columnwidth}
    \raggedright
%    \vspace{-0.5cm}
		\includegraphics[scale=0.7]{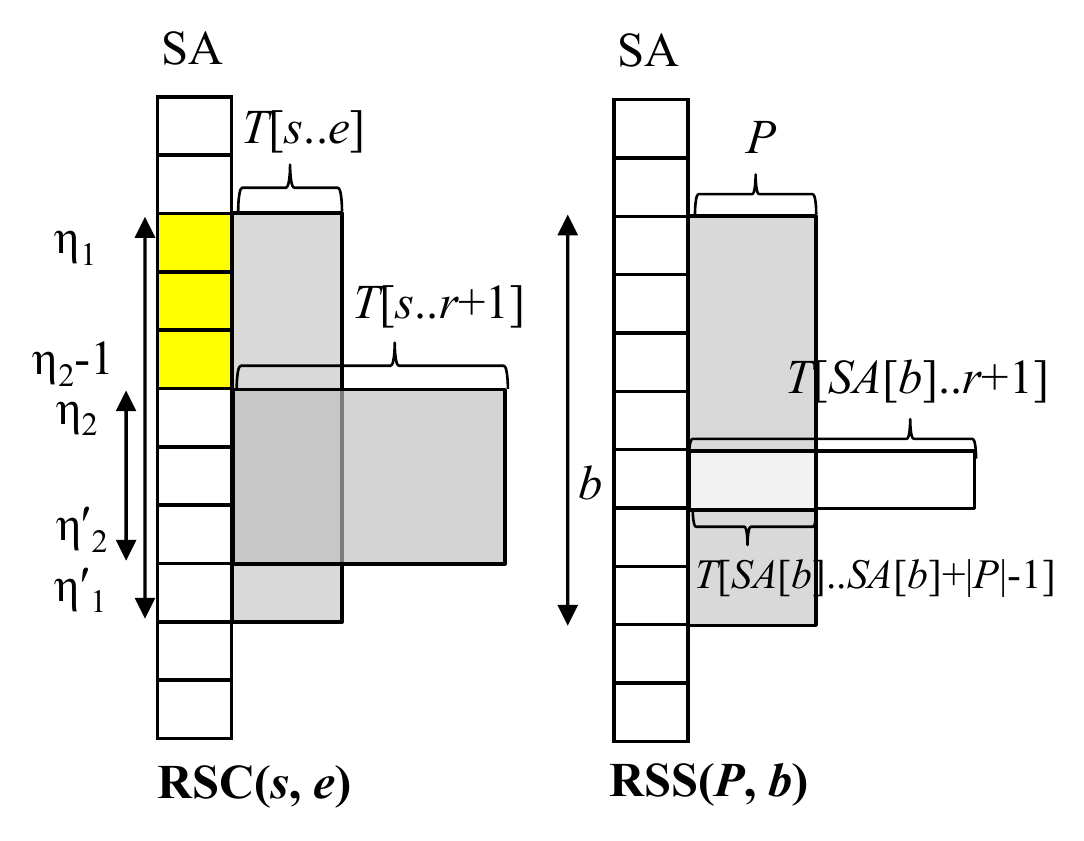}
    \caption{An illustration of RSC query $\RSCQ(s, e)$ (left) and RSS query $\RSSQ(P, b)$ (right). }
        \label{fig:intro_rsc_rss}
\end{minipage}
\begin{minipage}[b]{0.45\columnwidth}
    %\vspace{-3.5cm}
    \raggedright
	\includegraphics[scale=0.7]{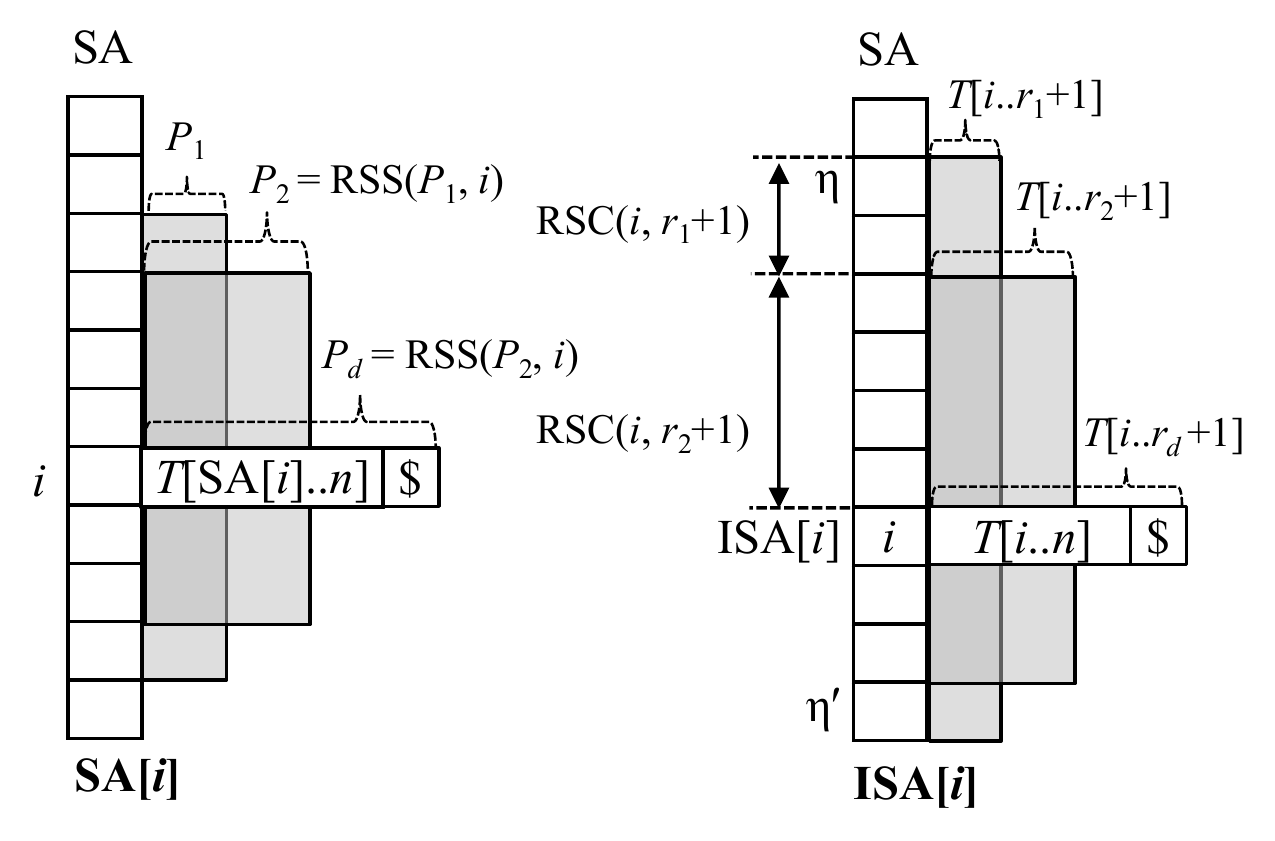}
    \caption{An illustration of Theorem~\ref{theo:intro_sa_query} (left) and Theorem~\ref{theo:intro_isa_query} (right).}
        \label{fig:intro_sa_isa}
\end{minipage}
\end{figure}

\emph{Restricted suffix count} (RSC) and \emph{restricted suffix search} (RSS) queries are ones that count and search suffixes associated with interval attractors.
%Let $u \in \mathcal{U}$ be a node whose associated partition $\Delta(u)$ includes an interval $[s,e] \in \Delta$. 
%Let $I(u) = ([p,q],[\ell,r])$ be the interval attractor associated with $u \in \mathcal{U}$.
\begin{definition}[RSC query]
    For a given interval $[s, e] \in \Delta$, there exists a node $u \in \mathcal{U}$ such that $[s, e] \in \Delta(u)$.
    Let $I(u)=([p,q], [\ell, r])$ be the interval attractor associated with $u$. 
    Consider the sa-intervals $[\eta_{1}, \eta^{\prime}_{1}]$ and $[\eta_{2}, \eta^{\prime}_{2}]$ of two substrings $T[s..e]$ and $T[s..r+1]$ on the suffix array $\SA$ of $T$, respectively, both starting at the same position $s$.
    For the given interval $[s, e]$, $\RSCQ(s, e) = \eta_{2} - \eta_{1}$, 
    which represents the number of suffixes within interval $[\eta_{1}, \eta_{2}-1]$ on the suffix array. 
\end{definition}

The figure on the left in Figure~\ref{fig:intro_rsc_rss} illustrates the RSC query.
The RSC query computes the number of suffixes that have the prefix $T[s,e]$ and is lexicographically smaller than $T[s..r+1]$, which is represented as positions colored yellow in SA in the figure. 

\begin{definition}[RSS query]
    RSS query $\RSSQ(P, b)$ takes as input a pair of a substring $P$ of $T$ with $|P| \geq 2$ and a position $b$ in the sa-interval of $P$ on $\SA$. 
    For the interval $[\SA[b], \SA[b]+|P|-1]$ with the prefix $T[\SA[b]..\SA[b]+|P|-1]$ of the suffix $T[\SA[b]..n]$, 
    there exists an interval attractor $I(u) = ([p,q], [\ell,r])$ for some $u \in \mathcal{U}$ such that $[\SA[b],\SA[b]+|P|-1] \in \Delta(u)$.
    Then, $\RSSQ(P, b)$ returns $T[\SA[b]..r+1]$.
\end{definition}

The figure on the right in Figure~\ref{fig:intro_rsc_rss} illustrates the RSS query.
The following theorems show that RSC query can be answered by counting interval attractors satisfying certain conditions. 
\begin{theorem}\label{theo:intro_rsc_query_formula}
    $\RSCQ(s, e)$ for an input $[s,e] \in \Delta$ can be computed as the number of nodes $u'$ with interval attractors $I(u^{\prime})$ that satisfy the following three conditions: 
        (i) the height of $u^{\prime}$ is the same as that of $u$ in the derivation tree;
        (ii) $T[p^\prime-1..\gamma^\prime-1]$ includes $T[s..\gamma-1]$ as a proper suffix;
        (iii) $T[\gamma^\prime..r^\prime+1]$ includes $T[\gamma..e]$ as a proper prefix;
        (iv) $T[\gamma^\prime..r^\prime+1]$ is lexicographically smaller than $T[\gamma..r+1]$.
%    \begin{enumerate}[itemsep=0pt, parsep=0pt,before=\vspace{-0.3\baselineskip}, after=\vspace{-0.3\baselineskip}, label=\textbf{(\roman*)}]
%        \item the height of $u^{\prime}$ is the same as that of $u$ in the derivation tree.
%        \item $T[p^\prime-1..\gamma^\prime-1]$ includes $T[s..\gamma-1]$ as a proper suffix.
%        \item $T[\gamma^\prime..r^\prime+1]$ includes $T[\gamma..e]$ as a proper prefix.
%        \item $T[\gamma^\prime..r^\prime+1]$ is lexicographically smaller than %$T[\gamma..r+1]$.
%    \end{enumerate}
\end{theorem}

The following theorem shows that RSS query can be answered using the RSC query. 
\begin{theorem}\label{theo:intro_rss_query_formula}
    For an input pair $(P, b)$ of $\RSSQ(P, b)$, 
    let $[\eta_{1}, \eta^{\prime}_{1}]$ be the sa-interval of $P$ on $\SA$ and let $b' \in [\eta_{1}, \eta^{\prime}_{1}]$.
    For any position $b' \in [\eta_{1}, \eta^{\prime}_{1}]$, 
    there exists the interval $[\SA[b^{\prime}], \SA[b^{\prime}]+|P|-1]$ with the prefix $T[\SA[b^{\prime}]..\SA[b^{\prime}]+|P|-1]$ of the suffix $T[\SA[b^{\prime}]..n]$. 
    Additionally there exists an interval attractor $I(u^{\prime}) = ([p^{\prime},q^{\prime}], [\ell^{\prime},r^{\prime}])$ for some $u^{\prime} \in \mathcal{U}$ such that $[\SA[b^{\prime}],\SA[b^{\prime}]+|P|-1] \in \Delta(u^{\prime})$.
    Then, $\RSSQ(P, b) = T[\SA[b^{\prime}]..r^{\prime}+1]$ for $b' \in [\eta_{1}, \eta^{\prime}_{1}]$ such that the interval $[\SA[b^{\prime}], \SA[b^{\prime}]+|P|-1]$ satisfies the following two conditions: (i) $\eta_{1} + \RSCQ(\SA[b^{\prime}], \SA[b^{\prime}]+|P|-1) \leq b$; (ii) $\eta_{1} + \RSCQ(\SA[b^{\prime}], \SA[b^{\prime}]+|P|-1) + |[\eta_{2}, \eta^{\prime}_{2}]| - 1 \geq b$, where $[\eta_{2}, \eta^{\prime}_{2}]$ is the sa-interval of $T[\SA[b^{\prime}]..r^{\prime}+1]$.
%    \end{enumerate}
%    \begin{enumerate}[itemsep=0pt, parsep=0pt,before=\vspace{-0.3\baselineskip}, %after=\vspace{-0.3\baselineskip}, label=\textbf{(\roman*)}]    
%        \item $\eta_{1} + \RSCQ(\SA[b^{\prime}], \SA[b^{\prime}]+|P|-1) \leq b$;
%        \item $\eta_{1} + \RSCQ(\SA[b^{\prime}], \SA[b^{\prime}]+|P|-1) + |[\eta_{2}, %\eta^{\prime}_{2}]| - 1 \geq b$, where $[\eta_{2}, \eta^{\prime}_{2}]$ is the sa-interval %of $T[\SA[b^{\prime}]..r^{\prime}+1]$.
%    \end{enumerate}
\end{theorem}
%\color{red}
%We explain Theorem~\ref{theo:intro_rss_query_formula} using Theorem~\ref{theo:intro_sa_intv_formula}. 
%According to Theorem~\ref{theo:intro_sa_intv_formula}, 
%the interval attractor $I(u^{\prime})$ corresponds to the suffix $T[b^{\prime}..n]$. 
%Using Theorem~\ref{theo:intro_sa_intv_formula}, 
%we can show that the substring returned by RSS query is 
%the same as substring $T[\SA[b^{\prime}]..r^{\prime}+1]$ if 
%the sa-interval $[\eta_{2}, \eta^{\prime}_{2}]$ of the substring $T[\SA[b^{\prime}]..r^{\prime}+1]$ contains position $b$. 
%The two conditions of Theorem~\ref{theo:intro_rss_query_formula} shows that 
%we can verify whether $b \in [\eta_{2}, \eta^{\prime}_{2}]$ or not using RSC query. 
%Therefore, we can answer RSS query using the two conditions of Theorem~\ref{theo:intro_rss_query_formula}.  

%the sa-interval $[\eta_{2}, \eta^{\prime}_{2}]$
%the interval attractor $I(u^{\prime})$ corresponds to a suffix in the sa-interval of $T[s..e]$ in Theorem~\ref{theo:intro_sa_intv_formula}. 
%In Condition (i) (respectively, Condition (ii)), 
%the left side of the inequality represents position $\eta_{2}$ (respectively, $\eta^{\prime}_{2}$). 
%\color{black}

    Both Theorem~\ref{theo:intro_rsc_query_formula} and Theorem~\ref{theo:intro_rss_query_formula} are proven by employing Theorem~\ref{theo:intro_sa_intv_formula}. 
    RSC and RSS queries are decomposed into seven and eight distinct subqueries, respectively, each of which can be efficiently solved on a grid constructed by mapping interval attractors. The solution to the original query is obtained by summing the results of these subqueries. Details of the subqueries are listed in Table~\ref{table:RSC_query_result} and 
    Table~\ref{table:RSS_query_result} and elaborated upon in Section~\ref{sec:RSC_query} and Section~\ref{sec:RSS_query}, respectively.

\section{SA and ISA Queries}
%\begin{figure}[t]
% \begin{center}
%		\includegraphics[scale=0.7]{figures/sa_query.pdf}
%		\includegraphics[scale=0.7]{figures/isa_query.pdf}
%
%	  \caption{ 
%    An illustration of Theorem~\ref{theo:intro_sa_query} (left) and Theorem~\ref{theo:intro_isa_query} (right). 
%	  }
%\label{fig:intro_sa_isa}
% \end{center}
%\end{figure}

SA and ISA queries can be addressed using RSS and RSC queries, as shown in Figure~\ref{fig:intro_rsc_rss} and described below.
%\begin{theorem}[The relationship between ISA and RSC queries]
\begin{theorem}\label{theo:intro_sa_query}
    %Let $i \in [1,|\SA|]$ be a position on $\SA$ such that $\SA[i] \neq n$. 
    Let $i \in [1,n]$ be an input to the SA query $\SA[i]$ such that $\SA[i] \neq n$.
    Define $P_{j}$ for $j \geq 1$ recursively as follows:
    $P_{1} = T[\SA[i]..\SA[i]+1]$; 
    $P_{j} = \RSSQ(P_{j-1}, i)$ for $j \geq 2$. 
    Then, $\SA[i] = n - |P_{d}| + 2$ for some $d \geq 1$ and the last character $\$$ of string $P_{d}$.
\end{theorem}

%\color{red}
\begin{theorem}\label{theo:intro_isa_query}
    Let $i \in [1,n-1]$ be an input to the ISA query $\ISA[i]$.
    Define $r_{j} \in [1, n]$ for $j \geq 1$ recursively as follows:     
    $r_{1} = i$; 
    for $j \geq 2$, 
    $r_{j} = r$ for the interval attractor $I(u) = ([p, q], [\ell, r])$ associated with the unique node $u \in \mathcal{U}$ such that $[i, r_{j-1} + 1] \in \Delta(u)$.  
    Then, for $d \geq 1$ such that $r_{d} = n$ and for the sa-interval $[\eta, \eta^{\prime}]$ of the substring $T[i..r_{1}+1]$, the ISA query is computed as $\ISA[i] = \eta + \sum_{j = 1}^{d-1} \RSCQ(i, r_{j}+1)$.    
    %Let $i \in [1,n-1]$ be an input to the ISA query $\ISA[i]$.
    %Define $u_{j} \in \mathcal{U}$ for $j \geq 1$ recursively as follows: 
    %$u_{1}$ is a node such that $[i, r_{0} + 1] \in \Delta(u_1)$ where $r_{0} = 1$; 
    %$u_{j}$ is a node such that $[i, r_{j-1} + 1] \in \Delta(u_j)$ for $j \geq 2$.
    %Then, for $d \geq 1$ such that $r_{d} = n$ and for the sa-interval $[\eta, \eta^{\prime}]$ of the substring $T[i..r_{0}+1]$, the ISA query is computed as $\ISA[i] = \eta + \sum_{j = 1}^{d} \RSCQ(i, r_{j-1}+1)$.    
    %Let $I(u_j) = ([p_j, q_j], [\ell_j, r_j])$ ($j \geq 1$) be an interval attractor for node $u_j \in \mathcal{U}$ such that $[i, r_{j-1} + 1] \in \Delta(u_j)$.
    %Then, for $d \geq 1$ such that $r_0=1$ and $r_{d} = n$ and for the sa-interval $[\eta, \eta^{\prime}]$ of the substring $T[i..i+1]$, the ISA query is computed as $\ISA[i] = \eta + \sum_{j = 1}^{d} \RSCQ(i, r_{j-1}+1)$.
\end{theorem}
%\color{black}

\begin{comment}
\color{blue}
\begin{theorem}\label{theo:intro_isa_query}
    Let $I(u_j) = ([p_j, q_j], [\ell_j, r_j])$ ($j \geq 1$) be an interval attractor for node $u_j \in \mathcal{U}$ such that $[i, r_{j-1} + 1] \in \Delta(u_j)$.
    Then, for $d \geq 1$ such that $r_0=1$ and $r_{d} = n$ and for the sa-interval $[\eta, \eta^{\prime}]$ of the substring $T[i..i+1]$, the ISA query is computed as $\ISA[i] = \eta + \sum_{j = 1}^{d} \RSCQ(i, r_{j-1}+1)$.
\end{theorem}
\color{black}
\end{comment}

\begin{comment}
\color{blue}
\begin{theorem}\label{theo:intro_isa_query}
    Consider a suffix $T[i..n]$ of length at least $2$ in $T$.  
    Positions $r_j$ for $j \geq 1$ on $T$ are defined as  
    $r_1 = i$; $r_j = r$ for $j \geq 2$, where $I(s^k)=([p,q],[\ell,r])$ for the tail $[s^k, e^k]$ of $A(i, r_{j-1} + 1)$.
    Let $d^{\prime} \geq 1$ be the integer such that $r_{d^{\prime}} = n$.
    Let $[\eta, \eta^{\prime}]$ be the sa-interval of string $T[i..i+1]$. 
    ISA query $\ISA[i]$ for a given position $i$ is computed as $\eta + \sum_{j = 1}^{d^{\prime}-1} \RSCQ(i, r_{j}+1)$.
\end{theorem}
\color{black}
\end{comment}

%\color{red}
%Figure~\ref{fig:intro_sa_isa} illustrates Theorem~\ref{theo:intro_sa_query} and %Theorem~\ref{theo:intro_isa_query}.
%\color{black}
Pseudo-codes for computing SA and ISA queries are presented in Algorithm~\ref{algo:light_sa_query} and Algorithm~\ref{algo:light_isa_query} in Section~\ref{sec:sa_and_isa_queries_with_optimal}. 
In the subsequent section, we present an overview of our compressed dynamic data structure for answering SA and ISA queries in $O(H^{3} \log^{2} n + H \log^{6} n)$ time and $O(H^{3} \log n + H \log^{4} n)$ time, respectively, 
using expected $O(\delta \log \frac{n \log \sigma}{\delta \log n} B)$ bits of space for machine word size $B = \Theta(\log n)$.

\section{Restricted Recompression}
Restricted recompression~\cite{9961143,DBLP:journals/corr/KociumakaRRW13} is a randomized algorithm developed to construct the RLSLP $\mathcal{G}^R=(\mathcal{V}, \Sigma, \mathcal{D}, E)$ 
with three distinct properties from $T$.
The first property is that restricted recompression generates an RLSLP whose derivation tree of height $H$ is \emph{height-balanced}. 
This means that the tree either (a) consists solely of a single node, namely the root, or (b) has all children of any given node at the same height. 
Thus, nonterminals in $\mathcal{V}$ or characters in $\Sigma$ at each height $h \in [0,H]$ in the derivation tree can be regarded as a sequence $S^h$, resulting in a total of $H+1$ sequences, denoted by $S^0$, $S^1$, $\ldots$, $S^H$. 
The second property is that if any two distinct nodes $u, u' \in \mathcal{U}$ at the same height in the derivation tree derive the same substring in $T$, they have the same nonterminal in $\mathcal{V}$ as a node label. 
%\color{red}
%Let $\mu(t)$ be a function that returns $(8/7)^{\lceil h/2 \rceil - 1}$ for integer $t$. 
%\color{black}
The third property is as follows: let $T[s..e]$ and $T[s^{\prime}..e^{\prime}]$ be two distinct occurrences of the same substring $P$ in $T$.  
If there exists a node $u \in \mathcal{U}$ at height $h$ that derives a substring starting at position $\gamma \in [s + 16\mu(h+1), e - 16\mu(h+1)]$ in $T$ where $\mu(h)=(8/7)^{\lceil h/2 \rceil - 1}$, 
then there exists another node $u' \in \mathcal{U}$ at the same height $h$ that derives a substring starting at position $s^{\prime} + |[s, \gamma]| - 1$~\cite{9961143}.

These three properties, especially the third property, allow us to present a dynamic data structure whose size and query time are dependent on the height $H$ of the derivation tree by the restricted recompression. Thus, we derive a new theorem, which states that the height $H$ of the derivation tree is $O(\log n)$ with a probabilistic guarantee.
\begin{theorem}\label{theo:tree_height}
$H \leq 2(w+1) \log_{8/7} (4n) + 2$ holds with probability $1 - (1/n^{w})$ 
for the height $H$ of the RLSLP built by the restricted recompression and for all integers $w \geq 1$. 
\end{theorem}

%\color{red}
%To determine the form of each production rule in $\mathcal{D}$, 
%the restricted recompression assigns to each nonterminal contained in $S^h$ for each $h %\in [0, H]$, at least one value among $\{-1, 0, +1\}$. 
%If the nonterminal generates a string of length greater than $(8/7)^{\lceil h/2 \rceil %- 1}$, then $-1$ is assigned to the nonterminal. Otherwise, $0$ or $1$ is uniformly and %randomly assigned to the nonterminal. 
%These randomly assigned integers need to preserved for updating dynamic data structures %on the RLSLP built by the restricted recompression.
%\color{black}

\section{Dynamic Compressed Representation of Interval Attractors} %in \texorpdfstring{$\delta$}{}-optimal space}
RSC and RSS queries can efficiently be answered in $O(\polylog(n))$ time by mapping interval attractors to a grid and computing solutions satisfying Theorem~\ref{theo:intro_rsc_query_formula} and Theorem~\ref{theo:intro_rss_query_formula} using a range reporting on the grid. 

\subsection{Reduction of the Number of Interval Attractors by Periodicity}
\begin{comment}
\color{red}
Periodicity within interval attractors enables reducing the number of interval attractors that need to be preserved. 
Consider the interval attractor $I(u) = ([p, q], [\ell, r])$ associated with a node $u$ of height $h$ deriving a substring $C$ starting at a position $\gamma$ on $T$. 
An interval attractor $I(u)$ is periodic if it meets the following two conditions: (i) $T[p-1..\gamma-1]$ is a suffix of a repeated sequence of $C$; (ii) $T[\gamma..r]$ begins with a repetition of $C$ that exceeds the length of $1 + \sum_{w=1}^{h+3} \lfloor \mu(w) \rfloor$.
If either of these conditions is not met, $I(u)$ is considered non-periodic.
\color{black}
\end{comment}
Periodicity within interval attractors enables reducing the number of interval attractors that need to be preserved. 
%\color{red}
%A periodic interval attractor is an interval attractor such that 
%the substring associated with the interval attractor contains 
%a sufficiently long repetition of the substring derived from the corresponding node. 
%Interval attractors can be classified as either periodic or non-periodic, defined as follows.
%\color{black} 
Consider an interval attractor, $I(u) = ([p, q], [\ell, r])$, for the node $u$ corresponding to a position $s \in [1,|S^h|]$ in $S^h$ at height $h \in [0, H]$. 
%Consider an interval attractor, $I(S^h[s]) = ([p, q], [\ell, r])$, for a position $s \in [1,|S^h|]$ in $S^h$ at $h \in [0, H]$. 
Let $C$ represent a substring derived from the nonterminal $S^h[s]$, starting from position $\gamma$ in $T$. %(i.e., $C = \text{val}(S^{h}[s^{h}])$). 
%Denote by $\gamma$ a starting position of the substring $C$ in $T$. 
An interval attractor $I(u)$ is periodic if it meets the following two conditions: (i) $T[p-1..\gamma-1]$ is a suffix of a repeated sequence of $C$; (ii) $T[\gamma..r]$ is a repetition of $C$ with a length more than or equal to $1 + \sum_{w=1}^{h+3} \lfloor \mu(w) \rfloor$.
If either of these conditions is not met, $I(u)$ is considered non-periodic.

The following theorem states that periodic interval attractors can be reconstructed from non-periodic interval attractors.
\begin{theorem}\label{theo:intro_periodicity}
    Consider the interval attractor $I(u_s) = ([p, q], [\ell, r])$ for the node $u_s$ at each position $s \in [1, |S^h|]$ at height $h \in [0, H]$. 
    Let $C$ be the substring derived from the nonterminal $S^h[s]$, starting from 
    position $\gamma$ in $T$.  
    Let $K$ be the length of the longest common prefix between two strings $T[\gamma..r]$ and $C^{n+1}$.
    For $d = \lfloor \frac{K - (2 + \sum_{w = 1}^{h+3} \lfloor \mu(w) \rfloor)}{|C|} \rfloor$ with $d \geq 1$,
    if $I(u_s)$ is non-periodic and $I(u_{s+1})$ is periodic, then the following three statements hold: 
        (i) $I(u_{s+j})$ for each $j \in [1, d]$ is periodic, 
        (ii) $I(u_{s+d+1})$ is non-periodic;
        (iii) for all $j \in [1, d]$, there exists a unique integer $\tau \in \mathbb{Z}$ such that $([q+1 + (j - 1)|C|, q + d|C|], [\ell + \tau + (j - 1)|C|, r])$ is the interval attractor associated with node $u_{s+j} \in \mathcal{U}$ (i.e., $I(u_{s+j}) = ([q+1 + (j - 1)|C|, q + d|C|], [\ell + \tau + (j - 1)|C|, r])$).    
       %(ii) there exists an integer $\tau \in \mathbb{Z}$ such that
        %$I(u_{s+1}) = ([q+1, q + |C|], [\ell + \tau, r])$, 
        %$I(u_{s+2}) = ([q+1 + |C|, q + 2|C|], [\ell + \tau + |C|, r])$, 
        %$\ldots$
        %$I(u_{s+d}) = ([q+1 + (d - 1)|C|, q + d|C|], [\ell + \tau + (d - 1)|C|, r])$.    
%
%    \begin{enumerate}[itemsep=0pt, parsep=0pt,before=\vspace{-0.5\baselineskip}, after=\vspace{-0.5\baselineskip}, label=\textbf{(\roman*)}]
%    %\begin{enumerate}[label=\textbf{(\roman*)}]
%        \item $d \geq 1$;
%        \item $I(u_{s+j})$ for any $j \in [1, d]$ is periodic, 
%        and $I(u_{s + d + 1})$ is non-periodic;
%        \item there exists an integer $\tau \in \mathbb{Z}$ such that
%        $I(u_{s+1}) = ([q+1, q + |C|], [\ell + \tau, r])$, 
%        $I(u_{s+2}) = ([q+1 + |C|, q + 2|C|], [\ell + \tau + |C|, r])$, 
%        $\ldots$
%        $I(u_{s+d}) = ([q+1 + (d - 1)|C|, q + d|C|], [\ell + \tau + (d - 1)|C|, r])$.    
%    \end{enumerate}
%    Here, $I(u_{s+1}), I(u_{s+2}), \ldots, I(u_{s + d})$ are called interval attractors %reconstructed from $I(u_s)$. 
\end{theorem}

%\color{red}
%Periodic interval attractors are used to reduce the number of interval attractors mapped to %a grid, which is used to answer RSC and RSS queries.  
%In other words, only non-periodic interval attractors are mapped to the grid. 
%By leveraging the periodicity of interval attractors, 
%we can answer RSC and RSS queries without storing periodic interval attractors. 
%The next subsection will explain how to map non-periodic interval attractors to a grid. 
%\color{black}

\subsection{Weighted Points Representation of Non-periodic Interval Attractors} %in \texorpdfstring{$\delta$}{}-optimal Space}
Let $\mathcal{X}$ and $\mathcal{Y}$ be linearly ordered sets. 
$\mathcal{J}^h$ denotes a finite set of weighted points on $\mathcal{X}$ and $\mathcal{Y}$. Each point within $\mathcal{J}^h$ represents the non-periodic interval attractor associated with a nonterminal in $S^h$ at height $h \in [0, H]$.
Each point in $\mathcal{J}^h$ for any $h \in [0, H]$ is associated with two integer weights, which quantify the number of aggregated points, detailed as follows.
%subject to (i) the periodicity of interval attractors and (ii) a specific condition pertaining to two sets $\mathcal{X}$ and $\mathcal{Y}$, as follows.

Consider two non-periodic interval attractors $I(u) = ([p, q], [\ell, r])$ and $I(u^\prime)=([p^\prime, q^\prime], [\ell^\prime, r^\prime])$ of nodes $u, u' \in \mathcal{U}$ corresponding to $s^h$-th and $s^{\prime h}$-th nonterminals $S^h[s^h]$ and $S^h[s^{\prime h}]$ in $S^h$ at $h \in [0, H]$. 
Let $\gamma$ and $\gamma^\prime$ denote the starting positions of the substrings derived from $S^h[s^h]$ and $S^{\prime h}[s^{\prime h}]$, respectively. 
$I(u)$ is considered lexicographically smaller than $I(u^\prime)$ on $\mathcal{X}$ if and only if
$\reverse(T[p-1..\gamma-1]) \prec \reverse(T[p^\prime-1..\gamma^\prime-1])$ for the two strings $\reverse(T[p-1..\gamma-1])$ and $\reverse(T[p^\prime-1..\gamma^\prime-1])$ obtained by reversing two substrings $T[p-1..\gamma-1]$ and $T[p^\prime-1..\gamma^\prime-1]$, respectively.
Similarly, $I(u)$ is considered lexicographically smaller than $I(u^{\prime})$ on $\mathcal{Y}$
if and only if $T[\gamma..r+1] \prec T[\gamma^\prime..r^\prime+1]$.

The first integer weight assigned to each non-periodic interval attractor, represented as a point in $\mathcal{J}^h$ for any $h \in [0, H]$, reflects the number of periodic interval attractors that can be reconstructed from a non-periodic interval attractor, as established by Theorem~\ref{theo:intro_periodicity}. 
The weight for each non-periodic interval attractor is defined as $1$ plus the total number of periodic interval attractors reconstructed by that non-periodic interval attractor. 

The second integer weight is set as the number of points in $\mathcal{J}^h$ that share the same order on $\mathcal{X}$ and $\mathcal{Y}$, i.e., the number of 
non-periodic interval attractors $I(u)= ([p,q], [\ell, r])$ with the same two substrings $T[p-1..\gamma-1]$ and $T[\gamma..r+1]$.
Such interval attractors are aggregated, and the second integer weight represents the number of aggregated interval attractors.

%\color{blue}
The following theorem concerning the total number of weighted points is as follows.
%\color{black}
%\color{red}
%As a result of aggregating interval attractors into weighted points, 
%these weighted points correspond to 
%a subset $\Psi$ of non-periodic interval attractors such that 
%for any pair of interval attractors $I(u) = ([p,q], [\ell, r]), I(u^{\prime}) = %%[p^\prime, q^\prime], [\ell^\prime, r^\prime]) \in \Psi$, 
%two substrings $T[p-1..r+1]$ and $T[p^{\prime}-1..r^{\prime}+1]$ are distinct (i.e., %$T[p-1..r+1] \neq T[p^{\prime}..r^{\prime}+1]$). 
%By leveraging properties of the restricted recompression, 
%we can construct partition $\Delta(u)$ so that the expected size of the subset $\Psi$ %is bounded by $O(\delta \log \frac{n \log \sigma}{\delta \log n})$, 
%leading to the following theorem for the total number of weighted points. 
%\color{black}

\begin{theorem}
    The expected value of the total number of weighted points in 
    $H+1$ sets $\mathcal{J}^{0}$, $\mathcal{J}^{1}$, $\ldots$, $\mathcal{J}^{H}$ is 
    $O(\delta \log \frac{n \log \sigma}{\delta \log n})$ 
    (i.e., $\mathbb{E}[\sum_{h=0}^{H} |\mathcal{J}^{h}|] = O(\delta \log \frac{n \log \sigma}{\delta \log n})$).     
\end{theorem}

\subsection{Directed Acyclic Graph Representation of Derivation Tree}% in \texorpdfstring{$\delta$}{}-optimal Space}
We represent the derivation tree built by restricted recompression as a directed acyclic graph (DAG), termed \emph{RR-DAG}.
The RR-DAG $\mathcal{DG}$ of RLSLP $\mathcal{G}^{R}$ is defined as a triplet $(\mathcal{U}_{\RR}, \mathcal{E}_{\RR}, L_{\RR})$, as follows.
%$\mathcal{U}_{\RR}$ is a set of nodes in the RR-DAG. 
Let $\mathcal{U}_{\RR}$ be a set of nodes in the RR-DAG, defined by assigning a value from $\{ -1, 0, 1 \}$ to each nonterminal in set $\mathcal{V}$ according to a rule in restricted recompression.
The start symbol and every nonterminal assigned $0$ or $1$ in the RLSLP $\mathcal{G}^{R}$ correspond one-to-one to a node in $\mathcal{U}_{\RR}$. Let the subscript $i$ of nonterminal $X_i \in \mathcal{V}$ correspond to the subscript $i$ of $v_{i} \in \mathcal{U}_{\RR}$.  
$\mathcal{E}_{\RR} \subseteq \mathcal{U}_{\RR} \times \mathcal{U}_{\RR}$ is a set of directed edges in $\mathcal{DG}$. A directed edge from a node labeled $X_i$ to another node labeled $X_j$ is included in $\mathcal{E}_{\RR}$ if either of the following conditions is met: (a) the node labeled $X_i$ is directly connected to the node labeled $X_j$ in the derivation tree of the RLSLP $\mathcal{G}^{R}$; or (b) there exists a path from the node labeled $X_i$ to the node labeled $X_j$ through intermediate nodes corresponding to nonterminals assigned $-1$. 
$L_{\RR}: \mathcal{U}_{\RR} \rightarrow \mathcal{V}$ is a label function that returns the nonterminal $X_{i} \in \mathcal{V}$ corresponding to a given node $v_{i} \in \mathcal{U}_{\RR}$. 
The expected total number of nodes and edges in $\mathcal{DG}$ is estimated as follows.
%can be upper-bounded by $O(\delta \log \frac{n \log \sigma}{\delta \log %n})$. 
\begin{lemma}\label{theo:rrdag_size}
    For the RR-DAG $\mathcal{DG} = (\mathcal{U}_{\RR}, \mathcal{E}_{\RR}, L_{\RR})$ of the RLSLP $\mathcal{G}^{R} = (\mathcal{V}, \Sigma, \mathcal{D}, E)$, 
    $\mathbb{E} [ |\mathcal{U}_{\RR}| + |\mathcal{E}_{\RR}|] = O(\delta \log \frac{n \log \sigma}{\delta \log n} )$ holds. 
\end{lemma}

In summary, we store (i) the data structure supporting \emph{range-sum} query~\cite{doi:10.1137/0217026} on weighted points representing non-periodic interval attractors and (ii) the RR-DAG representing the derivation tree by restricted recompression in expected $\delta$-optimal space.
Range-sum queries enable counting the number of interval attractors satisfying certain conditions in polylogarithmic time with respect to $n$ on data structures (i) and (ii). 
We obtain the following theorem. 
\begin{theorem}\label{theo:intro_rsc_rss}
%    Using the data structures of $\delta$-optimal space, 
%    On the weighted points representation of interval attractors, 
    We can answer the seven subqueries listed in Table~\ref{table:RSC_query_result} 
    in $O(H^{2} \log n + \log^{4} n)$ time on 
    a data structure of expected $O(\delta \log \frac{n \log \sigma}{\delta \log n} B)$ bits of space for machine word size $B$. 
    RSC query can be answered in the same time using these seven subqueries.     
    Similarly, 
    we can answer the seven subqueries in Table~\ref{table:RSS_query_result} 
    in $O(H^{2} \log^{2} n + \log^{6} n)$ time. 
    RSS query can be answered in the same time using these eight subqueries. 
    As a result, 
    SA and ISA queries can be answered in $O(H^{3} \log^{2} n + H \log^{6} n)$ time and $O(H^{3} \log n + H \log^{4} n)$ time, respectively.
\end{theorem}

\section{Update Operation}
This section presents the updates to our data structures, which consist of (i) a weighted points representation and (ii) a directed acyclic graph representing a derivation tree, in response to the insertion or deletion of a single character into or from $T$ at a position.

\subsection{Update of RR-DAG}
A character's insertion into, or deletion from, a string allows for the efficient update of the directed acyclic graph of parse trees by adding or removing a specified number of its nodes or edges.
Here, we consider changes in the nodes and edges of a RR-DAG according to changes in sequences of nonterminals in an RLSLP, which occur by inserting or deleting a character into or from a string.
Let $T^\prime$ be a string built by the insertion or deletion of a character from string $T$.
%In addition, 
Let $\mathcal{G^R} = (\mathcal{V}, \Sigma, \mathcal{D}, E)$ (respectively, $\mathcal{G^{\prime R}} = (\mathcal{V^\prime}, \Sigma^\prime, \mathcal{D^\prime}, E^\prime)$) denote an RLSLP of $T$ (respectively, $T^\prime$).
%Let \(S^h\) for \(h \in [0,H]\) (respectively, \(S^{\prime h}\)) be a sequence of nonterminals in \(\mathcal{V}\) (respectively, \(\mathcal{V^\prime}\)). 
Let \(S^h\) (respectively, \(S^{\prime h}\)) be 
the sequence of nonterminals at height $h$ in the derivation tree of RLSLP $\mathcal{G^R}$ (respectively $\mathcal{G^{\prime R}}$).
%, which are \(H+1\) sequences (respectively, \(H^\prime + 1\) sequences).
The following theorem guarantees that updating \(\mathcal{G^R}\) to \(\mathcal{G^{\prime R}}\) can be achieved by modifying a localized region within \(S^h\) of a bounded length \(O(\max\{H, H^\prime\})\) to obtain \(S^{\prime h}\).
\begin{theorem}\label{theo:replace}
For each $h \in [0, \max \{ H, H^{\prime} \}]$, 
there exist two segments $S^{h}[p^{h}..q^{h}]$ and $S^{\prime h}[p^{\prime h}..q^{\prime h}]$ of two sequences $S^{h}$ and $S^{\prime h}$, 
respectively, satisfying the following three conditions: 
    (i) $S^{\prime h} = S^{h}[1..p^{h}-1]S^{\prime h}[p^{\prime h}..q^{\prime h}]S^{h}[q^{h}+1..|S^{h}|]$;
    (ii) $|S^{h}[p^{h}..q^{h}]| = O(H)$;
    (iii) $|S^{\prime h}[p^{\prime h}..q^{\prime h}]| = O(H^{\prime})$.
%\begin{enumerate}[itemsep=0pt, parsep=0pt,before=\vspace{-0.5\baselineskip}, %after=\vspace{-0.5\baselineskip}, label=\textbf{(\roman*)}]
%%\begin{enumerate}[label=\textbf{(\roman*)}]
%    \item $S^{\prime h} = S^{h}[1..p^{h}-1]S^{\prime h}[p^{\prime h}..q^{\prime h}]S^{h}[q^{h}+1..|S^{h}|]$;
%    \item $|S^{h}[p^{h}..q^{h}]| = O(H)$;
%    \item $|S^{\prime h}[p^{\prime h}..q^{\prime h}]| = O(H^{\prime})$.
%\end{enumerate}
If \(h \geq H+1\), then $S^{h}$ is defined as the start symbol $E$ of $\mathcal{G}^{R}$. Similarly, if \(h \geq H^{\prime}+1\), then $S^{\prime h}$ is defined as $E^{\prime}$ as the start symbol $E^{\prime}$ of $\mathcal{G}^{\prime R}$.
\end{theorem}

Theorem~\ref{theo:replace} shows that each sequence $S^{\prime h}$ is obtained by replacing segment $S^{h}[p^{h}..q^{h}]$ of sequence $S^{h}$ with segment $S^{\prime h}[p^{\prime h}..q^{\prime h}]$ of sequence $S^{\prime h}$. 
Based on this modification, we update RR-DAG $\mathcal{DG}$ of RLSLP $\mathcal{G}^{R}$ as follows:
\begin{enumerate}[itemsep=0pt, parsep=0pt,before=\vspace{-0.5\baselineskip}, after=\vspace{-0.5\baselineskip}, label=\textbf{(\roman*)}]
%\begin{enumerate}[label=\textbf{(\roman*)}]
    \item  For all $i \in [p^{h}..q^{h}]$ where nonterminal $S^h[i]$ is unequal to any element of $S^{\prime h}$ (i.e., $S^{h}[i] \neq S^{\prime h}[j]$ for all $j \in \{ 1, 2, \ldots, |S^{\prime h}| \}$), all nodes labeled $S^h[i]$ and their adjacent edges in $\mathcal{DG}$ are removed;
%    if sequence $S^{\prime h}$ does not contain the nonterminal $S^{h}[i]$, 
%    and $DG$ has a node $v$ labeled the nonterminal, 
%    then the node $v$ and directed edges connected to the node $v$ are removed from the RR-DAG. 
    \item For all $i \in [p^{\prime h}..q^{\prime h}]$ that satisfy the following two conditions:
(a) the nonterminal $S^{\prime h}[i]$ is unequal to any element of $S^h$, and (b) either $S^{\prime h}[i]$ is the start symbol $E$ in $\mathcal{G}^{\prime R}$ or $S^{\prime h}[i]$ is assigned to $0$ or $1$, a node labeled $S^{\prime h}[i]$ is added to $\mathcal{DG}$, and directed edges connected to this node are created as appropriate from the definition of RR-DAG.
\end{enumerate}
We can update the RR-DAG in $O(\log^{2} (nHH^{\prime}))$ time for each nonterminal in $S^{h}[p^{h}..q^{h}]$ and $S^{\prime h}[p^{\prime h}..q^{\prime h}]$. 
Theorem~\ref{theo:replace} shows that segments $S^{0}[p^{0}..q^{0}]$, 
$S^{1}[p^{1}..q^{1}]$, $\ldots$, $S^{h}[p^{h}..q^{h}]$ and $S^{\prime 0}[p^{\prime 0}..q^{\prime 0}]$, 
$S^{\prime 1}[p^{\prime 1}..q^{\prime 1}]$, $\ldots$, $S^{\prime H^{\prime}}[p^{\prime H^{\prime}}..q^{\prime H^{\prime}}]$ 
contain $O(H^{2} + H^{\prime 2})$ nonterminals in total. 
Therefore, the update of RR-DAG takes $O(\log^{4} n)$ time if $H, H^{\prime} = O(\log n)$. 
Formally, we obtain the following theorem.

\begin{theorem}
Consider the scenario where a string $T^\prime$ is created by inserting or deleting a single character into or from the string $T$. Let $\mathcal{G}^{R}$ and $\mathcal{G}^{\prime R}$ be RLSLPs deriving $T$ and $T^\prime$, respectively, both constructed via restricted recompression. Then, the RR-DAG $\mathcal{DG}$ of $\mathcal{G}^{R}$ can be updated in $O((\max \{ H, H^{\prime}, \log (nHH^{\prime}) \})^{4})$ time, where $H$ and $H^{\prime}$ are the heights of the derivation trees of $\mathcal{G}^R$ and $\mathcal{G}^{\prime R}$, respectively.
\end{theorem}

\subsection{Update of Weighted Points of Non-periodic Interval Attractors}
Following the update of RR-DAG, 
a character's insertion into, or deletion from, a string allows for the efficient update of finite set $\mathcal{J}^{h}$ of weighted points associated with non-periodic interval attractors for $h \in [0, H]$. 
Here, we consider an update in the weighted points in each set $\mathcal{J}^{h}$ according to the modification of $\mathcal{G}^{R}$. 
Let $\Psi_{\RR}$ (respectively, $\Psi^{\prime}_{\RR}$) be the set of interval attractors obtained from 
$\mathcal{G}^{R}$ (respectively, $\mathcal{G}^{\prime R}$). 
The following theorem shows that a subset of set $\Psi^{\prime}_{\RR}$ can be obtained by 
modifying the set $\Psi_{\RR}$ of interval attractors. 

\begin{theorem}\label{theo:IA_change_formula}
    Consider the string $T^{\prime}$ built by inserting or deleting a single character into or from string $T$ at position $\lambda$.
    Let $\epsilon = 1$ and $\epsilon^{\prime} = 0$ if $T^\prime$ is updated from $T$ by an insertion; 
    otherwise $\epsilon = -1$ and $\epsilon^{\prime} = 1$.
    Then, 
    subset $\{ ([p, q], [\ell, r]) \in \Psi^{\prime}_{\RR} \mid [p-1, r+1] \cap [\lambda-1, \lambda + 1 - \epsilon^{\prime}] = \emptyset \}$ of set $\Psi^{\prime}_{\RR}$ is equal to the union of the following two sets: 
    (i) $\{ ([p, q], [\ell, r]) \in \Psi_{\RR} \mid r+3 \leq \lambda \}$;
    (ii) $\{ ([p+\epsilon, q+\epsilon], [\ell+\epsilon, r+\epsilon]) \mid ([p, q], [\ell, r]) \in \Psi_{\RR} \wedge \lambda \leq p-2 - \epsilon^{\prime} \}$.  
%    \begin{enumerate}[itemsep=0pt, parsep=0pt,before=\vspace{-0.5\baselineskip}, after=\vspace{-0.5\baselineskip}, label=\textbf{(\roman*)}]
%    \begin{enumerate}[label=\textbf{(\roman*)}]
%    \item $\{ ([p, q], [\ell, r]) \in \Psi_{\RR} \mid r+3 \leq \lambda \}$;
%    \item $\{ ([p+\epsilon, q+\epsilon], [\ell+\epsilon, r+\epsilon]) \mid ([p, q], [\ell, %r]) \in \Psi_{\RR} \wedge \lambda \leq p-2 - \epsilon^{\prime} \}$.  
%    \end{enumerate}
\end{theorem}

Consider the sets $\mathcal{J}^h \subseteq \mathcal{X} \times \mathcal{Y} \times \mathbb{Z} \times \mathbb{Z}$ and $\mathcal{J}^{\prime h} \subseteq \mathcal{X} \times \mathcal{Y} \times \mathbb{Z} \times \mathbb{Z}$ of weighted points for $h \in [0, H]$ and $h^\prime \in [0, H^\prime]$, respectively. Each element of $\mathcal{J}^h$ corresponds to a non-periodic interval attractor in $\Psi_{\RR}$, and each element of $\mathcal{J}^{\prime h}$ corresponds to a non-periodic interval attractor in $\Psi^\prime_{\RR}$.
Let $\mathcal{J}^{h} = \emptyset$ for $h \geq H+1$ and $\mathcal{J}^{\prime h} = \emptyset$ for $h \geq H+1$.

Define $\kappa(x, y)$ for any $(x, y) \in \mathcal{X} \times \mathcal{Y}$ as the number of non-periodic interval attractors in $\Psi_{\RR}$ that satisfies two specific conditions. For each $([p, q], [\ell, r]) \in \Psi_{\RR}$, the conditions are as follows:
(A) $[p-1, r+1] \cap [\lambda - 1, \lambda + \epsilon^{\prime}] \neq \emptyset$;
(B) the weighted point representing interval attractor $([p, q], [\ell, r])$ is located at the same $(x, y)$ in the $\mathcal{X} \times \mathcal{Y}$ plane.
Here, $\epsilon^{\prime}$ is the integer introduced in Theorem~\ref{theo:IA_change_formula}.  

Similarly, define $\kappa^\prime(x, y)$ for any point $(x, y) \in \mathcal{X} \times \mathcal{Y}$ as the number of non-periodic interval attractors in $\Psi^{\prime}_{\RR}$ that satisfies two specific conditions. For each $([p^{\prime}, q^{\prime}], [\ell^{\prime}, r^{\prime}]) \in \Psi^{\prime}_{\RR}$, the conditions are as follows:
(a) $[p^{\prime}-1, r^{\prime}+1] \cap [\lambda - 1, \lambda + 1 - \epsilon^{\prime}] \neq \emptyset$;
(b) the weighted point representing interval attractor $([p^{\prime}, q^{\prime}], [\ell^{\prime}, r^{\prime}])$ is located at the same $(x, y)$ in the $\mathcal{X} \times \mathcal{Y}$ plane.

Set $\mathcal{J}^{\prime h}$ can be obtained by updating set $\mathcal{J}^{h}$ as follows. 
\begin{theorem}\label{theo:intro_weighted_point_formula}
    Consider two sets $\mathcal{J}^{h}$ and $\mathcal{J}^{\prime h}$ of weighted points for an integer $h \in [0, \max \{ H, H^{\prime} \}]$. 
    Then, the following two statements hold. 
    \begin{enumerate}[itemsep=0pt, parsep=0pt,before=\vspace{-0.5\baselineskip}, after=\vspace{-0.5\baselineskip}, label=\textbf{(\roman*)}]
%    \begin{enumerate}[label=\textbf{(\roman*)}]
    \item For each weighted point $(x, y, \alpha, \beta)$ in set $\mathcal{J}^{h}$, 
    set $\mathcal{J}^{\prime h}$ contains a weighted point $(x^{\prime}, y^{\prime}, \alpha^{\prime}, \beta^{\prime})$ 
    as $x = x^{\prime}$, $y = y^{\prime}$,  $\alpha^{\prime} = \alpha$ and $\beta^{\prime} = \beta - \kappa(x, y) + \kappa^{\prime}(x, y)$ if and only if $\beta - \kappa(x, y) + \kappa^{\prime}(x, y) \geq 1$. 
%    Otherwise, $x = x^{\prime}$, $y = y^{\prime}$,  $\alpha^{\prime} = \alpha$
%    In addition, $\alpha^{\prime} = \alpha$ and $\beta^{\prime} = \beta - \kappa(x, y) + \kappa^{\prime}(x, y)$ hold 
%    if $\beta - \kappa(x, y) + \kappa^{\prime}(x, y) \geq 1$. 
    \item Consider a weighted point $(x^{\prime}, y^{\prime}, \alpha^{\prime}, \beta^{\prime})$ in set $\mathcal{J}^{\prime h}$ such that the same point $(x^{\prime}, y^{\prime})$ with any weights is not contained in $\mathcal{J}^h$.      
%    If set $\mathcal{J}^{h}$ does not contain any weighted point $(x, y, \alpha, \beta)$ satisfying $x = x^{\prime}$ and $y = y^{\prime}$ for a weighted point $(x^{\prime}, y^{\prime}, \alpha^{\prime}, \beta^{\prime})$ in set $\mathcal{J}^{\prime h}$,
    For any non-periodic interval attractor $([p^{\prime}, q^{\prime}], [\ell^{\prime}, r^{\prime}]) \in \Psi^{\prime}_{\RR}$ located at the same point $(x^\prime, y^\prime)$, condition $[p^{\prime}-1, r^{\prime}+1] \cap [\lambda - 1, \lambda + 1 - \epsilon^{\prime}] \neq \emptyset$ holds.
    \end{enumerate}
\end{theorem}

Theorem~\ref{theo:intro_weighted_point_formula} indicates that 
$(H^{\prime}+1)$ sets $\mathcal{J}^{\prime 0}$, $\mathcal{J}^{\prime 1}$, $\ldots$, $\mathcal{J}^{\prime H^{\prime}}$ 
of weighted points can be computed by 
modifying $(H+1)$ sets $\mathcal{J}^{0}$, $\mathcal{J}^{1}$, $\ldots$, $\mathcal{J}^{H}$.
The update of weighted points takes $O((\max \{ H, H^{\prime}$, $\log (nHH^{\prime}) \})^{7})$ time 
per weighted point representing either 
(A) non-periodic interval attractor $([p, q], [\ell, r]) \in \Psi_{\RR}$ satisfying $[p-1, r+1] \cap [\lambda - 1, \lambda + \epsilon^{\prime}] \neq \emptyset$ 
or (B) non-periodic interval attractor $([p^{\prime}, q^{\prime}], [\ell^{\prime}, r^{\prime}]) \in \Psi^{\prime}_{\RR}$ satisfying $[p^{\prime}-1, r^{\prime}+1] \cap [\lambda - 1, \lambda + 1 - \epsilon^{\prime}] \neq \emptyset$. 
The following theorem bounds the number of such weighted points by expected $O(H + H^{\prime} + \log n)$.

\begin{theorem}\label{theo:bound_overlapping_IA}
Let $W$ be the number of non-periodic interval attractors $([p, q], [\ell, r])$ in set $\Psi_{\RR}$ satisfying $[p-1, r+1] \cap [\lambda - 1, \lambda + \epsilon^{\prime}] \neq \emptyset$. 
Similarly, 
let $W^{\prime}$ be the number of non-periodic interval attractors $([p, q], [\ell, r])$ in set $\Psi^{\prime}_{\RR}$ satisfying $[p-1, r+1] \cap [\lambda - 1, \lambda + 1 - \epsilon^{\prime}] \neq \emptyset$. 
Then, $\mathbb{E}[W] = O(H + \log n)$ and $\mathbb{E}[W^{\prime}] = O(H^{\prime} + \log n)$. 
\end{theorem}

Finally, we can update weighted points representing non-periodic interval attractors in expected $O(\log^{8} n)$ time in total if $H, H^{\prime} = O(\log n)$. Formally, we obtain the following theorem. 
\begin{theorem}
    Consider the scenario where a string $T^\prime$ is created by inserting or deleting a single character into or from the string $T$. 
    Let $\mathcal{G}^{R}$ and $\mathcal{G}^{\prime R}$ be RLSLPs deriving $T$ and $T^\prime$, respectively, both constructed via restricted recompression. 
    Then, weighted points representing non-periodic interval attractors can be updated in 
    expected $O((\max \{ H, H^{\prime}, \log (nHH^{\prime}) \})^{8})$ time.
    %, where $H^{\prime}$ is the height of the derivation tree of %$\mathcal{G}^{\prime R}$.
\end{theorem}

\paragraph{Results on Queries with Polylogarithmic Time and $\delta$-Optimal Space.}
Our data structure's performance relies on the height $H$ of the derivation tree of RLSLP $\mathcal{G}^{R}$, because SA, ISA, and update operations run in $O((\max { H, H^{\prime}, \log (nHH^{\prime}) })^{8})$ time. To ensure polylogarithmic query and update time in $n$, we bound $H$ by $2(w+1) \log_{8/7} (4|T|) + 2$ for a constant $w \geq 2$. Update operations succeed if $H$ stays within this bound; otherwise, they fail, with failure probability $O(n^{-w})$.
Theorem~\ref{theo:tree_height} that guarantees $H \leq 2(w+1) \log_{8/7} (4|T|) + 2$ holds with probability $1 - (1/n^{2w})$.

In summary, our structure achieves expected $\delta$-optimal space, with SA and ISA queries supported in $O(\log^{7} n)$ and $O(\log^{5} n)$ time, and updates in expected $O(\log^{8} n)$ time. Additionally, RA and LCE queries can be performed in $O(\log n)$ time (see Section~\ref{sec:recompression}).

\paragraph{Construction.}
Our data structure can be built by updating it incrementally as we read through the input string $T$, character by character. 
This construction takes expected $O(n \log^{8} n)$ time and fails with probability $O(n^{-(w-2)/2})$ 
because each update runs in $O(\log^{8} n)$ time with failure probability $O(n^{-w})$. 
See Section~\ref{sec:summary} for the detailed analysis of our construction.

\addtocontents{toc}{\protect\setcounter{tocdepth}{3}}

\clearpage
\tableofcontents
\clearpage

\section{Related Works}\label{sec:related_works}
\renewcommand{\arraystretch}{1.2}
\begin{table}[h]
    \footnotesize
    \vspace{-0.5cm}
    \caption{
    Summary of state-of-the-art static and dynamic data structures supporting 
    suffix array~(SA), inverse suffix array~(ISA), random access~(RA), and longest common extension~(LCE) queries. 
    $T$ is an input string of alphabet size $\sigma$ and length $n$ with substring complexity $\delta$; 
    $r$ is the number of equal-letter runs in the BWT~\cite{burrows1994block} of the input string. 
    Here, $r = O(\delta \log^{2} n)$ between $r$ and $\delta$~\cite{DBLP:conf/focs/KempaK20}; 
    $\delta = O(n / \log_{\sigma} n)$ between $n$ and $\delta$~\cite{9961143}. 
%    In the second column, 
%    D means that the data structure is built by a deterministic algorithm. 
%    In contrast, R means that the data structure is built by a randomized algorithm.    
    The rightmost column shows the time needed to update data structures for the following two operations: 
    (i) insert a given character into string $T$ at a given position; 
    (ii) remove the character of string $T$ at a given position.
%    (iii) replace the character of string $T$ at a given position by a given character. 
%    The second method proposed in \cite{DBLP:journals/corr/abs-2112-12678} only supports the third operation for the update of their data structure. 
    Our data structure fails with probability $O(n^{-w})$ for a user-defined constant $w \geq 2$. 
    We assume that $B = \Theta(\log n)$ for machine word size $B$. 
    }
    \vspace{-5mm}    
    \label{table:result} 
    \center{
    \scalebox{0.85}{
    \begin{tabular}{r||c|cccccc}
 \multirow{2}{*}{Method} & Working space & \multicolumn{5}{|c}{Query time}  \\ 
        & (bits)        & SA & ISA & RA  & LCE & Update \\ \hline \hline 
Kempa and Kociumaka & \multirow{2}{*}{$O(n \polylog(n))$} & \multirow{2}{*}{$O(\log^{4} n)$} & \multirow{2}{*}{$O(\log^{5} n)$} & \multirow{2}{*}{$O(\log n)$}  & \multirow{2}{*}{Unsupported} & \multirow{2}{*}{$O(\log^{3} n (\log\log n)^{2})$}  \\  
\cite{DBLP:conf/stoc/KempaK22} & &  & & & &  \\ \hline
Amir and Boneh                   & \multirow{2}{*}{$O(n \polylog(n))$}  & \multirow{2}{*}{$O(\log^{5} n)$} & \multirow{2}{*}{Unsupported} & \multirow{2}{*}{$O(\log n)$} & \multirow{2}{*}{$O(\log n)$} & \multirow{2}{*}{$O(n^{2/3} \polylog(n))$}  \\  
\cite[Theorem 24]{DBLP:journals/corr/abs-2112-12678}                 &  &  &  &  & &  &   \\  \hline
Amir and Boneh              & \multirow{2}{*}{$O(n \polylog(n))$} & \multirow{2}{*}{Unsupported} & \multirow{2}{*}{$O(\log^{4} n)$} & \multirow{2}{*}{$O(\log n)$} & \multirow{2}{*}{$O(\log n)$} & \multirow{2}{*}{$O(n^{1/2} \polylog(n))$}  \\  
\cite[Theorem 21]{DBLP:journals/corr/abs-2112-12678}                   &  & &  &  &  &   \\ \hline 
$\delta$-SA~\cite{DBLP:journals/corr/abs-2308-03635,DBLP:conf/focs/KempaK23}   & $O(\delta \log \frac{n \log \sigma}{\delta \log n} \log n)$ & $O(\log^{4+\epsilon} n)$ & $O(\log^{4+\epsilon} n)$ & $O(\log n)$ & $O(\log n)$ & Unsupported \\ \hline 
Gagie et al.~\cite{DBLP:journals/jacm/GagieNP20}  & $O(r \log (n/r) \log n)$ & $O(\log (n/r))$ & $O(\log (n/r))$ & $O(\log (n/r))$ & $O(\log(n/r))$ & Unsupported \\ \hline \hline 
This study  & expected  & \multirow{2}{*}{$O(\log^{7} n)$}  & \multirow{2}{*}{$O(\log^{5} n)$}  & \multirow{2}{*}{$O(\log n)$}  & \multirow{2}{*}{$O(\log n)$}  & expected \\ 
 & $O(\delta \log \frac{n \log \sigma}{\delta \log n} \log n)$ &  &  &  & & $O(\log^{8} n)$ 
    \end{tabular} 
    }
    }
\end{table}

To date, various static and dynamic data structures have been developed to support SA and related queries, essential for implementing suffix trees. 
These data structures can be categorized based on their working space, estimated for either a general type of strings or highly repetitive strings. 
A summary of state-of-the-art static and dynamic data structures is presented in Table~\ref{table:result}.

For strings of a general type, the compressed suffix array (CSA)\cite{doi:10.1137/S0097539702402354} and the compressed suffix tree (CST)\cite{DBLP:conf/soda/Sadakane02} stand out as popular static compressed data structures, where input strings are not subject to alteration.
They operate in $\log{n}$ time and utilize $n\log{\sigma}$ bits of space for a string of length $n$ and an alphabet size of $\sigma$. 
While a dynamic data structure supporting the SA query and updates in $O(\polylog{(n)})$ time has been proposed~\cite{DBLP:conf/stoc/KempaK22}, it does not operate in a compressed space.
%While Kempa and Kociumaka~\cite{} also presented a dynamic data structure supporting the SA query and updates in %$O(\polylog{(n)})$ time, it does not operate in a compressed space. 

For highly repetitive strings, 
research for data structures has explored various dimensions, including linear-time construction~\cite{DBLP:journals/talg/BelazzouguiCKM20}, optimal time or space efficiency~\cite{DBLP:conf/soda/Kempa19,DBLP:conf/soda/KempaK23}, and implementation within BWT-runs bounded space~\cite{DBLP:journals/jacm/GagieNP20,DBLP:conf/spire/BoucherKP022}. 
A prevalent observation across these studies is the suboptimal space efficiency of existing compressed data structures.
Kociumaka, Navarro, and Prezza~\cite{9961143} introduced an asymptotically minimal space requirement, \(O\left(\delta \log\frac{n\log\sigma}{\delta\log n} \log n \right)\) bits, sufficient to represent any string $T$ of length \(n\), with an alphabet size of \(\sigma\), and substring complexity \(\delta\), serving as a measure of repetitiveness. 
The space is referred to as $\delta$-\emph{optimal space}~\cite{ViceVersa,DBLP:conf/focs/KempaK23}. 
Furthermore, they also presented a compressed data structure supporting random access and pattern-matching queries in $\delta$-optimal space. 
A compressed data structure supporting locate queries in near-optimal time and $\delta$-optimal space is also proposed~\cite{ViceVersa}.
Recently, Kempa and Kociumaka~\cite{DBLP:journals/corr/abs-2308-03635,DBLP:conf/focs/KempaK23} presented $\delta$-SA,  a static compressed data structure supporting the SA query in $O(\polylog{(n)})$ using $\delta$-optimal space. 
On the practical side, several compressed data structures have been developed to support queries on suffix trees~\cite{CACERES2022104749,DBLP:conf/alenex/BoucherCGHMNR21,DBLP:journals/algorithms/AbeliukCN13}. While these structures are expected to occupy a small space when applied to string data in practical scenarios, their theoretical space usage remains unestimated. 
Overall, these data structures are static, meaning that input strings are not subject to alteration.

%While Kempa and Kociumaka also presented a dynamic data structure supporting the SA query and updates in $O(\polylog{(n)})$ %time, it does not operate in a compressed space. 

Despite the critical significance of compressed dynamic data structures capable of supporting SA queries and updates, no prior studies have succeeded in presenting such structures within a $\delta$-optimal space.
Current data structures that support the SA query has a drawback of static or suboptimal working space. 
We present a dynamic data structure that supports important queries such as SA, ISA, RA, LCE and updates necessary for implementing suffix trees within polylogarithmic time, utilizing expected $\delta$-optimal space. 
Details of the proposed data structure are presented in the remaining of the paper.

%Despite the importance of scalable learning of interpretable
%linear models, no previous work has been able to achieve
%high prediction accuracy for classification/regression tasks
%and high interpretability of the learned models. We present
%a scalable learning algorithm that meets both these demands
%and is made possible by learning linear models on grammarcompressed data in the %framework of PLS. Details of the
%proposed method are presented in the next section.

\section{Preliminaries}\label{sec:preliminary}
\begin{figure}[t]
 \begin{center}
		\includegraphics[scale=0.7]{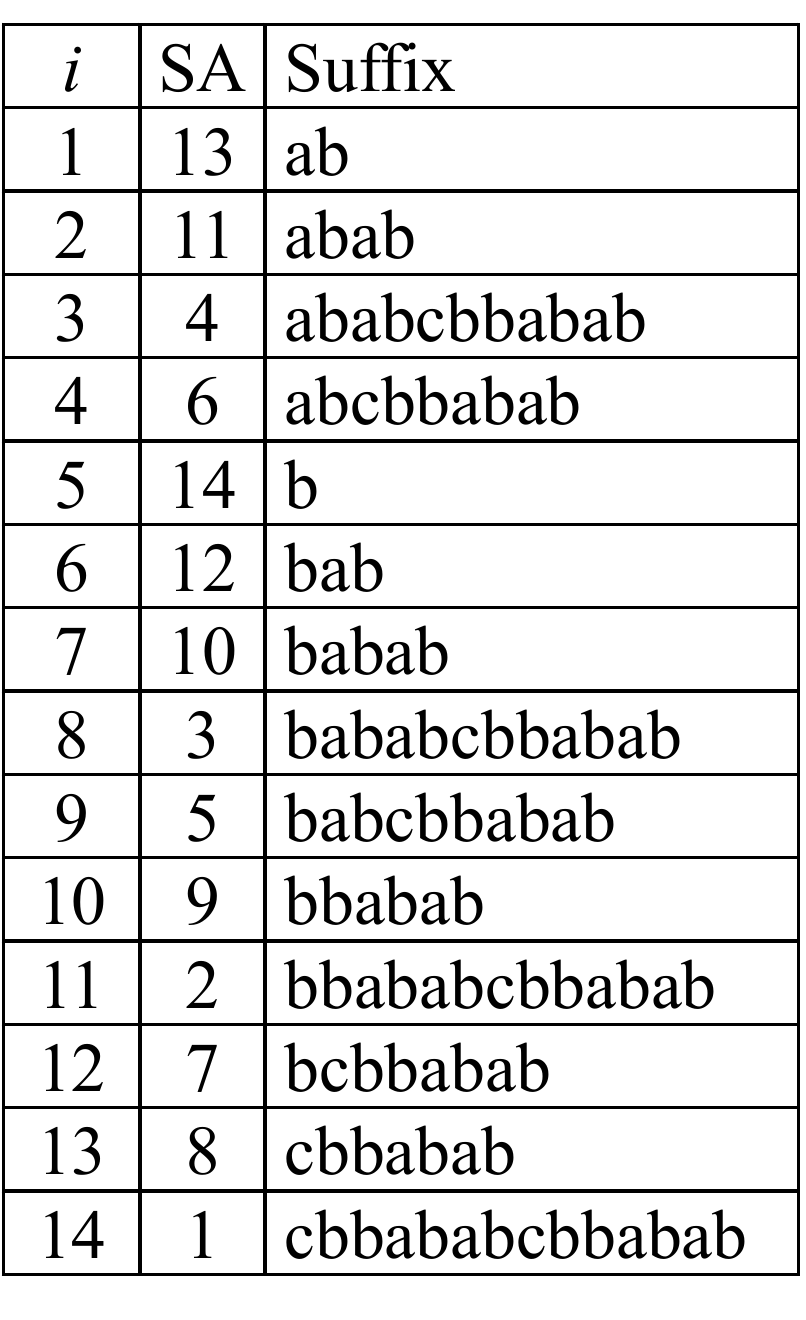}
	  \caption{ 
	  An illustration of the suffix array of string $T = \mathrm{cbb abab cbb abab}$.  
	  }
\label{fig:sa}
 \end{center}
\end{figure}

\subsection{Basic Notation.}
An \emph{interval} $[b, e]$ for two integers $b$ and $e$ represents the set $\{b, b+1, \ldots, e \}$ if $b \geq e$; 
otherwise, the interval $[b, e]$ represents an empty set $\emptyset$ (i.e., $[b, e] = \emptyset$). 
Let $\min \mathcal{S}$~(respectively, $\max \mathcal{S}$) be the smallest integer~(respectively, the largest integer) in a nonempty set $\mathcal{S}$ of integers. 
Let $\mathbb{N}_{0}$ be the set of non-negative integers and $\mathbb{N}_{>0}$ be the set of positive integers.

Let $T$ be a string of length $n \geq 1$ over an alphabet $\Sigma = \{ 1, 2, \ldots, n^{\mathcal{O}(1)} \}$ of size $\sigma$, and 
$|T|$ denotes the length of $T$ (i.e., $|T| = n$). 
We assume a total order over the alphabet $\Sigma$.
Let $\varepsilon$ be a string of length $0$. 
Let $T[i]$ be the $i$-th character of $T$ (i.e., $T = T[1], T[2], \ldots, T[n]$). 
Let $T[i..j]$ be the substring of $T$ that begins at position $i$
and ends at position $j$ (i.e., $T[i..j] = T[i], T[i+1], \ldots, T[j]$) if $i \leq j$; 
otherwise, let $T[i..j]$ be a string of length $0$ (i.e., $T[i..j] = \varepsilon$). 
A \emph{prefix} (respectively, \emph{suffix}) of $T$ is a substring that begins at position $1$ (respectively, ends at position $n$). 
A prefix of $T$ is called \emph{proper prefix} if it is shorter than $T$. 
Similarly, a suffix of $T$ is called \emph{proper suffix} if it is shorter than $T$. 
%A \emph{prefix} of string $T$ is a substring begins at position $1$, 
%and the prefix is called \emph{proper prefix} if the length of a prefix is less than $n$. 
%Similarly, 
%a \emph{suffix} of string $T$ is a substring ends at position $n$, 
%and the suffix is called \emph{proper suffix} if the length of a suffix is less than $n$. 
$\reverse(T)$ denotes the string obtained by reversing string $T$ (i.e., $\reverse(T) = T[n], T[n-1], \ldots, T[1]$). 
Let $\$$ and $\#$ be the smallest and largest characters in alphabet $\Sigma$, respectively, such that 
neither $\$$ nor $\#$ is contained in string $T$ (i.e., for all $i \in [1, n]$, both $T[i] \neq \$$ and $T[i] \neq \#$ hold).
We assume $T$ begins with $\$$ and ends with $\$$ 
($T[0] = \$$ and $T[n+1] = \$$) throughout this paper. 

For two characters $c, c^{\prime} \in \Sigma$, 
$c < c^{\prime}$ means that $c$ is smaller than $c^{\prime}$. 
%For a string $P$, $P[i] < P[j]$ means that the $i$-th character of $P$ is smaller than the $j$-th character of $P$. 
For a string $P$, $T \prec P$ holds if and only if either of the following two conditions holds: 
(i) there exists $i \in [1, n]$ such that $T[1..(i-1)] = P[1..(i-1)]$ and $T[i] < P[i]$ hold; (ii) $T = P[1..|T|]$ (i.e., $T$ is a prefix of $P$) and $|T| < |P|$ hold.  
In this case, $T$ is said to be lexicographically smaller than $P$.
$T \preceq P$ means that either $T = P$ or $T \prec P$ holds. 
%$T \prec P$ means that $T$ is lexicographically smaller than $P$.
%For two integers $0 \leq b \leq < $
A substring $T[i..j]$ of string $T$ is called \emph{occurrence} of string $P$ if 
the substring is equal to string $P$ (i.e., $T[i..j] = P$). 
Similarly, a position $i$ in string $T$ is called \emph{occurrence position} of string $P$ 
if substring $T[i..i + |P| - 1]$ is an occurrence of string $P$. 
Let $\Occ(T, P)$ be the set of all the occurrence positions of string $P$ in string $T$ (i.e., $\Occ(T, P) = \{ i \in [1, n-|P|+1] \mid T[i..(i+|P|-1)] = P \}$).
Let $\lcp(T, P)$ be the longest common prefix~(LCP) between two strings $T$ and $P$ (i.e., 
if $T[1] = P[1]$, then $\lcp(T, P) = T[1..1+d-1]$ for the largest integer $d \in [1, \min \{ |T|, |P| \}]$ satisfying $T[1..d] = P[1..d]$; 
otherwise, let $\lcp(T, P)$ be a string of length $0$). 
Similarly, let $\lcs(T, P)$ be the longest common suffix~(LCS) between two strings $T$ and $P$ (i.e., 
if $T[|T|] = P[|P|]$, then $\lcs(T, P) = T[(n-d+1)..n]$, where $d$ is the largest integer in the range $[1, \min \{ |T|, |P| \}]$ satisfying $T[(n-d+1)..n] = P[(|P|-d+1)..|P|]$; 
otherwise, $\lcs(T, P)$ is defined as a string of length $0$). 
Let $T \cdot P$ be a string representing the concatenation of two strings, $T$ and $P$ (i.e., $T \cdot P = T[1], T[2], \ldots, T[n], P[1], P[2], \ldots, P[|P|]$). 
Let $P^{k}$ be a string that consists of $k \geq 1$ repetitions of string $P$ (i.e., $P^{1} = P$, and $P^{k} = P \cdot (P^{k-1})$ for all integer $k \geq 2$).  

A suffix array~\cite{DBLP:journals/siamcomp/ManberM93} ($\SA$) of string $T$ is an integer array of size $n$ such that 
$\SA[i]$ stores the starting position of the $i$-th suffix of $T$ in lexicographical order. 
Formally, $\SA$ is a permutation of $n$ integers $1, 2, \ldots, n$ satisfying $T[\SA[1]..n] \prec \cdots \prec T[\SA[n]..n]$. 
The \emph{suffix array interval}~(\emph{sa-interval}) of string $P$ is an interval $[\eta, \eta^{\prime}] \subseteq \{ 1, 2, \ldots, n \}$ such that 
$\SA[\eta..\eta^{\prime}]$ represents all the occurrence positions of $P$ in string $T$ (i.e., $\forall i \in \{ 1, 2, \ldots, n \}, T[\SA[i]..\SA[i] + |P| - 1] = P \Leftrightarrow \eta \leq i \leq \eta^{\prime}$). 
If $P$ is the concatenation of a suffix $T[s..n]$ and character $\$$, 
then the sa-interval of $P$ is defined as $[s, s]$ for simplicity. 
An inverse suffix array ($\ISA$) of string $T$ is the inverse of suffix array $\SA$. 
That is, for any pair of integers $i, j \in \{ 1, 2, \ldots, n \}$, $\ISA[i] = j$ if and only if $\SA[j] = i$. 
Figure~\ref{fig:sa} illustrates the suffix array of a string. 

We will use base-2 logarithms throughout this paper unless otherwise indicated. 
Our computation model is a unit-cost word RAM~\cite{DBLP:conf/stacs/Hagerup98} with 
multiplication, randomization, and a machine word size of $B = \Omega(\log n)$ bits for input string $T$ of length $n$. 
We evaluate the space complexity of a data structure in terms of the number of machine words used by the data structure. 
A bitwise evaluation of space complexity can be obtained with a $B$ multiplicative factor. 

\subsection{Substring Complexity \texorpdfstring{$\delta$}{delta}~\texorpdfstring{\cite{DBLP:journals/algorithmica/RaskhodnikovaRRS13,DBLP:journals/talg/ChristiansenEKN21}}{}}\label{subsec:delta}
Substring complexity~\cite{DBLP:journals/algorithmica/RaskhodnikovaRRS13,DBLP:journals/talg/ChristiansenEKN21} is a measure of a repetitiveness for strings. 
$\mathcal{S}_{d}$ is defined as a set of substrings of length $d \geq 0$ in a string $T$ of length $n \geq 1$ (i.e., $\mathcal{S}_{d} = \{ T[i..(i+d-1)] \mid 1 \leq i \leq n - d + 1 \}$). Note that $\mathcal{S}_{1} = \sigma$ and $\mathcal{S}_{n} = 1$ hold.
The substring complexity $\delta$ of string $T$ is defined as $\max \{ |\mathcal{S}_{1}| / 1, |\mathcal{S}_{2}| / 2, \ldots, |\mathcal{S}_{n}| / n  \}$. 
Thus, the more frequently a shorter substring appears in string $T$, the larger the substring complexity $\delta$ is.
$1 \leq \delta \leq n$ always holds from the definition of $\delta$. In addition, 
$\delta \geq 2$ always holds for string $T$ containing at least two distinct characters. 

For string $T = \mathrm{abaaabbba}$, 
$\mathcal{S}_{1} = \{ a, b \}$;
$\mathcal{S}_{2} = \{ \mathrm{ab}, \mathrm{ba}, \mathrm{aa}, \mathrm{bb} \}$; 
$\mathcal{S}_{3} = \{ \mathrm{aaa}$, $\mathrm{aab}$, $\mathrm{aba}$, $\mathrm{abb}$, $\mathrm{baa}$; $\mathrm{bba}$, $\mathrm{bbb} \}$; 
$\mathcal{S}_{4} = \{ \mathrm{aaab}$, $\mathrm{aabb}$, $\mathrm{abaa}$, $\mathrm{abbb}$, $\mathrm{baaa}$; $\mathrm{bbba}  \}$; 
$\mathcal{S}_{5} = \{ \mathrm{aaabb}$, $\mathrm{aabbb}$, $\mathrm{abaaa}$, $\mathrm{abbba}$;
$\mathrm{baaab} \}$; 
$\mathcal{S}_{6} = \{ \mathrm{aaabbb}$, $\mathrm{aabbba}$, $\mathrm{abaaab}$, $\mathrm{baaabb} \}$;
$\mathcal{S}_{7} = \{ \mathrm{aaabbba}$, $\mathrm{abaaabb}$, $\mathrm{baaabbb} \}$; 
$\mathcal{S}_{8} = \{ \mathrm{abaaabbb}$, $\mathrm{baaabbba} \}$; 
$\mathcal{S}_{9} = \{ \mathrm{abaaabbba} \}$.
Therefore, we obtain $\delta = 7/3$.

$\delta$-optimal space is defined as a space of $\Theta(\delta \log \frac{n \log \sigma}{\delta \log n} \log n)$ bits for string $T$ of length $n$, with an alphabet size of $\sigma$, and a substring complexity of $\delta$. 
Any string $T$ can be encoded into a bit string of length $\mathcal{O}(\delta \log \frac{n \log \sigma}{\delta \log n} \log n)$~\cite{9961143}. This bound is tight for the three parameters $n$, $\sigma$, and $\delta$~\cite{9961143}.

%Recently, Kociumaka et al. showed upper and lower bounds on compression size of string $T$ with substring %complexity $\delta$ in \cite{9961143}. 
%The following lemma states an upper bound on compression size of string $T$. 
%\begin{lemma}[Theorem IV.8(1) in~\cite{9961143}]\label{lem:upper_bound_delta}
%Any string $T$ among strings of length $n$, with an alphabet size of $\sigma = n^{\mathcal{O}(1)}$, and a substring complexity of $\delta$ can be encoded into a bit string of length $\mathcal{O}(\delta \log \frac{n \log \sigma}{\delta \log n} \log n)$.
%\end{lemma}

%Similarly, the following lemma states an lower bound on compression size of string $T$.
%\begin{lemma}[Theorem IV.8(2) in~\cite{9961143}]\label{lem:lower_bound_delta}
%At least one string $T$ among strings of length $n$, with an alphabet size of $\sigma$, and a substring complexity of $\delta$ can be encoded into 
%a bit string of length $\Omega(\delta \log \frac{n \log \sigma}{\delta \log n} \log n)$.
%\end{lemma}

%$\delta$-optimal space is defined as a space of $\Theta(\delta \log \frac{n \log \sigma}{\delta \log n} \log n)$ bits. 
%Lemma~\ref{lem:upper_bound_delta} and lemma~\ref{lem:lower_bound_delta} 
%ensure that $\delta$-optimal space is worst-case optimal 
%to represent a given string of length $n$ with an alphabet size of $\sigma$ and a substring complexity of $\delta$. 

\subsection{Range-count and Range-sum Queries on Two-Dimensional Grid}\label{subsec:range_data_structure}
A \emph{grid} is a pair $(\mathcal{X}, \mathcal{Y})$ of two ordered sets, $\mathcal{X} = \{ L_{1}, L_{2}, \ldots, L_{d} \}$~($L_{1} < L_{2} < \cdots < L_{d}$) and 
$\mathcal{Y} = \{ R_{1}, R_{2}, \ldots, R_{d^{\prime}} \}$~($R_{1} < R_{2} < \cdots < R_{d^{\prime}}$). 
Here, each element of set $\mathcal{X}$ represents a $x$-coordinate on two-dimensional space. 
Similarly, each element of set $\mathcal{Y}$ represents a $y$-coordinate on two-dimensional space. 
A \emph{weighted point} on grid $(\mathcal{X}, \mathcal{Y})$ is defined as a 4-tuple $(x, y, w, e)$ of four elements $x \in \mathcal{X}$, $y \in \mathcal{Y}$, $w \geq 0$, 
and $e \in \mathcal{E}$ for a set $\mathcal{E}$ of elements. 
Here, the $x$-coordinate and $y$-coordinate of the weighted point are represented by the first element $x$ and second element $y$ of 4-tuple $(x, y, w, e)$, respectively;
the third element $w$ is a weight of the weighted point; the fourth element $e$ represents the identifier of the weight point.
The identifiers of weighted points are used to distinguish them.

\emph{Range-sum data structure}~\cite{doi:10.1137/0217026} is a data structure supporting the following two queries and two operations for a given set $\mathcal{J}$ of weighted points on grid $(\mathcal{X}, \mathcal{Y})$: 
(i) a range-sum query, denoted as \emph{range-sum query} $\rangesum(\mathcal{J}, L, L^{\prime}, R, R^{\prime})$, returns the sum of all the weights $w$ in the subset of weighted points $(x, y, w, e)$ in the set $\mathcal{J}$ which meet the condition that each pair of $x$- and $y$-coordinates satisfies $L \leq x \leq L^{\prime}$ and $R \leq y \leq R^{\prime}$ for given $L, L^{\prime} \in \mathcal{X}$ and $R, R^{\prime} \in \mathcal{Y}$ (i.e., $\rangesum(\mathcal{J}, L, L^{\prime}, R, R^{\prime}) = \sum_{(x, y, w, e) \in \mathcal{I}} w$ for set $\mathcal{I} = \{ (x, y, w, e) \in \mathcal{J} \mid (L \leq x \leq L^{\prime}) \land (R \leq y \leq R^{\prime}) \}$);
(ii) a range-count query, denoted as \emph{range-count query} $\rangecount(\mathcal{J}, L, L^{\prime}, R, R^{\prime})$, returns the number of weighted points $(x, y, w, e)$ in the set $\mathcal{J}$ which meet the condition that each pair of $x$- and $y$-coordinates satisfy $L \leq x \leq L^{\prime}$ and $R \leq y \leq R^{\prime}$ for given $L, L^{\prime} \in \mathcal{X}$ and $R, R^{\prime} \in \mathcal{Y}$ (i.e., $\rangecount(\mathcal{J}, L, L^{\prime}, R, R^{\prime}) = |\{ (x, y, w, e) \in \mathcal{J} \mid (L \leq x \leq L^{\prime}) \land (R \leq y \leq R^{\prime}) \}|$);
(iii) the insertion operation inserts a given weighted point $(x, y, w, e)$ into set $\mathcal{J}$; 
(iv) the deletion operation deletes a given weighted point $(x, y, w, e)$ from $\mathcal{J}$. 
The range-sum data structure supports (i) range-count query in $O(\log^{2} |\mathcal{J}|)$ time and (ii) the other queries in $O(\log^{4} |\mathcal{J}|)$ time with $O(|\mathcal{J}| B)$ bits of space for machine word size $B$. 
The running time of the above four queries can be achieved under the condition that 
any pair of elements, either from set $\mathcal{X}$ or from set $\mathcal{Y}$), can be compared in $O(1)$ time. 
For such $O(1)$ time comparisons, we build an \emph{order maintenance data structure}~\cite{DBLP:conf/stoc/DietzS87} on each set of $\mathcal{X}$ and $\mathcal{Y}$. 
The order maintenance data structure is detailed in Section~\ref{subsubsec:JA_X_ds}.  
%and the combination of order maintenance and range-sum data structures is detailed in Section~\ref{subsubsec:RA_ds}.

\subsection{Straight-Line Program~(SLP)~\texorpdfstring{\cite{DBLP:journals/njc/KarpinskiRS97}}{}}\label{subsec:slp}
\begin{figure}[t]
 \begin{center}
		\includegraphics[scale=0.7]{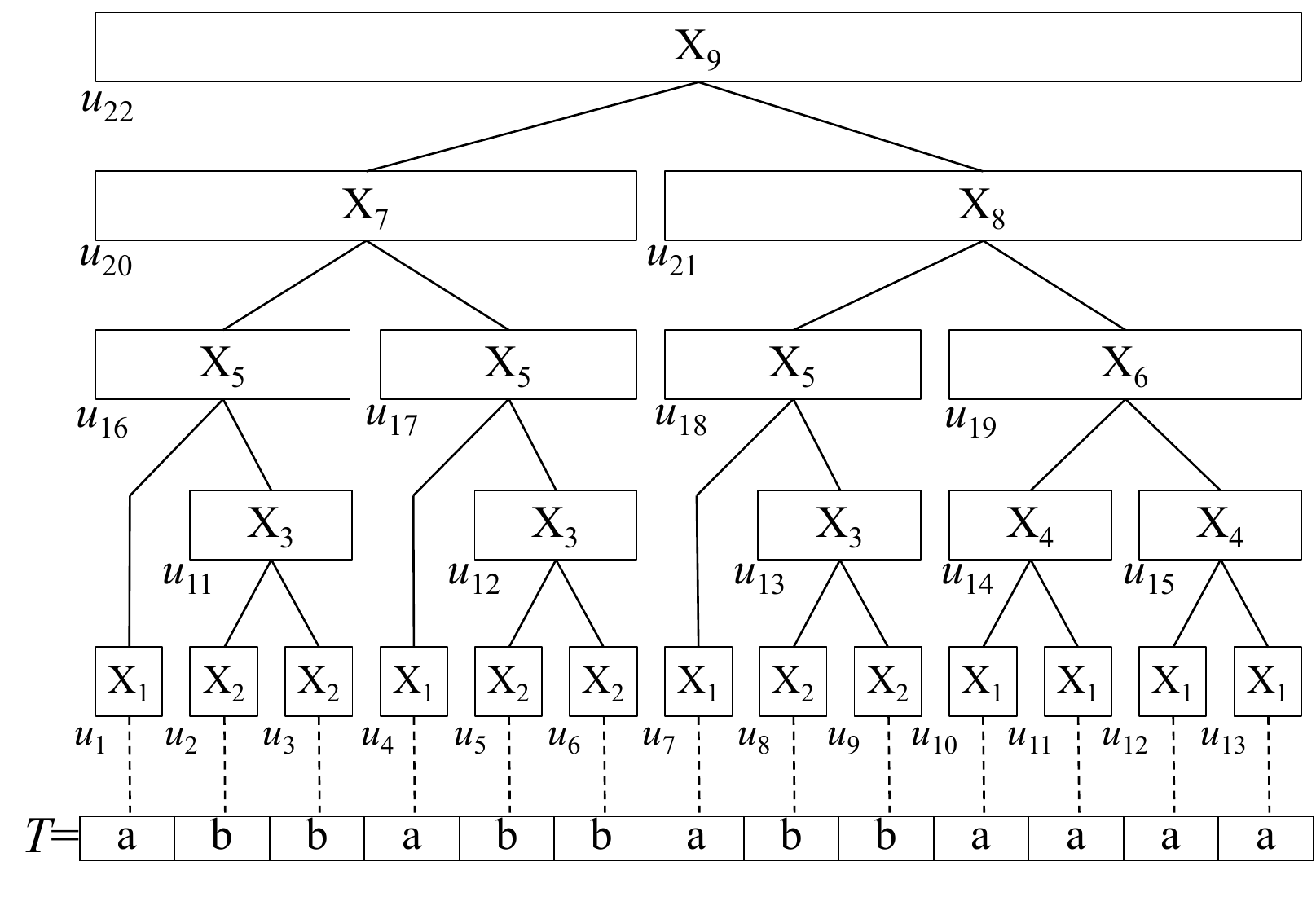}

	  \caption{An Illustration of the derivation tree $\mathcal{T}$ of SLP $\mathcal{G}$ 
	  with set $\{ u_{1}, u_{2}, \ldots, u_{21} \}$ of nodes.  
	  Each rectangle represents the node written in the lower left corner of the rectangle, 
	  and the nonterminal in the rectangle depicts the label of the corresponding node.} 
	  %(Bottom) The label function $\mathcal{L}$ of derivation tree $\mathcal{T}$.
\label{fig:slp_derivation_tree}
 \end{center}
\end{figure}
\emph{Straight-Line Program~(SLP)}~\cite{DBLP:journals/njc/KarpinskiRS97} is a grammar-based compression for strings, and it is a context-free grammar deriving a single string in \emph{Chomsky normal form}. 
SLP deriving a string $T$ of length $n \geq 1$ is defined as 4-tuple $\mathcal{G} = (\mathcal{V}, \Sigma, \mathcal{D}, E)$, where 
$\mathcal{V} = \{ X_{1}, X_{2}, \ldots, X_{g} \}$ is a set of \emph{nonterminals};
$\Sigma$ is an alphabet for string $T$;
$\mathcal{D} = \{ X_{1} \rightarrow \expr_{1}, X_{2} \rightarrow \expr_{2}, \ldots, X_{g} \rightarrow \expr_{g} \} \subseteq \mathcal{V} \times \{(\mathcal{V} \times \mathcal{V}) \cup \Sigma\}$ is a set of \emph{production rules}.
Each production rule $(X_{i} \rightarrow \expr_{i}) \in \mathcal{D}$ means that nonterminal $X_{i} \in \mathcal{V}$ in the left-hand side generates $\expr_{i} \in  (\mathcal{V} \times \mathcal{V}) \cup \Sigma$ as a pair of nonterminals $X_{j}X_{k} \in \mathcal{V} \times \mathcal{V}$ ($j, k < i$) or a character $c \in \Sigma$ in the right-hand side, and 
%it satisfies the condition $\expr_{i} \neq \expr_{i^\prime}$ for each pair of $\expr_{i}$ and $\expr_{i^\prime}$ in $(\mathcal{V} \times \mathcal{V}) \cup \Sigma$. 
it satisfies the condition $\expr_{i} \neq \expr_{i^\prime}$ for any pair of two production rules $(X_{i} \rightarrow \expr_{i})$ and 
$(X_{i^{\prime}} \rightarrow \expr_{i^{\prime}})$ in $\mathcal{D}$. 
%it satisfies the condition $\expr_{i} \neq \expr_{i^\prime}$ for each pair of $\expr_{i}$ and $\expr_{i^\prime}$ in $(\mathcal{V} \times \mathcal{V})$. 
%$E \in \mathcal{V}$ is the unique nonterminal called \emph{start symbol} which does not appear in the right-hand side of 
%any production rule in $\mathcal{D}$ (i.e., $\nexists \expr_{i} \in \{\expr_{1},\expr_{2},...,\expr_{g}\}$ such that %either of $\expr_{i}$ is $E$ for all $(X_{i} \rightarrow \expr_{i}) \in \mathcal{D}$).
$E \in \mathcal{V}$ is the unique nonterminal called \emph{start symbol} which does not appear in the right-hand side of 
any production rule in $\mathcal{D}$ (i.e., $\expr_{i} \in \mathcal{V}$ does not contain $E$ for all $(X_{i} \rightarrow \expr_{i}) \in \mathcal{D}$). 
%neither of the nonterminals in any pair $\expr_{i} \in \mathcal{V}$ is $E$ for all $(X_{i} \rightarrow \expr_{i}) \in \mathcal{D}$).
%In addition, the other nonterminals except the start symbol $E$ must appear at least once in the right-hand side of any %production rule (i.e., $\forall X_{i} \in \mathcal{V} \setminus \{ E \}$, $\exists (X_j \rightarrow \expr_j) \in %\mathcal{D}$ such that either of the pair $\expr_{j}$ of nonterminals is $X_j$).
Additionally, all nonterminals except the start symbol $E$ must appear at least once in the right-hand side of some production rule. In other words, for every $X_i \in \mathcal{V} \setminus { E }$, there exists a production rule $(X_j \rightarrow \expr_j) \in \mathcal{D}$ 
where either of the nonterminals in the pair $\expr_{j}$ is $X_{j}$.

In SLP, each nonterminal in $\mathcal{V}$ derives the substring in input string $T$, forming the complete binary tree called \emph{derivation tree} with the root labeled the nonterminal. 
The start symbol $E$ derives input string $T$.
Let $\val : \mathcal{V} \rightarrow \Sigma^{+}$ be a function that returns the substring in string $T$ derived by a given nonterminal in $\mathcal{V}$. 
%The complete binary tree deriving the input string $T$ for the start symbol $E$ is called \emph{derivation tree} of SLP $\mathcal{G}$. 
The complete binary tree deriving the input string $T$ is called \emph{derivation tree} of SLP $\mathcal{G}$. 
For the derivation tree of SLP $\mathcal{G}$, 
a node of the derivation tree is said to derive a substring $T[i..j]$ in string $T$ 
if the leftmost and rightmost leaves of the subtree with the node as the root are 
the $i$-th and $j$-th leaves of the derivation tree, respectively.

Figure~\ref{fig:slp_derivation_tree} illustrates the derivation tree of an SLP $\mathcal{G} = (\mathcal{V}, \Sigma, \mathcal{D}, E)$ 
deriving a string $T$. 
Here, (i) $T = \mathrm{abb abb abb aaaa}$;
(ii) $\mathcal{V} = \{ X_{1}, X_{2}, \ldots, X_{9} \}$; 
(iii) $\Sigma = \{ \mathrm{a}, \mathrm{b} \}$; 
(iv) $\mathcal{D} = \{ X_{1} \rightarrow \mathrm{a}$, 
$X_{2} \rightarrow \mathrm{b}$, 
$X_{3} \rightarrow (X_{2}, X_{2})$, 
$X_{4} \rightarrow (X_{1}, X_{1})$, 
$X_{5} \rightarrow (X_{1}, X_{3})$, 
$X_{6} \rightarrow (X_{4}, X_{4})$, 
$X_{7} \rightarrow (X_{5}, X_{5})$, 
$X_{8} \rightarrow (X_{5}, X_{6})$, 
$X_{9} \rightarrow (X_{7}, X_{8}) \}$; 
(v) $E = X_{9}$. 

\subsection{Run-Length SLP~(RLSLP)~\texorpdfstring{\cite{DBLP:conf/mfcs/TanimuraNBIT17}}{}}\label{subsec:run-length_slp}
%\begin{figure}[t]
% \begin{center}
%		\includegraphics[scale=0.7]{figures/rlslp_derivation_tree-crop.pdf}
%
%	  \caption{An illustration of the derivation tree $\mathcal{T}$ of RLSLP $\mathcal{G}_{2}$ 
%	  with set $\{ u_{1}, u_{2}, \ldots, u_{27} \}$ of nodes.  
%	  Each rectangle represents the node written in the lower left corner of the rectangle, 
%	  and the nonterminal in the rectangle depicts the label of the corresponding node.}
%\label{fig:rlslp_nonterminal_tree}
% \end{center}
%\end{figure}

An \emph{run-length SLP (RLSLP)} is an SLP with nonterminals that can produce a repetition of characters in an alphabet.
Several algorithms have been proposed to build an RLSLP for a string~(e.g., \emph{signature encoding}~\cite{DBLP:conf/mfcs/TanimuraNBIT17}, \emph{recompression}~\cite{DBLP:journals/talg/Jez15}, \emph{signature grammar}~\cite{DBLP:conf/latin/ChristiansenE18} and \emph{restricted block compression}~\cite{DBLP:conf/latin/KociumakaNO22}). 

%let $(X_{i})^{d}$ be a $d$ successive appearance of nonterminals.% $X_{i}, X_{i}, \ldots, X_{i}$.
RLSLP is defined as a 4-tuple $\mathcal{G}^R=(\mathcal{V}, \Sigma, \mathcal{D}, E)$ where the definitions of $\mathcal{V}$, $\Sigma$, and $E$ are the same as those in SLP $\mathcal{G}$. 
The set of production rules, denoted as $\mathcal{D}$, in RLSLP $\mathcal{G}^R$ differs from that in SLP $\mathcal{G}$.
Specifically, $\mathcal{D}$ is defined as a subset of $\mathcal{V} \times (\Sigma \cup \mathcal{V}^+)$; that is $\mathcal{D} \subseteq \mathcal{V} \times (\Sigma \cup \mathcal{V}^+)$. Each production rule in $\mathcal{D}$ takes one of the following forms: 
(i) $X_{i} \rightarrow c$ for $c \in \Sigma$; 
(ii) $X_{i} \rightarrow X_j, X_k$  ($j, k < i$) for two distinct nonterminals $X_j, X_k \in \mathcal{V}, X_j \neq X_k$;
(iii) $X_{i} \rightarrow X_j$ ($j < i$)  for $X_j \in \mathcal{V}$;
(iv) $X_{i} \rightarrow (X_{j})^{d}$  ($j < i$) for $(X_{j})^{d}$ representing $d(\geq 2)$ successive appearance of a nonterminal $X_j \in \mathcal{V}$.

A derivation tree corresponding to an RLSLP is said to be {\em height-balanced} if it meets one of the following conditions: 
(i) the derivation tree comprises only a single node (i.e., root); 
or (ii) all children of each node in the tree have the same height.
We define sequences $S^{0}, S^{1}, \ldots, S^{H}$, comprising a total of $H+1$ sequences, such that each $S^{h}$ maps to a sequence of nonterminal symbols. These symbols are utilized to denote the labels of consecutive nodes at height $h$ within the derivation tree of $\mathcal{G}^{R}$.

\section{Interval Attractors for SA and ISA Queries}\label{sec:simplicied_RASS}
This section introduces a new concept, termed \emph{interval attractors}, which is defined within the derivation tree of RLSLP, and explains how interval attractors can be used to answer SA and ISA queries. 
Section~\ref{subsec:simplicied_RASS} introduces interval attractors and explains their properties. 
Section~\ref{subsec:simplified_rss_rsc} introduces two queries and shows that 
the two queries can be answered using interval attractors.
Section~\ref{subsec:simplified_sa_and_isa_queries} shows that 
SA and ISA queries can be answered using the two queries of Section~\ref{subsec:simplified_rss_rsc}. 

\subsection{Interval Attractors}\label{subsec:simplicied_RASS}
Each node in the derivation tree of an RLSLP can be associated with a mutually exclusive and collectively exhaustive set of intervals (i.e., a partition) in the input string. An interval attractor is defined for each node based on this partition.

Throughout this paper, let $u, u^{\prime} \in \mathcal{U}$ be nodes at heights $h, h' \in \{0,1, \ldots, H\}$ in the derivation tree of RLSLP $\mathcal{G}^{R}$, 
deriving the substrings that start at positions $\gamma$ and $\gamma^{\prime}$ on $T$, respectively.
Let $\Delta$ denote the set of all possible intervals of length at least 2 on input string $T$ (i.e., $\Delta = \{[s, e] \mid 1 \leq s < e \leq n \}$) throughout this paper.

We can construct the partition $\Delta(u) \subseteq \Delta$ associated with each $u \in \mathcal{U}$ such that any pair of intervals $[s,e] \in \Delta(u)$ and $[s^{\prime}, e^{\prime}] \in \Delta(u^{\prime})$ for any $u^{\prime} \in \mathcal{U}$ satisfies the following conditions:

\begin{enumerate}[label=\textbf{(\roman*)}] 
    \item \label{condition:IA:include} $\gamma \in [s, e-1]$ and $\gamma^{\prime} \in [s^{\prime}, e^{\prime}-1]$; 
    \item \label{condition:IA:2} If the distance from $s$ to $\gamma$ is unequal to the distance from $s'$ to $\gamma'$ (i.e., $\gamma - s \neq \gamma' - s'$), then substrings $T[s..e]$ and $T[s'..e']$ are different (i.e., $T[s..e] \neq T[s'..e']$).
    \item \label{condition:IA:3} If $\gamma > \gamma'$ holds, then the substring $T[s..e]$ is not a prefix of $T[s'..e']$ (i.e., either $s \neq s^{\prime}$ or $s = s^{\prime} \leq e^{\prime} < e$).
    \item \label{condition:IA:4} If $h > h'$ holds, then $T[s..e]$ is not a substring of $T[s'..e']$ (i.e., $[s..e] \not \subseteq [s', e']$).
    \item \label{condition:IA:extension} Additionally, if $u = u'$, then for any pair $(s^{\prime\prime}, e^{\prime\prime})$ such that $s^{\prime \prime} \in [\min \{ s, s^{\prime}  \}, s ]$ and $e^{\prime \prime} \in [e, \max \{ e, e^{\prime} \}]$, it holds that $[s^{\prime\prime}, e^{\prime\prime}] \in \Delta(u)$.
\end{enumerate}
Note that for $\Delta(u)$ and $\Delta(u')$ associated with any pair of nodes $u, u' \in \mathcal{U}$, the conditions $\Delta(u) \cap \Delta(u^{\prime}) = \emptyset$ and $\bigcup_{u \in \mathcal{U}} \Delta(u) = \Delta$ hold.

Interval attractor $I(u)$ associated with $u \in \mathcal{U}$ is defined as a pair of intervals $([p,q],[\ell,r])$ such that $p = \min \{ s \}$, $q = \max \{ s \}$, $\ell = \min \{ e \}$, and $r = \max\{ e \}$ for all intervals $[s, e] \in \Delta(u)$.
If $\Delta(u)$ is empty, then the interval attractor $I(u)$ is undefined.

\subsubsection{Properties of Interval Attractors}
We explain properties of interval attractors and each set $\Delta(u)$. 
The following corollary follows from the conditions of set $\Delta(u)$.
\begin{corollary}\label{cor:IA_basic_corollary}
    Let $u \in \mathcal{U}$ be a node of height $h$ deriving a substring starting at position $\gamma$ in input string $T$.
    Similarly, 
    let $u^{\prime} \in \mathcal{U}$ be a node of height $h^{\prime}$ deriving a substring starting at position $\gamma^{\prime}$ in input string $T$. 
    For two intervals $[s, e] \in \Delta(u)$ and $[s^{\prime}, e^{\prime}] \in \Delta(u^{\prime})$, 
    the two nodes satisfy the following three conditions: 
    \begin{enumerate}[label=\textbf{(\roman*)}] 
        \item \label{enum:IA_basic_corollary:1} if $T[s..e] = T[s^{\prime}..e^{\prime}]$, then 
        $|[s, \gamma]| = |[s^{\prime}, \gamma^{\prime}]|$ and $h = h^{\prime}$;
        \item \label{enum:IA_basic_corollary:2} if $s = e$, $e < e^{\prime}$, and $h = h^{\prime}$, then $u = u^{\prime}$;
        \item \label{enum:IA_basic_corollary:3} if $[s, e] \subseteq [s^{\prime}, e^{\prime}]$, then $h \leq h^{\prime}$.
    \end{enumerate}    
\end{corollary}

The following corollary follows from the definition of interval attractors 
and Condition~\ref{condition:IA:include} of $\Delta(u)$. 
\begin{corollary}\label{cor:IA_exist_corollary}    
    For a node $u \in \mathcal{U}$ such that $\Delta(u) \neq \emptyset$,     
    let $I(u)=([p,q], [\ell, r])$ be an interval attractor associated with $u$ 
    and $\gamma$ be the starting position of the substring derived from $u$ on $T$. 
    Then, the following five statements hold: 
    \begin{enumerate}[label=\textbf{(\roman*)}]
        \item \label{enum:IA_exist_corollary:1} $[\gamma, \gamma+1] \subseteq [s, e]$ 
        for each interval $[s, e] \in \Delta(u)$;
        \item there exists an integer $e \in [\ell, r]$ 
        satisfying $[p, e] \in \Delta(u)$;
        \item there exists an integer $e \in [\ell, r]$ 
        satisfying $[q, e] \in \Delta(u)$;
        \item there exists an integer $s \in [p, q]$ 
        satisfying $[s, \ell] \in \Delta(u)$;
        \item there exists an integer $s \in [p, q]$ 
        satisfying $[s, r] \in \Delta(u)$.
    \end{enumerate}
\end{corollary}

The following lemma states the forms of intervals contained in set $\Delta(u)$.
\begin{lemma}\label{lem:IA_maximal_lemma}    
    For a node $u \in \mathcal{U}$ such that $\Delta(u) \neq \emptyset$,     
    let $I(u)=([p,q], [\ell, r])$ be an interval attractor associated with $u$. 
    Then, the following four statements hold: 
    \begin{enumerate}[label=\textbf{(\roman*)}]
    \item $[p, r], [p, \ell], [q, r] \in \Delta(u)$;
    \item consider an interval $[s, e]$ in set $\Delta(u)$. 
    Then, $[s^{\prime}, e^{\prime}] \in \Delta(u)$ for any pair of integers $s^{\prime} \in [p, s]$ and $e^{\prime} \in [e, r]$;
    \item $[s, r] \in \Delta(u)$ for any integer $s \in [p, q]$;
    \item $[p, e] \in \Delta(u)$ for any integer $e \in [\ell, r]$.
    \end{enumerate}
\end{lemma}
\begin{proof}
    This lemma can be proved using Condition~\ref{condition:IA:extension} of set $\Delta(u)$ 
    and Corollary~\ref{cor:IA_exist_corollary}. 
    The detailed proof is as follows. 

    \paragraph{Proof of Lemma~\ref{lem:IA_maximal_lemma}(i).}    
    Corollary~\ref{cor:IA_exist_corollary} shows that 
    there exist two integers $s$ and $e$ such that 
    satisfying $[s, r] \in \Delta(u)$;
    $[p, e] \in \Delta(u)$, respectively. 
    In this case, 
    $[p, r] \in \Delta(u)$ follows from Condition~\ref{condition:IA:extension} of $\Delta(u)$. 
    Similarly, 
    we can prove $[p, \ell], [q, r] \in \Delta(u)$ using the same approach. 

    \paragraph{Proof of Lemma~\ref{lem:IA_maximal_lemma}(ii).}    
    Since  $[p, r], [s, e] \in \Delta(u)$, 
    $[s^{\prime}, e^{\prime}] \in \Delta(u)$ follows from Condition~\ref{condition:IA:extension} of $\Delta(u)$.  

    \paragraph{Proof of Lemma~\ref{lem:IA_maximal_lemma}(iii).}    
    From Corollary~\ref{cor:IA_exist_corollary}, 
    there exists an integer $e \in [\ell, r]$ satisfying $[q, e] \in \Delta(u)$.
    Since  $[p, r], [q, e] \in \Delta(u)$, 
    $[s, r] \in \Delta(u)$ follows from Condition~\ref{condition:IA:extension} of $\Delta(u)$.  

    \paragraph{Proof of Lemma~\ref{lem:IA_maximal_lemma}(iv).}    
    From Corollary~\ref{cor:IA_exist_corollary}, 
    there exists an integer $s \in [p, q]$ satisfying $[s, \ell] \in \Delta(u)$.
    Since  $[p, r], [s, \ell] \in \Delta(u)$, 
    $[p, e] \in \Delta(u)$ follows from Condition~\ref{condition:IA:extension} of $\Delta(u)$.  
\end{proof}

%\begin{lemma}\label{lem:IA_matchX}    
%    Let $I(u)=([p,q], [\ell, r])$ be an interval attractor associated with 
%    a node $u \in \mathcal{U}$ such that $\Delta(u)$ contains an interval $[s, e]$. 
%    Then, $[s^{\prime}, e^{\prime}] \in \Delta(u)$ for any pair of integers $s^{\prime} \in [p, s]$ and $e^{\prime} \in [e, r]$. 
%\end{lemma}

The following lemma states basic properties of interval attractors. 
\begin{lemma}\label{lem:IA_super_basic_property}
For a node $u \in \mathcal{U}$ such that $\Delta(u) \neq \emptyset$,     
let $I(u)=([p,q], [\ell, r])$ be an interval attractor associated with $u$ 
and $\gamma$ be the starting position of the substring derived from $u$ on $T$. 
The following two statements hold:
\begin{enumerate}[label=\textbf{(\roman*)}]
\item \label{enum:IA_super_basic_property:1} $q \leq \gamma < \ell$;
\item \label{enum:IA_super_basic_property:3} 
$[p, r] \neq [p^{\prime}, r^{\prime}]$ and $I(u) \neq I(u^{\prime})$ for 
the interval attractor $I(u^{\prime})=([p^{\prime},q^{\prime}], [\ell^{\prime}, r^{\prime}])$ associated with 
a node $u^{\prime} \in \mathcal{U}$ ($u \neq u^{\prime}$). 
\end{enumerate}
\end{lemma}
\begin{proof}
The proof of Lemma~\ref{lem:IA_super_basic_property} is as follows. 

\paragraph{Proof of Lemma~\ref{lem:IA_super_basic_property}(i).}
From Condition~\ref{condition:IA:include}, 
every interval $[s, e]$ of set $\Delta(u)$ contains two positions $\gamma$ and $\gamma+1$. 
We obtain $q \leq \gamma$ because 
$q = \max \{ s \mid [s, e] \in \Delta(u) \}$ and $s \leq \gamma$. 
Similarly, 
we obtain $\gamma + 1 \leq \ell$ because 
$\ell = \min \{ e \mid [s, e] \in \Delta(u) \}$ and $\gamma + 1 \leq e$. 
Therefore, we obtain $q \leq \gamma < \ell$. 

\paragraph{Proof of Lemma~\ref{lem:IA_super_basic_property}(ii).}
We prove $[p, r] \neq [p^{\prime}, r^{\prime}]$ by contradiction. 
We assume that $[p, r] = [p^{\prime}, r^{\prime}]$ holds. 
Then, 
Lemma~\ref{lem:IA_maximal_lemma} shows that $[p, r] \in \Delta(u)$ and $[p, r] \in \Delta(u^{\prime})$, 
and hence, $\Delta(u) \cap \Delta(u^{\prime}) \neq \emptyset$.
On the other hand, $\Delta(u) \cap \Delta(u^{\prime}) = \emptyset$ follows from the definition of set $\Delta(u)$. 
The two facts yield a contradiction. 
Therefore, $[p, r] \neq [p^{\prime}, r^{\prime}]$ must hold. 
$I(u) \neq I(u^{\prime})$ follows from $[p, r] \neq [p^{\prime}, r^{\prime}]$. 
\end{proof}

The following corollary follows from Lemma~\ref{lem:IA_super_basic_property}~\ref{enum:IA_super_basic_property:1}, 
and this corollary shows that for two interval attractors $I(u)$ and $I(u^{\prime})$, 
we can verify whether $I(u) = I(u^{\prime})$ or not by comparing two nodes $u$ and $u^{\prime}$. 

\begin{corollary}\label{cor:IA_identify_corollary}
Consider the two interval attractors $I(u)=([p,q], [\ell, r])$ and $I(u^{\prime})=([p^{\prime},q^{\prime}], [\ell^{\prime}, r^{\prime}])$ 
associated with two nodes $u$ of height $h$ and $u^{\prime}$ of height $h^{\prime}$ in $\mathcal{U}$, respectively. 
Let $\gamma$ and $\gamma^{\prime}$ be the starting positions of the two substrings derived from $u$ 
and $u^{\prime}$, respectively, on $T$. 
Then, (i) $u = u^{\prime} \Leftrightarrow h = h^{\prime} \land \gamma = \gamma^{\prime}$ 
and (ii) $I(u) = I(u^{\prime}) \Leftrightarrow u = u^{\prime}$.
\end{corollary}

The following lemma states the number of interval attractors associated with nodes in the derivation tree of RLSLP. 
\begin{lemma}\label{lem:non_comp_IA_size}
Let $m$ be the number of interval attractors obtained from the derivation tree of RLSLP $\mathcal{G}^{R}$. 
Then, $m = O(n^{2})$.
\end{lemma}
\begin{proof}
    For each position $i \in [1, n]$ of $T$, 
    let $m_{i}$ be the number of interval attractors $I(u) = ([p, q], [\ell, r])$ such that 
    $p = i$. 
    Then, $m_{i} \leq n$ follows from Lemma~\ref{lem:IA_super_basic_property}~\ref{enum:IA_super_basic_property:3}. 
    Therefore, $m = O(n^{2})$ follows from $m = \sum_{i = 1}^{n} m_{i}$ and $m_{i} \leq n$. 
\end{proof}

The following lemma states the relationship between the heights of two nodes $u$ and $u^{\prime}$ 
using the interval attractor associated with $u$ and an interval contained in set $\Delta(u^{\prime})$. 

\begin{lemma}\label{lem:interval_extension_propertyX}
    For a node $u \in \mathcal{U}$ of height $h$ such that $\Delta(u) \neq \emptyset$,     
    let $I(u)=([p,q], [\ell, r])$ be an interval attractor associated with $u$. 
    Let $u^{\prime} \in \mathcal{U}$ be a node of height $h^{\prime}$ such that set $\Delta(u^{\prime})$ contains an interval $[s, e]$. 
    Then, the following three statements hold: 
    \begin{enumerate}[label=\textbf{(\roman*)}]
        \item \label{enum:interval_extension_propertyX:right}
        if $s \in [p, q]$ and $e \geq r+1$, then $h^{\prime} \geq h+1$; 
        \item \label{enum:interval_extension_propertyX:left}
        if $s \leq p-1$ and $e \in [\ell, n]$, 
        then $h^{\prime} \geq h$. 
        \item \label{enum:interval_extension_propertyX:low}
        if $s \in [p, q]$, $e \in [\ell, r]$, and $[s, e] \not \in \Delta(u)$, 
        then $h^{\prime} < h$. 
    \end{enumerate}
\end{lemma}
\begin{proof}
    The proof of Lemma~\ref{lem:interval_extension_propertyX} is as follows. 
    \paragraph{Proof of Lemma~\ref{lem:interval_extension_propertyX}(i).}
    $[q, r] \in \Delta(u)$ follows from Lemma~\ref{lem:IA_maximal_lemma}. 
    $[q, r+1] \not \in \Delta(u)$ follows from the definition of interval attractor. 
    Let $u^{\prime \prime} \in \mathcal{U}$ be a node of height $h^{\prime \prime}$ such that set $\Delta(u^{\prime \prime})$ contains an interval $[q, r+1]$. 
    Let $\gamma$ and $\gamma^{\prime}$ be the starting positions of the substrings derived from $u$ and $u^{\prime}$, respectively. 
    Since $[q, r] \subseteq [q, r+1]$, 
    $h \leq h^{\prime\prime}$ follows from 
    Condition~\ref{enum:IA_basic_corollary:3} of Corollary~\ref{cor:IA_basic_corollary}. 
    Similarly, 
    we obtain $h^{\prime\prime} \leq h^{\prime}$. 
    Therefore, we obtain Lemma~\ref{lem:interval_extension_propertyX}~\ref{enum:interval_extension_propertyX:left} 
    if $h \neq h^{\prime\prime}$

    We prove $h \neq h^{\prime\prime}$ by contradiction. 
    We assume that $h = h^{\prime\prime}$ holds. 
    Then, $u = u^{\prime\prime}$ follows from 
    Condition~\ref{enum:IA_basic_corollary:2} of Corollary~\ref{cor:IA_basic_corollary}. 
    On the other hand, $u \neq u^{\prime\prime}$ follows from $[q, r+1] \not \in \Delta(u)$. 
    The two facts $u = u^{\prime\prime}$ and $u \neq u^{\prime\prime}$ yield a contradiction, 
    and hence $h \neq h^{\prime\prime}$ must hold. 

    \paragraph{Proof of Lemma~\ref{lem:interval_extension_propertyX}(ii).}
    $[p, \ell] \in \Delta(u)$ follows from Lemma~\ref{lem:IA_maximal_lemma}. 
    Since $[p, \ell] \subseteq [s, e]$, 
    $h \leq h^{\prime}$ follows from 
    Condition~\ref{enum:IA_basic_corollary:3} of Corollary~\ref{cor:IA_basic_corollary}. 

    \paragraph{Proof of Lemma~\ref{lem:interval_extension_propertyX}(iii).}
    Consider interval $[s, r]$. 
    Lemma~\ref{lem:IA_maximal_lemma} shows that $[s, r] \in \Delta(u)$. 
    Since $[s, e] \subseteq [s, r]$, 
    $h^{\prime} \leq h$ follows from 
    Corollary~\ref{cor:IA_basic_corollary}~\ref{enum:IA_basic_corollary:3}. 
    Therefore, 
    $h^{\prime} < h$ holds if $h^{\prime} \neq h$. 

    We prove $h^{\prime} \neq h$ by contradiction. 
    We assume that $h^{\prime} = h$ holds. 
    Then, 
    Corollary~\ref{cor:IA_basic_corollary}~\ref{enum:IA_basic_corollary:2} shows that 
    $u = u^{\prime}$. 
    On the other hand,  
    $u \neq u^{\prime}$ follows from $[s, e] \not \in \Delta(u)$. 
    Therefore,  $h^{\prime} \neq h$ must hold.
\end{proof}

Consider two suffixes $T[\SA[i]..n]$ and $T[\SA[i^{\prime}]..n]$ within the sa-interval of a string $P$ 
on the suffix array $\SA$ of $T$. 
Then, two substrings $T[\SA[i]..\SA[i] + |P|-1]$ and $T[\SA[i^{\prime}]..\SA[i^{\prime}] + |P|-1]$ are 
occurrences of $P$ on $T$. 
The following theorem follows from the conditions of set $\Delta(u)$, 
and this theorem states the relationship between two nodes $u$ and $u^{\prime}$ such that $[\SA[i], \SA[i] + |P|-1] \in \Delta(u)$ and $T[\SA[i^{\prime}]..(\SA[i^{\prime}] + |P| - 1)] \in \Delta(u^{\prime})$. 

\begin{corollary}\label{cor:capture_gamma_corollary}
    For any two suffixes $T[\SA[i]..n]$ and $T[\SA[i^{\prime}]..n]$ ($i \neq i^{\prime}$) in the sa-interval of a string $P (|P| \geq 2)$ on the suffix array $\SA$ of $T$, 
    there exist nodes $u, u' \in \mathcal{U}$ such that $[\SA[i], \SA[i] + |P|-1] \in \Delta(u)$ and $T[\SA[i']..(\SA[i'] + |P| - 1)] \in \Delta(u')$.
    Also, there exist interval attractors $I(u)=([p,q], [\ell, r])$ and $I(u')=([p',q'], [\ell', r'])$ for such nodes $u, u' \in \mathcal{U}$. 
    Then, the following three statements hold: 
    \begin{enumerate}[label=\textbf{(\roman*)}]
    \item \label{enum:capture_gamma_corollary:1} the height of $u^{\prime}$ is the same as that of $u$ in the derivation tree; 
    \item \label{enum:capture_gamma_corollary:2} $T[\SA[i]..\gamma-1] = T[\SA[i^{\prime}]..\gamma^{\prime}-1]$ and $T[\gamma..\SA[i] + |P| -1] = T[\gamma^{\prime}..\SA[i^{\prime}] + |P| -1]$;
    \item \label{enum:capture_gamma_corollary:3} $u \neq u^{\prime}$.
    \end{enumerate}
\end{corollary}

Consider a node $u$ such that $\Delta(u)$ contains the interval 
representing an occurrence of $P$ in $T$. 
The following theorem states that for a given node $u^{\prime} \in \mathcal{U}$, 
we can verify whether set $\Delta(u^{\prime})$ contains the interval representing another occurrence of $P$ in $T$ or not by comparing the two interval attractors associated with $u$ and $u^{\prime}$.

\begin{theorem}\label{theo:SA_INTV_INTV_ATTR}    
    For a suffix $T[\SA[i]..n]$ in the sa-interval $[\eta, \eta^{\prime}]$ of a string $P (|P| \geq 2)$ on the suffix array $\SA$ of $T$, 
    an interval attractors $I(u)=([p,q], [\ell, r])$ is associated with a node $u \in \mathcal{U}$ such that $[\SA[i], \SA[i] + |P|-1] \in \Delta(u)$. 
    For a node $u^{\prime} \in \mathcal{U}$, 
    there exists a suffix $T[\SA[i^{\prime}]..n]$ in the sa-interval $[\eta, \eta^{\prime}]$ 
    satisfying $[\SA[i^{\prime}], \SA[i^{\prime}] + |P|-1] \in \Delta(u^{\prime})$ 
    if and only if 
    the interval attractor $I(u')=([p',q'], [\ell', r'])$ associated with $u^{\prime}$ satisfies 
    the following three conditions: 
    \begin{enumerate}[label=\textbf{(\roman*)}]
        \item the height of $u^{\prime}$ is the same as that of $u$ in the derivation tree.
        \item $T[p^\prime-1..\gamma^\prime-1]$ includes $T[\SA[i]..\gamma-1]$ as a proper suffix (i.e., $\lcs(T[p^\prime..\gamma^\prime-1], T[\SA[i]..\gamma-1]) = T[\SA[i]..\gamma-1]$).
        \item $T[\gamma^\prime..r^\prime+1]$ includes $T[\gamma..\SA[i] + |P|-1]$ as a proper prefix 
        (i.e., $\lcp(T[\gamma^\prime..r^\prime], T[\gamma..\SA[i] + |P|-1]) = T[\gamma..\SA[i] + |P|-1]$).
    \end{enumerate}
    \end{theorem}
    \begin{proof}
    Theorem~\ref{theo:SA_INTV_INTV_ATTR} follows from the following two statements: 
    \begin{enumerate}[label=\textbf{(\Alph*)}]
        \item if interval attractor $I(u')$ satisfies the three conditions of Theorem~\ref{theo:SA_INTV_INTV_ATTR}, 
        then suffix $T[\SA[i^{\prime}]..n]$ exists; 
        \item if suffix $T[\SA[i^{\prime}]..n]$ exists, 
        then interval attractor $I(u')$ satisfies the three conditions of Theorem~\ref{theo:SA_INTV_INTV_ATTR}. 
    \end{enumerate}

    \paragraph{Proof of statement (A).}
    We define $i^{\prime}$ as an integer satisfying $\SA[i^{\prime}] = \gamma^\prime - |[\SA[i], \gamma-1]|$. 
    Then, 
    suffix $T[\SA[i^{\prime}]..n]$ has string $P$ as a prefix, 
    and hence, this suffix is contained in the sa-interval of $P$. 

    There exists a node $u^{\prime \prime} \in \mathcal{U}$ of $h^{\prime}$ such that 
    interval $[\SA[i^{\prime}], \SA[i^{\prime}] + |P| - 1]$ is contained in set $\Delta(u^{\prime \prime})$. 
    Let $h$ and $h^{\prime}$ be the heights of two nodes $u$ and $u^{\prime}$, respectively. 
    Here, $h = h^{\prime}$ holds. 
    Let $\gamma^{\prime \prime}$ be the starting position of the substring derived from node $u^{\prime \prime}$.
    Since $P = T[\SA[i]..\SA[i] + |P|-1] = T[\SA[i^{\prime}]..\SA[i^{\prime}] + |P|-1]$, 
    $h = h^{\prime \prime}$ and $|[\SA[i], \gamma]| = |[\SA[i^{\prime}], \gamma^{\prime \prime}]|$ 
    follows from Condition~\ref{enum:IA_basic_corollary:1} of Corollary~\ref{cor:IA_basic_corollary}. 
    $\gamma^{\prime} = \gamma^{\prime \prime}$ follows from 
    $\SA[i^{\prime}] = \gamma^\prime - |[\SA[i], \gamma]| - 1$, $|[\SA[i], \gamma]| = |[\SA[i^{\prime}], \gamma^{\prime \prime}]|$. 
    Since $h^{\prime \prime} = h^{\prime}$ and $\gamma = \gamma^{\prime}$, 
    $u^{\prime}$ and $u^{\prime \prime}$ are the same node.  
    Therefore, 
    suffix $T[\SA[i^{\prime}]..n]$ is contained in the sa-interval $[\eta, \eta^{\prime}]$ 
    and $[\SA[i^{\prime}], \SA[i^{\prime}] + |P|-1] \in \Delta(u^{\prime})$ holds. 

    \paragraph{Proof of statement (B).}
    Since $P = T[\SA[i]..\SA[i] + |P|-1] = T[\SA[i^{\prime}]..\SA[i^{\prime}] + |P|-1]$, 
    $h = h^{\prime}$ and $|[\SA[i], \gamma]| = |[\SA[i^{\prime}], \gamma^{\prime}]|$ 
    follows from Condition~\ref{enum:IA_basic_corollary:1} of Corollary~\ref{cor:IA_basic_corollary}. 
    From the definition of interval attractor, 
    $[\SA[i^{\prime}], \SA[i^{\prime}] + |P|-1] \subseteq [p^{\prime}, r^{\prime}]$ always holds. 
    Therefore, 
    interval attractor $I(u')$ satisfies the three conditions of Theorem~\ref{theo:SA_INTV_INTV_ATTR}. 
\end{proof}

Consider 
the interval attractor $([p, p+\alpha_{1}], [p + \alpha_{2}, p+\alpha_{3}])$ associated with a node. 
If string $T$ contains another occurrence of substring $T[p-1..p+\alpha_{3}+1]$ as a substring $T[x-1..x+\alpha_{3}+1]$, then the following theorem ensures that pair $([x, x+\alpha_{1}], [x + \alpha_{2}, x+\alpha_{3}])$ of two intervals is an interval attractor with a node. 

\begin{theorem}\label{theo:IA_SYNC_X}
    Consider the interval attractor $I(u) = ([p, p+\alpha_{1}], [p + \alpha_{2}, p+\alpha_{3}])$ 
    associated with a node $u \in \mathcal{U}$ of height $h$ deriving a substring starting at position $\gamma$. 
    If there exists a position $x$ of input string $T$ such that 
    $T[x-1..x +\alpha_{3} + 1] = T[p-1..p +\alpha_{3} + 1]$, 
    then 
    consider a node $u^{\prime} \in \mathcal{U}$ such that $[x..x +\alpha_{3}] \in \Delta(u)$. 
    Then, this node $u^{\prime}$ satisfies the following three conditions: 
    \begin{enumerate}[label=\textbf{(\roman*)}]
        \item $I(u^{\prime}) = ([x, x+\alpha_{1}], [x + \alpha_{2}, x+\alpha_{3}])$;
        \item the height $h^{\prime}$ of $u^{\prime}$ is the same as that of $u$ in the derivation tree;
        \item $|[p, \gamma]| = |[x, \gamma^{\prime}]|$ 
        for the starting position $\gamma^{\prime}$ of the substring derived from $u^{\prime}$.
    \end{enumerate}
\end{theorem}
\begin{proof}
    $[p, p+\alpha_{3}] \in \Delta(u)$ follows from Lemma~\ref{lem:IA_maximal_lemma}. 
    Since $T[p..p+\alpha_{3}] = T[x..x+\alpha_{3}]$, 
    Corollary~\ref{cor:capture_gamma_corollary} shows that the following two statements hold: 
    \begin{itemize}
        \item the height $h^{\prime}$ of node $u^{\prime}$ is $h$;
        \item $|[p, \gamma]| = |[x, \gamma^{\prime}]|$ for the starting position $\gamma^{\prime}$ of the substring derived from $u^{\prime}$. 
    \end{itemize}
    Let $([p^{\prime}, q^{\prime}], [\ell^{\prime}, r^{\prime}])$ be the interval attractor $I(u^{\prime})$ associated with node $u^{\prime}$. 
    We already proved $h^{\prime} = h$ and $|[p, \gamma]| = |[x, \gamma^{\prime}]|$, 
    and hence, Theorem~\ref{theo:IA_SYNC_X} holds if 
    the following four statements holds. 
    \begin{enumerate}[label=\textbf{(\Alph*)}]
        \item $p^{\prime} = x$; 
        \item $r^{\prime} = x + \alpha_{3}$; 
        \item $q^{\prime} = p^{\prime} + \alpha_{1}$; 
        \item $\ell^{\prime} = p^{\prime} + \alpha_{2}$.
    \end{enumerate}    

    \textbf{Proof of statement (A).}
    We prove $p^{\prime} = x$ by contradiction. 
    We assume that $p^{\prime} \neq x$. 
    Then, $p^{\prime} < x$ holds 
    because $p^{\prime} \leq x$ follows from the definition of interval attractor. 
    $[x-1, x + \alpha_{3}] \in \Delta(u^{\prime})$ follows from Lemma~\ref{lem:IA_maximal_lemma}. 
    Let $u_{1} \in \mathcal{U}$ be a node such that $[p-1, p + \alpha_{3}] \in \Delta(u_{1})$. 
    We can apply Corollary~\ref{cor:capture_gamma_corollary} 
    to the two substrings $T[p-1..p + \alpha_{3}]$ and $T[x-1..x + \alpha_{3}]$ 
    because $T[p-1..p + \alpha_{3}] = T[x-1..x + \alpha_{3}]$. 
    Since $[p-1, p + \alpha_{3}] \in \Delta(u_{1})$ and $[x-1, x + \alpha_{3}] \in \Delta(u^{\prime})$, 
    this corollary shows that 
    the height of $u_{1}$ is $h^{\prime}$, and the node derives a substring starting at position $\gamma$ on $T$. 
    Since $h = h^{\prime}$, 
    this node $u_{1}$ is $u$. 
    In this case, $[p-1..p + \alpha_{3}] \in \Delta(u)$ holds. 
    On the other hand, $[p-1..p + \alpha_{3}] \not \in \Delta(u)$ follows from the definition of interval attractor. 
    The two facts $[p-1..p + \alpha_{3}] \in \Delta(u)$ and $[p-1..p + \alpha_{3}] \not \in \Delta(u)$ yield a contradiction. 
    Therefore, $p^{\prime} = x$ must hold. 
    
    \textbf{Proof of statement (B).}
    Statement (B) can be proved using the same approach used to prove statement (A). 

    \textbf{Proof of statement (C).}
    We prove $q^{\prime} = p^{\prime} + \alpha_{1}$ by contradiction. 
    We assume that $q^{\prime} \neq p^{\prime} + \alpha_{1}$ does not hold. 
    Then, $q^{\prime} > p^{\prime} + \alpha_{1}$ holds 
    because $q^{\prime} >= p^{\prime} + \alpha_{1}$ follows from the definition of interval attractor. 
    Corollary~\ref{cor:IA_exist_corollary} shows that 
    there exists an integer $e \in [\ell, r]$ 
    satisfying $[q, e] \in \Delta(u^{\prime})$. 
    $[p^{\prime} + \alpha_{1}+1, e] \in \Delta(u^{\prime})$ follows from Lemma~\ref{lem:IA_maximal_lemma}. 
    Let $u_{1} \in \mathcal{U}$ be a node such that $[p + \alpha_{1}+1, e] \in \Delta(u_{1})$. 
    Then, we can prove $u_{1} = u$ by the same approach used in the proof of statement (A). 
    In this case, $[p + \alpha_{1}+1, e] \in \Delta(u)$ holds. 
    On the other hand, $[p + \alpha_{1}+1, e] \not \in \Delta(u)$ follows from the definition of interval attractor. 
    Therefore, $q^{\prime} = p^{\prime} + \alpha_{1}$ must hold. 

    \textbf{Proof of statement (D).}
    Statement (D) can be proved using the same approach used to prove statement (C). 
\end{proof}

The following lemma states between two substrings $T[\gamma..r+1]$ and $T[\gamma^{\prime}..r^{\prime}+1]$ 
for the two interval attractors $I(u)$ and $I(u^{\prime})$ stated in Corollary~\ref{cor:capture_gamma_corollary}. 

\begin{lemma}\label{lem:proper_prefix_lemma}
Consider the two suffixes $T[\SA[i]..n]$ and $T[\SA[i']..n]$ of Corollary~\ref{cor:capture_gamma_corollary}. 
If $T[\SA[i]..r+1] \neq T[\SA[i^{\prime}]..r^{\prime}+1]$, 
then neither $T[\gamma..r+1]$ nor $T[\gamma^{\prime}..r^{\prime}+1]$ is a proper prefix of the other 
(i.e., $|\lcp(T[\gamma..r+1], T[\gamma^{\prime}..r^{\prime}+1])| < \min \{ |[\gamma, r+1]|, |[\gamma^{\prime}, r^{\prime}+1]| \}$). 
\end{lemma}
\begin{proof}
    We apply Corollary~\ref{cor:capture_gamma_corollary} to the two suffixes $T[\SA[i]..n]$ and $T[\SA[i^{\prime}]..n]$. 
    Then, this corollary shows that the two nodes $u$ and $u^{\prime}$ have the same height $h$ in the derivation tree of RLSLP.  

    We prove Lemma~\ref{lem:proper_prefix_lemma} by contradiction. 
    We assume that Lemma~\ref{lem:proper_prefix_lemma} does not hold. 
    Since $T[\gamma..r+1] \neq T[\gamma^{\prime}..r^{\prime}+1]$, 
    either of the following two conditions is satisfied: 
    (A) $T[\gamma..r+1]$ is a proper prefix of $T[\gamma^{\prime}..r^{\prime}+1]$; 
    (B) $T[\gamma^{\prime}..r^{\prime}+1]$ is a proper prefix of $T[\gamma..r+1]$. 
    
    For condition (A), 
    we obtain $\gamma^{\prime} + |[\gamma, r+1]| - 1 \in [\SA[i^{\prime}] + |P| - 1, r^{\prime}]$, 
    and two substrings $T[\SA[i]..r+1]$ and $T[\SA[i^{\prime}]..\gamma^{\prime} + |[\gamma, r+1]| - 1]$ are the same string. 
    Consider two interval attractors $I(u_{1})$ and $I(u_{2})$ such that 
    $[\SA[i], r+1] \in \Delta(u_{1})$ and $[\SA[i^{\prime}], \gamma^{\prime} + |[\gamma, r+1]| - 1] \in \Delta(u_{2})$, respectively. 
    We can apply Lemma~\ref{lem:IA_maximal_lemma} two intervals $[\SA[i^{\prime}], \SA[i^{\prime}] + |P| - 1]$ 
    and $[\SA[i^{\prime}], \gamma^{\prime} + |[\gamma, r+1]| - 1]$. 
    Then, this lemma shows that $u^{\prime} = u_{2}$. 
    Therefore, 
    the height of $u_{2}$ is $h$.

    On the other hand, Lemma~\ref{lem:interval_extension_propertyX}~\ref{enum:interval_extension_propertyX:right} shows that 
    the height of $u_{1}$ is at least $h+1$. 
    Since $T[\SA[i]..r+1] = T[\SA[i^{\prime}]..\gamma^{\prime} + |[\gamma, r+1]| - 1]$, 
    we can apply Corollary~\ref{cor:capture_gamma_corollary} to the two substrings $T[\SA[i]..r+1]$ and $T[\SA[i^{\prime}]..\gamma^{\prime} + |[\gamma, r+1]| - 1]$. 
    Then, this corollary shows that the node $u_{1}$ and $u_{2}$ have the same height in the derivation tree of RLSLP, 
    and hence, 
    the height of $u_{2}$ is at least $h+1$. 
    For the height $h_{2}$ of node $u_{2}$, 
    the two facts $h_{2} = h$ and $h_{2} \geq h+1$ yield a contradiction. 
    
    For condition (B), 
    we show that there exists a contradiction using the same approach as for condition (A). 
    Therefore, Lemma~\ref{lem:proper_prefix_lemma} must hold.     
\end{proof}

The lexicographical order between suffixes in the sa-interval of a string can be determined by comparing substrings associated with interval attractors as follows.

\begin{theorem}\label{theo:sa_intv_formulaX}
    For any two suffixes $T[\SA[i]..n]$ and $T[\SA[i']..n]$ in the sa-interval of a string $P (|P| \geq 2)$ on the suffix array $\SA$ of $T$, 
    there exist nodes $u, u' \in \mathcal{U}$ such that $[\SA[i], \SA[i] + |P|-1] \in \Delta(u)$ and $T[\SA[i']..(\SA[i'] + |P| - 1)] \in \Delta(u')$.
    Also, there exist interval attractors $I(u)=([p,q], [\ell, r])$ and $I(u')=([p',q'], [\ell', r'])$ for such nodes $u, u' \in \mathcal{U}$. 
    If $T[\SA[i]..r+1] \neq T[\SA[i^{\prime}]..r^{\prime}+1]$, 
    then the relation $i < i^{\prime} \Leftrightarrow T[\gamma..r+1] \prec T[\gamma^{\prime}..r^{\prime}+1]$ holds.
\end{theorem}
\begin{proof}
    $\SA[i] \in [p, q]$ and $\SA[i] + |P| - 1 \in [\ell, r]$ follow from the definition of interval attractor.
    Similarly, 
    $\SA[i^{\prime}] \in [p^{\prime}, q^{\prime}]$ and $\SA[i^{\prime}] + |P| - 1 \in [\ell^{\prime}, r^{\prime}]$ hold. 
    We apply Corollary~\ref{cor:capture_gamma_corollary} to the two suffixes $T[\SA[i]..n]$ and $T[\SA[i^{\prime}]..n]$. 
    Then, this corollary shows that 
    (i) the two nodes $u$ and $u^{\prime}$ have the same height $h$ in the derivation tree of RLSLP, 
    and (ii) $T[\SA[i]..\gamma-1] = T[\SA[i^{\prime}]..\gamma^{\prime}-1]$ and $T[\gamma..\SA[i] + |P| -1] = T[\gamma^{\prime}..\SA[i^{\prime}] + |P| -1]$. 

    Consider the longest prefix $Z$ between two strings $T[\gamma..r+1]$ and $T[\gamma^{\prime}..r^{\prime}+1]$ 
    (i.e., $Z = \lcp(T[\gamma..r+1], T[\gamma^{\prime}..r^{\prime}+1])$). 
    Since $T[\gamma..r+1] \neq T[\gamma^{\prime}..r^{\prime}+1]$, 
    Lemma~\ref{lem:proper_prefix_lemma} shows that neither $T[\gamma..r+1]$ nor $T[\gamma^{\prime}..r^{\prime}+1]$ is a proper prefix of the other 
    (i.e., $Z \neq T[\gamma..r+1]$ and $Z \neq T[\gamma^{\prime}..r^{\prime}+1]$). 

    The proof of Theorem~\ref{theo:sa_intv_formulaX} is as follows. 

    \textbf{Proof of $i < i^{\prime} \Rightarrow T[\gamma..r+1] \prec T[\gamma^{\prime}..r^{\prime}+1]$.}    
    $T[\gamma..r+1] \prec T[\gamma^{\prime}..r^{\prime}+1]$ holds if $T[\gamma + |Z|] < T[\gamma^{\prime} + |Z|]$. 
    We show that $T[\gamma + |Z|] < T[\gamma^{\prime} + |Z|]$. 
    Consider two suffixes $T[\SA[i]..n]$ and $T[\SA[i^{\prime}]..n]$. 
    Let $Z^{\prime}$ be the longest prefix between the two suffixes, 
    then $T[\SA[i] + |Z|] < T[\SA[i^{\prime}] + |Z|]$ follows from $i < i^{\prime}$. 
    The two characters $T[\SA[i] + |Z|]$ and $T[\SA[i^{\prime}] + |Z|]$ correspond with $T[\gamma + |Z|]$ and $T[\gamma^{\prime} + |Z|]$, respectively. 
    This is because $Z^{\prime} = T[\SA[i]..\gamma-1] \cdot Z = T[\SA[i^{\prime}]..\gamma^{\prime}-1] \cdot Z$ holds. 
    We obtain $T[\gamma + |Z|] < T[\gamma^{\prime} + |Z|]$, 
    and hence, $i < i^{\prime} \Rightarrow T[\gamma..r+1] \prec T[\gamma^{\prime}..r^{\prime}+1]$. 

    \textbf{Proof of $i < i^{\prime} \Leftarrow T[\gamma..r+1] \prec T[\gamma^{\prime}..r^{\prime}+1]$.}
    $i < i^{\prime}$ holds if $T[\gamma + |Z|] < T[\gamma^{\prime} + |Z|]$. 
    $T[\gamma + |Z|] < T[\gamma^{\prime} + |Z|]$ follows from $T[\gamma..r+1] \prec T[\gamma^{\prime}..r^{\prime}+1]$ 
    and $|Z| < \min \{ |[\gamma, r+1]|, |[\gamma^{\prime}, r^{\prime}+1]| \}$ (Lemma~\ref{lem:proper_prefix_lemma}). 
    Therefore, we obtain $i < i^{\prime} \Leftarrow T[\gamma..r+1] \prec T[\gamma^{\prime}..r^{\prime}+1]$. 

\end{proof}

The following corollary follows from Theorem~\ref{theo:sa_intv_formulaX} and Lemma~\ref{lem:proper_prefix_lemma}. 
\begin{corollary}\label{cor:sa_intv_corollary}
    Consider the two suffixes $T[\SA[i]..n]$ and $T[\SA[i']..n]$ of Theorem~\ref{theo:sa_intv_formulaX}. 
    For the sa-interval $[\eta_{1}, \eta^{\prime}_{1}]$ of substring $T[\SA[i]..\SA[i] + r+1]$, 
    $T[\SA[i]..r+1] = T[\SA[i^{\prime}]..r^{\prime}+1] \Leftrightarrow i^{\prime} \in [\eta_{1}, \eta^{\prime}_{1}]$. 
\end{corollary}
    
\subsection{RSC and RSS queries}\label{subsec:simplified_rss_rsc}
\begin{figure}[t]
 \begin{center}
		\includegraphics[scale=0.7]{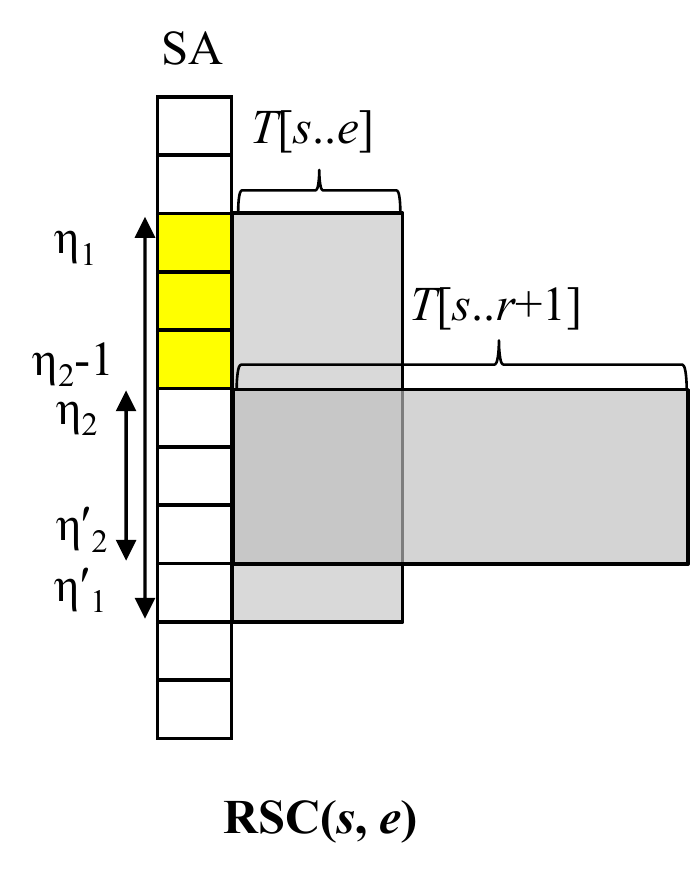}
		\includegraphics[scale=0.7]{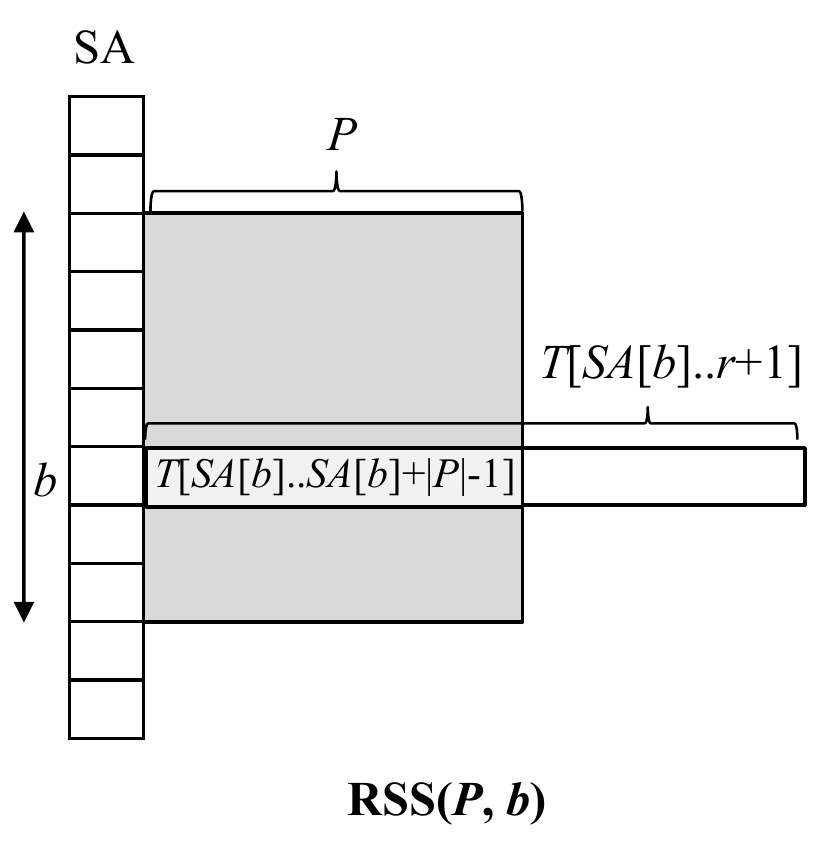}

	  \caption{ 
    An illustration of RSC query $\RSCQ(s, e)$ (left) and RSS query $\RSSQ(P, b)$ (right). 
    }
\label{fig:rsc_rss}
 \end{center}
\end{figure}

\emph{Restricted suffix count} (RSC) and \emph{restricted suffix search} (RSS) queries are ones that count and search suffixes associated with interval attractors.
%Let $u \in \mathcal{U}$ be a node whose associated partition $\Delta(u)$ includes an interval $[s,e] \in \Delta$. 
%Let $I(u) = ([p,q],[\ell,r])$ be the interval attractor associated with $u \in \mathcal{U}$.
\begin{definition}[RSC query]\label{def:simplified_rsc_query}
    For a given interval $[s, e] \in \Delta$, there exists a node $u \in \mathcal{U}$ such that $[s, e] \in \Delta(u)$.
    Let $I(u)=([p,q], [\ell, r])$ be the interval attractor associated with $u$. 
    Consider the sa-intervals $[\eta_{1}, \eta^{\prime}_{1}]$ and $[\eta_{2}, \eta^{\prime}_{2}]$ of two substrings $T[s..e]$ and $T[s..r+1]$ on the suffix array $\SA$ of $T$, respectively, both starting at the same position $s$.
    For the given interval $[s, e]$, $\RSCQ(s, e) = \eta_{2} - \eta_{1}$, 
    which represents the number of suffixes within interval $[\eta_{1}, \eta_{2}-1]$ on the suffix array. 
\end{definition}
The following corollary follows from the definitions RSC query and suffix array. 
\begin{corollary}\label{cor:rsc_corollary}
    For the interval attractor $I(u)=([p,q], [\ell, r])$ of Definition~\ref{def:simplified_rsc_query}, 
    $\RSCQ(s, e) = |\{ i \in [1, n] \mid T[s..e] \preceq T[i..n] \prec T[s..(\min \{ r+1, n \})] \}|$.
\end{corollary}

The figure on the left in Figure~\ref{fig:rsc_rss} illustrates the RSC query.
The RSC query computes the number of suffixes that have the prefix $T[s,e]$ and is lexicographically smaller than $T[s..r+1]$, which is represented as positions colored yellow in SA in the figure. 

\begin{definition}[RSS query]
    RSS query $\RSSQ(P, b)$ takes as input a pair of a substring $P$ of $T$ with $|P| \geq 2$ and a position $b$ in the sa-interval of $P$ on $\SA$. 
    For the interval $[\SA[b], \SA[b]+|P|-1]$ with the prefix $T[\SA[b]..\SA[b]+|P|-1]$ of the suffix $T[\SA[b]..n]$, 
    there exists an interval attractor $I(u) = ([p,q], [\ell,r])$ for some $u \in \mathcal{U}$ such that $[\SA[b],\SA[b]+|P|-1] \in \Delta(u)$.
    Then, $\RSSQ(P, b)$ returns $T[\SA[b]..r+1]$.
\end{definition}

The figure on the right in Figure~\ref{fig:rsc_rss} illustrates the RSS query.

The following theorems show that RSC query can be answered by counting interval attractors satisfying certain conditions. 
\begin{theorem}\label{theo:rsc_query_sub_formula_X}
    $\RSCQ(s, e)$ for an input $[s,e] \in \Delta$ can be computed as the number of interval attractors $I(u^{\prime})$ that satisfy the following three conditions.
    \begin{enumerate}[label=\textbf{(\roman*)}]   
        \item the height of $u^{\prime}$ is the same as that of $u$ in the derivation tree.
        \item $T[p^\prime-1..\gamma^\prime-1]$ includes $T[s..\gamma-1]$ as a proper suffix.
        \item $T[\gamma^\prime..r^\prime+1]$ includes $T[\gamma..e]$ as a proper prefix.
        \item $T[\gamma^\prime..r^\prime+1]$ is lexicographically smaller than $T[\gamma..r+1]$.
    \end{enumerate}
\end{theorem}
\begin{proof}
Let $\Psi$ be the set of interval attractors satisfying the four conditions of Theorem~\ref{theo:rsc_query_sub_formula_X}. 
Then, Theorem~\ref{theo:rsc_query_sub_formula_X} holds if 
$\eta_{2} - \eta_{1} = |\Psi|$. 
Here, $[\eta_{1}, \eta^{\prime}_{1}]$ and $[\eta_{2}, \eta^{\prime}_{2}]$ are the sa-intervals 
of $T[s..e]$ and $T[s..r+1]$, respectively, defined in Definition~\ref{def:simplified_rsc_query}. 
We prove $\eta_{2} - \eta_{1} = |\Psi|$ by $\eta_{2} - \eta_{1} \leq |\Psi|$ and 
$\eta_{2} - \eta_{1} \geq |\Psi|$. 
The two inequalities are proven as follows. 

\paragraph{Proof of $\eta_{2} - \eta_{1} \leq |\Psi|$.}
Consider a suffix $T[\SA[i]..n]$ starting at a position in interval $[\eta_{1}, \eta_{2}-1]$ on 
the suffix array of $T$ (i.e., $i \in [\eta_{1}, \eta_{2}-1]$). 
There exists a node $u^{\prime} \in \mathcal{U}$ such that $[\SA[i], \SA[i] + |[s, e]| - 1] \in \Delta(u^{\prime})$, 
and an interval attractor $I(u^{\prime}) = ([p^{\prime}, q^{\prime}], [\ell^{\prime}, r^{\prime}])$ is associated with the node $u^{\prime}$. 
Then, Theorem~\ref{theo:SA_INTV_INTV_ATTR} shows that 
This interval attractor $I(u^{\prime})$ satisfies Conditions (i)-(iii) of Theorem~\ref{theo:rsc_query_sub_formula_X}. 

Next, we show that $T[s..r+1] \neq T[\SA[i]..r^{\prime}+1]$ by contradiction. 
We assume that $T[s..r+1] = T[\SA[i]..r^{\prime}+1]$. 
Then, the suffix $T[\SA[i]..n]$ starts at a position in interval $[\eta_{2}, \eta^{\prime}_{2}]$ on 
the suffix array of $T$ (i.e., $i \in [\eta_{2}, \eta^{\prime}_{2}]$). 
The two facts $i \in [\eta_{2}, \eta^{\prime}_{2}]$ and $i < \eta_{2}$ yield a contradiction. 
Therefore, $T[s..r+1] \neq T[\SA[i]..r^{\prime}+1]$ must hold. 

We show that $I(u^{\prime})$ satisfies the fourth condition of Theorem~\ref{theo:rsc_query_sub_formula_X}. 
Let $i^{\prime}$ be the position on the suffix array satisfying $\SA[i^{\prime}] = s$. 
Then, $i < i^{\prime}$ follows from $i < \eta_{2}$ and $i^{\prime} \in [\eta_{2}, \eta^{\prime}_{2}]$.
Since $T[s..r+1] \neq T[\SA[i]..r^{\prime}+1]$, 
we can apply Theorem~\ref{theo:sa_intv_formulaX} to the two substrings $T[s..e]$ and $T[\SA[i]..\SA[i] + |[s, e]| - 1]$. 
Then, Theorem~\ref{theo:sa_intv_formulaX} shows that $T[\gamma^\prime..r^\prime+1] \prec T[\gamma..r+1]$ holds. 
Therefore, $I(u^{\prime})$ satisfies the third condition of Theorem~\ref{theo:rsc_query_sub_formula_X}. 

We showed that $I(u^{\prime})$ satisfies the four conditions of Theorem~\ref{theo:rsc_query_sub_formula_X} (i.e., $I(u^{\prime}) \in \Psi$). 
Therefore, we obtain $\eta_{2} - \eta_{1} \leq |\Psi|$. 

\paragraph{Proof of $\eta_{2} - \eta_{1} \geq |\Psi|$.}
Consider an interval attractor $I(u^{\prime}) \in \Psi$, 
and let $i$ be the position on the suffix array satisfying $\SA[i] = \gamma^\prime - |[s, \gamma-1]|$. 
Then, Theorem~\ref{theo:SA_INTV_INTV_ATTR} shows that 
the position $i$ is contained in the sa-interval $[\eta_{1}, \eta^{\prime}_{1}]$ of $T[s..e]$. 
Therefore, $\eta_{2} - \eta_{1} \geq |\Psi|$ holds if $i \in [\eta_{1}, \eta_{2}-1]$. 

We prove $i < i^{\prime}$ for the position $i^{\prime}$ used in the proof of $\eta_{2} - \eta_{1} \leq |\Psi|$. 
$T[s..r+1] \neq T[\SA[i]..r^{\prime}+1]$ and $T[\gamma^\prime..r^\prime+1] \prec T[\gamma..r+1]$ follow from the four conditions of Theorem~\ref{theo:rsc_query_sub_formula_X}. 
Since $T[s..r+1] \neq T[\SA[i]..r^{\prime}+1]$, 
we can apply Theorem~\ref{theo:sa_intv_formulaX} to the two substrings $T[s..e]$ and $T[\SA[i]..\SA[i] + |[s, e]| - 1]$. 
Then, this theorem shows that $i < i^{\prime}$. 

We prove $i \not \in [\eta_{2}, i^{\prime}]$ by contradiction. 
We assume that $i \in [\eta_{2}, i^{\prime}]$ holds. 
Then, string $T[\gamma^{\prime}..n+1]$ include two strings $T[\gamma..r+1]$ and $T[\gamma^{\prime}..r^{\prime}+1]$ as prefixes. 
Since $T[\gamma..r+1] \neq T[\gamma^{\prime}..r^{\prime}+1]$, 
this fact indicates that string $T[\gamma..r+1]$ is a proper prefix of string $T[\gamma^{\prime}..r^{\prime}+1]$, 
or $T[\gamma^{\prime}..r^{\prime}+1]$ is a proper prefix of string $T[\gamma..r+1]$. 
On the other hand, Lemma~\ref{lem:proper_prefix_lemma} shows that 
neither $T[\gamma..r+1]$ nor $T[\gamma^{\prime}..r^{\prime}+1]$ is a proper prefix of the other. 
There exists a contradiction. Therefore, $i \not \in [\eta_{2}, i^{\prime}]$ must hold.

Since $i \in [\eta_{1}, i^{\prime}]$, 
either $i \in [\eta_{1}, \eta_{2}-1]$ or $i \in [\eta_{2}, i^{\prime}]$ holds. 
We already proved $i \not \in [\eta_{2}, i^{\prime}]$, 
and hence, 
we obtain $i \in [\eta_{1}, \eta_{2}-1]$. 
Therefore, we obtain $\eta_{2} - \eta_{1} \geq |\Psi|$. 
\end{proof}

The following theorem shows that RSS query can be answered using the RSC query. 
\begin{theorem}\label{theo:rss_query_sub_formula_X}
    For an input pair $(P, b)$ of $\RSSQ(P, b)$, 
    let $[\eta_{1}, \eta^{\prime}_{1}]$ be the sa-interval of $P$ on $\SA$ and let $b' \in [\eta_{1}, \eta^{\prime}_{1}]$.
    For any position $b' \in [\eta_{1}, \eta^{\prime}_{1}]$, 
    there exists the interval $[\SA[b^{\prime}], \SA[b^{\prime}]+|P|-1]$ with the prefix $T[\SA[b^{\prime}]..\SA[b^{\prime}]+|P|-1]$ of the suffix $T[\SA[b^{\prime}]..n]$. 
    Additionally there exists an interval attractor $I(u^{\prime}) = ([p^{\prime},q^{\prime}], [\ell^{\prime},r^{\prime}])$ for some $u^{\prime} \in \mathcal{U}$ such that $[\SA[b^{\prime}],\SA[b^{\prime}]+|P|-1] \in \Delta(u^{\prime})$.
    Then, $\RSSQ(P, b) = T[\SA[b^{\prime}]..r^{\prime}+1]$ for $b' \in [\eta_{1}, \eta^{\prime}_{1}]$ such that the interval $[\SA[b^{\prime}], \SA[b^{\prime}]+|P|-1]$ satisfies the following two conditions: 
    \begin{enumerate}[itemsep=0pt, parsep=0pt,before=\vspace{-0.3\baselineskip}, after=\vspace{-0.3\baselineskip}, label=\textbf{(\roman*)}]    
        \item $\eta_{1} + \RSCQ(\SA[b^{\prime}], \SA[b^{\prime}]+|P|-1) \leq b$;
        \item $\eta_{1} + \RSCQ(\SA[b^{\prime}], \SA[b^{\prime}]+|P|-1) + |[\eta_{2}, \eta^{\prime}_{2}]| - 1 \geq b$, where $[\eta_{2}, \eta^{\prime}_{2}]$ is the sa-interval of $T[\SA[b^{\prime}]..r^{\prime}+1]$.
    \end{enumerate}
\end{theorem}
\begin{proof}
    Consider the interval $[\SA[b^{\prime}], \SA[b^{\prime}]+|P|-1]$ satisfying the two conditions of Theorem~\ref{theo:rss_query_sub_formula_X}. 
    From the definition of RSC query, 
    $\RSCQ(\SA[b^{\prime}]$, $\SA[b^{\prime}]+|P|-1)$ returns the length of interval $[\eta_{1}, \eta_{2} - 1]$ 
    for the sa-interval $[\eta_{2}, \eta^{\prime}_{2}]$ of substring $T[\SA[b^{\prime}]..r^{\prime}+1]$. 
    This fact indicates that     
    $\eta_{2} = \eta_{1} + \RSCQ(\SA[b^{\prime}]$, $\SA[b^{\prime}]+|P|-1)$ 
    and $\eta^{\prime}_{2} = \eta_{1} + \RSCQ(\SA[b^{\prime}], \SA[b^{\prime}]+|P|-1) + |[\eta_{2}, \eta^{\prime}_{2}]| - 1$ hold. 
    Therefore, the sa-interval of $T[\SA[b^{\prime}]..r^{\prime}+1]$ contains position $b$. 
    
    For the substring $T[b..r+1]$ returned by RSS query $\RSSQ(P, b)$, 
    let $[\eta_{3}, \eta^{\prime}_{3}]$ be the sa-interval of $T[b..r+1]$. 
    Since $b \in [\eta_{2}, \eta^{\prime}_{2}]$, 
    Corollary~\ref{cor:sa_intv_corollary} shows that 
    $[\eta_{3}, \eta^{\prime}_{3}] = [\eta_{2}, \eta^{\prime}_{2}]$ 
    and $T[b..r+1] = T[\SA[b^{\prime}]..r^{\prime}+1]$. 
    Therefore, we obtain Theorem~\ref{theo:rss_query_sub_formula_X}. 
\end{proof}

In the worst case, 
Lemma~\ref{lem:non_comp_IA_size} shows that $O(n^2)$ interval attractors are created for input string $T$ of length $n$. Since verifying whether an interval attractor satisfies the four conditions of Theorem~\ref{theo:rsc_query_sub_formula_X} straightforwardly takes $O(n)$ time and there are $O(n^2)$ interval attractors, an RSC query can be answered in $O(n^3)$ time. Similarly, 
since straightforwardly verifying whether an interval attractor satisfies the two conditions of Theorem~\ref{theo:rss_query_sub_formula_X} takes $O(n^3)$ time and there are $O(n^2)$ interval attractors, an RSS query can be answered in $O(n^5)$ time.

For supporting RSC and RSS queries with $O(\polylog (n))$ time and expected $\delta$-optimal space, 
RSC and RSS queries are decomposed into seven and eight distinct subqueries, respectively, each of which can be efficiently solved on a grid constructed by mapping interval attractors. The solution to the original query is obtained by summing the results of these subqueries. Details of the subqueries are listed in Table~\ref{table:RSC_query_result} and 
Table~\ref{table:RSS_query_result} and elaborated upon in Section~\ref{sec:RSC_query} and Section~\ref{sec:RSS_query}, respectively.

\subsection{SA and ISA Queries}\label{subsec:simplified_sa_and_isa_queries}
\begin{figure}[t]
 \begin{center}
		\includegraphics[scale=0.7]{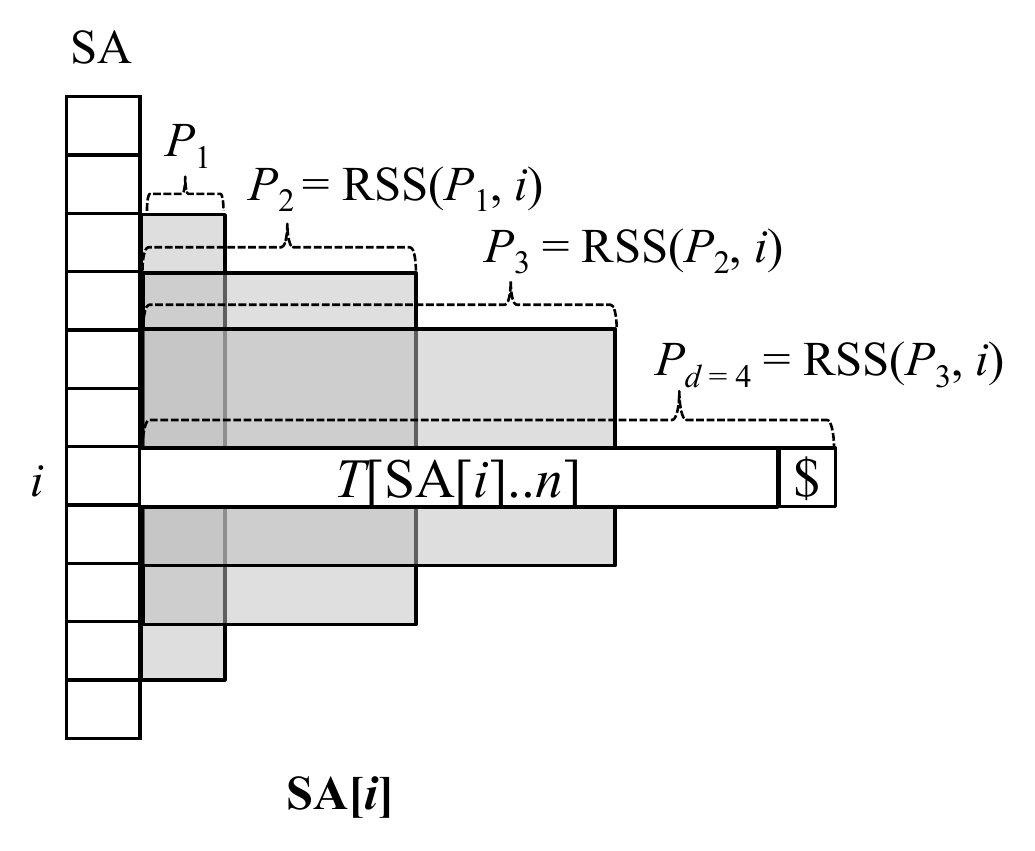}
		\includegraphics[scale=0.7]{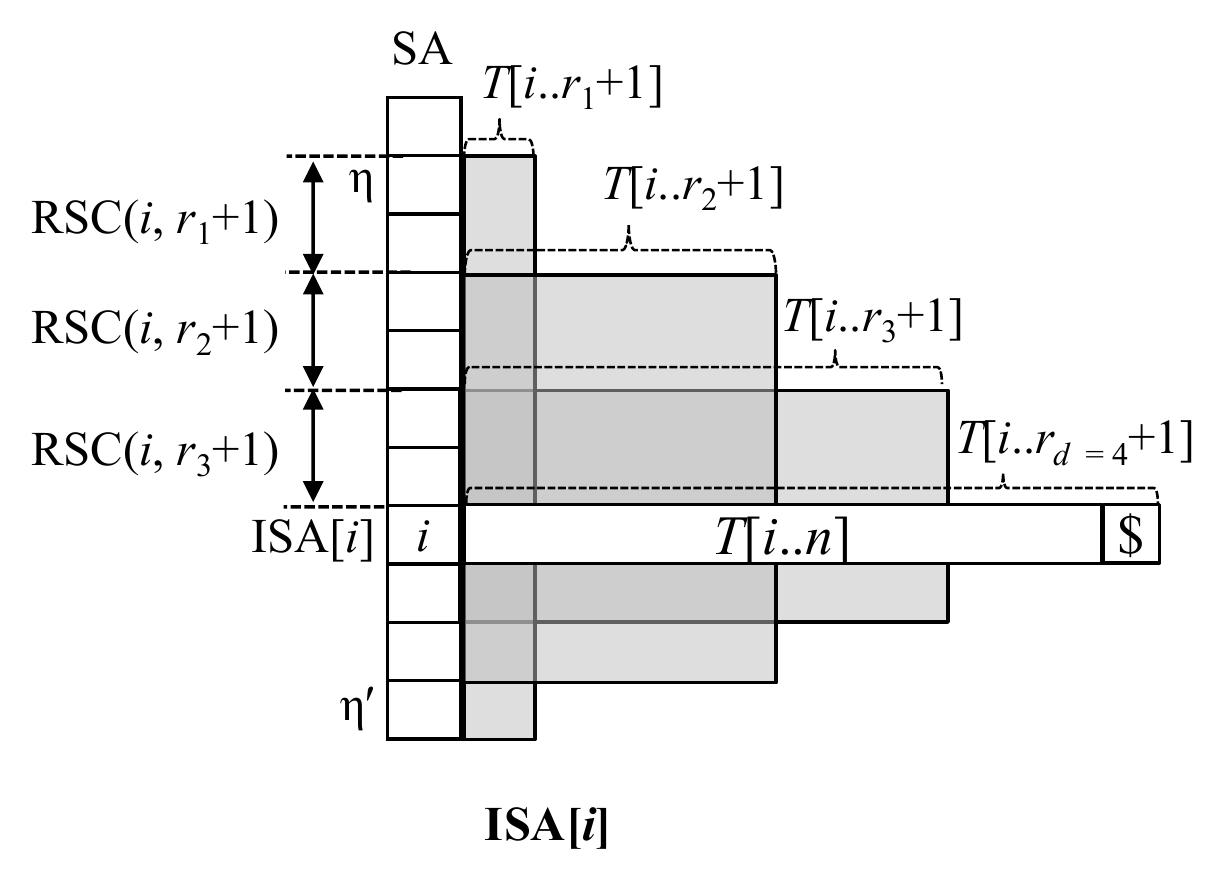}

	  \caption{ 
    An illustration of Theorem~\ref{theo:sa_query} (left) and Theorem~\ref{theo:isa_query} (right). 
	  }
\label{fig:sa_isa}
 \end{center}
\end{figure}

SA and ISA queries can be answered using RSS and RSC queries as follows.  
\begin{theorem}\label{theo:sa_query}
    Let $i \in [1,n]$ be an input to the SA query $\SA[i]$ except for $\SA[i] = n$.
    Define $P_{j}$ for $j \geq 1$ recursively as follows:
    $P_{1} = T[\SA[i]..\SA[i]+1]$; 
    $P_{j} = \RSSQ(P_{j-1}, i)$ for $j \geq 2$. 
    Then, $\SA[i] = n - |P_{d}| + 2$ for some $d \geq 1$ such that the last character of string $P_{d}$ is $\$$. 
\end{theorem}
\begin{proof}
    From the definition of RSS query, 
    the following three statements hold: 
    \begin{enumerate}[label=\textbf{(\roman*)}]
    \item $|P_{2}| < |P_{3}| < \cdots < |P_{d}|$;
    \item $P_{2}, P_{3}, \ldots, P_{d}$ are prefixes of string $T[\SA[i]..n+1]$;
    \item $P_{d} = T[\SA[i]..n+1]$.
    \end{enumerate}
    Therefore, $\SA[i] = n - |P_{d}| + 2$ follows from $P_{d} = T[\SA[i]..n+1]$.     
\end{proof}

\begin{theorem}\label{theo:isa_query}
    Let $i \in [1,n-1]$ be an input to the ISA query $\ISA[i]$.
    Define $r_{j} \in [1, n]$ for $j \geq 1$ recursively as follows:     
    $r_{1} = i$; 
    for $j \geq 2$, 
    $r_{j} = r$ for the interval attractor $I(u) = ([p, q], [\ell, r])$ associated with the unique node $u \in \mathcal{U}$ such that $[i, r_{j-1} + 1] \in \Delta(u)$.  
    Then, for $d \geq 1$ such that $r_{d} = n$ and for the sa-interval $[\eta, \eta^{\prime}]$ of the substring $T[i..r_{1}+1]$, the ISA query is computed as $\ISA[i] = \eta + \sum_{j = 1}^{d-1} \RSCQ(i, r_{j}+1)$.    
\end{theorem}
\begin{proof}
    $r_{1} < r_{2} < \cdots < r_{d^{\prime}} = n$ follows from the definition of interval attractors. 
    Let $[\eta_{j}, \eta^{\prime}_{j}]$ be the sa-interval of string $T[i..r_{j}+1]$ 
    for each integer $j \geq 1$. 
    Here, $\eta = \eta_{1}$ holds. 
    $[\eta_{d}, \eta^{\prime}_{d}] = [\ISA[i], \ISA[i]]$ 
    because $[\eta_{d}, \eta^{\prime}_{d}]$ is the sa-interval of $T[i..n+1]$, 
    and $T[i..n+1] = T[i..n]\$$ (see Section~\ref{sec:preliminary}). 

    We obtain $\eta = \eta_{1} \leq \eta_{2} \leq \cdots \leq \eta^{\prime}_{d} = \ISA[i]$ 
    because each $T[i..r_{j}+1]$ is a prefix of $T[i..r_{j+1}+1]$ for all $j \in [1, d-1]$. 
    This fact indicates that 
    $\ISA[i] = \eta + \sum_{j = 1}^{d-1} (\eta_{j+1} - \eta_{j})$. 
    The integer $(\eta_{j+1} - \eta_{j})$ can be obtained from $\RSCQ(i, r_{j} + 1)$. 
    Therefore, we obtain $\ISA[i] = \eta + \sum_{j = 1}^{d-1} \RSCQ(i, r_{j}+1)$.  
\end{proof}

Theorem~\ref{theo:sa_query} shows that SA query $\SA[i]$ can be answered using $d-1$ RSS queries 
except for $\SA[i] = n$. 
The following lemma shows this integer $d$ can be bounded by $H+2$. 
\begin{lemma}\label{lem:d_bound}
    Consider $d$ strings $P_{1}, P_{2}, \ldots, P_{d}$ introduced in Theorem~\ref{theo:intro_sa_query} 
    for SA query $\SA[i]$. 
    Then, $d \leq H+2$. 
\end{lemma}
\begin{proof}
    For each integer $j \in [1, d-1]$, 
    string $P_{j}$ occurs in $T$ as substring $T[\SA[i]..\SA[i] + |P_{j}| - 1]$. 
    Let $I(u_j) = ([p_j, q_j], [\ell_j, r_j])$ be an interval attractor for node $u_j \in \mathcal{U}$ such that $[\SA[i], \SA[i] + |P_{j}| - 1] \in \Delta(u_j)$.
    Then, $\SA[i] + |P_{j}| - 1 > r_{j-1}$ holds for $j \in [2, d]$. 

    Let $h_{j}$ be the height of node $u_{j}$ in the derivation tree of RLSLP $\mathcal{G}^{R}$. 
    Lemma~\ref{lem:interval_extension_propertyX}~\ref{enum:interval_extension_propertyX:right} 
    shows that $0 \leq h_{1} < h_{2} < \cdots < h_{d-1} \leq H$ 
    because 
    $\SA[i] + |P_{j}| - 1 > r_{j-1}$ and $[\SA[i], \SA[i] + |P_{j}| - 1] \in \Delta(u_j)$ hold. 
    Therefore, we obtain $d \leq H+2$. 
\end{proof}

Theorem~\ref{theo:isa_query} shows that ISA query $\ISA[i]$ can be answered using 
$d-1$ RSC queries except for $i = n$. 
The following lemma shows this integer $d$ can be bounded by $H+1$. 
\begin{lemma}\label{lem:d_prime_bound}
    Consider the $d$ interval attractors $r_{1}, r_{2}, \ldots, r_{d}$ introduced in Theorem~\ref{theo:intro_isa_query} for ISA query $\ISA[i]$. 
    Then, $d \leq H+1$. 
\end{lemma}
\begin{proof}
    For an integer $j \in [1, d-1]$, 
    let $I(u) = ([p, q], [\ell, r])$ be the interval attractor associated with 
    a node $u \in \mathcal{U}$ of height $h$
    such that $[i, r_{j}+1] \in \Delta(u)$. 
    Then, $r_{j+1}$ is defined as $r$.     
    $r_{j}+1 \leq r$ follows from the definition of interval attractor. 
    Therefore, $r_{j} < r_{r+1}$. 

    Let $u^{\prime} \in \mathcal{U}$ of height $h^{\prime}$
    such that $[i, r_{j+1}+1] \in \Delta(u)$. 
    Then, $d \leq H+1$ holds if $h < h^{\prime}$ 
    because $0 \leq h < h^{\prime} \leq H$ for the height $H$ of the derivation tree.          
    Since $r = r_{j+1}$, 
    we can apply Lemma~\ref{lem:interval_extension_propertyX}~\ref{enum:interval_extension_propertyX:right} 
    to the two nodes $u$ and $u^{\prime}$. 
    Then, Lemma~\ref{lem:interval_extension_propertyX}~\ref{enum:interval_extension_propertyX:right} shows that $h < h^{\prime}$. 
    Therefore, $d \leq H+1$. 
\end{proof}

Figure~\ref{fig:intro_sa_isa} illustrates Theorem~\ref{theo:sa_query} and Theorem~\ref{theo:isa_query}.
Pseudo-codes for computing SA and ISA queries are presented in Algorithm~\ref{algo:light_sa_query} and Algorithm~\ref{algo:light_isa_query} in Section~\ref{sec:sa_and_isa_queries_with_optimal}. 
In the subsequent section, we present an overview of our compressed dynamic data structure for answering SA and ISA queries in $O(H^{3} \log^{2} n + H \log^{6} n)$ time and $O(H^{3} \log n + H \log^{4} n)$ time, respectively, 
using expected $O(\delta \log \frac{n \log \sigma}{\delta \log n} B)$ bits of space for machine word size $B = \Theta(\log n)$.

\section{Restricted Recompression}\label{sec:recompression}
%\section{Restricted Recompression and Run-Length SLP~(RLSLP)}\label{sec:recompression}
This section begins by reviewing \emph{restricted recompression}~\cite{9961143}, a randomized algorithm developed for constructing RLSLP. 
While the space for representing RLSLP is not $\delta$-optimal, 
we introduce a novel representation for RLSLP, termed \emph{RR-DAG}, which encodes it within a $\delta$-optimal space. 
Furthermore, we highlight a unique property of the restricted recompression algorithm, which guarantees the construction of RLSLPs with probability $1$. These RLSLPs have derivation trees with heights logarithmically proportional to the input string lengths with high probability. 
We leverage this property to answer SA and ISA queries in $\delta$-optimal space in subsequent sections.

\subsection{Restricted Recompression~\texorpdfstring{\cite{9961143,DBLP:journals/corr/KociumakaRRW13}}{}}\label{subsec:recompression}
\begin{figure}[t]
 \begin{center}
		\includegraphics[scale=0.8]{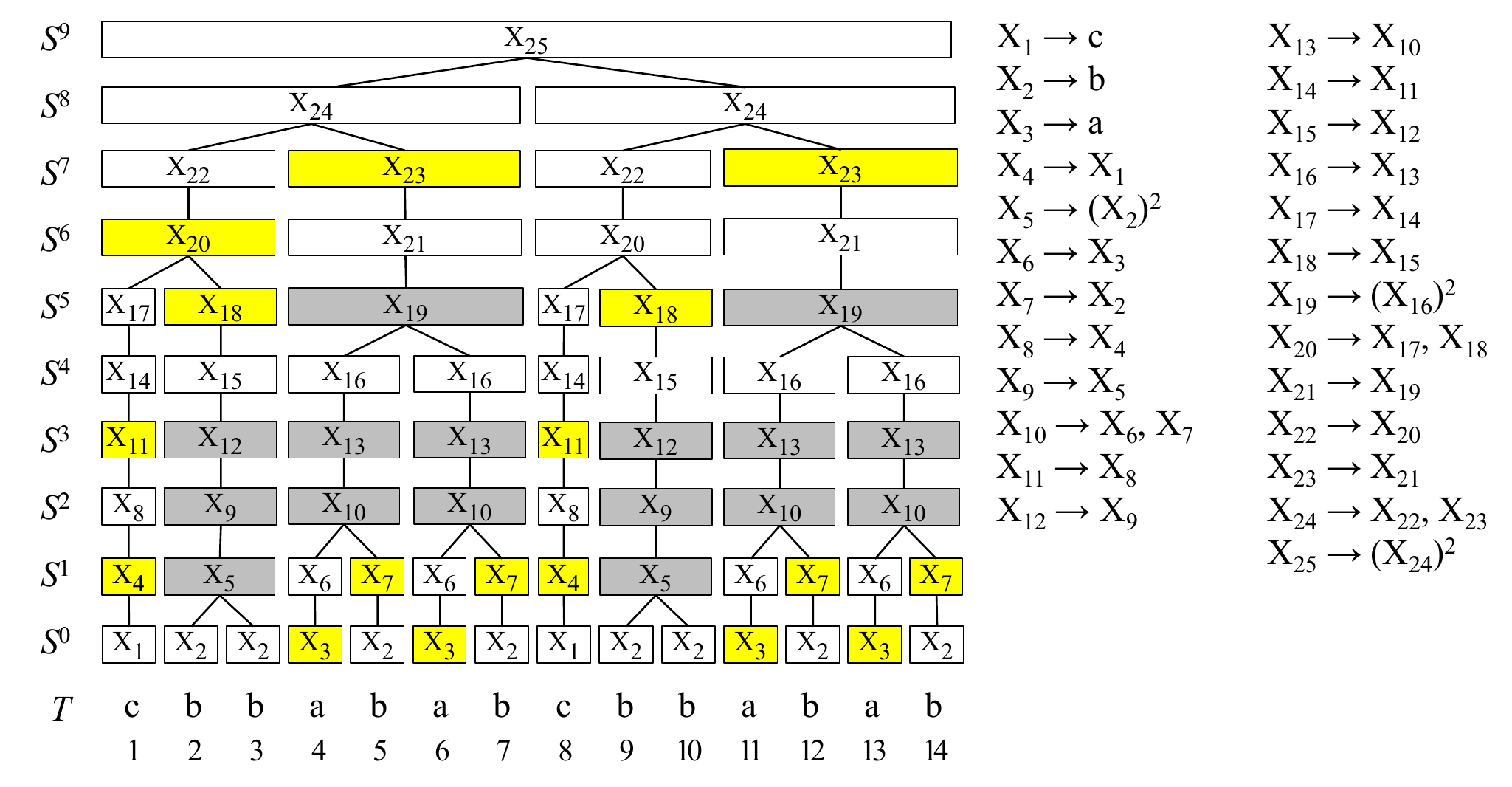}

	  \caption{ 
	  An illustration of (i) the derivation tree of an RLSLP $\mathcal{G}^{R}$ 
	   constructed using restricted recompression, and (ii) production rules of $\mathcal{G}^{R}$. 
        Each node is presented by a rectangle that encloses its corresponding nonterminal, denoted by $X_{i}$. 
        If the integer $1$ is assigned to the nonterminal $X_{i}$, then the rectangle is colored yellow.
        If the integer $-1$ is assigned to the nonterminal $X_{i}$, then the rectangle is colored gray.
        %The node's id, represented as $u_i$, is displayed in the rectangle's lower left corner.
	  }
\label{fig:restricted_recompression}
 \end{center}
\end{figure}

Restricted recompression as described in \cite{9961143,DBLP:journals/corr/KociumakaRRW13}, is a randomized algorithm designed to construct an RLSLP with a height-balanced derivation tree, which is defined in Section~\ref{subsec:run-length_slp}. 
As stated in Section~\ref{subsec:run-length_slp}, 
$S^h$ is a sequence of nonterminal symbols as the labels of consecutive nodes at height $h$ within the height-balanced derivation tree.  
These symbols are utilized to denote the labels of consecutive nodes at height $h$ within the derivation tree of $\mathcal{G}^{R}$.
Given an input string $T$ of length $n \geq 2$, it systematically builds the RLSLP from the bottom up, achieving this in an expected $O(n)$ time, 
as illustrated in Figure~\ref{fig:restricted_recompression}. 
Initially, restricted recompression constructs a sequence $S^{0}$ of nonterminals by transforming the characters of input string $T$ as follows. 
\begin{enumerate}[label=\textbf{Step} (\roman*)\\,leftmargin=1.5cm, align=left]
    \item For every character $c \in \Sigma$ that appears in $T$, a production rule of the form (i) $X_{i} \to c$ is established. Distinct nonterminals $X_j$ and $X_k$ (with $j \neq k$) are introduced for each pair of different characters $c$ and $c'$ where ($c \neq c'$).
    \item Each occurrence of the character $c$ in $T$ is substituted with the nonterminal $X_{i}$ that produces the character $c$.
\end{enumerate}
The aforementioned two steps yield a sequence $S^{0}$ of nonterminals with length $n$ situated at height $0$ in the height-balanced derivation tree.

Subsequently, for integer $h \geq 1$, the restricted recompression algorithm generates a sequence $S^{h}$ of nonterminals by substituting the nonterminals of sequence $S^{h-1}$ as follows:

\begin{enumerate}[label=\textbf{Step} (\roman*)\\,leftmargin=1.5cm, start=3, align=left]
    \item For a nonterminal $X_{i}$ appearing in sequence $S^{h-1}$, 
    it is assigned an integer from the set $\{ -1, 0, 1 \}$, denoted as $\assign(X_{i})$.
    Specifically, if the length of the string derived from $X_{i}$ is at most $\mu(h)$, where $\mu(h) = (8/7)^{\lceil h/2 \rceil - 1}$, 
    then $X_{i}$ is uniformly and randomly assigned an integer from $\{0, 1\}$ (i.e., $|\val(X_{i})| \leq \mu(h) \Leftrightarrow \assign(X_{i}) = 0$ or $1$). Otherwise $X_{i}$ is assigned the integer $-1$ (i.e., $|\val(X_{i})| > \mu(h) \Leftrightarrow \assign(X_{i}) = -1$).

    \item The sequence $S^{h-1}$ is partitioned into $m$ segments $S^{h-1}[\pi_{1}..(\pi_{2}-1)]$, $S^{h-1}[\pi_{2}..(\pi_{3}-1)]$, $\ldots$, $S^{h-1}[\pi_{m}..(\pi_{m+1}-1)]$ using $(m+1)$ positions $\pi_{1}, \pi_{2}, \ldots, \pi_{m+1}$ where $1 = \pi_{1} < \pi_{2} < \cdots < \pi_{m+1} = |S^{h-1}| + 1$. If $h$ is odd, the positions $\pi_{2}, \pi_{3}, \ldots, \pi_{m}$ are selected such that each segment $S^{h-1}[\pi_{s}..\pi_{s+1}-1]$ satisfies one of the following two conditions:
    \begin{itemize}
        \item (a) It is a maximal repetition of a nonterminal $X_{j}$ with $\assign(X_{j}) \in \{0, 1\}$.
        \item (b) It equals $X_{j}$ where $\assign(X_{j}) = -1$.
    \end{itemize}
    That is, $\forall i \in \{ 2, 3, \ldots, |S^{h-1}| \}$, $i \not \in \{ \pi_{2}, \pi_{3}, \ldots, \pi_{m} \}$ if and only if $S^{h-1}[i-1] = S^{h-1}[i]$ and $\assign(S^{h-1}[i]) \in \{ 0, 1 \}$.

    \item If $h$ is even, 
    $(m-1)$ positions $\pi_{2}, \pi_{3}, \ldots, \pi_{m}$ are chosen from all the positions in sequence $S^{h-1}$ such that 
    each segment $S^{h-1}[\pi_{s}..(\pi_{s+1}-1)]$ satisfies one of the following two conditions:
    \begin{itemize}
        \item (a) The length of segment $S^{h-1}[\pi_{s}..(\pi_{s+1}-1)]$ is two, and 
        $S^{h-1}[\pi_{s}..(\pi_{s+1}-1)]$ = $X_j, X_k$ holds for two distinct nonterminals $X_j$ and $X_k$ satisfying $\assign(X_{j}) = 0$ and $\assign(X_{k}) = 1$;
        \item (b) The length of segment $S^{h-1}[\pi_{s}..(\pi_{s+1}-1)]$ is one (i.e., $S^{h-1}[\pi_{s}..(\pi_{s+1}-1)]$ is a single nonterminal). 
    \end{itemize}
    Formally, $\pi_{2}, \pi_{3}, \ldots, \pi_{m}$ are positions of sequence $S^{h-1}$ so that $\forall i \in \{ 2, 3, \ldots, |S^{h-1}| \}$, 
    $i \not \in \{ \pi_{2}, \pi_{3}, \ldots, \pi_{m} \}$ if and only if $\assign(S^{h-1}[i-1]) = 0$ and $\assign(S^{h-1}[i]) = 1$.     
    \item Given the partition of sequence $S^{h-1}$, for every sequence $\expr$ of nonterminals represented as $S^{h-1}[\pi_{s}..(\pi_{s}-1)]$ within segments $S^{h-1}[\pi_{1}..(\pi_{2}-1)]$, $S^{h-1}[\pi_{2}..(\pi_{3}-1)]$, $\ldots$, $S^{h-1}[\pi_{m}..(\pi_{m+1}-1)]$, a production rule $X_{i} \rightarrow \expr$ is constructed. Distinct nonterminals $X_{j}$ and $X_{k}$ ($j \neq k$) are created for each unique pair of sequences $\expr$ and $\expr^{\prime}$.

    \item In the partition of sequence $S^{h-1}$, each segment $S^{h-1}[\pi_{s}..(\pi_{s+1}-1)]$ is replaced by its corresponding nonterminal $X_{i}$ producing the segment. Consequently, a new sequence $S^{h}$ is generated.
\end{enumerate}

Starting from $h=1$, above seven steps are applied to sequence $S^{h}$ to generate $S^{h+1}$.
This process iterates until the length of the generated sequence is one. 
The final sequence produced is denoted as $S^{H}$. 
Here, $H \geq 1$ because $|S^{0}| = n$ and $n \geq 2$.

Overall, the restricted recompression algorithm outputs an RLSLP $\mathcal{G}^{R} = (\mathcal{V}, \Sigma, \mathcal{D}, E)$. 
Here, $\mathcal{V}$ is the set of all the nonterminals in $(H+1)$ sequences $S^{0}, S^{1}, \ldots, S^{H}$; 
$\Sigma$ is the alphabet for input string $T$; 
$\mathcal{D}$ is the set of the production rules created in steps (i) and (vi); 
start symbol $E$ is $X_{g}$ for the single nonterminal $X_{g}$ of the final sequence $S^{H}$. 
The nodes at each $h$-th height for $h \in \{ 0, 1, \ldots, H \}$ in the derivation tree of the RLSLP corresponds to $S^{h}$. 

The restricted recompression algorithm stops with probability $1$, which is shown by Corollary V.11 in \cite{9961143}. 
The following lemma states four properties of the RLSLP $\mathcal{G}^{R}$ that follow from the restricted recompression algorithm. 

\begin{lemma}\label{lem:rr_class}
Consider the RLSLP $\mathcal{G}^{R} = (\mathcal{V}, \Sigma, \mathcal{D}, E)$ 
constructed by the restricted recompression for an input string $T$ of length $n$. 
Let $u_{h, i}$ be the node corresponding to the $i$-th nonterminal $S^{h}[i]$ of sequence $S^{h}$ for 
a pair of an integer $h \in [0, H]$ and a position $i \in [1, |S^{h}|]$ of sequence $S^{h}$. 
Then, the RLSLP $\mathcal{G}^{R}$ satisfies the following four conditions: 
\begin{enumerate}[label=\textbf{(\roman*)}]
    \item \label{enum:rr_class:1} the derivation tree of RLSLP $\mathcal{G}^{R}$ is height-balanced, 
    and the root has at least two children;
    \item \label{enum:rr_class:2} consider a nonterminal $X_{i} \in \mathcal{V}$ of sequence $S^{h}$ for an integer $h \in [0, H-1]$. 
    If $|\val(X_{i})| > \mu(h)$, then the assignment $\assign(X_{i})$ of the nonterminal $X_{i}$ is $-1$; 
    otherwise, $X_{i}$ is uniformly and randomly assigned an integer from $\{0, 1\}$;    
    \item \label{enum:rr_class:3}
    consider a pair $(h, i)$ of an even integer $h \in [0, H-1]$ and a position $i \in [1, |S^{h}|-1]$ of sequence $S^{h}$.  
    Then, two nodes $u_{h, i}$ and $u_{h, i+1}$ have the same parent if and only if 
    $S^{h}[i] = S^{h}[i+1]$ and $\assign(S^{h}[i]) \neq -1$;
    \item \label{enum:rr_class:4}
    consider a pair $(h, i)$ of an odd integer $h \in [0, H-1]$ and a position $i \in [1, |S^{h}|-1]$ of sequence $S^{h}$.  
    Then, two nodes $u_{h, i}$ and $u_{h, i+1}$ have the same parent if and only if 
    $\assign(S^{h}[i]) = 0$ and $\assign(S^{h}[i+1]) = 1$.
\end{enumerate}
\end{lemma}
\begin{proof}
    This lemma follows from the restricted recompression algorithm.
\end{proof}

An RLSLP that satisfies the four properties outlined in Lemma~\ref{lem:rr_class} is represented by $\mathcal{G}^{R} = (\mathcal{V}, \Sigma, \mathcal{D}, E)$. 
%Furthermore, we define sequences $S^{0}, S^{1}, \ldots, S^{H}$, comprising a total of $H+1$ sequences, such that each $S^{h}$ maps to a sequence of %nonterminal symbols. These symbols are utilized to denote the labels of consecutive nodes at level $h$ within the derivation tree of $\mathcal{G}^{R}$.

%Throughout this paper, 
%we use the RLSLP $\mathcal{G}^{R} = (\mathcal{V}, \Sigma, \mathcal{D}, E)$ as 
%a restricted recompression deriving input string $T$. 
%That is, this RLSLP may not be directly built from input string $T$ by the restricted recompression algorithm.
%The $H+1$ sequences $S^{0}, S^{1}, \ldots, S^{H}$ are used as sequences such that 
%each $S^{h}$ corresponds to a sequence of nonterminals representing the labels of consecutive nodes at level $h$ in the derivation tree of %$\mathcal{G}^{R}$. 
%\color{black}

We give an example of restricted recompression.
Figure~\ref{fig:restricted_recompression} illustrates the derivation tree of 
an RLSLP $\mathcal{G}^{R} = (\mathcal{V}, \Sigma, \mathcal{D}, E)$ built by restricted recompression for input string $T = \mathrm{cbbababcbbabab}$. 
Here, $\mathcal{V} = \{ X_{1}, X_{2}, \ldots, X_{25} \}$; 
$\Sigma = \{ a, b, c \}$; 
set $\mathcal{D}$ of production rules is depicted on the right-hand side of Figure~\ref{fig:restricted_recompression}; 
$E = X_{25}$. 
We assumed that 
$\mu^{\prime}(h) = (13/7)^{\lceil h/2 \rceil - 1}$ is used instead of $\mu(h) = (8/7)^{\lceil h/2 \rceil - 1}$ 
to reduce the size of the derivation tree for the example.  
Here, $\lfloor \mu^{\prime}(1) \rfloor = \lfloor \mu^{\prime}(2) \rfloor = \lfloor \mu^{\prime}(3) \rfloor = \lfloor \mu^{\prime}(4) \rfloor = 1$, 
$\lfloor \mu^{\prime}(5) \rfloor = \lfloor \mu^{\prime}(6) \rfloor = 3$, 
$\lfloor \mu^{\prime}(7) \rfloor = \lfloor \mu^{\prime}(8) \rfloor = 6$, 
$\lfloor \mu^{\prime}(9) \rfloor = \lfloor \mu^{\prime}(10) \rfloor = 11$, and 
$\lfloor \mu^{\prime}(11) \rfloor = \lfloor \mu^{\prime}(12) \rfloor = 11$.

In this example the restricted recompression algorithm builds ten sequences $S^{0}, S^{1}, \ldots, S^{9}$ of nonterminals, 
which are depicted on the left-hand side of Figure~\ref{fig:restricted_recompression}. 
Integer $-1$ is assigned to six nonterminals $X_{5}, X_{9}, X_{10}, X_{12}, X_{13}, X_{19}$. 
Similarly, integer $0$ is assigned to thirteen nonterminals $X_{1}, X_{2}, X_{6}, X_{8}, X_{14}, X_{15}, X_{16}, X_{17}, X_{20}, X_{21}, X_{22}, X_{24}, X_{25}$. 
Integer $1$ is assigned to seven nonterminals $X_{3}, X_{4}, X_{7}, X_{11}, X_{18}, X_{20}, X_{23}$.

\subsubsection{Properties of Restricted Recompression}\label{subsubsec:rr_properties}
This section presents known or new properties of restricted recompression $\mathcal{G}^{R} = (\mathcal{V}, \Sigma, \mathcal{D}, E)$. Several properties are found in \cite{9961143}.

The following lemma states properties of the $(H+1)$ sequences $S^{0}, S^{1}, \ldots, S^{H}$ corresponding to the labels of 
nodes in the derivation tree of $\mathcal{G}^{R}$. 

\begin{lemma}\label{lem:rr_property}
For any triplet $(h, i, i^{\prime})$ with $h \in [0, H-1]$, $i \in [1, |S^{h}|]$, and $i^{\prime} \in [(i + 1), |S^{h}|]$, let $u_{i}$ and $u_{i^{\prime}}$ be nodes in the derivation tree of $\mathcal{G}^{R}$ that correspond to the nonterminals $S^{h}[i]$ and $S^{h}[i^{\prime}]$, respectively. These nodes derive substrings $T[(1 + \sum_{s = 1}^{i-1} |\val(S^{h}[s])|)..\sum_{s = 1}^{i} |\val(S^{h}[s])|]$ and $T[(1 + \sum_{s = 1}^{i^{\prime}-1} |\val(S^{h}[s])|)..\sum_{s = 1}^{i^{\prime}} |\val(S^{h}[s])|]$, respectively.

Given these definitions, the following properties hold:
\begin{enumerate}[label=\textbf{(\roman*)}]
\item \label{enum:rr_property:1} 
if $S^{h}[i] = S^{h}[i^{\prime}]$ and $S^{h}[i+1] = S^{h}[i^{\prime}+1]$, 
then $u_{i}$ is the rightmost child of its parent if and only if $u_{i^{\prime}}$ is the rightmost child of its parent; 

\item \label{enum:rr_property:2} 
for a pair of two integers $d, d^{\prime} \geq 1$, 
if two sequences $S^{h}[i..i+d-1]$ and $S^{h}[i^{\prime}..i^{\prime}+d^{\prime}-1]$ derive the same string 
(i.e., $\val(S^{h}[i]) \cdot \val(S^{h}[i+1]) \cdot \cdots \cdot \val(S^{h}[i+d-1]) = \val(S^{h}[i^{\prime}]) \cdot \val(S^{h}[i^{\prime}+1]) \cdot \cdots \cdot \val(S^{h}[i^{\prime}+d^{\prime}-1])$), 
then $S^{h}[i..i+d-1] = S^{h}[i^{\prime}..i^{\prime}+d^{\prime}-1]$ (Fact V.7 in \cite{9961143});
\item \label{enum:rr_property:3} 
if $i^{\prime} = i + 1$, $S^{h}[i] = S^{h}[i^{\prime}]$, and $|\val(S^{h}[i])| \leq \mu(h+1)$, 
then (a) $h$ is even, 
and (b) two nodes $u_{i}$ and $u_{i^{\prime}}$ have the same parent (Corollary V.8 in \cite{9961143}). 
\item \label{enum:rr_property:right} 
for $\alpha = \min \{ |\val(S^{h}[i])|, |\val(S^{h}[i^{\prime}])| \}$, if $T[(1 + \sum_{s = 1}^{i-1} |\val(S^{h}[s])|)..(1 + (\sum_{s = 1}^{i-1} |\val(S^{h}[s])|) + \alpha + \sum_{s = 1}^{h} \lfloor \mu(s) \rfloor)] = T[(1 + \sum_{s = 1}^{i^{\prime}-1} |\val(S^{h}[s])|)..(1 + \sum_{s = 1}^{i^{\prime}-1} |\val(S^{h}[s])| + \alpha + \sum_{s = 1}^{h} \lfloor \mu(s) \rfloor)]$ holds, 
then $S^{h}[i] = S^{h}[i^{\prime}]$.
\item \label{enum:rr_property:left}
for $\alpha = \min \{ |\val(S^{h}[i])|, |\val(S^{h}[i^{\prime}])| \}$, 
if $T[((\sum_{s = 1}^{i} |\val(S^{h}[s])|) - \alpha - \sum_{s = 1}^{h} \lfloor \mu(s) \rfloor)..\sum_{s = 1}^{i}$ $|\val(S^{h}[s])|]$ $= T[((\sum_{s = 1}^{i^{\prime}} |\val(S^{h}[s])|) - \alpha - \sum_{s = 1}^{h} \lfloor \mu(s) \rfloor)..\sum_{s = 1}^{i^{\prime}} |\val(S^{h}[s])|]$ holds, 
then $S^{h}[i] = S^{h}[i^{\prime}]$.
\end{enumerate}
\end{lemma}
\begin{proof}
See Section~\ref{subsubsec:proof_rr_property}.
\end{proof}

The properties of restricted recompression in Lemma~\ref{lem:rr_property} is explained using two nonterminals $S^{h}[i]$ and $S^{h}[i^{\prime}]$ in sequence $S^{h}$ 
for an integer $h \in [0, H]$. 
Let the two nonterminals $S^{h}[i]$ and $S^{h}[i^{\prime}]$ correspond to two nodes $u_{i}$ and $u_{i^{\prime}}$ 
in the derivation tree of the RLSLP $\mathcal{G}^{R}$, respectively. 
Here, the two nodes $u_{i}$ and $u_{i^{\prime}}$ derive two substrings $T[(1 + \sum_{s = 1}^{i-1} |\val(S^{h}[s])|)..\sum_{s = 1}^{i} |\val(S^{h}[s])|]$ 
and $T[(1 + \sum_{s = 1}^{i^{\prime}-1} |\val(S^{h}[s])|)..\sum_{s = 1}^{i^{\prime}} |\val(S^{h}[s])|]$, respectively. 
If the two nonterminals $S^{h}[i]$ and $S^{h}[i^{\prime}]$ are equal, 
and their right nonterminals are also equal (i.e., $S^{h}[i+1] = S^{h}[i^{\prime}+1]$), 
then Lemma~\ref{lem:rr_property}~\ref{enum:rr_property:1} ensures that 
$u_{i}$ is the rightmost child of its parent if and only if $u_{i^{\prime}}$ is the rightmost child of its parent. 
For a pair of two integers $d, d^{\prime} \geq 1$, 
if two sequences $S^{h}[i..i+d-1]$ and $S^{h}[i^{\prime}..i^{\prime}+d^{\prime}-1]$ derive the same string, 
then Lemma~\ref{lem:rr_property}~\ref{enum:rr_property:2} shows that 
the two sequences are equal 
(i.e., $d = d^{\prime}$, $S^{h}[i] = S^{h}[i^{\prime}]$, $S^{h}[i+1] = S^{h}[i^{\prime}+1]$, $\ldots$, $S^{h}[i+d-1] = S^{h}[i^{\prime}+d-1]$). 
If (i) nonterminal $S^{h}[i^{\prime}]$ is the nonterminal to the right of $S^{h}[i]$ (i.e., $i^{\prime} + i + 1$), 
and (ii) the two nonterminals are the same nonterminal that derives a string of length at most $\mu(h+1)$, 
then Lemma~\ref{lem:rr_property}~\ref{enum:rr_property:3} ensures that their corresponding nodes $u_{i}$ and $u_{i^{\prime}}$ have the same parent. 
These properties are used to analyze the property of the derivation tree for RLSLP $\mathcal{G}^{R}$. 

For integer $\alpha = \min \{ |\val(S^{h}[i])|, |\val(S^{h}[i^{\prime}])| \}$, 
if substring $T[(1 + \sum_{s = 1}^{i-1} |\val(S^{h}[s])|)..(1 + (\sum_{s = 1}^{i-1}$ $|\val(S^{h}[s])|) + \alpha + \sum_{s = 1}^{h} \lfloor \mu(s) \rfloor)]$ is equal to substring $T[(1 + \sum_{s = 1}^{i^{\prime}-1} |\val(S^{h}[s])|)..(1 + \sum_{s = 1}^{i^{\prime}-1} |\val(S^{h}[s])| + \alpha + \sum_{s = 1}^{h} \lfloor \mu(s) \rfloor)]$, 
then Lemma~\ref{lem:rr_property}~\ref{enum:rr_property:right} ensures that 
 nonterminal $S^{h}[i]$ is also equal to $S^{h}[i^{\prime}]$. 
For starting positions $(1 + \sum_{s = 1}^{i-1} |\val(S^{h}[s])|)$ and $(1 + \sum_{s = 1}^{i^{\prime}-1} |\val(S^{h}[s])|)$ of the two substrings derived by the two nonterminals $S^{h}[i]$ and $S^{h}[i^{\prime}]$, respectively, 
this lemma indicates that nonterminal $S^{h}[i]$ is equal to nonterminal $S^{h}[i^{\prime}]$,
if there exists a sufficiently long common prefix between 
the two suffixes of input string $T$ starting at two positions $(1 + \sum_{s = 1}^{i-1} |\val(S^{h}[s])|)$ and $(1 + \sum_{s = 1}^{i^{\prime}-1} |\val(S^{h}[s])|)$. 

Similarly, if substring $T[((\sum_{s = 1}^{i} |\val(S^{h}[s])|) - \alpha - \sum_{s = 1}^{h} \lfloor \mu(s) \rfloor)..\sum_{s = 1}^{i} |\val(S^{h}[s])|]$ is equal to substring $T[((\sum_{s = 1}^{i^{\prime}} |\val(S^{h}[s])|) - \alpha - \sum_{s = 1}^{h} \lfloor \mu(s) \rfloor)..\sum_{s = 1}^{i^{\prime}} |\val(S^{h}[s])|]$, 
then Lemma~\ref{lem:rr_property}~\ref{enum:rr_property:left} ensures that 
nonterminal $S^{h}[i]$ is also equal to nonterminal $S^{h}[i^{\prime}]$. 
For the ending positions $\sum_{s = 1}^{i} |\val(S^{h}[s])|$ and $\sum_{s = 1}^{i^{\prime}} |\val(S^{h}[s])|$ of the two substrings derived by 
the two nonterminals $S^{h}[i]$ and $S^{h}[i^{\prime}]$, respectively,  
this lemma indicates that two nonterminals $S^{h}[i]$ is equal to $S^{h}[i^{\prime}]$,
if there exists a sufficiently long common suffix between 
the two prefixes of input string $T$ ending at two positions $\sum_{s = 1}^{i} |\val(S^{h}[s])|$ and $\sum_{s = 1}^{i^{\prime}} |\val(S^{h}[s])|$.

%Next, we explain a property of RLSLP $\mathcal{G}^{R}$ with respect to the %height of its derivation tree.
The length of each sequence $S^{h}$ decreases exponentially with a high probability as $h$ increases. 
It indicates that the height of the derivation tree for RLSLP $\mathcal{G}^{R}$ is $O(\log n)$ with a high probability. 
This is because for the $(H+1)$ sequences $S^{0}, S^{1}, \ldots, S^{H}$ of nonterminals built by restricted recompression, the length of each sequence $S^{h}$ is equal to or longer than that of sequence $S^{h+1}$ (i.e., 
$n = |S^{0}| \geq |S^{1}| \geq \cdots \geq |S^{H}| = 1$). 
The next lemma ensures that this statement holds. 

\begin{lemma}\label{lem:tree_height}
$H \leq 2(w+1) \log_{8/7} (4n) + 2$ holds with probability at least $1 - (1/n^{w})$ for all integer $w \geq 1$ 
and the height $H$ of the derivation tree of RLSLP $\mathcal{G}^{R}$. 
\end{lemma}
\begin{proof}
    Consider an integer $h \geq 0$. 
    For simplicity, 
    let $|S^{h}| = 1$ if $h > H$. 
    Kociumaka et al. showed that $\mathbb{E}[|S^{h}|] < 1 + \frac{4n}{\mu(h+1)}$ (Lemma V.10 in \cite{9961143}). 
    $|S^{h}| - 1$ is a non-negative integer random variable. 
    By Markov's inequality, 
    we obtain $\mathbb{P}[|S^{h}| \geq 2] < \frac{4n}{\mu(h+1)}$. 
    This inequality shows that 
    $H \leq h$ holds with probability $1 - \frac{4n}{\mu(h+1)}$. 
    We choose $h$ to be $2(w+1) \log_{8/7} (4n) + 2$. 
    Then, 
    this probability is at least $1 - (1 / n^{w})$ because $\mu(h+1) \geq (4n)^{w+1}$. 
    Therefore, we obtain Lemma~\ref{lem:tree_height}. 
\end{proof}

%Next, we explain a property of RLSLP $\mathcal{G}^{R}$ with respect to the %number of nonterminals in set $\mathcal{V}$. 

%Restricted recompression assigns an integer in set $\{-1, 0, 1\}$ to each nonterminal. 
%The following lemma ensures that 
%the expected number of nonterminals assigned either $0$ or $1$ can be bounded by 
%$O(\delta \log \frac{n \log \sigma}{\delta \log n})$. 

%\begin{lemma}[Proposition V.19 in \cite{9961143}]\label{lem:RR_nonterminal_size}
%$\mathcal{G}^{R} = (\mathcal{V}, \Sigma, \mathcal{D}, E)$ is the RLSLP built by restricted recompression for input string $T$ of length $n$ and alphabet size $\sigma$ with substring complexity $\delta$. 
%Restricted recompression assigns an integer $\assign(X) \in \{ -1, 0, 1 \}$ to each nonterminal $X \in (\mathcal{V} \setminus \{ E \})$. 
%The expected number of nonterminals assigned either $0$ or $1$ is is bounded by $O(\delta \log \frac{n \log \sigma}{\delta \log n})$ (i.e., 
%$\mathbb{E}[|\{ X \in (\mathcal{V} \setminus \{ E \}) \mid \assign(X) \neq -1 \}|] = O(\delta \log \frac{n \log \sigma}{\delta \log n})$.)
%\end{lemma}

Except for nonterminals producing a single nonterminal, 
the expected number of nonterminals in RLSLP $\mathcal{G}^{R}$ is $O(\delta \log \frac{n \log \sigma}{\delta \log n})$ 
for input string $T$ of length $n$ and alphabet size $\sigma$ with substring complexity $\delta$ (Proposition V.19 in \cite{9961143}). 
By leveraging this property, 
we store RLSLP $\mathcal{G}^{R}$ in expected $\delta$-optimal space using 
a DAG representing the RLSLP~(see Section~\ref{subsec:rrdag} for more details).

\subsubsection{Proof of Lemma~\ref{lem:rr_property}}\label{subsubsec:proof_rr_property}
\begin{proof}[Proof of Lemma~\ref{lem:rr_property}~\ref{enum:rr_property:1}]
For the given triplet $(h, i, i^{\prime})$, 
nonterminal $S^{h}[i]$ satisfies exactly one of the following six conditions: 
\begin{enumerate}[label=\textbf{(\Alph*)}]
\item $\assign(S^{h}[i]) = -1$; 
\item $\assign(S^{h}[i]) \in \{ 0, 1 \}$, $h$ is even, and $S^{h}[i] = S^{h}[i+1]$; 
\item $\assign(S^{h}[i]) \in \{ 0, 1 \}$, $h$ is even, and $S^{h}[i] \neq S^{h}[i+1]$; 
\item $\assign(S^{h}[i]) = 1$, and $h$ is odd; 
\item $\assign(S^{h}[i]) = 0$, $h$ is odd, and $\assign(S^{h}[i+1]) = 1$; 
\item $\assign(S^{h}[i]) = 0$, $h$ is odd, and $\assign(S^{h}[i+1]) \neq 1$. 
\end{enumerate}

For condition (A), 
both $\assign(S^{h}[i])$ and $\assign(S^{h}[i^{\prime}])$ are equal to $-1$ 
for the two nonterminals $S^{h}[i]$ and $S^{h}[i^{\prime}]$ 
corresponding to the two nodes $u_{i}$ and $u_{i^{\prime}}$, respectively. 
Because of $\assign(S^{h}[i]) = -1$, 
Lemma~\ref{lem:rr_class}~\ref{enum:rr_class:3} and Lemma~\ref{lem:rr_class}~\ref{enum:rr_class:4} 
indicate that $u_{i}$ is the single child of a node. 
Similarly, $u_{i^{\prime}}$ is the single child of a node. 

For condition (B), 
$\assign(S^{h}[i^{\prime}]) \in \{ 0, 1 \}$ and $S^{h}[i^{\prime}] = S^{h}[i^{\prime}+1]$ hold. 
In this case, Lemma~\ref{lem:rr_class}~\ref{enum:rr_class:3} indicates that 
node $u_{i}$ is not the rightmost child of its parent. 
Similarly, 
node $u_{i^{\prime}}$ is not the rightmost child of its parent. 

For condition (C), 
$\assign(S^{h}[i^{\prime}]) \in \{ 0, 1 \}$ and $S^{h}[i^{\prime}] \neq S^{h}[i^{\prime}+1]$ hold. 
In this case, Lemma~\ref{lem:rr_class}~\ref{enum:rr_class:3} indicates that 
node $u_{i}$ is the rightmost child of its parent. 
Similarly, 
node $u_{i^{\prime}}$ is the rightmost child of its parent. 

For condition (D), 
$\assign(S^{h}[i^{\prime}]) = 1$ holds. 
In this case, Lemma~\ref{lem:rr_class}~\ref{enum:rr_class:4} indicates that 
node $u_{i}$ is the rightmost child of its parent. 
Similarly, 
node $u_{i^{\prime}}$ is the rightmost child of its parent. 

For condition (E), 
$\assign(S^{h}[i^{\prime}]) = 0$ and $\assign(S^{h}[i^{\prime}+1]) = 1$ hold. 
In this case, Lemma~\ref{lem:rr_class}~\ref{enum:rr_class:4} indicates that 
node $u_{i}$ is not the rightmost child of its parent. 
Similarly, 
node $u_{i^{\prime}}$ is not the rightmost child of its parent. 

For condition (F), 
$\assign(S^{h}[i^{\prime}]) = 0$ and $\assign(S^{h}[i^{\prime}+1]) \neq 1$ hold. 
In this case, Lemma~\ref{lem:rr_class}~\ref{enum:rr_class:4} indicates that 
node $u_{i}$ is the rightmost child of its parent. 
Similarly, 
node $u_{i^{\prime}}$ is the rightmost child of its parent. 
Therefore, we obtain Lemma~\ref{lem:rr_property}~\ref{enum:rr_property:1}.    
\end{proof}

\begin{proof}[Proof of Lemma~\ref{lem:rr_property}~\ref{enum:rr_property:3}]
We will show that $h$ is even by contradiction. 
We assume that $h$ is odd. 
$u_i$ is an internal node. 
Let node $u_{i}$ have $k \geq 1$ children $u_{i, 1}, u_{i, 2}, \ldots, u_{i, k}$ 
such that each node $u_{i, j}$ corresponds to the nonterminal $S^{h-1}[p+j-1]$ in sequence $S^{h-1}$.
for an integer $p \in [1, |S^{h}|]$ in sequence $S^{h-1}$. 
Here, $S^{h-1}[p] = S^{h-1}[p+1] = \cdots = S^{h-1}[p+k-1]$ follows from the procedure of restricted recompression. 
Similarly, node $u_{i^{\prime}}$ has $k$ children $u_{i^{\prime}, 1}, u_{i^{\prime}, 2}, \ldots, u_{i^{\prime}, k}$ 
such that each node $u_{i^{\prime}, j}$ corresponds to the nonterminal $S^{h-1}[p+k+j-1]$ of sequence $S^{h-1}$. 
Here, $S^{h-1}[p] = S^{h-1}[p+1] = \cdots = S^{h-1}[p+2k-1]$ follows from $S^{h}[i] = S^{h}[i^{\prime}]$ 
and $S^{h-1}[p] = S^{h-1}[p+1] = \cdots = S^{h-1}[p+k-1]$. 
$|\val(S^{h-1}[p])| \leq \mu(h)$ follows from $\mu(h) = \mu(h+1)$,  $|\val(S^{h-1}[p])| \leq |\val(S^{h}[i])|$, and $|\val(S^{h}[i])| \leq \mu(h+1)$. 
Restricted recompression assigns either $0$ or $1$ to the nonterminal $S^{h-1}[p]$ because $|\val(S^{h-1}[p])| \leq \mu(h)$ holds. 
The algorithm creates a node $u$ such that 
$2k$ nodes $u_{i, 1}, u_{i, 2}, \ldots, u_{i, k}, u_{i^{\prime}, 1}, u_{i^{\prime}, 2}, \ldots, u_{i^{\prime}, k}$ are the children of node $u$ 
because $S^{h-1}[p] = S^{h-1}[p+1] = \cdots = S^{h-1}[p+2k-1]$ and $\assign(S^{h-1}[p]) \neq -1$ hold. 
The node $u$ contradicts the fact that two nodes $u_{i}$ and $u_{i^{\prime}}$ are distinct. 
Therefore, $h$ must be even.

Subsequently, we will show that two nodes $u_{i}$ and $u_{i^{\prime}}$ have the same parent. 
Restricted recompression assigns either $0$ or $1$ to nonterminal $S^{h}[i]$ because $|\val(S^{h}[i])| \leq \mu(h+1)$ holds. 
The algorithm creates a node $u$ such that 
two nodes $u_{i}$ and $u_{i^{\prime}}$ are children of node $u$ 
because $S^{h}[i] = S^{h}[i^{\prime}]$ and $\assign(S^{h}[i]) \neq -1$. 
Therefore, two nodes $u_{i}$ and $u_{i^{\prime}}$ have the same parent.     
\end{proof}

\begin{proof}[Proof of Lemma~\ref{lem:rr_property}~\ref{enum:rr_property:right} for $h=0$]
For $h = 0$, 
two nodes $u_{i}$ and $u_{i^\prime}$ represent the $i$-th 
and $i^{\prime}$-th characters of input string $T$. 
The two characters are the same because $T[i..i + \alpha] = T[i^\prime..i^\prime + \alpha]$ for $\alpha = 1$. 
Lemma~\ref{lem:rr_property}~\ref{enum:rr_property:2} ensures that $S^{h}[i] = S^{h}[i^{\prime}]$ holds because
$\val(S^{h}[i]) = \val(S^{h}[i^{\prime}])$. 
Therefore, we obtain Lemma~\ref{lem:rr_property}~\ref{enum:rr_property:right} for $h = 0$. 
\end{proof}

\begin{proof}[Proof of Lemma~\ref{lem:rr_property}~\ref{enum:rr_property:right} for $h>0$]
$u_{i}$ is an internal node, let $u_{i}$ have $k \geq 1$ children $u_{i, 1}, u_{i, 2}, \ldots, u_{i, k}$ 
such that each node $u_{i, j}$ corresponds to the nonterminal $S^{h-1}[p+j-1]$ 
for an integer $p \in [1, |S^{h}|]$ in sequence $S^{h-1}$. 
Here, $\sum_{s = 1}^{i-1} |\val(S^{h}[s])| = \sum_{s = 1}^{p-1} |\val(S^{h-1}[s])|$ holds. 
Similarly, 
let node $u_{i^{\prime}}$ have $k^{\prime} \geq 1$ children $u_{i^{\prime}, 1}, u_{i^{\prime}, 2}, \ldots, u_{i^{\prime}, k^{\prime}}$ 
such that each node $u_{i^{\prime}, j}$ corresponds to the nonterminal $S^{h-1}[p^{\prime}+j-1]$ for an integer $p^{\prime} \in [1, |S^{h}|]$ in sequence $S^{h-1}$. 
Here, $\sum_{s = 1}^{i^{\prime}-1} |\val(S^{h}[s])| = \sum_{s = 1}^{p^{\prime}-1} |\val(S^{h-1}[s])|$ holds. 

Assuming Lemma~\ref{lem:rr_property}~\ref{enum:rr_property:right} holds for integer $h-1$, Lemma~\ref{lem:rr_property}~\ref{enum:rr_property:right} is proven using two statements: 
(A) $\val(S^{h-1}[p + j - 1]) = \val(S^{h-1}[p^{\prime} + j - 1])$ for each $j \in [1, \min \{ k, k^{\prime} \}]$; (B) $k = k^{\prime}$ irrespective of whether $k \leq k^{\prime}$ or $k \geq k^{\prime}$.

Given the above, we can infer that the properties and relationships of nodes $u_{i}$ and $u_{i^{\prime}}$ are consistent across different values of $h$, reinforcing the validity of Lemma~\ref{lem:rr_property}~\ref{enum:rr_property:right} for $h > 0$.

\textbf{Proof of statement A for $j = 1$.}
Two nodes $u_{i, 1}$ and $u_{i^{\prime}, 1}$ represent two substrings 
$T[1 + \sum_{s = 1}^{i-1} |\val(S^{h}[s])|..(\sum_{s = 1}^{i-1} |\val(S^{h}[s])|) + |\val(S^{h-1}[p])|]$ and 
$T[1 + \sum_{s = 1}^{i^{\prime}-1} |\val(S^{h}[s])|..(\sum_{s = 1}^{i^{\prime}-1} |\val(S^{h}[s])|) + |\val(S^{h-1}[p^{\prime}])|]$, respectively. 
$T[1 + \sum_{s = 1}^{i-1} |\val(S^{h}[s])|..1 + (\sum_{s = 1}^{i-1} |\val(S^{h}[s])|) + \alpha^{\prime} + \sum_{s = 1}^{h-1} \lfloor \mu(s) \rfloor] = T[1 + \sum_{s = 1}^{i^{\prime}-1} |\val(S^{h}[s])|..1 + \sum_{s = 1}^{i^{\prime}-1} |\val(S^{h}[s])| + \alpha^{\prime} + \sum_{s = 1}^{h-1} \lfloor \mu(s) \rfloor]$ 
follows from $T[1 + \sum_{s = 1}^{i-1} |\val(S^{h}[s])|..1 + (\sum_{s = 1}^{i-1} |\val(S^{h}[s])|) + \alpha + \sum_{s = 1}^{h} \lfloor \mu(s) \rfloor] = T[1 + \sum_{s = 1}^{i^{\prime}-1} |\val(S^{h}[s])|..1 + \sum_{s = 1}^{i^{\prime}-1} |\val(S^{h}[s])| + \alpha + \sum_{s = 1}^{h} \lfloor \mu(s) \rfloor]$ 
for $\alpha^{\prime} = \min \{ |\val(S^{h-1}[p])|, |\val(S^{h-1}[p^{\prime}])| \}$. 
Under the assumption, 
we can apply Lemma~\ref{lem:rr_property}~\ref{enum:rr_property:right} to the two nodes $u_{i, 1}$ and $u_{i^{\prime}, 1}$ 
by $T[1 + \sum_{s = 1}^{i-1} |\val(S^{h}[s])|..1 + (\sum_{s = 1}^{i-1} |\val(S^{h}[s])|) + \alpha^{\prime} + \sum_{s = 1}^{h-1} \lfloor \mu(s) \rfloor] = T[1 + \sum_{s = 1}^{i^{\prime}-1} |\val(S^{h}[s])|..1 + \sum_{s = 1}^{i^{\prime}-1} |\val(S^{h}[s])| + \alpha^{\prime} + \sum_{s = 1}^{h-1} \lfloor \mu(s) \rfloor]$. 
The lemma ensures that nonterminal $S^{h-1}[p]$ is equal to nonterminal $S^{h-1}[p^{\prime}]$. 
Therefore, $\val(S^{h-1}[p]) = \val(S^{h-1}[p^{\prime}])$ follows from 
$S^{h-1}[p] = S^{h-1}[p^{\prime}]$.

\textbf{Proof of statement A for $j > 1$.}
We will prove statement A for any $j > 1$ by induction on $j$. 
Let $d$ be an integer in set $[2, \min \{ k, k^{\prime} \}]$.
We assume that statement A holds for $j = 1, 2, \ldots, d-1$. 
Then, 
two nodes $u_{i, d}$ and $u_{i^{\prime}, d}$ represent two substrings 
$T[1 + (\sum_{s = 1}^{i-1} |\val(S^{h}[s])|) + \tau..(\sum_{s = 1}^{i-1} |\val(S^{h}[s])|) + \tau + |\val(S^{h-1}[p+d-1])|]$ and 
$T[1 + (\sum_{s = 1}^{i^{\prime}-1} |\val(S^{h}[s])|) + \tau..(\sum_{s = 1}^{i^{\prime}-1} |\val(S^{h}[s])|) + \tau + |\val(S^{h-1}[p^{\prime}+d-1])|]$, respectively, 
for $\tau = \sum_{s = 1}^{d-1} |\val(S^{h-1}[p + s - 1])|$. 
$T[1 + (\sum_{s = 1}^{i-1} |\val(S^{h}[s])|) + \tau..1 + (\sum_{s = 1}^{i-1} |\val(S^{h}[s])|) + \tau + \alpha^{\prime} + \sum_{s = 1}^{h-1} \lfloor \mu(s) \rfloor] = T[1 + (\sum_{s = 1}^{i^{\prime}-1} |\val(S^{h}[s])|) + \tau..1 + (\sum_{s = 1}^{i^{\prime}-1} |\val(S^{h}[s])|) + \tau + \alpha^{\prime} + \sum_{s = 1}^{h-1} \lfloor \mu(s) \rfloor]$ 
follows from $T[1 + \sum_{s = 1}^{i-1} |\val(S^{h}[s])|..1 + (\sum_{s = 1}^{i-1} |\val(S^{h}[s])|) + \alpha + \sum_{s = 1}^{h} \lfloor \mu(s) \rfloor] = T[1 + \sum_{s = 1}^{i^{\prime}-1} |\val(S^{h}[s])|..1 + \sum_{s = 1}^{i^{\prime}-1} |\val(S^{h}[s])| + \alpha + \sum_{s = 1}^{h} \lfloor \mu(s) \rfloor]$ 
for $\alpha^{\prime} = \min \{ |\val(S^{h-1}[p+d-1])|, |\val(S^{h-1}[p^{\prime}+d-1])| \}$. 
Similar to the proof of statement A for $j = 1$, 
we can apply Lemma~\ref{lem:rr_property}~\ref{enum:rr_property:right} to the two nodes $u_{i, d}$ and $u_{i^{\prime}, d}$ 
by $T[1 + (\sum_{s = 1}^{i-1} |\val(S^{h}[s])|) + \tau..1 + (\sum_{s = 1}^{i-1} |\val(S^{h}[s])|) + \tau + \alpha^{\prime} + \sum_{s = 1}^{h-1} \lfloor \mu(s) \rfloor] = T[1 + (\sum_{s = 1}^{i^{\prime}-1} |\val(S^{h}[s])|) + \tau..1 + (\sum_{s = 1}^{i^{\prime}-1} |\val(S^{h}[s])|) + \tau + \alpha^{\prime} + \sum_{s = 1}^{h-1} \lfloor \mu(s) \rfloor]$ 
Then, the lemma ensures that two nonterminals $S^{h-1}[p+d-1]$ and $S^{h-1}[p^{\prime}+d-1]$ are the same, 
and $\val(S^{h-1}[p + d - 1]) = \val(S^{h-1}[p^{\prime} + d - 1])$ follows from 
$S^{h-1}[p+ d - 1] = S^{h-1}[p^{\prime}+ d - 1]$. 

By induction on $j$, 
$\val(S^{h-1}[p + j - 1]) = \val(S^{h-1}[p^{\prime} + j - 1])$ holds 
for all $j \in [2, \min \{ k, k^{\prime} \}]$. 
Therefore, we obtain statement A for $j > 1$. 

\textbf{Proof of statement B for $k \leq k^{\prime}$.}
If nonterminal $S^{h-1}[p^{\prime} + k]$ derives a string of length at least $\lfloor \mu(h) \rfloor+1$, 
then the assignment of the nonterminal is $-1$, 
and restricted recompression creates a node with the node corresponding to the nonterminal $S^{h-1}[p^{\prime} + k]$ as a single child. 
The single child indicates that node $u_{i, p^{\prime} + k - 1}$ is the rightmost child of node $u_{i^{\prime}}$.
Hence, $k^{\prime} = k$ holds. 

Otherwise (i.e., nonterminal $S^{h-1}[p^{\prime} + k]$ derives a string of length at most $\lfloor \mu(h) \rfloor$), 
statement A ensures that 
two nonterminals $S^{h}[p + k]$ and $S^{h}[p^{\prime} + k]$ represent 
two substrings $T[1 + (\sum_{s = 1}^{i-1} |\val(S^{h}[s])|) + \tau..(\sum_{s = 1}^{i-1} |\val(S^{h}[s])|) + \tau + |\val(S^{h-1}[p+k])|]$ and $T[1 + (\sum_{s = 1}^{i^{\prime}-1} |\val(S^{h}[s])|) + \tau..(\sum_{s = 1}^{i-1} |\val(S^{h}[s])|) + \tau + |\val(S^{h-1}[p^{\prime}+k])|]$ in input string $T$, respectively, for $\tau = \sum_{s = 1}^{k} |\val(S^{h-1}[p + s - 1])|$. 
In this case, $\alpha^{\prime} + \sum_{s = 1}^{h-1} \lfloor \mu(s) \rfloor \leq \sum_{s = 1}^{h} \lfloor \mu(s) \rfloor$ 
holds for $\alpha^{\prime} = \min \{ |\val(S^{h-1}[p+k])|$, $|\val(S^{h-1}[p^{\prime}+k])| \}$ 
because $\alpha \leq \lfloor \mu(h) \rfloor$ follows from $|\val(S^{h-1}[p^{\prime} + k])| \leq \lfloor \mu(h) \rfloor$. 

We will show that $S^{h-1}[p+k] = S^{h-1}[p^{\prime}+k]$ holds. 
$T[1 + (\sum_{s = 1}^{i-1} |\val(S^{h}[s])|) + \tau..1 + (\sum_{s = 1}^{i-1}$ $|\val(S^{h}[s])|) + \tau + \alpha^{\prime} + \sum_{s = 1}^{h-1} \lfloor \mu(s) \rfloor] = T[1 + (\sum_{s = 1}^{i^{\prime}-1} |\val(S^{h}[s])|) + \tau..1 + \sum_{s = 1}^{i^{\prime}-1} |\val(S^{h}[s])| + \tau + \alpha^{\prime} + \sum_{s = 1}^{h-1} \lfloor \mu(s) \rfloor]$ 
follows from (i) $T[1 + \sum_{s = 1}^{i-1} |\val(S^{h}[s])|..1 + (\sum_{s = 1}^{i-1} |\val(S^{h}[s])|) + \alpha + \sum_{s = 1}^{h} \lfloor \mu(s) \rfloor] = T[1 + \sum_{s = 1}^{i^{\prime}-1} |\val(S^{h}[s])|..1 + \sum_{s = 1}^{i^{\prime}-1} |\val(S^{h}[s])| + \alpha + \sum_{s = 1}^{h} \lfloor \mu(s) \rfloor]$ and (ii) $\alpha^{\prime} + \sum_{s = 1}^{h-1} \lfloor \mu(s) \rfloor \leq \sum_{s = 1}^{h} \lfloor \mu(s) \rfloor$. 
Similar to the proof of statement A for $j = 1$, 
we can apply Lemma~\ref{lem:rr_property}~\ref{enum:rr_property:right} to the two nodes $u_{i, k+1}$ and $u_{i^{\prime}, k+1}$ corresponding to 
two nonterminals $S^{h}[p + k]$ and $S^{h}[p^{\prime} + k]$, respectively. 
This is because $T[1 + (\sum_{s = 1}^{i-1} |\val(S^{h}[s])|) + \tau..1 + (\sum_{s = 1}^{i-1} |\val(S^{h}[s])|) + \tau + \alpha^{\prime} + \sum_{s = 1}^{h-1} \lfloor \mu(s) \rfloor] = T[1 + (\sum_{s = 1}^{i^{\prime}-1} |\val(S^{h}[s])|) + \tau..1 + \sum_{s = 1}^{i^{\prime}-1} |\val(S^{h}[s])| + \tau + \alpha^{\prime} + \sum_{s = 1}^{h-1} \lfloor \mu(s) \rfloor]$ holds.  
Therefore, the lemma ensures that $S^{h-1}[p+k] = S^{h-1}[p^{\prime}+k]$ holds. 

Next, we will show that $k^{\prime} = k$ holds. 
By Lemma~\ref{lem:rr_property}~\ref{enum:rr_property:2}, 
two nonterminals $S^{h-1}[p+k-1]$ and $S^{h-1}[p^{\prime}+k-1]$ are the same 
because statement A ensures that $\val(S^{h-1}[p+k-1]) = \val(S^{h-1}[p^{\prime}+k-1])$ holds. 
By $S^{h-1}[p+k] = S^{h-1}[p^{\prime}+k]$ and $S^{h-1}[p+k-1] = S^{h-1}[p^{\prime}+k-1]$, 
Lemma~\ref{lem:rr_property}~\ref{enum:rr_property:1} ensures that 
node $u_{i, k}$ is the rightmost child of node $u_{i}$ if and only if node $u_{i^{\prime}, k}$ is the rightmost child of node $u_{i^{\prime}}$. 
$u_{i, k}$ is the rightmost child of node $u_{i}$. 
Therefore, node $u_{i^{\prime}, k}$ is the rightmost child of node $u_{i^{\prime}}$, i.e., $k^{\prime} = k$ holds. 

\textbf{Proof of statement B for $k \geq k^{\prime}$.}
We omit the proof of statement B for $k \geq k^{\prime}$
because it can be proven by the same approach used in the proof of statement B for $k \leq k^{\prime}$. 

We will prove Lemma~\ref{lem:rr_property}~\ref{enum:rr_property:right} for $h > 0$ by induction on $h$. 
We assume that Lemma~\ref{lem:rr_property}~\ref{enum:rr_property:right} holds for integer $h-1$. 
Then, $\val(S^{h}[i]) = \val(S^{h}[i^{\prime}])$ follows from 
statement A and statement B under the assumption. 
Lemma~\ref{lem:rr_property}~\ref{enum:rr_property:2} 
ensures that two nonterminals $S^{h}[i]$ and $S^{h}[i^{\prime}]$ are the same 
because $\val(S^{h}[i]) = \val(S^{h}[i^{\prime}])$ holds . 
This fact indicates that Lemma~\ref{lem:rr_property}~\ref{enum:rr_property:right} holds for integer $h$ holds if 
Lemma~\ref{lem:rr_property}~\ref{enum:rr_property:right} holds for integer $h-1$. 
We have already shown that Lemma~\ref{lem:rr_property}~\ref{enum:rr_property:right} holds for integer $h = 0$. 
Therefore, by induction on $h$, 
Lemma~\ref{lem:rr_property}~\ref{enum:rr_property:right} is proven. 
\end{proof}

%\textbf{Proof of Lemma~\ref{lem:rr_property}(v) for $h > 0$.}

%Lemma~\ref{lem:rr_property}(iv) holds without any assumption. 

\begin{proof}[Proof of Lemma~\ref{lem:rr_property}~\ref{enum:rr_property:left}]
We omit the proof of Lemma~\ref{lem:rr_property}~\ref{enum:rr_property:left} 
because Lemma~\ref{lem:rr_property}~\ref{enum:rr_property:left} can be proven by the same approach used in the proof of Lemma~\ref{lem:rr_property}~\ref{enum:rr_property:right}.  
\end{proof}

%\textbf{Proof of Lemma~\ref{lem:rr_property}(v).}

\subsection{RR-DAG Representation of RLSLP in Expected \texorpdfstring{$\delta$}{delta}-Optimal Space}\label{subsec:rrdag}
\begin{figure}[t]
 \begin{center}
		\includegraphics[scale=0.8]{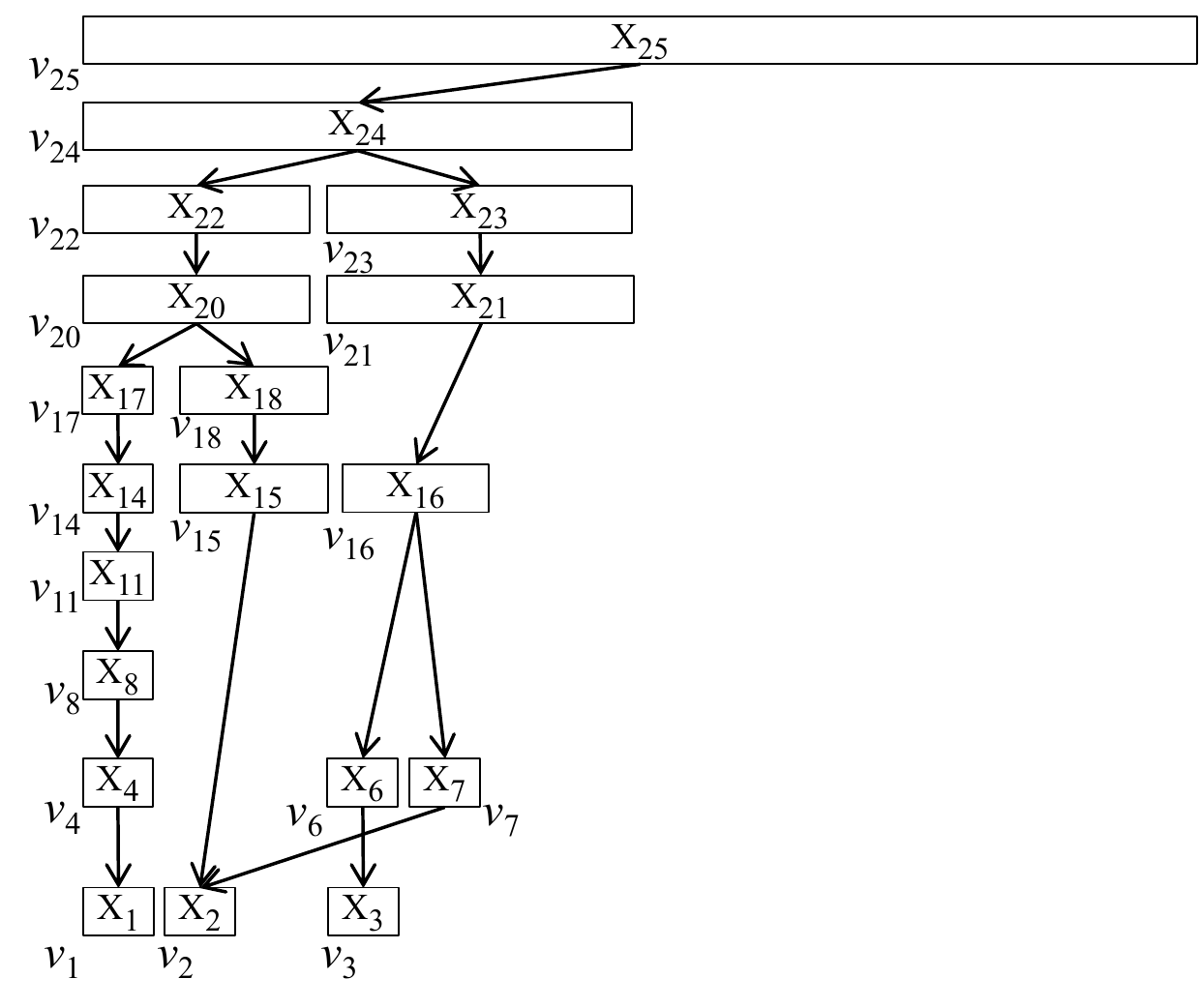}

	  \caption{ 
	  An illustration of the RR-DAG representation of the RLSLP $\mathcal{G}^{R}$ used in Figure~\ref{fig:restricted_recompression}.   
	  }
\label{fig:rrdag}
 \end{center}
\end{figure}

In this section, 
we present a directed acyclic graph (DAG) representation, termed \emph{RR-DAG}, for 
 RLSLP $\mathcal{G}^{R}$ using an expected $\delta$-optimal space. 
RR-DAG can be defined through the derivation tree of $\mathcal{G}^{R}$, and 
the RR-DAG of RLSLP $\mathcal{G}^{R}$ is defined as a triplet $(\mathcal{U}_{\RR}, \mathcal{E}_{\RR}, L_{\RR})$, as follows.
$\mathcal{U}_{\RR}$ is a set of nodes in the RR-DAG. The start symbol and every nonterminal assigned $0$ or $1$ in the RLSLP $\mathcal{G}^{R}$ correspond one-to-one to a node in $\mathcal{U}_{\RR}$. Let the subscript $i$ of nonterminal $X_i \in \mathcal{G}^R$ have the subscript $i$ of $v_{i} \in \mathcal{U}_{\RR}$.  
$\mathcal{E}_{\RR} \subseteq \mathcal{U}_{\RR} \times \mathcal{U}_{\RR}$ is a set of directed edges in the RR-DAG. A directed edge from a node labeled $X_i$ to another node labeled $X_j$ is included in $\mathcal{E}_{\RR}$ if either of the following conditions is met: (a) the node labeled $X_i$ is directly connected to the node labeled $X_j$ in the derivation tree of the RLSLP $\mathcal{G}^{R}$; or (b) there exists a path from the node labeled $X_i$ to the node labeled $X_j$ through intermediate nodes corresponding to nonterminals assigned $-1$.
$L_{\RR}: \mathcal{U}_{\RR} \rightarrow \mathcal{V}$ is a label function that returns the nonterminal $X_{i} \in \mathcal{V}$ corresponding to a given node $v_{i} \in \mathcal{U}_{\RR}$.

%We will use the RR-DAG for two purposes. 
%The first purpose is to store the RLSLP $\mathcal{G}^{R}$ in expected $\delta$-optimal space. 
%To achieve this purpose, 
%we introduce a label function $L_{\prule}: \mathcal{U}_{\RR} \rightarrow \mathcal{D}$. 
%This function returns the production rule with the nonterminal $X_{i} \in \mathcal{V}$ 
%corresponding to a given node $v_{i} \in \mathcal{U}_{\RR}$. 
%Here, if this label function returns a nonterminal $X_{j}$ assigned $-1$, 
%then the nonterminal is represented as a pair of (i) 

%Here, $\mathcal{U}_{\RR} = \{ v_{1}, v_{2}, \ldots, v_{16}\}$; 

%%%%%%%%%%%%%%%%

Six label functions $L_{\level}$, $L_{\prule}$, $L_{\pathP}$, $L_{\assign}$, $L_{\length}$, and $L_{\vOcc}$ are defined on the set $\mathcal{U}_{\RR}$ of nodes.
%A given node $v \in \mathcal{U}_{\RR}$ corresponds to a nonterminal $X \in %\mathcal{V}$ contained in sequence $S^{q}$ for an integer $q \in [0, H]$. 
%There exists the smallest integer $y$ in set $[0, t]$ satisfying either of the following conditions: (a) $y < t$, and function $\locus(\val(X_{i}), y)$ returns a nonterminal with assignment $-1$; 
%(b) (ii) $y = t$.
%Formally, let $y = \min (\{ t \} \cup \{ t^{\prime} \in [0, t-1] \mid (\locus(\val(X_{i}), t^{\prime}) \neq \perp) \land (\assign(\locus(\val(X_{i}), t^{\prime})) = -1) \})$. 
\begin{itemize}
\item $L_{\level}: \mathcal{U}_{\RR} \rightarrow [0, H]$ returns an integer $q$ such that sequence $S^q$ includes the nonterminal $X_{i} \in \mathcal{V}$ corresponding to a given node $v_{i} \in \mathcal{U}_{\RR}$. 
%\end{itemize}
%In the other function, let $q$ be the integer returned by $L_{\level}(v_{i})$ for a given node $v_{i}$. 

%\begin{itemize}
    \item $L_{\prule}: \mathcal{U}_{\RR} \rightarrow \mathcal{D}$ returns the production rule with $X_{i}$ on the left-hand side.     
    \item $L_{\pathP}: \mathcal{U}_{\RR} \rightarrow \mathcal{U}_{\RR}$ is defined using the longest path 
    $\mathbb{P} = v_{j_{1}} \rightarrow v_{j_{2}} \rightarrow \cdots \rightarrow v_{j_{k}}$ on the RR-DAG satisfying the following three conditions: 
    (i) it starts at node $v_{i}$ (i.e., $v_{j_{1}} = v_{i}$), 
    (ii) $v_{j_{x}}$ is a child of node $v_{j_{x-1}}$ for all $x \in [2, k]$, 
    and (iii) all the nodes of this path represent the same string (i.e., $\val(L_{\RR}(v_{j_{1}})) = \val(L_{\RR}(v_{j_{2}})) = \cdots = \val(L_{\RR}(v_{j_{k}}))$).
    If such path $\mathbb{P}$ exists, then the function $L_{\pathP}(v_{i})$ returns the last node $v_{j_{k}}$ of the path; 
    otherwise (i.e., $v_{i}$ is either a leaf or has at least two children), 
    this function returns the given node $v_{i}$. 
    \item $L_{\assign}: \mathcal{U}_{\RR} \rightarrow \{ 0, 1, 2 \}$ returns a value in $\{ 0, 1 \}$ as the assignment $\assign(X_{i})$ of nonterminal $X_{i}$ if the nonterminal is not the start symbol of RLSLP $\mathcal{G}^{R}$; otherwise, the label function returns $2$. 
    \item $L_{\length}: \mathcal{U}_{\RR} \rightarrow [1, n]$ returns the length of the string derived by the nonterminal $X_{i}$ (i.e., $L_{\length}(v_{i}) = |\val(X_{i})|$). 
    \item $L_{\vOcc}: \mathcal{U}_{\RR} \rightarrow \mathbb{N}_{>0}$ returns 
    the number of occurrences of nonterminal $X_{i}$ in the derivation tree of RLSLP $\mathcal{G}^{R}$ (i.e., 
    $L_{\vOcc}(v_{i}) = |\bigcup_{t = 0}^{H} \{ (t, j) \mid S^{t}[j] = X_{i} \text{ for } j \in [1, |S^t|]\}|$). 
\end{itemize}
    
%\begin{itemize}
%\item $L_{\bottom}: \mathcal{U}_{\RR} \rightarrow [-1, H]$ returns the largest integer $p \in [0, q-1]$ satisfying $\locus(\val(X), p) = \perp$ if 
%such integer $p$ exists; otherwise this function returns $-1$. 
%\item $L_{\prule}: \mathcal{U}_{\RR} \rightarrow \mathcal{D}$ returns the production rule with the nonterminal obtained by function $\locus(\val(X_{i}), p+1)$ on the left-hand side if $p+1 \leq q-1$; 
%otherwise, this function returns the production rule with $X_{i}$ on the left-hand side. 
%\item $L_{\assign}: \mathcal{U}_{\RR} \rightarrow \{ 0, 1, 2 \}$ returns a value in $\{ 0, 1 \}$ as the assignment $\assign(X_{i})$ of nonterminal $X_{i}$ if the nonterminal is not the start symbol of RLSLP $\mathcal{G}^{R}$; otherwise, the label function returns $2$. 
%\item $L_{\length}: \mathcal{U}_{\RR} \rightarrow [1, n]$ returns the length of the string derived by the nonterminal $X_{i}$ (i.e., $L_{\length}(v_{i}) = |\val(X_{i})|$). 
%\item $L_{\vOcc}: \mathcal{U}_{\RR} \rightarrow \mathbb{N}_{>0}$ returns 
%the number of occurrences of nonterminal $X_{i}$ in the derivation tree of RLSLP $\mathcal{G}^{R}$ (i.e., 
%$L_{\vOcc}(v_{i}) = |\bigcup_{t = 0}^{H} \{ (t, j) \mid S^{t}[j] = X_{i} \text{ for } j \in [1, |S^t|]\}|$). 
%\end{itemize}

Label function $L_{\prule}$ is used to recover the original RLSLP $\mathcal{G}^{R}$ from its RR-DAG representation; 
$L_{\assign}$ is used to obtain the assignments of nonterminals; 
$L_{\pathP}$ and $L_{\length}$ are used to traverse the derivation tree of RLSLP $\mathcal{G}^{R}$ using the RR-DAG;
$L_{\vOcc}$ is used to appropriately update the RR-DAG when the RLSLP $\mathcal{G}^{R}$ is changed~(see Section~\ref{subsec:rlslp_update}). 

%Four label functions $L_{\level}$, $L_{\bottom}$, $L_{\prule}$, and $L_{\assign}$ are used to 
%recover the original RLSLP $\mathcal{G}^{R}$ from its RR-DAG representation; $L_{\length}$ is used to traverse the derivation tree of RLSLP $\mathcal{G}^{R}$ using the RR-DAG;
%$L_{\vOcc}$ is used to appropriately update the RR-DAG when the RLSLP $\mathcal{G}^{R}$ is changed~(see Section~\ref{subsec:rlslp_update}). 

Figure~\ref{fig:rrdag} illustrates an example of RR-DAG of the RLSLP used in Figure~\ref{fig:restricted_recompression}. 
For example, $L_{\level}(v_{15}) = 4$, 
$L_{\prule}(v_{15}) = X_{5} \rightarrow (X_{2})^{2}$, 
$L_{\assign}(v_{15}) = 0$, 
$L_{\length}(v_{15}) = 2$, 
and $L_{\vOcc}(v_{15}) = 2$. 

\paragraph{Representation of nonterminals.}
In this paper, we represent nonterminals RLSLP $\mathcal{G}^{R}$ using nodes in the RR-DAG for the RLSLP. 
Consider a nonterminal $X_{i} \in \mathcal{V}$, and 
it appears in a sequence $S~{h}$. 
This nonterminal $X_{i}$ is represented as follows. 
If the nonterminal $X_{i}$ corresponds to a node $v_{i} \in \mathcal{U}_{\RR}$ in the RR-DAG (i.e., $L_{\RR}(v_{i}) = X_{i}$), 
then the nonterminal is represented as the node $v_{i}$. 
Otherwise (i.e., any node of the RR-DAG does not correspond to the nonterminal $X_{i}$), 
the nonterminal $X_{i}$ is assigned to $-1$, and let $h^{\prime} \geq 0$ be the smallest integer 
satisfying $|\val(X_{i})| \leq \mu(h^{\prime} + 1)$. 
Then, $h^{\prime} > h$ holds, and 
sequence $S^{h^{\prime}}$ contains a nonterminal $X_{j} \in \mathcal{V}$ such that 
$X_{j}$ and $X_{i}$ derive the same string (i.e., $\val(X_{i}) = \val(X_{j})$). 
This nonterminal $X_{j}$ is uniquely determined by Lemma~\ref{lem:rr_property}~\ref{enum:rr_property:2}, 
and the RR-DAG contains a node $v_{j}$ corresponding to the nonterminal $X_{j}$. 
In this case, the nonterminal $X_{i}$ is represented as a pair of the node $v_{j}$ and integer $h$. 

We show that the production rule $X_{i} \rightarrow \expr_{i}$ with $X_{i}$ on the left-hand side can be obtained from the RR-DAG with label functions. 
If the nonterminal $X_{i}$ is represented as node $v_{i}$, 
then we can obtain the production rule $X_{i} \rightarrow \expr_{i}$ from label function $L_{\prule}(v_{i})$. 
Otherwise, $X_{i}$ is represented as a pair of node $v_{j}$ and integer $h$. 
In this case, node $v_{j}$ has one or two children. 
If the node has only one child, then sequence $\expr_{i}$ can be obtained by the following lemma. 

\begin{lemma}\label{lem:node_representation_one}
    Consider a production rule $X_{i} \rightarrow \expr_{i} \in \mathcal{D}$ for RLSLP $\mathcal{G}^{R}$ 
    satisfying the following two conditions: 
    (A) $X_{i}$ is represented as a pair of a node $v_{j} \in \mathcal{U}_{\RR}$ and integer $h$; 
    (B) $v_{j}$ has only one child $v_{s} \in \mathcal{U}_{\RR}$. 
    Let $X_{j}$ and $X_{s}$ be the two nonterminals corresponding to $v_{j}$ and $v_{s}$, respectively.
    Then, the following two statements hold. 
    \begin{enumerate}[label=\textbf{(\roman*)}]
        \item $h > L_{\level}(v_{s})$; 
        \item if $h = L_{\level}(v_{s}) + 1$, 
        then $\expr_{i} = (X_{s})^{d}$ for $d = |\val(X_{j})| / |\val(X_{s})|$. 
        Otherwise, sequence $S^{h-1}$ contains a nonterminal $X_{s^{\prime}}$ assigned $-1$, 
        and $\expr_{i} = X_{s^{\prime}}$. Here, $X_{s^{\prime}}$ is represented as a pair of $v_{j}$ and $h-1$.
    \end{enumerate}
\end{lemma}
\begin{proof}
    Let $\mathbb{P} = u_{j_{1}} \rightarrow u_{j_{2}} \rightarrow \cdots \rightarrow u_{j_{k}}$ be 
    the longest path in the derivation tree of RLSLP $\mathcal{G}^{R}$ satisfying the following three conditions:
    \begin{enumerate}[label=\textbf{(\alph*)}]
        \item $u_{j_{1}}$ corresponds to nonterminal $X_{j}$; 
        \item node $u_{j_{x}}$ corresponds a nonterminal assigned $-1$ for all $x \in [2, k]$, 
        and $u_{j_{x}}$ is the single child of node $u_{j_{x-1}}$;
        \item node $u_{j_{k}}$ has at least two children, 
        and every child of the node $u_{j_{k}}$ corresponds to $X_{s}$. 
    \end{enumerate}

    From the definition of RR-DAG and the existence of $X_{i}$, 
    such path $\mathbb{P}$ exists, 
    a node $u_{j_{y}}$ corresponds to $X_{i}$. 
    Here, the height of the node $u_{j_{y}}$ is $h$. 

    \paragraph{Proof of Lemma~\ref{lem:node_representation_one}(i).}
    Consider a child $u$ of node $u_{j_{k}}$. 
    $h$ is larger than the height of the child $u$, 
    and the height of $u$ obtained from label function $L_{\level}(v_{s})$. 
    Therefore, $h > L_{\level}(v_{s})$ holds. 

    \paragraph{Proof of Lemma~\ref{lem:node_representation_one}(ii).}
    If $h = L_{\level}(v_{s}) + 1$, 
    then $y = k$ holds. 
    In this case, node $u_{j_{y}}$ has $d$ children, 
    and each child derives string $\val(X_{s})$. 
    Therefore, $\expr_{i} = (X_{s})^{d}$ holds. 

    Otherwise (i.e., $h > L_{\level}(v_{s}) + 1$), 
    $1 < y < k$ holds. 
    For each integer $x \in [2, k]$, 
    node $u_{j_{x}}$ corresponds a nonterminal $X_{j_{x}}$ assigned $-1$ 
    and the nonterminal and $X_{j}$ derive the same string. 
    Here, the nonterminal $X_{j_{x}}$ is represented as a pair of $v_{j}$ and $h^{\prime} - x + 1$ for the height $h^{\prime}$ of node $u_{j_{1}}$. 
    Therefore, we obtain Lemma~\ref{lem:node_representation_one}(ii). 
\end{proof}

If the node has only two children, then sequence $\expr_{i}$ can be obtained by the following lemma. 

\begin{lemma}\label{lem:node_representation_two}
    Consider a production rule $X_{i} \rightarrow \expr_{i} \in \mathcal{D}$ for RLSLP $\mathcal{G}^{R}$ 
    satisfying the following three conditions: 
    (A) $X_{i}$ is represented as a pair of a node $v_{j} \in \mathcal{U}_{\RR}$ and integer $h$; 
    (B) $v_{j}$ has two children $v_{s}$ and $v_{s^{\prime}} \in \mathcal{U}_{\RR}$. 
    Let $X_{j}, X_{s}$, and $X_{s^{\prime}}$ be the three nonterminals corresponding to $v_{j}, v_{s}$, and $v_{s^{\prime}}$, respectively. 
    For simplicity, we assume that $L_{\assign}(v_{s}) \leq L_{\assign}(v_{s^{\prime}})$ holds. 
    Then, the following three statements hold. 
    \begin{enumerate}[label=\textbf{(\roman*)}]
        \item $h > L_{\level}(v_{s})$ and $L_{\level}(v_{s}) = L_{\level}(v_{s^{\prime}})$; 
        \item $L_{\assign}(v_{s}) = 0$ and $L_{\assign}(v_{s^{\prime}}) = 1$; 
        \item if $h = L_{\level}(v_{s}) + 1$, 
        then $\expr_{i} = X_{s}, X_{s^{\prime}}$. 
        Otherwise, sequence $S^{h-1}$ contains a nonterminal $X_{s^{\prime \prime}}$ assigned $-1$, 
        and $\expr_{i} = X_{s^{\prime \prime}}$. Here, $X_{s^{\prime}}$ is represented as a pair of $v_{j}$ and $h-1$.
    \end{enumerate}
\end{lemma}
\begin{proof}
    Lemma~\ref{lem:node_representation_two} can be proved using the same approach used in the proof of Lemma~\ref{lem:node_representation_one}.
\end{proof}

%we can obtain the production rule by the following lemma. 
%\begin{lemma}\label{lem:node_representation}
%    Consider a production rule $X_{i} \rightarrow \expr_{i} \in \mathcal{D}$ for RLSLP $\mathcal{G}^{R}$ 
%    such that $X_{i}$ is represented as a pair of a node $v_{j} \in \mathcal{U}_{\RR}$ and integer $h$. 
%    Here, the node $v_{j}$ has at most two nodes $v_{x}$ and $v_{j} \in \mathcal{U}_{\RR}$. 
%    For simplicity, we assume that $v_{x}$ is a child of $v_{j}$ if $v_{j}$ has a child. 
%
%    \begin{enumerate}[label=\textbf{(\roman*)}]
%        \item $v_{j}$ has at least one child, and $h > L_{\level}(v_{x})$.
%        \item If $h = L_{\level}(v_{x}) + 1$
%        \item If $L_{\level}(v_{j^{\prime}}) = h + 1$, then $X_{i}$ is represented as node $v_{j^{\prime}}$ (i.e., $L_{\RR}(v_{j^{\prime}}) = X_{i}$); 
%        otherwise, $X_{i}$ is represented as a pair of node $v_{j^{\prime}}$ and integer $h+1$.
%    \end{enumerate}
%    Here, if node $v_{j}$ has only one child $v_{j^{\prime}}$, and the label function $L_{\level}(v_{j^{\prime}})$ returns $h-1$, 
%    then the nonterminal $X_{j^{\prime}}$ is represented as the node $v_{j^{\prime}}$; 
%    otherwise, the nonterminal $X_{j^{\prime}}$ is represented as a pair of node $v_{j}$ and integer $h-1$. 
%\end{lemma}
%\begin{proof}
%    This lemma follows from the definition of RR-DAG.
%\end{proof}

Lemma~\ref{lem:node_representation_one} and Lemma~\ref{lem:node_representation_two} indicate that 
we can recover RLSLP $\mathcal{G}^{R}$ from its RR-DAG with label functions. 
Therefore, the RR-DAG can be used to store RLSLP $\mathcal{G}^{R}$.

\paragraph{Four functions to obtain production rules.}
We introduce four functions $f_{\prule, A}$, $f_{\prule, B}$, $f_{\prule, C}$, and $f_{\prule, D}$ to obtain 
the left-hand side of a production rule from its right-hand side. 
These four functions are defined as follows.
\begin{itemize}
    \item $f_{\prule, A}: \mathcal{V} \rightarrow \mathcal{V} \cup \{ \perp \}$ returns $X_{j}$ for a nonterminal $X_{i} \in \mathcal{V}$ assigned $0$ or $1$ 
    if there exists a nonterminal $X_{j} \in \mathcal{V}$ satisfying $X_{j} \rightarrow X_{i} \in \mathcal{D}$; 
    otherwise, it returns a special character $\perp$.
    \item $f_{\prule, B}: \mathcal{V}^{2} \rightarrow \mathcal{V} \cup \{ \perp \}$ returns $X_{j}$ for a pair of nonterminals $X_{i}, X_{i^{\prime}} \in \mathcal{V}$ assigned $0$ or $1$ 
    if there exists a nonterminal $X_{j} \in \mathcal{V}$ satisfying $X_{j} \rightarrow X_{i}, X_{i^{\prime}} \in \mathcal{D}$; 
    otherwise, it returns a special character $\perp$. 
    \item $f_{\prule, C}: \mathcal{V} \times \mathbb{N}_{>0} \rightarrow \mathcal{V} \cup \{ \perp \}$ returns $X_{j}$ for a pair of (i) a nonterminal $X_{i} \in \mathcal{V}$ assigned $0$ or $1$ 
    and (ii) an integer $d \geq 1$ 
    if there exists a nonterminal $X_{j} \in \mathcal{V}$ satisfying $X_{j} \rightarrow (X_{i})^{d}$; 
    otherwise, it returns a special character $\perp$.
    \item $f_{\prule, D}: \Sigma \rightarrow \mathcal{V} \cup \{ \perp \}$ returns $X_{j}$ for a character $c \in \Sigma$ 
    if there exists a nonterminal $X_{j} \in \mathcal{V}$ satisfying $X_{j} \rightarrow c \in \mathcal{D}$; 
    otherwise, it returns a special character $\perp$.
\end{itemize}

The left-hand side of a production rule $X_{i} \rightarrow \expr_{i} \in \mathcal{D}$ can be obtained from 
the four functions $f_{\prule, A}$, $f_{\prule, B}$, $f_{\prule, C}$, or $f_{\prule, D}$ if the right-hand side of the production rule 
does not contain a nonterminal assigned $-1$. 
Similar to label function $L_{\vOcc}$, 
these four functions are used to appropriately update the RR-DAG when the RLSLP $\mathcal{G}^{R}$ is changed.

Each interval node of RR-DAG has at most two children, 
and hence, the number of edges in this DAG is linearly proportional to the number of nodes. 
The following lemma ensures that 
the expected total number of nodes in RR-DAG can be upper-bounded by $O(\delta \log \frac{n \log \sigma}{\delta \log n})$. 
\begin{lemma}\label{lem:rrdag_node_size}
    $\mathbb{E}[|\mathcal{U}_{\RR}|] = O(\delta \log \frac{n \log \sigma}{\delta \log n} )$ 
    for the RR-DAG $(\mathcal{U}_{\RR}, \mathcal{E}_{\RR}, L_{\RR})$ of the RLSLP $\mathcal{G}^{R} = (\mathcal{V}, \Sigma, \mathcal{D}, E)$.  
    Here, $n$ is the length of input string $T$; $\sigma$ is the alphabet size for $T$; 
    $\delta$ is the substring complexity of $T$. 
\end{lemma}
\begin{proof}
See Section~\ref{subsec:proof_rrdag_node_size}.
\end{proof}

The RLSLP $\mathcal{G}^{R}$ can be recovered from its RR-DAG, 
and the expected total number of nodes and edges in RR-DAG is upper-bounded by $O(\delta \log \frac{n \log \sigma}{\delta \log n})$. 
The two facts indicate that 
the RLSLP can be stored in expected $\delta$-optimal space. 
We then present several data structures to store the RR-DAG in expected $\delta$-optimal space.

%%%%%%%%%%%%%%%%%%%%%%%%%%%%%%%%%%%%%%%%%%%%%%%%%%%%%%%%%%%%%%%%%%%%%%%%%%%
\subsubsection{Dynamic Data Structures for RR-DAG}\label{subsubsec:rrdag_ds}
The RR-DAG of RLSLP $\mathcal{G}^{R}$ is represented using a doubly linked list storing several pieces of information on its nodes 
and four B-tree data structures (i.e., a type of self-balancing search tree). 
Each element in the doubly linked list stores the following nine pieces of information on a node $v_{i} \in \mathcal{U}_{\RR}$: (i) $L_{\RR}(v_{i})$, (ii) $L_{\level}(v_{i})$, 
(iii) $L_{\prule}(v_{i})$, (iv) $L_{\pathP}(v_{i})$, (v) $L_{\assign}(v_{i})$, 
(vi) $L_{\length}(v_{i})$, (vii) $L_{\vOcc}(v_{i})$, (viii) the subscript $i$ of the node $v_{i}$, 
(ix) the set of all directed edges starting at node $v_{i}$. The number of such directed edges is at most $2$. 
See Section~\ref{subsec:rrdag} for these label functions 
$L_{\RR}$, $L_{\level}$, $L_{\prule}$, $L_{\pathP}(v_{i})$, $L_{\assign}$, $L_{\length}$, and $L_{\vOcc}$. 

For the RR-DAG with $|\mathcal{U}_{\RR}|$ nodes, the doubly linked list consists of $|\mathcal{U}_{\RR}|$ elements in total. 
Each of the above nine values can be stored in $O(B)$ bits for machine word size $B$. 
Therefore, the doubly linked list requires $O(|\mathcal{U}_{\RR}| B)$ bits. 

Generally, a B-tree is a data structure storing pairs of key and value, and it can be used for 
referring to the value from a given key. 
Consider $k$ production rules $X_{i_{1}} \rightarrow \expr_{i_{1}}$, $X_{i_{2}} \rightarrow \expr_{i_{2}}$, 
$\ldots$, $X_{i_{k}} \rightarrow \expr_{i_{k}}$ in set $\mathcal{D}$ 
such that the sequence $\expr_{i_{x}}$ of each production rule $X_{i_{x}} \rightarrow \expr_{i_{x}}$ is a nonterminal $X_{j_{x}}$ assigned $0$ or $1$. 
For simplicity, we assume that $j_{1} < j_{2} < \cdots j_{k}$. 
For each production rule $X_{i_{x}} \rightarrow X_{j_{x}}$, 
its key and value are defined as $X_{j_{x}}$ and $X_{i_{x}}$, respectively. 
A B-tree storing these pairs of key and value is used to compute function $f_{\prule, A}$. 
Here, the keys $X_{j_{1}}$, $X_{j_{2}}$, $\ldots$, $X_{j_{k}}$ are sorted in increasing order of their subscripts $j_{1}, j_{2}, \ldots,j_{k}$. 
Similarly, we use three B-trees for computing $f_{\prule, B}$, $f_{\prule, C}$, and $f_{\prule, D}$. 
These four B-trees stores at most $|\mathcal{U}_{\RR}|$ pairs in total 
and require $O(|\mathcal{U}_{\RR}| B)$ bits.

The total space of the data structures for the RR-DAG of RLSLP $\mathcal{G}^{R}$ is $O(|\mathcal{U}_{\RR}| B)$ bits of space. 
Lemma~\ref{lem:rrdag_node_size} ensures that the expected value of $|\mathcal{U}_{\RR}|$ is $O(\delta \log \frac{n \log \sigma}{\delta \log n})$ 
for input string $T$ of length $n$, with an alphabet size of $\sigma$, and a substring complexity of $\delta$. 
Therefore, if the machine word size $B$ is $O(\log n)$, 
then the RR-DAG of RLSLP $\mathcal{G}^{R}$ can be stored in expected $\delta$-optimal space. 

\subsubsection{Queries}
We explain queries supported by the RR-DAG for RLSLP $\mathcal{G}^{R}$.  
The RR-DAG is used to traverse the derivation tree of the RLSLP. 
The following lemma states the running time of basic queries to traverse the derivation tree. 

\begin{lemma}\label{lem:basic_operations_on_dev_tree}
    Consider the RR-DAG $(\mathcal{U}_{\RR}, \mathcal{E}_{\RR}, L_{\RR})$ of the RLSLP $\mathcal{G}^{R} = (\mathcal{V}, \Sigma, \mathcal{D}, E)$ 
    and a node $u_{i}$ in the derivation tree of RLSLP $\mathcal{G}^{R}$. 
    Here, the node $u_{i}$ corresponds to a nonterminal $X_{i} \in \mathcal{V}$. 
    Let $\mathbb{P} = u_{j_{1}} \rightarrow u_{j_{2}} \rightarrow \cdots \rightarrow u_{j_{k}}$ be 
    the longest path on the derivation tree of RLSLP $\mathcal{G}^{R}$ satisfying two conditions: 
    (A) it starts at node $u_{i}$; 
    (B) $u_{j_{x}}$ is the single child of node $u_{j_{x-1}}$ for all $x \in [2, k]$.
    If such path $\mathbb{P}$ does not exist, then we define $\mathbb{P}$ as a path of length $0$ starting at node $u_{i}$.  
    Using the nonterminal $X_{i}$, 
    the RR-DAG can support the following five queries on the nonterminal $u$ in $O(1)$ time. 
    \begin{enumerate}[label=\textbf{(\roman*)}]
        \item return $|\val(X_{i})|$;
        \item return the height $h$ of node $u$;
        \item return the production rule with the nonterminal $X_{i}$ on the left-hand side;
        \item return the nonterminal corresponding to the $i$-th child of node $u_{i}$ for a given integer $i \in [1, m]$;
        \item return the nonterminal corresponding to the last node $u_{j_{k}}$ of the path $\mathbb{P}$. 
    \end{enumerate}
    Here, $m$ is the number of children of node $u_{i}$. 
\end{lemma}
\begin{proof}
    If $X_{i}$ corresponds to a node $v_{i}$ in the RR-DAG, then $X_{i}$ is represented as the node $v_{i}$; 
    otherwise, $X_{i}$ is represented as a pair of a node $v_{j} \in \mathcal{U}_{\RR}$ and integer $h$. 
    The proof of Lemma~\ref{lem:basic_operations_on_dev_tree} is as follows. 

    \paragraph{Proof of Lemma~\ref{lem:basic_operations_on_dev_tree}(i).}
    If $X_{i}$ corresponds to node $v_{i}$, then $L_{\length}(v_{i}) = |\val(X_{i})|$; 
    otherwise, $L_{\length}(v_{j}) = |\val(X_{i})|$ holds. 
    This label function $L_{\length}$ can be computed in $O(1)$ time.

    \paragraph{Proof of Lemma~\ref{lem:basic_operations_on_dev_tree}(ii).}
    If $X_{i}$ corresponds to node $v_{i}$, then $L_{\level}(v_{i}) = h$; 
    otherwise, $X_{i}$ is represented as pair $(v_{j}, h)$. 
    The label function $L_{\level}$ can be computed in $O(1)$ time.

    \paragraph{Proof of Lemma~\ref{lem:basic_operations_on_dev_tree}(iii).}
    Consider the production rule $X_{i} \rightarrow \expr_{i}$ with the nonterminal $X_{i}$ on the left-hand side.
    If $X_{i}$ corresponds to node $v_{i}$, then $L_{\prule}(v_{i}) = (X_{i} \rightarrow \expr_{i})$ holds, 
    and this label function can be computed in $O(1)$ time. 
    Otherwise, we can compute sequence $\expr_{i}$ using Lemma~\ref{lem:node_representation_one} and Lemma~\ref{lem:node_representation_two} in $O(1)$ time. 

    \paragraph{Proof of Lemma~\ref{lem:basic_operations_on_dev_tree}(iv).}
    The $i$-ht child of node $u_{i}$ has the $i$-th nonterminal of sequence $\expr_{i}$ as a label. 
    We already showed that this sequence can be computed in $O(1)$ time. 

    \paragraph{Proof of Lemma~\ref{lem:basic_operations_on_dev_tree}(v).}
    There exists three cases: 
    \begin{enumerate}[label=\textbf{(\alph*)}]
        \item $u_{i}$ is a leaf or has at least two children;
        \item $u_{i}$ has a single child, and $X_{i}$ is represented as node $v_{i}$;
        \item $u_{i}$ has a single child, and $X_{i}$ is represented as pair $(v_{j}, h)$.
    \end{enumerate}

    For case (a), 
    the $\mathbb{P}$ is defined as a path of length $0$ starting at node $u_{i}$. 
    In this case, the fifth query of Lemma~\ref{lem:basic_operations_on_dev_tree} returns $X_{i}$. 

    For case (b), the length of the path $\mathbb{P}$ is at least $1$. 
    If the last node $u_{j_{k}}$ of the path $\mathbb{P}$ is a leaf, 
    then the last node $u_{j_{k}}$ corresponds to a leaf $v_{j_{k}}$ in the RR-DAG. 
    The node $v_{i}$ is an ancestor of the leaf $v_{j_{k}}$. 
    %Let $v_{w}$ be a node in the RR-DAG such that 
    %it is a parent of $v_{j_{k}}$ and a descendant of $v_{i}$. 
    Let $\mathbb{P}_{1}$ be the shortest path starting at node $v_{i}$ and ending at node $v_{j_{k}}$ in the RR-DAG. 
    Then, every node of the path $\mathbb{P}_{1}$ has only one child because 
    every node of the path $\mathbb{P}$ has at most one child. 
    In this case, label function $L_{\pathP}(v_{i})$ returns node $v_{j_{k}}$. 
    Two nodes $u_{j_{k}}$ and $v_{j_{k}}$ has the same label.     
    We can compute the two label functions $L_{\pathP}(v_{i})$ and $L_{\RR}(v_{j_{k}})$ in $O(1)$ time. 
    Therefore, we can answer the fifth query of Lemma~\ref{lem:basic_operations_on_dev_tree} in $O(1)$ time. 

    Otherwise (i.e., the last node $u_{j_{k}}$ of the path $\mathbb{P}$ is not a leaf), 
    $u_{j_{k}}$ has at least two children. 
    Let $u_{j_{k+1}}$ be a child of the node $u_{j_{k}}$. 
    Then, the child $u_{j_{k+1}}$ corresponds to a node $v_{j_{k+1}}$ in the RR-DAG 
    because each child of $u_{j_{k}}$ is assigned $0$ or $1$. 
    Let $v_{w}$ be a node in the RR-DAG such that 
    it is a parent of $v_{j_{k+1}}$ and a descendant of $v_{i}$. 
    Then, label function $L_{\pathP}(v_{i})$ returns $v_{w}$ 
    because $v_{i}$ and $v_{w}$ represent the same string (i.e., $\val(L_{\RR}(v_{i})) = \val(L_{\RR}(v_{w}))$), 
    and $v_{w}$ and $v_{j_{k+1}}$ represent distinct strings. 
    Therefore, we can obtain the node $v_{w}$ in $O(1)$ time by the label function $L_{\pathP}(v_{i})$. 

    Let $X_{j_{k}}$ be the nonterminal corresponding to node $u_{j_{k}}$. 
    If $L_{\level}(v_{w}) = L_{\level}(v_{j_{k+1}}) + 1$ (i.e., node $u_{j_{k}}$ correspond to node $v_{w}$), 
    then the nonterminal $X_{j_{k}}$ is represented as $v_{w}$. 
    Otherwise, the nonterminal $X_{j_{k}}$ is represented as a pair of node $v_{w}$ and integer $L_{\level}(v_{j_{k+1}}) + 1$. 
    We can compute the label function $L_{\level}$ in $O(1)$ time. 
    Therefore, we can answer the fifth query of Lemma~\ref{lem:basic_operations_on_dev_tree} in $O(1)$ time. 

    For case (c),         
    let $u_{w^{\prime}}$ be the lowest ancestor of $u_{i}$ corresponding to a node $v_{w^{\prime}}$ in the RR-DAG. 
    Then, $v_{w^{\prime}} = v_{j}$ holds; 
    $u_{w^{\prime}}$ and $u_{i}$ represent the same string; 
    $v_{w^{\prime}}$ has a single child.  
    Let $\mathbb{P}^{\prime}$ be the path starting at node $u_{w^{\prime}}$ and ending at node $u_{j_{k}}$. 
    Then, for the node $u_{w^{\prime}}$, 
    the fifth query of Lemma~\ref{lem:basic_operations_on_dev_tree} returns the nonterminal corresponding to the last node of the path $\mathbb{P}^{\prime}$. 
    This query for $u_{w^{\prime}}$ can be answered by the algorithm used in the proof for case (b) 
    because $u_{w^{\prime}}$ has a single child, 
    and the nonterminal of $u_{w^{\prime}}$ is represented as $v_{j}$. 
    The nonterminal corresponding to $v_{j}$ can be obtained from label function $L_{\RR}(v_{j})$ in $O(1)$ time. 
    Therefore, we can answer the fifth query of Lemma~\ref{lem:basic_operations_on_dev_tree} in $O(1)$ time for case (c).
\end{proof}

\paragraph{Random access, LCE, and reversed LCE queries.}
Random access query $\RAQ(i)$ returns the $i$-th character $T[i]$ of input string $T$ for a given position $i \in [1, n]$ in input string $T$. 
LCE query $\LCEQ(i, j)$ returns the length $\ell$ of the longest common prefix between two suffixes $T[i..n]$ and $T[j..n]$ of for a given pair of two positions $i, j \in [1, n]$ in input string $T$~(i.e., $\ell = |\lcp(T[i..n], T[j..n])|$). 
Similarly, reversed LCE query $\rLCEQ(i, j)$ returns the length $\ell^{\prime}$ of the longest common suffix between two prefixes $T[1..i]$ and $T[1..j]$ for a given pair of two positions $i, j \in [1, n]$ in input string $T$ (i.e., $\ell^{\prime} = |\lcs(T[1..i], T[1..j])|$). 
Kempa and Kociumaka present algorithms answering random access, LCE, and reversed LCE queries in $O(H)$ time using 
the derivation tree of RLSLP $\mathcal{G}^{R}$ (Theorem 5.24 and Theorem 5.25 in \cite{DBLP:journals/corr/abs-2308-03635}). 
Here, $H$ is the height of the derivation tree, 
and their algorithms assumes that the five basic queries of Lemma~\ref{lem:basic_operations_on_dev_tree} can be executed in $O(1)$ time. 
Their algorithms use the RLSLP computed by the restricted recompression that runs in a deterministic way. 
The restricted recompression used in this paper is a randomized version, 
but their algorithms work on the derivation tree constructed by the randomized algorithm. 
Therefore, the RR-DAG can support random access, LCE, and reversed LCE queries in $O(H)$ time. 

The following theorem summarizes RLSLP $\mathcal{G}^{R}$ and its RR-DAG. 
\begin{theorem}\label{theo:rr_dag_summary}
    Consider the RR-DAG of the RLSLP $\mathcal{G}^{R} = (\mathcal{V}, \Sigma, \mathcal{D}, E)$ constructed by the restricted recompression for input string $T$ of length $n$ with an alphabet size of $\sigma$, and a substring complexity of $\delta$. 
    Let $H$ be the height of the derivation tree of RLSLP $\mathcal{G}^{R}$. 
    The following four statements hold: 
    \begin{enumerate}[label=\textbf{(\roman*)}]
    \item the restricted recompression algorithm runs in expected $O(n)$ time \cite{9961143}; 
    \item $H \leq 2(w+1) \log_{8/7} (4n) + 2$ holds with probability $1 - (1/n^{w})$ for all integer $w \geq 1$; 
    \item let $|\mathcal{U}_{\RR}|$ be the number of nodes in the RR-DAG of RLSLP $\mathcal{G}^{R}$. 
    Then, the data structures for the RR-DAG introduced in Section~\ref{subsubsec:rrdag_ds} requires $O(|\mathcal{U}_{\RR}| B)$ bits of space for machine word size $B$, 
    and $\mathbb{E}(|\mathcal{U}_{\RR}|) = O(\delta \log \frac{n \log \sigma}{\delta \log n})$ holds;
    \item using the data structures for the RR-DAG, 
    we can support the following three queries: 
    %(A) path query $\pathQ(i)$ in $O(1 + H)$ time, 
    (A) random access query $\RAQ(i)$ in $O(H)$ time, 
    (B) LCE query $\LCEQ(i, j)$ in $O(H)$ time, 
    and (C) reversed LCE query $\rLCEQ(i, j)$ in $O(H)$ time. 
\end{enumerate}
\end{theorem}

\subsubsection{Proof of Lemma~\ref{lem:rrdag_node_size}}\label{subsec:proof_rrdag_node_size}
We introduce a susbet $\mathcal{V}_{A}$ of set $\mathcal{V}$. 
This subset $\mathcal{V}_{A}$ consists of nonterminals such that 
the production rule with each nonterminal $X_{i} \in \mathcal{V}^{h}_{A}$ on the left-hand side 
takes form $X_{i} \rightarrow c$, $X_{i} \rightarrow X_{j}, X_{k}$, or $X_{i} \rightarrow (X_{j})^{d}$. 
In other words, the subset $\mathcal{V}_{A}$ consists of nonterminals such that 
each nonterminal does not produce a single nonterminal. 
Kociumaka et al. showed that the expected number of nonterminals in the subset $\mathcal{V}_{A}$ 
is $O(\delta \log \frac{n \log \sigma}{\delta \log n})$ (Proposition V.19 in \cite{9961143}).

Similarly, we introduce a susbet $\mathcal{V}^{h}_{B}$ of set $\mathcal{V}$ for each integer $h \in [0, H]$. 
This subset $\mathcal{V}^{h}_{B}$ consists of nonterminals such that 
each nonterminal $X_{i} \in \mathcal{V}^{h}_{B}$ satisfies the following three conditions: 
\begin{itemize}
    \item it appears in sequence $S^{h}$;
    \item it is assigned $0$ or $1$ (i.e., $\assign(X_{i}) \in \{ 0, 1 \}$); 
    \item it produce a single nonterminal (i.e., there exists a nonterminal $X_{j}$ satisfying $X_{i} \rightarrow X_{j} \in \mathcal{D}$).
\end{itemize}
The following proposition shows that 
the expected number of nodes in set $\bigcup_{h=0}^{H} \mathcal{V}^{h}_{B}$ can be bounded by $O(\delta \log \frac{n \log \sigma}{\delta \log n})$. 

\begin{proposition}\label{prop:vb_size}
    $\mathbb{E}[|\bigcup_{h=0}^{H} \mathcal{V}^{h}_{B}|] = O(\delta \log \frac{n \log \sigma}{\delta \log n})$. 
\end{proposition}
\begin{proof}
This proposition can be proved by modifying the proof of Corollary V.17 in \cite{9961143}. 
Let $\tau = \lceil 2\log_{8/7}\log_{\sigma} \delta \rceil$ 
and $\tau^{\prime} = \lceil 2\log_{8/7} \frac{n}{\delta} \rceil$. 
Then, Proposition~\ref{prop:vb_size} follows from the following three statements.
\begin{enumerate}[label=\textbf{(\roman*)}]
    \item $\sum_{h = 0}^{\tau-1} |\mathcal{V}^{h}_{B}| = O(\delta)$; 
    \item $\mathbb{E}[\sum_{h = \tau}^{\tau^{\prime}} |\mathcal{V}^{h}_{B}|] = O(\delta \log \frac{n \log \sigma}{\delta \log n})$;
    \item $\mathbb{E}[\sum_{h = \tau^{\prime}}^{H} |\mathcal{V}^{h}_{B}|] = O(\delta)$.
\end{enumerate}        

\paragraph{Proof of statement (i).}
Each nonterminal of set $\mathcal{V}^{h}_{B}$ derives a substring of length at most $\mu(h+1)$ on $T$ 
because the nonterminal is assigned $-1$. 
Lemma~\ref{lem:rr_property}~\ref{enum:rr_property:2} shows that 
distinct nonterminals of $\mathcal{V}^{h}_{B}$ derive distinct substrings. 
These two observations indicate that $|\mathcal{V}^{h}_{B}| \leq \sum_{t = 1}^{\lceil \mu(h+1) \rceil} \sigma^{t}$ 
for the alphabet $\Sigma$ of size $\sigma$. 
Here, $\sum_{t = 1}^{\lceil \mu(h+1) \rceil} \sigma^{t} = O(\sigma^{\lceil \mu(h+1) \rceil})$.
Therefore, 
\begin{equation*}
    \begin{split}
        \sum_{h = 0}^{\tau-1} |\mathcal{V}^{h}_{B}| & = \sum_{t = 0}^{\lceil 2\log_{8/7}\log_{\sigma} \delta \rceil - 1} O(\sigma^{\lceil \mu(t+1) \rceil})\\
        &= \sum_{i = 0}^{\lceil \log_{\sigma} \delta \rceil } O(\sigma^{i}) \\
        &= O(\delta).
    \end{split}
\end{equation*}

\paragraph{Proof of statement (ii).}
Kociumaka et al. showed that the expected number of nonterminals in the subset $|\mathcal{V}^{h}_{B}|$ 
is $O(\delta)$ for every $h \geq 0$ (Lemma V.14 in \cite{9961143}). 
We obtain $\tau^{\prime} - \tau + 1 = O( \log \frac{n \log \sigma}{\delta \log \delta})$. 
Kociumaka et al. showed that 
$\log \frac{n \log \sigma}{\delta \log \delta} = O(\log \frac{n \log \sigma}{\delta \log n})$ holds 
(the proof of Lemma II.4 in \cite{9961143}). 
Therefore, $\mathbb{E}[\sum_{h = \tau}^{\tau^{\prime}} |\mathcal{V}^{h}_{B}|] = (\tau^{\prime} - \tau + 1) \cdot O(\delta) = O(\delta \log \frac{n \log \sigma}{\delta \log n})$.
%\begin{equation*}
%    \begin{split}
%        \mathbb{E}[\sum_{h = \tau}^{\tau^{\prime}} |\mathcal{V}^{h}_{B}|] & = (\tau^{\prime} - \tau + 1) \cdot O(\delta) \\
%        &= O(\delta \log \frac{n \log \sigma}{\delta \log n}).
%    \end{split}
%\end{equation*}

\paragraph{Proof of statement (iii).}
We prove $\mathbb{E}[|\mathcal{V}^{h}_{B}|] = O(\frac{7}{8}^{h/2} n)$. 
$|\mathcal{V}^{h}_{B}| \leq |S^{h}|$ follows from the definition of the set $\mathcal{V}^{h}_{B}$. 
$|S^{h}| = 1 \Rightarrow |\mathcal{V}^{h}_{B}| = 0$ holds 
because (A) $h = H$ if $|S^{h}| = 1$, 
and (B) the root of the derivation tree of RLSLP $\mathcal{G}^{R}$ always has at least two children. 
These two observations indicate that $\mathbb{E}[|\mathcal{V}^{h}_{B}|] \leq \mathbb{E}[|S^{h}|] - \mathbb{P}[|S^{h}| = 1]$ holds. 
Kociumaka et al. showed that 
$\mathbb{E}[|S^{h}|] - \mathbb{P}[|S^{h}| = 1] \leq \frac{8n}{\mu(h+1)}$ holds (the proof of Corollary V.17 in \cite{9961143}). 
$\frac{8n}{\mu(h+1)} = O(\frac{7}{8}^{h/2} n)$ follows from the definition of function $\mu$.
Therefore, we obtain $\mathbb{E}[|\mathcal{V}^{h}_{B}|] = O(\frac{7}{8}^{h/2} n)$. 

Kociumaka et al. showed that $\sum_{h = \tau^{\prime}}^{\infty} \frac{7}{8}^{h/2} n = O(\delta)$ (the proof of Corollary V.17 in \cite{9961143}). 
%$\mathbb{E}[\sum_{h = \tau^{\prime}}^{H} |\mathcal{V}^{h}_{B}|] = O(\frac{7}{8}^{h/2} n)$
Therefore, $\mathbb{E}[\sum_{h = \tau^{\prime}}^{H} |\mathcal{V}^{h}_{B}|] = O(\sum_{h = \tau^{\prime}}^{\infty} \frac{7}{8}^{h/2} n) = O(\delta)$. 
\end{proof}

Each node of the RR-DAG corresponds to a nonterminal in two subsets $\mathcal{V}_{A}$ and $\bigcup_{h=0}^{H} \mathcal{V}^{h}_{B}$. 
We already showed that the expected number of nonterminals in the two subsets  
is $O(\delta \log \frac{n \log \sigma}{\delta \log n})$. 
Therefore, we obtain $\mathbb{E}[|\mathcal{U}_{\RR}|] = O(\delta \log \frac{n \log \sigma}{\delta \log n})$.

\section{Interval Attractors via Restricted Recompression}\label{sec:RASSO}
\begin{figure}[t]
 \begin{center}
		\includegraphics[scale=0.9]{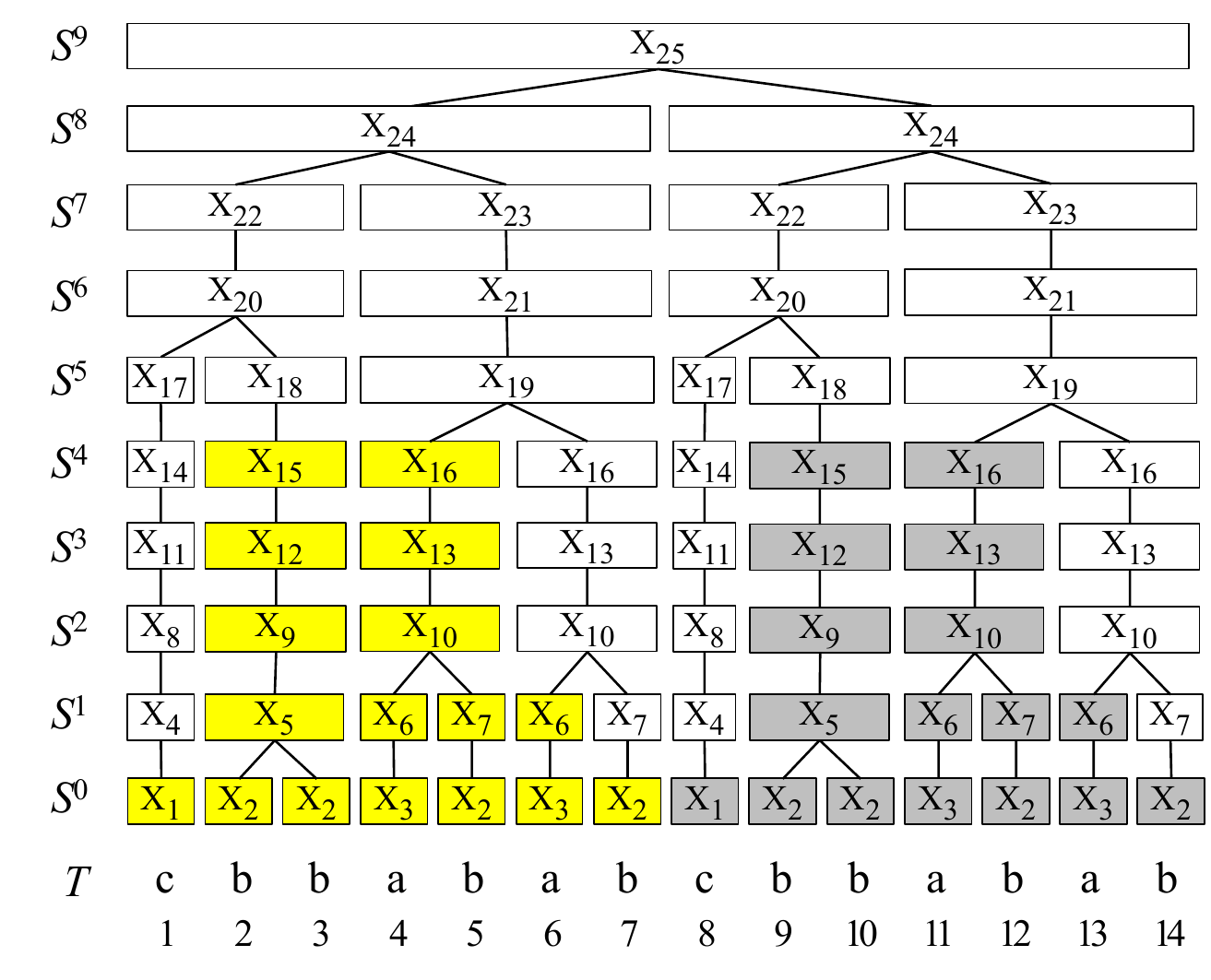}

	  \caption{ 
	  An illustration of the intervals of $A(1, 7)$ and $A(8, 14)$. 
   Here, we used the derivation tree of the RLSLP in Figure~\ref{fig:restricted_recompression}. 
   Yellow rectangles represent the nonterminals in the intervals $[1, 7]$, $[2, 5]$, $[2, 3]$, $[2, 3]$, $[2, 3]$ of $A(1, 7)$. 
   Gray rectangles represent the nonterminals in the intervals $[8, 14]$, $[8, 11]$, $[6, 7]$, $[6, 7]$, $[6, 7]$ of $A(8, 14)$.    
	  }
\label{fig:finterval}
 \end{center}
\end{figure}
This section shows that interval attractors can be constructed from the RLSLP $\mathcal{G}^{R}$ built by the restricted recompression. 

\subsection{Partition \texorpdfstring{$\Delta(u)$}{Delta} via Restricted Recompression}\label{subsec:RR_delta}
%\tnnote*{MARK: need to add an explanation}{

Consider the $(H+1)$ sequences $S^{0}, S^{1}, \ldots, S^{H}$ of nonterminals representing 
the derivation tree of RLSLP $\mathcal{G}^{R}$. 
To construct partition $\Delta(u)$, 
we introduce a function $f_{\interval}$, which is defined as a mapping from an interval $[s^h, e^h]$ on $S^h$ to an interval $[s^{h+1}, e^{h+1}]$ on $S^{h+1}$ for $h \in [0, H-1]$.
If such interval $[s^{h+1}, e^{h+1}]$ does not exist, $f_{\interval}$ returns $\perp$ for this input. 
Formally, $f_{\interval}$ is defined as follows. 
%}
\begin{definition}\label{def:f_interval}
Let $f_{\interval}$ be a recursive function defined on intervals $[s^h, e^h]$ on $S^h$ for $h \in [0,H-1]$.
It takes the interval $[s^h, e^h]$ as input and returns the interval $[s^{h+1}, e^{h+1}]$ as output for $h \in [0,H-1]$.
Let $s^{h}$-th and $e^{h}$-th nonterminals of $S^{h}$ derive  $T[x_{s}..y_{s}]$ and $T[x_{e}..y_{e}]$, respectively.
The position $s^{h+1}$ in $S^{h+1}$ is defined as the smallest $\tau$ where $x^\prime$, from the substring $T[x^{\prime}..y^{\prime}]$ derived by $S^{h+1}[\tau]$, satisfies at least one of the following conditions:
\begin{enumerate}[label=\textbf{(\roman*)}]
    \item $x^{\prime} \in [x_{s} + 1, x_{e} - 1]$; 
    \item There exists a position $\tau^{\prime} \in  [s^{h}, e^{h}-1]$ on $S^{h}$ such that $S^h[\tau^{\prime}]$ derives $T[x_{\tau^{\prime}}..y_{\tau^{\prime}}]$, meeting the following conditions:
    (a) $y_{\tau^{\prime}} - x_{\tau^{\prime}} + 1 > \lfloor \mu(h+1) \rfloor$ and (b) $x^{\prime} \in \{ x_{\tau^{\prime}}, y_{\tau^{\prime}} + 1 \}$; 
    \item $x^{\prime} = x_{e}$ and $e^{0} - x^{\prime} + 1 > \sum_{w = 1}^{h+1} \lfloor \mu(w) \rfloor$; 
    \item $x^{\prime} = x_{s}$ and $x^{\prime} - s^{0} > \sum_{w = 1}^{h+1} \lfloor \mu(w) \rfloor$. 
\end{enumerate}
Similarly, 
$e^{h+1}$ is the largest position in $S^{h+1}$ such that (i) $S^{h+1}[e^{h+1}]$ derives $T[x^{\prime}..y^{\prime}]$ and
(ii) $x^\prime$ satisfies at least one of above four conditions (i)-(iv). 
If such $x^\prime$ satisfying any of the four conditions for $s^{h+1}$ or $e^{h+1}$ does not exist, $f_{\interval}$ returns $\perp$.
\end{definition}

The function \(f_{\interval}\) can create a sequence of segments $[s^h, e^h]$ within the sequence $S^h$ for every $h \in [0, k]$, as follows:
(i) For the base case, $[s^0,e^0] = [s,e]$; 
(ii) For any $h \in [1, k]$, $[s^h,e^h] = f_{\interval}(s^{h-1},e^{h-1})$; (iii) The process concludes either when $k=H$ or $f_{\interval}(s^k,e^k) = \perp$.
For an interval $[s,e]$ on $T$, $A(s,e)$ denotes a sequence of segments 
created by \(f_{\interval}\). Formally, $A(s,e) = [s^0,e^0], [s^1,e^1], \ldots, [s^k,e^k]$. The last element $[s^k,e^k]$ of $A(s,e)$ is referred to as the \emph{tail} of $A(s,e)$. 

We explain a property of the interval obtained from the function $f_{\interval}$. 
For simplicity, let $\gamma_{i}$ represent the starting position of the substring on $T$ derived from each nonterminal $S^{h+1}[i]$ in sequence $S^{h+1}$. 
The purpose of the function $f_{\interval}$ is to recursively compute a sufficiently long interval $[s^{h+1}, e^{h+1}]$ on sequence $S^{h+1}$ such that 
each nonterminal $S^{h+1}[i]$ within segment $S^{h+1}[s^{h+1}..e^{h+1}]$ satisfies the following condition for any occurrence $T[s^{\prime}..e^{\prime}]$ of string $T[s..e]$ on $T$: 
there exists a nonterminal $S^{h+1}[i^{\prime}]$ in sequence $S^{h+1}$ such that the distance from $s'$ to $\gamma_{i^{\prime}}$ is equal to the distance from $s$ to $\gamma_{i}$. 
Such an interval $[s^{h+1}, e^{h+1}]$ can be computed using the substring $T[s..e]$ because 
the restricted recompression uses local contexts to determine the structure of the derivation tree. 
Similar ideas are used to efficiently compute various queries on the string $T$ represented as an RLSLP~\cite{DBLP:conf/cpm/I17,DBLP:journals/corr/GawrychowskiKKL15}, 
but the definition of function $f_{\interval}$ is more complicated for finding a sufficiently long interval.

%We introduce a new concept termed \emph{interval attractors} defined within the derivation tree by restricted recompression. 
%The interval attractors are defined as a set of intervals on string $T$, constructed based on the results of restricted recompression and the function $f_{\interval}$. 
%The function 

Figure~\ref{fig:finterval} illustrates two sequences $A(1, 7) = [1, 7]$, $[2, 5]$, $[2, 3]$, $[2, 3]$, $[2, 3]$ and $A(8, 14) = [8, 14]$, $[8, 11]$, $[6, 7]$, $[6, 7]$, $[6, 7]$. 
The tails of the two sequences $A(1, 7)$ and $A(8, 14)$ are $[2, 3]$ and $[6, 7]$, respectively. 
Consider the recursive function $f_{\interval}(s^{1}, e^{1})$ used to compute $A(1, 7)$. 
Here, $[s^{1}, e^{1}] = [2, 5]$, $x_{s} = 2$, $x_{e} = 6$, and $h = 1$. 
In this case, the second position of sequence $S^{2}$ is the smallest position satisfying 
at least one of the four conditions in Definition~\ref{def:f_interval} 
because $x^{\prime} \in [x_{s}+1, x_{e}-1]$ holds for 
the substring $T[x^{\prime}..y^{\prime}]$ derived by the second nonterminal of $S^{2}$. 
On the other hand, 
the third position of sequence $S^{2}$ is the largest position satisfying 
at least one of the four conditions in Definition~\ref{def:f_interval}. 
Therefore, the recursive function $f_{\interval}(s^{1}, e^{1})$ outputs interval $[2, 3]$ in sequence $S^{2}$. 

By using the function $f_{\interval}$, 
we construct partition $\Delta(u) \subseteq \Delta$ as follows. 
\begin{definition}\label{def:RR_Delta}
Consider the node $u \in \mathcal{U}$ corresponding to each position $s^{h}$ of sequence $S^{h}$ 
at $h \in [0, H]$. 
Then, partition $\Delta(u)$ is defined as a subset of set $\Delta$ such that 
for any $[s, e] \in \Delta(u)$, 
there exists the tail $[s^k,e^k]$ of $A(s, e)$ satisfying $h = k$ and $s^{k} = s^{h}$ hold. 
If such interval $[s, e]$ does not exist, then $\Delta(u) = \emptyset$. 
%For simplicity, $\Delta(s^h)$ is defined as $\Delta(u)$ for each position $s^{h}$ of sequence $S^{h}$, 
%which corresponds to a node $u \in \mathcal{U}$ in the derivation tree of RLSLP. 
%Then, partition $\Delta(s^h)$ is defined as 
%the subset $\Delta(s^h) \subseteq \Delta$ such that for any $[s, e] \in \Delta(s^h)$, 
%there exists the tail $[s^k,e^k]$ of $A(s, e)$ satisfying $h = k$ and $s^{k} = s^{h}$ hold. 
\end{definition}
For simplicity, 
$\Delta(s^h)$ is defined as $\Delta(u)$ for each position $s^{h}$ of sequence $S^{h}$, 
which corresponds to a node $u \in \mathcal{U}$ in the derivation tree of RLSLP. 
For clarity, 
$\Delta(h, s^{h})$ denotes $\Delta(s^{h})$. 
%Similarly, let $\Delta(h, s^{h}) = \Delta(s^h)$ for each position $s^{h}$ of sequence $S^{h}$. 

We provide an example of interval attractors using the partition of Definition~\ref{def:RR_Delta}.
in Figure~\ref{fig:restricted_recompression}. 
Let $s^{1}$ and $s^{\prime 1}$ be the third and ninth positions of sequence $S^{1}$. 
Then, we obtain $\Delta(s^{1}) = [2, 5], [2, 6], [2, 7], [3, 5], [3, 6], [3, 7]$ 
and $\Delta(s^{\prime 1}) = [9, 12], [9, 13], [9, 14], [10, 12], [10, 13], [10, 14]$. 
Therefore, two interval attractors $I(s^{1})$ and $I(s^{\prime 1})$ are 
$([2, 3], [5, 7])$ and $([9, 10], [12, 14])$, respectively.

The remaining part of this section explains properties of sequence $A(s, e)$ 
and shows that the partition $\Delta(u)$ of Definition~\ref{def:RR_Delta} 
satisfies the conditions of the partition introduced in Section~\ref{subsec:simplicied_RASS}.

\subsection{Properties of Sequence \texorpdfstring{$A(s, e)$}{}}\label{subsec:f_rec}
This subsection explain properties of sequence $A(s, e)$ for an interval $[s, e] \in \Delta$. 
%Here, $\Delta$ is the set of intervals introduced in Definition~\ref{def:IA}.
The following lemma states properties of sequence $A(s, e)$. 

\begin{lemma}\label{lem:rec_function_basic_relation}
Consider the sequence $A(s, e) = [s^{0}, e^{0}], [s^{1}, e^{1}], \ldots, [s^{k}, e^{k}]$ of intervals for an interval $[s, e]$ ($s < e$) on input string $T$. 
For each integer $h \in [0, k]$, 
let $T[x^{h}_{s}..y^{h}_{s}]$ and $T[x^{h}_{e}..y^{h}_{e}]$ be the two substrings derived from the $s^{h}$-th and $e^{h}$-th nonterminals of sequence $S^{h}$, respectively. 
The following three statements hold for the height $H \geq 1$ of RLSLP $\mathcal{G}^{R}$: 
\begin{enumerate}[label=\textbf{(\roman*)}]
    \item \label{enum:rec_function_basic_relation:1} $k < H$; 
    \item \label{enum:rec_function_basic_relation:4} 
    for each integer $h \in [1, k]$, 
    (A) $s+1 \leq x^{h}_{s} \leq x^{h}_{e} \leq e-1$, 
    (B) $s^{h} \geq 2$, 
    (C) $|[x^{h}_{e}, y^{h}_{e}]| \geq |[x^{h}_{e}, e]| - \sum_{w = 1}^{h} \lfloor \mu(w) \rfloor$, 
    and (D) $|[x^{h}_{s-1}, y^{h}_{s-1}]| \geq |[s, x^{h}_{s}]| - 1 - \sum_{w = 1}^{h} \lfloor \mu(w) \rfloor$ 
    for the substring $T[x^{h}_{s-1}..y^{h}_{s-1}]$ derived from the $(s^{h} - 1)$-th nonterminal of sequence $S^{h}$;    
    \item \label{enum:rec_function_basic_relation:2} if $|[s, e]| = 2$, then $k = 0$.
\end{enumerate}
\end{lemma}
\begin{proof}
See Section~\ref{subsubsec:rec_function_basic_relation_proof}. 
\end{proof}
    %\item \label{enum:rec_function_basic_relation:3} if $h \geq 1$ and $f_{\rec}(h-1, [i, j]) = \emptyset$, 
    %then $f_{\rec}(h, [i, j]) = \emptyset$;

The following lemma states properties of the tail $[s^{k}, e^{k}]$ of sequence $A(s, e)$. 

\begin{lemma}\label{lem:f_rec_top_property}
Consider the sequence $A(s, e) = [s^{0}, e^{0}], [s^{1}, e^{1}], \ldots, [s^{k}, e^{k}]$ of intervals for an interval $[s, e]$ ($s < e$) on input string $T$. 
Let $T[x^{k}_{s}..y^{k}_{s}]$ and $T[x^{k}_{e}..y^{k}_{e}]$ be the two substrings derived from the $s^{k}$-th and $e^{k}$-th nonterminals of sequence $S^{k}$, respectively, in input string $T$. 
For each integer $h \in [0, k]$, 
The following four statements hold: 
\begin{enumerate}[label=\textbf{(\roman*)}]
    \item \label{enum:f_rec_top_property:1} if $|[s^{k}, e^{k}]| \geq 2$, 
    then $|\val(S^{k}[i])| \leq \lfloor \mu(k+1) \rfloor$ for all $i \in [s^{k}, e^{k}-1]$; 
    \item \label{enum:f_rec_top_property:2} if $|[s^{k}, e^{k}]| \geq 4$, then $S^{k}[s^{k}] = S^{k}[s^{k}+1] = \cdots = S^{k}[e^{k}-1]$; 
    \item \label{enum:f_rec_top_property:3} if there exists an integer $d \geq 0$ satisfying 
    $|[x^{k}_{s}, e]| > d \lfloor \mu(k+1) \rfloor + \sum_{w = 1}^{k+1} \lfloor \mu(w) \rfloor$, 
    then $|[s^{k}, e^{k}]| \geq d + 2$; 
    \item \label{enum:f_rec_top_property:4} $|[s, x^{k}_{s}]| \leq 1 + \sum_{w = 1}^{k+1} \lfloor \mu(w) \rfloor$ 
    and $|[x^{k}_{e}, e]| \leq \sum_{w = 1}^{k+1} \lfloor \mu(w) \rfloor$. 
\end{enumerate}
\end{lemma}
\begin{proof}
See Section~\ref{subsubsec:proof_f_rec_top_property}.
\end{proof}

The following lemma states the relationship between two sequences $A(s, e)$ and $A(s^{\prime}, e^{\prime})$ for 
two intervals $[s, e]$ and $[s^{\prime}, e^{\prime}]$ on input string $T$.

\begin{lemma}\label{lem:intv_function_shift}
    Consider the sequence $A(s, e) = [s^{0}, e^{0}], [s^{1}, e^{1}], \ldots, [s^{k}, e^{k}]$ of intervals for an interval $[s, e]$ ($s < e$) on input string $T$. 
    Similarly, consider 
    the sequence $A(s^{\prime}, e^{\prime}) = [s^{\prime 0}, e^{\prime 0}]$, $[s^{\prime 1}, e^{\prime 1}]$, $\ldots$, $[s^{\prime k^{\prime}}, e^{\prime k^{\prime}}]$ of intervals for another interval $[s^{\prime}, e^{\prime}]$ ($s^{\prime} < e^{\prime}$) on input string $T$. 
    Then, the following five statements hold for each integer $h \in [0, k]$:
\begin{enumerate}[label=\textbf{(\roman*)}]
        \item \label{enum:intv_function_shift:1} 
        if $s^{\prime} \leq s$ and $e^{\prime} = e$, 
        then $k^{\prime} \geq h$, $[s^{h}, e^{h}] \subseteq [s^{\prime h}, e^{\prime h}]$, and $e^{h} = e^{\prime h}$;
        \item \label{enum:intv_function_shift:3} 
        if $s^{\prime} = s$ and $e \leq e^{\prime}$
        then $k^{\prime} \geq h$, $[s^{h}, e^{h}] \subseteq [s^{\prime h}, e^{\prime h}]$, and $s^{h} = s^{\prime h}$;
        \item \label{enum:intv_function_shift:5}
        if $s \leq s^{\prime}$ (respectively, $s \geq s^{\prime}$) and $k^{\prime} \geq h$, 
        then $s^{h} \leq s^{\prime h}$ (respectively, $s^{h} \geq s^{\prime h}$); 
        \item \label{enum:intv_function_shift:6}
        if $e \leq e^{\prime}$ (respectively, $e \geq e^{\prime}$) and $k^{\prime} \geq h$, 
        then $e^{h} \leq e^{\prime h}$ (respectively, $e^{h} \geq e^{\prime h}$); 
        \item \label{enum:intv_function_shift:7} 
        if $[s, e] \subseteq [s^{\prime}, e^{\prime}]$, 
        then $k^{\prime} \geq h$ 
        and $[s^{h}, e^{h}] \subseteq [s^{\prime h}, e^{\prime h}]$.
\end{enumerate}
\end{lemma}
\begin{proof}
See Section~\ref{subsubsec:rec_function_property_overview_proof1}.
\end{proof}

The next theorem ensures that $f_\interval$ creates the same sequences of segments $A(s, e)$ and $A(s^{\prime}, e^{\prime})$ from the same substrings $T[s..e]$ and $T[s^{\prime}..e^{\prime}]$ for different intervals $[s..e]$ and $[s^{\prime}, e^{\prime}]$ on $T$.

\begin{theorem}\label{theo:f_interval_syncro_property}
Let $[s, e]$ and $[s^{\prime}, e^{\prime}]$ be two intervals on input string $T$ satisfying $T[s..e] = T[s^{\prime}..e^{\prime}]$. 
Consider two sequences $A(s, e) = [s^{0}, e^{0}], [s^{1}, e^{1}], \ldots, [s^{k}, e^{k}]$ 
and $A(s^{\prime}, e^{\prime}) = [s^{\prime 0}, e^{\prime 0}], [s^{\prime 1}, e^{\prime 1}]$, $\ldots$, $[s^{\prime k^{\prime}}, e^{\prime k^{\prime}}]$ of intervals. 
For each integer $h \in [0, k]$, 
let $T[x^{h}_{s}..y^{h}_{s}]$ and $T[x^{h}_{e}..y^{h}_{e}]$ be 
the two substrings derived from the $s^{h}$-th and $e^{h}$-th nonterminals of sequence $S^{h}$, respectively, in input string $T$. 
Similarly, 
for each integer $h \in [0, k^{\prime}]$, 
let $T[x^{\prime h}_{s}..y^{\prime h}_{s}]$ and $T[x^{\prime h}_{e}..y^{\prime h}_{e}]$ be 
the two substrings derived from the $s^{\prime h}$-th and $e^{\prime h}$-th nonterminals of sequence $S^{h}$, respectively, in input string $T$. 
Then, the following two statements hold: 
\begin{enumerate}[label=\textbf{(\roman*)}]        
        \item \label{enum:rec_function_syncro_property:1} for each integer $h \in [0, \min \{ k, k^{\prime} \}]$, 
        (A) $S^{h}[s^{h}..e^{h}-1] = S^{h}[s^{\prime h}..e^{\prime h}-1]$,
        (B) $|[s, x^{h}_{s}]| = |[s^{\prime}, x^{\prime h}_{s}]|$, 
        and (C) $|[x^{h}_{e}, e]| = |[x^{\prime h}_{e}, e^{\prime}]|$;
        \item \label{enum:rec_function_syncro_property:2} $k = k^{\prime}$. 
\end{enumerate}
\end{theorem}
\begin{proof}
See Section~\ref{subsubsec:proof_rec_function_syncro_property}.
\end{proof}

We provide an example of Theorem~\ref{theo:f_interval_syncro_property} using 
the two sequences $A(1, 7) = [1, 7]$, $[2, 5]$, $[2, 3]$, $[2, 3]$, $[2, 3]$ and $A(8, 14) = [8, 14]$, $[8, 11]$, $[6, 7]$, $[6, 7]$, $[6, 7]$ shown in Figure~\ref{fig:finterval}. 
Here, the two substrings $T[1..7]$ and $T[8..14]$ are the same string $\mathrm{cbbabab}$. 
Let 
$[s, e] = [1, 7]$, 
$[s^{0}, e^{0}] = [1, 7]$, 
$[s^{1}, e^{1}] = [2, 5]$, 
$[s^{2}, e^{2}] = [2, 3]$, 
$[s^{3}, e^{3}] = [2, 3]$, 
and $[s^{4}, e^{4}] = [2, 3]$. 
Similarly, 
let 
$[s^{\prime}, e^{\prime}] = [8, 14]$, 
$[s^{\prime 0}, e^{\prime 0}] = [8, 14]$, 
$[s^{\prime 1}, e^{\prime 1}] = [8, 11]$, 
$[s^{\prime 2}, e^{\prime 2}] = [6, 7]$, 
$[s^{\prime 3}, e^{\prime 3}] = [6, 7]$, 
and $[s^{\prime 4}, e^{\prime 4}] = [6, 7]$. 
In this case, $x_{s}^{3} = 2$, $x_{e}^{3} = 4$, 
$x_{s}^{\prime 3} = 9$, and $x_{e}^{\prime 3} = 11$. 
Theorem~\ref{theo:f_interval_syncro_property}~\ref{enum:rec_function_syncro_property:1} shows that (A) $S^{3}[s^{3}..e^{3}-1] = S^{0}[s^{\prime 3}..e^{\prime 3}-1] = X_{12}$, 
(B) $|[s, x_{s}^{3}]| = |[s^{\prime}, x_{\prime s}^{3}]| = 2$, 
and (C) $|[x_{e}^{3}, e]| = |[x_{e}^{\prime 3}, e^{\prime}]| = 4$. 
Theorem~\ref{theo:f_interval_syncro_property}~\ref{enum:rec_function_syncro_property:2} shows that $|A(1, 7)| = |A(8, 14)|$ holds.

\subsubsection{Proof of Lemma~\ref{lem:rec_function_basic_relation}}\label{subsubsec:rec_function_basic_relation_proof}
The following three propositions are used to prove Lemma~\ref{lem:rec_function_basic_relation}
\begin{proposition}\label{prop:sync_set_sub_property1}
Consider the sequence $A(s, e) = [s^0, e^0], [s^1, e^1], \ldots, [s^k, e^k]$ of intervals for an interval $[s, e]$ ($s < e$) on input string $T$. 
For each integer $h \in [0, k]$, 
let $T[x^{h}_{s}..y^{h}_{s}]$ and $T[x^{h}_{e}..y^{h}_{e}]$ be the two substrings derived from the $s^{h}$-th and $e^{h}$-th nonterminals of sequence $S^{h}$, respectively. 
Then, the following four statements hold: 
\begin{enumerate}[label=\textbf{(\roman*)}]
    \item \label{enum:sync_set_sub_property1:1} $x^{h}_{s} \in [x^{h-1}_{s}, x^{h-1}_{e}]$ if $h \neq 0$. Otherwise (i.e., $h = 0$), $x^{h}_{s} = s$;
    \item \label{enum:sync_set_sub_property1:2} $x^{h}_{e} \in [x^{h-1}_{s}, x^{h-1}_{e}]$ if $h \neq 0$. Otherwise (i.e., $h = 0$), $x^{h}_{e} = e$;
    \item \label{enum:sync_set_sub_property1:3} $|[x^{h}_{e}, y^{h}_{e}]| \geq |[x^{h}_{e}, e]| - \sum_{w = 1}^{h} \lfloor \mu(w) \rfloor$;
    \item \label{enum:sync_set_sub_property1:4} 
    if $x^{h}_{s} \geq s+1$, then $s^{h} \geq 2$ and 
    $|[x^{h}_{s-1}, y^{h}_{s-1}]| \geq |[s, x^{h}_{s}]| - 1 - \sum_{w = 1}^{h} \lfloor \mu(w) \rfloor$ 
    for the substring $T[x^{h}_{s-1}..y^{h}_{s-1}]$ derived from the $(s^{h} - 1)$-th nonterminal of sequence $S^{h}$.    
\end{enumerate}
\end{proposition}
\begin{proof}
We prove Proposition~\ref{prop:sync_set_sub_property1} by induction on $h$. 
For the base case $h = 0$, 
Proposition~\ref{prop:sync_set_sub_property1} holds 
because $[x^{h}_{s}, y^{h}_{s}] = [s, s]$, $[x^{h}_{e}, y^{h}_{e}] = [e, e]$, and $\sum_{w = 1}^{0} \lfloor \mu(w) \rfloor = 0$.

For $1 \leq h \leq k$, 
the following four statements hold by the inductive assumption:
\begin{itemize}
    \item $x^{h-1}_{s} \in [s, e]$; 
    \item $x^{h-1}_{e} \in [s, e]$; 
    \item $|[x^{h-1}_{e}, y^{h-1}_{e}]| \geq |[x^{h-1}_{e}, e]| - \sum_{w = 1}^{h-1} \lfloor \mu(w) \rfloor$; 
    \item if $x^{h-1}_{s} \geq s+1$, then $s^{h-1} \geq 2$ and 
    $|[x^{h-1}_{s-1}, y^{h-1}_{s-1}]| \geq |[s, x^{h-1}_{s}]| - 1 - \sum_{w = 1}^{h-1} \lfloor \mu(w) \rfloor$ 
    for the substring $T[x^{h-1}_{s-1}..y^{h-1}_{s-1}]$ derived from the $(s^{h-1} - 1)$-th nonterminal of sequence $S^{h-1}$.
\end{itemize}

\textbf{Proof of $x^{h}_{s} \in [x^{h-1}_{s}, x^{h-1}_{e}]$.}
From the definition of sequence $A(s, e)$, 
the position $s^{h}$ satisfies at least one of the four conditions (i), (ii), (iii), and (iv) of Definition~\ref{def:f_interval}. 
This fact indicates that $x^{h}_{s} \in [x^{h-1}_{s}, x^{h-1}_{e}]$ holds. 

\textbf{Proof of $x^{h}_{e} \in [x^{h-1}_{s}, x^{h-1}_{e}]$.}
We can prove $x^{h}_{e} \in [x^{h-1}_{s}, x^{h-1}_{e}]$ using the same approach as for $x^{h}_{s} \in [x^{h-1}_{s}, x^{h-1}_{e}]$. 

\textbf{Proof of $|[x^{h}_{e}, y^{h}_{e}]| \geq |[x^{h}_{e}, e]| - \sum_{w = 1}^{h} \lfloor \mu(w) \rfloor$.}
At least one of the following three conditions is satisfied: 
\begin{enumerate}[label=\textbf{(\alph*)}]
    \item $|[x^{h}_{e}, e]| \leq \sum_{w = 1}^{h} \lfloor \mu(w) \rfloor$; 
    \item $|[x^{h}_{e}, e]| > \sum_{w = 1}^{h} \lfloor \mu(w) \rfloor$ and $x^{h}_{e} = x^{h-1}_{e}$; 
    \item $|[x^{h}_{e}, e]| > \sum_{w = 1}^{h} \lfloor \mu(w) \rfloor$ and $x^{h}_{e} < x^{h-1}_{e}$. 
\end{enumerate}

For condition (a), 
$|[x^{h}_{e}, y^{h}_{e}]| \geq 1$ 
and $|[x^{h}_{e}, e]| - \sum_{w = 1}^{h} \lfloor \mu(w) \rfloor \leq 0$ hold. 
Therefore, $|[x^{h}_{e}, y^{h}_{e}]| \geq |[x^{h}_{e}, e]| - \sum_{w = 1}^{h} \lfloor \mu(w) \rfloor$ holds. 

For condition (b), 
$[x^{h-1}_{e}, y^{h-1}_{e}] \subseteq [x^{h}_{e}, y^{h}_{e}]$ holds. 
Because of $|[x^{h-1}_{e}, y^{h-1}_{e}]| \geq |[x^{h-1}_{e}, e]| - \sum_{w = 1}^{h-1} \lfloor \mu(w) \rfloor$ and $x^{h}_{e} = x^{h-1}_{e}$, 
$|[x^{h}_{e}, y^{h}_{e}]| \geq |[x^{h}_{e}, e]| - \sum_{w = 1}^{h} \lfloor \mu(w) \rfloor$ follows from the following equation: 
\begin{equation*}
    \begin{split}
        |[x^{h}_{e}, y^{h}_{e}]| & \geq |[x^{h-1}_{e}, y^{h-1}_{e}]| \\
        &\geq |[x^{h-1}_{e}, e]| - \sum_{w = 1}^{h-1} \lfloor \mu(w) \rfloor \\
        &= |[x^{h}_{e}, e]| - \sum_{w = 1}^{h-1} \lfloor \mu(w) \rfloor \\
        &\geq |[x^{h}_{e}, e]| - \sum_{w = 1}^{h} \lfloor \mu(w) \rfloor.
    \end{split}
\end{equation*}

For condition (c), 
we prove at least one of $|[x^{h-1}_{e}, e]| \leq \sum_{w = 1}^{h} \lfloor \mu(w) \rfloor$ or   
$|[x^{h-1}_{e}, y^{h-1}_{e}]| \leq \lfloor \mu(h) \rfloor$ holds by contradiction. 
We assume that $|[x^{h-1}_{e}, e]| > \sum_{w = 1}^{h} \lfloor \mu(w) \rfloor$ 
and $|[x^{h-1}_{e}, y^{h-1}_{e}]| > \lfloor \mu(h) \rfloor$. 
Because of $|[x^{h-1}_{e}, y^{h-1}_{e}]| > \lfloor \mu(h) \rfloor$, 
the assignment $\assign(S^{h-1}[e])$ of the nonterminal $S^{h-1}[e]$ is $-1$. 
Because of $\assign(S^{h-1}[e]) = -1$, 
Lemma~\ref{lem:rr_class} indicates that 
sequence $S^{h}$ contains a position $g$ satisfying 
$[x^{h}_{g}, y^{h}_{g}] = [x^{h-1}_{e}, y^{h-1}_{e}]$ 
for the substring $T[x^{h}_{g}..y^{h}_{g}]$ derived from the $g$-th nonterminal of sequence $S^{h}$ in string $T$. 
In this case, $e^{h} = g$ holds 
because the position $g$ satisfies condition (iii) of Definition~\ref{def:f_interval}. 
$x^{h}_{e} = x^{h-1}_{e}$ follows from $x^{h-1}_{e} = x^{h}_{g}$ and $x^{h}_{g} = x^{h}_{e}$. 
The two facts $x^{h}_{e} < x^{h-1}_{e}$ and $x^{h}_{e} = x^{h-1}_{e}$ yield a contradiction. 
Therefore, at least one of $|[x^{h-1}_{e}, e]| \leq \sum_{w = 1}^{h} \lfloor \mu(w) \rfloor$ or   
$|[x^{h-1}_{e}, y^{h-1}_{e}]| \leq \lfloor \mu(h) \rfloor$ must hold. 

We prove $|[x^{h-1}_{e}, e]| \leq \sum_{w = 1}^{h} \lfloor \mu(w) \rfloor$ by contradiction. 
We assume that $|[x^{h-1}_{e}, e]| > \sum_{w = 1}^{h} \lfloor \mu(w) \rfloor$. 
Then, $|[x^{h-1}_{e}, y^{h-1}_{e}]| \leq \lfloor \mu(h) \rfloor$ 
because 
at least one of $|[x^{h-1}_{e}, e]| \leq \sum_{w = 1}^{h} \lfloor \mu(w) \rfloor$ or   
$|[x^{h-1}_{e}, y^{h-1}_{e}]| \leq \lfloor \mu(h) \rfloor$ holds. 
On the other hand, 
$|[x^{h-1}_{e}, y^{h-1}_{e}]| > \lfloor \mu(h) \rfloor$ follows from 
$|[x^{h-1}_{e}, y^{h-1}_{e}]| \geq |[x^{h-1}_{e}, e]| - \sum_{w = 1}^{h-1} \lfloor \mu(w) \rfloor$ 
and $|[x^{h-1}_{e}, e]| > \sum_{w = 1}^{h} \lfloor \mu(w) \rfloor$. 
The two facts $|[x^{h-1}_{e}, y^{h-1}_{e}]| \leq \lfloor \mu(h) \rfloor$ and $|[x^{h-1}_{e}, y^{h-1}_{e}]| > \lfloor \mu(h) \rfloor$ yield a contradiction. 
Therefore, $|[x^{h-1}_{e}, e]| \leq \sum_{w = 1}^{h} \lfloor \mu(w) \rfloor$ must hold. 

We prove $y^{h}_{e} \geq x^{h-1}_{e} - 1$ by contradiction. 
We assume that $y^{h}_{e} < x^{h-1}_{e} - 1$ holds. 
Then, the $(e^{h} + 1)$-th position of sequence $S^{h}$ 
satisfies $x^{h}_{e+1} \leq x^{h-1}_{e}-1$ 
for the substring $T[x^{h}_{e+1}..y^{h}_{e+1}]$ derived from the $(e^{h} + 1)$-th nonterminal of sequence $S^{h}$ in string $T$. 
$x^{h}_{e+1} \geq x^{h-1}_{s} + 1$ follows from 
$x^{h}_{e+1} = y^{h}_{e} + 1$, $x^{h}_{e} \leq y^{h}_{e}$, and $x^{h}_{e} \geq x^{h-1}_{s}$.
In this case, the $(e^{h} + 1)$-th position of sequence $S^{h}$ satisfies condition (i) of Definition~\ref{def:f_interval}. 
On the other hand, the $(e^{h} + 1)$-th position of sequence $S^{h}$ does not satisfy condition (i) of Definition~\ref{def:f_interval} 
because $e^{h}$ is the largest position of sequence $S^{h}$ satisfying at least one of the four conditions (i), (ii), (iii), and (iv). 
The two facts yield a contradiction. 
Therefore, $y^{h}_{e} \geq x^{h-1}_{e} - 1$ must hold. 

We prove $|[x^{h}_{e}, y^{h}_{e}]| \geq |[x^{h}_{e}, e]| - \sum_{w = 1}^{h} \lfloor \mu(w) \rfloor$ for condition (c). 
$|[x^{h}_{e}, x^{h-1}_{e}-1]| \geq |[x^{h}_{e}, e]| - \sum_{w = 1}^{h} \lfloor \mu(w) \rfloor$ 
follows from $|[x^{h}_{e}, x^{h-1}_{e}-1]| = |[x^{h}_{e}, e]| - |[x^{h-1}_{e}, e]|$ 
and $|[x^{h-1}_{e}, e]| \leq \sum_{w = 1}^{h} \lfloor \mu(w) \rfloor$.
$|[x^{h}_{e}, y^{h}_{e}]| \geq |[x^{h}_{e}, x^{h-1}_{e}-1]|$ because $y^{h}_{e} \geq x^{h-1}_{e} - 1$. 
Therefore, $|[x^{h}_{e}, y^{h}_{e}]| \geq |[x^{h}_{e}, e]| - \sum_{w = 1}^{h} \lfloor \mu(w) \rfloor$ 
follows from 
$|[x^{h}_{e}, y^{h}_{e}]| \geq |[x^{h}_{e}, x^{h-1}_{e}-1]|$ 
and $|[x^{h}_{e}, x^{h-1}_{e}-1]| \geq |[x^{h}_{e}, e]| - \sum_{w = 1}^{h} \lfloor \mu(w) \rfloor$. 

\textbf{Proof of $s^{h} \geq 2$ and $|[x^{h}_{s-1}, y^{h}_{s-1}]| \geq |[s, x^{h}_{s}]| - 1 - \sum_{w = 1}^{h} \lfloor \mu(w) \rfloor$ if $x^{h}_{s} \geq s+1$.}
$s^{h} \geq 2$ holds because of $x^{h}_{s} \geq s+1$ and $s \geq 1$.
We prove $|[x^{h}_{s-1}, y^{h}_{s-1}]| \geq |[s, x^{h}_{s}]| - 1 - \sum_{w = 1}^{h} \lfloor \mu(w) \rfloor$ using a similar approach as for $|[x^{h}_{e}, y^{h}_{e}]| \geq |[x^{h}_{e}, e]| - \sum_{w = 1}^{h} \lfloor \mu(w) \rfloor$. 
At least one of the following three conditions is satisfied: 
\begin{enumerate}[label=\textbf{(\arabic*)}]
    \item $|[s, x^{h}_{s}]| \leq 1 + \sum_{w = 1}^{h} \lfloor \mu(w) \rfloor$; 
    \item $|[s, x^{h}_{s}]| > 1 + \sum_{w = 1}^{h} \lfloor \mu(w) \rfloor$ and $x^{h-1}_{s} = x^{h}_{s}$; 
    \item $|[s, x^{h}_{s}]| > 1 + \sum_{w = 1}^{h} \lfloor \mu(w) \rfloor$ and $x^{h-1}_{s} < x^{h}_{s}$. 
\end{enumerate}

For condition (1), 
$|[x^{h}_{s-1}, y^{h}_{s-1}]| \geq 1$ 
and $|[s, x^{h}_{s}]| - 1 - \sum_{w = 1}^{h} \lfloor \mu(w) \rfloor \leq 0$ hold. 
Therefore, $|[x^{h}_{s-1}, y^{h}_{s-1}]| \geq |[s, x^{h}_{s}]| - 1 - \sum_{w = 1}^{h} \lfloor \mu(w) \rfloor$ holds. 

For condition (2), 
$x^{h-1}_{s} \geq s+1$ follows from $x^{h}_{s} = x^{h-1}_{s}$ and $x^{h}_{s} \geq s+1$. 
Because of $x^{h-1}_{s} \geq s+1$, 
$s^{h-1} \geq 2$ and $|[x^{h-1}_{s-1}, y^{h-1}_{s-1}]| \geq |[s, x^{h-1}_{s}]| - 1 - \sum_{w = 1}^{h-1} \lfloor \mu(w) \rfloor$ 
follows from the inductive assumption. 
In this case, 
$[x^{h-1}_{s-1}, y^{h-1}_{s-1}] \subseteq [x^{h}_{s-1}, y^{h}_{s-1}]$ holds 
because $y^{h}_{s-1} = y^{h-1}_{s-1}$ holds. 
$|[x^{h}_{s-1}, y^{h}_{s-1}]| \geq |[s, x^{h}_{s}]| - 1 - \sum_{w = 1}^{h} \lfloor \mu(w) \rfloor$ follows from the following equation: 
\begin{equation*}
    \begin{split}
        |[x^{h}_{s-1}, y^{h}_{s-1}]| & \geq |[x^{h-1}_{s-1}, y^{h-1}_{s-1}]| \\
        &\geq |[s, x^{h-1}_{s}]| - 1 - \sum_{w = 1}^{h-1} \lfloor \mu(w) \rfloor \\
        &= |[s, x^{h}_{s}]| - 1 - \sum_{w = 1}^{h-1} \lfloor \mu(w) \rfloor \\
        &\geq |[s, x^{h}_{s}]| - 1 - \sum_{w = 1}^{h} \lfloor \mu(w) \rfloor.
    \end{split}
\end{equation*}

For condition (3), 
we can prove the following statements using the same approach as for condition (c): 
\begin{itemize}
    \item at least one of $|[s, x^{h-1}_{s}]| \leq \sum_{w = 1}^{h} \lfloor \mu(w) \rfloor$ and $|[x^{h-1}_{s}, y^{h-1}_{s}]| \leq \lfloor \mu(h) \rfloor$;
    \item $|[s, x^{h-1}_{s}]| \leq 1 + \sum_{w = 1}^{h} \lfloor \mu(w) \rfloor$ (i.e., $|[s, x^{h-1}_{s} - 1]| \leq \sum_{w = 1}^{h} \lfloor \mu(w) \rfloor$).
\end{itemize}

We prove $x^{h}_{s-1} \leq x^{h-1}_{s}+1$ by contradiction. 
We assume that $x^{h}_{s-1} \geq x^{h-1}_{s}+2$ holds. 
$x^{h}_{s-1} \leq x^{h-1}_{e} - 1$ follows from 
$x^{h}_{s-1} \leq y^{h}_{s-1}$, $y^{h}_{s-1} = x^{h}_{s} - 1$, and $x^{h}_{s} \leq x^{h-1}_{e}$. 
In this case, the $(s^{h} - 1)$-th position of sequence $S^{h}$ satisfies condition (i) of Definition~\ref{def:f_interval}. 
On the other hand, the $(s^{h} + 1)$-th position of sequence $S^{h}$ does not satisfy condition (i) of Definition~\ref{def:f_interval} 
because $s^{h}$ is the smallest position of sequence $S^{h}$ satisfying at least one of the four conditions (i), (ii), (iii), and (iv). 
The two facts yield a contradiction. 
Therefore, $x^{h}_{s-1} \leq x^{h-1}_{s}+1$ must hold. 

We prove $|[x^{h}_{s-1}, y^{h}_{s-1}]| \geq |[s, x^{h}_{s}]| - 1 - \sum_{w = 1}^{h} \lfloor \mu(w) \rfloor$ for condition (3). 
$|[x^{h-1}_{s}+1, x^{h}_{s}-1]| \geq |[s, x^{h}_{s}-1]| - \sum_{w = 1}^{h} \lfloor \mu(w) \rfloor$ 
follows from $|[x^{h-1}_{s}+1, x^{h}_{s}-1]| = |[s, x^{h}_{s}-1]| - |[s, x^{h-1}_{s}-1]|$ 
and $|[s, x^{h-1}_{s}-1]| \leq \sum_{w = 1}^{h} \lfloor \mu(w) \rfloor$.
$|[x^{h}_{s-1}, y^{h}_{s-1}]| \geq |[x^{h-1}_{s}+1, x^{h}_{s}-1]|$ follows from $y^{h}_{s-1} = x^{h}_{s} - 1$ and $x^{h}_{s-1} \leq x^{h-1}_{s}+1$.
Therefore, $|[x^{h}_{s-1}, y^{h}_{s-1}]| \geq |[s, x^{h}_{s}]| - 1 - \sum_{w = 1}^{h} \lfloor \mu(w) \rfloor$ 
follows from the following equation: 
\begin{equation*}
    \begin{split}
        |[x^{h}_{s-1}, y^{h}_{s-1}]| & \geq |[x^{h-1}_{s}+1, x^{h}_{s}-1]| \\
        & \geq |[s, x^{h}_{s} - 1]| - \sum_{w = 1}^{h} \lfloor \mu(w) \rfloor \\
        & = |[s, x^{h}_{s}]| - 1 - \sum_{w = 1}^{h} \lfloor \mu(w) \rfloor.
    \end{split}
\end{equation*}

Finally, we obtain Proposition~\ref{prop:sync_set_sub_property1}.
\end{proof}

\begin{proof}[Proof of Lemma~\ref{lem:rec_function_basic_relation}]
The proof of Lemma~\ref{lem:rec_function_basic_relation} is as follows. 

\textbf{Proof of Lemma~\ref{lem:rec_function_basic_relation}(i).}
We prove $k < H$ by contradiction. 
We assume that $k = H$ holds. 
Then, $s^{H} = 1$, $e^{H} = 1$, $x^{H}_{s} = 1$, and $y^{H}_{s} = n$ hold because of $|S^{H}| = 1$. 
From the definition of function $f_{\interval}$, 
the position $s^{H}$ satisfies at least one of the four conditions (i), (ii), (iii), and (iv) of Definition~\ref{def:f_interval}. 
If the position $s^{H}$ satisfies condition (i), 
then $x^{H-1}_{s} + 1 \leq x^{H}_{s}$ holds. 
$x^{H}_{s} \geq 2$ follows from $x^{H-1}_{s} + 1 \leq x^{H}_{s}$ and $x^{H-1}_{s} \geq 1$. 
The two facts $x^{H}_{s} \geq 2$ and $x^{H}_{s} = 1$ yield a contradiction. 

If the position $s^{H}$ satisfies condition (ii), 
then $x^{H-1}_{s} = x^{H}_{s}$ and $|\val(S^{H-1}[s^{H-1}])| > \lfloor \mu(H) \rfloor$ hold 
because of $x^{H}_{s} = 1$. 
Because of $|\val(S^{H-1}[s^{H-1}])| > \lfloor \mu(H) \rfloor$, 
the assignment of nonterminal $S^{H-1}[s^{H-1}]$ is $-1$.

In the derivation tree of RLSLP $\mathcal{G}$, 
the root corresponds to the $s^{H}$-th nonterminal of sequence $S^{H}$ 
and has the node $u$ corresponding to the $s^{H-1}$-th nonterminal of sequence $S^{H-1}$ as a child. 
Because of $\assign(S^{H-1}[s^{H-1}]) = -1$, 
Lemma~\ref{lem:rr_class}~\ref{enum:rr_class:3} and Lemma~\ref{lem:rr_class}~\ref{enum:rr_class:4} indicate that 
the node $u$ is the single child of the root. 
On the other hand, 
Lemma~\ref{lem:rr_class}~\ref{enum:rr_class:1} indicates that the root has at least two children. 
Therefore, there exists a contradiction. 

If the position $s^{H}$ satisfies condition (iii), 
then $x^{H}_{s} = x^{H-1}_{e}$ and $|[x^{H-1}_{e}, e]| > \sum_{w = 1}^{H} \lfloor \mu(w) \rfloor$ hold. 
$|[x^{H-1}_{e}, y^{H-1}_{e}]| \geq |[x^{H-1}_{e}, e]| - \sum_{w = 1}^{H-1} \lfloor \mu(w) \rfloor$
follows from Proposition~\ref{prop:sync_set_sub_property1}~\ref{enum:sync_set_sub_property1:3}. 
$|[x^{H-1}_{e}, y^{H-1}_{e}]| > \lfloor \mu(H) \rfloor$ follows from 
$|[x^{H-1}_{e}, e]| > \sum_{w = 1}^{H} \lfloor \mu(w) \rfloor$ and 
$|[x^{H-1}_{e}, y^{H-1}_{e}]| \geq |[x^{H-1}_{e}, e]| - \sum_{w = 1}^{H-1} \lfloor \mu(w) \rfloor$. 
Because of $|\val(S^{H-1}[e^{H-1}])| > \lfloor \mu(H) \rfloor$, 
the assignment of nonterminal $S^{H-1}[e^{H-1}]$ is $-1$.

In this case, we can show that the root has a single child using the fact that 
$\assign(S^{H-1}[e^{H-1}]) = -1$ using the same approach as for condition (ii). 
On the other hand, 
Lemma~\ref{lem:rr_class}~\ref{enum:rr_class:1} indicates that the root has at least two children. 
Therefore, there exists a contradiction. 

If the position $s^{H}$ satisfies condition (iv), 
then $x^{H}_{s} = x^{H-1}_{s}$ and $|[s, x^{H-1}_{s}]| > 1 + \sum_{w = 1}^{H} \lfloor \mu(w) \rfloor$ hold. 
$|[s, x^{H}_{s}]| > 1$ follows from $x^{H}_{s} = x^{H-1}_{s}$ and $|[s, x^{H-1}_{s}]| > 1$. 
On the other hand, $|[s, x^{H}_{s}]| = 1$ follows from 
$1 \leq s \leq x^{H}_{s}$ (Proposition~\ref{prop:sync_set_sub_property1}~\ref{enum:sync_set_sub_property1:1}) and $x^{H}_{s} = 1$. 
The two facts $|[s, x^{H}_{s}]| > 1$ and $|[s, x^{H}_{s}]| = 1$ yield a contradiction. 

We showed that there always exists a contradiction under the assumption that $k = H$ holds. 
Therefore, $k < H$ must hold.

\textbf{Proof of Lemma~\ref{lem:rec_function_basic_relation}(ii).}
We prove Lemma~\ref{lem:rec_function_basic_relation}~\ref{enum:rec_function_basic_relation:4} using induction on $h$. 
For the base case $h = 1$, 
$[x^{1}_{s}, x^{1}_{e}] \subseteq [s, e]$ holds 
because $[x^{1}_{s}, x^{1}_{e}] \subseteq [x^{0}_{s}, x^{0}_{e}] = [s, e]$ follows from Proposition~\ref{prop:sync_set_sub_property1}~\ref{enum:sync_set_sub_property1:1} and Proposition~\ref{prop:sync_set_sub_property1}~\ref{enum:sync_set_sub_property1:2}. 

The position $s^{1}$ satisfies at least one of the four conditions (i), (ii), (iii), and (iv) of Definition~\ref{def:f_interval}. 
We show that the position $s^{1}$ satisfies condition (i). 
We assume that the position $s^{1}$ does not satisfy condition (i). 
Then, the position satisfies at least one of three conditions (ii), (iii), and (iv). 
If the position $s^{1}$ satisfies condition (ii), 
then there exists a position $i$ of sequence $S^{0}$ satisfying $|\val(S^{0}[i])| > \lfloor \mu(1) \rfloor$. 
$|\val(S^{0}[i])| \geq 2$ follows from $|\val(S^{0}[i])| > \lfloor \mu(1) \rfloor$ and $\lfloor \mu(1) \rfloor \geq 1$. 
On the other hand, $|\val(S^{0}[i])| = 1$ holds because the nonterminals of sequence $S^{0}$ corresponds to leaves in the derivation tree of RLSLP $\mathcal{G}^{R}$. 
The two facts $|\val(S^{0}[i])| \geq 2$ and $|\val(S^{0}[i])| = 1$ yields a contradiction. 

If the position $s^{1}$ satisfies condition (iii), 
then $1 > \sum_{w = 1}^{1} \lfloor \mu(w) \rfloor$ holds. 
On the other hand, $\sum_{w = 1}^{1} \lfloor \mu(w) \rfloor \geq 1$ holds. 
The two facts $1 > \sum_{w = 1}^{1} \lfloor \mu(w) \rfloor$ and $\sum_{w = 1}^{1} \lfloor \mu(w) \rfloor \geq 1$ yield a contradiction. 

If the position $s^{1}$ satisfies condition (iv), 
then $0 > \sum_{w = 1}^{1} \lfloor \mu(w) \rfloor$ holds. 
On the other hand, $\sum_{w = 1}^{1} \lfloor \mu(w) \rfloor \geq 0$ holds. 
The two facts $0 > \sum_{w = 1}^{1} \lfloor \mu(w) \rfloor$ and $\sum_{w = 1}^{1} \lfloor \mu(w) \rfloor \geq 0$ yield a contradiction. 
Therefore, the position $s^{1}$ must satisfy condition (i).

Similarly, 
the position $e^{1}$ satisfies at least one of the four conditions (i), (ii), (iii), and (iv) of Definition~\ref{def:f_interval}. 
We can show that $e^{1}$ satisfies condition (i) using the same approach as for the position $s^{1}$. 

We prove $s+1 \leq x^{1}_{s} \leq x^{1}_{e} \leq e-1$. 
$s+1 \leq x^{1}_{s}$ holds because the position $s^{1}$ satisfies condition (i). 
$x^{1}_{e} \geq e-1$ holds because the position $e^{1}$ satisfies condition (A). 
$x^{1}_{s} \leq x^{1}_{e}$ holds because $[s^{1}, e^{1}]$ is an interval on sequence $S^{1}$. 
Therefore, $s+1 \leq x^{1}_{s} \leq x^{1}_{e} \leq e-1$ holds. 

We prove $s^{1} \geq 2$ by contradiction. 
We assume that $s^{1} < 2$ (i.e., $s^{1} = 1$) holds. 
Then, $x^{1}_{s} = 1$ holds because $s^{1}$ is the first position of sequence $S^{1}$. 
On the other hand, $x^{1}_{s} \geq 2$ follows from $s \geq 1$ and $s+1 \leq x^{1}_{s}$. 
The two facts $x^{1}_{s} = 1$ and $x^{1}_{s} \geq 2$ yield a contradiction. 
Therefore, $s^{1} \geq 2$ must hold. 

$|[x^{1}_{e}, y^{1}_{e}]| \geq |[x^{1}_{e}, e]| - \sum_{w = 1}^{1} \lfloor \mu(w) \rfloor$ 
and $|[x^{1}_{s-1}, y^{1}_{s-1}]| \geq |[s, x^{1}_{s}]| - 1 - \sum_{w = 1}^{1} \lfloor \mu(w) \rfloor$ 
follow from Proposition~\ref{prop:sync_set_sub_property1}~\ref{enum:sync_set_sub_property1:3} and Proposition~\ref{prop:sync_set_sub_property1}~\ref{enum:sync_set_sub_property1:4}, respectively. 
Therefore, Lemma~\ref{lem:rec_function_basic_relation}~\ref{enum:rec_function_basic_relation:4} holds for the base case. 

For $h \geq 2$, 
$s+1 \leq x^{h-1}_{s} \leq x^{h-1}_{e} \leq e-1$ holds by the inductive assumption. 
$s+1 \leq x^{h}_{s} \leq x^{h}_{e} \leq e-1$ follows from 
$s+1 \leq x^{h-1}_{s} \leq x^{h-1}_{e} \leq e-1$, Proposition~\ref{prop:sync_set_sub_property1}~\ref{enum:sync_set_sub_property1:1}, 
and Proposition~\ref{prop:sync_set_sub_property1}~\ref{enum:sync_set_sub_property1:2}. 
$s^{h} \geq 2$ can be proved using the same approach as for $s^{1} \geq 2$. 
$|[x^{h}_{e}, y^{h}_{e}]| \geq |[x^{h}_{e}, e]| - \sum_{w = 1}^{h} \lfloor \mu(w) \rfloor$ 
and $|[x^{h}_{s-1}, y^{h}_{s-1}]| \geq |[s, x^{h}_{s}]| - 1 - \sum_{w = 1}^{h} \lfloor \mu(w) \rfloor$ 
follow from Proposition~\ref{prop:sync_set_sub_property1}~\ref{enum:sync_set_sub_property1:3} and Proposition~\ref{prop:sync_set_sub_property1}~\ref{enum:sync_set_sub_property1:4}, respectively. 
Therefore, Lemma~\ref{lem:rec_function_basic_relation}~\ref{enum:rec_function_basic_relation:4} holds for $h \geq 2$.

\textbf{Proof of Lemma~\ref{lem:rec_function_basic_relation}(iii).}
We prove Lemma~\ref{lem:rec_function_basic_relation}~\ref{enum:rec_function_basic_relation:2} by contradiction. 
We assume that $k \geq 1$ holds. 
Then, $s+1 \leq x^{1}_{s} \leq x^{1}_{e} \leq s$ follows from Lemma~\ref{lem:rec_function_basic_relation}~\ref{enum:rec_function_basic_relation:4} 
and $e = s+1$. 
The two facts $s < s+1$ and $s + 1 \leq s$ yield a contradiction. 
Therefore, Lemma~\ref{lem:rec_function_basic_relation}~\ref{enum:rec_function_basic_relation:2} must hold.

\end{proof}

%%%%%%%%%%%%%%%%%%%%%%%%%%%%%%%%%%%%%%%%%%%%%%%%%%%%%%%%%%%%
\subsubsection{Proof of Lemma~\ref{lem:f_rec_top_property}}\label{subsubsec:proof_f_rec_top_property}
The following two propositions are used to prove Lemma~\ref{lem:f_rec_top_property}. 

\begin{proposition}\label{prop:I_SYNC_top_property1}
Consider the sequence $A(s, e) = [s^{0}, e^{0}], [s^{1}, e^{1}], \ldots, [s^{k}, e^{k}]$ of intervals for an interval $[s, e]$ ($s < e$) on input string $T$. 
For even $h$ in set $[0, k]$, 
if there exists an integer $i \in [s^{h}+1, e^{h}-1]$ satisfying $S^{h}[i-1] \neq S^{h}[i]$, 
then $h < k$ holds. 
\end{proposition}
\begin{proof}
$k < H$ follows from Lemma~\ref{lem:rec_function_basic_relation}~\ref{enum:rec_function_basic_relation:1}. 
Let $T[x^{h}_{s}..y^{h}_{s}]$, $T[x^{h}_{e}..y^{h}_{e}]$, $T[x^{h}_{i}..y^{h}_{i}]$ be 
the three substrings derived from the $s^{h}$-th, $e^{h}$-th, and $i$-th nonterminals of sequence $S^{h}$, respectively, in input string $T$. 
Because of $S^{h}[i-1] \neq S^{h}[i]$, 
Lemma~\ref{lem:rr_class}~\ref{enum:rr_class:3} indicates that 
sequence $S^{h+1}$ contains a position $i^{\prime}$ satisfying 
$x^{h+1}_{i^{\prime}} = x^{h}_{i}$ for the substring $T[x^{h+1}_{i^{\prime}}..y^{h+1}_{i^{\prime}}]$ derived from the $i^{\prime}$-th nonterminal of sequence $S^{h+1}$. 
$x^{h}_{s} + 1 \leq x^{h+1}_{i^{\prime}}$ holds 
because $x^{h}_{s} < x^{h}_{i}$ and $x^{h}_{i} = x^{h+1}_{i^{\prime}}$ hold. 
Similarly, 
$x^{h+1}_{i^{\prime}} \leq x^{h}_{e} - 1$ holds 
because $x^{h}_{i} < x^{h}_{e}$ and $x^{h}_{i} = x^{h+1}_{i^{\prime}}$ hold. 
Because of $x^{h+1}_{i^{\prime}} \in [x^{h}_{s} + 1, x^{h}_{e} - 1]$, 
the position $i^{\prime}$ satisfies condition (i) of Definition~\ref{def:f_interval}. 
This fact indicates $f_{\interval}(s^{h}, e^{h}) \neq \perp$ holds, 
and hence, $h < k$ holds. 
\end{proof}

\begin{proposition}\label{prop:I_SYNC_top_property2}
Consider the sequence $A(s, e) = [s^{0}, e^{0}], [s^{1}, e^{1}], \ldots, [s^{k}, e^{k}]$ of intervals for an interval $[s, e]$ ($s < e$) on input string $T$. 
For odd $h$ in set $[0, k]$, 
if $|[s^{h}, e^{h}] \geq 4$, then $h < k$ holds. 
\end{proposition}
\begin{proof}
$k < H$ follows from Lemma~\ref{lem:rec_function_basic_relation}~\ref{enum:rec_function_basic_relation:1}. 
Let $T[x^{h}_{s}..y^{h}_{s}]$, $T[x^{h}_{e}..y^{h}_{e}]$, $T[x^{h}_{s+1}..y^{h}_{s+1}]$ be 
the three substrings derived from the $s^{h}$-th, $e^{h}$-th, and $(s^{h}+1)$-th nonterminals of sequence $S^{h}$, respectively, in input string $T$. 
If the assignment of $S^{h}[s^{h}+1] = 0$, 
then 
Lemma~\ref{lem:rr_class}~\ref{enum:rr_class:3} indicates that 
sequence $S^{h+1}$ contains a position $i$ satisfying $x^{h+1}_{i} = x^{h}_{s+1}$ 
for the substring $T[x^{h+1}_{i}..y^{h+1}_{i}]$ derived from the $i$-th nonterminal of sequence $S^{h+1}$. 
Otherwise, 
sequence $S^{h+1}$ contains a position $i^{\prime}$ satisfying $x^{h+1}_{i^{\prime}} = y^{h}_{s+1}+1$ 
for the substring $T[x^{h+1}_{i^{\prime}}..y^{h+1}_{i^{\prime}}]$ derived from the $i^{\prime}$-th nonterminal of sequence $S^{h+1}$. 
Therefore, 
sequence $S^{h+1}$ contains a position $j$ satisfying either $x^{h+1}_{j} = x^{h}_{s+1}$ or $x^{h+1}_{j} = y^{h}_{s+1}+1$ 
for the substring $T[x^{h+1}_{j}..y^{h+1}_{j}]$ derived from the $j$-th nonterminal of sequence $S^{h+1}$. 

We prove $x^{h+1}_{j} \in [x^{h}_{s} + 1, x^{h}_{e} - 1]$. 
$x^{h}_{s} + 1 \leq x^{h}_{s+1}$ holds because of $s^{h} < s^{h}+1$. 
$y^{h}_{s+1} + 1 \leq x^{h}_{e} - 1$ holds because of $s+1 \leq e^{h}-2$. 
Therefore, $x^{h+1}_{j} \in [x^{h}_{s} + 1, x^{h}_{e} - 1]$ follows from 
$x^{h}_{s} + 1 \leq x^{h}_{s+1}$, $y^{h}_{s+1} + 1 \leq x^{h}_{e} - 1$, and $x^{h+1}_{j} \in [x^{h}_{s+1}, y^{h}_{s+1} + 1]$. 

Because of $x^{h+1}_{j} \in [x^{h}_{s} + 1, x^{h}_{e} - 1]$, 
the position $j$ satisfies condition (i) of Definition~\ref{def:f_interval}. 
This fact indicates $f_{\interval}(s^{h}, e^{h}) \neq \perp$ holds, 
and hence, $h < k$ holds. 
\end{proof}

The proof of Lemma~\ref{lem:f_rec_top_property} is as follows. 

%The proof of 
%For each integer $s \in [1, |S^{h}|]$, 
%let $T[x_{s}..y_{s}]$ be the substring derived from the $s$-th nonterminal of sequence $S^{h}$. 
%Similarly, 
%for each integer $s \in [1, |S^{h+1}|]$, 
%let $T[x^{\prime}_{s}..y^{\prime}_{s}]$ be the substring derived from the $s$-th nonterminal of sequence $S^{h+1}$. 

\begin{proof}[Proof of Lemma~\ref{lem:f_rec_top_property}~\ref{enum:f_rec_top_property:1}]
We prove Lemma~\ref{lem:f_rec_top_property}~\ref{enum:f_rec_top_property:1} by contradiction. 
We assume that Lemma \ref{lem:f_rec_top_property}~\ref{enum:f_rec_top_property:1} does not hold. 
Then, there exists an integer $i \in [s^{k}, e^{k}-1]$ 
satisfying $|\val(S^{k}[i])| > \lfloor \mu(k+1) \rfloor$. 
Let $T[x^{k}_{i}..y^{k}_{i}]$ be the substring derived from the $i$-th nonterminal of sequence $S^{k}$. 
Because of $|\val(S^{k}[i])| > \lfloor \mu(k+1) \rfloor$, 
the assignment of the nonterminal $S^{k}[i]$ is $-1$. 
Because of $\assign(S^{k}[i]) = -1$, 
Lemma~\ref{lem:rr_class}~\ref{enum:rr_class:3} and Lemma~\ref{lem:rr_class}~\ref{enum:rr_class:4} indicate that 
sequence $S^{k+1}$ contains a position $i^{\prime}$ satisfying 
$x^{k+1}_{i^{\prime}} = x^{k}_{i}$ 
for the substring $T[x^{k+1}_{i^{\prime}}..y^{k+1}_{i^{\prime}}]$ derived from the $i^{\prime}$-th nonterminal of sequence $S^{k+1}$ in input string $T$. 
In this case, the position $i^{\prime}$ satisfies condition (ii) of Definition~\ref{def:f_interval}. 
By the existence of the position $i^{\prime}$, 
$f_{\interval}(s^{k}, e^{k}) \neq \perp$ holds. 
On the other hand, $f_{\interval}(s^{k}, e^{k}) = \perp$ follows from the definition of the sequence $S(s, e)$. 
The two facts $f_{\interval}(s^{k}, e^{k}) \neq \perp$ and $f_{\interval}(s^{k}, e^{k}) = \perp$ yield a contradiction. 
Therefore, Lemma~\ref{lem:f_rec_top_property}~\ref{enum:f_rec_top_property:1} must hold. 
\end{proof}

\begin{proof}[Proof of Lemma~\ref{lem:f_rec_top_property}~\ref{enum:f_rec_top_property:2}]
We prove Lemma~\ref{lem:f_rec_top_property}~\ref{enum:f_rec_top_property:2} by contradiction. 
We assume that Lemma \ref{lem:f_rec_top_property} \ref{enum:f_rec_top_property:2} does not hold. 
Then, there exists an integer $i \in [s^{k}, e^{k}-2]$ satisfying 
$S^{k}[i] \neq S^{k}[i+1]$. 
If $k$ is even, then Proposition~\ref{prop:I_SYNC_top_property1} shows that $S^{k}[i] = S^{k}[i+1]$. 
The two facts $S^{k}[i] \neq S^{k}[i+1]$ and $S^{k}[i] = S^{k}[i+1]$ yield a contradiction. 
Otherwise (i.e., $k$ is odd), 
Proposition~\ref{prop:I_SYNC_top_property2} shows that $|[s^{k}, e^{k}]| \leq 3$. 
The two facts $|[s^{k}, e^{k}]| \geq 4$ and $|[s^{k}, e^{k}]| \leq 3$ yield a contradiction. 
Therefore, Lemma~\ref{lem:f_rec_top_property}~\ref{enum:f_rec_top_property:2} must hold. 
\end{proof}

\begin{proof}[Proof of Lemma~\ref{lem:f_rec_top_property}~\ref{enum:f_rec_top_property:3}]
We prove Lemma~\ref{lem:f_rec_top_property}~\ref{enum:f_rec_top_property:3} by contradiction. 
We assume that Lemma \ref{lem:f_rec_top_property} \ref{enum:f_rec_top_property:3} does not hold. 
Then, $|[s^{k}, e^{k}]| \leq d + 1$ holds. 
$|[x^{k}_{s}, x^{k}_{e}-1]| \leq d \lfloor \mu(k+1) \rfloor$ 
follows from Lemma~\ref{lem:f_rec_top_property}~\ref{enum:f_rec_top_property:1}. 
$|[x^{k}_{e}, e]| > \sum_{w = 1}^{k+1} \lfloor \mu(w) \rfloor$ because 
(A) $|[x^{k}_{e}, e]| = |[x^{k}_{s}, e]| - |[x^{k}_{s}, x^{k}_{e}-1]|$, 
(B) $|[x^{k}_{s}, x^{k}_{e}-1]| \leq d \lfloor \mu(k+1) \rfloor$, and 
(C) $|[x^{k}_{s}, e]| > d \lfloor \mu(k+1) \rfloor + \sum_{w = 1}^{k+1} \lfloor \mu(w) \rfloor$. 
$|[x^{k}_{e}, y^{k}_{e}]| \geq |[x^{k}_{e}, e]| - \sum_{w = 1}^{k} \lfloor \mu(w) \rfloor$ 
follows from Lemma~\ref{lem:rec_function_basic_relation}~\ref{enum:rec_function_basic_relation:4}.

$|[x^{k}_{e}, y^{k}_{e}]| > \lfloor \mu(k+1) \rfloor$ 
follows from 
$|[x^{k}_{e}, y^{k}_{e}]| \geq |[x^{k}_{e}, e]| - \sum_{w = 1}^{k} \lfloor \mu(w) \rfloor$ 
and $|[x^{k}_{e}, e]| > \sum_{w = 1}^{k+1} \lfloor \mu(w) \rfloor$. 
Because of $|[x^{k}_{e}, y^{k}_{e}]| > \lfloor \mu(k+1) \rfloor$, 
the assignment $\assign(S^{k}[e])$ of the nonterminal $S^{k}[e]$ is $-1$. 
Because of $\assign(S^{k}[e]) = -1$, 
Lemma~\ref{lem:rr_class}~\ref{enum:rr_class:3} and Lemma~\ref{lem:rr_class}~\ref{enum:rr_class:4} indicate that 
sequence $S^{k+1}$ contains a position $i$ satisfying 
$x^{k+1}_{i} = x^{k}_{e}$ 
for the substring $T[x^{k+1}_{i}..y^{k+1}_{i}]$ derived from the $i$-th nonterminal of sequence $S^{k+1}$ in input string $T$. 

The position $i$ of sequence $S^{k+1}$ 
satisfies the condition (iii) of Definition~\ref{def:f_interval}.   
By the existence of the position $i$, 
$f_{\interval}(s^{k}, e^{k}) \neq \perp$ holds. 
On the other hand, $f_{\interval}(s^{k}, e^{k}) = \perp$ follows from the definition of the sequence $S(s, e)$. 
The two facts $f_{\interval}(s^{k}, e^{k}) \neq \perp$ and $f_{\interval}(s^{k}, e^{k}) = \perp$ yield a contradiction. 
Therefore, Lemma~\ref{lem:f_rec_top_property}~\ref{enum:f_rec_top_property:3} must hold. 
\end{proof}

\begin{proof}[Proof of Lemma~\ref{lem:f_rec_top_property}~\ref{enum:f_rec_top_property:4}]
We prove $|[s, x^{k}_{s}]| \leq 1 + \sum_{w = 1}^{k+1} \lfloor \mu(w) \rfloor$ by contradiction. 
We assume that $|[s, x^{k}_{s}]| > 1 + \sum_{w = 1}^{k+1} \lfloor \mu(w) \rfloor$ holds. 
For the substring $T[x^{k}_{s-1}..y^{k}_{s-1}]$ derived from the $(s^{k} - 1)$-th nonterminal of sequence $S^{k}$, 
$|[x^{k}_{s-1}, y^{k}_{s-1}]| \geq |[s, x^{k}_{s}]| - 1 - \sum_{w = 1}^{k} \lfloor \mu(w) \rfloor$ 
follows from Lemma~\ref{lem:rec_function_basic_relation}~\ref{enum:rec_function_basic_relation:4}. 
$|[x^{k}_{s-1}, y^{k}_{s-1}]| > \lfloor \mu(k+1) \rfloor$ follows from 
$|[x^{k}_{s-1}, y^{k}_{s-1}]| \geq |[s, x^{k}_{s}]| - 1 - \sum_{w = 1}^{k} \lfloor \mu(w) \rfloor$ 
and $|[s, x^{k}_{s}]| > 1 + \sum_{w = 1}^{k+1} \lfloor \mu(w) \rfloor$. 
Because of $|[x^{k}_{s-1}, y^{k}_{s-1}]| > \lfloor \mu(k+1) \rfloor$, 
the assignment of nonterminals $S^{k}[s-1]$ is $-1$. 

Because of $\assign(S^{k}[s-1]) = -1$, 
Lemma~\ref{lem:rr_class}~\ref{enum:rr_class:3} and Lemma~\ref{lem:rr_class}~\ref{enum:rr_class:4} indicate that 
sequence $S^{k+1}$ contains a position $i$ satisfying 
$x^{k+1}_{i} = x^{k}_{s}$ 
for the substring $T[x^{k+1}_{i}..y^{k+1}_{i}]$ derived from the $i$-th nonterminal of sequence $S^{k+1}$ in input string $T$. 
The position $i$ of sequence $S^{k+1}$ 
satisfies the condition (iv) of Definition~\ref{def:f_interval}.   
By the existence of the position $i$, 
$f_{\interval}(s^{k}, e^{k}) \neq \perp$ holds. 
On the other hand, $f_{\interval}(s^{k}, e^{k}) = \perp$ follows from the definition of the sequence $S(s, e)$. 
The two facts $f_{\interval}(s^{k}, e^{k}) \neq \perp$ and $f_{\interval}(s^{k}, e^{k}) = \perp$ yield a contradiction. 
Therefore, $|[s, x^{k}_{s}]| \leq 1 + \sum_{w = 1}^{k+1} \lfloor \mu(w) \rfloor$ must hold. 

Similarly, $|[x^{k}_{e}, e]| \leq \sum_{w = 1}^{k+1} \lfloor \mu(w) \rfloor$ can be proved using the same approach. 
\end{proof}

%%%%%%%%%%%%%%%%%%%%%%%%%%%%%%%%%%%%%%%%%%%%%%%%%%%%%%%%%%%%

\subsubsection{Proof of Lemma~\ref{lem:intv_function_shift}}\label{subsubsec:rec_function_property_overview_proof1}
For each integer $h \in [0, k]$, 
let $T[x^{h}_{s}..y^{h}_{s}]$ and $T[x^{h}_{e}..y^{h}_{e}]$ be 
the two substrings derived from the $s^{h}$-th and $e^{h}$-th nonterminals of sequence $S^{h}$, respectively, in input string $T$. 
Similarly, 
for each integer $h \in [0, k^{\prime}]$, 
let $T[x^{\prime h}_{s}..y^{\prime h}_{s}]$ and $T[x^{\prime h}_{e}..y^{\prime h}_{e}]$ be 
the two substrings derived from the $s^{\prime h}$-th and $e^{\prime h}$-th nonterminals of sequence $S^{h}$, respectively, in input string $T$. 

The following proposition is used to prove Lemma~\ref{lem:intv_function_shift}. 
\begin{proposition}\label{prop:rec_primitive_relation}
    Consider the two sequences $A(s, e)$ and $A(s^{\prime}, e^{\prime})$ introduced in Lemma~\ref{lem:intv_function_shift}. 
    The following two statements hold for each integer $h \in [0, k]$:     
    \begin{enumerate}[label=\textbf{(\roman*)}]
        \item \label{enum:rec_primitive_relation:1} 
        if $[s^{\prime}, e^{\prime}] = [s-1, e]$, 
        then $k^{\prime} \geq h$, $[s^{h}, e^{h}] \subseteq [s^{\prime h}, e^{\prime h}]$, and $e^{h} = e^{\prime h}$;
        \item \label{enum:rec_primitive_relation:2} 
        if $[s^{\prime}, e^{\prime}] = [s, e+1]$, 
        then $k^{\prime} \geq h$, $[s^{h}, e^{h}] \subseteq [s^{\prime h}, e^{\prime h}]$, and $s^{h} = s^{\prime h}$;
\end{enumerate}
\end{proposition}
\begin{proof}
The proof of Proposition~\ref{prop:rec_primitive_relation} is as follows.

\textbf{Proof of Proposition~\ref{prop:rec_primitive_relation}(i).}
We will prove Proposition~\ref{prop:rec_primitive_relation}~\ref{enum:rec_primitive_relation:1} by induction on $h$. 
For the base case $h = 0$, 
$[s^{0}, e^{0}] = [s, e]$ and $[s^{\prime 0}, e^{\prime 0}] = [s^{\prime}, e^{\prime}]$ hold. 
Therefore, $k^{\prime} \geq h$, $[s^{h}, e^{h}] \subseteq [s^{\prime h}, e^{\prime h}]$, and $e^{h} = e^{\prime h}$ holds. 

For the inductive step, 
consider $h \in [1, k]$. 
$k^{\prime} \geq h-1$, $[s^{h-1}, e^{h-1}] \subseteq [s^{\prime h-1}, e^{\prime h-1}]$, and $e^{h-1} = e^{\prime h-1}$ holds by the inductive assumption. 
We show that either position $s^{h}$ satisfies at least one of the four conditions (i), (ii), (iii), and (iv) of Definition~\ref{def:f_interval} 
for function $f_{\interval}(s^{\prime h-1}, e^{\prime h-1})$. 
From the definition of sequence $A(s, e)$, 
the position $s^{h}$ satisfies at least one of the four conditions (i), (ii), (iii), and (iv) of Definition~\ref{def:f_interval} 
for function $f_{\interval}(s^{h-1}, e^{h-1})$. 

If the position $s^{h}$ satisfies condition (i) for function $f_{\interval}(s^{h-1}, e^{h-1})$, 
then the position satisfies condition (i) for function $f_{\interval}(s^{\prime h-1}, e^{\prime h-1})$.
If the position $s^{h}$ satisfies condition (ii) for function $f_{\interval}(s^{h-1}, e^{h-1})$, 
then the position satisfies condition (ii) for function $f_{\interval}(s^{\prime h-1}, e^{\prime h-1})$.
If the position $s^{h}$ satisfies condition (iii) for function $f_{\interval}(s^{h-1}, e^{h-1})$, 
then the position satisfies condition (iii) for function $f_{\interval}(s^{\prime h-1}, e^{\prime h-1})$.
If the position $s^{h}$ satisfies condition (iv) for function $f_{\interval}(s^{h-1}, e^{h-1})$, 
and $s^{h-1} = s^{\prime h-1}$, 
then the position $s^{h}$ satisfies condition (iv) for function $f_{\interval}(s^{\prime h-1}, e^{\prime h-1})$.

If the position $s^{h}$ satisfies condition (iv) for function $f_{\interval}(s^{h-1}, e^{h-1})$, 
and $s^{h-1} \neq s^{\prime h-1}$ (i.e., $s^{\prime h-1} < s^{h-1}$), 
then $x^{h}_{s} = x^{h-1}_{s}$, $x^{\prime h-1}_{s} < x^{h-1}_{s}$, and $|[s, x^{h-1}_{s}]| > 1 + \sum_{w = 1}^{h} \lfloor \mu(w) \rfloor$ hold. 
In this case, $h \geq 2$ holds because 
$|[s, x^{0}_{s}]| > 1 + \sum_{w = 1}^{0} \lfloor \mu(w) \rfloor$ does not hold by $|[s, x^{0}_{s}]| = 1$. 
Because of $h \geq 2$, 
Lemma~\ref{lem:rec_function_basic_relation}~\ref{enum:rec_function_basic_relation:4} shows that 
$|[x^{h-1}_{s-1}, y^{h-1}_{s-1}]| \geq |[s, x^{h-1}_{s}]| - 1 - \sum_{w = 1}^{h-1} \lfloor \mu(w) \rfloor$ 
for the substring $T[x^{h-1}_{s-1}..y^{h-1}_{s-1}]$ derived from the $(s^{h-1} - 1)$-th nonterminal of sequence $S^{h-1}$
$|[x^{h-1}_{s-1}, y^{h-1}_{s-1}]| > \lfloor \mu(h) \rfloor$ follows from 
$|[x^{h-1}_{s-1}, y^{h-1}_{s-1}]| \geq |[s, x^{h-1}_{s}]| - 1 - \sum_{w = 1}^{h-1} \lfloor \mu(w) \rfloor$ and 
$|[s, x^{h-1}_{s}]| > 1 + \sum_{w = 1}^{h} \lfloor \mu(w) \rfloor$. 
The position $s^{h}$ satisfies condition (ii) for function $f_{\interval}(s^{\prime h-1}, e^{\prime h-1})$ 
because $|[x^{h-1}_{s-1}, y^{h-1}_{s-1}]| > \lfloor \mu(h) \rfloor$, $s^{h-1}-1 \in [s^{\prime h-1}, e^{\prime h-1}-1]$, 
and $x^{h}_{s-1} = y^{h-1}_{s-1}+1$. 
Therefore, the position $s^{h}$ always satisfies at least one of the four conditions (i), (ii), (iii), and (iv) of Definition~\ref{def:f_interval} for function $f_{\interval}(s^{\prime h-1}, e^{\prime h-1})$. 
Similarly, 
we can show that position $e^{h}$ always satisfies at least one of the four conditions (i), (ii), (iii), and (iv) of Definition~\ref{def:f_interval} for function $f_{\interval}(s^{\prime h-1}, e^{\prime h-1})$. 

The existence of the two positions $s^{h}$ and $e^{h}$ indicate that 
$f_{\interval}(s^{\prime h-1}, e^{\prime h-1}) = [s^{\prime h}, e^{\prime h}]$ and $[s^{h}, e^{h}] \subseteq [s^{\prime h}, e^{\prime h}]$ hold. 
$k^{\prime} \geq h$ follows from $f_{\interval}(s^{\prime h-1}, e^{\prime h-1}) = [s^{\prime h}, e^{\prime h}]$. 

We show that $s^{h-1} \leq e^{h-1}-1$ holds if $e^{\prime h} > e^{h}$ and $x^{\prime h}_{e} = x^{\prime h-1}_{e}$. 
$[s^{h-1}, e^{h-1}] \subseteq [s^{\prime h-1}, e^{\prime h-1}]$ holds by the inductive assumption. 
Sequence $S^{h-1}$ contains a position $i$ satisfying $x^{h-1}_{i} = x^{h}_{e}$ 
for the substring $T[x^{h-1}_{i}..y^{h-1}_{i}]$ derived from the $i$-th nonterminal of sequence $S^{h-1}$. 
$x^{h-1}_{i} \in [x^{h-1}_{s}, x^{h-1}_{e}]$ follows from  
$x^{h-1}_{i} = x^{h}_{e}$ and $x^{h}_{e} \in [x^{h-1}_{s}, x^{h-1}_{e}]$ (Proposition~\ref{prop:sync_set_sub_property1}~\ref{enum:sync_set_sub_property1:2}).
$x^{h-1}_{i} \neq x^{h-1}_{e}$ because 
$x^{h-1}_{i} = x^{h}_{e}$, $x^{h-1}_{e} < x^{\prime h}_{e}$, and $x^{\prime h}_{e} = x^{\prime h-1}_{e}$. 
$i \in [s^{h-1}, e^{h-1}-1]$ holds because of  
$x^{h-1}_{i} \in [x^{h-1}_{s}, x^{h-1}_{e}]$ and $x^{h-1}_{i} \neq x^{h-1}_{e}$. 
Therefore, $s^{h-1} \leq e^{h-1}-1$ follow from $i \in [s^{h-1}, e^{h-1}-1]$. 

We prove $e^{\prime h} = e^{h}$ by contradiction. 
We assume that $e^{\prime h} \neq e^{h}$ holds. 
Then, $e^{\prime h} > e^{h}$ follows from $e^{\prime h} \geq e^{h}$ and $e^{\prime h} \neq e^{h}$. 
$x^{h-1}_{s} \leq x^{h}_{e}$ and $x^{\prime h}_{e} \leq x^{\prime h-1}_{e}$ follows from Proposition~\ref{prop:sync_set_sub_property1}~\ref{enum:sync_set_sub_property1:2}.
$x^{h-1}_{s} + 1 \leq x^{\prime h}_{e}$ follows from $x^{h-1}_{s} \leq x^{h}_{e}$, $x^{h}_{e} < x^{\prime h}_{e}$. 
The position $e^{\prime h}$ satisfies at least one of 
the four conditions (i), (ii), (iii), and (iv) of Definition~\ref{def:f_interval} for function $f_{\interval}(s^{\prime h-1}, e^{\prime h-1})$. 
Therefore, one of the following four conditions is satisfied: 
\begin{enumerate}[label=\textbf{(\alph*)}]
    \item $x^{\prime h}_{e} \leq x^{\prime h-1}_{e} - 1$;
    \item $x^{\prime h}_{e} = x^{\prime h-1}_{e}$ holds, 
    and the position $e^{\prime h}$ satisfies condition (B) for function $f_{\interval}(s^{\prime h-1}, e^{\prime h-1})$;
    \item $x^{\prime h}_{e} = x^{\prime h-1}_{e}$ holds, 
    and the position $e^{\prime h}$ satisfies condition (C) for function $f_{\interval}(s^{\prime h-1}, e^{\prime h-1})$;
    \item $x^{\prime h}_{e} = x^{\prime h-1}_{e}$ holds, 
    and the position $e^{\prime h}$ satisfies condition (D) for function $f_{\interval}(s^{\prime h-1}, e^{\prime h-1})$.
\end{enumerate}

For condition (a), 
$x^{\prime h}_{e} \in [x^{h-1}_{s}+1, x^{h-1}_{e}-1]$ follows from 
$x^{h-1}_{s} + 1 \leq x^{\prime h}_{e}$, $x^{\prime h}_{e} \leq x^{\prime h-1}_{e}-1$, and $x^{\prime h-1}_{e} = x^{h-1}_{e}$. 
Because of $x^{\prime h}_{e} \in [x^{h-1}_{s}+1, x^{h-1}_{e}-1]$, 
the position $e^{\prime h}$ satisfies condition (A) for function $f_{\interval}(s^{h-1}, e^{h-1})$, 
which indicates that $e^{h} \geq e^{\prime h}$ holds. 
The two facts $e^{\prime h} > e^{h}$ and $e^{h} \geq e^{\prime h}$ yield a contradiction. 

For condition (b), 
the $(e^{\prime h-1}-1)$-th position of sequence $S^{h-1}$ satisfies 
$e^{\prime h-1}-1 \in [s^{\prime h-1}, e^{\prime h-1} - 1]$, 
$|\val(S^{h-1}[e^{\prime h-1}-1])| > \lfloor \mu(h) \rfloor$, 
and $y^{\prime h-1}_{e-1} + 1 = x^{\prime h}_{e}$ 
for the substring $T[x^{\prime h-1}_{e-1}..y^{\prime h-1}_{e-1}]$ derived from the $(e^{\prime h-1}-1)$-th nonterminal of sequence $S^{h-1}$. 
In this case, $s^{h-1} \leq e^{h-1}-1$ holds because of $e^{\prime h} > e^{h}$ and $x^{\prime h}_{e} = x^{\prime h-1}_{e}$. 
$e^{\prime h-1}-1 \in [s^{h-1}, e^{h-1} - 1]$ follows from 
$e^{\prime h-1} = e^{h-1}$, $s^{h-1} \leq e^{h-1}-1$, and $e^{\prime h-1}-1 \in [s^{\prime h-1}, e^{\prime h-1} - 1]$. 
The position $e^{\prime h}$ satisfies condition (B) for function $f_{\interval}(s^{h-1}, e^{h-1})$ 
because $e^{\prime h-1}-1 \in [s^{h-1}, e^{h-1} - 1]$, $|\val(S^{h-1}[e^{\prime h-1}-1])| > \lfloor \mu(h) \rfloor$, 
and $y^{\prime h-1}_{e-1} + 1 = x^{\prime h}_{e}$. 
The existence of the position $e^{\prime h}$ indicates that $e^{h} \geq e^{\prime h}$ holds, 
but the two facts $e^{\prime h} > e^{h}$ and $e^{h} \geq e^{\prime h}$ yield a contradiction. 

For condition (c), 
the position $e^{\prime h}$ satisfies condition (C) for function $f_{\interval}(s^{h-1}, e^{h-1})$ 
because $x^{\prime h}_{e} = x^{\prime h-1}_{e}$, $x^{\prime h-1}_{e} = x^{h-1}_{e}$, and $e = e^{\prime}$. 
The existence of the position $e^{\prime h}$ indicates that $e^{h} \geq e^{\prime h}$ holds, 
but the two facts $e^{\prime h} > e^{h}$ and $e^{h} \geq e^{\prime h}$ yield a contradiction. 

For condition (d), 
$s^{\prime h-1} = e^{\prime h-1}$ holds. 
$s^{h-1} = e^{h-1}$ follows from $s^{\prime h-1} = e^{\prime h-1}$ and $[s^{h-1}, e^{h-1}] \subseteq [s^{\prime h-1}, e^{\prime h-1}]$. 
On the other hand, 
$s^{h-1} \leq e^{h-1}-1$ holds because of $e^{\prime h} > e^{h}$ and $x^{\prime h}_{e} = x^{\prime h-1}_{e}$. 
The two facts $s^{h-1} = e^{h-1}$ and $s^{h-1} \leq e^{h-1}-1$ yield a contradiction. 

We showed that there exists a contradiction under the assumption that $e^{\prime h} \neq e^{h}$ holds. 
Therefore, $e^{\prime h} = e^{h}$ must hold. 
Finally, we obtain Proposition~\ref{prop:rec_primitive_relation}~\ref{enum:rec_primitive_relation:1} by induction on $h$. 

\textbf{Proof of Proposition~\ref{prop:rec_primitive_relation}(ii).}
Proposition~\ref{prop:rec_primitive_relation}~\ref{enum:rec_primitive_relation:2} is symmetric to 
Proposition~\ref{prop:rec_primitive_relation}~\ref{enum:rec_primitive_relation:1}. 
Therefore, 
Proposition~\ref{prop:rec_primitive_relation}~\ref{enum:rec_primitive_relation:2} can be proved using the same approach as for 
Proposition~\ref{prop:rec_primitive_relation}~\ref{enum:rec_primitive_relation:1}.
\end{proof}

The proof of Lemma~\ref{lem:intv_function_shift} is as follows. 

\begin{proof}[Proof of Lemma~\ref{lem:intv_function_shift}~\ref{enum:intv_function_shift:1}]
We prove Lemma~\ref{lem:intv_function_shift}~\ref{enum:intv_function_shift:1} by induction on $s^{\prime}$. 
For the base case $s^{\prime} = s$, 
$k^{\prime} \geq h$, $[s^{h}, e^{h}] \subseteq [s^{\prime h}, e^{\prime h}]$, and $e^{h} = e^{\prime h}$ hold 
because of $A(s, e) = A(s^{\prime}, e^{\prime})$.

For the inductive step, 
consider $s^{\prime} \in [1, s-1]$ 
and the sequence $A(s^{\prime} + 1, e^{\prime}) = [s^{\prime\prime 0}, e^{\prime\prime 0}]$, $[s^{\prime\prime 1}, e^{\prime\prime 1}]$, $\ldots$, 
$[s^{\prime\prime k^{\prime\prime}}, e^{\prime\prime k^{\prime\prime}}]$ of intervals for interval $[s^{\prime}+1, e^{\prime}]$ on input string $T$. 
Then, $k^{\prime\prime} \geq h$, $[s^{h}, e^{h}] \subseteq [s^{\prime\prime h}, e^{\prime\prime h}]$, and $e^{h} = e^{\prime\prime h}$ 
hold by the inductive assumption. 
We apply Proposition~\ref{prop:rec_primitive_relation}~\ref{enum:rec_primitive_relation:1} to the two sequences $A(s^{\prime}, e^{\prime})$ and $A(s^{\prime} + 1, e^{\prime})$. Then, $k^{\prime} \geq h$, $[s^{\prime\prime h}, e^{\prime\prime h}] \subseteq [s^{\prime h}, e^{\prime h}]$, and $e^{\prime\prime h} = e^{\prime h}$ hold. 
$[s^{h}, e^{h}] \subseteq [s^{\prime h}, e^{\prime h}]$ follows from $[s^{h}, e^{h}] \subseteq [s^{\prime\prime h}, e^{\prime\prime h}]$ and $[s^{\prime\prime h}, e^{\prime\prime h}] \subseteq [s^{\prime h}, e^{\prime h}]$. 
$e^{h} = e^{\prime h}$ follows from $e^{h} = e^{\prime\prime h}$ and $e^{\prime\prime h} = e^{\prime h}$. 
Therefore, $k^{\prime} \geq h$, $[s^{h}, e^{h}] \subseteq [s^{\prime h}, e^{\prime h}]$, and $e^{h} = e^{\prime h}$ hold. 

Finally, we obtain Lemma~\ref{lem:intv_function_shift}~\ref{enum:intv_function_shift:1} by induction on $s^{\prime}$. 
\end{proof}

\begin{proof}[Proof of Lemma~\ref{lem:intv_function_shift}~\ref{enum:intv_function_shift:3}]
We proved Lemma~\ref{lem:intv_function_shift}~\ref{enum:intv_function_shift:1} using Proposition~\ref{prop:rec_primitive_relation}~\ref{enum:rec_primitive_relation:1}. 
Similarly, Lemma \ref{lem:intv_function_shift}~\ref{enum:intv_function_shift:3} can be proved using Proposition~\ref{prop:rec_primitive_relation}~\ref{enum:rec_primitive_relation:2}. 
\end{proof}

\begin{proof}[Proof of Lemma~\ref{lem:intv_function_shift}~\ref{enum:intv_function_shift:5}]
We prove $s^{h} \leq s^{\prime h}$ for the case of $s \leq s^{\prime}$. 
If $e \leq e^{\prime}$, 
then 
consider the sequence $A(s, e^{\prime}) = [s^{\prime\prime 0}, e^{\prime\prime 0}]$, $[s^{\prime\prime 1}, e^{\prime\prime 1}]$, $\ldots$, 
$[s^{\prime\prime k^{\prime\prime}}, e^{\prime\prime k^{\prime\prime}}]$ of intervals for interval $[s, e^{\prime}]$ on input string $T$. 
We apply Lemma~\ref{lem:intv_function_shift}~\ref{enum:intv_function_shift:1} to the two sequences $A(s^{\prime}, e^{\prime})$ and $A(s, e^{\prime})$. 
Then, $s^{\prime\prime h} \leq s^{\prime h}$.
Similarly, 
we apply Lemma~\ref{lem:intv_function_shift}~\ref{enum:intv_function_shift:3} to the two sequences $A(s, e)$ and $A(s, e^{\prime})$. 
Then, $s^{h} = s^{\prime\prime h}$ holds. 
Therefore, $s^{h} \leq s^{\prime h}$ follows from $s^{h} = s^{\prime\prime h}$ and $s^{\prime\prime h} \leq s^{\prime h}$. 

Otherwise (i.e., $e > e^{\prime}$), 
consider the sequence $A(s^{\prime}, e) = [s^{\prime\prime 0}, e^{\prime\prime 0}]$, $[s^{\prime\prime 1}, e^{\prime\prime 1}]$, $\ldots$, 
$[s^{\prime\prime k^{\prime\prime}}, e^{\prime\prime k^{\prime\prime}}]$ of intervals for interval $[s^{\prime}, e]$ on input string $T$. 
We apply Lemma~\ref{lem:intv_function_shift}~\ref{enum:intv_function_shift:3} to the two sequences $A(s^{\prime}, e^{\prime})$ and $A(s^{\prime}, e)$. 
Then, $s^{\prime h} = s^{\prime\prime h}$ holds. 
Similarly, 
we apply Lemma~\ref{lem:intv_function_shift}~\ref{enum:intv_function_shift:1} to the two sequences $A(s^{\prime}, e)$ and $A(s, e)$. 
Then, $s^{h} \leq s^{\prime\prime h}$. 
Therefore, $s^{h} \leq s^{\prime h}$ follows from $s^{\prime h} = s^{\prime\prime h}$ and $s^{h} \leq s^{\prime\prime h}$. 

For the case of $s \geq s^{\prime}$, 
we can prove $s^{h} \geq s^{\prime h}$ using the same approach as for the case of $s \leq s^{\prime}$. 
Therefore, Lemma~\ref{lem:intv_function_shift}~\ref{enum:intv_function_shift:5} holds. 
\end{proof}

\begin{proof}[Proof of Lemma~\ref{lem:intv_function_shift}~\ref{enum:intv_function_shift:6}]
Lemma~\ref{lem:intv_function_shift}~\ref{enum:intv_function_shift:6} is symmetric to 
Lemma~\ref{lem:intv_function_shift}~\ref{enum:intv_function_shift:5}. 
We can prove Lemma \ref{lem:intv_function_shift}~\ref{enum:intv_function_shift:6} using the same approach as for 
Lemma~\ref{lem:intv_function_shift}~\ref{enum:intv_function_shift:5}. 
\end{proof}

\begin{proof}[Proof of Lemma~\ref{lem:intv_function_shift}~\ref{enum:intv_function_shift:7}]
Consider the sequence $A(s, e^{\prime}) = [s^{\prime\prime 0}, e^{\prime\prime 0}]$, $[s^{\prime\prime 1}, e^{\prime\prime 1}]$, $\ldots$, 
$[s^{\prime\prime k^{\prime\prime}}, e^{\prime\prime k^{\prime\prime}}]$ of intervals for interval $[s, e^{\prime}]$ on input string $T$. 
We apply Lemma~\ref{lem:intv_function_shift}~\ref{enum:intv_function_shift:3} to the two sequences $A(s, e)$ and $A(s, e^{\prime})$. 
Then, $s^{h} = s^{\prime\prime h}$ and $e^{h} \leq e^{\prime\prime h}$.
We apply Lemma~\ref{lem:intv_function_shift}~\ref{enum:intv_function_shift:1} to the two sequences $A(s, e^{\prime})$ and $A(s^{\prime}, e^{\prime})$. 
Then, $s^{\prime h} \leq s^{\prime\prime h}$ and $e^{\prime h} = e^{\prime\prime h}$.
$s^{\prime h} \leq s^{h}$ follows from $s^{\prime h} \leq s^{\prime\prime h}$ and $s^{h} = s^{\prime\prime h}$.
$e^{h} \leq e^{\prime h}$ follows from $e^{h} \leq e^{\prime\prime h}$ and $e^{\prime h} = e^{\prime\prime h}$. 
Therefore, Lemma~\ref{lem:intv_function_shift}~\ref{enum:intv_function_shift:7} holds. 
\end{proof}

\subsubsection{Proof of Theorem~\ref{theo:f_interval_syncro_property}}\label{subsubsec:proof_rec_function_syncro_property}
The following two propositions are used to prove Theorem~\ref{theo:f_interval_syncro_property}. 

\begin{proposition}\label{prop:rec_function_syncro_sub1}
    Consider the two sequences $A(s, e)$ and $A(s^{\prime}, e^{\prime})$ introduced in Theorem~\ref{theo:f_interval_syncro_property}. 
    For an integer $h \in [0, k-1]$, 
    we assume the following four statements hold: 
    \begin{itemize}        
        \item $k^{\prime} \geq h$;
        \item $|[s, x^{h}_{s}]| = |[s^{\prime}, x^{\prime h}_{s}]|$;
        \item $|[x^{h}_{e}, e]| = |[x^{\prime h}_{e}, e^{\prime}]|$;
        \item $S^{h}[s^{h}..e^{h}-1] = S^{h}[s^{\prime h}..e^{\prime h}-1]$.
    \end{itemize}
    Then, the following three statements hold:
    \begin{enumerate}[label=\textbf{(\roman*)}]        
        \item $k^{\prime} \geq h+1$;
        \item $|[s, x^{h+1}_{s}]| \geq |[s^{\prime}, x^{\prime h+1}_{s}]|$;
        \item $|[x^{h+1}_{e}, e]| \geq |[x^{\prime h+1}_{e}, e^{\prime}]|$.
    \end{enumerate}
\end{proposition}
\begin{proof}
    The following two statements are used to prove Proposition~\ref{prop:rec_function_syncro_sub1}: 
    \begin{enumerate}[label=\textbf{(\Alph*)}]        
        \item sequence $S^{h+1}$ contains a position $\alpha$ satisfying 
        $|[x^{h}_{s}, x^{h+1}_{s}]| = |[x^{\prime h}_{s}, x^{h+1}_{\alpha}]|$ for function $f_{\interval}$ $(s^{\prime h}, e^{\prime h})$, 
        and this position $\alpha$ satisfies at least one of the four conditions (i), (ii), (iii), and (iv) of Definition~\ref{def:f_interval} 
        for function $f_{\interval}(s^{\prime h}, e^{\prime h})$. 
        Here, $T[x^{h+1}_{\alpha}..y^{h+1}_{\alpha}]$ is 
        the substring derived from the $\alpha$-th nonterminal of sequence $S^{h+1}$ in $T$; 
        \item sequence $S^{h+1}$ contains a position $\beta$ satisfying 
        $|[x^{h}_{s}, x^{h+1}_{s}]| = |[x^{\prime h}_{s}, x^{h+1}_{\beta}]|$ for function $f_{\interval}$ $(s^{\prime h}, e^{\prime h})$, 
        and this position $\beta$ satisfies at least one of the four conditions (i), (ii), (iii), and (iv) of Definition~\ref{def:f_interval} 
        for function $f_{\interval}(s^{\prime h}, e^{\prime h})$.
        Here, $T[x^{h+1}_{\beta}..y^{h+1}_{\beta}]$ is 
        the substring derived from the $\beta$-th nonterminal of sequence $S^{h+1}$ in $T$.
    \end{enumerate}

    \textbf{Proof of statement (A).}
    $x^{h+1}_{s} \in [x^{h}_{s}, x^{h}_{e}]$ follows from Proposition~\ref{prop:sync_set_sub_property1}~\ref{enum:sync_set_sub_property1:1}. 
    Because of $x^{h+1}_{s} \in [x^{h}_{s}, x^{h}_{e}]$, 
    there exists an integer $d \in [0, e^{h} - s^{h}]$ 
    satisfying 
    $x^{h}_{s+d} = x^{h+1}_{s}$ for the substring $T[x^{h}_{s+d}..y^{h}_{s+d}]$ derived from the $(s^{h} + d)$-th nonterminal of sequence $S^{h}$ in $T$. 
    Because of $S^{h}[s^{h}..e^{h}-1] = S^{h}[s^{\prime h}..e^{\prime h}-1]$, 
    the $(s^{\prime h} + d)$-th nonterminal of sequence $S^{h}$ satisfies 
    $|[x^{h}_{s}, x^{h+1}_{s}]| = |[x^{\prime h}_{s}, x^{h}_{s^{\prime} + d}]|$ 
    for the substring $T[x^{h}_{s^{\prime}+d}..y^{h}_{s^{\prime}+d}]$ derived from the $(s^{\prime h} + d)$-th nonterminal of sequence $S^{h}$ in $T$. 
    In the derivation tree of RLSLP $\mathcal{G}^{R}$, 
    let $u$ and $u^{\prime}$ be the nodes corresponding to the $(s^{h}+d)$-th and $(s^{\prime h}+d)$-th nonterminals of sequence $S^{h}$, respectively. 
    Here, $u$ is the leftmost child of the node corresponding to the $s^{h+1}$-th nonterminal of sequence $S^{h+1}$.     
    If $u^{\prime}$ is the leftmost child of a node, 
    then position $\alpha$ exists, 
    and the parent of the node $u^{\prime}$ corresponds to the $\alpha$-th nonterminal of sequence $S^{h+1}$. 

    The position $s^{h+1}$ satisfies at least one of the four conditions (i), (ii), (iii), and (iv) of Definition~\ref{def:f_interval} 
    for function $f_{\interval}(s^{h}, e^{h})$.         
    If the position $s^{h+1}$ satisfies condition (i), 
    then $x^{h+1}_{s} \in [x^{h}_{s}+1, x^{h}_{e}-1]$ holds. 
    $x^{h}_{s+d} \in [x^{h}_{s}+1, x^{h}_{e}-1]$ follows from $x^{h}_{s+d} = x^{h+1}_{s}$ and $x^{h+1}_{s} \in [x^{h}_{s}+1, x^{h}_{e}-1]$.     
    Because of $x^{h}_{s+d} \in [x^{h}_{s}+1, x^{h}_{e}-1]$, 
    $s^{h} < s^{h}+d < e^{h}$ holds. 
    $S^{h}[s^{h}+d-1] = S^{h}[s^{\prime h}+d-1]$ and $S^{h}[s^{h}+d] = S^{h}[s^{\prime h}+d]$ 
    follow from $s^{h} < s^{h}+d < e^{h}$ and $S^{h}[s^{h}..e^{h}-1] = S^{h}[s^{\prime h}..e^{\prime h}-1]$. 
    In this case, 
    Lemma~\ref{lem:rr_property}~\ref{enum:rr_property:1} ensures that 
    $u^{\prime}$ is the leftmost child of a node because 
    $S^{h}[s^{h}+d-1..s^{h}+d] = S^{h}[s^{\prime h}+d-1..s^{\prime h}+d]$ 
    and $u$ is the leftmost child of a node. 
    This lemma indicates that position $\alpha$ exists. 
    The position $\alpha$ satisfies condition (i) of Definition~\ref{def:f_interval} 
    for function $f_{\interval}(s^{\prime h}, e^{\prime h})$ 
    because $x^{h+1}_{\alpha} = x^{h}_{s^{\prime}+d}$ and $x^{h}_{s^{\prime}+d} \in [x^{\prime h}_{s}+1, x^{\prime h}_{e}-1]$ hold. 

    Let $T[x^{h}_{s+d-1}..y^{h}_{s+d-1}]$ and $T[x^{h}_{s^{\prime}+d-1}..y^{h}_{s^{\prime}+d-1}]$ be 
    the two substrings derived from the $(s^{h} + d-1)$-th and $(s^{\prime h} + d-1)$-th nonterminals of sequence $S^{h}$ in $T$, respectively. 
    If the position $s^{h+1}$ satisfies condition (ii), 
    then at least one of the following two conditions is satisfied: 
    \begin{enumerate}[label=\textbf{(\alph*)}]        
        \item $s^{h}+d-1 \in [s^{h}, e^{h}-1]$ and $|\val(S^{h}[s^{h}+d-1])| > \lfloor \mu(h+1) \rfloor$;
        \item $s^{\prime h}+d \in [s^{h}, e^{h}-1]$ and $|\val(S^{h}[s^{h}+d])| > \lfloor \mu(h+1) \rfloor$. 
    \end{enumerate}
    If condition (a) is satisfied, 
    then $s^{\prime}+d-1 \in [s^{\prime h}, e^{\prime h}-1]$ and $|\val(S^{h}[s^{\prime h}+d-1])| > \lfloor \mu(h+1) \rfloor$ hold 
    because 
    $S^{h}[s^{h}+d-1] = S^{h}[s^{\prime h}+d-1]$  
    follows from $s^{h}+d-1 \in [s^{h}, e^{h}-1]$ and $S^{h}[s^{h}..e^{h}-1] = S^{h}[s^{\prime h}..e^{\prime h}-1]$. 
    In this case, 
    the position $\alpha$ exists 
    because Lemma~\ref{lem:rr_class}~\ref{enum:rr_class:3} and Lemma~\ref{lem:rr_class}~\ref{enum:rr_class:4} indicate that 
    $u^{\prime}$ is the leftmost child of a node. 
    The existence of the nonterminal $S^{h}[s^{\prime h}+d-1]$ indicates that 
    the position $\alpha$ satisfies condition (ii) of Definition~\ref{def:f_interval} 
    for function $f_{\interval}(s^{\prime h}, e^{\prime h})$. 

    Otherwise (i.e., condition (a) is not satisfied), 
    condition (b) is satisfied. 
    $s^{\prime}+d \in [s^{\prime h}, e^{\prime h}-1]$ and $|\val(S^{h}[s^{\prime h}+d])| > \lfloor \mu(h+1) \rfloor$ hold 
    because 
    $S^{h}[s^{h}+d] = S^{h}[s^{\prime h}+d]$  
    follows from $s^{h}+d \in [s^{h}, e^{h}-1]$ and $S^{h}[s^{h}..e^{h}-1] = S^{h}[s^{\prime h}..e^{\prime h}-1]$. 
    In this case, 
    the position $\alpha$ exists 
    because Lemma~\ref{lem:rr_class}~\ref{enum:rr_class:3} and Lemma~\ref{lem:rr_class}~\ref{enum:rr_class:4} indicate that 
    $u^{\prime}$ is the leftmost child of a node. 
    The existence of the nonterminal $S^{h}[s^{\prime h}+d]$ indicates that 
    the position $\alpha$ satisfies condition (ii) of Definition~\ref{def:f_interval} 
    for function $f_{\interval}(s^{\prime h}, e^{\prime h})$. 
    
    If the position $s^{h+1}$ satisfies condition (iii), 
    then $s^{h} + d = e^{h}$ and $|[x^{h}_{e}, e]| > \sum_{w = 1}^{h+1} \lfloor \mu(w) \rfloor$ hold. 
    $s^{\prime h} + d = e^{\prime h}$ follows from 
    $s^{h} + d = e^{h}$ and $|S^{h}[s^{h}..e^{h}-1]| = |S^{h}[s^{\prime h}..e^{\prime h}-1]|$.     
    $|[x^{\prime h}_{e}, e^{\prime}]| > \sum_{w = 1}^{h+1} \lfloor \mu(w) \rfloor$ follows from 
    $|[x^{h}_{e}, e]| > \sum_{w = 1}^{h+1} \lfloor \mu(w) \rfloor$ and $|[x^{h}_{e}, e]| = |[x^{\prime h}_{e}, e^{\prime}]|$. 
    $|[x^{\prime h}_{e}, y^{\prime h}_{e}]| \geq |[x^{\prime h}_{e}, e^{\prime}]| - \sum_{w = 1}^{h} \lfloor \mu(w) \rfloor$ 
    follows from Lemma~\ref{lem:rec_function_basic_relation}~\ref{enum:rec_function_basic_relation:4}. 
    $|[x^{\prime h}_{e}, y^{\prime h}_{e}]| > \lfloor \mu(h+1) \rfloor$ follows from 
    $|[x^{\prime h}_{e}, e^{\prime}]| > \sum_{w = 1}^{h+1} \lfloor \mu(w) \rfloor$ and 
    $|[x^{\prime h}_{e}, y^{\prime h}_{e}]| \geq |[x^{\prime h}_{e}, e^{\prime}]| - \sum_{w = 1}^{h} \lfloor \mu(w) \rfloor$. 
    $|\val(S^{h}[s^{\prime h}+d])| > \lfloor \mu(h+1) \rfloor$ follows from 
    $s^{\prime h} + d = e^{\prime h}$, 
    $|\val(S^{h}[e^{\prime h}])| = |[x^{\prime h}_{e}, y^{\prime h}_{e}]|$, 
    and $|[x^{\prime h}_{e}, y^{\prime h}_{e}]| > \lfloor \mu(h+1) \rfloor$. 
    Because of $|\val(S^{h}[s^{\prime h}+d])| > \lfloor \mu(h+1) \rfloor$, 
    Lemma~\ref{lem:rr_class}~\ref{enum:rr_class:3} and Lemma~\ref{lem:rr_class}~\ref{enum:rr_class:4} indicate that 
    $u^{\prime}$ is the leftmost child of a node, 
    and hence, the position $\alpha$ exists. 
    In this case, 
    the position $\alpha$ satisfies condition (iii) of Definition~\ref{def:f_interval} 
    for function $f_{\interval}(s^{\prime h}, e^{\prime h})$. 
    
    If the position $s^{h+1}$ satisfies condition (iv), 
    then $s^{h} + d = s^{h}$ (i.e., $d = 0$) and $|[s, x^{h}_{s}]| > 1 + \sum_{w = 1}^{h+1} \lfloor \mu(w) \rfloor$ hold. 
    $s^{\prime h} + d = s^{\prime h}$ follows from $d = 0$. 
    $|[s^{\prime}, x^{\prime h}_{s}]| > 1 + \sum_{w = 1}^{h+1} \lfloor \mu(w) \rfloor$ follows from 
    $|[s, x^{h}_{s}]| > 1 + \sum_{w = 1}^{h+1} \lfloor \mu(w) \rfloor$ and $|[s, x^{h}_{s}]| = |[s^{\prime}, x^{\prime h}_{s}]|$. 
    $|[x^{h}_{s^{\prime}+d-1}, y^{h}_{s^{\prime}+d-1}]| \geq |[s^{\prime}, x^{\prime h}_{s}]| - 1 - \sum_{w = 1}^{h} \lfloor \mu(w) \rfloor$ 
    follows from Lemma~\ref{lem:rec_function_basic_relation}~\ref{enum:rec_function_basic_relation:4}. 
    $|[x^{h}_{s^{\prime}+d-1}, y^{h}_{s^{\prime}+d-1}]| > \lfloor \mu(h+1) \rfloor$ follows from 
    $|[x^{h}_{s^{\prime}+d-1}, y^{h}_{s^{\prime}+d-1}]| \geq |[s^{\prime}, x^{\prime h}_{s}]| - 1 - \sum_{w = 1}^{h} \lfloor \mu(w) \rfloor$  
    and $|[s^{\prime}, x^{\prime h}_{s}]| > 1 + \sum_{w = 1}^{h+1} \lfloor \mu(w) \rfloor$. 
    $|\val(S^{h}[s^{\prime h}+d-1])| > \lfloor \mu(h+1) \rfloor$ follows from 
    $|\val(S^{h}[e^{\prime h}+d-1])| = |[x^{h}_{s^{\prime}+d-1}, y^{h}_{s^{\prime}+d-1}]|$, 
    and $|[x^{h}_{s^{\prime}+d-1}, y^{h}_{s^{\prime}+d-1}]| > \lfloor \mu(h+1) \rfloor$. 
    Because of $|\val(S^{h}[s^{\prime h}+d-1])| > \lfloor \mu(h+1) \rfloor$, 
    Lemma~\ref{lem:rr_class}~\ref{enum:rr_class:3} and Lemma~\ref{lem:rr_class}~\ref{enum:rr_class:4} indicate that 
    $u^{\prime}$ is the leftmost child of a node, 
    and hence, the position $\alpha$ exists. 
    In this case, 
    the position $\alpha$ satisfies condition (iv) of Definition~\ref{def:f_interval} 
    for function $f_{\interval}(s^{\prime h}, e^{\prime h})$. 
    
    Finally, statement (A) holds. 
    
    \textbf{Proof of statement (B).}
    Statement (B) can be proved using the same approach as for statement (A).

    \textbf{Proof of Proposition~\ref{prop:rec_function_syncro_sub1}(i).}
    $f_{\interval}(s^{\prime h}, e^{\prime h}) \neq \perp$ follows from statement (A) and statement (B). 
    Therefore, $k^{\prime} \geq h+1$ holds.

    \textbf{Proof of Proposition~\ref{prop:rec_function_syncro_sub1}(ii).}
    Statement (A) indicates that 
    $x^{\prime h+1}_{s} \leq x^{h+1}_{\alpha}$ holds. 
    $|[s^{\prime}$, $x^{\prime h+1}_{s}]| \leq |[s^{\prime}, x^{h+1}_{\alpha}]|$ follows from 
    $x^{\prime h+1}_{s} \leq x^{h+1}_{\alpha}$. 
    Therefore, 
    $|[s, x^{h+1}_{s}]| \geq |[s^{\prime}, x^{\prime h+1}_{s}]|$ follows from 
    $|[x^{h}_{s}, x^{h+1}_{s}]| = |[x^{\prime h}_{s}, x^{h+1}_{\alpha}]|$ and $|[s^{\prime}, x^{\prime h+1}_{s}]| \leq |[s^{\prime}, x^{h+1}_{\alpha}]|$. 

    \textbf{Proof of Proposition~\ref{prop:rec_function_syncro_sub1}(iii).}
    Statement (B) indicates that 
    $x^{\prime h+1}_{e} \geq x^{h+1}_{\beta}$ holds. 
    $|[x^{\prime h+1}_{e}$, $e^{\prime}]| \leq |[x^{h+1}_{\beta}, e^{\prime}]|$ 
    follows from $x^{\prime h+1}_{e} \geq x^{h+1}_{\beta}$. 
    Therefore, 
    $|[x^{h+1}_{e}, e]| \geq |[x^{\prime h+1}_{e}, e^{\prime}]|$ follows from 
    $|[x^{h}_{s}, x^{h+1}_{s}]| = |[x^{\prime h}_{s}, x^{h+1}_{\beta}]|$ and $|[x^{\prime h+1}_{e}, e^{\prime}]| \leq |[x^{h+1}_{\beta}, e^{\prime}]|$.     
\end{proof}

\begin{proposition}\label{prop:rec_function_syncro_sub2}
    Consider the two sequences $A(s, e)$ and $A(s^{\prime}, e^{\prime})$ introduced in Theorem~\ref{theo:f_interval_syncro_property}. 
    For an integer $h \in [0, k^{\prime}-1]$, 
    we assume the following four statements hold: 
    \begin{itemize}        
        \item $k \geq h$;
        \item $|[s, x^{h}_{s}]| = |[s^{\prime}, x^{\prime h}_{s}]|$;
        \item $|[x^{h}_{e}, e]| = |[x^{\prime h}_{e}, e^{\prime}]|$;
        \item $S^{h}[s^{h}..e^{h}-1] = S^{h}[s^{\prime h}..e^{\prime h}-1]$.
    \end{itemize}
    Then, the following three statements hold:
    \begin{enumerate}[label=\textbf{(\roman*)}]        
        \item $k \geq h+1$;
        \item $|[s, x^{h+1}_{s}]| \leq |[s^{\prime}, x^{\prime h+1}_{s}]|$;
        \item $|[x^{h+1}_{e}, e]| \leq |[x^{\prime h+1}_{e}, e^{\prime}]|$.
    \end{enumerate}
\end{proposition}
\begin{proof}
    Proposition~\ref{prop:rec_function_syncro_sub2} can be proved using the same approach as for 
    Proposition~\ref{prop:rec_function_syncro_sub1} 
    because 
    Proposition~\ref{prop:rec_function_syncro_sub2} is symmetric to Proposition~\ref{prop:rec_function_syncro_sub1}. 
\end{proof}

The proof of Theorem~\ref{theo:f_interval_syncro_property} is as follows. 

\begin{proof}[Proof of Theorem~\ref{theo:f_interval_syncro_property}~\ref{enum:rec_function_syncro_property:1}]
We prove Theorem~\ref{theo:f_interval_syncro_property}~\ref{enum:rec_function_syncro_property:1} by induction on $h$. 
For the base case $h = 0$, 
Theorem~\ref{theo:f_interval_syncro_property}~\ref{enum:rec_function_syncro_property:1} clearly holds. 

For $h \in [1, \min \{ k, k^{\prime} \}]$, 
the following three statements hold by the inductive assumption: 
\begin{itemize}
    \item $S^{h-1}[s^{h-1}..e^{h-1}-1] = S^{h-1}[s^{\prime h-1}..e^{\prime h-1}-1]$; 
    \item $|[s, x^{h-1}_{s}]| = |[s^{\prime}, x^{\prime h-1}_{s}]|$; 
    \item $|[x^{h-1}_{e}, e]| = |[x^{\prime h-1}_{e}, e^{\prime}]|$.
\end{itemize}

We prove $|[s, x^{h}_{s}]| = |[s^{\prime}, x^{\prime h}_{s}]|$ and $|[x^{h}_{e}, e]| = |[x^{\prime h}_{e}, e^{\prime}]|$. 
Proposition~\ref{prop:rec_function_syncro_sub1} shows that 
$|[s, x^{h}_{s}]| \geq |[s^{\prime}, x^{\prime h}_{s}]|$ and $|[x^{h}_{e}, e]| \geq |[x^{\prime h}_{e}, e^{\prime}]|$ hold. 
Proposition~\ref{prop:rec_function_syncro_sub2} shows that 
$|[s, x^{h}_{s}]| \leq |[s^{\prime}, x^{\prime h}_{s}]|$ and $|[x^{h}_{e}, e]| \leq |[x^{\prime h}_{e}, e^{\prime}]|$ hold. 
Therefore, $|[s, x^{h}_{s}]| = |[s^{\prime}, x^{\prime h}_{s}]|$ and $|[x^{h}_{e}, e]| = |[x^{\prime h}_{e}, e^{\prime}]|$ hold. 

We prove $S^{h}[s^{h}..e^{h}-1] = S^{h}[s^{\prime h}..e^{\prime h}-1]$. 
Two sequences $S^{h}[s^{h}..e^{h}-1]$ and $S^{h}[s^{\prime h}..e^{\prime h}-1]$ derive the same string 
(i.e., $T[x^{h}_{s}..x^{h}_{e}-1] = T[x^{\prime h}_{s}..x^{\prime h}_{e}-1]$) 
because $T[s..e] = T[s^{\prime}..e^{\prime}]$, $|[s, x^{h}_{s}]| = |[s^{\prime}, x^{\prime h}_{s}]|$, and $|[x^{h}_{e}, e]| = |[x^{\prime h}_{e}, e^{\prime}]|$ hold. 
We can apply Lemma~\ref{lem:rr_property}~\ref{enum:rr_property:2} to the two sequences $S^{h}[s^{h}..e^{h}-1]$ and $S^{h}[s^{\prime h}..e^{\prime h}-1]$. 
Then, the lemma shows that $S^{h}[s^{h}..e^{h}-1] = S^{h}[s^{\prime h}..e^{\prime h}-1]$ holds. 
Therefore, we obtain Theorem~\ref{theo:f_interval_syncro_property}~\ref{enum:rec_function_syncro_property:1} by induction on $h$. 
\end{proof}

\begin{proof}[Proof of Theorem~\ref{theo:f_interval_syncro_property}~\ref{enum:rec_function_syncro_property:2}]
We prove $k = k^{\prime}$ by contradiction. 
We assume that $k \neq k^{\prime}$ holds. 
Then, either $k < k^{\prime}$ or $k > k^{\prime}$ holds. 
If $k < k^{\prime}$, 
then the following three statements follows from Theorem~\ref{theo:f_interval_syncro_property}~\ref{enum:rec_function_syncro_property:1}: 
\begin{itemize}
    \item $S^{k}[s^{k}..e^{k}-1] = S^{k}[s^{\prime k}..e^{\prime k}-1]$; 
    \item $|[s, x^{k}_{s}]| = |[s^{\prime}, x^{\prime k}_{s}]|$; 
    \item $|[x^{k}_{e}, e]| = |[x^{\prime k}_{e}, e^{\prime}]|$.
\end{itemize}
In this case, 
Proposition~\ref{prop:rec_function_syncro_sub2} shows that $f_{\interval}(s^{k}, e^{k}) \neq \perp$ holds. 
On the other hand, $f_{\interval}(s^{k}$, $e^{k}) = \perp$ follows from the definition of sequence $A(s, e)$. 
The two facts $f_{\interval}(s^{k}, e^{k}) = \perp$ and $f_{\interval}(s^{k}, e^{k}) \neq \perp$ yield a contradiction. 

Otherwise (i.e., $k > k^{\prime}$), 
we can show that there exists a contradiction by Proposition~\ref{prop:rec_function_syncro_sub1}. 
Therefore, $k = k^{\prime}$ must hold. 
\end{proof}

\subsection{Verification of Partition \texorpdfstring{$\Delta(u)$}{}}
We show that interval attractors can be constructed from the partition $\Delta(u)$ of Definition~\ref{def:RR_Delta}. 
Formally, we prove the following theorem. 
\begin{theorem}\label{theo:delta_proof}
    The partition $\Delta(s^h)$ of Definition~\ref{def:RR_Delta} 
    satisfies the conditions of the partition introduced in Section~\ref{subsec:simplicied_RASS}. 
\end{theorem}

Consider two nodes $u, u^{\prime} \in \mathcal{U}$ such that 
two sets $\Delta(u)$ and $\Delta(u^{\prime})$ contains two intervals $[s, e]$ and $[s^{\prime}, e^{\prime}]$. 
Here, $u$ (respectively, $u^{\prime}$) corresponds to a position $s^{h}$ of sequence $S^{h}$ for some $h \geq 0$ 
(respectively, a position $s^{\prime h^{\prime}}$ of sequence $S^{h^{\prime}}$ for some $h^{\prime} \geq 0$). 
Let $\gamma$ and $\gamma^{\prime}$ be the starting positions of the substring 
derived from two nonterminals $S^{h}[s^{h}]$ and $S^{h^{\prime}}[s^{h^{\prime}}]$ on input string $T$. 
From Definition~\ref{def:RR_Delta}, 
sequence $A(s, e)$ consists of $h+1$ intervals $[s^{0}, e^{0}], [s^{1}, e^{1}], \ldots, [s^{h}, e^{h}]$. 
Similarly, 
sequence $A(s^{\prime}, e^{\prime})$ consists of $h+1$ intervals $[s^{\prime 0}, e^{\prime 0}], [s^{\prime 1}, e^{\prime 1}], \ldots, [s^{\prime h^{\prime}}, e^{\prime h^{\prime}}]$. 

Theorem~\ref{theo:delta_proof} holds if the following five statements hold:
\begin{enumerate}[label=\textbf{(\roman*)}] 
    \item $[\gamma, \gamma+1] \subseteq [s, e]$ and $[\gamma^{\prime}, \gamma^{\prime}+1] \subseteq [s^{\prime}, e^{\prime}]$; 
    \item If $\gamma - s \neq \gamma' - s'$, then $T[s..e] \neq T[s'..e']$.
    \item If $\gamma > \gamma'$ holds, then the substring $T[s..e]$ is not a prefix of $T[s'..e']$ (i.e., either $s \neq s^{\prime}$ or $s = s^{\prime} \leq e^{\prime} < e$).
    \item If $h > h'$ holds, then $[s..e] \not \subseteq [s', e']$.
    \item If $u = u'$, then for any pair $(s^{\prime\prime}, e^{\prime\prime})$ such that $s^{\prime \prime} \in [\min \{ s, s^{\prime}  \}, s ]$ and $e^{\prime \prime} \in [e, \max \{ e, e^{\prime} \}]$, it holds that $[s^{\prime\prime}, e^{\prime\prime}] \in \Delta(u)$.
\end{enumerate}

\paragraph{Proof of statement (i).}
If $h = 0$, then $\gamma = s$ follows from the definition of $f_{\interval}$. 
$[\gamma, \gamma+1] \subseteq [s, e]$ follows from $\gamma = s$ and $|[s, e]| \geq 2$. 
Otherwise, $\gamma \in [s+1, e-1]$ follows from Lemma~\ref{lem:rec_function_basic_relation}~\ref{enum:rec_function_basic_relation:4}. 
Therefore, we obtain $[\gamma, \gamma+1] \subseteq [s, e]$. 
Similarly, we can prove $[\gamma^{\prime}, \gamma^{\prime}+1] \subseteq [s^{\prime}, e^{\prime}]$ using the same approach.

\paragraph{Proof of statement (ii).}
Statement (ii) holds if $T[s..e] = T[s'..e'] \Rightarrow \gamma - s = \gamma' - s'$. 
Theorem~\ref{theo:f_interval_syncro_property}~\ref{enum:rec_function_syncro_property:1} shows that 
$T[s..e] = T[s'..e'] \Rightarrow \gamma - s = \gamma' - s'$ holds. 
Therefore, we obtain statement (ii). 

\paragraph{Proof of statement (iii).}
Statement (iii) holds if $(s = s^{\prime} \land e < e^{\prime}) \Rightarrow \gamma < \gamma^{\prime}$. 
We prove $(s = s^{\prime} \land e < e^{\prime}) \Rightarrow \gamma < \gamma^{\prime}$. 
Lemma~\ref{lem:intv_function_shift}~\ref{enum:intv_function_shift:3} shows that 
$h^{\prime} \geq h$ and $s^{h} \leq s^{\prime h}$. 
Let $\gamma^{\prime \prime}$ be the starting position of the substring derived from the $s^{\prime h}$-th nonterminal of sequence $S^{h}$ on $T$. 
Then, $\gamma \leq \gamma^{\prime \prime}$ follows from $s^{h} \leq s^{\prime h}$. 
Since $h^{\prime} \geq h$, 
Proposition~\ref{prop:sync_set_sub_property1}~\ref{enum:sync_set_sub_property1:1} shows that 
$\gamma^{\prime \prime} \leq \gamma^{\prime}$. 
$\gamma < \gamma^{\prime}$ follows from $\gamma \leq \gamma^{\prime \prime}$ and $\gamma^{\prime \prime} \leq \gamma^{\prime}$. 
Therefore, we obtain statement (iii). 

\paragraph{Proof of statement (iv).}
Statement (iv) holds if $[s, e] \subseteq [s^{\prime}, e^{\prime}] \Rightarrow h \leq h^{\prime}$. 
Lemma~\ref{lem:intv_function_shift}~\ref{enum:intv_function_shift:7} shows that 
$[s, e] \subseteq [s^{\prime}, e^{\prime}] \Rightarrow h \leq h^{\prime}$ holds. 
Therefore, we obtain statement (iv).  

\paragraph{Proof of statement (v).}
%In this case, $h = h^{\prime}$ and $\gamma = \gamma^{\prime}$. 
%We prove $|[s, e] \cap [s^{\prime}, e^{\prime}]| \geq 2$. 
%If $h = 0$, then $\gamma = s = s^{\prime}$ follows from the definition of function $f_{\interval}$. 
%$|[s, e] \cap [s^{\prime}, e^{\prime}]| \geq 2$ follows from $\gamma = s = s^{\prime}$, $|[s, e]| \geq 2$, and $|[s^{\prime}, e^{\prime}]| \geq 2$. 
%Otherwise (i.e., $h > 0$), $\gamma \in [s+1, e-1]$ and $\gamma \in [s^{\prime}+1, e^{\prime}-1]$ follows from Lemma~\ref{lem:rec_function_basic_relation}~\ref{enum:rec_function_basic_relation:4}. 
%Therefore, we obtain $|[s, e] \cap [s^{\prime}, e^{\prime}]| \geq 2$. 
We prove $[s^{\prime \prime}, e^{\prime \prime}] \in \Delta(u)$ by the following proposition.

\begin{proposition}\label{prop:last_set_inclusion}
    If $u = u^{\prime}$, 
    then 
    $[s^{\prime \prime}, e^{\prime \prime}] \in \Delta(u)$ for any pair of integers 
    $s^{\prime \prime} \in [\min \{ s, s^{\prime}  \}, s ]$ and $e^{\prime \prime} \in [e, \max \{ e, e^{\prime} \}]$.
\end{proposition}
\begin{proof}
Consider sequence $A(s^{\prime \prime}, e^{\prime \prime}) = [s^{\prime \prime 0}, e^{\prime \prime 0}], [s^{\prime \prime 1}, e^{\prime \prime 1}], \ldots, [s^{\prime \prime k}, e^{\prime \prime k}]$. 
Because of $[s, e] \subseteq [s^{\prime \prime}, e^{\prime \prime}]$, 
Lemma~\ref{lem:intv_function_shift}~\ref{enum:intv_function_shift:7} shows that 
$k \geq h$ and $[s^{h}, e^{h}] \subseteq [s^{\prime\prime h}, e^{\prime\prime h}]$ hold. 
    
The following seven statements are used to prove Proposition~\ref{prop:last_set_inclusion}. 

\begin{enumerate}[label=\textbf{(\roman*)}]
    \item $s^{h} = s^{\prime\prime h}$; 
    \item $k = h$ if $[s, e] \subseteq [s^{\prime}, e^{\prime}]$; 
    \item $k = h$ if $[s^{\prime}, e^{\prime}] \subseteq [s, e]$; 
    \item $k = h$ if $s^{\prime} > s$, $e^{\prime} > e$, $s^{\prime \prime} = s$, and $e^{\prime \prime} = e^{\prime}$; 
    \item $k = h$ if $s^{\prime} < s$, $e^{\prime} < e$, $s^{\prime \prime} = s^{\prime}$, and $e^{\prime \prime} = e$;
    \item $k = h$ if $s^{\prime} > s$ and $e^{\prime} > e$; 
    \item $k = h$ if $s^{\prime} < s$ and $e^{\prime} < e$.     
\end{enumerate}

\textbf{Proof of statement (i).}
If $s^{\prime} < s$, 
then 
we apply Lemma~\ref{lem:intv_function_shift}~\ref{enum:intv_function_shift:5} to the two intervals 
$[s, e]$ and $[s^{\prime \prime}, e^{\prime \prime}]$. 
The lemma shows that $s^{\prime\prime h} \leq s^{h}$ holds because $s^{\prime \prime} \leq s$. 
Similarly, 
we apply Lemma~\ref{lem:intv_function_shift}~\ref{enum:intv_function_shift:5} to the two intervals 
$[s^{\prime}, e^{\prime}]$ and $[s^{\prime \prime}, e^{\prime \prime}]$. 
The lemma shows that $s^{\prime\prime h} \geq s^{h}$ holds because $s^{\prime \prime} \geq s^{\prime}$. 
Therefore, $s^{h} = s^{\prime\prime h}$ follows from $s^{\prime\prime h} \leq s^{h}$ and $s^{\prime\prime h} \geq s^{h}$.

Otherwise (i.e., $s^{\prime} \geq s$), 
$s^{\prime \prime} = s$ follows from 
$s^{\prime \prime} \in [\min \{ s, s^{\prime} \}, s ]$ and $\min \{ s, s^{\prime} \} = s$. 
We can apply Lemma~\ref{lem:intv_function_shift}~\ref{enum:intv_function_shift:3} to two intervals $[s, e]$ and $[s^{\prime \prime}, e^{\prime \prime}]$. 
Then, the lemma shows that $s^{h} = s^{\prime\prime h}$ holds. 

\textbf{Proof of statement (ii).}
We prove $k = h$ by contradiction. 
We assume that $k \neq h$ holds. 
Then, $k \geq h+1$ holds. 
$[s^{\prime \prime}, e^{\prime \prime}] \subseteq [s^{\prime}, e^{\prime}]$ follows from 
$[s, e] \subseteq [s^{\prime}, e^{\prime}]$, $s^{\prime \prime} \in [\min \{ s, s^{\prime} \}, s ]$ and $e^{\prime \prime} \in [e, \max \{ e, e^{\prime} \}]$. 
Because of $[s^{\prime \prime}, e^{\prime \prime}] \subseteq [s^{\prime}, e^{\prime}]$, 
we can apply Lemma~\ref{lem:intv_function_shift}~\ref{enum:intv_function_shift:7} to two intervals $[s^{\prime}, e^{\prime}]$ and $[s^{\prime \prime}, e^{\prime \prime}]$. 
Then, the lemma shows that $k \leq h$ holds, 
but the two facts $k \geq h+1$ and $k \leq h$ yield a contradiction. 
Therefore, $k = h$ must hold. 

\textbf{Proof of statement (iii).}
$[s^{\prime \prime}, e^{\prime \prime}] = [s, e]$ follows from 
$[s^{\prime}, e^{\prime}] \subseteq [s, e]$, $s^{\prime \prime} \in [\min \{ s, s^{\prime} \}, s ]$ and $e^{\prime \prime} \in [e, \max \{ e, e^{\prime} \}]$. 
$A(s, e) = A(s^{\prime \prime}, e^{\prime \prime})$ follows from $[s^{\prime \prime}, e^{\prime \prime}] = [s, e]$. 
Therefore, we obtain $k = h$.

\textbf{Proof of statement (iv).}
We apply Lemma~\ref{lem:intv_function_shift} to the three intervals 
$[s, e]$, $[s^{\prime}, e^{\prime}]$, and $[s^{\prime \prime}, e^{\prime \prime}]$. 
Then, we obtain 
(A) $s^{\prime\prime h} \leq s^{\prime h}$, (B) $e^{\prime\prime h} = e^{\prime h}$, 
(C) $s^{\prime\prime h} = s^{h}$, (D) $e^{\prime\prime h} \geq e^{h}$, 
and (E) $e^{h} \leq e^{\prime h}$. 
Since $s^{h} = s^{\prime h}$, 
we obtain $s^{h} = s^{\prime h} = s^{\prime\prime h}$ and $e^{h} \leq e^{\prime h} = e^{\prime\prime h}$. 
 
We prove $k = h$ by contradiction. 
We assume that $k \neq h$ holds. 
Then, $k \geq h+1$ holds. 

Because of $f_{\interval}(s^{\prime\prime h}, e^{\prime\prime h}) = [s^{\prime\prime h+1}, e^{\prime\prime h+1}]$, 
the position $s^{\prime\prime h+1}$ of sequence $S^{h+1}$ satisfies at least one of the four conditions (i), (ii), (iii), and (iv) of Definition~\ref{def:f_interval}. 
If the position $s^{\prime\prime h+1}$ satisfies condition (i), 
then the position $s^{\prime\prime h+1}$ satisfies the condition (i) of Definition~\ref{def:f_interval} 
for function $f_{\interval}(s^{\prime h}, e^{\prime h})$. 
The existence of the position $s^{\prime\prime h+1}$ indicates that $f_{\interval}(s^{\prime h}, e^{\prime h}) \neq \perp$. 

If the position $s^{\prime\prime h+1}$ satisfies condition (ii), 
then the position $s^{\prime\prime h+1}$ satisfies the condition (ii) of Definition~\ref{def:f_interval} 
for function $f_{\interval}(s^{\prime h}, e^{\prime h})$. 
The existence of the position $s^{\prime\prime h+1}$ indicates that $f_{\interval}(s^{\prime h}, e^{\prime h}) \neq \perp$. 

If the position $s^{\prime\prime h+1}$ satisfies condition (iii), 
then the position $s^{\prime\prime h+1}$ satisfies the condition (iii) of Definition~\ref{def:f_interval} 
for function $f_{\interval}(s^{\prime h}, e^{\prime h})$ 
because $e^{\prime h} = e^{\prime\prime h}$ and $e^{\prime \prime} = e^{\prime}$. 
The existence of the position $s^{\prime\prime h+1}$ indicates that $f_{\interval}(s^{\prime h}, e^{\prime h}) \neq \perp$. 

If the position $s^{\prime\prime h+1}$ satisfies condition (iv), 
then the position $s^{\prime\prime h+1}$ satisfies the condition (iv) of Definition~\ref{def:f_interval} 
for function $f_{\interval}(s^{h}, e^{h})$ 
because $s^{h} = s^{\prime\prime h}$ and $s = s^{\prime \prime}$.
The existence of the position $s^{\prime\prime h+1}$ indicates that $f_{\interval}(s^{h}, e^{h}) \neq \perp$. 

We showed that at least one of $f_{\interval}(s^{\prime h}, e^{\prime h}) \neq \perp$ and $f_{\interval}(s^{h}, e^{h}) \neq \perp$ holds. 
On the other hand, $f_{\interval}(s^{\prime h}, e^{\prime h}) = \perp$ and $f_{\interval}(s^{h}, e^{h}) = \perp$ 
follows from the definitions of $A(s, e)$ and $A(s^{\prime}, e^{\prime})$. 
There exists a contradiction. Therefore, $k = h$ must hold. 

\textbf{Proof of statement (v).}
Statement (v) can be proved using the same approach as for statement (iv). 

\textbf{Proof of statement (vi).}
We prove $k = h$ by contradiction. 
We assume that $k \neq h$ holds. 
Then, $k \geq h+1$ holds. 
We apply Lemma~\ref{lem:intv_function_shift}~\ref{enum:intv_function_shift:7} to two intervals $[s^{\prime \prime}, e^{\prime \prime}]$ and $[s, e^{\prime}]$. 
The lemma shows that $|A(s, e^{\prime})| \geq h+2$ holds  
because $[s^{\prime \prime}, e^{\prime \prime}] \subseteq [s, e^{\prime}]$ and $k \geq h+1$. 
On the other hand, statement (iv) shows that $|A(s, e^{\prime})| = h+1$ holds. 
The two facts $|A(s, e^{\prime})| = h+1$ and $|A(s, e^{\prime})| \geq h+2$ yield a contradiction. 
Therefore, $k = h$ must hold. 

\textbf{Proof of statement (vii).}
We prove $k = h$ by contradiction. 
We assume that $k \neq h$ holds. 
Then, $k \geq h+1$ holds. 
We apply Lemma~\ref{lem:intv_function_shift}~\ref{enum:intv_function_shift:7} to two intervals $[s^{\prime \prime}, e^{\prime \prime}]$ and $[s^{\prime}, e]$. 
The lemma shows that $|A(s^{\prime}, e)| \geq h+2$ holds  
because $[s^{\prime \prime}, e^{\prime \prime}] \subseteq [s^{\prime}, e]$ and $k \geq h+1$. 
On the other hand, statement (v) shows that $|A(s^{\prime}, e)| = h+1$ holds. 
The two facts $|A(s^{\prime}, e)| = h+1$ and $|A(s^{\prime}, e)| \geq h+2$ yield a contradiction. 
Therefore, $k = h$ must hold. 

\textbf{Proof of Proposition~\ref{prop:last_set_inclusion}.}
$s^{h} = s^{\prime\prime h}$ follows from statement (i). 
We prove $k = h$. 
Because of $\gamma \in [s, e]$ and $\gamma \in [s^{\prime}, e^{\prime}]$, 
one of the following four conditions holds: 
\begin{enumerate}[label=\textbf{(\Alph*)}]
    \item $[s, e] \subseteq [s^{\prime}, e^{\prime}]$;
    \item $[s^{\prime}, e^{\prime}] \subseteq [s, e]$; 
    \item $s^{\prime} > s$ and $e^{\prime} > e$; 
    \item $s^{\prime} < s$ and $e^{\prime} < e$.     
\end{enumerate}
For condition (A), $h = k$ follows from statement (ii).
For condition (B), $h = k$ follows from statement (iii).
For condition (C), $h = k$ follows from statement (vi).
For condition (D), $h = k$ follows from statement (vii). 
Therefore, $k = h$ always holds. 

$s^{\prime \prime k} = s^{h}$ follows from $s^{h} = s^{\prime\prime h}$ and $k = h$. 
Therefore, we obtain $[s^{\prime \prime}, e^{\prime \prime}] \in \Delta(u)$. 
\end{proof}

We showed that the partition $\Delta(s^h)$ of Definition~\ref{def:RR_Delta} 
satisfies the conditions of the partition introduced in Section~\ref{subsec:simplicied_RASS}. 
Therefore, we obtain Theorem~\ref{theo:delta_proof}.

%\subsection{Properties of Interval Attractors}\label{subsec:RR_interval_properties}

%\input{4_interval_attractor/4_2_1_interval_attractor}
%\input{4_interval_attractor/4_2_2_interval_attractor_proof}

\section{Subsets of Interval Attractors}\label{sec:IA_subset_and_functions} 
This section introduces sixteen subsets of interval attractors and a function that returns a set of interval attractors for a given interval attractor. 
Thirteen queries on these subsets are defined in the later sections, and these queries are used to solve the SA and ISA queries. 
Some subsets and the function introduced in this section are used to show that the algorithms presented in the later sections work, and that the dynamic data structures introduced there can be stored in $\delta$-optimal space. 

The subsets and function introduced in this section are defined on 
the interval attractors constructed from the RLSLP $\mathcal{G}^R=(\mathcal{V}, \Sigma, \mathcal{D}, E)$ built by the restricted recompression, 
which are explained in Section~\ref{sec:RASSO}. 
Here, the height-balanced derivation tree of the RLSLP is represented as $H+1$ sequences 
$S^{0}, S^{1}, \ldots, S^{H}$. 
For simplicity, 
we introduce several notations. 
For the position $s^{h}$ of sequence $S^{h}$ corresponding to a node $u \in \mathcal{U}$, 
$I(s^{h})$ is defined as the interval attractor $I(u)$ associated with the node $u$ 
(i.e., $I(s^{h}) = I(u)$). 
$\Psi_{\RR}$ is defined the set of interval attractors obtained from RLSLP $\mathcal{G}^{R}$. 
Formally, let $\Psi_{\RR} = \{ I(s^{h}) \mid h \in [0, H] \land s^{h} \in [1, |S^{h}|] \land \Delta(s^{h}) \neq \emptyset \}$. 
Here, $\Delta(s^{h})$ is the set of Definition~\ref{def:RR_Delta}. 
For each interval $[s, e] \in \Delta$, 
$I_{\capture}(s, e)$ is defined as 
the interval attractor $I(u)$ with the node $u \in \mathcal{U}$ that satisfies 
$\Delta(u)$ contains the interval $[s, e]$. 
The substring $C$ derived from $u$ on $T$ is called \emph{associated string} of interval attractor $I(u)$. 
The starting position $\gamma$ of $C$ on $T$ is called \emph{attractor position} of interval attractor $I(u)$. 
The height $h$ of $u$ in the derivation tree is called \emph{level} of interval attractor $I(u)$.

\subsection{Twelve Subsets of Interval Attractors}\label{subsec:IA_subsets} 
\begin{figure}[p]
 \begin{center}
		\includegraphics[scale=0.8]{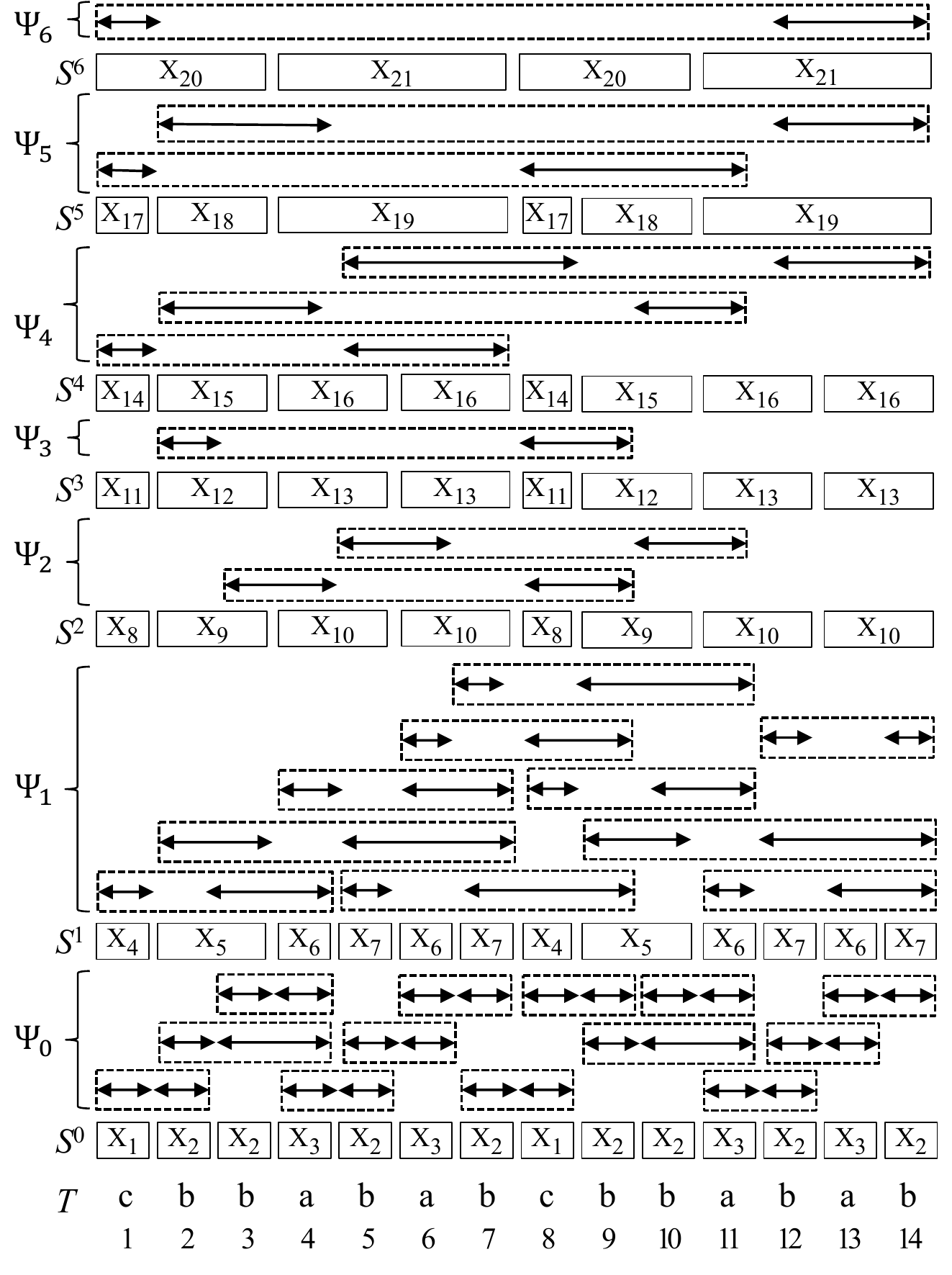}

	  \caption{ 
   An illustration of the level-wise interval attractors in the derivation tree of the RLSLP used in Figure~\ref{fig:restricted_recompression}. 
   Each dotted rectangle represents an interval attractor $([p, q], [\ell, r])$ in set $\Psi_{\RR}$. The left arrow in the dotted rectangle represents interval $[p, q]$.
   Similarly, the right arrow in the dotted rectangle represents interval $[\ell, r]$.    
	  }
\label{fig:psi_h}
 \end{center}
\end{figure}

This section introduces twelve subsets of interval attractors $\Psi_{\RR}$ and their properties. 
Thirteen queries on those subsets are defined in the later sections, and these queries are used to solve the SA and ISA queries. 

\paragraph{Level-wise Interval Attractors in Derivation tree.}
For each integer $h \in [0, H]$, 
subset $\Psi_{h}$ consist of interval attractors such that the level of each interval attractor is $h$. 
$\Psi_{\RR}$ is divided into $H+1$ disjoint sets $\Psi_{0}$, $\Psi_{1}$, $\ldots$, $\Psi_{H}$ (i.e., $\Psi_{h} \cap \Psi_{h^{\prime}} = \emptyset$ for any pair of distinct integers $h, h^{\prime} \in [0, H]$ and $\Psi_{\RR} = \bigcup_{h = 0}^{H} \Psi_{h}$).
We refer to each of these $H+1$ disjoint sets, $\Psi_{0}$, $\Psi_{1}$, $\ldots$, $\Psi_{H}$, as level-wise interval attractors in the derivation tree. Especially $\Psi_{h}$ is refereed to as $h$-th level interval attractors.

Figure~\ref{fig:psi_h} illustrates the level-wise interval attractors in the derivation tree of the RLSLP used in Figure~\ref{fig:restricted_recompression}. 
Here, 
\begin{itemize}
    \item $\Psi_{0} = \{ ([1, 1], [2, 2])$, $([2, 2], [3, 4])$, $([3, 3], [4, 4])$, $([4, 4], [5, 5])$, $([5, 5], [6, 6])$, $([6, 6], [7, 7])$, $([7, 7]$, $[8, 8])$, $([8, 8], [9, 9])$, $([9, 9], [10, 11])$, $([10, 10], [11, 11])$, $([11, 11], [12, 12])$, $([12, 12], [13, 13])$, $([13, 13]$, $[14, 14]) \}$;
    \item $\Psi_{1} = \{ ([1, 1], [3, 4]), ([2, 3], [5, 7]), ([4, 4], [6, 7]), ([5, 5], [7, 9]), ([6, 6], [8, 9]), ([7, 7], [9, 11]), ([8, 8]$, $[10, 11]), ([9, 10], [12, 14]), ([11, 11], [13, 14]), ([12, 12], [14, 14]) \}$;
    \item $\Psi_{2} = \{ ([3, 4], [8, 9]), ([5, 6], [10, 11]) \}$;
    \item $\Psi_{3} = \{ ([2, 2], [8, 9]) \}$;
    \item $\Psi_{4} = \{ ([1, 1], [5, 7]), ([2, 4], [10, 11]), ([5, 8], [12, 14]) \}$;
    \item $\Psi_{5} = \{ ([1, 1], [8, 11]), ([2, 4], [12, 14]) \}$;
    \item $\Psi_{6} = \{ ([1, 1], [12, 14]) \}$;
    \item $\Psi_{7} = \Psi_{8} = \Psi_{9} = \emptyset$. 
\end{itemize}
Let $s^{2}$ be the fifth position of sequence $S^{2}$. 
Then, $\Delta(s^{2}) = \{ [5, 10], [5, 11], [6, 10], [6, 11] \}$ 
and $I(s^{2}) = ([5, 6], [10, 11])$. 
Therefore, $I_{\capture}(6, 10) = ([5, 6], [10, 11])$ holds. 
The attractor position of the interval attractor $I_{\capture}(6, 10)$ is $8$ 
because the fifth nonterminal $X_{8}$ of sequence $S^{2}$ derives substring $T[8..8]$. 

\paragraph{Subset $\Psi_{\leftmost}$.}
Subset $\Psi_{\leftmost} \subseteq \Psi_{\RR}$ consists of interval attractors. 
Each interval attractor $([p, q], [\ell, r])$ $\in \Psi_{\leftmost}$ satisfies one of the following  three conditions: (i) $p=1$; (ii) $r=n$; (iii) for $p>1$ and $r<n$, substring $T[p-1..r+1]$ appears as the leftmost occurrence in $T$. 
Formally, 
$\Psi_{\leftmost} = \{ ([p, q], [\ell, r]) \in \Psi_{\RR} \mid (p = 1) \lor (r = n) \} \cup \{ ([p, q], [\ell, r]) \in \Psi_{\RR} \mid (p > 1) \land (r < n) \land (p - 1 = \min \Occ(T, T[p-1..r+1]) \}$. 

%%%%%%%%%%%%%%%%%%%%%%%%%%%%%%%%%%%%%%%%%%%%%%%%%

\paragraph{Subset $\Psi_{\centerset}(C)$.}
For a string $C \in \Sigma^{+}$, 
subset $\Psi_{\centerset}(C) \subseteq \Psi_{\RR}$ consists of interval attractors 
such that the associated string of each interval attractor is $C$. 

%as its a
%is defined as a set of interval attractors in $([p, q], [\ell, r]) \in \Psi_{\RR}$ associating with such $C$.
%For a string $C$ not associating with any interval attract in $\Psi_{\RR}$, $\Psi_{\centerset}(C) = \emptyset$.

%If the length of interval $[\gamma, r]$ is larger than $\sum_{w = 1}^{h+1} \lfloor \mu(w) \rfloor$ (i.e., $r - \gamma + 1 > \sum_{w = 1}^{h+1} \lfloor \mu(w) \rfloor$) 
%for the attractor position $\gamma$ of the interval attractor $([p, q], [\ell, r])$, 
%then $C$ is defined as the string derived from the $s^{h}$-th nonterminal of sequence $S^{h}$ (i.e., $C = \val(S^{h}[s^{h}])$). 
%Otherwise, $C$ is defined as substring $T[\gamma..r]$. 

%%%%%%%%%%%%%%%%%%%%%%%%%%%%%%%%%%%%%%%%%%%%%%%%%

\paragraph{Subset $\Psi_{\run}$.}
Periodicity within interval attractors enables reducing the number of interval attractors that need to be preserved. 
Consider an interval attractor $I(u) = ([p, q], [\ell, r]) \in \Psi_{\RR}$ of level $h$ 
with attractor position $\gamma$ and attractor string $C$.  
An interval attractor $I(u)$ is periodic string $T[p-1..r+1]$ has a sufficiently long repetition of the string $C$ as a prefix. 
Formally, $I(u)$ is periodic if it satisfies the following two conditions: 
\begin{enumerate}[label=\textbf{(\roman*)}]
\item \label{cond:Psi_HR:1} 
String $T[p-1..\gamma - 1]$ is a suffix of string $C^{n+1}$~(i.e., $\lcs(T[p-1..\gamma - 1], C^{n+1}) = T[p-1..\gamma - 1]$); 
\item \label{cond:Psi_HR:2} 
The length of the longest common prefix between two strings $T[\gamma..r]$ and $C^{n+1}$ is larger than $1 + \sum_{w = 1}^{h+3} \lfloor \mu(w) \rfloor$~(i.e., $|\lcp(T[\gamma..r], C^{n+1})| > 1 + \sum_{w = 1}^{h+3} \lfloor \mu(w) \rfloor$). 
\end{enumerate}

Subset $\Psi_{\run} \subseteq \Psi_{\RR}$ consists of 
periodic interval attractors in set $\Psi_{\RR}$. 
Therefore, the interval attractors of subset $\Psi_{\run}$ are periodic interval attractors. 
Conversely, interval attractors not contained in the subset $\Psi_{\run}$ are called \emph{non-periodic interval attractors}.

%%%%%%%%%%%%%%%%%%%%%%%%%%%%%%%%%

\paragraph{Subset $\Psi_{\source}$.}
For the attractor position $\gamma$ and associated string $C$ of each interval attractor $([p, q], [\ell, r]) \in \Psi_{\RR}$,
subset $\Psi_{\source} \subseteq \Psi_{\RR}$ consists of interval attractors $([p, q], [\ell, r]) \in \Psi_{\RR}$ satisfying two conditions:
\begin{enumerate}[label=\textbf{(\roman*)}]
    \item each $([p, q], [\ell, r])$ is non-periodic (i.e., $([p, q], [\ell, r]) \not \in \Psi_{\run}$); 
    \item for the $h$-th level interval attractors $\Psi_{h}$ containing $([p, q], [\ell, r]) \in \Psi_{\RR}$, 
    set $\Psi_{h} \cap \Psi_{\run} \cap \Psi_{\centerset}(C)$ contains 
    an interval attractor $([p^{\prime}, q^{\prime}], [\ell^{\prime}, r^{\prime}])$ 
    such that its attractor position $\gamma^{\prime}$ is equal to $\gamma + |C|$. 
\end{enumerate}

%%%%%%%%%%%%%%%%%%%%%%%

\paragraph{Subset $\Psi_{\str}(P)$.}
For a string $P \in \Sigma^{+}$, 
subset $\Psi_{\str}(P) \subseteq \Psi_{\RR}$ consists of interval attractors $([p, q], [\ell, r]) \in \Psi_{\RR}$ such that substring $T[p-1..r+1]$ is equal to string $P$. 
Formally, $\Psi_{\str}(P) = \{ ([p, q], [\ell, r]) \in \Psi_{\RR} \mid T[p-1..r+1] = P \}$. 

\paragraph{Subset $\Psi_{\CCP}(P)$.}
For a string $P \in \Sigma^{+}$, 
let $T[i_{1}..i_{1} + |P|-1]$, $T[i_{2}..i_{2} + |P|-1]$, $\ldots$, $T[i_{k}..i_{k} + |P|-1]$ be all the occurrences of the string $P$ in input string $T$ (i.e., $\{ i_{1}, i_{2}, \ldots, i_{k} \} = \Occ(T, P)$). 
Then, 
subset $\Psi_{\CCP}(P) \subseteq \Psi_{\RR}$ is defined as the set of $k$ interval attractors $I_{\capture}(i_{1}, i_{1} + |P|-1)$, 
$I_{\capture}(i_{2}, i_{2} + |P|-1)$, $\ldots$, $I_{\capture}(i_{k}, i_{k} + |P|-1)$ 
(i.e., $\Psi_{\CCP}(P) = \bigcup_{i \in \Occ(T, P)} I_{\capture}(i, i+|P|-1)$).

\paragraph{Subset $\Psi_{\lex}(P)$.}
For a string $P \in \Sigma^{+}$, 
$\Psi_{\lex}(P) \subseteq \Psi_{\RR}$ consists of interval attractors $([p, q], [\ell, r]) \in \Psi_{\RR}$ such that for the attractor position $\gamma$ of each interval attractor $([p, q], [\ell, r]) \in \Psi_{\RR}$, 
substring $T[\gamma..r+1]$ is lexicographically smaller than the string $P$.
Formally, $\Psi_{\lex}(P) = \{ ([p, q], [\ell, r]) \in \Psi_{\RR} \mid T[\gamma..r+1] \prec P \}$.

\paragraph{Subset $\Psi_{\lcp}(K)$.}
For an integer $K \geq 0$, 
$\Psi_{\lcp}(K) \subseteq \Psi_{\RR}$ consists of interval attractors $([p, q], [\ell, r]) \in \Psi_{\RR}$ such that for the attractor position $\gamma$ and associated string $C$ of each interval attractor $([p, q], [\ell, r]) \in \Psi_{\RR}$, 
the length of the longest common prefix between two strings $T[\gamma..r]$ and $C^{n+1}$ is $K$. 
Formally, $\Psi_{\lcp}(K) = \{ ([p, q], [\ell, r]) \in \Psi_{\RR} \mid |\lcp(T[\gamma..r], C^{n+1})| = K \}$. 

\paragraph{Subset $\Psi_{\modulo}(M)$.}
For an integer $M \geq 0$, 
subset $\Psi_{\modulo}(M) \subseteq \Psi_{\RR}$ consists of interval attractors $([p, q], [\ell, r]) \in \Psi_{\RR}$ such that each $([p,q], [\ell,r]) \in \Psi_{\RR}$ satisfies the following two conditions:
\begin{enumerate}[label=\textbf{(\roman*)}]
    \item 
    Let $\Psi_{h}$ be the $h$-th level interval attractors containing the interval attractor $([p, q], [\ell, r])$. 
    The interval attractor $([p,q],[\ell,r])$ is contained in subset $\Psi_{\lcp}(K)$ for an integer $K > 1 + \sum_{w = 1}^{h+3} \lfloor \mu(w) \rfloor$ (i.e., 
    $\exists K \in [2 + \sum_{w = 1}^{h+3} \lfloor \mu(w) \rfloor, n+1] \text{ s.t. } ([p, q], [\ell, r]) \in \Psi_{\lcp}(K)$); 
    \item $(K - (2 + \sum_{w = 1}^{h+3} \lfloor \mu(w) \rfloor) ) \mod |C| = M$ for the associated string $C$ of the interval attractor $([p, q], [\ell, r])$.     
\end{enumerate}
Formally, $\Psi_{\modulo}(M) = \bigcup_{h = 0}^{H} \bigcup_{C \in \Sigma^{+}} \bigcup_{K \geq 2 + \sum_{w = 1}^{h+3} \lfloor \mu(w) \rfloor}^{n+1} \{ ([p, q], [\ell, r]) \in \Psi_{h} \cap \Psi_{\centerset}(C) \cap \Psi_{\lcp}(K) \mid (K - 2 - \sum_{w = 1}^{h+3} \lfloor \mu(w) \rfloor) \mod |C| = M \}$. 

\paragraph{Subsets $\Psi_{\preceding}$ and $\Psi_{\succeeding}$.}
Subset $\Psi_{\preceding} \subseteq \Psi_{\RR}$ 
consists of interval attractors $([p,q], [\ell, r]) \in \Psi_{\RR}$ such that 
for the attractor position $\gamma$ and associated string $C$ of each interval attractor $([p, q], [\ell, r]) \in \Psi_{\RR}$, 
substring $T[\gamma..r+1]$ is lexicographically smaller than string $C^{n+1}$. 
Formally, 
$\Psi_{\preceding} = \{ ([p, q], [\ell, r]) \in \Psi_{\RR} \mid T[\gamma..r+1] \prec C^{n+1} \}$. 

In contrast, 
subset $\Psi_{\succeeding} \subseteq \Psi_{\RR}$ 
consists of interval attractors $([p,q], [\ell, r]) \in \Psi_{\RR}$ such that 
for the attractor position $\gamma$ and associated string $C$ of each interval attractor $([p, q], [\ell, r]) \in \Psi_{\RR}$, 
substring $T[\gamma..r+1]$ is lexicographically larger than string $C^{n+1}$. 
Formally, 
$\Psi_{\succeeding} = \{ ([p, q], [\ell, r]) \in \Psi_{\RR} \mid C^{n+1} \prec T[\gamma..r+1] \}$. 

Set $\Psi_{\RR}$ can be divided into the two subsets $\Psi_{\preceding}$ and $\Psi_{\succeeding}$ (i.e., $\Psi_{\RR} = \Psi_{\preceding} \cup \Psi_{\succeeding}$ and $\Psi_{\preceding} \cap \Psi_{\succeeding} = \emptyset$). 
This is because for the attractor position $\gamma$ and associated string $C$ of each interval attractor $([p, q], [\ell, r]) \in \Psi_{\RR}$, 
either $T[\gamma..r+1] \prec C^{n+1}$ or $C^{n+1} \prec T[\gamma..r+1]$ holds.

\subsubsection{Properties of Subsets}

The following lemma ensures that 
for each interval attractor $([p, q], [\ell, r]) \in \Psi_{\RR}$, 
there exists the only one interval attractor $([p^{\prime}, q^{\prime}], [\ell^{\prime}, r^{\prime}])$ 
satisfying $T[p - 1..r + 1] = T[p^{\prime} - 1..r^{\prime} + 1]$ in subset $\Psi_{\leftmost}$.

\begin{lemma}\label{lem:lm_basic_property}
The following two statements hold for subset $\Psi_{\leftmost}$. 
\begin{enumerate}[label=\textbf{(\roman*)}]
\item \label{enum:lm_basic_property:1} $T[p-1..r+1] \neq T[p^{\prime}-1..r^{\prime}+1]$ for any pair of two distinct interval attractors $([p, q], [\ell, r])$ and $([p^{\prime}, q^{\prime}], [\ell^{\prime}, r^{\prime}])$ in subset $\Psi_{\leftmost}$; 
\item \label{enum:lm_basic_property:2} for each interval attractor $([p, q], [\ell, r]) \in \Psi_{\RR}$, 
subset $\Psi_{\leftmost}$ contains an interval attractor $([p^{\prime}, q^{\prime}], [\ell^{\prime}, r^{\prime}])$ satisfying $T[p-1..r+1] = T[p^{\prime}-1..r^{\prime}+1]$.
\end{enumerate}
\end{lemma}
\begin{proof}
The proof of Lemma~\ref{lem:lm_basic_property} is as follows. 

\textbf{Proof of Lemma~\ref{lem:lm_basic_property}(i).}
We prove $T[p-1..r+1] \neq T[p^{\prime}-1..r^{\prime}+1]$ by contradiction. 
We assume that $T[p-1..r+1] = T[p^{\prime}-1..r^{\prime}+1]$ holds. 
Then, 
$[p, r] = [p^{\prime}, r^{\prime}]$ follows from the definition of the subset $\Psi_{\leftmost}$. 
On the other hand, $[p, r] \neq [p^{\prime}, r^{\prime}]$ follows from Lemma~\ref{lem:IA_super_basic_property}~\ref{enum:IA_super_basic_property:3}. 
The two facts $T[p-1..r+1] = T[p^{\prime}-1..r^{\prime}+1]$ and $[p, r] \neq [p^{\prime}, r^{\prime}]$ yield a contradiction. 
Therefore, $T[p-1..r+1] \neq T[p^{\prime}-1..r^{\prime}+1]$ must hold. 

\textbf{Proof of Lemma~\ref{lem:lm_basic_property}(ii).}
If $[p-1, r+1] \subseteq [1, n]$, 
then there exists an interval $[p^{\prime}, r^{\prime}] \subseteq [1, n]$ in input string $T$ 
satisfying $p^{\prime} - 1 = \min \Occ(T, T[p-1..r+1])$ and $T[p-1..r+1] = T[p^{\prime}-1..r^{\prime}+1]$. 
From Theorem~\ref{theo:IA_SYNC_X}, 
there exists a pair of two integers $q^{\prime}$ and $\ell^{\prime}$ satisfying 
$([p^{\prime}, q^{\prime}], [\ell^{\prime}, r^{\prime}]) \in \Psi_{\RR}$. 
The subset $\Psi_{\leftmost}$ contains the interval attractor $([p^{\prime}, q^{\prime}], [\ell^{\prime}, r^{\prime}])$. 

Otherwise (i.e., $p = 1$ or $r = n$), 
the subset $\Psi_{\leftmost}$ contains the interval attractor $([p, q], [\ell, r])$. 
Therefore, Lemma~\ref{lem:lm_basic_property}(ii) holds. 
\end{proof}

%%%%%%%%%%%%%%%%%%%%%%%%%%%%%%%%%%%%%%%%%%%%

The following two lemmas state properties of subset $\Psi_{\centerset}(C)$. 

\begin{lemma}\label{lem:associated_string_C}
    Consider two interval attractors $([p, q], [\ell, r])$ and $([p^{\prime}, q^{\prime}], [\ell^{\prime}, r^{\prime}])$ 
    in the $h$-th level interval attractors $\Psi_{h}$ for an integer $h \in [0, H]$. 
    Let $\gamma$ and $C$ be the attractor position and associated string of interval attractor $([p, q], [\ell, r])$, respectively. 
    Analogously, $\gamma^{\prime}$ and $C^{\prime}$ are defined for interval attractor $([p^{\prime}, q^{\prime}], [\ell^{\prime}, r^{\prime}])$.
    The following two statements hold: 
    \begin{enumerate}[label=\textbf{(\roman*)}]
        \item \label{enum:associated_string_C:1} if $|[\gamma, r]| > \sum_{w = 1}^{h+1} \lfloor \mu(w) \rfloor$, then 
        the associated string $C$ is a prefix of string $T[\gamma..r]$ (i.e., $\lcp(T[\gamma..r], C) = C$), 
        and $|C| \leq \lfloor \mu(h+1) \rfloor$; 
        \item \label{enum:associated_string_C:2} if the length of the longest common prefix between two strings $T[\gamma..r]$ and $T[\gamma^{\prime}..r^{\prime}]$ 
        is larger than $\sum_{w = 1}^{h+1} \lfloor \mu(w) \rfloor$~(i.e., $|\lcp(T[\gamma..r], T[\gamma^{\prime}..r^{\prime}])| > \sum_{w = 1}^{h+1} \lfloor \mu(w) \rfloor$), 
        then $C = C^{\prime}$. 
    \end{enumerate}    
\end{lemma}
\begin{proof}
Consider two interval attractors $I_{\capture}(p, r)$ and $I_{\capture}(p^{\prime}, r^{\prime})$. 
Then, Lemma~\ref{lem:IA_maximal_lemma} shows that $I_{\capture}(p, r) = ([p, q], [\ell, r])$ and $I_{\capture}(p^{\prime}, r^{\prime}) = ([p^{\prime}, q^{\prime}], [\ell^{\prime}, r^{\prime}])$.
Let $[s^{k}, e^{k}]$ and $[s^{\prime k^{\prime}}, e^{\prime k^{\prime}}]$ be the tails of sequences $A(p, r)$ and $A(p^{\prime}, r^{\prime})$, 
respectively. 
Then, 
the associated string $C$ is defined as 
the substring $T[x^{k}_{s}..y^{k}_{s}]$ derived from 
the $s^{k}$-th nonterminal of sequence $S^{k}$ in string $T$. 
Similarly, 
the associated string $C^{\prime}$ is defined as 
the substring $T[x^{\prime k^{\prime}}_{s}..y^{\prime k^{\prime}}_{s}]$ derived from 
the $s^{\prime k^{\prime}}$-th nonterminal of sequence $S^{k^{\prime}}$ in string $T$. 
Here, $\gamma = x^{k}_{s}$, $\gamma^{\prime} = x^{\prime k^{\prime}}_{s}$, and $h = k = k^{\prime}$ hold.

The proof of Lemma~\ref{lem:associated_string_C} is as follows. 

\paragraph{Proof of Lemma~\ref{lem:associated_string_C}(i).}
Lemma~\ref{lem:associated_string_C}(i) holds if 
$|C| \leq \lfloor \mu(h+1) \rfloor$ 
because $C$ is a prefix of suffix $T[\gamma..n]$. 
$|[s^{h}, e^{h}]| \geq 2$ follows from Lemma~\ref{lem:f_rec_top_property}~\ref{enum:f_rec_top_property:3} 
and $|[\gamma, r]| > \sum_{w = 1}^{h+1} \lfloor \mu(w) \rfloor$. 
$|C| \leq \lfloor \mu(h+1) \rfloor$ follows from 
Lemma~\ref{lem:f_rec_top_property}~\ref{enum:f_rec_top_property:1} and $|[s^{h}, e^{h}]| \geq 2$. 
Therefore, we obtain Lemma~\ref{lem:associated_string_C}(i). 

\paragraph{Proof of Lemma~\ref{lem:associated_string_C}(ii).}
Let $\alpha = \min \{ |C|, |C^{\prime}| \}$. 
Then, $\alpha \leq \lfloor \mu(h+1) \rfloor$ follows from Lemma~\ref{lem:associated_string_C}(i). 
We can apply Lemma~\ref{lem:rr_property}~\ref{enum:rr_property:right} to the two nonterminals 
$S^{h}[s^{h}]$ and $S^{h}[s^{\prime h}]$ 
if the following condition holds: 
$T[\gamma..\gamma + \alpha + \sum_{w = 1}^{h} \lfloor \mu(w) \rfloor] = T[\gamma^{\prime}..\gamma^{\prime} + \alpha + \sum_{w = 1}^{h} \lfloor \mu(w) \rfloor]$. 
Since $\alpha \leq \lfloor \mu(h+1) \rfloor$ and $|\lcp(T[\gamma..r], T[\gamma^{\prime}..r^{\prime}])| > \sum_{w = 1}^{h+1} \lfloor \mu(w) \rfloor$, 
this conditions is satisfied. 
Lemma~\ref{lem:rr_property}~\ref{enum:rr_property:right} shows that $C = C^{\prime}$. 
Therefore, we obtain Lemma~\ref{lem:associated_string_C}(ii). 

\end{proof}

The following two lemmas state properties of set $\Psi_{\str}(P)$. 
\begin{lemma}\label{lem:psi_str_property}
Consider two interval attractors $([p, q], [\ell, r]), ([p^{\prime}, q^{\prime}], [\ell^{\prime}, r^{\prime}]) \in \Psi_{\RR}$ 
satisfying $T[p-1..r+1] = T[p^{\prime}-1..r^{\prime}+1]$. 
Here, $([p, q], [\ell, r])$ and $([p^{\prime}, q^{\prime}], [\ell^{\prime}, r^{\prime}]) \in \Psi_{\str}(P)$ follow from 
the definition of the subset $\Psi_{\str}(P)$ for string $P = T[p-1..r+1]$. 
Then, the following three statements hold for the subset $\Psi_{\str}(P)$ of set $\Psi_{\RR}$: 
\begin{enumerate}[label=\textbf{(\roman*)}]
    \item \label{enum:psi_str_property:1} 
    let $\gamma$ and $\gamma^{\prime}$ be 
    the attractor positions of the two interval attractors $([p, q], [\ell, r])$ and $([p^{\prime}, q^{\prime}], [\ell^{\prime}, r^{\prime}])$, respectively. 
    Then, $|[p, \gamma-1]| = |[p^{\prime}, \gamma^{\prime}-1]|$ and $|[\gamma, r]| = |[\gamma^{\prime}, r^{\prime}]|$; 
    \item \label{enum:psi_str_property:2} 
    let $h$ and $h^{\prime}$ be the levels of the two interval attractors $([p, q], [\ell, r])$ and $([p^{\prime}, q^{\prime}], [\ell^{\prime}, r^{\prime}])$, respectively. 
    Then, $h = h^{\prime}$; 
    \item \label{enum:psi_str_property:3} 
    $|[p, q]| = |[p^{\prime}, q^{\prime}]|$ and $|[\ell, r]| = |[\ell^{\prime}, r^{\prime}]|$.  
\end{enumerate}

\end{lemma}
\begin{proof}
    Consider two interval attractors $I_{\capture}(p, r)$ and $I_{\capture}(p^{\prime}, r^{\prime})$. 
    Then, $I_{\capture}(p, r) = ([p, q], [\ell, r])$ and $I_{\capture}(p^{\prime}, r^{\prime}) = ([p^{\prime}, q^{\prime}], [\ell^{\prime}, r^{\prime}])$ follow from Lemma~\ref{lem:IA_maximal_lemma}. 

    The proof of Lemma~\ref{lem:psi_str_property} is as follows. 

    \textbf{Proof of Lemma~\ref{lem:psi_str_property}(i).}
    Corollary~\ref{cor:capture_gamma_corollary} shows that 
    $|[p, \gamma-1]| = |[p^{\prime}, \gamma^{\prime}-1]|$ and $|[\gamma, r]| = |[\gamma^{\prime}, r^{\prime}]|$  
    because $T[p..r] = T[p^{\prime}..r^{\prime}]$. 
    
    \textbf{Proof of Lemma~\ref{lem:psi_str_property}(ii).}
    Corollary~\ref{cor:capture_gamma_corollary} shows that 
    $h = h^{\prime}$ holds 
    because $T[p..r] = T[p^{\prime}..r^{\prime}]$. 
    
    \textbf{Proof of Lemma~\ref{lem:psi_str_property}(iii).}
    Because of $T[p-1..r+1] = T[p^{\prime}-1..r^{\prime}+1]$, 
    Theorem~\ref{theo:IA_SYNC_X} shows that 
    set $\Psi_{\RR}$ contains interval attractor 
    $([p^{\prime}, p^{\prime} + (q - p)], [r^{\prime} - (r - \ell), r^{\prime}])$. 
        
    We prove $|[p, q]| = |[p^{\prime}, q^{\prime}]|$ by contradiction. 
    We assume that $|[p, q]| \neq |[p^{\prime}, q^{\prime}]|$ holds. 
    Then, $([p^{\prime}, q^{\prime}], [\ell^{\prime}, r^{\prime}]) \neq ([p^{\prime}, p^{\prime} + (q - p)], [r^{\prime} - (r - \ell), r^{\prime}])$ holds. 
    Because of $([p^{\prime}, q^{\prime}], [\ell^{\prime}, r^{\prime}]) \neq ([p^{\prime}, p^{\prime} + (q - p)], [r^{\prime} - (r - \ell), r^{\prime}])$, 
    Lemma~\ref{lem:IA_super_basic_property}~\ref{enum:IA_super_basic_property:3} shows that 
    $[p^{\prime}, r^{\prime}] \neq [p^{\prime}, r^{\prime}]$ holds, 
    but it contradicts the fact that $[p^{\prime}, r^{\prime}] = [p^{\prime}, r^{\prime}]$ holds. 
    Therefore, $|[p, q]| = |[p^{\prime}, q^{\prime}]|$ must hold. 

    Similarly, 
    we can prove $|[\ell, r]| = |[\ell^{\prime}, r^{\prime}]|$ using the same approach used to prove $|[p, q]| = |[p^{\prime}, q^{\prime}]|$. 
    Therefore, $|[p, q]| = |[p^{\prime}, q^{\prime}]|$ and $|[\ell, r]| = |[\ell^{\prime}, r^{\prime}]|$ hold. 
    
\end{proof}

\begin{lemma}\label{lem:psi_str_occ_property}
Consider an interval attractor $([p, q], [\ell, r])$ in set $\Psi_{\RR}$. 
If $[p-1, r+1] \subseteq [1, n]$, 
then $|\Psi_{\str}(T[p-1..r+1])| = |\Occ(T, T[p-1..r+1])|$ and $\Occ(T, T[p-1..r+1]) = \{ p^{\prime} - 1 \mid ([p^{\prime}, q^{\prime}], [\ell^{\prime}, r^{\prime}]) \in \Psi_{\str}(T[p-1..r+1]) \}$. 
Otherwise, $\Psi_{\str}(T[p-1..r+1]) = \{ ([p, q], [\ell, r]) \}$ holds. 
\end{lemma}
\begin{proof}
    The proof of Lemma~\ref{lem:psi_str_occ_property} is as follows. 

    \textbf{Proof of Lemma~\ref{lem:psi_str_occ_property} for $[p-1, r+1] \subseteq [1, n]$.}
    Let $([p_{1}, q_{1}], [\ell_{1}, r_{1}])$, $([p_{2}, q_{2}], [\ell_{2}, r_{2}])$, 
    $\ldots$, $([p_{k}, q_{k}], [\ell_{k}, r_{k}])$ ($p_{1} \leq p_{2} \leq \cdots \leq p_{k}$) be the interval attractors of the subset $\Psi_{\str}(T[p-1..r+1])$.     
    Here, 
    $T[p_{s}-1..r_{s}+1] = T[p-1..r+1]$ follows from the definition of the subset $\Psi_{\str}(T[p-1..r+1])$ 
    for each integer $s \in [1, k]$. 
    Therefore, $\Occ(T, T[p-1..r+1]) \supseteq \{ p_{s} - 1 \mid s \in [1, k] \}$ holds. 

    We prove $p_{1} < p_{2} < \cdots < p_{k}$.     
    $r_{s} = p_{s} + (r - p)$ follows from $T[p_{s}-1..r_{s}+1] = T[p-1..r+1]$ for each integer $s \in [1, k]$. 
    Lemma~\ref{lem:IA_super_basic_property}~\ref{enum:IA_super_basic_property:3} shows that 
    $[p_{s}, p_{s} + (r - p)] \neq [p_{s^{\prime}}, p_{s^{\prime}} + (r - p)]$ holds 
    for any pair of two interval attractors 
    $([p_{s}, q_{s}], [\ell_{s}, r_{s}])$, $([p_{s^{\prime}}, q_{s^{\prime}}], [\ell_{s^{\prime}}, r_{s^{\prime}}]) \in \Psi_{\str}(T[p-1..r+1])$. 
    $p_{s} \neq p_{s^{\prime}}$ follows from $[p_{s}, p_{s} + (r - p)] \neq [p_{s^{\prime}}, p_{s^{\prime}} + (r - p)]$. 
    Therefore, $p_{1} < p_{2} < \cdots < p_{k}$ holds.  

    We prove $\Occ(T, T[p-1..r+1]) = \{ p_{s} - 1 \mid s \in [1, k] \}$ by contradiction. 
    We assume that $\Occ(T, T[p-1..r+1]) \neq \{ p_{s} - 1 \mid s \in [1, k] \}$ holds. 
    Then, there exists a position $i$ of input string $T$ satisfying 
    $i \in \Occ(T, T[p-1..r+1])$ and $i \not \in \{ p_{s} - 1 \mid s \in [1, k] \}$. 
    Because of $i \in \Occ(T, T[p-1..r+1])$, 
    Theorem~\ref{theo:IA_SYNC_X} shows that 
    set $\Psi_{\RR}$ contains an interval attractor $([p^{\prime}, q^{\prime}], [\ell^{\prime}, r^{\prime}])$ 
    satisfying $p^{\prime} - 1 = i$ and $T[p^{\prime}-1..r^{\prime}+1] = T[p-1..r+1]$. 
    Because of $T[p^{\prime}-1..r^{\prime}+1] = T[p-1..r+1]$, 
    $([p^{\prime}, q^{\prime}], [\ell^{\prime}, r^{\prime}]) \in \Psi_{\str}(T[p-1..r+1])$ holds. 
    $i \in \{ p_{s} - 1 \mid s \in [1, k] \}$ follows from 
    $([p^{\prime}, q^{\prime}], [\ell^{\prime}, r^{\prime}]) \in \Psi_{\str}(T[p-1..r+1])$ and $p^{\prime} - 1 = i$. 
    The two facts $i \not \in \{ p_{s} - 1 \mid s \in [1, k] \}$ and $i \in \{ p_{s} - 1 \mid s \in [1, k] \}$ yield a contradiction. 
    Therefore, $\Occ(T, T[p-1..r+1]) = \{ p_{s} - 1 \mid s \in [1, k] \}$ must hold. 

    Finally, Lemma~\ref{lem:psi_str_occ_property} follows from 
    $p_{1} < p_{2} < \cdots < p_{k}$ and $\Occ(T, T[p-1..r+1]) = \{ p_{s} - 1 \mid s \in [1, k] \}$. 
    
    \textbf{Proof of Lemma~\ref{lem:psi_str_occ_property} for $[p-1, r+1] \not \subseteq [1, n]$.}
    In this case, $p = 1$ or $r = n$ holds because $[p-1, r+1] \not \subseteq [1, n]$ and $[p-1, r+1] \subseteq [0, n+1]$ hold. 
    $([p, q], [\ell, r]) \in \Psi_{\str}(T[p-1..r+1])$ follows from the definition of the subset $\Psi_{\str}(T[p-1..r+1])$. 

    We prove $\Psi_{\str}(T[p-1..r+1]) = \{ ([p, q], [\ell, r]) \}$ by contradiction. 
    We assume that $\Psi_{\str}(T[p-1..r+1]) \neq \{ ([p, q], [\ell, r]) \}$ holds. 
    Then, the subset $\Psi_{\str}(T[p-1..r+1])$ contains an interval attractor $([p^{\prime}, q^{\prime}], [\ell^{\prime}, r^{\prime}])$ satisfying $([p^{\prime}, q^{\prime}], [\ell^{\prime}, r^{\prime}]) \neq ([p, q], [\ell, r])$. 
    Here, $T[p^{\prime}-1..r^{\prime}+1] = T[p-1..r+1]$ holds. 
    $|[p, r]| \neq |[p^{\prime}, r^{\prime}]|$ follows from $T[p^{\prime}-1..r^{\prime}+1] = T[p-1..r+1]$. 
    Lemma~\ref{lem:IA_super_basic_property}~\ref{enum:IA_super_basic_property:3} shows that 
    $[p, r] \neq [p^{\prime}, r^{\prime}]$ holds. 
    $p \neq p^{\prime}$ follows from $[p, r] \neq [p^{\prime}, r^{\prime}]$ and $|[p, r]| \neq |[p^{\prime}, r^{\prime}]|$. 

    If $p = 1$, 
    then $T[p-1] = \$$ holds. 
    $p^{\prime} \in [2, n]$ follows from $p \in [1, n]$ and $p \neq p^{\prime}$. 
    $T[p^{\prime}-1] \neq \$$ follows from $p^{\prime} \in [2, n]$. 
    On the other hand, $T[p^{\prime}-1] = \$$ follows from $T[p-1] = \$$ and $T[p^{\prime}-1..r^{\prime}+1] = T[p-1..r+1]$. 
    The two facts $T[p^{\prime}-1] = \$$ and $T[p^{\prime}-1] \neq \$$ yield a contradiction. 

    Otherwise (i.e., $p \neq 1$), 
    $r = n$ holds. 
    In this case, we can prove two facts 
    $T[r^{\prime}+1] = \$$ and $T[r^{\prime}+1] \neq \$$ using the same approach as for $p = 1$. 
    Therefore, $\Psi_{\str}(T[p-1..r+1]) = \{ ([p, q], [\ell, r]) \}$ must hold. 
\end{proof}

The following lemma states properties of interval attractors contained in subset $\Psi_{\run}$.

\begin{lemma}\label{lem:psi_run_basic_property}
Consider an interval attractor $([p, q], [\ell, r])$ in subset $\Psi_{\run}$. 
Here, the interval attractor $([p, q], [\ell, r])$ is 
contained in the $h$-th level interval attractors $\Psi_{h}$ for an integer $h \in [0, H]$; 
let $\gamma$ and $C$ be the attractor position and associated string of the interval attractor $([p, q], [\ell, r])$, respectively; 
let $K \geq 0$ be an integer satisfying $([p, q], [\ell, r]) \in \Psi_{\lcp}(K)$. 
The following nine statements hold: 
\begin{enumerate}[label=\textbf{(\roman*)}]
    \item \label{enum:psi_run_basic_property:1}
    $|C| \leq \lfloor \mu(h+1) \rfloor$, $\ell \leq \gamma + \sum_{w = 1}^{h} \lfloor \mu(w) \rfloor$, 
    $|[p, q]| = |C|$, 
    $|[p, \gamma-1]| \geq |C| - 1$, $p \geq 2$, and $|[\gamma, r]| > 1 + \sum_{w = 1}^{h+3} \lfloor \mu(w) \rfloor$;
    \item \label{enum:psi_run_basic_property:2}
    a pair $([p + |C|, q + |C|], [\ell + |C|, r])$ of intervals is contained in the set $\Psi_{h} \cap \Psi_{\centerset}(C) \cap \Psi_{\lcp}(K - |C|)$, 
    and $\gamma^{\prime} = \gamma + |C|$ holds for the attractor position $\gamma^{\prime}$ of the interval attractor $([p + |C|, q + |C|], [\ell + |C|, r])$; 
    \item \label{enum:psi_run_basic_property:3} 
    set $\Psi_{h}$ contains an interval attractor 
    $([p^{\prime}, q^{\prime}], [\ell^{\prime}, r^{\prime}]) \in \Psi_{\RR}$ satisfying  
    (a) $p^{\prime} < p$, 
    (b) $q^{\prime} = p-1$, 
    (c) $\ell^{\prime} \leq \ell$, 
    (d) $r^{\prime} = r$, 
    (e) $\gamma^{\prime} = \gamma - |C|$ for the attractor position $\gamma^{\prime}$ of the interval attractor $([p^{\prime}, q^{\prime}], [\ell^{\prime}, r^{\prime}])$, 
    (f) $([p^{\prime}, q^{\prime}], [\ell^{\prime}, r^{\prime}]) \in \Psi_{\centerset}(C) \cap \Psi_{\lcp}(K + |C|)$, 
    and (g) $([p^{\prime}, q^{\prime}], [\ell^{\prime}, r^{\prime}]) \in \Psi_{\run} \cup \Psi_{\source}$;
    \item \label{enum:psi_run_basic_property:4}
    there exists an integer $M \geq 0$ satisfying $([p, q], [\ell, r]) \in \Psi_{\modulo}(M)$;
    \item \label{enum:psi_run_basic_property:5}
    consider interval attractor $I_{\capture}(p-1, \ell) \in \Psi_{\RR}$.  
    Then, $I_{\capture}(p-1, \ell) = ([p^{\prime}, q^{\prime}], [\ell^{\prime}, r^{\prime}])$ 
    for the interval attractor $([p^{\prime}, q^{\prime}], [\ell^{\prime}, r^{\prime}])$ of statement (iii);
    \item \label{enum:psi_run_basic_property:6}
    consider the interval attractor $([p^{\prime}, q^{\prime}], [\ell^{\prime}, r^{\prime}])$ of statement (iii). 
    Then, $([p^{\prime}, q^{\prime}], [\ell^{\prime}, r^{\prime}]) \in \Psi_{\run} \Leftrightarrow |\lcs(T[1..\gamma-1], C^{n+1})| \geq |C| + |[p-1, \gamma-1]|$.
    \item \label{enum:psi_run_basic_property:7}
    consider an interval attractor $([\hat{p}, \hat{q}], [\hat{\ell}, \hat{r}])$ in set $\Psi_{h} \cap \Psi_{\run} \cap \Psi_{\centerset}(C)$.
    Then, $T[p-1..\gamma-1] = T[\hat{p}-1..\hat{\gamma}-1]$ 
    and $|\lcp(T[\gamma..r], T[\hat{\gamma}..\hat{r}])| > 1 + \sum_{w = 1}^{h+3} \lfloor \mu(w) \rfloor$ 
    for the attractor position $\hat{\gamma}$ of the interval attractor $([\hat{p}, \hat{q}], [\hat{\ell}, \hat{r}])$; 
    \item \label{enum:psi_run_basic_property:8}
    $|\lcp(T[\gamma..r+1], C^{n+1})| = K$; 
    \item \label{enum:psi_run_basic_property:9}
    consider interval attractor $I_{\capture}(i, j) \in \Psi_{\RR}$ 
    for an interval $[i, j]$ satisfying 
    $i \in [p, q]$ and $j \in [\gamma + \sum_{w = 1}^{h + 3} \lfloor \mu(w) \rfloor, r]$. 
    Then, $I_{\capture}(i, j) = ([p, q], [\ell, r])$.     
\end{enumerate}
\end{lemma}
\begin{proof}
See Section~\ref{subsubsec:proof_psi_run_basic_property}.
\end{proof}

The following two lemmas state properties of set $\Psi_{h} \cap \Psi_{\run} \cap \Psi_{\centerset}(C)$ 
for an integer $h \in [0, H]$ and a string $C \in \Sigma^{+}$. 

\begin{lemma}\label{lem:suffix_syncro}
    Consider an interval attractor $([p, q], [\ell, r]) \in \Psi_{\RR}$ satisfying the following two conditions: 
    \begin{itemize}
        \item let $\gamma$ and $C$ be the attractor position and associated string of the interval attractor $([p, q], [\ell, r])$, respectively. 
        Then, there exists an integer $\tau \in [p, q]$ such that 
        string $T[\tau..\gamma-1]$ is a suffix of string $C^{n+1}$ (i.e., $\lcs(T[\tau..\gamma-1], C^{n+1}) = T[\tau..\gamma-1]$); 
        \item 
        the length of the longest common prefix between two strings $T[\gamma..r]$ and $C^{n+1}$ is larger than $\sum_{w = 1}^{h+3} \lfloor \mu(w) \rfloor$ for the level $h$ of the interval attractor $([p, q], [\ell, r])$ 
        (i.e., $|\lcp(T[\gamma..r], C^{n+1})| > \sum_{w = 1}^{h+3} \lfloor \mu(w) \rfloor$).
    \end{itemize}
    For an interval attractor $([p^{\prime}, q^{\prime}], [\ell^{\prime}, r^{\prime}]) \in \Psi_{h} \cap \Psi_{\run} \cap \Psi_{\centerset}(C)$, 
    let $\gamma^{\prime}$ be the attractor position of the interval attractor $([p^{\prime}, q^{\prime}], [\ell^{\prime}, r^{\prime}])$. 
    Then, the string $T[\tau..\gamma-1]$ is a suffix of string $T[p^{\prime}..\gamma^{\prime}-1]$  
    (i.e., $\lcs(T[\tau..\gamma-1], T[p^{\prime}..\gamma^{\prime}-1]) = T[\tau..\gamma-1]$). 
\end{lemma}
\begin{proof}
    See Section~\ref{subsubsec:suffix_syncro_proof}.
\end{proof}

\begin{lemma}\label{lem:psi_HR_repetitive_C_property}
    Consider two strings $C_{1}, C_{2} \in \Sigma^{+}$ ($C_{1} \neq C_{2}$) satisfying 
    $\Psi_{h} \cap \Psi_{\run} \cap \Psi_{\centerset}(C_{1}) \neq \emptyset$ 
    and 
    $\Psi_{h} \cap \Psi_{\run} \cap \Psi_{\centerset}(C_{2}) \neq \emptyset$ 
    for an integer $h \in [0, H]$. 
    Then, $n+1 > 1 + \sum_{w = 1}^{h+3} \lfloor \mu(w) \rfloor$ 
    and 
    $C_{1}^{n+1}[1..2 + \sum_{w = 1}^{h+3} \lfloor \mu(w) \rfloor] \neq C_{2}^{n+1}[1..2 + \sum_{w = 1}^{h+3} \lfloor \mu(w) \rfloor]$ hold. 
\end{lemma}
\begin{proof}
    Let $([p_{1}, q_{1}], [\ell_{1}, r_{1}])$ and $([p_{2}, q_{2}], [\ell_{2}, r_{2}])$ be 
    two interval attractors in the two sets $\Psi_{h} \cap \Psi_{\run} \cap \Psi_{\centerset}(C_{1})$ and 
    $\Psi_{h} \cap \Psi_{\run} \cap \Psi_{\centerset}(C_{2})$, respectively.     
    Then, $|\lcp(T[\gamma_{1}..r_{1}], C_{1}^{n+1})| > 1 + \sum_{w = 1}^{h+3} \lfloor \mu(w) \rfloor$ follows from the definition of the subset $\Psi_{\run}$ 
    for the attractor position $\gamma_{1}$ of the interval attractor $([p_{1}, q_{1}], [\ell_{1}, r_{1}])$. 
    Similarly, $|\lcp(T[\gamma_{2}..r_{2}], C_{2}^{n+1})| > 1 + \sum_{w = 1}^{h+3} \lfloor \mu(w) \rfloor$ holds for the attractor position $\gamma_{2}$ of the interval attractor $([p_{2}, q_{2}], [\ell_{2}, r_{2}])$. 
    Therefore, $n+1 > 1 + \sum_{w = 1}^{h+3} \lfloor \mu(w) \rfloor$ follows from $|\lcp(T[\gamma_{1}..r_{1}], C_{1}^{n+1})| > 1 + \sum_{w = 1}^{h+3} \lfloor \mu(w) \rfloor$. 

    We prove $C_{1}^{n+1}[1..2 + \sum_{w = 1}^{h+3} \lfloor \mu(w) \rfloor] \neq C_{2}^{n+1}[1..2 + \sum_{w = 1}^{h+3} \lfloor \mu(w) \rfloor]$ by contradiction. 
    We assume that $C_{1}^{n+1}[1..2 + \sum_{w = 1}^{h+3} \lfloor \mu(w) \rfloor] = C_{2}^{n+1}[1..2 + \sum_{w = 1}^{h+3} \lfloor \mu(w) \rfloor]$. 
    Then, $|\lcp(T[\gamma_{1}..r_{1}], T[\gamma_{2}..r_{2}])| > \sum_{w = 1}^{h+1} \lfloor \mu(w) \rfloor$ holds 
    because (A) string $C_{1}^{n+1}[1..2 + \sum_{w = 1}^{h+3} \lfloor \mu(w) \rfloor]$ is a prefix of string $T[\gamma_{1}..r_{1}]$, 
    (B) string $C_{2}^{n+1}[1..2 + \sum_{w = 1}^{h+3} \lfloor \mu(w) \rfloor]$ is a prefix of string $T[\gamma_{2}..r_{2}]$, 
    and (C) $C_{1}^{n+1}[1..2 + \sum_{w = 1}^{h+3} \lfloor \mu(w) \rfloor] \neq C_{2}^{n+1}[1..2 + \sum_{w = 1}^{h+3} \lfloor \mu(w) \rfloor]$. 
    Because of $|\lcp(T[\gamma_{1}..r_{1}], T[\gamma_{2}..r_{2}])| > \sum_{w = 1}^{h+1} \lfloor \mu(w) \rfloor$, 
    Lemma~\ref{lem:associated_string_C}~\ref{enum:associated_string_C:2} shows that $C_{1} = C_{2}$ holds. 
    The two facts $C_{1} \neq C_{2}$ and $C_{1} = C_{2}$ yield a contradiction. 
    Therefore, $C_{1}^{n+1}[1..2 + \sum_{w = 1}^{h+3} \lfloor \mu(w) \rfloor] \neq C_{2}^{n+1}[1..2 + \sum_{w = 1}^{h+3} \lfloor \mu(w) \rfloor]$ must hold. 
\end{proof}

%\color{red}
The following lemma states properties of subset $\Psi_{\source}$.
\begin{lemma}\label{lem:light_source_basic_property}
    Consider an interval attractor $([p, q], [\ell, r])$ in set $\Psi_{\source}$. 
    Let $h, \gamma$, and $C$ be the level, attractor position, and associated string of the interval attractor $([p, q], [\ell, r])$, respectively;
    let $K = |\lcp(T[\gamma..r], C^{n+1})|$. 
    Then, (i) $|C| + 2 + \sum_{w = 1}^{h+3} \lfloor \mu(w) \rfloor \leq K \leq n$, 
    (ii) $|\lcs(T[p-1..\gamma-1], C^{n+1})| < |[p-1, \gamma-1]|$, 
    and (iii) $([p, q], [\ell, r]) \in \Psi_{\modulo}(M)$ for the integer $M = (K - (2 + \sum_{w = 1}^{h+3} \lfloor \mu(w) \rfloor) ) \mod |C|$.    
\end{lemma}
\begin{proof}
    From the definition of set $\Psi_{\source}$, 
    there exists an interval attractor $([p^{\prime}, q^{\prime}], [\ell^{\prime}, r^{\prime}]) \in \Psi_{h} \cap \Psi_{\run} \cap \Psi_{\centerset}(C)$ whose attractor position $\gamma^{\prime}$ is equal to $\gamma + |C|$ (i.e., $\gamma^{\prime} = \gamma + |C|$). 
    Let $K^{\prime} = |\lcp(T[\gamma^{\prime}..r^{\prime}], C^{n+1})|$. 
    Then, $K^{\prime} > 1 + \sum_{w = 1}^{h+3} \lfloor \mu(w) \rfloor$ follows from 
    the definition of periodic interval attractor. 
    $K = |C| + K^{\prime}$ follows from Lemma~\ref{lem:psi_run_basic_property}~\ref{enum:psi_run_basic_property:3}. 
    Therefore, we obtain $|C| + 2 + \sum_{w = 1}^{h+3} \lfloor \mu(w) \rfloor \leq K \leq n$.     

We prove $|\lcs(T[p-1..\gamma-1], C^{n+1})| < |[p-1, \gamma-1]|$ by contradiction. 
Here, $|\lcs(T[p-1..\gamma-1], C^{n+1})| \leq |[p-1, \gamma-1]|$ follows from the definition of longest common suffix. 
We assume that $|\lcs(T[p-1..\gamma-1], C^{n+1})| = |[p-1, \gamma-1]|$ holds. 
Then, $\lcs(T[p-1..\gamma-1], C^{n+1}) = T[p-1..\gamma-1]$ holds. 
$|\lcp(T[\gamma..r+1, r], C^{n+1})| > |C| + 2 + \sum_{w = 1}^{h+3} \lfloor \mu(w) \rfloor$ follows from 
$K \geq |C| + 2 + \sum_{w = 1}^{h+3} \lfloor \mu(w) \rfloor$. 
$([p, q], [\ell, r]) \in \Psi_{\run}$ follows from the definition of the subset $\Psi_{\run}$ 
because $\lcs(T[p-1..\gamma-1], C^{n+1}) = T[p-1..\gamma-1]$ 
and $|\lcp(T[\gamma..r], C^{n+1})| > 1 + \sum_{w = 1}^{h+3} \lfloor \mu(w) \rfloor$. 
On the other hand, $([p, q], [\ell, r]) \not \in \Psi_{\run}$ follows from the definition of the subset $\Psi_{\source}$. 
The two facts $([p, q], [\ell, r]) \in \Psi_{\run}$ and $([p, q], [\ell, r]) \not \in \Psi_{\run}$ yield a contradiction. 
Therefore, $|\lcs(T[p-1..\gamma-1], C^{n+1})| < |[p-1, \gamma-1]|$ must hold. 

    Since $K \geq |C| + 2 + \sum_{w = 1}^{h+3} \lfloor \mu(w) \rfloor$, 
    $([p, q], [\ell, r]) \in \Psi_{\modulo}(M)$ follows from the definition of the subset $\Psi_{\modulo}(M)$. 
\end{proof}
%\color{black}

The following lemma states properties of four subsets $\Psi_{\lcp}(K)$, $\Psi_{\modulo}(M)$, $\Psi_{\preceding}$, and $\Psi_{\succeeding}$
for two integers $K \geq 0$ and $M \geq 0$.

\begin{lemma}\label{lem:psi_LMPS_property}
    Let $C \in (\Sigma \setminus \{ \$, \# \})^{+}$ be a string, 
    and let $K$ and $K^{\prime}$ be two integers satisfying $1 \leq K < K^{\prime} \leq n$.     
    Then, the following four statements hold:
    \begin{enumerate}[label=\textbf{(\roman*)}]
    \item \label{enum:psi_LMPS_property:lcp:1} $\Psi_{\RR} = \bigcup_{\lambda = 1}^{n} \Psi_{\lcp}(\lambda)$;
    \item \label{enum:psi_LMPS_property:lcp:2} $\Psi_{\lcp}(\lambda) \cap \Psi_{\lcp}(\lambda^{\prime}) = \emptyset$ for any pair of two integers $0 \leq \lambda < \lambda^{\prime} \leq n$;    
    \item \label{enum:psi_LMPS_property:preceding:1} 
    consider two interval attractors $([p, q], [\ell, r]) \in \Psi_{\centerset}(C) \cap \Psi_{\run} \cap \Psi_{\lcp}(K) \cap \Psi_{\preceding}$ and $([p^{\prime}, q^{\prime}]$, $[\ell^{\prime}, r^{\prime}]) \in \Psi_{\centerset}(C) \cap \Psi_{\run} \cap \Psi_{\lcp}(K^{\prime}) \cap \Psi_{\preceding}$. 
    Then, $T[\gamma..r+1] \prec T[\gamma^{\prime}..r^{\prime}+1]$ holds 
    for the attractor position $\gamma$ (respectively, $\gamma^{\prime}$) of the interval attractor $([p, q], [\ell, r])$ (respectively, $([p^{\prime}, q^{\prime}]$, $[\ell^{\prime}, r^{\prime}])$);
    \item \label{enum:psi_LMPS_property:succeeding:1} 
    consider two interval attractors $([p, q], [\ell, r]) \in \Psi_{\centerset}(C) \cap \Psi_{\run} \cap \Psi_{\lcp}(K) \cap \Psi_{\succeeding}$ and $([p^{\prime}, q^{\prime}]$, $[\ell^{\prime}, r^{\prime}]) \in \Psi_{\centerset}(C) \cap \Psi_{\run} \cap \Psi_{\lcp}(K^{\prime}) \cap \Psi_{\succeeding}$. 
    Then, $T[\gamma^{\prime}..r^{\prime}+1] \prec T[\gamma..r+1]$ holds 
    for the attractor position $\gamma$ (respectively, $\gamma^{\prime}$) of the interval attractor $([p, q], [\ell, r])$ (respectively, $([p^{\prime}, q^{\prime}], [\ell^{\prime}, r^{\prime}])$).
    %\item \label{enum:psi_LMPS_property:mod:1} $\Psi_{\run} \cap \Psi_{\centerset}(C) = \bigcup_{\lambda = 0}^{|C|-1} \Psi_{\modulo}(\lambda)$.
    \end{enumerate}
\end{lemma}
\begin{proof}
    The proof of Lemma~\ref{lem:psi_LMPS_property} is as follows.

    \textbf{Proof of Lemma~\ref{lem:psi_LMPS_property}(i).}
    Lemma~\ref{lem:psi_LMPS_property}~\ref{enum:psi_LMPS_property:lcp:1} follows from the definition of subset $\Psi_{\lcp}(\lambda)$. 
    
    %We prove $\Psi_{\RR} \subseteq \bigcup_{\lambda = 1}^{n} \Psi_{\lcp}(\lambda)$. 
    %Let $\gamma$ and $C$ be the attractor position and associated string of an interval attractor $([p, q], [\ell, r])$ in 
    %set $\Psi_{\RR}$, respectively. 
    %Let $K$ be the length of the longest common prefix between two strings $T[\gamma..r]$ and $C^{n+1}$ (i.e., 
    %$K = |\lcp(T[\gamma..r], C^{n+1})|$). 
    %Then, $([p, q], [\ell, r]) \in \Psi_{\lcp}(K)$ follows from the definition of the subset $\Psi_{\lcp}(K)$. 
    %$K \leq n$ holds because $[\gamma, r] \subseteq [1, n]$ holds. 
    %$K \geq 1$ holds because $|C| \geq 1$ 
    %and 
    %Lemma~\ref{lem:C_prefix_property} shows that the associated string $C$ is a prefix of the string $T[\gamma..r]$. 
    %Therefore, $\Psi_{\RR} \subseteq \bigcup_{\lambda = 1}^{n} \Psi_{\lcp}(\lambda)$ holds. 

    %On the other hand, $\Psi_{\RR} \supseteq \bigcup_{\lambda = 1}^{n} \Psi_{\lcp}(\lambda)$ holds 
    %because $\Psi_{\lcp}(\lambda) \subseteq \Psi_{\RR}$ follows from the definition of each subset $\Psi_{\lcp}(\lambda)$. 
    %Therefore, $\Psi_{\RR} = \bigcup_{\lambda = 1}^{n} \Psi_{\lcp}(\lambda)$ holds. 
    
    \textbf{Proof of Lemma~\ref{lem:psi_LMPS_property}(ii).}
    $\Psi_{\lcp}(\lambda) \cap \Psi_{\lcp}(\lambda^{\prime}) = \emptyset$ follows from the definitions of the two subsets $\Psi_{\lcp}(\lambda)$ and $\Psi_{\lcp}(\lambda^{\prime})$.

    \textbf{Proof of Lemma~\ref{lem:psi_LMPS_property}(iii).}
    We prove $C^{n+1}[1..K^{\prime}] \prec T[\gamma^{\prime}..r^{\prime}+1]$. 
    Because of $([p^{\prime}, q^{\prime}], [\ell^{\prime}, r^{\prime}]) \in \Psi_{\centerset}(C) \cap \Psi_{\lcp}(K^{\prime})$, 
    $T[\gamma^{\prime}..r^{\prime}+1] = C^{n+1}[1..K^{\prime}] \cdot T[\gamma^{\prime} + K^{\prime}..r^{\prime}+1]$ follows from the definition of the subset $\Psi_{\lcp}(K^{\prime})$. 
    Therefore, $C^{n+1}[1..K^{\prime}] \prec T[\gamma^{\prime}..r^{\prime}+1]$ follows from 
    $T[\gamma^{\prime}..r^{\prime}+1] = C^{n+1}[1..K^{\prime}] \cdot T[\gamma^{\prime} + K^{\prime}..r^{\prime}+1]$. 

    Next, we prove $T[\gamma..r+1] \prec C^{n+1}[1..K+1]$. 
    Because of $([p, q], [\ell, r]) \in \Psi_{\centerset}(C) \cap \Psi_{\lcp}(K)$, 
    $|\lcp(T[\gamma..r], C^{n+1})|= K$ follows from the definition of the subset $\Psi_{\lcp}(K)$. 
    Because of $([p, q], [\ell, r]) \in \Psi_{\centerset}(C) \cap \Psi_{\run} \cap \Psi_{\lcp}(K)$, 
    Lemma~\ref{lem:psi_run_basic_property}~\ref{enum:psi_run_basic_property:8} shows that 
    $|\lcp(T[\gamma..r+1], C^{n+1})|= K$. 
    Because of $|\lcp(T[\gamma..r], C^{n+1})|= K$, 
    the string $T[\gamma..r+1]$ can be divided into two strings $T[\gamma..\gamma + K-1]$ and 
    $T[\gamma + K..r+1]$. 
    Here, $T[\gamma..\gamma + K-1] = C^{n+1}[1..K]$ holds. 
    Because of $|\lcp(T[\gamma..r+1], C^{n+1})|= K$, 
    $T[\gamma + K] \neq C^{n+1}[K+1]$ holds. 
    $T[\gamma..r+1] \prec C^{n+1}$ follows from the definition of the subset $\Psi_{\preceding}$ 
    because $([p, q], [\ell, r]) \in \Psi_{\centerset}(C) \cap \Psi_{\preceding}$. 
    $T[\gamma + K] < C^{n+1}[K+1]$ follows from 
    $T[\gamma..\gamma + K-1] = C^{n+1}[1..K]$, $T[\gamma + K] \neq C^{n+1}[K+1]$, 
    and $T[\gamma..r+1] \prec C^{n+1}$. 
    Therefore, $T[\gamma..r+1] \prec C^{n+1}[1..K+1]$ follows from 
    $T[\gamma..r+1] = T[\gamma..\gamma + K-1] \cdot T[\gamma + K..r+1]$, 
    $T[\gamma..\gamma + K-1] = C^{n+1}[1..K]$, 
    and $T[\gamma + K] < C^{n+1}[K+1]$.

    We prove $T[\gamma..r+1] \prec T[\gamma^{\prime}..r^{\prime}+1]$. 
    $C^{n+1}[1..K+1] \preceq C^{n+1}[1..K^{\prime}]$ holds 
    because $K < K^{\prime}$. 
    Therefore, $T[\gamma..r+1] \prec T[\gamma^{\prime}..r^{\prime}+1]$ follows from 
    $T[\gamma..r+1] \prec C^{n+1}[1..K+1]$, $C^{n+1}[1..K+1] \preceq C^{n+1}[1..K^{\prime}]$, 
    and $C^{n+1}[1..K^{\prime}] \prec T[\gamma^{\prime}..r^{\prime}+1]$. 

    \textbf{Proof of Lemma~\ref{lem:psi_LMPS_property}(iv).}
    We prove $C^{n+1}[1..K+1]\# \prec T[\gamma..r+1]$. 
    Because of $([p, q], [\ell, r]) \in \Psi_{\centerset}(C) \cap \Psi_{\lcp}(K)$, 
    $|\lcp(T[\gamma..r], C^{n+1})|= K$ follows from the definition of the subset $\Psi_{\lcp}(K)$. 
    Because of $([p, q], [\ell, r]) \in \Psi_{\centerset}(C) \cap \Psi_{\run} \cap \Psi_{\lcp}(K)$, 
    Lemma~\ref{lem:psi_run_basic_property}~\ref{enum:psi_run_basic_property:8} shows that 
    $|\lcp(T[\gamma..r+1], C^{n+1})|= K$. 
    Because of $|\lcp(T[\gamma..r], C^{n+1})|= K$, 
    the string $T[\gamma..r+1]$ can be divided into two strings $T[\gamma..\gamma + K-1]$ and 
    $T[\gamma + K..r+1]$. 
    Here, $T[\gamma..\gamma + K-1] = C^{n+1}[1..K]$ holds. 
    Because of $|\lcp(T[\gamma..r+1], C^{n+1})|= K$, 
    $T[\gamma + K] \neq C^{n+1}[K+1]$ holds. 
    $C^{n+1} \prec T[\gamma..r+1]$ follows from the definition of the subset $\Psi_{\succeeding}$ 
    because $([p, q], [\ell, r]) \in \Psi_{\centerset}(C) \cap \Psi_{\succeeding}$. 
    $C^{n+1}[K+1] < T[\gamma + K]$ follows from 
    $T[\gamma..\gamma + K-1] = C^{n+1}[1..K]$, $T[\gamma + K] \neq C^{n+1}[K+1]$, 
    and $T[\gamma..r+1] \prec C^{n+1}$. 
    Therefore, $C^{n+1}[1..K+1]\# \prec T[\gamma..r+1]$ follows from 
    $T[\gamma..r+1] = T[\gamma..\gamma + K-1] \cdot T[\gamma + K..r+1]$, 
    $T[\gamma..\gamma + K-1] = C^{n+1}[1..K]$, 
    and $C^{n+1}[K+1] < T[\gamma + K]$. 

    Next, we prove $T[\gamma^{\prime}..r^{\prime}+1] \prec C^{n+1}[1..K^{\prime}]\#$. 
    Similar to the string $T[\gamma..r+1]$, 
    the string $T[\gamma^{\prime}..r^{\prime}+1]$ satisfies the following two conditions: 
    \begin{enumerate}[label=\textbf{(\alph*)}]
        \item $T[\gamma^{\prime}..r^{\prime}+1] = T[\gamma^{\prime}..\gamma^{\prime} + K^{\prime}-1] \cdot T[\gamma^{\prime} + K^{\prime}..r^{\prime}+1]$;
        \item $T[\gamma^{\prime}..\gamma^{\prime} + K^{\prime}-1] = C^{n+1}[1..K^{\prime}]$.
    \end{enumerate}
    $T[\gamma^{\prime} + K^{\prime}] < \#$ holds because $T[\gamma^{\prime} + K^{\prime}] \neq \#$ and 
    $\#$ is the largest character in the alphabet $\Sigma$. 
    Therefore, $T[\gamma^{\prime}..r^{\prime}+1] \prec C^{n+1}[1..K^{\prime}]\#$ follows from 
    condition (a), condition (b), and $T[\gamma^{\prime} + K^{\prime}] < \#$. 

    We prove $T[\gamma^{\prime}..r^{\prime}+1] \prec T[\gamma..r+1]$. 
    $C^{n+1}[1..K^{\prime}]\# \preceq C^{n+1}[1..K+1]\#$ holds 
    because $K < K^{\prime}$ holds, 
    and the string $C$ does not contain the character $\#$. 
    Therefore, $T[\gamma^{\prime}..r^{\prime}+1] \prec T[\gamma..r+1]$ follows from 
    $T[\gamma^{\prime}..r^{\prime}+1] \prec C^{n+1}[1..K^{\prime}]\#$, 
    $C^{n+1}[1..K^{\prime}]\# \preceq C^{n+1}[1..K+1]\#$, 
    and $C^{n+1}[1..K+1]\# \prec T[\gamma..r+1]$. 
\end{proof}

The following two lemmas state properties of the subset $\Psi_{\CCP}(T[i..j])$ for a substring $T[i..j]$ in input string $T$. 

\begin{lemma}\label{lem:CCP_property}
Consider interval attractor $I_{\capture}(i, j) = ([p, q], [\ell, r]) \in \Psi_{\RR}$ for an interval $[i, j]$ ($i < j$) in input string $T$. 
Let $\gamma$, $C$, and $h$ be the attractor position, associated string, and level of the interval attractor $([p, q], [\ell, r])$, 
respectively. 
Then, the following six statements hold for the subset $\Psi_{\CCP}(T[i..j])$.
\begin{enumerate}[label=\textbf{(\roman*)}]
    \item \label{enum:CCP_property:1} $\Psi_{\CCP}(T[i..j]) \subseteq \Psi_{h}$; 
    \item \label{enum:CCP_property:2}
    $\reverse(T[i..\gamma-1]) \prec \reverse(T[p-1..\gamma-1]) \prec \reverse(\# T[i..\gamma-1])$ 
    and $T[\gamma..j] \prec T[\gamma..r + 1] \prec T[\gamma..j]\#$; 
    \item \label{enum:CCP_property:3}
    consider an interval attractor $([p^{\prime}, q^{\prime}], [\ell^{\prime}, r^{\prime}])$ in the $h$-th level interval attractors $\Psi_{h}$. 
    If $T[\gamma^{\prime} - |[i, \gamma-1]|..\gamma^{\prime} -1] = T[i..\gamma-1]$ and 
    $T[\gamma^{\prime}..\gamma^{\prime} + |[\gamma, j]| -1] = T[\gamma..j]$ hold for 
    the attractor position $\gamma^{\prime}$ of the interval attractor $([p^{\prime}, q^{\prime}], [\ell^{\prime}, r^{\prime}])$, 
    then $I_{\capture}(\gamma^{\prime} - |[i, \gamma-1]|, \gamma^{\prime} + |[\gamma, j]| -1) = ([p^{\prime}, q^{\prime}], [\ell^{\prime}, r^{\prime}])$ 
    and $([p^{\prime}, q^{\prime}], [\ell^{\prime}, r^{\prime}]) \in \Psi_{\CCP}(T[i..j])$; 
    \item \label{enum:CCP_property:4}
    $\Psi_{\CCP}(T[i..j]) = \{ ([p^{\prime}, q^{\prime}], [\ell^{\prime}, r^{\prime}]) \in \Psi_{h} \mid \reverse(T[i..\gamma-1]) \prec \reverse(T[p^{\prime}-1..\gamma^{\prime}-1]) \prec \reverse(\#T[i..\gamma-1]) \text{ and } T[\gamma..j] \prec T[\gamma^{\prime}..r^{\prime}+1] \prec T[\gamma..j]\# \}$. 
    Here, $\gamma^{\prime}$ is the attractor position of each interval attractor $([p^{\prime}, q^{\prime}], [\ell^{\prime}, r^{\prime}]) \in \Psi_{h}$; 
    \item \label{enum:CCP_property:5}
    if (a) $|[\gamma, j]| > 1 + \sum_{w = 1}^{h+3} \lfloor \mu(w) \rfloor$ 
    and (b) there exists a string $C^{\prime} \in \Sigma^{+}$ satisfying $\Psi_{\CCP}(T[i..j]) \cap \Psi_{\centerset}(C^{\prime}) \neq \emptyset$, then $C = C^{\prime}$; 
    \item \label{enum:CCP_property:6}
    consider an interval attractor $([p^{\prime}, q^{\prime}], [\ell^{\prime}, r^{\prime}]) \in \Psi_{\CCP}(T[i..j])$ 
    and its attractor position $\gamma^{\prime}$. 
    Then, string $T[\gamma^{\prime} - |[i, \gamma-1]|..\gamma^{\prime} + |[\gamma, j]| - 1]$ is an occurrence of 
    string $T[i..j]$ in input string $T$ (i.e., $T[i..j] = T[\gamma^{\prime} - |[i, \gamma-1]|..\gamma^{\prime} + |[\gamma, j]| - 1]$), 
    and $I_{\capture}(\gamma^{\prime} - |[i, \gamma-1]|, \gamma^{\prime} + |[\gamma, j]| - 1) = ([p^{\prime}, q^{\prime}], [\ell^{\prime}, r^{\prime}])$. 
\end{enumerate}
\end{lemma}
\begin{proof}
See Section~\ref{subsubsec:CCP_property_proof1}.
\end{proof}

\begin{lemma}\label{lem:CCP_special_property}
Consider interval attractor $I_{\capture}(i, j) = ([p, q], [\ell, r]) \in \Psi_{\RR}$ for an interval $[i, j]$ ($i < j$) in input string $T$. 
Let $\gamma$, $C$, and $h$ be the attractor position, associated string, and level of the interval attractor $([p, q], [\ell, r])$, 
respectively. 
Then, $\Psi_{h} \cap \Psi_{\run} \cap \Psi_{\centerset}(C) \cap (\bigcup_{\lambda = |[\gamma, j]|}^{n} \Psi_{\lcp}(\lambda)) \subseteq \Psi_{\CCP}(T[i..j])$ holds if the following two conditions are satisfied: 
\begin{enumerate}[label=\textbf{(\roman*)}]
    \item string $T[i..\gamma-1]$ is a suffix of string $C^{n+1}$ (i.e., $\lcs(T[i..\gamma-1], C^{n+1}) = T[i..\gamma-1]$);
    \item string $T[\gamma..j]$ is a prefix of string $C^{n+1}$ (i.e., $\lcp(T[\gamma..j], C^{n+1}) = T[\gamma..j]$).
\end{enumerate}
\end{lemma}
\begin{proof}
See Section~\ref{subsubsec:CCP_property_proof2}.    
\end{proof}

Consider two interval attractors $([p, q], [\ell, r]), ([p^{\prime}, q^{\prime}], [\ell^{\prime}, r^{\prime}]) \in \Psi_{\RR}$ 
satisfying $T[p-1..r+1] = T[p^{\prime}-1..r^{\prime}+1]$. 
For each of the following subsets, denoted as $\Psi$,
\begin{itemize}
    \item $\Psi_{\centerset}(C)$ for a string $C \in \Sigma^{+}$; 
    \item $\Psi_{\run}$, $\Psi_{\source}$, $\Psi_{\preceding}$, and $\Psi_{\succeeding}$; 
    \item $\Psi_{\str}(P)$, $\Psi_{\CCP}(P)$, and $\Psi_{\lex}(P)$ for a string $P \in \Sigma^{+}$; 
    \item $\Psi_{\lcp}(K)$ for an integer $K \geq 0$; 
    \item $\Psi_{\modulo}(M)$ for an integer $M \geq 0$.
\end{itemize}
If the subset $\Psi$ contains the interval attractor $([p, q], [\ell, r])$, 
then the following lemma ensures that 
$\Psi$ contains the other one $([p^{\prime}, q^{\prime}], [\ell^{\prime}, r^{\prime}])$. 

\begin{lemma}\label{lem:psi_equality_basic_property}
Consider two interval attractors $([p, q], [\ell, r]), ([p^{\prime}, q^{\prime}], [\ell^{\prime}, r^{\prime}]) \in \Psi_{\RR}$ 
satisfying $T[p-1..r+1] = T[p^{\prime}-1..r^{\prime}+1]$. 
Let $h$ be the level of the interval attractor $([p, q], [\ell, r])$. 
Here, $([p^{\prime}, q^{\prime}], [\ell^{\prime}, r^{\prime}]) \in \Psi_{h}$ follows from Lemma~\ref{lem:psi_str_property}~\ref{enum:psi_str_property:2}.
The following ten statements hold: 
\begin{enumerate}[label=\textbf{(\roman*)}]
    \item \label{enum:psi_equality_basic_property:1}    
    for a string $P \in \Sigma^{+}$, 
    $([p, q], [\ell, r]) \in \Psi_{\str}(P) \Leftrightarrow ([p^{\prime}, q^{\prime}], [\ell^{\prime}, r^{\prime}]) \in \Psi_{\str}(P)$;
    \item \label{enum:psi_equality_basic_property:2}
    for a string $P \in \Sigma^{+}$, 
    $([p, q], [\ell, r]) \in \Psi_{\CCP}(P) \Leftrightarrow ([p^{\prime}, q^{\prime}], [\ell^{\prime}, r^{\prime}]) \in \Psi_{\CCP}(P)$;
    \item \label{enum:psi_equality_basic_property:4}    
    $([p, q], [\ell, r]) \in \Psi_{\run} \Leftrightarrow ([p^{\prime}, q^{\prime}], [\ell^{\prime}, r^{\prime}]) \in \Psi_{\run}$;    
    \item \label{enum:psi_equality_basic_property:5}    
    $([p, q], [\ell, r]) \in \Psi_{\source} \Leftrightarrow ([p^{\prime}, q^{\prime}], [\ell^{\prime}, r^{\prime}]) \in \Psi_{\source}$;
    \item \label{enum:psi_equality_basic_property:center_set}    
    for a string $P \in \Sigma^{+}$, 
    $([p, q], [\ell, r]) \in \Psi_{\centerset}(P) \cap (\Psi_{\run} \cup \Psi_{\source}) \Leftrightarrow ([p^{\prime}, q^{\prime}], [\ell^{\prime}, r^{\prime}]) \in \Psi_{\centerset}(P) \cap (\Psi_{\run} \cup \Psi_{\source})$;

    \item \label{enum:psi_equality_basic_property:6}    
    $([p, q], [\ell, r]) \in \Psi_{\preceding} \cap (\Psi_{\run} \cup \Psi_{\source}) \Leftrightarrow ([p^{\prime}, q^{\prime}], [\ell^{\prime}, r^{\prime}]) \in \Psi_{\preceding} \cap (\Psi_{\run} \cup \Psi_{\source})$;
    \item \label{enum:psi_equality_basic_property:7}    
    $([p, q], [\ell, r]) \in \Psi_{\succeeding} \cap (\Psi_{\run} \cup \Psi_{\source}) \Leftrightarrow ([p^{\prime}, q^{\prime}], [\ell^{\prime}, r^{\prime}]) \in \Psi_{\succeeding} \cap (\Psi_{\run} \cup \Psi_{\source})$;
    \item \label{enum:psi_equality_basic_property:8}
    for a string $P \in \Sigma^{+}$, 
    $([p, q], [\ell, r]) \in \Psi_{\lex}(P) \Leftrightarrow ([p^{\prime}, q^{\prime}], [\ell^{\prime}, r^{\prime}]) \in \Psi_{\lex}(P)$;
    \item \label{enum:psi_equality_basic_property:9}
    for an integer $K > \sum_{w = 1}^{h+1} \lfloor \mu(w) \rfloor$, 
    $([p, q], [\ell, r]) \in \Psi_{\lcp}(K) \Leftrightarrow ([p^{\prime}, q^{\prime}], [\ell^{\prime}, r^{\prime}]) \in \Psi_{\lcp}(K)$;
    \item \label{enum:psi_equality_basic_property:10}
    for an integer $M \geq 0$, 
    $([p, q], [\ell, r]) \in \Psi_{\modulo}(M) \Leftrightarrow ([p^{\prime}, q^{\prime}], [\ell^{\prime}, r^{\prime}]) \in \Psi_{\modulo}(M)$.
\end{enumerate}
\end{lemma}
\begin{proof}
    $([p^{\prime}, q^{\prime}], [\ell^{\prime}, r^{\prime}]) = ([p^{\prime}, q^{\prime} + (q-p)], [r^{\prime} - (r-\ell), r^{\prime}])$ follows from Theorem~\ref{theo:IA_SYNC_X}.     
    Let $\gamma$ and $C$ be attractor position and associated string of 
    the interval attractor $([p, q], [\ell, r])$, respectively.
    Similarly, 
    let $\gamma^{\prime}$ and $C^{\prime}$ be attractor position and associated string of 
    the interval attractor $([p^{\prime}, q^{\prime}], [\ell^{\prime}, r^{\prime}])$, respectively. 
    Because of $T[p-1..r+1] = T[p^{\prime}-1..r^{\prime}+1]$, 
    $([p, q], [\ell, r]), ([p^{\prime}, q^{\prime}], [\ell^{\prime}, r^{\prime}]) \in \Psi_{\str}(T[p-1..r+1])$ 
    holds.     
    $|[p, \gamma]| = |[p^{\prime}, \gamma^{\prime}]|$ 
    follows from Lemma~\ref{lem:psi_str_property}~\ref{enum:psi_str_property:1}.

    \textbf{Proof of Lemma~\ref{lem:psi_equality_basic_property}(i).}
    Lemma~\ref{lem:psi_equality_basic_property}~\ref{enum:psi_equality_basic_property:1} follows from the definition of the subset $\Psi_{\str}(P)$. 
    
    \textbf{Proof of Lemma~\ref{lem:psi_equality_basic_property}(ii).}
    We prove $([p, q], [\ell, r]) \in \Psi_{\CCP}(P) \Rightarrow ([p^{\prime}, q^{\prime}], [\ell^{\prime}, r^{\prime}]) \in \Psi_{\CCP}(P)$. 
    Because of $([p, q], [\ell, r]) \in \Psi_{\CCP}(P)$, 
    there exists an occurrence position $i \in \Occ(T, P)$ of string $P$ 
    satisfying $I_{\capture}(i, i + |P|-1) = ([p, q], [\ell, r])$. 
    Here, $i \in [p, q]$ and $i + |P|-1 \in [\ell, r]$ follow from the definition of interval attractor.     
    $T[\gamma^{\prime} - |[i, \gamma-1]|..\gamma^{\prime} -1] = T[i..\gamma-1]$ and 
    $T[\gamma^{\prime}..\gamma^{\prime} + |[\gamma, i + |P|-1]| -1] = T[\gamma..i + |P|-1]$ 
    hold because $|[p, \gamma]| = |[p^{\prime}, \gamma^{\prime}]|$, 
    $T[p-1..r+1] = T[p^{\prime}-1..r^{\prime}+1]$, and $q \leq \gamma < \ell$ (Lemma~\ref{lem:IA_super_basic_property}~\ref{enum:IA_super_basic_property:1}). 
    In this case, Lemma~\ref{lem:CCP_property}~\ref{enum:CCP_property:3} shows that 
    $I_{\capture}(\gamma^{\prime} - |[i, \gamma-1]|, \gamma^{\prime} + |[\gamma, i + |P|-1]| -1) = ([p^{\prime}, q^{\prime}], [\ell^{\prime}, r^{\prime}])$ and 
    $I_{\capture}(\gamma^{\prime} - |[i, \gamma-1]|, \gamma^{\prime} + |[\gamma, i + |P|-1]| -1) \in \Psi_{\CCP}(P)$. 
    Therefore, we obtain $([p, q], [\ell, r]) \in \Psi_{\CCP}(P) \Rightarrow ([p^{\prime}, q^{\prime}], [\ell^{\prime}, r^{\prime}]) \in \Psi_{\CCP}(P)$. 
        
    Similarly, we can prove $([p, q], [\ell, r]) \in \Psi_{\CCP}(P) \Leftarrow ([p^{\prime}, q^{\prime}], [\ell^{\prime}, r^{\prime}]) \in \Psi_{\CCP}(P)$ using the same approach used to prove $([p, q], [\ell, r]) \in \Psi_{\CCP}(P) \Rightarrow ([p^{\prime}, q^{\prime}], [\ell^{\prime}, r^{\prime}]) \in \Psi_{\CCP}(P)$. 
    Therefore, we obtain Lemma~\ref{lem:psi_equality_basic_property}(ii). 

    \textbf{Proof of Lemma~\ref{lem:psi_equality_basic_property}(iii).}
    Lemma~\ref{lem:psi_equality_basic_property}(iii) can be proved using 
    the four facts $([p, q], [\ell, r])$, $([p^{\prime}, q^{\prime}]$, $[\ell^{\prime}, r^{\prime}]) \in \Psi_{h}$, 
    $C = C^{\prime}$, $T[p-1..\gamma-1] = T[p^{\prime}-1..\gamma^{\prime}-1]$, 
    and $T[\gamma..r+1] = T[\gamma^{\prime}..r^{\prime}+1]$, 
    but we have not yet proven that $C = C^{\prime}$. 
    We can apply Lemma~\ref{lem:associated_string_C}~\ref{enum:associated_string_C:2} to the two interval attractors 
    $([p, q], [\ell, r])$, $([p^{\prime}, q^{\prime}]$, $[\ell^{\prime}, r^{\prime}])$ 
    because $T[\gamma..r+1] = T[\gamma^{\prime}..r^{\prime}+1]$, 
    and $\max \{ |[\gamma, r+1]|, |[\gamma^{\prime}, r^{\prime}+1]|  \} > \sum_{w = 1}^{h+1} \lfloor \mu(w) \rfloor$ 
    follows from the definition of the subset $\Psi_{\run}$. 
    $C = C^{\prime}$ follows from Lemma~\ref{lem:associated_string_C}~\ref{enum:associated_string_C:2}. 
    Therefore, we obtain Lemma~\ref{lem:psi_equality_basic_property}(iii). 

    \textbf{Proof of $([p, q], [\ell, r]) \in \Psi_{\source} \Rightarrow ([p^{\prime}, q^{\prime}], [\ell^{\prime}, r^{\prime}]) \in \Psi_{\source}$.}
    From the definition of set $\Psi_{\source}$, 
    there exists an interval attractor $([p_{1}, q_{1}], [\ell_{1}, r_{1}]) \in \Psi_{h} \cap \Psi_{\run} \cap \Psi_{\centerset}(C)$ whose attractor position $\gamma_{1}$ is equal to $\gamma + |C|$ (i.e., $\gamma_{1} = \gamma + |C|$). 
    Lemma~\ref{lem:psi_run_basic_property}~\ref{enum:psi_run_basic_property:3} shows that 
    set $\Psi_{h}$ contains an interval attractor 
    $([p_{0}, q_{0}], [\ell_{0}, r_{0}])$ satisfying 
    (a) $p_{0} < p_{1}$, (b) $\gamma_{0} = \gamma_{1} - |C|$ for the attractor position $\gamma_{0}$ of the interval attractor $([p_{0}, q_{0}], [\ell_{0}, r_{0}])$, (c) $\ell_{0} \leq \ell_{1}$, and (d) $r_{0} = r_{1}$. 

    We prove $([p_{0}, q_{0}], [\ell_{0}, r_{0}]) = ([p, q], [\ell, r])$ by contradiction. 
    We assume that $([p_{0}, q_{0}], [\ell_{0}, r_{0}]) \neq ([p, q], [\ell, r])$ holds. 
    Then, $\gamma_{0} \neq \gamma$ follows from Corollary~\ref{cor:IA_identify_corollary}. 
    On the other hand, $\gamma_{0} = \gamma$ follows from $\gamma_{0} = \gamma_{1} - |C|$ and $\gamma = \gamma_{1} - |C|$. 
    The two facts $\gamma_{0} \neq \gamma$ and $\gamma_{0} = \gamma$ yield a contradiction. 
    Therefore, $([p_{0}, q_{0}], [\ell_{0}, r_{0}]) = ([p, q], [\ell, r])$ must hold. 

    Let $p^{\prime}_{1} = p^{\prime} + (p_{1}-p)$, $q^{\prime}_{1} = p^{\prime}_{1} + (q_{1} - p_{1})$, and $\ell^{\prime}_{1} = r^{\prime} - (r_{1} - \ell_{1})$. 
    Then, we prove $([p^{\prime}_{1}, q^{\prime}_{1}], [\ell^{\prime}_{1}, r^{\prime}]) \in \Psi_{h}$. 
    $T[p_{1}-1..r_{1}+1] = T[p^{\prime}_{1}-1..r^{\prime}+1]$ 
    follows from $T[p-1..r+1] = T[p^{\prime}-1..r^{\prime}+1]$, $p_{1} \in [p, r]$, 
    and $r = r_{1}$.
    Lemma~\ref{lem:psi_str_property}~\ref{enum:psi_str_property:2} shows that 
    the $h$-th level interval attractors $\Psi_{h}$ contains 
    interval attractor $([p^{\prime}_{1}, q^{\prime}_{1}], [\ell^{\prime}_{1}, r^{\prime}])$ 
    because $T[p_{1}-1..r_{1}+1] = T[p^{\prime}_{1}-1..r^{\prime}+1]$ and 
    $([p_{1}, q_{1}], [\ell_{1}, r_{1}]) \in \Psi_{h}$. 
    
    We prove $\gamma^{\prime}_{1} = \gamma^{\prime} + |C|$ for the attractor position $\gamma^{\prime}_{1}$ of the interval attractor $([p^{\prime}_{1}, q^{\prime}_{1}], [\ell^{\prime}_{1}, r^{\prime}])$. 
    Because of $T[p_{1}-1..r_{1}+1] = T[p^{\prime}_{1}-1..r^{\prime}+1]$, 
    $|[p^{\prime}_{1}, \gamma^{\prime}_{1}-1]| = |[p_{1}, \gamma_{1}-1]|$ follows from Lemma~\ref{lem:psi_str_property}~\ref{enum:psi_str_property:1}. 
    $\gamma^{\prime}_{1} = \gamma + |C| - p^{\prime} - p$ follows from $|[p^{\prime}_{1}, \gamma^{\prime}_{1}-1]| = |[p_{1}, \gamma_{1}-1]|$. 
    $\gamma = \gamma^{\prime} - p^{\prime} + p$ follows from $|[p, \gamma]| = |[p^{\prime}, \gamma^{\prime}]|$. 
    Therefore, $\gamma^{\prime}_{1} = \gamma^{\prime} + |C|$ follows from the equation:
\begin{equation*}
    \begin{split}
    \gamma^{\prime}_{1} &= \gamma_{1} - p_{1} + p^{\prime}_{1}  \\
    &= \gamma + |C| - p_{1} + p^{\prime}_{1} \\
    &= (\gamma^{\prime} - p^{\prime} + p) + |C| - p_{1} + p^{\prime}_{1} \\ 
    &= (\gamma^{\prime} - p^{\prime} + p) + |C| - p_{1} + (p^{\prime} + (p_{1}-p)) \\ 
    &= \gamma^{\prime} + |C|.
    \end{split}
\end{equation*}
    
    We prove $([p^{\prime}_{1}, q^{\prime}_{1}], [\ell^{\prime}_{1}, r^{\prime}]) \in \Psi_{\centerset}(C) \cap \Psi_{\run}$. 
    Lemma~\ref{lem:psi_equality_basic_property}~\ref{enum:psi_equality_basic_property:4} shows that 
    $([p^{\prime}_{1}, q^{\prime}_{1}], [\ell^{\prime}_{1}, r^{\prime}]) \in \Psi_{\run}$ holds 
    because $([p_{1}, q_{1}], [\ell_{1}, r_{1}]) \in \Psi_{\run}$. 
    In this case, $([p^{\prime}_{1}, q^{\prime}_{1}], [\ell^{\prime}_{1}, r^{\prime}]) \in \Psi_{\centerset}(C)$ also holds 
    (see the proof of Lemma~\ref{lem:psi_equality_basic_property}~\ref{enum:psi_equality_basic_property:4}). 

    We prove $([p^{\prime}, q^{\prime}], [\ell^{\prime}, r^{\prime}]) \not \in \Psi_{\run}$. 
    Because of $([p, q], [\ell, r]) \in \Psi_{\source}$, 
    $([p, q], [\ell, r]) \not \in \Psi_{\run}$ follows from the definition of the subset $\Psi_{\source}$. 
    $([p^{\prime}, q^{\prime}], [\ell^{\prime}, r^{\prime}]) \not \in \Psi_{\run}$ follows from 
    Lemma~\ref{lem:psi_equality_basic_property}~\ref{enum:psi_equality_basic_property:4} and 
    $([p, q], [\ell, r]) \not \in \Psi_{\run}$. 

    Finally, $([p^{\prime}, q^{\prime}], [\ell^{\prime}, r^{\prime}]) \in \Psi_{\source}$ 
    follows from (A) $([p^{\prime}, q^{\prime}], [\ell^{\prime}, r^{\prime}]) \not \in \Psi_{\run}$, 
    (B) $([p^{\prime}_{1}, q^{\prime}_{1}], [\ell^{\prime}_{1}, r^{\prime}]) \in \Psi_{h} \cap \Psi_{\centerset}(C) \cap \Psi_{\run}$, 
    and (C) $\gamma^{\prime}_{1} = \gamma^{\prime} + |C|$. 
    Therefore, $([p, q], [\ell, r]) \in \Psi_{\source} \Rightarrow ([p^{\prime}, q^{\prime}], [\ell^{\prime}, r^{\prime}]) \in \Psi_{\source}$ holds. 

    \textbf{Proof of Lemma~\ref{lem:psi_equality_basic_property}(iv).}
    We already proved $([p, q], [\ell, r]) \in \Psi_{\source} \Rightarrow ([p^{\prime}, q^{\prime}], [\ell^{\prime}, r^{\prime}]) \in \Psi_{\source}$. 
    We can prove 
    $([p, q], [\ell, r]) \in \Psi_{\source} \Leftarrow ([p^{\prime}, q^{\prime}], [\ell^{\prime}, r^{\prime}]) \in \Psi_{\source}$ 
    using the same approach used to prove $([p, q], [\ell, r]) \in \Psi_{\source} \Rightarrow ([p^{\prime}, q^{\prime}], [\ell^{\prime}, r^{\prime}]) \in \Psi_{\source}$. 
    Therefore, $([p, q], [\ell, r]) \in \Psi_{\source} \Leftrightarrow ([p^{\prime}, q^{\prime}], [\ell^{\prime}, r^{\prime}]) \in \Psi_{\source}$ follows from $([p, q], [\ell, r]) \in \Psi_{\source} \Rightarrow ([p^{\prime}, q^{\prime}], [\ell^{\prime}, r^{\prime}]) \in \Psi_{\source}$ 
    and $([p, q], [\ell, r]) \in \Psi_{\source} \Leftarrow ([p^{\prime}, q^{\prime}], [\ell^{\prime}, r^{\prime}]) \in \Psi_{\source}$.

    \textbf{Proof of Lemma~\ref{lem:psi_equality_basic_property}(v).}
    Similar to the proof of Lemma~\ref{lem:psi_equality_basic_property}~\ref{enum:psi_equality_basic_property:4}, 
    $C = C^{\prime}$ follows from Lemma~\ref{lem:associated_string_C}~\ref{enum:associated_string_C:2}. 
    This is because  
    $\max \{ |[\gamma, r+1]|, |[\gamma^{\prime}, r^{\prime}+1]|  \} > \sum_{w = 1}^{h+1} \lfloor \mu(w) \rfloor$ 
    follows from the definition of the subset $\Psi_{\run}$ and Lemma~\ref{lem:light_source_basic_property}. 
    Therefore, Lemma~\ref{lem:psi_equality_basic_property}(v) 
    follows from $C = C$, Lemma~\ref{lem:psi_equality_basic_property}(iii), and Lemma~\ref{lem:psi_equality_basic_property}(iv).

    \textbf{Proof of Lemma~\ref{lem:psi_equality_basic_property}(vi).}
    $C = C^{\prime}$ follows from Lemma~\ref{lem:psi_equality_basic_property}~\ref{enum:psi_bigcup_property:centerset}.      
    Lemma~\ref{lem:psi_equality_basic_property}(vi) can be proved using 
    the two facts $C = C^{\prime}$ and $T[\gamma..r+1] = T[\gamma^{\prime}..r^{\prime}+1]$. 

    \textbf{Proof of Lemma~\ref{lem:psi_equality_basic_property}(vii).}
    Similar to Lemma~\ref{lem:psi_equality_basic_property}(vi), 
    Lemma~\ref{lem:psi_equality_basic_property}(vii) can be proved using 
    the two facts $C = C^{\prime}$ and $T[\gamma..r+1] = T[\gamma^{\prime}..r^{\prime}+1]$. 

    \textbf{Proof of Lemma~\ref{lem:psi_equality_basic_property}(viii).}
    Lemma~\ref{lem:psi_equality_basic_property}(viii) can be proved using the fact $T[\gamma..r+1] = T[\gamma^{\prime}..r^{\prime}+1]$. 

    \textbf{Proof of Lemma~\ref{lem:psi_equality_basic_property}(ix).}
    We prove $([p, q], [\ell, r]) \in \Psi_{\lcp}(K) \Rightarrow ([p^{\prime}, q^{\prime}], [\ell^{\prime}, r^{\prime}]) \in \Psi_{\lcp}(K)$. 
    $K \leq |[\gamma, r]|$ follows from the definition of subset $\Psi_{\lcp}(K)$. 
    Consider the length $K^{\prime}$ the longest common prefix between 
    two strings $T[\gamma..r]$ and $T[\gamma^{\prime}..r^{\prime}]$. 
    Since $T[\gamma..r+1] = T[\gamma^{\prime}..r^{\prime}+1]$, 
    $K^{\prime} \geq |[\gamma, r]|$ holds. 
    $K^{\prime} \geq K$ follows from  $K \leq |[\gamma, r]|$ and $K^{\prime} \geq |[\gamma, r]|$.  
    $K^{\prime} > \sum_{w = 1}^{h+1} \lfloor \mu(w) \rfloor$ follows from 
    $K^{\prime} \geq K$ and $K > \sum_{w = 1}^{h+1} \lfloor \mu(w) \rfloor$. 
    Since $K^{\prime} \geq \sum_{w = 1}^{h+1} \lfloor \mu(w) \rfloor$, 
    $C = C^{\prime}$ follows from Lemma~\ref{lem:associated_string_C}~\ref{enum:associated_string_C:2}. 
    $([p^{\prime}, q^{\prime}], [\ell^{\prime}, r^{\prime}]) \in \Psi_{\lcp}(K)$ follows from 
    $C = C^{\prime}$ and $T[\gamma..r+1] = T[\gamma^{\prime}..r^{\prime}+1]$. 
    Therefore, 
    we obtain $([p, q], [\ell, r]) \in \Psi_{\lcp}(K) \Rightarrow ([p^{\prime}, q^{\prime}], [\ell^{\prime}, r^{\prime}]) \in \Psi_{\lcp}(K)$. 
    Similarly, 
    $([p, q], [\ell, r]) \in \Psi_{\lcp}(K) \Leftarrow ([p^{\prime}, q^{\prime}], [\ell^{\prime}, r^{\prime}]) \in \Psi_{\lcp}(K)$ 
    can be proved by the same approach. 

    \textbf{Proof of Lemma~\ref{lem:psi_equality_basic_property}(x).}
    We prove $([p, q], [\ell, r]) \in \Psi_{\modulo}(M) \Rightarrow ([p^{\prime}, q^{\prime}], [\ell^{\prime}, r^{\prime}]) \in \Psi_{\modulo}(M)$. 
    Consider the length $K$ the longest common prefix between 
    two strings $T[\gamma..r]$ and $C^{n+1}$. 
    Then, $K \geq 2 + \sum_{w = 1}^{h+3} \lfloor \mu(w) \rfloor$ follows from the definition of subset $\Psi_{\modulo}(M)$. 
    Consider the length $K^{\prime}$ the longest common prefix between 
    two strings $T[\gamma..r]$ and $T[\gamma^{\prime}..r^{\prime}]$. 
    Similar to the proof of Lemma~\ref{lem:psi_equality_basic_property}(ix), 
    we can prove $C = C^{\prime}$
    using the two facts $K^{\prime} \geq K$ and $K \geq 2 + \sum_{w = 1}^{h+3} \lfloor \mu(w) \rfloor$. 
    $([p^{\prime}, q^{\prime}], [\ell^{\prime}, r^{\prime}]) \in \Psi_{\lcp}(K)$ follows from Lemma~\ref{lem:psi_equality_basic_property}(ix). 
    $([p^{\prime}, q^{\prime}], [\ell^{\prime}, r^{\prime}]) \in \Psi_{\modulo}(M)$ can be obtained using 
    the three facts $C = C^{\prime}$, $([p^{\prime}, q^{\prime}], [\ell^{\prime}, r^{\prime}]) \in \Psi_{\lcp}(K)$, and $T[\gamma..r+1] = T[\gamma^{\prime}..r^{\prime}+1]$. 
    Therefore, we obtain $([p, q], [\ell, r]) \in \Psi_{\modulo}(M) \Rightarrow ([p^{\prime}, q^{\prime}], [\ell^{\prime}, r^{\prime}]) \in \Psi_{\modulo}(M)$. 
    Similarly, 
    we can prove $([p, q], [\ell, r]) \in \Psi_{\modulo}(M) \Leftarrow ([p^{\prime}, q^{\prime}], [\ell^{\prime}, r^{\prime}]) \in \Psi_{\modulo}(M)$ 
    by the same approach.

    %Lemma~\ref{lem:psi_equality_basic_property}(x) can be proved using the three facts 
    %$([p, q], [\ell, r])$, $([p^{\prime}, q^{\prime}], [\ell^{\prime}, r^{\prime}]) \in \Psi_{h}$, 
    %$C = C^{\prime}$, and $T[\gamma..r+1] = T[\gamma^{\prime}..r^{\prime}+1]$.    
    %\color{black}

\end{proof}

While $\Psi_{\leftmost}$ and $\Psi_{\run}$ can be stored in $O(n)$ space, which is not optimal, the set difference $\Psi_{\leftmost} \setminus \Psi_{\run}$ can be stored in expected $\delta$-optimal space. This is ensured by the subsequent theorem.
\begin{theorem}\label{theo:RR_Psi_set_size}
$\mathbb{E}[|\Psi_{\leftmost} \setminus \Psi_{\run}|] = O(\delta \log \frac{n \log \sigma}{\delta \log n})$ 
for input string $T$ of length $n$ with an alphabet size of $\sigma$, and a substring complexity of $\delta$. 
\end{theorem}
\begin{proof}
See Section~\ref{subsec:size_proof}. 
\end{proof}

\subsubsection{Proof of Lemma~\ref{lem:psi_run_basic_property}}\label{subsubsec:proof_psi_run_basic_property}
The following propositions are used to prove Lemma~\ref{lem:psi_run_basic_property}. 

%%%%%%%%%%%%%%%%%%%%%%%%%%%%%%%%%%%%%%%%%%%%%%%%%%%%%%%%%%%%%%%%%%%%%%%%%%%%%%%%%%%%%%%%%%%%%%%%

\begin{proposition}\label{prop:RB_IA_plus_right}
Consider two sequences $A(s, e) = [s^{0}, e^{0}], [s^{1}, e^{1}], \ldots, [s^{k}, e^{k}]$ and 
$A(s, e+1) = [s^{\prime 0}, e^{\prime 0}], [s^{\prime 1}, e^{\prime 1}]$, $\ldots, [s^{\prime k^{\prime}}, e^{\prime k^{\prime}}]$ of intervals. 
Here, $k \leq k^{\prime}$ follows from Lemma~\ref{lem:intv_function_shift}~\ref{enum:intv_function_shift:7}. 
Let $T[x^{k}_{s}..y^{k}_{s}]$ and $T[x^{k}_{e}..y^{k}_{e}]$ be the two substrings derived from the $s^{k}$-th and $e^{k}$-th nonterminals of sequence $S^{k}$, respectively, in input string $T$. 
Similarly, 
let $T[x^{\prime k}_{s}..y^{\prime k}_{s}]$ and $T[x^{\prime k}_{e}..y^{\prime k}_{e}]$ be the two substrings derived from the $s^{\prime k}$-th and $e^{\prime k}$-th nonterminals of sequence $S^{k}$, respectively, in input string $T$. 

If $k \neq k^{\prime}$, then at least one of the following four conditions is satisfied: 
\begin{enumerate}[label=\textbf{(\roman*)}]
    \item $|[s^{k}, e^{k}]| = 1$, $e^{k} < e^{\prime k}$, and 
    there exists a position $i$ of sequence $S^{k+1}$ satisfying 
    $x^{k+1}_{i} \in [x^{k}_{e} + 1, x^{\prime k}_{e} - 1]$ for 
    the substring $T[x^{k+1}_{i}..y^{k+1}_{i}]$ derived from the $i$-th position of sequence $S^{k+1}$;
    \item $|[s^{k}, e^{k}]| \geq 2$, $e^{k} < e^{\prime k}$, and 
    there exists a position $i^{\prime}$ of sequence $S^{k+1}$ satisfying 
    $x^{k+1}_{i^{\prime}} \in [x^{k}_{e}, x^{\prime k}_{e} - 1]$ for 
    the substring $T[x^{k+1}_{i^{\prime}}..y^{k+1}_{i^{\prime}}]$ derived from the $i$-th position of sequence $S^{k+1}$;
    \item $e^{k} < e^{\prime k}$, and there exists a position $j$ of sequence $S^{k}$ satisfying 
    $j \in [e^{k}, e^{\prime k} - 1]$ and $|\val(S^{k}[j])| > \lfloor \mu(k+1) \rfloor$;
    \item $e^{k} = e^{\prime k}$, $|[x^{k}_{e}, e]| = \sum_{w = 1}^{k+1} \lfloor \mu(w) \rfloor$, 
    and $|[x^{k}_{e}, y^{k}_{e}]| > \lfloor \mu(k+1) \rfloor$. 
\end{enumerate}
\end{proposition}
\begin{proof}
    Proposition~\ref{prop:RB_IA_plus_right} can be proved using Definition~\ref{def:f_interval} 
    and Lemma~\ref{lem:rec_function_basic_relation}~\ref{enum:rec_function_basic_relation:4}. 
\end{proof}

%%%%%%%%%%%%%%%%%%%%%%%%%%%%%%%%%%%%%%%%%%%%%%%%%%%%%%%%%%%%%%%%%%%%%%%%%%%%%%

\begin{proposition}\label{prop:RB_two_last_sets}
    For an integer $h \in [0, H]$, 
    consider two positions $b, b^{\prime} \in [1, |S^{h}|]$ ($b < b^{\prime}$) 
    in sequence $S^{h}$ 
    satisfying $\Delta(h, b) \neq \emptyset$ 
    and $\Delta(h, b^{\prime}) \neq \emptyset$, respectively. 
    Let $p = \min \{ s \mid [s, e] \in \Delta(h, b) \}$, 
    $p^{\prime} = \min \{ s \mid [s, e] \in \Delta(h, b^{\prime}) \}$, 
    $r = \max \{ e \mid [s, e] \in \Delta(h, b) \}$, 
    and $r^{\prime} = \max \{ e \mid [s, e] \in \Delta(h, b^{\prime}) \}$. 
    Here, $[p, r] \in \Delta(h, b)$ and $[p^{\prime}, r^{\prime}] \in \Delta(h, b^{\prime})$
    follow from Lemma~\ref{lem:IA_maximal_lemma}.

    Consider two sequences $A(p, r) = [s^{0}, e^{0}], [s^{1}, e^{1}], \ldots, [s^{h}, e^{h}]$ and 
    $A(p^{\prime}, r^{\prime}) = [s^{\prime 0}, e^{\prime 0}], [s^{\prime 1}, e^{\prime 1}]$, $\ldots, [s^{\prime h}, e^{\prime h}]$ of intervals. 
    Then, the following three statements hold: 
\begin{enumerate}[label=\textbf{(\roman*)}]
    \item $p < p^{\prime}$;
    \item $r \leq r^{\prime}$;
    \item $e^{h} \leq e^{\prime h}$.    
\end{enumerate}
\end{proposition}
\begin{proof}
    $s^{h} = b$ and $s^{\prime h} = b^{\prime}$ follow from the definition of the two sets $\Delta(h, b)$ 
    and $\Delta(h, b^{\prime})$ 
    because 
    $[p, r] \in \Delta(h, b)$ and $[p^{\prime}, r^{\prime}] \in \Delta(h, b^{\prime})$.     
    The proof of Proposition~\ref{prop:RB_two_last_sets} is as follows. 

    \textbf{Proof of Proposition~\ref{prop:RB_two_last_sets}(i).}
    We prove $p < p^{\prime}$ by contradiction. 
    We assume that $p \geq p^{\prime}$ holds. 
    Under the assumption, 
    we apply Lemma~\ref{lem:intv_function_shift}~\ref{enum:intv_function_shift:5} 
    to the two intervals $[p, r]$ and $[p^{\prime}, r^{\prime}]$. 
    Then, the lemma shows that 
    $s^{h} \geq s^{\prime h}$ (i.e., $b \geq b^{\prime}$) holds. 
    On the other hand, $b < b^{\prime}$ follows from the premise of Proposition~\ref{prop:RB_two_last_sets}. 
    The two facts $b \geq b^{\prime}$ and $b < b^{\prime}$ yield a contradiction. 
    Therefore, $p < p^{\prime}$ must hold. 
    
    \textbf{Proof of Proposition~\ref{prop:RB_two_last_sets}(ii).}    
    $([p, q], [\ell, r]) \in \Psi_{h}$ 
    follows from the definition of interval attractor 
    for the two integers $q = \max \{ s \mid [s, e] \in \Delta(h, b) \}$ 
    and $\ell = \min \{ e \mid [s, e] \in \Delta(h, b) \}$.
    Similarly, 
    $([p^{\prime}, q^{\prime}], [\ell^{\prime}, r^{\prime}]) \in \Psi_{h}$ 
    holds 
    for the two integers $q^{\prime} = \max \{ s \mid [s, e] \in \Delta(h, b^{\prime}) \}$ 
    and $\ell^{\prime} = \min \{ e \mid [s, e] \in \Delta(h, b^{\prime}) \}$.
    Here, $([p, q], [\ell, r]) \neq ([p^{\prime}, q^{\prime}], [\ell^{\prime}, r^{\prime}])$ 
    follows from Lemma~\ref{lem:IA_super_basic_property}~\ref{enum:IA_super_basic_property:3}. 
            
    We prove $r \leq r^{\prime}$ by contradiction. 
    We assume that $r > r^{\prime}$ holds. 
    Then, $[p^{\prime}, r^{\prime}+1] \subseteq [p, r]$ follows from 
    $p < p^{\prime}$ and $r > r^{\prime}$. 
    Let $h_{1}$ be the level of interval attractor $I_{\capture}(p^{\prime}, r^{\prime}+1) = ([p_{1}, q_{1}], [\ell_{1}, r_{1}])$.     
    Lemma~\ref{lem:IA_maximal_lemma} shows that 
    $I_{\capture}(p, r) = ([p, q], [\ell, r])$ and $I_{\capture}(p^{\prime}, r^{\prime}) = ([p^{\prime}, q^{\prime}], [\ell^{\prime}, r^{\prime}])$ hold.     
    $h \geq h_{1}$ follows from Corollary~\ref{cor:IA_basic_corollary}~\ref{enum:IA_basic_corollary:3} 
    because $[p^{\prime}, r^{\prime}+1] \subseteq [p, r]$, 
    and the level of the interval attractor $I_{\capture}(p, r)$ is $h$. 
    On the other hand, $h < h_{1}$ follows from Lemma~\ref{lem:interval_extension_propertyX}~\ref{enum:interval_extension_propertyX:left}
    because $[p^{\prime}, r^{\prime}] \subseteq [p^{\prime}, r^{\prime}+1]$, 
    and the level of the interval attractor $I_{\capture}(p^{\prime}, r^{\prime})$ is $h$. 
    The two facts $h \geq h_{1}$ and $h < h_{1}$ yield a contradiction. 
    Therefore, $r \leq r^{\prime}$ must hold. 
    
    \textbf{Proof of Proposition~\ref{prop:RB_two_last_sets}(iii).}
    We apply Lemma~\ref{lem:intv_function_shift}~\ref{enum:intv_function_shift:6} 
    to the two intervals $[p, r]$ and $[p^{\prime}, r^{\prime}]$. 
    The lemma shows that $e^{h} \leq e^{\prime h}$ holds 
    because $r \leq r^{\prime}$. 
\end{proof}

%%%%%%%%%%%%%%%%%%%%%%%%%%%%%%%%%%%%%%%%%%%%%%%%%%%%%%%%%%%%%%%%%%%%%%%%%%%%%%%%%%%%%%%%%%%%%%%%

\begin{proposition}\label{prop:adjacent_last_set}
    For an integer $h \in [0, H]$, 
    consider a position $b \in [1, |S^{h}|-1]$ in sequence $S^{h}$ 
    satisfying $\Delta(h, b) \neq \emptyset$ 
    and $\Delta(h, b+1) \neq \emptyset$. 
    Let $p = \min \{ s \mid [s, e] \in \Delta(h, b) \}$, 
    $p^{\prime} = \min \{ s \mid [s, e] \in \Delta(h, b+1) \}$, 
    $r = \max \{ e \mid [s, e] \in \Delta(h, b) \}$, 
    and $r^{\prime} = \max \{ e \mid [s, e] \in \Delta(h, b+1) \}$. 
    Here, $[p, r] \in \Delta(h, b)$ and $[p^{\prime}, r^{\prime}] \in \Delta(h, b+1)$
    follows from Lemma~\ref{lem:IA_maximal_lemma}. 

    Consider two sequences $A(p, r) = [s^{0}, e^{0}], [s^{1}, e^{1}], \ldots, [s^{h}, e^{h}]$ and 
    $A(p^{\prime}, r^{\prime}) = [s^{\prime 0}, e^{\prime 0}], [s^{\prime 1}, e^{\prime 1}]$, $\ldots, [s^{\prime h}, e^{\prime h}]$ of intervals. 
    If $|[s^{h}, e^{h}]| \geq 4$, 
    then the following four statements hold: 
\begin{enumerate}[label=\textbf{(\roman*)}]
    \item $S^{h}[s^{h}] = S^{h}[s^{h}+1] = \cdots = S^{h}[e^{\prime h}-1]$; 
    \item $r = r^{\prime}$; 
    \item $e^{h} = e^{\prime h}$; 
    \item $[p^{\prime}-1, r^{\prime}] \in \Delta(h, b)$. 
\end{enumerate}
\end{proposition}
\begin{proof}
    $s^{h} = b$ and $s^{\prime h} = b+1$ follow from the definition of the two sets $\Delta(h, b)$ 
    and $\Delta(h, b+1)$
    because 
    $[p, r] \in \Delta(h, b)$ and $[p^{\prime}, r^{\prime}] \in \Delta(h, b+1)$.     
    $p < p^{\prime}$, 
    $r \leq r^{\prime}$, 
    and $e^{h} \leq e^{\prime h}$
    follow from Proposition~\ref{prop:RB_two_last_sets}. 

    Consider sequence $A(p^{\prime}-1, r^{\prime}) = [s^{\prime\prime 0}, e^{\prime\prime 0}], [s^{\prime\prime 1}, e^{\prime\prime 1}], \ldots, [s^{\prime\prime k}, e^{\prime\prime k}]$.
    We apply Lemma~\ref{lem:intv_function_shift}~\ref{enum:intv_function_shift:1} 
    to the two intervals $[p^{\prime}-1, r^{\prime}]$ and $[p^{\prime}, r^{\prime}]$.
    Then, the lemma shows that 
    $k \geq h$, $s^{\prime\prime h} \leq s^{\prime h}$ (i.e., $s^{\prime\prime h} \leq b+1$), and $e^{\prime\prime h} = e^{\prime h}$. 
        
    Similarly, consider sequence $A(p, r+1) = [s^{\prime\prime\prime 0}, e^{\prime\prime\prime 0}], [s^{\prime\prime\prime 1}, e^{\prime\prime\prime 1}], \ldots, [s^{\prime\prime\prime k^{\prime}}, e^{\prime\prime\prime k^{\prime}}]$.
    We apply Lemma \ref{lem:intv_function_shift}~\ref{enum:intv_function_shift:3} 
    to the two intervals $[p, r]$ and $[p, r+1]$. 
    Then, the lemma shows that 
    $k^{\prime} \geq h$, $s^{\prime\prime\prime h} = s^{h}$ (i.e., $s^{\prime\prime\prime h} = b$), 
    and $e^{\prime\prime\prime h} \geq e^{h}$. 

    Let $T[x^{h}_{s}..y^{h}_{s}]$ and $T[x^{h}_{e}..y^{h}_{e}]$ be the two substrings derived from the $s^{h}$-th and $e^{h}$-th nonterminals of sequence $S^{h}$, respectively, in input string $T$. 
    Similarly, 
    let $T[x^{\prime h}_{s}..y^{\prime h}_{s}]$ and $T[x^{\prime h}_{e}..y^{\prime h}_{e}]$ be the two substrings derived from the $s^{\prime h}$-th and $e^{\prime h}$-th nonterminals of sequence $S^{h}$, respectively, in input string $T$. 

    \textbf{Proof of Proposition~\ref{prop:adjacent_last_set}(i).}
    We can apply Lemma~\ref{lem:f_rec_top_property}~\ref{enum:f_rec_top_property:2} 
    to the interval $[p, r]$ because $|[s^{h}, e^{h}]| \geq 4$. 
    The lemma shows that 
    $S^{h}[s^{h}] = S^{h}[s^{h}+1] = \cdots = S^{h}[e^{h}-1]$. 
    $e^{h} \leq e^{\prime h}$ follows from Proposition~\ref{prop:RB_two_last_sets}(iii). 
    If $e^{h} = e^{\prime h}$ holds, 
    then $S^{h}[s^{h}+1] = S^{h}[s^{h}+2] = \cdots = S^{h}[e^{\prime h}-1]$ holds. 
    Otherwise (i.e., $e^{h} < e^{\prime h}$), 
    $|[s^{\prime h}, e^{\prime h}-1]| \geq 4$ follows from 
    $s^{\prime h} = b+1$, $s^{h} = b$, $e^{h} < e^{\prime h}$, and $|[s^{h}, e^{h}]| \geq 4$.    
    We can apply Lemma~\ref{lem:f_rec_top_property}~\ref{enum:f_rec_top_property:2} 
    to the interval $[p^{\prime}, r^{\prime}]$. 
    The lemma shows that $S^{h}[s^{\prime h}] = S^{h}[s^{\prime h}+1] = \cdots = S^{h}[e^{\prime h}-1]$. 
    Therefore, Proposition~\ref{prop:adjacent_last_set}(i) follows from 
    $S^{h}[s^{h}] = S^{h}[s^{h}+1] = \cdots = S^{h}[e^{h}-1]$, $S^{h}[s^{\prime h}] = S^{h}[s^{\prime h}+1] = \cdots = S^{h}[e^{\prime h}-1]$, 
    and $s^{\prime h} = s^{h} + 1$. 

    \textbf{Proof of Proposition~\ref{prop:adjacent_last_set}(ii).}        
    We prove $e^{h} \in [s^{\prime h} + 1, e^{\prime h}]$. 
    $e^{h} \geq s^{\prime h} + 1$ follows from 
    $|[s^{h}, e^{h}]| \geq 4$, $s^{h} = b$, and $s^{\prime h} = b+1$. 
    We already proved $e^{h} \leq e^{\prime h}$. 
    Therefore, $e^{h} \in [s^{\prime h} + 1, e^{\prime h}]$ holds. 
    
    We prove $r = r^{\prime}$ by contradiction. 
    We assume that $r \neq r^{\prime}$ holds. 
    Then, $r < r^{\prime}$ follows from $r \leq r^{\prime}$ 
    and $r \neq r^{\prime}$. 
    We apply Lemma~\ref{lem:intv_function_shift}~\ref{enum:intv_function_shift:6} 
    to the two intervals $[p, r+1]$ and $[p^{\prime}, r^{\prime}]$. 
    Then, the lemma shows that $e^{\prime\prime\prime h} \leq e^{\prime h}$ 
    because $r+1 \leq r^{\prime}$. 
    
    We can apply Proposition~\ref{prop:RB_IA_plus_right} to the two intervals $[p, r]$ and $[p, r+1]$ 
    because $k^{\prime} > h$ holds. 
    This proposition shows that at least one of the four conditions (i), (ii), (iii), and (iv) in Proposition~\ref{prop:RB_IA_plus_right} is satisfied. 

    If condition (i) of Proposition~\ref{prop:RB_IA_plus_right} is satisfied, 
    then $|[s^{h}, e^{h}]| = 1$ holds, 
    but the two facts $|[s^{h}, e^{h}]| = 1$ and $|[s^{h}, e^{h}]| \geq 4$ yield a contradiction. 

    Let $T[x^{\prime\prime\prime h}_{e}..y^{\prime\prime\prime h}_{e}]$ be the substring derived from the $e^{\prime\prime\prime h}$-th nonterminals of sequence $S^{h}$ in input string $T$. 
    If condition (ii) of Proposition~\ref{prop:RB_IA_plus_right} is satisfied,     
    then $|[s^{h}, e^{h}]| \geq 2$, $e^{h} < e^{\prime\prime\prime h}$, and 
    there exists a position $i$ of sequence $S^{h+1}$ satisfying 
    $x^{h+1}_{i} \in [x^{h}_{e}, x^{\prime\prime\prime h}_{e} - 1]$ for 
    the substring $T[x^{h+1}_{i}..y^{h+1}_{i}]$ derived from the $i$-th position of sequence $S^{h+1}$. 
    $x^{h}_{e} \geq x^{\prime h}_{s} + 1$ holds because $s^{\prime h} + 1 \leq e^{h}$.     
    $x^{\prime\prime\prime h}_{e} \leq x^{\prime h}_{e}$ holds because $e^{\prime\prime\prime h} \leq e^{\prime h}$.     
    $x^{h+1}_{i} \geq x^{\prime h}_{s} + 1$ follows from 
    $x^{h+1}_{i} \geq x^{h}_{e}$ and $x^{h}_{e} \geq x^{\prime h}_{s} + 1$. 
    $x^{h+1}_{i} \leq x^{\prime h}_{e} - 1$ follows from  
    $x^{h+1}_{i} \leq x^{\prime\prime\prime h}_{e} - 1$ and $x^{\prime\prime\prime h}_{e} \leq x^{\prime h}_{e}$. 
    The position $i$ of sequence $S^{h+1}$ satisfies condition (i) of Definition~\ref{def:f_interval} for function $f_{\interval}(s^{\prime h}, e^{\prime h})$ because $x^{h+1}_{i} \in [x^{\prime h}_{s} + 1, x^{\prime h}_{e} - 1]$. 
    The existence of the position $i$ indicates that $f_{\interval}(s^{\prime h}, e^{\prime h}) \neq \perp$ holds. 
    On the other hand, $f_{\interval}(s^{\prime h}, e^{\prime h}) = \perp$ follows from the definition of sequence $A(p^{\prime}, r^{\prime})$. 
    The two facts $f_{\interval}(s^{\prime h}, e^{\prime h}) \neq \perp$ and $f_{\interval}(s^{\prime h}, e^{\prime h}) = \perp$ yield a contradiction. 

    If condition (iii) of Proposition~\ref{prop:RB_IA_plus_right} is satisfied,     
    then $e^{h} < e^{\prime\prime\prime h}$, and there exists a position $j$ of sequence $S^{h}$ satisfying 
    $j \in [e^{h}, e^{\prime\prime\prime h} - 1]$ and $|\val(S^{h}[j])| > \lfloor \mu(h+1) \rfloor$. 
    The assignment $\assign(S^{h}[j])$ of the nonterminal $S^{h}[j]$ is $-1$ because $|\val(S^{h}[j])| > \lfloor \mu(h+1) \rfloor$. 
    In this case, Lemma~\ref{lem:rr_class}~\ref{enum:rr_class:3} and Lemma~\ref{lem:rr_class}~\ref{enum:rr_class:4} indicate that 
    there exists a position $j^{\prime}$ of sequence $S^{h+1}$ satisfying 
    $x^{h+1}_{j^{\prime}} = x^{h}_{j}$. 
    Here, $T[x^{h+1}_{j^{\prime}}..y^{h+1}_{j^{\prime}}]$ (respectively, $T[x^{h}_{j}..y^{h}_{j}]$) is the substring derived from 
    the $j^{\prime}$-th nonterminal of sequence $S^{h+1}$ (respectively, the $j$-th nonterminal of sequence $S^{h}$). 
    The position $j^{\prime}$ of sequence $S^{h+1}$ satisfies condition (ii) of Definition~\ref{def:f_interval} for function $f_{\interval}(s^{\prime h}, e^{\prime h})$ because 
    $j \in [s^{\prime h}, e^{\prime h}-1]$ follows from $j \in [e^{h}, e^{\prime\prime\prime h} - 1]$, 
    $e^{h} \in [s^{\prime h} + 1, e^{\prime h}]$, and $e^{\prime\prime\prime h} \leq e^{\prime h}$. 
    The existence of the position $j^{\prime}$ indicates that $f_{\interval}(s^{\prime h}, e^{\prime h}) \neq \perp$ holds. 
    On the other hand, $f_{\interval}(s^{\prime h}, e^{\prime h}) = \perp$ follows from the definition of sequence $A(p^{\prime}, r^{\prime})$. 
    The two facts $f_{\interval}(s^{\prime h}, e^{\prime h}) \neq \perp$ and $f_{\interval}(s^{\prime h}, e^{\prime h}) = \perp$ yield a contradiction. 
    
    If condition (iv) of Proposition~\ref{prop:RB_IA_plus_right} is satisfied, 
    then $|[x^{h}_{e}, y^{h}_{e}]| > \lfloor \mu(h+1) \rfloor$ and $|[x^{h}_{e}, r]| = \sum_{w = 1}^{h+1} \lfloor \mu(w) \rfloor$ hold.
    The assignment $\assign(S^{h}[e^{h}])$ of the nonterminal $S^{h}[e^{h}]$ is $-1$ because $|[x^{h}_{e}, y^{h}_{e}]| > \lfloor \mu(h+1) \rfloor$. 
    In this case, Lemma~\ref{lem:rr_class}~\ref{enum:rr_class:3} and Lemma~\ref{lem:rr_class}~\ref{enum:rr_class:4} indicate that 
    there exists a position $i^{\prime}$ of sequence $S^{h+1}$ satisfying 
    $x^{h+1}_{i^{\prime}} = x^{h}_{e}$ 
    for the substring $T[x^{h+1}_{i^{\prime}}..y^{h+1}_{i^{\prime}}]$ derived from the $i^{\prime}$-th nonterminal of sequence $S^{h+1}$ in input string $T$. 
    Either $e^{h} = e^{\prime h}$ or $e^{h} < e^{\prime h}$ holds 
    because $e^{h} \leq e^{\prime\prime\prime h} \leq e^{\prime h}$. 
    If $e^{h} = e^{\prime h}$, 
    then 
    $|[x^{\prime h}_{e}, r^{\prime}]| > |[x^{h}_{e}, r]|$ follows from 
    $x^{h}_{e} = x^{\prime h}_{e}$ and $r < r^{\prime}$. 
    The position $i^{\prime}$ of sequence $S^{h+1}$ satisfies condition (iii) of Definition~\ref{def:f_interval} for function $f_{\interval}(s^{\prime h}, e^{\prime h})$ because     
    $|[x^{\prime h}_{e}, r^{\prime}]| > \sum_{w = 1}^{h+1} \lfloor \mu(w) \rfloor$ follows from 
    $|[x^{\prime h}_{e}, r^{\prime}]| > |[x^{h}_{e}, r]|$ and $|[x^{h}_{e}, r]| = \sum_{w = 1}^{h+1} \lfloor \mu(w) \rfloor$. 
    Otherwise (i.e., $e^{h} < e^{\prime h}$), 
    the position $i^{\prime}$ of sequence $S^{h+1}$ satisfies condition (ii) of Definition~\ref{def:f_interval} for function $f_{\interval}(s^{\prime h}, e^{\prime h})$ because 
    $e^{h} \in [s^{\prime h}, e^{\prime h}-1]$ and $|[x^{h}_{e}, y^{h}_{e}]| > \lfloor \mu(h+1) \rfloor$ hold. 
    The existence of the position $i^{\prime}$ indicates that $f_{\interval}(s^{\prime h}, e^{\prime h}) \neq \perp$ holds. 
    On the other hand, $f_{\interval}(s^{\prime h}, e^{\prime h}) = \perp$ follows from the definition of sequence $A(p^{\prime}, r^{\prime})$. 
    The two facts $f_{\interval}(s^{\prime h}, e^{\prime h}) \neq \perp$ and $f_{\interval}(s^{\prime h}, e^{\prime h}) = \perp$ yield a contradiction. 
    
    We showed that there exists a contradiction under the assumption that $r \neq r^{\prime}$ holds. 
    Therefore, $r = r^{\prime}$ must hold. 

    \textbf{Proof of Proposition~\ref{prop:adjacent_last_set}(iii).}
    We apply Lemma~\ref{lem:intv_function_shift}~\ref{enum:intv_function_shift:6} 
    to the two intervals $[p, r]$ and $[p^{\prime}, r^{\prime}]$. 
    Then, the lemma shows that 
    $e = e^{\prime}$ because $r = r^{\prime}$. 
    Therefore, Proposition~\ref{prop:adjacent_last_set}(iii) holds. 

    \textbf{Proof of Proposition~\ref{prop:adjacent_last_set}(iv).}
    We prove $k = h$ by contradiction for sequence $A(p^{\prime}-1, r^{\prime}) = [s^{\prime\prime 0}, e^{\prime\prime 0}], [s^{\prime\prime 1}, e^{\prime\prime 1}], \ldots, [s^{\prime\prime k}, e^{\prime\prime k}]$. 
    We assume that $k \neq h$. 
    Then, $k \geq h+1$ follows from $k \neq h$ and $k \geq h$. 
    $[p^{\prime}-1, r^{\prime}] \subseteq [p, r]$ follows from $p < p^{\prime}$ and $r = r^{\prime}$. 
    We can apply Lemma~\ref{lem:intv_function_shift}~\ref{enum:intv_function_shift:7} 
    to the two intervals $[p, r]$ and $[p^{\prime}-1, r^{\prime}]$. 
    The lemma shows that $h \geq k$ holds, 
    but the two facts $k \geq h+1$ and $h \geq k$ yield a contradiction. 
    Therefore, $k = h$ must hold. 

    We prove $s^{\prime\prime h} \neq b+1$ by contradiction. 
    We assume that $s^{\prime\prime h} = b+1$. 
    Then, $[p^{\prime}-1, r^{\prime}] \in \Delta(h, b+1)$ 
    follows from $k = h$ and $s^{\prime\prime h} = b+1$. 
    On the other hand, $[p^{\prime}-1, r^{\prime}] \not \in \Delta(h, b+1)$ 
    because $p^{\prime} = \min \{ s \mid [s, e] \in \Delta(h, b+1) \}$. 
    The two facts $[p^{\prime}-1, r^{\prime}] \in \Delta(h, b+1)$ and $[p^{\prime}-1, r^{\prime}] \not \in \Delta(h, b+1)$ yield a contradiction. 
    Therefore, $s^{\prime\prime h} \neq b+1$ must hold. 

    We prove $s^{\prime\prime h} = b$. 
    We can apply Lemma~\ref{lem:intv_function_shift}~\ref{enum:intv_function_shift:7} 
    to the two intervals $[p^{\prime}-1, r^{\prime}]$ and $[p, r]$ 
    because $[p^{\prime}-1, r^{\prime}] \subseteq [p, r]$. 
    The lemma shows that $[s^{\prime\prime h}, e^{\prime\prime h}] \subseteq [s^{h}, e^{h}]$. 
    $s^{\prime\prime h} \geq b$ follows from $[s^{\prime\prime h}, e^{\prime\prime h}] \subseteq [s^{h}, e^{h}]$ and $s^{h} = b$. 
    Therefore, $s^{\prime\prime h} = b$ follows from $s^{\prime\prime h} \geq b$, $s^{\prime\prime h} \neq b+1$, and $s^{\prime\prime h} \in [1, b+1]$. 
    
    Finally, $[p^{\prime}-1, r^{\prime}] \in \Delta(h, b)$ follows from 
    $k = h$ and $s^{\prime\prime h} = b$. 
\end{proof}

%%%%%%%%%%%%%%%%%%%%%%%%%%%%%%%%%%%%%%%%%%%%%%%%%%%%%%%%%%%%%%%%%%%%%%%%%%%%%%%%%%%%%%%%%%%%%%%%

\begin{proposition}\label{prop:f_rec_long_property}
    Consider two sequences $A(s, e) = [s^{0}, e^{0}], [s^{1}, e^{1}], \ldots, [s^{k}, e^{k}]$ 
    and $A(s, e^{\prime}) = [s^{\prime 0}, e^{\prime 0}], [s^{\prime 1}, e^{\prime 1}], \ldots, [s^{\prime k^{\prime}}, e^{\prime k^{\prime}}]$
    of intervals 
    for three positions $s, e$, and $e^{\prime}$ ($s < e^{\prime} \leq e$) in input string $T$. 
    For an integer $h \in [0, k-1]$, 
    let $T[x^{k}_{s}..y^{k}_{s}]$ be the substring derived from the $s^{k}$-th nonterminal of sequence $S^{k}$ in input string $T$. 
    If $|[x^{k}_{s}, e]| \geq \sum_{w = 1}^{k} \lfloor \mu(w) \rfloor$ and $e^{\prime} \in [x^{k}_{s} + \sum_{w = 1}^{k} \lfloor \mu(w) \rfloor, e]$, 
    then $k^{\prime} = k$. 
\end{proposition}
\begin{proof}
For each integer $h \in [0, k]$, 
let $T[x^{h}_{s}..y^{h}_{s}]$ and $T[x^{h}_{e}..y^{h}_{e}]$ be the two substrings derived from the $s^{h}$-th and $e^{h}$-th nonterminals of sequence $S^{h}$, respectively, in input string $T$. 
Similarly, for each integer $h \in [0, k^{\prime}]$, 
let $T[x^{\prime h}_{s}..y^{\prime h}_{s}]$ and $T[x^{\prime h}_{e}..y^{\prime h}_{e}]$ be the two substrings derived from the $s^{\prime h}$-th and $e^{\prime h}$-th nonterminals of sequence $S^{h}$, respectively, in input string $T$. 
For each integer $h \in [0, k]$, 
sequence $S^{h}$ contains a position $i^{h}$ satisfying $x^{h}_{i} = x^{k}_{s}$ 
for the substring $T[x^{h}_{i}..y^{h}_{i}]$ derived from the $i^{h}$-th nonterminal of sequence $S^{h}$. 
We apply Lemma~\ref{lem:intv_function_shift}~\ref{enum:intv_function_shift:3} to the two intervals $[s, e]$ and $[s, e^{\prime}]$. 
Then, the lemma shows that $k \geq k^{\prime}$, 
$s^{h} = s^{\prime h}$, and $e^{h} \geq e^{\prime h}$ for each integer $h \in [0, k^{\prime}]$. 

The following five statements are used to prove Proposition~\ref{prop:f_rec_long_property}.
\begin{enumerate}[label=\textbf{(\roman*)}]
    \item $x^{h-1}_{s} \leq x^{h}_{s} \leq x^{h}_{e} \leq x^{h-1}_{e}$ for each integer $h \in [1, k]$;  
    \item $x^{h}_{s} \leq x^{k}_{s}$ and $x^{k}_{s} \leq x^{h}_{e}$ for each integer $h \in [0, k]$;
    \item $i^{h} \in [s^{h}, e^{h}]$ for each integer $h \in [0, k]$;
    \item if $e^{\prime h} \leq i^{h}$ 
    for an integer $h \in [1, \min \{ k^{\prime}, k-1 \} ]$, 
    then $h < k^{\prime}$;
    \item if $e^{\prime h} > i^{h}$ 
    for an integer $h \in [1, \min \{ k^{\prime}, k-1 \} ]$, 
    then $h < k^{\prime}$.
\end{enumerate}

\textbf{Proof of statement (i).}
$x^{h-1}_{s} \leq x^{h}_{s} \leq x^{h}_{e} \leq x^{h-1}_{e}$ follow from Proposition~\ref{prop:sync_set_sub_property1}~\ref{enum:sync_set_sub_property1:1} and Proposition~\ref{prop:sync_set_sub_property1}~\ref{enum:sync_set_sub_property1:2}. 

\textbf{Proof of statement (ii).}
$x^{h}_{s} \leq x^{k}_{s}$ and $x^{k}_{s} \leq x^{h}_{e}$ follow from statement (i). 

\textbf{Proof of statement (iii).}
$s^{h} \leq i^{h}$ because $x^{h}_{s} \leq x^{k}_{s}$ and $x^{h}_{i} = x^{k}_{s}$. 
$i^{h} \leq e^{h}$ because $x^{k}_{s} \leq x^{h}_{e}$ and $x^{h}_{i} = x^{k}_{s}$. 
Therefore, $i^{h} \in [s^{h}, e^{h}]$ holds. 

\textbf{Proof of statement (iv).}
We prove $|[x^{\prime h}_{e}, y^{\prime h}_{e}]| > \lfloor \mu(h+1) \rfloor$. 
$x^{\prime h}_{e} \leq x^{h}_{i}$ follows from $e^{\prime h} \leq i^{h}$.
$x^{\prime h}_{e} \leq x^{k}_{s}$ follows from $x^{\prime h}_{e} \leq x^{h}_{i}$ and $x^{k}_{s} = x^{h}_{i}$. 
$|[x^{\prime h}_{e}, e^{\prime}]| > \sum_{w = 1}^{k} \lfloor \mu(w) \rfloor$ 
follows from $e^{\prime} \in [x^{k}_{s} + \sum_{w = 1}^{k} \lfloor \mu(w) \rfloor, e]$ 
and $x^{\prime h}_{e} \leq x^{k}_{s}$. 
We apply Lemma~\ref{lem:rec_function_basic_relation}~\ref{enum:rec_function_basic_relation:4} 
to the interval $[s, e^{\prime}]$. 
Then, the lemma shows that 
$|[x^{\prime h}_{e}, y^{\prime h}_{e}]| \geq |[x^{\prime h}_{e}, e^{\prime}]| - \sum_{w = 1}^{h} \lfloor \mu(w) \rfloor$ holds. 
Since $h \leq k-1$, 
the following equation holds: 
\begin{equation*}
    \begin{split}
        |[x^{\prime h}_{e}, y^{\prime h}_{e}]| &\geq |[x^{\prime h}_{e}, e^{\prime}]| - \sum_{w = 1}^{h} \lfloor \mu(w) \rfloor \\
        &> \sum_{w = 1}^{k} \lfloor \mu(w) \rfloor - \sum_{w = 1}^{h} \lfloor \mu(w) \rfloor \\
        &\geq \lfloor \mu(h+1) \rfloor.
    \end{split}
\end{equation*}

Since $|[x^{\prime h}_{e}, y^{\prime h}_{e}]| > \lfloor \mu(h+1) \rfloor$, 
the assignment $\assign(S^{h}[e^{\prime h}])$ of the nonterminal $S^{h}[e^{\prime h}]$ is $-1$. 
In this case, 
Lemma~\ref{lem:rr_class}~\ref{enum:rr_class:3} and Lemma~\ref{lem:rr_class}~\ref{enum:rr_class:4} indicate that 
there exists a position $g$ of sequence $S^{h+1}$ satisfying 
$x^{h+1}_{g} = x^{\prime h}_{e}$ for the substring $T[x^{h+1}_{g}..y^{h+1}_{g}]$ derived from the $g$-th nonterminal of sequence $S^{h+1}$.

We prove $h < k^{\prime}$. 
$|[x^{\prime h}_{e}, e^{\prime}]| > \sum_{w = 1}^{h+1} \lfloor \mu(w) \rfloor$ follows from 
$|[x^{\prime h}_{e}, e^{\prime}]| > \sum_{w = 1}^{k} \lfloor \mu(w) \rfloor$ and $k \geq h+1$. 
The position $g$ of sequence $S^{h+1}$ satisfies condition (iii) of 
Definition~\ref{def:f_interval} for function $f_{\interval}(s^{\prime h}, e^{\prime h})$. 
The existence of the position $g$ indicates that $f_{\interval}(s^{\prime h}, e^{\prime h}) \neq \perp$ holds. 
Therefore, $h < k^{\prime}$ follows from $f_{\interval}(s^{\prime h}, e^{\prime h}) \neq \perp$. 

\textbf{Proof of statement (v).}
$x^{h+1}_{s} \leq x^{h}_{i}$ follows from $x^{h+1}_{s} \leq x^{k}_{s}$ and $x^{h}_{i} = x^{k}_{s}$. 
$x^{h}_{i} < x^{\prime h}_{e}$ holds because $s^{\prime h} > i^{h}$. 
Therefore, $x^{h+1}_{s} < x^{\prime h}_{e}$ holds. 

Sequence $S^{h}$ contains a position $j$ satisfying $x^{h+1}_{s} = x^{h}_{j}$ 
for the substring $T[x^{h}_{j}..y^{h}_{j}]$ derived from the $j$-th nonterminal of sequence $S^{h}$. 
$s^{h} \leq j \leq i^{h}$ holds because $x^{h}_{s} \leq x^{h}_{j} \leq x^{h}_{i}$. 

The position $s^{h+1}$ of sequence $S^{h+1}$ satisfies at least one of four conditions (i), (ii), (iii), and (iv) of 
Definition~\ref{def:f_interval} for function $f_{\interval}(s^{h}, e^{h})$. 
If the position $s^{h+1}$ satisfies condition (i), 
then $x^{h+1}_{s} \in [x^{h}_{s}+1, x^{h}_{e} - 1]$ holds. 
$x^{h+1}_{s} \geq x^{\prime h}_{s}+1$ follows from $x^{h+1}_{s} \in [x^{h}_{s}+1, x^{h}_{e} - 1]$ and $s^{h} = s^{\prime h}$. 
$x^{h+1}_{s} \leq x^{\prime h}_{e}-1$ follows from $x^{h+1}_{s} < x^{\prime h}_{e}$. 
The position $s^{h+1}$ of sequence $S^{h+1}$ satisfies condition (i) of 
Definition~\ref{def:f_interval} for function $f_{\interval}(s^{\prime h}, e^{\prime h})$. 
The existence of the position $s^{h+1}$ indicates that $h < k^{\prime}$. 

If the position $s^{h+1}$ satisfies condition (ii) and $j = s^{h}$ (i.e., $j = s^{\prime h}$), 
then $|[x^{h}_{j}, y^{h}_{j}]| > \lfloor \mu(h+1) \rfloor$. 
In this case, 
the position $s^{h+1}$ of sequence $S^{h+1}$ satisfies condition (ii) of 
Definition~\ref{def:f_interval} for function $f_{\interval}(s^{\prime h}, e^{\prime h})$ 
because $j \in [s^{\prime h}, e^{\prime h} - 1]$ follows from $s^{\prime h} \leq j \leq i^{h}$ and $e^{\prime h} > i^{h}$.
The existence of the position $s^{h+1}$ indicates that $h < k^{\prime}$ holds. 

If the position $s^{h+1}$ satisfies condition (ii) and $j \neq s^{h}$ (i.e., $j > s^{\prime h}$), 
then $|[x^{h}_{j}, y^{h}_{j}]| > \lfloor \mu(h+1) \rfloor$ or $|[x^{h}_{j-1}, y^{h}_{j-1}]| > \lfloor \mu(h+1) \rfloor$ 
for the substring $T[x^{h}_{j-1}..y^{h}_{j-1}]$ derived from the $(j-1)$-th nonterminal of sequence $S^{h}$. 
In this case, 
the position $s^{h+1}$ of sequence $S^{h+1}$ satisfies condition (ii) of 
Definition~\ref{def:f_interval} for function $f_{\interval}(s^{\prime h}, e^{\prime h})$ 
because $j \in [s^{\prime h}+1, e^{\prime h} - 1]$ follows from $s^{\prime h} < j \leq i^{h}$ and $e^{\prime h} > i^{h}$.
The existence of the position $s^{h+1}$ indicates that $h < k^{\prime}$ holds. 

We show that the position $s^{h+1}$ does not satisfy condition (iii) by contradiction. 
We assume that the position $s^{h+1}$ satisfies condition (iii). 
Then, $x^{h+1}_{s} = x^{h}_{e}$ holds. 
$j = e^{h}$ follows from $x^{h+1}_{s} = x^{h}_{e}$. 
On the other hand, $j < e^{h}$ follows from $s^{h} \leq j \leq i^{h}$ and $i^{h} < e^{\prime h} \leq e^{h}$. 
The two facts $j = e^{h}$ and $j < e^{h}$ yield a contradiction. 
Therefore, the position $s^{h+1}$ does not satisfy condition (iii). 

If the position $s^{h+1}$ satisfies condition (iv), 
then $x^{h+1}_{s} = x^{h}_{s}$ and $|[s, x^{h}_{s}]| > 1 + \sum_{w = 1}^{h+1} \lfloor \mu(w) \rfloor$ hold. 
In this case, 
the position $s^{h+1}$ of sequence $S^{h+1}$ satisfies condition (iv) of 
Definition~\ref{def:f_interval} for function $f_{\interval}(s^{\prime h}, e^{\prime h})$ 
because $x^{h}_{s} = x^{\prime h}_{s}$.
The existence of the position $s^{h+1}$ indicates that $h < k^{\prime}$ holds. 

Finally, $h < k^{\prime}$ always holds. 

\textbf{Proof of Proposition~\ref{prop:f_rec_long_property}.}
$k^{\prime} \geq k$ follows from statement (iv) and statement (v). 
We already proved $k^{\prime} \leq k$. 
Therefore, $k = k^{\prime}$ holds. 
\end{proof}

%%%%%%%%%%%%%%%%%%%%%%%%%%%%%%%%%%%%%%%%%%%%%%%%%%%%%%%%%%%%%%%%%%%%%%%%%%%%%%%%%%%%%%%%%%%%%%%%

\begin{proposition}\label{prop:ILAST_set_long_property}
    Let $(h, b)$ be a pair of an integer $h \in [0, H]$ 
    and a position $b \in [1, |S^{h}|]$ of sequence $S^{h}$ 
    satisfying $\Delta(h, b) \neq \emptyset$. 
    Let 
    (A) $T[x^{h}_{b}..y^{h}_{b}]$ be the substring derived from the $b$-th nonterminal of sequence $S^{h}$ in input string $T$, 
    (B) $e_{\max} = \max \{ e \mid [s, e] \in \Delta(h, b) \}$, 
    and (C) $e_{\min} = \min \{ e \mid [s, e] \in \Delta(h, b) \}$.    
    If $|[x^{h}_{b}, e_{\max}]| \geq \sum_{w = 1}^{h} \lfloor \mu(w) \rfloor$, 
    then $e_{\min} \leq x^{h}_{b} + \sum_{w = 1}^{h} \lfloor \mu(w) \rfloor$. 
\end{proposition}
\begin{proof}
Let $s_{\min} = \min \{ s \mid [s, e] \in \Delta(h, b) \}$ 
and $s_{\max} = \max \{ s \mid [s, e] \in \Delta(h, b) \}$. 
Then, 
$[s_{\min}, e_{\max}] \in \Delta(h, b)$ follows from Lemma~\ref{lem:IA_maximal_lemma}. 
Consider sequence $A(s_{\min}, e_{\max}) = [s^{0}, e^{0}], [s^{1}, e^{1}], \ldots, [s^{k}, e^{k}]$. 
Here, $k = h$ and $s^{h} = b$ follows from the definition of set $\Delta(h, b)$. 

Let $e^{\prime} = x^{h}_{b} + \sum_{w = 1}^{h} \lfloor \mu(w) \rfloor$. 
Consider sequence $A(s_{\min}, e^{\prime}) = [s^{0}, e^{0}], [s^{1}, e^{1}], \ldots, [s^{\prime k^{\prime}}, e^{\prime k^{\prime}}]$. 
Then, $k^{\prime} = h$ follows from Proposition~\ref{prop:f_rec_long_property}. 
We apply Lemma~\ref{lem:intv_function_shift}~\ref{enum:intv_function_shift:3} to 
the two intervals $[s_{\min}, e_{\max}]$ and $[s_{\min}, e^{\prime}]$. 
Then, the lemma shows that $s^{h} = s^{\prime h}$ and $e^{\prime h} \leq e^{h}$. 

We prove $e_{\min} \leq x^{h}_{b} + \sum_{w = 1}^{h} \lfloor \mu(w) \rfloor$. 
$[s_{\min}, e^{\prime}] \in \Delta(h, b)$ follows from 
$k^{\prime} = h$ and $s^{\prime h} = s^{h}$ and $s^{h} = b$. 
$e_{\min} \leq e^{\prime}$ follows from $[s_{\min}, e^{\prime}] \in \Delta(h, b)$ 
and $e_{\min} = \min \{ e \mid [s, e] \in \Delta(h, b) \}$. 
Therefore, $e_{\min} \leq x^{h}_{b} + \sum_{w = 1}^{h} \lfloor \mu(w) \rfloor$ 
follows from $e_{\min} \leq e^{\prime}$ and $e^{\prime} = x^{h}_{b} + \sum_{w = 1}^{h} \lfloor \mu(w) \rfloor$. 

\end{proof}

Consider an interval attractor $([p, q], [\ell, r])$ in subset $\Psi_{\run}$. 
From Definition~\ref{def:RR_Delta} and Lemma~\ref{lem:rec_function_basic_relation}~\ref{enum:rec_function_basic_relation:1}, 
there exists a pair of an integer $h \in [0, H-1]$ and a position $b \in [1, |S^{h}|]$ in sequence $S^{h}$ 
satisfying 
$p = \min \{ s \mid [s, e] \in \Delta(h, b) \}$, 
$q = \max \{ s \mid [s, e] \in \Delta(h, b) \}$, 
$\ell = \min \{ e \mid [s, e] \in \Delta(h, b) \}$, 
and $r = \max \{ e \mid [s, e] \in \Delta(h, b) \}$. 
The following proposition states properties of the two sets $\Delta(h, b)$ 
and $\Delta(h, b+1)$.

%%%%%%%%%%%%%%%%%%%%%%%%%%%%%%%%%%%%%%%%%%%%%%%%%%%%%%%%%%%%%%%%%%%%%%%%%%%%%%%%%%%%%%%%%%%%%%%%

\begin{proposition}\label{prop:I_LAST_HR_proceeding_property}
Consider an interval attractor $([p, q], [\ell, r])$ in subset $\Psi_{\run}$ 
and sequence $A(p, r) = [s^{0}, e^{0}], [s^{1}, e^{1}], \ldots, [s^{k}, e^{k}]$. 
Let $\gamma$ and $C$ be the attractor position and associated string of 
the interval attractor $([p, q], [\ell, r])$, respectively. 
Here, 
\begin{itemize}
    \item from Definition~\ref{def:RR_Delta} and Lemma~\ref{lem:rec_function_basic_relation}~\ref{enum:rec_function_basic_relation:1}, 
    there exists a pair of an integer $h \in [0, H-1]$ and a position $b \in [1, |S^{h}|]$ in sequence $S^{h}$ 
    satisfying 
    $p = \min \{ s \mid [s, e] \in \Delta(h, b) \}$, 
    $q = \max \{ s \mid [s, e] \in \Delta(h, b) \}$, 
    $\ell = \min \{ e \mid [s, e] \in \Delta(h, b) \}$, 
    and $r = \max \{ e \mid [s, e] \in \Delta(h, b) \}$; 
    \item $[p, r] \in \Delta(h, b)$ follows from Lemma~\ref{lem:IA_maximal_lemma}; 
    \item $h = k$ and $b = s^{k}$ follows from $[p, r] \in \Delta(h, b)$. 
\end{itemize}

Let $T[x^{h}_{s}..y^{h}_{s}]$ be the substring derived from the $s^{h}$-th nonterminal of sequence $S^{h}$. 
Then, the following eight statements hold:
\begin{enumerate}[label=\textbf{(\roman*)}]
    \item \label{enum:I_LAST_HR_proceeding_property:1} $|[x^{h}_{s}, r]| > 1 + \sum_{w = 1}^{h+3} \lfloor \mu(w) \rfloor$ and $C = \val(S^{h}[b])$;
    \item \label{enum:I_LAST_HR_proceeding_property:2} $|[s^{h}, e^{h}]| \geq 4$, $S^{h}[s^{h}] = S^{h}[s^{h}+1] = \cdots = S^{h}[e^{h}-1]$, 
    $|\val(S^{h}[s^{h}])| \leq \lfloor \mu(h+1) \rfloor$, and $|C| \leq \lfloor \mu(h+1) \rfloor$; 
    \item \label{enum:I_LAST_HR_proceeding_property:3} $\ell \leq x^{h}_{s} +  \sum_{w = 1}^{h} \lfloor \mu(w) \rfloor$;
    \item \label{enum:I_LAST_HR_proceeding_property:4} $T[p-1..x^{h}_{s} + (\sum_{w = 1}^{h} \lfloor \mu(w) \rfloor) + 2 \lfloor \mu(h+1) \rfloor] = T[p-1 + |C|..x^{h}_{s} + (\sum_{w = 1}^{h} \lfloor \mu(w) \rfloor) + 2 \lfloor \mu(h+1) \rfloor + |C|]$;
    \item \label{enum:I_LAST_HR_proceeding_property:5} $[p, x^{h}_{s} + (\sum_{w = 1}^{h} \lfloor \mu(w) \rfloor) + |C|] \in \Delta(h, b)$;
    \item \label{enum:I_LAST_HR_proceeding_property:6} 
    for a pair of two integers $s \in [p-1, x^{h}_{s}-1]$ and $e \in [x^{h}_{s}, x^{h}_{s} + (\sum_{w = 1}^{h} \lfloor \mu(w) \rfloor) + 2|C|]$, 
    $[s + |C|, e + |C|] \in \Delta(h, b+1)$ 
    if $[s, e] \in \Delta(h, b)$;
    \item \label{enum:I_LAST_HR_proceeding_property:7} 
    for a pair of two integers $s \in [p-1, x^{h}_{s}-1]$ and $e \in [x^{h}_{s}, x^{h}_{s} + (\sum_{w = 1}^{h} \lfloor \mu(w) \rfloor) + 2|C|]$, 
    $[s, e] \in \Delta(h, b)$ 
    if $[s + |C|, e + |C|] \in \Delta(h, b+1)$; 
    \item \label{enum:I_LAST_HR_proceeding_property:8} $[p + |C|, x^{h}_{s} + (\sum_{w = 1}^{h} \lfloor \mu(w) \rfloor) + 2|C|] \in \Delta(h, b+1)$.
\end{enumerate}
\end{proposition}
\begin{proof}
The proof of Proposition~\ref{prop:I_LAST_HR_proceeding_property} is as follows. 

\textbf{Proof of Proposition~\ref{prop:I_LAST_HR_proceeding_property}(i).}
$\lcs(T[p-1..\gamma - 1], C^{n+1}) = T[p-1..\gamma - 1]$ 
and $|\lcp(T[\gamma..r]$, $C^{n+1})| > 1 + \sum_{w = 1}^{h+3} \lfloor \mu(w) \rfloor$ follow from 
the definition of the subset $\Psi_{\run}$. 
Here, $\gamma = x^{h}_{s}$ follows from the definition of attractor position. 
$|[x^{h}_{s}, r]| > 1 + \sum_{w = 1}^{h+3} \lfloor \mu(w) \rfloor$ follows from 
$|\lcp(T[\gamma..r], C^{n+1})| > 1 + \sum_{w = 1}^{h+3} \lfloor \mu(w) \rfloor$ 
and $\gamma = x^{h}_{s}$. 
Since $|[\gamma, r]| > \sum_{w = 1}^{h+3} \lfloor \mu(w) \rfloor$, 
the associated string $C$ is defined as the string derived from the $b$-th nonterminal of sequence $S^{h}$ 
(i.e., $C = \val(S^{h}[b])$).

\textbf{Proof of Proposition~\ref{prop:I_LAST_HR_proceeding_property}(ii).}
We prove $|[s^{h}, e^{h}]| \geq 4$. 
$2 \lfloor \mu(h+1) \rfloor + \sum_{w = 1}^{h+1} \lfloor \mu(w) \rfloor \leq \sum_{w = 1}^{h+3} \lfloor \mu(w) \rfloor$ 
because $\lfloor \mu(h+1) \rfloor \leq \lfloor \mu(h+2) \rfloor$ and $\lfloor \mu(h+2) \rfloor \leq \lfloor \mu(h+3) \rfloor$. 
$|[x^{h}_{s}, r]| > 2 \lfloor \mu(h+1) \rfloor + \sum_{w = 1}^{h+1} \lfloor \mu(w) \rfloor$ 
follows from $|[x^{h}_{s}, r]| > 1 + \sum_{w = 1}^{h+3} \lfloor \mu(w) \rfloor$ 
and $2 \lfloor \mu(h+1) \rfloor + \sum_{w = 1}^{h+1} \lfloor \mu(w) \rfloor \leq \sum_{w = 1}^{h+3} \lfloor \mu(w) \rfloor$. 
We apply Lemma~\ref{lem:f_rec_top_property}~\ref{enum:f_rec_top_property:3} to the interval $[p, r]$. 
Then, the lemma shows that $|[s^{h}, e^{h}]| \geq 4$ holds. 

We prove $S^{h}[s^{h}] = S^{h}[s^{h}+1] = \cdots = S^{h}[e^{h}-1]$, $|\val(S^{h}[s^{h}])| \leq \lfloor \mu(h+1) \rfloor$, and $|C| \leq \lfloor \mu(h+1) \rfloor$.
We can apply Lemma~\ref{lem:f_rec_top_property}~\ref{enum:f_rec_top_property:1} and Lemma~\ref{lem:f_rec_top_property}~\ref{enum:f_rec_top_property:2} to the interval $[p, r]$ 
because $|[s^{h}, e^{h}]| \geq 4$. 
The lemma shows that $S^{h}[s^{h}] = S^{h}[s^{h}+1] = \cdots = S^{h}[e^{h}-1]$. 
$|\val(S^{h}[s^{h}])| \leq \lfloor \mu(h+1) \rfloor$ follows from Lemma~\ref{lem:f_rec_top_property}~\ref{enum:f_rec_top_property:1}. 
$|C| \leq \lfloor \mu(h+1) \rfloor$ follows from $C = \val(S^{h}[s^{h}])$ and $|\val(S^{h}[s^{h}])| \leq \lfloor \mu(h+1) \rfloor$. 

\textbf{Proof of Proposition~\ref{prop:I_LAST_HR_proceeding_property}(iii).}
$\ell \leq x^{h}_{s} + \sum_{w = 1}^{h} \lfloor \mu(w) \rfloor$ follows from Proposition~\ref{prop:ILAST_set_long_property} 
and $|[x^{h}_{s}, r]| > 1 + \sum_{w = 1}^{h+3} \lfloor \mu(w) \rfloor$. 

\textbf{Proof of Proposition~\ref{prop:I_LAST_HR_proceeding_property}(iv).}
We show that string $T[x^{h}_{s}..x^{h}_{s} + (\sum_{w = 1}^{h} \lfloor \mu(w) \rfloor) + 2 \lfloor \mu(h+1) \rfloor + |C|]$ is 
a prefix of the string $C^{n+1}$. 
$(\sum_{w = 1}^{h} \lfloor \mu(w) \rfloor) + 2 \lfloor \mu(h+1) \rfloor + |C| \leq \sum_{w = 1}^{h+3} \lfloor \mu(w) \rfloor$ 
because $|C| \leq \lfloor \mu(h+1) \rfloor \leq \lfloor \mu(h+2) \rfloor \leq \lfloor \mu(h+3) \rfloor$. 
The string $T[x^{h}_{s}..x^{h}_{s} + (\sum_{w = 1}^{h} \lfloor \mu(w) \rfloor) + 2 \lfloor \mu(h+1) \rfloor + |C|]$ 
is a prefix of the longest common prefix $\lcp(T[x^{h}_{s}..r], C^{n+1})$ of 
two strings $T[x^{h}_{s}..r]$ and $C^{n+1}$ 
because $|\lcp(T[x^{h}_{s}..r], C^{n+1})| > 1 + \sum_{w = 1}^{h+3} \lfloor \mu(w) \rfloor$ 
and $|T[x^{h}_{s}..x^{h}_{s} + (\sum_{w = 1}^{h} \lfloor \mu(w) \rfloor) + 2 \lfloor \mu(h+1) \rfloor + |C|]| \leq 1 + \sum_{w = 1}^{h+3} \lfloor \mu(w) \rfloor$. 
Therefore, string $T[x^{h}_{s}..x^{h}_{s} + (\sum_{w = 1}^{h} \lfloor \mu(w) \rfloor) + 2 \lfloor \mu(h+1) \rfloor + |C|]$ is 
a prefix of the string $C^{n+1}$. 

We prove $T[p-1..x^{h}_{s} + (\sum_{w = 1}^{h} \lfloor \mu(w) \rfloor) + 2 \lfloor \mu(h+1) \rfloor] = T[p-1 + |C|..x^{h}_{s} + (\sum_{w = 1}^{h} \lfloor \mu(w) \rfloor) + 2 \lfloor \mu(h+1) \rfloor + |C|]$. 
$T[p-1..x^{h}_{s}-1] = T[p-1 + |C|..x^{h}_{s}-1 + |C|]$ follows from 
$\lcs(T[p-1..x^{h}_{s}-1], C^{n+1}) = T[p-1..x^{h}_{s}-1]$, 
$|\lcp(T[x^{h}_{s}..r+1], C^{n+1})| > 1 + \sum_{w = 1}^{h+3} \lfloor \mu(w) \rfloor$, 
and $|C| \leq \lfloor \mu(h+1) \rfloor$. 
$T[x^{h}_{s}..x^{h}_{s} + (\sum_{w = 1}^{h} \lfloor \mu(w) \rfloor) + 2 \lfloor \mu(h+1) \rfloor] = T[x^{h}_{s} + |C|..x^{h}_{s} + (\sum_{w = 1}^{h} \lfloor \mu(w) \rfloor) + 2 \lfloor \mu(h+1) \rfloor + |C|]$ holds 
because string $T[x^{h}_{s}..x^{h}_{s} + (\sum_{w = 1}^{h} \lfloor \mu(w) \rfloor) + 2 \lfloor \mu(h+1) \rfloor + |C|]$ is 
a prefix of the string $C^{n+1}$. 
Therefore, $T[p-1..x^{h}_{s} + (\sum_{w = 1}^{h} \lfloor \mu(w) \rfloor) + 2 \lfloor \mu(h+1) \rfloor] = T[p-1 + |C|..x^{h}_{s} + (\sum_{w = 1}^{h} \lfloor \mu(w) \rfloor) + 2 \lfloor \mu(h+1) \rfloor + |C|]$ follows from 
$T[p-1..x^{h}_{s}-1] = T[p-1 + |C|..x^{h}_{s}-1 + |C|]$ and $T[x^{h}_{s}..x^{h}_{s} + (\sum_{w = 1}^{h} \lfloor \mu(w) \rfloor) + 2 \lfloor \mu(h+1) \rfloor] = T[x^{h}_{s} + |C|..x^{h}_{s} + (\sum_{w = 1}^{h} \lfloor \mu(w) \rfloor) + 2 \lfloor \mu(h+1) \rfloor + |C|]$.

\textbf{Proof of Proposition~\ref{prop:I_LAST_HR_proceeding_property}(v).}
We prove $x^{h}_{s} + (\sum_{w = 1}^{h} \lfloor \mu(w) \rfloor) + |C| \in [\ell, r]$. 
$\ell \leq x^{h}_{s} + (\sum_{w = 1}^{h} \lfloor \mu(w) \rfloor) + |C|$ 
follows from $\ell \leq x^{h}_{s} +  \sum_{w = 1}^{h} \lfloor \mu(w) \rfloor$ and 
$x^{h}_{s} +  \sum_{w = 1}^{h} \lfloor \mu(w) \rfloor \leq x^{h}_{s} + (\sum_{w = 1}^{h} \lfloor \mu(w) \rfloor) + |C|$. 
$x^{h}_{s} + (\sum_{w = 1}^{h} \lfloor \mu(w) \rfloor) + |C| \leq r$ holds because 
$|[x^{h}_{s}, x^{h}_{s} + (\sum_{w = 1}^{h} \lfloor \mu(w) \rfloor) + |C|]| \leq 1 + (\sum_{w = 1}^{h+1} \lfloor \mu(w) \rfloor)$ 
and $|[x^{h}_{s}, r]| > 1 + \sum_{w = 1}^{h+3} \lfloor \mu(w) \rfloor$. 
Therefore, $x^{h}_{s} + (\sum_{w = 1}^{h} \lfloor \mu(w) \rfloor) + |C| \in [\ell, r]$ holds. 

We prove $[p, x^{h}_{s} + (\sum_{w = 1}^{h} \lfloor \mu(w) \rfloor) + |C|] \in \Delta(h, b)$. 
We apply Lemma~\ref{lem:IA_maximal_lemma} to the set $\Delta(h, b)$. 
Then, the lemma shows that 
$[p, x^{h}_{s} + (\sum_{w = 1}^{h} \lfloor \mu(w) \rfloor) + |C|] \in \Delta(h, b)$ 
because $x^{h}_{s} + (\sum_{w = 1}^{h} \lfloor \mu(w) \rfloor) + |C| \in [\ell, r]$. 

\textbf{Proof of Proposition~\ref{prop:I_LAST_HR_proceeding_property}(vi).}
Consider two sequences 
$A(s, e) = [s^{\prime 0}, e^{\prime 0}], [s^{\prime 1}, e^{\prime 1}], \ldots, [s^{\prime k^{\prime}}, e^{\prime k^{\prime}}]$ 
and 
$A(s + |C|, e + |C|) = [s^{\prime\prime 0}, e^{\prime\prime 0}], [s^{\prime\prime 1}, e^{\prime\prime 1}], \ldots, [s^{\prime\prime k^{\prime\prime}}, e^{\prime\prime k^{\prime\prime}}]$. 
Here, $k^{\prime} = h$ and $s^{\prime k^{\prime}} = b$ follows from $[s, e] \in \Delta(h, b)$. 
We can apply Corollary~\ref{cor:capture_gamma_corollary} to 
two intervals $[s, e]$ 
and $[s + |C|, e + |C|]$ 
because $T[s..e] = T[s + |C|..e + |C|]$ follows from Proposition~\ref{prop:I_LAST_HR_proceeding_property}~\ref{enum:I_LAST_HR_proceeding_property:4} and $|C| \leq \lfloor \mu(h+1) \rfloor$. 
Corollary~\ref{cor:capture_gamma_corollary} shows that $k^{\prime} = k^{\prime\prime}$ (i.e., $k^{\prime\prime} = h$) 
and $|[s, x^{\prime h}_{s}-1]| = |[s + |C|, x^{\prime\prime h}_{s}-1]|$ hold. 
Here, $T[x^{\prime h}_{s}..y^{\prime h}_{s}]$ and $T[x^{\prime\prime h}_{s}..y^{\prime\prime h}_{s}]$ 
are the two substrings derived from the $s^{\prime h}$-th and $s^{\prime\prime h}$-th nonterminals of sequence $S^{h}$, respectively. 
$x^{\prime h}_{s} = x^{h}_{s}$ holds because $s^{h} = s^{\prime h}$. 
$x^{\prime\prime h} = x^{h}_{s} + |C|$ follows from $|[s, x^{\prime h}_{s}-1]| = |[s + |C|, x^{\prime\prime h}_{s}-1]|$ 
and $x^{\prime h}_{s} = x^{h}_{s}$. 
$s^{\prime\prime h} = b+1$ holds because 
the $b$-th nonterminal of sequence $S^{h}$ derives substring $T[x^{h}_{s}..y^{h}_{s}]$, 
$C = T[x^{h}_{s}..y^{h}_{s}]$. 
Therefore, $[s + |C|, e + |C|] \in \Delta(h, b+1)$ follows from $k^{\prime\prime} = h$ and $s^{\prime\prime h} = b+1$. 

\textbf{Proof of Proposition~\ref{prop:I_LAST_HR_proceeding_property}(vii).}
Proposition~\ref{prop:I_LAST_HR_proceeding_property}(vii) can be proved using the same approach 
as for Proposition~\ref{prop:I_LAST_HR_proceeding_property}(vi). 

\textbf{Proof of Proposition~\ref{prop:I_LAST_HR_proceeding_property}(viii).}
Proposition~\ref{prop:I_LAST_HR_proceeding_property}(viii) follows from Proposition~\ref{prop:I_LAST_HR_proceeding_property}(v) 
and Proposition~\ref{prop:I_LAST_HR_proceeding_property}(vi). 
\end{proof}

The following proposition states the relationship among 
the eight integers $p$, $q$, $\ell$, $r$, 
$\min \{ s \mid [s, e] \in \Delta(h, b+1) \}$, 
$\max \{ s \mid [s, e] \in \Delta(h, b+1) \}$, 
$\min \{ e \mid [s, e] \in \Delta(h, b+1) \}$, 
and $\max \{ e \mid [s, e] \in \Delta(h, b+1) \}$. 

%%%%%%%%%%%%%%%%%%%%%%%%%%%%%%%%%%%%%%%%%%%%%%%%%%%%%%%%%%%%%%%%%%%%%%%%%%%%%%%%%%%%%%%%%%%%%%%%

\begin{proposition}\label{prop:I_LAST_HR_proceeding_RB}
Consider an interval attractor $([p, q], [\ell, r])$ in subset $\Psi_{\run}$ 
and sequence $A(p, r) = [s^{0}, e^{0}], [s^{1}, e^{1}], \ldots, [s^{k}, e^{k}]$. 
Let $\gamma$ and $C$ be the attractor position and associated string of 
the interval attractor $([p, q], [\ell, r])$, respectively. 
Here, 
\begin{itemize}
    \item from Definition~\ref{def:RR_Delta} and Lemma~\ref{lem:rec_function_basic_relation}~\ref{enum:rec_function_basic_relation:1}, 
    there exists a pair of an integer $h \in [0, H-1]$ and a position $b \in [1, |S^{h}|]$ in sequence $S^{h}$ 
    satisfying 
    $p = \min \{ s \mid [s, e] \in \Delta(h, b) \}$, 
    $q = \max \{ s \mid [s, e] \in \Delta(h, b) \}$, 
    $\ell = \min \{ e \mid [s, e] \in \Delta(h, b) \}$, 
    and $r = \max \{ e \mid [s, e] \in \Delta(h, b) \}$; 
    \item $[p, r] \in \Delta(h, b)$ follows from Lemma~\ref{lem:IA_maximal_lemma}; 
    \item $h = k$ and $b = s^{k}$ follows from $[p, r] \in \Delta(h, b)$;
    \item $\Delta(h, b+1) \neq \emptyset$ follows from Proposition~\ref{prop:I_LAST_HR_proceeding_property}~\ref{enum:I_LAST_HR_proceeding_property:8}. 
\end{itemize}
Let $s_{\min} = \min \{ s \mid [s, e] \in \Delta(h, b+1) \}$, 
$s_{\max} = \max \{ s \mid [s, e] \in \Delta(h, b+1) \}$,  
$e_{\min} = \min \{ e \mid [s, e] \in \Delta(h, b+1) \}$, and
$e_{\max} = \max \{ e \mid [s, e] \in \Delta(h, b+1) \}$. 
Then, the following two statements hold: 
\begin{enumerate}[label=\textbf{(\roman*)}]
    \item $s_{\min} = p + |C|$, $s_{\max} = q + |C|$, $e_{\min} = \ell + |C|$, and $e_{\max} = r$; 
    \item $q = p + |C| - 1$.
\end{enumerate}
\end{proposition}
\begin{proof}
Consider sequence $A(s_{\min}, e_{\max}) = [s^{\prime 0}, e^{\prime 0}], [s^{\prime 1}, e^{\prime 1}], \ldots, [s^{\prime k^{\prime}}, e^{\prime k^{\prime}}]$. 
$[s_{\min}, e_{\max}] \in \Delta(h, b+1)$ follows from Lemma~\ref{lem:IA_maximal_lemma}. 
$h = k^{\prime}$ and $b+1 = s^{\prime h}$ follows from $[s_{\min}, e_{\max}] \in \Delta(h, b+1)$;

Let $T[x^{h}_{s}..y^{h}_{s}]$ be the substring derived from the $s^{h}$-th nonterminal of sequence $S^{h}$. 
The following seven statements are used to prove Proposition~\ref{prop:I_LAST_HR_proceeding_RB}. 
\begin{enumerate}[label=\textbf{(\arabic*)}]
    \item $[s^{\prime h}, e^{\prime h}] = [s^{h}+1, e^{h}]$, $e_{\max} = r$, and $q \geq s_{\min}-1$; 
    \item $s_{\min} = p + |C|$; 
    \item $e_{\min} \leq \ell + |C|$;
    \item $e_{\min} \geq \ell + |C|$;
    \item $[q, x^{h}_{s} + (\sum_{w = 1}^{h} \lfloor \mu(w) \rfloor) + 2|C|] \in \Delta(h, b)$ and $q \leq p + |C| - 1$;
    \item $s_{\max} \geq q + |C|$; 
    \item $s_{\max} = q + |C|$. 
\end{enumerate}

\textbf{Proof of statement (1).} 
We can apply Proposition~\ref{prop:adjacent_last_set} to two sets $\Delta(h, b)$ and $\Delta(h, b+1)$ 
because $|[s^{h}, e^{h}]| \geq 4$ follows from Proposition~\ref{prop:I_LAST_HR_proceeding_property}~\ref{enum:I_LAST_HR_proceeding_property:2}. 
The proposition shows that $e^{h} = e^{\prime h}$, $r = e_{\max}$, and $[s_{\min}-1, e_{\max}] \in \Delta(h, b)$. 
$[s^{\prime h}, e^{\prime h}] = [s^{h}+1, e^{h}]$ follows from $s^{\prime h} = b+1$, $s^{h} = b$, and $e^{h} = e^{\prime h}$. 
$q \geq s_{\min}-1$ follows from $[s_{\min}-1, e_{\max}] \in \Delta(h, b)$ 
and $q = \max \{ s \mid [s, e] \in \Delta(h, b) \}$. 

\textbf{Proof of statement (2).} 
We prove $s_{\min} = p + |C|$ by contradiction. 
We assume that $s_{\min} \neq p + |C|$ holds. 
$[p + |C|, x^{h}_{s} + (\sum_{w = 1}^{h} \lfloor \mu(w) \rfloor) + 2|C|] \in \Delta(h, b+1)$ 
follows from Proposition~\ref{prop:I_LAST_HR_proceeding_property}~\ref{enum:I_LAST_HR_proceeding_property:8}. 
$s_{\min} \leq p + |C|$ follows from $[p + |C|, x^{h}_{s} + (\sum_{w = 1}^{h} \lfloor \mu(w) \rfloor) + 2|C|] \in \Delta(h, b+1)$. 
$s_{\min} < p + |C|$ follows from $s_{\min} \leq p + |C|$ and $s_{\min} \neq p + |C|$. 

We apply Lemma~\ref{lem:IA_maximal_lemma} to the interval 
$[p + |C|, x^{h}_{s} + (\sum_{w = 1}^{h} \lfloor \mu(w) \rfloor) + 2|C|]$. 
The proposition shows that 
$[p + |C| - 1, x^{h}_{s} + (\sum_{w = 1}^{h} \lfloor \mu(w) \rfloor) + 2|C|] \in \Delta(h, b+1)$ 
because $p + |C| - 1 \in [s_{\min}, s_{\max}]$. 
$[p - 1, x^{h}_{s} + (\sum_{w = 1}^{h} \lfloor \mu(w) \rfloor) + |C|] \in \Delta(h, b)$ 
follows from Proposition~\ref{prop:I_LAST_HR_proceeding_property}~\ref{enum:I_LAST_HR_proceeding_property:7} 
and $[p + |C| - 1, x^{h}_{s} + (\sum_{w = 1}^{h} \lfloor \mu(w) \rfloor) + 2|C|] \in \Delta(h, b+1)$. 

On the other hand, 
$[p - 1, x^{h}_{s} + (\sum_{w = 1}^{h} \lfloor \mu(w) \rfloor) + |C|] \not \in \Delta(h, b)$ 
because $p-1 \not \in [p, q]$. 
The two facts $[p - 1, x^{h}_{s} + (\sum_{w = 1}^{h} \lfloor \mu(w) \rfloor) + |C|] \in \Delta(h, b)$ 
and $[p - 1, x^{h}_{s} + (\sum_{w = 1}^{h} \lfloor \mu(w) \rfloor) + |C|] \not \in \Delta(h, b)$ 
yield a contradiction. 
Therefore, $s_{\min} = p + |C|$ must hold. 

\textbf{Proof of statement (3).} 
$\ell \leq x^{h}_{s} + (\sum_{w = 1}^{h} \lfloor \mu(w) \rfloor) + |C|$ 
follows from $[p, x^{h}_{s} + (\sum_{w = 1}^{h} \lfloor \mu(w) \rfloor) + |C|] \in \Delta(h, b)$ (Proposition~\ref{prop:I_LAST_HR_proceeding_property}~\ref{enum:I_LAST_HR_proceeding_property:5}). 
$[p, \ell] \in \Delta(h, b)$ follows from Lemma~\ref{lem:IA_maximal_lemma}. 
Since $[p, \ell] \in \Delta(h, b)$, 
$x^{h}_{s} \leq \ell$ follows from the definition of interval attractor. 
We can apply Proposition~\ref{prop:I_LAST_HR_proceeding_property}~\ref{enum:I_LAST_HR_proceeding_property:6} to 
the interval $[p, \ell]$ 
because $p \in [p-1, x^{h}_{s} - 1]$ 
and $\ell \in [x^{h}_{s}, x^{h}_{s} + (\sum_{w = 1}^{h} \lfloor \mu(w) \rfloor) + |C|]$. 
$[p + |C|, \ell + |C|] \in \Delta(h, b+1)$ follows from Proposition~\ref{prop:I_LAST_HR_proceeding_property}~\ref{enum:I_LAST_HR_proceeding_property:6}. 
Therefore, 
$e_{\min} \leq \ell + |C|$ follows from $[p + |C|, \ell + |C|] \in \Delta(h, b+1)$. 

\textbf{Proof of statement (4).} 
$e_{\min} \leq x^{h}_{s} + (\sum_{w = 1}^{h} \lfloor \mu(w) \rfloor) + 2|C|$ 
follows from $[p + |C|, x^{h}_{s} + (\sum_{w = 1}^{h} \lfloor \mu(w) \rfloor) + 2|C|] \in \Delta(h, b+1)$ (Proposition~\ref{prop:I_LAST_HR_proceeding_property}~\ref{enum:I_LAST_HR_proceeding_property:8}). 
$[s_{\max}, e_{\min}] \in \Delta(h, b+1)$ follows from Lemma~\ref{lem:IA_maximal_lemma}. 
$1 + y^{h}_{b} \leq e_{\min}$ follows from Corollary~\ref{cor:IA_exist_corollary}~\ref{enum:IA_exist_corollary:1}. 
Here, $y^{h}_{b} = x^{h}_{s} + |C| - 1$ because $\val(S^{h}[b]) = C$ follows from Proposition~\ref{prop:I_LAST_HR_proceeding_property}~\ref{enum:I_LAST_HR_proceeding_property:1}. 
$e_{\min} - |C| \in [x^{h}_{s}, x^{h}_{s} + (\sum_{w = 1}^{h} \lfloor \mu(w) \rfloor) + |C|]$ 
follows from $1 + y^{h}_{b} \leq e_{\min} \leq x^{h}_{s} + (\sum_{w = 1}^{h} \lfloor \mu(w) \rfloor) + 2|C|$ 
and $y^{h}_{b} = x^{h}_{s} + |C| - 1$. 
We can apply Proposition~\ref{prop:I_LAST_HR_proceeding_property}~\ref{enum:I_LAST_HR_proceeding_property:7} to 
the interval $[s_{\max}, e_{\min}]$ 
because $s_{\max} - |C| = p$ 
and $e_{\min} - |C| \in [x^{h}_{s}, x^{h}_{s} + (\sum_{w = 1}^{h} \lfloor \mu(w) \rfloor) + |C|]$. 
$[s_{\max} - |C|, e_{\min} - |C|] \in \Delta(h, b)$ follows from Proposition~\ref{prop:I_LAST_HR_proceeding_property}~\ref{enum:I_LAST_HR_proceeding_property:7}. 
Therefore, $\ell \leq e_{\min} - |C|$ (i.e., $e_{\min} \geq \ell + |C|$) follows from 
$[s_{\max} - |C|, e_{\min} - |C|] \in \Delta(h, b)$. 

\textbf{Proof of statement (5).} 
We prove $q \leq p + |C| - 1$ by contradiction. 
We assume that $q > p + |C| - 1$ holds. 
Then, $[p + |C|, r] \in \Delta(h, b)$ follows from Lemma~\ref{lem:IA_maximal_lemma} 
because $p + |C| \in [p, q]$. 
$[s_{\min}, e_{\max}] \in \Delta(h, b+1)$ follows from Lemma~\ref{lem:IA_maximal_lemma}. 
$[p + |C|, r] \in \Delta(h, b+1)$ follows from 
$[s_{\min}, e_{\max}] \in \Delta(h, b+1)$, $s_{\min} = p + |C|$, and $e_{\max} = r$. 
$\Delta(h, b) \cap \Delta(h, b+1) \neq \emptyset$ follows from 
$[p + |C|, r] \in \Delta(h, b)$ and $[p + |C|, r] \in \Delta(h, b+1)$. 

On the other hand, 
$\Delta(h, b) \cap \Delta(h, b+1) = \emptyset$ 
follows from the definition of sets $\Delta(h, b)$ and $\Delta(h, b+1)$. 
The two facts $\Delta(h, b) \cap \Delta(h, b+1) \neq \emptyset$ and 
$\Delta(h, b) \cap \Delta(h, b+1) = \emptyset$ yield a contradiction. 
Therefore, $q \leq p + |C| - 1$ must hold. 

Next, 
we consider two sequences 
$A(q, x^{h}_{s} + (\sum_{w = 1}^{h} \lfloor \mu(w) \rfloor) + 2|C|) = [s^{\prime\prime 0}, e^{\prime\prime 0}]$, $[s^{\prime\prime 1}, e^{\prime\prime 1}]$, $\ldots$, $[s^{\prime\prime k^{\prime\prime}}, e^{\prime\prime k^{\prime\prime}}]$ 
and 
$A(p + |C|, x^{h}_{s} + (\sum_{w = 1}^{h} \lfloor \mu(w) \rfloor) + 2|C|) = [s^{\prime\prime\prime 0}, e^{\prime\prime\prime 0}], [s^{\prime\prime\prime 1}, e^{\prime\prime\prime 1}], \ldots, [s^{\prime\prime\prime k^{\prime\prime\prime}}, e^{\prime\prime\prime k^{\prime\prime\prime}}]$. 
Here, $k^{\prime\prime\prime} = h$ and $s^{\prime\prime\prime h} = b+1$ follow from $[p + |C|, x^{h}_{s} + (\sum_{w = 1}^{h} \lfloor \mu(w) \rfloor) + 2|C|] \in \Delta(h, b+1)$. 
$x^{h}_{s} + (\sum_{w = 1}^{h} \lfloor \mu(w) \rfloor) + 2|C| \in [\ell, r]$ 
because (A) $\ell \leq x^{h}_{s} +  \sum_{w = 1}^{h} \lfloor \mu(w) \rfloor$~(Proposition~\ref{prop:I_LAST_HR_proceeding_property}~\ref{enum:I_LAST_HR_proceeding_property:2}), 
(B) $r \geq x^{h}_{s} + \sum_{w = 1}^{h+3} \lfloor \mu(w) \rfloor$~(Proposition~\ref{prop:I_LAST_HR_proceeding_property}~\ref{enum:I_LAST_HR_proceeding_property:1}), 
and (C) $|C| \leq \lfloor \mu(h+1) \rfloor$~(Proposition~\ref{prop:I_LAST_HR_proceeding_property}~\ref{enum:I_LAST_HR_proceeding_property:2}). 
$[p + |C|, x^{h}_{s} + (\sum_{w = 1}^{h} \lfloor \mu(w) \rfloor) + 2|C|] \subseteq [q, x^{h}_{s} + (\sum_{w = 1}^{h} \lfloor \mu(w) \rfloor) + 2|C|]$ follows from $q \leq p + |C| - 1$.

We apply Lemma~\ref{lem:intv_function_shift}~\ref{enum:intv_function_shift:7} to the two intervals $[q, x^{h}_{s} + (\sum_{w = 1}^{h} \lfloor \mu(w) \rfloor) + 2|C|]$ and $[p + |C|, x^{h}_{s} + (\sum_{w = 1}^{h} \lfloor \mu(w) \rfloor) + 2|C|]$. 
Then, the lemma shows that $k^{\prime\prime} \geq h$ and $[s^{\prime\prime\prime k^{\prime\prime\prime}}, e^{\prime\prime\prime h}] \subseteq [s^{\prime\prime h}, e^{\prime\prime h}]$ because $[p + |C|, x^{h}_{s} + (\sum_{w = 1}^{h} \lfloor \mu(w) \rfloor) + 2|C|] \subseteq [q, x^{h}_{s} + (\sum_{w = 1}^{h} \lfloor \mu(w) \rfloor) + 2|C|]$. 

We prove $(k^{\prime\prime}, s^{\prime\prime k^{\prime\prime}}) = (h, b)$ by contradiction. 
We assume that $(k^{\prime\prime}, s^{\prime\prime k^{\prime\prime}}) \neq (h, b)$. 
Then, set $\Delta(h, b)$ does not contain the interval $[q, x^{h}_{s} + (\sum_{w = 1}^{h} \lfloor \mu(w) \rfloor) + 2|C|]$. 
$k^{\prime\prime} < h$ follows from Lemma~\ref{lem:interval_extension_propertyX}~\ref{enum:interval_extension_propertyX:low}. 
The two facts $k^{\prime\prime} \geq h$ and $k^{\prime\prime} < h$ yield a contradiction. 
Therefore, $(k^{\prime\prime}, s^{\prime\prime k^{\prime\prime}}) = (h, b)$ must hold. 
$[q, x^{h}_{s} + (\sum_{w = 1}^{h} \lfloor \mu(w) \rfloor) + 2|C|] \in \Delta(h, b)$ follows from $(k^{\prime\prime}, s^{\prime\prime k^{\prime\prime}}) = (h, b)$.

\textbf{Proof of statement (6).}
We use interval $[q, x^{h}_{s} + (\sum_{w = 1}^{h} \lfloor \mu(w) \rfloor) + 2|C|]$ 
to prove $s_{\max} \geq q + |C|$. 
We can apply Proposition~\ref{prop:I_LAST_HR_proceeding_property}~\ref{enum:I_LAST_HR_proceeding_property:6} 
to the interval $[q, x^{h}_{s} + (\sum_{w = 1}^{h} \lfloor \mu(w) \rfloor) + 2|C|]$ 
because $q \in [p-1, x^{h}_{s} - 1]$ and 
$x^{h}_{s} + (\sum_{w = 1}^{h} \lfloor \mu(w) \rfloor) + 2|C| \in [x^{h}_{s}, x^{h}_{s} + (\sum_{w = 1}^{h} \lfloor \mu(w) \rfloor) + 2|C|]$. 
The proposition shows that 
$[q + |C|, x^{h}_{s} + (\sum_{w = 1}^{h} \lfloor \mu(w) \rfloor) + 3|C|] \in \Delta(h, b+1)$. 
Therefore, $s_{\max} \geq q + |C|$ follows from $[q + |C|, x^{h}_{s} + (\sum_{w = 1}^{h} \lfloor \mu(w) \rfloor) + 3|C|] \in \Delta(h, b+1)$. 

\textbf{Proof of statement (7).}
We use interval $[q + |C| + 1, x^{h}_{s} + (\sum_{w = 1}^{h} \lfloor \mu(w) \rfloor) + 3|C|]$ 
to prove $s_{\max} = q + |C|$. 
$e_{\min} \leq x^{h}_{s} + (\sum_{w = 1}^{h} \lfloor \mu(w) \rfloor) + 3|C|$ 
follows from $e_{\min} \leq \ell + |C|$ and 
$\ell \leq x^{h}_{s} +  \sum_{w = 1}^{h} \lfloor \mu(w) \rfloor$. 
On the other hand, $x^{h}_{s} + (\sum_{w = 1}^{h} \lfloor \mu(w) \rfloor) + 3|C| \leq e_{\max}$ holds 
because (A) $e_{\max} = r$, 
(B) $r > x^{h}_{s} + \sum_{w = 1}^{h+3} \lfloor \mu(w) \rfloor$, 
and (C) $|C| \leq \lfloor \mu(h+1) \rfloor$.
$q + |C| + 1 \leq p + 2|C|$ follows from 
$q \leq p + |C| - 1$. 
$[p + 2|C|, x^{h}_{s} + (\sum_{w = 1}^{h} \lfloor \mu(w) \rfloor) + 3|C|] \subseteq [q + |C| + 1, x^{h}_{s} + (\sum_{w = 1}^{h} \lfloor \mu(w) \rfloor) + 3|C|]$ follows from $q + |C| + 1 \leq p + 2|C|$. 

Consider three sequence 
$A(q + |C| + 1, x^{h}_{s} + (\sum_{w = 1}^{h} \lfloor \mu(w) \rfloor) + 3|C|) = [s^{\prime\prime 0}, e^{\prime\prime 0}]$, $[s^{\prime\prime 1}, e^{\prime\prime 1}]$, $\ldots$, $[s^{\prime\prime k^{\prime\prime}}, e^{\prime\prime k^{\prime\prime}}]$, 
$A(p + 2|C|, x^{h}_{s} + (\sum_{w = 1}^{h} \lfloor \mu(w) \rfloor) + 3|C|) = [s^{\prime\prime\prime 0}, e^{\prime\prime\prime 0}], [s^{\prime\prime\prime 1}, e^{\prime\prime\prime 1}], \ldots, [s^{\prime\prime\prime k^{\prime\prime\prime}}, e^{\prime\prime\prime k^{\prime\prime\prime}}]$, 
and 
$A(p, x^{h}_{s} + (\sum_{w = 1}^{h} \lfloor \mu(w) \rfloor) + |C|) = [s^{\prime\prime\prime\prime 0}, e^{\prime\prime\prime\prime 0}], [s^{\prime\prime\prime\prime 1}, e^{\prime\prime\prime\prime 1}], \ldots, [s^{\prime\prime\prime\prime k^{\prime\prime\prime\prime}}, e^{\prime\prime\prime\prime k^{\prime\prime\prime\prime}}]$. 
Here, $k^{\prime\prime\prime\prime} = h$ and $s^{\prime\prime\prime\prime h} = b$ hold 
because $[p, x^{h}_{s} + (\sum_{w = 1}^{h} \lfloor \mu(w) \rfloor) + |C|] \in \Delta(h, b)$ 
follows from Proposition~\ref{prop:I_LAST_HR_proceeding_property}~\ref{enum:I_LAST_HR_proceeding_property:5}. 

$T[p..x^{h}_{s} + (\sum_{w = 1}^{h} \lfloor \mu(w) \rfloor) + |C|] = T[p + 2|C|..x^{h}_{s} + (\sum_{w = 1}^{h} \lfloor \mu(w) \rfloor) + 3|C|]$ follows from $T[p-1..x^{h}_{s} + (\sum_{w = 1}^{h} \lfloor \mu(w) \rfloor) + 2 \lfloor \mu(h+1) \rfloor] = T[p-1 + |C|..x^{h}_{s} + (\sum_{w = 1}^{h} \lfloor \mu(w) \rfloor) + 2 \lfloor \mu(h+1) \rfloor + |C|]$~(Proposition~\ref{prop:I_LAST_HR_proceeding_property}~\ref{enum:I_LAST_HR_proceeding_property:4}) and $|C| \leq \lfloor \mu(h+1) \rfloor$. 
Since $T[p..x^{h}_{s} + (\sum_{w = 1}^{h} \lfloor \mu(w) \rfloor) + |C|] = T[p + 2|C|..x^{h}_{s} + (\sum_{w = 1}^{h} \lfloor \mu(w) \rfloor) + 3|C|]$, 
we can apply Theorem~\ref{theo:f_interval_syncro_property} to the two intervals 
$[p, x^{h}_{s} + (\sum_{w = 1}^{h} \lfloor \mu(w) \rfloor) + |C|]$ and 
$[p + 2|C|, x^{h}_{s} + (\sum_{w = 1}^{h} \lfloor \mu(w) \rfloor) + 3|C|]$. 
The lemma shows that $k^{\prime\prime\prime} = h$. 

We can apply Lemma~\ref{lem:intv_function_shift}~\ref{enum:intv_function_shift:7} to the two intervals 
$[p + 2|C|, x^{h}_{s} + (\sum_{w = 1}^{h} \lfloor \mu(w) \rfloor) + 3|C|]$ and $[q + |C| + 1, x^{h}_{s} + (\sum_{w = 1}^{h} \lfloor \mu(w) \rfloor) + 3|C|]$ because $[p + 2|C|, x^{h}_{s} + (\sum_{w = 1}^{h} \lfloor \mu(w) \rfloor) + 3|C|] \subseteq [q + |C| + 1, x^{h}_{s} + (\sum_{w = 1}^{h} \lfloor \mu(w) \rfloor) + 3|C|]$. 
Then, the lemma shows that $k^{\prime\prime} \geq h$ and 
$[s^{\prime\prime\prime h}, e^{\prime\prime\prime h}] \subseteq [s^{\prime\prime h}, e^{\prime\prime h}]$.

We prove $s_{\max} = q + |C|$ by contradiction. 
We assume that $s_{\max} \neq q + |C|$ holds. 
Then, $s_{\max} > q + |C|$ follows from 
$s_{\max} \neq q + |C|$ and $s_{\max} \geq q + |C|$. 
If $[q + |C| + 1, x^{h}_{s} + (\sum_{w = 1}^{h} \lfloor \mu(w) \rfloor) + 3|C|] \in \Delta(h, b+1)$, 
then $q + |C| + 1 \leq y^{h}_{s}$ follows from Corollary~\ref{cor:IA_exist_corollary}~\ref{enum:IA_exist_corollary:1}. 
Here, $y^{h}_{s} = x^{h}_{s} + |C| - 1$ holds. 
$q + 1 \leq x^{h}_{s} - 1$ follows from $q + |C| + 1 \leq y^{h}_{s}$ and $y^{h}_{s} = x^{h}_{s} + |C| - 1$. 
We can apply Proposition~\ref{prop:I_LAST_HR_proceeding_property}~\ref{enum:I_LAST_HR_proceeding_property:7} 
to the interval $[q + |C| + 1, x^{h}_{s} + (\sum_{w = 1}^{h} \lfloor \mu(w) \rfloor) + 3|C|]$ 
because 
$q + 1 \in [p-1, x^{h}_{s} - 1]$ and 
$x^{h}_{s} + (\sum_{w = 1}^{h} \lfloor \mu(w) \rfloor) + 2|C| \in [x^{h}_{s}, x^{h}_{s} + (\sum_{w = 1}^{h} \lfloor \mu(w) \rfloor) + 2|C|]$. 
The proposition shows that 
$[q + 1, x^{h}_{s} + (\sum_{w = 1}^{h} \lfloor \mu(w) \rfloor) + 2|C|] \in \Delta(h, b)$. 
On the other hand, $[q + 1, x^{h}_{s} + (\sum_{w = 1}^{h} \lfloor \mu(w) \rfloor) + 2|C|] \not \in \Delta(h, b)$ holds 
because $q+1 \not \in [p, q]$. 
The two facts $[q + 1, x^{h}_{s} + (\sum_{w = 1}^{h} \lfloor \mu(w) \rfloor) + 2|C|] \in \Delta(h, b)$ 
and $[q + 1, x^{h}_{s} + (\sum_{w = 1}^{h} \lfloor \mu(w) \rfloor) + 2|C|] \not \in \Delta(h, b)$ yield a contradiction. 

Otherwise (i.e., $[q + |C| + 1, x^{h}_{s} + (\sum_{w = 1}^{h} \lfloor \mu(w) \rfloor) + 3|C|] \not \in \Delta(h, b+1)$), 
$k^{\prime\prime} < h$ follows from Lemma~\ref{lem:interval_extension_propertyX}~\ref{enum:interval_extension_propertyX:low}. 
The two facts $k^{\prime\prime} \geq h$ and $k^{\prime\prime} < h$ yield a contradiction.
Therefore, $s_{\max} = q + |C|$ must hold. 

\textbf{Proof of Proposition~\ref{prop:I_LAST_HR_proceeding_RB}(i).} 
Proposition~\ref{prop:I_LAST_HR_proceeding_RB}(i) follows from statement (1), statement (2), 
statement (3), statement (4), and statement (7). 

\textbf{Proof of Proposition~\ref{prop:I_LAST_HR_proceeding_RB}(ii).} 
$q = p + |C| - 1$ follows from 
$q \leq p + |C| - 1$ (statement (5)), 
$q \geq s_{\min}-1$ (statement (1)), 
and $s_{\min} = p + |C|$ (statement (2)). 
\end{proof}

%%%%%%%%%%%%%%%%%%%%%%%%%%%%%%%%%%%%%%%%%%%%%%%%%%%%%%%%%%%%%%%%%%%%%%%%%%%%%%%%%%%%%%%%%%%%%%%%

The following proposition states properties of the two sets $\Delta(h, b)$ 
and $\Delta(h, b-1)$.

\begin{proposition}\label{prop:I_LAST_HR_back_property}
Consider an interval attractor $([p, q], [\ell, r])$ in subset $\Psi_{\run}$ 
and sequence $A(p, r) = [s^{0}, e^{0}], [s^{1}, e^{1}], \ldots, [s^{k}, e^{k}]$. 
Let $\gamma$ and $C$ be the attractor position and associated string of 
the interval attractor $([p, q], [\ell, r])$, respectively. 
Here, 
\begin{itemize}
    \item from Definition~\ref{def:RR_Delta} and Lemma~\ref{lem:rec_function_basic_relation}~\ref{enum:rec_function_basic_relation:1}, 
    there exists a pair of an integer $h \in [0, H-1]$ and a position $b \in [1, |S^{h}|]$ in sequence $S^{h}$ 
    satisfying 
    $p = \min \{ s \mid [s, e] \in \Delta(h, b) \}$, 
    $q = \max \{ s \mid [s, e] \in \Delta(h, b) \}$, 
    $\ell = \min \{ e \mid [s, e] \in \Delta(h, b) \}$, 
    and $r = \max \{ e \mid [s, e] \in \Delta(h, b) \}$; 
    \item $[p, r] \in \Delta(h, b)$ follows from Lemma~\ref{lem:IA_maximal_lemma}; 
    \item $h = k$ and $b = s^{k}$ follows from $[p, r] \in \Delta(h, b)$. 
\end{itemize}

Let $T[x^{h}_{s}..y^{h}_{s}]$ be the substring derived from the $s^{h}$-th nonterminal of sequence $S^{h}$. 
Then, the following three statements hold:
\begin{enumerate}[label=\textbf{(\roman*)}]
    \item \label{enum:I_LAST_HR_back_property:1} $[p-1 + |C|, x^{h}_{s} + 2\lfloor \mu(h+1) \rfloor + \sum_{w = 1}^{h+1} \lfloor \mu(w) \rfloor] \in \Delta(h, b)$;
    \item \label{enum:I_LAST_HR_back_property:3} $b \geq 2$ and $S^{h}[b-1] = S^{h}[b]$; 
    \item \label{enum:I_LAST_HR_back_property:5} $[p-1, x^{h}_{s} + 2 \lfloor \mu(h+1) \rfloor + (\sum_{w = 1}^{h+1} \lfloor \mu(w) \rfloor) - |C|] \in \Delta(h, b-1)$.
\end{enumerate}
\end{proposition}
\begin{proof}
The proof of Proposition~\ref{prop:I_LAST_HR_back_property} is as follows.

\textbf{Proof of Proposition~\ref{prop:I_LAST_HR_back_property}(i).} 
We prove $p - 1 + |C| \in [p, q]$ and $x^{h}_{s} + 2\lfloor \mu(h+1) \rfloor + \sum_{w = 1}^{h+1} \lfloor \mu(w) \rfloor \in [x^{h}_{s} + (\sum_{w = 1}^{h} \lfloor \mu(w) \rfloor) + 2|C| ,r]$. 
$q = p + |C| - 1$ follows from Proposition~\ref{prop:I_LAST_HR_proceeding_RB}. 
$p - 1 + |C| \in [p, q]$ follows from $q = p + |C| - 1$ and $|C| \geq 1$. 
$x^{h}_{s} + 2\lfloor \mu(h+1) \rfloor + \sum_{w = 1}^{h+1} \lfloor \mu(w) \rfloor \leq r$ 
because $r \geq x^{h}_{s} + \sum_{w = 1}^{h+3} \lfloor \mu(w) \rfloor$ (Proposition~\ref{prop:I_LAST_HR_proceeding_property}~\ref{enum:I_LAST_HR_proceeding_property:1}), 
and $2\lfloor \mu(h+1) \rfloor + \sum_{w = 1}^{h+1} \lfloor \mu(w) \rfloor \leq \sum_{w = 1}^{h+3} \lfloor \mu(w) \rfloor$. 
$x^{h}_{s} + 2\lfloor \mu(h+1) \rfloor + \sum_{w = 1}^{h+1} \lfloor \mu(w) \rfloor \geq x^{h}_{s} + (\sum_{w = 1}^{h} \lfloor \mu(w) \rfloor) + 2|C|$ because $|C| \leq \lfloor \mu(h+1) \rfloor$ follows from 
Proposition~\ref{prop:I_LAST_HR_proceeding_property}~\ref{enum:I_LAST_HR_proceeding_property:2}. 
Therefore, $x^{h}_{s} + 2\lfloor \mu(h+1) \rfloor + \sum_{w = 1}^{h+1} \lfloor \mu(w) \rfloor \in [x^{h}_{s} + (\sum_{w = 1}^{h} \lfloor \mu(w) \rfloor) + 2|C| ,r]$ holds. 

We prove $[p - 1 + |C|, x^{h}_{s} + 2\lfloor \mu(h+1) \rfloor + \sum_{w = 1}^{h+1} \lfloor \mu(w) \rfloor] \in \Delta(h, b)$. 
We already showed that $[q, x^{h}_{s} + (\sum_{w = 1}^{h} \lfloor \mu(w) \rfloor) + 2|C|] \in \Delta(h, b)$ holds 
in the proof of Proposition~\ref{prop:I_LAST_HR_proceeding_property}. 
We apply Lemma~\ref{lem:IA_maximal_lemma} to the interval $[q, x^{h}_{s} + (\sum_{w = 1}^{h} \lfloor \mu(w) \rfloor) + 2|C|]$. 
Then, the lemma shows that $[p - 1 + |C|, x^{h}_{s} + 2\lfloor \mu(h+1) \rfloor + \sum_{w = 1}^{h+1} \lfloor \mu(w) \rfloor] \in \Delta(h, b)$  
because $p - 1 + |C| \in [p, q]$ and $x^{h}_{s} + 2\lfloor \mu(h+1) \rfloor + \sum_{w = 1}^{h+1} \lfloor \mu(w) \rfloor \in [x^{h}_{s} + (\sum_{w = 1}^{h} \lfloor \mu(w) \rfloor) + 2|C|, r]$. 

\textbf{Proof of Proposition~\ref{prop:I_LAST_HR_back_property}(ii).} 
Consider two sequences $A(p-1, x^{h}_{s} + 2 \lfloor \mu(h+1) \rfloor + (\sum_{w = 1}^{h+1} \lfloor \mu(w) \rfloor) - |C|) = [s^{\prime 0}, e^{\prime 0}], [s^{\prime 1}, e^{\prime 1}], \ldots, [s^{\prime k^{\prime}}, e^{\prime k^{\prime}}]$ 
and $A(p-1+|C|, x^{h}_{s} + 2 \lfloor \mu(h+1) \rfloor + \sum_{w = 1}^{h+1} \lfloor \mu(w) \rfloor) = [s^{\prime\prime 0}, e^{\prime\prime 0}], [s^{\prime\prime 1}, e^{\prime\prime 1}], \ldots, [s^{\prime\prime k^{\prime\prime}}, e^{\prime\prime k^{\prime\prime}}]$. 
Here, $k^{\prime\prime} = h$ and $s^{\prime\prime h} = b$ follow from $[p-1+|C|, x^{h}_{s} + 2\lfloor \mu(h+1) \rfloor + \sum_{w = 1}^{h+1} \lfloor \mu(w) \rfloor] \in \Delta(h, b)$.
$T[p-1..x^{h}_{s} + 2 \lfloor \mu(h+1) \rfloor + (\sum_{w = 1}^{h+1} \lfloor \mu(w) \rfloor) - |C|] = T[p-1+|C|..x^{h}_{s} + 2 \lfloor \mu(h+1) \rfloor + (\sum_{w = 1}^{h+1} \lfloor \mu(w) \rfloor)]$ follows from 
Proposition~\ref{prop:I_LAST_HR_proceeding_property}~\ref{enum:I_LAST_HR_proceeding_property:4} 
and $|C| \leq \lfloor \mu(h+1) \rfloor$. 
Since $T[p-1..x^{h}_{s} + 2 \lfloor \mu(h+1) \rfloor + (\sum_{w = 1}^{h+1} \lfloor \mu(w) \rfloor) - |C|] = T[p -1 + |C|..x^{h}_{s} + 2 \lfloor \mu(h+1) \rfloor + \sum_{w = 1}^{h+1} \lfloor \mu(w) \rfloor]$, 
we can apply Corollary~\ref{cor:capture_gamma_corollary} to the two intervals $[p-1, x^{h}_{s} + 2 \lfloor \mu(h+1) \rfloor + (\sum_{w = 1}^{h+1} \lfloor \mu(w) \rfloor) - |C|]$ and $[p-1+|C|, x^{h}_{s} + 2 \lfloor \mu(h+1) \rfloor + \sum_{w = 1}^{h+1} \lfloor \mu(w) \rfloor]$. 
Corollary~\ref{cor:capture_gamma_corollary} shows that $k^{\prime} = h$ and $|[p-1, x^{\prime h}_{s}-1]| = |[p - 1 + |C|, x^{h}_{s} - 1]|$
for the substring $T[x^{\prime h}_{s}..y^{\prime h}_{s}]$ derived from the $s^{\prime h}$-th nonterminal of sequence $S^{h}$. 
$x^{\prime h}_{s} = x^{h}_{s} - |C|$ follows from $|[p-1, x^{\prime h}_{s}-1]| = |[p - 1 + |C|, x^{h}_{s} - 1]|$. 
This equation $x^{\prime h}_{s} = x^{h}_{s} - |C|$ indicates that $1 \leq s^{\prime h} < s^{h}$ holds. 
Therefore, $b \geq 2$ follows from $1 \leq s^{\prime h} < s^{h}$ and $s^{h} = b$. 

We prove $T[x^{\prime h}_{s}..x^{h}_{s} + 2 \lfloor \mu(h+1) \rfloor + (\sum_{w = 1}^{h+1} \lfloor \mu(w) \rfloor) - |C|] = T[x^{h}_{s}..x^{h}_{s} + 2\lfloor \mu(h+1) \rfloor + \sum_{w = 1}^{h+1} \lfloor \mu(w) \rfloor]$. 
Let $d = (x^{h}_{s} - 1 - |C|) - (p-1)$. 
$x^{h}_{s} - p \geq |C|$ because 
$q = p + |C| - 1$ (Proposition~\ref{prop:I_LAST_HR_proceeding_property})
and $q \leq x^{h}_{s} - 1$ (Corollary~\ref{cor:IA_exist_corollary}~\ref{enum:IA_exist_corollary:1}). 
$d \geq 0$ follows from $d = (x^{h}_{s} - 1 - |C|) - (p-1)$ and $x^{h}_{s} - 1 - |C| \geq p - 1$. 
$T[p + d..x^{h}_{s} + 2 \lfloor \mu(h+1) \rfloor + (\sum_{w = 1}^{h+1} \lfloor \mu(w) \rfloor) - |C|] = T[p + |C| + d..x^{h}_{s} + 2 \lfloor \mu(h+1) \rfloor + (\sum_{w = 1}^{h+1} \lfloor \mu(w) \rfloor)]$ 
follows from $T[p-1..x^{h}_{s} + 2 \lfloor \mu(h+1) \rfloor + (\sum_{w = 1}^{h+1} \lfloor \mu(w) \rfloor) - |C|] = T[p-1 + |C|..x^{h}_{s} + 2 \lfloor \mu(h+1) \rfloor + (\sum_{w = 1}^{h+1} \lfloor \mu(w) \rfloor)]$ and $d \geq 0$. 
$p + d = x^{\prime h}_{s}$ and $p + |C| + d = x^{h}_{s}$ 
follow from $d = (x^{h}_{s} - 1 - |C|) - (p-1)$ and $x^{\prime h}_{s} = x^{h}_{s} - |C|$. 
Therefore, $T[x^{\prime h}_{s}..x^{h}_{s} + 2 \lfloor \mu(h+1) \rfloor + (\sum_{w = 1}^{h+1} \lfloor \mu(w) \rfloor) - |C|] = T[x^{h}_{s}..x^{h}_{s} + 2\lfloor \mu(h+1) \rfloor + \sum_{w = 1}^{h+1} \lfloor \mu(w) \rfloor]$ follows from 
(A) $T[p + d..x^{h}_{s} + 2 \lfloor \mu(h+1) \rfloor + (\sum_{w = 1}^{h+1} \lfloor \mu(w) \rfloor) - |C|] = T[p + |C| + d..x^{h}_{s} + 2 \lfloor \mu(h+1) \rfloor + (\sum_{w = 1}^{h+1} \lfloor \mu(w) \rfloor)]$, 
(B) $p + d = x^{\prime h}_{s}$, 
and (C) $p + |C| + d = x^{h}_{s}$. 

We prove $S^{h}[s^{\prime h}] = S^{h}[b]$.
Let $\alpha = \min \{ |\val(S^{h}[b])|, |\val(S^{h}[s^{\prime h}])| \}$. 
Then, $\alpha \leq |C|$ because $\val(S^{h}[b]) = C$. 
$T[x^{\prime h}_{s}..x^{\prime h}_{s} + \alpha + \sum_{w = 1}^{h} \lfloor \mu(w) \rfloor)] = T[x^{h}_{s}..x^{h}_{s} + \alpha + \sum_{w = 1}^{h} \lfloor \mu(w) \rfloor]$ follows from 
(a) $\alpha \leq |C|$, (b) $|C| \leq \lfloor \mu(h+1) \rfloor$, 
and (c) $T[x^{\prime h}_{s}..x^{h}_{s} + 2 \lfloor \mu(h+1) \rfloor + (\sum_{w = 1}^{h+1} \lfloor \mu(w) \rfloor) - |C|] = T[x^{h}_{s}..x^{h}_{s} + 2\lfloor \mu(h+1) \rfloor + \sum_{w = 1}^{h+1} \lfloor \mu(w) \rfloor]$. 
Since $T[x^{\prime h}_{s}..x^{\prime h}_{s} + \alpha + \sum_{w = 1}^{h} \lfloor \mu(w) \rfloor] = T[x^{h}_{s}..x^{h}_{s} + \alpha + \sum_{w = 1}^{h} \lfloor \mu(w) \rfloor]$, 
we can apply Lemma~\ref{lem:rr_property}~\ref{enum:rr_property:right} to 
the two nonterminals $S^{h}[s^{\prime h}]$ and $S^{h}[b]$. 
Then, $S^{h}[s^{\prime h}] = S^{h}[b]$ follows from Lemma~\ref{lem:rr_property}~\ref{enum:rr_property:right}. 

We prove $s^{\prime h} = b - 1$. 
$s^{\prime h} = b - 1$ holds if $y^{\prime h}_{s} = x^{h}_{s} - 1$. 
$y^{\prime h}_{s} = x^{\prime h}_{s} + |C| - 1$ follows from 
$S^{h}[s^{\prime h}] = S^{h}[b]$ and $\val(S^{h}[b]) = C$. 
$y^{\prime h}_{s} = x^{h}_{s} - 1$ follows from 
$y^{\prime h}_{s} = x^{\prime h}_{s} + |C| - 1$ and $x^{\prime h}_{s} = x^{h}_{s} - |C|$.
Therefore, $s^{\prime h} = b - 1$ holds. 

Finally, $S^{h}[b-1] = S^{h}[b]$ follows from $S^{h}[s^{\prime h}] = S^{h}[b]$ and $s^{\prime h} = b - 1$.

\textbf{Proof of Proposition~\ref{prop:I_LAST_HR_back_property}(iii).} 
We showed that $[p-1, x^{h}_{s} + 2 \lfloor \mu(h+1) \rfloor + (\sum_{w = 1}^{h+1} \lfloor \mu(w) \rfloor) - |C|] \in \Delta(h, s^{\prime h})$ and $s^{\prime h} = b-1$ in the proof of Proposition~\ref{prop:I_LAST_HR_back_property}(ii). 
Therefore, $[p-1, x^{h}_{s} + 2 \lfloor \mu(h+1) \rfloor + (\sum_{w = 1}^{h+1} \lfloor \mu(w) \rfloor) - |C|] \in \Delta(h, b-1)$ holds. 

\end{proof}

The following proposition states the relationship among 
the eight integers $p$, $q$, $\ell$, $r$, 
$\min \{ s \mid [s, e] \in \Delta(h, b-1) \}$, 
$\max \{ s \mid [s, e] \in \Delta(h, b-1) \}$, 
$\min \{ e \mid [s, e] \in \Delta(h, b-1) \}$, 
and $\max \{ e \mid [s, e] \in \Delta(h, b-1) \}$. 

%%%%%%%%%%%%%%%%%%%%%%%%%%%%%%%%%%%%%%%%%%%%%%%%%%%%%%%%%%%%%%%%%%%%%%%%%%%%%%%%%%%%%%%%%%%%%%%%

\begin{proposition}\label{prop:I_LAST_HR_back_RB}
Consider an interval attractor $([p, q], [\ell, r])$ in subset $\Psi_{\run}$ 
and sequence $A(p, r) = [s^{0}, e^{0}], [s^{1}, e^{1}], \ldots, [s^{k}, e^{k}]$. 
Let $\gamma$ and $C$ be the attractor position and associated string of 
the interval attractor $([p, q], [\ell, r])$, respectively. 
Here, 
\begin{itemize}
    \item from Definition~\ref{def:RR_Delta} and Lemma~\ref{lem:rec_function_basic_relation}~\ref{enum:rec_function_basic_relation:1}, 
    there exists a pair of an integer $h \in [0, H-1]$ and a position $b \in [1, |S^{h}|]$ in sequence $S^{h}$ 
    satisfying 
    $p = \min \{ s \mid [s, e] \in \Delta(h, b) \}$, 
    $q = \max \{ s \mid [s, e] \in \Delta(h, b) \}$, 
    $\ell = \min \{ e \mid [s, e] \in \Delta(h, b) \}$, 
    and $r = \max \{ e \mid [s, e] \in \Delta(h, b) \}$; 
    \item $[p, r] \in \Delta(h, b)$ follows from Lemma~\ref{lem:IA_maximal_lemma}; 
    \item $h = k$ and $b = s^{k}$ follows from $[p, r] \in \Delta(h, b)$; 
    \item $\Delta(h, b-1) \neq \emptyset$ follows from Proposition~\ref{prop:I_LAST_HR_back_property}~\ref{enum:I_LAST_HR_back_property:5}.
\end{itemize}
Let $s_{\min} = \min \{ s \mid [s, e] \in \Delta(h, b-1) \}$, 
$s_{\max} = \max \{ s \mid [s, e] \in \Delta(h, b-1) \}$,  
$e_{\min} = \min \{ e \mid [s, e] \in \Delta(h, b-1) \}$, and 
$e_{\max} = \max \{ e \mid [s, e] \in \Delta(h, b-1) \}$. 
Then, 
$s_{\min} < p$, 
$s_{\max} = p-1$, 
$e_{\min} \leq \ell$, 
and $e_{\max} = r$.
\end{proposition}
\begin{proof}
Consider sequence $A(s_{\min}, e_{\max}) = [s^{\prime 0}, e^{\prime 0}], [s^{\prime 1}, e^{\prime 1}], \ldots, [s^{\prime k^{\prime}}, e^{\prime k^{\prime}}]$. 
$k^{\prime} = h$ and $s^{\prime h} = b-1$ follows from the definition of set $\Delta(h, b-1)$ 
because $[s_{\min}, e_{\max}] \in \Delta(h, b-1)$ follows from Lemma~\ref{lem:IA_maximal_lemma}. 

Let $T[x^{h}_{s}..y^{h}_{s}]$ and $T[x^{\prime h}_{s}..y^{\prime h}_{s}]$ 
be the two substrings derived from the $s^{h}$-th and $s^{\prime h}$-th nonterminals of sequence $S^{h}$ (i.e., the $b$-th and $(b-1)$-th nonterminals of sequence $S^{h}$), respectively. 
The following five statements are used to prove Proposition~\ref{prop:I_LAST_HR_back_RB}. 
\begin{enumerate}[label=\textbf{(\arabic*)}]
    \item $s_{\min} < p$; 
    \item $|[s^{\prime h}, e^{\prime h}]| \geq 4$;
    \item $e_{\max} = r$;
    \item $s_{\max} = p-1$;
    \item $e_{\min} \leq \ell$.
\end{enumerate}

\textbf{Proof of statement (1).} 
$[p-1, x^{h}_{s} + 2 \lfloor \mu(h+1) \rfloor + (\sum_{w = 1}^{h+1} \lfloor \mu(w) \rfloor) - |C|] \in \Delta(h, b-1)$ 
follows from Proposition~\ref{prop:I_LAST_HR_back_property}~\ref{enum:I_LAST_HR_back_property:5}.
$s_{\min} \leq p-1$ follows from $[p-1, x^{h}_{s} + 2 \lfloor \mu(h+1) \rfloor + (\sum_{w = 1}^{h+1} \lfloor \mu(w) \rfloor) - |C|] \in \Delta(h, b-1)$ and $s_{\min} = \min \{ s \mid [s, e] \in \Delta(h, b-1) \}$. 
Therefore, $s_{\min} < p$ holds. 

\textbf{Proof of statement (2).} 
We prove $|[x^{\prime h}_{s}, e_{\max}]| > 2 \lfloor \mu(h+1) \rfloor + \sum_{w = 1}^{h+1} \lfloor \mu(w) \rfloor$.  
Here, $x^{\prime h}_{s} = x^{\prime}_{s} - |C|$ holds 
because $s^{\prime h} = b-1$, $s^{h} = b$, 
$C = \val(S^{h}[b])$ (Proposition~\ref{prop:I_LAST_HR_proceeding_property}~\ref{enum:I_LAST_HR_proceeding_property:1}), 
and $S^{h}[b-1] = S^{h}[b]$ (Proposition~\ref{prop:I_LAST_HR_back_property}~\ref{enum:I_LAST_HR_back_property:3}).
$e_{\max} \geq x^{h}_{s} + 2 \lfloor \mu(h+1) \rfloor + (\sum_{w = 1}^{h+1} \lfloor \mu(w) \rfloor) - |C|$ follows from 
$[p-1, x^{h}_{s} + 2 \lfloor \mu(h+1) \rfloor + (\sum_{w = 1}^{h+1} \lfloor \mu(w) \rfloor) - |C|] \in \Delta(h, b-1)$ 
and $e_{\max} = \max \{ e \mid [s, e] \in \Delta(h, b-1) \}$. 
Therefore, 
$|[x^{\prime h}_{s}, e_{\max}]| > 2 \lfloor \mu(h+1) \rfloor + \sum_{w = 1}^{h+1} \lfloor \mu(w) \rfloor$ follows from 
(A) $|[x^{\prime h}_{s}, e_{\max}]| = e_{\max} - x^{\prime h}_{s} + 1$, 
(B) $e_{\max} \geq x^{h}_{s} + 2 \lfloor \mu(h+1) \rfloor + (\sum_{w = 1}^{h+1} \lfloor \mu(w) \rfloor) - |C|$, 
and (C) $x^{\prime h}_{s} = x^{h}_{s} - |C|$. 

We prove $|[s^{\prime h}, e^{\prime h}]| \geq 4$. 
We can apply Lemma~\ref{lem:f_rec_top_property}~\ref{enum:f_rec_top_property:3} to the interval $[s_{\min}, e_{\max}]$ 
because $|[x^{\prime h}_{s}, e_{\max}]| > 2 \lfloor \mu(h+1) \rfloor + \sum_{w = 1}^{h+1} \lfloor \mu(w) \rfloor$. 
The lemma shows that $|[s^{\prime h}, e^{\prime h}]| \geq 4$. 

\textbf{Proof of statement (3).} 
Since $|[s^{\prime h}, e^{\prime h}]| \geq 4$, 
we can apply Proposition~\ref{prop:adjacent_last_set} to 
the two sets $\Delta(h, b-1)$ and $\Delta(h, b)$. 
Then, $e_{\max} = r$ follows from Proposition~\ref{prop:adjacent_last_set}(ii). 

\textbf{Proof of statement (4).} 
We prove $s_{\max} = p-1$ by contradiction. 
We assume that $s_{\max} \neq p-1$ holds. 
$s_{\max} \geq p-1$ follows from 
$[p-1, x^{h}_{s} + 2 \lfloor \mu(h+1) \rfloor + (\sum_{w = 1}^{h+1} \lfloor \mu(w) \rfloor) - |C|] \in \Delta(h, b-1)$. 
$s_{\max} > p-1$ follows from $s_{\max} \neq p-1$ and $s_{\max} \geq p-1$. 

$[p, e_{\max}] \in \Delta(h, b-1)$ follows from Lemma~\ref{lem:IA_maximal_lemma} 
because $p \in [s_{\min}, s_{\max}]$. 
$[p, r] \in \Delta(h, b)$ follows from 
Lemma~\ref{lem:IA_maximal_lemma}. 
$\Delta(h, b-1) \cap \Delta(h, b) \neq \emptyset$ 
follows from $[p, e_{\max}] \in \Delta(h, b-1)$, 
$[p, r] \in \Delta(h, b)$, and $r = e_{\max}$. 

On the other hand, $\Delta(h, b-1) \cap \Delta(h, b) = \emptyset$ follows from 
the definition of sets $\Delta(h, b-1)$ and $\Delta(h, b)$.
The two facts $\Delta(h, b-1) \cap \Delta(h, b) \neq \emptyset$ and 
$\Delta(h, b-1) \cap \Delta(h, b) = \emptyset$ yield a contradiction. 
Therefore, $s_{\max} = p-1$ must hold. 

\textbf{Proof of statement (5).} 
Consider three sequences $A(p-1, \ell) = [s^{\prime\prime 0}, e^{\prime\prime 0}], [s^{\prime\prime 1}, e^{\prime\prime 1}]$, $\ldots$, $[s^{\prime\prime k^{\prime\prime}}, e^{\prime\prime k^{\prime\prime}}]$, 
$A(p, \ell) = [s^{\prime\prime\prime 0}, e^{\prime\prime\prime 0}]$, $[s^{\prime\prime\prime 1}, e^{\prime\prime\prime 1}]$, $\ldots$, $[s^{\prime\prime\prime k^{\prime\prime\prime}}, e^{\prime\prime\prime k^{\prime\prime\prime}}]$, 
and 
$A(p-1, r) = [s^{\prime\prime\prime\prime 0}, e^{\prime\prime\prime\prime 0}]$, $[s^{\prime\prime\prime\prime 1}, e^{\prime\prime\prime\prime 1}]$, $\ldots$, $[s^{\prime\prime\prime\prime k^{\prime\prime\prime\prime}}, e^{\prime\prime\prime\prime k^{\prime\prime\prime\prime}}]$. 
Here, $k^{\prime\prime\prime} = h$ and $s^{\prime\prime\prime h} = b$ follow from the definition of set $\Delta(h, b)$ 
because $[p, \ell] \in \Delta(h, b)$ follows from Lemma~\ref{lem:IA_maximal_lemma}. 
We apply Lemma~\ref{lem:intv_function_shift}~\ref{enum:intv_function_shift:7} to the two intervals $[p-1, \ell]$ and $[p, \ell]$. 
Then, the lemma shows that $k^{\prime\prime} \geq h$.

We prove $k^{\prime\prime} = h$ by contradiction. 
We assume that $k^{\prime\prime} \neq h$ holds. 
Then, $k^{\prime\prime} > h$ follows from $k^{\prime\prime} \neq h$ and $k^{\prime\prime} \geq h$. 
We apply Lemma~\ref{lem:intv_function_shift}~\ref{enum:intv_function_shift:7} to the two intervals $[p-1, \ell]$ and $[p-1, r]$. 
Then, the lemma shows that $k^{\prime\prime} \leq k^{\prime\prime\prime\prime}$ 
because $[p-1, \ell] \subseteq [p-1, r]$. 
$k^{\prime\prime\prime\prime} \geq h+1$ follows from $k^{\prime\prime} \leq k^{\prime\prime\prime\prime}$ and $k^{\prime\prime} > h$. 

On the other hand, $[p-1, r] \in \Delta(h, b-1)$ follows from Lemma~\ref{lem:IA_maximal_lemma} 
because $s_{\max} = p-1$ and $e_{\max} = r$. 
$k^{\prime\prime\prime\prime} = h$ follows from $[p-1, r] \in \Delta(h, b-1)$. 
The two facts $k^{\prime\prime\prime\prime} \geq h+1$ and $k^{\prime\prime\prime\prime} = h$ yield a contradiction. 
Therefore, $k^{\prime\prime} = h$ must hold. 

We prove $s^{\prime\prime h} = b-1$. 
We apply Lemma~\ref{lem:intv_function_shift}~\ref{enum:intv_function_shift:3} to the two intervals $[p-1, \ell]$ 
and $[p-1, r]$. 
Then, the lemma shows that $s^{\prime\prime h} = s^{\prime\prime\prime\prime h}$.
$s^{\prime\prime\prime\prime h} = b-1$ follows from $[p-1, r] \in \Delta(h, b-1)$. 
Therefore, $s^{\prime\prime h} = b-1$ holds. 

We prove $e_{\min} \leq \ell$. 
$[p-1, \ell] \in \Delta(h, b-1)$ holds 
because $k^{\prime\prime} = h$, and $s^{\prime\prime h} = b-1$. 
$e_{\min} \leq \ell$ follows from $[p-1, \ell] \in \Delta(h, b-1)$ 
and $e_{\min} = \min \{ e \mid [s, e] \in \Delta(h, b-1) \}$. 

\textbf{Proof of Proposition~\ref{prop:I_LAST_HR_back_RB}.}
Proposition~\ref{prop:I_LAST_HR_back_RB} follows from 
statement (1), statement (3), statement (4), and statement (5). 
\end{proof}

We prove Lemma~\ref{lem:psi_run_basic_property} using Proposition~\ref{prop:I_LAST_HR_proceeding_RB} 
and Proposition~\ref{prop:I_LAST_HR_back_RB}. 
From Definition~\ref{def:RR_Delta}, 
there exists a position $b \in [1, |S^{h}|]$ in sequence $S^{h}$ 
satisfying 
$p = \min \{ s \mid [s, e] \in \Delta(h, b) \}$, 
$q = \max \{ s \mid [s, e] \in \Delta(h, b) \}$, 
$\ell = \min \{ e \mid [s, e] \in \Delta(h, b) \}$, 
and $r = \max \{ e \mid [s, e] \in \Delta(h, b) \}$. 

Consider sequence $A(p, r) = [s^{0}, e^{0}], [s^{1}, e^{1}], \ldots, [s^{k}, e^{k}]$. 
Here, $h = k$ and $b = s^{k}$ follows from the definition of the set $\Delta(h, b)$ 
because $[p, r] \in \Delta(h, b)$ follows from Lemma~\ref{lem:IA_maximal_lemma}. 
Let $T[x^{h}_{s}..y^{h}_{s}]$ be the substring derived from 
the $s^{h}$-th nonterminal of sequence $S^{h}$. 
Then, $\gamma = x^{h}_{s}$ follows from the definition of the attractor position.  

%%%%%%%%%%%%%%%%%%%%%%%%%%%%%%%%%%%%%%%%%%%%%%%%%%%%%%%%%%%%%%%%%%%%%%%%%%%%%%%%%%%%%%%%%%%%%%%%

\begin{proof}[Proof of Lemma~\ref{lem:psi_run_basic_property}~\ref{enum:psi_run_basic_property:1}]
$|C| \leq \lfloor \mu(h+1) \rfloor$ follows from Proposition~\ref{prop:I_LAST_HR_proceeding_property}~\ref{enum:I_LAST_HR_proceeding_property:2}. 
$\ell \leq \gamma + \sum_{w = 1}^{h} \lfloor \mu(w) \rfloor$ follows from 
$\ell \leq x^{h}_{s} +  \sum_{w = 1}^{h} \lfloor \mu(w) \rfloor$ (Proposition~\ref{prop:I_LAST_HR_proceeding_property}~\ref{enum:I_LAST_HR_proceeding_property:3}) and $\gamma = x^{h}_{s}$. 
$|[\gamma, r]| > 1 + \sum_{w = 1}^{h+3} \lfloor \mu(w) \rfloor$ follows from 
$|[x^{h}_{s}, r]| > 1 + \sum_{w = 1}^{h+3} \lfloor \mu(w) \rfloor$ (Proposition~\ref{prop:I_LAST_HR_proceeding_property}~\ref{enum:I_LAST_HR_proceeding_property:1}) and $\gamma = x^{h}_{s}$. 
$|[p, q]| = |C|$ follows from Proposition~\ref{prop:I_LAST_HR_proceeding_RB}(ii). 
$|[p, \gamma-1]| \geq |C| - 1$ holds because 
$|[p, q]| = |C|$, 
and $|[p, \gamma]| \geq |[p, q]|$ follows from Lemma~\ref{lem:IA_super_basic_property}~\ref{enum:IA_super_basic_property:1}.  

We prove $p \geq 2$ by contradiction. 
We assume that $p < 2$ holds. 
Then, $p = 1$ follows from $p \in [1, n]$ and $p < 2$. 
$\lcs(T[p-1..\gamma-1], C^{n+1}) = T[p-1..\gamma-1]$ follows from the definition of the subset $\Psi_{\run}$. 
Since $\lcs(T[p-1..\gamma-1], C^{n+1}) = T[p-1..\gamma-1]$, 
$T[0]$ is a character of the associated string $C$. 
$T[0] \neq \$$ because 
the associated string $C$ does not contain character $\$$. 
On the other hand, 
$T[0]$ is defined as character $\$$. 
The two facts $T[0] \neq \$$ and $T[0] = \$$ yield a contradiction. 
Therefore, $p \geq 2$ must hold.
\end{proof}

%%%%%%%%%%%%%%%%%%%%%%%%%%%%%%%%%%%%%%%%%%%%%%%%%%%%%%%%%
\begin{proof}[Proof of Lemma~\ref{lem:psi_run_basic_property}~\ref{enum:psi_run_basic_property:2}]
We prove $([p + |C|, q + |C|], [\ell + |C|, r]) \in \Psi_{h}$. 
Let $p_{1} = \min \{ s \mid [s, e] \in \Delta(h, b+1) \}$, 
$q_{1} = \max \{ s \mid [s, e] \in \Delta(h, b+1) \}$, 
$\ell_{1} = \min \{ e \mid [s, e] \in \Delta(h, b+1) \}$, 
and $r_{1} = \max \{ e \mid [s, e] \in \Delta(h, b+1) \}$. 
Then, $([p_{1}, q_{1}], [\ell_{1}, r_{1}]) \in \Psi_{h}$ follows from the definition of interval attractor. 
Proposition~\ref{prop:I_LAST_HR_proceeding_RB} shows that 
$p_{1} = p + |C|$, 
$q_{1} = q + |C|$, 
$\ell_{1} = \ell + |C|$, 
and $r_{1} = r$. 
Therefore, $([p + |C|, q + |C|], [\ell + |C|, r]) \in \Psi_{h}$. 

We prove $\gamma_{1} = \gamma + |C|$ and $|[\gamma_{1}, r]| > \sum_{w = 1}^{h+1} \lfloor \mu(w) \rfloor$ for the attractor position $\gamma_{1}$ of the interval attractor $([p + |C|, q + |C|], [\ell + |C|, r])$. 
Here, $\gamma_{1} = x^{h}_{s} + |\val(S^{h}[b])|$ holds. 
$|C| = |\val(S^{h}[b])|$ follows from Proposition~\ref{prop:I_LAST_HR_proceeding_property}~\ref{enum:I_LAST_HR_proceeding_property:1}. 
$\gamma_{1} = \gamma + |C|$ follows from 
$\gamma = x^{h}_{s}$, $\gamma_{1} = x^{h}_{s} + |\val(S^{h}[b])|$, and $|C| = |\val(S^{h}[b])|$. 
$|[\gamma_{1}, r]| > \sum_{w = 1}^{h+1} \lfloor \mu(w) \rfloor$ follows from 
$|[x^{h}_{s}, r]| > 1 + \sum_{w = 1}^{h+3} \lfloor \mu(w) \rfloor$, $\gamma_{1} = \gamma + |C|$, and $|C| \leq \lfloor \mu(h+1) \rfloor$.

We prove $([p + |C|, q + |C|], [\ell + |C|, r]) \in \Psi_{\centerset}(C)$. 
Let $C_{1}$ be the associated string of the interval attractor $([p_{1}, q_{1}], [\ell_{1}, r_{1}])$. 
Because of $|[\gamma_{1}, r]| > \sum_{w = 1}^{h+1} \lfloor \mu(w) \rfloor$, 
the associated string $C_{1}$ is defined as the string $\val(S^{h}[b+1])$ derived from the $(b+1)$-th nonterminal of sequence $S^{h}$. 
$S^{h}[b] = S^{h}[b+1]$ follows from Proposition~\ref{prop:I_LAST_HR_proceeding_property}~\ref{enum:I_LAST_HR_proceeding_property:2}. 
$C_{1} = C$ follows from 
$C_{1} = \val(S^{h}[b+1])$, $\val(S^{h}[b]) = \val(S^{h}[b+1])$, and $C = \val(S^{h}[b])$. 
Therefore, $([p + |C|, q + |C|], [\ell + |C|, r]) \in \Psi_{\centerset}(C)$ follows from $C_{1} = C$. 

We prove $K > |C|$. 
Because of $([p, q], [\ell, r]) \in \Psi_{\lcp}(K)$, 
$K = |\lcp(T[\gamma..r], C^{n+1})|$ follows from the definition of the subset $\Psi_{\lcp}(K)$. 
$|\lcp(T[\gamma..r], C^{n+1})| > 1 + \sum_{w = 1}^{h+3} \lfloor \mu(w) \rfloor$ follows from the definition of the subset $\Psi_{\run}$. 
On the other hand, $|C| \leq \lfloor \mu(h+1) \rfloor$ follows from Lemma~\ref{lem:psi_run_basic_property}~\ref{enum:psi_run_basic_property:1}. 
Therefore, 
\begin{equation*}
    \begin{split}
        K &> 1 + \sum_{w = 1}^{h+3} \lfloor \mu(w) \rfloor \\
        &\geq \lfloor \mu(h+1) \rfloor \\
        &\geq |C|.
    \end{split}    
\end{equation*}

We prove $([p + |C|, q + |C|], [\ell + |C|, r]) \in \Psi_{\lcp}(K - |C|)$. 
Let $K_{1} \geq 0$ be an integer satisfying $([p + |C|, q + |C|], [\ell + |C|, r]) \in \Psi_{\lcp}(K_{1})$. 
Then, $K_{1} = |\lcp(T[\gamma_{1}..r], C^{n+1})|$ follows from the definition of the subset $\Psi_{\lcp}(K_{1})$. 
$|\lcp(T[\gamma_{1}..r], C^{n+1})| = |\lcp(T[\gamma + |C|..r], C^{n+1})|$ follows from $\gamma_{1} = \gamma + |C|$. 
$|\lcp(T[\gamma + |C|..r], C^{n+1})| = |\lcp(T[\gamma..r], C^{n+1})| - |C|$ 
because $|\lcp(T[\gamma..r], C^{n+1})| = K$ and $K > |C|$. 
$K_{1} = K - |C|$ follows from 
$K_{1} = |\lcp(T[\gamma..r], C^{n+1})| - |C|$ and $|\lcp(T[\gamma..r], C^{n+1})| = K$. 
Therefore, $([p + |C|, q + |C|], [\ell + |C|, r]) \in \Psi_{\lcp}(K - |C|)$ holds. 
\end{proof}

%%%%%%%%%%%%%%%%%%%%%%%%%%%%%%%%%%%%%%%%%%%%%%%%%%%%%%%%%
\begin{proof}[Proof of Lemma~\ref{lem:psi_run_basic_property}~\ref{enum:psi_run_basic_property:3}]
Let $p_{2} = \min \{ s \mid [s, e] \in \Delta(h, b-1) \}$, 
$q_{2} = \max \{ s \mid [s, e] \in \Delta(h, b-1) \}$, 
$\ell_{2} = \min \{ e \mid [s, e] \in \Delta(h, b-1) \}$, 
and $r_{2} = \max \{ e \mid [s, e] \in \Delta(h, b-1) \}$. 
Then, $([p_{2}, q_{2}], [\ell_{2}, r_{2}]) \in \Psi_{h}$ follows from the definition of interval attractor. 
$p_{2} < p$, $q_{2} = p-1$, $\ell_{2} \leq \ell$, and $r_{2} = r$ follow from Proposition~\ref{prop:I_LAST_HR_back_RB}. 

We prove $\gamma_{2} = \gamma - |C|$ for the attractor position $\gamma_{2}$ of the interval attractor $([p_{2}, q_{2}], [\ell_{2}, r_{2}])$. 
Here, $\gamma_{2} = x^{h}_{s} - |\val(S^{h}[b-1])|$ follows from the definition of the attractor position. 
$|\val(S^{h}[b-1])| = |C|$ follows from 
$\val(S^{h}[b]) = C$ (Proposition~\ref{prop:I_LAST_HR_proceeding_property}~\ref{enum:I_LAST_HR_proceeding_property:1}) 
and $S^{h}[b-1] = S^{h}[b]$ (Proposition~\ref{prop:I_LAST_HR_back_property}~\ref{enum:I_LAST_HR_back_property:3}). 
$\gamma_{2} = \gamma - |C|$ follows from 
$\gamma = x^{h}_{s}$, $\gamma_{2} = x^{h}_{s} - |\val(S^{h}[b-1])|$, and $|\val(S^{h}[b-1])| = |C|$. 

$|[\gamma_{2}, r_{2}]| > \sum_{w = 1}^{h+1} \lfloor \mu(w) \rfloor$ follows from the following equation:
Therefore, 
\begin{equation*}
    \begin{split}
        |[\gamma_{2}, r_{2}]| &= |[\gamma_{2}, r]| \\
        &\geq |[\gamma, r]| \\
        &> \sum_{w = 1}^{h+3} \lfloor \mu(w) \rfloor \\
        &\geq \sum_{w = 1}^{h+1} \lfloor \mu(w) \rfloor.
    \end{split}    
\end{equation*}

We prove $([p_{2}, q_{2}], [\ell_{2}, r_{2}]) \in \Psi_{\centerset}(C)$. 
Let $C_{2}$ be the associated string of the interval attractor $([p_{2}, q_{2}], [\ell_{2}, r_{2}])$. 
Since $|[\gamma_{2}, r_{2}]| > \sum_{w = 1}^{h+1} \lfloor \mu(w) \rfloor$, 
the associated string $C_{2}$ is defined as the string $\val(S^{h}[b-1])$ derived from the $(b-1)$-th nonterminal of sequence $S^{h}$. 
$C_{2} = C$ follows from 
$C_{2} = \val(S^{h}[b-1])$, $\val(S^{h}[b-1]) = \val(S^{h}[b])$, and $C = \val(S^{h}[b])$. 
Therefore, $([p_{2}, q_{2}], [\ell_{2}, r_{2}]) \in \Psi_{\centerset}(C)$ follows from $C_{2} = C$. 

We prove $([p_{2}, q_{2}], [\ell_{2}, r_{2}]) \in \Psi_{\lcp}(K + |C|)$. 
$([p_{2}, q_{2}], [\ell_{2}, r_{2}]) \in \Psi_{\lcp}(K + |C|)$ holds 
if $|\lcp(T[\gamma_{2}..r_{2}]$, $C^{n+1})| = K + |C|$. 
$T[\gamma_{2}..\gamma_{2}+|C| - 1] = C$ follows from $([p_{2}, q_{2}], [\ell_{2}, r_{2}]) \in \Psi_{\centerset}(C)$. 
Since $|[\gamma_{2}, r_{2}]| > \sum_{w = 1}^{h+1} \lfloor \mu(w) \rfloor$, 
we can apply Lemma~\ref{lem:associated_string_C}~\ref{enum:associated_string_C:1} to the interval attractor $([p_{2}, q_{2}], [\ell_{2}, r_{2}])$. 
This lemma shows that $\gamma_{2}+|C| - 1 \leq r_{2}$ holds. 
$|\lcp(T[\gamma_{2}..r_{2}], C^{n+1})| = |C| + |\lcp(T[\gamma_{2} + |C|..r_{2}], C^{n+1})|$ holds 
because $T[\gamma_{2}..\gamma_{2}+|C| - 1] = C$ and $\gamma_{2}+|C| - 1 \leq r_{2}$. 
$|\lcp(T[\gamma_{2} + |C|..r_{2}], C^{n+1})| = |\lcp(T[\gamma..r], C^{n+1})|$ 
because $\gamma = \gamma_{2} + |C|$ and $r = r_{2}$. 
$|\lcp(T[\gamma..r], C^{n+1})| = K$ follows from $([p, q], [\ell, r]) \in \Psi_{\lcp}(K)$. 
Therefore, 
\begin{equation*}
    \begin{split}
        |\lcp(T[\gamma_{2}..r_{2}], C^{n+1})| &= |C| + |\lcp(T[\gamma_{2} + |C|..r_{2}], C^{n+1})| \\
        &= |C| + |\lcp(T[\gamma..r], C^{n+1})| \\
        &= K + |C|.
    \end{split}    
\end{equation*}

%Because of $([p_{2}, q_{2}], [\ell_{2}, r_{2}]) \in \Psi_{\centerset}(C)$, 
%$T[\gamma_{2}..\gamma_{2}+|C| - 1] = C$ and $\gamma_{2}+|C| - 1 \leq r_{2}$ follow from Lemma~\ref{lem:C_prefix_property}. 

We prove $([p_{2}, q_{2}], [\ell_{2}, r_{2}]) \in \Psi_{\source} \cup \Psi_{\run}$. 
If $([p_{2}, q_{2}], [\ell_{2}, r_{2}]) \in \Psi_{\run}$, 
then $([p_{2}, q_{2}], [\ell_{2}, r_{2}]) \in \Psi_{\source} \cup \Psi_{\run}$ holds. 
Otherwise, $([p_{2}, q_{2}], [\ell_{2}, r_{2}]) \in \Psi_{\source}$ follows from the definition of the subset $\Psi_{\source}$
because (1) $([p_{2}, q_{2}], [\ell_{2}, r_{2}]) \not \in \Psi_{\run}$, 
(2) $([p_{2}, q_{2}], [\ell_{2}, r_{2}]) \in \Psi_{h} \cup \Psi_{\centerset}(C)$, 
and (3) the set $\Psi_{h} \cup \Psi_{\run} \cup \Psi_{\centerset}(C)$ contains interval attractor $([p, q], [\ell, r])$, 
and $\gamma = \gamma_{2} + |C|$. 
Therefore, $([p_{2}, q_{2}], [\ell_{2}, r_{2}]) \in \Psi_{\source} \cup \Psi_{\run}$ holds. 

We showed that $([p_{2}, q_{2}], [\ell_{2}, r_{2}]) \in \Psi_{h} \cap \Psi_{\centerset}(C) \cap (\Psi_{\source} \cup \Psi_{\run})$, 
$p_{2} < p$, $q_{2} = p-1$, $\ell_{2} \leq \ell$, $r_{2} = r$, and $\gamma_{2} = \gamma - |C|$. 
Therefore, Lemma~\ref{lem:psi_run_basic_property}(iii) holds. 
\end{proof}

%%%%%%%%%%%%%%%%%%%%%%%%%%%%%%%%%%%%%%%%%%%%%%%%%%%%%%%%%
\begin{proof}[Proof of Lemma~\ref{lem:psi_run_basic_property}~\ref{enum:psi_run_basic_property:4}]
Let $K \geq 0$ be an integer satisfying $([p, q], [\ell, r]) \in \Psi_{\lcp}(K)$. 
Then, $K = |\lcp(T[\gamma..r], C^{n+1})|$ follows from the definition of the subset $\Psi_{\lcp}(K)$. 
Here, $|\lcp(T[\gamma..r], C^{n+1})| > 1 + \sum_{w = 1}^{h+3} \lfloor \mu(w) \rfloor$ 
follows from the definition of the subset $\Psi_{\run}$. 
Because of $K \geq 2 + \sum_{w = 1}^{h+3} \lfloor \mu(w) \rfloor$, 
the subset $\Psi_{\modulo}((K - (2 + \sum_{w = 1}^{h+3} \lfloor \mu(w) \rfloor) ) \mod |C|)$ contains the interval attractor $([p, q], [\ell, r])$. 
\end{proof}

%%%%%%%%%%%%%%%%%%%%%%%%%%%%%%%%%%%%%%%%%%%%%%%%%%%%%%%%%
\begin{proof}[Proof of Lemma~\ref{lem:psi_run_basic_property}~\ref{enum:psi_run_basic_property:5}]
From Lemma~\ref{lem:psi_run_basic_property}~\ref{enum:psi_run_basic_property:3}, 
we obtain the following six statements: 
\begin{itemize}
\item $p^{\prime} < p$, 
\item $q^{\prime} = p-1$, 
\item $\ell^{\prime} \leq \ell$, 
\item $r^{\prime} = r$, 
\item $\gamma^{\prime} = \gamma - |C|$, 
\item $([p^{\prime}, q^{\prime}], [\ell^{\prime}, r^{\prime}]) \in \Psi_{\centerset}(C) \cap \Psi_{h} \cap \Psi_{\lcp}(K + |C|)$. 
\end{itemize}
In addition, we proved $([p^{\prime}, q^{\prime}], [\ell^{\prime}, r^{\prime}]) \in \Delta(h, b-1)$ in the proof of Lemma~\ref{lem:psi_run_basic_property}~\ref{enum:psi_run_basic_property:3}. 

Consider two sequences $A(p-1, \ell) = [s^{\prime 0}, e^{\prime 0}]$, $[s^{\prime 1}, e^{\prime 1}]$, $\ldots$, $[s^{\prime k^{\prime}}, e^{\prime k^{\prime}}]$ 
and $A(p, \ell) = [s^{\prime\prime 0}, e^{\prime\prime 0}]$, $[s^{\prime\prime 1}, e^{\prime\prime 1}]$, $\ldots$, $[s^{\prime\prime k^{\prime\prime}}, e^{\prime\prime k^{\prime\prime}}]$. 
$[p, \ell] \in \Delta(h, b)$ follows from Lemma~\ref{lem:IA_maximal_lemma}. 
$k^{\prime\prime} = h$ and $s^{\prime\prime h} = b$ follow from $[p, \ell] \in \Delta(h, b)$. 
We apply Lemma~\ref{lem:intv_function_shift}~\ref{enum:intv_function_shift:1} to the two intervals $[p, \ell]$ and $[p-1, \ell]$. 
Then, the lemma shows that $k^{\prime} \geq h$ and $s^{\prime h} \leq b$. 

We prove $I_{\capture}(p-1, \ell) = ([p^{\prime}, q^{\prime}], [\ell^{\prime}, r^{\prime}])$ by contradiction. 
We assume that $I_{\capture}(p-1, \ell) \neq ([p^{\prime}, q^{\prime}], [\ell^{\prime}, r^{\prime}])$ holds. 
Then, $(k^{\prime\prime}, s^{\prime h}) \neq (h, b-1)$ holds. 
In this case, sequence $\Delta(h, b-1)$ does not contain interval $[p-1, \ell]$. 
We can apply Lemma~\ref{lem:interval_extension_propertyX}~\ref{enum:interval_extension_propertyX:low} to the interval $[p-1, \ell]$ 
because $[p-1, \ell] \not \in \Delta(h, b-1)$, $p-1 \in [p^{\prime}, q^{\prime}]$, and $\ell \in [\ell^{\prime}, r^{\prime}]$. 
The proposition shows that $k^{\prime} < h$, 
but the two facts $k^{\prime} \geq h$ and $k^{\prime} < h$ yield a contradiction. 
Therefore, $I_{\capture}(p-1, \ell) = ([p^{\prime}, q^{\prime}], [\ell^{\prime}, r^{\prime}])$ must hold. 

\end{proof}

%%%%%%%%%%%%%%%%%%%%%%%%%%%%%%%%%%%%%%%%%%%%%%%%%%%%%%%%%
\begin{proof}[Proof of Lemma~\ref{lem:psi_run_basic_property}~\ref{enum:psi_run_basic_property:6}]
Lemma~\ref{lem:psi_run_basic_property}~\ref{enum:psi_run_basic_property:6} holds if the following two statements hold: 
(1) $([p^{\prime}, q^{\prime}], [\ell^{\prime}, r^{\prime}]) \in \Psi_{\run} \Rightarrow |\lcs(T[1..\gamma-1], C^{n+1})| \geq |C| + |[p-1, \gamma-1]|$; 
(2) $([p^{\prime}, q^{\prime}], [\ell^{\prime}, r^{\prime}]) \in \Psi_{\run} \Leftarrow |\lcs(T[1..\gamma-1], C^{n+1})| \geq |C| + |[p-1, \gamma-1]|$. 

\textbf{Proof of $([p^{\prime}, q^{\prime}], [\ell^{\prime}, r^{\prime}]) \in \Psi_{\run} \Rightarrow |\lcs(T[1..\gamma-1], C^{n+1})| \geq |C| + |[p-1, \gamma-1]|$.}
We apply Lemma~\ref{lem:psi_run_basic_property}~\ref{enum:psi_run_basic_property:2} to the interval attractor $([p^{\prime}, q^{\prime}], [\ell^{\prime}, r^{\prime}])$. 
Then, the lemma shows that 
$([p^{\prime}, q^{\prime}], [\ell^{\prime}, r^{\prime}]) \in \Psi_{\centerset}(C) \cap \Psi_{h}$, 
$p = p^{\prime} + |C|$, $q = q^{\prime} + |C|$, $\ell = \ell^{\prime} + |C|$, $r = r^{\prime}$, and $\gamma = \gamma^{\prime} + |C|$. 
$\lcs(T[p-1..\gamma-1], C^{n+1}) = T[p-1..\gamma-1]$ follows from the definition of the subset $\Psi_{\run}$. 
Similarly, 
$\lcs(T[p^{\prime}-1..\gamma^{\prime}-1], C^{n+1}) = T[p^{\prime}-1..\gamma^{\prime}-1]$ follows from the definition of the subset $\Psi_{\run}$. 

We prove $|\lcs(T[1..\gamma-1], C^{n+1})| = |\lcs(T[1..\gamma^{\prime} - 1], C^{n+1})| + |C|$. 
$T[\gamma - |C|..\gamma-1] = C$ holds 
because $\lcs(T[p-1..\gamma-1], C^{n+1}) = T[p-1..\gamma-1]$, 
and $|[p-1, \gamma-1]| \geq |C|$ follows from Lemma~\ref{lem:psi_run_basic_property}~\ref{enum:psi_run_basic_property:1}. 
Since $T[\gamma - |C|..\gamma-1] = C$, 
$|\lcs(T[1..\gamma-1], C^{n+1})| = |\lcs(T[1..\gamma - |C| - 1], C^{n+1})| + |C|$ holds. 
Here, $|\lcs(T[1..\gamma - |C| - 1], C^{n+1})| = |\lcs(T[1..\gamma^{\prime} - 1], C^{n+1})|$ 
because $\gamma = \gamma^{\prime} + |C|$. 
Therefore, $|\lcs(T[1..\gamma-1], C^{n+1})| = |\lcs(T[1..\gamma^{\prime} - 1], C^{n+1})| + |C|$ holds. 

We prove $|\lcs(T[1..\gamma-1], C^{n+1})| \geq |C| + |[p-1, \gamma-1]|$.
$|\lcs(T[1..\gamma^{\prime}-1], C^{n+1})| \geq |\lcs(T[p^{\prime}-1..\gamma^{\prime}-1], C^{n+1})|$ 
because $p^{\prime}-1 \geq 1$ follows from Lemma~\ref{lem:psi_run_basic_property}~\ref{enum:psi_run_basic_property:1}. 
$|\lcs(T[p^{\prime}-1..\gamma^{\prime}-1], C^{n+1})| = |[p-1, \gamma-1]|$ follows from the following equation: 
\begin{equation*}
    \begin{split}
        |\lcs(T[p^{\prime}-1..\gamma^{\prime}-1], C^{n+1})| &= |T[p^{\prime}-1..\gamma^{\prime}-1]| \\
        &= |[p^{\prime}-1, \gamma^{\prime}-1]| \\
        &= |[p - |C| -1, \gamma - |C| -1]| \\
        &= |[p-1, \gamma -1]|.
    \end{split}
\end{equation*}
Therefore, $|\lcs(T[1..\gamma-1], C^{n+1})| \geq |C| + |[p-1, \gamma-1]|$ follows from the following equation:
\begin{equation*}
    \begin{split}
        |\lcs(T[1..\gamma-1], C^{n+1})| &= |\lcs(T[1..\gamma^{\prime} - 1], C^{n+1})| + |C| \\
        &\geq |\lcs(T[p^{\prime}-1..\gamma^{\prime}-1], C^{n+1})| + |C| \\
        &= |[p-1, \gamma-1]| + |C|.
    \end{split}
\end{equation*}

\textbf{Proof of $([p^{\prime}, q^{\prime}], [\ell^{\prime}, r^{\prime}]) \in \Psi_{\run} \Leftarrow |\lcs(T[1..\gamma-1], C^{n+1})| \geq |C| + |[p-1, \gamma-1]|$.}
We prove $T[p-1..\ell] = T[p-1 - |C|..\ell - |C|]$. 
$\lcs(T[p-1-|C|..\gamma-1], C^{n+1}) = T[p-1-|C|..\gamma-1]$ follows from $|\lcs(T[1..\gamma-1], C^{n+1})| \geq |C| + |[p-1, \gamma-1]|$. 
$|\lcp(T[\gamma..r], C^{n+1})| > 1 + \sum_{w = 1}^{h+3} \lfloor \mu(w) \rfloor$ follows from the definition of the subset $\Psi_{\run}$. 
$T[p-1..\gamma + (\sum_{w = 1}^{h+3} \lfloor \mu(w) \rfloor)] = T[p-1 - |C|..\gamma + (\sum_{w = 1}^{h+3} \lfloor \mu(w) \rfloor) - |C|]$ 
follows from $\lcs(T[p-1-|C|..\gamma-1], C^{n+1}) = T[p-1-|C|..\gamma-1]$ 
and $|\lcp(T[\gamma..r], C^{n+1})| > 1 + \sum_{w = 1}^{h+3} \lfloor \mu(w) \rfloor$. 
$\ell \leq \gamma + \sum_{w = 1}^{h} \lfloor \mu(w) \rfloor$ follows from Lemma~\ref{lem:psi_run_basic_property}~\ref{enum:psi_run_basic_property:1}. 
Therefore, $T[p-1..\ell] = T[p-1 - |C|..\ell - |C|]$ follows from 
$T[p-1..\gamma + (\sum_{w = 1}^{h+3} \lfloor \mu(w) \rfloor)] = T[p-1 - |C|..\gamma + (\sum_{w = 1}^{h+3} \lfloor \mu(w) \rfloor) - |C|]$ 
and $\ell \leq \gamma + \sum_{w = 1}^{h} \lfloor \mu(w) \rfloor$.

We prove $I_{\capture}(p - |C|, \ell - |C|) = ([p^{\prime}, q^{\prime}], [\ell^{\prime}, r^{\prime}])$. 
$[p, \ell] \in \Delta(h, b)$ follows from Lemma~\ref{lem:IA_maximal_lemma}. 
$I_{\capture}(p, \ell) = ([p, q], [\ell, r])$ follows from $[p, \ell] \in \Delta(h, b)$. 
Let $([p_{1}, q_{1}], [\ell_{1}, r_{1}])$ be the interval attractor $I_{\capture}(p - |C|, \ell - |C|)$. 
Then, we can apply Corollary~\ref{cor:capture_gamma_corollary} to the two intervals $[p, \ell]$ and $[p - |C|, \ell - |C|]$ 
because $T[p..\ell] = T[p - |C|..\ell - |C|]$. 
Corollary~\ref{cor:capture_gamma_corollary} shows that 
$([p_{1}, q_{1}], [\ell_{1}, r_{1}]) \in \Psi_{h}$ 
and $|[p-|C|, \gamma_{1}]| = |[p, \gamma]|$ hold for the attractor position $\gamma_{1}$ of the interval attractor $([p_{1}, q_{1}], [\ell_{1}, r_{1}])$ 
because $I_{\capture}(p - |C|, \ell - |C|) = ([p_{1}, q_{1}], [\ell_{1}, r_{1}])$, 
$I_{\capture}(p, \ell) = ([p, q], [\ell, r])$, 
and $([p, q], [\ell, r]) \in \Psi_{h}$. 
$\gamma_{1} = \gamma - |C|$ follows from $|[p-|C|, \gamma_{1}]| = |[p, \gamma]|$. 
Lemma~\ref{lem:psi_run_basic_property}~\ref{enum:psi_run_basic_property:3} shows that 
(a) $p^{\prime} < p$, (b) $q^{\prime} = p-1$, (c) $\ell^{\prime} \leq \ell$, (d) $r^{\prime} = r$, 
(e) $\gamma^{\prime} = \gamma - |C|$, and 
(f) $([p^{\prime}, q^{\prime}], [\ell^{\prime}, r^{\prime}]) \in \Psi_{\centerset}(C) \cap \Psi_{h} \cap \Psi_{\lcp}(K + |C|)$. 
Corollary~\ref{cor:IA_identify_corollary} shows that 
$([p_{1}, q_{1}], [\ell_{1}, r_{1}]) = ([p^{\prime}, q^{\prime}], [\ell^{\prime}, r^{\prime}])$ holds 
because $\gamma_{1} = \gamma^{\prime}$ and $([p_{1}, q_{1}], [\ell_{1}, r_{1}]), ([p^{\prime}, q^{\prime}], [\ell^{\prime}, r^{\prime}]) \in \Psi_{h}$. 
Therefore, $I_{\capture}(p - |C|, \ell - |C|) = ([p^{\prime}, q^{\prime}], [\ell^{\prime}, r^{\prime}])$ holds. 

Next, we prove $p^{\prime} = p - |C|$ by contradiction. 
We assume that $p^{\prime} \neq p - |C|$ holds. 
Since $I_{\capture}(p - |C|, \ell - |C|) = ([p^{\prime}, q^{\prime}], [\ell^{\prime}, r^{\prime}])$, 
$p-|C| \in [p^{\prime}, q^{\prime}]$ and $\ell-|C| \in [\ell^{\prime}, r^{\prime}]$ follows from the definition of interval attractor. 
$p^{\prime} \leq p - |C|$ follows from $p-|C| \in [p^{\prime}, q^{\prime}]$. 
$p^{\prime} < p - |C|$ follows from $p^{\prime} \leq p - |C|$ and $p^{\prime} \neq p - |C|$. 

Consider two interval attractors $I_{\capture}(p-1, \ell) = ([p_{2}, q_{2}], [\ell_{2}, r_{2}])$ 
and $I_{\capture}(p-|C|-1, \ell-|C|)$. 
Here, Lemma~\ref{lem:IA_maximal_lemma} shows that $I_{\capture}(p-|C|-1, \ell-|C|) = ([p^{\prime}, q^{\prime}], [\ell^{\prime}, r^{\prime}])$ 
because $I_{\capture}(p - |C|, \ell - |C|) = ([p^{\prime}, q^{\prime}], [\ell^{\prime}, r^{\prime}])$ 
and $p^{\prime} < p - |C|$. 
Since $T[p-1..\ell] = T[p - |C| - 1..\ell - |C|]$, 
we can apply Corollary~\ref{cor:capture_gamma_corollary} to the two intervals $[p-1, \ell]$ and $[p - |C| - 1, \ell - |C|]$. 
Corollary~\ref{cor:capture_gamma_corollary} shows that 
$([p_{2}, q_{2}], [\ell_{2}, r_{2}]) \in \Psi_{h}$ 
and $|[p-|C|-1, \gamma^{\prime}]| = |[p-1, \gamma_{2}]|$ hold for the attractor position $\gamma_{2}$ of the interval attractor $([p_{2}, q_{2}], [\ell_{2}, r_{2}])$. 
$\gamma_{2} = \gamma$ follows from $|[p-|C|-1, \gamma^{\prime}]| = |[p-1, \gamma_{2}]|$ and $\gamma^{\prime} = \gamma - |C|$. 
Corollary~\ref{cor:IA_identify_corollary} shows that 
$([p_{2}, q_{2}], [\ell_{2}, r_{2}]) = ([p, q], [\ell, r])$ (i.e., $I_{\capture}(p-1, \ell) = ([p, q], [\ell, r])$) holds 
because $\gamma_{2} = \gamma$ and $([p_{2}, q_{2}], [\ell_{2}, r_{2}]), ([p, q], [\ell, r]) \in \Psi_{h}$. 
Since $I_{\capture}(p-1, \ell) = ([p, q], [\ell, r])$, 
$p-1 \in [p, q]$ follows from the definition of interval attractor. 
On the other hand, $p-1 \not \in [p, q]$ always holds. 
The two facts $p-1 \in [p, q]$ and $p-1 \not \in [p, q]$ yield a contradiction. 
Therefore, $p^{\prime} = p - |C|$ must hold.

We prove $\lcs(T[p^{\prime}-1..\gamma^{\prime}-1], C^{n+1}) = T[p^{\prime}-1..\gamma^{\prime}-1]$. 
$\gamma < \ell$ follows from Lemma~\ref{lem:IA_super_basic_property}~\ref{enum:IA_super_basic_property:1}. 
$T[p^{\prime}-1..\gamma^{\prime}-1] = T[p-1..\gamma-1]$ follows from 
$T[p-1 - |C|..\ell - |C|] = T[p-1..\ell]$, 
$p^{\prime} = p - |C|$, $\gamma < \ell$, and $\gamma^{\prime} = \gamma - |C|$. 
Because of $T[p^{\prime}-1..\gamma^{\prime}-1] = T[p-1..\gamma-1]$, 
$\lcs(T[p^{\prime}-1..\gamma^{\prime}-1], C^{n+1}) = \lcs(T[p-1..\gamma-1], C^{n+1})$ holds. 
$\lcs(T[p-1..\gamma-1], C^{n+1}) = T[p-1..\gamma-1]$ follows from the definition of the subset $\Psi_{\run}$. 
Therefore, $\lcs(T[p^{\prime}-1..\gamma^{\prime}-1], C^{n+1}) = T[p^{\prime}-1..\gamma^{\prime}-1]$ follows from 
$\lcs(T[p^{\prime}-1..\gamma^{\prime}-1], C^{n+1}) = \lcs(T[p-1..\gamma-1], C^{n+1})$, 
$\lcs(T[p-1..\gamma-1], C^{n+1}) = T[p-1..\gamma-1]$, 
and $T[p^{\prime}-1..\gamma^{\prime}-1] = T[p-1..\gamma-1]$. 

We prove $|\lcp(T[\gamma^{\prime}..r^{\prime}], C^{n+1})| > 1 + \sum_{w = 1}^{h+3} \lfloor \mu(w) \rfloor$. 
Because of $([p^{\prime}, q^{\prime}], [\ell^{\prime}, r^{\prime}]) \in \Psi_{\centerset}(C) \cap \Psi_{\lcp}(K + |C|)$, 
$|\lcp(T[\gamma^{\prime}..r^{\prime}], C^{n+1})| = K + |C|$ holds. 
Here, $K > 1 + \sum_{w = 1}^{h+3} \lfloor \mu(w) \rfloor$ follows from 
$K = |\lcp(T[\gamma..r], C^{n+1})|$ and $|\lcp(T[\gamma..r], C^{n+1})| > 1 + \sum_{w = 1}^{h+3} \lfloor \mu(w) \rfloor$. 
Therefore, $|\lcp(T[\gamma^{\prime}..r^{\prime}], C^{n+1})| > 1 + \sum_{w = 1}^{h+3} \lfloor \mu(w) \rfloor$ holds. 

Finally, $([p^{\prime}, q^{\prime}], [\ell^{\prime}, r^{\prime}]) \in \Psi_{\run}$ follows from 
(a) $\lcs(T[p^{\prime}-1..\gamma^{\prime}-1], C^{n+1}) = T[p^{\prime}-1..\gamma^{\prime}-1]$, 
(b) $|\lcp(T[\gamma^{\prime}..r^{\prime}], C^{n+1})| > 1 + \sum_{w = 1}^{h+3} \lfloor \mu(w) \rfloor$, 
and (c) $([p^{\prime}, q^{\prime}], [\ell^{\prime}, r^{\prime}]) \in \Psi_{\centerset}(C) \cap \Psi_{h}$. 

\end{proof}

%\begin{proof}[Proof of Lemma~\ref{lem:psi_run_basic_property}~\ref{enum:psi_run_basic_property:6}]

%Next, we prove $T[p-1..\gamma-1] = T[\hat{p}-1..\hat{\gamma}-1]$. 
%This equation holds if the following two statements hold: 
%(1) $([p^{\prime}, q^{\prime}], [\ell^{\prime}, r^{\prime}]) \in \Psi_{\run} \Rightarrow |\lcs(T[1..\gamma-1], C^{n+1})| \geq |C| + |[p-1, \gamma-1]|$; 
%(2) $([p^{\prime}, q^{\prime}], [\ell^{\prime}, r^{\prime}]) \in \Psi_{\run} \Leftarrow |\lcs(T[1..\gamma-1], C^{n+1})| \geq |C| + |[p-1, \gamma-1]|$. 
%Therefore, we prove statement (1) and statement (2). 
%\end{proof}

%%%%%%%%%%%%%%%%%%%%%%%%%%%%%%%%%%%%%%%%%%%%%%%%%%%%%%%%%

\begin{proof}[Proof of Lemma~\ref{lem:psi_run_basic_property}~\ref{enum:psi_run_basic_property:7}]
We prove $|\lcp(T[\gamma..r], T[\hat{\gamma}..\hat{r}])| > 1 + \sum_{w = 1}^{h+3} \lfloor \mu(w) \rfloor$. 
Since $([p, q], [\ell, r]) \in \Psi_{\run}$, 
$|\lcp(T[\gamma..r], C^{n+1})| > 1 + \sum_{w = 1}^{h+3} \lfloor \mu(w) \rfloor$ follows from the definition of the subset $\Psi_{\run}$. 
Similarly, 
$|\lcp(T[\hat{\gamma}..\hat{r}], C^{n+1})| > 1 + \sum_{w = 1}^{h+3} \lfloor \mu(w) \rfloor$ holds. 
Therefore, $|\lcp(T[\gamma..r], T[\hat{\gamma}..\hat{r}])| > 1 + \sum_{w = 1}^{h+3} \lfloor \mu(w) \rfloor$ follows from 
$|\lcp(T[\gamma..r], C^{n+1})| > 1 + \sum_{w = 1}^{h+3} \lfloor \mu(w) \rfloor$ and $|\lcp(T[\hat{\gamma}..\hat{r}], C^{n+1})| > 1 + \sum_{w = 1}^{h+3} \lfloor \mu(w) \rfloor$. 

Consider the longest common suffix $\lcs(T[p-1..\gamma-1], T[\hat{p}-1..\hat{\gamma}-1])$ between two strings $T[p-1..\gamma-1]$ and $T[\hat{p}-1..\hat{\gamma}-1]$. 
Then, $|\lcs(T[p-1..\gamma-1], T[\hat{p}-1..\hat{\gamma}-1])| = \min \{ |[p-1, \gamma-1]|, |[\hat{p}-1, \hat{\gamma}-1]| \}$ holds 
because the definition of the subset $\Psi_{\run}$ indicates that 
the two strings $T[p-1..\gamma-1]$ and $T[\hat{p}-1..\hat{\gamma}-1]$ are suffixes of string $C^{n+1}$. 

We prove $T[p-1..\gamma-1] = T[\hat{p}-1..\hat{\gamma}-1]$ by contradiction. 
We assume that $T[p-1..\gamma-1] \neq T[\hat{p}-1..\hat{\gamma}-1]$ holds. 
Since $|\lcs(T[p-1..\gamma-1], T[\hat{p}-1..\hat{\gamma}-1])| = \min \{ |[p-1, \gamma-1]|, |[\hat{p}-1, \hat{\gamma}-1]| \}$, 
either of the following two cases occurs: 
(A) $|[p-1, \gamma-1]| < |[\hat{p}-1, \hat{\gamma}-1]|$; 
(B) $|[p-1, \gamma-1]| > |[\hat{p}-1, \hat{\gamma}-1]|$. 

\textbf{Case (A).}
We prove $T[\gamma - |[p-1, \gamma-1]|..\gamma + \sum_{w = 1}^{h} \lfloor \mu(w) \rfloor] = T[\hat{\gamma} - |[p-1, \gamma-1]|..\hat{\gamma} + \sum_{w = 1}^{h} \lfloor \mu(w) \rfloor]$. 
$T[\gamma - |[p-1, \gamma-1]|..\gamma-1] = T[\hat{\gamma} - |[p-1, \gamma-1]|..\hat{\gamma}-1]$ follows from 
$|\lcs(T[p-1..\gamma-1], T[\hat{p}-1..\hat{\gamma}-1])| = |[p-1, \gamma-1]|$. 
$T[\gamma..\gamma + \sum_{w = 1}^{h} \lfloor \mu(w) \rfloor] = T[\hat{\gamma}..\hat{\gamma} + \sum_{w = 1}^{h} \lfloor \mu(w) \rfloor]$ holds 
because $|\lcp(T[\gamma..r], C^{n+1})| > 1 + \sum_{w = 1}^{h+3} \lfloor \mu(w) \rfloor$ and 
$|\lcp(T[\hat{\gamma}..\hat{r}], C^{n+1})| > 1 + \sum_{w = 1}^{h+3} \lfloor \mu(w) \rfloor$ follow from the definition of the subset $\Psi_{\run}$. 
Therefore, $T[\gamma - |[p-1, \gamma-1]|..\gamma + \sum_{w = 1}^{h} \lfloor \mu(w) \rfloor] = T[\hat{\gamma} - |[p-1, \gamma-1]|..\hat{\gamma} + \sum_{w = 1}^{h} \lfloor \mu(w) \rfloor]$ follows from $T[\gamma - |[p-1, \gamma-1]|..\gamma-1] = T[\hat{\gamma} - |[p-1, \gamma-1]|..\hat{\gamma}-1]$ 
and $T[\gamma..\gamma + \sum_{w = 1}^{h} \lfloor \mu(w) \rfloor] = T[\hat{\gamma}..\hat{\gamma} + \sum_{w = 1}^{h} \lfloor \mu(w) \rfloor]$. 

Consider two interval attractors $I_{\capture}(\gamma - |[p, \gamma - 1]|, \gamma + \sum_{w = 1}^{h} \lfloor \mu(w) \rfloor)$ 
and $I_{\capture}(\hat{\gamma} - |[p, \gamma - 1]|, \hat{\gamma} + \sum_{w = 1}^{h} \lfloor \mu(w) \rfloor)$. 
Lemma~\ref{lem:IA_maximal_lemma} shows that $I_{\capture}(\gamma - |[p, \gamma - 1]|, \gamma + \sum_{w = 1}^{h} \lfloor \mu(w) \rfloor) = ([p, q], [\ell, r])$ 
because (a) $\gamma - |[p, \gamma - 1]| = p$, 
and (b) $\gamma + \sum_{w = 1}^{h} \lfloor \mu(w) \rfloor \in [\ell, r]$ follows from Lemma~\ref{lem:psi_run_basic_property}~\ref{enum:psi_run_basic_property:1}. 

We prove $I_{\capture}(\hat{\gamma} - |[p, \gamma - 1]|, \hat{\gamma} + \sum_{w = 1}^{h} \lfloor \mu(w) \rfloor) = ([\hat{p}, \hat{q}], [\hat{\ell}, \hat{r}])$. 
We can apply Corollary~\ref{cor:capture_gamma_corollary} to the two intervals $[\gamma - |[p, \gamma - 1]|, \gamma + \sum_{w = 1}^{h} \lfloor \mu(w) \rfloor)]$ 
and $[\hat{\gamma} - |[p, \gamma - 1]|, \hat{\gamma} + \sum_{w = 1}^{h} \lfloor \mu(w) \rfloor]$ 
because $T[\gamma - |[p, \gamma-1]|..\gamma + \sum_{w = 1}^{h} \lfloor \mu(w) \rfloor] = T[\hat{\gamma} - |[p, \gamma-1]|..\hat{\gamma} + \sum_{w = 1}^{h} \lfloor \mu(w) \rfloor]$. 
Corollary~\ref{cor:capture_gamma_corollary}~\ref{enum:capture_gamma_corollary:1} shows that the level of the interval attractor $I_{\capture}(\hat{\gamma} - |[p, \gamma - 1]|, \hat{\gamma} + \sum_{w = 1}^{h} \lfloor \mu(w) \rfloor)$ is $h$. 
Corollary~\ref{cor:capture_gamma_corollary}~\ref{enum:capture_gamma_corollary:2} shows that 
$T[\gamma - |[p, \gamma-1]|..\gamma-1] = T[\hat{\gamma} - |[p, \gamma-1]|..\gamma_{A}-1]$ holds for 
the attractor position $\gamma_{A}$ of the interval attractor $I_{\capture}(\hat{\gamma} - |[p, \gamma - 1]|, \hat{\gamma} + \sum_{w = 1}^{h} \lfloor \mu(w) \rfloor)$.  
$|[\gamma - |[p, \gamma-1]|, \gamma-1]| = |[\hat{\gamma} - |[p, \gamma-1]|, \gamma_{A}-1]|$ follows from $T[\gamma - |[p, \gamma-1]|..\gamma-1] = T[\hat{\gamma} - |[p, \gamma-1]|..\gamma_{A}-1]$. 
$\gamma_{A} = \hat{\gamma}$ follows from $|[\gamma - |[p, \gamma-1]|, \gamma-1]| = |[\hat{\gamma} - |[p, \gamma-1]|, \gamma_{A}-1]|$. 
Corollary~\ref{cor:IA_identify_corollary} shows that 
$I_{\capture}(\hat{\gamma} - |[p, \gamma - 1]|, \hat{\gamma} + \sum_{w = 1}^{h} \lfloor \mu(w) \rfloor)  = ([\hat{p}, \hat{q}], [\hat{\ell}, \hat{r}])$ because 
$\gamma_{A} = \hat{\gamma}$ and $I_{\capture}(\hat{\gamma} - |[p, \gamma - 1]|, \hat{\gamma} + \sum_{w = 1}^{h} \lfloor \mu(w) \rfloor), ([\hat{p}, \hat{q}], [\hat{\ell}, \hat{r}]) \in \Psi_{h}$. 

We prove $I_{\capture}(\hat{\gamma} - |[p-1, \gamma - 1]|, \hat{\gamma} + \sum_{w = 1}^{h} \lfloor \mu(w) \rfloor) = ([\hat{p}, \hat{q}], [\hat{\ell}, \hat{r}])$ for interval attractor $I_{\capture}(\hat{\gamma} - |[p-1, \gamma - 1]|, \hat{\gamma} + \sum_{w = 1}^{h} \lfloor \mu(w) \rfloor)$. 
$\hat{\gamma} - |[p-1, \gamma - 1]| \geq \hat{p}$ follows from $|[p-1, \gamma-1]| < |[\hat{p}-1, \hat{\gamma}-1]|$. 
$\hat{\gamma} - |[p-1, \gamma - 1]| < \hat{\gamma} - |[p, \gamma - 1]|$ follows from $|[p-1, \gamma - 1]| > |[p, \gamma - 1]|$. 
Since $I_{\capture}(\hat{\gamma} - |[p, \gamma - 1]|, \hat{\gamma} + \sum_{w = 1}^{h} \lfloor \mu(w) \rfloor)  = ([\hat{p}, \hat{q}], [\hat{\ell}, \hat{r}])$, 
$\hat{\gamma} - |[p, \gamma - 1]| \in [\hat{p}, \hat{q}]$ and $\hat{\gamma} + \sum_{w = 1}^{h} \lfloor \mu(w) \rfloor \in [\hat{\ell}, \hat{r}]$ 
follow from the definition of interval attractor. 
$\hat{\gamma} - |[p-1, \gamma - 1]| \in [\hat{p}, \hat{q}]$ follows from $\hat{\gamma} - |[p, \gamma - 1]| \in [\hat{p}, \hat{q}]$ and $\hat{\gamma} - |[p-1, \gamma - 1]| \geq \hat{p}$. 
Lemma~\ref{lem:IA_maximal_lemma} shows that 
$I_{\capture}(\hat{\gamma} - |[p-1, \gamma - 1]|, \hat{\gamma} + \sum_{w = 1}^{h} \lfloor \mu(w) \rfloor) = ([\hat{p}, \hat{q}], [\hat{\ell}, \hat{r}])$ 
because 
(a) $I_{\capture}(\hat{\gamma} - |[p, \gamma - 1]|, \hat{\gamma} + \sum_{w = 1}^{h} \lfloor \mu(w) \rfloor)  = ([\hat{p}, \hat{q}], [\hat{\ell}, \hat{r}])$, (b) $\hat{\gamma} - |[p-1, \gamma - 1]| \in [\hat{p}, \hat{q}]$, 
and (c) $\hat{\gamma} - |[p-1, \gamma - 1]| < \hat{\gamma} - |[p, \gamma - 1]|$. 

We prove $I_{\capture}(\gamma - |[p-1, \gamma - 1]|, \gamma + \sum_{w = 1}^{h} \lfloor \mu(w) \rfloor) = ([p, q], [\ell, r])$ for interval attractor $I_{\capture}(\gamma - |[p-1, \gamma - 1]|, \gamma + \sum_{w = 1}^{h} \lfloor \mu(w) \rfloor)$. 
We can apply Corollary~\ref{cor:capture_gamma_corollary} to the two intervals $[\gamma - |[p-1, \gamma-1]|, \gamma + \sum_{w = 1}^{h} \lfloor \mu(w) \rfloor]$ 
and $[\hat{\gamma} - |[p-1, \gamma-1]|, \hat{\gamma} + \sum_{w = 1}^{h} \lfloor \mu(w) \rfloor]$ 
because $T[\gamma - |[p-1, \gamma-1]|..\gamma + \sum_{w = 1}^{h} \lfloor \mu(w) \rfloor] = T[\hat{\gamma} - |[p-1, \gamma-1]|..\hat{\gamma} + \sum_{w = 1}^{h} \lfloor \mu(w) \rfloor]$. 
Corollary~\ref{cor:capture_gamma_corollary}~\ref{enum:capture_gamma_corollary:1} shows that the level of the interval attractor $I_{\capture}(\gamma - |[p-1, \gamma - 1]|, \gamma + \sum_{w = 1}^{h} \lfloor \mu(w) \rfloor)$ is $h$. 
Corollary~\ref{cor:capture_gamma_corollary}~\ref{enum:capture_gamma_corollary:2} shows that 
$T[\gamma - |[p-1, \gamma-1]|..\gamma_{B}-1] = T[\hat{\gamma} - |[p-1, \gamma-1]|..\hat{\gamma}-1]$ holds for 
the attractor position $\gamma_{B}$ of the interval attractor $I_{\capture}(\gamma - |[p-1, \gamma - 1]|, \gamma + \sum_{w = 1}^{h} \lfloor \mu(w) \rfloor)$. 
$|[\gamma - |[p-1, \gamma-1]|, \gamma_{B}-1]| = |[\hat{\gamma} - |[p-1, \gamma-1]|, \hat{\gamma}-1]|$ follows from 
$T[\gamma - |[p-1, \gamma-1]|..\gamma_{B}-1] = T[\hat{\gamma} - |[p-1, \gamma-1]|..\hat{\gamma}-1]$. 
$\gamma_{B} = \gamma$ follows from $|[\gamma - |[p-1, \gamma-1]|, \gamma_{B}-1]| = |[\hat{\gamma} - |[p-1, \gamma-1]|, \hat{\gamma}-1]|$. 
Corollary~\ref{cor:IA_identify_corollary} shows that 
$I_{\capture}(\gamma - |[p-1, \gamma - 1]|, \gamma + \sum_{w = 1}^{h} \lfloor \mu(w) \rfloor) = ([p, q], [\ell, r])$ because 
$\gamma_{B} = \gamma$ and $I_{\capture}(\gamma - |[p-1, \gamma - 1]|, \gamma + \sum_{w = 1}^{h} \lfloor \mu(w) \rfloor), ([p, q], [\ell, r]) \in \Psi_{h}$. 

We show that there exists a contradiction. 
Since $I_{\capture}(\gamma - |[p-1, \gamma - 1]|, \gamma + \sum_{w = 1}^{h} \lfloor \mu(w) \rfloor) = ([p, q], [\ell, r])$, 
$\gamma - |[p-1, \gamma - 1]| \in [p, q]$ follows from the definition of interval attractor. 
On the other hand, $\gamma - |[p-1, \gamma - 1]| \not \in [p, q]$ holds 
because $\gamma - |[p-1, \gamma - 1]| = p-1$. 
The two facts $\gamma - |[p-1, \gamma - 1]| \in [p, q]$ and $\gamma - |[p-1, \gamma - 1]| \not \in [p, q]$ yield a contradiction. 
Therefore, $T[p-1..\gamma-1] = T[\hat{p}-1..\hat{\gamma}-1]$ must hold. 

\textbf{Case (B).}
Case (B) is symmetric to case (A). 
Therefore, we can show that there exists a contradiction using the same approach as for case (A). 
\end{proof}

\begin{proof}[Proof of Lemma~\ref{lem:psi_run_basic_property}~\ref{enum:psi_run_basic_property:8}]
Consider two interval attractors $I_{\capture}(p, r + 1 - |C|)$ and $I_{\capture}(p + |C|, r + 1)$. 
Here, $r+1-|C| \in [\ell, r]$ follows from 
$|C| \leq \lfloor \mu(h+1) \rfloor$, $\ell \leq \gamma + \sum_{w = 1}^{h} \lfloor \mu(w) \rfloor$, and $|[\gamma, r]| > 1 + \sum_{w = 1}^{h+3} \lfloor \mu(w) \rfloor$ (Lemma~\ref{lem:psi_run_basic_property}~\ref{enum:psi_run_basic_property:1}). 
Since $r + 1 - |C| \in [\ell, r]$, 
Lemma~\ref{lem:IA_maximal_lemma} shows that $I_{\capture}(p, r + 1 - |C|) = ([p, q], [\ell, r])$.

We prove $|\lcp(T[\gamma..r+1], C^{n+1})| \neq |[\gamma, r+1]|$ by contradiction.
We assume that $|\lcp(T[\gamma..r+1], C^{n+1})| = |[\gamma, r+1]|$ holds. 
Then, $T[p + |C|..r+1] = T[p..r + 1 - |C|]$ follows from 
$|\lcp(T[\gamma..r+1], C^{n+1})| = |[\gamma, r+1]|$, 
$\lcs(T[p-1..\gamma-1], C^{n+1}) = T[p-1..\gamma-1]$, and $|[p-1, \gamma-1]| \geq |C|$. 
Since $T[p + |C|..r+1] = T[p..r + 1 - |C|]$, 
we can apply Corollary~\ref{cor:capture_gamma_corollary} to the two intervals $[p + |C|, r + 1]$ and $[p, r + 1 - |C|]$. 
Corollary~\ref{cor:capture_gamma_corollary}~\ref{enum:capture_gamma_corollary:1} shows that 
$I_{\capture}(p + |C|, r + 1) \in \Psi_{h}$ holds. 
Corollary~\ref{cor:capture_gamma_corollary}~\ref{enum:capture_gamma_corollary:2} shows that 
$|[p, \gamma-1]| = |[p + |C|, \gamma_{A}-1]|$ holds for the attractor position $\gamma_{A}$ of $I_{\capture}(p + |C|, r + 1)$. 
$\gamma_{A} = \gamma + |C|$ follows from $|[p, \gamma-1]| = |[p+|C|, \gamma_{A}-1]|$. 

From Lemma~\ref{lem:psi_run_basic_property}~\ref{enum:psi_run_basic_property:2}, 
the $h$-th level interval attractors $\Psi_{h}$ contains interval attractor $([p + |C|, q + |C|], [\ell + |C|, r])$, 
and its attractor position $\gamma^{\prime}$ is equal to $\gamma + |C|$. 
Corollary~\ref{cor:IA_identify_corollary} shows that 
$I_{\capture}(p + |C|, r + 1) = ([p + |C|, q + |C|], [\ell + |C|, r])$ holds 
because $I_{\capture}(p + |C|, r + 1) \in \Psi_{h}$, $([p + |C|, q + |C|], [\ell + |C|, r]) \in \Psi_{h}$, 
and $\gamma^{\prime} = \gamma_{A}$. 
Since $I_{\capture}(p + |C|, r + 1) = ([p + |C|, q + |C|], [\ell + |C|, r])$, 
$r+1 \in [\ell + |C|, r]$ follows from the definition of interval attractor. 
On the other hand, $r+1 \not \in [\ell + |C|, r]$ follows from $r+1 \geq r$. 
The two facts $r+1 \in [\ell + |C|, r]$ and $r+1 \not \in [\ell + |C|, r]$ yield a contradiction. 
Therefore, $|\lcp(T[\gamma..r+1], C^{n+1})| \neq |[\gamma, r+1]|$ must hold. 

We prove $|\lcp(T[\gamma..r+1], C^{n+1})| = K$. 
Here, $K = |\lcp(T[\gamma..r], C^{n+1})|$ follows from the definition of the subset $\Psi_{\lcp}(K)$; 
$0 \leq |\lcp(T[\gamma..r], C^{n+1})| \leq |[\gamma, r]|$ holds. 
If $K < |[\gamma, r]|$, 
then $|\lcp(T[\gamma..r+1], C^{n+1})| = |\lcp(T[\gamma..r], C^{n+1})|$ holds. 
Therefore, $|\lcp(T[\gamma..r+1], C^{n+1})| = K$ follows from 
$K = |\lcp(T[\gamma..r], C^{n+1})|$ and $|\lcp(T[\gamma..r+1], C^{n+1})| = |\lcp(T[\gamma..r], C^{n+1})|$. 

Otherwise (i.e., $K = |[\gamma, r]|$), 
either $|\lcp(T[\gamma..r+1], C^{n+1})| = |[\gamma, r]|$ or $|\lcp(T[\gamma..r+1], C^{n+1})| = |[\gamma, r+1]|$ holds. 
Because of $|\lcp(T[\gamma..r+1], C^{n+1})| \neq |[\gamma, r+1]|$, 
$|\lcp(T[\gamma..r+1], C^{n+1})| = |[\gamma, r]|$ must hold. 
Therefore, $|\lcp(T[\gamma..r+1], C^{n+1})| = K$ follows from 
$K = |[\gamma, r]|$ and $|\lcp(T[\gamma..r+1], C^{n+1})| = |[\gamma, r]|$. 
\end{proof}

\begin{proof}[Proof of Lemma~\ref{lem:psi_run_basic_property}~\ref{enum:psi_run_basic_property:9}]
We prove $[q, \gamma + 2\lfloor \mu(h+1) \rfloor + \sum_{w = 1}^{h+1} \lfloor \mu(w) \rfloor] \in \Delta(h, b)$. 
Proposition~\ref{prop:I_LAST_HR_back_property}~\ref{enum:I_LAST_HR_back_property:1} shows that 
$[p - 1 + |C|, x^{h}_{s} + 2\lfloor \mu(h+1) \rfloor + \sum_{w = 1}^{h+1} \lfloor \mu(w) \rfloor] \in \Delta(h, b)$ holds 
for the substring $T[x^{h}_{s}..y^{h}_{s}]$ derived from the $s^{h}$-th nonterminal of sequence $S^{h}$. 
Here, $I(s^{h}) = ([p, q], [\ell, r])$ holds. 
$p - 1 + |C| = q$ follows from Lemma~\ref{lem:psi_run_basic_property}~\ref{enum:psi_run_basic_property:1}. 
$\gamma = x^{h}_{s}$ follows from the definition of the attractor position. 
Therefore, $[q, \gamma + 2\lfloor \mu(h+1) \rfloor + \sum_{w = 1}^{h+1} \lfloor \mu(w) \rfloor] \in \Delta(h, b)$ holds. 

We prove $I_{\capture}(i, j) = ([p, q], [\ell, r])$. 
$j \in [\gamma + 2\lfloor \mu(h+1) \rfloor + \sum_{w = 1}^{h+1} \lfloor \mu(w) \rfloor, r]$ holds 
because $j \in [\gamma + \sum_{w = 1}^{h + 3} \lfloor \mu(w) \rfloor, r]$ 
and $2\lfloor \mu(h+1) \rfloor + \sum_{w = 1}^{h+1} \lfloor \mu(w) \rfloor \leq \sum_{w = 1}^{h + 3} \lfloor \mu(w) \rfloor$. 
Lemma~\ref{lem:IA_maximal_lemma} shows that 
$[i, j] \in \Delta(h, b)$ holds 
because $i \in [p, q]$, $j \in [\gamma + 2\lfloor \mu(h+1) \rfloor + \sum_{w = 1}^{h+1} \lfloor \mu(w) \rfloor, r]$, 
and $[q, \gamma + 2\lfloor \mu(h+1) \rfloor + \sum_{w = 1}^{h+1} \lfloor \mu(w) \rfloor] \in \Delta(h, b)$. 
Since $b = s^{h}$, 
$I_{\capture}(i, j) = I(s^{h})$ follows from the definition of interval attractor $I_{\capture}(i, j)$.
Therefore, $I_{\capture}(i, j) = ([p, q], [\ell, r])$ follows from $I_{\capture}(i, j) = I(s^{h})$ and $I(s^{h}) = ([p, q], [\ell, r])$. 
\end{proof}

\subsubsection{Proof of Lemma~\ref{lem:suffix_syncro}}\label{subsubsec:suffix_syncro_proof}
\begin{proof}
We prove $|\lcs(T[\tau..\gamma-1], T[p^{\prime}-1..\gamma^{\prime}-1])| = \min \{ |[\tau, \gamma-1]|, |[p^{\prime}-1, \gamma^{\prime}-1]| \}$. 
$\lcs(T[\tau..\gamma-1], C^{n+1}) = T[\tau..\gamma-1]$ follows from the premise of Lemma~\ref{lem:suffix_syncro}. 
Because of $([p^{\prime}, q^{\prime}], [\ell^{\prime}, r^{\prime}]) \in \Psi_{h} \cap \Psi_{\run} \cap \Psi_{\centerset}(C)$, 
$\lcs(T[p^{\prime}-1..\gamma^{\prime}-1], C^{n+1}) = T[p^{\prime}-1..\gamma^{\prime}-1]$ follows from the definition of the subset $\Psi_{\run}$. 
Therefore, $|\lcs(T[\tau..\gamma-1], T[p^{\prime}-1..\gamma^{\prime}-1])| = \min \{ |[\tau, \gamma-1]|, |[p^{\prime}-1, \gamma^{\prime}-1]| \}$ 
follows from $\lcs(T[\tau..\gamma-1], C^{n+1}) = T[\tau..\gamma-1]$ and $\lcs(T[p^{\prime}-1..\gamma^{\prime}-1], C^{n+1}) = T[p^{\prime}-1..\gamma^{\prime}-1]$. 

We assume that $\lcs(T[\tau..\gamma-1], T[p^{\prime}..\gamma^{\prime}-1]) \neq T[\tau..\gamma-1]$ holds. 
Then, $|\lcs(T[\tau..\gamma-1], T[p^{\prime}..\gamma^{\prime}-1])| < |[\tau, \gamma-1]|$ holds. 
$|\lcs(T[\tau..\gamma-1], T[p^{\prime}..\gamma^{\prime}-1])| = \min \{ |[\tau, \gamma-1]|, |[p^{\prime}, \gamma^{\prime}-1]| \}$ 
follows from $|\lcs(T[\tau..\gamma-1], T[p^{\prime}-1..\gamma^{\prime}-1])| = \min \{ |[\tau, \gamma-1]|, |[p^{\prime}-1, \gamma^{\prime}-1]| \}$.
Therefore, $|[p^{\prime}, \gamma^{\prime}-1]| < |[\tau, \gamma-1]|$ follows from 
$|\lcs(T[\tau..\gamma-1], T[p^{\prime}..\gamma^{\prime}-1])| < |[\tau, \gamma-1]|$ and 
$|\lcs(T[\tau..\gamma-1], T[p^{\prime}..\gamma^{\prime}-1])| = \min \{ |[\tau, \gamma-1]|, |[p^{\prime}, \gamma^{\prime}-1]| \}$. 

The following five statements are used to show that there exists a contradiction under the assumption that $\lcs(T[\tau..\gamma-1], T[p^{\prime}..\gamma^{\prime}-1]) \neq T[\tau..\gamma-1]$ holds. 

\begin{enumerate}[label=\textbf{(\arabic*)}]
    \item $T[p^{\prime}-1..\gamma^{\prime} + \sum_{w = 1}^{h} \lfloor \mu(w) \rfloor] = T[\gamma - |[p^{\prime}-1, \gamma^{\prime} - 1]|..\gamma + \sum_{w = 1}^{h} \lfloor \mu(w) \rfloor]$; 
    \item $I_{\capture}(p^{\prime}, \gamma^{\prime} + \sum_{w = 1}^{h} \lfloor \mu(w) \rfloor) = ([p^{\prime}, q^{\prime}], [\ell^{\prime}, r^{\prime}])$;
    \item $I_{\capture}(\gamma - |[p^{\prime}, \gamma^{\prime} - 1]|, \gamma + \sum_{w = 1}^{h} \lfloor \mu(w) \rfloor) = ([p, q], [\ell, r])$;
    \item $I_{\capture}(\gamma - |[p^{\prime}-1, \gamma^{\prime} - 1]|, \gamma + \sum_{w = 1}^{h} \lfloor \mu(w) \rfloor) = ([p, q], [\ell, r])$;    
    \item $I_{\capture}(p^{\prime}-1, \gamma^{\prime} + \sum_{w = 1}^{h} \lfloor \mu(w) \rfloor) = ([p^{\prime}, q^{\prime}], [\ell^{\prime}, r^{\prime}])$.
\end{enumerate}

\textbf{Proof of statement (1).}
We prove $T[\gamma..\gamma + \sum_{w = 1}^{h} \lfloor \mu(w) \rfloor] = T[\gamma^{\prime}..\gamma^{\prime} + \sum_{w = 1}^{h} \lfloor \mu(w) \rfloor]$. 
Because of $K > 1 + \sum_{w = 1}^{h+3} \lfloor \mu(w) \rfloor$, 
$\lcp(T[\gamma..\gamma + \sum_{w = 1}^{h} \lfloor \mu(w) \rfloor], C^{n+1}) = T[\gamma..\gamma + \sum_{w = 1}^{h} \lfloor \mu(w) \rfloor]$ holds. 
Because of $([p^{\prime}, q^{\prime}], [\ell^{\prime}, r^{\prime}]) \in \Psi_{h} \cap \Psi_{\run} \cap \Psi_{\centerset}(C)$, 
$\lcp(T[\gamma^{\prime}..\gamma^{\prime} + \sum_{w = 1}^{h} \lfloor \mu(w) \rfloor], C^{n+1}) = T[\gamma^{\prime}..\gamma^{\prime} + \sum_{w = 1}^{h} \lfloor \mu(w) \rfloor]$ follows from the definition of the subset $\Psi_{\run}$. 
Therefore, $T[\gamma..\gamma + \sum_{w = 1}^{h} \lfloor \mu(w) \rfloor] = T[\gamma^{\prime}..\gamma^{\prime} + \sum_{w = 1}^{h} \lfloor \mu(w) \rfloor]$ follows from $\lcp(T[\gamma..\gamma + \sum_{w = 1}^{h} \lfloor \mu(w) \rfloor], C^{n+1}) = T[\gamma..\gamma + \sum_{w = 1}^{h} \lfloor \mu(w) \rfloor]$ and $\lcp(T[\gamma^{\prime}..\gamma^{\prime} + \sum_{w = 1}^{h} \lfloor \mu(w) \rfloor], C^{n+1}) = T[\gamma^{\prime}..\gamma^{\prime} + \sum_{w = 1}^{h} \lfloor \mu(w) \rfloor]$. 

We prove $T[p^{\prime}-1..\gamma^{\prime} + \sum_{w = 1}^{h} \lfloor \mu(w) \rfloor] = T[\gamma - |[p^{\prime}-1, \gamma^{\prime} - 1]|..\gamma + \sum_{w = 1}^{h} \lfloor \mu(w) \rfloor]$. 
$|[p^{\prime}-1, \gamma^{\prime}-1]| \leq |[\tau, \gamma-1]|$ follows from $|[p^{\prime}, \gamma^{\prime}-1]| < |[\tau, \gamma-1]|$. 
$|\lcs(T[\tau..\gamma-1], T[p^{\prime}-1..\gamma^{\prime}-1])| = |[p^{\prime}-1, \gamma^{\prime}-1]|$ 
follows from $|\lcs(T[\tau..\gamma-1], T[p^{\prime}-1..\gamma^{\prime}-1])| = \min \{ |[\tau, \gamma-1]|, |[p^{\prime}-1, \gamma^{\prime}-1]| \}$ 
and $|[p^{\prime}-1, \gamma^{\prime}-1]| \leq |[\tau, \gamma-1]|$. 
Therefore, $T[p^{\prime}-1..\gamma^{\prime} + \sum_{w = 1}^{h} \lfloor \mu(w) \rfloor] = T[\gamma - |[p^{\prime}-1, \gamma^{\prime} - 1]|..\gamma + \sum_{w = 1}^{h} \lfloor \mu(w) \rfloor]$ follows from 
$|\lcs(T[\tau..\gamma-1], T[p^{\prime}-1..\gamma^{\prime}-1])| = |[p^{\prime}-1, \gamma^{\prime}-1]|$ and 
$T[\gamma..\gamma + \sum_{w = 1}^{h} \lfloor \mu(w) \rfloor] = T[\gamma^{\prime}..\gamma^{\prime} + \sum_{w = 1}^{h} \lfloor \mu(w) \rfloor]$. 

\textbf{Proof of statement (2).}
Lemma~\ref{lem:psi_run_basic_property}~\ref{enum:psi_run_basic_property:1} shows that 
$\gamma^{\prime} + \sum_{w = 1}^{h} \lfloor \mu(w) \rfloor \in [\ell^{\prime}, r^{\prime}]$ holds. 
Because of $\gamma^{\prime} + \sum_{w = 1}^{h} \lfloor \mu(w) \rfloor \in [\ell^{\prime}, r^{\prime}]$, 
Lemma~\ref{lem:IA_maximal_lemma} shows that $I_{\capture}(p^{\prime}, \gamma^{\prime} + \sum_{w = 1}^{h} \lfloor \mu(w) \rfloor) = ([p^{\prime}, q^{\prime}], [\ell^{\prime}, r^{\prime}])$. 

\textbf{Proof of statement (3).}
We can apply Corollary~\ref{cor:capture_gamma_corollary} to the two intervals $[p^{\prime}, \gamma^{\prime} + \sum_{w = 1}^{h} \lfloor \mu(w) \rfloor]$ 
and $[\gamma - |[p^{\prime}, \gamma^{\prime} - 1]|, \gamma + \sum_{w = 1}^{h} \lfloor \mu(w) \rfloor]$ 
because $T[p^{\prime}..\gamma^{\prime} + \sum_{w = 1}^{h} \lfloor \mu(w) \rfloor] = T[\gamma - |[p^{\prime}, \gamma^{\prime} - 1]|..\gamma + \sum_{w = 1}^{h} \lfloor \mu(w) \rfloor]$ (statement (1)). 
Corollary~\ref{cor:capture_gamma_corollary}~\ref{enum:capture_gamma_corollary:1} shows that 
the level of the interval attractor $I_{\capture}(\gamma - |[p^{\prime}, \gamma^{\prime} - 1]|, \gamma + \sum_{w = 1}^{h} \lfloor \mu(w) \rfloor)$ is $h$ 
because $I_{\capture}(p^{\prime}, \gamma^{\prime} + \sum_{w = 1}^{h} \lfloor \mu(w) \rfloor) = ([p^{\prime}, q^{\prime}], [\ell^{\prime}, r^{\prime}])$ 
and $([p^{\prime}, q^{\prime}], [\ell^{\prime}, r^{\prime}]) \in \Psi_{h}$. 
Corollary~\ref{cor:capture_gamma_corollary}~\ref{enum:capture_gamma_corollary:2} shows that 
$|[p^{\prime}, \gamma^{\prime}-1]| = |[\gamma - |[p^{\prime}, \gamma^{\prime} - 1]|, \gamma_{A}-1]|$ holds 
for the attractor position $\gamma_{A}$ of the interval attractor $I_{\capture}(\gamma - |[p^{\prime}, \gamma^{\prime} - 1]|, \gamma + \sum_{w = 1}^{h} \lfloor \mu(w) \rfloor)$. 
$\gamma_{A} = \gamma$ follows from $|[p^{\prime}, \gamma^{\prime}-1]| = |[\gamma - |[p^{\prime}, \gamma^{\prime} - 1]|, \gamma_{A}-1]|$. 
Corollary~\ref{cor:IA_identify_corollary} shows that 
$I_{\capture}(\gamma - |[p^{\prime}, \gamma^{\prime} - 1]|, \gamma + \sum_{w = 1}^{h} \lfloor \mu(w) \rfloor) = ([p, q], [\ell, r])$ holds because 
$\gamma_{A} = \gamma$, $I_{\capture}(\gamma - |[p^{\prime}, \gamma^{\prime} - 1]|, \gamma + \sum_{w = 1}^{h} \lfloor \mu(w) \rfloor) \in \Psi_{h}$, 
and $([p, q], [\ell, r]) \in \Psi_{h}$. 

\textbf{Proof of statement (4).}
We prove $\gamma - |[p^{\prime}-1, \gamma^{\prime} - 1]| \in [p, \gamma - |[p^{\prime}, \gamma^{\prime} - 1]|]$. 
$\gamma - |[p^{\prime}-1, \gamma^{\prime} - 1]| < \gamma - |[p^{\prime}, \gamma^{\prime} - 1]|$ follows from 
$|[p^{\prime}-1, \gamma^{\prime} - 1]| > |[p^{\prime}, \gamma^{\prime} - 1]|$. 
$\gamma - |[p^{\prime}-1, \gamma^{\prime} - 1]| \geq \tau$ follows from $|[\tau, \gamma-1]| > |[p^{\prime}, \gamma^{\prime}-1]|$. 
$\gamma - |[p^{\prime}-1, \gamma^{\prime} - 1]| \geq p$ follows from 
$\tau \in [p, q]$ and $\gamma - |[p^{\prime}-1, \gamma^{\prime} - 1]| \geq \tau$. 
Therefore, $\gamma - |[p^{\prime}-1, \gamma^{\prime} - 1]| \in [p, \gamma - |[p^{\prime}, \gamma^{\prime} - 1]|]$ holds. 

Lemma~\ref{lem:IA_maximal_lemma} shows that 
$I_{\capture}(\gamma - |[p^{\prime}-1, \gamma^{\prime} - 1]|, \gamma + \sum_{w = 1}^{h} \lfloor \mu(w) \rfloor) = ([p, q], [\ell, r])$ 
because 
$\gamma - |[p^{\prime}-1, \gamma^{\prime} - 1]| \in [p, \gamma - |[p^{\prime}, \gamma^{\prime} - 1]|]$ 
and $I_{\capture}(\gamma - |[p^{\prime}, \gamma^{\prime} - 1]|, \gamma + \sum_{w = 1}^{h} \lfloor \mu(w) \rfloor) = ([p, q], [\ell, r])$. 
Therefore, statement (4) holds. 

\textbf{Proof of statement (5).}
We can apply Corollary~\ref{cor:capture_gamma_corollary} to two intervals $[p^{\prime}-1, \gamma^{\prime} + \sum_{w = 1}^{h} \lfloor \mu(w) \rfloor]$ 
and $[\gamma - |[p^{\prime}-1, \gamma^{\prime} - 1]|, \gamma + \sum_{w = 1}^{h} \lfloor \mu(w) \rfloor]$.
Corollary~\ref{cor:capture_gamma_corollary}~\ref{enum:capture_gamma_corollary:1} shows that 
the level of the interval attractor $I_{\capture}(p^{\prime}-1, \gamma^{\prime} + \sum_{w = 1}^{h} \lfloor \mu(w) \rfloor)$ is $h$ 
because $I_{\capture}(\gamma - |[p^{\prime}-1, \gamma^{\prime} - 1]|, \gamma + \sum_{w = 1}^{h} \lfloor \mu(w) \rfloor) = ([p, q], [\ell, r])$ 
and $([p, q], [\ell, r]) \in \Psi_{h}$. 
Corollary~\ref{cor:capture_gamma_corollary}~\ref{enum:capture_gamma_corollary:2} shows that 
$|[\gamma - |[p^{\prime}-1, \gamma^{\prime} - 1]|, \gamma-1]| = |[p^{\prime}-1, \gamma_{B}-1]|$ holds 
for the attractor position $\gamma_{B}$ of the interval attractor $I_{\capture}(p^{\prime}-1, \gamma^{\prime} + \sum_{w = 1}^{h} \lfloor \mu(w) \rfloor)$. 
$\gamma_{B} = \gamma^{\prime}$ follows from $|[\gamma - |[p^{\prime} - 1, \gamma^{\prime} - 1]|, \gamma-1]| = |[p^{\prime}-1, \gamma_{B}-1]|$. 
Corollary~\ref{cor:IA_identify_corollary} shows that 
$I_{\capture}(p^{\prime}-1, \gamma^{\prime} + \sum_{w = 1}^{h} \lfloor \mu(w) \rfloor) = ([p^{\prime}, q^{\prime}], [\ell^{\prime}, r^{\prime}])$ holds because 
$\gamma_{B} = \gamma^{\prime}$, $I_{\capture}(p^{\prime}-1, \gamma^{\prime} + \sum_{w = 1}^{h} \lfloor \mu(w) \rfloor) \in \Psi_{h}$, 
and $([p^{\prime}, q^{\prime}], [\ell^{\prime}, r^{\prime}]) \in \Psi_{h}$. 

\textbf{Proof of Lemma~\ref{lem:suffix_syncro}.}
We prove $\lcs(T[\tau..\gamma-1], T[p^{\prime}..\gamma^{\prime}-1]) = T[\tau..\gamma-1]$ by contradiction. 
Under the assumption that $\lcs(T[\tau..\gamma-1], T[p^{\prime}..\gamma^{\prime}-1]) \neq T[\tau..\gamma-1]$ holds, 
statement (5) shows that $I_{\capture}(p^{\prime}-1, \gamma^{\prime} + \sum_{w = 1}^{h} \lfloor \mu(w) \rfloor) = ([p^{\prime}, q^{\prime}], [\ell^{\prime}, r^{\prime}])$. 
Since $I_{\capture}(p^{\prime}-1, \gamma^{\prime} + \sum_{w = 1}^{h} \lfloor \mu(w) \rfloor) = ([p^{\prime}, q^{\prime}], [\ell^{\prime}, r^{\prime}])$, 
$p^{\prime}-1 \in [p^{\prime}, q^{\prime}]$ follows from the definition of interval attractor. 
On the other hand, $p^{\prime}-1 \not \in [p^{\prime}, q^{\prime}]$ because $p^{\prime}-1 < p^{\prime}$. 
The two facts $p^{\prime}-1 \in [p^{\prime}, q^{\prime}]$ and $p^{\prime}-1 \not \in [p^{\prime}, q^{\prime}]$ yield a contradiction. 
Therefore, $\lcs(T[\tau..\gamma-1], T[p^{\prime}..\gamma^{\prime}-1]) = T[\tau..\gamma-1]$ must hold. 

\end{proof}

\subsubsection{Proof of Lemma~\ref{lem:CCP_property}}\label{subsubsec:CCP_property_proof1}
The proof of Lemma~\ref{lem:CCP_property} is as follows.

\textbf{Proof of Lemma~\ref{lem:CCP_property}(i).}
Consider an interval attractor $([p^{\prime}, q^{\prime}], [\ell^{\prime}, r^{\prime}])$ in the subset $\Psi_{\CCP}(T[i..j])$. 
Then, there exists an occurrence $T[i^{\prime}..i^{\prime} + |[i, j]|-1]$ of string $T[i..j]$ in input string $T$ satisfying 
$I_{\capture}(i^{\prime}, i^{\prime} + |[i, j]|-1) = ([p^{\prime}, q^{\prime}], [\ell^{\prime}, r^{\prime}])$. 
Corollary~\ref{cor:capture_gamma_corollary} shows that 
$([p^{\prime}, q^{\prime}], [\ell^{\prime}, r^{\prime}]) \in \Psi_{h}$ because 
$I_{\capture}(i^{\prime}, i^{\prime} + |[i, j]|-1) = ([p^{\prime}, q^{\prime}], [\ell^{\prime}, r^{\prime}])$, 
$I_{\capture}(i, j) = ([p, q], [\ell, r]) \in \Psi_{h}$, 
and $T[i..j] = T[i^{\prime}..i^{\prime} + |[i, j]|-1]$. 
Therefore, $\Psi_{\CCP}(T[i..j]) \subseteq \Psi_{h}$ holds. 

\textbf{Proof of Lemma~\ref{lem:CCP_property}(ii).}
Because of $I_{\capture}(i, j) = ([p, q], [\ell, r])$, 
$i \in [p, q]$ and $j \in [\ell, r]$ follow from the definition of interval attractor.
$T[\gamma..j] \prec T[\gamma..r + 1] \prec T[\gamma..j]\#$ holds because 
(a) string $T[\gamma..j]$ is a proper prefix of string $T[\gamma..r + 1]$, 
(b) the character $\#$ does not occur in input string $T$, 
and (c) $T[n+1]$ is defined as character $\$$. 
Similarly, $\reverse(T[i..\gamma-1]) \prec \reverse(T[p-1..\gamma-1]) \prec \reverse(\# T[i..\gamma-1])$ holds 
because string $T[i..\gamma-1]$ is a proper suffix of string $T[p-1..\gamma-1]$. 

\textbf{Proof of Lemma~\ref{lem:CCP_property}(iii).}
We prove $I_{\capture}(\gamma^{\prime} - |[i, \gamma-1]|, \gamma^{\prime} + |[\gamma, j]| -1) = ([p^{\prime}, q^{\prime}], [\ell^{\prime}, r^{\prime}])$ by contradiction. 
We assume that $I_{\capture}(\gamma^{\prime} - |[i, \gamma-1]|, \gamma^{\prime} + |[\gamma, j]| -1) \neq ([p^{\prime}, q^{\prime}], [\ell^{\prime}, r^{\prime}])$. 
We can apply Corollary~\ref{cor:capture_gamma_corollary} to the two intervals $[i, j]$ and $[\gamma^{\prime} - |[i, \gamma-1]|, \gamma^{\prime} + |[\gamma, j]| -1]$ because $T[\gamma^{\prime} - |[i, \gamma-1]|..\gamma^{\prime} + |[\gamma, j]| -1] = T[i..j]$.
Corollary~\ref{cor:capture_gamma_corollary}~\ref{enum:capture_gamma_corollary:1} shows that 
$I_{\capture}(\gamma^{\prime} - |[i, \gamma-1]|, \gamma^{\prime} + |[\gamma, j]| -1) \in \Psi_{h}$ holds 
because $I_{\capture}(i, j) = ([p, q], [\ell, r]) \in \Psi_{h}$. 
Corollary~\ref{cor:capture_gamma_corollary}~\ref{enum:capture_gamma_corollary:2} shows that 
$|[i, \gamma-1]| = |[\gamma_{A} - |[i, \gamma-1]|, \gamma_{A}-1]|$ holds for the attractor position $\gamma_{A}$ of the interval attractor $I_{\capture}(\gamma^{\prime} - |[i, \gamma-1]|, \gamma^{\prime} + |[\gamma, j]| -1)$. 
$|[i, \gamma-1]| = |[\gamma^{\prime} - |[i, \gamma-1]|, \gamma^{\prime}-1]|$ follows from $T[i..\gamma-1] = T[\gamma^{\prime} - |[i, \gamma-1]|..\gamma^{\prime}-1]$. 
$\gamma_{A} = \gamma^{\prime}$ follows from $|[i, \gamma-1]| = |[\gamma^{\prime} - |[i, \gamma-1]|, \gamma^{\prime}-1]|$ and $|[i, \gamma-1]| = |[\gamma_{A} - |[i, \gamma-1]|, \gamma_{A}-1]|$. 

Corollary~\ref{cor:IA_identify_corollary} shows that 
$I_{\capture}(\gamma^{\prime} - |[i, \gamma-1]|, \gamma^{\prime} + |[\gamma, j]| -1) = ([p^{\prime}, q^{\prime}], [\ell^{\prime}, r^{\prime}])$ because 
$I_{\capture}(\gamma^{\prime} - |[i, \gamma-1]|, \gamma^{\prime} + |[\gamma, j]| -1) \Psi_{h}$, 
$([p^{\prime}, q^{\prime}]$, $[\ell^{\prime}, r^{\prime}]) \in \Psi_{h}$, and $\gamma_{A} = \gamma^{\prime}$. 
The two facts $I_{\capture}(\gamma^{\prime} - |[i, \gamma-1]|, \gamma^{\prime} + |[\gamma, j]| -1) = ([p^{\prime}, q^{\prime}], [\ell^{\prime}, r^{\prime}])$ and $I_{\capture}(\gamma^{\prime} - |[i, \gamma-1]|, \gamma^{\prime} + |[\gamma, j]| -1) \neq ([p^{\prime}, q^{\prime}], [\ell^{\prime}, r^{\prime}])$ yield a contradiction. 
Therefore, $I_{\capture}(\gamma^{\prime} - |[i, \gamma-1]|, \gamma^{\prime} + |[\gamma, j]| -1) = ([p^{\prime}, q^{\prime}], [\ell^{\prime}, r^{\prime}])$ must hold. 

$([p^{\prime}, q^{\prime}], [\ell^{\prime}, r^{\prime}]) \in \Psi_{\CCP}(T[i..j])$ follows from 
$I_{\capture}(\gamma^{\prime} - |[i, \gamma-1]|, \gamma^{\prime} + |[\gamma, j]| -1) = ([p^{\prime}, q^{\prime}], [\ell^{\prime}, r^{\prime}])$ and 
$T[\gamma^{\prime} - |[i, \gamma-1]|..\gamma^{\prime} + |[\gamma, j]|-1] = T[i..j]$. 
Therefore, Lemma~\ref{lem:CCP_property}~\ref{enum:CCP_property:3} holds.

\textbf{Proof of Lemma~\ref{lem:CCP_property}(iv).}
Let $\Psi_{A} = \{ ([p^{\prime}, q^{\prime}], [\ell^{\prime}, r^{\prime}]) \in \Psi_{h} \mid \reverse(T[i..\gamma-1]) \prec \reverse(T[p^{\prime}-1..\gamma^{\prime}-1]) \prec \reverse(\#T[i..\gamma-1]) \text{ and } T[\gamma..j] \prec T[\gamma^{\prime}..r^{\prime}+1] \prec T[\gamma..j]\# \}$ for simplicity. 
Then, Lemma~\ref{lem:CCP_property}~\ref{enum:CCP_property:4} holds if 
$\Psi_{\CCP}(T[i..j]) \subseteq \Psi_{A}$ and 
$\Psi_{\CCP}(T[i..j]) \supseteq \Psi_{A}$ hold.

\textbf{Proof of $\Psi_{\CCP}(T[i..j]) \subseteq \Psi_{A}$.}
Consider an interval attractor $([p_{A}, q_{A}], [\ell_{A}, r_{A}]) \in \Psi_{\CCP}(T[i..j])$. 
Then, 
there exists an occurrence $T[i_{A}..i_{A} + |[i, j]|-1]$ of string $T[i..j]$ in input string $T$ satisfying 
$I_{\capture}(i_{A}, i_{A} + |[i, j]|-1) = ([p_{A}, q_{A}], [\ell_{A}, r_{A}])$. 
Since $I_{\capture}(i_{A}, i_{A} + |[i, j]|-1) = ([p_{A}, q_{A}], [\ell_{A}, r_{A}])$, 
$i_{A} \in [p_{A}, q_{A}]$ and $i_{A} + |[i, j]|-1 \in [\ell_{A}, r_{A}]$ follows from the definition of interval attractor. 

We can apply Corollary~\ref{cor:capture_gamma_corollary} to the two intervals 
$[i, j]$ and $[i_{A}, i_{A} + |[i, j]|-1]$ 
because $T[i..j] = T[i_{A}..i_{A} + |[i, j]|-1]$. 
Corollary~\ref{cor:capture_gamma_corollary}~\ref{enum:capture_gamma_corollary:1} shows that 
$([p_{A}, q_{A}], [\ell_{A}, r_{A}]) \in \Psi_{h}$ 
because $I_{\capture}(i, j) = ([p, q], [\ell, r]) \in \Psi_{h}$. 
Corollary~\ref{cor:capture_gamma_corollary}~\ref{enum:capture_gamma_corollary:2} shows that 
$T[i..\gamma-1] = T[i_{A}..\gamma_{A}-1]$ and $T[\gamma..j] = T[\gamma_{A}..i_{A} + |[i, j]|-1]$ hold 
for the attractor position $\gamma_{A}$ of the interval attractor $([p_{A}, q_{A}], [\ell_{A}, r_{A}])$. 

We prove $T[\gamma..j] \prec T[\gamma_{A}..r_{A}+1] \prec T[\gamma..j]\#$ and $\reverse(T[i..\gamma-1]) \prec \reverse(T[p_{A}-1..\gamma_{A}-1]) \prec \reverse(\#T[i..\gamma-1])$. 
We can apply Lemma~\ref{lem:CCP_property}~\ref{enum:CCP_property:2} to the interval attractor $([p_{A}, q_{A}], [\ell_{A}, r_{A}])$ 
because $I_{\capture}(i_{A}, i_{A} + |[i, j]|-1) = ([p_{A}, q_{A}], [\ell_{A}, r_{A}])$. 
Lemma~\ref{lem:CCP_property}~\ref{enum:CCP_property:2} shows that 
$T[\gamma_{A}..i_{A} + |[i, j]|-1] \prec T[\gamma_{A}..r_{A}+1] \prec T[\gamma_{A}..i_{A} + |[i, j]|-1]\#$ and $\reverse(T[i_{A}..\gamma_{A}-1]) \prec \reverse(T[p_{A}-1..\gamma_{A}-1]) \prec \reverse(\#T[i_{A}..\gamma_{A}-1])$. 
We already proved $T[i..\gamma-1] = T[i_{A}..\gamma_{A}-1]$ and $T[\gamma..j] = T[\gamma_{A}..i_{A} + |[i, j]|-1]$. 
Therefore, 
$T[\gamma..j] \prec T[\gamma_{A}..r_{A}+1] \prec T[\gamma..j]\#$ 
and $\reverse(T[i..\gamma-1]) \prec \reverse(T[p_{A}-1..\gamma_{A}-1]) \prec \reverse(\#T[i..\gamma-1])$. 

$([p_{A}, q_{A}], [\ell_{A}, r_{A}]) \in \Psi_{A}$ follows from 
$([p_{A}, q_{A}], [\ell_{A}, r_{A}]) \in \Psi_{h}$, 
$T[\gamma..j] \prec T[\gamma_{A}..r_{A}+1] \prec T[\gamma..j]\#$, 
and $\reverse(T[i..\gamma-1]) \prec \reverse(T[p_{A}-1..\gamma_{A}-1]) \prec \reverse(\#T[i..\gamma-1])$. 
Therefore, $\Psi_{\CCP}(T[i..j]) \subseteq \Psi_{A}$ follows from the fact that 
$([p_{A}, q_{A}], [\ell_{A}, r_{A}]) \in \Psi_{A}$ holds for each interval attractor $([p_{A}, q_{A}], [\ell_{A}, r_{A}]) \in \Psi_{\CCP}(T[i..j])$. 

\textbf{Proof of $\Psi_{\CCP}(T[i..j]) \supseteq \Psi_{A}$.}
Consider an interval attractor $([p_{B}, q_{B}], [\ell_{B}, r_{B}]) \in \Psi_{A}$. 
Then, $([p_{B}, q_{B}], [\ell_{B}, r_{B}]) \in \Psi_{h}$, 
$\reverse(T[i..\gamma-1]) \prec \reverse(T[p_{B}-1..\gamma_{B}-1]) \prec \reverse(\#T[i..\gamma-1])$ 
and $T[\gamma..j] \prec T[\gamma_{B}..r_{B}+1] \prec T[\gamma..j]\#$ hold 
for the attractor position $\gamma_{B}$ of the interval attractor $([p_{B}, q_{B}], [\ell_{B}, r_{B}])$. 
The equation $T[\gamma..j] \prec T[\gamma_{B}..r_{B}+1] \prec T[\gamma..j]\#$ indicates that 
string $T[\gamma..j]$ is a proper prefix of string $T[\gamma_{B}..r_{B}+1]$. 
Similarly, 
the equation $\reverse(T[i..\gamma-1]) \prec \reverse(T[p_{B}-1..\gamma_{B}-1]) \prec \reverse(\#T[i..\gamma-1])$ indicates that string $T[i..\gamma-1]$ is a proper suffix of string $T[p_{B}-1..\gamma_{B}-1]$. 
We can apply Lemma~\ref{lem:CCP_property}~\ref{enum:CCP_property:3} to the interval attractor $([p_{B}, q_{B}], [\ell_{B}, r_{B}])$ 
because $([p_{B}, q_{B}], [\ell_{B}, r_{B}]) \in \Psi_{h}$, 
$T[\gamma_{B} - |[i, \gamma-1]|..\gamma_{B} -1] = T[i..\gamma-1]$, 
and $T[\gamma_{B}..\gamma_{B} + |[\gamma, j]| -1] = T[\gamma..j]$. 
Lemma~\ref{lem:CCP_property}~\ref{enum:CCP_property:3} shows that $([p_{B}, q_{B}], [\ell_{B}, r_{B}]) \in \Psi_{\CCP}(T[i..j])$ holds. 
Therefore, $\Psi_{\CCP}(T[i..j]) \supseteq \Psi_{A}$ follows from the fact that 
$([p_{B}, q_{B}], [\ell_{B}, r_{B}]) \in \Psi_{\CCP}(T[i..j])$ holds for each interval attractor $([p_{B}, q_{B}], [\ell_{B}, r_{B}]) \in \Psi_{A}$. 

\textbf{Proof of Lemma~\ref{lem:CCP_property}(v).}
Because of $\Psi_{\CCP}(T[i..j]) \cap \Psi_{\centerset}(C^{\prime}) \neq \emptyset$, 
the set $\Psi_{\CCP}(T[i..j]) \cap \Psi_{\centerset}(C^{\prime})$ contains an interval attractor $([p^{\prime}, q^{\prime}], [\ell^{\prime}, r^{\prime}])$. 
Here, $([p^{\prime}, q^{\prime}], [\ell^{\prime}, r^{\prime}]) \in \Psi_{h}$ follows from Lemma~\ref{lem:CCP_property}~\ref{enum:CCP_property:1}; 
there exists an occurrence $T[i^{\prime}..i^{\prime} + |[i, j]|-1]$ of string $T[i..j]$ in input string $T$ satisfying 
$I_{\capture}(i^{\prime}, i^{\prime} + |[i, j]|-1) = ([p^{\prime}, q^{\prime}], [\ell^{\prime}, r^{\prime}])$. 
Since $I_{\capture}(i^{\prime}, i^{\prime} + |[i, j]|-1) = ([p^{\prime}, q^{\prime}], [\ell^{\prime}, r^{\prime}])$, 
$i^{\prime} \in [p^{\prime}, q^{\prime}]$ and $i^{\prime} + |[i, j]|-1 \in [\ell^{\prime}, r^{\prime}]$ follow from 
the definition of interval attractor. 

We prove $([p^{\prime}, q^{\prime}], [\ell^{\prime}, r^{\prime}]) \in \Psi_{\centerset}(C)$. 
Corollary~\ref{cor:capture_gamma_corollary} shows that 
$T[\gamma..j] = T[\gamma^{\prime}..i^{\prime} + |[i, j]|-1]$ holds for the attractor position $\gamma^{\prime}$ of the interval attractor $([p^{\prime}, q^{\prime}], [\ell^{\prime}, r^{\prime}])$. 
The length of the longest common prefix of two strings $T[\gamma..r]$ and $T[\gamma^{\prime}..r^{\prime}]$ is larger than $\sum_{w = 1}^{h+3} \lfloor \mu(w) \rfloor$ 
because $T[\gamma..j] = T[\gamma^{\prime}..i^{\prime} + |[i, j]|-1]$, 
$|[\gamma, j]| \leq 1 + \sum_{w = 1}^{h+3} \lfloor \mu(w) \rfloor$, 
$j \leq r$, and $i^{\prime} + |[i, j]|-1 \leq r^{\prime}$. 
We can apply Lemma~\ref{lem:associated_string_C}~\ref{enum:associated_string_C:2} to the two interval attractors 
$([p, q], [\ell, r])$ and $([p^{\prime}, q^{\prime}], [\ell^{\prime}, r^{\prime}])$ 
because $([p, q], [\ell, r]) \in \Psi_{h}$, $([p^{\prime}, q^{\prime}], [\ell^{\prime}, r^{\prime}]) \in \Psi_{h}$, 
and $|\lcp(T[\gamma..r], T[\gamma^{\prime}..r^{\prime}])| > \sum_{w = 1}^{h+3} \lfloor \mu(w) \rfloor$. 
This lemma shows that $([p^{\prime}, q^{\prime}], [\ell^{\prime}, r^{\prime}]) \in \Psi_{\centerset}(C)$. 

We prove $C = C^{\prime}$ by contradiction. 
We assume that $C \neq C^{\prime}$ holds. 
Then, $\Psi_{\centerset}(C) \cap \Psi_{\centerset}(C^{\prime}) \neq \emptyset$ follows from 
$([p^{\prime}, q^{\prime}], [\ell^{\prime}, r^{\prime}]) \in \Psi_{\centerset}(C)$  and $([p^{\prime}, q^{\prime}], [\ell^{\prime}, r^{\prime}]) \in \Psi_{\centerset}(C^{\prime})$. 
On the other hand, $\Psi_{\centerset}(C) \cap \Psi_{\centerset}(C^{\prime}) = \emptyset$ follows from the definition of associated string. 
Therefore, $C = C^{\prime}$ must hold. 

\textbf{Proof of Lemma~\ref{lem:CCP_property}(vi).}
From the definition of the subset $\Psi_{\CCP}(T[i..j])$, 
there exists an occurrence $T[i^{\prime}..i^{\prime} + |[i, j]| - 1]$ of substring $T[i..j]$ in input string $T$ 
satisfying $I_{\capture}(i^{\prime}, i^{\prime} + |[i, j]| - 1) = ([p^{\prime}, q^{\prime}], [\ell^{\prime}, r^{\prime}])$.

We can apply Corollary~\ref{cor:capture_gamma_corollary} to the two intervals $[i, j]$ and $[i^{\prime}, i^{\prime} + |[i, j]| - 1]$ 
because $T[i^{\prime}..i^{\prime} + |[i, j]| - 1] = T[i..j]$. 
Corollary~\ref{cor:capture_gamma_corollary}~\ref{enum:capture_gamma_corollary:2} shows that 
$|[i^{\prime}, \gamma^{\gamma}-1]| = |[i, \gamma-1]|$ and $|[\gamma^{\prime}, i^{\prime} + |[i, j]| - 1]| = |[\gamma, j]|$ hold. 
$i^{\prime} = \gamma^{\gamma} - |[i, \gamma-1]|$ follows from $|[i^{\prime}, \gamma^{\gamma}-1]| = |[i, \gamma-1]|$. 
$i^{\prime} + |[i, j]| - 1 = \gamma^{\prime} + |[\gamma, j]| - 1$ follows from $|[\gamma^{\prime}, i^{\prime} + |[i, j]| - 1]| = |[\gamma, j]|$. 
Therefore, Lemma~\ref{lem:CCP_property}~\ref{enum:CCP_property:6} holds. 

%%%%%%%%%%%%%%%%%%%%%%%%%%%%%%%%%%%%%%%%%%%%%%%%%%%%%%%%%%%%%%%%%%%%%%%%%%%%%%%%

\subsubsection{Proof of Lemma~\ref{lem:CCP_special_property}}\label{subsubsec:CCP_property_proof2}
The following proposition states a property of the interval attractor $([p, q], [\ell, r])$ of Lemma~\ref{lem:CCP_special_property}.

\begin{proposition}\label{prop:CCP_special_left_property}
Consider the interval $[i, j]$ and interval attractor $([p, q], [\ell, r])$ of Lemma~\ref{lem:CCP_special_property}. 
Here, the interval attractor $([p, q], [\ell, r])$ satisfies the two conditions of Lemma~\ref{lem:CCP_special_property}; 
let $\gamma$, $C$, and $h$ be the attractor position, associated string, and level of the interval attractor $([p, q], [\ell, r])$, 
respectively. 
Let $\gamma^{\prime}$ be the attractor position of an interval attractor $([p^{\prime}, q^{\prime}], [\ell^{\prime}, r^{\prime}])$ in 
set $\Psi_{h} \cap \Psi_{\run} \cap \Psi_{\centerset}(C)$. 
Then, $|[p^{\prime}, \gamma^{\prime}-1]| \geq |[i, \gamma-1]|$ holds. 
\end{proposition}
\begin{proof}
    We assume that $|[p^{\prime}, \gamma^{\prime}-1]| < |[i, \gamma-1]|$ holds. 
    Then, $i \leq \gamma-1$ holds because $|[p^{\prime}, \gamma^{\prime}-1]| \geq 0$. 
    Under the assumption, 
    let $\tau \geq 0$ be the smallest integer satisfying $|[p^{\prime}, \gamma^{\prime}-1]| \geq |[i, \gamma-1]| - \tau |C|$. 
    Here, $\tau \geq 1$ holds because $|[p^{\prime}, \gamma^{\prime}-1]| < |[i, \gamma-1]|$. 

    The following six statements are used to there exists a contradiction under the assumption:     
\begin{enumerate}[label=\textbf{(\roman*)}]
    \item $\gamma^{\prime} - |[i, \gamma-1]| + \tau |C| \in [p^{\prime}, q^{\prime}]$;
    \item let $d \geq 1$ be the smallest integer satisfying 
    $I_{\capture}(\gamma^{\prime} - |[i, \gamma-1]| + \tau |C|, \gamma^{\prime} + d - 1) = ([p^{\prime}, q^{\prime}], [\ell^{\prime}, r^{\prime}])$. 
    Then, $d \leq 1 + \sum_{w = 1}^{h + 3} \lfloor \mu(w) \rfloor$; 
    \item 
    $i \leq \gamma - 1 - \tau |C|$ holds, 
    and string $T[i..\gamma - 1 - \tau |C|]$ is a suffix of string $T[p^{\prime}..\gamma^{\prime}-1]$ 
    (i.e., $\lcs(T[i..\gamma - 1 - \tau |C|], T[p^{\prime}..\gamma^{\prime}-1]) = T[i..\gamma - 1 - \tau |C|]$);
    \item 
    let $R$ be the length of the longest common prefix between two strings 
    $T[\gamma^{\prime}..r^{\prime}]$ and $T[\gamma - \tau |C|..j]$
    (i.e., $R = |\lcp(T[\gamma^{\prime}..r^{\prime}], T[\gamma - \tau |C|..j])|$). 
    Then, $R \geq \min \{ d, \tau |C| + |[\gamma, j]| \}$ 
    for the integer $d$ of statement (i);
    \item if $\tau |C| + |[\gamma, j]| < d$ for the integer $d$ of statement (i), 
    then $T[i..j] = T[\gamma^{\prime} - |[i, \gamma-1]| + \tau |C|..\gamma^{\prime} + \tau |C| + |[\gamma, j]| - 1]$ holds, 
    and $I_{\capture}(\gamma^{\prime} - |[i, \gamma-1]| + \tau |C|, \gamma^{\prime} + \tau |C| + |[\gamma, j]| - 1) \in \Psi_{h}$;
    \item if $\tau |C| + |[\gamma, j]| \geq d$ for the integer $d$ of statement (i), 
    then $T[i..\gamma - \tau |C| + d - 1] = T[\gamma^{\prime} - |[i, \gamma-1]| + \tau |C|..\gamma^{\prime} + d - 1]$ holds, 
    and $I_{\capture}(i, \gamma - \tau |C| + d - 1]) \in \Psi_{h}$. 
\end{enumerate}
    \textbf{Proof of statement (i).}
    We prove $\gamma^{\prime} - |[i, \gamma-1]| + \tau |C| \in [p^{\prime}, p^{\prime} + |C| - 1]$. 
    $|[p^{\prime}, \gamma^{\prime}-1]| - (|[i, \gamma-1]| - \tau |C|) \in \{ 0, 1, \ldots, |C|-1 \}$ holds 
    because $\tau \geq 1$ is the smallest integer satisfying $|[p^{\prime}, \gamma^{\prime}-1]| \geq |[i, \gamma-1]| - \tau |C|$. 
    $|[p^{\prime}, \gamma^{\prime}-1]| - (|C| - 1) \leq (|[i, \gamma-1]| - \tau |C|) \leq |[p^{\prime}, \gamma^{\prime}-1]|$ 
    follows from $|[p^{\prime}, \gamma^{\prime}-1]| - (|[i, \gamma-1]| - \tau |C|) \in \{ 0, 1, \ldots, |C|-1 \}$. 
    $\gamma^{\prime} - |[i, \gamma-1]| + \tau |C| \geq p^{\prime}$ follows from $(|[i, \gamma-1]| - \tau |C|) \leq |[p^{\prime}, \gamma-1]|$. 
    On the other hand, $\gamma^{\prime} - |[i, \gamma-1]| + \tau |C| \leq p^{\prime} + |C| - 1$ follows from 
    $|[p^{\prime}, \gamma^{\prime}-1]| - (|C| - 1) \leq (|[i, \gamma-1]| - \tau |C|)$. 
    Therefore, $\gamma^{\prime} - |[i, \gamma-1]| + \tau |C| \in [p^{\prime}, p^{\prime} + |C| - 1]$ holds. 

    We prove $\gamma^{\prime} - |[i, \gamma-1]| + \tau |C| \in [p^{\prime}, q^{\prime}]$. 
    Because of $([p^{\prime}, q^{\prime}], [\ell^{\prime}, r^{\prime}]) \in \Psi_{\run} \cap \Psi_{\centerset}(C)$, 
    Lemma~\ref{lem:psi_run_basic_property}~\ref{enum:psi_run_basic_property:1} shows that 
    $|[p^{\prime}, q^{\prime}]| = |C|$ holds. 
    $[p^{\prime}, p^{\prime} + |C| - 1] = [p^{\prime}, q^{\prime}]$ follows from $|[p^{\prime}, q^{\prime}]| = |C|$. 
    Therefore, $\gamma^{\prime} - |[i, \gamma-1]| + \tau |C| \in [p^{\prime}, q^{\prime}]$ follows from 
    $\gamma^{\prime} - |[i, \gamma-1]| + \tau |C| \in [p^{\prime}, p^{\prime} + |C| - 1]$ and $[p^{\prime}, p^{\prime} + |C| - 1] = [p^{\prime}, q^{\prime}]$. 

    \textbf{Proof of statement (ii).}    
    Because of $\gamma^{\prime} - |[i, \gamma-1]| + \tau |C| \in [p^{\prime}, q^{\prime}]$, 
    Lemma~\ref{lem:psi_run_basic_property}~\ref{enum:psi_run_basic_property:9} shows that 
    $I_{\capture}(\gamma^{\prime} - |[i, \gamma-1]| + \tau |C|, \gamma^{\prime} + \sum_{w = 1}^{h + 3} \lfloor \mu(w) \rfloor) = ([p^{\prime}, q^{\prime}], [\ell^{\prime}, r^{\prime}])$ holds. 
    Therefore, $d \leq 1 + \sum_{w = 1}^{h + 3} \lfloor \mu(w) \rfloor$ holds. 

    Lemma~\ref{lem:IA_maximal_lemma} shows that 
    $I_{\capture}(\gamma^{\prime} - |[i, \gamma-1]| + \tau |C|, r^{\prime}) = ([p^{\prime}, q^{\prime}], [\ell^{\prime}, r^{\prime}])$. 
    Here, $\gamma^{\prime} \leq r^{\prime}$ follows from $\gamma^{\prime} \leq \ell^{\prime}$ (Lemma~\ref{lem:IA_super_basic_property}~\ref{enum:IA_super_basic_property:1}) and $\ell^{\prime} \leq r^{\prime}$. 
    Therefore, $\gamma^{\prime} + d - 1 \leq r^{\prime}$ holds. 

    \textbf{Proof of statement (iii).}
    We prove $i \leq \gamma - \tau |C|$. 
    Because of $([p^{\prime}, q^{\prime}], [\ell^{\prime}, r^{\prime}]) \in \Psi_{\run} \cap \Psi_{\centerset}(C)$, 
    Lemma~\ref{lem:psi_run_basic_property}~\ref{enum:psi_run_basic_property:1} shows that 
    $|[p^{\prime}, \gamma^{\prime}-1]| \geq |C|-1$ holds. 
    $|[p^{\prime}, \gamma^{\prime}-1]| - |C| < |[i, \gamma-1]| - \tau |C|$ holds 
    because $\tau \geq 1$ is the smallest integer satisfying $|[p^{\prime}, \gamma^{\prime}-1]| \geq |[i, \gamma-1]| - \tau |C|$. 
    $|[i, \gamma-1]| - \tau |C| \geq 0$ follows from 
    $|[p^{\prime}, \gamma^{\prime}-1]| - |C| < |[i, \gamma-1]| - \tau |C|$ and $|[p^{\prime}, \gamma^{\prime}-1]| \geq |C| - 1$. 
    Therefore, $i \leq \gamma - \tau |C|$ follows from $|[i, \gamma-1]| - \tau |C| \geq 0$. 

    We prove $\lcs(T[i..\gamma - 1 - \tau |C|], T[p^{\prime}..\gamma^{\prime}-1]) = T[i..\gamma - 1 - \tau |C|]$. 
    $\lcs(T[i..\gamma-1], C^{n+1}) = T[i..\gamma-1]$ follows from the first condition of Lemma~\ref{lem:CCP_special_property}. 
    Since $i \leq \gamma - \tau |C|$, 
    either $i = \gamma - \tau |C|$ or $i \leq \gamma - 1 - \tau |C|$ holds. 
    If $i = \gamma - \tau |C|$, 
    then $\lcs(T[i..\gamma - 1 - \tau |C|], T[p^{\prime}..\gamma^{\prime}-1]) = T[i..\gamma - 1 - \tau |C|]$ holds 
    because $T[i..\gamma - 1 - \tau |C|] = \varepsilon$. 
    
    Otherwise (i.e., $i \leq \gamma - 1 - \tau |C|$), 
    $\lcs(T[i..\gamma-1 - \tau |C|], C^{n+1}) = T[i..\gamma-1 - \tau |C|]$ follows from     
    $\lcs(T[i..\gamma-1], C^{n+1}) = T[i..\gamma-1]$. 
    Because of $([p^{\prime}, q^{\prime}], [\ell^{\prime}, r^{\prime}]) \in \Psi_{\run} \cap \Psi_{\centerset}(C)$, 
    $\lcs(T[p^{\prime}..\gamma^{\prime}-1], C^{n+1}) = T[p^{\prime}..\gamma^{\prime}-1]$ follows from the definition of the subset $\Psi_{\run}$. 

    Let $n^{\prime} = |C^{n+1}|$ and $m = |[i, \gamma-1 - \tau |C|]|$ for simplicity. 
    Then, $T[i..\gamma-1 - \tau |C|] = C^{n+1}[n^{\prime} - m + 1..n^{\prime}]$ follows from 
    $\lcs(T[i..\gamma-1 - \tau |C|], C^{n+1}) = T[i..\gamma-1 - \tau |C|]$. 
    Similarly, $T[p^{\prime}..\gamma^{\prime}-1] = C^{n+1}[n^{\prime} - |[p^{\prime}, \gamma^{\prime}-1]| + 1..n^{\prime}]$ follows 
    from $\lcs(T[p^{\prime}..\gamma^{\prime}-1], C^{n+1}) = T[p^{\prime}..\gamma^{\prime}-1]$. 
    $|[p^{\prime}, \gamma^{\prime}-1]| \geq m$ follows from $|[p^{\prime}, \gamma^{\prime}-1]| \geq |[i, \gamma-1]| - \tau |C|$ and $m = |[i, \gamma-1 - \tau |C|]|$. 
    String $C^{n+1}[n^{\prime} - m + 1..n^{\prime}]$ is a suffix of 
    string $C^{n+1}[n^{\prime} - |[p^{\prime}, \gamma^{\prime}-1]| + 1..n^{\prime}]$     
    because $|[p^{\prime}, \gamma^{\prime}-1]| \geq m$. 
    $\lcs(T[i..\gamma - 1 - \tau |C|], T[p^{\prime}..\gamma^{\prime}-1]) = C^{n+1}[n^{\prime} - m + 1..n^{\prime}]$ holds 
    because (1) $T[i..\gamma-1 - \tau |C|] = C^{n+1}[n^{\prime} - m + 1..n^{\prime}]$, 
    (2) $T[p^{\prime}..\gamma^{\prime}-1] = C^{n+1}[n^{\prime} - |[p^{\prime}, \gamma^{\prime}-1]| + 1..n^{\prime}]$, 
    and (3) $C^{n+1}[n^{\prime} - m + 1..n^{\prime}]$ is a suffix of $C^{n+1}[n^{\prime} - |[p^{\prime}, \gamma^{\prime}-1]| + 1..n^{\prime}]$. 
    Therefore, 
    $\lcs(T[i..\gamma - 1 - \tau |C|], T[p^{\prime}..\gamma^{\prime}-1]) = T[i..\gamma - 1 - \tau |C|]$ 
    follows from $\lcs(T[i..\gamma - 1 - \tau |C|], T[p^{\prime}..\gamma^{\prime}-1]) = C^{n+1}[n^{\prime} - m + 1..n^{\prime}]$ 
    and $T[i..\gamma-1 - \tau |C|] = C^{n+1}[n^{\prime} - m + 1..n^{\prime}]$. 

    \textbf{Proof of statement (iv).}
    Let $K^{\prime}$ be the length of the longest common prefix between two strings $T[\gamma^{\prime}..r^{\prime}]$ and $C^{n+1}$ 
    (i.e., $K^{\prime} = |\lcp(T[\gamma^{\prime}..r^{\prime}], C^{n+1})|$). 
    Because of $([p^{\prime}, q^{\prime}], [\ell^{\prime}, r^{\prime}]) \in \Psi_{h} \cap \Psi_{\run} \cap \Psi_{\centerset}(C)$, 
    $K^{\prime} > 1 + \sum_{w = 1}^{h+3} \lfloor \mu(w) \rfloor$ follows from the definition of the subset $\Psi_{\run}$. 
    String $T[\gamma^{\prime}..r^{\prime}+1]$ can be divided into two strings $C^{n+1}[1..K^{\prime}]$ and $T[\gamma^{\prime} + K^{\prime}..r^{\prime}+1]$. 
    $d < K^{\prime}$ follows from $K^{\prime} > 1 + \sum_{w = 1}^{h+3} \lfloor \mu(w) \rfloor$ and 
    $d \leq 1 + \sum_{w = 1}^{h+3} \lfloor \mu(w) \rfloor$ (statement (ii)).

    We prove $T[\gamma - \tau |C|..j] = C^{n+1}[1..\tau |C| + |[\gamma, j]|]$. 
    $\lcs(T[i..\gamma-1], C^{n+1}) = T[i..\gamma-1]$ 
    and $\lcp(T[\gamma..j], C^{n+1}) = T[\gamma..j]$ 
    follow from the two conditions of Lemma~\ref{lem:CCP_special_property}.          
    $T[\gamma - \tau |C|..\gamma-1] = C^{\tau}$ follows from 
    $\lcs(T[i..\gamma-1], C^{n+1}) = T[i..\gamma-1]$ and $i \leq \gamma - \tau |C|$ (statement (iii)). 
    $T[\gamma..j] = C^{n+1}[1..|[\gamma, j]|]$ follows from $\lcp(T[\gamma..j], C^{n+1}) = T[\gamma..j]$. 
    Therefore, $T[\gamma - \tau |C|..j] = C^{n+1}[1..\tau |C| + |[\gamma, j]|]$ follows from the following equation: 
    \begin{equation*}
        \begin{split}
            T[\gamma - \tau |C|..j] &= T[\gamma - \tau |C|..\gamma-1] \cdot T[\gamma..j] \\
            &= C^{\tau} \cdot (C^{n+1}[1..|[\gamma, j]|]) \\
            &= C^{n+1}[1..\tau |C| + |[\gamma, j]|].
        \end{split}
    \end{equation*}

    We prove $R \geq \min \{ d, \tau |C| + |[\gamma, j]| \}$. 
    $R \geq \min \{ K^{\prime}, \tau |C| + |[\gamma, j]| \}$ follows from 
    (1) $R = |\lcp(T[\gamma^{\prime}..r^{\prime}], T[\gamma - \tau |C|..j])|$, 
    (2) $T[\gamma^{\prime}..r^{\prime}+1] = C^{n+1}[1..K^{\prime}] \cdot T[\gamma^{\prime} + K^{\prime}..r^{\prime}+1]$, 
    and (3) $T[\gamma - \tau |C|..j] = C^{n+1}[1..\tau |C| + |[\gamma, j]|]$. 
    $\{ K^{\prime}, \tau |C| + |[\gamma, j]| \} \geq \min \{ d, \tau |C| + |[\gamma, j]| \}$ holds 
    because $d < K^{\prime}$. 
    Therefore, $R \geq \min \{ d, \tau |C| + |[\gamma, j]| \}$ follows from 
    $R \geq \min \{ K^{\prime}, \tau |C| + |[\gamma, j]| \}$ and $\{ K^{\prime}, \tau |C| + |[\gamma, j]| \} \geq \min \{ d, \tau |C| + |[\gamma, j]| \}$. 

    \textbf{Proof of statement (v).}
    We prove $T[i..j] = T[\gamma^{\prime} - |[i, \gamma-1]| + \tau |C|..\gamma^{\prime} + \tau |C| + |[\gamma, j]| - 1]$. 
    In this case, $R \geq \tau |C| + |[\gamma, j]|$ follows from statement (iv) for the integer $R$ of statement (iv). 
    $T[\gamma^{\prime}..\gamma^{\prime} + \tau |C| + |[\gamma, j]| - 1] = T[\gamma - \tau |C|..j]$ follows from 
    $R = |\lcp(T[\gamma^{\prime}..r^{\prime}], T[\gamma - \tau |C|..j])|$ and $R \geq \tau |C| + |[\gamma, j]|$. 
    Statement (iii) shows that 
    $T[i..\gamma - 1 - \tau |C|] = T[\gamma^{\prime} - |[i, \gamma-1]| + \tau |C|..\gamma^{\prime} - 1]$ holds. 
    Therefore, 
    $T[i..j] = T[\gamma^{\prime} - |[i, \gamma-1]| + \tau |C|..\gamma^{\prime} + \tau |C| + |[\gamma, j]| - 1]$ follows from 
    $T[i..\gamma - 1 - \tau |C|] = T[\gamma^{\prime} - |[i, \gamma-1]| + \tau |C|..\gamma^{\prime} - 1]$ 
    and $T[\gamma^{\prime}..\gamma^{\prime} + \tau |C| + |[\gamma, j]| - 1] = T[\gamma - \tau |C|..j]$. 

    We can apply Corollary~\ref{cor:capture_gamma_corollary} to the two intervals $[i, j]$ and $[\gamma^{\prime} - |[i, \gamma-1]| + \tau |C|, \gamma^{\prime} + \tau |C| + |[\gamma, j]| - 1]$ 
    because $T[i..j] = T[\gamma^{\prime} - |[i, \gamma-1]| + \tau |C|..\gamma^{\prime} + \tau |C| + |[\gamma, j]| - 1]$. 
    Because of $([p, q], [\ell, r]) \in \Psi_{h}$, 
    Corollary~\ref{cor:capture_gamma_corollary}~\ref{enum:capture_gamma_corollary:1} shows that 
    $I_{\capture}(\gamma^{\prime} - |[i, \gamma-1]| + \tau |C|, \gamma^{\prime} + \tau |C| + |[\gamma, j]| - 1) \in \Psi_{h}$ 
    because $I_{\capture}(i, j) = ([p, q], [\ell, r]) \in \Psi_{h}$. 
    Therefore, statement (v) holds. 
    
    \textbf{Proof of statement (vi).}
    We prove $T[i..\gamma - \tau |C| + d - 1] = T[\gamma^{\prime} - |[i, \gamma-1]| + \tau |C|..\gamma^{\prime} + d - 1]$. 
    In this case, $R \geq d$ follows from statement (iv) for the integer $R$ of statement (iv). 
    $T[\gamma^{\prime}..\gamma^{\prime} + d - 1] = T[\gamma - \tau |C|..\gamma - \tau |C| + d - 1]$ follows from 
    $R = |\lcp(T[\gamma^{\prime}..r^{\prime}], T[\gamma - \tau |C|..j])|$ and $R \geq d$. 
    Statement (iii) shows that 
    $T[i..\gamma - 1 - \tau |C|] = T[\gamma^{\prime} - |[i, \gamma-1]| + \tau |C|..\gamma^{\prime} - 1]$ holds. 
    Therefore, $T[i..\gamma - \tau |C| + d - 1] = T[\gamma^{\prime} - |[i, \gamma-1]| + \tau |C|..\gamma^{\prime} + d - 1]$ 
    follows from 
    $T[i..\gamma - 1 - \tau |C|] = T[\gamma^{\prime} - |[i, \gamma-1]| + \tau |C|..\gamma^{\prime} - 1]$ 
    and $T[\gamma^{\prime}..\gamma^{\prime} + d - 1] = T[\gamma - \tau |C|..\gamma - \tau |C| + d - 1]$. 

    $I_{\capture}(\gamma^{\prime} - |[i, \gamma-1]| + \tau |C|, \gamma^{\prime} + d - 1) = ([p^{\prime}, q^{\prime}], [\ell^{\prime}, r^{\prime}])$ follows from the definition of the integer $d$. 
    We can apply Corollary~\ref{cor:capture_gamma_corollary} to the two intervals 
    $[\gamma^{\prime} - |[i, \gamma-1]| + \tau |C|, \gamma^{\prime} + d - 1]$ and $[i, \gamma - \tau |C| + d - 1]$ 
    because $T[i..\gamma - \tau |C| + d - 1] = T[\gamma^{\prime} - |[i, \gamma-1]| + \tau |C|..\gamma^{\prime} + d - 1]$. 
    Corollary~\ref{cor:capture_gamma_corollary}~\ref{enum:capture_gamma_corollary:1} shows that 
    $I_{\capture}(i, \gamma - \tau |C| + d - 1) \in \Psi_{h}$ holds 
    because $I_{\capture}(\gamma^{\prime} - |[i, \gamma-1]| + \tau |C|, \gamma^{\prime} + d - 1) = ([p^{\prime}, q^{\prime}], [\ell^{\prime}, r^{\prime}]) \in \Psi_{h}$. 
    Therefore, statement (vi) holds. 
    
    \textbf{Proof of Proposition~\ref{prop:CCP_special_left_property}.}
    We prove Proposition~\ref{prop:CCP_special_left_property} by contradiction, 
    i.e., we show that there exists a contradiction under the assumption that $|[p, \gamma-1]| < |[i, \gamma-1]|$ holds. 
    Consider two interval attractors $I_{\capture}(\gamma^{\prime} - |[i, \gamma-1]| + \tau |C|, \gamma^{\prime} + \tau |C| + |[\gamma, j]| - 1) = ([p_{A}, q_{A}], [\ell_{A}, r_{A}])$ 
    and $I_{\capture}(i, \gamma - \tau |C| + d - 1]) = ([p_{B}, q_{B}], [\ell_{B}, r_{B}])$. 
    From statement (v) and statement (vi), 
    one of the following six conditions is satisfied: 
    \begin{enumerate}[label=\textbf{(\Alph*)}]
    \item $\tau |C| + |[\gamma, j]| < d$ and $r_{A} < r^{\prime}$;
    \item $\tau |C| + |[\gamma, j]| < d$ and $r_{A} > r^{\prime}$;
    \item $\tau |C| + |[\gamma, j]| < d$ and $r_{A} = r^{\prime}$;
    \item $\tau |C| + |[\gamma, j]| \geq d$ and $r_{B} < r$;
    \item $\tau |C| + |[\gamma, j]| \geq d$ and $r_{B} > r$;
    \item $\tau |C| + |[\gamma, j]| \geq d$ and $r_{B} = r$.    
    \end{enumerate}

    For condition (A),     
    Lemma~\ref{lem:interval_extension_propertyX}~\ref{enum:interval_extension_propertyX:right} shows that 
    $I_{\capture}(\gamma^{\prime} - |[i, \gamma-1]| + \tau |C|, r_{A}+1) \in \bigcup_{t = h+1}^{H} \Psi_{t}$ 
    because $I_{\capture}(\gamma^{\prime} - |[i, \gamma-1]| + \tau |C|, \gamma^{\prime} + \tau |C| + |[\gamma, j]| - 1) = ([p_{A}, q_{A}], [\ell_{A}, r_{A}])$ and $([p_{A}, q_{A}], [\ell_{A}, r_{A}]) \in \Psi_{h}$ hold (statement (v)).
    Lemma~\ref{lem:IA_maximal_lemma} shows that 
    $I_{\capture}(\gamma^{\prime} - |[i, \gamma-1]| + \tau |C|, r^{\prime}) = ([p^{\prime}, q^{\prime}], [\ell^{\prime}, r^{\prime}])$ holds. 
    because $\gamma^{\prime} - |[i, \gamma-1]| + \tau |C| \in [p^{\prime}, q^{\prime}]$.
    Corollary~\ref{cor:IA_basic_corollary}~\ref{enum:IA_basic_corollary:3} shows that 
    $I_{\capture}(\gamma^{\prime} - |[i, \gamma-1]| + \tau |C|, r^{\prime}) \in \bigcup_{t = h+1}^{H} \Psi_{t}$ holds 
    because $I_{\capture}(\gamma^{\prime} - |[i, \gamma-1]| + \tau |C|, r_{A}+1) \subseteq \bigcup_{t = h+1}^{H} \Psi_{t}$ 
    and $[\gamma^{\prime} - |[i, \gamma-1]| + \tau |C|, r_{A}+1] \subseteq [\gamma^{\prime} - |[i, \gamma-1]| + \tau |C|, r^{\prime}]$. 
    $h^{\prime} \in [h+1, H]$ holds for the level $h^{\prime}$ of the interval attractor $([p^{\prime}, q^{\prime}], [\ell^{\prime}, r^{\prime}])$. 
    On the other hand, 
    $h^{\prime} = h$ holds because $([p^{\prime}, q^{\prime}], [\ell^{\prime}, r^{\prime}]) \in \Psi_{h}$ holds. 
    The two facts $h^{\prime} \in [h+1, H]$ and $h^{\prime} = h$ yield a contradiction.

    For condition (B), we can show that there exists a contradiction using the same approach as for condition (A).
    
    For condition (C), 
    consider two interval attractors 
    $I_{\capture}(\gamma^{\prime} - |[i, \gamma-1]| + \tau |C|, r^{\prime})$ and $I_{\capture}(\gamma^{\prime} - |[i, \gamma-1]| + \tau |C|, r_{A})$.
    Lemma~\ref{lem:IA_maximal_lemma} shows that 
    $I_{\capture}(\gamma^{\prime} - |[i, \gamma-1]| + \tau |C|, r^{\prime}) = ([p^{\prime}, q^{\prime}], [\ell^{\prime}, r^{\prime}])$ holds.     
    Lemma~\ref{lem:IA_maximal_lemma} shows that 
    $I_{\capture}(\gamma^{\prime} - |[i, \gamma-1]| + \tau |C|, r_{A}) = ([p_{A}, q_{A}]$, $[\ell_{A}, r_{A}]) $ holds 
    because $I_{\capture}(\gamma^{\prime} - |[i, \gamma-1]| + \tau |C|, \gamma^{\prime} + \tau |C| + |[\gamma, j]| - 1) = ([p_{A}, q_{A}]$, $[\ell_{A}, r_{A}])$. 
    $([p^{\prime}, q^{\prime}], [\ell^{\prime}, r^{\prime}]) = ([p_{A}, q_{A}], [\ell_{A}, r_{A}])$ follows from 
    $I_{\capture}(\gamma^{\prime} - |[i, \gamma-1]| + \tau |C|, r^{\prime}) = ([p^{\prime}, q^{\prime}], [\ell^{\prime}, r^{\prime}])$, 
    $I_{\capture}(\gamma^{\prime} - |[i, \gamma-1]| + \tau |C|, \gamma^{\prime} + \tau |C| + |[\gamma, j]| - 1) = ([p_{A}, q_{A}], [\ell_{A}, r_{A}])$, 
    and $r_{A} = r^{\prime}$. 
    In this case, $d \leq \tau |C| + |[\gamma, j]|$ follows from the definition of the integer $d$ 
    because 
    $I_{\capture}(\gamma^{\prime} - |[i, \gamma-1]| + \tau |C|, \gamma^{\prime} + \tau |C| + |[\gamma, j]| - 1) = ([p^{\prime}, q^{\prime}], [\ell^{\prime}, r^{\prime}])$ holds. 
    The two facts $\tau |C| + |[\gamma, j]| < d$ and $d \leq \tau |C| + |[\gamma, j]|$ yield a contradiction.

    For condition (D), 
    consider two interval attractors $I_{\capture}(i, r_{B}+1)$ and $I_{\capture}(i, r)$. 
    Since $I_{\capture}(i, j) = ([p, q], [\ell, r])$, 
    $i \in [p, q]$ follows from the definition of interval attractor. 
    Lemma~\ref{lem:IA_maximal_lemma} shows that $I_{\capture}(i, r) = ([p, q], [\ell, r])$. 
    
    Lemma~\ref{lem:interval_extension_propertyX}~\ref{enum:interval_extension_propertyX:right} shows that 
    $I_{\capture}(i, r_{B}+1) \in \bigcup_{t = h+1}^{H} \Psi_{t}$ 
    because $I_{\capture}(i, \gamma - \tau |C| + d - 1) = ([p_{B}, q_{B}], [\ell_{B}, r_{B}])$ and 
    $([p_{B}, q_{B}], [\ell_{B}, r_{B}]) \in \Psi_{h}$ hold (statement (vi)).
    Corollary~\ref{cor:IA_basic_corollary}~\ref{enum:IA_basic_corollary:3} shows that 
    $I_{\capture}(i, r) \in \bigcup_{t = h+1}^{H} \Psi_{t}$ holds 
    because $I_{\capture}(i, r_{B}+1) \in \bigcup_{t = h+1}^{H} \Psi_{t}$  
    and $[i, r_{B}+1] \subseteq [i, r]$. 
    On the other hand, $I_{\capture}(i, r) \not \in \bigcup_{t = h+1}^{H} \Psi_{t}$ 
    because $I_{\capture}(i, r) = ([p, q], [\ell, r])$ and $([p, q], [\ell, r]) \in \Psi_{h}$. 
    The two facts $I_{\capture}(i, r_{B}+1) \in \bigcup_{t = h+1}^{H} \Psi_{t}$ and 
    $I_{\capture}(i, r) \not \in \bigcup_{t = h+1}^{H} \Psi_{t}$ yield a contradiction.

    For condition (E), 
    we can show that there exists a contradiction using the same approach as for condition (D). 

    For condition (F), 
    we can show that $([p_{B}, q_{B}], [\ell_{B}, r_{B}]) = ([p, q], [\ell, r])$ holds using the same approach for condition (C). 
    In this case, 
    statement (iv) shows that 
    $T[i..\gamma - \tau |C| + d - 1] = T[\gamma^{\prime} - |[i, \gamma-1]| + \tau |C|..\gamma^{\prime} + d - 1]$ 
    and $I_{\capture}(i, \gamma - \tau |C| + d - 1) = ([p_{B}, q_{B}], [\ell_{B}, r_{B}])$ hold.     
    Here, $I_{\capture}(i, \gamma - \tau |C| + d - 1) = ([p, q], [\ell, r])$ holds 
    because $([p_{B}, q_{B}], [\ell_{B}, r_{B}]) = ([p, q], [\ell, r])$. 
    
    Consider interval attractor $I_{\capture}(\gamma^{\prime} - |[i, \gamma-1]| + \tau |C|, \gamma^{\prime} + d - 1)$. 
    Then, $I_{\capture}(\gamma^{\prime} - |[i, \gamma-1]| + \tau |C|, \gamma^{\prime} + d - 1) = ([p^{\prime}, q^{\prime}], [\ell^{\prime}, r^{\prime}])$ follows from the definition of the integer $d$. 
    We can apply Corollary~\ref{cor:capture_gamma_corollary} to the two intervals 
    $[i, \gamma - \tau |C| + d - 1]$ and $[\gamma^{\prime} - |[i, \gamma-1]| + \tau |C|, \gamma^{\prime} + d - 1]$ 
    because $T[i..\gamma - \tau |C| + d - 1] = T[\gamma^{\prime} - |[i, \gamma-1]| + \tau |C|..\gamma^{\prime} + d - 1]$. 
    Corollary~\ref{cor:capture_gamma_corollary} shows that 
    $|[i, \gamma-1]| = |[\gamma^{\prime} - |[i, \gamma-1]| + \tau |C|, \gamma^{\prime} - 1]|$ holds 
    because $I_{\capture}(i, \gamma - \tau |C| + d - 1) = ([p, q], [\ell, r])$ 
    and $I_{\capture}(\gamma^{\prime} - |[i, \gamma-1]| + \tau |C|, \gamma^{\prime} + d - 1) = ([p^{\prime}, q^{\prime}], [\ell^{\prime}, r^{\prime}])$.
    $\tau |C| = 0$ follows from $|[i, \gamma-1]| = |[\gamma^{\prime} - |[i, \gamma-1]| + \tau |C|, \gamma^{\prime} - 1]|$. 
    $|C| \geq 1$ holds because $C \in \Sigma^{+}$. 
    $\tau = 0$ follows from $\tau |C| = 0$ and $|C| \geq 1$. 
    This equation $\tau = 0$ indicates that 
    $|[p^{\prime}, \gamma^{\prime}-1]| \geq |[i, \gamma-1]|$ holds 
    because $\tau$ is the smallest integer satisfying $|[p^{\prime}, \gamma^{\prime}-1]| \geq |[i, \gamma-1]| - \tau |C|$. 
    On the other hand, we assumed that $|[p, \gamma-1]| < |[i, \gamma-1]|$ holds. 
    Therefore, the two facts $|[p^{\prime}, \gamma^{\prime}-1]| \geq |[i, \gamma-1]|$ and $|[p, \gamma-1]| < |[i, \gamma-1]|$ yield a contradiction. 

    We showed that there exists a contradiction under the assumption that $|[p, \gamma-1]| < |[i, \gamma-1]|$ holds. 
    Therefore, $|[p^{\prime}, \gamma^{\prime}-1]| \geq |[i, \gamma-1]|$ must hold. 
\end{proof}

\begin{proof}[Proof of Lemma~\ref{lem:CCP_special_property}]
    Consider an interval attractor $([p^{\prime}, q^{\prime}], [\ell^{\prime}, r^{\prime}])$ in set $\Psi_{h} \cap \Psi_{\run} \cap \Psi_{\centerset}(C) \cap (\bigcup_{\lambda = |[\gamma, j]|}^{n} \Psi_{\lcp}(\lambda))$. 
    We prove $T[i..\gamma-1] = T[\gamma^{\prime} - |[i, \gamma-1]|..\gamma^{\prime}-1]$ for the attractor position $\gamma^{\prime}$ 
    of the interval attractor $([p^{\prime}, q^{\prime}], [\ell^{\prime}, r^{\prime}])$. 
    Proposition~\ref{prop:CCP_special_left_property} shows that 
    $|[p^{\prime}, \gamma^{\prime}-1]| \geq |[i, \gamma-1]|$ holds.
    Because of $([p^{\prime}, q^{\prime}], [\ell^{\prime}, r^{\prime}]) \in \Psi_{\run} \cap \Psi_{\centerset}(C)$, 
    $\lcs(T[p^{\prime}-1..\gamma^{\prime} - 1], C^{n+1}) = T[p^{\prime}-1..\gamma^{\prime} - 1]$ follows from the definition of the subset $\Psi_{\run}$. 
    $\lcs(T[\gamma^{\prime} - |[i, \gamma-1]|..\gamma^{\prime} - 1], C^{n+1}) = T[\gamma^{\prime} - |[i, \gamma-1]|..\gamma^{\prime} - 1]$ 
    follows from $\lcs(T[p^{\prime}-1..\gamma^{\prime} - 1], C^{n+1}) = T[p^{\prime}-1..\gamma^{\prime} - 1]$ and $|[p^{\prime}, \gamma^{\prime}-1]| \geq |[i, \gamma-1]|$.     
    $\lcs(T[i..\gamma-1], C^{n+1}) = T[i..\gamma-1]$ follows from the first condition of Lemma~\ref{lem:CCP_special_property}. 

    Let $n^{\prime} = |C^{n+1}|$ for simplicity. 
    $T[\gamma^{\prime} - |[i, \gamma-1]|..\gamma^{\prime} - 1] = C^{n+1}[n^{\prime} - |[i, \gamma-1]|..n^{\prime}]$ follows from $\lcs(T[\gamma^{\prime} - |[i, \gamma-1]|..\gamma^{\prime} - 1], C^{n+1}) = T[\gamma^{\prime} - |[i, \gamma-1]|..\gamma^{\prime} - 1]$. 
    Similarly, 
    $T[i..\gamma-1] = C^{n+1}[n^{\prime} - |[i, \gamma-1]|..n^{\prime}]$ follows from $\lcs(T[i..\gamma-1], C^{n+1}) = T[i..\gamma-1]$.     
    Therefore, $T[i..\gamma-1] = T[\gamma^{\prime} - |[i, \gamma-1]|..\gamma^{\prime}-1]$ follows from 
    $T[\gamma^{\prime} - |[i, \gamma-1]|..\gamma^{\prime} - 1] = C^{n+1}[n^{\prime} - |[i, \gamma-1]|..n^{\prime}]$ and 
    $T[i..\gamma-1] = C^{n+1}[n^{\prime} - |[i, \gamma-1]|..n^{\prime}]$. 

    Next, we prove $T[\gamma..j] = T[\gamma^{\prime}..\gamma^{\prime} + |[\gamma, j]| - 1]$. 
    $\lcp(T[\gamma..j], C^{n+1}) = T[\gamma..j]$ follows from the second condition of Lemma~\ref{lem:CCP_special_property}. 
    $T[\gamma..j] = C^{n+1}[1..|[\gamma, j]|]$ follows from $\lcp(T[\gamma..j], C^{n+1}) = T[\gamma..j]$. 
    Because of $([p^{\prime}, q^{\prime}], [\ell^{\prime}, r^{\prime}]) \in \bigcup_{\lambda = |[\gamma, j]|}^{n} \Psi_{\lcp}(\lambda)$, 
    there exists an integer $K^{\prime} \in [|[\gamma, j]|, n]$ satisfying 
    $([p^{\prime}, q^{\prime}], [\ell^{\prime}, r^{\prime}]) \in \Psi_{\lcp}(K^{\prime})$. 
    $|\lcp(T[\gamma^{\prime}..r^{\prime}], C^{n+1})| = K^{\prime}$ follows from the definition of the subset $\Psi_{\lcp}(K^{\prime})$. 
    $T[\gamma^{\prime}..\gamma^{\prime} + |[\gamma, j]| - 1] = C^{n+1}[1..|[\gamma, j]|]$ follows from 
    $|\lcp(T[\gamma^{\prime}..r^{\prime}], C^{n+1})| = K^{\prime}$ and $K^{\prime} \geq |[\gamma, j]|$. 
    Therefore, $T[\gamma..j] = T[\gamma^{\prime}..\gamma^{\prime} + |[\gamma, j]| - 1]$ follows from 
    $T[\gamma..j] = C^{n+1}[1..|[\gamma, j]|]$ and $T[\gamma^{\prime}..\gamma^{\prime} + |[\gamma, j]| - 1] = C^{n+1}[1..|[\gamma, j]|]$. 

    We prove $([p^{\prime}, q^{\prime}], [\ell^{\prime}, r^{\prime}]) \in \Psi_{\CCP}(T[i..j])$. 
    $([p^{\prime}, q^{\prime}], [\ell^{\prime}, r^{\prime}]) \in \Psi_{h}$ holds, 
    and we showed that $T[i..\gamma-1] = T[\gamma^{\prime} - |[i, \gamma-1]|..\gamma^{\prime}-1]$ 
    and $T[\gamma..j] = T[\gamma^{\prime}..\gamma^{\prime} + |[\gamma, j]| - 1]$. 
    Therefore, Lemma~\ref{lem:CCP_property}~\ref{enum:CCP_property:3} shows that 
    $([p^{\prime}, q^{\prime}], [\ell^{\prime}, r^{\prime}]) \in \Psi_{\CCP}(T[i..j])$ holds. 

    We showed that $([p^{\prime}, q^{\prime}], [\ell^{\prime}, r^{\prime}]) \in \Psi_{\CCP}(T[i..j])$ holds 
    for each interval attractor $([p^{\prime}, q^{\prime}], [\ell^{\prime}, r^{\prime}])$ in set $\Psi_{h} \cap \Psi_{\run} \cap \Psi_{\centerset}(C) \cap (\bigcup_{\lambda = |[\gamma, j]|}^{n} \Psi_{\lcp}(\lambda))$. 
    Therefore, $\Psi_{h} \cap \Psi_{\run} \cap \Psi_{\centerset}(C) \cap (\bigcup_{\lambda = |[\gamma, j]|}^{n} \Psi_{\lcp}(\lambda)) \subseteq \Psi_{\CCP}(T[i..j])$ holds.     
\end{proof}

\subsection{Proof of Theorem~\ref{theo:RR_Psi_set_size}}\label{subsec:size_proof}
In this section, we will prove Theorem~\ref{theo:RR_Psi_set_size}. 
For proving the theorem, 
we introduce two sets of positions in input string $T$ and three subsets of interval attractors $\Psi_{\RR}$. 

\paragraph{$d$-cover.}
A set $\mathsf{C} \subseteq [1, n]$ is called a $d$-\emph{cover} of input string $T$ if 
(i) $[\max \{ 1, n - d + 1 \}, n] \subseteq \mathsf{C}$, and (ii) $[i, i + d - 1] \subseteq \mathsf{C}$ for every $i \in [1, n - d + 1]$ 
satisfying $i = \min \Occ(T, T[i..i + d - 1])$. 
We introduce a set $\mathsf{C}_{\cover}(d)$ as a $d$-cover of $T$.  
Let $k = 2^{\lceil \log d \rceil}$ and $\mathsf{I} = \{ i \in [1, n-2k] \mid i \mod k = 1 \mbox{ and } i = \min \Occ(T, T[i..i+2k-1]) \}$. 
Then, the set $\mathsf{C}_{\cover}(d)$ is defined as $\mathsf{C}_{\cover}(d) = \bigcup_{i \in \mathsf{I}} [i, i + 2k -1] \cup [\max \{1, n - 2k + 1 \}, n]$. 
Kempa and Kociumaka shows that this set $\mathsf{C}_{\cover}(d)$ is a $d$-cover of $T$~\cite{DBLP:conf/focs/KempaK23}. 
For two integers $d$ and $d^{\prime}$ satisfying $1 \leq d < d^{\prime}$, 
$\mathsf{C}_{\cover}(d) \subseteq \mathsf{C}_{\cover}(d^{\prime})$ follows from 
the definition of the $d$-cover.

The following lemma states properties of $d$-cover $\mathsf{C}_{\cover}(d)$.
\begin{lemma}\label{lem:m_cover_properties}[Lemma 4.7 in \cite{DBLP:journals/corr/abs-2308-03635}]
The following two statements hold for an integer $d \geq 0$: 
\begin{enumerate}[label=\textbf{(\roman*)}]
\item \label{enum:m_cover_properties:1} $|\mathsf{C}_{\cover}(d)| = |\mathcal{S}_{8d}| + 8d \leq 16 d \delta$ for the substring complexity $\delta$ of input string $T$ and the set $\mathcal{S}_{8d}$ of strings introduced in Section~\ref{sec:preliminary}; 
\item \label{enum:m_cover_properties:2} there exists a sequence of $k$ disjoint intervals $I_{1}, I_{2}, \ldots, I_{k} \subseteq [1, n]$ in input string $T$ 
such that (A) the union of the $k$ intervals is equal to the set $\mathsf{C}_{\cover}(d)$ (i.e., $\bigcup_{s = 1}^{k} I_{s} = \mathsf{C}_{\cover}(d)$), 
and (B) $k \leq \frac{|\mathcal{S}_{8d}|}{d} + 8 \leq 16 \delta$.
\end{enumerate}
\end{lemma}

The following lemma shows states the relationship between 
the substring complexity $\delta$ of $T$ and the size of the $\lceil \mu(t+1) \rceil$-cover $\mathsf{C}_{\cover}(\lceil \mu(t+1) \rceil)$ for an integer $t \geq 0$. 
\begin{lemma}\label{lem:d_cover_total_size}
$\sum_{t = 0}^{\infty} \frac{|\mathsf{C}_{\cover}(c\lceil \mu(t+1) \rceil)|}{\lceil \mu(t+1) \rceil} = O(c^{2} \delta \log \frac{n \log \sigma}{\delta \log n})$ 
for every integer $c \geq 1$. 
\end{lemma}
\begin{proof}
    Let $m_{t} = 2 \lfloor 16 \mu(t) \rfloor + \floor \mu(t+1) \rfloor$ for an integer $t \geq 0$. 
    Then, Kempa and Kociumaka showed that 
    $\sum_{t = 0}^{\infty} \frac{|\mathsf{C}_{\cover}(m_{t})|}{m_{t}} = O(\delta \log \frac{n \log \sigma}{\delta \log n})$ 
    (Lemma 5.18 in \cite{DBLP:journals/corr/abs-2308-03635}). 
    Because of $\frac{m_{t}}{\lceil \mu(t+1) \rceil} = O(1)$, 
    there exists a constant $c_{1} \geq 1$ satisfying $c\lceil \mu(t+1) \rceil \leq (c + c_{1}) m_{t}$.
    Similarly, 
    there exists a constant $c_{2} \geq 1$ satisfying $(c + c_{1}) m_{t} \leq (c + c_{2}) \lceil \mu(t+1) \rceil$. 
    We already showed that 
    $\mathsf{C}_{\cover}(d) \subseteq \mathsf{C}_{\cover}(d^{\prime})$ 
    for two integers $d$ and $d^{\prime}$ satisfying $1 \leq d < d^{\prime}$. 
    Therefore, $|\mathsf{C}_{\cover}(c\lceil \mu(t+1) \rceil)| \leq |\mathsf{C}_{\cover}((c + c_{1}) m_{t})|$ holds.

    By generalizing Lemma 5.18 in \cite{DBLP:journals/corr/abs-2308-03635}, 
    we obtain   
    $\sum_{t = 0}^{\infty} \frac{|\mathsf{C}_{\cover}((c + c_{1}) m_{t})|}{(c + c_{1})m_{t}} = O((c + c_{1}) \delta \log \frac{n \log \sigma}{\delta \log n})$. 
    For every integer $t \geq 0$, 
    $\frac{|\mathsf{C}_{\cover}(c\lceil \mu(t+1) \rceil)|}{(c+c_{2}) \lceil \mu(t+1) \rceil } \leq \frac{|\mathsf{C}_{\cover}((c + c_{1}) m_{t})|}{(c + c_{1}) m_{t}}$ 
    follows from $|\mathsf{C}_{\cover}(c\lceil \mu(t+1) \rceil)| \leq |\mathsf{C}_{\cover}((c + c_{1}) m_{t})|$ 
    and $(c + c_{1}) m_{t} \leq (c + c_{2}) \lceil \mu(t+1) \rceil$. 
    Therefore, 
    \begin{equation*}
        \begin{split}
            & \sum_{t = 0}^{\infty} \frac{|\mathsf{C}_{\cover}(c\lceil \mu(t+1) \rceil)|}{\lceil \mu(t+1) \rceil} \\
            &= (c + c_{2}) \sum_{t = 0}^{\infty} \frac{|\mathsf{C}_{\cover}(c\lceil \mu(t+1) \rceil)|}{(c + c_{2}) \lceil \mu(t+1) \rceil} \\
            &\leq (c + c_{2}) \sum_{t = 0}^{\infty} \frac{|\mathsf{C}_{\cover}((c + c_{1}) m_{t})|}{(c + c_{1})m_{t}} \\
            &= O(c^{2} \delta \log \frac{n \log \sigma}{\delta \log n}).
        \end{split}
    \end{equation*}
\end{proof}

\paragraph{Set $\mathcal{B}_{\bstart}(h)$ of positions.}
For an integer $h \in [0, H]$, 
let $\mathcal{B}_{\bstart}(h)$ be the set of the starting positions of the substring derived from each nonterminal of sequence $S^{h}$ in input string $T$. 
Formally, let $\mathcal{B}_{\bstart}(h) = \{ 1 + \sum_{w = 1}^{s-1} |\val(S^{h}[w])| \mid 1 \leq s \leq |S^{h}| \}$. 
$\mathcal{B}_{\bstart}(H) = \{ 1 \}$, $\mathcal{B}_{\bstart}(H) \subseteq \mathcal{B}_{\bstart}(H-1) \subseteq \cdots \subseteq \mathcal{B}_{\bstart}(0)$, and $\mathcal{B}_{\bstart}(0) = \{ 1, 2, \ldots, n \}$ follow from the definition of the set $\mathcal{B}_{\bstart}(h)$. 

The following proposition states properties of the set $\mathcal{B}_{\bstart}(h)$. 

\begin{proposition}\label{prop:b_set_properties}
The following two statements hold for each integer $h \in [0, H]$: 
\begin{enumerate}[label=\textbf{(\roman*)}]
\item \label{enum:b_set_properties:1} $\mathbb{E}[|\mathcal{B}_{\bstart}(h) \cap [i, j]|] < 1 + \frac{4 |[i, j]| }{\mu(h+1)}$ for an interval $[i, j] \subseteq [1, n]$ in input string $T$ of length $n$; 
%\item \label{enum:b_set_properties:5} $\mathbb{E}[|\mathcal{B}_{\bstart}(h) \cap [i, j]|] < 1 + \frac{4 |[i, j]| }{\mu(h+1)}$ for 
%an interval $[i, j] \subseteq [1, n]$ in input string $T$ of length $n$~(Lemma V.12 in~\cite{9961143})
\item \label{enum:b_set_properties:4} 
let $b = 1$ if $h > 0$; otherwise $b = n$. 
Then, set $\mathcal{B}_{\bstart} \setminus \{ b \}$ contains 
the attractor position of each interval attractor in the $h$-th level interval attractors $\Psi_{h}$.
\end{enumerate}
\end{proposition}
\begin{proof}
The proof of Proposition~\ref{prop:b_set_properties} is as follows.
\paragraph{Proof of Proposition~\ref{prop:b_set_properties}(i).}
Similar to set $\mathcal{B}_{\bstart}(h)$, 
let $\mathcal{B}_{\bend}(h)$ be the set of the ending positions of the substring derived from each nonterminal of sequence $S^{h}$ in input string $T$. 
Then, $\mathbb{E}[|\mathcal{B}_{\bend}(h) \cap [i, j]|] < 1 + \frac{4 |[i, j]| }{\mu(h+1)}$ holds~(Lemma V.12 in~\cite{9961143}).
Similarly, we can prove $\mathbb{E}[|\mathcal{B}_{\bstart}(h) \cap [i, j]|] < 1 + \frac{4 |[i, j]| }{\mu(h+1)}$ 
using the fact that $\mathcal{B}_{\bstart}(h) = \{ 1 \} \cup \{ x+1 \mid x \in \mathcal{B}_{\bstart}(h) \setminus \{ n \} \}$.

\paragraph{Proof of Proposition~\ref{prop:b_set_properties}(ii).}
We prove Proposition~\ref{prop:b_set_properties}~\ref{enum:b_set_properties:4}. 
Let $([p, q], [\ell, r])$ be an interval attractor in set $\Psi_{h}$.
We show that set $\mathcal{B}_{\bstart}$ contains 
the attractor position of the interval attractor. 
From Definition~\ref{def:RR_Delta}, 
there exists a position $s^{h} \in [1, |S^{h}|]$ of sequence $S^{h}$ 
satisfying $I(s^{h}) = ([p, q], [\ell, r])$. 
The attractor position $\gamma$ of the interval attractor is defined as $x^{h}_{s}$ 
for the substring $T[x^{h}_{s}..y^{h}_{s}]$ derived from the $s^{h}$-th nonterminal of sequence $S^{h}$ in $T$. 
Therefore, $\gamma \in \mathcal{B}_{\bstart}$. 

Next, we prove $\gamma \neq b$. 
Lemma~\ref{lem:IA_maximal_lemma} shows that $I_{\capture}(p, r) = ([p, q], [\ell, r])$ for interval attractor $I_{\capture}(p, r)$. 
Consider the tail $[s^{\prime k}, e^{\prime k}]$ of sequence $A(p, r)$. 
Then, $k = h$ and $s^{\prime h} = s^{h}$ hold. 
If $h \neq 0$, 
then Lemma~\ref{lem:rec_function_basic_relation}~\ref{enum:rec_function_basic_relation:4} shows that 
$x^{h}_{s} \geq p + 1$. 
$\gamma \neq b$ follows from $\gamma = x^{h}_{s}$, 
$x^{h}_{s} \geq p + 1$, $p \geq 1$, and $b = 1$. 

Otherwise (i.e., $h = 0$), 
$q \leq \gamma < \ell \leq n$ follows from Lemma~\ref{lem:IA_super_basic_property}~\ref{enum:IA_super_basic_property:1}.
$\gamma \neq b$ follows from $\gamma < n$ and $b = n$. 
Therefore, $\gamma \in \mathcal{B}_{\bstart}(h) \setminus \{ b \}$ follows from 
$\gamma \in \mathcal{B}_{\bstart}$ and $\gamma \neq b$. 
\end{proof}

The following lemma states the expected size of the union of intersection of the two sets $\mathcal{B}_{\bstart}(t)$ and $\mathsf{C}_{\cover}(d)$. 
\begin{lemma}\label{lem:d_cover_expected_size}
$\mathbb{E}[ |\mathcal{B}_{\bstart}(t) \cap \mathsf{C}_{\cover}(d)|] \leq 8 + \frac{|\mathcal{S}_{8d}|}{d} + \frac{4 |\mathsf{C}_{\cover}(d)| }{\mu(t+1)}$ for two integers $t \in [0, H]$ and $d \geq 1$. 
\end{lemma}
\begin{proof}
We can divide the set $\mathsf{C}_{\cover}(d)$ into $k$ disjoint intervals 
$I_{1}, I_{2}, \ldots, I_{k} \subseteq [1, n]$. 
$k \leq \frac{|\mathcal{S}_{8d}|}{d} + 8$ follows from Lemma~\ref{lem:m_cover_properties}~\ref{enum:m_cover_properties:2}. 
For each interval $I_{s}$, 
$\mathbb{E}[|\mathcal{B}_{\bstart}(t) \cap I_{s}|] < 1 + \frac{4 |I_{s}| }{\mu(t+1)}$ follows from 
Proposition~\ref{prop:b_set_properties}~\ref{enum:b_set_properties:1}. 
Therefore, 
\begin{equation*}
    \begin{split}
        \mathbb{E}[ |\mathcal{B}_{\bstart}(t) \cap \mathsf{C}_{\cover}(d)|] &= \sum_{s=1}^{k} \mathbb{E}[|\mathcal{B}_{\bstart}(t) \cap I_{s}|] \\
        &\leq \sum_{s=1}^{k} (1 + \frac{4 |I_{s}| }{\mu(t+1)}) \\
        &= k + \frac{4 |\mathsf{C}_{\cover}(d)| }{\mu(t+1)} \\
        &\leq 8 + \frac{|\mathcal{S}_{8d}|}{d} + \frac{4 |\mathsf{C}_{\cover}(d)| }{\mu(t+1)}.
    \end{split}
\end{equation*}
\end{proof}

\paragraph{Two subsets $\Psi_{\leftLen}(d, d^{\prime})$ and $\Psi_{\rightLen}(d, d^{\prime})$.}
For two real numbers $d$ and $d^{\prime}$, 
subset $\Psi_{\leftLen}(d, d^{\prime}) \subseteq \Psi_{\RR}$ consists of 
interval attractors such that 
each interval attractor $([p, q], [\ell, r]) \in \Psi_{\leftLen}(d, d^{\prime})$ with attractor position $\gamma$ satisfies 
$d < |[p-1, \gamma-1]| \leq d^{\prime}$ (i.e., $\Psi_{\leftLen}(d, d^{\prime}) = \{ ([p, q], [\ell, r]) \in \Psi_{\RR} \mid d < |[p-1, \gamma-1]| \leq d^{\prime} \}$). 
Similarly, 
subset $\Psi_{\rightLen}(d, d^{\prime}) \subseteq \Psi_{\RR}$ consists of 
interval attractors such that 
each interval attractor $([p, q], [\ell, r]) \in \Psi_{\rightLen}(d, d^{\prime})$ with attractor position $\gamma$ satisfies 
$d < |[\gamma, r+1]| \leq d^{\prime}$.

%each interval attractor $([p, q], [\ell, r]) \in \Psi_{\length}(t)$ with attractor position $\gamma$ satisfies 
%$\sum_{w = 1}^{t} \lfloor \mu(w) \rfloor < |[\gamma, r + 1]| \leq \sum_{w = 1}^{t+1} \lfloor \mu(w) \rfloor$. 
%Formally, let $\Psi_{\length}(t) = \{ ([p, q], [\ell, r]) \in \Psi_{\RR} \mid \sum_{w = 1}^{t} \lfloor \mu(w) \rfloor < |[\gamma, r + 1]| \leq \sum_{w = 1}^{t+1} \lfloor \mu(w) \rfloor \}$. 

\paragraph{Subset $\Psi_{\rightRun}$.}
Subset $\Psi_{\rightRun} \subseteq \Psi_{\RR}$ consists of 
interval attractors such that 
each interval attractor $([p, q], [\ell, r]) \in \Psi_{\rightRun}$ satisfies 
$|\lcp(T[\gamma..r], C^{n+1})| > 1 + \sum_{w = 1}^{h+3} \lfloor \mu(w) \rfloor$ for 
the attractor position $\gamma$, associated string $C$, and level $h$ of interval attractor $([p, q], [\ell, r])$. 

\paragraph{Subset $\Psi_{\group}(t, i)$.}
For a pair of integers $t \geq 3$ and $i \in [1, n]$, 
subset $\Psi_{\group}(t, i)$ consists of interval attractors such that 
each interval attractor $([p, q], [\ell, r]) \in \Psi_{\group}(t, i)$ with attractor position $\gamma$ satisfies 
$[i, i + \lfloor (7/8) \mu(t) \rfloor - 1 ] \subseteq [\gamma+1, \gamma + \lfloor \mu(t) \rfloor]$. 
Here, $\lfloor (7/8) \mu(t) \rfloor \geq 1$ and $\lfloor \mu(t) \rfloor \geq 1$ hold for all $t \geq 3$. 
Formally, $\Psi_{\group}(t, i) = \{ ([p, q], [\ell, r]) \in \Psi_{\RR} \mid [i, i + \lfloor (7/8) \mu(t) \rfloor - 1 ] \subseteq [\gamma+1, \gamma + \lfloor \mu(t) \rfloor] \}$.

%\subsection{Properties of three sets \texorpdfstring{$\mathcal{B}_{\bstart}(t)$}{Bt}, \texorpdfstring{$\mathsf{C}_{\cover}(d)$}{Bt}, and \texorpdfstring{$\mathcal{S}_{d}$}{Sd}}

\subsubsection{Properties of Interval Attractors and Function \texorpdfstring{$\mu$}{mu}}\label{subsubsec:sub_properties}
This section explains several properties of interval attractors and function $\mu$. These properties are used to prove Theorem~\ref{theo:RR_Psi_set_size}. 
For simplicity, 
let $\mu_{\SUM}(h) = \sum_{w = 1}^{h} \lfloor \mu(w) \rfloor$ for an integer $h \geq 0$. 
The following proposition gives an upper bound on the integer $\mu_{\SUM}(h)$. 
\begin{proposition}\label{prop:gamma_bound}
For an integer $h \geq 1$, 
let $b_{h} = 1$ if $h$ is even; otherwise $b_{h} = 0$. 
Then, $\mu_{\SUM}(h) \leq (15 + b_{h}) \mu(h)$.  
\end{proposition}
\begin{proof}
We will prove $\sum_{w = 1}^{h} \lfloor \mu(w) \rfloor \leq (15 + b_{h}) \mu(h)$ by induction on $h$. 
For the base case $h = 1$,
$\sum_{w = 1}^{h} \lfloor \mu(w) \rfloor = 1$ 
and $(15 + b_{h}) \mu(h) = 15$ hold 
because $\mu(1) = 1$. 
Therefore, $\sum_{w = 1}^{h} \lfloor \mu(w) \rfloor \leq (15 + b_{h}) \mu(h)$ holds. 

For the inductive step, 
consider $h \geq 2$. 
Then, $\sum_{w = 1}^{h-1} \lfloor \mu(w) \rfloor \leq (15 + b_{h-1}) \mu(h-1)$ holds. 
If $h$ is even, 
$\sum_{w = 1}^{h} \lfloor \mu(w) \rfloor \leq 15 \mu(h-1) + \mu(h)$ by $\sum_{w = 1}^{h-1} \lfloor \mu(w) \rfloor \leq 15 \mu(h-1)$ and $\lfloor \mu(h) \rfloor \leq \mu(h)$. 
$\sum_{w = 1}^{h} \lfloor \mu(w) \rfloor \leq (15 + b_{h}) \mu(h)$ 
because $15 \mu(h-1) + \mu(h) \leq (15 + b_{h}) \mu(h)$. 
Otherwise (i.e., $h$ is odd), 
$\sum_{w = 1}^{h} \lfloor \mu(w) \rfloor \leq 16 \mu(h-1) + \mu(h)$ because 
$\sum_{w = 1}^{h-1} \lfloor \mu(w) \rfloor \leq 16 \mu(h-1)$ and $\lfloor \mu(h) \rfloor \leq \mu(h)$. 
$16 \mu(h-1) = 14 \mu(h)$ follows from $\mu(h) = (8/7) \mu(h-1)$. 
$\sum_{w = 1}^{h} \lfloor \mu(w) \rfloor \leq (15 + b_{h-1}) \mu(h)$ follows from the following equation: 
\begin{equation*}
    \begin{split}
        \sum_{w = 1}^{h} \lfloor \mu(w) \rfloor &\leq 16 \mu(h-1) + \mu(h) \\
        &= 15\mu(h) \\
        &= (15 + b_{h}) \mu(h). 
    \end{split}
\end{equation*}
Therefore, 
$\sum_{w = 1}^{h} \lfloor \mu(w) \rfloor \leq (15 + b_{h}) \mu(h)$ holds for all $h \geq 1$. 
\end{proof}

The following proposition gives an upper bound on the length of the interval $[p-1, r+1]$ for 
each interval attractor $([p, q], [\ell, r]) \in \Psi_{\RR}$.  
\begin{proposition}\label{prop:psi_H_length}
    The following three statements hold for each integer $h \in [0, H]$. 
    \begin{enumerate}[label=\textbf{(\roman*)}]
    \item $\Psi_{h} \setminus \Psi_{\rightRun} \subseteq \Psi_{\rightLen}(0, 36 \mu(h+1))$; 
    \item $\Psi_{h} \subseteq \Psi_{\leftLen}(0, 16 \mu(h+1))$;    
    \item     
    $\Psi_{h} \cap \Psi_{\rightRun} \cap \Psi_{\rightLen}(d, n) \subseteq \bigcup_{K = d - \mu_{\SUM}(h+1)}^{n+1} \Psi_{\lcp}(K)$ for all integer $d \geq 0$.
        
\end{enumerate}
\end{proposition}
\begin{proof}
Consider an interval attractor $([p, q], [\ell, r]) \in \Psi_{\rightRun}$ of level $h$ with attractor position $\gamma$. 
From Definition~\ref{def:RR_Delta}, 
there exists a position $b \in [1, |S^{h}|]$ in sequence $S^{h}$ 
satisfying 
$p = \min \{ s \mid [s, e] \in \Delta(h, b) \}$, 
$q = \max \{ s \mid [s, e] \in \Delta(h, b) \}$, 
$\ell = \min \{ e \mid [s, e] \in \Delta(h, b) \}$, 
and $r = \max \{ e \mid [s, e] \in \Delta(h, b) \}$. 
Here, $\Delta(h, b)$ is the set of intervals in input string $T$ introduced in Section~\ref{subsec:RR_delta}; 
$[p, r] \in \Delta(h, b)$ follows from Lemma~\ref{lem:IA_maximal_lemma}. 
Consider sequence $A(p, r) = [s^{0}, e^{0}], [s^{1}, e^{1}], \ldots, [s^{k}, e^{k}]$. 
Here, $k = h$ and $s^{k} = b$ follow from $[p, r] \in \Delta(h, b)$. 
$\gamma = x^{h}_{s}$ follows from the definition of the attractor position 
for the substring $T[x^{h}_{s}..y^{h}_{s}]$ derived from 
the $s^{h}$-th nonterminal $S^{h}[s^{h}]$ of sequence $S^{h}$.
Let $T[x^{h}_{e}..y^{h}_{e}]$ be the substring derived from 
the $e^{h}$-th nonterminal $S^{h}[e^{h}]$ of sequence $S^{h}$. 

\paragraph{Proof of $|[\gamma, r+1]| \leq 1 + (\sum_{w = 1}^{h+3} \lfloor \mu(w) \rfloor) + (\sum_{w = 1}^{h+1} \lfloor \mu(w) \rfloor)$.}
We prove $|[\gamma, r+1]| \leq 1 + (\sum_{w = 1}^{h+3} \lfloor \mu(w) \rfloor) + (\sum_{w = 1}^{h+1} \lfloor \mu(w) \rfloor)$ by contradiction. 
We assume that $|[\gamma, r+1]| > 1 + (\sum_{w = 1}^{h+3} \lfloor \mu(w) \rfloor) + (\sum_{w = 1}^{h+1} \lfloor \mu(w) \rfloor)$ holds. 
Because of $|[\gamma, r+1]| > 1 + \sum_{w = 1}^{h+3} \lfloor \mu(w) \rfloor$, 
the associated string $C$ of the interval attractor $([p, q], [\ell, r])$ is defined as string $\val(S^{h}[s^{h}])$. 
$|[x^{h}_{s}, r]| > 2\lfloor \mu(h+1) \rfloor + \sum_{w = 1}^{h+1} \lfloor \mu(w) \rfloor$ holds 
because $\gamma = x^{h}_{s}$. 
Because of $|[x^{h}_{s}, r]| > 2\lfloor \mu(h+1) \rfloor + \sum_{w = 1}^{h+1} \lfloor \mu(w) \rfloor$, 
we can apply Lemma~\ref{lem:f_rec_top_property}~\ref{enum:f_rec_top_property:3} to 
interval $[p, r]$. 
Then, the lemma shows that $|[s^{h}, e^{h}]| \geq 4$. 
Because of $|[s^{h}, e^{h}]| \geq 4$, 
$|\val(S^{h}[s^{h}])| \leq \lfloor \mu(h+1) \rfloor$ and $S^{h}[s^{h}] = S^{h}[s^{h}+1] = \cdots = S^{h}[e^{h}-1]$ 
follow from Lemma~\ref{lem:f_rec_top_property}~\ref{enum:f_rec_top_property:1} and Lemma~\ref{lem:f_rec_top_property}~\ref{enum:f_rec_top_property:2}, respectively. 
$|\lcp(T[\gamma..r], C^{n+1})| \geq |[x^{h}_{s}, x^{h}_{e}-1]|$ holds because 
$C = \val(S^{h}[s^{h}])$, $S^{h}[s^{h}] = S^{h}[s^{h}+1] = \cdots = S^{h}[e^{h}-1]$, 
and $\gamma = x^{h}_{s}$. 

We apply Lemma~\ref{lem:f_rec_top_property}~\ref{enum:f_rec_top_property:4} to interval $[p, r]$. 
Then, the lemma shows that $|[x^{h}_{e}, r]| \leq \sum_{w = 1}^{h+1} \lfloor \mu(w) \rfloor$ holds. 
$|[x^{h}_{s}, x^{h}_{e}-1]| \geq 2 + \sum_{w = 1}^{h+3} \lfloor \mu(w) \rfloor$ because 
$|[x^{h}_{s}, x^{h}_{e}-1]| = |[x^{h}_{s}, r]| - |[x^{h}_{e}, r]|$, 
$\gamma = x^{h}_{s}$, $|[\gamma, r+1]| > 1 + (\sum_{w = 1}^{h+3} \lfloor \mu(w) \rfloor) + (\sum_{w = 1}^{h+1} \lfloor \mu(w) \rfloor)$, and $|[x^{h}_{e}, r]| \leq \sum_{w = 1}^{h+1} \lfloor \mu(w) \rfloor$.
$|\lcp(T[\gamma..r], C^{n+1})| > 1 + \sum_{w = 1}^{h+3} \lfloor \mu(w) \rfloor$ 
follows from $|\lcp(T[\gamma..r], C^{n+1})| \geq |[x^{h}_{s}, x^{h}_{e}-1]|$ and $|[x^{h}_{s}, x^{h}_{e}-1]| \geq 2 + \sum_{w = 1}^{h+3} \lfloor \mu(w) \rfloor$. 
On the other hand, $|\lcp(T[\gamma..r], C^{n+1})| \leq 1 + \sum_{w = 1}^{h+3} \lfloor \mu(w) \rfloor$ 
follows from $([p, q], [\ell, r]) \not \in \Psi_{\rightRun}$. 
The two facts $|\lcp(T[\gamma..r], C^{n+1})| > 1 + \sum_{w = 1}^{h+3} \lfloor \mu(w) \rfloor$ and $|\lcp(T[\gamma..r], C^{n+1})| \leq 1 + \sum_{w = 1}^{h+3} \lfloor \mu(w) \rfloor$ yield a contradiction. 
Therefore, $|[\gamma, r+1]| \leq 1 + (\sum_{w = 1}^{h+3} \lfloor \mu(w) \rfloor) + (\sum_{w = 1}^{h+1} \lfloor \mu(w) \rfloor)$ must hold. 

\textbf{Proof of Proposition~\ref{prop:psi_H_length}(i).}
We prove $|[\gamma, r+1]| \leq 36 \mu(h + 1)$ (i.e., $\Psi_{h} \setminus \Psi_{\rightRun} \subseteq \Psi_{\rightLen}(0, 36 \mu(h+1))$). 
$(\sum_{w = 1}^{h+3} \lfloor \mu(w) \rfloor) + (\sum_{w = 1}^{h+1} \lfloor \mu(w) \rfloor) \leq 16 \mu(h + 3) + 16 \mu(h + 1)$ follows from Proposition~\ref{prop:gamma_bound}. 
Here, $\mu(h + 3) = (8/7) \mu(h + 1)$ and $1 \leq \mu(h + 1)$ follows from the definition of the function $\mu$.
Therefore, 
\begin{equation*}
    \begin{split}
    |[\gamma, r+1]| &\leq (\sum_{w = 1}^{h+3} \lfloor \mu(w) \rfloor) + (\sum_{w = 1}^{h+1} \lfloor \mu(w) \rfloor) \\
    &\leq 16 \mu(h + 3) + 16 \mu(h + 1) \\
    &\leq 36 \mu(h + 1).
    \end{split}
\end{equation*}

\textbf{Proof of Proposition~\ref{prop:psi_H_length}(ii).}
Consider an interval attractor $([p^{\prime}, q^{\prime}], [\ell^{\prime}, r^{\prime}])$ of level $h$ with attractor position $\gamma^{\prime}$. 
We apply Lemma~\ref{lem:f_rec_top_property}~\ref{enum:f_rec_top_property:4} to interval $[p, r]$. 
Then, the lemma shows that $|[p-1, \gamma-1]| \leq \sum_{w = 1}^{h+1} \lfloor \mu(w) \rfloor$. 
$\sum_{w = 1}^{h+1} \lfloor \mu(w) \rfloor \leq 16 \mu(h + 1)$ follows from Proposition~\ref{prop:gamma_bound}. 
Therefore, $\Psi_{h} \subseteq \Psi_{\leftLen}(0, 16 \mu(h+1))$ follows from 
$|[p-1, \gamma-1]| \leq 16 \mu(h + 1)$ and $|[p-1, \gamma-1]| \geq 1$. 

\textbf{Proof of Proposition~\ref{prop:psi_H_length}(iii).}
For simplicity, we assume that $([p, q], [\ell, r]) \in \Psi_{\rightLen}(d, n) \cap \Psi_{\rightRun}$ holds. 
In this case, we already proved $|\lcp(T[\gamma..r], C^{n+1})| \geq |[\gamma, x^{h}_{e}-1]|$ 
and $|[x^{h}_{e}, r]| \leq \sum_{w = 1}^{h+1} \lfloor \mu(w) \rfloor$ 
in the proof of $|[\gamma, r+1]| \leq 1 + (\sum_{w = 1}^{h+3} \lfloor \mu(w) \rfloor) + (\sum_{w = 1}^{h+1} \lfloor \mu(w) \rfloor)$. 
$|[\gamma, x^{h}_{e}-1]| \geq d - \sum_{w = 1}^{h+1} \lfloor \mu(w) \rfloor$ follows from $|[\gamma, r+1]| > d$ 
and $|[x^{h}_{e}, r]| \leq \sum_{w = 1}^{h+1} \lfloor \mu(w) \rfloor$. 
$|\lcp(T[\gamma..r], C^{n+1})| \geq d - \sum_{w = 1}^{h+1} \lfloor \mu(w) \rfloor$ follows from 
$|\lcp(T[\gamma..r], C^{n+1})| \geq |[\gamma, x^{h}_{e}-1]|$ and $|[\gamma, x^{h}_{e}-1]| \geq d - \sum_{w = 1}^{h+1} \lfloor \mu(w) \rfloor$. 
Therefore, $([p, q], [\ell, r]) \in \bigcup_{K = d - \mu_{\SUM}(h+1)}^{n+1} \Psi_{\lcp}(K)$ follows from 
$|\lcp(T[\gamma..r], C^{n+1})| \geq d - \sum_{w = 1}^{h+1} \lfloor \mu(w) \rfloor$. 
\end{proof}

The following proposition states the expected size of the $h$-th level interval attractors $\Psi_{h}$ for each $h \in [0, H]$. 
\begin{proposition}\label{prop:psi_H_expected_size}
$\mathbb{E}[|\Psi_{h}|] < (32n/7) \frac{1}{(8/7)^{(h/2)}}$ for each integer $h \in [0, H]$.
\end{proposition}
\begin{proof}
We prove $|\Psi_{h}| \leq |\mathcal{B}_{\bstart}(h)| - 1$.
Let $I_{1}$, $I_{2}$, $\ldots$, $I_{k}$ be the interval attractors in the set $\Psi_{h}$, 
and let $\gamma_{i}$ be the attractor position of each interval attractor $I_{i}$. 
Then, $\gamma_{i} \in \mathcal{B}_{\bstart}(h) \setminus \{ b \}$ holds for the integer $b$ introduced in Proposition~\ref{prop:b_set_properties}~\ref{enum:b_set_properties:4}. 
$|\Psi_{h}| = |\{ \gamma_{1}, \gamma_{2}, \ldots, \gamma_{k} \}|$ because 
Corollary~\ref{cor:IA_identify_corollary} shows that 
$\gamma_{i} \neq \gamma_{i^{\prime}}$ for any pair of integers $1 \leq i < i^{\prime} < k$. 
Therefore, $|\Psi_{h}| \leq |\mathcal{B}_{\bstart}(h)| - 1$ follows from 
$|\Psi_{h}| = |\{ \gamma_{1}, \gamma_{2}, \ldots, \gamma_{k} \}|$, 
$\{ \gamma_{1}, \gamma_{2}, \ldots, \gamma_{k} \} \subseteq \mathcal{B}_{\bstart}(h) \setminus \{ b \}$, 
and $b \in \mathcal{B}_{\bstart}(h)$.

We prove $\mathbb{E}[|\Psi_{h}|] < (32n/7) \frac{1}{(8/7)^{(h/2)}}$. 
$\mathbb{E}[|\mathcal{B}_{\bstart}(h)|] < 1 + \frac{4n}{\mu(h+1)}$ because 
Proposition~\ref{prop:b_set_properties}~\ref{enum:b_set_properties:1} shows that 
$\mathbb{E}(|\mathcal{B}_{\bstart}(h) \cap [1, n]|) < 1 + \frac{4n}{\mu(h+1)}$ holds. 
Therefore, 
\begin{equation*}
    \begin{split}
    \mathbb{E}[|\Psi_{h}|] &\leq \mathbb{E}[|\mathcal{B}_{\bstart}(h)| - 1] \\
    &< \frac{4n}{\mu(h+1)} \\
    &= (32n/7) \frac{1}{(8/7)^{\lceil h/2 \rceil}} \\
    &\leq (32n/7)\frac{1}{(8/7)^{h/2}}.
    \end{split}
\end{equation*}
\end{proof}

The following proposition states properties of the subset $\Psi_{\leftmost}$. 

\begin{proposition}\label{prop:leftmost_susbet_sub_properties}
Let $t \in [0, H]$, $K_{1} \geq 1$, and $K_{2} \geq 1$ be three integers. 
The following three statements hold: 
\begin{enumerate}[label=\textbf{(\roman*)}]
    \item \label{enum:leftmost_susbet_sub_properties:1}
    consider an interval attractor $([p, q], [\ell, r]) \in \Psi_{\leftmost} \cap \Psi_{\leftLen}(0, K_{1}) \cap \Psi_{\rightLen}(0, K_{2})$ of level $t$ with attractor position $\gamma$. 
    Then,     
    $[p-1, r+1] \subseteq \mathsf{C}_{\cover}(K_{1} + K_{2}) \cup \{ 0, n+1 \}$ 
    and $\gamma \in \mathcal{B}_{\bstart}(t) \cap \mathsf{C}_{\cover}(K_{1} + K_{2})$;     
    %$[\gamma - K_{1}, \gamma + K_{2} - 1] \subseteq \mathsf{C}_{\cover}(K_{1} + K_{2})$, 
    \item \label{enum:leftmost_susbet_sub_properties:2}
    $|\Psi_{t} \cap \Psi_{\leftmost} \cap \Psi_{\leftLen}(0, K_{1}) \cap \Psi_{\rightLen}(0, K_{2})| \leq |\mathcal{B}_{\bstart}(t) \cap \mathsf{C}_{\cover}(K_{1} + K_{2})|$ for each integer $t \in [0, H]$;    
    \item \label{enum:leftmost_susbet_sub_properties:3} consider $k$ interval attractors in subset $\Psi_{\leftmost} \cap \Psi_{\leftLen}(0, K_{1}) \cap \Psi_{\rightLen}(0, K_{2})$ such that their attractor positions are distinct. 
    Then, $k \leq \sigma^{K_{1} + K_{2}}$.
\end{enumerate}
\end{proposition}
\begin{proof}
The proof of Proposition~\ref{prop:leftmost_susbet_sub_properties} is as follows. 

\paragraph{Proof of Proposition~\ref{prop:leftmost_susbet_sub_properties}(i).}
We prove $[p-1, r+1] \subseteq \mathsf{C}_{\cover}(K_{1} + K_{2}) \cup \{ 0, n+1 \}$. 
$p-1 = \min \Occ(T, T[p-1..r+1])$ 
or $p = 1$ follows from the definition of the subset $\Psi_{\leftmost}$. 
If $p-1 = \min \Occ(T, T[p-1..r+1])$, 
then $[p-1, r+1] \subseteq \mathsf{C}_{\cover}(K_{1} + K_{2})$ holds 
because $|[p-1, r+1]| \leq K_{1} + K_{2}$, 
and the leftmost occurrence of string $T[p-1..r+1]$ starts at position $p-1$ in $T$. 
Otherwise (i.e., $p = 1$), 
$[p, \min \{ r+1, m \}] \subseteq \mathsf{C}_{\cover}(K_{1} + K_{2})$ holds 
because the leftmost occurrence of string $T[1..\min \{ r+1, m \}]$ starts at position $1$ in $T$. 
Therefore, we obtain $[p-1, r+1] \subseteq \mathsf{C}_{\cover}(K_{1} + K_{2}) \cup \{ 0, n+1 \}$.

We prove $\gamma \in \mathcal{B}_{\bstart}(t) \cap \mathsf{C}_{\cover}(K_{1} + K_{2})$. 
$\gamma \in \mathcal{B}_{\bstart}(t)$ follows from the definition of the attractor position. 
$\gamma \in \mathsf{C}_{\cover}(K_{1} + K_{2})$ follows from 
$[p-1, r+1] \subseteq \mathsf{C}_{\cover}(K_{1} + K_{2}) \cup \{ 0, n+1 \}$ 
and $p \leq \gamma \leq r$. 
Therefore, we obtain $\gamma \in \mathcal{B}_{\bstart}(t) \cap \mathsf{C}_{\cover}(K_{1} + K_{2})$. 

\paragraph{Proof of Proposition~\ref{prop:leftmost_susbet_sub_properties}(ii).}
Let $\gamma$ and $\gamma^{\prime}$ be the attractor positions of two interval attractors in 
$\Psi_{t} \cap \Psi_{\leftmost} \cap \Psi_{\leftLen}(0, K_{1}) \cap \Psi_{\rightLen}(0, K_{2})$. 
$\gamma \neq \gamma^{\prime}$ follows from Corollary~\ref{cor:IA_identify_corollary}. 
$\gamma, \gamma^{\prime} \in \mathcal{B}_{\bstart}(t) \cap \mathsf{C}_{\cover}(K_{1} + K_{2})$ follows from 
Proposition~\ref{prop:leftmost_susbet_sub_properties}(i). 
Therefore, $|\Psi_{t} \cap \Psi_{\leftmost} \cap \Psi_{\leftLen}(0, K_{1}) \cap \Psi_{\rightLen}(0, K_{2})| \leq |\mathcal{B}_{\bstart}(t) \cap \mathsf{C}_{\cover}(K_{1} + K_{2})|$ 
follows from $\gamma \neq \gamma^{\prime}$ and $\gamma, \gamma^{\prime} \in \mathcal{B}_{\bstart}(t) \cap \mathsf{C}_{\cover}(K_{1} + K_{2})$. 

\paragraph{Proof of Proposition~\ref{prop:leftmost_susbet_sub_properties}(iii).}
Let $([p_{1}, q_{1}], [\ell_{1}, r_{1}])$, $([p_{2}, q_{2}], [\ell_{2}, r_{2}])$, $\ldots$, $([p_{k}, q_{k}]$, $[\ell_{k}, r_{k}])$ be $k$ interval attractors in $\Psi_{\leftmost} \cap \Psi_{\leftLen}(0, K_{1}) \cap \Psi_{\rightLen}(0, K_{2})$ such that their attractor positions are distinct. 
Let $\gamma_{i}$ be the attractor position of each interval attractor $([p_{i}, q_{i}], [\ell_{i}, r_{i}])$
For simplicity, 
we assume that $\gamma_{1} < \gamma_{2} < \cdots < \gamma_{k}$. 
Let $L_{i} = T[p_{i}-1..\gamma_{i}-1]$ and $R_{i} = T[\gamma_{i}..r_{i}-1]$.

For any pair of integers $1 \leq i < j \leq k$, 
we prove $T[\gamma_{i} - |L_{j}|..\gamma_{i} + |R_{j}| - 1] \neq T[p_{j}-1..r_{j}+1]$ by contradiction. 
We assume that $T[\gamma_{i} - |L_{j}|..\gamma_{i} + |R_{j}| - 1] = T[p_{j}-1..r_{j}+1]$ holds. 
In this case, $[p_{i}-1, r_{i}+1], [p_{j}-1, r_{j}+1] \subseteq [1, n]$ holds. 
$\gamma_{i} - |L_{j}| < \gamma_{j} - |L_{j}|$~(i.e., $\gamma_{i} - |L_{j}| < p_{j}-1$) follows from 
$\gamma_{i} < \gamma_{j}$. 
$p_{j}-1 \neq \min \Occ(T, T[p_{j}-1..r_{j}+1])$ holds because 
$T[\gamma_{i} - |L_{j}|..\gamma_{i} + |R_{j}| - 1] = T[p_{j}-1..r_{j}+1]$ 
and $\gamma_{i} - |L_{j}| < p_{j}-1$. 
On the other hand, 
$p_{j}-1 = \min \Occ(T, T[p_{j}-1..r_{j}+1])$ follows from the definition of the subset $\Psi_{\leftmost}$. 
The two facts $p_{j}-1 \neq \min \Occ(T, T[p_{j}-1..r_{j}+1])$ and $p_{j}-1 = \min \Occ(T, T[p_{j}-1..r_{j}+1])$ yield a contradiction. 
Therefore, $T[\gamma_{i} - |L_{j}|..\gamma_{i} + |R_{j}| - 1] \neq T[p_{j}-1..r_{j}+1]$ must hold. 

We prove $k \leq \sigma^{K_{1}+K_{2}}$. 
For simplicity, let $T[i] = \#$ for all $i \geq n+2$. 
Similarly, let $T[i] = \#$ for all $i \leq -1$. 
Here, $T[0] = \$$, $T[n+1] = \$$, and $T[i] \neq \$$ for all $i \in [1, n]$ (see Section~\ref{sec:preliminary}). 
For any pair of integers $1 \leq i < j \leq k$, 
$T[\gamma_{i} - K_{1}..\gamma_{i} + K_{2} - 1] \neq T[\gamma_{j} - K_{1}..\gamma_{j} + K_{2} - 1]$ 
holds because (i) $T[\gamma_{i} - |L_{j}|..\gamma_{i} + |R_{j}| - 1] \neq T[p_{j}-1..r_{j}+1]$, 
(ii) $|L_{i}|, |L_{j}| \leq K_{1}$, and (iii) $|R_{i}|, |R_{j}| \leq K_{2}$. 
Therefore, $k \leq \sigma^{K_{1}+K_{2}}$ follows from 
$k = |\{ T[\gamma_{i} - K_{1}..\gamma_{i} + K_{2} - 1] \mid s \in [1, k] \}|$ 
and $|\{ T[\gamma_{i} - K_{1}..\gamma_{i} + K_{2} - 1] \mid s \in [1, k] \}| \leq \sigma^{K_{1} + K_{2}}$.

\end{proof}

The following two propositions state properties of set $(\Psi_{\rightRun} \cap \Psi_{\rightLen}(\mu_{\SUM}(t), \mu_{\SUM}(t+1))) \setminus \Psi_{\run}$ for an integer $t \geq 1$. 

\begin{proposition}\label{prop:rightRun_property_H}
    Consider an interval attractor $([p, q], [\ell, r])$ of level $h$ 
    in set $(\Psi_{\rightRun} \cap \Psi_{\rightLen}(\mu_{\SUM}(t), \mu_{\SUM}(t+1))) \setminus \Psi_{\run}$ for an integer $t \geq 1$. 
    Let $\gamma$ be the attractor position of the interval attractor $([p, q], [\ell, r])$. 
    Then, 
    (i) $|[\gamma, r+1]| > 1 + \sum_{w = 1}^{h+3} \lfloor \mu(w) \rfloor$, 
    (ii) $h \leq t-3$, 
    (iii) $([p, q], [\ell, r]) \in \Psi_{\leftLen}(0, 16 \mu(t+1)) \cap \Psi_{\rightLen}(0, 17 \mu(t+1))$, 
    and (iv) $([p, q], [\ell, r]) \in \bigcup_{K = \lfloor \mu(t) \rfloor }^{n+1} \Psi_{\lcp}(K)$. 
\end{proposition}
\begin{proof}
We prove $|[\gamma, r+1]| > 1 + \sum_{w = 1}^{h+3} \lfloor \mu(w) \rfloor$ 
for the attractor position $\gamma$ of the interval attractor $([p, q], [\ell, r])$. 
For the associated string $C$ of the interval attractor $([p, q], [\ell, r])$, 
$|\lcp(T[\gamma..r], C^{n+1})| > 1 + \sum_{w = 1}^{h+3} \lfloor \mu(w) \rfloor$ follows from 
the definition of the subset $\Psi_{\rightRun}$. 
Therefore, 
we obtain $|[\gamma, r+1]| > 1 + \sum_{w = 1}^{h+3} \lfloor \mu(w) \rfloor$. 

We prove $h \leq t - 3$ by contradiction. 
We assume that $h > t - 3$ holds. 
Then, $|[\gamma, r+1]| > \sum_{w = 1}^{t+1} \lfloor \mu(w) \rfloor$ follows from 
$|[\gamma, r+1]| > 1 + \sum_{w = 1}^{h+3} \lfloor \mu(w) \rfloor$ and $h > t-3$. 
On the other hand, $|[\gamma, r + 1]| \leq \sum_{w = 1}^{t+1} \lfloor \mu(w) \rfloor$ follows from 
$([p, q], [\ell, r]) \in \Psi_{\rightLen}(\mu_{\SUM}(t), \mu_{\SUM}(t+1))$. 
The two facts $|[\gamma, r+1]| > \sum_{w = 1}^{t+1} \lfloor \mu(w) \rfloor$ and $|[\gamma, r + 1]| \leq \sum_{w = 1}^{t+1} \lfloor \mu(w) \rfloor$ yield a contradiction. 
Therefore, $h \leq t - 3$ must hold. 

We prove $([p, q], [\ell, r]) \in \Psi_{\leftLen}(0, 17 \mu(t+1)) \cap \Psi_{\rightLen}(0, 16 \mu(t+1))$. 
$([p, q], [\ell, r]) \in \Psi_{\leftLen}(0$, $17 \mu(t+1))$ follows from Proposition~\ref{prop:psi_H_length} 
and $h \leq t - 3$. 
$([p, q], [\ell, r]) \in \Psi_{\rightLen}(0, 16 \mu(t+1))$ follows from 
$|[\gamma, r+1]| > 1 + \sum_{w = 1}^{h+3} \lfloor \mu(w) \rfloor$, 
$h \leq t - 3$, and Proposition~\ref{prop:gamma_bound}. 

We prove $([p, q], [\ell, r]) \in \bigcup_{K = \lfloor \mu(t) \rfloor }^{n+1} \Psi_{\lcp}(K)$. 
$([p, q], [\ell, r]) \in \bigcup_{K = \mu_{\SUM}(t) - \mu_{\SUM}(h+1)}^{n+1} \Psi_{\lcp}(K)$ follows from Proposition~\ref{prop:psi_H_length}.
$\mu_{\SUM}(t) - \mu_{\SUM}(h+1) \geq \lfloor \mu(t) \rfloor$ follows from $h \leq t - 3$. 
Therefore, we obtain $([p, q], [\ell, r]) \in \bigcup_{K = \lfloor \mu(t) \rfloor }^{n+1} \Psi_{\lcp}(K)$. 
\end{proof}

\begin{proposition}\label{prop:rightRun_minor_property}
    Consider an interval attractor $([p, q], [\ell, r])$ of level $h$ 
    in set $(\Psi_{\rightRun} \cap \Psi_{\rightLen}(\mu_{\SUM}(t), \mu_{\SUM}(t+1))) \setminus \Psi_{\run}$ for an integer $t \geq 1$. 
    Let $\gamma$ and $C$ be the attractor position and associated string of the interval attractor $([p, q], [\ell, r])$, respectively. 
    Then, the following five statements hold: 
\begin{enumerate}[label=\textbf{(\roman*)}]
    %\item \label{enum:rightRun_property:1} $|[p-1, \gamma-1]| \leq 1 + \sum_{w = 1}^{h+1} \lfloor \mu(w) \rfloor$, 
    %$1 + \sum_{w = 1}^{h+1} \lfloor \mu(w) \rfloor \leq 17 \mu(t+1)$, and $|[\gamma, r+1]| \leq 16 \mu(t+1)$;
    \item \label{enum:rightRun_minor_property:1} $|\lcs(T[p-1..\gamma-1], C^{n+1})| < |[p-1, \gamma-1]|$;
    \item \label{enum:rightRun_minor_property:2} $1 \leq |C| \leq \lfloor (7/8) \mu(t) \rfloor$;
    \item \label{enum:rightRun_minor_property:3}
    consider a position $j \in [1, |S^{h}|]$ of sequence $S^{h}$ satisfying 
    $i \in [x^{h}_{j}+1, y^{h}_{j}]$ 
    for an integer $i \in [\gamma+1, \gamma + \lfloor \mu(t) \rfloor]$ 
    and the substring $T[x^{h}_{j}..y^{h}_{j}]$ derived from the $j$-th nonterminal of sequence $S^{h}$ in string $T$. 
    Then, $\val(S^{h}[j]) = C$, and $j \in [s^{k}, e^{k}-1]$ for the tail $[s^{k}, e^{k}]$ of sequence $A(p, r)$;    
    \item \label{enum:rightRun_minor_property:4} $\mathcal{B}_{\bstart}(h) \cap [i, i + \lfloor (7/8) \mu(t) \rfloor - 1 ] \neq \emptyset$ for all $i \in [\gamma+1, \gamma + \lfloor \mu(t) \rfloor]$;
    \item \label{enum:rightRun_minor_property:5} $\mathcal{B}_{\bstart}(h+1) \cap [\gamma+1, \gamma + \lfloor \mu(t) \rfloor] = \emptyset$.
    %substring $T[\gamma..\gamma + \mu_{\SUM}(t)]$
    %$\mathcal{B}_{\bstart}(h+1) \cap [\gamma+1, \gamma + \lfloor \mu(t) \rfloor] = \emptyset$.

    %\item \label{enum:rightRun_property:7} $h \leq t-3$.
\end{enumerate}
\end{proposition}
\begin{proof}
From Definition~\ref{def:RR_Delta}, 
there exists a position $b \in [1, |S^{h}|]$ in sequence $S^{h}$ 
satisfying 
$p = \min \{ i \mid [i, j] \in \Delta(h, b) \}$, 
$q = \max \{ i \mid [i, j] \in \Delta(h, b) \}$, 
$\ell = \min \{ j \mid [i, j] \in \Delta(h, b) \}$, 
and $r = \max \{ j \mid [i, j] \in \Delta(h, b) \}$. 
Here, $\Delta(h, b)$ is the set of intervals in input string $T$ introduced in Section~\ref{subsec:RR_delta}; 
$[p, r] \in \Delta(h, b)$ follows from Lemma~\ref{lem:IA_maximal_lemma}. 
Consider sequence $A(p, r) = [s^{0}, e^{0}], [s^{1}, e^{1}], \ldots, [s^{k}, e^{k}]$. 
Here, $k = h$ and $s^{k} = b$ follows from $[p, r] \in \Delta(h, b)$. 
$\gamma = x^{h}_{s}$ follows from the definition of the attractor position 
for the substring $T[x^{h}_{s}..y^{h}_{s}]$ derived from 
the $s^{h}$-th nonterminal $S^{h}[s^{h}]$ of sequence $S^{h}$.
Let $T[x^{h}_{e}..y^{h}_{e}]$ be the substring derived from 
the $e^{h}$-th nonterminal $S^{h}[e^{h}]$ of sequence $S^{h}$. 
Because of $|[\gamma, r+1]| > 1 + \sum_{w = 1}^{h+3} \lfloor \mu(w) \rfloor$ (Proposition~\ref{prop:rightRun_property_H}), 
the associated string $C$ is defined as the string $\val(S^{h}[s^{h}])$ derived from the 
$s^{h}$-th nonterminal of sequence $S^{h}$. 

We prove (A) $|[s^{h}, e^{h}]| \geq 4$, (B) $|\val(S^{h}[s^{h}])| \leq \lfloor \mu(h+1) \rfloor$, 
(C) $S^{h}[s^{h}] = S^{h}[s^{h}+1] = \cdots = S^{h}[e^{h}-1]$ 
and $C = \val(S^{h}[s^{h}])$, and (D) $|[\gamma, x^{h}_{e}-1]| = |C||[s^{h}, e^{h}-1]|$. 
$|[\gamma, r]| > 1 + 2\lfloor \mu(h+1) \rfloor + \sum_{w = 1}^{h+1} \lfloor \mu(w) \rfloor$ 
follows from $|[\gamma, r+1]| > 1 + \sum_{w = 1}^{h+3} \lfloor \mu(w) \rfloor$.  
In this case, 
we can apply Lemma~\ref{lem:f_rec_top_property}~\ref{enum:f_rec_top_property:3} to interval $[p, r]$, 
and it shows that $|[s^{h}, e^{h}]| \geq 4$. 
$|\val(S^{h}[s^{h}])| \leq \lfloor \mu(h+1) \rfloor$ 
and $S^{h}[s^{h}] = S^{h}[s^{h}+1] = \cdots = S^{h}[e^{h}-1]$ follow from 
Lemma~\ref{lem:f_rec_top_property}~\ref{enum:f_rec_top_property:1} and 
Lemma~\ref{lem:f_rec_top_property}~\ref{enum:f_rec_top_property:2}, respectively. 
$x^{h}_{e} = x^{h}_{s} + |C||[s^{h}, e^{h}-1]|$ follows from $S^{h}[s^{h}] = S^{h}[s^{h}+1] = \cdots = S^{h}[e^{h}-1]$ 
and $C = \val(S^{h}[s^{h}])$. 
Therefore, 
$|[\gamma, x^{h}_{e}-1]| = |C||[s^{h}, e^{h}-1]|$ follows from 
$x^{h}_{e} = x^{h}_{s} + |[s^{h}, e^{h}-1]||C|$ 
and $\gamma = x^{h}_{s}$. 

We prove (E) $|[\gamma, x^{h}_{e}-1]| > \lfloor \mu(t) \rfloor$.
$\sum_{w = 1}^{t} \lfloor \mu(w) \rfloor < |[\gamma, r+1]| \leq \sum_{w = 1}^{t+1} \lfloor \mu(w) \rfloor$ 
follows from $([p, q], [\ell, r]) \in \Psi_{\rightLen}(\mu_{\SUM}(t), \mu_{\SUM}(t+1))$.  
$|[x^{h}_{e}, r]| \leq \sum_{w = 1}^{h+1} \lfloor \mu(w) \rfloor$ follows from 
Lemma~\ref{lem:f_rec_top_property} \ref{enum:f_rec_top_property:4}. 
$|[x^{h}_{e}, r]| \leq \sum_{w = 1}^{t-2} \lfloor \mu(w) \rfloor$ follows from 
$|[x^{h}_{e}, r]| \leq \sum_{w = 1}^{h+1} \lfloor \mu(w) \rfloor$ and $h \leq t - 3$. 
$|[\gamma, x^{h}_{e}-1] = |[\gamma, r+1]| - (|[x^{h}_{e}, r]| + |[r+1, r+1]|)$. 
$|[\gamma, r+1]| - (|[x^{h}_{e}, r]| + |[r+1, r+1]|) \geq (\sum_{w = 1}^{t} \lfloor \mu(w) \rfloor) - (\sum_{w = 1}^{t-2} \lfloor \mu(w) \rfloor)$ follows from 
$|[\gamma, r+1]| \geq 1 + \sum_{w = 1}^{t} \lfloor \mu(w) \rfloor$, 
$|[r+1, r+1]| = 1$, and $|[x^{h}_{e}, r]| \leq \sum_{w = 1}^{t-2} \lfloor \mu(w) \rfloor$. 
$(\sum_{w = 1}^{t} \lfloor \mu(w) \rfloor) - (\sum_{w = 1}^{t-2} \lfloor \mu(w) \rfloor) > \lfloor \mu(t) \rfloor$. 
Therefore, 
\begin{equation*}
    \begin{split}
    |[\gamma, x^{h}_{e}-1]| &= |[\gamma, r+1]| - (|[x^{h}_{e}, r]| + |[r+1, r+1]|) \\
    &\geq (\sum_{w = 1}^{t} \lfloor \mu(w) \rfloor) - (\sum_{w = 1}^{t-2} \lfloor \mu(w) \rfloor) \\
    &> \lfloor \mu(t) \rfloor.
    \end{split}
\end{equation*}

%We prove (F) $|[s^{h}, e^{h}-1]| > \lfloor \frac{\lfloor \mu(t) \rfloor}{|C|} \rfloor$ by contradiction. 
%We assume that $|[s^{h}, e^{h}-1]| \leq \lfloor \frac{\lfloor \mu(t) \rfloor}{|C|} \rfloor$ holds. 
%Let $\alpha = \lfloor \mu(t) \rfloor \mod |C|$. 
%Then, $\frac{\lfloor \mu(t) \rfloor - \alpha}{|C|} = \lfloor \frac{\lfloor \mu(t) \rfloor}{|C|} \rfloor$. 
%$|[\gamma, x^{h}_{e}-1]| \leq \lfloor \mu(t) \rfloor - \alpha$ 
%follows from $|[\gamma, x^{h}_{e}-1]| = |C||[s^{h}, e^{h}-1]|$, 
%$|[s^{h}, e^{h}-1]| \leq \lfloor \frac{\lfloor \mu(t) \rfloor}{|C|} \rfloor$, 
%and $\frac{\lfloor \mu(t) \rfloor - \alpha}{|C|} = \lfloor \frac{\lfloor \mu(t) \rfloor}{|C|} \rfloor$. 
%$|[\gamma, x^{h}_{e}-1]| \leq \lfloor \mu(t) \rfloor$ 
%follows from $|[\gamma, x^{h}_{e}-1]| \leq \lfloor \mu(t) \rfloor - \alpha$, 
%$\alpha \in [0, |C| - 1]$. 
%On the other hand, we already proved $|[\gamma, x^{h}_{e}-1]| > \lfloor \mu(t) \rfloor$. 
%The two facts $|[\gamma, x^{h}_{e}-1]| \leq \lfloor \mu(t) \rfloor$ and $|[\gamma, x^{h}_{e}-1]| > \lfloor \mu(t) \rfloor$ yield a contradiction. 
%Therefore, $|[s^{h}, e^{h}-1]| > \lfloor \frac{\lfloor \mu(t) \rfloor}{|C|} \rfloor$ must hold. 

\textbf{Proof of Proposition~\ref{prop:rightRun_minor_property}(i).}
Proposition~\ref{prop:rightRun_minor_property}(i) follows from the definitions of two subsets $\Psi_{\run}$ and $\Psi_{\rightRun}$. 

\textbf{Proof of Proposition~\ref{prop:rightRun_minor_property}(ii).}
$|C| \geq 1$ follows from the definition of the associated string $C$. 
$|C| \leq \lfloor \mu(h+1) \rfloor$ follows from 
$C = \val(S^{h}[s^{h}])$ and $|\val(S^{h}[s^{h}])| \leq \lfloor \mu(h+1) \rfloor$. 
$\mu(h+1) \leq (7/8) \mu(h+3)$ holds because $\mu(h+3) = (8/7) \mu(h+1)$.
$\mu(h+3) \leq \mu(t)$ follows from Proposition~\ref{prop:rightRun_property_H}. 
Therefore, 
\begin{equation*}
    \begin{split}
    |C| &\leq \lfloor \mu(h+1) \rfloor \\
    &\leq \lfloor (7/8) \mu(h+3) \rfloor \\
    &\leq \lfloor (7/8) \mu(t) \rfloor.
    \end{split}
\end{equation*}

\textbf{Proof of Proposition~\ref{prop:rightRun_minor_property}(iii).}
$y^{h}_{j} > x^{h}_{s}$ follows from 
$\gamma < i$, $i \leq y^{h}_{j}$, and $\gamma = x^{h}_{s}$. 
This inequality $y^{h}_{j} > x^{h}_{s}$ indicates that $j \geq s^{h}$ 
because $x^{h}_{j} \geq x^{h}_{s}$ must hold. 
$x^{h}_{e} > \gamma + \lfloor \mu(t) \rfloor$ follows from $|[\gamma, x^{h}_{e}-1]| > \lfloor \mu(t) \rfloor$ (statement (E)). 
$x^{h}_{j} \leq \gamma + \lfloor \mu(t) \rfloor - 1$ 
follows from $i \in [x^{h}_{j}+1, y^{h}_{j}]$ 
and $i \in [\gamma+1, \gamma + \lfloor \mu(t) \rfloor]$. 
$x^{h}_{j} < x^{h}_{e}$ 
follows from 
$x^{h}_{j} \leq \gamma + \lfloor \mu(t) \rfloor - 1$ and $x^{h}_{e} > \gamma + \lfloor \mu(t) \rfloor$.
This inequality $x^{h}_{j} < x^{h}_{e}$ indicates that $j < e^{h}$. 
$j \in [s^{h}, e^{h}-1]$ follows from $j \geq s^{h}$ and $j < e^{h}$. 
Therefore, $\val(S^{h}[j]) = C$ follows from 
$j \in [s^{h}, e^{h}-1]$, $S^{h}[s^{h}] = S^{h}[s^{h}+1] = \cdots = S^{h}[e^{h}-1]$, 
and $C = \val(S^{h}[s^{h}])$. 

\textbf{Proof of Proposition~\ref{prop:rightRun_minor_property}(iv).}
We prove Proposition~\ref{prop:rightRun_minor_property}(iv) by contradiction. 
We assume that there exists an integer $i \in [\gamma+1, \gamma + \lfloor \mu(t) \rfloor]$ 
satisfying $\mathcal{B}_{\bstart}(h) \cap [i, i + \lfloor (7/8) \mu(t) \rfloor - 1 ] = \emptyset$. 
Then, there exists a position $j \in [1, |S^{h}|]$ of sequence $S^{h}$ satisfying 
$[i, i + \lfloor (7/8) \mu(t) \rfloor - 1 ] \subseteq [x^{h}_{j}+1, y^{h}_{j}]$ 
for the substring $T[x^{h}_{j}..y^{h}_{j}]$ derived from the $j$-th nonterminal of sequence $S^{h}$ in string $T$. 
$|[x^{h}_{j}, y^{h}_{j}]| \geq 1 + \lfloor (7/8) \mu(t) \rfloor$ 
follows from $|[x^{h}_{j}, y^{h}_{j}]| = 1 + [x^{h}_{j}+1, y^{h}_{j}]$, 
$|[x^{h}_{j}+1, y^{h}_{j}]| \geq |[i, i + \lfloor (7/8) \mu(t) \rfloor - 1 ]|$, 
and $|[i, i + \lfloor (7/8) \mu(t) \rfloor - 1 ]| = \lfloor (7/8) \mu(t) \rfloor$. 

On the other hand, 
$\val(S^{h}[j]) = C$ follows from Proposition~\ref{prop:rightRun_minor_property}(iii). 
$|[x^{h}_{j}, y^{h}_{j}]| < 1 + \lfloor (7/8) \mu(t) \rfloor$ follows from 
$|[x^{h}_{j}, y^{h}_{j}]| = |\val(S^{h}[j])|$, $\val(S^{h}[j]) = C$, 
and $1 \leq |C| \leq \lfloor (7/8) \mu(t) \rfloor$. 
The two facts $|[x^{h}_{j}, y^{h}_{j}]| \geq 1 + \lfloor (7/8) \mu(t) \rfloor$ and $|[x^{h}_{j}, y^{h}_{j}]| < 1 + \lfloor (7/8) \mu(t) \rfloor$ 
yield a contradiction. 
Therefore, Proposition~\ref{prop:rightRun_minor_property}(iv) must hold.

\textbf{Proof of Proposition~\ref{prop:rightRun_minor_property}(v).}
In the derivation tree, 
let $u_{j}$ be the node corresponding to the $(s^{h}+j-1)$-th nonterminal $S^{h}[s^{h}+j-1]$ of sequence $S^{h}$ 
for each $j \in [1, e^{h} - s^{h} - 1]$. 
Then, $\mathcal{B}_{\bstart}(h+1) \cap [x^{\prime}+1, y^{\prime}] = \emptyset$ 
holds 
for the substring $T[x^{\prime}..y^{\prime}]$ derived from the parent $u^{\prime}$ of $u_{1}$. 
Here, $x^{\prime} \leq x^{h}_{s}$ holds. 

We prove $[\gamma+1, x^{h}_{e} - 1] \subseteq [x^{\prime}+1, y^{\prime}]$. 
Let $d = e^{h} - s^{h} - 1$ for simplicity. 
Then, $d$ nodes $u_{1}, u_{2}, \ldots, u_{d}$ have the same nonterminal~(i.e., $S^{h}[s^{h}] = S^{h}[s^{h}+1] = \cdots = S^{h}[e^{h}-1]$). 
We can apply Lemma~\ref{lem:rr_property}~\ref{enum:rr_property:3} to 
the nonterminals $u_{1}, u_{2}, \ldots, u_{d}$ 
because $S^{h}[s^{h}] = S^{h}[s^{h}+1] = \cdots = S^{h}[e^{h}-1]$, 
and $|\val(S^{h}[s^{h}])| \leq \lfloor \mu(h+1) \rfloor$. 
The lemma ensures that 
the nonterminals $u_{1}, u_{2}, \ldots, u_{d}$ have the same parent $u^{\prime}$. 
This fact indicates that $y^{\prime} \geq x^{h}_{e} - 1$. 
Therefore,  
$[\gamma+1, x^{h}_{e} - 1] \subseteq [x^{\prime}+1, y^{\prime}]$ follows from 
$x^{h}_{s} = \gamma$, $x^{\prime} \leq x^{h}_{s}$, and $y^{\prime} \geq x^{h}_{e} - 1$. 

We prove Proposition~\ref{prop:rightRun_minor_property}(v). 
We already proved $|[\gamma, x^{h}_{e}-1]| > \lfloor \mu(t) \rfloor$ (statement (E)). 
Therefore, 
\begin{equation*}
    \begin{split}
    \mathcal{B}_{\bstart}(h+1) \cap [\gamma+1, \gamma + \lfloor \mu(t) \rfloor] &\subseteq \mathcal{B}_{\bstart}(h+1) \cap [\gamma+1, x^{h}_{e} - 1] \\
    &\subseteq \mathcal{B}_{\bstart}(h+1) \cap [x^{\prime}+1, y^{\prime}] \\
    &= \emptyset.
    \end{split}
\end{equation*}
\end{proof}

The following proposition states the relationship between two interval attractors in set 
$(\Psi_{\rightRun} \cap \Psi_{\rightLen}(\mu_{\SUM}(t), \mu_{\SUM}(t+1)) \cap \Psi_{\group}(t, i)) \setminus \Psi_{\run}$.

\begin{proposition}\label{prop:in_group_rightRun_property}
    Consider two distinct interval attractors $([p, q], [\ell, r])$ and $([p^{\prime}, q^{\prime}], [\ell^{\prime}, r^{\prime}])$ 
    in set $(\Psi_{\rightRun} \cap \Psi_{\rightLen}(\mu_{\SUM}(t), \mu_{\SUM}(t+1)) \cap \Psi_{\group}(t, i)) \setminus \Psi_{\run}$ 
    for a pair of integers $t \geq 3$ and $i \in [1, n]$. 
    Let $h, \gamma, C$ be the level, attractor position, and associated string of interval attractor $([p, q], [\ell, r])$, respectively. 
    Similarly, 
    let $h^{\prime}, \gamma^{\prime}, C^{\prime}$ be the level, attractor position, and associated string of interval attractor $([p^{\prime}, q^{\prime}], [\ell^{\prime}, r^{\prime}])$, 
    respectively.
    %%Here, . 
    Then, the following four statements hold: 
\begin{enumerate}[label=\textbf{(\roman*)}]
    \item \label{enum:in_group_rightRun_property:1} $h = h^{\prime}$;
    \item \label{enum:in_group_rightRun_property:2} $C = C^{\prime}$; 
    \item \label{enum:in_group_rightRun_property:3} $\gamma \neq \gamma^{\prime}$;    
    \item \label{enum:in_group_rightRun_property:5} $-16\mu(h+1) < \gamma^{\prime} - \gamma < 16\mu(h+1)$. 
    %\item \label{enum:in_group_rightRun_property:4} $\gamma - |\lcs(T[p-1..\gamma-1], C^{n+1})| = \gamma^{\prime} - |\lcs(T[p^{\prime}-1..\gamma^{\prime}-1], (C^{\prime})^{n+1})|$. 
\end{enumerate}
\end{proposition}
\begin{proof}
    The proof of Proposition~\ref{prop:in_group_rightRun_property} is as follows.
%There exists a position $b^{h}$ in sequence $S^{h}$ satisfying $I(b^{h}) = ([p, q], [\ell, r])$, 
%and Lemma~\ref{lem:IA_wide} shows that $I_{\capture}(p, r) = I(b^{h})$. 
%Consider sequence $A(p, r) = [s^{0}, e^{0}], [s^{1}, e^{1}], \ldots, [s^{k}, e^{k}]$. 
%Here, $k = h$ and $s^{h} = b^{h}$ follow from $I_{\capture}(p, r) = I(b^{h})$. 
%$\gamma = x^{h}_{s}$ follows from the definition of the attractor position 
%for the substring $T[x^{h}_{s}..y^{h}_{s}]$ derived from 
%the $s^{h}$-th nonterminal $S^{h}[s^{h}]$ of sequence $S^{h}$ in $T$.
%
%Similarly, 
%there exists a position $b^{\prime h^{\prime}}$ in sequence $S^{h^{\prime}}$ satisfying $I(b^{\prime h^{\prime}}) = ([p^{\prime}, q^{\prime}], [\ell^{\prime}, r^{\prime}])$. 
%Lemma \ref{lem:IA_wide} shows that $I_{\capture}(p^{\prime}, r^{\prime}) = I(b^{\prime h^{\prime}})$. 
%Consider sequence $A(p^{\prime}, r^{\prime}) = [s^{\prime 0}, e^{\prime 0}], [s^{\prime 1}, e^{\prime 1}], \ldots, [s^{\prime k^{\prime}}, e^{k^{\prime}}]$. 
%Then, $k^{\prime} = h^{\prime}$ and $s^{\prime h^{\prime}} = b^{\prime h^{\prime}}$. 
%The attractor position $\gamma^{\prime}$ is defined as $x^{\prime h^{\prime}}_{s}$ 
%for the substring $T[x^{\prime h^{\prime}}_{s}..y^{\prime h^{\prime}}_{s}]$ derived from 
%the $s^{\prime h^{\prime}}$-th nonterminal of sequence $S^{h^{\prime}}$ in $T$.

\textbf{Proof of Proposition~\ref{prop:in_group_rightRun_property}(i).}
We prove $\mathcal{B}_{\bstart}(h) \cap [i, i + \lfloor (7/8) \mu(t) \rfloor - 1 ] \neq \emptyset$. 
$i \in [\gamma+1, \gamma + \lfloor \mu(t) \rfloor]$ follows from the definition of the $\Psi_{\group}(t, i)$. 
In this case, $\mathcal{B}_{\bstart}(h) \cap [i, i + \lfloor (7/8) \mu(t) \rfloor - 1 ] \neq \emptyset$ follow from Proposition~\ref{prop:rightRun_minor_property}~\ref{enum:rightRun_minor_property:4}. 
Similarly, we obtain $\mathcal{B}_{\bstart}(h^{\prime}) \cap [i, i + \lfloor (7/8) \mu(t) \rfloor - 1 ] \neq \emptyset$. 

We prove $\mathcal{B}_{\bstart}(h+1) \cap [i, i + \lfloor (7/8) \mu(t) \rfloor - 1 ] = \emptyset$. 
$\mathcal{B}_{\bstart}(h+1) \cap [\gamma+1, \gamma + \lfloor \mu(t) \rfloor] = \emptyset$ 
follows from Proposition~\ref{prop:rightRun_minor_property}~\ref{enum:rightRun_minor_property:5}. 
$[i, i + \lfloor (7/8) \mu(t) \rfloor - 1 ] \subseteq [\gamma+1, \gamma + \lfloor \mu(t) \rfloor]$ 
follows from the definition of the $\Psi_{\group}(t, i)$. 
Therefore, we obtain $\mathcal{B}_{\bstart}(h+1) \cap [i, i + \lfloor (7/8) \mu(t) \rfloor - 1 ] = \emptyset$. 
Similarly, we obtain $\mathcal{B}_{\bstart}(h^{\prime}+1) \cap [i, i + \lfloor (7/8) \mu(t) \rfloor - 1 ] = \emptyset$.

    We prove $h = h^{\prime}$ by contradiction. 
    We assume that $h \neq h^{\prime}$ holds. 
    If $h < h^{\prime}$, 
    then $\mathcal{B}_{\bstart}(h^{\prime}) \cap [i, i + \lfloor (7/8) \mu(t) \rfloor - 1 ] = \emptyset$ 
    follows from 
    $\mathcal{B}_{\bstart}(h+1) \cap [i, i + \lfloor (7/8) \mu(t) \rfloor - 1 ] = \emptyset$ and 
    $\mathcal{B}_{\bstart}(h^{\prime}) \subseteq \mathcal{B}_{\bstart}(h^{\prime} - 1) \subseteq \cdots \subseteq \mathcal{B}_{\bstart}(h+1)$. 
    The two facts $\mathcal{B}_{\bstart}(h^{\prime}) \cap [i, i + \lfloor (7/8) \mu(t) \rfloor - 1 ] = \emptyset$ 
    and $\mathcal{B}_{\bstart}(h^{\prime}) \cap [i, i + \lfloor (7/8) \mu(t) \rfloor - 1 ] \neq \emptyset$ 
    yield a contradiction. 
    
    Otherwise (i.e., $h > h^{\prime}$), 
    we can show that there exists a contradiction using the same approach. 
    Therefore, Proposition~\ref{prop:in_group_rightRun_property}(i) must hold. 

\textbf{Proof of Proposition~\ref{prop:in_group_rightRun_property}(ii).}    
    We prove $C = C^{\prime}$. 
    Consider a position $j$ of sequence $S^{h}$ satisfying 
    $i \in [x^{h}_{j}+1, y^{h}_{j}]$ 
    for the substring $T[x^{h}_{j}..y^{h}_{j}]$ derived from 
    the $j$-th nonterminal $S^{h}[j]$ of sequence $S^{h}$ in input string $T$.     
    $i \in [\gamma+1, \gamma + \lfloor \mu(t) \rfloor]$ 
    and $i \in [\gamma^{\prime}+1, \gamma^{\prime} + \lfloor \mu(t) \rfloor]$ 
    follow from the definition of set $\Psi_{\group}(t, i)$. 
    We apply Proposition~\ref{prop:rightRun_minor_property}~\ref{enum:rightRun_minor_property:3} to 
    the interval attractor $([p, q], [\ell, r])$. 
    Then, the proposition shows that 
    $\val(S^{h}[j]) = C$ and $j \in [s^{h}, e^{h}-1]$. 
    Similarly, 
    we apply Proposition~\ref{prop:rightRun_minor_property}~\ref{enum:rightRun_minor_property:3} to 
    the interval attractor $([p^{\prime}, q^{\prime}], [\ell^{\prime}, r^{\prime}])$. 
    Then, $\val(S^{h}[j]) = C^{\prime}$ and $j \in [s^{\prime h}, e^{\prime h}-1]$ hold. 
    Therefore, $C = C^{\prime}$ holds. 
    
    \textbf{Proof of Proposition~\ref{prop:in_group_rightRun_property}(iii).}
    $\gamma \neq \gamma^{\prime}$ follows from $h = h^{\prime}$ and Corollary~\ref{cor:IA_identify_corollary}.

    \textbf{Proof of Proposition~\ref{prop:in_group_rightRun_property}(iv).}
    We prove $|\lcs(T[p^{\prime}-1, \gamma-1], (C^{\prime})^{n+1})| \leq 16\mu(h+1)$. 
    $|\lcs(T[p^{\prime}-1, \gamma-1], (C^{\prime})^{n+1})| < |[p^{\prime}-1, \gamma-1]|$ follows from Proposition~\ref{prop:rightRun_minor_property}~\ref{enum:rightRun_minor_property:1}. 
    $|[p^{\prime}-1, \gamma-1]| \leq 16\mu(h+1)$ follows from Proposition~\ref{prop:psi_H_length} and $h = h^{\prime}$. 
    Therefore, we obtain $|\lcs(T[p^{\prime}-1, \gamma-1], (C^{\prime})^{n+1})| < 16\mu(h+1)$. 

    We prove $-16\mu(h+1) < \gamma^{\prime} - \gamma < 16\mu(h+1)$. 
    If $\gamma \leq \gamma^{\prime}$, 
    then $\gamma < \gamma^{\prime} \leq i$ follows from $\gamma \leq \gamma^{\prime}$, 
    $\gamma \neq \gamma^{\prime}$, and the definition of set $\Psi_{\group}(t, i)$. 
    Proposition~\ref{prop:rightRun_property_H} shows that 
    substring $T[\gamma..i]$ is a prefix of string $C^{n+1}$. 
    This fact indicates that 
    $|\lcs(T[p^{\prime}-1, \gamma-1], (C^{\prime})^{n+1})| \geq \gamma^{\prime} - \gamma$ 
    because $C = C^{\prime}$ and $\gamma < \gamma^{\prime} \leq i$ hold. 
    Therefore, $\gamma^{\prime} - \gamma \leq 16\mu(h+1)$ follows from 
    $\gamma^{\prime} - \gamma \leq |\lcs(T[p^{\prime}-1, \gamma-1], (C^{\prime})^{n+1})|$  
    and $|\lcs(T[p^{\prime}-1, \gamma-1], (C^{\prime})^{n+1})| < 16\mu(h+1)$. 

    Otherwise (i.e., $\gamma > \gamma^{\prime}$), 
    we can prove $\gamma - \gamma^{\prime} \leq 16\mu(h+1)$ using the same approach. 
    Therefore, we obtain Proposition~\ref{prop:in_group_rightRun_property}(iv). 
\end{proof}

The following proposition states properties of 
the subset $\Psi_{\group}(t, i)$. 

\begin{proposition}\label{prop:group_super_properties}
    Let $d = \max \{ 1, \lfloor (1/8) \mu(t) \rfloor \}$ 
    and $M = \lceil \frac{n}{d} \rceil + 1$ for an integer $t \geq 3$. 
    Let $\mathcal{W}$ be a set of integers 
    such that each integer $j \in \mathcal{W}$ satisfies 
    $[1 + (j-1)d, (j-1)d + \lfloor (7/8) \mu(t) \rfloor] \subseteq \mathsf{C}_{\cover}(33 \mu(t+1))$.  
    (i.e., $\mathcal{W} = \{ j \geq 1 \mid [1 + (j-1)d, (j-1)d + \lfloor (7/8) \mu(t) \rfloor] \subseteq \mathsf{C}_{\cover}(33 \mu(t+1)) \}$).     
    The following four statements hold: 
\begin{enumerate}[label=\textbf{(\roman*)}]
    \item \label{enum:group_super_properties:1} $\mathbb{E}[|(\Psi_{\rightRun} \cap \Psi_{\rightLen}(\mu_{\SUM}(t), \mu_{\SUM}(t+1)) \cap \Psi_{\group}(t, i)) \setminus \Psi_{\run}|] = O(1)$ 
    for each position $i \in [1, n]$ of $T$; 
    \item \label{enum:group_super_properties:2} $(\Psi_{\rightRun} \cap \Psi_{\rightLen}(\mu_{\SUM}(t), \mu_{\SUM}(t+1))) \setminus \Psi_{\run} \subseteq \bigcup_{j \in [1, M]} \Psi_{\group}(t, 1 + (j-1)d)$; 
    \item \label{enum:group_super_properties:3} 
    $(\Psi_{\rightRun} \cap \Psi_{\leftmost} \cap \Psi_{\rightLen}(\mu_{\SUM}(t), \mu_{\SUM}(t+1))) \setminus \Psi_{\run} \subseteq \bigcup_{j \in \mathcal{W}} \Psi_{\group}(t, 1 + (j-1)d)$ 
    and $|\mathcal{W}| = O(\delta)$.

    %\item \label{enum:rightRun_size:4} $\mu(t+1) / d \leq 19$.
\end{enumerate}     
\end{proposition}
\begin{proof}
    The following two statements are used to prove Proposition~\ref{prop:group_super_properties}. 
\begin{enumerate}[label=\textbf{(\Alph*)}]
    \item consider an interval attractor $([p, q], [\ell, r])$ 
    in set $(\Psi_{\rightRun} \cap \Psi_{\rightLen}(\mu_{\SUM}(t), \mu_{\SUM}(t+1))) \setminus \Psi_{\run}$. 
    Let $j^{\prime} \geq 1$ be the smallest integer satisfying $1 + (j^{\prime}-1)d \geq \gamma$ 
    (i.e., $j^{\prime} = \min \{ j \geq 1 \mid 1 + (j-1)d \geq \gamma \}$) 
    for the attractor position $\gamma$ of the interval attractor $([p, q], [\ell, r])$. 
    Then, $j^{\prime} \leq M$ and $([p, q], [\ell, r]) \in \Psi_{\group}(t, 1 + (j^{\prime}-1)d)$ hold;
    \item $\mu(t+1) / d \leq 19$.
\end{enumerate}     

    \textbf{Proof of statement (A).}
    We prove $j^{\prime} \leq M$ by contradiction. 
    We assume that $j^{\prime} > M$. 
    Then, 
    $1 + (M-1)d < \gamma$ follows from the definition of the integer $j^{\prime}$. 
    On the other hand, 
    $1 + (M-1)d \geq \gamma$ follows from $M = \lceil \frac{n}{d} \rceil + 1$ 
    and $\gamma \leq n$. 
    The two facts $1 + (M-1)d < \gamma$ and $1 + (M-1)d \geq \gamma$ yield a contradiction. 
    Therefore, $j^{\prime} \leq M$ must hold.
    
    We prove $(j^{\prime}-1)d + \lfloor (7/8) \mu(t) \rfloor \leq \gamma + \lfloor \mu(t) \rfloor - 1$. 
    Since $t \geq 3$, 
    $1 \leq \lfloor (7/8) \mu(t) \rfloor \leq \lfloor \mu(t) \rfloor$ holds. 
    $\gamma \leq 1 + (j^{\prime}-1)d \leq \gamma + d - 1$ follows from 
    $j^{\prime} = \min \{ j \geq 1 \mid 1 + (j-1)d \geq \gamma \}$.     
    If $\lfloor (1/8) \mu(t) \rfloor \geq 1$, 
    then $d = \lfloor (1/8) \mu(t) \rfloor$ holds. 
    $(j^{\prime}-1)d + \lfloor (7/8) \mu(t) \rfloor \leq \gamma + \lfloor \mu(t) \rfloor - 1$ 
    follows from $1 + (j^{\prime}-1)d \leq \gamma + d - 1$ 
    and $\lfloor (1/8) \mu(t) \rfloor + \lfloor (7/8) \mu(t) \rfloor \leq \lfloor \mu(t) \rfloor$.     
    Otherwise (i.e., $\lfloor (1/8) \mu(t) \rfloor < 1$), 
    $d = 1$ holds. 
    $(j^{\prime}-1)d + \lfloor (7/8) \mu(t) \rfloor \leq \gamma + \lfloor \mu(t) \rfloor - 1$ 
    follows from $1 + (j^{\prime}-1)d \leq \gamma + d - 1$ 
    and $\lfloor (7/8) \mu(t) \rfloor \leq \lfloor \mu(t) \rfloor$. 

    We prove $([p, q], [\ell, r]) \in \Psi_{\group}(t, 1 + (j^{\prime}-1)d)$. 
    Let $i = 1 + (j^{\prime}-1)d$ for simplicity. 
    Then, 
    $([p, q], [\ell, r]) \in \Psi_{\group}(t, 1 + (j^{\prime}-1)d)$ holds if 
    $[i, i + \lfloor (7/8) \mu(t) \rfloor] \subseteq [\gamma, \gamma + \lfloor \mu(t) \rfloor - 1]$.
    $[i, i + \lfloor (7/8) \mu(t) \rfloor] \subseteq [\gamma, \gamma + \lfloor \mu(t) \rfloor - 1]$ 
    follows from $1 + (j^{\prime}-1)d \geq \gamma$, 
    $(j^{\prime}-1)d + \lfloor (7/8) \mu(t) \rfloor \leq \gamma + \lfloor \mu(t) \rfloor - 1$, 
    and $i = 1 + (j^{\prime}-1)d$.     
    Therefore, $([p, q], [\ell, r]) \in \Psi_{\group}(t, 1 + (j^{\prime}-1)d)$ holds. 
    
    \textbf{Proof of statement (B).}
    We consider the following three cases: 
    (a) $1 \leq \mu(t) < 8$, (b) $8 \leq \mu(t) < 16$, and (c) $\mu(t) \geq 16$. 
    For case (a), 
    $\mu(t+1) \leq 64/7$ follows from $\mu(t+1) \leq (8/7)\mu(t)$ and $\mu(t) < 8$. 
    $d = 1$ follows from $d = \max \{ 1, \lfloor (1/8) \mu(t) \rfloor \}$ 
    and $\lfloor (1/8) \mu(t) \rfloor = 0$. 
    Therefore, $\mu(t+1) / d \leq 19$. 

    For case (b), 
    $\mu(t+1) \leq 128/7$ follows from $\mu(t+1) \leq (8/7)\mu(t)$ and $\mu(t) < 16$. 
    $d \geq 1$ follows from $d = \max \{ 1, \lfloor (1/8) \mu(t) \rfloor \}$ 
    and $\lfloor (1/8) \mu(t) \rfloor \geq 1$. 
    Therefore, $\mu(t+1) / d \leq 19$. 

    For case (c), 
    $d = \lfloor (1/8) \mu(t) \rfloor$ holds. 
    There exists a rational number $0 \leq g < 8$ satisfying 
    $\lfloor (1/8) \mu(t) \rfloor = \frac{\mu(t) - g}{8}$. 
    $\frac{g}{\mu(t)} \leq (1/2)$ follows from $0 \leq g < 8$ and $\mu(t) \geq 16$. 
    Therefore, 
    \begin{equation*}
        \begin{split}
        \frac{\mu(t+1)}{d} &\leq (8/7) \frac{\mu(t)}{\frac{\mu(t) - g}{8}} \\
        &= (64/7) \frac{1}{1 - \frac{g}{\mu(t)}} \\
        &\leq 128/7 \\
        &\leq 19.
        \end{split}
    \end{equation*}

    %The proof of Proposition~\ref{prop:rightRun_size} is as follows.

    \textbf{Proof of Proposition~\ref{prop:group_super_properties}(i).}
    Let $\Psi = (\Psi_{\rightRun} \cap \Psi_{\rightLen}(\mu_{\SUM}(t), \mu_{\SUM}(t+1)) \cap \Psi_{\group}(t, i)) \setminus \Psi_{\run}$ for simplicity. 
    Proposition~\ref{prop:in_group_rightRun_property}~\ref{enum:in_group_rightRun_property:1} shows that 
    there exists an integer $h \in [0, H]$ satisfying 
    $\Psi \subseteq \Psi_{h}$. 
    Let $\Gamma$ be a set of positions obtained by collecting the attractor positions of 
    all the interval attractors in the set $\Psi$. 
    Then, $|\Gamma| = |\Psi|$ follows from 
    Proposition~\ref{prop:in_group_rightRun_property}~\ref{enum:in_group_rightRun_property:3}. 
    $\Gamma \subseteq \mathcal{B}_{\bstart}(h)$ holds because 
    each position of the set $\Gamma$ is the attractor position of an interval attractor of level $h$.
    Proposition~\ref{prop:in_group_rightRun_property}~\ref{enum:in_group_rightRun_property:5} shows that 
    there exists an interval $[s, s + \lceil 32\mu(h+1) \rceil - 1]$ of length $\lceil 32\mu(h+1) \rceil$ satisfying  
    $\Gamma \subseteq [s, s + \lceil 32\mu(h+1) \rceil - 1]$. 
    Proposition~\ref{prop:b_set_properties} shows that 
    $\mathbb{E}[|\mathcal{B}_{\bstart}(h) \cap [s, s + \lceil 32\mu(h+1) \rceil - 1]|] < 133$ holds. 
    The set $\Gamma$ is a subset of set $\mathcal{B}_{\bstart}(h) \cap [s, s + \lceil 32\mu(h+1) \rceil - 1]$. 
    Therefore, 
    \begin{equation*}
        \begin{split}
            \mathbb{E}[|\Psi|] &= \mathbb{E}[|\Gamma|] \\
            &\leq \mathbb{E}[|\mathcal{B}_{\bstart}(h) \cap [s, s + \lceil 32\mu(h+1) \rceil - 1]|] \\
            & < 133 \\
            &= O(1).
        \end{split}
    \end{equation*}
    
    \textbf{Proof of Proposition~\ref{prop:group_super_properties}(ii).}
    $(\Psi_{\rightRun} \cap \Psi_{\rightLen}(\mu_{\SUM}(t), \mu_{\SUM}(t+1))) \setminus \Psi_{\run} \subseteq \bigcup_{j \in [1, M]}$ $ \Psi_{\group}(t, 1 + (j-1)d)$ 
    follows from statement (A). 

    \textbf{Proof of Proposition~\ref{prop:group_super_properties}(iii).}        
    Consider an interval attractor $([p, q], [\ell, r])$ with attractor position $\gamma$ 
    in subset $(\Psi_{\rightRun} \cap \Psi_{\leftmost} \cap \Psi_{\rightLen}(\mu_{\SUM}(t), \mu_{\SUM}(t+1))) \setminus \Psi_{\run}$. 
    Let $j^{\prime} \geq 1$ be the smallest integer satisfying $1 + (j^{\prime}-1)d \geq \gamma$ 
    (i.e., $j^{\prime} = \min \{ j \geq 1 \mid 1 + (j-1)d \geq \gamma \}$). 
    Then, $j^{\prime} \in [1, M]$ and $([p, q], [\ell, r]) \in \Psi_{\group}(t, 1 + (j^{\prime}-1)d)$ 
    follow from statement (A). 

    We prove $(\Psi_{\rightRun} \cap \Psi_{\leftmost} \cap \Psi_{\rightLen}(\mu_{\SUM}(t), \mu_{\SUM}(t+1))) \setminus \Psi_{\run} \subseteq \bigcup_{j \in \mathcal{W}} \Psi_{\group}(t, 1 + (j-1)d)$. 
    $|[\gamma, r+1]| > \lfloor \mu(t) \rfloor$ follows from the definition of set $\Psi_{\rightLen}(\mu_{\SUM}(t), \mu_{\SUM}(t+1))$ 
    and $\mu_{\SUM}(t) \geq \lfloor \mu(t) \rfloor$. 
    $([p, q], [\ell, r]) \in \Psi_{\leftLen}(0, 16 \mu(t+1)) \cap \Psi_{\rightLen}(0, 17 \mu(t+1))$ follows from 
    Proposition~\ref{prop:rightRun_property_H}. 
    $[p-1, r+1] \subseteq \mathsf{C}_{\cover}(33 \mu(t+1)) \cup \{ 0, n+1 \}$ follows from 
    Proposition~\ref{prop:leftmost_susbet_sub_properties}~\ref{enum:leftmost_susbet_sub_properties:2}. 
    $\gamma \in [p, r]$ follows from the definition of interval attractor. 
    $[1 + (j^{\prime}-1)d, (j^{\prime}-1)d + \lfloor (7/8) \mu(t) \rfloor] \subseteq \mathsf{C}_{\cover}(33 \mu(t+1))$ follows from 
    the following equation: 
    \begin{equation*}
        \begin{split}
            [1 + (j^{\prime}-1)d, (j^{\prime}-1)d + \lfloor (7/8) \mu(t) \rfloor] &\subseteq [\gamma, \gamma + \lfloor \mu(t) \rfloor - 1] \\
            &\subseteq [\gamma, r] \\
            &\subseteq [p, r] \\
            &\subseteq \mathsf{C}_{\cover}(33 \mu(t+1)).
        \end{split}
    \end{equation*}
    $j^{\prime} \in \mathcal{W}$ follows from $[1 + (j^{\prime}-1)d, (j^{\prime}-1)d + \lfloor (7/8) \mu(t) \rfloor] \subseteq \mathsf{C}_{\cover}(33 \mu(t+1))$. 
    Therefore, we obtain $(\Psi_{\rightRun} \cap \Psi_{\leftmost} \cap \Psi_{\rightLen}(\mu_{\SUM}(t), \mu_{\SUM}(t+1))) \setminus \Psi_{\run} \subseteq \bigcup_{j \in \mathcal{W}} \Psi_{\group}(t, 1 + (j-1)d)$.

    Next, we prove $|\mathcal{W}| = O(\delta)$. 
    $|\mathcal{W}| \leq |\{ 1 + (j-1)d) \mid j \geq 1 \} \cap \mathsf{C}_{\cover}(33 \mu(t+1))|$ 
    follows from the definition of the set $\mathcal{W}$. 
    From Lemma~\ref{lem:m_cover_properties}~\ref{enum:m_cover_properties:2}, 
    the set $\mathsf{C}_{\cover}(33 \mu(t+1))$ 
    can be divided into $k$ disjoint intervals $I_{1}, I_{2}, \ldots, I_{k} \subseteq [1, n]$ in input string $T$ 
    satisfying $k \leq c \delta$ for a constant $c \geq 1$. 
    
    From Lemma~\ref{lem:m_cover_properties}~\ref{enum:m_cover_properties:1}, 
    there exists a constant $c^{\prime} \geq 1$ satisfying 
    $|\mathsf{C}_{\cover}(33 \mu(t+1))| \leq c^{\prime} \mu(t+1) \delta$. 
    $\mu(t+1) / d \leq 19$ follows from statement (B). 
    Therefore, $|\mathcal{W}| = O(\delta)$ follows from the following equation: 
\begin{equation*}
\begin{split}
|\mathcal{W}| &\leq |\{ 1 + (j-1)d) \mid j \geq 1 \} \cap \mathsf{C}_{\cover}(33 \mu(t+1))| \\
              &= \sum_{s = 1}^{k} |\{ 1 + (j-1)d) \mid j \geq 1 \} \cap I_{s}|  \\
              &\leq k + \sum_{s = 1}^{k} \frac{|I_{s}|}{d} \\
              &= k + \frac{|\mathsf{C}_{\cover}(33 \mu(t+1))|}{d} \\
              &\leq c \delta + \frac{c^{\prime} \mu(t+1) \delta}{d} \\
              &\leq c \delta + 19c^{\prime}\delta\\
              &=O(\delta).
\end{split}
\end{equation*}
\end{proof}

\subsubsection{Proof of Theorem~\ref{theo:RR_Psi_set_size}}
We prove $\mathbb{E}[|\Psi_{\leftmost} \setminus \Psi_{\run}|] = O(\delta \log \frac{n \log \sigma}{\delta \log n})$. 
The subset $\Psi_{\leftmost} \setminus \Psi_{\run}$ can be divided into two subsets $\Psi_{\leftmost} \setminus \Psi_{\rightRun}$ and $(\Psi_{\leftmost} \cap \Psi_{\rightRun}) \setminus \Psi_{\run}$ 
because $(\Psi_{\leftmost} \setminus \Psi_{\rightRun}) \cap \Psi_{\run} = \emptyset$. 
The following proposition gives an upper bound to the expected size of the subset $\Psi_{\leftmost} \setminus \Psi_{\run}$. 

\begin{proposition}\label{prop:expected_size_A}
$\mathbb{E}[|\Psi_{\leftmost} \setminus \Psi_{\rightRun}|] = O(\delta \log \frac{n \log \sigma}{\delta \log n})$. 
\end{proposition}
\begin{proof}
Let $\pi = \lfloor 2 \log_{8/7} \frac{\log_{\sigma} \delta}{53} \rfloor - 1$ and $\tau = \lceil 2 \log_{8/7} \frac{n}{\delta} \rceil$. 
The subset $\Psi_{\leftmost} \setminus \Psi_{\rightRun}$ 
is a subset of the union of three subsets $(\bigcup_{t = 0}^{\pi} \Psi_{t} \cap \Psi_{\leftmost}) \setminus \Psi_{\rightRun}$, 
$\Bigl(\bigcup_{t = \max \{ 0, \pi\}}^{\tau} \Psi_{t} \cap \Psi_{\leftmost} \Bigr) \setminus \Psi_{\rightRun}$, 
and $\bigcup_{t = \tau}^{H} \Psi_{t}$ because $\Psi_{\leftmost} \subseteq \bigcup_{t=0}^{H} \Psi_{t}$ holds. 
Therefore, 
Proposition~\ref{prop:expected_size_A} follows from the following three statements. 
\begin{enumerate}[label=\textbf{(\roman*)}]
\item $|(\bigcup_{t = 0}^{\max \{ 0, \pi\}} \Psi_{t} \cap \Psi_{\leftmost}) \setminus \Psi_{\rightRun}| = O(\delta)$; 
\item $\mathbb{E}[|\Bigl(\bigcup_{t = \max \{ 0, \pi\}}^{\tau} \Psi_{t} \cap \Psi_{\leftmost} \Bigr) \setminus \Psi_{\rightRun}|] = O(\delta \log \frac{n \log \sigma}{\delta \log n})$;
\item $\mathbb{E}[|\bigcup_{t = \tau}^{H} \Psi_{t}|] = O(\delta)$.
\end{enumerate}

\paragraph{Proof of statement (i).}
    We prove $|(\Psi_{t} \cap \Psi_{\leftmost}) \setminus \Psi_{\rightRun}| \leq \sigma^{52\mu(t+1)}$ 
    for each integer $t \in [0, H]$.     
    $(\Psi_{t} \cap \Psi_{\leftmost}) \setminus \Psi_{\rightRun} \subseteq \Psi_{\leftLen}(0, 16\mu(t+1)) \cap \Psi_{\rightLen}(0, 36\mu(t+1))$ follows from Proposition~\ref{prop:psi_H_length}. 
    We can apply Proposition~\ref{prop:leftmost_susbet_sub_properties}~\ref{enum:leftmost_susbet_sub_properties:3} to the set $(\Psi_{t} \cap \Psi_{\leftmost}) \setminus \Psi_{\rightRun}$ because 
    Corollary~\ref{cor:IA_identify_corollary} shows that 
    the attractor positions of the interval attractors in the set $\Psi_{t}$ are distinct. 
    $|(\Psi_{t} \cap \Psi_{\leftmost}) \setminus \Psi_{\rightRun}| \leq \sigma^{52\mu(t+1)}$ 
    follows from Proposition~\ref{prop:leftmost_susbet_sub_properties}~\ref{enum:leftmost_susbet_sub_properties:3} 
    and $(\Psi_{t} \cap \Psi_{\leftmost}) \setminus \Psi_{\rightRun} \subseteq \Psi_{\leftLen}(0, 16\mu(t+1)) \cap \Psi_{\rightLen}(0, 36\mu(t+1))$. 
        
    We prove $|(\bigcup_{t = 0}^{\pi} \Psi_{t} \cap \Psi_{\leftmost}) \setminus \Psi_{\rightRun}| = O(\delta)$. 
    We already proved $|(\Psi_{t} \cap \Psi_{\leftmost}) \setminus \Psi_{\rightRun}| \leq \sigma^{52\mu(t+1)}$. 
    Here, $52\mu(\pi+1) \leq \log_{\sigma} \delta$ holds. 
    $\sum_{i = 0}^{\lfloor \log_{\sigma} \delta \rfloor} \sigma^{i} \leq 2\delta$ follows from 
    $\sum_{i = 0}^{\lfloor \log_{\sigma} \delta \rfloor} \sigma^{i} \leq 2\sigma^{\log_{\sigma} \delta}$ and 
    $\sigma^{\log_{\sigma} \delta} = \delta$. 
    Therefore, 
\begin{equation*}
    \begin{split}
    |(\bigcup_{t = 0}^{\max \{ 0, \pi\}} \Psi_{t} \cap \Psi_{\leftmost}) \setminus \Psi_{\rightRun}| &\leq \sum_{t = 0}^{\max \{ 0, \pi\}} \sigma^{53\mu(t+1)} \\
    &= O(\sum_{i = 0}^{\lfloor \log_{\sigma} \delta \rfloor} \sigma^{i}) \\
    &= O(\delta).
    \end{split}
\end{equation*}
    
\paragraph{Proof of statement (ii).}
We will prove $\tau - \pi = O(\log \frac{n \log \sigma}{\delta \log \delta})$. 
$\tau \leq 2 \log_{8/7} \frac{n}{\delta} + 1$ follows from $\tau = \lceil 2 \log_{8/7} \frac{n}{\delta} \rceil$. 
$\pi \geq 2 \log_{8/7} \frac{\log_{\sigma} \delta}{53} - 2$ follows from $\pi = \lfloor 2 \log_{8/7} \frac{\log_{\sigma} \delta}{53} \rfloor - 1$. 
Therefore, $\tau - \pi = O(\log \frac{n \log \sigma}{\delta \log \delta})$ follows from the following equation. 
\begin{equation*}
\begin{split}
\tau - \pi &\leq (2 \log_{8/7} \frac{n}{\delta} + 1) - (2 \log_{8/7} \frac{\log_{\sigma} \delta}{53} - 2) \\
&= 2 \log_{8/7} \frac{53n}{\delta \log_{\sigma} \delta} + 3 \\
&= 2 \log_{8/7} \frac{53n \log \sigma}{\delta \log \delta} + 3 \\
&= O(\log \frac{n \log \sigma}{\delta \log \delta}).
\end{split}
\end{equation*}

We already proved $\log \frac{n \log \sigma}{\delta \log \delta} = O(\log \frac{n \log \sigma}{\delta \log n})$ in the proof of Proposition~\ref{prop:vb_size}.
Therefore, $\tau - \pi = O(\log \frac{n \log \sigma}{\delta \log n})$ holds. 

We prove statement (ii). 
We already proved $(\Psi_{t} \cap \Psi_{\leftmost}) \setminus \Psi_{\rightRun} \subseteq \Psi_{\leftLen}(0, 16\mu(t+1)) \cap \Psi_{\rightLen}(0, 36\mu(t+1))$. 
$|\Psi_{t} \cap \Psi_{\leftmost} \cap \Psi_{\leftLen}(0, 16\mu(t+1)) \cap \Psi_{\rightLen}(0, 36\mu(t+1))| \leq |\mathcal{B}_{\bstart}(t) \cap \mathsf{C}_{\cover}(52 \mu(t+1))|$ follows from Proposition~\ref{prop:leftmost_susbet_sub_properties}~\ref{enum:leftmost_susbet_sub_properties:2}. 
$\mathbb{E}[ |\mathcal{B}_{\bstart}(t) \cap \mathsf{C}_{\cover}(52 \mu(t+1))|] = 8 + \frac{|\mathcal{S}_{416 \mu(t+1)}|}{52 \mu(t+1)} + \frac{4 |\mathsf{C}_{\cover}(52 \mu(t+1))| }{\mu(t+1)}$ follows from Lemma~\ref{lem:d_cover_expected_size}. 
$\sum_{t = 0}^{\infty} \frac{|\mathsf{C}_{\cover}(52 \mu(t+1))| }{\mu(t+1)} = O(\delta \log \frac{n \log \sigma}{\delta \log n})$ 
follows from Lemma~\ref{lem:d_cover_total_size}. 
$\sum_{t = 0}^{\infty} \frac{|\mathcal{S}_{416 \mu(t+1)}|}{52 \mu(t+1)} = O(\delta \log \frac{n \log \sigma}{\delta \log n})$ holds because 
Kempa and Kociumaka showed that $\sum_{i = 0}^{\infty} \frac{1}{2^{i}}|\mathcal{S}_{c2^{i}}| = O(c \delta \log \frac{c n \log \sigma}{\delta \log n})$ 
for every $c \geq 1$~(Lemma 8.12 in \cite{DBLP:journals/corr/abs-2308-03635}).

Therefore, 
\begin{equation*}
    \begin{split}
    & \mathbb{E}[|\Bigl(\bigcup_{t = \max \{ 0, \pi\}}^{\tau} \Psi_{t} \cap \Psi_{\leftmost} \Bigr) \setminus \Psi_{\rightRun}|]  \\ 
    &\leq  \sum_{t = \max \{ 0, \pi\}}^{\tau} \mathbb{E}[|\Psi_{t} \cap \Psi_{\leftmost} \cap \Psi_{\leftLen}(0, 16\mu(t+1)) \cap \Psi_{\rightLen}(0, 36\mu(t+1))|]  \\
    &\leq \sum_{t = \max \{ 0, \pi\}}^{\tau} \mathbb{E}[|\mathcal{B}_{\bstart}(t) \cap \mathsf{C}_{\cover}(52 \mu(t+1))|]   \\
    &\leq \sum_{t = \max \{ 0, \pi\}}^{\tau} (8 + \frac{|\mathcal{S}_{416 \mu(t+1)}|}{52 \mu(t+1)} + \frac{4 |\mathsf{C}_{\cover}(52 \mu(t+1))| }{\mu(t+1)}) \\
    & \leq 8(\tau - \pi + 1) + \sum_{t = 0}^{\infty} \frac{|\mathcal{S}_{416 \mu(t+1)}|}{52 \mu(t+1)} + 4\sum_{t = 0}^{\infty} \frac{|\mathsf{C}_{\cover}(52 \mu(t+1))| }{\mu(t+1)} \\
    &= O(\delta \log \frac{n \log \sigma}{\delta \log n}). 
    \end{split}
\end{equation*}

%The following proposition states a property of the set $\mathcal{S}_{d}$ of strings. 
%\begin{lemma}\label{lem:substring_set_total_size}[Lemma 8.12 in \cite{DBLP:journals/corr/abs-2308-03635}]
%$\sum_{i = 0}^{\infty} \frac{1}{2^{i}}|\mathcal{S}_{c2^{i}}| = O(c \delta \log \frac{c n \log \sigma}{\delta \log n})$ 
%for every $c \geq 1$.
%\end{lemma}

\paragraph{Proof of statement (iii).}
$\sum_{t = \tau}^{H} \mathbb{E}[|\Psi_{t}|] \leq \sum_{t = \tau}^{H} (32n/7) \frac{1}{(8/7)^{(t/2)}}$ 
follows from Proposition~\ref{prop:psi_H_expected_size}. 
There exists a constant $\alpha$ satisfying $\sum_{i = 0}^{\infty} (7/8)^{i/2} \leq \alpha$. 
$\frac{1}{(8/7)^{(\tau/2)}} \leq \delta / n$ follows from $\tau = \lceil 2 \log_{8/7} \frac{n}{\delta} \rceil$. 
Therefore, 
\begin{equation*}
    \begin{split}
    \mathbb{E}[|\bigcup_{t = \tau}^{H} \Psi_{t}|] &= \sum_{t = \tau}^{H} \mathbb{E}[|\Psi_{t}|] \\
    &< \sum_{t = \tau}^{H} (32n/7) \frac{1}{(8/7)^{(t/2)}} \\
    &\leq (32n/7) \frac{1}{(8/7)^{(\tau/2)}} \sum_{i = 0}^{\infty} (7/8)^{i/2} \\
    &\leq \frac{32\alpha n}{7} \times \frac{\delta}{n} \\
    &= O(\delta).     
    \end{split}
\end{equation*}
\end{proof}

The following proposition gives an upper bound to the expected size of the subset $(\Psi_{\leftmost} \cap \Psi_{\rightRun}) \setminus \Psi_{\run}$. 

\begin{proposition}\label{prop:expected_size_B}
$\mathbb{E}[|(\Psi_{\leftmost} \cap \Psi_{\rightRun}) \setminus \Psi_{\run}|] = O(\delta \log \frac{n \log \sigma}{\delta \log n})$. 
\end{proposition}
\begin{proof}
Let $\pi$ and $\tau$ be the two integers introduced in the proof of Proposition~\ref{prop:expected_size_A}. 
Let $\mu_{\SUM}(t)$ be the integer introduced in Section~\ref{subsubsec:sub_properties}. 
$(\Psi_{\leftmost} \cap \Psi_{\rightRun}) \setminus \Psi_{\run} = \Bigl(\bigcup_{t = 0}^{\infty} \Psi_{\rightRun} \cap \Psi_{\leftmost} \cap \Psi_{\rightLen}(\mu_{\SUM}(t), \mu_{\SUM}(t+1)) \Bigr) \setminus \Psi_{\run}$ 
because $\Psi_{\RR} = \bigcup_{t = 0}^{\infty} \Psi_{\rightLen}(\mu_{\SUM}(t), \mu_{\SUM}(t+1))$. 
Therefore, Proposition~\ref{prop:expected_size_B} follows from the following three statements. 
\begin{enumerate}[label=\textbf{(\roman*)}]
\item $|\Bigl(\bigcup_{t = 0}^{\max \{ 0, \pi\}} \Psi_{\rightRun} \cap \Psi_{\leftmost} \cap \Psi_{\rightLen}(\mu_{\SUM}(t), \mu_{\SUM}(t+1)) \Bigr) \setminus \Psi_{\run}| = O(\delta)$;
\item $\mathbb{E}[|\Bigl(\bigcup_{t = \max \{ 0, \pi\}}^{\tau+2} \Psi_{\rightRun} \cap \Psi_{\leftmost} \cap \Psi_{\rightLen}(\mu_{\SUM}(t), \mu_{\SUM}(t+1)) \Bigr) \setminus \Psi_{\run}|] = O(\delta \log \frac{n \log \sigma}{\delta \log n})$;
\item $\mathbb{E}[|\Bigl(\bigcup_{t = \tau+3}^{\infty} \Psi_{\rightRun} \cap \Psi_{\leftmost} \cap \Psi_{\rightLen}(\mu_{\SUM}(t), \mu_{\SUM}(t+1)) \Bigr) \setminus \Psi_{\run}|] = O(\delta)$.
\end{enumerate}

\paragraph{Proof of $|(\Psi_{\rightRun} \cap \Psi_{\leftmost} \cap \Psi_{\rightLen}(\mu_{\SUM}(t), \mu_{\SUM}(t+1))) \setminus \Psi_{\run}| \leq \sigma^{33\mu(t+1)}$.}
We will prove $|(\Psi_{\rightRun} \cap \Psi_{\leftmost} \cap \Psi_{\rightLen}(\mu_{\SUM}(t), \mu_{\SUM}(t+1))) \setminus \Psi_{\run}| \leq \sigma^{33\mu(t+1)}$ 
for each integer $t \geq 0$. This statement is used to prove statement (i) and statement (ii). 

We show that $|(\Psi_{\rightRun} \cap \Psi_{\leftmost} \cap \Psi_{\rightLen}(\mu_{\SUM}(t), \mu_{\SUM}(t+1))) \setminus \Psi_{\run}| = 0$ by contradiction for $t = 0$. 
We assume that $(\Psi_{\rightRun} \cap \Psi_{\leftmost} \cap \Psi_{\rightLen}(\mu_{\SUM}(0), \mu_{\SUM}(1))) \setminus \Psi_{\run} \neq \emptyset$ holds. 
Then, the subset $\Psi_{\rightLen}(\mu_{\SUM}(0), \mu_{\SUM}(1))$ contains an interval attractor $([p, q], [\ell, r])$. 
Here, $\mu_{\SUM}(0) = 0$ and $\mu_{\SUM}(1) = 1$. 
$([p, q], [\ell, r]) \not \in \Psi_{\rightLen}(2, n+1)$ follows from $([p, q], [\ell, r]) \in \Psi_{\rightLen}(0, 1)$ and 
$\Psi_{\rightLen}(0, 1) \cap \Psi_{\rightLen}(2, n+1) = \emptyset$. 
On the other hand, 
$|[\gamma, r + 1]| \geq 2$ (i.e., $([p, q], [\ell, r]) \in \Psi_{\rightLen}(2, n+1)$) follows from the definition of interval attractor  
for the attractor position $\gamma$ of the interval attractor $([p, q], [\ell, r])$. 
The two facts $([p, q], [\ell, r]) \not \in \Psi_{\rightLen}(2, n+1)$ and $([p, q], [\ell, r]) \in \Psi_{\rightLen}(2, n+1)$ yield a contradiction. 
Therefore, $|(\Psi_{\rightRun} \cap \Psi_{\leftmost} \cap \Psi_{\rightLen}(\mu_{\SUM}(t), \mu_{\SUM}(t+1))) \setminus \Psi_{\run}| = 0$ must hold. 

Similarly, we show that $|(\Psi_{\rightRun} \cap \Psi_{\leftmost} \cap \Psi_{\rightLen}(\mu_{\SUM}(t), \mu_{\SUM}(t+1))) \setminus \Psi_{\run}| = 0$ by contradiction for $t \in \{ 1, 2 \}$. 
We assume that $(\Psi_{\rightRun} \cap \Psi_{\leftmost} \cap \Psi_{\rightLen}(\mu_{\SUM}(t), \mu_{\SUM}(t+1))) \setminus \Psi_{\run} \neq \emptyset$ holds. 
Then, the subset contains an interval attractor $([p, q], [\ell, r])$ with level $h \geq 0$, 
and $h \leq t-3$ follows from Proposition~\ref{prop:rightRun_property_H}. 
$h < 0$ follows from $h \leq t-3$ and $t \in \{ 1, 2 \}$, 
but this fact contradicts $h \geq 0$. 
Therefore, $|(\Psi_{\rightRun} \cap \Psi_{\leftmost} \cap \Psi_{\rightLen}(\mu_{\SUM}(t), \mu_{\SUM}(t+1))) \setminus \Psi_{\run}| = 0$ must hold. 

Next, we show that the attractor positions of the interval attractors in the subset $(\Psi_{\rightRun} \cap \Psi_{\leftmost} \cap \Psi_{\rightLen}(\mu_{\SUM}(t), \mu_{\SUM}(t+1))) \setminus \Psi_{\run}$ are distinct by contradiction for $t \geq 3$. 
We assume that the subset contains two interval attractors $([p_{1}, q_{1}], [\ell_{1}, r_{1}])$ 
and $([p_{2}, q_{2}], [\ell_{2}, r_{2}])$ such that their attractor positions are the same. 
Let $\gamma_{1}$ and 
$h_{1}$~(respectively, $\gamma_{2}$ and $h_{2}$) be the attractor position and level of the interval attractor $([p_{1}, q_{1}], [\ell_{1}, r_{1}])$~(respectively, $([p_{2}, q_{2}], [\ell_{2}, r_{2}])$). 
The subset $\Psi_{\group}(t, \gamma_{1})$ contains the two interval attractors $([p_{1}, q_{1}], [\ell_{1}, r_{1}])$ 
and $([p_{2}, q_{2}], [\ell_{2}, r_{2}])$. 
$h_{1} = h_{2}$ follows from Proposition~\ref{prop:in_group_rightRun_property}~\ref{enum:in_group_rightRun_property:1}. 
$\gamma_{1} \neq \gamma_{2}$ follows from $h_{1} = h_{2}$ and Corollary~\ref{cor:IA_identify_corollary}. 
This fact contradicts $\gamma_{1} = \gamma_{2}$. 
Therefore, the attractor positions of the interval attractors in the subset $(\Psi_{\rightRun} \cap \Psi_{\leftmost} \cap \Psi_{\rightLen}(\mu_{\SUM}(t), \mu_{\SUM}(t+1))) \setminus \Psi_{\run}$ must be distinct. 

We prove $|(\Psi_{\rightRun} \cap \Psi_{\leftmost} \cap \Psi_{\rightLen}(\mu_{\SUM}(t), \mu_{\SUM}(t+1))) \setminus \Psi_{\run}| \leq \sigma^{33\mu(t+1)}$ 
for $t \geq 3$.
$(\Psi_{\rightRun} \cap \Psi_{\leftmost} \cap \Psi_{\rightLen}(\mu_{\SUM}(t), \mu_{\SUM}(t+1))) \setminus \Psi_{\run} \subseteq \Psi_{\leftLen}(0, 17 \mu(t+1))) \cap \Psi_{\rightLen}(0, 16 \mu(t+1)))$ follows from Proposition~\ref{prop:psi_H_length}. 
We apply Proposition~\ref{prop:leftmost_susbet_sub_properties}~\ref{enum:leftmost_susbet_sub_properties:3} 
to $(\Psi_{\rightRun} \cap \Psi_{\leftmost} \cap \Psi_{\rightLen}(\mu_{\SUM}(t), \mu_{\SUM}(t+1))) \setminus \Psi_{\run}$. 
Then, the proposition shows that $|(\Psi_{\rightRun} \cap \Psi_{\leftmost} \cap \Psi_{\rightLen}(\mu_{\SUM}(t), \mu_{\SUM}(t+1))) \setminus \Psi_{\run}| \leq \sigma^{33\mu(t+1)}$ holds. 
Therefore, $|(\Psi_{\rightRun} \cap \Psi_{\leftmost} \cap \Psi_{\rightLen}(\mu_{\SUM}(t)$, $\mu_{\SUM}(t+1))) \setminus \Psi_{\run}| \leq \sigma^{33\mu(t+1)}$ holds for all $t \geq 0$. 

\paragraph{Proof of $\lceil (n/d) \rceil + 1 \leq 16\delta(7/8)^{t^{\prime}/2} + 2$.}
Let $d = \max \{ 1, \lfloor (1/8) \mu(t^{\prime} + \tau + 2) \rfloor \}$ for an integer $t^{\prime} \geq 1$. 
Then, we will prove $(\lceil (n/d) \rceil + 1) \leq 16\delta(7/8)^{t^{\prime}/2} + 2$. 
This statement is used to prove statement (iii). 

$\mu(t^{\prime} + \tau + 2) \geq (8/7)^{(t^{\prime}/2)} (n / \delta)$ follows from the following equation. 
\begin{equation*}
\begin{split}
\mu(t^{\prime} + \tau + 2) &= (8/7)^{\lceil (t^{\prime} + \tau + 2)/2 \rceil - 1} \\
&\geq (8/7)^{\lceil (t^{\prime}/2) + 1 + \log_{8/7} \frac{n}{\delta} \rceil - 1} \\
&\geq (8/7)^{(t^{\prime}/2) + \log_{8/7} \frac{n}{\delta}} \\
&= (8/7)^{(t^{\prime}/2)} (n / \delta).  
\end{split}
\end{equation*} 

If $1 \leq \mu(t^{\prime} + \tau + 2) < 16$, 
then $d = 1$ and $(8/7)^{(t^{\prime}/2)} (n / \delta) \leq \mu(t^{\prime} + \tau + 2) < 16$ hold. 
$n < 16\delta (7/8)^{t^{\prime}/2}$ follows from $(8/7)^{(t^{\prime}/2)} (n / \delta) < 16$ and $t^{\prime} \geq 1$. 
$\lceil (n/d) \rceil + 1 \leq 16\delta(7/8)^{t^{\prime}/2} + 2$ follows from the following equation. 
\begin{equation*}
\begin{split}
\lceil (n/d) \rceil + 1 &= n + 1 \\
&\leq 16\delta (7/8)^{b/2} + 1 \\
&\leq 16\delta(7/8)^{b/2} + 2.  
\end{split}
\end{equation*}    

%We prove $\lceil (n/d) \rceil + 1 \leq 16\delta(7/8)^{t^{\prime}/2} + 2$. 

Otherwise (i.e., $\mu(t^{\prime} + \tau + 2) \geq 16$), 
$d = \lfloor (1/8) \mu(t^{\prime} + \tau + 2) \rfloor$ holds, 
and there exists a rational number $0 \leq g < 8$ satisfying 
$\lfloor (1/8) \mu(t^{\prime} + \tau + 2) \rfloor = \frac{\mu(t^{\prime} + \tau + 2) - g}{8}$. 
$\frac{g}{\mu(t^{\prime} + \tau + 2)} \leq 1/2$ follows from $0 \leq g < 8$ 
and $\mu(t^{\prime} + \tau + 2) \geq 16$. 
Therefore, 
$\lceil (n/d) \rceil + 1 \leq 16\delta(7/8)^{t^{\prime}/2} + 2$ follows from the following equation. 
\begin{equation*}
\begin{split}
\lceil (n/d) \rceil + 1 &= \lceil \frac{8n}{\mu(t^{\prime} + \tau + 2) - g} \rceil + 1 \\
&= \lceil \frac{8n / \mu(t^{\prime} + \tau + 2)}{1 - (g/\mu(t^{\prime} + \tau + 2))} \rceil + 1 \\
&\leq \lceil 16\frac{n}{\mu(t^{\prime} + \tau + 2)} \rceil + 1 \\
&\leq \lceil 16\delta(7/8)^{t^{\prime}/2} \rceil + 1 \\
&\leq 16\delta(7/8)^{t^{\prime}/2} + 2.  
\end{split}
\end{equation*}

\paragraph{Proof of statement (i).}
We already proved $\sum_{t=0}^{\pi} \sigma^{53\mu(t+1)} = O(\delta)$ in the proof of Proposition~\ref{prop:expected_size_A}. 
Therefore, 
\begin{equation*}
    \begin{split}
    & |\Bigl(\bigcup_{t = 0}^{\max \{ 0, \pi\}} \Psi_{\rightRun} \cap \Psi_{\leftmost} \cap \Psi_{\rightLen}(\mu_{\SUM}(t), \mu_{\SUM}(t+1)) \Bigr) \setminus \Psi_{\run}| \\
    &\leq \sum_{t = 0}^{\pi} \Bigl( (\Psi_{\rightRun} \cap \Psi_{\leftmost} \cap \Psi_{\rightLen}(\mu_{\SUM}(t), \mu_{\SUM}(t+1)) ) \setminus \Psi_{\run}| \Bigr) \\
    &\leq \sum_{t = 0}^{\pi} \sigma^{33\mu(t+1)}\\
    &= O(\delta).     
    \end{split}
\end{equation*}

\paragraph{Proof of statement (ii).}
We prove $\mathbb{E}[|(\Psi_{\rightRun} \cap \Psi_{\leftmost} \cap \Psi_{\rightLen}(\mu_{\SUM}(t), \mu_{\SUM}(t+1))) \setminus \Psi_{\run}|] = O(\delta)$ for each integer $t \in [0, H]$. 
If $t < 3$, then $\mathbb{E}[|(\Psi_{\rightRun} \cap \Psi_{\leftmost} \cap \Psi_{\rightLen}(\mu_{\SUM}(t), \mu_{\SUM}(t+1))) \setminus \Psi_{\run}|] = 0$ because we already proved $|(\Psi_{\rightRun} \cap \Psi_{\leftmost} \cap \Psi_{\rightLen}(\mu_{\SUM}(t), \mu_{\SUM}(t+1))) \setminus \Psi_{\run}| = 0$. 
Otherwise, 
Proposition~\ref{prop:group_super_properties}~\ref{enum:group_super_properties:3} shows that 
there exists a set $\mathcal{W}^{\prime}$ of $O(\delta)$ integers 
satisfying $(\Psi_{\rightRun} \cap \Psi_{\leftmost} \cap \Psi_{\rightLen}(\mu_{\SUM}(t), \mu_{\SUM}(t+1))) \setminus \Psi_{\run} \subseteq \bigcup_{j \in \mathcal{W}^{\prime}} \Psi_{\group}(t, 1 + (j-1)d)$. 
Here, $d = \max \{ 1, \lfloor (1/8) \mu(t) \rfloor \}$.     
For each integer $j \in \mathcal{W}^{\prime}$, 
$\mathbb{E}[|\Psi_{\rightRun} \cap \Psi_{\leftmost} \cap \Psi_{\rightLen}(\mu_{\SUM}(t), \mu_{\SUM}(t+1)) \cap \Psi_{\group}(t, 1 + (j-1)d)) \setminus \Psi_{\run}|] = O(1)$ follows from Proposition~\ref{prop:group_super_properties}~\ref{enum:group_super_properties:1}. 
Therefore, 
\begin{equation*}
    \begin{split}
        & \mathbb{E}[|(\Psi_{\rightRun} \cap \Psi_{\leftmost} \cap \Psi_{\rightLen}(\mu_{\SUM}(t), \mu_{\SUM}(t+1))) \setminus \Psi_{\run}|] \\ 
        & \leq \mathbb{E}[\sum_{j \in \mathcal{W}^{\prime}} |\Psi_{\rightRun} \cap \Psi_{\leftmost} \cap \Psi_{\rightLen}(\mu_{\SUM}(t), \mu_{\SUM}(t+1)) \cap \Psi_{\group}(t, 1 + (j-1)d)) \setminus \Psi_{\run}|] \\
        &= \sum_{j \in \mathcal{W}^{\prime}} \mathbb{E}[|\Psi_{\rightRun} \cap \Psi_{\leftmost} \cap \Psi_{\rightLen}(\mu_{\SUM}(t), \mu_{\SUM}(t+1)) \cap \Psi_{\group}(t, 1 + (j-1)d)) \setminus \Psi_{\run}|] \\
        &= O(\delta).            
    \end{split}
\end{equation*}

We prove statement (ii). 
We already proved $\tau - \pi = O(\log \frac{n \log \sigma}{\delta \log n})$ in the proof of Proposition~\ref{prop:expected_size_B}. 
Therefore, 
\begin{equation*}
    \begin{split}
    & \mathbb{E}[| \Bigl(\bigcup_{t = \max\{ 0, \pi \} }^{\tau+2} (\Psi_{\rightRun} \cap \Psi_{\leftmost} \cap \Psi_{\rightLen}(\mu_{\SUM}(t), \mu_{\SUM}(t+1)) \Bigr) \setminus \Psi_{\run}|] \\
    &= \sum_{t = \max\{ 0, \pi \}}^{\tau+2} \mathbb{E}[|(\Psi_{\rightRun} \cap \Psi_{\leftmost} \cap \Psi_{\rightLen}(\mu_{\SUM}(t), \mu_{\SUM}(t+1))) \setminus \Psi_{\run}|] \\
    &= O(\delta \log \frac{n \log \sigma}{\delta \log n}).  
    \end{split}
\end{equation*}    

\paragraph{Proof of statement (iii).}
Proposition~\ref{prop:group_super_properties}~\ref{enum:group_super_properties:1} shows that 
there exists a constant $c \geq 1$ satisfying 
$\mathbb{E}[|(\Psi_{\rightRun} \cap \Psi_{\rightLen}(\mu_{\SUM}(t), \mu_{\SUM}(t+1)) \cap \Psi_{\group}(t, i)) \setminus \Psi_{\run}|] \leq c$ 
for any pair of integers $t \geq 3$ and $i \in [1, n]$. 
For simplicity, let $t = t^{\prime} + \tau + 2$ 
for an integer $t^{\prime} \geq 1$. 
We prove $\mathbb{E}[|( \Psi_{\rightRun} \cap \Psi_{\leftmost} \cap \Psi_{\rightLen}(\mu_{\SUM}(t), \mu_{\SUM}(t+1))) \setminus \Psi_{\run}|] \leq 16c\delta(7/8)^{t^{\prime}/2} + 2c$. 
Proposition~\ref{prop:group_super_properties}~\ref{enum:group_super_properties:2} shows that 
$(\Psi_{\rightRun} \cap \Psi_{\rightLen}(\mu_{\SUM}(t), \mu_{\SUM}(t+1))) \setminus \Psi_{\run} \subseteq \bigcup_{j \in [1, \lceil (n/d) \rceil + 1]} \Psi_{\group}(t, 1 + (j-1)d)$ 
for $d = \max \{ 1, \lfloor (1/8) \mu(t^{\prime} + \tau + 2) \rfloor \}$. 
We already proved $\lceil (n/d) \rceil + 1 \leq 16\delta(7/8)^{t^{\prime}/2} + 2$. 
Therefore, 
\begin{equation*}
\begin{split}
& \mathbb{E}[|(\Psi_{\rightRun} \cap \Psi_{\rightLen}(\mu_{\SUM}(t), \mu_{\SUM}(t+1)) \setminus \Psi_{\run}|] \\
&\leq \mathbb{E}[\sum_{j \in [1, \lceil (n/d) \rceil + 1]} |(\Psi_{\rightRun} \cap \Psi_{\rightLen}(\mu_{\SUM}(t), \mu_{\SUM}(t+1)) \cap \Psi_{\group}(t, 1 + (j-1)d)) \setminus \Psi_{\run}|] \\
&\leq \sum_{j \in [1, \lceil (n/d) \rceil + 1]} \mathbb{E}[|(\Psi_{\rightRun} \cap \Psi_{\rightLen}(\mu_{\SUM}(t), \mu_{\SUM}(t+1)) \cap \Psi_{\group}(t, 1 + (j-1)d)) \setminus \Psi_{\run}|] \\
&\leq c |[1, \lceil (n/d) \rceil + 1]| \\
&= c (\lceil (n/d) \rceil + 1) \\
&\leq 16c\delta(7/8)^{t^{\prime}/2} + 2c. 
\end{split}
\end{equation*} 

Statement (iii) follows from the following equation. 
\begin{equation*}
\begin{split}
& \mathbb{E}[|(\bigcup_{t = \tau+3}^{\infty} (\Psi_{\rightRun} \cap \Psi_{\leftmost} \cap \Psi_{\rightLen}(\mu_{\SUM}(t), \mu_{\SUM}(t+1))) \setminus \Psi_{\run}|] \\
&\leq \mathbb{E}[|(\bigcup_{t = \tau+3}^{\infty} (\Psi_{\rightRun} \cap \Psi_{\rightLen}(\mu_{\SUM}(t), \mu_{\SUM}(t+1))) \setminus \Psi_{\run}|] \\
&= \sum_{t = \tau+3}^{\infty} \mathbb{E}[|(\Psi_{\rightRun} \cap \Psi_{\rightLen}(\mu_{\SUM}(t), \mu_{\SUM}(t+1))) \setminus \Psi_{\run}|]  \\
&\leq \sum_{t^{\prime} = 1}^{\infty} (16c\delta(7/8)^{b/2} + 2c) \\
&\leq 18c\delta \sum_{i = 0}^{\infty} (7/8)^{i/2} \\
&= O(\delta).
\end{split}
\end{equation*}    
\end{proof}

Finally, Theorem~\ref{theo:RR_Psi_set_size} follows from Proposition~\ref{prop:expected_size_A} and Proposition~\ref{prop:expected_size_B}.

\subsection{Reconstructing Periodic Interval Attractors}\label{subsec:function_recover}
Periodicity within interval attractors enables reducing the number of interval attractors that need to be preserved. 
This section shows that any periodic interval attractor can be obtained from a non-periodic interval attractor using the periodicity within interval attractors. 

To leverage the periodicity within interval attractors, 
we use the following properties of the interval attractors obtained from the RLSLP $\mathcal{G}^{R}$, which is built by restricted recompression.
\begin{lemma}\label{lem:IA_reconstruct_condition}
Consider an interval attractor, $I(u) = ([p, q], [\ell, r]) \in \Psi_{\RR}$, for the node $u$ corresponding to a position $s \in [1,|S^h|]$ in $S^h$ at $h \in [0, H]$. 
Let $C$ and $\gamma$ be the associated string and attractor position of $I(u)$, respectively. 
Let $u_{s-1}$ and $u_{s+1} \in \mathcal{U}$ be the nodes corresponding to positions $(s-1)$ and $(s+1)$ in $S^h$, respectively. 
Then, the following three statements hold: 
\begin{enumerate}[label=\textbf{(\roman*)}]
    \item \label{lem:IA_reconstruct_condition:1} If a periodic interval attractor $([p^{\prime}, q^{\prime}], [\ell^{\prime}, r^{\prime}])$ is associated with $u_{s-1}$, then $I(u) = ([p^{\prime} + |C|, q^{\prime} + |C|], [\ell^{\prime} + |C|, r^{\prime}])$. 
    \item If interval attractor $I(u)$ is periodic 
    and a non-periodic interval attractor $([p^{\prime}, q^{\prime}], [\ell^{\prime}, r^{\prime}])$ is associated with $u_{s-1}$, then 
    $I(u) = ([q^{\prime} + 1, q^{\prime} + |C|], [\ell^{\prime} + \tau, r^{\prime}])$ 
    for some integer $\tau \geq 0$. 
    \item 
    If a periodic interval attractor is associated with $u_{s+1}$, 
    then interval attractor $I(u)$ is defined (i.e., $\Delta(u) \neq \emptyset$). 
\end{enumerate}
\end{lemma}
\begin{proof}
    The first statement is follows from Lemma~\ref{lem:psi_run_basic_property}~\ref{enum:psi_run_basic_property:2}. 
    The second and third statements follow from Lemma~\ref{lem:psi_run_basic_property}~\ref{enum:psi_run_basic_property:3}. 
\end{proof}

The following corollary follows from Lemma~\ref{lem:psi_run_basic_property}~\ref{enum:psi_run_basic_property:3}. 
\begin{corollary}\label{cor:IA_reconstruct_corollary}
    Consider the two interval attractors $I(u)$ and $([p^{\prime}, q^{\prime}], [\ell^{\prime}, r^{\prime}])$ of Lemma~\ref{lem:IA_reconstruct_condition}(ii). 
    Then, their associated strings are the same, 
    and the interval attractor $([p^{\prime}, q^{\prime}], [\ell^{\prime}, r^{\prime}])$ is contained in 
    set $\Psi_{\source}$. 
\end{corollary}

The following lemma states the relationship among interval attractors associated with nodes corresponding to consecutive nonterminals of sequence $S^{h}$ at $h \in [0, H]$. 
\begin{lemma}\label{lem:source_and_recover}
Consider an interval attractor, $I(u_{s}) = ([p, q], [\ell, r]) \in \Psi_{\source}$, for the node $u_{s}$ corresponding to a position $s \in [1,|S^h|]$ in $S^h$ at $h \in [0, H]$. 
Let $C$ and $\gamma$ be the associated string and attractor position of $I(u_{s})$, respectively. 
Let $u_{s-1}$ and $u_{s+1} \in \mathcal{U}$ be the nodes corresponding to positions $(s-1)$ and $(s+1)$ in $S^h$, respectively. 
Let $K = |\lcp(T[\gamma..r], C^{n+1})|$ 
and $k = \lfloor \frac{K - (2 + \sum_{w = 1}^{h+3} \lfloor \mu(w) \rfloor)}{|C|} \rfloor$.
From the definition of set $\Psi_{\source}$, 
there exists an interval attractor $([p^{\prime}, q^{\prime}], [\ell^{\prime}, r^{\prime}]) \in \Psi_{h} \cap \Psi_{\run} \cap \Psi_{\centerset}(C)$ whose attractor position $\gamma^{\prime}$ is equal to $\gamma + |C|$ (i.e., $\gamma^{\prime} = \gamma + |C|$). 
The following two statements hold: 
\begin{enumerate}[label=\textbf{(\roman*)}]
    \item \label{enum:source_and_recover:1} $1 \leq k \leq n$ and $k |C| + (2 + \sum_{w = 1}^{h+3} \lfloor \mu(w) \rfloor) \leq K <  (k+1) |C| + (2 + \sum_{w = 1}^{h+3} \lfloor \mu(w) \rfloor)$;
    \item \label{enum:source_and_recover:2} for each integer $j \in [1, k]$, 
    interval attractor $([p^{\prime} + (j - 1) |C|, q^{\prime} + (j - 1) |C|], [\ell^{\prime} + (j - 1) |C|, r])$ is 
    periodic and associated with the node $u_{s+j}$ corresponding to the $(s+j)$-th nonterminal of sequence $S^{h}$. 
    In addition, the associated string and attractor position of this interval attractor are $C$ and $\gamma + j|C|$, respectively. 
    \item \label{enum:source_and_recover:3} 
    interval attractor $([p^{\prime} + k|C|, q^{\prime} + k|C|], [\ell^{\prime} + k|C|, r])$ is 
    non-periodic and associated with the node $u_{s+k+1}$ corresponding to the $(s+k+1)$-th nonterminal of sequence $S^{h}$. 
    In addition, the associated string and attractor position of this interval attractor are $C$ and $\gamma + (k+1)|C|$, respectively. 

\end{enumerate}
\end{lemma}
\begin{proof}
    The proof of Lemma~\ref{lem:source_and_recover} is as follows.
    \paragraph{Proof of Lemma~\ref{lem:source_and_recover}(i).}
    Lemma~\ref{lem:light_source_basic_property} shows that 
    $|C| + 2 + \sum_{w = 1}^{h+3} \lfloor \mu(w) \rfloor \leq K \leq n$ holds. 
    Since $|C| \geq 1$, 
    we obtain $1 \leq \lfloor \frac{K - (2 + \sum_{w = 1}^{h+3} \lfloor \mu(w) \rfloor)}{|C|} \rfloor \leq n$. 
    Therefore, $1 \leq k \leq n$ holds. 

    $k |C| + (2 + \sum_{w = 1}^{h+3} \lfloor \mu(w) \rfloor) \leq K <  (k+1) |C| + (2 + \sum_{w = 1}^{h+3} \lfloor \mu(w) \rfloor)$ follows from 
    $k \geq 1$ and the definition of $k$. 

    \paragraph{Proof of Lemma~\ref{lem:source_and_recover}(ii).}
We prove Lemma~\ref{lem:source_and_recover}(ii) by induction on $j$. 
For the base case $j = 1$, 
interval attractor $([p^{\prime}, q^{\prime}], [\ell^{\prime}, r^{\prime}])$ is periodic and associated with the node $u_{s+1}$ 
because the level, associated string and attractor position of this interval attractor are $h$, $C$ and $\gamma + |C|$, respectively.

For the inductive step, 
consider $j \in [2, k]$. 
Then, 
interval attractor $([p^{\prime} + (j - 2) |C|, q^{\prime} + (j - 2) |C|], [\ell^{\prime} + (j - 2) |C|, r])$ is 
periodic and associated with node $u_{s+j-1}$. 
In addition, the associated string and attractor position of this interval attractor are $C$ and $\gamma + (j-1)|C|$, respectively. 
In this case, Lemma~\ref{lem:IA_reconstruct_condition}~\ref{lem:IA_reconstruct_condition:1} shows that 
interval attractor $([p^{\prime} + (j - 1) |C|, q^{\prime} + (j - 1) |C|], [\ell^{\prime} + (j - 1) |C|, r])$ is associated with node $u_{s+j}$. 
Lemma~\ref{lem:psi_run_basic_property}~\ref{enum:psi_run_basic_property:2} shows that 
the associated string and attractor position $\gamma^{\prime}$ of this interval attractor are $C$ and $\gamma + j|C|$, respectively. 

Let $K^{\prime} = |\lcp(T[\gamma^{\prime}..r], C^{n+1})|$. 
The interval attractor $([p^{\prime} + (j - 1) |C|, q^{\prime} + (j - 1) |C|], [\ell^{\prime} + (j - 1) |C|, r])$ is periodic 
if $K^{\prime} > 1 + \sum_{w = 1}^{h+3} \lfloor \mu(w) \rfloor$. 
$K^{\prime} = K - j |C|$ and $K^{\prime} > 1 + \sum_{w = 1}^{h+3} \lfloor \mu(w) \rfloor$ follow from $j \leq k$ and $K \geq k |C| + (2 + \sum_{w = 1}^{h+3} \lfloor \mu(w) \rfloor)$. 
This is because $T[\gamma..r]$ contains a long repetition of $C$ as a prefix.
Therefore, Lemma~\ref{lem:source_and_recover}(ii) holds for $j$. 
By induction on $j$, 
we obtain Lemma~\ref{lem:source_and_recover}(ii). 

    \paragraph{Proof of Lemma~\ref{lem:source_and_recover}(iii).}
    Lemma~\ref{lem:source_and_recover}(ii) shows that 
    interval attractor $([p^{\prime} + (k - 1) |C|, q^{\prime} + (k - 1) |C|], [\ell^{\prime} + (k - 1) |C|, r])$ is 
    periodic and associated with node $u_{s+k}$. 
    In addition, the associated string and attractor position of this interval attractor are $C$ and $\gamma + k|C|$, respectively. 
    
    In this case, Lemma~\ref{lem:IA_reconstruct_condition}~\ref{lem:IA_reconstruct_condition:1} shows that 
    interval attractor $([p^{\prime} + k |C|, q^{\prime} + k |C|], [\ell^{\prime} + k |C|, r])$ is associated with node $u_{s+k+1}$. 
    Lemma~\ref{lem:psi_run_basic_property}~\ref{enum:psi_run_basic_property:2} shows that 
    the associated string and attractor position $\gamma^{\prime}$ of this interval attractor are $C$ and $\gamma + (k+1)|C|$, respectively. 

    Let $K^{\prime} = |\lcp(T[\gamma^{\prime}..r], C^{n+1})|$. 
    The interval attractor $([p^{\prime} + k |C|, q^{\prime} + k |C|], [\ell^{\prime} + k |C|, r])$ is non-periodic 
    if $K^{\prime} \leq 1 + \sum_{w = 1}^{h+3} \lfloor \mu(w) \rfloor$. 
    $K^{\prime} = K - (k+1) |C|$ and $K^{\prime} \leq 1 + \sum_{w = 1}^{h+3} \lfloor \mu(w) \rfloor$ follow from $k |C| + (2 + \sum_{w = 1}^{h+3} \lfloor \mu(w) \rfloor) \leq K <  (k+1) |C| + (2 + \sum_{w = 1}^{h+3} \lfloor \mu(w) \rfloor)$. 
    Therefore, we obtain Lemma~\ref{lem:source_and_recover}(iii). 
\end{proof}

For an interval attractor $([p, q], [\ell, r])$ in the set $\Psi_{\source}$,
function $f_{\recover}: \Psi_{\source} \rightarrow 2^{\Psi_{\run}}$ returns a set of the $k$ periodic interval attractors stated in Lemma~\ref{lem:source_and_recover}~\ref{enum:source_and_recover:2}. 

\subsubsection{Properties of Function \texorpdfstring{$f_{\recover}$}{frec}}
The following lemma states basic properties of the function $f_{\recover}$. 
\begin{lemma}\label{lem:recover_basic_property}
Consider an interval attractor $([p, q], [\ell, r]) \in \Psi_{\source}$. 
Let $h, \gamma$, and $C$ be the level, attractor position, and associated string of the interval attractor $([p, q], [\ell, r])$, respectively;
let $\gamma_{s}$ be the attractor position of the interval attractor $([p_{s}, q_{s}]$, $[\ell_{s}, r_{s}])$ for each integer $s \in [1, k]$;
let $K = |\lcp(T[\gamma..r], C^{n+1})|$ and $M = (K - (2 + \sum_{w = 1}^{h+3} \lfloor \mu(w) \rfloor) ) \mod |C|$; 
let $([p_{1}, q_{1}]$, $[\ell_{1}, r_{1}])$, $([p_{2}, q_{2}]$, $[\ell_{2}, r_{2}])$, $\ldots$, $([p_{k}, q_{k}], [\ell_{k}, r_{k}])$ ($p_{1} < p_{2} < \ldots < p_{k}$) be the interval attractors obtained from the function $f_{\recover}(([p, q], [\ell, r]))$. 
Here, $([p, q], [\ell, r]) \in \Psi_{\lcp}(K)$ follows from the definition of the subset $\Psi_{\lcp}(K)$; 
$([p, q], [\ell, r]) \in \Psi_{\modulo}(M)$ follows from Lemma~\ref{lem:light_source_basic_property}. 
From the definition of set $\Psi_{\source}$, 
there exists an interval attractor $([p^{\prime}, q^{\prime}], [\ell^{\prime}, r^{\prime}]) \in \Psi_{h} \cap \Psi_{\run} \cap \Psi_{\centerset}(C)$ whose attractor position $\gamma^{\prime}$ is equal to $\gamma + |C|$ (i.e., $\gamma^{\prime} = \gamma + |C|$). 
The following six statements hold:
\begin{enumerate}[label=\textbf{(\roman*)}]
    \item \label{enum:recover_basic_property:1}
    $|C| + 2 + \sum_{w = 1}^{h+3} \lfloor \mu(w) \rfloor \leq K \leq n$, 
    $k = \lfloor \frac{K - (2 + \sum_{w = 1}^{h+3} \lfloor \mu(w) \rfloor)}{|C|} \rfloor$, and $1 \leq k \leq n$;
    \item \label{enum:recover_basic_property:2} 
    $([p_{s}, q_{s}], [\ell_{s}, r_{s}]) = ([q + 1 + (s - 1) |C|, q + s|C|], [\ell^{\prime} + (s - 1) |C|, r])$ 
    and $\gamma_{s} = \gamma + s |C|$ for each integer $s \in [1, k]$;
    \item \label{enum:recover_basic_property:3} 
    $([p_{1}, q_{1}], [\ell_{1}, r_{1}]) = ([p^{\prime}, q^{\prime}], [\ell^{\prime}, r^{\prime}])$, 
    $p < p^{\prime}$, $q = p^{\prime}-1$, $\ell \leq \ell^{\prime}$, and $r = r^{\prime}$;
    \item \label{enum:recover_basic_property:4} 
    $([p_{s}, q_{s}]$, $[\ell_{s}, r_{s}]) \in \Psi_{\run} \cap \Psi_{h} \cap \Psi_{\centerset}(C) \cap \Psi_{\modulo}(M) \cap \Psi_{\lcp}(K - s |C|)$ for each integer $s \in [1, k]$; 
    \item \label{enum:recover_basic_property:5}
    for each integer $s \in [1, k]$, 
    $([p, q], [\ell, r]) \in \Psi_{\preceding} \Leftrightarrow ([p_{s}, q_{s}]$, $[\ell_{s}, r_{s}]) \in \Psi_{\preceding}$;
    \item \label{enum:recover_basic_property:6}
    if $([p, q], [\ell, r]) \in \Psi_{\preceding}$, 
    then $T[\gamma_{k}..r_{k}+1] \prec T[\gamma_{k-1}..r_{k-1}+1] \prec \cdots \prec T[\gamma_{1}..r_{1}+1]$ holds. 
    Otherwise, 
    $T[\gamma_{1}..r_{1}+1] \prec T[\gamma_{2}..r_{2}+1] \prec \cdots \prec T[\gamma_{k}..r_{k}+1]$ holds.
\end{enumerate}
\end{lemma}
\begin{proof}
    For each integer $s \in [1, k]$, 
    $([p_{s}, q_{s}], [\ell_{s}, r_{s}]) = ([p^{\prime} + (s-1)|C|, q^{\prime} + (s-1)|C|], [\ell^{\prime} + (s-1)|C|, r])$ 
    follows from the definition of the function $f_{\recover}$.     
    The proof of Lemma~\ref{lem:recover_basic_property} is as follows. 

    \textbf{Proof of Lemma~\ref{lem:recover_basic_property}(i).}
    This statement follows from Lemma~\ref{lem:source_and_recover} and Lemma~\ref{lem:light_source_basic_property}.

    \textbf{Proof of Lemma~\ref{lem:recover_basic_property}(ii).}
    This statement follows from 
    $([p_{s}, q_{s}], [\ell_{s}, r_{s}]) = ([p^{\prime} + (s-1)|C|, q^{\prime} + (s-1)|C|], [\ell^{\prime} + (s-1)|C|, r])$ 
    and Corollary~\ref{cor:IA_reconstruct_corollary}. 

    \textbf{Proof of Lemma~\ref{lem:recover_basic_property}(iii).}
    $([p_{1}, q_{1}], [\ell_{1}, r_{1}]) = ([p^{\prime}, q^{\prime}], [\ell^{\prime}, r^{\prime}])$ 
    follows from the definition of $f_{\recover}$. 
    $p < p^{\prime}$, $q = p^{\prime}-1$, $\ell \leq \ell^{\prime}$, and $r = r^{\prime}$ follows from 
    Corollary~\ref{cor:IA_reconstruct_corollary} and $p \leq q$.

    \textbf{Proof of Lemma~\ref{lem:recover_basic_property}(iv).}
    We already showed that 
    $([p_{s}, q_{s}]$, $[\ell_{s}, r_{s}]) \in \Psi_{\run} \cap \Psi_{h} \cap \Psi_{\centerset}(C) \cap \Psi_{\lcp}(K - s |C|)$ 
    and $K - s |C| \geq 2 + \sum_{w = 1}^{h+3} \lfloor \mu(w) \rfloor$
    in the proof of Lemma~\ref{lem:source_and_recover}~\ref{enum:source_and_recover:2}.     
    $([p_{s}, q_{s}], [\ell_{s}, r_{s}]) \in \Psi_{\modulo}(M)$ follows from 
    $([p, q], [\ell, r]) \in \Psi_{\modulo}(M)$, $K - s |C| \geq 2 + \sum_{w = 1}^{h+3} \lfloor \mu(w) \rfloor$, 
    and $([p, q], [\ell, r]), ([p_{s}, q_{s}], [\ell_{s}, r_{s}]) \in \Psi_{h} \cap \Psi_{\centerset}(C)$. 
    
    \textbf{Proof of Lemma~\ref{lem:recover_basic_property}(v).}
    $T[\gamma..r+1] = C^{s} \cdot T[\gamma_{s}..r+1]$ holds 
    because $T[\gamma..r]$ contains a sufficiently long repetition of $C$ as a prefix.    
    We prove $([p, q], [\ell, r]) \in \Psi_{\preceding} \Rightarrow ([p_{s}, q_{s}]$, $[\ell_{s}, r_{s}]) \in \Psi_{\preceding}$. 
    $T[\gamma..r+1] \prec C^{n+1}$ follows from $([p, q], [\ell, r]) \in \Psi_{\preceding} \cap \Psi_{\centerset}(C)$. 
    $T[\gamma_{s}..r+1] \prec C^{n+1}$ follows from $T[\gamma..r+1] = C^{s} \cdot T[\gamma_{s}..r+1]$ and $T[\gamma..r+1] \prec C^{n+1}$. 
    Therefore, 
    we obtain $([p_{s}, q_{s}]$, $[\ell_{s}, r_{s}]) \in \Psi_{\preceding}$.  

    Similarly, we can prove $([p, q], [\ell, r]) \in \Psi_{\preceding} \Leftarrow ([p_{s}, q_{s}]$, $[\ell_{s}, r_{s}]) \in \Psi_{\preceding}$ 
    using the fact that $T[\gamma..r+1] = C^{s} \cdot T[\gamma_{s}..r+1]$. 
    Therefore, we obtain Lemma~\ref{lem:recover_basic_property}(v). 

    \textbf{Proof of Lemma~\ref{lem:recover_basic_property}(vi).}
    Consider an integer $s \in [1, k-1]$. 
    Then, Lemma~\ref{lem:recover_basic_property}~\ref{enum:recover_basic_property:4} shows that 
    $([p_{s}, q_{s}], [\ell_{s}, r_{s}]) \in \Psi_{\run} \cap \Psi_{\centerset}(C) \cap \Psi_{\lcp}(K - s |C|)$ 
    and 
    $([p_{s+1}, q_{s+1}], [\ell_{s+1}, r_{s+1}]) \in \Psi_{\run} \cap \Psi_{\centerset}(C) \cap \Psi_{\lcp}(K - (s+1) |C|)$ hold.     
    If $([p, q], [\ell, r]) \in \Psi_{\preceding}$, 
    then Lemma~\ref{lem:recover_basic_property}~\ref{enum:recover_basic_property:5} shows that 
    $([p_{s}, q_{s}], [\ell_{s}, r_{s}]), ([p_{s+1}, q_{s+1}], [\ell_{s+1}, r_{s+1}]) \in \Psi_{\preceding}$ holds. 
    In this case, Lemma~\ref{lem:psi_LMPS_property}~\ref{enum:psi_LMPS_property:preceding:1} shows that 
    $T[\gamma_{s+1}..r_{s+1}+1] \prec T[\gamma_{s}..r_{s}+1]$ holds. 

    Otherwise (i.e., $([p, q], [\ell, r]) \not \in \Psi_{\preceding}$), 
    Lemma~\ref{lem:recover_basic_property}~\ref{enum:recover_basic_property:5} shows that 
    $([p_{s}, q_{s}], [\ell_{s}, r_{s}]), ([p_{s+1}, q_{s+1}]$, $[\ell_{s+1}, r_{s+1}]) \not \in \Psi_{\preceding}$ holds. 
    In this case, 
    $([p_{s}, q_{s}], [\ell_{s}, r_{s}]), ([p_{s+1}, q_{s+1}], [\ell_{s+1}, r_{s+1}]) \in \Psi_{\succeeding}$ holds 
    because each interval attractor is contained in either $\Psi_{\preceding}$ or $\Psi_{\succeeding}$. 
    Lemma~\ref{lem:recover_basic_property}~\ref{enum:recover_basic_property:5} shows that 
    Lemma~\ref{lem:psi_LMPS_property}~\ref{enum:psi_LMPS_property:succeeding:1} shows that 
    $T[\gamma_{s}..r_{s}+1] \prec T[\gamma_{s+1}..r_{s+1}+1]$ holds. 
    Therefore, Lemma~\ref{lem:recover_basic_property}~\ref{enum:recover_basic_property:6} holds. 
\end{proof}

%let $h \in [0, H]$ and $K \geq 0$ be two integers satisfying $([p, q], [\ell, r]) \in \Psi_{h}$ 
%and $([p, q], [\ell, r]) \in \Psi_{\lcp}(K)$, respectively.
%Let $\gamma$ and $C$ be the attractor position and the associated string of the interval attractor $([p, q], [\ell, r])$, respectively. 
%From the definition of set $\Psi_{\source}$, 
%there exists an interval attractor $([p^{\prime}, q^{\prime}], [\ell^{\prime}, r^{\prime}]) \in \Psi_{\run}$ whose attractor position $\gamma^{\prime}$ is equal to $\gamma + |C|$ (i.e., $%\gamma^{\prime} = \gamma + |C|$). 
%Lemma~\ref{lem:source_basic_property}~\ref{enum:source_basic_property:4} ensures that 
%subset $\Psi_{\run}$ contains the $k$ interval attractors $([p^{\prime}, q^{\prime}], [\ell^{\prime}, r])$, 
%$([p^{\prime} + |C|, q^{\prime} + |C|], [\ell^{\prime} + |C|, r])$, $\ldots$, $([p^{\prime} + (k-1)|C|, q^{\prime} + (k-1)|C|], [\ell^{\prime} + (k-1)|C|, r])$, where $k = \lfloor \frac{K - (2 + \sum_{w = 1}^{h+3} \lfloor \mu(w) \rfloor)}{|C|} \rfloor$.
%Finally, $f_{\recover}$ is defined as the function that, given a interval attractor $([p, q], [\ell, r]) \in \Psi_{\source}$, returns the $k$ interval attractors $([p^{\prime}, q^{\prime}], [\ell^{\prime}, r])$, 
%$([p^{\prime} + |C|, q^{\prime} + |C|], [\ell^{\prime} + |C|, r])$, $\ldots$, $([p^{\prime} + (k-1)|C|, q^{\prime} + (k-1)|C|], [\ell^{\prime} + (k-1)|C|, r])$ (i.e., $f_{\recover}([p, q], [\ell, r]) = \{ ([p^{\prime} + (s-1)|C|, q^{\prime} + (s-1)|C|], [\ell^{\prime} + (s-1)|C|, r]) \mid s \in [1, k] \}$).

The following lemma ensures that 
the function $f_{\recover}$ makes a one-to-many correspondence between the two subsets $\Psi_{\source}$ and $\Psi_{\run}$. 

\begin{lemma}\label{lem:recover_division_property}
    Consider the subset $\Psi_{\source}$ consisting of $m$ interval attractors $([p_{1}, q_{1}]$, $[\ell_{1}, r_{1}])$, $([p_{2}, q_{2}]$, $[\ell_{2}, r_{2}])$, 
    $\ldots$, $([p_{m}, q_{m}], [\ell_{m}, r_{m}])$. 
    For each interval attractor $([p_{s}, q_{s}], [\ell_{s}, r_{s}]) \in \Psi_{\source}$, 
    let $\Psi_{\recover, s}$ be the set of interval attractors obtained from the function $f_{\recover}(([p_{s}, q_{s}], [\ell_{s}, r_{s}]))$.
    Then, subset $\Psi_{\run}$ can be divided into $m$ disjoint sets $\Psi_{\recover, 1}, \Psi_{\recover, 2}, \ldots, \Psi_{\recover, m}$, 
    i.e., the following two statements hold: 
    \begin{enumerate}[label=\textbf{(\roman*)}]
    \item \label{enum:recover_division_property:1} $\Psi_{\run} = \bigcup_{s = 1}^{m} \Psi_{\recover, s}$; 
    \item \label{enum:recover_division_property:2} $\Psi_{\recover, s} \cap \Psi_{\recover, s^{\prime}} = \emptyset$ for any pair of integers $1 \leq s < s^{\prime} \leq m$.
    \end{enumerate}
\end{lemma}
\begin{proof}
    The proof of Lemma~\ref{lem:recover_division_property} is as follows. 

    \textbf{Proof of Lemma~\ref{lem:recover_division_property}(i).}    
    We prove $\Psi_{\run} \supseteq \bigcup_{s = 1}^{m} \Psi_{\recover, s}$. 
    For each integer $s \in [1, k]$, 
    $\Psi_{\recover, 1} \subseteq \Psi_{\run}$ follows from the definition of $f_{\recover}$. 
    Therefore, $\Psi_{\run} \supseteq \bigcup_{s = 1}^{m} \Psi_{\recover, s}$ holds. 

    Next, we prove $\Psi_{\run} \subseteq \bigcup_{s = 1}^{m} \Psi_{\recover, s}$. 
    Consider a periodic interval attractor, $I(u_{x}) \in \Psi_{\run}$, for the node $u_{x}$ corresponding to a position $s \in [1,|S^h|]$ in $S^h$ at $h \in [0, H]$. 
    From Lemma~\ref{lem:IA_reconstruct_condition} and Corollary~\ref{cor:IA_reconstruct_corollary}, 
    there exists a position $x^{\prime} \in [1,|S^h|]$ in sequence $S^{h}$ such that $x^{\prime} < x$ satisfying the following two conditions: 
    (A) an interval attractor $I(u_{x^{\prime}})$ of set $\Psi_{\source}$ 
    is associated with the node $u_{x^{\prime}}$ corresponding to the position $x^{\prime}$; 
    (B) for each $j \in [x^{\prime}, x]$,  
    a periodic interval attractor is associated with the node corresponding to position $j$ of sequence $S^{h}$. 
    In this case, 
    Lemma~\ref{lem:source_and_recover} shows that $f_{\recover}(I(u_{x^{\prime}}))$ returns interval attractor $I(u_{x})$. 
    Therefore, we obtain $\Psi_{\run} \subseteq \bigcup_{s = 1}^{m} \Psi_{\recover, s}$. 
        
    \textbf{Proof of Lemma~\ref{lem:recover_division_property}(ii).}
    Consider the periodic interval attractor $I(u_{x})$ and non-periodic interval attractor $I(u_{x^{\prime}})$ used in the proof of Lemma~\ref{lem:recover_division_property}(i). 
    For each interval attractor $([p_{s}, q_{s}], [\ell_{s}, r_{s}]) \in \Psi_{\source} \setminus \{ I(u_{x^{\prime}}) \}$, 
    Lemma~\ref{lem:source_and_recover} shows that $I(u_{x}) \not \in f_{\recover}(([p_{s}, q_{s}], [\ell_{s}, r_{s}]))$ holds. 
    This fact indicates that Lemma~\ref{lem:recover_division_property}(ii) holds. 
\end{proof}

Consider two interval attractors $([p, q], [\ell, r]), ([p^{\prime}, q^{\prime}], [\ell^{\prime}, r^{\prime}]) \in \Psi_{\source}$ 
satisfying $T[p-1..r+1] = T[p^{\prime}-1..r^{\prime}+1]$. 
The following lemma states the relationship between two functions $f_{\recover}(([p, q]$, $[\ell, r]))$ and 
$f_{\recover}(([p, q], [\ell, r]))$. 

\begin{lemma}\label{lem:recover_super_property}
    Consider two interval attractors $([p, q], [\ell, r]), ([p^{\prime}, q^{\prime}], [\ell^{\prime}, r^{\prime}]) \in \Psi_{\source}$ 
    satisfying $T[p-1..r+1] = T[p^{\prime}-1..r^{\prime}+1]$. 
    Let $([p_{1}, q_{1}], [\ell_{1}, r_{1}])$, $([p_{2}, q_{2}], [\ell_{2}, r_{2}])$, $\ldots$, 
    $([p_{k}, q_{k}], [\ell_{k}, r_{k}])$ ($p_{1} < p_{2} < \cdots < p_{k}$) be the interval attractors obtained from 
    the function $f_{\recover}(([p, q], [\ell, r]))$. 
    Similarly, 
    let $([p^{\prime}_{1}, q^{\prime}_{1}], [\ell^{\prime}_{1}, r^{\prime}_{1}])$, 
    $([p^{\prime}_{2}, q^{\prime}_{2}], [\ell^{\prime}_{2}, r^{\prime}_{2}])$, $\ldots$, 
    $([p^{\prime}_{k^{\prime}}, q^{\prime}_{k^{\prime}}], [\ell^{\prime}_{k^{\prime}}, r^{\prime}_{k^{\prime}}])$ 
    ($p^{\prime}_{1} < p^{\prime}_{2} < \cdots < p^{\prime}_{k^{\prime}}$) be the interval attractors obtained from 
    the function $f_{\recover}(([p^{\prime}, q^{\prime}], [\ell^{\prime}, r^{\prime}]))$. 
    Then, the following two statements hold: 
    \begin{enumerate}[label=\textbf{(\roman*)}]
    \item \label{enum:recover_super_property:1} 
    $k = k^{\prime}$ (i.e., $|f_{\recover}(([p, q], [\ell, r]))| = |f_{\recover}(([p^{\prime}, q^{\prime}], [\ell^{\prime}, r^{\prime}]))|$);
    \item \label{enum:recover_super_property:2} 
    $T[p_{s}-1..r_{s}+1] = T[p^{\prime}_{s}-1..r^{\prime}_{s}+1]$ for each integer $s \in [1, k]$.
    \end{enumerate}
\end{lemma}
\begin{proof}
Let $\gamma$, $C$, and $h$ be the attractor position, associated string, and level of the interval attractor $([p, q], [\ell, r])$, 
respectively. 
Similarly, 
Let $\gamma^{\prime}$, $C^{\prime}$, and $h^{\prime}$ be the attractor position, associated string, and level of the interval attractor $([p^{\prime}, q^{\prime}], [\ell^{\prime}, r^{\prime}])$, respectively. 
We can apply Lemma~\ref{lem:psi_equality_basic_property} to the two interval attractors 
$([p, q], [\ell, r])$ and $([p^{\prime}, q^{\prime}], [\ell^{\prime}, r^{\prime}])$ 
because $T[p-1..r+1] = T[p^{\prime}-1..r^{\prime}+1]$. 
Lemma~\ref{lem:psi_equality_basic_property}~\ref{enum:psi_equality_basic_property:center_set} shows that 
$C = C^{\prime}$ holds. 
Similarly, 
we can apply Lemma~\ref{lem:psi_str_property} to the two interval attractors 
$([p, q], [\ell, r])$ and $([p^{\prime}, q^{\prime}], [\ell^{\prime}, r^{\prime}])$.
Lemma~\ref{lem:psi_str_property} shows that 
$h = h^{\prime}$ and $|[p, q]| = |[p^{\prime}, q^{\prime}]|$ hold.

Let $K$ be the length of the longest common prefix between two strings $T[\gamma..r]$ and $C^{n+1}$ 
(i.e., $K = |\lcp(T[\gamma..r], C^{n+1})|$). 
Here, $([p, q], [\ell, r]) \in \Psi_{\lcp}(K)$ follows from the definition of the subset $\Psi_{\lcp}(K)$, 
and $K \geq \sum_{w = 1}^{h+1} \lfloor \mu(w) \rfloor$ follows from Lemma~\ref{lem:recover_basic_property}~\ref{enum:recover_basic_property:1}. 
Lemma~\ref{lem:psi_equality_basic_property}~\ref{enum:psi_equality_basic_property:9} shows that 
$([p^{\prime}, q^{\prime}], [\ell^{\prime}, r^{\prime}]) \in \Psi_{\lcp}(K)$ holds.

The proof of Lemma~\ref{lem:recover_super_property} is as follows. 

\textbf{Proof of Lemma~\ref{lem:recover_super_property}(i).}
We apply Lemma~\ref{lem:recover_basic_property}~\ref{enum:recover_basic_property:1} to 
the two interval attractors $([p, q], [\ell, r])$ and $([p^{\prime}, q^{\prime}], [\ell^{\prime}, r^{\prime}])$. 
Then, 
$k = \lfloor \frac{K - (2 + \sum_{w = 1}^{h+3} \lfloor \mu(w) \rfloor)}{|C|} \rfloor$ 
and $k^{\prime} = \lfloor \frac{K - (2 + \sum_{w = 1}^{h+3} \lfloor \mu(w) \rfloor)}{|C|} \rfloor$ hold 
because 
$([p, q], [\ell, r]), ([p^{\prime}, q^{\prime}], [\ell^{\prime}, r^{\prime}]) \in \Psi_{h} \cap \Psi_{\centerset}(C) \cap \Psi_{\lcp}(K)$. 
Therefore, $k = k^{\prime}$ holds. 

\textbf{Proof of Lemma~\ref{lem:recover_super_property}(ii).}
We apply Lemma~\ref{lem:recover_basic_property}~\ref{enum:recover_basic_property:2} to 
the two interval attractors $([p, q], [\ell, r])$ and $([p^{\prime}, q^{\prime}], [\ell^{\prime}, r^{\prime}])$. 
Then, $[p_{s}-1, r_{s}+1] = [q + (s-1) |C|, r + 1]$ and $[p^{\prime}_{s}-1, r^{\prime}_{s}+1] = [q^{\prime} + (s-1) |C|, r^{\prime}+1]$ hold. 
$T[q + (s-1) |C|..r + 1] = T[q^{\prime} + (s-1) |C|..r^{\prime}+1]$ follows from 
$T[p-1..r+1] = T[p^{\prime}-1..r^{\prime}+1]$ and $|[p, q]| = |[p^{\prime}, q^{\prime}]|$. 
Therefore, $T[p_{s}-1..r_{s}+1] = T[p^{\prime}_{s}-1..r^{\prime}_{s}+1]$ follows from 
$[p_{s}-1, r_{s}+1] = [q + (s-1) |C|, r + 1]$, $[p^{\prime}_{s}-1, r^{\prime}_{s}+1] = [q^{\prime} + (s-1) |C|, r^{\prime}+1]$, 
and $T[q + (s-1) |C|..r + 1] = T[q^{\prime} + (s-1) |C|..r^{\prime}+1]$. 
\end{proof}

A periodic interval attractor can be reconstructed in $O(1)$ time by utilizing a non-periodic interval attractor 
and function $f_{\recover}$, as shown in the following theorem. 

\begin{theorem}\label{theo:simple_periodicity}
    Consider interval attractor $I(u_s) = ([p, q], [\ell, r])$ for the node $u_s$ at each position $s \in [1, |S^h|]$ at height $h \in [0, H]$. 
    Let $C$ be the substring derived from the nonterminal $S^h[u_s]$, and 
    let $\gamma$ be the starting position of the substring $C$ in $T$.  
    Define $K$ as the length of the longest common prefix between two strings $T[\gamma..r]$ and $C^{n+1}$.
    Set $d = \lfloor \frac{K - (2 + \sum_{w = 1}^{h+3} \lfloor \mu(w) \rfloor)}{|C|} \rfloor$.
    If interval attractor $I(u_s)$ is non-periodic and interval attractor $I(u_{s+1})$ is periodic, then the following three statements hold: 
    \begin{enumerate}[label=\textbf{(\roman*)}]
        \item $d \geq 1$;
        \item $I(u_{s+j})$ for any $j \in [1, d]$ is periodic, 
        and $I(u_{s + d + 1})$ is non-periodic;
        \item for all $j \in [1, d]$, there exists a unique integer $\tau \in \mathbb{Z}$ such that $([q+1 + (j - 1)|C|, q + d|C|], [\ell + \tau + (j - 1)|C|, r])$ is the interval attractor associated with node $u_{s+j} \in \mathcal{U}$ (i.e., $I(u_{s+j}) = ([q+1 + (j - 1)|C|, q + d|C|], [\ell + \tau + (j - 1)|C|, r])$).    
    \end{enumerate}
    Here, $I(u_{s+1}), I(u_{s+2}), \ldots, I(u_{s + d})$ are called interval attractors reconstructed from $I(u_s)$. 
\end{theorem}
\begin{proof}
    Corollary~\ref{cor:IA_reconstruct_corollary} shows that 
    set $\Psi_{\source}$ contains interval attractor $I(u_s)$. 
    Lemma~\ref{lem:IA_reconstruct_condition} shows that 
    (A) $f_{\recover}(I(u_s))$ returns the $d$ interval attractors $I(u_{s+1}), I(u_{s+2}), \ldots, I(u_{s+d})$, 
    (B) these interval attractors are periodic, and 
    (C) $I(u_{s+d+1})$ is non-periodic. 
    Lemma~\ref{lem:recover_basic_property}~\ref{enum:recover_basic_property:2} and Lemma~\ref{lem:recover_basic_property}~\ref{enum:recover_basic_property:3} show that 
    there exists an integer $\tau \geq 0$ satisfying 
    $I(u_{s+j}) = ([q+1 + (j - 1)|C|, q + j|C|], [\ell + \tau + (j - 1)|C|, r])$ for all $j \in [1, k]$. 
    Therefore, we obtain Theorem~\ref{theo:simple_periodicity}.      
\end{proof}

\subsection{Three Subsets of Interval Attractors Based on Function \texorpdfstring{$f_{\recover}$}{}}\label{subsubsec:recover_subsets}
This section introduces three subsets of set $\Psi_{\RR}$ using function $f_{\recover}$. 
These subsets are used to explain the relationship among interval attractors obtained from 
function $f_{\recover}$.

%We introduce two subsets $\Psi_{\sRecover}(([p, q], [\ell, r]))$ and 
%$\Psi_{\nRecover}(\lambda)$ of set $\Psi_{\RR}$ using function $f_{\recover}$ 
%for an interval attractor $([p, q], [\ell, r]) \in \Psi_{\source}$ 
%and integer $\lambda \geq 0$. 
%The 

\paragraph{Subset $\Psi_{\sRecover}(([p, q], [\ell, r]))$.}
For an interval attractor $([p, q], [\ell, r])$ in subset $\Psi_{\source}$, 
subset $\Psi_{\sRecover}(([p, q], [\ell, r]))$ of set $\Psi_{\RR}$ 
consists of the interval attractors such that 
for each interval attractor $([p^{\prime}, q^{\prime}], [\ell^{\prime}, r^{\prime}]) \in \Psi_{\sRecover}(([p, q], [\ell, r]))$, 
the subset $\Psi_{\source}$ contains an interval attractor $([\hat{p}, \hat{q}], [\hat{\ell}, \hat{r}])$ satisfying the following two conditions: 
\begin{itemize}
    \item $([p^{\prime}, q^{\prime}], [\ell^{\prime}, r^{\prime}]) \in f_{\recover}(([\hat{p}, \hat{q}], [\hat{\ell}, \hat{r}]))$;
    \item $T[\hat{p}-1..\hat{r}+1] = T[p-1..r+1]$.
\end{itemize}
Formally, let $\Psi_{\sRecover}(([p, q], [\ell, r])) = \{ ([p^{\prime}, q^{\prime}], [\ell^{\prime}, r^{\prime}]) \in \Psi_{\RR} \mid \exists ([\hat{p}, \hat{q}], [\hat{\ell}, \hat{r}]) \in \Psi_{\source} \text{ s.t. }$ $([p^{\prime}, q^{\prime}]$, $[\ell^{\prime}, r^{\prime}]) \in f_{\recover}(([\hat{p}, \hat{q}], [\hat{\ell}, \hat{r}])) \text{ and } T[\hat{p}-1..\hat{r}+1] = T[p-1..r+1]\}$.  

%\paragraph{Properties of two subsets $\Psi_{\sRecover}(([p, q], [\ell, r]))$ and $\Psi_{\nRecover}(\lambda)$.}

The following two lemmas state properties of the subset $\Psi_{\sRecover}(([p, q], [\ell, r]))$. 

\begin{lemma}\label{lem:sRecover_basic_property}
Consider an interval attractor $([p, q], [\ell, r])$ in subset $\Psi_{\source}$. 
Let $\gamma$, $C$, and $h$ be the attractor position, associated string, and level of the interval attractor $([p, q], [\ell, r])$, 
respectively; 
let $([p_{1}, q_{1}], [\ell_{1}, r_{1}])$, $([p_{2}, q_{2}], [\ell_{2}, r_{2}])$, $\ldots$, $([p_{k}, q_{k}], [\ell_{k}, r_{k}])$ 
be the interval attractors in set $\Psi_{\sRecover}(([p, q], [\ell, r]))$; 
let $\gamma_{s}$ be the attractor position of each interval attractor $([p_{s}, q_{s}], [\ell_{s}, r_{s}])$.
Then, the following five statements hold:

\begin{enumerate}[label=\textbf{(\roman*)}]
    \item \label{enum:sRecover_basic_property:str} 
    $\Psi_{\sRecover}(([p, q], [\ell, r])) = \bigcup_{([p^{\prime}, q^{\prime}], [\ell^{\prime}, r^{\prime}]) \in \Psi_{\str}(T[p-1..r+1])} f_{\recover}(([p^{\prime}, q^{\prime}], [\ell^{\prime}, r^{\prime}]))$;     
    \item \label{enum:sRecover_basic_property:centerset} 
    let $K = |\lcp(T[\gamma..r], C^{n+1})|$, 
    then, the following equation holds:
    \begin{equation*}
    \Psi_{\sRecover}(([p, q], [\ell, r])) \subseteq \Psi_{h} \cap \Psi_{\run} \cap \Psi_{\centerset}(C) \cap (\bigcup_{\lambda = 2+\sum_{w = 1}^{h+3} \lfloor \mu(w) \rfloor}^{K-|C|} \Psi_{\lcp}(\lambda));
    \end{equation*}    
    %$\Psi_{\sRecover}(([p, q], [\ell, r])) \subseteq \Psi_{h} \cap \Psi_{\run} \cap \Psi_{\centerset}(C) \cap (\bigcup_{\lambda = 2+\sum_{w = 1}^{h+3} \lfloor \mu(w) \rfloor}^{K-|C|} \Psi_{\lcp}(\lambda))$;     
    \item \label{enum:sRecover_basic_property:overlap} $\Psi_{\sRecover}(([p, q], [\ell, r])) \cap \Psi_{\sRecover}(([p^{\prime}, q^{\prime}], [\ell^{\prime}, r^{\prime}])) = \emptyset$ for each interval attractor $([p^{\prime}, q^{\prime}], [\ell^{\prime}$, $r^{\prime}]) \in \Psi_{\source}$ satisfying $T[p-1..r+1] \neq T[p^{\prime}-1..r^{\prime}+1]$;
    \item \label{enum:sRecover_basic_property:equality} $\Psi_{\sRecover}(([p, q], [\ell, r])) = \Psi_{\sRecover}(([p^{\prime}, q^{\prime}], [\ell^{\prime}, r^{\prime}]))$ for each interval attractor $([p^{\prime}, q^{\prime}], [\ell^{\prime}, r^{\prime}]) \in \Psi_{\source}$ satisfying $T[p-1..r+1] = T[p^{\prime}-1..r^{\prime}+1]$;
    \item \label{enum:sRecover_basic_property:lex} 
    if there exists an integer $s \in [1, k]$ satisfying $|\lcp(T[\gamma_{s}..r_{s}+1], P)| \leq 1 + \sum_{w = 1}^{h+3} \mu(w)$ for a string $P \in \Sigma^{+}$, 
    then for each integer $s^{\prime} \in [1, k]$, 
    (A) $T[\gamma_{s}..r_{s}+1] \prec P \Rightarrow T[\gamma_{s^{\prime}}..r_{s^{\prime}}+1] \prec P$ 
    and (B) $P \prec T[\gamma_{s}..r_{s}+1] \Rightarrow P \prec T[\gamma_{s^{\prime}}..r_{s^{\prime}}+1]$.
\end{enumerate}
\end{lemma}
\begin{proof}
The proof of Lemma~\ref{lem:sRecover_basic_property} is as follows. 

\textbf{Proof of Lemma~\ref{lem:sRecover_basic_property}(i).}
Let $\Psi = \bigcup_{([p^{\prime}, q^{\prime}], [\ell^{\prime}, r^{\prime}]) \in \Psi_{\str}(T[p-1..r+1])}$ $f_{\recover}(([p^{\prime}, q^{\prime}], [\ell^{\prime}, r^{\prime}]))$ for simplicity. 
We prove $\Psi_{\sRecover}(([p, q], [\ell, r])) \subseteq \Psi$. 
Consider an integer $s \in [1, k]$. 
From the definition of the subset $\Psi_{\sRecover}(([p, q], [\ell, r]))$, 
subset $\Psi_{\source}$ contains an interval attractor $([p_{A}, q_{A}], [\ell_{A}, r_{A}])$ 
satisfying $([p_{s}, q_{s}], [\ell_{s}, r_{s}]) \in f_{\recover}(([p_{A}, q_{A}], [\ell_{A}, r_{A}]))$ 
and $T[p_{A}-1..r_{A}+1] = T[p-1..r+1]$. 
Because of $T[p_{A}-1..r_{A}+1] = T[p-1..r+1]$, 
$([p_{A}, q_{A}], [\ell_{A}, r_{A}]) \in \Psi_{\str}(T[p-1..r+1])$ follows from the definition of the subset $\Psi_{\str}(T[p-1..r+1])$. 
$([p_{s}, q_{s}], [\ell_{s}, r_{s}]) \in \Psi$ follows from 
$([p_{s}, q_{s}], [\ell_{s}, r_{s}]) \in f_{\recover}(([p_{A}, q_{A}], [\ell_{A}, r_{A}]))$ and $([p_{A}, q_{A}], [\ell_{A}, r_{A}]) \in \Psi_{\str}(T[p-1..r+1])$. 
Therefore, $\Psi_{\sRecover}(([p, q], [\ell, r])) \subseteq \Psi$ holds. 

Next, we prove $\Psi_{\sRecover}(([p, q], [\ell, r])) \supseteq \Psi$. 
Consider an interval attractor $([p_{B}, q_{B}], [\ell_{B}, r_{B}])$ in set $\Psi$. 
Then, set $\Psi_{\str}(T[p-1..r+1])$ contains an interval attractor $([p_{C}, q_{C}], [\ell_{C}, r_{C}])$ satisfying 
$([p_{B}, q_{B}], [\ell_{B}, r_{B}]) \in f_{\recover}(([p_{C}, q_{C}], [\ell_{C}, r_{C}]))$. 
$([p_{B}, q_{B}], [\ell_{B}, r_{B}]) \in \Psi_{\sRecover}(([p, q], [\ell, r]))$ follows from the definition of the subset $\Psi_{\sRecover}(([p, q], [\ell, r]))$ 
because $T[p_{C}-1..r_{C}+1] = T[p-1..r+1]$ follows from the definition of the subset $\Psi_{\str}(T[p-1..r+1])$, 
and Lemma~\ref{lem:psi_equality_basic_property}~\ref{enum:psi_equality_basic_property:5} shows that 
$([p_{C}, q_{C}], [\ell_{C}, r_{C}]) \in \Psi_{\source}$ holds. 
Therefore, $\Psi_{\sRecover}(([p, q], [\ell, r])) \supseteq \Psi$ holds. 

We showed that $\Psi_{\sRecover}(([p, q], [\ell, r])) \subseteq \Psi$ and $\Psi_{\sRecover}(([p, q], [\ell, r])) \supseteq \Psi$ hold. 
Therefore, $\Psi_{\sRecover}(([p, q], [\ell, r])) = \Psi$ holds.

\textbf{Proof of Lemma~\ref{lem:sRecover_basic_property}(ii).}
Consider an integer $s \in [1, k]$. 
From the definition of the subset $\Psi_{\sRecover}(([p, q], [\ell, r]))$, 
the subset $\Psi_{\source}$ contains an interval attractor $([p^{\prime}, q^{\prime}], [\ell^{\prime}, r^{\prime}])$ satisfying 
$([p_{s}, q_{s}], [\ell_{s}, r_{s}]) \in f_{\recover}(([p^{\prime}, q^{\prime}], [\ell^{\prime}, r^{\prime}]))$ and 
$([p^{\prime}, q^{\prime}], [\ell^{\prime}, r^{\prime}]) \in \Psi_{\str}(([p, q], [\ell, r]))$. 
$T[p-1..r+1] = T[p^{\prime}-1..r^{\prime}+1]$ follows from the definition of the subset $\Psi_{\str}(([p, q], [\ell, r]))$. 
$([p, q], [\ell, r]) \in \Psi_{\lcp}(K)$ follows from the definition of the subset $\Psi_{\lcp}(K)$. 
$K \geq \sum_{w = 1}^{h+1} \lfloor \mu(w) \rfloor$ follows from Lemma~\ref{lem:recover_basic_property}~\ref{enum:recover_basic_property:1}. 
Because of $T[p-1..r+1] = T[p^{\prime}-1..r^{\prime}+1]$, 
we can apply Lemma~\ref{lem:psi_str_property} and Lemma~\ref{lem:psi_equality_basic_property} to the two interval attractors $([p, q], [\ell, r])$ and $([p^{\prime}, q^{\prime}], [\ell^{\prime}, r^{\prime}])$. 
This lemma shows that $([p^{\prime}, q^{\prime}], [\ell^{\prime}, r^{\prime}]) \in \Psi_{\centerset}(C) \cap \Psi_{h} \cap \Psi_{\lcp}(K)$ holds. 

%We prove $([p^{\prime}, q^{\prime}], [\ell^{\prime}, r^{\prime}]) \in \Psi_{\centerset}(C) \cap \Psi_{h} \cap \Psi_{\lcp}(K)$. 
%Because of $T[p-1..r+1] = T[p^{\prime}-1..r^{\prime}+1]$, 
%we can apply Lemma~\ref{lem:psi_str_property} and Lemma~\ref{lem:psi_equality_basic_property} to the two interval attractors $([p, q], [\ell, r])$ and $([p^{\prime}, q^{\prime}], [\ell^{\prime}, r^{\prime}])$. 
%Lemma~\ref{lem:psi_equality_basic_property}~\ref{enum:psi_equality_basic_property:center_set} shows that 
%$([p^{\prime}, q^{\prime}], [\ell^{\prime}, r^{\prime}]) \in \Psi_{\centerset}(C)$ holds. 
%Lemma~\ref{lem:psi_str_property}~\ref{enum:psi_str_property:2} shows that 
%$([p^{\prime}, q^{\prime}], [\ell^{\prime}, r^{\prime}]) \in \Psi_{h}$ holds. 
%Lemma \ref{lem:psi_equality_basic_property}~\ref{enum:psi_equality_basic_property:9} shows that 
%$([p^{\prime}, q^{\prime}], [\ell^{\prime}, r^{\prime}]) \in \Psi_{\lcp}(K)$ holds 
%because $([p, q], [\ell, r]) \in \Psi_{\lcp}(K)$ follows from the definition of the subset $\Psi_{\lcp}(K)$. 

We prove $([p_{s}, q_{s}], [\ell_{s}, r_{s}]) \in \Psi_{h} \cap \Psi_{\run} \cap \Psi_{\centerset}(C) \cap (\bigcup_{\lambda = 2+\sum_{w = 1}^{h+3} \lfloor \mu(w) \rfloor}^{K-|C|} \Psi_{\lcp}(\lambda))$. 
Let $d = |f_{\recover}(([p^{\prime}, q^{\prime}], [\ell^{\prime}, r^{\prime}]))|$ for simplicity. 
We apply Lemma~\ref{lem:recover_basic_property} to the interval attractor $([p^{\prime}, q^{\prime}], [\ell^{\prime}, r^{\prime}])$. 
Then, Lemma~\ref{lem:recover_basic_property}~\ref{enum:recover_basic_property:4} shows that 
there exists an integer $\tau \in [1, d]$ satisfying 
$([p_{s}, q_{s}]$, $[\ell_{s}, r_{s}]) \in \Psi_{\run} \cap \Psi_{h} \cap \Psi_{\centerset}(C) \cap \Psi_{\lcp}(K - \tau |C|)$. 
Let $K_{s} = |\lcp(T[\gamma_{s}..r_{s}], C^{n+1})|$. 
Because of $([p_{s}, q_{s}]$, $[\ell_{s}, r_{s}]) \in \Psi_{\run} \cap \Psi_{h} \cap \Psi_{\centerset}(C)$, 
$K_{s} \geq 2+\sum_{w = 1}^{h+3} \lfloor \mu(w) \rfloor$ follows from the definition of the subset $\Psi_{\run}$. 
On the other hand, 
$K_{s} = K - \tau |C|$ follows from the definition of the subset $\Psi_{\lcp}(K - \tau |C|)$ 
because $([p_{s}, q_{s}], [\ell_{s}, r_{s}]) \in \Psi_{\centerset}(C) \cap \Psi_{\lcp}(K - \tau |C|)$. 
$2+\sum_{w = 1}^{h+3} \lfloor \mu(w) \rfloor \leq K - \tau |C| \leq K - |C|$ follows from 
$K_{s} \geq 2+\sum_{w = 1}^{h+3} \lfloor \mu(w) \rfloor$, $K_{s} = K - \tau |C|$, and $\tau \geq 1$. 
Therefore, $([p_{s}, q_{s}], [\ell_{s}, r_{s}]) \in \Psi_{h} \cap \Psi_{\run} \cap \Psi_{\centerset}(C) \cap (\bigcup_{\lambda = 2+\sum_{w = 1}^{h+3} \lfloor \mu(w) \rfloor}^{K-|C|} \Psi_{\lcp}(\lambda))$ holds. 

We showed that $([p_{s}, q_{s}], [\ell_{s}, r_{s}]) \in \Psi_{h} \cap \Psi_{\run} \cap \Psi_{\centerset}(C) \cap (\bigcup_{\lambda = 2+\sum_{w = 1}^{h+3} \lfloor \mu(w) \rfloor}^{K-|C|} \Psi_{\lcp}(\lambda))$ holds for each integer $s \in [1, k]$. 
Therefore, Lemma~\ref{lem:sRecover_basic_property}~\ref{enum:sRecover_basic_property:centerset} holds. 

\textbf{Proof of Lemma~\ref{lem:sRecover_basic_property}(iii).}
We prove $\Psi_{\sRecover}(([p, q], [\ell, r])) \cap \Psi_{\sRecover}(([p^{\prime}, q^{\prime}], [\ell^{\prime}, r^{\prime}])) = \emptyset$ by contradiction. 
We assume that $\Psi_{\sRecover}(([p, q], [\ell, r])) \cap \Psi_{\sRecover}(([p^{\prime}, q^{\prime}], [\ell^{\prime}, r^{\prime}])) \neq \emptyset$ holds. 
Then, the set $\Psi_{\sRecover}(([p, q], [\ell, r])) \cap \Psi_{\sRecover}(([p^{\prime}, q^{\prime}], [\ell^{\prime}, r^{\prime}]))$ contains an interval attractor $([p_{s}, q_{s}]$, $[\ell_{s}, r_{s}])$. 

From the definition of the subset $\Psi_{\sRecover}(([p, q], [\ell, r]))$, 
the subset $\Psi_{\source}$ contains an interval attractor $([p_{A}, q_{A}], [\ell_{A}, r_{A}])$ satisfying 
$([p_{s}, q_{s}], [\ell_{s}, r_{s}]) \in f_{\recover}(([p_{A}, q_{A}], [\ell_{A}, r_{A}]))$ and 
$([p_{A}, q_{A}]$, $[\ell_{A}, r_{A}]) \in \Psi_{\str}(([p, q], [\ell, r]))$. 
Similarly, from the definition of the subset $\Psi_{\sRecover}(([p^{\prime}, q^{\prime}], [\ell^{\prime}, r^{\prime}]))$, 
the subset $\Psi_{\source}$ contains an interval attractor $([p_{B}, q_{B}], [\ell_{B}, r_{B}])$ satisfying 
$([p_{s}, q_{s}], [\ell_{s}, r_{s}]) \in f_{\recover}(([p_{B}, q_{B}], [\ell_{B}, r_{B}]))$ and 
$([p_{B}, q_{B}], [\ell_{B}, r_{B}]) \in \Psi_{\str}(([p^{\prime}, q^{\prime}], [\ell^{\prime}, r^{\prime}]))$. 

We prove $T[p_{A}-1..r_{A}+1] = T[p_{B}-1..r_{B}+1]$. 
Lemma~\ref{lem:recover_division_property}~\ref{enum:recover_division_property:2} shows that 
$([p_{A}, q_{A}], [\ell_{A}, r_{A}]) = ([p_{B}, q_{B}], [\ell_{B}, r_{B}])$ holds 
because $([p_{s}, q_{s}], [\ell_{s}, r_{s}]) \in f_{\recover}(([p_{A}, q_{A}], [\ell_{A}, r_{A}]))$ and 
$([p_{s}, q_{s}], [\ell_{s}, r_{s}]) \in f_{\recover}(([p_{B}, q_{B}], [\ell_{B}, r_{B}]))$. 
Therefore, $T[p_{A}-1..r_{A}+1] = T[p_{B}-1..r_{B}+1]$ follows from $([p_{A}, q_{A}]$, $[\ell_{A}, r_{A}]) = ([p_{B}, q_{B}], [\ell_{B}, r_{B}])$. 

On the other hand, $T[p_{A}-1..r_{A}+1] = T[p-1..r+1]$ follows from $([p_{A}, q_{A}], [\ell_{A}, r_{A}]) \in \Psi_{\str}(([p, q], [\ell, r]))$. Similarly, $T[p_{B}-1..r_{B}+1] = T[p^{\prime}-1..r^{\prime}+1]$ follows from $([p_{B}, q_{B}], [\ell_{B}, r_{B}]) \in \Psi_{\str}(([p^{\prime}, q^{\prime}], [\ell^{\prime}, r^{\prime}]))$. 
$T[p_{A}-1..r_{A}+1] \neq T[p_{B}-1..r_{B}+1]$ follows from (A) $T[p_{A}-1..r_{A}+1] = T[p-1..r+1]$, (B) $T[p_{B}-1..r_{B}+1] = T[p^{\prime}-1..r^{\prime}+1]$, and (C) $T[p-1..r+1] \neq T[p^{\prime}-1..r^{\prime}+1]$. 
The two facts $T[p_{A}-1..r_{A}+1] = T[p_{B}-1..r_{B}+1]$ and $T[p_{A}-1..r_{A}+1] \neq T[p_{B}-1..r_{B}+1]$ yield a contradiction. 
Therefore, $\Psi_{\sRecover}(([p, q], [\ell, r])) \cap \Psi_{\sRecover}(([p^{\prime}, q^{\prime}], [\ell^{\prime}, r^{\prime}])) = \emptyset$ must hold. 

\textbf{Proof of Lemma~\ref{lem:sRecover_basic_property}(iv).}
Lemma~\ref{lem:sRecover_basic_property}~\ref{enum:sRecover_basic_property:equality} follows from 
the definitions of the two subsets $\Psi_{\sRecover}(([p, q], [\ell, r]))$ and $\Psi_{\sRecover}(([p^{\prime}, q^{\prime}], [\ell^{\prime}, r^{\prime}]))$. 

\textbf{Proof of Lemma~\ref{lem:sRecover_basic_property}(v).}
We prove $T[\gamma_{s}..\gamma_{s} + 1 + \sum_{w = 1}^{h+3} \mu(w)] = T[\gamma_{s^{\prime}}..\gamma_{s^{\prime}} + 1 + \sum_{w = 1}^{h+3} \mu(w)]$. 
$([p_{s}, q_{s}], [\ell_{s}, r_{s}]), ([p_{s^{\prime}}, q_{s^{\prime}}], [\ell_{s^{\prime}}, r_{s^{\prime}}]) \in \Psi_{h} \cap \Psi_{\run} \cap \Psi_{\centerset}(C)$ follows from Lemma~\ref{lem:sRecover_basic_property}~\ref{enum:sRecover_basic_property:centerset}. 
Because of $([p_{s}, q_{s}], [\ell_{s}, r_{s}]) \in \Psi_{h} \cap \Psi_{\run} \cap \Psi_{\centerset}(C)$, 
$|\lcp(T[\gamma_{s}..r_{s} + 1], C^{n+1})| > 1 + \sum_{w = 1}^{h+3} \mu(w)$ follows from the definition of the subset $\Psi_{\run}$. 
Similarly, 
$|\lcp(T[\gamma_{s^{\prime}}..r_{s^{\prime}} + 1], C^{n+1})| > 1 + \sum_{w = 1}^{h+3} \mu(w)$ follows from the definition of the subset $\Psi_{\run}$. 
Therefore, $T[\gamma_{s}..\gamma_{s} + 1 + \sum_{w = 1}^{h+3} \mu(w)] = T[\gamma_{s^{\prime}}..\gamma_{s^{\prime}} + 1 + \sum_{w = 1}^{h+3} \mu(w)]$ follows from $|\lcp(T[\gamma_{s}..r_{s} + 1], C^{n+1})| > 1 + \sum_{w = 1}^{h+3} \mu(w)$ and $|\lcp(T[\gamma_{s^{\prime}}..r_{s^{\prime}} + 1], C^{n+1})| > 1 + \sum_{w = 1}^{h+3} \mu(w)$. 

We prove $T[\gamma_{s}..r_{s}+1] \prec P \Rightarrow T[\gamma_{s^{\prime}}..r_{s^{\prime}}+1] \prec P$. 
$T[\gamma_{s}..\gamma_{s} + 1 + \sum_{w = 1}^{h+3} \mu(w)] \prec P$ follows from 
$T[\gamma_{s}..r_{s}+1] \prec P$ and $|\lcp(T[\gamma_{s}..r_{s}+1], P)| \leq 1 + \sum_{w = 1}^{h+3} \mu(w)$. 
$T[\gamma_{s^{\prime}}..\gamma_{s^{\prime}} + 1 + \sum_{w = 1}^{h+3} \mu(w)] \prec P$ follows from 
$T[\gamma_{s}..\gamma_{s} + 1 + \sum_{w = 1}^{h+3} \mu(w)] = T[\gamma_{s^{\prime}}..\gamma_{s^{\prime}} + 1 + \sum_{w = 1}^{h+3} \mu(w)]$ and $T[\gamma_{s}..\gamma_{s} + 1 + \sum_{w = 1}^{h+3} \mu(w)] \prec P$. 
The string $T[\gamma_{s^{\prime}}..\gamma_{s^{\prime}} + 1 + \sum_{w = 1}^{h+3} \mu(w)]$ is a prefix of string $T[\gamma_{s}..r_{s}+1]$, 
and hence, 
$T[\gamma_{s^{\prime}}..\gamma_{s^{\prime}} + 1 + \sum_{w = 1}^{h+3} \mu(w)] \preceq T[\gamma_{s^{\prime}}..r_{s^{\prime}}+1]$ holds. 
Therefore, $T[\gamma_{s^{\prime}}..r_{s^{\prime}}+1] \prec P$ follows from 
$T[\gamma_{s^{\prime}}..\gamma_{s^{\prime}} + 1 + \sum_{w = 1}^{h+3} \mu(w)] \prec P$ and $T[\gamma_{s^{\prime}}..\gamma_{s^{\prime}} + 1 + \sum_{w = 1}^{h+3} \mu(w)] \preceq T[\gamma_{s^{\prime}}..r_{s^{\prime}}+1]$. 

Similarly, 
we can prove $P \prec T[\gamma_{s}..r_{s}+1] \Rightarrow P \prec T[\gamma_{s^{\prime}}..r_{s^{\prime}}+1]$ 
using the same approach used to prove $T[\gamma_{s}..r_{s}+1] \prec P \Rightarrow T[\gamma_{s^{\prime}}..r_{s^{\prime}}+1] \prec P$. 
Therefore, Lemma~\ref{lem:sRecover_basic_property}~\ref{enum:sRecover_basic_property:lex} holds.

\end{proof}

\begin{lemma}\label{lem:sRecover_size_property}
Consider an interval attractor $([p, q], [\ell, r])$ in subset $\Psi_{\source}$. 
Let $\gamma$, $C$, and $h$ be the attractor position, associated string, and level of the interval attractor $([p, q], [\ell, r])$, 
respectively; 
let $\zeta = 2 + \sum_{w = 1}^{h+3} \lfloor \mu(w) \rfloor$ for simplicity; 
let $K = |\lcp(T[\gamma..r], C^{n+1})|$ and $M = (K - \zeta) \mod |C|$. 
Then, the following four statements hold:

\begin{enumerate}[label=\textbf{(\roman*)}]
    \item \label{enum:sRecover_size_property:1}
    $|\Psi_{\sRecover}(([p, q], [\ell, r])) \cap \Psi| = \sum_{([p^{\prime}, q^{\prime}], [\ell^{\prime}, r^{\prime}]) \in \Psi_{\str}(T[p-1..r+1])} |f_{\recover}(([p^{\prime}, q^{\prime}], [\ell^{\prime}, r^{\prime}])) \cap \Psi|$    
    for a subset $\Psi$ of set $\Psi_{\RR}$; 
    \item \label{enum:sRecover_size_property:2} $|\Psi_{\sRecover}(([p, q], [\ell, r]))| = |\Psi_{\str}(T[p-1..r+1])| |f_{\recover}(([p, q], [\ell, r]))|$;
    \item \label{enum:sRecover_size_property:3}
    $|\Psi_{\sRecover}(([p, q], [\ell, r])) \cap (\bigcup_{\lambda = (y-1)|C| + \zeta}^{y|C| + \zeta - 1} \Psi_{\lcp}(\lambda))| = |\Psi_{\str}(T[p-1..r+1])|$ for each integer $y \in [1, \lfloor \frac{K - \zeta}{|C|} \rfloor]$;    
    \item \label{enum:sRecover_size_property:4}
    $|\Psi_{\sRecover}(([p, q], [\ell, r])) \cap \Psi_{\lcp}((y-1)|C| + \zeta + M)| = |\Psi_{\str}(T[p-1..r+1])|$ 
    for each integer $y \in [1, \lfloor \frac{K - \zeta}{|C|} \rfloor]$.    
\end{enumerate}
\end{lemma}
\begin{proof}
$K \geq |C| + \zeta$ follows from Lemma~\ref{lem:recover_basic_property}~\ref{enum:recover_basic_property:1}.
$([p, q], [\ell, r]) \in \Psi_{\lcp}(K) \cap \Psi_{\modulo}(M)$ follows from the definitions of the two subsets $\Psi_{\lcp}(K)$ 
and $\Psi_{\modulo}(M)$. 

Let $([p_{1}, q_{1}], [\ell_{1}, r_{1}])$, $([p_{2}, q_{2}], [\ell_{2}, r_{2}])$, $\ldots$, $([p_{k}, q_{k}], [\ell_{k}, r_{k}])$ 
be the interval attractors in the subset $\Psi_{\str}(T[p-1..r+1])$. 
Here, $T[p-1..r+1] = T[p_{s}-1..r_{s}+1]$ follows from the definition of the subset $\Psi_{\str}(T[p-1..r+1])$ 
for each integer $s \in [1, k]$. 
Lemma~\ref{lem:psi_equality_basic_property} shows that 
$([p_{s}, q_{s}], [\ell_{s}, r_{s}]) \in \Psi_{h} \cap \Psi_{\source} \cap \Psi_{\centerset}(C) \cap \Psi_{\lcp}(K) \cap \Psi_{\modulo}(M)$ holds 
because $([p, q], [\ell, r]) \in \Psi_{h} \cap \Psi_{\source} \cap \Psi_{\centerset}(C) \cap \Psi_{\lcp}(K) \cap \Psi_{\modulo}(M)$ 
and $T[p-1..r+1] = T[p_{s}-1..r_{s}+1]$. 

The following two statements are used to prove Lemma~\ref{lem:sRecover_size_property}: 
\begin{enumerate}[label=\textbf{(\Alph*)}]
    \item $|f_{\recover}(([p_{s}, q_{s}], [\ell_{s}, r_{s}])) \cap \Psi_{\lcp}((y-1)|C| + \zeta + M)| = 1$ for any pair of two integers $s \in [1, k]$ and $y \in [1, \lfloor \frac{K - \zeta}{|C|} \rfloor]$;
    \item $|f_{\recover}(([p_{s}, q_{s}], [\ell_{s}, r_{s}])) \cap (\bigcup_{\lambda = (y-1)|C| + \zeta}^{y|C| + \zeta - 1} \Psi_{\lcp}(\lambda))| = 1$ for any pair of two integers $s \in [1, k]$ and $y \in [1, \lfloor \frac{K - \zeta}{|C|} \rfloor]$. 
\end{enumerate}
\textbf{Proof of statement (A).}
For simplicity, let $m = |f_{\recover}(([p_{s}, q_{s}], [\ell_{s}, r_{s}]))|$.
We prove $y \leq m$. 
Lemma~\ref{lem:recover_basic_property}~\ref{enum:recover_basic_property:1} shows that 
$|f_{\recover}(([p, q], [\ell, r]))| = \lfloor \frac{K - \zeta}{|C|} \rfloor$ holds. 
Lemma~\ref{lem:recover_super_property}~\ref{enum:recover_super_property:1} shows that 
$|f_{\recover}(([p, q], [\ell, r]))| = m$ holds 
because $T[p-1..r+1] = T[p_{s}-1..r_{s}+1]$. 
Therefore, $y \leq m$ follows from 
$y \in [1, \lfloor \frac{K - \zeta}{|C|} \rfloor]$, $|f_{\recover}(([p, q], [\ell, r]))| = \lfloor \frac{K - \zeta}{|C|} \rfloor$, 
and $|f_{\recover}(([p, q], [\ell, r]))| = m$.

We prove $K - (m - y + 1)|C| = (y-1)|C| + \zeta + M$. 
$|f_{\recover}(([p, q], [\ell, r]))| = \lfloor \frac{K - \zeta}{|C|} \rfloor$ follows from Lemma~\ref{lem:recover_basic_property}~\ref{enum:recover_basic_property:1}. 
$\lfloor \frac{K - \zeta}{|C|} \rfloor = \frac{K - M - \zeta}{|C|}$ follows from 
$K \geq \zeta$ and $M = (K - \zeta) \mod |C|$. 
Therefore, $K - (m - y + 1)|C| = (y-1)|C| + \zeta + M$ follows from the following equation: 
\begin{equation*}
    \begin{split}
    K - (m - y + 1)|C| &= K - |f_{\recover}(([p, q], [\ell, r]))| |C| + (y - 1)|C|  \\
    &= K - \lfloor \frac{K - \zeta}{|C|} \rfloor |C| (y - 1)|C| \\
    &= K - (K - M - \zeta) (y - 1)|C| \\
    &= (y - 1)|C| + \zeta + M.
    \end{split}
\end{equation*}

Lemma~\ref{lem:recover_basic_property}~\ref{enum:recover_basic_property:4} shows that 
$|f_{\recover}(([p_{s}, q_{s}], [\ell_{s}, r_{s}])) \cap \Psi_{\lcp}(K - (m - y + 1)|C|)| = 1$ holds 
because $([p_{s}, q_{s}], [\ell_{s}, r_{s}]) \in \Psi_{\source} \cap \Psi_{\centerset}(C) \cap \Psi_{\lcp}(K)$ and $y \leq m$. 
Therefore, $|f_{\recover}(([p_{s}, q_{s}], [\ell_{s}, r_{s}])) \cap \Psi_{\lcp}((y-1)|C| + \zeta + M)| = 1$ follows from 
$|f_{\recover}(([p_{s}, q_{s}], [\ell_{s}, r_{s}])) \cap \Psi_{\lcp}(K - (m - y + 1)|C|)| = 1$ 
and $K - (m - y + 1)|C| = (y-1)|C| + \zeta + M$. 

\textbf{Proof of statement (B).}
Let $\mathcal{I} = [0, |C| - 1] \setminus \{ M \}$ for simplicity. 
Because of $M \in [0, |C| - 1]$,  
the following equation holds: 
\begin{equation}\label{eq:sRecover_size_property:1}
    \begin{split}
    f_{\recover}(([p_{s}, q_{s}], [\ell_{s}, r_{s}])) & \cap (\bigcup_{\lambda = (y-1)|C| + \zeta}^{y|C| + \zeta - 1} \Psi_{\lcp}(\lambda)) \\
    &= (f_{\recover}(([p_{s}, q_{s}], [\ell_{s}, r_{s}])) \cap \Psi_{\lcp}((y-1)|C| + \zeta + M))  \\
    &\cup (f_{\recover}(([p_{s}, q_{s}], [\ell_{s}, r_{s}])) \cap (\bigcup_{\lambda \in \mathcal{I}} \Psi_{\lcp}( (y-1)|C| + \zeta + \lambda))).
    \end{split}
\end{equation}

We prove $|f_{\recover}(([p_{s}, q_{s}], [\ell_{s}, r_{s}])) \cap (\bigcup_{\lambda = (y-1)|C| + \zeta}^{y|C| + \zeta - 1} \Psi_{\lcp}(\lambda))| = 1$ by contradiction. 
We assume that $|f_{\recover}(([p_{s}, q_{s}], [\ell_{s}, r_{s}])) \cap (\bigcup_{\lambda = (y-1)|C| + \zeta}^{y|C| + \zeta - 1} \Psi_{\lcp}(\lambda))| \neq 1$. 
Then, Equation~\ref{eq:sRecover_size_property:1} indicates that 
the set $f_{\recover}(([p_{s}, q_{s}], [\ell_{s}, r_{s}])) \cap (\bigcup_{\lambda \in \mathcal{I}} \Psi_{\lcp}( (y-1)|C| + \zeta + \lambda))$ 
contains an interval attractor $([p^{\prime}, q^{\prime}], [\ell^{\prime}, r^{\prime}])$ 
because $|f_{\recover}(([p_{s}, q_{s}], [\ell_{s}, r_{s}])) \cap \Psi_{\lcp}((y-1)|C| + \zeta + M)| = 1$ follows from statement (A). 
Here, Lemma~\ref{lem:recover_basic_property}~\ref{enum:recover_basic_property:4} shows that 
$([p^{\prime}, q^{\prime}], [\ell^{\prime}, r^{\prime}]) \in \Psi_{h} \cap \Psi_{\run} \cap \Psi_{\centerset}(C) \cap \Psi_{\modulo}(M)$ holds because $([p_{s}, q_{s}], [\ell_{s}, r_{s}]) \in \Psi_{h} \cap \Psi_{\source} \cap \Psi_{\centerset}(C) \cap \Psi_{\modulo}(M)$ 
and $([p^{\prime}, q^{\prime}], [\ell^{\prime}, r^{\prime}]) \in f_{\recover}(([p_{s}, q_{s}], [\ell_{s}, r_{s}]))$. 
The set $\mathcal{I}$ contains an integer $M^{\prime}$ 
satisfying $([p^{\prime}, q^{\prime}], [\ell^{\prime}, r^{\prime}]) \in \Psi_{\lcp}( (y-1)|C| + \zeta + M^{\prime})$. 

Let $\gamma^{\prime}$ be the attractor position of the interval attractor $([p^{\prime}, q^{\prime}], [\ell^{\prime}, r^{\prime}])$. 
Let $K^{\prime} = |\lcp(T[\gamma^{\prime}..r^{\prime}]$, $C^{n+1})|$. 
Because of $([p^{\prime}, q^{\prime}], [\ell^{\prime}, r^{\prime}]) \in \Psi_{\centerset}(C) \cap \Psi_{\lcp}( (y-1)|C| + \zeta + M^{\prime})$, 
$K^{\prime} = (y-1)|C| + \zeta + M^{\prime}$ follows from the definition of the subset $\Psi_{\lcp}( (y-1)|C| + \zeta + M^{\prime})$. 
Because of $([p^{\prime}, q^{\prime}], [\ell^{\prime}, r^{\prime}]) \in \Psi_{h} \cap \Psi_{\centerset}(C) \cap \Psi_{\modulo}(M)$, 
$(K^{\prime} - \zeta) \mod |C| = M$ follows from the definition of the subset $\Psi_{\modulo}(M)$. 
$M^{\prime} = M$ follows from 
$K^{\prime} = (y-1)|C| + \zeta + M^{\prime}$, $(K^{\prime} - \zeta) \mod |C| = M$, $y \geq 1$, and $M^{\prime} \in [0, |C| - 1]$. 
On the other hand, $M^{\prime} \neq M$ follows from $M^{\prime} \in \mathcal{I}$ and $\mathcal{I} = [0, |C| - 1] \setminus \{ M \}$. 
The two facts $M^{\prime} = M$ and $M^{\prime} \neq M$ yield a contradiction. 
Therefore, $|f_{\recover}(([p_{s}, q_{s}], [\ell_{s}, r_{s}])) \cap (\bigcup_{\lambda = (y-1)|C| + \zeta}^{y|C| + \zeta - 1} \Psi_{\lcp}(\lambda))| = 1$ must hold.

%The proof of Lemma~\ref{lem:sRecover_size_property} is as follows. 

\textbf{Proof of Lemma~\ref{lem:sRecover_size_property}(i).}
Lemma~\ref{lem:sRecover_basic_property}~\ref{enum:sRecover_basic_property:str} indicates that 
$|\Psi_{\sRecover}(([p, q], [\ell, r])) \cap \Psi| = |\bigcup_{s = 1}^{k}$ $(f_{\recover}(([p_{s}, q_{s}], [\ell_{s}, r_{s}])) \cap \Psi)|$ holds. 
$|\bigcup_{s = 1}^{k} (f_{\recover}(([p_{s}, q_{s}], [\ell_{s}, r_{s}])) \cap \Psi)| = \sum_{s = 1}^{k} |f_{\recover}(([p_{s}, q_{s}]$, $[\ell_{s}, r_{s}])) \cap \Psi|$ holds 
because 
Lemma~\ref{lem:recover_division_property}~\ref{enum:recover_division_property:2} shows that 
$f_{\recover}(([p_{s}, q_{s}], [\ell_{s}, r_{s}])) \cap f_{\recover}(([p_{s^{\prime}}, q_{s^{\prime}}]$, $[\ell_{s^{\prime}}, r_{s^{\prime}}])) = \emptyset$ 
for any pair of two integers $1 \leq s < s^{\prime} \leq k$. 
Therefore, $|\Psi_{\sRecover}(([p, q], [\ell, r])) \cap \Psi| = \sum_{s = 1}^{k} |f_{\recover}(([p_{s}, q_{s}], [\ell_{s}, r_{s}])) \cap \Psi|$ holds. 

\textbf{Proof of Lemma~\ref{lem:sRecover_size_property}(ii).}
$|\Psi_{\sRecover}(([p, q], [\ell, r]))| = |\Psi_{\sRecover}(([p, q], [\ell, r])) \cap \Psi_{\RR}|$ follows from 
$\Psi_{\sRecover}(([p, q], [\ell, r])) \subseteq \Psi_{\RR}$. 
$|\Psi_{\sRecover}(([p, q], [\ell, r])) \cap \Psi_{\RR}| = \sum_{s = 1}^{k} |f_{\recover}(([p_{s}, q_{s}], [\ell_{s}, r_{s}])) \cap \Psi_{\RR}|$ follows from Lemma~\ref{lem:sRecover_size_property}(i). 
Here, for each integer $s \in [1, k]$, 
$|f_{\recover}(([p_{s}, q_{s}], [\ell_{s}, r_{s}])) \cap \Psi_{\RR}| = |f_{\recover}(([p_{s}, q_{s}], [\ell_{s}, r_{s}]))|$ holds 
because $f_{\recover}(([p_{s}, q_{s}], [\ell_{s}, r_{s}])) \subseteq \Psi_{\RR}$. 
Lemma~\ref{lem:recover_super_property}~\ref{enum:recover_super_property:1} shows that 
$|f_{\recover}(([p, q], [\ell, r]))| = |f_{\recover}(([p_{s}, q_{s}], [\ell_{s}, r_{s}]))|$ holds. 
because 
$T[p-1..r+1] = T[p_{s}-1..r_{s}+1]$ follows from the definition of the subset $\Psi_{\str}(T[p-1..r+1])$. 
Therefore, $|\Psi_{\sRecover}(([p, q], [\ell, r]))| = |\Psi_{\str}(T[p-1..r+1])| |f_{\recover}(([p, q], [\ell, r]))|$ 
follows from the following equation: 
\begin{equation*}
    \begin{split}
    |\Psi_{\sRecover}(([p, q], [\ell, r]))| &= |\Psi_{\sRecover}(([p, q], [\ell, r])) \cap \Psi_{\RR}| \\
    &= \sum_{s = 1}^{k} |f_{\recover}(([p_{s}, q_{s}], [\ell_{s}, r_{s}])) \cap \Psi_{\RR}| \\ 
    &= \sum_{s = 1}^{k} |f_{\recover}(([p_{s}, q_{s}], [\ell_{s}, r_{s}]))| \\
    &= \sum_{s = 1}^{k} |f_{\recover}(([p, q], [\ell, r]))| \\ 
    &= k |f_{\recover}(([p, q], [\ell, r]))| \\
    &= |\Psi_{\str}(T[p-1..r+1])| |f_{\recover}(([p, q], [\ell, r]))|.
    \end{split}
\end{equation*}

\textbf{Proof of Lemma~\ref{lem:sRecover_size_property}(iii).}
The following equation follows from Lemma~\ref{lem:sRecover_size_property}(i): 
\begin{equation*}
    \begin{split}
    |\Psi_{\sRecover}(([p, q], [\ell, r])) & \cap \Bigl(\bigcup_{\lambda = (y-1)|C| + \zeta}^{y|C| + \zeta - 1} \Psi_{\lcp}(\lambda) \Bigr)| \\
    &= \sum_{s = 1}^{k} |f_{\recover}(([p_{s}, q_{s}], [\ell_{s}, r_{s}])) \cap \Bigl(\bigcup_{\lambda = (y-1)|C| + \zeta}^{y|C| + \zeta - 1} \Psi_{\lcp}(\lambda) \Bigr)|. 
    \end{split}
\end{equation*}
Statement (B) shows that 
$|f_{\recover}(([p_{s}, q_{s}], [\ell_{s}, r_{s}])) \cap (\bigcup_{\lambda = (y-1)|C| + \zeta}^{y|C| + \zeta - 1} \Psi_{\lcp}(\lambda))| = 1$ for each integer $s \in [1, k]$. 
Therefore, $|(\Psi_{\sRecover}(([p, q], [\ell, r])) \cap (\bigcup_{\lambda = (y-1)|C| + \zeta}^{y|C| + \zeta - 1} \Psi_{\lcp}(\lambda))| = |\Psi_{\str}(T[p-1..r+1])|$ holds. 

\textbf{Proof of Lemma~\ref{lem:sRecover_size_property}(iv).}
The following equation follows from Lemma~\ref{lem:sRecover_size_property}(i): 
\begin{equation*}
    \begin{split}
    |\Psi_{\sRecover}(([p, q], [\ell, r])) & \cap \Psi_{\lcp}((y-1)|C| + \zeta + M)| \\
    &= \sum_{s = 1}^{k} |f_{\recover}(([p_{s}, q_{s}], [\ell_{s}, r_{s}])) \cap \Psi_{\lcp}((y-1)|C| + \zeta + M)|.
    \end{split}
\end{equation*}
Statement (A) shows that $|(\Psi_{\sRecover}(([p, q], [\ell, r])) \cap \Psi_{\lcp}((y-1)|C| + \zeta + M)| = 1$ for each integer $s \in [1, k]$. Therefore, $|\Psi_{\sRecover}(([p, q], [\ell, r])) \cap \Psi_{\lcp}((y-1)|C| + \zeta + M)| = |\Psi_{\str}(T[p-1..r+1])|$ holds. 
\end{proof}

\paragraph{Subset $\Psi_{\nRecover}(\lambda)$.}
For an integer $\lambda \geq 0$, 
subset $\Psi_{\nRecover}(\lambda)$ of set $\Psi_{\RR}$ consists of the interval attractors such that 
for each interval attractor $([p, q], [\ell, r]) \in \Psi_{\nRecover}(\lambda)$, 
the subset $\Psi_{\source}$ contains an interval attractor $([p^{\prime}, q^{\prime}], [\ell^{\prime}, r^{\prime}])$ satisfying the following two conditions: 
\begin{itemize}
    \item $([p, q], [\ell, r]) \in f_{\recover}(([p^{\prime}, q^{\prime}], [\ell^{\prime}, r^{\prime}]))$;
    \item $|f_{\recover}(([p^{\prime}, q^{\prime}], [\ell^{\prime}, r^{\prime}]))| = \lambda$.
\end{itemize}
Formally, let $\Psi_{\nRecover}(\lambda) = \{ ([p, q], [\ell, r]) \in \Psi_{\RR} \mid \exists ([p^{\prime}, q^{\prime}], [\ell^{\prime}, r^{\prime}]) \in \Psi_{\source} \text{ s.t. } ([p, q]$, $[\ell, r]) \in f_{\recover}(([p^{\prime}, q^{\prime}], [\ell^{\prime}, r^{\prime}])) \text{ and } |f_{\recover}(([p^{\prime}, q^{\prime}], [\ell^{\prime}, r^{\prime}]))| = \lambda \}$.  

\begin{lemma}\label{lem:nRecover_basic_property}
    The following two statements hold: 
\begin{enumerate}[label=\textbf{(\roman*)}]
    \item \label{enum:nRecover_basic_property:1}
    $\Psi_{\run} = \bigcup_{\lambda = 1}^{n} \Psi_{\nRecover}(\lambda)$;
    \item \label{enum:nRecover_basic_property:2}
    $\Psi_{\nRecover}(\lambda) \cap \Psi_{\nRecover}(\lambda^{\prime}) = \emptyset$ 
    for any pair of two integers $1 \leq \lambda < \lambda^{\prime} \leq n$.
\end{enumerate}
\end{lemma}
\begin{proof}
    The proof of Lemma~\ref{lem:nRecover_basic_property} is as follows.

    \textbf{Proof of Lemma~\ref{lem:nRecover_basic_property}(i).}    
    We prove $\Psi_{\run} \subseteq \bigcup_{\lambda = 1}^{n} \Psi_{\nRecover}(\lambda)$.     
    Consider an interval attractor $([p, q], [\ell, r])$ in subset $\Psi_{\run}$. 
    Then, Lemma~\ref{lem:recover_division_property}~\ref{enum:recover_division_property:1} shows that 
    subset $\Psi_{\source}$ contains an interval attractor $([p^{\prime}, q^{\prime}], [\ell^{\prime}, r^{\prime}])$ 
    satisfying $([p, q], [\ell, r]) \in f_{\recover}(([p^{\prime}, q^{\prime}], [\ell^{\prime}, r^{\prime}]))$. 
    Let $k = |f_{\recover}(([p^{\prime}, q^{\prime}], [\ell^{\prime}, r^{\prime}]))|$ for simplicity. 
    Then, $([p, q], [\ell, r]) \in \Psi_{\nRecover}(k)$ follows from the definition of the subset $\Psi_{\nRecover}(k)$. 
    $1 \leq k \leq n$ follows from Lemma~\ref{lem:recover_basic_property}~\ref{enum:recover_basic_property:1}. 
    Therefore, $\Psi_{\run} \subseteq \bigcup_{\lambda = 1}^{n} \Psi_{\nRecover}(\lambda)$ holds. 

    On the other hand, $\Psi_{\run} \supseteq \bigcup_{\lambda = 1}^{n} \Psi_{\nRecover}(\lambda)$ holds 
    because $\Psi_{\run} \supseteq \Psi_{\nRecover}(\lambda)$ follows from the definition of the subset $\Psi_{\nRecover}(\lambda)$ 
    for each integer $\lambda \in [1, n]$. 
    Therefore, $\Psi_{\run} = \bigcup_{\lambda = 1}^{n} \Psi_{\nRecover}(\lambda)$ follows from 
    $\Psi_{\run} \subseteq \bigcup_{\lambda = 1}^{n} \Psi_{\nRecover}(\lambda)$ and $\Psi_{\run} \supseteq \bigcup_{\lambda = 1}^{n} \Psi_{\nRecover}(\lambda)$. 

    \textbf{Proof of Lemma~\ref{lem:nRecover_basic_property}(ii).}    
    We prove $\Psi_{\nRecover}(\lambda) \cap \Psi_{\nRecover}(\lambda^{\prime}) = \emptyset$ by contradiction. 
    We assume that $\Psi_{\nRecover}(\lambda) \cap \Psi_{\nRecover}(\lambda^{\prime}) \neq \emptyset$ holds. 
    Then, there exists an interval attractor $([p, q], [\ell, r]) \in \Psi_{\RR}$ satisfying 
    $([p, q], [\ell, r]) \in \Psi_{\nRecover}(\lambda) \cap \Psi_{\nRecover}(\lambda^{\prime})$. 
    Because of $([p, q], [\ell, r]) \in \Psi_{\nRecover}(\lambda)$, 
    set $\Psi_{\source}$ contains an interval attractor $([p_{A}, q_{A}], [\ell_{A}, r_{A}])$ satisfying 
    $([p, q], [\ell, r]) \in f_{\recover}(([p_{A}, q_{A}], [\ell_{A}, r_{A}]))$ and 
    $|f_{\recover}(([p_{A}, q_{A}], [\ell_{A}, r_{A}]))| = \lambda$. 
    Similarly, 
    set $\Psi_{\source}$ contains an interval attractor $([p_{B}, q_{B}], [\ell_{B}, r_{B}])$ satisfying 
    $([p, q], [\ell, r]) \in f_{\recover}(([p_{B}, q_{B}], [\ell_{B}, r_{B}]))$ and 
    $|f_{\recover}(([p_{B}, q_{B}], [\ell_{B}, r_{B}]))| = \lambda^{\prime}$. 
    $([p_{A}, q_{A}], [\ell_{A}, r_{A}]) \neq ([p_{B}, q_{B}], [\ell_{B}, r_{B}])$ holds 
    because $|f_{\recover}(([p_{A}$, $q_{A}], [\ell_{A}, r_{A}]))| \neq |f_{\recover}(([p_{B}, q_{B}], [\ell_{B}, r_{B}]))|$. 

    We show that there exists a contradiction. 
    $f_{\recover}(([p_{A}, q_{A}], [\ell_{A}, r_{A}])) \cap f_{\recover}(([p_{B}, q_{B}]$, $[\ell_{B}$, $r_{B}])) \neq \emptyset$ 
    follows from $([p, q], [\ell, r]) \in f_{\recover}(([p_{A}, q_{A}], [\ell_{A}, r_{A}]))$ and 
    $([p, q], [\ell, r]) \in f_{\recover}(([p_{B}, q_{B}]$, $[\ell_{B}, r_{B}]))$. 
    On the other hand, 
    Lemma~\ref{lem:recover_division_property}~\ref{enum:recover_division_property:2} shows that 
    $f_{\recover}(([p_{A}, q_{A}], [\ell_{A}, r_{A}])) \cap f_{\recover}(([p_{B}, q_{B}]$, $[\ell_{B}, r_{B}])) = \emptyset$ holds. 
    The two facts $f_{\recover}(([p_{A}, q_{A}], [\ell_{A}, r_{A}])) \cap f_{\recover}(([p_{B}, q_{B}], [\ell_{B}, r_{B}])) \neq \emptyset$ and $f_{\recover}(([p_{A}$, $q_{A}], [\ell_{A}, r_{A}])) \cap f_{\recover}(([p_{B}, q_{B}], [\ell_{B}, r_{B}])) = \emptyset$ yield a contradiction. 
    Therefore, $\Psi_{\nRecover}(\lambda) \cap \Psi_{\nRecover}(\lambda^{\prime}) = \emptyset$ must hold. 
\end{proof}

\paragraph{Subset $\Psi_{\mRecover}$.}
Subset $\Psi_{\mRecover}$ of set $\Psi_{\RR}$ consists of the interval attractors such that 
for each interval attractor $([p, q], [\ell, r]) \in \Psi_{\mRecover}$, 
the subset $\Psi_{\source}$ contains an interval attractor $([p^{\prime}, q^{\prime}], [\ell^{\prime}, r^{\prime}])$ satisfying the following two conditions: 
\begin{itemize}
    \item $([p, q], [\ell, r]) \in f_{\recover}(([p^{\prime}, q^{\prime}], [\ell^{\prime}, r^{\prime}]))$;
    \item $f_{\recover}(([p^{\prime}, q^{\prime}], [\ell^{\prime}, r^{\prime}])) \cap \Psi_{\lex}(T[\gamma..r+1]) = \emptyset$ for the attractor position $\gamma$ of the interval attractor $([p, q], [\ell, r])$.     
\end{itemize}
Formally, let $\Psi_{\mRecover} = \{ ([p, q], [\ell, r]) \in \Psi_{\RR} \mid \exists ([p^{\prime}, q^{\prime}], [\ell^{\prime}, r^{\prime}]) \in \Psi_{\source} \text{ s.t. } ([p, q]$, $[\ell, r]) \in f_{\recover}(([p^{\prime}, q^{\prime}], [\ell^{\prime}, r^{\prime}])) \text{ and } (f_{\recover}(([p^{\prime}, q^{\prime}], [\ell^{\prime}, r^{\prime}])) \cap \Psi_{\lex}(T[\gamma..r+1])) = \emptyset  \}$. 
$\Psi_{\mRecover} \subseteq \Psi_{\run}$ follows from the definition of the subset $\Psi_{\mRecover}$.

The following two lemmas state properties of the subset $\Psi_{\mRecover}$. 
\begin{lemma}\label{lem:mRecover_basic_property}
    Let $([p_{1}, q_{1}], [\ell_{1}, r_{1}])$, 
    $([p_{2}, q_{2}], [\ell_{2}, r_{2}])$, $\ldots$, $([p_{k}, q_{k}], [\ell_{k}, r_{k}])$ ($p_{1} < p_{2} < \ldots < p_{k}$)
    be the interval attractors in the set obtained from function $f_{\recover}(([p, q], [\ell, r]))$ 
    for an interval attractor $([p, q], [\ell, r])$ in subset $\Psi_{\source}$.     
    If $([p, q], [\ell, r]) \in \Psi_{\preceding}$, 
    then $f_{\recover}(([p, q], [\ell, r])) \cap \Psi_{\mRecover} = \{ ([p_{k}, q_{k}], [\ell_{k}, r_{k}]) \}$. 
    Otherwise, $f_{\recover}(([p, q], [\ell, r])) \cap \Psi_{\mRecover} = \{ ([p_{1}, q_{1}], [\ell_{1}, r_{1}]) \}$.

%    The following two statements hold for subset $\Psi_{\mRecover}$: 
%\begin{enumerate}[label=\textbf{(\roman*)}]
%    \item \label{enum:mRecover_basic_property:1X}
%    $\Psi_{\mRecover} \subseteq \Psi_{\run}$;
%    \item \label{enum:mRecover_basic_property:2}
%\end{enumerate}
\end{lemma}
\begin{proof}
    Let $\gamma_{s}$ be the attractor position of each interval attractor $([p_{s}, q_{s}]$, $[\ell_{s}, r_{s}]) \in f_{\recover}(([p, q]$, $[\ell, r]))$. 
    If $([p, q], [\ell, r]) \in \Psi_{\preceding}$, 
    then Lemma~\ref{lem:recover_basic_property}~\ref{enum:recover_basic_property:6} shows that 
    $T[\gamma_{k}..r_{k}+1] \prec T[\gamma_{k-1}..r_{k-1}+1] \prec \cdots \prec T[\gamma_{1}..r_{1}+1]$ holds. 
    In this case, 
    $([p_{k}, q_{k}], [\ell_{k}, r_{k}]) \in \Psi_{\mRecover}$ holds 
    because $f_{\recover}(([p, q]$, $[\ell, r])) \cap \Psi_{\lex}(T[\gamma_{k}..r_{k}+1]) = \emptyset$ holds. 
    In contrast, 
    $([p_{s}, q_{s}], [\ell_{s}, r_{s}]) \not \in \Psi_{\mRecover}$ holds for each integer $s \in [1, k-1]$ 
    because 
    $([p_{k}, q_{k}], [\ell_{k}, r_{k}]) \in f_{\recover}(([p, q]$, $[\ell, r])) \cap \Psi_{\lex}(T[\gamma_{s}..r_{s}+1])$ holds. 
    Therefore, $f_{\recover}(([p, q], [\ell, r])) \cap \Psi_{\mRecover} = \{ ([p_{k}, q_{k}], [\ell_{k}, r_{k}]) \}$ holds. 

    Otherwise (i.e., $([p, q], [\ell, r]) \not \in \Psi_{\preceding}$), 
    Lemma~\ref{lem:recover_basic_property}~\ref{enum:recover_basic_property:6} shows that 
    $T[\gamma_{1}..r_{1}+1] \prec T[\gamma_{2}..r_{2}+1] \prec \cdots \prec T[\gamma_{k}..r_{k}+1]$ holds. 
    In this case, 
    $([p_{1}, q_{1}]$, $[\ell_{1}, r_{1}]) \in \Psi_{\mRecover}$ holds 
    because $f_{\recover}(([p, q]$, $[\ell, r])) \cap \Psi_{\lex}(T[\gamma_{1}..r_{1}+1]) = \emptyset$ holds. 
    In contrast, 
    $([p_{s}, q_{s}], [\ell_{s}, r_{s}]) \not \in \Psi_{\mRecover}$ holds for each integer $s \in [2, k]$ 
    because 
    $([p_{1}, q_{1}], [\ell_{1}, r_{1}]) \in f_{\recover}(([p, q], [\ell, r])) \cap \Psi_{\lex}(T[\gamma_{s}..r_{s}+1])$ holds. 
    Therefore, $f_{\recover}(([p, q], [\ell, r])) \cap \Psi_{\mRecover} = \{ ([p_{1}, q_{1}], [\ell_{1}, r_{1}]) \}$ holds.     
\end{proof}

\begin{lemma}\label{lem:m_recover_equality}
    Consider two interval attractors $([p, q], [\ell, r]), ([p^{\prime}, q^{\prime}], [\ell^{\prime}, r^{\prime}])$ in set $\Psi_{\source}$ 
    satisfying $T[p-1..r+1] = T^{\prime}[p^{\prime}-1..r^{\prime}+1]$. 
    Let $([p_{1}, q_{1}], [\ell_{1}, r_{1}])$, 
    $([p_{2}, q_{2}], [\ell_{2}, r_{2}])$, $\ldots$, $([p_{k}, q_{k}], [\ell_{k}, r_{k}])$ ($p_{1} < p_{2} < \ldots < p_{k}$)
    be the interval attractors in the set obtained from function $f_{\recover}(([p, q], [\ell, r]))$ 
    Analogously, 
    $([p^{\prime}_{1}, q^{\prime}_{1}], [\ell^{\prime}_{1}, r^{\prime}_{1}])$, 
    $([p^{\prime}_{2}, q^{\prime}_{2}], [\ell^{\prime}_{2}, r^{\prime}_{2}])$, $\ldots$, $([p^{\prime}_{k^{\prime}}, q^{\prime}_{k^{\prime}}], [\ell^{\prime}_{k^{\prime}}, r^{\prime}_{k^{\prime}}])$ ($p^{\prime}_{1} < p^{\prime}_{2} < \ldots < p^{\prime}_{k^{\prime}}$)
    are defined for interval attractor $([p^{\prime}, q^{\prime}], [\ell^{\prime}, r^{\prime}])$. 
    Here, 
    Lemma~\ref{lem:mRecover_basic_property} shows that 
    there exists an integer $x \in [1, k]$ satisfying 
    $f_{\recover}(([p, q]$, $[\ell, r])) \cap \Psi_{\mRecover} = \{ ([p_{x}, q_{x}], [\ell_{x}, r_{x}]) \}$. 
    Similarly, 
    there exists an integer $y \in [1, k^{\prime}]$ satisfying 
    $f_{\recover}(([p^{\prime}, q^{\prime}]$, $[\ell^{\prime}, r^{\prime}])) \cap \Psi_{\mRecover} = \{ ([p^{\prime}_{y}, q^{\prime}_{y}], [\ell^{\prime}_{y}, r^{\prime}_{y}]) \}$. 
    Then, 
    $T[p_{x}-1..r_{x}+1] = T[p^{\prime}_{y}-1..r^{\prime}_{y}+1]$. 
\end{lemma}
\begin{lemma}
    The proof of Lemma~\ref{lem:m_recover_equality} is as follows. 
    %Let $([p_{1, 1}, q_{1, 1}], [\ell_{1, 1}, r_{1, 1}])$, 
    %$([p_{1, 2}, q_{1, 2}], [\ell_{1, 2}, r_{1, 2}])$, $\ldots$, $([p_{1, k}, q_{1, k}], [\ell_{1, k}, r_{1, k}])$ ($p_{1, 1} < p_{1, 2} < \ldots < p_{1, k}$)
    %be the interval attractors in the set obtained from function $f_{\recover}(([p_{1}, q_{1}], [\ell_{1}, r_{1}]))$. 
    %Analogously, 
    %$([p_{2, 1}, q_{2, 1}], [\ell_{2, 1}, r_{2, 1}])$, 
    %$([p_{2, 2}, q_{2, 2}], [\ell_{2, 2}, r_{2, 2}])$, $\ldots$, $([p_{2, k^{\prime}}, q_{2, k^{\prime}}], [\ell_{2, k^{\prime}}, r_{2, k^{\prime}}])$ 
    %($p_{2, 1} < p_{2, 2} < \ldots < p_{2, k^{\prime}}$) 
    %are defined for interval attractor $([p_{2}, q_{2}], [\ell_{2}, r_{2}])$. 

    \paragraph{Proof of Lemma~\ref{lem:m_recover_equality} for $([p, q], [\ell, r]) \in \Psi_{\preceding}$.}
    $([p^{\prime}, q^{\prime}], [\ell^{\prime}, r^{\prime}]) \in \Psi_{\preceding}$ follows from Lemma~\ref{lem:psi_equality_basic_property}~\ref{enum:psi_equality_basic_property:6}. 
    Since $([p, q], [\ell, r]), ([p^{\prime}, q^{\prime}], [\ell^{\prime}, r^{\prime}]) \in \Psi_{\preceding}$, 
    Lemma~\ref{lem:mRecover_basic_property} shows that $x = k$ and $y = k^{\prime}$. 
    Lemma~\ref{lem:recover_super_property} shows that $T[p_{x}-1..r_{x}+1] = T[p^{\prime}_{y}-1..r^{\prime}_{y}+1]$. 
    Therefore, we obtain 
    $T[p_{x}-1..r_{x}+1] = T[p^{\prime}_{y}-1..r^{\prime}_{y}+1]$. 

    \paragraph{Proof of Lemma~\ref{lem:m_recover_equality} for $([p_{1}, q_{1}], [\ell_{1}, r_{1}]) \not \in \Psi_{\preceding}$.}
    We can prove, $x = y = 1$ 
    and $T[p_{x}-1..r_{x}+1] = T[p^{\prime}_{y}-1..r^{\prime}_{y}+1]$. 
    by the same approach as for $([p, q], [\ell, r]) \in \Psi_{\preceding}$. 
\end{lemma}
    
\subsection{Sampling Interval Attractors}\label{subsec:sampling_subset}
We introduce a subset of set $\Psi_{\RR}$ called \emph{sampling subset} as follows. 
\begin{definition}\label{def:sampling_subset}
A sampling subset $\Psi_{\samp} = \{ ([p_{1}, q_{1}], [\ell_{1}, r_{1}])$, $([p_{2}, q_{2}], [\ell_{2}, r_{2}])$, $\ldots$, 
$([p_{k}, q_{k}], [\ell_{k}, r_{k}]) \}$ for RLSLP $\mathcal{G}^{R}$ 
is defined as a subset of set $\Psi_{\RR}$ 
satisfying the following two conditions: 
\begin{enumerate}[label=\textbf{(\roman*)}]
    \item \label{enum:sampling_subset:1} the strings represented by the interval attractors in the subset $\Psi_{\samp}$ are distinct 
    (i.e., $T[p_{s}-1..r_{s}+1] \neq T[p_{s^{\prime}}-1..r_{s^{\prime}}+1]$ for any pair of 
    two integers $1 \leq s < s^{\prime} \leq k$);
    \item \label{enum:sampling_subset:2} the strings represented by the interval attractors in the subset $\Psi_{\samp}$ are equal to 
    the strings represented by the interval attractors in the subset $\Psi_{\leftmost} \setminus \Psi_{\run}$ 
    (i.e., $\{ T[p_{s}-1..r_{s}+1] \mid s \in [1, k] \} = \{ T[p-1..r+1] \mid ([p, q], [\ell, r]) \in \Psi_{\leftmost} \setminus \Psi_{\run} \}$).    
\end{enumerate}

\end{definition}

As opposed to the twelve subsets introduced in Section~\ref{subsec:IA_subsets}, 
the sampling subset $\Psi_{\samp}$ is not unique. 
The following lemma states properties of the sampling subset $\Psi_{\samp}$. 

\begin{lemma}\label{lem:samp_basic_property}
Consider a sampling subset $\Psi_{\samp}$ of the set $\Psi_{\RR}$. 
Then, the following four statements hold. 
\begin{enumerate}[label=\textbf{(\roman*)}]
\item \label{enum:samp_basic_property:1} $|\Psi_{\samp}| = |\Psi_{\leftmost} \setminus \Psi_{\run}|$; 
\item \label{enum:samp_basic_property:2} for each interval attractor $([p, q], [\ell, r]) \in \Psi_{\RR} \setminus \Psi_{\run}$, 
the sampling subset $\Psi_{\samp}$ contains an interval attractor $([p^{\prime}, q^{\prime}], [\ell^{\prime}, r^{\prime}])$ 
satisfying $T[p-1..r+1] = T[p^{\prime}-1..r^{\prime}-1]$; 
\item \label{enum:samp_basic_property:3} $\Psi_{\samp} \cap \Psi_{\run} = \emptyset$;
\item \label{enum:samp_basic_property:4} $\mathbb{E}[|\Psi_{\samp}|] = O(\delta \log \frac{n \log \sigma}{\delta \log n})$.
\end{enumerate}
\end{lemma}
\begin{proof}
The proof of Lemma~\ref{lem:samp_basic_property} is as follows.

\textbf{Proof of Lemma~\ref{lem:samp_basic_property}(i).} 
Let $d$ be the number of strings in set $\{ T[p-1..r+1] \mid ([p, q], [\ell, r]) \in \Psi_{\samp} \}$ 
(i.e., $d = |\{ T[p-1..r+1] \mid ([p, q], [\ell, r]) \in \Psi_{\samp} \}|$). 
Similarly, 
let $d^{\prime}$ be the number of strings in set $\{ T[p-1..r+1] \mid ([p, q], [\ell, r]) \in \Psi_{\leftmost} \setminus \Psi_{\run} \}$. 
Then, $d = d^{\prime}$ holds 
because $\{ T[p-1..r+1] \mid ([p, q], [\ell, r]) \in \Psi_{\samp} \} = \{ T[p-1..r+1] \mid ([p, q], [\ell, r]) \in \Psi_{\leftmost} \setminus \Psi_{\run} \}$ follows from the definition of the sampling subset $\Psi_{\samp}$. 

We prove $|\Psi_{\samp}| = |\Psi_{\leftmost} \setminus \Psi_{\run}|$. 
$d = |\Psi_{\samp}|$ follows from the definition of the sampling subset $\Psi_{\samp}$. 
$d^{\prime} = |\Psi_{\leftmost} \setminus \Psi_{\run}|$ follows from Lemma~\ref{lem:lm_basic_property}~\ref{enum:lm_basic_property:1}. 
Therefore, $|\Psi_{\samp}| = |\Psi_{\leftmost} \setminus \Psi_{\run}|$ follows from 
$d = d^{\prime}$, $d = |\Psi_{\samp}|$, and $d^{\prime} = |\Psi_{\leftmost} \setminus \Psi_{\run}|$. 

\textbf{Proof of Lemma~\ref{lem:samp_basic_property}(ii).} 
Lemma~\ref{lem:lm_basic_property}~\ref{enum:lm_basic_property:2} shows that 
the subset $\Psi_{\leftmost}$ contains an interval attractor $([p_{1}, q_{1}], [\ell_{1}, r_{1}])$ 
satisfying $T[p_{1}-1..r_{1}+1] = T[p-1..r-1]$. 
Because of $([p, q], [\ell, r]) \not \in \Psi_{\run}$, 
$([p_{1}, q_{1}], [\ell_{1}, r_{1}]) \not \in \Psi_{\run}$ follows from Lemma~\ref{lem:psi_equality_basic_property}~\ref{enum:psi_equality_basic_property:4}. 
The sampling subset $\Psi_{\samp}$ contains an interval attractor $([p^{\prime}, q^{\prime}], [\ell^{\prime}, r^{\prime}])$ 
satisfying $T[p^{\prime}-1..r^{\prime}+1] = T[p_{1}-1..r_{1}-1]$ 
because $([p_{1}, q_{1}], [\ell_{1}, r_{1}]) \in \Psi_{\samp} \setminus \Psi_{\run}$. 
Therefore, Lemma~\ref{lem:samp_basic_property}(ii) holds. 

\textbf{Proof of Lemma~\ref{lem:samp_basic_property}(iii).} 
$\Psi_{\samp} \cap \Psi_{\run} = \emptyset$ holds 
if $([p, q], [\ell, r]) \not \in \Psi_{\run}$ holds for each interval attractor $([p, q], [\ell, r]) \in \Psi_{\samp}$. 
We prove $([p, q], [\ell, r]) \not \in \Psi_{\run}$. 
From the definition of the subset $\Psi_{\samp}$, 
the $\Psi_{\leftmost} \setminus \Psi_{\run}$ contains an interval attractor $([p^{\prime}, q^{\prime}], [\ell^{\prime}, r^{\prime}])$ 
satisfying $T[p-1..r+1] = T[p^{\prime}-1..r^{\prime}-1]$. 
Lemma~\ref{lem:psi_equality_basic_property}~\ref{enum:psi_equality_basic_property:4} shows that 
$([p, q], [\ell, r]) \not \in \Psi_{\run}$ holds because 
$T[p-1..r+1] = T[p^{\prime}-1..r^{\prime}-1]$ and $([p^{\prime}, q^{\prime}], [\ell^{\prime}, r^{\prime}]) \not \in \Psi_{\run}$. 
Therefore, $\Psi_{\samp} \cap \Psi_{\run} = \emptyset$ holds. 

\textbf{Proof of Lemma~\ref{lem:samp_basic_property}(iv).} 
$\mathbb{E}[|\Psi_{\samp}|] = O(\delta \log \frac{n \log \sigma}{\delta \log n})$ 
follows from $|\Psi_{\samp}| = |\Psi_{\leftmost} \setminus \Psi_{\run}|$ (Lemma~\ref{lem:samp_basic_property}~\ref{enum:samp_basic_property:1})  
and $\mathbb{E}[|\Psi_{\leftmost} \setminus \Psi_{\run}|] = O(\delta \log \frac{n \log \sigma}{\delta \log n})$ (Theorem~\ref{theo:RR_Psi_set_size}).
\end{proof}

%\section{OptDynSA}\label{sec:main_components}
\section{Queries on Interval Attractors \texorpdfstring{$\Psi_{\RR}$}{}}\label{sec:other_query}
\renewcommand{\arraystretch}{1.2}
\begin{table}[p]
    \normalsize
    \vspace{-0.5cm}
    \caption{
    Queries on the set $\Psi_{\RR}$ of interval attractors. 
    Here, 
    $H$ is the height of the derivation tree of RLSLP $\mathcal{G}^{R}$;     
    $n$ is the length of input string $T$;
    $([p, q], [\ell, r]) \in \Psi_{\RR}$ is a given interval attractor; 
    $h \in [0, H]$ is an integer satisfying $([p, q], [\ell, r]) \in \Psi_{h}$;
    $\gamma$ is the attractor position of the interval attractor $([p, q], [\ell, r])$;
    $C$ is the associated string of the interval attractor $([p, q], [\ell, r])$; 
    $s$ is an integer in set $\{ 1, 2, \ldots, |\Psi_{\samp}| \}$ for the sampling subset $\Psi_{\samp}$ introduced in Section~\ref{subsec:sampling_subset};
    $[i_{1}, j_{1}] \subseteq [1, n]$ ($i < j$) is a given interval in input string $T$; 
    $[i_{2}, j_{2}] \subseteq [1, n]$ is a given interval in input string $T$; 
    $P \in \Sigma^{+}$ is a given string of length $2$; 
    $b \in [1, n]$ is a position in the suffix array $\SA$ of input string $T$;    
    $m = \max \{ |[i_{2}, j_{2}]|, H, \log n \}$. 
    These queries are detailed in Section~\ref{sec:other_query}. 
    %The RSC and RSS queries in Table~\ref{table:IA_query_result} are detailed in 
    %Section~\ref{sec:RSC_query} and Section~\ref{sec:RSS_query}, respectively. 
    }
    %\vspace{-5mm}    
    \label{table:IA_query_result} 
    \center{
    \scalebox{0.85}{
    \begin{tabular}{l||l|l|l}
 Query & Description & Time complexity & Sections \\  \hline 
 $\levelQ(([p, q], [\ell, r]))$ & Return the level $h$  & $O(H^{2})$ & \ref{subsec:level_query} \\ \hline 
%  & the $h$-th level interval attractors $\Psi_{h}$ &  & \\ \hline 

 $\attrQ(([p, q], [\ell, r]))$ & Return the attractor position $\gamma$ & $O(H^{2})$ & \ref{subsec:attr_pos_query} \\ \hline
 $\clenQ(([p, q], [\ell, r]))$ & Return the length of the associated string $C$ & $O(H^{2})$ & \ref{subsec:C_length_query} \\ \hline
 $\clcpQ(([p, q], [\ell, r]))$ & Return $|\lcp(T[\gamma..r+1], C^{n+1})|$ & $O(H^{2})$ & \ref{subsec:C_lcp_query} \\ \hline
 $\clcsQ(([p, q], [\ell, r]))$ & Return $|\lcs(T[p-1..\gamma-1], C^{n+1})|$ & $O(H^{2})$ & \ref{subsec:C_lcs_query} \\ \hline
 $\CAPQ([i_{1}, j_{1}])$ & Return interval attractor $I_{\capture}(i_{1}, j_{1})$ & $O(H^{2} \log n)$ & \ref{subsec:capture_query} \\ \hline
 $\runQ(([p, q], [\ell, r]))$ & Verify whether $([p, q], [\ell, r]) \in \Psi_{\run}$ or not & $O(H^{2})$ & \ref{subsec:verify_run_query} \\ \hline
 $\precQ(([p, q], [\ell, r]))$ & Verify whether $([p, q], [\ell, r]) \in \Psi_{\preceding}$ or not & $O(H^{2})$ & \ref{subsec:verify_prec_query} \\ \hline
 $\sourceQ(([p, q], [\ell, r]))$ & Verify whether $([p, q], [\ell, r]) \in \Psi_{\source}$ or not & $O(H^{2} \log n)$ & \ref{subsec:verify_source_query} \\ \hline
 $\rsizeQ(([p, q], [\ell, r]))$ & Return $|f_{\recover}([p, q], [\ell, r])|$ if $([p, q], [\ell, r]) \in \Psi_{\source}$ & $O(H^{2})$ & \ref{subsec:rsize_query}   \\ \hline
 \multirow{2}{*}{$\OVQ([i_{2}, j_{2}])$} & Return all the interval attractors in set & expected & \multirow{2}{*}{\ref{subsec:overlap_query}}
 \\ 
& $\{ ([p^{\prime}, q^{\prime}], [\ell^{\prime}, r^{\prime}]) \in \Psi_{\RR} \setminus \Psi_{\run} \mid [p^{\prime}, r^{\prime}] \cap [i_{2}, j_{2}] \neq \emptyset \}$  & $O(m H^{3} \log n)$ &  \\ \hline
 \multirow{1}{*}{$\sampleQ(s)$} & Return the $s$-th interval attractor of $\Psi_{\samp}$ & \multirow{1}{*}{$O(\log n)$} & \multirow{1}{*}{\ref{subsec:sample_query}} \\ \hline
 \multirow{4}{*}{$\BiSQ(P)$} & Return (A) $|\{ x \in [1, n] \mid T[x..n] \prec P \}|$, & \multirow{4}{*}{$O((H + \log n) \log n)$} & \multirow{4}{*}{\ref{subsec:bigram_search_query}} \\ 
  & (B) $|\Occ(T, P)|$, and  & & \\ 
  & (C) an occurrence position of & &   \\ 
  & string $P$ in input string $T$ & &   \\ \hline
 $\BiAQ(i^{\prime})$ & Return string $T[\SA[i^{\prime}]..\SA[i^{\prime}]+1]$ & $O(H \log^{2} n + \log^{3} n)$ & \ref{subsec:bigram_access_query} \\
    \end{tabular} 
    }
    }
\end{table}

This section introduces fourteen queries on the set $\Psi_{\RR}$ of interval attractors, as summarized in Table~\ref{table:IA_query_result}. Along with these queries, we present efficient data structures and algorithms for responding to them. These queries are utilized in SA query, ISA query, and the update of our data structure in subsequent sections.

\subsection{Level Query}\label{subsec:level_query}
For a given interval attractor $([p, q], [\ell, r]) \in \Psi_{\RR}$,
the \emph{level} query $\levelQ(([p, q], [\ell, r]))$ returns the level $h$ of the interval attractor $([p, q], [\ell, r])$. 
The level $h$ can be determined by computing the length of sequence $A(p, r)$ of intervals 
because $h = |A(p, r)| - 1$ always holds. 
This computation of sequence $A(p, r)$ is performed using the data structure for the RR-DAG of RLSLP $\mathcal{G}^{R}$,
which is introduced in Section~\ref{subsec:rrdag}.
The following lemma presents two queries related to the sequence $A(p, r)$.

\begin{lemma}\label{lem:function_running_time}
For an interval $[s, e] \subseteq [1, n]$ ($s < e$) in input string $T$ of length $n$, 
consider sequence $A(s, e) = [s^{0}, e^{0}], [s^{1}, e^{1}], \ldots, [s^{k}, e^{k}]$. 
For each integer $h \in [0, k]$,  
let $u_{h}$ and $u^{\prime}_{h}$ be the two nodes corresponding to the $s^{h}$-th and $e^{h}$-th nonterminals of sequence $S^{h}$, respectively, in the derivation tree of RLSLP $\mathcal{G}^{R}$. 
Using the data structure introduced in Section~\ref{subsec:rrdag} for the RR-DAG of RLSLP $\mathcal{G}^{R}$, 
we can support the following two queries in $O(H^{2})$ time for the interval $[s, e]$: 
\begin{enumerate}[label=\textbf{(\roman*)}]
    \item It computes the integer $k$; 
    \item It computes $2(k+1)$ nodes $u_{0}, u^{\prime}_{0}$, $u_{1}, u^{\prime}_{1}$, $\ldots$, $u_{k}, u^{\prime}_{k}$, 
    and the starting positions of the substrings derived from these nodes on $T$.     
\end{enumerate}
\end{lemma}
\begin{proof}
See Section~\ref{subsubsec:proof_function_running_time}.
\end{proof}

Lemma~\ref{lem:function_running_time} ensures that 
we can compute the length of sequence $A(p, r)$ in $O(H^{2})$ time. 
Therefore, we can answer a given level query in $O(H^{2})$ time using the data structure for the RR-DAG of RLSLP $\mathcal{G}^{R}$. 

%%%%%%%%%%%%%%%%%%%%%%%%%%%%%%%%%%%%%%%%%%%%%%%%
\subsubsection{Proof of Lemma~\ref{lem:function_running_time}}\label{subsubsec:proof_function_running_time}
For each integer $h \in [0, k]$, 
let $T[x^{h}_{s}..y^{h}_{s}]$ and $T[x^{h}_{e}..y^{h}_{e}]$ be the two substrings derived from the $s^{h}$-th and $e^{h}$-th nonterminals of sequence $S^{h}$, respectively. 
The following proposition is used to prove Lemma~\ref{lem:function_running_time}. 

\begin{proposition}\label{prop:verify_I_sync_algorithm}
Consider the sequence $A(s, e) = [s^{0}, e^{0}], [s^{1}, e^{1}], \ldots, [s^{k}, e^{k}]$ of Lemma~\ref{lem:function_running_time}.  
For a given integer $h \in [0, k]$, 
we can support the following four queries in $O(H)$ time 
using (1) the data structure introduced in Section~\ref{subsec:rrdag} for the RR-DAG of RLSLP $\mathcal{G}^{R}$, 
and (2) the two intervals $[x^{h}_{s}, y^{h}_{s}]$ and $[x^{h}_{e}, y^{h}_{e}]$ in input string $T$: 
\begin{enumerate}[label=\textbf{(\roman*)}]
    \item let $g \in [1, |S^{h+1}|]$ be a position of sequence $S^{h+1}$ satisfying $i \in [x^{h+1}_{g}, y^{h+1}_{g}]$ 
    for a given position $i \in [1, n]$ in input string $T$ 
    and the substring $T[x^{h+1}_{g}..y^{h+1}_{g}]$ derived from the $g$-th nonterminal of sequence $S^{h+1}$. 
    Then, it verifies whether the position $g$ satisfies at least one of four conditions (i), (ii), (iii), and (iv) of Definition~\ref{def:f_interval} for function $f_{\interval}(s^{h}, e^{h})$ or not;
    \item it verifies whether $f_{\interval}(s^{h}, e^{h}) \neq \perp$ or not;
    \item     
    if $f_{\interval}(s^{h}, e^{h}) \neq \perp$, 
    then it returns interval $[x^{h+1}_{s}, y^{h+1}_{s}]$ 
    for the substring $T[x^{h+1}_{s}..y^{h+1}_{s}]$ derived from the $s^{h+1}$-th nonterminal of sequence $S^{h+1}$;
    \item 
    if $f_{\interval}(s^{h}, e^{h}) \neq \perp$, 
    then it returns interval $[x^{h+1}_{e}, y^{h+1}_{e}]$ 
    for the substring $T[x^{h+1}_{e}..y^{h+1}_{e}]$ derived from the $e^{h+1}$-th nonterminal of sequence $S^{h+1}$.
\end{enumerate}
\end{proposition}
\begin{proof}
Let $\lambda \in [1, |S^{h+1}|]$ be the smallest position of sequence $S^{h+1}$ satisfying 
$x^{h+1}_{\lambda} \geq x^{h}_{s}$ for the substring $T[x^{h+1}_{\lambda}..y^{h+1}_{\lambda}]$ derived from the $\lambda$-th nonterminal of sequence $S^{h+1}$. 
Similarly, 
let $\lambda^{\prime} \in [1, |S^{h+1}|]$ be the largest position of sequence $S^{h+1}$ satisfying 
$x^{h+1}_{\lambda^{\prime}} \leq x^{h}_{e}$ for the substring $T[x^{h+1}_{\lambda^{\prime}}..y^{h+1}_{\lambda^{\prime}}]$ derived from the $\lambda^{\prime}$-th nonterminal of sequence $S^{h+1}$. 
$\lambda \leq \lambda^{\prime}$ holds because $x^{h}_{s} \leq x^{h}_{e}$.

The following two statements are used to prove Proposition~\ref{prop:verify_I_sync_algorithm}:
\begin{enumerate}[label=\textbf{(\Alph*)}]
    \item $s^{h+1} \in \{ \lambda, \lambda+1 \}$ if $f_{\interval}(s^{h}, e^{h}) \neq \perp$; 
    \item $e^{h+1} \in \{ \lambda^{\prime}-1, \lambda^{\prime} \}$ if $f_{\interval}(s^{h}, e^{h}) \neq \perp$. 
\end{enumerate}

\textbf{Proof of statement (A).}
$\lambda \leq s^{h+1} \leq e^{h+1} \leq \lambda^{\prime}$ follows from the definition of function $f_{\interval}$. 
If $\lambda + 1 \geq \lambda^{\prime}$, 
then $s^{h+1} \in \{ \lambda, \lambda + 1\}$ follows from 
$\lambda \leq s^{h+1}$ and $s^{h+1} \leq \lambda^{\prime} \leq \lambda + 1$. 

Otherwise (i.e., $\lambda + 1 < \lambda^{\prime}$), 
$x^{h}_{s} + 1 \leq x^{h+1}_{\lambda+1} \leq x^{h}_{e} - 1$ holds 
for the substring $T[x^{h+1}_{\lambda+1}..y^{h+1}_{\lambda+1}]$ derived from the $(\lambda+1)$-th nonterminal of sequence $S^{h+1}$. 
In this case, $s^{h+1} \leq \lambda+1$ holds 
because the position $\lambda+1$ satisfies condition (i) of Definition~\ref{def:f_interval} for function $f_{\interval}(s^{h}, e^{h})$. 
Therefore, $s^{h+1} \in \{ \lambda, \lambda + 1\}$ follows from $\lambda \leq s^{h+1}$ and $s^{h+1} \leq \lambda+1$. 

\textbf{Proof of statement (B).}
Statement (B) can be proved using the same approach as for statement (A). 

\textbf{Proof of Proposition~\ref{prop:verify_I_sync_algorithm}(i).}
We verify the position $g$ in the following five phases. 
In the first phase, 
we find the node $u_{g}$ corresponding to the $g$-th nonterminal of sequence $S^{h+1}$ by traversing 
the derivation tree of RLSLP $\mathcal{G}^{R}$. 
This traversal can be executed in $O(H)$ time using the queries of Lemma~\ref{lem:basic_operations_on_dev_tree}. 
In addition, we can compute interval $[x^{h+1}_{g}, y^{h+1}_{g}]$ in the same time 
for the substring $T[x^{h+1}_{g}..y^{h+1}_{g}]$ derived from the $g$-th nonterminal of sequence $S^{h+1}$.

In the second phase, 
we verify whether the position $g$ satisfies condition (i) of Definition~\ref{def:f_interval} or not, 
i.e., we verify whether $x^{h+1}_{g} \in [x^{h}_{s} + 1, x^{h}_{e} - 1]$ or not. 
This verification can be executed in $O(1)$ time. 

In the third phase, 
we verify whether the position $g$ satisfies condition (ii) of Definition~\ref{def:f_interval} or not. 
Sequence $S^{h}$ contains a position $j$ satisfying $x^{h}_{j} = x^{h+1}_{g}$ 
for the substring $T[x^{h}_{j}..y^{h}_{j}]$ derived from the $j$-th nonterminal of sequence $S^{h}$. 
Let $T[x^{h}_{j-1}..y^{h}_{j-1}]$ be the substring derived from the $(j-1)$-th nonterminal of sequence $S^{h}$.
Then, the position $g$ satisfies condition (ii) if at least one of the following two statements: 
(a) $x^{h}_{s} \leq x^{h}_{j} < x^{h}_{e}$ and $|[x^{h}_{j}, y^{h}_{j}]| > \lfloor \mu(h+1) \rfloor$; 
(b) $x^{h}_{s} \leq x^{h}_{j-1} < x^{h}_{e}$ and $|[x^{h}_{j-1}, y^{h}_{j-1}]| > \lfloor \mu(h+1) \rfloor$. 

Similar to the first phase, 
the interval $[x^{h}_{j}, y^{h}_{j}]$ can be computed in $O(H)$ time by traversing 
the derivation tree of RLSLP $\mathcal{G}^{R}$. 
Similarly, 
the interval $[x^{h}_{j-1}, y^{h}_{j-1}]$ can be computed in the same time. 
The integer $\lfloor \mu(h+1) \rfloor$ can be computed in $O(H)$ time. 
Therefore, we can verify whether the position $g$ satisfies condition (ii) or not in $O(H)$ time. 

In the fourth phase, 
we verify whether the position $g$ satisfies condition (iii) of Definition~\ref{def:f_interval} or not. 
Here, the position $g$ satisfies condition (iii) if 
$x^{h+1}_{g} = x^{h}_{e}$ and 
$|[x^{h}_{e}, e]| > \sum_{w = 1}^{h+1} \lfloor \mu(w) \rfloor$. 
$\sum_{w = 1}^{h+1} \lfloor \mu(w) \rfloor$ can be computed in $O(H)$ time. 
Therefore, we can verify whether the position $g$ satisfies condition (iii) or not in $O(H)$ time. 

In the fifth phase, 
we verify whether the position $g$ satisfies condition (iv) of Definition~\ref{def:f_interval} or not. 
Similar to the third phase, 
we can verify whether the position $g$ satisfies condition (iv) or not in $O(H)$ time. 

Therefore, the five phases run in $O(H)$ time in total. 

\textbf{Proof of Proposition~\ref{prop:verify_I_sync_algorithm}(ii).}
From the definition of function $f_{\interval}$ and statement (A), 
at least one of two positions $\lambda$ and $\lambda+1$ in sequence $S^{h+1}$ 
satisfies at least one of four conditions (i), (ii), (iii), and (iv) of Definition~\ref{def:f_interval} for function $f_{\interval}(s^{h}, e^{h})$ 
if and only if $f_{\interval}(s^{h}, e^{h}) \neq \perp$. 
We can verify whether the two positions $\lambda$ and $\lambda+1$ satisfy at least one of the four conditions or not 
using Proposition~\ref{prop:verify_I_sync_algorithm}(i). 
Therefore, we can verify whether $f_{\interval}(s^{h}, e^{h}) \neq \perp$ or not in $O(H)$ time.

\textbf{Proof of Proposition~\ref{prop:verify_I_sync_algorithm}(iii).}
$s^{h+1} \in \{ \lambda, \lambda+1 \}$ follows from statement (A). 
If $s^{h+1} = \lambda$, 
then $[x^{h+1}_{s}, y^{h+1}_{s}] = [x^{h+1}_{\lambda}, y^{h+1}_{\lambda}]$ 
for the substring $T[x^{h+1}_{\lambda}..y^{h+1}_{\lambda}]$ derived from the $\lambda$-th nonterminal of sequence $S^{h+1}$. 
Otherwise, 
$[x^{h+1}_{s}, y^{h+1}_{s}] = [x^{h+1}_{\lambda+1}, y^{h+1}_{\lambda+1}]$ 
for the substring $T[x^{h+1}_{\lambda+1}..y^{h+1}_{\lambda+1}]$ derived from the $(\lambda+1)$-th nonterminal of sequence $S^{h+1}$. 
The two intervals $[x^{h+1}_{\lambda}, y^{h+1}_{\lambda}]$ and $[x^{h+1}_{\lambda+1}, y^{h+1}_{\lambda+1}]$ 
can be computed by traversing the derivation tree of RLSLP $\mathcal{G}^{R}$. 
This traversal can be executed in $O(H)$ time using the queries of Lemma~\ref{lem:basic_operations_on_dev_tree}. 
We can verify whether $s^{h+1} = \lambda$ or not by Proposition~\ref{prop:verify_I_sync_algorithm}(i). 
Therefore, the interval $[x^{h+1}_{s}, y^{h+1}_{s}]$ can be computed in $O(H)$ time. 

\textbf{Proof of Proposition~\ref{prop:verify_I_sync_algorithm}(iv).}
Proposition~\ref{prop:verify_I_sync_algorithm}(iv) can be proved using the same approach as for Proposition~\ref{prop:verify_I_sync_algorithm}(iii). 

\end{proof}

The proof of Lemma~\ref{lem:function_running_time} is as follows. 

\begin{proof}[Proof of Lemma~\ref{lem:function_running_time}]
We compute $2(k+1)$ nodes $u_{0}, u^{\prime}_{0}$, $u_{1}, u^{\prime}_{1}$, $\ldots$, $u_{k}, u^{\prime}_{k}$ 
in a bottom-up approach. 
The two nodes $u_{0}$ and $u^{\prime}_{0}$ can be obtained by traversing the derivation tree of RLSLP $\mathcal{G}^{R}$ 
because $u_{0}$ and $u^{\prime}_{0}$ derive the $s$-th and $e$-th characters of $T$, respectively. 
This traversal can be executed in $O(H)$ time using the queries of Lemma~\ref{lem:basic_operations_on_dev_tree}. 
For each integer $h \in [0, k-1]$, 
we compute two nodes $u_{h+1}$ and $u^{\prime}_{h+1}$ in $O(H)$ time by applying Proposition~\ref{prop:verify_I_sync_algorithm} to the two nodes $u_{h}$ and $u^{\prime}_{h}$. 
Proposition~\ref{prop:verify_I_sync_algorithm}(ii) can verify whether $h \leq k$ or not in $O(H)$ time. 
Therefore, we can answer the two queries of Lemma~\ref{lem:function_running_time} in $O(kH)$ time, 
and $kH = O(H^{2})$.

%We answer the two queries stated in Lemma~\ref{lem:function_running_time} in the following two phases. 
%In the first phase, 
%we compute $2(k+1)$ intervals $[x^{0}_{s}, y^{0}_{s}]$, $[x^{0}_{e}, y^{0}_{e}]$, $[x^{1}_{s}, y^{1}_{s}]$, $[x^{1}_{e}, y^{1}_{e}]$, 
%$\ldots$, $[x^{k}_{s}, y^{k}_{s}]$, $[x^{k}_{e}, y^{k}_{e}]$. 
%The first two intervals $[x^{0}_{s}, y^{0}_{s}]$ and $[x^{0}_{e}, y^{0}_{e}]$ can be computed in $O(1)$ time 
%because $x^{0}_{s} = s$, $y^{0}_{s} = s$, $x^{0}_{e} = e$, and $y^{0}_{e} = e$. 
%The other intervals can be computed in $O(kH)$ time by Proposition~\ref{prop:verify_I_sync_algorithm}. 
%After the first phase, 
%we can answer the first query of Lemma~\ref{lem:function_running_time}. 
%
%In the second phase, 
%we compute $2(k+1)$ triplets 
%$\repr(u_{0})$, $\repr(u^{\prime}_{0})$, 
%$\repr(u_{1})$, $\repr(u^{\prime}_{1})$, $\ldots$, 
%$\repr(u_{k})$, $\repr(u^{\prime}_{k})$. 
%For each integer $h \in [0, k]$, 
%the two triplets $\repr(u_{h})$ and $\repr(u^{\prime}_{h})$ can be obtained by 
%two path queries $\pathQ(x^{h}_{s})$ and $\pathQ(x^{h}_{e})$, respectively. 
%The two path queries can be executed in $O(H)$ time by Theorem~\ref{theo:rr_dag_summary}. 
%After the second phase, 
%we can answer the second query of Lemma~\ref{lem:function_running_time}. 
%
%The two phases take $O(kH)$ in total. 
%$k \leq H$, and hence, we obtain Lemma~\ref{lem:function_running_time}. 

\end{proof}
%%%%%%%%%%%%%%%%%%%%%%%%%%%%%%%%%%%%%%%%%%%%%%%%%%%%%%%%%

\subsection{Attractor Position Query}\label{subsec:attr_pos_query}
For a given interval attractor $([p, q], [\ell, r]) \in \Psi_{\RR}$, 
\emph{attractor position} query $\attrQ(([p, q], [\ell, r]))$ returns the attractor position $\gamma$ of the given interval attractor (i.e., 
$\attrQ(([p, q], [\ell, r])) = \gamma$). 
Consider sequence $A(p, r) = [s^{0}, e^{0}], [s^{1}, e^{1}], \ldots, [s^{k}, e^{k}]$ 
of intervals for an interval attractor $([p, q], [\ell, r]) \in \Psi_{\RR}$, 
and let $u$ be the node corresponding to the $s^{k}$-th nonterminal of sequence $S^{k}$ in the derivation tree of RLSLP $\mathcal{G}^{R}$. 
From the definition of attractor position, 
$\gamma$ is the starting position of the substring derived from the node $u$ on $T$. 
We can compute this starting position $\gamma$ in $O(H^{2})$ time by Lemma~\ref{lem:function_running_time}. 
Therefore, we can answer a given attractor position query runs in $O(H^{2})$ time.

\subsection{C-Length Query}\label{subsec:C_length_query}
\emph{C-length} query $\clenQ(([p, q], [\ell, r]))$ returns the length of the associated string $C$ of a given interval attractor $([p, q], [\ell, r]) \in \Psi_{\RR}$ (i.e., $\clenQ(([p, q], [\ell, r])) = |C|$).
Similar to attractor position query, 
the associated string $C$ can be obtained using sequence $A(p, r) = [s^{0}, e^{0}], [s^{1}, e^{1}], \ldots, [s^{k}, e^{k}]$ of intervals. 
Let $u$ be the node corresponding to the $s^{k}$-th nonterminal of sequence $S^{k}$ in the derivation tree of RLSLP $\mathcal{G}^{R}$. 
Then, $C$ is defined as the substring $T[\gamma..\gamma + |\val(S^{k}[s^{k}])| - 1]$. 
The length of this substring can be computed in $O(H^{2})$ time Lemma~\ref{lem:function_running_time}(ii) and the first query of Lemma~\ref{lem:basic_operations_on_dev_tree}. 
Therefore, C-length query can be answered in $O(H^{2})$ time.

%The algorithm for C-length query consists of the following five steps. 
%\begin{enumerate}[label=\textbf{(\roman*)}]
%    \item Compute the node $u$ and $\gamma$ in $O(H^{2})$ time by Lemma~\ref{lem:function_running_time}(ii). 
%    \item Compute the length $|\val(S^{k}[s^{k}])|$ in $O(1)$ time by the first query of Lemma~\ref{lem:basic_operations_on_dev_tree}. 
%    \item Compute $\sum_{w = 1}^{k+1} \lfloor \mu(w) \rfloor$ in $O(H)$ time. 
%    \item If $|[\gamma, r]| \geq \sum_{w = 1}^{k+1} \lfloor \mu(w) \rfloor$, 
%    return $|\val(S^{k}[s^{k}])|$ as the answer to the given C-length query.  
%    \item Otherwise, return $|[\gamma, r]|$ as the answer to the given C-length query. 
%\end{enumerate}
%
%The above algorithm for C-length query runs in $O(H^{2})$ time.

%%%%%%%%%%%%%%%%%%%%%%%%%%%%%%%%%%%%%%%%%%%%%%%%%%%%%%%%%
\subsection{C-LCP Query}\label{subsec:C_lcp_query}
For a given interval attractor $([p, q], [\ell, r]) \in \Psi_{\RR}$, 
\emph{C-LCP} query $\clcpQ(([p, q]$, $[\ell, r]))$ returns 
the length $K$ of the longest common prefix between two strings $T[\gamma..r+1]$ and $C^{n+1}$ (i.e., $\clcpQ(([p, q]$, $[\ell, r])) = |\lcp(T[\gamma..r+1], C^{n+1})|$). 
Here, $\gamma$ and $C$ are the attractor position and associated string of the given interval attractor, respectively. 
From the definition of the associated string, 
the length $K$ is equal to $\min \{ |[\gamma, r+1]|, |C| + \LCEQ(\gamma, \gamma + |C|) \}$; 
$\gamma$ and $|C|$ can be computed in $O(H^{2})$ time by attractor position and C-length queries. 
LCE query can be answered in $O(H)$ time. 
Therefore, C-LCP query can be answered in $O(H^{2})$ time. 

\subsection{C-LCS Query}\label{subsec:C_lcs_query}
For a given interval attractor $([p, q], [\ell, r]) \in \Psi_{\RR}$, 
C-LCS query $\clcsQ(([p, q]$, $[\ell, r]))$ returns the length $K^{\prime}$ of the longest common suffix between two strings $T[p-1..\gamma-1]$ and $C^{n+1}$ (i.e., $\clcsQ(([p, q], [\ell, r])) = |\lcs(T[p-1..\gamma-1], C^{n+1})|$). 
Here, $C$ is the associated string of the given interval attractor.
From the definition of the associated string, 
the length $K^{\prime}$ is equal to $\min \{ |[p-1, \gamma-1]|, \rLCEQ(\gamma-1, \gamma+|C|-1) \}$. 
Therefore, C-LCS query can be answered in $O(H^{2})$ time using the same approach as C-LCP query.

%The following lemma indicates that 
%the C-LCS query can be answered by the reversed LCE query on input string $T$. 

%\begin{lemma}\label{lem:clcs_query_property}
%For an interval attractor $([p, q], [\ell, r]) \in \Psi_{\RR}$, 
%let $\gamma$ and $C$ be the attractor position and associated string of the interval attractor, respectively. 
%Then, $|\lcs(T[p-1..\gamma-1], C^{n+1})| = \min \{ |[p-1, \gamma-1]|, \rLCEQ(\gamma-1, \gamma+|C|-1) \}$ holds. 
%\end{lemma}
%\begin{proof}
%    Lemma~\ref{lem:clcs_query_property} can be proved using a similar approach as for Lemma~\ref{lem:clcp_query_property}. 
%\end{proof}

%Similar to C-LCP query, 
%a C-LCS query can be computed using attractor position, C-length, and reversed LCE queries with the data structure for the RR-DAG of RLSLP $\mathcal{G}^{R}$ in $O(H^{2})$ time. 

%%%%%%%%%%%%%%%%%%%%%%%%%%%%%%%%%%%%%%%%%%%%%%%%%%%%%%%%
\subsection{Capture Query}\label{subsec:capture_query}
For a given interval $[s, e] \subseteq [1, n]$ of length at least $2$ in the input string $T$, \emph{capture} query $\CAPQ([s, e])$ returns interval attractor $I_{\capture}(s, e) = ([p, q], [\ell, r]) \in \Psi_{\RR}$. 
Here, $I(s^{k}) = ([p, q], [\ell, r])$ follows from the definition of interval attractor 
for sequence $A(s, e) = [s^{0}, e^{0}], [s^{1}, e^{1}], \ldots, [s^{k}, e^{k}]$ of intervals. 
We compute the four positions $p, q, \ell$, and $r$ in the interval attractor $([p, q], [\ell, r])$ 
using the data structure introduced in Section~\ref{subsec:rrdag} for the RR-DAG of RLSLP $\mathcal{G}^{R}$. 
The idea for computing the four positions $p, q, \ell$, and $r$ is as follows. 

\paragraph{Computation of position $r$.}
We use a sequence of $(n-e+1)$ integers $\pi_{1}, \pi_{2}, \ldots, \pi_{n-e+1} \in \{ 0, 1 \}$ for this subsection. 
Each integer $\pi_{i}$ is defined as $0$ if $I_{\capture}(s, e) = I_{\capture}(s, e + i - 1)$; 
otherwise, let $\pi_{i} = 1$. 
The following lemma states three properties of the sequence $\pi_{1}, \pi_{2}, \ldots, \pi_{n-e+1}$.

\begin{lemma}\label{lem:cap_query_r_properties}
The following three statements hold for sequence $\pi_{1}, \pi_{2}, \ldots, \pi_{n-e+1}$. 
\begin{enumerate}[label=\textbf{(\roman*)}]
    \item for an integer $i \in [1, n-e+1]$, $\pi_{i} = 0 \Leftrightarrow e + i - 1 \leq r$;
    \item the sequence $\pi_{1}, \pi_{2}, \ldots, \pi_{n-e+1}$ is a non-decreasing sequence 
    (i.e., $\pi_{1} \leq \pi_{2} \leq \cdots \leq \pi_{n-e+1}$);
    \item $r = r^{\prime} + e - 1$ for 
    the largest integer $r^{\prime}$ in set $[1, n-e+1]$ that 
    satisfies $\pi_{r^{\prime}} = 0$; 
\end{enumerate}
\end{lemma}
\begin{proof}
    The proof of Lemma~\ref{lem:cap_query_r_properties} is as follows.

    \textbf{Proof of Lemma~\ref{lem:cap_query_r_properties}(i).}        
    We prove $\pi_{i} = 0 \Rightarrow e + i - 1 \leq r$. 
    $I_{\capture}(s, e) = I_{\capture}(s, e + i - 1)$ follows from $\pi_{i} = 0$. 
    $I_{\capture}(s, e + i - 1) = ([p, q], [\ell, r])$ follows from 
    $I_{\capture}(s, e) = I_{\capture}(s, e + i - 1)$ and $I_{\capture}(s, e) = ([p, q], [\ell, r])$. 
    Since $I_{\capture}(s, e + i - 1) = ([p, q], [\ell, r])$, 
    $e + i - 1 \in [\ell, r]$ follows from the definition of interval attractor. 
    Therefore, $e + i - 1 \leq r$ holds. 

    Next, 
    we prove $\pi_{i} = 0 \Leftarrow e + i - 1 \leq r$. 
    $e + i - 1 \in [e, r]$ follows from $i \geq 1$ and $e + i - 1 \leq r$. 
    Lemma~\ref{lem:IA_maximal_lemma} shows that 
    $I_{\capture}(s, e) = I_{\capture}(s, e + i - 1)$ 
    because $I_{\capture}(s, e + i - 1) = ([p, q], [\ell, r])$ 
    and $e + i - 1 \in [e, r]$. 
    Therefore, $\pi_{i} = 0$ follows from $I_{\capture}(s, e) = I_{\capture}(s, e + i - 1)$. 
    
    \textbf{Proof of Lemma~\ref{lem:cap_query_r_properties}(ii).}
    Lemma~\ref{lem:cap_query_r_properties}(ii) follows from Lemma~\ref{lem:cap_query_r_properties}(i). 

    \textbf{Proof of Lemma~\ref{lem:cap_query_r_properties}(iii).}
    Lemma~\ref{lem:cap_query_r_properties}(iii) follows from Lemma~\ref{lem:cap_query_r_properties}(i). 

\end{proof}

We can compute each integer $\pi_{i}$ by the following lemma. 
\begin{lemma}\label{lem:verify_I_I}
For two intervals $[s, e]$ $(s < e)$ and $[s^{\prime}, e^{\prime}]$ $(s^{\prime} < e^{\prime})$ in input string $T$, 
we can verify whether $I_{\capture}(s, e) = I_{\capture}(s^{\prime}, e^{\prime})$ or not 
in $O(H^{2})$ time using 
the data structure introduced in Section~\ref{subsec:rrdag} for the RR-DAG of RLSLP $\mathcal{G}^{R}$.
\end{lemma}
\begin{proof}
    Consider two sequences $A(s, e) = [s^{0}, e^{0}], [s^{1}, e^{1}], \ldots, [s^{k}, e^{k}]$ 
    and $A(s^{\prime}, e^{\prime}) = [s^{\prime 0}, e^{\prime 0}]$, $[s^{\prime 1}, e^{\prime 1}]$, $\ldots$, $[s^{\prime k^{\prime}}, e^{\prime k^{\prime}}]$. 
    Let $T[x^{k}_{s}..y^{k}_{s}]$ and $T[x^{\prime k^{\prime}}_{s}..y^{\prime k^{\prime}}_{s}]$ be 
    the two substrings derived from the $s^{k}$-th and $s^{\prime k}$-th nonterminals of sequence $S^{k}$, respectively, in input string $T$. 
    From the definition of interval attractor, 
    $I_{\capture}(s, e) = I_{\capture}(s^{\prime}, e^{\prime})$ if and only if 
    $k = k^{\prime}$ and $x^{k}_{s} = x^{\prime k^{\prime}}_{s}$. 
    The four integers 
    $k, k^{\prime}, x^{k}_{s}$ and $x^{\prime k^{\prime}}_{s}$ can be computed in 
    $O(H^{2})$ time by Lemma~\ref{lem:function_running_time}. 
    Therefore, Lemma~\ref{lem:verify_I_I} holds. 
\end{proof}

Lemma~\ref{lem:cap_query_r_properties}(iii) indicates that 
the position $r$ of the interval attractor $([p, q], [\ell, r])$ can be computed by 
finding the largest integer $r^{\prime}$ in set $[1, n-e+1]$ 
that satisfies $\pi_{r^{\prime}} = 0$. 
The integer $r^{\prime}$ can be found by binary search on the sequence $\pi_{1}, \pi_{2}, \ldots, \pi_{n-e+1}$ 
because Lemma~\ref{lem:cap_query_r_properties}(ii) shows that 
the sequence $\pi_{1}, \pi_{2}, \ldots, \pi_{n-e+1}$ is a non-decreasing sequence. 
This binary search on the sequence $\pi_{1}, \pi_{2}, \ldots, \pi_{n-e+1}$ 
needs to access $O(\log n)$ integers of the sequence. 
Each of the $O(\log n)$ integers can be computed in $O(H^{2})$ time by Lemma~\ref{lem:verify_I_I}. 
Therefore, the position $r$ can be computed in $O(H^{2} \log n)$ time. 

\paragraph{Computing the position $p$.}
We use a sequence of $s$ integers $\tau_{1}, \tau_{2}, \ldots, \tau_{s} \in \{ 0, 1 \}$ for this subsection. 
Each integer $\tau_{i}$ is defined as $0$ if $I_{\capture}(s, e) = I_{\capture}(s-i+1, e)$; 
otherwise, let $\tau_{i} = 1$. 
The following lemma states three properties of the sequence $\tau_{1}, \tau_{2}, \ldots, \tau_{s}$.

\begin{lemma}\label{lem:cap_query_p_properties}
The following three statements hold for sequence $\tau_{1}, \tau_{2}, \ldots, \tau_{s}$. 
\begin{enumerate}[label=\textbf{(\roman*)}]
    \item for an integer $i \in [1, s]$, $\tau_{i} = 0 \Leftrightarrow s-i+1 \geq p$;
    \item the sequence $\tau_{1}, \tau_{2}, \ldots, \tau_{s}$ is a non-decreasing sequence;
    \item $p = s-p^{\prime}+1$ for 
    the largest integer $p^{\prime}$ in set $[1, s]$ that 
    satisfies $\tau_{p^{\prime}} = 0$. 
\end{enumerate}
\end{lemma}
\begin{proof}
    Lemma~\ref{lem:cap_query_p_properties} can be proved using the same approach as for Lemma~\ref{lem:cap_query_r_properties}.
\end{proof}

Lemma~\ref{lem:cap_query_p_properties}(iii) indicates that 
the position $p$ of the interval attractor $([p, q], [\ell, r])$ can be computed by 
finding the smallest integer $p^{\prime}$ in set $[1, s]$ 
that satisfies $\tau_{p^{\prime}} = 0$. 
Similar to the integer $r^{\prime}$, 
the integer $p^{\prime}$ can be found by binary search on the sequence $\tau_{1}, \tau_{2}, \ldots, \tau_{s}$. 
This binary search can be executed in $O(H^{2} \log n)$ time using Lemma~\ref{lem:verify_I_I}. 
Therefore, the position $p$ can be computed in $O(H^{2} \log n)$ time. 

\paragraph{Computing the position $\ell$.}
We use a sequence of $(e-p)$ integers $\pi^{\prime}_{1}, \pi^{\prime}_{2}, \ldots, \pi^{\prime}_{e-p} \in \{ 0, 1 \}$ for this subsection. 
Each integer $\pi^{\prime}_{i}$ is defined as $0$ if $I_{\capture}(p, e) = I_{\capture}(p, e - i + 1)$; 
otherwise, let $\pi^{\prime}_{i} = 1$. 
The following lemma states three properties of the sequence $\pi^{\prime}_{1}, \pi^{\prime}_{2}, \ldots, \pi^{\prime}_{e-p}$.

\begin{lemma}\label{lem:cap_query_ell_properties}
The following three statements hold for sequence $\pi^{\prime}_{1}, \pi^{\prime}_{2}, \ldots, \pi^{\prime}_{e-p}$. 
\begin{enumerate}[label=\textbf{(\roman*)}]
    \item for an integer $i \in [1, e-p]$, $\pi^{\prime}_{i} = 0 \Leftrightarrow e - i + 1 \geq \ell$;
    \item the sequence $\pi^{\prime}_{1}, \pi^{\prime}_{2}, \ldots, \pi^{\prime}_{e-p}$ is a non-decreasing sequence;
    \item $\ell = e - \ell^{\prime} + 1$ for 
    the largest integer $\ell^{\prime}$ in set $[1, e-p]$ that 
    satisfies $\pi^{\prime}_{\ell^{\prime}} = 0$; 
\end{enumerate}
\end{lemma}
\begin{proof}
    The proof of Lemma~\ref{lem:cap_query_ell_properties} is as follows.

    \textbf{Proof of Lemma~\ref{lem:cap_query_ell_properties}(i).}        
    We prove $\pi^{\prime}_{i} = 0 \Rightarrow e - i + 1 \geq \ell$. 
    $I_{\capture}(p, e) = I_{\capture}(p, e - i + 1)$ follows from $\pi^{\prime}_{i} = 0$. 
    Since $I_{\capture}(s, e) = ([p, q], [\ell, r])$, 
    $I_{\capture}(p, e) = ([p, q], [\ell, r])$ 
    follows from Lemma~\ref{lem:IA_maximal_lemma}.
    $I_{\capture}(p, e - i + 1) = ([p, q], [\ell, r])$ follows from 
    $I_{\capture}(p, e) = I_{\capture}(p, e - i + 1)$ and $I_{\capture}(p, e) = ([p, q], [\ell, r])$. 
    Since $I_{\capture}(p, e - i + 1) = ([p, q], [\ell, r])$, 
    $e - i + 1 \in [\ell, r]$ follows from the definition of interval attractor. 
    Therefore, $e - i + 1 \geq \ell$ holds. 

    Next, 
    we prove $\pi^{\prime}_{i} = 0 \Leftarrow e - i + 1 \geq \ell$. 
    $e - i + 1 \in [\ell, r]$ follows from 
    $e \in [\ell, r]$, 
    $i \geq 1$, and $e - i + 1 \geq \ell$. 
    Lemma~\ref{lem:IA_maximal_lemma} shows that 
    $I_{\capture}(p, e) = I_{\capture}(p, e - i + 1)$ 
    because $I_{\capture}(p, e) = ([p, q], [\ell, r])$ 
    and $e - i + 1 \in [\ell, r]$. 
    Therefore, $\pi^{\prime}_{i} = 0$ follows from $I_{\capture}(p, e) = I_{\capture}(p, e - i + 1)$. 

    \textbf{Proof of Lemma~\ref{lem:cap_query_ell_properties}(ii).}        
    Lemma~\ref{lem:cap_query_ell_properties}(ii) follows from Lemma~\ref{lem:cap_query_ell_properties}(i).

    \textbf{Proof of Lemma~\ref{lem:cap_query_ell_properties}(iii).}        
    Lemma~\ref{lem:cap_query_ell_properties}(iii) follows from Lemma~\ref{lem:cap_query_ell_properties}(i).
\end{proof}    

Lemma~\ref{lem:cap_query_ell_properties}(iii) indicates that 
the position $\ell$ of the interval attractor $([p, q], [\ell, r])$ can be computed by 
finding the smallest integer $\ell^{\prime}$ in set $[1, e-p]$ 
that satisfies $\pi^{\prime}_{\ell^{\prime}} = 0$. 
Similar to the integer $r^{\prime}$, 
the integer $\ell^{\prime}$ can be found by binary search on the sequence $\pi^{\prime}_{1}, \pi^{\prime}_{2}, \ldots, \pi^{\prime}_{e-p}$. 
This binary search can be executed in $O(H^{2} \log n)$ time using Lemma~\ref{lem:verify_I_I} 
after computing the position $p$. 
Therefore, the position $\ell$ can be computed in $O(H^{2} \log n)$ time.
    
\paragraph{Computing the position $q$.}
We use a sequence of $r-s$ integers $\tau^{\prime}_{1}, \tau^{\prime}_{2}, \ldots, \tau^{\prime}_{r-s} \in \{ 0, 1 \}$ for this subsection. 
Each integer $\tau^{\prime}_{i}$ is defined as $0$ if $I_{\capture}(s, r) = I_{\capture}(s+i-1, r)$; 
otherwise, let $\tau^{\prime}_{i} = 1$. 
The following lemma states three properties of the sequence $\tau^{\prime}_{1}, \tau^{\prime}_{2}, \ldots, \tau^{\prime}_{r-s}$.

\begin{lemma}\label{lem:cap_query_q_properties}
The following three statements hold for sequence $\tau^{\prime}_{1}, \tau^{\prime}_{2}, \ldots, \tau^{\prime}_{r-s}$. 
\begin{enumerate}[label=\textbf{(\roman*)}]
    \item for an integer $i \in [1, r-s]$, $\tau^{\prime}_{i} = 0 \Leftrightarrow s+i-1 \geq q$;
    \item the sequence $\tau^{\prime}_{1}, \tau^{\prime}_{2}, \ldots, \tau^{\prime}_{r-s}$ is a non-decreasing sequence;
    \item $q = s+q^{\prime}-1$ for 
    the largest integer $q^{\prime}$ in set $[1, r-s]$ that 
    satisfies $\tau^{\prime}_{q^{\prime}} = 0$. 
\end{enumerate}
\end{lemma}
\begin{proof}
    Lemma~\ref{lem:cap_query_q_properties} can be proved using the same approach as for Lemma~\ref{lem:cap_query_ell_properties}.
\end{proof}

Lemma~\ref{lem:cap_query_q_properties}(iii) indicates that 
the position $q$ of the interval attractor $([p, q], [\ell, r])$ can be computed by 
finding the largest integer $q^{\prime}$ in set $[1, r-s]$ 
that satisfies $\tau^{\prime}_{q^{\prime}} = 0$. 
Similar to the integer $r^{\prime}$, 
the integer $q^{\prime}$ can be found by binary search on the sequence $\tau^{\prime}_{1}, \tau^{\prime}_{2}, \ldots, \tau^{\prime}_{r-s}$. 
This binary search can be executed in $O(H^{2} \log n)$ time using Lemma~\ref{lem:verify_I_I} 
after computing the position $r$. 
Therefore, the position $q$ can be computed in $O(H^{2} \log n)$ time. 

We showed that 
we can compute the four positions $p, q, \ell$, and $r$ in the interval attractor $([p, q], [\ell, r])$ 
in $O(H^{2} \log n)$ time. 
Therefore, a given capture query can be answered in $O(H^{2} \log n)$ time. 

%%%%%%%%%%%%%%%%%%%%%%%%%%%%%%%%%%%%%%%%%%%%%%%%%%%%%%%%%
\subsection{Verify-Run Query}\label{subsec:verify_run_query}
For a given interval attractor $([p, q], [\ell, r]) \in \Psi_{\RR}$, 
\emph{verify-run} query $\runQ(([p, q], [\ell, r]))$ verifies whether $([p, q], [\ell, r]) \in \Psi_{\run}$ or not. 
If $([p, q], [\ell, r]) \in \Psi_{\run}$, then the verify-run query returns a boolean value, True; 
otherwise, this query returns False. 
From the definition of set $\Psi_{\run}$, 
$([p, q], [\ell, r]) \in \Psi_{\run}$ holds if and only if 
the given interval attractor satisfies the following two conditions: 
\begin{enumerate}[label=\textbf{(\roman*)}]
\item $\clcsQ(([p, q], [\ell, r])) = |[p-1, \gamma-1]|$ for the attractor position $\gamma$ obtained from attractor position query $\attrQ(([p, q], [\ell, r]))$; 
\item $\clcpQ(([p, q], [\ell, r])) > 1 + \sum_{w = 1}^{h+3} \lfloor \mu(w) \rfloor$ 
for the integer $h$ obtained from level query $\levelQ(([p$, $q], [\ell, r]))$.
\end{enumerate}

The C-LCP, C-LCS, attractor position, and level queries can be answered in $O(H^{2})$ time using the data structure for the RR-DAG of RLSLP $\mathcal{G}^{R}$. 
Therefore, the total computation time of verify-run query is $O(H^{2})$. 

%%%%%%%%%%%%%%%%%%%%%%%%%%%%%%%%%%%%%%%%%%%%%%%%%%%%%%%%%
\subsection{Verify-Prec Query}\label{subsec:verify_prec_query}
For a given interval attractor $([p, q], [\ell, r]) \in \Psi_{\RR}$, 
\emph{verify-prec} query $\precQ(([p, q], [\ell, r]))$ verifies whether $([p, q], [\ell, r]) \in \Psi_{\preceding}$ or not.
If $([p, q], [\ell, r]) \in \Psi_{\preceding}$, then the verify-prec query returns a boolean value, True; 
otherwise, this query returns False. 
Let $C$ and $\gamma$ be the associated string and attractor position of the interval attractor $([p, q], [\ell, r])$, respectively. 
Let $K$ be the length of the longest common prefix between two strings $T[\gamma..r+1]$ and $C^{n+1}$. 
From the definition of set $\Psi_{\preceding}$,  
$([p, q], [\ell, r]) \in \Psi_{\preceding}$ holds if and only if 
either of the following two conditions is satisfied: 
\begin{enumerate}[label=\textbf{(\roman*)}]
\item $K < |[\gamma, r+1]|$ and $T[\gamma + K] < T[\gamma + (K \mod |C|)]$;
\item $K = |[\gamma, r+1]|$.
\end{enumerate}
$\gamma$, $|C|$, and $K$ can be computed in $O(H^{2})$ time by attractor position, C-length, and C-LCP queries. 
We can access the two characters $T[\gamma + K]$ and $T[\gamma + (K \mod |C|)]$ in $O(H)$ time by random access query. 
Therefore, the total computation time of verify-prec query is $O(H^{2})$.

%$C[1 + (K \mod |C|)] = T[\gamma + (K \mod |C|)]$ holds 
%because Lemma~\ref{lem:C_prefix_property} shows that 
%$C$ is a prefix of string $T[\gamma..r]$. 
%Therefore, we can answer a given verify-prec query by computing the length $K$ 
%and two characters $T[\gamma + K]$ and $T[\gamma + (K \mod |C|)]$. 
%
%We answer verify-prec query $\precQ(([p, q], [\ell, r]))$ in the following four steps: 
%\begin{enumerate}[label=\textbf{(\Alph*)}]
%\item compute $\gamma$ and $|C|$ by  attractor position query $\attrQ(([p, q], [\ell, r]))$ and 
%C-length query $\clenQ(([p, q]$, $[\ell, r]))$, respectively;
%\item compute the length $K$ by C-LCP query $\clcpQ(([p, q], [\ell, r]))$;
%\item compute two characters $T[\gamma + K]$ and $T[\gamma + (K \mod |C|)]$ 
%by two random access queries $\RAQ(\gamma + K)$ and $\RAQ(\gamma + (K \mod |C|))$, respectively; 
%\item verify $([p, q], [\ell, r]) \in \Psi_{\preceding}$ or not using the length $K$ and two characters $T[\gamma + K]$ and $T[\gamma + (K \mod |C|)]$. 
%\end{enumerate}
%
%The four steps take $O(H^{2})$ time in total using the data structure for the RR-DAG of RLSLP $\mathcal{G}^{R}$. 
%Therefore, the total computation time of verify-prec query is $O(H^{2})$.

%%%%%%%%%%%%%%%%%%%%%%%%%%%%%%%%%%%%%%%%%%%%%%%%%%%%%%%%%
\subsection{Verify-Source Query}\label{subsec:verify_source_query}
For a given interval attractor $([p, q], [\ell, r]) \in \Psi_{\RR}$, 
\emph{verify-source} query $\sourceQ(([p, q], [\ell, r]))$ verifies whether $([p, q], [\ell, r]) \in \Psi_{\source}$ or not. 
This verification can be executed as follows. 
Let $h, \gamma$, and $C$ be the level, attractor position, and associated string of the interval attractor $([p, q], [\ell, r])$. 
Consider an interval attractor $([p^{\prime}, q^{\prime}], [\ell^{\prime}, r^{\prime}]) \in \Psi_{\RR}$ 
satisfying $h^{\prime} = h$ and $\gamma^{\prime} = \gamma + |C|$ for the level $h^{\prime}$ and attractor position $\gamma^{\prime}$ of interval attractor $([p^{\prime}, q^{\prime}], [\ell^{\prime}, r^{\prime}])$. 
From the definition of set $\Psi_{\source}$,  
$([p, q], [\ell, r]) \in \Psi_{\source}$ holds if and only if 
the following two conditions are satisfied: 
\begin{itemize}
\item $([p, q], [\ell, r]) \not \in \Psi_{\run}$; 
\item the interval attractor $([p^{\prime}, q^{\prime}], [\ell^{\prime}, r^{\prime}])$ exists, 
and $([p^{\prime}, q^{\prime}], [\ell^{\prime}, r^{\prime}]) \in \Psi_{\run} \cap \Psi_{\centerset}(C)$.
\end{itemize}

If $([p, q], [\ell, r]) \in \Psi_{\source}$, 
then the following lemma shows that the interval attractor $([p^{\prime}, q^{\prime}], [\ell^{\prime}, r^{\prime}])$ can be computed by capture query. 
\begin{lemma}\label{lem:source_query_succinct_property}
If $([p, q], [\ell, r]) \in \Psi_{\source}$, 
then the interval attractor $([p^{\prime}, q^{\prime}], [\ell^{\prime}, r^{\prime}])$ exists, 
and $I_{\capture}(q+1, r) = ([p^{\prime}, q^{\prime}], [\ell^{\prime}, r^{\prime}])$. 
\end{lemma}
\begin{proof}
    Lemma~\ref{lem:recover_basic_property}~\ref{enum:recover_basic_property:2} 
    and Lemma~\ref{lem:recover_basic_property}~\ref{enum:recover_basic_property:3} show that 
    $p^{\prime} = q + 1$ and $r^{\prime} = r$. 
    In this case, 
    Lemma~\ref{lem:IA_maximal_lemma} shows that $I_{\capture}(q+1, r) = ([p^{\prime}, q^{\prime}], [\ell^{\prime}, r^{\prime}])$. 
\end{proof}

We need to verify whether the interval attractor $([p^{\prime}, q^{\prime}], [\ell^{\prime}, r^{\prime}])$ is contained in subset $\Psi_{\centerset}(C)$ or not 
for answering the verify-source query. 
The next lemma can be used for this verification. 
\begin{lemma}\label{lem:verify_centerset}
For an interval attractor $([p, q], [\ell, r]) \in \Psi_{\RR}$, 
let $C$ be the associated string of the interval attractor. 
Then, we can verify whether a given interval attractor $([p^{\prime}, q^{\prime}], [\ell^{\prime}, r^{\prime}])$ 
is contained in subset $\Psi_{\centerset}(C)$ or not in $O(H^{2})$ time 
using the interval attractor $([p, q], [\ell, r])$ 
and data structure for the RR-DAG of RLSLP $\mathcal{G}^{R}$, which is introduced in Section~\ref{subsubsec:rrdag_ds}. 
\end{lemma}
\begin{proof}
$T[\gamma..\gamma + |C| - 1] = C$ follows from the definition of the assoicated string. 
for the attractor position $\gamma$ of the interval attractor $([p, q], [\ell, r])$. 
Similarly, we obtain $T[\gamma^{\prime}..\gamma^{\prime} + |C^{\prime}| - 1] = C^{\prime}$ 
for the attractor position $\gamma^{\prime}$ and associated string $C^{\prime}$ of the interval attractor $([p^{\prime}, q^{\prime}], [\ell^{\prime}, r^{\prime}])$. 
Therefore, 
$C = C^{\prime}$ holds if and only if 
$|C| = |C^{\prime}|$ and $\LCEQ(\gamma, \gamma^{\prime}) \geq |C|$. 

The lengths of the two associated strings $C$ and $C^{\prime}$ 
can be computed in $O(H^{2})$ time by two C-length queries $\clenQ(([p, q], [\ell, r]))$ and $\clenQ(([p^{\prime}, q^{\prime}], [\ell^{\prime}, r^{\prime}]))$, 
respectively. 
The two attractor positions $\gamma$ and $\gamma^{\prime}$ can be computed in 
$O(H^{2})$ time by two attractor position queries $\attrQ(([p, q], [\ell, r]))$ and $\attrQ(([p^{\prime}, q^{\prime}], [\ell^{\prime}, r^{\prime}]))$, 
respectively. 
LCE query can be executed in $O(H)$ time (Theorem~\ref{theo:rr_dag_summary}). 
Therefore, 
we can verify whether $([p^{\prime}, q^{\prime}], [\ell^{\prime}, r^{\prime}]) \in \Psi_{\centerset}(C)$ 
or not in $O(H^{2})$ time. 
\end{proof}

With the data structure for the RR-DAG of RLSLP $\mathcal{G}^{R}$, 
the algorithm for the verify-source query consists of the following nine steps. 
\begin{enumerate}[label=\textbf{(\roman*)}]
    \item compute interval attractor $I_{\capture}(q+1, r) = ([p^{\prime\prime}, q^{\prime\prime}], [\ell^{\prime\prime}, r^{\prime\prime}])$ by capture query $\CAPQ(q+1, r)$; 
    \item compute the level $h$ and attractor position $\gamma$ of interval attractor $([p, q], [\ell, r])$ by level-query $\levelQ(([p, q], [\ell, r]))$ and attractor position query $\attrQ(([p, q], [\ell, r]))$, respectively;
    \item compute the level $h^{\prime\prime}$ and attractor position $\gamma^{\prime\prime}$ of 
    interval attractor $([p^{\prime\prime}, q^{\prime\prime}], [\ell^{\prime\prime}, r^{\prime\prime}])$ by level-query $\levelQ(([p^{\prime\prime}, q^{\prime\prime}], [\ell^{\prime\prime}, r^{\prime\prime}]))$ and attractor position query $\attrQ(([p^{\prime\prime}, q^{\prime\prime}], [\ell^{\prime\prime}, r^{\prime\prime}]))$, respectively;
    \item compute the length $|C|$ of the associated string of interval attractor $([p, q], [\ell, r])$ 
    by C-length query $\clenQ(([p, q], [\ell, r]))$;     
    \item verify whether $([p^{\prime\prime}, q^{\prime\prime}], [\ell^{\prime\prime}, r^{\prime\prime}]) = ([p^{\prime}, q^{\prime}], [\ell^{\prime}, r^{\prime}])$ or not using five integers $\gamma$, $\gamma^{\prime}$, $|C|$, $h$, and $h^{\prime}$. If $([p^{\prime\prime}, q^{\prime\prime}], [\ell^{\prime\prime}, r^{\prime\prime}]) \neq ([p^{\prime}, q^{\prime}], [\ell^{\prime}, r^{\prime}])$, then return the boolean value False 
    as the answer to the given verify-source query;     
    \item Verify whether $([p, q], [\ell, r]) \not \in \Psi_{\run}$ or not by verify-run query $\runQ(([p, q], [\ell, r]))$; 
    \item Verify whether $([p^{\prime}, q^{\prime}], [\ell^{\prime}, r^{\prime}]) \in \Psi_{\run}$ or not by verify-run query 
    $\runQ(([p^{\prime}, q^{\prime}], [\ell^{\prime}, r^{\prime}]))$; 
    \item Verify whether $([p^{\prime}, q^{\prime}], [\ell^{\prime}, r^{\prime}]) \in \Psi_{\centerset}(C)$ or not by Lemma~\ref{lem:verify_centerset}; 
    \item Return the boolean value True as the answer to the given verify-source query 
    if $([p, q], [\ell, r]) \not \in \Psi_{\run}$ and $([p^{\prime}, q^{\prime}], [\ell^{\prime}, r^{\prime}]) \in \Psi_{\run} \cap \Psi_{\centerset}(C)$; 
    otherwise, return the boolean value False as the answer to the given verify-source query.
\end{enumerate}

Step (i) takes $O(H^{2} \log n)$ time. 
The other steps take $O(H^{2})$ time. 
Therefore, we can answer a given verify-source query in $O(H^{2} \log n)$ time.

%%%%%%%%%%%%%%%%%%%%%%%%%%%%%%%%%%%%%%%%%%%%%%%%%%%%%%%%%
\subsection{R-Size Query}\label{subsec:rsize_query}
For a given interval attractor $([p, q], [\ell, r]) \in \Psi_{\source}$, 
\emph{r-size} query $\rsizeQ(([p, q], [\ell, r]))$ returns the number of the interval attractors in the set obtained by function $f_{\recover}(([p, q], [\ell, r]))$ (i.e., $\rsizeQ(([p, q], [\ell, r])) = |f_{\recover}(([p, q], [\ell, r]))|$). 
The following lemma states the number of interval attractors obtained from the function $f_{\recover}(([p, q], [\ell, r]))$. 

\begin{lemma}\label{lem:recover_size_property}
For an interval attractor $([p, q], [\ell, r]) \in \Psi_{\source}$, 
let (A) $\gamma$ be the attractor position of the interval attractor, 
(B) $C$ be the associated string of the interval attractor, 
(C) $h$ be an integer satisfying $([p, q], [\ell, r]) \in \Psi_{h}$, 
and (D) $K$ be the length of the longest common prefix between two strings $T[\gamma..r+1]$ and $C^{n+1}$~(i.e., $K = |\lcp(T[\gamma..r], C^{n+1})|$).
Then, $|f_{\recover}(([p, q], [\ell, r]))| = \lfloor \frac{K - (2 + \sum_{w = 1}^{h+3} \lfloor \mu(w) \rfloor)}{|C|} \rfloor$. 
\end{lemma}
\begin{proof}
$|f_{\recover}(([p, q], [\ell, r]))| = \lfloor \frac{K - (2 + \sum_{w = 1}^{h+3} \lfloor \mu(w) \rfloor)}{|C|} \rfloor$ follows from 
the definition of the function $f_{\recover}$. 
\end{proof}

Lemma~\ref{lem:recover_size_property} indicates that 
we can compute $|f_{\recover}(([p, q], [\ell, r]))|$ using the following three queries: 
(i) attractor position query $\attrQ(([p, q], [\ell, r]))$, 
(ii) C-length query $\clenQ(([p, q], [\ell, r]))$, 
and (iii) C-LCP query $\clcsQ(([p, q], [\ell, r]))$.
These three queries can be answered in $O(H^{2})$ time using 
the data structure for the RR-DAG of RLSLP $\mathcal{G}^{R}$. 
Therefore, we can answer a given r-size query in $O(H^{2})$ time.

\subsection{Overlap Query}\label{subsec:overlap_query}
For a given interval $[i, j] \subseteq [1, n]$ in input string $T$, 
\emph{overlap} query $\OVQ([i, j])$ returns a set of non-periodic interval attractors; 
each interval attractor $([p, q], [\ell, r])$ in this set satisfies the condition that the intervals $[p, r]$ and $[i, j]$ overlap, meaning $[p, r] \cap [i, j] \neq \emptyset$. 
Formally, the overlap query returns set $\{ ([p, q], [\ell, r]) \in \Psi_{\RR} \setminus \Psi_{\run} \mid [p, r] \cap [i, j] \neq \emptyset \}$. 

%such that each non-periodic interval attractor satisfies 
%the condition that the intervals $[p, r]$ and $[i, j]$ overlap, meaning $[p, r] \cap [i, j] \neq \emptyset$. 
%interval attractors $([p, q], [\ell, r]) \in \Psi_{\RR} \setminus \Psi_{\run}$; 

For answering overlap query $\OVQ([i, j])$, 
we introduce a subset $\Psi_{\OVQ, C}$ of set $\Psi_{\RR}$, which is used in this subsection. 
The subset $\Psi_{\OVQ, C}$ consists of interval attractors such that 
each interval attractor $([p, q], [\ell, r])$ satisfies 
$q \leq i-1$ and $r \geq i+1$ 
(i.e., $\Psi_{\OVQ, C} = \{ ([p, q], [\ell, r]) \in \Psi_{\RR} \mid (q \leq i-1) \land (r \geq i+1) \}$).

The following lemma states the relationship between overlap query and the subset $\Psi_{\OVQ, C}$. 
\begin{lemma}\label{lem:RB_OVQ_and_Subset}
    For an interval $[i, j] \subseteq [1, n]$ in input string $T$,
    let $\Psi_{\OVQ, L} = \{ I_{\capture}(x, i) \mid x \in [1, i-1] \}$ 
    and $\Psi_{\OVQ, R} = \bigcup_{i^{\prime} = i}^{j} \{ I_{\capture}(i^{\prime}, j^{\prime}) \mid j^{\prime} \in [i^{\prime}+1, n] \}$. 
    Then, 
    $\OVQ([i, j]) = (\Psi_{\OVQ, C} \cap \Psi_{\OVQ, L} \cap \Psi_{\OVQ, R}) \setminus \Psi_{\run}$ holds. 
\end{lemma}
\begin{proof}
    Lemma~\ref{lem:RB_OVQ_and_Subset} follows from the following four statements: 
\begin{enumerate}[label=\textbf{(\roman*)}]    
    \item $\OVQ([i, j]) \subseteq (\Psi_{\OVQ, C} \cap \Psi_{\OVQ, L} \cap \Psi_{\OVQ, R}) \setminus \Psi_{\run}$;
    \item $\Psi_{\OVQ, C} \setminus \Psi_{\run} \subseteq \OVQ([i, j])$; 
    \item $\Psi_{\OVQ, L} \setminus \Psi_{\run} \subseteq \OVQ([i, j])$; 
    \item $\Psi_{\OVQ, R} \setminus \Psi_{\run} \subseteq \OVQ([i, j])$.
\end{enumerate} 
    
    \textbf{Proof of statement (i).}
    Consider an interval attractor $([p, q], [\ell, r]) \in \OVQ([i, j])$. 
    Then, $[p, r] \cap [i, j] \neq \emptyset$ and $([p, q], [\ell, r]) \not \in \Psi_{\run}$ follow from overlap query. 
    We prove $([p, q], [\ell, r]) \in \Psi_{\OVQ, C} \cup \Psi_{\OVQ, L} \cup \Psi_{\OVQ, R}$.
    Because of $[p, r] \cap [i, j] \neq \emptyset$, 
    the interval attractor $([p, q], [\ell, r])$ satisfies at least one of the following four conditions: 
\begin{enumerate}[label=\textbf{(\Alph*)}]    
    \item $i \in [\ell, r]$; 
    \item $i \in [q+1, \ell-1]$; 
    \item $i \in [p, q]$;
    \item $p \in [i, j]$.     
\end{enumerate} 
    
    For condition (A), 
    $I_{\capture}(p, i) = ([p, q], [\ell, r])$ follows from Lemma~\ref{lem:IA_maximal_lemma}. 
    Therefore, 
    $([p, q], [\ell, r]) \in \Psi_{\OVQ, L}$ follows from 
    $I_{\capture}(p, i) = ([p, q], [\ell, r])$ and $I_{\capture}(p, i) \in \Psi_{\OVQ, L}$. 
        
    For condition (B), 
    $q \leq i-1$ and $r \geq i+1$ hold. 
    The two inequalities indicates that $([p, q], [\ell, r]) \in \Psi_{\OVQ, C}$. 

    For condition (C), 
    $I_{\capture}(i, r) = ([p, q], [\ell, r])$ follows from Lemma~\ref{lem:IA_maximal_lemma}. 
    Therefore, $([p, q], [\ell, r]) \in \Psi_{\OVQ, R}$ 
    follows from $I_{\capture}(i, r) = ([p, q], [\ell, r])$ and $I_{\capture}(i, r) \in \Psi_{\OVQ, R}$. 

    For condition (D), 
    $I_{\capture}(p, r) = ([p, q], [\ell, r])$ follows from Lemma~\ref{lem:IA_maximal_lemma}. 
    $I_{\capture}(p, r) \in \Psi_{\OVQ, R}$ because 
    $p \in [i, j]$ and $r \in [p+1, n]$.     
    Therefore, $([p, q], [\ell, r]) \in \Psi_{\OVQ, R}$ follows from 
    $I_{\capture}(p, r) = ([p, q], [\ell, r])$ and $I_{\capture}(p, r) \in \Psi_{\OVQ, R}$. 
    
    We proved $([p, q], [\ell, r]) \in \Psi_{\OVQ, C} \cup \Psi_{\OVQ, L} \cup \Psi_{\OVQ, R}$. 
    $\OVQ([i, j]) \subseteq (\Psi_{\OVQ, C} \cup \Psi_{\OVQ, L} \cup \Psi_{\OVQ, R}) \setminus \Psi_{\run}$ follows from 
    $([p, q], [\ell, r]) \in \Psi_{\OVQ, C} \cup \Psi_{\OVQ, L} \cup \Psi_{\OVQ, R}$ and 
    $([p, q], [\ell, r]) \not \in \Psi_{\run}$. 
        
    \textbf{Proof of statement (ii).}
    Consider an interval attractor $([p, q], [\ell, r]) \in \Psi_{\OVQ, C} \setminus \Psi_{\run}$. 
    $q \leq i-1$ and $r \geq i+1$ follow from $([p, q], [\ell, r]) \in \Psi_{\OVQ, C}$. 
    $[p, r] \cap [i, j] \neq \emptyset$ follows from $p \leq q$, $q \leq i-1$, and $r \geq i+1$. 
    $([p, q], [\ell, r]) \in \OVQ([i, j])$ follows from $[p, r] \cap [i, j] \neq \emptyset$ and 
    $([p, q], [\ell, r]) \not \in \Psi_{\run}$. 
    Therefore, $\Psi_{\OVQ, C} \setminus \Psi_{\run} \subseteq \OVQ([i, j])$ holds. 
    
    \textbf{Proof of statement (iii).}
    Consider an interval attractor $([p, q], [\ell, r]) \in \Psi_{\OVQ, L} \setminus \Psi_{\run}$. 
    Since $([p, q], [\ell, r]) \in \Psi_{\OVQ, L}$, 
    there exists an integer $x \in [1, i-1]$ satisfying $I_{\capture}(x, i) = ([p, q], [\ell, r])$. 
    $x \in [p, q]$ and $i \in [\ell, r]$ follow from the definition of interval attractor.
    $i \in [p, r]$ follows from $p \leq \ell$ and $i \in [\ell, r]$. 
    $([p, q], [\ell, r]) \in \OVQ([i, j])$ follows from $i \in [p, r]$ and $([p, q], [\ell, r]) \not \in \Psi_{\run}$. 
    Therefore, $\Psi_{\OVQ, L} \setminus \Psi_{\run} \subseteq \OVQ([i, j])$ holds. 
        
    \textbf{Proof of statement (iv).}
    Consider an interval attractor $([p, q], [\ell, r]) \in \Psi_{\OVQ, R} \setminus \Psi_{\run}$. 
    Since $([p, q], [\ell, r]) \in \Psi_{\OVQ, R}$, 
    there exists a pair of two integers $x \in [i, j]$ and $y \in [x+1, n]$ 
    satisfying $I_{\capture}(x, y) = ([p, q], [\ell, r])$. 
    $x \in [p, q]$ and $y \in [\ell, r]$ follow from the definition of interval attractor.
    $x \in [p, r]$ follows from $x \in [p, q]$ and $q \leq r$. 
    $[i, j] \cap [p, r] \neq \emptyset$ follows from $x \in [p, r]$ and $x \in [i, j]$. 
    $([p, q], [\ell, r]) \in \OVQ([i, j])$ follows from $[i, j] \cap [p, r] \neq \emptyset$ and $([p, q], [\ell, r]) \not \in \Psi_{\run}$. 
    Therefore, $\Psi_{\OVQ, R} \setminus \Psi_{\run} \subseteq \OVQ([i, j])$ holds. 
\end{proof}

Lemma~\ref{lem:RB_OVQ_and_Subset} indicates that 
overlap query can be answered by computing the following three sets: 
(i) $\bigcup_{i^{\prime} = i}^{j} \{ I_{\capture}(i^{\prime}, j^{\prime}) \mid j^{\prime} \in [i^{\prime}+1, n] \} \setminus \Psi_{\run}$;
(ii) $\Psi_{\OVQ, C} \setminus \Psi_{\run}$; 
(iii) $\{ I_{\capture}(x, i) \mid x \in [1, i-1] \} \setminus \Psi_{\run}$. 
Therefore, we present the three algorithms for computing these sets using the data structure detailed in Section~\ref{subsubsec:rrdag_ds}, which is associated with the RR-DAG of the RLSLP $\mathcal{G}^{R}$.

\subsubsection{Computation of Set \texorpdfstring{$\bigcup_{i^{\prime} = i}^{j} \{ I_{\capture}(i^{\prime}, j^{\prime}) \mid j^{\prime} \in [i^{\prime}+1, n] \} \setminus \Psi_{\run}$}{}}\label{subsubsec:ovqr}
We explain the algorithm for computing set $\bigcup_{i^{\prime} = i}^{j} \{ I_{\capture}(i^{\prime}, j^{\prime}) \mid j^{\prime} \in [i^{\prime}+1, n] \} \setminus \Psi_{\run}$ of interval attractors. 
This algorithm consists of two phases, which are explained as follows. 

\paragraph{Phase 1: Computation of set $\{ I_{\capture}(i^{\prime}, j^{\prime}) \mid j^{\prime} \in [i^{\prime}+1, n] \}$ for each integer $i^{\prime} \in [i, j]$.}
In the first phase, we compute set $\{ I_{\capture}(i^{\prime}, j^{\prime}) \mid j^{\prime} \in [i^{\prime}+1, n] \}$ of interval attractors for each integer $i^{\prime} \in [i, j]$. 
This phase leverages the following lemma, which states 
four properties of the set $\{ I_{\capture}(i^{\prime}, j^{\prime}) \mid j^{\prime} \in [i^{\prime}+1, n] \}$. 

\begin{lemma}\label{lem:psi_OVR_conditions}
Let $([p_{1}, q_{1}], [\ell_{1}, r_{1}])$, $([p_{2}, q_{2}], [\ell_{2}, r_{2}])$, $\ldots$, 
$([p_{k}, q_{k}], [\ell_{k}, r_{k}])$ ($r_{1} \leq r_{2} \leq \cdots \leq r_{k}$) be the interval attractors in 
set $\{ I_{\capture}(i^{\prime}, j^{\prime}) \mid j^{\prime} \in [i^{\prime}+1, n] \}$ 
for a position $i^{\prime} \in [1, n]$ in input string $T$. 
Then, the following four statements hold: 
\begin{enumerate}[label=\textbf{(\roman*)}]
    \item $r_{1} < r_{2} < \cdots < r_{k} = n$; 
    \item $k \leq H+1$;
    \item if $i^{\prime} < n$, then $k \geq 1$ and $([p_{1}, q_{1}], [\ell_{1}, r_{1}]) = I_{\capture}(i^{\prime}, i^{\prime}+1)$; 
    otherwise $k = 0$;
    \item $([p_{s}, q_{s}], [\ell_{s}, r_{s}]) = I_{\capture}(i^{\prime}, r_{s-1}+1)$ for all $s \in [2, k]$.     
\end{enumerate}    
\end{lemma}
\begin{proof}
Let $j_{1} = i^{\prime}+1$ 
and $\{ j_{2}, j_{3}, \ldots, j_{m} \}$ ($j_{2} < j_{3} < \cdots < j_{m}$) be a subset of set $\{ i^{\prime}+3, i^{\prime}+2, \ldots, n \}$ 
such that each integer $j_{s}$ satisfies $I_{\capture}(i^{\prime}, j_{s}) \neq I_{\capture}(i^{\prime}, j_{s}-1)$ 
(i.e., $\{ j_{2}, j_{3}, \ldots, j_{m} \} = \{ j^{\prime} \in [i^{\prime}+3, n] \mid I_{\capture}(i^{\prime}, j^{\prime}) \neq I_{\capture}(i^{\prime}, j^{\prime}-1) \}$). 

For each integer $s \in [1, m]$, 
let $([p^{\prime}_{s}, q^{\prime}_{s}], [\ell^{\prime}_{s}, r^{\prime}_{s}])$ be interval attractor $I_{\capture}(i^{\prime}, j_{s})$. 
Then, 
$i^{\prime} \in [p^{\prime}_{s}, q^{\prime}_{s}]$, 
$j_{s} = r^{\prime}_{s-1}+1$, 
and $r^{\prime}_{m} = n$ can be proved using 
(A) the definition of interval attractor $I_{\capture}(i^{\prime}, j_{s})$, 
(B) $I_{\capture}(i^{\prime}, j_{s}) = I_{\capture}(i^{\prime}, j_{s} + 1) = \cdots = I_{\capture}(i^{\prime}, r^{\prime}_{s})$, 
and (C) $I_{\capture}(i^{\prime}, j_{s}) \neq I_{\capture}(i^{\prime}, j_{s}-1)$
In addition, $r^{\prime}_{s} > r^{\prime}_{s-1}$ follows from 
$j_{s-1} < j_{s}$, 
$j_{s-1} = r^{\prime}_{s-2}+1$, 
and $j_{s} = r^{\prime}_{s-1}+1$. 
$k = m$ holds because 
$\{ ([p_{s}, q_{s}], [\ell_{s}, r_{s}]) \mid s \in [1, k] \} = \{ I_{\capture}(i^{\prime}, j_{s}) \mid s \in [1, m] \}$ holds. 
$([p_{s}, q_{s}], [\ell_{s}, r_{s}]) = I_{\capture}(i^{\prime}, r_{s-1}+1)$ for all $s \in [1, k]$ 
because (A) $r_{1} \leq r_{2} \leq \cdots \leq r_{k}$, 
(B) $r^{\prime}_{1} < r^{\prime}_{2} < \cdots < r^{\prime}_{k}$, 
and (C) 
$\{ ([p_{s}, q_{s}], [\ell_{s}, r_{s}]) \mid s \in [1, k] \} = \{ I_{\capture}(i^{\prime}, j_{s}) \mid s \in [1, k] \}$ hold.

We show that $m \leq H+1$. 
Let $h_{s} \in [0, H]$ be the level of each interval attractor $I_{\capture}(i^{\prime}, j_{s})$. 
Then, $h_{s-1} < h_{s}$ follows from 
Lemma~\ref{lem:interval_extension_propertyX}~\ref{enum:interval_extension_propertyX:right}, 
$j_{s-1} < j_{s}$, and $j_{s} = r^{\prime}_{s-1}+1$. 
This fact indicates that $0 \leq h_{1} < h_{2} < \cdots < h_{m} \leq H$. 
Therefore, we obtain $m \leq H+1$.

\textbf{Proof of Lemma~\ref{lem:psi_OVR_conditions}(i).}
$r_{1} < r_{2} < \cdots < r_{k}$ holds 
because $r_{s} = r^{\prime}_{s}$ for all $s \in [1, k]$, 
and $r^{\prime}_{1} < r^{\prime}_{2} < \cdots < r^{\prime}_{k}$. 
$r_{k} = n$ follows from $r_{k} = r^{\prime}_{m}$ and $r^{\prime}_{m} = n$.  

\textbf{Proof of Lemma~\ref{lem:psi_OVR_conditions}(ii).}
$k \leq H+1$ follows from $k = m$ and $m \leq H+1$. 

\textbf{Proof of Lemma~\ref{lem:psi_OVR_conditions}(iii).}
If $i^{\prime} < n$, 
then $j_{1} = i^{\prime}+1$ and $m \geq 1$. 
$k \geq 1$ follows from $k = m$ and $m \geq 1$. 
$([p_{1}, q_{1}], [\ell_{1}, r_{1}]) = I_{\capture}(i^{\prime}, i^{\prime}+1)$ follows from 
$([p_{1}, q_{1}], [\ell_{1}, r_{1}]) = ([p^{\prime}_{1}, q^{\prime}_{1}], [\ell^{\prime}_{1}, r^{\prime}_{1}])$ 
and $([p^{\prime}_{1}, q^{\prime}_{1}], [\ell^{\prime}_{1}, r^{\prime}_{1}]) = I_{\capture}(i^{\prime}, i^{\prime}+1)$. 
Otherwise (i.e., $i^{\prime} = n$), 
$m = 0$ holds. 
$k = 0$ follows from $k = m$ and $m = 0$. 

\textbf{Proof of Lemma~\ref{lem:psi_OVR_conditions}(iv).}
$([p_{s}, q_{s}], [\ell_{s}, r_{s}]) = I_{\capture}(i^{\prime}, r_{s-1}+1)$ follows from 
$([p_{s}, q_{s}], [\ell_{s}, r_{s}])$ $= ([p^{\prime}_{s}, q^{\prime}_{s}], [\ell^{\prime}_{s}, r^{\prime}_{s}])$, 
$([p^{\prime}_{s}, q^{\prime}_{s}], [\ell^{\prime}_{s}, r^{\prime}_{s}]) = I_{\capture}(i^{\prime}, j_{s})$, 
$j_{s} = r^{\prime}_{s-1} + 1$, and $r^{\prime}_{s-1} = r_{s-1}$. 
\end{proof}

For each integer $i^{\prime} \in [i, j]$, 
we compute the $k$ interval attractors 
$([p_{1}, q_{1}], [\ell_{1}, r_{1}])$, $([p_{2}, q_{2}], [\ell_{2}, r_{2}])$, $\ldots$, 
$([p_{k}, q_{k}], [\ell_{k}, r_{k}])$ ($r_{1} \leq r_{2} \leq \cdots \leq r_{k}$) 
in set $\{ I_{\capture}(i^{\prime}, j^{\prime}) \mid j^{\prime} \in [i^{\prime}+1, n] \}$ 
by $k$ capture queries. 
Here, $r_{1} < r_{2} < \cdots < r_{k} = n$ follows from Lemma~\ref{lem:psi_OVR_conditions}(i). 
Lemma~\ref{lem:psi_OVR_conditions}(iii) shows that 
the first interval attractor $([p_{1}, q_{1}], [\ell_{1}, r_{1}])$ is $I_{\capture}(i^{\prime}, i^{\prime}+1)$, and 
it can be obtained by capture query $\CAPQ([i^{\prime}, i^{\prime}+1])$. 
For each integer $s \in [2, k]$, 
Lemma~\ref{lem:psi_OVR_conditions}(iv) shows that 
the $s$-th interval attractor is $I_{\capture}(i^{\prime}, r_{s-1}+1)$, 
and 
it can be obtained by capture query $\CAPQ([i^{\prime}, r_{s-1}+1])$. 

The $k$ capture queries take $O(kH^{2} \log n)$ time. 
$k \leq H+1$ because $k = |\{ I_{\capture}(i^{\prime}, j^{\prime}) \mid j^{\prime} \in [i^{\prime}+1, n] \}|$, 
and $|\{ I_{\capture}(i^{\prime}, j^{\prime}) \mid j^{\prime} \in [i^{\prime}+1, n] \}| \leq H+1$ follows from Lemma~\ref{lem:psi_OVR_conditions}(ii). 
Therefore, the first phase takes $O(|[i, j]| H^{3} \log n)$ time in total.

%Therefore, we execute $k$ capture queries for each integer $i^{\prime} \in [i, j]$ in the first phase. 
%the set $\{ I_{\capture}(i^{\prime}, j^{\prime}) \mid j^{\prime} \in [i^{\prime}+1, n] \}$ 
%can be computed by $k$ capture queries. 

\paragraph{Phase 2: Computation of set $\bigcup_{i^{\prime} = i}^{j} \{ I_{\capture}(i^{\prime}, j^{\prime}) \mid j^{\prime} \in [i^{\prime}+1, n] \} \setminus \Psi_{\run}$.} 
In the second phase, we compute set $\bigcup_{i^{\prime} = i}^{j} \{ I_{\capture}(i^{\prime}, j^{\prime}) \mid j^{\prime} \in [i^{\prime}+1, n] \} \setminus \Psi_{\run}$. 
We obtain the set $\bigcup_{i^{\prime} = i}^{j} \{ I_{\capture}(i^{\prime}, j^{\prime}) \mid j^{\prime} \in [i^{\prime}+1, n] \} \setminus \Psi_{\run}$ by removing the interval attractors of the subset $\Psi_{\run}$ from the set 
$\bigcup_{i^{\prime} = i}^{j} \{ I_{\capture}(i^{\prime}, j^{\prime}) \mid j^{\prime} \in [i^{\prime}+1, n] \}$, 
which is computed in the first phase. 
This removal needs to verify whether 
the subset $\Psi_{\run}$ contains each interval attractor of the set $\bigcup_{i^{\prime} = i}^{j} \{ I_{\capture}(i^{\prime}, j^{\prime}) \mid j^{\prime} \in [i^{\prime}+1, n] \}$ or not. 
We execute this verification by verify-run query, 
and each verify-run query takes $O(H^{2})$ time.
Lemma~\ref{lem:psi_OVR_conditions}(ii) shows that the set $\bigcup_{i^{\prime} = i}^{j} \{ I_{\capture}(i^{\prime}, j^{\prime}) \mid j^{\prime} \in [i^{\prime}+1, n] \}$ contains at most $|[i, j]|(H+1)$ interval attractors. 
Therefore, the second phase takes $O(|[i, j]| H^{3})$ time in total. 

We showed that the first and second phases take $O(|[i, j]| H^{3} \log n)$ time and $O(|[i, j]| H^{3})$ time, respectively. 
Therefore, the two phases take $O(|[i, j]| H^{3} \log n)$ time in total. 

%%%%%%%%%%%%%%%%%%%%%%%%%%%%%%%%%%%%%%%%%%%%%%%%%%%%%%%%%%%%%%%%%%%%%%%%%%%%%%%%%%%%%%%%%%

\subsubsection{Computation of Set \texorpdfstring{$\Psi_{\OVQ, C} \setminus \Psi_{\run}$}{}}\label{subsubsec:ovqc}
We introduce several notions for quickly computing the set $\Psi_{\OVQ, C} \setminus \Psi_{\run}$.

\paragraph{Left-extension of interval attractor.}
For two interval attractors $([p, q], [\ell, r]), ([p^{\prime}, q^{\prime}], [\ell^{\prime}, r^{\prime}]) \in \Psi_{\RR}$, 
the interval attractor $([p, q], [\ell, r])$ is called \emph{left-extension} of the interval attractor $([p^{\prime}, q^{\prime}], [\ell^{\prime}, r^{\prime}])$ if $q = p^{\prime}-1$ and $r \in [\ell^{\prime}, r^{\prime}]$ hold. 
The following lemma states four properties of left-extension of interval attractor. 

\begin{lemma}\label{lem:left_extension_property}
The following four statements hold for an interval attractor $([p, q], [\ell, r]) \in \Psi_{\RR}$: 
\begin{enumerate}[label=\textbf{(\roman*)}]
    \item \label{enum:left_extension_property:1} for an interval attractor $([p^{\prime}, q^{\prime}], [\ell^{\prime}, r^{\prime}]) \in \Psi_{\RR}$, 
    the interval attractor $([p, q], [\ell, r])$ is a left-extension of the interval attractor $([p^{\prime}, q^{\prime}], [\ell^{\prime}, r^{\prime}])$ if and only if $q + 1 < r$ and $([p^{\prime}, q^{\prime}], [\ell^{\prime}, r^{\prime}]) = I_{\capture}(q+1, r)$; 
    \item \label{enum:left_extension_property:2} the set $\Psi_{\RR}$ contains at most one interval attractor $([p^{\prime}, q^{\prime}], [\ell^{\prime}, r^{\prime}])$ 
    such that $([p, q], [\ell, r])$ is a left-extension of the interval attractor $([p^{\prime}, q^{\prime}], [\ell^{\prime}, r^{\prime}])$; 
    \item \label{enum:left_extension_property:3}
    if the subset $\Psi_{\run}$ contains the interval attractor $([p, q], [\ell, r])$, 
    then the interval attractor $([p, q], [\ell, r])$ is a left-extension of the interval attractor $([p + |C|, q + |C|], [\ell + |C|, r])$ 
    for the associated string $C$ of the interval attractor $([p, q], [\ell, r])$; 
    \item \label{enum:left_extension_property:4}
    let $([p^{\prime}, q^{\prime}], [\ell^{\prime}, r^{\prime}]) \in \Psi_{\RR}$ be an interval attractor such that 
    $([p, q], [\ell, r])$ is a left-extension of the interval attractor $([p^{\prime}, q^{\prime}], [\ell^{\prime}, r^{\prime}])$. 
    If the subset $\Psi_{\source}$ contains the interval attractor $([p, q], [\ell, r])$, 
    then $([p^{\prime}, q^{\prime}], [\ell^{\prime}, r^{\prime}]) = ([p_{1}, q_{1}], [\ell_{1}, r_{1}])$ 
    for the $k$ interval attractors $([p_{1}, q_{1}]$, $[\ell_{1}, r_{1}])$, $([p_{2}, q_{2}], [\ell_{2}, r_{2}])$, 
    $\ldots$, $([p_{k}, q_{k}], [\ell_{k}, r_{k}])$ ($p_{1} \leq p_{2} \leq \cdots \leq p_{k}$) obtained from the function $f_{\recover}(([p, q], [\ell, r]))$.     
\end{enumerate}
\end{lemma}
\begin{proof}
    The following two statements are used to prove Lemma~\ref{lem:left_extension_property}(i). 
\begin{enumerate}[label=\textbf{(\arabic*)}]
    \item $q + 1 < r$ and $([p^{\prime}, q^{\prime}], [\ell^{\prime}, r^{\prime}]) = I_{\capture}(q+1, r)$ hold if 
    the interval attractor $([p, q], [\ell, r])$ is a left-extension of the interval attractor $([p^{\prime}, q^{\prime}], [\ell^{\prime}, r^{\prime}])$ (i.e., $q = p^{\prime}-1$ and $r \in [\ell^{\prime}, r^{\prime}]$); 
    \item the interval attractor $([p, q], [\ell, r])$ is a left-extension of the interval attractor $([p^{\prime}, q^{\prime}], [\ell^{\prime}, r^{\prime}])$ if $q + 1 < r$ and $([p^{\prime}, q^{\prime}], [\ell^{\prime}, r^{\prime}]) = I_{\capture}(q+1, r)$.
\end{enumerate}

    \textbf{Proof of statement (1).}
    $p^{\prime} < \ell^{\prime}$ follows from the definition of interval attractor. 
    $q + 1 < r$ follows from $q = p^{\prime}-1$, $r \in [\ell^{\prime}, r^{\prime}]$, and $p^{\prime} < \ell^{\prime}$. 
    $([p^{\prime}, q^{\prime}], [\ell^{\prime}, r^{\prime}]) = I_{\capture}(q+1, r)$ holds because 
    (A) $[q+1, r] = [p^{\prime}, r]$ follows from $q = p^{\prime}-1$, 
    and (B) $I_{\capture}(p^{\prime}, r) = ([p^{\prime}, q^{\prime}], [\ell^{\prime}, r^{\prime}])$ 
    follows from Lemma~\ref{lem:IA_maximal_lemma}. 
    Therefore, $q + 1 < r$ and $([p^{\prime}, q^{\prime}], [\ell^{\prime}, r^{\prime}]) = I_{\capture}(q+1, r)$ hold. 

    \textbf{Proof of statement (2).}    
    We prove $q = p^{\prime}-1$ by contradiction. 
    We assume that $q \neq p^{\prime}-1$ holds. 
    $q + 1 \geq p^{\prime}$ and $r \in [\ell^{\prime}, r^{\prime}]$ follow from the definition of interval attractor 
    for the interval attractor $I_{\capture}(q+1, r) = ([p^{\prime}, q^{\prime}], [\ell^{\prime}, r^{\prime}])$. 
    $q \geq p^{\prime}$ follows from $q \neq p^{\prime}-1$ and $q + 1 \geq p^{\prime}$. 
    
    We consider interval attractor $I_{\capture}(q, r)$. 
    $I_{\capture}(q, r) = ([p, q], [\ell, r])$ follows from Lemma~\ref{lem:IA_maximal_lemma}. 
    On the other hand, $I_{\capture}(q, r) = ([p^{\prime}, q^{\prime}], [\ell^{\prime}, r^{\prime}])$ 
    follows from Lemma~\ref{lem:IA_maximal_lemma}
    because (A) $I_{\capture}(q+1, r) = ([p^{\prime}, q^{\prime}], [\ell^{\prime}, r^{\prime}])$ 
    and (B) $q \in [p^{\prime}, q+1]$. 
    The two facts $I_{\capture}(q, r) = ([p, q], [\ell, r])$
    and $I_{\capture}(q+1, r) = ([p^{\prime}, q^{\prime}], [\ell^{\prime}, r^{\prime}])$ yield a contradiction 
    because $([p, q], [\ell, r]) \neq ([p^{\prime}, q^{\prime}], [\ell^{\prime}, r^{\prime}])$. 
    Therefore, $q = p^{\prime}-1$ must hold. 

    We prove statement (2). 
    This statement holds if $q = p^{\prime}-1$ and $r \in [\ell^{\prime}, r^{\prime}]$. 
    We already proved $q = p^{\prime}-1$ and $r \in [\ell^{\prime}, r^{\prime}]$. 
    Therefore, statement (2) holds. 
        
    \textbf{Proof of Lemma~\ref{lem:left_extension_property}(i).}    
    Lemma~\ref{lem:left_extension_property}(i) follows from statement (1) and statement (2). 
        
    \textbf{Proof of Lemma~\ref{lem:left_extension_property}(ii).}    
    We prove Lemma~\ref{lem:left_extension_property}(ii) by contradiction. 
    We assume that Lemma~\ref{lem:left_extension_property}(ii) does not hold. 
    Then, the set $\Psi_{\RR}$ contains two interval attractors $([p_{1}, q_{1}], [\ell_{1}, r_{1}])$ 
    and $([p_{2}, q_{2}], [\ell_{2}, r_{2}])$ ($([p_{1}, q_{1}], [\ell_{1}, r_{1}]) \neq ([p_{2}, q_{2}], [\ell_{2}, r_{2}])$) 
    such that the interval attractor $([p, q], [\ell, r])$ is a left-extension of both $([p_{1}, q_{1}], [\ell_{1}, r_{1}])$ and $([p_{2}, q_{2}], [\ell_{2}, r_{2}])$. 
    In this case, 
    $([p_{1}, q_{1}], [\ell_{1}, r_{1}]) = ([p_{2}, q_{2}], [\ell_{2}, r_{2}])$ 
    because 
    $([p_{1}, q_{1}], [\ell_{1}, r_{1}]) = I_{\capture}(q+1, r)$ and $([p_{2}, q_{2}], [\ell_{2}, r_{2}]) = I_{\capture}(q+1, r)$ 
    follows from Lemma~\ref{lem:left_extension_property}(i). 
    The two facts $([p_{1}, q_{1}], [\ell_{1}, r_{1}]) = ([p_{2}, q_{2}], [\ell_{2}, r_{2}])$ and $([p_{1}, q_{1}], [\ell_{1}, r_{1}]) \neq ([p_{2}, q_{2}], [\ell_{2}, r_{2}])$ yield a contradiction. 
    Therefore, Lemma~\ref{lem:left_extension_property}(ii) must hold.

    \textbf{Proof of Lemma~\ref{lem:left_extension_property}(iii).}
    We apply Lemma~\ref{lem:psi_run_basic_property}~\ref{enum:psi_run_basic_property:1} and Lemma~\ref{lem:psi_run_basic_property}~\ref{enum:psi_run_basic_property:2} to the interval attractor 
    $([p, q], [\ell, r])$. 
    Lemma~\ref{lem:psi_run_basic_property}~\ref{enum:psi_run_basic_property:1} shows that 
    $q - p = |C| - 1$ holds. 
    Lemma~\ref{lem:psi_run_basic_property}~\ref{enum:psi_run_basic_property:2} shows that 
    the set $\Psi_{\RR}$ contains interval attractor $([p + |C|, q + |C|], [\ell + |C|, r])$. 
    $q+1 = p + |C|$ follows from $q - p = |C| - 1$. 
    Because of $q+1 = p + |C|$ and $r \in [\ell + |C|, r]$, 
    the interval attractor $([p, q], [\ell, r])$ is a left-extension of the interval attractor $([p + |C|, q + |C|], [\ell + |C|, r])$. 
        
    \textbf{Proof of Lemma~\ref{lem:left_extension_property}(iv).}    
    Lemma~\ref{lem:recover_basic_property}~\ref{enum:recover_basic_property:3} shows that 
    $p < p_{1}$, $q = p_{1}-1$, $\ell \leq \ell_{1}$, and $r = r_{1}$ hold. 
    $q + 1 = p_{1}$ follows from $q = p_{1}-1$. 
    $r \in [\ell_{1}, r_{1}]$ follows from $r = r_{1}$. 
    Because of $q+1 = p_{1}$ and $r \in [\ell_{1}, r_{1}]$, 
    the interval attractor $([p, q], [\ell, r])$ is a left-extension of the interval attractor $([p_{1}, q_{1}], [\ell_{1}, r_{1}])$. 

    We prove $([p_{1}, q_{1}], [\ell_{1}, r_{1}]) = ([p^{\prime}, q^{\prime}], [\ell^{\prime}, r^{\prime}])$ by contradiction. 
    We assume that $([p_{1}, q_{1}], [\ell_{1}, r_{1}]) \neq ([p^{\prime}, q^{\prime}], [\ell^{\prime}, r^{\prime}])$ holds. 
    Then, the interval attractor $([p, q], [\ell, r])$ is a left-extension of both 
    $([p_{1}, q_{1}]$, $[\ell_{1}, r_{1}])$ and $([p^{\prime}, q^{\prime}], [\ell^{\prime}, r^{\prime}])$.     
    This fact contradicts Lemma~\ref{lem:left_extension_property}~\ref{enum:left_extension_property:2}, 
    i.e., 
    the interval attractor $([p, q], [\ell, r])$ is a left-extension of at most one interval attractor. 
    Therefore, $([p_{1}, q_{1}], [\ell_{1}, r_{1}]) = ([p^{\prime}, q^{\prime}], [\ell^{\prime}, r^{\prime}])$ must hold. 
\end{proof}

The following lemma is used to compute all the left-extensions of a given interval attractor.

\begin{lemma}\label{lem:comp_left_extensions}
    Consider the set $\Psi$ of left-extensions of an interval attractor $([p, q], [\ell, r]) \in \Psi_{\RR}$ (i.e., 
    $\Psi = \{ ([p^{\prime}, q^{\prime}], [\ell^{\prime}, r^{\prime}]) \in \Psi_{\RR} \mid (q^{\prime}+1 = p) \land (r^{\prime} \in [\ell, r]) \}$). 
    Then, $\Psi = \{ ([p^{\prime}, q^{\prime}], [\ell^{\prime}, r^{\prime}]) \in \bigcup_{j^{\prime} = p}^{n} I_{\capture}(p-1, j^{\prime}) \mid r^{\prime} \in [\ell, r] \}$.  
\end{lemma}
\begin{proof}
    Lemma~\ref{lem:comp_left_extensions} holds if the following two statements: 
    (i) $\Psi \subseteq \{ ([p^{\prime}, q^{\prime}], [\ell^{\prime}, r^{\prime}]) \in \bigcup_{j^{\prime} = p}^{n} I_{\capture}(p-1, j^{\prime}) \mid r^{\prime} \in [\ell, r] \}$; 
    (ii) $\Psi \supseteq \{ ([p^{\prime}, q^{\prime}], [\ell^{\prime}, r^{\prime}]) \in \bigcup_{j^{\prime} = p}^{n} I_{\capture}(p-1, j^{\prime}) \mid r^{\prime} \in [\ell, r] \}$. 
    The proof of these statements are as follows. 

    \textbf{Proof of $\Psi \subseteq \{ ([p^{\prime}, q^{\prime}], [\ell^{\prime}, r^{\prime}]) \in \bigcup_{j^{\prime} = p}^{n} I_{\capture}(p-1, j^{\prime}) \mid r^{\prime} \in [\ell, r] \}$.}    
    We consider an interval attractor $([p_{1}, q_{1}], [\ell_{1}, r_{1}])$ in the subset $\Psi$. 
    Here, $q_{1} = p-1$ and $r_{1} \in [\ell, r]$ hold. 
    We prove $([p_{1}, q_{1}], [\ell_{1}, r_{1}]) \in \{ ([p^{\prime}, q^{\prime}], [\ell^{\prime}, r^{\prime}]) \in \bigcup_{j^{\prime} = p}^{n} I_{\capture}(p-1, j^{\prime}) \mid r^{\prime} \in [\ell, r] \}$ (i.e., $\Psi \subseteq \{ ([p^{\prime}, q^{\prime}], [\ell^{\prime}, r^{\prime}]) \in \bigcup_{j^{\prime} = p}^{n} I_{\capture}(p-1, j^{\prime}) \mid r^{\prime} \in [\ell, r] \}$). 
    $I_{\capture}(q_{1}, r_{1}) = ([p_{1}, q_{1}], [\ell_{1}, r_{1}])$ 
    follows from Lemma~\ref{lem:IA_maximal_lemma}. 
    $I_{\capture}(q_{1}, r_{1}) = I_{\capture}(p-1, r_{1})$ follows from $q_{1} = p-1$. 
    Therefore, $([p_{1}, q_{1}], [\ell_{1}, r_{1}]) \in \{ ([p^{\prime}, q^{\prime}]$, $[\ell^{\prime}, r^{\prime}]) \in \bigcup_{j^{\prime} = p}^{n} I_{\capture}(p-1, j^{\prime}) \mid r^{\prime} \in [\ell, r] \}$ 
    follows from 
    (A) $I_{\capture}(q_{1}, r_{1}) = I_{\capture}(p-1, r_{1})$, 
    (B) $I_{\capture}(q_{1}, r_{1}) = ([p_{1}, q_{1}], [\ell_{1}, r_{1}])$, 
    and (C) $r_{1} \in [\ell, r]$. 
    
    \textbf{Proof of $\Psi \supseteq \{ ([p^{\prime}, q^{\prime}], [\ell^{\prime}, r^{\prime}]) \in \bigcup_{j^{\prime} = p}^{n} I_{\capture}(p-1, j^{\prime}) \mid r^{\prime} \in [\ell, r] \}$.}    
    We consider an interval attractor $([p_{2}, q_{2}], [\ell_{2}, r_{2}])$ in the subset $\{ ([p^{\prime}, q^{\prime}], [\ell^{\prime},r^{\prime}]) \in \bigcup_{j^{\prime} = p}^{n} I_{\capture}(p-1, j^{\prime}) \mid r^{\prime} \in [\ell, r] \}$. 
    Then, $r_{2} \in [\ell, r]$ holds, 
    and there exists an integer $j^{\prime} \in [p, n]$ 
    satisfying $([p_{2}, q_{2}], [\ell_{2}, r_{2}]) = I_{\capture}(p-1, j^{\prime})$. 
    Since $([p_{2}, q_{2}], [\ell_{2}, r_{2}]) = I_{\capture}(p-1, j^{\prime})$, 
    $p-1 \in [p_{2}, q_{2}]$ follows from the definition of interval attractor. 
    $q_{2} \geq p-1$ follows from $p-1 \in [p_{2}, q_{2}]$. 

    We prove $q_{2} = p-1$ by contradiction. 
    We assume that $q_{2} \neq p-1$. 
    Then, $q_{2} > p-1$ follows from $q_{2} \geq p-1$ and $q_{2} \neq p-1$. 
    We consider interval attractor $I_{\capture}(p, r_{2})$. 
    Then, $I_{\capture}(p, r_{2}) = ([p_{2}, q_{2}], [\ell_{2}, r_{2}])$ follows from 
    Lemma~\ref{lem:IA_maximal_lemma} 
    because $p \in [p_{2}, q_{2}]$. 
    Similarly, 
    $I_{\capture}(p, r_{2}) = ([p, q], [\ell, r])$ follows from 
    Lemma~\ref{lem:IA_maximal_lemma} 
    because $r_{2} \in [\ell, r]$. 
    $([p, q], [\ell, r]) = ([p_{2}, q_{2}], [\ell_{2}, r_{2}])$ follows from $I_{\capture}(p, r_{2}) = ([p_{2}, q_{2}], [\ell_{2}, r_{2}])$ and $I_{\capture}(p, r_{2}) = ([p, q], [\ell, r])$. 
    On the other hand, $([p, q], [\ell, r]) \neq ([p_{2}, q_{2}], [\ell_{2}, r_{2}])$ follows from $p \neq p_{2}$. 
    The two facts $([p, q], [\ell, r]) = ([p_{2}, q_{2}], [\ell_{2}, r_{2}])$ and $([p, q], [\ell, r]) \neq ([p_{2}, q_{2}], [\ell_{2}, r_{2}])$ yield a contradiction. 
    Therefore, $q_{2} = p-1$ must hold. 

    The interval attractor $([p_{2}, q_{2}], [\ell_{2}, r_{2}])$ is contained in the subset $\Psi$ 
    because we already proved $q_{2} = p-1$ and $r_{2} \in [\ell, r]$. 
    Because of $([p_{2}, q_{2}], [\ell_{2}, r_{2}]) \in \Psi$, 
    $\Psi \supseteq \{ ([p^{\prime}, q^{\prime}], [\ell^{\prime}, r^{\prime}]) \in \bigcup_{j^{\prime} = p}^{n} I_{\capture}(p-1, j^{\prime}) \mid r^{\prime} \in [\ell, r] \}$ holds. 
\end{proof}

We explain the algorithm for computing all the left-extensions of a given interval attractor $([p, q], [\ell, r]) \in \Psi_{\RR}$ 
using Lemma~\ref{lem:comp_left_extensions}. 
Lemma~\ref{lem:comp_left_extensions} indicates that 
the left-extensions of a given interval attractor $([p, q], [\ell, r])$ can be obtained by 
verifying whether each interval attractor $([p^{\prime}, q^{\prime}], [\ell^{\prime}, r^{\prime}])$ 
of set $\bigcup_{j^{\prime} = p}^{n} I_{\capture}(p-1, j^{\prime})$ satisfies $r \in [\ell^{\prime}, r^{\prime}]$ or not. 
The set $\bigcup_{j^{\prime} = p}^{n} I_{\capture}(p-1, j^{\prime})$ can be computed in $O(H^{3} \log n)$ time by 
the first phase of the algorithm presented in Section~\ref{subsubsec:ovqr}. 
Therefore, all the left-extensions of a given interval attractor in $O(H^{3} \log n)$ time.

%%%%%%%%%%%%%%%%%%%%%%%%%%%%%%%%%%%%%%%%%%%%%%%%%%%%%%%%%%%%%%%%%
\paragraph{Run-extension of an interval attractor.}

For two interval attractors $([p, q], [\ell, r])$, $([p^{\prime}, q^{\prime}]$, $[\ell^{\prime}, r^{\prime}]) \in \Psi_{\RR}$, 
the interval attractor $([p, q], [\ell, r])$ is called \emph{run-extension} of the interval attractor $([p^{\prime}, q^{\prime}], [\ell^{\prime}, r^{\prime}])$ if $([p, q], [\ell, r]) \in \Psi_{\source}$ and $([p^{\prime}, q^{\prime}], [\ell^{\prime}, r^{\prime}]) \in f_{\recover}(([p, q], [\ell, r]))$ for the function $f_{\recover}$ introduced in Section~\ref{subsec:function_recover}. 
Here, $([p^{\prime}, q^{\prime}], [\ell^{\prime}, r^{\prime}]) \in \Psi_{\run}$ follows from the definition of the function $f_{\recover}(([p, q], [\ell, r]))$ if the interval attractor $([p, q], [\ell, r])$ is a run-extension of the interval attractor $([p^{\prime}, q^{\prime}], [\ell^{\prime}, r^{\prime}])$.

The following lemma is used to compute all the run-extensions of a given interval attractor.

\begin{lemma}\label{lem:run_extension_property}
    Consider the set $\Psi$ of run-extensions of an interval attractor $([p, q], [\ell, r]) \in \Psi_{\run}$ (i.e., 
    $\Psi = \{ ([p^{\prime}, q^{\prime}], [\ell^{\prime}, r^{\prime}]) \in \Psi_{\RR} \mid (([p^{\prime}, q^{\prime}], [\ell^{\prime}, r^{\prime}] \in \Psi_{\source}) \land (([p, q], [\ell, r]) \in f_{\recover}(([p^{\prime}, q^{\prime}], [\ell^{\prime}, r^{\prime}]))  \})$). 
    Here, 
    Lemma~\ref{lem:recover_division_property}~\ref{enum:recover_division_property:1} shows that the subset $\Psi_{\source}$ contains an interval attractor 
    $([p_{A}, q_{A}], [\ell_{A}, r_{A}])$ satisfying $([p, q], [\ell, r]) \in f_{\recover}(([p_{A}, q_{A}], [\ell_{A}, r_{A}]))$. 
    Let $\gamma$ and $C$ be the attractor position and associated string of the interval attractor $([p, q], [\ell, r])$, respectively; 
    let $m \geq 0$ be the smallest integer satisfying $|[p-1, \gamma-1]| + m|C| > |\lcs(T[1..\gamma-1], C^{n+1})|$ 
    (i.e., $m = 1 + \lfloor (|\lcs(T[1..\gamma-1], C^{n+1})| - |[p-1, \gamma-1]|) / |C| \rfloor$). 
    Then, the following three statements hold: 
    \begin{enumerate}[label=\textbf{(\roman*)}]
    \item \label{lem:run_extension_property:1} $\Psi = \{ ([p_{A}, q_{A}], [\ell_{A}, r_{A}]) \}$; 
    \item \label{lem:run_extension_property:2} $I_{\capture}(p - (m-1) |C| - 1, r) = ([p_{A}, q_{A}], [\ell_{A}, r_{A}])$; 
    \item \label{lem:run_extension_property:3} $m = 1 + \lfloor (\rLCEQ(\gamma-1, \gamma+|C|-1) - |[p-1, \gamma-1]|) / |C| \rfloor$.
    \end{enumerate}
\end{lemma}
\begin{proof}

    Let $([p_{1}, q_{1}], [\ell_{1}, r_{1}])$, $([p_{2}, q_{2}], [\ell_{2}, r_{2}])$, 
    $\ldots$, $([p_{k}, q_{k}], [\ell_{k}, r_{k}])$ ($p_{1} \leq p_{2} \leq \cdots \leq p_{k}$) be the interval attractors obtained from the function $f_{\recover}(([p_{A}, q_{A}], [\ell_{A}, r_{A}]))$. 
    Let $\gamma_{A}$ be the attractor position of the interval attractor $([p_{A}, q_{A}], [\ell_{A}, r_{A}])$. 
    Similarly, 
    let $\gamma_{s}$ be the attractor position of each interval attractor $([p_{s}, q_{s}], [\ell_{s}, r_{s}]) \in f_{\recover}(([p_{A}, q_{A}], [\ell_{A}, r_{A}]))$. 
    Because of $([p, q], [\ell, r]) \in f_{\recover}(([p_{A}, q_{A}], [\ell_{A}, r_{A}]))$, 
    there exists an integer $x \in [1, k]$ satisfying $([p_{x}, q_{x}], [\ell_{x}, r_{x}]) = ([p, q], [\ell, r])$ 
    and $\gamma_{x} = \gamma$. 
    Lemma~\ref{lem:recover_basic_property}~\ref{enum:recover_basic_property:4} shows that 
    $([p_{A}, q_{A}], [\ell_{A}, r_{A}])$, $([p_{1}, q_{1}], [\ell_{1}, r_{1}])$, $([p_{2}, q_{2}], [\ell_{2}, r_{2}])$, $\ldots$, $([p_{k}, q_{k}], [\ell_{k}, r_{k}]) \in \Psi_{\centerset}(C)$.

    The following six statements are used to prove Lemma~\ref{lem:run_extension_property}:
\begin{enumerate}[label=\textbf{(\Alph*)}]
    \item 
    (a) $([p_{s}, q_{s}], [\ell_{s}, r_{s}]) = ([q_{A} + 1 + (s-1)|C|, q_{A} + s|C|], [\ell_{1} + (s-1)|C|, r_{A}])$, 
    (b) $|[p_{s}, q_{s}]| = |C|$, 
    (c) $T[\gamma_{s}..\gamma_{s} + |C| - 1] = C$, 
    (d) $|[p_{s}-1, \gamma_{s} -1]| = |[q_{A}, \gamma_{A}]| + |C|$.
    and (e) $\gamma_{s} = \gamma_{A} + s|C|$ for each integer $s \in [1, k]$; 
    \item $T[\gamma_{A}..\gamma - 1] = C^{x}$; 
    \item $|\lcs(T[1..\gamma-1], C^{n+1})| = |\lcs(T[p_{A}-1..\gamma_{A}-1], C^{n+1})| + x|C|$ 
    and $|\lcs(T[1..\gamma_{1}-1], C^{n+1})| = |\lcs(T[p_{A}-1..\gamma_{A}-1], C^{n+1})| + |C|$;
    \item $|\lcs(T[p_{A}-1..\gamma_{A}-1], C^{n+1})| \geq |[p-1, \gamma-1]| - |C|$;
    \item $|\lcs(T[p_{A}-1..\gamma_{A}-1], C^{n+1})| < |[p-1, \gamma-1]|$;
    \item $m = x$.
\end{enumerate}
    \textbf{Proof of statement (A).}
    Because of $([p_{s}, q_{s}], [\ell_{s}, r_{s}]) \in \Psi_{\run} \cap \Psi_{\centerset}(C)$, 
    $|[p_{s}, q_{s}]| = |C|$ follows from Lemma~\ref{lem:psi_run_basic_property}~\ref{enum:psi_run_basic_property:1}. 
    $T[\gamma_{s}..\gamma_{s} + |C| - 1] = C$ follows from $([p_{s}, q_{s}], [\ell_{s}, r_{s}]) \in \Psi_{\centerset}(C)$. 
    $\gamma_{s} = \gamma_{A} + s |C|$ follows from Lemma~\ref{lem:source_and_recover}~\ref{enum:source_and_recover:2}. 
    
    $([p_{s}, q_{s}], [\ell_{s}, r_{s}]) = ([p_{1} + (s-1)|C|, q_{1} + (s-1)|C|], [\ell_{1} + (s-1)|C|, r_{A}])$ 
    follows from the definition of the function $f_{\recover}$. 
    $p_{A} < p_{1}$, $q_{A} = p_{1} - 1$, and $\ell_{A} \leq \ell_{1}$ follow from Lemma~\ref{lem:recover_basic_property}~\ref{enum:recover_basic_property:3}. 
    $p_{s} = q_{A} + 1 + (s-1)|C|$ follows from $p_{s} = p_{1} + (s-1)|C|$ and $q_{A} = p_{1} - 1$. 
    $q_{s} = q_{A} + s|C|$ follows from $p_{s} = q_{A} + 1 + (s-1)|C|$ and $|[p_{s}, q_{s}]| = |C|$.

    $|[p_{s}-1, \gamma_{s} -1]| = |[q_{A}, \gamma_{A}]| + |C|$ follows from 
    $p_{s} = q_{A} + 1 + (s-1)|C|$ and $\gamma_{s} = \gamma_{A} + s|C|$.    
    Therefore, statement (A) holds. 
    
    \textbf{Proof of statement (B).}
    $T[\gamma_{A}..\gamma_{A} + |C| - 1] = C$ follows from $([p_{A}, q_{A}]$, $[\ell_{A}, r_{A}]) \in \Psi_{\centerset}(C)$.
    $T[\gamma_{A} + |C|..\gamma_{A} + |C| - 1] = \cdots = T[\gamma_{A} + (k-1)|C|..\gamma_{A} + k|C| - 1] = C$ 
    holds because 
    $T[\gamma_{s}..\gamma_{s} + |C| - 1] = C$ for each integer $s \in [1, k]$, 
    and $T[\gamma_{s}..\gamma_{s} + |C| - 1] = T[\gamma_{A} + s|C|..\gamma_{A} + (s+1)|C| - 1]$ 
    follows from $\gamma_{s} = \gamma_{A} + s|C|$. 
    $T[\gamma_{A}..\gamma_{A} + x|C| - 1] = C^{x}$ follows from 
    $T[\gamma_{A}..\gamma_{A} + |C| - 1] = T[\gamma_{A} + |C|..\gamma_{A} + |C| - 1] = \cdots = T[\gamma_{A} + (x-1)|C|..\gamma_{A} + x|C| - 1] = C$.
    $\gamma_{A} + x|C| - 1 = \gamma - 1$ follows from $\gamma_{x} = \gamma_{A} + x|C|$ 
    and $\gamma_{x} = \gamma$. 
    Therefore, $T[\gamma_{A}..\gamma - 1] = C^{x}$ follows from 
    $T[\gamma_{A}..\gamma_{A} + x|C| - 1] = C^{x}$ and $\gamma_{A} + x|C| - 1 = \gamma - 1$. 
    
    \textbf{Proof of statement (C).}
    We prove $|\lcs(T[1..\gamma-1], C^{n+1})| = |\lcs(T[p_{A}-1..\gamma_{A}-1], C^{n+1})| + x|C|$. 
    $|\lcs(T[1..\gamma-1], C^{n+1})| = |\lcs(T[1..\gamma_{A}-1], C^{n+1})| + x|C|$ follows from $T[\gamma_{A}..\gamma - 1] = C^{x}$. 
    $|\lcs(T[1..\gamma_{A}-1], C^{n+1})| = |\lcs(T[p_{A}-1..\gamma_{A}-1], C^{n+1})|$ holds 
    because $|\lcs(T[p_{A}-1..\gamma_{A}-1], C^{n+1})| < |[p_{A}-1, \gamma_{A}-1]|$ 
    follows from Lemma~\ref{lem:light_source_basic_property}. 
    Therefore, $|\lcs(T[1..\gamma-1], C^{n+1})| = |\lcs(T[p_{A}-1..\gamma_{A}-1], C^{n+1})| + x|C|$ holds. 

    Next, we prove $|\lcs(T[1..\gamma_{1}-1], C^{n+1})| = |\lcs(T[p_{A}-1..\gamma_{A}-1], C^{n+1})| + |C|$. 
    $|\lcs(T[1..\gamma_{1}-1], C^{n+1})| = |\lcs(T[1..\gamma_{A}-1], C^{n+1})| + |C|$ because 
    $T[\gamma_{A}..\gamma_{A} + |C| - 1] = C$ and $\gamma_{1} = \gamma_{A} + |C|$. 
    $|\lcs(T[1..\gamma_{1}-1], C^{n+1})| = |\lcs(T[p_{A}-1..\gamma_{A}-1], C^{n+1})| + |C|$ follows from 
    $|\lcs(T[1..\gamma_{1}-1], C^{n+1})| = |\lcs(T[1..\gamma_{A}-1], C^{n+1})| + |C|$ and 
    $|\lcs(T[1..\gamma_{A}-1], C^{n+1})| = |\lcs(T[p_{A}-1..\gamma_{A}-1], C^{n+1})|$.

    \textbf{Proof of statement (D).}    
    Because of $T[\gamma_{A}..\gamma_{A} + |C| - 1] = C$, 
    $|\lcs(T[p_{A}-1..\gamma_{A} + |C| -1], C^{n+1})| = |\lcs(T[p_{A}-1..\gamma_{A}-1], C^{n+1})| + |C|$ holds. 
    $|\lcs(T[p_{A}-1..\gamma_{A} + |C| -1], C^{n+1})| \geq |\lcs(T[p_{1}-1..\gamma_{A} + |C| -1], C^{n+1})|$ 
    because $p_{A} < p_{1}$. 
    Here, $|\lcs(T[p_{1}-1..\gamma_{A} + |C| -1], C^{n+1})| = |\lcs(T[p_{1}-1..\gamma_{1} -1], C^{n+1})|$ holds 
    because $\gamma_{1} = \gamma_{A} + |C|$.
    Because of $([p_{1}, q_{1}], [\ell_{1}, r_{1}]) \in \Psi_{\run}$, 
    $|\lcs(T[p_{1}-1..\gamma_{1} -1], C^{n+1})| = |[p_{1}-1, \gamma_{1} -1]|$ follows from 
    the definition of the subset $\Psi_{\run}$. 
    $|[p_{1}-1, \gamma_{1} -1]| = |[p_{x}-1, \gamma_{x} -1]|$ follows from 
    $|[p_{1}-1, \gamma_{1} -1]| = |[q_{A}, \gamma_{A}]| + |C|$ and $|[p_{x}-1, \gamma_{x} -1]| = |[q_{A}, \gamma_{A}]| + |C|$. 
    $|[p_{x}-1, \gamma_{x} -1]| = |[p-1, \gamma -1]|$ follows from $([p_{x}, q_{x}], [\ell_{x}, r_{x}]) = ([p, q], [\ell, r])$. 
    Therefore, $|\lcs(T[p_{A}-1..\gamma_{A}-1], C^{n+1})| \geq |[p-1, \gamma-1]| - |C|$ follows from the following equation.
\begin{equation*}
\begin{split}
|\lcs(T[p_{A}-1..\gamma_{A}-1], C^{n+1})| &= |\lcs(T[p_{A}-1..\gamma_{A} + |C| -1], C^{n+1})| - |C| \\
              &\geq |\lcs(T[p_{1}-1..\gamma_{A} + |C| -1], C^{n+1})| - |C|  \\
              &= |\lcs(T[p_{1}-1..\gamma_{1} -1], C^{n+1})| - |C|  \\
              &= |[p_{x}-1, \gamma_{x} -1]| - |C|  \\
              &= |[p-1, \gamma -1]| - |C|.
\end{split}
\end{equation*}

    \textbf{Proof of statement (E).}    
    We prove $|\lcs(T[1..\gamma_{1}-1], C^{n+1})| < |C| + |[p_{1}-1, \gamma_{1}-1]|$. 
    We apply Lemma~\ref{lem:psi_run_basic_property}~\ref{enum:psi_run_basic_property:3} and Lemma~\ref{lem:psi_run_basic_property}~\ref{enum:psi_run_basic_property:6} to the interval attractor $([p_{1}, q_{1}], [\ell_{1}, r_{1}])$. 
    Lemma~\ref{lem:psi_run_basic_property}~\ref{enum:psi_run_basic_property:3} shows that 
    $([p_{A}, q_{A}], [\ell_{A}, r_{A}]) \in \Psi_{\run} \cup \Psi_{\source}$ holds. 
    Lemma~\ref{lem:psi_run_basic_property}~\ref{enum:psi_run_basic_property:6} shows that  
    $([p_{A}, q_{A}], [\ell_{A}, r_{A}]) \in \Psi_{\run} \Leftrightarrow |\lcs(T[1..\gamma_{1}-1], C^{n+1})| \geq |C| + |[p_{1}-1, \gamma_{1}-1]|$ holds. 
    Because of $([p_{A}, q_{A}], [\ell_{A}, r_{A}]) \in \Psi_{\source}$, 
    $|\lcs(T[1..\gamma_{1}-1], C^{n+1})| < |C| + |[p_{1}-1, \gamma_{1}-1]|$ holds. 

    We prove $|\lcs(T[p_{A}-1..\gamma_{A}-1], C^{n+1})| < |[p-1, \gamma-1]|$. 
    We already proved $|[p_{1}-1, \gamma_{1}-1]| = |[p_{x}-1, \gamma_{x}-1]|$ in the proof of statement (D). 
    $|\lcs(T[p_{A}-1..\gamma_{A}-1], C^{n+1})| < |[p-1, \gamma-1]|$ follows from 
    (a) $|\lcs(T[1..\gamma_{1}-1], C^{n+1})| < |C| + |[p_{1}-1, \gamma_{1}-1]|$, 
    (b) $|\lcs(T[1..\gamma_{1}-1], C^{n+1})| = |\lcs(T[p_{A}-1..\gamma_{A}-1], C^{n+1})| + |C|$ (statement (C)), 
    (c) $|[p_{1}-1, \gamma_{1}-1]| = |[p_{x}-1, \gamma_{x}-1]|$, 
    and (d) $|[p_{x}-1, \gamma_{x}-1]| = |[p-1, \gamma-1]|$.

    \textbf{Proof of statement (F).}    
    $m = 1 + x + \lfloor (|\lcs(T[1..\gamma-1], C^{n+1})| - |[p-1, \gamma-1]|) / |C| \rfloor$ 
    follows from 
    $m = 1 + \lfloor (|\lcs(T[1..\gamma-1], C^{n+1})| - |[p-1, \gamma-1]|) / |C| \rfloor$ 
    and $|\lcs(T[1..\gamma-1], C^{n+1})| = |\lcs(T[p_{A}-1..\gamma_{A}-1], C^{n+1})| + x|C|$ (statement (C)).     
    $-|C| \leq |\lcs(T[p_{A}-1..\gamma_{A}-1], C^{n+1})| - |[p-1, \gamma-1]| < 0$ 
    follows from statement (D) and (E). 
    $\lfloor (|\lcs(T[1..\gamma-1], C^{n+1})| - |[p-1, \gamma-1]|) / |C| \rfloor = -1$ 
    follows from $-|C| \leq |\lcs(T[p_{A}-1..\gamma_{A}-1], C^{n+1})| - |[p-1, \gamma-1]| < 0$. 
    Therefore, 
\begin{equation*}
\begin{split}
m &= 1 + x + \lfloor (|\lcs(T[1..\gamma-1], C^{n+1})| - |[p-1, \gamma-1]|) / |C| \rfloor \\
              &= 1 + x - 1  \\
              &= x.
\end{split}
\end{equation*}

    \textbf{Proof of Lemma~\ref{lem:run_extension_property}(i).}    
    We prove $\Psi = \{ ([p_{A}, q_{A}], [\ell_{A}, r_{A}]) \}$ by contradiction. 
    Here, $([p_{A}, q_{A}], [\ell_{A}, r_{A}]) \in \Psi$ follows from the definition of run-extension. 
    We assume that $\Psi \neq \{ ([p_{A}, q_{A}]$, $[\ell_{A}, r_{A}]) \}$ holds. 
    Then, the subset $\Psi_{\source}$ contains an interval attractor $([p_{B}, q_{B}], [\ell_{B}, r_{B}])$ 
    ($([p_{B}, q_{B}]$, $[\ell_{B}, r_{B}]) \neq ([p_{A}, q_{A}], [\ell_{A}, r_{A}])$)
    satisfying $([p, q], [\ell, r]) \in f_{\recover}(([p_{B}, q_{B}], [\ell_{B}, r_{B}]))$. 
    $f_{\recover}(([p_{A}, q_{A}]$, $[\ell_{A}, r_{A}])) \cap f_{\recover}(([p_{B}, q_{B}], [\ell_{B}, r_{B}])) \neq \emptyset$ 
    follows from $([p, q], [\ell, r]) \in f_{\recover}(([p_{A}, q_{A}], [\ell_{A}, r_{A}]))$ 
    and $([p, q], [\ell, r]) \in f_{\recover}(([p_{B}, q_{B}], [\ell_{B}, r_{B}]))$. 
    On the other hand, $f_{\recover}(([p_{A}, q_{A}], [\ell_{A}, r_{A}])) \cap f_{\recover}(([p_{B}, q_{B}], [\ell_{B}, r_{B}])) = \emptyset$ 
    follows from Lemma~\ref{lem:recover_division_property}~\ref{enum:recover_division_property:2}. 
    The two facts $f_{\recover}(([p_{A}, q_{A}]$, $[\ell_{A}, r_{A}])) \cap f_{\recover}(([p_{B}, q_{B}], [\ell_{B}, r_{B}])) \neq \emptyset$ 
    and $f_{\recover}(([p_{A}, q_{A}], [\ell_{A}, r_{A}])) \cap f_{\recover}(([p_{B}, q_{B}], [\ell_{B}, r_{B}])) = \emptyset$ yield a contradiction. 
    Therefore, $\Psi = \{ ([p_{A}, q_{A}], [\ell_{A}, r_{A}]) \}$ must hold. 
    
    \textbf{Proof of Lemma~\ref{lem:run_extension_property}(ii).}
    We prove $q_{A} = p - (m-1)|C| - 1$ and $r_{A} = r$. 
    $p_{x} = q_{A} + 1 + (x-1)|C|$ and $r_{x} = r_{A}$ follow from statement (A). 
    $q_{A} = p - (m-1)|C| - 1$ follows from 
    $p_{x} = q_{A} + 1 + (x-1)|C|$, $x = m$, and $p_{x} = p$.
    $r_{A} = r$ follows from $r_{x} = r_{A}$ and $r = r_{x}$. 

    We prove $I_{\capture}(p - (m-1) |C| - 1, r) = ([p_{A}, q_{A}], [\ell_{A}, r_{A}])$. 
    We apply Lemma~\ref{lem:IA_maximal_lemma} to the interval attractor $([p_{A}, q_{A}], [\ell_{A}, r_{A}])$. 
    Then, the lemma shows that 
    $I_{\capture}(q_{A}, r_{A}) = ([p_{A}, q_{A}], [\ell_{A}, r_{A}])$. 
    Therefore, $I_{\capture}(p - (m-1) |C| - 1, r) = ([p_{A}, q_{A}], [\ell_{A}, r_{A}])$ 
    follows from $q_{A} = p - (m-1)|C| - 1$ and $r_{A} = r$. 

    \textbf{Proof of Lemma~\ref{lem:run_extension_property}(iii).} 
    The length $|\lcs(T[1..\gamma-1], C^{n+1})|$ of the longest common suffix between 
    two strings $T[1..\gamma-1]$ and $C^{n+1}$ can be computed by reversed LCE query $\rLCEQ(\gamma-1, \gamma+|C|-1)$. 
    $m = 1 + \lfloor (\rLCEQ(\gamma-1, \gamma+|C|-1) - |[p-1, \gamma-1]|) / |C| \rfloor$ 
    follows from 
    $m = 1 + \lfloor (|\lcs(T[1..\gamma-1], C^{n+1})| - |[p-1, \gamma-1]|) / |C| \rfloor$ 
    and $|\lcs(T[1..\gamma-1], C^{n+1})| = \rLCEQ(\gamma-1, \gamma+|C|-1)$. 
    
\end{proof}

We explain the algorithm for computing all the run-extensions of a given interval attractor $([p, q], [\ell, r]) \in \Psi_{\RR}$ 
using Lemma~\ref{lem:run_extension_property}. 
If $([p, q], [\ell, r]) \not \in \Psi_{\run}$ holds, 
then the given interval attractor has no run-extensions because 
$([p, q], [\ell, r]) \not \in f_{\recover}(([p^{\prime}, q^{\prime}], [\ell^{\prime}, r^{\prime}]))$ holds
for each interval attractor $([p^{\prime}, q^{\prime}], [\ell^{\prime}, r^{\prime}]) \in \Psi_{\source}$. 
Otherwise (i.e., $([p, q], [\ell, r]) \in \Psi_{\run}$), 
Lemma~\ref{lem:run_extension_property}~\ref{lem:run_extension_property:1} shows that 
the subset $\Psi_{\source}$ contains the only one interval attractor $([p^{\prime}, q^{\prime}], [\ell^{\prime}, r^{\prime}]) \in \Psi_{\source}$ such that it is a run-extension of the given interval attractor. 
From Lemma~\ref{lem:run_extension_property}~\ref{lem:run_extension_property:2}, 
there exists an integer $m \geq 0$ satisfying $([p^{\prime}, q^{\prime}], [\ell^{\prime}, r^{\prime}]) = I_{\capture}(p - (m-1) |C| - 1, r)$. This interval attractor $I_{\capture}(p - (m-1) |C| - 1, r)$ can be obtained by capture query. 
Lemma~\ref{lem:run_extension_property}~\ref{lem:run_extension_property:3} shows that 
$m = 1 + \lfloor (\rLCEQ(\gamma-1, \gamma+|C|-1) - |[p-1, \gamma-1]|) / |C| \rfloor$ holds 
for the attractor position $\gamma$ and associated string $C$ of the given interval attractor. 
Therefore, all the run-extensions of a given interval attractor can be computed using capture query and reversed LCE query. 

We compute all the run-extensions of the interval attractor $([p, q], [\ell, r])$ in the following five steps: 

\begin{enumerate}[label=\textbf{(\roman*)}]
    \item verify whether $([p, q], [\ell, r]) \in \Psi_{\run}$ or not by verify-run query $\runQ(([p, q], [\ell, r]))$. 
    If $([p, q], [\ell, r]) \not \in \Psi_{\run}$, then the given interval attractor has no run-extensions; 
    otherwise, proceed to the next step;
    \item compute the attractor position $\gamma$ of the given interval attractor 
    by attractor position query $\attrQ(([p, q], [\ell, r]))$;
    \item compute the length $|C|$ of the associated string $C$ of the given interval attractor 
    by C-length query $\clenQ(([p, q], [\ell, r]))$;
    \item execute reversed LCE query $\rLCEQ(\gamma-1, \gamma+|C|-1)$ 
    and compute $m = 1 + \lfloor (|\lcs(T[1..\gamma-1], C^{n+1})| - |[p-1, \gamma-1]|) / |C| \rfloor$; 
    \item return the run-extension of the given interval attractor by capture query $\CAPQ([p - (m-1) |C| - 1, r])$.
\end{enumerate}

The bottleneck of this algorithm is capture query, which takes $O(H^{2} \log n)$ time. 
Therefore, we can compute all the run-extensions of a given interval attractor in $O(H^{2} \log n)$ time.

%%%%%%%%%%%%%%%%%%%%%%%%%%%%%%%%%%%%%%%%%%%%%%%%%%%%%%%%%%%%%%%%%

\paragraph{Two subsets $\Psi_{\OVQ, A}$ and $\Psi_{\OVQ, B}$ of set $\Psi_{\RR}$.}

For this subsection, 
we introduce two subsets $\Psi_{\OVQ, A}$ and $\Psi_{\OVQ, B}$ of the set $\Psi_{\RR}$. 
The two subsets $\Psi_{\OVQ, A}$ and $\Psi_{\OVQ, B}$ are defined as follows: 
\begin{description}
    \item[$\Psi_{\OVQ, A}$:] it consists of all the left-extensions of interval attractors in set $(\Psi_{\OVQ, C} \cup \{ I_{\capture}(i, j^{\prime}) \mid j^{\prime} \in [i+1, n] \} ) \setminus \Psi_{\run}$ (i.e., $\Psi_{\OVQ, A} = \{ ([p, q], [\ell, r]) \in \Psi_{\RR} \mid \exists ([p^{\prime}, q^{\prime}], [\ell^{\prime}, r^{\prime}]) \in (\Psi_{\OVQ, C} \cup \{ I_{\capture}(i, j^{\prime}) \mid j^{\prime} \in [i+1, n] \}) \setminus \Psi_{\run} \mbox{ s.t. } (q+1 = p^{\prime}) \land (r \in [\ell^{\prime}, r^{\prime}]) \}$); 
    \item[$\Psi_{\OVQ, B}$:]: it consists of all the run-extensions of interval attractors in set $\{ I_{\capture}(i, j^{\prime}) \mid j^{\prime} \in [i+1, n] \} \cup \{ ([p, q], [\ell, r]) \in \Psi_{\OVQ, A} \mid r \geq i+1 \}$ (i.e., $\Psi_{\OVQ, B} = \{ ([p, q], [\ell, r]) \in \Psi_{\RR} \mid ([p, q], [\ell, r]) \in \Psi_{\source} \mbox{ and } \exists ([p^{\prime}, q^{\prime}], [\ell^{\prime}, r^{\prime}]) \in \{ I_{\capture}(i, j^{\prime}) \mid j^{\prime} \in [i+1, n] \} \cup \Psi^{\prime}_{A} \mbox{ s.t. } ([p^{\prime}, q^{\prime}], [\ell^{\prime}, r^{\prime}]) \in f_{\recover}(([p, q], [\ell, r])) \}$ for the set $\Psi^{\prime}_{A} = \{ ([p, q], [\ell, r]) \in \Psi_{\OVQ, A} \mid r \geq i+1 \}$).
\end{description}

The following lemma states properties of the two subsets $\Psi_{\OVQ, A}$ and $\Psi_{\OVQ, B}$. 

\begin{lemma}\label{lem:ovqab_property}
The following three statements hold for an overlap query $\OVQ([i, j])$: 

\begin{enumerate}[label=\textbf{(\roman*)}]
    \item \label{enum:ovqab_property:1}
    consider an interval attractor $([p, q], [\ell, r]) \in \Psi_{\OVQ, A}$. 
    Then, $q \leq i-1$, and 
    the set $\Psi_{\RR} \setminus \Psi_{\run}$ contains 
    an interval attractor $([p^{\prime}, q^{\prime}], [\ell^{\prime}, r^{\prime}])$ such that 
    $([p, q], [\ell, r])$ is a left-extension of the interval attractor $([p^{\prime}, q^{\prime}], [\ell^{\prime}, r^{\prime}])$; 
    \item \label{enum:ovqab_property:2}
    consider an interval attractor $([p, q], [\ell, r]) \in \Psi_{\OVQ, B}$. 
    Then, $q \leq i-1$, $r \geq i+1$ and $([p, q], [\ell, r]) \in \Psi_{\source}$.
    \item \label{enum:ovqab_property:3} 
    $|\Psi_{\OVQ, A} \cap \Psi_{\run}| \leq |\{ ([p, q], [\ell, r]) \in \Psi_{\RR} \setminus \Psi_{\run} \mid i \in [p, r] \}|$.    
\end{enumerate}
\end{lemma}
\begin{proof}
    The proof of Lemma~\ref{lem:ovqab_property} is as follows. 

    \textbf{Proof of Lemma~\ref{lem:ovqab_property}(i).}
    Because of $([p, q], [\ell, r]) \in \Psi_{\OVQ, A}$, 
    the set $(\Psi_{\OVQ, C} \cup \{ I_{\capture}(i, j^{\prime}) \mid j^{\prime} \in [i+1, n] \}) \setminus \Psi_{\run}$ contains 
    an interval attractor $([p^{\prime}, q^{\prime}], [\ell^{\prime}, r^{\prime}])$ such that 
    $([p, q], [\ell, r])$ is a left-extension of the interval attractor $([p^{\prime}, q^{\prime}], [\ell^{\prime}, r^{\prime}])$ 
    (i.e., $q + 1 = p^{\prime}$ and $r \in [\ell^{\prime}, r^{\prime}]$). 
    $([p^{\prime}, q^{\prime}], [\ell^{\prime}, r^{\prime}]) \in \Psi_{\RR} \setminus \Psi_{\run}$ 
    follows from $([p^{\prime}, q^{\prime}], [\ell^{\prime}, r^{\prime}]) \in (\Psi_{\OVQ, C} \cup \{ I_{\capture}(i, j^{\prime}) \mid j^{\prime} \in [i+1, n] \}) \setminus \Psi_{\run}$. 
    If $p^{\prime} \leq i$ holds, 
    then $q \leq i-1$ follows from $q + 1 = p^{\prime}$.
    
    We prove $p^{\prime} \leq i$. 
    If $([p^{\prime}, q^{\prime}], [\ell^{\prime}, r^{\prime}]) \in \{ I_{\capture}(i, j^{\prime}) \mid j^{\prime} \in [i+1, n] \}$, 
    then there exists an integer $j^{\prime}$ satisfying $I_{\capture}(i, j^{\prime}) = ([p^{\prime}, q^{\prime}], [\ell^{\prime}, r^{\prime}])$. 
    $i \in [p^{\prime}, q^{\prime}]$) follows from the definition of interval attractor 
    for the interval attractor $I_{\capture}(i, j^{\prime}) = ([p^{\prime}, q^{\prime}], [\ell^{\prime}, r^{\prime}])$. 
    Therefore, $p^{\prime} \leq i$ follows from $i \in [p^{\prime}, q^{\prime}]$. 
    
    Otherwise (i.e., $([p^{\prime}, q^{\prime}], [\ell^{\prime}, r^{\prime}]) \not \in \{ I_{\capture}(i, j^{\prime}) \mid j^{\prime} \in [i+1, n] \}$), 
    the subset $\Psi_{\OVQ, C}$ contains the interval attractor $([p^{\prime}, q^{\prime}], [\ell^{\prime}, r^{\prime}])$. 
    $p^{\prime} \leq i-1$ follows from $([p^{\prime}, q^{\prime}], [\ell^{\prime}, r^{\prime}]) \in \Psi_{\OVQ, C}$. 
    Therefore, $p^{\prime} \leq i$ holds.
    
    \textbf{Proof of Lemma~\ref{lem:ovqab_property}(ii).}
    Because of $([p, q], [\ell, r]) \in \Psi_{\OVQ, B}$, 
    the set $\{ I_{\capture}(i, j^{\prime}) \mid j^{\prime} \in [i+1, n] \} \cup \{ ([p, q], [\ell, r]) \in \Psi_{\OVQ, A} \mid r \geq i+1 \}$ contains
    an interval attractor $([p^{\prime}, q^{\prime}], [\ell^{\prime}, r^{\prime}])$ such that 
    $([p, q], [\ell, r])$ is a run-extension of the interval attractor $([p^{\prime}, q^{\prime}], [\ell^{\prime}, r^{\prime}])$ 
    (i.e., $([p, q], [\ell, r]) \in \Psi_{\source}$, $([p^{\prime}, q^{\prime}], [\ell^{\prime}, r^{\prime}]) \in \Psi_{\run}$, and $([p^{\prime}, q^{\prime}], [\ell^{\prime}, r^{\prime}]) \in f_{\recover}(([p, q], [\ell, r]))$). 
    Therefore, $([p, q], [\ell, r]) \in \Psi_{\source}$ holds. 

    We prove $p^{\prime} \leq i$ and $r^{\prime} \geq i+1$. 
    If $([p^{\prime}, q^{\prime}], [\ell^{\prime}, r^{\prime}]) \in \{ I_{\capture}(i, j^{\prime}) \mid j^{\prime} \in [i+1, n] \}$, 
    then there exists an integer $j^{\prime} \in [i+1, n]$ satisfying $I_{\capture}(i, j^{\prime}) = ([p^{\prime}, q^{\prime}], [\ell^{\prime}, r^{\prime}])$. 
    $i \in [p^{\prime}, q^{\prime}]$ and $\ell^{\prime} \geq i+1$ follow from the definition of interval attractor 
    because 
    $I_{\capture}(i, j^{\prime}) = ([p^{\prime}, q^{\prime}], [\ell^{\prime}, r^{\prime}])$ 
    and $j^{\prime} \geq i+1$.
    $p^{\prime} \leq i$ follows from $i \in [p^{\prime}, q^{\prime}]$. 
    $r^{\prime} \geq i+1$ follows from $\ell^{\prime} \geq i+1$ and $\ell^{\prime} \leq r^{\prime}$. 

    Otherwise (i.e., $([p^{\prime}, q^{\prime}], [\ell^{\prime}, r^{\prime}]) \not \in \{ I_{\capture}(i, j^{\prime}) \mid j^{\prime} \in [i+1, n] \}$), 
    $([p^{\prime}, q^{\prime}], [\ell^{\prime}, r^{\prime}]) \in \{ ([p, q], [\ell, r]) \in \Psi_{\OVQ, A} \mid r \geq i+1 \}$ holds. 
    $q^{\prime} \leq i-1$ follows from statement (1). 
    $p^{\prime} \leq i$ follows from $p^{\prime} \leq q^{\prime}$ and $q^{\prime} \leq i-1$. 
    $r^{\prime} \geq i+1$ follows from $([p^{\prime}, q^{\prime}], [\ell^{\prime}, r^{\prime}]) \in \{ ([p, q], [\ell, r]) \in \Psi_{\OVQ, A} \mid r \geq i+1 \}$. 

    We prove $q \leq i-1$ and $r \geq i+1$. 
    Let $([p_{1}, q_{1}], [\ell_{1}, r_{1}])$, 
    $([p_{2}, q_{2}], [\ell_{2}, r_{2}])$, $\ldots$, $([p_{k}, q_{k}], [\ell_{k}, r_{k}])$ ($p_{1} \leq p_{2} \leq \cdots \leq p_{k}$) 
    be the interval attractors obtained from the function $f_{\recover}(([p, q], [\ell, r]))$.     
    For each integer $s \in [1, k]$, 
    $p_{1} \leq p_{s}$ and $r_{s} = r$ 
    follow from the definition of the function $f_{\recover}$. 
    Because of $([p^{\prime}, q^{\prime}], [\ell^{\prime}, r^{\prime}]) \in f_{\recover}(([p, q], [\ell, r]))$, 
    $p_{1} \leq p^{\prime}$ and $r^{\prime} = r$ hold.      
    $q = p_{1} - 1$ follows from Lemma~\ref{lem:recover_basic_property}~\ref{enum:recover_basic_property:3}. 
    Therefore, $q \leq i-1$ follows from 
    $q = p_{1} - 1$, $p_{1} \leq p^{\prime}$, and $p^{\prime} \leq i$. 
    Similarly, $r \geq i+1$ follows from $r^{\prime} = r$ and $r^{\prime} \geq i+1$. 
    
    \textbf{Proof of Lemma~\ref{lem:ovqab_property}(iii).}
    Let $([p_{1}, q_{1}], [\ell_{1}, r_{1}])$, 
    $([p_{2}, q_{2}], [\ell_{2}, r_{2}])$, $\ldots$, $([p_{k}, q_{k}], [\ell_{k}, r_{k}])$ 
    be the interval attractors in the set $\Psi_{\OVQ, A} \cap \Psi_{\run}$. 
    Let $C_{s}$ be the associated string of each interval attractor $([p_{s}, q_{s}], [\ell_{s}, r_{s}])$. 
    Then, Lemma~\ref{lem:left_extension_property}~\ref{enum:left_extension_property:3} shows that 
    the interval attractor $([p_{s}, q_{s}], [\ell_{s}, r_{s}])$ is a left-extension of 
    interval attractor $([p_{s} + |C_{s}|, q_{s} + |C_{s}|], [\ell_{s} + |C_{s}|, r_{s}])$ 
    (i.e., $q_{s} + 1 = p_{s} + |C_{s}|$ and $r_{s} \in [\ell_{s} + |C_{s}|, r_{s}]$). 

    We prove $k = |\{ ([p_{s} + |C_{s}|, q_{s} + |C_{s}|], [\ell_{s} + |C_{s}|, r_{s}]) \mid s \in [1, k] \}|$ by contradiction. 
    We assume that $k = |\{ ([p_{s} + |C_{s}|, q_{s} + |C_{s}|], [\ell_{s} + |C_{s}|, r_{s}]) \mid s \in [1, k] \}|$ does not hold. 
    Then, there exist two integers $1 \leq x \leq y \leq k$ satisfying 
    $([p_{x} + |C_{x}|, q_{x} + |C_{x}|], [\ell_{x} + |C_{x}|, r_{x}]) = ([p_{y} + |C_{y}|, q_{y} + |C_{y}|], [\ell_{y} + |C_{y}|, r_{y}])$. 
    $q_{x} + 1 = p_{y} + |C_{y}|$ follows from $q_{x} + 1 = p_{x} + |C_{x}|$ and $p_{x} + |C_{x}| = p_{y} + |C_{y}|$. 
    $r_{x} \in [\ell_{y} + |C_{y}|, r_{y}]$ follows from $r_{x} \in [\ell_{x} + |C_{x}|, r_{x}]$ 
    and $[\ell_{x} + |C_{x}|, r_{x}] = [\ell_{y} + |C_{y}|, r_{y}]$. 
    Because of $q_{x} + 1 = p_{y} + |C_{y}|$ and $r_{x} \in [\ell_{y} + |C_{y}|, r_{y}]$, 
    the interval attractor $([p_{x}, q_{x}], [\ell_{x}, r_{x}])$ is a left-extension of 
    interval attractor $([p_{y} + |C_{y}|, q_{y} + |C_{y}|], [\ell_{y} + |C_{y}|, r_{y}])$. 
    The interval attractor $([p_{x}, q_{x}], [\ell_{x}, r_{x}])$ is a left-extension of 
    both $([p_{x} + |C_{x}|, q_{x} + |C_{x}|], [\ell_{x} + |C_{x}|, r_{x}])$ and $([p_{y} + |C_{y}|, q_{y} + |C_{y}|], [\ell_{y} + |C_{y}|, r_{y}])$. 
    This fact contradicts Lemma~\ref{lem:left_extension_property}~\ref{enum:left_extension_property:2}, 
    i.e., 
    the interval attractor $([p_{x}, q_{x}], [\ell_{x}, r_{x}])$ is a left-extension of at most one interval attractor.  
    Therefore, $k = |\{ ([p_{s} + |C_{s}|, q_{s} + |C_{s}|], [\ell_{s} + |C_{s}|, r_{s}]) \mid s \in [1, k] \}|$ must hold. 
    
    We prove $\{ ([p_{s} + |C_{s}|, q_{s} + |C_{s}|], [\ell_{s} + |C_{s}|, r_{s}]) \mid s \in [1, k] \} \subseteq (\Psi_{\OVQ, C} \cup \{ I_{\capture}(i, j^{\prime}) \mid j^{\prime} \in [i+1, n] \}) \setminus \Psi_{\run}$ by contradiction. 
    We assume that $\{ ([p_{s} + |C_{s}|, q_{s} + |C_{s}|], [\ell_{s} + |C_{s}|, r_{s}]) \mid s \in [1, k] \} \subseteq (\Psi_{\OVQ, C} \cup \{ I_{\capture}(i, j^{\prime}) \mid j^{\prime} \in [i+1, n] \}) \setminus \Psi_{\run}$ does not hold. 
    Then, there exists an integer $s \in [1, k]$ satisfying 
    $([p_{s} + |C_{s}|, q_{s} + |C_{s}|], [\ell_{s} + |C_{s}|, r_{s}]) \not \in (\Psi_{\OVQ, C} \cup \{ I_{\capture}(i, j^{\prime}) \mid j^{\prime} \in [i+1, n] \}) \setminus \Psi_{\run}$. 
    Because of $([p_{s}, q_{s}], [\ell_{s}, r_{s}]) \in \Psi_{\OVQ, A}$, 
    the set $(\Psi_{\OVQ, C} \cup \{ I_{\capture}(i, j^{\prime}) \mid j^{\prime} \in [i+1, n] \}) \setminus \Psi_{\run}$ contains an interval attractor 
    $([p^{\prime}, q^{\prime}], [\ell^{\prime}, r^{\prime}])$ such that 
    the interval attractor $([p_{s}, q_{s}], [\ell_{s}, r_{s}])$ is a left-extension of 
    interval attractor $([p^{\prime}, q^{\prime}], [\ell^{\prime}, r^{\prime}])$ (see the definition of the subset $\Psi_{\OVQ, A}$). 
    The interval attractor $([p_{s}, q_{s}], [\ell_{s}, r_{s}])$ is a left-extension of two interval attractors, 
    but this fact contradicts Lemma~\ref{lem:left_extension_property}~\ref{enum:left_extension_property:2}. 
    Therefore, $\{ ([p_{s} + |C_{s}|, q_{s} + |C_{s}|], [\ell_{s} + |C_{s}|, r_{s}]) \mid s \in [1, k] \} \subseteq (\Psi_{\OVQ, C} \cup \{ I_{\capture}(i, j^{\prime}) \mid j^{\prime} \in [i+1, n] \}) \setminus \Psi_{\run}$ must hold. 

    We prove $\Psi_{\OVQ, C} \cup \{ I_{\capture}(i, j^{\prime}) \mid j^{\prime} \in [i+1, n] \} \subseteq \{ ([p, q], [\ell, r]) \in \Psi_{\RR} \mid i \in [p, r] \}$. 
    For each interval attractor $([p, q], [\ell, r]) \in \Psi_{\OVQ, C} \cup \{ I_{\capture}(i, j^{\prime}) \mid j^{\prime} \in [i+1, n] \}$, 
    if $([p, q], [\ell, r]) \in \Psi_{\OVQ, C}$, 
    then $q \leq i-1$ and $r \geq i+1$ follow from the definition of the subset $\Psi_{\OVQ, C}$. 
    $i \in [p, r]$ follows from $p \leq q$, $q \leq i-1$, and $r \geq i+1$. 
    Otherwise (i.e., $([p, q], [\ell, r]) \not \in \Psi_{\OVQ, C}$), 
    $([p, q], [\ell, r]) \in \{ I_{\capture}(i, j^{\prime}) \mid j^{\prime} \in [i+1, n] \}$ holds, 
    which indicates that there exists an integer $j^{\prime} \in [i+1, n]$ satisfying 
    $I_{\capture}(i, j^{\prime}) = ([p, q], [\ell, r])$. 
    Since $I_{\capture}(i, j^{\prime}) = ([p, q], [\ell, r])$, 
    $i \in [p, q]$ follows from the definition of interval attractor. 
    $i \in [p, r]$ follows from $i \in [p, q]$ and $[p, q] \subseteq [p, r]$. 
    Therefore, $\Psi_{\OVQ, C} \cup \{ I_{\capture}(i, j^{\prime}) \mid j^{\prime} \in [i+1, n] \} \subseteq \{ ([p, q], [\ell, r]) \in \Psi_{\RR} \mid i \in [p, r] \}$ holds. 

    We prove $|\Psi_{\OVQ, A} \cap \Psi_{\run}| \leq |\{ ([p, q], [\ell, r]) \in \Psi_{\RR} \setminus \Psi_{\run} \mid i \in [p, r] \}|$. 
    Here, $|\Psi_{\OVQ, A} \cap \Psi_{\run}| = k$ holds. 
    $\{ ([p_{s} + |C_{s}|, q_{s} + |C_{s}|], [\ell_{s} + |C_{s}|, r_{s}]) \mid s \in [1, k] \} \subseteq \{ ([p, q], [\ell, r]) \in \Psi_{\RR} \setminus \Psi_{\run} \mid i \in [p, r] \}$ follows from 
    $\{ ([p_{s} + |C_{s}|, q_{s} + |C_{s}|], [\ell_{s} + |C_{s}|, r_{s}]) \mid s \in [1, k] \} \subseteq (\Psi_{\OVQ, C} \cup \{ I_{\capture}(i, j^{\prime}) \mid j^{\prime} \in [i+1, n] \}) \setminus \Psi_{\run}$ and $\Psi_{\OVQ, C} \cup \{ I_{\capture}(i, j^{\prime}) \mid j^{\prime} \in [i+1, n] \} \subseteq \{ ([p, q], [\ell, r]) \in \Psi_{\RR} \mid i \in [p, r] \}$. 
    $k \leq |\{ ([p, q], [\ell, r]) \in \Psi_{\RR} \setminus \Psi_{\run} \mid i \in [p, r] \}|$ follows from $\{ ([p_{s} + |C_{s}|, q_{s} + |C_{s}|], [\ell_{s} + |C_{s}|, r_{s}]) \mid s \in [1, k] \} \subseteq \{ ([p, q], [\ell, r]) \in \Psi_{\RR} \setminus \Psi_{\run} \mid i \in [p, r] \}$. 
    Therefore, $|\Psi_{\OVQ, A} \cap \Psi_{\run}| \leq |\{ ([p, q], [\ell, r]) \in \Psi_{\RR} \setminus \Psi_{\run} \mid i \in [p, r] \}|$ follows from $|\Psi_{\OVQ, A} \cap \Psi_{\run}| = k$ and $k \leq |\{ ([p, q], [\ell, r]) \in \Psi_{\RR} \setminus \Psi_{\run} \mid i \in [p, r] \}|$. 
\end{proof}

The following lemma states the relationship among the three subsets $(\Psi_{\OVQ, C} \setminus \Psi_{\run})$, $\{ ([p, q]$, $[\ell, r]) \in \Psi_{\OVQ, A} \mid r \geq i+1 \}$, and $\Psi_{\OVQ, B}$. 

\begin{lemma}\label{lem:RB_left_extension}
$\Psi_{\OVQ, C} \setminus \Psi_{\run} \subseteq \{ ([p, q], [\ell, r]) \in \Psi_{\OVQ, A} \mid r \geq i+1 \} \cup \Psi_{\OVQ, B}$ 
for an overlap query $\OVQ([i, j])$. 
\end{lemma}
\begin{proof}
Consider an interval attractor $([p, q], [\ell, r])$ in the set $\Psi_{\OVQ, C} \setminus \Psi_{\run}$. 
Let $h \in [0, H]$ be an integer satisfying $([p, q], [\ell, r]) \in \Psi_{h}$; 
let $\gamma$ and $C$ be the attractor position and associated string of the interval attractor $([p, q], [\ell, r])$; 
let $([p^{\prime}, q^{\prime}], [\ell^{\prime}, r^{\prime}])$ be the interval attractor $I_{\capture}(q+1, r)$. 
Here, Lemma~\ref{lem:left_extension_property}~\ref{enum:left_extension_property:1} ensures that 
the interval attractor $([p, q], [\ell, r])$ is a left-extension of the interval attractor $([p^{\prime}, q^{\prime}], [\ell^{\prime}, r^{\prime}])$ because $q + 1 < r$ follows from $([p, q], [\ell, r]) \in \Psi_{\OVQ, C}$. 
If $([p, q], [\ell, r]) \in \Psi_{\source}$, 
then let $([p_{1}, q_{1}], [\ell_{1}, r_{1}])$, $([p_{2}, q_{2}], [\ell_{2}, r_{2}])$, 
$\ldots$, $([p_{k}, q_{k}], [\ell_{k}, r_{k}])$ ($p_{1} \leq p_{2} \leq \cdots \leq p_{k}$) be the interval attractors obtained from the function $f_{\recover}(([p, q], [\ell, r]))$. 

Lemma~\ref{lem:RB_left_extension} holds if $([p, q], [\ell, r]) \in \{ ([\hat{p}, \hat{q}], [\hat{\ell}, \hat{r}]) \in \Psi_{\OVQ, A} \mid \hat{r} \geq i+1 \} \cup \Psi_{\OVQ, B}$ holds. 
The following three statements are used to prove $([p, q], [\ell, r]) \in  \{ ([\hat{p}, \hat{q}], [\hat{\ell}, \hat{r}]) \in \Psi_{\OVQ, A} \mid \hat{r} \geq i+1 \} \cup \Psi_{\OVQ, B}$.

\begin{enumerate}[label=\textbf{(\arabic*)}]
    \item $([p^{\prime}, q^{\prime}], [\ell^{\prime}, r^{\prime}]) \in \Psi_{\OVQ, C} \cup \{ I_{\capture}(i, j^{\prime}) \mid j^{\prime} \in [i+1, n] \}$;
    \item $([p, q], [\ell, r]) \in \Psi_{\source}$ if $([p^{\prime}, q^{\prime}], [\ell^{\prime}, r^{\prime}]) \in \Psi_{\run}$;
    \item $([p_{k}, q_{k}], [\ell_{k}, r_{k}]) \in \Psi_{\OVQ, C} \cap \Psi_{\run} \cap \Psi_{\centerset}(C)$ 
    and $([p_{k} + |C|, q_{k} + |C|], [\ell_{k} + |C|, r_{k}]) \in (\Psi_{\OVQ, C} \cup \{ I_{\capture}(i, j^{\prime}) \mid j^{\prime} \in [i+1, n] \}) \setminus \Psi_{\run}$
    if $([p, q], [\ell, r]) \in \Psi_{\source}$ and $i \not \in [p_{s}, q_{s}]$ for all $s \in [1, k]$.
\end{enumerate}

\textbf{Proof of statement (1).}
$q \leq i-1$ and $r \geq i+1$ follow from $([p, q], [\ell, r]) \in \Psi_{\OVQ, C}$.
$p^{\prime} = q+1$ and $r \in [\ell^{\prime}, r^{\prime}]$ hold because 
the interval attractor $([p, q], [\ell, r])$ is a left-extension of the interval attractor $([p^{\prime}, q^{\prime}], [\ell^{\prime}, r^{\prime}])$. 
$r^{\prime} \geq i+1$ follows from $r \in [\ell^{\prime}, r^{\prime}]$ and $r \geq i+1$. 

We prove $([p^{\prime}, q^{\prime}], [\ell^{\prime}, r^{\prime}]) \in \Psi_{\OVQ, C} \cup \{ I_{\capture}(i, j^{\prime}) \mid j^{\prime} \in [i+1, n] \}$. 
If $q^{\prime} \leq i-1$, 
then $([p^{\prime}, q^{\prime}], [\ell^{\prime}, r^{\prime}]) \in \Psi_{\OVQ, C}$ 
follows from $q^{\prime} \leq i-1$ and $r^{\prime} \geq i+1$. 
Otherwise (i.e., $q^{\prime} \geq i$), 
Lemma~\ref{lem:IA_maximal_lemma} shows that 
$I_{\capture}(i, r^{\prime}) = ([p^{\prime}, q^{\prime}], [\ell^{\prime}, r^{\prime}])$ holds. 
$([p^{\prime}, q^{\prime}], [\ell^{\prime}, r^{\prime}]) \in \{ I_{\capture}(i, j^{\prime}) \mid j^{\prime} \in [i+1, n] \}$ holds 
because $I_{\capture}(i, r^{\prime}) = ([p^{\prime}, q^{\prime}], [\ell^{\prime}, r^{\prime}])$ and $r^{\prime} \geq i+1$. 
Therefore, $([p^{\prime}, q^{\prime}], [\ell^{\prime}, r^{\prime}]) \in \Psi_{\OVQ, C} \cup \{ I_{\capture}(i, j^{\prime}) \mid j^{\prime} \in [i+1, n] \}$ holds.

\textbf{Proof of statement (2).}
Let $h^{\prime} \in [0, H]$ be an integer satisfying $([p^{\prime}, q^{\prime}], [\ell^{\prime}, r^{\prime}]) \in \Psi_{h^{\prime}}$. 
We apply Lemma~\ref{lem:psi_run_basic_property}~\ref{enum:psi_run_basic_property:3} to the interval attractor $([p^{\prime}, q^{\prime}], [\ell^{\prime}, r^{\prime}])$. 
Then, the lemma shows that the set $\Psi_{h^{\prime}}$ contains 
interval attractor $([p^{\prime}_{1}, q^{\prime}_{1}], [\ell^{\prime}_{1}, r^{\prime}_{1}])$ satisfying 
$p^{\prime}_{1} < p^{\prime}$, $q^{\prime}_{1} = p^{\prime}-1$, $\ell^{\prime}_{1} \leq \ell^{\prime}$, $r^{\prime}_{1} = r^{\prime}$, and 
$([p^{\prime}_{1}, q^{\prime}_{1}], [\ell^{\prime}_{1}, r^{\prime}_{1}]) \in \Psi_{\run} \cup \Psi_{\source}$. 
$q \in [p^{\prime}_{1}, q^{\prime}_{1}]$ follows from 
$q = p^{\prime} - 1$, $p^{\prime}_{1} < p^{\prime}$, and $q^{\prime}_{1} = p^{\prime}-1$. 
$r \leq [\ell^{\prime}_{1}, r^{\prime}_{1}]$ follows from 
$r \in [\ell^{\prime}, r^{\prime}]$, $r^{\prime} = r^{\prime}_{1}$, and $\ell^{\prime}_{1} \leq \ell^{\prime}$. 

We prove $I_{\capture}(q, r) \in \bigcup_{t = h^{\prime}}^{H} \Psi_{t}$ 
for interval attractor $I_{\capture}(q, r)$. 
We consider interval attractor $I_{\capture}(q+1, r) = ([p^{\prime}, q^{\prime}], [\ell^{\prime}, r^{\prime}])$. 
$I_{\capture}(p^{\prime}-1, r) \in \bigcup_{t = h^{\prime}}^{H} \Psi_{t}$ follows from 
Lemma~\ref{lem:interval_extension_propertyX}~\ref{enum:interval_extension_propertyX:left}. 
Therefore, $I_{\capture}(q, r) \in \bigcup_{t = h^{\prime}}^{H} \Psi_{t}$ follows from 
$I_{\capture}(p^{\prime}-1, r) \in \bigcup_{t = h^{\prime}}^{H} \Psi_{t}$ and $q = p^{\prime} - 1$. 

We prove $I_{\capture}(q, r) = ([p^{\prime}_{1}, q^{\prime}_{1}], [\ell^{\prime}_{1}, r^{\prime}_{1}])$ by contradiction. 
We assume that $I_{\capture}(q, r) \neq ([p^{\prime}_{1}, q^{\prime}_{1}], [\ell^{\prime}_{1}, r^{\prime}_{1}])$. 
Consider sequence $A(q, r) = [s^{0}, e^{0}], [s^{1}, e^{1}], \ldots, [s^{k^{\prime\prime}}, e^{k^{\prime\prime}}]$ of intervals. 
Then, $[q, r] \in \Delta(k^{\prime\prime}, s^{k^{\prime\prime}})$ follows from 
the definition of the set $\Delta(k^{\prime\prime}, s^{k^{\prime\prime}})$ introduced in Section~\ref{subsec:RR_delta}. 
Here, $I_{\capture}(q, r) = I(s^{k^{\prime\prime}})$ holds, 
and the level of the interval attractor $I_{\capture}(q, r)$ is $k^{\prime\prime}$. 
$k^{\prime\prime} \geq h^{\prime}$ follows from $I_{\capture}(q, r) \subseteq \bigcup_{t = h^{\prime}}^{H} \Psi_{t}$. 

From Definition~\ref{def:RR_Delta}, 
there exists a position $b \in [1, |S^{h^{\prime}}|]$ in sequence $S^{h^{\prime}}$ 
satisfying 
$p^{\prime}_{1} = \min \{ s \mid [s, e] \in \Delta(h^{\prime}_{1}, b) \}$, 
$q^{\prime}_{1} = \max \{ s \mid [s, e] \in \Delta(h^{\prime}_{1}, b) \}$, 
$\ell^{\prime}_{1} = \min \{ e \mid [s, e] \in \Delta(h^{\prime}_{1}, b) \}$, 
and $r^{\prime}_{1} = \max \{ e \mid [s, e] \in \Delta(h^{\prime}_{1}, b) \}$. 
Since $I_{\capture}(q, r) \neq ([p^{\prime}_{1}, q^{\prime}_{1}], [\ell^{\prime}_{1}, r^{\prime}_{1}])$, 
$[q, r] \not \in \Delta(h^{\prime}_{1}, b)$ follows from the definition of the set $\Delta(k^{\prime\prime}, s^{k^{\prime\prime}})$. 
In this case, Lemma~\ref{lem:interval_extension_propertyX}~\ref{enum:interval_extension_propertyX:low} shows that 
$k^{\prime\prime} < h^{\prime}$ 
because $q \in [p^{\prime}_{1}, q^{\prime}_{1}]$ and $r \leq [\ell^{\prime}_{1}, r^{\prime}_{1}]$. 
The two facts $k^{\prime\prime} \geq h^{\prime}$ and $k^{\prime\prime} < h^{\prime}$ yield a contradiction. 
Therefore, $I_{\capture}(q, r) = ([p^{\prime}_{1}, q^{\prime}_{1}], [\ell^{\prime}_{1}, r^{\prime}_{1}])$ must hold. 

We prove $([p, q], [\ell, r]) \in \Psi_{\source}$. 
$I_{\capture}(q, r) = ([p, q], [\ell, r])$ follows from Lemma~\ref{lem:IA_maximal_lemma}. 
$([p, q]$, $[\ell, r]) = ([p^{\prime}_{1}, q^{\prime}_{1}], [\ell^{\prime}_{1}, r^{\prime}_{1}])$ 
follows from $I_{\capture}(q, r) = ([p, q], [\ell, r])$ and $I_{\capture}(q, r) = ([p^{\prime}_{1}, q^{\prime}_{1}], [\ell^{\prime}_{1}, r^{\prime}_{1}])$. 
$([p, q], [\ell, r]) \in \Psi_{\source} \cup \Psi_{\run}$ follows from 
$([p, q], [\ell, r]) = ([p^{\prime}_{1}, q^{\prime}_{1}], [\ell^{\prime}_{1}, r^{\prime}_{1}])$ and 
$([p^{\prime}_{1}, q^{\prime}_{1}], [\ell^{\prime}_{1}, r^{\prime}_{1}]) \in \Psi_{\run} \cup \Psi_{\source}$. 
On the other hand, $([p, q], [\ell, r]) \not \in \Psi_{\run}$ follows from $([p, q], [\ell, r]) \in \Psi_{\OVQ, C} \setminus \Psi_{\run}$. 
Therefore, $([p, q], [\ell, r]) \in \Psi_{\source}$ follows from 
$([p, q], [\ell, r]) \in \Psi_{\source} \cup \Psi_{\run}$ and $([p, q], [\ell, r]) \not \in \Psi_{\run}$.

\textbf{Proof of statement (3).}
For each integer $s \in [1, k]$, 
$([p_{s}, q_{s}], [\ell_{s}, r_{s}]) = ([p_{1} + (s-1)|C|, q_{1} + (s-1)|C|], [\ell_{1} + (s-1)|C|, r])$ 
follows from the definition of the function $f_{\recover}$. 
Let $K$ be the length of the longest common prefix of two strings $T[\gamma..r+1]$ and $C^{n+1}$ (i.e., $K = |\lcp(T[\gamma..r+1], C^{n+1})|$). 
Then, 
$([p_{1} + (s-1)|C|, q_{1} + (s-1)|C|], [\ell_{1} + (s-1)|C|, r]) \in \Psi_{\run} \cap \Psi_{h} \cap \Psi_{\centerset}(C) \cap \Psi_{\lcp}(K - s |C|)$ follows from Lemma~\ref{lem:recover_basic_property}~\ref{enum:recover_basic_property:4}. 
$p < p_{1}$, $q = p_{1} - 1$, and $\ell \leq \ell_{1}$ follow from Lemma~\ref{lem:recover_basic_property}~\ref{enum:recover_basic_property:3}. 

We prove $[p_{s}, q_{s}] = ([q + 1 + (s-1)|C|, q + s|C|]$ 
for each integer $s \in [1, k]$. 
$p_{s} = q + 1 + (s-1)|C|$ follows from $q = p_{1} - 1$ and $p_{s} = p_{1} + (s-1)|C|$. 
Because of $([p_{s}, q_{s}], [\ell_{s}, r_{s}]) \in \Psi_{\run} \cap \Psi_{\centerset}(C)$, 
$|[p_{s}, q_{s}]| = |C|$ follows from Lemma~\ref{lem:psi_run_basic_property}~\ref{enum:psi_run_basic_property:1}. 
Therefore, $q_{s} = q + s|C|$ follows from $p_{s} = q + 1 + (s-1)|C|$ and $|[p_{s}, q_{s}]| = |C|$. 

We prove $i > q_{k}$ by contradiction.
We assume that $i \leq q_{k}$ holds. 
Then, $i \in [q+1, q_{k}]$ holds 
because $q \leq i-1$ (i.e., $i > q$) follows from $([p, q], [\ell, r]) \in \Psi_{\OVQ, C}$. 
Under the assumption, 
there exists an integer $s^{\prime} \in [1, k]$ satisfying $i \in [p_{s^{\prime}}, q_{s^{\prime}}]$. 
This is because $\bigcup_{s = 1}^{k} [p_{s}, q_{s}] = [q+1, q_{k}]$ follows from 
$[p_{1}, q_{1}] = ([q + 1, q + |C|]$, $[p_{2}, q_{2}] = ([q + 1 + |C|, q + 2|C|]$, 
$\ldots$, $[p_{k}, q_{k}] = ([q + 1 + (k-1)|C|, q + k|C|]$. 
The existence of the integer $s^{\prime}$ contradicts the fact that $i \not \in [p_{s}, q_{s}]$ for all $s \in [1, k]$. 
Therefore, $i > q_{k}$ must hold. 

We prove $([p_{k}, q_{k}], [\ell_{k}, r_{k}]) \in \Psi_{\OVQ, C}$. 
Here, $([p_{k}, q_{k}], [\ell_{k}, r_{k}]) \in \Psi_{\OVQ, C}$ holds if $q_{k} \leq i-1$ and $r_{k} \geq i+1$. 
$q_{k} \leq i-1$ follows from $i > q_{k}$. 
$r_{k} \geq i+1$ holds because 
$r_{k} = r$, 
and $r \geq i+1$ follows from $([p, q], [\ell, r]) \in \Psi_{\OVQ, C}$. 
Therefore, $([p_{k}, q_{k}], [\ell_{k}, r_{k}]) \in \Psi_{\OVQ, C}$ holds.

Because of $([p_{k}, q_{k}], [\ell_{k}, r_{k}]) \in \Psi_{\run} \cap \Psi_{\centerset}(C) \cap \Psi_{\lcp}(K - k|C|)$, 
Lemma~\ref{lem:psi_run_basic_property}~\ref{enum:psi_run_basic_property:2} shows that 
the set $\Psi_{h} \cap \Psi_{\centerset}(C) \cap \Psi_{\lcp}(K - (k+1)|C|)$ contains interval attractor $([p_{k} + |C|, q_{k} + |C|], [\ell_{k} + |C|, r_{k}])$. 
Here, $[p_{k} + |C|, q_{k} + |C|] = [q_{k} + 1, q_{k} + |C|]$ holds because $|[p_{k}, q_{k}]| = |C|$.

We prove $([p_{k} + |C|, q_{k} + |C|], [\ell_{k} + |C|, r_{k}]) \in \Psi_{\OVQ, C} \cup \{ I_{\capture}(i, j^{\prime}) \mid j^{\prime} \in [i+1, n] \}$. 
If $i > q_{k} + |C|$, 
then $([p_{k} + |C|, q_{k} + |C|], [\ell_{k} + |C|, r_{k}]) \in \Psi_{\OVQ, C}$ holds 
because $q_{k} + |C| \leq i-1$ and $r_{k} \geq i+1$. 
Otherwise (i.e., $i \leq q_{k} + |C|$), 
$i \in [p_{k} + |C|, q_{k} + |C|]$ follows from $i > q_{k}$, $i \leq q_{k} + |C|$, and $|[p_{k}, q_{k}]| = |C|$. 
Because of $i \in [p_{k} + |C|, q_{k} + |C|]$, 
$I_{\capture}(i, r_{k}) = ([p_{k} + |C|, q_{k} + |C|], [\ell_{k} + |C|, r_{k}])$ follows from Lemma~\ref{lem:IA_maximal_lemma}. 
$([p_{k} + |C|, q_{k} + |C|], [\ell_{k} + |C|, r_{k}]) \in \{ I_{\capture}(i, j^{\prime}) \mid j^{\prime} \in [i+1, n] \}$ holds 
for the interval attractor $I_{\capture}(i, r_{k}) = ([p_{k} + |C|, q_{k} + |C|], [\ell_{k} + |C|, r_{k}])$. 
Therefore, $([p_{k} + |C|, q_{k} + |C|], [\ell_{k} + |C|, r_{k}]) \in \Psi_{\OVQ, C} \cup \{ I_{\capture}(i, j^{\prime}) \mid j^{\prime} \in [i+1, n] \}$ holds.

We prove $([p_{k} + |C|, q_{k} + |C|], [\ell_{k} + |C|, r_{k}]) \not \in \Psi_{\run}$. 
Because of $([p_{k} + |C|, q_{k} + |C|], [\ell_{k} + |C|, r_{k}]) \in \Psi_{h} \cap \Psi_{\lcp}(K - (k+1)|C|)$, 
$([p_{k} + |C|, q_{k} + |C|], [\ell_{k} + |C|, r_{k}]) \not \in \Psi_{\run}$ holds if 
$K - (k+1)|C| \leq 1 + \sum_{w = 1}^{h+3} \lfloor \mu(w) \rfloor$. 
$k = \lfloor \frac{K - (2 + \sum_{w = 1}^{h+3} \lfloor \mu(w) \rfloor)}{|C|} \rfloor$ follows from the definition of the function $f_{\recover}$. 
$K - (k+1)|C| \leq 1 + \sum_{w = 1}^{h+3} \lfloor \mu(w) \rfloor$ follows from $k = \lfloor \frac{K - (2 + \sum_{w = 1}^{h+3} \lfloor \mu(w) \rfloor)}{|C|} \rfloor$. 
Therefore, $([p_{k} + |C|, q_{k} + |C|], [\ell_{k} + |C|, r_{k}]) \not \in \Psi_{\run}$ holds. 

Finally, $([p_{k}, q_{k}], [\ell_{k}, r_{k}]) \in \Psi_{\OVQ, C} \cap \Psi_{\run} \cap \Psi_{\centerset}(C)$ 
follows from $([p_{k}, q_{k}], [\ell_{k}, r_{k}]) \in \Psi_{\OVQ, C}$ 
and $([p_{k}, q_{k}], [\ell_{k}, r_{k}]) \in \Psi_{\run} \cap \Psi_{h} \cap \Psi_{\centerset}(C) \cap \Psi_{\lcp}(K - k |C|)$. 
$([p_{k} + |C|, q_{k} + |C|], [\ell_{k} + |C|, r_{k}]) \in (\Psi_{\OVQ, C} \cup \{ I_{\capture}(i, j^{\prime}) \mid j^{\prime} \in [i+1, n] \}) \setminus \Psi_{\run}$ 
follows from $([p_{k} + |C|, q_{k} + |C|], [\ell_{k} + |C|, r_{k}]) \in \Psi_{\OVQ, C} \cup \{ I_{\capture}(i, j^{\prime}) \mid j^{\prime} \in [i+1, n] \}$ and $([p_{k} + |C|, q_{k} + |C|], [\ell_{k} + |C|, r_{k}]) \not \in \Psi_{\run}$. 

\textbf{Proof of $([p, q], [\ell, r]) \in  \{ ([\hat{p}, \hat{q}], [\hat{\ell}, \hat{r}]) \in \Psi_{\OVQ, A} \mid \hat{r} \geq i+1 \} \cup \Psi_{\OVQ, B}$.}
One of the following three conditions is satisfied: 
\begin{enumerate}[label=\textbf{(\Alph*)}]
    \item $([p, q], [\ell, r]) \in \Psi_{\source}$, and there exists an integer $s \in [1, k]$ satisfying $i \in [p_{s}, q_{s}]$; 
    \item $([p, q], [\ell, r]) \in \Psi_{\source}$, and $i \not \in [p_{s}, q_{s}]$ for all $s \in [1, k]$; 
    \item $([p, q], [\ell, r]) \not \in \Psi_{\source}$. 
\end{enumerate}

For condition (A), 
$([p_{s}, q_{s}], [\ell_{s}, r_{s}]) \in \{ I_{\capture}(i, j^{\prime}) \mid j^{\prime} \in [i+1, n] \}$ holds 
because $I_{\capture}(i, r_{s}) = ([p_{s}, q_{s}], [\ell_{s}, r_{s}])$ follows from Lemma~\ref{lem:IA_maximal_lemma}. 
Therefore, $([p, q], [\ell, r]) \in \Psi_{\OVQ, B}$ follows from 
$([p_{s}, q_{s}]$, $[\ell_{s}, r_{s}]) \in f_{\recover}(([p, q], [\ell, r]))$ 
and $([p_{s}, q_{s}], [\ell_{s}, r_{s}]) \in \{ I_{\capture}(i, j^{\prime}) \mid j^{\prime} \in [i+1, n] \}$. 

For condition (B), 
$([p_{k}, q_{k}], [\ell_{k}, r_{k}]) \in \Psi_{\OVQ, C} \cap \Psi_{\run} \cap \Psi_{\centerset}(C)$ 
and $([p_{k} + |C|, q_{k} + |C|], [\ell_{k} + |C|, r_{k}]) \in (\Psi_{\OVQ, C} \cup \{ I_{\capture}(i, j^{\prime}) \mid j^{\prime} \in [i+1, n] \}) \setminus \Psi_{\run}$ 
follow from statement (3). 
Here, $r_{k} = r$ follows from the definition of the function $f_{\recover}$; 
$r \geq i+1$ follows from $([p, q], [\ell, r]) \in \Psi_{\OVQ, C}$. 
$([p_{k}, q_{k}], [\ell_{k}, r_{k}]) \in \{ ([\hat{p}, \hat{q}], [\hat{\ell}, \hat{r}]) \in \Psi_{\OVQ, A} \mid \hat{r} \geq i+1 \}$ holds 
because (a) $([p_{k} + |C|, q_{k} + |C|], [\ell_{k} + |C|, r_{k}]) \in (\Psi_{\OVQ, C} \cup \{ I_{\capture}(i, j^{\prime}) \mid j^{\prime} \in [i+1, n] \}) \setminus \Psi_{\run}$, 
(b) $r_{k} \geq i+1$, 
and (c) Lemma~\ref{lem:left_extension_property}~\ref{enum:left_extension_property:3} shows that 
the interval attractor $([p_{k}, q_{k}], [\ell_{k}, r_{k}])$ is a left-extension of the interval attractor $([p_{k} + |C|, q_{k} + |C|], [\ell_{k} + |C|, r_{k}])$. 
Therefore, $([p, q], [\ell, r]) \in \Psi_{\OVQ, B}$ follows from 
$([p_{k}, q_{k}], [\ell_{k}, r_{k}]) \in f_{\recover}(([p, q], [\ell, r]))$ 
and $([p_{k}, q_{k}], [\ell_{k}, r_{k}]) \in \{ ([\hat{p}, \hat{q}], [\hat{\ell}, \hat{r}]) \in \Psi_{\OVQ, A} \mid \hat{r} \geq i+1 \}$. 

For condition (C), 
$([p^{\prime}, q^{\prime}], [\ell^{\prime}, r^{\prime}]) \not \in \Psi_{\run}$ follows from 
statement (2) and $([p, q], [\ell, r]) \not \in \Psi_{\source}$. 
$([p^{\prime}, q^{\prime}], [\ell^{\prime}, r^{\prime}]) \in \Psi_{\OVQ, C} \cup \{ I_{\capture}(i, j^{\prime}) \mid j^{\prime} \in [i+1, n] \}$ follows from statement (1). 
$([p^{\prime}, q^{\prime}], [\ell^{\prime}, r^{\prime}]) \in (\Psi_{\OVQ, C} \cup \{ I_{\capture}(i, j^{\prime}) \mid j^{\prime} \in [i+1, n] \}) \setminus \Psi_{\run}$ 
follows from $([p^{\prime}, q^{\prime}], [\ell^{\prime}, r^{\prime}]) \not \in \Psi_{\run}$ and $([p^{\prime}, q^{\prime}], [\ell^{\prime}, r^{\prime}]) \in \Psi_{\OVQ, C} \cup \{ I_{\capture}(i, j^{\prime}) \mid j^{\prime} \in [i+1, n] \}$. 
$([p, q], [\ell, r]) \in \Psi_{\OVQ, A}$ holds 
because the interval attractor $([p, q], [\ell, r])$ is a left-extension of the interval attractor $([p^{\prime}, q^{\prime}], [\ell^{\prime}, r^{\prime}])$. 
$r \geq i+1$ follows from $([p, q], [\ell, r]) \in \Psi_{\OVQ, C}$. 
Therefore, $([p, q], [\ell, r]) \in \{ ([\hat{p}, \hat{q}], [\hat{\ell}, \hat{r}]) \in \Psi_{\OVQ, A} \mid \hat{r} \geq i+1 \}$ holds. 

We showed that $([p, q], [\ell, r]) \in \{ ([\hat{p}, \hat{q}], [\hat{\ell}, \hat{r}]) \in \Psi_{\OVQ, A} \mid \hat{r} \geq i+1 \}$ or $([p, q], [\ell, r]) \in \Psi_{\OVQ, B}$ holds for 
each of the three conditions (A), (B), and (C). 
Therefore, $([p, q], [\ell, r]) \in \{ ([\hat{p}, \hat{q}], [\hat{\ell}, \hat{r}]) \in \Psi_{\OVQ, A} \mid \hat{r} \geq i+1 \} \cup \Psi_{\OVQ, B}$ holds. 

\end{proof}

Lemma~\ref{lem:RB_left_extension} indicates that 
we can obtain all the interval attractors in the subset $\Psi_{\OVQ, C} \setminus \Psi_{\run}$ by computing 
the two subsets $\{ ([p, q], [\ell, r]) \in \Psi_{\OVQ, A} \mid r \geq i+1 \}$ and $\Psi_{\OVQ, B}$. 
In the next paragraph, we introduce a directed graph $\mathcal{G}_{\OVQ}$ for quickly computing 
the two subsets $\{ ([p, q], [\ell, r]) \in \Psi_{\OVQ, A} \mid r \geq i+1 \}$ and $\Psi_{\OVQ, B}$. 

\paragraph{Directed graph $\mathcal{G}_{\OVQ}$.}
$\mathcal{G}_{\OVQ} = (\mathcal{U}_{\OVQ}, \mathcal{E}_{\OVQ})$ is a directed graph representing the relationship among interval attractors in 
set $\{ I_{\capture}(i, j^{\prime}) \mid j^{\prime} \in [i+1, n] \} \cup \{ ([p, q], [\ell, r]) \in \Psi_{\OVQ, A} \mid r \geq i+1 \} \cup \Psi_{\OVQ, B}$, 
where each directed edge represents a left-extension or run-extension of an interval attractor. 
Set $\mathcal{U}_{\OVQ}$ is a set of nodes such that each node is an interval attractor in the set $\{ I_{\capture}(i, j^{\prime}) \mid j^{\prime} \in [i+1, n] \} \cup \{ ([p, q], [\ell, r]) \in \Psi_{\OVQ, A} \mid r \geq i+1 \} \cup \Psi_{\OVQ, B}$ (i.e., $\mathcal{U}_{\OVQ} = \{ I_{\capture}(i, j^{\prime}) \mid j^{\prime} \in [i+1, n] \} \cup \{ ([p, q], [\ell, r]) \in \Psi_{\OVQ, A} \mid r \geq i+1 \} \cup \Psi_{\OVQ, B}$). 
Set $\mathcal{E}_{\OVQ} \subseteq \mathcal{U}_{\OVQ} \times \mathcal{U}_{\OVQ}$ is a set of directed edges. 
For a pair of two interval attractors $([p, q], [\ell, r]), ([p^{\prime}, q^{\prime}], [\ell^{\prime}, r^{\prime}])$ in set $\mathcal{U}_{\OVQ}$, 
the interval attractor $([p, q], [\ell, r])$ has a directed edge to the interval attractor $([p^{\prime}, q^{\prime}], [\ell^{\prime}, r^{\prime}])$ if either of the following two statements holds: 
\begin{itemize}
    \item $([p^{\prime}, q^{\prime}], [\ell^{\prime}, r^{\prime}])$ is a left-extension of $([p, q], [\ell, r])$, 
    and $([p, q], [\ell, r]) \not \in \Psi_{\run}$;
    \item $([p^{\prime}, q^{\prime}], [\ell^{\prime}, r^{\prime}])$ is a run-extension of $([p, q], [\ell, r])$, 
    and $([p, q], [\ell, r]) \in \Psi_{\run}$.
\end{itemize}

The following lemma shows that the directed graph $\mathcal{G}_{\OVQ}$ is a forest of $O(H + |\{ ([p, q], [\ell, r]) \in \Psi_{\RR} \setminus \Psi_{\run} \mid i \in [p, r] \}|)$ nodes.

\begin{lemma}\label{lem:forest_g_ovq_property}
    Consider the directed graph $\mathcal{G}_{\OVQ} = (\mathcal{U}_{\OVQ}, \mathcal{E}_{\OVQ})$ for an overlap query $\OVQ([i, j])$. 
    Then, the following two statements hold. 
    \begin{enumerate}[label=\textbf{(\roman*)}]
    \item \label{lem:forest_g_ovq_property:1} the in-degree of a node $([p, q], [\ell, r]) \in \mathcal{U}_{\OVQ}$ is $0$ if $([p, q], [\ell, r]) \in \{ I_{\capture}(i, j^{\prime}) \mid j^{\prime} \in [i+1, n] \}$; 
    otherwise, its in-degree is $1$;
    \item \label{lem:forest_g_ovq_property:2} the number of nodes in the directed graph $\mathcal{G}_{\OVQ}$ is 
    at most $H+1 + 3|\{ ([p, q], [\ell, r]) \in \Psi_{\RR} \setminus \Psi_{\run} \mid i \in [p, r] \}|$. 
    \end{enumerate}
\end{lemma}
\begin{proof}
    The following five statements are used to prove Lemma~\ref{lem:forest_g_ovq_property}: 
    \begin{enumerate}[label=\textbf{(\arabic*)}]
    \item 
    $|f_{\recover}(([p, q], [\ell, r])) \cap \mathcal{U}_{\OVQ}| \leq 1$ 
    for a node $([p, q], [\ell, r]) \in \mathcal{U}_{\OVQ}$ satisfying $([p, q], [\ell, r]) \in \Psi_{\source}$;    
    \item the in-degree of a node $([p, q], [\ell, r]) \in \mathcal{U}_{\OVQ}$ is at most $1$;
    \item for a node $([p, q], [\ell, r]) \in \mathcal{U}_{\OVQ}$ satisfying $([p, q], [\ell, r]) \in \{ I_{\capture}(i, j^{\prime}) \mid j^{\prime} \in [i+1, n] \}$, 
    its in-degree is $0$;
    \item $(\Psi_{\OVQ, C} \cup \{ I_{\capture}(i, j^{\prime}) \mid j^{\prime} \in [i+1, n] \}) \setminus \Psi_{\run} \subseteq \mathcal{U}_{\OVQ}$; 
    \item for a node $([p, q], [\ell, r]) \in \mathcal{U}_{\OVQ}$ satisfying $([p, q], [\ell, r]) \not \in \{ I_{\capture}(i, j^{\prime}) \mid j^{\prime} \in [i+1, n] \}$, 
    its in-degree at least $1$.
    \end{enumerate}

    \textbf{Proof of statement (1).} 
    Let $([p_{1}, q_{1}], [\ell_{1}, r_{1}])$, 
    $([p_{2}, q_{2}], [\ell_{2}, r_{2}])$, $\ldots$, $([p_{k}, q_{k}], [\ell_{k}, r_{k}])$ ($p_{1} \leq p_{2} \leq \cdots \leq p_{k}$) 
    be the interval attractors obtained from the function $f_{\recover}(([p, q], [\ell, r]))$. 
    For each integer $s \in [1, k]$, 
    $([p_{s}, q_{s}], [\ell_{s}, r_{s}]) = ([p_{1} + (s-1)|C|, q_{1} + (s-1)|C|], [\ell_{1} + (s-1)|C|, r^{\prime}])$ 
    follows from the definition of the function $f_{\recover}$ 
    for the associated string $C$ of the interval attractor $([p, q], [\ell, r])$.

    We prove $[p_{s}, q_{s}] = [q + 1 + (s-1)|C|, q + s|C|]$.
    $p < p_{1}$ and $q = p_{1} - 1$ follow from Lemma~\ref{lem:recover_basic_property}~\ref{enum:recover_basic_property:3}. 
    $([p_{s}, q_{s}], [\ell_{s}, r_{s}]) \in \Psi_{\run} \cap \Psi_{h} \cap \Psi_{\centerset}(C)$ follows from Lemma~\ref{lem:recover_basic_property}~\ref{enum:recover_basic_property:4}. 
    Because of $([p_{s}, q_{s}], [\ell_{s}, r_{s}]) \in \Psi_{\run} \cap \Psi_{\centerset}(C)$, 
    $|[p_{s}, q_{s}]| = |C|$ follows from Lemma~\ref{lem:psi_run_basic_property}~\ref{enum:psi_run_basic_property:1}. 
    $p_{s} = q + 1 + (s-1)|C|$ follows from $q = p_{1} - 1$ and $p_{s} = p_{1} + (s-1)|C|$. 
    $q_{s} = q + s|C|$ follows from $p_{s} = q + 1 + (s-1)|C|$ and $|[p_{s}, q_{s}]| = |C|$. 

    We prove $|f_{\recover}(([p, q], [\ell, r])) \cap \mathcal{U}_{\OVQ}| \leq 1$ by contradiction. 
    We assume that $|f_{\recover}(([p, q]$, $[\ell, r])) \cap \mathcal{U}_{\OVQ}| \geq 2$ holds. 
    Then, there exists two integers $1 \leq x < y \leq k$ 
    satisfying $([p_{x}, q_{x}], [\ell_{x}, r_{x}])$, $([p_{y}, q_{y}], [\ell_{y}, r_{y}]) \in \mathcal{U}_{\OVQ}$. 
    Here, $\mathcal{U}_{\OVQ} = \{ I_{\capture}(i, j^{\prime}) \mid j^{\prime} \in [i+1, n] \} \cup \{ ([p, q], [\ell, r]) \in \Psi_{\OVQ, A} \mid r \geq i+1 \} \cup \Psi_{\OVQ, B}$ 
    for the two subsets $\Psi_{\OVQ, A}$ and $\Psi_{\OVQ, B}$ defined in Lemma~\ref{lem:RB_left_extension}.
    Under the assumption, 
    one of the following three conditions are satisfied: 
    
    \begin{enumerate}[label=\textbf{(\Alph*)}]
    \item $([p_{x}, q_{x}], [\ell_{x}, r_{x}]), ([p_{y}, q_{y}], [\ell_{y}, r_{y}]) \in \{ I_{\capture}(i, j^{\prime}) \mid j^{\prime} \in [i+1, n] \}$; 
    \item $([p_{x}, q_{x}], [\ell_{x}, r_{x}]) \in \{ I_{\capture}(i, j^{\prime}) \mid j^{\prime} \in [i+1, n] \}$ and $([p_{y}, q_{y}], [\ell_{y}, r_{y}]) \not \in \{ I_{\capture}(i, j^{\prime}) \mid j^{\prime} \in [i+1, n] \}$;
    \item $([p_{x}, q_{x}], [\ell_{x}, r_{x}]) \not \in \{ I_{\capture}(i, j^{\prime}) \mid j^{\prime} \in [i+1, n] \}$.
    \end{enumerate}

    For condition (A), 
    there exist two integers $j^{\prime}_{1}, j^{\prime}_{2} \in [i+1, n]$ satisfying $I_{\capture}(i, j^{\prime}_{1}) = ([p_{x}, q_{x}]$, $[\ell_{x}, r_{x}])$ and $I_{\capture}(i, j^{\prime}_{2}) = ([p_{y}, q_{y}], [\ell_{y}, r_{y}])$, respectively. 
    $i \in [p_{x}, q_{x}]$ follows from the definition of interval attractor 
    for the interval attractor $I_{\capture}(i, j^{\prime}_{1}) = ([p_{x}, q_{x}], [\ell_{x}, r_{x}])$. 
    Similarly, $i \in [p_{y}, q_{y}]$ holds.     
    $[p_{x}, q_{x}] \cap [p_{y}, q_{y}] \neq \emptyset$ follows from $i \in [p_{x}, q_{x}]$ and $i \in [p_{y}, q_{y}]$. 
    On the other hand, $[p_{x}, q_{x}] \cap [p_{y}, q_{y}] = \emptyset$ follows from 
    $[p_{x}, q_{x}] = [q + 1 + (x-1)|C|, q + x|C|]$ and $[p_{y}, q_{y}] = [q + 1 + (y-1)|C|, q + y|C|]$. 
    The two facts $[p_{x}, q_{x}] \cap [p_{y}, q_{y}] \neq \emptyset$ and $[p_{x}, q_{x}] \cap [p_{y}, q_{y}] = \emptyset$ 
    yield a contradiction. 
    
    For condition (B), 
    there exists an integer $j^{\prime} \in [i+1, n]$ satisfying $I_{\capture}(i, j^{\prime}) = ([p_{x}, q_{x}], [\ell_{x}, r_{x}])$. 
    $i \in [p_{x}, q_{x}]$ follows from the definition of interval attractor 
    for the interval attractor $I_{\capture}(i, j^{\prime}) = ([p_{x}, q_{x}], [\ell_{x}, r_{x}])$. 
    The interval attractor $([p_{y}, q_{y}], [\ell_{y}, r_{y}])$ is contained in the set 
    $(\{ ([p, q], [\ell, r]) \in \Psi_{\OVQ, A} \mid r \geq i+1 \} \cup \Psi_{\OVQ, B}) \cap \Psi_{\run}$. 
    $([p_{y}, q_{y}], [\ell_{y}, r_{y}]) \in \{ ([p, q], [\ell, r]) \in \Psi_{\OVQ, A} \mid r \geq i+1 \}$ holds 
    because $\Psi_{\OVQ, B} \subseteq \Psi_{\source}$ follows from Lemma~\ref{lem:ovqab_property}~\ref{enum:ovqab_property:2}, 
    and $\Psi_{\source} \cap \Psi_{\run} = \emptyset$ holds. 
    Because of $([p_{y}, q_{y}], [\ell_{y}, r_{y}]) \in \Psi_{\OVQ, A}$, 
    $q_{y} \leq i-1$ follows from Lemma~\ref{lem:ovqab_property}~\ref{enum:ovqab_property:1}. 
    $q_{x} < q_{y}$ follows from $[p_{x}, q_{x}] = [q + 1 + (x-1)|C|, q + x|C|]$ and $[p_{y}, q_{y}] = [q + 1 + (y-1)|C|, q + y|C|]$. 
    $q_{x} \leq i-1$ follows from $q_{x} < q_{y}$ and $q_{y} \leq i-1$. 
    The two facts $i \in [p_{x}, q_{x}]$) and $q_{x} \leq i-1$ yield a contradiction. 
    
    For condition (C), 
    the interval attractor $([p_{x}, q_{x}], [\ell_{x}, r_{x}])$ is contained in the set $\{ ([p, q], [\ell, r]) \in \Psi_{\OVQ, A} \mid r \geq i+1 \}$, which is similar to the interval attractor $([p_{y}, q_{y}], [\ell_{y}, r_{y}])$ for condition (B).     
    Lemma~\ref{lem:ovqab_property}~\ref{enum:ovqab_property:1} shows that 
    the set $\Psi_{\RR} \setminus \Psi_{\run}$ contains 
    an interval attractor $([p^{\prime}, q^{\prime}], [\ell^{\prime}, r^{\prime}])$ such that 
    $([p_{x}, q_{x}], [\ell_{x}, r_{x}])$ is a left-extension of the interval attractor $([p^{\prime}, q^{\prime}], [\ell^{\prime}, r^{\prime}])$. 
    Lemma~\ref{lem:left_extension_property}~\ref{enum:left_extension_property:3} shows that 
    $([p_{x}, q_{x}], [\ell_{x}, r_{x}])$ is a left-extension of the interval attractor $([p_{x+1}, q_{x+1}], [\ell_{x+1}, r_{x+1}])$ 
    because $([p_{x+1}, q_{x+1}], [\ell_{x+1}, r_{x+1}]) = ([p_{x} + |C|, q_{x} + |C|], [\ell_{x} + |C|, r_{x}])$ holds. 
    $([p^{\prime}, q^{\prime}], [\ell^{\prime}, r^{\prime}]) \neq ([p_{x+1}, q_{x+1}], [\ell_{x+1}, r_{x+1}])$ 
    follows from $([p^{\prime}, q^{\prime}], [\ell^{\prime}, r^{\prime}]) \in \Psi_{\RR} \setminus \Psi_{\run}$ 
    and $([p_{x+1}, q_{x+1}], [\ell_{x+1}, r_{x+1}]) \in \Psi_{\run}$. 
    Because of $([p^{\prime}, q^{\prime}], [\ell^{\prime}, r^{\prime}]) \neq ([p_{x+1}, q_{x+1}], [\ell_{x+1}, r_{x+1}])$, 
    $([p_{x}, q_{x}], [\ell_{x}, r_{x}])$ is a left-extension of two interval attractors. 
    On the other hand, Lemma~\ref{lem:left_extension_property}~\ref{enum:left_extension_property:2} shows that 
    $([p_{x}, q_{x}], [\ell_{x}, r_{x}])$ is a left-extension of at most one interval attractor. 
    Therefore, there exists a contradiction. 

    We showed that there exists a contradiction for each of the three conditions (A), (B), and (C). 
    Therefore, $|f_{\recover}(([p, q], [\ell, r])) \cap \mathcal{U}_{\OVQ}| \leq 1$ must hold. 
    
    \textbf{Proof of statement (2).} 
    We prove statement (2) by contradiction.
    We assume that statement (2) does not hold. 
    Then, there exist two nodes $([p_{A}, q_{A}], [\ell_{A}, r_{A}]), ([p_{B}, q_{B}], [\ell_{B}, r_{B}]) \in \mathcal{U}_{\OVQ}$ 
    ($p_{A} \leq p_{B}$ and $([p_{A}, q_{A}], [\ell_{A}, r_{A}]) \neq ([p_{B}, q_{B}], [\ell_{B}, r_{B}])$)     
    such that each of the two nodes has a directed edge to the node $([p, q], [\ell, r])$. 
    Under the assumption, 
    one of the following four conditions are satisfied: 
    \begin{enumerate}[label=\textbf{(\alph*)}]
    \item $([p_{A}, q_{A}], [\ell_{A}, r_{A}]), ([p_{B}, q_{B}], [\ell_{B}, r_{B}]) \not \in \Psi_{\run}$; 
    \item $([p_{A}, q_{A}], [\ell_{A}, r_{A}]) \not \in \Psi_{\run}$ and $([p_{B}, q_{B}], [\ell_{B}, r_{B}]) \in \Psi_{\run}$;
    \item $([p_{A}, q_{A}], [\ell_{A}, r_{A}]) \in \Psi_{\run}$ and $([p_{B}, q_{B}], [\ell_{B}, r_{B}]) \not \in \Psi_{\run}$;
    \item $([p_{A}, q_{A}], [\ell_{A}, r_{A}]), ([p_{B}, q_{B}], [\ell_{B}, r_{B}]) \in \Psi_{\run}$. 
    \end{enumerate}

    For condition (a), 
    node $([p, q], [\ell, r])$ is a left-extension of node $([p_{A}, q_{A}], [\ell_{A}, r_{A}])$ 
    because the node $([p_{A}, q_{A}], [\ell_{A}, r_{A}])$ has a directed edge to the node $([p, q], [\ell, r])$, 
    and $([p_{A}, q_{A}], [\ell_{A}, r_{A}]) \not \in \Psi_{\run}$. 
    Similarly, 
    node $([p, q], [\ell, r])$ is a left-extension of node $([p_{B}, q_{B}], [\ell_{B}, r_{B}])$.
    In this case, 
    $([p_{A}, q_{A}], [\ell_{A}, r_{A}]) = ([p_{B}, q_{B}], [\ell_{B}, r_{B}])$ holds 
    because 
    Lemma~\ref{lem:left_extension_property}~\ref{enum:left_extension_property:1} shows that 
    $I_{\capture}(q+1, r) = ([p_{A}, q_{A}], [\ell_{A}, r_{A}])$ and $I_{\capture}(q+1, r) = ([p_{B}, q_{B}]$, $[\ell_{B}, r_{B}])$ hold. 
    The two facts $([p_{A}, q_{A}], [\ell_{A}, r_{A}]) \neq ([p_{B}, q_{B}]$, $[\ell_{B}, r_{B}])$ 
    and $([p_{A}, q_{A}], [\ell_{A}, r_{A}]) = ([p_{B}, q_{B}], [\ell_{B}, r_{B}])$ yield a contradiction.         

    For condition (b), 
    node $([p, q], [\ell, r])$ is a left-extension of node $([p_{A}, q_{A}], [\ell_{A}, r_{A}])$. 
    On the other hand, 
    node $([p, q], [\ell, r])$ is a run-extension of node $([p_{B}, q_{B}], [\ell_{B}, r_{B}])$ 
    because the node $([p_{B}, q_{B}], [\ell_{B}, r_{B}])$ has a directed edge to the node $([p, q], [\ell, r])$, 
    and $([p_{B}, q_{B}], [\ell_{B}, r_{B}]) \in \Psi_{\run}$. 
    $([p, q], [\ell, r]) \in \Psi_{\source}$ follows from the definition of run-extension. 
    Lemma~\ref{lem:left_extension_property}~\ref{enum:left_extension_property:4} shows that 
    $([p_{A}, q_{A}], [\ell_{A}, r_{A}]) \in f_{\recover}(([p, q], [\ell, r]))$ holds. 
    $([p_{A}, q_{A}], [\ell_{A}, r_{A}]) \in \Psi_{\run}$ holds. 
    This is because $f_{\recover}(([p, q], [\ell, r])) \subseteq \Psi_{\run}$ follows from the definition of the function $f_{\recover}$. 
    The two facts $([p_{A}, q_{A}], [\ell_{A}, r_{A}]) \not \in \Psi_{\run}$ and 
    $([p_{A}, q_{A}], [\ell_{A}, r_{A}]) \in \Psi_{\run}$ yield a contradiction. 
    Similarly, we can show that 
    there exists a contradiction for condition (c) using the same approach. 

    For condition (d), 
    node $([p, q], [\ell, r])$ is a run-extension of both $([p_{A}, q_{A}], [\ell_{A}, r_{A}])$ and $([p_{B}, q_{B}]$, $[\ell_{B}, r_{B}])$. 
    $([p, q], [\ell, r]) \in \Psi_{\source}$ and 
    $([p_{A}, q_{A}], [\ell_{A}, r_{A}]), ([p_{B}, q_{B}], [\ell_{B}, r_{B}]) \in f_{\recover}([p, q], [\ell, r])$    
    follow from the definition of run-extension. 
    $([p_{A}, q_{A}], [\ell_{A}, r_{A}]) = ([p_{B}, q_{B}], [\ell_{B}, r_{B}])$ holds 
    because $|f_{\recover}(([p, q], [\ell, r])) \cap \mathcal{U}_{\OVQ}| \leq 1$ follows from statement (1). 
    The two facts $([p_{A}, q_{A}], [\ell_{A}, r_{A}]) \neq ([p_{B}, q_{B}], [\ell_{B}, r_{B}])$ 
    and $([p_{A}, q_{A}], [\ell_{A}, r_{A}]) = ([p_{B}, q_{B}], [\ell_{B}, r_{B}])$ yield a contradiction.         
    
    We showed that there exists a contradiction for each of the four conditions (a), (b), (c) and (d). 
    Therefore, statement (2) must hold. 

    \textbf{Proof of statement (3).}
    We show that the in-degree of the node $([p, q], [\ell, r])$ is $0$ by contradiction. 
    We assume that the in-degree of the node $([p, q], [\ell, r])$ is not $0$. 
    Then, the forest $\mathcal{G}_{\OVQ}$ contains a node 
    $([p^{\prime}, q^{\prime}], [\ell^{\prime}, r^{\prime}])$ that has a directed edge to the node $([p, q], [\ell, r])$. 

    We prove $p^{\prime} \leq i$. 
    The node $([p^{\prime}, q^{\prime}], [\ell^{\prime}, r^{\prime}])$ satisfies 
    at least one of the following three conditions: 
    (I) $([p^{\prime}, q^{\prime}], [\ell^{\prime}, r^{\prime}]) \in \Psi_{\OVQ, B}$; 
    (II) $([p^{\prime}, q^{\prime}], [\ell^{\prime}, r^{\prime}]) \in \{ I_{\capture}(i, j^{\prime}) \mid j^{\prime} \in [i+1, n] \}$; 
    (III) $([p^{\prime}, q^{\prime}], [\ell^{\prime}, r^{\prime}]) \in \Psi_{\OVQ, A}$. 

    For condition (I), 
    $q^{\prime} \leq i-1$ follows from Lemma~\ref{lem:ovqab_property}~\ref{enum:ovqab_property:2}. 
    $p^{\prime} \leq i$ follows from $p^{\prime} \leq q^{\prime}$ and $q^{\prime} \leq i-1$. 
    For condition (II), 
    there exists an integer $j^{\prime} \in [i+1, n]$ satisfying $I_{\capture}(i, j^{\prime}) = ([p^{\prime}, q^{\prime}], [\ell^{\prime}, r^{\prime}])$. 
    Since $I_{\capture}(i, j^{\prime}) = ([p^{\prime}, q^{\prime}], [\ell^{\prime}, r^{\prime}])$, 
    $i \in [p^{\prime}, q^{\prime}]$ follows from the definition of interval attractor. 
    $p^{\prime} \leq i$ follows from $i \in [p^{\prime}, q^{\prime}]$. 
    For condition (III), 
    $q^{\prime} \leq i-1$ follows from statement Lemma~\ref{lem:ovqab_property}~\ref{enum:ovqab_property:1}. 
    $p^{\prime} \leq i$ follows from $p^{\prime} \leq q^{\prime}$ and $q^{\prime} \leq i-1$. 
    Therefore, $p^{\prime} \leq i$ holds. 

    We prove $p^{\prime} \geq i + 1$. 
    Because of $([p, q], [\ell, r]) \in \{ I_{\capture}(i, j^{\prime}) \mid j^{\prime} \in [i+1, n] \}$, 
    there exists an integer $j^{\prime} \in [i+1, n]$ satisfying $I_{\capture}(i, j^{\prime}) = ([p, q], [\ell, r])$. 
    Since $I_{\capture}(i, j^{\prime}) = ([p, q], [\ell, r])$, 
    $i \in [p, q]$ follows from the definition of interval attractor. 
    If $([p^{\prime}, q^{\prime}], [\ell^{\prime}, r^{\prime}]) \in \Psi_{\run}$, 
    then $([p, q], [\ell, r])$ is a run-extension of the interval attractor $([p^{\prime}, q^{\prime}], [\ell^{\prime}, r^{\prime}])$, 
    which indicates that $([p, q], [\ell, r]) \in \Psi_{\source}$ and $([p^{\prime}, q^{\prime}], [\ell^{\prime}, r^{\prime}]) \in f_{\recover}(([p, q], [\ell, r]))$. 
    Let $([p_{1}, q_{1}]$, $[\ell_{1}, r_{1}])$, 
    $([p_{2}, q_{2}], [\ell_{2}, r_{2}])$, $\ldots$, $([p_{k}, q_{k}], [\ell_{k}, r_{k}])$ ($p_{1} \leq p_{2} \leq \cdots \leq p_{k}$) 
    be the interval attractors obtained from the function $f_{\recover}(([p, q], [\ell, r]))$. 
    Then, $p_{1} \leq p^{\prime}$ follows from the definition of the function $f_{\recover}$.
    $q = p_{1} - 1$ follows from Lemma~\ref{lem:recover_basic_property}~\ref{enum:recover_basic_property:3}.
    $p^{\prime} \geq i + 1$ follows from 
    $i \leq q$, $q = p_{1} - 1$, and $p_{1} \leq p^{\prime}$. 

    Otherwise (i.e., $([p^{\prime}, q^{\prime}], [\ell^{\prime}, r^{\prime}]) \not \in \Psi_{\run}$), 
    then $([p, q], [\ell, r])$ is a left-extension of the interval attractor $([p^{\prime}, q^{\prime}], [\ell^{\prime}, r^{\prime}])$ 
    (i.e., $q = p^{\prime} - 1$ and $r \in [\ell^{\prime}, r^{\prime}]$). 
    $p^{\prime} \geq i + 1$ follows from $i \leq q$ and $q = p^{\prime} - 1$. 
    Therefore, $p^{\prime} \geq i + 1$ holds. 

    The two facts $p^{\prime} \leq i$ and $p^{\prime} \geq i + 1$ yield a contradiction. 
    Therefore, the in-degree of the node $([p, q], [\ell, r])$ must be $0$.

    \textbf{Proof of statement (4).} 
    $\{ I_{\capture}(i, j^{\prime}) \mid j^{\prime} \in [i+1, n] \} \subseteq \mathcal{U}_{\OVQ}$ follows from 
    $\mathcal{U}_{\OVQ} = \{ I_{\capture}(i, j^{\prime}) \mid j^{\prime} \in [i+1, n] \} \cup \{ ([p, q], [\ell, r]) \in \Psi_{\OVQ, A} \mid r \geq i+1 \} \cup \Psi_{\OVQ, B}$. 
    $\Psi_{\OVQ, C} \setminus \Psi_{\run} \subseteq \mathcal{U}_{\OVQ}$ follows from 
    $\Psi_{\OVQ, C} \setminus \Psi_{\run} \subseteq \{ ([p, q], [\ell, r]) \in \Psi_{\OVQ, A} \mid r \geq i+1 \} \cup \Psi_{\OVQ, B}$ 
    (Lemma~\ref{lem:RB_left_extension}) and 
    $\{ ([p, q], [\ell, r]) \in \Psi_{\OVQ, A} \mid r \geq i+1 \} \cup \Psi_{\OVQ, B} \subseteq \mathcal{U}_{\OVQ}$. 
    Therefore, $(\Psi_{\OVQ, C} \cup \{ I_{\capture}(i, j^{\prime}) \mid j^{\prime} \in [i+1, n] \}) \setminus \Psi_{\run} \subseteq \mathcal{U}_{\OVQ}$ holds. 

    \textbf{Proof of statement (5).}   
    Because of $\mathcal{U}_{\OVQ} = \{ I_{\capture}(i, j^{\prime}) \mid j^{\prime} \in [i+1, n] \} \cup \{ ([p, q], [\ell, r]) \in \Psi_{\OVQ, A} \mid r \geq i+1 \} \cup \Psi_{\OVQ, B}$,     
    $([p, q], [\ell, r]) \in \{ ([p, q], [\ell, r]) \in \Psi_{\OVQ, A} \mid r \geq i+1 \} \cup \Psi_{\OVQ, B}$ follows from 
    $([p, q], [\ell, r]) \in \mathcal{U}_{\OVQ}$ and $([p, q], [\ell, r]) \not \in \{ I_{\capture}(i, j^{\prime}) \mid j^{\prime} \in [i+1, n] \}$. 
    
    If $([p, q], [\ell, r]) \in \Psi_{\OVQ, A}$, 
    then the set $(\Psi_{\OVQ, C} \cup \{ I_{\capture}(i, j^{\prime}) \mid j^{\prime} \in [i+1, n] \}) \setminus \Psi_{\run}$ contains an interval attractor 
    $([p^{\prime}, q^{\prime}], [\ell^{\prime}, r^{\prime}])$ such that 
    $([p, q], [\ell, r])$ is a left-extension of the interval attractor $([p^{\prime}, q^{\prime}], [\ell^{\prime}, r^{\prime}])$. 
    Statement (4) shows that 
    the interval attractor $([p^{\prime}, q^{\prime}], [\ell^{\prime}, r^{\prime}])$ is a node of the forest $\mathcal{G}_{\OVQ}$. 
    Because of $([p^{\prime}, q^{\prime}], [\ell^{\prime}, r^{\prime}]) \not \in \Psi_{\run}$, 
    the node $([p^{\prime}, q^{\prime}], [\ell^{\prime}, r^{\prime}])$ has a directed edge to the node $([p, q], [\ell, r])$. 
    Therefore, the in-degree of the node $([p, q], [\ell, r])$ is at least $1$. 
    
    Otherwise (i.e., $([p, q], [\ell, r]) \not \in \Psi_{\OVQ, A}$), 
    $([p, q], [\ell, r]) \in \Psi_{\OVQ, B}$ holds. 
    This fact $([p, q], [\ell, r]) \in \Psi_{\OVQ, B}$ indicates that 
    the set $\{ I_{\capture}(i, j^{\prime}) \mid j^{\prime} \in [i+1, n] \} \cup \{ ([p, q], [\ell, r]) \in \Psi_{\OVQ, A} \mid r \geq i+1 \}$ contains an interval attractor 
    $([p^{\prime}, q^{\prime}], [\ell^{\prime}, r^{\prime}])$ such that 
    $([p, q], [\ell, r])$ is a run-extension of the interval attractor $([p^{\prime}, q^{\prime}], [\ell^{\prime}, r^{\prime}])$ 
    (see the definition of the subset $\Psi_{\OVQ, B}$). 
    Here, the interval attractor $([p^{\prime}, q^{\prime}], [\ell^{\prime}, r^{\prime}])$ is a node of the forest $\mathcal{G}_{\OVQ}$ 
    because $\{ I_{\capture}(i, j^{\prime}) \mid j^{\prime} \in [i+1, n] \} \cup \{ ([p, q], [\ell, r]) \in \Psi_{\OVQ, A} \mid r \geq i+1 \} \subseteq \mathcal{U}_{\OVQ}$. 
    $([p^{\prime}, q^{\prime}], [\ell^{\prime}, r^{\prime}]) \in \Psi_{\run}$ follows from the definition of run-extension. 
    Because of $([p^{\prime}, q^{\prime}], [\ell^{\prime}, r^{\prime}]) \in \Psi_{\run}$, 
    the node $([p^{\prime}, q^{\prime}], [\ell^{\prime}, r^{\prime}])$ has a directed edge to the node $([p, q], [\ell, r])$. 
    Therefore, the in-degree of the node $([p, q], [\ell, r])$ is at least $1$.

    \textbf{Proof of Lemma~\ref{lem:forest_g_ovq_property}(i).}    
    If $([p, q], [\ell, r]) \in \{ I_{\capture}(i, j^{\prime}) \mid j^{\prime} \in [i+1, n] \}$, 
    then statement (3) shows that the in-degree of the node $([p, q], [\ell, r])$ is $0$. 
    Otherwise, 
    statement (2) and statement (5) show that 
    the in-degree of the node $([p, q], [\ell, r])$ is $1$. 
    Therefore, Lemma~\ref{lem:forest_g_ovq_property}(i) holds. 
    
    \textbf{Proof of Lemma~\ref{lem:forest_g_ovq_property}(ii).} 
    We prove $\{ ([p, q], [\ell, r]) \in \Psi_{\OVQ, A} \mid r \geq i+1 \} \setminus \Psi_{\run} \subseteq \{ ([p, q], [\ell, r]) \in \Psi_{\RR} \setminus \Psi_{\run} \mid i \in [p, r] \}$.
    For each interval attractor $([p, q], [\ell, r]) \in \{ ([p, q], [\ell, r]) \in \Psi_{\OVQ, A} \mid r \geq i+1 \} \setminus \Psi_{\run}$, 
    $q \leq i-1$ follows from Lemma~\ref{lem:ovqab_property}~\ref{enum:ovqab_property:1}. 
    $i \in [p, r]$ follows from $p \leq q$, $q \leq i-1$, and $r \geq i+1$. 
    Therefore, $\{ ([p, q], [\ell, r]) \in \Psi_{\OVQ, A} \mid r \geq i+1 \} \setminus \Psi_{\run} \subseteq \{ ([p, q], [\ell, r]) \in \Psi_{\RR} \setminus \Psi_{\run} \mid i \in [p, r] \}$ holds. 
    
    %$[p, r] \cap [i, j] \neq \emptyset$ follows from 
    %$p \leq q$, $q \leq i-1$, and $r \geq i+1$. 
    %Therefore, $([p, q], [\ell, r]) \in \OVQ([i, j])$ follows from 
    %$[p, r] \cap [i, j] \neq \emptyset$ and $([p, q], [\ell, r]) \not \in \Psi_{\run}$. 
        
    We prove $\Psi_{\OVQ, B} \subseteq \{ ([p, q], [\ell, r]) \in \Psi_{\RR} \setminus \Psi_{\run} \mid i \in [p, r] \}$.
    For each interval attractor $([p, q], [\ell, r]) \in \Psi_{\OVQ, B}$, 
    $q-1 \leq i-1$, $r \geq i+1$, and $([p, q], [\ell, r]) \in \Psi_{\source}$ follows from Lemma~\ref{lem:ovqab_property}~\ref{enum:ovqab_property:2}. 
    $([p, q], [\ell, r]) \not \in \Psi_{\run}$ follows from 
    $([p, q], [\ell, r]) \in \Psi_{\source}$ and $\Psi_{\source} \cap \Psi_{\run} = \emptyset$. 
    $i \in [p, r]$ follows from $p \leq q$, $q-1 \leq i-1$, and $r \geq i+1$. 
    Therefore, $\Psi_{\OVQ, B} \subseteq \{ ([p, q], [\ell, r]) \in \Psi_{\RR} \setminus \Psi_{\run} \mid i \in [p, r] \}$ holds. 

    We prove $|\mathcal{U}_{\OVQ}| \leq H+1 + 3|\{ ([p, q], [\ell, r]) \in \Psi_{\RR} \setminus \Psi_{\run} \mid i \in [p, r] \}|$. 
    $|\mathcal{U}_{\OVQ}| \leq |\{ I_{\capture}(i, j^{\prime}) \mid j^{\prime} \in [i+1, n] \}| + |\{ ([p, q], [\ell, r]) \in \Psi_{\OVQ, A} \mid r \geq i+1 \}| + |\Psi_{\OVQ, B}|$ 
    follows from $\mathcal{U}_{\OVQ} = \{ I_{\capture}(i, j^{\prime}) \mid j^{\prime} \in [i+1, n] \} \cup \{ ([p, q], [\ell, r]) \in \Psi_{\OVQ, A} \mid r \geq i+1 \} \cup \Psi_{\OVQ, B}$. 
    $|\{ I_{\capture}(i, j^{\prime}) \mid j^{\prime} \in [i+1, n] \}| \leq H+1$ follows from Lemma~\ref{lem:psi_OVR_conditions}(ii). 
    $|\{ ([p, q], [\ell, r]) \in \Psi_{\OVQ, A} \mid r \geq i+1 \}| = |\{ ([p, q], [\ell, r]) \in \Psi_{\OVQ, A} \mid r \geq i+1 \} \cap \Psi_{\run}| + |\{ ([p, q], [\ell, r]) \in \Psi_{\OVQ, A} \mid r \geq i+1 \} \setminus \Psi_{\run}|$. 
    $|\{ ([p, q], [\ell, r]) \in \Psi_{\OVQ, A} \mid r \geq i+1 \} \cap \Psi_{\run}| \leq |\{ ([p, q], [\ell, r]) \in \Psi_{\RR} \setminus \Psi_{\run} \mid i \in [p, r] \}|$ 
    follows from statement Lemma~\ref{lem:ovqab_property}~\ref{enum:ovqab_property:3}. 
    $|\{ ([p, q], [\ell, r]) \in \Psi_{\OVQ, A} \mid r \geq i+1 \} \setminus \Psi_{\run}| \leq |\{ ([p, q], [\ell, r]) \in \Psi_{\RR} \setminus \Psi_{\run} \mid i \in [p, r] \}|$ 
    follows from $\{ ([p, q], [\ell, r]) \in \Psi_{\OVQ, A} \mid r \geq i+1 \} \setminus \Psi_{\run} \subseteq \{ ([p, q], [\ell, r]) \in \Psi_{\RR} \setminus \Psi_{\run} \mid i \in [p, r] \}$. 
    $|\Psi_{\OVQ, B}| \leq |\{ ([p, q], [\ell, r]) \in \Psi_{\RR} \setminus \Psi_{\run} \mid i \in [p, r] \}|$ 
    follows from $\Psi_{\OVQ, B} \subseteq \{ ([p, q], [\ell, r]) \in \Psi_{\RR} \setminus \Psi_{\run} \mid i \in [p, r] \}$. 
    Therefore, $|\mathcal{U}_{\OVQ}| \leq H+1 + 3|\{ ([p, q], [\ell, r]) \in \Psi_{\RR} \setminus \Psi_{\run} \mid i \in [p, r] \}|$ holds.
        
\end{proof}

%%%%%%%%%%%%%%%%%%%%%%%%%%%%%%%%%%%%%%%%%%%%%%%%%%

\paragraph{Computation of children of a node.}
For constructing the forest $\mathcal{G}_{\OVQ}$, 
we explain an algorithm for computing the children $([p_{1}, q_{1}], [\ell_{1}, r_{1}]), ([p_{2}, q_{2}], [\ell_{2}, r_{2}]), \ldots, ([p_{k}, q_{k}], [\ell_{k}, r_{k}])$ of a given node $([p, q], [\ell, r]) \in \mathcal{U}_{\OVQ}$. 
The following lemma states the relationship among the children, left-extensions, and run-extensions of a node in the forest $\mathcal{G}_{\OVQ}$. 

\begin{lemma}\label{lem:comp_ovrc_children}
    Consider a node $([p, q], [\ell, r]) \in \mathcal{U}_{\OVQ}$ 
    in the directed graph $\mathcal{G}_{\OVQ} = (\mathcal{U}_{\OVQ}, \mathcal{E}_{\OVQ})$ for an overlap query $\OVQ([i, j])$. 
    Let $\Psi$ and $\Psi^{\prime}$ be the two sets of the left-extensions and run-extensions of the node $([p, q], [\ell, r])$. 
    If $([p, q], [\ell, r]) \in \Psi_{\run}$, 
    then the children of the node $([p, q], [\ell, r])$ are all the run-extensions of the node $([p, q], [\ell, r])$. 
    Otherwise (i.e., $([p, q], [\ell, r]) \not \in \Psi_{\run}$), 
    the children of the node $([p, q], [\ell, r])$ are left-extensions of the node $([p, q], [\ell, r])$ 
    such that each left-extension $([p^{\prime}, q^{\prime}], [\ell^{\prime}, r^{\prime}])$ satisfies $r^{\prime} \geq i+1$.     
\end{lemma}
\begin{proof}
    Lemma~\ref{lem:comp_ovrc_children} holds if 
    the following four statements holds for 
    each interval attractor $([p^{\prime}, q^{\prime}]$, $[\ell^{\prime}, r^{\prime}]) \in \Psi_{\RR}$.

    \begin{enumerate}[label=\textbf{(\roman*)}]
    \item if $([p, q], [\ell, r]) \in \Psi_{\run}$, 
    and the interval attractor $([p^{\prime}, q^{\prime}], [\ell^{\prime}, r^{\prime}])$ is a run-extension of node $([p, q], [\ell, r])$, 
    then the interval attractor $([p^{\prime}, q^{\prime}], [\ell^{\prime}, r^{\prime}])$ is a node in the forest $\mathcal{G}_{\OVQ}$, 
    and the node $([p, q], [\ell, r])$ has a directed edge to the node $([p^{\prime}, q^{\prime}], [\ell^{\prime}, r^{\prime}])$;
    \item if (A) $([p, q], [\ell, r]) \in \Psi_{\run}$, 
    (B) the interval attractor $([p^{\prime}, q^{\prime}], [\ell^{\prime}, r^{\prime}])$ is a node in the forest $\mathcal{G}_{\OVQ}$, 
    and (C) the node $([p, q], [\ell, r])$ has a directed edge to the node $([p^{\prime}, q^{\prime}], [\ell^{\prime}, r^{\prime}])$, 
    then the node $([p^{\prime}, q^{\prime}], [\ell^{\prime}, r^{\prime}])$ is a run-extension of the node $([p, q], [\ell, r])$; 
    \item if (A) $([p, q], [\ell, r]) \not \in \Psi_{\run}$, (B) node $([p^{\prime}, q^{\prime}], [\ell^{\prime}, r^{\prime}])$ is a left-extension of node $([p, q], [\ell, r])$, and (C) $r^{\prime} \geq i+1$, 
    then the interval attractor $([p^{\prime}, q^{\prime}], [\ell^{\prime}, r^{\prime}])$ is a node in the forest $\mathcal{G}_{\OVQ}$, 
    and the node $([p, q], [\ell, r])$ has a directed edge to the node $([p^{\prime}, q^{\prime}], [\ell^{\prime}, r^{\prime}])$;
    \item if (A) $([p, q], [\ell, r]) \not \in \Psi_{\run}$, 
    (B) the interval attractor $([p^{\prime}, q^{\prime}], [\ell^{\prime}, r^{\prime}])$ is a node in the forest $\mathcal{G}_{\OVQ}$, 
    and (C) the node $([p, q], [\ell, r])$ has a directed edge to the node $([p^{\prime}, q^{\prime}], [\ell^{\prime}, r^{\prime}])$, 
    then the node $([p^{\prime}, q^{\prime}], [\ell^{\prime}, r^{\prime}])$ is a left-extension of the node $([p, q], [\ell, r])$, 
    and $r^{\prime} \geq i+1$ holds.     
    \end{enumerate}

    \textbf{Proof of statement (i).}
    We prove $([p, q], [\ell, r]) \in \{ I_{\capture}(i, j^{\prime}) \mid j^{\prime} \in [i+1, n] \} \cup \{ ([p, q], [\ell, r]) \in \Psi_{\OVQ, A} \mid r \geq i+1 \}$. 
    $\Psi_{\OVQ, B} \subseteq \Psi_{\source}$ holds 
    because each interval attractor in the subset $\Psi_{\OVQ, B}$ is a run-extension of an interval attractor. 
    $\Psi_{\OVQ, B} \cap \Psi_{\run} = \emptyset$ holds 
    because $\Psi_{\OVQ, B} \subseteq \Psi_{\source}$, 
    and $\Psi_{\source} \cap \Psi_{\run} = \emptyset$ follows from the definition of the subset $\Psi_{\source}$. 
    Therefore, $([p, q], [\ell, r]) \in \{ I_{\capture}(i, j^{\prime}) \mid j^{\prime} \in [i+1, n] \} \cup \{ ([p, q], [\ell, r]) \in \Psi_{\OVQ, A} \mid r \geq i+1 \}$ 
    follows from 
    $([p, q], [\ell, r]) \in \mathcal{U}_{\OVQ} \cap \Psi_{\run}$, 
    $\mathcal{U}_{\OVQ} = \{ I_{\capture}(i, j^{\prime}) \mid j^{\prime} \in [i+1, n] \} \cup \{ ([p, q], [\ell, r]) \in \Psi_{\OVQ, A} \mid r \geq i+1 \} \cup \Psi_{\OVQ, B}$, 
    and $\Psi_{\OVQ, B} \cap \Psi_{\run} = \emptyset$. 

    $([p^{\prime}, q^{\prime}], [\ell^{\prime}, r^{\prime}]) \in \Psi_{\OVQ, B}$ follows from the definition of the set $\Psi_{\OVQ, B}$ 
    because $([p, q], [\ell, r]) \in \{ I_{\capture}(i, j^{\prime}) \mid j^{\prime} \in [i+1, n] \} \cup \{ ([p, q], [\ell, r]) \in \Psi_{\OVQ, A} \mid r \geq i+1 \}$, 
    and the interval attractor $([p^{\prime}, q^{\prime}], [\ell^{\prime}, r^{\prime}])$ is a run-extension of node $([p, q], [\ell, r])$. 
    The interval attractor $([p^{\prime}, q^{\prime}], [\ell^{\prime}, r^{\prime}])$ is a node in the forest $\mathcal{G}_{\OVQ}$ 
    because $([p^{\prime}, q^{\prime}], [\ell^{\prime}, r^{\prime}]) \in \Psi_{\OVQ, B}$ and $\Psi_{\OVQ, B} \subseteq \mathcal{U}_{\OVQ}$. 
    The node $([p, q], [\ell, r])$ has a directed edge to the node $([p^{\prime}, q^{\prime}], [\ell^{\prime}, r^{\prime}])$ 
    because $([p, q], [\ell, r]) \in \Psi_{\run}$, 
    and the node $([p^{\prime}, q^{\prime}], [\ell^{\prime}, r^{\prime}])$ is a run-extension of node $([p, q], [\ell, r])$. 
        
    \textbf{Proof of statement (ii).}
    Statement (ii) follows from the definition of the set $\mathcal{E}_{\OVQ}$ of edges.

    \textbf{Proof of statement (iii).}
    We prove $([p, q], [\ell, r]) \in \Psi_{\OVQ, C} \cup \{ I_{\capture}(i, j^{\prime}) \mid j^{\prime} \in [i+1, n] \}$. 
    Because of $\mathcal{U}_{\OVQ} = \{ I_{\capture}(i, j^{\prime}) \mid j^{\prime} \in [i+1, n] \} \cup \{ ([p, q], [\ell, r]) \in \Psi_{\OVQ, A} \mid r \geq i+1 \} \cup \Psi_{\OVQ, B}$, 
    the node $([p, q], [\ell, r])$ satisfies 
    one of the following three conditions: 
    (a) $([p, q], [\ell, r]) \in \{ I_{\capture}(i, j^{\prime}) \mid j^{\prime} \in [i+1, n] \}$; 
    (b) $([p, q], [\ell, r]) \in \{ ([p, q], [\ell, r]) \in \Psi_{\OVQ, A} \mid r \geq i+1 \}$; 
    (c) $([p, q], [\ell, r]) \in \Psi_{\OVQ, B}$. 

    For condition (a), 
    $([p, q], [\ell, r]) \in \Psi_{\OVQ, C} \cup \{ I_{\capture}(i, j^{\prime}) \mid j^{\prime} \in [i+1, n] \}$ follows from $([p, q], [\ell, r]) \in \{ I_{\capture}(i, j^{\prime}) \mid j^{\prime} \in [i+1, n] \}$. 
    For condition (b), 
    $r \leq i+1$ holds. 
    $q \leq i-1$ follows from Lemma~\ref{lem:ovqab_property}~\ref{enum:ovqab_property:1}. 
    $([p, q], [\ell, r]) \in \Psi_{\OVQ, C}$ follows from $q \leq i-1$ and $r \leq i+1$. 
    For condition (c), 
    $q \leq i-1$ and $r \leq i+1$ follow from Lemma~\ref{lem:ovqab_property}~\ref{enum:ovqab_property:2}.
    $([p, q], [\ell, r]) \in \Psi_{\OVQ, C}$ follows from $q \leq i-1$ and $r \leq i+1$. 
    Therefore, $([p, q], [\ell, r]) \in \Psi_{\OVQ, C} \cup \{ I_{\capture}(i, j^{\prime}) \mid j^{\prime} \in [i+1, n] \}$ holds. 

    We prove statement (iii). 
    $([p, q], [\ell, r]) \in (\Psi_{\OVQ, C} \cup \{ I_{\capture}(i, j^{\prime}) \mid j^{\prime} \in [i+1, n] \}) \setminus \Psi_{\run}$ follows from 
    $([p, q], [\ell, r]) \in \Psi_{\OVQ, C} \cup \{ I_{\capture}(i, j^{\prime}) \mid j^{\prime} \in [i+1, n] \}$ and $([p, q], [\ell, r]) \not \in \Psi_{\run}$. 
    $([p^{\prime}, q^{\prime}], [\ell^{\prime}, r^{\prime}]) \in \Psi_{\OVQ, A}$ follows from the definition of the set $\Psi_{\OVQ, A}$ 
    because $([p, q], [\ell, r]) \in (\Psi_{\OVQ, C} \cup \{ I_{\capture}(i, j^{\prime}) \mid j^{\prime} \in [i+1, n] \}) \setminus \Psi_{\run}$, 
    and the interval attractor $([p^{\prime}, q^{\prime}], [\ell^{\prime}, r^{\prime}])$ is a left-extension of node $([p, q], [\ell, r])$. 
    $([p^{\prime}, q^{\prime}], [\ell^{\prime}, r^{\prime}]) \in \mathcal{U}_{\OVQ}$ follows from 
    $([p^{\prime}, q^{\prime}], [\ell^{\prime}, r^{\prime}]) \in \Psi_{\OVQ, A}$, 
    $r^{\prime} \geq i+1$, 
    and $\mathcal{U}_{\OVQ} = \{ I_{\capture}(i, j^{\prime}) \mid j^{\prime} \in [i+1, n] \} \cup \{ ([p, q], [\ell, r]) \in \Psi_{\OVQ, A} \mid r \geq i+1 \} \cup \Psi_{\OVQ, B}$.
    The node $([p, q], [\ell, r])$ has a directed edge to the node $([p^{\prime}, q^{\prime}], [\ell^{\prime}, r^{\prime}])$ 
    because $([p, q], [\ell, r]) \not \in \Psi_{\run}$, 
    and the node $([p^{\prime}, q^{\prime}], [\ell^{\prime}, r^{\prime}])$ is a left-extension of node $([p, q], [\ell, r])$. 

    \textbf{Proof of statement (iv).}
    From the definition of the set $\mathcal{E}_{\OVQ}$ of edges, 
    the node $([p^{\prime}, q^{\prime}], [\ell^{\prime}, r^{\prime}])$ is a left-extension of the node $([p, q], [\ell, r])$. 
    We prove $r^{\prime} \geq i+1$. 
    Because of $\mathcal{U}_{\OVQ} = \{ I_{\capture}(i, j^{\prime}) \mid j^{\prime} \in [i+1, n] \} \cup \{ ([p, q], [\ell, r]) \in \Psi_{\OVQ, A} \mid r \geq i+1 \} \cup \Psi_{\OVQ, B}$, 
    the node $([p^{\prime}, q^{\prime}], [\ell^{\prime}, r^{\prime}])$ satisfies 
    one of the following three conditions: 
    (a) $([p^{\prime}, q^{\prime}], [\ell^{\prime}, r^{\prime}]) \in \{ I_{\capture}(i, j^{\prime}) \mid j^{\prime} \in [i+1, n] \}$; 
    (b) $([p^{\prime}, q^{\prime}], [\ell^{\prime}, r^{\prime}]) \in \{ ([p, q], [\ell, r]) \in \Psi_{\OVQ, A} \mid r \geq i+1 \}$; 
    (c) $([p^{\prime}, q^{\prime}], [\ell^{\prime}, r^{\prime}]) \in \Psi_{\OVQ, B}$. 
    
    For condition (a), 
    there exists an integer $j^{\prime} \in [i+1, n]$ satisfying $I_{\capture}(i, j^{\prime}) = ([p^{\prime}, q^{\prime}], [\ell^{\prime}, r^{\prime}])$. 
    $i \in [p^{\prime}, q^{\prime}]$ and $\ell^{\prime} \geq i+1$ follow from the definition of interval attractor 
    because $I_{\capture}(i, j^{\prime}) = ([p^{\prime}, q^{\prime}], [\ell^{\prime}, r^{\prime}])$ and $j^{\prime} \geq i+1$. 
    $r^{\prime} \geq i+1$ follows from $\ell^{\prime} \geq i+1$ and $\ell^{\prime} \leq r^{\prime}$. 
    For condition (b), 
    $r^{\prime} \geq i+1$ follows from $([p^{\prime}, q^{\prime}], [\ell^{\prime}, r^{\prime}]) \in \{ ([p, q], [\ell, r]) \in \Psi_{\OVQ, A} \mid r \geq i+1 \}$. 
    For condition (c), 
    Lemma~\ref{lem:ovqab_property}~\ref{enum:ovqab_property:2} shows that $r^{\prime} \geq i+1$ holds. 
    Therefore, $r^{\prime} \geq i+1$ holds.     
\end{proof}

Lemma~\ref{lem:comp_ovrc_children} shows that the children of a node $([p, q], [\ell, r]) \in \mathcal{U}_{\OVQ}$ 
are all the run-extensions of the node if $([p, q], [\ell, r]) \in \Psi_{\run}$; 
otherwise, the children of the node $([p, q], [\ell, r])$ are left-extensions of the node 
such that each left-extension $([p^{\prime}, q^{\prime}], [\ell^{\prime}, r^{\prime}])$ satisfies $r^{\prime} \geq i+1$. 
Therefore, the children of a node can be obtained by computing left-extensions and run-extensions of the node. 

The algorithm for computing the children of node $([p, q], [\ell, r])$ consists of the following four steps: 
\begin{enumerate}[label=\textbf{(\roman*)}]
    \item verify whether $([p, q], [\ell, r]) \in \Psi_{\run}$ or not by verify-run query $\runQ(([p, q], [\ell, r]))$; 
    \item if $([p, q], [\ell, r]) \in \Psi_{\run}$, then 
    compute all the run-extensions of the node $([p, q], [\ell, r])$ 
    and return the run-extensions as the children of the node $([p, q], [\ell, r])$; 
    otherwise, proceed to the next step; 
    \item compute all the left-extensions of the node $([p, q], [\ell, r])$     
    \item return each left-extension $([p^{\prime}, q^{\prime}], [\ell^{\prime}, r^{\prime}])$ of the node $([p, q], [\ell, r])$ 
    satisfying $r^{\prime} \geq i+1$ as a child of the node.
\end{enumerate}

The bottleneck of this algorithm is computing all the left-extensions of the node $([p, q], [\ell, r])$, 
which takes $O(H^{3} \log n)$ time. 
Therefore, we can compute the children of a given node in $O(H^{3} \log n)$ time. 

\paragraph{Construction of forest $\mathcal{G}_{\OVQ}$.}
We construct the forest $\mathcal{G}_{\OVQ}$ in two phases. 
Let $([p_{1}, q_{1}], [\ell_{1}, r_{1}])$, $([p_{2}, q_{2}], [\ell_{2}, r_{2}])$, $\ldots$, $([p_{k}, q_{k}], [\ell_{k}, r_{k}])$ be the roots of the forest $\mathcal{G}_{\OVQ}$. 
Here, $\{ ([p_{s}, q_{s}], [\ell_{s}, r_{s}]) \mid s \in [1, k] \} = \{ I_{\capture}(i, j^{\prime}) \mid j^{\prime} \in [i+1, n] \}$ follows from Lemma~\ref{lem:forest_g_ovq_property}~\ref{lem:forest_g_ovq_property:1}. 

In the first phase, we obtain the $k$ roots $([p_{1}, q_{1}], [\ell_{1}, r_{1}])$, $([p_{2}, q_{2}], [\ell_{2}, r_{2}])$, $\ldots$, $([p_{k}, q_{k}], [\ell_{k}, r_{k}])$ by computing the set $\{ I_{\capture}(i, j^{\prime}) \mid j^{\prime} \in [i+1, n] \}$ of interval attractors. This set can be computed by the first phase of the algorithm presented in Section~\ref{subsubsec:ovqr}. 
Therefore, we can obtain the $k$ roots in $O(H^{3} \log n)$ time. 

In the second phase, we obtain all the descendants of the $k$ roots $([p_{1}, q_{1}], [\ell_{1}, r_{1}])$, $([p_{2}, q_{2}], [\ell_{2}, r_{2}])$, $\ldots$, $([p_{k}, q_{k}], [\ell_{k}, r_{k}])$. 
These descendants can be obtained by computing the children of each node. 
Computing the children of each node takes $O(H^{3} \log n)$ time. 
Therefore, all the descendants of the $k$ roots can be obtained in $O(m H^{3} \log n)$ time 
for the number $m$ of nodes in the forest $\mathcal{G}_{\OVQ}$. 

After executing the second phase, we obtain the forest $\mathcal{G}_{\OVQ}$. 
This is because all the nodes in the forest can be found in the two phases. 
Therefore, 
the construction of the forest $\mathcal{G}_{\OVQ}$ takes $O(m H^{3} \log n)$ time in total. 

\paragraph{Computation of set $\Psi_{\OVQ, C} \setminus \Psi_{\run}$.}
For computing set $\Psi_{\OVQ, C} \setminus \Psi_{\run}$, 
we leverage the fact that 
each interval attractor of the set $\Psi_{\OVQ, C} \setminus \Psi_{\run}$ 
occurs in the forest $\mathcal{G}_{\OVQ}$ as a node. 
Here, a node $([p, q], [\ell, r])$ of the forest is contained in the set $\Psi_{\OVQ, C} \setminus \Psi_{\run}$ 
if (A) $(q \leq i-1)$, (B) $(r \leq i+1)$, and (C) $([p, q], [\ell, r]) \not \in \Psi_{\run}$. 
Therefore, we can obtain the set $\Psi_{\OVQ, C} \setminus \Psi_{\run}$ 
by verifying whether each node of the forest $\mathcal{G}_{\OVQ}$ 
is contained in the set $\Psi_{\OVQ, C} \setminus \Psi_{\run}$ or not. 

The algorithm for computing the set $\Psi_{\OVQ, C} \setminus \Psi_{\run}$ consists of two phases. 
In the first phase, we construct the forest $\mathcal{G}_{\OVQ}$, 
which takes $O(m H^{3} \log n)$ time for the number $m$ of nodes in the forest $\mathcal{G}_{\OVQ}$. 

In the second phase, find all the interval attractors in the set $\Psi_{\OVQ, C} \setminus \Psi_{\run}$ 
by verifying whether each node of the forest $\mathcal{G}_{\OVQ}$ 
is contained in the set $\Psi_{\OVQ, C} \setminus \Psi_{\run}$ or not. 
This verification need to execute verify-run query, which takes $O(H^{2})$ time. 
Therefore, the second phase takes $O(m H^{2})$ time. 

The computation of the set $\Psi_{\OVQ, C} \setminus \Psi_{\run}$ takes $O(m H^{3} \log n)$ time in total. 
Here, Lemma~\ref{lem:forest_g_ovq_property} shows that 
the number $m$ can be bounded by 
$m = O(H + |\{ ([p, q], [\ell, r]) \in \Psi_{\RR} \setminus \Psi_{\run} \mid i \in [p, r] \}|)$. 

%%%%%%%%%%%%%%%%%%%%%%%%%%%%%%%%%%%%%%%%%%%%%%%%%%%%%%%%%%%%%%%%%%%%%%%%%%%%%%%%%%%%%%%%%%%%%%%

\subsubsection{Computation of Set \texorpdfstring{$\{ I_{\capture}(x, i) \mid x \in [1, i-1] \} \setminus \Psi_{\run}$}{}}\label{subsubsec:ovql}
The following lemma is used to compute the set $\{ I_{\capture}(x, i) \mid x \in [1, i-1] \} \setminus \Psi_{\run}$. 

\begin{lemma}\label{lem:psi_OVL_conditions}
Consider the $k$ interval attractors $([p_{1}, q_{1}], [\ell_{1}, r_{1}])$, $([p_{2}, q_{2}], [\ell_{2}, r_{2}])$, $\ldots$, 
$([p_{k}, q_{k}]$, $[\ell_{k}, r_{k}])$ ($p_{1} \geq p_{2} \geq \cdots \geq p_{k}$) in 
set $\{ I_{\capture}(x, i) \mid x \in [1, i-1] \} \setminus \Psi_{\run}$ 
for a position $i \in [1, n]$ in input string $T$ satisfying $i \geq 2$.
Then, the following four statements hold: 
\begin{enumerate}[label=\textbf{(\roman*)}]
    \item \label{lem:psi_OVL_conditions:1} $p_{1} > p_{2} > \cdots > p_{k} = 1$; 
    \item \label{lem:psi_OVL_conditions:2} $1 \leq k \leq |\{ ([p, q], [\ell, r]) \in \Psi_{\RR} \setminus \Psi_{\run} \mid i \in [p, r] \}|$;
    \item \label{lem:psi_OVL_conditions:3}
    consider interval attractor $I_{\capture}(i-1, i) = ([p, q], [\ell, r])$. 
    If this interval attractor is contained in set $\Psi_{\run}$, 
    then the interval attractor $([p_{1}, q_{1}], [\ell_{1}, r_{1}])$ is a run-extension of the interval attractor $([p, q], [\ell, r])$; 
    otherwise $([p_{1}, q_{1}], [\ell_{1}, r_{1}]) = ([p, q], [\ell, r])$; 
    \item \label{lem:psi_OVL_conditions:4} consider $I_{\capture}(p_{s-1}-1, i) = ([p, q], [\ell, r])$ for an integer $s \in [2, k]$.
    If $([p, q], [\ell, r]) \in \Psi_{\run}$, 
    then the interval attractor $([p_{s}, q_{s}], [\ell_{s}, r_{s}])$ is a run-extension of the interval attractor $([p, q], [\ell, r])$; 
    otherwise $([p_{s}, q_{s}], [\ell_{s}, r_{s}]) = ([p, q], [\ell, r])$. 
\end{enumerate}     
\end{lemma}
\begin{proof}
Let $\{ x_{1}, x_{2}, \ldots, x_{m} \}$ ($x_{1} > x_{2} > \cdots > x_{m}$) be a subset of set $\{ 1, 2, \ldots, i-1 \}$ 
such that each integer $x_{s}$ satisfies either of the following two conditions: 
(A) $x_{s} = i-1$; 
(B) $x_{s} < i-1$ and $I_{\capture}(x_{s}, i) \neq I_{\capture}(x_{s}+1, i)$. 
Because of $i \geq 2$, 
$m \geq 1$ and $x_{1} = i-1$ hold. 
Let $\{ x^{\prime}_{1}, x^{\prime}_{2}, \ldots, x^{\prime}_{m^{\prime}} \}$ 
($x^{\prime}_{1} > x^{\prime}_{2} > \cdots > x^{\prime}_{m^{\prime}}$) be a subset of set $\{ x_{1}, x_{2}, \ldots, x_{m} \}$ 
such that each integer $x^{\prime}_{s}$ satisfies $I_{\capture}(x^{\prime}_{s}, i) \not \in \Psi_{\run}$ 
(i.e., $\{ x^{\prime}_{1}, x^{\prime}_{2}, \ldots, x^{\prime}_{m^{\prime}} \} = \{ x_{s} \in [1, m] \mid I_{\capture}(x_{s}, i) \not \in \Psi_{\run} \}$). 
For each integer $s \in [1, m]$, 
let $([p^{\prime}_{s}, q^{\prime}_{s}], [\ell^{\prime}_{s}, r^{\prime}_{s}])$ be interval attractor $I_{\capture}(x_{s}, i)$. 

The following four statements are used to prove Lemma~\ref{lem:psi_OVL_conditions}: 

\begin{enumerate}[label=\textbf{(\arabic*)}]
    \item      
    $x_{s} \in [p^{\prime}_{s}, q^{\prime}_{s}]$, $i \in [\ell^{\prime}_{s}, r^{\prime}_{s}]$, 
    and $I_{\capture}(x_{s}, i) = I_{\capture}(x_{s}-1, i) = \cdots = I_{\capture}(p^{\prime}_{s}, i)$ for each integer $s \in [1, m]$;     
    \item  
    $x_{s} = p^{\prime}_{s-1} - 1$ and $p^{\prime}_{s} < p^{\prime}_{s-1}$ for each integer $s \in [2, m]$;
    \item 
    let $\tau$ be the smallest integer in set $\{ s+1, s+2, \ldots, m \}$ 
    that satisfies $([p^{\prime}_{\tau}, q^{\prime}_{\tau}], [\ell^{\prime}_{\tau}, r^{\prime}_{\tau}]) \not \in \Psi_{\run}$ for an integer $s \in [1, m]$. 
    If $([p^{\prime}_{s}, q^{\prime}_{s}], [\ell^{\prime}_{s}, r^{\prime}_{s}]) \in \Psi_{\run}$, 
    then 
    (A) $s \leq m-1$, 
    (B) the smallest integer $\tau$ exists, and 
    (C) the interval attractor $([p^{\prime}_{\tau}, q^{\prime}_{\tau}], [\ell^{\prime}_{\tau}, r^{\prime}_{\tau}])$ 
    is a run-extension of the interval attractor $([p^{\prime}_{s}, q^{\prime}_{s}], [\ell^{\prime}_{s}, r^{\prime}_{s}])$;    
    \item $k = m^{\prime}$, and 
    $([p_{s}, q_{s}], [\ell_{s}, r_{s}]) = I_{\capture}(x^{\prime}_{s}, i)$ for all $s \in [1, k]$.
\end{enumerate}    

\textbf{Proof of statement (1).}
$x_{s} \in [p^{\prime}_{s}, q^{\prime}_{s}]$ and $i \in [\ell^{\prime}_{s}, r^{\prime}_{s}]$ follow from the definition of interval attractor for the interval attractor $I_{\capture}(x_{s}, i) = ([p^{\prime}_{s}, q^{\prime}_{s}], [\ell^{\prime}_{s}, r^{\prime}_{s}])$. 
Lemma~\ref{lem:IA_maximal_lemma} shows that 
$I_{\capture}(x_{s}, i) = I_{\capture}(x_{s}-1, i) = \cdots = I_{\capture}(p^{\prime}_{s}, i)$. 

\textbf{Proof of statement (2).}
We prove $x_{s} = p^{\prime}_{s-1}-1$ by contradiction. 
Here, $x_{s} \leq p^{\prime}_{s-1}-1$ holds 
because $I_{\capture}(x_{s-1}, i) = I_{\capture}(x_{s-1}-1, i) = \cdots = I_{\capture}(p^{\prime}_{s-1}, i)$ follows from statement (1). 
We assume that $x_{s} \neq p^{\prime}_{s-1}-1$ holds. 
Then, $x_{s} < p^{\prime}_{s-1}-1$ follows from $x_{s} \neq p^{\prime}_{s-1}-1$ and $x_{s} \leq p^{\prime}_{s-1}-1$. 
This inequality $x_{s} < p^{\prime}_{s-1}-1$ indicates that $I_{\capture}(p^{\prime}_{s-1}, i) = I_{\capture}(p^{\prime}_{s-1}-1, i)$ holds. 
$I_{\capture}(p^{\prime}_{s-1}-1, i) = ([p^{\prime}_{s-1}, q^{\prime}_{s-1}], [\ell^{\prime}_{s-1}, r^{\prime}_{s-1}])$ 
follows from 
$I_{\capture}(x_{s-1}, i) = I_{\capture}(x_{s-1}-1, i) = \cdots = I_{\capture}(p^{\prime}_{s-1}-1, i)$ 
and $I_{\capture}(x_{s-1}, i) = ([p^{\prime}_{s-1}, q^{\prime}_{s-1}], [\ell^{\prime}_{s-1}, r^{\prime}_{s-1}])$. 
$p^{\prime}_{s-1}-1 \in [p^{\prime}_{s-1}, q^{\prime}_{s-1}]$ and $i \in [\ell^{\prime}_{s-1}, r^{\prime}_{s-1}]$ follow from the definition of interval attractor for 
the interval attractor $I_{\capture}(p^{\prime}_{s-1}-1, i) = ([p^{\prime}_{s-1}, q^{\prime}_{s-1}], [\ell^{\prime}_{s-1}, r^{\prime}_{s-1}])$. 
On the other hand, $p^{\prime}_{s-1}-1 \not \in [p^{\prime}_{s-1}, q^{\prime}_{s-1}]$ follows from $p^{\prime}_{s-1}-1 < p^{\prime}_{s-1}$. 
The two facts $p^{\prime}_{s-1}-1 \in [p^{\prime}_{s-1}, q^{\prime}_{s-1}]$ and $p^{\prime}_{s-1}-1 \not \in [p^{\prime}_{s-1}, q^{\prime}_{s-1}]$ yield a contradiction. 
Therefore, $x_{s} = p^{\prime}_{s-1}-1$ must hold. 

$p^{\prime}_{s} < p^{\prime}_{s-1}$ follows from $x_{s} = p^{\prime}_{s-1}-1$ and $x_{s} \in [p^{\prime}_{s}, q^{\prime}_{s}]$. Therefore, statement (2) holds. 

\textbf{Proof of statement (3).}
Lemma~\ref{lem:recover_division_property}~\ref{enum:recover_division_property:1} shows that the subset $\Psi_{\source}$ contains an interval attractor 
$([p, q], [\ell, r])$ satisfying $([p^{\prime}_{s}, q^{\prime}_{s}], [\ell^{\prime}_{s}, r^{\prime}_{s}]) \in f_{\recover}(([p, q], [\ell, r]))$. 
Consider $d$ interval attractors 
$([p_{A, 1}, q_{A, 1}], [\ell_{A, 1}, r_{A, 1}])$, 
$([p_{A, 2}, q_{A, 2}], [\ell_{A, 2}, r_{A, 2}])$, $\ldots$, 
$([p_{A, d}, q_{A, d}], [\ell_{A, d}, r_{A, d}])$ ($p_{A, 1} \leq p_{A, 2} \leq \cdots \leq p_{A, d}$) obtained from the function $f_{\recover}(([p, q], [\ell, r]))$. 
Because of $([p^{\prime}_{s}, q^{\prime}_{s}], [\ell^{\prime}_{s}, r^{\prime}_{s}]) \in f_{\recover}(([p, q], [\ell, r]))$, 
there exists an integer $g \in [1, d]$ satisfying $([p_{A, g}, q_{A, g}], [\ell_{A, g}, r_{A, g}]) = ([p^{\prime}_{s}, q^{\prime}_{s}]$, $[\ell^{\prime}_{s}, r^{\prime}_{s}])$. 
Let $C$ be the associated string of the interval attractor $([p, q], [\ell, r])$. 
Then, for each integer $b \in [1, g]$, 
$([p_{A, b}, q_{A, b}], [\ell_{A, b}, r_{A, b}]) = ([q + 1 + (b - 1) |C|, q + b|C|], [\ell_{A, 1} + (b - 1) |C|, r])$ 
and $([p_{A, b}, q_{A, b}], [\ell_{A, b}, r_{A, b}]) \in \Psi_{\run}$ 
follow from the definition of the function $f_{\recover}$ and Lemma~\ref{lem:recover_basic_property}~\ref{enum:recover_basic_property:2}. 

For each integer $b \in [1, g]$, 
we prove $([p_{A, b}, q_{A, b}], [\ell_{A, b}, r_{A, b}]) = ([p^{\prime}_{s + g - b}, q^{\prime}_{s + g - b}], [\ell^{\prime}_{s + g - b}, r^{\prime}_{s + g - b}])$ by induction on $b$. 
For the base case $b = g$, 
$([p_{A, b}, q_{A, b}], [\ell_{A, b}, r_{A, b}]) = ([p^{\prime}_{s + g - b}, q^{\prime}_{s + g - b}], [\ell^{\prime}_{s + g - b}, r^{\prime}_{s + g - b}])$ holds 
because $([p_{A, g}, q_{A, g}], [\ell_{A, g}, r_{A, g}]) = ([p^{\prime}_{s}, q^{\prime}_{s}], [\ell^{\prime}_{s}, r^{\prime}_{s}])$. 

For the inductive step, 
consider $b \in [1, g-1]$. 
Then, we assume that 
$([p_{A, b+1}, q_{A, b+1}]$, $[\ell_{A, b+1}$, $r_{A, b+1}]) = ([p^{\prime}_{s + g - b - 1}$, $q^{\prime}_{s + g - b - 1}], [\ell^{\prime}_{s + g - b - 1}, r^{\prime}_{s + g - b - 1}])$. 
Statement (1) shows that 
$i \in [\ell_{A, b+1}, r_{A, b+1}]$ and 
$I_{\capture}(x_{s + g - b - 1}, i) = I_{\capture}(x_{s + g - b - 1}-1, i) = \cdots = I_{\capture}(p^{\prime}_{s + g - b - 1}, i)$ hold. 
Because of $I_{\capture}(x_{s + g - b - 1}, i) = I_{\capture}(x_{s + g - b - 1}-1, i) = \cdots = I_{\capture}(p^{\prime}_{s + g - b - 1}, i)$, 
$([p_{A, b}, q_{A, b}], [\ell_{A, b}, r_{A, b}]) = ([p^{\prime}_{s + g - b}, q^{\prime}_{s + g - b}], [\ell^{\prime}_{s + g - b}, r^{\prime}_{s + g - b}])$ holds 
if $I_{\capture}(p^{\prime}_{s + g - b - 1} - 1, i) = ([p_{A, b}, q_{A, b}], [\ell_{A, b}, r_{A, b}])$. 
Here, $I_{\capture}(p^{\prime}_{s + g - b - 1} - 1, i) = I_{\capture}(p_{A, b+1} - 1, i)$ follows from 
$([p_{A, b+1}, q_{A, b+1}], [\ell_{A, b+1}, r_{A, b+1}]) = ([p^{\prime}_{s + g - b - 1}, q^{\prime}_{s + g - b - 1}], [\ell^{\prime}_{s + g - b - 1}, r^{\prime}_{s + g - b - 1}])$.

We prove $I_{\capture}(p_{A, b+1} - 1, i) = ([p_{A, b}, q_{A, b}], [\ell_{A, b}, r_{A, b}])$ under the assumption. 
Consider interval attractor $I_{\capture}(p_{A, b+1}-1, \ell_{A, b+1})$. 
Since $([p_{A, b+1}, q_{A, b+1}], [\ell_{A, b+1}, r_{A, b+1}]) \in \Psi_{\run}$, 
$I_{\capture}(p_{A, b+1}-1, \ell_{A, b+1}) = ([p_{A, b}$, $q_{A, b}], [\ell_{A, b}, r_{A, b}])$ 
follows from Lemma~\ref{lem:psi_run_basic_property}~\ref{enum:psi_run_basic_property:5}. 
$i \in [\ell_{A, b+1}, r_{A, b}]$ follows from $i \in [\ell_{A, b+1}, r_{A, b+1}]$ and $r_{A, b} = r_{A, b+1}$. 
Lemma~\ref{lem:IA_maximal_lemma} shows that 
$I_{\capture}(p_{A, b+1}-1, i) = ([p_{A, b}$, $q_{A, b}], [\ell_{A, b}, r_{A, b}])$ 
because 
$i \in [\ell_{A, b+1}, r_{A, b}]$ and $I_{\capture}(p_{A, b+1}-1, \ell_{A, b+1}) = ([p_{A, b}$, $q_{A, b}], [\ell_{A, b}, r_{A, b}])$.

$I_{\capture}(p^{\prime}_{s + g - b - 1} - 1, i) = ([p_{A, b}, q_{A, b}], [\ell_{A, b}, r_{A, b}])$ 
follows from $I_{\capture}(p^{\prime}_{s + g - b - 1} - 1, i) = I_{\capture}(p_{A, b+1} - 1, i)$ and 
$I_{\capture}(p_{A, b+1} - 1, i) = ([p_{A, b}, q_{A, b}], [\ell_{A, b}, r_{A, b}])$. 
This fact indicates that $([p_{A, b}, q_{A, b}], [\ell_{A, b}, r_{A, b}]) = ([p^{\prime}_{s + g - b}, q^{\prime}_{s + g - b}], [\ell^{\prime}_{s + g - b}, r^{\prime}_{s + g - b}])$ holds. 

Finally, by induction on $b$, 
$([p_{A, b}, q_{A, b}], [\ell_{A, b}, r_{A, b}]) = ([p^{\prime}_{s + g - b}, q^{\prime}_{s + g - b}], [\ell^{\prime}_{s + g - b}, r^{\prime}_{s + g - b}])$ holds for all $b \in [1, g]$. 
Similarly, we can prove $([p, q], [\ell, r]) = ([p^{\prime}_{s + g}, q^{\prime}_{s + g}], [\ell^{\prime}_{s + g}, r^{\prime}_{s + g}])$ using the same approach. 

We prove $s \leq m-1$. 
Because of $([p, q], [\ell, r]) = ([p^{\prime}_{s + g}, q^{\prime}_{s + g}], [\ell^{\prime}_{s + g}, r^{\prime}_{s + g}])$, 
$s+g \leq m$ holds. 
$s \leq m-1$ follows from $g \in [1, d]$ and $s+g \leq m$. 

We prove $\tau = s+g$. 
For each integer $b \in [1, g]$, 
$([p^{\prime}_{s + g - b}, q^{\prime}_{s + g - b}], [\ell^{\prime}_{s + g - b}, r^{\prime}_{s + g - b}]) \in \Psi_{\run}$ holds 
because $([p^{\prime}_{s + g - b}, q^{\prime}_{s + g - b}], [\ell^{\prime}_{s + g - b}, r^{\prime}_{s + g - b}]) = ([p_{A, b}, q_{A, b}], [\ell_{A, b}, r_{A, b}])$ and $([p_{A, b}, q_{A, b}], [\ell_{A, b}, r_{A, b}]) \in \Psi_{\run}$. 
In contrast, 
$([p^{\prime}_{s + g}, q^{\prime}_{s + g}], [\ell^{\prime}_{s + g}, r^{\prime}_{s + g}]) \not \in \Psi_{\run}$ holds 
because 
(a) $([p^{\prime}_{s + g}, q^{\prime}_{s + g}], [\ell^{\prime}_{s + g}, r^{\prime}_{s + g}]) = ([p, q], [\ell, r])$, 
(b) $([p, q], [\ell, r]) \in \Psi_{\source}$, 
and (c) $\Psi_{\source} \cap \Psi_{\run} = \emptyset$. 
Therefore, $\tau = s+g$ follows from $([p^{\prime}_{s + g}, q^{\prime}_{s + g}], [\ell^{\prime}_{s + g}, r^{\prime}_{s + g}]) \not \in \Psi_{\run}$ and 
$\{ ([p^{\prime}_{s + g - b}, q^{\prime}_{s + g - b}], [\ell^{\prime}_{s + g - b}, r^{\prime}_{s + g - b}]) \mid b \in [1, g] \} \subseteq \Psi_{\run}$.

We show that the interval attractor $([p^{\prime}_{\tau}, q^{\prime}_{\tau}], [\ell^{\prime}_{\tau}, r^{\prime}_{\tau}])$ 
is a run-extension of the interval attractor $([p^{\prime}_{s}, q^{\prime}_{s}], [\ell^{\prime}_{s}, r^{\prime}_{s}])$. 
Because of $([p^{\prime}_{s}, q^{\prime}_{s}], [\ell^{\prime}_{s}, r^{\prime}_{s}]) \in f_{\recover}(([p, q], [\ell, r]))$, 
the interval attractor $([p, q], [\ell, r])$ is a run-extension of the interval attractor $([p^{\prime}_{s}, q^{\prime}_{s}], [\ell^{\prime}_{s}, r^{\prime}_{s}])$. 
Because of $([p, q], [\ell, r]) = ([p^{\prime}_{\tau}, q^{\prime}_{\tau}], [\ell^{\prime}_{\tau}, r^{\prime}_{\tau}])$, 
the interval attractor $([p^{\prime}_{\tau}, q^{\prime}_{\tau}], [\ell^{\prime}_{\tau}, r^{\prime}_{\tau}])$ 
is a run-extension of the interval attractor $([p^{\prime}_{s}, q^{\prime}_{s}], [\ell^{\prime}_{s}, r^{\prime}_{s}])$. 

\textbf{Proof of statement (4).}
We prove $k = m^{\prime}$.
$\{ ([p_{s}, q_{s}], [\ell_{s}, r_{s}]) \mid s \in [1, k] \} = \{ ([p^{\prime}_{s}, q^{\prime}_{s}], [\ell^{\prime}_{s}, r^{\prime}_{s}]) \mid s \in [1, m] \} \subseteq \Psi_{\run}$ follows from the definition of the subset $\{ x_{1}, x_{2}, \ldots, x_{m} \}$. 
Let $\mathcal{X} = \{ x^{\prime}_{1}$, $x^{\prime}_{2}$, $\ldots$, $x^{\prime}_{m^{\prime}} \}$ for simplicity. 
Then, 
$\{ ([p^{\prime}_{s}, q^{\prime}_{s}], [\ell^{\prime}_{s}, r^{\prime}_{s}]) \mid s \in [1, m] \} \subseteq \Psi_{\run} = \{ ([p^{\prime}_{s}, q^{\prime}_{s}], [\ell^{\prime}_{s}, r^{\prime}_{s}]) \mid s \in \mathcal{X} \}$ follows from 
the subset of $\{ x^{\prime}_{1}, x^{\prime}_{2}, \ldots, x^{\prime}_{m^{\prime}} \}$. 
$|\{ ([p_{s}, q_{s}], [\ell_{s}, r_{s}]) \mid s \in [1, k] \}| = |\{ ([p^{\prime}_{s}, q^{\prime}_{s}], [\ell^{\prime}_{s}, r^{\prime}_{s}]) \mid s \in \mathcal{X} \}|$ follows from 
$\{ ([p_{s}, q_{s}], [\ell_{s}, r_{s}]) \mid s \in [1, k] \} = \{ ([p^{\prime}_{s}, q^{\prime}_{s}], [\ell^{\prime}_{s}, r^{\prime}_{s}]) \mid s \in [1, m] \} \subseteq \Psi_{\run}$ 
and $\{ ([p^{\prime}_{s}, q^{\prime}_{s}], [\ell^{\prime}_{s}, r^{\prime}_{s}]) \mid s \in [1, m] \} \subseteq \Psi_{\run} = \{ ([p^{\prime}_{s}, q^{\prime}_{s}], [\ell^{\prime}_{s}, r^{\prime}_{s}]) \mid s \in \mathcal{X} \}$. 
$|\{ ([p_{s}, q_{s}], [\ell_{s}, r_{s}]) \mid s \in [1, k] \}| = k$ 
and $|\{ ([p^{\prime}_{s}, q^{\prime}_{s}], [\ell^{\prime}_{s}, r^{\prime}_{s}]) \mid s \in \mathcal{X} \}| = m^{\prime}$. 
Therefore, $k = m^{\prime}$ holds. 

We prove $([p_{s}, q_{s}], [\ell_{s}, r_{s}]) = I_{\capture}(x^{\prime}_{s}, i)$ for all $s \in [1, k]$. 
From the definition of the subset $\{ x^{\prime}_{1}, x^{\prime}_{2}, \ldots, x^{\prime}_{m^{\prime}} \}$, 
there exists an integer $g_{s} \in [1, m]$ satisfying 
$I_{\capture}(x^{\prime}_{s}, i) = ([p^{\prime}_{g_{s}}, q^{\prime}_{g_{s}}], [\ell^{\prime}_{g_{s}}, r^{\prime}_{g_{s}}])$. 
Here, $g_{1} < g_{2} < \cdots < g_{k}$ holds. 
$p^{\prime}_{g_{1}} > p^{\prime}_{g_{2}} > \cdots > p^{\prime}_{x^{\prime}_{g_{k}}}$ 
follows from $p^{\prime}_{1} > p^{\prime}_{2} > \cdots > p^{\prime}_{x^{\prime}_{m^{\prime}}}$ 
and $g_{1} < g_{2} < \cdots < g_{k}$. 
This order $p^{\prime}_{g_{1}} > p^{\prime}_{g_{2}} > \cdots > p^{\prime}_{x^{\prime}_{g_{k}}}$ indicates that 
$([p_{s}, q_{s}], [\ell_{s}, r_{s}]) = I_{\capture}(x^{\prime}_{s}, i)$ holds for all $s \in [1, k]$ 
because $p_{1} \geq p_{2} \geq \cdots \geq p_{k}$ holds. 

\textbf{Proof of Lemma~\ref{lem:psi_OVL_conditions}(i).}
We prove $p_{1} > p_{2} > \cdots > p_{k}$ using the $k$ integers $g_{1}, g_{2}, \ldots, g_{k}$ used in 
the proof of statement (4). 
Here, $g_{1} < g_{2} < \cdots < g_{k}$ holds,
and $p_{s} = p^{\prime}_{g_{s}}$ follows from statement (4) for all $s \in [1, k]$. 
$p^{\prime}_{g_{1}} > p^{\prime}_{g_{2}} > \cdots > p^{\prime}_{g_{k}}$ follows from statement (2). 
Therefore, $p_{1} > p_{2} > \cdots > p_{k}$ follows from 
$p^{\prime}_{g_{1}} > p^{\prime}_{g_{2}} > \cdots > p^{\prime}_{g_{k}}$, 
$p_{1} = p^{\prime}_{g_{1}}$, 
$p_{2} = p^{\prime}_{g_{2}}$, $\ldots$, $p_{k} = p^{\prime}_{g_{k}}$.

Consider interval attractor $I_{\capture}(1, i) = ([p, q], [\ell, r])$. 
Then, $1 \in [p, q]$ and $i \in [\ell, r]$ follow from the definition of interval attractor. 
$p = 1$ follows from $1 \in [p, q]$. 
The set $\{ I_{\capture}(x, i) \mid x \in [1, i-1] \}$ contains the interval attractor $([p, q], [\ell, r])$. 

We prove $([p, q], [\ell, r]) \not \in \Psi_{\run}$ by contradiction. 
We assume that $([p, q], [\ell, r]) \in \Psi_{\run}$ holds. 
Then, Lemma~\ref{lem:psi_run_basic_property} shows that $p \geq 2$ holds. 
The two facts $p = 1$ and $p \geq 2$ yield a contradiction. 
Therefore, $([p, q], [\ell, r]) \not \in \Psi_{\run}$ must hold. 

We prove $p_{k} = 1$. 
$([p, q], [\ell, r]) \in \{ I_{\capture}(x, i) \mid x \in [1, i-1] \} \setminus \Psi_{\run}$ follows from 
$([p, q], [\ell, r]) \in \{ I_{\capture}(x, i) \mid x \in [1, i-1] \}$ and $([p, q], [\ell, r]) \not \in \Psi_{\run}$. 
Because of $([p, q], [\ell, r]) \in \{ I_{\capture}(x, i) \mid x \in [1, i-1] \} \setminus \Psi_{\run}$, 
$p \in \{ p_{1}, p_{2}, \ldots, p_{k} \}$ holds. 
$([p, q], [\ell, r]) = ([p_{k}, q_{k}], [\ell_{k}, r_{k}])$ follows from 
(a) $p \in \{ p_{1}, p_{2}, \ldots, p_{k} \}$, 
(b) $p_{1} > p_{2} > \cdots > p_{k} \geq 1$, 
and (c) $p = 1$. 
Therefore, $p_{k} = 1$ holds. 

\textbf{Proof of Lemma~\ref{lem:psi_OVL_conditions}(ii).}
We prove $k \geq 1$. 
We already showed that $I_{\capture}(1, i-1) \in \{ I_{\capture}(x, i) \mid x \in [1, i-1] \} \setminus \Psi_{\run}$ holds 
in the proof of Lemma~\ref{lem:psi_OVL_conditions}(i). 
Here, $k = |\{ I_{\capture}(x, i) \mid x \in [1, i-1] \} \setminus \Psi_{\run}|$ holds. 
Therefore, $k \geq 1$ follows from $k = |\{ I_{\capture}(x, i) \mid x \in [1, i-1] \} \setminus \Psi_{\run}|$ 
and $I_{\capture}(1, i-1) \in \{ I_{\capture}(x, i) \mid x \in [1, i-1] \} \setminus \Psi_{\run}$. 

We prove $k \leq |\{ ([p, q], [\ell, r]) \in \Psi_{\RR} \setminus \Psi_{\run} \mid i \in [p, r] \}|$. 
For each integer $s \in [1, k]$, 
$i \in [p_{s}, r_{s}]$ holds because 
(a) there exists an integer $g \in [1, m]$ satisfying $([p_{s}, q_{s}], [\ell_{s}, r_{s}]) = ([p^{\prime}_{g}, q^{\prime}_{g}], [\ell^{\prime}_{g}, r^{\prime}_{g}])$, 
(b) $i \in [\ell^{\prime}_{g}, r^{\prime}_{g}]$ follows from statement (1). 
$([p_{1}, q_{1}], [\ell_{1}, r_{1}]), ([p_{2}, q_{2}], [\ell_{2}, r_{2}]), \ldots, ([p_{k}, q_{k}], [\ell_{k}, r_{k}]) \in \{ ([p, q], [\ell, r]) \in \Psi_{\RR} \setminus \Psi_{\run} \mid i \in [p, r] \}$ holds 
because $i \in [p_{s}, r_{s}]$ and 
$([p_{s}, q_{s}], [\ell_{s}, r_{s}]) \not \in \Psi_{\run}$ for all $s \in [1, k]$. 
Therefore, $k \leq |\{ ([p, q], [\ell, r]) \in \Psi_{\RR} \setminus \Psi_{\run} \mid i \in [p, r] \}|$ 
follows from $([p_{1}, q_{1}], [\ell_{1}, r_{1}])$, $([p_{2}, q_{2}], [\ell_{2}, r_{2}])$, $\ldots$, $([p_{k}, q_{k}], [\ell_{k}, r_{k}]) \in \{ ([p, q], [\ell, r]) \in \Psi_{\RR} \setminus \Psi_{\run} \mid i \in [p, r] \}$. 

\textbf{Proof of Lemma~\ref{lem:psi_OVL_conditions}(iii).}
$([p, q], [\ell, r]) = ([p^{\prime}_{1}, q^{\prime}_{1}], [\ell^{\prime}_{1}, r^{\prime}_{1}])$ holds 
because $([p, q], [\ell, r]) = I_{\capture}(i-1, i)$, 
$([p^{\prime}_{1}, q^{\prime}_{1}], [\ell^{\prime}_{1}, r^{\prime}_{1}]) = I_{\capture}(x_{1}, i)$, 
and $x_{1} = i-1$. 
Consider the smallest integer $\tau$ in set $\{ 2, 3, \ldots, m \}$ 
that satisfies $([p^{\prime}_{\tau}, q^{\prime}_{\tau}], [\ell^{\prime}_{\tau}, r^{\prime}_{\tau}]) \not \in \Psi_{\run}$. 
If $([p, q], [\ell, r]) \in \Psi_{\run}$, 
then statement (3) shows that the interval attractor $([p^{\prime}_{\tau}, q^{\prime}_{\tau}], [\ell^{\prime}_{\tau}, r^{\prime}_{\tau}])$ is a run-extension of the interval attractor $([p^{\prime}_{1}, q^{\prime}_{1}], [\ell^{\prime}_{1}, r^{\prime}_{1}])$. 
In this case, $x_{\tau} = x^{\prime}_{1}$ holds. Statement (4) shows that 
$([p_{1}, q_{1}], [\ell_{1}, r_{1}]) = I_{\capture}(x^{\prime}_{1}, i)$ holds, 
which indicates that the interval attractor $([p_{1}, q_{1}], [\ell_{1}, r_{1}])$ 
is a run-extension of the interval attractor $([p, q], [\ell, r])$. 

Otherwise (i.e., $([p, q], [\ell, r]) \not \in \Psi_{\run}$), 
$x_{1} = x^{\prime}_{1}$ holds. 
$([p_{1}, q_{1}], [\ell_{1}, r_{1}]) = ([p, q], [\ell, r])$ follows from 
$([p_{1}, q_{1}], [\ell_{1}, r_{1}]) = I_{\capture}(x^{\prime}_{1}, i)$, 
$I_{\capture}(x^{\prime}_{1}, i) = I_{\capture}(x_{1}, i)$, 
and $I_{\capture}(x_{1}, i) = ([p, q], [\ell, r])$. 
Therefore, Lemma~\ref{lem:psi_OVL_conditions}(iii) holds. 

\textbf{Proof of Lemma~\ref{lem:psi_OVL_conditions}(iv).}
$([p_{s-1}, q_{s-1}], [\ell_{s-1}, r_{s-1}]) = I_{\capture}(x^{\prime}_{s-1}, i)$ follows from statement (4). 
Because of $\{ x^{\prime}_{1}, x^{\prime}_{2}, \ldots, x^{\prime}_{k} \} \subseteq \{ x_{1}, x_{2}, \ldots, x_{m} \}$, 
there exists an integer $g \in [1, m]$ satisfying $x_{g} = x^{\prime}_{s-1}$. 
Therefore, $([p_{s-1}, q_{s-1}], [\ell_{s-1}, r_{s-1}]) = ([p^{\prime}_{g}, q^{\prime}_{g}], [\ell^{\prime}_{g}, r^{\prime}_{g}])$ holds. 

We prove $([p, q], [\ell, r]) = ([p^{\prime}_{g+1}, q^{\prime}_{g+1}], [\ell^{\prime}_{g+1}, r^{\prime}_{g+1}])$. 
$x_{g+1} = p^{\prime}_{g} - 1$ follows from statement (2). 
$p_{s-1} = p^{\prime}_{g}$ follows from $([p_{s-1}, q_{s-1}], [\ell_{s-1}, r_{s-1}]) = ([p^{\prime}_{g}, q^{\prime}_{g}], [\ell^{\prime}_{g}, r^{\prime}_{g}])$. 
$x_{g+1} = p_{s-1} - 1$ follows from 
$x_{g+1} = p^{\prime}_{g} - 1$ and $p_{s-1} = p^{\prime}_{g}$. 
Because of $x_{g+1} = p_{s-1} - 1$, 
$I_{\capture}(x_{g+1}, i) = I_{\capture}(p^{\prime}_{s-1} - 1, i)$ holds. 
Therefore, $([p, q], [\ell, r]) = ([p^{\prime}_{g+1}, q^{\prime}_{g+1}], [\ell^{\prime}_{g+1}, r^{\prime}_{g+1}])$ 
follows from $I_{\capture}(x_{g+1}, i) = ([p^{\prime}_{g+1}, q^{\prime}_{g+1}], [\ell^{\prime}_{g+1}, r^{\prime}_{g+1}])$ 
and $I_{\capture}(p^{\prime}_{s-1} - 1, i) = ([p, q], [\ell, r])$. 

We prove Lemma~\ref{lem:psi_OVL_conditions}(iv). 
Consider the smallest integer $\tau$ in set $\{ g+2, g+3, \ldots, m \}$ 
that satisfies $([p^{\prime}_{\tau}, q^{\prime}_{\tau}], [\ell^{\prime}_{\tau}, r^{\prime}_{\tau}]) \not \in \Psi_{\run}$. 
If $([p, q], [\ell, r]) \in \Psi_{\run}$ (i.e., $([p^{\prime}_{g+1}, q^{\prime}_{g+1}], [\ell^{\prime}_{g+1}, r^{\prime}_{g+1}]) \in \Psi_{\run}$), 
then statement (3) shows that the interval attractor $([p^{\prime}_{\tau}, q^{\prime}_{\tau}], [\ell^{\prime}_{\tau}, r^{\prime}_{\tau}])$ is a run-extension of the interval attractor $([p^{\prime}_{g+1}, q^{\prime}_{g+1}], [\ell^{\prime}_{g+1}, r^{\prime}_{g+1}])$. 
In this case, $x^{\prime}_{s} = x_{\tau}$ holds 
because (a) $x^{\prime}_{s-1} = x_{g}$, 
(b) $\{ ([p^{\prime}_{b}, q^{\prime}_{b}], [\ell^{\prime}_{b}, r^{\prime}_{b}]) \mid b \in [g+1, \tau] \} \subseteq \Psi_{\run}$, 
and (c) $([p^{\prime}_{\tau}, q^{\prime}_{\tau}], [\ell^{\prime}_{\tau}, r^{\prime}_{\tau}]) \not \in \Psi_{\run}$. 
Because of $x^{\prime}_{s} = x_{\tau}$, 
statement (4) shows that $([p_{s}, q_{s}], [\ell_{s}, r_{s}]) = ([p^{\prime}_{\tau}, q^{\prime}_{\tau}], [\ell^{\prime}_{\tau}, r^{\prime}_{\tau}])$ holds. 
Because $([p, q], [\ell, r]) = ([p^{\prime}_{g+1}, q^{\prime}_{g+1}], [\ell^{\prime}_{g+1}, r^{\prime}_{g+1}])$, 
the interval attractor $([p_{s}, q_{s}], [\ell_{s}, r_{s}])$ is a run-extension of the interval attractor $([p, q], [\ell, r])$. 

Otherwise (i.e., $([p^{\prime}_{g+1}, q^{\prime}_{g+1}], [\ell^{\prime}_{g+1}, r^{\prime}_{g+1}]) \not \in \Psi_{\run}$), 
$x^{\prime}_{s} = x_{g+1}$ holds. 
Because of $x^{\prime}_{s} = x_{g+1}$, 
statement (4) shows that $([p_{s}, q_{s}], [\ell_{s}, r_{s}]) = ([p^{\prime}_{g+1}, q^{\prime}_{g+1}], [\ell^{\prime}_{g+1}, r^{\prime}_{g+1}])$ holds. 
Therefore, $([p_{s}, q_{s}], [\ell_{s}, r_{s}]) = ([p, q], [\ell, r])$ follows from 
$([p, q], [\ell, r]) = ([p^{\prime}_{g+1}, q^{\prime}_{g+1}], [\ell^{\prime}_{g+1}, r^{\prime}_{g+1}])$ 
and $([p_{s}, q_{s}], [\ell_{s}, r_{s}]) = ([p^{\prime}_{g+1}, q^{\prime}_{g+1}]$, $[\ell^{\prime}_{g+1}, r^{\prime}_{g+1}])$. 

\end{proof}

Consider $k$ interval attractors 
$([p_{1}, q_{1}], [\ell_{1}, r_{1}])$, $([p_{2}, q_{2}], [\ell_{2}, r_{2}])$, $\ldots$, 
$([p_{k}, q_{k}], [\ell_{k}, r_{k}])$ ($p_{1} \geq p_{2} \geq \cdots \geq p_{k}$) 
in set $\{ I_{\capture}(x, i) \mid x \in [1, i-1] \} \setminus \Psi_{\run}$. 
We compute these $k$ interval attractors in the following three phases. 

\paragraph{Phase 1.}
In the first phase, we verify whether $k \geq 1$ or not. 
If $i = 1$, then $k = 0$ because 
$\{ I_{\capture}(x, i) \mid x \in [1, i-1] \} = \emptyset$ holds.  
Otherwise (i.e., $i \in [2, n]$), 
Lemma~\ref{lem:psi_OVL_conditions}~\ref{lem:psi_OVL_conditions:2} shows that $k \geq 1$ holds. 
Therefore, we can verify whether $k \geq 1$ or not in $O(1)$ time. 

If $k = 0$, then the algorithm stops; 
otherwise, the algorithm proceeds to the second phase. 

\paragraph{Phase 2.}
In the second phase, we compute the first interval attractor $([p_{1}, q_{1}], [\ell_{1}, r_{1}])$ 
using interval attractor $I_{\capture}(i-1, i) = ([p, q], [\ell, r])$. 
Lemma~\ref{lem:psi_OVL_conditions}~\ref{lem:psi_OVL_conditions:3} shows that 
$([p_{1}, q_{1}], [\ell_{1}, r_{1}]) = ([p, q], [\ell, r])$ holds if $([p, q], [\ell, r]) \not \in \Psi_{\run}$; 
otherwise, the interval attractor $([p_{1}, q_{1}], [\ell_{1}, r_{1}])$ is a run-extension of the interval attractor $([p, q], [\ell, r])$. 
Therefore, the first interval attractor $([p_{1}, q_{1}], [\ell_{1}, r_{1}])$ can be obtained by the following three steps: 
\begin{enumerate}[label=\textbf{(\roman*)}]
    \item obtain the interval attractor $([p, q], [\ell, r])$ by capture query $\CAPQ([i-1, i])$;
    \item verify whether $([p, q], [\ell, r]) \not \in \Psi_{\run}$ by verify-run query $\runQ(([p, q], [\ell, r]))$. 
    If $([p, q]$, $[\ell, r]) \not \in \Psi_{\run}$, then return the interval attractor $([p, q], [\ell, r])$ as the interval attractor $([p_{1}, q_{1}]$, $[\ell_{1}, r_{1}])$. 
    Otherwise, proceed to the next step; 
    \item compute all the run-extensions of the interval attractor $([p, q], [\ell, r])$. 
    Here, Lemma~\ref{lem:run_extension_property} shows that 
    the interval attractor $([p, q], [\ell, r])$ has only one run-extension. 
    Return the found run-extension as the $([p_{1}, q_{1}], [\ell_{1}, r_{1}])$. 
\end{enumerate}

The bottleneck of the second phase is capture query in the first step, which takes $O(H^{2} \log n)$ time. 
Therefore, the second phase takes $O(H^{2} \log n)$ time in total. 

\paragraph{Phase 3.}
In the third phase, we compute $k-1$ interval attractors $([p_{2}, q_{2}], [\ell_{2}, r_{2}])$, $([p_{3}, q_{3}], [\ell_{3}, r_{3}])$, $\ldots$, $([p_{k}, q_{k}], [\ell_{k}, r_{k}])$. 
Consider interval attractor $I_{\capture}(p_{s-1}-1, i) = ([p^{\prime}, q^{\prime}], [\ell^{\prime}, r^{\prime}])$ for each integer $s \in [2, k]$. 
Then, Lemma~\ref{lem:psi_OVL_conditions}~\ref{lem:psi_OVL_conditions:4} shows that 
$([p_{s}, q_{s}], [\ell_{s}, r_{s}]) = ([p^{\prime}, q^{\prime}], [\ell^{\prime}, r^{\prime}])$ holds if $([p^{\prime}, q^{\prime}], [\ell^{\prime}, r^{\prime}]) \not \in \Psi_{\run}$; 
otherwise, the interval attractor $([p_{s}, q_{s}], [\ell_{s}, r_{s}])$ is a run-extension of the interval attractor $([p^{\prime}, q^{\prime}], [\ell^{\prime}, r^{\prime}])$. 
Therefore, each interval attractor $([p_{s}, q_{s}], [\ell_{s}, r_{s}])$ can be obtained by the following three steps: 
\begin{enumerate}[label=\textbf{(\Alph*)}]
    \item obtain the interval attractor $([p^{\prime}, q^{\prime}], [\ell^{\prime}, r^{\prime}])$ by capture query $\CAPQ([p_{s-1}-1, i])$;
    \item verify whether $([p^{\prime}, q^{\prime}], [\ell^{\prime}, r^{\prime}]) \not \in \Psi_{\run}$ by verify-run query $\runQ(([p^{\prime}, q^{\prime}], [\ell^{\prime}, r^{\prime}]))$. 
    If $([p^{\prime}, q^{\prime}]$, $[\ell^{\prime}, r^{\prime}]) \not \in \Psi_{\run}$, then return the interval attractor $([p^{\prime}, q^{\prime}], [\ell^{\prime}, r^{\prime}])$ as the interval attractor $([p_{s}, q_{s}], [\ell_{s}, r_{s}])$. 
    Otherwise, proceed to the next step; 
    \item compute all the run-extensions of the interval attractor $([p^{\prime}, q^{\prime}], [\ell^{\prime}, r^{\prime}])$. 
    Here, Lemma~\ref{lem:run_extension_property} shows that 
    the interval attractor $([p^{\prime}, q^{\prime}], [\ell^{\prime}, r^{\prime}])$ has only one run-extension. 
    Return the found run-extension as the $([p_{s}, q_{s}], [\ell_{s}, r_{s}])$. 
\end{enumerate}

Similar to the second phase, 
each interval attractor can be obtained in $O(H^{2} \log n)$ time. 
Therefore, the third phase takes $O(k H^{2} \log n)$ time in total.

Finally, we can compute the $k$ interval attractors $([p_{1}, q_{1}], [\ell_{1}, r_{1}])$, $([p_{2}, q_{2}], [\ell_{2}, r_{2}])$, $\ldots$, $([p_{k}, q_{k}]$, $[\ell_{k}, r_{k}])$ in $O(1 + k H^{2} \log n)$ time. 
Here, $k \leq |\{ ([p, q], [\ell, r]) \in \Psi_{\RR} \setminus \Psi_{\run} \mid i \in [p, r] \}|$ follows from  Lemma~\ref{lem:psi_OVL_conditions}~\ref{lem:psi_OVL_conditions:2}. 
Therefore, the total running time is $O(1 + (H^{2} \log n) |\{ ([p, q], [\ell, r]) \in \Psi_{\RR} \setminus \Psi_{\run} \mid i \in [p, r] \}|)$. 

\subsubsection{Algorithm of Overlap Query}
We explain the algorithm for answering a given overlap query $\OVQ([i, j])$ using the data structure detailed in Section~\ref{subsubsec:rrdag_ds}. 
Lemma~\ref{lem:RB_OVQ_and_Subset} shows that 
the set of interval attractors obtained from the overlap query $\OVQ([i, j])$ is equal to the union of 
the following three sets: 
(i) $\bigcup_{i^{\prime} = i}^{j} \{ I_{\capture}(i^{\prime}, j^{\prime}) \mid j^{\prime} \in [i^{\prime}+1, n] \} \setminus \Psi_{\run}$.
(ii) $\Psi_{\OVQ, C} \setminus \Psi_{\run}$, 
and (iii) $\{ I_{\capture}(x, i) \mid x \in [1, i-1] \} \setminus \Psi_{\run}$. 

The first set $\bigcup_{i^{\prime} = i}^{j} \{ I_{\capture}(i^{\prime}, j^{\prime}) \mid j^{\prime} \in [i^{\prime}+1, n] \} \setminus \Psi_{\run}$ can be computed in $O(|[i, j]| H^{3} \log n))$ 
time by the algorithm presented in Section~\ref{subsubsec:ovqr}. 
The second set $\Psi_{\OVQ, C} \setminus \Psi_{\run}$ can be computed in $O((H + m^{\prime}) H^{3} \log n)$ time by 
the algorithm presented in Section~\ref{subsubsec:ovqc} 
for the number $m^{\prime}$ of interval attractors in set $\{ ([p, q], [\ell, r]) \in \Psi_{\RR} \setminus \Psi_{\run} \mid i \in [p, r] \}$. 
The third set $\{ I_{\capture}(x, i) \mid x \in [1, i-1] \} \setminus \Psi_{\run}$ can be computed in 
$O(1 + m^{\prime} H^{2} \log n)$ time by the algorithm presented in Section~\ref{subsubsec:ovql}. 
Therefore, the above three sets can be obtained in $O((|[i, j]| + m^{\prime}) H^{3} \log n))$ time. 

The following lemma shows that 
the expected value of $m^{\prime}$ can be bounded by $O(H + \log n)$. 

\begin{lemma}\label{lem:overlap_count}
$\mathbb{E}(|\{ ([p, q], [\ell, r]) \in \Psi_{\RR} \setminus \Psi_{\run} \mid i \in [p, r] \}|) = O(H + \log n)$ 
for a position $i \in [1, n]$ in input string $T$. 
\end{lemma}
\begin{proof}
See Section~\ref{subsubsec:overlap_count_proof}.
\end{proof}

Therefore, we can answer a given overlap query in expected $O(\max \{ |[i, j]|, H, \log n \} H^{3} \log n)$ time.

\subsubsection{Proof of Lemma~\ref{lem:overlap_count}}\label{subsubsec:overlap_count_proof}
We prove Lemma~\ref{lem:overlap_count}. 
The set $\{ ([p, q], [\ell, r]) \in \Psi_{\RR} \setminus \Psi_{\run} \mid i \in [p, r] \}$ 
can be divided into two subsets 
(i) $\{ ([p, q], [\ell, r]) \in \Psi_{\RR} \setminus \Psi_{\rightRun} \mid i \in [p, r] \}$ 
and (ii) $\{ ([p, q], [\ell, r]) \in \Psi_{\rightRun} \setminus \Psi_{\run} \mid i \in [p, r] \}$. 
Here, $\Psi_{\rightRun}$ is the subset introduced in Section~\ref{subsec:size_proof}. 
Therefore, Lemma~\ref{lem:overlap_count} follows from the following two propositions. 

\begin{proposition}\label{prop:overlap_proof_H1}
$\mathbb{E}[|\{ ([p, q], [\ell, r]) \in \Psi_{\RR} \setminus \Psi_{\rightRun} \mid i \in [p, r] \}|] = O(H)$ for each integer $i \in [1, n]$. 
\end{proposition}
\begin{proof}
    Consider an integer $h \in [0, H]$, 
    and let $([p_{1}, q_{1}], [\ell_{1}, r_{1}])$, $([p_{2}, q_{2}], [\ell_{2}, r_{2}])$, $\ldots$, 
    $([p_{k}, q_{k}], [\ell_{k}, r_{k}])$ be the interval attractors in 
    the set $\{ ([p, q], [\ell, r]) \in \Psi_{h} \setminus \Psi_{\rightRun} \mid i \in [p, r] \}$. 
    Let $\gamma_{s}$ be the attractor position of each interval attractor $([p_{s}, q_{s}], [\ell_{s}, r_{s}])$. 
    Then, $k = \{ \gamma_{s} \mid s \in [1, k] \}$ 
    because Corollary~\ref{cor:IA_identify_corollary} shows that their attractor positions are distinct. 
    $\{ \gamma_{s} \mid s \in [1, k] \} \subseteq \mathcal{B}_{\bstart}$ follows from 
    the definition of the attractor position 
    for the set $\mathcal{B}_{\bstart}(h)$ of positions introduced in Section~\ref{subsec:size_proof}.  

    We prove $\{ \gamma_{s} \mid s \in [1, k] \} \subseteq [i - 53\mu(h+1), i + 53\mu(h+1)]$. 
    $\Psi_{h} \setminus \Psi_{\rightRun} \subseteq \Psi_{\leftLen}(0, 16 \mu(h+1)) \cap \Psi_{\rightLen}(0, 36 \mu(h+1))$ follows from Proposition~\ref{prop:psi_H_length} 
    for the two subsets $\Psi_{\leftLen}(0, 16 \mu(h+1))$ and $\Psi_{\rightLen}(0, 36 \mu(h+1))$ of set $\Psi_{\RR}$ 
    introduced in Section~\ref{subsec:size_proof}. 
    This fact indicates that 
    $[p_{s}-1, r_{s}+1] \subseteq [i - 53\mu(h+1), i + 53\mu(h+1)]$ holds 
    for each interval attractor $([p_{s}, q_{s}], [\ell_{s}, r_{s}])$. 
    Similarly, $\gamma_{s} \in [i - 53\mu(h+1), i + 53\mu(h+1)]$ holds 
    because $\gamma_{s} \in [p_{s}-1, r_{s}+1]$ follows from the definition of interval attractor. 
    Therefore, we obtain $\{ \gamma_{s} \mid s \in [1, k] \} \subseteq [i - 53\mu(h+1), i + 53\mu(h+1)]$. 

    We prove $\mathbb{E}[|\{ ([p, q], [\ell, r]) \in \Psi_{h} \setminus \Psi_{\rightRun} \mid i \in [p, r] \}|] < 429$. 
    $|\{ ([p, q], [\ell, r]) \in \Psi_{h} \setminus \Psi_{\rightRun} \mid i \in [p, r] \}| \leq |\mathcal{B}_{\bstart}(h) \cap [i - 53\mu(h+1), i + 53\mu(h+1)]|$ follows from 
    $\{ \gamma_{s} \mid s \in [1, k] \} \subseteq \mathcal{B}_{\bstart}$ and $\{ \gamma_{s} \mid s \in [1, k] \} \subseteq [i - 53\mu(h+1), i + 53\mu(h+1)]$. 
    Proposition~\ref{prop:b_set_properties}~\ref{enum:b_set_properties:1} shows that 
    $|\{ \gamma_{s} \mid s \in [1, k] \} \subseteq [i - 53\mu(h+1), i + 53\mu(h+1)]| < 429$. 
    Therefore, we obtain $\mathbb{E}[|\{ ([p, q], [\ell, r]) \in \Psi_{h} \setminus \Psi_{\rightRun} \mid i \in [p, r] \}|] < 429$. 

    Finally, Proposition~\ref{prop:overlap_proof_H1} follows from the following equation: 
\begin{equation*}
\begin{split}
&\mathbb{E}[|\{ ([p, q], [\ell, r]) \in \Psi_{\RR} \setminus \Psi_{\rightRun} \mid i \in [p, r] \}|]\\
&= \mathbb{E}[|\bigcup_{h = 0}^{H} \{ ([p, q], [\ell, r]) \in \Psi_{h} \setminus \Psi_{\rightRun} \mid i \in [p, r] \}|] \\
&\leq \sum_{h=0}^{H} \mathbb{E}[|\{ ([p, q], [\ell, r]) \in \Psi_{h} \setminus \Psi_{\rightRun} \mid i \in [p, r] \}|] \\
              &< \sum_{h=0}^{H} 429  \\
              &= O(H).
\end{split}
\end{equation*}    
\end{proof}

\begin{proposition}\label{prop:overlap_proof_H2}
$\mathbb{E}[| \{ ([p, q], [\ell, r]) \in \Psi_{\rightRun} \setminus \Psi_{\run} \mid i \in [p, r] \}|] = O(\log n)$ for each integer $i \in [1, n]$. 
\end{proposition}
\begin{proof}
    Set $\{ ([p, q], [\ell, r]) \in \Psi_{\rightRun} \setminus \Psi_{\run} \mid i \in [p, r] \}$ is equal to 
    set $\bigcup_{t = 0}^{\infty} \{ ([p, q], [\ell, r]) \in (\Psi_{\rightRun} \cap \Psi_{\rightLen}(\mu_{\SUM}(t), \mu_{\SUM}(t+1))) \setminus \Psi_{\run} \mid i \in [p, r] \}$. 
    Here, $\mu_{\SUM}(t)$ is the integer introduced in Section~\ref{subsec:size_proof}. 
    Therefore, Proposition~\ref{prop:overlap_proof_H2} follows from the following three statements: 
    \begin{enumerate}[label=\textbf{(\roman*)}]
    \item $(\Psi_{\rightRun} \cap \Psi_{\rightLen}(\mu_{\SUM}(t), \mu_{\SUM}(t+1))) \setminus \Psi_{\run} = \emptyset$ for $t \in \{ 0, 1, 2 \}$; 
    \item $\mathbb{E}[|\{ ([p, q], [\ell, r]) \in (\Psi_{\rightRun} \cap \Psi_{\rightLen}(\mu_{\SUM}(t), \mu_{\SUM}(t+1))) \setminus \Psi_{\run} \mid i \in [p, r] \}|] = O(1)$ for $t \geq 3$;
    \item $(\Psi_{\rightRun} \cap \Psi_{\rightLen}(\mu_{\SUM}(t), \mu_{\SUM}(t+1))) \setminus \Psi_{\run} = \emptyset$ for $t \geq 2 + \lceil 2 \log_{8/7} (n+1) \rceil$.
    \end{enumerate}

    \paragraph{Proof of statement (i).}
    We already proved 
    $|(\Psi_{\rightRun} \cap \Psi_{\leftmost} \cap \Psi_{\rightLen}(\mu_{\SUM}(t), \mu_{\SUM}(t+1))) \setminus \Psi_{\run}| = 0$ 
    in the proof of Proposition~\ref{prop:expected_size_B}. 
    We can prove statement (i) using the same approach. 
    
    \paragraph{Proof of statement (ii).}
    Let $([p_{1}, q_{1}], [\ell_{1}, r_{1}])$, $([p_{2}, q_{2}], [\ell_{2}, r_{2}])$, $\ldots$, 
$([p_{k}, q_{k}], [\ell_{k}, r_{k}])$ be the interval attractors in 
the set $\{ ([p, q], [\ell, r]) \in (\Psi_{\rightRun} \cap \Psi_{\rightLen}(\mu_{\SUM}(t), \mu_{\SUM}(t+1))) \setminus \Psi_{\run} \mid i \in [p, r] \}$. 
Let $\gamma_{s}$ be the attractor position of each interval attractor $([p_{s}, q_{s}], [\ell_{s}, r_{s}])$, 
let $d = \max \{ 1, \lfloor (1/8) \mu(t) \rfloor \}$, 
and let $j_{s} \geq 1$ be the smallest integer satisfying $1 + (j_{s}-1)d \geq \gamma_{s}$ 
(i.e., $j_{s} = \min \{ j \geq 1 \mid 1 + (j-1)d \geq \gamma_{s} \}$). 
Then, Proposition~\ref{prop:group_super_properties}~\ref{enum:group_super_properties:2} shows that 
$([p_{s}, q_{s}], [\ell_{s}, r_{s}]) \in \Psi_{\group}(t, 1 + (j_{s}-1)d)$ holds 
for the subset $\Psi_{\group}(t, 1 + (j_{s}-1)d)$ of set $\Psi_{\RR}$ introduced in Section~\ref{subsec:size_proof}.

Let $\mathcal{W}^{\prime}$ be a set of integers 
such that each integer $j \in \mathcal{W}^{\prime}$ satisfies 
$i \in [1 + (j-1)d - 17 \lfloor \mu(t+1) \rfloor, (j-1)d + 18 \lfloor \mu(t+1) \rfloor - 1]$. 
We prove $\{ ([p, q], [\ell, r]) \in \Psi_{\rightRun} \setminus \Psi_{\run} \mid i \in [p, r] \} \subseteq \bigcup_{j \in \mathcal{W}^{\prime}} \Psi_{\group}(t, 1 + (j-1)d)$. 
For each interval attractor $([p_{s}, q_{s}], [\ell_{s}, r_{s}])$, 
$([p_{s}, q_{s}], [\ell_{s}, r_{s}]) \in \Psi_{\leftLen}(0, 16 \mu(t+1)) \cap \Psi_{\rightLen}(0, 17 \mu(t+1))$ 
follows from Proposition~\ref{prop:rightRun_property_H}. 
$[p_{s}, r_{s}] \subseteq [\gamma_{s} - 16 \mu(t+1), \gamma + 17 \mu(t+1) - 1]$ follows from 
$([p_{s}, q_{s}], [\ell_{s}, r_{s}]) \in \Psi_{\leftLen}(0, 16 \mu(t+1)) \cap \Psi_{\rightLen}(0, 17 \mu(t+1))$. 
$[1 + (j_{s}-1)d, 1 + (j_{s}-1)d + \lfloor (7/8) \mu(t) \rfloor - 1 ] \subseteq [\gamma_{s}+1, \gamma_{s} + \lfloor \mu(t) \rfloor]$ 
follows from the definition of the subset $\Psi_{\group}(t, 1 + (j_{s}-1)d)$. 
$[\gamma_{s} - 16 \mu(t+1), \gamma + 17 \mu(t+1) - 1] \subseteq [1 + (j_{s}-1)d - 17 \lfloor \mu(t+1) \rfloor, (j_{s}-1)d + 18 \lfloor \mu(t+1) \rfloor - 1]$ follows from $[1 + (j_{s}-1)d, 1 + (j_{s}-1)d + \lfloor (7/8) \mu(t) \rfloor - 1 ] \subseteq [\gamma_{s}+1, \gamma_{s} + \lfloor \mu(t) \rfloor]$. 
$j_{s} \in \mathcal{W}^{\prime}$ follows from the following equation. 
\begin{equation*}
\begin{split}
i &\in [p_{s}, r_{s}] \\
  &\subseteq [\gamma_{s} - 16 \mu(t+1), \gamma + 17 \mu(t+1) - 1] \\
  &\subseteq [1 + (j_{s}-1)d - 17 \lfloor \mu(t+1) \rfloor, (j_{s}-1)d + 18 \lfloor \mu(t+1) \rfloor - 1]. 
\end{split}
\end{equation*}
Therefore, we obtain $\{ ([p, q], [\ell, r]) \in \Psi_{\rightRun} \setminus \Psi_{\run} \mid i \in [p, r] \} \subseteq \bigcup_{j \in \mathcal{W}^{\prime}} \Psi_{\group}(t, 1 + (j-1)d)$.

We prove statement (ii). 
Proposition~\ref{prop:group_super_properties}~\ref{enum:group_super_properties:1} shows that 
$\mathbb{E}[|(\Psi_{\rightRun} \cap \Psi_{\rightLen}(\mu_{\SUM}(t), \mu_{\SUM}(t+1)) \cap \Psi_{\group}(t, 1 + (j_{s}-1)d)) \setminus \Psi_{\run}|] = O(1)$ for each integer $s \in [1, k]$. 
The set $\mathcal{W}^{\prime}$ contains $O(1)$ integers. 
Therefore, 
\begin{equation*}
\begin{split}
&\mathbb{E}[|\{ ([p, q], [\ell, r]) \in (\Psi_{\rightRun} \cap \Psi_{\rightLen}(\mu_{\SUM}(t), \mu_{\SUM}(t+1))) \setminus \Psi_{\run} \mid i \in [p, r] \}|] \\
&\leq \mathbb{E}[|\bigcup_{j \in \mathcal{W}^{\prime}} (\Psi_{\rightRun} \cap \Psi_{\rightLen}(\mu_{\SUM}(t), \mu_{\SUM}(t+1)) \cap \Psi_{\group}(t, 1 + (j-1)d)) \setminus \Psi_{\run}|] \\
&\leq \sum_{j \in \mathcal{W}^{\prime}} \mathbb{E}[|(\Psi_{\rightRun} \cap \Psi_{\rightLen}(\mu_{\SUM}(t), \mu_{\SUM}(t+1)) \cap \Psi_{\group}(t, 1 + (j-1)d)) \setminus \Psi_{\run}|] \\
&= O(|\mathcal{W}^{\prime}|) \\
&= O(1).
\end{split}
\end{equation*}

\paragraph{Proof of statement (iii).}
We prove statement (iii) by contradiction. 
We assume that statement (iii) does not hold, i.e., 
the set $(\Psi_{\rightRun} \cap \Psi_{\rightLen}(\mu_{\SUM}(t), \mu_{\SUM}(t+1))) \setminus \Psi_{\run}$ contains an interval attractor $([p, q], [\ell, r])$ with attractor position $\gamma$. 
Then, $|[\gamma, r+1]| > \sum_{w = 1}^{t} \lfloor \mu(w) \rfloor$ 
follows from the definition of the subset $\Psi_{\rightLen}(\mu_{\SUM}(t), \mu_{\SUM}(t+1)))$. 
Since $t \geq 2 + \lceil 2 \log_{8/7} (n+1) \rceil$, 
$|[\gamma, r+1]| > n+1$ follows from the following equation. 
\begin{equation*}
\begin{split}
|[\gamma, r+1]| &> \sum_{w = 1}^{t} \lfloor \mu(w) \rfloor \\
&\geq \lfloor \mu(\lceil 2 \log_{8/7} (n+1) \rceil + 2) \rfloor \\
&= \lfloor (8/7)^{\lceil (\lceil 2 \log_{8/7} (n+1) \rceil + 2)/2 \rceil - 1} \rfloor \\
&\geq \lfloor (8/7)^{\log_{8/7} (n+1)} \rfloor \\
&= \lfloor n+1 \rfloor \\
&= n+1.
\end{split}
\end{equation*}    

On the other hand, $|[\gamma, r+1]| \leq n+1$ follows from $[\gamma, r] \subseteq [1, n]$. 
The two facts $|[\gamma, r+1]| > n+1$ and $|[\gamma, r+1]| \leq n+1$ yield a contradiction. 
Therefore, statement (iii) must hold. 
\end{proof}

\subsection{Sample Query}\label{subsec:sample_query}
Consider the interval attractors $([p_{1}, q_{1}]$, $[\ell_{1}, r_{1}])$, $([p_{2}, q_{2}], [\ell_{2}, r_{2}])$, $\ldots$, $([p_{k}, q_{k}], [\ell_{k}, r_{k}])$ of the sampling subset $\Psi_{\samp}$ introduced in Section~\ref{subsec:sampling_subset}. 
Here, the $k$ interval attractors are sorted in lexicographical order of the $k$ strings 
$T[p_{1}-1..r_{1}+1]$, $T[p_{2}-1..r_{2}+1]$, $\ldots$, $T[p_{k}-1..r_{k}+1]$ 
(i.e., $T[p_{1}-1..r_{1}+1] \preceq T[p_{2}-1..r_{2}+1] \preceq \cdots \preceq T[p_{k}-1..r_{k}+1]$). 
The order of the $k$ interval attractors is unique. 
This is because 
$T[p_{s}-1..r_{s}+1] \neq T[p_{s^{\prime}}-1..r_{s^{\prime}}+1]$ 
follows from the definition of the sampling subset 
for any pair of two interval attractors $([p_{s}, q_{s}], [\ell_{s}, r_{s}])$, $([p_{s^{\prime}}, q_{s^{\prime}}], [\ell_{s^{\prime}}, r_{s^{\prime}}]) \in \Psi_{\samp}$. 
For a given integer $s \in [1, k]$, 
\emph{sample query} $\sampleQ(i)$ returns the $i$-th interval attractor $([p_{s}, q_{s}], [\ell_{s}, r_{s}])$ of 
the sampling subset $\Psi_{\samp}$. 

\paragraph{Permutation $\phi_{1}, \phi_{2}, \ldots, \phi_{k}$ representing the order of interval attractors.}
For this subsection, 
we introduce a permutation $\phi_{1}, \phi_{2}, \ldots, \phi_{k}$ of $k$ integers $1, 2, \ldots, k$ 
representing an order of the $k$ interval attractors of the sampling subset $\Psi_{\samp}$. 
This permutation is obtained by sorting the $k$ interval attractors in increasing order of the $k$ positions $p_{1}, p_{2}, \ldots, p_{k}$.
If the sampling subset $\Psi_{\samp}$ contains two distinct interval attractors 
$([p_{s}, q_{s}], [\ell_{s}, r_{s}])$ and $([p_{s^{\prime}}, q_{s^{\prime}}], [\ell_{s^{\prime}}, r_{s^{\prime}}])$ 
satisfying $p_{s} = p_{s^{\prime}}$, 
then the interval attractor $([p_{s}, q_{s}], [\ell_{s}, r_{s}])$ precedes 
the interval attractor $([p_{s^{\prime}}, q_{s^{\prime}}], [\ell_{s^{\prime}}, r_{s^{\prime}}])$ if and only if $r_{s} < r_{s^{\prime}}$. 
Formally, let $\phi_{1}, \phi_{2}, \ldots, \phi_{k}$ be a permutation of $k$ integers $1, 2, \ldots, k$ such that 
each integer $s \in [1, k-1]$ satisfies either of the following two conditions: 
\begin{itemize}
    \item $p_{\phi_{s}} < p_{\phi_{s+1}}$; 
    \item $p_{\phi_{s}} = p_{\phi_{s+1}}$ and $r_{\phi_{s}} < r_{\phi_{s+1}}$.
\end{itemize}

Lemma~\ref{lem:IA_super_basic_property}~\ref{enum:IA_super_basic_property:3} shows that 
$p_{s} \neq p_{s^{\prime}}$ or $r_{s} \neq r_{s^{\prime}}$ holds 
for any pair of two integers $1 \leq s < s^{\prime} \leq k$. 
Therefore, 
the permutation $\phi_{1}, \phi_{2}, \ldots, \phi_{k}$ is unique.

\paragraph{Two Sequences $\mathbf{Q}_{\samp}$ and $\mathbf{Q}_{\ofs}$.}
We represent the sampling subset $\Psi_{\samp}$ as a sequence $\mathbf{Q}_{\samp}$ of $k$ nodes 
such that each node $u_{s}$ corresponds to the interval attractor $([p_{\phi_{s}}, q_{\phi_{s}}], [\ell_{\phi_{s}}, r_{\phi_{s}}])$. 
A label function $\mathcal{L}_{\samp}: \{ u_{1}$, $u_{2}$, $\ldots$, $u_{k}  \} \rightarrow \mathcal{N}_{0} \times \mathcal{N}_{0} \times \mathcal{N}_{0}$ 
is defined on the set of $k$ nodes $u_{1}$, $u_{2}$, $\ldots$, $u_{k}$. 
The label function $\mathcal{L}_{\samp}(u_{s})$ returns a triplet $(q_{\phi_{s}} - p_{\phi_{s}}, \ell_{\phi_{s}} - p_{\phi_{s}}, r_{\phi_{s}} - p_{\phi_{s}})$ of 
three non-negative integers $(q_{\phi_{s}} - p_{\phi_{s}})$, $(\ell_{\phi_{s}} - p_{\phi_{s}})$, and $(r_{\phi_{s}} - p_{\phi_{s}})$. 

Sequence $\mathbf{Q}_{\ofs}$ consists of $k$ non-negative integer 
$g_{1}, g_{2}, \ldots, g_{k}$ 
such that 
each integer $g_{s}$ is defined as $(p_{\phi_{s}} - p_{\phi_{s-1}})$. 
Here, let $p_{\phi_{0}} = 0$ for simplicity. 
These two sequences $\Psi_{\samp}$ and $\mathbf{Q}_{\ofs}$ 
are used to store the sampling subset $\Psi_{\samp}$. 

\subsubsection{Dynamic Data Structures for Sample Query}\label{subsubsec:sample_ds}
Sequence $\mathbf{Q}_{\samp} = u_{1}$, $u_{2}$, $\ldots$, $u_{k}$ is 
represented using a doubly linked list $\mathbf{L}_{\samp, 1}$ of $k$ elements 
such that each $s$-th element corresponds to the $s$-th node $u_{s}$. 
The $s$-th element of the doubly linked list stores the triplet obtained from 
function $\mathcal{L}_{\samp}(u_{s})$. 
A \emph{list indexing data structure}~\cite{DBLP:conf/wads/Dietz89} is used for quickly accessing to the $k$ elements of the doubly linked list representing $\mathbf{Q}_{\samp}$. 
For machine word size $B$, 
the list indexing data structure requires $O(kB)$ bits of space  
and supports the following four operations in $O(\log k)$ time: 
\begin{itemize}
\item return the $i$-th element of the doubly linked list for a given integer $i \in [1, k]$; 
\item return the index $i$ of a given $i$-th element of the doubly linked list; 
\item insert a new element after a given element of the doubly linked list; 
\item delete a given element from the doubly linked list. 
\end{itemize}

The order of the $k$ interval attractors $([p_{1}, q_{1}]$, $[\ell_{1}, r_{1}])$, $([p_{2}, q_{2}], [\ell_{2}, r_{2}])$, $\ldots$, $([p_{k}, q_{k}], [\ell_{k}, r_{k}])$ is stored in 
a doubly linked list $\mathbf{L}_{\samp, 2}$ of $k$ elements. 
For each integer $s \in [1, k]$, 
the $\phi_{s}$-th element of this doubly linked list stores a pointer to the $s$-th element of the doubly linked list $\mathbf{L}_{\samp, 1}$. 
A list indexing data structure is used for quickly accessing to the $k$ elements of the doubly linked list $\mathbf{L}_{\samp, 2}$. 

Sequence $\mathbf{Q}_{\ofs} = g_{1}, g_{2}, \ldots, g_{k}$ is represented using a doubly linked list $\mathbf{L}_{\samp, 3}$ of $k$ elements 
such that each $s$-th element stores the $s$-th integer $g_{s}$. 
A \emph{partial sum data structure}~\cite{DBLP:journals/algorithmica/BilleCCGSVV18} is built on 
the sequence $\mathbf{Q}_{\ofs}$. 
The partial sum data structure requires $O(kB)$ bits of space 
and supports the following four operations in $O(\log k)$ time: 
\begin{itemize}
\item return $\sum_{w = 1}^{i} g_{w}$ for a given integer $i \in [1, k]$; 
\item return the smallest integer $i$ in set $[1, k]$ satisfying $\sum_{w = 1}^{i} g_{w} \geq m$ for a given integer $m \geq 0$;
\item insert integer $b$ into the sequence $\mathbf{Q}_{\ofs}$ as the $i$-th integer for a pair of two integers $b \geq 0$ and $i \in [1, k+1]$;
\item delete the $i$-th from the sequence $\mathbf{Q}_{\ofs}$ for a given integer $i \in [1, k]$.
\end{itemize}
These data structures requires $O(kB)$ bits of space in total (i.e., $O(|\Psi_{\samp}|B)$ bits). 

\subsubsection{Computation of the \texorpdfstring{$\phi_{s}$}{}-th Interval Attractor}\label{subsubsec:computation_delta_samp}
For a given integer $s \in [1, k]$, 
we provide the algorithm computing the $\phi_{s}$-th interval attractor $([p_{\phi_{s}}, q_{\phi_{s}}], [\ell_{\phi_{s}}, r_{\phi_{s}}])$ using the dynamic data structures for sample query. 
This algorithm is used to answer sample query. 

For computing the $\phi_{s}$-th interval attractor $([p_{\phi_{s}}, q_{\phi_{s}}], [\ell_{\phi_{s}}, r_{\phi_{s}}])$, 
we leverage the following two observations: 
\begin{enumerate}[label=\textbf{(\roman*)}]
    \item $p_{\phi_{s}}$ is equal to $\sum_{w = 1}^{s} g_{w}$ for sequence $\mathbf{Q}_{\ofs} = g_{1}, g_{2}, \ldots, g_{k}$; 
    \item three integers $(q_{\phi_{s}} - p_{\phi_{s}}, \ell_{\phi_{s}} - p_{\phi_{s}}, r_{\phi_{s}} - p_{\phi_{s}})$ can be obtained from the label function $\mathcal{L}_{\samp}(u_{s})$. 
\end{enumerate}
Therefore, we can obtain the interval attractor $([p_{\phi_{s}}, q_{\phi_{s}}], [\ell_{\phi_{s}}, r_{\phi_{s}}])$ by computing $\sum_{w = 1}^{s} g_{w}$ and $\mathcal{L}_{\samp}(u_{s})$. 
The sum $\sum_{w = 1}^{s} g_{w}$ can be computed in $O(\log k)$ time using the partial sum data structure built on sequence $\mathbf{Q}_{\ofs}$. 
The result of the function $\mathcal{L}_{\samp}(u_{s})$ is stored in the $s$-th element of doubly linked list $\mathbf{L}_{\samp, 1}$. We can access this element in $O(\log k)$ time using the list indexing data structure built on $\mathbf{L}_{\samp, 1}$. 

The total running time of this algorithm is $O(\log k)$. 
Here, $k = O(n^{2})$ because 
$k \leq |\Psi_{\RR}|$ holds, 
and $|\Psi_{\RR}| = O(n^{2})$ follows from Lemma~\ref{lem:non_comp_IA_size}. 
Therefore, the interval attractor $([p_{\phi_{s}}, q_{\phi_{s}}], [\ell_{\phi_{s}}, r_{\phi_{s}}])$ can be computed in $O(\log n)$ time. 

\subsubsection{Algorithm for Sample Query}\label{subsubsec:algo_sample_query}
We present the algorithm answering a given sample query $\sampleQ(s)$ 
using the dynamic data structures for sample query. 
Let $s^{\prime}$ be an integer in set $\{ 1, 2, \ldots, k \}$ satisfying 
$\phi_{s^{\prime}} = s$ for permutation $\phi_{1}, \phi_{2}, \ldots, \phi_{k}$. 
Then, sample query $\sampleQ(s)$ returns 
the interval attractor $([p_{\phi_{s^{\prime}}}, q_{\phi_{s^{\prime}}}], [\ell_{\phi_{s^{\prime}}}, r_{\phi_{s^{\prime}}}])$. 
Therefore, the sample query can be answered 
using the integer $s^{\prime}$ and 
the algorithm provided in Section~\ref{subsubsec:computation_delta_samp}. 

We compute the interval attractor $([p_{\phi_{s^{\prime}}}, q_{\phi_{s^{\prime}}}], [\ell_{\phi_{s^{\prime}}}, r_{\phi_{s^{\prime}}}])$ in four steps: 
\begin{enumerate}[label=\textbf{(\roman*)}]
    \item access the $s$-th element of doubly linked list $\mathbf{L}_{\samp, 2}$ 
    using the list indexing data structure built on $\mathbf{L}_{\samp, 2}$;
    \item access the $s^{\prime}$-th element of doubly linked list $\mathbf{L}_{\samp, 1}$ 
    using the pointer stored in the $s$-th element of doubly linked list $\mathbf{L}_{\samp, 2}$;
    \item compute the integer $s^{\prime}$ by the list indexing data structure built on $\mathbf{L}_{\samp, 1}$;
    \item compute the interval attractor $([p_{\phi_{s^{\prime}}}, q_{\phi_{s^{\prime}}}], [\ell_{\phi_{s^{\prime}}}, r_{\phi_{s^{\prime}}}])$ by the algorithm provided in Section~\ref{subsubsec:computation_delta_samp}.
\end{enumerate}

These four steps take $O(\log k)$ time in total. 
Here, $O(\log k) = O(\log n)$ because of $k = O(n^{2})$. 
Therefore, a given sample query can be answered in $O(\log n)$ time. 

\subsection{Bigram Search Query}\label{subsec:bigram_search_query}
For a given string $P \in \Sigma^{+}$ of length $2$, 
\emph{bigram search query} $\BiSQ(P)$ returns a triplet $(d, g, j)$ of the following three integers $d$, $g$, and $j$: 
\begin{description}
    \item [$d$:] the number of suffixes of input string $T$ and lexicographically smaller than string $P$ 
    (i.e., $d = |\{ x \in [1, n] \mid T[x..n] \prec P \}|$); 
    \item [$g$:] the number of occurrences of string $P$ in input string $T$ (i.e., $g = |\Occ(T, P)|$);
    \item [$j$:] an occurrence position of string $P$ in string $T$ (i.e., $j \in \Occ(T, P)$)
    if the string $P$ occurs in $T$; 
    otherwise, $j = -1$.
\end{description}

For answering a given bigram search query, 
we leverage the $k$ interval attractors 
$([p_{1}, q_{1}], [\ell_{1}, r_{1}])$, 
$([p_{2}, q_{2}], [\ell_{2}, r_{2}])$, $\ldots$, $([p_{k}, q_{k}], [\ell_{k}, r_{k}])$ 
in subset $\Psi_{0} \cap \Psi_{\samp}$. 

\paragraph{Permutation $\varrho_{1}, \varrho_{2}, \ldots, \varrho_{k}$.}
For this subsection, 
we introduce a permutation $\varrho_{1}, \varrho_{2}, \ldots, \varrho_{k}$ of $k$ integers $1, 2, \ldots, k$. 
This permutation represents the order of the $k$ interval attractors $([p_{1}, q_{1}], [\ell_{1}, r_{1}])$, 
$([p_{2}, q_{2}], [\ell_{2}, r_{2}])$, $\ldots$, $([p_{k}, q_{k}], [\ell_{k}, r_{k}])$ obtained by 
sorting these interval attractors in lexicographical order of the $k$ strings 
$T[q_{1}..q_{1}+1]$, $T[q_{2}..q_{2}+1]$, $\ldots$, $T[q_{k}..q_{k}+1]$. 
Here, if the subset $\Psi_{0} \cap \Psi_{\samp}$ contains two distinct interval attractors 
$([p_{s}, q_{s}], [\ell_{s}, r_{s}])$ and $([p_{s^{\prime}}, q_{s^{\prime}}], [\ell_{s^{\prime}}, r_{s^{\prime}}])$ 
satisfying $T[q_{s}..q_{s}+1] = T[q_{s^{\prime}}..q_{s^{\prime}}+1]$, 
then the interval attractor $([p_{s}, q_{s}], [\ell_{s}, r_{s}])$ precedes the interval attractor $([p_{s^{\prime}}, q_{s^{\prime}}], [\ell_{s^{\prime}}, r_{s^{\prime}}])$ if and only if $T[p_{s}-1..r_{s}+1] \prec T[p_{s^{\prime}}-1..r_{s^{\prime}}+1]$. 
Formally, let $\varrho_{1}, \varrho_{2}, \ldots, \varrho_{k}$ be a permutation of $k$ integers $1, 2, \ldots, k$ such that 
each integer $s \in [1, k-1]$ satisfies either of the following two conditions: 
\begin{itemize}
    \item $T[q_{\varrho_{s}}..q_{\varrho_{s}}+1] \prec T[q_{\varrho_{s+1}}..q_{\varrho_{s+1}}+1]$; 
    \item $T[q_{\varrho_{s}}..q_{\varrho_{s}}+1] = T[q_{\varrho_{s+1}}..q_{\varrho_{s+1}}+1]$ and 
    $T[p_{\varrho_{s}}-1..r_{\varrho_{s}}+1] \prec T[p_{\varrho_{s+1}}-1..r_{\varrho_{s+1}}+1]$.     
\end{itemize}
The permutation $\varrho_{1}, \varrho_{2}, \ldots, \varrho_{k}$ is unique because 
$T[p-1..r+1] \neq T[p^{\prime}-1..r^{\prime}+1]$ follows from the definition of the sampling subset $\Psi_{\samp}$
for any pair of two distinct interval attractors $([p, q], [\ell, r])$, $([p^{\prime}, q^{\prime}], [\ell^{\prime}, r^{\prime}]) \in \Psi_{\samp}$.

\paragraph{Sequence $\mathbf{Q}_{\BiSQ}$ of nodes.}
We represent the sorted $k$ interval attractors 
$([p_{\varrho_{1}}, q_{\varrho_{1}}], [\ell_{\varrho_{1}}, r_{\varrho_{1}}])$, 
$([p_{\varrho_{2}}, q_{\varrho_{2}}], [\ell_{\varrho_{2}}, r_{\varrho_{2}}])$, $\ldots$, 
$([p_{\varrho_{k}}, q_{\varrho_{k}}], [\ell_{\varrho_{k}}, r_{\varrho_{k}}])$ as 
a sequence $\mathbf{Q}_{\BiSQ}$ of $k$ nodes $u_{1}, u_{2}, \ldots, u_{k}$ such that 
each node $u_{s}$ corresponds to the $s$-th interval attractor$([p_{\varrho_{s}}, q_{\varrho_{s}}], [\ell_{\varrho_{s}}, r_{\varrho_{s}}])$. 
A label function $\mathcal{L}_{\BiSQ}: \{ u_{1}$, $u_{2}$, $\ldots$, $u_{k}  \} \rightarrow \mathcal{N}_{0}$ 
is defined on the set of $k$ nodes $u_{1}$, $u_{2}$, $\ldots$, $u_{k}$. 
For a given node $u_{s}$ of sequence $\mathbf{Q}_{\BiSQ}$, 
the result of the function $\mathcal{L}_{\BiSQ}(u_{s})$ is defined as follows: 
if $([p_{\varrho_{s}}, q_{\varrho_{s}}], [\ell_{\varrho_{s}}, r_{\varrho_{s}}]) \in \Psi_{\source}$ holds, 
then the function $\mathcal{L}_{\BiSQ}(u_{s})$ returns $(1+m_{s})W_{s}$ 
for two integers $m_{s} = |f_{\recover}(([p_{\varrho_{s}}, q_{\varrho_{s}}], [\ell_{\varrho_{s}}, r_{\varrho_{s}}]))|$ and 
$W_{s} = |\Psi_{\str}(([p_{\varrho_{s}}, q_{\varrho_{s}}], [\ell_{\varrho_{s}}, r_{\varrho_{s}}]))|$; 
otherwise, the function $\mathcal{L}_{\BiSQ}(u_{s})$ returns the integer $W_{s}$. 
Here, $f_{\recover}$ and $\Psi_{\str}$ are the function and subset introduced in 
Section~\ref{subsec:function_recover} and Section~\ref{subsec:IA_subsets}, respectively. 

The following lemma states the relationship between the bigram search query and label function $\mathcal{L}_{\BiSQ}$. 

\begin{lemma}\label{lem:bis_property}
Consider the sequence $\mathbf{Q}_{\BiSQ}$ of $k$ nodes $u_{1}, u_{2}, \ldots, u_{k}$ 
such that each node $u_{s}$ corresponds to the $\varrho_{s}$-th interval attractor 
$([p_{\varrho_{s}}, q_{\varrho_{s}}], [\ell_{\varrho_{s}}, r_{\varrho_{s}}])$ in the subset 
$\Psi_{0} \cap \Psi_{\samp} = \{ ([p_{\varrho_{1}}, q_{\varrho_{1}}]$, $[\ell_{\varrho_{1}}, r_{\varrho_{1}}])$, 
$([p_{\varrho_{2}}, q_{\varrho_{2}}], [\ell_{\varrho_{2}}, r_{\varrho_{2}}])$, $\ldots$, 
$([p_{\varrho_{k}}, q_{\varrho_{k}}], [\ell_{\varrho_{k}}, r_{\varrho_{k}}]) \}$. 
For a string $P \in \Sigma^{+}$ of length $2$, 
we define three integers $\tau$, $\tau^{\prime}$, and $\lambda$ as follows: 
\begin{description}
    \item [$\tau$:] the largest integer $s$ in set $[1, k]$ satisfying $T[q_{\varrho_{s}}..q_{\varrho_{s}}+1] \prec P$ 
    if such integer $s$ exists; otherwise, let $\tau = 0$;
    \item [$\tau^{\prime}$:] the largest integer $s$ in set $[1, k]$ satisfying $T[q_{\varrho_{s}}..q_{\varrho_{s}}+1] \preceq P$ 
    if such integer $s$ exists; otherwise, let $\tau^{\prime} = 0$;
    \item [$\lambda$:] $\lambda = 1$ if $T[n..n] \prec P$; otherwise, let $\lambda = 0$.
\end{description}
The following three statements hold for the triplet $(d, g, j)$ obtained by bigram search query $\BiSQ(P)$: 
\begin{enumerate}[label=\textbf{(\roman*)}]
    \item $d = \lambda + \sum_{w = 1}^{\tau} \mathcal{L}_{\BiSQ}(u_{w})$;
    \item $g = (\sum_{w = 1}^{\tau^{\prime}} \mathcal{L}_{\BiSQ}(u_{w})) - (\sum_{w = 1}^{\tau} \mathcal{L}_{\BiSQ}(u_{w}))$;
    \item if $\tau < \tau^{\prime}$, 
    then $q_{\varrho_{\tau+1}} \in \Occ(T, P)$; 
    otherwise, $\Occ(T, P) = \emptyset$. 
\end{enumerate}
\end{lemma}
\begin{proof}
See Section~\ref{subsubsec:bis_property_proof}.
\end{proof}

Lemma~\ref{lem:bis_property} indicates that 
we can answer bigram search query $\BiSQ(P)$ using the label function $\mathcal{L}_{\BiSQ}$. 
The algorithm for answering a given bigram search query is detailed in Section~\ref{subsubsec:bis_query_algo}.

%%%%%%%%%%%%%%%%%%%%%%%%%%%%%%%%%%%%%%%%%%%%%%%%%%%%%%%%%%%%%%%%%%%%%%%%%%%
\subsubsection{Proof of Lemma~\ref{lem:bis_property}}\label{subsubsec:bis_property_proof}
The following proposition states five properties of the $0$-th level interval attractors $\Psi_{0}$.

\begin{proposition}\label{prop:bis1}
    The following five statements hold for the $0$-th level interval attractors $\Psi_{0}$: 
\begin{enumerate}[label=\textbf{(\roman*)}]
    \item \label{enum:bis1:1} consider interval attractor $I_{\capture}(q, q+1)$  
    for an interval attractor $([p, q], [\ell, r]) \in \Psi_{0}$. 
    Then, $I_{\capture}(q, q+1) = ([p, q], [\ell, r])$ holds;
    \item \label{enum:bis1:2} $\ell = q + 1$ for each interval attractor $([p, q], [\ell, r]) \in \Psi_{0}$;    
    \item \label{enum:bis1:3} $q \neq q^{\prime}$ for any pair of two distinct interval attractors 
    $([p, q], [\ell, r]), ([p^{\prime}, q^{\prime}], [\ell^{\prime}, r^{\prime}]) \in \Psi_{0}$;
    \item \label{enum:bis1:4} consider interval attractor $I_{\capture}(i, i+1)$ 
    for a position $i \in [1, n]$ in input string $T$ satisfying $i \leq n-1$. 
    Then, $I_{\capture}(i, i+1) \in \Psi_{0}$ holds;    
    \item \label{enum:bis1:5} $\{ q \mid ([p, q], [\ell, r]) \in \Psi_{0} \} = [1, n-1]$.
    
\end{enumerate}
\end{proposition}
\begin{proof}
The proof of Proposition~\ref{prop:bis1} is as follows. 

\textbf{Proof of Proposition~\ref{prop:bis1}(i).}
We prove $I_{\capture}(q, q+1) = ([p, q], [\ell, r])$ by contradiction. 
We assume that $I_{\capture}(q, q+1) \neq ([p, q], [\ell, r])$ holds. 
Then, there exists an interval attractor $([p^{\prime}, q^{\prime}], [\ell^{\prime}, r^{\prime}]) \in \Psi_{\RR}$ 
($([p^{\prime}, q^{\prime}], [\ell^{\prime}, r^{\prime}]) \neq ([p, q], [\ell, r])$)
satisfying $I_{\capture}(q, q+1) = ([p^{\prime}, q^{\prime}], [\ell^{\prime}, r^{\prime}])$. 
Since $I_{\capture}(q, q+1) = ([p^{\prime}, q^{\prime}], [\ell^{\prime}, r^{\prime}])$, 
$q \in [p^{\prime}, q^{\prime}]$ and $q+1 \in [\ell^{\prime}, r^{\prime}]$ follow from the definition of interval attractor. 
Let $h^{\prime}$ be the level of interval attractor $([p^{\prime}, q^{\prime}], [\ell^{\prime}, r^{\prime}])$. 

Consider interval attractor $I_{\capture}(q, r)$. 
Then, Lemma~\ref{lem:IA_maximal_lemma} shows that $I_{\capture}(q, r) = ([p, q], [\ell, r])$. 
If $r^{\prime} \geq r$, 
then Lemma~\ref{lem:IA_maximal_lemma} shows that $I_{\capture}(q, r) = ([p^{\prime}, q^{\prime}], [\ell^{\prime}, r^{\prime}])$ 
because $I_{\capture}(q, q+1) = ([p^{\prime}, q^{\prime}], [\ell^{\prime}, r^{\prime}])$, 
$q+1 \leq r$, and $r \in [\ell^{\prime}, r^{\prime}]$. 
$([p, q], [\ell, r]) = ([p^{\prime}, q^{\prime}], [\ell^{\prime}, r^{\prime}])$ follows from 
$I_{\capture}(q, r) = ([p, q], [\ell, r])$ and $I_{\capture}(q, r) = ([p^{\prime}, q^{\prime}], [\ell^{\prime}, r^{\prime}])$. 
The two facts $([p^{\prime}, q^{\prime}], [\ell^{\prime}, r^{\prime}]) \neq ([p, q], [\ell, r])$ 
and $([p, q], [\ell, r]) = ([p^{\prime}, q^{\prime}], [\ell^{\prime}, r^{\prime}])$ yield a contradiction. 

Otherwise (i.e., $r^{\prime} < r$), 
$I_{\capture}(q, r^{\prime}+1) \in \bigcup_{h = h^{\prime}+1}^{H} \Psi_{h}$ follows from Lemma~\ref{lem:interval_extension_propertyX}~\ref{enum:interval_extension_propertyX:right}. 
Since $I_{\capture}(q, r^{\prime}+1) \in \bigcup_{h = h^{\prime}+1}^{H} \Psi_{h}$, 
Corollary~\ref{cor:IA_basic_corollary}~\ref{enum:IA_basic_corollary:3} shows that 
$I_{\capture}(q, j^{\prime}) \in \bigcup_{h = h^{\prime}+1}^{H} \Psi_{h}$ holds 
for any integer $j^{\prime} \in [r^{\prime}+1, n]$.
$I_{\capture}(q, r) \in \bigcup_{h = h^{\prime}+1}^{H} \Psi_{h}$ 
follows from $I_{\capture}(q, j^{\prime}) \in \bigcup_{h = h^{\prime}+1}^{H} \Psi_{h}$ 
and $r \in [r^{\prime}+1, n]$. 
The level $h^{\prime\prime}$ of the interval attractor $([p, q], [\ell, r])$ is larger than $h^{\prime}+1$ 
because 
$I_{\capture}(q, r) \in \bigcup_{h = h^{\prime}+1}^{H} \Psi_{h}$ 
and $I_{\capture}(q, r) = ([p, q], [\ell, r])$. 
$h^{\prime\prime} \geq 1$ follows from $h^{\prime\prime} \geq h^{\prime}+1$ and $h^{\prime} \geq 0$. 
On the other hand, $h^{\prime\prime} = 0$ follows from $([p, q], [\ell, r]) \in \Psi_{0}$. 
The two facts $h^{\prime\prime} \geq 1$ and $h^{\prime\prime} = 0$ yield a contradiction. 

We showed that there exists a contradiction under the assumption that $I_{\capture}(q, q+1) \neq ([p, q], [\ell, r])$. 
Therefore, $I_{\capture}(q, q+1) = ([p, q], [\ell, r])$ must hold.

\textbf{Proof of Proposition~\ref{prop:bis1}(ii).}
Since $I_{\capture}(q, q+1) = ([p, q], [\ell, r])$, 
$q \in [p, q]$ and $q+1 \in [\ell, r]$ follow from the definition of interval attractor. 
$q < \ell$ follows from Lemma~\ref{lem:IA_super_basic_property}~\ref{enum:IA_super_basic_property:1}. 
Therefore, $\ell = q+1$ follows from $q+1 \in [\ell, r]$ and $q < \ell$. 

\textbf{Proof of Proposition~\ref{prop:bis1}(iii).}
We prove $q \neq q^{\prime}$ by contradiction. 
We assume that $q = q^{\prime}$ holds. 
Then, $I_{\capture}(q, q+1) = ([p, q], [\ell, r])$ and 
$I_{\capture}(q, q+1) = ([p^{\prime}, q^{\prime}], [\ell^{\prime}, r^{\prime}])$ follows from 
Proposition~\ref{prop:bis1}~\ref{enum:bis1:1}. 
$([p, q], [\ell, r]) = ([p^{\prime}, q^{\prime}], [\ell^{\prime}, r^{\prime}])$ follows from 
$I_{\capture}(q, q+1) = ([p, q], [\ell, r])$ and $I_{\capture}(q, q+1) = ([p^{\prime}, q^{\prime}], [\ell^{\prime}, r^{\prime}])$. 
$q = q^{\prime}$ follows from $([p, q], [\ell, r]) = ([p^{\prime}, q^{\prime}], [\ell^{\prime}, r^{\prime}])$. 
The two facts $q = q^{\prime}$ and $q \neq q^{\prime}$ yield a contradiction. 
Therefore, $q \neq q^{\prime}$ must hold. 

\textbf{Proof of Proposition~\ref{prop:bis1}(iv).}
Consider sequence $A(i, i+1) = [s^{0}, e^{0}], [s^{1}, e^{1}], \ldots, [s^{0 k^{\prime}}, e^{0 k^{\prime}}]$ of intervals 
and the set $\Delta(0, i)$ of intervals, which is introduced in Section~\ref{subsec:RR_delta}. 
Here, $[s^{0}, e^{0}] = [i, i+1]$ holds, 
and $k^{\prime} = 0$ follows from Lemma~\ref{lem:rec_function_basic_relation}~\ref{enum:rec_function_basic_relation:2}. 
Since $s^{0} = i$ and $k^{\prime} = 0$, 
$[i, i+1] \in \Delta(0, i)$ follows from the definition of the set $\Delta(0, i)$. 
Let $p = \min \{ i^{\prime} \mid [i^{\prime}, j^{\prime}] \in \Delta(0, i) \}$, 
$q = \max \{ i^{\prime} \mid [i^{\prime}, j^{\prime}] \in \Delta(0, i) \}$, 
$\ell = \min \{ j^{\prime} \mid [i^{\prime}, j^{\prime}] \in \Delta(0, i) \}$, 
and $r = \max \{ j^{\prime} \mid [i^{\prime}, j^{\prime}] \in \Delta(0, i) \}$. 
From Definition~\ref{def:RR_Delta}, 
the $0$-th level interval attractors $\Psi_{0}$ contains interval attractor $([p, q], [\ell, r])$. 

We prove $I_{\capture}(i, i+1) \in \Psi_{0}$. 
$I_{\capture}(i, i+1) = ([p, q], [\ell, r])$ hold because $[i, i+1] \in \Delta(0, i)$.
Therefore, $I_{\capture}(i, i+1) \in \Psi_{0}$ follows from 
$I_{\capture}(i, i+1) = ([p, q], [\ell, r])$ and $([p, q], [\ell, r]) \in \Psi_{0}$. 

\textbf{Proof of Proposition~\ref{prop:bis1}(v).}
We prove $\{ q \mid ([p, q], [\ell, r]) \in \Psi_{0} \} \subseteq [1, n-1]$. 
For each interval attractor $([p, q], [\ell, r]) \in \Psi_{0}$, 
$\ell = q + 1$ follows from Proposition~\ref{prop:bis1}(ii). 
$q \in [1, n-1]$ follows from $q, \ell \in [1, n]$ and $\ell = q + 1$. 
Therefore, $\{ q \mid ([p, q], [\ell, r]) \in \Psi_{0} \} \subseteq [1, n-1]$ holds. 

We prove $\{ q \mid ([p, q], [\ell, r]) \in \Psi_{0} \} = [1, n-1]$ by contradiction. 
We assume that $\{ q \mid ([p, q], [\ell, r]) \in \Psi_{0} \} \neq [1, n-1]$ holds. 
Then, there exists an integer $i \in [1, n-1]$ satisfying $i \not \in \{ q \mid ([p, q], [\ell, r]) \in \Psi_{0} \}$ 
because $\{ q \mid ([p, q], [\ell, r]) \in \Psi_{0} \} \subseteq [1, n-1]$. 

Consider interval attractor $I_{\capture}(i, i+1) = ([p^{\prime}, q^{\prime}], [\ell^{\prime}, r^{\prime}])$. 
Since $I_{\capture}(i, i+1) = ([p^{\prime}, q^{\prime}], [\ell^{\prime}, r^{\prime}])$, 
$i \in [p^{\prime}, q^{\prime}]$ and $i+1 \in [\ell^{\prime}, r^{\prime}]$ follow from the definition of interval attractor. 
$q^{\prime} < \ell^{\prime}$ follows from Lemma~\ref{lem:IA_super_basic_property}~\ref{enum:IA_super_basic_property:1}. 
Therefore, $i = q^{\prime}$ and $i+1 = \ell^{\prime}$ follow from $i \in [p^{\prime}, q^{\prime}]$, $i+1 \in [\ell^{\prime}, r^{\prime}]$, and $q^{\prime} < \ell^{\prime}$.

Let $h^{\prime}$ be the level of interval attractor $I_{\capture}(i, i+1)$. 
Then, $h^{\prime} = 0$ holds 
because $I_{\capture}(i, i+1) \in \Psi_{0}$ follows from Proposition~\ref{prop:bis1}~\ref{enum:bis1:4}.  
$i \in \{ q \mid ([p, q], [\ell, r]) \in \Psi_{0} \}$ follows from 
$I_{\capture}(i, i+1) \in \Psi_{0}$ and $i = q^{\prime}$. 
The two facts $i \in \{ q \mid ([p, q], [\ell, r]) \in \Psi_{0} \}$ and $i \not \in \{ q \mid ([p, q], [\ell, r]) \in \Psi_{0} \}$ yield a contradiction. 
Therefore, $\{ q \mid ([p, q], [\ell, r]) \in \Psi_{0} \} = [1, n-1]$ must hold. 
\end{proof}

Consider the subset $\Psi_{0} \cap \Psi_{\samp} = \{ ([p_{1}, q_{1}], [\ell_{1}, r_{1}]), ([p_{2}, q_{2}], [\ell_{2}, r_{2}]), \ldots, ([p_{k}, q_{k}], [\ell_{k}, r_{k}]) \}$. 
Here, each interval attractor $([p_{s}, q_{s}], [\ell_{s}, r_{s}])$ is contained in 
the subset $\Psi_{\str}(T[p_{s}-1..r_{s}+1])$, which is introduced in Section~\ref{subsec:IA_subsets}. 
For this subsection, 
we introduce $k$ subsets $\Psi^{\BiSQ}_{1}$, $\Psi^{\BiSQ}_{2}$, $\ldots$, $\Psi^{\BiSQ}_{k}$ of the set $\Psi_{\RR}$. 
For each integer $s \in [1, k]$, 
if $([p_{s}, q_{s}], [\ell_{s}, r_{s}]) \in \Psi_{\source}$ holds, 
then Lemma~\ref{lem:psi_equality_basic_property}~\ref{enum:psi_equality_basic_property:5} shows that 
the subset $\Psi_{\str}(T[p_{s}-1..r_{s}+1])$ is a subset of the set $\Psi_{\source}$ (i.e., $\Psi_{\str}(T[p_{s}-1..r_{s}+1]) \subseteq \Psi_{\source}$). 
In this case, the subset $\Psi^{\BiSQ}_{s}$ is defined as the union of 
two sets $\Psi^{\BiSQ}_{s} = \Psi_{\str}(T[p_{s}-1..r_{s}+1])$ and $\bigcup_{([p, q], [\ell, r]) \in \Psi_{\str}(T[p_{s}-1..r_{s}+1])} f_{\recover}(([p, q], [\ell, r]))$, i.e., 
$$\Psi^{\BiSQ}_{s} = \Psi_{\str}(T[p_{s}-1..r_{s}+1]) \cup \Bigl(\bigcup_{([p, q], [\ell, r]) \in \Psi_{\str}(T[p_{s}-1..r_{s}+1])} f_{\recover}(([p, q], [\ell, r])) \Bigr)$$ 
Otherwise (i.e., $([p_{s}, q_{s}], [\ell_{s}, r_{s}]) \not \in \Psi_{\source}$), 
let $\Psi^{\BiSQ}_{s} = \Psi_{\str}(T[p_{s}-1..r_{s}+1])$. 

The following proposition states properties of the $k$ subsets $\Psi^{\BiSQ}_{1}$, $\Psi^{\BiSQ}_{2}$, $\ldots$, $\Psi^{\BiSQ}_{k}$.  

\begin{proposition}\label{prop:bis2}
    Consider the $k$ subsets $\Psi^{\BiSQ}_{1}$, $\Psi^{\BiSQ}_{2}$, $\ldots$, $\Psi^{\BiSQ}_{k}$ of set $\Psi_{\RR}$ 
    for the subset $\Psi_{0} \cap \Psi_{\samp} = \{ ([p_{1}, q_{1}], [\ell_{1}, r_{1}])$, $([p_{2}, q_{2}], [\ell_{2}, r_{2}])$, $\ldots$, $([p_{k}, q_{k}], [\ell_{k}, r_{k}]) \}$. 
    Here, each node $u_{s}$ of the sequence $\mathbf{Q}_{\BiSQ}$ corresponds to 
    the $\varrho_{s}$-th interval attractor $([p_{\varrho_{s}}, q_{\varrho_{s}}], [\ell_{\varrho_{s}}, r_{\varrho_{s}}])$.    
    The following six statements hold: 
\begin{enumerate}[label=\textbf{(\roman*)}]
    \item for each interval attractor $([p, q], [\ell, r]) \in \Psi_{0}$, 
    there exists an integer $s \in [1, k]$ satisfying $([p, q], [\ell, r]) \in \Psi^{\BiSQ}_{s}$;
    \item $\Psi^{\BiSQ}_{s} \subseteq \Psi_{0}$ for all integer $s \in [1, k]$;
    \item $\mathcal{L}_{\BiSQ}(u_{s}) = |\Psi^{\BiSQ}_{\varrho_{s}}|$ for all integer $s \in [1, k]$;     
    \item consider an interval attractor $([p, q], [\ell, r]) \in \Psi^{\BiSQ}_{s}$ for an integer $s \in [1, s]$. 
    Then, $T[q_{s}..q_{s}+1] = T[q..q+1]$ holds;
    \item $\Psi^{\BiSQ}_{s} \cap \Psi^{\BiSQ}_{s^{\prime}} = \emptyset$ for any pair of two integers $1 \leq s < s^{\prime} \leq k$;
    \item $|\Psi^{\BiSQ}_{s}| = |\{ q \mid ([p, q], [\ell, r]) \in \Psi^{\BiSQ}_{s} \}|$ for all integer $s \in [1, k]$.
\end{enumerate}
\end{proposition}
\begin{proof}
The following three statements are used to prove Proposition~\ref{prop:bis2}: 
\begin{enumerate}[label=\textbf{(\Alph*)}]
    \item consider two interval attractors $([p, q], [\ell, r]), ([p^{\prime}, q^{\prime}], [\ell^{\prime}, r^{\prime}]) \in \Psi_{\RR}$ satisfying $([p^{\prime}, q^{\prime}], [\ell^{\prime}, r^{\prime}]) \in \Psi_{\str}(T[p-1..r+1])$. 
    Then, $T[q..q+1] = T[q^{\prime}..q^{\prime}+1]$ holds;    
    \item consider two interval attractors $([p, q], [\ell, r]), ([p^{\prime}, q^{\prime}], [\ell^{\prime}, r^{\prime}]) \in \Psi_{\RR}$ satisfying $([p, q], [\ell, r]) \in \Psi_{0} \cap \Psi_{\source}$ and $([p^{\prime}, q^{\prime}], [\ell^{\prime}, r^{\prime}]) \in f_{\recover}(([p, q], [\ell, r]))$, respectively. 
    Then, $T[q] = T[q^{\prime}] = (C^{n+1})[1]$ and $T[q+1] = T[q^{\prime}+1] = (C^{n+1})[2]$ hold
    for the associated string $C$ of the interval attractor $([p, q], [\ell, r])$; 
    \item consider an interval attractor $([p, q], [\ell, r])$ in the subset $\Psi^{\BiSQ}_{s}$ for an integer $s \in [1, k]$. 
    If $([p, q], [\ell, r]) \not \in \Psi_{\str}(T[p_{s}-1..r_{s}+1])$ holds, 
    then $([p_{s}, q_{s}], [\ell_{s}, r_{s}]) \in \Psi_{\source}$, 
    $([p, q], [\ell, r]) \in \Psi_{\run}$, and 
    the subset $\Psi_{\str}(T[p_{s}-1..r_{s}+1])$ contains an interval attractor $([p^{\prime}, q^{\prime}], [\ell^{\prime}, r^{\prime}])$ satisfying $([p, q], [\ell, r]) \in f_{\recover}(([p^{\prime}, q^{\prime}], [\ell^{\prime}, r^{\prime}]))$. 
\end{enumerate}

\textbf{Proof of statement (A).}
$T[p-1..r+1] = T[p^{\prime}-1..r^{\prime}+1]$ follows from $([p^{\prime}, q^{\prime}], [\ell^{\prime}, r^{\prime}]) \in \Psi_{\str}(T[p-1..r+1])$. 
Because of $T[p-1..r+1] = T[p^{\prime}-1..r^{\prime}+1]$, 
$|[p, q]| = |[p^{\prime}, q^{\prime}]|$ follows from Lemma~\ref{lem:psi_str_property}~\ref{enum:psi_str_property:1}. 
Therefore, 
$T[q..q+1] = T[q^{\prime}..q^{\prime}+1]$ follows from 
(1) $T[p-1..r+1] = T[p^{\prime}-1..r^{\prime}+1]$, 
(2) $|[p, q]| = |[p^{\prime}, q^{\prime}]|$, 
and (3) $q+1 \leq r+1$. 

\textbf{Proof of statement (B).}
Let $([p^{\prime}_{1}, q^{\prime}_{1}], [\ell^{\prime}_{1}, r^{\prime}_{1}])$, 
$([p^{\prime}_{2}, q^{\prime}_{2}], [\ell^{\prime}_{2}, r^{\prime}_{2}])$, $\ldots$, 
$([p^{\prime}_{m}, q^{\prime}_{m}], [\ell^{\prime}_{m}, r^{\prime}_{m}])$ ($p^{\prime}_{1} < p^{\prime}_{2} < \cdots < p^{\prime}_{m}$) be the interval attractors obtained from the function $f_{\recover}(([p, q], [\ell, r]))$. 

We prove $T[q^{\prime}_{s}..q^{\prime}_{s}+1] = (C^{n+1})[1..2]$ for each integer $s \in [1, m]$. 
Because of $([p, q], [\ell, r]) \in \Psi_{0} \cap \Psi_{\centerset}(C)$, 
$([p^{\prime}_{s}, q^{\prime}_{s}], [\ell^{\prime}_{s}, r^{\prime}_{s}]) \in \Psi_{0} \cap \Psi_{\centerset}(C) \cap \Psi_{\run}$ follows from 
Lemma~\ref{lem:recover_basic_property}~\ref{enum:recover_basic_property:4}. 
Because of $([p^{\prime}_{s}, q^{\prime}_{s}], [\ell^{\prime}_{s}, r^{\prime}_{s}]) \in \Psi_{0}$, 
$\ell^{\prime}_{s} = q^{\prime}_{s} + 1$ follows from Proposition~\ref{prop:bis1}(ii). 
Let $\gamma^{\prime}_{s}$ be the attractor position of the interval attractor $([p^{\prime}_{s}, q^{\prime}_{s}], [\ell^{\prime}_{s}, r^{\prime}_{s}])$. 
Then, $q^{\prime}_{s} \leq \gamma^{\prime}_{s} < \ell^{\prime}_{s}$ follows from Lemma~\ref{lem:IA_super_basic_property}~\ref{enum:IA_super_basic_property:1}. 
$\gamma^{\prime}_{s} = q^{\prime}_{s}$ follows from 
$q^{\prime}_{s} \leq \gamma^{\prime}_{s} < \ell^{\prime}_{s}$ and $\ell^{\prime}_{s} = q^{\prime}_{s} + 1$. 

Because of $([p^{\prime}_{s}, q^{\prime}_{s}], [\ell^{\prime}_{s}, r^{\prime}_{s}]) \in \Psi_{0} \cap \Psi_{\centerset}(C) \cap \Psi_{\run}$, 
$\lcs(T[p^{\prime}_{s}-1..\gamma^{\prime}_{s} - 1], C^{n+1}) = T[p^{\prime}_{s}-1..\gamma^{\prime}_{s} - 1]$ 
and $|\lcp(T[\gamma^{\prime}_{s}..r^{\prime}_{s}], C^{n+1})| > 1 + \sum_{w = 1}^{3} \lfloor \mu(w) \rfloor$ 
follow from the definition of the subset $\Psi_{\run}$. 
Therefore, $T[q^{\prime}_{s}..q^{\prime}_{s}+1] = (C^{n+1})[1..2]$ holds  
because (a) $T[\gamma^{\prime}_{s}..\gamma^{\prime}_{s}+1] = (C^{n+1})[1..2]$ follows from 
$|\lcp(T[\gamma^{\prime}_{s}..r^{\prime}_{s}], C^{n+1})| > 1 + \sum_{w = 1}^{3} \lfloor \mu(w) \rfloor$, and 
(b) $\gamma^{\prime}_{s} = q^{\prime}_{s}$. 

%$T[q^{\prime}_{s}] = C[|C|]$ holds because 
%(a) $T[\gamma^{\prime}_{s} - 1] = C[|C|]$ follows from $\lcs(T[p^{\prime}_{s}-1..\gamma^{\prime}_{s} - 1], C^{n+1}) = T[p^{\prime}_{s}-1..\gamma^{\prime}_{s} - 1]$, 
%and (b) $\gamma^{\prime}_{s} - 1 = q^{\prime}_{s}$. 
%Therefore, $T[q^{\prime}_{s}] = C[|C|]$ and $T[q^{\prime}_{s}+1] = C[1]$ hold. 

Next, we prove $T[q..q+1] = (C^{n+1})[1..2]$. 
Let $K \geq 0$ be an integer satisfying $([p, q], [\ell, r]) \in \Psi_{\lcp}(K)$. 
Then, Lemma~\ref{lem:recover_basic_property}~\ref{enum:recover_basic_property:1} shows that $K \geq 2$ holds. 
Let $\gamma$ be the attractor position of the interval attractor $([p, q], [\ell, r])$. 
Then, we can prove $\gamma = q$ using the same approach used to prove $\gamma^{\prime}_{s} = q^{\prime}_{s}$. 
Because of $K \geq 2$,
$|\lcp(T[\gamma..r], C^{n+1})| \geq 2$ holds (see the definition of the subset $\Psi_{\lcp}(K)$). 
$T[\gamma..\gamma+1] = (C^{n+1})[1..2]$ follows from $|\lcp(T[\gamma..r+1], C^{n+1})| \geq 2$. 
Therefore, $T[q..q+1] = (C^{n+1})[1..2]$ follows from 
$T[\gamma..\gamma+1] = (C^{n+1})[1..2]$ and $\gamma = q$. 

Finally, $T[q] = T[q^{\prime}] = (C^{n+1})[1]$ and $T[q+1] = T[q^{\prime}+1] = (C^{n+1})[2]$ follow from 
$T[q^{\prime}_{s}..q^{\prime}_{s}+1] = (C^{n+1})[1..2]$ and $T[q..q+1] = (C^{n+1})[1..2]$. 

\textbf{Proof of statement (C).}
We prove $([p_{s}, q_{s}], [\ell_{s}, r_{s}]) \in \Psi_{\source}$ by contradiction. 
We assume that $([p_{s}, q_{s}], [\ell_{s}, r_{s}]) \not \in \Psi_{\source}$ holds. 
Then, the subset $\Psi^{\BiSQ}_{s} = \Psi_{\str}(T[p_{s}-1..r_{s}+1])$ follows from the definition of the subset $\Psi^{\BiSQ}_{s}$. 
$([p_{s}, q_{s}], [\ell_{s}, r_{s}]) \not \in \Psi^{\BiSQ}_{s}$ follows from 
$\Psi^{\BiSQ}_{s} = \Psi_{\str}(T[p_{s}-1..r_{s}+1])$ and $([p, q], [\ell, r]) \not \in \Psi_{\str}(T[p_{s}-1..r_{s}+1])$. 
The two facts $([p_{s}, q_{s}], [\ell_{s}, r_{s}]) \not \in \Psi^{\BiSQ}_{s}$ and $([p_{s}, q_{s}], [\ell_{s}, r_{s}]) \in \Psi^{\BiSQ}_{s}$ yield a contradiction. 
Therefore, $([p_{s}, q_{s}], [\ell_{s}, r_{s}]) \in \Psi_{\source}$ must hold. 

Because of $([p_{s}, q_{s}], [\ell_{s}, r_{s}]) \in \Psi_{\source}$, 
the subset $\Psi^{\BiSQ}_{s}$ is defined as the union of two sets $\Psi_{\str}(T[p_{s}-1..r_{s}+1])$ 
and $\bigcup_{([p_{A}, q_{A}], [\ell_{A}, r_{A}]) \in \Psi_{\str}(T[p_{s}-1..r_{s}+1])} f_{\recover}(([p_{A}, q_{A}], [\ell_{A}, r_{A}]))$. 
Because of $([p, q], [\ell, r]) \not \in \Psi_{\str}(T[p_{s}-1..r_{s}+1])$, 
the subset $\Psi_{\str}(T[p_{s}-1..r_{s}+1])$ contains an interval attractor $([p^{\prime}, q^{\prime}], [\ell^{\prime}, r^{\prime}])$ satisfying $([p, q], [\ell, r]) \in f_{\recover}(([p^{\prime}, q^{\prime}], [\ell^{\prime}, r^{\prime}]))$. 
Because of $([p, q], [\ell, r]) \in f_{\recover}(([p^{\prime}, q^{\prime}], [\ell^{\prime}, r^{\prime}]))$, 
$([p, q], [\ell, r]) \in \Psi_{\run}$ follows from the definition of the function $f_{\recover}$. 
Therefore, statement (C) holds. 

\textbf{Proof of Proposition~\ref{prop:bis2}(i).}
We prove Proposition~\ref{prop:bis2}(i) holds if $([p, q], [\ell, r]) \not \in \Psi_{\run}$. 
In this case, Lemma~\ref{lem:lm_basic_property}~\ref{enum:lm_basic_property:2} shows that 
the subset $\Psi_{\leftmost}$ contains an interval attractor $([p_{A}, q_{A}], [\ell_{A}, r_{A}])$ 
satisfying $T[p_{A}-1..r_{A}+1] = T[p-1..r+1]$. 
Because of $([p, q], [\ell, r]) \not \in \Psi_{\run}$, 
$([p_{A}, q_{A}], [\ell_{A}, r_{A}]) \not \in \Psi_{\run}$ follows from Lemma~\ref{lem:psi_equality_basic_property}~\ref{enum:psi_equality_basic_property:4}. 
Because of $([p_{A}, q_{A}], [\ell_{A}, r_{A}]) \in \Psi_{\leftmost} \setminus \Psi_{\run}$, 
the subset $\Psi_{\samp}$ contains an interval attractor $([p_{B}, q_{B}], [\ell_{B}, r_{B}])$ 
satisfying $T[p_{B}-1..r_{B}+1] = T[p_{A}-1..r_{A}+1]$ (see the definition of the subset $\Psi_{\samp}$). 
Here, $T[p_{B}-1..r_{B}+1] = T[p-1..r+1]$ follows from 
$T[p_{B}-1..r_{B}+1] = T[p_{A}-1..r_{A}+1]$ and $T[p_{A}-1..r_{A}+1] = T[p-1..r+1]$. 
Lemma~\ref{lem:psi_str_property}~\ref{enum:psi_str_property:2} shows that 
$([p_{B}, q_{B}], [\ell_{B}, r_{B}]) \in \Psi_{0}$ holds 
because $T[p_{B}-1..r_{B}+1] = T[p-1..r+1]$ and $([p, q], [\ell, r]) \in \Psi_{0}$. 
Because of $([p_{B}, q_{B}], [\ell_{B}, r_{B}]) \in \Psi_{\samp} \cap \Psi_{0}$, 
there exists an integer $s \in [1, k]$ satisfying $([p_{s}, q_{s}], [\ell_{s}, r_{s}]) = ([p_{B}, q_{B}], [\ell_{B}, r_{B}])$. 
$([p, q], [\ell, r]) \in \Psi_{\str}(T[p_{s}-1..r_{s}+1])$ follows from 
$T[p_{B}-1..r_{B}+1] = T[p-1..r+1]$ and $([p_{s}, q_{s}], [\ell_{s}, r_{s}]) = ([p_{B}, q_{B}], [\ell_{B}, r_{B}])$. 
$\Psi_{\str}(T[p_{s}-1..r_{s}+1]) \subseteq \Psi^{\BiSQ}_{s}$ follows from the definition of the subset $\Psi^{\BiSQ}_{s}$. 
Therefore, $([p, q], [\ell, r]) \in \Psi^{\BiSQ}_{s}$ follows from 
$([p, q], [\ell, r]) \in \Psi_{\str}(T[p_{s}-1..r_{s}+1])$ and $\Psi_{\str}(T[p_{s}-1..r_{s}+1]) \subseteq \Psi^{\BiSQ}_{s}$. 

Next, we prove Proposition~\ref{prop:bis2}(i) holds if $([p, q], [\ell, r]) \in \Psi_{\run}$. 
Lemma~\ref{lem:recover_division_property} shows that 
the subset $\Psi_{\source}$ contains an interval attractor $([p_{C}, q_{C}], [\ell_{C}, r_{C}])$ 
satisfying $([p, q], [\ell, r]) \in f_{\recover}(([p_{C}, q_{C}]$, $[\ell_{C}, r_{C}]))$. 
Here, $([p_{C}, q_{C}], [\ell_{C}, r_{C}]) \in \Psi_{0}$ follows from 
Lemma~\ref{lem:recover_basic_property}~\ref{enum:recover_basic_property:4} 
because $([p, q], [\ell, r]) \in \Psi_{0}$. 

Lemma~\ref{lem:lm_basic_property}~\ref{enum:lm_basic_property:2} shows that 
the subset $\Psi_{\leftmost}$ contains an interval attractor $([p_{D}, q_{D}], [\ell_{D}, r_{D}])$ 
satisfying $T[p_{D}-1..r_{D}+1] = T[p_{C}-1..r_{C}+1]$. 
Because of $([p_{C}, q_{C}], [\ell_{C}, r_{C}]) \in \Psi_{\source}$, 
$([p_{D}, q_{D}]$, $[\ell_{D}, r_{D}]) \in \Psi_{\source}$ follows from Lemma~\ref{lem:psi_equality_basic_property}~\ref{enum:psi_equality_basic_property:5}. 
$([p_{D}, q_{D}], [\ell_{D}, r_{D}]) \not \in \Psi_{\run}$ holds 
because $([p_{C}, q_{C}]$, $[\ell_{C}, r_{C}]) \in \Psi_{\source}$ and $\Psi_{\source} \cap \Psi_{\run} = \emptyset$. 
Because of $([p_{D}, q_{D}], [\ell_{D}, r_{D}]) \in \Psi_{\leftmost} \setminus \Psi_{\run}$, 
the subset $\Psi_{\samp}$ contains an interval attractor $([p_{E}, q_{E}], [\ell_{E}, r_{E}])$ 
satisfying $T[p_{E}-1..r_{E}+1] = T[p_{D}-1..r_{D}+1]$. 
Here, $T[p_{E}-1..r_{E}+1] = T[p_{C}-1..r_{C}+1]$ follows from 
$T[p_{E}-1..r_{E}+1] = T[p_{D}-1..r_{D}+1]$ and $T[p_{D}-1..r_{D}+1] = T[p_{C}-1..r_{C}+1]$. 
Lemma~\ref{lem:psi_str_property}~\ref{enum:psi_str_property:2} shows that 
$([p_{E}, q_{E}], [\ell_{E}, r_{E}]) \in \Psi_{0}$ holds 
because $T[p_{E}-1..r_{E}+1] = T[p_{C}-1..r_{C}+1]$ and $([p_{C}, q_{C}], [\ell_{C}, r_{C}]) \in \Psi_{0}$. 
Because of $([p_{E}, q_{E}], [\ell_{E}, r_{E}]) \in \Psi_{\samp} \cap \Psi_{0}$, 
there exists an integer $s \in [1, k]$ satisfying $([p_{s}, q_{s}], [\ell_{s}, r_{s}]) = ([p_{E}, q_{E}], [\ell_{E}, r_{E}])$. 

Because of $([p_{C}, q_{C}], [\ell_{C}, r_{C}]) \in \Psi_{\source}$, 
$([p_{s}, q_{s}], [\ell_{s}, r_{s}]) \in \Psi_{\source}$ follows from Lemma~\ref{lem:psi_equality_basic_property}~\ref{enum:psi_equality_basic_property:5}. 
Because of $([p_{s}, q_{s}], [\ell_{s}, r_{s}]) \in \Psi_{\source}$, 
$\bigcup_{([p^{\prime}, q^{\prime}], [\ell^{\prime}, r^{\prime}]) \in \Psi_{\str}(T[p_{s}-1..r_{s}+1])} f_{\recover}(([p^{\prime}, q^{\prime}], [\ell^{\prime}, r^{\prime}])) \subseteq \Psi^{\BiSQ}_{s}$ follows from the definition of the subset $\Psi^{\BiSQ}_{s}$. 
Here, $([p_{C}, q_{C}], [\ell_{C}, r_{C}]) \in \Psi_{\str}(T[p_{s}-1..r_{s}+1])$ follows from 
$T[p_{E}-1..r_{E}+1] = T[p_{C}-1..r_{C}+1]$ and $([p_{s}, q_{s}], [\ell_{s}, r_{s}]) = ([p_{E}, q_{E}], [\ell_{E}, r_{E}])$. 
Therefore, $([p, q], [\ell, r]) \in \Psi^{\BiSQ}_{s}$ follows from the following equation:
\begin{equation*}
\begin{split}
([p, q], [\ell, r]) &\in f_{\recover}(([p_{C}, q_{C}], [\ell_{C}, r_{C}])) \\
&\subseteq \bigcup_{([p^{\prime}, q^{\prime}], [\ell^{\prime}, r^{\prime}]) \in \Psi_{\str}(T[p_{s}-1..r_{s}+1])} f_{\recover}(([p^{\prime}, q^{\prime}], [\ell^{\prime}, r^{\prime}])) \\
&\subseteq \Psi^{\BiSQ}_{s}.
\end{split}
\end{equation*}

We showed that Proposition~\ref{prop:bis2}(i) whether $([p, q], [\ell, r]) \in \Psi_{\run}$ or not. 
Therefore, Proposition~\ref{prop:bis2}(i) holds.

\textbf{Proof of Proposition~\ref{prop:bis2}(ii).}
Consider an interval attractor $([p, q], [\ell, r])$ in the subset $\Psi^{\BiSQ}_{s}$. 
Then, we prove $([p, q], [\ell, r]) \in \Psi_{0}$. 
One of the following three conditions is satisfied: 
\begin{enumerate}[label=\textbf{(\alph*)}]
    \item $([p_{s}, q_{s}], [\ell_{s}, r_{s}]) \not \in \Psi_{\source}$;
    \item $([p_{s}, q_{s}], [\ell_{s}, r_{s}]) \in \Psi_{\source}$ 
    and $([p, q], [\ell, r]) \in \Psi_{\str}(T[p_{s}-1..r_{s}+1])$;
    \item $([p_{s}, q_{s}], [\ell_{s}, r_{s}]) \in \Psi_{\source}$ 
    and $([p, q], [\ell, r]) \not \in \Psi_{\str}(T[p_{s}-1..r_{s}+1])$.
\end{enumerate}

For condition (a), 
$\Psi^{\BiSQ}_{s} = \Psi_{\str}(T[p_{s}-1..r_{s}+1])$ follows from the definition of the subset $\Psi^{\BiSQ}_{s}$. 
$([p, q], [\ell, r]) \in \Psi_{\str}(T[p_{s}-1..r_{s}+1])$ follows from 
$([p, q], [\ell, r]) \in \Psi^{\BiSQ}_{s}$ and $\Psi^{\BiSQ}_{s} = \Psi_{\str}(T[p_{s}-1..r_{s}+1])$. 
$T[p-1..r+1] = T[p_{s}-1..r_{s}+1]$ follows from $([p, q], [\ell, r]) \in \Psi_{\str}(T[p_{s}-1..r_{s}+1])$. 
Lemma~\ref{lem:psi_str_property}~\ref{enum:psi_str_property:2} shows that 
$([p, q], [\ell, r]) \in \Psi_{0}$ holds 
because $T[p-1..r+1] = T[p_{s}-1..r_{s}+1]$ and $([p_{s}, q_{s}], [\ell_{s}, r_{s}]) \in \Psi_{0}$. 

For condition (b), 
$T[p-1..r+1] = T[p_{s}-1..r_{s}+1]$ follows from $([p, q], [\ell, r]) \in \Psi_{\str}(T[p_{s}-1..r_{s}+1])$. 
Lemma~\ref{lem:psi_str_property}~\ref{enum:psi_str_property:2} shows that  
$([p, q], [\ell, r]) \in \Psi_{0}$ holds 
because $T[p-1..r+1] = T[p_{s}-1..r_{s}+1]$ and $([p_{s}, q_{s}], [\ell_{s}, r_{s}]) \in \Psi_{0}$. 

For condition (c), 
statement (C) shows that 
the subset $\Psi_{\str}(T[p_{s}-1..r_{s}+1])$ contains an interval attractor $([p^{\prime}, q^{\prime}], [\ell^{\prime}, r^{\prime}])$ satisfying $([p, q], [\ell, r]) \in f_{\recover}(([p^{\prime}, q^{\prime}], [\ell^{\prime}, r^{\prime}]))$. 
$T[p^{\prime}-1..r^{\prime}+1] = T[p_{s}-1..r_{s}+1]$ follows from $([p^{\prime}, q^{\prime}], [\ell^{\prime}, r^{\prime}]) \in \Psi_{\str}(T[p_{s}-1..r_{s}+1])$. 
Lemma~\ref{lem:psi_str_property}~\ref{enum:psi_str_property:2} shows that 
$([p^{\prime}, q^{\prime}], [\ell^{\prime}, r^{\prime}]) \in \Psi_{0}$ holds 
because $T[p^{\prime}-1..r^{\prime}+1] = T[p_{s}-1..r_{s}+1]$ and $([p_{s}, q_{s}], [\ell_{s}, r_{s}]) \in \Psi_{0}$. 
$([p, q], [\ell, r]) \in \Psi_{0}$ follows from 
Lemma~\ref{lem:recover_basic_property}~\ref{enum:recover_basic_property:4} 
because $([p^{\prime}, q^{\prime}], [\ell^{\prime}, r^{\prime}]) \in \Psi_{0}$ 
and $([p, q], [\ell, r]) \in f_{\recover}(([p^{\prime}, q^{\prime}], [\ell^{\prime}, r^{\prime}]))$.

%the subset $\Psi_{\str}(T[p_{s}-1..r_{s}+1])$ is a subset of the subset $\Psi_{\source}$, 
%and the subset $\Psi_{\str}(T[p_{s}-1..r_{s}+1])$ contains 
%an interval attractor $([p^{\prime}, q^{\prime}], [\ell^{\prime}, r^{\prime}])$ 
%satisfying $([p, q], [\ell, r]) \in f_{\recover}(([p^{\prime}, q^{\prime}], [\ell^{\prime}, r^{\prime}]))$ 
%(see the definition of the subset $\Psi^{\BiSQ}_{s}$). 

We showed that $([p, q], [\ell, r]) \in \Psi_{0}$ holds for 
each interval attractor $([p, q], [\ell, r])$ in the subset $\Psi^{\BiSQ}_{s}$. 
Therefore, $\Psi^{\BiSQ}_{s} \subseteq \Psi_{0}$ holds. 

\textbf{Proof of Proposition~\ref{prop:bis2}(iii).}
If $([p_{\varrho_{s}}, q_{\varrho_{s}}], [\ell_{\varrho_{s}}, r_{\varrho_{s}}]) \not \in \Psi_{\source}$, 
then $\mathcal{L}_{\BiSQ}(u_{s}) = |\Psi_{\str}(T[p_{\varrho_{s}}-1..r_{\varrho_{s}}+1])|$ follows from 
the definition of the label function $\mathcal{L}_{\BiSQ}$. 
Similarly, $\Psi^{\BiSQ}_{\varrho_{s}} = \Psi_{\str}(T[p_{\varrho_{s}}-1..r_{\varrho_{s}}+1])$ follows from 
the definition of the subset $\Psi^{\BiSQ}_{\varrho_{s}}$. 
Therefore, $\mathcal{L}_{\BiSQ}(u_{s}) = |\Psi^{\BiSQ}_{\varrho_{s}}|$ holds. 

Otherwise, (i.e., $([p_{\varrho_{s}}, q_{\varrho_{s}}], [\ell_{\varrho_{s}}, r_{\varrho_{s}}]) \in \Psi_{\source}$), 
let $\Psi_{A} = \Psi_{\str}(([p_{\varrho_{s}}, q_{\varrho_{s}}], [\ell_{\varrho_{s}}, r_{\varrho_{s}}]))$ 
and $m = |f_{\recover}(([p_{\varrho_{s}}, q_{\varrho_{s}}], [\ell_{\varrho_{s}}, r_{\varrho_{s}}]))|$ for simplicity. 
Then, $\mathcal{L}_{\BiSQ}(u_{s}) = (1+m)|\Psi_{A}|$ follows from 
the definition of the label function $\mathcal{L}_{\BiSQ}$. 
Similarly, $\Psi^{\BiSQ}_{\varrho_{s}} = \Psi_{A} \cup (\bigcup_{([p, q], [\ell, r]) \in \Psi_{A}} f_{\recover}(([p, q], [\ell, r])))$ follows from the definition of the subset $\Psi^{\BiSQ}_{\varrho_{s}}$. 
In this case, $\mathcal{L}_{\BiSQ}(u_{s}) = |\Psi^{\BiSQ}_{\varrho_{s}}|$ holds 
if the following three statements hold: 
\begin{enumerate}[label=\textbf{(\alph*)}]
    \item $f_{\recover}(([p, q], [\ell, r])) \cap f_{\recover}(([p^{\prime}, q^{\prime}], [\ell^{\prime}, r^{\prime}]) = \emptyset$ 
    for any pair of two distinct interval attractors $([p, q], [\ell, r]), ([p^{\prime}, q^{\prime}], [\ell^{\prime}, r^{\prime}]) \in \Psi_{A}$;
    \item $\Psi_{A} \cap f_{\recover}(([p, q], [\ell, r])) = \emptyset$ for each interval attractor $([p, q], [\ell, r]) \in \Psi_{A}$;
    \item $m = |f_{\recover}(([p, q], [\ell, r]))|$ for each interval attractor $([p, q], [\ell, r]) \in \Psi_{A}$.
\end{enumerate}
Statement (a) follows from Lemma~\ref{lem:recover_division_property}. 
Statement (b) holds because 
(1) $\Psi_{A} \subseteq \Psi_{\source}$, 
(2) $f_{\recover}(([p, q], [\ell, r])) \subseteq \Psi_{\run}$ follows from the definition of the function $f_{\recover}$, 
and (3) $\Psi_{\source} \cap \Psi_{\run} = \emptyset$. 
We prove statement (c). 
$T[p-1..r+1] = T[p_{s}-1..r_{s}+1]$ follows from $([p, q], [\ell, r]) \in \Psi_{A}$. 
Because of $T[p-1..r+1] = T[p_{s}-1..r_{s}+1]$, 
$|f_{\recover}(([p_{s}, q_{s}], [\ell_{s}, r_{s}]))| = |f_{\recover}(([p, q], [\ell, r]))|$ 
follows from Lemma~\ref{lem:recover_super_property}~\ref{enum:recover_super_property:1}. 
Here, $m = |f_{\recover}(([p_{s}, q_{s}], [\ell_{s}, r_{s}]))|$ holds. 
Therefore, statement (c) holds. 

We proved three statements (a), (b), and (c). 
Therefore, $\mathcal{L}_{\BiSQ}(u_{s}) = |\Psi^{\BiSQ}_{\varrho_{s}}|$ holds. 

\textbf{Proof of Proposition~\ref{prop:bis2}(iv).}
One of the following three conditions is satisfied: 
\begin{enumerate}[label=\textbf{(\alph*)}]
    \item $([p_{s}, q_{s}], [\ell_{s}, r_{s}]) \not \in \Psi_{\source}$;
    \item $([p_{s}, q_{s}], [\ell_{s}, r_{s}]) \in \Psi_{\source}$ 
    and $([p, q], [\ell, r]) \in \Psi_{\str}(T[p_{s}-1..r_{s}+1])$;
    \item $([p_{s}, q_{s}], [\ell_{s}, r_{s}]) \in \Psi_{\source}$ 
    and $([p, q], [\ell, r]) \not \in \Psi_{\str}(T[p_{s}-1..r_{s}+1])$.
\end{enumerate}

For condition (a), 
$\Psi^{\BiSQ}_{s} = \Psi_{\str}(T[p_{s}-1..r_{s}+1])$ follows from the definition of the subset $\Psi^{\BiSQ}_{s}$. 
$([p, q], [\ell, r]) \in \Psi_{\str}(T[p_{s}-1..r_{s}+1])$ follows from 
$([p, q], [\ell, r]) \in \Psi^{\BiSQ}_{s}$ and $\Psi^{\BiSQ}_{s} = \Psi_{\str}(T[p_{s}-1..r_{s}+1])$. 
Because of $([p, q], [\ell, r]) \in \Psi_{\str}(T[p_{s}-1..r_{s}+1])$, 
$T[q..q+1] = T[q_{s}..q_{s}+1]$ follows from statement (A). 

For condition (b), 
$T[q..q+1] = T[q_{s}..q_{s}+1]$ follows from statement (A) 
because $([p, q], [\ell, r]) \in \Psi_{\str}(T[p_{s}-1..r_{s}+1])$. 

For condition (c), 
statement (C) shows that 
the subset $\Psi_{\str}(T[p_{s}-1..r_{s}+1])$ contains an interval attractor $([p^{\prime}, q^{\prime}], [\ell^{\prime}, r^{\prime}])$ satisfying $([p, q], [\ell, r]) \in f_{\recover}(([p^{\prime}, q^{\prime}], [\ell^{\prime}, r^{\prime}]))$. 
Because of $([p^{\prime}, q^{\prime}], [\ell^{\prime}$, $r^{\prime}]) \in \Psi_{\str}(T[p_{s}-1..r_{s}+1])$, 
$T[q^{\prime}..q^{\prime}+1] = T[q_{s}..q_{s}+1]$ follows from statement (A). 
Lemma~\ref{lem:psi_str_property}~\ref{enum:psi_str_property:2} shows that 
$([p^{\prime}, q^{\prime}], [\ell^{\prime}, r^{\prime}]) \in \Psi_{0}$ holds 
because (1) $T[p^{\prime}-1..r^{\prime}+1] = T[p_{s}-1..r_{s}+1]$ follows from $([p^{\prime}, q^{\prime}], [\ell^{\prime}, r^{\prime}]) \in \Psi_{\str}(T[p_{s}-1..r_{s}+1])$, 
and (2) $([p_{s}, q_{s}], [\ell_{s}, r_{s}]) \in \Psi_{0}$. 
Because of $([p^{\prime}, q^{\prime}], [\ell^{\prime}, r^{\prime}]) \in \Psi_{0} \cap \Psi_{\source}$, 
$T[q^{\prime}..q^{\prime}+1] = T[q..q+1]$ follows from statement (B). 
Therefore, 
$T[q..q+1] = T[q_{s}..q_{s}+1]$ follows from 
$T[q^{\prime}..q^{\prime}+1] = T[q_{s}..q_{s}+1]$ and $T[q^{\prime}..q^{\prime}+1] = T[q..q+1]$.

%the subset $\Psi_{\str}(T[p_{s}-1..r_{s}+1])$ is a subset of the subset $\Psi_{\source}$, 
%and the subset $\Psi_{\str}(T[p_{s}-1..r_{s}+1])$ contains 
%an interval attractor $([p^{\prime}, q^{\prime}], [\ell^{\prime}, r^{\prime}])$ 
%satisfying $([p, q], [\ell, r]) \in f_{\recover}(([p^{\prime}, q^{\prime}], [\ell^{\prime}, r^{\prime}]))$. 

\textbf{Proof of Proposition~\ref{prop:bis2}(v).}
We prove $\Psi^{\BiSQ}_{s} \cap \Psi^{\BiSQ}_{s^{\prime}} = \emptyset$ by contradiction. 
Here, $\Psi^{\BiSQ}_{s}$, $\Psi^{\BiSQ}_{s^{\prime}} \subseteq \Psi_{0}$ follows from Proposition~\ref{prop:bis2}(ii). 
We assume that $\Psi^{\BiSQ}_{s} \cap \Psi^{\BiSQ}_{s^{\prime}} \neq \emptyset$ holds. 
Then, the $0$-th level interval attractors $\Psi_{0}$ contains an interval attractor $([p, q], [\ell, r])$ 
satisfying $([p, q], [\ell, r]) \in \Psi^{\BiSQ}_{s}$ and $([p, q], [\ell, r]) \in \Psi^{\BiSQ}_{s^{\prime}}$. 
The interval attractor $([p, q], [\ell, r])$ satisfies one of the following four conditions: 

\begin{enumerate}[label=\textbf{(\alph*)}]
    \item $([p, q], [\ell, r]) \in \Psi_{\str}(T[p_{s}-1..r_{s}+1])$ and $([p, q], [\ell, r]) \in \Psi_{\str}(T[p_{s^{\prime}}-1..r_{s^{\prime}}+1])$;
    \item $([p, q], [\ell, r]) \in \Psi_{\str}(T[p_{s}-1..r_{s}+1])$ and $([p, q], [\ell, r]) \not \in \Psi_{\str}(T[p_{s^{\prime}}-1..r_{s^{\prime}}+1])$;
    \item $([p, q], [\ell, r]) \not \in \Psi_{\str}(T[p_{s}-1..r_{s}+1])$ and $([p, q], [\ell, r]) \in \Psi_{\str}(T[p_{s^{\prime}}-1..r_{s^{\prime}}+1])$;
    \item $([p, q], [\ell, r]) \not \in \Psi_{\str}(T[p_{s}-1..r_{s}+1])$ and $([p, q], [\ell, r]) \not \in \Psi_{\str}(T[p_{s^{\prime}}-1..r_{s^{\prime}}+1])$.
\end{enumerate}

For condition (a), 
$T[p-1..r+1] = T[p_{s}-1..r_{s}+1]$ and $T[p-1..r+1] = T[p_{s^{\prime}}-1..r_{s^{\prime}}+1]$ follow from 
$([p, q], [\ell, r]) \in \Psi_{\str}(T[p_{s}-1..r_{s}+1])$ and 
$([p, q], [\ell, r]) \in \Psi_{\str}(T[p_{s^{\prime}}-1..r_{s^{\prime}}+1])$, respectively. 
$T[p_{s}-1..r_{s}+1] = T[p_{s^{\prime}}-1..r_{s^{\prime}}+1]$ follows from 
$T[p-1..r+1] = T[p_{s}-1..r_{s}+1]$ and $T[p-1..r+1] = T[p_{s^{\prime}}-1..r_{s^{\prime}}+1]$. 
On the other hand, 
$T[p_{s}-1..r_{s}+1] \neq T[p_{s^{\prime}}-1..r_{s^{\prime}}+1]$ follows from 
the definition of the sampling subset $\Psi_{\samp}$. 
The two facts $T[p_{s}-1..r_{s}+1] = T[p_{s^{\prime}}-1..r_{s^{\prime}}+1]$ 
and $T[p_{s}-1..r_{s}+1] \neq T[p_{s^{\prime}}-1..r_{s^{\prime}}+1]$ yield a contradiction.

For condition (b), 
$T[p-1..r+1] = T[p_{s}-1..r_{s}+1]$ follows from $([p, q], [\ell, r]) \in \Psi_{\str}(T[p_{s}-1..r_{s}+1])$. 
$([p_{s}, q_{s}], [\ell_{s}, r_{s}]) \not \in \Psi_{\run}$ follows from Lemma~\ref{lem:samp_basic_property}~\ref{enum:samp_basic_property:3}. 
Lemma~\ref{lem:psi_equality_basic_property}~\ref{enum:psi_equality_basic_property:4} shows that 
$([p, q], [\ell, r]) \not \in \Psi_{\run}$ holds 
because $T[p-1..r+1] = T[p_{s}-1..r_{s}+1]$ and $([p_{s}, q_{s}], [\ell_{s}, r_{s}]) \not \in \Psi_{\run}$. 
On the other hand, statement (C) shows that $([p, q], [\ell, r]) \in \Psi_{\run}$ holds 
because $([p, q], [\ell, r]) \not \in \Psi_{\str}(T[p_{s^{\prime}}-1..r_{s^{\prime}}+1])$. 
The two facts $([p, q], [\ell, r]) \in \Psi_{\run}$ and $([p, q], [\ell, r]) \not \in \Psi_{\run}$ yield a contradiction. 

For condition (c), 
we can show that there exists a contradiction using the same approach as for condition (b). 

For condition (d), 
statement (C) shows that 
the subset $\Psi_{\str}(T[p_{s}-1..r_{s}+1])$ contains an interval attractor $([p_{A}, q_{A}], [\ell_{A}, r_{A}])$ satisfying $([p, q], [\ell, r]) \in f_{\recover}(([p_{A}, q_{A}], [\ell_{A}, r_{A}]))$ 
because $([p, q], [\ell, r]) \not \in \Psi_{\str}(T[p_{s}-1..r_{s}+1])$. 
Similarly, 
statement (C) shows that 
the subset $\Psi_{\str}(T[p_{s^{\prime}}-1..r_{s^{\prime}}+1])$ contains an interval attractor $([p_{B}, q_{B}]$, $[\ell_{B}, r_{B}])$ satisfying $([p, q], [\ell, r]) \in f_{\recover}(([p_{B}, q_{B}]$, $[\ell_{B}, r_{B}]))$. 
$T[p_{A}-1..r_{A}+1] = T[p_{s}-1..r_{s}+1]$~(respectively, $T[p_{B}-1..r_{B}+1] = T[p_{s^{\prime}}-1..r_{s^{\prime}}+1]$) follows from $([p_{A}, q_{A}], [\ell_{A}, r_{A}]) \in \Psi_{\str}(T[p_{s}-1..r_{s}+1])$ (respectively, $([p_{B}, q_{B}], [\ell_{B}, r_{B}]) \in \Psi_{\str}(T[p_{s^{\prime}}-1..r_{s^{\prime}}+1])$). 

If $([p_{A}, q_{A}], [\ell_{A}, r_{A}]) = ([p_{B}, q_{B}], [\ell_{B}, r_{B}])$ holds, 
then $T[p_{A}-1..r_{A}+1] = T[p_{B}-1..r_{B}+1]$. 
$T[p_{s}-1..r_{s}+1] = T[p_{s^{\prime}}-1..r_{s^{\prime}}+1]$ follows from 
$T[p_{A}-1..r_{A}+1] = T[p_{s}-1..r_{s}+1]$, $T[p_{B}-1..r_{B}+1] = T[p_{s^{\prime}}-1..r_{s^{\prime}}+1]$, 
and $T[p_{A}-1..r_{A}+1] = T[p_{B}-1..r_{B}+1]$. 
On the other hand, 
$T[p_{s}-1..r_{s}+1] \neq T[p_{s^{\prime}}-1..r_{s^{\prime}}+1]$ follows from 
the definition of the sampling subset $\Psi_{\samp}$. 
The two facts $T[p_{s}-1..r_{s}+1] = T[p_{s^{\prime}}-1..r_{s^{\prime}}+1]$ 
and $T[p_{s}-1..r_{s}+1] \neq T[p_{s^{\prime}}-1..r_{s^{\prime}}+1]$ yield a contradiction.

Otherwise (i.e., $([p_{A}, q_{A}], [\ell_{A}, r_{A}]) \neq ([p_{B}, q_{B}], [\ell_{B}, r_{B}])$), 
$f_{\recover}(([p_{A}, q_{A}], [\ell_{A}, r_{A}])) \cap f_{\recover}(([p_{B}$, $q_{B}], [\ell_{B}, r_{B}])) \neq \emptyset$ 
follows from $([p, q], [\ell, r]) \in f_{\recover}(([p_{A}, q_{A}], [\ell_{A}, r_{A}]))$ and $([p, q], [\ell, r]) \in f_{\recover}(([p_{B}$, $q_{B}], [\ell_{B}, r_{B}]))$. 
On the other hand, $f_{\recover}(([p_{A}, q_{A}], [\ell_{A}, r_{A}])) \cap f_{\recover}(([p_{B}, q_{B}], [\ell_{B}, r_{B}])) = \emptyset$ follows from Lemma~\ref{lem:recover_division_property}~\ref{enum:recover_division_property:2}. 
The two facts $f_{\recover}(([p_{A}, q_{A}], [\ell_{A}, r_{A}])) \cap f_{\recover}(([p_{B}, q_{B}], [\ell_{B}, r_{B}])) \neq \emptyset$ and $f_{\recover}(([p_{A}, q_{A}], [\ell_{A}, r_{A}])) \cap f_{\recover}(([p_{B}, q_{B}], [\ell_{B}, r_{B}])) = \emptyset$ yield a contradiction. 

We showed that there exists a contradiction for each of four conditions (a), (b), (c), and (d). 
Therefore, $\Psi^{\BiSQ}_{s} \cap \Psi^{\BiSQ}_{s^{\prime}} = \emptyset$ must hold. 

\textbf{Proof of Proposition~\ref{prop:bis2}(vi).}
We prove $|\Psi^{\BiSQ}_{s}| = |\{ q \mid ([p, q], [\ell, r]) \in \Psi^{\BiSQ}_{s} \}|$ by contradiction. 
We assume that $|\Psi^{\BiSQ}_{s}| \neq |\{ q \mid ([p, q], [\ell, r]) \in \Psi^{\BiSQ}_{s} \}|$ holds. 
Then, the subset $\Psi^{\BiSQ}_{s}$ contains two distinct interval attractors 
$([p, q], [\ell, r])$ and $([p^{\prime}, q^{\prime}], [\ell^{\prime}, r^{\prime}])$ satisfying $q = q^{\prime}$. 
Here, $([p, q], [\ell, r]), ([p^{\prime}, q^{\prime}], [\ell^{\prime}, r^{\prime}]) \in \Psi_{0}$ follows from 
Proposition~\ref{prop:bis2}(ii). 
Proposition~\ref{prop:bis1}(iii) shows that $q \neq q^{\prime}$ holds. 
The two facts $q = q^{\prime}$ and $q \neq q^{\prime}$ yield a contradiction. 
Therefore, $|\Psi^{\BiSQ}_{s}| = |\{ q \mid ([p, q], [\ell, r]) \in \Psi^{\BiSQ}_{s} \}|$ must hold. 
\end{proof}

The following proposition states properties of 
set $\bigcup_{s = 1}^{k} \{ q \mid ([p, q], [\ell, r]) \in \Psi^{\BiSQ}_{s} \}$ of integers. 

\begin{proposition}\label{prop:bis3}
Consider the $k$ subsets $\Psi^{\BiSQ}_{1}$, $\Psi^{\BiSQ}_{2}$, $\ldots$, $\Psi^{\BiSQ}_{k}$ of set $\Psi_{\RR}$. 
Then, the following two statements hold: 
\begin{enumerate}[label=\textbf{(\roman*)}]
    \item $\{ q \mid ([p, q], [\ell, r]) \in \Psi^{\BiSQ}_{s} \} \cap \{ q \mid ([p, q], [\ell, r]) \in \Psi^{\BiSQ}_{s^{\prime}} \} = \emptyset$ for any pair of two integers $1 \leq s < s^{\prime} \leq k$;
    \item $\bigcup_{s = 1}^{k} \{ q \mid ([p, q], [\ell, r]) \in \Psi^{\BiSQ}_{s} \} = [1, n]$.
\end{enumerate}
\end{proposition}
\begin{proof}
The proof of Proposition~\ref{prop:bis3} is as follows. 

\textbf{Proof of Proposition~\ref{prop:bis3}(i).}
We prove $\{ q \mid ([p, q], [\ell, r]) \in \Psi^{\BiSQ}_{s} \} \cap \{ q \mid ([p, q], [\ell, r]) \in \Psi^{\BiSQ}_{s^{\prime}} \} = \emptyset$ by contradiction. 
Here, $\Psi^{\BiSQ}_{s}, \Psi^{\BiSQ}_{s^{\prime}} \subseteq \Psi_{0}$ follows from Proposition~\ref{prop:bis2}(ii). 
We assume that $\{ q \mid ([p, q], [\ell, r]) \in \Psi^{\BiSQ}_{s} \} \cap \{ q \mid ([p, q], [\ell, r]) \in \Psi^{\BiSQ}_{s^{\prime}} \} \neq \emptyset$ holds. 
Then, the $0$-th level interval attractors $\Psi_{0}$ contains 
two interval attractors $([p, q], [\ell, r]), ([p^{\prime}, q^{\prime}], [\ell^{\prime}, r^{\prime}])$ satisfying the following three conditions: 
(A) $q = q^{\prime}$, (B) $([p, q], [\ell, r]) \in \Psi^{\BiSQ}_{s}$, and (C) $([p^{\prime}, q^{\prime}], [\ell^{\prime}, r^{\prime}]) \in \Psi^{\BiSQ}_{s}$. 
$([p, q], [\ell, r]) = ([p^{\prime}, q^{\prime}], [\ell^{\prime}, r^{\prime}])$ follows from Lemma~\ref{prop:bis1}(iii) 
because $q = q^{\prime}$ and $([p, q], [\ell, r]), ([p^{\prime}, q^{\prime}], [\ell^{\prime}, r^{\prime}]) \in \Psi_{0}$. 
$\Psi^{\BiSQ}_{s} \cap \Psi^{\BiSQ}_{s^{\prime}} \neq \emptyset$ follows from 
$([p, q], [\ell, r]) = ([p^{\prime}, q^{\prime}], [\ell^{\prime}, r^{\prime}])$, 
$([p, q], [\ell, r]) \in \Psi^{\BiSQ}_{s}$, 
and $([p^{\prime}, q^{\prime}], [\ell^{\prime}, r^{\prime}]) \in \Psi^{\BiSQ}_{s}$. 
On the other hand, $\Psi^{\BiSQ}_{s} \cap \Psi^{\BiSQ}_{s^{\prime}} = \emptyset$ follows from Proposition~\ref{prop:bis2}(v). 
The two facts $\Psi^{\BiSQ}_{s} \cap \Psi^{\BiSQ}_{s^{\prime}} \neq \emptyset$ and $\Psi^{\BiSQ}_{s} \cap \Psi^{\BiSQ}_{s^{\prime}} = \emptyset$ yield a contradiction. 
Therefore, $\{ q \mid ([p, q], [\ell, r]) \in \Psi^{\BiSQ}_{s} \} \cap \{ q \mid ([p, q], [\ell, r]) \in \Psi^{\BiSQ}_{s^{\prime}} \} = \emptyset$ must hold. 

\textbf{Proof of Proposition~\ref{prop:bis3}(ii).}
We prove $\bigcup_{s = 1}^{k} \{ q \mid ([p, q], [\ell, r]) \in \Psi^{\BiSQ}_{s} \} \subseteq [1, n]$. 
For each interval attractor $([p, q], [\ell, r]) \in \bigcup_{s = 1}^{k} \Psi^{\BiSQ}_{s}$, 
$([p, q], [\ell, r]) \in \Psi_{0}$ holds because 
$\bigcup_{s = 1}^{k} \Psi^{\BiSQ}_{s} \subseteq \Psi_{0}$ follows from Proposition~\ref{prop:bis2}(ii). 
Because of $([p, q], [\ell, r]) \in \Psi_{0}$, 
$q \in [1, n-1]$ follows from Proposition~\ref{prop:bis1}(v). 
Therefore, $\bigcup_{s = 1}^{k} \{ q \mid ([p, q], [\ell, r]) \in \Psi^{\BiSQ}_{s} \} \subseteq [1, n]$ holds. 

On the other hand, 
we prove $\bigcup_{s = 1}^{k} \{ q \mid ([p, q], [\ell, r]) \in \Psi^{\BiSQ}_{s} \} \supseteq [1, n]$. 
For each integer $x \in [1, n-1]$, 
Proposition~\ref{prop:bis1}(v) shows that 
the $0$-th level interval attractors $\Psi_{0}$ contains an interval attractor $([p, q], [\ell, r])$ satisfying $q = x$. 
Proposition~\ref{prop:bis2}(i) shows that 
there exists an integer $s \in [1, k]$ satisfying 
$([p, q], [\ell, r]) \in \Psi^{\BiSQ}_{s}$. 
The existence of the integer $s$ indicates that $x \in \bigcup_{s = 1}^{k} \{ q \mid ([p, q], [\ell, r]) \in \Psi^{\BiSQ}_{s} \}$ holds. 
Therefore, $\bigcup_{s = 1}^{k} \{ q \mid ([p, q], [\ell, r]) \in \Psi^{\BiSQ}_{s} \} \supseteq [1, n]$ holds. 

Finally, Proposition~\ref{prop:bis3}(ii) follows from 
$\bigcup_{s = 1}^{k} \{ q \mid ([p, q], [\ell, r]) \in \Psi^{\BiSQ}_{s} \} \subseteq [1, n]$ 
and $\bigcup_{s = 1}^{k} \{ q \mid ([p, q], [\ell, r]) \in \Psi^{\BiSQ}_{s} \} \supseteq [1, n]$.
\end{proof}

Next, 
we consider a position $x$ of input string $T$ satisfying $x \leq n-1$. 
From Proposition~\ref{prop:bis3}, 
there exists the only one integer $s \in [1, k]$ satisfying 
$x \in \{ q \mid ([p, q], [\ell, r]) \in \Psi^{\BiSQ}_{s} \}$. 
For the given bigram search query $\BiSQ(P)$, 
the following proposition states the relationship among 
three strings $T[x..n]$, $T[q_{s}..q_{s}+1]$, and $P$. 

\begin{proposition}\label{prop:bis4}
    Consider (A) a position $x$ of input string $T$ satisfying $x \leq n-1$ (i.e., $x \in [1, n-1]$), 
    (B) the subset $\Psi_{0} \cap \Psi_{\samp} = \{ ([p_{1}, q_{1}], [\ell_{1}, r_{1}]), ([p_{2}, q_{2}], [\ell_{2}, r_{2}]), \ldots, ([p_{k}, q_{k}], [\ell_{k}, r_{k}]) \}$, 
    and (C) the $k$ subsets $\Psi^{\BiSQ}_{1}$, $\Psi^{\BiSQ}_{2}$, $\ldots$, $\Psi^{\BiSQ}_{k}$ of set $\Psi_{\RR}$.     
    From Proposition~\ref{prop:bis3}, 
    there exists the only one integer $s \in [1, k]$ satisfying 
    $x \in \{ q \mid ([p, q], [\ell, r]) \in \Psi^{\BiSQ}_{s} \}$. 
    The following two statements hold for a string $P \in \Sigma^{+}$ of length $2$:
\begin{enumerate}[label=\textbf{(\roman*)}]
    \item $T[x..n] \prec P \Leftrightarrow T[q_{s}..q_{s}+1] \prec P$;
    \item $x \in \Occ(T, P) \Leftrightarrow T[q_{s}..q_{s}+1] = P$.    
\end{enumerate}
\end{proposition}
\begin{proof}
We prove $T[x..x+1] = T[q_{s}..q_{s}+1]$. 
Because of $x \in \{ q \mid ([p, q], [\ell, r]) \in \Psi^{\BiSQ}_{s} \}$, 
the set $\Psi^{\BiSQ}_{s}$ contains an interval attractor $([p^{\prime}, q^{\prime}], [\ell^{\prime}, r^{\prime}])$ 
satisfying $q^{\prime} = x$. 
Lemma~\ref{prop:bis2}(iv) shows that $T[q_{s}..q_{s}+1] = T[q^{\prime}..q^{\prime}+1]$ holds. 
Therefore, $T[x..x+1] = T[q_{s}..q_{s}+1]$ follows from 
$T[q_{s}..q_{s}+1] = T[q^{\prime}..q^{\prime}+1]$ and $q^{\prime} = x$. 

Proposition~\ref{prop:bis4} holds if the following four statements hold: 
(A) $T[x..n] \prec P \Rightarrow T[q_{s}..q_{s}+1] \prec P$; 
(B) $T[x..n] \prec P \Leftarrow T[q_{s}..q_{s}+1] \prec P$; 
(C) $x \in \Occ(T, P) \Rightarrow T[q_{s}..q_{s}+1] = P$;
(D) $x \in \Occ(T, P) \Leftarrow T[q_{s}..q_{s}+1] = P$. 

\textbf{Proof of $T[x..n] \prec P \Rightarrow T[q_{s}..q_{s}+1] \prec P$.}
$T[x..x+1] \prec P$ holds because 
$T[x..x+1] \preceq T[x..n]$ and $T[x..n] \prec P$. 
$T[q_{s}..q_{s}+1] \prec P$ follows from 
$T[x..x+1] = T[q_{s}..q_{s}+1]$ and $T[x..x+1] \prec P$. 

\textbf{Proof of $T[x..n] \prec P \Leftarrow T[q_{s}..q_{s}+1] \prec P$.}
$T[x..x+1] \prec P$ follows from 
$T[x..x+1] = T[q_{s}..q_{s}+1]$ and $T[q_{s}..q_{s}+1] \prec P$. 
$T[x..n] \prec P$ holds 
because (1) $T[x..n]$ is a prefix of string $T[x..x+1]$, 
(2) $T[x..x+1] \prec P$, 
and (3) $T[x..x+1]$ and $P$ have the same length. 

\textbf{Proof of $x \in \Occ(T, P) \Rightarrow T[q_{s}..q_{s}+1] = P$.}
$T[x..x+1] = P$ follows from $x \in \Occ(T, P)$. 
$T[q_{s}..q_{s}+1] = P$ follows from 
$T[x..x+1] = T[q_{s}..q_{s}+1]$ and $T[x..x+1] = P$. 

\textbf{Proof of $x \in \Occ(T, P) \Leftarrow T[q_{s}..q_{s}+1] = P$.}
$T[x..x+1] = P$ follows from 
$T[x..x+1] = T[q_{s}..q_{s}+1]$ and $T[q_{s}..q_{s}+1] = P$. 
$x \in \Occ(T, P)$ follows from 
$T[x..x+1] = P$ and $[x..x+|P|-1] \subseteq [1, n]$. 
\end{proof}

We prove Lemma~\ref{lem:bis_property} using Proposition~\ref{prop:bis2}, Proposition~\ref{prop:bis3}, and Proposition~\ref{prop:bis4}.

\begin{proof}[Proof of Lemma~\ref{lem:bis_property}]
For this proof, 
we introduce two sets $\mathcal{J}, \mathcal{J}^{\prime}$ of integers in set $\{ 1, 2, \ldots, k \}$. 
The former set $\mathcal{J}$ consists of integers such that 
each integer $s \in \mathcal{J}$ satisfies $T[q_{s}..q_{s}+1] \prec P$ 
(i.e., $\mathcal{J} = \{ s \in [1, k] \mid T[q_{s}..q_{s}+1] \prec P \}$). 
The latter set $\mathcal{J}$ consists of integers such that 
each integer $s \in \mathcal{J}$ satisfies $T[q_{s}..q_{s}+1] = P$ 
(i.e., $\mathcal{J}^{\prime} = \{ s \in [1, k] \mid T[q_{s}..q_{s}+1] = P \}$). 
Here, $\mathcal{J} = \{ \varrho_{w} \mid w \in [1, \tau] \}$ and $\mathcal{J}^{\prime} = \{ \varrho_{w} \mid w \in [\tau+1, \tau^{\prime}] \}$ 
hold because 
$T[q_{\varrho_{1}}..q_{\varrho_{1}}+1] \preceq T[q_{\varrho_{2}}..q_{\varrho_{2}}+1] \preceq \cdots \preceq T[q_{\varrho_{k}}..q_{\varrho_{k}}+1]$ 
follows from the definition of the permutation $\varrho_{1}, \varrho_{2}, \ldots, \varrho_{k}$. 

The following four statements are used to prove Lemma~\ref{lem:bis_property}. 
\begin{enumerate}[label=\textbf{(\Alph*)}]
    \item $\{ x \in [1, n-1] \mid T[x..n] \prec P \} \subseteq \bigcup_{s \in \mathcal{J}} \{ q \mid ([p, q], [\ell, r]) \in \Psi^{\BiSQ}_{s} \}$;
    \item $\{ x \in [1, n-1] \mid T[x..n] \prec P \} \supseteq \bigcup_{s \in \mathcal{J}} \{ q \mid ([p, q], [\ell, r]) \in \Psi^{\BiSQ}_{s} \}$;
    \item $\Occ(T, P) \subseteq \bigcup_{s \in \mathcal{J}^{\prime}} \{ q \mid ([p, q], [\ell, r]) \in \Psi^{\BiSQ}_{s} \}$;
    \item $\Occ(T, P) \supseteq \bigcup_{s \in \mathcal{J}^{\prime}} \{ q \mid ([p, q], [\ell, r]) \in \Psi^{\BiSQ}_{s} \}$.
\end{enumerate}

\textbf{Proof of statement (A).}
Consider an integer $x^{\prime} \in \{ x \in [1, n-1] \mid T[x..n] \prec P \}$. 
From Proposition~\ref{prop:bis3}, 
there exists the only one integer $s \in [1, k]$ satisfying 
$x^{\prime} \in \{ q \mid ([p, q], [\ell, r]) \in \Psi^{\BiSQ}_{s} \}$. 
Because of $T[x^{\prime}..n] \prec P$, 
Proposition~\ref{prop:bis4}(i) shows that 
$T[q_{s}..q_{s}+1] \prec P$ holds. 
$s \in \mathcal{J}$ follows from $T[q_{s}..q_{s}+1] \prec P$. 
Therefore, $\{ x \in [1, n-1] \mid T[x..n] \prec P \} \subseteq \bigcup_{s \in \mathcal{J}} \{ q \mid ([p, q], [\ell, r]) \in \Psi^{\BiSQ}_{s} \}$ holds. 

\textbf{Proof of statement (B).}
Consider an integer $x^{\prime} \in \{ q \mid ([p, q], [\ell, r]) \in \Psi^{\BiSQ}_{s} \}$ 
for an integer $s \in \mathcal{J}$. 
Then $x^{\prime} \in [1, n-1]$ follows from Proposition~\ref{prop:bis3}(ii). 
Because of $T[q_{s}..q_{s}+1] \prec P$, 
Proposition~\ref{prop:bis4}(i) shows that 
$T[x^{\prime}..n] \prec P$ holds. 
$x^{\prime} \in \{ x \in [1, n-1] \mid T[x..n] \prec P \}$ 
follows from  $x^{\prime} \in [1, n-1]$ and $T[x^{\prime}..n] \prec P$. 
Therefore, $\{ x \in [1, n-1] \mid T[x..n] \prec P \} \supseteq \bigcup_{s \in \mathcal{J}} \{ q \mid ([p, q], [\ell, r]) \in \Psi^{\BiSQ}_{s} \}$ holds. 

\textbf{Proof of statement (C).}
Consider an integer $x \in \Occ(T, P)$. 
Then, $T[x..x+1] = P$ and $x \in [1, n-1]$ hold. 
From Proposition~\ref{prop:bis3}, 
there exists the only one integer $s \in [1, k]$ satisfying 
$x \in \{ q \mid ([p, q], [\ell, r]) \in \Psi^{\BiSQ}_{s} \}$. 
Because of $x \in \Occ(T, P)$, 
Proposition~\ref{prop:bis4}(ii) shows that 
$T[q_{s}..q_{s}+1] = P$ holds. 
$s \in \mathcal{J}^{\prime}$ follows from $T[q_{s}..q_{s}+1] = P$. 
Therefore, 
$\Occ(T, P) \subseteq \bigcup_{s \in \mathcal{J}^{\prime}} \{ q \mid ([p, q], [\ell, r]) \in \Psi^{\BiSQ}_{s} \}$ holds. 

\textbf{Proof of statement (D).}
Consider an integer $x \in \{ q \mid ([p, q], [\ell, r]) \in \Psi^{\BiSQ}_{s} \}$ 
for an integer $s \in \mathcal{J}^{\prime}$. 
Then $x \in [1, n-1]$ follows from Proposition~\ref{prop:bis3}(ii). 
Because of $T[q_{s}..q_{s}+1] = P$, 
Proposition~\ref{prop:bis4}(ii) shows that 
$x \in \Occ(T, P)$ holds. 
Therefore, $\Occ(T, P) \supseteq \bigcup_{s \in \mathcal{J}^{\prime}} \{ q \mid ([p, q], [\ell, r]) \in \Psi^{\BiSQ}_{s} \}$ holds. 

\textbf{Proof of Lemma~\ref{lem:bis_property}(i).}
We prove $|\{ x \in [1, n-1] \mid T[x..n] \prec P \}| = \sum_{s = 1}^{\tau} \mathcal{L}_{\BiSQ}(u_{s})$. 
$|\{ x \in [1, n-1] \mid T[x..n] \prec P \}| = |\bigcup_{s \in \mathcal{J}} \{ q \mid ([p, q], [\ell, r]) \in \Psi^{\BiSQ}_{s} \}|$ holds because 
$\{ x \in [1, n-1] \mid T[x..n] \prec P \} = \bigcup_{s \in \mathcal{J}} \{ q \mid ([p, q], [\ell, r]) \in \Psi^{\BiSQ}_{s} \}$ 
follows from statement (A) and statement (B). 
$|\bigcup_{s \in \mathcal{J}} \{ q \mid ([p, q], [\ell, r]) \in \Psi^{\BiSQ}_{s} \}| = \sum_{s \in \mathcal{J}} |\{ q \mid ([p, q], [\ell, r]) \in \Psi^{\BiSQ}_{s} \}|$ holds 
because $\{ q \mid ([p, q], [\ell, r]) \in \Psi^{\BiSQ}_{s} \} \cap \{ q \mid ([p, q], [\ell, r]) \in \Psi^{\BiSQ}_{s^{\prime}} \} = \emptyset$ follows from Proposition~\ref{prop:bis3}(i) for any pair of two integers $s, s^{\prime} \in \mathcal{J}$. 
For each integer $s^{\prime} \in \mathcal{J}$, 
$|\{ q \mid ([p, q], [\ell, r]) \in \Psi^{\BiSQ}_{s} \}| = |\Psi^{\BiSQ}_{s}|$ 
follows from Proposition~\ref{prop:bis2}(vi). 
$\sum_{s \in \mathcal{J}} |\Psi^{\BiSQ}_{s}| = \sum_{w = 1}^{\tau} |\mathcal{L}_{\BiSQ}(u_{w})|$ holds 
because (a) $|\mathcal{L}_{\BiSQ}(u_{w})| = |\Psi^{\BiSQ}_{\varrho_{w}}|$ follows from Proposition~\ref{prop:bis2}(iii) 
for each integer $w \in [1, \tau]$, 
and (b) $\mathcal{J} = \{ \varrho_{w} \mid w \in [1, \tau] \}$. 
Therefore, $|\{ x \in [1, n-1] \mid T[x..n] \prec P \}| = \sum_{w = 1}^{\tau} \mathcal{L}_{\BiSQ}(u_{w})$ follows from the following equation: 

\begin{equation*}
\begin{split}
|\{ x \in [1, n-1] \mid T[x..n] \prec P \}| &= |\bigcup_{s \in \mathcal{J}} \{ q \mid ([p, q], [\ell, r]) \in \Psi^{\BiSQ}_{s} \}| \\
&= \sum_{s \in \mathcal{J}} |\{ q \mid ([p, q], [\ell, r]) \in \Psi^{\BiSQ}_{s} \}| \\
&= \sum_{s \in \mathcal{J}} |\Psi^{\BiSQ}_{s}| \\
&= \sum_{w = 1}^{\tau} |\mathcal{L}_{\BiSQ}(u_{w})|.
\end{split}
\end{equation*}

We prove $d = \lambda + \sum_{w = 1}^{\tau} \mathcal{L}_{\BiSQ}(u_{w})$. 
$d = \lambda + |\{ x \in [1, n-1] \mid T[x..n] \prec P \}|$ follows from the definition of bigram search query. 
We already proved $|\{ x \in [1, n-1] \mid T[x..n] \prec P \}| = \sum_{w = 1}^{\tau} \mathcal{L}_{\BiSQ}(u_{w})$. 
Therefore, $d = \lambda + \sum_{w = 1}^{\tau} \mathcal{L}_{\BiSQ}(u_{w})$ holds. 

\textbf{Proof of Lemma~\ref{lem:bis_property}(ii).}
We prove $g = (\sum_{w = 1}^{\tau^{\prime}} \mathcal{L}_{\BiSQ}(u_{w})) - (\sum_{w = 1}^{\tau} \mathcal{L}_{\BiSQ}(u_{w}))$. 
$g = |\Occ(T, P)|$ follows from the definition of bigram search query. 
$|\Occ(T, P)| = |\bigcup_{s \in \mathcal{J}^{\prime}} \{ q \mid ([p, q]$, $[\ell, r]) \in \Psi^{\BiSQ}_{s} \}|$ follows from statement (C) and statement (D). 
For each integer $s^{\prime} \in \mathcal{J}^{\prime}$, 
$|\{ q \mid ([p, q], [\ell, r]) \in \Psi^{\BiSQ}_{s} \}| = |\Psi^{\BiSQ}_{s}|$ 
follows from Proposition~\ref{prop:bis2}(vi). 
$\sum_{s \in \mathcal{J}^{\prime}} |\Psi^{\BiSQ}_{s}| = \sum_{w = \tau+1}^{\tau^{\prime}} |\mathcal{L}_{\BiSQ}(u_{w})|$ holds 
because (a) $|\mathcal{L}_{\BiSQ}(u_{w})| = |\Psi^{\BiSQ}_{\varrho_{w}}|$ follows from Proposition~\ref{prop:bis2}(iii) 
for each integer $w \in [\tau+1, \tau^{\prime}]$, 
and (b) $\mathcal{J}^{\prime} = \{ \varrho_{w} \mid w \in [\tau+1, \tau^{\prime}] \}$. 
Therefore, 
$g = (\sum_{w = 1}^{\tau^{\prime}} \mathcal{L}_{\BiSQ}(u_{w})) - (\sum_{w = 1}^{\tau} \mathcal{L}_{\BiSQ}(u_{w}))$ 
follows from the following equation: 
\begin{equation*}
\begin{split}
g &= |\Occ(T, P)| \\
&= |\bigcup_{s \in \mathcal{J}^{\prime}} \{ q \mid ([p, q], [\ell, r]) \in \Psi^{\BiSQ}_{s} \}| \\
&= \sum_{s \in \mathcal{J}^{\prime}} |\Psi^{\BiSQ}_{s}|  \\
&= \sum_{w = \tau+1}^{\tau^{\prime}} |\mathcal{L}_{\BiSQ}(u_{w})| \\ 
&= (\sum_{w = 1}^{\tau^{\prime}} \mathcal{L}_{\BiSQ}(u_{w})) - (\sum_{w = 1}^{\tau} \mathcal{L}_{\BiSQ}(u_{w})).
\end{split}
\end{equation*}

\textbf{Proof of Lemma~\ref{lem:bis_property}(iii).}
We prove $q_{s} \in \{ q \mid ([p, q], [\ell, r]) \in \Psi^{\BiSQ}_{s} \}$ for each integer $s \in [1, k]$.
$\Psi_{\str}(T[p_{s}-1..r_{s}+1]) \subseteq \Psi^{\BiSQ}_{s}$ follows from the definition of the subset $\Psi^{\BiSQ}_{s}$. 
$([p_{s}, q_{s}], [\ell_{s}, r_{s}]) \in \Psi_{\str}(T[p_{s}-1..r_{s}+1])$ follows from the definition of the subset $\Psi_{\str}$. 
$([p_{s}, q_{s}], [\ell_{s}, r_{s}]) \in \Psi^{\BiSQ}_{s}$ follows from 
$([p_{s}, q_{s}], [\ell_{s}, r_{s}]) \in \Psi_{\str}(T[p_{s}-1..r_{s}+1])$ and $\Psi_{\str}(T[p_{s}-1..r_{s}+1]) \subseteq \Psi^{\BiSQ}_{s}$. 
Therefore, $q_{s} \in \{ q \mid ([p, q], [\ell, r]) \in \Psi^{\BiSQ}_{s} \}$ holds.

$\Occ(T, P) = \bigcup_{s \in \mathcal{J}^{\prime}} \{ q \mid ([p, q], [\ell, r]) \in \Psi^{\BiSQ}_{s} \}$ follows from statement (C) and statement (D). 
If $\tau < \tau^{\prime}$, 
then $\varrho_{\tau+1} \in \mathcal{J}^{\prime}$ 
follows from $\mathcal{J}^{\prime} = \{ \varrho_{w} \mid w \in [\tau+1, \tau^{\prime}] \}$. 
We already proved $q_{\varrho_{\tau+1}} \in \{ q \mid ([p, q], [\ell, r]) \in \Psi^{\BiSQ}_{\varrho_{\tau+1}} \}$ holds. 
Therefore, $q_{\varrho_{\tau+1}} \in \Occ(T, P)$ follows from the following equation: 
\begin{equation*}
\begin{split}
q_{\varrho_{\tau+1}} &\in \{ q \mid ([p, q], [\ell, r]) \in \Psi^{\BiSQ}_{\varrho_{\tau+1}} \} \\
&\subseteq \bigcup_{s \in \mathcal{J}^{\prime}} \{ q \mid ([p, q], [\ell, r]) \in \Psi^{\BiSQ}_{s} \} \\
&= \Occ(T, P).
\end{split}
\end{equation*}

Otherwise (i.e., $\tau \geq \tau^{\prime}$), 
$\mathcal{J}^{\prime} = \emptyset$ holds 
because $\mathcal{J}^{\prime} = \{ \varrho_{w} \mid w \in [\tau+1, \tau^{\prime}] \}$ and 
$\{ \varrho_{w} \mid w \in [\tau+1, \tau^{\prime}] \} = \emptyset$. 
$\Occ(T, P) = \emptyset$ follows from 
$\Occ(T, P) = \bigcup_{s \in \mathcal{J}^{\prime}} \{ q \mid ([p, q], [\ell, r]) \in \Psi^{\BiSQ}_{s} \}$ 
and $\mathcal{J}^{\prime} = \emptyset$. 
\end{proof}

\subsubsection{Dynamic Data Structures for Bigram Search Query}\label{subsubsec:bis_ds}
The sequence $\mathbf{Q}_{\BiSQ} = u_{1}, u_{2}, \ldots, u_{k}$ is represented using a doubly linked list $\mathbf{L}_{\BiSQ}$ of $k$ elements 
such that each $s$-th element corresponds to the $s$-th node $u_{s}$. 
The $s$-th element of the doubly linked list $\mathbf{L}_{\BiSQ}$ stores the following two pieces of information on its corresponding node $u_{s}$: 
\begin{itemize}
    \item the integer obtained from $\mathcal{L}_{\BiSQ}(u_{s})$; 
    \item consider the doubly linked list $\mathbf{L}_{\samp, 1}$ representing the sequence $\mathbf{Q}_{\samp}$ of nodes, which is introduced in Section~\ref{subsubsec:sample_ds}. 
    The doubly linked list $\mathbf{L}_{\samp, 1}$ has an element $e$ corresponding to 
    the interval attractor $([p_{\varrho_{s}}, q_{\varrho_{s}}], [\ell_{\varrho_{s}}, r_{\varrho_{s}}])$. 
    The $s$-th element of the doubly linked list $\mathbf{L}_{\BiSQ}$ stores a pointer to the element $e$. 
\end{itemize}

List indexing and partial sum data structures are built on the doubly linked list $\mathbf{L}_{\BiSQ}$. 
Here, the list indexing and partial sum data structures are introduced in Section~\ref{subsubsec:sample_ds}. 
Using the partial sum data structure, 
we can compute the sum $\sum_{w = 1}^{i} \mathcal{L}_{\BiSQ}(u_{w})$ in $O(\log k)$ time for a given integer $i \in [1, k]$. 
The total size of these data structures is $O(k B)$ bits of space for machine word size $B$. 
Here, $k \leq |\Psi_{\samp}|$ holds because 
$k = |\Psi_{0} \cap \Psi_{\samp}|$ and 
$|\Psi_{0} \cap \Psi_{\samp}| \leq |\Psi_{\samp}|$. 

\subsubsection{Answering Bigram Search Query}\label{subsubsec:bis_query_algo}
We explain the algorithm for a given bigram search query $\BiSQ(P)$. 
This algorithm leverages Lemma~\ref{lem:bis_property}, which states 
the relationship the relationship between the bigram search query and label function $\mathcal{L}_{\BiSQ}$. 
For answering the given bigram search query, 
we use the following three data structures: 
\begin{itemize}
    \item the data structure detailed in Section~\ref{subsubsec:rrdag_ds}, which is associated with the RR-DAG of the RLSLP $\mathcal{G}^{R}$;
    \item the data structure detailed in Section~\ref{subsubsec:sample_ds} for sample query;
    \item the data structure detailed in Section~\ref{subsubsec:bis_ds} for bigram search query.
\end{itemize}

The algorithm for answering bigram search query consists of four phases. 
In the first phase, 
the algorithm computes the integer $\tau$ of Lemma~\ref{lem:bis_property} by binary search on 
the $k$ nodes $u_{1}, u_{2}, \ldots, u_{k}$ of sequence $\mathbf{Q}_{\BiSQ}$. 
Here, $\tau$ is defined as the largest integer $s$ in set $[1, k]$ satisfying $T[q_{\varrho_{s}}..q_{\varrho_{s}}+1] \prec P$ 
if such integer $s$ exists; otherwise, $\tau = 0$. 
The binary search of this phase access $O(\log k)$ nodes of the sequence $\mathbf{Q}_{\BiSQ}$. 
For each accessed node $u_{s}$, 
we verify whether $T[q_{\varrho_{s}}..q_{\varrho_{s}}+1] \prec P$ or not in the following three steps: 
\begin{enumerate}[label=\textbf{(\roman*)}]
    \item obtain the interval attractor $([p_{\varrho_{s}}, q_{\varrho_{s}}], [\ell_{\varrho_{s}}, r_{\varrho_{s}}])$ 
    by the algorithm presented in Section~\ref{subsubsec:computation_delta_samp};
    \item obtain the substring $T[q_{\varrho_{s}}..q_{\varrho_{s}}+1]$ by two random access queries $\RAQ(q_{\varrho_{s}})$ and $\RAQ(q_{\varrho_{s}}+1)$;
    \item verify whether $T[q_{\varrho_{s}}..q_{\varrho_{s}}+1] \prec P$ or not.
\end{enumerate}
The above three steps take $O(H + \log n)$ time, 
and we need $O(\log^{2} k)$ time to access $O(\log k)$ nodes of the sequence $\mathbf{Q}_{\BiSQ}$. 
Therefore, the first phase takes $O((H + \log n + \log k)\log k)$ time in total. 

In the second phase, 
the algorithm computes the integer $\tau^{\prime}$ of Lemma~\ref{lem:bis_property}. 
The definition of the integer $\tau^{\prime}$ is similar to 
that of the integer $\tau$, and hence, 
the integer $\tau^{\prime}$ can be computed by binary search on the sequence $\mathbf{Q}_{\BiSQ}$. 
We compute the integer $\tau^{\prime}$ using the approach used in the first phase. 
Therefore, the second phase takes $O((H + \log n + \log k)\log k)$ time. 

In the third phase, 
the algorithm computes the integer $\lambda$ of Lemma~\ref{lem:bis_property}. 
Here, $\lambda = 1$ if $T[n..n] \prec P$; otherwise, $\lambda = 0$. 
The last character $T[n]$ of input string $T$ can be computed by random access query $\RAQ(n)$, 
which takes $O(H)$ time. 
Therefore, we can compute the integer $\lambda$ in $O(H)$ time. 

In the fourth phase, 
the algorithm computes the triplet $(d, g, j)$ obtained by the bigram search query $\BiSQ(P)$. 
Lemma~\ref{lem:bis_property}(i) shows that $d = \lambda + \sum_{w = 1}^{\tau} \mathcal{L}_{\BiSQ}(u_{w})$ holds. 
The sum $\sum_{w = 1}^{\tau} \mathcal{L}_{\BiSQ}(u_{w})$ can be computed in $O(\log k)$ time 
by the partial sum data structure built on the doubly linked list representing the sequence $\mathbf{Q}_{\BiSQ}$. 
Lemma~\ref{lem:bis_property}(ii) shows that $g = (\sum_{w = 1}^{\tau^{\prime}} \mathcal{L}_{\BiSQ}(u_{w})) - (\sum_{w = 1}^{\tau} \mathcal{L}_{\BiSQ}(u_{w}))$ holds. 
Similar to the sum $\sum_{w = 1}^{\tau} \mathcal{L}_{\BiSQ}(u_{w})$, 
the sum $\sum_{w = 1}^{\tau^{\prime}} \mathcal{L}_{\BiSQ}(u_{w})$ can be computed in $O(\log k)$ time. 
Lemma~\ref{lem:bis_property}(iii) shows that 
we can return the integer $q_{\varrho_{\tau+1}}$ of the interval attractor 
$([p_{\tau+1}, q_{\tau+1}]$, $[\ell_{\tau+1}, r_{\tau+1}])$ as the integer $j$ 
if $\tau < \tau^{\prime}$; 
otherwise, $j = -1$ holds. 
We obtain the interval attractor $([p_{\tau+1}, q_{\tau+1}], [\ell_{\tau+1}, r_{\tau+1}])$ by the algorithm presented in Section~\ref{subsubsec:computation_delta_samp}, which takes $O(\log n + \log k)$ time. 
Therefore, the fourth phase takes $O(\log n + \log k)$ time in total. 

The above four phases take $O((H + \log n + \log k)\log k)$ time in total. 
Here, $k \leq |\Psi_{\samp}|$ holds, and 
we proved $|\Psi_{\samp}| = O(n^{2})$ in Section~\ref{subsubsec:computation_delta_samp}. 
Therefore, 
we can answer a given bigram search query in $O((H + \log n)\log n)$ time.

\subsection{Bigram Access Query}\label{subsec:bigram_access_query}
For a given position $x \in [1, n]$ in the suffix array $\SA$ of input string $T$, 
\emph{bigram access} query $\BiAQ(x)$ return string $T[\SA[x]..\SA[x] + 1]$ of length $2$. 
Here, $T[\SA[x]..\SA[x] + 1] = T[n]\$$ if $\SA[x] = n$. 

For answering bigram access query, 
we leverage bigram search query introduced in Section~\ref{subsec:bigram_search_query}. 
Let $P_{1}, P_{2}, \ldots, P_{k}$ ($P_{1} \prec P_{2} \cdots \prec P_{k}$) be the strings in set $\Sigma^{+}$ 
such that the length of each string $P_{s}$ is $2$ (i.e., $\{ P_{1}, P_{2}, \ldots, P_{k} \} = \{ P \in \Sigma^{+} \mid |P| = 2 \}$). 
For each string $P_{s}$, 
let $(d_{s}, g_{s}, j_{s})$ be the triplet obtained by bigram search query $\BiSQ(P_{s})$. 
Because of $P_{1} \prec P_{2} \cdots \prec P_{k}$, 
$d_{1} \leq d_{2} \leq \cdots \leq d_{k}$ holds. 
The following lemma states the relationship among the string $T[\SA[x]..\SA[x]+1]$ and the $k$ triplets $(d_{1}, g_{1}, j_{1})$, $(d_{2}, g_{2}, j_{2})$, $\ldots$, $(d_{k}, g_{k}, j_{k})$. 

\begin{lemma}\label{lem:inverse_BiSQ_property}
Let $\lambda$ be the largest integer in set $[1, k]$ satisfying $d_{\lambda} \leq x - 1$ 
for a position $x \in [1, n]$ in the suffix array of input string $T$. 
If $x \leq d_{\lambda} + g_{\lambda}$, 
then $\SA[x] \neq n$ and $P_{\lambda} = T[\SA[x]..\SA[x]+1]$; 
otherwise, $\SA[x] = n$ and $P_{\lambda+1} = T[\SA[x]..\SA[x]+1]$.      
\end{lemma}
\begin{proof}
    The following two statements are used to prove Lemma~\ref{lem:inverse_BiSQ_property}: 
    \begin{enumerate}[label=\textbf{(\roman*)}]
    \item $\SA[x] \neq n \Rightarrow (x \leq d_{\lambda} + g_{\lambda}) \land (P_{\lambda} = T[\SA[x]..\SA[x]+1])$;
    \item $\SA[x] = n \Rightarrow (x > d_{\lambda} + g_{\lambda}) \land (P_{\lambda+1} = T[\SA[x]..\SA[x]+1])$.    
    \end{enumerate}

    \textbf{Proof of statement (i).}
    Because of $\SA[x] \neq n$, 
    there exists an integer $\alpha \in [1, k]$ satisfying $P_{\alpha} = T[\SA[x]..\SA[x]+1]$.     
    $\alpha < k$ holds 
    because $P_{k} = \# \#$ for the largest character $\#$ in alphabet $\Sigma$, 
    and input string $T$ does not contain the character $\#$. 
    The sa-interval of the string $P_{\alpha}$ is interval $[1 + d_{\alpha}, d_{\alpha} + g_{\alpha}]$. 
    $x \in [1 + d_{\alpha}, d_{\alpha} + g_{\alpha}]$ holds because 
    The $\SA[x]$-th suffix of string $T$ contains string $P_{\alpha}$ as a prefix. 

    We prove $d_{\alpha} + g_{\alpha} \leq d_{\alpha+1}$. 
    $T[\SA[1]..n] \prec T[\SA[2]..n] \prec \cdots \prec T[\SA[n]..n]$ follows from the definition of suffix array. 
    $T[\SA[d_{\alpha} + g_{\alpha}]..n] \prec P_{\alpha}\#$ holds 
    because $[1 + d_{\alpha}, d_{\alpha} + g_{\alpha}]$ is the sa-interval of the string $P_{\alpha}$. 
    $T[\SA[1]..n] \prec T[\SA[2]..n] \prec \cdots T[\SA[d_{\alpha} + g_{\alpha}]..n] \prec P_{\alpha+1}$ 
    follows from $T[\SA[1]..n] \prec T[\SA[2]..n] \prec \cdots T[\SA[d_{\alpha} + g_{\alpha}]..n] \prec P_{\alpha}\#$ 
    and $P_{\alpha}\# \prec P_{\alpha+1}$. 
    Therefore, $d_{\alpha} + g_{\alpha} \leq d_{\alpha+1}$ holds. 

    We prove $\lambda = \alpha$. 
    $d_{\alpha} \leq x - 1$ follows from $x \in [1 + d_{\alpha}, d_{\alpha} + g_{\alpha}]$. 
    $d_{\alpha+1} \leq x - 1$ does not hold 
    because $x \leq d_{\alpha} + g_{\alpha}$ and $d_{\alpha} + g_{\alpha} \leq d_{\alpha+1}$. 
    Because of $d_{1} \leq d_{2} \leq \cdots \leq d_{k}$, 
    $\alpha$ is the largest integer in set $[1, k]$ satisfying $d_{\alpha} \leq x - 1$, 
    i.e., $\lambda = \alpha$ holds. 

    We prove $x \leq d_{\lambda} + g_{\lambda}$ and $P_{\lambda} = T[\SA[x]..\SA[x]+1]$. 
    $x \leq d_{\lambda} + g_{\lambda}$ follows from $\lambda = \alpha$ and $x \in [1 + d_{\alpha}, d_{\alpha} + g_{\alpha}]$. 
    $P_{\lambda} = T[\SA[x]..\SA[x]+1]$ follows from $\lambda = \alpha$ and $P_{\alpha} = T[\SA[x]..\SA[x]+1]$. 
    Therefore, statement (i) holds. 

    \textbf{Proof of statement (ii).}
    Let $c_{1}, c_{2}, \ldots, c_{\sigma}$ ($c_{1} < c_{2} < \cdots < c_{\sigma}$) be the characters in alphabet $\Sigma$ (i.e., 
    $\Sigma = \{ c_{1}, c_{2}, \ldots, c_{\sigma} \}$). 
    Here, $c_{1} = \$$ and $c_{\sigma} = \#$ hold. 
    The two characters $\$$ and $\#$ are not contained in input string $T$. 
    The last character $T[n]$ of the string $T$ is contained in alphabet $\Sigma$ as a character $c_{e}$ satisfying $e \geq 2$.
    
    Because of $\SA[x] = n$, 
    $T[\SA[x]..\SA[x]+1] = c_{e}\$$ holds. 
    The string $c_{e}\$$ is contained in set $\{ P_{1}, P_{2}, \ldots, P_{k} \}$ as a string $P_{\beta}$. 
    $P_{\beta-1} = c_{e-1}\#$ holds because $\$$ is the smallest character in set $\Sigma$. 

    $T[\SA[1]..n] \prec T[\SA[2]..n] \prec \cdots \prec T[\SA[n]..n]$ follows from the definition of suffix array. 
    $P_{\beta-1} \prec T[\SA[x]..n] \prec P_{\beta}$ holds 
    because $T[\SA[x]..n] = c_{e}$.     
    $T[\SA[1]..n] \prec T[\SA[2]..n] \prec \cdots \prec T[\SA[x-1]..n] \prec P_{\beta-1}$ because 
    $T[\SA[i]] \leq c_{e-1}$ and $T[\SA[i]+1] < \#$ hold for all integer $i \in [1, x-1]$. 
    On the other hand, $P_{\beta} \prec T[\SA[x+1]..n] \prec T[\SA[x+2]..n] \prec \cdots \prec T[\SA[n]..n]$ holds.

    We prove $\lambda = \beta - 1$. 
    $d_{\beta} = x$ holds 
    because $T[\SA[1]..n] \prec T[\SA[2]..n] \prec \cdots \prec T[\SA[x]..n] \prec P_{\beta}$ 
    and $P_{\beta} \prec T[\SA[x+1]..n] \prec T[\SA[x+2]..n] \prec \cdots \prec T[\SA[n]..n]$. 
    $d_{\beta-1} = x-1$ holds 
    because $T[\SA[1]..n] \prec T[\SA[2]..n] \prec \cdots \prec T[\SA[x-1]..n] \prec P_{\beta-1}$ 
    and $P_{\beta-1} \prec T[\SA[x]..n] \prec T[\SA[x+1]..n] \prec \cdots \prec T[\SA[n]..n]$. 
    Therefore, $\lambda = \beta - 1$ follows from 
    $d_{\beta-1} = x-1$, $d_{\beta} = x$, and $d_{1} \leq d_{2} \leq \cdots \leq d_{k}$. 

    We prove $x > d_{\lambda} + g_{\lambda}$ and $P_{\lambda+1} = T[\SA[x]..\SA[x]+1]$. 
    $g_{\beta-1} = 0$ because $P_{\beta-1}$ contains character $\#$, 
    and string $T$ does not contain the character $\#$. 
    $x > d_{\lambda} + g_{\lambda}$ follows from 
    $\lambda = \beta - 1$, $d_{\beta-1} = x-1$, and $g_{\beta-1} = 0$.     
    $P_{\lambda+1} = T[\SA[x]..\SA[x]+1]$ follows from $\lambda = \beta - 1$ and $P_{\beta} = T[\SA[x]..\SA[x]+1]$. 
    Therefore, $x > d_{\lambda} + g_{\lambda}$ and $P_{\lambda+1} = T[\SA[x]..\SA[x]+1]$ hold, i.e., 
    statement (ii) holds.

    \textbf{Proof of Lemma~\ref{lem:inverse_BiSQ_property}.}
    Lemma~\ref{lem:inverse_BiSQ_property} follows from statement (i) and statement (ii).
\end{proof}

For the integer $\lambda$ defined in Lemma~\ref{lem:inverse_BiSQ_property},  
Lemma~\ref{lem:inverse_BiSQ_property} shows that 
$P_{\lambda} = T[\SA[x]..\SA[x]+1]$ holds if $x \leq d_{\lambda} + g_{\lambda}$. 
Otherwise, $P_{\lambda+1} = T[\SA[x]..\SA[x]+1]$ holds. 
Therefore, we can compute the string $T[\SA[x]..\SA[x]+1]$ using the integer $\lambda$. 

\paragraph{Algorithm.}
We answer a given bigram access query using the dynamic data structures of Section~\ref{subsubsec:bis_query_algo}, 
which are used to answer bigram search query. 
The algorithm answering bigram access query consists of four steps: 
\begin{enumerate}[label=\textbf{(\roman*)}]
    \item compute the integer $\lambda$ by binary search on the $k$ integers $d_{1}, d_{2}, \ldots, d_{k}$;
    \item compute two strings $P_{\lambda}$ and $P_{\lambda+1}$;
    \item obtain $(d_{\lambda}, g_{\lambda}, j_{\lambda})$ by bigram search query $\BiSQ(P_{\lambda})$;
    \item if $x \leq d_{\lambda} + g_{\lambda}$, then return string $P_{\lambda}$ as the answer to 
    the given bigram access query. 
    Otherwise, return string $P_{\lambda+1}$ as the answer to the given bigram access query.    
\end{enumerate}

We show that the first step takes $O(H \log^{2} n + \log^{3} n)$ time.  
The binary search of the first step executes $O(\log k)$ bigram search queries 
because we need to execute bigram search query $\BiSQ(P_{s})$ for computing each integer $d_{s}$. 
$k = n^{O(1)}$ holds 
because $k = \sigma^{2}$ and $\sigma = n^{O(1)}$ (see Section~\ref{sec:preliminary}) 
for the size $\sigma$ of alphabet $\Sigma$. 
Therefore, the first step takes $O(H \log^{2} n + \log^{3} n)$ time. 

The bottleneck of the algorithm answering bigram access query is the first step. 
Therefore, we can answer a given bigram access query in $O(H \log^{2} n + \log^{3} n)$ time.

\section{RSC Query with Expected \texorpdfstring{$\delta$}{}-optimal Space}\label{sec:RSC_query}

The goal of this section is to answer a given RSC query $\RSCQ(i, j)$ using 
a dynamic data structure of $O((|\Psi_{\samp}| + |\mathcal{U}_{\RR}|) B)$ bits of space for machine word size $B = \Theta(\log n)$. 
Here, $|\mathcal{U}_{\RR}|$ is the number of nodes in the RR-DAG of RLSLP $\mathcal{G}^{R}$. 
This dynamic data structure achieves expected $\delta$-optimal space 
because (i) $\mathbb{E}[|\mathcal{U}_{\RR}|] = O(\delta \log \frac{n \log \sigma}{\delta \log n})$ (Lemma~\ref{lem:rrdag_node_size}), 
and (ii) $\mathbb{E}[|\Psi_{\samp}|] = O(\delta \log \frac{n \log \sigma}{\delta \log n})$ (Lemma~\ref{lem:samp_basic_property}~\ref{enum:samp_basic_property:4}). 
The following lemma states the summary of this section. 

\begin{theorem}\label{theo:rsc_query_summary}
Using a dynamic data structure of $O((|\Psi_{\samp}| + |\mathcal{U}_{\RR}|) B)$ bits of space for machine word size $B$, 
we can answer a given RSC query in $O(H^{2} \log n + \log^{4} n)$ time. 
\end{theorem}
\begin{proof}
See Section~\ref{subsec:answer_rsc_query}.
\end{proof}

For this section, 
$h_{Q}$, $\gamma_{Q}$, and $C_{Q}$ are defined as the level, attractor position, and associated string of 
interval attractor $I_{\capture}(i, j) = ([p_{Q}, q_{Q}], [\ell_{Q}, r_{Q}])$, respectively. 
$K_{Q}$ is defined as the length of the longest common prefix between two strings $T[\gamma_{Q}..r_{Q}]$ and $C_{Q}^{n+1}$~(i.e., $K_{Q} = |\lcp(T[\gamma_{Q}..r_{Q}], C_{Q}^{n+1})|$). 
$M_{Q}$ is defined as $M_{Q} = (K_{Q} - (2 + \sum_{w = 1}^{h_{Q}+3} \lfloor \mu(w) \rfloor) ) \mod |C_{Q}|$. 

We already showed that RSC query can be answered by counting interval attractors satisfying 
the conditions of Theorem~\ref{theo:rsc_query_sub_formula_X}. 
The following theorem shows that 
the set of such interval attractors is the intersection of the three subsets 
$\Psi_{\CCP}(T[i..j])$, $\Psi_{\lex}(T[\gamma_{Q}..r_{Q}+1])$, and $\Psi_{h_{Q}}$. 

\begin{theorem}\label{theo:rsc_query_sub_formula}
%Consider interval attractor $I_{\capture}(s, e) = ([p, q], [\ell, r])$ with level $h$ and attractor position $\gamma$. 
%Then, 
$\RSCQ(i, j) = |\Psi_{\CCP}(T[i..j]) \cap \Psi_{\lex}(T[\gamma_{Q}..r_{Q}+1])|$ and $\Psi_{\CCP}(T[i..j]) \cap \Psi_{\lex}(T[\gamma_{Q}..r_{Q}+1]) \subseteq \Psi_{h_{Q}}$ hold. 
\end{theorem}
\begin{proof}
Let $\Psi \subseteq \Psi_{\RR}$ be the set of all the interval attractors satisfying the conditions of Theorem~\ref{theo:rsc_query_sub_formula_X} in set $\Psi_{\RR}$. 
Then, 
$\RSCQ(i, j) = |\Psi_{\CCP}(T[i..j]) \cap \Psi_{\lex}(T[\gamma_{Q}..r_{Q}+1])|$ holds if 
$\Psi \subseteq \Psi_{\CCP}(T[i..j]) \cap \Psi_{\lex}(T[\gamma_{Q}..r_{Q}+1])$ 
and $\Psi \supseteq \Psi_{\CCP}(T[i..j]) \cap \Psi_{\lex}(T[\gamma_{Q}..r_{Q}+1])$. 

\paragraph{Proof of $\Psi \subseteq \Psi_{\CCP}(T[i..j]) \cap \Psi_{\lex}(T[\gamma_{Q}..r_{Q}+1])$.}
Consider an interval attractor $([p, q], [\ell, r])$ in set $\Psi$. 
Theorem~\ref{theo:SA_INTV_INTV_ATTR} shows that 
$([p, q], [\ell, r]) \in \Psi_{\CCP}(T[i..j])$ follows from Conditions (i)-(iii) of Theorem~\ref{theo:rsc_query_sub_formula_X}. 
$([p, q], [\ell, r]) \in \Psi_{\lex}(T[\gamma_{Q}..r_{Q}+1])$ follows from 
the fourth condition of Theorem~\ref{theo:rsc_query_sub_formula_X}. 
Therefore, we obtain $\Psi \subseteq \Psi_{\CCP}(T[i..j]) \cap \Psi_{\lex}(T[\gamma_{Q}..r_{Q}+1])$. 

\paragraph{Proof of $\Psi \supseteq \Psi_{\CCP}(T[i..j]) \cap \Psi_{\lex}(T[\gamma_{Q}..r_{Q}+1])$.}
Consider an interval attractor $([p, q], [\ell, r])$ in set $\Psi_{\CCP}(T[i..j]) \cap \Psi_{\lex}(T[\gamma_{Q}..r_{Q}+1])$. 
Theorem~\ref{theo:SA_INTV_INTV_ATTR} shows that 
each interval attractor of set $\Psi_{\CCP}(T[i..j])$ satisfies Conditions (i)-(iii) of Theorem~\ref{theo:rsc_query_sub_formula_X}. 
From the definition of set $\Psi_{\lex}(T[\gamma_{Q}..r_{Q}+1])$, 
each interval attractor of set $\Psi_{\lex}(T[\gamma_{Q}..r_{Q}+1])$ satisfies 
the fourth condition of Theorem~\ref{theo:rsc_query_sub_formula_X}. 
Therefore, we obtain $\Psi \supseteq \Psi_{\CCP}(T[i..j]) \cap \Psi_{\lex}(T[\gamma_{Q}..r_{Q}+1])$. 

\paragraph{Proof of $\Psi_{\CCP}(T[i..j]) \cap \Psi_{\lex}(T[\gamma_{Q}..r_{Q}+1]) \subseteq \Psi_{h_{Q}}$.}
We already showed that 
each interval attractor of set $\Psi_{\CCP}(T[i..j]) \cap \Psi_{\lex}(T[\gamma_{Q}..r_{Q}+1])$ 
satisfies the conditions of Theorem~\ref{theo:rsc_query_sub_formula_X}. 
$\Psi_{\CCP}(T[i..j]) \cap \Psi_{\lex}(T[\gamma_{Q}..r_{Q}+1]) \subseteq \Psi_{h_{Q}}$ follows from 
the first condition of Theorem~\ref{theo:rsc_query_sub_formula_X}. 
\end{proof}

\subsection{Seven Subqueries}\label{subsec:rsc_sub}
\begin{table}[t]
    \normalsize
    \vspace{-0.5cm}
    \caption{
    Seven subqueries used to answer RSC query $\RSCQ(i, j)$. 
    $\gamma_{Q}$ and $C_{Q}$ are the attractor position and associated string of 
    interval attractor $I_{\capture}(i, j) = ([p_{Q}, q_{Q}], [\ell_{Q}, r_{Q}])$, respectively. 
    $K_{Q}$ is the length of the longest common prefix between two strings $T[\gamma_{Q}..r_{Q}]$ and $C_{Q}^{n+1}$. 
    }
    \vspace{1mm}    
    \label{table:RSC_query_result} 
    \center{
    \scalebox{0.85}{
    \begin{tabular}{l||l|l}
 Subquery & Description & Sections \\  \hline 
 $\RSCQA(i, j)$ & Return $|(\Psi_{\CCP}(T[i..j]) \cap \Psi_{\lex}(T[\gamma_{Q}..r_{Q}+1])) \setminus \Psi_{\run}|$ & \ref{subsec:RSC_comp_A} \\ \hline
 $\RSCQBX(i, j)$ & Return $|\Psi_{\CCP}(T[i..j]) \cap \Psi_{\lex}(T[\gamma_{Q}..r_{Q}+1]) \cap \Psi_{\run}|$ & \ref{subsec:RSC_comp_B} \\ \hline
 $\RSCQBY(i, j)$ & Return $|(\Psi_{\CCP}(T[i..j]) \cap \Psi_{\lex}(T[\gamma_{Q}..r_{Q}+1]) \cap \Psi_{\run}) \setminus \Psi_{\centerset}(C_{Q})|$ & \ref{subsec:RSC_comp_B} \\ \hline 
 $\RSCQCX(i, j)$ & Return $|\Psi_{\CCP}(T[i..j]) \cap \Psi_{\lex}(T[\gamma_{Q}..r_{Q}+1]) \cap \Psi_{\run} \cap \Psi_{\centerset}(C_{Q}) \cap \Psi_{\lcp}(K_{Q}) \cap \Psi_{\preceding}|$ & \ref{subsec:RSC_comp_C1} \\ \hline
 $\RSCQCY(i, j)$ & Return $|\Psi_{\CCP}(T[i..j]) \cap \Psi_{\lex}(T[\gamma_{Q}..r_{Q}+1]) \cap \Psi_{\run} \cap \Psi_{\centerset}(C_{Q}) \cap \Psi_{\lcp}(K_{Q}) \cap \Psi_{\succeeding}|$ & \ref{subsec:RSC_comp_C2} \\ \hline
 $\RSCQDX(i, j)$ & Return $|(\Psi_{\CCP}(T[i..j]) \cap \Psi_{\lex}(T[\gamma_{Q}..r_{Q}+1]) \cap \Psi_{\run} \cap \Psi_{\centerset}(C_{Q}) \cap \Psi_{\preceding}) \setminus \Psi_{\lcp}(K_{Q})|$  & \ref{subsec:RSC_comp_D1} \\ \hline
 $\RSCQDY(i, j)$ & Return $|(\Psi_{\CCP}(T[i..j]) \cap \Psi_{\lex}(T[\gamma_{Q}..r_{Q}+1]) \cap \Psi_{\run} \cap \Psi_{\centerset}(C_{Q}) \cap \Psi_{\succeeding}) \setminus \Psi_{\lcp}(K_{Q})|$  & \ref{subsec:RSC_comp_D2} \\ \hline
    \end{tabular} 
    }
    }
\end{table}

As already explained in Section~\ref{subsec:simplified_rss_rsc}, 
RSC query $\RSCQ(i, j)$ is decomposed into seven subqueries. 
The seven subqueries are defined as follows (see also Table~\ref{table:RSC_query_result}): 
\begin{description}
    \item[$\RSCQA(i, j)$:] returns the number of non-periodic interval attractors satisfying the following two conditions: 
    (1) they are contained in set $\Psi_{\CCP}(T[i..j])$; 
    (2) they are contained in set $\Psi_{\lex}(T[\gamma_{Q}..r_{Q}+1])$. 
    Formally, $$\RSCQA(i, j) = |(\Psi_{\CCP}(T[i..j]) \cap \Psi_{\lex}(T[\gamma_{Q}..r_{Q}+1])) \setminus \Psi_{\run}|.$$ 
    \item[$\RSCQBX(i, j)$:]  
    returns the number of periodic interval attractors satisfying the following two conditions: 
    (1) they are contained in set $\Psi_{\CCP}(T[i..j])$; 
    (2) they are contained in set $\Psi_{\lex}(T[\gamma_{Q}..r_{Q}+1])$. 
    Formally, $$\RSCQBX(i, j) = |\Psi_{\CCP}(T[i..j]) \cap \Psi_{\lex}(T[\gamma_{Q}..r_{Q}+1]) \cap \Psi_{\run}|.$$ 
    \item[$\RSCQBY(i, j)$:]  
    returns the number of periodic interval attractors satisfying the following three conditions: 
    (1) they are contained in set $\Psi_{\CCP}(T[i..j])$; 
    (2) they are contained in set $\Psi_{\lex}(T[\gamma_{Q}..r_{Q}+1])$; 
    (3) each interval attractor does not have string $C_{Q}$ as its associated string. 
    Formally, 
    $$\RSCQBY(i, j) = |(\Psi_{\CCP}(T[i..j]) \cap \Psi_{\lex}(T[\gamma_{Q}..r_{Q}+1]) \cap \Psi_{\run}) \setminus \Psi_{\centerset}(C_{Q})|.$$     
    \item[$\RSCQCX(i, j)$:] 
    returns the number of periodic interval attractors satisfying the following five conditions: 
    (1) they are contained in set $\Psi_{\CCP}(T[i..j])$; 
    (2) they are contained in set $\Psi_{\lex}(T[\gamma_{Q}..r_{Q}+1])$; 
    (3) they are contained in set $\Psi_{\lcp}(K_{Q})$; 
    (4) they are contained in set $\Psi_{\preceding}$;
    (5) each interval attractor has string $C_{Q}$ as its associated string.  
    Formally, $$\RSCQCX(i, j) = |\Psi_{\CCP}(T[i..j]) \cap \Psi_{\lex}(T[\gamma_{Q}..r_{Q}+1]) \cap \Psi_{\run} \cap \Psi_{\centerset}(C_{Q}) \cap \Psi_{\lcp}(K_{Q}) \cap \Psi_{\preceding}|.$$ 
    \item[$\RSCQCY(i, j)$:] 
    returns the number of periodic interval attractors satisfying the following five conditions: 
    (1) they are contained in set $\Psi_{\CCP}(T[i..j])$; 
    (2) they are contained in set $\Psi_{\lex}(T[\gamma_{Q}..r_{Q}+1])$; 
    (3) they are contained in set $\Psi_{\lcp}(K_{Q})$; 
    (4) they are contained in set $\Psi_{\succeeding}$;
    (5) each interval attractor has string $C_{Q}$ as its associated string.  
    Formally, $$\RSCQCY(i, j) = |\Psi_{\CCP}(T[i..j]) \cap \Psi_{\lex}(T[\gamma_{Q}..r_{Q}+1]) \cap \Psi_{\run} \cap \Psi_{\centerset}(C_{Q}) \cap \Psi_{\lcp}(K_{Q}) \cap \Psi_{\succeeding}|.$$ 
    \item[$\RSCQDX(i, j)$:] 
    returns the number of periodic interval attractors satisfying the following five conditions: 
    (1) they are contained in set $\Psi_{\CCP}(T[i..j])$; 
    (2) they are contained in set $\Psi_{\lex}(T[\gamma_{Q}..r_{Q}+1])$; 
    (3) they are not contained in set $\Psi_{\lcp}(K_{Q})$; 
    (4) they are contained in set $\Psi_{\preceding}$;
    (5) each interval attractor has string $C_{Q}$ as its associated string.  
    Formally, $$\RSCQDX(i, j) = |(\Psi_{\CCP}(T[i..j]) \cap \Psi_{\lex}(T[\gamma_{Q}..r_{Q}+1]) \cap \Psi_{\run} \cap \Psi_{\centerset}(C_{Q}) \cap \Psi_{\preceding}) \setminus \Psi_{\lcp}(K_{Q})|.$$ 
    \item[$\RSCQDY(i, j)$:] 
    returns the number of periodic interval attractors satisfying the following five conditions: 
    (1) they are contained in set $\Psi_{\CCP}(T[i..j])$; 
    (2) they are contained in set $\Psi_{\lex}(T[\gamma_{Q}..r_{Q}+1])$; 
    (3) they are not contained in set $\Psi_{\lcp}(K_{Q})$; 
    (4) they are contained in set $\Psi_{\preceding}$;
    (5) each interval attractor has string $C_{Q}$ as its associated string.  
    Formally, $$\RSCQDY(i, j) = |(\Psi_{\CCP}(T[i..j]) \cap \Psi_{\lex}(T[\gamma_{Q}..r_{Q}+1]) \cap \Psi_{\run} \cap \Psi_{\centerset}(C_{Q}) \cap \Psi_{\succeeding}) \setminus \Psi_{\lcp}(K_{Q})|.$$ 
\end{description}

%\begin{description}
%    \item[$\RSCQA(i, j)$:] returns $|(\Psi_{\CCP}(T[i..j]) \cap \Psi_{\lex}(T[\gamma_{Q}..r_{Q}+1])) \setminus \Psi_{\run}|$.
%    \item[$\RSCQBX(i, j)$:] returns $|\Psi_{\CCP}(T[i..j]) \cap \Psi_{\lex}(T[\gamma_{Q}..r_{Q}+1]) \cap \Psi_{\run}|$.
%    \item[$\RSCQBY(i, j)$:] returns $|(\Psi_{\CCP}(T[i..j]) \cap \Psi_{\lex}(T[\gamma_{Q}..r_{Q}+1]) \cap \Psi_{\run}) \setminus \Psi_{\centerset}(C_{Q})|$.
%    \item[$\RSCQCX(i, j)$:] returns $|\Psi_{\CCP}(T[i..j]) \cap \Psi_{\lex}(T[\gamma_{Q}..r_{Q}+1]) \cap \Psi_{\run} \cap \Psi_{\centerset}(C_{Q}) \cap \Psi_{\lcp}(K_{Q}) \cap \Psi_{\preceding}|$.
%    \item[$\RSCQCY(i, j)$:] returns $|\Psi_{\CCP}(T[i..j]) \cap \Psi_{\lex}(T[\gamma_{Q}..r_{Q}+1]) \cap \Psi_{\run} \cap \Psi_{\centerset}(C_{Q}) \cap \Psi_{\lcp}(K_{Q}) \cap \Psi_{\succeeding}|$.
%    \item[$\RSCQDX(i, j)$:] returns $|(\Psi_{\CCP}(T[i..j]) \cap \Psi_{\lex}(T[\gamma_{Q}..r_{Q}+1]) \cap \Psi_{\run} \cap \Psi_{\centerset}(C_{Q}) \cap \Psi_{\preceding}) \setminus \Psi_{\lcp}(K_{Q})|$.
%    \item[$\RSCQDY(i, j)$:] returns $|(\Psi_{\CCP}(T[i..j]) \cap \Psi_{\lex}(T[\gamma_{Q}..r_{Q}+1]) \cap \Psi_{\run} \cap \Psi_{\centerset}(C_{Q}) \cap \Psi_{\succeeding}) \setminus \Psi_{\lcp}(K_{Q})|$.
%\end{description}

We show that RSC query can be computed by the above seven subqueries. 
We already showed that 
RSC query can be computed as the number of interval attractors in set $\Psi_{\CCP}(T[i..j]) \cap \Psi_{\lex}(T[\gamma_{Q}..r_{Q}+1])$ (Theorem~\ref{theo:rsc_query_sub_formula}). 

For quickly counting the interval attractors in the set $\Psi_{\CCP}(T[i..j]) \cap \Psi_{\lex}(T[\gamma_{Q}..r_{Q}+1])$, 
we consider the following four conditions of RSC query $\RSCQ(i, j)$: 
\begin{description}
    \item[Condition A] $K_{Q} \leq 1 + \sum_{w = 1}^{h_{Q}+3} \lfloor \mu(w) \rfloor$;
    \item[Condition B] (a) $K_{Q} > 1 + \sum_{w = 1}^{h_{Q}+3} \lfloor \mu(w) \rfloor$ and (b) string $T[i..\gamma_{Q}-1]$ is not a suffix of string $C_{Q}^{n+1}$;
    \item[Condition C] (a) $K_{Q} > 1 + \sum_{w = 1}^{h_{Q}+3} \lfloor \mu(w) \rfloor$, (b) string $T[i..\gamma_{Q}-1]$ is a suffix of string $C_{Q}^{n+1}$, and (c) $|[\gamma_{Q}, j]| > K_{Q}$; 
    \item[Condition D] (a) $K_{Q} > 1 + \sum_{w = 1}^{h_{Q}+3} \lfloor \mu(w) \rfloor$, (b) string $T[i..\gamma_{Q}-1]$ is a suffix of string $C_{Q}^{n+1}$, and (c) $|[\gamma_{Q}, j]| \leq K_{Q}$.
\end{description}
The given RSC query $\RSCQ(i, j)$ satisfies one of the above four conditions (A), (B), (C), and (D). 
The following lemma shows that the set $\Psi_{\CCP}(T[i..j]) \cap \Psi_{\lex}(T[\gamma_{Q}..r_{Q}+1])$ 
can be divided into at most six disjoint sets. 

\begin{lemma}\label{lem:Psi_capture_division_property}
Consider the four conditions (A), (B), (C), and (D) of RSC query $\RSCQ(i, j)$. 
The following four statements hold:
\begin{enumerate}[label=\textbf{(\roman*)}]
    \item for condition A, 
    \begin{equation*}
    \begin{split}
    |\Psi_{\CCP}(T[i..j]) \cap \Psi_{\lex}(T[\gamma_{Q}..r_{Q}+1])| &= |(\Psi_{\CCP}(T[i..j]) \cap \Psi_{\lex}(T[\gamma_{Q}..r_{Q}+1])) \setminus \Psi_{\run}| \\ 
    & + |\Psi_{\CCP}(T[i..j]) \cap \Psi_{\lex}(T[\gamma_{Q}..r_{Q}+1]) \cap \Psi_{\run}|;
    \end{split}
    \end{equation*}
    \item for condition B, 
    \begin{equation*}
    \begin{split}
    |\Psi_{\CCP}(T[i..j]) \cap & \Psi_{\lex}(T[\gamma_{Q}..r_{Q}+1])| = |(\Psi_{\CCP}(T[i..j]) \cap \Psi_{\lex}(T[\gamma_{Q}..r_{Q}+1])) \setminus \Psi_{\run}| \\ 
    & + |(\Psi_{\CCP}(T[i..j]) \cap \Psi_{\lex}(T[\gamma_{Q}..r_{Q}+1]) \cap \Psi_{\run}) \setminus \Psi_{\centerset}(C_{Q})|;
    \end{split}
    \end{equation*}
    \item for condition C, 
    \begin{equation*}
    \begin{split}
    & |\Psi_{\CCP}(T[i..j]) \cap \Psi_{\lex}(T[\gamma_{Q}..r_{Q}+1])| \\
    & = |(\Psi_{\CCP}(T[i..j]) \cap \Psi_{\lex}(T[\gamma_{Q}..r_{Q}+1])) \setminus \Psi_{\run}| \\
    & + |(\Psi_{\CCP}(T[i..j]) \cap \Psi_{\lex}(T[\gamma_{Q}..r_{Q}+1]) \cap \Psi_{\run}) \setminus \Psi_{\centerset}(C_{Q})| \\
    & + |\Psi_{\CCP}(T[i..j]) \cap \Psi_{\lex}(T[\gamma_{Q}..r_{Q}+1]) \cap \Psi_{\run} \cap \Psi_{\centerset}(C_{Q}) \cap \Psi_{\lcp}(K_{Q}) \cap \Psi_{\preceding}| \\
    & + |\Psi_{\CCP}(T[i..j]) \cap \Psi_{\lex}(T[\gamma_{Q}..r_{Q}+1]) \cap \Psi_{\run} \cap \Psi_{\centerset}(C_{Q}) \cap \Psi_{\lcp}(K_{Q}) \cap \Psi_{\succeeding}|;
    \end{split}
    \end{equation*}
    \item for condition D, 
    \begin{equation*}
    \begin{split}
    & |\Psi_{\CCP}(T[i..j]) \cap \Psi_{\lex}(T[\gamma_{Q}..r_{Q}+1])| \\ 
    & = |(\Psi_{\CCP}(T[i..j]) \cap \Psi_{\lex}(T[\gamma_{Q}..r_{Q}+1])) \setminus \Psi_{\run}| \\
    & + |(\Psi_{\CCP}(T[i..j]) \cap \Psi_{\lex}(T[\gamma_{Q}..r_{Q}+1]) \cap \Psi_{\run}) \setminus \Psi_{\centerset}(C_{Q})| \\
    & + |\Psi_{\CCP}(T[i..j]) \cap \Psi_{\lex}(T[\gamma_{Q}..r_{Q}+1]) \cap \Psi_{\run} \cap \Psi_{\centerset}(C_{Q}) \cap \Psi_{\lcp}(K_{Q}) \cap \Psi_{\preceding}| \\
    & + |\Psi_{\CCP}(T[i..j]) \cap \Psi_{\lex}(T[\gamma_{Q}..r_{Q}+1]) \cap \Psi_{\run} \cap \Psi_{\centerset}(C_{Q}) \cap \Psi_{\lcp}(K_{Q}) \cap \Psi_{\succeeding}| \\
    & + |(\Psi_{\CCP}(T[i..j]) \cap \Psi_{\lex}(T[\gamma_{Q}..r_{Q}+1]) \cap \Psi_{\run} \cap \Psi_{\centerset}(C_{Q}) \cap \Psi_{\preceding}) \setminus \Psi_{\lcp}(K_{Q})| \\
    & + |(\Psi_{\CCP}(T[i..j]) \cap \Psi_{\lex}(T[\gamma_{Q}..r_{Q}+1]) \cap \Psi_{\run} \cap \Psi_{\centerset}(C_{Q}) \cap \Psi_{\succeeding}) \setminus \Psi_{\lcp}(K_{Q})|. 
    \end{split}
    \end{equation*}
\end{enumerate}
\end{lemma}
\begin{proof}
The following five equations are used to prove Lemma~\ref{lem:Psi_capture_division_property}:
\begin{equation}\label{eq:Psi_capture_division_property:1}
    \begin{split}
     |\Psi_{\CCP}(T[i..j]) & \cap \Psi_{\lex}(T[\gamma_{Q}..r_{Q}+1])| = |(\Psi_{\CCP}(T[i..j]) \cap \Psi_{\lex}(T[\gamma_{Q}..r_{Q}+1])) \setminus \Psi_{\run}| \\ 
    & + |(\Psi_{\CCP}(T[i..j]) \cap \Psi_{\lex}(T[\gamma_{Q}..r_{Q}+1]) \cap \Psi_{\run}) \setminus \Psi_{\centerset}(C_{Q})| \\
    & + |\Psi_{\CCP}(T[i..j]) \cap \Psi_{\lex}(T[\gamma_{Q}..r_{Q}+1]) \cap \Psi_{\run} \cap \Psi_{\centerset}(C_{Q})|;
    \end{split}
\end{equation}
\begin{equation}\label{eq:Psi_capture_division_property:2}
    \begin{split}
    & |\Psi_{\CCP}(T[i..j]) \cap \Psi_{\lex}(T[\gamma_{Q}..r_{Q}+1]) \cap \Psi_{\run} \cap \Psi_{\centerset}(C_{Q})| \\ 
    &= |\Psi_{\CCP}(T[i..j]) \cap \Psi_{\lex}(T[\gamma_{Q}..r_{Q}+1]) \cap \Psi_{\run} \cap \Psi_{\centerset}(C_{Q}) \cap \Psi_{\lcp}(K_{Q}) \cap \Psi_{\preceding})| \\
    & + |\Psi_{\CCP}(T[i..j]) \cap \Psi_{\lex}(T[\gamma_{Q}..r_{Q}+1]) \cap \Psi_{\run} \cap \Psi_{\centerset}(C_{Q}) \cap \Psi_{\lcp}(K_{Q}) \cap \Psi_{\succeeding})| \\ 
    & + |(\Psi_{\CCP}(T[i..j]) \cap \Psi_{\lex}(T[\gamma_{Q}..r_{Q}+1]) \cap \Psi_{\run} \cap \Psi_{\centerset}(C_{Q})) \setminus \Psi_{\lcp}(K_{Q})|;
    \end{split}
\end{equation}
\begin{equation}\label{eq:Psi_capture_division_property:3}
    \begin{split}
    & |(\Psi_{\CCP}(T[i..j]) \cap \Psi_{\lex}(T[\gamma_{Q}..r_{Q}+1]) \cap \Psi_{\run} \cap \Psi_{\centerset}(C_{Q})) \setminus \Psi_{\lcp}(K_{Q})| \\ 
    &= |(\Psi_{\CCP}(T[i..j]) \cap \Psi_{\lex}(T[\gamma_{Q}..r_{Q}+1]) \cap \Psi_{\run} \cap \Psi_{\centerset}(C_{Q}) \cap \Psi_{\preceding}) \setminus \Psi_{\lcp}(K_{Q})| \\
    & + |(\Psi_{\CCP}(T[i..j]) \cap \Psi_{\lex}(T[\gamma_{Q}..r_{Q}+1]) \cap \Psi_{\run} \cap \Psi_{\centerset}(C_{Q}) \cap \Psi_{\succeeding}) \setminus \Psi_{\lcp}(K_{Q})|;
    \end{split}
\end{equation}
\begin{equation}\label{eq:Psi_capture_division_property:4}
    |\Psi_{\CCP}(T[i..j]) \cap \Psi_{\lex}(T[\gamma_{Q}..r_{Q}+1]) \cap \Psi_{\run} \cap \Psi_{\centerset}(C_{Q})| = 0 \mbox { for condition B};
\end{equation}
\begin{equation}\label{eq:Psi_capture_division_property:5}
    |(\Psi_{\CCP}(T[i..j]) \cap \Psi_{\lex}(T[\gamma_{Q}..r_{Q}+1]) \cap \Psi_{\run} \cap \Psi_{\centerset}(C_{Q})) \setminus \Psi_{\lcp}(K_{Q})| = 0 \mbox { for condition C}.
\end{equation}

\textbf{Proof of Equation~\ref{eq:Psi_capture_division_property:1}.}
Set $\Psi_{\CCP}(T[i..j]) \cap \Psi_{\lex}(T[\gamma_{Q}..r_{Q}+1])$ can be divided into two disjoint sets 
$(\Psi_{\CCP}(T[i..j]) \cap \Psi_{\lex}(T[\gamma_{Q}..r_{Q}+1])) \setminus \Psi_{\run}$ 
and $\Psi_{\CCP}(T[i..j]) \cap \Psi_{\lex}(T[\gamma_{Q}..r_{Q}+1]) \cap \Psi_{\run}$. 
Similarly, set $\Psi_{\CCP}(T[i..j]) \cap \Psi_{\lex}(T[\gamma_{Q}..r_{Q}+1]) \cap \Psi_{\run}$ 
can be divided into two disjoint sets $(\Psi_{\CCP}(T[i..j]) \cap \Psi_{\lex}(T[\gamma_{Q}..r_{Q}+1]) \cap \Psi_{\run}) \setminus \Psi_{\centerset}(C_{Q})$ 
and $\Psi_{\CCP}(T[i..j]) \cap \Psi_{\lex}(T[\gamma_{Q}..r_{Q}+1]) \cap \Psi_{\run} \cap \Psi_{\centerset}(C_{Q})$. 
Therefore, Equation~\ref{eq:Psi_capture_division_property:1} follows from the following equation: 
\begin{equation*}
    \begin{split}
    |\Psi_{\CCP}(T[i..j]) \cap & \Psi_{\lex}(T[\gamma_{Q}..r_{Q}+1])| \\ 
    &= |(\Psi_{\CCP}(T[i..j]) \cap \Psi_{\lex}(T[\gamma_{Q}..r_{Q}+1])) \setminus \Psi_{\run}| \\
    &+ |\Psi_{\CCP}(T[i..j]) \cap \Psi_{\lex}(T[\gamma_{Q}..r_{Q}+1]) \cap \Psi_{\run}| \\
    &= |(\Psi_{\CCP}(T[i..j]) \cap \Psi_{\lex}(T[\gamma_{Q}..r_{Q}+1])) \setminus \Psi_{\run}| \\
    &+ |(\Psi_{\CCP}(T[i..j]) \cap \Psi_{\lex}(T[\gamma_{Q}..r_{Q}+1]) \cap \Psi_{\run}) \setminus \Psi_{\centerset}(C_{Q})| \\
    &+ |\Psi_{\CCP}(T[i..j]) \cap \Psi_{\lex}(T[\gamma_{Q}..r_{Q}+1]) \cap \Psi_{\run} \cap \Psi_{\centerset}(C_{Q})|.
    \end{split}
\end{equation*}

\textbf{Proof of Equation~\ref{eq:Psi_capture_division_property:2}.}
Set $\Psi_{\CCP}(T[i..j]) \cap \Psi_{\lex}(T[\gamma_{Q}..r_{Q}+1]) \cap \Psi_{\run} \cap \Psi_{\centerset}(C_{Q})$ 
can be divided into two disjoint sets $\Psi_{\CCP}(T[i..j]) \cap \Psi_{\lex}(T[\gamma_{Q}..r_{Q}+1]) \cap \Psi_{\run} \cap \Psi_{\centerset}(C_{Q}) \cap \Psi_{\lcp}(K_{Q})$ and 
$(\Psi_{\CCP}(T[i..j]) \cap \Psi_{\lex}(T[\gamma_{Q}..r_{Q}+1]) \cap \Psi_{\run} \cap \Psi_{\centerset}(C_{Q})) \setminus \Psi_{\lcp}(K_{Q})$. 
Set $\Psi_{\CCP}(T[i..j]) \cap \Psi_{\lex}(T[\gamma_{Q}..r_{Q}+1]) \cap \Psi_{\run} \cap \Psi_{\centerset}(C_{Q}) \cap \Psi_{\lcp}(K_{Q})$ 
can be divided into two disjoint sets $\Psi_{\CCP}(T[i..j]) \cap \Psi_{\lex}(T[\gamma_{Q}..r_{Q}+1]) \cap \Psi_{\run} \cap \Psi_{\centerset}(C_{Q}) \cap \Psi_{\lcp}(K_{Q}) \cap \Psi_{\preceding})$ 
and $\Psi_{\CCP}(T[i..j]) \cap \Psi_{\lex}(T[\gamma_{Q}..r_{Q}+1]) \cap \Psi_{\run} \cap \Psi_{\centerset}(C_{Q}) \cap \Psi_{\lcp}(K_{Q}) \cap \Psi_{\succeeding})$. 
This is because $\Psi_{\preceding} \cap \Psi_{\succeeding} = \emptyset$ 
and $\Psi_{\RR} = \Psi_{\preceding} \cup \Psi_{\succeeding}$ hold 
(see the definition of the two subsets $\Psi_{\preceding}$ and $\Psi_{\succeeding}$ in Section~\ref{subsec:IA_subsets}). 
Here, $\Psi_{\CCP}(T[i..j]) \cap \Psi_{\lex}(T[\gamma_{Q}..r_{Q}+1]) \cap \Psi_{\run} \cap \Psi_{\centerset}(C_{Q}) \cap \Psi_{\lcp}(K_{Q}) \subseteq \Psi_{\RR}$ holds. 
Therefore, Equation~\ref{eq:Psi_capture_division_property:2} follows from the following equation. 

\begin{equation*}
    \begin{split}
    |\Psi_{\CCP}(T[i..j]) & \cap \Psi_{\lex}(T[\gamma_{Q}..r_{Q}+1]) \cap \Psi_{\run} \cap \Psi_{\centerset}(C_{Q})|  \\ 
    &= |\Psi_{\CCP}(T[i..j]) \cap \Psi_{\lex}(T[\gamma_{Q}..r_{Q}+1]) \cap \Psi_{\run} \cap \Psi_{\centerset}(C_{Q}) \cap \Psi_{\lcp}(K_{Q})| \\
    &+ |(\Psi_{\CCP}(T[i..j]) \cap \Psi_{\lex}(T[\gamma_{Q}..r_{Q}+1]) \cap \Psi_{\run} \cap \Psi_{\centerset}(C_{Q})) \setminus \Psi_{\lcp}(K_{Q})| \\
    &= |\Psi_{\CCP}(T[i..j]) \cap \Psi_{\lex}(T[\gamma_{Q}..r_{Q}+1]) \cap \Psi_{\run} \cap \Psi_{\centerset}(C_{Q}) \cap \Psi_{\lcp}(K_{Q}) \cap \Psi_{\preceding})| \\
    &+ |\Psi_{\CCP}(T[i..j]) \cap \Psi_{\lex}(T[\gamma_{Q}..r_{Q}+1]) \cap \Psi_{\run} \cap \Psi_{\centerset}(C_{Q}) \cap \Psi_{\lcp}(K_{Q}) \cap \Psi_{\succeeding})| \\
    &+ |(\Psi_{\CCP}(T[i..j]) \cap \Psi_{\lex}(T[\gamma_{Q}..r_{Q}+1]) \cap \Psi_{\run} \cap \Psi_{\centerset}(C_{Q})) \setminus \Psi_{\lcp}(K_{Q})|.
    \end{split}
\end{equation*}

\textbf{Proof of Equation~\ref{eq:Psi_capture_division_property:3}.}
Equation~\ref{eq:Psi_capture_division_property:3} follows from 
(a) $(\Psi_{\CCP}(T[i..j]) \cap \Psi_{\lex}(T[\gamma_{Q}..r_{Q}+1]) \cap \Psi_{\run} \cap \Psi_{\centerset}(C_{Q})) \setminus \Psi_{\lcp}(K_{Q}) \subseteq \Psi_{\RR}$, 
(b) $\Psi_{\preceding} \cap \Psi_{\succeeding} = \emptyset$, 
and (c) $\Psi_{\RR} = \Psi_{\preceding} \cup \Psi_{\succeeding}$.

\textbf{Proof of Equation~\ref{eq:Psi_capture_division_property:4}.}
We will prove Equation~\ref{eq:Psi_capture_division_property:4} by contradiction. 
We assume that Equation~\ref{eq:Psi_capture_division_property:4} does not hold for condition (B). 
Then, set $\Psi_{\CCP}(T[i..j]) \cap \Psi_{\lex}(T[\gamma_{Q}..r_{Q}+1]) \cap \Psi_{\run} \cap \Psi_{\centerset}(C_{Q})$ contains 
an interval attractor $([p, q], [\ell, r])$. 
Let $\gamma$ be the attractor position of the interval attractor $([p, q], [\ell, r])$. 
Because of $([p, q], [\ell, r]) \in \Psi_{\CCP}(T[i..j])$, 
there exists an occurrence $T[i^{\prime}..i^{\prime}+|[i, j]|-1]$ of string $T[i..j]$ in input string $T$ 
satisfying 
$I_{\capture}(i^{\prime}, i^{\prime}+|[i, j]|-1) = ([p, q], [\ell, r])$. 
Since $I_{\capture}(i^{\prime}, i^{\prime}+|[i, j]|-1) = ([p, q], [\ell, r])$, 
$i^{\prime} \in [p, q]$ and $i^{\prime}+|[i, j]|-1 \in [\ell, r]$ follow from the definition of interval attractor. 
Corollary~\ref{cor:capture_gamma_corollary} shows that 
$T[i^{\prime}..\gamma-1] = T[i..\gamma_{Q}-1]$ holds 
because 
$I_{\capture}(i, j) = ([p_{Q}, q_{Q}], [\ell_{Q}, r_{Q}])$, 
$I_{\capture}(i^{\prime}, i^{\prime}+|[i, j]|-1) = ([p, q], [\ell, r])$, 
and $T[i..j] = T[i^{\prime}..i^{\prime}+|[i, j]|-1]$. 

We will show that string $T[i..\gamma_{Q}-1]$ is a suffix of string $C_{Q}^{n+1}$ under the assumption. 
From the definition of the subset $\Psi_{\run}$, 
string $T[p-1..\gamma-1]$ is a suffix of string $C_{Q}^{n+1}$ 
because $([p, q], [\ell, r]) \in \Psi_{\run} \cap \Psi_{\centerset}(C_{Q})$. 
Because of $i^{\prime} \in [p, q]$, 
the substring $T[i^{\prime}..\gamma-1]$ is a suffix of $T[p-1..\gamma-1]$, 
which indicates that 
$T[i^{\prime}..\gamma-1]$ is also a suffix of string $C_{Q}^{n+1}$. 
Because of $T[i^{\prime}..\gamma-1] = T[i..\gamma_{Q}-1]$, 
$T[i..\gamma_{Q}-1]$ is also a suffix of string $C_{Q}^{n+1}$. 

We will showed that string $T[i..\gamma_{Q}-1]$ is a suffix of string $C_{Q}^{n+1}$ under the assumption. 
On the other hand, 
the condition (B) of RSC query ensures that 
string $T[i..\gamma_{Q}-1]$ is not a suffix of string $C_{Q}^{n+1}$. 
The two facts yield a contradiction. 
Therefore, Equation~\ref{eq:Psi_capture_division_property:4} must hold. 

\textbf{Proof of Equation~\ref{eq:Psi_capture_division_property:5}.}
We will prove Equation~\ref{eq:Psi_capture_division_property:5} by contradiction. 
We assume that Equation~\ref{eq:Psi_capture_division_property:5} does not hold for condition (C). 
Then, set $(\Psi_{\CCP}(T[i..j]) \cap \Psi_{\lex}(T[\gamma_{Q}..r_{Q}+1]) \cap \Psi_{\run} \cap \Psi_{\centerset}(C_{Q})) \setminus \Psi_{\lcp}(K_{Q})$ contains an interval attractor $([p, q], [\ell, r])$. 
Let $\gamma$ be the attractor position of the interval attractor $([p, q], [\ell, r])$. 
Because of $([p, q], [\ell, r]) \in \Psi_{\CCP}(T[i..j])$, 
there exists an occurrence $T[i^{\prime}..i^{\prime}+|[i, j]|-1]$ of string $T[i..j]$ in input string $T$ 
satisfying $([p, q], [\ell, r]) = I_{\capture}(i^{\prime}, i^{\prime}+|[i, j]|-1)$. 
Since $([p, q], [\ell, r]) = I_{\capture}(i^{\prime}, i^{\prime}+|[i, j]|-1)$, 
$i^{\prime} \in [p, q]$ and $i^{\prime}+|[i, j]|-1 \in [\ell, r]$ follow from the definition of interval attractor. 
Corollary~\ref{cor:capture_gamma_corollary} shows that 
$T[i^{\prime}..\gamma-1] = T[i..\gamma_{Q}-1]$ and $T[\gamma..i^{\prime} + |[i, j]| - 1] = T[\gamma_{Q}..j]$ hold.  

Let $K^{\prime}$ be the length of the longest common prefix between two strings $T[\gamma..r]$ and $C_{Q}^{n+1}$~(i.e., $K^{\prime} = |\lcp(T[\gamma..r], C_{Q}^{n+1})|$). 
We will prove $K^{\prime} = K_{Q}$ under the assumption. 
Here, $|[\gamma_{Q}, j]| > K_{Q}$ follows from the condition (C) of RSC query. 
$|\lcp(T[\gamma..r], T[\gamma_{Q}..r_{Q}])| > K_{Q}$ holds 
because (a) string $T[\gamma..j]$ is a prefix of string $T[\gamma..r]$, 
(b) $T[\gamma_{Q}..j]$ is a prefix of string $T[\gamma_{Q}..r_{Q}]$, 
(c) $T[\gamma..i^{\prime} + |[i, j]| - 1] = T[\gamma_{Q}..j]$, 
and (d) $|[\gamma_{Q}, j]| > K_{Q}$. 
$|\lcp(T[\gamma_{Q}..r_{Q}], C_{Q}^{n+1})| = K_{Q}$ follows from $([p_{Q}, q_{Q}], [\ell_{Q}, r_{Q}]) \in \Psi_{\centerset}(C_{Q}) \cap \Psi_{\lcp}(K_{Q})$.
$|\lcp(T[\gamma..r], C_{Q}^{n+1})| = K_{Q}$ follows from 
$|\lcp(T[\gamma..r], T[\gamma_{Q}..r_{Q}])| > K_{Q}$ and $|\lcp(T[\gamma_{Q}..r_{Q}], C_{Q}^{n+1})| = K_{Q}$. 
Therefore, $K^{\prime} = K_{Q}$ follows from $K^{\prime} = |\lcp(T[\gamma..r], C_{Q}^{n+1})|$ and $K = |\lcp(T[\gamma..r], C_{Q}^{n+1})|$. 

We proved $K^{\prime} = K_{Q}$. 
On the other hand, $K^{\prime} \neq K_{Q}$ follows from $([p, q], [\ell, r]) \in \Psi_{\centerset}(C_{Q})$ and $([p, q], [\ell, r]) \not \in \Psi_{\lcp}(K_{Q})$. 
The two facts $K^{\prime} \neq K_{Q}$ and $K^{\prime} = K_{Q}$ yield a contradiction. 
Therefore, Equation~\ref{eq:Psi_capture_division_property:5} must hold. 

\textbf{Proof of Lemma~\ref{lem:Psi_capture_division_property}(i).}
Set $\Psi_{\CCP}(T[i..j]) \cap \Psi_{\lex}(T[\gamma_{Q}..r_{Q}+1])$ 
can be divided into two disjoint sets $(\Psi_{\CCP}(T[i..j]) \cap \Psi_{\lex}(T[\gamma_{Q}..r_{Q}+1])) \setminus \Psi_{\run}$ 
and $\Psi_{\CCP}(T[i..j]) \cap \Psi_{\lex}(T[\gamma_{Q}..r_{Q}+1]) \cap \Psi_{\run}$. 
This division indicates that Lemma~\ref{lem:Psi_capture_division_property}(i) holds. 

\textbf{Proof of Lemma~\ref{lem:Psi_capture_division_property}(ii).}
This statement follows from Equation~\ref{eq:Psi_capture_division_property:1} and Equation~\ref{eq:Psi_capture_division_property:4}. 

\textbf{Proof of Lemma~\ref{lem:Psi_capture_division_property}(iii).}
This statement follows from Equation~\ref{eq:Psi_capture_division_property:1}, Equation~\ref{eq:Psi_capture_division_property:2}, and Equation~\ref{eq:Psi_capture_division_property:5}. 

\textbf{Proof of Lemma~\ref{lem:Psi_capture_division_property}(iv).}
This statement follows from Equation~\ref{eq:Psi_capture_division_property:1}, Equation~\ref{eq:Psi_capture_division_property:2}, and Equation~\ref{eq:Psi_capture_division_property:3}. 
\end{proof}

Combining Theorem~\ref{theo:rsc_query_sub_formula}, Lemma~\ref{lem:Psi_capture_division_property}, and seven subqueries, 
we obtain the following corollary. 

\begin{corollary}\label{cor:RB_rsc_subqueries}
Consider the four conditions (A), (B), (C), and (D) of RSC query $\RSCQ(i, j)$. 
The following four statements hold:
\begin{enumerate}[label=\textbf{(\roman*)}]
    \item $\RSCQ(i, j) = \RSCQA(i, j) + \RSCQBX(i, j)$ for condition (A);
    \item $\RSCQ(i, j) = \RSCQA(i, j) + \RSCQBY(i, j)$ for condition (B);
    \item $\RSCQ(i, j) = \RSCQA(i, j) + \RSCQBY(i, j) + \RSCQCX(i, j) + \RSCQCY(i, j)$ for condition (C);
    \item $\RSCQ(i, j) = \RSCQA(i, j) + \RSCQBY(i, j) + \RSCQCX(i, j) + \RSCQCY(i, j) + \RSCQDX(i, j) + \RSCQDY(i, j)$ 
    for condition (D).
\end{enumerate}
\end{corollary}
Corollary~\ref{cor:RB_rsc_subqueries} shows that 
RSC query can be computed as a sum of seven subqueries. 
Therefore, RSC query can be answered using the seven subqueries. 

In the next subsections, we present dynamic data structures supporting these seven subqueries.

\subsection{Weighted Points Representations of Non-periodic Interval Attractors}\label{subsec:weighted_point}
The seven subqueries for RSC query can efficiently be answered in $O(\polylog(n))$ time by 
mapping non-periodic interval attractors to a grid and using range-sum queries on the grid. 
This subsection explains the basic idea for mapping non-periodic interval attractors to a grid. 

Let $\mathcal{X}$ and $\mathcal{Y}$ be linearly ordered sets. 
$\mathcal{J}^{h}$ denotes a finite set of weighted points on $\mathcal{X}$ and $\mathcal{Y}$. 
Each point within $\mathcal{J}^{h}$ corresponds to a non-periodic interval attractor in the $h$-th level interval attractors $\Psi_{h}$.
Each point in $\mathcal{J}^{h}$ for any $h \in [0, H]$ is associated with two integer weights, which quantify the number of aggregated points, subject to (i) the periodicity of interval attractors and (ii) a specific condition pertaining to two sets $\mathcal{X}$ and $\mathcal{Y}$.

Consider two non-periodic interval attractors $([p, q], [\ell, r])$ and $([p^{\prime}, q^{\prime}], [\ell^{\prime}, r^{\prime}])$ in subset $\Psi_{h}$. 
Let $\gamma$ and $\gamma^{\prime}$ be the two attractor positions of the two interval attractors $([p, q], [\ell, r])$ and $([p^{\prime}, q^{\prime}], [\ell^{\prime}, r^{\prime}])$, respectively. 
$([p, q], [\ell, r])$ is considered lexicographically smaller than $([p^{\prime}, q^{\prime}], [\ell^{\prime}, r^{\prime}])$ on $\mathcal{X}$ if and only if
$\reverse(T[p-1..\gamma-1]) \prec \reverse(T[p^\prime-1..\gamma^\prime-1])$.
Similarly, $([p, q], [\ell, r])$ is considered lexicographically smaller than $([p^{\prime}, q^{\prime}], [\ell^{\prime}, r^{\prime}])$ on $\mathcal{Y}$ if and only if $T[\gamma..r+1] \prec T[\gamma^\prime..r^\prime+1]$.

The first integer weight assigned to each non-periodic interval attractor, represented as a point in $\mathcal{J}^{h}$ for any $h \in [0, H]$, reflects the number of periodic interval attractors that can be reconstructed from a single non-periodic interval attractor, as established by Theorem~\ref{theo:simple_periodicity}. 
The weight for each non-periodic interval attractor is defined as $1$ plus the total number of periodic interval attractors reconstructed by that non-periodic interval attractor. 

The second integer weight is set as the number of points in $\mathcal{J}^{h}$ that shares the same order on $\mathcal{X}$ and $\mathcal{Y}$, i.e., the number of 
non-periodic interval attractors $([p, q], [\ell, r])$ with the same two substrings $T[p-1..\gamma-1]$ and $T[\gamma..r+1]$.
Such interval attractors are aggregated, and the second integer weight represents the number of aggregated interval attractors. 

%We say that a weighted point 
%\mathcal{X} \times \mathcal{Y}$ 

The following lemma states properties of weighted points in each set $\mathcal{J}^{h}$. 
\begin{lemma}\label{lem:weight_point_property}
    The following three statements hold:
    \begin{enumerate}[label=\textbf{(\roman*)}]
    \item \label{enum:weight_point_property:1} consider two non-periodic interval attractors 
    $([p, q], [\ell, r]), ([p^{\prime}, q^{\prime}], [\ell^{\prime}, r^{\prime}]) \in \Psi_{\RR} \setminus \Psi_{\run}$. 
    $T[p-1..r+1] = T[p^{\prime}-1..r^{\prime}+1]$ if and only if 
    the two non-periodic interval attractors are aggregated; 
    \item \label{enum:weight_point_property:2} for an integer $h \in [0, H]$, 
    each interval attractor of set $\Psi_{h} \cap \Psi_{\samp}$ is aggregated into 
    a distinct weighted point in set $\mathcal{J}^{h}$;
    \item \label{enum:weight_point_property:3} for an integer $h \in [0, H]$, 
    consider a weighted point in set $\mathcal{J}^{h}$. 
    Then, there exists an interval attractor in set $\Psi_{h} \cap \Psi_{\samp}$ 
    that aggregated into the weighted point. 
    \end{enumerate}
\end{lemma}
\begin{proof}
    The proof of Lemma~\ref{lem:weight_point_property} is as follows. 

    \textbf{Proof of Lemma~\ref{lem:weight_point_property}(i).}
    Let $\gamma$ and $\gamma^{\prime}$ be the two attractor positions of two interval attractors $([p, q], [\ell, r])$ and $([p^{\prime}, q^{\prime}], [\ell^{\prime}, r^{\prime}])$, respectively.     
    If two non-periodic interval attractors $([p, q], [\ell, r])$ and 
    $([p^{\prime}, q^{\prime}], [\ell^{\prime}, r^{\prime}])$ are aggregated, 
    then $T[p-1..\gamma-1] = T[p^{\prime}-1..\gamma^{\prime}-1]$ and 
    $T[\gamma..r+1] = T[\gamma^{\prime}..r^{\prime}+1]$ hold. 
    $T[p-1..r+1] = T[p^{\prime}-1..r^{\prime}+1]$ follows from 
    $T[p-1..\gamma-1] = T[p^{\prime}-1..\gamma^{\prime}-1]$ and 
    $T[\gamma..r+1] = T[\gamma^{\prime}..r^{\prime}+1]$. 

    On the other hand, 
    if $T[p-1..r+1] = T[p^{\prime}-1..r^{\prime}+1]$ holds, 
    then Lemma~\ref{lem:psi_str_property} shows that 
    (A) $T[p-1..\gamma-1] = T[p^{\prime}-1..\gamma^{\prime}-1]$ and 
    $T[\gamma..r+1] = T[\gamma^{\prime}..r^{\prime}+1]$, 
    and (B) there exists an integer $h \in [0, H]$ satisfying 
    $([p, q], [\ell, r]), ([p^{\prime}, q^{\prime}], [\ell^{\prime}, r^{\prime}]) \in \Psi_{h}$. 
    Therefore, the two non-periodic interval attractors are aggregated. 

    \textbf{Proof of Lemma~\ref{lem:weight_point_property}(ii).}
    Consider two interval attractors 
    $([p, q], [\ell, r])$ and $([p^{\prime}, q^{\prime}], [\ell^{\prime}, r^{\prime}])$ in set $\Psi_{h} \cap \Psi_{\samp}$. 
    Let $\gamma$ and $\gamma^{\prime}$ be the two attractor positions of two interval attractors $([p, q], [\ell, r])$ and $([p^{\prime}, q^{\prime}], [\ell^{\prime}, r^{\prime}])$, respectively. 
    Then, at least one of $T[p-1..\gamma-1] \neq T[p^{\prime}-1..\gamma^{\prime}-1]$ and 
    $T[\gamma..r+1] \neq T[\gamma^{\prime}..r^{\prime}+1]$ holds because 
    $T[p-1..r+1] = T[p^{\prime}-1..r^{\prime}+1]$ follows from the definition of sampling subset $\Psi_{\samp}$ (Definition~\ref{def:sampling_subset}). 
    Lemma~\ref{lem:samp_basic_property}~\ref{enum:samp_basic_property:3} shows that 
    the two interval attractors $([p, q], [\ell, r])$ and $([p^{\prime}, q^{\prime}], [\ell^{\prime}, r^{\prime}])$ 
    are non-periodic. 
    Therefore, the two non-periodic interval attractors are aggregated into two distinct weighted points in set $\Psi_{h}$. 

    \textbf{Proof of Lemma~\ref{lem:weight_point_property}(iii).}
    Consider a point in set $\Psi_{h}$ representing 
    a non-periodic interval attractor $([p, q], [\ell, r])$ in the $h$-th level interval attractors $\Psi_{h}$. 
    Lemma~\ref{lem:samp_basic_property}~\ref{enum:samp_basic_property:2} shows that 
    sampling subset $\Psi_{\samp}$ contains an interval attractor $([p^{\prime}, q^{\prime}], [\ell^{\prime}, r^{\prime}])$ satisfying $T[p-1..r+1] = T[p^{\prime}-1..r^{\prime}+1]$. 
    The interval attractor $([p^{\prime}, q^{\prime}], [\ell^{\prime}, r^{\prime}])$ is 
    contained in set $\Psi_{h} \cap \Psi_{\samp}$ 
    because     
    Lemma~\ref{lem:psi_str_property} shows that 
    $([p^{\prime}, q^{\prime}], [\ell^{\prime}, r^{\prime}]) \Psi_{h}$. 
    Lemma~\ref{lem:weight_point_property}~\ref{enum:weight_point_property:1} shows that 
    the two interval attractors $([p, q], [\ell, r])$ and $([p^{\prime}, q^{\prime}], [\ell^{\prime}, r^{\prime}])$ are aggregated. 
    Therefore, Lemma~\ref{lem:weight_point_property}~\ref{enum:weight_point_property:3}.
\end{proof}

Lemma~\ref{lem:weight_point_property}~\ref{enum:weight_point_property:2} and Lemma~\ref{lem:weight_point_property}~\ref{enum:weight_point_property:3} indicate that 
there is a one-to-one correspondence between (i) set $\mathcal{J}^{h}$ of weighted points and 
(ii) set $\Psi_{\samp} \cap \Psi_{h}$ of interval attractors. 

The following lemma states properties of the weights of each point. 
\begin{lemma}\label{lem:weight}
    Let $\alpha$ and $\beta$ be the first and second integer weights of a weighted point $v$ in set $\mathcal{J}^{h}$, respectively, for an integer $h \in [0, H]$. 
    Here, set $\Psi_{h} \cap \Psi_{\samp}$ contains an interval attractor $([p, q], [\ell, r])$ corresponding to 
    the weighted point. 
    The following two statements hold:
    \begin{enumerate}[label=\textbf{(\roman*)}]
    \item \label{enum:weight:1} if $([p, q], [\ell, r]) \in \Psi_{\source}$, 
    then $\alpha = 1 + |f_{\recover}(([p, q], [\ell, r]))|$; 
    otherwise $\alpha = 1$;
    \item \label{enum:weight:2} $\beta = |\Psi_{\str}(T[p-1..r+1])|$.
    \end{enumerate}
\end{lemma}
\begin{proof}
    There exists a position $s^{h}$ of sequence $S^{h}$ such that 
    the weighted point $v$ represents non-periodic interval attractor 
    $I(s^{h}) = ([p^{\prime}, q^{\prime}], [\ell^{\prime}, r^{\prime}])$ in set $\Psi_{h}$. 
    If $I(s^{h}+1)$ is a periodic interval attractor, 
    then $\alpha$ is defined as $1 + d$ for the integer $d$ used in Theorem~\ref{theo:simple_periodicity}. 
    Otherwise, $\alpha = 1$. 
    For simplicity, let $P = T[p^{\prime}-1..r^{\prime}+1]$. 
    Then, 
    Lemma~\ref{lem:weight_point_property}~\ref{enum:weight_point_property:1} shows that 
    the following two conditions: 
    \begin{itemize}
        \item $T[p-1..r+1] = P$;
        \item $\beta = |\{ ([p^{\prime\prime}, q^{\prime\prime}], [\ell^{\prime\prime}, r^{\prime\prime}]) \in \Psi_{h} \mid T[p^{\prime\prime}-1..r^{\prime\prime}+1] = P \}|$. 
    \end{itemize} 

    \textbf{Proof of Lemma~\ref{lem:weight}(i).}
    Lemma~\ref{lem:psi_equality_basic_property}~\ref{enum:psi_equality_basic_property:5} shows that 
    $([p, q], [\ell, r]) \in \Psi_{\source} \Leftrightarrow ([p^{\prime}, q^{\prime}], [\ell^{\prime}, r^{\prime}]) \in \Psi_{\source}$. 
    If $([p, q], [\ell, r]) \not \in \Psi_{\source}$, 
    then $\alpha = 1$. 
    Otherwise, $\alpha = 1 + d$. 
    %Theorem~\ref{theo:periodicity}~\ref{enum:periodicity:2} shows that 
    $d = |f_{\recover}(([p^{\prime}, q^{\prime}], [\ell^{\prime}, r^{\prime}]))|$ holds (see the proof of Theorem~\ref{theo:simple_periodicity}). 
    Since $T[p-1..r+1] = P$, 
    Lemma~\ref{lem:recover_super_property}~\ref{enum:recover_super_property:1} shows that 
    $|f_{\recover}(([p, q], [\ell, r]))| = |f_{\recover}(([p^{\prime}, q^{\prime}], [\ell^{\prime}, r^{\prime}]))|$. 
    Therefore, $\alpha = 1 + |f_{\recover}(([p, q], [\ell, r]))|$. 
    
    \textbf{Proof of Lemma~\ref{lem:weight}(ii).}
    Lemma~\ref{lem:psi_str_property}~\ref{enum:psi_str_property:2} indicates that 
    $\Psi_{\str}(P) = \{ ([p^{\prime\prime}, q^{\prime\prime}], [\ell^{\prime\prime}, r^{\prime\prime}]) \in \Psi_{h} \mid T[p^{\prime\prime}-1..r^{\prime\prime}+1] = P \}|$. 
    $\Psi_{\str}(P) = \Psi_{\str}(T[p-1..r+1])$ follows from $T[p-1..r+1] = P$. 
    Therefore, $\beta = |\Psi_{\str}(T[p-1..r+1])|$ holds. 
\end{proof}

The following theorem concerning the total number of weighted points is as follows.
\begin{theorem}\label{theo:weighted_point_formular}
    The expected value of the total number of weighted points in 
    $H+1$ sets $\mathcal{J}^{0}, \mathcal{J}^{1}, \ldots, \mathcal{J}^{H}$ is 
    $O(\delta \log \frac{n \log \sigma}{\delta \log n})$ 
    (i.e., $\mathbb{E}[\sum_{h=0}^{H} |\mathcal{J}^{h}|] = O(\delta \log \frac{n \log \sigma}{\delta \log n})$).     
\end{theorem}
\begin{proof}
    Lemma~\ref{lem:weight_point_property}~\ref{enum:weight_point_property:2} and Lemma~\ref{lem:weight_point_property}~\ref{enum:weight_point_property:3} indicate that 
    $|\mathcal{J}^{h}| = |\Psi_{h} \cap \Psi_{\samp}|$ holds. 
    Lemma~\ref{lem:samp_basic_property} \ref{enum:samp_basic_property:4} shows that 
    $\mathbb{E}[|\Psi_{\samp}|] = O(\delta \log \frac{n \log \sigma}{\delta \log n})$. 
    Therefore, Theorem~\ref{theo:weighted_point_formular} follows from the following equation: 
    \begin{equation*}
        \begin{split}
            \mathbb{E}[\sum_{h=0}^{H} |\mathcal{J}^{h}|]  &= \mathbb{E}[\sum_{h=0}^{H} |\Psi_{h} \cap \Psi_{\samp}|] \\
            &= \mathbb{E}[|\Psi_{\samp}|] \\
            &= O(\delta \log \frac{n \log \sigma}{\delta \log n}).
        \end{split}
    \end{equation*}    
\end{proof}

Each subquery can be answered by $O(1)$ range-sum queries on weighted points obtained by transforming 
weighted points in each set $\mathcal{J}^{h}$. 
Such range-sum queries can be supported by range-sum data structures introduced in Section~\ref{subsec:range_data_structure}, 
and these data structures can be stored in expected $\delta$-optimal space. 
This is because 
Theorem~\ref{theo:weighted_point_formular} shows that 
the expected total number of weighted points can be bounded by $O(\delta \log \frac{n \log \sigma}{\delta \log n})$. 
We show that each subquery can be answered by $O(1)$ range-sum queries in subsequent subsections.

\subsection{Subquery \texorpdfstring{$\RSCQA(i, j)$}{RSCA(i, j)}}\label{subsec:RSC_comp_A}
The goal of this subsection is to answer subquery $\RSCQA(i, j)$. 
The following lemma states the summary of this subsection. 

\begin{lemma}\label{lem:RSC_subquery_A_summary}
Using a dynamic data structure of $O((|\Psi_{\samp}| + |\mathcal{U}_{\RR}|) B)$ bits of space for machine word size $B$, 
we can answer $\RSCQA(i, j)$ in $O(H^{2} \log n + \log^{4} n)$ time. 
That is, we can compute $|(\Psi_{\CCP}(T[i..j]) \cap \Psi_{\lex}(T[\gamma_{Q}..r_{Q}+1])) \setminus \Psi_{\run}|$ in $O(H^{2} \log n + \log^{4} n)$ time. 
Here, $|\mathcal{U}_{\RR}|$ is the number of nodes in the RR-DAG of RLSLP $\mathcal{G}^{R}$. 
\end{lemma}
\begin{proof}
See Section~\ref{subsubsec:JA_subquery_ds}.
\end{proof}

For answering subquery $\RSCQA(i, j)$, 
we leverage range-sum query on weighted points corresponding to the interval attractors in 
set $\Psi_{h_{Q}} \cap \Psi_{\samp}$. 
Here, range-sum query and weighted points are detailed in Section~\ref{subsec:range_data_structure}. 
For this purpose, 
we introduce a set $\mathcal{J}_{A}(h)$ of weighted points on a grid $(\mathcal{X}_{A}(h), \mathcal{Y}_{A}(h))$ 
for each integer $h \in [0, H]$. 
Here, the set $\mathcal{J}_{A}(h)$ and grid $(\mathcal{X}_{A}(h), \mathcal{Y}_{A}(h))$ are defined using 
set $\Psi_{h} \cap \Psi_{\samp}$ of $k$ interval attractors 
$([p_{1}, q_{1}], [\ell_{1}, r_{1}]), ([p_{2}, q_{2}], [\ell_{2}, r_{2}])$, 
$\ldots$, $([p_{k}, q_{k}], [\ell_{k}, r_{k}])$. 
Let $\gamma_{s}$ be the attractor position of each interval attractor $([p_{s}, q_{s}], [\ell_{s}, r_{s}]) \in \Psi_{h} \cap \Psi_{\samp}$. 

\paragraph{Grid $(\mathcal{X}_{A}(h), \mathcal{Y}_{A}(h))$.}
$\mathcal{X}_{A}(h) \subseteq \Sigma^{*}$ is 
the ordered set of strings defined as the union of two sets 
$\{ \varepsilon, \# \}$ and $\{ \reverse(T[p_{s}-1..\gamma_{s}-1]) \mid s \in [1, k] \}$ of strings
(i.e., $\mathcal{X}_{A}(h) = \{ \varepsilon, \# \} \cup \{ \reverse(T[p_{s}-1..\gamma_{s}-1]) \mid s \in [1, k] \}$). 
Here, $\varepsilon$ is the string of length $0$, 
and $\#$ is the largest character in the alphabet $\Sigma$ (see Section~\ref{sec:preliminary}). 
This ordered set $\mathcal{X}_{A}(h)$ consists of $d$ strings $L_{1}, L_{2}, \ldots, L_{d}$ 
that are sorted in lexicographical order (i.e., $L_{1} \prec L_{2} \prec \cdots \prec L_{d}$). 
$L_{1} = \varepsilon$ and $L_{d} = \#$ always hold because 
every string of the set $\{ \reverse(T[p_{s}-1..\gamma_{s}-1]) \mid s \in [1, k] \}$ does not contain the character $\#$. 

Similarly, $\mathcal{Y}_{A}(h) \subseteq \Sigma^{*}$ is the ordered set of strings defined as the union of two sets 
$\{ \varepsilon, \# \}$ and $\{ T[\gamma_{s}..r_{s}+1] \mid s \in [1, k] \}$ of strings. 
This ordered set $\mathcal{Y}_{A}(h)$ consists of $d^{\prime}$ strings $R_{1}, R_{2}, \ldots, R_{d^{\prime}}$ 
that are sorted in lexicographical order (i.e., $R_{1} \prec R_{2} \prec \cdots \prec R_{d^{\prime}}$). 
Similar to the ordered set $\mathcal{X}_{A}(h)$, 
$R_{1} = \varepsilon$ and $R_{d^{\prime}} = \#$ always hold. 

Grid $(\mathcal{X}_{A}(h), \mathcal{Y}_{A}(h))$ consists of two ordered sets $\mathcal{X}_{A}(h)$ and $\mathcal{Y}_{A}(h)$. 
Each string of the ordered set $\mathcal{X}_{A}(h)$ represents x-coordinate on two dimensional space. 
Similarly, each string of the ordered set $\mathcal{Y}_{A}(h)$ represents y-coordinate on two dimensional space.

\paragraph{Set $\mathcal{J}_{A}(h)$ of Weighted Points.}
Set $\mathcal{J}_{A}(h)$ consists of $k$ weighted points $(\reverse(T[p_{1}-1..\gamma_{1}-1]), T[\gamma_{1}..r_{1}+1], |\Psi_{\str}(T[p_{1}-1..r_{1}+1])|, T[p_{1}-1..r_{1}+1])$, $(\reverse(T[p_{2}-1..\gamma_{2}-1]), T[\gamma_{2}..r_{2}+1], |\Psi_{\str}(T[p_{2}-1..r_{2}+1])|, T[p_{2}-1..r_{2}+1])$, 
$\ldots$, $(\reverse(T[p_{k}-1..\gamma_{k}-1]), T[\gamma_{k}..r_{k}+1], |\Psi_{\str}(T[p_{k}-1..r_{k}+1])|, T[p_{k}-1..r_{k}+1])$ 
on grid $(\mathcal{X}_{A}(h), \mathcal{Y}_{A}(h))$. 
Each weighted point $(\reverse(T[p_{s}-1..\gamma_{s}-1]), T[\gamma_{s}..r_{s}+1], |\Psi_{\str}(T[p_{s}-1..r_{s}+1])|, T[p_{s}-1..r_{s}+1])$ corresponds to interval attractor $([p_{s}, q_{s}], [\ell_{s}, r_{s}])$ in set $\Psi_{h} \cap \Psi_{\samp}$. 
The details of the four elements of the weighted point $(\reverse(T[p_{s}-1..\gamma_{s}-1]), T[\gamma_{s}..r_{s}+1], |\Psi_{\str}(T[p_{s}-1..r_{s}+1])|, T[p_{s}-1..r_{s}+1])$ are as follows:
\begin{itemize}
    \item the first string $\reverse(T[p_{s}-1..\gamma_{s}-1])$ is the x-coordinate of this weighted point; 
    \item the second string $T[\gamma_{s}..r_{s}+1]$ is the y-coordinate of this weighted point; 
    \item the third integer $|\Psi_{\str}(T[p_{s}-1..r_{s}+1])|$ is the weight of this weighted point;
    \item the fourth string $T[p_{s}-1..r_{s}+1]$ is the identifier of this weighted point.
\end{itemize}
From the definition of the sampling subset $\Psi_{\samp}$, 
the identifiers $T[p_{1}-1..r_{1}+1], T[p_{2}-1..r_{2}+1], \ldots, T[p_{k}-1..r_{k}+1]$ of all the weighted points in 
the set $\mathcal{J}_{A}(h)$ are different. 

The following lemma states the sizes of the three sets $\mathcal{X}_{A}(h)$, $\mathcal{Y}_{A}(h)$, and $\mathcal{J}_{A}(h)$. 

\begin{lemma}\label{lem:JA_size}
    The following three statements hold: 
    \begin{enumerate}[label=\textbf{(\roman*)}]
    \item \label{enum:JA_size:1} $|\mathcal{X}_{A}(h)|, |\mathcal{Y}_{A}(h)| \leq 2 + |\mathcal{J}_{A}(h)|$ for each integer $h \in [0, H]$;
    \item \label{enum:JA_size:2} $\sum_{h = 0}^{H} |\mathcal{J}_{A}(h)| = |\Psi_{\samp}|$;
    \item \label{enum:JA_size:3} $|\mathcal{X}_{A}(h)|, |\mathcal{Y}_{A}(h)|, |\mathcal{J}_{A}(h)| = O(n^{2})$.
    \end{enumerate}
\end{lemma}
\begin{proof}
The proof of Lemma~\ref{lem:JA_size} is as follows. 

\textbf{Proof of Lemma~\ref{lem:JA_size}(i).}
$|\mathcal{X}_{A}(h)|, |\mathcal{Y}_{A}(h)| \leq 2 + |\mathcal{J}_{A}(h)|$ follows from the definitions of the two ordered sets 
$\mathcal{X}_{A}(h)$ and $\mathcal{Y}_{A}(h)$. 

\textbf{Proof of Lemma~\ref{lem:JA_size}(ii).}
For each integer $h \in [0, H]$, 
there exists a one-to-one correspondence between the weighted points of set $\mathcal{J}_{A}(h)$ 
and the interval attractors of set $\Psi_{h} \cap \Psi_{\samp}$. 
The sampling subset $\Psi_{\samp}$ can be divided into $(1 + H)$ sets 
$\Psi_{0} \cap \Psi_{\samp}$, $\Psi_{1} \cap \Psi_{\samp}$, $\ldots$, $\Psi_{H} \cap \Psi_{\samp}$. 
Therefore, $\sum_{h = 0}^{H} |\mathcal{J}_{A}(h)| = |\Psi_{\samp}|$ follows from the equation: 
\begin{equation*}
    \begin{split}
        \sum_{h = 0}^{H} |\mathcal{J}_{A}(h)| &= \sum_{h = 0}^{H} |\Psi_{h} \cap \Psi_{\samp}| \\
        &= |\bigcup_{h=0}^{H} (\Psi_{h} \cap \Psi_{\samp})| \\
        &= |\Psi_{\samp}|.
    \end{split}
\end{equation*}    

\textbf{Proof of Lemma~\ref{lem:JA_size}(iii).}
$|\Psi_{\samp}| = O(n^{2})$ follows from 
$\Psi_{\samp} \subseteq \Psi_{\RR}$ and 
$|\Psi_{\RR}| = O(n^{2})$ (Lemma~\ref{lem:non_comp_IA_size}). 
Therefore, 
$|\mathcal{X}_{A}(h)|, |\mathcal{Y}_{A}(h)|, |\mathcal{J}_{A}(h)| = O(n^{2})$ follows from 
Lemma~\ref{lem:JA_size}~\ref{enum:JA_size:1}, Lemma~\ref{lem:JA_size}~\ref{enum:JA_size:2}, 
and $|\Psi_{\samp}| = O(n^{2})$. 
\end{proof}

The following lemma shows that 
the interval attractors in the set $(\Psi_{\CCP}(T[i..j]) \cap \Psi_{\lex}(T[\gamma_{Q}..r_{Q}+1])) \setminus \Psi_{\run}$ 
can be counted by one range-sum query on the set $\mathcal{J}_{A}(h_{Q})$ of weighted points. 

\begin{lemma}\label{lem:JA_main_lemma}
For RSC query $\RSCQ(i, j)$, 
consider the set $\mathcal{J}_{A}(h_{Q})$ of weighted points on grid $(\mathcal{X}_{A}(h_{Q}), \mathcal{Y}_{A}(h_{Q}))$. 
Here, let $\gamma_{Q}$ and $h_{Q}$ be the attractor position and level of the interval attractor $I_{\capture}(i, j) = ([p_{Q}, q_{Q}], [\ell_{Q}, r_{Q}])$, 
respectively; 
$\mathcal{X}_{A}(h_{Q}) = \{ L_{1}, L_{2}, \ldots, L_{d} \}$~($L_{1} \prec L_{2} \prec \cdots \prec L_{d}$); 
$\mathcal{Y}_{A}(h_{Q}) = \{ R_{1}, R_{2}, \ldots, R_{d^{\prime}} \}$~($R_{1} \prec R_{2} \prec \cdots \prec R_{d^{\prime}}$). 
Let $x, x^{\prime}, y$, and $y^{\prime}$ be the four integers defined as follows: 
\begin{itemize}
\item $x = \min \{ s \in [1, d] \mid \reverse(T[i..\gamma_{Q}-1]) \prec L_{s} \}$;
\item $x^{\prime} = \max \{ s \in [1, d] \mid L_{s} \prec \reverse(\# T[i..\gamma_{Q}-1]) \}$;
\item $y = \min \{ s \in [1, d^{\prime}] \mid T[\gamma_{Q}..j]  \prec R_{s} \}$;
\item $y^{\prime} = \max \{ s \in [1, d^{\prime}] \mid R_{s} \prec T[\gamma_{Q}..r_{Q} + 1] \}$. 
\end{itemize}
Then, the following equation holds: 
\begin{equation*}
|(\Psi_{\CCP}(T[i..j]) \cap \Psi_{\lex}(T[\gamma_{Q}..r_{Q}+1])) \setminus \Psi_{\run}| = \rangesum(\mathcal{J}_{A}(h_{Q}), L_{x}, L_{x^{\prime}}, R_{y}, R_{y^{\prime}}).
\end{equation*}
Here, $\rangesum(\mathcal{J}_{A}(h_{Q}), L_{x}, L_{x^{\prime}}, R_{y}, R_{y^{\prime}})$ is the range-sum query introduced in Section~\ref{subsec:range_data_structure}. 
\end{lemma}
\begin{proof}
See Section~\ref{subsubsec:RA_main_lemma_proof}.
\end{proof}

\subsubsection{Proof of Lemma~\ref{lem:JA_main_lemma}}\label{subsubsec:RA_main_lemma_proof}

For a string $P \in \Sigma^{+}$, 
the following proposition states properties of the subset $\Psi_{\str}(P)$ introduced in Section~\ref{subsec:IA_subsets}. 
\begin{proposition}\label{prop:psi_bigcup_property}
The following three statements hold: 
\begin{enumerate}[label=\textbf{(\roman*)}]
    \item \label{enum:psi_bigcup_property:level}
    $\Psi_{h} = \bigcup_{([p, q], [\ell, r]) \in \Psi_{h}} \Psi_{\str}(T[p-1..r+1])$ for each integer $h \in [0, H]$;
    \item \label{enum:psi_bigcup_property:run}
    $\Psi_{\run} = \bigcup_{([p, q], [\ell, r]) \in \Psi_{\run}} \Psi_{\str}(T[p-1..r+1])$; 
    \item \label{enum:psi_bigcup_property:CCP}
    $\Psi_{\CCP}(P) = \bigcup_{([p, q], [\ell, r]) \in \Psi_{\CCP}(P)} \Psi_{\str}(T[p-1..r+1])$ for a string $P \in \Sigma^{+}$;    
    \item \label{enum:psi_bigcup_property:lex}
    $\Psi_{\lex}(P) = \bigcup_{([p, q], [\ell, r]) \in \Psi_{\lex}(P)} \Psi_{\str}(T[p-1..r+1])$ for a string $P \in \Sigma^{+}$; 
    \item \label{enum:psi_bigcup_property:centerset}
    $\Psi_{\centerset}(P) \cap \Psi_{\run} = \bigcup_{([p, q], [\ell, r]) \in \Psi_{\centerset}(P) \cap \Psi_{\run}} \Psi_{\str}(T[p-1..r+1])$ 
    for a string $P \in \Sigma^{+}$; 
    \item \label{enum:psi_bigcup_property:samp}
    let $\Psi = \bigcup_{Z \in \mathcal{Z}} \Psi_{\str}(Z)$ for a set $\mathcal{Z} \subseteq \Sigma^{+}$ of strings. 
    Then, $|\Psi \setminus \Psi_{\run}| = \sum_{([p, q], [\ell, r]) \in (\Psi \setminus \Psi_{\run}) \cap \Psi_{\samp}}$ $|\Psi_{\str}(T[p-1..r+1])|$.
\end{enumerate}
\end{proposition}
\begin{proof}
    The proof of Proposition~\ref{prop:psi_bigcup_property} is as follows: 

    \textbf{Proof of Proposition~\ref{prop:psi_bigcup_property}(i).}
    Proposition~\ref{prop:psi_bigcup_property}(i) holds if 
    $\Psi_{h} \subseteq \bigcup_{([p, q], [\ell, r]) \in \Psi_{h}} \Psi_{\str}(T[p-1..r+1])$ 
    and $\Psi_{h} \supseteq \bigcup_{([p, q], [\ell, r]) \in \Psi_{h}} \Psi_{\str}(T[p-1..r+1])$.

    \textbf{Proof of $\Psi_{h} \subseteq \bigcup_{([p, q], [\ell, r]) \in \Psi_{h}} \Psi_{\str}(T[p-1..r+1])$.}
    For each interval $([p^{\prime}, q^{\prime}], [\ell^{\prime}, r^{\prime}]) \in \Psi_{h}$, 
    $([p^{\prime}, q^{\prime}], [\ell^{\prime}, r^{\prime}]) \in \Psi_{\str}(T[p^{\prime}-1..r^{\prime}+1])$ 
    follows from the definition of the subset $\Psi_{\str}(T[p^{\prime}-1..r^{\prime}+1])$. 
    Therefore, $\Psi_{h} \subseteq \bigcup_{([p, q], [\ell, r]) \in \Psi_{h}} \Psi_{\str}(T[p-1..r+1])$ holds. 

    \textbf{Proof of $\Psi_{h} \supseteq \bigcup_{([p, q], [\ell, r]) \in \Psi_{h}} \Psi_{\str}(T[p-1..r+1])$.}
    Consider an interval attractor $([p_{A}, q_{A}]$, $[\ell_{A}, r_{A}]) \in \bigcup_{([p, q], [\ell, r]) \in \Psi_{h}} \Psi_{\str}(T[p-1..r+1])$. 
    Then, the $h$-th interval attractors $\Psi_{h}$ contains an interval attractor $([p_{B}, q_{B}], [\ell_{B}, r_{B}])$ 
    satisfying $([p_{A}, q_{A}], [\ell_{A}, r_{A}]) \in \Psi_{\str}(T[p_{B}-1..r_{B}+1])$. 

    We prove $([p_{A}, q_{A}], [\ell_{A}, r_{A}]) \in \Psi_{h}$. 
    Because of $([p_{A}, q_{A}], [\ell_{A}, r_{A}]) \in \Psi_{\str}(T[p_{B}-1..r_{B}+1])$, 
    $T[p_{A}-1..r_{A}+1] = T[p_{B}-1..r_{B}+1]$ follows from the definition of the subset $\Psi_{\str}(T[p_{B}-1..r_{B}+1])$. 
    Lemma~\ref{lem:psi_str_property}~\ref{enum:psi_str_property:2} shows that 
    $([p_{A}, q_{A}], [\ell_{A}, r_{A}]) \in \Psi_{h}$ holds 
    because $T[p_{A}-1..r_{A}+1] = T[p_{B}-1..r_{B}+1]$ and $([p_{B}, q_{B}], [\ell_{B}, r_{B}]) \in \Psi_{h}$. 

    We showed that $([p_{A}, q_{A}], [\ell_{A}, r_{A}]) \in \Psi_{h}$ holds for each interval attractor $([p_{A}, q_{A}], [\ell_{A}, r_{A}]) \in \bigcup_{([p, q], [\ell, r]) \in \Psi_{h}} \Psi_{\str}(T[p-1..r+1])$. 
    Therefore, $\Psi_{h} \supseteq \bigcup_{([p, q], [\ell, r]) \in \Psi_{h}} \Psi_{\str}(T[p-1..r+1])$ holds. 

    \textbf{Proof of Proposition~\ref{prop:psi_bigcup_property}(ii).}
    Proposition~\ref{prop:psi_bigcup_property}(ii) follows from Lemma~\ref{lem:psi_equality_basic_property}~\ref{enum:psi_equality_basic_property:4}. 

    \textbf{Proof of Proposition~\ref{prop:psi_bigcup_property}(iii).}
    Proposition~\ref{prop:psi_bigcup_property}(iii) follows from Lemma~\ref{lem:psi_equality_basic_property}~\ref{enum:psi_equality_basic_property:2}. 

    \textbf{Proof of Proposition~\ref{prop:psi_bigcup_property}(iv).}    
    Proposition~\ref{prop:psi_bigcup_property}(iv) follows from Lemma~\ref{lem:psi_equality_basic_property}~\ref{enum:psi_equality_basic_property:8}. 

    \textbf{Proof of Proposition~\ref{prop:psi_bigcup_property}(v).}    
    Proposition~\ref{prop:psi_bigcup_property}(v) follows from Lemma~\ref{lem:psi_equality_basic_property}~\ref{enum:psi_equality_basic_property:center_set} 
    and Lemma~\ref{lem:psi_equality_basic_property}~\ref{enum:psi_equality_basic_property:4}. 

    \textbf{Proof of Proposition~\ref{prop:psi_bigcup_property}(vi).}    
    Proposition~\ref{prop:psi_bigcup_property}(vi) holds if 
    the following three statements hold: 
\begin{enumerate}[label=\textbf{(\Alph*)}]
    \item $\Psi \setminus \Psi_{\run} \subseteq \bigcup_{([p, q], [\ell, r]) \in (\Psi \setminus \Psi_{\run}) \cap \Psi_{\samp}} \Psi_{\str}(T[p-1..r+1])$; 
    \item $\Psi \setminus \Psi_{\run} \supseteq \bigcup_{([p, q], [\ell, r]) \in (\Psi \setminus \Psi_{\run}) \cap \Psi_{\samp}} \Psi_{\str}(T[p-1..r+1])$; 
    \item $\Psi_{\str}(T[p_{1}-1..r_{1}+1]) \cap \Psi_{\str}(T[p_{2}-1..r_{2}+1]) = \emptyset$ 
    for any pair of two interval attractors $([p_{1}, q_{1}], [\ell_{1}, r_{1}]), ([p_{2}, q_{2}], [\ell_{2}, r_{2}]) \in (\Psi \setminus \Psi_{\run}) \cap \Psi_{\samp}$.
\end{enumerate}

    \textbf{Proof of statement (A).}    
    Consider an interval attractor $([p_{A}, q_{A}], [\ell_{A}, r_{A}])$ in set $\Psi \setminus \Psi_{\run}$. 
    Because of $([p_{A}, q_{A}], [\ell_{A}, r_{A}]) \not \in \Psi_{\run}$, 
    Lemma~\ref{lem:samp_basic_property}~\ref{enum:samp_basic_property:2} shows that 
    the sampling subset $\Psi_{\samp}$ contains an interval attractor $([p_{B}, q_{B}], [\ell_{B}, r_{B}])$ satisfying 
    $T[p_{B}-1..r_{B}+1] = T[p_{A}-1..r_{A}+1]$. 

    We prove $([p_{A}, q_{A}], [\ell_{A}, r_{A}]) \in \bigcup_{([p, q], [\ell, r]) \in (\Psi \setminus \Psi_{\run}) \cap \Psi_{\samp}} \Psi_{\str}(T[p-1..r+1])$. 
    From the definition of the subset $\Psi$, 
    the set $\mathcal{Z}$ contains string $T[p_{A}-1..r_{A}+1]$.     
    Because of $T[p_{B}-1..r_{B}+1] = T[p_{A}-1..r_{A}+1]$, 
    $([p_{B}, q_{B}], [\ell_{B}, r_{B}]) \in \Psi_{\str}(T[p_{A}-1..r_{A}+1])$ follows from the definition of the subset $\Psi_{\str}(T[p_{A}-1..r_{A}+1])$. 
    Similarly,  $([p_{A}, q_{A}], [\ell_{A}, r_{A}]) \in \Psi_{\str}(T[p_{B}-1..r_{B}+1])$ holds. 
    $([p_{B}, q_{B}], [\ell_{B}, r_{B}]) \in \Psi$ follows from 
    $([p_{B}, q_{B}], [\ell_{B}, r_{B}]) \in \Psi_{\str}(T[p_{A}-1..r_{A}+1])$ and $T[p_{A}-1..r_{A}+1] \in \mathcal{Z}$. 
    $([p_{B}, q_{B}], [\ell_{B}, r_{B}]) \not \in \Psi_{\run}$ 
    follows from Lemma~\ref{lem:psi_equality_basic_property}~\ref{enum:psi_equality_basic_property:4} 
    because $([p_{A}, q_{A}], [\ell_{A}, r_{A}]) \not \in \Psi_{\run}$ and $T[p_{B}-1..r_{B}+1] = T[p_{A}-1..r_{A}+1]$.     
    $([p_{B}, q_{B}], [\ell_{B}, r_{B}]) \in (\Psi \setminus \Psi_{\run}) \cap \Psi_{\samp}$ 
    follows from $([p_{B}, q_{B}], [\ell_{B}, r_{B}]) \in \Psi$, 
    $([p_{B}, q_{B}], [\ell_{B}, r_{B}]) \in \Psi_{\samp}$, 
    and $([p_{B}, q_{B}], [\ell_{B}, r_{B}]) \not \in \Psi_{\run}$. 
    Therefore, $([p_{A}, q_{A}], [\ell_{A}, r_{A}]) \in \bigcup_{([p, q], [\ell, r]) \in (\Psi \setminus \Psi_{\run}) \cap \Psi_{\samp}} \Psi_{\str}(T[p-1..r+1])$ 
    follows from $([p_{B}, q_{B}], [\ell_{B}, r_{B}]) \in (\Psi \setminus \Psi_{\run}) \cap \Psi_{\samp}$ 
    and $([p_{A}, q_{A}], [\ell_{A}, r_{A}]) \in \Psi_{\str}(T[p_{B}-1..r_{B}+1])$. 

    We showed that $([p_{A}, q_{A}], [\ell_{A}, r_{A}]) \in \bigcup_{([p, q], [\ell, r]) \in (\Psi \setminus \Psi_{\run}) \cap \Psi_{\samp}} \Psi_{\str}(T[p-1..r+1])$ 
    for each interval attractor $([p_{A}, q_{A}], [\ell_{A}, r_{A}])$ in set $\Psi \setminus \Psi_{\run}$. 
    Therefore, statement (A) holds. 

    \textbf{Proof of statement (B).}    
    Similar to the proof of statement (A), 
    we consider an interval attractor $([p_{A}, q_{A}], [\ell_{A}, r_{A}])$ in set $\bigcup_{([p, q], [\ell, r]) \in (\Psi \setminus \Psi_{\run}) \cap \Psi_{\samp}} \Psi_{\str}(T[p-1..r+1])$. 
    Then, the set $(\Psi \setminus \Psi_{\run}) \cap \Psi_{\samp}$ contains an interval attractor $([p_{B}, q_{B}], [\ell_{B}, r_{B}])$ satisfying 
    $([p_{A}, q_{A}], [\ell_{A}, r_{A}]) \in \Psi_{\str}(T[p_{B}-1..r_{B}+1])$.
    Because of $([p_{A}, q_{A}], [\ell_{A}, r_{A}]) \in \Psi_{\str}(T[p_{B}-1..r_{B}+1])$, 
    $T[p_{B}-1..r_{B}+1] = T[p_{A}-1..r_{A}+1]$ follows from the definition of the subset $\Psi_{\str}(T[p_{B}-1..r_{B}+1])$.  

    We prove $([p_{A}, q_{A}], [\ell_{A}, r_{A}]) \not \in \Psi_{\run}$. 
    $(\Psi \setminus \Psi_{\run}) \cap \Psi_{\samp} = (\Psi \cap \Psi_{\samp}) \setminus \Psi_{\run}$ holds 
    because $\Psi_{\samp} \cap \Psi_{\run} = \emptyset$ follows from Lemma~\ref{lem:samp_basic_property}~\ref{enum:samp_basic_property:3}.     
    $([p_{B}, q_{B}], [\ell_{B}, r_{B}]) \not \in \Psi_{\run}$ follows from 
    $([p_{B}, q_{B}], [\ell_{B}, r_{B}]) \in (\Psi \setminus \Psi_{\run}) \cap \Psi_{\samp}$ and $(\Psi \setminus \Psi_{\run}) \cap \Psi_{\samp} = (\Psi \cap \Psi_{\samp}) \setminus \Psi_{\run}$. 
    $([p_{A}, q_{A}], [\ell_{A}, r_{A}]) \not \in \Psi_{\run}$ 
    follows from Lemma~\ref{lem:psi_equality_basic_property}~\ref{enum:psi_equality_basic_property:4} 
    because $([p_{B}, q_{B}], [\ell_{B}, r_{B}]) \not \in \Psi_{\run}$ and $T[p_{B}-1..r_{B}+1] = T[p_{A}-1..r_{A}+1]$.     

    We prove $([p_{A}, q_{A}], [\ell_{A}, r_{A}]) \in \Psi$. 
    Because of $([p_{B}, q_{B}], [\ell_{B}, r_{B}]) \in \Psi$, 
    the set $\mathcal{Z}$ contains string $T[p_{B}-1..r_{B}+1]$. 
    The subset $\Psi$ contains the interval attractor $([p_{A}, q_{A}], [\ell_{A}, r_{A}])$ 
    because (a) $T[p_{B}-1..r_{B}+1] \in \mathcal{Z}$, 
    (b) $T[p_{B}-1..r_{B}+1] = T[p_{A}-1..r_{A}+1]$, 
    and (c) $([p_{A}, q_{A}], [\ell_{A}, r_{A}]) \in \Psi_{\str}(T[p_{A}-1..r_{A}+1])$ hold. 

    For each interval attractor $([p_{A}, q_{A}], [\ell_{A}, r_{A}]) \in \bigcup_{([p, q], [\ell, r]) \in (\Psi \setminus \Psi_{\run}) \cap \Psi_{\samp}} \Psi_{\str}(T[p-1..r+1])$, 
    $([p_{A}, q_{A}], [\ell_{A}, r_{A}]) \in \Psi \setminus \Psi_{\run}$ holds 
    because we showed that $([p_{A}, q_{A}], [\ell_{A}, r_{A}]) \not \in \Psi_{\run}$ and $([p_{A}, q_{A}]$, $[\ell_{A}, r_{A}]) \in \Psi$ hold. 
    Therefore, statement (B) holds. 
    
    \textbf{Proof of statement (C).}
    We prove $\Psi_{\str}(T[p_{1}-1..r_{1}+1]) \cap \Psi_{\str}(T[p_{2}-1..r_{2}+1]) = \emptyset$ by contradiction. 
    We assume that $\Psi_{\str}(T[p_{1}-1..r_{1}+1]) \cap \Psi_{\str}(T[p_{2}-1..r_{2}+1]) \neq \emptyset$ holds. 
    Then, the set $\Psi_{\str}(T[p_{1}-1..r_{1}+1]) \cap \Psi_{\str}(T[p_{2}-1..r_{2}+1])$ contains 
    an interval attractor $([p, q], [\ell, r])$. 
    $T[p_{1}-1..r_{1}+1] = T[p_{2}-1..r_{2}+1]$ holds 
    because 
    $T[p-1..r+1] = T[p_{1}-1..r_{1}+1]$ and $T[p-1..r+1] = T[p_{2}-1..r_{2}+1]$ follow from $([p, q], [\ell, r]) \in \Psi_{\str}(T[p_{1}-1..r_{1}+1])$ and 
    $([p, q], [\ell, r]) \in \Psi_{\str}(T[p_{2}-1..r_{2}+1])$, respectively. 
    On the other hand, 
    $T[p_{1}-1..r_{1}+1] \neq T[p_{2}-1..r_{2}+1]$ follows from the definition of the sampling subset $\Psi_{\samp}$. 
    The two facts $T[p_{1}-1..r_{1}+1] = T[p_{2}-1..r_{2}+1]$ and $T[p_{1}-1..r_{1}+1] \neq T[p_{2}-1..r_{2}+1]$ yield a contradiction. 
    Therefore, $\Psi_{\str}(T[p_{1}-1..r_{1}+1]) \cap \Psi_{\str}(T[p_{2}-1..r_{2}+1]) = \emptyset$ must hold. 

\end{proof}

%%%%%%%%%%%%%%%%%%%%%%%%%%%%%%%%%%%%%%%%%%%%%%%%%%%%
We prove Lemma~\ref{lem:JA_main_lemma} using Proposition~\ref{prop:psi_bigcup_property} and Lemma~\ref{lem:CCP_property}. 

\begin{proof}[Proof of Lemma~\ref{lem:JA_main_lemma}]
Consider the $k$ interval attractors 
$([p_{1}, q_{1}], [\ell_{1}, r_{1}])$, $([p_{2}, q_{2}], [\ell_{2}, r_{2}])$, $\ldots$, $([p_{k}$, $q_{k}]$, $[\ell_{k}, r_{k}])$ 
in set $\Psi_{h_{Q}} \cap \Psi_{\samp}$. 
Let $\gamma_{s}$ of the attractor position of each interval attractor $([p_{s}, q_{s}]$, $[\ell_{s}, r_{s}])$. 
Set $\mathcal{J}_{A}(h_{Q})$ consists of $k$ weighted points $v_{1}, v_{2}, \ldots, v_{k}$ on grid $(\mathcal{X}_{A}(h), \mathcal{Y}_{A}(h))$. 
Here, for each integer $s \in [1, k]$, 
let $v_{s} = (\reverse(T[p_{s}-1..\gamma_{s}-1]), T[\gamma_{s}..r_{s}+1], |\Psi_{\str}(T[p_{s}-1..r_{s}+1])|, T[p_{s}-1..r_{s}+1])$, 
and the weighted point $v_{s}$ corresponds to the interval attractor $([p_{s}, q_{s}], [\ell_{s}, r_{s}])$. 

We introduce a set $\mathcal{I}_{A} \subseteq [1, k]$ of integers for this proof. 
This set $\mathcal{I}_{A}$ consists of integers in set $\{ 1, 2, \ldots, k \}$ such that 
each integer $s$ satisfies $L_{x} \preceq \reverse(T[p_{s}-1..\gamma_{s}-1]) \preceq L_{x^{\prime}}$ 
and $R_{y} \preceq T[\gamma_{s}..r_{s}+1] \preceq R_{y^{\prime}}$~(i.e.,  
$\mathcal{I}_{A} = \{ s \in [1, k] \mid L_{x} \preceq \reverse(T[p_{s}-1..\gamma_{s}-1]) \preceq L_{x^{\prime}} \text{ and } R_{y} \preceq T[\gamma_{s}..r_{s}+1] \preceq R_{y^{\prime}} \}$). 
Let $\Psi_{\answer} = (\Psi_{\CCP}(T[i..j]) \cap \Psi_{\lex}(T[\gamma_{Q}..r_{Q}+1])) \setminus \Psi_{\run}$ for simplicity. 

The following four statements are used to prove Lemma~\ref{lem:JA_main_lemma}. 

\begin{enumerate}[label=\textbf{(\roman*)}]
    \item $\{ ([p_{s}, q_{s}], [\ell_{s}, r_{s}]) \mid s \in \mathcal{I}_{A} \} \subseteq \Psi_{\CCP}(T[i..j]) \cap \Psi_{\lex}(T[\gamma_{Q}..r_{Q}+1]) \cap \Psi_{\samp}$;
    \item $\{ ([p_{s}, q_{s}], [\ell_{s}, r_{s}]) \mid s \in \mathcal{I}_{A} \} \supseteq \Psi_{\CCP}(T[i..j]) \cap \Psi_{\lex}(T[\gamma_{Q}..r_{Q}+1]) \cap \Psi_{\samp}$;
    \item $\{ ([p_{s}, q_{s}], [\ell_{s}, r_{s}]) \mid s \in \mathcal{I}_{A} \} = \Psi_{\answer} \cap \Psi_{\samp}$; 
    \item $|\Psi_{\answer}| = \sum_{([p, q], [\ell, r]) \in \Psi_{\answer} \cap \Psi_{\samp}} |\Psi_{\str}(T[p-1..r+1])|$.

    %\item $\rangesum(\mathcal{J}_{A}(h_{Q}), L_{x}, L_{x^{\prime}}, R_{y}, R_{y^{\prime}}) = \sum_{([p, q], [\ell, r]) \in \Psi_{\answer} \cap \Psi_{\samp}} |\Psi_{\str}(T[p-1..r+1])|$. 
\end{enumerate}

\paragraph{Proof of statement (i).}
Consider an interval attractor $([p_{s}, q_{s}], [\ell_{s}, r_{s}])$ in set $\{ ([p_{s}, q_{s}], [\ell_{s}, r_{s}]) \mid s \in \mathcal{I}_{A} \}$. 
Then, $L_{x} \preceq \reverse(T[p_{s}-1..\gamma_{s}-1]) \preceq L_{x^{\prime}}$, 
$R_{y} \preceq T[\gamma_{s}..r_{s}+1] \preceq R_{y^{\prime}}$, 
and $([p_{s}, q_{s}], [\ell_{s}, r_{s}]) \in \Psi_{h} \cap \Psi_{\samp}$ hold. 
$\reverse(T[i..\gamma_{Q}-1]) \prec \reverse(T[p_{s}-1..\gamma_{s}-1]) \prec \reverse(\# T[i..\gamma_{Q}-1])$ follows from $L_{x} \preceq \reverse(T[p_{s}-1..\gamma_{s}-1]) \preceq L_{x^{\prime}}$. 
Similarly, $T[\gamma_{Q}..j] \prec T[\gamma_{s}..r_{s}+1] \prec T[\gamma_{Q}..r_{Q} + 1]$ follows from $R_{y} \preceq T[\gamma_{s}..r_{s}+1] \preceq R_{y^{\prime}}$. 

We prove $T[\gamma_{s}..r_{s}+1] \prec (T[\gamma_{Q}..j] \cdot \#)$. 
For the interval attractor $I_{\capture}(i, j) = ([p_{Q}, q_{Q}], [\ell_{Q}, r_{Q}])$, 
Lemma~\ref{lem:CCP_property} \ref{enum:CCP_property:2} shows that 
$T[\gamma_{Q}..r_{Q} + 1] \prec T[\gamma_{Q}..j]\#$ holds. 
Therefore, $T[\gamma_{s}..r_{s}+1] \prec T[\gamma_{Q}..j]\#$ follows from 
$T[\gamma_{s}..r_{s}+1] \prec T[\gamma_{Q}..r_{Q} + 1]$ and $T[\gamma_{Q}..r_{Q} + 1] \prec T[\gamma_{Q}..j]\#$. 

We prove $([p_{s}, q_{s}], [\ell_{s}, r_{s}]) \in \Psi_{\CCP}(T[i..j]) \cap \Psi_{\lex}(T[\gamma_{Q}..r_{Q}+1]) \cap \Psi_{\samp}$. 
Lemma~\ref{lem:CCP_property}~\ref{enum:CCP_property:4} shows that 
$([p_{s}, q_{s}], [\ell_{s}, r_{s}]) \in \Psi_{\CCP}(T[i..j])$ holds 
because (a) $\reverse(T[i..\gamma_{Q}-1]) \prec \reverse(T[p_{s}-1..\gamma_{s}-1]) \prec \reverse(\# T[i..\gamma_{Q}-1])$, 
(b) $T[\gamma_{Q}..j] \prec T[\gamma_{s}..r_{s}+1] \prec T[\gamma_{Q}..j]\#$, 
and (c) $([p_{s}, q_{s}], [\ell_{s}, r_{s}]) \in \Psi_{h_{Q}}$. 
Because of $T[\gamma_{s}..r_{s}+1] \prec T[\gamma_{Q}..r_{Q} + 1]$, 
$([p_{s}, q_{s}], [\ell_{s}, r_{s}]) \in \Psi_{\lex}(T[\gamma_{Q}..r_{Q}+1])$ follows from the definition of the subset $\Psi_{\lex}(T[\gamma_{Q}..r_{Q}+1])$. 
Therefore, $([p_{s}, q_{s}], [\ell_{s}, r_{s}]) \in \Psi_{\CCP}(T[i..j]) \cap \Psi_{\lex}(T[\gamma_{Q}..r_{Q}+1]) \cap \Psi_{\samp}$ 
follows from $([p_{s}, q_{s}], [\ell_{s}, r_{s}]) \in \Psi_{\CCP}(T[i..j])$, $([p_{s}, q_{s}], [\ell_{s}, r_{s}]) \in \Psi_{\lex}(T[\gamma_{Q}..r_{Q}+1])$, 
and $([p_{s}, q_{s}], [\ell_{s}, r_{s}]) \in \Psi_{\samp}$.

We showed that $([p_{s}, q_{s}], [\ell_{s}, r_{s}]) \in \Psi_{\CCP}(T[i..j]) \cap \Psi_{\lex}(T[\gamma_{Q}..r_{Q}+1]) \cap \Psi_{\samp}$ holds 
for each interval attractor $([p_{s}, q_{s}], [\ell_{s}, r_{s}])$ in set $\{ ([p_{s}, q_{s}], [\ell_{s}, r_{s}]) \mid s \in \mathcal{I}_{A} \}$. 
Therefore, $\{ ([p_{s}, q_{s}], [\ell_{s}, r_{s}]) \mid s \in \mathcal{I}_{A} \} \subseteq \Psi_{\CCP}(T[i..j]) \cap \Psi_{\lex}(T[\gamma_{Q}..r_{Q}+1]) \cap \Psi_{\samp}$ holds.

\paragraph{Proof of statement (ii).}
Consider an interval attractor $([p, q], [\ell, r])$ in set $\Psi_{\CCP}(T[i..j]) \cap \Psi_{\lex}$ $(T[\gamma_{Q}..r_{Q}+1]) \cap \Psi_{\samp}$.
$T[\gamma..r+1] \prec T[\gamma_{Q}..r_{Q} + 1]$ follows from $([p, q], [\ell, r]) \in \Psi_{\lex}(T[\gamma_{Q}..r_{Q}+1])$. 
Because of $([p, q], [\ell, r]) \in \Psi_{\CCP}(T[i..j])$, 
Lemma~\ref{lem:CCP_property}~\ref{enum:CCP_property:1} shows that 
the level of the interval attractor $([p, q], [\ell, r])$ is $h_{Q}$. 
Because of $([p, q], [\ell, r]) \in \Psi_{h} \cap \Psi_{\samp}$, 
there exists an integer $s \in [1, k]$ satisfying $([p_{s}, q_{s}, [\ell_{s}, r_{s}]) = ([p, q], [\ell, r])$. 

We prove $L_{x} \preceq \reverse(T[p_{s}-1..\gamma_{s}-1]) \preceq L_{x^{\prime}}$. 
Lemma~\ref{lem:CCP_property}~\ref{enum:CCP_property:4} shows that 
$\reverse(T[i..\gamma_{Q}-1]) \prec \reverse(T[p_{s}-1..\gamma_{s}-1]) \prec \reverse(\# T[i..\gamma_{Q}-1])$ holds. 
$L_{x} \preceq \reverse(T[p_{s}-1..\gamma_{s}-1])$ follows from $\reverse(T[i..\gamma_{Q}-1]) \prec \reverse(T[p_{s}-1..\gamma_{s}-1])$ 
and $x = \min \{ s \in [1, d] \mid \reverse(T[i..\gamma_{Q}-1]) \prec L_{s} \}$. 
Similarly, $\reverse(T[p_{s}-1..\gamma_{s}-1]) \preceq L_{x^{\prime}}$ follows from 
$\reverse(T[p_{s}-1..\gamma_{s}-1]) \prec \reverse(\# T[i..\gamma_{Q}-1])$ and 
$x^{\prime} = \max \{ s \in [1, d] \mid L_{s} \prec \reverse(\# T[i..\gamma_{Q}-1]) \}$. 
Therefore, $L_{x} \preceq \reverse(T[p_{s}-1..\gamma_{s}-1]) \preceq L_{x^{\prime}}$ holds. 

Next, we prove $R_{y} \preceq T[\gamma_{s}..r_{s}+1] \preceq R_{y^{\prime}}$. 
Lemma~\ref{lem:CCP_property}~\ref{enum:CCP_property:4} shows that 
$T[\gamma_{Q}..j] \prec T[\gamma_{s}..r_{s}+1]$ holds.  
$T[\gamma_{s}..r_{s}+1] \prec T[\gamma_{Q}..r_{Q}+1]$ follows from $([p_{s}, q_{s}, [\ell_{s}, r_{s}]) \in \Psi_{\lex}(T[\gamma_{Q}..r_{Q}+1])$. 
$R_{y} \preceq T[\gamma_{s}..r_{s}+1]$ follows from 
$T[\gamma_{Q}..j] \prec T[\gamma_{s}..r_{s}+1]$ and $y = \min \{ s \in [1, d^{\prime}] \mid T[\gamma_{Q}..j]  \prec R_{s} \}$. 
$T[\gamma_{s}..r_{s}+1] \preceq R_{y^{\prime}}$ follows from 
$T[\gamma_{s}..r_{s}+1] \prec T[\gamma_{Q}..r_{Q}+1]$ and $y^{\prime} = \max \{ s \in [1, d^{\prime}] \mid R_{s} \prec T[\gamma_{Q}..r_{Q} + 1] \}$. 
Therefore, $R_{y} \preceq T[\gamma_{s}..r_{s}+1] \preceq R_{y^{\prime}}$ holds. 

We prove $([p, q], [\ell, r]) \in \{ ([p_{s}, q_{s}], [\ell_{s}, r_{s}]) \mid s \in \mathcal{I}_{A} \}$. 
$s \in \mathcal{I}_{A}$ follows from $L_{x} \preceq \reverse(T[p_{s}-1..\gamma_{s}-1]) \preceq L_{x^{\prime}}$ and $R_{y} \preceq T[\gamma_{s}..r_{s}+1] \preceq R_{y^{\prime}}$. 
$([p, q], [\ell, r]) \in \{ ([p_{s}, q_{s}], [\ell_{s}, r_{s}]) \mid s \in \mathcal{I}_{A} \}$ follows from 
$([p_{s}, q_{s}, [\ell_{s}, r_{s}]) = ([p, q], [\ell, r])$ and $s \in \mathcal{I}_{A}$. 

We showed that $([p, q], [\ell, r]) \in \{ ([p_{s}, q_{s}], [\ell_{s}, r_{s}]) \mid s \in \mathcal{I}_{A} \}$ holds for 
each interval attractor $([p, q], [\ell, r])$ in set $\Psi_{\CCP}(T[i..j]) \cap \Psi_{\lex}(T[\gamma_{Q}..r_{Q}+1]) \cap \Psi_{\samp}$. 
Therefore, $\{ ([p_{s}, q_{s}], [\ell_{s}, r_{s}]) \mid s \in \mathcal{I}_{A} \} \supseteq \Psi_{\CCP}(T[i..j]) \cap \Psi_{\lex}(T[\gamma_{Q}..r_{Q}+1]) \cap \Psi_{\samp}$ holds. 

\paragraph{Proof of statement (iii).}
$\{ ([p_{s}, q_{s}], [\ell_{s}, r_{s}]) \mid s \in \mathcal{I}_{A} \} = \Psi_{\CCP}(T[i..j]) \cap \Psi_{\lex}(T[\gamma_{Q}..r_{Q}+1]) \cap \Psi_{\samp}$ 
follows from statement (i) and statement (ii). 
Here, $\Psi_{\samp} \cap \Psi_{\run} = \emptyset$ follows from Lemma~\ref{lem:samp_basic_property}~\ref{enum:samp_basic_property:3}. 
Because of $\Psi_{\samp} \cap \Psi_{\run} = \emptyset$, 
$\Psi_{\CCP}(T[i..j]) \cap \Psi_{\lex}(T[\gamma_{Q}..r_{Q}+1]) \cap \Psi_{\samp} = ((\Psi_{\CCP}(T[i..j]) \cap \Psi_{\lex}(T[\gamma_{Q}..r_{Q}+1])) \setminus \Psi_{\run}) \cap \Psi_{\samp}$ holds. 
$\Psi_{\answer} = (\Psi_{\CCP}(T[i..j]) \cap \Psi_{\lex}(T[\gamma_{Q}..r_{Q}+1])) \setminus \Psi_{\run}$, 
and hence, statement (iii) follows from the following equation: 
\begin{equation*}
    \begin{split}
    \{ ([p_{s}, q_{s}], [\ell_{s}, r_{s}]) \mid s \in \mathcal{I}_{A} \} &= \Psi_{\CCP}(T[i..j]) \cap \Psi_{\lex}(T[\gamma_{Q}..r_{Q}+1]) \cap \Psi_{\samp} \\ 
    &= ((\Psi_{\CCP}(T[i..j]) \cap \Psi_{\lex}(T[\gamma_{Q}..r_{Q}+1])) \setminus \Psi_{\run}) \cap \Psi_{\samp} \\
    &= \Psi_{\answer} \cap \Psi_{\samp}.  
    \end{split}
\end{equation*}

\paragraph{Proof of statement (iv).}
Let $\mathcal{Z} = \{ T[p-1..r+1] \mid ([p, q], [\ell, r]) \in \Psi_{\CCP}(T[i..j]) \cap \Psi_{\lex}(T[\gamma_{Q}..r_{Q}+1]) \}$. 
Then, $\Psi_{\CCP}(T[i..j]) \cap \Psi_{\lex}(T[\gamma_{Q}..r_{Q}+1]) = \bigcup_{Z \in \mathcal{Z}} \Psi_{\str}(Z)$ follows from Proposition~\ref{prop:psi_bigcup_property}~\ref{enum:psi_bigcup_property:CCP} and Proposition~\ref{prop:psi_bigcup_property}~\ref{enum:psi_bigcup_property:lex}. 
We apply Proposition~\ref{prop:psi_bigcup_property}~\ref{enum:psi_bigcup_property:samp} to the set $\mathcal{Z}$ of strings. 
Then, Proposition~\ref{prop:psi_bigcup_property}~\ref{enum:psi_bigcup_property:samp} shows that the following equation holds: 
\begin{equation*}
    \begin{split}
    & |(\Psi_{\CCP}(T[i..j]) \cap  \Psi_{\lex}(T[\gamma_{Q}..r_{Q}+1])) \setminus \Psi_{\run}| \\ 
    &= \sum_{([p, q], [\ell, r]) \in ((\Psi_{\CCP}(T[i..j]) \cap \Psi_{\lex}(T[\gamma_{Q}..r_{Q}+1])) \setminus \Psi_{\run}) \cap \Psi_{\samp}} |\Psi_{\str}(T[p-1..r+1])|.
    \end{split}
\end{equation*}
Here, $\Psi_{\answer} = (\Psi_{\CCP}(T[i..j]) \cap \Psi_{\lex}(T[\gamma_{Q}..r_{Q}+1])) \setminus \Psi_{\run}$ holds. 
Therefore, statement (iv) holds. 

\paragraph{Proof of Lemma~\ref{lem:JA_main_lemma}.}
Statement (iv) shows that $|\Psi_{\answer}| = \sum_{([p, q], [\ell, r]) \in \Psi_{\answer} \cap \Psi_{\samp}} |\Psi_{\str}(T[p-1..r+1])|$ holds. 
Statement (iii) indicates that $\sum_{([p, q], [\ell, r]) \in \Psi_{\answer} \cap \Psi_{\samp}} |\Psi_{\str}(T[p-1..r+1])| = \sum_{s \in \mathcal{I}_{A}} |\Psi_{\str}(T[p_{s}-1..r_{s}+1])|$ holds. 
$\rangesum(\mathcal{J}_{A}(h_{Q}), L_{x}, L_{x^{\prime}}, R_{y}, R_{y^{\prime}}) = \sum_{s \in \mathcal{I}_{A}} |\Psi_{\str}(T[p_{s}-1..r_{s}+1])|$ 
follows from the definitions of the set $\mathcal{I}_{A}$ and range-sum query. 
Therefore, Lemma~\ref{lem:JA_main_lemma} follows from the following lemma:
\begin{equation*}
    \begin{split}
    |(\Psi_{\CCP}(T[i..j]) \cap  \Psi_{\lex}(T[\gamma_{Q}..r_{Q}+1])) \setminus \Psi_{\run}| &= |\Psi_{\answer}| \\ 
    &= \sum_{([p, q], [\ell, r]) \in \Psi_{\answer} \cap \Psi_{\samp}} |\Psi_{\str}(T[p-1..r+1])| \\ 
    &= \sum_{s \in \mathcal{I}_{A}} |\Psi_{\str}(T[p_{s}-1..r_{s}+1])| \\ 
    &= \rangesum(\mathcal{J}_{A}(h_{Q}), L_{x}, L_{x^{\prime}}, R_{y}, R_{y^{\prime}}).     
    \end{split}
\end{equation*}
\end{proof}

\subsubsection{Dynamic Data Structures for Ordered Set \texorpdfstring{$\mathcal{X}_{A}(h)$}{}}\label{subsubsec:JA_X_ds}
We present dynamic data structures to store the ordered set $\mathcal{X}_{A}(h)$ for each integer $h \in [0, H]$. 
Here, the ordered set $\mathcal{X}_{A}(h)$ consists of $d$ strings $L_{1}, L_{2}, \ldots, L_{d}$~($L_{1} \prec L_{2} \prec \cdots \prec L_{d}$). 
For each string $L_{b} \in \mathcal{X}_{A}(h)$, 
we introduce a sequence $\mathbf{Q}^{X}_{A}(h, L_{b})$ of $m$ weighted points $(x_{1}, y_{1}, w_{1}, e_{1})$, 
$(x_{2}, y_{2}, w_{2}, e_{2})$, $\ldots$, $(x_{m}, y_{m}, w_{m}, e_{m})$ in set $\mathcal{J}_{A}(h)$ of weighted points 
such that each weighted point $(x_{s}, y_{s}, w_{s}, e_{s})$ contains the string $L_{b}$ as its x-coordinate 
(i.e., $\{ (x_{s}, y_{s}, w_{s}, e_{s}) \mid s \in [1, m] \} = \{ (x, y, w, e) \in \mathcal{J}_{A}(h) \mid x = L_{b} \}$). 
The $m$ weighted points are sorted in lexicographic order of their y-coordinates $y_{1}, y_{2}, \ldots, y_{m} \in \Sigma^{+}$. 
If the sequence $\mathbf{Q}^{X}_{A}(h, L_{b})$ contains two weighted points of the same y-coordinate, 
then the two weighted points are sorted in lexicographical order of their identifiers. 
Formally, either of the following two conditions is satisfied for each integer $s \in [1, m-1]$: 
\begin{itemize}
    \item $y_{s} \prec y_{s+1}$;
    \item $y_{s} = y_{s+1}$ and $e_{s} \prec e_{s+1}$.
\end{itemize}

We store the ordered set $\mathcal{X}_{A}(h)$ using $d+1$ doubly linked lists $\mathbf{X}_{A}(h, L_{1}), \mathbf{X}_{A}(h, L_{2})$, $\ldots$, $\mathbf{X}_{A}(h, L_{d})$, and $\mathbf{L}^{X}_{A}(h)$. 
For each integer $b \in [1, d]$, 
the doubly linked list $\mathbf{X}_{A}(h, L_{b})$ consists of $m$ elements 
for the $m$ weighted points in sequence $\mathbf{Q}^{X}_{A}(h, L_{b})$. 
Each $s$-th element of $\mathbf{X}_{A}(h, L_{b})$ corresponds to the $s$-th weighted point $(x_{s}, y_{s}, w_{s}, e_{s})$ of $\mathbf{Q}^{X}_{A}(h, L_{b})$. 
Here, the following three statements hold: 
\begin{itemize}
    \item from the definition of set $\mathcal{J}_{A}(h)$, 
    the $s$-th weighted point $(x_{s}, y_{s}, w_{s}, e_{s})$ corresponds to an interval attractor $([p, q], [\ell, r])$ in the sampling subset $\Psi_{\samp}$;
    \item the interval attractor $([p, q], [\ell, r])$ corresponds to a node $u$ of the sequence $\mathbf{Q}_{\samp}$ introduced in Section~\ref{subsec:sample_query}; 
    \item the node $u$ is represented as an element $v$ of the doubly linked list $\mathbf{L}_{\samp, 1}$ introduced in Section~\ref{subsubsec:sample_ds}. 
\end{itemize}
The $s$-th element of the doubly linked list $\mathbf{X}_{A}(h, L_{b})$ stores a pointer to the element $v$ corresponding to the $s$-th weighted point $(x_{s}, y_{s}, w_{s}, e_{s})$. 
A list indexing data structure is used for quickly accessing to the elements of the doubly linked list $\mathbf{X}_{A}(h, L_{b})$. 
Here, the list indexing data structure is introduced in Section~\ref{subsubsec:sample_ds}. 

The last doubly-linked list $\mathbf{L}^{X}_{A}(h)$ consists of $d$ elements such that 
each $b$-th element corresponds to the $b$-th string $L_{b}$ of the ordered set $\mathcal{X}_{A}(h)$. 
Here, the $b$-th element stores a pointer to the $b$-th doubly linked list $\mathbf{X}_{A}(h, L_{b})$. 
Similar to the doubly linked list $\mathbf{X}_{A}(h, L_{b})$, 
a list indexing data structure is used for quickly accessing to the elements of the doubly linked list $\mathbf{L}^{X}_{A}(h)$. 
An order maintenance data structure~\cite{DBLP:conf/stoc/DietzS87} is used for comparing two elements of the doubly linked list $\mathbf{L}^{X}_{A}(h)$ with the total order of the doubly linked list. 
This data structure supports the following three operations: 
\begin{itemize}
\item the order operation determines whether or not an element precedes another element in doubly linked list $\mathbf{L}^{X}_{A}(h)$; 
\item the insertion operation inserts a new element right after a given element of doubly linked list $\mathbf{L}^{X}_{A}(h)$; 
\item the deletion operation deletes a given element from doubly linked list $\mathbf{L}^{X}_{A}(h)$.
\end{itemize}
The data structure supports these three operations in $O(1)$ time with $O(d B)$ bits of space for machine word size $B$. 
This order maintenance data structure is used to compare two given strings $L_{b}$ and $L_{b^{\prime}}$ 
in ordered set $\mathcal{X}_{A}(h)$ in $O(1)$ time. 

Overall, these $d+1$ doubly linked lists $\mathbf{X}_{A}(h, L_{1}), \mathbf{X}_{A}(h, L_{2})$, $\ldots$, $\mathbf{X}_{A}(h, L_{d})$, and $\mathbf{L}^{X}_{A}(h)$ require $O((d + |\mathcal{J}_{A}(h)|) B)$ bits of space in total for machine word size $B$. 
This is because 
there exists a one-to-one correspondence between (i) the weighted points of the set $\mathcal{J}_{A}(h)$ 
and (ii) the elements of the $d$ doubly linked lists $\mathbf{X}_{A}(h, L_{1}), \mathbf{X}_{A}(h, L_{2})$, $\ldots$, $\mathbf{X}_{A}(h, L_{d})$. 
$d \leq 2 + |\mathcal{J}_{A}(h)|$ follows from Lemma~\ref{lem:JA_size}~\ref{enum:JA_size:1}. 
Therefore, the dynamic data structures for the ordered set $\mathcal{X}_{A}(h)$ can be stored in $O(|\mathcal{J}_{A}(h)| B)$ bits of space. 

The following lemma states queries supported by the dynamic data structures for the ordered set $\mathcal{X}_{A}(h)$. 

\begin{lemma}\label{lem:JA_X_queries}
    Let $L_{1}, L_{2}, \ldots, L_{d}$~($L_{1} \prec L_{2} \prec \cdots \prec L_{d}$) be the $d$ strings of 
    the ordered set $\mathcal{X}_{A}(h)$ introduced in Section~\ref{subsec:RSC_comp_A} for an integer $h \in [0, H]$. 
    Consider the $d+1$ doubly linked lists $\mathbf{X}_{A}(h, L_{1}), \mathbf{X}_{A}(h, L_{2})$, $\ldots$, $\mathbf{X}_{A}(h, L_{d})$, $\mathbf{L}^{X}_{A}(h)$. 
    Using these $d+1$ doubly linked lists and the dynamic data structures of Section~\ref{subsubsec:rrdag_ds} 
    and Section~\ref{subsubsec:sample_ds}, 
    we can support the following three queries: 
    \begin{enumerate}[label=\textbf{(\roman*)}]
    \item \label{enum:JA_X_queries:1} 
    verify whether $L_{s} \prec L_{s^{\prime}}$ or not in $O(1)$ time 
    for the given $s$-th and $s^{\prime}$-th elements of the doubly linked list $\mathbf{L}^{X}_{A}(h)$;
    \item \label{enum:JA_X_queries:2} 
    for a given integer $s \in [1, d]$, 
    return an interval $[g, g + |L_{s}| - 1]$ in input string $T$ satisfying $\reverse(T[g..g + |L_{s}| - 1]) = L_{s}$ 
    in $O(H^{2} + \log n)$ time if $L_{s} \not \in \{ \varepsilon, \# \}$; 
    otherwise return the string $L_{s}$ in $O(1)$ time;
    \item \label{enum:JA_X_queries:3} 
    consider a given pair of an integer $s \in [1, d]$ and interval $[\alpha, \beta] \subseteq [1, n]$ in input string $T$. 
    Then, verify the following three conditions in $O(H^{2} + \log n)$ time: 
    (A) $\reverse(T[\alpha..\beta]) \prec L_{s}$, (B) $\reverse(T[\alpha..\beta]) = L_{s}$, and (C) $L_{s} \prec \reverse(\#T[\alpha..\beta])$.    
    \end{enumerate}
\end{lemma}
\begin{proof}
    The proof of Lemma~\ref{lem:JA_X_queries} is as follows:

    \textbf{Proof of Lemma~\ref{lem:JA_X_queries}(i).}
    We can verify whether $L_{s} \prec L_{s^{\prime}}$ or not in $O(1)$ time by the order maintenance data structure built on 
    doubly linked list $\mathbf{L}^{X}_{A}(h)$. 

    \textbf{Proof of Lemma~\ref{lem:JA_X_queries}(ii).}
    One of the following three conditions is satisfied: 
    (a) $s = 1$; (b) $s = d$; (c) $1 < s < d$. 
    For case (a), 
    $L_{s} = \varepsilon$ follows from the definition of the ordered set $\mathcal{X}_{A}(h)$.
    In this case, we can return the character $\varepsilon$ in $O(1)$ time. 
    For case (b), 
    $L_{s} = \#$ follows from the definition of the ordered set $\mathcal{X}_{A}(h)$.
    In this case, we can return the character $\#$ in $O(1)$ time. 
    
    For case (c), 
    the $s$-th doubly linked list $\mathbf{X}_{A}(h, L_{s})$ contains at least one element. 
    Let $(x_{1}, y_{1}, w_{1}, e_{1})$ be the weighted point corresponding to the first element of the doubly linked list $\mathbf{X}_{A}(h, L_{s})$. 
    Then, $x_{1} = L_{s}$ holds. 
    This weighted point $(x_{1}, y_{1}, w_{1}, e_{1})$ corresponds to an interval attractor $([p_{1}, q_{1}], [\ell_{1}, r_{1}])$ 
    in set $\Psi_{h} \cap \Psi_{\samp}$ satisfying 
    $\reverse(T[p_{1}-1..\gamma_{1}-1]) = x_{1}$ for the attractor position $\gamma_{1}$ of the $([p_{1}, q_{1}], [\ell_{1}, r_{1}])$. 
    Here, the interval attractor $([p_{1}, q_{1}], [\ell_{1}, r_{1}])$ is represented as a node $u$ of 
    the sequence $\mathbf{Q}_{\samp}$ introduced in Section~\ref{subsec:sample_query}.     
    $\reverse(T[p_{1}-1..\gamma_{1}-1]) = L_{s}$ holds, 
    and hence, we can return interval $[p_{1}-1, \gamma_{1}-1]$ as the answer to the query of Lemma~\ref{lem:JA_X_queries}~\ref{enum:JA_X_queries:2}.

    In this case, we answer the query of Lemma~\ref{lem:JA_X_queries}~\ref{enum:JA_X_queries:2} in two phases. 
    In the first phase, we obtain the node $u$ of the sequence $\mathbf{Q}_{\samp}$. 
    The $s$-th element of doubly linked list $\mathbf{L}^{X}_{A}(h)$ stores 
    a pointer to the doubly linked list $\mathbf{X}_{A}(h, L_{s})$.     
    The first element of the doubly linked list $\mathbf{X}_{A}(h, L_{s})$ stores 
    a pointer to the element representing the node $u$ 
    in the doubly linked list of Section~\ref{subsubsec:sample_ds}. 
    The $s$-th element can be accessed in $O(\log d)$ time by the list indexing data structure built on the doubly linked list $\mathbf{L}^{X}_{A}(h)$. 
    Here, $d = O(n^{2})$ follows from Lemma~\ref{lem:JA_size}~\ref{enum:JA_size:3}. 
    Therefore, we can obtain the node $u$ in $O(\log n)$ time. 

    In the second phase, we return interval $[p_{1}-1, \gamma_{1}-1]$ as the answer to the query of Lemma~\ref{lem:JA_X_queries}~\ref{enum:JA_X_queries:2}.     
    We recover the interval attractor $([p_{1}, q_{1}], [\ell_{1}, r_{1}])$ from the node $u$ 
    in $O(\log n)$ time by the algorithm presented in Section~\ref{subsubsec:computation_delta_samp}. 
    The attractor position $\gamma_{1}$ can be computed in $O(H^{2})$ time 
    by attractor position query $\attrQ(([p_{1}, q_{1}], [\ell_{1}, r_{1}]))$. 
    Therefore, the second phase takes $O(H^{2} + \log n)$ time. 

    Finally, Lemma~\ref{lem:JA_X_queries}~\ref{enum:JA_X_queries:2} holds. 
    
    \textbf{Proof of Lemma~\ref{lem:JA_X_queries}(iii).}
    One of the following three conditions is satisfied: 
    (a) $L_{s} = \varepsilon$; (b) $L_{s} = \#$; (c) $L_{s} \not \in \{ \varepsilon, \# \}$. 
    For case (a), 
    $\reverse(T[\alpha..\beta]) \prec L_{s}$ does not hold. 
    For case (b)
    $\reverse(T[\alpha..\beta]) \prec L_{s}$ holds because the substring $T[\alpha..\beta]$ does not contain the character $\#$. 
    For case (c),     
    there exists an interval $[g, g + |L_{s}| - 1]$ in input string $T$ 
    satisfying $\reverse(T[g..g + |L_{s}| - 1]) = L_{s}$. 
    Here, this interval $[g, g + |L_{s}| - 1]$ can be obtained by the query of Lemma~\ref{lem:JA_X_queries}~\ref{enum:JA_X_queries:2}. 
    In this case, 
    we know that two intervals representing two strings $\reverse(T[\alpha..\beta])$ and $L_{s}$. 
    Therefore, we can verify whether $\reverse(T[\alpha..\beta]) \prec L_{s}$ or not 
    using one reversed LCE query and $O(1)$ random access queries. 
    Similarly, 
    we can verify 
    $\reverse(T[\alpha..\beta]) = L_{s}$ and $L_{s} \prec \reverse(\#T[\alpha..\beta])$ 
    using one reversed LCE query and $O(1)$ random access queries. 

    The query of Lemma~\ref{lem:JA_X_queries}~\ref{enum:JA_X_queries:2} takes $O(H^{2} + \log n)$ time. 
    Theorem~\ref{theo:rr_dag_summary} shows that 
    the reversed LCE query takes $O(H)$ time. 
    Similarly, 
    $O(1)$ random access queries take $O(H)$ time. 
    Therefore, 
    The query of Lemma~\ref{lem:JA_X_queries}~\ref{enum:JA_X_queries:3} can be answered in $O(H^{2} + \log n)$ time. 
            
\end{proof}

\subsubsection{Dynamic Data Structures for Ordered Set \texorpdfstring{$\mathcal{Y}_{A}(h)$}{}}\label{subsubsec:JA_Y_ds}
For each integer $h \in [0, H]$, 
we present dynamic data structures to store the ordered set $\mathcal{Y}_{A}(h)$ 
using an approach similar to the one used for storing the ordered set $\mathcal{X}_{A}(h)$. 
Here, the ordered set $\mathcal{Y}_{A}(h)$ consists of $d^{\prime}$ strings $R_{1}, R_{2}, \ldots, R_{d^{\prime}}$~($R_{1} \prec R_{2} \prec \cdots \prec R_{d^{\prime}}$). 
For each string $R_{b} \in \mathcal{Y}_{A}(h)$, 
we introduce a sequence $\mathbf{Q}^{Y}_{A}(h, R_{b})$ of $m$ weighted points $(x_{1}, y_{1}, w_{1}, e_{1})$, 
$(x_{2}, y_{2}, w_{2}, e_{2})$, $\ldots$, $(x_{m}, y_{m}, w_{m}, e_{m})$ in set $\mathcal{J}_{A}(h)$ of weighted points 
such that each weighted point $(x_{s}, y_{s}, w_{s}, e_{s})$ contains the string $R_{b}$ as its y-coordinate 
(i.e., $\{ (x_{s}, y_{s}, w_{s}, e_{s}) \mid s \in [1, m] \} = \{ (x, y, w, e) \in \mathcal{J}_{A}(h) \mid y = R_{b} \}$). 
The $m$ weighted points are sorted in lexicographic order of their x-coordinates $x_{1}, x_{2}, \ldots, x_{m} \in \Sigma^{+}$. 
If the sequence $\mathbf{Q}^{Y}_{A}(h, R_{b})$ contains two weighted points of the same x-coordinate, 
then the two weighted points are sorted in lexicographical order of their identifiers. 
Formally, either of the following two conditions is satisfied for each integer $s \in [1, m-1]$: 
\begin{itemize}
    \item $x_{s} \prec x_{s+1}$;
    \item $x_{s} = x_{s+1}$ and $e_{s} \prec e_{s+1}$.
\end{itemize}

We store the ordered set $\mathcal{Y}_{A}(h)$ using $d^{\prime}+1$ doubly linked lists $\mathbf{Y}_{A}(h, R_{1}), \mathbf{Y}_{A}(h, R_{2})$, $\ldots$, $\mathbf{Y}_{A}(h, R_{d^{\prime}})$, and $\mathbf{L}^{Y}_{A}(h)$. 
For each integer $b \in [1, d^{\prime}]$, 
the doubly linked list $\mathbf{Y}_{A}(h, R_{b})$ consists of $m$ elements 
for the $m$ weighted points in sequence $\mathbf{Q}^{Y}_{A}(h, R_{b})$. 
Each $s$-th element of $\mathbf{Y}_{A}(h, R_{b})$ corresponds to the $s$-th weighted point $(x_{s}, y_{s}, w_{s}, e_{s})$ of $\mathbf{Q}^{Y}_{A}(h, R_{b})$. 
Similar to Section~\ref{subsubsec:JA_X_ds}, 
the $s$-th weighted point $(x_{s}, y_{s}, w_{s}, e_{s})$ corresponds to an element $v$ of the doubly linked list introduced in Section~\ref{subsubsec:sample_ds}. 
The $s$-th element of $\mathbf{Y}_{A}(h, R_{b})$ stores a pointer to the element $v$ corresponding to the $s$-th weighted point 
$(x_{s}, y_{s}, w_{s}, e_{s})$. 
A list indexing data structure is used for quickly accessing to the elements of the doubly linked list $\mathbf{Y}_{A}(h, R_{b})$. 

The last doubly-linked list $\mathbf{L}^{Y}_{A}(h)$ consists of $d^{\prime}$ elements such that 
each $b$-th element corresponds to the $b$-th string $R_{b}$ of the ordered set $\mathcal{Y}_{A}(h)$. 
Here, the $b$-th element stores a pointer to the $b$-th doubly linked list $\mathbf{Y}_{A}(h, R_{b})$. 
Similar to the doubly linked list $\mathbf{L}^{Y}_{A}(h)$ of Section~\ref{subsubsec:JA_X_ds}, 
list indexing and order maintenance data structures are built on the doubly-linked list $\mathbf{L}^{Y}_{A}(h)$. 
These $d^{\prime}+1$ doubly linked lists $\mathbf{Y}_{A}(h, R_{1}), \mathbf{Y}_{A}(h, R_{2})$, $\ldots$, $\mathbf{Y}_{A}(h, R_{d^{\prime}})$, $\mathbf{L}^{Y}_{A}(h)$ require 
$O((d^{\prime} + |\mathcal{J}_{A}(h)|) B)$ bits of space in total for machine word size $B$. 
$d^{\prime} \leq 2 + |\mathcal{J}_{A}(h)|$ follows from Lemma~\ref{lem:JA_size}~\ref{enum:JA_size:1}. 
Therefore, the dynamic data structures for the ordered set $\mathcal{Y}_{A}(h)$ can be stored in $O(|\mathcal{J}_{A}(h)| B)$ bits of space.

The following lemma states queries supported by the dynamic data structures for the ordered set $\mathcal{Y}_{A}(h)$. 

\begin{lemma}\label{lem:JA_Y_queries}
    Let $L_{1}, L_{2}, \ldots, L_{d}$~($L_{1} \prec L_{2} \prec \cdots \prec L_{d}$) be the $d$ strings of 
    the ordered set $\mathcal{X}_{A}(h)$ introduced in Section~\ref{subsec:RSC_comp_A} for an integer $h \in [0, H]$. 
    Similarly, 
    let $R_{1}, R_{2}, \ldots, R_{d^{\prime}}$~($R_{1} \prec R_{2} \prec \cdots \prec R_{d^{\prime}}$) be 
    the $d^{\prime}$ strings of the ordered set $\mathcal{Y}_{A}(h)$ introduced in Section~\ref{subsec:RSC_comp_A}. 
    Consider the $d^{\prime}+1$ doubly linked lists $\mathbf{Y}_{A}(h, R_{1}), \mathbf{Y}_{A}(h, R_{2})$, $\ldots$, $\mathbf{Y}_{A}(h, R_{d^{\prime}})$, $\mathbf{L}^{Y}_{A}(h)$. 
    Using these $d^{\prime}+1$ doubly linked lists and the dynamic data structures of Section~\ref{subsubsec:rrdag_ds} 
    and Section~\ref{subsubsec:sample_ds}, 
    we can support the following four queries: 
    \begin{enumerate}[label=\textbf{(\roman*)}]
    \item \label{enum:JA_Y_queries:1} 
    verify whether $R_{s} \prec R_{s^{\prime}}$ or not in $O(1)$ time 
    for the given $s$-th and $s^{\prime}$-th elements of the doubly linked list $\mathbf{L}^{Y}_{A}(h)$;
    \item \label{enum:JA_Y_queries:2} 
    for a given integer $s \in [1, d^{\prime}]$, 
    return an interval $[g, g + |R_{s}| - 1]$ in input string $T$ satisfying $T[g..g + |R_{s}| - 1] = R_{s}$ 
    in $O(H^{2} + \log n)$ time if $R_{s} \not \in \{ \varepsilon, \# \}$; 
    otherwise return the string $R_{s}$ in $O(1)$ time;
    \item \label{enum:JA_Y_queries:3} 
    consider a given pair of an integer $s \in [1, d^{\prime}]$ and interval $[\alpha, \beta] \subseteq [1, n]$ in input string $T$. 
    Then, verify the following four conditions in $O(H^{2} + \log n)$ time: 
    (A) $T[\alpha..\beta] \prec R_{s}$, 
    (B) $T[\alpha..\beta] = R_{s}$, 
    (C) $R_{s} \prec T[\alpha..\beta]$, 
    and (D) $R_{s} \prec T[\alpha..\beta]\#$;
    \item \label{enum:JA_Y_queries:4} 
    consider a given triplet of three integers $\tau, \tau^{\prime} \in [1, d]$ ($\tau \leq \tau^{\prime}$) 
    and $s \in [1, d^{\prime}]$. 
    If set $\mathcal{J}_{A}(h)$ contains a weighed point $(x, y, w, e)$ 
    satisfying $L_{\tau} \preceq x \preceq L_{\tau^{\prime}}$ and $y = R_{s}$, 
    then return the interval attractor corresponding to the weighed point $(x, y, w, e)$ 
    in $O(H^{2} \log n + \log^{2} n)$ time.
    For answering this query, 
    we need the dynamic data structures for the ordered set $\mathcal{X}_{A}(h)$ introduced in Section~\ref{subsubsec:JA_X_ds}.     
    \end{enumerate}
\end{lemma}
\begin{proof}
    The proof of Lemma~\ref{lem:JA_Y_queries} is as follows. 

    \textbf{Proof of Lemma~\ref{lem:JA_Y_queries}(i).}
    We can verify whether $L_{s} \prec L_{s^{\prime}}$ or not in $O(1)$ time by the order maintenance data structure built on 
    doubly linked list $\mathbf{L}^{Y}_{A}(h)$. 

    \textbf{Proof of Lemma~\ref{lem:JA_Y_queries}(ii).}
    Lemma~\ref{lem:JA_Y_queries}~\ref{enum:JA_Y_queries:2} corresponds to Lemma~\ref{lem:JA_X_queries}~\ref{enum:JA_X_queries:2}. 
    Therefore, 
    Lemma~\ref{lem:JA_Y_queries}~\ref{enum:JA_Y_queries:2} can be proved using 
    a similar approach as for Lemma~\ref{lem:JA_X_queries}~\ref{enum:JA_X_queries:2}. 

    \textbf{Proof of Lemma~\ref{lem:JA_Y_queries}(iii).}
    Lemma~\ref{lem:JA_Y_queries}~\ref{enum:JA_Y_queries:3} can be proved using 
    a similar approach as for Lemma~\ref{lem:JA_X_queries}~\ref{enum:JA_X_queries:3}. 
    The detailed proof of Lemma~\ref{lem:JA_Y_queries}~\ref{enum:JA_Y_queries:3} is as follows. 
    
    One of the following three conditions is satisfied: 
    (a) $R_{s} = \varepsilon$; (b) $R_{s} = \#$; (c) $R_{s} \not \in \{ \varepsilon, \# \}$. 
    For case (a), 
    $T[\alpha..\beta] \prec R_{s}$ does not hold. 
    For case (b)
    $T[\alpha..\beta] \prec R_{s}$ holds because the substring $T[\alpha..\beta]$ does not contain the character $\#$. 
    For case (c),     
    there exists an interval $[g, g + |R_{s}| - 1]$ in input string $T$ 
    satisfying $T[g..g + |R_{s}| - 1] = R_{s}$. 
    Here, this interval $[g, g + |R_{s}| - 1]$ can be obtained by the query of Lemma~\ref{lem:JA_Y_queries}~\ref{enum:JA_Y_queries:2}. 
    In this case, 
    we can verify whether $T[\alpha..\beta] \prec R_{s}$ or not 
    using one LCE query and at most two random access queries 
    because we know that two intervals representing two strings $T[\alpha..\beta]$ and $R_{s}$. 
    Similarly, we can verify 
    $T[\alpha..\beta] = R_{s}$, 
    $R_{s} \prec T[\alpha..\beta]$, 
    and $R_{s} \prec T[\alpha..\beta]\#$ using the same approach.     

    The query of Lemma~\ref{lem:JA_Y_queries}~\ref{enum:JA_Y_queries:2} takes $O(H^{2} + \log n)$ time. 
    Theorem~\ref{theo:rr_dag_summary} shows that 
    the LCE query takes $O(H)$ time. 
    Similarly, 
    $O(1)$ random access queries take $O(H^{2})$ time. 
    Therefore, the query of Lemma~\ref{lem:JA_Y_queries}~\ref{enum:JA_Y_queries:3} can be answered in $O(H^{2} + \log n)$ time.

    \textbf{Proof of Lemma~\ref{lem:JA_Y_queries}(iv).}
    We leverage the doubly linked list $\mathbf{Y}_{A}(h, R_{s})$ for this query. 
    Let $(x_{1}, y_{1}, w_{1}, e_{1})$, $(x_{2}, y_{2}, w_{2}, e_{2})$, $\ldots$, 
    $(x_{m}, y_{m}, w_{m}, e_{m})$ ($x_{1} \preceq x_{2} \preceq \cdots \preceq x_{m}$) be 
    the weighted points in set $\{ (x, y, w, e) \in \mathcal{J}_{A}(h) \mid y = R_{s} \}$. 
    Then, each $b$-th weighted point $(x_{b}, y_{b}, w_{b}, e_{b})$ corresponds to the $b$-th element of the doubly linked list $\mathbf{Y}_{A}(h, R_{s})$. 
    On the other hand, the $b$-th weighted point $(x_{b}, y_{b}, w_{b}, e_{b})$ corresponds to 
    an interval attractor $([p_{b}, q_{b}], [\ell_{b}, r_{b}])$ in set $\Psi_{h} \cap \Psi_{\samp}$.     
    Let $\lambda$ be the smallest integer in set $[1, m]$ satisfying 
    $L_{\tau} \preceq x_{\lambda}$. 
    Then, the $\lambda$-th weighted point $(x_{\lambda}, y_{\lambda}, w_{\lambda}, e_{\lambda})$ 
    satisfies $L_{\tau} \preceq x_{\lambda} \preceq L_{\tau^{\prime}}$ and $y_{\lambda} = R_{s}$. 
    Therefore, we can return the interval attractor corresponding to the weighted point $(x_{\lambda}, y_{\lambda}, w_{\lambda}, e_{\lambda})$ as the answer to the query of Lemma~\ref{lem:JA_Y_queries}~\ref{enum:JA_Y_queries:4}. 

    We find the smallest integer $\lambda$ by binary search on the doubly linked list $\mathbf{Y}_{A}(h, R_{s})$. 
    This binary search needs to verify whether $L_{\tau} \preceq x_{b}$ or not for an integer $b \in [1, m]$. 
    We execute this verification in the following four steps: 
    \begin{enumerate}[label=\textbf{(\arabic*)}]
    \item access the $b$-th element of the doubly linked list $\mathbf{Y}_{A}(h, R_{s})$ by 
    the list indexing data structure built on the doubly linked list. 
    Here, the $b$-th element stores a pointer to the element $v$ 
    corresponding to the interval attractor $([p_{b}, q_{b}], [\ell_{b}, r_{b}])$ 
    in the doubly linked list introduced in Section~\ref{subsubsec:sample_ds};
    \item recover the interval attractor $([p_{b}, q_{b}], [\ell_{b}, r_{b}])$ from the element $v$ 
    in $O(\log n)$ time by the algorithm of Section~\ref{subsubsec:computation_delta_samp};
    \item compute the attractor position $\gamma_{b}$ of the $([p_{b}, q_{b}], [\ell_{b}, r_{b}])$ 
    in $O(H^{2})$ time by 
    attractor position query $\attrQ(([p_{b}, q_{b}], [\ell_{b}, r_{b}]))$. 
    Here, $x_{b} = \reverse(T[p_{b}-1..\gamma_{b}-1])$ holds;
    \item verify whether $L_{\tau} \preceq x_{b}$ or not 
    in $O(H^{2} + \log n)$ time by Lemma~\ref{lem:JA_X_queries}~\ref{enum:JA_X_queries:3}.    
    \end{enumerate}
    The verification takes $O(H^{2} + \log n)$ time, 
    and the binary search executes this verification $O(\log m)$ times. 
    $m \leq |\mathcal{J}_{A}(h)|$ holds, 
    and Lemma~\ref{lem:JA_size}~\ref{enum:JA_size:3} shows that $|\mathcal{J}_{A}(h)| = O(n^{2})$ holds. 
    Therefore, this binary search takes $O(H^{2} \log n + \log^{2} n)$ time in total. 

    After executing the binary search, 
    we return the interval attractor $([p_{\lambda}, q_{\lambda}], [\ell_{\lambda}, r_{\lambda}])$ 
    as the answer to the query of Lemma~\ref{lem:JA_Y_queries}~\ref{enum:JA_Y_queries:4}. 
    This interval attractor can be obtained by the algorithm of Section~\ref{subsubsec:computation_delta_samp}. 
    Therefore, Lemma~\ref{lem:JA_Y_queries}~\ref{enum:JA_Y_queries:4} holds.    
\end{proof}

\subsubsection{Dynamic Data Structures for Set \texorpdfstring{$\mathcal{J}_{A}(h)$}{} of Weighted Points}\label{subsubsec:JA_ds}
For each integer $h \in [0, H]$, 
we present dynamic data structures to support range-sum query on set $\mathcal{J}_{A}(h)$ of weighted points. 
The set $\mathcal{J}_{A}(h)$ consists of $k$ weighted points $(x_{1}, y_{1}, w_{1}, e_{1})$, 
$(x_{2}, y_{2}, w_{2}, e_{2})$, $\ldots$, $(x_{k}, y_{k}, w_{k}, e_{k})$ ($e_{1} \prec e_{2} \prec \cdots e_{k}$). 
Here, the following two statements hold for each weighted point $(x_{s}, y_{s}, w_{s}, e_{s}) \in \mathcal{J}_{A}(h)$: 
\begin{itemize}
    \item the doubly linked list $\mathbf{X}_{A}(h, x_{s})$ of Section~\ref{subsubsec:JA_X_ds} 
    contains an element $v_{s}$ representing the weighted point $(x_{s}, y_{s}, w_{s}, e_{s})$; 
    \item the doubly linked list $\mathbf{Y}_{A}(h, y_{s})$ of Section~\ref{subsubsec:JA_Y_ds} 
    contains an element $v^{\prime}_{s}$ representing the weighted point $(x_{s}, y_{s}, w_{s}, e_{s})$. 
\end{itemize}

We store the set $\mathcal{J}_{A}(h)$ using a doubly linked list $\mathbf{L}_{A}(h)$ of $k$ elements. 
For each integer $s \in [1, k]$, 
the $s$-th element of the doubly linked list $\mathbf{L}_{A}(h)$ corresponds to the $s$-th weighted point $(x_{s}, y_{s}, w_{s}, e_{s})$. This element stores (i) the weight $w_{s}$ and (ii) two pointers to the two elements $v_{s}$ and $v^{\prime}_{s}$. 
List indexing and range-sum data structures are built on doubly linked list $\mathbf{L}_{A}(h)$. 
Here, the range-sum data structure is introduced in Section~\ref{sec:preliminary}, 
and we use this range-sum data structure to support range-count and range-sum queries on the set $\mathcal{J}_{A}(h)$ of weighted points. 
These dynamic data structures require $O(|\mathcal{J}_{A}(h)| B)$ bits of space in total for machine word size $B$.

As already explained in Section~\ref{sec:preliminary}, 
the range-sum data structure for the set $\mathcal{J}_{A}(h)$ requires 
the condition that 
any pair of elements, either from set $\mathcal{X}_{A}(h)$ or from set $\mathcal{Y}_{A}(h)$, can be compared in $O(1)$ time. 
This condition can be achieved using Lemma~\ref{lem:JA_X_queries}, Lemma~\ref{lem:JA_Y_queries}, and the pointers stored in the doubly linked list $\mathbf{L}_{A}(h)$. 

\subsubsection{Dynamic Data Structures and Algorithm for \texorpdfstring{$\RSCQA(i, j)$}{RSCA(i, j)}}\label{subsubsec:JA_subquery_ds}
We prove Lemma~\ref{lem:RSC_subquery_A_summary}, i.e., 
we show that $\RSCQA(i, j)$ can be answered in $O(H^{2} \log n + \log^{4} n)$ time using dynamic data structures of $O((H + |\mathcal{U}_{\RR}| + |\Psi_{\samp}|)B)$ bits of space for machine word size $B$. 
Here, $|\mathcal{U}_{\RR}|$ is the number of nodes in the RR-DAG of RLSLP $\mathcal{G}^{R}$. 

\paragraph{Data Structures.}
We answer $\RSCQA(i, j)$ using the following dynamic data structures: 
\begin{itemize}
    \item the dynamic data structures of $O(|\mathcal{U}_{\RR}|B)$ bits of space 
    for the RR-DAG of RLSLP $\mathcal{G}^{R}$ (Section~\ref{subsubsec:rrdag_ds}). 
    \item the dynamic data structures of $O(|\Psi_{\samp}|B)$ bits of space for sample query (Section~\ref{subsubsec:sample_ds});
    \item the dynamic data structures of $O(\sum_{h = 0}^{H} |\mathcal{J}_{A}(h)| B)$ bits of space for $(1 + H)$ sets $\mathcal{X}_{A}(0)$, $\mathcal{X}_{A}(1)$, $\ldots$, $\mathcal{X}_{A}(H)$ 
    (Section~\ref{subsubsec:JA_X_ds});
    \item the dynamic data structures of $O(\sum_{h = 0}^{H} |\mathcal{J}_{A}(h)| B)$ bits of space 
    for $(1 + H)$ sets $\mathcal{Y}_{A}(0)$, $\mathcal{Y}_{A}(1)$, $\ldots$, $\mathcal{Y}_{A}(H)$ 
    (Section~\ref{subsubsec:JA_Y_ds});
    \item the dynamic data structures of $O(\sum_{h = 0}^{H} |\mathcal{J}_{A}(h)| B)$ bits of space 
    for $(1 + H)$ sets $\mathcal{J}_{A}(0)$, $\mathcal{J}_{A}(1)$, $\ldots$, $\mathcal{J}_{A}(H)$ 
    (Section~\ref{subsubsec:JA_ds}).
\end{itemize}
$\sum_{h = 0}^{H} |\mathcal{J}_{A}(h)| = |\Psi_{\samp}|$ follows from Lemma~\ref{lem:JA_size}~\ref{enum:JA_size:2}. 
Therefore, these dynamic data structures can be stored in $O((|\mathcal{U}_{\RR}| + |\Psi_{\samp}|) B)$ bits of space. 

\paragraph{Algorithm.}
The algorithm for $\RSCQA(i, j)$ computes $|(\Psi_{\CCP}(T[i..j]) \cap \Psi_{\lex}(T[\gamma_{Q}..r_{Q}+1])) \setminus \Psi_{\run}|$. 
This algorithm leverages Lemma~\ref{lem:JA_main_lemma}, which shows that 
the size of set $(\Psi_{\CCP}(T[i..j]) \cap \Psi_{\lex}(T[\gamma_{Q}..r_{Q}+1])) \setminus \Psi_{\run}$ can be computed 
by one range-sum query on set $\mathcal{J}_{A}(h_{Q})$ of weighted points 
for the level $h_{Q}$ of interval attractor $([p_{Q}, q_{Q}], [\ell_{Q}, r_{Q}])$. 

The algorithm for $\RSCQA(i, j)$ consists of three phases. 
In the first phase, 
we compute the level $h_{Q}$ and attractor position $\gamma_{Q}$ of interval attractor $([p_{Q}, q_{Q}], [\ell_{Q}, r_{Q}])$. 
The interval attractor $([p_{Q}, q_{Q}], [\ell_{Q}, r_{Q}])$ can be obtained by capture query $\CAPQ([i, j])$. 
The level and attractor position of the interval attractor $([p_{Q}, q_{Q}], [\ell_{Q}, r_{Q}])$ 
can be computed by level query $\levelQ(([p_{Q}, q_{Q}]$, $[\ell_{Q}, r_{Q}]))$ and attractor position query $\attrQ(([p_{Q}, q_{Q}], [\ell_{Q}, r_{Q}]))$, respectively. 
The bottleneck of the first phase is capture query, which takes $O(H^{2} \log n)$ time. 

In the second phase, 
we compute four integers $x, x^{\prime}, y$, and $y^{\prime}$ of Lemma~\ref{lem:JA_main_lemma}. 
Here, 
\begin{itemize}
\item $x = \min \{ s \in [1, d] \mid \reverse(T[i..\gamma_{Q}-1]) \prec L_{s} \}$ 
for the $d$ strings $L_{1}, L_{2}, \ldots, L_{d}$~($L_{1} \prec L_{2} \prec \cdots \prec L_{d}$) in the ordered set $\mathcal{X}_{A}(h_{Q})$;
\item $x^{\prime} = \max \{ s \in [1, d] \mid L_{s} \prec \reverse(\# T[i..\gamma_{Q}-1]) \}$;
\item $y = \min \{ s \in [1, d^{\prime}] \mid T[\gamma_{Q}..j]  \prec R_{s} \}$ 
for the $d^{\prime}$ strings $R_{1}, R_{2}, \ldots, R_{d^{\prime}}$~($R_{1} \prec R_{2} \prec \cdots \prec R_{d^{\prime}}$) in the ordered set $\mathcal{Y}_{A}(h_{Q})$;
\item $y^{\prime} = \max \{ s \in [1, d^{\prime}] \mid R_{s} \prec T[\gamma_{Q}..r_{Q} + 1] \}$. 
\end{itemize}
We compute the two integers $x$ and $x^{\prime}$ by binary search on the $d$ strings $L_{1}, L_{2}, \ldots, L_{d}$. 
This binary search can be executed in $O((H^{2} + \log n)\log d)$ time using Lemma~\ref{lem:JA_X_queries}~\ref{enum:JA_X_queries:3}. 
Similarly, we compute the two integers $y$ and $y^{\prime}$ by binary search on the $d^{\prime}$ strings $R_{1}, R_{2}, \ldots, R_{d^{\prime}}$. 
This binary search can be executed in $O((H^{2} + \log n)\log d^{\prime})$ time using Lemma~\ref{lem:JA_Y_queries}~\ref{enum:JA_X_queries:3}. 
Lemma~\ref{lem:JA_size}~\ref{enum:JA_size:3} shows that $d + d^{\prime} = O(n^{2})$ holds. 
Therefore, the second phase takes $O((H^{2} + \log n) \log n)$ time. 

In the third phase, we compute $|(\Psi_{\CCP}(T[i..j]) \cap  \Psi_{\lex}(T[\gamma_{Q}..r_{Q}+1])) \setminus \Psi_{\run}|$ by 
range-sum query $\rangesum(\mathcal{J}_{A}(h_{Q}), L_{x}, L_{x^{\prime}}$, $R_{y}, R_{y^{\prime}})$. 
For executing this range-sum query, 
we need to access the following elements of two doubly linked lists $\mathbf{L}^{X}_{A}(h_{Q})$ and $\mathbf{L}^{Y}_{A}(h_{Q})$: 
\begin{itemize}
    \item the $x$-th and $x^{\prime}$-th elements of $\mathbf{L}^{X}_{A}(h_{Q})$ corresponding to 
    the two strings $L_{x}$ and $L_{x^{\prime}}$, respectively;
    \item the $y$-th and $y^{\prime}$-th elements of $\mathbf{L}^{Y}_{A}(h_{Q})$ corresponding to 
    the two strings $R_{y}$ and $R_{y^{\prime}}$, respectively. 
\end{itemize}
The $x$-th and $x^{\prime}$-th elements of $\mathbf{L}^{X}_{A}(h_{Q})$ can be accessed in $O(\log d)$ time by 
the list indexing data structure built on the doubly linked list $\mathbf{L}^{X}_{A}(h_{Q})$. 
Similarly, 
the $y$-th and $y^{\prime}$-th elements of $\mathbf{L}^{Y}_{A}(h_{Q})$ can be accessed in $O(\log d^{\prime})$ time. 
The range-sum query $\rangesum(\mathcal{J}_{A}(h_{Q}), L_{x}, L_{x^{\prime}}$, $R_{y}, R_{y^{\prime}})$ takes 
$O(\log^{4} k)$ time for the number $k$ of weighted points in set $\mathcal{J}_{A}(h_{Q})$. 
Lemma~\ref{lem:JA_size}~\ref{enum:JA_size:3} shows that $d, d^{\prime}, k = O(n^{2})$ holds. 
Therefore, the third phases takes $O(\log^{4} n)$ time. 

Finally, the algorithm for $\RSCQA(i, j)$ can be executed in $O(H^{2} \log n + \log^{4} n)$ time in total. 
Therefore, Lemma~\ref{lem:RSC_subquery_A_summary} holds.

\subsection{Two Subqueries \texorpdfstring{$\RSCQBX(i, j)$}{RSCB1(i, j)} and \texorpdfstring{$\RSCQBY(i, j)$}{RSCB2(i, j)}}\label{subsec:RSC_comp_B}
The goal of this subsection is to answer two subqueries $\RSCQBX(i, j)$ and $\RSCQBY(i, j)$. 
The following lemma states the summary of this subsection. 

\begin{lemma}\label{lem:RSC_subquery_B_summary}
The following two statements hold:
\begin{enumerate}[label=\textbf{(\roman*)}]
    \item 
    Using a dynamic data structure of $O((|\Psi_{\samp}| + |\mathcal{U}_{\RR}|) B)$ bits of space for machine word size $B$, 
    we can answer $\RSCQBX(i, j)$ (i.e., computing $|\Psi_{\CCP}(T[i..j]) \cap \Psi_{\lex}(T[\gamma_{Q}..r_{Q}+1]) \cap \Psi_{\run}|$) in $O(H^{2} \log n + \log^{4} n)$ time if the given RSC query $\RSCQ(i, j)$ satisfies condition (A) of RSC query.     
    \item 
    Using a dynamic data structure of $O((|\Psi_{\samp}| + |\mathcal{U}_{\RR}|) B)$ bits of space for machine word size $B$, 
    we can answer $\RSCQBY(i, j)$ (i.e., computing $|(\Psi_{\CCP}(T[i..j]) \cap \Psi_{\lex}(T[\gamma_{Q}..r_{Q}+1]) \cap \Psi_{\run}) \setminus \Psi_{\centerset}(C_{Q})|$) in $O(H^{2} \log n + \log^{4} n)$ time if the given RSC query $\RSCQ(i, j)$ satisfies any one of three conditions (B), (C), and (D) of RSC query. 
\end{enumerate}
Here, $|\mathcal{U}_{\RR}|$ is the number of nodes in the RR-DAG of RLSLP $\mathcal{G}^{R}$. 
\end{lemma}
\begin{proof}
See Section~\ref{subsubsec:JB_subquery_ds}.
\end{proof}

For answering two subqueries $\RSCQBX(i, j)$ and $\RSCQBY(i, j)$, 
we leverage range-sum query on weighted points corresponding to the interval attractors in 
set $\Psi_{h_{Q}} \cap \Psi_{\source} \cap \Psi_{\samp}$. 
For this purpose, 
we introduce a set $\mathcal{J}_{B}(h)$ of weighted points on a grid $(\mathcal{X}_{B}(h), \mathcal{Y}_{B}(h))$ 
for each integer $h \in [0, H]$. 
Here, the set $\mathcal{J}_{B}(h)$ and grid $(\mathcal{X}_{B}(h), \mathcal{Y}_{B}(h))$ are defined using 
set $\Psi_{h} \cap \Psi_{\source} \cap \Psi_{\samp}$ of $k$ interval attractors 
$([p_{1}, q_{1}], [\ell_{1}, r_{1}]), ([p_{2}, q_{2}], [\ell_{2}, r_{2}])$, 
$\ldots$, $([p_{k}, q_{k}], [\ell_{k}, r_{k}])$. 
For each interval attractor $([p_{s}, q_{s}], [\ell_{s}, r_{s}]) \in \Psi_{h} \cap \Psi_{\source} \cap \Psi_{\samp}$, 
Lemma~\ref{lem:mRecover_basic_property} shows that 
set $f_{\recover}(([p_{s}, q_{s}], [\ell_{s}, r_{s}])) \cap \Psi_{\mRecover}$ consists of an interval attractor 
$([\hat{p}_{s}, \hat{q}_{s}], [\hat{\ell}_{s}, \hat{r}_{s}])$ (i.e., $f_{\recover}(([p_{s}, q_{s}], [\ell_{s}, r_{s}])) \cap \Psi_{\mRecover} = \{ ([\hat{p}_{s}, \hat{q}_{s}], [\hat{\ell}_{s}, \hat{r}_{s}]) \}$). 
Let $\hat{\gamma}_{s}$ be the attractor position of the interval attractor $([\hat{p}_{s}, \hat{q}_{s}], [\hat{\ell}_{s}, \hat{r}_{s}])$.

\paragraph{Grid $(\mathcal{X}_{B}(h), \mathcal{Y}_{B}(h))$.}
$\mathcal{X}_{B}(h) \subseteq \Sigma^{*}$ is 
the ordered set of strings defined as the union of two sets 
$\{ \varepsilon, \# \}$ and $\{ \reverse(T[\hat{p}_{s}-1..\hat{\gamma}_{s}-1]) \mid s \in [1, k] \}$ of strings
(i.e., $\mathcal{X}_{B}(h) = \{ \varepsilon, \# \} \cup \{ \reverse(T[\hat{p}_{s}-1..\hat{\gamma}_{s}-1]) \mid s \in [1, k] \}$). 
$\varepsilon$ is the string of length $0$, 
and $\#$ is the largest character in the alphabet $\Sigma$ (see Section~\ref{sec:preliminary}). 
This ordered set $\mathcal{X}_{B}(h)$ consists of $d$ strings $L_{1}, L_{2}, \ldots, L_{d}$ 
that are sorted in lexicographical order (i.e., $L_{1} \prec L_{2} \prec \cdots \prec L_{d}$). 
$L_{1} = \varepsilon$ and $L_{d} = \#$ always hold because 
every string of the set $\{ \reverse(T[\hat{p}_{s}-1..\hat{\gamma}_{s}-1]) \mid s \in [1, k] \}$ does not contain the character $\#$. 

Similarly, $\mathcal{Y}_{B}(h) \subseteq \Sigma^{*}$ is the ordered set of strings defined as the union of two sets 
$\{ \varepsilon, \# \}$ and $\{ T[\hat{\gamma}_{s}..\hat{r}_{s}+1] \mid s \in [1, k] \}$ of strings. 
This ordered set $\mathcal{Y}_{B}(h)$ consists of $d^{\prime}$ strings $R_{1}, R_{2}, \ldots, R_{d^{\prime}}$ 
that are sorted in lexicographical order (i.e., $R_{1} \prec R_{2} \prec \cdots \prec R_{d^{\prime}}$). 
Similar to the ordered set $\mathcal{X}_{B}(h)$, 
$R_{1} = \varepsilon$ and $R_{d^{\prime}} = \#$ always hold. 

Grid $(\mathcal{X}_{B}(h), \mathcal{Y}_{B}(h))$ consists of two ordered sets $\mathcal{X}_{B}(h)$ and $\mathcal{Y}_{B}(h)$. 
Each string of the ordered set $\mathcal{X}_{B}(h)$ represents x-coordinate on two dimensional space. 
Similarly, each string of the ordered set $\mathcal{Y}_{B}(h)$ represents y-coordinate on two dimensional space.

\paragraph{Set $\mathcal{J}_{B}(h)$ of Weighted Points.}
Set $\mathcal{J}_{B}(h)$ consists of $k$ weighted points 
$(\reverse(T[\hat{p}_{1}-1..\hat{\gamma}_{1}-1]), T[\hat{\gamma}_{1}..\hat{r}_{1}+1], |\Psi_{\str}(T[p_{1}-1..r_{1}+1])| |f_{\recover}(([p_{1}, q_{1}], [\ell_{1}, r_{1}]))|, T[p_{1}-1..r_{1}+1])$, 
$(\reverse(T[\hat{p}_{2}-1..\hat{\gamma}_{2}-1]), T[\hat{\gamma}_{2}..\hat{r}_{2}+1], |\Psi_{\str}(T[p_{2}-1..r_{2}+1])||f_{\recover}(([p_{2}, q_{2}], [\ell_{2}, r_{2}]))| , T[p_{2}-1..r_{2}+1])$, 
$\ldots$, 
$(\reverse(T[\hat{p}_{k}-1..\hat{\gamma}_{k}-1]), T[\hat{\gamma}_{k}..\hat{r}_{k}+1], |\Psi_{\str}(T[p_{k}-1..r_{k}+1])| |f_{\recover}(([p_{k}, q_{k}], [\ell_{k}, r_{k}]))|, T[p_{k}-1..r_{k}+1])$ 
on grid $(\mathcal{X}_{B}(h), \mathcal{Y}_{B}(h))$. 
Each weighted point $(\reverse(T[\hat{p}_{s}-1..\hat{\gamma}_{s}-1]), T[\hat{\gamma}_{s}..\hat{r}_{s}+1], |\Psi_{\str}(T[p_{s}-1..r_{s}+1])|, T[p_{s}-1..r_{s}+1])$ corresponds to interval attractor $([p_{s}, q_{s}], [\ell_{s}, r_{s}])$ in set $\Psi_{h} \cap \Psi_{\source} \cap \Psi_{\samp}$. 
The details of the four elements of the weighted point $(\reverse(T[\hat{p}_{s}-1..\hat{\gamma}_{s}-1]), T[\hat{\gamma}_{s}..\hat{r}_{s}+1], |\Psi_{\str}(T[p_{s}-1..r_{s}+1])|, T[p_{s}-1..r_{s}+1])$ are as follows:
\begin{itemize}
    \item the first string $\reverse(T[\hat{p}_{s}-1..\hat{\gamma}_{s}-1])$ is the x-coordinate of this weighted point; 
    \item the second string $T[\hat{\gamma}_{s}..\hat{r}_{s}+1]$ is the y-coordinate of this weighted point; 
    \item the third integer $|\Psi_{\str}(T[p_{s}-1..r_{s}+1])| |f_{\recover}(([p_{s}, q_{s}], [\ell_{s}, r_{s}]))|$ is the weight of this weighted point;
    \item the fourth string $T[p_{s}-1..r_{s}+1]$ is the identifier of this weighted point.
\end{itemize}

From the definition of the sampling subset $\Psi_{\samp}$, 
the identifiers $T[p_{1}-1..r_{1}+1], T[p_{2}-1..r_{2}+1], \ldots, T[p_{k}-1..r_{k}+1]$ of all the weighted points in 
the set $\mathcal{J}_{B}(h)$ are different. 

The following lemma states the sizes of the three sets $\mathcal{X}_{A}(B)$, $\mathcal{Y}_{B}(h)$, and $\mathcal{J}_{B}(h)$. 

\begin{lemma}\label{lem:JB_size}
    The following three statements hold: 
    \begin{enumerate}[label=\textbf{(\roman*)}]
    \item \label{enum:JB_size:1} $|\mathcal{X}_{B}(h)|, |\mathcal{Y}_{B}(h)| \leq 2 + |\mathcal{J}_{B}(h)|$ for each integer $h \in [0, H]$;
    \item \label{enum:JB_size:2} $\sum_{h = 0}^{H} |\mathcal{J}_{B}(h)| \leq |\Psi_{\samp}|$;
    \item \label{enum:JB_size:3} $|\mathcal{X}_{B}(h)|, |\mathcal{Y}_{B}(h)|, |\mathcal{J}_{B}(h)| = O(n^{2})$.
    \end{enumerate}
\end{lemma}
\begin{proof}
The proof of Lemma~\ref{lem:JB_size} is as follows. 

\textbf{Proof of Lemma~\ref{lem:JB_size}(i).}
$|\mathcal{X}_{B}(h)|, |\mathcal{Y}_{B}(h)| \leq 2 + |\mathcal{J}_{B}(h)|$ follows from the definitions of the two ordered sets 
$\mathcal{X}_{B}(h)$ and $\mathcal{Y}_{B}(h)$. 

\textbf{Proof of Lemma~\ref{lem:JB_size}(ii).}
For each integer $h \in [0, H]$, 
there exists a one-to-one correspondence between the weighted points of set $\mathcal{J}_{B}(h)$ 
and the interval attractors of set $\Psi_{h} \cap \Psi_{\source} \cap \Psi_{\samp}$. 
The sampling subset $\Psi_{\samp}$ can be divided into $(1 + H)$ sets 
$\Psi_{0} \cap \Psi_{\samp}$, $\Psi_{1} \cap \Psi_{\samp}$, $\ldots$, $\Psi_{H} \cap \Psi_{\samp}$. 
Therefore, $\sum_{h = 0}^{H} |\mathcal{J}_{B}(h)| \leq |\Psi_{\samp}|$ follows from the equation: 
\begin{equation*}
    \begin{split}
        \sum_{h = 0}^{H} |\mathcal{J}_{B}(h)| &= \sum_{h = 0}^{H} |\Psi_{h} \cap \Psi_{\source} \cap \Psi_{\samp}| \\
        &= |\bigcup_{h=0}^{H} (\Psi_{h} \cap \Psi_{\source} \cap \Psi_{\samp})| \\
        &\leq |\bigcup_{h=0}^{H} (\Psi_{h} \cap \Psi_{\samp})| \\
        &= |\Psi_{\samp}|.
    \end{split}
\end{equation*}    

\textbf{Proof of Lemma~\ref{lem:JB_size}(iii).}
$|\mathcal{X}_{B}(h)|, |\mathcal{Y}_{B}(h)|, |\mathcal{J}_{B}(h)| = O(n^{2})$ can be proved using the same approach as for 
Lemma~\ref{lem:JA_size}~\ref{enum:JA_size:3}.    
\end{proof}

The following lemma shows that 
we can count the interval attractors in 
the two sets $\Psi_{\CCP}(T[i..j]) \cap \Psi_{\lex}(T[\gamma_{Q}..r_{Q}+1]) \cap \Psi_{\run}$ 
and $(\Psi_{\CCP}(T[i..j]) \cap \Psi_{\lex}(T[\gamma_{Q}..r_{Q}+1]) \cap \Psi_{\run}) \setminus \Psi_{\centerset}(C_{Q})$ 
by two range-sum queries on the set $\mathcal{J}_{B}(h_{Q})$ of weighted points.

\begin{lemma}\label{lem:JB_main_lemma}
For RSC query $\RSCQ(i, j)$, 
consider the set $\mathcal{J}_{B}(h_{Q})$ of weighted points on grid $(\mathcal{X}_{B}(h_{Q}), \mathcal{Y}_{B}(h_{Q}))$. 
Here, let $([p_{Q}, q_{Q}], [\ell_{Q}, r_{Q}])$ be interval attractor $I_{\capture}(i, j)$; 
let $\gamma_{Q}$, $C_{Q}, $and $h_{Q}$ be the attractor position, associated string, and level of the interval attractor $([p_{Q}, q_{Q}], [\ell_{Q}, r_{Q}])$, 
respectively; 
$\mathcal{X}_{B}(h_{Q}) = \{ L_{1}, L_{2}, \ldots, L_{d} \}$~($L_{1} \prec L_{2} \prec \cdots \prec L_{d}$); 
$\mathcal{Y}_{B}(h_{Q}) = \{ R_{1}, R_{2}, \ldots, R_{d^{\prime}} \}$~($R_{1} \prec R_{2} \prec \cdots \prec R_{d^{\prime}}$). 
Let $x, x^{\prime}, y, y^{\prime}$, and $\hat{y}$ be the five integers defined as follows: 
\begin{itemize}
\item $x = \min \{ s \in [1, d] \mid \reverse(T[i..\gamma_{Q}-1]) \prec L_{s} \}$; 
\item $x^{\prime} = \max \{ s \in [1, d] \mid L_{s} \prec \reverse(\# \cdot T[i..\gamma_{Q}-1]) \}$; 
\item $y = \min \{ s \in [1, d^{\prime}] \mid T[\gamma_{Q}..j] \prec R_{s} \}$; 
\item $y^{\prime} = \max \{ s \in [1, d^{\prime}] \mid R_{s} \prec T[\gamma_{Q}..r_{Q} + 1] \}$;  
\item $\hat{y} = \max \{ s \in [1, d^{\prime}] \mid R_{s} \prec T[\gamma_{Q}..\gamma_{Q} + 1 + \sum_{w = 1}^{h_{Q}+3} \mu(w)] \}$.
\end{itemize}

If $|\lcp(T[\gamma_{Q}..r_{Q} + 1], C_{Q}^{n+1})| \leq 1 + \sum_{w = 1}^{h_{Q}+3} \mu(w)$ holds, then 
the following equation holds: 
\begin{equation}\label{eq:x_set_RB_property:1}
      |\Psi_{\CCP}(T[i..j]) \cap \Psi_{\lex}(T[\gamma_{Q}..r_{Q}+1]) \cap \Psi_{\run}| = \rangesum(\mathcal{J}_{B}(h_{Q}), L_{x}, L_{x^{\prime}}, R_{y}, R_{y^{\prime}}).  
\end{equation}    
Otherwise~(i.e., $|\lcp(T[\gamma_{Q}..r_{Q} + 1], C_{Q}^{n+1})| > 1 + \sum_{w = 1}^{h_{Q}+3} \mu(w)$), 
the following equation holds: 
\begin{equation}\label{eq:x_set_RB_property:2}
      |(\Psi_{\CCP}(T[i..j]) \cap \Psi_{\lex}(T[\gamma_{Q}..r_{Q}+1]) \cap \Psi_{\run}) \setminus \Psi_{\centerset}(C_{Q})| = \rangesum(\mathcal{J}_{B}(h_{Q}), L_{x}, L_{x^{\prime}}, R_{y}, R_{\hat{y}}).
\end{equation}

\end{lemma}
\begin{proof}
See Section~\ref{subsubsec:RB_main_lemma_proof}.
\end{proof}

\subsubsection{Proof of Lemma~\ref{lem:JB_main_lemma}}\label{subsubsec:RB_main_lemma_proof}
Consider the $k$ interval attractors 
$([p_{1}, q_{1}], [\ell_{1}, r_{1}])$, $([p_{2}, q_{2}], [\ell_{2}, r_{2}])$, $\ldots$, $([p_{k}, q_{k}], [\ell_{k}, r_{k}])$ 
in set $(\Psi_{h} \cap \Psi_{\source}) \cap \Psi_{\samp}$. 
Let $\gamma_{s}$ and $C_{s}$ of the attractor position and associate string of each interval attractor $([p_{s}, q_{s}], [\ell_{s}, r_{s}])$, respectively. 
From the definition of the subset $\Psi_{\source}$, 
there exists an interval attractor $([p^{\prime}_{s}, q^{\prime}_{s}], [\ell^{\prime}_{s}, r^{\prime}_{s}]) \in \Psi_{\run} \cap \Psi_{h} \cap \Psi_{\centerset}(C_{s})$ 
such that its attractor position $\gamma^{\prime}_{s}$ is equal to $\gamma_{s} + |C_{s}|$.

We introduce two sets $\mathcal{I}_{B}, \mathcal{I}^{\prime}_{B} \subseteq [1, k]$ of integers for this proof. 
The former set $\mathcal{I}_{B}$ consists of integers in set $\{ 1, 2, \ldots, k \}$ such that 
each integer $s$ satisfies $L_{x} \preceq \reverse(T[p^{\prime}_{s}-1..\gamma^{\prime}_{s}-1]) \preceq L_{x^{\prime}}$ 
and $R_{y} \preceq T[\gamma^{\prime}_{s}..r^{\prime}_{s}+1] \preceq R_{y^{\prime}}$~(i.e., 
$\mathcal{I}_{B} = \{ s \in [1, k] \mid L_{x} \preceq \reverse(T[p^{\prime}_{s}-1..\gamma^{\prime}_{s}-1]) \preceq L_{x^{\prime}} \text{ and } R_{y} \preceq T[\gamma^{\prime}_{s}..r^{\prime}_{s}+1] \preceq R_{y^{\prime}} \}$). 
Similarly, 
the latter set $\mathcal{I}^{\prime}_{B}$ consists of integers in set $\{ 1, 2, \ldots, k \}$ such that 
each integer $s$ satisfies $L_{x} \preceq \reverse(T[p^{\prime}_{s}-1..\gamma^{\prime}_{s}-1]) \preceq L_{x^{\prime}}$ 
and $R_{y} \preceq T[\gamma^{\prime}_{s}..r^{\prime}_{s}+1] \preceq R_{\hat{y}}$. 

The following four propositions are used to prove Lemma~\ref{lem:JB_main_lemma}. 

%%%%%%%%%%%%%%%%%%%%%%%%%%%%

\begin{proposition}\label{prop:set_JB1_subseq}
If $|\lcp(T[\gamma_{Q}..r_{Q} + 1], C_{Q}^{n+1})| \leq 1 + \sum_{w = 1}^{h_{Q}+3} \mu(w)$, 
then the following equation holds: 
\begin{equation*}
\bigcup_{s \in \mathcal{I}_{B}} \Psi_{\sRecover}(([p_{s}, q_{s}], [\ell_{s}, r_{s}])) \subseteq \Psi_{\CCP}(T[i..j]) \cap \Psi_{\lex}(T[\gamma_{Q}..r_{Q}+1]) \cap \Psi_{\run};
\end{equation*}
\end{proposition}
\begin{proof}
Let $\zeta = 1 + \sum_{w = 1}^{h_{Q}+3} \mu(w)$ for simplicity. 
Consider an interval attractor $([p_{A}, q_{A}], [\ell_{A}, r_{A}]) \in \bigcup_{s \in \mathcal{I}_{B}}$ $\Psi_{\sRecover}(([p_{s}, q_{s}], [\ell_{s}, r_{s}]))$. 
Then, set $\mathcal{I}_{B}$ contains an integer $b$ satisfying $([p_{A}, q_{A}], [\ell_{A}, r_{A}]) \in \Psi_{\sRecover}(([p_{b}, q_{b}], [\ell_{b}, r_{b}]))$. 
Here, $([p^{\prime}_{s}, q^{\prime}_{s}], [\ell^{\prime}_{s}, r^{\prime}_{s}]) \in \Psi_{\sRecover}(([p_{b}, q_{b}], [\ell_{b}, r_{b}]))$;     
$([p_{A}, q_{A}], [\ell_{A}, r_{A}]) \in \Psi_{h_{Q}} \cap \Psi_{\run} \cap \Psi_{\centerset}(C_{b})$ 
follows from Lemma~\ref{lem:sRecover_basic_property}~\ref{enum:sRecover_basic_property:centerset}. 
Let $\gamma_{A}$ be the attractor position of the interval attractor $([p_{A}, q_{A}], [\ell_{A}, r_{A}])$. 

The following five statements are used to prove Proposition~\ref{prop:set_JB1_subseq}. 

\begin{enumerate}[label=\textbf{(\roman*)}]
    \item $|\lcp(T[\gamma^{\prime}_{b}..r^{\prime}_{b} + 1], T[\gamma_{Q}..r_{Q}+1])| \leq \zeta$; 
    \item $T[\gamma_{Q}..j] \prec T[\gamma_{A}..r_{A} + 1] \prec T[\gamma_{Q}..j]\#$ and $T[\gamma_{A}..r_{A} + 1] \prec T[\gamma_{Q}..r_{Q}+1]$; 
    \item $\reverse(T[i..\gamma_{Q}-1]) \prec \reverse(T[p_{A}-1..\gamma_{A}-1]) \prec \reverse(\#T[i..\gamma_{Q}-1])$; 
    \item $([p_{A}, q_{A}], [\ell_{A}, r_{A}]) \in \Psi_{\CCP}(T[i..j])$; 
    \item $([p_{A}, q_{A}], [\ell_{A}, r_{A}]) \in \Psi_{\lex}(T[\gamma_{Q}..r_{Q}+1])$.
\end{enumerate}
\textbf{Proof of statement (i).}
We prove $|\lcp(T[\gamma^{\prime}_{b}..r^{\prime}_{b} + 1], T[\gamma_{Q}..r_{Q}+1])| \leq \zeta$ by contradiction. 
We assume that $|\lcp(T[\gamma^{\prime}_{b}..r^{\prime}_{b} + 1], T[\gamma_{Q}..r_{Q}+1])| > \zeta$ holds. 
Because of $([p^{\prime}_{b}, q^{\prime}_{b}], [\ell^{\prime}_{b}, r^{\prime}_{b}]) \in \Psi_{h_{Q}} \cap \Psi_{\run} \cap \Psi_{\centerset}(C_{b})$, 
$|\lcp(T[\gamma^{\prime}_{b}..r^{\prime}_{b} + 1], C_{b}^{n+1})| > \zeta$ 
follows from the definition of the subset $\Psi_{\run}$. 
If $C_{b} = C_{Q}$, 
then $|\lcp(T[\gamma_{Q}..r_{Q} + 1], C_{Q}^{n+1})| > \zeta$ follows from 
$|\lcp(T[\gamma^{\prime}_{b}..r^{\prime}_{b} + 1], T[\gamma_{Q}..r_{Q}+1])| > \zeta$ and $|\lcp(T[\gamma^{\prime}_{b}..r^{\prime}_{b} + 1], C_{b}^{n+1})| > \zeta$. 
On the other hand, $|\lcp(T[\gamma_{Q}..r_{Q} + 1], C_{Q}^{n+1})| \leq \zeta$ follows from the premise of Proposition~\ref{prop:set_JB1_subseq}. 
The two facts $|\lcp(T[\gamma_{Q}..r_{Q} + 1], C_{Q}^{n+1})| > \zeta$ and $|\lcp(T[\gamma_{Q}..r_{Q} + 1], C_{Q}^{n+1})| \leq \zeta$ yield a contradiction. 

Otherwise (i.e., $C_{b} \neq C_{Q}$), 
we can apply Lemma~\ref{lem:associated_string_C}~\ref{enum:associated_string_C:2} to the two interval attractors 
$([p^{\prime}_{b}, q^{\prime}_{b}], [\ell^{\prime}_{b}, r^{\prime}_{b}])$ and 
$([p_{Q}, q_{Q}], [\ell_{Q}, r_{Q}])$ because 
$([p^{\prime}_{b}, q^{\prime}_{b}], [\ell^{\prime}_{b}, r^{\prime}_{b}]), ([p_{Q}, q_{Q}], [\ell_{Q}, r_{Q}]) \in \Psi_{h_{Q}}$ and $|\lcp(T[\gamma^{\prime}_{b}..r^{\prime}_{b} + 1]$, $T[\gamma_{Q}..r_{Q}+1])| > \zeta$. 
This lemma shows that $C_{b} = C_{Q}$ holds, but it contradicts $C_{b} \neq C_{Q}$. 
Therefore, $|\lcp(T[\gamma^{\prime}_{b}..r^{\prime}_{b} + 1], T[\gamma_{Q}..r_{Q}+1])| \leq \zeta$ must hold. 

\textbf{Proof of statement (ii).}
We prove $T[\gamma_{A}..r_{A} + 1] \prec T[\gamma_{Q}..r_{Q}+1]$. 
$R_{y} \preceq T[\gamma^{\prime}_{b}..r^{\prime}_{b}+1] \preceq R_{y^{\prime}}$ follows from $b \in \mathcal{I}_{B}$. 
$T[\gamma_{Q}..j] \prec T[\gamma^{\prime}_{b}..r^{\prime}_{b}+1] \prec T[\gamma_{Q}..r_{Q}+1]$ 
follows from $R_{y} \preceq T[\gamma^{\prime}_{b}..r^{\prime}_{b}+1] \preceq R_{y^{\prime}}$. 
Lemma~\ref{lem:sRecover_basic_property}~\ref{enum:sRecover_basic_property:lex} shows that 
$T[\gamma_{A}..r_{A} + 1] \prec T[\gamma_{Q}..r_{Q}+1]$ 
because $T[\gamma^{\prime}_{b}..r^{\prime}_{b}+1] \prec T[\gamma_{Q}..r_{Q}+1]$, 
$|\lcp(T[\gamma^{\prime}_{b}..r^{\prime}_{b} + 1], T[\gamma_{Q}..r_{Q}+1])| \leq \zeta$ (statement (i)), 
and 
$([p^{\prime}_{b}, q^{\prime}_{b}], [\ell^{\prime}_{b}, r^{\prime}_{b}]), ([p_{A}, q_{A}], [\ell_{A}, r_{A}]) \in \Psi_{\sRecover}(([p_{b}, q_{b}], [\ell_{b}, r_{b}]))$. 

We prove $T[\gamma_{Q}..j] \prec T[\gamma_{A}..r_{A} + 1]$. 
The lexicographic order $T[\gamma_{Q}..j] \prec T[\gamma^{\prime}_{b}..r^{\prime}_{b}+1] \prec T[\gamma_{Q}..r_{Q}+1]$ indicates that 
$j < r_{Q} + 1$ holds. 
$|\lcp(T[\gamma^{\prime}_{b}..r^{\prime}_{b} + 1], T[\gamma_{Q}..j])| \leq \zeta$ follows from 
$|\lcp(T[\gamma^{\prime}_{b}..r^{\prime}_{b} + 1], T[\gamma_{Q}..r_{Q}+1])| \leq \zeta$ and $j < r_{Q} + 1$. 
Lemma~\ref{lem:sRecover_basic_property}~\ref{enum:sRecover_basic_property:lex} shows that 
$T[\gamma_{A}..r_{A} + 1] \prec T[\gamma_{Q}..j]$ 
because $T[\gamma_{Q}..j] \prec T[\gamma^{\prime}_{b}..r^{\prime}_{b}+1]$, 
$|\lcp(T[\gamma^{\prime}_{b}..r^{\prime}_{b} + 1], T[\gamma_{Q}..j])| \leq \zeta$, 
and 
$([p^{\prime}_{b}, q^{\prime}_{b}], [\ell^{\prime}_{b}, r^{\prime}_{b}])$, 
$([p_{A}, q_{A}], [\ell_{A}, r_{A}]) \in \Psi_{\sRecover}(([p_{b}, q_{b}], [\ell_{b}, r_{b}]))$. 

We prove $T[\gamma_{A}..r_{A} + 1] \prec T[\gamma_{Q}..j]\#$. 
The string $T[\gamma_{Q}..j]$ is a prefix of string $T[\gamma^{\prime}_{b}..r^{\prime}_{b} + 1]$ 
(i.e., $\lcp(T[\gamma^{\prime}_{b}..r^{\prime}_{b} + 1], T[\gamma_{Q}..j]) = T[\gamma_{Q}..j]$) 
because $T[\gamma_{Q}..j] \prec T[\gamma^{\prime}_{b}..r^{\prime}_{b}+1] \prec T[\gamma_{Q}..r_{Q}+1]$ 
and $j < r_{Q} + 1$. 
$T[\gamma^{\prime}_{b}..r^{\prime}_{b} + 1] \prec T[\gamma_{Q}..j]\#$ follows from 
$\lcp(T[\gamma^{\prime}_{b}..r^{\prime}_{b} + 1], T[\gamma_{Q}..j]) = T[\gamma_{Q}..j]$. 
$|\lcp(T[\gamma^{\prime}_{b}..r^{\prime}_{b} + 1], T[\gamma_{Q}..j]\#)| \leq \zeta$ 
follows from $|\lcp(T[\gamma^{\prime}_{b}..r^{\prime}_{b} + 1], T[\gamma_{Q}..j])| \leq \zeta$. 
Lemma~\ref{lem:sRecover_basic_property}~\ref{enum:sRecover_basic_property:lex} shows that 
$T[\gamma_{A}..r_{A} + 1] \prec T[\gamma_{Q}..j]\#$
because $T[\gamma^{\prime}_{b}..r^{\prime}_{b} + 1] \prec T[\gamma_{Q}..j]\#$, 
$|\lcp(T[\gamma^{\prime}_{b}..r^{\prime}_{b} + 1], T[\gamma_{Q}..j])| \leq \zeta$, 
and 
$([p^{\prime}_{b}, q^{\prime}_{b}], [\ell^{\prime}_{b}, r^{\prime}_{b}]), ([p_{A}, q_{A}], [\ell_{A}, r_{A}]) \in \Psi_{\sRecover}(([p_{b}, q_{b}], [\ell_{b}, r_{b}]))$. 

\textbf{Proof of statement (iii).}
$L_{x} \preceq \reverse(T[p^{\prime}_{b}-1..\gamma^{\prime}_{b}-1]) \preceq L_{x^{\prime}}$ follows from 
$b \in \mathcal{I}_{B}$. 
$\reverse(T[i..\gamma_{Q}-1]) \prec \reverse(T[p^{\prime}_{b}-1..\gamma^{\prime}_{b}-1]) \prec \reverse(\#T[i..\gamma_{Q}-1])$ follows from $L_{x} \preceq \reverse(T[p^{\prime}_{b}-1..\gamma^{\prime}_{b}-1]) \preceq L_{x^{\prime}}$. 
Lemma~\ref{lem:psi_run_basic_property}~\ref{enum:psi_run_basic_property:7} shows that 
$\reverse(T[p^{\prime}_{b}-1..\gamma^{\prime}_{b}-1]) = \reverse(T[p_{A}-1..\gamma_{A}-1])$ holds. 
Therefore, $\reverse(T[i..\gamma_{Q}-1]) \prec \reverse(T[p_{A}-1..\gamma_{A}-1]) \prec \reverse(\#T[i..\gamma_{Q}-1])$ 
follows from $\reverse(T[i..\gamma_{Q}-1]) \prec \reverse(T[p^{\prime}_{b}-1..\gamma^{\prime}_{b}-1]) \prec \reverse(\#T[i..\gamma_{Q}-1])$ and $\reverse(T[p^{\prime}_{b}-1..\gamma^{\prime}_{b}-1]) = \reverse(T[p_{A}-1..\gamma_{A}-1])$.

\textbf{Proof of statement (iv).}
$\Psi_{\CCP}(T[i..j]) = \{ ([p, q], [\ell, r]) \in \Psi_{h_{Q}} \mid \reverse(T[i..\gamma_{Q}-1]) \prec \reverse(T[p-1..\gamma-1]) \prec \reverse(\#T[i..\gamma_{Q}-1]) \text{ and } T[\gamma_{Q}..j] \prec T[\gamma..r+1] \prec T[\gamma_{Q}..j]\# \}$
follows from Lemma~\ref{lem:CCP_property}~\ref{enum:CCP_property:4}. 
We already proved $([p_{A}, q_{A}], [\ell_{A}, r_{A}]) \in \Psi_{h_{Q}}$, 
$\reverse(T[i..\gamma_{Q}-1]) \prec \reverse(T[p_{A}-1..\gamma_{A}-1]) \prec \reverse(\#T[i..\gamma_{Q}-1])$ (statement (iii)), 
and $T[\gamma_{Q}..j] \prec T[\gamma_{A}..r_{A}+1] \prec T[\gamma_{Q}..j]\#$ (statement (ii)). 
Therefore, $([p_{A}, q_{A}], [\ell_{A}, r_{A}]) \in \Psi_{\CCP}(T[i..j])$ holds.

\textbf{Proof of statement (v).}
$T[\gamma_{A}..r_{A} + 1] \prec T[\gamma_{Q}..r_{Q}+1]$ follows from statement (ii). 
Therefore, 
$([p_{A}, q_{A}], [\ell_{A}, r_{A}]) \in \Psi_{\lex}(T[\gamma_{Q}..r_{Q}+1])$ follows from 
the definition of the subset $\Psi_{\lex}(T[\gamma_{Q}..r_{Q}+1])$. 

\textbf{Proof of Proposition~\ref{prop:set_JB1_subseq}.}
$([p_{A}, q_{A}], [\ell_{A}, r_{A}]) \in \Psi_{\CCP}(T[i..j]) \cap \Psi_{\lex}(T[\gamma_{Q}..r_{Q}+1]) \cap \Psi_{\run}$ 
follows from $([p_{A}, q_{A}], [\ell_{A}, r_{A}]) \in \Psi_{\run}$, 
$([p_{A}, q_{A}], [\ell_{A}, r_{A}]) \in \Psi_{\CCP}(T[i..j])$ (statement (iv)), 
and $([p_{A}, q_{A}], [\ell_{A}, r_{A}]) \in \Psi_{\lex}(T[\gamma_{Q}..r_{Q}+1])$ (statement (iii)).
Therefore, Proposition~\ref{prop:set_JB1_subseq} follows from the fact that 
$([p_{A}, q_{A}], [\ell_{A}, r_{A}]) \in \Psi_{\CCP}(T[i..j]) \cap \Psi_{\lex}(T[\gamma_{Q}..r_{Q}+1]) \cap \Psi_{\run}$ for each interval attractor $([p_{A}, q_{A}], [\ell_{A}, r_{A}]) \in \bigcup_{s \in \mathcal{I}_{B}}$ $\Psi_{\sRecover}(([p_{s}$, $q_{s}], [\ell_{s}, r_{s}]))$. 

\end{proof}

\begin{proposition}\label{prop:set_JB1_supseq}
If $|\lcp(T[\gamma_{Q}..r_{Q} + 1], C_{Q}^{n+1})| \leq 1 + \sum_{w = 1}^{h_{Q}+3} \mu(w)$, 
then the following equation holds: 
\begin{equation*}
\bigcup_{s \in \mathcal{I}_{B}} \Psi_{\sRecover}(([p_{s}, q_{s}], [\ell_{s}, r_{s}])) \supseteq \Psi_{\CCP}(T[i..j]) \cap \Psi_{\lex}(T[\gamma_{Q}..r_{Q}+1]) \cap \Psi_{\run}.
\end{equation*}
\end{proposition}
\begin{proof}
Let $\zeta = 1 + \sum_{w = 1}^{h_{Q}+3} \mu(w)$ for simplicity. 
Consider an interval attractor $([p_{A}, q_{A}], [\ell_{A}, r_{A}]) \in \Psi_{\CCP}(T[i..j]) \cap \Psi_{\lex}(T[\gamma_{Q}..r_{Q}+1]) \cap \Psi_{\run}$. 
Then, we show that 
the sampling subset $\Psi_{\samp}$ contains an interval attractor $([p_{C}, q_{C}], [\ell_{C}, r_{C}])$ satisfying 
$([p_{A}, q_{A}], [\ell_{A}, r_{A}]) \in \Psi_{\sRecover}(([p_{C}, q_{C}], [\ell_{C}, r_{C}]))$.     
Because of $([p_{A}, q_{A}], [\ell_{A}, r_{A}]) \in \Psi_{\run}$, 
Lemma~\ref{lem:recover_division_property}~\ref{enum:recover_division_property:1} shows that 
subset $\Psi_{\source}$ contains an interval attractor $([p_{B}, q_{B}], [\ell_{B}, r_{B}])$ satisfying 
$([p_{A}, q_{A}], [\ell_{A}, r_{A}]) \in f_{\recover}(([p_{B}, q_{B}], [\ell_{B}, r_{B}]))$. 
Here, $([p_{A}, q_{A}], [\ell_{A}, r_{A}]) \in \Psi_{\sRecover}(([p_{B}, q_{B}], [\ell_{B}, r_{B}]))$ follows from 
the definition of the subset $\Psi_{\sRecover}(([p_{B}$, $q_{B}], [\ell_{B}, r_{B}]))$; 
$([p_{B}, q_{B}], [\ell_{B}, r_{B}]) \not \in \Psi_{\run}$ holds 
because $\Psi_{\source} \cap \Psi_{\run} = \emptyset$. 
From the definition of the sampling subset $\Psi_{\samp}$, 
the sampling subset contains an interval attractor $([p_{C}, q_{C}], [\ell_{C}, r_{C}])$ satisfying 
$T[p_{C}-1..r_{C}+1] = T[p_{B}-1..r_{B}+1]$. 
Here, $([p_{C}, q_{C}], [\ell_{C}, r_{C}]) \in \Psi_{\source}$ follows from 
Lemma~\ref{lem:psi_equality_basic_property}~\ref{enum:psi_equality_basic_property:5}. 
Because of $T[p_{C}-1..r_{C}+1] = T[p_{B}-1..r_{B}+1]$, 
Lemma~\ref{lem:sRecover_basic_property}~\ref{enum:sRecover_basic_property:equality} shows that 
$\Psi_{\sRecover}(([p_{B}, q_{B}], [\ell_{B}, r_{B}])) = \Psi_{\sRecover}(([p_{C}, q_{C}], [\ell_{C}, r_{C}]))$ holds. 
Therefore, $([p_{A}, q_{A}], [\ell_{A}, r_{A}]) \in \Psi_{\sRecover}(([p_{C}, q_{C}], [\ell_{C}, r_{C}]))$ 
follows from 
$\Psi_{\sRecover}(([p_{B}$, $q_{B}], [\ell_{B}, r_{B}])) = \Psi_{\sRecover}(([p_{C}$, $q_{C}], [\ell_{C}, r_{C}]))$ 
and $([p_{A}, q_{A}], [\ell_{A}, r_{A}]) \in \Psi_{\sRecover}(([p_{B}, q_{B}], [\ell_{B}, r_{B}]))$.

We show that there exists an integer $b \in [1, k]$ satisfying 
$([p_{b}, q_{b}], [\ell_{b}, r_{b}]) = ([p_{C}, q_{C}], [\ell_{C}, r_{C}])$. 
Lemma~\ref{lem:CCP_property}~\ref{enum:CCP_property:1} shows that 
the level of the interval attractor $([p_{A}, q_{A}], [\ell_{A}, r_{A}])$ is $h_{Q}$ 
because $([p_{A}, q_{A}], [\ell_{A}, r_{A}]) \in \Psi_{\CCP}(T[i..j])$ and 
$([p_{Q}, q_{Q}], [\ell_{Q}, r_{Q}]) \in \Psi_{h_{Q}}$. 
Lemma~\ref{lem:sRecover_basic_property}~\ref{enum:sRecover_basic_property:centerset} shows that 
$([p_{C}, q_{C}], [\ell_{C}, r_{C}]) \in \Psi_{h_{Q}}$ holds 
because $([p_{A}, q_{A}], [\ell_{A}, r_{A}]) \in \Psi_{\sRecover}(([p_{C}, q_{C}], [\ell_{C}, r_{C}]))$ 
and $([p_{A}$, $q_{A}], [\ell_{A}, r_{A}]) \in \Psi_{h_{Q}}$. 
$([p_{C}, q_{C}], [\ell_{C}, r_{C}]) \in \Psi_{h_{Q}} \cap \Psi_{\source} \cap \Psi_{\samp}$ 
follows from $([p_{C}, q_{C}], [\ell_{C}, r_{C}]) \in \Psi_{h_{Q}}$,
$([p_{C}, q_{C}], [\ell_{C}, r_{C}]) \in \Psi_{\source}$, 
and $([p_{C}, q_{C}], [\ell_{C}, r_{C}]) \in \Psi_{\samp}$.
Because of $\Psi_{h_{Q}} \cap \Psi_{\source} \cap \Psi_{\samp} = \{ ([p_{s}, q_{s}], [\ell_{s}, r_{s}]) \mid s \in [1, k] \}$, 
there exists an integer $b \in [1, k]$ satisfying 
$([p_{b}, q_{b}], [\ell_{b}, r_{b}]) = ([p_{C}, q_{C}], [\ell_{C}, r_{C}])$. 

The following three statements are used to prove Proposition~\ref{prop:set_JB1_supseq}. 

\begin{enumerate}[label=\textbf{(\roman*)}]
    \item $L_{x} \preceq \reverse(T[p^{\prime}_{b}-1..\gamma^{\prime}_{b}-1]) \preceq L_{x^{\prime}}$; 
    \item $|\lcp(T[\gamma_{A}..r_{A}+1], T[\gamma_{Q}..r_{Q}+1])| \leq \zeta$; 
    \item $R_{y} \preceq T[\gamma^{\prime}_{b}..r^{\prime}_{b}+1] \preceq R_{y^{\prime}}$.
\end{enumerate}

\textbf{Proof of statement (i).}
We prove $\reverse(T[p^{\prime}_{b}-1..\gamma^{\prime}_{b}-1]) = \reverse(T[p_{A}-1..\gamma_{A}-1])$ 
and $T[\gamma^{\prime}_{b}..\gamma^{\prime}_{b} + \zeta] = T[\gamma_{A}..\gamma_{A} + \zeta]$.     
Lemma~\ref{lem:sRecover_basic_property}~\ref{enum:sRecover_basic_property:centerset} shows that 
$([p_{A}, q_{A}], [\ell_{A}, r_{A}]) \in \Psi_{h_{Q}} \cap \Psi_{\run} \cap \Psi_{\centerset}(C_{b})$ holds 
because $([p_{A}, q_{A}], [\ell_{A}, r_{A}]) \in \Psi_{\sRecover}(([p_{b}, q_{b}], [\ell_{b}, r_{b}]))$ 
and $([p_{b}, q_{b}], [\ell_{b}, r_{b}]) \in \Psi_{h_{Q}} \cap \Psi_{\source} \cap \Psi_{\centerset}(C_{b})$. 
Lemma~\ref{lem:psi_run_basic_property}~\ref{enum:psi_run_basic_property:7} shows that 
$\reverse(T[p^{\prime}_{b}-1..\gamma^{\prime}_{b}-1]) = \reverse(T[p_{A}-1..\gamma_{A}-1])$ 
and $T[\gamma^{\prime}_{b}..\gamma^{\prime}_{b} + \zeta] = T[\gamma_{A}..\gamma_{A} + \zeta]$     
because $([p^{\prime}_{b}, q^{\prime}_{b}], [\ell^{\prime}_{b}, r^{\prime}_{b}]), ([p_{A}, q_{A}], [\ell_{A}, r_{A}]) \in \Psi_{h_{Q}} \cap \Psi_{\run} \cap \Psi_{\centerset}(C_{b})$. 

We prove $L_{x} \preceq \reverse(T[p^{\prime}_{b}-1..\gamma^{\prime}_{b}-1]) \preceq L_{x^{\prime}}$. 
$\Psi_{\CCP}(T[i..j]) = \{ ([p, q], [\ell, r]) \in \Psi_{h_{Q}} \mid \reverse(T[i..\gamma_{Q}-1]) \prec \reverse(T[p-1..\gamma-1]) \prec \reverse(\#T[i..\gamma_{Q}-1]) \text{ and } T[\gamma_{Q}..j] \prec T[\gamma..r+1] \prec T[\gamma_{Q}..j]\# \}$
follows from Lemma~\ref{lem:CCP_property}~\ref{enum:CCP_property:4}. 
Because of $([p_{A}, q_{A}], [\ell_{A}, r_{A}]) \in \Psi_{\CCP}(T[i..j])$, 
$\reverse(T[i..\gamma_{Q}-1]) \prec \reverse(T[p_{A}-1..\gamma_{A}-1]) \prec \reverse(\#T[i..\gamma_{Q}-1])$ holds. 
$\reverse(T[i..\gamma_{Q}-1]) \prec \reverse(T[p^{\prime}_{b}-1..\gamma^{\prime}_{b}-1]) \prec \reverse(\#T[i..\gamma_{Q}-1])$ follows from $\reverse(T[i..\gamma_{Q}-1]) \prec \reverse(T[p_{A}-1..\gamma_{A}-1]) \prec \reverse(\#T[i..\gamma_{Q}-1])$ and $\reverse(T[p^{\prime}_{b}-1..\gamma^{\prime}_{b}-1]) = \reverse(T[p_{A}-1..\gamma_{A}-1])$. 
Therefore, $L_{x} \preceq \reverse(T[p^{\prime}_{b}-1..\gamma^{\prime}_{b}-1]) \preceq L_{x^{\prime}}$ 
follows from 
(A) $\reverse(T[i..\gamma_{Q}-1]) \prec \reverse(T[p^{\prime}_{b}-1..\gamma^{\prime}_{b}-1]) \prec \reverse(\#T[i..\gamma_{Q}-1])$, 
(B) $x = \min \{ s \in [1, d] \mid \reverse(T[i..\gamma_{Q}-1]) \prec L_{s} \}$, 
and (C) $x^{\prime} = \max \{ s \in [1, d] \mid L_{s} \prec \reverse(\# \cdot T[i..\gamma_{Q}-1]) \}$.

\textbf{Proof of statement (ii).}
We prove $|\lcp(T[\gamma_{A}..r_{A} + 1], T[\gamma_{Q}..r_{Q}+1])| \leq \zeta$ by contradiction. 
We assume that $|\lcp(T[\gamma_{A}..r_{A} + 1], T[\gamma_{Q}..r_{Q}+1])| > \zeta$ holds. 
Because of $([p_{A}, q_{A}], [\ell_{A}, r_{A}]) \in \Psi_{h_{Q}} \cap \Psi_{\run} \cap \Psi_{\centerset}(C_{b})$, 
$|\lcp(T[\gamma_{A}..r_{A} + 1], C_{b}^{n+1})| > \zeta$ 
follows from the definition of the subset $\Psi_{\run}$. 
If $C_{b} = C_{Q}$, 
then $|\lcp(T[\gamma_{Q}..r_{Q} + 1], C_{Q}^{n+1})| > \zeta$ follows from 
$|\lcp(T[\gamma_{A}..r_{A} + 1], T[\gamma_{Q}..r_{Q}+1])| > \zeta$ and $|\lcp(T[\gamma_{A}..r_{A} + 1], C_{b}^{n+1})| > \zeta$. 
On the other hand, $|\lcp(T[\gamma_{Q}..r_{Q} + 1], C_{Q}^{n+1})| \leq \zeta$ follows from the premise of Proposition~\ref{prop:set_JB1_supseq}. 
The two facts $|\lcp(T[\gamma_{Q}..r_{Q} + 1], C_{Q}^{n+1})| > \zeta$ and $|\lcp(T[\gamma_{Q}..r_{Q} + 1], C_{Q}^{n+1})| \leq \zeta$ yield a contradiction. 

Otherwise (i.e., $C_{b} \neq C_{Q}$), 
we can apply Lemma~\ref{lem:associated_string_C}~\ref{enum:associated_string_C:2} to the two interval attractors 
$([p_{A}, q_{A}], [\ell_{A}$, $r_{A}])$ and 
$([p_{Q}, q_{Q}], [\ell_{Q}, r_{Q}])$ because 
$([p_{A}, q_{A}], [\ell_{A}, r_{A}]), ([p_{Q}, q_{Q}], [\ell_{Q}, r_{Q}]) \in \Psi_{h_{Q}}$ and $|\lcp(T[\gamma_{A}..r_{A} + 1], T[\gamma_{Q}..r_{Q}+1])| > \zeta$. 
This lemma shows that $C_{b} = C_{Q}$ holds, but it contradicts $C_{b} \neq C_{Q}$. 
Therefore, $|\lcp(T[\gamma_{A}..r_{A} + 1], T[\gamma_{Q}..r_{Q}+1])| \leq \zeta$ must hold. 

\textbf{Proof of statement (iii).}
We prove $T[\gamma_{Q}..j] \prec T[\gamma_{A}..r_{A}+1] \prec T[\gamma_{Q}..r_{Q}+1]$ and $j < r_{Q} + 1$. 
$\Psi_{\CCP}(T[i..j]) = \{ ([p, q], [\ell, r]) \in \Psi_{h_{Q}} \mid \reverse(T[i..\gamma_{Q}-1]) \prec \reverse(T[p-1..\gamma-1]) \prec \reverse(\#T[i..\gamma_{Q}-1]) \text{ and } T[\gamma_{Q}..j] \prec T[\gamma..r+1] \prec T[\gamma_{Q}..j]\# \}$
follows from Lemma~\ref{lem:CCP_property}~\ref{enum:CCP_property:4}. 
Because of $([p_{A}, q_{A}]$, $[\ell_{A}, r_{A}]) \in \Psi_{\CCP}(T[i..j])$, 
$T[\gamma_{Q}..j] \prec T[\gamma_{A}..r_{A}+1] \prec T[\gamma_{Q}..j]\#$. 
On the other hand, $T[\gamma_{A}..r_{A}+1] \prec T[\gamma_{Q}..r_{Q}+1]$ follows from 
$([p_{A}, q_{A}], [\ell_{A}, r_{A}]) \in \Psi_{\lex}(T[\gamma_{Q}..r_{Q}+1])$. 
Therefore, $T[\gamma_{Q}..j] \prec T[\gamma_{A}..r_{A}+1] \prec T[\gamma_{Q}..r_{Q}+1]$ holds, 
and this lexicographic order indicates that $j < r_{Q} + 1$ holds. 

We prove $T[\gamma_{Q}..j] \prec T[\gamma^{\prime}_{b}..r^{\prime}_{b}+1] \prec T[\gamma_{Q}..r_{Q}+1]$. 
Lemma~\ref{lem:sRecover_basic_property}~\ref{enum:sRecover_basic_property:lex} shows that 
$T[\gamma^{\prime}_{b}..r^{\prime}_{b} + 1] \prec T[\gamma_{Q}..r_{Q}+1]$ holds 
because $T[\gamma_{A}..r_{A} + 1] \prec T[\gamma_{Q}..r_{Q}+1]$, 
$|\lcp(T[\gamma_{A}..r_{A} + 1], T[\gamma_{Q}..r_{Q}+1])| \leq \zeta$ (statement (ii)), 
and 
$([p^{\prime}_{b}, q^{\prime}_{b}], [\ell^{\prime}_{b}, r^{\prime}_{b}]), ([p_{A}, q_{A}], [\ell_{A}, r_{A}]) \in \Psi_{\sRecover}(([p_{b}, q_{b}], [\ell_{b}, r_{b}]))$. 
$|\lcp(T[\gamma_{A}..r_{A} + 1], T[\gamma_{Q}..j])| \leq \zeta$ follows from 
$|\lcp(T[\gamma_{A}..r_{A} + 1], T[\gamma_{Q}..r_{Q}+1])| \leq \zeta$ and $j < r_{Q} + 1$. 
Lemma~\ref{lem:sRecover_basic_property}~\ref{enum:sRecover_basic_property:lex} shows that 
$T[\gamma_{Q}..j] \prec T[\gamma^{\prime}_{b}..r^{\prime}_{b}+1]$ holds 
because $T[\gamma_{Q}..j] \prec T[\gamma_{A}..r_{A}+1]$ and  
$|\lcp(T[\gamma_{A}..r_{A} + 1], T[\gamma_{Q}..j])| \leq \zeta$. 
Therefore, $T[\gamma_{Q}..j] \prec T[\gamma^{\prime}_{b}..r^{\prime}_{b}+1] \prec T[\gamma_{Q}..r_{Q}+1]$ holds. 

Finally, $R_{y} \preceq T[\gamma^{\prime}_{b}..r^{\prime}_{b}+1] \preceq R_{y^{\prime}}$ 
follows from 
(A) $T[\gamma_{Q}..j] \prec T[\gamma^{\prime}_{b}..r^{\prime}_{b}+1] \prec T[\gamma_{Q}..r_{Q} + 1]$, 
(B) $T[\gamma^{\prime}_{b}..r^{\prime}_{b}+1] \in \{ R_{1}, R_{2}, \ldots, R_{d^{\prime}} \}$, 
(C) $y = \min \{ s \in [1, d^{\prime}] \mid T[\gamma_{Q}..j] \prec R_{s} \}$, 
and (D) $y^{\prime} = \max \{ s \in [1, d^{\prime}] \mid R_{s} \prec T[\gamma_{Q}..r_{Q} + 1] \}$.

\textbf{Proof of Proposition~\ref{prop:set_JB1_supseq}.}
$b \in \mathcal{I}_{B}$ follows from 
$L_{x} \preceq \reverse(T[p^{\prime}_{b}-1..\gamma^{\prime}_{b}-1]) \preceq L_{x^{\prime}}$ 
and $R_{y} \preceq T[\gamma^{\prime}_{b}..r^{\prime}_{b}+1] \preceq R_{\hat{y}}$. 
$([p_{A}, q_{A}], [\ell_{A}, r_{A}]) \in \bigcup_{s \in \mathcal{I}_{B}} \Psi_{\sRecover}(([p_{s}, q_{s}], [\ell_{s}, r_{s}]))$ follows from $b \in \mathcal{I}_{B}$ and 
$([p_{A}, q_{A}], [\ell_{A}, r_{A}]) \in \Psi_{\sRecover}(([p_{b}, q_{b}], [\ell_{b}, r_{b}]))$. 
Therefore, Proposition~\ref{prop:set_JB1_supseq} follows from the fact that 
$([p_{A}, q_{A}], [\ell_{A}, r_{A}]) \in \Psi_{\sRecover}(([p_{b}, q_{b}], [\ell_{b}, r_{b}]))$ holds for each interval attractor $([p_{A}, q_{A}], [\ell_{A}, r_{A}]) \in \Psi_{\CCP}(T[i..j]) \cap \Psi_{\lex}(T[\gamma_{Q}..r_{Q}+1]) \cap \Psi_{\run}$.

\end{proof}

\begin{proposition}\label{prop:set_JB2_subseq}
If $|\lcp(T[\gamma_{Q}..r_{Q} + 1], C_{Q}^{n+1})| > 1 + \sum_{w = 1}^{h_{Q}+3} \mu(w)$, 
then the following equation holds: 
\begin{equation*}
\bigcup_{s \in \mathcal{I}^{\prime}_{B}} \Psi_{\sRecover}(([p_{s}, q_{s}], [\ell_{s}, r_{s}])) \subseteq (\Psi_{\CCP}(T[i..j]) \cap \Psi_{\lex}(T[\gamma_{Q}..r_{Q}+1]) \cap \Psi_{\run})  \setminus \Psi_{\centerset}(C_{Q}). 
\end{equation*}
\end{proposition}
\begin{proof}
Let $\zeta = 1 + \sum_{w = 1}^{h_{Q}+3} \mu(w)$ for simplicity. 
Consider an interval attractor $([p_{A}, q_{A}], [\ell_{A}, r_{A}]) \in \bigcup_{s \in \mathcal{I}^{\prime}_{B}}$ $\Psi_{\sRecover}(([p_{s}$, $q_{s}], [\ell_{s}, r_{s}]))$. 
Then, set $\mathcal{I}^{\prime}_{B}$ contains an integer $b$ satisfying $([p_{A}, q_{A}], [\ell_{A}, r_{A}]) \in \Psi_{\sRecover}(([p_{b}, q_{b}], [\ell_{b}, r_{b}]))$. 
Here, $([p_{A}, q_{A}], [\ell_{A}, r_{A}]) \in \Psi_{h_{Q}} \cap \Psi_{\run} \cap \Psi_{\centerset}(C_{b})$ 
follows from Lemma \ref{lem:sRecover_basic_property} \ref{enum:sRecover_basic_property:centerset}. 
Let $\gamma_{A}$ be the attractor position of the interval attractor $([p_{A}, q_{A}], [\ell_{A}, r_{A}])$. 

The following five statements are used to prove Proposition~\ref{prop:set_JB2_subseq}. 

\begin{enumerate}[label=\textbf{(\roman*)}]
    \item $T[\gamma_{Q}..j] \prec T[\gamma_{A}..r_{A} + 1] \prec T[\gamma_{Q}..j]\#$, $T[\gamma_{A}..r_{A} + 1] \prec T[\gamma_{Q}..\gamma_{Q} + \zeta]$, and $T[\gamma_{A}..r_{A} + 1] \prec T[\gamma_{Q}..r_{Q}+1]$; 
    \item $\reverse(T[i..\gamma_{Q}-1]) \prec \reverse(T[p_{A}-1..\gamma_{A}-1]) \prec \reverse(\#T[i..\gamma_{Q}-1])$; 
    \item $C_{b} \neq C_{Q}$ and $([p_{A}, q_{A}], [\ell_{A}, r_{A}]) \not \in \Psi_{\centerset}(C_{Q})$; 
    \item $([p_{A}, q_{A}], [\ell_{A}, r_{A}]) \in \Psi_{\CCP}(T[i..j])$; 
    \item $([p_{A}, q_{A}], [\ell_{A}, r_{A}]) \in \Psi_{\lex}(T[\gamma_{Q}..r_{Q}+1])$.
\end{enumerate}

\textbf{Proof of statement (i).}
We prove $T[\gamma_{Q}..j] \prec T[\gamma^{\prime}_{b}..r^{\prime}_{b}+1] \prec T[\gamma_{Q}..\gamma_{Q} + \zeta]$ 
and $j < \gamma_{Q} + \zeta$. 
$R_{y} \preceq T[\gamma^{\prime}_{b}..r^{\prime}_{b}+1] \preceq R_{\hat{y}}$ follows from $b \in \mathcal{I}^{\prime}_{B}$. 
$T[\gamma_{Q}..j] \prec T[\gamma^{\prime}_{b}..r^{\prime}_{b}+1] \prec T[\gamma_{Q}..\gamma_{Q} + \zeta]$ 
follows from $R_{y} \preceq T[\gamma^{\prime}_{b}..r^{\prime}_{b}+1] \preceq R_{\hat{y}}$. 
This lexicographic order $T[\gamma_{Q}..j] \prec T[\gamma^{\prime}_{b}..r^{\prime}_{b}+1] \prec T[\gamma_{Q}..\gamma_{Q} + \zeta]$ indicates that $j < \gamma_{Q} + \zeta$ holds. 

We prove $T[\gamma_{A}..r_{A} + 1] \prec T[\gamma_{Q}..\gamma_{Q} + \zeta]$. 
$|\lcp(T[\gamma^{\prime}_{b}..r^{\prime}_{b} + 1], T[\gamma_{Q}..\gamma_{Q} + \zeta])| \leq \zeta$ follows from 
$T[\gamma^{\prime}_{b}..r^{\prime}_{b}+1] \prec T[\gamma_{Q}..\gamma_{Q} + \zeta]$. 
Lemma~\ref{lem:sRecover_basic_property}~\ref{enum:sRecover_basic_property:lex} shows that 
$T[\gamma_{A}..r_{A} + 1] \prec T[\gamma_{Q}..\gamma_{Q} + \zeta]$ holds 
because $T[\gamma^{\prime}_{b}..r^{\prime}_{b}+1] \prec T[\gamma_{Q}..\gamma_{Q} + \zeta]$, 
$|\lcp(T[\gamma^{\prime}_{b}..r^{\prime}_{b} + 1], T[\gamma_{Q}..\gamma_{Q} + \zeta])| \leq \zeta$, 
and 
$([p^{\prime}_{b}, q^{\prime}_{b}], [\ell^{\prime}_{b}, r^{\prime}_{b}])$, $([p_{A}, q_{A}]$, $[\ell_{A}, r_{A}]) \in \Psi_{\sRecover}(([p_{b}, q_{b}], [\ell_{b}, r_{b}]))$. 

We prove $T[\gamma_{A}..r_{A} + 1] \prec T[\gamma_{Q}..r_{Q} + 1]$. 
$|[\gamma_{Q}, r_{Q} + 1]| > \zeta$ follows from $|\lcp(T[\gamma_{Q}..r_{Q} + 1], C_{Q}^{n+1})| > \zeta$. 
$T[\gamma_{Q}..\gamma_{Q} + \zeta] \preceq T[\gamma_{Q}..r_{Q} + 1]$ follows from $|[\gamma_{Q}, r_{Q} + 1]| > \zeta$. 
Therefore, $T[\gamma_{A}..r_{A} + 1] \prec T[\gamma_{Q}..r_{Q} + 1]$ follows from 
$T[\gamma_{A}..r_{A} + 1] \prec T[\gamma_{Q}..\gamma_{Q} + \zeta]$ and $T[\gamma_{Q}..\gamma_{Q} + \zeta] \preceq T[\gamma_{Q}..r_{Q} + 1]$. 

We prove $T[\gamma_{Q}..j] \prec T[\gamma_{A}..r_{A} + 1]$. 
$|\lcp(T[\gamma^{\prime}_{b}..r^{\prime}_{b} + 1], T[\gamma_{Q}..j])| \leq \zeta$ follows from 
$|\lcp(T[\gamma^{\prime}_{b}..r^{\prime}_{b} + 1], T[\gamma_{Q}..\gamma_{Q} + \zeta])| \leq \zeta$ and $j < \gamma_{Q} + \zeta$. 
Lemma~\ref{lem:sRecover_basic_property}~\ref{enum:sRecover_basic_property:lex} shows that 
$T[\gamma_{A}..r_{A} + 1] \prec T[\gamma_{Q}..j]$ holds 
because $T[\gamma_{Q}..j] \prec T[\gamma^{\prime}_{b}..r^{\prime}_{b}+1]$ 
and $|\lcp(T[\gamma^{\prime}_{b}..r^{\prime}_{b} + 1], T[\gamma_{Q}..j])| \leq \zeta$. 

We prove $T[\gamma_{A}..r_{A} + 1] \prec T[\gamma_{Q}..j]\#$. 
The string $T[\gamma_{Q}..j]$ is a prefix of string $T[\gamma^{\prime}_{b}..r^{\prime}_{b} + 1]$ 
(i.e., $\lcp(T[\gamma^{\prime}_{b}..r^{\prime}_{b} + 1], T[\gamma_{Q}..j]) = T[\gamma_{Q}..j]$) 
because $T[\gamma_{Q}..j] \prec T[\gamma^{\prime}_{b}..r^{\prime}_{b}+1] \prec T[\gamma_{Q}..\gamma_{Q} + \zeta]$ 
and $j < \gamma_{Q} + \zeta$. 
$T[\gamma^{\prime}_{b}..r^{\prime}_{b} + 1] \prec T[\gamma_{Q}..j]\#$ follows from 
$\lcp(T[\gamma^{\prime}_{b}..r^{\prime}_{b} + 1], T[\gamma_{Q}..j]) = T[\gamma_{Q}..j]$. 
$|\lcp(T[\gamma^{\prime}_{b}..r^{\prime}_{b} + 1], T[\gamma_{Q}..j]\#)| \leq \zeta$ 
follows from $|\lcp(T[\gamma^{\prime}_{b}..r^{\prime}_{b} + 1], T[\gamma_{Q}..j])| \leq \zeta$. 
Lemma~\ref{lem:sRecover_basic_property}~\ref{enum:sRecover_basic_property:lex} shows that 
$T[\gamma_{A}..r_{A} + 1] \prec T[\gamma_{Q}..j]\#$ holds 
because $T[\gamma^{\prime}_{b}..r^{\prime}_{b} + 1] \prec T[\gamma_{Q}..j]\#$ 
and $|\lcp(T[\gamma^{\prime}_{b}..r^{\prime}_{b} + 1], T[\gamma_{Q}..j])| \leq \zeta$. 

\textbf{Proof of statement (ii).}
This statement can be proved using the same approach as for statement (iii) in the proof of Proposition~\ref{prop:set_JB1_subseq}. 

\textbf{Proof of statement (iii).}
We prove $C_{b} \neq C_{Q}$ by contradiction. 
We assume that $C_{b} = C_{Q}$ holds. 
Because of $([p_{A}, q_{A}], [\ell_{A}, r_{A}]) \in \Psi_{h_{Q}} \cap \Psi_{\run} \cap \Psi_{\centerset}(C_{Q})$, 
$|\lcp(T[\gamma_{A}..r_{A} + 1], C_{Q}^{n+1})| > \zeta$ follows from 
the definition of the subset $\Psi_{\run}$. 
$T[\gamma_{Q}..\gamma_{Q} + \zeta] = T[\gamma_{A}..\gamma_{A} + \zeta]$ 
follows from $|\lcp(T[\gamma_{Q}..r_{Q} + 1], C_{Q}^{n+1})| > \zeta$ 
and $|\lcp(T[\gamma_{A}..r_{A} + 1], C_{Q}^{n+1})| > \zeta$. 
On the other hand, $T[\gamma_{Q}..\gamma_{Q} + \zeta] \neq T[\gamma_{A}..\gamma_{A} + \zeta]$ follows from 
$T[\gamma_{A}..r_{A} + 1] \prec T[\gamma_{Q}..\gamma_{Q} + \zeta]$ (statement (i)). 
The two facts $T[\gamma_{Q}..\gamma_{Q} + \zeta] = T[\gamma_{A}..\gamma_{A} + \zeta]$ and 
$T[\gamma_{Q}..\gamma_{Q} + \zeta] \neq T[\gamma_{A}..\gamma_{A} + \zeta]$ yield a contradiction. 
Therefore, $C_{b} \neq C_{Q}$ must hold. 

$([p_{A}, q_{A}], [\ell_{A}, r_{A}]) \not \in \Psi_{\centerset}(C_{Q})$ follows from 
$C_{b} \neq C_{Q}$ and $([p_{A}, q_{A}], [\ell_{A}, r_{A}]) \in \Psi_{h_{Q}} \cap \Psi_{\run} \cap \Psi_{\centerset}(C_{b})$.     

\textbf{Proof of statement (iv).}
We prove $([p_{A}, q_{A}], [\ell_{A}, r_{A}]) \in \Psi_{\CCP}(T[i..j])$. 
$\Psi_{\CCP}(T[i..j]) = \{ ([p, q], [\ell, r]) \in \Psi_{h_{Q}} \mid \reverse(T[i..\gamma_{Q}-1]) \prec \reverse(T[p-1..\gamma-1]) \prec \reverse(\#T[i..\gamma_{Q}-1]) \text{ and } T[\gamma_{Q}..j] \prec T[\gamma..r+1] \prec T[\gamma_{Q}..j]\# \}$
follows from Lemma~\ref{lem:CCP_property}~\ref{enum:CCP_property:4}.    
We already proved 
(A) $T[\gamma_{Q}..j] \prec T[\gamma_{A}..r_{A} + 1] \prec T[\gamma_{Q}..j]\#$ (statement (i)), 
(B) $\reverse(T[i..\gamma_{Q}-1]) \prec \reverse(T[p_{A}-1..\gamma_{A}-1]) \prec \reverse(\#T[i..\gamma_{Q}-1])$ (statement (ii)), 
and $([p_{A}, q_{A}], [\ell_{A}, r_{A}]) \in \Psi_{h_{Q}}$. 
Therefore, $([p_{A}, q_{A}], [\ell_{A}, r_{A}]) \in \Psi_{\CCP}(T[i..j])$ holds. 

\textbf{Proof of statement (v).}
$T[\gamma_{A}..r_{A} + 1] \prec T[\gamma_{Q}..r_{Q}+1]$ follows from statement (i).
Therefore, 
$([p_{A}, q_{A}], [\ell_{A}, r_{A}]) \in \Psi_{\lex}(T[\gamma_{Q}..r_{Q}+1])$ follows from 
the definition of the subset $\Psi_{\lex}(T[\gamma_{Q}..r_{Q}+1])$. 

\textbf{Proof of Proposition~\ref{prop:set_JB2_subseq}.}
$([p_{A}, q_{A}], [\ell_{A}, r_{A}]) \in (\Psi_{\CCP}(T[i..j]) \cap \Psi_{\lex}(T[\gamma_{Q}..r_{Q}+1]) \cap \Psi_{\run})  \setminus \Psi_{\centerset}(C_{Q})$ follows from $([p_{A}, q_{A}], [\ell_{A}, r_{A}]) \in \Psi_{\CCP}(T[i..j])$ (statement (iv)), 
$([p_{A}, q_{A}], [\ell_{A}, r_{A}]) \in \Psi_{\lex}(T[\gamma_{Q}..r_{Q}+1])$ (statement (v)), 
and $([p_{A}, q_{A}], [\ell_{A}, r_{A}]) \not \in \Psi_{\centerset}(C_{Q})$ (statement (iii)). 
Therefore, Proposition~\ref{prop:set_JB2_subseq} follows from the fact that 
$([p_{A}, q_{A}], [\ell_{A}, r_{A}]) \in (\Psi_{\CCP}(T[i..j]) \cap \Psi_{\lex}(T[\gamma_{Q}..r_{Q}+1]) \cap \Psi_{\run})  \setminus \Psi_{\centerset}(C_{Q})$ for each interval attractor $([p_{A}, q_{A}], [\ell_{A}, r_{A}]) \in \bigcup_{s \in \mathcal{I}^{\prime}_{B}}$ $\Psi_{\sRecover}(([p_{s}$, $q_{s}], [\ell_{s}, r_{s}]))$. 
    
\end{proof}

\begin{proposition}\label{prop:set_JB2_supseq}
If $|\lcp(T[\gamma_{Q}..r_{Q} + 1], C_{Q}^{n+1})| > 1 + \sum_{w = 1}^{h_{Q}+3} \mu(w)$, 
then the following equation holds: 
\begin{equation*}
\bigcup_{s \in \mathcal{I}^{\prime}_{B}} \Psi_{\sRecover}(([p_{s}, q_{s}], [\ell_{s}, r_{s}])) \supseteq (\Psi_{\CCP}(T[i..j]) \cap \Psi_{\lex}(T[\gamma_{Q}..r_{Q}+1]) \cap \Psi_{\run})  \setminus \Psi_{\centerset}(C_{Q}).
\end{equation*}
\end{proposition}
\begin{proof}
Let $\zeta = 1 + \sum_{w = 1}^{h_{Q}+3} \mu(w)$ for simplicity. 
Consider an interval attractor $([p_{A}, q_{A}], [\ell_{A}, r_{A}]) \in (\Psi_{\CCP}(T[i..j]) \cap \Psi_{\lex}(T[\gamma_{Q}..r_{Q}+1]) \cap \Psi_{\run}) \setminus \Psi_{\centerset}(C_{Q})$. 
Because of $([p_{A}, q_{A}], [\ell_{A}, r_{A}]) \in \Psi_{\CCP}(T[i..j]) \cap \Psi_{\lex}(T[\gamma_{Q}..r_{Q}+1]) \cap \Psi_{\run}$, 
the sampling subset $\Psi_{\samp}$ contains an interval attractor $([p_{C}, q_{C}], [\ell_{C}, r_{C}])$ satisfying 
$([p_{A}, q_{A}], [\ell_{A}, r_{A}]) \in \Psi_{\sRecover}(([p_{C}, q_{C}], [\ell_{C}, r_{C}]))$ (see the proof of Proposition~\ref{prop:set_JB1_supseq}).     
Here, there exists an integer $b \in [1, k]$ satisfying 
$([p_{b}, q_{b}], [\ell_{b}, r_{b}]) = ([p_{C}, q_{C}], [\ell_{C}, r_{C}])$ (see the proof of Proposition~\ref{prop:set_JB1_supseq}). 
Because of $([p_{A}, q_{A}], [\ell_{A}, r_{A}]) \in \Psi_{\sRecover}(([p_{b}, q_{b}]$, 
$[\ell_{b}, r_{b}]))$, 
Lemma~\ref{lem:sRecover_basic_property}~\ref{enum:psi_bigcup_property:centerset} shows that 
$([p_{A}, q_{A}], [\ell_{A}, r_{A}]) \in \Psi_{h_{Q}} \cap \Psi_{\run} \cap \Psi_{\centerset}(C_{b})$ holds.

The following four statements are used to prove Proposition~\ref{prop:set_JB2_supseq}. 

\begin{enumerate}[label=\textbf{(\roman*)}]
    \item $L_{x} \preceq \reverse(T[p^{\prime}_{b}-1..\gamma^{\prime}_{b}-1]) \preceq L_{x^{\prime}}$; 
    \item $C_{b} \neq C_{Q}$;
    \item $|\lcp(T[\gamma_{A}..r_{A}+1], T[\gamma_{Q}..r_{Q}+1])| \leq \zeta$; 
    \item $R_{y} \preceq T[\gamma^{\prime}_{b}..r^{\prime}_{b}+1] \preceq R_{\hat{y}}$.
\end{enumerate}

\textbf{Proof of statement (i).}
This statement can be proved using the same approach as for statement (i) in the proof of Proposition~\ref{prop:set_JB1_supseq}. 

\textbf{Proof of statement (ii).}
We prove $C_{b} \neq C_{Q}$. 
We already proved $([p_{A}, q_{A}], [\ell_{A}, r_{A}]) \in \Psi_{\centerset}(C_{b})$. 
On the other hand, $([p_{A}, q_{A}], [\ell_{A}, r_{A}]) \not \in \Psi_{\centerset}(C_{Q})$ follows from 
$([p_{A}, q_{A}], [\ell_{A}, r_{A}]) \in (\Psi_{\CCP}(T[i..j]) \cap \Psi_{\lex}(T[\gamma_{Q}..r_{Q}+1]) \cap \Psi_{\run}) \setminus \Psi_{\centerset}(C_{Q})$. 
Therefore, $C_{b} \neq C_{Q}$ follows from $([p_{A}, q_{A}]$, $[\ell_{A}, r_{A}]) \in \Psi_{\centerset}(C_{b})$ 
and $([p_{A}, q_{A}], [\ell_{A}, r_{A}]) \not \in \Psi_{\centerset}(C_{Q})$. 

\textbf{Proof of statement (iii).}
We prove $|\lcp(T[\gamma_{A}..r_{A}+1], T[\gamma_{Q}..r_{Q}+1])| \leq \zeta$ by contradiction. 
We assume that $|\lcp(T[\gamma_{A}..r_{A}+1], T[\gamma_{Q}..r_{Q}+1])| > \zeta$ holds. 
We can apply Lemma~\ref{lem:associated_string_C}~\ref{enum:associated_string_C:2} to the two interval attractors 
$([p_{A}, q_{A}], [\ell_{A}, r_{A}])$ and 
$([p_{Q}, q_{Q}], [\ell_{Q}, r_{Q}])$ 
because $([p_{A}, q_{A}], [\ell_{A}, r_{A}]) \in \Psi_{h_{Q}} \cap \Psi_{\centerset}(C_{b})$, 
$([p_{Q}, q_{Q}], [\ell_{Q}, r_{Q}]) \in \Psi_{h_{Q}} \cap \Psi_{\centerset}(C_{Q})$, 
and $|\lcp(T[\gamma_{A}..r_{A}+1], T[\gamma_{Q}..r_{Q}+1])| > \zeta$. 
This lemma shows that $C_{b} = C_{Q}$ holds. 
On the other hand, we already proved $C_{b} \neq C_{Q}$. 
The two facts $C_{b} = C_{Q}$ and $C_{b} \neq C_{Q}$ yield a contradiction. 
Therefore, $|\lcp(T[\gamma_{A}..r_{A}+1], T[\gamma_{Q}..r_{Q}+1])| \leq \zeta$ must hold.

\textbf{Proof of statement (iv).}
We prove $T[\gamma_{Q}..j] \prec T[\gamma_{A}..r_{A}+1] \prec T[\gamma_{Q}..\gamma_{Q} + \zeta]$ and $j < \gamma_{Q} + \zeta$. 
$T[\gamma_{Q}..j] \prec T[\gamma_{A}..r_{A}+1]$ follows from 
Lemma~\ref{lem:CCP_property}~\ref{enum:CCP_property:4} and $([p_{A}, q_{A}], [\ell_{A}, r_{A}]) \in \Psi_{\CCP}(T[i..j])$. 
Because of $([p_{A}, q_{A}], [\ell_{A}, r_{A}]) \in \Psi_{\lex}(T[\gamma_{Q}..r_{Q}+1])$, 
$T[\gamma_{A}..r_{A}+1] \prec T[\gamma_{Q}..r_{Q}+1]$ follows from the definition of the subset $\Psi_{\lex}(T[\gamma_{Q}..r_{Q}+1])$. 
$T[\gamma_{A}..r_{A}+1] \prec T[\gamma_{Q}..\gamma_{Q} + \zeta]$ follows from 
$T[\gamma_{A}..r_{A}+1] \prec T[\gamma_{Q}..r_{Q}+1]$ and $|\lcp(T[\gamma_{A}..r_{A}+1], T[\gamma_{Q}..r_{Q}+1])| \leq \zeta$ (statement (iii)). 
This lexicographical order $T[\gamma_{Q}..j] \prec T[\gamma_{A}..r_{A}+1] \prec T[\gamma_{Q}..\gamma_{Q} + \zeta]$ 
indicates that $j < \gamma_{Q} + \zeta$ holds. 

We prove $T[\gamma_{Q}..j] \prec T[\gamma^{\prime}_{b}..r^{\prime}_{b}+1] \prec T[\gamma_{Q}..\gamma_{Q} + \zeta]$. 
$|\lcp(T[\gamma_{A}..r_{A} + 1], T[\gamma_{Q}..\gamma_{Q} + \zeta])| \leq \zeta$ follows from 
$T[\gamma_{A}..r_{A}+1] \prec T[\gamma_{Q}..\gamma_{Q} + \zeta]$. 
Lemma~\ref{lem:sRecover_basic_property}~\ref{enum:sRecover_basic_property:lex} shows that 
$T[\gamma^{\prime}_{b}..r^{\prime}_{b} + 1] \prec T[\gamma_{Q}..\gamma_{Q} + \zeta]$ holds 
because $T[\gamma_{A}..r_{A} + 1] \prec T[\gamma_{Q}..\gamma_{Q} + \zeta]$, 
$|\lcp(T[\gamma_{A}..r_{A} + 1], T[\gamma_{Q}..\gamma_{Q} + \zeta])| \leq \zeta$, 
and 
$([p^{\prime}_{b}, q^{\prime}_{b}], [\ell^{\prime}_{b}, r^{\prime}_{b}]), ([p_{A}, q_{A}], [\ell_{A}, r_{A}]) \in \Psi_{\sRecover}(([p_{b}, q_{b}], [\ell_{b}, r_{b}]))$. 
$|\lcp(T[\gamma_{A}..r_{A} + 1], T[\gamma_{Q}..j])| \leq \zeta$ follows from 
$|\lcp(T[\gamma_{A}..r_{A} + 1], T[\gamma_{Q}..\gamma_{Q} + \zeta])| \leq \zeta$ and $j < \gamma_{Q} + \zeta$. 
Lemma~\ref{lem:sRecover_basic_property}~\ref{enum:sRecover_basic_property:lex} shows that 
$T[\gamma_{Q}..j] \prec T[\gamma^{\prime}_{b}..r^{\prime}_{b}+1]$ holds 
because $T[\gamma_{Q}..j] \prec T[\gamma_{A}..r_{A}+1]$ and  
$|\lcp(T[\gamma_{A}..r_{A} + 1], T[\gamma_{Q}..j])| \leq \zeta$. 
Therefore, $T[\gamma_{Q}..j] \prec T[\gamma^{\prime}_{b}..r^{\prime}_{b}+1] \prec T[\gamma_{Q}..\gamma_{Q} + \zeta]$ holds. 

Finally, $R_{y} \preceq T[\gamma^{\prime}_{b}..r^{\prime}_{b}+1] \preceq R_{\hat{y}}$ 
follows from 
(A) $T[\gamma_{Q}..j] \prec T[\gamma^{\prime}_{b}..r^{\prime}_{b}+1] \prec T[\gamma_{Q}..\gamma_{Q} + \zeta]$, 
(B) $T[\gamma^{\prime}_{b}..r^{\prime}_{b}+1] \in \{ R_{1}, R_{2}, \ldots, R_{d^{\prime}} \}$
(C) $y = \min \{ s \in [1, d^{\prime}] \mid T[\gamma_{Q}..j] \prec R_{s} \}$, 
and (D) $\hat{y} = \max \{ s \in [1, d^{\prime}] \mid R_{s} \prec T[\gamma_{Q}..\gamma_{Q} + \zeta] \}$. 

\textbf{Proof of Proposition~\ref{prop:set_JB2_supseq}.}
$b \in \mathcal{I}^{\prime}_{B}$ follows from 
$L_{x} \preceq \reverse(T[p^{\prime}_{b}-1..\gamma^{\prime}_{b}-1]) \preceq L_{x^{\prime}}$ (statement (i))
and $R_{y} \preceq T[\gamma^{\prime}_{b}..r^{\prime}_{b}+1] \preceq R_{\hat{y}}$ (statement (iv)). 
$([p_{A}, q_{A}], [\ell_{A}, r_{A}]) \in \bigcup_{s \in \mathcal{I}^{\prime}_{B}} \Psi_{\sRecover}(([p_{s}, q_{s}], [\ell_{s}, r_{s}]))$ follows from $b \in \mathcal{I}^{\prime}_{B}$ and 
$([p_{A}, q_{A}], [\ell_{A}, r_{A}]) \in \Psi_{\sRecover}(([p_{b}, q_{b}], [\ell_{b}, r_{b}]))$. 
Therefore, Proposition~\ref{prop:set_JB2_supseq} follows from the fact that 
$([p_{A}, q_{A}], [\ell_{A}, r_{A}]) \in \Psi_{\sRecover}(([p_{b}, q_{b}], [\ell_{b}, r_{b}]))$ holds for each interval attractor $([p_{A}, q_{A}], [\ell_{A}, r_{A}]) \in (\Psi_{\CCP}(T[i..j]) \cap \Psi_{\lex}(T[\gamma_{Q}..r_{Q}+1]) \cap \Psi_{\run}) \setminus \Psi_{\centerset}(C_{Q})$. 
\end{proof}

We prove Equation~\ref{eq:x_set_RB_property:1} in Lemma~\ref{lem:JB_main_lemma} using 
Proposition~\ref{prop:set_JB1_subseq} and Proposition~\ref{prop:set_JB1_supseq}. 

\begin{proof}[Proof of Equation~\ref{eq:x_set_RB_property:1} in Lemma~\ref{lem:JB_main_lemma}]
$|\Psi_{\CCP}(T[i..j]) \cap \Psi_{\lex}(T[\gamma_{Q}..r_{Q}+1]) \cap \Psi_{\run}| = |\bigcup_{s \in \mathcal{I}_{B}} \Psi_{\sRecover}$ $(([p_{s}, q_{s}], [\ell_{s}, r_{s}]))|$ follows from Proposition~\ref{prop:set_JB1_subseq} and Proposition~\ref{prop:set_JB1_supseq}. 
Lemma~\ref{lem:sRecover_basic_property}~\ref{enum:sRecover_basic_property:overlap} shows that 
$\Psi_{\sRecover}(([p_{s}, q_{s}], [\ell_{s}, r_{s}])) \cap \Psi_{\sRecover}(([p_{s^{\prime}}, q_{s^{\prime}}], [\ell_{s^{\prime}}, r_{s^{\prime}}])) = \emptyset$ for any pair of two integers $1 \leq s < s^{\prime} \leq k$ 
because $T[p_{s}-1..r_{s}+1] \neq T[p_{s^{\prime}}-1..r_{s^{\prime}}+1]$ follows from the definition of the set $\mathcal{J}_{B}(h_{Q})$. 
Because of $\Psi_{\sRecover}(([p_{s}, q_{s}], [\ell_{s}, r_{s}])) \cap \Psi_{\sRecover}(([p_{s^{\prime}}, q_{s^{\prime}}], [\ell_{s^{\prime}}, r_{s^{\prime}}])) = \emptyset$, 
$|\bigcup_{s \in \mathcal{I}_{B}} \Psi_{\sRecover}(([p_{s}, q_{s}], [\ell_{s}, r_{s}]))| = \sum_{s \in \mathcal{I}_{B}} |\Psi_{\sRecover}(([p_{s}, q_{s}], [\ell_{s}, r_{s}]))|$ holds. 
Lemma~\ref{lem:sRecover_size_property}~\ref{enum:sRecover_size_property:2} shows that 
$|\Psi_{\sRecover}(([p_{s}, q_{s}], [\ell_{s}, r_{s}]))| = |\Psi_{\str}(T[p_{s}-1..r_{s}+1])| |f_{\recover}(([p_{s}, q_{s}], [\ell_{s}, r_{s}]))|$ for each integer $s \in \mathcal{I}_{B}$. 
$\rangesum(\mathcal{J}_{B}(h_{Q}), L_{x}, L_{x^{\prime}}, R_{y}$, $R_{y^{\prime}}) = \sum_{s \in \mathcal{I}_{B}} |\Psi_{\str}(T[p_{s}-1..r_{s}+1])| |f_{\recover}(([p_{s}, q_{s}], [\ell_{s}, r_{s}]))|$ 
follows from the definitions of the set $\mathcal{I}_{B}$ and range-sum query. 
Therefore, Equation~\ref{eq:x_set_RB_property:1} follows from the following equation: 
\begin{equation*}
    \begin{split}
    |\Psi_{\CCP}(T[i..j]) \cap \Psi_{\lex}(T[\gamma_{Q}..r_{Q}+1]) & \cap \Psi_{\run}| \\
    &= |\bigcup_{s \in \mathcal{I}_{B}} \Psi_{\sRecover}(([p_{s}, q_{s}], [\ell_{s}, r_{s}]))| \\ 
    &= \sum_{s \in \mathcal{I}_{B}} |\Psi_{\sRecover}(([p_{s}, q_{s}], [\ell_{s}, r_{s}]))| \\ 
    &= \sum_{s \in \mathcal{I}_{B}} |\Psi_{\str}(T[p_{s}-1..r_{s}+1])| |f_{\recover}(([p_{s}, q_{s}], [\ell_{s}, r_{s}]))| \\ 
    &= \rangesum(\mathcal{J}_{B}(h_{Q}), L_{x}, L_{x^{\prime}}, R_{y}, R_{y^{\prime}}).     
    \end{split}
\end{equation*}
\end{proof}

Next, we prove Equation~\ref{eq:x_set_RB_property:2} in Lemma~\ref{lem:JB_main_lemma} using 
Proposition~\ref{prop:set_JB2_subseq} and Proposition~\ref{prop:set_JB2_supseq}. 

\begin{proof}[Proof of Equation~\ref{eq:x_set_RB_property:2} in Lemma~\ref{lem:JB_main_lemma}]
From Proposition~\ref{prop:set_JB2_subseq} and Proposition~\ref{prop:set_JB2_supseq}, 
$|(\Psi_{\CCP}(T[i..j]) \cap \Psi_{\lex}(T[\gamma_{Q}..r_{Q}+1]) \cap \Psi_{\run}) \setminus \Psi_{\centerset}(C_{Q})| = |\bigcup_{s \in \mathcal{I}^{\prime}_{B}} \Psi_{\sRecover}(([p_{s}, q_{s}], [\ell_{s}, r_{s}]))|$ holds. 
$|\bigcup_{s \in \mathcal{I}^{\prime}_{B}} \Psi_{\sRecover}$ $(([p_{s}, q_{s}], [\ell_{s}, r_{s}]))| = \sum_{s \in \mathcal{I}^{\prime}_{B}} |\Psi_{\sRecover}(([p_{s}, q_{s}], [\ell_{s}, r_{s}]))|$ holds because 
$\Psi_{\sRecover}(([p_{s}, q_{s}], [\ell_{s}, r_{s}])) \cap \Psi_{\sRecover}(([p_{s^{\prime}}$, $q_{s^{\prime}}], [\ell_{s^{\prime}}, r_{s^{\prime}}])) = \emptyset$ for any pair of two integers $1 \leq s < s^{\prime} \leq k$ (see the proof of Equation~\ref{eq:x_set_RB_property:1}). 
Lemma~\ref{lem:sRecover_size_property}~\ref{enum:sRecover_size_property:2} shows that 
$|\Psi_{\sRecover}(([p_{s}, q_{s}], [\ell_{s}, r_{s}]))| = |\Psi_{\str}(T[p_{s}-1..r_{s}+1])| |f_{\recover}(([p_{s}, q_{s}], [\ell_{s}, r_{s}]))|$ for each integer $s \in \mathcal{I}^{\prime}_{B}$. 
$\rangesum(\mathcal{J}_{B}(h_{Q}), L_{x}, L_{x^{\prime}}, R_{y}$, $R_{\hat{y}}) = \sum_{s \in \mathcal{I}^{\prime}_{B}}$ $|\Psi_{\str}(T[p_{s}-1..r_{s}+1])| |f_{\recover}(([p_{s}$, $q_{s}], [\ell_{s}, r_{s}]))|$ 
follows from the definitions of the set $\mathcal{I}^{\prime}_{B}$ and range-sum query. 
Therefore, Equation~\ref{eq:x_set_RB_property:2} follows from the following equation: 
\begin{equation*}
    \begin{split}
    |(\Psi_{\CCP}(T[i..j]) \cap \Psi_{\lex}(T[\gamma_{Q}..r_{Q}+1]) \cap & \Psi_{\run}) \setminus \Psi_{\centerset}(C_{Q})| \\
    &= |\bigcup_{s \in \mathcal{I}^{\prime}_{B}} \Psi_{\sRecover}(([p_{s}, q_{s}], [\ell_{s}, r_{s}]))| \\ 
    &= \sum_{s \in \mathcal{I}^{\prime}_{B}} |\Psi_{\sRecover}(([p_{s}, q_{s}], [\ell_{s}, r_{s}]))| \\ 
    &= \sum_{s \in \mathcal{I}^{\prime}_{B}} |\Psi_{\str}(T[p_{s}-1..r_{s}+1])| |f_{\recover}(([p_{s}, q_{s}], [\ell_{s}, r_{s}]))| \\ 
    &= \rangesum(\mathcal{J}_{B}(h_{Q}), L_{x}, L_{x^{\prime}}, R_{y}, R_{\hat{y}}).     
    \end{split}
\end{equation*}

\end{proof}

\subsubsection{Dynamic Data Structures for Ordered Set \texorpdfstring{$\mathcal{X}_{B}(h)$}{}}\label{subsubsec:JB_X_ds}
For each integer $h \in [0, H]$, 
we present dynamic data structures to store the ordered set $\mathcal{X}_{B}(h)$, 
which are similar to the dynamic data structures for the ordered set $\mathcal{X}_{A}(h)$ presented in Section~\ref{subsubsec:JA_X_ds}. 
Here, the ordered set $\mathcal{X}_{B}(h)$ consists of $d$ strings $L_{1}, L_{2}, \ldots, L_{d}$~($L_{1} \prec L_{2} \prec \cdots \prec L_{d}$). 
For each string $L_{b} \in \mathcal{X}_{B}(h)$, 
we introduce a sequence $\mathbf{Q}^{X}_{B}(h, L_{b})$ of $m$ weighted points $(x_{1}, y_{1}, w_{1}, e_{1})$, 
$(x_{2}, y_{2}, w_{2}, e_{2})$, $\ldots$, $(x_{m}, y_{m}, w_{m}, e_{m})$ in set $\mathcal{J}_{B}(h)$ of weighted points 
such that each weighted point $(x_{s}, y_{s}, w_{s}, e_{s})$ contains the string $L_{b}$ as its x-coordinate 
(i.e., $\{ (x_{s}, y_{s}, w_{s}, e_{s}) \mid s \in [1, m] \} = \{ (x, y, w, e) \in \mathcal{J}_{B}(h) \mid x = L_{b} \}$). 
The $m$ weighted points of sequence $\mathbf{Q}^{X}_{B}(h, L_{b})$ are sorted in 
a similar way as that of sequence $\mathbf{Q}^{X}_{A}(h, L_{b})$ introduced in Section~\ref{subsubsec:JA_X_ds}.

We store the ordered set $\mathcal{X}_{B}(h)$ using $d+1$ doubly linked lists $\mathbf{X}_{B}(h, L_{1}), \mathbf{X}_{B}(h, L_{2})$, $\ldots$, $\mathbf{X}_{B}(h, L_{d})$, and $\mathbf{L}^{X}_{B}(h)$. 
For each integer $b \in [1, d]$, 
the doubly linked list $\mathbf{X}_{B}(h, L_{b})$ consists of $m$ elements 
for the $m$ weighted points in sequence $\mathbf{Q}^{X}_{B}(h, L_{b})$. 
Each $s$-th element of $\mathbf{X}_{B}(h, L_{b})$ corresponds to the $s$-th weighted point $(x_{s}, y_{s}, w_{s}, e_{s})$ of $\mathbf{Q}^{X}_{B}(h, L_{b})$. 
Here, the following three statements hold: 
\begin{itemize}
    \item from the definition of set $\mathcal{J}_{B}(h)$, 
    the $s$-th weighted point $(x_{s}, y_{s}, w_{s}, e_{s})$ corresponds to an interval attractor $([p, q], [\ell, r])$ in the sampling subset $\Psi_{\samp}$;
    \item the interval attractor $([p, q], [\ell, r])$ corresponds to a node $u$ of the sequence $\mathbf{Q}_{\samp}$ introduced in Section~\ref{subsec:sample_query}; 
    \item the node $u$ is represented as an element $v$ of the doubly linked list introduced in Section~\ref{subsubsec:sample_ds}. 
\end{itemize}
The $s$-th element of the doubly linked list $\mathbf{X}_{B}(h, L_{b})$ stores a pointer to the element $v$ corresponding to the $s$-th weighted point $(x_{s}, y_{s}, w_{s}, e_{s})$. 
A list indexing data structure is used for quickly accessing to the elements of the doubly linked list $\mathbf{X}_{B}(h, L_{b})$. 
Here, the list indexing data structure is introduced in Section~\ref{subsubsec:sample_ds}. 

The last doubly-linked list $\mathbf{L}^{X}_{B}(h)$ consists of $d$ elements such that 
each $b$-th element corresponds to the $b$-th string $L_{b}$ of the ordered set $\mathcal{X}_{B}(h)$. 
Here, the $b$-th element stores a pointer to the $b$-th doubly linked list $\mathbf{X}_{B}(h, L_{b})$. 
A list indexing data structure is used for quickly accessing to the elements of the doubly linked list $\mathbf{L}^{X}_{B}(h)$. 
An order maintenance data structure~\cite{DBLP:conf/stoc/DietzS87} is used for comparing two elements of the doubly linked list $\mathbf{L}^{X}_{B}(h)$ with the total order of the doubly linked list. 
Here, the order maintenance data structure is introduced in Section~\ref{subsubsec:JA_X_ds}, 
and this order maintenance data structure is used to compare two given strings $L_{b}$ and $L_{b^{\prime}}$ 
in ordered set $\mathcal{X}_{B}(h)$ in $O(1)$ time. 

Overall, these $d+1$ doubly linked lists $\mathbf{X}_{B}(h, L_{1}), \mathbf{X}_{B}(h, L_{2})$, $\ldots$, $\mathbf{X}_{B}(h, L_{d})$, and $\mathbf{L}^{X}_{B}(h)$ require $O((d + |\mathcal{J}_{B}(h)|) B)$ bits of space in total for machine word size $B$. 
Similar to the dynamic data structure for the ordered set $\mathcal{X}_{A}(h)$, 
the dynamic data structures for the ordered set $\mathcal{X}_{B}(h)$ can be stored in $O(|\mathcal{J}_{B}(h)| B)$ bits of space. 

The following lemma states queries supported by the dynamic data structures for the ordered set $\mathcal{X}_{A}(h)$. 

\begin{lemma}\label{lem:JB_X_queries}
    Let $L_{1}, L_{2}, \ldots, L_{d}$~($L_{1} \prec L_{2} \prec \cdots \prec L_{d}$) be the $d$ strings of 
    the ordered set $\mathcal{X}_{B}(h)$ introduced in Section~\ref{subsec:RSC_comp_B} 
    for an integer $h \in [0, H]$. 
    Consider the $d+1$ doubly linked lists $\mathbf{X}_{B}(h, L_{1}), \mathbf{X}_{B}(h, L_{2})$, $\ldots$, $\mathbf{X}_{B}(h, L_{d})$, $\mathbf{L}^{X}_{B}(h)$. 
    Using these $d+1$ doubly linked lists and the dynamic data structures of Section~\ref{subsubsec:rrdag_ds} 
    and Section~\ref{subsubsec:sample_ds}, 
    we can support the following three queries: 
    \begin{enumerate}[label=\textbf{(\roman*)}]
    \item \label{enum:JB_X_queries:1} 
    verify whether $L_{s} \prec L_{s^{\prime}}$ or not in $O(1)$ time 
    for the given $s$-th and $s^{\prime}$-th elements of the doubly linked list $\mathbf{L}^{X}_{B}(h)$;
    \item \label{enum:JB_X_queries:2} 
    for a given integer $s \in [1, d]$, 
    return an interval $[g, g + |L_{s}| - 1]$ in input string $T$ satisfying $\reverse(T[g..g + |L_{s}| - 1]) = L_{s}$ 
    in $O(H^{2} + \log n)$ time if $L_{s} \not \in \{ \varepsilon, \# \}$; 
    otherwise return the string $L_{s}$ in $O(1)$ time;
    \item \label{enum:JB_X_queries:3} 
    consider a given pair of an integer $s \in [1, d]$ and interval $[\alpha, \beta] \subseteq [1, n]$ in input string $T$. 
    Then, verify the following three conditions in $O(H^{2} + \log n)$ time: 
    (A) $\reverse(T[\alpha..\beta]) \prec L_{s}$, 
    (B) $\reverse(T[\alpha..\beta]) = L_{s}$, 
    and (C) $L_{s} \prec \reverse(\#T[\alpha..\beta])$.    
    \end{enumerate}
\end{lemma}

The following lemma is used to prove Lemma~\ref{lem:JB_X_queries}. 
\begin{lemma}\label{lem:mRecover_query}
    Consider an interval attractor $([p, q], [\ell, r])$ in subset $\Psi_{\source}$. 
    Here, Lemma~\ref{lem:mRecover_basic_property} shows that 
    there exists an interval attractor $([p^{\prime}, q^{\prime}], [\ell^{\prime}, r^{\prime}]) \in \Psi_{\RR}$ 
    satisfying $f_{\recover}(([p, q], [\ell, r])) \cap \Psi_{\mRecover} = \{ ([p^{\prime}, q^{\prime}], [\ell^{\prime}, r^{\prime}]) \}$. 
    Let $\gamma^{\prime}$ be the attractor position of the interval attractor $([p^{\prime}, q^{\prime}], [\ell^{\prime}, r^{\prime}])$. 
    Then, 
    we can compute interval $[p^{\prime}, r^{\prime}]$ and the attractor position $\gamma^{\prime}$ 
    in $O(H^{2})$ time 
    using the interval attractor $([p, q], [\ell, r])$ and the dynamic data structures of Section~\ref{subsubsec:rrdag_ds}. 
\end{lemma}
\begin{proof}
    Let $([p_{1}, q_{1}], [\ell_{1}, r_{1}])$, 
    $([p_{2}, q_{2}], [\ell_{2}, r_{2}])$, $\ldots$, $([p_{k}, q_{k}], [\ell_{k}, r_{k}])$ ($p_{1} < p_{2} < \ldots < p_{k}$)
    be the interval attractors in the set obtained from function $f_{\recover}(([p, q], [\ell, r]))$. 
    Let $\gamma$ and $C$ be the attractor position and associated string of the interval attractor $([p, q], [\ell, r])$, respectively. 

    If $([p, q], [\ell, r]) \in \Psi_{\preceding}$, 
    then 
    Lemma~\ref{lem:mRecover_basic_property} shows that 
    $([p^{\prime}, q^{\prime}], [\ell^{\prime}, r^{\prime}]) = ([p_{k}, q_{k}], [\ell_{k}, r_{k}])$ holds. 
    In this case, 
    Lemma~\ref{lem:recover_basic_property}~\ref{enum:recover_basic_property:2} shows that 
    $[p^{\prime}, r^{\prime}] = [q + 1 + (k - 1) |C|, r]$ 
    and $\gamma^{\prime} = \gamma + k |C|$ hold. 
    Otherwise (i.e., $([p, q], [\ell, r]) \not \in \Psi_{\preceding}$), 
    Lemma~\ref{lem:mRecover_basic_property} shows that 
    $([p^{\prime}, q^{\prime}], [\ell^{\prime}, r^{\prime}]) = ([p_{1}, q_{1}], [\ell_{1}, r_{1}])$ holds. 
    In this case, 
    Lemma~\ref{lem:recover_basic_property}~\ref{enum:recover_basic_property:2} shows that 
    $[p^{\prime}, r^{\prime}] = [q + 1, r]$ and $\gamma^{\prime} = \gamma + |C|$ hold. 

    We can verify whether $([p, q], [\ell, r]) \in \Psi_{\preceding}$ or not by verify-prec query $\precQ(([p, q], [\ell, r]))$.
    The attractor position $\gamma$ can be obtained by attractor position query $\attrQ(([p, q], [\ell, r]))$.
    The length of the associated string $C$ can be obtained by C-length query $\clenQ(([p, q], [\ell, r]))$.
    The integer $k$ can be obtained by r-size query $\rsizeQ(([p, q], [\ell, r]))$.
    These four queries take $O(H^{2})$ time. 
    Therefore, we can compute interval $[p^{\prime}, r^{\prime}]$ and the attractor position $\gamma^{\prime}$ 
    in $O(H^{2})$ time.      
\end{proof}

The proof of Lemma~\ref{lem:JB_X_queries} is as follows.
\begin{proof}[Proof of Lemma~\ref{lem:JB_X_queries}~\ref{enum:JB_X_queries:1}]
We can verify whether $L_{s} \prec L_{s^{\prime}}$ or not in $O(1)$ time by the order maintenance data structure built on 
doubly linked list $\mathbf{L}^{X}_{B}(h)$. 
\end{proof}

\begin{proof}[Proof of Lemma~\ref{lem:JB_X_queries}~\ref{enum:JB_X_queries:2}]
Lemma~\ref{lem:JB_X_queries} \ref{enum:JB_X_queries:2} can be proved using 
a similar approach as for Lemma \ref{lem:JA_X_queries} \ref{enum:JA_X_queries:2}. 
The detailed proof of Lemma \ref{lem:JB_X_queries}~\ref{enum:JB_X_queries:2} is as follows. 

One of the following three conditions is satisfied: 
(a) $s = 1$; (b) $s = d$; (c) $1 < s < d$. 
For case (a), 
$L_{s} = \varepsilon$ follows from the definition of the ordered set $\mathcal{X}_{B}(h)$.
In this case, we can return the character $\varepsilon$ in $O(1)$ time. 
For case (b), 
$L_{s} = \#$ follows from the definition of the ordered set $\mathcal{X}_{B}(h)$.
In this case, we can return the character $\#$ in $O(1)$ time. 
    
For case (c), 
the $s$-th doubly linked list $\mathbf{X}_{B}(h, L_{s})$ contains at least one element. 
Let $(x_{1}, y_{1}, w_{1}$, $e_{1})$ be the weighted point corresponding to the first element of the doubly linked list $\mathbf{X}_{B}(h, L_{s})$. 
Then, $x_{1} = L_{s}$ holds. 
Let $([p_{1}, q_{1}], [\ell_{1}, r_{1}])$ be the interval attractor corresponding to the weighted point $(x_{1}, y_{1}, w_{1}, e_{1})$. 
Then, 
Lemma~\ref{lem:mRecover_basic_property} shows that 
set $f_{\recover}(([p_{1}, q_{1}], [\ell_{1}, r_{1}])) \cap \Psi_{\mRecover}$ contains 
an interval attractor $([\hat{p}_{1}, \hat{q}_{1}], [\hat{\ell}_{1}, \hat{r}_{1}])$ 
satisfying 
$f_{\recover}(([p_{1}, q_{1}], [\ell_{1}, r_{1}])) \cap \Psi_{\mRecover} = \{ ([\hat{p}_{1}, \hat{q}_{1}]$, $[\hat{\ell}_{1}, \hat{r}_{1}]) \}$).
From the definition of the set $\mathcal{J}_{B}(h)$, 
$x_{1} = \reverse(T[\hat{p}_{1}-1..\hat{\gamma}_{1}-1])$ holds 
for the attractor position $\hat{\gamma}_{1}$ of the interval attractor $([\hat{p}_{s}, \hat{q}_{s}], [\hat{\ell}_{s}, \hat{r}_{s}])$. 
$L_{s} = \reverse(T[\hat{p}_{1}-1..\hat{\gamma}_{1}-1])$ follows from $x_{1} = L_{s}$ and $x_{1} = \reverse(T[\hat{p}_{1}-1..\hat{\gamma}_{1}-1])$. 
Therefore, we can return interval $[\hat{p}_{1}-1, \hat{\gamma}_{1}-1]$ as the answer to the query of Lemma~\ref{lem:JA_X_queries}~\ref{enum:JA_X_queries:2}.

In this case, we answer the query of Lemma~\ref{lem:JB_X_queries}~\ref{enum:JB_X_queries:2} in two phases. 
In the first phase, we obtain the node $u$ of the sequence $\mathbf{Q}_{\samp}$. 
The $s$-th element of doubly linked list $\mathbf{L}^{X}_{B}(h)$ stores 
a pointer to the doubly linked list $\mathbf{X}_{B}(h, L_{s})$.     
The first element of the doubly linked list $\mathbf{X}_{B}(h, L_{s})$ stores 
a pointer to the element representing the node $u$ 
in the doubly linked list of Section~\ref{subsubsec:sample_ds}. 
The $s$-th element can be accessed in $O(\log d)$ time by 
the list indexing data structure built on the doubly linked list $\mathbf{L}^{X}_{B}(h)$. 
Here, $d = O(n^{2})$ follows from Lemma~\ref{lem:JB_size}~\ref{enum:JB_size:3}. 
Therefore, we can obtain the node $u$ in $O(\log n)$ time. 

In the second phase, we return interval $[\hat{p}_{1}-1, \hat{\gamma}_{1}-1]$ as the answer to the query of Lemma~\ref{lem:JA_X_queries}~\ref{enum:JA_X_queries:2}.     
We recover the interval attractor $([p_{1}, q_{1}], [\ell_{1}, r_{1}])$ from the node $u$ 
in $O(\log n)$ time by the algorithm of Section~\ref{subsubsec:computation_delta_samp}. 
Lemma~\ref{lem:mRecover_query} shows that 
we can compute the interval $[\hat{p}_{1}-1, \hat{\gamma}_{1}-1]$ in $O(H^{2})$ time using 
the interval attractor $([p_{1}, q_{1}], [\ell_{1}, r_{1}])$ and the dynamic data structures of Section~\ref{subsubsec:rrdag_ds}. 
Therefore, the second phase takes $O(H^{2} + \log n)$ time. 

Finally, Lemma~\ref{lem:JB_X_queries}~\ref{enum:JB_X_queries:2} holds. 
\end{proof}

\begin{proof}[Proof of Lemma~\ref{lem:JB_X_queries}~\ref{enum:JB_X_queries:3}]
Lemma~\ref{lem:JB_X_queries}~\ref{enum:JB_X_queries:3} corresponds to 
Lemma~\ref{lem:JA_X_queries}~\ref{enum:JA_X_queries:3}.
We proved Lemma~\ref{lem:JA_X_queries} \ref{enum:JA_X_queries:3} using 
the query of Lemma~\ref{lem:JA_X_queries}~\ref{enum:JA_X_queries:2}, one reversed LCE query, and random access queries. 
Lemma~\ref{lem:JB_X_queries} \ref{enum:JB_X_queries:2} corresponds to Lemma~\ref{lem:JA_X_queries}~\ref{enum:JA_X_queries:2}. 
Therefore, Lemma~\ref{lem:JB_X_queries}~\ref{enum:JB_X_queries:3} can be proved using the same approach as for Lemma~\ref{lem:JA_X_queries}~\ref{enum:JA_X_queries:3}. 
\end{proof}

\subsubsection{Dynamic Data Structures for Ordered Set \texorpdfstring{$\mathcal{Y}_{B}(h)$}{}}\label{subsubsec:JB_Y_ds}
For each integer $h \in [0, H]$, 
we present dynamic data structures to store the ordered set $\mathcal{Y}_{B}(h)$, 
which are similar to the dynamic data structures for the ordered set $\mathcal{Y}_{A}(h)$ presented in Section~\ref{subsubsec:JA_Y_ds}. 
Here, the ordered set $\mathcal{Y}_{B}(h)$ consists of $d^{\prime}$ strings $R_{1}, R_{2}, \ldots, R_{d^{\prime}}$~($R_{1} \prec R_{2} \prec \cdots \prec R_{d^{\prime}}$). 
For each string $R_{b} \in \mathcal{Y}_{B}(h)$, 
we introduce a sequence $\mathbf{Q}^{Y}_{B}(h, R_{b})$ of $m$ weighted points $(x_{1}, y_{1}, w_{1}, e_{1})$, 
$(x_{2}, y_{2}, w_{2}, e_{2})$, $\ldots$, $(x_{m}, y_{m}, w_{m}, e_{m})$ in set $\mathcal{J}_{B}(h)$ of weighted points 
such that each weighted point $(x_{s}, y_{s}, w_{s}, e_{s})$ contains the string $R_{b}$ as its y-coordinate. 
The $m$ weighted points of sequence $\mathbf{Q}^{Y}_{B}(h, Y_{b})$ are sorted in 
a similar way as that of sequence $\mathbf{Q}^{Y}_{A}(h, Y_{b})$ introduced in Section~\ref{subsubsec:JA_Y_ds}.

We store the ordered set $\mathcal{Y}_{B}(h)$ using $d^{\prime}+1$ doubly linked lists $\mathbf{Y}_{B}(h, R_{1}), \mathbf{Y}_{B}(h, R_{2})$, $\ldots$, $\mathbf{Y}_{B}(h, R_{d^{\prime}})$, and $\mathbf{L}^{Y}_{B}(h)$. 
For each integer $b \in [1, d^{\prime}]$, 
the doubly linked list $\mathbf{Y}_{B}(h, R_{b})$ consists of $m$ elements 
for the $m$ weighted points in sequence $\mathbf{Q}^{Y}_{B}(h, R_{b})$. 
Each $s$-th element of $\mathbf{Y}_{B}(h, R_{b})$ corresponds to the $s$-th weighted point $(x_{s}, y_{s}, w_{s}, e_{s})$ of $\mathbf{Q}^{Y}_{B}(h, R_{b})$. 
Similar to Section~\ref{subsubsec:JB_X_ds}, 
the $s$-th weighted point $(x_{s}, y_{s}, w_{s}, e_{s})$ corresponds to an element $v$ of the doubly linked list introduced in Section~\ref{subsubsec:sample_ds}. 
The $s$-th element of $\mathbf{Y}_{B}(h, R_{b})$ stores a pointer to the element $v$ corresponding to the $s$-th weighted point 
$(x_{s}, y_{s}, w_{s}, e_{s})$. 
A list indexing data structure is used for quickly accessing to the elements of the doubly linked list $\mathbf{Y}_{B}(h, R_{b})$. 

The last doubly-linked list $\mathbf{L}^{Y}_{B}(h)$ consists of $d^{\prime}$ elements such that 
each $b$-th element corresponds to the $b$-th string $R_{b}$ of the ordered set $\mathcal{Y}_{B}(h)$. 
Here, the $b$-th element stores a pointer to the $b$-th doubly linked list $\mathbf{Y}_{B}(h, R_{b})$. 
Similar to the doubly linked list $\mathbf{L}^{X}_{B}(h)$ of Section~\ref{subsubsec:JB_X_ds}, 
list indexing and order maintenance data structures are built on the doubly-linked list $\mathbf{L}^{Y}_{B}(h)$. 
These $d^{\prime}+1$ doubly linked lists $\mathbf{Y}_{B}(h, R_{1}), \mathbf{Y}_{B}(h, R_{2})$, $\ldots$, $\mathbf{Y}_{B}(h, R_{d^{\prime}})$, $\mathbf{L}^{Y}_{B}(h)$ require 
$O((d^{\prime} + |\mathcal{J}_{B}(h)|) B)$ bits of space in total for machine word size $B$. 
$d^{\prime} \leq 2 + |\mathcal{J}_{B}(h)|$ follows from Lemma~\ref{lem:JB_size}~\ref{enum:JB_size:1}. 
Therefore, the dynamic data structures for the ordered set $\mathcal{Y}_{B}(h)$ can be stored in $O(|\mathcal{J}_{B}(h)| B)$ bits of space. 

The following lemma states queries supported by the dynamic data structures for the ordered set $\mathcal{Y}_{B}(h)$. 

\begin{lemma}\label{lem:JB_Y_queries}
    Let $L_{1}, L_{2}, \ldots, L_{d}$~($L_{1} \prec L_{2} \prec \cdots \prec L_{d}$) be the $d$ strings of 
    the ordered set $\mathcal{X}_{A}(h)$ introduced in Section~\ref{subsec:RSC_comp_A} for an integer $h \in [0, H]$. 
    Similarly, 
    let $R_{1}, R_{2}, \ldots, R_{d^{\prime}}$~($R_{1} \prec R_{2} \prec \cdots \prec R_{d^{\prime}}$) be 
    the $d^{\prime}$ strings of the ordered set $\mathcal{Y}_{B}(h)$ introduced in Section~\ref{subsec:RSC_comp_B}. 
    Consider the $d^{\prime}+1$ doubly linked lists $\mathbf{Y}_{B}(h, R_{1}), \mathbf{Y}_{B}(h, R_{2})$, $\ldots$, $\mathbf{Y}_{B}(h, R_{d^{\prime}})$, $\mathbf{L}^{Y}_{B}(h)$. 
    Using these $d^{\prime}+1$ doubly linked lists and the dynamic data structures of Section~\ref{subsubsec:rrdag_ds} 
    and Section~\ref{subsubsec:sample_ds}, 
    we can support the following four queries: 
    \begin{enumerate}[label=\textbf{(\roman*)}]
    \item \label{enum:JB_Y_queries:1} 
    verify whether $R_{s} \prec R_{s^{\prime}}$ or not in $O(1)$ time 
    for the given $s$-th and $s^{\prime}$-th elements of the doubly linked list $\mathbf{L}^{Y}_{B}(h)$;    
    \item \label{enum:JB_Y_queries:2} 
    for a given integer $s \in [1, d^{\prime}]$, 
    return an interval $[g, g + |R_{s}| - 1]$ in input string $T$ satisfying $T[g..g + |R_{s}| - 1] = R_{s}$ 
    in $O(H^{2} + \log n)$ time if $R_{s} \not \in \{ \varepsilon, \# \}$; 
    otherwise return the string $R_{s}$ in $O(1)$ time;
    \item \label{enum:JB_Y_queries:3} 
    consider a given pair of an integer $s \in [1, d^{\prime}]$ and interval $[\alpha, \beta] \subseteq [1, n]$ in input string $T$. 
    Then, verify the following four conditions in $O(H^{2} + \log n)$ time: 
    (A) $T[\alpha..\beta] \prec R_{s}$, 
    (B) $T[\alpha..\beta] = R_{s}$, 
    (C) $R_{s} \prec T[\alpha..\beta]$, 
    and (D) $R_{s} \prec T[\alpha..\beta]\#$;
    \item \label{enum:JB_Y_queries:4} 
    consider a given triplet of three integers $\tau, \tau^{\prime} \in [1, d]$ ($\tau \leq \tau^{\prime}$) 
    and $s \in [1, d^{\prime}]$. 
    If set $\mathcal{J}_{B}(h)$ contains a weighed point $(x, y, w, e)$ 
    satisfying $L_{\tau} \preceq x \preceq L_{\tau^{\prime}}$ and $y = R_{s}$, 
    then return the interval attractor corresponding to the weighed point $(x, y, w, e)$ 
    in $O(H^{2} \log n + \log^{2} n)$ time.
    For answering this query, 
    we need the dynamic data structures for the ordered set $\mathcal{X}_{B}(h)$ introduced in Section~\ref{subsubsec:JB_X_ds}.     
    \end{enumerate}
\end{lemma}
\begin{proof}
    The proof of Lemma~\ref{lem:JB_Y_queries} is as follows. 

    \textbf{Proof of Lemma~\ref{lem:JB_Y_queries}(i).}
    We can verify whether $R_{s} \prec R_{s^{\prime}}$ or not in $O(1)$ time by the order maintenance data structure built on 
    doubly linked list $\mathbf{L}^{Y}_{B}(h)$. 

    \textbf{Proof of Lemma~\ref{lem:JB_Y_queries}(ii).}
    Lemma~\ref{lem:JB_Y_queries}~\ref{enum:JB_Y_queries:2} corresponds to 
    Lemma~\ref{lem:JB_X_queries}~\ref{enum:JB_X_queries:2}. 
    We proved Lemma~\ref{lem:JB_X_queries}~\ref{enum:JB_X_queries:2} using Lemma~\ref{lem:mRecover_query}. 
    Similarly, 
    Lemma~\ref{lem:JB_Y_queries}~\ref{enum:JB_Y_queries:2} can be proved using Lemma~\ref{lem:mRecover_query}. 

    \textbf{Proof of Lemma~\ref{lem:JB_Y_queries}(iii).}
    Lemma~\ref{lem:JB_Y_queries}~\ref{enum:JB_Y_queries:3} corresponds to 
    Lemma~\ref{lem:JA_Y_queries}~\ref{enum:JA_Y_queries:3}.
    We proved Lemma~\ref{lem:JA_Y_queries}~\ref{enum:JA_Y_queries:3} using 
    the query of Lemma~\ref{lem:JA_Y_queries}~\ref{enum:JA_Y_queries:2}, one LCE query, and random access queries. 
    Lemma~\ref{lem:JB_Y_queries}~\ref{enum:JB_Y_queries:2} corresponds to Lemma~\ref{lem:JA_Y_queries}~\ref{enum:JA_Y_queries:2}. 
    Therefore, Lemma~\ref{lem:JB_Y_queries}~\ref{enum:JB_Y_queries:3} can be proved using the same approach as for Lemma~\ref{lem:JA_Y_queries}~\ref{enum:JA_Y_queries:3}.     

    \textbf{Proof of Lemma~\ref{lem:JB_Y_queries}(iv).}
    Lemma~\ref{lem:JB_Y_queries}~\ref{enum:JB_Y_queries:4} corresponds to 
    Lemma~\ref{lem:JA_Y_queries}~\ref{enum:JA_Y_queries:4}. 
    Lemma~\ref{lem:JB_Y_queries} \ref{enum:JB_Y_queries:4} can be proved using the same approach as for Lemma~\ref{lem:JA_Y_queries}~\ref{enum:JA_Y_queries:4}.     

\end{proof}

\subsubsection{Dynamic Data Structures for Set \texorpdfstring{$\mathcal{J}_{B}(h)$}{} of Weighted Points}\label{subsubsec:JB_ds}
For each integer $h \in [0, H]$, 
we present dynamic data structures to support range-sum query on set $\mathcal{J}_{B}(h)$ of weighted points,
which are similar to the dynamic data structures for the ordered set $\mathcal{J}_{A}(h)$ presented in Section~\ref{subsubsec:JA_ds}. 
Let $(x_{1}, y_{1}, w_{1}, e_{1})$, $(x_{2}, y_{2}, w_{2}, e_{2})$, $\ldots$, $(x_{k}, y_{k}, w_{k}, e_{k})$ ($e_{1} \prec e_{2} \prec \cdots e_{k}$) be the weighted points in the set $\mathcal{J}_{B}(h)$. 
Here, the following two statements hold for each weighted point $(x_{s}, y_{s}, w_{s}, e_{s}) \in \mathcal{J}_{B}(h)$: 
\begin{itemize}
    \item the doubly linked list $\mathbf{X}_{B}(h, x_{s})$ of Section~\ref{subsubsec:JB_X_ds} 
    contains an element $v_{s}$ representing the weighted point $(x_{s}, y_{s}, w_{s}, e_{s})$; 
    \item the doubly linked list $\mathbf{Y}_{B}(h, y_{s})$ of Section~\ref{subsubsec:JB_Y_ds} 
    contains an element $v^{\prime}_{s}$ representing the weighted point $(x_{s}, y_{s}, w_{s}, e_{s})$. 
\end{itemize}

We store the set $\mathcal{J}_{B}(h)$ using a doubly linked list $\mathbf{L}_{B}(h)$ of $k$ elements. 
For each integer $s \in [1, k]$, 
the $s$-th element of the doubly linked list $\mathbf{L}_{B}(h)$ corresponds to the $s$-th weighted point $(x_{s}, y_{s}, w_{s}, e_{s})$. This element stores (i) the weight $w_{s}$ and (ii) two pointers to the two elements $v_{s}$ and $v^{\prime}_{s}$. 
List indexing and range-sum data structures are built on doubly linked list $\mathbf{L}_{B}(h)$. 
This range-sum data structure is used to support range-count and range-sum queries on the set $\mathcal{J}_{B}(h)$ of weighted points. 
These dynamic data structures require $O(|\mathcal{J}_{B}(h)| B)$ bits of space in total for machine word size $B$.

\subsubsection{Dynamic Data Structures and Algorithm for \texorpdfstring{$\RSCQBX(i, j)$}{RSCB1(i, j)} and \texorpdfstring{$\RSCQBY(i, j)$}{RSCB2(i, j)}}\label{subsubsec:JB_subquery_ds}
We prove Lemma~\ref{lem:RSC_subquery_B_summary}, i.e., 
we show that $\RSCQBX(i, j)$ and $\RSCQBY(i, j)$ can be answered in 
$O(H^{2} \log n + \log^{4} n)$ time using dynamic data structures of 
$O((H + |\mathcal{U}_{\RR}| + |\Psi_{\samp}|)B)$ bits of space for machine word size $B$. 

\paragraph{Data Structures.}
We answer $\RSCQBX(i, j)$ and $\RSCQBY(i, j)$ using the following dynamic data structures: 
\begin{itemize}
    \item the dynamic data structures of $O(|\mathcal{U}_{\RR}|B)$ bits of space 
    for the RR-DAG of RLSLP $\mathcal{G}^{R}$ (Section~\ref{subsubsec:rrdag_ds}). 
    \item the dynamic data structures of $O(|\Psi_{\samp}|B)$ bits of space 
    for sample query (Section~\ref{subsubsec:sample_ds});
    \item the dynamic data structures of $O(\sum_{h = 0}^{H} |\mathcal{J}_{B}(h)| B)$ bits of space for $(1 + H)$ sets $\mathcal{X}_{B}(0)$, $\mathcal{X}_{B}(1)$, $\ldots$, $\mathcal{X}_{B}(H)$ 
    (Section~\ref{subsubsec:JB_X_ds});
    \item the dynamic data structures of $O(\sum_{h = 0}^{H} |\mathcal{J}_{B}(h)| B)$ bits of space 
    for $(1 + H)$ sets $\mathcal{Y}_{B}(0)$, $\mathcal{Y}_{B}(1)$, $\ldots$, $\mathcal{Y}_{B}(H)$ 
    (Section~\ref{subsubsec:JB_Y_ds});
    \item the dynamic data structures of $O(\sum_{h = 0}^{H} |\mathcal{J}_{B}(h)| B)$ bits of space 
    for $(1 + H)$ sets $\mathcal{J}_{B}(0)$, $\mathcal{J}_{B}(1)$, $\ldots$, $\mathcal{J}_{B}(H)$ 
    (Section~\ref{subsubsec:JB_ds}).
\end{itemize}
$\sum_{h = 0}^{H} |\mathcal{J}_{B}(h)| \leq |\Psi_{\samp}|$ follows from Lemma~\ref{lem:JB_size}~\ref{enum:JB_size:2}. 
Therefore, these dynamic data structures can be stored in $O((|\mathcal{U}_{\RR}| + |\Psi_{\samp}|) B)$ bits of space. 

\paragraph{Algorithm for subquery $\RSCQBX(i, j)$.}
The algorithm for $\RSCQBX(i, j)$ computes $|\Psi_{\CCP}(T[i..j]) \cap \Psi_{\lex}(T[\gamma_{Q}..r_{Q}+1]) \cap \Psi_{\run}|$ 
under the condition that $K_{Q} \leq 1 + \sum_{w = 1}^{h_{Q}+3} \lfloor \mu(w) \rfloor$ for the length $K_{Q}$ of the longest common prefix between two strings $T[\gamma_{Q}..r_{Q}+1]$ and $C_{Q}^{n+1}$. 
This algorithm leverages Lemma~\ref{lem:JB_main_lemma}, which shows that 
the size of set $\Psi_{\CCP}(T[i..j]) \cap \Psi_{\lex}(T[\gamma_{Q}..r_{Q}+1]) \cap \Psi_{\run}$ can be computed 
by one range-sum query on set $\mathcal{J}_{B}(h_{Q})$ of weighted points 
for the level $h_{Q}$ of interval attractor $([p_{Q}, q_{Q}], [\ell_{Q}, r_{Q}])$. 

The algorithm for $\RSCQBX(i, j)$ consists of three phases. 
In the first phase, 
we compute the level $h_{Q}$ and attractor position $\gamma_{Q}$ of interval attractor $([p_{Q}, q_{Q}], [\ell_{Q}, r_{Q}])$. 
Similar to the first phase of the algorithm for subquery $\RSCQA(i, j)$, 
this phase takes $O(H^{2} \log n)$ time. 

In the second phase, 
we compute four integers $x, x^{\prime}, y$, and $y^{\prime}$ of Lemma~\ref{lem:JB_main_lemma}. 
Here, 
\begin{itemize}
\item $x = \min \{ s \in [1, d] \mid \reverse(T[i..\gamma_{Q}-1]) \prec L_{s} \}$ for the $d$ strings $L_{1}, L_{2}, \ldots, L_{d}$~($L_{1} \prec L_{2} \prec \cdots \prec L_{d}$) in the ordered set $\mathcal{X}_{B}(h_{Q})$; 
\item $x^{\prime} = \max \{ s \in [1, d] \mid L_{s} \prec \reverse(\# \cdot T[i..\gamma_{Q}-1]) \}$; 
\item $y = \min \{ s \in [1, d^{\prime}] \mid T[\gamma_{Q}..j] \prec R_{s} \}$ for the $d^{\prime}$ strings $R_{1}, R_{2}, \ldots, R_{d^{\prime}}$~($R_{1} \prec R_{2} \prec \cdots \prec R_{d^{\prime}}$) in the ordered set $\mathcal{Y}_{B}(h_{Q})$;
\item $y^{\prime} = \max \{ s \in [1, d^{\prime}] \mid R_{s} \prec T[\gamma_{Q}..r_{Q} + 1] \}$. 
\end{itemize}
We compute the two integers $x$ and $x^{\prime}$ by binary search on the $d$ strings $L_{1}, L_{2}, \ldots, L_{d}$. 
This binary search can be executed in $O((H^{2} + \log n)\log d)$ time using Lemma~\ref{lem:JB_X_queries}~\ref{enum:JB_X_queries:3}. 
Similarly, we compute the two integers $y$ and $y^{\prime}$ by binary search on the $d^{\prime}$ strings $R_{1}, R_{2}, \ldots, R_{d^{\prime}}$. 
This binary search can be executed in $O((H^{2} + \log n)\log d^{\prime})$ time using Lemma~\ref{lem:JB_Y_queries}~\ref{enum:JB_X_queries:3}. 
Lemma~\ref{lem:JB_size}~\ref{enum:JB_size:3} shows that $d + d^{\prime} = O(n^{2})$ holds. 
Therefore, the second phase takes $O((H^{2} + \log n) \log n)$ time. 

In the third phase, we compute $|\Psi_{\CCP}(T[i..j]) \cap \Psi_{\lex}(T[\gamma_{Q}..r_{Q}+1]) \cap \Psi_{\run}|$ by 
range-sum query $\rangesum(\mathcal{J}_{B}(h_{Q}), L_{x}, L_{x^{\prime}}, R_{y}, R_{y^{\prime}})$. 
Similar to the third phase of the algorithm for subquery $\RSCQA(i, j)$, 
this phase takes $O(\log d + \log d^{\prime} + \log^{4} k)$ time for the number $k$ of weighted points in set $\mathcal{J}_{A}(h_{Q})$. 
Lemma~\ref{lem:JA_size}~\ref{enum:JA_size:3} shows that $d, d^{\prime}, k = O(n^{2})$ holds. 
Therefore, the running time of the third phases can be bounded by $O(\log^{4} n)$. 

Finally, the algorithm for subquery $\RSCQBX(i, j)$ can be executed in $O(H^{2} \log n + \log^{4} n)$ time in total. 

\paragraph{Algorithm for $\RSCQBY(i, j)$.}
We explain the algorithm answering $\RSCQBY(i, j)$. 
This algorithm computes $|(\Psi_{\CCP}(T[i..j]) \cap \Psi_{\lex}(T[\gamma_{Q}..r_{Q}+1]) \cap \Psi_{\run}) \setminus \Psi_{\centerset}(C_{Q})|$ under the condition that $K_{Q} > 1 + \sum_{w = 1}^{h_{Q}+3} \lfloor \mu(w) \rfloor$ holds. 
Similar to subquery $\RSCQBX(i, j)$, 
this algorithm leverages Lemma~\ref{lem:JB_main_lemma}, which shows that 
the size of set $(\Psi_{\CCP}(T[i..j]) \cap \Psi_{\lex}(T[\gamma_{Q}..r_{Q}+1]) \cap \Psi_{\run}) \setminus \Psi_{\centerset}(C_{Q})$ can be computed by one range-sum query on set $\mathcal{J}_{B}(h_{Q})$ of weighted points. 

The algorithm for $\RSCQBY(i, j)$ consists of three phases. 
In the first phase, 
we compute the level $h_{Q}$ and attractor position $\gamma_{Q}$ of interval attractor $([p_{Q}, q_{Q}], [\ell_{Q}, r_{Q}])$. 
Similar to the first phase of the algorithm for $\RSCQBX(i, j)$, 
this phase takes $O(H^{2} \log n)$ time. 

In the second phase, 
we compute four integers $x, x^{\prime}, y$, and $\hat{y}$ of Lemma~\ref{lem:JB_main_lemma}. 
Here, $\hat{y} = \max \{ s \in [1, d^{\prime}] \mid R_{s} \prec T[\gamma_{Q}..\gamma_{Q} + 1 + \sum_{w = 1}^{h_{Q}+3} \mu(w)] \}$. 
Similar to $\RSCQBX(i, j)$, 
the four four integers $x, x^{\prime}, y$, and $\hat{y}$ can be computed in $O((H^{2} + \log n) \log n)$ time. 

In the third phase, we compute $|(\Psi_{\CCP}(T[i..j]) \cap \Psi_{\lex}(T[\gamma_{Q}..r_{Q}+1]) \cap \Psi_{\run}) \setminus \Psi_{\centerset}(C_{Q})|$ by 
range-sum query $\rangesum(\mathcal{J}_{B}(h_{Q}), L_{x}, L_{x^{\prime}}, R_{y}, R_{\hat{y}})$. 
Similar to the third phase of the algorithm for $\RSCQBX(i, j)$, 
this phase takes $O(\log^{4} n)$ time. 

Finally, the algorithm for $\RSCQBY(i, j)$ can be executed in $O(H^{2} \log n + \log^{4} n)$ time in total. 
Therefore, Lemma~\ref{lem:RSC_subquery_B_summary} holds.

\subsection{Subquery \texorpdfstring{$\RSCQCX(i, j)$}{RSCC1(i, j)}}\label{subsec:RSC_comp_C1}
The goal of this subsection is to answer $\RSCQCX(i, j)$. 
The following lemma states the summary of this subsection. 

\begin{lemma}\label{lem:RSC_subquery_C1_summary}
%Consider the two integers $K_{Q}$ and $M_{Q}$ introduced in Section~\ref{sec:RSC_query}. 
Using a dynamic data structure of $O((|\Psi_{\samp}| + |\mathcal{U}_{\RR}|) B)$ bits of space for machine word size $B$, 
we can answer subquery $\RSCQCX(i, j)$ (i.e., computing $|\Psi_{\CCP}(T[i..j]) \cap \Psi_{\lex}(T[\gamma_{Q}..r_{Q}+1]) \cap \Psi_{\run} \cap \Psi_{\centerset}(C_{Q}) \cap \Psi_{\lcp}(K_{Q}) \cap \Psi_{\preceding}|$) in $O(H^{2} \log n + \log^{4} n)$ time if the given RSC query $\RSCQ(i, j)$ satisfies either condition (C) or (D).
Here, $|\mathcal{U}_{\RR}|$ is the number of nodes in the RR-DAG of RLSLP $\mathcal{G}^{R}$.
\end{lemma}
\begin{proof}
See Section~\ref{subsubsec:JC1_subquery_ds}.
\end{proof}

For answering $\RSCQCX(i, j)$, 
we leverage range-sum query on weighted points corresponding to the interval attractors in 
set $\Psi_{h_{Q}} \cap \Psi_{\source} \cap \Psi_{\centerset}(C_{Q}) \cap \Psi_{\modulo}(M_{Q}) \cap \Psi_{\preceding} \cap \Psi_{\samp}$. 
Here, $M_{Q}$ is the integer introduced in Section~\ref{sec:RSC_query}.
For this purpose, 
we introduce a set $\mathcal{J}_{C}(h, C, M)$ of weighted points on a grid $([1, n], \mathcal{Y}_{C}(h, C, M))$ 
for a triplet $(h, C, M)$ of 
an integer $h \in [0, H]$, a string $C \in \Sigma^{+}$, and a non-negative integer $M \in \mathbb{N}_{0}$. 
Here, the set $\mathcal{J}_{C}(h)$ and grid $([1, n], \mathcal{Y}_{C}(h, C, M))$ are defined using 
set $\Psi_{h} \cap \Psi_{\source} \cap \Psi_{\centerset}(C) \cap \Psi_{\modulo}(M) \cap \Psi_{\preceding} \cap \Psi_{\samp}$ of $k$ interval attractors 
$([p_{1}, q_{1}], [\ell_{1}, r_{1}]), ([p_{2}, q_{2}], [\ell_{2}, r_{2}])$, 
$\ldots$, $([p_{k}, q_{k}], [\ell_{k}, r_{k}])$. 
For each integer $s \in [1, k]$, 
let $\gamma_{s}$ of the attractor position of each interval attractor $([p_{s}, q_{s}], [\ell_{s}, r_{s}])$; 
let $K_{s}$ be the length of the longest common prefix between two strings $T[\gamma_{s}..r_{s}]$ and $C^{n+1}$ 
(i.e., $K_{s} = |\lcp(T[\gamma_{s}..r_{s}], C^{n+1})|$). 
Here, $([p_{s}, q_{s}], [\ell_{s}, r_{s}]) \in \Psi_{\lcp}(K_{s})$ follows from the definition of subset $\Psi_{\lcp}(K_{s})$. 

\paragraph{Grid $([1, n], \mathcal{Y}_{C}(h, C, M))$.}
$\mathcal{Y}_{C}(h, C, M) \subseteq \Sigma^{*}$ is 
the ordered set of strings defined as the union of two sets 
$\{ \varepsilon, \# \}$ and $\{ T[\gamma_{s} + K_{s}..r_{s} + 1] \mid s \in [1, k] \}$ 
(i.e., $\mathcal{Y}_{C}(h, C, M) = \{ \varepsilon, \# \} \cup \{ T[\gamma_{s} + K_{s}..r_{s} + 1] \mid s \in [1, k] \}$). 
Here, $\varepsilon$ is the string of length $0$, 
and $\#$ is the largest character in the alphabet $\Sigma$ (see Section~\ref{sec:preliminary}). 
This ordered set $\mathcal{Y}_{C}(h, C, M)$ consists of $d$ strings $R_{1}, R_{2}, \ldots, R_{d}$ 
that are sorted in lexicographical order (i.e., $R_{1} \prec R_{2} \prec \cdots \prec R_{d}$). 
$R_{1} = \varepsilon$ and $R_{d} = \#$ always hold because 
every string of the set $\{ T[\gamma_{s} + K_{s}..r_{s} + 1] \mid s \in [1, k] \}$ does not contain the character $\#$. 

Grid $([1, n], \mathcal{Y}_{C}(h, C, M))$ consists of two ordered sets $[1, n]$ and $\mathcal{Y}_{C}(h, C, M)$. 
Each integer of the ordered set $[1, n]$ represents x-coordinate on two dimensional space. 
Similarly, each string of the ordered set $\mathcal{Y}_{C}(h, C, M)$ represents y-coordinate on two dimensional space. 

\paragraph{Set $\mathcal{J}_{C}(h, C, M)$ of Weighted Points.}
Set $\mathcal{J}_{C}(h, C, M)$ consists of $k$ weighted points 
$(|f_{\recover}(([p_{1}$, $q_{1}], [\ell_{1}, r_{1}]))|$, $T[\gamma_{1} + K_{1}..r_{1} + 1]$, $|\Psi_{\str}(T[p_{1}-1..r_{1}+1])|, T[p_{1}-1..r_{1}+1])$, 
$(|f_{\recover}(([p_{2}$, $q_{2}], [\ell_{2}, r_{2}]))|$, $T[\gamma_{2} + K_{2}..r_{2} + 1]$, $|\Psi_{\str}(T[p_{2}-1..r_{2}+1])|, T[p_{2}-1..r_{2}+1])$, $\ldots$, 
$(|f_{\recover}(([p_{k}, q_{k}], [\ell_{k}, r_{k}]))|, T[\gamma_{k} + K_{k}..r_{k} + 1], |\Psi_{\str}(T[p_{k}-1..r_{k}+1])|, T[p_{k}-1..r_{k}+1])$ on grid $([1, n], \mathcal{Y}_{C}(h, C, M))$. 
Each weighted point $(|f_{\recover}(([p_{s}, q_{s}], [\ell_{s}, r_{s}]))|, T[\gamma_{s} + K_{s}..r_{s} + 1], |\Psi_{\str}(T[p_{s}-1..r_{s}+1])|, T[p_{s}-1..r_{s}+1])$ corresponds to interval attractor $([p_{s}, q_{s}], [\ell_{s}, r_{s}])$ in set $\Psi_{h} \cap \Psi_{\source} \cap \Psi_{\centerset}(C) \cap \Psi_{\modulo}(M) \cap \Psi_{\preceding} \cap \Psi_{\samp}$. 
The details of the four elements of the weighted point $(|f_{\recover}(([p_{s}, q_{s}], [\ell_{s}, r_{s}]))|, T[\gamma_{s} + K_{s}..r_{s} + 1], |\Psi_{\str}(T[p_{s}-1..r_{s}+1])|, T[p_{s}-1..r_{s}+1])$ are as follows:
\begin{itemize}
    \item the first integer $|f_{\recover}(([p_{s}, q_{s}], [\ell_{s}, r_{s}]))|$ is the x-coordinate of this weighted point; 
    \item the second string $T[\gamma_{s} + K_{s}..r_{s} + 1]$ is the y-coordinate of this weighted point; 
    \item the third integer $|\Psi_{\str}(T[p_{s}-1..r_{s}+1])|$ is the weight of this weighted point;
    \item the fourth string $T[p_{s}-1..r_{s}+1]$ is the identifier of this weighted point.
\end{itemize}

From the definition of the sampling subset $\Psi_{\samp}$, 
the identifiers $T[p_{1}-1..r_{1}+1], T[p_{2}-1..r_{2}+1], \ldots, T[p_{k}-1..r_{k}+1]$ of all the weighted points in 
the set $\mathcal{J}_{C}(h, C, M)$ are different. 

The following lemma states the sizes of two sets $\mathcal{Y}_{C}(h, C, M)$ and $\mathcal{J}_{C}(h, C, M)$. 

\begin{lemma}\label{lem:JC1_size}
    The following two statements hold for a triplet of 
    an integer $h \in [0, H]$, a string $C \in \Sigma^{+}$, and a non-negative integer $M \in \mathbb{N}_{0}$: 
    \begin{enumerate}[label=\textbf{(\roman*)}]
    \item \label{enum:JC1_size:1} $|\mathcal{Y}_{C}(h, C, M)| \leq 2 + |\mathcal{J}_{C}(h, C, M)|$;
    \item \label{enum:JC1_size:2} $|\mathcal{Y}_{C}(h, C, M)|, |\mathcal{J}_{C}(h, C, M)| = O(n^{2})$.
    \end{enumerate}
\end{lemma}
\begin{proof}
    The proof of Lemma~\ref{lem:JC1_size} is as follows. 

    \textbf{Proof of Lemma~\ref{lem:JC1_size}(i).}
    $|\mathcal{Y}_{C}(h, C, M)| \leq 2 + |\mathcal{J}_{C}(h, C, M)|$ follows from the definition of the ordered set 
    $\mathcal{Y}_{C}(h, C, M)$. 

    \textbf{Proof of Lemma~\ref{lem:JC1_size}(ii).}
    We prove $|\mathcal{J}_{C}(h, C, M)| = O(n^{2})$. 
    $|\mathcal{J}_{C}(h, C, M)| \leq |\Psi_{h} \cap \Psi_{\source} \cap \Psi_{\centerset}(C) \cap \Psi_{\modulo}(M) \cap \Psi_{\preceding} \cap \Psi_{\samp}|$ holds 
    because 
    there exists a one-to-one correspondence between the weighted points of set $\mathcal{J}_{C}(h, C, M)$ 
    and the interval attractors of set $\Psi_{h} \cap \Psi_{\source} \cap \Psi_{\centerset}(C) \cap \Psi_{\modulo}(M) \cap \Psi_{\preceding} \cap \Psi_{\samp}$. 
    $|\Psi_{h} \cap \Psi_{\source} \cap \Psi_{\centerset}(C) \cap \Psi_{\modulo}(M) \cap \Psi_{\preceding} \cap \Psi_{\samp}| \leq |\Psi_{\samp}|$ because 
    the set $\Psi_{h} \cap \Psi_{\source} \cap \Psi_{\centerset}(C) \cap \Psi_{\modulo}(M) \cap \Psi_{\preceding} \cap \Psi_{\samp}$ is a subset of the sampling subset $\Psi_{\samp}$. 
    $|\Psi_{\samp}| = O(n^{2})$ follows from 
    $\Psi_{\samp} \subseteq \Psi_{\RR}$ and 
    $|\Psi_{\RR}| = O(n^{2})$ (Lemma~\ref{lem:non_comp_IA_size}). 
    Therefore, $|\mathcal{J}_{C}(h, C, M)| = O(n^{2})$ holds. 

    We prove $|\mathcal{Y}_{C}(h, C, M)| = O(n^{2})$. 
    $|\mathcal{Y}_{C}(h, C, M)| \leq 2 + |\mathcal{J}_{C}(h, C, M)|$ follows from Lemma~\ref{lem:JC1_size}~\ref{enum:JC1_size:1}. 
    We already proved $|\mathcal{J}_{C}(h, C, M)| = O(n^{2})$. 
    Therefore, $|\mathcal{Y}_{C}(h, C, M)| = O(n^{2})$ holds. 
    
\end{proof}

The following lemma shows that 
we can count the interval attractors in set $\Psi_{\CCP}(T[i..j]) \cap \Psi_{\lex}(T[\gamma_{Q}..r_{Q}+1]) \cap \Psi_{\run} \cap \Psi_{\centerset}(C_{Q}) \cap \Psi_{\lcp}(K_{Q}) \cap \Psi_{\preceding}$ 
by one range-sum query on the set $\mathcal{J}_{C}(h_{Q}, C_{Q}, M_{Q})$ of weighted points 
for the integer $K_{Q}$ of Lemma~\ref{lem:RSC_subquery_C1_summary}. 

\begin{lemma}\label{lem:JC1_main_lemma}
Consider RSC query $\RSCQ(i, j)$ satisfying either condition (C) or (D) of RSC query stated in Section~\ref{subsec:rsc_sub}. 
Let $K_{Q}$ and $M_{Q}$ be the two integers of Lemma~\ref{lem:RSC_subquery_C1_summary}; 
let $R_{1}, R_{2}, \ldots, R_{d}$ ($R_{1} \prec R_{2} \prec \cdots \prec R_{d}$) be 
the $d$ strings in the ordered set $\mathcal{Y}_{C}(h_{Q}, C_{Q}, M_{Q})$; 
let $x, y$, and $y^{\prime}$ be the three integers defined as follows: 
\begin{itemize}
\item $x =  1 + \lfloor \frac{K_{Q} - (2 + \sum_{w = 1}^{h_{Q}+3} \lfloor \mu(w) \rfloor)}{|C_{Q}|} \rfloor$; 
\item $y = \min \{ s \in [1, d] \mid T[\gamma_{Q} + K_{Q}..j] \prec R_{s} \}$ if $|[\gamma_{Q}, j]| > K_{Q}$. 
Otherwise, let $y = 1$; 
\item $y^{\prime} = \max \{ s \in [1, d] \mid R_{s} \prec T[\gamma_{Q} + K_{Q}..r_{Q} + 1] \}$.
\end{itemize}

Then, the following equation holds: 
\begin{equation*}
    \begin{split}
    |\Psi_{\CCP}(T[i..j]) \cap \Psi_{\lex}(T[\gamma_{Q}..r_{Q}+1]) \cap & \Psi_{\run} \cap \Psi_{\centerset}(C_{Q}) \cap \Psi_{\lcp}(K_{Q}) \cap \Psi_{\preceding})| \\
    &= \rangesum(\mathcal{J}_{C}(h_{Q}, C_{Q}, M_{Q}), x, n, R_{y}, R_{y^{\prime}}).
    \end{split}
\end{equation*}    
\end{lemma}
\begin{proof}
See Section~\ref{subsubsec:RC_main_lemma_proof}.
\end{proof}

\subsubsection{Proof of Lemma~\ref{lem:JC1_main_lemma}}\label{subsubsec:RC_main_lemma_proof}

For proving Lemma~\ref{lem:JC1_main_lemma}, 
we leverage the subset $\Psi_{\sRecover}(([p, q], [\ell, r]))$ of set $\Psi_{\RR}$ introduced in Section~\ref{subsubsec:recover_subsets} for an interval attractor $([p, q], [\ell, r]) \in \Psi_{\source}$. 
The following lemma states a property of the subset $\Psi_{\sRecover}(([p, q], [\ell, r]))$. 

\begin{proposition}\label{prop:psi_str_recover_lcp}
Consider an interval attractor $([p, q], [\ell, r])$ in subset $\Psi_{\source}$. 
Let $\gamma$, $C$, and $h$ be the attractor position, associated string, and level of the interval attractor $([p, q], [\ell, r])$, 
respectively; 
let $K$ be the length of the longest common prefix between two strings $T[\gamma..r]$ and $C^{n+1}$ 
(i.e., $K = |\lcp(T[\gamma..r], C^{n+1})|$). 
Here, $([p, q], [\ell, r]) \in \Psi_{\lcp}(K)$ follows from the definition of the subset $\Psi_{\lcp}(K)$. 
Then, $|\Psi_{\sRecover}(([p, q], [\ell, r])) \cap \Psi_{\lcp}(K - \tau |C|)| = |\Psi_{\str}(T[p-1..r+1])|$ holds 
for all integer $\tau \in [1, |f_{\recover}(([p, q], [\ell, r]))|]$. 
\end{proposition}
\begin{proof}
    Let $([p_{1}, q_{1}], [\ell_{1}, r_{1}])$, $([p_{2}, q_{2}], [\ell_{2}, r_{2}])$, $\ldots$, $([p_{k}, q_{k}], [\ell_{k}, r_{k}])$ 
    be the interval attractors in the subset $\Psi_{\str}(T[p-1..r+1])$. 
    Here, $T[p_{s}-1..r_{s}+1] = T[p-1..r+1]$ follows from the definition of the subset $\Psi_{\str}(T[p-1..r+1])$ for each interval attractor $([p_{s}, q_{s}], [\ell_{s}, r_{s}])$.     
    Lemma~\ref{lem:psi_equality_basic_property} shows that 
    $([p_{s}, q_{s}], [\ell_{s}, r_{s}]) \in \Psi_{\source} \cap \Psi_{\centerset}(C) \cap \Psi_{\lcp}(K)$ holds 
    because $T[p_{s}-1..r_{s}+1] = T[p-1..r+1]$ and $([p, q], [\ell, r]) \in \Psi_{\source} \cap \Psi_{\centerset}(C) \cap \Psi_{\lcp}(K)$. 
    Lemma~\ref{lem:psi_str_property}~\ref{enum:psi_str_property:2} shows that 
    the level of the interval attractor $([p_{s}, q_{s}], [\ell_{s}, r_{s}])$ is $h$ 
    because $T[p_{s}-1..r_{s}+1] = T[p-1..r+1]$ and $([p, q], [\ell, r]) \in \Psi_{h}$. 
    
    We prove $|f_{\recover}(([p_{s}, q_{s}], [\ell_{s}, r_{s}])) \cap \Psi_{\lcp}(K - \tau |C|)| = 1$. 
    Lemma~\ref{lem:recover_super_property}~\ref{enum:recover_super_property:1} shows that 
    $|f_{\recover}(([p_{s}$, $q_{s}]$, $[\ell_{s}, r_{s}]))| = |f_{\recover}(([p, q], [\ell, r]))|$ holds.          
    Lemma~\ref{lem:recover_basic_property}~\ref{enum:recover_basic_property:4} shows that 
    $|f_{\recover}(([p_{s}, q_{s}], [\ell_{s}, r_{s}])) \cap \Psi_{\lcp}(K - \tau |C|)| = 1$ holds 
    because $([p_{s}, q_{s}], [\ell_{s}, r_{s}]) \in \Psi_{\source} \cap \Psi_{\centerset}(C) \cap \Psi_{\lcp}(K)$ 
    and $\tau \in [1, |f_{\recover}(([p_{s}, q_{s}], [\ell_{s}, r_{s}]))|]$. 

    We prove $|\Psi_{\sRecover}(([p, q], [\ell, r])) \cap \Psi_{\lcp}(K - \tau |C|)| = |\Psi_{\str}(T[p-1..r+1])|$. 
    $|\Psi_{\sRecover}(([p, q]$, $[\ell, r])) \cap \Psi_{\lcp}(K - \tau |C|)| = |\bigcup_{s = 1}^{k} f_{\recover}(([p_{s}, q_{s}], [\ell_{s}, r_{s}])) \cap \Psi_{\lcp}(K - \tau |C|)|$ follows from the definition of the subset $\Psi_{\sRecover}(([p, q], [\ell, r]))$. 
    $|\bigcup_{s = 1}^{k} f_{\recover}(([p_{s}, q_{s}], [\ell_{s}, r_{s}])) \cap \Psi_{\lcp}(K - \tau |C|)| = \sum_{s = 1}^{k} |f_{\recover}(([p_{s}$, $q_{s}], [\ell_{s}, r_{s}])) \cap \Psi_{\lcp}(K - \tau |C|)|$ holds. 
    This is because $f_{\recover}(([p_{s}, q_{s}], [\ell_{s}, r_{s}])) \cap f_{\recover}(([p_{s^{\prime}}, q_{s^{\prime}}]$, $[\ell_{s^{\prime}}, r_{s^{\prime}}])) = \emptyset$ follows from Lemma~\ref{lem:recover_division_property}~\ref{enum:recover_division_property:2} 
    for any pair of two integers $1 \leq s < s^{\prime} \leq k$.     
    Therefore, $|\Psi_{\sRecover}(([p, q], [\ell, r])) \cap \Psi_{\lcp}(K - \tau |C|)| = |\Psi_{\str}(T[p-1..r+1])|$ follows from the following equation:
\begin{equation*}
\begin{split}
    |\Psi_{\sRecover}(([p, q], [\ell, r])) \cap \Psi_{\lcp}(K - \tau |C|)| &= |\bigcup_{s = 1}^{k} f_{\recover}(([p_{s}, q_{s}], [\ell_{s}, r_{s}])) \cap \Psi_{\lcp}(K - \tau |C|)| \\
    &= \sum_{s = 1}^{k} |f_{\recover}(([p_{s}, q_{s}], [\ell_{s}, r_{s}])) \cap \Psi_{\lcp}(K - \tau |C|)| \\ 
    &= \sum_{s = 1}^{k} 1 \\ 
    &= |\Psi_{\str}(T[p-1..r+1])|. 
\end{split}
\end{equation*}
\end{proof}

Consider the $k$ interval attractors 
$([p_{1}, q_{1}], [\ell_{1}, r_{1}])$, $([p_{2}, q_{2}], [\ell_{2}, r_{2}])$, $\ldots$, $([p_{k}, q_{k}], [\ell_{k}, r_{k}])$ 
in set $\Psi_{h_{Q}} \cap \Psi_{\source} \cap \Psi_{\centerset}(C_{Q}) \cap \Psi_{\modulo}(M_{Q}) \cap \Psi_{\preceding} \cap \Psi_{\samp}$. 
Let $\gamma_{s}$ be the attractor position of each interval attractor $([p_{s}, q_{s}], [\ell_{s}, r_{s}])$. 
Let $K_{s} = |\lcp(T[\gamma_{s}..r_{s}], C_{Q}^{n+1})|$ for simplicity. 
From the definition of the subset $\Psi_{\source}$, 
there exists an interval attractor $([p^{\prime}_{s}, q^{\prime}_{s}], [\ell^{\prime}_{s}, r^{\prime}_{s}]) \in \Psi_{\run} \cap \Psi_{h_{Q}} \cap \Psi_{\centerset}(C_{Q})$ 
such that its attractor position $\gamma^{\prime}_{s}$ is equal to $\gamma_{s} + |C_{Q}|$.

We introduce a set $\mathcal{I}_{C} \subseteq [1, k]$ of integers for this proof. 
The set $\mathcal{I}_{C}$ consists of integers in set $\{ 1, 2, \ldots, k \}$ such that 
each integer $s$ satisfies 
$x \leq |f_{\recover}(([p_{s}, q_{s}], [\ell_{s}, r_{s}]))| \leq n$ 
and $R_{y} \preceq T[\gamma_{s} + K_{s}..r_{s} + 1] \preceq R_{y^{\prime}}$~(i.e., 
$\mathcal{I}_{C} = \{ s \in [1, k] \mid x \leq |f_{\recover}(([p_{s}, q_{s}], [\ell_{s}, r_{s}]))| \leq n \text{ and } R_{y} \preceq T[\gamma_{s} + K_{s}..r_{s} + 1] \preceq R_{y^{\prime}} \}$). 

The following two propositions are used to prove Lemma~\ref{lem:JC1_main_lemma}. 

\begin{proposition}\label{prop:set_JC_subseq}
The following equation holds: 
\begin{equation*}
\begin{split}
    &\bigcup_{s \in \mathcal{I}_{C}} \Psi_{\sRecover}(([p_{s}, q_{s}], [\ell_{s}, r_{s}])) \cap \Psi_{\lcp}(K_{Q}) \\
    &\subseteq \Psi_{\CCP}(T[i..j]) \cap \Psi_{\lex}(T[\gamma_{Q}..r_{Q}+1]) \cap \Psi_{\run} \cap \Psi_{\centerset}(C_{Q}) \cap \Psi_{\lcp}(K_{Q}) \cap \Psi_{\preceding}.
\end{split}
\end{equation*}
\end{proposition}
\begin{proof}
Consider an interval attractor $([p_{A}, q_{A}], [\ell_{A}, r_{A}]) \in \bigcup_{s \in \mathcal{I}_{C}} \Psi_{\sRecover}(([p_{s}, q_{s}], [\ell_{s}, r_{s}])) \cap \Psi_{\lcp}(K_{Q})$. 
Then, set $\mathcal{I}_{C}$ contains an integer $b$ satisfying $([p_{A}, q_{A}], [\ell_{A}, r_{A}]) \in \Psi_{\sRecover}(([p_{b}, q_{b}], [\ell_{b}, r_{b}]))$. 
Here, $([p^{\prime}_{s}, q^{\prime}_{s}], [\ell^{\prime}_{s}, r^{\prime}_{s}]) \in \Psi_{\sRecover}(([p_{b}, q_{b}], [\ell_{b}, r_{b}]))$;     
$([p_{A}, q_{A}], [\ell_{A}, r_{A}]) \in \Psi_{h_{Q}} \cap \Psi_{\run} \cap \Psi_{\centerset}(C_{Q})$ 
follows from Lemma~\ref{lem:sRecover_basic_property}~\ref{enum:sRecover_basic_property:centerset}. 
From the definition of the function $f_{\recover}$, 
there exists an integer $\tau \in [1, |f_{\recover}(([p_{b}, q_{b}], [\ell_{b}, r_{b}]))|]$ satisfying 
$([p_{A}, q_{A}], [\ell_{A}, r_{A}]) = ([p^{\prime}_{b} + (\tau-1)|C_{Q}|, q^{\prime}_{b} + (\tau-1)|C_{Q}|], [\ell^{\prime}_{b} + (\tau-1)|C_{Q}|, r_{b}])$. 
From Lemma~\ref{lem:source_and_recover} and Lemma~\ref{lem:recover_basic_property}, 
the following three statements hold: 
\begin{enumerate}[label=\textbf{(\roman*)}]
    \item $\gamma_{A} = \gamma_{b} + \tau |C_{Q}|$ for the attractor position $\gamma_{A}$ of the interval attractor $([p_{A}, q_{A}], [\ell_{A}, r_{A}])$ (Lemma~\ref{lem:source_and_recover}~\ref{enum:source_and_recover:2}); 
    \item $([p_{A}, q_{A}], [\ell_{A}, r_{A}]) \in \Psi_{\run} \cap \Psi_{h_{Q}} \cap \Psi_{\centerset}(C_{Q}) \cap \Psi_{\modulo}(M_{Q}) \cap \Psi_{\lcp}(K_{b} - \tau |C_{Q}|)$ (Lemma~\ref{lem:recover_basic_property}~\ref{enum:recover_basic_property:4});
    \item $([p_{A}, q_{A}], [\ell_{A}, r_{A}]) \in \Psi_{\preceding}$ (Lemma~\ref{lem:recover_basic_property}~\ref{enum:recover_basic_property:5}).
\end{enumerate}

The following six statements are used to prove Proposition~\ref{prop:set_JB1_subseq}. 

\begin{enumerate}[label=\textbf{(\roman*)}]
\setcounter{enumi}{3}
    \item $K_{b} - \tau |C_{Q}| = K_{Q}$;
    \item $T[\gamma_{A}..\gamma_{A} + K_{Q} - 1] = T[\gamma_{Q}..\gamma_{Q} + K_{Q} - 1]$ and $T[\gamma_{b} + K_{b}..r_{b} + 1] = T[\gamma_{A} + K_{Q}..r_{A} + 1]$; 
    \item $T[\gamma_{A}..r_{A} + 1] \prec T[\gamma_{Q}..r_{Q}+1]$ and $T[\gamma_{Q}..j] \prec T[\gamma_{A}..r_{A} + 1] \prec T[\gamma_{Q}..j]\#$; 
    \item $\reverse(T[i..\gamma_{Q}-1]) \prec \reverse(T[p_{A}-1..\gamma_{A}-1]) \prec \reverse(\#T[i..\gamma_{Q}-1])$; 
    \item $([p_{A}, q_{A}], [\ell_{A}, r_{A}]) \in \Psi_{\CCP}(T[i..j])$; 
    \item $([p_{A}, q_{A}], [\ell_{A}, r_{A}]) \in \Psi_{\lex}(T[\gamma_{Q}..r_{Q}+1])$.
\end{enumerate}

\textbf{Proof of statement (iv).}
Because of $([p_{A}, q_{A}], [\ell_{A}, r_{A}]) \in \Psi_{\centerset}(C_{Q}) \cap \Psi_{\lcp}(K_{b} - \tau |C_{Q}|)$, 
$|\lcp(T[\gamma_{A}..r_{A}], C_{Q}^{n+1})| = K_{b} - \tau |C_{Q}|$ follows from the definition of the subset $\Psi_{\lcp}(K_{b} - \tau |C_{Q}|)$. 
On the other hand, $|\lcp(T[\gamma_{A}..r_{A}], C_{Q}^{n+1})| = K_{Q}$ 
follows from the definition of the subset $\Psi_{\lcp}(K_{Q})$ because 
$([p_{A}, q_{A}], [\ell_{A}, r_{A}]) \in \Psi_{\centerset}(C_{Q}) \cap \Psi_{\lcp}(K_{Q})$. 
Therefore, $K_{Q} = K_{b} - \tau |C_{Q}|$ follows from $|\lcp(T[\gamma_{A}..r_{A}], C_{Q}^{n+1})| = K_{b} - \tau |C_{Q}|$ 
and $|\lcp(T[\gamma_{A}..r_{A}], C_{Q}^{n+1})| = K_{Q}$.

\textbf{Proof of statement (v).}
We prove $T[\gamma_{A}..\gamma_{A} + K_{Q} - 1] = T[\gamma_{Q}..\gamma_{Q} + K_{Q} - 1]$. 
Because of $([p_{Q}, q_{Q}], [\ell_{Q}, r_{Q}]) \in \Psi_{\centerset}(C_{Q}) \cap \Psi_{\lcp}(K_{Q})$, 
$|\lcp(T[\gamma_{Q}..r_{Q}], C_{Q}^{n+1})| = K_{Q}$ follows from the definition of the subset $\Psi_{\lcp}(K_{Q})$. 
Similarly, $|\lcp(T[\gamma_{A}..r_{A}], C_{Q}^{n+1})| = K_{Q}$ 
follows from the definition of the subset $\Psi_{\lcp}(K_{Q})$ because 
$([p_{A}, q_{A}], [\ell_{A}, r_{A}]) \in \Psi_{\centerset}(C_{Q}) \cap \Psi_{\lcp}(K_{Q})$. 
Therefore, $T[\gamma_{A}..\gamma_{A} + K_{Q} - 1] = T[\gamma_{Q}..\gamma_{Q} + K_{Q} - 1]$ follows from 
$|\lcp(T[\gamma_{Q}..r_{Q}], C_{Q}^{n+1})| = K_{Q}$ and $|\lcp(T[\gamma_{A}..r_{A}], C_{Q}^{n+1})| = K_{Q}$.

We prove $T[\gamma_{b} + K_{b}..r_{b} + 1] = T[\gamma_{A} + K_{Q}..r_{A} + 1]$. 
$\gamma_{b} + K_{b} = \gamma_{A} + K_{Q}$ follows from 
$\gamma_{A} = \gamma_{b} + \tau |C_{Q}|$ and $K_{b} - \tau |C_{Q}| = K_{Q}$. 
$r_{b} + 1 = r_{A} + 1$ follows from $r_{b} = r_{A}$. 
Therefore, $T[\gamma_{b} + K_{b}..r_{b} + 1] = T[\gamma_{A} + K_{Q}..r_{A} + 1]$ holds. 

\textbf{Proof of statement (vi).}
We prove $T[\gamma_{A}..r_{A} + 1] \prec T[\gamma_{Q}..r_{Q}+1]$. 
Because of $b \in \mathcal{I}_{C}$, 
$R_{y} \preceq T[\gamma_{b} + K_{b}..r_{b} + 1] \preceq R_{y^{\prime}}$ follows from the definition of the set $\mathcal{I}_{C}$. 
$T[\gamma_{b} + K_{b}..r_{b} + 1] \prec T[\gamma_{Q} + K_{Q}..r_{Q} + 1]$ 
follows from 
$T[\gamma_{b} + K_{b}..r_{b} + 1] \preceq R_{y^{\prime}}$ and 
$y^{\prime} = \max \{ s \in [1, d] \mid R_{s} \prec T[\gamma_{Q} + K_{Q}..r_{Q} + 1] \}$. 
$T[\gamma_{A} + K_{Q}..r_{A} + 1] \prec T[\gamma_{Q} + K_{Q}..r_{Q} + 1]$ follows from 
$T[\gamma_{b} + K_{b}..r_{b} + 1] \prec T[\gamma_{Q} + K_{Q}..r_{Q} + 1]$ and $T[\gamma_{b} + K_{b}..r_{b} + 1] = T[\gamma_{A} + K_{Q}..r_{A} + 1]$. 
Therefore, $T[\gamma_{A}..r_{A} + 1] \prec T[\gamma_{Q}..r_{Q}+1]$ follows from 
the following four equations: 
\begin{itemize}
    \item $T[\gamma_{A}..r_{A} + 1] = T[\gamma_{A}..\gamma_{A} + K_{Q} - 1] \cdot T[\gamma_{A} + K_{Q}..r_{A} + 1]$;
    \item $T[\gamma_{Q}..r_{Q} + 1] = T[\gamma_{Q}..\gamma_{Q} + K_{Q} - 1] \cdot T[\gamma_{Q} + K_{Q}..r_{Q} + 1]$;
    \item $T[\gamma_{A}..\gamma_{A} + K_{Q} - 1] = T[\gamma_{Q}..\gamma_{Q} + K_{Q} - 1]$;
    \item $T[\gamma_{A} + K_{Q}..r_{A} + 1] \prec T[\gamma_{Q} + K_{Q}..r_{Q} + 1]$. 
\end{itemize}

We prove $T[\gamma_{Q}..j] \prec T[\gamma_{A}..r_{A} + 1]$. 
If $|[\gamma_{Q}, j]| > K_{Q}$ holds, 
then $T[\gamma_{Q} + K_{Q}..j] \prec T[\gamma_{b} + K_{b}..r_{b} + 1]$ follows from 
$R_{y} \preceq T[\gamma_{b} + K_{b}..r_{b} + 1]$ 
and $y = \min \{ s \in [1, d] \mid T[\gamma_{Q} + K_{Q}..j] \prec R_{s} \}$. 
Here, $T[\gamma_{Q}..j] = T[\gamma_{Q}..\gamma_{Q} + K_{Q} - 1] \cdot T[\gamma_{Q} + K_{Q}..j]$ 
follows from $|[\gamma_{Q}, j]| > K_{Q}$. 
$T[\gamma_{Q} + K_{Q}..j] \prec T[\gamma_{A} + K_{Q}..r_{A} + 1]$ follows from 
$T[\gamma_{Q} + K_{Q}..j] \prec T[\gamma_{b} + K_{b}..r_{b} + 1]$ and 
$T[\gamma_{b} + K_{b}..r_{b} + 1] = T[\gamma_{A} + K_{Q}..r_{A} + 1]$. 
Therefore, $T[\gamma_{Q}..j] \prec T[\gamma_{A}..r_{A} + 1]$ follows from the four equations: 
\begin{itemize}
    \item $T[\gamma_{Q}..j] = T[\gamma_{Q}..\gamma_{Q} + K_{Q} - 1] \cdot T[\gamma_{Q} + K_{Q}..j]$;
    \item $T[\gamma_{A}..r_{A} + 1] = T[\gamma_{A}..\gamma_{A} + K_{Q} - 1] \cdot T[\gamma_{A} + K_{Q}..r_{A} + 1]$;
    \item $T[\gamma_{A}..\gamma_{A} + K_{Q} - 1] = T[\gamma_{Q}..\gamma_{Q} + K_{Q} - 1]$;
    \item $T[\gamma_{Q} + K_{Q}..j] \prec T[\gamma_{A} + K_{Q}..r_{A} + 1]$. 
\end{itemize}
Otherwise (i.e., $|[\gamma_{Q}, j]| \leq K_{Q}$), 
$T[\gamma_{Q}..j] \prec T[\gamma_{A}..r_{A} + 1]$ follows from 
(A) $T[\gamma_{Q}..j] \preceq T[\gamma_{Q}..\gamma_{Q} + K_{Q} - 1]$, 
(B) $T[\gamma_{Q}..\gamma_{Q} + K_{Q} - 1] = T[\gamma_{A}..\gamma_{A} + K_{Q} - 1]$, 
and (C) $T[\gamma_{A}..\gamma_{A} + K_{Q} - 1] \prec T[\gamma_{A}..r_{A} + 1]$.

We prove $T[\gamma_{A}..r_{A} + 1] \prec T[\gamma_{Q}..j]\#$. 
Since $I_{\capture}(i, j) = ([p_{Q}, q_{Q}], [\ell_{Q}, r_{Q}])$, 
$j < r_{Q} + 1$ follows from the definition of interval attractor. 
$T[\gamma_{Q}..r_{Q}+1] \prec T[\gamma_{Q}..j]\#$ holds 
because $j < r_{Q} + 1$. 
Therefore, $T[\gamma_{A}..r_{A} + 1] \prec T[\gamma_{Q}..j]\#$ follows from 
$T[\gamma_{A}..r_{A} + 1] \prec T[\gamma_{Q}..r_{Q}+1]$ and $T[\gamma_{Q}..r_{Q}+1] \prec T[\gamma_{Q}..j]\#$. 

\textbf{Proof of statement (vii).}
We can apply Lemma~\ref{lem:suffix_syncro} to the interval attractor $([p_{Q}, q_{Q}], [\ell_{Q}, r_{Q}])$ 
because (A) string $T[i..\gamma_{Q}-1]$ is a suffix of string $C_{Q}^{n+1}$, 
and (B) $K_{Q} > 1 + \sum_{w = 1}^{h_{Q}+3} \lfloor \mu(w) \rfloor$. 
Lemma~\ref{lem:suffix_syncro} shows that 
the string $T[i..\gamma_{Q}-1]$ is a suffix of string $T[p_{A}..\gamma_{A}-1]$ 
(i.e., $\lcs(T[i..\gamma_{Q}-1], T[p_{A}..\gamma_{A}-1]) = T[i..\gamma-1]$) 
because $([p_{A}, q_{A}], [\ell_{A}, r_{A}]) \in \Psi_{h_{Q}} \cap \Psi_{\run} \cap \Psi_{\centerset}(C_{Q})$. 
Because of $\lcs(T[i..\gamma_{Q}-1], T[p_{A}..\gamma_{A}-1]) = T[i..\gamma-1]$, 
$\reverse(T[i..\gamma_{Q}-1]) \prec \reverse(T[p_{A}-1..\gamma_{A}-1]) \prec \reverse(\#T[i..\gamma_{Q}-1])$ holds. 

\textbf{Proof of statement (viii).}
$\Psi_{\CCP}(T[i..j]) = \{ ([p, q], [\ell, r]) \in \Psi_{h_{Q}} \mid \reverse(T[i..\gamma_{Q}-1]) \prec \reverse(T[p-1..\gamma-1]) \prec \reverse(\#T[i..\gamma_{Q}-1]) \text{ and } T[\gamma_{Q}..j] \prec T[\gamma..r+1] \prec T[\gamma_{Q}..j]\# \}$
follows from Lemma~\ref{lem:CCP_property}~\ref{enum:CCP_property:4}. 
We already proved $([p_{A}, q_{A}], [\ell_{A}, r_{A}]) \in \Psi_{h_{Q}}$, 
$\reverse(T[i..\gamma_{Q}-1]) \prec \reverse(T[p_{A}-1..\gamma_{A}-1]) \prec \reverse(\#T[i..\gamma_{Q}-1])$ (statement (vii)), 
and $T[\gamma_{Q}..j] \prec T[\gamma_{A}..r_{A}+1] \prec T[\gamma_{Q}..j]\#$ (statement (vi)). 
Therefore, $([p_{A}, q_{A}], [\ell_{A}, r_{A}]) \in \Psi_{\CCP}(T[i..j])$ holds.

\textbf{Proof of statement (ix).}
$T[\gamma_{A}..r_{A} + 1] \prec T[\gamma_{Q}..r_{Q}+1]$ follows from statement (vi). 
Therefore, 
$([p_{A}, q_{A}], [\ell_{A}, r_{A}]) \in \Psi_{\lex}(T[\gamma_{Q}..r_{Q}+1])$ follows from 
the definition of the subset $\Psi_{\lex}(T[\gamma_{Q}..r_{Q}+1])$. 

\textbf{Proof of Proposition~\ref{prop:set_JC_subseq}.}
$([p_{A}, q_{A}], [\ell_{A}, r_{A}]) \in \Psi_{\CCP}(T[i..j]) \cap \Psi_{\lex}(T[\gamma_{Q}..r_{Q}+1]) \cap \Psi_{\run} \cap \Psi_{\centerset}(C_{Q}) \cap \Psi_{\lcp}(K_{Q}) \cap \Psi_{\preceding}$ 
follows from statement (ii), statement (iii), statement (iv), statement (viii), and statement (ix). 
Therefore, Proposition~\ref{prop:set_JB1_subseq} follows from the fact that 
$([p_{A}, q_{A}], [\ell_{A}, r_{A}]) \in \Psi_{\CCP}(T[i..j]) \cap \Psi_{\lex}(T[\gamma_{Q}..r_{Q}+1]) \cap \Psi_{\run} \cap \Psi_{\centerset}(C_{Q}) \cap \Psi_{\lcp}(K_{Q}) \cap \Psi_{\preceding}$ for each interval attractor $([p_{A}, q_{A}], [\ell_{A}, r_{A}]) \in \bigcup_{s \in \mathcal{I}_{C}} \Psi_{\sRecover}(([p_{s}, q_{s}], [\ell_{s}, r_{s}])) \cap \Psi_{\lcp}(K_{Q})$. 

\end{proof}

\begin{proposition}\label{prop:set_JC_supseq}
The following equation holds: 
\begin{equation*}
\begin{split}
    &\bigcup_{s \in \mathcal{I}_{C}} \Psi_{\sRecover}(([p_{s}, q_{s}], [\ell_{s}, r_{s}])) \cap \Psi_{\lcp}(K_{Q}) \\
    &\supseteq \Psi_{\CCP}(T[i..j]) \cap \Psi_{\lex}(T[\gamma_{Q}..r_{Q}+1]) \cap \Psi_{\run} \cap \Psi_{\centerset}(C_{Q}) \cap \Psi_{\lcp}(K_{Q}) \cap \Psi_{\preceding}.
\end{split}
\end{equation*}
\end{proposition}
\begin{proof}
Consider an interval attractor $([p_{A}, q_{A}], [\ell_{A}, r_{A}]) \in \Psi_{\CCP}(T[i..j]) \cap \Psi_{\lex}(T[\gamma_{Q}..r_{Q}+1]) \cap \Psi_{\run} \cap \Psi_{\centerset}(C_{Q}) \cap \Psi_{\lcp}(K_{Q}) \cap \Psi_{\preceding}$. 
Here, Lemma~\ref{lem:CCP_property}~\ref{enum:CCP_property:1} shows that 
the level of the interval attractor $([p_{A}, q_{A}], [\ell_{A}, r_{A}])$ is $h_{Q}$ 
because $([p_{A}, q_{A}], [\ell_{A}, r_{A}]) \in \Psi_{\CCP}(T[i..j])$ and 
$([p_{Q}, q_{Q}], [\ell_{Q}, r_{Q}]) \in \Psi_{h_{Q}}$. 
Because of $([p_{A}, q_{A}], [\ell_{A}, r_{A}]) \in \Psi_{\run}$, 
Lemma~\ref{lem:recover_division_property}~\ref{enum:recover_division_property:1} shows that 
subset $\Psi_{\source}$ contains an interval attractor $([p_{B}, q_{B}], [\ell_{B}, r_{B}])$ satisfying 
$([p_{A}, q_{A}], [\ell_{A}, r_{A}]) \in f_{\recover}(([p_{B}, q_{B}], [\ell_{B}, r_{B}]))$. 
Here, $([p_{A}, q_{A}], [\ell_{A}, r_{A}]) \in \Psi_{\sRecover}(([p_{B}, q_{B}], [\ell_{B}, r_{B}]))$ follows from 
the definition of the subset $\Psi_{\sRecover}(([p_{B}$, $q_{B}], [\ell_{B}, r_{B}]))$; 
$([p_{B}, q_{B}]$, $[\ell_{B}, r_{B}]) \not \in \Psi_{\run}$ holds 
because $\Psi_{\source} \cap \Psi_{\run} = \emptyset$; 
$([p_{B}, q_{B}]$, $[\ell_{B}, r_{B}]) \in \Psi_{h_{Q}} \cap \Psi_{\centerset}(C_{Q})$ follows from 
Lemma~\ref{lem:recover_basic_property}~\ref{enum:recover_basic_property:4}. 
Let $K_{B} = |\lcp(T[\gamma_{B}..r_{B}], C_{Q}^{n+1})|$ and $M_{B} = (K_{B} - (2 + \sum_{w = 1}^{h_{Q}+3} \lfloor \mu(w) \rfloor) ) \mod |C_{Q}|$ for the attractor position $\gamma_{B}$ of the interval attractor $([p_{B}, q_{B}], [\ell_{B}, r_{B}])$. 
Then, $([p_{B}, q_{B}], [\ell_{B}, r_{B}]) \in \Psi_{\lcp}(K_{B})$ 
and $([p_{B}, q_{B}], [\ell_{B}, r_{B}]) \in \Psi_{\modulo}(M_{B})$ follows from 
the definitions of the two subsets $\Psi_{\lcp}(K_{B})$ and $\Psi_{\modulo}(M_{B})$, respectively. 

From the definition of the subset $\Psi_{\source}$, 
there exists an interval attractor $([p^{\prime}_{B}, q^{\prime}_{B}], [\ell^{\prime}_{B}, r^{\prime}_{B}]) \in \Psi_{\run} \cap \Psi_{h_{Q}} \cap \Psi_{\centerset}(C_{Q})$ 
such that its attractor position $\gamma^{\prime}_{B}$ is equal to $\gamma_{B} + |C_{Q}|$.
From the definition of the function $f_{\recover}$, 
there exists an integer $\tau \in [1, |f_{\recover}(([p_{B}, q_{B}], [\ell_{B}, r_{B}]))|]$ satisfying 
$([p_{A}, q_{A}], [\ell_{A}, r_{A}]) = ([p^{\prime}_{B} + (\tau-1)|C_{Q}|, q^{\prime}_{B} + (\tau-1)|C_{Q}|], [\ell^{\prime}_{B} + (\tau-1)|C_{Q}|, r_{B}])$. 
From Lemma~\ref{lem:source_and_recover} and Lemma~\ref{lem:recover_basic_property}, 
the following three statements hold: 
\begin{enumerate}[label=\textbf{(\roman*)}]
    \item $\gamma_{A} = \gamma_{B} + \tau |C_{Q}|$ (Lemma~\ref{lem:source_and_recover}~\ref{enum:source_and_recover:2}) 
    for the attractor position $\gamma_{A}$ of the interval attractor $([p_{A}, q_{A}], [\ell_{A}, r_{A}])$; 
    \item $([p_{A}, q_{A}], [\ell_{A}, r_{A}]) \in \Psi_{\lcp}(K_{B} - \tau |C_{Q}|) \cap \Psi_{\modulo}(M_{B})$ (Lemma~\ref{lem:recover_basic_property}~\ref{enum:recover_basic_property:4});
    \item $([p_{B}, q_{B}], [\ell_{B}, r_{B}]) \in \Psi_{\preceding}$ (Lemma~\ref{lem:recover_basic_property}~\ref{enum:recover_basic_property:5}).
\end{enumerate}

We show that 
the sampling subset $\Psi_{\samp}$ contains an interval attractor $([p_{C}, q_{C}], [\ell_{C}, r_{C}])$ satisfying 
$([p_{A}, q_{A}], [\ell_{A}, r_{A}]) \in \Psi_{\sRecover}(([p_{C}, q_{C}], [\ell_{C}, r_{C}]))$.     
From the definition of the sampling subset $\Psi_{\samp}$, 
the sampling subset contains an interval attractor $([p_{C}, q_{C}], [\ell_{C}, r_{C}])$ satisfying 
$T[p_{C}-1..r_{C}+1] = T[p_{B}-1..r_{B}+1]$. 
Here, $([p_{C}, q_{C}], [\ell_{C}, r_{C}]) \in \Psi_{\source}$ follows from 
Lemma~\ref{lem:psi_equality_basic_property}~\ref{enum:psi_equality_basic_property:5}. 
Because of $T[p_{C}-1..r_{C}+1] = T[p_{B}-1..r_{B}+1]$, 
Lemma~\ref{lem:sRecover_basic_property}~\ref{enum:sRecover_basic_property:equality} shows that 
$\Psi_{\sRecover}(([p_{B}, q_{B}], [\ell_{B}, r_{B}])) = \Psi_{\sRecover}(([p_{C}, q_{C}], [\ell_{C}, r_{C}]))$ holds. 
Therefore, $([p_{A}, q_{A}], [\ell_{A}, r_{A}]) \in \Psi_{\sRecover}(([p_{C}, q_{C}], [\ell_{C}, r_{C}]))$ 
follows from $([p_{A}, q_{A}], [\ell_{A}, r_{A}]) \in \Psi_{\sRecover}(([p_{B}, q_{B}], [\ell_{B}, r_{B}]))$ 
and $\Psi_{\sRecover}(([p_{B}$, $q_{B}], [\ell_{B}, r_{B}])) = \Psi_{\sRecover}(([p_{C}$, $q_{C}], [\ell_{C}, r_{C}]))$.

Let $\gamma_{C}$ be the attractor position of the interval attractor $([p_{C}, q_{C}], [\ell_{C}, r_{C}])$. 
The following five statements are used to prove Proposition~\ref{prop:set_JC_supseq}: 

\begin{enumerate}[label=\textbf{(\roman*)}]
\setcounter{enumi}{3}
    \item $K_{B} - \tau |C_{Q}| = K_{Q}$ and $M_{B} = M_{Q}$;
    \item $([p_{C}, q_{C}], [\ell_{C}, r_{C}]) \in \Psi_{h_{Q}} \cap \Psi_{\centerset}(C_{Q}) \cap \Psi_{\lcp}(K_{Q} + \tau |C_{Q}|) \cap \Psi_{\modulo}(M_{Q}) \cap \Psi_{\preceding}$; 
    \item $x \leq |f_{\recover}(([p_{C}, q_{C}], [\ell_{C}, r_{C}]))| \leq n$; 
    \item $T[\gamma_{A}..\gamma_{A} + K_{Q} - 1] = T[\gamma_{Q}..\gamma_{Q} + K_{Q} - 1]$ and $T[\gamma_{C} + K_{B}..r_{C} + 1] = T[\gamma_{A} + K_{Q}..r_{A} + 1]$; 
    \item there exists an integer $b \in [1, k]$ satisfying $([p_{b}, q_{b}], [\ell_{b}, r_{b}]) = ([p_{C}, q_{C}], [\ell_{C}, r_{C}])$ 
    and $R_{y} \preceq T[\gamma_{b}..K_{b}..r_{b} + 1] \preceq R_{y^{\prime}}$.
    %\item $T[\gamma_{Q}..j] \prec T[\gamma_{C} + K_{B}..r_{C} + 1] \prec T[\gamma_{Q} + K_{Q}..r_{Q}+1]$; 
    %\item $R_{y} \preceq T[\gamma_{C}..K_{C}..r_{C} + 1] \preceq R_{y^{\prime}}$. 
\end{enumerate}

\textbf{Proof of statement (iv).}
We prove $K_{B} - \tau |C_{Q}| = K_{Q}$. 
Because of $([p_{A}, q_{A}], [\ell_{A}, r_{A}]) \in \Psi_{\centerset}(C_{Q}) \cap \Psi_{\lcp}(K_{B} - \tau |C_{Q}|)$, 
$|\lcp(T[\gamma_{A}..r_{A}], C_{Q}^{n+1})| = K_{B} - \tau |C_{Q}|$ follows from the definition of the subset $\Psi_{\lcp}(K_{B} - \tau |C_{Q}|)$. 
On the other hand, $|\lcp(T[\gamma_{A}..r_{A}], C_{Q}^{n+1})| = K_{Q}$ 
follows from the definition of the subset $\Psi_{\lcp}(K_{Q})$ because 
$([p_{A}, q_{A}], [\ell_{A}, r_{A}]) \in \Psi_{\centerset}(C_{Q}) \cap \Psi_{\lcp}(K_{Q})$. 
Therefore, $K_{Q} = K_{b} - \tau |C_{Q}|$ follows from $|\lcp(T[\gamma_{A}..r_{A}], C_{Q}^{n+1})| = K_{B} - \tau |C_{Q}|$ 
and $|\lcp(T[\gamma_{A}..r_{A}], C_{Q}^{n+1})| = K_{Q}$. 

We prove $M_{B} = M_{Q}$. 
Here, the two integers $M_{B}$ and $M_{Q}$ are defined as 
$(K_{B} - (2 + \sum_{w = 1}^{h_{Q}+3} \lfloor \mu(w) \rfloor) ) \mod |C_{Q}|$ and 
$(K_{Q} - (2 + \sum_{w = 1}^{h_{Q}+3} \lfloor \mu(w) \rfloor)) \mod |C_{Q}|$, respectively. 
We already proved $K_{B} - \tau |C_{Q}| = K_{Q}$. 
Therefore, $M_{B} = M_{Q}$ follows from the following equation: 
\begin{equation*}
    \begin{split}
    M_{B} &= (K_{B} - (2 + \sum_{w = 1}^{h_{Q}+3} \lfloor \mu(w) \rfloor) ) \mod |C_{Q}| \\
    &= ((K_{Q} + \tau |C_{Q}|) - (2 + \sum_{w = 1}^{h_{Q}+3} \lfloor \mu(w) \rfloor)) \mod |C_{Q}| \\
    &= (K_{Q} - (2 + \sum_{w = 1}^{h_{Q}+3} \lfloor \mu(w) \rfloor)) \mod |C_{Q}| \\
    &= M_{Q}.     
    \end{split}
\end{equation*}

\textbf{Proof of statement (v).}
Lemma~\ref{lem:sRecover_basic_property}~\ref{enum:sRecover_basic_property:centerset} shows that 
$([p_{C}, q_{C}], [\ell_{C}, r_{C}]) \in \Psi_{h_{Q}}$ holds 
because $([p_{A}, q_{A}], [\ell_{A}, r_{A}]) \in \Psi_{\sRecover}(([p_{C}, q_{C}], [\ell_{C}, r_{C}]))$ 
and $([p_{A}, q_{A}], [\ell_{A}, r_{A}]) \in \Psi_{h_{Q}}$. 
We can apply Lemma~\ref{lem:psi_equality_basic_property} to the two interval attractors 
$([p_{B}, q_{B}], [\ell_{B}, r_{B}])$ and $([p_{C}, q_{C}], [\ell_{C}, r_{C}])$ 
because $T[p_{B}-1..r_{B}+1] = T[p_{C}-1..r_{C}+1]$. 
Because of $([p_{B}, q_{B}], [\ell_{B}, r_{B}]) \in \Psi_{\centerset}(C_{Q}) \cap \Psi_{\lcp}(K_{B}) \cap \Psi_{\modulo}(M_{B}) \cap \Psi_{\preceding}$, 
Lemma~\ref{lem:psi_equality_basic_property} shows that 
$([p_{C}, q_{C}], [\ell_{C}, r_{C}]) \in \Psi_{\centerset}(C_{Q}) \cap \Psi_{\lcp}(K_{B}) \cap \Psi_{\modulo}(M_{B}) \cap \Psi_{\preceding}$ holds.  
$K_{B} = K_{Q} + \tau |C_{Q}|$ and $M_{B} = M_{Q}$ follows from statement (iv). 
Therefore, $([p_{C}, q_{C}], [\ell_{C}, r_{C}]) \in \Psi_{h_{Q}} \cap \Psi_{\centerset}(C_{Q}) \cap \Psi_{\lcp}(K_{Q} + \tau |C_{Q}|) \cap \Psi_{\modulo}(M_{Q}) \cap \Psi_{\preceding}$ holds.

\textbf{Proof of statement (vi).}
We prove $|f_{\recover}(([p_{C}, q_{C}], [\ell_{C}, r_{C}]))| \geq x$. 
From the definition of the function $f_{\recover}$, 
$|f_{\recover}(([p_{C}, q_{C}], [\ell_{C}, r_{C}]))| = \lfloor \frac{(K_{Q} + \tau |C_{Q}|) - (2 + \sum_{w = 1}^{h_{Q}+3} \lfloor \mu(w) \rfloor)}{|C_{Q}|} \rfloor$ holds 
because $([p_{C}, q_{C}]$, $[\ell_{C}, r_{C}]) \in \Psi_{h_{Q}} \cap \Psi_{\centerset}(C_{Q}) \cap \Psi_{\lcp}(K_{Q} + \tau |C_{Q}|)$. 
%Because of , 
%$|f_{\recover}(([p_{C}, q_{C}], [\ell_{C}, r_{C}]))| = \lfloor \frac{(K_{Q} + \tau |C_{Q}|) - (2 + \sum_{w = 1}^{h_{Q}+3} \lfloor \mu(w) \rfloor)}{|C_{Q}|} \rfloor$ follows from the definition of the function $f_{\recover}$. 
Here, $\lfloor \frac{(K_{Q} + \tau |C_{Q}|) - (2 + \sum_{w = 1}^{h_{Q}+3} \lfloor \mu(w) \rfloor)}{|C_{Q}|} \rfloor = \tau + \lfloor \frac{K_{Q} - (2 + \sum_{w = 1}^{h_{Q}+3} \lfloor \mu(w) \rfloor)}{|C_{Q}|} \rfloor$; 
$\tau \geq 1$; 
$x = 1 + \lfloor \frac{K_{Q} - (2 + \sum_{w = 1}^{h_{Q}+3} \lfloor \mu(w) \rfloor)}{|C_{Q}|} \rfloor$. 
Therefore, $|f_{\recover}(([p_{C}, q_{C}]$, $[\ell_{C}, r_{C}]))| \geq x$ follows from the following equation: 
\begin{equation*}
    \begin{split}
    |f_{\recover}(([p_{C}, q_{C}], [\ell_{C}, r_{C}]))| &= \lfloor \frac{(K_{Q} + \tau |C_{Q}|) - (2 + \sum_{w = 1}^{h_{Q}+3} \lfloor \mu(w) \rfloor)}{|C_{Q}|} \rfloor \\
    &= \tau + \lfloor \frac{K_{Q} - (2 + \sum_{w = 1}^{h_{Q}+3} \lfloor \mu(w) \rfloor)}{|C_{Q}|} \rfloor \\
    &= \tau + (x - 1) \\
    &\geq x.     
    \end{split}
\end{equation*}

We prove $|f_{\recover}(([p_{C}, q_{C}], [\ell_{C}, r_{C}]))| \leq n$. 
$K_{Q} + \tau |C_{Q}| \leq n$ follows from the definition of the subset $\Psi_{\lcp}(K_{Q} + \tau |C_{Q}|)$. 
$|C_{Q}| \geq 1$ follows from $C_{Q} \in \Sigma^{+}$. 
Therefore, $|f_{\recover}(([p_{C}, q_{C}], [\ell_{C}, r_{C}]))| \leq n$ follows from 
(A) $|f_{\recover}(([p_{C}, q_{C}], [\ell_{C}, r_{C}]))| = \lfloor \frac{(K_{Q} + \tau |C_{Q}|) - (2 + \sum_{w = 1}^{h_{Q}+3} \lfloor \mu(w) \rfloor)}{|C_{Q}|} \rfloor$, (B) $K_{Q} + \tau |C_{Q}| \leq n$, 
and (C) $|C_{Q}| \geq 1$. 

\textbf{Proof of statement (vii).}
We prove $T[\gamma_{A}..\gamma_{A} + K_{Q} - 1] = T[\gamma_{Q}..\gamma_{Q} + K_{Q} - 1]$. 
Because of $([p_{Q}, q_{Q}], [\ell_{Q}, r_{Q}]) \in \Psi_{\centerset}(C_{Q}) \cap \Psi_{\lcp}(K_{Q})$, 
$|\lcp(T[\gamma_{Q}..r_{Q}], C_{Q}^{n+1})| = K_{Q}$ follows from the definition of the subset $\Psi_{\lcp}(K_{Q})$. 
Similarly, $|\lcp(T[\gamma_{A}..r_{A}], C_{Q}^{n+1})| = K_{Q}$ 
follows from the definition of the subset $\Psi_{\lcp}(K_{Q})$ because 
$([p_{A}, q_{A}], [\ell_{A}, r_{A}]) \in \Psi_{\centerset}(C_{Q}) \cap \Psi_{\lcp}(K_{Q})$. 
Therefore, $T[\gamma_{A}..\gamma_{A} + K_{Q} - 1] = T[\gamma_{Q}..\gamma_{Q} + K_{Q} - 1]$ follows from 
$|\lcp(T[\gamma_{Q}..r_{Q}], C_{Q}^{n+1})| = K_{Q}$ and $|\lcp(T[\gamma_{A}..r_{A}], C_{Q}^{n+1})| = K_{Q}$. 

We prove $T[\gamma_{C} + K_{B}..r_{C} + 1] = T[\gamma_{A} + K_{Q}..r_{A} + 1]$. 
$\gamma_{B} + K_{B} = \gamma_{A} + K_{Q}$ follows from 
$\gamma_{A} = \gamma_{B} + \tau |C_{Q}|$ and $K_{B} - \tau |C_{Q}| = K_{Q}$. 
$T[\gamma_{B} + K_{B}..r_{B} + 1] = T[\gamma_{A} + K_{Q}..r_{A} + 1]$ follows from 
$\gamma_{B} + K_{B} = \gamma_{A} + K_{Q}$ and $r_{B} = r_{A}$. 
Because of $T[p_{C}-1..r_{C}+1] = T[p_{B}-1..r_{B}+1]$, 
Lemma~\ref{lem:psi_str_property}~\ref{enum:psi_str_property:1} shows that 
$T[\gamma_{C}..r_{C} + 1] = T[\gamma_{B}..r_{B} + 1]$ holds. 
$T[\gamma_{C} + K_{B}..r_{C} + 1] = T[\gamma_{B} + K_{B}..r_{B} + 1]$ follows from 
$T[\gamma_{C}..r_{C} + 1] = T[\gamma_{B}..r_{B} + 1]$. 
Therefore, $T[\gamma_{C} + K_{B}..r_{C} + 1] = T[\gamma_{A} + K_{Q}..r_{A} + 1]$ 
follows from $T[\gamma_{C} + K_{B}..r_{C} + 1] = T[\gamma_{B} + K_{B}..r_{B} + 1]$ 
and $T[\gamma_{B} + K_{B}..r_{B} + 1] = T[\gamma_{A} + K_{Q}..r_{A} + 1]$. 

\textbf{Proof of statement (viii).}
$([p_{C}, q_{C}], [\ell_{C}, r_{C}]) \in \Psi_{h_{Q}} \cap \Psi_{\source} \cap \Psi_{\centerset}(C_{Q}) \cap \Psi_{\modulo}(M_{Q}) \cap \Psi_{\preceding} \cap \Psi_{\samp}$ follows from $([p_{C}, q_{C}], [\ell_{C}, r_{C}]) \in \cap \Psi_{\source} \cap \Psi_{\samp}$ and statement (v). 
Therefore, there exists an integer $b \in [1, k]$ satisfying $([p_{b}, q_{b}], [\ell_{b}, r_{b}]) = ([p_{C}, q_{C}], [\ell_{C}, r_{C}])$. 

We prove $T[\gamma_{C} + K_{B}..r_{C} + 1] \prec T[\gamma_{Q} + K_{Q}..r_{Q}+1]$. 
$T[\gamma_{A}..r_{A} + 1] \prec T[\gamma_{Q}..r_{Q}+1]$ follows from 
$([p_{A}, q_{A}], [\ell_{A}, r_{A}]) \in \Psi_{\lex}(T[\gamma_{Q}..r_{Q}+1])$. 
$T[\gamma_{A} + K_{Q}..r_{A} + 1] \prec T[\gamma_{Q} + K_{Q}..r_{Q}+1]$ follows from the following four equations: 
\begin{itemize}
    \item $T[\gamma_{A}..r_{A} + 1] = T[\gamma_{A}..\gamma_{A} + K_{Q} - 1] \cdot T[\gamma_{A} + K_{Q}..r_{A} + 1]$;
    \item $T[\gamma_{Q}..r_{Q} + 1] = T[\gamma_{Q}..\gamma_{Q} + K_{Q} - 1] \cdot T[\gamma_{Q} + K_{Q}..r_{Q} + 1]$;
    \item $T[\gamma_{A}..\gamma_{A} + K_{Q} - 1] = T[\gamma_{Q}..\gamma_{Q} + K_{Q} - 1]$;
    \item $T[\gamma_{A}..r_{A} + 1] \prec T[\gamma_{Q}..r_{Q}+1]$. 
\end{itemize}
Therefore, $T[\gamma_{C} + K_{B}..r_{C} + 1] \prec T[\gamma_{Q} + K_{Q}..r_{Q}+1]$ follows from 
$T[\gamma_{C} + K_{B}..r_{C} + 1] = T[\gamma_{A} + K_{Q}..r_{A} + 1]$ (statement (vii)) and $T[\gamma_{A} + K_{Q}..r_{A} + 1] \prec T[\gamma_{Q} + K_{Q}..r_{Q}+1]$.

We prove $T[\gamma_{b}..K_{b}..r_{b} + 1] \preceq R_{y^{\prime}}$. 
$K_{b} = K_{B}$ holds because 
(A) $([p_{b}, q_{b}], [\ell_{b}, r_{b}]) = ([p_{C}, q_{C}]$, $[\ell_{C}, r_{C}])$, 
(B) $([p_{C}, q_{C}], [\ell_{C}, r_{C}]) \in \Psi_{\lcp}(K_{Q} + \tau |C_{Q}|)$, 
and (C) $K_{Q} + \tau |C_{Q}| = K_{B}$. 
$\gamma_{b} = \gamma_{C}$ holds because $([p_{b}, q_{b}], [\ell_{b}, r_{b}]) = ([p_{C}, q_{C}], [\ell_{C}, r_{C}])$. 
$T[\gamma_{b} + K_{b}..r_{b} + 1] = T[\gamma_{C} + K_{B}..r_{C} + 1]$ follows from 
$\gamma_{b} = \gamma_{C}$, $K_{b} = K_{B}$, and $r_{b} = r_{C}$. 
$T[\gamma_{b}..K_{b}..r_{b} + 1] \prec T[\gamma_{Q} + K_{Q}..r_{Q}+1]$ 
follows from $T[\gamma_{b} + K_{b}..r_{b} + 1] = T[\gamma_{C} + K_{B}..r_{C} + 1]$ and $T[\gamma_{C} + K_{B}..r_{C} + 1] \prec T[\gamma_{Q} + K_{Q}..r_{Q}+1]$. 
Therefore, $T[\gamma_{b}..K_{b}..r_{b} + 1] \preceq R_{y^{\prime}}$ follows from 
$T[\gamma_{b}..K_{b}..r_{b} + 1] \prec T[\gamma_{Q} + K_{Q}..r_{Q}+1]$ and $y^{\prime} = \max \{ s \in [1, d] \mid R_{s} \prec T[\gamma_{Q} + K_{Q}..r_{Q} + 1] \}$. 

We show that string $T[\gamma_{Q}..j]$ is a prefix of string $T[\gamma_{A}..r_{A}]$ (i.e., $\lcp(T[\gamma_{Q}..j], T[\gamma_{A}..r_{A}]) = T[\gamma_{Q}..j]$). 
$\Psi_{\CCP}(T[i..j]) = \{ ([p, q], [\ell, r]) \in \Psi_{h_{Q}} \mid \reverse(T[i..\gamma_{Q}-1]) \prec \reverse(T[p-1..\gamma-1]) \prec \reverse(\#T[i..\gamma_{Q}-1]) \text{ and } T[\gamma_{Q}..j] \prec T[\gamma..r+1] \prec T[\gamma_{Q}..j]\# \}$
follows from Lemma~\ref{lem:CCP_property}~\ref{enum:CCP_property:4}. 
Because of $([p_{A}, q_{A}], [\ell_{A}, r_{A}]) \in \Psi_{\CCP}(T[i..j])$, 
$T[\gamma_{Q}..j] \prec T[\gamma_{A}..r_{A}+1] \prec T[\gamma_{Q}..j]\#$ holds. 
This lexicographic order $T[\gamma_{Q}..j] \prec T[\gamma_{A}..r_{A}+1] \prec T[\gamma_{Q}..j]\#$ indicates that 
string $T[\gamma_{Q}..j]$ is a prefix of string $T[\gamma_{A}..r_{A}]$. 

We prove $R_{y} \preceq T[\gamma_{b}..K_{b}..r_{b} + 1]$. 
If $|[\gamma_{Q}, j]| > K_{Q}$, 
then $T[\gamma_{Q} + K_{Q}..j] \prec T[\gamma_{A} + K_{Q}..r_{A}+1]$ follows from the following four equations: 
\begin{itemize}
    \item $T[\gamma_{A}..r_{A} + 1] = T[\gamma_{A}..\gamma_{A} + K_{Q} - 1] \cdot T[\gamma_{A} + K_{Q}..r_{A} + 1]$;
    \item $T[\gamma_{Q}..j] = T[\gamma_{Q}..\gamma_{Q} + K_{Q} - 1] \cdot T[\gamma_{Q} + K_{Q}..j]$;
    \item $T[\gamma_{A}..\gamma_{A} + K_{Q} - 1] = T[\gamma_{Q}..\gamma_{Q} + K_{Q} - 1]$;
    \item $T[\gamma_{Q}..j] \prec T[\gamma_{A}..r_{A}+1]$. 
\end{itemize}
$T[\gamma_{Q} + K_{Q}..j] \prec T[\gamma_{b} + K_{b}..r_{b}+1]$ follows from 
$T[\gamma_{Q} + K_{Q}..j] \prec T[\gamma_{A} + K_{Q}..r_{A}+1]$, 
$T[\gamma_{A} + K_{Q}..r_{A} + 1] = T[\gamma_{C} + K_{B}..r_{C} + 1]$, 
and $T[\gamma_{C} + K_{B}..r_{C} + 1] = T[\gamma_{b} + K_{b}..r_{b} + 1]$. 
Therefore, $R_{y} \preceq T[\gamma_{b}..K_{b}..r_{b} + 1]$ follows from 
$T[\gamma_{Q} + K_{Q}..j] \prec T[\gamma_{b} + K_{b}..r_{b}+1]$ and $y = \min \{ s \in [1, d] \mid T[\gamma_{Q} + K_{Q}..j] \prec R_{s} \}$. 
Otherwise (i.e., $|[\gamma_{Q}, j]| \leq K_{Q}$), 
the integer $y$ is defined as $1$, and $R_{1}$ is a string of length $0$~(i.e., $R_{1} = \varepsilon$). 
$R_{y} \preceq T[\gamma_{b}..K_{b}..r_{b} + 1]$ holds because $|R_{y}| = 0$. 

\textbf{Proof of Proposition~\ref{prop:set_JC_supseq}.}
We prove $([p_{A}, q_{A}], [\ell_{A}, r_{A}]) \in \bigcup_{s \in \mathcal{I}_{C}} \Psi_{\sRecover}(([p_{s}, q_{s}], [\ell_{s}, r_{s}])) \cap \Psi_{\lcp}(K_{Q})$. 
Statement (viii) shows that there exists an integer $b \in [1, k]$ satisfying $([p_{b}, q_{b}], [\ell_{b}, r_{b}]) = ([p_{C}, q_{C}], [\ell_{C}, r_{C}])$ and $R_{y} \preceq T[\gamma_{b}..K_{b}..r_{b} + 1] \preceq R_{y^{\prime}}$. 
$x \leq |f_{\recover}(([p_{b}, q_{b}], [\ell_{b}, r_{b}]))| \leq n$ follows from 
$x \leq |f_{\recover}(([p_{C}, q_{C}], [\ell_{C}, r_{C}]))| \leq n$ (statement (vi)) 
and $([p_{b}, q_{b}], [\ell_{b}, r_{b}]) = ([p_{C}, q_{C}], [\ell_{C}, r_{C}])$. 
$b \in \mathcal{I}_{C}$ follows from $x \leq |f_{\recover}(([p_{C}, q_{C}], [\ell_{C}, r_{C}]))| \leq n$ and 
$R_{y} \preceq T[\gamma_{b}..K_{b}..r_{b} + 1] \preceq R_{y^{\prime}}$. 
$([p_{A}, q_{A}], [\ell_{A}, r_{A}]) \in \Psi_{\sRecover}(([p_{b}, q_{b}], [\ell_{b}, r_{b}]))$ follows from 
$([p_{A}, q_{A}], [\ell_{A}, r_{A}]) \in \Psi_{\sRecover}(([p_{B}, q_{B}]$, $[\ell_{B}, r_{B}]))$, 
$\Psi_{\sRecover}(([p_{B}, q_{B}], [\ell_{B}, r_{B}])) = \Psi_{\sRecover}(([p_{C}, q_{C}], [\ell_{C}, r_{C}]))$, 
and $([p_{b}, q_{b}], [\ell_{b}, r_{b}]) = ([p_{C}$, $q_{C}], [\ell_{C}, r_{C}])$. 
Therefore, $([p_{A}, q_{A}], [\ell_{A}, r_{A}]) \in \bigcup_{s \in \mathcal{I}_{C}} \Psi_{\sRecover}(([p_{s}, q_{s}], [\ell_{s}, r_{s}])) \cap \Psi_{\lcp}(K_{Q})$ follows from 
(A) $b \in \mathcal{I}_{C}$, (B) $([p_{A}, q_{A}], [\ell_{A}, r_{A}]) \in \Psi_{\sRecover}(([p_{b}, q_{b}], [\ell_{b}, r_{b}]))$, 
and $([p_{A}, q_{A}], [\ell_{A}, r_{A}]) \in \Psi_{\lcp}(K_{Q})$. 

Finally, Proposition~\ref{prop:set_JC_supseq} follows from the fact that 
$([p_{A}, q_{A}], [\ell_{A}, r_{A}]) \in \bigcup_{s \in \mathcal{I}_{C}} \Psi_{\sRecover}(([p_{s}, q_{s}]$, $[\ell_{s}, r_{s}])) \cap \Psi_{\lcp}(K_{Q})$ for each interval attractor $([p_{A}, q_{A}], [\ell_{A}, r_{A}]) \in \Psi_{\CCP}(T[i..j]) \cap \Psi_{\lex}(T[\gamma_{Q}..r_{Q}+1]) \cap \Psi_{\run} \cap \Psi_{\centerset}(C_{Q}) \cap \Psi_{\lcp}(K_{Q}) \cap \Psi_{\preceding}$.

\end{proof}

We prove Lemma~\ref{lem:JC1_main_lemma} using Proposition~\ref{prop:psi_str_recover_lcp}, Proposition~\ref{prop:set_JC_subseq}, 
and Proposition~\ref{prop:set_JC_supseq}.

\begin{proof}[Proof of Lemma~\ref{lem:JC1_main_lemma}]
We prove $K_{Q} = (x-1) |C_{Q}| + (2 + \sum_{w = 1}^{h_{Q}+3} \lfloor \mu(w) \rfloor) + M_{Q}$. 
$\lfloor \frac{K_{Q} - (2 + \sum_{w = 1}^{h_{Q}+3} \lfloor \mu(w) \rfloor)}{|C_{Q}|} \rfloor = \frac{K_{Q} - (2 + \sum_{w = 1}^{h_{Q}+3} \lfloor \mu(w) \rfloor) - M_{Q}}{|C_{Q}|}$ follows from $M_{Q} = (K_{Q} - (2 + \sum_{w = 1}^{h_{Q}+3} \lfloor \mu(w) \rfloor) ) \mod |C_{Q}|$. 
Therefore, $K_{Q} = (x-1) |C_{Q}| + (2 + \sum_{w = 1}^{h_{Q}+3} \lfloor \mu(w) \rfloor) + M_{Q}$ 
follows from $x = 1 + \lfloor \frac{K_{Q} - (2 + \sum_{w = 1}^{h_{Q}+3} \lfloor \mu(w) \rfloor)}{|C_{Q}|} \rfloor$ 
and $\lfloor \frac{K_{Q} - (2 + \sum_{w = 1}^{h_{Q}+3} \lfloor \mu(w) \rfloor)}{|C_{Q}|} \rfloor = \frac{K_{Q} - (2 + \sum_{w = 1}^{h_{Q}+3} \lfloor \mu(w) \rfloor) - M_{Q}}{|C_{Q}|}$. 

Let $\tau = |f_{\recover}(([p_{s}, q_{s}], [\ell_{s}, r_{s}]))| - (x-1)$ for an integer $s$ in set $\mathcal{I}_{C}$. 
Then, we prove $K_{Q} = K_{s} - \tau |C_{Q}|$. 
Because of $([p_{s}, q_{s}], [\ell_{s}, r_{s}]) \in \Psi_{\source} \cap \Psi_{\lcp}(K_{s}) \cap \Psi_{h_{Q}} \cap \Psi_{\centerset}(C_{Q})$, 
Lemma~\ref{lem:recover_basic_property}~\ref{enum:recover_basic_property:1} shows that 
$|f_{\recover}(([p_{s}, q_{s}], [\ell_{s}, r_{s}]))| = \lfloor \frac{K_{s} - (2 + \sum_{w = 1}^{h_{Q}+3} \lfloor \mu(w) \rfloor)}{|C_{Q}|} \rfloor$ holds. 
Because of $([p_{s}, q_{s}], [\ell_{s}, r_{s}]) \in \Psi_{\modulo}(M_{Q}) \cap \Psi_{h_{Q}} \cap \Psi_{\centerset}(C_{Q})$, 
$M_{Q} = (K_{s} - (2 + \sum_{w = 1}^{h_{Q}+3} \lfloor \mu(w) \rfloor)) \mod |C_{Q}|$ follows from the definition of the subset $\Psi_{\modulo}(M_{Q})$. 
$\lfloor \frac{K_{s} - (2 + \sum_{w = 1}^{h_{Q}+3} \lfloor \mu(w) \rfloor)}{|C_{Q}|} \rfloor = \frac{K_{s} - (2 + \sum_{w = 1}^{h_{Q}+3} \lfloor \mu(w) \rfloor) - M_{Q}}{|C_{Q}|}$ follows from $M_{Q} = (K_{s} - (2 + \sum_{w = 1}^{h_{Q}+3} \lfloor \mu(w) \rfloor)) \mod |C_{Q}|$. 
$K_{s} = |f_{\recover}(([p_{s}, q_{s}], [\ell_{s}, r_{s}]))| |C_{Q}| + (2 + \sum_{w = 1}^{h_{Q}+3} \lfloor \mu(w) \rfloor) + M_{Q}$ 
follows from $|f_{\recover}(([p_{s}, q_{s}], [\ell_{s}, r_{s}]))| = \lfloor \frac{K_{s} - (2 + \sum_{w = 1}^{h_{Q}+3} \lfloor \mu(w) \rfloor)}{|C_{Q}|} \rfloor$ and $\lfloor \frac{K_{s} - (2 + \sum_{w = 1}^{h_{Q}+3} \lfloor \mu(w) \rfloor)}{|C_{Q}|} \rfloor = \frac{K_{s} - (2 + \sum_{w = 1}^{h_{Q}+3} \lfloor \mu(w) \rfloor) - M_{Q}}{|C_{Q}|}$. 
Therefore, $K_{Q} = K_{s} - \tau |C_{Q}|$ follows from the following equation: 
\begin{equation*}
    \begin{split}
    K_{Q} &= (x-1) |C_{Q}| + (2 + \sum_{w = 1}^{h_{Q}+3} \lfloor \mu(w) \rfloor) + M_{Q} \\
    &= (|f_{\recover}(([p_{s}, q_{s}], [\ell_{s}, r_{s}]))| - \tau) |C_{Q}| + (2 + \sum_{w = 1}^{h_{Q}+3} \lfloor \mu(w) \rfloor) + M_{Q} \\ 
    &= K_{s} - \tau |C_{Q}|.
    \end{split}
\end{equation*}

Next, we prove $|\Psi_{\sRecover}(([p_{s}, q_{s}], [\ell_{s}, r_{s}])) \cap \Psi_{\lcp}(K_{Q})| = |\Psi_{\str}(T[p_{s}-1..r_{s}+1])|$. 
From the definition of the set $\mathcal{I}_{C}$, 
$x \leq |f_{\recover}(([p_{s}, q_{s}]$, $[\ell_{s}, r_{s}]))|$ holds.
$x \geq 1$ follows from 
$x =  1 + \lfloor \frac{K_{Q} - (2 + \sum_{w = 1}^{h_{Q}+3} \lfloor \mu(w) \rfloor)}{|C_{Q}|} \rfloor$ 
and $K_{Q} \geq 2 + \sum_{w = 1}^{h_{Q}+3} \lfloor \mu(w) \rfloor$. 
$\tau \in [1, |f_{\recover}(([p_{s}, q_{s}], [\ell_{s}, r_{s}]))|]$ follows from 
$\tau = |f_{\recover}(([p_{s}, q_{s}]$, $[\ell_{s}, r_{s}]))| - (x-1)$ and $1 \leq x \leq |f_{\recover}(([p_{s}, q_{s}], [\ell_{s}, r_{s}]))|$. 
We can apply Proposition~\ref{prop:psi_str_recover_lcp} to the set $\Psi_{\sRecover}(([p_{s}, q_{s}], [\ell_{s}, r_{s}])) \cap \Psi_{\lcp}(K_{Q})$ because 
$\tau \in [1, |f_{\recover}(([p_{s}, q_{s}], [\ell_{s}, r_{s}]))|]$ and $K_{Q} = K_{s} - \tau |C_{Q}|$. 
Therefore, Proposition~\ref{prop:psi_str_recover_lcp} shows that $|\Psi_{\sRecover}(([p_{s}, q_{s}], [\ell_{s}, r_{s}])) \cap \Psi_{\lcp}(K_{Q})| = |\Psi_{\str}(T[p_{s}-1..r_{s}+1])|$ holds. 

We prove Lemma~\ref{lem:JC1_main_lemma}. 
$|\Psi_{\CCP}(T[i..j]) \cap \Psi_{\lex}(T[\gamma_{Q}..r_{Q}+1]) \cap \Psi_{\run} \cap \Psi_{\centerset}(C_{Q}) \cap \Psi_{\lcp}(K_{Q}) \cap \Psi_{\preceding}| = |\bigcup_{s \in \mathcal{I}_{C}} \Psi_{\sRecover}(([p_{s}, q_{s}], [\ell_{s}, r_{s}])) \cap \Psi_{\lcp}(K_{Q})|$ follows from Proposition~\ref{prop:set_JC_subseq} and Proposition~\ref{prop:set_JC_supseq}. 
$|\bigcup_{s \in \mathcal{I}_{C}} \Psi_{\sRecover}(([p_{s}, q_{s}], [\ell_{s}, r_{s}])) \cap \Psi_{\lcp}(K_{Q})| = \sum_{s \in \mathcal{I}_{C}} |\Psi_{\sRecover}(([p_{s}, q_{s}], [\ell_{s}, r_{s}])) \cap \Psi_{\lcp}(K_{Q})|$ holds. 
This is because the following two statements hold for any pair of two integers $1 \leq s < s^{\prime} \leq k$: 
(A) $T[p_{s}-1..r_{s}+1] \neq T[p_{s^{\prime}}-1..r_{s^{\prime}}+1]$; 
(B) $\Psi_{\sRecover}(([p_{s}, q_{s}], [\ell_{s}, r_{s}])) \cap \Psi_{\sRecover}(([p_{s^{\prime}}, q_{s^{\prime}}], [\ell_{s^{\prime}}, r_{s^{\prime}}])) = \emptyset$ follows from Lemma~\ref{lem:sRecover_basic_property}~\ref{enum:sRecover_basic_property:overlap}.  
We already showed that 
$|\Psi_{\sRecover}(([p_{s}$, $q_{s}], [\ell_{s}, r_{s}])) \cap \Psi_{\lcp}(K_{Q})| = |\Psi_{\str}(T[p_{s}-1..r_{s}+1])|$ holds 
for each integer $s \in \mathcal{I}_{C}$. 
$\rangesum(\mathcal{J}_{C}(h_{Q}, C_{Q}$, $M_{Q}), x, n, R_{y}, R_{y^{\prime}}) = \sum_{s \in \mathcal{I}_{C}} |\Psi_{\str}(T[p_{s}-1..r_{s}+1])|$ 
follows from the definitions of the set $\mathcal{I}_{C}$ and range-sum query. 
Therefore, Lemma~\ref{lem:JC1_main_lemma} follows from the following equation: 
\begin{equation*}
    \begin{split}
    |\Psi_{\CCP}(T[i..j]) \cap \Psi_{\lex}(T[\gamma_{Q}..r_{Q}+1]) \cap & \Psi_{\run} \cap \Psi_{\centerset}(C_{Q}) \cap \Psi_{\lcp}(K_{Q}) \cap \Psi_{\preceding}| \\
    &= |\bigcup_{s \in \mathcal{I}_{C}} \Psi_{\sRecover}(([p_{s}, q_{s}], [\ell_{s}, r_{s}])) \cap \Psi_{\lcp}(K_{Q})| \\ 
    &= \sum_{s \in \mathcal{I}_{C}} |\Psi_{\sRecover}(([p_{s}, q_{s}], [\ell_{s}, r_{s}])) \cap \Psi_{\lcp}(K_{Q})| \\ 
    &= \sum_{s \in \mathcal{I}_{C}} |\Psi_{\str}(T[p_{s}-1..r_{s}+1])| \\ 
    &= \rangesum(\mathcal{J}_{C}(h_{Q}, C_{Q}, M_{Q}), x, n, R_{y}, R_{y^{\prime}})     
    \end{split}
\end{equation*}
\end{proof}

%%%%%%%%%%%%%%%%%%%%%%%%%%%%%%%%%%%%%%%%%%%%%%%%%%%%%%%%%%%%%%%%%%%%%%%%%%%%%%%%

\subsubsection{Dynamic Data Structures for Ordered Set \texorpdfstring{$\mathcal{Y}_{C}(h, C, M)$}{}}\label{subsubsec:JC1_Y_ds}
Consider a triplet of an integer $h \in [0, H]$, a string $C \in \Sigma^{+}$, and a non-negative integer $M \in \mathbb{N}_{0}$. 
We present dynamic data structures to store the ordered set $\mathcal{Y}_{C}(h, C, M)$, 
which are similar to the dynamic data structures for the ordered set $\mathcal{Y}_{A}(h)$ presented in Section~\ref{subsubsec:JA_Y_ds}. 
Let $R_{1}, R_{2}, \ldots, R_{d}$~($R_{1} \prec R_{2} \prec \cdots \prec R_{d}$) be 
the strings in the ordered set $\mathcal{Y}_{C}(h, C, M)$. 
For each string $R_{b} \in \mathcal{Y}_{C}(h, C, M)$, 
we introduce a sequence $\mathbf{Q}^{Y}_{C}(h, C, M, R_{b})$ of $m$ weighted points $(x_{1}, y_{1}, w_{1}, e_{1})$, 
$(x_{2}, y_{2}, w_{2}, e_{2})$, $\ldots$, $(x_{m}, y_{m}, w_{m}, e_{m})$ in set $\mathcal{J}_{C}(h, C, M)$ of weighted points 
such that each weighted point $(x_{s}, y_{s}, w_{s}, e_{s})$ contains the string $R_{b}$ as its y-coordinate. 
The $m$ weighted points are sorted in increasing order of their x-coordinates $x_{1}, x_{2}, \ldots, x_{m} \in \mathbb{N}_{0}$. 
If the sequence $\mathbf{Q}^{Y}_{C}(h, C, M, R_{b})$ contains two weighted points of the same x-coordinate, 
then the two weighted points are sorted in lexicographical order of their identifiers. 
Formally, either of the following two conditions is satisfied for each integer $s \in [1, k-1]$: 
\begin{itemize}
    \item $x_{s} < x_{s+1}$;
    \item $x_{s} = x_{s+1}$ and $e_{s} \prec e_{s+1}$.
\end{itemize}

We store the ordered set $\mathcal{Y}_{C}(h, C, M)$ using $d+1$ doubly linked lists $\mathbf{Y}_{C}(h, C, M, R_{1})$, $\mathbf{Y}_{C}(h, C$, $M, R_{2})$, $\ldots$, $\mathbf{Y}_{C}(h, C, M, R_{d})$, and $\mathbf{L}^{Y}_{C}(h, C, M)$. 
For each integer $b \in [1, d]$, 
the doubly linked list $\mathbf{Y}_{C}(h, C, M, R_{b})$ consists of $m$ elements 
for the $m$ weighted points in sequence $\mathbf{Q}^{Y}_{C}(h, C, M, R_{b})$. 
Each $s$-th element of $\mathbf{Y}_{C}(h, C, M, R_{b})$ corresponds to the $s$-th weighted point $(x_{s}, y_{s}, w_{s}, e_{s})$ of $\mathbf{Q}^{Y}_{C}(h, C, M, R_{b})$. 
Here, the following three statements hold: 
\begin{itemize}
    \item from the definition of set $\mathcal{J}_{C}(h, C, M)$, 
    the $s$-th weighted point $(x_{s}, y_{s}, w_{s}, e_{s})$ corresponds to an interval attractor $([p, q], [\ell, r])$ in the sampling subset $\Psi_{\samp}$;
    \item the interval attractor $([p, q], [\ell, r])$ corresponds to a node $u$ of the sequence $\mathbf{Q}_{\samp}$ introduced in Section~\ref{subsec:sample_query}; 
    \item the node $u$ is represented as an element $v$ of the doubly linked list introduced in Section~\ref{subsubsec:sample_ds}. 
\end{itemize}
The $s$-th element of $\mathbf{Y}_{C}(h, C, M, R_{b})$ stores a pointer to the element $v$ corresponding to the $s$-th weighted point 
$(x_{s}, y_{s}, w_{s}, e_{s})$. 
A list indexing data structure is used for quickly accessing to the elements of the doubly linked list $\mathbf{Y}_{C}(h, C, M, R_{b})$. 

The last doubly-linked list $\mathbf{L}^{Y}_{C}(h, C, M)$ consists of $d$ elements such that 
each $b$-th element corresponds to the $b$-th string $R_{b}$ of the ordered set $\mathcal{Y}_{C}(h, C, M)$. 
Here, the $b$-th element stores a pointer to the $b$-th doubly linked list $\mathbf{Y}_{C}(h, C, M, R_{b})$. 
Similar to the doubly linked list $\mathbf{L}^{Y}_{A}(h)$ of Section~\ref{subsubsec:JA_Y_ds}, 
list indexing and order maintenance data structures are built on the doubly-linked list $\mathbf{L}^{Y}_{C}(h, C, M)$. 
These $d+1$ doubly linked lists $\mathbf{Y}_{C}(h, C, M, R_{1}), \mathbf{Y}_{C}(h, C, M, R_{2})$, $\ldots$, $\mathbf{Y}_{C}(h, C, M, R_{d})$, $\mathbf{L}^{Y}_{C}(h, C, M)$ require 
$O((d + |\mathcal{J}_{C}(h, C, M)|) B)$ bits of space in total for machine word size $B$. 
$d \leq 2 + |\mathcal{J}_{C}(h, C, M)|$ follows from Lemma~\ref{lem:JC1_size}~\ref{enum:JC1_size:1}. 
Therefore, the dynamic data structures for the ordered set $\mathcal{Y}_{C}(h, C, M)$ can be stored in $O(|\mathcal{J}_{C}(h, C, M)| B)$ bits of space. 

The following lemma states queries supported by the dynamic data structures for the ordered set $\mathcal{Y}_{C}(h, C, M)$. 

\begin{lemma}\label{lem:JC1_Y_queries}
    Let $R_{1}, R_{2}, \ldots, R_{d}$~($R_{1} \prec R_{2} \prec \cdots \prec R_{d}$) be 
    the $d$ strings of the ordered set $\mathcal{Y}_{C}(h, C, M)$ introduced in Section~\ref{subsec:RSC_comp_C1} 
    for a triplet of an integer $h \in [0, H]$, a string $C \in \Sigma^{+}$, and a non-negative integer $M \in \mathbb{N}_{0}$. 
    Consider the $d+1$ doubly linked lists $\mathbf{Y}_{C}(h, C, M, R_{1}), \mathbf{Y}_{C}(h, C, M$, $R_{2})$, $\ldots$, $\mathbf{Y}_{C}(h, C, M, R_{d})$, and $\mathbf{L}^{Y}_{C}(h, C, M)$.  
    Using these $d+1$ doubly linked lists and the dynamic data structures of Section~\ref{subsubsec:rrdag_ds} 
    and Section~\ref{subsubsec:sample_ds}, 
    we can support the following four queries: 
    \begin{enumerate}[label=\textbf{(\roman*)}]
    \item \label{enum:JC1_Y_queries:1} 
    verify whether $R_{s} \prec R_{s^{\prime}}$ or not in $O(1)$ time 
    for the given $s$-th and $s^{\prime}$-th elements of the doubly linked list $\mathbf{L}^{Y}_{C}(h, C, M)$;    
    \item \label{enum:JC1_Y_queries:2} 
    for a given integer $s \in [1, d]$, 
    return an interval $[g, g + |R_{s}| - 1]$ in input string $T$ satisfying $T[g..g + |R_{s}| - 1] = R_{s}$ 
    in $O(H^{2} + \log n)$ time if $R_{s} \not \in \{ \varepsilon, \# \}$; 
    otherwise return the string $R_{s}$ in $O(1)$ time;
    \item \label{enum:JC1_Y_queries:3} 
    consider a given pair of an integer $s \in [1, d]$ and interval $[\alpha, \beta] \subseteq [1, n]$ in input string $T$. 
    Then, verify the following four conditions in $O(H^{2} + \log n)$ time: 
    (A) $T[\alpha..\beta] \prec R_{s}$, 
    (B) $T[\alpha..\beta] = R_{s}$, 
    (C) $R_{s} \prec T[\alpha..\beta]$, 
    and (D) $R_{s} \prec T[\alpha..\beta]\#$;
    \item \label{enum:JC1_Y_queries:4} 
    consider a given triplet of three integers $\tau, \tau^{\prime} \in [1, n]$ ($\tau \leq \tau^{\prime}$) 
    and $s \in [1, d]$. 
    If set $\mathcal{J}_{C}(h, C, M)$ contains a weighed point $(x, y, w, e)$ 
    satisfying $\tau \leq x \leq \tau^{\prime}$ and $y = R_{s}$, 
    then return the interval attractor corresponding to the weighed point $(x, y, w, e)$ 
    in $O(H^{2} \log n + \log^{2} n)$ time.
    \end{enumerate}
\end{lemma}
\begin{proof}
    The proof of Lemma~\ref{lem:JB_Y_queries} is as follows. 

    \textbf{Proof of Lemma~\ref{lem:JC1_Y_queries}(i).}
    We can verify whether $R_{s} \prec R_{s^{\prime}}$ or not in $O(1)$ time by the order maintenance data structure built on 
    doubly linked list $\mathbf{L}^{Y}_{C}(h, C, M)$. 

    \textbf{Proof of Lemma~\ref{lem:JC1_Y_queries}(ii).}
Lemma~\ref{lem:JC1_Y_queries}~\ref{enum:JC1_Y_queries:2} can be proved using 
a similar approach as for Lemma~\ref{lem:JA_Y_queries}~\ref{enum:JA_Y_queries:2}. 
The detailed proof of Lemma~\ref{lem:JC1_Y_queries}~\ref{enum:JC1_Y_queries:2} is as follows. 

One of the following three conditions is satisfied: 
(A) $s = 1$; (B) $s = d$; (C) $1 < s < d$. 
For case (A), 
$R_{s} = \varepsilon$ follows from the definition of the ordered set $\mathcal{X}_{C}(h, C, M)$.
In this case, we can return the character $\varepsilon$ in $O(1)$ time. 
For case (B), 
$R_{s} = \#$ follows from the definition of the ordered set $\mathcal{X}_{C}(h, C, M)$.
In this case, we can return the character $\#$ in $O(1)$ time. 
    
For case (C), 
the $s$-th doubly linked list $\mathbf{Y}_{C}(h, C, M, R_{s})$ contains at least one element. 
Let $(x_{1}, y_{1}, w_{1}, e_{1})$ be the weighted point corresponding to the first element of the doubly linked list $\mathbf{Y}_{C}(h, C, M, R_{s})$. 
Then, $y_{1} = R_{s}$ holds. 
This weighted point $(x_{1}, y_{1}, w_{1}, e_{1})$ corresponds to an interval attractor $([p_{1}, q_{1}], [\ell_{1}, r_{1}])$ 
in set $\Psi_{\samp} \cap \Psi_{\centerset}(C)$ satisfying 
$T[\gamma_{1} + K_{1}..r_{1}+1] = y_{1}$ for 
(1) the attractor position $\gamma_{1}$ of the $([p_{1}, q_{1}], [\ell_{1}, r_{1}])$, 
and (2) the length $K_{1}$ of the longest common prefix between $T[\gamma_{1}..r_{1}]$ and $C^{n+1}$.
Here, the interval attractor $([p_{1}, q_{1}], [\ell_{1}, r_{1}])$ is represented as a node $u$ of 
the sequence $\mathbf{Q}_{\samp}$ introduced in Section~\ref{subsec:sample_query}.     
$T[\gamma_{1} + K_{1}..r_{1}+1] = R_{s}$ follows from 
$T[\gamma_{1} + K_{1}..r_{1}+1] = y_{1}$ and $y_{1} = R_{s}$. 
Therefore, we can return interval $[\gamma_{1} + K_{1}, r_{1}+1]$ as the answer to the query of Lemma~\ref{lem:JC1_Y_queries}~\ref{enum:JC1_Y_queries:2}.

In this case, we answer the query of Lemma~\ref{lem:JC1_Y_queries}~\ref{enum:JC1_Y_queries:2} in two phases. 
In the first phase, we obtain the node $u$ of the sequence $\mathbf{Q}_{\samp}$. 
The $s$-th element of doubly linked list $\mathbf{L}^{Y}_{C}(h, C, M)$ stores 
a pointer to the doubly linked list $\mathbf{Y}_{C}(h, C, M, R_{s})$.     
The first element of the doubly linked list $\mathbf{Y}_{C}(h, C, M, R_{s})$ stores 
a pointer to the element representing the node $u$ in the doubly linked list of Section~\ref{subsubsec:sample_ds}. 
The $s$-th element can be accessed in $O(\log d)$ time by the list indexing data structure built on the doubly linked list $\mathbf{L}^{Y}_{C}(h, C, M)$. 
Here, $d = O(n^{2})$ follows from Lemma~\ref{lem:JC1_size}~\ref{enum:JC1_size:2}. 
Therefore, we can obtain the node $u$ in $O(\log n)$ time. 

In the second phase, we return interval $[\gamma_{1} + K_{1}, r_{1}+1]$ as the answer to the query of 
Lemma~\ref{lem:JC1_Y_queries} \ref{enum:JC1_Y_queries:2}.     
We recover the interval attractor $([p_{1}, q_{1}], [\ell_{1}, r_{1}])$ from the node $u$ in $O(\log n)$ time by the algorithm of Section~\ref{subsubsec:computation_delta_samp}. 
The attractor position $\gamma_{1}$ can be computed in $O(H^{2})$ time 
by attractor position query $\attrQ(([p_{1}, q_{1}], [\ell_{1}, r_{1}]))$.
The length $K_{1}$ can be computed in $O(H^{2})$ time by C-LCP query $\clcpQ(([p_{1}, q_{1}], [\ell_{1}, r_{1}]))$. 
Therefore, the second phase takes in $O(H^{2} + \log n)$ time. 

Finally, Lemma~\ref{lem:JC1_Y_queries}~\ref{enum:JC1_Y_queries:2} holds.

\textbf{Proof of Lemma~\ref{lem:JC1_Y_queries}(iii).}
Lemma~\ref{lem:JC1_Y_queries}~\ref{enum:JC1_Y_queries:3} corresponds to 
Lemma~\ref{lem:JA_Y_queries}~\ref{enum:JA_Y_queries:3}.
We proved Lemma~\ref{lem:JA_Y_queries}~\ref{enum:JA_Y_queries:3} using 
the query of Lemma~\ref{lem:JA_Y_queries}~\ref{enum:JA_Y_queries:2}, one LCE query, and random access queries. 
Lemma~\ref{lem:JC1_Y_queries}~\ref{enum:JC1_Y_queries:2} corresponds to Lemma~\ref{lem:JA_Y_queries}~\ref{enum:JA_Y_queries:2}. 
Therefore, Lemma~\ref{lem:JC1_Y_queries}~\ref{enum:JC1_Y_queries:3} can be proved using the same approach as for Lemma~\ref{lem:JA_Y_queries}~\ref{enum:JA_Y_queries:3}.     

\textbf{Proof of Lemma~\ref{lem:JC1_Y_queries}(iv).}
Lemma~\ref{lem:JC1_Y_queries}~\ref{enum:JC1_Y_queries:4} can be proved using the same approach as for Lemma~\ref{lem:JA_Y_queries}~\ref{enum:JA_Y_queries:4}. 
The detailed proof of Lemma~\ref{lem:JC1_Y_queries}~\ref{enum:JC1_Y_queries:4} is as follows. 

We leverage the doubly linked list $\mathbf{Y}_{C}(h, C, M, R_{s})$ for this query. 
Let $(x_{1}, y_{1}, w_{1}, e_{1})$, $(x_{2}, y_{2}, w_{2}$, $e_{2})$, $\ldots$, 
$(x_{m}, y_{m}, w_{m}, e_{m})$ ($x_{1} \leq x_{2} \leq \cdots \leq x_{m}$) be 
the weighted points in set $\{ (x, y, w, e) \in \mathcal{J}_{C}(h, C, M) \mid y = R_{s} \}$. 
Then, each $b$-th weighted point $(x_{b}, y_{b}, w_{b}, e_{b})$ corresponds to the $b$-th element of the doubly linked list $\mathbf{Y}_{C}(h, C, M, R_{s})$. 
On the other hand, the $b$-th weighted point $(x_{b}, y_{b}, w_{b}, e_{b})$ corresponds to 
an interval attractor $([p_{b}, q_{b}], [\ell_{b}, r_{b}])$ in set $\Psi_{h} \cap \Psi_{\samp}$.     
Let $\lambda$ be the smallest integer in set $[1, m]$ satisfying 
$\tau \preceq x_{\lambda}$. 
Then, the $\lambda$-th weighted point $(x_{\lambda}, y_{\lambda}, w_{\lambda}, e_{\lambda})$ 
satisfies $\tau \leq x_{\lambda} \leq \tau^{\prime}$ and $y_{\lambda} = R_{s}$. 
Therefore, we can return the interval attractor corresponding to the weighted point $(x_{\lambda}, y_{\lambda}, w_{\lambda}, e_{\lambda})$ as the answer to the query of Lemma~\ref{lem:JC1_Y_queries}~\ref{enum:JC1_Y_queries:4}. 

We find the smallest integer $\lambda$ by binary search on the doubly linked list $\mathbf{Y}_{C}(h, C, M, R_{s})$. 
This binary search needs to verify whether $\tau \leq x_{b}$ or not for an integer $b \in [1, m]$. 
Here, $x_{b} = |f_{\recover}((x_{b}, y_{b}, w_{b}, e_{b}))|$ follows from the definition of the set $\mathcal{J}_{C}(h, C, M)$.
The x-coordinate $x_{b}$ can be computed $O(H^{2} + \log n)$ time in the following three steps: 
\begin{enumerate}[label=\textbf{(\arabic*)}]
\item access the $b$-th element of the doubly linked list $\mathbf{Y}_{C}(h, C, M, R_{s})$ by 
the list indexing data structure built on the doubly linked list. 
Here, the $b$-th element stores a pointer to the element $v$ 
corresponding to the interval attractor $([p_{b}, q_{b}], [\ell_{b}, r_{b}])$ 
in the doubly linked list introduced in Section~\ref{subsubsec:sample_ds};
\item recover the interval attractor $([p_{b}, q_{b}], [\ell_{b}, r_{b}])$ from the element $v$ 
in $O(\log n)$ time by the algorithm of Section~\ref{subsubsec:computation_delta_samp};
\item compute $|f_{\recover}((x_{b}, y_{b}, w_{b}, e_{b}))|$ by r-size query $\rsizeQ(([p_{s}, q_{s}], [\ell_{s}, r_{s}]))$. 
\end{enumerate}
The binary search takes $O((H^{2} + \log n)\log m)$ time. 
$m \leq |\mathcal{J}_{C}(h, C, M)|$ holds, 
and Lemma~\ref{lem:JC1_size}~\ref{enum:JC1_size:2} shows that $|\mathcal{J}_{C}(h, C, M)| = O(n^{2})$ holds. 
Therefore, the smallest integer $\lambda$ can be found in $O(H^{2} \log n + \log^{2} n)$ time. 

After executing the binary search, 
we return the interval attractor $([p_{\lambda}, q_{\lambda}], [\ell_{\lambda}, r_{\lambda}])$ 
as the answer to the query of Lemma~\ref{lem:JC1_Y_queries}~\ref{enum:JC1_Y_queries:4}. 
This interval attractor can be obtained by the algorithm of Section~\ref{subsubsec:computation_delta_samp}. 
Therefore, Lemma~\ref{lem:JC1_Y_queries}~\ref{enum:JC1_Y_queries:4} holds.    

\end{proof}

\subsubsection{Dynamic Data Structures for Set \texorpdfstring{$\mathcal{J}_{C}(h, C, M)$}{} of Weighted Points}\label{subsubsec:JC1_ds}
Consider a triplet of an integer $h \in [0, H]$, a string $C \in \Sigma^{+}$, and a non-negative integer $M \in \mathbb{N}_{0}$. 
We present dynamic data structures to support range-sum query on set $\mathcal{J}_{C}(h, C, M)$ of weighted points,
which are similar to the dynamic data structures for the ordered set $\mathcal{J}_{A}(h)$ presented in Section~\ref{subsubsec:JA_ds}. 
Let $(x_{1}, y_{1}, w_{1}, e_{1})$, $(x_{2}, y_{2}, w_{2}, e_{2})$, $\ldots$, $(x_{k}, y_{k}, w_{k}, e_{k})$ ($e_{1} \prec e_{2} \prec \cdots e_{k}$) be the weighted points in the set $\mathcal{J}_{C}(h, C, M)$. 
For each weighted point $(x_{s}, y_{s}, w_{s}, e_{s}) \in \mathcal{J}_{C}(h, C, M)$, 
the doubly linked list $\mathbf{Y}_{C}(h, C, M, y_{s})$ of Section~\ref{subsubsec:JC1_Y_ds} 
contains an element $v_{s}$ representing the weighted point $(x_{s}, y_{s}, w_{s}, e_{s})$. 

We store the set $\mathcal{J}_{C}(h, C, M)$ using a doubly linked list $\mathbf{L}_{C}(h, C, M)$ of $k$ elements. 
For each integer $s \in [1, k]$, 
the $s$-th element of the doubly linked list $\mathbf{L}_{C}(h, C, M)$ corresponds to the $s$-th weighted point $(x_{s}, y_{s}, w_{s}, e_{s})$. This element stores (i) the x-coordinate $x_{s}$, 
(ii) the weight $w_{s}$, and (iii) a pointer to the element $v_{s}$. 
List indexing and range-sum data structures are built on doubly linked list $\mathbf{L}_{C}(h, C, M)$. 
This range-sum data structure is used to support range-count and range-sum queries on the set $\mathcal{J}_{C}(h, C, M)$ of 
weighted points. 
These dynamic data structures require $O(|\mathcal{J}_{C}(h, C, M)| B)$ bits of space in total for machine word size $B$. 

\subsubsection{Dynamic Data Structures for Ordered Set \texorpdfstring{$\mathcal{T}_{C}$}{} of Triplets}\label{subsubsec:TC1_ds}
We introduce an ordered Set $\mathcal{T}_{C} \subseteq [0, H] \times \Sigma^{+} \times \mathbb{N}_{0}$ such that 
each triplet of the set $\mathcal{T}_{C}$ consists of an integer $h \in [0, H]$, a string $C \in \Sigma^{+}$, and a non-negative integer $M \in \mathbb{N}_{0}$ satisfying $\mathcal{J}_{C}(h, C, M) \neq \emptyset$ 
(i.e., $\mathcal{T}_{C} = \{ (h, C, M) \in [0, H] \times \Sigma^{+} \times \mathbb{N}_{0} \mid \mathcal{J}_{C}(h, C, M) \neq \emptyset \}$). 
Let $(h_{1}, C_{1}, M_{1})$, $(h_{2}, C_{2}, M_{2})$, $\ldots$, $(h_{m}, C_{m}, M_{m})$ be the triplets in the ordered set $\mathcal{T}_{C}$. 
In the ordered set $\mathcal{T}_{C}$, 
a triplet $(h_{s}, C_{s}, M_{s})$ precedes another triplet $(h_{s^{\prime}}, C_{s^{\prime}}, M_{s^{\prime}})$ if and only if 
one of the following three conditions is satisfied: 
\begin{itemize}
    \item $h_{s} < h_{s^{\prime}}$;
    \item $h_{s} = h_{s^{\prime}}$ and $C_{s} \prec C_{s^{\prime}}$;
    \item $h_{s} = h_{s^{\prime}}$, $C_{s} = C_{s^{\prime}}$, and $M_{s} < M_{s^{\prime}}$.
\end{itemize}
This ordered set $\mathcal{T}_{C}$ is used to verify whether set $\mathcal{J}_{C}(h, C, M)$ is empty or not for 
a given triplet of an integer $h \in [0, H]$, a string $C \in \Sigma^{+}$, and a non-negative integer $M \in \mathbb{N}_{0}$. 
The following lemma states three properties of the ordered set $\mathcal{T}_{C}$. 

\begin{lemma}\label{lem:TC1_size}
    The following three statements hold for ordered set $\mathcal{T}_{C}$: 
    \begin{enumerate}[label=\textbf{(\roman*)}]
    \item \label{enum:TC1_size:1} 
    $\sum_{(h, C, M) \in \mathcal{T}_{C}} |\mathcal{J}_{C}(h, C, M)| \leq |\Psi_{\samp}|$;
    \item \label{enum:TC1_size:2} $|\mathcal{T}_{C}| \leq |\Psi_{\samp}|$;
    \item \label{enum:TC1_size:3} $|\mathcal{T}_{C}| = O(n^{2})$. 
    \end{enumerate}
\end{lemma}
\begin{proof}
    The proof of Lemma~\ref{lem:TC1_size} is as follows. 

    \textbf{Proof of Lemma~\ref{lem:TC1_size}(i).}
    For a triplet of an integer $h \in [0, H]$, a string $C \in \Sigma^{+}$, and a non-negative integer $M \in \mathbb{N}_{0}$, 
    there exists a one-to-one correspondence between the weighted points of set $\mathcal{J}_{C}(h, C, M)$ 
    and the interval attractors of set $\Psi_{h} \cap \Psi_{\source} \cap \Psi_{\centerset}(C) \cap \Psi_{\modulo}(M) \cap \Psi_{\preceding} \cap \Psi_{\samp}$. 
    Therefore, the following two equations hold: 
    \begin{equation*}
    |\mathcal{J}_{C}(h, C, M)| = |\Psi_{h} \cap \Psi_{\source} \cap \Psi_{\centerset}(C) \cap \Psi_{\modulo}(M) \cap \Psi_{\preceding} \cap \Psi_{\samp}|.
    \end{equation*}    
    \begin{equation*}
    |\Psi_{h} \cap \Psi_{\source} \cap \Psi_{\centerset}(C) \cap \Psi_{\modulo}(M) \cap \Psi_{\preceding} \cap \Psi_{\samp}| \neq 0 \Leftrightarrow (h, C, M) \in \mathcal{T}_{C}.
    \end{equation*} 

    Let $(h_{1}, C_{1}, M_{1}), (h_{2}, C_{2}, M_{2}), \ldots, (h_{k}, C_{k}, M_{k})$ be the triplets in the set $\mathcal{T}_{C}$. 
    Then, 
    set $\Psi_{\source} \cap \Psi_{\preceding} \cap \Psi_{\samp}$ can be divided into 
    $k$ sets 
    $(\Psi_{h_{1}} \cap \Psi_{\source} \cap \Psi_{\centerset}(C_{1}) \cap \Psi_{\modulo}(M_{1}) \cap \Psi_{\preceding} \cap \Psi_{\samp})$,
    $(\Psi_{h_{2}} \cap \Psi_{\source} \cap \Psi_{\centerset}(C_{2}) \cap \Psi_{\modulo}(M_{2}) \cap \Psi_{\preceding} \cap \Psi_{\samp})$,
    $\ldots$, 
    $(\Psi_{h_{k}} \cap \Psi_{\source} \cap \Psi_{\centerset}(C_{k}) \cap \Psi_{\modulo}(M_{k}) \cap \Psi_{\preceding} \cap \Psi_{\samp})$. 
    This is because the following three statements hold: 
    \begin{itemize}
        \item $\Psi_{h} \cap \Psi_{h^{\prime}} = \emptyset$ for any pair of two integers $0 \leq h < h^{\prime} \leq H$;
        \item $\Psi_{\centerset}(C) \cap \Psi_{\centerset}(C^{\prime}) = \emptyset$ for any pair of two integers $0 \leq h < h^{\prime} \leq H$;
        \item $\Psi_{\modulo}(M) \cap \Psi_{\modulo}(M^{\prime}) = \emptyset$ for any pair of two integers $0 \leq M < M^{\prime}$.        
    \end{itemize}

    $\Psi_{\source} \cap \Psi_{\preceding} \cap \Psi_{\samp} \subseteq \Psi_{\samp}$ holds. 
    Therefore, $\sum_{(h, C, M) \in \mathcal{T}_{C}} |\mathcal{J}_{C}(h, C, M)| \leq |\Psi_{\samp}|$ follows from 
    the following equation: 
    \begin{equation*}
    \begin{split}
        \sum_{(h, C, M) \in \mathcal{T}_{C}} & |\mathcal{J}_{C}(h, C, M)| \\
        & = \sum_{(h, C, M) \in \mathcal{T}_{C}} |\Psi_{h} \cap \Psi_{\source} \cap \Psi_{\centerset}(C) \cap \Psi_{\modulo}(M) \cap \Psi_{\preceding} \cap \Psi_{\samp}| \\
        &= |\Psi_{\source} \cap \Psi_{\preceding} \cap \Psi_{\samp}| \\
        &\leq |\Psi_{\samp}|.
    \end{split}
    \end{equation*}    

    \textbf{Proof of Lemma~\ref{lem:TC1_size}(ii).}
    $|\mathcal{T}_{C}| \leq \sum_{(h, C, M) \in \mathcal{T}_{C}} |\mathcal{J}_{C}(h, C, M)|$ holds 
    because $|\mathcal{J}_{C}(h, C, M)| \geq 1$ for each triplet $(h, C, M) \in \mathcal{T}_{C}$. 
    We already proved $\sum_{(h, C, M) \in \mathcal{T}_{C}} |\mathcal{J}_{C}(h, C, M)| \leq |\Psi_{\samp}|$. 
    Therefore, $|\mathcal{T}_{C}| \leq |\Psi_{\samp}|$ holds. 

    \textbf{Proof of Lemma~\ref{lem:TC1_size}(iii).}
    $|\Psi_{\samp}| = O(n^{2})$ follows from 
    $\Psi_{\samp} \subseteq \Psi_{\RR}$ and 
    $|\Psi_{\RR}| = O(n^{2})$ (Lemma~\ref{lem:non_comp_IA_size}). 
    $|\mathcal{T}_{C}| = O(n^{2})$ follows from 
    $|\mathcal{T}_{C}| \leq |\Psi_{\samp}|$ and $|\Psi_{\samp}| = O(n^{2})$.

\end{proof}

We store the ordered set $\mathcal{T}_{C}$ using a doubly linked list of $m$ elements. 
For each integer $s \in [1, m]$, 
the $s$-th element of this doubly linked list corresponds to the $s$-th triplet $(h_{s}, C_{s}, M_{s})$ of the ordered set $\mathcal{T}_{C}$. 
The $s$-th element stores two pointers to the two doubly linked lists $\mathbf{L}_{C}(h, C, M)$ and $\mathbf{L}^{Y}_{C}(h, C, M)$. 
A list indexing data structure is used for quickly accessing to the elements of the doubly linked list storing the ordered set $\mathcal{T}_{C}$. 
These dynamic data structures require $O(m B)$ bits of space in total for machine word size $B$. 

\subsubsection{Dynamic Data Structures and Algorithm for \texorpdfstring{$\RSCQCX(i, j)$}{RSCC1(i, j)}}\label{subsubsec:JC1_subquery_ds}
We prove Lemma~\ref{lem:RSC_subquery_C1_summary}, i.e., 
we show that $\RSCQCX(i, j)$ can be answered in $O(H^{2} \log n + \log^{4} n)$ time using 
dynamic data structures of $O((|\mathcal{U}_{\RR}| + |\Psi_{\samp}|)B)$ bits of space for machine word size $B$. 
Here, $|\mathcal{U}_{\RR}|$ is the number of nodes in the RR-DAG of RLSLP $\mathcal{G}^{R}$. 
Let $(h_{1}, C_{1}, M_{1})$, $(h_{2}, C_{2}, M_{2})$, $\ldots$, $(h_{m}, C_{m}, M_{m})$ be the triplets in the ordered set $\mathcal{T}_{C}$. 

\paragraph{Data Structures.}
We answer $\RSCQCX(i, j)$ using the following dynamic data structures: 
\begin{itemize}
    \item the dynamic data structures of $O(|\mathcal{U}_{\RR}|B)$ bits of space 
    for the RR-DAG of RLSLP $\mathcal{G}^{R}$ (Section~\ref{subsubsec:rrdag_ds}). 
    \item the dynamic data structures of $O(|\Psi_{\samp}|B)$ bits of space for 
    sample query (Section~\ref{subsubsec:sample_ds});
    \item the dynamic data structures of $O(\sum_{s = 1}^{m} |\mathcal{J}_{C}(h_{s}, C_{s}, M_{s})| B)$ bits of space 
    for $m$ sets $\mathcal{Y}_{C}(h_{1}$, $C_{1}, M_{1})$, $\mathcal{Y}_{C}(h_{2}, C_{2}, M_{2})$, $\ldots$, $\mathcal{Y}_{C}(h_{m}, C_{m}, M_{m})$ 
    (Section~\ref{subsubsec:JC1_Y_ds});
    \item the dynamic data structures of $O(\sum_{s = 1}^{m} |\mathcal{J}_{C}(h_{s}, C_{s}, M_{s})| B)$ bits of space 
    for $m$ sets $\mathcal{J}_{C}(h_{1}$, $C_{1}, M_{1})$, $\mathcal{J}_{C}(h_{2}, C_{2}, M_{2})$, $\ldots$, $\mathcal{J}_{C}(h_{m}, C_{m}, M_{m})$ (Section~\ref{subsubsec:JC1_ds}); 
    \item the dynamic data structures of $O(m B)$ bits of space for the ordered set $\mathcal{T}_{C}$ (Section~\ref{subsubsec:TC1_ds}).
\end{itemize}
$\sum_{s = 1}^{m} |\mathcal{J}_{C}(h_{s}, C_{s}, M_{s})| \leq |\Psi_{\samp}|$ 
and $m \leq |\Psi_{\samp}|$ follow from Lemma~\ref{lem:TC1_size}. 
Therefore, these dynamic data structures can be stored in $O((|\mathcal{U}_{\RR}| + |\Psi_{\samp}|) B)$ bits of space. 

The following lemma states three queries supported by these dynamic data structures. 
\begin{lemma}\label{lem:TC1_queries}
    Let $m = |\mathcal{T}_{C}|$ 
    and $(h_{s}, C_{s}, M_{s})$ be the $s$-th triplet of the ordered set $\mathcal{T}_{C}$ for each integer $s \in [1, m]$.     
    Using the dynamic data structures of Section~\ref{subsubsec:JC1_subquery_ds}, 
    we can answer the following three queries:
    \begin{enumerate}[label=\textbf{(\roman*)}]
    \item \label{enum:TC1_queries:1}
    for a given integer $s \in [1, m]$, 
    return a triplet of an integer $h \in [0, H]$, an interval $[\alpha, \beta] \subseteq [1, n]$ in input string $T$, 
    and a non-negative integer $M \in \mathbb{N}_{0}$ 
    satisfying $(h, T[\alpha..\beta], M) = (h_{s}, C_{s}, M_{s})$ in $O(H^{2} + \log n)$ time; 
    \item \label{enum:TC1_queries:2}
    consider a given 4-tuple of an integer $s \in [1, m]$, an integer $h \in [0, H]$, an interval $[\alpha, \beta] \subseteq [1, n]$ in input string $T$, and a non-negative integer $M \in \mathbb{N}_{0}$. 
    Then, verify whether the $s$-th triplet of the ordered set $\mathcal{T}_{C}$ satisfies one of 
    the following three conditions: 
    (A) $h_{s} < h$; (B) $h_{s} = h$ and $C_{s} \prec T[\alpha..\beta]$; (C) $h_{s} = h$, $C_{s} = T[\alpha..\beta]$, and $M_{s} \leq M$. 
    This verification takes $O(H^{2} + \log n)$ time;
    \item \label{enum:TC1_queries:3}
    verify whether $(h, T[\alpha..\beta], M) \in \mathcal{T}_{C}$ or not in $O(H^{2} \log n + \log^{2} n)$ time  
    for a given triplet of an integer $h \in [0, H]$, an interval $[\alpha, \beta] \subseteq [1, n]$ in input string $T$, 
    and a non-negative integer $M \in \mathbb{N}_{0}$. 
    If $(h, T[\alpha..\beta], M) \in \mathcal{T}_{C}$ holds, 
    then return an integer $s \in [1, m]$ satisfying 
    $(h_{s}, C_{s}, M_{s}) = (h, T[\alpha..\beta], M)$ in the same time. 
    \end{enumerate}
\end{lemma}
\begin{proof}
    The proof of Lemma~\ref{lem:TC1_queries} is as follows: 

    \textbf{Proof of Lemma~\ref{lem:TC1_queries}(i).}
    $\mathcal{J}_{C}(h_{s}, C_{s}, M_{s}) \neq \emptyset$ follows from 
    the definition of the ordered set $\mathcal{T}_{C}$. 
    Let $v$ be the first element of doubly linked list $\mathbf{L}_{C}(h_{s}, C_{s}, M_{s})$. 
    Then, the following four statements hold: 
    \begin{itemize}
        \item the element $v$ corresponds to a weighted point $(x, y, w, e)$ in set $\mathcal{J}_{C}(h_{s}, C_{s}, M_{s})$;
        \item let $([p, q], [\ell, r])$ be the interval attractor corresponding to the weighted point $(x, y, w, e)$. 
        Then, $([p, q], [\ell, r]) \in \Psi_{h_{s}} \cap \Psi_{\centerset}(C_{s}) \cap \Psi_{\modulo}(M_{s})$ holds; 
        \item 
        let $v^{\prime}$ be the element of the doubly linked list $\mathbf{Y}_{C}(h_{s}, C_{s}, M_{s}, y)$ 
        corresponding to the weighted point $(x, y, w, e)$. 
        Then, the element $v$ of the doubly linked list $\mathbf{L}_{C}(h_{s}, C_{s}, M_{s})$ 
        stores a pointer to the element $v^{\prime}$;
        \item 
        for the sequence $\mathbf{Q}_{\samp}$ introduced in Section~\ref{subsec:sample_query}, 
        let $u$ be the node of the sequence $\mathbf{Q}_{\samp}$ corresponding to  
        the interval attractor $([p, q], [\ell, r])$. 
        Then, the element $v^{\prime}$ of the doubly linked list $\mathbf{Y}_{C}(h_{s}, C_{s}, M_{s}, y)$ 
        stores a pointer to the element representing the node $u$ in the doubly linked list of Section~\ref{subsubsec:sample_ds}.         
    \end{itemize}
    Let $h$, $\gamma$, and $C$ be the level, attractor position, and associated string of 
    the interval attractor $([p, q], [\ell, r])$. 
    Here, $C = T[\gamma..\gamma + |C| - 1]$ follows from the definition of the associated string $C$.
    $h = h_{s}$ and $C = C_{s}$ follow from $([p, q], [\ell, r]) \in \Psi_{h_{s}} \cap \Psi_{\centerset}(C_{s})$. 
    Let $K$ be the length of the longest common prefix between two strings $T[\gamma..r]$ and $C^{n+1}$. 
    Then, $M_{s} = (K - (2 + \sum_{w = 1}^{h+3} \lfloor \mu(w) \rfloor) ) \mod |C|$ 
    follows from the definition of the subset $\Psi_{\modulo}(M_{s})$ 
    because $([p, q], [\ell, r]) \in \Psi_{h} \cap \Psi_{\centerset}(C) \cap \Psi_{\modulo}(M_{s})$. 
    Therefore, 
    we can answer the query of Lemma~\ref{lem:TC1_queries}~\ref{enum:TC1_queries:1} using  
    the four integers $h$, $\gamma$, $|C|$, and $K$. 

    We answer the query of Lemma~\ref{lem:TC1_queries}~\ref{enum:TC1_queries:1} in two phases: 
    In the first phase, 
    we obtain the interval attractor $([p, q], [\ell, r])$ in two steps. 
    \begin{enumerate}[label=\textbf{(\Alph*)}]
    \item access the doubly linked list $\mathbf{L}_{C}(h_{s}, C_{s}, M_{s})$ 
    using the pointer stored in 
    the $s$-th element of the doubly linked list representing the ordered set $\mathcal{T}_{C}$; 
    \item access the element $v^{\prime}$ of the doubly linked list $\mathbf{Y}_{C}(h_{s}, C_{s}, M_{s}, y)$ 
    using the pointer stored in the element $v$ of the doubly linked list $\mathbf{L}_{C}(h_{s}, C_{s}, M_{s})$;
    \item obtain the node $u$ by the pointer stored in the element $v^{\prime}$;
    \item recover the interval attractor $([p, q], [\ell, r])$ from the node $u$ in $O(\log n)$ time by the algorithm of Section~\ref{subsubsec:computation_delta_samp}.     
    \end{enumerate}
    The first and fourth steps take $O(\log m)$ time and $O(\log n)$ time, respectively. 
    Here, $m = O(n^{2})$ follows from Lemma~\ref{lem:TC1_size}~\ref{enum:TC1_size:3}. 
    The second and third steps takes $O(1)$ time. 
    Therefore, the first phase takes $O(\log n)$ time. 

    In the second phase, 
    we compute the four integers $h$, $\gamma$, $|C|$, and $K$ 
    by level-query $\levelQ(([p, q]$, $[\ell, r]))$, attractor position query $\attrQ(([p, q], [\ell, r]))$, 
    C-length query $\clenQ(([p, q], [\ell, r]))$, and C-LCP query $\clcpQ(([p, q], [\ell, r]))$, 
    respectively. 
    The second phase takes $O(H^{2})$ time. 

    In the second phase, 
    we return triplet $(h, [\gamma, \gamma + |C| - 1], M_{s})$ 
    as the answer to the query of Lemma~\ref{lem:TC1_queries}~\ref{enum:TC1_queries:1}. 
    The integer $M_{s}$ can be computed in $O(H)$ time using the three integers $h$, $|C|$, and $K$. 
    Therefore, the third phase takes $O(H)$ time. 

    Finally, we can answer the query of Lemma~\ref{lem:TC1_queries}~\ref{enum:TC1_queries:1} in $O(H^{2} + \log n)$ time. 

    \textbf{Proof of Lemma~\ref{lem:TC1_queries}(ii).}    
    Let $[\alpha_{s}, \beta_{s}]$ be an interval in input string $T$ 
    satisfying $T[\alpha_{s}..\beta_{s}] = C_{s}$. 
    Then, we can compute triplet $(h_{s}, [\alpha_{s}, \beta_{s}], M_{s})$ by the query of Lemma~\ref{lem:TC1_queries}~\ref{enum:TC1_queries:1}. 
    We can verify whether $C_{s} \prec T[\alpha..\beta]$ or not by one LCE query and at most two random access queries. 
    Similarly, 
    we can verify whether $C_{s} = T[\alpha..\beta]$ or not by LCE query. 
    Therefore, we can answer the query of Lemma~\ref{lem:TC1_queries}~\ref{enum:TC1_queries:2} 
    using (1) the query of Lemma~\ref{lem:TC1_queries}~\ref{enum:TC1_queries:1}, 
    (2) two LCE queries, and (3) at most two random access queries. 
    These queries take $O(H^{2} + \log n)$ time in total. 

    \textbf{Proof of Lemma~\ref{lem:TC1_queries}(iii).}    
    Let $x$ be the largest integer in set $[1, m]$ satisfying 
    one of the following three conditions: 
    \begin{enumerate}[label=\textbf{(\Alph*)}]
        \item $h_{x} < h$;
        \item $h_{x} = h$ and $C_{x} \prec T[\alpha..\beta]$;
        \item $h_{x} = h$, $C_{x} = T[\alpha..\beta]$, and $M_{x} \leq M$. 
    \end{enumerate}
    Then, the ordered set $\mathcal{T}_{C}$ contains triplet $(h, T[\alpha..\beta], M)$ 
    if and only if $(h_{x}, C_{x}, M_{x}) = (h, T[\alpha..\beta], M)$ holds. 

    We answer the query of Lemma~\ref{lem:TC1_queries}~\ref{enum:TC1_queries:3} in three phases. 
    In the first phase, 
    we find the largest integer $x$ by binary search on the $m$ triplets of ordered set $\mathcal{T}_{C}$. 
    This binary search can be executed using $O(\log m)$ queries of Lemma~\ref{lem:TC1_queries}~\ref{enum:TC1_queries:2}. 
    Here, $m = O(n^{2})$ follows from Lemma~\ref{lem:TC1_size}~\ref{enum:TC1_size:3}. 
    If the largest integer $x$ does not exist, 
    then we report that the ordered set $\mathcal{T}_{C}$ does not contain triplet $(h, T[\alpha..\beta], M)$. 
    Otherwise, proceed to the next phase. 
    The first phase takes $O(H^{2} \log n + \log^{2} n)$ time. 

    In the second phase, 
    we verify whether $(h_{x}, C_{x}, M_{x}) = (h, T[\alpha..\beta], M)$ or not by 
    the query of Lemma~\ref{lem:TC1_queries}~\ref{enum:TC1_queries:2}. 
    If $(h_{x}, C_{x}, M_{x}) = (h, T[\alpha..\beta], M)$ holds, 
    then we return the integer $x$ as the answer to the query of Lemma~\ref{lem:TC1_queries}~\ref{enum:TC1_queries:3}. 
    Otherwise, we report that the ordered set $\mathcal{T}_{C}$ does not contain triplet $(h, T[\alpha..\beta], M)$. 
    The second phase takes $O(H^{2} + \log n)$ time. 

    Finally, we can answer the query of Lemma~\ref{lem:TC1_queries}~\ref{enum:TC1_queries:3} in $O(H^{2} \log n + \log^{2} n)$ time.     
\end{proof}

\paragraph{Algorithm.}
The algorithm for $\RSCQCX(i, j)$ computes $|\Psi_{\CCP}(T[i..j]) \cap \Psi_{\lex}(T[\gamma_{Q}..r_{Q}+1]) \cap \Psi_{\run} \cap \Psi_{\centerset}(C_{Q}) \cap \Psi_{\lcp}(K_{Q}) \cap \Psi_{\preceding}|$ 
under the condition that 
RSC query $\RSCQ(i, j)$ satisfies either condition (C) or (D) of RSC query stated in Section~\ref{subsec:rsc_sub}. 
This algorithm leverages Lemma~\ref{lem:JC1_main_lemma}, which shows that 
the size of set $|\Psi_{\CCP}(T[i..j]) \cap \Psi_{\lex}(T[\gamma_{Q}..r_{Q}+1]) \cap \Psi_{\run} \cap \Psi_{\centerset}(C_{Q}) \cap \Psi_{\lcp}(K_{Q}) \cap \Psi_{\preceding}|$ can be computed 
by one range-sum query on set $\mathcal{J}_{C}(h_{Q}, C_{Q}, M_{Q})$ of weighted points. 

The algorithm for $\RSCQCX(i, j)$ consists of five phases. 
In the first phase, 
we compute interval attractor $([p_{Q}, q_{Q}], [\ell_{Q}, r_{Q}])$ 
and five integers $h_{Q}$, $|C_{Q}|$, $\gamma_{Q}$, $K_{Q}$, and $M_{Q}$. 
The interval attractor $([p_{Q}, q_{Q}], [\ell_{Q}, r_{Q}])$ can be obtained by capture query $\CAPQ([i, j])$. 
The four integers $h_{Q}$, $\gamma_{Q}$, $|C_{Q}|$, and $K_{Q}$ can be computed by 
level-query $\levelQ(([p_{Q}, q_{Q}], [\ell_{Q}, r_{Q}]))$, attractor position query $\attrQ(([p_{Q}, q_{Q}], [\ell_{Q}, r_{Q}]))$, 
C-length query $\clenQ(([p_{Q}, q_{Q}], [\ell_{Q}, r_{Q}]))$, and C-LCP query $\clcpQ(([p_{Q}, q_{Q}], [\ell_{Q}, r_{Q}]))$, respectively. 
The integer $M_{Q}$ can be computed in $O(H)$ time using three integers $h_{Q}$, $|C_{Q}|$, and $K_{Q}$. 
The bottleneck of the first phase is capture query, which takes $O(H^{2} \log n)$ time. 

In the second phase, 
we verify whether $(h_{Q}, C_{Q}, M_{Q}) \in \mathcal{T}_{C}$ or not by Lemma~\ref{lem:TC1_queries}~\ref{enum:TC1_queries:3}. 
Here, let $(h_{1}, C_{1}, M_{1})$, $(h_{2}, C_{2}, M_{2})$, $\ldots$, $(h_{m}, C_{m}, M_{m})$ be the triplets in the ordered set $\mathcal{T}_{C}$. 
If $(h_{Q}, C_{Q}, M_{Q}) \not \in \mathcal{T}_{C}$, 
then Lemma~\ref{lem:JC1_main_lemma} indicates that 
$|\Psi_{\CCP}(T[i..j]) \cap \Psi_{\lex}(T[\gamma_{Q}..r_{Q}+1]) \cap \Psi_{\run} \cap \Psi_{\centerset}(C_{Q}) \cap \Psi_{\lcp}(K_{Q}) \cap \Psi_{\preceding}| = 0$ holds because $\mathcal{J}_{C}(h_{Q}, C_{Q}, M_{Q}) = \emptyset$. 
In this case, we stop this algorithm and report that the set $\Psi_{\CCP}(T[i..j]) \cap \Psi_{\lex}(T[\gamma_{Q}..r_{Q}+1]) \cap \Psi_{\run} \cap \Psi_{\centerset}(C_{Q}) \cap \Psi_{\lcp}(K_{Q}) \cap \Psi_{\preceding}$ is empty. 
Otherwise (i.e., $(h_{Q}, C_{Q}, M_{Q}) \in \mathcal{T}_{C}$), 
there exists an integer $s \in [1, m]$ satisfying 
$(h_{s}, C_{s}, M_{s}) = (h_{Q}, C_{Q}, M_{Q})$, 
and this integer $s$ can be obtained by the query of Lemma~\ref{lem:TC1_queries}~\ref{enum:TC1_queries:3}. 
The query of Lemma~\ref{lem:TC1_queries}~\ref{enum:TC1_queries:3} can be executed using 
triplet $(h_{Q}, [\gamma_{Q}, \gamma_{Q} + |C_{Q}| - 1], M_{Q})$ 
because $C_{Q} = T[\gamma_{Q}..\gamma_{Q} + |C_{Q}| - 1]$ follows from the definition of the associated string $C_{Q}$. 
Therefore, the second phase takes $O(H^{2} \log n + \log^{2} n)$ time. 

In the third phase, 
we access the two doubly linked lists $\mathbf{L}_{C}(h_{Q}, C_{Q}, M_{Q})$ and $\mathbf{L}^{Y}_{C}(h_{Q}, C_{Q}, M_{Q})$. 
The $s$-th element of the doubly linked list representing the ordered set $\mathcal{T}_{C}$ stores 
two pointers to the two doubly linked lists $\mathbf{L}_{C}(h_{Q}, C_{Q}, M_{Q})$ and $\mathbf{L}^{Y}_{C}(h_{Q}, C_{Q}, M_{Q})$. 
Therefore, the third phase can be executed in $O(\log m)$ time 
by the list indexing data structure built on the doubly linked list representing the ordered set $\mathcal{T}_{C}$. 
Here, $m = O(n^{2})$ follows from Lemma~\ref{lem:TC1_size}~\ref{enum:TC1_size:3}.

In the fourth phase, 
we compute three integers $x, y$, and $y^{\prime}$ of Lemma~\ref{lem:JC1_main_lemma}. 
Here, 
\begin{itemize}
\item $x =  1 + \lfloor \frac{K_{Q} - (2 + \sum_{w = 1}^{h_{Q}+3} \lfloor \mu(w) \rfloor)}{|C_{Q}|} \rfloor$; 
\item $y = \min \{ s \in [1, d] \mid T[\gamma_{Q} + K_{Q}..j] \prec R_{s} \}$ 
for the $d$ strings $R_{1}, R_{2}, \ldots, R_{d}$~($R_{1} \prec R_{2} \prec \cdots \prec R_{d}$) in the ordered set $\mathcal{Y}_{C}(h_{Q}, C_{Q}, M_{Q})$ if $|[\gamma_{Q}, j]| > K_{Q}$. 
Otherwise, let $y = 1$; 
\item $y^{\prime} = \max \{ s \in [1, d] \mid R_{s} \prec T[\gamma_{Q} + K_{Q}..r_{Q} + 1] \}$.
\end{itemize}
The integer $x$ can be computed in $O(H)$ time using three integers $h_{Q}$, $K_{Q}$, and $|C_{Q}|$. 
We compute the two integers $y$ and $y^{\prime}$ by binary search on the $d$ strings $R_{1}, R_{2}, \ldots, R_{d}$. 
This binary search can be executed in $O((H^{2} + \log n)\log d)$ time using Lemma~\ref{lem:JC1_Y_queries}~\ref{enum:JC1_Y_queries:3}. 
Lemma~\ref{lem:JC1_size}~\ref{enum:JC1_size:2} shows that $d = O(n^{2})$ holds. 
Therefore, the fourth phase takes $O((H^{2} + \log n) \log n)$ time. 

In the fifth phase, we compute $|\Psi_{\CCP}(T[i..j]) \cap \Psi_{\lex}(T[\gamma_{Q}..r_{Q}+1]) \cap \Psi_{\run} \cap \Psi_{\centerset}(C_{Q}) \cap \Psi_{\lcp}(K_{Q}) \cap \Psi_{\preceding}|$ by 
range-sum query $\rangesum(\mathcal{J}_{C}(h_{Q}, C_{Q}, M_{Q}), x, n, R_{y}, R_{y^{\prime}})$. 
For executing this range-sum query, 
we need to access 
the $y$-th and $y^{\prime}$-th elements of doubly linked list $\mathbf{L}^{Y}_{C}(h_{Q}, C_{Q}, M_{Q})$, 
which corresponding to two strings $R_{y}$ and $R_{y^{\prime}}$, respectively. 
The $y$-th and $y^{\prime}$-th elements of $\mathbf{L}^{Y}_{C}(h_{Q}, C_{Q}$, $M_{Q})$ can be accessed in $O(\log d)$ time by 
the list indexing data structure built on the doubly linked list $\mathbf{L}^{Y}_{C}(h_{Q}, C_{Q}, M_{Q})$. 
The range-sum query $\rangesum(\mathcal{J}_{C}(h_{Q}, C_{Q}, M_{Q}), x, n, R_{y}, R_{y^{\prime}})$ takes 
$O(\log^{4} k)$ time for the number $k$ of weighted points in set $\mathcal{J}_{C}(h_{Q}, C_{Q}, M_{Q})$. 
Lemma~\ref{lem:JC1_size}~\ref{enum:JC1_size:2} shows that $d, k = O(n^{2})$ holds. 
Therefore, the third phases takes $O(\log^{4} n)$ time. 

Finally, the algorithm for $\RSCQCX(i, j)$ can be executed in $O(H^{2} \log n + \log^{4} n)$ time in total. 
Therefore, Lemma~\ref{lem:RSC_subquery_C1_summary} holds.

\subsection{Subquery \texorpdfstring{$\RSCQCY(i, j)$}{RSCC2(i, j)}}\label{subsec:RSC_comp_C2}

The goal of this subsection is to answer subquery $\RSCQCY(i, j)$. 
The following lemma states the summary of this subsection. 

\begin{lemma}\label{lem:RSC_subquery_C2_summary}
Using a dynamic data structure of $O((|\Psi_{\samp}| + |\mathcal{U}_{\RR}|) B)$ bits of space for machine word size $B$, 
we can answer $\RSCQCY(i, j)$ in $O(H^{2} \log n + \log^{4} n)$ time 
if the given RSC query $\RSCQ(i, j)$ satisfies either condition (C) or (D). 
Here, $|\mathcal{U}_{\RR}|$ is the number of nodes in the RR-DAG of RLSLP $\mathcal{G}^{R}$.
\end{lemma}
\begin{proof}
See Section~\ref{subsubsec:JC2_subquery_ds}.
\end{proof}

Subquery $\RSCQCY(i, j)$ can be answered using the same approach as for $\RSCQCX(i, j)$, which explained in Section~\ref{subsec:RSC_comp_C1}. 
For answering $\RSCQCY(i, j)$, 
we leverage range-sum query on weighted points corresponding to the interval attractors in 
set $\Psi_{h_{Q}} \cap \Psi_{\source} \cap \Psi_{\centerset}(C_{Q}) \cap \Psi_{\modulo}(M_{Q}) \cap \Psi_{\succeeding} \cap \Psi_{\samp}$.
For this purpose, 
we introduce a set $\mathcal{J}_{C^{\prime}}(h, C, M)$ of weighted points on a grid $([1, n], \mathcal{Y}_{C^{\prime}}(h, C, M))$ 
for a triplet $(h, C, M)$ of 
an integer $h \in [0, H]$, a string $C \in \Sigma^{+}$, and a non-negative integer $M \in \mathbb{N}_{0}$. 
Here, the set $\mathcal{J}_{C^{\prime}}(h)$ and grid $([1, n], \mathcal{Y}_{C^{\prime}}(h, C, M))$ are defined using 
set $\Psi_{h} \cap \Psi_{\source} \cap \Psi_{\centerset}(C) \cap \Psi_{\modulo}(M) \cap \Psi_{\succeeding} \cap \Psi_{\samp}$ of $k$ interval attractors 
$([p_{1}, q_{1}], [\ell_{1}, r_{1}]), ([p_{2}, q_{2}]$, $[\ell_{2}, r_{2}])$, 
$\ldots$, $([p_{k}, q_{k}], [\ell_{k}, r_{k}])$. 
For each integer $s \in [1, k]$, 
let $\gamma_{s}$ of the attractor position of each interval attractor $([p_{s}, q_{s}], [\ell_{s}, r_{s}])$; 
let $K_{s}$ be the length of the longest common prefix between two strings $T[\gamma_{s}..r_{s}]$ and $C^{n+1}$ 
(i.e., $K_{s} = |\lcp(T[\gamma_{s}..r_{s}], C^{n+1})|$). 
Here, $([p_{s}, q_{s}], [\ell_{s}, r_{s}]) \in \Psi_{\lcp}(K_{s})$ follows from the definition of subset $\Psi_{\lcp}(K_{s})$. 

\paragraph{Grid $([1, n], \mathcal{Y}_{C^{\prime}}(h, C, M))$.}
$\mathcal{Y}_{C^{\prime}}(h, C, M) \subseteq \Sigma^{*}$ is 
the ordered set of strings defined as the union of two sets 
$\{ \varepsilon, \# \}$ and $\{ T[\gamma_{s} + K_{s}..r_{s} + 1] \mid s \in [1, k] \}$ 
(i.e., $\mathcal{Y}_{C^{\prime}}(h, C, M) = \{ \varepsilon, \# \} \cup \{ T[\gamma_{s} + K_{s}..r_{s} + 1] \mid s \in [1, k] \}$). 
Here, $\varepsilon$ is the string of length $0$, 
and $\#$ is the largest character in the alphabet $\Sigma$ (see Section~\ref{sec:preliminary}). 
This ordered set $\mathcal{Y}_{C^{\prime}}(h, C, M)$ consists of $d$ strings $R_{1}, R_{2}, \ldots, R_{d}$ 
that are sorted in lexicographical order (i.e., $R_{1} \prec R_{2} \prec \cdots \prec R_{d}$). 
$R_{1} = \varepsilon$ and $R_{d} = \#$ always hold because 
every string of the set $\{ T[\gamma_{s} + K_{s}..r_{s} + 1] \mid s \in [1, k] \}$ does not contain the character $\#$. 

Grid $([1, n], \mathcal{Y}_{C^{\prime}}(h, C, M))$ consists of two ordered sets $[1, n]$ and $\mathcal{Y}_{C^{\prime}}(h, C, M)$. 
Each integer of the ordered set $[1, n]$ represents x-coordinate on two dimensional space. 
Similarly, each string of the ordered set $\mathcal{Y}_{C^{\prime}}(h, C, M)$ represents y-coordinate on two dimensional space. 

\paragraph{Set $\mathcal{J}_{C^{\prime}}(h, C, M)$ of Weighted Points.}
Set $\mathcal{J}_{C^{\prime}}(h, C, M)$ consists of $k$ weighted points 
$(|f_{\recover}(([p_{1}$, $q_{1}], [\ell_{1}, r_{1}]))|$, $T[\gamma_{1} + K_{1}..r_{1} + 1]$, $|\Psi_{\str}(T[p_{1}-1..r_{1}+1])|, T[p_{1}-1..r_{1}+1])$, 
$(|f_{\recover}(([p_{2}$, $q_{2}], [\ell_{2}, r_{2}]))|$, $T[\gamma_{2} + K_{2}..r_{2} + 1]$, $|\Psi_{\str}(T[p_{2}-1..r_{2}+1])|, T[p_{2}-1..r_{2}+1])$, $\ldots$, 
$(|f_{\recover}(([p_{k}, q_{k}], [\ell_{k}, r_{k}]))|, T[\gamma_{k} + K_{k}..r_{k} + 1], |\Psi_{\str}(T[p_{k}-1..r_{k}+1])|, T[p_{k}-1..r_{k}+1])$ on grid $([1, n], \mathcal{Y}_{C^{\prime}}(h, C, M))$. 
Each weighted point $(|f_{\recover}(([p_{s}, q_{s}], [\ell_{s}, r_{s}]))|, T[\gamma_{s} + K_{s}..r_{s} + 1], |\Psi_{\str}(T[p_{s}-1..r_{s}+1])|, T[p_{s}-1..r_{s}+1])$ corresponds to interval attractor $([p_{s}, q_{s}], [\ell_{s}, r_{s}])$ in set $\Psi_{h} \cap \Psi_{\source} \cap \Psi_{\centerset}(C) \cap \Psi_{\modulo}(M) \cap \Psi_{\succeeding} \cap \Psi_{\samp}$. 
The details of the four elements of the weighted point $(|f_{\recover}(([p_{s}, q_{s}], [\ell_{s}, r_{s}]))|, T[\gamma_{s} + K_{s}..r_{s} + 1], |\Psi_{\str}(T[p_{s}-1..r_{s}+1])|, T[p_{s}-1..r_{s}+1])$ are as follows:
\begin{itemize}
    \item the first integer $|f_{\recover}(([p_{s}, q_{s}], [\ell_{s}, r_{s}]))|$ is the x-coordinate of this weighted point; 
    \item the second string $T[\gamma_{s} + K_{s}..r_{s} + 1]$ is the y-coordinate of this weighted point; 
    \item the third integer $|\Psi_{\str}(T[p_{s}-1..r_{s}+1])|$ is the weight of this weighted point;
    \item the fourth string $T[p_{s}-1..r_{s}+1]$ is the identifier of this weighted point.
\end{itemize}

The following lemma states the sizes of two sets $\mathcal{Y}_{C^{\prime}}(h, C, M)$ and $\mathcal{J}_{C^{\prime}}(h, C, M)$. 

\begin{lemma}\label{lem:JC2_size}
    The following two statements hold for a triplet of 
    an integer $h \in [0, H]$, a string $C \in \Sigma^{+}$, and a non-negative integer $M \in \mathbb{N}_{0}$: 
    \begin{enumerate}[label=\textbf{(\roman*)}]
    \item \label{enum:JC2_size:1} $|\mathcal{Y}_{C^{\prime}}(h, C, M)| \leq 2 + |\mathcal{J}_{C^{\prime}}(h, C, M)|$;
    \item \label{enum:JC2_size:2} $|\mathcal{Y}_{C^{\prime}}(h, C, M)|, |\mathcal{J}_{C^{\prime}}(h, C, M)| = O(n^{2})$.
    \end{enumerate}
\end{lemma}
\begin{proof}
    This lemma can be proved using the same approach as for Lemma~\ref{lem:JC1_size}.
\end{proof}

The following lemma shows that 
we can count the interval attractors in set $\Psi_{\CCP}(T[i..j]) \cap \Psi_{\lex}(T[\gamma_{Q}..r_{Q}+1]) \cap \Psi_{\run} \cap \Psi_{\centerset}(C_{Q}) \cap \Psi_{\lcp}(K_{Q}) \cap \Psi_{\succeeding}$ 
by one range-sum query on the set $\mathcal{J}_{C^{\prime}}(h_{Q}, C_{Q}, M_{Q})$ of weighted points 
for the integer $K_{Q}$ of Lemma~\ref{lem:RSC_subquery_C2_summary}. 

\begin{lemma}\label{lem:JC2_main_lemma}
Consider RSC query $\RSCQ(i, j)$ satisfying either condition (C) or (D) of RSC query stated in Section~\ref{subsec:rsc_sub}. 
Let $K_{Q}$ and $M_{Q}$ be the two integers of Lemma~\ref{lem:RSC_subquery_C2_summary}; 
let $R_{1}, R_{2}, \ldots, R_{d}$ ($R_{1} \prec R_{2} \prec \cdots \prec R_{d}$) be 
the $d$ strings in the ordered set $\mathcal{Y}_{C^{\prime}}(h_{Q}, C_{Q}, M_{Q})$; 
let $x, y$, and $y^{\prime}$ be the three integers defined as follows: 
\begin{itemize}
\item $x =  1 + \lfloor \frac{K_{Q} - (2 + \sum_{w = 1}^{h_{Q}+3} \lfloor \mu(w) \rfloor)}{|C_{Q}|} \rfloor$; 
\item $y = \min \{ s \in [1, d] \mid T[\gamma_{Q} + K_{Q}..j] \prec R_{s} \}$ if $|[\gamma_{Q}, j]| > K_{Q}$. 
Otherwise, let $y = 1$; 
\item $y^{\prime} = \max \{ s \in [1, d] \mid R_{s} \prec T[\gamma_{Q} + K_{Q}..r_{Q} + 1] \}$.
\end{itemize}

Then, the following equation holds: 
\begin{equation*}
    \begin{split}
    |\Psi_{\CCP}(T[i..j]) \cap \Psi_{\lex}(T[\gamma_{Q}..r_{Q}+1]) \cap & \Psi_{\run} \cap \Psi_{\centerset}(C_{Q}) \cap \Psi_{\lcp}(K_{Q}) \cap \Psi_{\succeeding})| \\
    &= \rangesum(\mathcal{J}_{C^{\prime}}(h_{Q}, C_{Q}, M_{Q}), x, n, R_{y}, R_{y^{\prime}}).
    \end{split}
\end{equation*}    
\end{lemma}
\begin{proof}
Lemma~\ref{lem:JC2_main_lemma} corresponds to Lemma~\ref{lem:JC1_main_lemma} in Section~\ref{subsec:RSC_comp_C1}. 
The definition of the ordered set $\mathcal{Y}_{C^{\prime}}(h_{Q}, C_{Q}, M_{Q})$ is symmetric to 
that of the ordered set $\mathcal{Y}_{C}(h_{Q}, C_{Q}, M_{Q})$ introduced in Section~\ref{subsec:RSC_comp_C1}.
Similarly, 
the definition of the set $\mathcal{J}_{C^{\prime}}(h_{Q}, C_{Q}, M_{Q})$ is symmetric to 
that of the set $\mathcal{J}_{C}(h_{Q}, C_{Q}, M_{Q})$ introduced in Section~\ref{subsec:RSC_comp_C1}. 
Therefore, 
Lemma~\ref{lem:JC2_main_lemma} can be proved using the same approach as for Lemma~\ref{lem:JC1_main_lemma}. 
\end{proof}

\subsubsection{Dynamic Data Structures for Ordered Set \texorpdfstring{$\mathcal{Y}_{C^{\prime}}(h, C, M)$}{}}\label{subsubsec:JC2_Y_ds}
Consider a triplet of an integer $h \in [0, H]$, a string $C \in \Sigma^{+}$, and a non-negative integer $M \in \mathbb{N}_{0}$. 
We present dynamic data structures to store the ordered set $\mathcal{Y}_{C^{\prime}}(h, C, M)$, 
which are similar to the dynamic data structures for the ordered set $\mathcal{Y}_{C}(h, C, M)$ presented in Section~\ref{subsubsec:JC1_Y_ds}. 
Let $R_{1}, R_{2}, \ldots, R_{d}$~($R_{1} \prec R_{2} \prec \cdots \prec R_{d}$) be 
the strings in the ordered set $\mathcal{Y}_{C^{\prime}}(h, C, M)$. 

We store the ordered set $\mathcal{Y}_{C^{\prime}}(h, C, M)$ using $d+1$ doubly linked lists $\mathbf{Y}_{C^{\prime}}(h, C, M, R_{1})$, $\mathbf{Y}_{C^{\prime}}(h, C$, $M, R_{2})$, $\ldots$, $\mathbf{Y}_{C^{\prime}}(h, C, M, R_{d})$, and $\mathbf{L}^{Y}_{C^{\prime}}(h, C, M)$. 
For each integer $b \in [1, d]$, 
the doubly linked list $\mathbf{Y}_{C^{\prime}}(h, C, M, R_{b})$ is defined in a similar way as 
the doubly linked list $\mathbf{Y}_{C}(h, C, M, R_{b})$ introduced in Section~\ref{subsubsec:JC1_Y_ds}. 
A list indexing data structure is used for quickly accessing to the elements of the doubly linked list $\mathbf{Y}_{C^{\prime}}(h, C, M, R_{b})$. 

Similarly, 
the last doubly linked list $\mathbf{L}^{Y}_{C^{\prime}}(h, C, M)$ is defined in a similar way as the doubly linked list $\mathbf{L}^{Y}_{C}(h, C, M)$ introduced in Section~\ref{subsubsec:JC1_Y_ds}. 
List indexing and order maintenance data structures are built on the doubly-linked list $\mathbf{L}^{Y}_{C^{\prime}}(h, C, M)$. 
These $d+1$ doubly linked lists $\mathbf{Y}_{C^{\prime}}(h, C, M, R_{1}), \mathbf{Y}_{C^{\prime}}(h, C, M, R_{2})$, $\ldots$, $\mathbf{Y}_{C^{\prime}}(h, C, M, R_{d})$, $\mathbf{L}^{Y}_{C}(h, C, M)$ require 
$O((d + |\mathcal{J}_{C^{\prime}}(h$, $C, M)|) B)$ bits of space in total for machine word size $B$. 
$d \leq 2 + |\mathcal{J}_{C^{\prime}}(h, C, M)|$ follows from Lemma~\ref{lem:JC2_size}~\ref{enum:JC2_size:1}. 
Therefore, the dynamic data structures for the ordered set $\mathcal{Y}_{C^{\prime}}(h, C, M)$ can be stored in $O(|\mathcal{J}_{C^{\prime}}(h$, $C, M)| B)$ bits of space.

The following lemma states queries supported by the dynamic data structures for the ordered set $\mathcal{Y}_{C^{\prime}}(h, C, M)$.

\begin{lemma}\label{lem:JC2_Y_queries}
    Let $R_{1}, R_{2}, \ldots, R_{d}$~($R_{1} \prec R_{2} \prec \cdots \prec R_{d}$) be 
    the $d$ strings of the ordered set $\mathcal{Y}_{C^{\prime}}(h, C, M)$ introduced in Section~\ref{subsec:RSC_comp_C2} 
    for a triplet of an integer $h \in [0, H]$, a string $C \in \Sigma^{+}$, and a non-negative integer $M \in \mathbb{N}_{0}$. 
    Consider the $d+1$ doubly linked lists $\mathbf{Y}_{C^{\prime}}(h, C, M, R_{1}), \mathbf{Y}_{C^{\prime}}(h, C, M$, $R_{2})$, $\ldots$, $\mathbf{Y}_{C^{\prime}}(h, C, M, R_{d})$, and $\mathbf{L}^{Y}_{C^{\prime}}(h, C, M)$.  
    Using these $d+1$ doubly linked lists and the dynamic data structures of Section~\ref{subsubsec:rrdag_ds} 
    and Section~\ref{subsubsec:sample_ds}, 
    we can support the following three queries: 
    \begin{enumerate}[label=\textbf{(\roman*)}]
    \item \label{enum:JC2_Y_queries:1} 
    verify whether $R_{s} \prec R_{s^{\prime}}$ or not in $O(1)$ time 
    for the given $s$-th and $s^{\prime}$-th elements of the doubly linked list $\mathbf{L}^{Y}_{C^{\prime}}(h, C, M)$;    
    \item \label{enum:JC2_Y_queries:2} 
    for a given integer $s \in [1, d]$, 
    return an interval $[g, g + |R_{s}| - 1]$ in input string $T$ satisfying $T[g..g + |R_{s}| - 1] = R_{s}$ 
    in $O(H^{2} + \log n)$ time if $R_{s} \not \in \{ \varepsilon, \# \}$; 
    otherwise return the string $R_{s}$ in $O(1)$ time;
    \item \label{enum:JC2_Y_queries:3} 
    verify whether $T[i..j] \prec R_{s}$ or not in $O(H^{2} + \log n)$ time 
    for a given pair of an integer $s \in [1, d]$ 
    and interval $[i, j] \subseteq [1, n]$ in input string $T$. 
    Similarly, we can verify whether $R_{s} \prec T[i..j]$ or not in the same time.
    \end{enumerate}
\end{lemma}
\begin{proof}
    This lemma can be proved using the same approach as for Lemma~\ref{lem:JC2_Y_queries}.
\end{proof}

\subsubsection{Dynamic Data Structures for Set \texorpdfstring{$\mathcal{J}_{C^{\prime}}(h, C, M)$}{} of Weighted Points}\label{subsubsec:JC2_ds}
Consider a triplet of an integer $h \in [0, H]$, a string $C \in \Sigma^{+}$, and a non-negative integer $M \in \mathbb{N}_{0}$. 
We present dynamic data structures to support range-sum query on set $\mathcal{J}_{C^{\prime}}(h, C, M)$ of weighted points,
which are similar to the dynamic data structures for the ordered set $\mathcal{J}_{C}(h, C, M)$ presented in Section~\ref{subsubsec:JC1_ds}. 
Let $(x_{1}, y_{1}, w_{1}, e_{1})$, $(x_{2}, y_{2}, w_{2}, e_{2})$, $\ldots$, $(x_{k}, y_{k}, w_{k}, e_{k})$ ($e_{1} \prec e_{2} \prec \cdots e_{k}$) be the weighted points in the set $\mathcal{J}_{C^{\prime}}(h, C, M)$. 
For each weighted point $(x_{s}, y_{s}, w_{s}, e_{s}) \in \mathcal{J}_{C^{\prime}}(h, C, M)$, 
the doubly linked list $\mathbf{Y}_{C^{\prime}}(h, C, M, y_{s})$ of Section~\ref{subsubsec:JC2_Y_ds} 
contains an element $v_{s}$ representing the weighted point $(x_{s}, y_{s}, w_{s}, e_{s})$. 

We store the set $\mathcal{J}_{C^{\prime}}(h, C, M)$ using a doubly linked list $\mathbf{L}_{C^{\prime}}(h, C, M)$ of $k$ elements. 
For each integer $s \in [1, k]$, 
the $s$-th element of the doubly linked list $\mathbf{L}_{C^{\prime}}(h, C, M)$ corresponds to the $s$-th weighted point $(x_{s}, y_{s}, w_{s}, e_{s})$. This element stores (i) the x-coordinate $x_{s}$, 
(ii) the weight $w_{s}$, and (iii) a pointer to the element $v_{s}$. 
List indexing and range-sum data structures are built on doubly linked list $\mathbf{L}_{C^{\prime}}(h, C, M)$. 
This range-sum data structure is used to support range-count and range-sum queries on the set $\mathcal{J}_{C^{\prime}}(h, C, M)$ of 
weighted points. 
These dynamic data structures require $O(|\mathcal{J}_{C^{\prime}}(h, C, M)| B)$ bits of space in total for machine word size $B$. 

\subsubsection{Dynamic Data Structures for Ordered Set \texorpdfstring{$\mathcal{T}_{C^{\prime}}$}{} of Triplets}\label{subsubsec:TC2_ds}
We introduce an ordered Set $\mathcal{T}_{C^{\prime}} \subseteq [0, H] \times \Sigma^{+} \times \mathbb{N}_{0}$ such that 
each triplet of the set $\mathcal{T}_{C^{\prime}}$ consists of an integer $h \in [0, H]$, a string $C \in \Sigma^{+}$, and a non-negative integer $M \in \mathbb{N}_{0}$ satisfying $\mathcal{J}_{C^{\prime}}(h, C, M) \neq \emptyset$. 
The triplets of this ordered set are sorted in a similar way as 
the triplets of the ordered set $\mathcal{T}_{C}$ introduced in Section~\ref{subsubsec:TC1_ds}.

The following lemma states three properties of the ordered set $\mathcal{T}_{C^{\prime}}$. 

\begin{lemma}\label{lem:TC2_size}
    The following three statements hold for ordered set $\mathcal{T}_{C^{\prime}}$: 
    \begin{enumerate}[label=\textbf{(\roman*)}]
    \item \label{enum:TC2_size:1} 
    $\sum_{(h, C, M) \in \mathcal{T}_{C^{\prime}}} |\mathcal{J}_{C^{\prime}}(h, C, M)| \leq |\Psi_{\samp}|$;
    \item \label{enum:TC2_size:2} $|\mathcal{T}_{C^{\prime}}| \leq |\Psi_{\samp}|$;
    \item \label{enum:TC2_size:3} $|\mathcal{T}_{C^{\prime}}| = O(n^{2})$. 
    \end{enumerate}
\end{lemma}
\begin{proof}
    This lemma can be proved using the same approach as for Lemma~\ref{lem:TC1_size}. 
\end{proof}

We store the ordered set $\mathcal{T}_{C^{\prime}}$ using a doubly linked list of $m$ elements 
for the $m$ triplets $(h_{1}, C_{1}, M_{1})$, $(h_{2}, C_{2}, M_{2})$, $\ldots$, $(h_{m}, C_{m}, M_{m})$ 
in the ordered set $\mathcal{T}_{C^{\prime}}$. 
For each integer $s \in [1, m]$, 
the $s$-th element of this doubly linked list corresponds to the $s$-th triplet $(h_{s}, C_{s}, M_{s})$ of the ordered set $\mathcal{T}_{C^{\prime}}$. 
The $s$-th element stores two pointers to the two doubly linked lists $\mathbf{L}_{C^{\prime}}(h, C, M)$ and $\mathbf{L}^{Y}_{C^{\prime}}(h, C, M)$. 
A list indexing data structure is used for quickly accessing to the elements of the doubly linked list storing the ordered set $\mathcal{T}_{C^{\prime}}$. 
These dynamic data structures require $O(m B)$ bits of space in total for machine word size $B$.

\subsubsection{Dynamic Data Structures and Algorithm for \texorpdfstring{$\RSCQCY(i, j)$}{RSCC2(i, j)}}\label{subsubsec:JC2_subquery_ds}
We prove Lemma~\ref{lem:RSC_subquery_C2_summary}, i.e., 
we show that subquery $\RSCQCY(i, j)$ can be answered in $O(H^{2} \log n + \log^{4} n)$ time using 
dynamic data structures of $O((|\mathcal{U}_{\RR}| + |\Psi_{\samp}|)B)$ bits of space for machine word size $B$. 
Here, $|\mathcal{U}_{\RR}|$ is the number of nodes in the RR-DAG of RLSLP $\mathcal{G}^{R}$. 
Let $(h_{1}, C_{1}, M_{1})$, $(h_{2}, C_{2}, M_{2})$, $\ldots$, $(h_{m}, C_{m}, M_{m})$ be the triplets in the ordered set $\mathcal{T}_{C^{\prime}}$. 

\paragraph{Data Structures.}
We answer $\RSCQCY(i, j)$ using the following dynamic data structures: 
\begin{itemize}
    \item the dynamic data structures of $O(|\mathcal{U}_{\RR}|B)$ bits of space 
    for the RR-DAG of RLSLP $\mathcal{G}^{R}$ (Section~\ref{subsubsec:rrdag_ds}). 
    \item the dynamic data structures of $O(|\Psi_{\samp}|B)$ bits of space for 
    sample query (Section~\ref{subsubsec:sample_ds});
    \item the dynamic data structures of $O(\sum_{s = 1}^{m} |\mathcal{J}_{C^{\prime}}(h_{s}, C_{s}, M_{s})| B)$ bits of space 
    for $m$ sets $\mathcal{Y}_{C^{\prime}}(h_{1}$, $C_{1}, M_{1})$, $\mathcal{Y}_{C^{\prime}}(h_{2}, C_{2}, M_{2})$, $\ldots$, $\mathcal{Y}_{C^{\prime}}(h_{m}, C_{m}, M_{m})$ 
    (Section~\ref{subsubsec:JC2_Y_ds});
    \item the dynamic data structures of $O(\sum_{s = 1}^{m} |\mathcal{J}_{C^{\prime}}(h_{s}, C_{s}, M_{s})| B)$ bits of space 
    for $m$ sets $\mathcal{J}_{C^{\prime}}(h_{1}$, $C_{1}, M_{1})$, $\mathcal{J}_{C^{\prime}}(h_{2}, C_{2}, M_{2})$, $\ldots$, $\mathcal{J}_{C^{\prime}}(h_{m}, C_{m}, M_{m})$ (Section~\ref{subsubsec:JC2_ds}); 
    \item the dynamic data structures of $O(m B)$ bits of space for the ordered set $\mathcal{T}_{C^{\prime}}$ (Section~\ref{subsubsec:TC2_ds}).
\end{itemize}
$\sum_{s = 1}^{m} |\mathcal{J}_{C^{\prime}}(h_{s}, C_{s}, M_{s})| \leq |\Psi_{\samp}|$ 
and $m \leq |\Psi_{\samp}|$ follow from Lemma~\ref{lem:TC2_size}. 
Therefore, these dynamic data structures can be stored in $O((|\mathcal{U}_{\RR}| + |\Psi_{\samp}|) B)$ bits of space. 

The following lemma states three queries supported by these dynamic data structures. 
\begin{lemma}\label{lem:TC2_queries}
    Let $m = |\mathcal{T}_{C^{\prime}}|$ 
    and $(h_{s}, C_{s}, M_{s})$ be the $s$-th triplet of the ordered set $\mathcal{T}_{C^{\prime}}$ for each integer $s \in [1, m]$.     
    Using the dynamic data structures of Section~\ref{subsubsec:JC2_subquery_ds}, 
    we can answer the following three queries:
    \begin{enumerate}[label=\textbf{(\roman*)}]
    \item \label{enum:TC2_queries:1}
    for a given integer $s \in [1, m]$, 
    return a triplet of an integer $h \in [0, H]$, an interval $[\alpha, \beta] \subseteq [1, n]$ in input string $T$, 
    and a non-negative integer $M \in \mathbb{N}_{0}$ 
    satisfying $(h, T[\alpha..\beta], M) = (h_{s}, C_{s}, M_{s})$ in $O(H^{2} + \log n)$ time; 
    \item \label{enum:TC2_queries:2}
    consider a given 4-tuple of an integer $s \in [1, m]$, an integer $h \in [0, H]$, an interval $[\alpha, \beta] \subseteq [1, n]$ in input string $T$, and a non-negative integer $M \in \mathbb{N}_{0}$. 
    Then, verify whether the $s$-th triplet of the ordered set $\mathcal{T}_{C^{\prime}}$ satisfies one of 
    the following three conditions: 
    (A) $h_{s} < h$; (B) $h_{s} = h$ and $C_{s} \prec T[\alpha..\beta]$; (C) $h_{s} = h$, $C_{s} = T[\alpha..\beta]$, and $M_{s} \leq M$. 
    This verification takes $O(H^{2} + \log n)$ time;
    \item \label{enum:TC2_queries:3}
    verify whether $(h, T[\alpha..\beta], M) \in \mathcal{T}_{C}$ or not in $O(H^{2} \log n + \log^{2} n)$ time  
    for a given triplet of an integer $h \in [0, H]$, an interval $[\alpha, \beta] \subseteq [1, n]$ in input string $T$, 
    and a non-negative integer $M \in \mathbb{N}_{0}$. 
    If $(h, T[\alpha..\beta], M) \in \mathcal{T}_{C^{\prime}}$ holds, 
    then return an integer $s \in [1, m]$ satisfying 
    $(h_{s}, C_{s}, M_{s}) = (h, T[\alpha..\beta], M)$ in the same time. 
    \end{enumerate}
\end{lemma}
\begin{proof}
    This lemma can be proved using the same approach as for Lemma~\ref{lem:TC2_queries}. 
\end{proof}

\paragraph{Algorithm.}
The algorithm for $\RSCQCY(i, j)$ computes $|\Psi_{\CCP}(T[i..j]) \cap \Psi_{\lex}(T[\gamma_{Q}..r_{Q}+1]) \cap \Psi_{\run} \cap \Psi_{\centerset}(C_{Q}) \cap \Psi_{\lcp}(K_{Q}) \cap \Psi_{\succeeding}|$ 
under the condition that 
RSC query $\RSCQ(i, j)$ satisfies either condition (C) or (D) of RSC query stated in Section~\ref{subsec:rsc_sub}. 
This algorithm leverages Lemma~\ref{lem:JC2_main_lemma}, which shows that 
the size of set $|\Psi_{\CCP}(T[i..j]) \cap \Psi_{\lex}(T[\gamma_{Q}..r_{Q}+1]) \cap \Psi_{\run} \cap \Psi_{\centerset}(C_{Q}) \cap \Psi_{\lcp}(K_{Q}) \cap \Psi_{\succeeding}|$ can be computed 
by one range-sum query on set $\mathcal{J}_{C^{\prime}}(h_{Q}, C_{Q}, M_{Q})$ of weighted points. 
Subqeury $\RSCQCY(i, j)$ can be answered in the same time (i.e., $O(H^{2} \log n + \log^{4} n)$ time) 
by modifying the algorithm for subquery $\RSCQCY(i, j)$ explained in Section~\ref{subsubsec:JC2_subquery_ds} 
because Lemma~\ref{lem:JC2_main_lemma} corresponds to Lemma~\ref{lem:JC1_main_lemma}. 
Therefore, Lemma~\ref{lem:RSC_subquery_C2_summary} can be proved using the same approach as for Lemma~\ref{lem:RSC_subquery_C1_summary}.

\subsection{Subquery \texorpdfstring{$\RSCQDX(i, j)$}{RSCD1(i, j)}}\label{subsec:RSC_comp_D1}
The goal of this subsection is to answer subquery $\RSCQDX(i, j)$. 
The following lemma states the summary of this subsection. 

\begin{lemma}\label{lem:RSC_subquery_D1_summary}
Using a dynamic data structure of $O((|\Psi_{\samp}| + |\mathcal{U}_{\RR}|) B)$ bits of space for machine word size $B$, 
we can answer $\RSCQDX(i, j)$ (i.e., computing $|(\Psi_{\CCP}(T[i..j]) \cap \Psi_{\lex}(T[\gamma_{Q}..r_{Q}+1]) \cap \Psi_{\run} \cap \Psi_{\centerset}(C_{Q}) \cap \Psi_{\preceding}) \setminus \Psi_{\lcp}(K_{Q})|$) in $O(H^{2} \log n + \log^{4} n)$ time if the given RSC query $\RSCQ(i, j)$ satisfies condition (D). 
Here, $|\mathcal{U}_{\RR}|$ is the number of nodes in the RR-DAG of RLSLP $\mathcal{G}^{R}$.
\end{lemma}
\begin{proof}
See Section~\ref{subsubsec:JD1_subquery_ds}.
\end{proof}

For answering $\RSCQDX(i, j)$., 
we leverage range-sum query on weighted points corresponding to the interval attractors in 
set $\Psi_{h_{Q}} \cap \Psi_{\source} \cap \Psi_{\centerset}(C_{Q}) \cap \Psi_{\preceding} \cap \Psi_{\samp}$. 
For this purpose, 
we introduce two sets $\mathcal{J}_{D}(h, C)$ and $\mathcal{J}_{E}(h, C)$ of weighted points 
on grid $([1, n], [-1, |C|])$ for a pair of an integer $h \in [0, H]$ and a string $C \in \Sigma^{+}$. 
Here, the two sets $\mathcal{J}_{D}(h, C)$ and $\mathcal{J}_{E}(h, C)$ are defined using 
set $\Psi_{h} \cap \Psi_{\source} \cap \Psi_{\centerset}(C) \cap \Psi_{\preceding} \cap \Psi_{\samp}$ of $k$ interval attractors 
$([p_{1}, q_{1}], [\ell_{1}, r_{1}]), ([p_{2}, q_{2}], [\ell_{2}, r_{2}])$, 
$\ldots$, $([p_{k}, q_{k}], [\ell_{k}, r_{k}])$. 
For each integer $s \in [1, k]$, 
let $\gamma_{s}$ of the attractor position of the interval attractor $([p_{s}, q_{s}], [\ell_{s}, r_{s}])$; 
let $g_{s} = |f_{\recover}(([p_{s}, q_{s}], [\ell_{s}, r_{s}]))|$, 
$K_{s} = |\lcp(T[\gamma_{s}..r_{s}], C^{n+1})|$, and $M_{s} = (K_{s} - (2 + \sum_{w = 1}^{h+3} \lfloor \mu(w) \rfloor) ) \mod |C|$. 
Here, $([p_{s}, q_{s}], [\ell_{s}, r_{s}]) \in \Psi_{\lcp}(K_{s})$ 
and $([p_{s}, q_{s}], [\ell_{s}, r_{s}]) \in \Psi_{\modulo}(M_{s})$ follow from 
the definitions of the two subsets $\Psi_{\lcp}(K_{s})$ and $\Psi_{\modulo}(M_{s})$, respectively. 

\paragraph{Set $\mathcal{J}_{D}(h, C)$ of Weighted Points.}
Set $\mathcal{J}_{D}(h, C)$ consists of $k$ weighted points 
$(g_{1}, M_{1}$, $|\Psi_{\str}(T[p_{1}-1..r_{1}+1])|, T[p_{1}-1..r_{1}+1]))$, 
$(g_{2}, M_{2}, |\Psi_{\str}(T[p_{2}-1..r_{2}+1])|, T[p_{2}-1..r_{2}+1]))$, $\ldots$, 
$(g_{k}, M_{k}, |\Psi_{\str}(T[p_{k}-1..r_{k}+1])|, T[p_{k}-1..r_{k}+1]))$. 
Each weighted point $(g_{s}, M_{s}, |\Psi_{\str}(T[p_{s}-1..r_{s}+1])|, T[p_{s}-1..r_{s}+1]))$ corresponds to interval attractor $([p_{s}, q_{s}], [\ell_{s}, r_{s}])$ in set $\Psi_{h} \cap \Psi_{\source} \cap \Psi_{\centerset}(C) \cap \Psi_{\preceding} \cap \Psi_{\samp}$. 
The details of the four elements of the weighted point $(g_{s}, M_{s}, |\Psi_{\str}(T[p_{s}-1..r_{s}+1])|, T[p_{s}-1..r_{s}+1]))$ are as follows:
\begin{itemize}
    \item the first integer $g_{s}$ is the x-coordinate of this weighted point; 
    \item the second integer $M_{s}$ is the y-coordinate of this weighted point; 
    \item the third integer $|\Psi_{\str}(T[p_{s}-1..r_{s}+1])|$ is the weight of this weighted point;
    \item the fourth string $T[p_{s}-1..r_{s}+1]$ is the identifier of this weighted point.
\end{itemize}

From the definition of the sampling subset $\Psi_{\samp}$, 
the identifiers $T[p_{1}-1..r_{1}+1], T[p_{2}-1..r_{2}+1], \ldots, T[p_{k}-1..r_{k}+1]$ of all the weighted points in 
the set $\mathcal{J}_{D}(h, C)$ are different. 

\paragraph{Set $\mathcal{J}_{E}(h, C)$ of Weighted Points.}
Set $\mathcal{J}_{E}(h, C)$ consists of $k$ weighted points 
$(g_{1}, M_{1}, g_{1}|\Psi_{\str}(T[p_{1}-1..r_{1}+1])|, T[p_{1}-1..r_{1}+1]))$, 
$(g_{2}, M_{2}, g_{2}|\Psi_{\str}(T[p_{2}-1..r_{2}+1])|, T[p_{2}-1..r_{2}+1]))$, $\ldots$, 
$(g_{k}, M_{k}$, $g_{k}|\Psi_{\str}(T[p_{k}-1..r_{k}+1])|, T[p_{k}-1..r_{k}+1]))$. 
Similar to the set $\mathcal{J}_{D}(h, C)$, 
each weighted point $(g_{s}, M_{s}, g_{s}|\Psi_{\str}(T[p_{s}-1..r_{s}+1])|, T[p_{s}-1..r_{s}+1]))$ corresponds to interval attractor $([p_{s}, q_{s}], [\ell_{s}, r_{s}])$ in set $\Psi_{h} \cap \Psi_{\source} \cap \Psi_{\centerset}(C) \cap \Psi_{\preceding} \cap \Psi_{\samp}$. 
From the definition of the sampling subset $\Psi_{\samp}$, 
the identifiers $T[p_{1}-1..r_{1}+1], T[p_{2}-1..r_{2}+1], \ldots, T[p_{k}-1..r_{k}+1]$ of all the weighted points in 
the set $\mathcal{J}_{E}(h, C)$ are different.

The following lemma shows that 
we can count the interval attractors in five subsets of set $\Psi_{\RR}$ by range-sum query on 
the two sets $\mathcal{J}_{D}(h, C)$ and $\mathcal{J}_{E}(h, C)$. 

\begin{lemma}\label{lem:JD1_sum}
Consider a triplet of an integer $h \in [0, H]$, a string $C \in \Sigma^{+}$, 
and an integer $K \geq 0$. 
Let $b$ and $M$ be the two integers defined as follows: 
\begin{itemize}
    \item $b = 1 + \lfloor \frac{K - (2 + \sum_{w = 1}^{h+3} \lfloor \mu(w) \rfloor)}{|C|} \rfloor$ if $K > 1 + \sum_{w = 1}^{h+3} \lfloor \mu(w) \rfloor$. Otherwise, let $b = 1$;
    \item $M = (K - (2 + \sum_{w = 1}^{h+3} \lfloor \mu(w) \rfloor) ) \mod |C|$ if $K > 1 + \sum_{w = 1}^{h+3} \lfloor \mu(w) \rfloor$. 
    Otherwise, let $M = 0$. 
\end{itemize}
Then, the following four equations hold:
\begin{equation}\label{eq:JD1_sum:1}
    \begin{split}
    |\Psi_{h} \cap \Psi_{\run} \cap \Psi_{\centerset}(C) \cap \Psi_{\preceding}| = \rangesum(\mathcal{J}_{E}(h, C), 1, n, 0, |C|-1).
    \end{split}
\end{equation}
\begin{equation}\label{eq:JD1_sum:2}
    \begin{split}
    |\Psi_{h} \cap \Psi_{\run} \cap \Psi_{\centerset}(C) \cap \Psi_{\preceding} & \cap  \Bigl(\bigcup_{\lambda = b}^{n} \Psi_{\nRecover}(\lambda) \Bigr) \cap \Bigl(\bigcup_{\lambda = 0}^{K - M - 1} \Psi_{\lcp}(\lambda) \Bigr)| \\
    &= (b-1) \rangesum(\mathcal{J}_{D}(h, C), b, n, 0, |C| - 1);
    \end{split}
\end{equation}
\begin{equation}\label{eq:JD1_sum:3}
    \begin{split}
    |\Psi_{h} \cap \Psi_{\run} \cap \Psi_{\centerset}(C) \cap \Psi_{\preceding} & \cap \Bigl(\bigcup_{\lambda = b}^{n} \Psi_{\nRecover}(\lambda) \Bigr) \cap \Bigl(\bigcup_{\lambda = K-M}^{K - 1} \Psi_{\lcp}(\lambda) \Bigr)| \\ 
    &= \rangesum(\mathcal{J}_{D}(h, C), b, n, 0, M - 1);
    \end{split}
\end{equation}
\begin{equation}\label{eq:JD1_sum:4}
    \begin{split}
    |\Psi_{h} \cap \Psi_{\run} \cap \Psi_{\centerset}(C) \cap \Psi_{\preceding} & \cap \Bigl(\bigcup_{\lambda = 1}^{b-1} \Psi_{\nRecover}(\lambda) \Bigr) \cap \Bigl(\bigcup_{\lambda = 0}^{K - 1} \Psi_{\lcp}(\lambda) \Bigr)| \\
    &= \rangesum(\mathcal{J}_{E}(h, C), 1, b-1, 0, |C|-1).
    \end{split}
\end{equation}
If $K > 1 + \sum_{w = 1}^{h+3} \lfloor \mu(w) \rfloor$, then the following equation holds:
\begin{equation}\label{eq:JD1_sum:5}
    |\Psi_{h} \cap \Psi_{\run} \cap \Psi_{\centerset}(C) \cap \Psi_{\preceding} \cap \Psi_{\lcp}(K)| = \rangesum(\mathcal{J}_{D}(h, C), b, n, M, M).
\end{equation}
\end{lemma}
\begin{proof}
    See Section~\ref{subsubsec:JD1_sum_proof}. 
\end{proof}

The following lemma states a property of the set $(\Psi_{\CCP}(T[i..j]) \cap \Psi_{\lex}(T[\gamma_{Q}..r_{Q}+1]) \cap \Psi_{\run} \cap  \Psi_{\centerset}(C_{Q}) \cap \Psi_{\preceding}) \setminus \Psi_{\lcp}(K_{Q})$ for the integer $K_{Q}$ of Lemma~\ref{lem:RSC_subquery_D1_summary}.
 
\begin{lemma}\label{lem:JD1_division}
Consider RSC query $\RSCQ(i, j)$ satisfying condition (D) of RSC query stated in Section~\ref{subsec:rsc_sub}. 
Here, let $([p_{Q}, q_{Q}], [\ell_{Q}, r_{Q}])$ be interval attractor $I_{\capture}(i, j)$; 
let $h_{Q}$, $\gamma_{Q}$, and $C_{Q}$ be the level, attractor position, and associated string of the interval attractor $([p_{Q}, q_{Q}], [\ell_{Q}, r_{Q}])$, respectively; 
let $K_{Q} = |\lcp(T[\gamma_{Q}..r_{Q}], C_{Q}^{n+1})|$ and $M_{Q} = (K_{Q} - (2 + \sum_{w = 1}^{h_{Q}+3} \lfloor \mu(w) \rfloor) ) \mod |C_{Q}|$; 
let $x, x^{\prime}$, and $M_{Q}^{\prime}$ be the three integers defined as follows: 
\begin{itemize}
    \item $x = 1 + \lfloor \frac{K_{Q} - (2 + \sum_{w = 1}^{h_{Q}+3} \lfloor \mu(w) \rfloor)}{|C_{Q}|} \rfloor$;
    \item $x^{\prime} = 1 + \lfloor \frac{|[\gamma_{Q}, j]| - (2 + \sum_{w = 1}^{h_{Q}+3} \lfloor \mu(w) \rfloor)}{|C_{Q}|} \rfloor$ 
    if $|[\gamma_{Q}, j]| > 1 + \sum_{w = 1}^{h_{Q}+3} \lfloor \mu(w) \rfloor$. 
    Otherwise, let $x^{\prime} = 1$; 
    \item $M_{Q}^{\prime} = (|[\gamma_{Q}, j]| - (2 + \sum_{w = 1}^{h_{Q}+3} \lfloor \mu(w) \rfloor)) \mod |C_{Q}|$ if $|[\gamma_{Q}, j]| > 1 + \sum_{w = 1}^{h_{Q}+3} \lfloor \mu(w) \rfloor$. 
    Otherwise, let $M_{Q}^{\prime} = 0$. 
\end{itemize}

If $([p_{Q}, q_{Q}], [\ell_{Q}, r_{Q}]) \in \Psi_{\preceding}$, 
then the following equation holds: 
\begin{equation}\label{eq:JD1_division:1}
    \begin{split}
& |(\Psi_{\CCP}(T[i..j]) \cap \Psi_{\lex}(T[\gamma_{Q}..r_{Q}+1]) \cap \Psi_{\run} \cap \Psi_{\centerset}(C_{Q}) \cap \Psi_{\preceding}) \setminus \Psi_{\lcp}(K_{Q})| \\
&= |\Psi_{h_{Q}} \cap \Psi_{\run} \cap \Psi_{\centerset}(C_{Q}) \cap \Psi_{\preceding} \cap \Bigl(\bigcup_{\lambda = x}^{n} \Psi_{\nRecover}(\lambda) \Bigr) \cap \Bigl(\bigcup_{\lambda = 0}^{K_{Q} - M_{Q} - 1} \Psi_{\lcp}(\lambda) \Bigr)| \\
&+ |\Psi_{h_{Q}} \cap \Psi_{\run} \cap \Psi_{\centerset}(C_{Q}) \cap \Psi_{\preceding} \cap \Bigl(\bigcup_{\lambda = x}^{n} \Psi_{\nRecover}(\lambda) \Bigr) \cap \Bigl(\bigcup_{\lambda = K_{Q} - M_{Q}}^{K_{Q}-1} \Psi_{\lcp}(\lambda) \Bigr)| \\
&+ |\Psi_{h_{Q}} \cap \Psi_{\run} \cap \Psi_{\centerset}(C_{Q}) \cap \Psi_{\preceding} \cap \Bigl(\bigcup_{\lambda = 1}^{x-1} \Psi_{\nRecover}(\lambda) \Bigr) \cap \Bigl(\bigcup_{\lambda = 0}^{K_{Q} - 1} \Psi_{\lcp}(\lambda) \Bigr)| \\
&- |\Psi_{h_{Q}} \cap \Psi_{\run} \cap \Psi_{\centerset}(C_{Q}) \cap \Psi_{\preceding} \cap \Bigl(\bigcup_{\lambda = x^{\prime}}^{n} \Psi_{\nRecover}(\lambda) \Bigr) \cap \Bigl(\bigcup_{\lambda = 0}^{|[\gamma_{Q}, j]| - M_{Q}^{\prime} - 1} \Psi_{\lcp}(\lambda) \Bigr)| \\
&- |\Psi_{h_{Q}} \cap \Psi_{\run} \cap \Psi_{\centerset}(C_{Q}) \cap \Psi_{\preceding} \cap \Bigl(\bigcup_{\lambda = x^{\prime}}^{n} \Psi_{\nRecover}(\lambda) \Bigr) \cap \Bigl(\bigcup_{\lambda = |[\gamma_{Q}, j]| - M_{Q}^{\prime}}^{|[\gamma_{Q}, j]| - 1} \Psi_{\lcp}(\lambda) \Bigr)| \\
&- |\Psi_{h_{Q}} \cap \Psi_{\run} \cap \Psi_{\centerset}(C_{Q}) \cap \Psi_{\preceding} \cap \Bigl(\bigcup_{\lambda = 1}^{x^{\prime}-1} \Psi_{\nRecover}(\lambda) \Bigr) \cap \Bigl(\bigcup_{\lambda = 0}^{|[\gamma_{Q}, j]| - 1} \Psi_{\lcp}(\lambda) \Bigr)|.
    \end{split}
\end{equation}
Otherwise (i.e., $([p_{Q}, q_{Q}], [\ell_{Q}, r_{Q}]) \not \in \Psi_{\preceding}$), 
the following equation holds: 
\begin{equation}\label{eq:JD1_division:2}
    \begin{split}
& |(\Psi_{\CCP}(T[i..j]) \cap \Psi_{\lex}(T[\gamma_{Q}..r_{Q}+1]) \cap \Psi_{\run} \cap \Psi_{\centerset}(C_{Q}) \cap \Psi_{\preceding}) \setminus \Psi_{\lcp}(K_{Q})| \\
&= |\Psi_{h_{Q}} \cap \Psi_{\run} \cap \Psi_{\centerset}(C_{Q}) \cap \Psi_{\preceding}| \\
&- |\Psi_{h_{Q}} \cap \Psi_{\run} \cap \Psi_{\centerset}(C_{Q}) \cap \Psi_{\preceding} \cap \Psi_{\lcp}(K_{Q})| \\
&- |\Psi_{h_{Q}} \cap \Psi_{\run} \cap \Psi_{\centerset}(C_{Q}) \cap \Psi_{\preceding} \cap \Bigl(\bigcup_{\lambda = x^{\prime}}^{n} \Psi_{\nRecover}(\lambda) \Bigr) \cap \Bigl(\bigcup_{\lambda = 0}^{|[\gamma_{Q}, j]| - M_{Q}^{\prime} - 1} \Psi_{\lcp}(\lambda) \Bigr)| \\
&- |\Psi_{h_{Q}} \cap \Psi_{\run} \cap \Psi_{\centerset}(C_{Q}) \cap \Psi_{\preceding} \cap \Bigl(\bigcup_{\lambda = x^{\prime}}^{n} \Psi_{\nRecover}(\lambda) \Bigr) \cap \Bigl(\bigcup_{\lambda = |[\gamma_{Q}, j]| - M_{Q}^{\prime}}^{|[\gamma_{Q}, j]| - 1} \Psi_{\lcp}(\lambda) \Bigr)| \\
&- |\Psi_{h_{Q}} \cap \Psi_{\run} \cap \Psi_{\centerset}(C_{Q}) \cap \Psi_{\preceding} \cap \Bigl(\bigcup_{\lambda = 1}^{x^{\prime}-1} \Psi_{\nRecover}(\lambda) \Bigr) \cap \Bigl(\bigcup_{\lambda = 0}^{|[\gamma_{Q}, j]| - 1} \Psi_{\lcp}(\lambda) \Bigr)|.
    \end{split}
\end{equation}
\end{lemma}
\begin{proof}
    See Section~\ref{subsubsec:JD1_division_proof}.
\end{proof}

Lemma~\ref{lem:JD1_division} shows that 
the set $(\Psi_{\CCP}(T[i..j]) \cap \Psi_{\lex}(T[\gamma_{Q}..r_{Q}+1]) \cap \Psi_{\run} \cap  \Psi_{\centerset}(C_{Q}) \cap \Psi_{\preceding}) \setminus \Psi_{\lcp}(K_{Q})$ can be divided into at most six subsets of Lemma~\ref{lem:JD1_sum}. 
The size of each subset of Lemma~\ref{lem:JD1_sum} can be computed by range-sum query on 
the two sets $\mathcal{J}_{D}(h_{Q}, C_{Q})$ and $\mathcal{J}_{E}(h_{Q}, C_{Q})$. 
Therefore, we obtain the following lemma by combining Lemma~\ref{lem:JD1_division} and Lemma~\ref{lem:JD1_sum}. 

\begin{lemma}\label{lem:JD1_main_lemma}
Consider RSC query $\RSCQ(i, j)$ satisfying condition (D) of RSC query stated in Section~\ref{subsec:rsc_sub}. 
Here, let $([p_{Q}, q_{Q}], [\ell_{Q}, r_{Q}])$ be interval attractor $I_{\capture}(i, j)$; 
let $h_{Q}$, $\gamma_{Q}$, and $C_{Q}$ be the level, attractor position, and associated string of the interval attractor $([p_{Q}, q_{Q}], [\ell_{Q}, r_{Q}])$, respectively; 
let $K_{Q}, M_{Q}, x, x^{\prime}$ and $M_{Q}^{\prime}$ be the five integers defined in Lemma~\ref{lem:JD1_division}. 
If $([p_{Q}, q_{Q}], [\ell_{Q}, r_{Q}]) \in \Psi_{\preceding}$ holds, 
then the following equation holds: 
\begin{equation}\label{eq:JD1_main_lemma:1}
    \begin{split}
    |(\Psi_{\CCP}(T[i..j]) \cap \Psi_{\lex}(T[\gamma_{Q}..r_{Q}+1]) & \cap \Psi_{\run} \cap \Psi_{\centerset}(C_{Q}) \cap \Psi_{\preceding}) \setminus \Psi_{\lcp}(K_{Q})| \\
    &= (x-1)(\rangesum(\mathcal{J}_{D}(h_{Q}, C_{Q}), x, n, -1, |C_{Q}|)) \\ 
    &+ \rangesum(\mathcal{J}_{E}(h_{Q}, C_{Q}), 1, x-1, 0, |C_{Q}| - 1) \\
    &+ \rangesum(\mathcal{J}_{D}(h_{Q}, C_{Q}), x, n, 0, M_{Q}-1) \\
    &-(x^{\prime}-1)(\rangesum(\mathcal{J}_{D}(h_{Q}, C_{Q}), x^{\prime}, n, 0, |C_{Q}| - 1)) \\
    &- \rangesum(\mathcal{J}_{E}(h_{Q}, C_{Q}), 1, x^{\prime}-1, 0, |C_{Q}| - 1) \\
    &- \rangesum(\mathcal{J}_{D}(h_{Q}, C_{Q}), x^{\prime}, n, 0, M_{Q}^{\prime} - 1).
    \end{split}
\end{equation}
Otherwise~(i.e., $([p_{Q}, q_{Q}], [\ell_{Q}, r_{Q}]) \not \in \Psi_{\preceding}$), 
the following equation holds: 
\begin{equation}\label{eq:JD1_main_lemma:2}
    \begin{split}
    |(\Psi_{\CCP}(T[i..j]) \cap \Psi_{\lex}(T[\gamma_{Q}..r_{Q}+1]) & \cap \Psi_{\run} \cap \Psi_{\centerset}(C_{Q}) \cap \Psi_{\preceding}) \setminus \Psi_{\lcp}(K_{Q})| \\
    &= \rangesum(\mathcal{J}_{E}(h_{Q}, C_{Q}), 1, n, 0, |C_{Q}|) \\ 
    &- \rangesum(\mathcal{J}_{D}(h_{Q}, C_{Q}), x, n, M_{Q}, M_{Q}) \\
    &- (x^{\prime}-1)(\rangesum(\mathcal{J}_{D}(h_{Q}, C_{Q}), x^{\prime}, n, 0, |C_{Q}| - 1)) \\ 
    &- \rangesum(\mathcal{J}_{E}(h_{Q}, C_{Q}), 1, x^{\prime}-1, 0, |C_{Q}| - 1) \\
    &- \rangesum(\mathcal{J}_{D}(h_{Q}, C_{Q}), x^{\prime}, n, 0, M_{Q}^{\prime} - 1).
    \end{split}
\end{equation}
\end{lemma}
\begin{proof}
See Section~\ref{subsubsec:JD1_main_lemma_proof}.
\end{proof}

%Lemma~\ref{lem:JD1_main_lemma} shows that 
%the size of the set $(\Psi_{\CCP}(T[i..j]) \cap \Psi_{\lex}(T[\gamma_{Q}..r_{Q}+1]) \cap \Psi_{\run} \cap  \Psi_{\centerset}(C_{Q}) \cap \Psi_{\preceding}) \setminus \Psi_{\lcp}(K_{Q})$ can be computed by 
%at most six range-sum queries on the two sets $\mathcal{J}_{D}(h_{Q}, C_{Q})$ and $\mathcal{J}_{E}(h_{Q}, C_{Q})$ of weighted points. 
%Therefore, we can compute 

\subsubsection{Proof of Lemma~\ref{lem:JD1_sum}}\label{subsubsec:JD1_sum_proof}

For this proof, 
we introduce five sets 
$\mathcal{I}_{D, 1}$, $\mathcal{I}_{D, 2}$, $\mathcal{I}_{D, 3}$, $\mathcal{I}_{E, 1}$, and $\mathcal{I}_{E, 2}$ 
of integers in set $\{ 1, 2, \ldots, k \}$. 
These sets are defined as follows: 
\begin{itemize}
    \item $\mathcal{I}_{D, 1} = \{ s \in [1, k] \mid b \leq g_{s} \leq n \text{ and } 0 \leq M_{s} \leq |C|-1 \}$; 
    \item $\mathcal{I}_{D, 2} = \{ s \in [1, k] \mid b \leq g_{s} \leq n \text{ and } 0 \leq M_{s} \leq M-1 \}$; 
    \item $\mathcal{I}_{D, 3} = \{ s \in [1, k] \mid b \leq g_{s} \leq n \text{ and } M_{s} = M \}$;
    \item $\mathcal{I}_{E, 1} = \{ s \in [1, k] \mid 1 \leq g_{s} \leq n \text{ and } 0 \leq M_{s} \leq |C|-1 \}$; 
    \item $\mathcal{I}_{E, 2} = \{ s \in [1, k] \mid 1 \leq g_{s} \leq b-1 \text{ and } 0 \leq M_{s} \leq |C|-1 \}$. 
\end{itemize}

\begin{proposition}\label{prop:JD_sum_set_subseteq}
The following five equations hold:

\begin{equation}\label{eq:JD_sum_set_subseteq:E1}
    \Psi_{h} \cap \Psi_{\run} \cap \Psi_{\centerset}(C) \cap \Psi_{\preceding} \subseteq \bigcup_{s \in \mathcal{I}_{E, 1}} \Psi_{\sRecover}(([p_{s}, q_{s}], [\ell_{s}, r_{s}]));
\end{equation}
\begin{equation}\label{eq:JD_sum_set_subseteq:D1}
    \begin{split}
    \Psi_{h} \cap \Psi_{\run} \cap \Psi_{\centerset}(C) \cap \Psi_{\preceding} & \cap  (\bigcup_{\lambda = b}^{n} \Psi_{\nRecover}(\lambda)) \cap (\bigcup_{\lambda = 0}^{K - M - 1} \Psi_{\lcp}(\lambda)) \\ 
    &\subseteq \bigcup_{s \in \mathcal{I}_{D, 1}} \bigcup_{\lambda = 0}^{K - M - 1} (\Psi_{\sRecover}(([p_{s}, q_{s}], [\ell_{s}, r_{s}])) \cap \Psi_{\lcp}(\lambda));
    \end{split}
\end{equation}
\begin{equation}\label{eq:JD_sum_set_subseteq:D2}
    \begin{split}
    \Psi_{h} \cap \Psi_{\run} \cap \Psi_{\centerset}(C) \cap \Psi_{\preceding} & \cap (\bigcup_{\lambda = b}^{n} \Psi_{\nRecover}(\lambda)) \cap (\bigcup_{\lambda = K-M}^{K - 1} \Psi_{\lcp}(\lambda)) \\ 
    &\subseteq \bigcup_{s \in \mathcal{I}_{D, 2}} \bigcup_{\lambda = K-M}^{K - 1} (\Psi_{\sRecover}(([p_{s}, q_{s}], [\ell_{s}, r_{s}])) \cap \Psi_{\lcp}(\lambda));
    \end{split}
\end{equation}
\begin{equation}\label{eq:JD_sum_set_subseteq:E2}
    \begin{split}
    \Psi_{h} \cap \Psi_{\run} \cap \Psi_{\centerset}(C) \cap \Psi_{\preceding} & \cap (\bigcup_{\lambda = 1}^{b-1} \Psi_{\nRecover}(\lambda)) \cap (\bigcup_{\lambda = 0}^{K - 1} \Psi_{\lcp}(\lambda)) \\ 
    &\subseteq \bigcup_{s \in \mathcal{I}_{E, 2}} \Psi_{\sRecover}(([p_{s}, q_{s}], [\ell_{s}, r_{s}]));
    \end{split}
\end{equation}
\begin{equation}\label{eq:JD_sum_set_subseteq:D3}
    \begin{split}
    \Psi_{h} \cap \Psi_{\run} \cap \Psi_{\centerset}(C) \cap \Psi_{\preceding} & \cap \Psi_{\lcp}(K) \\ 
    &\subseteq \bigcup_{s \in \mathcal{I}_{D, 3}} (\Psi_{\sRecover}(([p_{s}, q_{s}], [\ell_{s}, r_{s}])) \cap \Psi_{\lcp}(K)).
    \end{split}
\end{equation}
\end{proposition}
\begin{proof}
The following four statements are used to prove Proposition~\ref{prop:JD_sum_set_subseteq}: 
\begin{enumerate}[label=\textbf{(\roman*)}]
    \item consider an interval attractor $([p_{A}, q_{A}], [\ell_{A}, r_{A}])$ in set $\Psi_{h} \cap \Psi_{\run} \cap \Psi_{\centerset}(C) \cap \Psi_{\preceding}$. 
    Then, the subset $\Psi_{\source}$ contains an interval attractor $([p_{B}, q_{B}], [\ell_{B}, r_{B}])$ satisfying 
    $([p_{A}, q_{A}], [\ell_{A}, r_{A}]) \in f_{\recover}(([p_{B}, q_{B}], [\ell_{B}, r_{B}]))$ 
    and $([p_{B}, q_{B}], [\ell_{B}, r_{B}]) \in \Psi_{h} \cap \Psi_{\centerset}(C) \cap \Psi_{\preceding}$; 
    \item for the two interval attractors $([p_{A}, q_{A}], [\ell_{A}, r_{A}])$ and $([p_{B}, q_{B}], [\ell_{B}, r_{B}])$ of statement (i), 
    there exists an integer $\tau \in [1, k]$ satisfying 
    $([p_{A}, q_{A}], [\ell_{A}, r_{A}]) \in \Psi_{\sRecover}(([p_{\tau}, q_{\tau}], [\ell_{\tau}, r_{\tau}]))$ 
    and $T[p_{\tau}-1..r_{\tau}+1] = T[p_{B}-1..r_{B}+1]$; 
    \item for the interval attractor $([p_{A}, q_{A}], [\ell_{A}, r_{A}])$ of statement (i), 
    if there exist two integers $1 \leq x_{1} \leq x_{2} \leq n$ satisfying 
    $([p_{A}, q_{A}], [\ell_{A}, r_{A}]) \in \bigcup_{\lambda = x_{1}}^{x_{2}} \Psi_{\nRecover}(\lambda)$, 
    then $x_{1} \leq g_{\tau} \leq x_{2}$ holds for the integer $\tau$ of statement (ii);
    \item consider an integer $K_{A} \geq 0$ satisfying $([p_{A}, q_{A}], [\ell_{A}, r_{A}]) \in \Psi_{\lcp}(K_{A})$ for the interval attractor $([p_{A}, q_{A}], [\ell_{A}, r_{A}])$ of statement (i). 
    Let $M_{A} = (K_{A} - (2 + \sum_{w = 1}^{h+3} \lfloor \mu(w) \rfloor) ) \mod |C|$, 
    then 
    $K_{A} > 1 + \sum_{w = 1}^{h+3} \lfloor \mu(w) \rfloor$ 
    and $M_{\tau} = M_{A}$ for the integer $\tau$ of statement (ii). 
\end{enumerate}

\textbf{Proof of statement (i).}
Lemma~\ref{lem:recover_division_property}~\ref{enum:recover_division_property:1} shows that 
the subset $\Psi_{\source}$ contains an interval attractor $([p_{B}, q_{B}], [\ell_{B}, r_{B}])$ satisfying 
$([p_{A}, q_{A}], [\ell_{A}, r_{A}]) \in f_{\recover}(([p_{B}, q_{B}], [\ell_{B}, r_{B}]))$. 
Because of $([p_{A}, q_{A}], [\ell_{A}, r_{A}]) \in \Psi_{h} \cap \Psi_{\centerset}(C)$, 
Lemma~\ref{lem:recover_basic_property}~\ref{enum:recover_basic_property:4} shows that 
$([p_{B}, q_{B}], [\ell_{B}, r_{B}]) \in \Psi_{h} \cap \Psi_{\centerset}(C)$ holds. 
Because of $([p_{A}, q_{A}], [\ell_{A}, r_{A}]) \in \Psi_{\preceding}$, 
Lemma~\ref{lem:recover_basic_property}~\ref{enum:recover_basic_property:5} shows that $([p_{B}, q_{B}], [\ell_{B}, r_{B}]) \in \Psi_{\preceding}$. 
Therefore, $([p_{B}, q_{B}], [\ell_{B}, r_{B}]) \in \Psi_{h} \cap \Psi_{\centerset}(C) \cap \Psi_{\preceding}$ holds. 

\textbf{Proof of statement (ii).}
$([p_{B}, q_{B}], [\ell_{B}, r_{B}]) \not \in \Psi_{\run}$ follows from 
$([p_{B}, q_{B}], [\ell_{B}, r_{B}]) \in \Psi_{\source}$ and $\Psi_{\source} \cap \Psi_{\run} = \emptyset$. 
Because of $([p_{B}, q_{B}], [\ell_{B}, r_{B}]) \not \in \Psi_{\run}$, 
Lemma~\ref{lem:samp_basic_property}~\ref{enum:samp_basic_property:2} shows that 
the sampling subset $\Psi_{\samp}$ contains an interval attractor $([p_{C}, q_{C}], [\ell_{C}, r_{C}])$ satisfying 
$T[p_{B}-1..r_{B}+1] = T[p_{C}-1..r_{C}+1]$. 
Because of $([p_{B}, q_{B}], [\ell_{B}, r_{B}]) \in \Psi_{h} \cap \Psi_{\source} \cap \Psi_{\centerset}(C) \cap \Psi_{\preceding}$, 
Lemma~\ref{lem:psi_equality_basic_property} shows that 
$([p_{C}, q_{C}], [\ell_{C}, r_{C}]) \in \Psi_{h} \cap \Psi_{\source} \cap \Psi_{\centerset}(C) \cap \Psi_{\preceding}$ holds. 
Because of $([p_{C}, q_{C}], [\ell_{C}, r_{C}]) \in \Psi_{h} \cap \Psi_{\source} \cap \Psi_{\centerset}(C) \cap \Psi_{\preceding} \cap \Psi_{\samp}$, 
there exists an integer $\tau \in [1, k]$ satisfying $([p_{\tau}, q_{\tau}], [\ell_{\tau}, r_{\tau}]) = ([p_{C}, q_{C}], [\ell_{C}, r_{C}])$. 
$([p_{A}, q_{A}], [\ell_{A}, r_{A}]) \in \Psi_{\sRecover}(([p_{\tau}, q_{\tau}], [\ell_{\tau}, r_{\tau}]))$ 
follows from the definition of the subset $\Psi_{\sRecover}(([p_{\tau}, q_{\tau}], [\ell_{\tau}, r_{\tau}]))$. 
Therefore, statement (ii) holds. 

\textbf{Proof of statement (iii).}
Let $g_{B} = |f_{\recover}(([p_{B}, q_{B}], [\ell_{B}, r_{B}]))|$ for 
the interval attractor $([p_{B}, q_{B}], [\ell_{B}, r_{B}])$ of statement (i). 
Then, $([p_{A}, q_{A}], [\ell_{A}, r_{A}]) \in \Psi_{\nRecover}(g_{B})$ follows from the definition of the subset $\Psi_{\nRecover}(g_{B})$. 
$x_{1} \leq g_{B} \leq x_{2}$ follows from 
$([p_{A}, q_{A}], [\ell_{A}, r_{A}]) \in \Psi_{\nRecover}(g_{B})$ and 
$([p_{A}, q_{A}], [\ell_{A}, r_{A}]) \in \bigcup_{\lambda = x_{1}}^{x_{2}} \Psi_{\nRecover}(\lambda)$. 
For the integer $\tau$ of statement (ii), 
$|f_{\recover}(([p_{B}, q_{B}]$, $[\ell_{B}, r_{B}]))| = |f_{\recover}(([p_{\tau}, q_{\tau}], [\ell_{\tau}, r_{\tau}]))|$ 
follows from Lemma~\ref{lem:recover_super_property}~\ref{enum:recover_basic_property:1}. 
Therefore, 
$x_{1} \leq g_{\tau} \leq x_{2}$ follows from $x_{1} \leq g_{B} \leq x_{2}$, 
$g_{B} = |f_{\recover}(([p_{B}, q_{B}], [\ell_{B}, r_{B}]))|$, 
and $g_{\tau} = |f_{\recover}(([p_{\tau}, q_{\tau}], [\ell_{\tau}, r_{\tau}]))|$. 

\textbf{Proof of statement (iv).}
Because of $([p_{A}, q_{A}], [\ell_{A}, r_{A}]) \in \Psi_{\lcp}(K_{A}) \cap \Psi_{\centerset}(C)$, 
$|\lcp(T[\gamma_{A}..r_{A}]$, $C^{n+1})| = K_{A}$ follows from the definition of the subset $\Psi_{\lcp}(K_{A})$ 
for the attractor position $\gamma_{A}$ of the interval attractor $([p_{A}, q_{A}], [\ell_{A}, r_{A}])$. 
Because of $([p_{A}, q_{A}], [\ell_{A}, r_{A}]) \in \Psi_{h} \cap \Psi_{\run} \cap \Psi_{\centerset}(C)$, 
$|\lcp(T[\gamma_{A}..r_{A}], C^{n+1})| > 1 + \sum_{w = 1}^{h+3} \lfloor \mu(w) \rfloor$ follows from the definition of the subset $\Psi_{\run}$. 
$([p_{A}, q_{A}]$, $[\ell_{A}, r_{A}]) \in \Psi_{\modulo}(M_{A})$ holds 
because $K_{A} > 1 + \sum_{w = 1}^{h+3} \lfloor \mu(w) \rfloor$ 
and $([p_{A}, q_{A}], [\ell_{A}, r_{A}]) \in \Psi_{h} \cap \Psi_{\run} \cap \Psi_{\centerset}(C)$. 
Because of $([p_{A}, q_{A}], [\ell_{A}, r_{A}]) \in \Psi_{\modulo}(M_{A})$, 
Lemma~\ref{lem:recover_basic_property}~\ref{enum:recover_basic_property:4} shows that 
$([p_{B}, q_{B}]$, $[\ell_{B}, r_{B}]) \in \Psi_{\modulo}(M_{A})$ for the interval attractor $([p_{B}, q_{B}]$, $[\ell_{B}, r_{B}])$ of statement (i). 
For the integer $\tau$ of statement (ii), 
Lemma~\ref{lem:psi_equality_basic_property}~\ref{enum:psi_equality_basic_property:10} shows that 
$([p_{\tau}, q_{\tau}], [\ell_{\tau}, r_{\tau}]) \in \Psi_{\modulo}(M_{A})$ holds. 
On the other hand, $([p_{\tau}, q_{\tau}], [\ell_{\tau}, r_{\tau}]) \in \Psi_{\modulo}(M_{s})$ holds. 
Therefore, $M_{s} = M_{A}$ follows from $([p_{\tau}, q_{\tau}], [\ell_{\tau}, r_{\tau}]) \in \Psi_{\modulo}(M_{A})$ 
and $([p_{\tau}, q_{\tau}], [\ell_{\tau}, r_{\tau}]) \in \Psi_{\modulo}(M_{s})$. 

\textbf{Proof of Equation~\ref{eq:JD_sum_set_subseteq:E1}.}
Consider an interval attractor $([p_{A}, q_{A}], [\ell_{A}, r_{A}])$ in set $\Psi_{h} \cap \Psi_{\run} \cap \Psi_{\centerset}(C) \cap \Psi_{\preceding}$. 
Then, statement (ii) shows that 
there exists an integer $\tau \in [1, k]$ satisfying 
$([p_{A}, q_{A}], [\ell_{A}, r_{A}]) \in \Psi_{\sRecover}(([p_{\tau}, q_{\tau}], [\ell_{\tau}, r_{\tau}]))$. 

We prove $([p_{A}, q_{A}], [\ell_{A}, r_{A}]) \in \bigcup_{s \in \mathcal{I}_{E, 1}} \Psi_{\sRecover}(([p_{s}, q_{s}], [\ell_{s}, r_{s}]))$.  
$1 \leq g_{\tau} \leq n$ because 
$g_{\tau} = |f_{\recover}([p_{\tau}, q_{\tau}], [\ell_{\tau}, r_{\tau}])|$ holds, 
and $1 \leq |f_{\recover}([p_{\tau}, q_{\tau}], [\ell_{\tau}, r_{\tau}])| \leq n$ follows from Lemma~\ref{lem:recover_basic_property}~\ref{enum:recover_basic_property:1}. 
$0 \leq M_{\tau} \leq |C| - 1$ follows from $M_{\tau} = (K_{\tau} - (2 + \sum_{w = 1}^{h+3} \lfloor \mu(w) \rfloor) ) \mod |C|$. 
$\tau \in \mathcal{I}_{E, 1}$ follows from $1 \leq g_{\tau} \leq n$ and $0 \leq M_{\tau} \leq |C| - 1$. 
Therefore, $([p_{A}, q_{A}], [\ell_{A}, r_{A}]) \in \bigcup_{s \in \mathcal{I}_{E, 1}} \Psi_{\sRecover}(([p_{s}, q_{s}], [\ell_{s}, r_{s}]))$ follows from $([p_{A}, q_{A}], [\ell_{A}, r_{A}]) \in \Psi_{\sRecover}(([p_{\tau}, q_{\tau}], [\ell_{\tau}, r_{\tau}]))$ and $\tau \in \mathcal{I}_{E, 1}$. 

We showed that $([p_{A}, q_{A}], [\ell_{A}, r_{A}]) \in \bigcup_{s \in \mathcal{I}_{E, 1}} \Psi_{\sRecover}(([p_{s}, q_{s}], [\ell_{s}, r_{s}]))$ holds for each interval attractor $([p_{A}, q_{A}], [\ell_{A}, r_{A}])$ in set $\Psi_{h} \cap \Psi_{\run} \cap \Psi_{\centerset}(C) \cap \Psi_{\preceding}$. 
Therefore, Equation~\ref{eq:JD_sum_set_subseteq:E1} holds. 

\textbf{Proof of Equation~\ref{eq:JD_sum_set_subseteq:D1}.}
Consider an interval attractor $([p_{A}, q_{A}], [\ell_{A}, r_{A}])$ in set $\Psi_{h} \cap \Psi_{\run} \cap \Psi_{\centerset}(C) \cap \Psi_{\preceding} \cap  (\bigcup_{\lambda = b}^{n} \Psi_{\nRecover}(\lambda)) \cap (\bigcup_{\lambda = 0}^{K - M - 1} \Psi_{\lcp}(\lambda))$. 
Then, statement (ii) shows that 
there exists an integer $\tau \in [1, k]$ satisfying 
$([p_{A}, q_{A}], [\ell_{A}, r_{A}]) \in \Psi_{\sRecover}(([p_{\tau}, q_{\tau}], [\ell_{\tau}, r_{\tau}]))$. 

We prove $([p_{A}, q_{A}], [\ell_{A}, r_{A}]) \in \bigcup_{s \in \mathcal{I}_{D, 1}} \bigcup_{\lambda = 0}^{K - M - 1} (\Psi_{\sRecover}(([p_{s}, q_{s}], [\ell_{s}, r_{s}])) \cap \Psi_{\lcp}(\lambda))$. 
Because of $([p_{A}, q_{A}], [\ell_{A}, r_{A}]) \in \bigcup_{\lambda = b}^{n} \Psi_{\nRecover}(\lambda)$, 
$b \leq g_{\tau} \leq n$ follows from statement (iii). 
$0 \leq M_{\tau} \leq |C| - 1$ follows from $M_{\tau} = (K_{\tau} - (2 + \sum_{w = 1}^{h+3} \lfloor \mu(w) \rfloor) ) \mod |C|$. 
$\tau \in \mathcal{I}_{D, 1}$ follows from $b \leq g_{\tau} \leq n$ and $0 \leq M_{\tau} \leq |C|-1$. 
Therefore, $([p_{A}, q_{A}], [\ell_{A}, r_{A}]) \in \bigcup_{s \in \mathcal{I}_{D, 1}} \bigcup_{\lambda = 0}^{K - M - 1} (\Psi_{\sRecover}(([p_{s}, q_{s}], [\ell_{s}, r_{s}])) \cap \Psi_{\lcp}(\lambda))$ follows from 
$\tau \in \mathcal{I}_{D, 1}$ and 
$([p_{A}, q_{A}], [\ell_{A}, r_{A}]) \in \Psi_{\sRecover}(([p_{\tau}, q_{\tau}], [\ell_{\tau}, r_{\tau}])) \cap (\bigcup_{\lambda = 0}^{K - M - 1} \Psi_{\lcp}(\lambda))$. 

We showed that $([p_{A}, q_{A}], [\ell_{A}, r_{A}]) \in \bigcup_{s \in \mathcal{I}_{D, 1}} \bigcup_{\lambda = 0}^{K - M - 1} (\Psi_{\sRecover}(([p_{s}, q_{s}], [\ell_{s}, r_{s}])) \cap \Psi_{\lcp}(\lambda))$ holds for each interval attractor $([p_{A}, q_{A}], [\ell_{A}, r_{A}])$ in set $\Psi_{h} \cap \Psi_{\run} \cap \Psi_{\centerset}(C) \cap \Psi_{\preceding} \cap  (\bigcup_{\lambda = b}^{n}$ $\Psi_{\nRecover}(\lambda)) \cap (\bigcup_{\lambda = 0}^{K - M - 1} \Psi_{\lcp}(\lambda))$. 
Therefore, Equation~\ref{eq:JD_sum_set_subseteq:D1} holds. 

\textbf{Proof of Equation~\ref{eq:JD_sum_set_subseteq:D2}.}
Consider an interval attractor $([p_{A}, q_{A}], [\ell_{A}, r_{A}])$ in set $\Psi_{h} \cap \Psi_{\run} \cap \Psi_{\centerset}(C) \cap \Psi_{\preceding} \cap (\bigcup_{\lambda = b}^{n} \Psi_{\nRecover}(\lambda)) \cap (\bigcup_{\lambda = K-M}^{K - 1} \Psi_{\lcp}(\lambda))$. 
Then, statement (ii) shows that 
there exists an integer $\tau \in [1, k]$ satisfying 
$([p_{A}, q_{A}], [\ell_{A}, r_{A}]) \in \Psi_{\sRecover}(([p_{\tau}, q_{\tau}], [\ell_{\tau}, r_{\tau}]))$. 

Because of $([p_{A}, q_{A}], [\ell_{A}, r_{A}]) \in \bigcup_{\lambda = K-M}^{K - 1} \Psi_{\lcp}(\lambda)$, 
there exists an integer $K_{A} \in [K - M , K - 1]$ satisfying $([p_{A}, q_{A}], [\ell_{A}, r_{A}]) \in \Psi_{\lcp}(K_{A})$. 
Here, $M \geq 1$ follows from $K_{A} \in [K - M , K - 1]$.

Let $M_{A} = (K_{A} - (2 + \sum_{w = 1}^{h+3} \lfloor \mu(w) \rfloor) ) \mod |C|$. 
Then, we prove $0 \leq M_{A} \leq M-1$. 
$K > 1 + \sum_{w = 1}^{h+3} \lfloor \mu(w) \rfloor$ holds 
because $K > K_{A}$ holds, 
and $K_{A} > 1 + \sum_{w = 1}^{h+3} \lfloor \mu(w) \rfloor$ follows from statement (iv). 
Because of $K > 1 + \sum_{w = 1}^{h+3} \lfloor \mu(w) \rfloor$, 
the integer $M$ is defined as $M = (K - (2 + \sum_{w = 1}^{h+3} \lfloor \mu(w) \rfloor) ) \mod |C|$. 
$0 \leq (K_{A} - (2 + \sum_{w = 1}^{h+3} \lfloor \mu(w) \rfloor) ) \mod |C| \leq M-1$ holds 
because 
(A) $(K_{A} - (2 + \sum_{w = 1}^{h+3} \lfloor \mu(w) \rfloor) ) \mod |C| = ((K - (K - K_{A})) - (2 + \sum_{w = 1}^{h+3} \lfloor \mu(w) \rfloor) ) \mod |C|$, 
(B) $(K - (2 + \sum_{w = 1}^{h+3} \lfloor \mu(w) \rfloor) ) \mod |C| = M$, 
(C) $M \leq K - K_{A} \leq 1$, 
and (D) $M \geq 1$. 
Therefore, $0 \leq M_{A} \leq M-1$ follows from $M_{A} = (K_{A} - (2 + \sum_{w = 1}^{h+3} \lfloor \mu(w) \rfloor) ) \mod |C|$ 
and $0 \leq (K_{A} - (2 + \sum_{w = 1}^{h+3} \lfloor \mu(w) \rfloor) ) \mod |C| \leq M-1$. 

We prove $([p_{A}, q_{A}], [\ell_{A}, r_{A}]) \in \bigcup_{s \in \mathcal{I}_{D, 2}} \bigcup_{\lambda = K-M}^{K - 1} (\Psi_{\sRecover}(([p_{s}, q_{s}], [\ell_{s}, r_{s}])) \cap \Psi_{\lcp}(\lambda))$. 
Because of $([p_{A}, q_{A}], [\ell_{A}, r_{A}]) \in \bigcup_{\lambda = b}^{n} \Psi_{\nRecover}(\lambda)$, 
$b \leq g_{\tau} \leq n$ follows from statement (iii). 
$0 \leq M_{\tau} \leq M-1$ holds because 
$0 \leq M_{A} \leq M-1$ holds, 
and $M_{\tau} = M_{A}$ follows from statement (iv). 
$\tau \in \mathcal{I}_{D, 2}$ follows from $b \leq g_{\tau} \leq n$ and $0 \leq M_{\tau} \leq M-1$. 
Therefore, $([p_{A}, q_{A}], [\ell_{A}, r_{A}]) \in \bigcup_{s \in \mathcal{I}_{D, 2}} \bigcup_{\lambda = K-M}^{K - 1} (\Psi_{\sRecover}(([p_{s}, q_{s}], [\ell_{s}, r_{s}])) \cap \Psi_{\lcp}(\lambda))$ follows from 
$\tau \in \mathcal{I}_{D, 2}$ 
and $([p_{A}, q_{A}], [\ell_{A}, r_{A}]) \in \Psi_{\sRecover}(([p_{\tau}, q_{\tau}], [\ell_{\tau}, r_{\tau}])) \cap (\bigcup_{\lambda = K-M}^{K - 1} \Psi_{\lcp}(\lambda))$. 

We showed that $([p_{A}, q_{A}], [\ell_{A}, r_{A}]) \in \bigcup_{s \in \mathcal{I}_{D, 2}} \bigcup_{\lambda = K-M}^{K - 1} (\Psi_{\sRecover}(([p_{s}, q_{s}], [\ell_{s}, r_{s}])) \cap \Psi_{\lcp}(\lambda))$ holds for each interval attractor $([p_{A}, q_{A}], [\ell_{A}, r_{A}])$ in set $\Psi_{h} \cap \Psi_{\run} \cap \Psi_{\centerset}(C) \cap \Psi_{\preceding} \cap (\bigcup_{\lambda = b}^{n}$ $\Psi_{\nRecover}(\lambda)) \cap (\bigcup_{\lambda = K-M}^{K - 1} \Psi_{\lcp}(\lambda))$. 
Therefore, Equation~\ref{eq:JD_sum_set_subseteq:D2} holds.

\textbf{Proof of Equation~\ref{eq:JD_sum_set_subseteq:E2}.}
Consider an interval attractor $([p_{A}, q_{A}], [\ell_{A}, r_{A}])$ in set $\Psi_{h} \cap \Psi_{\run} \cap \Psi_{\centerset}(C) \cap \Psi_{\preceding} \cap (\bigcup_{\lambda = 1}^{b-1} \Psi_{\nRecover}(\lambda)) \cap (\bigcup_{\lambda = 0}^{K - 1} \Psi_{\lcp}(\lambda))$. 
Then, statement (ii) shows that 
there exists an integer $\tau \in [1, k]$ satisfying 
$([p_{A}, q_{A}], [\ell_{A}, r_{A}]) \in \Psi_{\sRecover}(([p_{\tau}, q_{\tau}], [\ell_{\tau}, r_{\tau}]))$. 

We prove $([p_{A}, q_{A}], [\ell_{A}, r_{A}]) \in \bigcup_{s \in \mathcal{I}_{E, 2}} \Psi_{\sRecover}(([p_{s}, q_{s}], [\ell_{s}, r_{s}]))$. 
Because of $([p_{A}, q_{A}], [\ell_{A}, r_{A}]) \in \bigcup_{\lambda = 1}^{b-1} \Psi_{\nRecover}(\lambda)$, 
$1 \leq g_{\tau} \leq b-1$ follows from statement (iii). 
$0 \leq M_{\tau} \leq |C| - 1$ follows from $M_{\tau} = (K_{\tau} - (2 + \sum_{w = 1}^{h+3} \lfloor \mu(w) \rfloor) ) \mod |C|$. 
$\tau \in \mathcal{I}_{E, 2}$ follows from $1 \leq g_{\tau} \leq b-1$ and $0 \leq M_{\tau} \leq |C|-1$. 
Therefore, $([p_{A}, q_{A}], [\ell_{A}, r_{A}]) \in \bigcup_{s \in \mathcal{I}_{E, 2}} \Psi_{\sRecover}(([p_{s}, q_{s}], [\ell_{s}, r_{s}]))$ follows from 
$\tau \in \mathcal{I}_{E, 2}$ and 
$([p_{A}, q_{A}], [\ell_{A}, r_{A}]) \in \Psi_{\sRecover}(([p_{\tau}, q_{\tau}], [\ell_{\tau}, r_{\tau}]))$. 

We showed that $([p_{A}, q_{A}], [\ell_{A}, r_{A}]) \in \bigcup_{s \in \mathcal{I}_{E, 2}} \Psi_{\sRecover}(([p_{s}, q_{s}], [\ell_{s}, r_{s}]))$ holds for each interval attractor $([p_{A}, q_{A}], [\ell_{A}, r_{A}])$ in set $\Psi_{h} \cap \Psi_{\run} \cap \Psi_{\centerset}(C) \cap \Psi_{\preceding} \cap (\bigcup_{\lambda = 1}^{b-1} \Psi_{\nRecover}(\lambda)) \cap (\bigcup_{\lambda = 0}^{K - 1} \Psi_{\lcp}(\lambda))$. 
Therefore, Equation~\ref{eq:JD_sum_set_subseteq:E2} holds.

\textbf{Proof of Equation~\ref{eq:JD_sum_set_subseteq:D3}.}
Consider an interval attractor $([p_{A}, q_{A}], [\ell_{A}, r_{A}])$ in set $\Psi_{h} \cap \Psi_{\run} \cap \Psi_{\centerset}(C) \cap \Psi_{\preceding} \cap \Psi_{\lcp}(K)$. 
Then, statement (ii) shows that 
there exists an integer $\tau \in [1, k]$ satisfying 
$([p_{A}, q_{A}], [\ell_{A}, r_{A}]) \in \Psi_{\sRecover}(([p_{\tau}, q_{\tau}], [\ell_{\tau}, r_{\tau}]))$. 
Because of $([p_{A}, q_{A}], [\ell_{A}, r_{A}]) \in \Psi_{\lcp}(K)$, 
$K > 1 + \sum_{w = 1}^{h+3} \lfloor \mu(w) \rfloor$ follows from statement (iii). 
Because of $K > 1 + \sum_{w = 1}^{h+3} \lfloor \mu(w) \rfloor$, 
the integer $M$ is defined as $M = (K - (2 + \sum_{w = 1}^{h+3} \lfloor \mu(w) \rfloor) ) \mod |C|$. 
Similarly, the integer $b$ is defined as $b = 1 + \lfloor \frac{K - (2 + \sum_{w = 1}^{h+3} \lfloor \mu(w) \rfloor)}{|C|} \rfloor$. 

Statement (i) shows that 
the subset $\Psi_{\source}$ contains an interval attractor $([p_{B}, q_{B}], [\ell_{B}, r_{B}])$ satisfying 
$([p_{A}, q_{A}], [\ell_{A}, r_{A}]) \in f_{\recover}(([p_{B}, q_{B}], [\ell_{B}, r_{B}]))$ 
and $\Psi_{\source} \in \Psi_{h} \cap \Psi_{\centerset}(C) \cap \Psi_{\preceding}$. 
Here, $T[p_{\tau}-1..r_{\tau}+1] = T[p_{B}-1..r_{B}+1]$ follows from statement (ii). 

Let $K_{B} \geq 0$ be an integer satisfying $([p_{B}, q_{B}], [\ell_{B}, r_{B}]) \in \Psi_{\lcp}(K_{B})$. 
We prove $K_{\tau} = K_{B}$. 
Because of $([p_{\tau}, q_{\tau}], [\ell_{\tau}, r_{\tau}]) \in \Psi_{\source}$, 
$K_{\tau} \geq 1 + \sum_{w = 1}^{h+3} \lfloor \mu(w) \rfloor$ follows from Lemma~\ref{lem:recover_basic_property}~\ref{enum:recover_basic_property:1}. 
We can apply Lemma~\ref{lem:psi_equality_basic_property}~\ref{enum:psi_equality_basic_property:9} 
to the two intervals $([p_{\tau}, q_{\tau}], [\ell_{\tau}, r_{\tau}])$ and $([p_{B}, q_{B}], [\ell_{B}, r_{B}])$ 
because $T[p_{\tau}-1..r_{\tau}+1] = T[p_{B}-1..r_{B}+1]$ and $K_{\tau} \geq 1 + \sum_{w = 1}^{h+3} \lfloor \mu(w) \rfloor$. 
Therefore, $K_{\tau} = K_{B}$ follows from this lemma. 

%This lemma shows that $K_{B}$ 
%Then, we prove $K_{\tau} = K_{B}$. 
%Lemma~\ref{lem:psi_equality_basic_property}~\ref{enum:psi_equality_basic_property:9} shows that 
%$([p_{\tau}, q_{\tau}], [\ell_{\tau}, r_{\tau}]) \in \Psi_{\lcp}(K_{B})$ holds 
%because $T[p_{\tau}-1..r_{\tau}+1] = T[p_{B}-1..r_{B}+1]$ and $([p_{B}, q_{B}], [\ell_{B}, r_{B}]) \in \Psi_{\lcp}(K_{B})$. 
%$K_{\tau} = K_{B}$ holds because $([p_{\tau}, q_{\tau}], [\ell_{\tau}, r_{\tau}]) \in \Psi_{\lcp}(K_{B})$ 
%and $([p_{\tau}, q_{\tau}], [\ell_{\tau}, r_{\tau}]) \in \Psi_{\lcp}(K_{\tau})$. 

We prove $K + |C| \leq K_{\tau} \leq n$. 
Because of $([p_{A}, q_{A}], [\ell_{A}, r_{A}]) \in f_{\recover}(([p_{B}, q_{B}], [\ell_{B}, r_{B}]))$, 
Lemma~\ref{lem:recover_basic_property}~\ref{enum:recover_basic_property:4} shows that 
there exists an integer $z \geq 1$ satisfying $([p_{A}, q_{A}], [\ell_{A}, r_{A}]) \in \Psi_{\lcp}(K_{B} - z|C|)$. 
$K_{B} - z|C| = K$ holds because 
$([p_{A}, q_{A}], [\ell_{A}, r_{A}]) \in \Psi_{\lcp}(K_{B} - z|C|)$ and $([p_{A}, q_{A}], [\ell_{A}, r_{A}]) \in \Psi_{\lcp}(K)$. 
$K_{\tau} \geq K + |C|$ follows from 
$K_{\tau} = K_{B}$, $K_{B} - z|C| = K$, and $z \geq 1$. 
On the other hand, $K_{\tau} \leq n$ follows from the $K_{\tau} = |\lcp(T[\gamma_{\tau}..r_{\tau}], C^{n+1})|$. 
Therefore, $K + |C| \leq K_{\tau} \leq n$ holds. 

We prove $b \leq g_{\tau} \leq n$. 
Because of $([p_{\tau}, q_{\tau}], [\ell_{\tau}, r_{\tau}]) \in \Psi_{h} \cap \Psi_{\centerset}(C) \cap \Psi_{\lcp}(K_{\tau})$,  
Lemma~\ref{lem:recover_basic_property}~\ref{enum:recover_basic_property:1} shows that 
$g_{\tau} = \lfloor \frac{K_{\tau} - (2 + \sum_{w = 1}^{h+3} \lfloor \mu(w) \rfloor)}{|C|} \rfloor$ and $g_{\tau} \leq n$. 
Because of $b = 1 + \lfloor \frac{K - (2 + \sum_{w = 1}^{h+3} \lfloor \mu(w) \rfloor)}{|C|} \rfloor$, 
$g_{\tau} \geq b$ follows from the following equation: 
\begin{equation*}
    \begin{split}
    g_{\tau} &= \lfloor \frac{K_{\tau} - (2 + \sum_{w = 1}^{h+3} \lfloor \mu(w) \rfloor)}{|C|} \rfloor \\
    &\geq \lfloor \frac{K + |C| - (2 + \sum_{w = 1}^{h+3} \lfloor \mu(w) \rfloor)}{|C|} \rfloor \\
    &= 1 + \lfloor \frac{K - (2 + \sum_{w = 1}^{h+3} \lfloor \mu(w) \rfloor)}{|C|} \rfloor \\
    &= b.
    \end{split}
\end{equation*}

We prove $([p_{A}, q_{A}], [\ell_{A}, r_{A}]) \in \bigcup_{s \in \mathcal{I}_{D, 3}} (\Psi_{\sRecover}(([p_{s}, q_{s}], [\ell_{s}, r_{s}])) \cap \Psi_{\lcp}(K))$. 
Statement (iii) shows that $M_{\tau} = M$ holds 
because $([p_{A}, q_{A}], [\ell_{A}, r_{A}]) \in \Psi_{\lcp}(K)$ and $M = (K - (2 + \sum_{w = 1}^{h+3} \lfloor \mu(w) \rfloor) ) \mod |C|$. 
$\tau \in \mathcal{I}_{D, 3}$ follows from $b \leq g_{\tau} \leq n$ and $M_{\tau} = M$. 
Therefore, $([p_{A}, q_{A}], [\ell_{A}, r_{A}]) \in \bigcup_{s \in \mathcal{I}_{D, 3}} (\Psi_{\sRecover}(([p_{s}, q_{s}], [\ell_{s}, r_{s}])) \cap \Psi_{\lcp}(K))$ follows from 
$([p_{A}, q_{A}], [\ell_{A}, r_{A}]) \in \Psi_{\sRecover}(([p_{\tau}, q_{\tau}]$, $[\ell_{\tau}, r_{\tau}])) \cap \Psi_{\lcp}(K)$ 
and $\tau \in \mathcal{I}_{D, 3}$.

We showed that $([p_{A}, q_{A}], [\ell_{A}, r_{A}]) \in \bigcup_{s \in \mathcal{I}_{D, 3}} (\Psi_{\sRecover}(([p_{s}, q_{s}], [\ell_{s}, r_{s}])) \cap \Psi_{\lcp}(K))$ holds for each interval attractor $([p_{A}, q_{A}], [\ell_{A}, r_{A}])$ in set $\Psi_{h} \cap \Psi_{\run} \cap \Psi_{\centerset}(C) \cap \Psi_{\preceding} \cap \Psi_{\lcp}(K)$. 
Therefore, Equation~\ref{eq:JD_sum_set_subseteq:D3} holds. 
\end{proof}

\begin{proposition}\label{prop:JD_sum_set_supseteq}
The following five equations hold:
\begin{equation}\label{eq:JD_sum_set_supseteq:E1}
    \Psi_{h} \cap \Psi_{\run} \cap \Psi_{\centerset}(C) \cap \Psi_{\preceding} \supseteq \bigcup_{s \in \mathcal{I}_{E, 1}} \Psi_{\sRecover}(([p_{s}, q_{s}], [\ell_{s}, r_{s}]));
\end{equation}
\begin{equation}\label{eq:JD_sum_set_supseteq:D1}
    \begin{split}
    \Psi_{h} \cap \Psi_{\run} \cap \Psi_{\centerset}(C) \cap \Psi_{\preceding} & \cap  (\bigcup_{\lambda = b}^{n} \Psi_{\nRecover}(\lambda)) \cap (\bigcup_{\lambda = 0}^{K - M - 1} \Psi_{\lcp}(\lambda)) \\ 
    &\supseteq \bigcup_{s \in \mathcal{I}_{D, 1}} \bigcup_{\lambda = 0}^{K - M - 1} (\Psi_{\sRecover}(([p_{s}, q_{s}], [\ell_{s}, r_{s}])) \cap \Psi_{\lcp}(\lambda))
    \end{split}
\end{equation}
\begin{equation}\label{eq:JD_sum_set_supseteq:D2}
    \begin{split}
    \Psi_{h} \cap \Psi_{\run} \cap \Psi_{\centerset}(C) \cap \Psi_{\preceding} & \cap (\bigcup_{\lambda = b}^{n} \Psi_{\nRecover}(\lambda)) \cap (\bigcup_{\lambda = K-M}^{K - 1} \Psi_{\lcp}(\lambda)) \\ 
    &\supseteq \bigcup_{s \in \mathcal{I}_{D, 2}} \bigcup_{\lambda = K-M}^{K - 1} (\Psi_{\sRecover}(([p_{s}, q_{s}], [\ell_{s}, r_{s}])) \cap \Psi_{\lcp}(\lambda));
    \end{split}
\end{equation}
\begin{equation}\label{eq:JD_sum_set_supseteq:E2}
    \begin{split}
    \Psi_{h} \cap \Psi_{\run} \cap \Psi_{\centerset}(C) \cap \Psi_{\preceding} & \cap (\bigcup_{\lambda = 1}^{b-1} \Psi_{\nRecover}(\lambda)) \cap (\bigcup_{\lambda = 0}^{K - 1} \Psi_{\lcp}(\lambda)) \\ 
    &\supseteq \bigcup_{s \in \mathcal{I}_{E, 2}} \Psi_{\sRecover}(([p_{s}, q_{s}], [\ell_{s}, r_{s}]));
    \end{split}
\end{equation}
\begin{equation}\label{eq:JD_sum_set_supseteq:D3}
    \begin{split}
    \Psi_{h} \cap \Psi_{\run} \cap \Psi_{\centerset}(C) \cap \Psi_{\preceding} & \cap \Psi_{\lcp}(K) \\ 
    &\supseteq \bigcup_{s \in \mathcal{I}_{D, 3}} (\Psi_{\sRecover}(([p_{s}, q_{s}], [\ell_{s}, r_{s}])) \cap \Psi_{\lcp}(K)).
    \end{split}
\end{equation}
\end{proposition}
\begin{proof}
The following six statements are used to prove Proposition~\ref{prop:JD_sum_set_supseteq}: 
\begin{enumerate}[label=\textbf{(\roman*)}]
    \item 
    $\bigcup_{s \in \mathcal{I}} \Psi_{\sRecover}(([p_{s}, q_{s}], [\ell_{s}, r_{s}])) \subseteq \Psi_{h} \cap \Psi_{\run} \cap \Psi_{\centerset}(C) \cap \Psi_{\preceding}$ holds for a subset $\mathcal{I}$ of set $\{ 1, 2, \ldots, k \}$; 
    \item consider an integer $s \in [1, k]$. 
    Then, $\Psi_{\sRecover}(([p_{s}, q_{s}], [\ell_{s}, r_{s}])) \subseteq \bigcup_{\lambda = x_{1}}^{x_{2}} \Psi_{\nRecover}(\lambda)$ 
    for any pair of two integers $x_{1}, x_{2} \in [1, n]$ satisfying $x_{1} \leq g_{s} \leq x_{2}$; 
    \item $\bigcup_{s \in \mathcal{I}_{D, 1}} \Psi_{\sRecover}(([p_{s}, q_{s}], [\ell_{s}, r_{s}])) \subseteq \bigcup_{\lambda = b}^{n} \Psi_{\nRecover}(\lambda)$; 
    \item $\bigcup_{s \in \mathcal{I}_{D, 2}} \Psi_{\sRecover}(([p_{s}, q_{s}], [\ell_{s}, r_{s}])) \subseteq \bigcup_{\lambda = b}^{n} \Psi_{\nRecover}(\lambda)$; 
    \item $\bigcup_{s \in \mathcal{I}_{E, 2}} \Psi_{\sRecover}(([p_{s}, q_{s}], [\ell_{s}, r_{s}])) \subseteq \bigcup_{\lambda = 1}^{b-1} \Psi_{\nRecover}(\lambda)$; 
    \item $\bigcup_{s \in \mathcal{I}_{E, 2}} \Psi_{\sRecover}(([p_{s}, q_{s}], [\ell_{s}, r_{s}])) \subseteq \bigcup_{\lambda = 0}^{K - 1} \Psi_{\lcp}(\lambda)$. 
\end{enumerate}

\textbf{Proof of statement (i).}
Consider an interval attractor $([p_{A}, q_{A}], [\ell_{A}, r_{A}]) \in \bigcup_{s \in \mathcal{I}} \Psi_{\sRecover}(([p_{s}$, $q_{s}], [\ell_{s}, r_{s}]))$. 
Then, the set $\mathcal{I}$ contains an integer $\tau$ satisfying 
$([p_{A}, q_{A}], [\ell_{A}, r_{A}]) \in \Psi_{\sRecover}(([p_{\tau}, q_{\tau}]$, $[\ell_{\tau}, r_{\tau}]))$. 
Because of $([p_{A}, q_{A}], [\ell_{A}, r_{A}]) \in \Psi_{\sRecover}(([p_{\tau}, q_{\tau}], [\ell_{\tau}, r_{\tau}]))$, 
the subset $\Psi_{\source}$ contains an interval attractor $([p_{B}, q_{B}], [\ell_{B}, r_{B}])$ satisfying 
$([p_{A}, q_{A}], [\ell_{A}, r_{A}]) \in f_{\recover}(([p_{B}, q_{B}], [\ell_{B}, r_{B}]))$ 
and $T[p_{B}-1..r_{B}+1] = T[p_{\tau}-1..r_{\tau}+1]$. 
Because of $([p_{\tau}, q_{\tau}], [\ell_{\tau}, r_{\tau}]) \in \Psi_{h} \cap \Psi_{\centerset}(C) \cap \Psi_{\preceding}$, 
Lemma~\ref{lem:psi_equality_basic_property} shows that 
$([p_{B}, q_{B}], [\ell_{B}, r_{B}]) \in \Psi_{h} \cap \Psi_{\centerset}(C) \cap \Psi_{\preceding}$ holds. 
Lemma~\ref{lem:recover_basic_property}~\ref{enum:recover_basic_property:4} shows that 
$([p_{A}, q_{A}], [\ell_{A}, r_{A}]) \in \Psi_{h} \cap \Psi_{\run} \cap \Psi_{\centerset}(C)$ holds 
because $([p_{A}, q_{A}], [\ell_{A}, r_{A}]) \in f_{\recover}(([p_{B}, q_{B}]$, $[\ell_{B}, r_{B}]))$ 
and $([p_{B}, q_{B}], [\ell_{B}, r_{B}]) \in \Psi_{h} \cap \Psi_{\centerset}(C)$. 
Similarly, Lemma~\ref{lem:recover_basic_property}~\ref{enum:recover_basic_property:5} shows that 
$([p_{A}, q_{A}]$, $[\ell_{A}, r_{A}]) \in \Psi_{\preceding}$ holds 
because $([p_{B}, q_{B}], [\ell_{B}, r_{B}]) \in \Psi_{\preceding}$. 
Therefore, $([p_{A}, q_{A}], [\ell_{A}, r_{A}]) \in \Psi_{h} \cap \Psi_{\run} \cap \Psi_{\centerset}(C) \cap \Psi_{\preceding}$ holds. 

We showed that $([p_{A}, q_{A}], [\ell_{A}, r_{A}]) \in \Psi_{h} \cap \Psi_{\run} \cap \Psi_{\centerset}(C) \cap \Psi_{\preceding}$ holds for each interval attractor $([p_{A}, q_{A}], [\ell_{A}, r_{A}]) \in \bigcup_{s \in \mathcal{I}} \Psi_{\sRecover}(([p_{s}, q_{s}], [\ell_{s}, r_{s}]))$. 
Therefore, $\bigcup_{s \in \mathcal{I}} \Psi_{\sRecover}(([p_{s}, q_{s}], [\ell_{s}, r_{s}])) \subseteq \Psi_{h} \cap \Psi_{\run} \cap \Psi_{\centerset}(C) \cap \Psi_{\preceding}$ holds. 

\textbf{Proof of statement (ii).}
Consider an interval attractor $([p_{A}, q_{A}], [\ell_{A}, r_{A}]) \in \Psi_{\sRecover}(([p_{s}$, $q_{s}], [\ell_{s}, r_{s}]))$. 
Then, 
the subset $\Psi_{\source}$ contains an interval attractor $([p_{B}, q_{B}], [\ell_{B}, r_{B}])$ satisfying 
$([p_{A}, q_{A}], [\ell_{A}, r_{A}]) \in f_{\recover}(([p_{B}, q_{B}], [\ell_{B}, r_{B}]))$ 
and $T[p_{B}-1..r_{B}+1] = T[p_{s}-1..r_{s}+1]$. 
Let $g_{B} = |f_{\recover}(([p_{B}, q_{B}], [\ell_{B}, r_{B}]))|$. 
Then, $([p_{A}, q_{A}], [\ell_{A}, r_{A}]) \in \Psi_{\nRecover}(g_{B})$ follows from the definition of the subset $\Psi_{\nRecover}(g_{B})$. 
Because of $T[p_{B}-1..r_{B}+1] = T[p_{s}-1..r_{s}+1]$, 
Lemma~\ref{lem:recover_super_property}~\ref{enum:recover_super_property:1} shows that 
$g_{B} = |f_{\recover}(([p_{s}, q_{s}], [\ell_{s}, r_{s}]))|$ holds. 
$x_{1} \leq g_{B} \leq x_{2}$ follows from 
$g_{B} = |f_{\recover}(([p_{s}, q_{s}], [\ell_{s}, r_{s}]))|$, 
$g_{s} = |f_{\recover}(([p_{s}, q_{s}], [\ell_{s}, r_{s}]))|$, 
and $x_{1} \leq g_{s} \leq x_{2}$. 
Therefore, $([p_{A}, q_{A}], [\ell_{A}, r_{A}]) \in \bigcup_{\lambda = x_{1}}^{x_{2}} \Psi_{\nRecover}(\lambda)$ 
follows from $([p_{A}, q_{A}], [\ell_{A}, r_{A}]) \in \Psi_{\nRecover}(g_{B})$ 
and $x_{1} \leq g_{B} \leq x_{2}$. 

We showed that $([p_{A}, q_{A}], [\ell_{A}, r_{A}]) \in \bigcup_{\lambda = x_{1}}^{x_{2}} \Psi_{\nRecover}(\lambda)$ holds for each interval attractor $([p_{A}, q_{A}]$, $[\ell_{A}, r_{A}]) \in \Psi_{\sRecover}(([p_{s}$, $q_{s}], [\ell_{s}, r_{s}]))$. 
Therefore, $\Psi_{\sRecover}(([p_{s}, q_{s}], [\ell_{s}, r_{s}])) \subseteq \bigcup_{\lambda = x_{1}}^{x_{2}} \Psi_{\nRecover}(\lambda)$ holds. 

\textbf{Proof of statement (iii).}
For each integer $s \in \mathcal{I}_{D, 1}$, 
$b \leq g_{s} \leq n$ follows from the definition of the set $\mathcal{I}_{D, 1}$. 
Because of $b \leq g_{s} \leq n$, 
statement (ii) shows that 
$\Psi_{\sRecover}(([p_{s}, q_{s}], [\ell_{s}, r_{s}])) \subseteq \bigcup_{\lambda = b}^{n} \Psi_{\nRecover}(\lambda)$ holds. 
Therefore, $\bigcup_{s \in \mathcal{I}_{D, 1}} \Psi_{\sRecover}(([p_{s}, q_{s}], [\ell_{s}, r_{s}])) \subseteq \bigcup_{\lambda = b}^{n} \Psi_{\nRecover}(\lambda)$ holds.

\textbf{Proof of statement (iv).}
Statement (iv) can be proved using the same approach as for statement (iii).

\textbf{Proof of statement (v).}
Statement (v) can be proved using the same approach as for statement (iii).

\textbf{Proof of statement (vi).}
Consider an interval attractor $([p_{A}, q_{A}], [\ell_{A}, r_{A}])$ in set $\bigcup_{s \in \mathcal{I}_{E, 2}}$ $\Psi_{\sRecover}(([p_{s}, q_{s}], [\ell_{s}, r_{s}]))$. 
Then, the set $\mathcal{I}_{E, 2}$ contains an integer $\tau$ satisfying 
$([p_{A}, q_{A}], [\ell_{A}, r_{A}]) \in \Psi_{\sRecover}(([p_{\tau}, q_{\tau}]$, $[\ell_{\tau}, r_{\tau}]))$. 
Here, $1 \leq g_{\tau} \leq b-1$ follows from the definition of the set $\mathcal{I}_{E, 2}$. 
$b \geq 2$ follows from $1 \leq g_{\tau} \leq b-1$. 
In this case, $K > 1 + \sum_{w = 1}^{h+3} \lfloor \mu(w) \rfloor$ must hold 
because the integer $b$ is defined as $1$ if $K \leq 1 + \sum_{w = 1}^{h+3} \lfloor \mu(w) \rfloor$. 

We prove $K_{\tau} \leq K + |C| - 1$ by contradiction. 
We assume that $K_{\tau} \geq K + |C|$ holds. 
Because of $([p_{\tau}, q_{\tau}], [\ell_{\tau}, r_{\tau}]) \in \Psi_{h} \cap \Psi_{\source} \cap \Psi_{\centerset}(C) \cap \Psi_{\lcp}(K_{\tau})$, 
Lemma~\ref{lem:recover_basic_property}~\ref{enum:recover_basic_property:1} shows that 
$g_{\tau} = \lfloor \frac{K_{\tau} - (2 + \sum_{w = 1}^{h+3} \lfloor \mu(w) \rfloor)}{|C|} \rfloor$ holds. 
Because of $K > 1 + \sum_{w = 1}^{h+3} \lfloor \mu(w) \rfloor$, 
the integer $b$ is defined as $b = 1 + \lfloor \frac{K - (2 + \sum_{w = 1}^{h+3} \lfloor \mu(w) \rfloor)}{|C|} \rfloor$. 
Therefore, $g_{\tau} \geq b$ follows from the following equation: 
\begin{equation*}
    \begin{split}
    g_{\tau} &= \lfloor \frac{K_{\tau} - (2 + \sum_{w = 1}^{h+3} \lfloor \mu(w) \rfloor)}{|C|} \rfloor \\
    &\geq \lfloor \frac{(K + |C|) - (2 + \sum_{w = 1}^{h+3} \lfloor \mu(w) \rfloor)}{|C|} \rfloor \\
    &= 1 + \lfloor \frac{K - (2 + \sum_{w = 1}^{h+3} \lfloor \mu(w) \rfloor)}{|C|} \rfloor \\
    &= b.
    \end{split}
\end{equation*}
On the other hand, $g_{\tau} \leq b-1$ follows from the definition of the set $\mathcal{I}_{E, 2}$. 
The two facts $g_{\tau} \geq b$ and $g_{\tau} \leq b-1$ yield a contradiction. 
Therefore, $K_{\tau} \leq K + |C| - 1$ must hold. 

We prove $([p_{A}, q_{A}], [\ell_{A}, r_{A}]) \in \bigcup_{\lambda = 0}^{K - 1} \Psi_{\lcp}(\lambda)$. 
Because of $([p_{A}, q_{A}], [\ell_{A}, r_{A}]) \in \Psi_{\sRecover}(([p_{\tau}, q_{\tau}]$, $[\ell_{\tau}, r_{\tau}]))$, 
the subset $\Psi_{\source}$ contains an interval attractor $([p_{B}, q_{B}], [\ell_{B}, r_{B}])$ satisfying 
$([p_{A}, q_{A}]$, $[\ell_{A}, r_{A}]) \in f_{\recover}(([p_{B}, q_{B}], [\ell_{B}, r_{B}]))$ 
and $T[p_{B}-1..r_{B}+1] = T[p_{\tau}-1..r_{\tau}+1]$. 
Lemma~\ref{lem:psi_equality_basic_property} shows that 
$([p_{B}, q_{B}], [\ell_{B}, r_{B}]) \in \Psi_{h} \cap \Psi_{\source} \cap \Psi_{\centerset}(C) \cap \Psi_{\lcp}(K_{\tau})$ holds 
because $T[p_{B}-1..r_{B}+1] = T[p_{\tau}-1..r_{\tau}+1]$ and 
$([p_{\tau}, q_{\tau}], [\ell_{\tau}, r_{\tau}]) \in \Psi_{h} \cap \Psi_{\source} \cap \Psi_{\centerset}(C) \cap \Psi_{\lcp}(K_{\tau})$. 
Because of $([p_{\tau}, q_{\tau}], [\ell_{\tau}, r_{\tau}]) \in \Psi_{h} \cap \Psi_{\source} \cap \Psi_{\centerset}(C) \cap \Psi_{\lcp}(K_{\tau})$, 
Lemma~\ref{lem:recover_basic_property}~\ref{enum:recover_basic_property:4} shows that 
there exists an integer $z \geq 1$ satisfying $([p_{A}, q_{A}], [\ell_{A}, r_{A}]) \in \Psi_{\lcp}(K_{\tau} - z|C|)$. 
$K_{\tau} - z|C| \leq K-1$ follows from $K_{\tau} \leq K + |C| - 1$ and $z \geq 1$. 
Therefore, $([p_{A}, q_{A}], [\ell_{A}, r_{A}]) \in \bigcup_{\lambda = 0}^{K - 1} \Psi_{\lcp}(\lambda)$ holds. 

We showed that $([p_{A}, q_{A}], [\ell_{A}, r_{A}]) \in \bigcup_{\lambda = 0}^{K - 1} \Psi_{\lcp}(\lambda)$ for each interval attractor $([p_{A}, q_{A}], [\ell_{A}, r_{A}]) \in \bigcup_{s \in \mathcal{I}_{E, 2}} \Psi_{\sRecover}(([p_{s}, q_{s}], [\ell_{s}, r_{s}]))$. 
Therefore, $\bigcup_{s \in \mathcal{I}_{E, 2}} \Psi_{\sRecover}(([p_{s}, q_{s}], [\ell_{s}, r_{s}])) \subseteq \bigcup_{\lambda = 0}^{K - 1} \Psi_{\lcp}(\lambda)$ holds. 

\textbf{Proof of Equation~\ref{eq:JD_sum_set_supseteq:E1}.}
Equation~\ref{eq:JD_sum_set_supseteq:E1} follows from statement (i). 

\textbf{Proof of Equation~\ref{eq:JD_sum_set_supseteq:D1}.}
Equation~\ref{eq:JD_sum_set_supseteq:D1} follows from statement (i), statement (iii), and 
$\bigcup_{\lambda = 0}^{K - M - 1}$ $(\Psi_{\sRecover}(([p_{s}, q_{s}], [\ell_{s}, r_{s}])) \cap \Psi_{\lcp}(\lambda)) \subseteq \bigcup_{\lambda = 0}^{K - M - 1} \Psi_{\lcp}(\lambda)$.

\textbf{Proof of Equation~\ref{eq:JD_sum_set_supseteq:D2}.}
Equation~\ref{eq:JD_sum_set_supseteq:D2} follows from statement (i), statement (iv), and 
$\bigcup_{s \in \mathcal{I}_{D, 2}}$ $\bigcup_{\lambda = K-M}^{K - 1}(\Psi_{\sRecover}(([p_{s}, q_{s}], [\ell_{s}, r_{s}])) \cap \Psi_{\lcp}(\lambda)) \subseteq \bigcup_{\lambda = K-M}^{K - 1} \Psi_{\lcp}(\lambda)$.

\textbf{Proof of Equation~\ref{eq:JD_sum_set_supseteq:E2}.}
Equation~\ref{eq:JD_sum_set_supseteq:E2} follows from statement (i), statement (v), and statement (vi). 

\textbf{Proof of Equation~\ref{eq:JD_sum_set_supseteq:D3}.}
Equation~\ref{eq:JD_sum_set_supseteq:D3} follows from statement (i) and 
$\bigcup_{s \in \mathcal{I}_{D, 3}} (\Psi_{\sRecover}(([p_{s}, q_{s}]$, $[\ell_{s}, r_{s}])) \cap \Psi_{\lcp}(K)) \subseteq \Psi_{\lcp}(K)$. 
\end{proof}

%%%%%%%%%%%%%%%%%%%%%%%%%%%%%%%%%%%%%%%%%%%%%%%%%%%%

\begin{proposition}\label{prop:JD1_sum_sub}
The following five statements hold:
\begin{enumerate}[label=\textbf{(\roman*)}]
    \item \label{enum:JD1_sum_sub:1} $|(\bigcup_{s \in \mathcal{I}} \Psi_{\sRecover}(([p_{s}, q_{s}], [\ell_{s}, r_{s}]))) \cap \Psi| = \sum_{s \in \mathcal{I}} |\Psi_{\sRecover}(([p_{s}, q_{s}], [\ell_{s}, r_{s}])) \cap \Psi|$ for 
    two subsets $\mathcal{I} \subseteq \{ 1, 2, \ldots, k \}$ 
    and $\Psi \subseteq \Psi_{\RR}$; 
    \item \label{enum:JD1_sum_sub:2} 
    $|\bigcup_{s \in \mathcal{I}} \Psi_{\sRecover}(([p_{s}, q_{s}], [\ell_{s}, r_{s}]))| = \sum_{s \in \mathcal{I}} g_{s} |\Psi_{\str}(T[p_{s}-1..r_{s}+1])|$ for a subset $\mathcal{I}$ of set $\{ 1, 2, \ldots, k \}$;
    \item \label{enum:JD1_sum_sub:3} $|\Psi_{\sRecover}(([p_{s}, q_{s}], [\ell_{s}, r_{s}])) \cap (\bigcup_{\lambda = 0}^{K - M - 1} \Psi_{\lcp}(\lambda))| = (b-1) |\Psi_{\str}(T[p_{s}-1..r_{s}+1])|$ 
    for each integer $s \in \mathcal{I}_{D, 1}$;
    \item \label{enum:JD1_sum_sub:4}
    $|\Psi_{\sRecover}(([p_{s}, q_{s}], [\ell_{s}, r_{s}])) \cap (\bigcup_{\lambda = K-M}^{K - 1} \Psi_{\lcp}(\lambda))| = |\Psi_{\str}(T[p_{s}-1..r_{s}+1])|$ for each integer $s \in \mathcal{I}_{D, 2}$;  
    \item \label{enum:JD1_sum_sub:5}
    if $K > 1 + \sum_{w = 1}^{h+3} \lfloor \mu(w) \rfloor$, 
    then $|\Psi_{\sRecover}(([p_{s}, q_{s}], [\ell_{s}, r_{s}])) \cap \Psi_{\lcp}(K)| = |\Psi_{\str}(T[p_{s}-1..r_{s}+1])|$ holds 
    for each integer $s \in \mathcal{I}_{D, 3}$.
\end{enumerate}
\end{proposition}
\begin{proof}
Let $\zeta = 2 + \sum_{w = 1}^{h+3} \lfloor \mu(w) \rfloor$ for simplicity. 
The proof of Proposition~\ref{prop:JD1_sum_sub} is as follows. 

\textbf{Proof of Proposition~\ref{prop:JD1_sum_sub}(i).}
For any pair of two integers $s, s^{\prime} \in \mathcal{I}$, 
$T[p_{s}-1..r_{s}+1] \neq T[p_{s^{\prime}}-1..r_{s^{\prime}}+1]$ follows from the definition of the sampling subset $\Psi_{\samp}$. 
Because of $T[p_{s}-1..r_{s}+1] \neq T[p_{s^{\prime}}-1..r_{s^{\prime}}+1]$, 
Lemma~\ref{lem:sRecover_basic_property}~\ref{enum:sRecover_basic_property:overlap} shows that 
$\Psi_{\sRecover}(([p_{s}, q_{s}], [\ell_{s}, r_{s}])) \cap \Psi_{\sRecover}(([p_{s^{\prime}}, q_{s^{\prime}}]$, $[\ell_{s^{\prime}}, r_{s^{\prime}}])) = \emptyset$ holds. 
Therefore, 
$|(\bigcup_{s \in \mathcal{I}} \Psi_{\sRecover}(([p_{s}, q_{s}], [\ell_{s}, r_{s}]))) \cap \Psi| = \sum_{s \in \mathcal{I}} |\Psi_{\sRecover}(([p_{s}, q_{s}]$, $[\ell_{s}, r_{s}])) \cap \Psi|$ holds. 

\textbf{Proof of Proposition~\ref{prop:JD1_sum_sub}(ii).}
$|\bigcup_{s \in \mathcal{I}} \Psi_{\sRecover}(([p_{s}, q_{s}], [\ell_{s}, r_{s}]))| = |\bigcup_{s \in \mathcal{I}} (\Psi_{\sRecover}(([p_{s}, q_{s}]$, $[\ell_{s}, r_{s}])) \cap \Psi_{\RR})|$ holds because 
$\Psi_{\sRecover}(([p_{s}, q_{s}], [\ell_{s}, r_{s}])) \subseteq \Psi_{\RR}$ holds for each integer $s \in \mathcal{I}$. 
Proposition~\ref{prop:JD1_sum_sub}(i) shows that 
$|(\bigcup_{s \in \mathcal{I}} (\Psi_{\sRecover}(([p_{s}, q_{s}], [\ell_{s}, r_{s}]))) \cap \Psi_{\RR})| = \sum_{s \in \mathcal{I}} |\Psi_{\sRecover}(([p_{s}, q_{s}]$, $[\ell_{s}, r_{s}])) \cap \Psi_{\RR}|$ holds. 
Here, 
$|\Psi_{\sRecover}(([p_{s}, q_{s}], [\ell_{s}, r_{s}])) \cap \Psi_{\RR}| = |\Psi_{\sRecover}(([p_{s}, q_{s}], [\ell_{s}, r_{s}]))|$ holds for each integer $s \in \mathcal{I}$. 
Lemma~\ref{lem:sRecover_size_property}~\ref{enum:sRecover_size_property:2} shows that 
$|\Psi_{\sRecover}(([p_{s}, q_{s}], [\ell_{s}, r_{s}]))| = g_{s} |\Psi_{\str}(T[p_{s}-1..r_{s}+1])|$ holds. 
Therefore, Proposition~\ref{prop:JD1_sum_sub}(ii) follows from the following equation: 
\begin{equation*}
    \begin{split}
        |\bigcup_{s \in \mathcal{I}} \Psi_{\sRecover}(([p_{s}, q_{s}], [\ell_{s}, r_{s}]))| &= |\bigcup_{s \in \mathcal{I}} (\Psi_{\sRecover}(([p_{s}, q_{s}], [\ell_{s}, r_{s}])) \cap \Psi_{\RR})| \\
        &= \sum_{s \in \mathcal{I}} |\Psi_{\sRecover}(([p_{s}, q_{s}], [\ell_{s}, r_{s}])) \cap \Psi_{\RR}| \\
        &= \sum_{s \in \mathcal{I}} |\Psi_{\sRecover}(([p_{s}, q_{s}], [\ell_{s}, r_{s}]))| \\
        &= \sum_{s \in \mathcal{I}} g_{s} |\Psi_{\str}(T[p_{s}-1..r_{s}+1])|.
    \end{split}
\end{equation*}

\textbf{Proof of Proposition~\ref{prop:JD1_sum_sub}(iii).}
Lemma~\ref{lem:sRecover_basic_property}~\ref{enum:sRecover_basic_property:centerset} shows that 
$\Psi_{\sRecover}(([p_{s}, q_{s}], [\ell_{s}, r_{s}])) \subseteq \bigcup_{\lambda = \zeta}^{K_{s}-|C|} \Psi_{\lcp}(\lambda)$ holds. 
This equation indicates that 
$\Psi_{\sRecover}(([p_{s}, q_{s}], [\ell_{s}, r_{s}])) \cap \bigcup_{\lambda = 0}^{\zeta-1} \Psi_{\lcp}(\lambda) = \emptyset$ holds 
because $\lcp(\lambda) \cap \lcp(\lambda^{\prime}) = \emptyset$ holds for any pair of two distinct integers $\lambda, \lambda^{\prime} \geq 0$. 

One of the following three conditions is satisfied: 
(a) $K < \zeta$; (b) $\zeta \leq K < \zeta + |C|$; $K \geq \zeta + |C|$. 
For condition (a), 
the two integers $b$ and $M$ are defined as $1$ and $0$, respectively. 
$|(\Psi_{\sRecover}(([p_{s}, q_{s}]$, $[\ell_{s}, r_{s}])) \cap (\bigcup_{\lambda = 0}^{K - M - 1} \Psi_{\lcp}(\lambda)))| = 0$ follows from 
the following equation: 
\begin{equation}\label{eq:JD1_sum_sub:1}
    \begin{split}
        (\Psi_{\sRecover}(([p_{s}, q_{s}], [\ell_{s}, r_{s}])) & \cap (\bigcup_{\lambda = 0}^{K - M - 1} \Psi_{\lcp}(\lambda)))  \\
        &\subseteq \Psi_{\sRecover}(([p_{s}, q_{s}], [\ell_{s}, r_{s}])) \cap \bigcup_{\lambda = 0}^{\zeta-1} \Psi_{\lcp}(\lambda) \\
        &= \emptyset.
    \end{split}
\end{equation}
Therefore, $|(\Psi_{\sRecover}(([p_{s}, q_{s}], [\ell_{s}, r_{s}])) \cap (\bigcup_{\lambda = 0}^{K - M - 1} \Psi_{\lcp}(\lambda)))| = (b-1) |\Psi_{\str}(T[p_{s}-1..r_{s}+1])|$ follows from $b = 1$ and $|(\Psi_{\sRecover}(([p_{s}, q_{s}], [\ell_{s}, r_{s}])) \cap (\bigcup_{\lambda = 0}^{K - M - 1} \Psi_{\lcp}(\lambda)))| = 0$. 

For condition (b), 
the two integers $b$ and $M$ are defined as $1 + \lfloor \frac{K - \zeta}{|C|} \rfloor$ and $(K - \zeta) \mod |C|$, respectively. 
$b = 1$ follows from $b = 1 + \lfloor \frac{K - \zeta}{|C|} \rfloor$ and $\zeta \leq K < \zeta + |C|$. 
Similarly, $K - M - 1 \leq \zeta-1$ follows from $M = (K - \zeta) \mod |C|$ and $\zeta \leq K < \zeta + |C|$. 
Similar to condition (a), 
Equation~\ref{eq:JD1_sum_sub:1} holds, which indicates that 
$|(\Psi_{\sRecover}(([p_{s}, q_{s}]$, $[\ell_{s}, r_{s}])) \cap (\bigcup_{\lambda = 0}^{K - M - 1} \Psi_{\lcp}(\lambda)))| = 0$ holds. 
Therefore, $|(\Psi_{\sRecover}(([p_{s}, q_{s}], [\ell_{s}, r_{s}])) \cap (\bigcup_{\lambda = 0}^{K - M - 1} \Psi_{\lcp}(\lambda)))| = (b-1) |\Psi_{\str}(T[p_{s}-1..r_{s}+1])|$ follows from $b = 1$ and $|(\Psi_{\sRecover}(([p_{s}, q_{s}], [\ell_{s}, r_{s}])) \cap (\bigcup_{\lambda = 0}^{K - M - 1} \Psi_{\lcp}(\lambda)))| = 0$. 

For condition (c), 
the two integers $b$ and $M$ are defined as $1 + \lfloor \frac{K - \zeta}{|C|} \rfloor$ and $(K - \zeta) \mod |C|$, respectively. 
$\lfloor \frac{K - \zeta}{|C|} \rfloor = \frac{K - M - \zeta}{|C|}$ follows from 
$K \geq \zeta$ and $M = (K - \zeta) \mod |C|$. 
$K - M = (b-1)|C| + \zeta$ follows from $b = 1 + \lfloor \frac{K - \zeta}{|C|} \rfloor$ and 
$\lfloor \frac{K - \zeta}{|C|} \rfloor = \frac{K - M - \zeta}{|C|}$. 
$b \geq 2$ follows from $b = 1 + \lfloor \frac{K - \zeta}{|C|} \rfloor$ and $K \geq \zeta + |C|$. 
We already proved $\Psi_{\sRecover}(([p_{s}, q_{s}], [\ell_{s}, r_{s}])) \cap \bigcup_{\lambda = 0}^{\zeta-1} \Psi_{\lcp}(\lambda) = \emptyset$. Therefore, $|\Psi_{\sRecover}(([p_{s}, q_{s}], [\ell_{s}, r_{s}])) \cap (\bigcup_{\lambda = 0}^{K - M - 1} \Psi_{\lcp}(\lambda))| = |\bigcup_{y = 1}^{b-1} (\Psi_{\sRecover}(([p_{s}, q_{s}], [\ell_{s}, r_{s}])) \cap \bigcup_{\lambda = (y-1)|C| + \zeta}^{y|C| + \zeta - 1} \Psi_{\lcp}(\lambda))|$ follows from the following equation: 
\begin{equation}\label{eq:JD1_sum_sub:2}
    \begin{split}
        \Psi_{\sRecover}(([p_{s}, q_{s}], [\ell_{s}, r_{s}])) & \cap (\bigcup_{\lambda = 0}^{K - M - 1} \Psi_{\lcp}(\lambda))  \\
        &= (\Psi_{\sRecover}(([p_{s}, q_{s}], [\ell_{s}, r_{s}])) \cap (\bigcup_{\lambda = 0}^{(b-1)|C| + \zeta - 1} \Psi_{\lcp}(\lambda))) \\
        &= (\Psi_{\sRecover}(([p_{s}, q_{s}], [\ell_{s}, r_{s}])) \cap \bigcup_{\lambda = 0}^{\zeta-1} \Psi_{\lcp}(\lambda)) \\
        &\cup (\Psi_{\sRecover}(([p_{s}, q_{s}], [\ell_{s}, r_{s}])) \cap \bigcup_{\lambda = \zeta}^{(b-1)|C| + \zeta - 1} \Psi_{\lcp}(\lambda)) \\
        &= \bigcup_{y = 1}^{b-1} (\Psi_{\sRecover}(([p_{s}, q_{s}], [\ell_{s}, r_{s}])) \cap \bigcup_{\lambda = (y-1)|C| + \zeta}^{y|C| + \zeta - 1} \Psi_{\lcp}(\lambda)).
    \end{split}
\end{equation}

We prove $b \leq \lfloor \frac{K_{s} - \zeta}{|C|} \rfloor$. 
%Here, $\lfloor \frac{K_{s} - \zeta}{|C|} \rfloor = \frac{K_{s} - M_{s} - \zeta}{|C|}$ follows from 
%$M_{s} = (K_{s} - \zeta) \mod |C|$. 
Because of $s \in \mathcal{I}_{D, 1}$, 
$g_{s} \geq b$ follows from the definition of the set $\mathcal{I}_{D, 1}$. 
Lemma~\ref{lem:recover_basic_property}~\ref{enum:recover_basic_property:1} shows that 
$g_{s} = \lfloor \frac{K_{s} - \zeta}{|C|} \rfloor$ holds. 
Therefore, $b \leq \lfloor \frac{K_{s} - \zeta}{|C|} \rfloor$ follows from 
$g_{s} \geq b$ and $g_{s} = \lfloor \frac{K_{s} - \zeta}{|C|} \rfloor$. 

For each integer $y \in [1, b-1]$, 
Lemma~\ref{lem:sRecover_size_property}~\ref{enum:sRecover_size_property:3} shows that 
$|\Psi_{\sRecover}(([p_{s}, q_{s}], [\ell_{s}, r_{s}])) \cap \bigcup_{\lambda = (y-1)|C| + \zeta}^{y|C| + \zeta - 1}$ $\Psi_{\lcp}(\lambda)| = |\Psi_{\str}(([p_{s}, q_{s}], [\ell_{s}, r_{s}]))|$ holds 
because $b-1 \leq \lfloor \frac{K_{s} - \zeta}{|C|} \rfloor$. 
Therefore, 
$|(\Psi_{\sRecover}$ $(([p_{s}, q_{s}]$, $[\ell_{s}, r_{s}])) \cap (\bigcup_{\lambda = 0}^{K - M - 1} \Psi_{\lcp}(\lambda)))| = (b-1) |\Psi_{\str}(T[p_{s}-1..r_{s}+1])|$ follows from the following equation: 
\begin{equation*}
    \begin{split}
        |\Psi_{\sRecover}(([p_{s}, q_{s}], [\ell_{s}, r_{s}])) & \cap \bigcup_{\lambda = (y-1)|C| + \zeta}^{y|C| + \zeta - 1} \Psi_{\lcp}(\lambda)|  \\
        &= |\bigcup_{y = 1}^{b-1} (\Psi_{\sRecover}(([p_{s}, q_{s}], [\ell_{s}, r_{s}])) \cap \bigcup_{\lambda = (y-1)|C| + \zeta}^{y|C| + \zeta - 1} \Psi_{\lcp}(\lambda))| \\
        &= \sum_{y = 1}^{b-1} |(\Psi_{\sRecover}(([p_{s}, q_{s}], [\ell_{s}, r_{s}])) \cap \bigcup_{\lambda = (y-1)|C| + \zeta}^{y|C| + \zeta - 1} \Psi_{\lcp}(\lambda))| \\
        &= \sum_{y = 1}^{b-1} |\Psi_{\str}(([p_{s}, q_{s}], [\ell_{s}, r_{s}]))| \\
        &= (b-1) |\Psi_{\str}(([p_{s}, q_{s}], [\ell_{s}, r_{s}]))|. 
    \end{split}
\end{equation*}

\textbf{Proof of Proposition~\ref{prop:JD1_sum_sub}(iv).}
Because of $s \in \mathcal{I}_{D, 2}$, 
$b \leq g_{s} \leq n$ and $0 \leq M_{s} \leq M-1$ follow from the definition of the set $\mathcal{I}_{D, 2}$. 
$M \geq 1$ follows from $0 \leq M_{s} \leq M-1$. 
Because of $M \geq 1$, 
$K \geq \zeta$ and $M = (K - \zeta) \mod |C|$ hold (see the definition of the integer $M$).
$K > \zeta$ follows from $M \neq 0$, $K \geq \zeta$, and $M = (K - \zeta) \mod |C|$. 
Because of $K > \zeta$, 
the integer $b$ is defined as $1 + \lfloor \frac{K - \zeta}{|C|} \rfloor$. 

The following four equations are used to prove Proposition~\ref{prop:JD1_sum_sub}(iv): 
\begin{equation}\label{eq:JD1_sum_sub:3}
    \begin{split}
    |\Psi_{\sRecover}(([p_{s}, q_{s}], [\ell_{s}, r_{s}])) & \cap (\bigcup_{\lambda = K-M}^{K - 1} \Psi_{\lcp}(\lambda))| \\
    &= |\Psi_{\sRecover}(([p_{s}, q_{s}], [\ell_{s}, r_{s}])) \cap (\bigcup_{\lambda = (b-1)|C| + \zeta}^{(b-1)|C| + \zeta + M - 1} \Psi_{\lcp}(\lambda)))|;
    \end{split}
\end{equation}
\begin{equation}\label{eq:JD1_sum_sub:4}
    \begin{split}
    |\Psi_{\sRecover}(([p_{s}, q_{s}], [\ell_{s}, r_{s}])) & \cap (\bigcup_{\lambda = (b-1)|C| + \zeta}^{(b-1)|C| + \zeta + M - 1} \Psi_{\lcp}(\lambda)))| \\
    & \leq |\Psi_{\sRecover}(([p_{s}, q_{s}], [\ell_{s}, r_{s}])) \cap (\bigcup_{\lambda = (b-1)|C| + \zeta}^{b|C| + \zeta - 1} \Psi_{\lcp}(\lambda))|;
    \end{split}
\end{equation}
\begin{equation}\label{eq:JD1_sum_sub:5}
    |\Psi_{\sRecover}(([p_{s}, q_{s}], [\ell_{s}, r_{s}])) \cap (\bigcup_{\lambda = (b-1)|C| + \zeta}^{b|C| + \zeta - 1} \Psi_{\lcp}(\lambda))| = |\Psi_{\str}(([p_{s}, q_{s}], [\ell_{s}, r_{s}]))|;
\end{equation}
\begin{equation}\label{eq:JD1_sum_sub:6}
    |\Psi_{\sRecover}(([p_{s}, q_{s}], [\ell_{s}, r_{s}])) \cap (\bigcup_{\lambda = (b-1)|C| + \zeta}^{(b-1)|C| + \zeta + M - 1} \Psi_{\lcp}(\lambda)))| \geq |\Psi_{\str}(([p_{s}, q_{s}], [\ell_{s}, r_{s}]))|.
\end{equation}

We prove Equation~\ref{eq:JD1_sum_sub:3} and Equation~\ref{eq:JD1_sum_sub:4}.  
$\lfloor \frac{K - \zeta}{|C|} \rfloor = \frac{K - M - \zeta}{|C|}$ follows from 
$K \geq \zeta$ and $M = (K - \zeta) \mod |C|$. 
$K - M = (b-1)|C| + \zeta$ follows from $b = 1 + \lfloor \frac{K - \zeta}{|C|} \rfloor$ and $\lfloor \frac{K - \zeta}{|C|} \rfloor = \frac{K - M - \zeta}{|C|}$. 
$K - 1 = (b-1)|C| + \zeta + M - 1$ follows from $K - M = (b-1)|C| + \zeta$. 
$(b-1)|C| + \zeta + M - 1 \leq b|C| + \zeta - 1$ because $M \in [1, |C|-1]$. 
Therefore, Equation~\ref{eq:JD1_sum_sub:3} and Equation~\ref{eq:JD1_sum_sub:4} follow from the following equation: 
\begin{equation*}
    \begin{split}
        \Psi_{\sRecover}(([p_{s}, q_{s}], [\ell_{s}, r_{s}])) & \cap (\bigcup_{\lambda = K-M}^{K - 1} \Psi_{\lcp}(\lambda))  \\
        &= \Psi_{\sRecover}(([p_{s}, q_{s}], [\ell_{s}, r_{s}])) \cap (\bigcup_{\lambda = (b-1)|C| + \zeta}^{(b-1)|C| + \zeta + M - 1} \Psi_{\lcp}(\lambda))) \\
        &\subseteq \Psi_{\sRecover}(([p_{s}, q_{s}], [\ell_{s}, r_{s}])) \cap (\bigcup_{\lambda = (b-1)|C| + \zeta}^{b|C| + \zeta - 1} \Psi_{\lcp}(\lambda)).
    \end{split}
\end{equation*}

We prove Equation~\ref{eq:JD1_sum_sub:5}. 
Lemma~\ref{lem:recover_basic_property}~\ref{enum:recover_basic_property:1} shows that 
$g_{s} = \lfloor \frac{K_{s} - \zeta}{|C|} \rfloor$ holds. 
$b \leq \lfloor \frac{K_{s} - \zeta}{|C|} \rfloor$ follows from 
$g_{s} \geq b$ and $g_{s} = \lfloor \frac{K_{s} - \zeta}{|C|} \rfloor$. 
Lemma~\ref{lem:sRecover_size_property}~\ref{enum:sRecover_size_property:3} shows that 
$|\Psi_{\sRecover}(([p_{s}, q_{s}], [\ell_{s}, r_{s}])) \cap (\bigcup_{\lambda = (b-1)|C| + \zeta}^{b|C| + \zeta - 1} \Psi_{\lcp}(\lambda))| = |\Psi_{\str}(([p_{s}, q_{s}], [\ell_{s}, r_{s}]))|$ holds 
because $1 \leq b \leq \lfloor \frac{K_{s} - \zeta}{|C|} \rfloor$. 
Therefore, Equation~\ref{eq:JD1_sum_sub:5} hols.

Next, we prove Equation~\ref{eq:JD1_sum_sub:6}. 
Because of $1 \leq b \leq \lfloor \frac{K_{s} - \zeta}{|C|} \rfloor$, 
Lemma~\ref{lem:sRecover_size_property}~\ref{enum:sRecover_size_property:4} shows that 
$|\Psi_{\sRecover}(([p_{s}, q_{s}], [\ell_{s}, r_{s}])) \cap \Psi_{\lcp}((b-1)|C| + \zeta + M_{s})| = |\Psi_{\str}(T[p_{s}-1..r_{s}+1])|$ 
holds. 
$(b-1)|C| + \zeta \leq (b-1)|C| + \zeta + M_{s} \leq (b-1)|C| + \zeta + M - 1$ follows from $0 \leq M_{s} \leq M-1$. 
This inequality $(b-1)|C| + \zeta \leq (b-1)|C| + \zeta + M_{s} \leq (b-1)|C| + \zeta + M - 1$ indicates that 
$\Psi_{\lcp}((b-1)|C| + \zeta + M_{s}) \subseteq \bigcup_{\lambda = (b-1)|C| + \zeta}^{(b-1)|C| + \zeta + M - 1} \Psi_{\lcp}(\lambda)$ holds.
Therefore, Equation~\ref{eq:JD1_sum_sub:6} follows from the following equation: 
\begin{equation*}
    \begin{split}
        |\Psi_{\sRecover}(([p_{s}, q_{s}], [\ell_{s}, r_{s}])) & \cap (\bigcup_{\lambda = (b-1)|C| + \zeta}^{(b-1)|C| + \zeta + M - 1} \Psi_{\lcp}(\lambda)))|  \\
        &\geq |\Psi_{\sRecover}(([p_{s}, q_{s}], [\ell_{s}, r_{s}])) \cap \Psi_{\lcp}((b-1)|C| + \zeta + M_{s})| \\
        &= |\Psi_{\str}(([p_{s}, q_{s}], [\ell_{s}, r_{s}]))|.
    \end{split}
\end{equation*}

Finally, $|(\Psi_{\sRecover}(([p_{s}, q_{s}], [\ell_{s}, r_{s}])) \cap (\bigcup_{\lambda = K-M}^{K - 1} \Psi_{\lcp}(\lambda)))| = |\Psi_{\str}(T[p_{s}-1..r_{s}+1])|$ follows from Equation~\ref{eq:JD1_sum_sub:3}, Equation~\ref{eq:JD1_sum_sub:4}, Equation~\ref{eq:JD1_sum_sub:5}, and Equation~\ref{eq:JD1_sum_sub:6}.

\textbf{Proof of Proposition~\ref{prop:JD1_sum_sub}(v).}
Because of $s \in \mathcal{I}_{D, 3}$, 
$b \leq g_{s} \leq n$ and $M_{s} = M$ follow from the definition of the set $\mathcal{I}_{D, 3}$. 
Because of $K \geq \zeta$, 
the two integers $b$ and $M$ are defined as $1 + \lfloor \frac{K - \zeta}{|C|} \rfloor$ and $(K - \zeta) \mod |C|$, respectively. 
$\lfloor \frac{K - \zeta}{|C|} \rfloor = \frac{K - M - \zeta}{|C|}$ follows from 
$K \geq \zeta$ and $M = (K - \zeta) \mod |C|$. 
$K - M = (b-1)|C| + \zeta$ follows from $b = 1 + \lfloor \frac{K - \zeta}{|C|} \rfloor$ and $\lfloor \frac{K - \zeta}{|C|} \rfloor = \frac{K - M - \zeta}{|C|}$. 

Lemma~\ref{lem:recover_basic_property}~\ref{enum:recover_basic_property:1} shows that 
$g_{s} = \lfloor \frac{K_{s} - \zeta}{|C|} \rfloor$ holds. 
$b \leq \lfloor \frac{K_{s} - \zeta}{|C|} \rfloor$ follows from 
$g_{s} \geq b$ and $g_{s} = \lfloor \frac{K_{s} - \zeta}{|C|} \rfloor$. 
Because of $1 \leq b \leq \lfloor \frac{K_{s} - \zeta}{|C|} \rfloor$, 
Lemma~\ref{lem:sRecover_size_property}~\ref{enum:sRecover_size_property:4} shows that 
$|\Psi_{\sRecover}(([p_{s}, q_{s}], [\ell_{s}, r_{s}])) \cap \Psi_{\lcp}((b-1)|C| + \zeta + M_{s})| = |\Psi_{\str}(T[p_{s}-1..r_{s}+1])|$ holds. 
$(b-1)|C| + \zeta + M_{s} = K$ follows from 
$K - M = (b-1)|C| + \zeta$ and $M_{s} = M$. 
Therefore, Proposition~\ref{prop:JD1_sum_sub}(v) follows from the following equation: 
\begin{equation*}
    \begin{split}
        |\Psi_{\sRecover}(([p_{s}, q_{s}], [\ell_{s}, r_{s}])) & \cap \Psi_{\lcp}(K)|  \\
        &= |\Psi_{\sRecover}(([p_{s}, q_{s}], [\ell_{s}, r_{s}])) \cap \Psi_{\lcp}((b-1)|C| + \zeta + M_{s})| \\
        &= |\Psi_{\str}(T[p_{s}-1..r_{s}+1])|.
    \end{split}
\end{equation*}

\end{proof}

We prove Lemma~\ref{lem:JD1_sum} using Proposition~\ref{prop:JD_sum_set_subseteq}, 
Proposition~\ref{prop:JD_sum_set_supseteq}, and Proposition~\ref{prop:JD1_sum_sub}. 

\begin{proof}[Proof of Lemma~\ref{lem:JD1_sum}]
The proof of Lemma~\ref{lem:JD1_sum} is as follows.
    
\paragraph{Proof of Equation~\ref{eq:JD1_sum:1}.}
$\sum_{s \in \mathcal{I}_{E, 1}} g_{s} |\Psi_{\str}(T[p_{s}-1..r_{s}+1])| = \rangesum(\mathcal{J}_{E}(h, C), 1, n, 0, |C|-1)$ follows from the definitions of the set $\mathcal{I}_{E, 1}$ and range-sum query. 
The following equation follows from Equation~\ref{eq:JD_sum_set_subseteq:E1}, Equation~\ref{eq:JD_sum_set_supseteq:E1}, 
Proposition~\ref{prop:JD1_sum_sub}~\ref{enum:JD1_sum_sub:2}, and $\sum_{s \in \mathcal{I}_{E, 1}} g_{s} |\Psi_{\str}(T[p_{s}-1..r_{s}+1])| = \rangesum(\mathcal{J}_{E}(h, C), 1, n, 0, |C|-1)$:
\begin{equation*}
    \begin{split}
    |\Psi_{h} \cap \Psi_{\run} & \cap \Psi_{\centerset}(C) \cap \Psi_{\preceding}| \\
    &= |\bigcup_{s \in \mathcal{I}_{E, 1}} \Psi_{\sRecover}(([p_{s}, q_{s}], [\ell_{s}, r_{s}]))| \\
    &= \sum_{s \in \mathcal{I}_{E, 1}} g_{s} |\Psi_{\str}(T[p_{s}-1..r_{s}+1])| \\
    &= \rangesum(\mathcal{J}_{E}(h, C), 1, n, 0, |C|-1).
    \end{split}
\end{equation*}
Therefore, Equation~\ref{eq:JD1_sum:1} holds.

\paragraph{Proof of Equation~\ref{eq:JD1_sum:2}.}
$\sum_{s \in \mathcal{I}_{D, 1}} |\Psi_{\str}(T[p_{s}-1..r_{s}+1])| = \rangesum(\mathcal{J}_{D}(h, C), b, n, 0, |C| - 1)$ follows from the definitions of the set $\mathcal{I}_{D, 1}$ and range-sum query. 
The following equation follows from 
Equation~\ref{eq:JD_sum_set_subseteq:D1}, Equation~\ref{eq:JD_sum_set_supseteq:D1}, 
Proposition~\ref{prop:JD1_sum_sub}~\ref{enum:JD1_sum_sub:1}, Proposition~\ref{enum:JD1_sum_sub:3}, 
and $\sum_{s \in \mathcal{I}_{D, 1}} |\Psi_{\str}(T[p_{s}-1..r_{s}+1])| = \rangesum(\mathcal{J}_{D}(h, C), b, n, 0, |C| - 1)$: 
\begin{equation*}
    \begin{split}
    |\Psi_{h} \cap \Psi_{\run} & \cap \Psi_{\centerset}(C) \cap \Psi_{\preceding} \cap  (\bigcup_{\lambda = b}^{n} \Psi_{\nRecover}(\lambda)) \cap (\bigcup_{\lambda = 0}^{K - M - 1} \Psi_{\lcp}(\lambda))| \\
    &= |\bigcup_{s \in \mathcal{I}_{D, 1}} \bigcup_{\lambda = 0}^{K - M - 1} (\Psi_{\sRecover}(([p_{s}, q_{s}], [\ell_{s}, r_{s}])) \cap \Psi_{\lcp}(\lambda))| \\
    &= |\bigcup_{s \in \mathcal{I}_{D, 1}} (\Psi_{\sRecover}(([p_{s}, q_{s}], [\ell_{s}, r_{s}])) \cap (\bigcup_{\lambda = 0}^{K - M - 1} \Psi_{\lcp}(\lambda)))| \\
    &= \sum_{s \in \mathcal{I}_{D, 1}} |(\Psi_{\sRecover}(([p_{s}, q_{s}], [\ell_{s}, r_{s}])) \cap (\bigcup_{\lambda = 0}^{K - M - 1} \Psi_{\lcp}(\lambda)))| \\
    &= (b-1) \sum_{s \in \mathcal{I}_{D, 1}} |\Psi_{\str}(T[p_{s}-1..r_{s}+1])| \\
    &= (b-1) \rangesum(\mathcal{J}_{D}(h, C), b, n, 0, |C| - 1).
    \end{split}
\end{equation*}
Therefore, Equation~\ref{eq:JD1_sum:2} holds. 

\paragraph{Proof of Equation~\ref{eq:JD1_sum:3}.}
$\sum_{s \in \mathcal{I}_{D, 2}} |\Psi_{\str}(T[p_{s}-1..r_{s}+1])| = \rangesum(\mathcal{J}_{D}(h, C), b, n, 0, M - 1)$ follows from the definitions of the set $\mathcal{I}_{D, 2}$ and range-sum query. 
The following equation follows from Equation~\ref{eq:JD_sum_set_subseteq:D2}, Equation~\ref{eq:JD_sum_set_supseteq:D2}, 
Proposition~\ref{prop:JD1_sum_sub}~\ref{enum:JD1_sum_sub:1}, Proposition~\ref{prop:JD1_sum_sub}~\ref{enum:JD1_sum_sub:4}, 
and $\sum_{s \in \mathcal{I}_{D, 2}} |\Psi_{\str}(T[p_{s}-1..r_{s}+1])| = \rangesum(\mathcal{J}_{D}(h, C), b, n, 0, M - 1)$:

\begin{equation*}
    \begin{split}
    |\Psi_{h} \cap \Psi_{\run} & \cap \Psi_{\centerset}(C) \cap \Psi_{\preceding} \cap (\bigcup_{\lambda = b}^{n} \Psi_{\nRecover}(\lambda)) \cap (\bigcup_{\lambda = K-M}^{K - 1} \Psi_{\lcp}(\lambda))| \\
    &= |\bigcup_{s \in \mathcal{I}_{D, 2}} \bigcup_{\lambda = K-M}^{K - 1} (\Psi_{\sRecover}(([p_{s}, q_{s}], [\ell_{s}, r_{s}])) \cap \Psi_{\lcp}(\lambda))| \\
    &= |\bigcup_{s \in \mathcal{I}_{D, 2}} (\Psi_{\sRecover}(([p_{s}, q_{s}], [\ell_{s}, r_{s}]))) \cap (\bigcup_{\lambda = K-M}^{K - 1} \Psi_{\lcp}(\lambda))| \\
    &= \sum_{s \in \mathcal{I}_{D, 2}} |\Psi_{\sRecover}(([p_{s}, q_{s}], [\ell_{s}, r_{s}])) \cap (\bigcup_{\lambda = K-M}^{K - 1} \Psi_{\lcp}(\lambda))| \\
    &= \sum_{s \in \mathcal{I}_{D, 2}} |\Psi_{\str}(T[p_{s}-1..r_{s}+1])| \\
    &= \rangesum(\mathcal{J}_{D}(h, C), b, n, 0, M - 1).
    \end{split}
\end{equation*}
Therefore, Equation~\ref{eq:JD1_sum:3} holds.

\paragraph{Proof of Equation~\ref{eq:JD1_sum:4}.}
$\sum_{s \in \mathcal{I}_{E, 2}} g_{s} |\Psi_{\str}(T[p_{s}-1..r_{s}+1])| = \rangesum(\mathcal{J}_{E}(h, C), 1, b-1, 0, |C|-1)$ follows from the definitions of the set $\mathcal{I}_{E, 2}$ and range-sum query. 
The following equation follows from 
Equation~\ref{eq:JD_sum_set_subseteq:E2}, Equation~\ref{eq:JD_sum_set_supseteq:E2}, 
Proposition~\ref{prop:JD1_sum_sub}~\ref{enum:JD1_sum_sub:2}, and $\sum_{s \in \mathcal{I}_{E, 2}} g_{s} |\Psi_{\str}(T[p_{s}-1..r_{s}+1])| = \rangesum(\mathcal{J}_{E}(h, C), 1, b-1, 0, |C|-1)$: 
\begin{equation*}
    \begin{split}
    |\Psi_{h} \cap \Psi_{\run} & \cap \Psi_{\centerset}(C) \cap \Psi_{\preceding} \cap (\bigcup_{\lambda = 1}^{b-1} \Psi_{\nRecover}(\lambda)) \cap (\bigcup_{\lambda = 0}^{K - 1} \Psi_{\lcp}(\lambda))| \\
    &= |\bigcup_{s \in \mathcal{I}_{E, 2}} \Psi_{\sRecover}(([p_{s}, q_{s}], [\ell_{s}, r_{s}]))| \\
    &= \sum_{s \in \mathcal{I}_{E, 2}} g_{s} |\Psi_{\str}(T[p_{s}-1..r_{s}+1])| \\
    &= \rangesum(\mathcal{J}_{E}(h, C), 1, b-1, 0, |C|-1).
    \end{split}
\end{equation*}
Therefore, Equation~\ref{eq:JD1_sum:4} holds. 

\paragraph{Proof of Equation~\ref{eq:JD1_sum:5}.}
$\sum_{s \in \mathcal{I}_{D, 3}} |\Psi_{\str}(T[p_{s}-1..r_{s}+1])| = \rangesum(\mathcal{J}_{D}(h, C), b, n, M, M)$ follows from 
the definitions of the set $\mathcal{I}_{D, 3}$ and range-sum query. 
The following equation follows from 
Equation~\ref{eq:JD_sum_set_subseteq:D3}, Equation~\ref{eq:JD_sum_set_supseteq:D3}, 
Proposition~\ref{prop:JD1_sum_sub}~\ref{enum:JD1_sum_sub:1}, Proposition~\ref{prop:JD1_sum_sub}~\ref{enum:JD1_sum_sub:5}, and 
$\sum_{s \in \mathcal{I}_{D, 3}} |\Psi_{\str}(T[p_{s}-1..r_{s}+1])| = \rangesum(\mathcal{J}_{D}(h, C), b, n, M, M)$: 
\begin{equation*}
    \begin{split}
    |\Psi_{h} \cap \Psi_{\run} \cap \Psi_{\centerset}(C) \cap \Psi_{\preceding} & \cap \Psi_{\lcp}(K)| \\ 
    &= |\bigcup_{s \in \mathcal{I}_{D, 3}} (\Psi_{\sRecover}(([p_{s}, q_{s}], [\ell_{s}, r_{s}])) \cap \Psi_{\lcp}(K))| \\
    &= \sum_{s \in \mathcal{I}_{D, 3}} |(\Psi_{\sRecover}(([p_{s}, q_{s}], [\ell_{s}, r_{s}])) \cap \Psi_{\lcp}(K))| \\
    &= \sum_{s \in \mathcal{I}_{D, 3}} |\Psi_{\str}(T[p_{s}-1..r_{s}+1])| \\
    &= \rangesum(\mathcal{J}_{D}(h, C), b, n, M, M).
    \end{split}
\end{equation*}
Therefore, Equation~\ref{eq:JD1_sum:5} holds. 
\end{proof}

\subsubsection{Proof of Lemma~\ref{lem:JD1_division}}\label{subsubsec:JD1_division_proof}
The following propositions are used to prove Lemma~\ref{lem:JD1_division}. 

\begin{proposition}\label{prop:psi_PS_basic_property}
Let $\gamma$ and $C$ be the attractor position and associated string of an interval attractor $([p, q], [\ell, r]) \in \Psi_{\RR}$, respectively. 
Here, $([p, q], [\ell, r]) \in \Psi_{\lcp}(K)$ holds for the length $K$ of 
the longest common prefix between two strings $T[\gamma..r]$ and $C^{n+1}$ 
(i.e., $K = |\lcp(T[\gamma..r], C^{n+1})|$). 
The following four statements hold: 
\begin{enumerate}[label=\textbf{(\roman*)}]
    \item \label{enum:psi_PS_basic_property:1}
    $\Psi_{\run} \cap \Psi_{\centerset}(C) \cap \Psi_{\preceding} \cap (\bigcup_{\lambda = 0}^{K-1} \Psi_{\lcp}(\lambda)) \subseteq \Psi_{\lex}(T[\gamma..r+1])$; 
    \item \label{enum:psi_PS_basic_property:2}
    if $([p, q], [\ell, r]) \in \Psi_{\preceding}$, then 
    $\Psi_{\run} \cap \Psi_{\centerset}(C) \cap \Psi_{\succeeding} \cap \Psi_{\lex}(T[\gamma..r+1]) = \emptyset$; 
    \item \label{enum:psi_PS_basic_property:3}
    if $([p, q], [\ell, r]) \not \in \Psi_{\preceding}$, then 
    $\Psi_{\run} \cap \Psi_{\centerset}(C) \cap \Psi_{\preceding} \subseteq \Psi_{\lex}(T[\gamma..r+1])$;
    \item \label{enum:psi_PS_basic_property:4}
    $\Psi_{\run} \cap \Psi_{\centerset}(C) \cap \Psi_{\succeeding} \cap (\bigcup_{\lambda = 0}^{K-1} \Psi_{\lcp}(\lambda)) \cap \Psi_{\lex}(T[\gamma..r+1]) = \emptyset$.
\end{enumerate}
\end{proposition}
\begin{proof}
    
    The proof of Proposition~\ref{prop:psi_PS_basic_property} is as follows. 

    \textbf{Proof of Proposition~\ref{prop:psi_PS_basic_property}(i).}
    Consider an interval attractor $([p^{\prime}, q^{\prime}], [\ell^{\prime}, r^{\prime}])$ in 
    the set $\Psi_{\run} \cap \Psi_{\centerset}(C) \cap \Psi_{\preceding} \cap (\bigcup_{\lambda = 0}^{K-1} \Psi_{\lcp}(\lambda))$. 
    Let $\gamma^{\prime}$ be the attractor position of the interval attractor $([p^{\prime}, q^{\prime}], [\ell^{\prime}, r^{\prime}])$. 
    Let $K^{\prime} = |\lcp(T[\gamma^{\prime}..r^{\prime}], C^{n+1})|$. 
    Then, $([p^{\prime}, q^{\prime}], [\ell^{\prime}, r^{\prime}]) \in \Psi_{\lcp}(K^{\prime})$ follows from the definition of the subset $\Psi_{\lcp}(K^{\prime})$. 
    $0 \leq K^{\prime} \leq K-1$ follows from $([p^{\prime}, q^{\prime}], [\ell^{\prime}, r^{\prime}]) \in \bigcup_{\lambda = 0}^{K-1} \Psi_{\lcp}(\lambda)$. 
    Because of $([p^{\prime}, q^{\prime}], [\ell^{\prime}, r^{\prime}]) \in \Psi_{\run}$, 
    Lemma~\ref{lem:psi_run_basic_property}~\ref{enum:psi_run_basic_property:8} shows that 
    $|\lcp(T[\gamma^{\prime}..r^{\prime} + 1], C^{n+1})| = K^{\prime}$ holds.
    
    We prove $T[\gamma^{\prime}..r^{\prime}+1] \prec C^{n+1}[1..K]$. 
    Because of $([p^{\prime}, q^{\prime}], [\ell^{\prime}, r^{\prime}]) \in \Psi_{\centerset}(C) \cap \Psi_{\preceding}$, 
    $T[\gamma^{\prime}..r^{\prime}+1] \prec C^{n+1}$ follows from the definition of the subset $\Psi_{\preceding}$. 
    $T[\gamma^{\prime}..r^{\prime}+1] \prec C^{n+1}[1..K^{\prime}+1]$ follows from 
    $|\lcp(T[\gamma^{\prime}..r^{\prime} + 1], C^{n+1})| = K^{\prime}$ and $T[\gamma^{\prime}..r^{\prime}+1] \prec C^{n+1}$. 
    $C^{n+1}[1..K^{\prime}+1] \preceq C^{n+1}[1..K]$ holds because $K^{\prime} \leq K-1$. 
    Therefore, $T[\gamma^{\prime}..r^{\prime}+1] \prec C^{n+1}[1..K]$ follows from 
    $T[\gamma^{\prime}..r^{\prime}+1] \prec C^{n+1}[1..K^{\prime}+1]$ and $C^{n+1}[1..K^{\prime}+1] \preceq C^{n+1}[1..K]$. 

    We prove $([p^{\prime}, q^{\prime}], [\ell^{\prime}, r^{\prime}]) \in \Psi_{\lex}(T[\gamma..r+1])$. 
    $C^{n+1}[1..K] = \lcp(T[\gamma..r], C^{n+1})$ follows from $K = |\lcp(T[\gamma..r], C^{n+1})|$.
    $C^{n+1}[1..K] \preceq T[\gamma..r]$ follows from $C^{n+1}[1..K] = \lcp(T[\gamma..r], C^{n+1})$. 
    $T[\gamma^{\prime}..r^{\prime}+1] \prec T[\gamma..r+1]$ follows from 
    $T[\gamma^{\prime}..r^{\prime}+1] \prec C^{n+1}[1..K]$, 
    $C^{n+1}[1..K] \preceq T[\gamma..r]$, and $T[\gamma..r] \prec T[\gamma..r+1]$. 
    Because of $T[\gamma^{\prime}..r^{\prime}+1] \prec T[\gamma..r+1]$, 
    $([p^{\prime}, q^{\prime}], [\ell^{\prime}, r^{\prime}]) \in \Psi_{\lex}(T[\gamma..r+1])$ 
    follows from the definition of the subset $\Psi_{\lex}(T[\gamma..r+1])$. 

    We showed that $([p^{\prime}, q^{\prime}], [\ell^{\prime}, r^{\prime}]) \in \Psi_{\lex}(T[\gamma..r+1])$ holds 
    for each interval attractor $([p^{\prime}, q^{\prime}]$, $[\ell^{\prime}, r^{\prime}]) \in \Psi_{\run} \cap \Psi_{\centerset}(C) \cap \Psi_{\preceding} \cap (\bigcup_{\lambda = 0}^{K-1} \Psi_{\lcp}(\lambda))$. 
    Therefore, $\Psi_{\run} \cap \Psi_{\centerset}(C) \cap \Psi_{\preceding} \cap (\bigcup_{\lambda = 0}^{K-1}$ $\Psi_{\lcp}(\lambda)) \subseteq \Psi_{\lex}(T[\gamma..r+1])$ holds. 

    \textbf{Proof of Proposition~\ref{prop:psi_PS_basic_property}(ii).}    
    We prove $\Psi_{\run} \cap \Psi_{\centerset}(C) \cap \Psi_{\succeeding} \cap \Psi_{\lex}(T[\gamma..r+1]) = \emptyset$ by contradiction. 
    We assume that 
    the set $\Psi_{\run} \cap \Psi_{\centerset}(C) \cap \Psi_{\succeeding} \cap \Psi_{\lex}(T[\gamma..r+1])$ 
    contains an interval attractor $([p^{\prime}, q^{\prime}], [\ell^{\prime}, r^{\prime}])$. 
    Because of $([p^{\prime}, q^{\prime}], [\ell^{\prime}, r^{\prime}]) \in \Psi_{\centerset}(C) \cap \Psi_{\succeeding}$, 
    $C^{n+1} \prec T[\gamma^{\prime}..r^{\prime}+1]$ follows from the definition of the subset $\Psi_{\succeeding}$ 
    for the attractor position $\gamma^{\prime}$ of the interval attractor $([p^{\prime}, q^{\prime}], [\ell^{\prime}, r^{\prime}])$. 

    On the other hand, $T[\gamma^{\prime}..r^{\prime}+1] \prec T[\gamma..r+1]$ follows from 
    $([p^{\prime}, q^{\prime}], [\ell^{\prime}, r^{\prime}]) \in \Psi_{\lex}(T[\gamma..r+1])$. 
    Because of $([p, q], [\ell, r]) \in \Psi_{\centerset}(C) \cap \Psi_{\preceding}$, 
    $T[\gamma..r+1] \prec C^{n+1}$ follows from the definition of the subset $\Psi_{\preceding}$. 
    $T[\gamma^{\prime}..r^{\prime}+1] \prec C^{n+1}$ follows from 
    $T[\gamma^{\prime}..r^{\prime}+1] \prec T[\gamma..r+1]$ and $T[\gamma..r+1] \prec C^{n+1}$. 
    The two facts $C^{n+1} \prec T[\gamma^{\prime}..r^{\prime}+1]$ and $T[\gamma^{\prime}..r^{\prime}+1] \prec C^{n+1}$ yield a contradiction. 
    Therefore, $\Psi_{\run} \cap \Psi_{\centerset}(C) \cap \Psi_{\succeeding} \cap \Psi_{\lex}(T[\gamma..r+1]) = \emptyset$ must hold. 
    
    \textbf{Proof of Proposition~\ref{prop:psi_PS_basic_property}(iii).}
    $([p, q], [\ell, r]) \in \Psi_{\succeeding}$ follows from 
    $\Psi_{\RR} = \Psi_{\preceding} \cup \Psi_{\succeeding}$ and $([p, q], [\ell, r]) \not \in \Psi_{\preceding}$. 
    Because of $([p, q], [\ell, r]) \in \Psi_{\centerset}(C) \cap \Psi_{\succeeding}$, 
    $C^{n+1} \prec T[\gamma..r+1]$ follows from the definition of the subset $\Psi_{\succeeding}$.

    We prove $([p^{\prime}, q^{\prime}], [\ell^{\prime}, r^{\prime}]) \in \Psi_{\lex}(T[\gamma..r+1])$ for 
    each interval attractor $([p^{\prime}, q^{\prime}], [\ell^{\prime}, r^{\prime}])$ in the set $\Psi_{\run} \cap \Psi_{\centerset}(C) \cap \Psi_{\preceding}$. 
    Because of $([p^{\prime}, q^{\prime}], [\ell^{\prime}, r^{\prime}]) \in \Psi_{\centerset}(C) \cap \Psi_{\preceding}$, 
    $T[\gamma^{\prime}..r^{\prime}+1] \prec C^{n+1}$ follows from the definition of the subset $\Psi_{\preceding}$ 
    for the attractor position $\gamma^{\prime}$ of the interval attractor $([p^{\prime}, q^{\prime}], [\ell^{\prime}, r^{\prime}])$. 
    $T[\gamma^{\prime}..r^{\prime}+1] \prec T[\gamma..r+1]$ follows from 
    $T[\gamma^{\prime}..r^{\prime}+1] \prec C^{n+1}$ and $C^{n+1} \prec T[\gamma..r+1]$. 
    Therefore, $([p^{\prime}, q^{\prime}], [\ell^{\prime}, r^{\prime}]) \in \Psi_{\lex}(T[\gamma..r+1])$ 
    follows from $T[\gamma^{\prime}..r^{\prime}+1] \prec T[\gamma..r+1]$. 

    \textbf{Proof of Proposition~\ref{prop:psi_PS_basic_property}(iv).} 
    Consider an interval attractor $([p^{\prime}, q^{\prime}], [\ell^{\prime}, r^{\prime}])$ in 
    the set $\Psi_{\run} \cap \Psi_{\centerset}(C) \cap \Psi_{\succeeding} \cap (\bigcup_{\lambda = 0}^{K-1} \Psi_{\lcp}(\lambda))$. 
    Let $\gamma^{\prime}$ be the attractor position of the interval attractor $([p^{\prime}, q^{\prime}], [\ell^{\prime}, r^{\prime}])$. 
    Let $K^{\prime} = |\lcp(T[\gamma^{\prime}..r^{\prime}], C^{n+1})|$. 
    Then, $([p^{\prime}, q^{\prime}], [\ell^{\prime}, r^{\prime}]) \in \Psi_{\lcp}(K^{\prime})$ follows from the definition of the subset $\Psi_{\lcp}(K^{\prime})$. 
    $0 \leq K^{\prime} \leq K-1$ follows from $([p^{\prime}, q^{\prime}], [\ell^{\prime}, r^{\prime}]) \in \bigcup_{\lambda = 0}^{K-1} \Psi_{\lcp}(\lambda)$. 
    Because of $([p^{\prime}, q^{\prime}], [\ell^{\prime}, r^{\prime}]) \in \Psi_{\run}$, 
    Lemma~\ref{lem:psi_run_basic_property}~\ref{enum:psi_run_basic_property:8} shows that 
    $|\lcp(T[\gamma^{\prime}..r^{\prime} + 1], C^{n+1})| = K^{\prime}$ holds.

    We prove $C^{n+1}[1..K]\# \prec T[\gamma^{\prime}..r^{\prime} + 1]$. 
    Because of $([p^{\prime}, q^{\prime}], [\ell^{\prime}, r^{\prime}]) \in \Psi_{\centerset}(C) \cap \Psi_{\succeeding}$, 
    $C^{n+1} \prec T[\gamma^{\prime}..r^{\prime}+1]$ follows from the definition of the subset $\Psi_{\succeeding}$. 
    $C^{n+1}[1..K^{\prime}+1]\# \prec T[\gamma^{\prime}..r^{\prime}+1]$ follows from 
    $|\lcp(T[\gamma^{\prime}..r^{\prime} + 1], C^{n+1})| = K^{\prime}$ and $C^{n+1} \prec T[\gamma^{\prime}..r^{\prime}+1]$. 
    The string $C^{n+1}[1..K^{\prime}+1]$ is a proper prefix of string $C^{n+1}[1..K]\#$ 
    because $K^{\prime}+1 \leq K$. 
    $C^{n+1}[1..K]\# \preceq C^{n+1}[1..K^{\prime}+1]\#$ follows from 
    the fact that the string $C^{n+1}[1..K^{\prime}+1]$ is a proper prefix of string $C^{n+1}[1..K]\#$. 
    Therefore, $C^{n+1}[1..K]\# \prec T[\gamma^{\prime}..r^{\prime} + 1]$ follows from 
    $C^{n+1}[1..K]\# \preceq C^{n+1}[1..K^{\prime}+1]\#$ and $C^{n+1}[1..K^{\prime}+1]\# \prec T[\gamma^{\prime}..r^{\prime}+1]$. 

    We prove $([p^{\prime}, q^{\prime}], [\ell^{\prime}, r^{\prime}]) \not \in \Psi_{\lex}(T[\gamma..r+1])$. 
    $T[\gamma..r+1] \preceq C^{n+1}[1..K]\#$ follows from $K = |\lcp(T[\gamma..r], C^{n+1})|$. 
    $T[\gamma..r+1] \prec T[\gamma^{\prime}..r^{\prime} + 1]$ follows from 
    $T[\gamma..r+1] \preceq C^{n+1}[1..K]\#$ and $C^{n+1}[1..K]\# \prec T[\gamma^{\prime}..r^{\prime} + 1]$. 
    Therefore, $([p^{\prime}, q^{\prime}], [\ell^{\prime}, r^{\prime}]) \not \in \Psi_{\lex}(T[\gamma..r+1])$ 
    follows from $T[\gamma..r+1] \prec T[\gamma^{\prime}..r^{\prime} + 1]$. 

    We showed that $([p^{\prime}, q^{\prime}], [\ell^{\prime}, r^{\prime}]) \not \in \Psi_{\lex}(T[\gamma..r+1])$ holds 
    for each interval attractor $([p^{\prime}, q^{\prime}], [\ell^{\prime}, r^{\prime}])$ in 
    the set $\Psi_{\run} \cap \Psi_{\centerset}(C) \cap \Psi_{\succeeding} \cap (\bigcup_{\lambda = 0}^{K-1} \Psi_{\lcp}(\lambda))$. 
    Therefore, $\Psi_{\run} \cap \Psi_{\centerset}(C) \cap \Psi_{\succeeding} \cap (\bigcup_{\lambda = 0}^{K-1} \Psi_{\lcp}(\lambda)) \cap \Psi_{\lex}(T[\gamma..r+1]) = \emptyset$ holds. 
\end{proof}

\begin{proposition}\label{prop:psi_condiiton_D_super_property}
Consider the RSC query $\RSCQ(i, j)$ satisfying condition (D) of RSC query stated in Section~\ref{subsec:rsc_sub}.  
If $\Psi_{h_{Q}} \cap \Psi_{\run} \cap \Psi_{\centerset}(C_{Q}) \cap (\bigcup_{\lambda = |[\gamma_{Q}, j]|}^{n} \Psi_{\lcp}(\lambda)) \neq \emptyset$, 
then $|\lcp(T[\gamma_{Q}..r_{Q}+1], C_{Q}^{n+1})| = K_{Q}$. 
\end{proposition}
\begin{proof}
Consider an interval attractor $([p, q], [\ell, r])$ in the set $\Psi_{h_{Q}} \cap \Psi_{\run} \cap \Psi_{\centerset}(C_{Q}) \cap (\bigcup_{\lambda = K_{Q}+1}^{n}$ $\Psi_{\lcp}(\lambda))$. 
Let $K = |\lcp(T[\gamma..r], C_{Q}^{n+1})|$ for the attractor position $\gamma$ of the interval attractor $([p, q]$, $[\ell, r])$. 
Then, $K_{Q} + 1 \leq K \leq n$ follows from 
$([p, q], [\ell, r]) \in \Psi_{\centerset}(C_{Q}) \cap (\bigcup_{\lambda = K_{Q}+1}^{n} \Psi_{\lcp}(\lambda))$. 

The following five statements are used to prove Proposition~\ref{prop:psi_condiiton_D_super_property}: 
\begin{enumerate}[label=\textbf{(\roman*)}]
    \item $T[i..j] = T[\gamma - |[i, \gamma_{Q}-1]|..\gamma + |[\gamma_{Q}, j]| - 1]$; 
    \item $I_{\capture}(\gamma - |[i, \gamma_{Q}-1]|, \gamma + |[\gamma_{Q}, j]| - 1) = ([p, q], [\ell, r])$ 
    for interval attractor $I_{\capture}(\gamma - |[i, \gamma_{Q}-1]|, \gamma + |[\gamma_{Q}, j]| - 1)$; 
    \item $I_{\capture}(\gamma - |[i, \gamma_{Q}-1]|, \gamma + K_{Q}) = ([p, q], [\ell, r])$ 
    for interval attractor $I_{\capture}(\gamma - |[i, \gamma_{Q}-1]|, \gamma + K_{Q})$;
    \item if $|\lcp(T[\gamma_{Q}..r_{Q}+1], C_{Q}^{n+1})| = |[\gamma_{Q}, r_{Q}+1]|$, 
    then $K_{Q} = |[\gamma_{Q}, r_{Q}]|$  
    and $I_{\capture}(i, \gamma_{Q} + K_{Q}) = ([p_{Q}, q_{Q}], [\ell_{Q}, r_{Q}])$ 
    for interval attractor $I_{\capture}(i, \gamma_{Q} + K_{Q})$;    
    \item $|\lcp(T[\gamma_{Q}..r_{Q}+1], C_{Q}^{n+1})| < |[\gamma_{Q}, r_{Q}+1]|$.
\end{enumerate}

    \textbf{Proof of statement (i).}
    We prove $T[\gamma_{Q}..j] = T[\gamma - |[i, \gamma_{Q}-1]|..\gamma + |[\gamma_{Q}, j]| - 1]$.
    $T[\gamma_{Q}..\gamma_{Q} + K_{Q}-1] = T[\gamma..\gamma + K_{Q}-1]$ 
    follows from 
    $|\lcp(T[\gamma_{Q}..r_{Q}], C_{Q}^{n+1})| = K_{Q}$
    $|\lcp(T[\gamma..r], C_{Q}^{n+1})| = K$, 
    and $K_{Q} + 1 \leq K$. 
    $T[\gamma_{Q}..\gamma_{Q} + |[\gamma_{Q}, j]|-1] = T[\gamma..\gamma + |[\gamma_{Q}, j]|-1]$ 
    follows from 
    $T[\gamma_{Q}..\gamma_{Q} + K_{Q}-1] = T[\gamma..\gamma + K_{Q}-1]$ 
    and $|[\gamma_{Q}, j]| \leq K_{Q}$. 
    $\gamma_{Q} + |[\gamma_{Q}, j]|-1 = j$. 
    Therefore, $T[\gamma_{Q}..j] = T[\gamma - |[i, \gamma_{Q}-1]|..\gamma + |[\gamma_{Q}, j]| - 1]$ holds. 

    We prove $T[i..j] = T[\gamma - |[i, \gamma_{Q}-1]|..\gamma + |[\gamma_{Q}, j]| - 1]$. 
    We can apply Lemma~\ref{lem:suffix_syncro} to the two interval attractors 
    $([p_{Q}, q_{Q}], [\ell_{Q}, r_{Q}])$ and $([p, q], [\ell, r])$. 
    This lemma shows that $T[i..\gamma_{Q}-1] = T[\gamma - |[i, \gamma_{Q}-1]|..\gamma-1]$ holds. 
    Therefore, $T[i..j] = T[\gamma - |[i, \gamma_{Q}-1]|..\gamma + |[\gamma_{Q}, j]| - 1]$ 
    follows from 
    $T[i..\gamma_{Q}-1] = T[\gamma - |[i, \gamma_{Q}-1]|..\gamma-1]$ and 
    $T[\gamma_{Q}..j] = T[\gamma - |[i, \gamma_{Q}-1]|..\gamma + |[\gamma_{Q}, j]| - 1]$. 

    \textbf{Proof of statement (ii).}    
    We prove $I_{\capture}(\gamma - |[i, \gamma_{Q}-1]|, \gamma + |[\gamma_{Q}, j]| - 1) \in \Psi_{h_{Q}}$ and 
    $\gamma_{A} = \gamma$ for the attractor position $\gamma_{A}$ of 
    the interval attractor $I_{\capture}(\gamma - |[i, \gamma_{Q}-1]|, \gamma + K_{Q}])$.     
    Consider interval attractor $I_{\capture}(i, j) = ([p_{Q}, q_{Q}], [\ell_{Q}, r_{Q}])$. 
    We can apply Corollary~\ref{cor:capture_gamma_corollary} to two interval attractors  
    $I_{\capture}(i, j)$ and $I_{\capture}(\gamma - |[i, \gamma_{Q}-1]|, \gamma + |[\gamma_{Q}, j]| - 1)$  
    because $T[i..j] = T[\gamma - |[i, \gamma_{Q}-1]|..\gamma + |[\gamma_{Q}, j]| - 1]$ holds (statement (i)).
    Corollary~\ref{cor:capture_gamma_corollary} shows that 
    $I_{\capture}(\gamma - |[i, \gamma_{Q}-1]|, \gamma + |[\gamma_{Q}, j]| - 1) \in \Psi_{h_{Q}}$ 
    and $|[i, \gamma_{Q}-1]| = |[\gamma_{A} - |[i, \gamma_{Q}-1]|, \gamma_{A}-1]|$ hold. 
    We already proved $T[i..\gamma_{Q}-1] = T[\gamma - |[i, \gamma_{Q}-1]|..\gamma-1]$ in the proof of statement (i).
    $|[i, \gamma_{Q}-1]| = |[\gamma - |[i, \gamma_{Q}-1]|, \gamma-1]|$ follows from 
    $T[i..\gamma_{Q}-1] = T[\gamma - |[i, \gamma_{Q}-1]|..\gamma-1]$. 
    $|[\gamma_{A} - |[i, \gamma_{Q}-1]|, \gamma_{A}-1]| = |[\gamma - |[i, \gamma_{Q}-1]|, \gamma-1]|$ follows from 
    $|[i, \gamma_{Q}-1]| = |[\gamma_{A} - |[i, \gamma_{Q}-1]|, \gamma_{A}-1]|$ and $|[i, \gamma_{Q}-1]| = |[\gamma - |[i, \gamma_{Q}-1]|, \gamma-1]|$. 
    Therefore, $\gamma_{A} = \gamma$ follows from $|[\gamma_{A} - |[i, \gamma_{Q}-1]|, \gamma_{A}-1]| = |[\gamma - |[i, \gamma_{Q}-1]|, \gamma-1]|$. 
    
    Corollary~\ref{cor:IA_identify_corollary} shows that 
    $I_{\capture}(\gamma - |[i, \gamma_{Q}-1]|, \gamma + |[\gamma_{Q}, j]| - 1) = ([p, q], [\ell, r])$ holds 
    because $I_{\capture}(\gamma - |[i, \gamma_{Q}-1]|, \gamma + |[\gamma_{Q}, j]| - 1) \in \Psi_{h_{Q}}$ and $\gamma_{A} = \gamma$. 
    Therefore, statement (ii) holds. 
    
    \textbf{Proof of statement (iii).}
    We prove $\ell \leq \gamma + |[\gamma_{Q}, j]| - 1 \leq \gamma + K_{Q} \leq r$. 
    Since $I_{\capture}(\gamma - |[i, \gamma_{Q}-1]|, \gamma + |[\gamma_{Q}, j]| - 1) = ([p, q], [\ell, r])$, 
    $\gamma - |[i, \gamma_{Q}-1]| \in [p, q]$ and $\gamma + |[\gamma_{Q}, j]| - 1 \in [\ell, r]$ follow from 
    the definition of interval attractor. 
    $\gamma + |[\gamma_{Q}, j]| - 1 \leq \gamma + K_{Q}$ follows from $|[\gamma_{Q}, j]| \leq K_{Q}$. 
    $\gamma + K - 1 \leq r$ follows from $K = |\lcp(T[\gamma..r], C_{Q}^{n+1})|$. 
    $\gamma + K_{Q} \leq \gamma + K - 1$ follows from $K_{Q} + 1 \leq K$. 
    Therefore, $\ell \leq \gamma + |[\gamma_{Q}, j]| - 1 \leq \gamma + K_{Q} \leq r$ 
    follows from $\gamma + |[\gamma_{Q}, j]| - 1 \in [\ell, r]$, 
    $\gamma + |[\gamma_{Q}, j]| - 1 \leq \gamma + K_{Q}$, 
    $\gamma + K_{Q} \leq \gamma + K - 1$, and $\gamma + K - 1 \leq r$. 

    We prove 
    $I_{\capture}(\gamma - |[i, \gamma_{Q}-1]|, \gamma + K_{Q}) = ([p, q], [\ell, r])$.
    We can apply Lemma~\ref{lem:IA_maximal_lemma} to the two interval attractors 
    $I_{\capture}(\gamma - |[i, \gamma_{Q}-1]|, \gamma + K_{Q})$ 
    and $I_{\capture}(\gamma - |[i, \gamma_{Q}-1]|, \gamma + |[\gamma_{Q}, j]| - 1)$ 
    because $I_{\capture}(\gamma - |[i, \gamma_{Q}-1]|, \gamma + |[\gamma_{Q}, j]| - 1) = ([p, q], [\ell, r])$ 
    and $\ell \leq \gamma + |[\gamma_{Q}, j]| - 1 \leq \gamma + K_{Q} \leq r$. 
    This lemma shows that $I_{\capture}(\gamma - |[i, \gamma_{Q}-1]|, \gamma + K_{Q}) = ([p, q], [\ell, r])$. 

    \textbf{Proof of statement (iv).}
    We prove $K_{Q} = |[\gamma_{Q}, r_{Q}]|$. 
    $|\lcp(T[\gamma_{Q}..r_{Q}], C_{Q}^{n+1})| = |[\gamma_{Q}, r_{Q}+1]| - 1$ follows from 
    $|\lcp(T[\gamma_{Q}..r_{Q}+1], C_{Q}^{n+1})| = |[\gamma_{Q}, r_{Q}+1]|$. 
    Therefore, $K_{Q} = |[\gamma_{Q}, r_{Q}]|$ follows from 
    $|\lcp(T[\gamma_{Q}..r_{Q}], C_{Q}^{n+1})| = K_{Q}$ and 
    $|\lcp(T[\gamma_{Q}..r_{Q}], C_{Q}^{n+1})| = |[\gamma_{Q}, r_{Q}+1]| - 1$. 
    
    We prove $T[i..\gamma_{Q} + K_{Q}] = T[\gamma - |[i, \gamma_{Q}-1]|..\gamma + K_{Q}]$. 
    $|\lcp(T[\gamma_{Q}..r_{Q}+1], C_{Q}^{n+1})| = K_{Q} + 1$ follows from 
    $|\lcp(T[\gamma_{Q}..r_{Q}+1], C_{Q}^{n+1})| = |[\gamma_{Q}, r_{Q}+1]|$ and $K_{Q} = |[\gamma_{Q}, r_{Q}]|$. 
    $T[\gamma_{Q}..\gamma_{Q} + K_{Q}] = T[\gamma..\gamma + K_{Q}]$ 
    follows from 
    $|\lcp(T[\gamma_{Q}..r_{Q}+1], C_{Q}^{n+1})| = K_{Q} + 1$, 
    $|\lcp(T[\gamma..r], C_{Q}^{n+1})| = K$, 
    and $K_{Q} + 1 \leq K$. 
    We already showed that $T[i..\gamma_{Q}-1] = T[\gamma - |[i, \gamma_{Q}-1]|..\gamma-1]$ holds 
    in the proof of statement (i). 
    Therefore, $T[i..\gamma_{Q} + K_{Q}] = T[\gamma - |[i, \gamma_{Q}-1]|..\gamma + K_{Q}]$ 
    follows from $T[i..\gamma_{Q}-1] = T[\gamma - |[i, \gamma_{Q}-1]|..\gamma-1]$ and $T[\gamma_{Q}..\gamma_{Q} + K_{Q}] = T[\gamma..\gamma + K_{Q}]$. 

    Consider interval attractor $I_{\capture}(i, \gamma_{Q} + K_{Q})$.     
    We prove $I_{\capture}(i, \gamma_{Q} + K_{Q}) \in \Psi_{h_{Q}}$ and 
    $\gamma_{A} = \gamma_{Q}$ for the attractor position $\gamma_{A}$ of 
    the interval attractor $I_{\capture}(i, \gamma_{Q} + K_{Q})$. 
    We can apply Corollary~\ref{cor:capture_gamma_corollary} to 
    the two interval attractors $I_{\capture}(i, \gamma_{Q} + K_{Q})$ 
    and $I_{\capture}(\gamma - |[i, \gamma_{Q}-1]|, \gamma + K_{Q})$
    because $T[i..\gamma_{Q} + K_{Q}] = T[\gamma - |[i, \gamma_{Q}-1]|..\gamma + K_{Q}]$ holds. 
    The theorem shows that 
    $I_{\capture}(i, \gamma_{Q} + K_{Q}) \in \Psi_{h_{Q}}$ and 
    $|[\gamma - |[i, \gamma_{Q}-1]|, \gamma-1]| = |[i, \gamma_{A}-1]|$ hold 
    for the attractor position $\gamma_{A}$ of the interval attractor $([p_{A}, q_{A}], [\ell_{A}, r_{A}])$. 
    Therefore, 
    $\gamma_{Q} = \gamma_{A}$ follows from $|[\gamma - |[i, \gamma_{Q}-1]|, \gamma-1]| = |[i, \gamma_{A}-1]|$. 

    Corollary~\ref{cor:IA_identify_corollary} shows that 
    $I_{\capture}(i, \gamma_{Q} + K_{Q}) = ([p_{Q}, q_{Q}], [\ell_{Q}, r_{Q}])$ holds 
    because $I_{\capture}(i, \gamma_{Q} + K_{Q}), ([p_{Q}, q_{Q}], [\ell_{Q}, r_{Q}]) \in \Psi_{h_{Q}}$ and $\gamma_{A} = \gamma_{Q}$. 
    Therefore, statement (iv) holds. 

    \textbf{Proof of statement (v).}
    We prove $|\lcp(T[\gamma_{Q}..r_{Q}+1], C_{Q}^{n+1})| < |[\gamma_{Q}, r_{Q}+1]|$ by contradiction. 
    Here, $|\lcp(T[\gamma_{Q}..r_{Q}+1], C_{Q}^{n+1})| \leq |[\gamma_{Q}, r_{Q}+1]|$ holds. 
    We assume that $|\lcp(T[\gamma_{Q}..r_{Q}+1], C_{Q}^{n+1})| \geq |[\gamma_{Q}, r_{Q}+1]|$ holds. 
    Then, $|\lcp(T[\gamma_{Q}..r_{Q}+1], C_{Q}^{n+1})| = |[\gamma_{Q}, r_{Q}+1]|$ follows from 
    $|\lcp(T[\gamma_{Q}..r_{Q}+1], C_{Q}^{n+1})| \geq |[\gamma_{Q}, r_{Q}+1]|$ and $|\lcp(T[\gamma_{Q}..r_{Q}+1], C_{Q}^{n+1})| \leq |[\gamma_{Q}, r_{Q}+1]|$. 
    Statement (iv) shows that 
    $K_{Q} = |[\gamma_{Q}, r_{Q}]|$ and 
    $I_{\capture}(i, \gamma_{Q} + K_{Q}) = ([p_{Q}, q_{Q}], [\ell_{Q}, r_{Q}])$ hold. 
    Since $I_{\capture}(i, \gamma_{Q} + K_{Q}) = ([p_{Q}, q_{Q}], [\ell_{Q}, r_{Q}])$, 
    $i \in [p_{Q}, q_{Q}]$ and $\gamma_{Q} + K_{Q} \in [\ell_{Q}, r_{Q}]$) follow from the definition of interval attractor. 
    On the other hand, $\gamma_{Q} + K_{Q} = r_{Q} + 1$ follows from $K_{Q} = |[\gamma_{Q}, r_{Q}]|$. 
    $\gamma_{Q} + K_{Q} \not \in [\ell_{Q}, r_{Q}]$ follows from $\gamma_{Q} + K_{Q} = r_{Q} + 1$. 
    The two facts $\gamma_{Q} + K_{Q} \in [\ell_{Q}, r_{Q}]$ and $\gamma_{Q} + K_{Q} \not \in [\ell_{Q}, r_{Q}]$ yield a contradiction. 
    Therefore, $|\lcp(T[\gamma_{Q}..r_{Q}+1], C_{Q}^{n+1})| < |[\gamma_{Q}, r_{Q}+1]|$ must hold.

    \textbf{Proof of Proposition~\ref{prop:psi_succ_super_property}.}
    $K \leq |[\gamma_{Q}, r_{Q}]|$ follows from $|\lcp(T[\gamma..r], C_{Q}^{n+1})| = K$. 
    If $K < |[\gamma_{Q}, r_{Q}]|$, 
    then $|\lcp(T[\gamma..r], C_{Q}^{n+1})| = |\lcp(T[\gamma..r+1], C_{Q}^{n+1})|$ holds. 
    Therefore, $|\lcp(T[\gamma..r+1]$, $C_{Q}^{n+1})| = K$ follows from 
    $|\lcp(T[\gamma..r], C_{Q}^{n+1})| = |\lcp(T[\gamma..r+1], C_{Q}^{n+1})|$ and $|\lcp(T[\gamma..r], C_{Q}^{n+1})| = K$. 

    Otherwise (i.e., $K = |[\gamma_{Q}, r_{Q}]|$), 
    either $|\lcp(T[\gamma..r+1], C_{Q}^{n+1})| = K$ or $|\lcp(T[\gamma..r+1], C_{Q}^{n+1})| = K+1$ holds. 
    In this case, 
    statement (v) shows that $|\lcp(T[\gamma..r+1], C_{Q}^{n+1})| < K+1$ holds. 
    Therefore, $|\lcp(T[\gamma..r+1], C_{Q}^{n+1})| = K$ holds. 
\end{proof}

\begin{proposition}\label{prop:psi_condition_D_prec_super_property}
Consider the RSC query $\RSCQ(i, j)$ satisfying condition (D) of RSC query stated in Section~\ref{subsec:rsc_sub}. 
If $([p_{Q}, q_{Q}], [\ell_{Q}, r_{Q}]) \in \Psi_{\preceding}$, then 
$\Psi_{h_{Q}} \cap \Psi_{\run} \cap \Psi_{\centerset}(C_{Q}) \cap (\bigcup_{\lambda = K_{Q}+1}^{n}$ $\Psi_{\lcp}(\lambda)) \cap \Psi_{\lex}(T[\gamma_{Q}..r_{Q}+1]) = \emptyset$ holds.  
\end{proposition}
\begin{proof}
If $\Psi_{h_{Q}} \cap \Psi_{\run} \cap \Psi_{\centerset}(C_{Q}) \cap (\bigcup_{\lambda = K_{Q}+1}^{n} \Psi_{\lcp}(\lambda)) = \emptyset$, 
then $\Psi_{h_{Q}} \cap \Psi_{\run} \cap \Psi_{\centerset}(C_{Q}) \cap (\bigcup_{\lambda = K_{Q}+1}^{n} \Psi_{\lcp}(\lambda)) \cap \Psi_{\lex}(T[\gamma_{Q}..r_{Q}+1]) = \emptyset$ holds. 
Otherwise, 
the set $\Psi_{h_{Q}} \cap \Psi_{\run} \cap \Psi_{\centerset}(C_{Q}) \cap (\bigcup_{\lambda = K_{Q}+1}^{n} \Psi_{\lcp}(\lambda))$ contains an interval attractor $([p, q], [\ell, r])$. 
Let $K = |\lcp(T[\gamma..r], C_{Q}^{n+1})|$ for the attractor position of the interval attractor $([p, q], [\ell, r])$. 
Then, $K_{Q}+1 \leq K \leq n$ 
follows from $([p, q], [\ell, r]) \in \Psi_{\centerset}(C_{Q}) \cap (\bigcup_{\lambda = K_{Q}+1}^{n} \Psi_{\lcp}(\lambda))$.

We prove $T[\gamma_{Q}..r_{Q}+1] \prec C_{Q}^{n+1}[1..K_{Q}+1]$. 
Because of $\Psi_{h_{Q}} \cap \Psi_{\run} \cap \Psi_{\centerset}(C_{Q}) \cap (\bigcup_{\lambda = K_{Q}+1}^{n}$ $\Psi_{\lcp}(\lambda)) \neq \emptyset$, 
Proposition~\ref{prop:psi_condiiton_D_super_property} shows that $|\lcp(T[\gamma_{Q}..r_{Q}+1], C_{Q}^{n+1})| = K_{Q}$ holds. 
Because of $([p_{Q}, q_{Q}], [\ell_{Q}, r_{Q}]) \in \Psi_{\centerset}(C_{Q}) \cap \Psi_{\preceding}$, 
$T[\gamma_{Q}..r_{Q}+1] \prec C_{Q}^{n+1}$ follows from the definition of the subset $\Psi_{\preceding}$. 
Therefore, $T[\gamma_{Q}..r_{Q}+1] \prec C_{Q}^{n+1}[1..K_{Q}+1]$ follows from 
$|\lcp(T[\gamma_{Q}..r_{Q}+1], C_{Q}^{n+1})| = K_{Q}$ and $T[\gamma_{Q}..r_{Q}+1] \prec C_{Q}^{n+1}$. 

We prove $([p, q], [\ell, r]) \not \in \Psi_{\lex}(T[\gamma_{Q}..r_{Q}+1])$. 
$C_{Q}^{n+1}[1..K] \prec T[\gamma..r+1]$ follows from $K = |\lcp(T[\gamma..r], C_{Q}^{n+1})|$. 
$C_{Q}^{n+1}[1..K_{Q}+1] \preceq C_{Q}^{n+1}[1..K]$ holds because $K_{Q}+1 \leq K$. 
$T[\gamma_{Q}..r_{Q}+1] \prec T[\gamma..r+1]$ follows from 
$T[\gamma_{Q}..r_{Q}+1] \prec C_{Q}^{n+1}[1..K_{Q}+1]$, $C_{Q}^{n+1}[1..K_{Q}+1] \preceq C_{Q}^{n+1}[1..K]$, 
and $C_{Q}^{n+1}[1..K] \prec T[\gamma..r+1]$. 
Therefore, $([p, q], [\ell, r]) \not \in \Psi_{\lex}(T[\gamma_{Q}..r_{Q}+1])$ follows from $T[\gamma_{Q}..r_{Q}+1] \prec T[\gamma..r+1]$. 

We showed that $([p, q], [\ell, r]) \not \in \Psi_{\lex}(T[\gamma_{Q}..r_{Q}+1])$ holds for each interval attractor 
$([p, q], [\ell, r]) \in \Psi_{h_{Q}} \cap \Psi_{\run} \cap \Psi_{\centerset}(C_{Q}) \cap (\bigcup_{\lambda = K_{Q}+1}^{n} \Psi_{\lcp}(\lambda))$. 
Therefore, we obtain $\Psi_{h_{Q}} \cap \Psi_{\run} \cap \Psi_{\centerset}(C_{Q}) \cap (\bigcup_{\lambda = K_{Q}+1}^{n} \Psi_{\lcp}(\lambda)) \cap \Psi_{\lex}(T[\gamma_{Q}..r_{Q}+1]) = \emptyset$. 
\end{proof}

\begin{proposition}\label{prop:set_h_HR_center_prec_lcp_sum}
Consider a triplet of an integer $h \in [0, H]$, a string $C \in \Sigma^{+}$, 
and an integer $K \in [1, n]$. 
Let $b$ and $M$ be the two integers defined as follows: 
\begin{itemize}
    \item $b = 1 + \lfloor \frac{K - (2 + \sum_{w = 1}^{h+3} \lfloor \mu(w) \rfloor)}{|C|} \rfloor$ if $K > 1 + \sum_{w = 1}^{h+3} \lfloor \mu(w) \rfloor$. Otherwise, let $b = 1$;
    \item $M = (K - (2 + \sum_{w = 1}^{h+3} \lfloor \mu(w) \rfloor) ) \mod |C|$ if $K > 1 + \sum_{w = 1}^{h+3} \lfloor \mu(w) \rfloor$. 
    Otherwise, let $M = 0$. 
\end{itemize}
Then, the following equation holds: 
\begin{equation*}
    \begin{split}
    |\Psi_{h} \cap \Psi_{\run} & \cap \Psi_{\centerset}(C) \cap \Psi_{\preceding} \cap (\bigcup_{\lambda = 0}^{K - 1} \Psi_{\lcp}(\lambda))| \\
    &= |\Psi_{h} \cap \Psi_{\run} \cap \Psi_{\centerset}(C) \cap \Psi_{\preceding} \cap (\bigcup_{\lambda = 0}^{K - M - 1} \Psi_{\lcp}(\lambda)) \cap (\bigcup_{\lambda = b}^{n} \Psi_{\nRecover}(\lambda))| \\
    &+ |\Psi_{h} \cap \Psi_{\run} \cap \Psi_{\centerset}(C) \cap \Psi_{\preceding} \cap (\bigcup_{\lambda = K-M}^{K-1} \Psi_{\lcp}(\lambda)) \cap (\bigcup_{\lambda = b}^{n} \Psi_{\nRecover}(\lambda))| \\
    &+ |\Psi_{h} \cap \Psi_{\run} \cap \Psi_{\centerset}(C) \cap \Psi_{\preceding} \cap (\bigcup_{\lambda = 0}^{K - 1} \Psi_{\lcp}(\lambda)) \cap (\bigcup_{\lambda = 1}^{b-1} \Psi_{\nRecover}(\lambda))|.
    \end{split}
\end{equation*}
\end{proposition}
\begin{proof}
    We prove $b \in [1, n]$. 
    If $K > 1 + \sum_{w = 1}^{h+3} \lfloor \mu(w) \rfloor$, 
    then $b \in [1, n]$ follows from $\lfloor \frac{K - (2 + \sum_{w = 1}^{h+3} \lfloor \mu(w) \rfloor)}{|C|} \rfloor \in [0, n-1]$. 
    Otherwise, $b = 1$. 

    Next, we prove $M \in [0, K-1]$. 
    If $K > 1 + \sum_{w = 1}^{h+3} \lfloor \mu(w) \rfloor$, 
    then $M \in [0, K-1]$ holds because 
    $M \leq K-1$ follows from $M = (K - (2 + \sum_{w = 1}^{h+3} \lfloor \mu(w) \rfloor) ) \mod |C|$. 
    Otherwise (i.e., $K \in [1, 1 + \sum_{w = 1}^{h+3} \lfloor \mu(w) \rfloor]$), 
    $M \leq [0, K-1]$ follows from 
    $M = 0$ and $K \geq 1$. 

    Proposition~\ref{prop:set_h_HR_center_prec_lcp_sum} follows from the following four equations: 

\begin{equation}\label{eq:set_h_HR_center_prec_lcp_sum:1}
    \begin{split}
    \Psi_{h} \cap \Psi_{\run} & \cap \Psi_{\centerset}(C) \cap \Psi_{\preceding} \cap (\bigcup_{\lambda = 0}^{K - 1} \Psi_{\lcp}(\lambda)) \\
    &= (\Psi_{h} \cap \Psi_{\run} \cap \Psi_{\centerset}(C) \cap \Psi_{\preceding} \cap (\bigcup_{\lambda = 0}^{K - M - 1} \Psi_{\lcp}(\lambda)) \cap (\bigcup_{\lambda = b}^{n} \Psi_{\nRecover}(\lambda))) \\
    & \cup (\Psi_{h} \cap \Psi_{\run} \cap \Psi_{\centerset}(C) \cap \Psi_{\preceding} \cap (\bigcup_{\lambda = K-M}^{K-1} \Psi_{\lcp}(\lambda)) \cap (\bigcup_{\lambda = b}^{n} \Psi_{\nRecover}(\lambda))) \\
    & \cup (\Psi_{h} \cap \Psi_{\run} \cap \Psi_{\centerset}(C) \cap \Psi_{\preceding} \cap (\bigcup_{\lambda = 0}^{K - 1} \Psi_{\lcp}(\lambda)) \cap (\bigcup_{\lambda = 1}^{b-1} \Psi_{\nRecover}(\lambda))).    
    \end{split}
    \end{equation}
    \begin{equation}\label{eq:set_h_HR_center_prec_lcp_sum:2}
    \begin{split}
    (\Psi_{h} \cap \Psi_{\run} & \cap \Psi_{\centerset}(C) \cap \Psi_{\preceding} \cap (\bigcup_{\lambda = 0}^{K - M - 1} \Psi_{\lcp}(\lambda)) \cap (\bigcup_{\lambda = b}^{n} \Psi_{\nRecover}(\lambda))) \\
    & \cap (\Psi_{h} \cap \Psi_{\run} \cap \Psi_{\centerset}(C) \cap \Psi_{\preceding} \cap (\bigcup_{\lambda = K-M}^{K-1} \Psi_{\lcp}(\lambda)) \cap (\bigcup_{\lambda = b}^{n} \Psi_{\nRecover}(\lambda))) = \emptyset. 
    \end{split}
\end{equation}

\begin{equation}\label{eq:set_h_HR_center_prec_lcp_sum:3}
    \begin{split}
    (\Psi_{h} \cap \Psi_{\run} & \cap \Psi_{\centerset}(C) \cap \Psi_{\preceding} \cap (\bigcup_{\lambda = K-M}^{K-1} \Psi_{\lcp}(\lambda)) \cap (\bigcup_{\lambda = b}^{n} \Psi_{\nRecover}(\lambda)))  \\
    & \cap (\Psi_{h} \cap \Psi_{\run} \cap \Psi_{\centerset}(C) \cap \Psi_{\preceding} \cap (\bigcup_{\lambda = 0}^{K - 1} \Psi_{\lcp}(\lambda)) \cap (\bigcup_{\lambda = 1}^{b-1} \Psi_{\nRecover}(\lambda))) = \emptyset. 
    \end{split}
\end{equation}

\begin{equation}\label{eq:set_h_HR_center_prec_lcp_sum:4}
    \begin{split}
    (\Psi_{h} \cap \Psi_{\run} & \cap \Psi_{\centerset}(C) \cap \Psi_{\preceding} \cap (\bigcup_{\lambda = 0}^{K - 1} \Psi_{\lcp}(\lambda)) \cap (\bigcup_{\lambda = 1}^{b-1} \Psi_{\nRecover}(\lambda)))  \\
    & \cap (\Psi_{h} \cap \Psi_{\run} \cap \Psi_{\centerset}(C) \cap \Psi_{\preceding} \cap (\bigcup_{\lambda = 0}^{K - M - 1} \Psi_{\lcp}(\lambda)) \cap (\bigcup_{\lambda = b}^{n} \Psi_{\nRecover}(\lambda))) = \emptyset. 
    \end{split}
\end{equation}

\textbf{Proof of Equation~\ref{eq:set_h_HR_center_prec_lcp_sum:1}.}
Because of $b \in [1, n]$, 
Lemma~\ref{lem:nRecover_basic_property}~\ref{enum:nRecover_basic_property:1} shows that 
the following equation holds: 
\begin{equation}\label{eq:set_h_HR_center_prec_lcp_sum:5}
    \begin{split}
    \Psi_{h} \cap \Psi_{\run} & \cap \Psi_{\centerset}(C) \cap \Psi_{\preceding} \cap (\bigcup_{\lambda = 0}^{K - 1} \Psi_{\lcp}(\lambda)) \\
    &= (\Psi_{h} \cap \Psi_{\run} \cap \Psi_{\centerset}(C) \cap \Psi_{\preceding} \cap (\bigcup_{\lambda = 0}^{K - 1} \Psi_{\lcp}(\lambda)) \cap (\bigcup_{\lambda = b}^{n} \Psi_{\nRecover}(\lambda))) \\
    & \cup (\Psi_{h} \cap \Psi_{\run} \cap \Psi_{\centerset}(C) \cap \Psi_{\preceding} \cap (\bigcup_{\lambda = 0}^{K - 1} \Psi_{\lcp}(\lambda)) \cap (\bigcup_{\lambda = 1}^{b-1} \Psi_{\nRecover}(\lambda))).
    \end{split}
\end{equation}

Because of $M \in [0, K-1]$, 
the following equation holds: 
\begin{equation}\label{eq:set_h_HR_center_prec_lcp_sum:6}
    \begin{split}
    \Psi_{h} \cap \Psi_{\run} & \cap \Psi_{\centerset}(C) \cap \Psi_{\preceding} \cap (\bigcup_{\lambda = 0}^{K - 1} \Psi_{\lcp}(\lambda)) \cap (\bigcup_{\lambda = b}^{n} \Psi_{\nRecover}(\lambda)) \\
    &= (\Psi_{h} \cap \Psi_{\run} \cap \Psi_{\centerset}(C) \cap \Psi_{\preceding} \cap (\bigcup_{\lambda = 0}^{K - M - 1} \Psi_{\lcp}(\lambda)) \cap (\bigcup_{\lambda = b}^{n} \Psi_{\nRecover}(\lambda))) \\
    &\cup (\Psi_{h} \cap \Psi_{\run} \cap \Psi_{\centerset}(C) \cap \Psi_{\preceding} \cap (\bigcup_{\lambda = K-M}^{K-1} \Psi_{\lcp}(\lambda)) \cap (\bigcup_{\lambda = b}^{n} \Psi_{\nRecover}(\lambda))).
    \end{split}
\end{equation}

Therefore, Equation~\ref{eq:set_h_HR_center_prec_lcp_sum:1} follows from Equation~\ref{eq:set_h_HR_center_prec_lcp_sum:5} 
and Equation~\ref{eq:set_h_HR_center_prec_lcp_sum:6}.

\textbf{Proof of Equation~\ref{eq:set_h_HR_center_prec_lcp_sum:2}.}
Lemma~\ref{lem:psi_LMPS_property}~\ref{enum:psi_LMPS_property:lcp:2} shows that 
$(\bigcup_{\lambda = 0}^{K - M - 1} \Psi_{\lcp}(\lambda)) \cap (\bigcup_{\lambda = K-M}^{K-1} \Psi_{\lcp}(\lambda)) = \emptyset$ holds. 
Therefore, Equation~\ref{eq:set_h_HR_center_prec_lcp_sum:2} holds. 

\textbf{Proof of Equation~\ref{eq:set_h_HR_center_prec_lcp_sum:3}.}
Lemma~\ref{lem:nRecover_basic_property}~\ref{enum:nRecover_basic_property:2} shows that 
$(\bigcup_{\lambda = b}^{n} \Psi_{\nRecover}(\lambda)) \cap (\bigcup_{\lambda = 1}^{b-1} \Psi_{\nRecover}(\lambda)) = \emptyset$ holds. 
Therefore, Equation~\ref{eq:set_h_HR_center_prec_lcp_sum:3} holds. 

\textbf{Proof of Equation~\ref{eq:set_h_HR_center_prec_lcp_sum:4}.}
Similar to Equation~\ref{eq:set_h_HR_center_prec_lcp_sum:3}, 
Equation~\ref{eq:set_h_HR_center_prec_lcp_sum:4} follows from the fact that 
$(\bigcup_{\lambda = b}^{n} \Psi_{\nRecover}(\lambda)) \cap (\bigcup_{\lambda = 1}^{b-1} \Psi_{\nRecover}(\lambda)) = \emptyset$ holds. 

\end{proof}

\begin{proposition}\label{prop:set_JD_property_big_sum}
The following equation holds: 
\begin{equation*}
    \begin{split}
    |(\Psi_{\CCP}(T[i..j]) \cap & \Psi_{\lex}(T[\gamma_{Q}..r_{Q}+1]) \cap \Psi_{\run} \cap \Psi_{\centerset}(C_{Q}) \cap \Psi_{\preceding}) \setminus \Psi_{\lcp}(K_{Q})| \\
    &= |(\Psi_{h_{Q}} \cap \Psi_{\lex}(T[\gamma_{Q}..r_{Q}+1]) \cap \Psi_{\run} \cap \Psi_{\centerset}(C_{Q}) \cap \Psi_{\preceding}) \setminus \Psi_{\lcp}(K_{Q})| \\
    &- |\Psi_{h_{Q}} \cap \Psi_{\run} \cap \Psi_{\centerset}(C_{Q}) \cap \Psi_{\preceding} \cap (\bigcup_{\lambda = 0}^{|[\gamma_{Q}, j]|-1} \Psi_{\lcp}(\lambda))|. 
    \end{split}
\end{equation*}
\end{proposition}
\begin{proof}
The following three equations are used to prove Proposition~\ref{prop:set_JD_property_big_sum}.

\begin{equation}\label{eq:set_JD_property_big_sum:1}
    \begin{split}
    (\Psi_{\CCP}(T[i..j]) \cap & \Psi_{\lex}(T[\gamma_{Q}..r_{Q}+1]) \cap \Psi_{\run} \cap \Psi_{\centerset}(C_{Q}) \cap \Psi_{\preceding}) \setminus \Psi_{\lcp}(K_{Q}) \\
    &\subseteq (\Psi_{h_{Q}} \cap \Psi_{\lex}(T[\gamma_{Q}..r_{Q}+1]) \cap \Psi_{\run} \cap \Psi_{\centerset}(C_{Q}) \cap \Psi_{\preceding}) \\ 
    &\setminus (\Psi_{\lcp}(K_{Q}) \cup (\bigcup_{\lambda = 0}^{|[\gamma_{Q}, j]|-1} \Psi_{\lcp}(\lambda)));
    \end{split}
\end{equation}
\begin{equation}\label{eq:set_JD_property_big_sum:2}
    \begin{split}
    (\Psi_{\CCP}(T[i..j]) \cap & \Psi_{\lex}(T[\gamma_{Q}..r_{Q}+1]) \cap \Psi_{\run} \cap \Psi_{\centerset}(C_{Q}) \cap \Psi_{\preceding}) \setminus \Psi_{\lcp}(K_{Q}) \\
    &\supseteq (\Psi_{h_{Q}} \cap \Psi_{\lex}(T[\gamma_{Q}..r_{Q}+1]) \cap \Psi_{\run} \cap \Psi_{\centerset}(C_{Q}) \cap \Psi_{\preceding}) \\ 
    &\setminus (\Psi_{\lcp}(K_{Q}) \cup (\bigcup_{\lambda = 0}^{|[\gamma_{Q}, j]|-1} \Psi_{\lcp}(\lambda)));
    \end{split}
\end{equation}
\begin{equation}\label{eq:set_JD_property_big_sum:3}
    \begin{split}
    \Psi_{h_{Q}} \cap \Psi_{\run} \cap & \Psi_{\centerset}(C_{Q}) \cap \Psi_{\preceding} \cap (\bigcup_{\lambda = 0}^{|[\gamma_{Q}, j]|-1} \Psi_{\lcp}(\lambda)) \\
    & \subseteq (\Psi_{h_{Q}} \cap \Psi_{\lex}(T[\gamma_{Q}..r_{Q}+1]) \cap \Psi_{\run} \cap \Psi_{\centerset}(C_{Q}) \cap \Psi_{\preceding}) \setminus \Psi_{\lcp}(K_{Q}).
    \end{split}
\end{equation}

\textbf{Proof of Equation~\ref{eq:set_JD_property_big_sum:1}.}
Consider an interval attractor $([p, q], [\ell, r])$ in the set 
$(\Psi_{\CCP}(T[i..j]) \cap \Psi_{\lex}(T[\gamma_{Q}..r_{Q}+1]) \cap \Psi_{\run} \cap \Psi_{\centerset}(C_{Q}) \cap \Psi_{\preceding}) \setminus \Psi_{\lcp}(K_{Q})$. 
Lemma~\ref{lem:CCP_property}~\ref{enum:CCP_property:1} shows that $([p, q], [\ell, r]) \in \Psi_{h_{Q}}$ holds 
because $([p, q], [\ell, r]) \in \Psi_{\CCP}(T[i..j])$ and $([p_{Q}, q_{Q}], [\ell_{Q}, r_{Q}]) \in \Psi_{h_{Q}}$. 
$([p, q], [\ell, r]) \not \in \bigcup_{\lambda = 0}^{|[\gamma_{Q}, j]|-1} \Psi_{\lcp}(\lambda)$ holds 
because $([p, q], [\ell, r]) \in \Psi_{\lcp}(K_{Q})$ and $|[\gamma_{Q}, j]| \leq K$. 
Therefore, $([p, q], [\ell$, $r]) \in (\Psi_{h_{Q}} \cap \Psi_{\lex}(T[\gamma_{Q}..r_{Q}+1]) \cap \Psi_{\run} \cap \Psi_{\centerset}(C_{Q}) \cap \Psi_{\preceding}) \setminus (\Psi_{\lcp}(K_{Q}) \cup (\bigcup_{\lambda = 0}^{|[\gamma_{Q}, j]|-1} \Psi_{\lcp}(\lambda)))$ follows from the following three equations: 
\begin{itemize}
    \item $([p, q], [\ell, r]) \in (\Psi_{\lex}(T[\gamma_{Q}..r_{Q}+1]) \cap \Psi_{\run} \cap \Psi_{\centerset}(C_{Q}) \cap \Psi_{\preceding}) \setminus \Psi_{\lcp}(K_{Q})$; 
    \item $([p, q], [\ell, r]) \in \Psi_{h_{Q}}$;
    \item $([p, q], [\ell, r]) \not \in \bigcup_{\lambda = 0}^{|[\gamma_{Q}, j]|-1} \Psi_{\lcp}(\lambda)$.
\end{itemize}

We showed that $([p, q], [\ell, r]) \in (\Psi_{h_{Q}} \cap \Psi_{\lex}(T[\gamma_{Q}..r_{Q}+1]) \cap \Psi_{\run} \cap \Psi_{\centerset}(C_{Q}) \cap \Psi_{\preceding}) \setminus (\Psi_{\lcp}(K_{Q}) \cup (\bigcup_{\lambda = 0}^{|[\gamma_{Q}, j]|-1} \Psi_{\lcp}(\lambda)))$ for each interval attractor $([p, q], [\ell, r]) \in (\Psi_{\CCP}(T[i..j]) \cap \Psi_{\lex}(T[\gamma_{Q}..r_{Q}+1]) \cap \Psi_{\run} \cap \Psi_{\centerset}(C_{Q}) \cap \Psi_{\preceding}) \setminus \Psi_{\lcp}(K_{Q})$. 
Therefore, Equation~\ref{eq:set_JD_property_big_sum:1} holds. 

\textbf{Proof of Equation~\ref{eq:set_JD_property_big_sum:2}.}
Consider an interval attractor $([p, q], [\ell, r])$ in set 
$(\Psi_{h_{Q}} \cap \Psi_{\lex}(T[\gamma_{Q}..r_{Q}+1]) \cap \Psi_{\run} \cap \Psi_{\centerset}(C_{Q}) \cap \Psi_{\preceding}) \setminus (\Psi_{\lcp}(K_{Q}) \cup (\bigcup_{\lambda = 0}^{|[\gamma_{Q}, j]|-1} \Psi_{\lcp}(\lambda)))$. 
Let $K = |\lcp(T[\gamma..r], C_{Q}^{n+1})|$ for the attractor position $\gamma$ of the interval attractor $([p, q], [\ell, r])$. 
Then, $([p, q], [\ell, r]) \in \Psi_{\lcp}(K)$ follows from the definition of the subset $\Psi_{\lcp}(K)$. 
$K \geq |[\gamma_{Q}, j]|$ holds because $([p, q], [\ell, r]) \not \in \bigcup_{\lambda = 0}^{|[\gamma_{Q}, j]|-1} \Psi_{\lcp}(\lambda)$. 
Similarly, $K \neq K_{Q}$ holds because $([p, q], [\ell, r]) \not \in \Psi_{\lcp}(K_{Q})$. 

We prove $T[\gamma_{Q}..j] \prec T[\gamma..r+1] \prec T[\gamma_{Q}..j]\#$. 
The string $T[\gamma_{Q}..j]$ is a prefix of string $C_{Q}^{n+1}$ (i.e., $\lcp(T[\gamma_{Q}..j], C_{Q}^{n+1}) = T[\gamma_{Q}..j]$) 
because $|[\gamma_{Q}, j]| \leq K_{Q}$.
The string $T[\gamma_{Q}..j]$ is a prefix of the string $T[\gamma..r]$ (i.e., $\lcp(T[\gamma_{Q}..j], T[\gamma..r]) = T[\gamma_{Q}..j]$) because $\lcp(T[\gamma_{Q}..j], C_{Q}^{n+1}) = T[\gamma_{Q}..j]$ and $K \geq |[\gamma_{Q}, j]|$. 
Therefore, $T[\gamma_{Q}..j] \prec T[\gamma..r+1] \prec T[\gamma_{Q}..j]\#$ follows from $\lcp(T[\gamma_{Q}..j], T[\gamma..r]) = T[\gamma_{Q}..j]$. 

We prove $\reverse(T[i..\gamma_{Q}-1]) \prec \reverse(T[p-1..\gamma-1]) \prec \reverse(\# T[i..\gamma_{Q}-1])$. 
We can apply Lemma~\ref{lem:suffix_syncro} to the interval attractor $([p_{Q}, q_{Q}], [\ell_{Q}, r_{Q}])$ 
because (A) string $T[i..\gamma_{Q}-1]$ is a suffix of string $C_{Q}^{n+1}$, 
and (B) $K_{Q} > 1 + \sum_{w = 1}^{h_{Q}+3} \lfloor \mu(w) \rfloor$ holds. 
Lemma~\ref{lem:suffix_syncro} shows that the string $T[i..\gamma_{Q}-1]$ is a suffix of string $T[p..\gamma-1]$ 
(i.e., $\lcs(T[i..\gamma_{Q}-1], T[p..\gamma-1]) = T[i..\gamma_{Q}-1]$) because 
$([p, q], [\ell, r]) \in \Psi_{h_{Q}} \cap \Psi_{\run} \cap \Psi_{\centerset}(C_{Q})$. 
Therefore, $\reverse(T[i..\gamma_{Q}-1]) \prec \reverse(T[p-1..\gamma-1]) \prec \reverse(\# T[i..\gamma_{Q}-1])$ follows from $\lcs(T[i..\gamma_{Q}-1], T[p..\gamma-1]) = T[i..\gamma_{Q}-1]$. 

We prove $([p, q], [\ell, r]) \in (\Psi_{\CCP}(T[i..j]) \cap \Psi_{\lex}(T[\gamma_{Q}..r_{Q}+1]) \cap \Psi_{\run} \cap \Psi_{\centerset}(C_{Q}) \cap \Psi_{\preceding}) \setminus \Psi_{\lcp}(K_{Q})$. 
Lemma~\ref{lem:CCP_property}~\ref{enum:CCP_property:4} shows that 
$\Psi_{\CCP}(T[i..j]) = \{ ([p^{\prime}, q^{\prime}], [\ell^{\prime}, r^{\prime}]) \in \Psi_{h_{Q}} \mid \reverse(T[i..\gamma_{Q}-1]) \prec \reverse(T[p^{\prime}-1..\gamma^{\prime}-1]) \prec \reverse(\#T[i..\gamma_{Q}-1]) \text{ and } T[\gamma_{Q}..j] \prec T[\gamma^{\prime}..r^{\prime}+1] \prec T[\gamma_{Q}..j]\# \}$ holds. 
Here, $\gamma^{\prime}$ is the attractor position of each interval attractor $([p^{\prime}, q^{\prime}], [\ell^{\prime}, r^{\prime}]) \in \Psi_{h_{Q}}$. 
$([p, q], [\ell, r]) \in \Psi_{\CCP}(T[i..j])$ follows from $([p, q], [\ell, r]) \in \Psi_{h_{Q}}$, 
$\reverse(T[i..\gamma_{Q}-1]) \prec \reverse(T[p-1..\gamma-1]) \prec \reverse(\# T[i..\gamma_{Q}-1])$, and 
$T[\gamma_{Q}..j] \prec T[\gamma..r+1] \prec T[\gamma_{Q}..j]\#$. 
Therefore, $([p, q], [\ell, r]) \in (\Psi_{\CCP}(T[i..j]) \cap \Psi_{\lex}(T[\gamma_{Q}..r_{Q}+1]) \cap \Psi_{\run} \cap \Psi_{\centerset}(C_{Q}) \cap \Psi_{\preceding}) \setminus \Psi_{\lcp}(K_{Q})$ follows from 
$([p, q], [\ell, r]) \in \Psi_{\CCP}(T[i..j])$ and 
$([p, q], [\ell, r]) \in (\Psi_{\lex}(T[\gamma_{Q}..r_{Q}+1]) \cap \Psi_{\run} \cap \Psi_{\centerset}(C_{Q}) \cap \Psi_{\preceding}) \setminus \Psi_{\lcp}(K_{Q})$.

We showed that $([p, q], [\ell, r]) \in (\Psi_{\CCP}(T[i..j]) \cap \Psi_{\lex}(T[\gamma_{Q}..r_{Q}+1]) \cap \Psi_{\run} \cap \Psi_{\centerset}(C_{Q}) \cap \Psi_{\preceding}) \setminus \Psi_{\lcp}(K_{Q})$ holds for each interval attractor $([p, q], [\ell, r]) \in (\Psi_{h_{Q}} \cap \Psi_{\lex}(T[\gamma_{Q}..r_{Q}+1]) \cap \Psi_{\run} \cap \Psi_{\centerset}(C_{Q}) \cap \Psi_{\preceding}) \setminus (\Psi_{\lcp}(K_{Q}) \cup (\bigcup_{\lambda = 0}^{|[\gamma_{Q}, j]|-1} \Psi_{\lcp}(\lambda)))$. 
Therefore, Equation~\ref{eq:set_JD_property_big_sum:2} holds. 

\textbf{Proof of Equation~\ref{eq:set_JD_property_big_sum:3}.}
Because of $|[\gamma_{Q}, j]| \leq K_{Q}$, 
Lemma~\ref{lem:psi_LMPS_property}~\ref{enum:psi_LMPS_property:lcp:2} indicates that 
the following equation holds: 
\begin{equation}\label{eq:set_JD_property_big_sum:4}
    \Psi_{h_{Q}} \cap \Psi_{\run} \cap \Psi_{\centerset}(C_{Q}) \cap \Psi_{\preceding} \cap (\bigcup_{\lambda = 0}^{|[\gamma_{Q}, j]|-1} \Psi_{\lcp}(\lambda)) \cap \Psi_{\lcp}(K_{Q}) = \emptyset. 
\end{equation}
The following equation follows from Proposition~\ref{prop:psi_PS_basic_property}~\ref{enum:psi_PS_basic_property:1}. 
\begin{equation}\label{eq:set_JD_property_big_sum:5}
    \begin{split}
    \Psi_{h_{Q}} \cap \Psi_{\run} \cap & \Psi_{\centerset}(C_{Q}) \cap \Psi_{\preceding} \cap (\bigcup_{\lambda = 0}^{|[\gamma_{Q}, j]|-1} \Psi_{\lcp}(\lambda)) \\
    & \subseteq \Psi_{h_{Q}} \cap \Psi_{\run} \cap \Psi_{\centerset}(C_{Q}) \cap \Psi_{\preceding} \cap (\bigcup_{\lambda = 0}^{K_{Q}-1} \Psi_{\lcp}(\lambda)) \\
    & \subseteq \Psi_{\lex}(T[\gamma_{Q}..r_{Q}+1]).
    \end{split}
\end{equation}
Therefore, Equation~\ref{eq:set_JD_property_big_sum:3} follows from Equation~\ref{eq:set_JD_property_big_sum:4} and Equation~\ref{eq:set_JD_property_big_sum:5}. 

\textbf{Proof of Proposition~\ref{prop:set_JD_property_big_sum}.}
The following equation follows from Equation~\ref{eq:set_JD_property_big_sum:1} and Equation~\ref{eq:set_JD_property_big_sum:2}: 
\begin{equation}\label{eq:set_JD_property_big_sum:6}
    \begin{split}
    (\Psi_{\CCP}(T[i..j]) \cap & \Psi_{\lex}(T[\gamma_{Q}..r_{Q}+1]) \cap \Psi_{\run} \cap \Psi_{\centerset}(C_{Q}) \cap \Psi_{\preceding}) \setminus \Psi_{\lcp}(K_{Q}) \\
    &= (\Psi_{h_{Q}} \cap \Psi_{\lex}(T[\gamma_{Q}..r_{Q}+1]) \cap \Psi_{\run} \cap \Psi_{\centerset}(C_{Q}) \cap \Psi_{\preceding}) \\ 
    &\setminus (\Psi_{\lcp}(K_{Q}) \cup (\bigcup_{\lambda = 0}^{|[\gamma_{Q}, j]|-1} \Psi_{\lcp}(\lambda))) \\
    &= ((\Psi_{h_{Q}} \cap \Psi_{\lex}(T[\gamma_{Q}..r_{Q}+1]) \cap \Psi_{\run} \cap \Psi_{\centerset}(C_{Q}) \cap \Psi_{\preceding}) \setminus \Psi_{\lcp}(K_{Q})) \\
    &\setminus \bigcup_{\lambda = 0}^{|[\gamma_{Q}, j]|-1} \Psi_{\lcp}(\lambda) \\
    &= ((\Psi_{h_{Q}} \cap \Psi_{\lex}(T[\gamma_{Q}..r_{Q}+1]) \cap \Psi_{\run} \cap \Psi_{\centerset}(C_{Q}) \cap \Psi_{\preceding}) \setminus \Psi_{\lcp}(K_{Q})) \\
    &\setminus (\Psi_{h_{Q}} \cap \Psi_{\run} \cap \Psi_{\centerset}(C_{Q}) \cap \Psi_{\preceding} \cap (\bigcup_{\lambda = 0}^{|[\gamma_{Q}, j]|-1} \Psi_{\lcp}(\lambda))).
    \end{split}
\end{equation}
Therefore, Proposition~\ref{prop:set_JD_property_big_sum} follows from Equation~\ref{eq:set_JD_property_big_sum:6} and 
Equation~\ref{eq:set_JD_property_big_sum:3}. 
\end{proof}

\begin{proposition}\label{prop:set_JD_property_small_prec}
If $([p_{Q}, q_{Q}], [\ell_{Q}, r_{Q}]) \in \Psi_{\preceding}$, 
then the following equation holds: 
\begin{equation}
    \begin{split}
    |(\Psi_{h_{Q}} \cap \Psi_{\lex}(T[\gamma_{Q}..r_{Q}+1]) & \cap \Psi_{\run} \cap \Psi_{\centerset}(C_{Q}) \cap \Psi_{\preceding}) \setminus \Psi_{\lcp}(K_{Q})| \\
    &= |\Psi_{h_{Q}} \cap \Psi_{\run} \cap \Psi_{\centerset}(C_{Q}) \cap \Psi_{\preceding} \cap (\bigcup_{\lambda = 0}^{K_{Q} - 1} \Psi_{\lcp}(\lambda))|;
    \end{split}
\end{equation}
\end{proposition}
\begin{proof}
$\Psi_{h_{Q}} \cap \Psi_{\lex}(T[\gamma_{Q}..r_{Q}+1]) \cap \Psi_{\run} \cap \Psi_{\centerset}(C_{Q}) \cap \Psi_{\preceding} \subseteq \bigcup_{\lambda = 1}^{n} \Psi_{\lcp}(\lambda)$ follows from Lemma \ref{lem:psi_LMPS_property}~\ref{enum:psi_LMPS_property:lcp:1}. 
Proposition~\ref{prop:psi_PS_basic_property}~\ref{enum:psi_PS_basic_property:2} shows that 
the set $\Psi_{h_{Q}} \cap \Psi_{\lex}(T[\gamma_{Q}..r_{Q}+1]) \cap \Psi_{\run} \cap \Psi_{\centerset}(C_{Q}) \cap \Psi_{\preceding} \cap \bigcup_{\lambda = 1}^{K_{Q}-1} \Psi_{\lcp}(\lambda)$ is equal to set 
$\Psi_{h_{Q}} \cap \Psi_{\run} \cap \Psi_{\centerset}(C_{Q}) \cap \Psi_{\preceding} \cap \bigcup_{\lambda = 1}^{K_{Q}-1} \Psi_{\lcp}(\lambda)$. 
Proposition~\ref{prop:psi_condition_D_prec_super_property} shows that 
the set $\Psi_{h_{Q}} \cap \Psi_{\lex}(T[\gamma_{Q}..r_{Q}+1]) \cap \Psi_{\run} \cap \Psi_{\centerset}(C_{Q}) \cap \Psi_{\preceding} \cap \bigcup_{\lambda = K_{Q}+1}^{n} \Psi_{\lcp}(\lambda)$ is empty. 
Therefore, the following equation holds: 
\begin{equation}\label{eq:set_JD_property_small_prec}
    \begin{split}
    (\Psi_{h_{Q}} & \cap \Psi_{\lex}(T[\gamma_{Q}..r_{Q}+1]) \cap \Psi_{\run} \cap \Psi_{\centerset}(C_{Q}) \cap \Psi_{\preceding}) \setminus \Psi_{\lcp}(K_{Q}) \\
    &= (\Psi_{h_{Q}} \cap \Psi_{\lex}(T[\gamma_{Q}..r_{Q}+1])  \cap \Psi_{\run} \cap \Psi_{\centerset}(C_{Q}) \cap \Psi_{\preceding} \cap (\bigcup_{\lambda = 1}^{K_{Q}-1} \Psi_{\lcp}(\lambda))) \\
    & \cup (\Psi_{h_{Q}} \cap \Psi_{\lex}(T[\gamma_{Q}..r_{Q}+1])  \cap \Psi_{\run} \cap \Psi_{\centerset}(C_{Q}) \cap \Psi_{\preceding} \cap (\bigcup_{\lambda = K_{Q}+1}^{n} \Psi_{\lcp}(\lambda))) \\
    &= \Psi_{h_{Q}} \cap \Psi_{\run} \cap \Psi_{\centerset}(C_{Q}) \cap \Psi_{\preceding} \cap (\bigcup_{\lambda = 1}^{K_{Q}-1} \Psi_{\lcp}(\lambda)).
    \end{split}
\end{equation}
Finally, Proposition~\ref{prop:set_JD_property_small_prec} follows from Equation~\ref{eq:set_JD_property_small_prec}.
\end{proof}

\begin{proposition}\label{prop:set_JD_property_small_succ_sum}
If $([p_{Q}, q_{Q}], [\ell_{Q}, r_{Q}]) \not \in \Psi_{\preceding}$, 
then the following equation holds: 
\begin{equation}
    \begin{split}
    |(\Psi_{h_{Q}} \cap \Psi_{\lex}(T[\gamma_{Q}..r_{Q}+1]) & \cap \Psi_{\run} \cap \Psi_{\centerset}(C_{Q}) \cap \Psi_{\preceding}) \setminus \Psi_{\lcp}(K_{Q})| \\
    &= |\Psi_{h_{Q}} \cap \Psi_{\run} \cap \Psi_{\centerset}(C) \cap \Psi_{\preceding}| \\ 
    &- |\Psi_{h_{Q}} \cap \Psi_{\run} \cap \Psi_{\centerset}(C) \cap \Psi_{\preceding} \cap \Psi_{\lcp}(K_{Q})|.
    \end{split}
\end{equation}
\end{proposition}
\begin{proof}
Lemma~\ref{lem:psi_LMPS_property}~\ref{enum:psi_LMPS_property:lcp:1} indicates that 
the following equation holds: 
\begin{equation*}
    \begin{split}
    |(\Psi_{h_{Q}} & \cap \Psi_{\lex}(T[\gamma_{Q}..r_{Q}+1]) \cap \Psi_{\run} \cap \Psi_{\centerset}(C_{Q}) \cap \Psi_{\preceding}) \setminus \Psi_{\lcp}(K_{Q})| \\
    &= |\Psi_{h_{Q}} \cap \Psi_{\lex}(T[\gamma_{Q}..r_{Q}+1]) \cap \Psi_{\run} \cap \Psi_{\centerset}(C_{Q}) \cap \Psi_{\preceding} \cap (\bigcup_{\lambda = 1}^{n} \Psi_{\lcp}(\lambda))| \\ 
    &- |\Psi_{h_{Q}} \cap \Psi_{\lex}(T[\gamma_{Q}..r_{Q}+1]) \cap \Psi_{\run} \cap \Psi_{\centerset}(C_{Q}) \cap \Psi_{\preceding} \cap \Psi_{\lcp}(K_{Q})|\\
    &= |\Psi_{h_{Q}} \cap \Psi_{\lex}(T[\gamma_{Q}..r_{Q}+1]) \cap \Psi_{\run} \cap \Psi_{\centerset}(C_{Q}) \cap \Psi_{\preceding}| \\ 
    &- |\Psi_{h_{Q}} \cap \Psi_{\lex}(T[\gamma_{Q}..r_{Q}+1]) \cap \Psi_{\run} \cap \Psi_{\centerset}(C_{Q}) \cap \Psi_{\preceding} \cap \Psi_{\lcp}(K_{Q})|.
    \end{split}
\end{equation*} 
Proposition~\ref{prop:psi_PS_basic_property}~\ref{enum:psi_PS_basic_property:3} indicates that 
$\Psi_{h_{Q}} \cap \Psi_{\lex}(T[\gamma_{Q}..r_{Q}+1]) \cap \Psi_{\run} \cap \Psi_{\centerset}(C_{Q}) \cap \Psi_{\preceding} = \Psi_{h_{Q}} \cap \Psi_{\run} \cap \Psi_{\centerset}(C_{Q}) \cap \Psi_{\preceding}$ and 
$\Psi_{h_{Q}} \cap \Psi_{\lex}(T[\gamma_{Q}..r_{Q}+1]) \cap \Psi_{\run} \cap \Psi_{\centerset}(C_{Q}) \cap \Psi_{\preceding} \cap \Psi_{\lcp}(K_{Q}) = \Psi_{h_{Q}} \cap \Psi_{\run} \cap \Psi_{\centerset}(C_{Q}) \cap \Psi_{\preceding} \cap \Psi_{\lcp}(K_{Q})$ hold. 
Therefore, Proposition~\ref{prop:set_JD_property_small_succ_sum} holds. 
\end{proof}

We prove Lemma~\ref{lem:JD1_division}. 
\paragraph{Proof of Equation~\ref{eq:JD1_division:1} in Lemma~\ref{lem:JD1_division}.}
We prove Equation~\ref{eq:JD1_division:1} using 
Proposition~\ref{prop:set_JD_property_big_sum}, Proposition~\ref{prop:set_JD_property_small_prec}, 
and Proposition~\ref{prop:set_h_HR_center_prec_lcp_sum}. 
Equation~\ref{eq:JD1_division:1} follows from the following three equations: 
\begin{equation}\label{eq:JD1_division:1:A}
    \begin{split}
    |(\Psi_{h_{Q}} & \cap \Psi_{\lex}(T[\gamma_{Q}..r_{Q}+1]) \cap \Psi_{\run} \cap \Psi_{\centerset}(C_{Q}) \cap \Psi_{\preceding}) \setminus \Psi_{\lcp}(K_{Q})| \\
    &= |\Psi_{h_{Q}} \cap \Psi_{\run} \cap \Psi_{\centerset}(C_{Q}) \cap \Psi_{\preceding} \cap (\bigcup_{\lambda = 0}^{K_{Q} - 1} \Psi_{\lcp}(\lambda))| \\
    &- |\Psi_{h_{Q}} \cap \Psi_{\run} \cap \Psi_{\centerset}(C_{Q}) \cap \Psi_{\preceding} \cap (\bigcup_{\lambda = 0}^{|[\gamma_{Q}, j]|-1} \Psi_{\lcp}(\lambda))|;
    \end{split}
\end{equation}
\begin{equation}\label{eq:JD1_division:1:B}
    \begin{split}
    |\Psi_{h_{Q}} & \cap \Psi_{\run} \cap \Psi_{\centerset}(C_{Q}) \cap \Psi_{\preceding} \cap (\bigcup_{\lambda = 0}^{K_{Q} - 1} \Psi_{\lcp}(\lambda))| \\
    &= |\Psi_{h_{Q}} \cap \Psi_{\run} \cap \Psi_{\centerset}(C_{Q}) \cap \Psi_{\preceding} \cap (\bigcup_{\lambda = 0}^{K_{Q} - M_{Q} - 1} \Psi_{\lcp}(\lambda)) \cap (\bigcup_{\lambda = x}^{n} \Psi_{\nRecover}(\lambda))| \\
    &+ |\Psi_{h_{Q}} \cap \Psi_{\run} \cap \Psi_{\centerset}(C_{Q}) \cap \Psi_{\preceding} \cap (\bigcup_{\lambda = K_{Q} - M_{Q}}^{K_{Q}-1} \Psi_{\lcp}(\lambda)) \cap (\bigcup_{\lambda = x}^{n} \Psi_{\nRecover}(\lambda))| \\
    &+ |\Psi_{h_{Q}} \cap \Psi_{\run} \cap \Psi_{\centerset}(C_{Q}) \cap \Psi_{\preceding} \cap (\bigcup_{\lambda = 0}^{K_{Q} - 1} \Psi_{\lcp}(\lambda)) \cap (\bigcup_{\lambda = 1}^{x-1} \Psi_{\nRecover}(\lambda))|;
    \end{split}
\end{equation}
\begin{equation}\label{eq:JD1_division:1:C}
    \begin{split}
    |\Psi_{h_{Q}} & \cap \Psi_{\run} \cap \Psi_{\centerset}(C_{Q}) \cap \Psi_{\preceding} \cap (\bigcup_{\lambda = 0}^{|[\gamma_{Q}, j]|-1} \Psi_{\lcp}(\lambda))| \\
    &= |\Psi_{h_{Q}} \cap \Psi_{\run} \cap \Psi_{\centerset}(C_{Q}) \cap \Psi_{\preceding} \cap (\bigcup_{\lambda = 0}^{|[\gamma_{Q}, j]| - M_{Q}^{\prime} - 1} \Psi_{\lcp}(\lambda)) \cap (\bigcup_{\lambda = x^{\prime}}^{n} \Psi_{\nRecover}(\lambda))| \\
    &+ |\Psi_{h_{Q}} \cap \Psi_{\run} \cap \Psi_{\centerset}(C_{Q}) \cap \Psi_{\preceding} \cap (\bigcup_{\lambda = |[\gamma_{Q}, j]| - M_{Q}^{\prime}}^{|[\gamma_{Q}, j]| - 1} \Psi_{\lcp}(\lambda)) \cap (\bigcup_{\lambda = x^{\prime}}^{n} \Psi_{\nRecover}(\lambda))| \\
    &+ |\Psi_{h_{Q}} \cap \Psi_{\run} \cap \Psi_{\centerset}(C_{Q}) \cap \Psi_{\preceding} \cap (\bigcup_{\lambda = 0}^{|[\gamma_{Q}, j]| - 1} \Psi_{\lcp}(\lambda)) \cap (\bigcup_{\lambda = 1}^{x^{\prime}-1} \Psi_{\nRecover}(\lambda))|.
    \end{split}
\end{equation}

Equation~\ref{eq:JD1_division:1:A} follows from Proposition~\ref{prop:set_JD_property_big_sum} and Proposition~\ref{prop:set_JD_property_small_prec}.
We obtain Equation~\ref{eq:JD1_division:1:B} by applying 
Proposition~\ref{prop:set_h_HR_center_prec_lcp_sum} to the set $\Psi_{h_{Q}} \cap \Psi_{\run} \cap \Psi_{\centerset}(C_{Q}) \cap \Psi_{\preceding} \cap (\bigcup_{\lambda = 0}^{K_{Q} - 1} \Psi_{\lcp}(\lambda))$. 
Similarly, we obtain Equation~\ref{eq:JD1_division:1:C} by applying 
Proposition~\ref{prop:set_h_HR_center_prec_lcp_sum} to the set $\Psi_{h_{Q}} \cap \Psi_{\run} \cap \Psi_{\centerset}(C_{Q}) \cap \Psi_{\preceding} \cap (\bigcup_{\lambda = 0}^{|[\gamma_{Q}, j]|-1} \Psi_{\lcp}(\lambda))$. 
Therefore, Equation~\ref{eq:JD1_division:1} holds. 

%\begin{proof}[Proof of Equation~\ref{eq:JD1_division:1}]
%\end{proof}

%%%%%%%%%%%%%%%%%%%%%%%%%%%
\paragraph{Proof of Equation~\ref{eq:JD1_division:2} in Lemma~\ref{lem:JD1_division}.}
We prove Equation~\ref{eq:JD1_division:2} using Proposition~\ref{prop:set_JD_property_big_sum},  Proposition~\ref{prop:set_JD_property_small_succ_sum}, and Equation~\ref{eq:JD1_division:1:C}. 
Here, Equation~\ref{eq:JD1_division:1:C} holds even if $([p_{Q}, q_{Q}], [\ell_{Q}, r_{Q}]) \not \in \Psi_{\preceding}$. 
The following equation follows from 
Proposition~\ref{prop:set_JD_property_big_sum} and Proposition~\ref{prop:set_JD_property_small_succ_sum}. 
\begin{equation}\label{eq:JD1_division:2:A}
    \begin{split}
    |(\Psi_{\CCP}(T[i..j]) & \cap \Psi_{\lex}(T[\gamma_{Q}..r_{Q}+1]) \cap \Psi_{\run} \cap \Psi_{\centerset}(C_{Q}) \cap \Psi_{\preceding}) \setminus \Psi_{\lcp}(K_{Q})| \\
    &= |(\Psi_{h_{Q}} \cap \Psi_{\lex}(T[\gamma_{Q}..r_{Q}+1]) \cap \Psi_{\run} \cap \Psi_{\centerset}(C_{Q}) \cap \Psi_{\preceding}) \setminus \Psi_{\lcp}(K_{Q})| \\
    &- |\Psi_{h_{Q}} \cap \Psi_{\run} \cap \Psi_{\centerset}(C_{Q}) \cap \Psi_{\preceding} \cap (\bigcup_{\lambda = 0}^{|[\gamma_{Q}, j]|-1} \Psi_{\lcp}(\lambda))| \\    
    &= |\Psi_{h_{Q}} \cap \Psi_{\run} \cap \Psi_{\centerset}(C) \cap \Psi_{\preceding}| \\
    &- |\Psi_{h_{Q}} \cap \Psi_{\run} \cap \Psi_{\centerset}(C) \cap \Psi_{\preceding} \cap \Psi_{\lcp}(K_{Q})| \\
    &- |\Psi_{h_{Q}} \cap \Psi_{\run} \cap \Psi_{\centerset}(C_{Q}) \cap \Psi_{\preceding} \cap (\bigcup_{\lambda = 0}^{|[\gamma_{Q}, j]|-1} \Psi_{\lcp}(\lambda))|.
    \end{split}
\end{equation}
Therefore, Equation~\ref{eq:JD1_division:2} follows from Equation~\ref{eq:JD1_division:2:A} and Equation~\ref{eq:JD1_division:1:C}.

%%%%%%%%%%%%%%%%%%%%%%%%%%%%%%%%%%%%%%%%%%%%%%%%%%%%%%%%%%%%%%%

\subsubsection{Proof of Lemma~\ref{lem:JD1_main_lemma}}\label{subsubsec:JD1_main_lemma_proof}
Equation~\ref{eq:JD1_sum:2} of Lemma~\ref{lem:JD1_sum} indicates that 
the following two equations hold: 
\begin{equation}\label{eq:JD1_main_lemma:3}
    \begin{split}
    |\Psi_{h_{Q}} \cap \Psi_{\run} \cap \Psi_{\centerset}(C_{Q}) & \cap \Psi_{\preceding} \cap (\bigcup_{\lambda = x}^{n} \Psi_{\nRecover}(\lambda)) \cap (\bigcup_{\lambda = 0}^{K_{Q} - M_{Q} - 1} \Psi_{\lcp}(\lambda))| \\
    &= (x-1) \rangesum(\mathcal{J}_{D}(h_{Q}, C_{Q}), x, n, 0, |C_{Q}| - 1);
    \end{split}
\end{equation}
\begin{equation}\label{eq:JD1_main_lemma:4}
    \begin{split}
    |\Psi_{h_{Q}} \cap \Psi_{\run} \cap \Psi_{\centerset}(C_{Q}) & \cap \Psi_{\preceding} \cap (\bigcup_{\lambda = x^{\prime}}^{n} \Psi_{\nRecover}(\lambda)) \cap (\bigcup_{\lambda = 0}^{|[\gamma_{Q}, j]| - M_{Q}^{\prime} - 1} \Psi_{\lcp}(\lambda))| \\
    &= (x^{\prime}-1) \rangesum(\mathcal{J}_{D}(h_{Q}, C_{Q}), x^{\prime}, n, 0, |C_{Q}| - 1).
    \end{split}
\end{equation}

Equation~\ref{eq:JD1_sum:3} of Lemma~\ref{lem:JD1_sum} indicates that 
the following two equations hold: 
\begin{equation}\label{eq:JD1_main_lemma:5}
    \begin{split}
    |\Psi_{h_{Q}} \cap \Psi_{\run} \cap \Psi_{\centerset}(C_{Q}) & \cap \Psi_{\preceding} \cap (\bigcup_{\lambda = x}^{n} \Psi_{\nRecover}(\lambda)) \cap (\bigcup_{\lambda = K_{Q} - M_{Q}}^{K_{Q}-1} \Psi_{\lcp}(\lambda))| \\
    &= \rangesum(\mathcal{J}_{D}(h_{Q}, C_{Q}), x, n, 0, M_{Q} - 1); 
    \end{split}
\end{equation}
\begin{equation}\label{eq:JD1_main_lemma:6}
    \begin{split}
    |\Psi_{h_{Q}} \cap \Psi_{\run} \cap \Psi_{\centerset}(C_{Q}) & \cap \Psi_{\preceding} \cap (\bigcup_{\lambda = x^{\prime}}^{n} \Psi_{\nRecover}(\lambda)) \cap (\bigcup_{\lambda = |[\gamma_{Q}, j]| - M_{Q}^{\prime}}^{|[\gamma_{Q}, j]| - 1} \Psi_{\lcp}(\lambda))| \\
    &= \rangesum(\mathcal{J}_{D}(h_{Q}, C_{Q}), x^{\prime}, n, 0, M^{\prime}_{Q} - 1). 
    \end{split}
\end{equation}

Equation~\ref{eq:JD1_sum:4} of Lemma~\ref{lem:JD1_sum} indicates that 
the following two equations hold: 
\begin{equation}\label{eq:JD1_main_lemma:7}
    \begin{split}
    |\Psi_{h_{Q}} \cap \Psi_{\run} \cap \Psi_{\centerset}(C_{Q}) & \cap \Psi_{\preceding} \cap (\bigcup_{\lambda = 1}^{x-1} \Psi_{\nRecover}(\lambda)) \cap (\bigcup_{\lambda = 0}^{K_{Q} - 1} \Psi_{\lcp}(\lambda))| \\
    &= \rangesum(\mathcal{J}_{E}(h_{Q}, C_{Q}), 1, x-1, 0, |C_{Q}|-1); 
    \end{split}
\end{equation}
\begin{equation}\label{eq:JD1_main_lemma:8}
    \begin{split}
    |\Psi_{h_{Q}} \cap \Psi_{\run} \cap \Psi_{\centerset}(C_{Q}) & \cap \Psi_{\preceding} \cap (\bigcup_{\lambda = 1}^{x^{\prime}-1} \Psi_{\nRecover}(\lambda)) \cap (\bigcup_{\lambda = 0}^{|[\gamma_{Q}, j]| - 1} \Psi_{\lcp}(\lambda))| \\
    &= \rangesum(\mathcal{J}_{E}(h_{Q}, C_{Q}), 1, x^{\prime}-1, 0, |C_{Q}|-1). 
    \end{split}
\end{equation}

Because of $K_{Q} > 1 + \sum_{w = 1}^{h_{Q}+3} \lfloor \mu(w) \rfloor$, 
Equation~\ref{eq:JD1_sum:5} of Lemma~\ref{lem:JD1_sum} indicates that 
the following equation holds: 
\begin{equation}\label{eq:JD1_main_lemma:9}
    \begin{split}
    |\Psi_{h_{Q}} \cap \Psi_{\run} \cap \Psi_{\centerset}(C_{Q}) & \cap \Psi_{\preceding} \cap \Psi_{\lcp}(K_{Q})| \\
    &= \rangesum(\mathcal{J}_{D}(h_{Q}, C_{Q}), x, n, M_{Q}, M_{Q}). 
    \end{split}
\end{equation}

We prove Equation~\ref{eq:JD1_main_lemma:1} and Equation~\ref{eq:JD1_main_lemma:2}. 
Equation~\ref{eq:JD1_main_lemma:1} follows from Equation~\ref{eq:JD1_division:1}, 
Equation~\ref{eq:JD1_main_lemma:3}, 
Equation~\ref{eq:JD1_main_lemma:4}, 
Equation~\ref{eq:JD1_main_lemma:5}, 
Equation~\ref{eq:JD1_main_lemma:6}, 
Equation~\ref{eq:JD1_main_lemma:7}, 
and Equation~\ref{eq:JD1_main_lemma:8}. 
Equation~\ref{eq:JD1_main_lemma:2} follows from 
Equation~\ref{eq:JD1_division:2}, 
Equation~\ref{eq:JD1_sum:1}, 
Equation~\ref{eq:JD1_main_lemma:4}, 
Equation~\ref{eq:JD1_main_lemma:6}, 
Equation~\ref{eq:JD1_main_lemma:8}, 
and Equation~\ref{eq:JD1_main_lemma:9}.

\subsubsection{Dynamic Data Structures for Two Sets \texorpdfstring{$\mathcal{J}_{D}(h, C)$}{} and \texorpdfstring{$\mathcal{J}_{E}(h, C)$}{} of Weighted Points}\label{subsubsec:JD1_ds}
Consider a pair of an integer $h \in [0, H]$ and a string $C \in \Sigma^{+}$. 
We present dynamic data structures to support range-sum query on set $\mathcal{J}_{D}(h, C)$ of weighted points,
which are similar to the dynamic data structures for the ordered set $\mathcal{J}_{A}(h)$ presented in Section~\ref{subsubsec:JA_ds}. 
Let $(x_{1}, y_{1}, w_{1}, e_{1})$, $(x_{2}, y_{2}, w_{2}, e_{2})$, $\ldots$, $(x_{k}, y_{k}, w_{k}, e_{k})$ ($e_{1} \prec e_{2} \prec \cdots e_{k}$) be the weighted points in the set $\mathcal{J}_{D}(h, C)$. 
Here, the following three statements hold for each integer $s \in [1, k]$: 
\begin{itemize}
    \item from the definition of set $\mathcal{J}_{D}(h, C)$, 
    the $s$-th weighted point $(x_{s}, y_{s}, w_{s}, e_{s})$ corresponds to an interval attractor $([p_{s}, q_{s}], [\ell_{s}, r_{s}])$ in the sampling subset $\Psi_{\samp}$;
    \item the interval attractor $([p_{s}, q_{s}], [\ell_{s}, r_{s}])$ corresponds to 
    a node $u_{s}$ of the sequence $\mathbf{Q}_{\samp}$ introduced in Section~\ref{subsec:sample_query}; 
    \item the node $u_{s}$ is represented as an element $v_{s}$ of the doubly linked list introduced in Section~\ref{subsubsec:sample_ds}. 
\end{itemize}

We store the set $\mathcal{J}_{D}(h, C)$ using a doubly linked list $\mathbf{L}_{D}(h, C)$ of $k$ elements. 
For each integer $s \in [1, k]$, 
the $s$-th element of the doubly linked list $\mathbf{L}_{D}(h, C)$ corresponds to 
the $s$-th weighted point $(x_{s}, y_{s}, w_{s}, e_{s})$. 
This element stores (i) the x-coordinate $x_{s}$, (ii) the y-coordinate $y_{s}$, (iii) the weight $w_{s}$, 
and (iv) a pointer to the element $v_{s}$ of the doubly linked list introduced in Section~\ref{subsubsec:sample_ds}. 
List indexing and range-sum data structures are built on doubly linked list $\mathbf{L}_{D}(h, C)$. 
This range-sum data structure is used to support range-count and range-sum queries on the set $\mathcal{J}_{D}(h, C)$ of 
weighted points. 

Similarly, 
we present dynamic data structures to support range-sum query on set $\mathcal{J}_{E}(h, C)$ of weighted points.
We store the set $\mathcal{J}_{E}(h, C)$ using a doubly linked list $\mathbf{L}_{E}(h, C)$ of $|\mathbf{L}_{E}(h, C)|$ elements 
in a similar way as the set $\mathcal{J}_{D}(h, C)$.
Similar to the doubly linked list $\mathbf{L}_{D}(h, C)$, 
list indexing and range-sum data structures are built on doubly linked list $\mathbf{L}_{E}(h, C)$. 
These dynamic data structures require $O((|\mathcal{J}_{D}(h, C)| + |\mathcal{J}_{E}(h, C)|) B)$ bits of space in total for machine word size $B$. 
Here, $|\mathcal{J}_{D}(h, C)| = |\mathcal{J}_{E}(h, C)|$ follows from the definitions of the two sets 
$\mathcal{J}_{D}(h, C)$ and $\mathbf{L}_{E}(h, C)$.

\subsubsection{Dynamic Data Structures for Ordered Set \texorpdfstring{$\mathcal{T}_{D}$}{} of Pairs}\label{subsubsec:TD1_ds}
We introduce an ordered set $\mathcal{T}_{D} \subseteq [0, H] \times \Sigma^{+}$ such that 
each pair of the set $\mathcal{T}_{D}$ consists of an integer $h \in [0, H]$ and a string $C \in \Sigma^{+}$ 
satisfying $\mathcal{J}_{D}(h, C) \neq \emptyset$ 
(i.e., $\mathcal{T}_{D} = \{ (h, C) \in [0, H] \times \Sigma^{+} \mid \mathcal{J}_{C}(h, D) \neq \emptyset \}$). 
Let $(h_{1}, C_{1})$, $(h_{2}, C_{2})$, $\ldots$, $(h_{m}, C_{m})$ be the pairs in the ordered set $\mathcal{T}_{D}$. 
In the ordered set $\mathcal{T}_{D}$, 
a triplet $(h_{s}, C_{s})$ precedes another pair $(h_{s^{\prime}}, C_{s^{\prime}})$ if and only if 
either (i) $h_{s} < h_{s^{\prime}}$ or (ii) $h_{s} = h_{s^{\prime}}$ and $C_{s} \prec C_{s^{\prime}}$ holds. 
This ordered set $\mathcal{T}_{D}$ is used to verify whether two sets $\mathcal{J}_{D}(h, C)$ and $\mathcal{J}_{E}(h, C)$ 
are empty or not for a given pair of an integer $h \in [0, H]$ and a string $C \in \Sigma^{+}$. 

The following lemma states properties of two sets $\mathcal{T}_{D}$ and $\mathcal{J}_{D}(h, C)$. 

\begin{lemma}\label{lem:TD1_size}
    The following four statements hold: 
    \begin{enumerate}[label=\textbf{(\roman*)}]
    \item \label{enum:TD1_size:1} 
    $\sum_{(h, C) \in \mathcal{T}_{D}} |\mathcal{J}_{D}(h, C)| \leq |\Psi_{\samp}|$;
    \item \label{enum:TD1_size:2} $|\mathcal{T}_{D}| \leq |\Psi_{\samp}|$;
    \item \label{enum:TD1_size:3} $|\mathcal{T}_{D}| = O(n^{2})$;
    \item \label{enum:TD1_size:4} 
    $|\mathcal{J}_{D}(h, C)| \leq |\Psi_{\samp}|$ 
    and 
    $|\mathcal{J}_{D}(h, C)| = O(n^{2})$ for a pair of an integer $h \in [0, H]$ and a string $C \in \Sigma^{+}$. 
    \end{enumerate}
\end{lemma}
\begin{proof}
    The proof of Lemma~\ref{lem:TD1_size} is as follows. 

    \textbf{Proof of Lemma~\ref{lem:TD1_size}(i).}
    Lemma~\ref{lem:TD1_size}~\ref{enum:TD1_size:1} can be proved using 
    the same approach as for Lemma~\ref{lem:TC1_size}~\ref{enum:TC1_size:1}.

    \textbf{Proof of Lemma~\ref{lem:TD1_size}(ii).}
    $|\mathcal{T}_{D}| \leq \sum_{(h, C) \in \mathcal{T}_{D}} |\mathcal{J}_{D}(h, C)|$ holds 
    because $|\mathcal{J}_{D}(h, C)| \geq 1$ for each pair $(h, C) \in \mathcal{T}_{D}$. 
    We already proved $\sum_{(h, C) \in \mathcal{T}_{D}} |\mathcal{J}_{D}(h, C)| \leq |\Psi_{\samp}|$. 
    Therefore, $|\mathcal{T}_{D}| \leq |\Psi_{\samp}|$ holds. 

    \textbf{Proof of Lemma~\ref{lem:TD1_size}(iii).}
    $|\Psi_{\samp}| = O(n^{2})$ follows from 
    $\Psi_{\samp} \subseteq \Psi_{\RR}$ and 
    $|\Psi_{\RR}| = O(n^{2})$ (Lemma~\ref{lem:non_comp_IA_size}). 
    $|\mathcal{T}_{D}| = O(n^{2})$ follows from 
    $|\mathcal{T}_{D}| \leq |\Psi_{\samp}|$ and $|\Psi_{\samp}| = O(n^{2})$.

    \textbf{Proof of Lemma~\ref{lem:TD1_size}(iv).}
    $|\mathcal{J}_{D}(h, C)| \leq |\Psi_{h} \cap \Psi_{\source} \cap \Psi_{\centerset}(C) \cap \Psi_{\preceding} \cap \Psi_{\samp}|$ holds 
    because 
    there exists a one-to-one correspondence between the weighted points of set $\mathcal{J}_{D}(h, C)$ 
    and the interval attractors of set $\Psi_{h} \cap \Psi_{\source} \cap \Psi_{\centerset}(C) \cap \Psi_{\preceding} \cap \Psi_{\samp}$. 
    $|\Psi_{h} \cap \Psi_{\source} \cap \Psi_{\centerset}(C) \cap \Psi_{\preceding} \cap \Psi_{\samp}| \leq |\Psi_{\samp}|$ because 
    the set $\Psi_{h} \cap \Psi_{\source} \cap \Psi_{\centerset}(C) \cap \cap \Psi_{\preceding} \cap \Psi_{\samp}$ is a subset of the sampling subset $\Psi_{\samp}$. 
    Therefore, $|\mathcal{J}_{D}(h, C)| \leq |\Psi_{\samp}|$ holds. 
    $|\mathcal{J}_{D}(h, C)| = O(n^{2})$ follows from 
    $|\mathcal{J}_{D}(h, C)| \leq |\Psi_{\samp}|$ and 
    $|\Psi_{\samp}| = O(n^{2})$. 
\end{proof}

We store the ordered set $\mathcal{T}_{D}$ using a doubly linked list of $m$ elements. 
For each integer $s \in [1, m]$, 
the $s$-th element of this doubly linked list corresponds to the $s$-th triplet $(h_{s}, C_{s})$ of the ordered set $\mathcal{T}_{D}$. 
The $s$-th element stores two pointers to 
two doubly linked lists $\mathbf{L}_{D}(h, C)$ and $\mathbf{L}_{E}(h, C)$. 
A list indexing data structure is used for quickly accessing to the elements of the doubly linked list storing the ordered set $\mathcal{T}_{D}$. 
These dynamic data structures require $O(m B)$ bits of space in total for machine word size $B$.

\subsubsection{Dynamic Data Structures and Algorithm for \texorpdfstring{$\RSCQDX(i, j)$}{RSCD1(i, j)}}\label{subsubsec:JD1_subquery_ds}
We prove Lemma~\ref{lem:RSC_subquery_D1_summary}, i.e., 
we show that subquery $\RSCQDX(i, j)$ can be answered in $O(H^{2} \log n + \log^{4} n)$ time using 
dynamic data structures of $O((|\mathcal{U}_{\RR}| + |\Psi_{\samp}|)B)$ bits of space for machine word size $B$. 
Here, $|\mathcal{U}_{\RR}|$ is the number of nodes in the RR-DAG of RLSLP $\mathcal{G}^{R}$. 
Let $(h_{1}, C_{1})$, $(h_{2}, C_{2})$, $\ldots$, $(h_{m}, C_{m})$ be the triplets in the ordered set $\mathcal{T}_{D}$.

\paragraph{Data Structures.}
We answer $\RSCQDX(i, j)$ using the following dynamic data structures: 
\begin{itemize}
    \item the dynamic data structures of $O(|\mathcal{U}_{\RR}|B)$ bits of space 
    for the RR-DAG of RLSLP $\mathcal{G}^{R}$ (Section~\ref{subsubsec:rrdag_ds}). 
    \item the dynamic data structures of $O(|\Psi_{\samp}|B)$ bits of space 
    for sample query (Section~\ref{subsubsec:sample_ds});
    \item the dynamic data structures of $O(\sum_{s = 1}^{m} |\mathcal{J}_{D}(h_{s}, C_{s})| B)$ bits of space 
    for $2m$ sets $\mathcal{J}_{D}(h_{1}, C_{1})$, $\mathcal{J}_{E}(h_{1}, C_{1})$, 
    $\mathcal{J}_{D}(h_{2}, C_{2})$, $\mathcal{J}_{E}(h_{2}, C_{2})$, $\ldots$,
    $\mathcal{J}_{D}(h_{m}, C_{m})$, $\mathcal{J}_{E}(h_{m}, C_{m})$ (Section~\ref{subsubsec:JD1_ds}); 
    \item the dynamic data structures of $O(m B)$ bits of space for the ordered set $\mathcal{T}_{D}$ (Section~\ref{subsubsec:TD1_ds}).
\end{itemize}
$\sum_{s = 1}^{m} |\mathcal{J}_{D}(h_{s}, C_{s})| \leq |\Psi_{\samp}|$ 
and $m \leq |\Psi_{\samp}|$ follow from Lemma~\ref{lem:TD1_size}. 
Therefore, these dynamic data structures can be stored in $O((|\mathcal{U}_{\RR}| + |\Psi_{\samp}|) B)$ bits of space. 

The following lemma states three queries supported by these dynamic data structures. 
\begin{lemma}\label{lem:TD1_queries}
    Let $m = |\mathcal{T}_{D}|$ 
    and $(h_{s}, C_{s})$ be the $s$-th triplet of the ordered set $\mathcal{T}_{D}$ for each integer $s \in [1, m]$.     
    Using the dynamic data structures of Section~\ref{subsubsec:JD1_subquery_ds}, 
    we can answer the following three queries:
    \begin{enumerate}[label=\textbf{(\roman*)}]
    \item \label{enum:TD1_queries:1}
    for a given integer $s \in [1, m]$, 
    return a pair of an integer $h \in [0, H]$ and a string $C \in \Sigma^{+}$ 
    satisfying $(h, T[\alpha..\beta]) = (h_{s}, C_{s})$ in $O(H^{2} + \log n)$ time; 
    \item \label{enum:TD1_queries:2}
    consider a given triplet of an integer $s \in [1, m]$, an integer $h \in [0, H]$, and an interval $[\alpha, \beta] \subseteq [1, n]$ in input string $T$. 
    Then, verify whether the $s$-th triplet of the ordered set $\mathcal{T}_{D}$ satisfies 
    either (A) $h_{s} < h$ or (B) $h_{s} = h$ and $C_{s} \preceq T[\alpha..\beta]$. 
    This verification takes $O(H^{2} + \log n)$ time;
    \item \label{enum:TD1_queries:3}
    verify whether $(h, T[\alpha..\beta]) \in \mathcal{T}_{D}$ or not in $O(H^{2} \log n + \log^{2} n)$ time  
    for a given pair of an integer $h \in [0, H]$ and interval $[\alpha, \beta] \subseteq [1, n]$ in input string $T$. 
    If $(h, T[\alpha..\beta]) \in \mathcal{T}_{D}$ holds, 
    then return an integer $s \in [1, m]$ satisfying 
    $(h_{s}, C_{s}) = (h, T[\alpha..\beta])$ in the same time. 
    \end{enumerate}
\end{lemma}
\begin{proof}
    This lemma can be proved using the same approach as for Lemma~\ref{lem:TC1_queries}.
\end{proof}

\paragraph{Algorithm.}
The algorithm for $\RSCQDX(i, j)$ computes $|(\Psi_{\CCP}(T[i..j]) \cap \Psi_{\lex}(T[\gamma_{Q}..r_{Q}+1]) \cap \Psi_{\run} \cap \Psi_{\centerset}(C_{Q}) \cap \Psi_{\preceding}) \setminus \Psi_{\lcp}(K_{Q})|$ 
under the condition that RSC query $\RSCQ(i, j)$ satisfies condition (D) of RSC query stated in Section~\ref{subsec:rsc_sub}. 
Here, $K_{Q} = |\lcp(T[\gamma_{Q}..r_{Q}], C_{Q}^{n+1})|$. 
This algorithm leverages Lemma~\ref{lem:JD1_main_lemma}, which shows that 
the size of set $|(\Psi_{\CCP}(T[i..j]) \cap \Psi_{\lex}(T[\gamma_{Q}..r_{Q}+1]) \cap \Psi_{\run} \cap \Psi_{\centerset}(C_{Q}) \cap \Psi_{\preceding}) \setminus \Psi_{\lcp}(K_{Q})|$ can be computed 
by at most six range-sum query on two sets $\mathcal{J}_{D}(h_{Q}, C_{Q})$ and $\mathcal{J}_{E}(h_{Q}, C_{Q})$ 
of weighted points.

The algorithm for $\RSCQDX(i, j)$ consists of five phases. 
In the first phase, 
we compute interval attractor $([p_{Q}, q_{Q}], [\ell_{Q}, r_{Q}])$ 
and four integers $h_{Q}$, $|C_{Q}|$, $\gamma_{Q}$, and $K_{Q}$ 
using the first phase of the algorithm for $\RSCQCX(i, j)$. 
This phase takes $O(H^{2} \log n)$ time.

In the second phase, 
we verify whether $(h_{Q}, C_{Q}) \in \mathcal{T}_{D}$ or not by Lemma~\ref{lem:TD1_queries}~\ref{enum:TD1_queries:3}. 
Here, let $(h_{1}, C_{1})$, $(h_{2}, C_{2})$, $\ldots$, $(h_{m}, C_{m})$ be the pairs in the ordered set $\mathcal{T}_{D}$. 
If $(h_{Q}, C_{Q}) \not \in \mathcal{T}_{D}$, 
then Lemma~\ref{lem:JD1_main_lemma} indicates that 
$|(\Psi_{\CCP}(T[i..j]) \cap \Psi_{\lex}(T[\gamma_{Q}..r_{Q}+1]) \cap \Psi_{\run} \cap \Psi_{\centerset}(C_{Q}) \cap \Psi_{\preceding}) \setminus \Psi_{\lcp}(K_{Q})| = 0$ holds 
because $\mathcal{J}_{D}(h_{Q}, C_{Q}) = \emptyset$ and $\mathcal{J}_{E}(h_{Q}, C_{Q}) = \emptyset$ hold. 
In this case, we stop this algorithm and report that the set $(\Psi_{\CCP}(T[i..j]) \cap \Psi_{\lex}(T[\gamma_{Q}..r_{Q}+1]) \cap \Psi_{\run} \cap \Psi_{\centerset}(C_{Q}) \cap \Psi_{\preceding}) \setminus \Psi_{\lcp}(K_{Q})$ is empty. 
Otherwise (i.e., $(h_{Q}, C_{Q}) \in \mathcal{T}_{D}$), 
there exists an integer $s \in [1, m]$ satisfying 
$(h_{s}, C_{s}) = (h_{Q}, C_{Q})$, 
and this integer $s$ can be obtained by the query of Lemma~\ref{lem:TD1_queries}~\ref{enum:TD1_queries:3}. 
The query of Lemma~\ref{lem:TD1_queries}~\ref{enum:TD1_queries:3} can be executed using 
triplet $(h_{Q}, [\gamma_{Q}, \gamma_{Q} + |C_{Q}| - 1])$ 
because $C_{Q} = T[\gamma_{Q}..\gamma_{Q} + |C_{Q}| - 1]$ follows from the definition of the associated string $C_{Q}$. 
Therefore, the second phase takes $O(H^{2} \log n + \log^{2} n)$ time. 

In the third phase, 
we access the two doubly linked list $\mathbf{L}_{D}(h_{Q}, C_{Q})$. 
The $s$-th element of the doubly linked list representing the ordered set $\mathcal{T}_{D}$ stores 
a pointer to the doubly linked lists $\mathbf{L}_{D}(h_{Q}, C_{Q})$. 
Therefore, the third phase can be executed in $O(\log m)$ time 
by the list indexing data structure built on the doubly linked list representing the ordered set $\mathcal{T}_{D}$. 
Here, $m = O(n^{2})$ follows from Lemma~\ref{lem:TD1_size}~\ref{enum:TD1_size:3}.

In the fourth phase, 
we compute three integers $x, x^{\prime}$, and $M^{\prime}_{Q}$ of Lemma~\ref{lem:JD1_division}. 
Here, 
\begin{itemize}
    \item $x = 1 + \lfloor \frac{K_{Q} - (2 + \sum_{w = 1}^{h_{Q}+3} \lfloor \mu(w) \rfloor)}{|C_{Q}|} \rfloor$;
    \item $x^{\prime} = 1 + \lfloor \frac{|[\gamma_{Q}, j]| - (2 + \sum_{w = 1}^{h_{Q}+3} \lfloor \mu(w) \rfloor)}{|C_{Q}|} \rfloor$ 
    if $|[\gamma_{Q}, j]| > 1 + \sum_{w = 1}^{h_{Q}+3} \lfloor \mu(w) \rfloor$. 
    Otherwise, let $x^{\prime} = 1$; 
    \item $M_{Q}^{\prime} = (|[\gamma_{Q}, j]| - (2 + \sum_{w = 1}^{h_{Q}+3} \lfloor \mu(w) \rfloor)) \mod |C_{Q}|$ if $|[\gamma_{Q}, j]| > 1 + \sum_{w = 1}^{h_{Q}+3} \lfloor \mu(w) \rfloor$. 
    Otherwise, let $M_{Q}^{\prime} = 0$. 
\end{itemize}
The three integers $x, x^{\prime}$, and $M^{\prime}_{Q}$ can be computed in $O(H)$ time using 
four integers $h_{Q}$, $\gamma_{Q}$, $K_{Q}$, and $|C_{Q}|$.
Therefore, the fourth phase takes $O(H)$ time. 

In the fifth phase, we compute $|(\Psi_{\CCP}(T[i..j]) \cap \Psi_{\lex}(T[\gamma_{Q}..r_{Q}+1]) \cap \Psi_{\run} \cap \Psi_{\centerset}(C_{Q}) \cap \Psi_{\preceding}) \setminus \Psi_{\lcp}(K_{Q})|$ by 
the range-sum queries stated in Lemma~\ref{lem:JD1_main_lemma}. 
We need to verify whether $([p_{Q}, q_{Q}], [\ell_{Q}, r_{Q}]) \in \Psi_{\preceding}$ or not to 
compute $|(\Psi_{\CCP}(T[i..j]) \cap \Psi_{\lex}(T[\gamma_{Q}..r_{Q}+1]) \cap \Psi_{\run} \cap \Psi_{\centerset}(C_{Q}) \cap \Psi_{\preceding}) \setminus \Psi_{\lcp}(K_{Q})|$ using Lemma~\ref{lem:JD1_main_lemma}. 
This verification can be executed by verify-prec query $\precQ(([p, q], [\ell, r]))$, 
which takes $O(H^{2})$ time. 
We can execute range-sum query on the two sets $\mathcal{J}_{D}(h_{Q}, C_{Q})$ and $\mathcal{J}_{E}(h_{Q}, C_{Q})$ of weighted points 
in $O(\log^{4} k)$ time for the number $k$ of weighted points in the set $\mathcal{J}_{D}(h_{Q}, C_{Q})$. 
$k = O(n^{2})$ follows from Lemma~\ref{lem:TD1_size}~\ref{enum:TD1_size:4}. 
Therefore, the fifth phase takes $O(H^{2} + \log^{4} n)$ time. 

Finally, the algorithm for $\RSCQDX(i, j)$ can be executed in $O(H^{2} \log n + \log^{4} n)$ time in total. 
Therefore, Lemma~\ref{lem:RSC_subquery_D1_summary} holds.

\subsection{Solution of \texorpdfstring{$\RSCQDY(i, j)$}{RSCD2(i, j)}}\label{subsec:RSC_comp_D2}
The goal of this subsection is to answer $\RSCQDY(i, j)$. 
The following lemma states the summary of this subsection. 

\begin{lemma}\label{lem:RSC_subquery_D2_summary}
Using a dynamic data structure of $O((|\Psi_{\samp}| + |\mathcal{U}_{\RR}|) B)$ bits of space for machine word size $B$, 
we can answer $\RSCQDY(i, j)$ (i.e., computing $|(\Psi_{\CCP}(T[i..j]) \cap \Psi_{\lex}(T[\gamma_{Q}..r_{Q}+1]) \cap \Psi_{\run} \cap \Psi_{\centerset}(C_{Q}) \cap \Psi_{\succeeding}) \setminus \Psi_{\lcp}(K_{Q})|$) in $O(H^{2} \log n + \log^{4} n)$ time if the given RSC query $\RSCQ(i, j)$ satisfies condition (D). 
Here, $|\mathcal{U}_{\RR}|$ is the number of nodes in the RR-DAG of RLSLP $\mathcal{G}^{R}$.
\end{lemma}
\begin{proof}
See Section~\ref{subsubsec:JD2_subquery_ds}.
\end{proof}

Subquery $\RSCQDY(i, j)$ can be answered using the same approach as for $\RSCQDX(i, j)$, which explained in Section~\ref{subsec:RSC_comp_D1}. 
For answering $\RSCQDY(i, j)$, 
we leverage range-sum query on weighted points corresponding to the interval attractors in 
set $\Psi_{h_{Q}} \cap \Psi_{\source} \cap \Psi_{\centerset}(C_{Q}) \cap \Psi_{\succeeding} \cap \Psi_{\samp}$. 
For this purpose, 
we introduce two sets $\mathcal{J}_{D^{\prime}}(h, C)$ and $\mathcal{J}_{E^{\prime}}(h, C)$ of weighted points 
on grid $([1, n], [-1, |C|])$ for a pair of an integer $h \in [0, H]$ and a string $C \in \Sigma^{+}$. 
Here, the two sets $\mathcal{J}_{D^{\prime}}(h, C)$ and $\mathcal{J}_{E^{\prime}}(h, C)$ are defined using 
set $\Psi_{h} \cap \Psi_{\source} \cap \Psi_{\centerset}(C) \cap \Psi_{\succeeding} \cap \Psi_{\samp}$ of $k$ interval attractors 
$([p_{1}, q_{1}], [\ell_{1}, r_{1}]), ([p_{2}, q_{2}], [\ell_{2}, r_{2}])$, 
$\ldots$, $([p_{k}, q_{k}], [\ell_{k}, r_{k}])$. 
For each integer $s \in [1, k]$, 
let $\gamma_{s}$ of the attractor position of the interval attractor $([p_{s}, q_{s}], [\ell_{s}, r_{s}])$; 
let $g_{s} = |f_{\recover}(([p_{s}, q_{s}], [\ell_{s}, r_{s}]))|$, 
$K_{s} = |\lcp(T[\gamma_{s}..r_{s}], C^{n+1})|$, and $M_{s} = (K_{s} - (2 + \sum_{w = 1}^{h+3} \lfloor \mu(w) \rfloor) ) \mod |C|$. 
Here, $([p_{s}, q_{s}], [\ell_{s}, r_{s}]) \in \Psi_{\lcp}(K_{s})$ 
and $([p_{s}, q_{s}], [\ell_{s}, r_{s}]) \in \Psi_{\modulo}(M_{s})$ follow from 
the definitions of the two subsets $\Psi_{\lcp}(K_{s})$ and $\Psi_{\modulo}(M_{s})$, respectively. 

\paragraph{Set $\mathcal{J}_{D^{\prime}}(h, C)$ of Weighted Points.}
Set $\mathcal{J}_{D^{\prime}}(h, C)$ is defined in a similar way as the set $\mathcal{J}_{D}(h, C)$ introduced in Section~\ref{subsec:RSC_comp_D1}, i.e., 
the set $\mathcal{J}_{D^{\prime}}(h, C)$ consists of $k$ weighted points 
$(g_{1}, M_{1}, |\Psi_{\str}(T[p_{1}-1..r_{1}+1])|, T[p_{1}-1..r_{1}+1]))$, 
$(g_{2}, M_{2}, |\Psi_{\str}(T[p_{2}-1..r_{2}+1])|, T[p_{2}-1..r_{2}+1]))$, $\ldots$, 
$(g_{k}, M_{k}$, $|\Psi_{\str}(T[p_{k}-1..r_{k}+1])|, T[p_{k}-1..r_{k}+1]))$. 
Each weighted point $(g_{s}, M_{s}, |\Psi_{\str}(T[p_{s}-1..r_{s}+1])|, T[p_{s}-1..r_{s}+1]))$ corresponds to interval attractor $([p_{s}, q_{s}], [\ell_{s}, r_{s}])$ in set $\Psi_{h} \cap \Psi_{\source} \cap \Psi_{\centerset}(C) \cap \Psi_{\succeeding} \cap \Psi_{\samp}$. 

\paragraph{Set $\mathcal{J}_{E^{\prime}}(h, C)$ of Weighted Points.}
Set $\mathcal{J}_{E^{\prime}}(h, C)$ is defined in a similar way as the set $\mathcal{J}_{E}(h, C)$ introduced in Section~\ref{subsec:RSC_comp_D1}, i.e., 
the set $\mathcal{J}_{E^{\prime}}(h, C)$ consists of $k$ weighted points 
$(g_{1}, M_{1}, g_{1}|\Psi_{\str}(T[p_{1}-1..r_{1}+1])|, T[p_{1}-1..r_{1}+1]))$, 
$(g_{2}, M_{2}, g_{2}|\Psi_{\str}(T[p_{2}-1..r_{2}+1])|, T[p_{2}-1..r_{2}+1]))$, $\ldots$, 
$(g_{k}, M_{k}$, $g_{k}|\Psi_{\str}(T[p_{k}-1..r_{k}+1])|, T[p_{k}-1..r_{k}+1]))$. 
Similar to the set $\mathcal{J}_{D^{\prime}}(h, C)$, 
each weighted point $(g_{s}, M_{s}, g_{s}|\Psi_{\str}(T[p_{s}-1..r_{s}+1])|, T[p_{s}-1..r_{s}+1]))$ corresponds to 
interval attractor $([p_{s}, q_{s}], [\ell_{s}, r_{s}])$ in set $\Psi_{h} \cap \Psi_{\source} \cap \Psi_{\centerset}(C) \cap \Psi_{\succeeding} \cap \Psi_{\samp}$. 

The following lemma shows that 
we can count the interval attractors in four subsets of set $\Psi_{\RR}$ by range-sum query on 
the two sets $\mathcal{J}_{D^{\prime}}(h, C)$ and $\mathcal{J}_{E^{\prime}}(h, C)$. 

\begin{lemma}\label{lem:JD2_sum}
Consider a triplet of an integer $h \in [0, H]$, a string $C \in \Sigma^{+}$, 
and an integer $K \geq 0$. 
Let $b$ and $M$ be the two integers defined as follows: 
\begin{itemize}
    \item $b = 1 + \lfloor \frac{K - (2 + \sum_{w = 1}^{h+3} \lfloor \mu(w) \rfloor)}{|C|} \rfloor$ if $K > 1 + \sum_{w = 1}^{h+3} \lfloor \mu(w) \rfloor$. Otherwise, let $b = 1$;
    \item $M = (K - (2 + \sum_{w = 1}^{h+3} \lfloor \mu(w) \rfloor) ) \mod |C|$ if $K > 1 + \sum_{w = 1}^{h+3} \lfloor \mu(w) \rfloor$. 
    Otherwise, let $M = 0$. 
\end{itemize}
Then, the following four equations hold:
\begin{equation}\label{eq:JD2_sum:1}
    \begin{split}
    |\Psi_{h} \cap \Psi_{\run} \cap \Psi_{\centerset}(C) \cap \Psi_{\succeeding}| = \rangesum(\mathcal{J}_{E^{\prime}}(h, C), 1, n, 0, |C|-1).
    \end{split}
\end{equation}
\begin{equation}\label{eq:JD2_sum:2}
    \begin{split}
    |\Psi_{h} \cap \Psi_{\run} \cap \Psi_{\centerset}(C) \cap \Psi_{\succeeding} & \cap \Bigl(\bigcup_{\lambda = b}^{n} \Psi_{\nRecover}(\lambda) \Bigr) \cap \Bigl(\bigcup_{\lambda = 0}^{K - M - 1} \Psi_{\lcp}(\lambda) \Bigr)| \\
    &= (b-1) \rangesum(\mathcal{J}_{D^{\prime}}(h, C), b, n, 0, |C| - 1);
    \end{split}
\end{equation}
\begin{equation}\label{eq:JD2_sum:3}
    \begin{split}
    |\Psi_{h} \cap \Psi_{\run} \cap \Psi_{\centerset}(C) \cap \Psi_{\succeeding} & \cap \Bigl(\bigcup_{\lambda = b}^{n} \Psi_{\nRecover}(\lambda) \Bigr) \cap \Bigl(\bigcup_{\lambda = K-M}^{K - 1} \Psi_{\lcp}(\lambda) \Bigr)| \\ 
    &= \rangesum(\mathcal{J}_{D^{\prime}}(h, C), b, n, 0, M - 1);
    \end{split}
\end{equation}
\begin{equation}\label{eq:JD2_sum:4}
    \begin{split}
    |\Psi_{h} \cap \Psi_{\run} \cap \Psi_{\centerset}(C) \cap \Psi_{\succeeding} & \cap \Bigl(\bigcup_{\lambda = 1}^{b-1} \Psi_{\nRecover}(\lambda) \Bigr) \cap \Bigl(\bigcup_{\lambda = 0}^{K - 1} \Psi_{\lcp}(\lambda) \Bigr)| \\
    &= \rangesum(\mathcal{J}_{E^{\prime}}(h, C), 1, b-1, 0, |C|-1).
    \end{split}
\end{equation}
\end{lemma}
\begin{proof}
    Lemma~\ref{lem:JD2_sum} can be proved by the same approach for proving Lemma~\ref{lem:JD1_sum}. 
    This is because 
    Equation~\ref{eq:JD2_sum:1}, Equation~\ref{eq:JD2_sum:2}, 
    Equation~\ref{eq:JD2_sum:3}, and Equation~\ref{eq:JD2_sum:4} are 
    symmetric to 
    Equation~\ref{eq:JD1_sum:1}, Equation~\ref{eq:JD1_sum:2}, 
    Equation~\ref{eq:JD1_sum:3}, and Equation~\ref{eq:JD1_sum:4} of Lemma~\ref{lem:JD1_sum}, 
    respectively. 
\end{proof}

The following lemma states a property of the set $(\Psi_{\CCP}(T[i..j]) \cap \Psi_{\lex}(T[\gamma_{Q}..r_{Q}+1]) \cap \Psi_{\run} \cap  \Psi_{\centerset}(C_{Q}) \cap \Psi_{\succeeding}) \setminus \Psi_{\lcp}(K_{Q})$ 
for the integer $K_{Q}$ of Lemma~\ref{lem:RSC_subquery_D2_summary}.

\begin{lemma}\label{lem:JD2_division}
Consider RSC query $\RSCQ(i, j)$ satisfying condition (D) of RSC query stated in Section~\ref{subsec:rsc_sub}. 
Here, let $([p_{Q}, q_{Q}], [\ell_{Q}, r_{Q}])$ be interval attractor $I_{\capture}(i, j)$; 
let $h_{Q}$, $\gamma_{Q}$, and $C_{Q}$ be the level, attractor position, and associated string of the interval attractor $([p_{Q}, q_{Q}], [\ell_{Q}, r_{Q}])$, respectively; 
let $K_{Q} = |\lcp(T[\gamma_{Q}..r_{Q}], C_{Q}^{n+1})|$; 
let $x^{\prime}$ and $M_{Q}^{\prime}$ be the two integers defined as follows: 
\begin{itemize}
    \item $x^{\prime} = 1 + \lfloor \frac{(K_{Q} + 1) - (2 + \sum_{w = 1}^{h_{Q}+3} \lfloor \mu(w) \rfloor)}{|C_{Q}|} \rfloor$;
    \item $M^{\prime}_{Q} = ((K_{Q} + 1) - (2 + \sum_{w = 1}^{h_{Q}+3} \lfloor \mu(w) \rfloor) ) \mod |C_{Q}|$. 
\end{itemize}

If $([p_{Q}, q_{Q}], [\ell_{Q}, r_{Q}]) \in \Psi_{\succeeding}$, 
then the following equation holds: 
\begin{equation}\label{eq:JD2_division:1}
    \begin{split}
& |(\Psi_{\CCP}(T[i..j]) \cap \Psi_{\lex}(T[\gamma_{Q}..r_{Q}+1]) \cap \Psi_{\run} \cap \Psi_{\centerset}(C_{Q}) \cap \Psi_{\succeeding}) \setminus \Psi_{\lcp}(K_{Q})| \\
&= |\Psi_{h_{Q}} \cap \Psi_{\run} \cap \Psi_{\centerset}(C_{Q}) \cap \Psi_{\succeeding}| \\
&- |\Psi_{h_{Q}} \cap \Psi_{\run} \cap \Psi_{\centerset}(C_{Q}) \cap \Psi_{\succeeding} \cap \Bigl(\bigcup_{\lambda = x^{\prime}}^{n} \Psi_{\nRecover}(\lambda) \Bigr) \cap \Bigl(\bigcup_{\lambda = 0}^{K_{Q} - M_{Q}^{\prime}} \Psi_{\lcp}(\lambda) \Bigr)| \\
&- |\Psi_{h_{Q}} \cap \Psi_{\run} \cap \Psi_{\centerset}(C_{Q}) \cap \Psi_{\succeeding} \cap \Bigl(\bigcup_{\lambda = x^{\prime}}^{n} \Psi_{\nRecover}(\lambda) \Bigr) \cap \Bigl(\bigcup_{\lambda = K_{Q} + 1 - M_{Q}^{\prime}}^{K_{Q}} \Psi_{\lcp}(\lambda) \Bigr)| \\
&- |\Psi_{h_{Q}} \cap \Psi_{\run} \cap \Psi_{\centerset}(C_{Q}) \cap \Psi_{\succeeding} \cap \Bigl(\bigcup_{\lambda = 1}^{x^{\prime}-1} \Psi_{\nRecover}(\lambda) \Bigr) \cap \Bigl(\bigcup_{\lambda = 0}^{K_{Q}} \Psi_{\lcp}(\lambda) \Bigr)|.
    \end{split}
\end{equation}
Otherwise (i.e., $([p_{Q}, q_{Q}], [\ell_{Q}, r_{Q}]) \not \in \Psi_{\succeeding}$), 
the following equation holds: 
\begin{equation}\label{eq:JD2_division:2}
    |(\Psi_{\CCP}(T[i..j]) \cap \Psi_{\lex}(T[\gamma_{Q}..r_{Q}+1]) \cap \Psi_{\run} \cap \Psi_{\centerset}(C_{Q}) \cap \Psi_{\succeeding}) \setminus \Psi_{\lcp}(K_{Q})| = 0.
\end{equation}
\end{lemma}
\begin{proof}
    See Section~\ref{subsubsec:JD2_division_proof}.
\end{proof}

Lemma~\ref{lem:JD2_division} shows that 
the set $(\Psi_{\CCP}(T[i..j]) \cap \Psi_{\lex}(T[\gamma_{Q}..r_{Q}+1]) \cap \Psi_{\run} \cap  \Psi_{\centerset}(C_{Q}) \cap \Psi_{\succeeding}) \setminus \Psi_{\lcp}(K_{Q})$ can be divided into at most four subsets of Lemma~\ref{lem:JD2_sum}. 
The size of each subset of Lemma~\ref{lem:JD2_sum} can be computed by range-sum query on 
the two sets $\mathcal{J}_{D^{\prime}}(h_{Q}, C_{Q})$ and $\mathcal{J}_{E^{\prime}}(h_{Q}, C_{Q})$. 
Therefore, we obtain the following lemma by combining Lemma~\ref{lem:JD2_division} and Lemma~\ref{lem:JD2_sum}. 

\begin{lemma}\label{lem:JD2_main_lemma}
Consider RSC query $\RSCQ(i, j)$ satisfying condition (D) of RSC query stated in Section~\ref{subsec:rsc_sub}. 
Here, let $([p_{Q}, q_{Q}], [\ell_{Q}, r_{Q}])$ be interval attractor $I_{\capture}(i, j)$; 
let $h_{Q}$, $\gamma_{Q}$, and $C_{Q}$ be the level, attractor position, and associated string of the interval attractor $([p_{Q}, q_{Q}], [\ell_{Q}, r_{Q}])$, respectively; 
let $K_{Q}, x^{\prime}$, and $M_{Q}^{\prime}$ be the three integers defined in Lemma~\ref{lem:JD2_division}. 
If $([p_{Q}, q_{Q}], [\ell_{Q}, r_{Q}]) \in \Psi_{\succeeding}$ holds, 
then the following equation holds: 
\begin{equation}\label{eq:JD2_main_lemma:1}
    \begin{split}
    |(\Psi_{\CCP}(T[i..j]) \cap \Psi_{\lex}(T[\gamma_{Q}..r_{Q}+1]) & \cap \Psi_{\run} \cap \Psi_{\centerset}(C_{Q}) \cap \Psi_{\succeeding}) \setminus \Psi_{\lcp}(K_{Q})| \\
    &= \rangesum(\mathcal{J}_{E^{\prime}}(h_{Q}, C_{Q}), 1, n, 0, |C_{Q}|-1) \\
    &- (x^{\prime}-1)\rangesum(\mathcal{J}_{D^{\prime}}(h_{Q}, C_{Q}), x^{\prime}, n, 0, |C_{Q}| - 1) \\
    &- \rangesum(\mathcal{J}_{D^{\prime}}(h_{Q}, C_{Q}), x^{\prime}, n, 0, M_{Q}^{\prime} - 1) \\
    &- \rangesum(\mathcal{J}_{E^{\prime}}(h_{Q}, C_{Q}), 1, x^{\prime}-1, 0, |C_{Q}| - 1).
    \end{split}
\end{equation}
Otherwise~(i.e., $([p_{Q}, q_{Q}], [\ell_{Q}, r_{Q}]) \not \in \Psi_{\succeeding}$), 
the following equation holds: 
\begin{equation}\label{eq:JD2_main_lemma:2}
    |(\Psi_{\CCP}(T[i..j]) \cap \Psi_{\lex}(T[\gamma_{Q}..r_{Q}+1]) \cap \Psi_{\run} \cap \Psi_{\centerset}(C_{Q}) \cap \Psi_{\succeeding}) \setminus \Psi_{\lcp}(K_{Q})| = 0.
\end{equation}
\end{lemma}
\begin{proof}
See Section~\ref{subsubsec:JD2_main_lemma_proof}.
\end{proof}

\subsubsection{Proof of Lemma~\ref{lem:JD2_division}}\label{subsubsec:JD2_division_proof}
\begin{proposition}\label{prop:psi_succ_super_property}
Consider the RSC query $\RSCQ(i, j)$ satisfying condition (D) of RSC query stated in Section~\ref{subsec:rsc_sub}. 
If $([p_{Q}, q_{Q}], [\ell_{Q}, r_{Q}]) \in \Psi_{\succeeding}$, then 
$\Psi_{h_{Q}} \cap \Psi_{\run} \cap \Psi_{\centerset}(C_{Q}) \cap (\bigcup_{\lambda = K_{Q}+1}^{n}$ $\Psi_{\lcp}(\lambda)) \subseteq \Psi_{\lex}(T[\gamma_{Q}..r_{Q}+1])$ holds.  
\end{proposition}
\begin{proof}
If $\Psi_{h_{Q}} \cap \Psi_{\run} \cap \Psi_{\centerset}(C_{Q}) \cap (\bigcup_{\lambda = K_{Q}+1}^{n} \Psi_{\lcp}(\lambda)) = \emptyset$, 
then $\Psi_{h_{Q}} \cap \Psi_{\run} \cap \Psi_{\centerset}(C_{Q}) \cap (\bigcup_{\lambda = K_{Q}+1}^{n} \Psi_{\lcp}(\lambda)) \cap \Psi_{\lex}(T[\gamma_{Q}..r_{Q}+1]) \subseteq \Psi_{\lex}(T[\gamma_{Q}..r_{Q}+1])$ holds. 
Otherwise, 
the set $\Psi_{h_{Q}} \cap \Psi_{\run} \cap \Psi_{\centerset}(C_{Q}) \cap (\bigcup_{\lambda = K_{Q}+1}^{n} \Psi_{\lcp}(\lambda))$ contains an interval attractor $([p, q], [\ell, r])$. 
Let $K = |\lcp(T[\gamma..r], C_{Q}^{n+1})|$ for the attractor position of the interval attractor $([p, q], [\ell, r])$. 
Then, $K_{Q}+1 \leq K \leq n$ 
follows from $([p, q], [\ell, r]) \in \Psi_{\centerset}(C_{Q}) \cap (\bigcup_{\lambda = K_{Q}+1}^{n} \Psi_{\lcp}(\lambda))$. 

We prove $C_{Q}^{n+1}[1..K_{Q}+1]\# \prec T[\gamma_{Q}..r_{Q}+1]$. 
Because of $\Psi_{h_{Q}} \cap \Psi_{\run} \cap \Psi_{\centerset}(C_{Q}) \cap (\bigcup_{\lambda = K_{Q}+1}^{n}$ $\Psi_{\lcp}(\lambda)) \neq \emptyset$, 
Proposition~\ref{prop:psi_condiiton_D_super_property} shows that $|\lcp(T[\gamma_{Q}..r_{Q}+1], C_{Q}^{n+1})| = K_{Q}$ holds. 
Because of $([p_{Q}, q_{Q}], [\ell_{Q}, r_{Q}]) \in \Psi_{\centerset}(C_{Q}) \cap \Psi_{\succeeding}$, 
$C_{Q}^{n+1} \prec T[\gamma_{Q}..r_{Q}+1]$ follows from the definition of the subset $\Psi_{\succeeding}$. 
Therefore, $C_{Q}^{n+1}[1..K_{Q}+1]\# \prec T[\gamma_{Q}..r_{Q}+1]$ follows from 
$|\lcp(T[\gamma_{Q}..r_{Q}+1], C_{Q}^{n+1})| = K_{Q}$ and $C_{Q}^{n+1} \prec T[\gamma_{Q}..r_{Q}+1]$.  

We prove $([p, q], [\ell, r]) \in \Psi_{\lex}(T[\gamma_{Q}..r_{Q}+1])$. 
$C_{Q}^{n+1}[1..K] \prec T[\gamma..r+1] \prec C_{Q}^{n+1}[1..K]\#$ follows from $K = |\lcp(T[\gamma..r], C_{Q}^{n+1})|$. 
$C_{Q}^{n+1}[1..K]\# \preceq C_{Q}^{n+1}[1..K_{Q}+1]\#$ holds because $K_{Q}+1 \leq K$. 
$T[\gamma..r+1] \prec T[\gamma_{Q}..r_{Q}+1]$ follows from 
$T[\gamma..r+1] \prec C_{Q}^{n+1}[1..K]\#$, $C_{Q}^{n+1}[1..K]\# \preceq C_{Q}^{n+1}[1..K_{Q}+1]\#$, 
and $C_{Q}^{n+1}[1..K_{Q}+1]\# \prec T[\gamma_{Q}..r_{Q}+1]$. 
Therefore, $([p, q], [\ell, r]) \in \Psi_{\lex}(T[\gamma_{Q}..r_{Q}+1])$ follows from $T[\gamma..r+1] \prec T[\gamma_{Q}..r_{Q}+1]$. 

We showed that $([p, q], [\ell, r]) \in \Psi_{\lex}(T[\gamma_{Q}..r_{Q}+1])$ holds for each interval attractor 
$([p, q], [\ell, r]) \in \Psi_{h_{Q}} \cap \Psi_{\run} \cap \Psi_{\centerset}(C_{Q}) \cap (\bigcup_{\lambda = K_{Q}+1}^{n} \Psi_{\lcp}(\lambda))$. 
Therefore, $\Psi_{h_{Q}} \cap \Psi_{\run} \cap \Psi_{\centerset}(C_{Q}) \cap (\bigcup_{\lambda = K_{Q}+1}^{n}$ $\Psi_{\lcp}(\lambda)) \subseteq \Psi_{\lex}(T[\gamma_{Q}..r_{Q}+1])$ holds. 
\end{proof}

\begin{proposition}\label{prop:set_JD_prime_property_big_sum}
If $([p_{Q}, q_{Q}], [\ell_{Q}, r_{Q}]) \in \Psi_{\succeeding}$, 
then the following equation holds: 
\begin{equation*}
    \begin{split}
    |(\Psi_{\CCP}(T[i..j]) & \cap \Psi_{\lex}(T[\gamma_{Q}..r_{Q}+1]) \cap \Psi_{\run} \cap \Psi_{\centerset}(C_{Q}) \cap \Psi_{\succeeding}) \setminus \Psi_{\lcp}(K_{Q})| \\
    &= |\Psi_{h_{Q}} \cap \Psi_{\run} \cap \Psi_{\centerset}(C_{Q}) \cap \Psi_{\succeeding}| \\
    &- |\Psi_{h_{Q}} \cap \Psi_{\run} \cap \Psi_{\centerset}(C_{Q}) \cap \Psi_{\succeeding} \cap (\bigcup_{\lambda = 0}^{K_{Q}} \Psi_{\lcp}(\lambda))|. 
    \end{split}
\end{equation*}
\end{proposition}
\begin{proof}
The following two equations are used to prove Proposition~\ref{prop:set_JD_prime_property_big_sum}.
\begin{equation}\label{eq:set_JD_prime_property_big_sum:1}
    \begin{split}
    (\Psi_{\CCP}(T[i..j]) & \cap \Psi_{\lex}(T[\gamma_{Q}..r_{Q}+1]) \cap \Psi_{\run} \cap \Psi_{\centerset}(C_{Q}) \cap \Psi_{\succeeding}) \setminus \Psi_{\lcp}(K_{Q}) \\
    &\subseteq \Psi_{h_{Q}} \cap \Psi_{\run} \cap \Psi_{\centerset}(C_{Q}) \cap \Psi_{\succeeding} \cap (\bigcup_{\lambda = K_{Q}+1}^{n} \Psi_{\lcp}(\lambda));
    \end{split}
\end{equation}
\begin{equation}\label{eq:set_JD_prime_property_big_sum:2}
    \begin{split}
    (\Psi_{\CCP}(T[i..j]) & \cap \Psi_{\lex}(T[\gamma_{Q}..r_{Q}+1]) \cap \Psi_{\run} \cap \Psi_{\centerset}(C_{Q}) \cap \Psi_{\succeeding}) \setminus \Psi_{\lcp}(K_{Q}) \\
    &\supseteq \Psi_{h_{Q}} \cap \Psi_{\run} \cap \Psi_{\centerset}(C_{Q}) \cap \Psi_{\succeeding} \cap (\bigcup_{\lambda = K_{Q}+1}^{n} \Psi_{\lcp}(\lambda)).
    \end{split}
\end{equation}

\textbf{Proof of Equation~\ref{eq:set_JD_prime_property_big_sum:1}.}
We prove $([p, q], [\ell, r]) \in \Psi_{h_{Q}} \cap \Psi_{\run} \cap \Psi_{\centerset}(C_{Q}) \cap \Psi_{\succeeding} \cap (\bigcup_{\lambda = K_{Q}+1}^{n}$ $\Psi_{\lcp}(\lambda))$ for each interval attractor $([p, q], [\ell, r])$ in set $(\Psi_{\CCP}(T[i..j]) \cap \Psi_{\lex}(T[\gamma_{Q}..r_{Q}+1]) \cap \Psi_{\run} \cap \Psi_{\centerset}(C_{Q}) \cap \Psi_{\succeeding}) \setminus \Psi_{\lcp}(K_{Q})$. 
Because of $([p, q], [\ell, r]) \in \Psi_{\CCP}(T[i..j])$, 
Lemma~\ref{lem:CCP_property}~\ref{enum:CCP_property:1} shows that 
$([p, q], [\ell, r]) \in \Psi_{h_{Q}}$ holds. 
Let $K = |\lcp(T[\gamma..r], C_{Q}^{n+1})|$ for the attractor position $\gamma$ of the interval attractor $([p, q], [\ell, r])$. 
Then, $([p, q], [\ell, r]) \in \Psi_{\lcp}(K)$ follows from the definition of the subset $\Psi_{\lcp}(K)$. 
Lemma~\ref{lem:psi_LMPS_property}~\ref{enum:psi_LMPS_property:lcp:1} shows that $1 \leq K \leq n$ holds. 
Proposition~\ref{prop:psi_PS_basic_property}~\ref{enum:psi_PS_basic_property:4} shows that 
$([p, q], [\ell, r]) \not \in (\bigcup_{\lambda = 1}^{K_{Q}-1} \Psi_{\lcp}(\lambda))$ holds. 
$K_{Q}+1 \leq K \leq n$ follows from 
$1 \leq K \leq n$, 
$([p, q], [\ell, r]) \not \in \Psi_{\lcp}(K_{Q})$, and $([p, q], [\ell, r]) \not \in (\bigcup_{\lambda = 1}^{K_{Q}-1} \Psi_{\lcp}(\lambda))$. 
Therefore, $([p, q], [\ell, r]) \in \Psi_{h_{Q}} \cap \Psi_{\run} \cap \Psi_{\centerset}(C_{Q}) \cap \Psi_{\succeeding} \cap (\bigcup_{\lambda = K_{Q}+1}^{n} \Psi_{\lcp}(\lambda))$ holds. 

We showed that $([p, q], [\ell, r]) \in \Psi_{h_{Q}} \cap \Psi_{\run} \cap \Psi_{\centerset}(C_{Q}) \cap \Psi_{\succeeding} \cap (\bigcup_{\lambda = K_{Q}+1}^{n} \Psi_{\lcp}(\lambda))$ holds for each interval attractor $([p, q], [\ell, r])$ in set $(\Psi_{\CCP}(T[i..j]) \cap \Psi_{\lex}(T[\gamma_{Q}..r_{Q}+1]) \cap \Psi_{\run} \cap \Psi_{\centerset}(C_{Q}) \cap \Psi_{\succeeding}) \setminus \Psi_{\lcp}(K_{Q})$. 
Therefore, Equation~\ref{eq:set_JD_prime_property_big_sum:1} holds. 

\textbf{Proof of Equation~\ref{eq:set_JD_prime_property_big_sum:2}.}
Consider an interval attractor $([p, q], [\ell, r])$ in set $\Psi_{h_{Q}} \cap \Psi_{\run} \cap \Psi_{\centerset}(C_{Q}) \cap \Psi_{\succeeding} \cap (\bigcup_{\lambda = K_{Q}+1}^{n} \Psi_{\lcp}(\lambda))$. 
Let $K = |\lcp(T[\gamma..r], C_{Q}^{n+1})|$ for the attractor position $\gamma$ of the interval attractor $([p, q], [\ell, r])$. 
Then, $([p, q], [\ell, r]) \in \Psi_{\lcp}(K)$ follows from the definition of the subset $\Psi_{\lcp}(K)$. 
$K_{Q}+1 \leq K \leq n$ follows from $([p, q], [\ell, r]) \in \bigcup_{\lambda = K_{Q}+1}^{n} \Psi_{\lcp}(\lambda)$. 

We prove $T[\gamma_{Q}..j] \prec T[\gamma..r+1] \prec T[\gamma_{Q}..j]\#$. 
The string $T[\gamma_{Q}..j]$ is a prefix of string $C_{Q}^{n+1}$ (i.e., $\lcp(T[\gamma_{Q}..j], C_{Q}^{n+1}) = T[\gamma_{Q}..j]$) 
because $|[\gamma_{Q}, j]| \leq K_{Q}$.
The string $T[\gamma_{Q}..j]$ is a prefix of the string $T[\gamma..r]$ (i.e., $\lcp(T[\gamma_{Q}..j], T[\gamma..r]) = T[\gamma_{Q}..j]$) because $\lcp(T[\gamma_{Q}..j], C_{Q}^{n+1}) = T[\gamma_{Q}..j]$, 
$K \geq K_{Q} + 1$, and $K_{Q} \geq |[\gamma_{Q}, j]|$. 
Therefore, $T[\gamma_{Q}..j] \prec T[\gamma..r+1] \prec T[\gamma_{Q}..j]\#$ follows from $\lcp(T[\gamma_{Q}..j], T[\gamma..r]) = T[\gamma_{Q}..j]$. 

We prove $\reverse(T[i..\gamma_{Q}-1]) \prec \reverse(T[p-1..\gamma-1]) \prec \reverse(\# T[i..\gamma_{Q}-1])$. 
We can apply Lemma~\ref{lem:suffix_syncro} to the interval attractor $([p_{Q}, q_{Q}], [\ell_{Q}, r_{Q}])$ 
because (A) string $T[i..\gamma_{Q}-1]$ is a suffix of string $C_{Q}^{n+1}$, 
and (B) $K_{Q} > 1 + \sum_{w = 1}^{h_{Q}+3} \lfloor \mu(w) \rfloor$ holds. 
Lemma~\ref{lem:suffix_syncro} shows that the string $T[i..\gamma_{Q}-1]$ is a suffix of string $T[p..\gamma-1]$ 
(i.e., $\lcs(T[i..\gamma_{Q}-1], T[p..\gamma-1]) = T[i..\gamma_{Q}-1]$) because 
$([p, q], [\ell, r]) \in \Psi_{h_{Q}} \cap \Psi_{\run} \cap \Psi_{\centerset}(C_{Q})$. 
Therefore, $\reverse(T[i..\gamma_{Q}-1]) \prec \reverse(T[p-1..\gamma-1]) \prec \reverse(\# T[i..\gamma_{Q}-1])$ follows from $\lcs(T[i..\gamma_{Q}-1], T[p..\gamma-1]) = T[i..\gamma_{Q}-1]$. 

We prove $([p, q], [\ell, r]) \in \Psi_{\CCP}(T[i..j])$. 
Lemma~\ref{lem:CCP_property}~\ref{enum:CCP_property:4} shows that 
$\Psi_{\CCP}(T[i..j]) = \{ ([p^{\prime}, q^{\prime}]$, $[\ell^{\prime}, r^{\prime}]) \in \Psi_{h_{Q}} \mid \reverse(T[i..\gamma_{Q}-1]) \prec \reverse(T[p^{\prime}-1..\gamma^{\prime}-1]) \prec \reverse(\#T[i..\gamma_{Q}-1]) \text{ and } T[\gamma_{Q}..j] \prec T[\gamma^{\prime}..r^{\prime}+1] \prec T[\gamma_{Q}..j]\# \}$ holds. 
Here, $\gamma^{\prime}$ is the attractor position of each interval attractor $([p^{\prime}, q^{\prime}], [\ell^{\prime}, r^{\prime}]) \in \Psi_{h_{Q}}$. 
Therefore, $([p, q], [\ell, r]) \in \Psi_{\CCP}(T[i..j])$ follows from $([p, q], [\ell, r]) \in \Psi_{h_{Q}}$, 
$\reverse(T[i..\gamma_{Q}-1]) \prec \reverse(T[p-1..\gamma-1]) \prec \reverse(\# T[i..\gamma_{Q}-1])$, and 
$T[\gamma_{Q}..j] \prec T[\gamma..r+1] \prec T[\gamma_{Q}..j]\#$. 

We prove $([p, q], [\ell, r]) \in (\Psi_{\CCP}(T[i..j]) \cap \Psi_{\lex}(T[\gamma_{Q}..r_{Q}+1]) \cap \Psi_{\run} \cap \Psi_{\centerset}(C_{Q}) \cap \Psi_{\succeeding}) \setminus \Psi_{\lcp}(K_{Q})$. 
$([p, q], [\ell, r]) \not \in \Psi_{\lcp}(K_{Q})$ follows from 
$([p, q], [\ell, r]) \in \Psi_{\lcp}(K)$ and $K \geq K_{Q} + 1$. 
Because of $([p, q], [\ell, r]) \in \Psi_{h_{Q}} \cap \Psi_{\run} \cap \Psi_{\centerset}(C_{Q}) \cap \Psi_{\succeeding} \cap (\bigcup_{\lambda = K_{Q}+1}^{n} \Psi_{\lcp}(\lambda))$, 
Proposition~\ref{prop:psi_succ_super_property} shows that 
$([p, q], [\ell, r]) \in \Psi_{\lex}(T[\gamma_{Q}..r_{Q}+1])$ holds. 
Therefore, $([p, q], [\ell, r]) \in (\Psi_{\CCP}(T[i..j]) \cap \Psi_{\lex}(T[\gamma_{Q}..r_{Q}+1]) \cap \Psi_{\run} \cap \Psi_{\centerset}(C_{Q}) \cap \Psi_{\succeeding}) \setminus \Psi_{\lcp}(K_{Q})$ follows from 
(A) $([p, q], [\ell, r]) \in \Psi_{\CCP}(T[i..j])$, 
(B) $([p, q], [\ell, r]) \in \Psi_{\lex}(T[\gamma_{Q}..r_{Q}+1])$, 
(C) $([p, q], [\ell, r]) \in \Psi_{\run} \cap \Psi_{\centerset}(C_{Q}) \cap \Psi_{\succeeding}$, 
and (D) $([p, q]$, $[\ell, r]) \not \in \Psi_{\lcp}(K_{Q})$. 

We showed that $([p, q], [\ell, r]) \in (\Psi_{\CCP}(T[i..j]) \cap \Psi_{\lex}(T[\gamma_{Q}..r_{Q}+1]) \cap \Psi_{\run} \cap \Psi_{\centerset}(C_{Q}) \cap \Psi_{\succeeding}) \setminus \Psi_{\lcp}(K_{Q})$ holds for each interval attractor $([p, q], [\ell, r])$ in set $\Psi_{h_{Q}} \cap \Psi_{\run} \cap \Psi_{\centerset}(C_{Q}) \cap \Psi_{\succeeding} \cap (\bigcup_{\lambda = K_{Q}+1}^{n} \Psi_{\lcp}(\lambda))$. 
Therefore, Equation~\ref{eq:set_JD_prime_property_big_sum:2} holds. 

\textbf{Proof of Proposition~\ref{prop:set_JD_prime_property_big_sum}.}
The following equation follows from Equation~\ref{eq:set_JD_prime_property_big_sum:1}, 
Equation~\ref{eq:set_JD_prime_property_big_sum:2}, Lemma~\ref{lem:psi_LMPS_property}~\ref{enum:psi_LMPS_property:lcp:1}, 
and Lemma~\ref{lem:psi_LMPS_property}~\ref{enum:psi_LMPS_property:lcp:1}:
\begin{equation}\label{eq:set_JD_prime_property_big_sum:3}
    \begin{split}
    (\Psi_{\CCP}(T[i..j]) & \cap \Psi_{\lex}(T[\gamma_{Q}..r_{Q}+1]) \cap \Psi_{\run} \cap \Psi_{\centerset}(C_{Q}) \cap \Psi_{\succeeding}) \setminus \Psi_{\lcp}(K_{Q}) \\
    &= \Psi_{h_{Q}} \cap \Psi_{\run} \cap \Psi_{\centerset}(C_{Q}) \cap \Psi_{\succeeding} \cap (\bigcup_{\lambda = K_{Q}+1}^{n} \Psi_{\lcp}(\lambda)) \\
    &= (\Psi_{h_{Q}} \cap \Psi_{\run} \cap \Psi_{\centerset}(C_{Q}) \cap \Psi_{\succeeding} \cap (\bigcup_{\lambda = 1}^{n} \Psi_{\lcp}(\lambda))) \\
    &\setminus (\Psi_{h_{Q}} \cap \Psi_{\run} \cap \Psi_{\centerset}(C_{Q}) \cap \Psi_{\succeeding} \cap (\bigcup_{\lambda = 1}^{K_{Q}} \Psi_{\lcp}(\lambda))) \\
    &= (\Psi_{h_{Q}} \cap \Psi_{\run} \cap \Psi_{\centerset}(C_{Q}) \cap \Psi_{\succeeding}) \\
    &\setminus (\Psi_{h_{Q}} \cap \Psi_{\run} \cap \Psi_{\centerset}(C_{Q}) \cap \Psi_{\succeeding} \cap (\bigcup_{\lambda = 1}^{K_{Q}} \Psi_{\lcp}(\lambda))).    
    \end{split}
\end{equation}
Therefore, Proposition~\ref{prop:set_JD_prime_property_big_sum} follows from Equation~\ref{eq:set_JD_prime_property_big_sum:3} and 
$\Psi_{h_{Q}} \cap \Psi_{\run} \cap \Psi_{\centerset}(C_{Q}) \cap \Psi_{\succeeding} \cap (\bigcup_{\lambda = 1}^{K_{Q}} \Psi_{\lcp}(\lambda)) \subseteq \Psi_{h_{Q}} \cap \Psi_{\run} \cap \Psi_{\centerset}(C_{Q}) \cap \Psi_{\succeeding}$. 
\end{proof}

\begin{proposition}\label{prop:set_h_HR_center_succ_lcp_sum}
The following equation holds: 
\begin{equation*}
    \begin{split}
    |\Psi_{h_{Q}} \cap \Psi_{\run} & \cap \Psi_{\centerset}(C_{Q}) \cap \Psi_{\succeeding} \cap (\bigcup_{\lambda = 0}^{K_{Q}} \Psi_{\lcp}(\lambda))| \\
&= |\Psi_{h_{Q}} \cap \Psi_{\run} \cap \Psi_{\centerset}(C_{Q}) \cap \Psi_{\succeeding} \cap (\bigcup_{\lambda = x^{\prime}}^{n} \Psi_{\nRecover}(\lambda)) \cap (\bigcup_{\lambda = 0}^{K_{Q} - M_{Q}^{\prime}} \Psi_{\lcp}(\lambda))| \\
&+ |\Psi_{h_{Q}} \cap \Psi_{\run} \cap \Psi_{\centerset}(C_{Q}) \cap \Psi_{\succeeding} \cap (\bigcup_{\lambda = x^{\prime}}^{n} \Psi_{\nRecover}(\lambda)) \cap (\bigcup_{\lambda = K_{Q} + 1 - M_{Q}^{\prime}}^{K_{Q}} \Psi_{\lcp}(\lambda))| \\
&+ |\Psi_{h_{Q}} \cap \Psi_{\run} \cap \Psi_{\centerset}(C_{Q}) \cap \Psi_{\succeeding} \cap (\bigcup_{\lambda = 1}^{x^{\prime}-1} \Psi_{\nRecover}(\lambda)) \cap (\bigcup_{\lambda = 0}^{K_{Q}} \Psi_{\lcp}(\lambda))|.
    \end{split}
\end{equation*}
\end{proposition}
\begin{proof}
    $x^{\prime} \in [1, n]$ follows form 
    $x^{\prime} = 1 + \lfloor \frac{(K_{Q} + 1) - (2 + \sum_{w = 1}^{h_{Q}+3} \lfloor \mu(w) \rfloor)}{|C_{Q}|} \rfloor$, 
    $1 + \sum_{w = 1}^{h_{Q}+3} \lfloor \mu(w) \rfloor < K_{Q} \leq n$, and $1 \leq |C_{Q}| \leq n$. 
    Similarly, $M_{Q}^{\prime} \in [0, K_{Q}]$ follows from 
    $M_{Q}^{\prime} = ((K_{Q}+1) - (2 + \sum_{w = 1}^{h_{Q}+3} \lfloor \mu(w) \rfloor) ) \mod |C_{Q}|$. 

    Proposition~\ref{prop:set_h_HR_center_succ_lcp_sum} follows from the following four equations: 

\begin{equation}\label{eq:set_h_HR_center_succ_lcp_sum:1}
    \begin{split}
    \Psi_{h_{Q}} & \cap \Psi_{\run} \cap \Psi_{\centerset}(C_{Q}) \cap \Psi_{\succeeding} \cap (\bigcup_{\lambda = 0}^{K_{Q}} \Psi_{\lcp}(\lambda)) \\
&= (\Psi_{h_{Q}} \cap \Psi_{\run} \cap \Psi_{\centerset}(C_{Q}) \cap \Psi_{\succeeding} \cap (\bigcup_{\lambda = x^{\prime}}^{n} \Psi_{\nRecover}(\lambda)) \cap (\bigcup_{\lambda = 0}^{K_{Q} - M_{Q}^{\prime}} \Psi_{\lcp}(\lambda))) \\
&\cup (\Psi_{h_{Q}} \cap \Psi_{\run} \cap \Psi_{\centerset}(C_{Q}) \cap \Psi_{\succeeding} \cap (\bigcup_{\lambda = x^{\prime}}^{n} \Psi_{\nRecover}(\lambda)) \cap (\bigcup_{\lambda = K_{Q} + 1 - M_{Q}^{\prime}}^{K_{Q}} \Psi_{\lcp}(\lambda))) \\
&\cup (\Psi_{h_{Q}} \cap \Psi_{\run} \cap \Psi_{\centerset}(C_{Q}) \cap \Psi_{\succeeding} \cap (\bigcup_{\lambda = 1}^{x^{\prime}-1} \Psi_{\nRecover}(\lambda)) \cap (\bigcup_{\lambda = 0}^{K_{Q}} \Psi_{\lcp}(\lambda))).
    \end{split}
    \end{equation}
    \begin{equation}\label{eq:set_h_HR_center_succ_lcp_sum:2}
    \begin{split}
    (\Psi_{h_{Q}} & \cap \Psi_{\run} \cap \Psi_{\centerset}(C_{Q}) \cap \Psi_{\succeeding} \cap (\bigcup_{\lambda = x^{\prime}}^{n} \Psi_{\nRecover}(\lambda)) \cap (\bigcup_{\lambda = 0}^{K_{Q} - M_{Q}^{\prime}} \Psi_{\lcp}(\lambda))) \\
    &\cap (\Psi_{h_{Q}} \cap \Psi_{\run} \cap \Psi_{\centerset}(C_{Q}) \cap \Psi_{\succeeding} \cap (\bigcup_{\lambda = x^{\prime}}^{n} \Psi_{\nRecover}(\lambda)) \cap (\bigcup_{\lambda = K_{Q} + 1 - M_{Q}^{\prime}}^{K_{Q}} \Psi_{\lcp}(\lambda))) = \emptyset. 
    \end{split}
\end{equation}

\begin{equation}\label{eq:set_h_HR_center_succ_lcp_sum:3}
    \begin{split}
    (\Psi_{h_{Q}} & \cap \Psi_{\run} \cap \Psi_{\centerset}(C_{Q}) \cap \Psi_{\succeeding} \cap (\bigcup_{\lambda = x^{\prime}}^{n} \Psi_{\nRecover}(\lambda)) \cap (\bigcup_{\lambda = K_{Q} + 1 - M_{Q}^{\prime}}^{K_{Q}} \Psi_{\lcp}(\lambda)))  \\
    &\cap (\Psi_{h_{Q}} \cap \Psi_{\run} \cap \Psi_{\centerset}(C_{Q}) \cap \Psi_{\succeeding} \cap (\bigcup_{\lambda = 1}^{x^{\prime}-1} \Psi_{\nRecover}(\lambda)) \cap (\bigcup_{\lambda = 0}^{K_{Q}} \Psi_{\lcp}(\lambda))) = \emptyset. 
    \end{split}
\end{equation}

\begin{equation}\label{eq:set_h_HR_center_succ_lcp_sum:4}
    \begin{split}
    (\Psi_{h_{Q}} & \cap \Psi_{\run} \cap \Psi_{\centerset}(C_{Q}) \cap \Psi_{\succeeding} \cap (\bigcup_{\lambda = 1}^{x^{\prime}-1} \Psi_{\nRecover}(\lambda)) \cap (\bigcup_{\lambda = 0}^{K_{Q}} \Psi_{\lcp}(\lambda)))  \\
    & \cap (\Psi_{h_{Q}} \cap \Psi_{\run} \cap \Psi_{\centerset}(C_{Q}) \cap \Psi_{\succeeding} \cap (\bigcup_{\lambda = x^{\prime}}^{n} \Psi_{\nRecover}(\lambda)) \cap (\bigcup_{\lambda = 0}^{K_{Q} - M_{Q}^{\prime}} \Psi_{\lcp}(\lambda))) = \emptyset. 
    \end{split}
\end{equation}

\textbf{Proof of Equation~\ref{eq:set_h_HR_center_succ_lcp_sum:1}.}
Because of $x^{\prime} \in [1, n]$, 
Lemma~\ref{lem:nRecover_basic_property}~\ref{enum:nRecover_basic_property:1} shows that 
the following equation holds: 
\begin{equation}\label{eq:set_h_HR_center_succ_lcp_sum:5}
    \begin{split}
    \Psi_{h_{Q}} \cap \Psi_{\run} & \cap \Psi_{\centerset}(C_{Q}) \cap \Psi_{\succeeding} \cap (\bigcup_{\lambda = 0}^{K_{Q}} \Psi_{\lcp}(\lambda)) \\
    &= (\Psi_{h_{Q}} \cap \Psi_{\run} \cap \Psi_{\centerset}(C_{Q}) \cap \Psi_{\succeeding} \cap (\bigcup_{\lambda = 0}^{K_{Q}} \Psi_{\lcp}(\lambda)) \cap (\bigcup_{\lambda = x^{\prime}}^{n} \Psi_{\nRecover}(\lambda))) \\
    & \cup (\Psi_{h_{Q}} \cap \Psi_{\run} \cap \Psi_{\centerset}(C_{Q}) \cap \Psi_{\succeeding} \cap (\bigcup_{\lambda = 0}^{K_{Q}} \Psi_{\lcp}(\lambda)) \cap (\bigcup_{\lambda = 1}^{x^{\prime}-1} \Psi_{\nRecover}(\lambda))).
    \end{split}
\end{equation}

Because of $M_{Q}^{\prime} \in [0, K_{Q}]$, 
the following equation holds: 
\begin{equation}\label{eq:set_h_HR_center_succ_lcp_sum:6}
    \begin{split}
    \Psi_{h_{Q}} & \cap \Psi_{\run} \cap \Psi_{\centerset}(C_{Q}) \cap \Psi_{\succeeding} \cap (\bigcup_{\lambda = 0}^{K_{Q}} \Psi_{\lcp}(\lambda)) \cap (\bigcup_{\lambda = x^{\prime}}^{n} \Psi_{\nRecover}(\lambda)) \\
    &= (\Psi_{h_{Q}} \cap \Psi_{\run} \cap \Psi_{\centerset}(C_{Q}) \cap \Psi_{\succeeding} \cap (\bigcup_{\lambda = 0}^{K_{Q} - M_{Q}^{\prime}} \Psi_{\lcp}(\lambda)) \cap (\bigcup_{\lambda = x^{\prime}}^{n} \Psi_{\nRecover}(\lambda))) \\
    &\cup (\Psi_{h_{Q}} \cap \Psi_{\run} \cap \Psi_{\centerset}(C_{Q}) \cap \Psi_{\succeeding} \cap (\bigcup_{\lambda = K_{Q} + 1 -M_{Q}^{\prime}}^{K_{Q}} \Psi_{\lcp}(\lambda)) \cap (\bigcup_{\lambda = x^{\prime}}^{n} \Psi_{\nRecover}(\lambda))).
    \end{split}
\end{equation}

Therefore, Equation~\ref{eq:set_h_HR_center_succ_lcp_sum:1} follows from Equation~\ref{eq:set_h_HR_center_succ_lcp_sum:5} 
and Equation~\ref{eq:set_h_HR_center_succ_lcp_sum:6}.

\textbf{Proof of Equation~\ref{eq:set_h_HR_center_succ_lcp_sum:2}.}
Lemma~\ref{lem:psi_LMPS_property}~\ref{enum:psi_LMPS_property:lcp:2} shows that 
$(\bigcup_{\lambda = 0}^{K_{Q} - M_{Q}^{\prime}} \Psi_{\lcp}(\lambda)) \cap (\bigcup_{\lambda = K_{Q} + 1 - M_{Q}^{\prime}}^{K_{Q}}$ $\Psi_{\lcp}(\lambda)) = \emptyset$ holds. 
Therefore, Equation~\ref{eq:set_h_HR_center_succ_lcp_sum:2} holds. 

\textbf{Proof of Equation~\ref{eq:set_h_HR_center_succ_lcp_sum:3}.}
Lemma~\ref{lem:nRecover_basic_property}~\ref{enum:nRecover_basic_property:2} shows that 
$(\bigcup_{\lambda = x^{\prime}}^{n} \Psi_{\nRecover}(\lambda)) \cap (\bigcup_{\lambda = 1}^{x^{\prime}-1} \Psi_{\nRecover}(\lambda)) = \emptyset$ holds. 
Therefore, Equation~\ref{eq:set_h_HR_center_succ_lcp_sum:3} holds. 

\textbf{Proof of Equation~\ref{eq:set_h_HR_center_succ_lcp_sum:4}.}
Similar to Equation~\ref{eq:set_h_HR_center_succ_lcp_sum:3}, 
Equation~\ref{eq:set_h_HR_center_succ_lcp_sum:4} follows from the fact that 
$(\bigcup_{\lambda = x^{\prime}}^{n} \Psi_{\nRecover}(\lambda)) \cap (\bigcup_{\lambda = 1}^{x^{\prime}-1} \Psi_{\nRecover}(\lambda)) = \emptyset$ holds. 

\end{proof}

\paragraph{Proof of Equation~\ref{eq:JD2_division:1} in Lemma~\ref{lem:JD2_division}.}
Equation~\ref{eq:JD2_division:1} follows from Proposition~\ref{prop:set_JD_prime_property_big_sum} and 
Proposition~\ref{prop:set_h_HR_center_succ_lcp_sum}. 

\paragraph{Proof of Equation~\ref{eq:JD2_division:2} in Lemma~\ref{lem:JD2_division}.}
$([p_{Q}, q_{Q}], [\ell_{Q}, r_{Q}]) \in \Psi_{\preceding}$ holds 
because $\Psi_{\RR} = \Psi_{\preceding} \cup \Psi_{\succeeding}$ 
and $([p_{Q}, q_{Q}], [\ell_{Q}, r_{Q}]) \not \in \Psi_{\succeeding}$. 
In this case, 
Proposition~\ref{prop:psi_PS_basic_property}~\ref{enum:psi_PS_basic_property:2} shows that 
set $\Psi_{\run} \cap \Psi_{\centerset}(C_{Q}) \cap \Psi_{\succeeding} \cap \Psi_{\lex}(T[\gamma_{Q}..r_{Q}+1])$ is empty. 
Therefore, Equation~\ref{eq:JD2_division:2} holds. 
\subsubsection{Proof of Lemma~\ref{lem:JD2_main_lemma}}\label{subsubsec:JD2_main_lemma_proof}
\paragraph{Proof of Equation~\ref{eq:JD2_main_lemma:1}.}
Equation~\ref{eq:JD2_sum:2} of Lemma~\ref{lem:JD2_sum} indicates that 
the following equation holds: 
\begin{equation}\label{eq:JD2_main_lemma:3}
    \begin{split}
    |\Psi_{h_{Q}} \cap \Psi_{\run} \cap \Psi_{\centerset}(C_{Q}) & \cap \Psi_{\succeeding} \cap (\bigcup_{\lambda = x^{\prime}}^{n} \Psi_{\nRecover}(\lambda)) \cap (\bigcup_{\lambda = 0}^{K_{Q} - M_{Q}^{\prime}} \Psi_{\lcp}(\lambda))| \\
    &= (x^{\prime}-1) \rangesum(\mathcal{J}_{D}(h_{Q}, C_{Q}), x^{\prime}, n, 0, |C_{Q}| - 1).
    \end{split}
\end{equation}

Equation~\ref{eq:JD2_sum:3} of Lemma~\ref{lem:JD2_sum} indicates that 
the following equation holds: 
\begin{equation}\label{eq:JD2_main_lemma:4}
    \begin{split}
    |\Psi_{h_{Q}} \cap \Psi_{\run} \cap \Psi_{\centerset}(C_{Q}) & \cap \Psi_{\succeeding} \cap (\bigcup_{\lambda = x^{\prime}}^{n} \Psi_{\nRecover}(\lambda)) \cap (\bigcup_{\lambda = K_{Q} + 1 - M_{Q}^{\prime}}^{K_{Q}} \Psi_{\lcp}(\lambda))| \\
    &= \rangesum(\mathcal{J}_{D}(h_{Q}, C_{Q}), x^{\prime}, n, 0, M^{\prime}_{Q} - 1). 
    \end{split}
\end{equation}

Equation~\ref{eq:JD2_sum:4} of Lemma~\ref{lem:JD2_sum} indicates that 
the following equation holds: 
\begin{equation}\label{eq:JD2_main_lemma:5}
    \begin{split}
    |\Psi_{h_{Q}} \cap \Psi_{\run} \cap \Psi_{\centerset}(C_{Q}) & \cap \Psi_{\succeeding} \cap (\bigcup_{\lambda = 1}^{x^{\prime}-1} \Psi_{\nRecover}(\lambda)) \cap (\bigcup_{\lambda = 0}^{K_{Q}} \Psi_{\lcp}(\lambda))| \\
    &= \rangesum(\mathcal{J}_{E}(h_{Q}, C_{Q}), 1, x^{\prime}-1, 0, |C_{Q}|-1). 
    \end{split}
\end{equation}

Therefore, Equation~\ref{eq:JD2_main_lemma:1} follows from Equation~\ref{eq:JD2_division:1}, 
Equation~\ref{eq:JD2_sum:1}, Equation~\ref{eq:JD2_main_lemma:3}, 
Equation~\ref{eq:JD2_main_lemma:4}, and Equation~\ref{eq:JD2_main_lemma:5}. 

\paragraph{Proof of Equation~\ref{eq:JD2_main_lemma:2}.}
This equation follows from Equation~\ref{eq:JD2_division:2} of Lemma~\ref{lem:JD2_division}.

\subsubsection{Dynamic Data Structures for Two Sets \texorpdfstring{$\mathcal{J}_{D^{\prime}}(h, C)$}{} and \texorpdfstring{$\mathcal{J}_{E^{\prime}}(h, C)$}{} of Weighted Points}\label{subsubsec:JD2_ds}
Consider a pair of an integer $h \in [0, H]$ and a string $C \in \Sigma^{+}$. 
We present dynamic data structures to support range-sum query on set $\mathcal{J}_{D^{\prime}}(h, C)$ of weighted points. 
We store the set $\mathcal{J}_{D^{\prime}}(h, C)$ using a doubly linked list $\mathbf{L}_{D^{\prime}}(h, C)$ 
in a similar way as the set $\mathcal{J}_{D}(h, C)$ introduced in Section~\ref{subsec:RSC_comp_D1}.
List indexing and range-sum data structures are built on doubly linked list $\mathbf{L}_{D^{\prime}}(h, C)$ of $|\mathbf{L}_{D^{\prime}}(h, C)|$ elements. 
This range-sum data structure is used to support range-count and range-sum queries on the set $\mathcal{J}_{D^{\prime}}(h, C)$ of 
weighted points. 

Similarly, 
we present dynamic data structures to support range-sum query on set $\mathcal{J}_{E^{\prime}}(h, C)$ of weighted points. 
We store the set $\mathcal{J}_{E^{\prime}}(h, C)$ using a doubly linked list $\mathbf{L}_{E^{\prime}}(h, C)$ of $|\mathbf{L}_{E^{\prime}}(h, C)|$ elements 
in a similar way as the set $\mathcal{J}_{D^{\prime}}(h, C)$.
Similar to the doubly linked list $\mathbf{L}_{D^{\prime}}(h, C)$, 
list indexing and range-sum data structures are built on doubly linked list $\mathbf{L}_{E^{\prime}}(h, C)$. 
These dynamic data structures require $O((|\mathcal{J}_{D^{\prime}}(h, C)| + |\mathcal{J}_{E^{\prime}}(h, C)|) B)$ bits of space in total for machine word size $B$. 
Here, $|\mathcal{J}_{D^{\prime}}(h, C)| = |\mathcal{J}_{E^{\prime}}(h, C)|$ follows from the definitions of the two sets 
$\mathcal{J}_{D^{\prime}}(h, C)$ and $\mathbf{L}_{E^{\prime}}(h, C)$. 

\subsubsection{Dynamic Data Structures for Ordered Set \texorpdfstring{$\mathcal{T}_{D^{\prime}}$}{} of Pairs}\label{subsubsec:TD2_ds}
We introduce an ordered set $\mathcal{T}_{D^{\prime}} \subseteq [0, H] \times \Sigma^{+}$ such that 
each pair of the set $\mathcal{T}_{D^{\prime}}$ consists of an integer $h \in [0, H]$ and a string $C \in \Sigma^{+}$ 
satisfying $\mathcal{J}_{D^{\prime}}(h, C) \neq \emptyset$. 
The pairs of the ordered set $\mathcal{T}_{D^{\prime}}$ are sorted in a similar way as 
that of the ordered set $\mathcal{T}_{D}$ introduced in Section~\ref{subsubsec:TD1_ds}. 

The following lemma states properties of two sets $\mathcal{T}_{D^{\prime}}$ and $\mathcal{J}_{D^{\prime}}(h, C)$. 

\begin{lemma}\label{lem:TD2_size}
    The following four statements hold: 
    \begin{enumerate}[label=\textbf{(\roman*)}]
    \item \label{enum:TD2_size:1} 
    $\sum_{(h, C) \in \mathcal{T}_{D^{\prime}}} |\mathcal{J}_{D^{\prime}}(h, C)| \leq |\Psi_{\samp}|$;
    \item \label{enum:TD2_size:2} $|\mathcal{T}_{D^{\prime}}| \leq |\Psi_{\samp}|$;
    \item \label{enum:TD2_size:3} $|\mathcal{T}_{D^{\prime}}| = O(n^{2})$;
    \item \label{enum:TD2_size:4} 
    $|\mathcal{J}_{D^{\prime}}(h, C)| \leq |\Psi_{\samp}|$ 
    and 
    $|\mathcal{J}_{D^{\prime}}(h, C)| = O(n^{2})$ for a pair of an integer $h \in [0, H]$ and a string $C \in \Sigma^{+}$. 
    \end{enumerate}
\end{lemma}
\begin{proof}
    This lemma can be proved using the same approach as for Lemma~\ref{lem:TD1_size}.
\end{proof}

We store the ordered set $\mathcal{T}_{D^{\prime}}$ of $m$ pairs using a doubly linked list of $m$ elements. 
For each integer $s \in [1, m]$, 
the $s$-th element of this doubly linked list corresponds to the $s$-th triplet $(h_{s}, C_{s})$ of the ordered set $\mathcal{T}_{D^{\prime}}$. 
The $s$-th element stores two pointers to two doubly linked lists $\mathbf{L}_{D^{\prime}}(h, C)$ and $\mathbf{L}_{E^{\prime}}(h, C)$. 
A list indexing data structure is used for quickly accessing to the elements of the doubly linked list storing the ordered set $\mathcal{T}_{D^{\prime}}$. 
These dynamic data structures require $O(m B)$ bits of space in total for machine word size $B$. 

\subsubsection{Dynamic Data Structures and Algorithm for \texorpdfstring{$\RSCQDY(i, j)$}{RSCD2(i, j)}}\label{subsubsec:JD2_subquery_ds}
We prove Lemma~\ref{lem:RSC_subquery_D2_summary}, i.e., 
we show that $\RSCQDY(i, j)$ can be answered in $O(H^{2} \log n + \log^{4} n)$ time using 
dynamic data structures of $O((|\mathcal{U}_{\RR}| + |\Psi_{\samp}|)B)$ bits of space for machine word size $B$. 
Here, $|\mathcal{U}_{\RR}|$ is the number of nodes in the RR-DAG of RLSLP $\mathcal{G}^{R}$. 
Let $(h_{1}, C_{1})$, $(h_{2}, C_{2})$, $\ldots$, $(h_{m}, C_{m})$ be the triplets in the ordered set $\mathcal{T}_{D^{\prime}}$.

\paragraph{Data Structures.}
We answer $\RSCQDY(i, j)$ using the following dynamic data structures: 
\begin{itemize}
    \item the dynamic data structures of $O(|\mathcal{U}_{\RR}|B)$ bits of space 
    for the RR-DAG of RLSLP $\mathcal{G}^{R}$ (Section~\ref{subsubsec:rrdag_ds}). 
    \item the dynamic data structures of $O(|\Psi_{\samp}|B)$ bits of space for 
    sample query (Section~\ref{subsubsec:sample_ds});
    \item the dynamic data structures of $O(\sum_{s = 1}^{m} |\mathcal{J}_{D^{\prime}}(h_{s}, C_{s})| B)$ bits of space 
    for $2m$ sets $\mathcal{J}_{D^{\prime}}(h_{1}, C_{1})$, $\mathcal{J}_{E^{\prime}}(h_{1}, C_{1})$, 
    $\mathcal{J}_{D^{\prime}}(h_{2}, C_{2})$, $\mathcal{J}_{E^{\prime}}(h_{2}, C_{2})$, $\ldots$,
    $\mathcal{J}_{D^{\prime}}(h_{m}, C_{m})$, $\mathcal{J}_{E^{\prime}}(h_{m}, C_{m})$ (Section~\ref{subsubsec:JD2_ds}); 
    \item the dynamic data structures of $O(m B)$ bits of space for the ordered set $\mathcal{T}_{D^{\prime}}$ (Section~\ref{subsubsec:TD2_ds}).
\end{itemize}
$\sum_{s = 1}^{m} |\mathcal{J}_{D^{\prime}}(h_{s}, C_{s})| \leq |\Psi_{\samp}|$ 
and $m \leq |\Psi_{\samp}|$ follow from Lemma~\ref{lem:TD2_size}. 
Therefore, these dynamic data structures can be stored in $O((|\mathcal{U}_{\RR}| + |\Psi_{\samp}|) B)$ bits of space. 

The following lemma states three queries supported by these dynamic data structures. 
\begin{lemma}\label{lem:TD2_queries}
    Let $m = |\mathcal{T}_{D^{\prime}}|$ 
    and $(h_{s}, C_{s})$ be the $s$-th triplet of the ordered set $\mathcal{T}_{D^{\prime}}$ for each integer $s \in [1, m]$.     
    Using the dynamic data structures of Section~\ref{subsubsec:JD2_subquery_ds}, 
    we can answer the following three queries:
    \begin{enumerate}[label=\textbf{(\roman*)}]
    \item \label{enum:TD2_queries:1}
    for a given integer $s \in [1, m]$, 
    return a pair of an integer $h \in [0, H]$ and a string $C \in \Sigma^{+}$ 
    satisfying $(h, T[\alpha..\beta]) = (h_{s}, C_{s})$ in $O(H^{2} + \log n)$ time; 
    \item \label{enum:TD2_queries:2}
    consider a given triplet of an integer $s \in [1, m]$, an integer $h \in [0, H]$, and an interval $[\alpha, \beta] \subseteq [1, n]$ in input string $T$. 
    Then, verify whether the $s$-th triplet of the ordered set $\mathcal{T}_{D^{\prime}}$ satisfies 
    either (A) $h_{s} < h$ or (B) $h_{s} = h$ and $C_{s} \preceq T[\alpha..\beta]$. 
    This verification takes $O(H^{2} + \log n)$ time;
    \item \label{enum:TD2_queries:3}
    verify whether $(h, T[\alpha..\beta]) \in \mathcal{T}_{D^{\prime}}$ or not in $O(H^{2} \log n + \log^{2} n)$ time  
    for a given pair of an integer $h \in [0, H]$ and interval $[\alpha, \beta] \subseteq [1, n]$ in input string $T$. 
    If $(h, T[\alpha..\beta]) \in \mathcal{T}_{D^{\prime}}$ holds, 
    then return an integer $s \in [1, m]$ satisfying 
    $(h_{s}, C_{s}) = (h, T[\alpha..\beta])$ in the same time. 
    \end{enumerate}
\end{lemma}
\begin{proof}
    This lemma can be proved using the same approach as for Lemma~\ref{lem:TC1_queries}.
\end{proof}

\paragraph{Algorithm.}
The algorithm for $\RSCQDY(i, j)$ computes $|(\Psi_{\CCP}(T[i..j]) \cap \Psi_{\lex}(T[\gamma_{Q}..r_{Q}+1]) \cap \Psi_{\run} \cap \Psi_{\centerset}(C_{Q}) \cap \Psi_{\succeeding}) \setminus \Psi_{\lcp}(K_{Q})|$ 
under the condition that 
RSC query $\RSCQ(i, j)$ satisfies condition (D) of RSC query stated in Section~\ref{subsec:rsc_sub}.  
This algorithm leverages Lemma~\ref{lem:JD2_main_lemma}, which shows that 
the size of set $|(\Psi_{\CCP}(T[i..j]) \cap \Psi_{\lex}(T[\gamma_{Q}..r_{Q}+1]) \cap \Psi_{\run} \cap \Psi_{\centerset}(C_{Q}) \cap \Psi_{\succeeding}) \setminus \Psi_{\lcp}(K_{Q})|$ can be computed 
by at most four range-sum query on two sets $\mathcal{J}_{D^{\prime}}(h_{Q}, C_{Q})$ and $\mathcal{J}_{E^{\prime}}(h_{Q}, C_{Q})$ 
of weighted points. 
Subquery $\RSCQDY(i, j)$ can be answered in the same time (i.e., $O(H^{2} \log n + \log^{4} n)$ time) by modifying the algorithm for $\RSCQDX(i, j)$ explained in Section~\ref{subsubsec:JC1_subquery_ds} 
because Lemma~\ref{lem:JD2_main_lemma} corresponds to Lemma~\ref{lem:JD1_main_lemma}. 
Therefore, Lemma~\ref{lem:RSC_subquery_D2_summary} can be proved using the same approach as for Lemma~\ref{lem:RSC_subquery_D1_summary}.

\subsection{Answering RSC Query}\label{subsec:answer_rsc_query}
We prove Theorem~\ref{theo:rsc_query_summary}, i.e., 
we show that 
we can answer a given RSC query in $O(H^{2} \log n + \log^{4} n)$ time using 
a dynamic data structure of $O((|\Psi_{\samp}| + |\mathcal{U}_{\RR}|) B)$ bits of space for machine word size $B$. 
Here, $|\mathcal{U}_{\RR}|$ is the number of nodes in the RR-DAG of RLSLP $\mathcal{G}^{R}$. 

\paragraph{Data Structures.}
We answer a given RSC query using the following dynamic data structures: 
\begin{itemize}
    \item the dynamic data structures of $O(|\mathcal{U}_{\RR}|B)$ bits of space 
    for the RR-DAG of RLSLP $\mathcal{G}^{R}$ (Section~\ref{subsubsec:rrdag_ds}). 
    \item the dynamic data structures of $O(|\Psi_{\samp}|B)$ bits of space for 
    sample query (Section~\ref{subsubsec:sample_ds});
    \item the dynamic data structures of $O((|\Psi_{\samp}| + |\mathcal{U}_{\RR}|) B)$ bits of space 
    for seven queries $\RSCQA(i, j)$, $\RSCQBX(i, j)$, $\RSCQBY(i, j)$, $\RSCQCX(i, j)$, $\RSCQCY(i, j)$, 
    $\RSCQDX(i, j)$, and $\RSCQDY(i, j)$
    (see Section~\ref{subsubsec:JA_subquery_ds}, Section~\ref{subsubsec:JB_subquery_ds}, Section~\ref{subsubsec:JC1_subquery_ds}, Section~\ref{subsubsec:JC2_subquery_ds}, Section~\ref{subsubsec:JD1_subquery_ds}, and Section~\ref{subsubsec:JD2_subquery_ds}).
\end{itemize}

\paragraph{Algorithm.}
We leverage the fact that 
$\RSCQ(i, j) = |\Psi_{\CCP}(T[i..j]) \cap \Psi_{\lex}(T[\gamma_{Q}..r_{Q}+1])|$ holds 
for the attractor position $\gamma_{Q}$ of interval attractor $([p_{Q}, q_{Q}], [\ell_{Q}, r_{Q}])$ 
(Theorem~\ref{theo:rsc_query_sub_formula}).
The algorithm for RSC query consists of three phases. 
In the first phase, 
we compute the interval attractor $([p_{Q}, q_{Q}], [\ell_{Q}, r_{Q}])$ by capture query $\CAPQ([i, j])$. 
The first phase takes $O(H^{2} \log n)$ time. 

The given RSC query $\RSCQ(i, j)$ satisfies one of four conditions (A), (B), (C), and (D) introduced in Section~\ref{subsec:rsc_sub}.  
In the second phase, 
we determine the condition satisfied by the given RSC query. 
This phase can be executed using level query $\levelQ(([p_{Q}, q_{Q}], [\ell_{Q}, r_{Q}]))$, 
attractor position query $\attrQ(([p_{Q}, q_{Q}], [\ell_{Q}, r_{Q}]))$, 
C-LCP query $\clcpQ(([p_{Q}, q_{Q}], [\ell_{Q}, r_{Q}]))$, 
and C-LCS query $\clcsQ(([p_{Q}, q_{Q}], [\ell_{Q}, r_{Q}]))$. 
Therefore, the second phase takes $O(H^{2})$ time. 

In the third phase, 
we answer the given RSC query using seven queries $\RSCQA(i, j)$, $\RSCQBX(i, j)$, $\RSCQBY(i, j)$, $\RSCQCX(i, j)$, $\RSCQCY(i, j)$, $\RSCQDX(i, j)$, and $\RSCQDY(i, j)$. 
Corollary~\ref{cor:RB_rsc_subqueries} shows that RSC query can be computed as a sum of the seven queries. 
We already showed that the seven queries can be answered in $O(H^{2} \log n + \log^{4} n)$ time. 
Therefore, the third phase can be executed in the same time. 

Finally, the algorithm for RSC query can be executed in $O(H^{2} \log n + \log^{4} n)$ time in total. 
Therefore, Theorem~\ref{theo:rsc_query_summary} holds. 

\section{RSS Query with Expected \texorpdfstring{$\delta$}{}-optimal Space}\label{sec:RSS_query}

%\begin{table}[t]
%    \normalsize
%    \centering
%    \vspace{-0.5cm}
%    \caption{
%    The relationship among subqueries and subsubqueries for RSS query.     
%    Second column represents subsubqueries used to answer each subquery.
%    }
%    \label{table:RSS_query_dependencies} 
%    \vspace{0.2cm}
%    \scalebox{0.85}{
%    \begin{tabular}{l||l|l|l|l|l|l}
% Subquery & $\levelQ$ & $\attrQ$ & $\CAPQ$ & $\clenQ$ & $\clcpQ$ & $\precQ$   \\  \hline 
%  $\RSSQ\text{-}\mathbf{A}$ & \checkmark & \checkmark & \checkmark & & & \\ \hline
%  $\RSSQ\text{-}\mathbf{B1}$ & \checkmark & \checkmark & \checkmark & \checkmark & \checkmark &   \\ \hline
%  $\RSSQ\text{-}\mathbf{B2}$ & \checkmark & \checkmark & \checkmark & \checkmark & \checkmark &   \\ \hline
%  $\RSSQ\text{-}\mathbf{C1}$ & \checkmark & \checkmark & \checkmark & \checkmark & \checkmark &   \\ \hline
%  $\RSSQ\text{-}\mathbf{C2}$ & \checkmark & \checkmark & \checkmark & \checkmark & \checkmark &  \\ \hline
%  $\RSSQ\text{-}\mathbf{D1}$ & \checkmark & \checkmark & \checkmark & \checkmark & \checkmark &   \\ \hline
%  $\RSSQ\text{-}\mathbf{D2}$ & \checkmark & \checkmark & \checkmark & \checkmark & \checkmark &  \\ \hline
%    \end{tabular}
%    }
%\end{table}

The goal of this section is to answer a given RSS query $\RSSQ(T[i..j], b)$ using 
the dynamic data structures for RSC query introduced in Section~\ref{subsec:answer_rsc_query}.
The following lemma states the summary of this section. 

\begin{theorem}\label{theo:rss_query_summary}
We can compute the string $F$ obtained by 
a given RSS query $\RSSQ(T[i..j], b)$ in $O(H^{2} \log^{2} n + \log^{6} n)$ time 
using (i) the data structures for RSC query, 
(ii) interval $[i, j]$, 
and (iii) the starting position $\eta$ of the sa-interval $[\eta, \eta^{\prime}]$ of $T[i..j]$. 
Here, $F$ is represented as an interval $[g, g + |F| - 1]$ satisfying $T[g..g + |F| - 1] = F$. 
\end{theorem}
\begin{proof}
See Section~\ref{subsec:answer_rss_query}.
\end{proof}

For this section, 
we consider two interval attractors $I_{\capture}(i, j) = ([p_{Q}, q_{Q}], [\ell_{Q}, r_{Q}])$ and $I_{\capture}(\SA[b], \SA[b] + |[i, j]| - 1) = ([p_{Q}^{\prime}, q_{Q}^{\prime}], [\ell_{Q}^{\prime}, r_{Q}^{\prime}])$. 
Here, $T[\SA[b]..\SA[b] + |[i, j]| - 1]$ is an occurrence of substring $T[i..j]$ in input string, 
and RSS query returns string $T[\SA[b]..r_{Q}^{\prime}+1]$. 
Let $h_{Q}$, $\gamma_{Q}$, and $C_{Q}$ be the level, attractor position, and associated string of the interval attractor $([p_{Q}, q_{Q}], [\ell_{Q}, r_{Q}])$, respectively. 
$[\eta, \eta^{\prime}]$ is defined as the sa-interval of string $T[i..j]$. 
Here, $\eta^{\prime} = \eta + |\Occ(T, T[i..j])| - 1$ follows from the definition of sa-interval. 
For simplicity, string $T[\gamma - |[i, \gamma_{Q}-1]|..r+1]$ is called \emph{RSS-suffix} of 
an interval attractor $([p, q], [\ell, r]) \in \Psi_{\RR}$ with attractor position $\gamma$ 
if the interval attractor is contained in set $\Psi_{\CCP}(T[i..j])$. 
The RSS-suffix has a property such that it has string $T[i..j]$ as a proper prefix. 
This is because Lemma~\ref{lem:CCP_property}~\ref{enum:CCP_property:6} shows that 
string $T[\gamma - |[i, \gamma_{Q}-1]|..\gamma + |[\gamma_{Q}, j]| - 1]$ is an occurrence of string $T[i..j]$ in input string $T$. 

We introduce three sets of strings to explain the key idea behind answering RSS query. 

\paragraph{Set $\mathcal{F}_{\suffix}(\Psi)$ of strings.}
Consider a subset $\Psi$ of set $\Psi_{\CCP}(T[i..j])$. 
Then, set $\mathcal{F}_{\suffix}(\Psi) \subseteq \Sigma^{+}$ consists of the RSS-suffixes of the interval attractors in the subset $\Psi$ (i.e., $\mathcal{F}_{\suffix}(\Psi) = \{ T[\gamma - |[i, \gamma_{Q}-1]|..r+1] \mid ([p, q], [\ell, r]) \in \Psi \}$ for each interval attractor $([p, q], [\ell, r]) \in \Psi$ with attractor position $\gamma$).

%for each string $F \in \mathcal{F}_{\suffix}(\Psi)$, 
%the subset $\Psi$ contains an interval attractor $([p, q], [\ell, r])$ satisfying 
%$F = T[\gamma - |[i, \gamma_{Q}-1]|..r+1]$. 
%Here, $\gamma$ and $\gamma_{Q}$ are the two attractor positions of the two interval attractor $([p, q], [\ell, r])$ and $([p_{Q}, q_{Q}], [\ell_{Q}, r_{Q}])$, respectively. 

%Formally, let $([p_{1}, q_{1}], [\ell_{1}, r_{1}])$, $([p_{2}, q_{2}], [\ell_{2}, r_{2}])$, $\ldots$, $([p_{k}, q_{k}], [\ell_{k}, r_{k}])$ 
%be the interval attractors in the subset $\Psi$; 
%let $\gamma_{s}$ be the attractor position of each interval attractor $([p_{s}, q_{s}], [\ell_{s}, r_{s}])$. 
%Then, let $\mathcal{F}_{\suffix}(\Psi) = \{ F \in \Sigma^{+} \mid \exists s \in [1, k] \text{ s.t. } F = T[\gamma_{s} - |[i, \gamma_{Q}-1]|..r_{s}+1] \}$. 
%The set $\mathcal{F}_{\suffix}(\Psi)$ has a property such that 
%string $T[i..j]$ is a proper prefix of each string $F \in \mathcal{F}_{\suffix}(\Psi)$. 

\paragraph{Set $\mathcal{F}_{\SA}$ of strings.}
Set $\mathcal{F}_{\SA} \subseteq \Sigma^{+}$ consists of the RSS-suffixes of the interval attractors in set $\Psi_{\CCP}(T[i..j])$ such that each RSS-suffix is equal to or lexicographically smaller than string $T[\SA[b]..r_{Q}^{\prime}+1]$. 
Formally, let $\mathcal{F}_{\SA} = \{ F \in \mathcal{F}_{\suffix}(\Psi_{\CCP}(T[i..j])) \mid F \preceq T[\SA[b]..r_{Q}^{\prime}+1] \}$. 
$\mathcal{F}_{\SA} \subseteq \mathcal{F}_{\suffix}(\Psi_{\CCP}(T[i..j]))$ follows from the definition of the set $\mathcal{F}_{\SA}$. 
The following lemma states properties of two sets $\mathcal{F}_{\SA}$ and $\mathcal{F}_{\suffix}(\Psi)$ for a subset $\Psi$ of set $\Psi_{\CCP}(T[i..j])$.

\begin{lemma}\label{lem:F_suffix_basic_property}
Let $\gamma_{Q}$ and $\gamma^{\prime}_{Q}$ be the two attractor positions of the two interval attractors $([p_{Q}, q_{Q}]$, $[\ell_{Q}, r_{Q}])$ 
and $([p^{\prime}_{Q}, q^{\prime}_{Q}], [\ell^{\prime}_{Q}, r^{\prime}_{Q}])$, respectively, for RSS query $\RSSQ(T[i..j], b)$. 
Then, the following four statements hold: 
\begin{enumerate}[label=\textbf{(\roman*)}]
    \item \label{enum:F_suffix_basic_property:1}  
    $T[\gamma^{\prime}_{Q} - |[i, \gamma_{Q}-1]|..r^{\prime}_{Q}+1] \in \mathcal{F}_{\SA}$ 
    and $\gamma^{\prime}_{Q} - |[i, \gamma_{Q}-1]| = \SA[b]$ 
    for the sa-interval $[\eta, \eta^{\prime}]$ of string $T[i..j]$;
    \item \label{enum:F_suffix_basic_property:2} 
    RSS query $\RSSQ(T[i..j], b)$ returns the lexicographically largest string in set $\mathcal{F}_{\SA}$ 
    (i.e., the lexicographically largest string in set $\mathcal{F}_{\SA}$ is $T[\SA[b]..r_{Q}^{\prime}+1]$);
    \item \label{enum:F_suffix_basic_property:3} $\mathcal{F}_{\SA} = \mathcal{F}_{\SA} \cap \mathcal{F}_{\suffix}(\Psi_{\CCP}(T[i..j]))$;
    \item \label{enum:F_suffix_basic_property:4} 
    $\mathcal{F}_{\suffix}(\emptyset) = \emptyset$;
    \item \label{enum:F_suffix_basic_property:5} 
    $\mathcal{F}_{\suffix}(\Psi_{A} \cup \Psi_{B}) = \mathcal{F}_{\suffix}(\Psi_{A}) \cup \mathcal{F}_{\suffix}(\Psi_{B})$ 
    for two subsets $\Psi_{A}$ and $\Psi_{B}$ of set $\Psi_{\CCP}(T[i..j])$;
    \item \label{enum:F_suffix_basic_property:6} 
    consider an interval attractor $([p, q], [\ell, r]) \in \Psi_{\CCP}(T[i..j])$ with attractor position $\gamma$. 
    If set $\mathcal{F}_{\SA}$ contains a string $F$ satisfying $T[\gamma - |[i, \gamma_{Q}-1]|..r+1] \preceq F$, 
    then $T[\gamma - |[i, \gamma_{Q}-1]|..r+1] \in \mathcal{F}_{\SA}$ holds. 
\end{enumerate}
\end{lemma}
\begin{proof}
The proof of Lemma~\ref{lem:F_suffix_basic_property} is as follows.

\textbf{Proof of Lemma~\ref{lem:F_suffix_basic_property}(i).}
We prove $T[\gamma^{\prime}_{Q} - |[i, \gamma_{Q}-1]|..r^{\prime}_{Q}+1] \in \mathcal{F}_{\suffix}(\Psi_{\CCP}(T[i..j]))$. 
Consider interval attractor $I_{\capture}(\SA[b], \SA[b] + |[i, j]| - 1) = ([p_{Q}^{\prime}, q_{Q}^{\prime}], [\ell_{Q}^{\prime}, r_{Q}^{\prime}])$. 
Here, $T[\SA[b]..\SA[b] + |[i, j]| - 1] = T[i..j]$ holds (see the definition of RSS query).
The set $\Psi_{\CCP}(T[i..j])$ contains the interval attractor $([p_{Q}^{\prime}, q_{Q}^{\prime}], [\ell_{Q}^{\prime}, r_{Q}^{\prime}])$ 
because $I_{\capture}(\SA[b], \SA[b] + |[i, j]| - 1) = ([p_{Q}^{\prime}, q_{Q}^{\prime}], [\ell_{Q}^{\prime}, r_{Q}^{\prime}])$, and $T[\SA[b]..\SA[b] + |[i, j]| - 1] = T[i..j]$. 
Because of $([p_{Q}^{\prime}, q_{Q}^{\prime}], [\ell_{Q}^{\prime}, r_{Q}^{\prime}]) \in \Psi_{\CCP}(T[i..j])$, 
$T[\gamma^{\prime}_{Q} - |[i, \gamma_{Q}-1]|..r^{\prime}_{Q}+1] \in \mathcal{F}_{\suffix}(\Psi_{\CCP}(T[i..j]))$ follows from 
the definition of the set $\mathcal{F}_{\suffix}(\Psi_{\CCP}(T[i..j]))$. 

We prove $\gamma^{\prime}_{Q} - |[i, \gamma_{Q}-1]| = \SA[b]$. 
We can apply Corollary~\ref{cor:capture_gamma_corollary} to the two interval attractors 
$I_{\capture}(i, j)$ and $I_{\capture}(\SA[b], \SA[b] + |[i, j]| - 1)$ because $T[\SA[b]..\SA[b] + |[i, j]| - 1] = T[i..j]$. 
Corollary~\ref{cor:capture_gamma_corollary} shows that 
$|[\SA[b], \gamma_{Q}^{\prime}-1]| = |[i, \gamma_{Q}-1]|$ holds. 
Therefore, $\gamma^{\prime}_{Q} - |[i, \gamma_{Q}-1]| = \SA[b]$ follows from $|[\SA[b], \gamma_{Q}^{\prime}-1]| = |[i, \gamma_{Q}-1]|$. 

We prove $T[\gamma^{\prime}_{Q} - |[i, \gamma_{Q}-1]|..r^{\prime}_{Q}+1] \in \mathcal{F}_{\SA}$. 
From the definition of the set $\mathcal{F}_{\SA}$, 
$T[\gamma^{\prime}_{Q} - |[i, \gamma_{Q}-1]|..r^{\prime}_{Q}+1] \in \mathcal{F}_{\SA}$ holds 
if $T[\gamma^{\prime}_{Q} - |[i, \gamma_{Q}-1]|..r^{\prime}_{Q}+1] \in \mathcal{F}_{\suffix}(\Psi_{\CCP}(T[i..j]))$ 
and $T[\gamma^{\prime}_{Q} - |[i, \gamma_{Q}-1]|..r^{\prime}_{Q}+1] = T[\SA[b]..r_{Q}^{\prime}+1]$. 
We already proved $T[\gamma^{\prime}_{Q} - |[i, \gamma_{Q}-1]|..r^{\prime}_{Q}+1] \in \mathcal{F}_{\suffix}(\Psi_{\CCP}(T[i..j]))$. 
$T[\gamma^{\prime}_{Q} - |[i, \gamma_{Q}-1]|..r^{\prime}_{Q}+1] = T[\SA[b]..r_{Q}^{\prime}+1]$ holds 
because $\gamma^{\prime}_{Q} - |[i, \gamma_{Q}-1]| = \SA[b]$. 
Therefore, $T[\gamma^{\prime}_{Q} - |[i, \gamma_{Q}-1]|..r^{\prime}_{Q}+1] \in \mathcal{F}_{\SA}$ holds. 

\textbf{Proof of Lemma~\ref{lem:F_suffix_basic_property}(ii).} 
We prove Lemma~\ref{lem:F_suffix_basic_property}~\ref{enum:F_suffix_basic_property:2} by contradiction. 
We assume that Lemma~\ref{lem:F_suffix_basic_property}~\ref{enum:F_suffix_basic_property:2} does not hold. 
Then, the set $\mathcal{F}_{\SA}$ contains a string $F$ satisfying $T[\SA[b]..r_{Q}^{\prime}+1] \prec F$. 
On the other hand, $F \preceq T[\SA[b]..r_{Q}^{\prime}+1]$ follows from the definition of the set $\mathcal{F}_{\SA}$. 
The two facts $T[\SA[b]..r_{Q}^{\prime}+1] \prec F$ and $F \preceq T[\SA[b]..r_{Q}^{\prime}+1]$ yield a contradiction. 
Therefore, Lemma~\ref{lem:F_suffix_basic_property}~\ref{enum:F_suffix_basic_property:2} must hold. 

\textbf{Proof of Lemma~\ref{lem:F_suffix_basic_property}(iii).} 
Lemma~\ref{lem:F_suffix_basic_property}(iii) follows from the definition of the set $\mathcal{F}_{\SA}$.

\textbf{Proof of Lemma~\ref{lem:F_suffix_basic_property}(iv).} 
Lemma~\ref{lem:F_suffix_basic_property}(iv) follows from the definition of the set $\mathcal{F}_{\suffix}(\emptyset)$.

\textbf{Proof of Lemma~\ref{lem:F_suffix_basic_property}(v).} 
Lemma~\ref{lem:F_suffix_basic_property}(v) follows from the definition of the set $\mathcal{F}_{\suffix}(\Psi_{A} \cup \Psi_{B})$.

\textbf{Proof of Lemma~\ref{lem:F_suffix_basic_property}(vi).} 
From the definition of the set $\mathcal{F}_{\SA}$, 
$F \in \mathcal{F}_{\suffix}(\Psi_{\CCP}(T[i..j]))$ 
and $F \preceq T[\SA[b]..r_{Q}^{\prime}+1]$ hold. 
$T[\gamma - |[i, \gamma_{Q}-1]|..r+1] \preceq T[\SA[b]..r_{Q}^{\prime}+1]$ follows from 
$T[\gamma - |[i, \gamma_{Q}-1]|..r+1] \preceq F$ and $F \preceq T[\SA[b]..r_{Q}^{\prime}+1]$. 
$T[\gamma - |[i, \gamma_{Q}-1]|..r+1] \in \mathcal{F}_{\suffix}(\Psi_{\CCP}(T[i..j]))$ follows from 
$([p, q], [\ell, r]) \in \Psi_{\CCP}(T[i..j])$. 
Therefore, $T[\gamma - |[i, \gamma_{Q}-1]|..r+1] \in \mathcal{F}_{\SA}$ follows from 
$T[\gamma - |[i, \gamma_{Q}-1]|..r+1] \in \mathcal{F}_{\suffix}(\Psi_{\CCP}(T[i..j]))$ and $T[\gamma - |[i, \gamma_{Q}-1]|..r+1] \preceq T[\SA[b]..r_{Q}^{\prime}+1]$. 

\end{proof}

The following lemma states the relationship between set $\mathcal{F}_{\SA}$ and RSC query. 

\begin{lemma}\label{lem:F_SA_formula}
Consider the interval attractors $([p_{1}, q_{1}], [\ell_{1}, r_{1}])$, $([p_{2}, q_{2}], [\ell_{2}, r_{2}])$, $\ldots$, $([p_{k}, q_{k}], [\ell_{k}, r_{k}])$ in set $\Psi_{\CCP}(T[i..j])$ for RSS query $\RSSQ(T[i..j], b)$. 
Let $\gamma_{s}$ be the attractor position of each interval attractor $([p_{s}, q_{s}], [\ell_{s}, r_{s}])$. 
Then, $\mathcal{F}_{\SA} = \{ T[\gamma_{s} - |[i, \gamma_{Q}-1]|..r_{s}+1] \mid s \in [1, k] \text{ s.t. } \RSCQ(\gamma_{s} - |[i, \gamma_{Q}-1]|, \gamma_{s} + |[\gamma_{Q}, j]| - 1) < b - \eta + 1 \}$. 
\end{lemma}
\begin{proof}
    %Let $h_{Q}$ be the level of the interval attractor $([p_{Q}, q_{Q}], [\ell_{Q}, r_{Q}])$. 
    For each integer $s \in [1, k]$, 
    Lemma~\ref{lem:CCP_property}~\ref{enum:CCP_property:6} shows that 
    the substring $T[\gamma_{s} - |[i, \gamma_{Q}-1]|..\gamma_{s} + |[\gamma_{Q}, j]| - 1]$ is an occurrence of string $T[i..j]$ in input string $T$ 
    (i.e., $T[\gamma_{s} - |[i, \gamma_{Q}-1]|..\gamma_{s} + |[\gamma_{Q}, j]| - 1] = T[i..j]$), 
    and $I_{\capture}(\gamma_{s} - |[i, \gamma_{Q}-1]|, \gamma_{s} + |[\gamma_{Q}, j]| - 1) = ([p_{s}, q_{s}], [\ell_{s}, r_{s}])$ holds. 
    For simplicity, we assume that 
    $k$ suffixes $T[\gamma_{1} - |[i, \gamma_{Q}-1]|..n]$, $T[\gamma_{2} - |[i, \gamma_{Q}-1]|..n]$, $\ldots$, 
    $T[\gamma_{k} - |[i, \gamma_{Q}-1]|..n]$ are sorted in lexicographical order. 
    Then, $T[\gamma_{1} - |[i, \gamma_{Q}-1]|..n]\$ \prec T[\gamma_{2} - |[i, \gamma_{Q}-1]|..n]\$ \prec \cdots \prec T[\gamma_{k} - |[i, \gamma_{Q}-1]|..n]$ holds. 
    
    Let $[\eta, \eta^{\prime}]$ be the sa-interval of string $T[i..j]$.     
    $\SA[\eta] = \gamma_{1} - |[i, \gamma_{Q}-1]|$ holds 
    because 
    $\eta$ is the smallest position satisfying $T[i..j] \preceq T[\SA[\eta]..n]\$$ in the suffix array.

    The following three statements are used to prove Lemma~\ref{lem:F_SA_formula}:
    \begin{enumerate}[label=\textbf{(\Alph*)}]
    \item $T[\gamma_{1} - |[i, \gamma_{Q}-1]|..r_{1}+1] \preceq T[\gamma_{2} - |[i, \gamma_{Q}-1]|..r_{2}+1] \preceq \cdots \preceq T[\gamma_{k} - |[i, \gamma_{Q}-1]|..r_{k}+1]$;
    \item $SA[\eta] = \gamma_{1} - |[i, \gamma_{Q}-1]|$, $SA[\eta+1] = \gamma_{2} - |[i, \gamma_{Q}-1]|$, 
    $\ldots$, $SA[\eta^{\prime}] = \gamma_{k} - |[i, \gamma_{Q}-1]|$, 
    and $([p_{b-\eta+1}, q_{b-\eta+1}], [\ell_{b-\eta+1}, r_{b-\eta+1}]) = ([p_{Q}^{\prime}, q_{Q}^{\prime}], [\ell_{Q}^{\prime}, r_{Q}^{\prime}])$;     
    \item 
    for an integer $s \in [1, k]$, 
    let $g$ be the smallest integer in set $[1, k]$ satisfying 
    $T[\gamma_{g} - |[i, \gamma_{Q}-1]|..r_{g}+1] = T[\gamma_{s} - |[i, \gamma_{Q}-1]|..r_{s}+1]$. 
    Then, $\RSCQ(\gamma_{s} - |[i, \gamma_{Q}-1]|, \gamma_{s} + |[\gamma_{Q}, j]| - 1) = g - 1$.     
    \end{enumerate}

    \textbf{Proof of statement (A).}
    Consider two integers $s_{1}$ and $s_{2} \in [1, k]$ ($s_{1} \leq s_{2}$). 
    Let $b_{1}$ and $b_{2}$ be two positions in the suffix array of $T$ satisfying 
    $\SA[b_{1}] = s_{1}$ and $\SA[b_{2}] = s_{2}$. 
    Then, $b_{1} < b_{2}$ follows from $T[\gamma_{s_{1}} - |[i, \gamma_{Q}-1]|..n] \prec T[\gamma_{s_{2}} - |[i, \gamma_{Q}-1]|..n]$. 
    We can apply Theorem~\ref{theo:sa_intv_formulaX} to 
    two interval attractors $I_{\capture}(\gamma_{s_{1}} - |[i, \gamma_{Q}-1]|, \gamma_{s_{1}} + |[\gamma_{Q}, j]| - 1)$ 
    and $I_{\capture}(\gamma_{s_{2}} - |[i, \gamma_{Q}-1]|, \gamma_{s_{2}} + |[\gamma_{Q}, j]| - 1)$. 
    Because of $b_{1} < b_{2}$, 
    this theorem indicates that 
    $T[\gamma_{s_{1}} - |[i, \gamma_{Q}-1]|..r_{s_{1}}+1] \prec T[\gamma_{s_{2}} - |[i, \gamma_{Q}-1]|..r_{s_{2}}+1]$ 
    or 
    $T[\gamma_{s_{1}} - |[i, \gamma_{Q}-1]|..r_{s_{1}}+1] = T[\gamma_{s_{2}} - |[i, \gamma_{Q}-1]|..r_{s_{2}}+1]$. 
    Therefore, we obtain statement (A). 

    \textbf{Proof of statement (B).}
    For a suffix $T[x..n]$ of string $T$, 
    the suffix contains string $T[i..j]$ as a prefix if and only if 
    the suffix is one of $k$ suffixes $T[\gamma_{1} - |[i, \gamma_{Q}-1]|..n]$, $T[\gamma_{2} - |[i, \gamma_{Q}-1]|..n]$, $\ldots$, 
    $T[\gamma_{k} - |[i, \gamma_{Q}-1]|..n]$. 
    The $k$ suffixes are sorted in lexicographical order. 
    Therefore, we obtain $SA[\eta] = \gamma_{1} - |[i, \gamma_{Q}-1]|$, $SA[\eta+1] = \gamma_{2} - |[i, \gamma_{Q}-1]|$, 
    $\ldots$, $SA[\eta^{\prime}] = \gamma_{k} - |[i, \gamma_{Q}-1]|$. 

    Set $\Psi_{\CCP}(T[i..j])$ contains interval attractor $I_{\capture}(\SA[b], \SA[b] + |[i, j]| - 1)$ 
    because $T[i..j] = T[\SA[b]..\SA[b] + |[i, j]| - 1]$. 
    This fact indicates that there exists an integer $g \in [1, k]$ satisfying 
    $([p_{g}, q_{g}], [\ell_{g}, r_{g}]) = ([p_{Q}^{\prime}, q_{Q}^{\prime}], [\ell_{Q}^{\prime}, r_{Q}^{\prime}])$. 
    $SA[\eta+g-1] = \gamma_{g} - |[i, \gamma_{Q}-1]|$ and $\SA[b] = \gamma_{g} - |[i, \gamma_{Q}-1]|$. 
    Therefore, we obtain $g = b - \eta + 1$. 
        
    \textbf{Proof of statement (C).}
    We prove $\RSCQ(\gamma_{s} - |[i, \gamma_{Q}-1]|, \gamma_{s} + |[\gamma_{Q}, j]| - 1) = |\{ x \in [1, n] \mid T[i..j] \preceq T[x..n]\$ \prec T[\gamma_{s} - |[i, \gamma_{Q}-1]|..r_{s}+1] \}|$.     
    $\RSCQ(\gamma_{s} - |[i, \gamma_{Q}-1]|, \gamma_{s} + |[\gamma_{Q}, j]| - 1) = |\{ x \in [1, n] \mid T[\gamma_{s} - |[i, \gamma_{Q}-1]|..\gamma_{s} + |[\gamma_{Q}, j]| - 1] \preceq T[x..n] \prec T[\gamma_{s} - |[i, \gamma_{Q}-1]|..\min\{ n, r_{s} + 1\}] \}|$ follows from the definition of RSC query (see Corollary~\ref{cor:rsc_corollary}). 
    Here, $T[i..j] = T[\gamma_{s} - |[i, \gamma_{Q}-1]|..\gamma_{s} + |[\gamma_{Q}, j]| - 1]$. 
    $T[x..n] \prec T[\gamma_{s} - |[i, \gamma_{Q}-1]|..\min\{ n, r_{s} + 1\}] \} \Leftrightarrow T[x..n]\$ \prec T[\gamma_{s} - |[i, \gamma_{Q}-1]|..r_{s} + 1] \}$ can be proved using the two facts that (a) $T[n+1] = \$$, 
    and (b) $T[1..n]$ does not contain the character $\$$. 
    Therefore, we obtain $\RSCQ(\gamma_{s} - |[i, \gamma_{Q}-1]|, \gamma_{s} + |[\gamma_{Q}, j]| - 1) = |\{ x \in [1, n] \mid T[i..j] \preceq T[x..n]\$ \prec T[\gamma_{s} - |[i, \gamma_{Q}-1]|..r_{s}+1] \}|$. 

    We prove statement (C). 
    As already explained in the proof of statement (B). 
    For a suffix $T[x..n]$ of string $T$, 
    the suffix contains string $T[i..j]$ as a prefix if and only if 
    the suffix is one of $k$ suffixes $T[\gamma_{1} - |[i, \gamma_{Q}-1]|..n]$, $T[\gamma_{2} - |[i, \gamma_{Q}-1]|..n]$, $\ldots$, 
    $T[\gamma_{k} - |[i, \gamma_{Q}-1]|..n]$. 
    If the suffix $T[x..n]$ is counted by RSC query $\RSCQ(\gamma_{s} - |[i, \gamma_{Q}-1]|, \gamma_{s} + |[\gamma_{Q}, j]| - 1)$, 
    then the suffix contains string $T[i..j]$ as a prefix.     
    Therefore, we obtain the following equation: 
    \begin{equation*}
    \begin{split}
        & \RSCQ(\gamma_{s} - |[i, \gamma_{Q}-1]|, \gamma_{s} + |[\gamma_{Q}, j]| - 1) \\
        &= |\{ x \in [1, n] \mid T[i..j] \preceq T[x..n]\$ \prec T[\gamma_{s} - |[i, \gamma_{Q}-1]|..r_{s}+1] \}| \\
        &= |\{ s^{\prime} \in [1, k] \mid T[\gamma_{s^{\prime}} - |[i, \gamma_{Q}-1]|..n]\$ \prec T[\gamma_{s} - |[i, \gamma_{Q}-1]|..r_{s}+1] \}| \\
        &= |\{ s^{\prime} \in [1, k] \mid T[\gamma_{s^{\prime}} - |[i, \gamma_{Q}-1]|..n]\$ \prec T[\gamma_{g} - |[i, \gamma_{Q}-1]|..r_{g}+1] \}|.
    \end{split}
    \end{equation*} 

    For an integer $s^{\prime} \in [1, k]$, 
    we prove $T[\gamma_{s^{\prime}} - |[i, \gamma_{Q}-1]|..n]\$ \prec T[\gamma_{g} - |[i, \gamma_{Q}-1]|..r_{g}+1] \Leftrightarrow s^{\prime} < g$. 
    We consider the following three cases: 
    (a) $s^{\prime} < g$, (b) $s^{\prime} \geq g$ and $T[\gamma_{s^{\prime}} - |[i, \gamma_{Q}-1]|..r_{s^{\prime}}+1] \neq T[\gamma_{g} - |[i, \gamma_{Q}-1]|..r_{g}+1]$, 
    and (c) $s^{\prime} \geq g$ and $T[\gamma_{s^{\prime}} - |[i, \gamma_{Q}-1]|..r_{s^{\prime}}+1] = T[\gamma_{g} - |[i, \gamma_{Q}-1]|..r_{g}+1]$. 

    For case (a), 
    $T[\gamma_{s^{\prime}} - |[i, \gamma_{Q}-1]|..r_{s^{\prime}}+1] \prec T[\gamma_{g} - |[i, \gamma_{Q}-1]|..r_{g}+1]$ 
    follows from statement (A) and the definition of the integer $g$. 
    Lemma~\ref{lem:proper_prefix_lemma} indicates that 
    string $T[\gamma_{s^{\prime}} - |[i, \gamma_{Q}-1]|..r_{s^{\prime}}+1]$ is not a prefix of string $T[\gamma_{g} - |[i, \gamma_{Q}-1]|..r_{g}+1]$. 
    This fact indicates that $T[\gamma_{s^{\prime}} - |[i, \gamma_{Q}-1]|..n]\$ \prec T[\gamma_{g} - |[i, \gamma_{Q}-1]|..r_{g}+1]$ holds. 
    Similarly, 
    we can prove $T[\gamma_{g} - |[i, \gamma_{Q}-1]|..r_{g}+1] \prec T[\gamma_{s^{\prime}} - |[i, \gamma_{Q}-1]|..n]\$$ using Lemma~\ref{lem:proper_prefix_lemma} for case (b). 
    For case (c), $T[\gamma_{g} - |[i, \gamma_{Q}-1]|..r_{g}+1] \preceq T[\gamma_{s^{\prime}} - |[i, \gamma_{Q}-1]|..n]\$$ holds 
    because string $T[\gamma_{s^{\prime}} - |[i, \gamma_{Q}-1]|..r_{s^{\prime}}+1]$ is a prefix of $T[\gamma_{s^{\prime}} - |[i, \gamma_{Q}-1]|..n]\$$. 
    Therefore, we obtain $T[\gamma_{s^{\prime}} - |[i, \gamma_{Q}-1]|..n]\$ \prec T[\gamma_{g} - |[i, \gamma_{Q}-1]|..r_{g}+1] \Leftrightarrow s^{\prime} < g$. 

    Finally, statement (C) follows from 
    $\RSCQ(\gamma_{s} - |[i, \gamma_{Q}-1]|, \gamma_{s} + |[\gamma_{Q}, j]| - 1) = |\{ s^{\prime} \in [1, k] \mid T[\gamma_{s^{\prime}} - |[i, \gamma_{Q}-1]|..n]\$ \prec T[\gamma_{g} - |[i, \gamma_{Q}-1]|..r_{g}+1] \}|$ 
    and $T[\gamma_{s^{\prime}} - |[i, \gamma_{Q}-1]|..n]\$ \prec T[\gamma_{g} - |[i, \gamma_{Q}-1]|..r_{g}+1] \Leftrightarrow s^{\prime} < g$. 
        
    \textbf{Proof of Lemma~\ref{lem:F_SA_formula}.}
    We prove $\mathcal{F}_{\SA} \subseteq \{ T[\gamma_{s} - |[i, \gamma_{Q}-1]|..r_{s}+1] \mid s \in [1, k] \text{ s.t. } \RSCQ(\gamma_{s} - |[i, \gamma_{Q}-1]|, \gamma_{s} + |[\gamma_{Q}, j]| - 1) < b - \eta + 1 \}$. 
    Consider a string $F \in \mathcal{F}_{\SA}$. 
    From the definition of the set $\mathcal{F}_{\SA}$, 
    $F \in \mathcal{F}_{\suffix}(\Psi_{\CCP}(T[i..j]))$ 
    and $F \preceq T[\SA[b]..r_{Q}^{\prime}+1]$ hold. 
    Since $F \in \mathcal{F}_{\suffix}(\Psi_{\CCP}(T[i..j]))$, 
    there exists an integer $s \in [1, k]$ satisfying 
    $F = T[\gamma_{s} - |[i, \gamma_{Q}-1]|..r_{s}+1]$. 
    Let $g$ be the smallest integer in set $[1, k]$ satisfying 
    $T[\gamma_{g} - |[i, \gamma_{Q}-1]|..r_{g}+1] = F$. 
    Similarly, 
    let $g^{\prime}$ be the smallest integer in set $[1, k]$ satisfying 
    $T[\gamma_{g^{\prime}} - |[i, \gamma_{Q}-1]|..r_{g^{\prime}}+1] = T[\gamma_{b-\eta+1} - |[i, \gamma_{Q}-1]|..r_{b-\eta+1}+1]$. 
    Here, $T[\gamma_{b-\eta+1} - |[i, \gamma_{Q}-1]|..r_{b-\eta+1}+1] = T[\SA[b]..r_{Q}^{\prime}+1]$ follows from statement (B). 
    Statement (C) shows that 
    $\RSCQ(\gamma_{s} - |[i, \gamma_{Q}-1]|, \gamma_{s} + |[\gamma_{Q}, j]| - 1) = g-1$.
    $g-1 < b-\eta+1$ holds 
    because $g \leq g^{\prime} \leq b-\eta+1$ follows from statement (A) and $T[\gamma_{b-\eta+1} - |[i, \gamma_{Q}-1]|..r_{b-\eta+1}+1] = T[\SA[b]..r_{Q}^{\prime}+1]$. 
    We obtain $\RSCQ(\gamma_{s} - |[i, \gamma_{Q}-1]|, \gamma_{s} + |[\gamma_{Q}, j]| - 1) < b-\eta+1$, 
    and hence, $\mathcal{F}_{\SA} \subseteq \{ T[\gamma_{s} - |[i, \gamma_{Q}-1]|..r_{s}+1] \mid s \in [1, k] \text{ s.t. } \RSCQ(\gamma_{s} - |[i, \gamma_{Q}-1]|, \gamma_{s} + |[\gamma_{Q}, j]| - 1) < b - \eta + 1 \}$ holds. 

    Next, we prove $\mathcal{F}_{\SA} \supseteq \{ T[\gamma_{s} - |[i, \gamma_{Q}-1]|..r_{s}+1] \mid s \in [1, k] \text{ s.t. } \RSCQ(\gamma_{s} - |[i, \gamma_{Q}-1]|, \gamma_{s} + |[\gamma_{Q}, j]| - 1) < b - \eta + 1 \}$.    
    Consider a string $F^{\prime} \in \{ T[\gamma_{s} - |[i, \gamma_{Q}-1]|..r_{s}+1] \mid s \in [1, k] \text{ s.t. }$ $\RSCQ(\gamma_{s} - |[i, \gamma_{Q}-1]|, \gamma_{s} + |[\gamma_{Q}, j]| - 1) < b - \eta + 1 \}$. 
    Then, there exists an integer $s^{\prime} \in [1, k]$ satisfying 
    $F^{\prime} = T[\gamma_{s^{\prime}} - |[i, \gamma_{Q}-1]|..r_{s^{\prime}}+1]$ and $\RSCQ(\gamma_{s^{\prime}} - |[i, \gamma_{Q}-1]|, \gamma_{s^{\prime}} + |[\gamma_{Q}, j]| - 1) < b - \eta + 1$.     
    Let $g^{\prime\prime} \in [1, k]$ be the smallest integer satisfying $F^{\prime} = T[\gamma_{g^{\prime\prime}} - |[i, \gamma_{Q}-1]|..r_{s^{\prime\prime}}+1]$. 
    Then, $\RSCQ(\gamma_{s^{\prime}} - |[i, \gamma_{Q}-1]|, \gamma_{s^{\prime}} + |[\gamma_{Q}, j]| - 1) = g^{\prime\prime}-1$. 
    $g^{\prime\prime} \leq b - \eta + 1$ follows from 
    $\RSCQ(\gamma_{s^{\prime}} - |[i, \gamma_{Q}-1]|, \gamma_{s^{\prime}} + |[\gamma_{Q}, j]| - 1) = g^{\prime\prime}-1$ and 
    $\RSCQ(\gamma_{s^{\prime}} - |[i, \gamma_{Q}-1]|, \gamma_{s^{\prime}} + |[\gamma_{Q}, j]| - 1) < b - \eta + 1$. 

    Let $m^{\prime}$ be the largest integer in set $[1, k]$ satisfying 
    $T[\gamma_{m^{\prime}} - |[i, \gamma_{Q}-1]|..r_{m^{\prime}}+1] = T[\gamma_{b-\eta+1} - |[i, \gamma_{Q}-1]|..r_{b-\eta+1}+1]$.  
    Then, $b-\eta+1 \in [g^{\prime}, m^{\prime}]$ holds because 
    the $k$ strings 
    $T[\gamma_{1} - |[i, \gamma_{Q}-1]|..r_{1}+1], T[\gamma_{2} - |[i, \gamma_{Q}-1]|..r_{2}+1], \ldots, T[\gamma_{k} - |[i, \gamma_{Q}-1]|..r_{k}+1]$ are sorted in lexicographical order (statement (A)). 
    
    Similarly, 
    let $m^{\prime\prime}$ be the largest integer in set $[1, k]$ satisfying 
    $T[\gamma_{m^{\prime\prime}} - |[i, \gamma_{Q}-1]|..r_{m^{\prime\prime}}+1] = F^{\prime}$. 
    Then, $m^{\prime\prime} \leq m^{\prime}$ holds because 
    $b-\eta+1 \in [g^{\prime}, m^{\prime}]$ and $g^{\prime\prime} \leq b - \eta + 1$. 
    This fact indicates that $F^{\prime} \preceq T[\gamma_{b-\eta+1} - |[i, \gamma_{Q}-1]|..r_{b-\eta+1}+1]$ (i.e., $F^{\prime} \preceq T[\SA[b]..r_{Q}^{\prime}+1]$). 

    $F^{\prime} \in \mathcal{F}_{\suffix}(\Psi_{\CCP}(T[i..j]))$ follows from 
    $F^{\prime} = T[\gamma_{s^{\prime}} - |[i, \gamma_{Q}-1]|..r_{s^{\prime}}+1]$ 
    and 
    $([p_{s^{\prime}}, q_{s^{\prime}}], [\ell_{s^{\prime}}, r_{s^{\prime}}]) \in \Psi_{\CCP}(T[i..j])$.    
    $F^{\prime} \in \mathcal{F}_{\SA}$ follows from 
    $F^{\prime} \in \mathcal{F}_{\suffix}(\Psi_{\CCP}(T[i..j]))$ and $F^{\prime} \preceq T[\gamma_{b-\eta+1} - |[i, \gamma_{Q}-1]|..r_{b-\eta+1}+1]$. 
    Therefore, $\mathcal{F}_{\SA} \supseteq \{ T[\gamma_{s} - |[i, \gamma_{Q}-1]|..r_{s}+1] \mid s \in [1, k] \text{ s.t. } \RSCQ(\gamma_{s} - |[i, \gamma_{Q}-1]|, \gamma_{s} + |[\gamma_{Q}, j]| - 1) < b - \eta + 1 \}$ holds. 

    Finally, Lemma~\ref{lem:F_SA_formula} follows from 
    $\mathcal{F}_{\SA} \subseteq \{ T[\gamma_{s} - |[i, \gamma_{Q}-1]|..r_{s}+1] \mid s \in [1, k] \text{ s.t. } \RSCQ(\gamma_{s} - |[i, \gamma_{Q}-1]|, \gamma_{s} + |[\gamma_{Q}, j]| - 1) < b - \eta + 1 \}$ 
    and $\mathcal{F}_{\SA} \supseteq \{ T[\gamma_{s} - |[i, \gamma_{Q}-1]|..r_{s}+1] \mid s \in [1, k] \text{ s.t. } \RSCQ(\gamma_{s} - |[i, \gamma_{Q}-1]|, \gamma_{s} + |[\gamma_{Q}, j]| - 1) < b - \eta + 1 \}$.

\end{proof}

\paragraph{Set $\mathcal{C}_{\run}$ of strings.}
Set $\mathcal{C}_{\run} \subseteq \Sigma^{+}$ consists of strings such that 
each string $C \in \mathcal{C}_{\run}$ satisfies 
$\mathcal{F}_{\SA} \cap \mathcal{F}_{\suffix}(\Psi_{\CCP}(T[i..j]) \cap \Psi_{\run} \cap \Psi_{\centerset}(C)) \neq \emptyset$ 
(i.e., $\mathcal{C}_{\run} = \{ C \in \Sigma^{+} \mid \mathcal{F}_{\SA} \cap \mathcal{F}_{\suffix}(\Psi_{\CCP}(T[i..j]) \cap \Psi_{\run} \cap \Psi_{\centerset}(C)) \neq \emptyset \}$). 

The following lemma states the relationship between two strings in the set $\mathcal{C}_{\run}$.

\begin{lemma}\label{lem:C_run_property}
    The following four statements hold for two strings $C_{1}, C_{2} \in \mathcal{C}_{\run}$ ($C_{1} \neq C_{2}$): 
    \begin{enumerate}[label=\textbf{(\roman*)}]
    \item \label{enum:C_run_property:1} $\Psi_{h_{Q}} \cap \Psi_{\run} \cap \Psi_{\centerset}(C_{1}) \neq \emptyset$ and $\Psi_{h_{Q}} \cap \Psi_{\run} \cap \Psi_{\centerset}(C_{2}) \neq \emptyset$;
    \item \label{enum:C_run_property:2} $C_{1}^{n+1}[1..2 + \sum_{w = 1}^{h_{Q}+3} \lfloor \mu(w) \rfloor] \neq C_{2}^{n+1}[1..2 + \sum_{w = 1}^{h_{Q}+3} \lfloor \mu(w) \rfloor]$;
    \item \label{enum:C_run_property:3} $T[i..\gamma_{Q}-1] \cdot (C_{1}^{n+1}[1..2 + \sum_{w = 1}^{h_{Q}+3} \lfloor \mu(w) \rfloor])$ is a prefix of each string $F_{1}$ in set $\mathcal{F}_{\suffix}(\Psi_{\CCP}(T[i..j]) \cap \Psi_{\run} \cap \Psi_{\centerset}(C_{1}))$. 
    Similarly, $T[i..\gamma_{Q}-1] \cdot (C_{2}^{n+1}[1..2 + \sum_{w = 1}^{h_{Q}+3} \lfloor \mu(w) \rfloor])$ is a prefix of each string $F_{2}$ in set $\mathcal{F}_{\suffix}(\Psi_{\CCP}(T[i..j]) \cap \Psi_{\run} \cap \Psi_{\centerset}(C_{2}))$; 
    \item \label{enum:C_run_property:4} 
    consider two strings $F_{1}$ and $F_{2}$ in the two sets $\mathcal{F}_{\suffix}(\Psi_{\CCP}(T[i..j]) \cap \Psi_{\run} \cap \Psi_{\centerset}(C_{1}))$ and $\mathcal{F}_{\suffix}(\Psi_{\CCP}(T[i..j]) \cap \Psi_{\run} \cap \Psi_{\centerset}(C_{2}))$. 
    Then, $F_{1} \prec F_{2} \Leftrightarrow C_{1}^{n+1}[1..2 + \sum_{w = 1}^{h_{Q}+3} \lfloor \mu(w) \rfloor] \prec C_{2}^{n+1}[1..2 + \sum_{w = 1}^{h_{Q}+3} \lfloor \mu(w) \rfloor]$. 
    \end{enumerate}
\end{lemma}
\begin{proof}
    The proof of Lemma~\ref{lem:C_run_property} is as follows.

    \textbf{Proof of Lemma~\ref{lem:C_run_property}(i).}
    We prove $\Psi_{h_{Q}} \cap \Psi_{\run} \cap \Psi_{\centerset}(C_{1}) \neq \emptyset$. 
    $\mathcal{F}_{\suffix}(\Psi_{\CCP}(T[i..j]) \cap \Psi_{\run} \cap \Psi_{\centerset}(C_{1})) \neq \emptyset$ 
    follows from the definition of the set $\mathcal{C}_{\run}$. 
    Because of $\mathcal{F}_{\suffix}(\Psi_{\CCP}(T[i..j]) \cap \Psi_{\run} \cap \Psi_{\centerset}(C_{1})) \neq \emptyset$, 
    set $\Psi_{\CCP}(T[i..j]) \cap \Psi_{\run} \cap \Psi_{\centerset}(C_{1})$ contains an interval attractor $([p, q], [\ell, r])$. 
    Lemma~\ref{lem:CCP_property}~\ref{enum:CCP_property:1} shows that $([p, q], [\ell, r]) \in \Psi_{h_{Q}}$ holds. 
    Therefore, $\Psi_{h_{Q}} \cap \Psi_{\run} \cap \Psi_{\centerset}(C_{1}) \neq \emptyset$ holds. 

    Similarly, we can prove $\Psi_{h_{Q}} \cap \Psi_{\run} \cap \Psi_{\centerset}(C_{2}) \neq \emptyset$ 
    using the same approach used to prove $\Psi_{h_{Q}} \cap \Psi_{\run} \cap \Psi_{\centerset}(C_{1}) \neq \emptyset$. 

    \textbf{Proof of Lemma~\ref{lem:C_run_property}(ii).}
    Lemma~\ref{lem:C_run_property}(ii) follows from Lemma~\ref{lem:psi_HR_repetitive_C_property} and Lemma~\ref{lem:C_run_property}(i). 

    \textbf{Proof of Lemma~\ref{lem:C_run_property}(iii).}
    We show that $T[i..\gamma_{Q}-1] \cdot (C_{1}^{n+1}[1..2 + \sum_{w = 1}^{h_{Q}+3} \lfloor \mu(w) \rfloor])$ is a prefix of the string $F_{1}$. 
    Because of $F_{1} \in \mathcal{F}_{\suffix}(\Psi_{\CCP}(T[i..j]) \cap \Psi_{\run} \cap \Psi_{\centerset}(C_{1}))$, 
    the set $\Psi_{\CCP}(T[i..j]) \cap \Psi_{\run} \cap \Psi_{\centerset}(C_{1})$ 
    contains an interval attractor $([p, q], [\ell, r])$ satisfying 
    $T[\gamma - |[i, \gamma_{Q}-1]|..r+1] = F_{1}$ for the interval attractor $\gamma$ of the interval attractor $([p, q], [\ell, r])$. 
    Here, Lemma~\ref{lem:CCP_property}~\ref{enum:CCP_property:1} shows that $([p, q], [\ell, r]) \in \Psi_{h_{Q}}$ holds (see the proof of Lemma~\ref{lem:C_run_property}(i)). 
    Because of $([p, q], [\ell, r]) \in \Psi_{\CCP}(T[i..j])$, 
    Lemma~\ref{lem:CCP_property}~\ref{enum:CCP_property:6} shows that 
    $T[\gamma - |[i, \gamma_{Q}-1]|..\gamma-1] = T[i..\gamma_{Q}-1]$ holds. 
    Because of $([p, q], [\ell, r]) \in \Psi_{\run} \cap \Psi_{h_{Q}} \cap \Psi_{\centerset}(C_{1})$, 
    $|\lcp(T[\gamma_{1}..r_{1}], C_{1}^{n+1})| > 1 + \sum_{w = 1}^{h_{Q}+3} \lfloor \mu(w) \rfloor$ follows from the definition of the subset $\Psi_{\run}$. 
    $C_{1}^{n+1}[1..2 + \sum_{w = 1}^{h_{Q}+3} \lfloor \mu(w) \rfloor] = T[\gamma_{1}..\gamma_{1} + 1 + \sum_{w = 1}^{h_{Q}+3} \lfloor \mu(w) \rfloor]$ follows from $|\lcp(T[\gamma_{1}..r_{1}], C_{1}^{n+1})| > 1 + \sum_{w = 1}^{h_{Q}+3} \lfloor \mu(w) \rfloor$. 
    Therefore, $T[i..\gamma_{Q}-1] \cdot (C_{1}^{n+1}[1..2 + \sum_{w = 1}^{h_{Q}+3} \lfloor \mu(w) \rfloor])$ is a prefix of the string $F_{1}$ 
    because 
    (A) $T[\gamma - |[i, \gamma_{Q}-1]|..\gamma-1] \cdot T[\gamma_{1}..\gamma_{1} + 1 + \sum_{w = 1}^{h_{Q}+3} \lfloor \mu(w) \rfloor]$ is a prefix of the string $F_{1}$, 
    (B) $T[\gamma - |[i, \gamma_{Q}-1]|..\gamma-1] = T[i..\gamma_{Q}-1]$, 
    and (C) $C_{1}^{n+1}[1..2 + \sum_{w = 1}^{h_{Q}+3} \lfloor \mu(w) \rfloor] = T[\gamma_{1}..\gamma_{1} + 1 + \sum_{w = 1}^{h_{Q}+3} \lfloor \mu(w) \rfloor]$. 
    
    \textbf{Proof of Lemma~\ref{lem:C_run_property}(iv).}
    Lemma~\ref{lem:C_run_property}(iv) follows from Lemma~\ref{lem:C_run_property}(ii) and Lemma~\ref{lem:C_run_property}(iii). 
\end{proof}

The following lemma states a property of the set $\mathcal{C}_{\run}$.
\begin{lemma}\label{lem:c_run_property}
    If $\mathcal{C}_{\run} \neq \emptyset$ and $|[\gamma_{Q}, j]| > 1 + \sum_{w = 1}^{h_{Q}+3} \lfloor \mu(w) \rfloor$, 
    then $\mathcal{C}_{\run} = \{ C_{Q} \}$ holds.
\end{lemma}
\begin{proof}
Consider a string $C$ in set $\mathcal{C}_{\run}$. 
Then, 
$\mathcal{F}_{\SA} \cap \mathcal{F}_{\suffix}(\Psi_{\CCP}(T[i..j]) \cap \Psi_{\run} \cap \Psi_{\centerset}(C)) \neq \emptyset$ follows from 
the definition of the set $\mathcal{C}_{\run}$. 
Because of $\mathcal{F}_{\suffix}(\Psi_{\CCP}(T[i..j]) \cap \Psi_{\run} \cap \Psi_{\centerset}(C)) \neq \emptyset$, 
$\Psi_{\CCP}(T[i..j]) \cap \Psi_{\run} \cap \Psi_{\centerset}(C) \neq \emptyset$ follows from 
the definition of the set $\mathcal{F}_{\suffix}(\Psi_{\CCP}(T[i..j]) \cap \Psi_{\run} \cap \Psi_{\centerset}(C))$. 
We can apply Lemma~\ref{lem:CCP_property}~\ref{enum:CCP_property:5} to the string $C$ 
because $|[\gamma_{Q}, j]| > 1 + \sum_{w = 1}^{h_{Q}+3} \lfloor \mu(w) \rfloor$ and 
$\Psi_{\CCP}(T[i..j]) \cap \Psi_{\centerset}(C) \neq \emptyset$ hold. 
Lemma~\ref{lem:CCP_property}~\ref{enum:CCP_property:5} shows that $C_{Q} = C$ holds. 
Therefore, $\mathcal{C}_{\run} = \{ C_{Q} \}$ holds. 
\end{proof}
%%%%%%%%%%%%%%%%%%%%%

\subsection{Eight Subqueries}\label{subsec:rss_subqueries}
\begin{table}[t]
    \normalsize
    \vspace{-0.5cm}
    \caption{
    Eight subqueries used to answer RSS query $\RSSQ(T[i..j], b)$. 
    $h_{Q}$, $\gamma_{Q}$, and $C_{Q}$ are the level, attractor position, and associated string of 
    interval attractor $I_{\capture}(i, j) = ([p_{Q}, q_{Q}], [\ell_{Q}, r_{Q}])$, respectively. 
    $\hat{K}$ is the length of the longest common prefix between two strings 
    $T[\gamma_{Q}..j]$ and $C_{Q}^{n+1}$. 
    $K$ is a given integer in $[\max\{ |[\gamma_{Q}, j]|, 2 + \sum_{w = 1}^{h_{Q}+3} \lfloor \mu(w) \rfloor \}, n]$.
    $C_{\max}$ is the string in set $\mathcal{C}_{\run}$ satisfying 
    $C^{n+1}[1..2 + \sum_{w = 1}^{h_{Q}+3} \lfloor \mu(w) \rfloor] \prec C^{n+1}_{\max}[1..2 + \sum_{w = 1}^{h_{Q}+3} \lfloor \mu(w) \rfloor]$ 
    for each string $C \in \mathcal{C}_{\run} \setminus \{ C_{\max} \}$.
    }
    \vspace{1mm}    
    \label{table:RSS_query_result} 
    \center{
    \scalebox{0.85}{
    \begin{tabular}{l||l|l}
 Subquery & Description & Sections \\  \hline 
 \multirow{2}{*}{$\RSSQA(T[i..j], b)$} & Return the lexicographically largest string  & \multirow{2}{*}{\ref{subsec:GammaA}} \\ 
  &  in set $\mathcal{F}_{\SA} \cap \mathcal{F}_{\suffix}(\Psi_{\CCP}(T[i..j]) \setminus \Psi_{\run})$ & \\ \hline  
 \multirow{2}{*}{$\RSSQB(T[i..j], b)$} & Verify whether $\mathcal{C}_{\run} = \emptyset$ or not. If $\mathcal{C}_{\run} \neq \emptyset$, then return an interval attractor & \multirow{2}{*}{\ref{subsec:GammaB}} \\ 
  & contained in set $\Psi_{\CCP}(T[i..j]) \cap \Psi_{\run} \cap \Psi_{\centerset}(C_{\max})$  & \\ \hline  
 \multirow{2}{*}{$\RSSQCX(T[i..j], b)$} & Return the lexicographically largest string & \multirow{2}{*}{\ref{subsec:GammaC1}} \\ 
  & in set $\mathcal{F}_{\SA} \cap \mathcal{F}_{\suffix}(\Psi_{\CCP}(T[i..j]) \cap \Psi_{\run} \cap \Psi_{\centerset}(C_{Q}) \cap \Psi_{\lcp}(\hat{K}) \cap \Psi_{\preceding})$ & \\ \hline 
 \multirow{2}{*}{$\RSSQCY(T[i..j], b)$} & Return the lexicographically largest string & \multirow{2}{*}{\ref{subsec:GammaC2}} \\ 
  & in set $\mathcal{F}_{\SA} \cap \mathcal{F}_{\suffix}(\Psi_{\CCP}(T[i..j]) \cap \Psi_{\run} \cap \Psi_{\centerset}(C_{Q}) \cap \Psi_{\lcp}(\hat{K}) \cap \Psi_{\succeeding})$ & \\  \hline
 \multirow{2}{*}{$\RSSQDX(T[i..j], b, K)$} & Return the lexicographically largest string & \multirow{2}{*}{\ref{subsec:GammaD1}} \\ 
  & in set $\mathcal{F}_{\SA} \cap \mathcal{F}_{\suffix}(\Psi_{\CCP}(T[i..j]) \cap \Psi_{\run} \cap \Psi_{\centerset}(C_{Q}) \cap \Psi_{\lcp}(K) \cap \Psi_{\preceding})$ & \\  \hline 
 \multirow{2}{*}{$\RSSQDY(T[i..j], b, K)$} & Return the lexicographically largest string & \multirow{2}{*}{\ref{subsec:GammaD2}} \\ 
  & in set $\mathcal{F}_{\SA} \cap \mathcal{F}_{\suffix}(\Psi_{\CCP}(T[i..j]) \cap \Psi_{\run} \cap \Psi_{\centerset}(C_{Q}) \cap \Psi_{\lcp}(K) \cap \Psi_{\succeeding})$ & \\  \hline 
 \multirow{2}{*}{$\RSSQEX(T[i..j], b)$} & Return the lexicographically largest string & \multirow{2}{*}{\ref{subsec:GammaE1}} \\ 
  & in set $\mathcal{F}_{\SA} \cap \mathcal{F}_{\suffix}(\Psi_{\CCP}(T[i..j]) \cap \Psi_{\run} \cap \Psi_{\centerset}(C_{Q}) \cap \Psi_{\preceding})$ & \\  \hline 
 \multirow{2}{*}{$\RSSQEY(T[i..j], b)$} & Return the lexicographically largest string & \multirow{2}{*}{\ref{subsec:GammaE2}} \\ 
  & in set $\mathcal{F}_{\SA} \cap \mathcal{F}_{\suffix}(\Psi_{\CCP}(T[i..j]) \cap \Psi_{\run} \cap \Psi_{\centerset}(C_{Q}) \cap \Psi_{\succeeding})$ & \\ \hline
    \end{tabular} 
    }
    }
\end{table}

As already explained in Section~\ref{subsec:simplified_rss_rsc}, 
RSS query $\RSSQ(T[i..j], b)$ is decomposed into eight subqueries. 
Let $\hat{K}$ be the length of the longest common prefix between two strings 
$T[\gamma_{Q}..j]$ and $C_{Q}^{n+1}$ (i.e., $\hat{K} = |\lcp(T[\gamma_{Q}..j], C_{Q}^{n+1})|$). 
Let $C_{\max}$ be the string in set $\mathcal{C}_{\run}$ satisfying 
$C^{n+1}[1..2 + \sum_{w = 1}^{h_{Q}+3} \lfloor \mu(w) \rfloor] \prec C^{n+1}_{\max}[1..2 + \sum_{w = 1}^{h_{Q}+3} \lfloor \mu(w) \rfloor]$ 
for each string $C \in \mathcal{C}_{\run} \setminus \{ C_{\max} \}$. 
Here, the string $C_{\max}$ exists only if the set $\mathcal{C}_{\run}$ is not empty. 
Then, the eight subqueries are defined as follows (see also Table~\ref{table:RSS_query_result}): 
\begin{description}
    \item{$\RSSQA(T[i..j], b)$:} 
    returns the lexicographically largest string among strings in set $\mathcal{F}_{\SA}$ such that 
    each string is the RSS-suffix of a non-periodic interval attractor in set $\Psi_{\CCP}(T[i..j])$. 
    Formally, this subquery returns the lexicographically largest string in set 
    $\mathcal{F}_{\SA} \cap \mathcal{F}_{\suffix}(\Psi_{\CCP}(T[i..j]) \setminus \Psi_{\run})$.     
    \item{$\RSSQB(T[i..j], b)$:} verifies whether set $\mathcal{C}_{\run}$ is empty or not. 
    If the set is not empty,     
    then this subquery returns a periodic interval attractor satisfying the following two conditions: 
    (1) it is contained in set $\Psi_{\CCP}(T[i..j])$; 
    (2) it has string $C_{\max}$ as its associated string 
    (i.e., this subquery returns an interval attractors in set $\Psi_{\CCP}(T[i..j]) \cap \Psi_{\run} \cap \Psi_{\centerset}(C_{\max})$). 
    \item{$\RSSQCX(T[i..j], b)$:} returns the lexicographically largest string among strings in set $\mathcal{F}_{\SA}$ 
    such that each string is the RSS-suffix of a periodic interval attractor satisfying the following four conditions: 
    (1) it is contained in set $\Psi_{\CCP}(T[i..j])$; 
    (2) it is contained in set $\Psi_{\lcp}(\hat{K})$; 
    (3) it is contained in set $\Psi_{\preceding}$; 
    (4) it has string $C_{Q}$ as its associated string. 
    Formally, this subquery returns the lexicographically largest string in set 
    $\mathcal{F}_{\SA} \cap \mathcal{F}_{\suffix}(\Psi_{\CCP}(T[i..j]) \cap \Psi_{\run} \cap \Psi_{\centerset}(C_{Q}) \cap \Psi_{\lcp}(\hat{K}) \cap \Psi_{\preceding})$.     
    \item{$\RSSQCY(T[i..j], b)$:} 
    returns the lexicographically largest string among strings in set $\mathcal{F}_{\SA}$ such that 
    each string is the RSS-suffix of a periodic interval attractor satisfying the following four conditions: 
    (1) it is contained in set $\Psi_{\CCP}(T[i..j])$; 
    (2) it is contained in set $\Psi_{\lcp}(\hat{K})$; 
    (3) it is contained in set $\Psi_{\succeeding}$; 
    (4) it has string $C_{Q}$ as its associated string. 
    Formally, this subquery returns the lexicographically largest string in set 
    $\mathcal{F}_{\SA} \cap \mathcal{F}_{\suffix}(\Psi_{\CCP}(T[i..j]) \cap \Psi_{\run} \cap \Psi_{\centerset}(C_{Q}) \cap \Psi_{\lcp}(\hat{K}) \cap \Psi_{\succeeding})$.     
    \item{$\RSSQDX(T[i..j], b, K)$:} 
    returns the lexicographically largest string among strings in set $\mathcal{F}_{\SA}$ such that 
    each string is the RSS-suffix of a periodic interval attractor satisfying the following four conditions: 
    (1) it is contained in set $\Psi_{\CCP}(T[i..j])$; 
    (2) it is contained in set $\Psi_{\lcp}(K)$ for a given integer $K \in [\max\{ |[\gamma_{Q}, j]|, 2 + \sum_{w = 1}^{h_{Q}+3} \lfloor \mu(w) \rfloor \}, n]$; 
    (3) it is contained in set $\Psi_{\preceding}$; 
    (4) it has string $C_{Q}$ as its associated string. 
    Formally, this subquery returns the lexicographically largest string in set 
    $\mathcal{F}_{\SA} \cap \mathcal{F}_{\suffix}(\Psi_{\CCP}(T[i..j]) \cap \Psi_{\run} \cap \Psi_{\centerset}(C_{Q}) \cap \Psi_{\lcp}(K) \cap \Psi_{\preceding})$. 
    \item{$\RSSQDY(T[i..j], b, K)$:} 
    returns the lexicographically largest string among strings in set $\mathcal{F}_{\SA}$ such that 
    each string is the RSS-suffix of a periodic interval attractor satisfying the following four conditions: 
    (1) it is contained in set $\Psi_{\CCP}(T[i..j])$; 
    (2) it is contained in set $\Psi_{\lcp}(K)$ for a given integer $K \in [\max\{ |[\gamma_{Q}, j]|, 2 + \sum_{w = 1}^{h_{Q}+3} \lfloor \mu(w) \rfloor \}, n]$; 
    (3) it is contained in set $\Psi_{\succeeding}$; 
    (4) it has string $C_{Q}$ as its associated string. 
    Formally, this subquery returns the lexicographically largest string in set 
    $\mathcal{F}_{\SA} \cap \mathcal{F}_{\suffix}(\Psi_{\CCP}(T[i..j]) \cap \Psi_{\run} \cap \Psi_{\centerset}(C_{Q}) \cap \Psi_{\lcp}(K) \cap \Psi_{\succeeding})$.
    \item{$\RSSQEX(T[i..j], b)$:} 
    returns the lexicographically largest string among strings in set $\mathcal{F}_{\SA}$ such that 
    each string is the RSS-suffix of a periodic interval attractor satisfying the following three conditions: 
    (1) it is contained in set $\Psi_{\CCP}(T[i..j])$; 
    (2) it is contained in set $\Psi_{\preceding}$; 
    (3) it has string $C_{Q}$ as its associated string. 
    Formally, this subquery returns the lexicographically largest string in set 
    $\mathcal{F}_{\SA} \cap \mathcal{F}_{\suffix}(\Psi_{\CCP}(T[i..j]) \cap \Psi_{\run} \cap \Psi_{\centerset}(C_{Q}) \cap \Psi_{\preceding})$.
    \item{$\RSSQEY(T[i..j], b)$:} 
    returns the lexicographically largest string among strings in set $\mathcal{F}_{\SA}$ such that 
    each string is the RSS-suffix of a periodic interval attractor satisfying the following three conditions: 
    (1) it is contained in set $\Psi_{\CCP}(T[i..j])$; 
    (2) it is contained in set $\Psi_{\succeeding}$; 
    (3) it has string $C_{Q}$ as its associated string. 
    Formally, this subquery returns the lexicographically largest string in set 
    $\mathcal{F}_{\SA} \cap \mathcal{F}_{\suffix}(\Psi_{\CCP}(T[i..j]) \cap \Psi_{\run} \cap \Psi_{\centerset}(C_{Q}) \cap \Psi_{\succeeding})$.
\end{description}

We show that RSS query can be computed by the above seven subqueries. 
We already showed that 
RSC query can be computed as the lexicographically largest string in set $\mathcal{F}_{\SA}$ (Lemma~\ref{lem:F_suffix_basic_property}~\ref{enum:F_suffix_basic_property:2}). 
For quickly finding the lexicographically largest string in set $\mathcal{F}_{\SA}$, 
we consider the following four conditions of RSS query: 

\begin{description}
 \item [Condition (A)] $\lcs(T[i..\gamma_{Q}-1], C_{Q}^{n+1}) = T[i..\gamma_{Q}-1]$ and 
 $1 + \sum_{w = 1}^{h_{Q}+3}$ $\lfloor \mu(w) \rfloor < \hat{K} < |[\gamma_{Q}, j]|$;
 \item [Condition (B)] 
 $\lcs(T[i..\gamma_{Q}-1], C_{Q}^{n+1}) = T[i..\gamma_{Q}-1]$ and $\hat{K} = |[\gamma_{Q}, j]|$;
 \item [Condition (C)] $\lcs(T[i..\gamma_{Q}-1], C_{Q}^{n+1}) \neq T[i..\gamma_{Q}-1]$;
 \item [Condition (D)] $\hat{K} < |[\gamma_{Q}, j]|$ and $\hat{K} \leq 1 + \sum_{w = 1}^{h_{Q}+3}$ $\lfloor \mu(w) \rfloor$.
\end{description}
RSS query satisfies at least one of the above four conditions 
because $\hat{K} \leq |[\gamma_{Q}, j]|$ follows from the definition of the length $\hat{K}$. 
The following three lemmas show that 
the lexicographically largest string in set $\mathcal{F}_{\SA}$ is equal to 
the lexicographically largest string in the union of at most three sets of strings. 

\begin{lemma}\label{lem:RSS_query_condition_A}
Assume that either $\mathcal{C}_{\run} = \emptyset$ or $C_{Q} = C_{\max}$ holds. 
If RSS query $\RSSQ(T[i..j], b)$ satisfies condition (A), 
then the lexicographically largest string in set $\mathcal{F}_{\SA}$ is equal to 
the lexicographically largest string in the union of the following three sets: 
\begin{itemize}
    \item $\mathcal{F}_{\SA} \cap \mathcal{F}_{\suffix}(\Psi_{\CCP}(T[i..j]) \setminus \Psi_{\run})$;
    \item $\mathcal{F}_{\SA} \cap \mathcal{F}_{\suffix}(\Psi_{\CCP}(T[i..j]) \cap \Psi_{\run} \cap \Psi_{\centerset}(C_{Q}) \cap \Psi_{\lcp}(\hat{K}) \cap \Psi_{\preceding})$;
    \item $\mathcal{F}_{\SA} \cap \mathcal{F}_{\suffix}(\Psi_{\CCP}(T[i..j]) \cap \Psi_{\run} \cap \Psi_{\centerset}(C_{Q}) \cap \Psi_{\lcp}(\hat{K}) \cap \Psi_{\succeeding})$.
\end{itemize}
\end{lemma}
\begin{proof}
    See Section~\ref{subsubsec:RSS_query_condition_A_proof}.
\end{proof}

\begin{lemma}\label{lem:RSS_query_condition_B}
Assume that either $\mathcal{C}_{\run} = \emptyset$ or $C_{Q} = C_{\max}$ holds. 
If RSS query $\RSSQ(T[i..j], b)$ satisfies condition (B), 
then the lexicographically largest string in set $\mathcal{F}_{\SA}$ is equal to 
the lexicographically largest string in the union of the following three sets: 
\begin{itemize}
    \item $\mathcal{F}_{\SA} \cap \mathcal{F}_{\suffix}(\Psi_{\CCP}(T[i..j]) \setminus \Psi_{\run})$;
    \item $\mathcal{F}_{\SA} \cap \mathcal{F}_{\suffix}(\Psi_{\CCP}(T[i..j]) \cap \Psi_{\run} \cap \Psi_{\centerset}(C_{Q}) \cap \Psi_{\preceding})$;
    \item $\mathcal{F}_{\SA} \cap \mathcal{F}_{\suffix}(\Psi_{\CCP}(T[i..j]) \cap \Psi_{\run} \cap \Psi_{\centerset}(C_{Q}) \cap \Psi_{\succeeding})$.
\end{itemize}
\end{lemma}
\begin{proof}
    See Section~\ref{subsubsec:RSS_query_condition_B_proof}.
\end{proof}

\begin{lemma}\label{lem:RSS_query_condition_CD}
Assume that either $\mathcal{C}_{\run} = \emptyset$ or $C_{Q} = C_{\max}$ holds. 
If RSS query $\RSSQ(T[i..j], b)$ satisfies at least one of condition (C) and condition (D), 
then 
the lexicographically largest string in set $\mathcal{F}_{\SA}$ 
is equal to the lexicographically largest string in set 
$\mathcal{F}_{\SA} \cap \mathcal{F}_{\suffix}(\Psi_{\CCP}(T[i..j]) \setminus \Psi_{\run})$. 
\end{lemma}
\begin{proof}
    See Section~\ref{subsubsec:RSS_query_condition_CD_proof}.
\end{proof}

Combining Lemma~\ref{lem:F_suffix_basic_property}~\ref{enum:F_suffix_basic_property:2}, 
Lemma~\ref{lem:RSS_query_condition_A}, 
Lemma~\ref{lem:RSS_query_condition_B}, 
Lemma~\ref{lem:RSS_query_condition_CD}, and eight subqueries, 
we obtain the following corollary. 

\begin{corollary}\label{cor:RB_rss_subqueries}
Assume that either $\mathcal{C}_{\run} = \emptyset$ or $C_{Q} = C_{\max}$ holds. 
Consider the four conditions (A), (B), (C), and (D) of RSS query $\RSSQ(T[i..j], b)$. 
The following three statements hold: 
\begin{enumerate}[label=\textbf{(\roman*)}]
    \item $\RSSQ(T[i..j], b)$ can be computed as the lexicographically largest string among the strings 
    obtained from three subqueries $\RSSQA(T[i..j], b)$, $\RSSQCX(T[i..j], b)$, and $\RSSQCY(T[i..j], b)$
    if condition (A) holds;
    \item $\RSSQ(T[i..j], b)$ can be computed as the lexicographically largest string among the strings 
    obtained from three subqueries $\RSSQA(T[i..j], b)$, $\RSSQEX(T[i..j], b)$, and $\RSSQEY(T[i..j], b)$
    if condition (B) holds;
    \item $\RSSQ(T[i..j], b)$ can be computed as the string obtained from $\RSSQA(T[i..j], b)$ if at least one of two conditions (C) and (D) holds.
\end{enumerate}
\end{corollary}

Corollary~\ref{cor:RB_rss_subqueries} shows that 
RSS query can be computed by five subqueries 
$\RSSQA(T[i..j], b)$, $\RSSQCX(T[i..j], b)$, $\RSSQCY(T[i..j], b)$, $\RSSQEX(T[i..j], b)$, and $\RSSQEY(T[i..j], b)$. 
The other subqueries $\RSSQB(T[i..j], b)$, $\RSSQDX(T[i..j], b, K)$, and $\RSSQDY(T[i..j], b, K)$ are not used to directly answer RSS query. 
Subquery $\RSSQB(T[i..j], b)$ is used to verify whether the assumption of Corollary~\ref{cor:RB_rss_subqueries} is satisfied or not. 
Two subqueries $\RSSQDX(T[i..j], b, K)$ and $\RSSQDY(T[i..j], b, K)$ are used to 
answer $\RSSQEX(T[i..j], b)$ and $\RSSQEY(T[i..j], b)$, respectively. 

In the next subsections, we present algorithms for these eight subqueries using 
the data structure for RSC query.

%\subsection{Key Idea behind Answering RSS Query}\label{subsec:rss_idea}

\subsubsection{Proof of Lemma~\ref{lem:RSS_query_condition_A}}\label{subsubsec:RSS_query_condition_A_proof}
The following two proposition are used to prove Lemma~\ref{subsubsec:RSS_query_condition_A_proof}.

\begin{proposition}\label{prop:CCP_division}
The following two equations hold:
\begin{equation}\label{eq:CCP_division:1}
    \begin{split}
    \Psi_{\CCP}(T[i..j]) &= \Psi_{\CCP}(T[i..j]) \setminus \Psi_{\run} \\ 
    &\cup (\bigcup_{C \in \Sigma^{+} \setminus \{ C_{Q} \}} \Psi_{\CCP}(T[i..j]) \cap \Psi_{\run} \cap \Psi_{\centerset}(C)) \\ 
    &\cup (\Psi_{\CCP}(T[i..j]) \cap \Psi_{\run} \cap \Psi_{\centerset}(C_{Q})). 
    \end{split}
\end{equation}
\begin{equation}\label{eq:CCP_division:2}
    \begin{split}
    \Psi_{\CCP}(T[i..j]) &= (\Psi_{\CCP}(T[i..j]) \setminus \Psi_{\run}) \\ 
    &\cup (\bigcup_{C \in \Sigma^{+} \setminus \{ C_{Q} \}} \Psi_{\CCP}(T[i..j]) \cap \Psi_{\run} \cap \Psi_{\centerset}(C)) \\ 
    &\cup ((\bigcup_{\lambda = 1}^{n} \Psi_{\CCP}(T[i..j]) \cap \Psi_{\run} \cap \Psi_{\centerset}(C_{Q}) \cap \Psi_{\lcp}(\lambda)) \setminus \Psi_{\lcp}(\hat{K})) \\ 
    &\cup (\Psi_{\CCP}(T[i..j]) \cap \Psi_{\run} \cap \Psi_{\centerset}(C_{Q}) \cap \Psi_{\lcp}(\hat{K}) \cap \Psi_{\preceding}) \\ 
    &\cup (\Psi_{\CCP}(T[i..j]) \cap \Psi_{\run} \cap \Psi_{\centerset}(C_{Q}) \cap \Psi_{\lcp}(\hat{K}) \cap \Psi_{\succeeding}). 
    \end{split}
\end{equation}
\end{proposition}
\begin{proof}
The proof of Proposition~\ref{prop:CCP_division} is as follows.

\textbf{Proof of Equation~\ref{eq:CCP_division:1}.}
Set $\Psi_{\CCP}(T[i..j])$ can be divided into two sets $\Psi_{\CCP}(T[i..j]) \setminus \Psi_{\run}$ and $\Psi_{\CCP}(T[i..j]) \cap \Psi_{\run}$. 
The latter set $\Psi_{\CCP}(T[i..j]) \cap \Psi_{\run}$ is equal to 
set $\bigcup_{C \in \Sigma^{+}} \Psi_{\CCP}(T[i..j]) \cap \Psi_{\run} \cap \Psi_{\centerset}(C)$ 
because each interval attractor is contained in set $\Psi_{\centerset}(C)$ for 
the associated string $C$ of the interval attractor.  
The set $\bigcup_{C \in \Sigma^{+}} \Psi_{\CCP}(T[i..j]) \cap \Psi_{\run} \cap \Psi_{\centerset}(C)$ 
is equal to the union of two sets (i) $\bigcup_{C \in \Sigma^{+} \setminus \{ C_{Q} \}} \Psi_{\CCP}(T[i..j]) \cap \Psi_{\run} \cap \Psi_{\centerset}(C)$ 
and (ii) $\Psi_{\CCP}(T[i..j]) \cap \Psi_{\run} \cap \Psi_{\centerset}(C_{Q})$. 
Therefore, Equation~\ref{eq:CCP_division:1} follows from the following equation: 
\begin{equation}\label{eq:CCP_division:3}
    \begin{split}
    \Psi_{\CCP}(T[i..j]) &= (\Psi_{\CCP}(T[i..j]) \setminus \Psi_{\run}) \\ 
    &\cup (\Psi_{\CCP}(T[i..j]) \cap \Psi_{\run}) \\ 
    &= (\Psi_{\CCP}(T[i..j]) \setminus \Psi_{\run}) \\
    &\cup (\bigcup_{C \in \Sigma^{+}} \Psi_{\CCP}(T[i..j]) \cap \Psi_{\run} \cap \Psi_{\centerset}(C)) \\
    &= (\Psi_{\CCP}(T[i..j]) \setminus \Psi_{\run}) \\
    &\cup (\bigcup_{C \in \Sigma^{+} \setminus \{ C_{Q} \}} \Psi_{\CCP}(T[i..j]) \cap \Psi_{\run} \cap \Psi_{\centerset}(C)) \\
    &\cup (\Psi_{\CCP}(T[i..j]) \cap \Psi_{\run} \cap \Psi_{\centerset}(C_{Q})). 
    \end{split}
\end{equation}

\textbf{Proof of Equation~\ref{eq:CCP_division:2}.}
Lemma~\ref{lem:psi_LMPS_property}~\ref{enum:psi_LMPS_property:lcp:1} indicates that 
set $\Psi_{\CCP}(T[i..j]) \cap \Psi_{\run} \cap \Psi_{\centerset}(C_{Q})$ is equal to set 
$\bigcup_{\lambda = 1}^{n} \Psi_{\CCP}(T[i..j]) \cap \Psi_{\run} \cap \Psi_{\centerset}(C_{Q}) \cap \Psi_{\lcp}(\lambda)$.  
The set $\bigcup_{\lambda = 1}^{n} \Psi_{\CCP}(T[i..j]) \cap \Psi_{\run} \cap \Psi_{\centerset}(C_{Q}) \cap \Psi_{\lcp}(\lambda)$ 
can be divided into two sets 
$(\bigcup_{\lambda = 1}^{n} \Psi_{\CCP}(T[i..j]) \cap \Psi_{\run} \cap \Psi_{\centerset}(C_{Q}) \cap \Psi_{\lcp}(\lambda)) \setminus \Psi_{\lcp}(\hat{K})$ 
and $\Psi_{\CCP}(T[i..j]) \cap \Psi_{\run} \cap \Psi_{\centerset}(C_{Q}) \cap \Psi_{\lcp}(\hat{K})$. 
The latter set $\Psi_{\CCP}(T[i..j]) \cap \Psi_{\run} \cap \Psi_{\centerset}(C_{Q}) \cap \Psi_{\lcp}(\hat{K})$ can be divided into 
two sets $\Psi_{\CCP}(T[i..j]) \cap \Psi_{\run} \cap \Psi_{\centerset}(C_{Q}) \cap \Psi_{\lcp}(\hat{K}) \cap \Psi_{\preceding}$ 
and $\Psi_{\CCP}(T[i..j]) \cap \Psi_{\run} \cap \Psi_{\centerset}(C_{Q}) \cap \Psi_{\lcp}(\hat{K}) \cap \Psi_{\succeeding}$. 
This is because each interval attractor of set $\Psi_{\RR}$ is contained in either the set $\Psi_{\preceding}$ or $\Psi_{\succeeding}$. 
Therefore, the following equation holds: 
\begin{equation}\label{eq:CCP_division:4}
    \begin{split}
    \Psi_{\CCP}(T[i..j]) \cap \Psi_{\run} & \cap \Psi_{\centerset}(C_{Q}) \\ 
    &= \bigcup_{\lambda = 1}^{n} \Psi_{\CCP}(T[i..j]) \cap \Psi_{\run} \cap \Psi_{\centerset}(C_{Q}) \cap \Psi_{\lcp}(\lambda) \\
    &= ((\bigcup_{\lambda = 1}^{n} \Psi_{\CCP}(T[i..j]) \cap \Psi_{\run} \cap \Psi_{\centerset}(C_{Q}) \cap \Psi_{\lcp}(\lambda)) \setminus \Psi_{\lcp}(\hat{K})) \\
    &\cup (\Psi_{\CCP}(T[i..j]) \cap \Psi_{\run} \cap \Psi_{\centerset}(C_{Q}) \cap \Psi_{\lcp}(\hat{K})) \\
    &= ((\bigcup_{\lambda = 1}^{n} \Psi_{\CCP}(T[i..j]) \cap \Psi_{\run} \cap \Psi_{\centerset}(C_{Q}) \cap \Psi_{\lcp}(\lambda)) \setminus \Psi_{\lcp}(\hat{K})) \\
    &\cup (\Psi_{\CCP}(T[i..j]) \cap \Psi_{\run} \cap \Psi_{\centerset}(C_{Q}) \cap \Psi_{\lcp}(\hat{K}) \cap \Psi_{\preceding}) \\
    &\cup (\Psi_{\CCP}(T[i..j]) \cap \Psi_{\run} \cap \Psi_{\centerset}(C_{Q}) \cap \Psi_{\lcp}(\hat{K}) \cap \Psi_{\succeeding}).
    \end{split}
\end{equation}
Finally, Equation~\ref{eq:CCP_division:2} follows from Equation~\ref{eq:CCP_division:1} and Equation~\ref{eq:CCP_division:4}. 
\end{proof}

\begin{proposition}\label{prop:empty_FSA_condition_A}
If RSS query $\RSSQ(T[i..j], b)$ satisfies condition (A) of RSS query, 
then the following two equations hold:
\begin{equation}\label{eq:empty_FSA_condition_A:1}
    \bigcup_{C \in \Sigma^{+} \setminus \{ C_{Q} \}} \Psi_{\CCP}(T[i..j]) \cap \Psi_{\run} \cap \Psi_{\centerset}(C) = \emptyset;
\end{equation}
\begin{equation}\label{eq:empty_FSA_condition_A:2}
    (\bigcup_{\lambda = 1}^{n} \Psi_{\CCP}(T[i..j]) \cap \Psi_{\run} \cap \Psi_{\centerset}(C_{Q}) \cap \Psi_{\lcp}(\lambda)) \setminus \Psi_{\lcp}(\hat{K}) = \emptyset.
\end{equation}
\end{proposition}
\begin{proof}
The proof of Proposition~\ref{prop:empty_FSA_condition_A} is as follows. 

\textbf{Proof of Equation~\ref{eq:empty_FSA_condition_A:1}.}
We prove Equation~\ref{eq:empty_FSA_condition_A:1} by contradiction. 
We assume that Equation~\ref{eq:empty_FSA_condition_A:1} does not hold.  
Then, there exists a string $C \in \Sigma^{+} \setminus \{ C_{Q} \}$ satisfying $\Psi_{\CCP}(T[i..j]) \cap \Psi_{\run} \cap \Psi_{\centerset}(C) \neq \emptyset$. 
We can apply Lemma~\ref{lem:CCP_property}~\ref{enum:CCP_property:5} to the string $C$ 
because (i) $\Psi_{\CCP}(T[i..j]) \cap \Psi_{\run} \cap \Psi_{\centerset}(C) \neq \emptyset$, 
and (ii) $|[\gamma_{Q}, j]| > 1 + \sum_{w = 1}^{h_{Q}+3} \lfloor \mu(w) \rfloor$ follows from $\hat{K} > 1 + \sum_{w = 1}^{h_{Q}+3} \lfloor \mu(w) \rfloor$. 
The lemma ensures that $C = C_{Q}$ holds. 
On the other hand, $C \neq C_{Q}$ follows from $C \in \Sigma^{+} \setminus \{ C_{Q} \}$.
The two facts $C = C_{Q}$ and $C \neq C_{Q}$ yield a contradiction. 
Therefore, Equation~\ref{eq:empty_FSA_condition_A:1} must hold. 

\textbf{Proof of Equation~\ref{eq:empty_FSA_condition_A:2}.}
We prove Equation~\ref{eq:empty_FSA_condition_A:2} by contradiction. 
We assume that Equation~\ref{eq:empty_FSA_condition_A:2} does not hold. 
Then, the set $(\bigcup_{\lambda = 1}^{n} \Psi_{\CCP}(T[i..j]) \cap \Psi_{\run} \cap \Psi_{\centerset}(C_{Q}) \cap \Psi_{\lcp}(\lambda)) \setminus \Psi_{\lcp}(\hat{K})$ contains an interval attractor $([p, q], [\ell, r])$. 
Let $K = |\lcp(T[\gamma..r], C_{Q}^{n+1})|$ for the attractor position $\gamma$ of the interval attractor $([p, q], [\ell, r])$. 
Then, $K \in [1, n] \setminus \{ \hat{K} \}$ holds 
because $([p, q], [\ell, r]) \in (\Psi_{\centerset}(C_{Q}) \cup (\bigcup_{\lambda = 1}^{n} \Psi_{\lcp}(\lambda))) \setminus \Psi_{\lcp}(\hat{K})$. 
Therefore, $K \neq \hat{K}$ follows from $K \in [1, n] \setminus \{ \hat{K} \}$. 

On the other hand, 
Lemma~\ref{lem:psi_run_basic_property}~\ref{enum:psi_run_basic_property:8} shows that 
$|\lcp(T[\gamma..r+1], C_{Q}^{n+1})| = K$ holds. 
Lemma \ref{lem:CCP_property}~\ref{enum:CCP_property:4} indicates that 
string $T[\gamma_{Q}..j]$ is a prefix of string $T[\gamma..r+1]$ 
(i.e., $\lcp(T[\gamma_{Q}..j], T[\gamma..r+1]) = T[\gamma_{Q}..j]$). 
$|\lcp(T[\gamma..r+1], C_{Q}^{n+1})| = \hat{K}$ holds 
because string $T[\gamma_{Q}..j]$ is a prefix of string $T[\gamma..r+1]$, 
$|\lcp(T[\gamma_{Q}..j], C_{Q}^{n+1})| = \hat{K}$ holds, 
and $\hat{K} < |[\gamma_{Q}, j]|$ holds. 
$K = \hat{K}$ follows from $|\lcp(T[\gamma..r+1], C_{Q}^{n+1})| = \hat{K}$ and $|\lcp(T[\gamma..r+1], C_{Q}^{n+1})| = K$. 
The two facts $K \neq \hat{K}$ and $K = \hat{K}$ yield a contradiction. 
Therefore, Equation~\ref{eq:empty_FSA_condition_A:2} must hold. 
\end{proof}

We prove Lemma~\ref{lem:RSS_query_condition_A}. 
\begin{proof}[Proof of Lemma~\ref{lem:RSS_query_condition_A}]
The following equation follows from 
Lemma~\ref{lem:F_suffix_basic_property}~\ref{enum:F_suffix_basic_property:5} and Equation~\ref{eq:CCP_division:2} of Proposition~\ref{prop:CCP_division}: 
\begin{equation}\label{eq:RSS_query_condition_A:1}
    \begin{split}
    \mathcal{F}_{\SA} & \cap \mathcal{F}_{\suffix}(\Psi_{\CCP}(T[i..j])) \\
    &= (\mathcal{F}_{\SA} \cap \mathcal{F}_{\suffix}(\Psi_{\CCP}(T[i..j]) \setminus \Psi_{\run})) \\
    &\cup (\mathcal{F}_{\SA} \cap \mathcal{F}_{\suffix}(\bigcup_{C \in \Sigma^{+} \setminus \{ C_{Q} \}} \Psi_{\CCP}(T[i..j]) \cap \Psi_{\run} \cap \Psi_{\centerset}(C))) \\
    &\cup (\mathcal{F}_{\SA} \cap \mathcal{F}_{\suffix}((\bigcup_{\lambda = 1}^{n} \Psi_{\CCP}(T[i..j]) \cap \Psi_{\run} \cap \Psi_{\centerset}(C_{Q}) \cap \Psi_{\lcp}(\lambda)) \setminus \Psi_{\lcp}(\hat{K}))) \\
    &\cup (\mathcal{F}_{\SA} \cap \mathcal{F}_{\suffix}(\Psi_{\CCP}(T[i..j]) \cap \Psi_{\run} \cap \Psi_{\centerset}(C_{Q}) \cap \Psi_{\lcp}(\hat{K}) \cap \Psi_{\preceding})) \\
    &\cup (\mathcal{F}_{\SA} \cap \mathcal{F}_{\suffix}(\Psi_{\CCP}(T[i..j]) \cap \Psi_{\run} \cap \Psi_{\centerset}(C_{Q}) \cap \Psi_{\lcp}(\hat{K}) \cap \Psi_{\succeeding})).
    \end{split}
\end{equation}

The following equation follows from Lemma~\ref{lem:F_suffix_basic_property}~\ref{enum:F_suffix_basic_property:4} and Equation~\ref{eq:empty_FSA_condition_A:1} of Proposition~\ref{prop:empty_FSA_condition_A}. 
\begin{equation}\label{eq:RSS_query_condition_A:2}
    \mathcal{F}_{\SA} \cap \mathcal{F}_{\suffix}(\bigcup_{C \in \Sigma^{+} \setminus \{ C_{Q} \}} \Psi_{\CCP}(T[i..j]) \cap \Psi_{\run} \cap \Psi_{\centerset}(C)) = \emptyset.
\end{equation}

Similarly, the following equation follows from Lemma~\ref{lem:F_suffix_basic_property}~\ref{enum:F_suffix_basic_property:4} and Equation~\ref{eq:empty_FSA_condition_A:2} of Proposition~\ref{prop:empty_FSA_condition_A}. 
\begin{equation}\label{eq:RSS_query_condition_A:3}
    \mathcal{F}_{\SA} \cap \mathcal{F}_{\suffix}((\bigcup_{\lambda = 1}^{n} \Psi_{\CCP}(T[i..j]) \cap \Psi_{\run} \cap \Psi_{\centerset}(C_{Q}) \cap \Psi_{\lcp}(\lambda)) \setminus \Psi_{\lcp}(\hat{K})) = \emptyset.
\end{equation}

Therefore, Lemma~\ref{lem:RSS_query_condition_A} follows from 
Lemma~\ref{lem:F_suffix_basic_property}~\ref{enum:F_suffix_basic_property:3}, 
Equation~\ref{eq:RSS_query_condition_A:1}, Equation~\ref{eq:RSS_query_condition_A:2}, and Equation~\ref{eq:RSS_query_condition_A:3}.    
\end{proof}

%%%%%%%%%%%%%%%%%%%%%%%%%%%%%%%%%%%%%%%%%%%%%%%%
\subsubsection{Proof of Lemma~\ref{lem:RSS_query_condition_B}}\label{subsubsec:RSS_query_condition_B_proof}
The following proposition is used to prove Lemma~\ref{lem:RSS_query_condition_B}. 

\begin{proposition}\label{prop:RSS_partiiton_C_lex_property}
Assume that either $\mathcal{C}_{\run} = \emptyset$ or $C_{Q} = C_{\max}$ holds. 
Consider the RSS query $\RSSQ(T[i..j], b)$ satisfying condition (B) of RSS query. 
If set $\mathcal{F}_{\SA} \cap \mathcal{F}_{\suffix}$ $(\bigcup_{C \in \Sigma^{+} \setminus \{ C_{Q} \}} \Psi_{\CCP}(T[i..j]) \cap \Psi_{\run} \cap \Psi_{\centerset}(C))$ is not empty, 
then the following the two statements hold: 
    \begin{enumerate}[label=\textbf{(\roman*)}]
    \item $\mathcal{F}_{\SA} \cap \mathcal{F}_{\suffix}(\Psi_{\CCP}(T[i..j]) \cap \Psi_{\run} \cap \Psi_{\centerset}(C_{Q})) \neq \emptyset$;
    \item 
    $F_{1} \prec F_{2}$ for any pair of two strings $F_{1}$ and $F_{2}$ in 
    the two sets $\mathcal{F}_{\SA} \cap \mathcal{F}_{\suffix}$ $(\bigcup_{C \in \Sigma^{+} \setminus \{ C_{Q} \}}$ $\Psi_{\CCP}(T[i..j]) \cap \Psi_{\run} \cap \Psi_{\centerset}(C))$ 
    and $\mathcal{F}_{\SA} \cap \mathcal{F}_{\suffix}(\Psi_{\CCP}(T[i..j]) \cap \Psi_{\run} \cap \Psi_{\centerset}(C_{Q}))$, 
    respectively. 
    \end{enumerate}
\end{proposition}
\begin{proof}
    Because of $\mathcal{F}_{\SA} \cap \mathcal{F}_{\suffix}$ $(\bigcup_{C \in \Sigma^{+} \setminus \{ C_{Q} \}} \Psi_{\CCP}(T[i..j]) \cap \Psi_{\run} \cap \Psi_{\centerset}(C)) \neq \emptyset$, 
    there exists a string $C^{\prime} \in \Sigma^{+} \setminus \{ C_{Q} \}$ satisfying 
    $\mathcal{F}_{\SA} \cap \mathcal{F}_{\suffix}$ $(\Psi_{\CCP}(T[i..j]) \cap \Psi_{\run} \cap \Psi_{\centerset}(C^{\prime})) \neq \emptyset$. 
    Here, $C^{\prime} \in \mathcal{C}_{\run}$ holds for set $\mathcal{C}_{\run}$ of strings. 
    In this case, 
    the string $C_{\max}$ is defined as the string in set $\mathcal{C}_{\run}$ satisfying 
    $C^{n+1}[1..2 + \sum_{w = 1}^{h_{Q}+3} \lfloor \mu(w) \rfloor] \prec C^{n+1}_{\max}[1..2 + \sum_{w = 1}^{h_{Q}+3} \lfloor \mu(w) \rfloor]$ 
    for each string $C \in \mathcal{C}_{\run} \setminus \{ C_{\max} \}$. 
    Here, $C_{Q} = C_{\max}$ holds because RSS query $\RSSQ(T[i..j], b)$ satisfies condition (B). 
            
    \textbf{Proof of Proposition~\ref{prop:RSS_partiiton_C_lex_property}(i).}
    $\mathcal{F}_{\SA} \cap \mathcal{F}_{\suffix}(\Psi_{\CCP}(T[i..j]) \cap \Psi_{\run} \cap \Psi_{\centerset}(C_{Q})) \neq \emptyset$ follows from the definition of the set $\mathcal{C}_{\run}$ 
    because $C_{Q} \in \mathcal{C}_{\run}$. 

    \textbf{Proof of Proposition~\ref{prop:RSS_partiiton_C_lex_property}(ii).}
    The set $\Sigma^{+} \setminus \{ C_{Q} \}$ contains a string $C_{1}$ satisfying 
    $F_{1} \in \mathcal{F}_{\SA} \cap \mathcal{F}_{\suffix}(\Psi_{\CCP}(T[i..j]) \cap \Psi_{\run} \cap \Psi_{\centerset}(C_{1}))$. 
    Similar to the string $C^{\prime}$, 
    $C_{1} \in \mathcal{C}_{\run}$ holds. 
    In this case, $C_{\max} = C_{Q}$ holds by the assumption. 
    $C_{1} \neq C_{\max}$ follows from $C_{1} \neq C_{Q}$ and $C_{\max} = C_{Q}$. 
    Therefore, 
    $C_{1}^{n+1}[1..2 + \sum_{w = 1}^{h_{Q}+3} \lfloor \mu(w) \rfloor] \prec C_{\max}^{n+1}[1..2 + \sum_{w = 1}^{h_{Q}+3} \lfloor \mu(w) \rfloor]$ follows from the definition of the string $C_{\max}$. 

    We prove $F_{1} \prec F_{2}$.     
    $C_{1}^{n+1}[1..2 + \sum_{w = 1}^{h_{Q}+3} \lfloor \mu(w) \rfloor] \prec C_{Q}^{n+1}[1..2 + \sum_{w = 1}^{h_{Q}+3} \lfloor \mu(w) \rfloor]$ 
    follows from $C_{1}^{n+1}[1..2 + \sum_{w = 1}^{h_{Q}+3} \lfloor \mu(w) \rfloor] \prec C_{\max}^{n+1}[1..2 + \sum_{w = 1}^{h_{Q}+3} \lfloor \mu(w) \rfloor]$ and $C_{Q} = C_{\max}$. 
    Therefore, $F_{1} \prec F_{2}$ follows from Lemma~\ref{lem:C_run_property}~\ref{enum:C_run_property:4} and 
    $C_{1}^{n+1}[1..2 + \sum_{w = 1}^{h_{Q}+3} \lfloor \mu(w) \rfloor] \prec C_{Q}^{n+1}[1..2 + \sum_{w = 1}^{h_{Q}+3} \lfloor \mu(w) \rfloor]$. 
\end{proof}

We prove Lemma~\ref{lem:RSS_query_condition_B}. 

\begin{proof}[Proof of Lemma~\ref{lem:RSS_query_condition_B}]
For simplicity, 
let $\mathcal{F}_{1} = \mathcal{F}_{\SA} \cap \mathcal{F}_{\suffix}(\Psi_{\CCP}(T[i..j]) \setminus \Psi_{\run})$, 
$\mathcal{F}_{2} = \mathcal{F}_{\SA} \cap \mathcal{F}_{\suffix}(\bigcup_{C \in \Sigma^{+} \setminus \{ C_{Q} \}} \Psi_{\CCP}(T[i..j]) \cap \Psi_{\run} \cap \Psi_{\centerset}(C))$, 
$\mathcal{F}_{3} = \mathcal{F}_{\SA} \cap \mathcal{F}_{\suffix}(\Psi_{\CCP}(T[i..j]) \cap \Psi_{\run} \cap \Psi_{\centerset}(C_{Q}))$, 
$\mathcal{F}_{4} = \mathcal{F}_{\SA} \cap \mathcal{F}_{\suffix}(\Psi_{\CCP}(T[i..j]) \cap \Psi_{\run} \cap \Psi_{\centerset}(C_{Q}) \cap \Psi_{\preceding})$, 
and 
$\mathcal{F}_{5} = \mathcal{F}_{\SA} \cap \mathcal{F}_{\suffix}(\Psi_{\CCP}(T[i..j]) \cap \Psi_{\run} \cap \Psi_{\centerset}(C_{Q}) \cap \Psi_{\succeeding})$. 
Here, $\Psi_{\CCP}(T[i..j]) \cap \Psi_{\run} \cap \Psi_{\centerset}(C_{Q}) = \Psi_{\CCP}(T[i..j]) \cap \Psi_{\run} \cap \Psi_{\centerset}(C_{Q}) \cap (\Psi_{\preceding} \cup \Psi_{\succeeding})$ holds because $\Psi_{\RR} = \Psi_{\preceding} \cup \Psi_{\succeeding}$ follows from the definitions of the two subsets $\Psi_{\preceding}$ and $\Psi_{\succeeding}$. 

We show that the lexicographically largest string in the set $\mathcal{F}_{3}$ is equal to the lexicographically largest string in the union of the two sets $\mathcal{F}_{4}$ and $\mathcal{F}_{5}$. 
Because of $\Psi_{\CCP}(T[i..j]) \cap \Psi_{\run} \cap \Psi_{\centerset}(C_{Q}) = \Psi_{\CCP}(T[i..j]) \cap \Psi_{\run} \cap \Psi_{\centerset}(C_{Q}) \cap (\Psi_{\preceding} \cup \Psi_{\succeeding})$, 
Lemma~\ref{lem:F_suffix_basic_property}~\ref{enum:F_suffix_basic_property:5} shows that 
$\mathcal{F}_{3} = \mathcal{F}_{4} \cup \mathcal{F}_{5}$ holds. 
Therefore, the lexicographically largest string in the set $\mathcal{F}_{3}$ is equal to the lexicographically largest string in the union of the two sets $\mathcal{F}_{4}$ and $\mathcal{F}_{5}$.  

We show that 
the lexicographically largest string in set $\mathcal{F}_{\SA} \cap \mathcal{F}_{\suffix}(\Psi_{\CCP}(T[i..j]))$ 
is equal to the lexicographically largest string in the union of the three sets $\mathcal{F}_{1}$, $\mathcal{F}_{4}$, and $\mathcal{F}_{5}$. 
The following equation follows from 
Lemma~\ref{lem:F_suffix_basic_property}~\ref{enum:F_suffix_basic_property:5} and Equation~\ref{eq:CCP_division:1} of Proposition~\ref{prop:CCP_division}: 
\begin{equation}\label{eq:RSS_query_condition_B:1}
    \mathcal{F}_{\SA} \cap \mathcal{F}_{\suffix}(\Psi_{\CCP}(T[i..j])) = \mathcal{F}_{1} \cup \mathcal{F}_{2} \cup \mathcal{F}_{3}.
\end{equation}
By combining Equation~\ref{eq:RSS_query_condition_B:1} and Proposition~\ref{prop:RSS_partiiton_C_lex_property}, 
we obtain the fact that 
the lexicographically largest string in set $\mathcal{F}_{\SA} \cap \mathcal{F}_{\suffix}(\Psi_{\CCP}(T[i..j]))$ 
is equal to the lexicographically largest string in the union of the two sets $\mathcal{F}_{1}$ and $\mathcal{F}_{3}$. 
We already showed that 
the lexicographically largest string in the set $\mathcal{F}_{3}$ is equal to the lexicographically largest string in the union of the two sets $\mathcal{F}_{4}$ and $\mathcal{F}_{5}$. 
Therefore, the lexicographically largest string in set $\mathcal{F}_{\SA} \cap \mathcal{F}_{\suffix}(\Psi_{\CCP}(T[i..j]))$ 
is equal to the lexicographically largest string in the union of the three sets $\mathcal{F}_{1}$, $\mathcal{F}_{4}$, and $\mathcal{F}_{5}$. 

We prove Lemma~\ref{lem:RSS_query_condition_B}. 
Lemma~\ref{lem:F_suffix_basic_property}~\ref{enum:F_suffix_basic_property:3} 
shows that $\mathcal{F}_{\SA} = \mathcal{F}_{\SA} \cap \mathcal{F}_{\suffix}(\Psi_{\CCP}(T[i..j]))$. 
We already showed that the lexicographically largest string in set $\mathcal{F}_{\SA} \cap \mathcal{F}_{\suffix}(\Psi_{\CCP}(T[i..j]))$ 
is equal to the lexicographically largest string in the union of the three sets $\mathcal{F}_{1}$, $\mathcal{F}_{4}$, and $\mathcal{F}_{5}$. 
Therefore, Lemma~\ref{lem:RSS_query_condition_B} holds. 
\end{proof}

%%%%%%%%%%%%%%%%%%
\subsubsection{Proof of Lemma~\ref{lem:RSS_query_condition_CD}}\label{subsubsec:RSS_query_condition_CD_proof}
The following proposition is used to prove Lemma~\ref{lem:RSS_query_condition_CD}. 

\begin{proposition}\label{prop:RSS_query_condition_CD_property}
    Assume that either $\mathcal{C}_{\run} = \emptyset$ or $C_{Q} = C_{\max}$ holds.
    Consider the RSS query $\RSSQ(T[i..j], b)$ satisfying either condition (C) or condition (D) or both. 
    Then, $\mathcal{C}_{\run} = \emptyset$ holds for set $\mathcal{C}_{\run}$ of strings.
\end{proposition}
\begin{proof}
We assume that $\mathcal{C}_{\run} \neq \emptyset$ holds. 
Then, the string $C_{\max}$ is defined as the string in set $\mathcal{C}_{\run}$ satisfying 
$C^{n+1}[1..2 + \sum_{w = 1}^{h_{Q}+3} \lfloor \mu(w) \rfloor] \prec C^{n+1}_{\max}[1..2 + \sum_{w = 1}^{h_{Q}+3} \lfloor \mu(w) \rfloor]$ 
for each string $C \in \mathcal{C}_{\run} \setminus \{ C_{\max} \}$. 
$C_{Q} \in \mathcal{C}_{\run}$ follows from $C_{Q} = C_{\max}$ and $C_{\max} \in \mathcal{C}_{\run}$. 
Because of $C_{Q} \in \mathcal{C}_{\run}$, 
$\mathcal{F}_{\SA} \cap \mathcal{F}_{\suffix}(\Psi_{\CCP}(T[i..j]) \cap \Psi_{\run} \cap \Psi_{\centerset}(C_{Q})) \neq \emptyset$ 
follows from the definition of the set $\mathcal{C}_{\run}$. 

Because of $\mathcal{F}_{\suffix}(\Psi_{\CCP}(T[i..j]) \cap \Psi_{\run} \cap \Psi_{\centerset}(C_{Q})) \neq \emptyset$, 
the set $\Psi_{\CCP}(T[i..j]) \cap \Psi_{\run} \cap \Psi_{\centerset}(C_{Q})$ contains an interval attractor $([p, q], [\ell, r])$. 
Here, Lemma~\ref{lem:CCP_property}~\ref{enum:CCP_property:1} shows that 
the level of the interval attractor $([p, q], [\ell, r])$ is $h_{Q}$. 
Let $\gamma$ be the attractor position of the interval attractor $([p, q], [\ell, r])$. 
Under the assumption that $\mathcal{C}_{\run} \neq \emptyset$ holds, 
we prove the following three statements: 
\begin{enumerate}[label=\textbf{(\roman*)}]
    \item $|\lcp(T[\gamma..r], C_{Q}^{n+1})| > 1 + \sum_{w = 1}^{h_{Q}+3} \lfloor \mu(w) \rfloor$;
    \item $\hat{K} = \min \{ |[\gamma_{Q}, j]|, |\lcp(T[\gamma..r], C_{Q}^{n+1})| \}$;
    \item $\lcs(T[i..\gamma_{Q}-1], C_{Q}^{n+1}) = T[i..\gamma_{Q}-1]$.
\end{enumerate}

\textbf{Proof of statement (i).}
Because of $([p, q], [\ell, r]) \in \Psi_{h_{Q}} \cap \Psi_{\run} \cap \Psi_{\centerset}(C_{Q})$, 
statement (i) follows from the definition of the subset $\Psi_{\run}$. 

\textbf{Proof of statement (ii).}
Because of $([p, q], [\ell, r]) \in \Psi_{\CCP}(T[i..j])$, 
Lemma~\ref{lem:CCP_property}~\ref{enum:CCP_property:6} shows that 
$T[i..\gamma_{Q}-1] = T[\gamma - |[i, \gamma_{Q}-1]|..\gamma-1]$, 
$T[\gamma_{Q}..j] = T[\gamma..\gamma + |[\gamma_{Q}, j]| - 1]$, 
and $I_{\capture}(\gamma - |[i, \gamma_{Q}-1]|, \gamma + |[\gamma_{Q}, j]| - 1) = ([p, q], [\ell, r])$. 
Since $I_{\capture}(\gamma - |[i, \gamma_{Q}-1]|, \gamma + |[\gamma_{Q}, j]| - 1) = ([p, q], [\ell, r])$, 
$\gamma - |[i, \gamma_{Q}-1]| \in [p, q]$ and $\gamma + |[\gamma_{Q}, j]| - 1 \in [\ell, r]$ follow from the definition of interval attractor. 

Let $K = |\lcp(T[\gamma..\gamma + |[\gamma_{Q}, j]| - 1], C_{Q}^{n+1})|$. 
Then, $K = \min \{ |[\gamma, \gamma + |[\gamma_{Q}, j]| - 1]|, |\lcp(T[\gamma..r]$, $C_{Q}^{n+1})| \}$ 
because $\gamma + |[\gamma_{Q}, j]| - 1 \leq r$. 
$K = \hat{K}$ follows from 
$K = |\lcp(T[\gamma..\gamma + |[\gamma_{Q}, j]| - 1], C_{Q}^{n+1})|$, 
$\hat{K} = |\lcp(T[\gamma_{Q}..j], C_{Q}^{n+1})|$, 
and $T[\gamma_{Q}..j] = T[\gamma..\gamma + |[\gamma_{Q}, j]| - 1]$. 
If $|[\gamma, \gamma + |[\gamma_{Q}, j]| - 1]| < |\lcp(T[\gamma..r], C_{Q}^{n+1})|$, 
then $\hat{K} = |[\gamma_{Q}, j]|$ follows from 
$\hat{K} = K$, $K = |[\gamma, \gamma + |[\gamma_{Q}, j]| - 1]|$, and $|[\gamma, \gamma + |[\gamma_{Q}, j]| - 1]| = |[\gamma_{Q}, j]|$. 
Otherwise (i.e., $|[\gamma, \gamma + |[\gamma_{Q}, j]| - 1]| \geq |\lcp(T[\gamma..r], C_{Q}^{n+1})|$), 
$\hat{K} = |\lcp(T[\gamma..r], C_{Q}^{n+1})|$ follows from  
$\hat{K} = K$ and $K = |\lcp(T[\gamma..r], C_{Q}^{n+1})|$. 
Therefore, $\hat{K} = \min \{ |[\gamma_{Q}, j]|, |\lcp(T[\gamma..r], C_{Q}^{n+1})| \}$ holds. 

\textbf{Proof of statement (iii).}
Because of $([p, q], [\ell, r]) \in \Psi_{h_{Q}} \cap \Psi_{\run} \cap \Psi_{\centerset}(C_{Q})$, 
$\lcs(T[p-1..\gamma-1], C_{Q}^{n+1}) = T[p-1..\gamma-1]$ follows from the definition of the subset $\Psi_{\run}$. 
$\lcs(T[\gamma - |[i, \gamma_{Q}-1]|..\gamma-1], C_{Q}^{n+1}) = T[\gamma - |[i, \gamma_{Q}-1]|..\gamma-1]$ follows from 
$\lcs(T[p-1..\gamma-1], C_{Q}^{n+1}) = T[p-1..\gamma-1]$ and $\gamma - |[i, \gamma_{Q}-1]| \in [p, q]$. 
We already proved $T[i..\gamma_{Q}-1] = T[\gamma - |[i, \gamma_{Q}-1]|..\gamma-1]$. 
Therefore, $\lcs(T[i..\gamma_{Q}-1], C_{Q}^{n+1}) = T[i..\gamma_{Q}-1]$ holds. 

\textbf{Proof of Proposition~\ref{prop:RSS_query_condition_CD_property}.}
We prove $\mathcal{C}_{\run} = \emptyset$ by contradiction. 
We assume that $\mathcal{C}_{\run} \neq \emptyset$ holds. 
If the RSS query $\RSSQ(T[i..j], b)$ satisfies condition (C) of RSS query, 
then $\lcs(T[i..\gamma_{Q}-1], C_{Q}^{n+1}) \neq T[i..\gamma_{Q}-1]$ holds. 
On the other hand, $\lcs(T[i..\gamma_{Q}-1], C_{Q}^{n+1}) = T[i..\gamma_{Q}-1]$ follows from statement (iii). 
The two facts $\lcs(T[i..\gamma_{Q}-1], C_{Q}^{n+1}) \neq T[i..\gamma_{Q}-1]$ and $\lcs(T[i..\gamma_{Q}-1], C_{Q}^{n+1}) = T[i..\gamma_{Q}-1]$ yield a contradiction. 

Otherwise (i.e., the RSS query $\RSSQ(T[i..j], b)$ does not satisfy condition (C)), 
the RSS query $\RSSQ(T[i..j], b)$ satisfies condition (D). 
In this case, $\hat{K} < |[\gamma_{Q}, j]|$ and $\hat{K} \leq 1 + \sum_{w = 1}^{h_{Q}+3}$ $\lfloor \mu(w) \rfloor$ hold. 
On the other hand, 
$\hat{K} = |[\gamma_{Q}, j]|$ follows from 
$\hat{K} = \min \{ |[\gamma_{Q}, j]|, |\lcp(T[\gamma..r]$, $C_{Q}^{n+1})| \}$ (statement (ii)), 
$|\lcp(T[\gamma..r], C_{Q}^{n+1})| > 1 + \sum_{w = 1}^{h_{Q}+3} \lfloor \mu(w) \rfloor$ (statement (i)), 
and $\hat{K} \leq 1 + \sum_{w = 1}^{h_{Q}+3} \lfloor \mu(w) \rfloor$. 
The two facts $\hat{K} < |[\gamma_{Q}, j]|$ and $\hat{K} = |[\gamma_{Q}, j]|$ yield a contradiction. 
Therefore, $\mathcal{C}_{\run} = \emptyset$ must hold. 
\end{proof}

We prove Lemma~\ref{lem:RSS_query_condition_CD}.

\begin{proof}[Proof of Lemma~\ref{lem:RSS_query_condition_CD}]
We prove $\mathcal{F}_{\SA} \cap \mathcal{F}_{\suffix}(\Psi_{\CCP}(T[i..j]) \cap \Psi_{\run}) = \emptyset$ by contradiction. 
We assume that $\mathcal{F}_{\SA} \cap \mathcal{F}_{\suffix}(\Psi_{\CCP}(T[i..j]) \cap \Psi_{\run}) \neq \emptyset$ holds. 
Then, the set $\Psi_{\CCP}(T[i..j]) \cap \Psi_{\run}$ contains an interval attractor $([p, q], [\ell, r])$ satisfying 
$T[\gamma - |[i, \gamma_{Q}-1]|..\gamma + |[\gamma_{Q}, j]| - 1] \in \mathcal{F}_{\SA} \cap \mathcal{F}_{\suffix}(\Psi_{\CCP}(T[i..j]) \cap \Psi_{\run})$ for the attractor position $\gamma$ of the interval attractor $([p, q], [\ell, r])$. 
Let $C$ be the associated string of the interval attractor $([p, q], [\ell, r])$. 
Then, $T[\gamma - |[i, \gamma_{Q}-1]|..\gamma + |[\gamma_{Q}, j]| - 1] \in \mathcal{F}_{\suffix}(\Psi_{\CCP}(T[i..j]) \cap \Psi_{\run} \cap \Psi_{\centerset}(C))$ follows from the definition of the set $\mathcal{F}_{\suffix}(\Psi_{\CCP}(T[i..j]) \cap \Psi_{\run} \cap \Psi_{\centerset}(C))$. 
$C \in \mathcal{C}_{\run}$ follows from the definition of the set $\mathcal{C}_{\run}$ 
because $\mathcal{F}_{\SA} \cap \mathcal{F}_{\suffix}(\Psi_{\CCP}(T[i..j]) \cap \Psi_{\run} \cap \Psi_{\centerset}(C)) \neq \emptyset$. 
On the other hand, Proposition~\ref{prop:RSS_query_condition_CD_property} shows that $\mathcal{C}_{\run} = \emptyset$ holds. 
The two facts $C \in \mathcal{C}_{\run}$ and $\mathcal{C}_{\run} = \emptyset$ yield a contradiction. 
Therefore, $\mathcal{F}_{\SA} \cap \mathcal{F}_{\suffix}(\Psi_{\CCP}(T[i..j]) \cap \Psi_{\run}) = \emptyset$ must hold.     

We prove Lemma~\ref{lem:RSS_query_condition_CD}. 
Because of $\mathcal{F}_{\SA} \cap \mathcal{F}_{\suffix}(\Psi_{\CCP}(T[i..j]) \cap \Psi_{\run}) = \emptyset$, 
the following equation follows from Lemma~\ref{lem:F_suffix_basic_property}~\ref{enum:F_suffix_basic_property:5}. 
\begin{equation}\label{eq:RSS_query_condition_CD:1}
    \begin{split}
    \mathcal{F}_{\SA} \cap \mathcal{F}_{\suffix}(\Psi_{\CCP}(T[i..j])) &= (\mathcal{F}_{\SA} \cap \mathcal{F}_{\suffix}(\Psi_{\CCP}(T[i..j]) \setminus \Psi_{\run})) \\
    &\cup (\mathcal{F}_{\SA} \cap \mathcal{F}_{\suffix}(\Psi_{\CCP}(T[i..j]) \cap \Psi_{\run})) \\ 
    &= \mathcal{F}_{\SA} \cap \mathcal{F}_{\suffix}(\Psi_{\CCP}(T[i..j]) \setminus \Psi_{\run}).
    \end{split}
\end{equation}
Therefore, Lemma~\ref{lem:RSS_query_condition_CD} follows from 
Lemma~\ref{lem:F_suffix_basic_property}~\ref{enum:F_suffix_basic_property:3} 
and Equation~\ref{eq:RSS_query_condition_CD:1}.      
\end{proof}

\subsection{Subquery \texorpdfstring{$\RSSQA(T[i..j], b)$}{RSSA(T[i..j], b)}}\label{subsec:GammaA}
The goal of this subsection is to solve subquery $\RSSQA(T[i..j], b)$ 
under the assumption that either $\mathcal{C}_{\run} = \emptyset$ or $C_{Q} = C_{\max}$ holds. 
The following lemma states the summary of this subsection. 

\begin{lemma}\label{lem:GammaA_algorithm}
Assume that either $\mathcal{C}_{\run} = \emptyset$ or $C_{Q} = C_{\max}$ holds. 
We can answer subquery $\RSSQA(T[i..j], b)$ in $O(H^{2} \log^{2} n + \log^{5} n)$ time 
using (A) the data structures for RSC query, 
(B) interval $[i, j]$, 
and (C) the starting position $\eta$ of the sa-interval $[\eta, \eta^{\prime}]$ of $T[i..j]$. 
If the subquery returns a string $F$, 
then $F$ is represented as an interval $[g, g + |F| - 1]$ satisfying $T[g..g + |F| - 1] = F$. 
\end{lemma}
\begin{proof}
See Section~\ref{subsubsec:gamma_A_algorithm}.
\end{proof}

We leverage the set $\mathcal{J}_{A}(h_{Q})$ of weighted points on grid $(\mathcal{X}_{A}(h_{Q}), \mathcal{Y}_{A}(h_{Q}))$ introduced in Section~\ref{subsec:RSC_comp_A}. 
Here, $h_{Q}$ is the level of interval attractor $([p_{Q}, q_{Q}], [\ell_{Q}, r_{Q}])$; 
the set $\mathcal{J}_{A}(h_{Q})$ and grid $(\mathcal{X}_{A}(h_{Q}), \mathcal{Y}_{A}(h_{Q}))$ are defined using 
the interval attractors $([p_{1}, q_{1}], [\ell_{1}, r_{1}]), ([p_{2}, q_{2}], [\ell_{2}, r_{2}])$, 
$\ldots$, $([p_{k}, q_{k}], [\ell_{k}, r_{k}])$ in set $\Psi_{h_{Q}} \cap \Psi_{\samp}$. 
Let $\gamma_{s}$ be the attractor position of each interval attractor $([p_{s}, q_{s}], [\ell_{s}, r_{s}]) \in \Psi_{h_{Q}} \cap \Psi_{\samp}$. 
The summary of the set $\mathcal{J}_{A}(h_{Q})$ and $(\mathcal{X}_{A}(h_{Q}), \mathcal{Y}_{A}(h_{Q}))$ is as follows 
(see Section~\ref{subsec:RSC_comp_A} for the details of the set and grid):
\begin{itemize}
    \item the ordered set $\mathcal{X}_{A}(h_{Q})$ consists of $d$ strings $L_{1}, L_{2}, \ldots, L_{d}$ ($L_{1} \prec L_{2} \prec \cdots \prec L_{d}$);
    \item the ordered set $\mathcal{Y}_{A}(h_{Q})$ consists of $d^{\prime}$ strings $R_{1}, R_{2}, \ldots, R_{d^{\prime}}$ ($R_{1} \prec R_{2} \prec \ldots \prec R_{d^{\prime}}$);
    \item the set $\mathcal{J}_{A}(h_{Q})$ contains the weighted point $(\reverse(T[p_{s}-1..\gamma_{s}-1]), T[\gamma_{s}..r_{s}+1], |\Psi_{\str}(T[p_{s}-1..r_{s}+1])|, T[p_{s}-1..r_{s}+1])$ corresponding to each interval attractor $([p_{s}, q_{s}], [\ell_{s}, r_{s}]) \in \Psi_{h_{Q}} \cap \Psi_{\samp}$. 
\end{itemize}

For this subsection, 
we define four integers $x, x^{\prime}, y$, and $y^{\prime}$ as follows: 
\begin{itemize}
    \item $x = \min \{ s \in [1, d] \mid \reverse(T[i..\gamma_{Q}-1]) \prec L_{s} \}$;
    \item $x^{\prime} = \max \{ s \in [1, d] \mid L_{s} \prec \reverse(\# \cdot T[i..\gamma_{Q}-1]) \}$;
    \item $y = \min \{ s \in [1, d^{\prime}] \mid T[\gamma_{Q}..j] \prec R_{s} \}$;
    \item $y^{\prime} = \max \{ s \in [1, d^{\prime}] \mid R_{s} \prec T[\gamma_{Q}..j]\# \}$. 
\end{itemize}

In the next paragraphs, 
we introduce $(y^{\prime} - y + 1)$ subsets of set $\Psi_{\RR}$ 
and two sequences of integers to explain the relationship between the set $\mathcal{J}_{A}(h_{Q})$ and subquery $\RSSQA(T[i..j], b)$. 

\paragraph{Subset $\Psi^{A}(t)$.}
For each integer $t \in [1, y^{\prime} - y + 1]$, 
subset $\Psi^{A}(t) \subseteq \Psi_{\RR}$ consists of interval attractors such that 
each interval attractor $([p, q], [\ell, r]) \in \Psi^{A}(t)$ satisfies the following three conditions: 
\begin{itemize}
    \item $([p, q], [\ell, r]) \in \Psi_{h_{Q}} \setminus \Psi_{\run}$;
    \item $L_{x} \preceq \reverse(T[p-1..\gamma-1]) \preceq L_{x^{\prime}}$ for the attractor position $\gamma$ of the interval attractor $([p, q], [\ell, r])$;
    \item $R_{y + t - 1} \preceq T[\gamma..r+1] \preceq R_{y^{\prime}}$.
\end{itemize}
Formally, the subset $\Psi^{A}(t)$ is defined as follows: 
\begin{equation*}
    \begin{split}
    \Psi^{A}(t) &= \{ ([p, q], [\ell, r]) \in \Psi_{\RR} \mid (([p, q], [\ell, r]) \in \Psi_{h_{Q}} \setminus \Psi_{\run}) \\
    &\land (L_{x} \preceq \reverse(T[p-1..\gamma-1]) \preceq L_{x^{\prime}}) \\
    &\land (R_{y + t - 1} \preceq T[\gamma..r+1] \preceq R_{y^{\prime}}) \}.
    \end{split}
\end{equation*}

\paragraph{Two sequences $\Gamma_{A}$ and $\Gamma_{A, \sub}$.}
Sequence $\Gamma_{A}$ consists of $(y^{\prime} - y + 1)$ integers $u_{1}, u_{2}, \ldots, u_{y^{\prime} - y + 1} \in \{ 0, 1 \}$. 
Here, each integer $u_{t}$ is $1$ if $\mathcal{F}_{\SA} \cap \mathcal{F}_{\suffix}(\Psi_{\CCP}(T[i..j]) \cap \Psi^{A}(t)) \neq \emptyset$; 
otherwise $u_{t}$ is $0$. 

The following lemma states four properties of sequence $\Gamma_{A}$. 

\begin{lemma}\label{lem:GammaA_property}
Let $\kappa$ be the largest integer in set $[1, y^{\prime} - y + 1]$ satisfying $u_{\kappa} = 1$ for sequence $\Gamma_{A} = u_{1}, u_{2}, \ldots, u_{y^{\prime} - y + 1}$. 
Then, the following four statements hold: 
\begin{enumerate}[label=\textbf{(\roman*)}]
    \item \label{enum:GammaA_property:1} if the largest integer $\kappa$ exists, then 
    string $T[i..\gamma_{Q} - 1] \cdot R_{y + \kappa - 1}$ is 
    the lexicographically largest string in set $\mathcal{F}_{\SA} \cap \mathcal{F}_{\suffix}(\Psi_{\CCP}(T[i..j]) \setminus \Psi_{\run})$. 
    Otherwise, the set $\mathcal{F}_{\SA} \cap \mathcal{F}_{\suffix}(\Psi_{\CCP}(T[i..j]) \setminus \Psi_{\run})$ is empty; 
    \item \label{enum:GammaA_property:2} if the largest integer $\kappa$ exists, 
    then set $\mathcal{J}_{A}(h_{Q})$ contains a weighted point 
    $(\reverse(T[p_{s}-1..\gamma_{s}-1]), T[\gamma_{s}..r_{s}+1], |\Psi_{\str}(T[p_{s}-1..r_{s}+1])|, T[p_{s}-1..r_{s}+1])$ 
    satisfying $L_{x} \preceq \reverse(T[p_{s}-1..\gamma_{s}-1] \preceq L_{x^{\prime}}$ 
    and $T[\gamma_{s}..r_{s}+1] = R_{y + \kappa - 1}$;
    \item \label{enum:GammaA_property:3} 
    $T[\gamma_{s} - |[i, \gamma_{Q}]| + 1..r_{s} + 1] = T[i..\gamma_{Q} - 1] \cdot R_{y + \kappa - 1}$ for 
    the interval attractor $([p_{s}, q_{s}], [\ell_{s}, r_{s}])$ corresponding to 
    the weighted point $(\reverse(T[p_{s}-1..\gamma_{s}-1]), T[\gamma_{s}..r_{s}+1], |\Psi_{\str}(T[p_{s}-1..r_{s}+1])|, T[p_{s}-1..r_{s}+1])$ 
    of Lemma~\ref{lem:GammaA_property}~\ref{enum:GammaA_property:2};
    \item \label{enum:GammaA_property:4}
    sequence $\Gamma_{A}$ is non-increasing~(i.e., $u_{1} \geq u_{2} \geq \cdots \geq u_{y^{\prime} - y + 1}$). 
\end{enumerate}
\end{lemma}
\begin{proof}
See Section~\ref{subsubsec:GammaA_proof1}.
\end{proof}

Next, sequence $\Gamma_{A, \sub}$ consists of $(y^{\prime} - y + 1)$ integers 
$\alpha_{1}, \alpha_{2}, \ldots, \alpha_{y^{\prime} - y + 1} \in \mathbb{N}_{0}$. 
Here, each integer $\alpha_{t}$ is defined as 
$\alpha_{t} = \rangecount(\mathcal{J}_{A}(h_{Q}), L_{x}, L_{x^{\prime}}$, $R_{y+t-1}, R_{y^{\prime}})$. 
Here, $\rangecount$ is the range-count query introduced in Section~\ref{subsec:range_data_structure}. 

The following lemma states three properties of sequence $\Gamma_{A, \sub}$. 

\begin{lemma}\label{lem:GammaA_sub_property}
The following three statements hold for two sequences $\Gamma_{A} = u_{1}, u_{2}, \ldots, u_{y^{\prime} - y + 1}$ and $\Gamma_{A, \sub} = \alpha_{1}, \alpha_{2}, \ldots, \alpha_{y^{\prime} - y + 1}$: 
\begin{enumerate}[label=\textbf{(\roman*)}]
    \item \label{enum:GammaA_sub_property:1} 
    sequence $\Gamma_{A, \sub}$ is non-increasing (i.e., $\alpha_{1} \geq \alpha_{2} \geq \cdots \geq \alpha_{y^{\prime} - y + 1}$);    
    \item \label{enum:GammaA_sub_property:2} 
    for an integer $t \in [1, y^{\prime} - y + 1]$, 
    consider the largest integer $t^{\prime}$ in set $[t, y^{\prime} - y + 1]$ satisfying 
    $\alpha_{t} = \alpha_{t^{\prime}}$. 
    If $\alpha_{t^{\prime}} \geq 1$, 
    then set $\mathcal{J}_{A}(h_{Q})$ contains a weighted point 
    $(\reverse(T[p_{s}-1..\gamma_{s}-1]), T[\gamma_{s}..r_{s}+1], |\Psi_{\str}(T[p_{s}-1..r_{s}+1])|, T[p_{s}-1..r_{s}+1])$ satisfying 
    $L_{x} \preceq \reverse(T[p_{s}-1..\gamma_{s}-1]) \preceq L_{x^{\prime}}$ and $T[\gamma_{s}..r_{s}+1] = R_{y + t^{\prime} - 1}$; 
    \item \label{enum:GammaA_sub_property:3}
    consider the three integers $t, t^{\prime}$, and $s$ of Lemma~\ref{lem:GammaA_sub_property}~\ref{enum:GammaA_sub_property:2}. 
    Then, 
    $u_{t} = 1 \Leftrightarrow (\alpha_{t^{\prime}} \geq 1) \land (\RSCQ(\gamma_{s} - |[i, \gamma_{Q}-1]|, \gamma_{s} + |[\gamma_{Q}, j]| - 1) < b - \eta + 1)$. 
\end{enumerate}
\end{lemma}
\begin{proof}
See Section~\ref{subsubsec:GammaA_proof2}.
\end{proof}

\subsubsection{Proof of Lemma~\ref{lem:GammaA_property}}\label{subsubsec:GammaA_proof1}

The following proposition states a property of set $\Psi_{h} \setminus \Psi_{\run}$ for an integer $h \in [0, H]$. 

\begin{proposition}\label{prop:JA_correspondence_property}
Consider an interval attractor $([p, q], [\ell, r]) \in \Psi_{h} \setminus \Psi_{\run}$ for an integer $h \in [0, H]$. 
Then, set $\Psi_{h} \cap \Psi_{\samp}$ contains an interval attractor $([p^{\prime}, q^{\prime}], [\ell^{\prime}, r^{\prime}])$ 
satisfying 
$T[p-1..\gamma-1] = T[p^{\prime}-1..\gamma^{\prime}-1]$ and $T[\gamma..r+1] = T[\gamma^{\prime}..r^{\prime}+1]$. 
Here, $\gamma$ and $\gamma^{\prime}$ are the attractor positions of the two interval attractors $([p, q], [\ell, r])$ 
and $([p^{\prime}, q^{\prime}], [\ell^{\prime}, r^{\prime}])$, respectively. 
\end{proposition}
\begin{proof}
    Lemma~\ref{lem:samp_basic_property}~\ref{enum:samp_basic_property:2} shows that 
    the sampling subset $\Psi_{\samp}$ contains an interval attractor $([p^{\prime}, q^{\prime}]$, $[\ell^{\prime}, r^{\prime}])$ 
    satisfying $T[p-1..r+1] = T[p^{\prime}-1..r^{\prime}+1]$. 
    Because of $T[p-1..r+1] = T[p^{\prime}-1..r^{\prime}+1]$, 
    Lemma~\ref{lem:psi_str_property}~\ref{enum:psi_str_property:1} shows that 
    $T[p-1..\gamma-1] = T[p^{\prime}-1..\gamma^{\prime}-1]$ and $T[\gamma..r+1] = T[\gamma^{\prime}..r^{\prime}+1]$ hold. 
    Similarly, Lemma~\ref{lem:psi_str_property}~\ref{enum:psi_str_property:2} shows that     
    $([p^{\prime}, q^{\prime}], [\ell^{\prime}, r^{\prime}]) \in \Psi_{h}$ holds.
    Therefore, Proposition~\ref{prop:JA_correspondence_property} holds.     
\end{proof}

The following proposition states three properties of set $\Psi^{A}(t)$ for each integer $t \in [1, y^{\prime} - y + 1]$.

\begin{proposition}\label{prop:Psi_A_Property}
The following three statements hold: 
\begin{enumerate}[label=\textbf{(\roman*)}]
    \item \label{enum:Psi_A_Property:1} $\Psi^{A}(1) = \Psi_{\CCP}(T[i..j]) \setminus \Psi_{\run}$;
    \item \label{enum:Psi_A_Property:2} $\Psi^{A}(t) \supseteq \Psi^{A}(t+1)$ for each integer $t \in [1, y^{\prime} - y + 1]$;
    \item \label{enum:Psi_A_Property:3}
    consider an interval attractor $([p, q], [\ell, r])$ in set $\Psi^{A}(t) \setminus \Psi^{A}(t+1)$ for an integer $t \in [1, y^{\prime} - y + 1]$. 
    Then, $T[\gamma - |[i, \gamma_{Q}-1]|..\gamma-1] \cdot T[\gamma..r+1] = T[i..\gamma_{Q}-1] \cdot R_{y+t-1}$ holds 
    for the attractor position $\gamma$ of the interval attractor $([p, q], [\ell, r])$.
\end{enumerate}
Here, let $\Psi^{A}(y^{\prime} - y + 2)$ be an empty set for simplicity. 
\end{proposition}
\begin{proof}
The following two statements are used to prove Proposition~\ref{prop:Psi_A_Property}:
\begin{enumerate}[label=\textbf{(\Alph*)}]
    \item $\Psi^{A}(1) \subseteq \Psi_{\CCP}(T[i..j]) \setminus \Psi_{\run}$;
    \item $\Psi^{A}(1) \supseteq \Psi_{\CCP}(T[i..j]) \setminus \Psi_{\run}$.    
\end{enumerate}

\textbf{Proof of statement (A).}
Consider an interval attractor $([p, q], [\ell, r]) \in \Psi^{A}(1)$.
The following three statements follow from the definition of the subset $\Psi^{A}(1)$: 
\begin{itemize}
    \item $([p, q], [\ell, r]) \in \Psi_{h_{Q}} \setminus \Psi_{\run}$;
    \item $L_{x} \preceq \reverse(T[p-1..\gamma-1]) \preceq L_{x^{\prime}}$ for the attractor position $\gamma$ of the interval attractor $([p, q], [\ell, r])$;
    \item $R_{y} \preceq T[\gamma..r+1] \preceq R_{y^{\prime}}$.
\end{itemize}

We prove $([p, q], [\ell, r]) \in \Psi_{\CCP}(T[i..j])$. 
$\reverse(T[i..\gamma_{Q}-1]) \prec \reverse(T[p-1..\gamma-1]) \prec \reverse(\#T[i..\gamma_{Q}-1])$ follows from 
(a) $L_{x} \preceq \reverse(T[p-1..\gamma-1]) \preceq L_{x^{\prime}}$, 
(b) $x = \min \{ s \in [1, d] \mid \reverse(T[i..\gamma_{Q}-1]) \prec L_{s} \}$, and 
(c) $x^{\prime} = \max \{ s \in [1, d] \mid L_{s} \prec \reverse(\# \cdot T[i..\gamma_{Q}-1]) \}$. 
Similarly, 
$T[\gamma_{Q}..j] \prec T[\gamma..r+1] \prec T[\gamma_{Q}..j]\#$ follows from 
(1) $R_{y} \preceq T[\gamma..r+1] \preceq R_{y^{\prime}}$, 
(2) $y = \min \{ s \in [1, d^{\prime}] \mid T[\gamma_{Q}..j] \prec R_{s} \}$, 
and (3) $y^{\prime} = \max \{ s \in [1, d^{\prime}] \mid R_{s} \prec (T[\gamma_{Q}..j] \cdot \#) \}$. 
Lemma~\ref{lem:CCP_property}~\ref{enum:CCP_property:4} shows that 
$\Psi_{\CCP}(T[i..j]) = \{ ([p^{\prime}, q^{\prime}], [\ell^{\prime}, r^{\prime}]) \in \Psi_{h_{Q}} \mid \reverse(T[i..\gamma_{Q}-1]) \prec \reverse(T[p^{\prime}-1..\gamma^{\prime}-1]) \prec \reverse(\#T[i..\gamma_{Q}-1]) \text{ and } T[\gamma_{Q}..j] \prec T[\gamma^{\prime}..r^{\prime}+1] \prec T[\gamma_{Q}..j]\# \}$ holds. 
Therefore, $([p, q], [\ell, r]) \in \Psi_{\CCP}(T[i..j])$ holds. 

We showed that $([p, q], [\ell, r]) \in \Psi_{\CCP}(T[i..j])$ and $([p, q], [\ell, r]) \not \in \Psi_{\run}$ for each interval attractor $([p, q], [\ell, r]) \in \Psi^{A}(1)$. 
Therefore, $\Psi^{A}(1) \subseteq \Psi_{\CCP}(T[i..j]) \setminus \Psi_{\run}$ holds. 

\textbf{Proof of statement (B).}
Consider an interval attractor $([p, q], [\ell, r]) \in \Psi_{\CCP}(T[i..j]) \setminus \Psi_{\run}$. 
From the definition of the subset $\Psi^{A}(1)$, 
$([p, q], [\ell, r]) \in \Psi^{A}(1)$ holds if 
the following three conditions are satisfied: 
\begin{itemize}
    \item $([p, q], [\ell, r]) \in \Psi_{h_{Q}} \setminus \Psi_{\run}$;
    \item $L_{x} \preceq \reverse(T[p-1..\gamma-1]) \preceq L_{x^{\prime}}$ for the attractor position $\gamma$ of the interval attractor $([p, q], [\ell, r])$;
    \item $R_{y} \preceq T[\gamma..r+1] \preceq R_{y^{\prime}}$.
\end{itemize}

We prove $([p, q], [\ell, r]) \in \Psi_{h_{Q}} \setminus \Psi_{\run}$.
Because of $([p, q], [\ell, r]) \in \Psi_{\CCP}(T[i..j])$, 
Lemma~\ref{lem:CCP_property}~\ref{enum:CCP_property:1} shows that 
$([p, q], [\ell, r]) \in \Psi_{h_{Q}}$ holds. 
Therefore, $([p, q], [\ell, r]) \in \Psi_{h_{Q}} \setminus \Psi_{\run}$ follows from 
$([p, q], [\ell, r]) \in \Psi_{h_{Q}}$ and $([p, q], [\ell, r]) \not \in \Psi_{\run}$. 

Because of $([p, q], [\ell, r]) \in \Psi_{\CCP}(T[i..j])$, 
Lemma~\ref{lem:CCP_property}~\ref{enum:CCP_property:4} shows that 
$\reverse(T[i..\gamma_{Q}-1]) \prec \reverse(T[p-1..\gamma-1]) \prec \reverse(\#T[i..\gamma_{Q}-1])$ 
and $T[\gamma_{Q}..j] \prec T[\gamma..r+1] \prec T[\gamma_{Q}..j]\#$ hold. 
Proposition~\ref{prop:JA_correspondence_property} shows that 
the set $\Psi_{h_{Q}} \cap \Psi_{\samp}$ contains an interval attractor $([p_{s}, q_{s}], [\ell_{s}, r_{s}])$ 
satisfying $T[p_{s}-1..\gamma_{s}-1] = T[p-1..\gamma-1]$ and $T[\gamma_{s}..r_{s}+1] = T[\gamma..r+1]$. 
Here, $\reverse(T[p_{s}-1..\gamma_{s}-1]) \in \mathcal{X}_{A}(h_{Q})$ and 
$T[\gamma_{s}..r_{s}+1] \in \mathcal{Y}_{A}(h_{Q})$ hold. 

We prove $L_{x} \preceq \reverse(T[p-1..\gamma-1]) \preceq L_{x^{\prime}}$ and $R_{y} \preceq T[\gamma..r+1] \preceq R_{y^{\prime}}$. 
$L_{x} \preceq \reverse(T[p-1..\gamma-1]) \preceq L_{x^{\prime}}$ follows from 
(1) $\reverse(T[i..\gamma_{Q}-1]) \prec \reverse(T[p-1..\gamma-1]) \prec \reverse(\#T[i..\gamma_{Q}-1])$,  
(2) $T[p_{s}-1..\gamma_{s}-1] = T[p-1..\gamma-1]$, 
(3) $x = \min \{ s \in [1, d] \mid \reverse(T[i..\gamma_{Q}-1]) \prec L_{s} \}$, 
(4) $x^{\prime} = \max \{ s \in [1, d] \mid L_{s} \prec \reverse(\# \cdot T[i..\gamma_{Q}-1]) \}$, 
and (5) $\reverse(T[p_{s}-1..\gamma_{s}-1]) \in \mathcal{X}_{A}(h_{Q})$. 
Similarly, $R_{y} \preceq T[\gamma..r+1] \preceq R_{y^{\prime}}$ follows from 
(a) $T[\gamma_{Q}..j] \prec T[\gamma..r+1] \prec T[\gamma_{Q}..j]\#$,  
(b) $T[\gamma_{s}..r_{s}+1] = T[\gamma..r+1]$, 
(c) $y = \min \{ s \in [1, d^{\prime}] \mid T[\gamma_{Q}..j] \prec R_{s} \}$, 
(d) $y^{\prime} = \max \{ s \in [1, d^{\prime}] \mid R_{s} \prec (T[\gamma_{Q}..j] \cdot \#) \}$, 
and (e) $\reverse(T[p_{s}-1..\gamma_{s}-1]) \in \mathcal{X}_{A}(h_{Q})$. 

We proved $([p, q], [\ell, r]) \in \Psi_{h_{Q}} \setminus \Psi_{\run}$, 
$L_{x} \preceq \reverse(T[p-1..\gamma-1]) \preceq L_{x^{\prime}}$, and 
$R_{y} \preceq T[\gamma..r+1] \preceq R_{y^{\prime}}$. 
Therefore, $([p, q], [\ell, r]) \in \Psi^{A}(1)$ holds for each interval attractor $([p, q], [\ell, r]) \in \Psi_{\CCP}(T[i..j]) \setminus \Psi_{\run}$. 
Finally, $\Psi^{A}(1) \supseteq \Psi_{\CCP}(T[i..j]) \setminus \Psi_{\run}$ holds. 

\textbf{Proof of Proposition~\ref{prop:Psi_A_Property}(i).}
Proposition~\ref{prop:Psi_A_Property}(i) follows from statement (A) and statement (B). 

\textbf{Proof of Proposition~\ref{prop:Psi_A_Property}(ii).}
Proposition~\ref{prop:Psi_A_Property}(ii) follows from the definitions of the two subsets $\Psi^{A}(t)$ and $\Psi^{A}(t+1)$. 

\textbf{Proof of Proposition~\ref{prop:Psi_A_Property}(iii).}
We prove $T[\gamma - |[i, \gamma_{Q}-1]|..\gamma-1] = T[i..\gamma_{Q}-1]$. 
Because of $([p, q], [\ell, r]) \in \Psi^{A}(t)$
$([p, q], [\ell, r]) \in \Psi_{\CCP}(T[i..j])$ follows from 
Proposition~\ref{prop:Psi_A_Property}~\ref{enum:Psi_A_Property:1} and Proposition~\ref{prop:Psi_A_Property}~\ref{enum:Psi_A_Property:2}. 
Because of $([p, q], [\ell, r]) \in \Psi_{\CCP}(T[i..j])$, 
Lemma~\ref{lem:CCP_property}~\ref{enum:CCP_property:6} shows that 
$T[\gamma - |[i, \gamma_{Q}-1]|..\gamma-1] = T[i..\gamma_{Q}-1]$ holds. 

We prove $T[\gamma..r+1] = R_{y+t-1}$. 
Because of $([p, q], [\ell, r]) \in \Psi^{A}(t) \setminus \Psi^{A}(t+1)$, 
the following three statements follow from the definitions of the two subsets $\Psi^{A}(t)$ and $\Psi^{A}(t+1)$: 
\begin{itemize}
    \item $([p, q], [\ell, r]) \in \Psi_{h_{Q}} \setminus \Psi_{\run}$;
    \item $L_{x} \preceq \reverse(T[p-1..\gamma-1]) \preceq L_{x^{\prime}}$ for the attractor position $\gamma$ of the interval attractor $([p, q], [\ell, r])$;
    \item $R_{y + t - 1} \preceq T[\gamma..r+1] \prec R_{y + t}$.
\end{itemize}
Because of $([p, q], [\ell, r]) \in \Psi_{h_{Q}} \setminus \Psi_{\run}$, 
Proposition~\ref{prop:JA_correspondence_property} shows that 
the set $\Psi_{h_{Q}} \cap \Psi_{\samp}$ contains an interval attractor $([p_{s}, q_{s}], [\ell_{s}, r_{s}])$ 
satisfying $T[p_{s}-1..\gamma_{s}-1] = T[p-1..\gamma-1]$ and $T[\gamma_{s}..r_{s}+1] = T[\gamma..r+1]$. 
Here, $\reverse(T[p_{s}-1..\gamma_{s}-1]) \in \mathcal{X}_{A}(h_{Q})$ and 
$T[\gamma_{s}..r_{s}+1] \in \mathcal{Y}_{A}(h_{Q})$ hold. 
Because of $T[\gamma_{s}..r_{s}+1] \in \mathcal{Y}_{A}(h_{Q})$, 
there exists an integer $\tau \in [1, d^{\prime}]$ satisfying 
$T[\gamma_{s}..r_{s}+1] = R_{\tau}$. 
$R_{y+t-1} \preceq R_{\tau} \prec R_{y+t}$ follows from 
$R_{y + t - 1} \preceq T[\gamma..r+1] \prec R_{y + t}$, $T[\gamma_{s}..r_{s}+1] = T[\gamma..r+1]$, and $T[\gamma_{s}..r_{s}+1] = R_{\tau}$. 
$\tau = y+t-1$ follows from $R_{y+t-1} \preceq R_{\tau} \prec R_{y+t}$ 
and $R_{1} \prec R_{2} \prec \cdots \prec R_{d^{\prime}}$. 
Therefore, $T[\gamma..r+1] = R_{y+t-1}$ holds. 

Finally, $T[\gamma - |[i, \gamma_{Q}-1]|..\gamma-1] \cdot T[\gamma..r+1] = T[i..\gamma_{Q}-1] \cdot R_{y+t-1}$ follows from 
$T[\gamma - |[i, \gamma_{Q}-1]|..\gamma-1] = T[i..\gamma_{Q}-1]$ and $T[\gamma..r+1] = R_{y+t-1}$. 

\end{proof}

For proving Lemma~\ref{lem:GammaA_property}, 
we introduce a set $\mathcal{I}^{A}$ of integers in set $\{ 1, 2, \ldots, y^{\prime} - y + 1 \}$. 
This set $\mathcal{I}^{A}$ consists of integers such that 
for each integer $t \in \mathcal{I}^{A}$, 
set $\Psi^{A}(t) \setminus \Psi^{A}(t+1)$ contains an interval attractor $([p, q], [\ell, r])$ satisfying 
$\RSCQ(\gamma - |[i, \gamma_{Q}-1]|, \gamma + |[\gamma_{Q}, j]| - 1) < b - \eta + 1$ for the attractor position $\gamma$ of the interval attractor $([p, q], [\ell, r])$. 
Formally, $\mathcal{I}^{A} = \{ t \in [1, y^{\prime} - y + 1] \mid \exists ([p, q], [\ell, r]) \in \Psi^{A}(t) \setminus \Psi^{A}(t+1) \text{ s.t. } \RSCQ(\gamma - |[i, \gamma_{Q}-1]|, \gamma + |[\gamma_{Q}, j]| - 1) < b - \eta + 1 \}$. 
Here, $\Psi^{A}(y^{\prime} - y + 2) = \emptyset$ for simplicity. 

The following proposition states three properties of the set $\mathcal{I}^{A}$. 

\begin{proposition}\label{prop:Set_IA_Property}
    The following three statements hold for set $\mathcal{I}^{A}$ and sequence $\Gamma_{A} = u_{1}, u_{2}$, $\ldots$, $u_{y^{\prime} - y + 1 }$: 
\begin{enumerate}[label=\textbf{(\roman*)}]
    \item \label{enum:Set_IA_Property:1} $\mathcal{F}_{\SA} \cap \mathcal{F}_{\suffix}(\Psi_{\CCP}(T[i..j]) \setminus \Psi_{\run}) = \{ T[i..\gamma_{Q} - 1] \cdot R_{y + t - 1} \mid t \in \mathcal{I}^{A} \}$ for the attractor position $\gamma_{Q}$ of interval attractor $([p_{Q}, q_{Q}], [\ell_{Q}, r_{Q}])$; 
    \item \label{enum:Set_IA_Property:2} $t \in \mathcal{I}^{A}$ for each integer $t \in [1, y^{\prime} - y + 1]$ satisfying $u_{t} = 1$ and $u_{t+1} = 0$;
    \item \label{enum:Set_IA_Property:3} $u_{t^{\prime}} = 1$ for 
    any pair of two integers $t \in \mathcal{I}^{A}$ and $t^{\prime} \in [1, t]$. 
\end{enumerate}
Here, let $u_{y^{\prime} - y + 2} = 0$ and $\Psi^{A}(y^{\prime} - y + 2) = \emptyset$ for simplicity. 
\end{proposition}
\begin{proof}
The following two statements are used to prove Proposition~\ref{prop:Set_IA_Property}: 
\begin{enumerate}[label=\textbf{(\Alph*)}]
    \item $\mathcal{F}_{\SA} \cap \mathcal{F}_{\suffix}(\Psi_{\CCP}(T[i..j]) \setminus \Psi_{\run}) \subseteq \{ T[i..\gamma_{Q} - 1] \cdot R_{y + t - 1} \mid t \in \mathcal{I}^{A} \}$;
    \item $\mathcal{F}_{\SA} \cap \mathcal{F}_{\suffix}(\Psi_{\CCP}(T[i..j]) \setminus \Psi_{\run}) \supseteq \{ T[i..\gamma_{Q} - 1] \cdot R_{y + t - 1} \mid t \in \mathcal{I}^{A} \}$. 
\end{enumerate}

\textbf{Proof of statement (A).}
The following equation follows from Proposition~\ref{prop:Psi_A_Property}~\ref{enum:Psi_A_Property:1}, 
Proposition~\ref{prop:Psi_A_Property}~\ref{enum:Psi_A_Property:2}, and $\Psi^{A}(y^{\prime} - y + 2) = \emptyset$: 
\begin{equation}\label{eq:Set_A_Property:1}
    \Psi_{\CCP}(T[i..j]) \setminus \Psi_{\run} = \bigcup_{t = 1}^{y^{\prime} - y + 1} \Psi^{A}(t) \setminus \Psi^{A}(t+1).
\end{equation}

Consider a string $F$ in set $\mathcal{F}_{\SA} \cap \mathcal{F}_{\suffix}(\Psi_{\CCP}(T[i..j]) \setminus \Psi_{\run})$. 
Because of $F \in \mathcal{F}_{\suffix}(\Psi_{\CCP}(T[i..j]) \setminus \Psi_{\run})$, 
set $\Psi_{\CCP}(T[i..j]) \setminus \Psi_{\run}$ contains an interval attractor $([p, q], [\ell, r])$ satisfying 
$T[\gamma - |[i, \gamma_{Q}-1]|..r+1] = F$ for the attractor position $\gamma$ of the interval attractor $([p, q], [\ell, r])$. 
Equation~\ref{eq:Set_A_Property:1} indicates that 
there exists an integer $t \in [1, y^{\prime} - y + 1]$ satisfying $([p, q], [\ell, r]) \in \Psi^{A}(t) \setminus \Psi^{A}(t+1)$. 

We prove $F \in \{ T[i..\gamma_{Q} - 1] \cdot R_{y + t - 1} \mid t \in \mathcal{I}^{A} \}$. 
Because of $F \in \mathcal{F}_{\SA}$, 
Lemma~\ref{lem:F_SA_formula} shows that 
$\RSCQ(\gamma - |[i, \gamma_{Q}-1]|, \gamma + |[\gamma_{Q}, j]| - 1) < b - \eta + 1$ holds. 
$t \in \mathcal{I}^{A}$ follows from $([p, q], [\ell, r]) \in \Psi^{A}(t) \setminus \Psi^{A}(t+1)$ 
and $\RSCQ(\gamma - |[i, \gamma_{Q}-1]|, \gamma + |[\gamma_{Q}, j]| - 1) < b - \eta + 1$. 
Proposition~\ref{prop:Psi_A_Property}~\ref{enum:Psi_A_Property:3} shows that 
$F = T[i..\gamma_{Q}-1] \cdot R_{y+t-1}$ holds. 
Therefore, $F \in \{ T[i..\gamma_{Q} - 1] \cdot R_{y + t - 1} \mid t \in \mathcal{I}^{A} \}$ follows from 
$F = T[i..\gamma_{Q}-1] \cdot R_{y+t-1}$ and $t \in \mathcal{I}^{A}$. 

We showed that $F \in \{ T[i..\gamma_{Q} - 1] \cdot R_{y + t - 1} \mid t \in \mathcal{I}^{A} \}$ for each string $F \in \mathcal{F}_{\SA} \cap \mathcal{F}_{\suffix}(\Psi_{\CCP}(T[i..j]) \setminus \Psi_{\run})$. 
Therefore, statement (A) holds. 

\textbf{Proof of statement (B).}
Consider an integer $t$ in set $\mathcal{I}^{A}$. 
From the definition of the set $\mathcal{I}^{A}$, 
set $\Psi^{A}(t) \setminus \Psi^{A}(t+1)$ contains an interval attractor $([p, q], [\ell, r])$ satisfying 
$\RSCQ(\gamma - |[i, \gamma_{Q}-1]|, \gamma + |[\gamma_{Q}, j]| - 1) < b - \eta + 1$ for the attractor position $\gamma$ of the interval attractor $([p, q], [\ell, r])$. 
Because of $([p, q], [\ell, r]) \in \Psi^{A}(t)$, 
$([p, q], [\ell, r]) \in \Psi_{\CCP}(T[i..j]) \setminus \Psi_{\run}$ follows from 
Proposition~\ref{prop:Psi_A_Property}~\ref{enum:Psi_A_Property:1} 
and Proposition~\ref{prop:Psi_A_Property}~\ref{enum:Psi_A_Property:2}. 

We prove $T[i..\gamma_{Q}-1] \cdot R_{y+t-1} \in \mathcal{F}_{\SA} \cap \mathcal{F}_{\suffix}(\Psi_{\CCP}(T[i..j]) \setminus \Psi_{\run})$. 
Because of $([p, q], [\ell, r]) \in \Psi_{\CCP}(T[i..j]) \setminus \Psi_{\run}$, 
$T[\gamma - |[i, \gamma_{Q}-1]|..r+1] \in \mathcal{F}_{\suffix}(\Psi_{\CCP}(T[i..j]) \setminus \Psi_{\run})$ follows from the definition of the set $\mathcal{F}_{\suffix}(\Psi_{\CCP}(T[i..j]) \setminus \Psi_{\run})$. 
Because of $\RSCQ(\gamma - |[i, \gamma_{Q}-1]|, \gamma + |[\gamma_{Q}, j]| - 1) < b - \eta + 1$, 
Lemma~\ref{lem:F_SA_formula} shows that 
$T[\gamma - |[i, \gamma_{Q}-1]|..r+1] \in \mathcal{F}_{\SA}$ holds. 
Proposition~\ref{prop:Psi_A_Property}~\ref{enum:Psi_A_Property:3} shows that 
$T[\gamma - |[i, \gamma_{Q}-1]|..\gamma-1] \cdot T[\gamma..r+1] = T[i..\gamma_{Q}-1] \cdot R_{y+t-1}$ holds. 
Therefore, $T[i..\gamma_{Q}-1] \cdot R_{y+t-1} \in \mathcal{F}_{\SA} \cap \mathcal{F}_{\suffix}(\Psi_{\CCP}(T[i..j]) \setminus \Psi_{\run})$ 
follows from (a) $T[\gamma - |[i, \gamma_{Q}-1]|..r+1] \in \mathcal{F}_{\suffix}(\Psi_{\CCP}(T[i..j]) \setminus \Psi_{\run})$, 
(b) $T[\gamma - |[i, \gamma_{Q}-1]|..r+1] \in \mathcal{F}_{\SA}$, 
and (c) $T[\gamma - |[i, \gamma_{Q}-1]|..\gamma-1] \cdot T[\gamma..r+1] = T[i..\gamma_{Q}-1] \cdot R_{y+t-1}$. 

We showed that $T[i..\gamma_{Q}-1] \cdot R_{y+t-1} \in \mathcal{F}_{\SA} \cap \mathcal{F}_{\suffix}(\Psi_{\CCP}(T[i..j]) \setminus \Psi_{\run})$ holds for each integer $t \in \mathcal{I}^{A}$. 
Therefore, statement (B) holds. 

\textbf{Proof of Proposition~\ref{prop:Set_IA_Property}(i).}
Proposition~\ref{prop:Set_IA_Property}(i) follows from statement (A) and statement (B).
 
\textbf{Proof of Proposition~\ref{prop:Set_IA_Property}(ii).}
$\mathcal{F}_{\SA} \cap \mathcal{F}_{\suffix}(\Psi_{\CCP}(T[i..j]) \cap \Psi^{A}(t)) \neq \emptyset$ follows from $u_{t} = 1$. 
Let $F$ be a string in set $\mathcal{F}_{\SA} \cap \mathcal{F}_{\suffix}(\Psi_{\CCP}(T[i..j]) \cap \Psi^{A}(t))$. 
Then, set $\Psi_{\CCP}(T[i..j]) \cap \Psi^{A}(t)$ contains an interval attractor $([p, q], [\ell, r])$ satisfying 
$T[\gamma - |[i, \gamma_{Q}-1]|..r+1] = F$ for the attractor position $\gamma$ of the interval attractor $([p, q], [\ell, r])$. 

We prove $([p, q], [\ell, r]) \not \in \Psi^{A}(t+1)$ by contradiction. 
We assume that $([p, q], [\ell, r]) \in \Psi^{A}(t+1)$ holds. 
Because of $([p, q], [\ell, r]) \in \Psi_{\CCP}(T[i..j]) \cap \Psi^{A}(t+1)$, 
$F \in \mathcal{F}_{\suffix}(\Psi_{\CCP}(T[i..j]) \cap \Psi^{A}(t+1))$ follows from the definition of the definition of the set $\mathcal{F}_{\suffix}(\Psi_{\CCP}(T[i..j]) \cap \Psi^{A}(t+1))$. 
$\mathcal{F}_{\SA} \cap \mathcal{F}_{\suffix}(\Psi_{\CCP}(T[i..j]) \cap \Psi^{A}(t+1)) \neq \emptyset$ follows from 
$F \in \mathcal{F}_{\SA}$ and $F \in \mathcal{F}_{\suffix}(\Psi_{\CCP}(T[i..j]) \cap \Psi^{A}(t+1))$. 
$u_{t+1} = 1$ follows from $\mathcal{F}_{\SA} \cap \mathcal{F}_{\suffix}(\Psi_{\CCP}(T[i..j]) \cap \Psi^{A}(t+1)) \neq \emptyset$. 
The two facts $u_{t+1} = 0$ and $u_{t+1} = 1$ yield a contradiction. 
Therefore, $([p, q], [\ell, r]) \not \in \Psi^{A}(t+1)$ must hold. 

We prove Proposition~\ref{prop:Set_IA_Property}(ii).  
Because of $F \in \mathcal{F}_{\SA}$, 
Lemma~\ref{lem:F_SA_formula} shows that 
$\RSCQ(\gamma - |[i, \gamma_{Q}-1]|, \gamma + |[\gamma_{Q}, j]| - 1) < b - \eta + 1$ holds. 
$t \in \mathcal{I}^{A}$ follows from 
$([p, q], [\ell, r]) \in \Psi^{A}(t) \setminus \Psi^{A}(t+1)$ and $\RSCQ(\gamma - |[i, \gamma_{Q}-1]|, \gamma + |[\gamma_{Q}, j]| - 1) < b - \eta + 1$. 
Therefore, Proposition~\ref{prop:Set_IA_Property}(ii) holds. 

\textbf{Proof of Proposition~\ref{prop:Set_IA_Property}(iii).}
From the definition of the set $\mathcal{I}^{A}$, 
set $\Psi^{A}(t) \setminus \Psi^{A}(t+1)$ contains an interval attractor $([p, q], [\ell, r])$ satisfying 
$\RSCQ(\gamma - |[i, \gamma_{Q}-1]|, \gamma + |[\gamma_{Q}, j]| - 1) < b - \eta + 1$ for the attractor position $\gamma$ of the interval attractor $([p, q], [\ell, r])$. 
Here, $([p, q], [\ell, r]) \in \Psi_{\CCP}(T[i..j]) \setminus \Psi_{\run}$ holds 
because (1) $([p, q], [\ell, r]) \in \Psi^{A}(t)$, 
(2) $\Psi^{A}(t) \subseteq \Psi^{A}(t-1) \subseteq \cdots \subseteq \Psi^{A}(1)$ (Proposition~\ref{prop:Psi_A_Property}~\ref{enum:Psi_A_Property:2}), 
and (3) $\Psi^{A}(1) = \Psi_{\CCP}(T[i..j]) \setminus \Psi_{\run}$ (Proposition~\ref{prop:Psi_A_Property}~\ref{enum:Psi_A_Property:1}).
Because of $\RSCQ(\gamma - |[i, \gamma_{Q}-1]|, \gamma + |[\gamma_{Q}, j]| - 1) < b - \eta + 1$, 
Lemma~\ref{lem:F_SA_formula} shows that 
$T[\gamma - |[i, \gamma_{Q}-1]|..r+1] \in \mathcal{F}_{\SA}$ holds. 

We prove Proposition~\ref{prop:Set_IA_Property}(iii). 
$([p, q], [\ell, r]) \in \Psi^{A}(t^{\prime})$ holds 
because (1) $\Psi^{A}(t^{\prime}) \supseteq \Psi^{A}(t^{\prime}+1) \supseteq \cdots \supseteq \Psi^{A}(t)$ (Proposition~\ref{prop:Psi_A_Property}~\ref{enum:Psi_A_Property:2}), 
and (2) $([p, q], [\ell, r]) \in \Psi^{A}(t)$. 
Because of $([p, q], [\ell, r]) \in \Psi_{\CCP}(T[i..j]) \cap \Psi^{A}(t^{\prime})$, 
$T[\gamma - |[i, \gamma_{Q}-1]|..r+1] \in \mathcal{F}_{\suffix}(\Psi_{\CCP}(T[i..j]) \cap \Psi^{A}(t^{\prime}))$ follows from the definition of the set $\Psi_{\CCP}(T[i..j]) \cap \Psi^{A}(t^{\prime})$. 
$\mathcal{F}_{\SA} \cap \mathcal{F}_{\suffix}(\Psi_{\CCP}(T[i..j]) \cap \Psi^{A}(t^{\prime})) \neq \emptyset$ follows from 
$T[\gamma - |[i, \gamma_{Q}-1]|..r+1] \in \mathcal{F}_{\SA}$ and $T[\gamma - |[i, \gamma_{Q}-1]|..r+1] \in \mathcal{F}_{\suffix}(\Psi_{\CCP}(T[i..j]) \cap \Psi^{A}(t^{\prime}))$. 
Because of $\mathcal{F}_{\SA} \cap \mathcal{F}_{\suffix}(\Psi_{\CCP}(T[i..j]) \cap \Psi^{A}(t^{\prime})) \neq \emptyset$, 
$u_{t^{\prime}} = 1$ follows from the definition of sequence $\Gamma_{A}$. 
Therefore, Proposition~\ref{prop:Set_IA_Property}(iii) holds. 
\end{proof}

%%%%%%%%%%%%%%%%%%%%%%%%%%%%%%%%

We prove Lemma~\ref{lem:GammaA_property} using Proposition~\ref{prop:JA_correspondence_property}, Proposition~\ref{prop:Psi_A_Property}, and Proposition~\ref{prop:Set_IA_Property}. 

\begin{proof}[Proof of Lemma~\ref{lem:GammaA_property}~\ref{enum:GammaA_property:1}]
Let $t_{1}, t_{2}, \ldots, t_{m}$ ($t_{1} < t_{2} < \ldots < t_{m}$) be the integers in set $\mathcal{I}^{A}$. 
If the integer $\kappa$ exists, 
then $\kappa = t_{m}$ follows from 
Proposition~\ref{prop:Set_IA_Property}~\ref{enum:Set_IA_Property:2} and Proposition~\ref{prop:Set_IA_Property}~\ref{enum:Set_IA_Property:3}. 
Proposition~\ref{prop:Set_IA_Property}~\ref{enum:Set_IA_Property:1} shows that 
$\mathcal{F}_{\SA} \cap \mathcal{F}_{\suffix}(\Psi_{\CCP}(T[i..j]) \setminus \Psi_{\run}) = \{ T[i..\gamma_{Q} - 1] \cdot R_{y + t_{\tau} - 1} \mid \tau \in [1, m] \}$ holds. 
Here, $T[i..\gamma_{Q} - 1] \cdot R_{y + t_{1} - 1} \prec T[i..\gamma_{Q} - 1] \cdot R_{y + t_{2} - 1} \prec \cdots \prec T[i..\gamma_{Q} - 1] \cdot R_{y + t_{m} - 1}$ holds 
because $R_{1} \prec R_{2} \prec \cdots \prec R_{d^{\prime}}$. 
Therefore, string $T[i..\gamma_{Q} - 1] \cdot R_{y + \kappa - 1}$ is 
the lexicographically largest string in set $\mathcal{F}_{\SA} \cap \mathcal{F}_{\suffix}(\Psi_{\CCP}(T[i..j]) \setminus \Psi_{\run})$. 

Otherwise (i.e., the integer $\kappa$ does not exist), 
Proposition~\ref{prop:Set_IA_Property}~\ref{enum:Set_IA_Property:3} indicates that 
$\mathcal{I}^{A} = \emptyset$ holds (i.e., $m = 0$).
In this case, 
Proposition~\ref{prop:Set_IA_Property}~\ref{enum:Set_IA_Property:1} shows that 
$\mathcal{F}_{\SA} \cap \mathcal{F}_{\suffix}(\Psi_{\CCP}(T[i..j]) \setminus \Psi_{\run}) = \emptyset$ holds. 
Therefore, Lemma~\ref{lem:GammaA_property}~\ref{enum:GammaA_property:1} holds. 
\end{proof}

\begin{proof}[Proof of Lemma~\ref{lem:GammaA_property}~\ref{enum:GammaA_property:2}]
We prove $\Psi^{A}(\kappa) \setminus \Psi^{A}(\kappa+1) \neq \emptyset$. 
$\kappa = \kappa^{\prime}$ follows from Proposition~\ref{prop:Set_IA_Property}~\ref{enum:Set_IA_Property:2} 
and Proposition~\ref{prop:Set_IA_Property}~\ref{enum:Set_IA_Property:3} 
for the largest integer $\kappa^{\prime}$ in the set $\mathcal{I}^{A}$. 
$\Psi^{A}(\kappa^{\prime}) \setminus \Psi^{A}(\kappa^{\prime}+1) \neq \emptyset$ follows from the definition of the set $\mathcal{I}^{A}$. 
Therefore, $\Psi^{A}(\kappa) \setminus \Psi^{A}(\kappa+1) \neq \emptyset$ holds. 

Consider an interval attractor $([p, q], [\ell, r])$ in set $\Psi^{A}(\kappa) \setminus \Psi^{A}(\kappa+1)$. 
Then, we prove $([p, q], [\ell, r]) \in \Psi_{h_{Q}} \setminus \Psi_{\run}$, 
$L_{x} \preceq \reverse(T[p-1..\gamma-1]) \preceq L_{x^{\prime}}$, 
and $T[\gamma..r+1] = R_{y + \kappa - 1}$ for the attractor position $\gamma$ of the interval attractor $([p, q], [\ell, r])$. 
Because of $([p, q], [\ell, r]) \in \Psi^{A}(\kappa)$, 
$([p, q], [\ell, r]) \in \Psi_{h_{Q}} \setminus \Psi_{\run}$ 
and $L_{x} \preceq \reverse(T[p-1..\gamma-1]) \preceq L_{x^{\prime}}$ follow from the definition of the subset $\Psi^{A}(\kappa)$. 
Proposition~\ref{prop:Psi_A_Property}~\ref{enum:Psi_A_Property:3} shows that $T[\gamma..r+1] = R_{y + \kappa - 1}$ holds. 

We prove Lemma~\ref{lem:GammaA_property}~\ref{enum:GammaA_property:2}. 
Because of $([p, q], [\ell, r]) \in \Psi_{h_{Q}} \setminus \Psi_{\run}$, 
Proposition~\ref{prop:JA_correspondence_property} shows that 
the set $\Psi_{h_{Q}} \cap \Psi_{\samp}$ contains an interval attractor $([p_{s}, q_{s}], [\ell_{s}, r_{s}])$ 
satisfying $T[p_{s}-1..\gamma_{s}-1] = T[p-1..\gamma-1]$ and $T[\gamma_{s}..r_{s}+1] = T[\gamma..r+1]$. 
The interval attractor $([p_{s}, q_{s}], [\ell_{s}, r_{s}])$ corresponds to 
the weighted point $(\reverse(T[p_{s}-1..\gamma_{s}-1]), T[\gamma_{s}..r_{s}+1], |\Psi_{\str}(T[p_{s}-1..r_{s}+1])|, T[p_{s}-1..r_{s}+1])$ 
in set $\mathcal{J}_{A}(h_{Q})$. 
$L_{x} \preceq \reverse(T[p_{s}-1..\gamma_{s}-1]) \preceq L_{x^{\prime}}$ follows from 
$L_{x} \preceq \reverse(T[p-1..\gamma-1]) \preceq L_{x^{\prime}}$ and $T[p_{s}-1..\gamma_{s}-1] = T[p-1..\gamma-1]$. 
$T[\gamma_{s}..r_{s}+1] = R_{y + \kappa - 1}$ follows from 
$T[\gamma_{s}..r_{s}+1] = T[\gamma..r+1]$ and $T[\gamma..r+1] = R_{y + \kappa - 1}$. 
Therefore, Lemma~\ref{lem:GammaA_property}~\ref{enum:GammaA_property:2} holds. 
\end{proof}

\begin{proof}[Proof of Lemma~\ref{lem:GammaA_property}~\ref{enum:GammaA_property:3}]
$T[\gamma_{s} - |[i, \gamma-1]|..\gamma_{s}-1] = T[i..\gamma_{Q} - 1]$ follows from 
$L_{x} \preceq \reverse(T[p_{s}-1..\gamma_{s}-1]) \preceq L_{x^{\prime}}$. 
Therefore, $T[\gamma_{s} - |[i, \gamma_{Q}]| + 1..r_{s} + 1] = T[i..\gamma_{Q} - 1] \cdot R_{y + \kappa - 1}$ 
follows from $T[\gamma_{s} - |[i, \gamma-1]|..\gamma_{s}-1] = T[i..\gamma_{Q} - 1]$ and $T[\gamma_{s}..r_{s}+1] = R_{y + \kappa - 1}$. 
\end{proof}

\begin{proof}[Proof of Lemma~\ref{lem:GammaA_property}~\ref{enum:GammaA_property:4}]
Let $u_{y^{\prime} - y + 2} = 0$ for simplicity. 
We prove Lemma~\ref{lem:GammaA_property}~\ref{enum:GammaA_property:4} by contradiction. 
We assume that Lemma~\ref{lem:GammaA_property}~\ref{enum:GammaA_property:4} does not hold. 
Then, there exists an integer $t \in [1, y^{\prime} - y]$ satisfying 
$u_{t} = 0$ and $u_{t+1} = 1$. 
Because of $u_{y^{\prime} - y + 2} = 0$, 
there exists an integer $t^{\prime} \in [t, y^{\prime} - y + 1]$ 
satisfying $u_{t^{\prime}} = 1$ and $u_{t^{\prime}+1} = 0$. 
Because of $u_{t^{\prime}} = 1$ and $u_{t^{\prime}+1} = 0$, 
Proposition~\ref{prop:Set_IA_Property}~\ref{enum:Set_IA_Property:2} shows that 
$t^{\prime} \in \mathcal{I}^{A}$ holds. 
Because of $t^{\prime} \in \mathcal{I}^{A}$, 
Proposition~\ref{prop:Set_IA_Property}~\ref{enum:Set_IA_Property:3} shows that 
$u_{t} = 1$ holds. 
The two facts $u_{t} = 0$ and $u_{t} = 1$ yield a contradiction. 
Therefore, Lemma~\ref{lem:GammaA_property}~\ref{enum:GammaA_property:4} must hold. 

\end{proof}

\subsubsection{Proof of Lemma~\ref{lem:GammaA_sub_property}}\label{subsubsec:GammaA_proof2}
The following proposition states the relationship between set $\Psi^{A}(t) \setminus \Psi^{A}(t+1)$ 
and range-count query on set $\mathcal{J}_{A}(h_{Q})$ for each integer $t \in [1, \hat{y} - y + 1]$. 

\begin{proposition}\label{prop:GammaA_Grid}
For an $t \in [1, \hat{y} - y + 1]$, 
$\rangecount(\mathcal{J}_{A}(h_{Q}), L_{x}, L_{x^{\prime}}, R_{y + t - 1}, R_{y + t - 1}) \geq 1 \Leftrightarrow \Psi^{A}(t) \setminus \Psi^{A}(t+1) \neq \emptyset$. 
Here, let $\Psi^{A}(\hat{y} - y + 2) = \emptyset$ for simplicity. 
\end{proposition}
\begin{proof}
Proposition~\ref{prop:GammaA_Grid} follows from the following two statements: 
\begin{enumerate}[label=\textbf{(\roman*)}]
    \item $\rangecount(\mathcal{J}_{A}(h_{Q}), L_{x}, L_{x^{\prime}}, R_{y + t - 1}, R_{y + t - 1}) \geq 1 \Rightarrow \Psi^{A}(t) \setminus \Psi^{A}(t+1) \neq \emptyset$;
    \item $\rangecount(\mathcal{J}_{A}(h_{Q}), L_{x}, L_{x^{\prime}}, R_{y + t - 1}, R_{y + t - 1}) \geq 1 \Leftarrow \Psi^{A}(t) \setminus \Psi^{A}(t+1) \neq \emptyset$.
\end{enumerate}

\textbf{Proof of statement (i).}
Because of $\rangecount(\mathcal{J}_{A}(h_{Q}), L_{x}, L_{x^{\prime}}, R_{y + t - 1}, R_{y + t - 1}) \geq 1$, 
set $\mathcal{J}_{A}(h_{Q})$ contains a weighted point $(\reverse(T[p_{s}-1..\gamma_{s}-1]), T[\gamma_{s}..r_{s}+1], |\Psi_{\str}(T[p_{s}-1..r_{s}+1])|, T[p_{s}-1..r_{s}+1])$ satisfying 
$L_{x} \preceq \reverse(T[p_{s}-1..\gamma_{s}-1]) \preceq L_{x^{\prime}}$ 
and $T[\gamma_{s}..r_{s}+1] = R_{y + t - 1}$. 
This weighted point corresponds to the interval attractor $([p_{s}, q_{s}], [\ell_{s}, r_{s}])$ in set $\Psi_{h_{Q}} \cap \Psi_{\samp}$. 

We prove $([p_{s}, q_{s}], [\ell_{s}, r_{s}]) \in \Psi^{A}(t) \setminus \Psi^{A}(t+1)$. 
$([p_{s}, q_{s}], [\ell_{s}, r_{s}]) \not \in \Psi_{\run}$ follows from 
Lemma~\ref{lem:samp_basic_property} \ref{enum:samp_basic_property:3}. 
$([p_{s}, q_{s}], [\ell_{s}, r_{s}]) \in \Psi^{A}(t)$ follows from the definition of the subset $\Psi^{A}(t)$ 
because $([p_{s}, q_{s}], [\ell_{s}, r_{s}]) \in \Psi_{h_{Q}} \setminus \Psi_{\run}$, 
$L_{x} \preceq \reverse(T[p_{s}-1..\gamma_{s}-1]) \preceq L_{x^{\prime}}$, 
and $T[\gamma_{s}..r_{s}+1] = R_{y + t - 1}$. 
On the other hand, $([p_{s}, q_{s}], [\ell_{s}, r_{s}]) \not \in \Psi^{A}(t+1)$ follows from the definition of the subset $\Psi^{A}(t+1)$ 
because $T[\gamma_{s}..r_{s}+1] = R_{y + t - 1}$. 
Therefore, $([p_{s}, q_{s}], [\ell_{s}, r_{s}]) \in \Psi^{A}(t) \setminus \Psi^{A}(t+1)$ holds. 

$\Psi^{A}(t) \setminus \Psi^{A}(t+1) \neq \emptyset$ follows from $([p_{s}, q_{s}], [\ell_{s}, r_{s}]) \in \Psi^{A}(t) \setminus \Psi^{A}(t+1)$. 
Therefore, statement (i) holds. 

\textbf{Proof of statement (ii).}
Consider an interval attractor $([p, q], [\ell, r])$ in set $\Psi^{A}(t) \setminus \Psi^{A}(t+1)$. 
Then, $([p, q], [\ell, r]) \in \Psi_{h_{Q}} \setminus \Psi_{\run}$ and 
$L_{x} \preceq \reverse(T[p-1..\gamma-1]) \preceq L_{x^{\prime}}$ follow from the definition of the subset $\Psi^{A}(t)$ 
for the attractor position $\gamma$ of the interval attractor $([p, q], [\ell, r])$. 
Proposition~\ref{prop:Psi_A_Property}~\ref{enum:Psi_A_Property:3} shows that 
$T[\gamma..r+1] = R_{y+t-1}$ holds. 
Because of $([p, q], [\ell, r]) \in \Psi_{h_{Q}} \setminus \Psi_{\run}$, 
Proposition~\ref{prop:JA_correspondence_property} shows that 
set $\Psi_{h_{Q}} \cap \Psi_{\samp}$ contains an interval attractor $([p_{s}, q_{s}], [\ell_{s}, r_{s}])$ satisfying 
$T[p-1..\gamma-1] = T[p_{s}-1..\gamma_{s}-1]$ 
and $T[\gamma..r+1] = T[\gamma_{s}..r_{s}+1]$. 
This interval attractor $([p_{s}, q_{s}], [\ell_{s}, r_{s}])$ corresponds to the weighted point $(\reverse(T[p_{s}-1..\gamma_{s}-1]), T[\gamma_{s}..r_{s}+1], |\Psi_{\str}(T[p_{s}-1..r_{s}+1])|, T[p_{s}-1..r_{s}+1])$ in set $\mathcal{J}_{A}(h_{Q})$. 
$L_{x} \preceq \reverse(T[p_{s}-1..\gamma_{s}-1]) \preceq L_{x^{\prime}}$ follows from 
$L_{x} \preceq \reverse(T[p-1..\gamma-1]) \preceq L_{x^{\prime}}$ 
and $T[p-1..\gamma-1] = T[p_{s}-1..\gamma_{s}-1]$. 
$T[\gamma_{s}..r_{s}+1] = R_{y + t - 1}$ follows from 
$T[\gamma..r+1] = R_{y+t-1}$ and $T[\gamma..r+1] = T[\gamma_{s}..r_{s}+1]$. 
The existence of the weighted point $(\reverse(T[p_{s}-1..\gamma_{s}-1]), T[\gamma_{s}..r_{s}+1], |\Psi_{\str}(T[p_{s}-1..r_{s}+1])|, T[p_{s}-1..r_{s}+1])$ indicates that 
$\rangecount(\mathcal{J}_{A}(h_{Q}), L_{x}, L_{x^{\prime}}, R_{y + t - 1}, R_{y + t - 1}) \geq 1$ holds. 
Therefore, statement (ii) holds. 

\end{proof}

We prove Lemma~\ref{lem:GammaA_sub_property} using Proposition~\ref{prop:Psi_A_Property}, Proposition~\ref{prop:Set_IA_Property}, 
and Proposition~\ref{prop:GammaA_Grid}. 
Here, let $u_{\hat{y} - y + 2} = 0$, $\alpha_{\hat{y} - y + 2} = 0$, 
and $\Psi^{A}(\hat{y} - y + 2) = \emptyset$ for simplicity. 

\begin{proof}[Proof of Lemma~\ref{lem:GammaA_sub_property}~\ref{enum:GammaA_sub_property:1}]
    The following equation follows from the definition of range-count query. 
    \begin{equation*}
    \begin{split}
    \rangecount(\mathcal{J}_{A}(h_{Q}), & L_{x}, L_{x^{\prime}}, R_{1}, R_{\hat{y}}) \\
    & \geq \rangecount(\mathcal{J}_{A}(h_{Q}), L_{x}, L_{x^{\prime}}, R_{2}, R_{\hat{y}}) \\ 
    & \geq \cdots \\
    & \geq \rangecount(\mathcal{J}_{A}(h_{Q}), L_{x}, L_{x^{\prime}}, R_{\hat{y}}, R_{\hat{y}}). 
    \end{split}
    \end{equation*}
    Therefore, $\alpha_{1} \geq \alpha_{2} \geq \cdots \geq \alpha_{\hat{y} - y + 1}$ holds. 
\end{proof}
\begin{proof}[Proof of Lemma~\ref{lem:GammaA_sub_property}~\ref{enum:GammaA_sub_property:2}]
    Let $\mathcal{J}$ be the subset of the set $\mathcal{J}_{A}(h_{Q})$ such that 
    each weighted point 
    $(\reverse(T[p_{s}-1..\gamma_{s}-1]), T[\gamma_{s}..r_{s}+1], |\Psi_{\str}(T[p_{s}-1..r_{s}+1])|, T[p_{s}-1..r_{s}+1]) \in \mathcal{J}$ 
    satisfies $L_{x} \preceq \reverse(T[p_{s}-1..\gamma_{s}-1]) \preceq L_{x^{\prime}}$ and $T[\gamma_{s}..r_{s}+1] = R_{y + t^{\prime} - 1}$. 
    Then, the following equation holds from the definition of range-count query: 
\begin{equation}\label{eq:GammaA_sub_property:1}
    \begin{split}
    |\mathcal{J}| &= \rangecount(\mathcal{J}_{A}(h_{Q}), L_{x}, L_{x^{\prime}}, R_{y+t^{\prime}-1}, R_{\hat{y}}) \\
    &- \rangecount(\mathcal{J}_{A}(h_{Q}), L_{x}, L_{x^{\prime}}, R_{y+t^{\prime}}, R_{\hat{y}}).
    \end{split}
\end{equation}

    We prove Lemma~\ref{lem:GammaA_sub_property}~\ref{enum:GammaA_sub_property:2} by contradiction. 
    We assume that Lemma~\ref{lem:GammaA_sub_property}~\ref{enum:GammaA_sub_property:2} does not hold. 
    Then, $\mathcal{J} = \emptyset$ holds. 
    If $t^{\prime} < \hat{y} - y + 1$, 
    then $\alpha_{t^{\prime}} = \alpha_{t^{\prime}+1}$ follows from 
    Equation~\ref{eq:GammaA_sub_property:1}, 
    $\alpha_{t^{\prime}} = \rangecount(\mathcal{J}_{A}(h_{Q}), L_{x}, L_{x^{\prime}}, R_{y+t^{\prime}-1}, R_{\hat{y}})$, 
    and $\alpha_{t^{\prime}+1} = \rangecount(\mathcal{J}_{A}(h_{Q}), L_{x}, L_{x^{\prime}}, R_{y+t^{\prime}}, R_{\hat{y}})$. 
    $\alpha_{t} = \alpha_{t^{\prime}+1}$ follows from 
    $\alpha_{t} = \alpha_{t^{\prime}}$ and $\alpha_{t^{\prime}} = \alpha_{t^{\prime}+1}$. 
    On the other hand, $\alpha_{t} \neq \alpha_{t^{\prime}+1}$ follows from the definition of the integer $t^{\prime}$. 
    The two facts $\alpha_{t} = \alpha_{t^{\prime}+1}$ and $\alpha_{t} \neq \alpha_{t^{\prime}+1}$ yield a contradiction. 

    Otherwise (i.e., $t^{\prime} = \hat{y} - y + 1$), 
    $|\mathcal{J}| = \rangecount(\mathcal{J}_{A}(h_{Q}), L_{x}, L_{x^{\prime}}, R_{\hat{y}}, R_{\hat{y}})$ 
    follows from the definition of range-query count. 
    $\alpha_{t^{\prime}} \geq 1$ follows from $\alpha_{t} = \alpha_{t^{\prime}}$ and $\alpha_{t} \geq 1$. 
    $|\mathcal{J}| \geq 1$ follows from 
    (A) $|\mathcal{J}| = \rangecount(\mathcal{J}_{A}(h_{Q}), L_{x}, L_{x^{\prime}}, R_{\hat{y}}, R_{\hat{y}})$, 
    (B) $\rangecount(\mathcal{J}_{A}(h_{Q}), L_{x}, L_{x^{\prime}}, R_{\hat{y}}, R_{\hat{y}}) = \alpha_{t^{\prime}}$, 
    and (C) $\alpha_{t^{\prime}} \geq 1$. 
    The two facts $\mathcal{J} = \emptyset$ and $|\mathcal{J}| \geq 1$ yield a contradiction. 
    Therefore, Lemma~\ref{lem:GammaA_sub_property}~\ref{enum:GammaA_sub_property:2} must hold. 
\end{proof}
\begin{proof}[Proof of Lemma~\ref{lem:GammaA_sub_property}~\ref{enum:GammaA_sub_property:3}]
    The following three statements are used to prove Lemma~\ref{lem:GammaA_sub_property}~\ref{enum:GammaA_sub_property:3}: 
    \begin{enumerate}[label=\textbf{(\Alph*)}]
    \item $u_{t} = 1 \Leftarrow (\alpha_{t^{\prime}} \geq 1) \land (\RSCQ(\gamma_{s} - |[i, \gamma_{Q}-1]|, \gamma_{s} + |[\gamma_{Q}, j]| - 1) < b - \eta + 1)$;
    \item 
    If $u_{t} = 1$, then $t^{\prime} = t_{A}$ and $\alpha_{t^{\prime}} \geq 1$ for the smallest integer $t_{A}$ in set $[t, \hat{y} - y + 1]$ satisfying  
    $\Psi^{A}(t_{A}) \setminus \Psi^{A}(t_{A} + 1) \neq \emptyset$; 
    \item $u_{t} = 1 \Rightarrow (\alpha_{t^{\prime}} \geq 1) \land (\RSCQ(\gamma_{s} - |[i, \gamma_{Q}-1]|, \gamma_{s} + |[\gamma_{Q}, j]| - 1) < b - \eta + 1)$.    
    \end{enumerate}

    \textbf{Proof of statement (A).}
    Consider the interval attractor $([p_{s}, q_{s}], [\ell_{s}, r_{s}])$ corresponding to the weighted point $(\reverse(T[p_{s}-1..\gamma_{s}-1]), T[\gamma_{s}..r_{s}+1], |\Psi_{\str}(T[p_{s}-1..r_{s}+1])|, T[p_{s}-1..r_{s}+1])$. 
    This interval attractor is contained in set $\Psi^{A}(t^{\prime}) \setminus \Psi^{A}(t^{\prime}+1)$ 
    because $L_{x} \preceq \reverse(T[p_{s}-1..\gamma_{s}-1]) \preceq L_{x^{\prime}}$ and $T[\gamma_{s}..r_{s}+1] = R_{y + t^{\prime} - 1}$. 
    For the set $\mathcal{I}^{A}$ of integers introduced in Section~\ref{subsubsec:GammaA_proof1}, 
    $t^{\prime} \in \mathcal{I}^{A}$ holds 
    because $([p_{s}, q_{s}], [\ell_{s}, r_{s}]) \in \Psi^{A}(t^{\prime}) \setminus \Psi^{A}(t^{\prime}+1)$ 
    and $\RSCQ(\gamma_{s} - |[i, \gamma_{Q}-1]|, \gamma_{s} + |[\gamma_{Q}, j]| - 1) < b - \eta + 1$. 
    Because of $t^{\prime} \in \mathcal{I}^{A}$ and $t \leq t^{\prime}$, 
    Proposition~\ref{prop:Set_IA_Property}~\ref{enum:Set_IA_Property:3} shows that 
    $u_{t} = 1$ holds. 
    Therefore, statement (A) holds. 

    \textbf{Proof of statement (B).}    
    We show that the smallest integer $t_{A}$ exists. 
    Because of $u_{t} = 1$, 
    $\mathcal{F}_{\SA} \cap \mathcal{F}_{\suffix}(\Psi_{\CCP}(T[i..j]) \cap \Psi^{A}(t)) \neq \emptyset$ follows from the definition of the sequence $\Gamma_{A}$. 
    Because of $\mathcal{F}_{\SA} \cap \mathcal{F}_{\suffix}(\Psi_{\CCP}(T[i..j]) \cap \Psi^{A}(t)) \neq \emptyset$, 
    $\Psi^{A}(t) \neq \emptyset$ holds. 
    Proposition~\ref{prop:Psi_A_Property}~\ref{enum:Psi_A_Property:2} indicates that 
    $\Psi^{A}(t) = \bigcup_{\lambda = t}^{\hat{y} - y + 1} \Psi^{A}(\lambda) \setminus \Psi^{A}(\lambda+1)$ holds. 
    Therefore, the existence of the smallest integer $t_{A}$ follows from 
    $\Psi^{A}(t) \neq \emptyset$ and $\Psi^{A}(t) = \bigcup_{\lambda = t}^{\hat{y} - y + 1} \Psi^{A}(\lambda) \setminus \Psi^{A}(\lambda+1)$. 

    We prove $\alpha_{t} = \alpha_{\tau}$ for each integer $\tau \in [t, t_{A}]$. 
    Because of $t \leq \tau$, 
    the following equation holds from the definition of range-count query:     
\begin{equation}\label{eq:GammaA_sub_property:2}
    \begin{split}
    \rangecount(\mathcal{J}_{A}(h_{Q}), L_{x}, L_{x^{\prime}}, & R_{y+t-1}, R_{\hat{y}}) \\
    &= (\sum_{\lambda = t}^{\tau-1} \rangecount(\mathcal{J}_{A}(h_{Q}), L_{x}, L_{x^{\prime}}, R_{y+\lambda-1}, R_{y+\lambda-1})) \\
    &+ \rangecount(\mathcal{J}_{A}(h_{Q}), L_{x}, L_{x^{\prime}}, R_{y+\tau - 1}, R_{\hat{y}}).
    \end{split}
\end{equation}
    Proposition~\ref{prop:GammaA_Grid} indicates that 
    $\sum_{\lambda = t}^{\tau-1} \rangecount(\mathcal{J}_{A}(h_{Q}), L_{x}, L_{x^{\prime}}, R_{y+\lambda-1}, R_{y+\lambda-1}) = 0$ holds 
    because $\Psi^{A}(\lambda) \setminus \Psi^{A}(\lambda+1) = \emptyset$ holds for all integer $\lambda \in [t, t_{A}-1]$. 
    Therefore, $\alpha_{t} = \alpha_{\tau}$ follows from 
    Equation~\ref{eq:GammaA_sub_property:2}, 
    $\alpha_{t} = \rangecount(\mathcal{J}_{A}(h_{Q}), L_{x}, L_{x^{\prime}}, R_{y+t-1}, R_{\hat{y}})$, 
    and $\alpha_{\tau} = \rangecount(\mathcal{J}_{A}(h_{Q}), L_{x}, L_{x^{\prime}}, R_{y+\tau-1}, R_{\hat{y}})$. 

    Next, we prove $\alpha_{t_{A}} > \alpha_{t_{A}+1}$.     
    The following equation holds from the definition of range-count query:     
\begin{equation}\label{eq:GammaA_sub_property:3}
    \begin{split}
    \rangecount(\mathcal{J}_{A}(h_{Q}), L_{x}, L_{x^{\prime}}, & R_{y+t_{A}-1}, R_{\hat{y}}) \\
    &= \rangecount(\mathcal{J}_{A}(h_{Q}), L_{x}, L_{x^{\prime}}, R_{y+t_{A} -1}, R_{y+t_{A}-1}) \\
    &+ \rangecount(\mathcal{J}_{A}(h_{Q}), L_{x}, L_{x^{\prime}}, R_{y+t_{A}}, R_{\hat{y}}).
    \end{split}
\end{equation}
    Because of $\Psi^{A}(t_{A}) \setminus \Psi^{A}(t_{A}+1) \neq \emptyset$, 
    Proposition~\ref{prop:GammaA_Grid} shows that 
    $\rangecount(\mathcal{J}_{A}(h_{Q}), L_{x}, L_{x^{\prime}}, R_{y+t_{A} -1}$, $R_{y+t_{A}-1}) \geq 1$ holds. 

    If $t_{A} < \hat{y} - y + 1$, then 
    $\alpha_{t_{A}} > \alpha_{t_{A}+1}$ follows from 
    Equation~\ref{eq:GammaA_sub_property:3}, 
    $\alpha_{t_{A}} = \rangecount(\mathcal{J}_{A}(h_{Q}), L_{x}, L_{x^{\prime}}$, $R_{y+t_{A}-1}, R_{\hat{y}})$, 
    and $\alpha_{t_{A}+1} = \rangecount(\mathcal{J}_{A}(h_{Q}), L_{x}, L_{x^{\prime}}, R_{y+t_{A}}, R_{\hat{y}})$. 
    Otherwise (i.e., $t_{A} = \hat{y} - y + 1$)
    $\alpha_{t_{A}+1} = 0$ holds. 
    $\alpha_{t_{A}} \geq 1$ follows from Equation~\ref{eq:GammaA_sub_property:3} and 
    $\alpha_{t_{A}} = \rangecount(\mathcal{J}_{A}(h_{Q}), L_{x}, L_{x^{\prime}}, R_{y+t_{A}-1}, R_{\hat{y}})$.
    Therefore, $\alpha_{t_{A}} > \alpha_{t_{A}+1}$ holds. 

    Finally, $t^{\prime} = t_{A}$ follows from 
    $\alpha_{t} = \alpha_{t+1} = \cdots = \alpha_{t_{A}}$ 
    and $\alpha_{t_{A}} > \alpha_{t_{A}+1}$. 
    $\alpha_{t^{\prime}} \geq 1$ follows from 
    $t^{\prime} = t_{A}$, $\alpha_{t_{A}} > \alpha_{t_{A}+1}$, and $\alpha_{t_{A}+1} \geq 0$. 
    
    \textbf{Proof of statement (C).}    
    Because of $u_{t} = 1$, 
    $\mathcal{F}_{\SA} \cap \mathcal{F}_{\suffix}(\Psi_{\CCP}(T[i..j]) \cap \Psi^{A}(t)) \neq \emptyset$ follows from the definition of the sequence $\Gamma_{A}$. 
    Because of $\mathcal{F}_{\SA} \cap \mathcal{F}_{\suffix}(\Psi_{\CCP}(T[i..j]) \cap \Psi^{A}(t)) \neq \emptyset$, 
    the set $\Psi_{\CCP}(T[i..j]) \cap \Psi^{A}(t)$ contains an interval attractor $([p_{B}, q_{B}], [\ell_{B}, r_{B}])$ satisfying 
    $T[\gamma_{B} - |[i, \gamma_{Q}-1]|..\gamma_{B}-1] \cdot T[\gamma_{B}..r_{B} + 1] \in \mathcal{F}_{\SA} \cap \mathcal{F}_{\suffix}(\Psi_{\CCP}(T[i..j]) \cap \Psi^{A}(t))$ for the attractor position $\gamma_{B}$ of the interval attractor $([p_{B}, q_{B}], [\ell_{B}, r_{B}])$. 
    Similar to Equation~\ref{eq:Set_A_Property:1}, 
    the following equation follows from Proposition~\ref{prop:Psi_A_Property}~\ref{enum:Psi_A_Property:2} and $\Psi^{A}(y^{\prime} - y + 2) = \emptyset$:
    \begin{equation*}
    \Psi^{A}(t) = \bigcup_{\lambda = t}^{\hat{y} - y + 1} \Psi^{A}(\lambda) \setminus \Psi^{A}(\lambda+1).
    \end{equation*}
    Therefore, there exists an integer $t_{B} \in [t, \hat{y} - y + 1]$ satisfying 
    $([p_{B}, q_{B}], [\ell_{B}, r_{B}]) \in \Psi^{A}(t_{B}) \setminus \Psi^{A}(t_{B}+1)$. 
    Here, $t_{A} \leq t_{B}$ holds for the smallest integer $t_{A}$ in set $[t, \hat{y} - y + 1]$ satisfying  
    $\Psi^{A}(t_{A}) \setminus \Psi^{A}(t_{A} + 1) \neq \emptyset$. 
    
    We prove $([p_{s}, q_{s}], [\ell_{s}, r_{s}]) \in \Psi^{A}(t^{\prime}) \setminus \Psi^{A}(t^{\prime}+1)$. 
    Because of $u_{t} = 1$, 
    statement (B) shows that $t^{\prime} = t_{A}$ and $\alpha_{t^{\prime}} \geq 1$ holds. 
    Lemma~\ref{lem:GammaA_sub_property}~\ref{enum:GammaA_sub_property:2} shows that 
    the weighted point 
    $(\reverse(T[p_{s}-1..\gamma_{s}-1]), T[\gamma_{s}..r_{s}+1], |\Psi_{\str}(T[p_{s}-1..r_{s}+1])|, T[p_{s}-1..r_{s}+1])$ satisfies  
    $L_{x} \preceq \reverse(T[p_{s}-1..\gamma_{s}-1]) \preceq L_{x^{\prime}}$ and $T[\gamma_{s}..r_{s}+1] = R_{y + t^{\prime} - 1}$. 
    Because of $([p_{s}, q_{s}], [\ell_{s}, r_{s}]) \in \Psi_{h_{Q}} \cap \Psi_{\samp}$, 
    Lemma~\ref{lem:samp_basic_property}~\ref{enum:samp_basic_property:3} shows that 
    $([p_{s}, q_{s}], [\ell_{s}, r_{s}]) \in \Psi_{h_{Q}} \setminus \Psi_{\run}$ holds. 
    We proved $([p_{s}, q_{s}], [\ell_{s}, r_{s}]) \in \Psi_{h_{Q}} \setminus \Psi_{\run}$, 
    $L_{x} \preceq \reverse(T[p_{s}-1..\gamma_{s}-1]) \preceq L_{x^{\prime}}$, 
    and $T[\gamma_{s}..r_{s}+1] = R_{y + t^{\prime} - 1}$. 
    Therefore, $([p_{s}, q_{s}], [\ell_{s}, r_{s}]) \in \Psi^{A}(t^{\prime}) \setminus \Psi^{A}(t^{\prime}+1)$ follows from 
    the definitions of the two subsets $\Psi^{A}(t^{\prime})$ and $\Psi^{A}(t^{\prime}+1)$.
    
    We prove $T[\gamma_{s} - |[i, \gamma_{Q}-1]|..\gamma_{s}-1] \cdot T[\gamma_{s}..r_{s} + 1] \preceq T[\gamma_{B} - |[i, \gamma_{Q}-1]|..\gamma_{B}-1] \cdot T[\gamma_{B}..r_{B} + 1]$. 
    Because of $([p_{s}, q_{s}], [\ell_{s}, r_{s}]) \in \Psi^{A}(t^{\prime}) \setminus \Psi^{A}(t^{\prime}+1)$, 
    Proposition~\ref{prop:Psi_A_Property}~\ref{enum:Psi_A_Property:3} shows that 
    $T[\gamma_{s} - |[i, \gamma_{Q}-1]|..\gamma_{s}-1] \cdot T[\gamma_{s}..r_{s} + 1] = T[i..\gamma_{Q}-1] \cdot R_{y + t^{\prime} - 1}$ holds. 
    Similarly, 
    Proposition~\ref{prop:Psi_A_Property}~\ref{enum:Psi_A_Property:3} shows that 
    $T[\gamma_{B} - |[i, \gamma_{Q}-1]|..\gamma_{B}-1] \cdot T[\gamma_{B}..r_{B} + 1] = T[i..\gamma_{Q}-1] \cdot R_{y + t_{B} - 1}$ holds. 
    $R_{y + t^{\prime} - 1} \preceq R_{y + t_{B} - 1}$ follows from 
    $t^{\prime} \leq t_{B}$ and $R_{1} \prec R_{2} \prec \cdots \prec R_{d^{\prime}}$. 
    Therefore, $T[\gamma_{s} - |[i, \gamma_{Q}-1]|..\gamma_{s}-1] \cdot T[\gamma_{s}..r_{s} + 1] \preceq T[\gamma_{B} - |[i, \gamma_{Q}-1]|..\gamma_{B}-1] \cdot T[\gamma_{B}..r_{B} + 1]$ follows from 
    (1) $T[\gamma_{s} - |[i, \gamma_{Q}-1]|..\gamma_{s}-1] \cdot T[\gamma_{s}..r_{s} + 1] = T[i..\gamma_{Q}-1] \cdot R_{y + t^{\prime} - 1}$, 
    (2) $T[\gamma_{B} - |[i, \gamma_{Q}-1]|..\gamma_{B}-1] \cdot T[\gamma_{B}..r_{B} + 1] = T[i..\gamma_{Q}-1] \cdot R_{y + t_{B} - 1}$, 
    and (3) $R_{y + t^{\prime} - 1} \preceq R_{y + t_{B} - 1}$. 

    We prove $\RSCQ(\gamma_{s} - |[i, \gamma_{Q}-1]|, \gamma_{s} + |[\gamma_{Q}, j]| - 1) < b - \eta + 1$. 
    Because of $([p_{s}, q_{s}], [\ell_{s}, r_{s}]) \in \Psi^{A}(t^{\prime})$, 
    $([p_{s}, q_{s}], [\ell_{s}, r_{s}]) \in \Psi_{\CCP}(T[i..])$ follows from 
    Proposition~\ref{prop:Psi_A_Property}~\ref{enum:Psi_A_Property:1} and Proposition~\ref{prop:Psi_A_Property}~\ref{enum:Psi_A_Property:2}.     
    Lemma~\ref{lem:F_suffix_basic_property}~\ref{enum:F_suffix_basic_property:6} shows that 
    $T[\gamma_{s} - |[i, \gamma_{Q}-1]|..\gamma_{s}-1] \cdot T[\gamma_{s}..r_{s} + 1] \in \mathcal{F}_{\SA}$ holds 
    because 
    (a) $([p_{s}, q_{s}], [\ell_{s}, r_{s}]), ([p_{B}, q_{B}], [\ell_{B}, r_{B}]) \in \Psi_{\CCP}(T[i..])$, 
    (b) $T[\gamma_{s} - |[i, \gamma_{Q}-1]|..\gamma_{s}-1] \cdot T[\gamma_{s}..r_{s} + 1] \preceq T[\gamma_{B} - |[i, \gamma_{Q}-1]|..\gamma_{B}-1] \cdot T[\gamma_{B}..r_{B} + 1]$, 
    and (c) $T[\gamma_{B} - |[i, \gamma_{Q}-1]|..\gamma_{B}-1] \cdot T[\gamma_{B}..r_{B} + 1] \in \mathcal{F}_{\SA}$.  
    $\RSCQ(\gamma_{s} - |[i, \gamma_{Q}-1]|, \gamma_{s} + |[\gamma_{Q}, j]| - 1) < b - \eta + 1$ 
    follows from Lemma~\ref{lem:F_SA_formula} and $T[\gamma_{s} - |[i, \gamma_{Q}-1]|..\gamma_{s}-1] \cdot T[\gamma_{s}..r_{s} + 1] \in \mathcal{F}_{\SA}$. 
    
    Finally, $u_{t} = 1 \Rightarrow (\alpha_{t^{\prime}} \geq 1) \land (\RSCQ(\gamma_{s} - |[i, \gamma_{Q}-1]|, \gamma_{s} + |[\gamma_{Q}, j]| - 1) < z)$ follows from statement (B) and $\RSCQ(\gamma_{s} - |[i, \gamma_{Q}-1]|, \gamma_{s} + |[\gamma_{Q}, j]| - 1) < z$. 

    \textbf{Proof of Lemma~\ref{lem:GammaA_sub_property}~\ref{enum:GammaA_sub_property:3}.}
    Lemma~\ref{lem:GammaA_sub_property}~\ref{enum:GammaA_sub_property:3} follows from statement (A) and statement (C).
\end{proof}

\subsubsection{Algorithm}\label{subsubsec:gamma_A_algorithm}
We prove Lemma~\ref{lem:GammaA_algorithm}, i.e., 
we show that subquery $\RSSQA(T[i..j], b)$ can be answered 
in $O(H^{2} \log^{2} n + \log^{6} n)$ time using the data structures for RSC query 
and interval $[i, j]$. 

\paragraph{Computation of each integer $u_{t}$ in sequence $\Gamma_{A}$.}
Consider the non-increasing sequence $\Gamma_{A} = u_{1}, u_{2}, \ldots, u_{y^{\prime} - y + 1} \in \{ 0, 1 \}$ 
for the two integers $y$ and $y^{\prime}$ introduced in Section~\ref{subsec:GammaA}. 
We show that 
the $t$-th integer $u_{t}$ can be computed in $O(H^{2} \log n + \log^{4} n)$ time for a given integer $t \in [1, y^{\prime} - y + 1]$. 

We leverage sequence $\Gamma_{A, \sub} = \alpha_{1}, \alpha_{2}, \ldots, \alpha_{y^{\prime} - y + 1}$ for computing the $t$-th integer $u_{t}$. 
Let $t^{\prime}$ be the largest integer in set $[t, y^{\prime} - y + 1]$ satisfying $\alpha_{t} = \alpha_{t^{\prime}}$. 
If $\alpha_{t^{\prime}} \geq 1$, 
then Lemma~\ref{lem:GammaA_sub_property}~\ref{enum:GammaA_sub_property:2} shows that 
set $\mathcal{J}_{A}(h_{Q})$ contains a weighted point 
$(\reverse(T[p_{s}-1..\gamma_{s}-1]), T[\gamma_{s}..r_{s}+1], |\Psi_{\str}(T[p_{s}-1..r_{s}+1])|, T[p_{s}-1..r_{s}+1])$ satisfying 
$L_{x} \preceq \reverse(T[p_{s}-1..\gamma_{s}-1]) \preceq L_{x^{\prime}}$ and $T[\gamma_{s}..r_{s}+1] = R_{y + t^{\prime} - 1}$. 
Here, $x$ and $x^{\prime}$ are the two integers introduced in Section~\ref{subsec:GammaA}. 
Lemma~\ref{lem:GammaA_sub_property}~\ref{enum:GammaA_sub_property:3} shows that 
$u_{t} = 1 \Leftrightarrow (\alpha_{t^{\prime}} \geq 1) \land (\RSCQ(\gamma_{s} - |[i, \gamma_{Q}-1]|, \gamma_{s} + |[\gamma_{Q}, j]| - 1) < b - \eta + 1)$ holds. 
Therefore, the $t$-th integer $u_{t}$ can be computed using the $t^{\prime}$-th integer $\alpha_{t^{\prime}}$ of the sequence $\Gamma_{A, \sub}$ and RSC query $\RSCQ(\gamma_{s} - |[i, \gamma_{Q}-1]|, \gamma_{s} + |[\gamma_{Q}, j]| - 1)$. 

The algorithm computing the $t$-th integer $u_{t}$ consists of five phases. 
In the first phase, 
we compute interval attractor $([p_{Q}, q_{Q}], [\ell_{Q}, r_{Q}])$, 
its level $h_{Q}$, and its attractor position $\gamma_{Q}$. 
The interval attractor $([p_{Q}, q_{Q}], [\ell_{Q}, r_{Q}])$ can be obtained by capture query $\CAPQ([i, j])$. 
The level $h_{Q}$ and attractor position $\gamma_{Q}$ can be obtained by 
level-query $\levelQ(([p_{Q}, q_{Q}], [\ell_{Q}, r_{Q}]))$ and attractor position query $\attrQ(([p_{Q}, q_{Q}], [\ell_{Q}, r_{Q}]))$, 
respectively. 
Therefore, the first phase takes $O(H^{2} \log n)$ time. 

In the second phase, 
we compute the four integers $x, x^{\prime}, y$, and $y^{\prime}$. 
The two integers $x$ and $x^{\prime}$ are obtained by binary search on the $d$ strings of the ordered set $\mathcal{X}_{A}(h_{Q})$. 
This binary search can be executed in $O((H^{2} + \log n) \log d)$ time using Lemma~\ref{lem:JA_X_queries}~\ref{enum:JA_X_queries:3}. 
Similarly, 
the two integers $y$ and $y^{\prime}$ are obtained by binary search on the $d^{\prime}$ strings 
of the ordered set $\mathcal{Y}_{A}(h_{Q})$. 
This binary search can be executed in $O((H^{2} + \log n)\log d^{\prime})$ time 
using Lemma~\ref{lem:JA_Y_queries}~\ref{enum:JA_X_queries:3}. 
$d, d^{\prime} = O(n^{2})$ follows from Lemma~\ref{lem:JA_size}~\ref{enum:JA_size:3}. 
Therefore, the second phase takes $O(H^{2} \log n + \log^{2} n)$ time.

In the third phase, 
we find the integer $t^{\prime}$ by binary search on the sequence $\Gamma_{A, \sub}$. 
This binary search can be executed by computing $O(\log (y^{\prime} - y + 1))$ integers of the sequence $\Gamma_{A, \sub}$ 
because Lemma~\ref{lem:GammaA_sub_property}~\ref{enum:GammaA_sub_property:1} shows that 
the sequence $\Gamma_{A, \sub}$ is non-increasing. 
Each integer of the sequence $\Gamma_{A, \sub}$ can be computed by one range-count query on the set $\mathcal{J}_{A}(h_{Q})$ of weighted points. 
This range-count query takes $O(\log^{2} |\mathcal{J}_{A}(h_{Q})|)$ time using the dynamic data structures of Section~\ref{subsubsec:JA_subquery_ds}. 
$y^{\prime} - y + 1 = O(n^{2})$ because $y^{\prime} - y + 1 \leq d^{\prime}$ and $d^{\prime} = O(n^{2})$. 
$|\mathcal{J}_{A}(h_{Q})| = O(n^{2})$ follows from Lemma~\ref{lem:JA_size}~\ref{enum:JA_size:3}. 
Therefore, the third phase takes $O(\log^{3} n)$ time. 

The integer $\alpha_{t^{\prime}}$ is obtained by the binary search of the third phase. 
The fourth phase is executed if $\alpha_{t^{\prime}} \geq 1$ holds. 
In the fourth phase, we find the interval attractor $([p_{s}, q_{s}], [\ell_{s}, r_{s}])$ corresponding to 
the weighted point $(\reverse(T[p_{s}-1..\gamma_{s}-1]), T[\gamma_{s}..r_{s}+1], |\Psi_{\str}(T[p_{s}-1..r_{s}+1])|, T[p_{s}-1..r_{s}+1])$. 
This interval attractor can be found in $O(H^{2} \log n + \log^{2} n)$ time 
using Lemma~\ref{lem:JA_Y_queries}~\ref{enum:JA_Y_queries:4}. 

In the fifth phase, 
we compute the $t$-th integer $u_{t}$ of sequence $\Gamma_{A}$ by verifying 
$\alpha_{t^{\prime}} \geq 1$ and $\RSCQ(\gamma_{s} - |[i, \gamma_{Q}-1]|, \gamma_{s} + |[\gamma_{Q}, j]| - 1) < b - \eta + 1$. 
The attractor position $\gamma_{s}$ is obtained in $O(H^{2})$ time 
by attractor position query $\attrQ(([p_{s}, q_{s}], [\ell_{s}, r_{s}]))$. 
The RSC query $\RSCQ(\gamma_{s} - |[i, \gamma_{Q}-1]|, \gamma_{s} + |[\gamma_{Q}, j]| - 1)$ takes $O(H^{2} \log n + \log^{4} n)$ time. 
Therefore, the fifth phase takes $O(H^{2} \log n + \log^{4} n)$ time. 

Finally, the algorithm computing the $t$-th integer $u_{t}$ takes $O(H^{2} \log n + \log^{4} n)$ time in total. 

\paragraph{Computation of the largest integer $\kappa$.}
Consider the largest integer $\kappa$ in set $[1, y^{\prime} - y + 1]$ satisfying $u_{\kappa} = 1$. 
We find the largest integer $\kappa$ by binary search on the non-increasing sequence $\Gamma_{A}$. 
This binary search can be executed by computing $O(\log (y^{\prime} - y + 1))$ integers of the sequence $\Gamma_{A}$. 
Each integer of the sequence $\Gamma_{A}$ can be computed in $O(H^{2} \log n + \log^{4} n)$ time. 
The two integers $y$ and $y^{\prime}$ can be obtained by the algorithm computing each integer $u_{t}$. 
Therefore, this binary search takes $O((H^{2} \log n + \log^{4} n) \log (y^{\prime} - y + 1))$ time 
(i.e., $O(H^{2} \log^{2} n + \log^{5} n)$ time).

\paragraph{Verification of $\mathcal{F}_{\SA} \cap \mathcal{F}_{\suffix}(\Psi_{\CCP}(T[i..j]) \setminus \Psi_{\run}) = \emptyset$.}
We verify whether $\mathcal{F}_{\SA} \cap \mathcal{F}_{\suffix}(\Psi_{\CCP}(T[i..j]) \setminus \Psi_{\run}) = \emptyset$ or not for answering subquery $\RSSQA(T[i..j], b)$. 
Lemma~\ref{lem:GammaA_property}~\ref{enum:GammaA_property:1} indicates that   
$\mathcal{F}_{\SA} \cap \mathcal{F}_{\suffix}(\Psi_{\CCP}(T[i..j]) \setminus \Psi_{\run}) \neq \emptyset$ holds if and only if 
the largest integer $\kappa$ exists. 
We can verify whether the largest integer $\kappa$ exists or not by the algorithm computing the largest integer $\kappa$. 
The verification of $\mathcal{F}_{\SA} \cap \mathcal{F}_{\suffix}(\Psi_{\CCP}(T[i..j]) \setminus \Psi_{\run}) = \emptyset$ can be executed in $O(H^{2} \log^{2} n + \log^{5} n)$ time. 

\paragraph{Algorithm for $\RSSQA(T[i..j], b)$.}
The algorithm for Subquery $\RSSQA(T[i..j], b)$ returns 
the lexicographically largest string $F$ in 
set $\mathcal{F}_{\SA} \cap \mathcal{F}_{\suffix}(\Psi_{\CCP}(T[i..j]) \setminus \Psi_{\run})$. 
This algorithm is executed only if $\mathcal{F}_{\SA} \cap \mathcal{F}_{\suffix}(\Psi_{\CCP}(T[i..j]) \setminus \Psi_{\run}) \neq \emptyset$ holds. 
Here, the verification of $\mathcal{F}_{\SA} \cap \mathcal{F}_{\suffix}(\Psi_{\CCP}(T[i..j]) \setminus \Psi_{\run}) = \emptyset$ can be executed in $O(H^{2} \log^{2} n + \log^{5} n)$ time. 
Since $\mathcal{F}_{\SA} \cap \mathcal{F}_{\suffix}(\Psi_{\CCP}(T[i..j]) \setminus \Psi_{\run}) \neq \emptyset$, 
Lemma~\ref{lem:GammaA_property}~\ref{enum:GammaA_property:1} shows that the largest integer $\kappa$ exists. 

The algorithm for $\RSSQA(T[i..j], b)$ leverages the largest integer $\kappa$. 
Lemma~\ref{lem:GammaA_property}~\ref{enum:GammaA_property:1} shows that 
$F = T[i..\gamma_{Q} - 1] \cdot R_{y + \kappa - 1}$ holds. 
Lemma~\ref{lem:GammaA_property}~\ref{enum:GammaA_property:2} shows that 
set $\mathcal{J}_{A}(h_{Q})$ contains 
a weighted point $(\reverse(T[p_{s}-1..\gamma_{s}-1]), T[\gamma_{s}..r_{s}+1], |\Psi_{\str}(T[p_{s}-1..r_{s}+1])|, T[p_{s}-1..r_{s}+1])$ 
satisfying $L_{x} \preceq \reverse(T[p_{s}-1..\gamma_{s}-1] \preceq L_{x^{\prime}}$ and $T[\gamma_{s}..r_{s}+1] = R_{y + \kappa - 1}$. 
Lemma~\ref{lem:GammaA_property}~\ref{enum:GammaA_property:3} shows that 
$T[\gamma_{s} - |[i, \gamma_{Q}]| + 1..r_{s} + 1] = T[i..\gamma_{Q} - 1] \cdot R_{y + \kappa - 1}$ holds. 
Therefore, we can return string $T[\gamma_{s} - |[i, \gamma_{Q}]| + 1..r_{s} + 1]$ as the answer to subquery $\RSSQA(T[i..j], b)$. 

The algorithm for $\RSSQA(T[i..j], b)$ consists of four phases.
In the first phase, 
we compute interval attractor $([p_{Q}, q_{Q}], [\ell_{Q}, r_{Q}])$, 
its level $h_{Q}$, and its attractor position $\gamma_{Q}$. 
This phase can be executed in $O(H^{2} \log n)$ time by the first phase of the algorithm computing an integer of sequence $\Gamma_{A}$. 

In the second phase, 
we compute the five integers $x, x^{\prime}, y, y^{\prime}$, and $\kappa$.
The four integers $x, x^{\prime}, y$, and $y^{\prime}$ can be computed 
in $O(H^{2} \log n + \log^{2} n)$ time by 
the second phase of the algorithm computing an integer of sequence $\Gamma_{A}$. 
The computation of the largest integer $\kappa$ takes $O(H^{2} \log^{2} n + \log^{5} n)$ time. 
Therefore, the second phase takes $O(H^{2} \log^{2} n + \log^{5} n)$ time.

In the third phase, 
we compute the interval attractor $([p_{s}, q_{s}], [\ell_{s}, r_{s}])$ corresponding to 
the weighted point $(\reverse(T[p_{s}-1..\gamma_{s}-1]), T[\gamma_{s}..r_{s}+1], |\Psi_{\str}(T[p_{s}-1..r_{s}+1])|, T[p_{s}-1..r_{s}+1])$. 
This interval attractor can be found in $O(H^{2} \log n + \log^{2} n)$ time 
using Lemma~\ref{lem:JA_Y_queries}~\ref{enum:JA_Y_queries:4}. 

In the fourth phase, 
we return string $T[\gamma_{s} - |[i, \gamma_{Q}]| + 1..r_{s} + 1]$ as the answer to subquery $\RSSQA(T[i..j], b)$. 
Here, the string is represented as interval $[\gamma_{s} - |[i, \gamma_{Q}]| + 1, r_{s} + 1]$. 
The attractor position $\gamma_{s}$ can be obtained by attractor position query $\attrQ(([p_{s}, q_{s}], [\ell_{s}, r_{s}]))$. 
Therefore, the fourth phase takes $O(H^{2})$ time. 

The four phases take $O(H^{2} \log^{2} n + \log^{5} n)$ time in total. 
Therefore, Lemma~\ref{lem:GammaA_algorithm} holds.

\subsection{Subquery \texorpdfstring{$\RSSQB(T[i..j], b)$}{RSSB(T[i..j], b)}}\label{subsec:GammaB}
The goal of this subsection is to answer subquery $\RSSQB(T[i..j], b)$ 
under the assumption that $|[\gamma_{Q}, j]| \leq 1 + \sum_{w = 1}^{h_{Q}+3} \lfloor \mu(w) \rfloor$ holds. 
The following lemma states the summary of this subsection. 

\begin{lemma}\label{lem:GammaB_algorithm}
We assume that $|[\gamma_{Q}, j]| \leq 1 + \sum_{w = 1}^{h_{Q}+3} \lfloor \mu(w) \rfloor$ holds 
for RSS query $\RSSQ(T[i..j], b)$. 
We can answer subquery $\RSSQB(T[i..j], b)$ in $O(H^{2} \log^{2} n + \log^{5} n)$ time 
using (A) the data structures for RSC query, 
(B) interval $[i, j]$, 
and (C) the starting position $\eta$ of the sa-interval $[\eta, \eta^{\prime}]$ of $T[i..j]$.  
\end{lemma}
\begin{proof}
See Section~\ref{subsubsec:gamma_B_algorithm}.
\end{proof}

For the level $h_{Q}$ of interval attractor $([p_{Q}, q_{Q}], [\ell_{Q}, r_{Q}])$, 
we use set $\Psi_{h_{Q}} \cap \Psi_{\source} \cap \Psi_{\samp}$ to explain the idea behind answering subquery $\RSSQB(T[i..j], b)$. 
Let $([p_{1}, q_{1}], [\ell_{1}, r_{1}]), ([p_{2}, q_{2}], [\ell_{2}, r_{2}])$, 
$\ldots$, $([p_{k}, q_{k}], [\ell_{k}, r_{k}])$ be the interval attractors in set $\Psi_{h_{Q}} \cap \Psi_{\source} \cap \Psi_{\samp}$. 
For each interval attractor $([p_{s}, q_{s}], [\ell_{s}, r_{s}]) \in \Psi_{h} \cap \Psi_{\source} \cap \Psi_{\samp}$, 
Lemma~\ref{lem:mRecover_basic_property} shows that 
set $f_{\recover}(([p_{s}, q_{s}], [\ell_{s}, r_{s}])) \cap \Psi_{\mRecover}$ consists of an interval attractor 
$([\hat{p}_{s}, \hat{q}_{s}], [\hat{\ell}_{s}, \hat{r}_{s}])$ (i.e., $f_{\recover}(([p_{s}, q_{s}], [\ell_{s}, r_{s}])) \cap \Psi_{\mRecover} = \{ ([\hat{p}_{s}, \hat{q}_{s}], [\hat{\ell}_{s}, \hat{r}_{s}]) \}$). 
Let $\hat{\gamma}_{s}$ be the attractor position of the interval attractor $([\hat{p}_{s}, \hat{q}_{s}], [\hat{\ell}_{s}, \hat{r}_{s}])$.

We leverage the set $\mathcal{J}_{B}(h_{Q})$ of weighted points on grid $(\mathcal{X}_{B}(h_{Q}), \mathcal{Y}_{B}(h_{Q}))$ introduced in 
Section~\ref{subsec:RSC_comp_B}. 
The summary of the set $\mathcal{J}_{B}(h_{Q})$ and $(\mathcal{X}_{B}(h_{Q}), \mathcal{Y}_{B}(h_{Q}))$ is as follows 
(see Section~\ref{subsec:RSC_comp_B} for the details of the set and grid):
\begin{itemize}
    \item the ordered set $\mathcal{X}_{B}(h_{Q})$ consists of $d$ strings $L_{1}, L_{2}, \ldots, L_{d}$ ($L_{1} \prec L_{2} \prec \cdots \prec L_{d}$);
    \item the ordered set $\mathcal{Y}_{B}(h_{Q})$ consists of $d^{\prime}$ strings $R_{1}, R_{2}, \ldots, R_{d^{\prime}}$ ($R_{1} \prec R_{2} \prec \ldots \prec R_{d^{\prime}}$);
    \item the set $\mathcal{J}_{B}(h_{Q})$ contains the weighted point 
    $(\reverse(T[\hat{p}_{s}-1..\hat{\gamma}_{s}-1]), T[\hat{\gamma}_{s}..\hat{r}_{s}+1], |\Psi_{\str}(T[p_{s}-1..r_{s}+1])| |f_{\recover}(([p_{s}, q_{s}], [\ell_{s}, r_{s}]))|, T[p_{s}-1..r_{s}+1])$ 
    corresponding to each interval attractor $([p_{s}, q_{s}], [\ell_{s}, r_{s}]) \in \Psi_{h_{Q}} \cap \Psi_{\source} \cap \Psi_{\samp}$. 
\end{itemize}
For this subsection, 
we define four integers $x, x^{\prime}, y$, and $y^{\prime}$ as follows: 
\begin{itemize}
    \item $x = \min \{ s \in [1, d] \mid \reverse(T[i..\gamma_{Q}-1]) \prec L_{s} \}$;
    \item $x^{\prime} = \max \{ s \in [1, d] \mid L_{s} \prec \reverse(\# \cdot T[i..\gamma_{Q}-1]) \}$;
    \item $y = \min \{ s \in [1, d^{\prime}] \mid T[\gamma_{Q}..j] \prec R_{s} \}$;
    \item $y^{\prime} = \max \{ s \in [1, d^{\prime}] \mid R_{s} \prec (T[\gamma_{Q}..j] \cdot \#) \}$. 
\end{itemize}

In the next paragraphs, 
we introduce $(y^{\prime} - y + 1)$ subsets of set $\Psi_{\RR}$ 
and two sequences of integers to explain the relationship between the set $\mathcal{J}_{B}(h_{Q})$ and subquery $\RSSQB(T[i..j], b)$. 

\paragraph{Subset $\Psi^{B}(t)$.}
For each integer $t \in [1, y^{\prime} - y + 1]$, 
subset $\Psi^{B}(t) \subseteq \Psi_{\RR}$ consists of interval attractors such that 
each interval attractor $([p, q], [\ell, r]) \in \Psi^{B}(t)$ satisfies the following three conditions: 
\begin{itemize}
    \item there exists an integer $s \in [1, k]$ satisfying $([p, q], [\ell, r]) = ([\hat{p}_{s}, \hat{q}_{s}], [\hat{\ell}_{s}, \hat{r}_{s}])$; 
    \item $L_{x} \preceq \reverse(T[p-1..\gamma-1]) \preceq L_{x^{\prime}}$ for the attractor position $\gamma$ of the interval attractor $([p, q], [\ell, r])$;
    \item $R_{y + t - 1} \preceq T[\gamma..r+1] \preceq R_{y^{\prime}}$.
\end{itemize}
Formally, the subset $\Psi^{B}(t)$ is defined as follows: 
\begin{equation*}
    \begin{split}
    \Psi^{B}(t) &= \{ ([\hat{p}_{s}, \hat{q}_{s}], [\hat{\ell}_{s}, \hat{r}_{s}]) \mid s \in [1, k] \text{ s.t. } \\
    &L_{x} \preceq \reverse(T[\hat{p}_{s}-1..\hat{\gamma}_{s}-1]) \preceq L_{x^{\prime}} \text{ and } R_{y + t - 1} \preceq T[\hat{\gamma}_{s}..\hat{r}_{s}+1] \preceq R_{y^{\prime}} \}.
    \end{split}
\end{equation*}

\paragraph{Two sequences $\Gamma_{B}$ and $\Gamma_{B, \sub}$.}
Sequence $\Gamma_{B}$ consists of $(y^{\prime} - y + 1)$ integers $u_{1}, u_{2}, \ldots, u_{y^{\prime} - y + 1} \in \{ 0, 1 \}$. 
Here, each integer $u_{t}$ is $1$ if $\mathcal{F}_{\SA} \cap \mathcal{F}_{\suffix}(\Psi_{\CCP}(T[i..j]) \cap \Psi^{B}(t)) \neq \emptyset$; 
otherwise $u_{t}$ is $0$. 

The following lemma states four properties of sequence $\Gamma_{B}$. 

\begin{lemma}\label{lem:GammaB_property}
Consider the case that $|[\gamma_{Q}, j]| \leq 1 + \sum_{w = 1}^{h_{Q}+3} \lfloor \mu(w) \rfloor$ holds 
for the given RSS query $\RSSQ(T[i..j], b)$. 
Let $\kappa$ be the largest integer in set $[1, y^{\prime} - y + 1]$ satisfying $u_{\kappa} = 1$ for sequence $\Gamma_{B} = u_{1}, u_{2}, \ldots, u_{y^{\prime} - y + 1}$. 
Then, the following four statements hold: 
\begin{enumerate}[label=\textbf{(\roman*)}]
    \item \label{enum:GammaB_property:1} if the largest integer $\kappa$ does not exist, 
    then $\mathcal{C}_{\run} = \emptyset$; 
    \item \label{enum:GammaB_property:2} if the largest integer $\kappa$ exists, 
    then set $\mathcal{J}_{B}(h_{Q})$ contains a weighted point 
    $(\reverse(T[\hat{p}_{s}-1..\hat{\gamma}_{s}-1]), T[\hat{\gamma}_{s}..\hat{r}_{s}+1], |\Psi_{\str}(T[p_{s}-1..r_{s}+1])| |f_{\recover}(([p_{s}, q_{s}], [\ell_{s}, r_{s}]))|, T[p_{s}-1..r_{s}+1])$  
    satisfying $L_{x} \preceq \reverse(T[\hat{p}_{s}-1..\hat{\gamma}_{s}-1]) \preceq L_{x^{\prime}}$ 
    and $T[\hat{\gamma}_{s}..\hat{r}_{s}+1] = R_{y + \kappa - 1}$;
    \item \label{enum:GammaB_property:3} 
    Let $C_{s}$ be the associated string of the interval attractor $([p_{s}, q_{s}], [\ell_{s}, r_{s}])$ corresponding to 
    the weighted point $(\reverse(T[\hat{p}_{s}-1..\hat{\gamma}_{s}-1]), T[\hat{\gamma}_{s}..\hat{r}_{s}+1]$, $|\Psi_{\str}(T[p_{s}-1..r_{s}+1])| |f_{\recover}(([p_{s}, q_{s}], [\ell_{s}$, $r_{s}]))|$, $T[p_{s}-1..r_{s}+1])$ of Lemma~\ref{lem:GammaB_property}~\ref{enum:GammaB_property:2}. 
    Then, $C_{s} \in \mathcal{C}_{\run}$, 
    $C_{s} = C_{\max}$ and $([\hat{p}_{s}, \hat{q}_{s}], [\hat{\ell}_{s}, \hat{r}_{s}]) \in \Psi_{\CCP}(T[i..j]) \cap \Psi_{\run} \cap \Psi_{\centerset}(C_{\max})$; 
    \item \label{enum:GammaB_property:4}
    sequence $\Gamma_{B}$ is non-increasing~(i.e., $u_{1} \geq u_{2} \geq \cdots \geq u_{y^{\prime} - y + 1}$). 
\end{enumerate}
\end{lemma}
\begin{proof}
See Section~\ref{subsubsec:GammaB_proof1}.
\end{proof}

Next, sequence $\Gamma_{B, \sub}$ consists of $(y^{\prime} - y + 1)$ integers 
$\alpha_{1}, \alpha_{2}, \ldots, \alpha_{y^{\prime} - y + 1} \in \mathbb{N}_{0}$. 
Here, each integer $\alpha_{t}$ is defined as 
$\alpha_{t} = \rangecount(\mathcal{J}_{B}(h_{Q}), L_{x}, L_{x^{\prime}}$, $R_{y+t-1}, R_{y^{\prime}})$. 
Here, $\rangecount$ is the range-count query introduced in Section~\ref{subsec:range_data_structure}. 

The following lemma states three properties of sequence $\Gamma_{B, \sub}$.

\begin{lemma}\label{lem:GammaB_sub_property}
Consider the case that $|[\gamma_{Q}, j]| \leq 1 + \sum_{w = 1}^{h_{Q}+3} \lfloor \mu(w) \rfloor$ holds 
for the given RSS query $\RSSQ(T[i..j], b)$. 
The following three statements hold for two sequences $\Gamma_{B} = u_{1}, u_{2}, \ldots, u_{y^{\prime} - y + 1}$ and $\Gamma_{B, \sub} = \alpha_{1}, \alpha_{2}, \ldots, \alpha_{y^{\prime} - y + 1}$: 
\begin{enumerate}[label=\textbf{(\roman*)}]
    \item \label{enum:GammaB_sub_property:1} 
    sequence $\Gamma_{B, \sub}$ is non-increasing (i.e., $\alpha_{1} \geq \alpha_{2} \geq \cdots \geq \alpha_{y^{\prime} - y + 1}$);    
    \item \label{enum:GammaB_sub_property:2} 
    for an integer $t \in [1, y^{\prime} - y + 1]$, 
    consider the largest integer $t^{\prime}$ in set $[t, y^{\prime} - y + 1]$ satisfying 
    $\alpha_{t} = \alpha_{t^{\prime}}$. 
    If $\alpha_{t^{\prime}} \geq 1$, 
    then set $\mathcal{J}_{B}(h_{Q})$ contains a weighted point 
    $(\reverse(T[\hat{p}_{s}-1..\hat{\gamma}_{s}-1]), T[\hat{\gamma}_{s}..\hat{r}_{s}+1], |\Psi_{\str}(T[p_{s}-1..r_{s}+1])| |f_{\recover}(([p_{s}, q_{s}], [\ell_{s}, r_{s}]))|, T[p_{s}-1..r_{s}+1])$ satisfying 
    $L_{x} \preceq \reverse(T[\hat{p}_{s}-1..\hat{\gamma}_{s}-1]) \preceq L_{x^{\prime}}$ and $T[\hat{\gamma}_{s}..\hat{r}_{s}+1] = R_{y + t^{\prime} - 1}$; 
    \item \label{enum:GammaB_sub_property:3}
    consider the three integers $t, t^{\prime}$, and $s$ of Lemma~\ref{lem:GammaB_sub_property}~\ref{enum:GammaB_sub_property:2}. 
    Then, 
    $u_{t} = 1 \Leftrightarrow (\alpha_{t^{\prime}} \geq 1) \land (\RSCQ(\hat{\gamma}_{s} - |[i, \gamma_{Q}-1]|, \hat{\gamma}_{s} + |[\gamma_{Q}, j]| - 1) < b - \eta + 1)$. 
\end{enumerate}
\end{lemma}
\begin{proof}
See Section~\ref{subsubsec:GammaB_proof2}.
\end{proof}

\subsubsection{Proof of Lemma~\ref{lem:GammaB_property}}\label{subsubsec:GammaB_proof1}

The following proposition states a property of set $\Psi_{h} \cap \Psi_{\run}$ for an integer $h \in [0, H]$. 

\begin{proposition}\label{prop:JB_correspondence_property}
Consider an interval attractor $([p_{A}, q_{A}], [\ell_{A}, r_{A}]) \in \Psi_{h} \cap \Psi_{\source}$ for an integer $h \in [0, H]$. 
Here, Lemma~\ref{lem:mRecover_basic_property} shows that 
there exists an interval attractor $([p_{B}, q_{B}], [\ell_{B}, r_{B}]) \in \Psi_{\run}$ satisfying 
$f_{\recover}(([p_{A}, q_{A}], [\ell_{A}, r_{A}])) \cap \Psi_{\mRecover} = \{ ([p_{B}, q_{B}], [\ell_{B}, r_{B}]) \}$. 
Then, set $\Psi_{h} \cap \Psi_{\source} \cap \Psi_{\samp}$ contains an interval attractor $([p_{C}, q_{C}], [\ell_{C}, r_{C}])$ 
satisfying $T[p_{A}-1..r_{A}-1] = T[p_{C}-1..r_{C}+1]$, 
and the interval attractor $([p_{C}, q_{C}], [\ell_{C}, r_{C}])$ satisfies the following two conditions: 
\begin{enumerate}[label=\textbf{(\roman*)}]
    \item $([p_{C}, q_{C}], [\ell_{C}, r_{C}]) \in \Psi_{\centerset}(C)$ 
    for the associated string $C$ of the interval attractor $([p_{A}, q_{A}]$, $[\ell_{A}, r_{A}])$;
    \item consider an interval attractor $([p_{D}, q_{D}], [\ell_{D}, r_{D}])$ 
    in set $f_{\recover}(([p_{C}, q_{C}], [\ell_{C}, r_{C}])) \cap \Psi_{\mRecover}$. 
    Then, $T[p_{B}-1..\gamma_{B}-1] = T[p_{D}-1..\gamma_{D}-1]$ and $T[\gamma_{B}..r_{B}+1] = T[\gamma_{D}..r_{D}+1]$. 
    Here, $\gamma_{B}$ and $\gamma_{D}$ are the attractor positions of the two interval attractors 
    $([p_{B}, q_{B}], [\ell_{B}, r_{B}])$ and $([p_{D}, q_{D}], [\ell_{D}, r_{D}])$, respectively. 
\end{enumerate}
\end{proposition}
\begin{proof}
$([p_{A}, q_{A}], [\ell_{A}, r_{A}]) \not \in \Psi_{\run}$ holds 
because $([p_{A}, q_{A}], [\ell_{A}, r_{A}]) \in \Psi_{\source}$ and $\Psi_{\source} \cap \Psi_{\run} = \emptyset$. 
Because of $([p_{A}, q_{A}], [\ell_{A}, r_{A}]) \not \in \Psi_{\run}$, 
Lemma~\ref{lem:samp_basic_property}~\ref{enum:samp_basic_property:2} shows that 
the sampling subset $\Psi_{\samp}$ contains an interval attractor $([p_{C}, q_{C}], [\ell_{C}, r_{C}])$ satisfying 
$T[p_{A}-1..r_{A}-1] = T[p_{C}-1..r_{C}+1]$. 
Lemma~\ref{lem:psi_str_property}~\ref{enum:psi_str_property:2} shows that 
$([p_{C}, q_{C}], [\ell_{C}, r_{C}]) \in \Psi_{h}$ holds. 
Lemma~\ref{lem:psi_equality_basic_property}~\ref{enum:psi_equality_basic_property:center_set} 
and Lemma~\ref{lem:psi_equality_basic_property}~\ref{enum:psi_equality_basic_property:5} 
show that 
$([p_{C}, q_{C}], [\ell_{C}, r_{C}]) \in \Psi_{\centerset}(C)$ and $([p_{C}, q_{C}], [\ell_{C}, r_{C}]) \in \Psi_{\source}$ hold, 
respectively. 
Therefore, $([p_{C}, q_{C}], [\ell_{C}, r_{C}]) \in \Psi_{h} \cap \Psi_{\source} \cap \Psi_{\samp} \cap \Psi_{\centerset}(C)$ holds. 

Let $([p_{A, 1}, q_{A, 1}]$, $[\ell_{A, 1}, r_{A, 1}])$, $([p_{A, 2}, q_{A, 2}], [\ell_{A, 2}, r_{A, 2}])$, $\ldots$, $([p_{A, m}, q_{A, m}], [\ell_{A, m}, r_{A, m}])$ ($p_{A, 1} < p_{A, 2} < \cdots < p_{A, m}$) be the interval attractors obtained from function $f_{\recover}([p_{A}, q_{A}], [\ell_{A}, r_{A}])$. 
Similarly, 
Let $([p_{C, 1}, q_{C, 1}], [\ell_{C, 1}, r_{C, 1}]), ([p_{C, 2}, q_{C, 2}], [\ell_{C, 2}, r_{C, 2}]), \ldots, ([p_{C, m^{\prime}}, q_{C, m^{\prime}}], [\ell_{C, m^{\prime}}, r_{C, m^{\prime}}])$ ($p_{C, 1} < p_{C, 2} < \cdots < p_{C, m}$) be the interval attractors obtained from function $f_{\recover}([p_{C}, q_{C}], [\ell_{C}, r_{C}])$. 
Then, Lemma~\ref{lem:recover_super_property}~\ref{enum:recover_super_property:1} shows that $m = m^{\prime}$ holds. 

Let $\gamma_{A, s}$ and $\gamma_{C, s}$ be the attractor positions of the two interval attractors 
$([p_{A, s}, q_{A, s}], [\ell_{A, s}, r_{A, s}])$ and $([p_{C, s}, q_{C, s}], [\ell_{C, s}, r_{C, s}])$, respectively, 
for an integer $s \in [1, m]$. 
Then, we prove $T[p_{A, s}-1..\gamma_{A, s}-1] = T[p_{C, s}-1..\gamma_{C, s}-1]$ and $T[\gamma_{A, s}..r_{A, s}+1] = T[\gamma_{C, s}..r_{C, s}+1]$. 
Lemma~\ref{lem:recover_super_property}~\ref{enum:recover_super_property:2} shows that 
$T[p_{A, s}-1..r_{A, s}+1] = T[p_{C, s}-1..r_{C, s}+1]$. 
Because of $T[p_{A, s}-1..r_{A, s}+1] = T[p_{C, s}-1..r_{C, s}+1]$, 
Lemma~\ref{lem:psi_str_property}~\ref{enum:psi_str_property:1} shows that 
$T[p_{A, s}-1..\gamma_{A, s}-1] = T[p_{C, s}-1..\gamma_{C, s}-1]$ and $T[\gamma_{A, s}..r_{A, s}+1] = T[\gamma_{C, s}..r_{C, s}+1]$ hold. 

We prove $T[p_{B}-1..\gamma_{B}-1] = T[p_{D}-1..\gamma_{D}-1]$ and $T[\gamma_{B}..r_{B}+1] = T[\gamma_{D}..r_{D}+1]$. 
If $([p_{A}, q_{A}], [\ell_{A}, r_{A}]) \in \Psi_{\preceding}$, 
then Lemma~\ref{lem:psi_equality_basic_property}~\ref{enum:psi_equality_basic_property:6} shows that 
$([p_{C}, q_{C}], [\ell_{C}, r_{C}]) \in \Psi_{\preceding}$ holds. 
Because of $([p_{A}, q_{A}], [\ell_{A}, r_{A}]) \in \Psi_{\preceding}$, 
Lemma~\ref{lem:mRecover_basic_property} shows that 
$([p_{B}, q_{B}], [\ell_{B}, r_{B}]) = ([p_{A, m}, q_{A, m}]$, $[\ell_{A, m}, r_{A, m}])$ holds. 
Similarly, 
Lemma~\ref{lem:mRecover_basic_property} shows that 
$([p_{D}, q_{D}], [\ell_{D}, r_{D}]) = ([p_{C, m}, q_{C, m}], [\ell_{C, m}, r_{C, m}])$ holds. 
Therefore, $T[p_{B}-1..\gamma_{B}-1] = T[p_{D}-1..\gamma_{D}-1]$ and $T[\gamma_{B}..r_{B}+1] = T[\gamma_{D}..r_{D}+1]$ 
follow from (A) $([p_{B}, q_{B}], [\ell_{B}, r_{B}]) = ([p_{A, m}, q_{A, m}], [\ell_{A, m}, r_{A, m}])$, 
(B) $([p_{D}, q_{D}], [\ell_{D}, r_{D}]) = ([p_{C, m}, q_{C, m}], [\ell_{C, m}, r_{C, m}])$, 
(C) $T[p_{A, m}-1..\gamma_{A, m}-1] = T[p_{C, m}-1..\gamma_{C, m}-1]$, 
and (D) $T[\gamma_{A, m}..r_{A, m}+1] = T[\gamma_{C, m}..r_{C, m}+1]$. 

Otherwise (i.e., $([p_{A}, q_{A}], [\ell_{A}, r_{A}]) \not \in \Psi_{\preceding}$), 
$([p_{A}, q_{A}], [\ell_{A}, r_{A}]) \in \Psi_{\succeeding}$ holds 
because $\Psi_{\RR} = \Psi_{\preceding} \cup \Psi_{\succeeding}$ follows from the definitions of the two subsets $\Psi_{\preceding}$ 
and $\Psi_{\succeeding}$. 
In this case, 
Lemma~\ref{lem:psi_equality_basic_property}~\ref{enum:psi_equality_basic_property:7} shows that 
$([p_{C}, q_{C}], [\ell_{C}, r_{C}]) \in \Psi_{\succeeding}$ holds. 
Because of $([p_{A}, q_{A}], [\ell_{A}, r_{A}]) \in \Psi_{\succeeding}$, 
Lemma~\ref{lem:mRecover_basic_property} shows that 
$([p_{B}, q_{B}], [\ell_{B}, r_{B}]) = ([p_{A, 1}, q_{A, 1}], [\ell_{A, 1}, r_{A, 1}])$ holds. 
Similarly, 
Lemma~\ref{lem:mRecover_basic_property} shows that 
$([p_{D}, q_{D}], [\ell_{D}, r_{D}]) = ([p_{C, 1}, q_{C, 1}], [\ell_{C, 1}, r_{C, 1}])$ holds. 
Therefore, we can prove $T[p_{B}-1..\gamma_{B}-1] = T[p_{D}-1..\gamma_{D}-1]$ and $T[\gamma_{B}..r_{B}+1] = T[\gamma_{D}..r_{D}+1]$ 
using the same approach as for $([p_{A}, q_{A}], [\ell_{A}, r_{A}]) \in \Psi_{\preceding}$. 

Finally, Proposition~\ref{prop:JB_correspondence_property} holds. 
\end{proof}

\begin{proposition}\label{prop:Psi_B_Property}
We assume that $|[\gamma_{Q}, j]| \leq 1 + \sum_{w = 1}^{h_{Q}+3} \lfloor \mu(w) \rfloor$ holds for RSS query $\RSSQ(T[i..j], b)$.
The following four statements hold: 
\begin{enumerate}[label=\textbf{(\roman*)}]
    \item \label{enum:Psi_B_Property:1} $\Psi^{B}(1) \subseteq \Psi_{\CCP}(T[i..j]) \cap \Psi_{\run}$;
    \item \label{enum:Psi_B_Property:2} $\Psi^{B}(t) \supseteq \Psi^{B}(t+1)$ for each integer $t \in [1, y^{\prime} - y + 1]$;
    \item \label{enum:Psi_B_Property:3}
    consider an interval attractor $([p, q], [\ell, r])$ in set $\Psi^{B}(t) \setminus \Psi^{B}(t+1)$ for an integer $t \in [1, y^{\prime} - y + 1]$. 
    Then, $T[\gamma - |[i, \gamma_{Q}-1]|..\gamma-1] \cdot T[\gamma..r+1] = T[i..\gamma_{Q}-1] \cdot R_{y+t-1}$ holds 
    for the attractor position $\gamma$ of the interval attractor $([p, q], [\ell, r])$. 
\end{enumerate}
Here, let $\Psi^{B}(y^{\prime} - y + 2) = \emptyset$ for simplicity. 
\end{proposition}
\begin{proof}
The proof of Proposition~\ref{prop:Psi_B_Property} is as follows. 

\textbf{Proof of Proposition~\ref{prop:Psi_B_Property}(i).}
Consider an interval attractor $([p, q], [\ell, r]) \in \Psi^{B}(1)$.
From the definition of the subset $\Psi^{B}(1)$, 
there exists an integer $s \in [1, k]$ satisfying the following three conditions: 
\begin{itemize}
    \item $([\hat{p}_{s}, \hat{q}_{s}], [\hat{\ell}_{s}, \hat{r}_{s}]) = ([p, q], [\ell, r])$;
    \item $L_{x} \preceq \reverse(T[\hat{p}_{s}-1..\hat{\gamma}_{s}-1]) \preceq L_{x^{\prime}}$;
    \item $R_{y} \preceq T[\hat{\gamma}_{s}..\hat{r}_{s}+1] \preceq R_{y^{\prime}}$.
\end{itemize}

We prove $([\hat{p}_{s}, \hat{q}_{s}], [\hat{\ell}_{s}, \hat{r}_{s}]) \in \Psi_{\CCP}(T[i..j]) \cap \Psi_{\run}$. 
Lemma~\ref{lem:recover_basic_property}~\ref{enum:recover_basic_property:4} shows that 
$([\hat{p}_{s}, \hat{q}_{s}], [\hat{\ell}_{s}, \hat{r}_{s}]) \in \Psi_{h_{Q}} \cap \Psi_{\run}$ holds 
because $([p_{s}, q_{s}], [\ell_{s}, r_{s}]) \in \Psi_{h_{Q}} \cap \Psi_{\source}$ 
and $([\hat{p}_{s}, \hat{q}_{s}], [\hat{\ell}_{s}, \hat{r}_{s}]) \in f_{\recover}(([p_{s}, q_{s}]$, $[\ell_{s}, r_{s}]))$. 
$\reverse(T[i..\gamma_{Q}-1]) \prec \reverse(T[p-1..\gamma-1]) \prec \reverse(\#T[i..\gamma_{Q}-1])$ follows from 
$L_{x} \preceq \reverse(T[\hat{p}_{s}-1..\hat{\gamma}_{s}-1]) \preceq L_{x^{\prime}}$, 
$x = \min \{ s \in [1, d] \mid \reverse(T[i..\gamma_{Q}-1]) \prec L_{s} \}$, 
and $x^{\prime} = \max \{ s \in [1, d] \mid L_{s} \prec \reverse(\# \cdot T[i..\gamma_{Q}-1]) \}$. 
$T[\gamma_{Q}..j] \prec T[\hat{\gamma}_{s}..\hat{r}_{s}+1] \prec T[\gamma_{Q}..j]\#$ follows from 
$R_{y} \preceq T[\hat{\gamma}_{s}..\hat{r}_{s}+1] \preceq R_{y^{\prime}}$, 
$y = \min \{ s \in [1, d^{\prime}] \mid T[\gamma_{Q}..j] \prec R_{s} \}$, 
and $y^{\prime} = \max \{ s \in [1, d^{\prime}] \mid R_{s} \prec (T[\gamma_{Q}..j] \cdot \#) \}$. 
Lemma~\ref{lem:CCP_property}~\ref{enum:CCP_property:4} shows that 
$\Psi_{\CCP}(T[i..j]) = \{ ([p^{\prime}, q^{\prime}], [\ell^{\prime}, r^{\prime}]) \in \Psi_{h_{Q}} \mid \reverse(T[i..\gamma_{Q}-1]) \prec \reverse(T[p^{\prime}-1..\gamma^{\prime}-1]) \prec \reverse(\#T[i..\gamma_{Q}-1]) \text{ and } T[\gamma_{Q}..j] \prec T[\gamma^{\prime}..r^{\prime}+1] \prec T[\gamma_{Q}..j]\# \}$ holds. 
Therefore, $([\hat{p}_{s}, \hat{q}_{s}], [\hat{\ell}_{s}, \hat{r}_{s}]) \in \Psi_{\CCP}(T[i..j])$. 

We showed that $([\hat{p}_{s}, \hat{q}_{s}], [\hat{\ell}_{s}, \hat{r}_{s}]) \in \Psi_{\CCP}(T[i..j]) \cap \Psi_{\run}$ 
(i.e., $([p, q], [\ell, r]) \in \Psi_{\CCP}(T[i..j]) \cap \Psi_{\run}$) 
for each interval attractor $([p, q], [\ell, r]) \in \Psi^{B}(1)$. 
Therefore, $\Psi^{B}(1) \subseteq \Psi_{\CCP}(T[i..j]) \cap \Psi_{\run}$ holds. 

\textbf{Proof of Proposition~\ref{prop:Psi_B_Property}(ii).}
Proposition~\ref{prop:Psi_B_Property}(ii) follows from the definitions of the two subsets $\Psi^{B}(t)$ and $\Psi^{B}(t+1)$. 

\textbf{Proof of Proposition~\ref{prop:Psi_B_Property}(iii).}
We prove $T[\gamma - |[i, \gamma_{Q}-1]|..\gamma-1] = T[i..\gamma_{Q}-1]$. 
Because of $([p, q], [\ell, r]) \in \Psi^{A}(t)$, 
$([p, q], [\ell, r]) \in \Psi_{\CCP}(T[i..j])$ follows from 
Proposition~\ref{prop:Psi_A_Property}~\ref{enum:Psi_A_Property:1} and Proposition~\ref{prop:Psi_A_Property}~\ref{enum:Psi_A_Property:2}. 
Because of $([p, q], [\ell, r]) \in \Psi_{\CCP}(T[i..j])$, 
Lemma~\ref{lem:CCP_property}~\ref{enum:CCP_property:6} shows that 
$T[\gamma - |[i, \gamma_{Q}-1]|..\gamma-1] = T[i..\gamma_{Q}-1]$ holds. 

We prove $T[\gamma..r+1] = R_{y+t-1}$. 
From the definition of the two subsets $\Psi^{B}(t)$ and $\Psi^{B}(t+1)$, 
there exists an integer $s \in [1, k]$ satisfying the following three conditions: 
\begin{itemize}
    \item $([\hat{p}_{s}, \hat{q}_{s}], [\hat{\ell}_{s}, \hat{r}_{s}]) = ([p, q], [\ell, r])$;
    \item $L_{x} \preceq \reverse(T[\hat{p}_{s}-1..\hat{\gamma}_{s}-1]) \preceq L_{x^{\prime}}$;
    \item $R_{y+t-1} \preceq T[\hat{\gamma}_{s}..\hat{r}_{s}+1] \prec R_{y+t}$.
\end{itemize}
Because of $T[\hat{\gamma}_{s}..\hat{r}_{s}+1] \in \mathcal{Y}_{B}(h_{Q})$, 
there exists an integer $\tau \in [1, d^{\prime}]$ satisfying 
$R_{\tau} = T[\hat{\gamma}_{s}..\hat{r}_{s}+1]$. 
$R_{y+t-1} \preceq R_{\tau} \prec R_{y+t}$ follows from 
$R_{y+t-1} \preceq T[\hat{\gamma}_{s}..\hat{r}_{s}+1] \prec R_{y+t}$ and $R_{\tau} = T[\hat{\gamma}_{s}..\hat{r}_{s}+1]$. 
$\tau = y+t-1$ follows from 
$R_{y+t-1} \preceq R_{\tau} \prec R_{y+t}$ and $R_{1} \prec R_{2} \prec \cdots \prec R_{d^{\prime}}$. 
Therefore, $T[\gamma..r+1] = R_{y+t-1}$ follows from 
$R_{y+t-1} = T[\hat{\gamma}_{s}..\hat{r}_{s}+1]$, $\hat{\gamma}_{s} = \gamma$, and $\hat{r}_{s} = r$. 

Finally, $T[\gamma - |[i, \gamma_{Q}-1]|..\gamma-1] \cdot T[\gamma..r+1] = T[i..\gamma_{Q}-1] \cdot R_{y+t-1}$ follows from 
$T[\gamma - |[i, \gamma_{Q}-1]|..\gamma-1] = T[i..\gamma_{Q}-1]$ and $T[\gamma..r+1] = R_{y+t-1}$.

\end{proof}

For proving Lemma~\ref{lem:GammaB_property}, 
we introduce a set $\mathcal{I}^{B}$ of integers in set $\{ 1, 2, \ldots, y^{\prime} - y + 1 \}$. 
This set $\mathcal{I}^{B}$ consists of integers such that 
for each integer $t \in \mathcal{I}^{B}$, 
set $\Psi^{B}(t) \setminus \Psi^{B}(t+1)$ contains an interval attractor $([p, q], [\ell, r])$ satisfying 
$\RSCQ(\gamma - |[i, \gamma_{Q}-1]|, \gamma + |[\gamma_{Q}, j]| - 1) < b - \eta + 1$ for the attractor position $\gamma$ of the interval attractor $([p, q], [\ell, r])$. 
Formally, $\mathcal{I}^{B} = \{ t \in [1, y^{\prime} - y + 1] \mid \exists ([p, q], [\ell, r]) \in \Psi^{B}(t) \setminus \Psi^{B}(t+1) \text{ s.t. } \RSCQ(\gamma - |[i, \gamma_{Q}-1]|, \gamma + |[\gamma_{Q}, j]| - 1) < b - \eta + 1 \}$. 
Here, $\Psi^{B}(y^{\prime} - y + 2) = \emptyset$ for simplicity. 

The following proposition states four properties of the set $\mathcal{I}^{B}$. 

\begin{proposition}\label{prop:Set_IB_Property}
We assume that $|[\gamma_{Q}, j]| \leq 1 + \sum_{w = 1}^{h_{Q}+3} \lfloor \mu(w) \rfloor$ holds for RSS query $\RSSQ(T[i..j], b)$.
The following four statements hold for set $\mathcal{I}^{B}$ and sequence $\Gamma_{B} = u_{1}, u_{2}, \ldots, u_{y^{\prime} - y + 1 }$: 
\begin{enumerate}[label=\textbf{(\roman*)}]
    \item \label{enum:Set_IB_Property:1} 
    $\mathcal{F}_{\SA} \cap \mathcal{F}_{\suffix}(\Psi_{\CCP}(T[i..j]) \cap \Psi_{\run}) \supseteq \{ T[i..\gamma_{Q} - 1] \cdot R_{y + t - 1} \mid t \in \mathcal{I}^{B} \}$; 
    \item \label{enum:Set_IB_Property:2}
    if there exists a string $C \in \Sigma^{+}$ satisfying $\mathcal{F}_{\SA} \cap \mathcal{F}_{\suffix}(\Psi_{\CCP}(T[i..j]) \cap \Psi_{\run} \cap \Psi_{\centerset}(C)) \neq \emptyset$, 
    then there exists a pair of two integers $s \in [1, k]$ and $t \in \mathcal{I}^{B}$  
    satisfying $([p_{s}, q_{s}], [\ell_{s}, r_{s}]) \in \Psi_{\centerset}(C)$ 
    and $([\hat{p}_{s}, \hat{q}_{s}], [\hat{\ell}_{s}, \hat{r}_{s}]) \in \Psi^{B}(t) \setminus \Psi^{B}(t+1)$;  
    \item \label{enum:Set_IB_Property:3} $t \in \mathcal{I}^{B}$ for each integer $t \in [1, y^{\prime} - y + 1]$ satisfying $u_{t} = 1$ and $u_{t+1} = 0$;
    \item \label{enum:Set_IB_Property:4} $u_{t^{\prime}} = 1$ for 
    any pair of two integers $t \in \mathcal{I}^{B}$ and $t^{\prime} \in [1, t]$. 
\end{enumerate}
Here, let $u_{y^{\prime} - y + 2} = 0$ and $\Psi^{B}(y^{\prime} - y + 2) = \emptyset$ for simplicity. 
\end{proposition}
\begin{proof}
The proof of Proposition~\ref{prop:Set_IB_Property} is as follows. 

\textbf{Proof of Proposition~\ref{prop:Set_IB_Property}(i).}
Consider an integer $t$ in set $\mathcal{I}^{B}$. 
From the definition of the set $\mathcal{I}^{B}$, 
set $\Psi^{B}(t) \setminus \Psi^{B}(t+1)$ contains an interval attractor $([p, q], [\ell, r])$ satisfying 
$\RSCQ(\gamma - |[i, \gamma_{Q}-1]|, \gamma + |[\gamma_{Q}, j]| - 1) < b - \eta + 1$ for the attractor position $\gamma$ of the interval attractor $([p, q], [\ell, r])$. 

We prove $T[\gamma - |[i, \gamma_{Q}-1]|..r+1] \in \mathcal{F}_{\SA} \cap \mathcal{F}_{\suffix}(\Psi_{\CCP}(T[i..j]) \cap \Psi_{\run})$. 
Because of $([p, q], [\ell, r]) \in \Psi^{B}(\kappa)$, 
$([p, q], [\ell, r]) \in \Psi_{\CCP}(T[i..j]) \cap \Psi_{\run}$ follows from 
Proposition~\ref{prop:Psi_B_Property}~\ref{enum:Psi_B_Property:1} and Proposition~\ref{prop:Psi_B_Property}~\ref{enum:Psi_B_Property:2}. 
Lemma~\ref{lem:F_SA_formula} shows that 
$T[\gamma - |[i, \gamma_{Q}-1]|..r+1] \in \mathcal{F}_{\SA}$ holds 
because $([p, q], [\ell, r]) \in \Psi_{\CCP}(T[i..j])$ and $\RSCQ(\gamma - |[i, \gamma_{Q}-1]|, \gamma + |[\gamma_{Q}, j]| - 1) < b - \eta + 1$. 
Because of $([p, q], [\ell, r]) \in \Psi_{\CCP}(T[i..j]) \cap \Psi_{\run}$, 
$T[\gamma - |[i, \gamma_{Q}-1]|..r+1] \in \mathcal{F}_{\suffix}(\Psi_{\CCP}(T[i..j]) \cap \Psi_{\run})$ follows from 
the definition of the set $\mathcal{F}_{\suffix}(\Psi_{\CCP}(T[i..j]) \cap \Psi_{\run})$. 
Therefore, $\mathcal{F}_{\SA} \cap \mathcal{F}_{\suffix}(\Psi_{\CCP}(T[i..j]) \cap \Psi_{\run}) \neq \emptyset$ follows from 
$T[\gamma - |[i, \gamma_{Q}-1]|..r+1] \in \mathcal{F}_{\SA} \cap \mathcal{F}_{\suffix}(\Psi_{\CCP}(T[i..j]) \cap \Psi_{\run})$. 

We prove $T[i..\gamma_{Q} - 1] \cdot R_{y + t - 1} \in \mathcal{F}_{\SA} \cap \mathcal{F}_{\suffix}(\Psi_{\CCP}(T[i..j]) \cap \Psi_{\run})$. 
Because of $([p, q], [\ell, r]) \in \Psi^{B}(t) \setminus \Psi^{B}(t+1)$, 
Proposition~\ref{prop:Psi_B_Property}~\ref{enum:Psi_B_Property:3} shows that 
$T[\gamma - |[i, \gamma_{Q}-1]|..r+1] = T[i..\gamma_{Q} - 1] \cdot R_{y + t - 1}$ holds. 
Therefore, $T[i..\gamma_{Q} - 1] \cdot R_{y + t - 1} \in \mathcal{F}_{\SA} \cap \mathcal{F}_{\suffix}(\Psi_{\CCP}(T[i..j]) \cap \Psi_{\run})$ 
follows from $T[\gamma - |[i, \gamma_{Q}-1]|..r+1] \in \mathcal{F}_{\SA} \cap \mathcal{F}_{\suffix}(\Psi_{\CCP}(T[i..j]) \cap \Psi_{\run})$ 
and $T[\gamma - |[i, \gamma_{Q}-1]|..r+1] = T[i..\gamma_{Q} - 1] \cdot R_{y + t - 1}$. 

We showed that $T[i..\gamma_{Q} - 1] \cdot R_{y + t - 1} \in \mathcal{F}_{\SA} \cap \mathcal{F}_{\suffix}(\Psi_{\CCP}(T[i..j]) \cap \Psi_{\run})$ holds for each integer $t \in \mathcal{I}^{B}$. 
Therefore, $\mathcal{F}_{\SA} \cap \mathcal{F}_{\suffix}(\Psi_{\CCP}(T[i..j]) \cap \Psi_{\run}) \supseteq \{ T[i..\gamma_{Q} - 1] \cdot R_{y + t - 1} \mid t \in \mathcal{I}^{B} \}$ holds. 

\textbf{Proof of Proposition~\ref{prop:Set_IB_Property}(ii).}
Consider a string $F$ in set $\mathcal{F}_{\SA} \cap \mathcal{F}_{\suffix}(\Psi_{\CCP}(T[i..j]) \cap \Psi_{\run} \cap \Psi_{\centerset}(C))$. 
Then, the set $\Psi_{\CCP}(T[i..j]) \cap \Psi_{\run} \cap \Psi_{\centerset}(C)$ contains an interval attractor $([p_{A}, q_{A}], [\ell_{A}, r_{A}])$ satisfying 
$T[\gamma_{A} - |[i, \gamma_{Q}-1]|..r_{A}+1] = F$ for the attractor position $\gamma_{A}$ of the interval attractor $([p_{A}, q_{A}], [\ell_{A}, r_{A}])$. 
Because of $([p_{A}, q_{A}], [\ell_{A}, r_{A}]) \in \Psi_{\CCP}(T[i..j])$, 
Lemma~\ref{lem:CCP_property}~\ref{enum:CCP_property:4} shows that 
$\reverse(T[i..\gamma_{Q}-1]) \prec \reverse(T[p_{A}-1..\gamma_{A}-1]) \prec \reverse(\#T[i..\gamma_{Q}-1])$ 
and $T[\gamma_{Q}..j] \prec T[\gamma_{A}..r_{A}+1] \prec T[\gamma_{Q}..j]\#$.  
Lemma~\ref{lem:CCP_property}~\ref{enum:CCP_property:1} shows that 
$([p_{A}, q_{A}], [\ell_{A}, r_{A}]) \in \Psi_{h_{Q}}$ holds. 

Lemma~\ref{lem:recover_division_property}~\ref{enum:recover_division_property:1} shows that 
subset $\Psi_{\source}$ contains an interval attractor $([p_{B}, q_{B}], [\ell_{B}, r_{B}])$ satisfying 
$([p_{A}, q_{A}], [\ell_{A}, r_{A}]) \in f_{\recover}(([p_{B}, q_{B}], [\ell_{B}, r_{B}]))$. 
Lemma~\ref{lem:recover_basic_property}~\ref{enum:recover_basic_property:4} shows that 
$([p_{B}, q_{B}], [\ell_{B}, r_{B}]) \in \Psi_{h_{Q}} \cap \Psi_{\centerset}(C)$ holds. 
Lemma~\ref{lem:mRecover_basic_property} shows that 
there exists an interval attractor $([p_{C}, q_{C}], [\ell_{C}, r_{C}]) \in \Psi_{\run}$ satisfying 
$f_{\recover}(([p_{B}, q_{B}], [\ell_{B}, r_{B}])) \cap \Psi_{\mRecover} = \{ ([p_{C}, q_{C}], [\ell_{C}, r_{C}]) \}$. 
Because of $([p_{B}, q_{B}]$, $[\ell_{B}, r_{B}]) \in \Psi_{h_{Q}} \cap \Psi_{\source} \cap \Psi_{\centerset}(C)$, 
Proposition~\ref{prop:JB_correspondence_property} shows that 
there exists an integer $s \in [1, k]$ satisfying the following three conditions: 
\begin{itemize}
    \item $T[p_{s}-1..r_{s}-1] = T[p_{B}-1..r_{B}+1]$;
    \item $([p_{s}, q_{s}], [\ell_{s}, r_{s}]) \in \Psi_{\centerset}(C)$;
    \item $T[\hat{p}_{s}-1..\hat{\gamma}_{s}-1] = T[p_{C}-1..\gamma_{C}-1]$ and $T[\hat{\gamma}_{s}..\hat{r}_{s}+1] = T[\gamma_{C}..r_{C}+1]$ for the attractor position $\gamma_{C}$ of the interval attractor $([p_{C}, q_{C}], [\ell_{C}, r_{C}])$.
\end{itemize}

We prove $L_{x} \preceq \reverse(T[\hat{p}_{s}-1..\hat{\gamma}_{s}-1]) \preceq L_{x^{\prime}}$.
Lemma~\ref{lem:recover_basic_property}~\ref{enum:recover_basic_property:4} shows that 
$([\hat{p}_{s}, \hat{q}_{s}], [\hat{\ell}_{s}, \hat{r}_{s}]) \in \Psi_{h_{Q}} \cap \Psi_{\run} \cap \Psi_{\centerset}(C)$ holds 
because $([p_{s}, q_{s}], [\ell_{s}, r_{s}]) \in \Psi_{h_{Q}} \cap \Psi_{\centerset}(C)$ 
and $([\hat{p}_{s}, \hat{q}_{s}], [\hat{\ell}_{s}, \hat{r}_{s}]) \in f_{\recover}(([p_{s}, q_{s}], [\ell_{s}, r_{s}]))$. 
We can apply Lemma~\ref{lem:psi_run_basic_property}~\ref{enum:psi_run_basic_property:7} to 
the two interval attractors $([p_{A}, q_{A}]$, $[\ell_{A}, r_{A}])$ and $([\hat{p}_{s}, \hat{q}_{s}], [\hat{\ell}_{s}, \hat{r}_{s}])$ 
because $([p_{A}, q_{A}], [\ell_{A}, r_{A}]), ([\hat{p}_{s}, \hat{q}_{s}], [\hat{\ell}_{s}, \hat{r}_{s}]) \in \Psi_{h_{Q}} \cap \Psi_{\run} \cap \Psi_{\centerset}(C)$. 
Lemma~\ref{lem:psi_run_basic_property}~\ref{enum:psi_run_basic_property:7} shows that 
$T[\hat{p}_{s}-1..\hat{\gamma}_{s}-1] = T[p_{A}-1..\gamma_{A}-1]$ holds. 
Because of $\reverse(T[\hat{p}_{s}-1..\hat{\gamma}_{s}-1]) \in \mathcal{X}_{B}(h_{Q})$, 
there exists an integer $\tau \in [1, d]$ satisfying 
$L_{\tau} = \reverse(T[\hat{p}_{s}-1..\hat{\gamma}_{s}-1])$. 
$\reverse(T[i..\gamma_{Q}-1]) \prec L_{\tau} \prec \reverse(\#T[i..\gamma_{Q}-1])$ follows from 
$L_{\tau} = \reverse(T[\hat{p}_{s}-1..\hat{\gamma}_{s}-1])$, $T[\hat{p}_{s}-1..\hat{\gamma}_{s}-1] = T[p_{A}-1..\gamma_{A}-1]$, 
and $\reverse(T[i..\gamma_{Q}-1]) \prec \reverse(T[p_{A}-1..\gamma_{A}-1]) \prec \reverse(\#T[i..\gamma_{Q}-1])$. 
$L_{x} \preceq L_{\tau} \preceq L_{x^{\prime}}$ follows from 
$x = \min \{ s \in [1, d] \mid \reverse(T[i..\gamma_{Q}-1]) \prec L_{s} \}$, 
$x^{\prime} = \max \{ s \in [1, d] \mid L_{s} \prec \reverse(\# \cdot T[i..\gamma_{Q}-1]) \}$, 
and $\reverse(T[i..\gamma_{Q}-1]) \prec L_{\tau} \prec \reverse(\#T[i..\gamma_{Q}-1])$. 
Therefore, $L_{x} \preceq \reverse(T[\hat{p}_{s}-1..\hat{\gamma}_{s}-1]) \preceq L_{x^{\prime}}$ follows from 
$L_{x} \preceq L_{\tau} \preceq L_{x^{\prime}}$ and $L_{\tau} = \reverse(T[\hat{p}_{s}-1..\hat{\gamma}_{s}-1])$. 

We prove $T[\gamma_{Q}..j] \prec T[\hat{\gamma}_{s}..\hat{r}_{s}+1] \prec T[\gamma_{Q}..j]\#$.
Let $\xi = 2 + \sum_{w = 1}^{h_{Q}+3} \lfloor \mu(w) \rfloor$ for simplicity. 
Because of $([p_{A}, q_{A}], [\ell_{A}, r_{A}]) \in \Psi_{h_{Q}} \cap \Psi_{\run} \cap \Psi_{\centerset}(C)$, 
$T[\gamma_{A}..r_{A} + 1] = C^{n+1}[1..\xi] \cdot T[\gamma_{A} + \xi..r_{A} + 1]$ follows from 
the definition of the subset $\Psi_{\run}$. 
Similarly, 
$T[\hat{\gamma}_{s}..\hat{r}_{s}+1] = C^{n+1}[1..\xi] \cdot T[\hat{\gamma}_{s} + \xi..\hat{r}_{s} + 1]$ follows from 
the definition of the subset $\Psi_{\run}$. 
$T[\gamma_{Q}..j] \prec C^{n+1}[1..\xi] \prec T[\gamma_{Q}..j]\#$ follows from 
$T[\gamma_{Q}..j] \prec T[\gamma_{A}..r_{A}+1] \prec T[\gamma_{Q}..j]\#$, 
$T[\gamma_{A}..r_{A} + 1] = C^{n+1}[1..\xi] \cdot T[\gamma_{A} + \xi..r_{A} + 1]$, 
and $|[\gamma_{Q}, j]| < \xi$. 
Therefore, $T[\gamma_{Q}..j] \prec T[\hat{\gamma}_{s}..\hat{r}_{s}+1] \prec T[\gamma_{Q}..j]\#$ follows from 
$T[\gamma_{Q}..j] \prec C^{n+1}[1..\xi] \prec T[\gamma_{Q}..j]\#$ 
and $T[\hat{\gamma}_{s}..\hat{r}_{s}+1] = C^{n+1}[1..\xi] \cdot T[\hat{\gamma}_{s} + \xi..\hat{r}_{s} + 1]$. 

Because of $T[\hat{\gamma}_{s}..\hat{r}_{s}+1] \in \mathcal{Y}_{B}(h_{Q})$, 
there exists an integer $\tau^{\prime} \in [1, d^{\prime}]$ satisfying 
$R_{\tau^{\prime}} = T[\hat{\gamma}_{s}..\hat{r}_{s}+1]$. 
We prove $\tau^{\prime} \in [y, y^{\prime}]$ 
and $R_{y} \preceq T[\hat{\gamma}_{s}..\hat{r}_{s}+1] \preceq R_{y^{\prime}}$. 
$T[\gamma_{Q}..j] \prec R_{\tau^{\prime}} \prec T[\gamma_{Q}..j]\#$ follows from 
$T[\gamma_{Q}..j] \prec T[\hat{\gamma}_{s}..\hat{r}_{s}+1] \prec T[\gamma_{Q}..j]\#$ and 
$R_{\tau^{\prime}} = T[\hat{\gamma}_{s}..\hat{r}_{s}+1]$. 
$R_{y} \preceq R_{\tau^{\prime}} \preceq R_{y^{\prime}}$ follows from 
$y = \min \{ s \in [1, d^{\prime}] \mid T[\gamma_{Q}..j] \prec R_{s} \}$, 
$y^{\prime} = \max \{ s \in [1, d^{\prime}] \mid R_{s} \prec (T[\gamma_{Q}..j] \cdot \#) \}$, 
and $T[\gamma_{Q}..j] \prec R_{\tau^{\prime}} \prec T[\gamma_{Q}..j]\#$. 
Therefore, $R_{y} \preceq T[\hat{\gamma}_{s}..\hat{r}_{s}+1] \preceq R_{y^{\prime}}$ follows from 
$R_{y} \preceq R_{\tau^{\prime}} \preceq R_{y^{\prime}}$ and $R_{\tau^{\prime}} = T[\hat{\gamma}_{s}..\hat{r}_{s}+1]$. 
$\tau^{\prime} \in [y, y^{\prime}]$ follows from 
$R_{y} \preceq R_{\tau^{\prime}} \preceq R_{y^{\prime}}$ and $R_{1} \prec R_{2} \prec \cdots \prec R_{d^{\prime}}$. 

We prove $([\hat{p}_{s}, \hat{q}_{s}], [\hat{\ell}_{s}, \hat{r}_{s}]) \in \Psi^{B}(\tau^{\prime} - y + 1) \setminus \Psi^{B}(\tau^{\prime} - y + 2)$. 
$([\hat{p}_{s}, \hat{q}_{s}], [\hat{\ell}_{s}, \hat{r}_{s}]) \in \Psi^{B}(\tau^{\prime} - y + 1)$ follows from 
$L_{x} \preceq \reverse(T[\hat{p}_{s}-1..\hat{\gamma}_{s}-1]) \preceq L_{x^{\prime}}$ 
and $R_{\tau^{\prime}} \preceq T[\hat{\gamma}_{s}..\hat{r}_{s}+1] \preceq R_{y^{\prime}}$. 
On the other hand, $([\hat{p}_{s}, \hat{q}_{s}], [\hat{\ell}_{s}, \hat{r}_{s}]) \not \in \Psi^{B}(\tau^{\prime} - y + 2)$ holds 
because $T[\hat{\gamma}_{s}..\hat{r}_{s}+1] = R_{\tau^{\prime}}$. 
Therefore, $([\hat{p}_{s}, \hat{q}_{s}], [\hat{\ell}_{s}, \hat{r}_{s}]) \in \Psi^{B}(\tau^{\prime} - y + 1) \setminus \Psi^{B}(\tau^{\prime} - y + 2)$ holds. 

We prove $T[\hat{\gamma}_{s} - |[i, \gamma_{Q}-1]|..\hat{\gamma}_{s} - 1] \prec T[\gamma_{A} - |[i, \gamma_{Q}-1]|..\gamma_{A} - 1]$. 
Let $t = \tau^{\prime} - y + 1$ for simplicity. 
Because of $([\hat{p}_{s}, \hat{q}_{s}], [\hat{\ell}_{s}, \hat{r}_{s}]) \in \Psi^{B}(t) \setminus \Psi^{B}(t+1)$, 
Proposition~\ref{prop:Psi_B_Property}~\ref{enum:Psi_B_Property:3} shows that 
$T[\hat{\gamma}_{s} - |[i, \gamma_{Q}-1]|..\hat{\gamma}_{s} - 1] = T[i..\gamma_{Q} - 1]$ holds. 
Because of $([p_{A}, q_{A}], [\ell_{A}, r_{A}]) \in \Psi_{\CCP}(T[i..j])$, 
Lemma~\ref{lem:CCP_property}~\ref{enum:CCP_property:6} shows that 
$T[\gamma_{A} - |[i, \gamma_{Q}-1]|..\gamma_{A} - 1] = T[i..\gamma_{Q} - 1]$ holds. 
Therefore, 
$T[\hat{\gamma}_{s} - |[i, \gamma_{Q}-1]|..\hat{\gamma}_{s} - 1] = T[\gamma_{A} - |[i, \gamma_{Q}-1]|..\gamma_{A} - 1]$ follows from 
$T[\hat{\gamma}_{s} - |[i, \gamma_{Q}-1]|..\hat{\gamma}_{s} - 1] = T[i..\gamma_{Q} - 1]$ 
and $T[\gamma_{A} - |[i, \gamma_{Q}-1]|..\gamma_{A} - 1] = T[i..\gamma_{Q} - 1]$.

We prove $T[\hat{\gamma}_{s}..\hat{r}_{s}+1] \prec T[\gamma_{A}..r_{A}+1]$. 
Because of $f_{\recover}(([p_{B}, q_{B}], [\ell_{B}, r_{B}])) \cap \Psi_{\mRecover} = \{ ([p_{C}, q_{C}], [\ell_{C}, r_{C}]) \}$, 
$f_{\recover}(([p_{B}, q_{B}], [\ell_{B}, r_{B}])) \cap \Psi_{\lex}(T[\gamma_{C}..r_{C}+1]) = \emptyset$ follows from 
the definition of the subset $\Psi_{\mRecover}$. 
$T[\gamma_{A}..r_{A}+1] \prec T[\gamma_{C}..r_{C}+1]$ does not hold (i.e., $T[\gamma_{C}..r_{C}+1] \preceq T[\gamma_{A}..r_{A}+1]$) because 
$f_{\recover}(([p_{B}, q_{B}], [\ell_{B}, r_{B}])) \cap \Psi_{\lex}(T[\gamma_{C}..r_{C}+1]) = \emptyset$ 
and $([p_{A}, q_{A}], [\ell_{A}, r_{A}]) \in f_{\recover}(([p_{B}, q_{B}], [\ell_{B}, r_{B}]))$. 
Therefore, $T[\hat{\gamma}_{s}..\hat{r}_{s}+1] \preceq T[\gamma_{A}..r_{A}+1]$ 
follows from $T[\hat{\gamma}_{s}..\hat{r}_{s}+1] = T[\gamma_{C}..r_{C}+1]$ and $T[\gamma_{C}..r_{C}+1] \preceq T[\gamma_{A}..r_{A}+1]$. 

We prove $t \in \mathcal{I}^{B}$. 
$T[\hat{\gamma}_{s} - |[i, \gamma_{Q}-1]|..\hat{\gamma}_{s} - 1] \cdot T[\hat{\gamma}_{s}..\hat{r}_{s}+1] \prec T[\gamma_{A} - |[i, \gamma_{Q}-1]|..\gamma_{A} - 1] \cdot T[\gamma_{A}..r_{A}+1]$ follows from 
$T[\hat{\gamma}_{s} - |[i, \gamma_{Q}-1]|..\hat{\gamma}_{s} - 1] \prec T[\gamma_{A} - |[i, \gamma_{Q}-1]|..\gamma_{A} - 1]$ and 
$T[\hat{\gamma}_{s}..\hat{r}_{s}+1] \prec T[\gamma_{A}..r_{A}+1]$. 
Because of $([\hat{p}_{s}, \hat{q}_{s}], [\hat{\ell}_{s}, \hat{r}_{s}]) \in \Psi^{B}(t)$, 
$([\hat{p}_{s}, \hat{q}_{s}], [\hat{\ell}_{s}, \hat{r}_{s}]) \in \Psi_{\CCP}(T[i..j])$ follows from 
Proposition~\ref{prop:Psi_B_Property}~\ref{enum:Psi_B_Property:1} and Proposition~\ref{prop:Psi_B_Property}~\ref{enum:Psi_B_Property:2}. 
Lemma~\ref{lem:F_suffix_basic_property}~\ref{enum:F_suffix_basic_property:6} shows that 
$T[\hat{\gamma}_{s} - |[i, \gamma_{Q}-1]|..\hat{\gamma}_{s} - 1] \cdot T[\hat{\gamma}_{s}..\hat{r}_{s}+1] \in \mathcal{F}_{\SA}$ holds because 
(a) $([\hat{p}_{s}, \hat{q}_{s}], [\hat{\ell}_{s}, \hat{r}_{s}]), ([p_{A}, q_{A}], [\ell_{A}, r_{A}]) \in \Psi_{\CCP}(T[i..])$, 
(b) $T[\hat{\gamma}_{s} - |[i, \gamma_{Q}-1]|..\hat{\gamma}_{s} - 1] \cdot T[\hat{\gamma}_{s}..\hat{r}_{s}+1] \prec T[\gamma_{A} - |[i, \gamma_{Q}-1]|..\gamma_{A} - 1] \cdot T[\gamma_{A}..r_{A}+1]$, 
and (c) $T[\gamma_{A} - |[i, \gamma_{Q}-1]|..\gamma_{A} - 1] \cdot T[\gamma_{A}..r_{A}+1] \in \mathcal{F}_{\SA}$. 
$\RSCQ(\hat{\gamma}_{s} - |[i, \gamma_{Q}-1]|, \hat{\gamma}_{s} + |[\gamma_{Q}, j]| - 1) < b - \eta + 1$ 
follows from Lemma~\ref{lem:F_SA_formula} and $T[\hat{\gamma}_{s} - |[i, \gamma_{Q}-1]|..\hat{\gamma}_{s} - 1] \cdot T[\hat{\gamma}_{s}..\hat{r}_{s}+1] \in \mathcal{F}_{\SA}$. 
Therefore, $t \in \mathcal{I}^{B}$ follows from $([\hat{p}_{s}, \hat{q}_{s}], [\hat{\ell}_{s}, \hat{r}_{s}]) \in \Psi^{B}(t) \setminus \Psi^{B}(t+1)$ and $\RSCQ(\hat{\gamma}_{s} - |[i, \gamma_{Q}-1]|, \hat{\gamma}_{s} + |[\gamma_{Q}, j]| - 1) < b - \eta + 1$. 

Finally, Proposition~\ref{prop:Set_IB_Property}~\ref{enum:Set_IB_Property:2} follows from 
$([p_{s}, q_{s}], [\ell_{s}, r_{s}]) \in \Psi_{\centerset}(C)$, $t \in \mathcal{I}^{B}$, and $([\hat{p}_{s}, \hat{q}_{s}]$, $[\hat{\ell}_{s}, \hat{r}_{s}]) \in \Psi^{B}(t) \setminus \Psi^{B}(t+1)$. 

\textbf{Proof of Proposition~\ref{prop:Set_IB_Property}(iii).}
Proposition~\ref{prop:Set_IB_Property}~\ref{enum:Set_IB_Property:3} can be proved using the same approach as for Proposition~\ref{prop:Set_IA_Property}~\ref{enum:Set_IA_Property:2}. 

\textbf{Proof of Proposition~\ref{prop:Set_IB_Property}(iv).}
Proposition~\ref{prop:Set_IB_Property}~\ref{enum:Set_IB_Property:4} can be proved using the same approach as for Proposition~\ref{prop:Set_IA_Property}~\ref{enum:Set_IA_Property:3}. 
\end{proof}

%%%%%%%%%%%%%%%%%%%%%%%%%%%%%%%%

We prove Lemma~\ref{lem:GammaB_property} using Proposition~\ref{prop:JB_correspondence_property}, Proposition~\ref{prop:Psi_B_Property}, and Proposition~\ref{prop:Set_IB_Property}. 

\begin{proof}[Proof of Lemma~\ref{lem:GammaB_property}~\ref{enum:GammaB_property:1}]
%Otherwise (i.e., the integer $\kappa$ does not exist), 
Proposition~\ref{prop:Set_IB_Property}~\ref{enum:Set_IB_Property:4} indicates that 
$\mathcal{I}^{B} = \emptyset$ holds.
In this case, 
Proposition~\ref{prop:Set_IB_Property}~\ref{enum:Set_IB_Property:2} shows that 
$\mathcal{F}_{\SA} \cap \mathcal{F}_{\suffix}(\Psi_{\CCP}(T[i..j]) \cap \Psi_{\run} \cap \Psi_{\centerset}(C)) = \emptyset$ holds 
for any string $C \in \Sigma^{+}$. 
Therefore, $\mathcal{C}_{\run} = \emptyset$ follows from 
$\mathcal{C}_{\run} = \{ C \in \Sigma^{+} \mid \mathcal{F}_{\SA} \cap \mathcal{F}_{\suffix}(\Psi_{\CCP}(T[i..j]) \cap \Psi_{\run} \cap \Psi_{\centerset}(C)) \neq \emptyset \}$ and 
$\{ C \in \Sigma^{+} \mid \mathcal{F}_{\SA} \cap \mathcal{F}_{\suffix}(\Psi_{\CCP}(T[i..j]) \cap \Psi_{\run} \cap \Psi_{\centerset}(C)) \neq \emptyset \} = \emptyset$. 
\end{proof}

\begin{proof}[Proof of Lemma~\ref{lem:GammaB_property}~\ref{enum:GammaB_property:2}]
We prove $\Psi^{B}(\kappa) \setminus \Psi^{B}(\kappa+1) \neq \emptyset$. 
$\kappa = \kappa^{\prime}$ follows from Proposition~\ref{prop:Set_IB_Property}~\ref{enum:Set_IB_Property:3} 
and Proposition~\ref{prop:Set_IB_Property}~\ref{enum:Set_IB_Property:4} 
for the largest integer $\kappa^{\prime}$ in the set $\mathcal{I}^{B}$. 
$\Psi^{B}(\kappa^{\prime}) \setminus \Psi^{B}(\kappa^{\prime}+1) \neq \emptyset$ follows from the definition of the set $\mathcal{I}^{B}$. 
Therefore, $\Psi^{B}(\kappa) \setminus \Psi^{B}(\kappa+1) \neq \emptyset$ holds. 

Because of $\Psi^{B}(\kappa) \setminus \Psi^{B}(\kappa+1) \neq \emptyset$, 
set $\Psi^{B}(\kappa) \setminus \Psi^{B}(\kappa+1)$ contains an interval attractor $([p, q], [\ell, r])$. 
From the definition of the subset $\Psi^{B}(\kappa)$, 
there exists an integer $s \in [1, k]$ satisfying 
$([\hat{p}_{s}, \hat{q}_{s}], [\hat{\ell}_{s}, \hat{r}_{s}]) = ([p, q], [\ell, r])$. 
Proposition~\ref{prop:Psi_B_Property}~\ref{enum:Psi_B_Property:3} shows that 
$T[\hat{\gamma}_{s} - |[i, \gamma_{Q}-1]|..\hat{\gamma}_{s}-1] \cdot T[\hat{\gamma}_{s}..\hat{r}_{s}+1] = T[i..\gamma_{Q}-1] \cdot R_{y+\kappa-1}$. 
Therefore, Lemma~\ref{lem:GammaB_property}~\ref{enum:GammaB_property:2} holds. 
\end{proof}

\begin{proof}[Proof of Lemma~\ref{lem:GammaB_property}~\ref{enum:GammaB_property:3}]
The following two statements are used to prove Lemma~\ref{lem:GammaB_property}~\ref{enum:GammaB_property:3}. 
\begin{enumerate}[label=\textbf{(\Alph*)}]
    \item $C_{s} \in \mathcal{C}_{\run}$. 
    \item if $C_{s} \neq C_{\max}$ holds, 
    then there exists an integer $t \in [\kappa+1, y^{\prime} - y + 1]$ satisfying $u_{t} = 1$.
\end{enumerate}

\textbf{Proof of statement (A).}
We prove $([\hat{p}_{s}, \hat{q}_{s}], [\hat{\ell}_{s}, \hat{r}_{s}]) \in \Psi_{\CCP}(T[i..j]) \cap \Psi_{\run} \cap \Psi_{\centerset}(C_{s})$. 
Here, $([\hat{p}_{s}, \hat{q}_{s}], [\hat{\ell}_{s}, \hat{r}_{s}]) \in \Psi^{B}(\kappa) \setminus \Psi^{B}(\kappa+1)$ 
and $\kappa = \kappa^{\prime}$ hold for the largest integer $\kappa^{\prime}$ in the set $\mathcal{I}^{B}$ 
(see the proof of Lemma~\ref{lem:GammaB_property}~\ref{enum:GammaB_property:2}). 
Because of $([p_{s}, q_{s}], [\ell_{s}, r_{s}]) \in \Psi_{h_{Q}} \cap \Psi_{\centerset}(C_{s})$, 
Lemma~\ref{lem:recover_basic_property}~\ref{enum:recover_basic_property:4} shows that 
$([\hat{p}_{s}, \hat{q}_{s}], [\hat{\ell}_{s}, \hat{r}_{s}]) \in \Psi_{h_{Q}} \cap \Psi_{\run} \cap \Psi_{\centerset}(C_{s})$ holds. 
Because of $([\hat{p}_{s}, \hat{q}_{s}], [\hat{\ell}_{s}, \hat{r}_{s}]) \in \Psi^{B}(\kappa)$, 
$([\hat{p}_{s}, \hat{q}_{s}], [\hat{\ell}_{s}, \hat{r}_{s}]) \in \Psi_{\CCP}(T[i..j])$ follows from 
Proposition~\ref{prop:Psi_B_Property}~\ref{enum:Psi_B_Property:1} and Proposition~\ref{prop:Psi_B_Property}~\ref{enum:Psi_B_Property:2}. 
Therefore, $([\hat{p}_{s}, \hat{q}_{s}], [\hat{\ell}_{s}, \hat{r}_{s}]) \in \Psi_{\CCP}(T[i..j]) \cap \Psi_{\run} \cap \Psi_{\centerset}(C_{s})$ holds. 

We prove $C_{s} \in \mathcal{C}_{\run}$. 
Because of $([\hat{p}_{s}, \hat{q}_{s}], [\hat{\ell}_{s}, \hat{r}_{s}]) \in \Psi_{\CCP}(T[i..j]) \cap \Psi_{\run} \cap \Psi_{\centerset}(C_{s})$, 
$T[\hat{\gamma}_{s} - |[i, \gamma_{Q}-1]|..\hat{\gamma}_{s}-1] \cdot T[\hat{\gamma}_{s}..\hat{r}_{s} + 1] \in \mathcal{F}_{\suffix}(\Psi_{\CCP}(T[i..j]) \cap \Psi_{\run} \cap \Psi_{\centerset}(C_{s}))$ follows from the definition of the set $\mathcal{F}_{\suffix}(\Psi_{\CCP}(T[i..j]) \cap \Psi_{\run} \cap \Psi_{\centerset}(C_{s}))$. 
Because of $([\hat{p}_{s}, \hat{q}_{s}], [\hat{\ell}_{s}, \hat{r}_{s}]) \in \Psi^{B}(\kappa) \setminus \Psi^{B}(\kappa+1)$, 
Proposition~\ref{prop:Psi_B_Property}~\ref{enum:Psi_B_Property:3} shows that 
$T[\hat{\gamma}_{s} - |[i, \gamma_{Q}-1]|..\hat{\gamma}_{s}-1] \cdot T[\hat{\gamma}_{s}..\hat{r}_{s} + 1] = T[i..\gamma_{Q}-1] \cdot R_{y + \kappa - 1}$ holds. 
Proposition~\ref{prop:Set_IB_Property}~\ref{enum:Set_IB_Property:1} shows that 
$T[i..\gamma_{Q}-1] \cdot R_{y + \kappa - 1} \in \mathcal{F}_{\SA}$ holds. 
$T[\hat{\gamma}_{s} - |[i, \gamma_{Q}-1]|..\hat{\gamma}_{s}-1] \cdot T[\hat{\gamma}_{s}..\hat{r}_{s} + 1] \in \mathcal{F}_{\SA} \cap \mathcal{F}_{\suffix}(\Psi_{\CCP}(T[i..j]) \cap \Psi_{\run} \cap \Psi_{\centerset}(C_{s}))$ follows from 
(1) $T[\hat{\gamma}_{s} - |[i, \gamma_{Q}-1]|..\hat{\gamma}_{s}-1] \cdot T[\hat{\gamma}_{s}..\hat{r}_{s} + 1] \in \mathcal{F}_{\suffix}(\Psi_{\CCP}(T[i..j]) \cap \Psi_{\run} \cap \Psi_{\centerset}(C_{s}))$, 
(2) $T[i..\gamma_{Q}-1] \cdot R_{y + \kappa - 1} \in \mathcal{F}_{\SA}$, 
and (3) $T[\hat{\gamma}_{s} - |[i, \gamma_{Q}-1]|..\hat{\gamma}_{s}-1] \cdot T[\hat{\gamma}_{s}..\hat{r}_{s} + 1] = T[i..\gamma_{Q}-1] \cdot R_{y + \kappa - 1}$. 
Therefore, $C_{s} \in \mathcal{C}_{\run}$ follows from 
$\mathcal{C}_{\run} = \{ C \in \Sigma^{+} \mid \mathcal{F}_{\SA} \cap \mathcal{F}_{\suffix}(\Psi_{\CCP}(T[i..j]) \cap \Psi_{\run} \cap \Psi_{\centerset}(C)) \neq \emptyset \}$ and $\mathcal{F}_{\SA} \cap \mathcal{F}_{\suffix}(\Psi_{\CCP}(T[i..j]) \cap \Psi_{\run} \cap \Psi_{\centerset}(C_{s})) \neq \emptyset$. 

\textbf{Proof of statement (B).}
Let $\xi = 2 + \sum_{w = 1}^{h_{Q}+3} \lfloor \mu(w) \rfloor$ for simplicity. 
Because of $C_{s} \in \mathcal{C}_{\run}$ and $C_{s} \neq C_{\max}$, 
the set $\mathcal{C}_{\run}$ contains a string $C$ satisfying the following three conditions: 
\begin{itemize}
    \item $C \neq C_{s}$; 
    \item $\mathcal{F}_{\SA} \cap \mathcal{F}_{\suffix}(\Psi_{\CCP}(T[i..j]) \cap \Psi_{\run} \cap \Psi_{\centerset}(C)) \neq \emptyset$;
    \item $C_{s}^{n+1}[1..\xi] \prec C^{n+1}[1..\xi]$. 
\end{itemize}
Because of $\mathcal{F}_{\SA} \cap \mathcal{F}_{\suffix}(\Psi_{\CCP}(T[i..j]) \cap \Psi_{\run} \cap \Psi_{\centerset}(C)) \neq \emptyset$, 
Proposition~\ref{prop:Set_IB_Property}~\ref{enum:Set_IB_Property:2} shows that 
there exists a pair of two integers $s^{\prime} \in [1, k]$ and $t \in \mathcal{I}^{B}$  
satisfying $([p_{s^{\prime}}, q_{s^{\prime}}], [\ell_{s^{\prime}}, r_{s^{\prime}}]) \in \Psi_{\centerset}(C)$ 
and $([\hat{p}_{s^{\prime}}, \hat{q}_{s^{\prime}}], [\hat{\ell}_{s^{\prime}}, \hat{r}_{s^{\prime}}]) \in \Psi^{B}(t) \setminus \Psi^{B}(t+1)$. 
Proposition~\ref{prop:Set_IB_Property}~\ref{enum:Set_IB_Property:4} shows that 
$u_{t} = 1$ holds.

We prove $T[\hat{\gamma}_{s}..\hat{r}_{s} + 1] = C_{s}^{n+1}[1..\xi] \cdot T[\hat{\gamma}_{s} + \xi..\hat{r}_{s} + 1]$ 
and $T[\hat{\gamma}_{s^{\prime}}..\hat{r}_{s^{\prime}} + 1] = C^{n+1}[1..\xi] \cdot T[\hat{\gamma}_{s^{\prime}} + \xi..\hat{r}_{s^{\prime}} + 1]$. 
Because of $([p_{s}, q_{s}], [\ell_{s}, r_{s}]) \in \Psi_{h_{Q}} \cap \Psi_{\centerset}(C_{s})$, 
Lemma~\ref{lem:recover_basic_property}~\ref{enum:recover_basic_property:4} shows that 
$([\hat{p}_{s}, \hat{q}_{s}], [\hat{\ell}_{s}, \hat{r}_{s}]) \in \Psi_{h_{Q}} \cap \Psi_{\run} \cap \Psi_{\centerset}(C_{s})$ holds. 
Therefore, 
$T[\hat{\gamma}_{s}..\hat{r}_{s} + 1] = C_{s}^{n+1}[1..\xi] \cdot T[\hat{\gamma}_{s} + \xi..\hat{r}_{s} + 1]$ follows from 
the definition of the subset $\Psi_{\run}$. 
Similarly, 
$([\hat{p}_{s^{\prime}}, \hat{q}_{s^{\prime}}], [\hat{\ell}_{s^{\prime}}, \hat{r}_{s^{\prime}}]) \in \Psi_{h_{Q}} \cap \Psi_{\run} \cap \Psi_{\centerset}(C)$ can be proved, 
and $T[\hat{\gamma}_{s^{\prime}}..\hat{r}_{s^{\prime}} + 1] = C^{n+1}[1..\xi] \cdot T[\hat{\gamma}_{s^{\prime}} + \xi..\hat{r}_{s^{\prime}} + 1]$ follows from 
the definition of the subset $\Psi_{\run}$. 

We prove $\kappa < t$. 
Because of $([\hat{p}_{s}, \hat{q}_{s}], [\hat{\ell}_{s}, \hat{r}_{s}]) \in \Psi^{B}(\kappa) \setminus \Psi^{B}(\kappa+1)$, 
Proposition~\ref{prop:Psi_B_Property}~\ref{enum:Psi_B_Property:3} shows that 
$T[\hat{\gamma}_{s}..\hat{r}_{s} + 1] = R_{y + \kappa - 1}$ holds. 
Similarly, 
Proposition~\ref{prop:Psi_B_Property}~\ref{enum:Psi_B_Property:3} shows that 
$T[\hat{\gamma}_{s^{\prime}}..\hat{r}_{s^{\prime}} + 1] = R_{y + t - 1}$ holds. 
$T[\hat{\gamma}_{s}..\hat{r}_{s} + 1] \prec T[\hat{\gamma}_{s^{\prime}}..\hat{r}_{s^{\prime}} + 1]$ follows from 
$T[\hat{\gamma}_{s}..\hat{r}_{s} + 1] = C_{s}^{n+1}[1..\xi] \cdot T[\hat{\gamma}_{s} + \xi..\hat{r}_{s} + 1]$, 
$T[\hat{\gamma}_{s^{\prime}}..\hat{r}_{s^{\prime}} + 1] = C^{n+1}[1..\xi] \cdot T[\hat{\gamma}_{s^{\prime}} + \xi..\hat{r}_{s^{\prime}} + 1]$, 
and $C_{s}^{n+1}[1..\xi] \prec C^{n+1}[1..\xi]$. 
$R_{y + \kappa - 1} \prec R_{y + t - 1}$ follows from 
$T[\hat{\gamma}_{s}..\hat{r}_{s} + 1] = R_{y + \kappa - 1}$, 
$T[\hat{\gamma}_{s^{\prime}}..\hat{r}_{s^{\prime}} + 1] = R_{y + t - 1}$, 
and $T[\hat{\gamma}_{s}..\hat{r}_{s} + 1] \prec T[\hat{\gamma}_{s^{\prime}}..\hat{r}_{s^{\prime}} + 1]$. 
Therefore, $\kappa < t$ follows from 
$R_{y + \kappa - 1} \prec R_{y + t - 1}$ and $R_{1} \prec R_{2} \prec \cdots R_{d^{\prime}}$. 

Finally, statement (B) follows from $u_{t} = 1$, $\kappa < t$, and $t \in \mathcal{I}^{B} \subseteq [1, y^{\prime} - y + 1]$. 

\textbf{Proof of Lemma~\ref{lem:GammaB_property}~\ref{enum:GammaB_property:3}.}
$C_{s} \in \mathcal{C}_{\run}$ follows from statement (A). 
We prove $C_{s} = C_{\max}$ by contradiction. 
We assume that $C_{s} \neq C_{\max}$ holds. 
Then, statement (B) shows that 
there exists an integer $t \in [\kappa+1, y^{\prime} - y + 1]$ satisfying $u_{t} = 1$. 
On the other hand, 
the existence of the largest integer $\kappa$ indicates that 
$u_{t} = 0$ holds. 
The two facts $u_{t} = 1$ and $u_{t} = 0$ yield a contradiction. 
Therefore, $C_{s} = C_{\max}$ must hold. 

We already proved $([\hat{p}_{s}, \hat{q}_{s}], [\hat{\ell}_{s}, \hat{r}_{s}]) \in \Psi_{\CCP}(T[i..j]) \cap \Psi_{\run} \cap \Psi_{\centerset}(C_{s})$ (i.e., $([\hat{p}_{s}, \hat{q}_{s}], [\hat{\ell}_{s}, \hat{r}_{s}]) \in \Psi_{\CCP}(T[i..j]) \cap \Psi_{\run} \cap \Psi_{\centerset}(C_{\max})$) in the proof of statement (A). 
Therefore, Lemma~\ref{lem:GammaB_property}~\ref{enum:GammaB_property:3} holds.
\end{proof}

\begin{proof}[Proof of Lemma~\ref{lem:GammaB_property}~\ref{enum:GammaB_property:4}] 
We can prove Lemma~\ref{lem:GammaB_property}~\ref{enum:GammaB_property:4} 
using the same approach used to prove Lemma~\ref{lem:GammaA_property}~\ref{enum:GammaA_property:4}. 

\end{proof}

\subsubsection{Proof of Lemma~\ref{lem:GammaB_sub_property}}\label{subsubsec:GammaB_proof2}

The following proposition states the relationship between set $\Psi^{B}(t) \setminus \Psi^{B}(t+1)$ 
and range-count query on set $\mathcal{J}_{B}(h_{Q})$ for each integer $t \in [1, \hat{y} - y + 1]$. 

\begin{proposition}\label{prop:GammaB_Grid}
We assume that $|[\gamma_{Q}, j]| \leq 1 + \sum_{w = 1}^{h_{Q}+3} \lfloor \mu(w) \rfloor$ holds for RSS query $\RSSQ(T[i..j], b)$.
For an $t \in [1, \hat{y} - y + 1]$, 
$\rangecount(\mathcal{J}_{B}(h_{Q}), L_{x}, L_{x^{\prime}}, R_{y + t - 1}, R_{y + t - 1}) \geq 1 \Leftrightarrow \Psi^{B}(t) \setminus \Psi^{B}(t+1) \neq \emptyset$. 
Here, let $\Psi^{B}(\hat{y} - y + 2) = \emptyset$ for simplicity. 
\end{proposition}
\begin{proof}
Proposition~\ref{prop:GammaB_Grid} follows from the following two statements: 
\begin{enumerate}[label=\textbf{(\roman*)}]
    \item $\rangecount(\mathcal{J}_{B}(h_{Q}), L_{x}, L_{x^{\prime}}, R_{y + t - 1}, R_{y + t - 1}) \geq 1 \Rightarrow \Psi^{B}(t) \setminus \Psi^{B}(t+1) \neq \emptyset$;
    \item $\rangecount(\mathcal{J}_{B}(h_{Q}), L_{x}, L_{x^{\prime}}, R_{y + t - 1}, R_{y + t - 1}) \geq 1 \Leftarrow \Psi^{B}(t) \setminus \Psi^{B}(t+1) \neq \emptyset$.
\end{enumerate}

\textbf{Proof of statement (i).}
Because of $\rangecount(\mathcal{J}_{B}(h_{Q}), L_{x}, L_{x^{\prime}}, R_{y + t - 1}, R_{y + t - 1}) \geq 1$, 
set $\mathcal{J}_{B}(h_{Q})$ contains a weighted point $(\reverse(T[\hat{p}_{s}-1..\hat{\gamma}_{s}-1]), T[\hat{\gamma}_{s}..\hat{r}_{s}+1], |\Psi_{\str}(T[p_{s}-1..r_{s}+1])| |f_{\recover}(([p_{s}, q_{s}]$, $[\ell_{s}, r_{s}]))|, T[p_{s}-1..r_{s}+1])$ satisfying 
$L_{x} \preceq \reverse(T[\hat{p}_{s}-1..\hat{\gamma}_{s}-1]) \preceq L_{x^{\prime}}$ 
and $T[\hat{\gamma}_{s}..\hat{r}_{s}+1] = R_{y + t - 1}$. 
This weighted point corresponds to the interval attractor $([p_{s}, q_{s}], [\ell_{s}, r_{s}])$ in set $\Psi_{h_{Q}} \cap \Psi_{\source} \cap \Psi_{\samp}$. 
$([\hat{p}_{s}, \hat{q}_{s}], [\hat{\ell}_{s}, \hat{r}_{s}]) \in \Psi^{B}(t)$ follows from 
$L_{x} \preceq \reverse(T[\hat{p}_{s}-1..\hat{\gamma}_{s}-1]) \preceq L_{x^{\prime}}$ 
and $T[\hat{\gamma}_{s}..\hat{r}_{s}+1] = R_{y + t - 1}$. 
On the other hand, $([\hat{p}_{s}, \hat{q}_{s}], [\hat{\ell}_{s}, \hat{r}_{s}]) \not \in \Psi^{B}(t+1)$ holds 
because $T[\hat{\gamma}_{s}..\hat{r}_{s}+1] = R_{y + t - 1}$. 
Therefore, $\Psi^{B}(t) \setminus \Psi^{B}(t+1) \neq \emptyset$ follows from 
$([\hat{p}_{s}, \hat{q}_{s}], [\hat{\ell}_{s}, \hat{r}_{s}]) \in \Psi^{B}(t)$ 
and $([\hat{p}_{s}, \hat{q}_{s}], [\hat{\ell}_{s}, \hat{r}_{s}]) \not \in \Psi^{B}(t+1)$. 

\textbf{Proof of statement (ii).}
Let $s \in [1, k]$ be an integer satisfying $([\hat{p}_{s}, \hat{q}_{s}], [\hat{\ell}_{s}, \hat{r}_{s}]) \in \Psi^{B}(t) \setminus \Psi^{B}(t+1)$. 
Then, $L_{x} \preceq \reverse(T[\hat{p}_{s}-1..\hat{\gamma}_{s}-1]) \preceq L_{x^{\prime}}$ follow from the definition of the subset $\Psi^{B}(t)$. 
Proposition~\ref{prop:Psi_B_Property}~\ref{enum:Psi_B_Property:3} shows that 
$T[\hat{\gamma}_{s}..\hat{r}_{s}+1] = R_{y+t-1}$ holds. 
The interval attractor $([p_{s}, q_{s}], [\ell_{s}, r_{s}])$ corresponds to 
the weighted point $(\reverse(T[\hat{p}_{s}-1..\hat{\gamma}_{s}-1]), T[\hat{\gamma}_{s}..\hat{r}_{s}+1], |\Psi_{\str}(T[p_{s}-1..r_{s}+1])| |f_{\recover}(([p_{s}, q_{s}], [\ell_{s}, r_{s}]))|, T[p_{s}-1..r_{s}+1])$. 
The existence of this weighted point indicates that 
$\rangecount(\mathcal{J}_{B}(h_{Q}), L_{x}, L_{x^{\prime}}, R_{y + t - 1}, R_{y + t - 1}) \geq 1$ holds. 
Therefore, statement (ii) holds. 

\end{proof}

We prove Lemma~\ref{lem:GammaB_sub_property} using Proposition~\ref{prop:Psi_B_Property}, Proposition~\ref{prop:Set_IB_Property}, 
and Proposition~\ref{prop:GammaB_Grid}. 
Here, let $u_{\hat{y} - y + 2} = 0$, $\alpha_{\hat{y} - y + 2} = 0$, 
and $\Psi^{B}(\hat{y} - y + 2) = \emptyset$ for simplicity. 

\begin{proof}[Proof of Lemma~\ref{lem:GammaB_sub_property}~\ref{enum:GammaB_sub_property:1}]
    Lemma~\ref{lem:GammaB_sub_property}~\ref{enum:GammaB_sub_property:1} 
    can be proved using the same approach as for Lemma~\ref{lem:GammaA_sub_property} \ref{enum:GammaA_sub_property:1}. 
\end{proof}
\begin{proof}[Proof of Lemma~\ref{lem:GammaB_sub_property}~\ref{enum:GammaB_sub_property:2}]
    Lemma~\ref{lem:GammaB_sub_property}~\ref{enum:GammaB_sub_property:2} 
    can be proved using the same approach as for Lemma~\ref{lem:GammaA_sub_property} \ref{enum:GammaA_sub_property:2}. 
\end{proof}
\begin{proof}[Proof of Lemma~\ref{lem:GammaB_sub_property}~\ref{enum:GammaB_sub_property:3}]
    The following three statements are used to prove Lemma~\ref{lem:GammaB_sub_property}~\ref{enum:GammaB_sub_property:3}: 
    \begin{enumerate}[label=\textbf{(\Alph*)}]
    \item $u_{t} = 1 \Leftarrow (\alpha_{t^{\prime}} \geq 1) \land (\RSCQ(\hat{\gamma}_{s} - |[i, \gamma_{Q}-1]|, \hat{\gamma}_{s} + |[\gamma_{Q}, j]| - 1) < b - \eta + 1)$;
    \item 
    If $u_{t} = 1$, then $t^{\prime} = t_{A}$ and $\alpha_{t^{\prime}} \geq 1$ for the smallest integer $t_{A}$ in set $[t, \hat{y} - y + 1]$ satisfying  
    $\Psi^{B}(t_{A}) \setminus \Psi^{B}(t_{A} + 1) \neq \emptyset$; 
    \item $u_{t} = 1 \Rightarrow (\alpha_{t^{\prime}} \geq 1) \land (\RSCQ(\hat{\gamma}_{s} - |[i, \gamma_{Q}-1]|, \hat{\gamma}_{s} + |[\gamma_{Q}, j]| - 1) < b - \eta + 1)$.    
    \end{enumerate}

    \textbf{Proof of statement (A).}
    This statement can be proved using the same approach as for statement (A) in the proof of Lemma~\ref{lem:GammaA_sub_property}~\ref{enum:GammaA_sub_property:3}.

    \textbf{Proof of statement (B).}
    This statement corresponds to statement (B) in the proof of Lemma~\ref{lem:GammaA_sub_property}~\ref{enum:GammaA_sub_property:3}. 
    We proved statement (B) in the proof of Lemma~\ref{lem:GammaA_sub_property}~\ref{enum:GammaA_sub_property:3} using 
    Proposition~\ref{prop:Psi_A_Property}~\ref{enum:Psi_A_Property:2} and Proposition~\ref{prop:GammaA_Grid}. 
    Proposition~\ref{prop:Psi_B_Property}~\ref{enum:Psi_B_Property:2} corresponds to 
    Proposition~\ref{prop:Psi_A_Property}~\ref{enum:Psi_A_Property:2}. 
    Similarly, Proposition~\ref{prop:GammaB_Grid} corresponds to Proposition~\ref{prop:GammaA_Grid}. 
    Therefore, 
    statement (B) can be proved using the same approach as for statement (B) 
    in the proof of Lemma~\ref{lem:GammaA_sub_property}~\ref{enum:GammaA_sub_property:3}.

    \textbf{Proof of statement (C).}    
    This statement can be proved using the same approach as for statement (C) 
    in the proof of Lemma~\ref{lem:GammaA_sub_property}~\ref{enum:GammaA_sub_property:3}. 
    The detailed proof of statement (C) is as follows. 

    Because of $u_{t} = 1$, 
    $\mathcal{F}_{\SA} \cap \mathcal{F}_{\suffix}(\Psi_{\CCP}(T[i..j]) \cap \Psi^{B}(t)) \neq \emptyset$ follows from the definition of the sequence $\Gamma_{B}$. 
    Because of $\mathcal{F}_{\SA} \cap \mathcal{F}_{\suffix}(\Psi_{\CCP}(T[i..j]) \cap \Psi^{B}(t)) \neq \emptyset$, 
    there exists a string $F \in \Sigma^{+}$ satisfying $F \in \mathcal{F}_{\SA} \cap \mathcal{F}_{\suffix}(\Psi_{\CCP}(T[i..j]) \cap \Psi^{B}(t))$. 
    Because of $F \in \mathcal{F}_{\suffix}(\Psi_{\CCP}(T[i..j]) \cap \Psi^{B}(t))$, 
    there exists an integer $s^{\prime} \in [1, k]$ satisfying 
    $T[\hat{\gamma}_{s^{\prime}} - |[i, \gamma_{Q}-1]|..\hat{\gamma}_{s^{\prime}}-1] \cdot T[\hat{\gamma}_{s^{\prime}}..\hat{r}_{s^{\prime}} + 1] = F$ 
    and 
    $([\hat{p}_{s^{\prime}}, \hat{q}_{s^{\prime}}], [\hat{\ell}_{s^{\prime}}, \hat{r}_{s^{\prime}}]) \in \Psi_{\CCP}(T[i..j]) \cap \Psi^{B}(t)$.  
    Proposition~\ref{prop:Psi_B_Property}~\ref{enum:Psi_B_Property:2} indicates that 
    $\Psi^{B}(t) = \bigcup_{\lambda = t}^{\hat{y} - y + 1} \Psi^{B}(\lambda) \setminus \Psi^{B}(\lambda+1)$ holds. 
    Therefore, there exists an integer $t_{B} \in [t, \hat{y} - y + 1]$ satisfying 
    $([\hat{p}_{s^{\prime}}, \hat{q}_{s^{\prime}}], [\hat{\ell}_{s^{\prime}}, \hat{r}_{s^{\prime}}]) \in \Psi^{B}(t_{B}) \setminus \Psi^{B}(t_{B}+1)$. 
    Here, $t_{A} \leq t_{B}$ holds for the smallest integer $t_{A}$ in set $[t, \hat{y} - y + 1]$ satisfying  
    $\Psi^{B}(t_{A}) \setminus \Psi^{B}(t_{A} + 1) \neq \emptyset$. 

    We prove $([\hat{p}_{s}, \hat{q}_{s}], [\hat{\ell}_{s}, \hat{r}_{s}]) \in \Psi^{B}(t^{\prime}) \setminus \Psi^{B}(t^{\prime}+1)$. 
    Because of $u_{t} = 1$, 
    statement (B) shows that $t^{\prime} = t_{A}$ and $\alpha_{t^{\prime}} \geq 1$ holds. 
    Lemma~\ref{lem:GammaB_sub_property}~\ref{enum:GammaB_sub_property:2} shows that 
    the weighted point 
    $(\reverse(T[\hat{p}_{s}-1..\hat{\gamma}_{s}-1]), T[\hat{\gamma}_{s}..\hat{r}_{s}+1], |\Psi_{\str}(T[p_{s}-1..r_{s}+1])| |f_{\recover}(([p_{s}, q_{s}], [\ell_{s}, r_{s}]))|, T[p_{s}-1..r_{s}+1])$ satisfies  
    $L_{x} \preceq \reverse(T[\hat{p}_{s}-1..\hat{\gamma}_{s}-1]) \preceq L_{x^{\prime}}$ and $T[\hat{\gamma}_{s}..\hat{r}_{s}+1] = R_{y + t^{\prime} - 1}$. 
    Because of $([p_{s}, q_{s}], [\ell_{s}, r_{s}]) \in \Psi_{h_{Q}}$, 
    Lemma~\ref{lem:recover_basic_property}~\ref{enum:recover_basic_property:4} shows that 
    $([\hat{p}_{s}, \hat{q}_{s}], [\hat{\ell}_{s}, \hat{r}_{s}]) \in \Psi_{h_{Q}} \cap \Psi_{\run}$ holds.     
    We proved $([\hat{p}_{s}, \hat{q}_{s}], [\hat{\ell}_{s}, \hat{r}_{s}]) \in \Psi_{h_{Q}} \cap \Psi_{\run}$, 
    $L_{x} \preceq \reverse(T[\hat{p}_{s}-1..\hat{\gamma}_{s}-1]) \preceq L_{x^{\prime}}$, and $T[\hat{\gamma}_{s}..\hat{r}_{s}+1] = R_{y + t^{\prime} - 1}$. 
    Therefore, $([\hat{p}_{s}, \hat{q}_{s}], [\hat{\ell}_{s}, \hat{r}_{s}]) \in \Psi^{B}(t^{\prime}) \setminus \Psi^{B}(t^{\prime}+1)$ 
    follows from the definitions of the two subsets $\Psi^{B}(t^{\prime})$ and $\Psi^{B}(t^{\prime}+1)$. 
    
    We prove $T[\hat{\gamma}_{s} - |[i, \gamma_{Q}-1]|..\hat{\gamma}_{s}-1] \cdot T[\hat{\gamma}_{s}..\hat{r}_{s} + 1] = T[i..\gamma_{Q}-1] \cdot R_{y + t^{\prime} - 1}$. 
    Because of $([\hat{p}_{s}, \hat{q}_{s}], [\hat{\ell}_{s}, \hat{r}_{s}]) \in \Psi^{B}(t^{\prime}) \setminus \Psi^{B}(t^{\prime}+1)$, 
    Proposition~\ref{prop:Psi_B_Property}~\ref{enum:Psi_B_Property:3} shows that 
    $T[\hat{\gamma}_{s} - |[i, \gamma_{Q}-1]|..\hat{\gamma}_{s}-1] = T[i..\gamma_{Q}-1]$ holds. 
    Therefore, $T[\hat{\gamma}_{s} - |[i, \gamma_{Q}-1]|..\hat{\gamma}_{s}-1] \cdot T[\hat{\gamma}_{s}..\hat{r}_{s} + 1] = T[i..\gamma_{Q}-1] \cdot R_{y + t^{\prime} - 1}$ follows from 
    $T[\hat{\gamma}_{s} - |[i, \gamma_{Q}-1]|..\hat{\gamma}_{s}-1] = T[i..\gamma_{Q}-1]$ 
    and $T[\hat{\gamma}_{s}..\hat{r}_{s}+1] = R_{y + t^{\prime} - 1}$. 

    We prove $T[\hat{\gamma}_{s} - |[i, \gamma_{Q}-1]|..\hat{\gamma}_{s}-1] \cdot T[\hat{\gamma}_{s}..\hat{r}_{s} + 1] \preceq T[\hat{\gamma}_{s^{\prime}} - |[i, \gamma_{Q}-1]|..\hat{\gamma}_{s^{\prime}}-1] \cdot T[\hat{\gamma}_{s^{\prime}}..\hat{r}_{s^{\prime}} + 1]$. 
    $R_{y + t^{\prime} - 1} \preceq R_{y + t_{B} - 1}$ follows from 
    $t^{\prime} \leq t_{B}$ and $R_{1} \prec R_{2} \prec \cdots \prec R_{d^{\prime}}$. 
    Because of $([\hat{p}_{s^{\prime}}, \hat{q}_{s^{\prime}}], [\hat{\ell}_{s^{\prime}}, \hat{r}_{s^{\prime}}]) \in \Psi^{B}(t_{B}) \setminus \Psi^{B}(t_{B}+1)$, 
    Proposition~\ref{prop:Psi_B_Property}~\ref{enum:Psi_B_Property:3} shows that 
    $T[\hat{\gamma}_{s^{\prime}} - |[i, \gamma_{Q}-1]|..\hat{\gamma}_{s^{\prime}}-1] \cdot T[\hat{\gamma}_{s^{\prime}}..\hat{r}_{s^{\prime}} + 1] = T[i..\gamma_{Q}-1] \cdot R_{y + t_{B} - 1}$ holds. 
    Therefore, 
    $T[\hat{\gamma}_{s} - |[i, \gamma_{Q}-1]|..\hat{\gamma}_{s}-1] \cdot T[\hat{\gamma}_{s}..\hat{r}_{s} + 1] \preceq T[\hat{\gamma}_{s^{\prime}} - |[i, \gamma_{Q}-1]|..\hat{\gamma}_{s^{\prime}}-1] \cdot T[\hat{\gamma}_{s^{\prime}}..\hat{r}_{s^{\prime}} + 1]$ follows from 
    (1) $T[\hat{\gamma}_{s} - |[i, \gamma_{Q}-1]|..\hat{\gamma}_{s}-1] \cdot T[\hat{\gamma}_{s}..\hat{r}_{s} + 1] = T[i..\gamma_{Q}-1] \cdot R_{y + t^{\prime} - 1}$, 
    (2) $T[\hat{\gamma}_{s^{\prime}} - |[i, \gamma_{Q}-1]|..\hat{\gamma}_{s^{\prime}}-1] \cdot T[\hat{\gamma}_{s^{\prime}}..\hat{r}_{s^{\prime}} + 1] = T[i..\gamma_{Q}-1] \cdot R_{y + t_{B} - 1}$, 
    and (3) $R_{y + t^{\prime} - 1} \preceq R_{y + t_{B} - 1}$. 
    
    We prove $\RSCQ(\hat{\gamma}_{s} - |[i, \gamma_{Q}-1]|, \hat{\gamma}_{s} + |[\gamma_{Q}, j]| - 1) < b - \eta + 1$. 
    Because of $([\hat{p}_{s}, \hat{q}_{s}], [\hat{\ell}_{s}, \hat{r}_{s}]) \in \Psi^{B}(t^{\prime})$, 
    $([\hat{p}_{s}, \hat{q}_{s}], [\hat{\ell}_{s}, \hat{r}_{s}]), ([\hat{p}_{s^{\prime}}, \hat{q}_{s^{\prime}}], [\hat{\ell}_{s^{\prime}}, \hat{r}_{s^{\prime}}]) \in \Psi_{\CCP}(T[i..])$ 
    follows from Proposition~\ref{prop:Psi_B_Property}~\ref{enum:Psi_B_Property:1} and Proposition~\ref{prop:Psi_B_Property}~\ref{enum:Psi_B_Property:2}.    
    Lemma~\ref{lem:F_suffix_basic_property}~\ref{enum:F_suffix_basic_property:6} shows that 
    $T[\hat{\gamma}_{s} - |[i, \gamma_{Q}-1]|..\hat{\gamma}_{s}-1] \cdot T[\hat{\gamma}_{s}..\hat{r}_{s} + 1] \in \mathcal{F}_{\SA}$ 
    because 
    (a) $([\hat{p}_{s}, \hat{q}_{s}], [\hat{\ell}_{s}, \hat{r}_{s}]), ([\hat{p}_{s^{\prime}}, \hat{q}_{s^{\prime}}], [\hat{\ell}_{s^{\prime}}, \hat{r}_{s^{\prime}}]) \in \Psi_{\CCP}(T[i..])$, 
    (b) $T[\hat{\gamma}_{s} - |[i, \gamma_{Q}-1]|..\hat{\gamma}_{s}-1] \cdot T[\hat{\gamma}_{s}..\hat{r}_{s} + 1] \preceq T[\hat{\gamma}_{s^{\prime}} - |[i, \gamma_{Q}-1]|..\hat{\gamma}_{s^{\prime}}-1] \cdot T[\hat{\gamma}_{s^{\prime}}..\hat{r}_{s^{\prime}} + 1]$, 
    and (c) $T[\hat{\gamma}_{s^{\prime}} - |[i, \gamma_{Q}-1]|..\hat{\gamma}_{s^{\prime}}-1] \cdot T[\hat{\gamma}_{s^{\prime}}..\hat{r}_{s^{\prime}} + 1] \in \mathcal{F}_{\SA}$. 
    Therefore, 
    $\RSCQ(\hat{\gamma}_{s} - |[i, \gamma_{Q}-1]|, \hat{\gamma}_{s} + |[\gamma_{Q}, j]| - 1) < b - \eta + 1$ follows from 
    Lemma~\ref{lem:F_SA_formula} and $T[\hat{\gamma}_{s} - |[i, \gamma_{Q}-1]|..\hat{\gamma}_{s}-1] \cdot T[\hat{\gamma}_{s}..\hat{r}_{s} + 1] \in \mathcal{F}_{\SA}$. 

    Finally, $u_{t} = 1 \Rightarrow (\alpha_{t^{\prime}} \geq 1) \land (\RSCQ(\hat{\gamma}_{s} - |[i, \gamma_{Q}-1]|, \hat{\gamma}_{s} + |[\gamma_{Q}, j]| - 1) < b - \eta + 1)$ follows from statement (B) and $\RSCQ(\hat{\gamma}_{s} - |[i, \gamma_{Q}-1]|, \hat{\gamma}_{s} + |[\gamma_{Q}, j]| - 1) < b - \eta + 1$. 

    \textbf{Proof of Lemma~\ref{lem:GammaB_sub_property}~\ref{enum:GammaB_sub_property:3}.}
    Lemma~\ref{lem:GammaB_sub_property}~\ref{enum:GammaB_sub_property:3} follows from statement (A) and statement (C).
\end{proof}

\subsubsection{Algorithm}\label{subsubsec:gamma_B_algorithm}
We prove Lemma~\ref{lem:GammaB_algorithm}, i.e., 
we show that subquery $\RSSQB(T[i..j], b)$ can be answered 
in $O(H^{2} \log^{2} n + \log^{6} n)$ time using the data structures for RSC query 
and interval $[i, j]$. 

\paragraph{Computation of each integer $u_{t}$ in sequence $\Gamma_{B}$.}
Consider the non-increasing sequence $\Gamma_{B} = u_{1}, u_{2}, \ldots, u_{y^{\prime} - y + 1} \in \{ 0, 1 \}$ 
for the two integers $y$ and $y^{\prime}$ introduced in Section~\ref{subsec:GammaB}. 
We show that 
the $t$-th integer $u_{t}$ can be computed in $O(H^{2} \log n + \log^{4} n)$ time for a given integer $t \in [1, y^{\prime} - y + 1]$. 

We leverage sequence $\Gamma_{B, \sub} = \alpha_{1}, \alpha_{2}$, $\ldots$, $\alpha_{y^{\prime} - y + 1}$ for computing the $t$-th integer $u_{t}$. 
Let $t^{\prime}$ be the largest integer in set $[t, y^{\prime} - y + 1]$ satisfying $\alpha_{t} = \alpha_{t^{\prime}}$. 
If $\alpha_{t^{\prime}} \geq 1$, 
then Lemma~\ref{lem:GammaB_sub_property}~\ref{enum:GammaB_sub_property:2} shows that 
set $\mathcal{J}_{B}(h_{Q})$ contains a weighted point 
$(\reverse(T[\hat{p}_{s}-1..\hat{\gamma}_{s}-1]), T[\hat{\gamma}_{s}..\hat{r}_{s}+1], |\Psi_{\str}(T[p_{s}-1..r_{s}+1])| |f_{\recover}(([p_{s}, q_{s}], [\ell_{s}, r_{s}]))|, T[p_{s}-1..r_{s}+1])$ satisfying 
$L_{x} \preceq \reverse(T[\hat{p}_{s}-1..\hat{\gamma}_{s}-1]) \preceq L_{x^{\prime}}$ and $T[\hat{\gamma}_{s}..\hat{r}_{s}+1] = R_{y + t^{\prime} - 1}$. 
Here, $x$ and $x^{\prime}$ are the two integers introduced in Section~\ref{subsec:GammaB}. 
Lemma~\ref{lem:GammaB_sub_property}~\ref{enum:GammaB_sub_property:3} shows that 
$u_{t} = 1 \Leftrightarrow (\alpha_{t^{\prime}} \geq 1) \land (\RSCQ(\hat{\gamma}_{s} - |[i, \gamma_{Q}-1]|, \hat{\gamma}_{s} + |[\gamma_{Q}, j]| - 1) < b - \eta + 1)$. 
Therefore, the $t$-th integer $u_{t}$ can be computed using the $t^{\prime}$-th integer $\alpha_{t^{\prime}}$ of the sequence $\Gamma_{B, \sub}$ and RSC query $\RSCQ(\hat{\gamma}_{s} - |[i, \gamma_{Q}-1]|, \hat{\gamma}_{s} + |[\gamma_{Q}, j]| - 1)$. 

The algorithm computing the $t$-th integer $u_{t}$ consists of six phases. 
In the first phase, 
we compute interval attractor $([p_{Q}, q_{Q}], [\ell_{Q}, r_{Q}])$, 
its level $h_{Q}$, and its attractor position $\gamma_{Q}$. 
The interval attractor $([p_{Q}, q_{Q}], [\ell_{Q}, r_{Q}])$ can be obtained by capture query $\CAPQ([i, j])$. 
The level $h_{Q}$ and attractor position $\gamma_{Q}$ can be obtained by 
level-query $\levelQ(([p_{Q}, q_{Q}], [\ell_{Q}, r_{Q}]))$ and attractor position query $\attrQ(([p_{Q}, q_{Q}], [\ell_{Q}, r_{Q}]))$, 
respectively. 
Therefore, the first phase takes $O(H^{2} \log n)$ time. 

In the second phase, 
we compute the four integers $x, x^{\prime}, y$, and $y^{\prime}$. 
The two integers $x$ and $x^{\prime}$ are obtained by binary search on the $d$ strings of the ordered set $\mathcal{X}_{B}(h_{Q})$. 
This binary search can be executed in $O((H^{2} + \log n) \log d)$ time using Lemma~\ref{lem:JB_X_queries}~\ref{enum:JB_X_queries:3}. 
Similarly, 
the two integers $y$ and $y^{\prime}$ are obtained by binary search on the $d^{\prime}$ strings 
of the ordered set $\mathcal{Y}_{B}(h_{Q})$. 
This binary search can be executed in $O((H^{2} + \log n)\log d^{\prime})$ time 
using Lemma~\ref{lem:JB_Y_queries}~\ref{enum:JB_X_queries:3}. 
$d, d^{\prime} = O(n^{2})$ follows from Lemma~\ref{lem:JB_size}~\ref{enum:JB_size:3}. 
Therefore, the second phase takes $O(H^{2} \log n + \log^{2} n)$ time.

In the third phase, 
we find the integer $t^{\prime}$ by binary search on the sequence $\Gamma_{B, \sub}$. 
This binary search can be executed by computing $O(\log (y^{\prime} - y + 1))$ integers of the sequence $\Gamma_{B, \sub}$ 
because the sequence $\Gamma_{B, \sub}$ is non-increasing (Lemma~\ref{lem:GammaB_sub_property}~\ref{enum:GammaB_sub_property:1}). 
Each integer of the sequence $\Gamma_{B, \sub}$ can be computed by one range-count query on the set $\mathcal{J}_{B}(h_{Q})$ of weighted points. 
This range-count query takes $O(\log^{2} |\mathcal{J}_{B}(h_{Q})|)$ time using the dynamic data structures of Section~\ref{subsubsec:JB_subquery_ds}. 
$y^{\prime} - y + 1 = O(n^{2})$ because $y^{\prime} - y + 1 \leq d^{\prime}$ and $d^{\prime} = O(n^{2})$. 
$|\mathcal{J}_{B}(h_{Q})| = O(n^{2})$ follows from Lemma~\ref{lem:JB_size}~\ref{enum:JB_size:3}. 
Therefore, the third phase takes $O(\log^{3} n)$ time. 

The integer $\alpha_{t^{\prime}}$ is obtained by the binary search of the third phase. 
The fourth phase is executed if $\alpha_{t^{\prime}} \geq 1$ holds. 
In the fourth phase, we find the interval attractor $([p_{s}, q_{s}], [\ell_{s}, r_{s}])$ corresponding to 
the weighted point $(\reverse(T[\hat{p}_{s}-1..\hat{\gamma}_{s}-1]), T[\hat{\gamma}_{s}..\hat{r}_{s}+1], |\Psi_{\str}(T[p_{s}-1..r_{s}+1])| |f_{\recover}(([p_{s}, q_{s}], [\ell_{s}, r_{s}]))|$, $T[p_{s}-1..r_{s}+1])$. 
This interval attractor can be found in $O(H^{2} \log n + \log^{2} n)$ time 
using Lemma~\ref{lem:JB_Y_queries} \ref{enum:JB_Y_queries:4}. 

In the fifth phase, 
we obtain the interval attractor $([\hat{p}_{s}, \hat{q}_{s}], [\hat{\ell}_{s}, \hat{r}_{s}])$ 
by applying Lemma~\ref{lem:mRecover_query} to the interval attractor $([p_{s}, q_{s}], [\ell_{s}, r_{s}])$.  
This phase takes $O(H^{2})$ time. 

In the sixth phase, 
we compute the $t$-th integer $u_{t}$ of sequence $\Gamma_{B}$ by verifying 
$\alpha_{t^{\prime}} \geq 1$ and $\RSCQ(\hat{\gamma}_{s} - |[i, \gamma_{Q}-1]|, \hat{\gamma}_{s} + |[\gamma_{Q}, j]| - 1) < b - \eta + 1$. 
The attractor position $\hat{\gamma}_{s}$ is obtained in $O(H^{2})$ time 
by attractor position query $\attrQ(([\hat{p}_{s}, \hat{q}_{s}], [\hat{\ell}_{s}, \hat{r}_{s}]))$. 
The RSC query $\RSCQ(\hat{\gamma}_{s} - |[i, \gamma_{Q}-1]|, \hat{\gamma}_{s} + |[\gamma_{Q}, j]| - 1)$ takes $O(H^{2} \log n + \log^{4} n)$ time. 
Therefore, the fifth phase takes $O(H^{2} \log n + \log^{4} n)$ time. 

Finally, the algorithm computing the $t$-th integer $u_{t}$ takes $O(H^{2} \log n + \log^{4} n)$ time in total. 

\paragraph{Computation of the largest integer $\kappa$.}
Consider the largest integer $\kappa$ in set $[1, y^{\prime} - y + 1]$ satisfying $u_{\kappa} = 1$. 
We find the largest integer $\kappa$ by binary search on the non-increasing sequence $\Gamma_{B}$. 
This binary search can be executed by computing $O(\log (y^{\prime} - y + 1))$ integers of the sequence $\Gamma_{B}$. 
Each integer of the sequence $\Gamma_{B}$ can be computed in $O(H^{2} \log n + \log^{4} n)$ time. 
The two integers $y$ and $y^{\prime}$ can be obtained by the algorithm computing each integer $u_{t}$. 
Therefore, this binary search takes $O((H^{2} \log n + \log^{4} n) \log (y^{\prime} - y + 1))$ time 
(i.e., $O(H^{2} \log^{2} n + \log^{5} n)$ time).

\paragraph{Verification of $\mathcal{C}_{\run} = \emptyset$.}
We verify whether $\mathcal{C}_{\run} = \emptyset$ or not for set $\mathcal{C}_{\run}$ of strings 
for answering subquery $\RSSQB(T[i..j], b)$. 
From Lemma~\ref{lem:GammaB_property}~\ref{enum:GammaB_property:1} and Lemma~\ref{lem:GammaB_property}~\ref{enum:GammaB_property:3}, 
the largest integer $\kappa$ exists if and only if $\mathcal{C}_{\run} \neq \emptyset$ holds. 
We can verify whether the largest integer $\kappa$ exists or not by the algorithm computing the largest integer $\kappa$. 
Therefore, the verification of $\mathcal{C}_{\run} = \emptyset$ takes $O(H^{2} \log^{2} n + \log^{5} n)$ time. 

\paragraph{Algorithm for subquery $\RSSQB(T[i..j], b)$.}
The algorithm for subquery $\RSSQB(T[i..j], b)$  
returns an interval attractor in set $\Psi_{\CCP}(T[i..j]) \cap \Psi_{\run} \cap \Psi_{\centerset}(C_{\max})$ for the string $C_{\max}$. 
This algorithm is executed only if $\mathcal{C}_{\run} \neq \emptyset$ holds; 
otherwise, subquery $\RSSQB(T[i..j], b)$ can be answered by verifying whether $\mathcal{C}_{\run} = \emptyset$ or not in $O(H^{2} \log^{2} n + \log^{5} n)$ time. 

Lemma~\ref{lem:GammaB_property}~\ref{enum:GammaB_property:2} shows that 
set $\mathcal{J}_{B}(h_{Q})$ contains a weighted point 
$(\reverse(T[\hat{p}_{s}-1..\hat{\gamma}_{s}-1]), T[\hat{\gamma}_{s}..\hat{r}_{s}+1], |\Psi_{\str}(T[p_{s}-1..r_{s}+1])| |f_{\recover}(([p_{s}, q_{s}], [\ell_{s}, r_{s}]))|, T[p_{s}-1..r_{s}+1])$ satisfying $L_{x} \preceq \reverse(T[\hat{p}_{s}-1..\hat{\gamma}_{s}-1]) \preceq L_{x^{\prime}}$ and $T[\hat{\gamma}_{s}..\hat{r}_{s}+1] = R_{y + \kappa - 1}$. 
Consider the associated string $C_{s}$ of the interval attractor $([p_{s}, q_{s}], [\ell_{s}, r_{s}])$ 
corresponding to this weighted point. 
Then, 
Lemma~\ref{lem:GammaB_property}~\ref{enum:GammaB_property:3} shows that 
$C_{s} = C_{\max}$ and 
$([\hat{p}_{s}, \hat{q}_{s}], [\hat{\ell}_{s}, \hat{r}_{s}]) \in \Psi_{\CCP}(T[i..j]) \cap \Psi_{\run} \cap \Psi_{\centerset}(C_{\max})$ hold. 
Therefore, we can return the interval attractor $([\hat{p}_{s}, \hat{q}_{s}], [\hat{\ell}_{s}, \hat{r}_{s}])$ as the answer to $\RSSQB(T[i..j], b)$. 

The algorithm for $\RSSQB(T[i..j], b)$ consists of four phases.
In the first phase, 
we compute interval attractor $([p_{Q}, q_{Q}], [\ell_{Q}, r_{Q}])$, 
its level $h_{Q}$, and its attractor position $\gamma_{Q}$. 
This phase can be executed in $O(H^{2} \log n)$ time by the first phase of the algorithm computing an integer of sequence $\Gamma_{B}$. 

In the second phase, 
we compute the five integers $x, x^{\prime}, y, y^{\prime}$, and $\kappa$.
The four integers $x, x^{\prime}, y$, and $y^{\prime}$ can be computed 
in $O(H^{2} \log n + \log^{2} n)$ time by 
the second phase of the algorithm computing an integer of sequence $\Gamma_{B}$. 
The computation of the largest integer $\kappa$ takes $O(H^{2} \log^{2} n + \log^{5} n)$ time. 
Therefore, the second phase takes $O(H^{2} \log^{2} n + \log^{5} n)$ time.

In the third phase, 
we compute the interval attractor $([p_{s}, q_{s}], [\ell_{s}, r_{s}])$ corresponding to 
the weighted point $(\reverse(T[\hat{p}_{s}-1..\hat{\gamma}_{s}-1]), T[\hat{\gamma}_{s}..\hat{r}_{s}+1], |\Psi_{\str}(T[p_{s}-1..r_{s}+1])| |f_{\recover}(([p_{s}, q_{s}], [\ell_{s}, r_{s}]))|$, $T[p_{s}-1..r_{s}+1])$. 
This interval attractor can be found in $O(H^{2} \log n + \log^{2} n)$ time 
using Lemma~\ref{lem:JB_Y_queries} \ref{enum:JB_Y_queries:4}. 

In the fourth phase, 
we return the interval attractor $([\hat{p}_{s}, \hat{q}_{s}], [\hat{\ell}_{s}, \hat{r}_{s}])$ as the answer to $\RSSQB(T[i..j], b)$. 
This interval attractor can be obtained by applying Lemma~\ref{lem:mRecover_query} to the interval attractor $([p_{s}, q_{s}], [\ell_{s}, r_{s}])$.  
This phase takes $O(H^{2})$ time. 

Finally, we can answer $\RSSQB(T[i..j], b)$ in $O(H^{2} \log^{2} n + \log^{5} n)$ time. 
Therefore, Lemma~\ref{lem:GammaB_algorithm} holds.

\subsection{Subquery \texorpdfstring{$\RSSQCX(T[i..j], b)$}{RSSC1(T[i..j], b)}}\label{subsec:GammaC1}
The goal of this subsection is to answer subquery $\RSSQCX(T[i..j], b)$ under the assumption 
that (i) either $\mathcal{C}_{\run} = \emptyset$ or $C_{Q} = C_{\max}$ holds, and (ii) condition (A) of RSS query is satisfied. 
For this subsection, 
let $\hat{K} = |\lcp(T[\gamma_{Q}..j], C_{Q}^{n+1})|$ 
and $\hat{M} = (\hat{K} - (2 + \sum_{w = 1}^{h_{Q}+3} \lfloor \mu(w) \rfloor) ) \mod |C_{Q}|$. 
The following lemma states the summary of this subsection. 

\begin{lemma}\label{lem:GammaC1_algorithm}
We assume that (i) either $\mathcal{C}_{\run} = \emptyset$ or $C_{Q} = C_{\max}$ holds, and (ii) condition (A) of RSS query is satisfied. 
We can answer $\RSSQCX(T[i..j], b)$ in $O(H^{2} \log^{2} n + \log^{5} n)$ time 
using (A) the data structures for RSC query, 
(B) interval $[i, j]$, 
and (C) the starting position $\eta$ of the sa-interval $[\eta, \eta^{\prime}]$ of $T[i..j]$. 
If the subquery returns a string $F$, 
then $F$ is represented as an interval $[g, g + |F| - 1]$ satisfying $T[g..g + |F| - 1] = F$. 
\end{lemma}
\begin{proof}
See Section~\ref{subsubsec:gamma_C1_algorithm}.
\end{proof}

We use set $\Psi_{h_{Q}} \cap \Psi_{\source} \cap \Psi_{\centerset}(C_{Q}) \cap \Psi_{\modulo}(\hat{M}) \cap \Psi_{\preceding} \cap \Psi_{\samp}$ to explain the idea behind answering $\RSSQCX(T[i..j], b)$. 
Let $([p_{1}, q_{1}], [\ell_{1}, r_{1}]), ([p_{2}, q_{2}], [\ell_{2}, r_{2}])$, 
$\ldots$, $([p_{k}, q_{k}], [\ell_{k}, r_{k}])$ be the interval attractors in the set $\Psi_{h_{Q}} \cap \Psi_{\source} \cap \Psi_{\centerset}(C_{Q}) \cap \Psi_{\modulo}(\hat{M}) \cap \Psi_{\preceding} \cap \Psi_{\samp}$. 
Let $\gamma_{s}$ be the attractor position of each interval attractor $([p_{s}, q_{s}], [\ell_{s}, r_{s}]) \in \Psi_{h_{Q}} \cap \Psi_{\source} \cap \Psi_{\centerset}(C_{Q}) \cap \Psi_{\modulo}(\hat{M}) \cap \Psi_{\preceding} \cap \Psi_{\samp}$. 
Let $K_{s} = |\lcp(T[\gamma_{s}..r_{s}], C_{Q}^{n+1})|$ for simplicity.  

We leverage the set $\mathcal{J}_{C}(h_{Q}, C_{Q}, \hat{M})$ of weighted points on grid $([1, n], \mathcal{Y}_{C}(h_{Q}, C_{Q}, \hat{M}))$ introduced in Section~\ref{subsec:RSC_comp_C1}. 
The summary of the set $\mathcal{J}_{C}(h_{Q}, C_{Q}, \hat{M})$ and ordered set $\mathcal{Y}_{C}(h_{Q}, C_{Q}, \hat{M})$ is as follows 
(see Section~\ref{subsec:RSC_comp_C1} for the details of the set and ordered set):
\begin{itemize}
    \item the ordered set $\mathcal{Y}_{C}(h_{Q}, C_{Q}, \hat{M})$ consists of $d$ strings $R_{1}, R_{2}, \ldots, R_{d}$ ($R_{1} \prec R_{2} \prec \cdots \prec R_{d}$);
    \item the set $\mathcal{J}_{C}(h_{Q}, C_{Q}, \hat{M})$ contains the weighted point 
    $(|f_{\recover}(([p_{s}, q_{s}], [\ell_{s}, r_{s}]))|$, $T[\gamma_{s} + K_{s}..r_{s} + 1]$, $|\Psi_{\str}(T[p_{s}-1..r_{s}+1])|$, $T[p_{s}-1..r_{s}+1])$ 
    corresponding to each interval attractor $([p_{s}, q_{s}]$, $[\ell_{s}, r_{s}])$ in set $\Psi_{h_{Q}} \cap \Psi_{\source} \cap \Psi_{\centerset}(C_{Q}) \cap \Psi_{\modulo}(\hat{M}) \cap \Psi_{\preceding} \cap \Psi_{\samp}$. 
\end{itemize}
For this subsection, 
we define three integers $x, y$ and $y^{\prime}$ as follows: 
\begin{itemize}
    \item $x = 1 + \lfloor \frac{\hat{K} - (2 + \sum_{w = 1}^{h_{Q}+3} \lfloor \mu(w) \rfloor)}{|C_{Q}|} \rfloor$;
    \item $y = \min \{ s \in [1, d] \mid T[\gamma_{Q} + \hat{K}..j] \prec R_{s} \}$;
    \item $y^{\prime} = \max \{ s \in [1, d] \mid R_{s} \prec T[\gamma_{Q} + \hat{K}..j]\# \}$. 
\end{itemize}

In the next paragraphs, 
we introduce $(y^{\prime} - y + 1)$ subsets of set $\Psi_{\RR}$ 
and two sequences of integers to explain the relationship between the set $\mathcal{Y}_{C}(h_{Q}, C_{Q}, \hat{M})$ and subquery $\RSSQCX(T[i..j], b)$. 

\paragraph{Subset $\Psi^{C}(t)$.}
For each integer $t \in [1, y^{\prime} - y + 1]$, 
subset $\Psi^{C}(t) \subseteq \Psi_{\RR}$ consists of interval attractors such that 
each interval attractor $([p, q], [\ell, r]) \in \Psi^{C}(t)$ satisfies the following two conditions: 
\begin{itemize}
    \item $([p, q], [\ell, r]) \in \Psi_{h_{Q}} \cap \Psi_{\run} \cap \Psi_{\centerset}(C_{Q}) \cap \Psi_{\lcp}(\hat{K}) \cap \Psi_{\preceding}$;
    \item $R_{y + t - 1} \preceq T[\gamma + \hat{K}..r + 1] \preceq R_{y^{\prime}}$ for the attractor position $\gamma$ of the interval attractor $([p, q], [\ell, r])$.
\end{itemize}
Formally, the subset $\Psi^{C}(t)$ is defined as follows: 
\begin{equation*}
    \begin{split}
    \Psi^{C}(t) &= \{ ([p, q], [\ell, r]) \in \Psi_{\RR} \mid (([p, q], [\ell, r]) \in \Psi_{h_{Q}} \cap \Psi_{\run} \cap \Psi_{\centerset}(C_{Q}) \cap \Psi_{\lcp}(\hat{K}) \cap \Psi_{\preceding}) \\
    &\land (R_{y + t - 1} \preceq T[\gamma + \hat{K}..r + 1] \preceq R_{y^{\prime}}) \}.
    \end{split}
\end{equation*}

\paragraph{Two sequences $\Gamma_{C}$ and $\Gamma_{C, \sub}$.}
The sequence $\Gamma_{C}$ consists of $(y^{\prime} - y + 1)$ integers $u_{1}$, $u_{2}$, $\ldots$, $u_{y^{\prime} - y + 1}$. 
Each integer $u_{t}$ is $1$ if $\mathcal{F}_{\SA} \cap \mathcal{F}_{\suffix}(\Psi_{\CCP}(T[i..j]) \cap \Psi^{C}(t)) \neq \emptyset$; 
otherwise $u_{t}$ is $0$. 
The following lemma states four properties of sequence $\Gamma_{C}$. 

\begin{lemma}\label{lem:GammaC1_property}
Let $\kappa$ be the largest integer in set $[1, y^{\prime} - y + 1]$ satisfying $u_{\kappa} = 1$ for sequence $\Gamma_{C} = u_{1}, u_{2}, \ldots, u_{y^{\prime} - y + 1}$. 
Then, the following four statements hold: 
\begin{enumerate}[label=\textbf{(\roman*)}]
    \item \label{enum:GammaC1_property:1} if the largest integer $\kappa$ exists, then 
    string $T[i..\gamma_{Q} - 1] \cdot C_{Q}^{n+1}[1..\hat{K}] \cdot R_{y + \kappa - 1}$ is 
    the lexicographically largest string in set $\mathcal{F}_{\SA} \cap \mathcal{F}_{\suffix}(\Psi_{\CCP}(T[i..j]) \cap \Psi_{\run} \cap \Psi_{\centerset}(C_{Q}) \cap \Psi_{\lcp}(\hat{K}) \cap \Psi_{\preceding})$. 
    Otherwise, the set $\mathcal{F}_{\SA} \cap \mathcal{F}_{\suffix}(\Psi_{\CCP}(T[i..j]) \cap \Psi_{\run} \cap \Psi_{\centerset}(C_{Q}) \cap \Psi_{\lcp}(\hat{K}) \cap \Psi_{\preceding})$ is empty; 
    \item \label{enum:GammaC1_property:2} if the largest integer $\kappa$ exists, 
    then set $\mathcal{J}_{C}(h_{Q}, C_{Q}, \hat{M})$ contains a weighted point 
    $(|f_{\recover}(([p_{s}$, $q_{s}]$, $[\ell_{s}, r_{s}]))|$, $T[\gamma_{s} + K_{s}..r_{s} + 1]$, $|\Psi_{\str}(T[p_{s}-1..r_{s}+1])|$, $T[p_{s}-1..r_{s}+1])$  
    satisfying $x \leq |f_{\recover}(([p_{s}, q_{s}], [\ell_{s}, r_{s}]))| \leq n$ 
    and $T[\gamma_{s} + K_{s}..r_{s}+1] = R_{y + \kappa - 1}$;
    \item \label{enum:GammaC1_property:3} 
    $T[\gamma_{s} + K_{s} - \hat{K} - |[i, \gamma_{Q}-1]|..r_{s} + 1] = T[i..\gamma_{Q} - 1] \cdot C_{Q}^{n+1}[1..\hat{K}] \cdot R_{y + \kappa - 1}$ for 
    the interval attractor $([p_{s}, q_{s}], [\ell_{s}, r_{s}])$ corresponding to 
    the weighted point $(|f_{\recover}(([p_{s}, q_{s}], [\ell_{s}, r_{s}]))|$, $T[\gamma_{s} + K_{s}..r_{s} + 1]$, $|\Psi_{\str}(T[p_{s}-1..r_{s}+1])|$, $T[p_{s}-1..r_{s}+1])$ 
    of Lemma~\ref{lem:GammaC1_property}~\ref{enum:GammaC1_property:2};
    \item \label{enum:GammaC1_property:4}
    sequence $\Gamma_{C}$ is non-increasing~(i.e., $u_{1} \geq u_{2} \geq \cdots \geq u_{y^{\prime} - y + 1}$). 
\end{enumerate}
\end{lemma}
\begin{proof}
See Section~\ref{subsubsec:GammaC1_property_proof}.
\end{proof}

Next, sequence $\Gamma_{C, \sub}$ consists of $(y^{\prime} - y + 1)$ integers 
$\alpha_{1}, \alpha_{2}, \ldots, \alpha_{y^{\prime} - y + 1} \in \mathbb{N}_{0}$. 
Here, each integer $\alpha_{t}$ is defined as 
$\alpha_{t} = \rangecount(\mathcal{J}_{C}(h_{Q}, C_{Q}, \hat{M}), x, n, R_{y + t - 1}, R_{y^{\prime}})$. 
Here, $\rangecount$ is the range-count query introduced in Section~\ref{subsec:range_data_structure}. 
The following lemma states three properties of sequence $\Gamma_{C, \sub}$.

\begin{lemma}\label{lem:GammaC1_sub_property}
The following three statements hold for two sequences $\Gamma_{C} = u_{1}, u_{2}, \ldots, u_{y^{\prime} - y + 1}$ and $\Gamma_{C, \sub} = \alpha_{1}$, $\alpha_{2}$, $\ldots$, $\alpha_{y^{\prime} - y + 1}$: 
\begin{enumerate}[label=\textbf{(\roman*)}]
    \item \label{enum:GammaC1_sub_property:1} 
    sequence $\Gamma_{C, \sub}$ is non-increasing (i.e., $\alpha_{1} \geq \alpha_{2} \geq \cdots \geq \alpha_{y^{\prime} - y + 1}$);    
    \item \label{enum:GammaC1_sub_property:2} 
    for an integer $t \in [1, y^{\prime} - y + 1]$, 
    consider the largest integer $t^{\prime}$ in set $[t, y^{\prime} - y + 1]$ satisfying 
    $\alpha_{t} = \alpha_{t^{\prime}}$. 
    If $\alpha_{t^{\prime}} \geq 1$, 
    then set $\mathcal{J}_{C}(h_{Q}, C_{Q}, \hat{M})$ contains a weighted point 
    $(|f_{\recover}(([p_{s}, q_{s}], [\ell_{s}, r_{s}]))|$, $T[\gamma_{s} + K_{s}..r_{s} + 1]$, $|\Psi_{\str}(T[p_{s}-1..r_{s}+1])|$, $T[p_{s}-1..r_{s}+1])$ satisfying 
    $x \leq |f_{\recover}(([p_{s}, q_{s}]$, $[\ell_{s}, r_{s}]))| \leq n$ 
    and $T[\gamma_{s} + K_{s}..r_{s}+1] = R_{y + t^{\prime} - 1}$;
    \item \label{enum:GammaC1_sub_property:3}
    consider the three integers $t, t^{\prime}$, and $s$ of Lemma~\ref{lem:GammaC1_sub_property}~\ref{enum:GammaC1_sub_property:2}. 
    Let $\upsilon = \gamma_{s} + K_{s} - \hat{K}$.
    Then, 
    $u_{t} = 1 \Leftrightarrow (\alpha_{t^{\prime}} \geq 1) \land (\RSCQ(\upsilon - |[i, \gamma_{Q}-1]|, \upsilon + |[\gamma_{Q}, j]| - 1) < b - \eta + 1)$. 
\end{enumerate}
\end{lemma}
\begin{proof}
See Section~\ref{subsubsec:GammaC1_sub_property_proof}.
\end{proof}

\subsubsection{Proof of Lemma~\ref{lem:GammaC1_property}}\label{subsubsec:GammaC1_property_proof}
The following two propositions state the relationship between 
two sets $\Psi_{h_{Q}} \cap \Psi_{\run} \cap \Psi_{\centerset}(C_{Q}) \cap \Psi_{\lcp}(\hat{K}) \cap \Psi_{\preceding}$ 
and $\Psi_{h_{Q}} \cap \Psi_{\source} \cap \Psi_{\centerset}(C_{Q}) \cap \Psi_{\modulo}(\hat{M}) \cap \Psi_{\preceding} \cap \Psi_{\samp}$. 

\begin{proposition}\label{prop:JC1_correspondence_property}
Consider an interval attractor $([p, q], [\ell, r]) \in \Psi_{h_{Q}} \cap \Psi_{\run} \cap \Psi_{\centerset}(C_{Q}) \cap \Psi_{\lcp}(\hat{K}) \cap \Psi_{\preceding}$. 
Then, there exists an integer $s \in [1, k]$ satisfying the following two conditions: 
\begin{enumerate}[label=\textbf{(\roman*)}]
    \item $x \leq |f_{\recover}(([p_{s}, q_{s}], [\ell_{s}, r_{s}]))| \leq n$; 
    \item $T[\gamma_{s} + K_{s}..r_{s}+1] = T[\gamma + \hat{K}..r+1]$ 
    for the attractor position $\gamma$ of the interval attractor $([p, q], [\ell, r])$.
\end{enumerate}
\end{proposition}
\begin{proof}
    Lemma~\ref{lem:recover_division_property}~\ref{enum:recover_division_property:1} shows that 
    subset $\Psi_{\source}$ contains an interval attractor $([p_{A}, q_{A}], [\ell_{A}, r_{A}])$ satisfying 
    $([p, q], [\ell, r]) \in f_{\recover}(([p_{A}, q_{A}], [\ell_{A}, r_{A}]))$. 
    Here, $([p_{A}, q_{A}], [\ell_{A}, r_{A}]) \not \in \Psi_{\run}$ holds because 
    $\Psi_{\source} \cap \Psi_{\run} = \emptyset$ follows from the definition of the subset $\Psi_{\source}$. 
    Because of $([p_{A}, q_{A}], [\ell_{A}, r_{A}]) \not \in \Psi_{\run}$, 
    Lemma~\ref{lem:samp_basic_property}~\ref{enum:samp_basic_property:2} shows that 
    the sampling subset $\Psi_{\samp}$ contains an interval attractor $([p_{B}, q_{B}]$, $[\ell_{B}, r_{B}])$ satisfying 
    $T[p_{B}-1..r_{B}+1] = T[p_{A}-1..r_{A}+1]$. 

    We show that there exists an integer $s \in [1, k]$ 
    satisfying $([p_{s}, q_{s}], [\ell_{s}, r_{s}]) = ([p_{B}, q_{B}], [\ell_{B}, r_{B}])$.  
    Because of $([p, q], [\ell, r]) \in \Psi_{h_{Q}} \cap \Psi_{\centerset}(C_{Q}) \cap \Psi_{\lcp}(\hat{K})$, 
    $([p, q], [\ell, r]) \in \Psi_{\modulo}(\hat{M})$ follows from the definition of the subset $\Psi_{\modulo}(\hat{M})$. 
    Because of $([p, q], [\ell, r]) \in \Psi_{h_{Q}} \cap \Psi_{\centerset}(C_{Q}) \cap \Psi_{\modulo}(\hat{M})$,     
    Lemma~\ref{lem:recover_basic_property}~\ref{enum:recover_basic_property:4} shows that     
    $([p_{A}, q_{A}], [\ell_{A}, r_{A}]) \in \Psi_{h_{Q}} \cap \Psi_{\centerset}(C_{Q}) \cap \Psi_{\modulo}(\hat{M})$ holds. 
    Because of $([p, q], [\ell, r]) \in \Psi_{\preceding}$, 
    Lemma~\ref{lem:recover_basic_property}~\ref{enum:recover_basic_property:5} shows that     
    $([p_{A}, q_{A}], [\ell_{A}, r_{A}]) \in \Psi_{\preceding}$ holds. 
    Lemma~\ref{lem:psi_equality_basic_property} shows that 
    $([p_{B}, q_{B}], [\ell_{B}, r_{B}]) \in \Psi_{h_{Q}} \cap \Psi_{\source} \cap \Psi_{\centerset}(C_{Q}) \cap \Psi_{\modulo}(\hat{M}) \cap \Psi_{\preceding}$ holds 
    because 
    $([p_{A}, q_{A}], [\ell_{A}, r_{A}]) \in \Psi_{h_{Q}} \cap \Psi_{\source} \cap \Psi_{\centerset}(C_{Q}) \cap \Psi_{\modulo}(\hat{M}) \cap \Psi_{\preceding}$ and $T[p_{B}-1..r_{B}+1] = T[p_{A}-1..r_{A}+1]$. 
    We showed that $([p_{B}, q_{B}], [\ell_{B}, r_{B}]) \in \Psi_{h_{Q}} \cap \Psi_{\source} \cap \Psi_{\centerset}(C_{Q}) \cap \Psi_{\modulo}(\hat{M}) \cap \Psi_{\preceding} \cap \Psi_{\samp}$ holds. 
    Therefore, there exists an integer $s \in [1, k]$ satisfying $([p_{s}, q_{s}], [\ell_{s}, r_{s}]) = ([p_{B}, q_{B}], [\ell_{B}, r_{B}])$. 

    Let $K_{A} = |\lcp(T[\gamma_{A}..r_{A}], C_{Q}^{n+1})|$ for the attractor position $\gamma_{A}$ of the interval attractor $([p_{A}, q_{A}]$, $[\ell_{A}, r_{A}])$. 
    Because of $([p_{A}, q_{A}], [\ell_{A}, r_{A}]) \in \Psi_{\centerset}(C_{Q})$, 
    $([p_{A}, q_{A}], [\ell_{A}, r_{A}]) \in \Psi_{\lcp}(K_{A})$ follows from the definition of the subset $\Psi_{\lcp}(K_{A})$. 
    Let $m = |f_{\recover}(([p_{A}, q_{A}], [\ell_{A}, r_{A}]))|$ for simplicity. 
    Because of $([p, q], [\ell, r]) \in \Psi_{\centerset}(C_{Q}) \cap \Psi_{\lcp}(\hat{K})$,
    Lemma~\ref{lem:recover_basic_property}~\ref{enum:recover_basic_property:4} shows that 
    there exists an integer $\tau \in [1, m]$ satisfying 
    $K_{A} = \hat{K} + \tau |C_{Q}|$. 
    Lemma~\ref{lem:recover_basic_property}~\ref{enum:recover_basic_property:2} shows that 
    $\gamma = \gamma_{A} + \tau |C_{Q}|$ and $r = r_{A}$ hold. 

    We prove $x \leq |f_{\recover}(([p_{s}, q_{s}], [\ell_{s}, r_{s}]))| \leq n$. 
    Lemma~\ref{lem:recover_basic_property}~\ref{enum:recover_basic_property:1} shows that 
    $m = \lfloor \frac{K_{A} - (2 + \sum_{w = 1}^{h_{Q}+3} \lfloor \mu(w) \rfloor)}{|C_{Q}|} \rfloor$ and $1 \leq m \leq n$ hold. 
    Because of $K_{A} = \hat{K} + \tau |C_{Q}|$ and $x = 1 + \lfloor \frac{\hat{K} - (2 + \sum_{w = 1}^{h_{Q}+3} \lfloor \mu(w) \rfloor)}{|C_{Q}|} \rfloor$, 
    $m = \tau + x - 1$ follows from the following equation: 
\begin{equation*}
    \begin{split}
    m &= \lfloor \frac{K_{A} - (2 + \sum_{w = 1}^{h_{Q}+3} \lfloor \mu(w) \rfloor)}{|C_{Q}|} \rfloor \\
    &= \tau + \lfloor \frac{\hat{K} - (2 + \sum_{w = 1}^{h_{Q}+3} \lfloor \mu(w) \rfloor)}{|C_{Q}|} \rfloor \\
    &= \tau + x - 1.
    \end{split}
\end{equation*}
    Because of $T[p_{s}-1..r_{s}+1] = T[p_{A}-1..r_{A}+1]$, 
    Lemma~\ref{lem:recover_super_property}~\ref{enum:recover_super_property:1} shows that 
    $|f_{\recover}(([p_{s}, q_{s}], [\ell_{s}, r_{s}]))| = m$ holds. 
    Therefore, $x \leq |f_{\recover}(([p_{s}, q_{s}], [\ell_{s}, r_{s}]))| \leq n$ 
    follows from 
    (1) $|f_{\recover}(([p_{s}, q_{s}], [\ell_{s}, r_{s}]))| = m$, 
    (2) $m = \tau + x - 1$, 
    (3) $\tau \in [1, m]$, 
    and (4) $1 \leq m \leq n$. 

    We prove $T[\gamma_{s} + K_{s}..r_{s}+1] = T[\gamma + \hat{K}..r+1]$. 
    $T[\gamma_{A} + K_{A}..r_{A}+1] = T[\gamma + \hat{K}..r+1]$ follows from 
    $\gamma_{A} = \gamma - \tau |C_{Q}|$, $K_{A} = \hat{K} + \tau |C_{Q}|$, 
    and $r_{A} = r$. 
    Lemma~\ref{lem:psi_equality_basic_property}~\ref{enum:psi_equality_basic_property:8} shows that 
    $K_{s} = K_{A}$ because 
    $([p_{s}, q_{s}], [\ell_{s}, r_{s}]) \in \Psi_{\lcp}(K_{s})$, 
    $([p_{A}, q_{A}], [\ell_{A}, r_{A}]) \in \Psi_{\lcp}(K_{A})$, 
    and $T[p_{s}-1..r_{s}+1] = T[p_{A}-1..r_{A}+1]$. 
    Lemma~\ref{lem:psi_str_property}~\ref{enum:psi_str_property:1} shows that 
    $T[\gamma_{s}..r_{s}+1] = T[\gamma_{A}..r_{A}+1]$ holds. 
    Therefore, $T[\gamma_{s} + K_{s}..r_{s}+1] = T[\gamma + \hat{K}..r+1]$ follows from 
    (1) $T[\gamma_{A} + K_{A}..r_{A}+1] = T[\gamma + \hat{K}..r+1]$, 
    (2) $T[\gamma_{s}..r_{s}+1] = T[\gamma_{A}..r_{A}+1]$, 
    and (3) $K_{s} = K_{A}$. 

    We showed that $x \leq |f_{\recover}(([p_{s}, q_{s}], [\ell_{s}, r_{s}]))| \leq n$ and $T[\gamma_{s} + K_{s}..r_{s}+1] = T[\gamma + \hat{K}..r+1]$ hold. 
    Therefore, Proposition~\ref{prop:JC1_correspondence_property} holds. 
\end{proof}

\begin{proposition}\label{prop:JC1_inverse_correspondence_property}
For an integer $s \in [1, k]$, 
if $x \leq |f_{\recover}(([p_{s}, q_{s}], [\ell_{s}, r_{s}]))| \leq n$ holds, 
then set $\Psi_{h_{Q}} \cap \Psi_{\run} \cap \Psi_{\centerset}(C_{Q}) \cap \Psi_{\lcp}(\hat{K}) \cap \Psi_{\preceding}$ contains an interval attractor $([p, q], [\ell, r])$ satisfying 
$T[\gamma_{s} + K_{s}..r_{s}+1] = T[\gamma + \hat{K}..r+1]$ and $\gamma = \gamma_{s} + K_{s} - \hat{K}$ 
for the attractor position $\gamma$ of the interval attractor $([p, q], [\ell, r])$.
\end{proposition}
\begin{proof}
    Let $m = |f_{\recover}(([p_{s}, q_{s}], [\ell_{s}, r_{s}]))|$ for simplicity.     
    We prove $K_{s} = \hat{K} + (m-x+1) |C_{Q}|$. 
    Because of $\hat{M} = (\hat{K} - (2 + \sum_{w = 1}^{h_{Q}+3} \lfloor \mu(w) \rfloor) ) \mod |C_{Q}|$, 
    the following equation holds: 
\begin{equation}\label{eq:JC1_inverse_correspondence_property:1}
    \begin{split}
    \hat{K} - (2 + \sum_{w = 1}^{h_{Q}+3} \lfloor \mu(w) \rfloor) &= \hat{M} + |C_{Q}| \lfloor \frac{\hat{K} - (2 + \sum_{w = 1}^{h_{Q}+3} \lfloor \mu(w) \rfloor)}{|C_{Q}|} \rfloor \\
    &= \hat{M} + (x - 1) |C_{Q}|.        
    \end{split}
\end{equation}
    Because of $([p_{s}, q_{s}], [\ell_{s}, r_{s}]) \in \Psi_{h_{Q}} \cap \Psi_{\centerset}(C_{Q}) \cap \Psi_{\lcp}(K_{s})$, 
    Lemma~\ref{lem:recover_basic_property}~\ref{enum:recover_basic_property:1} shows that 
    $m = \lfloor \frac{K_{s} - (2 + \sum_{w = 1}^{h_{Q}+3} \lfloor \mu(w) \rfloor)}{|C_{Q}|} \rfloor$ holds.     
    Because of $([p_{s}, q_{s}], [\ell_{s}, r_{s}]) \in \Psi_{h_{Q}} \cap \Psi_{\centerset}(C_{Q}) \cap \Psi_{\modulo}(\hat{M})$, 
    $(K_{s} - (2 + \sum_{w = 1}^{h_{Q}+3} \lfloor \mu(w) \rfloor) ) \mod |C_{Q}| = \hat{M}$ follows from the definition of the subset $\Psi_{\modulo}(\hat{M})$. 
    The following equation follows from $(K_{s} - (2 + \sum_{w = 1}^{h_{Q}+3} \lfloor \mu(w) \rfloor) ) \mod |C_{Q}| = \hat{M}$: 
\begin{equation}\label{eq:JC1_inverse_correspondence_property:2}
    \begin{split}
    K_{s} - (2 + \sum_{w = 1}^{h_{Q}+3} \lfloor \mu(w) \rfloor) &= \hat{M} + |C_{Q}| \lfloor \frac{K_{s} - (2 + \sum_{w = 1}^{h_{Q}+3} \lfloor \mu(w) \rfloor)}{|C_{Q}|} \rfloor \\
    &= \hat{M} + m |C_{Q}|.        
    \end{split}
\end{equation}
    Therefore, $K_{s} = \hat{K} + (m-x+1) |C_{Q}|$ follows from 
    Equation~\ref{eq:JC1_inverse_correspondence_property:1} and Equation~\ref{eq:JC1_inverse_correspondence_property:2}.

    Because of $m-x+1 \in [1, m]$, 
    Lemma~\ref{lem:recover_basic_property}~\ref{enum:recover_basic_property:2} shows that 
    there exists an interval attractor $([p, q], [\ell, r])$ in set $f_{\recover}(([p_{s}, q_{s}], [\ell_{s}, r_{s}]))$ satisfying 
    the following three conditions: 
    \begin{itemize}
        \item $p = q_{s} + 1 + (m-x) |C_{Q}|$;
        \item $r = r_{s}$;
        \item $\gamma = \gamma_{s} + (m-x+1)|C_{Q}|$ for the attractor position $\gamma$ of the interval attractor $([p, q], [\ell, r])$.
    \end{itemize}
    Lemma~\ref{lem:recover_basic_property}~\ref{enum:recover_basic_property:4} shows that 
    $([p, q], [\ell, r]) \in \Psi_{h_{Q}} \cap \Psi_{\run} \cap \Psi_{\centerset}(C_{Q}) \cap \Psi_{\modulo}(\hat{M}) \cap \Psi_{\lcp}(K_{s} - (m-x+1) |C_{Q}|)$ holds. 
    Here, $([p, q], [\ell, r]) \in \Psi_{\lcp}(\hat{K})$ follows from $([p, q], [\ell, r]) \in \Psi_{\lcp}(K_{s} - (m-x+1) |C_{Q}|)$ 
    and $K_{s} = \hat{K} + (m-x+1) |C_{Q}|$. 
    Because of $([p_{s}, q_{s}], [\ell_{s}, r_{s}]) \in \Psi_{\preceding}$, 
    Lemma~\ref{lem:recover_basic_property}~\ref{enum:recover_basic_property:5} shows that 
    $([p, q], [\ell, r]) \in \Psi_{\preceding}$ holds. 
    
    We prove $T[\gamma_{s} + K_{s}..r_{s}+1] = T[\gamma + \hat{K}..r+1]$ and $\gamma = \gamma_{s} + K_{s} - \hat{K}$. 
    $\gamma = \gamma_{s} + K_{s} - \hat{K}$ follows from $\gamma = \gamma_{s} + (m-x+1)|C_{Q}|$ and $K_{s} = \hat{K} + (m-x+1) |C_{Q}|$. 
    $T[\gamma_{s} + K_{s}..r_{s}+1] = T[\gamma + \hat{K}..r+1]$ follows from 
    $\gamma = \gamma_{s} + K_{s} - \hat{K}$ and $r = r_{s}$. 

    We proved $([p, q], [\ell, r]) \in \Psi_{h_{Q}} \cap \Psi_{\run} \cap \Psi_{\centerset}(C_{Q}) \cap \Psi_{\lcp}(\hat{K}) \cap \Psi_{\preceding}$, $T[\gamma_{s} + K_{s}..r_{s}+1] = T[\gamma + \hat{K}..r+1]$, and $\gamma = \gamma_{s} + K_{s} - \hat{K}$. 
    Therefore, Proposition~\ref{prop:JC1_inverse_correspondence_property} holds.     
\end{proof}

The following proposition states three properties of subset $\Psi^{C}(t)$ for each integer $t \in [1, y^{\prime} - y + 1]$. 
\begin{proposition}\label{prop:Psi_C1_Property}
Consider condition (A) of RSS query for the given RSS query $\RSSQ(T[i..j], b)$.
The following three statements hold: 
\begin{enumerate}[label=\textbf{(\roman*)}]
    \item \label{enum:Psi_C1_Property:1} 
    $\Psi^{C}(1) = \Psi_{\CCP}(T[i..j]) \cap \Psi_{\run} \cap \Psi_{\centerset}(C_{Q}) \cap \Psi_{\lcp}(\hat{K}) \cap \Psi_{\preceding}$;
    \item \label{enum:Psi_C1_Property:2} 
    $\Psi^{C}(t) \supseteq \Psi^{C}(t+1)$ for each integer $t \in [1, y^{\prime} - y + 1]$;
    \item \label{enum:Psi_C1_Property:3}
    consider an interval attractor $([p, q], [\ell, r])$ in set $\Psi^{C}(t) \setminus \Psi^{C}(t+1)$ for an integer $t \in [1, y^{\prime} - y + 1]$. 
    Then, $T[\gamma - |[i, \gamma_{Q}-1]|..\gamma-1] \cdot T[\gamma..r+1] = T[i..\gamma_{Q}-1] \cdot C_{Q}^{n+1}[1..\hat{K}] \cdot R_{y + t - 1}$ holds 
    for the attractor position $\gamma$ of the interval attractor $([p, q], [\ell, r])$.
\end{enumerate}
Here, let $\Psi^{C}(y^{\prime} - y + 2) = \emptyset$ for simplicity. 
\end{proposition}
\begin{proof}
The following two statements are used to prove Proposition~\ref{prop:Psi_C1_Property}:
\begin{enumerate}[label=\textbf{(\Alph*)}]
    \item $\Psi^{C}(1) \subseteq \Psi_{\CCP}(T[i..j]) \cap \Psi_{\run} \cap \Psi_{\centerset}(C_{Q}) \cap \Psi_{\lcp}(\hat{K}) \cap \Psi_{\preceding}$;
    \item $\Psi^{C}(1) \supseteq \Psi_{\CCP}(T[i..j]) \cap \Psi_{\run} \cap \Psi_{\centerset}(C_{Q}) \cap \Psi_{\lcp}(\hat{K}) \cap \Psi_{\preceding}$.    
\end{enumerate}

\textbf{Proof of statement (A).}
Consider an interval attractor $([p, q], [\ell, r]) \in \Psi^{C}(1)$. 
Then, 
$([p, q]$, $[\ell, r]) \in \Psi_{h_{Q}} \cap \Psi_{\run} \cap \Psi_{\centerset}(C_{Q}) \cap \Psi_{\lcp}(\hat{K}) \cap \Psi_{\preceding}$ 
and $R_{y} \preceq T[\gamma + \hat{K}..r + 1] \preceq R_{y^{\prime}}$ follow from the definition of the subset $\Psi^{C}(1)$ 
for the attractor position $\gamma$ of the interval attractor $([p, q], [\ell, r])$.

We prove $\reverse(T[i..\gamma_{Q}-1]) \prec \reverse(T[p-1..\gamma-1]) \prec \reverse(\#T[i..\gamma_{Q}-1])$. 
We can apply Lemma~\ref{lem:suffix_syncro} to the two interval attractors 
$([p_{Q}, q_{Q}], [\ell_{Q}, r_{Q}])$ and $([p, q], [\ell, r])$ because 
the following three statements hold: 
\begin{itemize}
    \item $([p, q], [\ell, r]) \in \Psi_{h_{Q}} \cap \Psi_{\run} \cap \Psi_{\centerset}(C_{Q})$;
    \item $i \in [p, q]$, 
    and $\lcs(T[i..\gamma_{Q}-1], C_{Q}^{n+1}) = T[i..\gamma_{Q}-1]$ for condition (A) of RSS query; 
    \item $|\lcp(T[\gamma_{Q}..r_{Q}], C_{Q}^{n+1})| > \sum_{w = 1}^{h_{Q}+3} \lfloor \mu(w) \rfloor$ for condition (A) of RSS query.
\end{itemize}
Lemma~\ref{lem:suffix_syncro} shows that 
the string $T[i..\gamma_{Q}-1]$ is a suffix of string $T[p..\gamma-1]$. 
Therefore, $\reverse(T[i..\gamma_{Q}-1]) \prec \reverse(T[p-1..\gamma-1]) \prec \reverse(\#T[i..\gamma_{Q}-1])$ holds. 

We prove $T[\gamma_{Q}..j] \prec T[\gamma..r+1] \prec T[\gamma_{Q}..j]\#$. 
Because of $([p, q], [\ell, r]) \in \Psi_{\centerset}(C_{Q}) \cap \Psi_{\lcp}(\hat{K})$, 
$T[\gamma..r+1] = C_{Q}^{n+1}[1..\hat{K}] \cdot T[\gamma + \hat{K}..r+1]$ follows from the definition of the subset $\Psi_{\lcp}(\hat{K})$. 
$T[\gamma_{Q} + \hat{K}..j] \prec T[\gamma + \hat{K}..r+1] \prec T[\gamma_{Q} + \hat{K}..j]\#$ 
follows from 
$R_{y} \preceq T[\gamma + \hat{K}..r + 1] \preceq R_{y^{\prime}}$, 
$y = \min \{ s \in [1, d] \mid T[\gamma_{Q} + \hat{K}..j] \prec R_{s} \}$, 
and $y^{\prime} = \max \{ s \in [1, d] \mid R_{s} \prec T[\gamma_{Q} + \hat{K}..j]\# \}$. 
$1 + \sum_{w = 1}^{h_{Q}+3} \lfloor \mu(w) \rfloor < \hat{K} < |[\gamma_{Q}, j]|$ follows from condition (A) of RSS query. 
$T[\gamma_{Q}..j] = C_{Q}^{n+1}[1..\hat{K}] \cdot T[\gamma_{Q} + \hat{K}..j]$ follows from $\hat{K} < |[\gamma_{Q}, j]|$. 
Therefore, $T[\gamma_{Q}..j] \prec T[\gamma..r+1] \prec T[\gamma_{Q}..j]\#$ follows from 
(1) $T[\gamma_{Q} + \hat{K}..j] \prec T[\gamma + \hat{K}..r+1] \prec T[\gamma_{Q} + \hat{K}..j]\#$, 
(2) $T[\gamma_{Q}..j] = C_{Q}^{n+1}[1..\hat{K}] \cdot T[\gamma_{Q} + \hat{K}..j]$, 
(3) $T[\gamma..r+1] = C_{Q}^{n+1}[1..\hat{K}] \cdot T[\gamma + \hat{K}..r+1]$, 
and (4) $T[\gamma_{Q} + \hat{K}..j] \prec T[\gamma + \hat{K}..r+1] \prec T[\gamma_{Q} + \hat{K}..j]\#$. 

We prove $([p, q], [\ell, r]) \in \Psi_{\CCP}(T[i..j])$. 
Lemma~\ref{lem:CCP_property}~\ref{enum:CCP_property:4} shows that 
$([p, q], [\ell, r]) \in \Psi_{\CCP}(T[i..j])$ holds if 
$([p, q], [\ell, r]) \in \Psi_{h_{Q}}$, 
$\reverse(T[i..\gamma_{Q}-1]) \prec \reverse(T[p-1..\gamma-1]) \prec \reverse(\#T[i..\gamma_{Q}-1])$, 
and $T[\gamma_{Q}..j] \prec T[\gamma..r+1] \prec T[\gamma_{Q}..j]\#$ hold. 
We already showed that  
$([p, q], [\ell, r]) \in \Psi_{h_{Q}}$, 
$\reverse(T[i..\gamma_{Q}-1]) \prec \reverse(T[p-1..\gamma-1]) \prec \reverse(\#T[i..\gamma_{Q}-1])$, 
and $T[\gamma_{Q}..j] \prec T[\gamma..r+1] \prec T[\gamma_{Q}..j]\#$ hold. 
Therefore, $([p, q], [\ell, r]) \in \Psi_{\CCP}(T[i..j])$ holds. 

We showed that $([p, q], [\ell, r]) \in \Psi_{\CCP}(T[i..j]) \cap \Psi_{\run} \cap \Psi_{\centerset}(C_{Q}) \cap \Psi_{\lcp}(\hat{K}) \cap \Psi_{\preceding}$ for each interval attractor $([p, q], [\ell, r]) \in \Psi^{C}(1)$. 
Therefore, $\Psi^{C}(1) \subseteq \Psi_{\CCP}(T[i..j]) \cap \Psi_{\run} \cap \Psi_{\centerset}(C_{Q}) \cap \Psi_{\lcp}(\hat{K}) \cap \Psi_{\preceding}$ holds. 

\textbf{Proof of statement (B).}
Consider an interval attractor $([p, q], [\ell, r]) \in \Psi_{\CCP}(T[i..j]) \cap \Psi_{\run} \cap \Psi_{\centerset}(C_{Q}) \cap \Psi_{\lcp}(\hat{K}) \cap \Psi_{\preceding}$. 
Because of $([p, q], [\ell, r]) \in \Psi_{\CCP}(T[i..j])$, 
Lemma~\ref{lem:CCP_property}~\ref{enum:CCP_property:4} shows that 
$T[\gamma_{Q}..j] \prec T[\gamma..r+1] \prec T[\gamma_{Q}..j]\#$ holds for the attractor position $\gamma$ of the interval attractor $([p, q], [\ell, r])$. 
Lemma~\ref{lem:CCP_property}~\ref{enum:CCP_property:1} shows that $([p, q], [\ell, r]) \in \Psi_{h_{Q}}$ holds. 
Because of $([p, q], [\ell, r]) \in \Psi_{h_{Q}} \cap \Psi_{\run} \cap \Psi_{\centerset}(C_{Q}) \cap \Psi_{\lcp}(\hat{K}) \cap \Psi_{\preceding}$, 
Proposition~\ref{prop:JC1_correspondence_property} shows that 
there exists an integer $s \in [1, k]$ satisfying 
$T[\gamma_{s} + K_{s}..r_{s}+1] = T[\gamma + \hat{K}..r+1]$.

We prove $T[\gamma_{Q} + \hat{K}..j] \prec T[\gamma + \hat{K}..r+1] \prec T[\gamma_{Q} + \hat{K}..j]\#$. 
Because of $([p, q], [\ell, r]) \in \Psi_{\centerset}(C_{Q}) \cap \Psi_{\lcp}(\hat{K})$, 
$T[\gamma..r+1] = C_{Q}^{n+1}[1..\hat{K}] \cdot T[\gamma + \hat{K}..r+1]$ follows from the definition of the subset $\Psi_{\lcp}(\hat{K})$. 
$1 + \sum_{w = 1}^{h_{Q}+3} \lfloor \mu(w) \rfloor < \hat{K} < |[\gamma_{Q}, j]|$ follows from condition (A) of RSS query. 
$T[\gamma_{Q}..j] = C_{Q}^{n+1}[1..\hat{K}] \cdot T[\gamma_{Q} + \hat{K}..j]$ follows from $\hat{K} < |[\gamma_{Q}, j]|$. 
Therefore, $T[\gamma_{Q} + \hat{K}..j] \prec T[\gamma + \hat{K}..r+1] \prec T[\gamma_{Q} + \hat{K}..j]\#$ follows from 
$T[\gamma_{Q}..j] \prec T[\gamma..r+1] \prec T[\gamma_{Q}..j]\#$, 
$T[\gamma..r+1] = C_{Q}^{n+1}[1..\hat{K}] \cdot T[\gamma + \hat{K}..r+1]$, 
and $T[\gamma_{Q}..j] = C_{Q}^{n+1}[1..\hat{K}] \cdot T[\gamma_{Q} + \hat{K}..j]$. 

We prove $R_{y} \preceq T[\gamma + \hat{K}..r + 1] \preceq R_{y^{\prime}}$. 
Because of $T[\gamma_{s} + K_{s}..r_{s}+1] \in \mathcal{Y}_{C}(h_{Q}, C_{Q}, \hat{M}))$, 
there exists an integer $\tau \in [1, d]$ satisfying $R_{\tau} = T[\gamma_{s} + K_{s}..r_{s}+1]$. 
$T[\gamma_{Q} + \hat{K}..j] \prec R_{\tau} \prec T[\gamma_{Q} + \hat{K}..j]\#$ follows from 
$T[\gamma_{Q} + \hat{K}..j] \prec T[\gamma + \hat{K}..r+1] \prec T[\gamma_{Q} + \hat{K}..j]\#$, 
$T[\gamma_{s} + K_{s}..r_{s}+1] = T[\gamma + \hat{K}..r+1]$, and $R_{\tau} = T[\gamma_{s} + K_{s}..r_{s}+1]$. 
$R_{y} \preceq R_{\tau} \preceq R_{y^{\prime}}$ follows from 
$T[\gamma_{Q} + \hat{K}..j] \prec R_{\tau} \prec T[\gamma_{Q} + \hat{K}..j]\#$, 
$y = \min \{ s \in [1, d] \mid T[\gamma_{Q} + \hat{K}..j] \prec R_{s} \}$, 
and $y^{\prime} = \max \{ s \in [1, d] \mid R_{s} \prec T[\gamma_{Q} + \hat{K}..j]\# \}$. 
Therefore, $R_{y} \preceq T[\gamma + \hat{K}..r + 1] \preceq R_{y^{\prime}}$ follows from 
$R_{y} \preceq R_{\tau} \preceq R_{y^{\prime}}$, $R_{\tau} = T[\gamma_{s} + K_{s}..r_{s}+1]$, 
and $T[\gamma_{s} + K_{s}..r_{s}+1] = T[\gamma + \hat{K}..r+1]$. 

We prove $([p, q], [\ell, r]) \in \Psi^{C}(1)$. 
From the definition of the subset $\Psi^{C}(1)$, 
$([p, q], [\ell, r]) \in \Psi^{C}(1)$ holds if 
$([p, q], [\ell, r]) \in \Psi_{h_{Q}} \cap \Psi_{\run} \cap \Psi_{\centerset}(C_{Q}) \cap \Psi_{\lcp}(\hat{K}) \cap \Psi_{\preceding}$ 
and $R_{y} \preceq T[\gamma + \hat{K}..r + 1] \preceq R_{y^{\prime}}$. 
We already proved $([p, q], [\ell, r]) \in \Psi_{h_{Q}} \cap \Psi_{\run} \cap \Psi_{\centerset}(C_{Q}) \cap \Psi_{\lcp}(\hat{K}) \cap \Psi_{\preceding}$ and $R_{y} \preceq T[\gamma + \hat{K}..r + 1] \preceq R_{y^{\prime}}$. 
Therefore, $([p, q], [\ell, r]) \in \Psi^{C}(1)$ holds. 

We proved $([p, q], [\ell, r]) \in \Psi^{C}(1)$ for each interval attractor $([p, q], [\ell, r]) \in \Psi_{\CCP}(T[i..j]) \cap \Psi_{\run} \cap \Psi_{\centerset}(C_{Q}) \cap \Psi_{\lcp}(\hat{K}) \cap \Psi_{\preceding}$. 
Therefore, $\Psi^{C}(1) \supseteq \Psi_{\CCP}(T[i..j]) \cap \Psi_{\run} \cap \Psi_{\centerset}(C_{Q}) \cap \Psi_{\lcp}(\hat{K}) \cap \Psi_{\preceding}$ holds. 

\textbf{Proof of Proposition~\ref{prop:Psi_C1_Property}(i).}
Proposition~\ref{prop:Psi_C1_Property}(i) follows from statement (A) and statement (B). 

\textbf{Proof of Proposition~\ref{prop:Psi_C1_Property}(ii).}
Proposition~\ref{prop:Psi_C1_Property}(ii) follows from the definitions of the two subsets $\Psi^{C}(t)$ and $\Psi^{C}(t+1)$. 

\textbf{Proof of Proposition~\ref{prop:Psi_C1_Property}(iii).}
We prove $T[\gamma - |[i, \gamma_{Q}-1]|..\gamma-1] = T[i..\gamma_{Q}-1]$. 
Because of $([p, q], [\ell, r]) \in \Psi^{C}(t)$
$([p, q], [\ell, r]) \in \Psi_{\CCP}(T[i..j])$ follows from 
Proposition~\ref{prop:Psi_C1_Property}~\ref{enum:Psi_C1_Property:1} and Proposition~\ref{prop:Psi_A_Property}~\ref{enum:Psi_A_Property:2}. 
Because of $([p, q], [\ell, r]) \in \Psi_{\CCP}(T[i..j])$, 
Lemma~\ref{lem:CCP_property}~\ref{enum:CCP_property:6} shows that 
$T[\gamma - |[i, \gamma_{Q}-1]|..\gamma-1] = T[i..\gamma_{Q}-1]$ holds. 

Because of $([p, q], [\ell, r]) \in \Psi^{C}(t) \setminus \Psi^{C}(t+1)$, 
$([p, q], [\ell, r]) \in \Psi_{h_{Q}} \cap \Psi_{\run} \cap \Psi_{\centerset}(C_{Q}) \cap \Psi_{\lcp}(\hat{K}) \cap \Psi_{\preceding}$ 
and $R_{y + t - 1} \preceq T[\gamma + \hat{K}..r + 1] \prec R_{y+t}$ follow from 
the definitions of the two subsets $\Psi^{C}(t)$ and $\Psi^{C}(t+1)$. 
Because of $([p, q], [\ell, r]) \in \Psi_{h_{Q}} \cap \Psi_{\run} \cap \Psi_{\centerset}(C_{Q}) \cap \Psi_{\lcp}(\hat{K}) \cap \Psi_{\preceding}$, 
Proposition~\ref{prop:JC1_correspondence_property} shows that 
there exists an integer $s \in [1, k]$ satisfying 
$T[\gamma_{s} + K_{s}..r_{s}+1] = T[\gamma + \hat{K}..r+1]$. 
Because of $T[\gamma_{s} + K_{s}..r_{s}+1] \in \mathcal{Y}_{C}(h_{Q}, C_{Q}, \hat{M}))$, 
there exists an integer $\tau \in [1, d]$ satisfying $R_{\tau} = T[\gamma_{s} + K_{s}..r_{s}+1]$. 

We prove $\tau = y+t-1$. 
$R_{y + t - 1} \preceq R_{\tau} \prec R_{y+t}$ follows from 
$R_{y + t - 1} \preceq T[\gamma + \hat{K}..r + 1] \prec R_{y+t}$, 
$T[\gamma_{s} + K_{s}..r_{s}+1] = T[\gamma + \hat{K}..r+1]$, and $R_{\tau} = T[\gamma_{s} + K_{s}..r_{s}+1]$. 
Therefore, $\tau = y+t-1$ follows from 
$R_{y + t - 1} \preceq R_{\tau} \prec R_{y+t}$ and $R_{1} \prec R_{2} \prec \cdots \prec R_{d}$. 

We prove $T[\gamma - |[i, \gamma_{Q}-1]|..\gamma-1] \cdot T[\gamma..r+1] = T[i..\gamma_{Q}-1] \cdot C_{Q}^{n+1}[1..\hat{K}] \cdot R_{y + t - 1}$. 
Because of $([p, q], [\ell, r]) \in \Psi_{\centerset}(C_{Q}) \cap \Psi_{\lcp}(\hat{K})$, 
$T[\gamma..r+1] = C_{Q}^{n+1}[1..\hat{K}] \cdot T[\gamma + \hat{K}..r+1]$ follows from the definition of the subset $\Psi_{\lcp}(\hat{K})$. 
$T[\gamma + \hat{K}..r+1] = R_{y + t - 1}$ follows from 
$T[\gamma + \hat{K}..r+1] = T[\gamma_{s} + K_{s}..r_{s}+1]$, $T[\gamma_{s} + K_{s}..r_{s}+1] = R_{\tau}$, 
and $\tau = y+t-1$. 
Therefore, $T[\gamma - |[i, \gamma_{Q}-1]|..\gamma-1] \cdot T[\gamma..r+1] = T[i..\gamma_{Q}-1] \cdot C_{Q}^{n+1}[1..\hat{K}] \cdot T[\gamma_{s} + K_{s}..r_{s}+1]$ follows from 
$T[\gamma - |[i, \gamma_{Q}-1]|..\gamma-1] = T[i..\gamma_{Q}-1]$, 
$T[\gamma..r+1] = C_{Q}^{n+1}[1..\hat{K}] \cdot T[\gamma + \hat{K}..r+1]$, 
and $T[\gamma + \hat{K}..r+1] = R_{y + t - 1}$. 
\end{proof}

For proving Lemma~\ref{lem:GammaC1_property}, 
we introduce a set $\mathcal{I}^{C}$ of integers in set $\{ 1, 2, \ldots, y^{\prime} - y + 1 \}$. 
This set $\mathcal{I}^{C}$ consists of integers such that 
for each integer $t \in \mathcal{I}^{C}$, 
set $\Psi^{C}(t) \setminus \Psi^{C}(t+1)$ contains an interval attractor $([p, q], [\ell, r])$ satisfying 
$\RSCQ(\gamma - |[i, \gamma_{Q}-1]|, \gamma + |[\gamma_{Q}, j]| - 1) < b - \eta + 1$ for the attractor position $\gamma$ of the interval attractor $([p, q], [\ell, r])$. 
Formally, $\mathcal{I}^{C} = \{ t \in [1, y^{\prime} - y + 1] \mid \exists ([p, q], [\ell, r]) \in \Psi^{C}(t) \setminus \Psi^{C}(t+1) \text{ s.t. } \RSCQ(\gamma - |[i, \gamma_{Q}-1]|, \gamma + |[\gamma_{Q}, j]| - 1) < b - \eta + 1 \}$. 
Here, $\Psi^{C}(y^{\prime} - y + 2) = \emptyset$ for simplicity.

The following proposition states three properties of the set $\mathcal{I}^{C}$. 

\begin{proposition}\label{prop:Set_IC1_Property}
    Consider condition (A) of RSS query for the given RSS query $\RSSQ(T[i..j], b)$.
    The following three statements hold for set $\mathcal{I}^{C}$ and sequence $\Gamma_{C} = u_{1}, u_{2}, \ldots, u_{y^{\prime} - y + 1 }$: 
\begin{enumerate}[label=\textbf{(\roman*)}]
    \item \label{enum:Set_IC1_Property:1} $\mathcal{F}_{\SA} \cap \mathcal{F}_{\suffix}(\Psi_{\CCP}(T[i..j]) \cap \Psi_{\run} \cap \Psi_{\centerset}(C_{Q}) \cap \Psi_{\lcp}(\hat{K}) \cap \Psi_{\preceding}) = \{ T[i..\gamma_{Q} - 1] \cdot C_{Q}^{n+1}[1..\hat{K}] \cdot R_{y + t - 1} \mid t \in \mathcal{I}^{C} \}$ for the attractor position $\gamma_{Q}$ of interval attractor $([p_{Q}, q_{Q}], [\ell_{Q}, r_{Q}])$; 
    \item \label{enum:Set_IC1_Property:2} $t \in \mathcal{I}^{C}$ for each integer $t \in [1, y^{\prime} - y + 1]$ satisfying $u_{t} = 1$ and $u_{t+1} = 0$;
    \item \label{enum:Set_IC1_Property:3} $u_{t^{\prime}} = 1$ for 
    any pair of two integers $t \in \mathcal{I}^{C}$ and $t^{\prime} \in [1, t]$. 
\end{enumerate}
Here, let $u_{y^{\prime} - y + 2} = 0$ and $\Psi^{C}(y^{\prime} - y + 2) = \emptyset$ for simplicity. 
\end{proposition}
\begin{proof}
The following two statements are used to prove Proposition~\ref{prop:Set_IC1_Property}: 
\begin{enumerate}[label=\textbf{(\Alph*)}]
    \item $\mathcal{F}_{\SA} \cap \mathcal{F}_{\suffix}(\Psi_{\CCP}(T[i..j]) \cap \Psi_{\run} \cap \Psi_{\centerset}(C_{Q}) \cap \Psi_{\lcp}(\hat{K}) \cap \Psi_{\preceding}) \subseteq \{ T[i..\gamma_{Q} - 1] \cdot C_{Q}^{n+1}[1..\hat{K}] \cdot R_{y + t - 1} \mid t \in \mathcal{I}^{C} \}$;
    \item $\mathcal{F}_{\SA} \cap \mathcal{F}_{\suffix}(\Psi_{\CCP}(T[i..j]) \cap \Psi_{\run} \cap \Psi_{\centerset}(C_{Q}) \cap \Psi_{\lcp}(\hat{K}) \cap \Psi_{\preceding}) \supseteq \{ T[i..\gamma_{Q} - 1] \cdot C_{Q}^{n+1}[1..\hat{K}] \cdot R_{y + t - 1} \mid t \in \mathcal{I}^{C} \}$.
\end{enumerate}

\textbf{Proof of statement (A).}
We can prove this statement using the same approach as for statement (A) in the proof of Proposition~\ref{prop:Set_IA_Property}. 
The detailed proof of this statement is as follows. 

The following equation follows from Proposition~\ref{prop:Psi_C1_Property}~\ref{enum:Psi_C1_Property:1}, 
Proposition~\ref{prop:Psi_C1_Property}~\ref{enum:Psi_C1_Property:2}, and $\Psi^{C}(y^{\prime} - y + 2) = \emptyset$: 
\begin{equation}\label{eq:Set_IC1_Property:1}
    \begin{split}
    \Psi_{\CCP}(T[i..j]) \cap \Psi_{\run} \cap \Psi_{\centerset}(C_{Q}) & \cap \Psi_{\lcp}(\hat{K}) \cap \Psi_{\preceding} \\
    &= \bigcup_{t = 1}^{y^{\prime} - y + 1} \Psi^{C}(t) \setminus \Psi^{C}(t+1). 
    \end{split}
\end{equation}

Consider a string $F$ in set $\mathcal{F}_{\SA} \cap \mathcal{F}_{\suffix}(\Psi_{\CCP}(T[i..j]) \cap \Psi_{\run} \cap \Psi_{\centerset}(C_{Q}) \cap \Psi_{\lcp}(\hat{K}) \cap \Psi_{\preceding})$. 
Because of $F \in \mathcal{F}_{\suffix}(\Psi_{\CCP}(T[i..j]) \cap \Psi_{\run} \cap \Psi_{\centerset}(C_{Q}) \cap \Psi_{\lcp}(\hat{K}) \cap \Psi_{\preceding})$, 
set $\Psi_{\CCP}(T[i..j]) \cap \Psi_{\run} \cap \Psi_{\centerset}(C_{Q}) \cap \Psi_{\lcp}(\hat{K}) \cap \Psi_{\preceding}$ contains an interval attractor $([p, q], [\ell, r])$ satisfying 
$T[\gamma - |[i, \gamma_{Q}-1]|..r+1] = F$ for the attractor position $\gamma$ of the interval attractor $([p, q], [\ell, r])$. 
Equation~\ref{eq:Set_IC1_Property:1} indicates that 
there exists an integer $t \in [1, y^{\prime} - y + 1]$ satisfying $([p, q], [\ell, r]) \in \Psi^{C}(t) \setminus \Psi^{C}(t+1)$. 

We prove $F \in \{ T[i..\gamma_{Q} - 1] \cdot C_{Q}^{n+1}[1..\hat{K}] \cdot R_{y + t - 1} \mid t \in \mathcal{I}^{C} \}$. 
Because of $F \in \mathcal{F}_{\SA}$, 
Lemma~\ref{lem:F_SA_formula} shows that 
$\RSCQ(\gamma - |[i, \gamma_{Q}-1]|, \gamma + |[\gamma_{Q}, j]| - 1) < b - \eta + 1$ holds. 
$t \in \mathcal{I}^{C}$ follows from $([p, q], [\ell, r]) \in \Psi^{C}(t) \setminus \Psi^{C}(t+1)$ 
and $\RSCQ(\gamma - |[i, \gamma_{Q}-1]|, \gamma + |[\gamma_{Q}, j]| - 1) < b - \eta + 1$. 
Proposition~\ref{prop:Psi_A_Property}~\ref{enum:Psi_A_Property:3} shows that 
$F = T[i..\gamma_{Q} - 1] \cdot C_{Q}^{n+1}[1..\hat{K}] \cdot R_{y + t - 1}$ holds. 
Therefore, $F \in \{ T[i..\gamma_{Q} - 1] \cdot C_{Q}^{n+1}[1..\hat{K}] \cdot R_{y + t - 1} \mid t \in \mathcal{I}^{C} \}$ follows from 
$F = T[i..\gamma_{Q} - 1] \cdot C_{Q}^{n+1}[1..\hat{K}] \cdot R_{y + t - 1}$ and $t \in \mathcal{I}^{C}$. 

We showed that $F \in \{ T[i..\gamma_{Q} - 1] \cdot C_{Q}^{n+1}[1..\hat{K}] \cdot R_{y + t - 1} \mid t \in \mathcal{I}^{C} \}$ for each string $F$ in set $\mathcal{F}_{\SA} \cap \mathcal{F}_{\suffix}(\Psi_{\CCP}(T[i..j]) \cap \Psi_{\run} \cap \Psi_{\centerset}(C_{Q}) \cap \Psi_{\lcp}(\hat{K}) \cap \Psi_{\preceding})$. 
Therefore, statement (A) holds.

\textbf{Proof of statement (B).}
This statement corresponds to statement (B) in the proof of Proposition~\ref{prop:Set_IA_Property}. 
We proved statement (B) in the proof of Proposition~\ref{prop:Set_IA_Property} using 
Proposition~\ref{prop:Psi_A_Property} and Lemma~\ref{lem:F_SA_formula}. 
Proposition~\ref{prop:Psi_C1_Property} corresponds to Proposition~\ref{prop:Psi_A_Property}. 
Therefore, 
statement (B) can be proved using the same approach as for statement (B) in the proof of Proposition~\ref{prop:Set_IA_Property}. 

\textbf{Proof of Proposition~\ref{prop:Set_IC1_Property}(i).}
Proposition~\ref{prop:Set_IC1_Property}~\ref{enum:Set_IC1_Property:1} follows from statement (A) and statement (B).
 
\textbf{Proof of Proposition~\ref{prop:Set_IC1_Property}(ii).}
Proposition~\ref{prop:Set_IC1_Property}~\ref{enum:Set_IC1_Property:2} can be proved using the same approach as for Proposition~\ref{prop:Set_IA_Property}~\ref{enum:Set_IA_Property:2}. 

\textbf{Proof of Proposition~\ref{prop:Set_IC1_Property}(iii).}
Proposition~\ref{prop:Set_IC1_Property}~\ref{enum:Set_IC1_Property:3} can be proved using the same approach as for Proposition~\ref{prop:Set_IA_Property}~\ref{enum:Set_IA_Property:3}. 
\end{proof}

We prove Lemma~\ref{lem:GammaC1_property} using Proposition~\ref{prop:JC1_correspondence_property}, 
Proposition~\ref{prop:JC1_inverse_correspondence_property}, Proposition~\ref{prop:Psi_C1_Property}, and Proposition~\ref{prop:Set_IC1_Property}. 

\begin{proof}[Proof of Lemma~\ref{lem:GammaC1_property}~\ref{enum:GammaC1_property:1}]
Lemma~\ref{lem:GammaC1_property}~\ref{enum:GammaC1_property:1} corresponds to Lemma~\ref{lem:GammaA_property}~\ref{enum:GammaA_property:1}. 
We proved Lemma \ref{lem:GammaA_property}~\ref{enum:GammaA_property:1} using 
Proposition \ref{prop:Set_IA_Property}. 
Proposition \ref{prop:Set_IC1_Property} corresponds to Proposition~\ref{prop:Set_IA_Property}. 
Therefore, Lemma~\ref{lem:GammaC1_property}~\ref{enum:GammaC1_property:1} can be proved using the same approach as for Lemma~\ref{lem:GammaA_property}~\ref{enum:GammaA_property:1}. 
\end{proof}

\begin{proof}[Proof of Lemma~\ref{lem:GammaC1_property}~\ref{enum:GammaC1_property:2}]
We can prove Lemma~\ref{lem:GammaC1_property}~\ref{enum:GammaC1_property:2} using 
the same approach as for Lemma \ref{lem:GammaA_property} \ref{enum:GammaA_property:2}. 
The detailed proof of Lemma~\ref{lem:GammaC1_property}~\ref{enum:GammaC1_property:2} is as follows. 

We prove $\Psi^{C}(\kappa) \setminus \Psi^{C}(\kappa+1) \neq \emptyset$. 
$\kappa = \kappa^{\prime}$ follows from Proposition~\ref{prop:Set_IC1_Property}~\ref{enum:Set_IC1_Property:2} 
and Proposition~\ref{prop:Set_IC1_Property}~\ref{enum:Set_IC1_Property:3} 
for the largest integer $\kappa^{\prime}$ in the set $\mathcal{I}^{C}$. 
$\Psi^{C}(\kappa^{\prime}) \setminus \Psi^{C}(\kappa^{\prime}+1) \neq \emptyset$ follows from the definition of the set $\mathcal{I}^{C}$. 
Therefore, $\Psi^{C}(\kappa) \setminus \Psi^{C}(\kappa+1) \neq \emptyset$ holds. 

Consider an interval attractor $([p, q], [\ell, r])$ in set $\Psi^{C}(\kappa) \setminus \Psi^{C}(\kappa+1)$. 
Then, we prove $([p, q], [\ell, r]) \in \Psi_{h_{Q}} \cap \Psi_{\source} \cap \Psi_{\centerset}(C_{Q}) \cap \Psi_{\lcp}(\hat{K}) \cap \Psi_{\preceding}$ and $T[\gamma + \hat{K}..r_+1] = R_{y + \kappa - 1}$ 
for the attractor position $\gamma$ of the interval attractor $([p, q], [\ell, r])$. 
Because of $([p, q], [\ell, r]) \in \Psi^{C}(\kappa)$, 
$([p, q], [\ell, r]) \in \Psi_{h_{Q}} \cap \Psi_{\run} \cap \Psi_{\centerset}(C_{Q}) \cap \Psi_{\lcp}(\hat{K}) \cap \Psi_{\preceding}$ 
follows from the definition of the subset $\Psi^{C}(\kappa)$. 
Proposition~\ref{prop:Psi_C1_Property}~\ref{enum:Psi_C1_Property:3} shows that $T[\gamma + \hat{K}..r+1] = R_{y + \kappa - 1}$ holds. 

We prove Lemma~\ref{lem:GammaC1_property}~\ref{enum:GammaC1_property:2}. 
Because of $([p, q], [\ell, r]) \in \Psi_{h_{Q}} \cap \Psi_{\source} \cap \Psi_{\centerset}(C_{Q}) \cap \Psi_{\lcp}(\hat{K}) \cap \Psi_{\preceding}$, 
Proposition~\ref{prop:JC1_correspondence_property} shows that 
the set $\Psi_{h_{Q}} \cap \Psi_{\source} \cap \Psi_{\centerset}(C_{Q}) \cap \Psi_{\lcp}(\hat{K}) \cap \Psi_{\preceding} \cap \Psi_{\samp}$ contains an interval attractor $([p_{s}, q_{s}], [\ell_{s}, r_{s}])$ 
satisfying 
$x \leq |f_{\recover}(([p_{s}, q_{s}], [\ell_{s}, r_{s}]))| \leq n$ 
and $T[\gamma_{s} + K_{s}..r_{s}+1] = T[\gamma + \hat{K}..r+1]$. 
The interval attractor $([p_{s}, q_{s}], [\ell_{s}, r_{s}])$ corresponds to 
the weighted point $(|f_{\recover}(([p_{s}, q_{s}]$, $[\ell_{s}, r_{s}]))|$, $T[\gamma_{s} + K_{s}..r_{s} + 1]$, $|\Psi_{\str}(T[p_{s}-1..r_{s}+1])|$, $T[p_{s}-1..r_{s}+1])$ in set $\mathcal{J}_{C}(h_{Q}, C_{Q}, \hat{M})$. 
$T[\gamma_{s} + K_{s}..r_{s}+1] = R_{y + \kappa - 1}$ follows from 
$T[\gamma_{s} + K_{s}..r_{s}+1] = T[\gamma + \hat{K}..r+1]$ and $T[\gamma + \hat{K}..r+1] = R_{y + \kappa - 1}$. 
Therefore, Lemma~\ref{lem:GammaC1_property}~\ref{enum:GammaC1_property:2} holds. 
\end{proof}

\begin{proof}[Proof of Lemma~\ref{lem:GammaC1_property}~\ref{enum:GammaC1_property:3}]
Because of $x \leq |f_{\recover}(([p_{s}, q_{s}], [\ell_{s}, r_{s}]))| \leq n$, 
Proposition~\ref{prop:JC1_inverse_correspondence_property} shows that 
set $\Psi_{h_{Q}} \cap \Psi_{\run} \cap \Psi_{\centerset}(C_{Q}) \cap \Psi_{\lcp}(\hat{K}) \cap \Psi_{\preceding}$ contains an interval attractor $([p, q], [\ell, r])$ satisfying 
$T[\gamma_{s} + K_{s}..r_{s}+1] = T[\gamma + \hat{K}..r+1]$ and $\gamma = \gamma_{s} + K_{s} - \hat{K}$ 
for the attractor position $\gamma$ of the interval attractor $([p, q], [\ell, r])$. 
Here, $T[\gamma + \hat{K}..r+1] = R_{y + \kappa - 1}$ follows from 
$T[\gamma_{s} + K_{s}..r_{s}+1] = T[\gamma + \hat{K}..r+1]$ and $T[\gamma_{s} + K_{s}..r_{s}+1] = R_{y + \kappa - 1}$. 

We prove $([p, q], [\ell, r]) \in \Psi^{C}(\kappa) \setminus \Psi^{C}(\kappa+1)$. 
$([p, q], [\ell, r]) \in \Psi^{C}(\kappa)$ follows from 
$([p, q], [\ell, r]) \in \Psi_{h_{Q}} \cap \Psi_{\run} \cap \Psi_{\centerset}(C_{Q}) \cap \Psi_{\lcp}(\hat{K}) \cap \Psi_{\preceding}$ 
and $T[\gamma + \hat{K}..r+1] = R_{y + \kappa - 1}$. 
On the other hand, $([p, q], [\ell, r]) \not \in \Psi^{C}(\kappa+1)$ holds 
because $T[\gamma + \hat{K}..r+1] = R_{y + \kappa - 1}$. 
Therefore, $([p, q], [\ell, r]) \in \Psi^{C}(\kappa) \setminus \Psi^{C}(\kappa+1)$ holds. 

We prove $T[\gamma_{s} + K_{s} - \hat{K} - |[i, \gamma_{Q}-1]|..r_{s} + 1] = T[i..\gamma_{Q} - 1] \cdot C_{Q}^{n+1}[1..\hat{K}] \cdot R_{y + \kappa - 1}$. 
Because of $([p, q], [\ell, r]) \in \Psi^{C}(\kappa) \setminus \Psi^{C}(\kappa+1)$, 
Proposition~\ref{prop:Psi_C1_Property}~\ref{enum:Psi_C1_Property:3} shows that 
$T[\gamma - |[i, \gamma_{Q}-1]|..r+1] = T[i..\gamma_{Q}-1] \cdot C_{Q}^{n+1}[1..\hat{K}] \cdot R_{y + \kappa - 1}$ holds. 
$\gamma - |[i, \gamma_{Q}-1]| = \gamma_{s} + K_{s} - \hat{K} - |[i, \gamma_{Q}-1]|$ follows from $\gamma = \gamma_{s} + K_{s} - \hat{K}$. 
Therefore, $T[\gamma_{s} + K_{s} - \hat{K} - |[i, \gamma_{Q}-1]|..r_{s} + 1] = T[i..\gamma_{Q} - 1] \cdot C_{Q}^{n+1}[1..\hat{K}] \cdot R_{y + \kappa - 1}$ holds. 
\end{proof}

\begin{proof}[Proof of Lemma~\ref{lem:GammaC1_property}~\ref{enum:GammaC1_property:4}]
Lemma~\ref{lem:GammaC1_property}~\ref{enum:GammaC1_property:4} corresponds to 
Lemma~\ref{lem:GammaA_property}~\ref{enum:GammaA_property:4}. 
We proved Lemma~\ref{lem:GammaA_property} \ref{enum:GammaA_property:4} using 
Proposition~\ref{prop:Set_IA_Property}. 
Proposition~\ref{prop:Set_IC1_Property} corresponds to Proposition~\ref{prop:Set_IA_Property}. 
Therefore, 
Lemma~\ref{lem:GammaC1_property} \ref{enum:GammaC1_property:4} can be proved using the same approach as for Lemma~\ref{lem:GammaA_property}~\ref{enum:GammaA_property:4}. 
\end{proof}

\subsubsection{Proof of Lemma~\ref{lem:GammaC1_sub_property}}\label{subsubsec:GammaC1_sub_property_proof}

The following proposition states the relationship between set $\Psi^{C}(t) \setminus \Psi^{C}(t+1)$ 
and range-count query on set $\mathcal{Y}_{C}(h_{Q}, C_{Q}, \hat{M})$ for each integer $t \in [1, \hat{y} - y + 1]$. 

\begin{proposition}\label{prop:GammaC1_Grid}
Consider condition (A) of RSS query for the given RSS query $\RSSQ(T[i..j], b)$.
For an $t \in [1, \hat{y} - y + 1]$, 
$\rangecount(\mathcal{Y}_{C}(h_{Q}, C_{Q}, \hat{M}), x, n, R_{y + t - 1}, R_{y + t - 1}) \geq 1 \Leftrightarrow \Psi^{C}(t) \setminus \Psi^{C}(t+1) \neq \emptyset$. 
Here, let $\Psi^{C}(\hat{y} - y + 2) = \emptyset$ for simplicity. 
\end{proposition}
\begin{proof}
Proposition~\ref{prop:GammaC1_Grid} follows from the following two statements: 
\begin{enumerate}[label=\textbf{(\roman*)}]
    \item $\rangecount(\mathcal{Y}_{C}(h_{Q}, C_{Q}, \hat{M}), x, n, R_{y + t - 1}, R_{y + t - 1}) \geq 1 \Rightarrow \Psi^{C}(t) \setminus \Psi^{C}(t+1) \neq \emptyset$;
    \item $\rangecount(\mathcal{Y}_{C}(h_{Q}, C_{Q}, \hat{M}), x, n, R_{y + t - 1}, R_{y + t - 1}) \geq 1 \Leftarrow \Psi^{C}(t) \setminus \Psi^{C}(t+1) \neq \emptyset$.
\end{enumerate}

\textbf{Proof of statement (i).}
Because of $\rangecount(\mathcal{Y}_{C}(h_{Q}, C_{Q}, \hat{M}), x, n, R_{y + t - 1}, R_{y + t - 1}) \geq 1$, 
set $\mathcal{J}_{A}(h_{Q})$ contains a weighted point $(|f_{\recover}(([p_{s}, q_{s}], [\ell_{s}, r_{s}]))|$, $T[\gamma_{s} + K_{s}..r_{s} + 1]$, $|\Psi_{\str}(T[p_{s}-1..r_{s}+1])|$, $T[p_{s}-1..r_{s}+1])$ satisfying 
$x \leq |f_{\recover}(([p_{s}, q_{s}], [\ell_{s}, r_{s}]))| \leq n$ and $T[\gamma_{s} + K_{s}..r_{s}+1] = R_{y + t - 1}$. 
This weighted point corresponds to the interval attractor $([p_{s}, q_{s}], [\ell_{s}, r_{s}])$ in set $\Psi_{h_{Q}} \cap \Psi_{\source} \cap \Psi_{\centerset}(C_{Q}) \cap \Psi_{\modulo}(\hat{M}) \cap \Psi_{\preceding} \cap \Psi_{\samp}$. 

Because of $x \leq |f_{\recover}(([p_{s}, q_{s}], [\ell_{s}, r_{s}]))| \leq n$, 
Proposition~\ref{prop:JC1_inverse_correspondence_property} shows that 
set $\Psi_{h_{Q}} \cap \Psi_{\run} \cap \Psi_{\centerset}(C_{Q}) \cap \Psi_{\lcp}(\hat{K}) \cap \Psi_{\preceding}$ contains an interval attractor $([p, q], [\ell, r])$ satisfying 
$T[\gamma_{s} + K_{s}..r_{s}+1] = T[\gamma + \hat{K}..r+1]$ 
for the attractor position $\gamma$ of the interval attractor $([p, q], [\ell, r])$. 
Here, $T[\gamma + \hat{K}..r+1] = R_{y + t - 1}$ follows from 
$T[\gamma_{s} + K_{s}..r_{s}+1] = T[\gamma + \hat{K}..r+1]$ and $T[\gamma_{s} + K_{s}..r_{s}+1] = R_{y + t - 1}$. 
$([p, q], [\ell, r]) \in \Psi^{C}(t)$ follows from 
$([p, q], [\ell, r]) \in \Psi_{h_{Q}} \cap \Psi_{\run} \cap \Psi_{\centerset}(C_{Q}) \cap \Psi_{\lcp}(\hat{K}) \cap \Psi_{\preceding}$ 
and $T[\gamma + \hat{K}..r+1] = R_{y + t - 1}$. 
On the other hand, $([p, q], [\ell, r]) \not \in \Psi^{C}(t+1)$ holds 
because $T[\gamma + \hat{K}..r+1] = R_{y + t - 1}$. 
Therefore, $\Psi^{C}(t) \setminus \Psi^{C}(t+1) \neq \emptyset$ follows from 
$([p, q], [\ell, r]) \in \Psi^{C}(t)$ and $([p, q], [\ell, r]) \not \in \Psi^{C}(t+1)$. 

\textbf{Proof of statement (ii).}
Consider an interval attractor $([p, q], [\ell, r])$ in set $\Psi^{C}(t) \setminus \Psi^{C}(t+1)$. 
Then, 
$([p, q], [\ell, r]) \in \Psi_{h_{Q}} \cap \Psi_{\run} \cap \Psi_{\centerset}(C_{Q}) \cap \Psi_{\lcp}(\hat{K}) \cap \Psi_{\preceding}$ 
follows from the definition of the subset $\Psi^{C}(t)$. 
Proposition~\ref{prop:Psi_C1_Property}~\ref{enum:Psi_C1_Property:3} shows that 
$T[\gamma + \hat{K}..r + 1] = R_{y+t-1}$ holds 
for the attractor position $\gamma$ of the interval attractor $([p, q], [\ell, r])$. 
We apply Proposition~\ref{prop:JC1_correspondence_property} to the interval attractor $([p, q], [\ell, r])$. 
Then, Proposition~\ref{prop:JC1_correspondence_property} shows that 
set $\mathcal{J}_{C}(h_{Q}, C_{Q}, \hat{M})$ contains a weighted point $e = (|f_{\recover}(([p_{s}, q_{s}], [\ell_{s}, r_{s}]))|$, $T[\gamma_{s} + K_{s}..r_{s} + 1]$, $|\Psi_{\str}(T[p_{s}-1..r_{s}+1])|$, $T[p_{s}-1..r_{s}+1])$ satisfying 
$x \leq |f_{\recover}(([p_{s}, q_{s}], [\ell_{s}, r_{s}]))| \leq n$ 
and $T[\gamma_{s} + K_{s}..r_{s}+1] = T[\gamma + \hat{K}..r + 1]$. 
Here, $T[\gamma_{s} + K_{s}..r_{s}+1] = R_{y+t-1}$ follows from 
$T[\gamma_{s} + K_{s}..r_{s}+1] = T[\gamma + \hat{K}..r + 1]$ and $T[\gamma + \hat{K}..r + 1] = R_{y+t-1}$. 
The existence of the weighted point $e$ indicates that 
$\rangecount(\mathcal{Y}_{C}(h_{Q}, C_{Q}, \hat{M}), x, n, R_{y + t - 1}, R_{y + t - 1}) \geq 1$ holds. 
Therefore, statement (ii) holds. 
\end{proof}

We prove Lemma~\ref{lem:GammaC1_sub_property} using 
Proposition~\ref{prop:JC1_inverse_correspondence_property}, Proposition~\ref{prop:Psi_C1_Property},  Proposition~\ref{prop:Set_IC1_Property}, and Proposition~\ref{prop:GammaC1_Grid}. 

\begin{proof}[Proof of Lemma~\ref{lem:GammaC1_sub_property}~\ref{enum:GammaC1_sub_property:1}]
Lemma~\ref{lem:GammaC1_sub_property}~\ref{enum:GammaC1_sub_property:1} can be proved using 
the same approach as for Lemma \ref{lem:GammaA_sub_property} \ref{enum:GammaA_sub_property:1}.
\end{proof}

\begin{proof}[Proof of Lemma~\ref{lem:GammaC1_sub_property}~\ref{enum:GammaC1_sub_property:2}]
Lemma~\ref{lem:GammaC1_sub_property}~\ref{enum:GammaC1_sub_property:2} 
can be proved using the same approach as for Lemma~\ref{lem:GammaA_sub_property} \ref{enum:GammaA_sub_property:2}. 
\end{proof}

\begin{proof}[Proof of Lemma~\ref{lem:GammaC1_sub_property}~\ref{enum:GammaC1_sub_property:3}]
The following three statements are used to prove Lemma~\ref{lem:GammaC1_sub_property}~\ref{enum:GammaC1_sub_property:3}: 
\begin{enumerate}[label=\textbf{(\Alph*)}]
    \item $u_{t} = 1 \Leftarrow (\alpha_{t^{\prime}} \geq 1) \land (\RSCQ(\upsilon - |[i, \gamma_{Q}-1]|, \upsilon + |[\gamma_{Q}, j]| - 1) < b - \eta + 1)$;
    \item 
    If $u_{t} = 1$, then $t^{\prime} = t_{A}$ and $\alpha_{t^{\prime}} \geq 1$ for the smallest integer $t_{A}$ in set $[t, \hat{y} - y + 1]$ satisfying  
    $\Psi^{C}(t_{A}) \setminus \Psi^{C}(t_{A} + 1) \neq \emptyset$; 
    \item $u_{t} = 1 \Rightarrow (\alpha_{t^{\prime}} \geq 1) \land (\RSCQ(\upsilon - |[i, \gamma_{Q}-1]|, \upsilon + |[\gamma_{Q}, j]| - 1) < b - \eta + 1)$.    
\end{enumerate}

    \textbf{Proof of statement (A).}
    Consider the interval attractor $([p_{s}, q_{s}], [\ell_{s}, r_{s}])$ corresponding to the weighted point $(|f_{\recover}(([p_{s}, q_{s}], [\ell_{s}, r_{s}]))|$, $T[\gamma_{s} + K_{s}..r_{s} + 1]$, $|\Psi_{\str}(T[p_{s}-1..r_{s}+1])|$, $T[p_{s}-1..r_{s}+1])$. 
    Because of $x \leq |f_{\recover}(([p_{s}, q_{s}], [\ell_{s}, r_{s}]))| \leq n$, 
    Proposition~\ref{prop:JC1_inverse_correspondence_property} shows that 
    set $\Psi_{h_{Q}} \cap \Psi_{\run} \cap \Psi_{\centerset}(C_{Q}) \cap \Psi_{\lcp}(\hat{K}) \cap \Psi_{\preceding}$ contains an interval attractor $([p, q], [\ell, r])$ satisfying 
    $T[\gamma_{s} + K_{s}..r_{s}+1] = T[\gamma + \hat{K}..r+1]$ and $\gamma = \gamma_{s} + K_{s} - \hat{K}$ 
    for the attractor position $\gamma$ of the interval attractor $([p, q], [\ell, r])$. 
    Here, $T[\gamma + \hat{K}..r+1] = R_{y + t^{\prime} - 1}$ follows from 
    $T[\gamma_{s} + K_{s}..r_{s}+1] = T[\gamma + \hat{K}..r+1]$ and $T[\gamma_{s} + K_{s}..r_{s}+1] = R_{y + t^{\prime} - 1}$. 

    We prove $([p, q], [\ell, r]) \in \Psi^{C}(t^{\prime}) \setminus \Psi^{C}(t^{\prime}+1)$.
    $([p, q], [\ell, r]) \in \Psi^{C}(t^{\prime})$ follows from 
    $([p, q], [\ell, r]) \in \Psi_{h_{Q}} \cap \Psi_{\run} \cap \Psi_{\centerset}(C_{Q}) \cap \Psi_{\lcp}(\hat{K}) \cap \Psi_{\preceding}$ 
    and $T[\gamma + \hat{K}..r+1] = R_{y + t^{\prime} - 1}$. 
    On the other hand, $([p, q], [\ell, r]) \not \in \Psi^{C}(t^{\prime}+1)$ holds 
    because $T[\gamma + \hat{K}..r+1] = R_{y + t^{\prime} - 1}$. 
    Therefore, $([p, q], [\ell, r]) \in \Psi^{C}(t^{\prime}) \setminus \Psi^{C}(t^{\prime}+1)$ follows from 
    $([p, q], [\ell, r]) \in \Psi^{C}(t^{\prime})$ and $([p, q], [\ell, r]) \not \in \Psi^{C}(t^{\prime}+1)$. 

    We prove $t^{\prime} \in \mathcal{I}^{C}$. 
    $\upsilon = \gamma$ follows from 
    $\upsilon = \gamma_{s} + K_{s} - \hat{K}$ and $\gamma = \gamma_{s} + K_{s} - \hat{K}$. 
    $\RSCQ(\gamma - |[i, \gamma_{Q}-1]|, \gamma + |[\gamma_{Q}, j]| - 1) < b - \eta + 1$ follows from 
    $\upsilon = \gamma$ and $\RSCQ(\upsilon - |[i, \gamma_{Q}-1]|, \upsilon + |[\gamma_{Q}, j]| - 1) < b - \eta + 1$. 
    Therefore, $t^{\prime} \in \mathcal{I}^{C}$ follows from 
    $([p, q], [\ell, r]) \in \Psi^{C}(t^{\prime}) \setminus \Psi^{C}(t^{\prime}+1)$ 
    and $\RSCQ(\gamma - |[i, \gamma_{Q}-1]|, \gamma + |[\gamma_{Q}, j]| - 1) < b - \eta + 1$. 

    Because of $t^{\prime} \in \mathcal{I}^{C}$ and $t \leq t^{\prime}$, 
    Proposition~\ref{prop:Set_IC1_Property}~\ref{enum:Set_IC1_Property:3} shows that 
    $u_{t} = 1$ holds. 
    Therefore, statement (A) holds. 

    \textbf{Proof of statement (B).}
    This statement corresponds to statement (B) in the proof of Lemma~\ref{lem:GammaA_sub_property}~\ref{enum:GammaA_sub_property:3}. 
    We proved statement (B) in the proof of Lemma~\ref{lem:GammaA_sub_property}~\ref{enum:GammaA_sub_property:3} using 
    Proposition~\ref{prop:Psi_A_Property}~\ref{enum:Psi_A_Property:2} and Proposition~\ref{prop:GammaA_Grid}. 
    Proposition~\ref{prop:Psi_C1_Property}~\ref{enum:Psi_C1_Property:2} corresponds to 
    Proposition~\ref{prop:Psi_A_Property}~\ref{enum:Psi_A_Property:2}. 
    Similarly, Proposition~\ref{prop:GammaC1_Grid} corresponds to Proposition~\ref{prop:GammaA_Grid}. 
    Therefore, 
    statement (B) can be proved using the same approach as for statement (B) 
    in the proof of Lemma~\ref{lem:GammaA_sub_property}~\ref{enum:GammaA_sub_property:3}.     

    \textbf{Proof of statement (C).}
    Statement (C) can be proved using the same approach as for 
    statement (C) in the proof of Lemma~\ref{lem:GammaA_sub_property}~\ref{enum:GammaA_sub_property:3}. 
    The detailed proof of statement (C) is as follows. 

    Because of $u_{t} = 1$, 
    $\mathcal{F}_{\SA} \cap \mathcal{F}_{\suffix}(\Psi_{\CCP}(T[i..j]) \cap \Psi^{C}(t)) \neq \emptyset$ follows from the definition of the sequence $\Gamma_{C}$. 
    Because of $\mathcal{F}_{\SA} \cap \mathcal{F}_{\suffix}(\Psi_{\CCP}(T[i..j]) \cap \Psi^{C}(t)) \neq \emptyset$, 
    the set $\Psi_{\CCP}(T[i..j]) \cap \Psi^{C}(t)$ contains an interval attractor $([p_{B}, q_{B}], [\ell_{B}, r_{B}])$ satisfying 
    $T[\gamma_{B} - |[i, \gamma_{Q}-1]|..\gamma_{B}-1] \cdot T[\gamma_{B}..r_{B} + 1] \in \mathcal{F}_{\SA} \cap \mathcal{F}_{\suffix}(\Psi_{\CCP}(T[i..j]) \cap \Psi^{C}(t))$ for the attractor position $\gamma_{B}$ of the interval attractor $([p_{B}, q_{B}], [\ell_{B}, r_{B}])$. 
    Similar to Equation~\ref{eq:Set_IC1_Property:1}, 
    the following equation follows from Proposition~\ref{prop:Psi_C1_Property}~\ref{enum:Psi_C1_Property:2} and $\Psi^{C}(y^{\prime} - y + 2) = \emptyset$:
    \begin{equation*}
    \Psi^{C}(t) = \bigcup_{\lambda = t}^{\hat{y} - y + 1} \Psi^{C}(\lambda) \setminus \Psi^{C}(\lambda+1).
    \end{equation*}
    Therefore, there exists an integer $t_{B} \in [t, \hat{y} - y + 1]$ satisfying 
    $([p_{B}, q_{B}], [\ell_{B}, r_{B}]) \in \Psi^{C}(t_{B}) \setminus \Psi^{C}(t_{B}+1)$. 
    Here, $t_{A} \leq t_{B}$ holds for the smallest integer $t_{A}$ in set $[t, \hat{y} - y + 1]$ satisfying  
    $\Psi^{C}(t_{A}) \setminus \Psi^{C}(t_{A} + 1) \neq \emptyset$. 

    Because of $u_{t} = 1$, 
    statement (B) shows that $t^{\prime} = t_{A}$ and $\alpha_{t^{\prime}} \geq 1$ holds. 
    Lemma~\ref{lem:GammaC1_sub_property}~\ref{enum:GammaC1_sub_property:2} shows that 
    the weighted point 
    $(|f_{\recover}(([p_{s}, q_{s}], [\ell_{s}, r_{s}]))|$, $T[\gamma_{s} + K_{s}..r_{s} + 1]$, $|\Psi_{\str}(T[p_{s}-1..r_{s}+1])|$, $T[p_{s}-1..r_{s}+1])$ satisfies $x \leq |f_{\recover}(([p_{s}, q_{s}], [\ell_{s}, r_{s}]))| \leq n$ 
    and $T[\gamma_{s} + K_{s}..r_{s}+1] = R_{y + t^{\prime} - 1}$. 
    Because of $x \leq |f_{\recover}(([p_{s}, q_{s}], [\ell_{s}, r_{s}]))| \leq n$, 
    Proposition~\ref{prop:JC1_inverse_correspondence_property} shows that 
    set $\Psi_{h_{Q}} \cap \Psi_{\run} \cap \Psi_{\centerset}(C_{Q}) \cap \Psi_{\lcp}(\hat{K}) \cap \Psi_{\preceding}$ contains an interval attractor $([p, q], [\ell, r])$ satisfying 
    $T[\gamma_{s} + K_{s}..r_{s}+1] = T[\gamma + \hat{K}..r+1]$ and $\gamma = \gamma_{s} + K_{s} - \hat{K}$ 
    for the attractor position $\gamma$ of the interval attractor $([p, q], [\ell, r])$. 
    Here, $T[\gamma + \hat{K}..r+1] = R_{y + t^{\prime} - 1}$ follows from 
    $T[\gamma_{s} + K_{s}..r_{s}+1] = T[\gamma + \hat{K}..r+1]$ and $T[\gamma_{s} + K_{s}..r_{s}+1] = R_{y + t^{\prime} - 1}$. 
    
    We prove $([p, q], [\ell, r]) \in \Psi^{C}(t^{\prime}) \setminus \Psi^{C}(t^{\prime}+1)$. 
    $([p, q], [\ell, r]) \in \Psi^{C}(t^{\prime})$ follows from 
    $([p, q], [\ell, r]) \in \Psi_{h_{Q}} \cap \Psi_{\run} \cap \Psi_{\centerset}(C_{Q}) \cap \Psi_{\lcp}(\hat{K}) \cap \Psi_{\preceding}$ 
    and $T[\gamma + \hat{K}..r+1] = R_{y + t^{\prime} - 1}$. 
    On the other hand, $([p, q], [\ell, r]) \not \in \Psi^{C}(t^{\prime}+1)$ holds 
    because $T[\gamma + \hat{K}..r+1] = R_{y + t^{\prime} - 1}$. 
    Therefore, $([p, q], [\ell, r]) \in \Psi^{C}(t^{\prime}) \setminus \Psi^{C}(t^{\prime}+1)$ holds. 

    We prove $T[\gamma - |[i, \gamma_{Q}-1]|..\gamma-1] \cdot T[\gamma..r + 1] \preceq T[\gamma_{B} - |[i, \gamma_{Q}-1]|..\gamma_{B}-1] \cdot T[\gamma_{B}..r_{B} + 1]$. 
    Because of $([p, q], [\ell, r]) \in \Psi^{C}(t^{\prime}) \setminus \Psi^{C}(t^{\prime}+1)$, 
    Proposition~\ref{prop:Psi_C1_Property}~\ref{enum:Psi_C1_Property:3} shows that 
    $T[\gamma - |[i, \gamma_{Q}-1]|..\gamma-1] \cdot T[\gamma..r+1] = T[i..\gamma_{Q}-1] \cdot C_{Q}^{n+1}[1..\hat{K}] \cdot R_{y + t^{\prime} - 1}$ holds.
    Similarly, 
    Proposition~\ref{prop:Psi_C1_Property}~\ref{enum:Psi_C1_Property:3} shows that 
    $T[\gamma_{B} - |[i, \gamma_{Q}-1]|..\gamma_{B}-1] \cdot T[\gamma_{B}..r_{B}+1] = T[i..\gamma_{Q}-1] \cdot C_{Q}^{n+1}[1..\hat{K}] \cdot R_{y + t_{B} - 1}$ holds.
    $R_{y + t^{\prime} - 1} \preceq R_{y + t_{B} - 1}$ follows from 
    $t^{\prime} \leq t_{B}$ and $R_{1} \prec R_{2} \prec \cdots \prec R_{d}$. 
    Therefore, $T[\gamma - |[i, \gamma_{Q}-1]|..\gamma-1] \cdot T[\gamma..r + 1] \preceq T[\gamma_{B} - |[i, \gamma_{Q}-1]|..\gamma_{B}-1] \cdot T[\gamma_{B}..r_{B} + 1]$ 
    follows from (1) $T[\gamma - |[i, \gamma_{Q}-1]|..\gamma-1] \cdot T[\gamma..r+1] = T[i..\gamma_{Q}-1] \cdot C_{Q}^{n+1}[1..\hat{K}] \cdot R_{y + t^{\prime} - 1}$, 
    (2) $T[\gamma_{B} - |[i, \gamma_{Q}-1]|..\gamma_{B}-1] \cdot T[\gamma_{B}..r_{B}+1] = T[i..\gamma_{Q}-1] \cdot C_{Q}^{n+1}[1..\hat{K}] \cdot R_{y + t_{B} - 1}$, 
    and (3) $R_{y + t^{\prime} - 1} \preceq R_{y + t_{B} - 1}$. 

    We prove $\RSCQ(\upsilon - |[i, \gamma_{Q}-1]|, \upsilon + |[\gamma_{Q}, j]| - 1) < b - \eta + 1$. 
    Because of $([p, q], [\ell, r]) \in \Psi^{C}(t^{\prime})$, 
    $([p, q], [\ell, r]) \in \Psi_{\CCP}(T[i..])$ follows from 
    Proposition~\ref{prop:Psi_C1_Property}~\ref{enum:Psi_C1_Property:1} and Proposition~\ref{prop:Psi_C1_Property}~\ref{enum:Psi_C1_Property:2}. 
    Lemma~\ref{lem:F_suffix_basic_property}~\ref{enum:F_suffix_basic_property:6} shows that 
    $T[\gamma - |[i, \gamma_{Q}-1]|..\gamma-1] \cdot T[\gamma..r + 1] \in \mathcal{F}_{\SA}$ 
    because 
    (a) $([p, q], [\ell, r])$, $([p_{B}, q_{B}], [\ell_{B}, r_{B}]) \in \Psi_{\CCP}(T[i..])$, 
    (b) $T[\gamma - |[i, \gamma_{Q}-1]|..\gamma-1] \cdot T[\gamma..r + 1] \preceq T[\gamma_{B} - |[i, \gamma_{Q}-1]|..\gamma_{B}-1] \cdot T[\gamma_{B}..r_{B} + 1]$, 
    and (c) $T[\gamma_{B} - |[i, \gamma_{Q}-1]|..\gamma_{B}-1] \cdot T[\gamma_{B}..r_{B} + 1] \in \mathcal{F}_{\SA}$. 
    $\RSCQ(\gamma - |[i, \gamma_{Q}-1]|, \gamma + |[\gamma_{Q}, j]| - 1) < b - \eta + 1$ 
    follows from 
    Lemma~\ref{lem:F_SA_formula} and $T[\gamma - |[i, \gamma_{Q}-1]|..\gamma-1] \cdot T[\gamma..r + 1] \in \mathcal{F}_{\SA}$. 
    $\upsilon = \gamma$ follows from 
    $\upsilon = \gamma_{s} + K_{s} - \hat{K}$ and $\gamma = \gamma_{s} + K_{s} - \hat{K}$. 
    Therefore, 
    $\RSCQ(\upsilon - |[i, \gamma_{Q}-1]|, \upsilon + |[\gamma_{Q}, j]| - 1) < b - \eta + 1$ follows from 
    $\RSCQ(\gamma - |[i, \gamma_{Q}-1]|, \gamma + |[\gamma_{Q}, j]| - 1) < b - \eta + 1$ and $\upsilon = \gamma$. 

    Finally, $u_{t} = 1 \Rightarrow (\alpha_{t^{\prime}} \geq 1) \land (\RSCQ(\upsilon - |[i, \gamma_{Q}-1]|, \upsilon + |[\gamma_{Q}, j]| - 1) < b - \eta + 1)$ holds. 
    
    \textbf{Proof of Lemma~\ref{lem:GammaC1_sub_property}~\ref{enum:GammaC1_sub_property:3}.}
    Lemma~\ref{lem:GammaC1_sub_property}~\ref{enum:GammaC1_sub_property:3} follows from statement (A) and statement (C).

\end{proof}

\subsubsection{Algorithm}\label{subsubsec:gamma_C1_algorithm}
We prove Lemma~\ref{lem:GammaC1_algorithm}, i.e., 
we show that subquery $\RSSQCX(T[i..j], b)$ can be answered 
in $O(H^{2} \log^{2} n + \log^{6} n)$ time using the data structures for RSC query 
and interval $[i, j]$. 

For answering subquery $\RSSQCX(T[i..j], b)$, 
we use the ordered set $\mathcal{T}_{C} = \{ (h_{1}, C_{1}, M_{1})$, $(h_{2}, C_{2}, M_{2})$, $\ldots$, $(h_{m}, C_{m}, M_{m}) \}$ of triplets introduced in Section~\ref{subsubsec:TC1_ds}. 
Let $\lambda$ be an integer in set $[1, m]$ satisfying $(h_{\lambda}, C_{\lambda}, M_{\lambda}) = (h_{Q}, C_{Q}, \hat{M})$. 
This integer $\lambda$ exists if and only if the ordered set $\mathcal{T}_{C}$ contains the triplet $(h_{Q}, C_{Q}, \hat{M})$.

\paragraph{Computation of integer $\lambda$.}
We compute the integer $\lambda$ in three phases. 
In the first phase, 
we compute interval attractor $([p_{Q}, q_{Q}], [\ell_{Q}, r_{Q}])$, 
its level $h_{Q}$, its attractor position $\gamma_{Q}$, and the length $|C_{Q}|$ of its associated string $C_{Q}$. 
The interval attractor $([p_{Q}, q_{Q}], [\ell_{Q}, r_{Q}])$ can be obtained by capture query $\CAPQ([i, j])$. 
The level $h_{Q}$ and attractor position $\gamma_{Q}$ can be obtained by 
level-query $\levelQ(([p_{Q}, q_{Q}], [\ell_{Q}, r_{Q}]))$ and attractor position query $\attrQ(([p_{Q}, q_{Q}], [\ell_{Q}, r_{Q}]))$, 
respectively. 
The length $|C_{Q}|$ can be obtained by C-length query $\clenQ(([p_{Q}, q_{Q}], [\ell_{Q}, r_{Q}]))$. 
Therefore, the first phase takes $O(H^{2} \log n)$ time. 

In the second phase, 
we compute two integers $\hat{K}$ and $\hat{M}$. 
Let $K_{Q}$ be the length of the longest common prefix between two strings $T[\gamma_{Q}..r_{Q}]$ and $C_{Q}^{n+1}$. 
Then, $\hat{K} = \min \{ K_{Q}, |[\gamma_{Q}, j]| \}$ holds, 
and the length $K_{Q}$ can be obtained in $O(H^{2})$ time by C-LCP query $\clcpQ(([p_{Q}, q_{Q}]$, $[\ell_{Q}, r_{Q}]))$. 
The integer $\hat{M}$ can be computed in $O(H)$ time using three integers $\hat{K}$, $h_{Q}$, and $|C_{Q}|$. 
Therefore, the second phase takes $O(H^{2})$ time. 

In the third phase, 
we compute the integer $\lambda$ using the query of Lemma~\ref{lem:TC1_queries}~\ref{enum:TC1_queries:3}. 
For answering this query, 
we need to know an interval $[\beta, \beta + |C_{Q}| - 1]$ in input string $T$ satisfying $T[\beta..\beta + |C_{Q}| - 1] = C_{Q}$. 
because $C_{Q} = T[\gamma_{Q}..\gamma_{Q} + |C_{Q}| - 1]$ follows from the definition of the associated string $C_{Q}$. 
Therefore, the query of Lemma~\ref{lem:TC1_queries}~\ref{enum:TC1_queries:3} can be answered in $O(H^{2} \log n + \log^{2} n)$ time. 

The three phases take $O(H^{2} \log n + \log^{2} n)$ time in total. 
Therefore, the integer $\lambda$ can be computed in $O(H^{2} \log n + \log^{2} n)$ time. 

\paragraph{Accessing the dynamic data structures for two sets $\mathcal{Y}_{C}(h_{\lambda}, C_{\lambda}, M_{\lambda})$ and $\mathcal{J}_{C}(h_{\lambda}, C_{\lambda}, M_{\lambda})$.}
For answering subquery $\RSSQCX(T[i..j], b)$, 
we use the dynamic data structures for two sets $\mathcal{Y}_{C}(h_{\lambda}, C_{\lambda}, M_{\lambda})$ and $\mathcal{J}_{C}(h_{\lambda}, C_{\lambda}, M_{\lambda})$ introduced in Section~\ref{subsubsec:JC1_Y_ds} and Section~\ref{subsubsec:JC1_ds}. 
For accessing theses dynamic data structures, 
we leverage the doubly linked list of $m$ elements for the ordered set $\mathcal{T}_{C}$, which is introduced in Section~\ref{subsubsec:TC1_ds}. 
The $\lambda$-th element of this doubly linked list stores two pointers to 
the dynamic data structures for the two sets $\mathcal{Y}_{C}(h_{\lambda}, C_{\lambda}, M_{\lambda})$ and $\mathcal{J}_{C}(h_{\lambda}, C_{\lambda}, M_{\lambda})$. 
We can access the $\lambda$-th element in $O(\log m)$ time by the list indexing data structure built on the doubly linked list for the ordered set $\mathcal{T}_{C}$. 
Here, $m = O(n^{2})$ follows from Lemma~\ref{lem:TC1_queries}~\ref{enum:TC1_queries:3}.
Therefore, we can access the dynamic data structures for two sets $\mathcal{Y}_{C}(h_{\lambda}, C_{\lambda}, M_{\lambda})$ and $\mathcal{J}_{C}(h_{\lambda}, C_{\lambda}, M_{\lambda})$ in $O(\log n)$ time if we know the integer $\lambda$. 

\paragraph{Computation of three integers $x$, $y$, and $y^{\prime}$.}
Consider the three integers $x$, $y$, and $y^{\prime}$ introduced in Section~\ref{subsec:GammaC1}. 
We compute the three integers in three phases. 

In the first phase, we compute the integer $\lambda$. 
This computation takes $O(H^{2} \log n + \log^{2} n)$ time. 

In the second phase, we compute the integer $x$. 
This integer $x$ can be computed in $O(H)$ time using three integers $\hat{K}$, $h_{Q}$, and $|C_{Q}|$. 
The three integers $\hat{K}$, $h_{Q}$, and $|C_{Q}|$ can be obtained by the algorithm computing the integer $\lambda$. 
Therefore, the second phase can be executed in $O(H)$ time. 

In the third phase, we compute the two integers $y$ and $y^{\prime}$. 
If the integer $\lambda$ exists, 
then 
we obtain the two integers $y$ and $y^{\prime}$ are obtained by binary search on the $d$ strings 
of the ordered set $\mathcal{Y}_{C}(h_{Q}, C_{Q}, \hat{M})$. 
This binary search can be executed in $O((H^{2} + \log n)\log d)$ time 
using the query of Lemma~\ref{lem:JA_Y_queries}~\ref{enum:JA_X_queries:3}, the integer $\hat{K}$, 
and the attractor position $\gamma_{Q}$ of the interval attractor $([p_{Q}, q_{Q}], [\ell_{Q}, r_{Q}])$. 
Here, the integer $\hat{K}$ and attractor position $\gamma_{Q}$ can be obtained by the algorithm computing the integer $\lambda$; 
$d = O(n^{2})$ follows from Lemma~\ref{enum:JC1_size:2}; 
for answering the query of Lemma~\ref{lem:JA_Y_queries}~\ref{enum:JA_X_queries:3}, 
we need to access the dynamic data structures for the ordered set $\mathcal{Y}_{C}(h_{\lambda}, C_{\lambda}, M_{\lambda})$ 
(i.e., $\mathcal{Y}_{C}(h_{Q}, C_{Q}, \hat{M})$) in $O(\log n)$ time.
Therefore, the two integers $y$ and $y^{\prime}$ can be computed in $O(H^{2} \log n + \log^{2} n)$ time. 

Otherwise (i.e., the integer $\lambda$ does not exist), 
$\mathcal{J}_{C}(h_{Q}, C_{Q}, \hat{M}) = \emptyset$ follows from the definition of the ordered set $\mathcal{T}_{C}$. 
$\mathcal{Y}_{C}(h_{Q}, C_{Q}, \hat{M}) = \{ R_{1}, R_{2} \}$, $R_{1} = \varepsilon$, and $R_{2} = \#$ hold 
because the set $\mathcal{J}_{C}(h_{Q}, C_{Q}, \hat{M})$. 
In this case, $y = y^{\prime} = 2$ holds. 
Therefore, the third phase takes $O(H^{2} \log n + \log^{2} n)$ time. 

The three phases take $O(H^{2} \log n + \log^{2} n)$ time in total. 
Therefore, we can compute the three integers $x$, $y$, and $y^{\prime}$ in $O(H^{2} \log n + \log^{2} n)$ time. 

\paragraph{Computation of each integer $u_{t}$ in sequence $\Gamma_{C}$.}
Consider the non-increasing sequence $\Gamma_{C} = u_{1}, u_{2}, \ldots, u_{y^{\prime} - y + 1} \in \{ 0, 1 \}$. 
We show that the $t$-th integer $u_{t}$ can be computed in $O(H^{2} \log n + \log^{4} n)$ time for a given integer $t \in [1, y^{\prime} - y + 1]$. 

We leverage sequence $\Gamma_{C, \sub} = \alpha_{1}$, $\alpha_{2}$, $\ldots$, $\alpha_{y^{\prime} - y + 1}$ for computing the $t$-th integer $u_{t}$. 
Let $t^{\prime}$ be the largest integer in set $[t, y^{\prime} - y + 1]$ satisfying $\alpha_{t} = \alpha_{t^{\prime}}$. 
If $\alpha_{t^{\prime}} \geq 1$, 
then Lemma~\ref{lem:GammaC1_sub_property}~\ref{enum:GammaC1_sub_property:2} shows that 
set $\mathcal{J}_{C}(h_{Q}, C_{Q}, \hat{M})$ contains a weighted point 
$(|f_{\recover}(([p_{s}, q_{s}], [\ell_{s}, r_{s}]))|$, $T[\gamma_{s} + K_{s}..r_{s} + 1]$, $|\Psi_{\str}(T[p_{s}-1..r_{s}+1])|$, $T[p_{s}-1..r_{s}+1])$ satisfying 
$x \leq |f_{\recover}(([p_{s}, q_{s}], [\ell_{s}, r_{s}]))| \leq n$ 
and $T[\gamma_{s} + K_{s}..r_{s}+1] = R_{y + t^{\prime} - 1}$. 
For the integer $\upsilon = \gamma_{s} + K_{s} - \hat{K}$, 
Lemma~\ref{lem:GammaC1_sub_property}~\ref{enum:GammaC1_sub_property:3} shows that 
$u_{t} = 1 \Leftrightarrow (\alpha_{t^{\prime}} \geq 1) \land (\RSCQ(\upsilon - |[i, \gamma_{Q}-1]|, \upsilon + |[\gamma_{Q}, j]| - 1) < b - \eta + 1)$ holds. 
Therefore, the $t$-th integer $u_{t}$ can be computed using the $t^{\prime}$-th integer $\alpha_{t^{\prime}}$ of the sequence $\Gamma_{C, \sub}$ and RSC query $\RSCQ(\upsilon - |[i, \gamma_{Q}-1]|, \upsilon + |[\gamma_{Q}, j]| - 1)$. 

The algorithm computing the $t$-th integer $u_{t}$ consists of four phases. 
In the first phase, we compute the four integers $\lambda$, $x$, $y$, and $y^{\prime}$. 
This computation takes $O(H^{2} \log n + \log^{2} n)$ time. 

In the second phase, 
we find the integer $t^{\prime}$ by binary search on the sequence $\Gamma_{C, \sub}$. 
This binary search can be executed by computing $O(\log (y^{\prime} - y + 1))$ integers of the sequence $\Gamma_{C, \sub}$ 
because Lemma~\ref{lem:GammaC1_sub_property}~\ref{enum:GammaC1_sub_property:1} shows that 
the sequence $\Gamma_{C, \sub}$ is non-increasing. 
Each integer of the sequence $\Gamma_{D, \sub}$ can be computed by one range-count query on the set $\mathcal{J}_{C}(h_{Q}, C_{Q}, \hat{M})$ of weighted points. 

If the integer $\lambda$ does not exist, 
$\mathcal{J}_{C}(h_{Q}, C_{Q}, \hat{M}) = \emptyset$ follows from the definition of the ordered set $\mathcal{T}_{C}$. 
In this case, any range-count query returns $0$ on the set $\mathcal{J}_{C}(h_{Q}, C_{Q}, \hat{M})$ 
Otherwise (i.e., the integer $\lambda$ exists), 
each range-count query can be executed by the dynamic data structures for the set $\mathcal{J}_{C}(h_{\lambda}, C_{\lambda}, M_{\lambda})$. 
Accessing these dynamic data structures takes $O(\log n)$ time, 
and each range-count query takes $O(\log^{2} k)$ time 
for the number $k$ of weighted points in the set $\mathcal{J}_{C}(h_{\lambda}, C_{\lambda}, M_{\lambda})$.
$y^{\prime} - y + 1 = O(n^{2})$ because $y^{\prime} - y + 1 \leq d$ and $d = O(n^{2})$. 
$k = O(n^{2})$ follows from Lemma~\ref{lem:JC1_size}~\ref{enum:JC1_size:2}. 
Therefore, the third phase takes $O(\log^{3} n)$ time. 

The integer $\alpha_{t^{\prime}}$ is obtained by the binary search of the second phase. 
The third phase is executed if $\alpha_{t^{\prime}} \geq 1$ holds. 
In this case, the integer $\lambda$ exists. 
This is because $\alpha_{t^{\prime}} = 0$ holds if the integer $\lambda$ does not exist. 
In the third phase, we find the interval attractor $([p_{s}, q_{s}], [\ell_{s}, r_{s}])$ corresponding to 
the weighted point $(|f_{\recover}(([p_{s}, q_{s}], [\ell_{s}, r_{s}]))|$, $T[\gamma_{s} + K_{s}..r_{s} + 1]$, $|\Psi_{\str}(T[p_{s}-1..r_{s}+1])|$, $T[p_{s}-1..r_{s}+1])$.  
This interval attractor can be found in $O(H^{2} \log n + \log^{2} n)$ time using the query of 
Lemma~\ref{lem:JC1_Y_queries}~\ref{enum:JC1_Y_queries:4}. 
This query can be answered after accessing the dynamic data structures for the ordered set $\mathcal{Y}_{C}(h_{\lambda}, C_{\lambda}, M_{\lambda})$ in $O(\log n)$ time. 
Therefore, the third phase takes $O(H^{2} \log n + \log^{2} n)$ time. 

In the fourth phase, 
we compute the $t$-th integer $u_{t}$ of sequence $\Gamma_{C}$ by verifying 
$\alpha_{t^{\prime}} \geq 1$ and $\RSCQ(\upsilon - |[i, \gamma_{Q}-1]|, \upsilon + |[\gamma_{Q}, j]| - 1) < b - \eta + 1$. 
The attractor position $\gamma_{s}$ is obtained in $O(H^{2})$ time 
by attractor position query $\attrQ(([p_{s}, q_{s}], [\ell_{s}, r_{s}]))$. 
The integer $\upsilon$ can be computed in $O(1)$ time using the three integers $\gamma_{s}$, $K_{s}$, and $\hat{K}$.
The integer $K_{s}$ can be obtained by C-LCP query $\clcpQ(([p_{s}, q_{s}], [\ell_{s}, r_{s}]))$. 
The integer $\hat{K}$ and attractor position $\gamma_{Q}$ are obtained by the algorithm computing the integer $\lambda$. 
The RSC query $\RSCQ(\upsilon - |[i, \gamma_{Q}-1]|, \upsilon + |[\gamma_{Q}, j]| - 1)$ takes $O(H^{2} \log n + \log^{4} n)$ time. 
Therefore, the fourth phase takes $O(H^{2} \log n + \log^{4} n)$ time. 

Finally, the algorithm computing the $t$-th integer $u_{t}$ takes $O(H^{2} \log n + \log^{4} n)$ time in total. 

\paragraph{Computation of the largest integer $\kappa$.}
Consider the largest integer $\kappa$ in set $[1, y^{\prime} - y + 1]$ satisfying $u_{\kappa} = 1$. 
We find the largest integer $\kappa$ by binary search on the non-increasing sequence $\Gamma_{C}$. 
This binary search can be executed by computing $O(\log (y^{\prime} - y + 1))$ integers of the sequence $\Gamma_{C}$. 
Each integer of the sequence $\Gamma_{C}$ can be computed in $O(H^{2} \log n + \log^{4} n)$ time. 
The two integers $y$ and $y^{\prime}$ can be computed in $O(H^{2} \log n + \log^{2} n)$ time. 
Therefore, this binary search takes $O((H^{2} \log n + \log^{5} n) \log (y^{\prime} - y + 1))$ time 
(i.e., $O(H^{2} \log^{2} n + \log^{5} n)$ time).

\paragraph{Verification $\mathcal{F}_{\SA} \cap \mathcal{F}_{\suffix}(\Psi_{\CCP}(T[i..j]) \cap \Psi_{\run} \cap \Psi_{\centerset}(C_{Q}) \cap \Psi_{\lcp}(\hat{K}) \cap \Psi_{\preceding}) = \emptyset$.}
We verify whether $\mathcal{F}_{\SA} \cap \mathcal{F}_{\suffix}(\Psi_{\CCP}(T[i..j]) \cap \Psi_{\run} \cap \Psi_{\centerset}(C_{Q}) \cap \Psi_{\lcp}(\hat{K}) \cap \Psi_{\preceding}) = \emptyset$ or not 
for answering $\RSSQCX(T[i..j], b)$. 
From Lemma~\ref{lem:GammaC1_property}~\ref{enum:GammaC1_property:1},  
$\mathcal{F}_{\SA} \cap \mathcal{F}_{\suffix}(\Psi_{\CCP}(T[i..j]) \cap \Psi_{\run} \cap \Psi_{\centerset}(C_{Q}) \cap \Psi_{\lcp}(\hat{K}) \cap \Psi_{\preceding}) = \emptyset$ holds if and only if 
the largest integer $\kappa$ exists. 
We can verify whether the largest integer $\kappa$ exists or not by the algorithm computing the largest integer $\kappa$. 
Therefore, the verification of $\mathcal{F}_{\SA} \cap \mathcal{F}_{\suffix}(\Psi_{\CCP}(T[i..j]) \cap \Psi_{\run} \cap \Psi_{\centerset}(C_{Q}) \cap \Psi_{\lcp}(\hat{K}) \cap \Psi_{\preceding}) = \emptyset$ 
takes $O(H^{2} \log^{2} n + \log^{5} n)$ time. 

\paragraph{Algorithm for subquery $\RSSQCX(T[i..j], b)$.}
The algorithm for $\RSSQCX(T[i..j], b)$ 
returns the lexicographically largest string $F$ in 
set $\mathcal{F}_{\SA} \cap \mathcal{F}_{\suffix}(\Psi_{\CCP}(T[i..j]) \cap \Psi_{\run} \cap \Psi_{\centerset}(C_{Q}) \cap \Psi_{\lcp}(\hat{K}) \cap \Psi_{\preceding})$. 
This algorithm is executed only if $\mathcal{F}_{\SA} \cap \mathcal{F}_{\suffix}(\Psi_{\CCP}(T[i..j]) \cap \Psi_{\run} \cap \Psi_{\centerset}(C_{Q}) \cap \Psi_{\lcp}(\hat{K}) \cap \Psi_{\preceding}) \neq \emptyset$; 
otherwise, subquery $\RSSQCX(T[i..j], b)$ can be answered by verifying whether $\mathcal{F}_{\SA} \cap \mathcal{F}_{\suffix}(\Psi_{\CCP}(T[i..j]) \cap \Psi_{\run} \cap \Psi_{\centerset}(C_{Q}) \cap \Psi_{\lcp}(\hat{K}) \cap \Psi_{\preceding}) = \emptyset$ or not in $O(H^{2} \log^{2} n + \log^{5} n)$ time. 

The algorithm for $\RSSQCX(T[i..j], b)$ leverages the largest integer $\kappa$. 
Lemma~\ref{lem:GammaC1_property}~\ref{enum:GammaC1_property:1} shows that 
$F = T[i..\gamma_{Q} - 1] \cdot C_{Q}^{n+1}[1..\hat{K}] \cdot R_{y + \kappa - 1}$ holds. 
Lemma~\ref{lem:GammaC1_property}~\ref{enum:GammaC1_property:2} shows that 
set $\mathcal{J}_{C}(h_{Q}, C_{Q}, \hat{M})$ contains 
a weighted point $(|f_{\recover}(([p_{s}, q_{s}]$, $[\ell_{s}, r_{s}]))|$, $T[\gamma_{s} + K_{s}..r_{s} + 1]$, $|\Psi_{\str}(T[p_{s}-1..r_{s}+1])|$, $T[p_{s}-1..r_{s}+1])$ satisfying $x \leq |f_{\recover}(([p_{s}, q_{s}], [\ell_{s}, r_{s}]))| \leq n$ and $T[\gamma_{s} + K_{s}..r_{s}+1] = R_{y + \kappa - 1}$. 
Lemma~\ref{lem:GammaC1_property}~\ref{enum:GammaC1_property:3} shows that 
$T[\gamma_{s} + K_{s} - \hat{K} - |[i, \gamma_{Q}-1]|..r_{s} + 1] = T[i..\gamma_{Q} - 1] \cdot C_{Q}^{n+1}[1..\hat{K}] \cdot R_{y + \kappa - 1}$ holds. 
Therefore, we can return string $T[\gamma_{s} + K_{s} - \hat{K} - |[i, \gamma_{Q}-1]|..r_{s} + 1]$ as the answer to subquery $\RSSQCX(T[i..j], b)$. 

The algorithm for $\RSSQCX(T[i..j], b)$ consists of three phases.
In the first phase, 
we compute the five integers $\lambda, x, y, y^{\prime}$, and $\kappa$.
The computation of the integer $\lambda$ takes $O(H^{2} \log n)$ time. 
The computation of the three integers $x, y$, and $y^{\prime}$ takes $O(H^{2} \log n + \log^{2} n)$ time. 
The computation of the largest integer $\kappa$ takes $O(H^{2} \log^{2} n + \log^{5} n)$ time. 
Therefore, the first phase takes $O(H^{2} \log^{2} n + \log^{5} n)$ time. 

In the second phase, 
we compute the interval attractor $([p_{s}, q_{s}], [\ell_{s}, r_{s}])$ corresponding to 
the weighted point $(|f_{\recover}(([p_{s}, q_{s}]$, $[\ell_{s}, r_{s}]))|$, $T[\gamma_{s} + K_{s}..r_{s} + 1]$, $|\Psi_{\str}(T[p_{s}-1..r_{s}+1])|$, $T[p_{s}-1..r_{s}+1])$. 
This interval attractor can be found in $O(H^{2} \log n + \log^{2} n)$ time 
using the query of Lemma~\ref{lem:JC1_Y_queries}~\ref{enum:JC1_Y_queries:4}. 
This query can be answered after accessing the dynamic data structures for the ordered set $\mathcal{Y}_{C}(h_{\lambda}, C_{\lambda}, M_{\lambda})$ in $O(\log n)$ time. 
Therefore, the second phase takes $O(H^{2} \log n + \log^{2} n)$ time. 

In the third phase, 
we return string $T[\gamma_{s} + K_{s} - \hat{K} - |[i, \gamma_{Q}-1]|..r_{s} + 1]$ as 
the answer to  subquery $\RSSQCX(T[i..j], b)$. 
Here, the string $T[\gamma_{s} + K_{s} - \hat{K} - |[i, \gamma_{Q}-1]|..r_{s} + 1]$ is represented as interval $[\gamma_{s} + K_{s} - \hat{K} - |[i, \gamma_{Q}-1]|, r_{s} + 1]$. 
The attractor position $\gamma_{s}$ can be obtained by attractor position query $\attrQ(([p_{s}, q_{s}], [\ell_{s}, r_{s}]))$. 
The integer $K_{s}$ can be obtained by C-LCP query $\clcpQ(([p_{s}, q_{s}]$, $[\ell_{s}, r_{s}]))$. 
The attractor position $\gamma_{Q}$ and $\hat{K}$ can be obtained by the algorithm computing the integer $\lambda$. 
Therefore, the third phase takes $O(H^{2})$ time. 

The three phases take $O(H^{2} \log^{2} n + \log^{5} n)$ time in total. 
Therefore, Lemma~\ref{lem:GammaC1_algorithm} holds.

\subsection{Subquery \texorpdfstring{$\RSSQCY(T[i..j], b)$}{RSSC2(T[i..j], b)}}\label{subsec:GammaC2}

The goal of this subsection is to answer subquery $\RSSQCY(T[i..j], b)$ under the assumption 
that (i) either $\mathcal{C}_{\run} = \emptyset$ or $C_{Q} = C_{\max}$ holds, and (ii) condition (A) of RSS query is satisfied. 
For this subsection, 
let $\hat{K} = |\lcp(T[\gamma_{Q}..j], C_{Q}^{n+1})|$ 
and $\hat{M} = (\hat{K} - (2 + \sum_{w = 1}^{h_{Q}+3} \lfloor \mu(w) \rfloor) ) \mod |C_{Q}|$. 
The following lemma states the summary of this subsection. 

\begin{lemma}\label{lem:GammaC2_algorithm}
We assume that (i) either $\mathcal{C}_{\run} = \emptyset$ or $C_{Q} = C_{\max}$ holds, and (ii) condition (A) of RSS query is satisfied. 
We can answer $\RSSQCY(T[i..j], b)$ in $O(H^{2} \log^{2} n + \log^{5} n)$ time 
using (A) the data structures for RSC query, 
(B) interval $[i, j]$, 
and (C) the starting position $\eta$ of the sa-interval $[\eta, \eta^{\prime}]$ of $T[i..j]$. 
If the subquery returns a string $F$, 
then $F$ is represented as an interval $[g, g + |F| - 1]$ satisfying $T[g..g + |F| - 1] = F$. 
\end{lemma}
\begin{proof}
See Section~\ref{subsubsec:gamma_C2_algorithm}.
\end{proof}

We use set $\Psi_{h_{Q}} \cap \Psi_{\source} \cap \Psi_{\centerset}(C_{Q}) \cap \Psi_{\modulo}(\hat{M}) \cap \Psi_{\succeeding} \cap \Psi_{\samp}$ to explain the idea behind solving subquery $\RSSQCY(T[i..j], b)$. 
Let $([p_{1}, q_{1}], [\ell_{1}, r_{1}]), ([p_{2}, q_{2}], [\ell_{2}, r_{2}])$, 
$\ldots$, $([p_{k}, q_{k}], [\ell_{k}, r_{k}])$ be the interval attractors in the set $\Psi_{h_{Q}} \cap \Psi_{\source} \cap \Psi_{\centerset}(C_{Q}) \cap \Psi_{\modulo}(\hat{M}) \cap \Psi_{\succeeding} \cap \Psi_{\samp}$. 
Let $\gamma_{s}$ be the attractor position of each interval attractor $([p_{s}, q_{s}], [\ell_{s}, r_{s}]) \in \Psi_{h_{Q}} \cap \Psi_{\source} \cap \Psi_{\centerset}(C_{Q}) \cap \Psi_{\modulo}(\hat{M}) \cap \Psi_{\succeeding} \cap \Psi_{\samp}$. 
Let $K_{s} = |\lcp(T[\gamma_{s}..r_{s}], C_{Q}^{n+1})|$ for simplicity.  

We leverage the set $\mathcal{J}_{C^{\prime}}(h_{Q}, C_{Q}, \hat{M})$ of weighted points on grid $([1, n], \mathcal{Y}_{C^{\prime}}(h_{Q}, C_{Q}, \hat{M}))$ introduced in Section~\ref{subsec:RSC_comp_C2}. 
The summary of the set $\mathcal{J}_{C^{\prime}}(h_{Q}, C_{Q}, \hat{M})$ and ordered set $\mathcal{Y}_{C^{\prime}}(h_{Q}, C_{Q}, \hat{M})$ is as follows 
(see Section~\ref{subsec:RSC_comp_C2} for the details of the set and ordered set):
\begin{itemize}
    \item the ordered set $\mathcal{Y}_{C^{\prime}}(h_{Q}, C_{Q}, \hat{M})$ consists of $d$ strings $R_{1}, R_{2}, \ldots, R_{d}$ ($R_{1} \prec R_{2} \prec \cdots \prec R_{d}$);
    \item the set $\mathcal{J}_{C^{\prime}}(h_{Q}, C_{Q}, \hat{M})$ contains the weighted point 
    $(|f_{\recover}(([p_{s}, q_{s}], [\ell_{s}, r_{s}]))|$, $T[\gamma_{s} + K_{s}..r_{s} + 1]$, $|\Psi_{\str}(T[p_{s}-1..r_{s}+1])|$, $T[p_{s}-1..r_{s}+1])$ 
    corresponding to each interval attractor $([p_{s}, q_{s}]$, $[\ell_{s}, r_{s}])$ in set $\Psi_{h_{Q}} \cap \Psi_{\source} \cap \Psi_{\centerset}(C_{Q}) \cap \Psi_{\modulo}(\hat{M}) \cap \Psi_{\succeeding} \cap \Psi_{\samp}$. 
\end{itemize}
For this subsection, 
we define three integers $x, y$ and $y^{\prime}$ as follows: 
\begin{itemize}
    \item $x = 1 + \lfloor \frac{\hat{K} - (2 + \sum_{w = 1}^{h_{Q}+3} \lfloor \mu(w) \rfloor)}{|C_{Q}|} \rfloor$;
    \item $y = \min \{ s \in [1, d] \mid T[\gamma_{Q} + \hat{K}..j] \prec R_{s} \}$;
    \item $y^{\prime} = \max \{ s \in [1, d] \mid R_{s} \prec T[\gamma_{Q} + \hat{K}..j]\# \}$. 
\end{itemize}

In the next paragraphs, 
we introduce $(y^{\prime} - y + 1)$ subsets of set $\Psi_{\RR}$ 
and two sequences of integers to explain the relationship between the set $\mathcal{Y}_{C^{\prime}}(h_{Q}, C_{Q}, \hat{M})$ and subquery $\RSSQCY(T[i..j], b)$. 

\paragraph{Subset $\Psi^{C^{\prime}}(t)$.}
For each integer $t \in [1, y^{\prime} - y + 1]$, 
subset $\Psi^{C^{\prime}}(t) \subseteq \Psi_{\RR}$ consists of interval attractors such that 
each interval attractor $([p, q], [\ell, r]) \in \Psi^{C^{\prime}}(t)$ satisfies the following two conditions: 
\begin{itemize}
    \item $([p, q], [\ell, r]) \in \Psi_{h_{Q}} \cap \Psi_{\run} \cap \Psi_{\centerset}(C_{Q}) \cap \Psi_{\lcp}(\hat{K}) \cap \Psi_{\succeeding}$;
    \item $R_{y + t - 1} \preceq T[\gamma + \hat{K}..r + 1] \preceq R_{y^{\prime}}$ for the attractor position $\gamma$ of the interval attractor $([p, q], [\ell, r])$.
\end{itemize}
Formally, the subset $\Psi^{C^{\prime}}(t)$ is defined as follows: 
\begin{equation*}
    \begin{split}
    \Psi^{C^{\prime}}(t) &= \{ ([p, q], [\ell, r]) \in \Psi_{\RR} \mid (([p, q], [\ell, r]) \in \Psi_{h_{Q}} \cap \Psi_{\run} \cap \Psi_{\centerset}(C_{Q}) \cap \Psi_{\lcp}(\hat{K}) \cap \Psi_{\succeeding}) \\
    &\land (R_{y + t - 1} \preceq T[\gamma + \hat{K}..r + 1] \preceq R_{y^{\prime}}) \}.
    \end{split}
\end{equation*}

\paragraph{Two sequences $\Gamma_{C^{\prime}}$ and $\Gamma_{C^{\prime}, \sub}$.}
The sequence $\Gamma_{C^{\prime}}$ consists of $(y^{\prime} - y + 1)$ integers $u_{1}$, $u_{2}$, $\ldots$, $u_{y^{\prime} - y + 1}$. 
Each integer $u_{t}$ is $1$ if $\mathcal{F}_{\SA} \cap \mathcal{F}_{\suffix}(\Psi_{\CCP}(T[i..j]) \cap \Psi^{C^{\prime}}(t)) \neq \emptyset$; 
otherwise $u_{t}$ is $0$. 
The following lemma states four properties of sequence $\Gamma_{C^{\prime}}$. 

\begin{lemma}\label{lem:GammaC2_property}
Consider condition (A) of RSS query for the given RSS query $\RSSQ(T[i..j], b)$.
Let $\kappa$ be the largest integer in set $[1, y^{\prime} - y + 1]$ satisfying $u_{\kappa} = 1$ for sequence $\Gamma_{C^{\prime}} = u_{1}, u_{2}, \ldots, u_{y^{\prime} - y + 1}$. 
Then, the following four statements hold: 
\begin{enumerate}[label=\textbf{(\roman*)}]
    \item \label{enum:GammaC2_property:1} if the largest integer $\kappa$ exists, then 
    string $T[i..\gamma_{Q} - 1] \cdot C_{Q}^{n+1}[1..\hat{K}] \cdot R_{y + \kappa - 1}$ is 
    the lexicographically largest string in set $\mathcal{F}_{\SA} \cap \mathcal{F}_{\suffix}(\Psi_{\CCP}(T[i..j]) \cap \Psi_{\run} \cap \Psi_{\centerset}(C_{Q}) \cap \Psi_{\lcp}(\hat{K}) \cap \Psi_{\succeeding})$. 
    Otherwise, the set $\mathcal{F}_{\SA} \cap \mathcal{F}_{\suffix}(\Psi_{\CCP}(T[i..j]) \cap \Psi_{\run} \cap \Psi_{\centerset}(C_{Q}) \cap \Psi_{\lcp}(\hat{K}) \cap \Psi_{\succeeding})$ is empty; 
    \item \label{enum:GammaC2_property:2} if the largest integer $\kappa$ exists, 
    then set $\mathcal{J}_{C}(h_{Q}, C_{Q}, \hat{M})$ contains a weighted point 
    $(|f_{\recover}(([p_{s}$, $q_{s}]$, $[\ell_{s}, r_{s}]))|$, $T[\gamma_{s} + K_{s}..r_{s} + 1]$, $|\Psi_{\str}(T[p_{s}-1..r_{s}+1])|$, $T[p_{s}-1..r_{s}+1])$  
    satisfying $x \leq |f_{\recover}(([p_{s}, q_{s}], [\ell_{s}, r_{s}]))| \leq n$ 
    and $T[\gamma_{s} + K_{s}..r_{s}+1] = R_{y + \kappa - 1}$;
    \item \label{enum:GammaC2_property:3} 
    $T[\gamma_{s} + K_{s} - \hat{K} - |[i, \gamma_{Q}-1]|..r_{s} + 1] = T[i..\gamma_{Q} - 1] \cdot C_{Q}^{n+1}[1..\hat{K}] \cdot R_{y + \kappa - 1}$ for 
    the interval attractor $([p_{s}, q_{s}], [\ell_{s}, r_{s}])$ corresponding to 
    the weighted point $(|f_{\recover}(([p_{s}, q_{s}], [\ell_{s}, r_{s}]))|$, $T[\gamma_{s} + K_{s}..r_{s} + 1]$, $|\Psi_{\str}(T[p_{s}-1..r_{s}+1])|$, $T[p_{s}-1..r_{s}+1])$ 
    of Lemma~\ref{lem:GammaC2_property}~\ref{enum:GammaC2_property:2};
    \item \label{enum:GammaC2_property:4}
    sequence $\Gamma_{C^{\prime}}$ is non-increasing~(i.e., $u_{1} \geq u_{2} \geq \cdots \geq u_{y^{\prime} - y + 1}$). 
\end{enumerate}
\end{lemma}
\begin{proof}
This lemma can be proved using the same approach as for Lemma~\ref{lem:GammaC1_property}.
\end{proof}

Next, sequence $\Gamma_{C^{\prime}, \sub}$ consists of $(y^{\prime} - y + 1)$ integers 
$\alpha_{1}, \alpha_{2}, \ldots, \alpha_{y^{\prime} - y + 1} \in \mathbb{N}_{0}$. 
Here, each integer $\alpha_{t}$ is defined as 
$\alpha_{t} = \rangecount(\mathcal{J}_{C}(h_{Q}, C_{Q}, \hat{M}), x, n, R_{y + t - 1}, R_{y^{\prime}})$. 
Here, $\rangecount$ is the range-count query introduced in Section~\ref{subsec:range_data_structure}. 
The following lemma states three properties of sequence $\Gamma_{C^{\prime}, \sub}$.

\begin{lemma}\label{lem:GammaC2_sub_property}
Consider condition (A) of RSS query for the given RSS query $\RSSQ(T[i..j], b)$.
The following three statements hold for two sequences $\Gamma_{C^{\prime}} = u_{1}, u_{2}, \ldots, u_{y^{\prime} - y + 1}$ and $\Gamma_{C^{\prime}, \sub} = \alpha_{1}$, $\alpha_{2}$, $\ldots$, $\alpha_{y^{\prime} - y + 1}$: 
\begin{enumerate}[label=\textbf{(\roman*)}]
    \item \label{enum:GammaC2_sub_property:1} 
    sequence $\Gamma_{C^{\prime}, \sub}$ is non-increasing (i.e., $\alpha_{1} \geq \alpha_{2} \geq \cdots \geq \alpha_{y^{\prime} - y + 1}$);    
    \item \label{enum:GammaC2_sub_property:2} 
    for an integer $t \in [1, y^{\prime} - y + 1]$, 
    consider the largest integer $t^{\prime}$ in set $[t, y^{\prime} - y + 1]$ satisfying 
    $\alpha_{t} = \alpha_{t^{\prime}}$. 
    If $\alpha_{t^{\prime}} \geq 1$, 
    then set $\mathcal{J}_{C}(h_{Q}, C_{Q}, \hat{M})$ contains a weighted point 
    $(|f_{\recover}(([p_{s}, q_{s}], [\ell_{s}, r_{s}]))|$, $T[\gamma_{s} + K_{s}..r_{s} + 1]$, $|\Psi_{\str}(T[p_{s}-1..r_{s}+1])|$, $T[p_{s}-1..r_{s}+1])$ satisfying 
    $x \leq |f_{\recover}(([p_{s}, q_{s}]$, $[\ell_{s}, r_{s}]))| \leq n$ 
    and $T[\gamma_{s} + K_{s}..r_{s}+1] = R_{y + t^{\prime} - 1}$;
    \item \label{enum:GammaC2_sub_property:3}
    consider the three integers $t, t^{\prime}$, and $s$ of Lemma~\ref{lem:GammaC2_sub_property}~\ref{enum:GammaC2_sub_property:2}. 
    Let $\upsilon = \gamma_{s} + K_{s} - \hat{K}$.
    Then, 
    $u_{t} = 1 \Leftrightarrow (\alpha_{t^{\prime}} \geq 1) \land (\RSCQ(\upsilon - |[i, \gamma_{Q}-1]|, \upsilon + |[\gamma_{Q}, j]| - 1) < b - \eta + 1)$. 
\end{enumerate}
\end{lemma}
\begin{proof}
This lemma can be proved using the same approach as for Lemma~\ref{lem:GammaC1_sub_property}.
\end{proof}

\subsubsection{Algorithm}\label{subsubsec:gamma_C2_algorithm}
We prove Lemma~\ref{lem:GammaC2_algorithm}, i.e., 
we show that subquery $\RSSQCY(T[i..j], b)$ can be answered 
in $O(H^{2} \log^{2} n + \log^{6} n)$ time using the data structures for RSC query 
and interval $[i, j]$. 

We answer subquery $\RSSQCY(T[i..j], b)$ by modifying 
the algorithm for subquery $\RSSQCX(T[i..j], b)$. 
In Section~\ref{subsubsec:gamma_C1_algorithm}, 
we presented the algorithm answering subquery $\RSSQCX(T[i..j], b)$ 
using two sequences $\Gamma_{C}$ and $\Gamma_{C, \sub}$. 
Two sequences $\Gamma_{C^{\prime}}$ and $\Gamma_{C^{\prime}, \sub}$ corresponds to 
the two sequences $\Gamma_{C}$ and $\Gamma_{C, \sub}$, respectively. 
Lemma~\ref{lem:GammaC2_property} states properties of the sequence $\Gamma_{C^{\prime}}$. 
This lemma corresponds to Lemma~\ref{lem:GammaC1_property}, which states properties of the sequence $\Gamma_{C}$. 
Similarly, 
Lemma~\ref{lem:GammaC2_sub_property} states properties of the sequence $\Gamma_{C^{\prime}, \sub}$. 
This lemma corresponds to Lemma~\ref{lem:GammaC1_sub_property}, which states properties of the sequence $\Gamma_{C, \sub}$. 
Therefore, subquery $\RSSQCY(T[i..j], b)$ can be answered in $O(H^{2} \log^{2} n + \log^{5} n)$ time 
using the same approach as for subquery $\RSSQCX(T[i..j], b)$.

\subsection{Subquery \texorpdfstring{$\RSSQDX(T[i..j], b, K)$}{RSSD1(T[i..j], b, K)}}\label{subsec:GammaD1}
The goal of this subsection is to answer subquery $\RSSQDX(T[i..j], b, K)$ under the assumption 
that (i) either $\mathcal{C}_{\run} = \emptyset$ or $C_{Q} = C_{\max}$ holds, and (ii) condition (B) of RSS query is satisfied. 
For this subsection, 
let $M = (K - (2 + \sum_{w = 1}^{h_{Q}+3} \lfloor \mu(w) \rfloor) ) \mod |C_{Q}|$ 
and $x = 1 + \lfloor \frac{K - (2 + \sum_{w = 1}^{h_{Q}+3} \lfloor \mu(w) \rfloor)}{|C_{Q}|} \rfloor$. 
The following lemma states the summary of this subsection. 

\begin{lemma}\label{lem:GammaD1_algorithm}
We assume that (i) either $\mathcal{C}_{\run} = \emptyset$ or $C_{Q} = C_{\max}$ holds, and (ii) condition (B) of RSS query is satisfied. 
We can answer $\RSSQDX(T[i..j], b, K)$ in $O(H^{2} \log^{2} n + \log^{5} n)$ time 
using (A) the data structures for RSC query, 
(B) interval $[i, j]$, 
and (C) the starting position $\eta$ of the sa-interval $[\eta, \eta^{\prime}]$ of $T[i..j]$. 
If the subquery returns a string $F$, 
then $F$ is represented as an interval $[g, g + |F| - 1]$ satisfying $T[g..g + |F| - 1] = F$; 
otherwise (i.e., $\mathcal{F}_{\SA} \cap \mathcal{F}_{\suffix}(\Psi_{\CCP}(T[i..j]) \cap \Psi_{\run} \cap \Psi_{\centerset}(C_{Q}) \cap \Psi_{\lcp}(K) \cap \Psi_{\preceding}) = \emptyset$), the running time of subquery $\RSSQDX(T[i..j], b, K)$ can be reduced to $O(H^{2} + \log^{4} n)$ time. 
\end{lemma}
\begin{proof}
See Section~\ref{subsubsec:gamma_D1_algorithm}.
\end{proof}

We use set $\Psi_{h_{Q}} \cap \Psi_{\source} \cap \Psi_{\centerset}(C_{Q}) \cap \Psi_{\modulo}(M) \cap \Psi_{\preceding} \cap \Psi_{\samp}$ to explain the idea behind solving $\RSSQDX(T[i..j], b, K)$. 
Let $([p_{1}, q_{1}], [\ell_{1}, r_{1}]), ([p_{2}, q_{2}], [\ell_{2}, r_{2}])$, 
$\ldots$, $([p_{k}, q_{k}], [\ell_{k}, r_{k}])$ be the interval attractors in the set $\Psi_{h_{Q}} \cap \Psi_{\source} \cap \Psi_{\centerset}(C_{Q}) \cap \Psi_{\modulo}(M) \cap \Psi_{\preceding} \cap \Psi_{\samp}$. 
Let $\gamma_{s}$ be the attractor position of each interval attractor $([p_{s}, q_{s}], [\ell_{s}, r_{s}]) \in \Psi_{h_{Q}} \cap \Psi_{\source} \cap \Psi_{\centerset}(C_{Q}) \cap \Psi_{\modulo}(M) \cap \Psi_{\preceding} \cap \Psi_{\samp}$. 
Let $K_{s} = |\lcp(T[\gamma_{s}..r_{s}], C_{Q}^{n+1})|$ for simplicity.  

Similar to subquery $\RSSQCX(T[i..j], b)$, 
we leverage the set $\mathcal{J}_{C}(h_{Q}, C_{Q}, M)$ of weighted points on grid $([1, n], \mathcal{Y}_{C}(h_{Q}$, $C_{Q}, M))$ introduced in Section~\ref{subsec:RSC_comp_C1}. 
The summary of the set $\mathcal{J}_{C}(h_{Q}, C_{Q}, M)$ and ordered set $\mathcal{Y}_{C}(h_{Q}, C_{Q}, M)$ is as follows 
(see Section~\ref{subsec:RSC_comp_C1} for the details of the set and ordered set):
\begin{itemize}
    \item the ordered set $\mathcal{Y}_{C}(h_{Q}, C_{Q}, M))$ consists of $d$ strings $R_{1}, R_{2}, \ldots, R_{d}$ ($R_{1} \prec R_{2} \prec \cdots \prec R_{d}$);
    \item the set $\mathcal{J}_{C}(h_{Q}, C_{Q}, M)$ contains the weighted point 
    $(|f_{\recover}(([p_{s}, q_{s}], [\ell_{s}, r_{s}]))|$, $T[\gamma_{s} + K_{s}..r_{s} + 1]$, $|\Psi_{\str}(T[p_{s}-1..r_{s}+1])|$, $T[p_{s}-1..r_{s}+1])$ 
    corresponding to each interval attractor $([p_{s}, q_{s}]$, $[\ell_{s}, r_{s}])$ in set $\Psi_{h_{Q}} \cap \Psi_{\source} \cap \Psi_{\centerset}(C_{Q}) \cap \Psi_{\modulo}(M) \cap \Psi_{\preceding} \cap \Psi_{\samp}$. 
\end{itemize}

In the next paragraphs, 
we introduce $d$ subsets of set $\Psi_{\RR}$ 
and two sequences of integers to explain the relationship between the set $\mathcal{Y}_{C}(h_{Q}, C_{Q}, M))$ and subquery $\RSSQDX(T[i..j], b, K)$. 

\paragraph{Subset $\Psi^{D}(t)$.}
For each integer $t \in [1, d]$, 
subset $\Psi^{D}(t) \subseteq \Psi_{\RR}$ consists of interval attractors such that 
each interval attractor $([p, q], [\ell, r]) \in \Psi^{D}(t)$ satisfies the following two conditions: 
\begin{itemize}
    \item $([p, q], [\ell, r]) \in \Psi_{h_{Q}} \cap \Psi_{\run} \cap \Psi_{\centerset}(C_{Q}) \cap \Psi_{\lcp}(K) \cap \Psi_{\preceding}$;
    \item $R_{t} \preceq T[\gamma + K..r + 1] \preceq R_{d}$ for the attractor position $\gamma$ of the interval attractor $([p, q], [\ell, r])$.
\end{itemize}
Formally, the subset $\Psi^{D}(t)$ is defined as follows: 
\begin{equation*}
    \begin{split}
    \Psi^{D}(t) &= \{ ([p, q], [\ell, r]) \in \Psi_{\RR} \mid (([p, q], [\ell, r]) \in \Psi_{h_{Q}} \cap \Psi_{\run} \cap \Psi_{\centerset}(C_{Q}) \cap \Psi_{\lcp}(K) \cap \Psi_{\preceding}) \\
    &\land (R_{t} \preceq T[\gamma + K..r + 1] \preceq R_{d}) \}.
    \end{split}
\end{equation*}

\paragraph{Two sequences $\Gamma_{D}$ and $\Gamma_{D, \sub}$.}
The sequence $\Gamma_{D}$ consists of $d$ integers $u_{1}, u_{2}, \ldots, u_{d}$. 
Each integer $u_{t}$ is $1$ if $\mathcal{F}_{\SA} \cap \mathcal{F}_{\suffix}(\Psi_{\CCP}(T[i..j]) \cap \Psi^{D}(t)) \neq \emptyset$; 
otherwise $u_{t}$ is $0$. 
The following lemma states four properties of sequence $\Gamma_{D}$.

\begin{lemma}\label{lem:GammaD1_property}
Consider condition (B) of RSS query for the given RSS query $\RSSQ(T[i..j], b)$.
Let $\kappa$ be the largest integer in set $[1, d]$ satisfying $u_{\kappa} = 1$ for sequence $\Gamma_{D} = u_{1}, u_{2}, \ldots, u_{d}$. 
Then, the following four statements hold: 
\begin{enumerate}[label=\textbf{(\roman*)}]
    \item \label{enum:GammaD1_property:1} if the largest integer $\kappa$ exists, then 
    string $T[i..\gamma_{Q} - 1] \cdot C_{Q}^{n+1}[1..K] \cdot R_{\kappa}$ is 
    the lexicographically largest string in set $\mathcal{F}_{\SA} \cap \mathcal{F}_{\suffix}(\Psi_{\CCP}(T[i..j]) \cap \Psi_{\run} \cap \Psi_{\centerset}(C_{Q}) \cap \Psi_{\lcp}(K) \cap \Psi_{\preceding})$. 
    Otherwise, the set $\mathcal{F}_{\SA} \cap \mathcal{F}_{\suffix}(\Psi_{\CCP}(T[i..j]) \cap \Psi_{\run} \cap \Psi_{\centerset}(C_{Q}) \cap \Psi_{\lcp}(K) \cap \Psi_{\preceding})$ is empty; 
    \item \label{enum:GammaD1_property:2} if the largest integer $\kappa$ exists, 
    then set $\mathcal{J}_{C}(h_{Q}, C_{Q}, M)$ contains a weighted point 
    $(|f_{\recover}(([p_{s}$, $q_{s}]$, $[\ell_{s}, r_{s}]))|$, $T[\gamma_{s} + K_{s}..r_{s} + 1]$, $|\Psi_{\str}(T[p_{s}-1..r_{s}+1])|$, $T[p_{s}-1..r_{s}+1])$  
    satisfying $x \leq |f_{\recover}(([p_{s}, q_{s}], [\ell_{s}, r_{s}]))| \leq n$ 
    and $T[\gamma_{s} + K_{s}..r_{s}+1] = R_{\kappa}$;
    \item \label{enum:GammaD1_property:3} 
    $T[\gamma_{s} + K_{s} - K - |[i, \gamma_{Q}-1]|..r_{s} + 1] = T[i..\gamma_{Q} - 1] \cdot C_{Q}^{n+1}[1..K] \cdot R_{\kappa}$ for 
    the interval attractor $([p_{s}, q_{s}], [\ell_{s}, r_{s}])$ corresponding to 
    the weighted point $(|f_{\recover}(([p_{s}, q_{s}], [\ell_{s}, r_{s}]))|$, $T[\gamma_{s} + K_{s}..r_{s} + 1]$, $|\Psi_{\str}(T[p_{s}-1..r_{s}+1])|$, $T[p_{s}-1..r_{s}+1])$ 
    of Lemma~\ref{lem:GammaD1_property}~\ref{enum:GammaD1_property:2};
    \item \label{enum:GammaD1_property:4}
    sequence $\Gamma_{D}$ is non-increasing~(i.e., $u_{1} \geq u_{2} \geq \cdots \geq u_{d}$). 
\end{enumerate}
\end{lemma}
\begin{proof}
See Section~\ref{subsubsec:GammaD1_property_proof}.
\end{proof}

Next, sequence $\Gamma_{D, \sub}$ consists of $d$ integers 
$\alpha_{1}, \alpha_{2}, \ldots, \alpha_{d} \in \mathbb{N}_{0}$. 
Here, each integer $\alpha_{t}$ is defined as 
$\alpha_{t} = \rangecount(\mathcal{J}_{C}(h_{Q}, C_{Q}, M), x, n, R_{t}, R_{d})$. 
Here, $\rangecount$ is the range-count query introduced in Section~\ref{subsec:range_data_structure}. 
The following lemma states three properties of sequence $\Gamma_{D, \sub}$. 

\begin{lemma}\label{lem:GammaD1_sub_property}
Consider condition (B) of RSS query for the given RSS query $\RSSQ(T[i..j], b)$.
The following three statements hold for two sequences $\Gamma_{D} = u_{1}, u_{2}, \ldots, u_{d}$ and $\Gamma_{D, \sub} = \alpha_{1}, \alpha_{2}, \ldots, \alpha_{d}$: 
\begin{enumerate}[label=\textbf{(\roman*)}]
    \item \label{enum:GammaD1_sub_property:1} 
    sequence $\Gamma_{D, \sub}$ is non-increasing (i.e., $\alpha_{1} \geq \alpha_{2} \geq \cdots \geq \alpha_{d}$);    
    \item \label{enum:GammaD1_sub_property:2} 
    for an integer $t \in [1, d]$, 
    consider the largest integer $t^{\prime}$ in set $[t, d]$ satisfying 
    $\alpha_{t} = \alpha_{t^{\prime}}$. 
    If $\alpha_{t^{\prime}} \geq 1$, 
    then set $\mathcal{J}_{C}(h_{Q}, C_{Q}, M)$ contains a weighted point 
    $(|f_{\recover}(([p_{s}, q_{s}], [\ell_{s}, r_{s}]))|$, $T[\gamma_{s} + K_{s}..r_{s} + 1]$, $|\Psi_{\str}(T[p_{s}-1..r_{s}+1])|$, $T[p_{s}-1..r_{s}+1])$ satisfying 
    $x \leq |f_{\recover}(([p_{s}, q_{s}], [\ell_{s}, r_{s}]))| \leq n$ 
    and $T[\gamma_{s} + K_{s}..r_{s}+1] = R_{t^{\prime}}$;
    \item \label{enum:GammaD1_sub_property:3}
    consider the three integers $t, t^{\prime}$, and $s$ of Lemma~\ref{lem:GammaD1_sub_property}~\ref{enum:GammaD1_sub_property:2}. 
    Let $\upsilon = \gamma_{s} + K_{s} - K$.
    Then, 
    $u_{t} = 1 \Leftrightarrow (\alpha_{t^{\prime}} \geq 1) \land (\RSCQ(\upsilon - |[i, \gamma_{Q}-1]|, \upsilon + |[\gamma_{Q}, j]| - 1) < b - \eta + 1)$. 
\end{enumerate}
\end{lemma}
\begin{proof}
See Section~\ref{subsubsec:GammaD1_sub_property_proof}.
\end{proof}

\subsubsection{Proof of Lemma~\ref{lem:GammaD1_property}}\label{subsubsec:GammaD1_property_proof}
The following two propositions state the relationship between 
two sets $\Psi_{h_{Q}} \cap \Psi_{\run} \cap \Psi_{\centerset}(C_{Q}) \cap \Psi_{\lcp}(K) \cap \Psi_{\preceding}$ 
and $\Psi_{h_{Q}} \cap \Psi_{\source} \cap \Psi_{\centerset}(C_{Q}) \cap \Psi_{\modulo}(M) \cap \Psi_{\preceding} \cap \Psi_{\samp}$. 

\begin{proposition}\label{prop:JD1_correspondence_property}
Consider an interval attractor $([p, q], [\ell, r]) \in \Psi_{h_{Q}} \cap \Psi_{\run} \cap \Psi_{\centerset}(C_{Q}) \cap \Psi_{\lcp}(K) \cap \Psi_{\preceding}$. 
Then, there exists an integer $s \in [1, k]$ satisfying the following two conditions: 
\begin{enumerate}[label=\textbf{(\roman*)}]
    \item $x \leq |f_{\recover}(([p_{s}, q_{s}], [\ell_{s}, r_{s}]))| \leq n$; 
    \item $T[\gamma_{s} + K_{s}..r_{s}+1] = T[\gamma + K..r+1]$ 
    for the attractor position $\gamma$ of the interval attractor $([p, q], [\ell, r])$.
\end{enumerate}
\end{proposition}
\begin{proof}
    Proposition~\ref{prop:JD1_correspondence_property} can be proved using the same approach as for Proposition~\ref{prop:JC1_correspondence_property}.
\end{proof}

\begin{proposition}\label{prop:JD1_inverse_correspondence_property}
For an integer $s \in [1, k]$, 
if $x \leq |f_{\recover}(([p_{s}, q_{s}], [\ell_{s}, r_{s}]))| \leq n$ holds, 
then set $\Psi_{h_{Q}} \cap \Psi_{\run} \cap \Psi_{\centerset}(C_{Q}) \cap \Psi_{\lcp}(K) \cap \Psi_{\preceding}$ contains an interval attractor $([p, q], [\ell, r])$ satisfying 
$T[\gamma_{s} + K_{s}..r_{s}+1] = T[\gamma + K..r+1]$ and $\gamma = \gamma_{s} + K_{s} - K$ 
for the attractor position $\gamma$ of the interval attractor $([p, q], [\ell, r])$.
\end{proposition}
\begin{proof}
    Proposition~\ref{prop:JD1_inverse_correspondence_property} can be proved using the same approach as for Proposition~\ref{prop:JC1_inverse_correspondence_property}.
\end{proof}

The following proposition states three properties of subset $\Psi^{D}(t)$ for each integer $t \in [1, d]$. 

\begin{proposition}\label{prop:Psi_D1_Property}
Consider condition (B) of RSS query for the given RSS query $\RSSQ(T[i..j], b)$.
The following three statements hold: 
\begin{enumerate}[label=\textbf{(\roman*)}]
    \item \label{enum:Psi_D1_Property:1} 
    $\Psi^{D}(1) = \Psi_{\CCP}(T[i..j]) \cap \Psi_{\run} \cap \Psi_{\centerset}(C_{Q}) \cap \Psi_{\lcp}(K) \cap \Psi_{\preceding}$;
    \item \label{enum:Psi_D1_Property:2} 
    $\Psi^{D}(t) \supseteq \Psi^{D}(t+1)$ for each integer $t \in [1, d]$;
    \item \label{enum:Psi_D1_Property:3}
    consider an interval attractor $([p, q], [\ell, r])$ in set $\Psi^{D}(t) \setminus \Psi^{D}(t+1)$ for an integer $t \in [1, d]$. 
    Then, $T[\gamma - |[i, \gamma_{Q}-1]|..\gamma-1] \cdot T[\gamma..r+1] = T[i..\gamma_{Q}-1] \cdot C_{Q}^{n+1}[1..K] \cdot R_{t}$ holds 
    for the attractor position $\gamma$ of the interval attractor $([p, q], [\ell, r])$.
\end{enumerate}
Here, let $\Psi^{D}(d + 1) = \emptyset$ for simplicity. 
\end{proposition}
\begin{proof}
The following two statements are used to prove Proposition~\ref{prop:Psi_D1_Property}:
\begin{enumerate}[label=\textbf{(\Alph*)}]
    \item $\Psi^{D}(1) \subseteq \Psi_{\CCP}(T[i..j]) \cap \Psi_{\run} \cap \Psi_{\centerset}(C_{Q}) \cap \Psi_{\lcp}(K) \cap \Psi_{\preceding}$;
    \item $\Psi^{D}(1) \supseteq \Psi_{\CCP}(T[i..j]) \cap \Psi_{\run} \cap \Psi_{\centerset}(C_{Q}) \cap \Psi_{\lcp}(K) \cap \Psi_{\preceding}$.    
\end{enumerate}

\textbf{Proof of statement (A).}
We can use Lemma~\ref{lem:CCP_special_property} because 
$\lcs(T[i..\gamma_{Q}-1], C_{Q}^{n+1}) = T[i..\gamma_{Q}-1]$ 
and $\lcp(T[\gamma_{Q}..j], C_{Q}^{n+1}) = T[\gamma_{Q}..j]$ follow from 
condition (B) of RSS query. 
Lemma~\ref{lem:CCP_special_property} shows that 
$\Psi_{h_{Q}} \cap \Psi_{\run} \cap \Psi_{\centerset}(C_{Q}) \cap (\bigcup_{\lambda = |[\gamma_{Q}, j]|}^{n} \Psi_{\lcp}(\lambda)) \subseteq \Psi_{\CCP}(T[i..j])$ holds. 
$\Psi^{D}(1) \subseteq \Psi_{\run} \cap \Psi_{\centerset}(C_{Q}) \cap \Psi_{\lcp}(K) \cap \Psi_{\preceding}$ 
and $\Psi_{\run} \cap \Psi_{\centerset}(C_{Q}) \cap \Psi_{\lcp}(K) \cap \Psi_{\preceding} \subseteq \Psi_{h_{Q}} \cap \Psi_{\run} \cap \Psi_{\centerset}(C_{Q}) \cap (\bigcup_{\lambda = |[\gamma_{Q}, j]|}^{n} \Psi_{\lcp}(\lambda)) \subseteq \Psi_{\CCP}(T[i..j])$ hold. 
Therefore, $\Psi^{D}(1) \subseteq \Psi_{\CCP}(T[i..j]) \cap \Psi_{\run} \cap \Psi_{\centerset}(C_{Q}) \cap \Psi_{\lcp}(K) \cap \Psi_{\preceding}$ holds. 

\textbf{Proof of statement (B).}
Consider an interval attractor $([p, q], [\ell, r]) \in \Psi_{\CCP}(T[i..j]) \cap \Psi_{\run} \cap \Psi_{\centerset}(C_{Q}) \cap \Psi_{\lcp}(K) \cap \Psi_{\preceding}$. 
Because of $([p, q], [\ell, r]) \in \Psi_{\CCP}(T[i..j])$, 
Lemma~\ref{lem:CCP_property}~\ref{enum:CCP_property:4} shows that 
$T[\gamma_{Q}..j] \prec T[\gamma..r+1] \prec T[\gamma_{Q}..j]\#$ holds for the attractor position $\gamma$ of the interval attractor $([p, q], [\ell, r])$. 
Lemma~\ref{lem:CCP_property}~\ref{enum:CCP_property:1} shows that $([p, q], [\ell, r]) \in \Psi_{h_{Q}}$ holds. 
Because of $([p, q], [\ell, r]) \in \Psi_{h_{Q}} \cap \Psi_{\run} \cap \Psi_{\centerset}(C_{Q}) \cap \Psi_{\lcp}(K) \cap \Psi_{\preceding}$, 
Proposition~\ref{prop:JD1_correspondence_property} shows that 
there exists an integer $s \in [1, k]$ satisfying 
$T[\gamma_{s} + K_{s}..r_{s}+1] = T[\gamma + K..r+1]$. 
Because of $T[\gamma_{s} + K_{s}..r_{s}+1] \in \mathcal{Y}_{C}(h_{Q}, C_{Q}, M))$, 
there exists an integer $\tau \in [1, d]$ satisfying $R_{\tau} = T[\gamma_{s} + K_{s}..r_{s}+1]$. 

We prove $([p, q], [\ell, r]) \in \Psi^{D}(1)$. 
From the definition of the subset $\Psi^{D}(1)$, 
$([p, q], [\ell, r]) \in \Psi^{D}(1)$ holds if 
$([p, q], [\ell, r]) \in \Psi_{h_{Q}} \cap \Psi_{\run} \cap \Psi_{\centerset}(C_{Q}) \cap \Psi_{\lcp}(K) \cap \Psi_{\preceding}$ 
and $R_{1} \preceq T[\gamma + K..r + 1] \preceq R_{d}$. 
We already proved $([p, q], [\ell, r]) \in \Psi_{h_{Q}} \cap \Psi_{\run} \cap \Psi_{\centerset}(C_{Q}) \cap \Psi_{\lcp}(K) \cap \Psi_{\preceding}$. 
$R_{1} \preceq T[\gamma + K..r + 1] \preceq R_{d}$ follows from 
$R_{\tau} = T[\gamma_{s} + K_{s}..r_{s}+1]$, $T[\gamma_{s} + K_{s}..r_{s}+1] = T[\gamma + K..r+1]$, and $\tau \in [1, d]$. 
Therefore, $([p, q], [\ell, r]) \in \Psi^{D}(1)$ holds. 

We proved $([p, q], [\ell, r]) \in \Psi^{D}(1)$ for each interval attractor $([p, q], [\ell, r]) \in \Psi_{\CCP}(T[i..j]) \cap \Psi_{\run} \cap \Psi_{\centerset}(C_{Q}) \cap \Psi_{\lcp}(K) \cap \Psi_{\preceding}$. 
Therefore, $\Psi^{D}(1) \supseteq \Psi_{\CCP}(T[i..j]) \cap \Psi_{\run} \cap \Psi_{\centerset}(C_{Q}) \cap \Psi_{\lcp}(K) \cap \Psi_{\preceding}$ holds. 

\textbf{Proof of Proposition~\ref{prop:Psi_D1_Property}(i).}
Proposition~\ref{prop:Psi_D1_Property}(i) follows from statement (A) and statement (B). 

\textbf{Proof of Proposition~\ref{prop:Psi_D1_Property}(ii).}
Proposition~\ref{prop:Psi_D1_Property}(ii) follows from the definitions of the two subsets $\Psi^{D}(t)$ and $\Psi^{D}(t+1)$. 

\textbf{Proof of Proposition~\ref{prop:Psi_D1_Property}(iii).}
Proposition~\ref{prop:Psi_D1_Property}(iii) can be proved using the same approach as for Proposition~\ref{prop:Psi_C1_Property}(iii).

\end{proof}

For proving Lemma~\ref{lem:GammaD1_property}, 
we introduce a set $\mathcal{I}^{D}$ of integers in set $\{ 1, 2, \ldots, d \}$. 
This set $\mathcal{I}^{D}$ consists of integers such that 
for each integer $t \in \mathcal{I}^{D}$, 
set $\Psi^{D}(t) \setminus \Psi^{D}(t+1)$ contains an interval attractor $([p, q], [\ell, r])$ satisfying 
$\RSCQ(\gamma - |[i, \gamma_{Q}-1]|, \gamma + |[\gamma_{Q}, j]| - 1) < b - \eta + 1$ for the attractor position $\gamma$ of the interval attractor $([p, q], [\ell, r])$. 
Formally, $\mathcal{I}^{D} = \{ t \in [1, d] \mid \exists ([p, q], [\ell, r]) \in \Psi^{D}(t) \setminus \Psi^{D}(t+1) \text{ s.t. } \RSCQ(\gamma - |[i, \gamma_{Q}-1]|, \gamma + |[\gamma_{Q}, j]| - 1) < b - \eta + 1 \}$. 
Here, $\Psi^{D}(d + 1) = \emptyset$ for simplicity. 

The following proposition states three properties of the set $\mathcal{I}^{D}$. 

\begin{proposition}\label{prop:Set_ID1_Property}
    Consider condition (B) of RSS query for the given RSS query $\RSSQ(T[i..j], b)$.
    The following three statements hold for set $\mathcal{I}^{D}$ and sequence $\Gamma_{D} = u_{1}, u_{2}, \ldots, u_{d}$: 
\begin{enumerate}[label=\textbf{(\roman*)}]
    \item \label{enum:Set_ID1_Property:1} $\mathcal{F}_{\SA} \cap \mathcal{F}_{\suffix}(\Psi_{\CCP}(T[i..j]) \cap \Psi_{\run} \cap \Psi_{\centerset}(C_{Q}) \cap \Psi_{\lcp}(K) \cap \Psi_{\preceding}) = \{ T[i..\gamma_{Q} - 1] \cdot C_{Q}^{n+1}[1..K] \cdot R_{t} \mid t \in \mathcal{I}^{D} \}$ for the attractor position $\gamma_{Q}$ of interval attractor $([p_{Q}, q_{Q}], [\ell_{Q}, r_{Q}])$; 
    \item \label{enum:Set_ID1_Property:2} $t \in \mathcal{I}^{D}$ for each integer $t \in [1, d]$ satisfying $u_{t} = 1$ and $u_{t+1} = 0$;
    \item \label{enum:Set_ID1_Property:3} $u_{t^{\prime}} = 1$ for 
    any pair of two integers $t \in \mathcal{I}^{D}$ and $t^{\prime} \in [1, t]$. 
\end{enumerate}
Here, let $u_{d + 1} = 0$ and $\Psi^{D}(d + 1) = \emptyset$ for simplicity. 
\end{proposition}
\begin{proof}
    Proposition~\ref{prop:Set_ID1_Property} can be proved using the same approach as for Proposition~\ref{prop:Set_IC1_Property}.
\end{proof}

We prove Lemma~\ref{lem:GammaD1_property} using Proposition~\ref{prop:JD1_correspondence_property}, 
Proposition~\ref{prop:JD1_inverse_correspondence_property}, Proposition~\ref{prop:Psi_D1_Property}, and Proposition~\ref{prop:Set_ID1_Property}. 

\begin{proof}[Proof of Lemma~\ref{lem:GammaD1_property}~\ref{enum:GammaD1_property:1}]
Lemma~\ref{lem:GammaD1_property}~\ref{enum:GammaD1_property:1} corresponds to Lemma~\ref{lem:GammaA_property}~\ref{enum:GammaA_property:1}. 
We proved Lemma \ref{lem:GammaA_property}~\ref{enum:GammaA_property:1} using 
Proposition~\ref{prop:Set_IA_Property}. 
Proposition~\ref{prop:Set_IC1_Property} corresponds to Proposition~\ref{prop:Set_IA_Property}. 
Therefore, Lemma~\ref{lem:GammaD1_property}~\ref{enum:GammaD1_property:1} can be proved using the same approach as for Lemma~\ref{lem:GammaA_property}~\ref{enum:GammaA_property:1}. 
\end{proof}

\begin{proof}[Proof of Lemma~\ref{lem:GammaD1_property}~\ref{enum:GammaD1_property:2}]
Lemma~\ref{lem:GammaD1_property}~\ref{enum:GammaD1_property:2} corresponds to 
Lemma~\ref{lem:GammaC1_property}~\ref{enum:GammaC1_property:2}. 
We proved Lemma \ref{lem:GammaC1_property} \ref{enum:GammaC1_property:2} 
using Proposition~\ref{prop:JC1_correspondence_property}, Proposition~\ref{prop:Psi_C1_Property}, and Proposition~\ref{prop:Set_IC1_Property}. 
Proposition~\ref{prop:JD1_correspondence_property}, Proposition~\ref{prop:Psi_D1_Property}, and Proposition~\ref{prop:Set_ID1_Property} correspond to Proposition~\ref{prop:JC1_correspondence_property}, Proposition~\ref{prop:Psi_C1_Property}, and Proposition~\ref{prop:Set_IC1_Property}, respectively. 
Therefore, Lemma~\ref{lem:GammaD1_property}~\ref{enum:GammaD1_property:2} can be proved using the same approach as for Lemma~\ref{lem:GammaC1_property}~\ref{enum:GammaC1_property:2}. 
\end{proof}

\begin{proof}[Proof of Lemma~\ref{lem:GammaD1_property}~\ref{enum:GammaD1_property:3}]
Lemma~\ref{lem:GammaD1_property}~\ref{enum:GammaD1_property:3} corresponds to 
Lemma~\ref{lem:GammaC1_property} \ref{enum:GammaC1_property:3}. 
We proved Lemma~\ref{lem:GammaC1_property} \ref{enum:GammaC1_property:2} 
using Proposition~\ref{prop:JC1_inverse_correspondence_property} and Proposition~\ref{prop:Psi_C1_Property}. 
Proposition~\ref{prop:JD1_inverse_correspondence_property} and Proposition~\ref{prop:Psi_D1_Property} correspond to 
Proposition~\ref{prop:JC1_inverse_correspondence_property} and Proposition~\ref{prop:Psi_C1_Property}, respectively. 
Therefore, Lemma \ref{lem:GammaD1_property}~\ref{enum:GammaD1_property:3} can be proved using the same approach as for Lemma~\ref{lem:GammaC1_property}~\ref{enum:GammaC1_property:3}. 
\end{proof}

\begin{proof}[Proof of Lemma~\ref{lem:GammaD1_property}~\ref{enum:GammaD1_property:4}]
Lemma~\ref{lem:GammaD1_property}~\ref{enum:GammaD1_property:4} corresponds to 
Lemma~\ref{lem:GammaA_property}~\ref{enum:GammaA_property:4}. 
We proved Lemma~\ref{lem:GammaA_property} \ref{enum:GammaA_property:4} using 
Proposition~\ref{prop:Set_IA_Property}. 
Proposition~\ref{prop:Set_ID1_Property} corresponds to Proposition~\ref{prop:Set_IA_Property}. 
Therefore, 
Lemma~\ref{lem:GammaD1_property} \ref{enum:GammaD1_property:4} can be proved using the same approach as for Lemma~\ref{lem:GammaA_property}~\ref{enum:GammaA_property:4}. 
\end{proof}

\subsubsection{Proof of Lemma~\ref{lem:GammaD1_sub_property}}\label{subsubsec:GammaD1_sub_property_proof}

The following proposition states the relationship between set $\Psi^{D}(t) \setminus \Psi^{D}(t+1)$ 
and range-count query on set $\mathcal{Y}_{C}(h_{Q}, C_{Q}, M)$ for each integer $t \in [1, d]$. 

\begin{proposition}\label{prop:GammaD1_Grid}
Consider condition (B) of RSS query for the given RSS query $\RSSQ(T[i..j], b)$.
For an $t \in [1, d]$, 
$\rangecount(\mathcal{Y}_{C}(h_{Q}, C_{Q}, M), x, n, R_{t}, R_{t}) \geq 1 \Leftrightarrow \Psi^{D}(t) \setminus \Psi^{D}(t+1) \neq \emptyset$. 
Here, let $\Psi^{D}(d + 1) = \emptyset$ for simplicity. 
\end{proposition}
\begin{proof}
Proposition~\ref{prop:GammaD1_Grid} can be proved using the same approach as for Proposition~\ref{prop:GammaC1_Grid}.
\end{proof}

We prove Lemma~\ref{lem:GammaD1_sub_property} using 
Proposition~\ref{prop:JD1_inverse_correspondence_property}, Proposition~\ref{prop:Psi_D1_Property},  Proposition~\ref{prop:Set_ID1_Property}, and Proposition~\ref{prop:GammaD1_Grid}. 

\begin{proof}[Proof of Lemma~\ref{lem:GammaD1_sub_property}~\ref{enum:GammaD1_sub_property:1}]
Lemma~\ref{lem:GammaD1_sub_property}~\ref{enum:GammaD1_sub_property:1} can be proved using 
the same approach as for Lemma \ref{lem:GammaA_sub_property} \ref{enum:GammaA_sub_property:1}.
\end{proof}

\begin{proof}[Proof of Lemma~\ref{lem:GammaD1_sub_property}~\ref{enum:GammaD1_sub_property:2}]
Lemma~\ref{lem:GammaD1_sub_property}~\ref{enum:GammaD1_sub_property:2} 
can be proved using the same approach as for Lemma~\ref{lem:GammaA_sub_property} \ref{enum:GammaA_sub_property:2}. 
\end{proof}

\begin{proof}[Proof of Lemma~\ref{lem:GammaD1_sub_property}~\ref{enum:GammaD1_sub_property:3}]
Lemma~\ref{lem:GammaD1_sub_property}~\ref{enum:GammaD1_sub_property:3} corresponds to 
Lemma~\ref{lem:GammaC1_sub_property}~\ref{enum:GammaC1_sub_property:3}. 
We showed that Lemma~\ref{lem:GammaC1_sub_property}~\ref{enum:GammaC1_sub_property:3} holds using 
Proposition~\ref{prop:JC1_inverse_correspondence_property}, Proposition~\ref{prop:Psi_C1_Property},  Proposition~\ref{prop:Set_IC1_Property}, and Proposition~\ref{prop:GammaC1_Grid}. 
Proposition~\ref{prop:JD1_inverse_correspondence_property}, Proposition~\ref{prop:Psi_D1_Property},  Proposition~\ref{prop:Set_ID1_Property}, and Proposition~\ref{prop:GammaD1_Grid} correspond to 
Proposition~\ref{prop:JC1_inverse_correspondence_property}, Proposition~\ref{prop:Psi_C1_Property},  Proposition~\ref{prop:Set_IC1_Property}, and Proposition~\ref{prop:GammaC1_Grid}, respectively. 
Therefore, Lemma~\ref{lem:GammaD1_sub_property}~\ref{enum:GammaD1_sub_property:3} can be proved using the same approach as for Lemma~\ref{lem:GammaC1_sub_property}~\ref{enum:GammaC1_sub_property:3}. 
\end{proof}

\subsubsection{Algorithm}\label{subsubsec:gamma_D1_algorithm}
We prove Lemma~\ref{lem:GammaD1_algorithm}, i.e., 
we show that subquery $\RSSQDX(T[i..j], b, K)$ can be answered 
in $O(H^{2} \log^{2} n + \log^{6} n)$ time using the data structures for RSC query 
and interval $[i, j]$.

For answering subquery $\RSSQDX(T[i..j], b, K)$, 
we use the ordered set $\mathcal{T}_{C} = \{ (h_{1}, C_{1}, M_{1})$, $(h_{2}, C_{2}, M_{2})$, $\ldots$, $(h_{m}, C_{m}, M_{m}) \}$ of triplets introduced in Section~\ref{subsubsec:TC1_ds}. 
Let $\lambda$ be an integer in set $[1, m]$ satisfying $(h_{\lambda}, C_{\lambda}, M_{\lambda}) = (h_{Q}, C_{Q}, M)$. 
This integer $\lambda$ exists if and only if the ordered set $\mathcal{T}_{C}$ contains the triplet $(h_{Q}, C_{Q}, M)$. 

\paragraph{Computation of integer $\lambda$.}
Consider the two integers 
$\hat{K} = |\lcp(T[\gamma_{Q}..j], C_{Q}^{n+1})|$ 
and $\hat{M} = (\hat{K} - (2 + \sum_{w = 1}^{h_{Q}+3} \lfloor \mu(w) \rfloor) ) \mod |C_{Q}|$ 
used in Section~\ref{subsec:GammaC1}. 
In Section~\ref{subsubsec:gamma_C1_algorithm}, 
we presented the algorithm computing an integer $\lambda^{\prime} \in [1, m]$ satisfying 
$(h_{\lambda^{\prime}}, C_{\lambda^{\prime}}, M_{\lambda^{\prime}}) = (h_{Q}, C_{Q}, \hat{M})$ 
in $O(H^{2} \log n + \log^{2} n)$ time. 
The integer $\lambda$ can be computed in the same time using the same approach as for the integer $\lambda^{\prime}$. 

\paragraph{Accessing the dynamic data structures for two sets $\mathcal{Y}_{C}(h_{\lambda}, C_{\lambda}, M_{\lambda})$ and $\mathcal{J}_{C}(h_{\lambda}, C_{\lambda}, M_{\lambda})$.}
For answering subquery $\RSSQDX(T[i..j], b, K)$, 
we use the dynamic data structures for two sets $\mathcal{Y}_{C}(h_{\lambda}, C_{\lambda}, M_{\lambda})$ and $\mathcal{J}_{C}(h_{\lambda}, C_{\lambda}, M_{\lambda})$ introduced in Section~\ref{subsubsec:JC1_Y_ds} and Section~\ref{subsubsec:JC1_ds}. 
In Section~\ref{subsubsec:gamma_C1_algorithm}, 
we presented the algorithm accessing the dynamic data structures for two sets $\mathcal{Y}_{C}(h_{\lambda^{\prime}}$, $C_{\lambda^{\prime}}, M_{\lambda^{\prime}})$ and $\mathcal{J}_{C}(h_{\lambda^{\prime}}, C_{\lambda^{\prime}}, M_{\lambda^{\prime}})$ 
in $O(\log n)$ time. 
Similarly, 
we can access the dynamic data structures for two sets $\mathcal{Y}_{C}(h_{\lambda}, C_{\lambda}, M_{\lambda})$ and $\mathcal{J}_{C}(h_{\lambda}, C_{\lambda}, M_{\lambda})$ in the same time. 

\paragraph{Computation of each integer $u_{t}$ in sequence $\Gamma_{D}$.}
Consider the non-increasing sequence $\Gamma_{D} = u_{1}, u_{2}, \ldots, u_{d} \in \{ 0, 1 \}$. 
We show that the $t$-th integer $u_{t}$ can be computed in $O(H^{2} \log n + \log^{4} n)$ time for a given integer $t \in [1, d]$. 

We leverage sequence $\Gamma_{D, \sub} = \alpha_{1}$, $\alpha_{2}$, $\ldots$, $\alpha_{d}$ for computing the $t$-th integer $u_{t}$. 
Let $t^{\prime}$ be the largest integer in set $[t, d]$ satisfying $\alpha_{t} = \alpha_{t^{\prime}}$. 
If $\alpha_{t^{\prime}} \geq 1$, 
then Lemma~\ref{lem:GammaD1_sub_property}~\ref{enum:GammaD1_sub_property:2} shows that 
set $\mathcal{J}_{C}(h_{Q}, C_{Q}, M)$ contains a weighted point 
$(|f_{\recover}(([p_{s}, q_{s}], [\ell_{s}, r_{s}]))|$, $T[\gamma_{s} + K_{s}..r_{s} + 1]$, $|\Psi_{\str}(T[p_{s}-1..r_{s}+1])|$, $T[p_{s}-1..r_{s}+1])$ satisfying 
$x \leq |f_{\recover}(([p_{s}, q_{s}], [\ell_{s}, r_{s}]))| \leq n$ 
and $T[\gamma_{s} + K_{s}..r_{s}+1] = R_{t^{\prime}}$. 
Here, $x$ is the integer introduced in Section~\ref{subsec:GammaD1}. 
For the integer $\upsilon = \gamma_{s} + K_{s} - K$, 
Lemma~\ref{lem:GammaD1_sub_property}~\ref{enum:GammaD1_sub_property:3} shows that 
$u_{t} = 1 \Leftrightarrow (\alpha_{t^{\prime}} \geq 1) \land (\RSCQ(\upsilon - |[i, \gamma_{Q}-1]|, \upsilon + |[\gamma_{Q}, j]| - 1) < b - \eta + 1)$ holds. 
Therefore, the $t$-th integer $u_{t}$ can be computed using the $t^{\prime}$-th integer $\alpha_{t^{\prime}}$ of the sequence $\Gamma_{D, \sub}$ and RSC query $\RSCQ(\upsilon - |[i, \gamma_{Q}-1]|, \upsilon + |[\gamma_{Q}, j]| - 1)$. 

The algorithm computing the $t$-th integer $u_{t}$ consists of four phases. 
In the first phase, we compute the three integers $\lambda$, $M$ and $x$. 
The computation of the integer $\lambda$ takes $O(H^{2} \log n + \log^{2} n)$ time. 
The two integers $M$ and $x$ can be computed in $O(H)$ time using the three integers $h_{Q}$, $|C_{Q}|$, and $K$. 
The two integers $h_{Q}$ and $|C_{Q}|$ can be obtained by the algorithm computing the integer $\lambda$ (see Section~\ref{subsubsec:gamma_C1_algorithm}). 
Therefore, the first phase takes $O(H^{2} \log n + \log^{2} n)$ time. 

In the second phase, 
we find the integer $t^{\prime}$ by binary search on the sequence $\Gamma_{D, \sub}$. 
This binary search can be executed by computing $O(\log d)$ integers of the sequence $\Gamma_{D, \sub}$ 
because the sequence $\Gamma_{D, \sub}$ is non-increasing (Lemma~\ref{lem:GammaD1_sub_property}~\ref{enum:GammaD1_sub_property:1}). 
Each integer of the sequence $\Gamma_{D, \sub}$ can be computed by one range-count query on the set $\mathcal{J}_{C}(h_{Q}, C_{Q}, M)$ of weighted points. 

If the integer $\lambda$ does not exist, 
then $\mathcal{J}_{C}(h_{Q}, C_{Q}, M) = \emptyset$ follows from the definition of the ordered set $\mathcal{T}_{C}$. 
In this case, any range-count query returns $0$ on the set $\mathcal{J}_{C}(h_{Q}, C_{Q}, M)$.  
Otherwise (i.e., the integer $\lambda$ exists), 
each range-count query can be executed by the dynamic data structures for the set $\mathcal{J}_{C}(h_{\lambda}, C_{\lambda}, M_{\lambda})$. 
Accessing these dynamic data structures takes $O(\log n)$ time, 
and each range-count query takes $O(\log^{2} k)$ time 
for the number $k$ of weighted points in the set $\mathcal{J}_{C}(h_{\lambda}, C_{\lambda}, M_{\lambda})$.
$d, k = O(n^{2})$ follows from Lemma~\ref{lem:JC1_size}~\ref{enum:JC1_size:2}. 
Therefore, the second phase takes $O(\log^{3} n)$ time. 

The integer $\alpha_{t^{\prime}}$ is obtained by the binary search of the second phase. 
The third phase is executed if $\alpha_{t^{\prime}} \geq 1$ holds. 
In this case, the integer $\lambda$ exists. 
This is because $\alpha_{t^{\prime}} = 0$ holds if the integer $\lambda$ does not exist. 
In the third phase, we find the interval attractor $([p_{s}, q_{s}], [\ell_{s}, r_{s}])$ corresponding to 
the weighted point $(|f_{\recover}(([p_{s}, q_{s}], [\ell_{s}, r_{s}]))|$, $T[\gamma_{s} + K_{s}..r_{s} + 1]$, $|\Psi_{\str}(T[p_{s}-1..r_{s}+1])|$, $T[p_{s}-1..r_{s}+1])$.  
This interval attractor can be found in $O(H^{2} \log n + \log^{2} n)$ time using the query of 
Lemma~\ref{lem:JC1_Y_queries}~\ref{enum:JC1_Y_queries:4}. 
This query can be answered after accessing the dynamic data structures for the ordered set $\mathcal{Y}_{C}(h_{\lambda}, C_{\lambda}, M_{\lambda})$ in $O(\log n)$ time. 
Therefore, the third phase takes $O(H^{2} \log n + \log^{2} n)$ time. 

In the fourth phase, 
we compute the $t$-th integer $u_{t}$ of sequence $\Gamma_{D}$ by verifying 
$\alpha_{t^{\prime}} \geq 1$ and $\RSCQ(\upsilon - |[i, \gamma_{Q}-1]|, \upsilon + |[\gamma_{Q}, j]| - 1) < b - \eta + 1$. 
The attractor position $\gamma_{s}$ is obtained in $O(H^{2})$ time 
by attractor position query $\attrQ(([p_{s}, q_{s}], [\ell_{s}, r_{s}]))$. 
The integer $\upsilon$ can be computed in $O(1)$ time using the three integers $\gamma_{s}$, $K_{s}$, and $K$.
The integer $K_{s}$ can be obtained by C-LCP query $\clcpQ(([p_{s}, q_{s}], [\ell_{s}, r_{s}]))$. 
The attractor position $\gamma_{Q}$ can be obtained by the algorithm computing the integer $\lambda$. 
The RSC query $\RSCQ(\upsilon - |[i, \gamma_{Q}-1]|, \upsilon + |[\gamma_{Q}, j]| - 1)$ takes $O(H^{2} \log n + \log^{4} n)$ time. 
Therefore, the fourth phase takes $O(H^{2} \log n + \log^{4} n)$ time. 

Finally, the algorithm computing the $t$-th integer $u_{t}$ takes $O(H^{2} \log n + \log^{4} n)$ time in total. 

\paragraph{Computation of the largest integer $\kappa$.}
Consider the largest integer $\kappa$ in set $[1, d]$ satisfying $u_{\kappa} = 1$. 
We find the largest integer $\kappa$ by binary search on the non-increasing sequence $\Gamma_{D}$. 
This binary search can be executed by computing $O(\log d)$ integers of the sequence $\Gamma_{D}$. 
Each integer of the sequence $\Gamma_{D}$ can be computed in $O(H^{2} \log n + \log^{4} n)$ time. 
Therefore, this binary search takes $O((H^{2} \log n + \log^{4} n) \log d)$ time 
(i.e., $O(H^{2} \log^{2} n + \log^{5} n)$ time).

\paragraph{Verification of $\mathcal{F}_{\SA} \cap \mathcal{F}_{\suffix}(\Psi_{\CCP}(T[i..j]) \cap \Psi_{\run} \cap \Psi_{\centerset}(C_{Q}) \cap \Psi_{\lcp}(K) \cap \Psi_{\preceding}) = \emptyset$.}
We verify whether $\mathcal{F}_{\SA} \cap \mathcal{F}_{\suffix}(\Psi_{\CCP}(T[i..j]) \cap \Psi_{\run} \cap \Psi_{\centerset}(C_{Q}) \cap \Psi_{\lcp}(K) \cap \Psi_{\preceding}) = \emptyset$ or not 
for answering $\RSSQDX(T[i..j], b, K)$.
From Lemma~\ref{lem:GammaD1_property}~\ref{enum:GammaD1_property:1},  
$\mathcal{F}_{\SA} \cap \mathcal{F}_{\suffix}(\Psi_{\CCP}(T[i..j]) \cap \Psi_{\run} \cap \Psi_{\centerset}(C_{Q}) \cap \Psi_{\lcp}(K) \cap \Psi_{\preceding}) = \emptyset$ holds if and only if 
the largest integer $\kappa$ exists. 
The largest integer $\kappa$ exists if and only if $u_{1} = 1$ holds. 
The first integer $u_{1}$ of the sequence $\Gamma_{D}$ is can be computed in $O(H^{2} \log n + \log^{4} n)$ time. 
Therefore, the verification of $\mathcal{F}_{\SA} \cap \mathcal{F}_{\suffix}(\Psi_{\CCP}(T[i..j]) \cap \Psi_{\run} \cap \Psi_{\centerset}(C_{Q}) \cap \Psi_{\lcp}(K) \cap \Psi_{\preceding}) = \emptyset$ 
takes $O(H^{2} \log n + \log^{4} n)$ time. 

\paragraph{Algorithm for subquery $\RSSQDX(T[i..j], b, K)$.}
The algorithm for subquery $\RSSQDX(T[i..j]$, $b, K)$ 
returns the lexicographically largest string $F$ in 
set $\mathcal{F}_{\SA} \cap \mathcal{F}_{\suffix}(\Psi_{\CCP}(T[i..j]) \cap \Psi_{\run} \cap \Psi_{\centerset}(C_{Q}) \cap \Psi_{\lcp}(K) \cap \Psi_{\preceding})$. 
This algorithm is executed only if $\mathcal{F}_{\SA} \cap \mathcal{F}_{\suffix}(\Psi_{\CCP}(T[i..j]) \cap \Psi_{\run} \cap \Psi_{\centerset}(C_{Q}) \cap \Psi_{\lcp}(K) \cap \Psi_{\preceding}) \neq \emptyset$ holds; 
otherwise, subquery $\RSSQDX(T[i..j], b, K)$  can be answered by verifying whether $\mathcal{F}_{\SA} \cap \mathcal{F}_{\suffix}(\Psi_{\CCP}(T[i..j]) \cap \Psi_{\run} \cap \Psi_{\centerset}(C_{Q}) \cap \Psi_{\lcp}(K) \cap \Psi_{\preceding}) = \emptyset$ or not in $O(H^{2} \log n + \log^{4} n)$ time. 

Similar to the algorithm for subquery $\RSSQCX(T[i..j], b)$, 
the algorithm for Lemma~\ref{lem:GammaD1_algorithm}~\ref{enum:GammaD1_property:2} 
leverages the largest integer $\kappa$. 
Lemma~\ref{lem:GammaD1_property}~\ref{enum:GammaD1_property:1} shows that 
$F = T[i..\gamma_{Q} - 1] \cdot C_{Q}^{n+1}[1..K] \cdot R_{\kappa}$ holds. 
Lemma~\ref{lem:GammaD1_property}~\ref{enum:GammaD1_property:2} shows that 
set $\mathcal{J}_{C}(h_{Q}, C_{Q}, M)$ contains a weighted point 
$(|f_{\recover}(([p_{s}, q_{s}]$, $[\ell_{s}, r_{s}]))|$, $T[\gamma_{s} + K_{s}..r_{s} + 1]$, $|\Psi_{\str}(T[p_{s}-1..r_{s}+1])|$, $T[p_{s}-1..r_{s}+1])$ satisfying $x \leq |f_{\recover}(([p_{s}, q_{s}]$, $[\ell_{s}, r_{s}]))| \leq n$ 
and $T[\gamma_{s} + K_{s}..r_{s}+1] = R_{\kappa}$. 
Lemma~\ref{lem:GammaD1_property}~\ref{enum:GammaD1_property:3} shows that 
$T[\gamma_{s} + K_{s} - K - |[i, \gamma_{Q}-1]|..r_{s} + 1] = T[i..\gamma_{Q} - 1] \cdot C_{Q}^{n+1}[1..K] \cdot R_{\kappa}$ holds. 
Therefore, we can return string $T[\gamma_{s} + K_{s} - K - |[i, \gamma_{Q}-1]|..r_{s} + 1]$ as the answer to subquery $\RSSQCX(T[i..j], b)$.

The algorithm for subquery $\RSSQCX(T[i..j], b)$ consists of three phases.
In the first phase, we compute the four integers $\lambda$, $M$, $x$, and $\kappa$. 
Similar to the first phase of the algorithm computing each integer of sequence $\Gamma_{D}$, 
we can compute the three integers $\lambda$, $M$ and $x$ in $O(H^{2} \log n + \log^{2} n)$ time. 
The computation of the largest integer $\kappa$ takes $O(H^{2} \log^{2} n + \log^{5} n)$ time. 
Therefore, this phase $O(H^{2} \log^{2} n + \log^{5} n)$ time.

In the second phase, 
we compute the interval attractor $([p_{s}, q_{s}], [\ell_{s}, r_{s}])$ corresponding to 
the weighted point $(|f_{\recover}(([p_{s}, q_{s}]$, $[\ell_{s}, r_{s}]))|$, $T[\gamma_{s} + K_{s}..r_{s} + 1]$, $|\Psi_{\str}(T[p_{s}-1..r_{s}+1])|$, $T[p_{s}-1..r_{s}+1])$. 
This interval attractor can be found in $O(H^{2} \log n + \log^{2} n)$ time 
using the query of Lemma~\ref{lem:JC1_Y_queries}~\ref{enum:JC1_Y_queries:4}. 
This query can be answered after accessing the dynamic data structures for the ordered set $\mathcal{Y}_{C}(h_{\lambda}, C_{\lambda}, M_{\lambda})$ in $O(\log n)$ time. 
Therefore, the second phase takes $O(H^{2} \log n + \log^{2} n)$ time. 

In the third phase, 
we return string $T[\gamma_{s} + K_{s} - K - |[i, \gamma_{Q}-1]|..r_{s} + 1]$ as the answer to subquery $\RSSQCX(T[i..j], b)$. 
Here, the string $T[\gamma_{s} + K_{s} - K - |[i, \gamma_{Q}-1]|..r_{s} + 1]$ is represented as 
interval $[\gamma_{s} + K_{s} - K - |[i, \gamma_{Q}-1]|, r_{s} + 1]$. 
The attractor position $\gamma_{s}$ can be obtained by attractor position query $\attrQ(([p_{s}, q_{s}], [\ell_{s}, r_{s}]))$. 
The integer $K_{s}$ can be obtained by C-LCP query $\clcpQ(([p_{s}, q_{s}]$, $[\ell_{s}, r_{s}]))$. 
The attractor position $\gamma_{Q}$ can be obtained by the algorithm computing the integer $\lambda$. 
Therefore, the third phase takes $O(H^{2})$ time. 

The three phases take $O(H^{2} \log^{2} n + \log^{5} n)$ time in total. 
Therefore, Lemma~\ref{lem:GammaD1_algorithm} holds.

\subsection{Subquery \texorpdfstring{$\RSSQDY(T[i..j], b, K)$}{RSSD2(T[i..j], b, K)}}\label{subsec:GammaD2}
The goal of this subsection is to answer subquery $\RSSQDY(T[i..j], b, K)$ under the assumption 
that (i) either $\mathcal{C}_{\run} = \emptyset$ or $C_{Q} = C_{\max}$ holds, and (ii) condition (B) of RSS query is satisfied. 
For this subsection, 
let $M = (K - (2 + \sum_{w = 1}^{h_{Q}+3} \lfloor \mu(w) \rfloor) ) \mod |C_{Q}|$ 
and $x = 1 + \lfloor \frac{K - (2 + \sum_{w = 1}^{h_{Q}+3} \lfloor \mu(w) \rfloor)}{|C_{Q}|} \rfloor$. 
The following lemma states the summary of this subsection. 

\begin{lemma}\label{lem:GammaD2_algorithm}
We assume that (i) either $\mathcal{C}_{\run} = \emptyset$ or $C_{Q} = C_{\max}$ holds, and (ii) condition (B) of RSS query is satisfied. 
We can answer $\RSSQDY(T[i..j], b, K)$ in $O(H^{2} \log^{2} n + \log^{5} n)$ time 
using (A) the data structures for RSC query, 
(B) interval $[i, j]$, 
and (C) the starting position $\eta$ of the sa-interval $[\eta, \eta^{\prime}]$ of $T[i..j]$. 
If the subquery returns a string $F$, 
then $F$ is represented as an interval $[g, g + |F| - 1]$ satisfying $T[g..g + |F| - 1] = F$; 
otherwise (i.e., $\mathcal{F}_{\SA} \cap \mathcal{F}_{\suffix}(\Psi_{\CCP}(T[i..j]) \cap \Psi_{\run} \cap \Psi_{\centerset}(C_{Q}) \cap \Psi_{\lcp}(K) \cap \Psi_{\succeeding}) = \emptyset$), the running time of subquery $\RSSQDY(T[i..j], b, K)$ can be reduced to $O(H^{2} + \log^{4} n)$ time. 
\end{lemma}
\begin{proof}
See Section~\ref{subsubsec:gamma_D2_algorithm}.
\end{proof}

We use set $\Psi_{h_{Q}} \cap \Psi_{\source} \cap \Psi_{\centerset}(C_{Q}) \cap \Psi_{\modulo}(M) \cap \Psi_{\succeeding} \cap \Psi_{\samp}$ to explain the idea behind solving subquery $\RSSQDY(T[i..j], b, K)$. 
Let $([p_{1}, q_{1}], [\ell_{1}, r_{1}]), ([p_{2}, q_{2}], [\ell_{2}, r_{2}])$, 
$\ldots$, $([p_{k}, q_{k}], [\ell_{k}, r_{k}])$ be the interval attractors in the set $\Psi_{h_{Q}} \cap \Psi_{\source} \cap \Psi_{\centerset}(C_{Q}) \cap \Psi_{\modulo}(M) \cap \Psi_{\succeeding} \cap \Psi_{\samp}$. 
Let $\gamma_{s}$ be the attractor position of each interval attractor $([p_{s}, q_{s}], [\ell_{s}, r_{s}]) \in \Psi_{h_{Q}} \cap \Psi_{\source} \cap \Psi_{\centerset}(C_{Q}) \cap \Psi_{\modulo}(M) \cap \Psi_{\succeeding} \cap \Psi_{\samp}$. 
Let $K_{s} = |\lcp(T[\gamma_{s}..r_{s}], C_{Q}^{n+1})|$ for simplicity.  

Similar to subquery $\RSSQCY(T[i..j], b)$, 
we leverage the set $\mathcal{J}_{C^{\prime}}(h_{Q}, C_{Q}, M)$ of weighted points on grid $([1, n]$, $\mathcal{Y}_{C^{\prime}}(h_{Q}, C_{Q}, M))$ introduced in Section~\ref{subsec:RSC_comp_C2}. 
The summary of the set $\mathcal{J}_{C^{\prime}}(h_{Q}, C_{Q}, M)$ and ordered set $\mathcal{Y}_{C^{\prime}}(h_{Q}, C_{Q}, M)$ is as follows 
(see Section~\ref{subsec:RSC_comp_C2} for the details of the set and ordered set):
\begin{itemize}
    \item the ordered set $\mathcal{Y}_{C^{\prime}}(h_{Q}, C_{Q}, M))$ consists of $d$ strings $R_{1}, R_{2}, \ldots, R_{d}$ ($R_{1} \prec R_{2} \prec \cdots \prec R_{d}$);
    \item the set $\mathcal{J}_{C^{\prime}}(h_{Q}, C_{Q}, M)$ contains the weighted point 
    $(|f_{\recover}(([p_{s}, q_{s}], [\ell_{s}, r_{s}]))|$, $T[\gamma_{s} + K_{s}..r_{s} + 1]$, $|\Psi_{\str}(T[p_{s}-1..r_{s}+1])|$, $T[p_{s}-1..r_{s}+1])$ 
    corresponding to each interval attractor $([p_{s}, q_{s}]$, $[\ell_{s}, r_{s}])$ in set $\Psi_{h_{Q}} \cap \Psi_{\source} \cap \Psi_{\centerset}(C_{Q}) \cap \Psi_{\modulo}(M) \cap \Psi_{\succeeding} \cap \Psi_{\samp}$. 
\end{itemize}

In the next paragraphs, 
we introduce $d$ subsets of set $\Psi_{\RR}$ 
and two sequences of integers to explain the relationship between the set $\mathcal{Y}_{C^{\prime}}(h_{Q}, C_{Q}, M))$ and subquery $\RSSQDY(T[i..j], b, K)$. 

\paragraph{Subset $\Psi^{D^{\prime}}(t)$.}
For each integer $t \in [1, d]$, 
subset $\Psi^{D^{\prime}}(t) \subseteq \Psi_{\RR}$ consists of interval attractors such that 
each interval attractor $([p, q], [\ell, r]) \in \Psi^{D^{\prime}}(t)$ satisfies the following two conditions: 
\begin{itemize}
    \item $([p, q], [\ell, r]) \in \Psi_{h_{Q}} \cap \Psi_{\run} \cap \Psi_{\centerset}(C_{Q}) \cap \Psi_{\lcp}(K) \cap \Psi_{\succeeding}$;
    \item $R_{t} \preceq T[\gamma + K..r + 1] \preceq R_{d}$ for the attractor position $\gamma$ of the interval attractor $([p, q], [\ell, r])$.
\end{itemize}
Formally, the subset $\Psi^{D^{\prime}}(t)$ is defined as follows: 
\begin{equation*}
    \begin{split}
    \Psi^{D^{\prime}}(t) &= \{ ([p, q], [\ell, r]) \in \Psi_{\RR} \mid (([p, q], [\ell, r]) \in \Psi_{h_{Q}} \cap \Psi_{\run} \cap \Psi_{\centerset}(C_{Q}) \cap \Psi_{\lcp}(K) \cap \Psi_{\succeeding}) \\
    &\land (R_{t} \preceq T[\gamma + K..r + 1] \preceq R_{d}) \}.
    \end{split}
\end{equation*}

\paragraph{Two sequences $\Gamma_{D^{\prime}}$ and $\Gamma_{D^{\prime}, \sub}$.}
The sequence $\Gamma_{D^{\prime}}$ consists of $d$ integers $u_{1}, u_{2}, \ldots, u_{d}$. 
Each integer $u_{t}$ is $1$ if $\mathcal{F}_{\SA} \cap \mathcal{F}_{\suffix}(\Psi_{\CCP}(T[i..j]) \cap \Psi^{D^{\prime}}(t)) \neq \emptyset$; 
otherwise $u_{t}$ is $0$. 
The following lemma states four properties of sequence $\Gamma_{D^{\prime}}$.

\begin{lemma}\label{lem:GammaD2_property}
Consider condition (B) of RSS query for the given RSS query $\RSSQ(T[i..j], b)$.
Let $\kappa$ be the largest integer in set $[1, d]$ satisfying $u_{\kappa} = 1$ for sequence $\Gamma_{D^{\prime}} = u_{1}, u_{2}, \ldots, u_{d}$. 
Then, the following four statements hold: 
\begin{enumerate}[label=\textbf{(\roman*)}]
    \item \label{enum:GammaD2_property:1} if the largest integer $\kappa$ exists, then 
    string $T[i..\gamma_{Q} - 1] \cdot C_{Q}^{n+1}[1..K] \cdot R_{\kappa}$ is 
    the lexicographically largest string in set $\mathcal{F}_{\SA} \cap \mathcal{F}_{\suffix}(\Psi_{\CCP}(T[i..j]) \cap \Psi_{\run} \cap \Psi_{\centerset}(C_{Q}) \cap \Psi_{\lcp}(K) \cap \Psi_{\succeeding})$. 
    Otherwise, the set $\mathcal{F}_{\SA} \cap \mathcal{F}_{\suffix}(\Psi_{\CCP}(T[i..j]) \cap \Psi_{\run} \cap \Psi_{\centerset}(C_{Q}) \cap \Psi_{\lcp}(K) \cap \Psi_{\succeeding})$ is empty; 
    \item \label{enum:GammaD2_property:2} if the largest integer $\kappa$ exists, 
    then set $\mathcal{J}_{C^{\prime}}(h_{Q}, C_{Q}, M)$ contains a weighted point 
    $(|f_{\recover}(([p_{s}$, $q_{s}]$, $[\ell_{s}, r_{s}]))|$, $T[\gamma_{s} + K_{s}..r_{s} + 1]$, $|\Psi_{\str}(T[p_{s}-1..r_{s}+1])|$, $T[p_{s}-1..r_{s}+1])$  
    satisfying $x \leq |f_{\recover}(([p_{s}, q_{s}], [\ell_{s}, r_{s}]))| \leq n$ 
    and $T[\gamma_{s} + K_{s}..r_{s}+1] = R_{\kappa}$;
    \item \label{enum:GammaD2_property:3} 
    $T[\gamma_{s} + K_{s} - K - |[i, \gamma_{Q}-1]|..r_{s} + 1] = T[i..\gamma_{Q} - 1] \cdot C_{Q}^{n+1}[1..K] \cdot R_{\kappa}$ for 
    the interval attractor $([p_{s}, q_{s}], [\ell_{s}, r_{s}])$ corresponding to 
    the weighted point $(|f_{\recover}(([p_{s}, q_{s}], [\ell_{s}, r_{s}]))|$, $T[\gamma_{s} + K_{s}..r_{s} + 1]$, $|\Psi_{\str}(T[p_{s}-1..r_{s}+1])|$, $T[p_{s}-1..r_{s}+1])$ 
    of Lemma~\ref{lem:GammaD2_property}~\ref{enum:GammaD2_property:2};
    \item \label{enum:GammaD2_property:4}
    sequence $\Gamma_{D^{\prime}}$ is non-increasing~(i.e., $u_{1} \geq u_{2} \geq \cdots \geq u_{d}$). 
\end{enumerate}
\end{lemma}
\begin{proof}
This lemma can be proved using the same approach as for Lemma~\ref{lem:GammaD1_property}.
\end{proof}

Next, sequence $\Gamma_{D^{\prime}, \sub}$ consists of $d$ integers 
$\alpha_{1}, \alpha_{2}, \ldots, \alpha_{d} \in \mathbb{N}_{0}$. 
Here, each integer $\alpha_{t}$ is defined as 
$\alpha_{t} = \rangecount(\mathcal{J}_{C^{\prime}}(h_{Q}, C_{Q}, M), x, n, R_{t}, R_{d})$. 
Here, $\rangecount$ is the range-count query introduced in Section~\ref{subsec:range_data_structure}. 
The following lemma states three properties of sequence $\Gamma_{D^{\prime}, \sub}$. 

\begin{lemma}\label{lem:GammaD2_sub_property}
Consider condition (B) of RSS query for the given RSS query $\RSSQ(T[i..j], b)$.
The following three statements hold for two sequences $\Gamma_{D^{\prime}} = u_{1}, u_{2}, \ldots, u_{d}$ and $\Gamma_{D^{\prime}, \sub} = \alpha_{1}, \alpha_{2}, \ldots, \alpha_{d}$: 
\begin{enumerate}[label=\textbf{(\roman*)}]
    \item \label{enum:GammaD2_sub_property:1} 
    sequence $\Gamma_{D^{\prime}, \sub}$ is non-increasing (i.e., $\alpha_{1} \geq \alpha_{2} \geq \cdots \geq \alpha_{d}$);    
    \item \label{enum:GammaD2_sub_property:2} 
    for an integer $t \in [1, d]$, 
    consider the largest integer $t^{\prime}$ in set $[t, d]$ satisfying 
    $\alpha_{t} = \alpha_{t^{\prime}}$. 
    If $\alpha_{t^{\prime}} \geq 1$, 
    then set $\mathcal{J}_{C^{\prime}}(h_{Q}, C_{Q}, M)$ contains a weighted point 
    $(|f_{\recover}(([p_{s}, q_{s}], [\ell_{s}, r_{s}]))|$, $T[\gamma_{s} + K_{s}..r_{s} + 1]$, $|\Psi_{\str}(T[p_{s}-1..r_{s}+1])|$, $T[p_{s}-1..r_{s}+1])$ satisfying 
    $x \leq |f_{\recover}(([p_{s}, q_{s}], [\ell_{s}, r_{s}]))| \leq n$ 
    and $T[\gamma_{s} + K_{s}..r_{s}+1] = R_{y + t^{\prime} - 1}$;
    \item \label{enum:GammaD2_sub_property:3}
    consider the three integers $t, t^{\prime}$, and $s$ of Lemma~\ref{lem:GammaD2_sub_property}~\ref{enum:GammaD2_sub_property:2}. 
    Let $\upsilon = \gamma_{s} + K_{s} - K$.
    Then, 
    $u_{t} = 1 \Leftrightarrow (\alpha_{t^{\prime}} \geq 1) \land (\RSCQ(\upsilon - |[i, \gamma_{Q}-1]|, \upsilon + |[\gamma_{Q}, j]| - 1) < b - \eta + 1)$. 
\end{enumerate}
\end{lemma}
\begin{proof}
This lemma can be proved using the same approach as for Lemma~\ref{lem:GammaD1_sub_property}.
\end{proof}
\subsubsection{Algorithm}\label{subsubsec:gamma_D2_algorithm}
We prove Lemma~\ref{lem:GammaD2_algorithm}, i.e., 
we show that subquery $\RSSQDY(T[i..j], b, K)$ can be answered 
in $O(H^{2} \log^{2} n + \log^{5} n)$ time using the data structures for RSC query 
and interval $[i, j]$. 

We answer $\RSSQDY(T[i..j], b, K)$ by modifying 
the algorithm answering $\RSSQDX(T[i..j], b, K)$. 
In Section~\ref{subsubsec:gamma_D1_algorithm}, 
we presented the algorithm answering $\RSSQDY(T[i..j], b, K)$ 
using two sequences $\Gamma_{D}$ and $\Gamma_{D, \sub}$. 
Two sequences $\Gamma_{D^{\prime}}$ and $\Gamma_{D^{\prime}, \sub}$ correspond to 
the two sequences $\Gamma_{D}$ and $\Gamma_{D, \sub}$, respectively. 
Lemma~\ref{lem:GammaD2_property} states properties of the sequence $\Gamma_{D^{\prime}}$. 
This lemma corresponds to Lemma~\ref{lem:GammaD1_property}, which states properties of the sequence $\Gamma_{D}$. 
Similarly, 
Lemma~\ref{lem:GammaD2_sub_property} states properties of the sequence $\Gamma_{D^{\prime}, \sub}$. 
This lemma corresponds to Lemma~\ref{lem:GammaD1_sub_property}, which states properties of the sequence $\Gamma_{D, \sub}$. 
Therefore, subquery $\RSSQDY(T[i..j], b, K)$ can be answered 
using the same approach as for subquery $\RSSQDX(T[i..j], b, K)$. 
Therefore, we obtain Lemma~\ref{lem:GammaD2_algorithm}.

\subsection{Subquery \texorpdfstring{$\RSSQEX(T[i..j], b)$}{RSSE1(T[i..j], b)}}\label{subsec:GammaE1}
The goal of this subsection is to solve subquery $\RSSQEX(T[i..j], b)$ 
under the assumption that 
(i) either $\mathcal{C}_{\run} = \emptyset$ or $C_{Q} = C_{\max}$ holds, 
and (ii) condition (B) of RSS query is satisfied. 
The following lemma states the summary of this subsection. 

\begin{lemma}\label{lem:GammaE1_algorithm}
We assume that (i) either $\mathcal{C}_{\run} = \emptyset$ or $C_{Q} = C_{\max}$ holds, and (ii) condition (B) of RSS query is satisfied. 
We can answer $\RSSQEX(T[i..j], b)$ in $O(H^{2} \log^{2} n + \log^{6} n)$ time 
using (A) the data structures for RSC query, 
(B) interval $[i, j]$, 
and (C) the starting position $\eta$ of the sa-interval $[\eta, \eta^{\prime}]$ of $T[i..j]$. 
If the subquery returns a string $F$, 
then $F$ is represented as an interval $[g, g + |F| - 1]$ satisfying $T[g..g + |F| - 1] = F$. 
\end{lemma}
\begin{proof}
See Section~\ref{subsubsec:gamma_E1_algorithm}.
\end{proof}

We leverage the two sets 
$\mathcal{J}_{D}(h_{Q}, C_{Q})$ and $\mathcal{J}_{E}(h_{Q}, C_{Q})$ of weighted points introduced in Section~\ref{subsec:RSC_comp_D1}. 
Here, the weighted points of the two sets $\mathcal{J}_{D}(h_{Q}, C_{Q})$ and $\mathcal{J}_{E}(h_{Q}, C_{Q})$ 
are located on grid $([1, n], [-1, |C_{Q}|])$ (see Section~\ref{subsec:RSC_comp_D1} for the details of the set and ordered set).
In the next paragraphs, 
we introduce two sequences of integers to explain the relationship among 
subquery $\RSSQEX(T[i..j], b)$ and the two sets $\mathcal{J}_{D}(h_{Q}, C_{Q})$ and $\mathcal{J}_{E}(h_{Q}, C_{Q})$. 

\paragraph{Two sequences $\Gamma_{E}$ and $\Gamma_{E, \sub}$.}
The sequence $\Gamma_{E}$ consists of $|[1, n - \theta + 1]|$ integers $u_{1}$, $u_{2}$, $\ldots$, $u_{n - \theta + 1} \in \{ 0, 1 \}$.  
Each integer $u_{t}$ is $1$ if the following equation holds: 
\begin{equation}\label{eq:GammaE1:1}
    \begin{split}
    \mathcal{F}_{\SA} \cap \mathcal{F}_{\suffix}(\Psi_{\CCP}(T[i..j]) \cap \Psi_{h_{Q}} \cap \Psi_{\run} \cap \Psi_{\centerset}(C_{Q}) \cap \Bigl(\bigcup_{\lambda = t + \theta - 1}^{n} \Psi_{\lcp}(\lambda) \Bigr) \cap \Psi_{\preceding}) \neq \emptyset.
    \end{split}
\end{equation}
Otherwise (i.e., Equation~\ref{eq:GammaE1:1} does not hold),  
$u_{t}$ is defined as $0$. 
The following lemma states two properties of sequence $\Gamma_{E}$. 

\begin{lemma}\label{lem:GammaE1_property}
Consider condition (B) of RSS query for the given RSS query $\RSSQ(T[i..j], b)$.
Let $\kappa$ be the largest integer in set $[1, n - \theta + 1]$ satisfying $u_{\kappa} = 1$ for sequence $\Gamma_{E} = u_{1}, u_{2}, \ldots, u_{n - \theta + 1}$. 
Then, the following four statements hold: 
\begin{enumerate}[label=\textbf{(\roman*)}]
    \item \label{enum:GammaE1_property:1} 
    let $F^{\prime}$ be the lexicographically largest string in set $\mathcal{F}_{\SA} \cap \mathcal{F}_{\suffix}(\Psi_{\CCP}(T[i..j]) \cap \Psi_{\run} \cap \Psi_{\centerset}(C_{Q}) \cap \Psi_{\lcp}(\kappa + \theta - 1) \cap \Psi_{\preceding})$. 
    If the largest integer $\kappa$ exists, then 
    the string $F^{\prime}$ is the lexicographically largest string in set $\mathcal{F}_{\SA} \cap \mathcal{F}_{\suffix}(\Psi_{\CCP}(T[i..j]) \cap \Psi_{h_{Q}} \cap \Psi_{\run} \cap \Psi_{\centerset}(C_{Q}) \cap \Psi_{\preceding})$. 
    Otherwise, the set $\mathcal{F}_{\SA} \cap \mathcal{F}_{\suffix}(\Psi_{\CCP}(T[i..j]) \cap \Psi_{\run} \cap \Psi_{\centerset}(C_{Q}) \cap \Psi_{\preceding})$ is empty;
    \item \label{enum:GammaE1_property:2}
    sequence $\Gamma_{E}$ is non-increasing~(i.e., $u_{1} \geq u_{2} \geq \cdots \geq u_{n - \theta + 1}$). 
\end{enumerate}
\end{lemma}
\begin{proof}
See Section~\ref{subsubsec:GammaE1_property_proof}.
\end{proof}

Next, sequence $\Gamma_{E, \sub}$ consists of $|[1, n - \theta + 1]|$ integers 
$\alpha_{1}, \alpha_{2}, \ldots, \alpha_{d} \in \mathbb{N}_{0}$. 
Each integer $\alpha_{t}$ is defined as follows: 
\begin{equation*}
    \alpha_{t} = |\Psi_{h_{Q}} \cap \Psi_{\run} \cap \Psi_{\centerset}(C_{Q}) \cap \Bigl(\bigcup_{\lambda = t + \theta - 1}^{n} \Psi_{\lcp}(\lambda) \Bigr) \cap \Psi_{\preceding}|.
\end{equation*}
The following lemma states three properties of sequence $\Gamma_{E, \sub}$.

\begin{lemma}\label{lem:GammaE1_sub_property}
Consider condition (B) of RSS query for the given RSS query $\RSSQ(T[i..j], b)$.
The following three statements hold for two sequences $\Gamma_{E} = u_{1}, u_{2}, \ldots, u_{n-\theta+1}$ and $\Gamma_{E, \sub} = \alpha_{1}$, $\alpha_{2}$, $\ldots$, $\alpha_{n-\theta+1}$: 
\begin{enumerate}[label=\textbf{(\roman*)}]
    \item \label{enum:GammaE1_sub_property:1} 
    sequence $\Gamma_{E, \sub}$ is non-increasing (i.e., $\alpha_{1} \geq \alpha_{2} \geq \cdots \geq \alpha_{n-\theta+1}$);    
    \item \label{enum:GammaE1_sub_property:2} 
    for an integer $t \in [1, n-\theta+1]$, 
    consider the largest integer $t^{\prime}$ in set $[t, n-\theta+1]$ satisfying 
    $\alpha_{t} = \alpha_{t^{\prime}}$. 
    Then, 
    $u_{t} = 1 \Leftrightarrow (\alpha_{t^{\prime}} \geq 1) \land (\mathcal{F}_{\SA} \cap \mathcal{F}_{\suffix}(\Psi_{\CCP}(T[i..j]) \cap \Psi_{\run} \cap \Psi_{\centerset}(C_{Q}) \cap \Psi_{\lcp}(t^{\prime} + \theta - 1) \cap \Psi_{\preceding}) \neq \emptyset)$;
    \item \label{enum:GammaE1_sub_property:3}
    For an integer $t \in [1, n - \theta + 1]$, 
    let $x_{t} = 1 + \lfloor \frac{(t + \theta - 1) - (2 + \sum_{w = 1}^{h_{Q}+3} \lfloor \mu(w) \rfloor)}{|C_{Q}|} \rfloor$ 
    and $M_{t} = ((t + \theta - 1) - (2 + \sum_{w = 1}^{h_{Q}+3} \lfloor \mu(w) \rfloor) ) \mod |C_{Q}|$. 
    Then, the following equation holds: 
    \begin{equation*}
    \begin{split}
    \alpha_{t} &= \rangesum(\mathcal{J}_{E}(h_{Q}, C_{Q}), 1, n, 0, |C_{Q}|-1) \\
    &- (x_{t}-1) \rangesum(\mathcal{J}_{D}(h_{Q}, C_{Q}), x_{t}, n, 0, |C_{Q}| - 1) \\
    &- \rangesum(\mathcal{J}_{D}(h_{Q}, C_{Q}), x_{t}, n, 0, M_{t} - 1) \\
    &- \rangesum(\mathcal{J}_{E}(h_{Q}, C_{Q}), 1, x_{t}-1, 0, |C_{Q}|-1). 
    \end{split}
    \end{equation*}
    Here, $\rangesum$ is the range-sum query introduced in Section~\ref{subsec:range_data_structure}.     
\end{enumerate}
\end{lemma}
\begin{proof}
See Section~\ref{subsubsec:GammaE1_sub_property_proof}.
\end{proof}

\subsubsection{Proof of Lemma~\ref{lem:GammaE1_property}}\label{subsubsec:GammaE1_property_proof}
The following proposition states properties of 
two sets $\Psi_{h_{Q}} \cap \Psi_{\run} \cap \Psi_{\centerset}(C_{Q}) \cap \Psi_{\lcp}(t + \theta - 1) \cap \Psi_{\preceding}$ 
and $\Psi_{\CCP}(T[i..j]) \cap \Psi_{\run} \cap \Psi_{\centerset}(C_{Q}) \cap \Psi_{\lcp}(t + \theta - 1) \cap \Psi_{\preceding}$ for each integer $t \in [1, n - \theta + 1]$. 

\begin{proposition}\label{prop:Psi_E1_Property}
Consider condition (B) of RSS query for the given RSS query $\RSSQ(T[i..j], b)$.
The following two statements hold: 
\begin{enumerate}[label=\textbf{(\roman*)}]
    \item \label{enum:Psi_E1_Property:1}
    $\Psi_{h_{Q}} \cap \Psi_{\run} \cap \Psi_{\centerset}(C_{Q}) \cap \Psi_{\lcp}(t + \theta - 1) \cap \Psi_{\preceding} \subseteq \Psi_{\CCP}(T[i..j])$ for each integer $t \in [1, n - \theta + 1]$;     
    \item \label{enum:Psi_E1_Property:2}
    consider two integers $t$ and $t^{\prime}$ satisfying $1 \leq t < t^{\prime} \leq n - \theta + 1$. 
    Then, $F \prec F^{\prime}$ holds for any pair of two strings $F \in \mathcal{F}_{\suffix}(\Psi_{\CCP}(T[i..j]) \cap \Psi_{\run} \cap \Psi_{\centerset}(C_{Q}) \cap \Psi_{\lcp}(t + \theta - 1) \cap \Psi_{\preceding})$ 
    and $F^{\prime} \in \mathcal{F}_{\suffix}(\Psi_{\CCP}(T[i..j]) \cap \Psi_{\run} \cap \Psi_{\centerset}(C_{Q}) \cap \Psi_{\lcp}(t^{\prime} + \theta - 1) \cap \Psi_{\preceding})$.
\end{enumerate}
\end{proposition}
\begin{proof}
The proof of Proposition~\ref{prop:Psi_E1_Property} is as follows. 

\textbf{Proof of Proposition~\ref{prop:Psi_E1_Property}(i).}
We can use Lemma~\ref{lem:CCP_special_property} because 
$\lcs(T[i..\gamma_{Q}-1], C_{Q}^{n+1}) = T[i..\gamma_{Q}-1]$ 
and $\lcp(T[\gamma_{Q}..j], C_{Q}^{n+1}) = T[\gamma_{Q}..j]$ follow from 
condition (B) of RSS query. 
Lemma~\ref{lem:CCP_special_property} shows that 
$\Psi_{h_{Q}} \cap \Psi_{\run} \cap \Psi_{\centerset}(C_{Q}) \cap (\bigcup_{\lambda = |[\gamma_{Q}, j]|}^{n} \Psi_{\lcp}(\lambda)) \subseteq \Psi_{\CCP}(T[i..j])$ holds. 
Therefore, 
$\Psi_{h_{Q}} \cap \Psi_{\run} \cap \Psi_{\centerset}(C_{Q}) \cap \Psi_{\lcp}(t + \theta - 1) \cap \Psi_{\preceding} \subseteq \Psi_{\CCP}(T[i..j])$ holds. 

\textbf{Proof of Proposition~\ref{prop:Psi_E1_Property}(ii).}
From the definition of the set $\mathcal{F}_{\suffix}(\Psi_{\CCP}(T[i..j]) \cap \Psi_{\run} \cap \Psi_{\centerset}(C_{Q}) \cap \Psi_{\lcp}(t + \theta - 1) \cap \Psi_{\preceding})$, 
set $\Psi_{\CCP}(T[i..j]) \cap \Psi_{\run} \cap \Psi_{\centerset}(C_{Q}) \cap \Psi_{\lcp}(t + \theta - 1) \cap \Psi_{\preceding}$ contains 
an interval attractor $([p, q], [\ell, r])$ satisfying 
$T[\gamma - |[i, \gamma_{Q}-1]|..\gamma-1] \cdot T[\gamma..r+1] = F$ for the attractor position $\gamma$ of the interval attractor 
$([p, q], [\ell, r])$. 
Similarly, 
set $\Psi_{\CCP}(T[i..j]) \cap \Psi_{\run} \cap \Psi_{\centerset}(C_{Q}) \cap \Psi_{\lcp}(t^{\prime} + \theta - 1) \cap \Psi_{\preceding}$ contains 
an interval attractor $([p^{\prime}, q^{\prime}], [\ell^{\prime}, r^{\prime}])$ satisfying 
$T[\gamma^{\prime} - |[i, \gamma_{Q}-1]|..\gamma^{\prime}-1] \cdot T[\gamma^{\prime}..r^{\prime}+1] = F^{\prime}$ for the attractor position $\gamma^{\prime}$ of the interval attractor $([p^{\prime}, q^{\prime}], [\ell^{\prime}, r^{\prime}])$. 

We prove $T[\gamma - |[i, \gamma_{Q}-1]|..\gamma-1] = T[\gamma^{\prime} - |[i, \gamma_{Q}-1]|..\gamma^{\prime}-1]$. 
Because of $([p, q], [\ell, r]) \in \Psi_{\CCP}(T[i..j])$, 
Lemma~\ref{lem:CCP_property}~\ref{enum:CCP_property:6} shows that 
$T[\gamma - |[i, \gamma_{Q}-1]|..\gamma-1] = T[i..\gamma_{Q}]$ holds. 
Similarly, 
Lemma~\ref{lem:CCP_property}~\ref{enum:CCP_property:6} shows that 
$T[\gamma^{\prime} - |[i, \gamma_{Q}-1]|..\gamma^{\prime}-1] = T[i..\gamma_{Q}]$ holds. 
Therefore, $T[\gamma - |[i, \gamma_{Q}-1]|..\gamma-1] = T[\gamma^{\prime} - |[i, \gamma_{Q}-1]|..\gamma^{\prime}-1]$ holds. 

We prove $T[\gamma..r+1] \prec T[\gamma^{\prime}..r^{\prime}+1]$. 
We can apply Lemma~\ref{lem:psi_LMPS_property}~\ref{enum:psi_LMPS_property:preceding:1} to the two interval attractors 
$([p, q], [\ell, r])$ and $([p^{\prime}, q^{\prime}], [\ell^{\prime}, r^{\prime}])$ 
because $([p, q], [\ell, r]) \in \Psi_{\run} \cap \Psi_{\centerset}(C_{Q}) \cap \Psi_{\lcp}(t + \theta - 1) \cap \Psi_{\preceding}$ 
and $([p^{\prime}, q^{\prime}], [\ell^{\prime}, r^{\prime}]) \in \Psi_{\run} \cap \Psi_{\centerset}(C_{Q}) \cap \Psi_{\lcp}(t^{\prime} + \theta - 1) \cap \Psi_{\preceding}$. 
Because of $t + \theta - 1 < t^{\prime} + \theta - 1$, 
Lemma~\ref{lem:psi_LMPS_property}~\ref{enum:psi_LMPS_property:preceding:1} shows that 
$T[\gamma..r+1] \prec T[\gamma^{\prime}..r^{\prime}+1]$ holds. 

Finally, $F \prec F^{\prime}$ follows from 
(1) $T[\gamma - |[i, \gamma_{Q}-1]|..\gamma-1] \cdot T[\gamma..r+1] = F$, 
(2) $T[\gamma^{\prime} - |[i, \gamma_{Q}-1]|..\gamma^{\prime}-1] \cdot T[\gamma^{\prime}..r^{\prime}+1] = F^{\prime}$, 
(3) $T[\gamma - |[i, \gamma_{Q}-1]|..\gamma-1] = T[\gamma^{\prime} - |[i, \gamma_{Q}-1]|..\gamma^{\prime}-1]$, 
and (4) $T[\gamma..r+1] \prec T[\gamma^{\prime}..r^{\prime}+1]$. 
\end{proof}

For proving Lemma~\ref{lem:GammaE1_property}, 
we introduce a set $\mathcal{I}^{E}$ of integers in set $\{ 1, 2, \ldots, n - \theta + 1 \}$. 
This set $\mathcal{I}^{E}$ consists of integers such that 
each integer $t \in \mathcal{I}^{E}$ satisfies 
$\mathcal{F}_{\SA} \cap \mathcal{F}_{\suffix}(\Psi_{\CCP}(T[i..j]) \cap \Psi_{\run} \cap \Psi_{\centerset}(C_{Q}) \cap \Psi_{\lcp}(t + \theta - 1) \cap \Psi_{\preceding}) \neq \emptyset$. 
Formally, the set $\mathcal{I}^{E}$ is defined as follows:
\begin{equation*}
    \begin{split}
    \mathcal{I}^{E} &= \{ t \in [1, n - \theta + 1] \mid \\
    & \mathcal{F}_{\SA} \cap \mathcal{F}_{\suffix}(\Psi_{\CCP}(T[i..j]) \cap \Psi_{\run} \cap \Psi_{\centerset}(C_{Q}) \cap \Psi_{\lcp}(t + \theta - 1) \cap \Psi_{\preceding}) \neq \emptyset \}.
    \end{split}
\end{equation*}

The following proposition states three properties of the set $\mathcal{I}^{E}$. 

\begin{proposition}\label{prop:Set_IE1_Property}
    Consider condition (B) of RSS query for the given RSS query $\RSSQ(T[i..j], b)$.
    The following three statements hold for set $\mathcal{I}^{E}$ and sequence $\Gamma_{E} = u_{1}, u_{2}, \ldots, u_{n - \theta + 1}$: 
\begin{enumerate}[label=\textbf{(\roman*)}]
    \item \label{enum:Set_IE1_Property:1} 
    the following equation holds: 
    \begin{equation*}
    \begin{split}
    & \mathcal{F}_{\SA} \cap \mathcal{F}_{\suffix}(\Psi_{\CCP}(T[i..j]) \cap \Psi_{\run} \cap \Psi_{\centerset}(C_{Q}) \cap \Psi_{\preceding}) \\
    &= \mathcal{F}_{\SA} \cap (\bigcup_{t \in \mathcal{I}^{E}} \mathcal{F}_{\suffix}(\Psi_{\CCP}(T[i..j]) \cap \Psi_{\run} \cap \Psi_{\centerset}(C_{Q}) \cap \Psi_{\lcp}(t + \theta - 1) \cap \Psi_{\preceding}));
    \end{split}
\end{equation*}    
    \item \label{enum:Set_IE1_Property:2} $t \in \mathcal{I}^{E}$ for each integer $t \in [1, n - \theta + 1]$ satisfying $u_{t} = 1$ and $u_{t+1} = 0$;
    \item \label{enum:Set_IE1_Property:3} $u_{t^{\prime}} = 1$ for 
    any pair of two integers $t \in \mathcal{I}^{E}$ and $t^{\prime} \in [1, t]$. 
\end{enumerate}
Here, let $u_{n - \theta + 2} = 0$ for simplicity. 
\end{proposition}
\begin{proof}
The following two statements are used to prove Proposition~\ref{prop:Set_IE1_Property}. 
\begin{enumerate}[label=\textbf{(\Alph*)}]
    \item $\Psi_{\CCP}(T[i..j]) \cap \Psi_{\run} \cap \Psi_{\centerset}(C_{Q}) \cap (\bigcup_{\lambda = 1}^{\theta-1} \Psi_{\lcp}(\lambda))\cap \Psi_{\preceding} = \emptyset$;
    \item the following equation holds: 
\begin{equation*}
    \begin{split}
    & \mathcal{F}_{\SA} \cap \mathcal{F}_{\suffix}(\Psi_{\CCP}(T[i..j]) \cap \Psi_{\run} \cap \Psi_{\centerset}(C_{Q}) \cap (\bigcup_{\lambda = \theta}^{n} \Psi_{\lcp}(\lambda)) \cap \Psi_{\preceding}) \\
    &= \mathcal{F}_{\SA} \cap (\bigcup_{t \in \mathcal{I}^{E}} \mathcal{F}_{\suffix}(\Psi_{\CCP}(T[i..j]) \cap \Psi_{\run} \cap \Psi_{\centerset}(C_{Q}) \cap \Psi_{\lcp}(t + \theta - 1) \cap \Psi_{\preceding})).
    \end{split}
\end{equation*}    
\end{enumerate}

\textbf{Proof of statement (A).}
We prove $\Psi_{\CCP}(T[i..j]) \cap \Psi_{\run} \cap \Psi_{\centerset}(C_{Q}) \cap (\bigcup_{\lambda = 1}^{\theta-1} \Psi_{\lcp}(\lambda))\cap \Psi_{\preceding} = \emptyset$ by contradiction. 
We assume that $\Psi_{\CCP}(T[i..j]) \cap \Psi_{\run} \cap \Psi_{\centerset}(C_{Q}) \cap (\bigcup_{\lambda = 1}^{\theta-1} \Psi_{\lcp}(\lambda))\cap \Psi_{\preceding} \neq \emptyset$ holds. 
Then, there exists an integer $K \in [1, \theta - 1]$ satisfying 
$\Psi_{\CCP}(T[i..j]) \cap \Psi_{\run} \cap \Psi_{\centerset}(C_{Q}) \cap \Psi_{\lcp}(K) \cap \Psi_{\preceding} \neq \emptyset$. 
Let $([p, q], [\ell, r])$ be an interval attractor in the set $\Psi_{\CCP}(T[i..j]) \cap \Psi_{\run} \cap \Psi_{\centerset}(C_{Q}) \cap \Psi_{\lcp}(K) \cap \Psi_{\preceding}$. 
Because of $([p, q], [\ell, r]) \in \Psi_{\centerset}(C_{Q}) \cap \Psi_{\lcp}(K)$, 
$|\lcp(T[\gamma..r], C_{Q}^{n+1})| = K$ follows from the definition of the subset $\Psi_{\lcp}(K)$ for the attractor position $\gamma$ of the interval attractor $([p, q], [\ell, r])$. 
$|\lcp(T[\gamma..r], C_{Q}^{n+1})| \leq \theta - 1$ follows from $|\lcp(T[\gamma..r], C_{Q}^{n+1})| = K$ and $K \in [1, \theta - 1]$. 

Under the assumption that $\Psi_{\CCP}(T[i..j]) \cap \Psi_{\run} \cap \Psi_{\centerset}(C_{Q}) \cap (\bigcup_{\lambda = 1}^{\theta-1} \Psi_{\lcp}(\lambda))\cap \Psi_{\preceding} \neq \emptyset$ holds, 
we prove $|\lcp(T[\gamma..r], C_{Q}^{n+1})| \geq 2 + \sum_{w = 1}^{h_{Q}+3} \lfloor \mu(w) \rfloor$. 
Because of $([p, q], [\ell, r]) \in \Psi_{\CCP}(T[i..j])$, 
Lemma~\ref{lem:CCP_property}~\ref{enum:CCP_property:1} shows that 
Because of $([p, q], [\ell, r]) \in \Psi_{h_{Q}} \cap \Psi_{\run} \cap \Psi_{\centerset}(C_{Q})$, 
$|\lcp(T[\gamma..r], C_{Q}^{n+1})| \geq 2 + \sum_{w = 1}^{h_{Q}+3} \lfloor \mu(w) \rfloor$ follows from the definition of the subset $\Psi_{\run}$. 

Under the assumption that $\Psi_{\CCP}(T[i..j]) \cap \Psi_{\run} \cap \Psi_{\centerset}(C_{Q}) \cap (\bigcup_{\lambda = 1}^{\theta-1} \Psi_{\lcp}(\lambda)) \cap \Psi_{\preceding} \neq \emptyset$ holds, 
we prove $|\lcp(T[\gamma..r], C_{Q}^{n+1})| \geq |[\gamma, j]|$. 
Let $\hat{K} = |\lcp(T[\gamma_{Q}..j], C_{Q}^{n+1})|$. 
Then, $\hat{K} = |[\gamma, j]|$ follows from condition (B) of RSS query. 
Because of $([p, q], [\ell, r]) \in \Psi_{\CCP}(T[i..j])$, 
Lemma~\ref{lem:CCP_property}~\ref{enum:CCP_property:6} shows that 
$I_{\capture}(\gamma - |[i, \gamma_{Q}-1]|, \gamma + |[\gamma_{Q}, j]| - 1) = ([p, q], [\ell, r])$ 
and $T[\gamma..j] = T[\gamma..\gamma + |[\gamma_{Q}, j]| - 1]$ hold. 
$|\lcp(T[\gamma..\gamma + |[\gamma_{Q}, j]| - 1], C_{Q}^{n+1})| = \hat{K}$ follows from 
$\hat{K} = |\lcp(T[\gamma_{Q}..j], C_{Q}^{n+1})|$ and $T[\gamma..j] = T[\gamma..\gamma + |[\gamma_{Q}, j]| - 1]$. 

Since $I_{\capture}(\gamma - |[i, \gamma_{Q}-1]|, \gamma + |[\gamma_{Q}, j]| - 1) = ([p, q], [\ell, r])$, 
$\gamma - |[i, \gamma_{Q}-1]| \in [p, q]$ and $\gamma + |[\gamma_{Q}, j]| - 1 \in [\ell, r]$ follow from the definition of interval attractor. 
$|[\gamma, r]| \geq |[\gamma_{Q}, j]|$ follows from $\gamma + |[\gamma_{Q}, j]| - 1 \in [\ell, r]$. 
$|\lcp(T[\gamma..r], C_{Q}^{n+1})| \geq |\lcp(T[\gamma..\gamma + |[\gamma_{Q}, j]| - 1], C_{Q}^{n+1})|$ holds 
because $\gamma + |[\gamma_{Q}, j]| - 1 \leq r$. 
Therefore, $|\lcp(T[\gamma..r], C_{Q}^{n+1})| \geq |[\gamma, j]|$ follows from 
$|\lcp(T[\gamma..r], C_{Q}^{n+1})| \geq |\lcp(T[\gamma..\gamma + |[\gamma_{Q}, j]| - 1], C_{Q}^{n+1})|$, 
$|\lcp(T[\gamma..\gamma + |[\gamma_{Q}, j]| - 1], C_{Q}^{n+1})| = \hat{K}$, 
and $\hat{K} = |[\gamma, j]|$.

$|\lcp(T[\gamma..r], C_{Q}^{n+1})| \geq \theta$ holds 
because (1) $|\lcp(T[\gamma..r], C_{Q}^{n+1})| > 1 + \sum_{w = 1}^{h_{Q}+3} \lfloor \mu(w) \rfloor$, 
(2) $|\lcp(T[\gamma..r]$, $C_{Q}^{n+1})| \geq |[\gamma, j]|$, 
and (3) $\theta = \max\{ |[\gamma_{Q}, j]|, 2 + \sum_{w = 1}^{h_{Q}+3} \lfloor \mu(w) \rfloor \}$. 
The two facts $|\lcp(T[\gamma..r]$, $C_{Q}^{n+1})| \leq \theta - 1$ and $|\lcp(T[\gamma..r], C_{Q}^{n+1})| \geq \theta$ yield a contradiction. 
Therefore, $\Psi_{\CCP}(T[i..j]) \cap \Psi_{\run} \cap \Psi_{\centerset}(C_{Q}) \cap (\bigcup_{\lambda = 1}^{\theta-1} \Psi_{\lcp}(\lambda))\cap \Psi_{\preceding} = \emptyset$ must hold. 

\textbf{Proof of statement (B).}
Lemma~\ref{lem:F_suffix_basic_property}~\ref{enum:F_suffix_basic_property:5} indicates that 
the following equation holds: 
\begin{equation*}
    \begin{split}
    & \mathcal{F}_{\SA} \cap \mathcal{F}_{\suffix}(\Psi_{\CCP}(T[i..j]) \cap \Psi_{\run} \cap \Psi_{\centerset}(C_{Q}) \cap (\bigcup_{\lambda = \theta}^{n} \Psi_{\lcp}(\lambda)) \cap \Psi_{\preceding}) \\
    &= \mathcal{F}_{\SA} \cap (\bigcup_{\lambda = \theta}^{n} \mathcal{F}_{\suffix}(\Psi_{\CCP}(T[i..j]) \cap \Psi_{\run} \cap \Psi_{\centerset}(C_{Q}) \cap \Psi_{\lcp}(\lambda) \cap \Psi_{\preceding})) \\
    &= \mathcal{F}_{\SA} \cap (\bigcup_{t = 1}^{n - \theta + 1} \mathcal{F}_{\suffix}(\Psi_{\CCP}(T[i..j]) \cap \Psi_{\run} \cap \Psi_{\centerset}(C_{Q}) \cap \Psi_{\lcp}(t + \theta - 1) \cap \Psi_{\preceding})).    
    \end{split}
\end{equation*}   
For each integer $t \in [1, n - \theta + 1] \setminus \mathcal{I}^{E}$
$\mathcal{F}_{\SA} \cap \mathcal{F}_{\suffix}(\Psi_{\CCP}(T[i..j]) \cap \Psi_{\run} \cap \Psi_{\centerset}(C_{Q}) \cap \Psi_{\lcp}(t + \theta - 1) \cap \Psi_{\preceding}) = \emptyset$ follows from the definition of the set $\mathcal{I}^{E}$. 
Therefore, statement (B) holds. 

\textbf{Proof of Proposition~\ref{prop:Set_IE1_Property}(i).}
Lemma~\ref{lem:psi_LMPS_property}~\ref{enum:psi_LMPS_property:lcp:1} indicates that 
set $\Psi_{\CCP}(T[i..j]) \cap \Psi_{\run} \cap \Psi_{\centerset}(C_{Q}) \cap \Psi_{\preceding}$ is equal to the union of 
two sets $\Psi_{\CCP}(T[i..j]) \cap \Psi_{\run} \cap \Psi_{\centerset}(C_{Q}) \cap (\bigcup_{\lambda = 1}^{\theta-1} \Psi_{\lcp}(\lambda)) \cap \Psi_{\preceding}$ and $\Psi_{\CCP}(T[i..j]) \cap \Psi_{\run} \cap \Psi_{\centerset}(C_{Q}) \cap (\bigcup_{\lambda = \theta}^{n} \Psi_{\lcp}(\lambda))\cap \Psi_{\preceding}$. 
Set $\mathcal{F}_{\suffix}(\Psi_{\CCP}(T[i..j]) \cap \Psi_{\run} \cap \Psi_{\centerset}(C_{Q}) \cap (\bigcup_{\lambda = 1}^{\theta-1} \Psi_{\lcp}(\lambda)) \cap \Psi_{\preceding})$ is empty 
because (1) $\Psi_{\CCP}(T[i..j]) \cap \Psi_{\run} \cap \Psi_{\centerset}(C_{Q}) \cap (\bigcup_{\lambda = 1}^{\theta-1} \Psi_{\lcp}(\lambda)) \cap \Psi_{\preceding} = \emptyset$ follows from statement (A), 
and (2) $\mathcal{F}_{\suffix}(\emptyset) = \emptyset$ follows from Lemma~\ref{lem:F_suffix_basic_property} \ref{enum:F_suffix_basic_property:4}. 
Therefore, Proposition~\ref{prop:Set_IE1_Property}~\ref{enum:Set_IE1_Property:1} follows from 
Lemma~\ref{lem:F_suffix_basic_property}~\ref{enum:F_suffix_basic_property:5}, statement (A), and statement (B). 
Formally, Proposition~\ref{prop:Set_IE1_Property}~\ref{enum:Set_IE1_Property:1} follows from the following equation: 
\begin{equation*}
    \begin{split}
    & \mathcal{F}_{\SA} \cap \mathcal{F}_{\suffix}(\Psi_{\CCP}(T[i..j]) \cap \Psi_{\run} \cap \Psi_{\centerset}(C_{Q}) \cap (\bigcup_{\lambda = \theta}^{n} \Psi_{\lcp}(\lambda)) \cap \Psi_{\preceding}) \\
    &= (\mathcal{F}_{\SA} \cap \mathcal{F}_{\suffix}(\Psi_{\CCP}(T[i..j]) \cap \Psi_{\run} \cap \Psi_{\centerset}(C_{Q}) \cap (\bigcup_{\lambda = 1}^{\theta-1} \Psi_{\lcp}(\lambda)) \cap \Psi_{\preceding})) \\
    &\cup (\mathcal{F}_{\SA} \cap \mathcal{F}_{\suffix}(\Psi_{\CCP}(T[i..j]) \cap \Psi_{\run} \cap \Psi_{\centerset}(C_{Q}) \cap (\bigcup_{\lambda = \theta}^{n} \Psi_{\lcp}(\lambda)) \cap \Psi_{\preceding})) \\
    &= (\mathcal{F}_{\SA} \cap \emptyset) \\
    &\cup (\mathcal{F}_{\SA} \cap (\bigcup_{t \in \mathcal{I}^{E}} \mathcal{F}_{\suffix}(\Psi_{\CCP}(T[i..j]) \cap \Psi_{\run} \cap \Psi_{\centerset}(C_{Q}) \cap \Psi_{\lcp}(t + \theta - 1) \cap \Psi_{\preceding}))) \\
    &= \mathcal{F}_{\SA} \cap (\bigcup_{t \in \mathcal{I}^{E}} \mathcal{F}_{\suffix}(\Psi_{\CCP}(T[i..j]) \cap \Psi_{\run} \cap \Psi_{\centerset}(C_{Q}) \cap \Psi_{\lcp}(t + \theta - 1) \cap \Psi_{\preceding})).
    \end{split}
\end{equation*} 

\textbf{Proof of Proposition~\ref{prop:Set_IE1_Property}(ii).}
Because of $u_{t} = 1$, 
the following equation follows from the definition of sequence $\gamma_{E}$: 
\begin{equation}\label{eq:Set_IE1_Property:1}
    \begin{split}
    \mathcal{F}_{\SA} \cap \mathcal{F}_{\suffix}(\Psi_{\CCP}(T[i..j]) \cap \Psi_{h_{Q}} \cap \Psi_{\run} \cap \Psi_{\centerset}(C_{Q}) \cap (\bigcup_{\lambda = t + \theta - 1}^{n} \Psi_{\lcp}(\lambda)) \cap \Psi_{\preceding}) \neq \emptyset.
    \end{split}
\end{equation}
If $t = n-\theta+1$, 
then $\mathcal{F}_{\SA} \cap \mathcal{F}_{\suffix}(\Psi_{\CCP}(T[i..j]) \cap \Psi_{h_{Q}} \cap \Psi_{\run} \cap \Psi_{\centerset}(C_{Q}) \cap \Psi_{\lcp}(t + \theta - 1) \cap \Psi_{\preceding}) \neq \emptyset$ follows from Equation~\ref{eq:Set_IE1_Property:1}. 
Therefore, $t \in \mathcal{I}^{E}$ holds. 

Otherwise (i.e., $t < n-\theta+1$), 
we leverage the $(t+1)$-th integer $u_{t+1}$ of the sequence $\gamma_{E}$. 
Because of $u_{t+1} = 0$, 
the following equation follows from the definition of sequence $\gamma_{E}$: 
\begin{equation}\label{eq:Set_IE1_Property:2}
    \begin{split}
    \mathcal{F}_{\SA} \cap \mathcal{F}_{\suffix}(\Psi_{\CCP}(T[i..j]) \cap \Psi_{h_{Q}} \cap \Psi_{\run} \cap \Psi_{\centerset}(C_{Q}) \cap (\bigcup_{\lambda = t + \theta}^{n} \Psi_{\lcp}(\lambda)) \cap \Psi_{\preceding}) = \emptyset.
    \end{split}
\end{equation}
$\mathcal{F}_{\SA} \cap \mathcal{F}_{\suffix}(\Psi_{\CCP}(T[i..j]) \cap \Psi_{h_{Q}} \cap \Psi_{\run} \cap \Psi_{\centerset}(C_{Q}) \cap \Psi_{\lcp}(t + \theta - 1) \cap \Psi_{\preceding}) \neq \emptyset$ follows from Equation~\ref{eq:Set_IE1_Property:1} 
and Equation~\ref{eq:Set_IE1_Property:2}. 
Therefore, $t \in \mathcal{I}^{E}$ follows from the definition of the set $\mathcal{I}^{E}$. 

\textbf{Proof of Proposition~\ref{prop:Set_IE1_Property}(iii).}
Proposition~\ref{prop:Set_IE1_Property}~\ref{enum:Set_IE1_Property:3} follows from 
the definitions of sequence $\gamma_{E}$ and set $\mathcal{I}^{E}$.
\end{proof}

We prove Lemma~\ref{lem:GammaE1_property} using Proposition~\ref{prop:Psi_E1_Property} and Proposition~\ref{prop:Set_IE1_Property}.

\begin{proof}[Proof of Lemma~\ref{lem:GammaE1_property}~\ref{enum:GammaE1_property:1}]
Lemma~\ref{lem:GammaE1_property}~\ref{enum:GammaE1_property:1} can be proved using the same approach as for 
Lemma \ref{lem:GammaA_property} \ref{enum:GammaA_property:1}. 
The detailed proof of Lemma~\ref{lem:GammaE1_property}~\ref{enum:GammaE1_property:1} is as follows. 

Consider the integers $t_{1}, t_{2}, \ldots, t_{m}$ ($t_{1} < t_{2} < \ldots < t_{m}$) in set $\mathcal{I}^{E}$. 
Let $F_{\tau}$ be the lexicographically largest string in set 
$\mathcal{F}_{\SA} \cap \mathcal{F}_{\suffix}(\Psi_{\CCP}(T[i..j]) \cap \Psi_{\run} \cap \Psi_{\centerset}(C_{Q}) \cap \Psi_{\lcp}(t_{\tau} + \theta - 1) \cap \Psi_{\preceding})$ for each integer $\tau \in [1, m]$. 
From the definition of the set $\mathcal{I}^{E}$, 
each string $F_{\tau}$ exists. 

If the integer $\kappa$ exists, 
then $\kappa = t_{m}$ follows from 
Proposition~\ref{prop:Set_IE1_Property}~\ref{enum:Set_IE1_Property:2} and 
Proposition~\ref{prop:Set_IE1_Property} \ref{enum:Set_IE1_Property:3}. 
Proposition~\ref{prop:Set_IE1_Property}~\ref{enum:Set_IE1_Property:1} indicates that 
the lexicographically largest string in set $\mathcal{F}_{\SA} \cap \mathcal{F}_{\suffix}$ $(\Psi_{\CCP}(T[i..j]) \cap \Psi_{\run} \cap \Psi_{\centerset}(C_{Q}) \cap \Psi_{\preceding})$ is equal to the lexicographically largest string in set $\{ F_{1}, F_{2}, \ldots, F_{m} \}$. 
Proposition~\ref{prop:Psi_E1_Property}~\ref{enum:Psi_E1_Property:2} indicates that 
$F_{1} \prec F_{2} \prec \ldots \prec F_{m}$ holds. 
$F^{\prime} = F_{m}$ follows from $\kappa = t_{m}$. 
Therefore, the string $F^{\prime}$ is the lexicographically largest string in set $\mathcal{F}_{\SA} \cap \mathcal{F}_{\suffix}(\Psi_{\CCP}(T[i..j]) \cap \Psi_{\run} \cap \Psi_{\centerset}(C_{Q}) \cap \Psi_{\preceding})$. 

Otherwise~(i.e., the integer $\kappa$ does not exist), 
Proposition~\ref{prop:Set_IE1_Property}~\ref{enum:Set_IE1_Property:3} indicates that 
$\mathcal{I}^{E} = \emptyset$ holds (i.e., $m = 0$).
In this case, 
Proposition~\ref{prop:Set_IE1_Property}~\ref{enum:Set_IE1_Property:1} shows that 
$\mathcal{F}_{\SA} \cap \mathcal{F}_{\suffix}(\Psi_{\CCP}(T[i..j]) \cap \Psi_{\run} \cap \Psi_{\centerset}(C_{Q}) \cap \Psi_{\preceding}) = \emptyset$ holds. 
Therefore, Lemma~\ref{lem:GammaE1_property}~\ref{enum:GammaE1_property:1} holds. 
\end{proof}

\begin{proof}[Proof of Lemma~\ref{lem:GammaE1_property}~\ref{enum:GammaE1_property:2}]
Lemma~\ref{lem:GammaE1_property}~\ref{enum:GammaE1_property:2} corresponds to 
Lemma~\ref{lem:GammaA_property}~\ref{enum:GammaA_property:4}. 
We showed that Lemma~\ref{lem:GammaA_property}~\ref{enum:GammaA_property:4} holds using 
Proposition~\ref{prop:Set_IA_Property}~\ref{enum:Set_IA_Property:2} and 
Proposition~\ref{prop:Set_IA_Property}~\ref{enum:Set_IA_Property:3}. 
Proposition~\ref{prop:Set_IE1_Property}~\ref{enum:Set_IE1_Property:2} corresponds to Proposition~\ref{prop:Set_IA_Property}~\ref{enum:Set_IA_Property:2}. 
Similarly, 
Proposition~\ref{prop:Set_IE1_Property}~\ref{enum:Set_IE1_Property:3} corresponds to Proposition~\ref{prop:Set_IA_Property}~\ref{enum:Set_IA_Property:3}. 
Therefore, Lemma~\ref{lem:GammaE1_property}~\ref{enum:GammaE1_property:2} can be proved using the same approach as for 
Lemma~\ref{lem:GammaA_property} \ref{enum:GammaA_property:4}. 
\end{proof}

\subsubsection{Proof of Lemma~\ref{lem:GammaE1_sub_property}}\label{subsubsec:GammaE1_sub_property_proof}
We prove Lemma~\ref{lem:GammaE1_sub_property} using Proposition~\ref{prop:Psi_E1_Property} and Proposition~\ref{prop:Set_IE1_Property}.

\begin{proof}[Proof of Lemma~\ref{lem:GammaE1_sub_property}~\ref{enum:GammaE1_sub_property:1}]
We prove $\alpha_{t} \geq \alpha_{t+1}$ for each integer $t \in [1, n-\theta]$. 
Lemma~\ref{lem:psi_LMPS_property}~\ref{enum:psi_LMPS_property:lcp:2} shows that 
$\Psi_{\lcp}(\lambda) \cap \Psi_{\lcp}(\lambda^{\prime}) = \emptyset$ holds 
for any pair of two integers $0 \leq \lambda < \lambda^{\prime} \leq n$. 
Therefore, following equation holds: 
\begin{equation*}
    \begin{split}
    |\Psi_{h_{Q}} \cap \Psi_{\run} & \cap \Psi_{\centerset}(C_{Q}) \cap (\bigcup_{\lambda = t + \theta - 1}^{n} \Psi_{\lcp}(\lambda)) \cap \Psi_{\preceding}| \\
    &= |\Psi_{h_{Q}} \cap \Psi_{\run} \cap \Psi_{\centerset}(C_{Q}) \cap \Psi_{\lcp}(t + \theta - 1) \cap \Psi_{\preceding}| \\
    &+ |\Psi_{h_{Q}} \cap \Psi_{\run} \cap \Psi_{\centerset}(C_{Q}) \cap (\bigcup_{\lambda = t + \theta}^{n} \Psi_{\lcp}(\lambda)) \cap \Psi_{\preceding}|.
    \end{split}
\end{equation*}

$\alpha_{t} \geq \alpha_{t+1}$ follows from the following equation:
\begin{equation}\label{eq:GammaE1_sub_property:1}
    \begin{split}
    \alpha_{t} &= |\Psi_{h_{Q}} \cap \Psi_{\run} \cap \Psi_{\centerset}(C_{Q}) \cap (\bigcup_{\lambda = t + \theta - 1}^{n} \Psi_{\lcp}(\lambda)) \cap \Psi_{\preceding}| \\
    &= |\Psi_{h_{Q}} \cap \Psi_{\run} \cap \Psi_{\centerset}(C_{Q}) \cap \Psi_{\lcp}(t + \theta - 1) \cap \Psi_{\preceding}| \\
    &+ |\Psi_{h_{Q}} \cap \Psi_{\run} \cap \Psi_{\centerset}(C_{Q}) \cap (\bigcup_{\lambda = t + \theta}^{n} \Psi_{\lcp}(\lambda)) \cap \Psi_{\preceding}| \\
    &= |\Psi_{h_{Q}} \cap \Psi_{\run} \cap \Psi_{\centerset}(C_{Q}) \cap \Psi_{\lcp}(t + \theta - 1) \cap \Psi_{\preceding}| + \alpha_{t+1} \\
    &\geq \alpha_{t+1}.
    \end{split}
\end{equation}

Therefore, $\alpha_{1} \geq \alpha_{2} \geq \cdots \geq \alpha_{n-\theta+1}$ holds. 
\end{proof}

\begin{proof}[Proof of Lemma~\ref{lem:GammaE1_sub_property}~\ref{enum:GammaE1_sub_property:2}]
Let $\alpha_{n-\theta+2} = 0$ for simplicity. 
The following three statements are used to prove Lemma~\ref{lem:GammaE1_sub_property}~\ref{enum:GammaE1_sub_property:2}:
\begin{enumerate}[label=\textbf{(\Alph*)}]
    \item $u_{t} = 1 \Leftarrow (\alpha_{t^{\prime}} \geq 1) \land (\mathcal{F}_{\SA} \cap \mathcal{F}_{\suffix}(\Psi_{\CCP}(T[i..j]) \cap \Psi_{\run} \cap \Psi_{\centerset}(C_{Q}) \cap \Psi_{\lcp}(t^{\prime} + \theta - 1) \cap \Psi_{\preceding}) \neq \emptyset)$;
    \item 
    if $u_{t} = 1$, then 
    $t^{\prime} = t_{A}$ and $\alpha_{t^{\prime}} \geq 1$ for the smallest integer $t_{A}$ in set $[t, n-\theta+1]$ 
    satisfying $\mathcal{F}_{\SA} \cap \mathcal{F}_{\suffix}(\Psi_{\CCP}(T[i..j]) \cap \Psi_{h_{Q}} \cap \Psi_{\run} \cap \Psi_{\centerset}(C_{Q}) \cap \Psi_{\lcp}(t_{A} + \theta - 1) \cap \Psi_{\preceding}) \neq \emptyset$;
    \item $u_{t} = 1 \Rightarrow (\alpha_{t^{\prime}} \geq 1) \land (\mathcal{F}_{\SA} \cap \mathcal{F}_{\suffix}(\Psi_{\CCP}(T[i..j]) \cap \Psi_{\run} \cap \Psi_{\centerset}(C_{Q}) \cap \Psi_{\lcp}(t^{\prime} + \theta - 1) \cap \Psi_{\preceding}) \neq \emptyset)$.
\end{enumerate}

\textbf{Proof of statement (A).}
Consider a string $F$ in set $\mathcal{F}_{\SA} \cap \mathcal{F}_{\suffix}(\Psi_{\CCP}(T[i..j]) \cap \Psi_{\run} \cap \Psi_{\centerset}(C_{Q}) \cap \Psi_{\lcp}(t^{\prime} + \theta - 1) \cap \Psi_{\preceding})$. 
Because of $F \in \mathcal{F}_{\suffix}(\Psi_{\CCP}(T[i..j]) \cap \Psi_{\run} \cap \Psi_{\centerset}(C_{Q}) \cap \Psi_{\lcp}(t^{\prime} + \theta - 1) \cap \Psi_{\preceding})$, 
set $\Psi_{\CCP}(T[i..j]) \cap \Psi_{\run} \cap \Psi_{\centerset}(C_{Q}) \cap \Psi_{\lcp}(t^{\prime} + \theta - 1) \cap \Psi_{\preceding}$ contains an interval attractor $([p, q], [\ell, r])$ satisfying 
$T[\gamma - |[i, \gamma_{Q}-1]|..r+1] = F$ for the attractor position $\gamma$ of the interval attractor $([p, q], [\ell, r])$. 
The existence of this interval attractor $([p, q], [\ell, r])$ indicates that 
Equation~\ref{eq:GammaE1:1} holds. Therefore, $u_{t} = 1$ holds. 

\textbf{Proof of statement (B).}
Because of $u_{t} = 1$, 
Equation~\ref{eq:GammaE1:1} indicates that the smallest integer $t_{A}$ exists. 
For an integer $\tau \in [t, t_{A}]$, 
we prove $\alpha_{t} = \alpha_{\tau}$ by contradiction. 
Here, $\alpha_{t} \geq \alpha_{\tau}$ follows from Lemma~\ref{lem:GammaE1_sub_property}~\ref{enum:GammaE1_sub_property:1}. 
We assume that $\alpha_{t} \neq \alpha_{\tau}$ holds. 
Then, $\alpha_{t} > \alpha_{\tau}$ follows from $\alpha_{t} \neq \alpha_{\tau}$ and $\alpha_{t} \geq \alpha_{\tau}$. 
In this case, $t < \tau$ holds, 
and there exists an integer $\tau^{\prime} \in [t, \tau-1]$ satisfying 
$\alpha_{\tau^{\prime}} > \alpha_{\tau^{\prime}+1}$. 
We apply Equation~\ref{eq:GammaE1_sub_property:1} to the two integers $\alpha_{\tau^{\prime}}$ and $\alpha_{\tau^{\prime}+1}$. 
Then, this equation indicates that 
$\alpha_{\tau^{\prime}} = |\Psi_{h_{Q}} \cap \Psi_{\run} \cap \Psi_{\centerset}(C_{Q}) \cap \Psi_{\lcp}(\tau^{\prime} + \theta - 1) \cap \Psi_{\preceding}| + \alpha_{\tau^{\prime}+1}$ holds. 
$\Psi_{h_{Q}} \cap \Psi_{\run} \cap \Psi_{\centerset}(C_{Q}) \cap \Psi_{\lcp}(\tau^{\prime} + \theta - 1) \cap \Psi_{\preceding} \neq \emptyset$ follows from $\alpha_{\tau^{\prime}} > \alpha_{\tau^{\prime}+1}$ and $\alpha_{\tau^{\prime}} = |\Psi_{h_{Q}} \cap \Psi_{\run} \cap \Psi_{\centerset}(C_{Q}) \cap \Psi_{\lcp}(\tau^{\prime} + \theta - 1) \cap \Psi_{\preceding}| + \alpha_{\tau^{\prime}+1}$. 
Let $([p, q], [\ell, r])$ be an interval attractor in the set $\Psi_{h_{Q}} \cap \Psi_{\run} \cap \Psi_{\centerset}(C_{Q}) \cap \Psi_{\lcp}(\tau^{\prime} + \theta - 1) \cap \Psi_{\preceding}$. 
Then, $([p, q], [\ell, r]) \in \Psi_{\CCP}(T[i..j])$ follows from Proposition~\ref{prop:Psi_E1_Property}~\ref{enum:Psi_E1_Property:1}. 
Let $F = T[\gamma - |[i, \gamma_{Q}-1]|..\gamma-1] \cdot T[\gamma..r+1]$ for simplicity. 
Because of $([p, q], [\ell, r]) \in \Psi_{\CCP}(T[i..j]) \cap \Psi_{h_{Q}} \cap \Psi_{\run} \cap \Psi_{\centerset}(C_{Q}) \cap \Psi_{\lcp}(\tau^{\prime} + \theta - 1) \cap \Psi_{\preceding}$, 
$F \in \mathcal{F}_{\suffix}(\Psi_{\CCP}(T[i..j]) \cap \Psi_{h_{Q}} \cap \Psi_{\run} \cap \Psi_{\centerset}(C_{Q}) \cap \Psi_{\lcp}(\tau^{\prime} + \theta - 1) \cap \Psi_{\preceding})$ follows from the definition of the subset $\mathcal{F}_{\suffix}(\Psi_{\CCP}(T[i..j]) \cap \Psi_{h_{Q}} \cap \Psi_{\run} \cap \Psi_{\centerset}(C_{Q}) \cap \Psi_{\lcp}(\tau^{\prime} + \theta - 1) \cap \Psi_{\preceding})$. 

We prove $F \in \mathcal{F}_{\SA}$. 
Because of $\mathcal{F}_{\SA} \cap \mathcal{F}_{\suffix}(\Psi_{\CCP}(T[i..j]) \cap \Psi_{h_{Q}} \cap \Psi_{\run} \cap \Psi_{\centerset}(C_{Q}) \cap \Psi_{\lcp}(t_{A} + \theta - 1) \cap \Psi_{\preceding}) \neq \emptyset$, 
the set $\mathcal{F}_{\SA} \cap \mathcal{F}_{\suffix}(\Psi_{\CCP}(T[i..j]) \cap \Psi_{h_{Q}} \cap \Psi_{\run} \cap \Psi_{\centerset}(C_{Q}) \cap \Psi_{\lcp}(t_{A} + \theta - 1) \cap \Psi_{\preceding})$ contains a string $F^{\prime}$. 
Because of $\tau^{\prime} < t_{A}$, 
Proposition~\ref{prop:Psi_E1_Property}~\ref{enum:Psi_E1_Property:2} shows that 
$F \prec F^{\prime}$ holds. 
Therefore, $F \in \mathcal{F}_{\SA}$ follows from Lemma~\ref{lem:F_suffix_basic_property}~\ref{enum:F_suffix_basic_property:6}, 
$F \prec F^{\prime}$, and $F^{\prime} \in \mathcal{F}_{\SA}$. 

$\mathcal{F}_{\SA} \cap \mathcal{F}_{\suffix}(\Psi_{\CCP}(T[i..j]) \cap \Psi_{h_{Q}} \cap \Psi_{\run} \cap \Psi_{\centerset}(C_{Q}) \cap \Psi_{\lcp}(\tau^{\prime} + \theta - 1) \cap \Psi_{\preceding}) \neq \emptyset$ follows from $F \in \mathcal{F}_{\SA} \cap \mathcal{F}_{\suffix}(\Psi_{\CCP}(T[i..j]) \cap \Psi_{h_{Q}} \cap \Psi_{\run} \cap \Psi_{\centerset}(C_{Q}) \cap \Psi_{\lcp}(\tau^{\prime} + \theta - 1) \cap \Psi_{\preceding})$. 
On the other hand, $\mathcal{F}_{\SA} \cap \mathcal{F}_{\suffix}(\Psi_{\CCP}(T[i..j]) \cap \Psi_{h_{Q}} \cap \Psi_{\run} \cap \Psi_{\centerset}(C_{Q}) \cap \Psi_{\lcp}(\tau^{\prime} + \theta - 1) \cap \Psi_{\preceding}) = \emptyset$ holds 
because $\tau^{\prime} < t_{A}$. 
The two facts $\mathcal{F}_{\SA} \cap \mathcal{F}_{\suffix}(\Psi_{\CCP}(T[i..j]) \cap \Psi_{h_{Q}} \cap \Psi_{\run} \cap \Psi_{\centerset}(C_{Q}) \cap \Psi_{\lcp}(\tau^{\prime} + \theta - 1) \cap \Psi_{\preceding}) \neq \emptyset$ and $\mathcal{F}_{\SA} \cap \mathcal{F}_{\suffix}(\Psi_{\CCP}(T[i..j]) \cap \Psi_{h_{Q}} \cap \Psi_{\run} \cap \Psi_{\centerset}(C_{Q}) \cap \Psi_{\lcp}(\tau^{\prime} + \theta - 1) \cap \Psi_{\preceding}) = \emptyset$ yield a contradiction. 
Therefore, $\alpha_{t} = \alpha_{\tau}$ must hold. 

Next, we prove $\alpha_{t_{A}} \geq 1$ and $\alpha_{t_{A}} > \alpha_{t_{A}+1}$. 
Because of $\mathcal{F}_{\suffix}(\Psi_{\CCP}(T[i..j]) \cap \Psi_{h_{Q}} \cap \Psi_{\run} \cap \Psi_{\centerset}(C_{Q}) \cap \Psi_{\lcp}(t_{A} + \theta - 1) \cap \Psi_{\preceding}) \neq \emptyset$, 
set $\Psi_{\CCP}(T[i..j]) \cap \Psi_{h_{Q}} \cap \Psi_{\run} \cap \Psi_{\centerset}(C_{Q}) \cap \Psi_{\lcp}(t_{A} + \theta - 1) \cap \Psi_{\preceding}$ contains an interval attractor $([p^{\prime}, q^{\prime}], [\ell^{\prime}, r^{\prime}])$. 
If $\tau_{A} = n - \theta + 1$, 
then $\alpha_{t_{A}} = |\Psi_{h_{Q}} \cap \Psi_{\run} \cap \Psi_{\centerset}(C_{Q}) \cap \Psi_{\lcp}(n) \cap \Psi_{\preceding}|$ 
and $\alpha_{t_{A}+1} = 0$ follows from the definitions of the two integers $\alpha_{t_{A}}$ and $\alpha_{t_{A}+1}$. 
Therefore, $\alpha_{t_{A}} \geq 1$ and $\alpha_{t_{A}} > \alpha_{t_{A}+1}$ because 
$([p^{\prime}, q^{\prime}], [\ell^{\prime}, r^{\prime}]) \in \Psi_{h_{Q}} \cap \Psi_{\run} \cap \Psi_{\centerset}(C_{Q}) \cap \Psi_{\lcp}(n) \cap \Psi_{\preceding}$ and $\alpha_{t_{A}+1} = 0$. 

Otherwise (i.e., $\tau_{A} < n - \theta + 1$), 
we apply Equation~\ref{eq:GammaE1_sub_property:1} to the two integers $\alpha_{\tau_{A}}$ and $\alpha_{\tau_{A}+1}$. 
Then, this equation indicates that 
$\alpha_{\tau_{A}} = |\Psi_{h_{Q}} \cap \Psi_{\run} \cap \Psi_{\centerset}(C_{Q}) \cap \Psi_{\lcp}(\tau_{A} + \theta - 1) \cap \Psi_{\preceding}| + \alpha_{\tau_{A}+1}$ holds. 
$|\Psi_{h_{Q}} \cap \Psi_{\run} \cap \Psi_{\centerset}(C_{Q}) \cap \Psi_{\lcp}(\tau_{A} + \theta - 1) \cap \Psi_{\preceding}| \geq 1$ holds 
because $([p^{\prime}, q^{\prime}], [\ell^{\prime}, r^{\prime}]) \in \Psi_{h_{Q}} \cap \Psi_{\run} \cap \Psi_{\centerset}(C_{Q}) \cap \Psi_{\lcp}(\tau_{A} + \theta - 1) \cap \Psi_{\preceding}$. 
Therefore, $\alpha_{t_{A}} \geq 1$ and $\alpha_{t_{A}} > \alpha_{t_{A}+1}$ hold. 

We prove statement (B). 
$t^{\prime} = t_{A}$ holds because 
we showed that $\alpha_{t} = \alpha_{t+1} = \cdots = \alpha_{t_{A}}$ and $\alpha_{t_{A}} > \alpha_{t_{A}+1}$. 
$\alpha_{t^{\prime}} \geq 1$ follows from $t^{\prime} = t_{A}$ and $\alpha_{t_{A}} \geq 1$. 
Therefore, statement (B) holds.

\textbf{Proof of statement (C).}
This statement follows from statement (B),

\textbf{Proof of Lemma~\ref{lem:GammaE1_sub_property}~\ref{enum:GammaE1_sub_property:2}.}
Lemma~\ref{lem:GammaE1_sub_property}~\ref{enum:GammaE1_sub_property:2} follows from statement (A) and statement (C).

\end{proof}

\begin{proof}[Proof of Lemma~\ref{lem:GammaE1_sub_property}~\ref{enum:GammaE1_sub_property:3}]
Let $K_{t} = t + \theta - 1$ for simplicity. 
Lemma~\ref{lem:psi_LMPS_property}~\ref{enum:psi_LMPS_property:lcp:2} shows that 
$\Psi_{\lcp}(\lambda) \cap \Psi_{\lcp}(\lambda^{\prime}) = \emptyset$ holds 
for any pair of two integers $0 \leq \lambda < \lambda^{\prime} \leq n$. 
Therefore, the following equation holds:
\begin{equation}\label{eq:GammaE1_sub_property:2}
    \begin{split}
    \alpha_{t} &= |\Psi_{h_{Q}} \cap \Psi_{\run} \cap \Psi_{\centerset}(C_{Q}) \cap (\bigcup_{\lambda = K_{t}}^{n} \Psi_{\lcp}(\lambda)) \cap \Psi_{\preceding}| \\
    &= |\Psi_{h_{Q}} \cap \Psi_{\run} \cap \Psi_{\centerset}(C_{Q}) \cap (\bigcup_{\lambda = 0}^{n} \Psi_{\lcp}(\lambda)) \cap \Psi_{\preceding}| \\
    &- |\Psi_{h_{Q}} \cap \Psi_{\run} \cap \Psi_{\centerset}(C_{Q}) \cap (\bigcup_{\lambda = 0}^{K_{t} - 1} \Psi_{\lcp}(\lambda)) \cap \Psi_{\preceding}|.
    \end{split}
\end{equation}

Lemma~\ref{lem:psi_LMPS_property}~\ref{enum:psi_LMPS_property:lcp:1} indicates that 
set $\Psi_{h_{Q}} \cap \Psi_{\run} \cap \Psi_{\centerset}(C_{Q}) \cap (\bigcup_{\lambda = 0}^{n} \Psi_{\lcp}(\lambda)) \cap \Psi_{\preceding}$ 
is equal to set $\Psi_{h_{Q}} \cap \Psi_{\run} \cap \Psi_{\centerset}(C_{Q}) \cap \Psi_{\preceding}$. 
Lemma~\ref{lem:JD1_sum} shows that 
the number of interval attractors in the set $\Psi_{h_{Q}} \cap \Psi_{\run} \cap \Psi_{\centerset}(C_{Q}) \cap \Psi_{\preceding}$ 
can be computed by range-sum query $\rangesum(\mathcal{J}_{E}(h_{Q}, C_{Q}), 1, n, 0, |C_{Q}|-1)$. 
Therefore, the following equation holds: 
\begin{equation}\label{eq:GammaE1_sub_property:3}
    \begin{split}
    |\Psi_{h_{Q}} & \cap \Psi_{\run} \cap \Psi_{\centerset}(C_{Q}) \cap (\bigcup_{\lambda = 0}^{n} \Psi_{\lcp}(\lambda)) \cap \Psi_{\preceding}| \\
    &= |\Psi_{h_{Q}} \cap \Psi_{\run} \cap \Psi_{\centerset}(C_{Q}) \cap \Psi_{\preceding}| \\
    &= \rangesum(\mathcal{J}_{E}(h_{Q}, C_{Q}), 1, n, 0, |C_{Q}|-1).        
    \end{split}
\end{equation}

Lemma~\ref{lem:nRecover_basic_property}~\ref{enum:nRecover_basic_property:1} indicates that 
set $\Psi_{h_{Q}} \cap \Psi_{\run} \cap \Psi_{\centerset}(C_{Q}) \cap (\bigcup_{\lambda = 0}^{K_{t} - 1} \Psi_{\lcp}(\lambda)) \cap \Psi_{\preceding}$ is equal to set $\Psi_{h_{Q}} \cap \Psi_{\run} \cap \Psi_{\centerset}(C_{Q}) \cap (\bigcup_{\lambda = 1}^{n} \Psi_{\nRecover}(\lambda)) \cap (\bigcup_{\lambda = 0}^{K_{t} - 1} \Psi_{\lcp}(\lambda)) \cap \Psi_{\preceding}$. 
Lemma~\ref{lem:nRecover_basic_property}~\ref{enum:nRecover_basic_property:2} indicates that 
the set $\Psi_{h_{Q}} \cap \Psi_{\run} \cap \Psi_{\centerset}(C_{Q}) \cap (\bigcup_{\lambda = 1}^{n} \Psi_{\nRecover}(\lambda)) \cap (\bigcup_{\lambda = 0}^{K_{t} - 1} \Psi_{\lcp}(\lambda)) \cap \Psi_{\preceding}$ can be divided into two sets 
$\Psi_{h_{Q}} \cap \Psi_{\run} \cap \Psi_{\centerset}(C_{Q}) \cap \Psi_{\preceding} \cap (\bigcup_{\lambda = x_{t}}^{n} \Psi_{\nRecover}(\lambda)) \cap (\bigcup_{\lambda = 0}^{K_{t} - 1} \Psi_{\lcp}(\lambda))$ and $\Psi_{h_{Q}} \cap \Psi_{\run} \cap \Psi_{\centerset}(C_{Q}) \cap \Psi_{\preceding} \cap (\bigcup_{\lambda = 1}^{x_{t}-1} \Psi_{\nRecover}(\lambda)) \cap (\bigcup_{\lambda = 0}^{K_{t} - 1} \Psi_{\lcp}(\lambda))$. 
Therefore, the following equation holds:
\begin{equation}\label{eq:GammaE1_sub_property:4}
    \begin{split}
    |\Psi_{h_{Q}} & \cap \Psi_{\run} \cap \Psi_{\centerset}(C_{Q}) \cap (\bigcup_{\lambda = 0}^{K_{t} - 1} \Psi_{\lcp}(\lambda)) \cap \Psi_{\preceding}| \\
    &= |\Psi_{h_{Q}} \cap \Psi_{\run} \cap \Psi_{\centerset}(C_{Q}) \cap (\bigcup_{\lambda = 1}^{n} \Psi_{\nRecover}(\lambda)) \cap (\bigcup_{\lambda = 0}^{K_{t} - 1} \Psi_{\lcp}(\lambda)) \cap \Psi_{\preceding}| \\
    &= |\Psi_{h_{Q}} \cap \Psi_{\run} \cap \Psi_{\centerset}(C_{Q}) \cap \Psi_{\preceding} \cap (\bigcup_{\lambda = x_{t}}^{n} \Psi_{\nRecover}(\lambda)) \cap (\bigcup_{\lambda = 0}^{K_{t} - 1} \Psi_{\lcp}(\lambda))| \\
    &+ |\Psi_{h_{Q}} \cap \Psi_{\run} \cap \Psi_{\centerset}(C_{Q}) \cap \Psi_{\preceding} \cap (\bigcup_{\lambda = 1}^{x_{t}-1} \Psi_{\nRecover}(\lambda)) \cap (\bigcup_{\lambda = 0}^{K_{t} - 1} \Psi_{\lcp}(\lambda))|. 
    \end{split}
\end{equation}

The set $\Psi_{h_{Q}} \cap \Psi_{\run} \cap \Psi_{\centerset}(C_{Q}) \cap \Psi_{\preceding} \cap (\bigcup_{\lambda = x_{t}}^{n} \Psi_{\nRecover}(\lambda)) \cap (\bigcup_{\lambda = 0}^{K_{t} - 1} \Psi_{\lcp}(\lambda))$ can be divided into 
two sets $\Psi_{h_{Q}} \cap \Psi_{\run} \cap \Psi_{\centerset}(C_{Q}) \cap \Psi_{\preceding} \cap (\bigcup_{\lambda = x_{t}}^{n} \Psi_{\nRecover}(\lambda)) \cap (\bigcup_{\lambda = 0}^{K_{t} - M_{t} - 1} \Psi_{\lcp}(\lambda))$ 
and $\Psi_{h_{Q}} \cap \Psi_{\run} \cap \Psi_{\centerset}(C_{Q}) \cap \Psi_{\preceding} \cap (\bigcup_{\lambda = x_{t}}^{n} \Psi_{\nRecover}(\lambda)) \cap (\bigcup_{\lambda = K_{t} - M_{t}}^{K_{t} - 1} \Psi_{\lcp}(\lambda))$.
Equation~\ref{eq:JD1_sum:2} of Lemma~\ref{lem:JD1_sum} indicates that 
the number of interval attractors in the set $\Psi_{h_{Q}} \cap \Psi_{\run} \cap \Psi_{\centerset}(C_{Q}) \cap \Psi_{\preceding} \cap (\bigcup_{\lambda = x_{t}}^{n} \Psi_{\nRecover}(\lambda)) \cap (\bigcup_{\lambda = 0}^{K_{t} - M_{t} - 1} \Psi_{\lcp}(\lambda))$ can be computed by range-sum query $(x_{t}-1) \rangesum(\mathcal{J}_{D}(h_{Q}, C_{Q}), x_{t}, n, 0, |C_{Q}| - 1)$. 
Similarly, 
Equation~\ref{eq:JD1_sum:3} of Lemma~\ref{lem:JD1_sum} indicates that 
the number of interval attractors in the set $\Psi_{h_{Q}} \cap \Psi_{\run} \cap \Psi_{\centerset}(C_{Q}) \cap \Psi_{\preceding} \cap (\bigcup_{\lambda = x_{t}}^{n} \Psi_{\nRecover}(\lambda)) \cap (\bigcup_{\lambda = K_{t} - M_{t}}^{K_{t} - 1} \Psi_{\lcp}(\lambda))$ can be computed by range-sum query $\rangesum(\mathcal{J}_{D}(h_{Q}, C_{Q}), x_{t}, n, 0, M_{t} - 1)$. 
Therefore, the following equation holds: 
\begin{equation}\label{eq:GammaE1_sub_property:5}
    \begin{split}
    |\Psi_{h_{Q}} & \cap \Psi_{\run} \cap \Psi_{\centerset}(C_{Q}) \cap \Psi_{\preceding} \cap (\bigcup_{\lambda = x_{t}}^{n} \Psi_{\nRecover}(\lambda)) \cap (\bigcup_{\lambda = 0}^{K_{t} - 1} \Psi_{\lcp}(\lambda))| \\
    &= |\Psi_{h_{Q}} \cap \Psi_{\run} \cap \Psi_{\centerset}(C_{Q}) \cap \Psi_{\preceding} \cap (\bigcup_{\lambda = x_{t}}^{n} \Psi_{\nRecover}(\lambda)) \cap (\bigcup_{\lambda = 0}^{K_{t} - M_{t} - 1} \Psi_{\lcp}(\lambda))| \\
    &+ |\Psi_{h_{Q}} \cap \Psi_{\run} \cap \Psi_{\centerset}(C_{Q}) \cap \Psi_{\preceding} \cap (\bigcup_{\lambda = x_{t}}^{n} \Psi_{\nRecover}(\lambda)) \cap (\bigcup_{\lambda = K_{t} - M_{t}}^{K_{t} - 1} \Psi_{\lcp}(\lambda))| \\
    &= (x_{t}-1) \rangesum(\mathcal{J}_{D}(h_{Q}, C_{Q}), x_{t}, n, 0, |C_{Q}| - 1) \\
    &+ \rangesum(\mathcal{J}_{D}(h_{Q}, C_{Q}), x_{t}, n, 0, M_{t} - 1).
    \end{split}
\end{equation}

Equation~\ref{eq:JD1_sum:4} of Lemma~\ref{lem:JD1_sum} indicates that 
the number of interval attractors in the set $\Psi_{h_{Q}} \cap \Psi_{\run} \cap \Psi_{\centerset}(C_{Q}) \cap \Psi_{\preceding} \cap (\bigcup_{\lambda = 1}^{x_{t}-1} \Psi_{\nRecover}(\lambda)) \cap (\bigcup_{\lambda = 0}^{K_{t} - 1} \Psi_{\lcp}(\lambda))$ can be computed by range-sum query $\rangesum(\mathcal{J}_{E}(h_{Q}, C_{Q}), 1, x_{t}-1, 0, |C_{Q}|-1)$. 
Therefore, the following equation holds: 
\begin{equation}\label{eq:GammaE1_sub_property:6}
    \begin{split}
    |\Psi_{h_{Q}} \cap \Psi_{\run} \cap \Psi_{\centerset}(C_{Q}) \cap \Psi_{\preceding} & \cap (\bigcup_{\lambda = 1}^{x_{t}-1} \Psi_{\nRecover}(\lambda)) \cap (\bigcup_{\lambda = 0}^{K_{t} - 1} \Psi_{\lcp}(\lambda))| \\
    &= \rangesum(\mathcal{J}_{E}(h_{Q}, C_{Q}), 1, x_{t}-1, 0, |C_{Q}|-1).
    \end{split}
\end{equation}

Finally, Lemma~\ref{lem:GammaE1_sub_property}~\ref{enum:GammaE1_sub_property:3} follows from Equation~\ref{eq:GammaE1_sub_property:2}, Equation~\ref{eq:GammaE1_sub_property:3}, Equation~\ref{eq:GammaE1_sub_property:4}, Equation~\ref{eq:GammaE1_sub_property:5}, and Equation~\ref{eq:GammaE1_sub_property:6}.
\end{proof}

\subsubsection{Algorithm}\label{subsubsec:gamma_E1_algorithm}
We prove Lemma~\ref{lem:GammaE1_algorithm}, i.e., 
we show that subquery $\RSSQEX(T[i..j], b)$ can be answered 
in $O(H^{2} \log^{2} n + \log^{6} n)$ time using the data structures for RSC query 
and interval $[i, j]$. 

For answering subquery $\RSSQEX(T[i..j], b)$, 
we use the ordered set $\mathcal{T}_{D} = \{ (h_{1}, C_{1})$, $(h_{2}, C_{2})$, $\ldots$, $(h_{m}, C_{m}) \}$ of pairs introduced in Section~\ref{subsubsec:TD1_ds}. 
Let $\lambda$ be an integer in set $[1, m]$ satisfying $(h_{\lambda}, C_{\lambda}) = (h_{Q}, C_{Q})$. 
This integer $\lambda$ exists if and only if the ordered set $\mathcal{T}_{D}$ contains the pair $(h_{Q}, C_{Q})$.

\paragraph{Computation of integer $\lambda$.}
We compute the integer $\lambda$ in two phases. 
In the first phase, 
we compute interval attractor $([p_{Q}, q_{Q}], [\ell_{Q}, r_{Q}])$, 
its level $h_{Q}$, its attractor position $\gamma_{Q}$, and the length $|C_{Q}|$ of its associated string $C_{Q}$. 
The interval attractor $([p_{Q}, q_{Q}], [\ell_{Q}, r_{Q}])$ can be obtained by capture query $\CAPQ([i, j])$. 
The level $h_{Q}$ and attractor position $\gamma_{Q}$ can be obtained by 
level-query $\levelQ(([p_{Q}, q_{Q}], [\ell_{Q}, r_{Q}]))$ and attractor position query $\attrQ(([p_{Q}, q_{Q}], [\ell_{Q}, r_{Q}]))$, 
respectively. 
The length $|C_{Q}|$ can be obtained by C-length query $\clenQ(([p_{Q}, q_{Q}], [\ell_{Q}, r_{Q}]))$. 
Therefore, the first phase takes $O(H^{2} \log n)$ time. 

In the second phase, 
we compute the integer $\lambda$ using the query of Lemma~\ref{lem:TD1_queries}~\ref{enum:TD1_queries:3}. 
For answering this query, 
we need to know an interval $[\beta, \beta + |C_{Q}| - 1]$ in input string $T$ satisfying $T[\beta..\beta + |C_{Q}| - 1] = C_{Q}$. 
$C_{Q} = T[\gamma_{Q}..\gamma_{Q} + |C_{Q}| - 1]$ follows from the definition of the associated string $C_{Q}$. 
Therefore, the query of Lemma~\ref{lem:TD1_queries}~\ref{enum:TD1_queries:3} can be answered in $O(H^{2} \log n + \log^{2} n)$ time. 

The three phases take $O(H^{2} \log n + \log^{2} n)$ time in total. 
Therefore, the integer $\lambda$ can be computed in $O(H^{2} \log n + \log^{2} n)$ time.

\paragraph{Accessing the dynamic data structures for two sets $\mathcal{J}_{D}(h_{\lambda}, C_{\lambda})$ and $\mathcal{J}_{E}(h_{\lambda}, C_{\lambda})$.}
For answering subquery $\RSSQEX(T[i..j], b)$, 
we use the dynamic data structures for two sets $\mathcal{J}_{D}(h_{\lambda}, C_{\lambda})$ and $\mathcal{J}_{E}(h_{\lambda}, C_{\lambda})$ introduced in Section~\ref{subsubsec:JD1_ds}. 
For accessing theses dynamic data structures, 
we leverage the doubly linked list of $m$ elements for the ordered set $\mathcal{T}_{D}$, which is introduced in Section~\ref{subsubsec:TD1_ds}. 
The $\lambda$-th element of this doubly linked list stores a pointer to 
the dynamic data structures for the two sets $\mathcal{J}_{D}(h_{\lambda}, C_{\lambda})$ and $\mathcal{J}_{E}(h_{\lambda}, C_{\lambda})$. 
We can access the $\lambda$-th element in $O(\log m)$ time by the list indexing data structure built on the doubly linked list for the ordered set $\mathcal{T}_{D}$. 
Here, $m = O(n^{2})$ follows from Lemma~\ref{lem:TD1_size}~\ref{enum:TD1_size:3}.
Therefore, we can access the dynamic data structures for two sets $\mathcal{J}_{D}(h_{\lambda}, C_{\lambda})$ and $\mathcal{J}_{E}(h_{\lambda}, C_{\lambda})$ in $O(\log n)$ time if we know the integer $\lambda$. 

\paragraph{Computation of each integer $u_{t}$ in sequence $\Gamma_{E}$.}
Consider the non-increasing sequence $\Gamma_{E} = u_{1}, u_{2}, \ldots, u_{n-\theta-1} \in \{ 0, 1 \}$. 
We show that the $t$-th integer $u_{t}$ can be computed in $O(H^{2} \log n + \log^{5} n)$ time for a given integer $t \in [1, n-\theta-1]$. 

We leverage sequence $\Gamma_{E, \sub} = \alpha_{1}$, $\alpha_{2}$, $\ldots$, $\alpha_{n-\theta-1}$ for computing the $t$-th integer $u_{t}$. 
Let $t^{\prime}$ be the largest integer in set $[t, n-\theta-1]$ satisfying $\alpha_{t} = \alpha_{t^{\prime}}$. 
Then, 
Lemma~\ref{lem:GammaE1_sub_property}~\ref{enum:GammaE1_sub_property:2} shows that 
$u_{t} = 1 \Leftrightarrow (\alpha_{t^{\prime}} \geq 1) \land (\mathcal{F}_{\SA} \cap \mathcal{F}_{\suffix}(\Psi_{\CCP}(T[i..j]) \cap \Psi_{\run} \cap \Psi_{\centerset}(C_{Q}) \cap \Psi_{\lcp}(t^{\prime} + \theta - 1) \cap \Psi_{\preceding}) \neq \emptyset)$ holds. 
Therefore, the $t$-th integer $u_{t}$ can be computed using the $t^{\prime}$-th integer $\alpha_{t^{\prime}}$ of the sequence $\Gamma_{E, \sub}$ and set $\mathcal{F}_{\SA} \cap \mathcal{F}_{\suffix}(\Psi_{\CCP}(T[i..j]) \cap \Psi_{\run} \cap \Psi_{\centerset}(C_{Q}) \cap \Psi_{\lcp}(t^{\prime} + \theta - 1) \cap \Psi_{\preceding})$. 

The algorithm computing the $t$-th integer $u_{t}$ consists of three phases. 
In the first phase, we compute the two integers $\lambda$, $\theta$, and the length $|C_{Q}|$ of the associated string of $([p_{Q}, q_{Q}], [\ell_{Q}, r_{Q}])$. 
The computation of the integer $\lambda$ takes $O(H^{2} \log n + \log^{2} n)$ time. 
The integer $\theta$ can be computed in $O(H)$ time using the level $h_{Q}$ and attractor position $\gamma_{Q}$ 
of the interval attractor $([p_{Q}, q_{Q}], [\ell_{Q}, r_{Q}])$. 
The level $h_{Q}$ and attractor position $\gamma_{Q}$ can be obtained by the algorithm computing the integer $\lambda$. 
Similarly, the length $|C_{Q}|$ can be obtained by the algorithm computing the integer $\lambda$. 
Therefore, this phase takes $O(H^{2} \log n + \log^{2} n)$ time. 

In the second phase, 
we find the integer $t^{\prime}$ by binary search on the sequence $\Gamma_{E, \sub}$. 
This binary search can be executed by computing $O(\log (n-\theta-1))$ integers of the sequence $\Gamma_{E, \sub}$ 
because Lemma~\ref{lem:GammaE1_sub_property}~\ref{enum:GammaE1_sub_property:1} shows that 
the sequence $\Gamma_{E, \sub}$ is non-increasing. 
Lemma~\ref{lem:GammaE1_sub_property}~\ref{enum:GammaE1_sub_property:3} shows that 
each integer $\alpha_{b}$ of the sequence $\Gamma_{E, \sub}$ can be computed by four range-sum queries 
on the two sets $\mathcal{J}_{D}(h_{Q}, C_{Q})$ and $\mathcal{J}_{E}(h_{Q}, C_{Q})$ of weighted points. 
The input of the four range-sum queries contains the two integers $x_{b}$ and $M_{b}$ introduced in 
Lemma~\ref{lem:GammaE1_sub_property}~\ref{enum:GammaE1_sub_property:3}. 
The two integers $x_{b}$ and $M_{b}$ can be computed in $O(H)$ time using three integers $\theta$, $h_{Q}$, and $|C_{Q}|$. 

If the integer $\lambda$ does not exist, 
$\mathcal{J}_{D}(h_{Q}, C_{Q}) = \emptyset$ follows from the definition of the ordered set $\mathcal{T}_{D}$. 
Similarly, $\mathcal{J}_{E}(h_{Q}, C_{Q}) = \emptyset$ holds because $|\mathcal{J}_{D}(h_{Q}, C_{Q})| = |\mathcal{J}_{E}(h_{Q}, C_{Q})|$. 
In this case, $\alpha_{b} = 0$ 
because any range-sum query returns $0$ on the two sets $\mathcal{J}_{D}(h_{Q}, C_{Q})$ and $\mathcal{J}_{E}(h_{Q}, C_{Q})$. 
Otherwise (i.e., the integer $\lambda$ exists), 
we can support range-sum query on the two sets $\mathcal{J}_{D}(h_{Q}, C_{Q})$ and $\mathcal{J}_{E}(h_{Q}, C_{Q})$ 
by the dynamic data structures for the two sets $\mathcal{J}_{D}(h_{\lambda}, C_{\lambda})$ and $\mathcal{J}_{E}(h_{\lambda}, C_{\lambda})$. 
Accessing these dynamic data structures takes $O(\log n)$ time, 
and each range-sum query takes $O(\log^{4} k)$ time 
for the number $k$ of weighted points in the set $\mathcal{J}_{D}(h_{\lambda}, C_{\lambda})$.
$k = O(n^{2})$ follows from Lemma~\ref{lem:TD1_size}~\ref{enum:TD1_size:4}. 
Therefore, the third phase takes $O(H + \log^{5} n)$ time. 

In the third phase, 
we compute the $t$-th integer $u_{t}$ of sequence $\Gamma_{E}$ by verifying 
$\alpha_{t^{\prime}} \geq 1$ and 
$\mathcal{F}_{\SA} \cap \mathcal{F}_{\suffix}(\Psi_{\CCP}(T[i..j]) \cap \Psi_{\run} \cap \Psi_{\centerset}(C_{Q}) \cap \Psi_{\lcp}(t^{\prime} + \theta - 1) \cap \Psi_{\preceding}) \neq \emptyset$. 
Here, the integer $\alpha_{t^{\prime}}$ is obtained by the binary search of the second phase. 
We verify whether $\mathcal{F}_{\SA} \cap \mathcal{F}_{\suffix}(\Psi_{\CCP}(T[i..j]) \cap \Psi_{\run} \cap \Psi_{\centerset}(C_{Q}) \cap \Psi_{\lcp}(t^{\prime} + \theta - 1) \cap \Psi_{\preceding}) \neq \emptyset$ or not by 
the first query of Lemma~\ref{lem:GammaD1_algorithm}. 
This query takes $O(H^{2} + \log^{4} n)$ time. 
Therefore, the third phase takes $O(H^{2} + \log^{4} n)$ time. 

Finally, the algorithm computing the $t$-th integer $u_{t}$ takes $O(H^{2} \log n + \log^{5} n)$ time in total. 

\paragraph{Computation of the largest integer $\kappa$.}
Consider the largest integer $\kappa$ in set $[1, n-\theta+1]$ satisfying $u_{\kappa} = 1$. 
We find the largest integer $\kappa$ by binary search on the non-increasing sequence $\Gamma_{E}$. 
This binary search can be executed by computing $O(\log (n-\theta+1))$ integers of the sequence $\Gamma_{E}$. 
Each integer of the sequence $\Gamma_{E}$ can be computed in $O(H^{2} \log n + \log^{5} n)$ time. 
Therefore, this binary search takes $O((H^{2} \log n + \log^{5} n) \log (n-\theta+1))$ time 
(i.e., $O(H^{2} \log^{2} n + \log^{6} n)$ time).

\paragraph{Verification of $\mathcal{F}_{\SA} \cap \mathcal{F}_{\suffix}(\Psi_{\CCP}(T[i..j]) \cap \Psi_{\run} \cap \Psi_{\centerset}(C_{Q}) \cap \Psi_{\preceding}) = \emptyset$.} 
We verify whether $\mathcal{F}_{\SA} \cap \mathcal{F}_{\suffix}(\Psi_{\CCP}(T[i..j]) \cap \Psi_{\run} \cap \Psi_{\centerset}(C_{Q}) \cap \Psi_{\preceding}) = \emptyset$ or not 
for answering subquery $\RSSQEX(T[i..j], b)$. 
From Lemma~\ref{lem:GammaE1_property}~\ref{enum:GammaE1_property:1},  
$\mathcal{F}_{\SA} \cap \mathcal{F}_{\suffix}(\Psi_{\CCP}(T[i..j]) \cap \Psi_{\run} \cap \Psi_{\centerset}(C_{Q}) \cap \Psi_{\lcp}(\hat{K}) \cap \Psi_{\preceding}) = \emptyset$ holds if and only if 
the largest integer $\kappa$ exists. 
We can verify whether the largest integer $\kappa$ exists or not by the algorithm computing the largest integer $\kappa$. 
Therefore, the verification of $\mathcal{F}_{\SA} \cap \mathcal{F}_{\suffix}(\Psi_{\CCP}(T[i..j]) \cap \Psi_{\run} \cap \Psi_{\centerset}(C_{Q}) \cap \Psi_{\preceding}) = \emptyset$ 
takes $O(H^{2} \log^{2} n + \log^{6} n)$ time.

\paragraph{Algorithm for subquery $\RSSQEX(T[i..j], b)$.}
The algorithm for subquery $\RSSQEX(T[i..j], b)$ 
returns the lexicographically largest string $F$ in 
set $\mathcal{F}_{\SA} \cap \mathcal{F}_{\suffix}(\Psi_{\CCP}(T[i..j]) \cap \Psi_{\run} \cap \Psi_{\centerset}(C_{Q}) \cap \Psi_{\preceding})$. 
This algorithm is executed only if $\mathcal{F}_{\SA} \cap \mathcal{F}_{\suffix}(\Psi_{\CCP}(T[i..j]) \cap \Psi_{\run} \cap \Psi_{\centerset}(C_{Q}) \cap \Psi_{\preceding}) \neq \emptyset$ holds; 
otherwise, subquery $\RSSQEX(T[i..j], b)$ can be answered by verifying whether $\mathcal{F}_{\SA} \cap \mathcal{F}_{\suffix}(\Psi_{\CCP}(T[i..j]) \cap \Psi_{\run} \cap \Psi_{\centerset}(C_{Q}) \cap \Psi_{\preceding}) = \emptyset$ or not in $O(H^{2} \log^{2} n + \log^{6} n)$ time. 

The algorithm for subquery $\RSSQEX(T[i..j], b)$ leverages the largest integer $\kappa$. 
Consider the lexicographically largest string $F^{\prime}$ in set $\mathcal{F}_{\SA} \cap \mathcal{F}_{\suffix}(\Psi_{\CCP}(T[i..j]) \cap \Psi_{\run} \cap \Psi_{\centerset}(C_{Q}) \cap \Psi_{\lcp}(\kappa + \theta - 1) \cap \Psi_{\preceding})$. 
Then, Lemma~\ref{lem:GammaE1_property}~\ref{enum:GammaE1_property:1} shows that 
the string $F^{\prime}$ is the lexicographically largest string in set $\mathcal{F}_{\SA} \cap \mathcal{F}_{\suffix}(\Psi_{\CCP}(T[i..j]) \cap \Psi_{h_{Q}} \cap \Psi_{\run} \cap \Psi_{\centerset}(C_{Q}) \cap \Psi_{\preceding})$. 
We can return a string $T[g^{\prime}..g^{\prime} + |F^{\prime}| -1]$ in input string $T$ 
satisfying $T[g^{\prime}..g^{\prime} + |F^{\prime}| -1] = F^{\prime}$ as the answer to subquery $\RSSQEX(T[i..j], b)$. 
The string $[g^{\prime}, g^{\prime} + |F^{\prime}| -1]$ can be computed by 
subquery $\RSSQDX(T[i..j], b, \kappa + \theta - 1)$
if we know the two integers $\kappa$ and $\theta$. 

The algorithm for subquery $\RSSQEX(T[i..j], b)$ consists of three steps.
In the first step, 
we compute the two integers $\kappa$ and $\theta$ in $O(H^{2} \log^{2} n + \log^{6} n)$ time. 
In the second step, 
we compute string $T[g^{\prime}..g^{\prime} + |F^{\prime}| -1]$ by $\RSSQDX(T[i..j], b, \kappa + \theta - 1)$ in $O(H^{2} \log^{2} n + \log^{5} n)$ time. 
Here, the string is represented as interval $[g^{\prime}, g^{\prime} + |F^{\prime}| -1]$. 
In the third step, 
we return string $T[g^{\prime}..g^{\prime} + |F^{\prime}| -1]$ as the answer to subquery $\RSSQEX(T[i..j], b)$. 

The three steps takes $O(H^{2} \log^{2} n + \log^{6} n)$ time in total. 
Therefore, Lemma~\ref{lem:GammaE1_algorithm} holds.

\subsection{Subquery \texorpdfstring{$\RSSQEY(T[i..j], b)$}{RSSE2(T[i..j], b)}}\label{subsec:GammaE2}
The goal of this subsection is to solve subquery $\RSSQEY(T[i..j], b)$ 
under the assumption that 
(i) either $\mathcal{C}_{\run} = \emptyset$ or $C_{Q} = C_{\max}$ holds, 
and (ii) condition (B) of RSS query is satisfied. 

\begin{lemma}\label{lem:GammaE2_algorithm}
We assume that (i) either $\mathcal{C}_{\run} = \emptyset$ or $C_{Q} = C_{\max}$ holds, and (ii) condition (B) of RSS query is satisfied. 
We can answer $\RSSQEY(T[i..j], b)$ in $O(H^{2} \log^{2} n + \log^{6} n)$ time 
using (A) the data structures for RSC query, 
(B) interval $[i, j]$, 
and (C) the starting position $\eta$ of the sa-interval $[\eta, \eta^{\prime}]$ of $T[i..j]$. 
If the subquery returns a string $F$, 
then $F$ is represented as an interval $[g, g + |F| - 1]$ satisfying $T[g..g + |F| - 1] = F$. 
\end{lemma}
\begin{proof}
See Section~\ref{subsubsec:gamma_E2_algorithm}.
\end{proof}

We leverage the two sets 
$\mathcal{J}_{D^{\prime}}(h_{Q}, C_{Q})$ and $\mathcal{J}_{E^{\prime}}(h_{Q}, C_{Q})$ of weighted points introduced in Section~\ref{subsec:RSC_comp_D1}. 
Here, the weighted points of the two sets $\mathcal{J}_{D^{\prime}}(h_{Q}, C_{Q})$ and $\mathcal{J}_{E^{\prime}}(h_{Q}, C_{Q})$ 
are located on grid $([1, n], [-1, |C_{Q}|])$ (see Section~\ref{subsec:RSC_comp_D2} for the details of the set and ordered set).
In the next paragraphs, 
we introduce two sequences of integers to explain the relationship among 
$\RSSQEY(T[i..j], b)$ and the two sets $\mathcal{J}_{D^{\prime}}(h_{Q}, C_{Q})$ and $\mathcal{J}_{E^{\prime}}(h_{Q}, C_{Q})$. 

\paragraph{Two sequences $\Gamma_{E^{\prime}}$ and $\Gamma_{E^{\prime}, \sub}$.}
The sequence $\Gamma_{E^{\prime}}$ consists of $|[1, n - \theta + 1]|$ integers $u_{1}$, $u_{2}$, $\ldots$, $u_{n - \theta + 1} \in \{ 0, 1 \}$.  
Each integer $u_{t}$ is $1$ if the following equation holds: 
\begin{equation}\label{eq:GammaE2:1}
    \begin{split}
    \mathcal{F}_{\SA} \cap \mathcal{F}_{\suffix}(\Psi_{\CCP}(T[i..j]) \cap \Psi_{h_{Q}} \cap \Psi_{\run} \cap \Psi_{\centerset}(C_{Q}) \cap (\bigcup_{\lambda = \theta}^{t + \theta - 1} \Psi_{\lcp}(\lambda)) \cap \Psi_{\succeeding}) \neq \emptyset.
    \end{split}
\end{equation}
Otherwise (i.e., Equation~\ref{eq:GammaE2:1} does not hold),  
$u_{t}$ is defined as $0$. 
The following lemma states four properties of sequence $\Gamma_{E^{\prime}}$. 

\begin{lemma}\label{lem:GammaE2_property}
Consider condition (B) of RSS query for the given RSS query $\RSSQ(T[i..j], b)$.
Let $\kappa$ be the smallest integer in set $[1, n - \theta + 1]$ satisfying $u_{\kappa} = 1$ for sequence $\Gamma_{E^{\prime}} = u_{1}, u_{2}, \ldots, u_{n - \theta + 1}$. 
Then, the following four statements hold: 
\begin{enumerate}[label=\textbf{(\roman*)}]
    \item \label{enum:GammaE2_property:1} 
    let $F^{\prime}$ be the lexicographically largest string in set $\mathcal{F}_{\SA} \cap \mathcal{F}_{\suffix}(\Psi_{\CCP}(T[i..j]) \cap \Psi_{\run} \cap \Psi_{\centerset}(C_{Q}) \cap \Psi_{\lcp}(\kappa + \theta - 1) \cap \Psi_{\succeeding})$. 
    If the largest integer $\kappa$ exists, then 
    the string $F^{\prime}$ is the lexicographically largest string in set $\mathcal{F}_{\SA} \cap \mathcal{F}_{\suffix}(\Psi_{\CCP}(T[i..j]) \cap \Psi_{h_{Q}} \cap \Psi_{\run} \cap \Psi_{\centerset}(C_{Q}) \cap \Psi_{\succeeding})$. 
    Otherwise, the set $\mathcal{F}_{\SA} \cap \mathcal{F}_{\suffix}(\Psi_{\CCP}(T[i..j]) \cap \Psi_{\run} \cap \Psi_{\centerset}(C_{Q}) \cap \Psi_{\succeeding})$ is empty;
    \item \label{enum:GammaE2_property:2}
    sequence $\Gamma_{E^{\prime}}$ is non-decreasing~(i.e., $u_{1} \leq u_{2} \leq \cdots \leq u_{n - \theta + 1}$). 
\end{enumerate}
\end{lemma}
\begin{proof}
See Section~\ref{subsubsec:GammaE2_property_proof}.
\end{proof}

Next, sequence $\Gamma_{E^{\prime}, \sub}$ consists of $|[1, n - \theta + 1]|$ integers 
$\alpha_{1}, \alpha_{2}, \ldots, \alpha_{d} \in \mathbb{N}_{0}$. 
Each integer $\alpha_{t}$ is defined as follows: 
\begin{equation*}
    \alpha_{t} = |\Psi_{h_{Q}} \cap \Psi_{\run} \cap \Psi_{\centerset}(C_{Q}) \cap (\bigcup_{\lambda = \theta}^{t + \theta - 1} \Psi_{\lcp}(\lambda)) \cap \Psi_{\succeeding}|.
\end{equation*}
The following lemma states three properties of sequence $\Gamma_{E^{\prime}, \sub}$. 

\begin{lemma}\label{lem:GammaE2_sub_property}
Consider condition (B) of RSS query for the given RSS query $\RSSQ(T[i..j], b)$.
The following three statements hold for two sequences $\Gamma_{E} = u_{1}, u_{2}, \ldots, u_{n-\theta+1}$ and $\Gamma_{E, \sub} = \alpha_{1}$, $\alpha_{2}$, $\ldots$, $\alpha_{n-\theta+1}$: 
\begin{enumerate}[label=\textbf{(\roman*)}]
    \item \label{enum:GammaE2_sub_property:1} 
    sequence $\Gamma_{E, \sub}$ is non-decreasing (i.e., $\alpha_{1} \leq \alpha_{2} \leq \cdots \leq \alpha_{n-\theta+1}$);    
    \item \label{enum:GammaE2_sub_property:2} 
    for an integer $t \in [1, n-\theta+1]$, 
    consider the smallest integer $t^{\prime}$ in set $[1, t]$ satisfying 
    $\alpha_{t} = \alpha_{t^{\prime}}$. 
    Then, 
    $u_{t} = 1 \Leftrightarrow (\alpha_{t^{\prime}} \geq 1) \land (\mathcal{F}_{\SA} \cap \mathcal{F}_{\suffix}(\Psi_{\CCP}(T[i..j]) \cap \Psi_{\run} \cap \Psi_{\centerset}(C_{Q}) \cap \Psi_{\lcp}(t^{\prime} + \theta - 1) \cap \Psi_{\succeeding}) \neq \emptyset)$;
    \item \label{enum:GammaE2_sub_property:3}
    the following equation holds for each integer $t \in [1, n - \theta + 1]$: 
    \begin{equation*}
    \begin{split}
    \alpha_{t} &= (x_{t}-1) \rangesum(\mathcal{J}_{D^{\prime}}(h_{Q}, C_{Q}), x_{t}, n, 0, |C_{Q}| - 1) \\
    &+ \rangesum(\mathcal{J}_{D^{\prime}}(h_{Q}, C_{Q}), x_{t}, n, 0, M_{t} - 1) \\
    &+ \rangesum(\mathcal{J}_{E^{\prime}}(h_{Q}, C_{Q}), 1, x_{t}-1, 0, |C_{Q}|-1) \\
    &- (x^{\prime}_{t}-1) \rangesum(\mathcal{J}_{D^{\prime}}(h_{Q}, C_{Q}), x^{\prime}_{t}, n, 0, |C_{Q}| - 1) \\
    &- \rangesum(\mathcal{J}_{D^{\prime}}(h_{Q}, C_{Q}), x^{\prime}_{t}, n, 0, M^{\prime}_{t} - 1) \\
    &- \rangesum(\mathcal{J}_{E^{\prime}}(h_{Q}, C_{Q}), 1, x^{\prime}_{t}-1, 0, |C_{Q}|-1).
    \end{split}
    \end{equation*}
    Here, $\rangesum$ is the range-sum query introduced in Section~\ref{subsec:range_data_structure}; 
    the four integers $x_{t}, M_{t}, x^{\prime}_{t}$, and $M^{\prime}_{t}$ are defined as follows: 
    \begin{itemize}
    \item $x_{t} = 1 + \lfloor \frac{(t + \theta) - (2 + \sum_{w = 1}^{h_{Q}+3} \lfloor \mu(w) \rfloor)}{|C_{Q}|} \rfloor$;
    \item $M_{t} = ((t + \theta) - (2 + \sum_{w = 1}^{h_{Q}+3} \lfloor \mu(w) \rfloor) ) \mod |C_{Q}|$;
    \item $x^{\prime} = 1 + \lfloor \frac{\theta - (2 + \sum_{w = 1}^{h_{Q}+3} \lfloor \mu(w) \rfloor)}{|C_{Q}|} \rfloor$;
    \item $M^{\prime}_{t} = (\theta - (2 + \sum_{w = 1}^{h_{Q}+3} \lfloor \mu(w) \rfloor) ) \mod |C_{Q}|$.    
    \end{itemize}
\end{enumerate}
\end{lemma}
\begin{proof}
See Section~\ref{subsubsec:GammaE2_sub_property_proof}.
\end{proof}

\subsubsection{Proof of Lemma~\ref{lem:GammaE2_property}}\label{subsubsec:GammaE2_property_proof}
The following proposition states properties of 
two sets $\Psi_{h_{Q}} \cap \Psi_{\run} \cap \Psi_{\centerset}(C_{Q}) \cap \Psi_{\lcp}(t + \theta - 1) \cap \Psi_{\succeeding}$ 
and $\Psi_{\CCP}(T[i..j]) \cap \Psi_{\run} \cap \Psi_{\centerset}(C_{Q}) \cap \Psi_{\lcp}(t + \theta - 1) \cap \Psi_{\succeeding}$ for each integer $t \in [1, n - \theta + 1]$. 

\begin{proposition}\label{prop:Psi_E2_Property}
Consider condition (B) of RSS query for the given RSS query $\RSSQ(T[i..j], b)$.
The following two statements hold: 
\begin{enumerate}[label=\textbf{(\roman*)}]
    \item \label{enum:Psi_E2_Property:1}
    $\Psi_{h_{Q}} \cap \Psi_{\run} \cap \Psi_{\centerset}(C_{Q}) \cap \Psi_{\lcp}(t + \theta - 1) \cap \Psi_{\succeeding} \subseteq \Psi_{\CCP}(T[i..j])$ for each integer $t \in [1, n - \theta + 1]$;     
    \item \label{enum:Psi_E2_Property:2}
    consider two integers $t$ and $t^{\prime}$ satisfying $1 \leq t < t^{\prime} \leq n - \theta + 1$. 
    Then, $F^{\prime} \prec F$ holds for any pair of two strings $F \in \mathcal{F}_{\suffix}(\Psi_{\CCP}(T[i..j]) \cap \Psi_{\run} \cap \Psi_{\centerset}(C_{Q}) \cap \Psi_{\lcp}(t + \theta - 1) \cap \Psi_{\succeeding})$ 
    and $F^{\prime} \in \mathcal{F}_{\suffix}(\Psi_{\CCP}(T[i..j]) \cap \Psi_{\run} \cap \Psi_{\centerset}(C_{Q}) \cap \Psi_{\lcp}(t^{\prime} + \theta - 1) \cap \Psi_{\succeeding})$.
\end{enumerate}
\end{proposition}
\begin{proof}
The proof of Proposition~\ref{prop:Psi_E2_Property} is as follows. 

\textbf{Proof of Proposition~\ref{prop:Psi_E2_Property}(i).}
Proposition~\ref{prop:Psi_E2_Property}(i) can be proved using the same approach for Proposition~\ref{prop:Psi_E1_Property}(i).

\textbf{Proof of Proposition~\ref{prop:Psi_E2_Property}(ii).}
From the definition of the set $\mathcal{F}_{\suffix}(\Psi_{\CCP}(T[i..j]) \cap \Psi_{\run} \cap \Psi_{\centerset}(C_{Q}) \cap \Psi_{\lcp}(t + \theta - 1) \cap \Psi_{\succeeding})$, 
set $\Psi_{\CCP}(T[i..j]) \cap \Psi_{\run} \cap \Psi_{\centerset}(C_{Q}) \cap \Psi_{\lcp}(t + \theta - 1) \cap \Psi_{\succeeding})$ contains 
an interval attractor $([p, q], [\ell, r])$ satisfying 
$T[\gamma - |[i, \gamma_{Q}-1]|..\gamma-1] \cdot T[\gamma..r+1] = F$ for the attractor position $\gamma$ of the interval attractor $([p, q], [\ell, r])$. 
Similarly, 
set $\Psi_{\CCP}(T[i..j]) \cap \Psi_{\run} \cap \Psi_{\centerset}(C_{Q}) \cap \Psi_{\lcp}(t^{\prime} + \theta - 1) \cap \Psi_{\succeeding})$ contains 
an interval attractor $([p^{\prime}, q^{\prime}], [\ell^{\prime}, r^{\prime}])$ satisfying 
$T[\gamma^{\prime} - |[i, \gamma_{Q}-1]|..\gamma^{\prime}-1] \cdot T[\gamma^{\prime}..r^{\prime}+1] = F^{\prime}$ for the attractor position $\gamma^{\prime}$ of the interval attractor $([p^{\prime}, q^{\prime}], [\ell^{\prime}, r^{\prime}])$. 

We prove $T[\gamma - |[i, \gamma_{Q}-1]|..\gamma-1] = T[\gamma^{\prime} - |[i, \gamma_{Q}-1]|..\gamma^{\prime}-1]$. 
Because of $([p, q], [\ell, r]) \in \Psi_{\CCP}(T[i..j])$, 
Lemma~\ref{lem:CCP_property}~\ref{enum:CCP_property:6} shows that 
$T[\gamma - |[i, \gamma_{Q}-1]|..\gamma-1] = T[i..\gamma_{Q}]$ holds. 
Similarly, 
Lemma~\ref{lem:CCP_property}~\ref{enum:CCP_property:6} shows that 
$T[\gamma^{\prime} - |[i, \gamma_{Q}-1]|..\gamma^{\prime}-1] = T[i..\gamma_{Q}]$ holds. 
Therefore, $T[\gamma - |[i, \gamma_{Q}-1]|..\gamma-1] = T[\gamma^{\prime} - |[i, \gamma_{Q}-1]|..\gamma^{\prime}-1]$ holds. 

We prove $T[\gamma^{\prime}..r^{\prime}+1] \prec T[\gamma..r+1]$. 
We can apply Lemma~\ref{lem:psi_LMPS_property}~\ref{enum:psi_LMPS_property:succeeding:1} to the two interval attractors 
$([p, q], [\ell, r])$ and $([p^{\prime}, q^{\prime}], [\ell^{\prime}, r^{\prime}])$ 
because $([p, q], [\ell, r]) \in \Psi_{\run} \cap \Psi_{\centerset}(C_{Q}) \cap \Psi_{\lcp}(t + \theta - 1) \cap \Psi_{\succeeding}$ 
and $([p^{\prime}, q^{\prime}], [\ell^{\prime}, r^{\prime}]) \in \Psi_{\run} \cap \Psi_{\centerset}(C_{Q}) \cap \Psi_{\lcp}(t^{\prime} + \theta - 1) \cap \Psi_{\succeeding}$. 
Because of $t + \theta - 1 < t^{\prime} + \theta - 1$, 
Lemma~\ref{lem:psi_LMPS_property}~\ref{enum:psi_LMPS_property:preceding:1} shows that 
$T[\gamma^{\prime}..r^{\prime}+1] \prec T[\gamma..r+1]$ holds. 

Finally, $F^{\prime} \prec F$ follows from 
(1) $T[\gamma - |[i, \gamma_{Q}-1]|..\gamma-1] \cdot T[\gamma..r+1] = F$, 
(2) $T[\gamma^{\prime} - |[i, \gamma_{Q}-1]|..\gamma^{\prime}-1] \cdot T[\gamma^{\prime}..r^{\prime}+1] = F^{\prime}$, 
(3) $T[\gamma - |[i, \gamma_{Q}-1]|..\gamma-1] = T[\gamma^{\prime} - |[i, \gamma_{Q}-1]|..\gamma^{\prime}-1]$, 
and (4) $T[\gamma^{\prime}..r^{\prime}+1] \prec T[\gamma..r+1]$. 
\end{proof}

For proving Lemma~\ref{lem:GammaE2_property}, 
we introduce a set $\mathcal{I}^{E^{\prime}}$ of integers in set $\{ 1, 2, \ldots, n - \theta + 1 \}$. 
This set $\mathcal{I}^{E^{\prime}}$ consists of integers such that 
each integer $t \in \mathcal{I}^{E^{\prime}}$ satisfies 
$\mathcal{F}_{\SA} \cap \mathcal{F}_{\suffix}(\Psi_{\CCP}(T[i..j]) \cap \Psi_{\run} \cap \Psi_{\centerset}(C_{Q}) \cap \Psi_{\lcp}(t + \theta - 1) \cap \Psi_{\succeeding}) \neq \emptyset$. 
Formally, the set $\mathcal{I}^{E^{\prime}}$ is defined as follows:
\begin{equation*}
    \begin{split}
    \mathcal{I}^{E^{\prime}} &= \{ t \in [1, n - \theta + 1] \mid \\
    & \mathcal{F}_{\SA} \cap \mathcal{F}_{\suffix}(\Psi_{\CCP}(T[i..j]) \cap \Psi_{\run} \cap \Psi_{\centerset}(C_{Q}) \cap \Psi_{\lcp}(t + \theta - 1) \cap \Psi_{\succeeding}) \neq \emptyset \}.
    \end{split}
\end{equation*}

The following proposition states three properties of the set $\mathcal{I}^{E^{\prime}}$. 

\begin{proposition}\label{prop:Set_IE2_Property}
    Consider condition (B) of RSS query for the given RSS query $\RSSQ(T[i..j], b)$.
    The following three statements hold for set $\mathcal{I}^{E^{\prime}}$ and sequence $\Gamma_{E^{\prime}} = u_{1}, u_{2}, \ldots, u_{n - \theta + 1}$: 
\begin{enumerate}[label=\textbf{(\roman*)}]
    \item \label{enum:Set_IE2_Property:1} 
    the following equation holds: 
    \begin{equation*}
    \begin{split}
    & \mathcal{F}_{\SA} \cap \mathcal{F}_{\suffix}(\Psi_{\CCP}(T[i..j]) \cap \Psi_{\run} \cap \Psi_{\centerset}(C_{Q}) \cap \Psi_{\succeeding}) \\
    &= \mathcal{F}_{\SA} \cap (\bigcup_{t \in \mathcal{I}^{E^{\prime}}} \mathcal{F}_{\suffix}(\Psi_{\CCP}(T[i..j]) \cap \Psi_{\run} \cap \Psi_{\centerset}(C_{Q}) \cap \Psi_{\lcp}(t + \theta - 1) \cap \Psi_{\succeeding}));
    \end{split}
\end{equation*}    
    \item \label{enum:Set_IE2_Property:2} $t \in \mathcal{I}^{E^{\prime}}$ for each integer $t \in [1, n - \theta + 1]$ satisfying $u_{t-1} = 0$ and $u_{t} = 1$;
    \item \label{enum:Set_IE2_Property:3} $u_{t^{\prime}} = 1$ for 
    any pair of two integers $t \in \mathcal{I}^{E^{\prime}}$ and $t^{\prime} \in [t, n - \theta + 1]$. 
\end{enumerate}
Here, let $u_{0} = 0$ for simplicity. 
\end{proposition}
\begin{proof}
    The proof of Proposition~\ref{prop:Set_IE2_Property} is as follows.

    \textbf{Proof of Proposition~\ref{prop:Set_IE2_Property}(i).}
    Proposition~\ref{prop:Set_IE2_Property}~\ref{enum:Set_IE2_Property:1} can be proved using the same approach as for Proposition~\ref{prop:Set_IE1_Property}~\ref{enum:Set_IE1_Property:1}.

    \textbf{Proof of Proposition~\ref{prop:Set_IE2_Property}(ii).}
    Because of $u_{t} = 1$, 
    the following equation follows from the definition of sequence $\Gamma_{E^{\prime}}$: 
    \begin{equation}\label{eq:Set_IE2_Property:1}
        \begin{split}
        \mathcal{F}_{\SA} \cap \mathcal{F}_{\suffix}(\Psi_{\CCP}(T[i..j]) \cap \Psi_{h_{Q}} \cap \Psi_{\run} \cap \Psi_{\centerset}(C_{Q}) \cap (\bigcup_{\lambda = \theta}^{t + \theta - 1} \Psi_{\lcp}(\lambda)) \cap \Psi_{\succeeding}) \neq \emptyset.
        \end{split}
    \end{equation}
    If $t = 1$, 
    then $\mathcal{F}_{\SA} \cap \mathcal{F}_{\suffix}(\Psi_{\CCP}(T[i..j]) \cap \Psi_{\run} \cap \Psi_{\centerset}(C_{Q}) \cap \Psi_{\lcp}(t + \theta - 1) \cap \Psi_{\succeeding}) \neq \emptyset$ follows from Equation~\ref{eq:Set_IE2_Property:1}. 
    Therefore, $t \in \mathcal{I}^{E^{\prime}}$ holds. 

    Otherwise (i.e., $t > 1$), 
    we leverage the $(t-1)$-th integer $u_{t-1}$ of sequence $\Gamma_{E^{\prime}}$.     
    Because of $u_{t-1} = 0$, 
    the following equation follows from the definition of sequence $\Gamma_{E^{\prime}}$: 
    \begin{equation}\label{eq:Set_IE2_Property:2}
        \begin{split}
        \mathcal{F}_{\SA} \cap \mathcal{F}_{\suffix}(\Psi_{\CCP}(T[i..j]) \cap \Psi_{h_{Q}} \cap \Psi_{\run} \cap \Psi_{\centerset}(C_{Q}) \cap (\bigcup_{\lambda = \theta}^{t + \theta} \Psi_{\lcp}(\lambda)) \cap \Psi_{\succeeding}) = \emptyset.
        \end{split}
    \end{equation}
    $\mathcal{F}_{\SA} \cap \mathcal{F}_{\suffix}(\Psi_{\CCP}(T[i..j]) \cap \Psi_{h_{Q}} \cap \Psi_{\run} \cap \Psi_{\centerset}(C_{Q}) \cap \Psi_{\lcp}(t + \theta - 1) \cap \Psi_{\succeeding}) \neq \emptyset$ follows from Equation~\ref{eq:Set_IE2_Property:1} 
    and Equation~\ref{eq:Set_IE2_Property:2}. 
    Therefore, $t \in \mathcal{I}^{E}$ follows from the definition of the set $\mathcal{I}^{E}$. 

    \textbf{Proof of Proposition~\ref{prop:Set_IE2_Property}(iii).}
    Proposition~\ref{prop:Set_IE2_Property}~\ref{enum:Set_IE2_Property:3} follows from 
    the definitions of sequence $\gamma_{E^{\prime}}$ and set $\mathcal{I}^{E^{\prime}}$.    
\end{proof}

We prove Lemma~\ref{lem:GammaE2_property} using Proposition~\ref{prop:Psi_E2_Property} and Proposition~\ref{prop:Set_IE2_Property}.

\begin{proof}[Proof of Lemma~\ref{lem:GammaE2_property}~\ref{enum:GammaE2_property:1}]
Lemma~\ref{lem:GammaE2_property}~\ref{enum:GammaE1_property:1} can be proved using the same approach as for 
Lemma \ref{lem:GammaE1_property} \ref{enum:GammaE1_property:1}. 
The detailed proof of Lemma~\ref{lem:GammaE2_property}~\ref{enum:GammaE2_property:1} is as follows. 

Consider the integers $t_{1}, t_{2}, \ldots, t_{m}$ ($t_{1} < t_{2} < \ldots < t_{m}$) in set $\mathcal{I}^{E^{\prime}}$. 
Let $F_{\tau}$ be the lexicographically largest string in set 
$\mathcal{F}_{\SA} \cap \mathcal{F}_{\suffix}(\Psi_{\CCP}(T[i..j]) \cap \Psi_{\run} \cap \Psi_{\centerset}(C_{Q}) \cap \Psi_{\lcp}(t_{\tau} + \theta - 1) \cap \Psi_{\succeeding})$ for each integer $\tau \in [1, m]$. 
From the definition of the set $\mathcal{I}^{E^{\prime}}$, 
each string $F_{\tau}$ exists. 

If the integer $\kappa$ exists, 
then $\kappa = t_{1}$ follows from 
Proposition~\ref{prop:Set_IE2_Property}~\ref{enum:Set_IE2_Property:2} and Proposition~\ref{prop:Set_IE2_Property}~\ref{enum:Set_IE2_Property:3}. 
Proposition~\ref{prop:Set_IE2_Property}~\ref{enum:Set_IE2_Property:1} indicates that 
the lexicographically largest string in set $\mathcal{F}_{\SA} \cap \mathcal{F}_{\suffix}(\Psi_{\CCP}(T[i..j]) \cap \Psi_{\run} \cap \Psi_{\centerset}(C_{Q}) \cap \Psi_{\succeeding})$ is equal to the lexicographically largest string in set $\{ F_{1}, F_{2}, \ldots, F_{m} \}$. 
Proposition~\ref{prop:Psi_E2_Property}~\ref{enum:Psi_E2_Property:2} indicates that 
$F_{m} \prec F_{m-1} \prec \ldots \prec F_{1}$ holds. 
$F^{\prime} = F_{1}$ follows from $\kappa = t_{1}$. 
Therefore, the string $F^{\prime}$ is the lexicographically largest string in set $\mathcal{F}_{\SA} \cap \mathcal{F}_{\suffix}(\Psi_{\CCP}(T[i..j]) \cap \Psi_{\run} \cap \Psi_{\centerset}(C_{Q}) \cap \Psi_{\succeeding})$. 

Otherwise~(i.e., the integer $\kappa$ does not exist), 
Proposition~\ref{prop:Set_IE2_Property}~\ref{enum:Set_IE2_Property:3} indicates that 
$\mathcal{I}^{E^{\prime}} = \emptyset$ holds (i.e., $m = 0$).
In this case, 
Proposition~\ref{prop:Set_IE2_Property}~\ref{enum:Set_IE2_Property:1} shows that 
$\mathcal{F}_{\SA} \cap \mathcal{F}_{\suffix}(\Psi_{\CCP}(T[i..j]) \cap \Psi_{\run} \cap \Psi_{\centerset}(C_{Q}) \cap \Psi_{\succeeding}) = \emptyset$ holds. 
Therefore, Lemma~\ref{lem:GammaE2_property}~\ref{enum:GammaE2_property:1} holds. 
\end{proof}

\begin{proof}[Proof of Lemma~\ref{lem:GammaE2_property}~\ref{enum:GammaE2_property:2}]
Let $u_{0} = 0$ for simplicity. 
We prove Lemma~\ref{lem:GammaE2_property}~\ref{enum:GammaE2_property:2} by contradiction. 
We assume that Lemma~\ref{lem:GammaE2_property}~\ref{enum:GammaE2_property:2} does not hold. 
Then, there exists an integer $t \in [2, n - \theta + 1]$ satisfying 
$u_{t-1} = 1$ and $u_{t} = 0$. 
Because of $u_{0} = 0$, 
there exists an integer $t^{\prime} \in [1, t-1]$ 
satisfying $u_{t^{\prime}-1} = 0$ and $u_{t^{\prime}} = 1$. 
Because of $u_{t^{\prime}-1} = 0$ and $u_{t^{\prime}} = 1$, 
Proposition~\ref{prop:Set_IE2_Property}~\ref{enum:Set_IE2_Property:2} shows that 
$t^{\prime} \in \mathcal{I}^{E^{\prime}}$ holds. 
Because of $t^{\prime} \in \mathcal{I}^{E^{\prime}}$, 
Proposition~\ref{prop:Set_IE2_Property}~\ref{enum:Set_IE2_Property:3} shows that 
$u_{t} = 1$ holds. 
The two facts $u_{t} = 0$ and $u_{t} = 1$ yield a contradiction. 
Therefore, Lemma~\ref{lem:GammaE2_property}~\ref{enum:GammaE2_property:2} must hold. 

\end{proof}

\subsubsection{Proof of Lemma~\ref{lem:GammaE2_sub_property}}\label{subsubsec:GammaE2_sub_property_proof}
We prove Lemma~\ref{lem:GammaE2_sub_property} using Proposition~\ref{prop:Psi_E2_Property} and Proposition~\ref{prop:Set_IE2_Property}.

\begin{proof}[Proof of Lemma~\ref{lem:GammaE2_sub_property}~\ref{enum:GammaE2_sub_property:1}]
Lemma~\ref{lem:GammaE2_sub_property}~\ref{enum:GammaE2_sub_property:1} can be proved using the same approach as for 
Lemma \ref{lem:GammaE1_sub_property} \ref{enum:GammaE1_sub_property:1}. 
The detailed proof of Lemma~\ref{lem:GammaE2_sub_property}~\ref{enum:GammaE2_sub_property:1} is as follows. 

We prove $\alpha_{t} \leq \alpha_{t+1}$ for each integer $t \in [1, n-\theta]$. 
Lemma~\ref{lem:psi_LMPS_property}~\ref{enum:psi_LMPS_property:lcp:2} shows that 
$\Psi_{\lcp}(\lambda) \cap \Psi_{\lcp}(\lambda^{\prime}) = \emptyset$ holds 
for any pair of two integers $0 \leq \lambda < \lambda^{\prime} \leq n$. 
Therefore, following equation holds: 
\begin{equation*}
    \begin{split}
    |\Psi_{h_{Q}} \cap \Psi_{\run} & \cap \Psi_{\centerset}(C_{Q}) \cap (\bigcup_{\lambda = t + \theta - 1}^{n} \Psi_{\lcp}(\lambda)) \cap \Psi_{\succeeding}| \\
    &= |\Psi_{h_{Q}} \cap \Psi_{\run} \cap \Psi_{\centerset}(C_{Q}) \cap \Psi_{\lcp}(t + \theta - 1) \cap \Psi_{\succeeding}| \\
    &+ |\Psi_{h_{Q}} \cap \Psi_{\run} \cap \Psi_{\centerset}(C_{Q}) \cap (\bigcup_{\lambda = t + \theta}^{n} \Psi_{\lcp}(\lambda)) \cap \Psi_{\succeeding}|.
    \end{split}
\end{equation*}

$\alpha_{t} \leq \alpha_{t+1}$ follows from the following equation:
\begin{equation}\label{eq:GammaE2_sub_property:1}
    \begin{split}
    \alpha_{t+1} &= |\Psi_{h_{Q}} \cap \Psi_{\run} \cap \Psi_{\centerset}(C_{Q}) \cap (\bigcup_{\lambda = \theta}^{t + \theta} \Psi_{\lcp}(\lambda)) \cap \Psi_{\succeeding}| \\
    &= |\Psi_{h_{Q}} \cap \Psi_{\run} \cap \Psi_{\centerset}(C_{Q}) \cap \Psi_{\lcp}(t + \theta) \cap \Psi_{\succeeding}| \\
    &+ |\Psi_{h_{Q}} \cap \Psi_{\run} \cap \Psi_{\centerset}(C_{Q}) \cap (\bigcup_{\lambda = \theta}^{t + \theta - 1} \Psi_{\lcp}(\lambda)) \cap \Psi_{\succeeding}| \\
    &= |\Psi_{h_{Q}} \cap \Psi_{\run} \cap \Psi_{\centerset}(C_{Q}) \cap \Psi_{\lcp}(t + \theta) \cap \Psi_{\succeeding}| + \alpha_{t} \\
    &\geq \alpha_{t}.
    \end{split}
\end{equation}

Therefore, $\alpha_{1} \leq \alpha_{2} \leq \cdots \leq \alpha_{n-\theta+1}$ holds.

\end{proof}

\begin{proof}[Proof of Lemma~\ref{lem:GammaE2_sub_property}~\ref{enum:GammaE2_sub_property:2}]
Let $\alpha_{0} = 0$ for simplicity. 
The following three statements are used to prove Lemma~\ref{lem:GammaE2_sub_property}~\ref{enum:GammaE2_sub_property:2}:
\begin{enumerate}[label=\textbf{(\Alph*)}]
    \item $u_{t} = 1 \Leftarrow (\alpha_{t^{\prime}} \geq 1) \land (\mathcal{F}_{\SA} \cap \mathcal{F}_{\suffix}(\Psi_{\CCP}(T[i..j]) \cap \Psi_{\run} \cap \Psi_{\centerset}(C_{Q}) \cap \Psi_{\lcp}(t^{\prime} + \theta - 1) \cap \Psi_{\succeeding}) \neq \emptyset)$;
    \item 
    if $u_{t} = 1$, then 
    $t^{\prime} = t_{A}$ and $\alpha_{t^{\prime}} \geq 1$ for the largest integer $t_{A}$ in set $[1, t]$ 
    satisfying $\mathcal{F}_{\SA} \cap \mathcal{F}_{\suffix}(\Psi_{\CCP}(T[i..j]) \cap \Psi_{h_{Q}} \cap \Psi_{\run} \cap \Psi_{\centerset}(C_{Q}) \cap \Psi_{\lcp}(t_{A} + \theta - 1) \cap \Psi_{\succeeding}) \neq \emptyset$;
    \item $u_{t} = 1 \Rightarrow (\alpha_{t^{\prime}} \geq 1) \land (\mathcal{F}_{\SA} \cap \mathcal{F}_{\suffix}(\Psi_{\CCP}(T[i..j]) \cap \Psi_{\run} \cap \Psi_{\centerset}(C_{Q}) \cap \Psi_{\lcp}(t^{\prime} + \theta - 1) \cap \Psi_{\succeeding}) \neq \emptyset)$.
\end{enumerate}

\textbf{Proof of statement (A).}
Consider a string $F$ in set $\mathcal{F}_{\SA} \cap \mathcal{F}_{\suffix}(\Psi_{\CCP}(T[i..j]) \cap \Psi_{\run} \cap \Psi_{\centerset}(C_{Q}) \cap \Psi_{\lcp}(t^{\prime} + \theta - 1) \cap \Psi_{\succeeding})$. 
Because of $F \in \mathcal{F}_{\suffix}(\Psi_{\CCP}(T[i..j]) \cap \Psi_{\run} \cap \Psi_{\centerset}(C_{Q}) \cap \Psi_{\lcp}(t^{\prime} + \theta - 1) \cap \Psi_{\succeeding})$, 
set $\Psi_{\CCP}(T[i..j]) \cap \Psi_{\run} \cap \Psi_{\centerset}(C_{Q}) \cap \Psi_{\lcp}(t^{\prime} + \theta - 1) \cap \Psi_{\succeeding}$ contains an interval attractor $([p, q], [\ell, r])$ satisfying 
$T[\gamma - |[i, \gamma_{Q}-1]|..r+1] = F$ for the attractor position $\gamma$ of the interval attractor $([p, q], [\ell, r])$. 
The existence of this interval attractor $([p, q], [\ell, r])$ indicates that 
Equation~\ref{eq:GammaE2:1} holds. Therefore, $u_{t} = 1$ holds. 

\textbf{Proof of statement (B).}
Because of $u_{t} = 1$, 
Equation~\ref{eq:GammaE2:1} indicates that the largest integer $t_{A}$ exists. 
For an integer $\tau \in [t_{A}, t]$, 
we prove $\alpha_{\tau} = \alpha_{t}$ by contradiction. 
Here, $\alpha_{\tau} \leq \alpha_{t}$ follows from Lemma~\ref{lem:GammaE2_sub_property}~\ref{enum:GammaE2_sub_property:1}. 
We assume that $\alpha_{\tau} \neq \alpha_{t}$ holds. 
Then, $\alpha_{\tau} < \alpha_{t}$ follows from $\alpha_{\tau} \neq \alpha_{t}$ and $\alpha_{\tau} \leq \alpha_{t}$. 
In this case, $\tau < t$ holds, 
and there exists an integer $\tau^{\prime} \in [\tau + 1, t]$ satisfying 
$\alpha_{\tau^{\prime}-1} < \alpha_{\tau^{\prime}}$. 
We apply Equation~\ref{eq:GammaE2_sub_property:1} to the two integers $\alpha_{\tau^{\prime}-1}$ and $\alpha_{\tau^{\prime}}$. 
Then, this equation indicates that 
$\alpha_{\tau^{\prime}} = |\Psi_{h_{Q}} \cap \Psi_{\run} \cap \Psi_{\centerset}(C_{Q}) \cap \Psi_{\lcp}(\tau^{\prime} + \theta - 1) \cap \Psi_{\succeeding}| + \alpha_{\tau^{\prime}-1}$ holds. 
$\Psi_{h_{Q}} \cap \Psi_{\run} \cap \Psi_{\centerset}(C_{Q}) \cap \Psi_{\lcp}(\tau^{\prime} + \theta - 1) \cap \Psi_{\succeeding} \neq \emptyset$ follows from $\alpha_{\tau^{\prime}-1} < \alpha_{\tau^{\prime}}$ and $\alpha_{\tau^{\prime}} = |\Psi_{h_{Q}} \cap \Psi_{\run} \cap \Psi_{\centerset}(C_{Q}) \cap \Psi_{\lcp}(\tau^{\prime} + \theta - 1) \cap \Psi_{\succeeding}| + \alpha_{\tau^{\prime}-1}$. 
Let $([p, q], [\ell, r])$ be an interval attractor in the set $\Psi_{h_{Q}} \cap \Psi_{\run} \cap \Psi_{\centerset}(C_{Q}) \cap \Psi_{\lcp}(\tau^{\prime} + \theta - 1) \cap \Psi_{\succeeding}$. 
Then, $([p, q], [\ell, r]) \in \Psi_{\CCP}(T[i..j])$ follows from Proposition~\ref{prop:Psi_E2_Property}~\ref{enum:Psi_E2_Property:1}. 
Let $F = T[\gamma - |[i, \gamma_{Q}-1]|..\gamma-1] \cdot T[\gamma..r+1]$ for simplicity. 
Because of $([p, q], [\ell, r]) \in \Psi_{\CCP}(T[i..j]) \cap \Psi_{h_{Q}} \cap \Psi_{\run} \cap \Psi_{\centerset}(C_{Q}) \cap \Psi_{\lcp}(\tau^{\prime} + \theta - 1) \cap \Psi_{\succeeding}$, 
$F \in \mathcal{F}_{\suffix}(\Psi_{\CCP}(T[i..j]) \cap \Psi_{h_{Q}} \cap \Psi_{\run} \cap \Psi_{\centerset}(C_{Q}) \cap \Psi_{\lcp}(\tau^{\prime} + \theta - 1) \cap \Psi_{\succeeding})$ follows from the definition of the subset $\mathcal{F}_{\suffix}(\Psi_{\CCP}(T[i..j]) \cap \Psi_{h_{Q}} \cap \Psi_{\run} \cap \Psi_{\centerset}(C_{Q}) \cap \Psi_{\lcp}(\tau^{\prime} + \theta - 1) \cap \Psi_{\succeeding})$. 

We prove $F \in \mathcal{F}_{\SA}$. 
Because of $\mathcal{F}_{\SA} \cap \mathcal{F}_{\suffix}(\Psi_{\CCP}(T[i..j]) \cap \Psi_{h_{Q}} \cap \Psi_{\run} \cap \Psi_{\centerset}(C_{Q}) \cap \Psi_{\lcp}(t_{A} + \theta - 1) \cap \Psi_{\succeeding}) \neq \emptyset$, 
the set $\mathcal{F}_{\SA} \cap \mathcal{F}_{\suffix}(\Psi_{\CCP}(T[i..j]) \cap \Psi_{h_{Q}} \cap \Psi_{\run} \cap \Psi_{\centerset}(C_{Q}) \cap \Psi_{\lcp}(t_{A} + \theta - 1) \cap \Psi_{\succeeding})$ contains a string $F^{\prime}$. 
Because of $t_{A} < \tau^{\prime}$, 
Proposition~\ref{prop:Psi_E2_Property}~\ref{enum:Psi_E2_Property:2} shows that 
$F \prec F^{\prime}$ holds. 
Therefore, $F \in \mathcal{F}_{\SA}$ follows from Lemma~\ref{lem:F_suffix_basic_property}~\ref{enum:F_suffix_basic_property:6}, 
$F \prec F^{\prime}$, and $F^{\prime} \in \mathcal{F}_{\SA}$. 

$\mathcal{F}_{\SA} \cap \mathcal{F}_{\suffix}(\Psi_{\CCP}(T[i..j]) \cap \Psi_{h_{Q}} \cap \Psi_{\run} \cap \Psi_{\centerset}(C_{Q}) \cap \Psi_{\lcp}(\tau^{\prime} + \theta - 1) \cap \Psi_{\succeeding}) \neq \emptyset$ follows from $F \in \mathcal{F}_{\SA} \cap \mathcal{F}_{\suffix}(\Psi_{\CCP}(T[i..j]) \cap \Psi_{h_{Q}} \cap \Psi_{\run} \cap \Psi_{\centerset}(C_{Q}) \cap \Psi_{\lcp}(\tau^{\prime} + \theta - 1) \cap \Psi_{\succeeding})$. 
On the other hand, $\mathcal{F}_{\SA} \cap \mathcal{F}_{\suffix}(\Psi_{\CCP}(T[i..j]) \cap \Psi_{h_{Q}} \cap \Psi_{\run} \cap \Psi_{\centerset}(C_{Q}) \cap \Psi_{\lcp}(\tau^{\prime} + \theta - 1) \cap \Psi_{\succeeding}) = \emptyset$ holds 
because $t_{A} < \tau^{\prime}$. 
The two facts $\mathcal{F}_{\SA} \cap \mathcal{F}_{\suffix}(\Psi_{\CCP}(T[i..j]) \cap \Psi_{h_{Q}} \cap \Psi_{\run} \cap \Psi_{\centerset}(C_{Q}) \cap \Psi_{\lcp}(\tau^{\prime} + \theta - 1) \cap \Psi_{\succeeding}) \neq \emptyset$ and $\mathcal{F}_{\SA} \cap \mathcal{F}_{\suffix}(\Psi_{\CCP}(T[i..j]) \cap \Psi_{h_{Q}} \cap \Psi_{\run} \cap \Psi_{\centerset}(C_{Q}) \cap \Psi_{\lcp}(\tau^{\prime} + \theta - 1) \cap \Psi_{\succeeding}) = \emptyset$ yield a contradiction. 
Therefore, $\alpha_{\tau} = \alpha_{t}$ must hold. 

Next, we prove $\alpha_{t_{A}} \geq 1$ and $\alpha_{t_{A}-1} < \alpha_{t_{A}}$. 
Because of $\mathcal{F}_{\suffix}(\Psi_{\CCP}(T[i..j]) \cap \Psi_{h_{Q}} \cap \Psi_{\run} \cap \Psi_{\centerset}(C_{Q}) \cap \Psi_{\lcp}(t_{A} + \theta - 1) \cap \Psi_{\succeeding}) \neq \emptyset$, 
set $\Psi_{\CCP}(T[i..j]) \cap \Psi_{h_{Q}} \cap \Psi_{\run} \cap \Psi_{\centerset}(C_{Q}) \cap \Psi_{\lcp}(t_{A} + \theta - 1) \cap \Psi_{\succeeding}$ contains an interval attractor $([p^{\prime}, q^{\prime}], [\ell^{\prime}, r^{\prime}])$. 
If $\tau_{A} = 1$, 
then $\alpha_{t_{A}} = |\Psi_{h_{Q}} \cap \Psi_{\run} \cap \Psi_{\centerset}(C_{Q}) \cap \Psi_{\lcp}(\theta) \cap \Psi_{\succeeding}|$ 
and $\alpha_{t_{A}-1} = 0$ follows from the definitions of the two integers $\alpha_{t_{A}-1}$ and $\alpha_{t_{A}}$. 
$\alpha_{t_{A}} \geq 1$ 
because $([p^{\prime}, q^{\prime}], [\ell^{\prime}, r^{\prime}]) \in \Psi_{h_{Q}} \cap \Psi_{\run} \cap \Psi_{\centerset}(C_{Q}) \cap \Psi_{\lcp}(1) \cap \Psi_{\succeeding}$. 
Therefore, $\alpha_{t_{A}} \geq 1$ and $\alpha_{t_{A}-1} < \alpha_{t_{A}}$ hold. 

Otherwise (i.e., $\tau_{A} > 1$), 
we apply Equation~\ref{eq:GammaE2_sub_property:1} to the two integers $\alpha_{\tau_{A}-1}$ and $\alpha_{\tau_{A}}$. 
Then, this equation indicates that 
$\alpha_{\tau_{A}} = |\Psi_{h_{Q}} \cap \Psi_{\run} \cap \Psi_{\centerset}(C_{Q}) \cap \Psi_{\lcp}(\tau_{A} + \theta - 1) \cap \Psi_{\succeeding}| + \alpha_{\tau_{A}-1}$ holds. 
$|\Psi_{h_{Q}} \cap \Psi_{\run} \cap \Psi_{\centerset}(C_{Q}) \cap \Psi_{\lcp}(\tau_{A} + \theta - 1) \cap \Psi_{\succeeding}| \geq 1$ holds 
because $([p^{\prime}, q^{\prime}], [\ell^{\prime}, r^{\prime}]) \in \Psi_{h_{Q}} \cap \Psi_{\run} \cap \Psi_{\centerset}(C_{Q}) \cap \Psi_{\lcp}(\tau_{A} + \theta - 1) \cap \Psi_{\succeeding}$. 
Therefore, $\alpha_{t_{A}} \geq 1$ and $\alpha_{t_{A}} > \alpha_{t_{A}+1}$ hold. 

We prove statement (B). 
$t^{\prime} = t_{A}$ holds because 
we showed that $\alpha_{t_{A}} = \alpha_{t_{A}+1} = \cdots = \alpha_{t}$ and $\alpha_{t_{A}-1} < \alpha_{t_{A}}$. 
$\alpha_{t^{\prime}} \geq 1$ follows from $t^{\prime} = t_{A}$ and $\alpha_{t_{A}} \geq 1$. 
Therefore, statement (B) holds. 

\textbf{Proof of statement (C).}
This statement follows from statement (B),

\textbf{Proof of Lemma~\ref{lem:GammaE2_sub_property}~\ref{enum:GammaE2_sub_property:2}.}
Lemma~\ref{lem:GammaE2_sub_property}~\ref{enum:GammaE2_sub_property:2} follows from statement (A) and statement (C).
\end{proof}

\begin{proof}[Proof of Lemma~\ref{lem:GammaE2_sub_property}~\ref{enum:GammaE2_sub_property:3}]
Let $K_{t} = t + \theta$ for simplicity. 
Lemma~\ref{lem:psi_LMPS_property}~\ref{enum:psi_LMPS_property:lcp:2} shows that 
$\Psi_{\lcp}(\lambda) \cap \Psi_{\lcp}(\lambda^{\prime}) = \emptyset$ holds 
for any pair of two integers $0 \leq \lambda < \lambda^{\prime} \leq n$. 
Therefore, the following equation holds:
\begin{equation}\label{eq:GammaE2_sub_property:2}
    \begin{split}
    \alpha_{t} &= |\Psi_{h_{Q}} \cap \Psi_{\run} \cap \Psi_{\centerset}(C_{Q}) \cap (\bigcup_{\lambda = \theta}^{K_{t}-1} \Psi_{\lcp}(\lambda)) \cap \Psi_{\succeeding}| \\
    &= |\Psi_{h_{Q}} \cap \Psi_{\run} \cap \Psi_{\centerset}(C_{Q}) \cap (\bigcup_{\lambda = 0}^{K_{t}-1} \Psi_{\lcp}(\lambda)) \cap \Psi_{\succeeding}| \\
    &- |\Psi_{h_{Q}} \cap \Psi_{\run} \cap \Psi_{\centerset}(C_{Q}) \cap (\bigcup_{\lambda = 0}^{\theta - 1} \Psi_{\lcp}(\lambda)) \cap \Psi_{\succeeding}|.
    \end{split}
\end{equation}

Lemma~\ref{lem:nRecover_basic_property}~\ref{enum:nRecover_basic_property:1} indicates that 
set $\Psi_{h_{Q}} \cap \Psi_{\run} \cap \Psi_{\centerset}(C_{Q}) \cap (\bigcup_{\lambda = 0}^{K_{t} - 1} \Psi_{\lcp}(\lambda)) \cap \Psi_{\succeeding}$ is equal to set $\Psi_{h_{Q}} \cap \Psi_{\run} \cap \Psi_{\centerset}(C_{Q}) \cap (\bigcup_{\lambda = 1}^{n} \Psi_{\nRecover}(\lambda)) \cap (\bigcup_{\lambda = 0}^{K_{t} - 1} \Psi_{\lcp}(\lambda)) \cap \Psi_{\succeeding}$. 
Lemma~\ref{lem:nRecover_basic_property}~\ref{enum:nRecover_basic_property:2} indicates that 
the set $\Psi_{h_{Q}} \cap \Psi_{\run} \cap \Psi_{\centerset}(C_{Q}) \cap (\bigcup_{\lambda = 1}^{n} \Psi_{\nRecover}(\lambda)) \cap (\bigcup_{\lambda = 0}^{K_{t} - 1} \Psi_{\lcp}(\lambda)) \cap \Psi_{\succeeding}$ can be divided into two sets 
$\Psi_{h_{Q}} \cap \Psi_{\run} \cap \Psi_{\centerset}(C_{Q}) \cap \Psi_{\succeeding} \cap (\bigcup_{\lambda = x_{t}}^{n} \Psi_{\nRecover}(\lambda)) \cap (\bigcup_{\lambda = 0}^{K_{t} - 1} \Psi_{\lcp}(\lambda))$ and $\Psi_{h_{Q}} \cap \Psi_{\run} \cap \Psi_{\centerset}(C_{Q}) \cap \Psi_{\succeeding} \cap (\bigcup_{\lambda = 1}^{x_{t}-1} \Psi_{\nRecover}(\lambda)) \cap (\bigcup_{\lambda = 0}^{K_{t} - 1} \Psi_{\lcp}(\lambda))$. 
The set $\Psi_{h_{Q}} \cap \Psi_{\run} \cap \Psi_{\centerset}(C_{Q}) \cap \Psi_{\succeeding} \cap (\bigcup_{\lambda = x_{t}}^{n} \Psi_{\nRecover}(\lambda)) \cap (\bigcup_{\lambda = 0}^{K_{t} - 1} \Psi_{\lcp}(\lambda))$ can be divided into 
two sets $\Psi_{h_{Q}} \cap \Psi_{\run} \cap \Psi_{\centerset}(C_{Q}) \cap \Psi_{\succeeding} \cap (\bigcup_{\lambda = x_{t}}^{n} \Psi_{\nRecover}(\lambda)) \cap (\bigcup_{\lambda = 0}^{K_{t} - M_{t} - 1} \Psi_{\lcp}(\lambda))$ 
and $\Psi_{h_{Q}} \cap \Psi_{\run} \cap \Psi_{\centerset}(C_{Q}) \cap \Psi_{\succeeding} \cap (\bigcup_{\lambda = x_{t}}^{n} \Psi_{\nRecover}(\lambda)) \cap (\bigcup_{\lambda = K_{t} - M_{t}}^{K_{t} - 1} \Psi_{\lcp}(\lambda))$. 
Equation~\ref{eq:JD2_sum:2} of Lemma~\ref{lem:JD2_sum} indicates that 
the number of interval attractors in the set $\Psi_{h_{Q}} \cap \Psi_{\run} \cap \Psi_{\centerset}(C_{Q}) \cap \Psi_{\succeeding} \cap (\bigcup_{\lambda = x_{t}}^{n} \Psi_{\nRecover}(\lambda)) \cap (\bigcup_{\lambda = 0}^{K_{t} - M_{t} - 1} \Psi_{\lcp}(\lambda))$ can be computed by range-sum query $(x_{t}-1) \rangesum(\mathcal{J}_{D^{\prime}}(h_{Q}, C_{Q}), x_{t}, n, 0, |C_{Q}| - 1)$. 
Equation~\ref{eq:JD2_sum:3} of Lemma~\ref{lem:JD2_sum} indicates that 
the number of interval attractors in the set $\Psi_{h_{Q}} \cap \Psi_{\run} \cap \Psi_{\centerset}(C_{Q}) \cap \Psi_{\succeeding} \cap (\bigcup_{\lambda = x_{t}}^{n} \Psi_{\nRecover}(\lambda)) \cap (\bigcup_{\lambda = K_{t} - M_{t}}^{K_{t} - 1} \Psi_{\lcp}(\lambda))$ can be computed by range-sum query $\rangesum(\mathcal{J}_{D^{\prime}}(h_{Q}, C_{Q}), x_{t}, n, 0, M_{t} - 1)$. 
Equation~\ref{eq:JD2_sum:4} of Lemma~\ref{lem:JD2_sum} indicates that 
the number of interval attractors in the set $\Psi_{h_{Q}} \cap \Psi_{\run} \cap \Psi_{\centerset}(C_{Q}) \cap \Psi_{\succeeding} \cap (\bigcup_{\lambda = 1}^{x_{t}-1} \Psi_{\nRecover}(\lambda)) \cap (\bigcup_{\lambda = 0}^{K_{t} - 1} \Psi_{\lcp}(\lambda))$ can be computed by range-sum query $\rangesum(\mathcal{J}_{E^{\prime}}(h_{Q}, C_{Q}), 1, x_{t}-1, 0, |C_{Q}|-1)$. 
Therefore, the following equation holds: 
\begin{equation}\label{eq:GammaE2_sub_property:3}
    \begin{split}
    |\Psi_{h_{Q}} & \cap \Psi_{\run} \cap \Psi_{\centerset}(C_{Q}) \cap (\bigcup_{\lambda = 0}^{K_{t} - 1} \Psi_{\lcp}(\lambda)) \cap \Psi_{\succeeding}| \\
    &= |\Psi_{h_{Q}} \cap \Psi_{\run} \cap \Psi_{\centerset}(C_{Q}) \cap (\bigcup_{\lambda = 1}^{n} \Psi_{\nRecover}(\lambda)) \cap (\bigcup_{\lambda = 0}^{K_{t} - 1} \Psi_{\lcp}(\lambda)) \cap \Psi_{\succeeding}| \\
    &= |\Psi_{h_{Q}} \cap \Psi_{\run} \cap \Psi_{\centerset}(C_{Q}) \cap \Psi_{\succeeding} \cap (\bigcup_{\lambda = x_{t}}^{n} \Psi_{\nRecover}(\lambda)) \cap (\bigcup_{\lambda = 0}^{K_{t} - 1} \Psi_{\lcp}(\lambda))| \\
    &+ |\Psi_{h_{Q}} \cap \Psi_{\run} \cap \Psi_{\centerset}(C_{Q}) \cap \Psi_{\succeeding} \cap (\bigcup_{\lambda = 1}^{x_{t}-1} \Psi_{\nRecover}(\lambda)) \cap (\bigcup_{\lambda = 0}^{K_{t} - 1} \Psi_{\lcp}(\lambda))| \\
    &= |\Psi_{h_{Q}} \cap \Psi_{\run} \cap \Psi_{\centerset}(C_{Q}) \cap \Psi_{\succeeding} \cap (\bigcup_{\lambda = x_{t}}^{n} \Psi_{\nRecover}(\lambda)) \cap (\bigcup_{\lambda = 0}^{K_{t} - M_{t} - 1} \Psi_{\lcp}(\lambda))| \\
    &+ |\Psi_{h_{Q}} \cap \Psi_{\run} \cap \Psi_{\centerset}(C_{Q}) \cap \Psi_{\succeeding} \cap (\bigcup_{\lambda = x_{t}}^{n} \Psi_{\nRecover}(\lambda)) \cap (\bigcup_{\lambda = K_{t} - M_{t}}^{K_{t} - 1} \Psi_{\lcp}(\lambda))| \\ 
    &+ |\Psi_{h_{Q}} \cap \Psi_{\run} \cap \Psi_{\centerset}(C_{Q}) \cap \Psi_{\succeeding} \cap (\bigcup_{\lambda = 1}^{x_{t}-1} \Psi_{\nRecover}(\lambda)) \cap (\bigcup_{\lambda = 0}^{K_{t} - 1} \Psi_{\lcp}(\lambda))| \\
    &= (x_{t}-1) \rangesum(\mathcal{J}_{D^{\prime}}(h_{Q}, C_{Q}), x_{t}, n, 0, |C_{Q}| - 1) \\
    &+ \rangesum(\mathcal{J}_{D^{\prime}}(h_{Q}, C_{Q}), x_{t}, n, 0, M_{t} - 1) \\
    &+ \rangesum(\mathcal{J}_{E^{\prime}}(h_{Q}, C_{Q}), 1, x_{t}-1, 0, |C_{Q}|-1).
    \end{split}
\end{equation}

Similarly, 
set $\Psi_{h_{Q}} \cap \Psi_{\run} \cap \Psi_{\centerset}(C_{Q}) \cap (\bigcup_{\lambda = 0}^{\theta - 1} \Psi_{\lcp}(\lambda)) \cap \Psi_{\succeeding}$ can be divided into the following three sets: 
\begin{itemize}
    \item $\Psi_{h_{Q}} \cap \Psi_{\run} \cap \Psi_{\centerset}(C_{Q}) \cap \Psi_{\succeeding} \cap (\bigcup_{\lambda = x^{\prime}_{t}}^{n} \Psi_{\nRecover}(\lambda)) \cap (\bigcup_{\lambda = 0}^{\theta - M^{\prime}_{t} - 1} \Psi_{\lcp}(\lambda))$;
    \item $\Psi_{h_{Q}} \cap \Psi_{\run} \cap \Psi_{\centerset}(C_{Q}) \cap \Psi_{\succeeding} \cap (\bigcup_{\lambda = x^{\prime}_{t}}^{n} \Psi_{\nRecover}(\lambda)) \cap (\bigcup_{\lambda = \theta - M^{\prime}_{t}}^{\theta - 1} \Psi_{\lcp}(\lambda))$;
    \item $\Psi_{h_{Q}} \cap \Psi_{\run} \cap \Psi_{\centerset}(C_{Q}) \cap \Psi_{\succeeding} \cap (\bigcup_{\lambda = 1}^{x^{\prime}_{t}-1} \Psi_{\nRecover}(\lambda)) \cap (\bigcup_{\lambda = 0}^{\theta - 1} \Psi_{\lcp}(\lambda))$.
\end{itemize}
Equation~\ref{eq:JD2_sum:2} of Lemma~\ref{lem:JD2_sum} indicates that 
the number of interval attractors in the set $\Psi_{h_{Q}} \cap \Psi_{\run} \cap \Psi_{\centerset}(C_{Q}) \cap \Psi_{\succeeding} \cap (\bigcup_{\lambda = x^{\prime}_{t}}^{n} \Psi_{\nRecover}(\lambda)) \cap (\bigcup_{\lambda = 0}^{\theta - M^{\prime}_{t} - 1} \Psi_{\lcp}(\lambda))$ can be computed by range-sum query $(x^{\prime}_{t}-1) \rangesum(\mathcal{J}_{D^{\prime}}(h_{Q}, C_{Q}), x^{\prime}_{t}, n, 0, |C_{Q}| - 1)$. 
Equation~\ref{eq:JD2_sum:3} of Lemma~\ref{lem:JD2_sum} indicates that 
the number of interval attractors in the set $\Psi_{h_{Q}} \cap \Psi_{\run} \cap \Psi_{\centerset}(C_{Q}) \cap \Psi_{\succeeding} \cap (\bigcup_{\lambda = x_{t}}^{n} \Psi_{\nRecover}(\lambda)) \cap (\bigcup_{\lambda = \theta - M^{\prime}_{t}}^{\theta - 1} \Psi_{\lcp}(\lambda))$ can be computed by range-sum query $\rangesum(\mathcal{J}_{D^{\prime}}(h_{Q}, C_{Q}), x^{\prime}_{t}, n, 0, M^{\prime}_{t} - 1)$. 
Equation~\ref{eq:JD2_sum:4} of Lemma~\ref{lem:JD2_sum} indicates that 
the number of interval attractors in the set $\Psi_{h_{Q}} \cap \Psi_{\run} \cap \Psi_{\centerset}(C_{Q}) \cap \Psi_{\succeeding} \cap (\bigcup_{\lambda = 1}^{x^{\prime}_{t}-1} \Psi_{\nRecover}(\lambda)) \cap (\bigcup_{\lambda = 0}^{\theta - 1} \Psi_{\lcp}(\lambda))$ can be computed by range-sum query $\rangesum(\mathcal{J}_{E^{\prime}}(h_{Q}, C_{Q}), 1, x^{\prime}_{t}-1, 0, |C_{Q}|-1)$. 
Therefore, the following equation holds: 
\begin{equation}\label{eq:GammaE2_sub_property:4}
    \begin{split}
    |\Psi_{h_{Q}} & \cap \Psi_{\run} \cap \Psi_{\centerset}(C_{Q}) \cap (\bigcup_{\lambda = 0}^{\theta - 1} \Psi_{\lcp}(\lambda)) \cap \Psi_{\succeeding}| \\
    &= |\Psi_{h_{Q}} \cap \Psi_{\run} \cap \Psi_{\centerset}(C_{Q}) \cap \Psi_{\succeeding} \cap (\bigcup_{\lambda = x^{\prime}_{t}}^{n} \Psi_{\nRecover}(\lambda)) \cap (\bigcup_{\lambda = 0}^{\theta - M^{\prime}_{t} - 1} \Psi_{\lcp}(\lambda))| \\
    &+ |\Psi_{h_{Q}} \cap \Psi_{\run} \cap \Psi_{\centerset}(C_{Q}) \cap \Psi_{\succeeding} \cap (\bigcup_{\lambda = x^{\prime}_{t}}^{n} \Psi_{\nRecover}(\lambda)) \cap (\bigcup_{\lambda = \theta - M^{\prime}_{t}}^{\theta - 1} \Psi_{\lcp}(\lambda))| \\ 
    &+ |\Psi_{h_{Q}} \cap \Psi_{\run} \cap \Psi_{\centerset}(C_{Q}) \cap \Psi_{\succeeding} \cap (\bigcup_{\lambda = 1}^{x^{\prime}_{t}-1} \Psi_{\nRecover}(\lambda)) \cap (\bigcup_{\lambda = 0}^{\theta - 1} \Psi_{\lcp}(\lambda))| \\
    &= (x^{\prime}_{t}-1) \rangesum(\mathcal{J}_{D^{\prime}}(h_{Q}, C_{Q}), x^{\prime}_{t}, n, 0, |C_{Q}| - 1) \\
    &+ \rangesum(\mathcal{J}_{D^{\prime}}(h_{Q}, C_{Q}), x^{\prime}_{t}, n, 0, M^{\prime}_{t} - 1) \\
    &+ \rangesum(\mathcal{J}_{E^{\prime}}(h_{Q}, C_{Q}), 1, x^{\prime}_{t}-1, 0, |C_{Q}|-1).
    \end{split}
\end{equation}

Finally, Lemma~\ref{lem:GammaE2_sub_property}~\ref{enum:GammaE1_sub_property:3} follows from Equation~\ref{eq:GammaE2_sub_property:2}, Equation~\ref{eq:GammaE2_sub_property:3}, and Equation~\ref{eq:GammaE2_sub_property:4}.

\end{proof}

\subsubsection{Algorithm}\label{subsubsec:gamma_E2_algorithm}
We prove Lemma~\ref{lem:GammaE2_algorithm}, i.e., 
we show that subquery $\RSSQEY(T[i..j], b)$ can be answered 
in $O(H^{2} \log^{2} n + \log^{6} n)$ time using the data structures for RSC query 
and interval $[i, j]$. 

We answer $\RSSQEY(T[i..j], b)$ by modifying 
the algorithm answering subquery $\RSSQEX(T[i..j], b)$. 
In Section~\ref{subsubsec:gamma_E1_algorithm}, 
we presented the algorithm answering subquery $\RSSQEX(T[i..j], b)$ 
using subquery $\RSSQDX(T[i..j], b, K)$ and two sequences $\Gamma_{E}$ and $\Gamma_{E, \sub}$. 
Subquery $\RSSQDY(T[i..j], b, K)$ corresponds to $\RSSQDX(T[i..j], b, K)$. 
Two sequences $\Gamma_{E^{\prime}}$ and $\Gamma_{E^{\prime}, \sub}$ corresponds 
the two sequences $\Gamma_{E}$ and $\Gamma_{E, \sub}$, respectively. 
Lemma~\ref{lem:GammaE2_property} states properties of the sequence $\Gamma_{E^{\prime}}$. 
This lemma corresponds to Lemma~\ref{lem:GammaE1_property}, which states properties of the sequence $\Gamma_{E}$. 
Similarly, 
Lemma~\ref{lem:GammaE2_sub_property} states properties of the sequence $\Gamma_{E^{\prime}, \sub}$. 
This lemma corresponds to Lemma~\ref{lem:GammaE1_sub_property}, which states properties of the sequence $\Gamma_{E, \sub}$. 
Therefore, subquery $\RSSQEY(T[i..j], b)$ can be answered in $O(H^{2} \log^{2} n + \log^{6} n)$ time 
using the same approach as for subquery $\RSSQEX(T[i..j], b)$.

\subsection{Answering RSS query}\label{subsec:answer_rss_query}
We prove Theorem~\ref{theo:rss_query_summary}, i.e., 
we show that 
a given RSS query $\RSSQ(T[i..j], b)$ can be answered in $O(H^{2} \log^{2} n + \log^{6} n)$ time 
using (i) the data structures for RSC query, 
(ii) interval $[i, j]$, 
and (iii) the starting position $\eta$ of the sa-interval $[\eta, \eta^{\prime}]$ of $T[i..j]$. 

We already showed that 
the string obtained by the given RSS query is computed as 
the lexicographically largest string among the strings 
obtained from at most five subqueries 
$\RSSQA(T[i..j], b)$, $\RSSQCX(T[i..j], b)$, $\RSSQCY(T[i..j], b)$, $\RSSQEX(T[i..j], b)$, and $\RSSQEY(T[i..j], b)$ (Corollary~\ref{cor:RB_rss_subqueries}) 
if either (A) $\mathcal{C}_{\run} = \emptyset$ or 
(B) there exists an occurrence $T[i^{\prime}..j^{\prime}]$ of string $T[i..i]$ in input string $T$ such that 
$C^{\prime} = C_{\max}$ for the associated string of interval attractor $I_{\capture}(i^{\prime}, j^{\prime})$. 
The following lemma shows that such occurrence $T[i^{\prime}..j^{\prime}]$ always exists, 
and the interval attractor $I_{\capture}(i^{\prime}, j^{\prime})$ can be obtained by $\RSSQB(T[i..j], b)$. 

\begin{lemma}\label{lem:RSS_query_condition_E}
If $\mathcal{C}_{\run} \neq \emptyset$ and $C_{Q} \neq C_{\max}$, 
then $|[\gamma_{Q}, j]| \leq 1 + \sum_{w = 1}^{h_{Q}+3} \lfloor \mu(w) \rfloor$ holds, 
and there exists an occurrence $T[i^{\prime}..j^{\prime}]$ of string $T[i..j]$ in input string $T$ such that interval attractor $I_{\capture}(i^{\prime}, j^{\prime})$ satisfies the following two conditions: 
\begin{enumerate}[label=\textbf{(\roman*)}]
    \item \label{enum:RSS_query_condition_E:1} 
    $I_{\capture}(i^{\prime}, j^{\prime})$ is obtained by subquery $\RSSQB(T[i..j], b)$;
    \item \label{enum:RSS_query_condition_E:2}
    $[i^{\prime}, j^{\prime}] = [\gamma^{\prime} - |[i, \gamma_{Q}-1]|, \gamma^{\prime} + |[\gamma_{Q}, j]| - 1]$ for the attractor position $\gamma^{\prime}$ of 
    interval attractor $I_{\capture}(i^{\prime}, j^{\prime})$;
    \item \label{enum:RSS_query_condition_E:3}
    $C^{\prime} = C_{\max}$ for the associated string $C^{\prime}$ of interval attractor $I_{\capture}(i^{\prime}, j^{\prime})$.    
\end{enumerate}
\end{lemma}
\begin{proof}
We prove $|[\gamma_{Q}, j]| \leq 1 + \sum_{w = 1}^{h_{Q}+3} \lfloor \mu(w) \rfloor$ by contradiction. 
We assume that $|[\gamma_{Q}, j]| > 1 + \sum_{w = 1}^{h_{Q}+3} \lfloor \mu(w) \rfloor$ holds. 
Then, Lemma~\ref{lem:c_run_property} shows that 
$\mathcal{C}_{\run} = \emptyset$ or $\mathcal{C}_{\run} = \{ C_{Q} \}$ holds 
for set $\mathcal{C}_{\run}$ of strings. 
In this case, $C_{Q} = C_{\max}$ follows from the definition of the string $C_{\max}$ 
because $\mathcal{C}_{\run} \neq \emptyset$. 
On the other hand, $C_{Q} \neq C_{\max}$ follows from the premise of Lemma~\ref{lem:RSS_query_condition_E}. 
The two facts $C_{Q} = C_{\max}$ and $C_{Q} \neq C_{\max}$ yield a contradiction. 
Therefore, $|[\gamma_{Q}, j]| \leq 1 + \sum_{w = 1}^{h_{Q}+3} \lfloor \mu(w) \rfloor$ must hold. 

Since $\mathcal{C}_{\run} \neq \emptyset$, 
$C_{\max}$ is defined as a string in set $\mathcal{C}_{\run}$. 
$\Psi_{\CCP}(T[i..j]) \cap \Psi_{\run} \cap \Psi_{\centerset}(C_{\max}) \neq \emptyset$ 
follows from the definition of set $\mathcal{C}_{\run}$. 
In this case, 
subquery $\RSSQB(T[i..j], b)$ returns an interval attractor $([p^{\prime}, q^{\prime}], [\ell^{\prime}, r^{\prime}])$ in set 
$\Psi_{\CCP}(T[i..j]) \cap \Psi_{\run} \cap \Psi_{\centerset}(C_{\max})$. 
Because of $([p^{\prime}, q^{\prime}], [\ell^{\prime}, r^{\prime}]) \in \Psi_{\CCP}(T[i..j])$, 
Lemma~\ref{lem:CCP_property}~\ref{enum:CCP_property:6} shows that 
$T[i..j] = T[\gamma^{\prime} - |[i, \gamma_{Q}-1]|..\gamma^{\prime} + |[\gamma_{Q}, j]| - 1]$ 
and 
$I_{\capture}(\gamma^{\prime} - |[i, \gamma_{Q}-1]|, \gamma^{\prime} + |[\gamma_{Q}, j]| - 1) = ([p^{\prime}, q^{\prime}], [\ell^{\prime}, r^{\prime}])$ 
for the attractor position $\gamma$ of interval attractor $([p^{\prime}, q^{\prime}], [\ell^{\prime}, r^{\prime}])$. 
Since $([p^{\prime}, q^{\prime}], [\ell^{\prime}, r^{\prime}]) \in \Psi_{\centerset}(C_{\max})$, 
the associated string of this interval attractor is $C_{\max}$. 
Therefore, Lemma~\ref{lem:RSS_query_condition_E} holds. 
\end{proof}

The algorithm for RSS query consists of six phases: 
\paragraph{Phase (i).}
In the first phase, 
we compute interval attractor $I_{\capture}(i, j) = ([p_{Q}, q_{Q}], [\ell_{Q}, r_{Q}])$, 
its level $h_{Q}$, and its attractor position $\gamma_{Q}$. 
The interval attractor $([p_{Q}, q_{Q}], [\ell_{Q}, r_{Q}])$ can be obtained in $O(H^{2} \log n)$ time by capture query $\CAPQ([i, j])$. 
Since $\CAPQ([i, j]) = ([p_{Q}, q_{Q}], [\ell_{Q}, r_{Q}])$, 
$i \in [p_{Q}, q_{Q}]$ and $j \in [\ell_{Q}, r_{Q}]$ follow from the definition of interval attractor. 
The level $h_{Q}$ and attractor position $\gamma_{Q}$ can be obtained in $O(H^{2})$ time 
by level query $\levelQ(([p_{Q}, q_{Q}], [\ell_{Q}, r_{Q}]))$ 
and attractor position query $\attrQ(([p_{Q}, q_{Q}], [\ell_{Q}, r_{Q}]))$, respectively. 
Therefore, the first phase takes $O(H^{2} \log n)$ time. 

\paragraph{Phase (ii).}
In the second phase, 
we verify whether at least one of $\mathcal{C}_{\run} = \emptyset$ and 
$C_{Q} = C_{\max}$ holds or not. 
If $|[\gamma_{Q}, j]| > 1 + \sum_{w = 1}^{h_{Q}+3} \lfloor \mu(w) \rfloor$, 
then Lemma~\ref{lem:c_run_property} shows that 
$\mathcal{C}_{\run} = \emptyset$ or $\mathcal{C}_{\run} = \{ C_{Q} \}$ holds. 
In this case, either $\mathcal{C}_{\run} = \emptyset$ or $C_{Q} = C_{\max}$ holds 
because $C_{\max}$ is defined as $C_{Q}$. 

Otherwise (i.e., $|[\gamma_{Q}, j]| \leq 1 + \sum_{w = 1}^{h_{Q}+3} \lfloor \mu(w) \rfloor$), 
we can use subquery $\RSSQB(T[i..j], b)$. 
We can verify whether $\mathcal{C}_{\run} = \emptyset$ or not by the subquery. 
If $\mathcal{C}_{\run} \neq \emptyset$, 
the subquery returns an interval attractor $([p, q], [\ell, r])$ such that 
its associated string is $C_{\max}$. 
In this case, we can verify $C_{Q} = C_{\max}$ or not using C-LCP and LCE queries. 
Therefore, the second phase takes $O(H^{2} \log^{2} n + \log^{6} n)$ time. 

\paragraph{Phase (iii).}
The third phase is executed if $\mathcal{C}_{\run} \neq \emptyset$ and $C_{Q} \neq C_{\max}$. 
In the third phase, we compute the occurrence $T[i^{\prime}..j^{\prime}]$ of string $T[i..j]$ stated in 
Lemma~\ref{lem:RSS_query_condition_E} using subquery $\RSSQB(T[i..j], b)$. 
After that, 
we restart this algorithm from the beginning using the occurrence $T[i^{\prime}..j^{\prime}]$ instead of string $T[i..j]$. 
Here, Lemma~\ref{lem:RSS_query_condition_E}~\ref{enum:RSS_query_condition_E:3} ensures that 
the third step is executed only once within this algorithm. 
Therefore, the third phase takes $O(H^{2} \log^{2} n + \log^{6} n)$ time. 

\paragraph{Phase (iv).}
In the fourth phase, the integer $\hat{K} = |\lcp(T[\gamma_{Q}..j], C_{Q}^{n+1})|$. 
$\hat{K} = \min \{ |[\gamma_{Q}, j]|, |\lcp(T[\gamma_{Q}..r_{Q}]$, $C_{Q}^{n+1})| \}$ holds 
because $j \in [\ell_{Q}, r_{Q}]$. 
The integer $|\lcp(T[\gamma_{Q}..r_{Q}], C_{Q}^{n+1})|$ can be obtained in $O(H^{2})$ time 
by C-LCP query $\clcpQ(([p_{Q}, q_{Q}], [\ell_{Q}, r_{Q}]))$. 
Therefore, the fourth phase takes $O(H^{2})$ time. 

\paragraph{Phase (v).}
In the fifth phase, 
we verify whether $\lcs(T[i..\gamma_{Q}-1], C_{Q}^{n+1}) = T[i..\gamma_{Q}-1]$ or not. 
Because of $i \in [p_{Q}, q_{Q}]$, 
$\lcs(T[i..\gamma_{Q}-1], C_{Q}^{n+1}) = T[i..\gamma_{Q}-1]$ holds if and only if 
$|\lcs(T[p_{Q}-1..\gamma_{Q}-1], C_{Q}^{n+1})| \geq |[i, \gamma_{Q}-1]|$ holds. 
The integer $|\lcs(T[p_{Q}-1..\gamma_{Q}-1], C_{Q}^{n+1})|$ can be obtained in $O(H^{2})$ time by C-LCS query $\clcsQ(([p_{Q}, q_{Q}], [\ell_{Q}, r_{Q}]))$. 
Therefore, the fifth phase takes $O(H^{2})$ time. 

\paragraph{Phase (vi).}
In the sixth phase, 
we compute the string $F$ obtained by RSS query $\RSSQ(T[i..j], b)$ using Corollary~\ref{cor:RB_rss_subqueries}. 
If the given RSS query satisfies condition (A), 
then $F$ is the lexicographically largest string among the strings 
obtained from three subqueries $\RSSQA(T[i..j], b)$, $\RSSQCX(T[i..j], b)$, and $\RSSQCY(T[i..j], b)$. 
If the given RSS query satisfies condition (B), 
then $F$ is the lexicographically largest string among the strings 
obtained from three subqueries $\RSSQA(T[i..j], b)$, $\RSSQEX(T[i..j], b)$, and $\RSSQEY(T[i..j], b)$. 
If the given RSS query satisfies condition (C) or (D), 
then $F$ is the string obtained from $\RSSQA(T[i..j], b)$. 
The five subqueries $\RSSQA(T[i..j], b)$, $\RSSQCX(T[i..j], b)$, $\RSSQCY(T[i..j], b)$, $\RSSQEX(T[i..j], b)$, and $\RSSQEY(T[i..j], b)$ can be computed in $O(H^{2} \log^{2} n + \log^{6} n)$ time. 
We can sort the strings obtained from the five subqueries in lexicographic order using 
$O(1)$ LCE and RA queries 
because each string $F^{\prime}$ is represented as an interval $[g, g + |F^{\prime}|-1]$ satisfying 
$F^{\prime} = T[g..g + |F^{\prime}|-1]$. 
Therefore, the sixth phase takes $O(H^{2} \log^{2} n + \log^{6} n)$ time. 

Finally, the six phases take $O(H^{2} \log^{2} n + \log^{6} n)$ time in total. 
Therefore, Theorem~\ref{theo:rss_query_summary} holds, 
i.e., RSS query can be answered in $O(H^{2} \log^{2} n + \log^{6} n)$ time.

\section{SA and ISA Queries with Expected \texorpdfstring{$\delta$}{}-optimal Space}\label{sec:sa_and_isa_queries_with_optimal}
We already showed that SA and ISA queries can be answered using 
RSS and RSC queries (see Theorem~\ref{theo:sa_query} and Theorem~\ref{theo:isa_query}) in Section~\ref{subsec:simplified_sa_and_isa_queries}. 
This section presents two algorithms answering SA and ISA queries using RSS and RSC queries 
in expected $\delta$-optimal space. 
These algorithms requires the dynamic data structures 
provided in Section~\ref{subsubsec:rrdag_ds}, Section~\ref{subsubsec:sample_ds}), Section~\ref{subsubsec:bis_ds}, and Section~\ref{subsec:answer_rsc_query}. 

\paragraph{SA query algorithm.}
%\begin{minipage}[t]{0.43\columnwidth}
%\begin{minipage}[t]{0.45\linewidth}
%\vspace{2pt}
 \begin{algorithm}[H]
\caption{SA query $\SA[i]$.}\label{algo:light_sa_query}
\begin{algorithmic}[1]
\Function{SAQuery}{$i$}
        \State{Compute $\ISA[n]$}            
        \If{$i \neq \ISA[n]$}
            \State{Compute string $T[\SA[i]..\SA[i]+1]$ by query $\BiAQ(i)$}
            \State{$P_{1} \leftarrow T[\SA[i]..\SA[i]+1]$}
            \State{$j \leftarrow 1$}
            \While{$P_{j}[|P_{j}|] \neq \$$}
                \State{$P_{j+1} \leftarrow \RSSQ(P_{j}, i)$}
                \State{$j \leftarrow j+1$}
            \EndWhile
            \State{$d \leftarrow j$}                        
            \State{\Return $n - |P_{d}| + 2$}           
            \Comment{Thm.~\ref{theo:sa_query}}
        \Else \Comment{$i = \ISA[n]$}
            \State{\Return $n$}
        \EndIf
    \EndFunction
\end{algorithmic}
\end{algorithm}
If $\SA[i] \neq n$ for SA query $\SA[i]$, 
then SA query can be straightforwardly answered using $(d-1)$ RSS queries for the integer $d$ used in Theorem~\ref{theo:intro_sa_query}. 
Otherwise (i.e., $\SA[i] \neq n$), 
SA query return $n$. 

Pseudo-codes for computing SA query is presented in Algorithm~\ref{algo:light_sa_query}. 
This algorithm uses bigram access query $\BiAQ(i)$ for computing the string $P_{1}$ of Theorem~\ref{theo:intro_sa_query}. 
For each integer $j \in [2, d]$, 
string $P_{j}$ is computed by RSS query $\RSSQ(P_{j-1}, i)$. 
We answer this RSS query by Theorem~\ref{theo:rss_query_summary}. 
For answering $\RSSQ(P_{j-1}, i)$, 
Theorem~\ref{theo:rss_query_summary} requires 
(i) an occurrence $T[s_{j-1}..s_{j-1} + |P_{j-1}| - 1]$ of string $P_{j-1}$ in input string $T$ 
and (ii) the starting position $\eta_{j-1}$ of the sa-interval $[\eta_{j-1}, \eta^{\prime}_{j-1}]$ of 
string $P_{j-1}$. 
In addition, we need to compute (iii) $\ISA[n]$ to verify $\SA[i] = n$ or not. 
The following lemma shows that these values can be computed in $O(H^{2} \log^{2} n + \log^{6} n)$ time. 

\begin{lemma}\label{lem:sa_query_sub}
    Consider $d$ strings $P_{1}, P_{2}, \ldots, P_{d}$ introduced in Theorem~\ref{theo:intro_sa_query} 
    for SA query $\SA[i]$.
    Let $T[s_{j}..s_{j} + |P_{j}| - 1]$ be an occurrence of each string $P_{j}$. 
    Let $[\eta_{j}, \eta^{\prime}_{j}]$ be the sa-interval of string $P_{j}$.     
    Using the dynamic data structures for bigram search query (Section~\ref{subsubsec:bis_query_algo}) and RSC query (Section~\ref{subsec:answer_rsc_query}), 
    we can support the following three queries: 
\begin{enumerate}[label=\textbf{(\roman*)}]
    \item compute two intervals $[s_{1}, s_{1} + |P_{1}| - 1]$ 
    and $[\eta_{1}, \eta^{\prime}_{1}]$ in $O(H \log n + \log^{2} n)$ time using string $P_{1}$; 
    \item for $j \in [2, d-1]$,     
    compute interval $[s_{j}, s_{j} + |P_{j}| - 1]$ 
    and position $\eta_{j}$ in $O(H^{2} \log^{2} n + \log^{6} n)$ time using 
    interval $[s_{j-1}, s_{j-1} + |P_{j-1}| - 1]$ and position $\eta_{j-1}$; 
    \item compute $\ISA[n]$ in $O((H + \log n) \log n)$ time.
\end{enumerate}
\end{lemma}
\begin{proof}
    Let $([p_{j}, q_{j}], [\ell_{j}, r_{j}])$ be interval attractor $I_{\capture}(\SA[i], \SA[i] + |P_{j-1}| - 1)$ for each integer $j \in [2, d-1]$. 
    Similarly, let $([p^{\prime}_{j}, q^{\prime}_{j}], [\ell^{\prime}_{j}, r^{\prime}_{j}])$ be interval attractor $I_{\capture}(s_{j}, s_{j} + |P_{j-1}| - 1)$.

    From the definition of RSS query, 
    the following statements hold: 
    \begin{itemize}
    \item $|P_{1}| < |P_{2}| < |P_{3}| < \cdots < |P_{d}|$;
    \item $r_{1} < r_{2} < \cdots < r_{d} = n$;
    \item $P_{1}, P_{2}, P_{3}, \ldots, P_{d-1}$ are prefixes of string $T[\SA[i]..n]$; 
    \item $T[\SA[i]..r_{j}+1] = P_{j}$ for all $j \in [2, d-1]$.
    \end{itemize}

    We show that $T[\SA[i]..r_{j}+1] = T[s_{j}..r^{\prime}_{j}+1]$ for each integer $j \in [2, d-1]$. 
    We can apply Corollary~\ref{cor:sa_intv_corollary} to the two substrings 
    $T[\SA[i]..\SA[i] + |P_{j-1}| -1]$ and $T[s_{j}..s_{j} + |P_{j-1}| - 1)]$ 
    because $P_{j-1} = T[\SA[i]..\SA[i] + |P_{j-1}| -1] = T[s_{j}..s_{j} + |P_{j-1}| - 1)]$. 
    Since $T[s_{j}..s_{j} + |P_{j}| - 1)] = P_{j}$, 
    Corollary~\ref{cor:sa_intv_corollary} shows that 
    $T[\SA[i]..r_{j}+1] = T[s_{j}..r^{\prime}_{j}+1]$ holds. 
    
    \textbf{Proof of Lemma~\ref{lem:sa_query_sub}(i).}
    Consider the triplet $(m, g, e)$ obtained by bigram search query $\BiSQ(P_{1})$. 
    Then, the sa-interval $[\eta_{1}, \eta^{\prime}_{1}]$ of $P_{1}$ is equal to $[1+m, m+g]$. 
    $e$ is an occurrence position of string $P_{1}$ (i.e., $s_{1} = e$). 
    Therefore, two intervals $[s_{1}, s_{1} + |P_{1}| - 1]$ 
    and $[\eta_{1}, \eta^{\prime}_{1}]$ can be computed by bigram search query $\BiSQ(P_{1})$. 
    This bigram search query takes $O(H \log n + \log^{2} n)$ time. 

    \textbf{Proof of Lemma~\ref{lem:sa_query_sub}(ii).}
    We can execute RSS query $\RSSQ(P_{j-1}, i)$ by Theorem~\ref{theo:rss_query_summary} 
    because we know interval $[s_{j-1}, s_{j-1} + |P_{j-1}| - 1]$ and position $\eta_{j-1}$. 
    Then, the RSS query returns $P_{j}$, 
    and it is represented as interval $[s_{j}, s_{j} + |P_{j}| - 1]$. 
    Therefore, the interval $[s_{j}, s_{j} + |P_{j}| - 1]$ can be computed in $O(H^{2} \log^{2} n + \log^{6} n)$ time. 

    Next, we compute position $\eta_{j}$ by RSC query and position $\eta_{j-1}$. 
    From the definition of RSC query, 
    we obtain $\eta_{j} = \eta_{j-1} + \RSCQ(\SA[i], \SA[i] + |P_{j-1}| - 1)$. 
    $\RSCQ(\SA[i], \SA[i] + |P_{j-1}| - 1) = \RSCQ(s_{j}, s_{j} + |P_{j-1}| - 1)$ follows from 
    the definition of RSC query  
    because $T[\SA[i]..\SA[i] + |P_{j-1}| - 1] = T[s_{j}..s_{j} + |P_{j-1}| - 1]$ 
    and $T[\SA[i]..r_{j} + 1] = T[s_{j}..r^{\prime}_{j} + 1]$. 
    The RSC query $\RSCQ(s_{j}, s_{j} + |P_{j-1}| - 1)$ can be computed in $O(H^{2} \log n + \log^{4} n)$ time by Theorem~\ref{theo:rsc_query_summary}. 
    Therefore, we obtain Lemma~\ref{lem:sa_query_sub}(ii). 

    \textbf{Proof of Lemma~\ref{lem:sa_query_sub}(iii).}
Bigram search query $\BiSQ(T[n]\$)$ returns a triplet $(m, g, e)$ such that 
$\SA[1+m] = n$ in $O((H + \log n) \log n)$ time. 
The last character of $T$ can be computed in $O(H)$ time by random access query. 
Therefore, we can compute $\ISA[n]$ in $O((H + \log n) \log n)$ time.

\end{proof}

We analyze the running time of the algorithm for SA query. 
This algorithm uses (A) bigram access query $\BiAQ(i)$, (B) $(d-1)$ RSS queries, 
and (C) $O(d)$ queries stated in Lemma~\ref{lem:sa_query_sub}. 
The bottleneck of the algorithm for SA query is RSS query, which takes $O(H^{2} \log^{2} n + \log^{6} n)$ time. 
The integer $d$ can be bounded by $O(H)$ (Lemma~\ref{lem:d_bound}).
Therefore, the algorithm for SA query takes $O(H^{3} \log^{2} n + H \log^{6} n)$ time in total.

\paragraph{ISA query algorithm.}
 \begin{algorithm}[H]
\caption{ISA query $\ISA[i]$. $[\eta, \eta^{\prime}]$: the sa-interval $[\eta, \eta^{\prime}]$ of $T[i..i+1]$}\label{algo:light_isa_query}
\begin{algorithmic}[1]
\Function{ISAQuery}{$i$}
        \If{$i \neq n$}
            \State{Access substring $T[i..i+1]$ by two queries $\RAQ(i)$ and $\RAQ(i+1)$}
            \State{Execute query $\BiSQ(T[i..i+1]) = (m, g, e)$}
            \State{$\eta \leftarrow 1 + m$}
            \Comment{$[\eta, \eta^{\prime}] = [1 + m, m + g]$}
            \State{$r_{1} \leftarrow i$}
            \State{$j \leftarrow 1$}
            \While{$r_{j} \neq n$}
                \State{compute interval attractor $([p, q], [\ell, r])$ by query $\CAPQ(i, r_{j} + 1)$}
                \State{$r_{j} \leftarrow r$}
                \State{$j \leftarrow j+1$}
            \EndWhile
            \State{$d \leftarrow j$}            
            \State{\Return $\eta + \sum_{j=1}^{d-1} \RSCQ(i, r_{j}+1)$}           
            \Comment{Thm.~\ref{theo:isa_query}}
        \Else \Comment{$i = n$}
            \State{Access $T[i]$ by query $\RAQ(i)$}
            \State{Execute query $\BiSQ(T[n]\$) = (m, g, e)$}            
            \Comment{$\SA[1+m] = i$}
            \State{\Return $1+m$}
        \EndIf
\EndFunction
\end{algorithmic}
\end{algorithm}
If the input $i$ of ISA query $\ISA[i]$ smaller than $n$, 
then this query can be straightforwardly answered using $d$ RSC queries 
for the integer $d$ used in Theorem~\ref{theo:isa_query}. 
Otherwise (i.e., $i = n$), 
we can compute $\ISA[n]$ by the third query of Lemma~\ref{lem:sa_query_sub}(iii) in 
$O((H + \log n) \log n)$ time. 

Pseudo-codes for computing ISA query is presented in Algorithm~\ref{algo:light_isa_query}. 
The algorithm for ISA query uses (i) two random access queries, (ii) bigram search query, 
(iii) $O(d)$ capture queries, and (iv) $O(d)$ RSC queries. 
The two random access queries are used to compute string $T[i..i+1]$ for a given ISA query $\ISA[i]$. 
The bigram search query is used to compute the sa-interval $[\eta, \eta^{\prime}]$ of string $T[i..i+1]$. 
The capture queries are used to compute $d-1$ integers  
$r_{1}$, $r_{2}$, $\ldots$, $I(r_{d})$ used in Theorem~\ref{theo:isa_query}. 
The bottleneck of the algorithm for ISA query is RSC query, which takes $O(H^{2} \log n + \log^{4} n)$ time. 
The integer $d$ can be bounded by $O(H)$ (Lemma~\ref{lem:d_prime_bound}).
Therefore, we can answer ISA query in $O(H^{3} \log n + H \log^{4} n)$ time. 

Finally, we obtain the following theorem. 
\begin{theorem}\label{theo:sa_isa_queries}
We can support SA and ISA queries in $O(H^{3} \log^{2} n + H \log^{6} n)$ time and 
$O(H^{3} \log n + H \log^{4} n)$ time, respectively 
using the dynamic data structures for RR-DAG (Section~\ref{subsubsec:rrdag_ds}), 
sample query (Section~\ref{subsubsec:sample_ds}), 
bigram search query (Section~\ref{subsubsec:bis_ds}), 
and RSC query (Section~\ref{subsec:answer_rsc_query}). 
These data structures require $O((|\Psi_{\samp}| + |\mathcal{U}_{\RR}|) B)$ bits of space. 
Here, (i) $B$ is machine word size, 
(ii) $\Psi_{\samp}$ is the sampling subset introduced in Section~\ref{subsec:sampling_subset}, 
and (iii) $|\mathcal{U}_{\RR}|$ is the number of nodes in the RR-DAG of RLSLP $\mathcal{G}^{R}$. 
\end{theorem}

\section{Insertion Operation}\label{sec:insertion}
\begin{table}[t]
    \normalsize
    \vspace{-0.5cm}
    \caption{
    The time needed to update the proposed data structures for answering SA, ISA, RA, and LCE queries. 
    Here, a character is inserted into string $T$ of length $n$ by an insertion operation.   
    $n$ is the length of string $T$. Here, $T$ is changed by inserting a given character 
    after executing an insertion operation string $T$ of length $n$. 
    $H$ is the height of the derivation tree of RLSLP $\mathcal{G}^{R}$. 
    $H^{\prime}$ is the height of the derivation tree of RLSLP $\mathcal{G}^{R}$ after updating the RLSLP.     
    }
    \vspace{1mm}    
    \label{table:insertion_operation_summary} 
    \center{
    \scalebox{0.85}{
    \begin{tabular}{l||l|l}
 Data Structure & Update (time) & Sections \\  \hline 
 Data Structures for RR-DAG (Section~\ref{subsubsec:rrdag_ds}) & $O((\max \{ H, H^{\prime}, \log (nHH^{\prime}) \})^{3})$ & \ref{subsubsec:update_rr_dag} \\ \hline
 Data Structures for sample query (Section~\ref{subsubsec:sample_ds}) & expected $O((\max \{H, H^{\prime}, \log (nHH^{\prime}) \})^{4})$ & 
 \ref{subsec:update_sample_query} \\ \hline
 Data Structures for bigram query (Section~\ref{subsubsec:bis_ds}) & expected $O((\max \{H, H^{\prime}, \log (nHH^{\prime}) \})^{4})$ & 
 \ref{subsec:bis_update} \\ \hline
 Data Structures for RSC-A subquery (Section~\ref{subsubsec:JA_subquery_ds}) & expected $O((\max \{H, H^{\prime}, \log (nHH^{\prime}) \})^{5})$ & 
 \ref{subsec:ra_update} \\ \hline
 Data Structures for RSC-B1 and RSC-B2 subqueries & \multirow{2}{*}{expected $O((\max \{H, H^{\prime}, \log (nHH^{\prime}) \})^{5})$} & 
\multirow{2}{*}{\ref{subsec:rb_update}} \\
(Section~\ref{subsubsec:JB_subquery_ds}) &  & 
 \\ \hline

 Data Structures for RSC-C1 subquery (Section~\ref{subsubsec:JC1_subquery_ds}) & expected $O((\max \{H, H^{\prime}, \log (nHH^{\prime}) \})^{5})$ & 
 \ref{subsec:rc1_update} \\ \hline
 Data Structures for RSC-C2 subquery (Section~\ref{subsubsec:JC2_subquery_ds}) & expected $O((\max \{H, H^{\prime}, \log (nHH^{\prime}) \})^{5})$ & 
 \ref{subsec:rc2_update} \\ \hline
 Data Structures for RSC-D1 subquery (Section~\ref{subsubsec:JD1_subquery_ds}) & expected $O((\max \{H, H^{\prime}, \log (nHH^{\prime}) \})^{5})$ & 
 \ref{subsec:rd1_update} \\ \hline
 Data Structures for RSC-D2 subquery (Section~\ref{subsubsec:JD2_subquery_ds}) & expected $O((\max \{H, H^{\prime}, \log (nHH^{\prime}) \})^{5})$ & 
 \ref{subsec:rd2_update} \\ \hline
 Data Structures for SA, ISA, RA, and LCE queries & \multirow{2}{*}{expected $O((\max \{H, H^{\prime}, \log (nHH^{\prime}) \})^{8})$} & 
 \multirow{2}{*}{\ref{subsec:update_dyn_ins}} \\ 
  (summary) &  & 
  \\ \hline

    \end{tabular} 
    }
    }
\end{table}

\begin{table}[t]
    \normalsize
    %\vspace{-0.5cm}
    \caption{
    The running time of the two algorithms used to update the data structures listed in Table~\ref{table:insertion_operation_summary}. 
    }
    \vspace{1mm}    
    \label{table:algorithm_for_insertion_operation_summary} 
    \center{
    \scalebox{0.85}{
    \begin{tabular}{l||l|l}
 Algorithm & Running time & Sections \\  \hline 
 Computing nonterminals added to/removed from & \multirow{2}{*}{$O(H^{2} H^{\prime} + H^{\prime 3} \log (nH))$} & \multirow{2}{*}{\ref{subsec:rlslp_update}} \\ 
 RLSLP $\mathcal{G}^{R}$ for updating the RLSLP &  &  \\ \hline
 
 Computing interval attractors used to update & \multirow{2}{*}{expected $O((\max \{H, H^{\prime}, \log (nHH^{\prime}) \})^{8})$} & 
 \multirow{2}{*}{\ref{subsec:comp_IA_for_update}} \\ 
  the dynamic data structures built on set $\Psi_{\RR}$ &  & \\ \hline
    \end{tabular} 
    }
    }
\end{table}

An insertion operation inserts a given character $c \in \Sigma \setminus \{ \$, \# \}$ into input string $T$ at a given position $\lambda \in \{ 2, 3, \ldots, n + 1 \}$. 
In this section, 
let $T^{\prime}$ be the string obtained by applying the insertion operation to the string $T$ 
(i.e., $T^{\prime} = T[1..\lambda-1] \cdot c \cdot T[\lambda..n]$ if $\lambda \neq n+1$; otherwise, $T^{\prime} = Tc$). 

This section explains how to update the dynamic data structures 
for RR-DAG (Section~\ref{subsubsec:rrdag_ds}), 
sample query (Section~\ref{subsubsec:sample_ds}), 
bigram search query (Section~\ref{subsubsec:bis_ds}), 
and RSC query (Section~\ref{subsec:answer_rsc_query}) 
after executing the insertion operation. 
Table~\ref{table:insertion_operation_summary} summarizes the time needed to update these data structures. 
Table~\ref{table:algorithm_for_insertion_operation_summary} summarizes the running time 
of the algorithms used to update the data structures listed in Table~\ref{table:insertion_operation_summary}.

Section~\ref{subsec:rlslp_update} presents an algorithm changing RLSLP $\mathcal{G}^{R}$ to an RLSLP deriving string $T^{\prime}$. 
After that,  
we present algorithms updating each of the dynamic data structures for SA and ISA queries based on the changes from $\mathcal{G}^{R}$ to the RLSLP deriving string $T^{\prime}$.

\subsection{Update of Restricted Recompression}\label{subsec:rlslp_update}
This subsection presents the algorithm 
changing RLSLP $\mathcal{G}^{R} = (\mathcal{V}, \Sigma, \mathcal{D}, E)$ 
to an RLSLP $\mathcal{G}^{R}_{\ins} = (\mathcal{V}^{\prime}, \Sigma, \mathcal{D}^{\prime}, E^{\prime})$ 
deriving string $T^{\prime}$ 
and satisfying the conditions of restricted recompression (i.e., the four conditions of the Lemma~\ref{lem:rr_class}). 
This algorithm is called \emph{dynamic restricted recompression}, 
and it requires the dynamic data structures for the RR-DAG of the RLSLP $\mathcal{G}^{R}$ (Section~\ref{sec:recompression}). 

This algorithm partially reconstructs the derivation tree of RLSLP $\mathcal{G}^{R}$, resulting in the RLSLP $\mathcal{G}^{R}_{\ins}$ that derives string $T^{\prime}$. 
This reconstruction is performed by removing nodes surrounding the $\lambda$-th character of $T$ from the derivation tree, 
and appropriately adding new nodes to the tree. Since the structure of the derivation tree is determined using local contexts, 
the number of removed or added nodes is bounded by $O(H^{2})$ where $H$ is the height of the derivation tree.

This algorithm requires shared randomness between the two RLSLPs $\mathcal{G}^{R}$ and $\mathcal{G}^{R}_{\ins}$. 
Under this requirement,
if the RLSLP $\mathcal{G}^{R}_{\ins}$ contains a nonterminal $X_{i}$ of RLSLP $\mathcal{G}^{R}$, and it is uniformly and randomly assigned an integer, 
then the nonterminal $X_{i}$ of the RLSLP $\mathcal{G}^{R}_{\ins}$ is assigned the same integer. 
The random integers assigned to nonterminals are stored in the RR-DAG for $\mathcal{G}^{R}$. 
Therefore,  this requirement can be satisfied.

%The following theorem is the summary of this subsection. 
The details of the dynamic restricted recompression are as follows.
Consider the $H+1$ sequences $S^{0}$, $S^{1}$, $\ldots$, $S^{H}$ of nonterminals introduced in Section~\ref{sec:recompression}. 
Similar to sequence $S^{h}$, 
$S^{h}_{\ins}$ denotes a sequence of nonterminal symbols as the labels of consecutive nodes at height $h$ within the height-balanced derivation tree of $\mathcal{G}^{R}_{\ins}$. 
The dynamic restricted recompression builds the RLSLP $\mathcal{G}^{R}_{\ins}$ from the bottom up, 
and each sequence $S^{h}_{\ins}$ is obtained by replacing a segment $S^{h}[p^{h}..q^{h}]$ of sequence $S^{h}$ with a sequence $Q^{h}$ of nonterminals. 
Here, for simplicity, let $S^{h}$ be a sequence of length $1$ that consists of start symbol $E$ (i.e., $S^{h} = E$) if $h \geq H+1$. 
For the insertion position $\lambda$, 
the starting position $p^{h}$ and ending position $q^{h}$ of the segment $S^{h}[p^{h}..q^{h}]$ are defined as follows: 
\begin{itemize}
    \item for $h = 0$, let $p^{h} = \lambda-1$ and $q^{h} = \lambda-1$;
    \item for $1 \leq h \leq H$, 
    let $u^{h-1}$ (respectively, $u^{\prime h-1}$) be the node corresponding to the $p^{h-1}$-th nonterminal (respectively, the $q^{h-1}$-th nonterminal) of sequence $S^{h-1}$. 
    Here, the parent of the node $u^{h-1}$ (respectively, $u^{\prime h-1}$) corresponds to a nonterminal $S^{h}[s]$~(respectively, $S^{h}[s^{\prime}]$) of sequence $S^{h}$. 
    Then, let $p^{h} = \max \{ s-1, 1 \}$ and $q^{h} = \min \{ s^{\prime} + 1, |S^{h}| \}$;
    \item for $h \geq H+1$, let $p^{h} = 1$ and $q^{h} = 1$. 
\end{itemize}

For the derivation tree of RLSLP $\mathcal{G}^{R}_{\ins}$ with height $H^{\prime}$, 
the dynamic restricted recompression constructs  
$H^{\prime}+1$ sequences $Q^{0}, Q^{1}, \ldots, Q^{H^{\prime}}$ in seven phases, which are detailed as follows. 

\paragraph{Phase (i).}
In the first phase, 
we compute the segment $S^{0}[p^{0}..q^{0}]$ of sequence $S^{0}$ in two steps. 
In the first step, 
we find the two nodes $u^{0}$ and $u^{\prime 0}$ corresponding to the $p^{0}$-th and $q^{0}$-th nonterminals of sequence $S^{0}$. 
This traversal can be executed in $O(H)$ time using the queries of Lemma~\ref{lem:basic_operations_on_dev_tree}. 
In the second step, 
we compute the segment $S^{0}[p^{0}..q^{0}]$ in $O(H)$ time using the two nodes $u^{0}$ and $u^{\prime 0}$. 
Therefore, the first phase takes $O(H)$ time.

%by traversing the derivation tree of RLSLP $\mathcal{G}^{R}$. 
%This traversal can be executed in $O(H)$ time using the queries of Lemma~\ref{lem:basic_operations_on_dev_tree} 
%ecause the $p^{0}$-th and $q^{0}$-th nonterminals of sequence $S^{0}$ derive the $\lambda-1$-th character of $T$. 

%in two steps. 
%In the first step, 
%we find the two nodes $u^{0}$ and $u^{\prime 0}$ corresponding to the $p^{0}$-th and $q^{0}$-th nonterminals of sequence $S^{0}$. 
%The two nodes can be found by traversing the derivation tree of RLSLP $\mathcal{G}^{R}$
%because of $p^{0} = \lambda-1$ and $q^{0} = \lambda-1$. 
%This traversal can be executed in $O(H)$ time using the queries of Lemma~\ref{lem:basic_operations_on_dev_tree}.  
%%The path query takes $O(H)$ time by Theorem~\ref{theo:rr_dag_summary}. 
%
%In the second step, 
%we compute the segment $S^{0}[p^{0}..q^{0}]$ applying Lemma~\ref{lem:basic_operation_on_derivation_tree} to the two nodes $u^{0}$ and $u^{\prime 0}$. 
%The second step takes $O(1)$ time. 

\paragraph{Phase (ii).}
In the second phase, 
the dynamic restricted recompression constructs sequence $Q^{0}$ of nonterminals.  
The sequence $Q^{0}$ consists of the $(\lambda-1)$-th nonterminal of sequence $S^{0}$ 
and a nonterminal $X$ producing the given character $c$ (i.e., $Q^{0} = S^{0}[\lambda-1], X$). 
If set $\mathcal{V}$ does not contain the nonterminal $X$, 
then a new production rule $X \rightarrow c$ is constructed. 

The nonterminal $S^{0}[\lambda-1]$ is obtained from the first phase 
because $p^{0} = \lambda-1$. 
We verify whether $X \in \mathcal{V}$ or not using the function $f_{\prule, D}$ introduced in Section~\ref{subsec:rrdag}. 
This function can be computed in $O(|\mathcal{U}_{\RR}|)$ time by the B-tree for this function, 
which is a component of the data structures for the RR-DAG. 
Therefore, the second phase takes $O(\log |\mathcal{U}_{\RR}|)$ time. 

Subsequently, for integer $h \geq 1$, the dynamic restricted recompression constructs sequence $Q^{h}$ of nonterminals. 
The sequence $Q^{h}$ is obtained by substituting the nonterminals of a sequence $Q^{h-1}_{\free}$. 
If $h \geq H+1$, 
then the sequence $Q^{h-1}_{\free}$ is defined as sequence $Q^{h-1}$ (i.e., $Q^{h-1}_{\free} = Q^{h-1}$). 
Otherwise (i.e., $h \leq H$), 
the sequence $Q^{h-1}_{\free}$ is defined as the concatenation of three sequences $S^{h-1}[\alpha^{h-1}..p^{h-1}-1]$, $Q^{h-1}$, and $S^{h-1}[q^{h-1}+1..\beta^{h-1}]$ 
for the segment $S^{h-1}[p^{h-1}..q^{h-1}]$ of sequence $S^{h-1}$. 
Here, for simplicity, 
let $S^{h-1}[\alpha^{h-1}..p^{h-1}-1]$ be a sequence of length $0$ if $\alpha^{h-1} > p^{h-1}-1$ (respectively, $q^{h-1}+1 > \beta^{h-1}$). 
The two positions $\alpha^{h-1}$ and $\beta^{h-1}$ of sequence $S^{h-1}$ are defined as follows: 
\begin{itemize}
    \item for the node $u^{h}$ corresponding to the $p^{h}$-th nonterminal of sequence $S^{h}$, 
    $\alpha^{h-1}$ is defined as a position of sequence $S^{h-1}$ such that 
    the leftmost child of node $u^{h}$ corresponds to the $\alpha^{h-1}$-th nonterminal of sequence $S^{h-1}$; 
    \item for the node $u^{\prime h}$ corresponding to the $q^{h}$-th nonterminal of sequence $S^{h}$, 
    $\beta^{h-1}$ is defined as a position of sequence $S^{h-1}$ such that 
    the rightmost child of node $v^{\prime}$ corresponds to the $\beta^{h-1}$-th nonterminal of sequence $S^{h-1}$. 
\end{itemize}

The dynamic restricted recompression constructs sequence $Q^{h}$ in the following five phases. 

\paragraph{Phase (iii).}
In the third phase, 
we compute the segment $S^{h}[p^{h}..q^{h}]$ of sequence $S^{h}$. 
If $h > H$, 
then the segment $S^{h}[p^{h}..q^{h}]$ can be obtained in $O(1)$ time 
because $S^{h}[p^{h}..q^{h}] = E$ holds for the start symbol $E$ of RLSLP $\mathcal{G}^{R}$. 

Otherwise (i.e., $h \leq H$), 
we compute the segment $S^{h}[p^{h}..q^{h}]$ in two steps. 
In the first step, 
we find the two nodes $u^{h}$ and $u^{\prime h}$ corresponding to the $p^{h}$-th and $q^{h}$-th nonterminals of sequence $S^{h}$ 
in the derivation tree of RLSLP $\mathcal{G}^{R}$. 
The two nodes $u^{h}$ and $u^{\prime h}$ can be found by traversing the derivation tree. 
This traversal can be executed in $O(H)$ time using  the queries of Lemma~\ref{lem:basic_operations_on_dev_tree}, 
$u^{h-1}$, and $u^{\prime h-1}$. 

In the second step, 
we obtain the segment $S^{h}[p^{h}..q^{h}]$ by traversing the derivation tree. 
This traversal can be executed in $O(|[p^{h}, q^{h}]| H)$ time using the queries of Lemma~\ref{lem:basic_operations_on_dev_tree}, 
$u^{h}$, and $u^{\prime h}$. 
The following lemma shows that the length of the segment $S^{h}[p^{h}..q^{h}]$ is at most $2h + 1$.

%by path query $\pathQ(x + \sum_{w = 1}^{i-1} |\val(S^{h}[w])|)$ 
%for the starting position $x$ of the substring derived from the $p^{h}$-th nonterminal of sequence $S^{h}$ in input string $T$. 
%The starting position $x$ can be computed by $O(1)$ time by Lemma~\ref{lem:basic_operation_on_derivation_tree}. 
%The integer $|\val(S^{h}[i])|$ can be computed in $O(1)$ time by Lemma~\ref{lem:basic_operation_on_derivation_tree} 
%after executing path query $\pathQ(x + \sum_{w = 1}^{i-1} |\val(S^{h}[w])|)$. 
%The second step takes $O(|[p^{h}, q^{h}]| H)$ time in total, 
%and 

\begin{lemma}\label{lem:change_sequence_property}
$|[p^{h}, q^{h}]| \leq 2h + 1$ for all integer $h \geq 0$. 
\end{lemma}
\begin{proof}
We prove Lemma~\ref{lem:change_sequence_property} by induction on $h$. 
For the base case $h = 0$, 
$|[p^{0}, q^{0}]| \leq 1$ holds 
because of $p^{0} = \lambda-1$ and $q^{0} = \lambda-1$. 

For $1 \leq h \leq H$, 
$|[p^{h-1}, q^{h-1}]| \leq 2(h-1) + 1$ holds by the inductive assumption. 
The parent of node $u^{h-1}$ (respectively, $u^{\prime h-1}$) corresponds to a nonterminal $S^{h}[s]$~(respectively, $S^{h}[s^{\prime}]$) of sequence $S^{h}$. 
Because of $|[p^{h-1}, q^{h-1}]| \leq 2(h-1) + 1$, 
$|[s, s^{\prime}]| \leq 2h+1$ holds. 
In this case, $p^{h} = \max \{ s-1, 1 \}$ and $q^{h} = \min \{ s^{\prime} + 1, |S^{h}| \}$. 
Therefore, $|[p^{h}, q^{h}]| \leq 2h + 1$. 

For $h \geq H+1$, 
$|[p^{h}, q^{h}]| \leq 2h + 1$ holds 
because of $p^{h} = 1$ and $q^{h} = 1$. 
Therefore, Lemma~\ref{lem:change_sequence_property} holds. 
\end{proof}

Finally, the third phase takes $O(H^{2})$ time in total. 

\paragraph{Phase (iv).}
The fourth phase is executed if $h \leq H$.
In the fourth phase, 
we find the two nodes $v$ and $v^{\prime}$ corresponding to the $\alpha^{h-1}$-th and $\beta^{h-1}$-th nonterminals of sequence $S^{h-1}$, 
respectively, in the derivation tree of RLSLP $\mathcal{G}^{R}$. 
The two nodes $v$ and $v^{\prime}$ can be found by traversing the derivation tree. 
This traversal can be executed in $O(H)$ time using  the queries of Lemma~\ref{lem:basic_operations_on_dev_tree}, 
$u^{h}$, and $u^{\prime h}$. 

\paragraph{Phase (v).}
For a nonterminal $X_{i}$ appearing in sequence $Q^{h-1}$, 
if the nonterminal $X_{i}$ is not contained in set $\mathcal{V}$, 
then it is assigned an integer from the set $\{ -1, 0, 1 \}$ using 
the same approach as for the third step of the restricted recompression algorithm presented in Section~\ref{subsec:recompression}. 
That is, if the length of the string derived from $X_{i}$ is at most $\mu(h)$, 
$X_{i}$ is uniformly and randomly assigned an integer from $\{ 0, 1 \}$. 
Otherwise $X_{i}$ is assigned the integer $-1$. 
The fifth phase takes $O(|Q^{h-1}|)$ time.

\paragraph{Phase (vi).}
In the sixth phase, 
the sequence $Q^{h-1}_{\free}$ is partitioned into $m$ segments $Q^{h-1}_{\free}[\pi_{1}..(\pi_{2}-1)]$, $Q^{h-1}_{\free}[\pi_{2}..(\pi_{3}-1)]$, $\ldots$, $Q^{h-1}_{\free}[\pi_{m}..(\pi_{m+1}-1)]$ using $(m+1)$ positions $\pi_{1}, \pi_{2}, \ldots, \pi_{m+1}$ where $1 = \pi_{1} < \pi_{2} < \cdots < \pi_{m+1} = |Q^{h-1}_{\free}| + 1$. 
If $h$ is odd, the positions $\pi_{2}, \pi_{3}, \ldots, \pi_{m}$ are selected 
using the same approach as for the fourth step of the restricted recompression algorithm. 
That is, the positions $\pi_{2}, \pi_{3}, \ldots, \pi_{m}$ are selected such that 
each segment $Q^{h-1}_{\free}[\pi_{s}..(\pi_{s+1}-1)]$ satisfies one of the following two conditions:
\begin{itemize}
    \item It is a maximal repetition of a nonterminal $X_{j}$ with $\assign(X_{j}) \in \{0, 1\}$.
    \item It equals a nonterminal $X_{j}$ where $\assign(X_{j}) = -1$.
\end{itemize}

Otherwise (i.e., $h$ is even), 
the $(m-1)$ positions $\pi_{2}, \pi_{3}, \ldots, \pi_{m}$ are chosen from all the positions in sequence $Q^{h-1}_{\free}$ using the same approach as for the fifth step of the restricted recompression algorithm. 
That is, 
the $(m-1)$ positions $\pi_{2}, \pi_{3}, \ldots, \pi_{m}$ are selected such that each segment $Q^{h-1}_{\free}[\pi_{s}..(\pi_{s+1}-1)]$ satisfies one of the following two conditions:
\begin{itemize}
    \item The length of segment $Q^{h-1}_{\free}[\pi_{s}..(\pi_{s+1}-1)]$ is two, and 
        $Q^{h-1}_{\free}[\pi_{s}..(\pi_{s+1}-1)]$ = $X_j, X_k$ holds 
        for two distinct nonterminals $X_j$ and $X_k$ satisfying $\assign(X_{j}) = 0$ and $\assign(X_{k}) = 1$.
    \item The length of segment $Q^{h-1}_{\free}[\pi_{s}..(\pi_{s+1}-1)]$ is one. 
\end{itemize}

The $m$ segments $Q^{h-1}_{\free}[\pi_{1}..(\pi_{2}-1)]$, $Q^{h-1}_{\free}[\pi_{2}..(\pi_{3}-1)]$, $\ldots$, $Q^{h-1}_{\free}[\pi_{m}..(\pi_{m+1}-1)]$ 
can be computed by traversing the derivation tree. 
This traversal can be executed in $O(mH)$ time using (A) the queries of Lemma~\ref{lem:basic_operations_on_dev_tree} 
and (B) four nodes $u^{h-1}$, $u^{\prime h-1}$, $v$, and $v^{\prime}$. 
Therefore, the sixth phase takes $O(m H)$ time.

%in $O(m H)$ time using (A) path query, (B) Lemma~\ref{lem:basic_operation_on_derivation_tree}, 
%and (C) four nodes $u^{h-1}$, $u^{\prime h-1}$, $v$, and $v^{\prime}$. 

\paragraph{Phase (vii).}
Given the partition of sequence $Q^{h-1}_{\free}$, for every sequence $\expr$ of nonterminals represented as $Q^{h-1}_{\free}[\pi_{s}..(\pi_{s}-1)]$ within segments $Q^{h-1}_{\free}[\pi_{1}..(\pi_{2}-1)]$, $Q^{h-1}_{\free}[\pi_{2}..(\pi_{3}-1)]$, $\ldots$, $Q^{h-1}_{\free}[\pi_{m}..(\pi_{m+1}-1)]$, 
a production rule $X_{i} \rightarrow \expr$ is constructed. 
If set $\mathcal{V}$ contains a nonterminal $X$ satisfying $X \rightarrow \expr \in \mathcal{D}$, 
then $X_{i}$ is defined as the nonterminal $X$;
otherwise, the nonterminal $X_{i}$ is a new nonterminal (i.e., $X_{i} \not \in \mathcal{V}$). 
In addition, distinct nonterminals $X_{j}$ and $X_{k}$ ($j \neq k$) are created for each unique pair of sequences $\expr$ and $\expr^{\prime}$. 

The following lemma is used to find the nonterminal $X$ in set $\mathcal{V}$. 
\begin{lemma}\label{lem:find_rule_lemma}
    Consider a sequence $\expr$ of nonterminals represented as $Q^{h-1}_{\free}[\pi_{s}..(\pi_{s}-1)]$ within segments $Q^{h-1}_{\free}[\pi_{1}..(\pi_{2}-1)]$, $Q^{h-1}_{\free}[\pi_{2}..(\pi_{3}-1)]$, $\ldots$, $Q^{h-1}_{\free}[\pi_{m}..(\pi_{m+1}-1)]$. 
    We can verify whether set $\mathcal{V}$ contains a nonterminal $X$ satisfying $X \rightarrow \expr \in \mathcal{D}$ or not 
    in $O(\log |\mathcal{U}_{\RR}|)$ time 
    using the dynamic data structures for the RR-DAG $(\mathcal{U}_{\RR}, \mathcal{E}_{\RR}, L_{\RR})$ of 
    the RLSLP $\mathcal{G}^{R}$ (Section~\ref{sec:recompression}).     
    If the nonterminal $X$ exists, then 
    we can return the nonterminal $X$ in the same time. 
\end{lemma}
\begin{proof}
    We can verify whether $X \rightarrow \expr \in \mathcal{D}$ or not using the three functions 
    $f_{\prule, A}$, $f_{\prule, B}$, and $f_{\prule, C}$ introduced in Section~\ref{subsec:rrdag}. 
    The segment $Q^{h-1}_{\free}[\pi_{s}..(\pi_{s}-1)]$ satisfies one of the following five conditions: 
    \begin{enumerate}[label=\textbf{(\roman*)}]
    \item segment $Q^{h-1}_{\free}[\pi_{s}..(\pi_{s}-1)]$ contains a new nonterminal generated by the dynamic restricted recompression 
    (i.e., there exists an integer $j \in [\pi_{s}, (\pi_{s}-1)]$ satisfying $Q^{h-1}_{\free}[j] \not \in \mathcal{V}$);
    \item segment $Q^{h-1}_{\free}[\pi_{s}..(\pi_{s}-1)]$ contains no new nonterminals, 
    and it is a repetition of a nonterminal $X_{j}$ with $\assign(X_{j}) \in \{ 0, 1 \}$; 
    \item segment $Q^{h-1}_{\free}[\pi_{s}..(\pi_{s}-1)]$ contains no new nonterminals, 
    and $Q^{h-1}_{\free}[\pi_{s}..(\pi_{s+1}-1)]$ = $X_j, X_k$ holds 
    for two distinct nonterminals $X_j$ and $X_k$ satisfying $\assign(X_{j}) = 0$ and $\assign(X_{k}) = 1$;
    \item segment $Q^{h-1}_{\free}[\pi_{s}..(\pi_{s}-1)]$ contains no new nonterminals, 
    and it equals a nonterminal $X_{j}$ with $\assign(X_{j}) \in \{ 0, 1 \}$;
    \item segment $Q^{h-1}_{\free}[\pi_{s}..(\pi_{s}-1)]$ contains no new nonterminals, 
    and it equals a nonterminal $X_{j}$ with $\assign(X_{j}) = -1$.
    \end{enumerate}

    \textbf{Proof of Lemma~\ref{lem:find_rule_lemma} for condition (i).}
    The nonterminal $X$ does not exist in set $\mathcal{V}$ for condition (i). 
    Therefore, we can verify whether set $X \in \mathcal{V}$ or not in $O(1)$ time. 

    \textbf{Proof of Lemma~\ref{lem:find_rule_lemma} for condition (ii).}
    If the nonterminal $X$ exists, 
    then $f_{\prule, C}(X_{j}, d) = X$ for the number $d$ of $X_{j}$ in segment $Q^{h-1}_{\free}[\pi_{s}..(\pi_{s}-1)]$; 
    otherwise, $f_{\prule, C}(X_{j}, d) = \perp$. 
    This function can be computed in $O(\log |\mathcal{U}_{\RR}|)$ time using the B-trees stored in 
    the data structures for the RR-DAG. 

    \textbf{Proof of Lemma~\ref{lem:find_rule_lemma} for condition (iii).}
    Similar to condition (ii), 
    the nonterminal $X$ can be computed in function $f_{\prule, B}$ in $O(\log |\mathcal{U}_{\RR}|)$ time. 
    
    \textbf{Proof of Lemma~\ref{lem:find_rule_lemma} for condition (iv).}
    Similar to condition (ii), 
    the nonterminal $X$ can be computed in function $f_{\prule, A}$ in $O(\log |\mathcal{U}_{\RR}|)$ time. 

    \textbf{Proof of Lemma~\ref{lem:find_rule_lemma} for condition (v).}
    Because of $\assign(X_{j}) = -1$, 
    the nonterminal $X_{j}$ is represented as a pair of a node $v \in \mathcal{U}_{\RR}$ in the RR-DAG for RLSLP $\mathcal{G}^{R}$ 
    and integer $h-1$ (see Section~\ref{subsec:rrdag}). 
    Here, $L_{\level}(v) \geq h$ holds for the label function $L_{\level}$ of RR-DAG. 
    From the definition of RR-DAG, 
    if $L_{\level}(v) = h$, then 
    the nonterminal $X$ is represented as the node $v$; 
    otherwise, it is represented as a pair of $v$ and $h$. 
    Therefore, the nonterminal $X$ can be computed in $O(1)$ time. 
\end{proof}

We can construct the production rule $X_{i} \rightarrow \expr$ in $O(\log |\mathcal{U}_{\RR}|)$ time using Lemma~\ref{lem:find_rule_lemma}. 
Therefore, the seventh phase takes $O(m \log |\mathcal{U}_{\RR}|)$ time in total. 

\paragraph{Phase (viii).}
In the partition of sequence $Q^{h-1}_{\free}$, each segment $Q^{h-1}_{\free}[\pi_{s}..(\pi_{s+1}-1)]$ is replaced by its corresponding nonterminal $X_{i}$ producing the segment. Consequently, a new sequence $Q^{h}$ is generated. 

Starting from $h=1$, above four phases are applied to sequence $Q^{h-1}_{\free}$ to generate sequence $Q^{h}$. 
Each sequence $S^{h}_{\ins}$ is obtained by modifying sequence $S^{h}$ using sequence $Q^{h}$, 
and this process iterates until the length of sequence $S^{h}_{\ins}$ is one.  
Consequently, $H^{\prime}+1$ sequences $S^{0}_{\ins}, S^{1}_{\ins}, \ldots, S^{H^{\prime}}_{\ins}$ are constructed. 

The following lemma shows that the length of each sequence $Q^{h}$ is at most $6(h+1)$. 
\begin{lemma}\label{lem:Q_sequence_length}
$|Q^{h}| \leq 6(h+1)$ for all integer $h \in [0, H^{\prime}]$. 
\end{lemma}
\begin{proof}
    For $h \in [0, H - 1]$, 
    let $\mathcal{B}^{h}$ be a set of integers in set $[\alpha^{h} + 1, p^{h}-1]$ 
    such that each integer $j \in \mathcal{B}^{h}$ satisfies $S^{h}[j] \neq S^{h}[j-1]$ 
    (i.e., $\mathcal{B}^{h} = \{ j \in [\alpha^{h} + 1, p^{h}-1] \mid S^{h}[j] \neq S^{h}[j-1] \}$). 
    Similarly, 
    let $\mathcal{B}^{\prime h}$ be a set of integers in set $[q^{h}+1, \beta^{h}]$ 
    such that each integer $j \in \mathcal{B}^{\prime h}$ satisfies $S^{h}[j] \neq S^{h}[j-1]$. 

    We prove $|\mathcal{B}^{h}| \leq 2$.     
    For the node $u^{h}$ corresponding to the $p^{h}$-th nonterminal of sequence $S^{h}$, 
    the parent of the node $u^{h}$ corresponds to a nonterminal $S^{h}[s]$ of sequence $S^{h}$. 
    Then, integer $p^{h+1}$ is defined as $\max \{ s-1, 1 \}$, 
    and node $u^{h+1}$ corresponds to the $p^{h+1}$-th nonterminal of sequence $S^{h+1}$. 
    If $s = 1$, 
    then $p^{h+1} = s$ and $\alpha^{h} = p^{h}$ hold. 
    $\mathcal{B}^{h} = \emptyset$ follows from $[\alpha^{h} + 1, p^{h}-1] = \emptyset$. 
    Therefore, $|\mathcal{B}^{h}| = 0$ holds. 

    Otherwise (i.e., $s \geq 2$), 
    $p^{h+1} = s - 1$ and $\alpha^{h} < p^{h}$ holds. 
    In this case, 
    there exists an integer $x \in [\alpha^{h}, p^{h}-1]$ such that 
    the rightmost child of node $u^{h+1}$ corresponds to the $x$-th nonterminal of sequence $S^{h}$. 
    From the definition of RLSLP, 
    one of the following three statements holds: 
    \begin{enumerate}[label=\textbf{(\roman*)}]
    \item $\alpha^{h} = x$;
    \item $\alpha^{h} = x-1$ and $S^{h}[x] \neq S^{h}[x-1]$;
    \item $\alpha^{h} \leq x-1$ and $S^{h}[\alpha^{h}] = S^{h}[\alpha^{h}+1] = \cdots = S^{h}[x]$.
    \end{enumerate}
    $|\{ j \in [\alpha^{h} + 1, x] \mid S^{h}[j] \neq S^{h}[j-1] \}| \leq 1$ 
    follows from the above three statements. 

    Similarly, 
    one of the following three statements holds: 
    \begin{enumerate}[label=\textbf{(\Alph*)}]
    \item $x = p^{h}-1$;
    \item $x = p^{h}-2$ and $S^{h}[p^{h}-1] \neq S^{h}[p^{h}]$;
    \item $x \leq p^{h}-2$ and $S^{h}[x+1] = S^{h}[x+2] = \cdots = S^{h}[p^{h}]$.
    \end{enumerate}
    $|\{ j \in [x+1, p^{h}-1] \mid S^{h}[j] \neq S^{h}[j-1] \}| \leq 1$ 
    follows from the above three statements. 
    Therefore, $|\mathcal{B}^{h}| \leq 2$ follows from the following equation: 
    \begin{equation*}
        \begin{split}
            |\mathcal{B}^{h}| &= |\{ j \in [\alpha^{h} + 1, p^{h}-1] \mid S^{h}[j] \neq S^{h}[j-1] \}| \\
            &= |\{ j \in [\alpha^{h} + 1, x] \mid S^{h}[j] \neq S^{h}[j-1] \}| + |\{ j \in [x+1, p^{h}-1]] \mid S^{h}[j] \neq S^{h}[j-1] \}| \\
            &\leq 2. 
        \end{split}
    \end{equation*}
    Similarly, $|\mathcal{B}^{\prime h}| \leq 2$ can be proved using the same approach. 

    We prove Lemma~\ref{lem:Q_sequence_length} by induction on $h$. 
    For the base case $h = 0$, 
    $|Q^{0}| = 2$ holds.  

    For $1 \leq h \leq H$, 
    $|Q^{h-1}| \leq 6h$ holds by the inductive assumption.     
    Sequence $Q^{h-1}_{\free}$ is defined as 
    the concatenation of three sequences $S^{h-1}[\alpha^{h-1}..p^{h-1}-1]$, $Q^{h-1}$, and $S^{h-1}[q^{h-1}+1..\beta^{h-1}]$. 
    Then, the length of sequence $Q^{h}$ is at most $2 + |\mathcal{B}^{h-1}| + |\mathcal{B}^{\prime h-1}| + |Q^{h-1}|$. 
    We already proved $|\mathcal{B}^{h-1}| \leq 2$ and $|\mathcal{B}^{\prime h-1}| \leq 2$.
    Therefore, $|Q^{h}| \leq 6(h+1)$ holds. 
    
    For $h \geq H+1$, 
    $|Q^{h-1}| \leq 6h$ holds by the inductive assumption.     
    Sequence $Q^{h-1}_{\free}$ is defined as sequence $Q^{h-1}$, 
    and the length of sequence $Q^{h}$ is at most $|Q^{h-1}|$.
    Therefore, $|Q^{h}| \leq 6(h+1)$ holds. 

    Finally, Lemma~\ref{lem:Q_sequence_length} holds. 
\end{proof}

The dynamic restricted recompression runs in $O(H^{\prime} (H^{2} + (\sum_{h = 1}^{H^{\prime}} |Q^{h}|) \log |\mathcal{U}_{\RR}|))$ time in total. 
Here, $\sum_{h = 1}^{H^{\prime}} |Q^{h}| = O(H^{\prime 2})$ follows from Lemma~\ref{lem:Q_sequence_length}. 
$|\mathcal{U}_{\RR}| = O(n H)$ holds 
because (A) each node of the RR-DAG corresponds a node in the derivation tree of RLSLP $\mathcal{G}^{R}$, 
and (B) the derivation tree has $O(n H)$ nodes. 
Therefore, the running time of the dynamic restricted recompression can be bounded by $O(H^{2} H^{\prime} + H^{\prime 3} \log (nH))$. 

In addition, 
for a given integer $w \geq 1$, 
we can verify whether $H^{\prime} \leq w$ or not in $O(H^{2} w + w^{3} \log (nH))$ time. 
This is because the first $w$ sequences $Q^{0}, Q^{1}, \ldots, Q^{w-1}$ 
can be constructed in $O(H^{2} w + w^{3} \log (nH))$ time by the dynamic restricted recompression. 

The following lemma shows that 
RLSLP $\mathcal{G}^{R}_{\ins}$ is equal to an RLSLP built by restricted recompression. 
\begin{lemma}\label{lem:DRR_ensure}
    In the derivation tree of $\mathcal{G}^{R}_{\ins}$, 
    let $u_{h, i}$ be the node corresponding to the $i$-th nonterminal $S^{h}_{\ins}[i]$ of sequence $S^{h}_{\ins}$ for 
    a pair of an integer $h \in [0, H^{\prime}]$ and a position $i \in [1, |S^{h}_{\ins}|]$ of sequence $S^{h}_{\ins}$. 
    RLSLP $\mathcal{G}^{R}_{\ins}$ satisfies the four conditions of Lemma~\ref{lem:rr_class}, 
    i.e., the following four statements hold: 
    \begin{enumerate}[label=\textbf{(\roman*)}]
    \item the derivation tree of RLSLP $\mathcal{G}^{R}_{\ins}$ is height-balanced;
    \item consider a nonterminal $X_{i} \in \mathcal{V}^{\prime}$ of sequence $S^{h}_{\ins}$ for an integer $h \in [0, H^{\prime}-1]$. 
    If $|\val(X_{i})| > \mu(h)$, then the assignment $\assign(X_{i})$ of the nonterminal $X_{i}$ is $-1$; 
    otherwise, $X_{i}$ is uniformly and randomly assigned an integer from $\{0, 1\}$;    
    \item 
    consider a pair $(h, i)$ of an even integer $h \in [0, H^{\prime}-1]$ and a position $i \in [1, |S^{h}_{\ins}|-1]$ of sequence $S^{h}_{\ins}$.  
    Then, two nodes $u_{h, i}$ and $u_{h, i+1}$ have the same parent if and only if 
    $S^{h}_{\ins}[i] = S^{h}_{\ins}[i+1]$ and $\assign(S^{h}_{\ins}[i]) \neq -1$;
    \item 
    consider a pair $(h, i)$ of an odd integer $h \in [0, H^{\prime}-1]$ and a position $i \in [1, |S^{h}_{\ins}|-1]$ of sequence $S^{h}$.  
    Then, two nodes $u_{h, i}$ and $u_{h, i+1}$ have the same parent if and only if 
    $\assign(S^{h}_{\ins}[i]) = 0$ and $\assign(S^{h}_{\ins}[i+1]) = 1$.
    \end{enumerate}
\end{lemma}
\begin{proof}
    The proof of Lemma~\ref{lem:DRR_ensure} is as follows. 

    \textbf{Proof of Lemma~\ref{lem:DRR_ensure}(i).}
    Lemma~\ref{lem:DRR_ensure}(i) follows from the procedure of the dynamic restricted recompression.     

    \textbf{Proof of Lemma~\ref{lem:DRR_ensure}(ii).}
    The nonterminal $X_{i}$ satisfies one of the following three conditions: 
    \begin{enumerate}[label=\textbf{(\Alph*)}]
    \item $X_{i} \in \mathcal{V}$ and $|\val(X_{i})| > \mu(h)$;
    \item $X_{i} \in \mathcal{V}$ and $|\val(X_{i})| \leq \mu(h)$;
    \item $X_{i} \not \in \mathcal{V}$ and $|\val(X_{i})| > \mu(h)$;
    \item $X_{i} \not \in \mathcal{V}$ and $|\val(X_{i})| \leq \mu(h)$.
    \end{enumerate}

    For condition (A), 
    $\assign(X_{i}) = -1$ follows from Lemma~\ref{lem:rr_class}~\ref{enum:rr_class:2}. 
    Similarly, 
    for condition (B), 
    Lemma~\ref{lem:rr_class}~\ref{enum:rr_class:2} shows that 
    $X_{i}$ is uniformly and randomly assigned an integer from $\{0, 1\}$. 

    For condition (C), 
    $X_{i}$ is assigned $-1$ in the fifth phase of the dynamic restricted recompression. 
    Similarly, 
    for condition (D), 
    $X_{i}$ is uniformly and randomly assigned an integer from $\{0, 1\}$ 
    in the fifth phase of the dynamic restricted recompression. 
    Therefore, Lemma~\ref{lem:DRR_ensure}(ii) holds. 

    \textbf{Proof of Lemma~\ref{lem:DRR_ensure}(iii).}
    If $h \geq H$, 
    then $S^{h}_{\ins} = Q^{h}_{\free}$ holds. 
    In this case, the sixth phase of the dynamic restricted recompression ensures that 
    two nodes $u_{h, i}$ and $u_{h, i+1}$ have the same parent if and only if 
    $S^{h}_{\ins}[i] = S^{h}_{\ins}[i+1]$ and $\assign(S^{h}[i]) \neq -1$. 

    Otherwise (i.e., $h \leq H-1$), 
    sequence $S^{h}_{\ins}$ can be divided into 
    three segments $S^{h}[1..\alpha^{h}-1]$, $Q^{h}_{\free}$, and $S^{h}[\beta^{h}+1..|S^{h}|]$. 
    Here, $|Q^{h}_{\free}| \geq 1$ follows from the procedure of the dynamic restricted recompression.     

    The integer $i$ satisfies one of the following five conditions: 
    \begin{enumerate}[label=\textbf{(\alph*)}]
    \item $\alpha^{h} \leq i \leq \alpha^{h} + |Q^{h}_{\free}| - 2$;
    \item $i \leq \alpha^{h} - 2$;
    \item $i \geq \alpha^{h} + |Q^{h}_{\free}|$;
    \item $i = \alpha^{h} - 1$;
    \item $i = \alpha^{h} + |Q^{h}_{\free}| - 1$.
    \end{enumerate}

    For condition (a), 
    two nonterminals $S^{h}_{\ins}[i]$ and $S^{h}_{\ins}[i+1]$ are contained in segment $Q^{h}_{\free}$. 
    In this case, the sixth phase of the dynamic restricted recompression ensures that 
    two nodes $u_{h, i}$ and $u_{h, i+1}$ have the same parent if and only if 
    $S^{h}_{\ins}[i] = S^{h}_{\ins}[i+1]$ and $\assign(S^{h}[i]) \neq -1$. 

    For condition (b)
    two nonterminals $S^{h}_{\ins}[i]$ and $S^{h}_{\ins}[i+1]$ are contained in segment $S^{h}[1..\alpha^{h}-1]$. 
    In this case, 
    Lemma~\ref{lem:rr_class}~\ref{enum:rr_class:3} indicates that 
    two nodes $u_{h, i}$ and $u_{h, i+1}$ have the same parent if and only if 
    $S^{h}_{\ins}[i] = S^{h}_{\ins}[i+1]$ and $\assign(S^{h}_{\ins}[i]) \neq -1$. 
    This is because 
    the parents of the two nodes $u_{h, i}$ and $u_{h, i+1}$ are not changed by the dynamic restricted recompression. 
    
    For condition (c)
    two nonterminals $S^{h}_{\ins}[i]$ and $S^{h}_{\ins}[i+1]$ are contained in segment $S^{h}[\beta^{h}+1..|S^{h}|]$. 
    Similar to condition (b), 
    Lemma~\ref{lem:rr_class}~\ref{enum:rr_class:3} indicates that 
    two nodes $u_{h, i}$ and $u_{h, i+1}$ have the same parent if and only if 
    $S^{h}_{\ins}[i] = S^{h}_{\ins}[i+1]$ and $\assign(S^{h}_{\ins}[i]) \neq -1$. 
     
    For condition (d), 
    two nonterminals $S^{h}_{\ins}[i]$ and $S^{h}_{\ins}[i+1]$ are contained in segments $S^{h}[1..\alpha^{h}-1]$ and $Q^{h}_{\free}$, respectively. 
    In this case, $Q^{h}_{\free}$ can be divided into 
    three segments $S^{h}[\alpha^{h}..p^{h}-1]$, $Q^{h}$, and $S^{h}[q^{h}+1..\beta^{h}]$. 
    Here, the length of segment $S^{h}[\alpha^{h}..p^{h}-1]$ is at least $1$ (i.e., $\alpha^{h} \leq p^{h}-1$).
    The nonterminal $S^{h}_{\ins}[i+1]$ is the first nonterminal of segment $S^{h}[\alpha^{h}..p^{h}-1]$ 
    (i.e., $i+1 = \alpha^{h}$). 
    
    From the procedure of the dynamic restricted recompression, 
    two nodes $u_{h, i}$ and $u_{h, i+1}$ do not have the same parent in the derivation tree of RLSLP $\mathcal{G}^{R}_{\ins}$. 
    Therefore, Lemma~\ref{lem:DRR_ensure}(iii) holds if 
    $S^{h}_{\ins}[i] \neq S^{h}_{\ins}[i+1]$ or $\assign(S^{h}_{\ins}[i]) = -1$ holds. 

    We show that $S^{h}_{\ins}[i] \neq S^{h}_{\ins}[i+1]$ or $\assign(S^{h}[i]) = -1$ holds. 
    The two nodes $u_{h, i}$ and $u_{h, i+1}$ are contained in the derivation tree of RLSLP $\mathcal{G}^{R}$ 
    (i.e., the two nodes are not removed from the derivation tree of RLSLP $\mathcal{G}^{R}$ by dynamic restricted recompression).
    In the derivation tree of RLSLP $\mathcal{G}^{R}$, 
    two nodes $u_{h, i}$ and $u_{h, i+1}$ correspond to the $i$-th and $i+1$-th nonterminals of sequence $S^{h}$, respectively. 
    This fact indicates that 
    $S^{h}[i] = S^{h}_{\ins}[i]$ and $S^{h}[i+1] = S^{h}_{\ins}[i+1]$ hold.     
    Lemma~\ref{lem:rr_class}~\ref{enum:rr_class:3} indicates that 
    two nodes $u_{h, i}$ and $u_{h, i+1}$ have the same parent if and only if 
    $S^{h}[i] = S^{h}[i+1]$ and $\assign(S^{h}[i]) \neq -1$. 

    Let $v$ be the parent of node $u_{h, i+1}$ in the derivation tree of RLSLP $\mathcal{G}^{R}$.     
    From the definition of the integer $\alpha^{h}$, 
    $u_{h, i+1}$ is the leftmost child of the node $v$. 
    In the derivation tree of RLSLP $\mathcal{G}^{R}$, 
    two nodes $u_{h, i}$ and $u_{h, i+1}$ do not have the same parent. 
    If $S^{h}_{\ins}[i] = S^{h}_{\ins}[i+1]$, 
    then $S^{h}[i] = S^{h}[i+1]$ holds, and $\assign(S^{h}[i]) = -1$ follows from Lemma~\ref{lem:rr_class}~\ref{enum:rr_class:3}. 
    Otherwise, $S^{h}[i] \neq S^{h}[i+1]$ holds. 
    Therefore, Lemma~\ref{lem:DRR_ensure}(iii) holds. 

    For condition (e), 
    we can prove Lemma~\ref{lem:DRR_ensure}(iii) using the same approach as for condition (c). 
    Therefore, Lemma~\ref{lem:DRR_ensure}(iii) always holds. 

    \textbf{Proof of Lemma~\ref{lem:DRR_ensure}(iv).}
    Lemma~\ref{lem:DRR_ensure}(iv) can be proved using the same approach as for Lemma~\ref{lem:DRR_ensure}(iii). 

\end{proof}

The following theorem states the summary of the dynamic restricted recompression. 

\begin{theorem}\label{theo:update1}
For an insertion operation, 
consider the dynamic restricted recompression algorithm changing 
RLSLP $\mathcal{G}^{R} = (\mathcal{V}, \Sigma, \mathcal{D}, E)$ 
to an RLSLP $\mathcal{G}^{R}_{\ins} = (\mathcal{V}^{\prime}, \Sigma, \mathcal{D}^{\prime}, E^{\prime})$. 
Then, the following four statements hold: 
\begin{enumerate}[label=\textbf{(\roman*)}]
    \item \label{enum:update1:1} RLSLP $\mathcal{G}^{R}_{\ins}$ derives string $T^{\prime}$ and satisfies the four conditions of the Lemma~\ref{lem:rr_class}); 
    \item \label{enum:update1:2} the dynamic restricted recompression runs in $O(H^{2} H^{\prime} + H^{\prime 3} \log (nH))$ time 
    using the dynamic data structures for the RR-DAG of the RLSLP $\mathcal{G}^{R}$ (Section~\ref{sec:recompression}). 
    Here, $H^{\prime}$ is the height of the derivation tree of $\mathcal{G}^{R}_{\ins}$;
    \item \label{enum:update1:3} the dynamic restricted recompression can output the following sequences of nonterminals: 
    (A) the segment $S^{h}[p^{h}..q^{h}]$ of sequence $S^{h}$ for each $h \in [0, H]$, 
    and (B) sequence $Q^{h}$ for each $h \in [0, H^{\prime}]$; 
    \item \label{enum:update1:4}
    for a given integer $w \geq 1$, 
    we can verify whether $H^{\prime} \leq w$ or not in $O(H^{2} w + w^{3} \log (nH))$ time. 
\end{enumerate}
\end{theorem}

\subsubsection{Updating RR-DAG}\label{subsubsec:update_rr_dag}
We present the algorithm updating the data structures for the RR-DAG $(\mathcal{U}_{\RR}, \mathcal{E}_{\RR}, L_{\RR})$ of the RLSLP $\mathcal{G}^{R}$ 
based on the changes from RLSLP $\mathcal{G}^{R}$ to $\mathcal{G}^{R}_{\ins}$. 
This algorithms requires the output of the dynamic restricted recompression, 
i.e., $H+1$ segments $S^{0}[p^{0}..q^{0}]$, $S^{1}[p^{1}..q^{1}]$, $\ldots$, $S^{H}[p^{H}..q^{H}]$ 
and $H^{\prime}+1$ sequences $Q^{0}$, $Q^{1}$, $\ldots$, $Q^{H^{\prime}}$. 

By the dynamic restricted recompression, 
each segment $S^{h}[p^{h}..q^{h}]$ of sequence $S^{h}$ is replaced with sequence $Q^{h}$ 
for each integer $h \in [0, H^{\prime}]$. 
After these replacements, 
the RR-DAG is modified as follows:

\begin{description}
    \item [Modification 1:]     
    For each segment $S^{h}[p^{h}..q^{h}]$ and each integer $i \in [p^{h}, q^{h}]$, 
    let $u$ be the node corresponding to the $i$-th nonterminal of sequence $S^{h}$ in the derivation tree of RLSLP $\mathcal{G}^{R}$. 
    Then, the node $u$ is removed from the derivation tree. 
    If either $S^{h}[i] = S$ or $\assign(S^{h}[i]) \in \{ 0, 1 \}$, then 
    the RR-DAG of RLSLP $\mathcal{G}^{R}$ 
    contains a node $v$ labeled $S^{h}[i]$, 
    and the output $W$ of the label function $L_{\vOcc}(v)$ is changed to $W-1$. 
    Here, $L_{\vOcc}$ is the label function introduced in Section~\ref{subsec:rrdag}.
    
    \item [Modification 2:] 
    For each nonterminal $X \in \mathcal{V} \setminus \mathcal{V}^{\prime}$, 
    if either $X = S$ or $\assign(X) \in \{ 0, 1 \}$, then 
    the RR-DAG of RLSLP $\mathcal{G}^{R}$ 
    contains a node $v$ labeled $X$. 
    This node $v$ is removed from the RR-DAG;
    
    \item [Modification 3:] 
    For each nonterminal $X \in \mathcal{V}^{\prime} \setminus \mathcal{V}$, 
    if either $X = S^{\prime}$ or $\assign(X) \in \{ 0, 1 \}$, then 
    a node $v$ is created, which corresponds to the nonterminal $X$. 
    This node $v$ is added to the RR-DAG of RLSLP $\mathcal{G}^{R}$. 
    
    \item [Modification 4:]
    For each sequence $Q^{h}$ and each integer $i \in [1, |Q^{h}|]$, 
    a node $u$ is created, which corresponds to the $i$-th nonterminal of sequence $Q^{h}$. 
    This node $u$ is added to the derivation tree of RLSLP $\mathcal{G}^{R}$.
    If either $Q^{h}[i] = S^{\prime}$ or $\assign(Q^{h}[i]) \in \{ 0, 1 \}$, then 
    the RR-DAG of RLSLP $\mathcal{G}^{R}$ 
    contains a node $v$ labeled $Q^{h}[i]$, 
    and the output $W$ of the label function $L_{\vOcc}(v)$ is changed to $W+1$. 
\end{description}

We update the data structures for the RR-DAG based on these modifications. 
The algorithm updating the data structures for the RR-DAG consists of the following six phases. 

\paragraph{Phase (i).}
In the first phase, 
we compute set $\mathcal{V} \setminus \mathcal{V}^{\prime}$ of nonterminals. 
The nonterminals of set $\mathcal{V} \setminus \mathcal{V}^{\prime}$ 
are contained in $(H+1)$ segments $S^{0}[p^{0}..q^{0}]$, $S^{1}[p^{1}..q^{1}]$, $\ldots$, $S^{H}[p^{H}..q^{H}]$. 
Therefore, 
we can compute the set $\mathcal{V} \setminus \mathcal{V}^{\prime}$ by verifying whether each nonterminal of the $(H+1)$ segments is contained in set $\mathcal{V} \setminus \mathcal{V}^{\prime}$. 

For this verification, 
we leverage the following relationship among 
set $\mathcal{V} \setminus \mathcal{V}^{\prime}$ and the $(H+1)$ segments $S^{0}[p^{0}..q^{0}]$, $S^{1}[p^{1}..q^{1}]$, $\ldots$, $S^{H}[p^{H}..q^{H}]$. 

\begin{lemma}\label{lem:remove_nonteminal_detection}
Consider a pair $(h, i)$ of two integers $h \in [0, H]$ and $i \in [p^{h}, q^{h}]$. 
Let $\kappa$ be the number of nonterminal $S^{h}[i]$ occurring in segment $S^{h}[p^{h}..q^{h}]$ 
(i.e., $\kappa = |\{ j \in [p^{h}, q^{h}] \mid S^{h}[i] = S^{h}[j] \}|$). 
Similarly, 
let $\kappa^{\prime}$ be the number of nonterminal $S^{h}[i]$ occurring in sequence $Q^{h}$. 
Let $u$ be the node corresponding to the $i$-th nonterminal of sequence $S^{h}$ in the derivation tree of RLSLP $\mathcal{G}^{R}$. 
Here, the node $u$ is represented as a node $v \in \mathcal{U}_{\RR}$ in the RR-DAG for RLSLP $\mathcal{G}^{R}$ 
or a pair of node $v$ and integer $h$ (See Section~\ref{subsec:rrdag}). 
%Here, the node $u$ is represented as a triplet $\repr(u) = (x^{\prime}, h, v)$ introduced in Section~\ref{subsubsec:node_representation}.  
Then, $S^{h}[i] \in \mathcal{V} \setminus \mathcal{V}^{\prime} \Leftrightarrow L_{\vOcc}(v) - \kappa + \kappa^{\prime} = 0$ holds 
for the label function $L_{\vOcc}$ introduced in Section~\ref{subsec:rrdag}.
\end{lemma}
\begin{proof}
    The following three statements are used to prove Lemma~\ref{lem:remove_nonteminal_detection}:
    \begin{enumerate}[label=\textbf{(\roman*)}]
    \item let $\kappa_{\ins}$ be the number of nonterminal $S^{h}[i]$ occurring in sequence $S^{h}_{\ins}$. 
    Then, $\kappa_{\ins} = L_{\vOcc}(v) - \kappa + \kappa^{\prime}$;
    \item $S^{h}[i] \in \mathcal{V} \setminus \mathcal{V}^{\prime} \Rightarrow L_{\vOcc}(v) - \kappa + \kappa^{\prime} = 0$;
    \item $S^{h}[i] \in \mathcal{V} \setminus \mathcal{V}^{\prime} \Leftarrow L_{\vOcc}(v) - \kappa + \kappa^{\prime} = 0$.    
    \end{enumerate}

    \textbf{Proof of statement (i).}    
    We show that the number of nonterminal $S^{h}[i]$ occurring in sequence $S^{h}$ is $L_{\vOcc}(v)$. 
    If the node $u$ is represented as node $v$, 
    then the two nodes $u$ and $v$ have the same label $S^{h}[i]$, 
    and the label function $L_{\vOcc}(v)$ returns the number of nonterminal $S^{h}[i]$ occurring in the derivation tree of RLSLP $\mathcal{G}^{R}$. 
    Sequence $S^{h}$ contains each occurrence of the nonterminal in the derivation tree, 
    and hence, the number of nonterminal $S^{h}[i]$ occurring in sequence $S^{h}$ is $L_{\vOcc}(v)$. 

    Otherwise (i.e., the node $u$ is represented as a pair of node $v$ and integer $h$), 
    the node $v$ has a nonterminal $X$ ($X \neq S^{h}[i]$) such that 
    (A) $X$ and $S^{h}[i]$ derive the same string (i.e., $\val(X) = \val(S^{h}[i])$), 
    and (B) $X$ is contained in a sequence $S^{h^{\prime}}$ satisfying $h^{\prime} > h$. 
    In this case, the number of nonterminal $S^{h}[i]$ occurring in the derivation tree of RLSLP $\mathcal{G}^{R}$ 
    is equal to the number of nonterminal $X$ occurring in the derivation tree of RLSLP $\mathcal{G}^{R}$. 
    This is because the following two statements follows from the definition of RR-DAG. 
    \begin{itemize}
        \item In the derivation tree, each node labeled with nonterminal $S^{h}[i]$ has an ancestor node with labeled with $X$. 
        \item Similarly, each node labeled with nonterminal $X$ has a descendant node with labeled with $S^{h}[i]$. 
    \end{itemize}
    Therefore, the number of nonterminal $S^{h}[i]$ occurring in sequence $S^{h}$ is $L_{\vOcc}(v)$.

    %For the label $X$ of node $v$,  
    %the label function $L_{\vOcc}(v)$ returns the number of $X$ occurring in the derivation tree of RLSLP $\mathcal{G}^{R}$. 
    %Lemma~\ref{lem:nonterminal_function} shows that 
    %the number of $S^{h}[i]$ occurring in the derivation tree of RLSLP $\mathcal{G}^{R}$ is equal to that of nonterminal $X$ in the tree. 
    %The number of $S^{h}[i]$ occurring in the derivation tree of RLSLP $\mathcal{G}^{R}$ 
    %is equal to that of $S^{h}[i]$ in sequence $S^{h}$ 
    %because the derivation tree of RLSLP $\mathcal{G}^{R}$ is balanced. 
    %Therefore, the number of nonterminal $S^{h}[i]$ occurring in sequence $S^{h}$ is $L_{\vOcc}(v)$. 

    We prove statement (i). 
    From the procedure of the dynamic restricted recompression, 
    sequence $S^{h}_{\ins}$ can be divided into three segments 
    $S^{h}[1..p^{h}-1]$, $Q^{h}$, and $S^{h}[q^{h}+1..|S^{h}|]$. 
    The sequence $S^{h}$ can be divided into three segments 
    $S^{h}[1..p^{h}-1]$, $S^{h}[p^{h}..q^{h}]$, and $S^{h}[q^{h}+1..|S^{h}|]$. 
    Therefore, statement (i) follows from the following equation:
    \begin{equation*}
        \begin{split}
        \kappa_{\ins} &= |\{ j \in [1, p^{h}-1] \mid S^{h}[i] = S^{h}[j] \}| + |\{ j \in [1, |Q^{h}|] \mid S^{h}[i] = Q^{h}[j] \}| \\
        &+ |\{ j \in [q^{h}+1, |S^{h}|] \mid S^{h}[i] = S^{h}[j] \}| \\
        &= (|\{ j \in [1, |S^{h}|] \mid S^{h}[i] = S^{h}[j] \}| -  |\{ j \in [p^{h}, q^{h}] \mid S^{h}[i] = S^{h}[j] \}|) \\
        &+ |\{ j \in [1, |Q^{h}|] \mid S^{h}[i] = Q^{h}[j] \}| \\
        &= L_{\vOcc}(v) - \kappa + \kappa^{\prime}.
        \end{split}        
    \end{equation*}

    \textbf{Proof of statement (ii).}
    Sequence $S^{h}_{\ins}$ does not contain nonterminal $S^{h}[i]$ 
    because $S^{h}[i] \not \in \mathcal{V}^{\prime}$ 
    follows from $S^{h}[i] \in \mathcal{V} \setminus \mathcal{V}^{\prime}$. 
    Statement (i) shows that $L_{\vOcc}(v) - \kappa + \kappa^{\prime} = 0$ holds. 
    Therefore, statement (ii) holds. 

    \textbf{Proof of statement (iii).}
    Because of $L_{\vOcc}(v) - \kappa + \kappa^{\prime} = 0$, 
    statement (i) shows that sequence $S^{h}_{\ins}$ does not contain nonterminal $S^{h}[i]$. 
    In this case, set $\mathcal{V}^{\prime}$ does not contain nonterminal $S^{h}[i]$. 
    $S^{h}[i] \in \mathcal{V} \setminus \mathcal{V}^{\prime}$ follows from 
    $S^{h}[i] \in \mathcal{V}$ and $S^{h}[i] \not \in \mathcal{V}^{\prime}$. 
    Therefore, statement (iii) holds. 

    \textbf{Proof of Lemma~\ref{lem:remove_nonteminal_detection}.}
    Lemma~\ref{lem:remove_nonteminal_detection} follows from statement (ii) and statement (iii). 
\end{proof}

The first phase consists of two steps for each pair $(h, i)$ of two integers $h \in [0, H]$ and $i \in [1, |Q^{h}|]$. 
Let $u$ be the node corresponding to the $i$-th nonterminal $S^{h}[i]$ of sequence $S^{h}$ in the derivation tree of RLSLP $\mathcal{G}^{R}$. 
Here, the node $u$ is represented as a node $v \in \mathcal{U}_{\RR}$ in the RR-DAG for RLSLP $\mathcal{G}^{R}$ 
or a pair of node $v$ and integer $h$ (See Section~\ref{subsec:rrdag}). 
In the first step, 
we obtain the node $u$ by traversing the derivation tree. 
This traversal is executed in the third phase of the dynamic restricted recompression. 
Therefore, the first step can be executed in $O(1)$ time. 

%In the first step, 
%we compute the node representation $\repr(u) = (x^{\prime}, h, v)$ by path query $\pathQ(x + \sum_{w = 1}^{i-1} |\val(S^{h}[w])|)$ 
%for the starting position $x$ of the substring derived from the $p^{h}$-th nonterminal of sequence $S^{h}$ in input string $T$. 
%This path query 

In the second step, we verify whether the nonterminal $S^{h}[i]$ is contained in set $\mathcal{V} \setminus \mathcal{V}^{\prime}$ 
by Lemma~\ref{lem:remove_nonteminal_detection}. 
The label function $L_{\vOcc}(v)$ can be computed in $O(1)$ time using the RR-DAG.
%because the node representation $\repr(u)$ contains a pointer to an element of the doubly linked list representing 
%the RR-DAG (see Section~\ref{subsubsec:rrdag_ds}), 
%and the element stores the result of the label function $L_{\vOcc}(v)$. 
We can compute the number of nonterminal $S^{h}[i]$ occurring in segment $S^{h}[p^{h}..q^{h}]$ 
in $O(q^{h} - p^{h} + 1)$ time. 
Here, $q^{h} - p^{h} = O(H)$ follows from Lemma~\ref{lem:change_sequence_property}. 
Similarly, 
we can compute the number of nonterminal $S^{h}[i]$ occurring in sequence $Q^{h}$ in $O(|Q^{h}|)$ time. 
Here, $|Q^{h}| = O(H^{\prime})$ follows from Lemma~\ref{lem:Q_sequence_length}. 
Therefore, the second step takes $O(H^{\prime} + H)$ time. 

The first phase takes $O((\sum_{h=0}^{H} |[p^{h}, q^{h}]|)(H^{\prime} + H))$ in total. 
Here, $\sum_{h=0}^{H} |[p^{h}, q^{h}]| = O(H^{2})$ follows from Lemma~\ref{lem:change_sequence_property}. 
Therefore, the running time of the first phase can be bounded by $O(H^{2} H^{\prime} + H^{3})$.  

\paragraph{Phase (ii).}
In the second phase, 
we compute set $\mathcal{V}^{\prime} \setminus \mathcal{V}$ of nonterminals. 
Here, the nonterminals of set $\mathcal{V}^{\prime} \setminus \mathcal{V}$ 
are contained in $(H^{\prime}+1)$ sequences $Q^{0}$, $Q^{1}$, $\ldots$, $Q^{H^{\prime}}$. 
Consider a nonterminal $X$ in a sequence $Q^{h}$. 
Then, set $\mathcal{V}^{\prime} \setminus \mathcal{V}$ contains the nonterminal $X$ if and only if 
the nonterminal is created in the second or seventh phase of the dynamic restricted recompression. 

We compute set $\mathcal{V}^{\prime} \setminus \mathcal{V}$ by verifying whether  
each nonterminal of the $(H^{\prime}+1)$ sequences $Q^{0}$, $Q^{1}$, $\ldots$, $Q^{H^{\prime}}$ 
is a new nonterminal or not. 
This verification takes $O(1)$ time using the result of the dynamic restricted recompression. 
Lemma~\ref{lem:Q_sequence_length} indicates that 
the $(H^{\prime}+1)$ sequences $Q^{0}$, $Q^{1}$, $\ldots$, $Q^{H^{\prime}}$ contains 
$O(H^{\prime 2})$ nonterminals. 
Therefore, the second phase takes $O(H^{\prime 2})$ time in total. 

\paragraph{Phase (iii).}
In the third phase, 
we update the data structures for the RR-DAG based on Modification 1 
for the $(H+1)$ segments $S^{0}[p^{0}..q^{0}]$, $S^{1}[p^{1}..q^{1}]$, $\ldots$, $S^{H}[p^{H}..q^{H}]$. 
%Consider a pair $(h, i)$ of two integers $h \in [0, H]$ and 
%$i \in [p^{h}, q^{h}]$ satisfying either $S^{h}[i] = S$ or $\assign(S^{h}[i]) \in \{ 0, 1 \}$. 
%Then, the RR-DAG of RLSLP $\mathcal{G}^{R}$ contains a node $v$ labeled $S^{h}[i]$. 
%Here, 
%$\repr(u) = (x + \sum_{w = 1}^{i-1} |\val(S^{h}[w])|, h, v)$ holds 
%for the node representation $\repr(u)$ of node $u$ and 
%the starting position $x$ of the substring derived from the $p^{h}$-th nonterminal of sequence $S^{h}$ in input string $T$. 

The third phase consists of two steps for each pair $(h, i)$ of two integers $h \in [0, H]$ and 
$i \in [p^{h}, q^{h}]$ satisfying either $S^{h}[i] = S$ or $\assign(S^{h}[i]) \in \{ 0, 1 \}$. 
Let $u$ be the node corresponding to the $i$-th nonterminal of sequence $S^{h}$ 
in the derivation tree of RLSLP $\mathcal{G}^{R}$. 
In the first step, 
we obtain the node $u$ by traversing the derivation tree. 
This traversal is executed in the third phase of the dynamic restricted recompression. 
Therefore, the first step can be executed in $O(1)$ time. 

The node $u$ is represented as a node $v \in \mathcal{U}_{\RR}$ in the RR-DAG. 
This node $v$ is stored as an element of the doubly linked list representing the RR-DAG (see Section~\ref{subsubsec:rrdag_ds}), 
and the element stores the integer $W$ obtained from label function $L_{\vOcc}(v)$. 
In the second step, 
we change the integer $W$ to $W-1$. 
The second step can be executed in $O(1)$ time. 
Therefore, the third phase takes $O(\sum_{h=0}^{H} |[p^{h}, q^{h}]|)$ (i.e., $O(H^{2})$) time in total. 

%The node $v$ is represented as an element of the doubly linked list representing 
%the RR-DAG (see Section~\ref{subsubsec:rrdag_ds}), 
%and the element stores the integer $W$ obtained from label function $L_{\vOcc}(v)$. 

\paragraph{Phase (iv).}
In the fourth phase, 
we update the data structures for the RR-DAG based on Modification 2 
for each nonterminal $X \in \mathcal{V} \setminus \mathcal{V}^{\prime}$ 
satisfying either $X = S$ or $\assign(X) \in \{ 0, 1 \}$. 
From the procedure of the first phase, 
there exists a pair of two integers $h \in [0, H]$ and $i \in [p^{h}, q^{h}]$ satisfying $X = S^{h}[i]$. 

The fourth phase consists of three steps for each nonterminal $X$. 
Let $u$ be the node corresponding to the $i$-th nonterminal of sequence $S^{h}$. 
This node $u$ is represented as a node $v \in \mathcal{U}_{\RR}$ in the RR-DAG. 
In the first step, 
we obtain the node $u$ by traversing the derivation tree of RLSLP $\mathcal{G}^{R}$. 
Similar to the first step of the third phase, 
this step can be executed in $O(1)$ time. 

From the definition of RR-DAG, 
the derivation tree of RLSLP $\mathcal{G}^{R}$ contains the longest path 
$\mathbb{P} = u_{1} \rightarrow u_{2} \rightarrow \cdots \rightarrow u_{k}$ 
satisfying two conditions: 
(A) it starts at node $u$ (i.e., $u_{1} = u$); 
(B) $u_{j}$ is the single child of node $u_{j-1}$ for all $x \in [2, k]$, 
and any node of the RR-DAG does not correspond to the node $u_{j}$ (i.e,. the nonterminal of $u_{j}$ is assigned $-1$). 
For the last node $u_{k}$ labeled with $X_{k}$ in the path $\mathbb{P}$,  
the pair of key and value corresponding to the nonterminal $X_{k}$ is stored in a B-tree, which is a component of 
the data structure for the RR-DAG. 
In the second step, we remove this pair from the B-tree in $O(\log |\mathcal{U}_{\RR}|)$ time. 
The key corresponding the nonterminal $X_{k}$ can be computed in $O(1)$ time using 
the children of the node $v$ of the RR-DAG. 
The second step takes $O(\log |\mathcal{U}_{\RR}|)$ time in total. 

The node $v$ is stored as an element of the doubly linked list representing the RR-DAG. 
In the third step, we remove the node $v$ from the doubly linked list. 
This removal takes $O(\log |\mathcal{U}_{\RR}|)$ time. 

The fourth phase takes $O(|\mathcal{V} \setminus \mathcal{V}^{\prime}| \log |\mathcal{U}_{\RR}|)$ time in total. 
$|\mathcal{V} \setminus \mathcal{V}^{\prime}| = O(H^{2})$ follows from 
$|\mathcal{V} \setminus \mathcal{V}^{\prime}| \leq \sum_{h=0}^{H} |[p^{h}, q^{h}]|$ 
and $\sum_{h=0}^{H} |[p^{h}, q^{h}]| = O(H^{2})$. 
We already proved $\log |\mathcal{U}_{\RR}| = O(\log nH)$. 
Therefore, the running time of the fourth phase can be bounded by $O(H^{2} \log n H)$. 

\paragraph{Phase (v).}
In the fifth phase, 
we update the data structures for the RR-DAG based on Modification 3 
for each nonterminal $X \in \mathcal{V}^{\prime} \setminus \mathcal{V}$ 
satisfying either $X = S^{\prime}$ or $\assign(X) \in \{ 0, 1 \}$. 
From the procedure of the second phase, 
there exists a pair of two integers $h \in [0, H^{\prime}]$ and $i \in [1, |Q^{h}|]$ satisfying $X = Q^{h}[i]$. 
Let $u$ be the node corresponding to the $i$-th nonterminal of sequence $Q^{h}$ 
in the derivation tree of RLSLP $\mathcal{G}^{R}_{\ins}$. 

The fifth phase consists of three steps for each pair $(h, i)$.
In the first step, 
we insert a new element into the doubly linked list representing the RR-DAG as the last element. 
This element represents a node $v$ labeled $X$ in the RR-DAG. 
The last element of the doubly linked list stores the results of the label functions 
$L_{\RR}(v)$, $L_{\level}(v)$, $L_{\prule}(v)$, $L_{\pathP}(v)$, $L_{\assign}(v)$, and $L_{\length}(v)$ introduced in Section~\ref{subsec:rrdag}. 
The results of these label functions can be computed in $O(H^{\prime})$ time. 
In addition, 
the last element of the doubly linked list temporarily stores the integer zero as the result of the label function $L_{\vOcc}(v)$. 
This value is appropriately updated in the next phase. 

In the second step, 
we create directed edges starting at node $v$. 
These directed edges are stored in the last element of the doubly linked list representing the RR-DAG. 
This step can be executed in $O(H^{\prime})$ time. 

Similar to the fourth phase, 
the derivation tree of RLSLP $\mathcal{G}^{R}_{\ins}$ contains the longest path 
$\mathbb{P} = u_{1} \rightarrow u_{2} \rightarrow \cdots \rightarrow u_{k}$ 
satisfying two conditions: 
(A) it starts at node $u$; 
(B) $u_{j}$ is the single child of node $u_{j-1}$ for all $x \in [2, k]$, 
and any node of the RR-DAG does not correspond to the node $u_{j}$. 
In the third step, 
for the last node $u_{k}$ labeled with $X_{k}$ in the path $\mathbb{P}$,  
the pair of key and value corresponding to the nonterminal $X_{k}$ is inserted into an appropriate B-tree in the data structure for the RR-DAG. 
This pair can be computed in $O(1)$ time using the children of the node $v$ of the RR-DAG. 
The third step takes $O(\log |\mathcal{U}^{\prime}_{\RR}|)$ time 
for the number $|\mathcal{U}^{\prime}_{\RR}|$ of nodes in the RR-DAG of RLSLP $\mathcal{G}^{R}_{\ins}$. 
Here, $|\mathcal{U}^{\prime}_{\RR}| = O(nH^{\prime})$ holds, which is similar to $|\mathcal{U}_{\RR}| = O(nH)$.

The fifth phase takes $O(|\mathcal{V}^{\prime} \setminus \mathcal{V}| (H^{\prime} + \log (nH^{\prime})))$ time in total. 
$|\mathcal{V}^{\prime} \setminus \mathcal{V}| = O(H^{\prime 2})$ holds 
because $|\mathcal{V}^{\prime} \setminus \mathcal{V}| \leq \sum_{h=0}^{H^{\prime}} |Q^{h}|$ holds, 
and $\sum_{h=0}^{H^{\prime}} |Q^{h}| = O(H^{\prime 2})$ follows from Lemma~\ref{lem:Q_sequence_length}. 
Therefore, the running time of the fifth phase is $O(H^{\prime 3} + H^{\prime 2} \log (nH))$. 

After executing the fifth phase, 
we can support the queries of Lemma~\ref{lem:basic_operations_on_dev_tree} on the derivation tree of RLSLP $\mathcal{G}^{R}_{\ins}$. 

\paragraph{Phase (vi).}
In the sixth phase, 
we update the data structures for the RR-DAG based on Modification 4 
for the $(H^{\prime}+1)$ sequences $Q^{0}$, $Q^{1}$, $\ldots$, $Q^{H^{\prime}}$. 
The third phase consists of two steps for each pair $(h, i)$ of two integers $h \in [0, H^{\prime}]$ and 
$i \in [1, |Q^{h}|]$ satisfying either $Q^{h}[i] = S^{\prime}$ or $\assign(Q^{h}[i]) \in \{ 0, 1 \}$. 
Let $u$ be the node corresponding to the $i$-th nonterminal of sequence $Q^{h}$ 
in the derivation tree of RLSLP $\mathcal{G}^{R}_{\ins}$. 
In the first step, 
we obtain the node $u$ by traversing the derivation tree. 
This traversal can be executed in $O(H^{\prime})$ time by the queries of Lemma~\ref{lem:basic_operations_on_dev_tree}. 

The node $u$ is represented as a node $v \in \mathcal{U}_{\RR}$ in the RR-DAG. 
This node $v$ is stored as an element of the doubly linked list representing the RR-DAG, 
and the element stores the integer $W$ obtained from label function $L_{\vOcc}(v)$. 
In the second step, 
we change the integer $W$ to $W+1$. 
The second step can be executed in $O(1)$ time. 
Therefore, 
the sixth phase takes $O(\sum_{h=0}^{H^{\prime}} |Q^{h}|)$ time in total (i.e., $O(H^{\prime 2})$ time).

%the third phase of the dynamic restricted recompression. 
%Therefore, the first step can be executed in $O(1)$ time. 
%the triplet $\repr(u)$ by path query $\pathQ(x + \sum_{w = 1}^{i-1} |\val(Q^{h}[w])|)$ on the derivation tree of RLSLP $\mathcal{G}^{R}_{\ins}$. 
%This path query takes $O(H^{\prime})$ time. 

%Then, $u$ is represented as 
%Consider a pair $(h, i)$ of two integers $h \in [0, H^{\prime}]$ and 
%$i \in [1, |Q^{h}|]$ satisfying either $Q^{h}[i] = S^{\prime}$ or $\assign(Q^{h}[i]) \in \{ 0, 1 \}$. 
%Then, the RR-DAG of RLSLP $\mathcal{G}^{R}$ contains a node $v$ labeled $Q^{h}[i]$. 
%Let $u$ be the node corresponding to the $i$-th nonterminal of sequence $Q^{h}$ 
%in the derivation tree of RLSLP $\mathcal{G}^{R}_{\ins}$. 
%Here, $\repr(u) = (x + \sum_{w = 1}^{i-1} |\val(Q^{h}[w])|, h, v)$ holds 
%for the node representation $\repr(u)$ of node $u$ and 
%the starting position $x$ of the substring derived from the $p^{h}$-th nonterminal of sequence $S^{h}$ in input string $T$. 
%The third phase consists of two steps for each pair $(h, i)$.

%The node $v$ is represented as an element of the doubly linked list representing 
%the RR-DAG, 
%and the element stores the integer $W$ obtained from label function $L_{\vOcc}(v)$. 
%In the second step, 
%we change the integer $W$ to $W+1$. 
%The triplet $\repr(u)$ contains a pointer to the node $v$. 
%The second step can be executed in $O(1)$ time. 
%Therefore, 
%the sixth phase takes $O(\sum_{h=0}^{H^{\prime}} |Q^{h}|)$ time in total (i.e., $O(H^{\prime 2})$ time). 

Finally, we can update the data structures for the RR-DAG of the RLSLP $\mathcal{G}^{R}$ in 
$O((\max \{ H$, $H^{\prime}, \log (nHH^{\prime}) \})^{3})$ time. 
Conversely, 
we can obtain the data structures for the RR-DAG of the RLSLP $\mathcal{G}^{R}_{\ins}$ 
by modifying the data structures for the RR-DAG of the RLSLP $\mathcal{G}^{R}$ in the same time 
if we store the differences between the two RR-DAGs of $\mathcal{G}^{R}$ and 
$\mathcal{G}^{R}_{\ins}$ in $O((\max \{ H, H^{\prime}, \log (nHH^{\prime}) \})^{3} B)$ bits of space for machine word size $B$. 
Therefore, we obtain the following lemma. 

\begin{lemma}\label{lem:dynamic_rrdag_summary}
Consider the two RLSLPs $\mathcal{G}^{R}$ and $\mathcal{G}^{R}_{\ins}$ of Theorem~\ref{theo:update1}, which derive input string $T$ and string $T^{\prime}$, respectively. 
After executing the dynamic restricted recompression, 
we can update the dynamic data structures for the RR-DAG of RLSLP $\mathcal{G}^{R}$ (Section~\ref{subsubsec:rrdag_ds}) 
based on the changes from RLSLP $\mathcal{G}^{R}$ to $\mathcal{G}^{R}_{\ins}$. 
This update takes $O((\max \{ H, H^{\prime}, \log (nHH^{\prime}) \})^{3})$ time. 
If necessary, 
we can obtain the data structures for the RR-DAG of the RLSLP $\mathcal{G}^{R}_{\ins}$ 
by modifying the data structures for the RR-DAG of the RLSLP $\mathcal{G}^{R}$ in the same time. 
This modification needs additional $O((\max \{ H, H^{\prime}, \log (nHH^{\prime}) \})^{3} B)$ bits of space.
\end{lemma}

\subsubsection{Relationship between two RLSLPs \texorpdfstring{$\mathcal{G}^{R}$}{} and \texorpdfstring{$\mathcal{G}^{R}_{\ins}$}{}}
We explain the relationship between two RLSLPs $\mathcal{G}^{R}$ and $\mathcal{G}^{R}_{\ins}$. 
The following lemma shows that 
Lemma~\ref{lem:rr_property} holds between $\mathcal{G}^{R}$ and $\mathcal{G}^{R}_{\ins}$.

\begin{lemma}\label{lem:dynamic_rr_property}
For any triplet $(h, i, i^{\prime})$ with $h \in [0, \min \{ H, H^{\prime} \} - 1]$, $i \in [1, |S^{h}|]$, and $i^{\prime} \in [1, |S^{h}_{\ins}|]$, 
let $u_{i}$ be the node in the derivation tree of $\mathcal{G}^{R}$ that 
correspond to the nonterminals $S^{h}[i]$. 
Similarly, let $u_{\ins, i^{\prime}}$ be the node in the derivation tree of $\mathcal{G}^{R}_{\ins}$ that 
correspond to the nonterminals $S^{h}_{\ins}[i^{\prime}]$. 
Let $T[x_{i}..y_{i}]$ and $T^{\prime}[x_{\ins, i^{\prime}}..y_{\ins, i^{\prime}}]$ be 
the two substrings derived from the two nodes $u_{i}$ and $u_{\ins, i^{\prime}}$, respectively. 
Let $\alpha = \min \{ |\val(S^{h}[i])|, |\val(S^{h}_{\ins}[i^{\prime}])| \}$. 

Given these definitions, the following properties hold:
\begin{enumerate}[label=\textbf{(\roman*)}]
\item \label{enum:dynamic_rr_property:1} 
if $S^{h}[i] = S^{h}_{\ins}[i^{\prime}]$ and $S^{h}[i+1] = S^{h}_{\ins}[i^{\prime}+1]$, 
then $u_{i}$ is the rightmost child of its parent if and only if 
$u_{\ins, i^{\prime}}$ is the rightmost child of its parent; 
\item \label{enum:dynamic_rr_property:2} 
for a pair of two integers $d, d^{\prime} \geq 1$, 
if two sequences $S^{h}[i..i+d-1]$ and $S^{h}_{\ins}[i^{\prime}..i^{\prime}+d^{\prime}-1]$ derive the same string 
(i.e., $\val(S^{h}[i]) \cdot \val(S^{h}[i+1]) \cdot \cdots \cdot \val(S^{h}[i+d-1]) = \val(S^{h}_{\ins}[i^{\prime}]) \cdot \val(S^{h}_{\ins}[i^{\prime}+1]) \cdot \cdots \cdot \val(S^{h}_{\ins}[i^{\prime}+d^{\prime}-1])$), 
then $S^{h}[i..i+d-1] = S^{h}_{\ins}[i^{\prime}..i^{\prime}+d^{\prime}-1]$;
\item \label{enum:dynamic_rr_property:right} 
if $T[x_{i}..x_{i} + \alpha + (\sum_{t = 1}^{h} \lfloor \mu(t) \rfloor)] = T[x_{\ins, i^{\prime}}..x_{\ins, i^{\prime}} + \alpha + (\sum_{t = 1}^{h} \lfloor \mu(t) \rfloor)]$ holds, 
then $S^{h}[i] = S^{h}_{\ins}[i^{\prime}]$.
\item \label{enum:dynamic_rr_property:left}
if $T[y_{i} - \alpha - (\sum_{t = 1}^{h} \lfloor \mu(t) \rfloor)..y_{i}] = T[y_{\ins, i^{\prime}} - \alpha - (\sum_{t = 1}^{h} \lfloor \mu(t) \rfloor)..y_{\ins, i^{\prime}}]$ holds, 
then $S^{h}[i] = S^{h}_{\ins}[i^{\prime}]$.
\end{enumerate}
\end{lemma}
\begin{proof}
See Section~\ref{subsubsec:proof_dynamic_rr_property}.
\end{proof}

%We explain properties of the RLSLP $\mathcal{G}^{R}_{\ins} = (\mathcal{V}^{\prime}, \Sigma, \mathcal{D}^{\prime}, S^{\prime})$ obtained by the dynamic restricted recompression algorithm. 

%\begin{lemma}\label{lem:RB_GR_GRINS}
%    The following three statements hold for two RLSLPs 
%    $\mathcal{G}^{R} = (\mathcal{V}, \Sigma, \mathcal{D}, S)$ and 
%    $\mathcal{G}^{R}_{\ins} = (\mathcal{V}^{\prime}, \Sigma, \mathcal{D}^{\prime}, S^{\prime})$. 
%
%    \begin{enumerate}[label=\textbf{(\roman*)}]
%    \item \label{enum:RB_GR_GRINS:1} $\mathcal{V}^{\prime} = (\mathcal{V} \cup \mathcal{V}_{+}) \setminus \mathcal{V}_{-}$;
%   \item \label{enum:RB_GR_GRINS:2} $\mathcal{D}^{\prime} = (\mathcal{D} \cup \mathcal{D}_{+}) \setminus \mathcal{D}_{-}$;
%    \item \label{enum:RB_GR_GRINS:3} consider two production rules $X \rightarrow \expr \in \mathcal{D}$ 
%    and $X^{\prime} \rightarrow \expr^{\prime} \in \mathcal{D}^{\prime}$. 
%    Then, $X = X^{\prime} \Leftrightarrow \expr = \expr^{\prime}$.
%    \end{enumerate}
%\end{lemma}
%\begin{proof}
%    \unproven
%\end{proof}

Consider the function $f_{\interval}$ and sequence $A(s, e)$ of intervals introduced in Section~\ref{sec:RASSO} for an interval $[s, e] \in \Delta$ in input string $T$. 
Let $f^{\prime}_{\interval}$ be the function $f_{\interval}$ defined using RLSLP $\mathcal{G}^{R}_{\ins}$ instead of $\mathcal{G}^{R}$. 
Similarly, 
let $A^{\prime}(s, e)$ be sequence $A(s, e)$ of intervals defined using function $f^{\prime}_{\interval}$ instead of 
function $f_{\interval}$. 

The following two lemmas state the relationship between two sequences $A(s, e)$ and $A^{\prime}(s, e)$. 

\begin{lemma}\label{lem:dynamicP_rec_function_syncro_property}
Consider two intervals $[s, e] \subseteq [1, n]$ ($s < e$) and $[s^{\prime}, e^{\prime}] \subseteq [1, n+1]$ ($s^{\prime} < e^{\prime}$) in two strings $T$ and $T^{\prime}$, respectively. 
Let $[s^{0}, e^{0}]$, $[s^{1}, e^{1}]$, $\ldots$, $[s^{k}, e^{k}]$ and 
$[s^{\prime 0}, e^{\prime 0}]$, $[s^{\prime 1}, e^{\prime 1}]$, $\ldots$, $[s^{\prime k^{\prime}}, e^{\prime k^{\prime}}]$ 
be two sequences $A(s, e)$ and $A^{\prime}(s^{\prime}, e^{\prime})$ of intervals, respectively. 
For each integer $h \in [0, k]$, 
let $T[x^{h}_{s}..y^{h}_{s}]$ and $T[x^{h}_{e}..y^{h}_{e}]$ be the two substrings 
derived from the $s^{h}$-th and $e^{h}$-th nonterminals of sequence $S^{h}$ in input string $T$, respectively. 
Similarly, 
for each integer $h \in [0, k^{\prime}]$, 
let $T^{\prime}[x^{\prime h}_{s}..y^{\prime h}_{s}]$ and $T^{\prime}[x^{\prime h}_{e}..y^{\prime h}_{e}]$ be the two substrings 
derived from the $s^{\prime h}$-th and $e^{\prime h}$-th nonterminals of sequence $S^{h}_{\ins}$ 
in input string $T^{\prime}$, respectively.     
If $T[s..e] = T^{\prime}[s^{\prime}..e^{\prime}]$, 
then then the following two statements hold: 
\begin{enumerate}[label=\textbf{(\roman*)}]
    \item \label{enum:dynamicP_rec_function_syncro_property:1}     
    $S^{h}[s^{h}+1..e^{h}-1] = S^{h}_{\ins}[s^{\prime h}+1..e^{\prime h}-1]$, 
    $|[s, x^{h}_{s}]| = |[s^{\prime}, x^{\prime h}_{s}]|$, 
    and $|[x^{h}_{e}, e]| = |[x^{\prime h}_{e}, e^{\prime}]|$
    for each integer $h \in [0, \min \{ k, k^{\prime} \}]$;
    \item \label{enum:dynamicP_rec_function_syncro_property:2} $k = k^{\prime}$.
\end{enumerate}
\end{lemma}
\begin{proof}
    See Section~\ref{subsubsec:proof_dynamicP_rec_function_syncro_property}.
\end{proof}

\subsubsection{Proof of Lemma~\ref{lem:dynamic_rr_property}}\label{subsubsec:proof_dynamic_rr_property}
\begin{proof}[Proof of Lemma~\ref{lem:dynamic_rr_property}~\ref{enum:dynamic_rr_property:1}]
Lemma~\ref{lem:dynamic_rr_property}~\ref{enum:dynamic_rr_property:1} corresponds to 
Lemma~\ref{lem:rr_property}~\ref{enum:rr_property:1}. 
We can prove Lemma~\ref{lem:dynamic_rr_property} \ref{enum:dynamic_rr_property:1} using 
the same approach as for Lemma~\ref{lem:rr_property}~\ref{enum:rr_property:1}. 
\end{proof}

\begin{proof}[Proof of Lemma~\ref{lem:dynamic_rr_property}~\ref{enum:dynamic_rr_property:2}]
We prove Lemma~\ref{lem:dynamic_rr_property}~\ref{enum:dynamic_rr_property:2} by induction on $h$. 
Let $P$ be the string derived by sequence $S^{h}[i..i+d-1]$ 
(i.e., $P = \val(S^{h}[i]) \cdot \val(S^{h}[i+1]) \cdot \cdots \cdot \val(S^{h}[i+d-1])$). 
For the base case $h = 0$, 
the length of the string $P$ is $d$. 
This is because each nonterminal of sequence $S^{0}$ derives a character. 
$d^{\prime} = d$ holds 
because sequence $S^{0}_{\ins}[i^{\prime}..i^{\prime}+d^{\prime}-1]$ derive the string $P$. 
For each integer $s \in [1, d]$, 
the two nonterminals $S^{0}[i+s-1]$ and $S^{0}_{\ins}[i^{\prime} + s - 1]$ derive the same character $P[s]$. 
That is, 
two sets $\mathcal{D}$ and $\mathcal{D}^{\prime}$ contain two production rules 
$S^{0}[i+s-1] \rightarrow P[s]$ and $S^{0}_{\ins}[i^{\prime} + s - 1] \rightarrow P[s]$, respectively. 
$S^{0}[i+s-1] = S^{0}_{\ins}[i^{\prime} + s - 1]$ follows from 
$S^{0}[i+s-1] \rightarrow P[s]$ and $S^{0}_{\ins}[i^{\prime} + s - 1] \rightarrow P[s]$. 
Therefore, $S^{h}[i..i+d-1] = S^{0}_{\ins}[i^{\prime}..i^{\prime}+d^{\prime}-1]$ holds. 

For the inductive step, consider $h \in [1, \min \{ H, H^{\prime} \} - 1]$. 
Let $u_{i+d-1}$ be the node in the derivation tree of $\mathcal{G}^{R}$ that 
correspond to the nonterminals $S^{h}[i+d-1]$. 
Similarly, let 
let $u_{\ins, i^{\prime}+d-1}$ be the node in the derivation tree of $\mathcal{G}^{R}_{\ins}$ that 
correspond to the nonterminals $S^{h}_{\ins}[i^{\prime}]$. 
For this proof, 
we use four positions $p$, $q$, $p^{\prime}$, and $q^{\prime}$ defined as follows: 
\begin{itemize}
    \item $p$ is a position of sequence $S^{h-1}$ such that 
    the $p$-th nonterminal of sequence $S^{h-1}$ corresponds to the leftmost child of node $u_{i}$;
    \item $q$ is a position of sequence $S^{h-1}$ such that 
    the $q$-th nonterminal of sequence $S^{h-1}$ corresponds to the rightmost child of node $u_{i+d-1}$;
    \item $p^{\prime}$ is a position of sequence $S^{h-1}_{\ins}$ such that 
    the $p^{\prime}$-th nonterminal of sequence $S^{h-1}_{\ins}$ corresponds to the leftmost child of node $u_{\ins, i^{\prime}}$;
    \item $q^{\prime}$ is a position of sequence $S^{h-1}_{\ins}$ such that 
    the $q^{\prime}$-th nonterminal of sequence $S^{h-1}_{\ins}$ corresponds to the rightmost child of node $u_{\ins, i^{\prime} + d^{\prime} - 1}$.    
\end{itemize}
Let $k_{s}$ be the number of children of the node corresponding to the $(i+s-1)$-th nonterminal $S^{h}[i+s-1]$ 
of sequence $S^{h}$ for each integer $s \in [1, d]$. 
Then, 
(A) $S^{h-1}[p..q]$ derives string $P$, 
(B) $k_{1} + k_{2} + \cdots k_{d} = |[p, q]|$ holds, 
and 
(C) set $\mathcal{D}$ contains production rule $S^{h}[i+s-1] \rightarrow S^{h-1}[p + (\sum_{w = 1}^{s-1} k_{w})..p + (\sum_{w = 1}^{s} k_{w}) - 1]$ for each integer $s \in [1, d]$. 
This is because 
sequence $S^{h-1}[p..q]$ corresponds to the children of the nodes corresponding to the sequence $S^{h}[i..i+d-1]$. 

Similarly, 
let $k^{\prime}_{s}$ be the number of children of the node corresponding to the $(i^{\prime}+s-1)$-th nonterminal $S^{h}_{\ins}[i^{\prime}+s-1]$ of sequence $S^{h}_{\ins}$ for each integer $s \in [1, d^{\prime}]$. 
Then, 
(a) sequence $S^{h-1}_{\ins}[p^{\prime}..q^{\prime}]$ derives string $P$, 
(b) $k^{\prime}_{1} + k^{\prime}_{2} + \cdots k^{\prime}_{d^{\prime}} = |[p^{\prime}, q^{\prime}]|$ holds, 
and (c) $\mathcal{D}^{\prime}$ contains production rule $S^{h}_{\ins}[i^{\prime}+s-1] \rightarrow S^{h-1}_{\ins}[p^{\prime} + (\sum_{w = 1}^{s-1} k^{\prime}_{w})..p + (\sum_{w = 1}^{s} k^{\prime}_{w}) - 1]$ for each integer $s \in [1, d^{\prime}]$. 

We can apply Lemma~\ref{lem:dynamic_rr_property}~\ref{enum:dynamic_rr_property:2} 
to the two sequences $S^{h-1}[p..q]$ and $S^{h-1}_{\ins}[p^{\prime}..q^{\prime}]$ 
because the two sequences derive the same string. 
Then, this lemma shows that two sequences $S^{h-1}[p..q]$ and $S^{h-1}_{\ins}[p^{\prime}..q^{\prime}]$ are equal. 
Let $m = |[p, q]|$ for simplicity. 
Because of $S^{h-1}[p..q] = S^{h-1}_{\ins}[p^{\prime}..q^{\prime}]$, 
we can apply Lemma~\ref{lem:dynamic_rr_property}~\ref{enum:dynamic_rr_property:2} to 
two nonterminals $S^{h-1}[p+s-1]$ and $S^{h-1}_{\ins}[p^{\prime}+s-1]$ for each integer $s \in [1, m-1]$. 
Then, Lemma~\ref{lem:dynamic_rr_property}~\ref{enum:dynamic_rr_property:2} indicates that 
$d = d^{\prime}$, $k_{1} = k^{\prime}_{1}$, $k_{2} = k^{\prime}_{2}$, $\ldots$, $k_{d} = k^{\prime}_{d}$. 

For each integer $s \in [1, d]$, 
let $\expr_{s} = S^{h-1}[p + (\sum_{w = 1}^{s-1} k_{w})..p + (\sum_{w = 1}^{s} k_{w}) - 1]$. 
Then, $S^{h}[i+s-1] \rightarrow \expr_{s} \in \mathcal{D}$ 
and $S^{h}_{\ins}[i^{\prime}+s-1] \rightarrow \expr_{s} \in \mathcal{D}^{\prime}$ hold. 
$S^{h}[i+s-1] = S^{h}_{\ins}[i^{\prime}+s-1]$ follows from 
$S^{h}[i+s-1] \rightarrow \expr_{s}$ and $S^{h}_{\ins}[i^{\prime}+s-1] \rightarrow \expr_{s}$. 
Therefore, $S^{h}[i..i+d-1] = S^{0}_{\ins}[i^{\prime}..i^{\prime}+d^{\prime}-1]$ holds for the inductive step. 
Finally, we obtain Lemma~\ref{lem:dynamic_rr_property}~\ref{enum:dynamic_rr_property:2} without any assumption. 
\end{proof}

\begin{proof}[Proof of Lemma~\ref{lem:dynamic_rr_property}~\ref{enum:dynamic_rr_property:right}]
Lemma~\ref{lem:dynamic_rr_property}~\ref{enum:dynamic_rr_property:right} corresponds to 
Lemma~\ref{lem:rr_property}~\ref{enum:rr_property:right}. 
We proved Lemma \ref{lem:rr_property}~\ref{enum:rr_property:right} using Lemma~\ref{lem:rr_property}~\ref{enum:rr_property:2}. 
Lemma~\ref{lem:dynamic_rr_property}~\ref{enum:dynamic_rr_property:2} corresponds to Lemma~\ref{lem:rr_property}~\ref{enum:rr_property:2}. 
Therefore, Lemma~\ref{lem:dynamic_rr_property}~\ref{enum:dynamic_rr_property:right} can be proved using the same approach as for Lemma~\ref{lem:rr_property}~\ref{enum:rr_property:right}. 
\end{proof}

\begin{proof}[Proof of Lemma~\ref{lem:dynamic_rr_property}~\ref{enum:dynamic_rr_property:left}]
Lemma~\ref{lem:dynamic_rr_property}~\ref{enum:dynamic_rr_property:left} can be proven by the same approach used in the proof of Lemma~\ref{lem:dynamic_rr_property}~\ref{enum:dynamic_rr_property:right}.
\end{proof}

\subsubsection{Proof of Lemma~\ref{lem:dynamicP_rec_function_syncro_property}}\label{subsubsec:proof_dynamicP_rec_function_syncro_property}
The proof of Lemma~\ref{lem:dynamicP_rec_function_syncro_property} is as follows.

\begin{proof}[Proof of Lemma~\ref{lem:dynamicP_rec_function_syncro_property}~\ref{enum:dynamicP_rec_function_syncro_property:1}]
We prove Lemma~\ref{lem:dynamicP_rec_function_syncro_property}~\ref{enum:dynamicP_rec_function_syncro_property:1} by induction on $h$. 
For the base case $h = 0$, 
$[s^{0}, e^{0}] = [s, e]$, $[s^{\prime 0}, e^{\prime 0}] = [s^{\prime}, e^{\prime}]$ 
and $S^{h}[s^{0}+1..e^{0}] = S^{h}_{\ins}[s^{\prime h}+1..e^{\prime h}]$ 
follow from the definitions of two sequence $A(s, e)$ and $A^{\prime}(s^{\prime}, e^{\prime})$. 
Therefore, Lemma~\ref{lem:dynamicP_rec_function_syncro_property}~\ref{enum:dynamicP_rec_function_syncro_property:1} holds. 

For the inductive step, 
consider $h \in [1, \min \{ k, k^{\prime} \}]$. 
Then, the following three statements hold by the inductive assumption: 
\begin{itemize}
    \item $|[s, x^{h-1}_{s}]| = |[s^{\prime}, x^{\prime h-1}_{s}]|$;
    \item $|[x^{h-1}_{e}, e]| = |[x^{\prime h-1}_{e}, e^{\prime}]|$;
    \item $S^{h-1}[s^{h-1}+1..e^{h-1}-1] = S^{h-1}_{\ins}[s^{\prime h-1}+1..e^{\prime h-1}-1]$.
\end{itemize}

The following four statements are used to prove Lemma~\ref{lem:dynamicP_rec_function_syncro_property}~\ref{enum:dynamicP_rec_function_syncro_property:1}: 
\begin{itemize}
    \item $|[x^{h-1}_{s}, x^{h}_{s}]| \leq |[x^{\prime h-1}_{s}, x^{\prime h}_{s}]|$;
    \item $|[x^{h-1}_{s}, x^{h}_{s}]| \geq |[x^{\prime h-1}_{s}, x^{\prime h}_{s}]|$;
    \item $|[x^{h}_{e}, x^{h-1}_{e}]| \leq |[x^{\prime h}_{e}, x^{\prime h-1}_{e}]|$;
    \item $|[x^{h}_{e}, x^{h-1}_{e}]| \geq |[x^{\prime h}_{e}, x^{\prime h-1}_{e}]|$.
\end{itemize}

\textbf{Proof of $|[x^{h-1}_{s}, x^{h}_{s}]| \leq |[x^{\prime h-1}_{s}, x^{\prime h}_{s}]|$.}
Let $\alpha$ be a position of sequence $S^{h}$ satisfying 
$|[s, x^{h}_{\alpha}]| = |[s^{\prime}, x^{\prime h}_{s}]|$ 
for the substring $T[x^{h}_{\alpha}..y^{h}_{\alpha}]$ 
derived from the $\alpha$-th nonterminal of sequence $S^{h}$ in input string $T$. 

Proposition~\ref{prop:sync_set_sub_property1} shows that 
$x^{\prime h-1}_{s} \leq x^{\prime h}_{s} \leq x^{\prime h-1}_{e}$ holds. 
Since $x^{\prime h-1}_{s} \leq x^{\prime h}_{s} \leq x^{\prime h-1}_{e}$, 
there exists an integer $d^{\prime} \in [0, e^{\prime h-1} - s^{\prime h-1}]$ 
satisfying 
$x^{\prime h-1}_{s+d} = x^{\prime h}_{s+d}$ 
for the substring $T^{\prime}[x^{\prime h-1}_{s+d}..y^{\prime h-1}_{s+d}]$ 
derived from the $(s^{\prime h} + d^{\prime})$-th nonterminal of sequence $S^{h}_{\ins}$ in input string $T^{\prime}$. 
Let $d \in [0, e^{h-1} - s^{h-1}]$ be an integer satisfying 
$|[x^{h-1}_{s}, x^{h-1}_{s+d}]| = |[x^{\prime h-1}_{s}, x^{\prime h-1}_{s+d}]|$ 
for the substring $T[x^{h-1}_{s+d}..y^{h-1}_{s+d}]$ 
derived from the $(s^{h} + d)$-th nonterminal of sequence $S^{h}$ in input string $T$. 
Such integer $d$ exists, and $d = d^{\prime}$ hold 
because $S^{h-1}[s^{h-1}+1..e^{h-1}-1] = S^{h-1}_{\ins}[s^{\prime h-1}+1..e^{\prime h-1}-1]$ holds. 

$|[x^{h-1}_{s}, x^{h}_{s}]| \leq |[x^{\prime h-1}_{s}, x^{\prime h}_{s}]|$ holds 
if the position $\alpha$ exists and satisfies at least one of four conditions (i), (ii), (iii), and (iv) of Definition~\ref{def:f_interval} for function $f_{\interval}(s^{h-1}, e^{h-1})$. 
From the definition of sequence $A^{\prime}(s^{\prime}, e^{\prime})$, 
the position $s^{\prime h}$ satisfies at least one of four conditions (i), (ii), (iii), and (iv) of Definition~\ref{def:f_interval} for function $f^{\prime}_{\interval}(s^{\prime h-1}, e^{\prime h-1})$. 

If the position $s^{\prime h}$ satisfies condition (i), 
then $x^{\prime h}_{s} \in [x^{\prime h-1}_{s} + 1, x^{\prime h-1}_{e} - 1]$. 
In this case, $1 \leq d^{\prime} \leq e^{\prime h-1} - s^{\prime h-1} - 1$ holds. 
$S^{h-1}[s^{h-1} + d - 1] = S^{h-1}_{\ins}[s^{\prime h-1} + d^{\prime} - 1]$ and 
$S^{h-1}[s^{h-1} + d] = S^{h-1}_{\ins}[s^{\prime h-1} + d^{\prime}]$ hold 
because $d = d^{\prime}$, $1 \leq d^{\prime} \leq e^{\prime h-1} - s^{\prime h-1} - 1$, 
and $S^{h-1}[s^{h-1}+1..e^{h-1}-1] = S^{h-1}_{\ins}[s^{\prime h-1}+1..e^{\prime h-1}-1]$. 
Since $S^{h-1}[s^{h-1} + d - 1..s^{h-1} + d] = S^{h-1}_{\ins}[s^{\prime h-1} + d^{\prime} - 1..s^{\prime h-1} + d^{\prime}]$, 
Lemma~\ref{lem:dynamic_rr_property}~\ref{enum:dynamic_rr_property:1} shows that 
the position $\alpha$ exists. 
$x^{h}_{\alpha} \in [x^{h-1}_{s} + 1, x^{h-1}_{e} - 1]$ holds 
because $x^{h}_{\alpha} = x^{h-1}_{s+d}$ and $x^{h-1}_{s+d} \in [x^{h-1}_{s} + 1, x^{h-1}_{e} - 1]$ hold.  
Therefore, 
the position $\alpha$ satisfies condition (i) of Definition~\ref{def:f_interval} for function $f_{\interval}(s^{h-1}, e^{h-1})$.

If the position $s^{\prime h}$ satisfies condition (ii) and $d^{\prime} = 0$, 
then $d^{\prime} \leq e^{\prime h-1} - s^{\prime h-1} - 1$ 
and $|[x^{\prime h-1}_{s+d}, y^{\prime h-1}_{s+d}]| > \lfloor \mu(h) \rfloor$ hold. 
$S^{h-1}[s^{h-1} + d] = S^{h-1}_{\ins}[s^{\prime h-1} + d^{\prime}]$ 
follows from $S^{h-1}[s^{h-1}+1..e^{h-1}-1] = S^{h-1}_{\ins}[s^{\prime h-1}+1..e^{\prime h-1}-1]$ 
and $d^{\prime} \leq e^{\prime h-1} - s^{\prime h-1} - 1$. 
$|[x^{h-1}_{s+d}, y^{h-1}_{s+d}]| > \lfloor \mu(h) \rfloor$ follows from 
$S^{h-1}[s^{h-1} + d] = S^{h-1}_{\ins}[s^{\prime h-1} + d^{\prime}]$ 
and $|[x^{\prime h-1}_{s+d}, y^{\prime h-1}_{s+d}]| > \lfloor \mu(h) \rfloor$. 
Since $|[x^{h-1}_{s+d}, y^{h-1}_{s+d}]| > \lfloor \mu(h) \rfloor$, 
the assignment of the nonterminal $S^{h-1}[s^{h-1} + d]$ is $-1$. 
Since $\assign(S^{h-1}[s^{h-1} + d]) = -1$, 
Lemma~\ref{lem:rr_class}~\ref{enum:rr_class:3} and Lemma~\ref{lem:rr_class}~\ref{enum:rr_class:4} indicate that 
the position $\alpha$ exists. 
The position $\alpha$ satisfies condition (ii) of Definition~\ref{def:f_interval} for function $f_{\interval}(s^{h-1}, e^{h-1})$ because $|[x^{h-1}_{s+d}, y^{h-1}_{s+d}]| > \lfloor \mu(h) \rfloor$ and $s^{h}+d \in [s^{h}, e^{h}-1]$ hold.

If the position $s^{\prime h}$ satisfies condition (ii) and $d^{\prime} \neq 0$, 
then $d^{\prime} \leq e^{\prime h-1} - s^{\prime h-1} - 1$ holds. 
Similar to condition (i), 
we can show that 
the position $\alpha$ exists and satisfies condition (i) of Definition~\ref{def:f_interval} for function $f_{\interval}(s^{h-1}, e^{h-1})$ because $1 \leq d^{\prime} \leq e^{\prime h-1} - s^{\prime h-1} - 1$ holds.

If the position $s^{\prime h}$ satisfies condition (iii), 
then $d^{\prime} = e^{\prime h-1} - s^{\prime h-1}$ 
and $|[x^{\prime h-1}_{s+d}, e^{\prime}]| > \sum_{w = 1}^{h} \lfloor \mu(w) \rfloor$ hold. 
$d = e^{h-1} - s^{h-1}$ follows from 
$e^{h-1} - s^{h-1} = e^{\prime h-1} - s^{\prime h-1}$ and $d^{\prime} = e^{\prime h-1} - s^{\prime h-1}$. 
$s^{\prime h-1} + d^{\prime} = e^{\prime h-1}$ follows from $d^{\prime} = e^{\prime h-1} - s^{\prime h-1}$. 
Similarly, $s^{h-1} + d = e^{h-1}$ holds. 
$|[x^{h-1}_{s+d}, e]| > \sum_{w = 1}^{h} \lfloor \mu(w) \rfloor$ follows from the following equation: 
\begin{equation*}
    \begin{split}
        |[x^{h-1}_{s+d}, e]| &= |[x^{h-1}_{e}, e]| \\
        &= |[x^{\prime h-1}_{e}, e^{\prime}]| \\
        &= |[x^{\prime h-1}_{s+d}, e^{\prime}]| \\
        &> \sum_{w = 1}^{h} \lfloor \mu(w) \rfloor.
    \end{split}
\end{equation*}
Lemma~\ref{lem:rec_function_basic_relation}~\ref{enum:rec_function_basic_relation:4} shows that 
$|[x^{h-1}_{e}, y^{h-1}_{e}]| \geq |[x^{h-1}_{e}, e]| - \sum_{w = 1}^{h-1} \lfloor \mu(w) \rfloor$ holds. 
$|[x^{h-1}_{e}, y^{h-1}_{e}]| > \lfloor \mu(h) \rfloor$ follows from 
$|[x^{h-1}_{e}, y^{h-1}_{e}]| \geq |[x^{h-1}_{e}, e]| - \sum_{w = 1}^{h-1} \lfloor \mu(w) \rfloor$, 
$|[x^{h-1}_{s+d}, e]| > \sum_{w = 1}^{h} \lfloor \mu(w) \rfloor$, 
and $|[x^{h-1}_{e}, e]| = |[x^{h-1}_{s+d}, e]|$. 
Since $|[x^{h-1}_{e}, y^{h-1}_{e}]| > \lfloor \mu(h) \rfloor$, 
the assignment of the nonterminal $S^{h-1}[e^{h-1}]$ is $-1$. 
Lemma~\ref{lem:rr_class}~\ref{enum:rr_class:3} and Lemma~\ref{lem:rr_class}~\ref{enum:rr_class:4} indicate that 
the position $\alpha$ exists because $\assign(S^{h-1}[e^{h-1}]) = -1$ and $s^{h-1} + d = e^{h-1}$. 
The position $\alpha$ satisfies condition (iii) of Definition~\ref{def:f_interval} for function $f_{\interval}(s^{h-1}, e^{h-1})$ because $x^{h}_{\alpha} = x^{h-1}_{e}$ 
and $|[x^{h-1}_{e}, e]| > \sum_{w = 1}^{h} \lfloor \mu(w) \rfloor$ hold. 

If the position $s^{\prime h}$ satisfies condition (iv), 
then $d^{\prime} = 0$ 
and $|[s^{\prime}, x^{\prime h-1}_{s+d}]| > 1 + \sum_{w = 1}^{h} \lfloor \mu(w) \rfloor$ hold. 
Here, $s^{h} + d = s^{h}$ and $s^{\prime h} + d^{\prime} = s^{\prime h}$ hold
because $d^{\prime} = 0$ and $d = d^{\prime}$. 
$|[s, x^{h-1}_{s+d}]| > 1 + \sum_{w = 1}^{h} \lfloor \mu(w) \rfloor$ follows from the following equation: 
\begin{equation*}
    \begin{split}
        |[s, x^{h-1}_{s+d}]| &= |[s, x^{h-1}_{s}]| \\
        &= |[s^{\prime}, x^{\prime h-1}_{s}]| \\
        &= |[s^{\prime}, x^{\prime h-1}_{s+d}]| \\
        &> 1 + \sum_{w = 1}^{h} \lfloor \mu(w) \rfloor.
    \end{split}
\end{equation*}
Lemma~\ref{lem:rec_function_basic_relation}~\ref{enum:rec_function_basic_relation:4} shows that 
$|[x^{h-1}_{s-1}, y^{h}_{s-1}]| \geq |[s, x^{h-1}_{s-1}]| - 1 - \sum_{w = 1}^{h-1} \lfloor \mu(w) \rfloor$ 
for the substring $T[x^{h-1}_{s-1}..y^{h-1}_{s-1}]$ derived from the $(s^{h-1} - 1)$-th nonterminal of sequence $S^{h}$. 
$|[x^{h-1}_{s-1}, y^{h}_{s-1}]| > \lfloor \mu(h) \rfloor$ 
follows from 
$|[x^{h-1}_{s-1}, y^{h}_{s-1}]| \geq |[s, x^{h-1}_{s-1}]| - 1 - \sum_{w = 1}^{h-1} \lfloor \mu(w) \rfloor$, 
$|[s, x^{h-1}_{s+d}]| > 1 + \sum_{w = 1}^{h} \lfloor \mu(w) \rfloor$, 
and $|[s, x^{h-1}_{s+d}]| = |[s, x^{h-1}_{s}]|$. 
Since $|[x^{h-1}_{s-1}, y^{h-1}_{s-1}]| > \lfloor \mu(h) \rfloor$, 
the assignment of the nonterminal $S^{h-1}[s^{h-1}-1]$ is $-1$. 
Lemma~\ref{lem:rr_class}~\ref{enum:rr_class:3} and Lemma~\ref{lem:rr_class}~\ref{enum:rr_class:4} indicate that 
the position $\alpha$ exists 
because $\assign(S^{h-1}[s^{h-1}-1]) = -1$ and $s^{h-1} + d = s^{h-1}$. 
The position $\alpha$ satisfies condition (iv) of Definition~\ref{def:f_interval} for function $f_{\interval}(s^{h-1}, e^{h-1})$ because $x^{h}_{\alpha} = x^{h-1}_{s}$ 
and $|[s, x^{h-1}_{s}]| > 1 + \sum_{w = 1}^{h} \lfloor \mu(w) \rfloor$ hold. 

We showed that the position $\alpha$ always exists and satisfies at least one of four conditions (i), (ii), (iii), and (iv) of Definition~\ref{def:f_interval} for function $f_{\interval}(s^{h-1}, e^{h-1})$. 
Therefore, $|[x^{h-1}_{s}, x^{h}_{s}]| \leq |[x^{\prime h-1}_{s}, x^{\prime h}_{s}]|$ holds. 

\textbf{Proof of $|[x^{h-1}_{s}, x^{h}_{s}]| \geq |[x^{\prime h-1}_{s}, x^{\prime h}_{s}]|$.}
$|[x^{h-1}_{s}, x^{h}_{s}]| \geq |[x^{\prime h-1}_{s}, x^{\prime h}_{s}]|$ can be proved using 
the same approach as for $|[x^{h-1}_{s}, x^{h}_{s}]| \leq |[x^{\prime h-1}_{s}, x^{\prime h}_{s}]|$. 

\textbf{Proof of $|[x^{h}_{e}, x^{h-1}_{e}]| \leq |[x^{\prime h}_{e}, x^{\prime h-1}_{e}]|$.}
$|[x^{h}_{e}, x^{h-1}_{e}]| \leq |[x^{\prime h}_{e}, x^{\prime h-1}_{e}]|$ can be proved using 
the same approach as for $|[x^{h-1}_{s}, x^{h}_{s}]| \leq |[x^{\prime h-1}_{s}, x^{\prime h}_{s}]|$. 

\textbf{Proof of $|[x^{h}_{e}, x^{h-1}_{e}]| \geq |[x^{\prime h}_{e}, x^{\prime h-1}_{e}]|$.}
$|[x^{h}_{e}, x^{h-1}_{e}]| \geq |[x^{\prime h}_{e}, x^{\prime h-1}_{e}]|$ can be proved using 
the same approach as for $|[x^{h-1}_{s}, x^{h}_{s}]| \leq |[x^{\prime h-1}_{s}, x^{\prime h}_{s}]|$. 

\textbf{Proof of $|[s, x^{h}_{s}]| = |[s^{\prime}, x^{\prime h}_{s}]|$ for $h \in [1, \min \{ k, k^{\prime} \}]$.}
$|[x^{h-1}_{s}, x^{h}_{s}]| = |[x^{\prime h-1}_{s}, x^{\prime h}_{s}]|$ 
follows from $|[x^{h-1}_{s}, x^{h}_{s}]| \geq |[x^{\prime h-1}_{s}, x^{\prime h}_{s}]|$ and 
$|[x^{h-1}_{s}, x^{h}_{s}]| \leq |[x^{\prime h-1}_{s}, x^{\prime h}_{s}]|$. 
Therefore, $|[s, x^{h}_{s}]| = |[s^{\prime}, x^{\prime h}_{s}]|$ follows from the following equation: 
\begin{equation*}
    \begin{split}
        |[s, x^{h}_{s}]| &= |[s, x^{h-1}_{s}]| + |[x^{h-1}_{s}, x^{h}_{s}]| - 1 \\
        &= |[s^{\prime}, x^{\prime h-1}_{s}]| + |[x^{\prime h-1}_{s}, x^{\prime h}_{s}]| - 1 \\
        &= |[s^{\prime}, x^{\prime h}_{s}]|. 
    \end{split}
\end{equation*}

\textbf{Proof of $|[x^{h}_{e}, e]| = |[x^{\prime h}_{e}, e^{\prime}]|$ for $h \in [1, \min \{ k, k^{\prime} \}]$.}
$|[x^{h}_{e}, x^{h-1}_{e}]| = |[x^{\prime h}_{e}, x^{\prime h-1}_{e}]|$ 
follows from $|[x^{h}_{e}, x^{h-1}_{e}]| \geq |[x^{\prime h}_{e}, x^{\prime h-1}_{e}]|$ 
and $|[x^{h}_{e}, x^{h-1}_{e}]| \leq |[x^{\prime h}_{e}, x^{\prime h-1}_{e}]|$.
Therefore, $|[x^{h}_{e}, e]| = |[x^{\prime h}_{e}, e^{\prime}]|$ follows from the following equation: 
\begin{equation*}
    \begin{split}
        |[x^{h}_{e}, e]| &= |[x^{h}_{e}, x^{h-1}_{e}]| + |[x^{h-1}_{e}, e]| - 1 \\
        &= |[x^{\prime h}_{e}, x^{\prime h-1}_{e}]| + |[x^{\prime h-1}_{e}, e^{\prime}]| - 1 \\
        &= |[x^{\prime h}_{e}, e^{\prime}]|. 
    \end{split}
\end{equation*}

\textbf{Proof of $S^{h}[s^{h}+1..e^{h}-1] = S^{h}_{\ins}[s^{\prime h}+1..e^{\prime h}-1]$ for $h \in [1, \min \{ k, k^{\prime} \}]$.}
Two Sequences $S^{h}[s^{h}+1..e^{h}-1]$ and $S^{h}_{\ins}[s^{\prime h}+1..e^{\prime h}-1]$ 
derive two strings $T[x^{h}_{s}..x^{h}_{e}-1]$ and $T^{\prime}[x^{\prime h}_{s}..x^{\prime h}_{e}-1]$, 
respectively. 
$T[x^{h}_{s}..x^{h}_{e}-1] = T^{\prime}[x^{\prime h}_{s}..x^{\prime h}_{e}-1]$ holds 
because 
$T[s..e] = T^{\prime}[s^{\prime}..e^{\prime}]$, 
$s \leq x^{h}_{s} \leq x^{h}_{e} \leq e$, 
$s^{\prime} \leq x^{\prime h}_{s} \leq x^{\prime h}_{e} \leq e^{\prime}$, 
$|[s, x^{h}_{s}]| = |[s^{\prime}, x^{\prime h}_{s}]|$, 
and $|[x^{h}_{e}, e]| = |[x^{\prime h}_{e}, e^{\prime}]|$. 
Since $T[x^{h}_{s}..x^{h}_{e}-1] = T^{\prime}[x^{\prime h}_{s}..x^{\prime h}_{e}-1]$, 
Lemma~\ref{lem:dynamic_rr_property}~\ref{enum:dynamic_rr_property:2} shows that 
$S^{h}[s^{h}+1..e^{h}-1] = S^{h}_{\ins}[s^{\prime h}+1..e^{\prime h}-1]$ hold. 

We proved $|[s, x^{h}_{s}]| = |[s^{\prime}, x^{\prime h}_{s}]|$, 
$|[x^{h}_{e}, e]| = |[x^{\prime h}_{e}, e^{\prime}]|$, and $S^{h}[s^{h}+1..e^{h}-1] = S^{h}_{\ins}[s^{\prime h}+1..e^{\prime h}-1]$ for $h \in [1, \min \{ k, k^{\prime} \}]$. 
Therefore, Lemma~\ref{lem:dynamicP_rec_function_syncro_property}~\ref{enum:dynamicP_rec_function_syncro_property:1}. 
\end{proof}

\begin{proof}[Proof of Lemma~\ref{lem:dynamicP_rec_function_syncro_property}~\ref{enum:dynamicP_rec_function_syncro_property:2}]
We prove $k = k^{\prime}$ by contradiction. 
Without loss of generality, 
we assume that $k < k^{\prime}$ holds. 
From Lemma~\ref{lem:dynamicP_rec_function_syncro_property}~\ref{enum:dynamicP_rec_function_syncro_property:1}, 
the following three statements hold: 
\begin{itemize}
    \item $|[s, x^{k}_{s}]| = |[s^{\prime}, x^{\prime k}_{s}]|$;
    \item $|[x^{k}_{e}, e]| = |[x^{\prime k}_{e}, e^{\prime}]|$;
    \item $S^{k}[s^{k}+1..e^{k}-1] = S^{k}_{\ins}[s^{\prime k}+1..e^{\prime k}-1]$.
\end{itemize}
In this case, 
we can show that sequence $S^{k+1}$ contains a position $\alpha$ 
satisfying at least one of four conditions (i), (ii), (iii), and (iv) of Definition~\ref{def:f_interval} for function $f^{\prime}_{\interval}(s^{k}, e^{k})$ using a similar approach as for Lemma~\ref{lem:dynamicP_rec_function_syncro_property}~\ref{enum:dynamicP_rec_function_syncro_property:1}. 
The existence of the position $\alpha$ indicates that 
$f^{\prime}_{\interval}(s^{k}, e^{k}) \neq \perp$ holds. 
On the other hand, 
$f^{\prime}_{\interval}(s^{k}, e^{k}) = \perp$ follows from the definition of sequence $A(s, e)$. 
The two facts $f^{\prime}_{\interval}(s^{k}, e^{k}) \neq \perp$ and $f^{\prime}_{\interval}(s^{k}, e^{k}) = \perp$ yield a contradiction. 
Therefore, $k = k^{\prime}$ must hold. 
\end{proof}

\subsection{Changes of Interval Attractors}\label{subsec:relationship_old_and_new}
Interval attractors in set $\Psi_{\RR}$ are changed by updating RLSLP $\mathcal{G}^{R}$. 
Let $\Psi_{\RR}$ (respectively, $\Psi^{\prime}_{\RR}$) be the set of interval attractors obtained from 
$\mathcal{G}^{R}$ (respectively, $\mathcal{G}^{\prime R}$). 
This subsection explains the relationship between the two sets $\Psi_{\RR}$ and $\Psi^{\prime}_{\RR}$. 

\subsubsection{The Relationship between Two Sets \texorpdfstring{$\Psi_{\RR}$}{Psi1} and \texorpdfstring{$\Psi^{\prime}_{\RR}$}{Psi2}}
Consider an interval attractor $([p, q], [\ell, r])$ in set $\Psi_{\RR}$. 
If there exists an integer $\epsilon \in \mathbb{Z}$ satisfying $T[p-1..r+1] = T^{\prime}[p-1+\epsilon..r+1+\epsilon]$, 
then the following lemma shows that interval attractor $([p+\epsilon, q+\epsilon], [\ell+\epsilon, r+\epsilon])$ is contained in set $\Psi^{\prime}_{\RR}$. 

\begin{lemma}\label{lem:RB_IA_proceeding_formula}
Consider an interval attractor $([p, q], [\ell, r])$ in set $\Psi_{\RR}$ 
and an interval $[i^{\prime}, j^{\prime}]$ on input string $T^{\prime}$ satisfying 
$T[p-1..r+1] = T^{\prime}[i^{\prime}-1, j^{\prime}+1]$. 
Then, set $\Psi^{\prime}_{\RR}$ contains interval attractor $([i^{\prime}, i^{\prime} + |[p, q]| - 1], [j^{\prime} - |[\ell, r]| + 1, j^{\prime}])$. 
\end{lemma}
\begin{proof}
See Section~\ref{subsubsec:proof_RB_IA_proceeding_formula}.
\end{proof}

Similarly, 
consider an interval attractor $([p^{\prime}, q^{\prime}], [\ell^{\prime}, r^{\prime}])$ in set $\Psi_{\RR}$. 
If there exists an integer $\epsilon \in \mathbb{Z}$ satisfying $T[p^{\prime}-1+\epsilon..r^{\prime}+1+\epsilon] = T^{\prime}[p^{\prime}-1..r^{\prime}+1]$, 
then the following lemma shows that interval attractor $([p^{\prime}+\epsilon, q^{\prime}+\epsilon], [\ell^{\prime}+\epsilon, r^{\prime}+\epsilon])$ is contained in set $\Psi_{\RR}$. 

\begin{lemma}\label{lem:RB_IA_back_formula}
Consider an interval attractor $([p^{\prime}, q^{\prime}], [\ell^{\prime}, r^{\prime}])$ in set $\Psi^{\prime}_{\RR}$ 
and an interval $[i, j]$ on input string $T$ satisfying 
$T[i-1..j+1] = T^{\prime}[p^{\prime}-1, r^{\prime}+1]$. 
Then, set $\Psi_{\RR}$ contains interval attractor $([i, i + |[p^{\prime}, q^{\prime}]| - 1], [j - |[\ell^{\prime}, r^{\prime}]| + 1, j])$. 
\end{lemma}
\begin{proof}
Lemma~\ref{lem:RB_IA_back_formula} can be proved using 
the same approach as for Lemma~\ref{lem:RB_IA_proceeding_formula}. 
\end{proof}

We introduce three subsets $\Psi_{\LEFT}$, $\Psi_{\RIGHT}$, and $\Psi_{\OLD}$ of set $\Psi_{\RR}$. 
The subset $\Psi_{\LEFT}$ consists of interval attractors such that 
each interval attractor $([p, q], [\ell, r]) \in \Psi_{\LEFT}$ satisfies $r+1 < \lambda - 1$ for the insertion position $\lambda$ (i.e., $\Psi_{\LEFT} = \{ ([p, q], [\ell, r]) \in \Psi_{\RR} \mid r+1 < \lambda - 1 \}$). 
In contrast, 
the subset $\Psi_{\RIGHT}$ consists of interval attractors such that 
each interval attractor $([p, q], [\ell, r]) \in \Psi_{\RIGHT}$ satisfies $p-1 > \lambda$ 
(i.e., $\Psi_{\RIGHT} = \{ ([p, q], [\ell, r]) \in \Psi_{\RR} \mid p-1 > \lambda \}$). 
The subset $\Psi_{\OLD}$ consists of interval attractors such that 
each interval attractor $([p, q], [\ell, r]) \in \Psi_{\OLD}$ satisfies 
$[p-1, r+1] \cap [\lambda-1, \lambda] \neq \emptyset$.
The following lemma states the relationship among three subsets $\Psi_{\LEFT}$, $\Psi_{\OLD}$, and $\Psi_{\RIGHT}$. 

\begin{lemma}\label{lem:psi_left_center_right}
The following two statements hold: 
\begin{enumerate}[label=\textbf{(\roman*)}]
    \item \label{enum:psi_left_center_right:1} 
    $\Psi_{\RR} = \Psi_{\LEFT} \cup \Psi_{\OLD} \cup \Psi_{\RIGHT}$; 
    \item \label{enum:psi_left_center_right:2} 
    three sets $\Psi_{\LEFT}, \Psi_{\OLD}, \Psi_{\RIGHT}$ are disjoint from each other 
    (i.e., $\Psi_{\LEFT} \cap \Psi_{\OLD} = \emptyset$, $\Psi_{\OLD} \cap \Psi_{\RIGHT} = \emptyset$, and $\Psi_{\LEFT} \cap \Psi_{\RIGHT} = \emptyset$). 
\end{enumerate}
\end{lemma}
\begin{proof}
    Lemma~\ref{lem:psi_left_center_right} follows from the definition of the three sets $\Psi_{\LEFT}$, $\Psi_{\RIGHT}$, and $\Psi_{\OLD}$.
\end{proof}

Because of $T[1..\lambda-1] = T^{\prime}[1..\lambda]$ and 
$T[\lambda..n] = T^{\prime}[\lambda+1..n+1]$, 
the following corollary follows from Lemma~\ref{lem:RB_IA_proceeding_formula}. 

\begin{corollary}\label{cor:RB_IA_before_after}
Consider an interval attractor $([p, q], [\ell, r])$ in set $\Psi_{\RR}$ 
satisfying $([p, q], [\ell, r]) \in \Psi_{\LEFT} \cup \Psi_{\RIGHT}$. 
Let $\epsilon = 0$ if $([p, q], [\ell, r]) \in \Psi_{\LEFT}$; 
otherwise, let $\epsilon = 1$. 
Then, the following two statements hold: 
\begin{enumerate}[label=\textbf{(\roman*)}]
    \item \label{enum:RB_IA_before_after:1} $T[p-1..r+1] = T^{\prime}[p + \epsilon -1..r + \epsilon +1]$;
    \item \label{enum:RB_IA_before_after:2} $([p+\epsilon, q+\epsilon], [\ell+\epsilon, r+\epsilon]) \in \Psi^{\prime}_{\RR}$. 
\end{enumerate}
\end{corollary}

Next, we introduce three subsets $\Psi^{\prime}_{\NEW}$, $\Psi^{\prime}_{\LEFT}$, $\Psi^{\prime}_{\RIGHT}$ of set $\Psi^{\prime}_{\RR}$. 
The subset $\Psi^{\prime}_{\NEW}$ consists of interval attractors such that 
each interval attractor $([p, q], [\ell, r]) \in \Psi^{\prime}_{\NEW}$ satisfies 
$[p-1, r+1] \cap [\lambda - 1, \lambda + 1] \neq \emptyset$ (i.e., 
$\Psi^{\prime}_{\NEW} = \{ ([p, q], [\ell, r]) \in \Psi^{\prime}_{\RR} \mid [p-1, r+1] \cap [\lambda - 1, \lambda + 1] \neq \emptyset \}$).
The subset $\Psi^{\prime}_{\LEFT}$ consists of interval attractors such that 
each interval attractor $([p, q], [\ell, r]) \in \Psi^{\prime}_{\LEFT}$ satisfies 
$r+1 < \lambda - 1$.  
Similarly, 
the subset $\Psi^{\prime}_{\RIGHT}$ consists of interval attractors such that 
each interval attractor $([p, q], [\ell, r]) \in \Psi^{\prime}_{\RIGHT}$ satisfies 
$p-1 > \lambda + 1$.  
Similar to the three subsets $\Psi_{\LEFT}$, $\Psi_{\OLD}$, and $\Psi_{\RIGHT}$, 
the following two relationships among $\Psi^{\prime}_{\LEFT}$, $\Psi^{\prime}_{\NEW}$, and $\Psi^{\prime}_{\RIGHT}$: 
\begin{enumerate}[label=\textbf{(\roman*)}]
    \item  
    $\Psi^{\prime}_{\RR} = \Psi^{\prime}_{\LEFT} \cup \Psi^{\prime}_{\NEW} \cup \Psi^{\prime}_{\RIGHT}$; 
    \item  
    three sets $\Psi^{\prime}_{\LEFT}, \Psi^{\prime}_{\NEW}, \Psi^{\prime}_{\RIGHT}$ are disjoint from each other. 
\end{enumerate}

%We define set $\Psi^{\prime}_{\RIGHT}$ as $\{ ([p + 1, q + 1], [\ell + 1, r + 1]) \mid ([p, q], [\ell, r]) \in \Psi_{\RIGHT} \}$ 
%using the subset $\Psi_{\RIGHT}$. 
%Then, Corollary~\ref{cor:RB_IA_after_before} shows that 
%two sets $\Psi_{\LEFT}$ and $\Psi^{\prime}_{\RIGHT}$ are subsets of set $\Psi^{\prime}_{\RR}$. 

Similar to Corollary~\ref{cor:RB_IA_before_after}, 
the following corollary follows from Lemma~\ref{lem:RB_IA_back_formula}. 
\begin{corollary}\label{cor:RB_IA_after_before}
Consider an interval attractor $([p, q], [\ell, r])$ in set $\Psi^{\prime}_{\RR}$ 
satisfying $([p, q], [\ell, r]) \in \Psi^{\prime}_{\LEFT} \cup \Psi^{\prime}_{\RIGHT}$. 
Let $\epsilon = 0$ if $([p, q], [\ell, r]) \in \Psi^{\prime}_{\LEFT}$; 
otherwise, let $\epsilon = 1$. 
Then, the following two statements hold: 
\begin{enumerate}[label=\textbf{(\roman*)}]
    \item \label{enum:RB_IA_after_before:1} $T[p-\epsilon-1..r-\epsilon+1] = T^{\prime}[p-1..r +1]$;
    \item \label{enum:RB_IA_after_before:2} $([p-\epsilon, q-\epsilon], [\ell-\epsilon, r-\epsilon]) \in \Psi_{\RR}$. 
\end{enumerate}
\end{corollary}

The following theorem states the relationship among four sets $\Psi_{\LEFT}$, $\Psi^{\prime}_{\LEFT}$, $\Psi_{\RIGHT}$, and $\Psi^{\prime}_{\RIGHT}$.  

\begin{theorem}\label{theo:RS_LEFT_RIGHT}
$\Psi_{\LEFT} = \Psi^{\prime}_{\LEFT}$ and 
$\Psi^{\prime}_{\RIGHT} = \{ ([p + 1, q + 1], [\ell + 1, r + 1]) \mid ([p, q], [\ell, r]) \in \Psi_{\RIGHT} \}$ hold 
for the four subsets $\Psi_{\LEFT}$, $\Psi^{\prime}_{\LEFT}$, $\Psi_{\RIGHT}$, and $\Psi^{\prime}_{\RIGHT}$. 
\end{theorem}
\begin{proof}
Theorem~\ref{theo:RS_LEFT_RIGHT} follows from Corollary~\ref{cor:RB_IA_before_after} and Corollary~\ref{cor:RB_IA_after_before}.
\end{proof}
Theorem~\ref{theo:RS_LEFT_RIGHT} shows that Theorem~\ref{theo:IA_change_formula} holds for 
inserting a single character into string $T$.

\subsubsection{The Relationship between Two Interval Attractors in \texorpdfstring{$\Psi_{\RR}$}{Psi1} and \texorpdfstring{$\Psi^{\prime}_{\RR}$}{Psi2}}
Consider two interval attractors $([p, q], [\ell, r]) \in \Psi_{\RR}$ and 
$([p^{\prime}, q^{\prime}], [\ell^{\prime}, r^{\prime}]) \in \Psi^{\prime}_{\RR}$ 
satisfying $T[p-1..r+1] = T^{\prime}[p^{\prime}-1..r^{\prime}+1]$. 
The following lemma states the relationship between the two interval attractors. 

\begin{lemma}\label{lem:dynamic_IA_super_correspondence}
Consider two interval attractors $([p, q], [\ell, r]) \in \Psi_{\RR}$ and 
$([p^{\prime}, q^{\prime}], [\ell^{\prime}, r^{\prime}]) \in \Psi^{\prime}_{\RR}$ 
satisfying $T[p-1..r+1] = T^{\prime}[p^{\prime}-1..r^{\prime}+1]$. 
Let $h$, $\gamma$, and $C$ be the level, attractor position, and associated string of interval attractor $([p, q], [\ell, r])$, respectively. 
Similarly, 
let $h^{\prime}$, $\gamma^{\prime}$, and $C^{\prime}$ be the level, attractor position, and associated string of interval attractor $([p^{\prime}, q^{\prime}], [\ell^{\prime}, r^{\prime}])$, respectively. 
Then the following three statements hold. 
\begin{enumerate}[label=\textbf{(\roman*)}]
    \item $|[p, q]| = |[p^{\prime}, q^{\prime}]|$ and $|[\ell, r]| = |[\ell^{\prime}, r^{\prime}]|$;
    \item $h = h^{\prime}$;
    \item $T[p-1..\gamma-1] = T^{\prime}[p^{\prime}-1..\gamma^{\prime}-1]$ 
    and $T[\gamma..r+1] = T^{\prime}[\gamma^{\prime}..r^{\prime}+1]$; 
    \item $C = C^{\prime}$ if $|[\gamma, r]| > \sum_{w = 1}^{h+1} \lfloor \mu(w) \rfloor$.
\end{enumerate}
\end{lemma}
\begin{proof}
See Section~\ref{subsubsec:proof_dynamic_IA_super_correspondence}.
\end{proof}

The following lemma is the dynamic version of Lemma~\ref{lem:psi_equality_basic_property}. 
\begin{lemma}\label{lem:dynamic_RR_subset}
Consider two interval attractors $([p, q], [\ell, r])$ in set $\Psi_{\RR}$ 
and $([p^{\prime}, q^{\prime}], [\ell^{\prime}, r^{\prime}])$ in set $\Psi^{\prime}_{\RR}$ 
satisfying $T[p-1..r+1] = T^{\prime}[p^{\prime}-1..r^{\prime}+1]$. 
Let $h$, $C$, and $\gamma$ be the level, associated string, and attractor position of the interval attractor $([p, q], [\ell, r])$, respectively. 
Analogously, $h$, $C$, and $\gamma$ are defined for interval attractor $([p^{\prime}, q^{\prime}], [\ell^{\prime}, r^{\prime}])$. 
Then, the following ten statements hold: 
\begin{enumerate}[label=\textbf{(\roman*)}]
    \item \label{enum:dynamic_RR_subset:H}    
    for an integer $t \geq 0$, 
    $([p, q], [\ell, r]) \in \Psi_{t} \Leftrightarrow ([p^{\prime}, q^{\prime}], [\ell^{\prime}, r^{\prime}]) \in \Psi^{\prime}_{t}$;
    \item \label{enum:dynamic_RR_subset:str}    
    for a string $P \in \Sigma^{+}$, 
    $([p, q], [\ell, r]) \in \Psi_{\str}(P) \Leftrightarrow ([p^{\prime}, q^{\prime}], [\ell^{\prime}, r^{\prime}]) \in \Psi^{\prime}_{\str}(P)$;   
    \item \label{enum:dynamic_RR_subset:run}    
    $([p, q], [\ell, r]) \in \Psi_{\run} \Leftrightarrow ([p^{\prime}, q^{\prime}], [\ell^{\prime}, r^{\prime}]) \in \Psi^{\prime}_{\run}$;
    \item \label{enum:dynamic_RR_subset:source}    
    $([p, q], [\ell, r]) \in \Psi_{\source} \Leftrightarrow ([p^{\prime}, q^{\prime}], [\ell^{\prime}, r^{\prime}]) \in \Psi^{\prime}_{\source}$;
    \item \label{enum:dynamic_RR_subset:C}    
    for a string $P \in \Sigma^{+}$, 
    $([p, q], [\ell, r]) \in \Psi_{\centerset}(P) \cap (\Psi_{\run} \cup \Psi_{\source}) \Leftrightarrow ([p^{\prime}, q^{\prime}], [\ell^{\prime}, r^{\prime}]) \in \Psi^{\prime}_{\centerset}(P) \cap (\Psi^{\prime}_{\run} \cup \Psi^{\prime}_{\source})$;
    \item \label{enum:dynamic_RR_subset:preceding}    
    $([p, q], [\ell, r]) \in \Psi_{\preceding} \cap (\Psi_{\run} \cup \Psi_{\source}) \Leftrightarrow ([p^{\prime}, q^{\prime}], [\ell^{\prime}, r^{\prime}]) \in \Psi^{\prime}_{\preceding} \cap (\Psi^{\prime}_{\run} \cup \Psi^{\prime}_{\source})$;
    \item \label{enum:dynamic_RR_subset:succeeding}    
    $([p, q], [\ell, r]) \in \Psi_{\succeeding} \cap (\Psi_{\run} \cup \Psi_{\source}) \Leftrightarrow ([p^{\prime}, q^{\prime}], [\ell^{\prime}, r^{\prime}]) \in \Psi^{\prime}_{\succeeding} \cap (\Psi^{\prime}_{\run} \cup \Psi^{\prime}_{\source})$;
    %\item \label{enum:dynamic_RR_subset:lex}
    %for a string $P \in \Sigma^{+}$, 
    %$([p, q], [\ell, r]) \in \Psi_{\lex}(P) \Leftrightarrow ([p^{\prime}, q^{\prime}], [\ell^{\prime}, r^{\prime}]) \in \Psi^{\prime}_{\lex}(P)$;
    \item \label{enum:dynamic_RR_subset:lcp}
    for an integer $K > \sum_{w = 1}^{h+1} \lfloor \mu(w) \rfloor$, 
    $([p, q], [\ell, r]) \in \Psi_{\lcp}(K) \Leftrightarrow ([p^{\prime}, q^{\prime}], [\ell^{\prime}, r^{\prime}]) \in \Psi^{\prime}_{\lcp}(K)$;
    \item \label{enum:dynamic_RR_subset:modulo}
    for an integer $M \geq 0$, 
    $([p, q], [\ell, r]) \in \Psi_{\modulo}(M) \Leftrightarrow ([p^{\prime}, q^{\prime}], [\ell^{\prime}, r^{\prime}]) \in \Psi^{\prime}_{\modulo}(M)$.
\end{enumerate}
\end{lemma}
\begin{proof}
Lemma~\ref{lem:dynamic_RR_subset} is the dynamic version of Lemma~\ref{lem:psi_equality_basic_property}. 
Lemma~\ref{lem:psi_equality_basic_property} is proved using the following three facts: 
\begin{itemize}
    \item $h = h^{\prime}$; 
    \item $T[p-1..\gamma-1] = T^{\prime}[p^{\prime}-1..\gamma^{\prime}-1]$ and $T[\gamma..r+1] = T^{\prime}[\gamma^{\prime}..r^{\prime}+1]$; 
    \item $C = C^{\prime}$ if $|[\gamma, r]| > \sum_{w = 1}^{h+1} \lfloor \mu(w) \rfloor$. 
\end{itemize}
For the dynamic setting, 
the above facts can be obtained by Lemma~\ref{lem:dynamic_IA_super_correspondence}. 
Therefore, we can prove Lemma~\ref{lem:dynamic_RR_subset} by the same approach used to prove Lemma~\ref{lem:psi_equality_basic_property}. 
\end{proof}

The following lemma states the relationship between two subsets $\Psi_{\str}(P)$ and $\Psi^{\prime}_{\str}(P)$. 

\begin{lemma}\label{lem:dynamic_and_basic_str_formula}
    Let $Z = T^{\prime}[p-1..r+1]$ for 
    an interval attractor $([p, q], [\ell, r])$ in set $\Psi^{\prime}_{\RR}$. 
    Then, the following two statements hold:
    \begin{enumerate}[label=\textbf{(\roman*)}]
    \item \label{enum:dynamic_and_basic_str_formula:1} $|\Psi^{\prime}_{\str}(Z)| = |\Psi_{\str}(Z)| - |\Psi_{\OLD} \cap \Psi_{\str}(Z)| + |\Psi^{\prime}_{\NEW} \cap \Psi^{\prime}_{\str}(Z)|$;
    \item \label{enum:dynamic_and_basic_str_formula:2} $|\Psi^{\prime}_{\str}(Z)| = |\Psi^{\prime}_{\NEW} \cap \Psi^{\prime}_{\str}(Z)|$ if $(\Psi_{\RR} \setminus \Psi_{\OLD}) \cap \Psi_{\str}(Z) = \emptyset$.
    \end{enumerate}
\end{lemma}
\begin{proof}
    The proof of Lemma~\ref{lem:dynamic_and_basic_str_formula} is as follows.

    \textbf{Proof of Lemma~\ref{lem:dynamic_and_basic_str_formula}(i).}    
    The set $\Psi^{\prime}_{\str}(Z)$ can be divided into three subsets 
    $(\Psi^{\prime}_{\LEFT} \cap \Psi^{\prime}_{\str}(Z))$, 
    $(\Psi^{\prime}_{\NEW} \cap \Psi^{\prime}_{\str}(Z))$, 
    and $(\Psi^{\prime}_{\RIGHT} \cap \Psi^{\prime}_{\str}(Z))$ 
    because set $\Psi^{\prime}$ can be divided into 
    three subsets $\Psi^{\prime}_{\LEFT}$, $\Psi^{\prime}_{\NEW}$, and $\Psi^{\prime}_{\RIGHT}$. 
    Similarly, 
    the set $\Psi_{\str}(Z)$ can be divided into three subsets 
    $(\Psi_{\LEFT} \cap \Psi_{\str}(Z))$, 
    $(\Psi_{\OLD} \cap \Psi_{\str}(Z))$, 
    and $(\Psi_{\RIGHT} \cap \Psi_{\str}(Z))$.  
    
    $\Psi^{\prime}_{\LEFT} \cap \Psi^{\prime}_{\str}(Z) = \Psi_{\LEFT} \cap \Psi_{\str}(Z)$ 
    holds because 
    (a) $\Psi_{\LEFT} = \Psi^{\prime}_{\LEFT}$ (Theorem~\ref{theo:RS_LEFT_RIGHT}) 
    and (b) $T[p-1..r+1] = T^{\prime}[p-1..r+1]$ for 
    each interval attractor $([p, q], [\ell, r]) \in \Psi_{\LEFT}$. 
    Similarly, 
    $\Psi^{\prime}_{\RIGHT} \cap \Psi^{\prime}_{\str}(Z) = \{ ([p+1, q+1], [\ell+1, r+1]) \mid ([p+1, q+1], [\ell+1, r+1]) \in \Psi_{\RIGHT}\} \cap \Psi_{\str}(Z)$ holds. 
    Therefore, Lemma~\ref{enum:dynamic_and_basic_str_formula:1} follows from the following equation: 
\begin{equation*}
        \begin{split}
            |\Psi^{\prime}_{\str}(Z)| &= \Psi^{\prime} \cap \Psi^{\prime}_{\str}(Z)  \\ 
            &= |\Psi^{\prime}_{\LEFT} \cap \Psi^{\prime}_{\str}(Z)| + |\Psi^{\prime}_{\RIGHT} \cap \Psi^{\prime}_{\str}(Z)| + |\Psi^{\prime}_{\NEW} \cap \Psi^{\prime}_{\str}(Z)| \\
            &= |\Psi_{\LEFT} \cap \Psi_{\str}(Z)| + |\Psi_{\RIGHT} \cap \Psi_{\str}(Z)| + |\Psi^{\prime}_{\NEW} \cap \Psi^{\prime}_{\str}(Z)| \\
            &= (|\Psi_{\str}(Z)| - |\Psi_{\OLD} \cap \Psi_{\str}(Z)|) + |\Psi^{\prime}_{\NEW} \cap \Psi^{\prime}_{\str}(Z)|.
        \end{split}
\end{equation*}

    \textbf{Proof of Lemma~\ref{lem:dynamic_and_basic_str_formula}(ii).}
    Lemma~\ref{lem:dynamic_and_basic_str_formula}(ii) follows from 
    Lemma~\ref{lem:dynamic_and_basic_str_formula}(i) and $\Psi_{\RR} \setminus \Psi_{\OLD} = \Psi_{\LEFT} \cup \Psi_{\RIGHT}$. 
\end{proof}

The following corollary is follows from Lemma~\ref{lem:dynamic_and_basic_str_formula}, 
Lemma~\ref{lem:dynamic_RR_subset}~\ref{enum:dynamic_RR_subset:H}, 
and Lemma~\ref{lem:dynamic_RR_subset}~\ref{enum:dynamic_RR_subset:run}. 

\begin{corollary}\label{cor:dynamic_str_formula}
    Let $Z = T^{\prime}[p-1..r+1]$ for 
    an interval attractor $([p, q], [\ell, r])$ in set $\Psi^{\prime}_{\RR}$ satisfying $([p, q], [\ell, r]) \not \in \Psi^{\prime}_{\run}$. 
    Then, the following two statements hold for the level $h$ of the interval attractor $([p, q], [\ell, r])$:
    \begin{enumerate}[label=\textbf{(\roman*)}]
    \item \label{enum:dynamic_str_formula:1} $|\Psi^{\prime}_{\str}(Z)| = |\Psi_{\str}(Z)| - |\Psi_{h} \cap  (\Psi_{\OLD} \setminus \Psi_{\run}) \cap \Psi_{\str}(Z)| + |\Psi^{\prime}_{h} \cap (\Psi^{\prime}_{\NEW} \setminus \Psi^{\prime}_{\run}) \cap \Psi^{\prime}_{\str}(Z)|$;
    \item \label{enum:dynamic_str_formula:2} $|\Psi^{\prime}_{\str}(Z)| = |\Psi^{\prime}_{h} \cap (\Psi^{\prime}_{\NEW} \setminus \Psi^{\prime}_{\run}) \cap \Psi^{\prime}_{\str}(Z)|$ if $(\Psi_{\RR} \setminus \Psi_{\OLD}) \cap \Psi_{\str}(Z) = \emptyset$.
    \end{enumerate}
\end{corollary}

Let $f^{\prime}_{\recover}: \Psi^{\prime}_{\source} \rightarrow 2^{\Psi^{\prime}_{\run}}$ be the function $f_{\recover}$ defined using set $\Psi^{\prime}_{\RR}$ instead of set $\Psi_{\RR}$. 
Similarly, let $\Psi^{\prime}_{\mRecover} \subseteq \Psi^{\prime}_{\RR}$ be the subset $\Psi_{\mRecover}$ defined using set $\Psi^{\prime}_{\RR}$ instead of set $\Psi_{\RR}$. 
Here, $f_{\recover}$ and $\Psi_{\mRecover}$ the function and subset introduced in Section~\ref{subsec:function_recover}. 
The following lemma states the relationship among $f_{\recover}$, $f^{\prime}_{\recover}$, $\Psi_{\mRecover}$, and $\Psi^{\prime}_{\mRecover}$.

\begin{lemma}\label{lem:dynamic_f_recover}
    Consider two interval attractors $([p, q], [\ell, r]) \in \Psi_{\source}$ 
    and $([p^{\prime}, q^{\prime}], [\ell^{\prime}, r^{\prime}]) \in \Psi^{\prime}_{\source}$ 
    satisfying $T[p-1..r+1] = T^{\prime}[p^{\prime}-1..r^{\prime}+1]$. 
    Let $([p_{1}, q_{1}], [\ell_{1}, r_{1}])$, 
    $([p_{2}, q_{2}], [\ell_{2}, r_{2}])$, $\ldots$, $([p_{k}, q_{k}], [\ell_{k}, r_{k}])$ ($p_{1} < p_{2} < \ldots < p_{k}$)
    be the interval attractors in the set obtained from function $f_{\recover}(([p, q], [\ell, r]))$ 
    Analogously, 
    $([p^{\prime}_{1}, q^{\prime}_{1}], [\ell^{\prime}_{1}, r^{\prime}_{1}])$, 
    $([p^{\prime}_{2}, q^{\prime}_{2}], [\ell^{\prime}_{2}, r^{\prime}_{2}])$, $\ldots$, $([p^{\prime}_{k^{\prime}}, q^{\prime}_{k^{\prime}}], [\ell^{\prime}_{k^{\prime}}, r^{\prime}_{k^{\prime}}])$ ($p^{\prime}_{1} < p^{\prime}_{2} < \ldots < p^{\prime}_{k^{\prime}}$)
    are defined for function $f_{\recover}(([p^{\prime}, q^{\prime}], [\ell^{\prime}, r^{\prime}]))$. 
    Here, 
    Lemma~\ref{lem:mRecover_basic_property} shows that 
    there exists an integer $x \in [1, k]$ satisfying 
    $f_{\recover}(([p, q]$, $[\ell, r])) \cap \Psi_{\mRecover} = \{ ([p_{x}, q_{x}], [\ell_{x}, r_{x}]) \}$. 
    Similarly, 
    there exists an integer $y \in [1, k^{\prime}]$ satisfying 
    $f^{\prime}_{\recover}(([p^{\prime}, q^{\prime}]$, $[\ell^{\prime}, r^{\prime}])) \cap \Psi^{\prime}_{\mRecover} = \{ ([p^{\prime}_{y}, q^{\prime}_{y}], [\ell^{\prime}_{y}, r^{\prime}_{y}]) \}$. 

    Then, the following three statements hold:
\begin{enumerate}[label=\textbf{(\roman*)}]
    \item \label{enum:dynamic_f_recover:X} $k = k^{\prime}$;
    \item \label{enum:dynamic_f_recover:Y} $T[p_{s}-1..r_{s}+1] = T^{\prime}[p^{\prime}_{s}-1..r^{\prime}_{s}+1]$ for each integer $s \in [1, k]$;
    \item \label{enum:dynamic_f_recover:Z} $T[p_{x}-1..r_{x}+1] = T^{\prime}[p^{\prime}_{y}-1..r^{\prime}_{y}+1]$.
\end{enumerate}
\end{lemma}
\begin{proof}
    Lemma~\ref{lem:dynamic_f_recover}~\ref{enum:dynamic_f_recover:X}, 
    Lemma~\ref{lem:dynamic_f_recover}~\ref{enum:dynamic_f_recover:Y}, 
    and Lemma~\ref{lem:dynamic_f_recover}~\ref{enum:dynamic_f_recover:Z} are 
    the dynamic versions of Lemma~\ref{lem:recover_super_property}~\ref{enum:recover_super_property:1}, 
    Lemma~\ref{lem:recover_super_property}~\ref{enum:recover_super_property:2}, 
    and Lemma~\ref{lem:m_recover_equality}, respectively. 
    Therefore, Lemma~\ref{lem:dynamic_f_recover} can be proved by the approach used to 
    prove Lemma~\ref{lem:recover_super_property} and Lemma~\ref{lem:m_recover_equality}. 
    
    Let $h$, $C$, and $\gamma$ be the level, associated string, and attractor position of interval attractor $([p, q], [\ell, r])$. 
    Analogously, $h^{\prime}$, $C^{\prime}$, and $\gamma^{\prime}$ are defined for interval attractor $([p^{\prime}, q^{\prime}], [\ell^{\prime}, r^{\prime}])$. 
    Let $K \geq 0$ and $K^{\prime} \geq 0$ be two integers satisfying 
    $([p, q], [\ell, r]) \in \Psi_{\lcp}(K)$ 
    and 
    $([p, q], [\ell$, $r]) \in \Psi^{\prime}_{\lcp}(K^{\prime})$, respectively. 
    If we prove Lemma~\ref{lem:dynamic_f_recover} can be proved by the approach used to 
    Lemma~\ref{lem:recover_super_property} and Lemma~\ref{lem:m_recover_equality}, 
    then we need to prove $h = h^{\prime}$, $C = C^{\prime}$, 
    $T[p-1..\gamma-1] = T^{\prime}[p^{\prime}-1..\gamma^{\prime}-1]$, $T[\gamma..r+1] = T^{\prime}[\gamma^{\prime}..r^{\prime}+1]$, and $K = K^{\prime}$. 
    $h = h^{\prime}$, $T[p-1..\gamma-1] = T^{\prime}[p^{\prime}-1..\gamma^{\prime}-1]$, and $T[\gamma..r+1] = T^{\prime}[\gamma^{\prime}..r^{\prime}+1]$ 
    follow from Lemma~\ref{lem:dynamic_IA_super_correspondence}. 
    $C = C^{\prime}$ follows from Lemma~\ref{lem:dynamic_RR_subset}~\ref{enum:dynamic_RR_subset:C}. 
    We can apply Lemma~\ref{lem:dynamic_RR_subset}~\ref{enum:dynamic_RR_subset:lcp} to the two interval attractors 
    $([p, q], [\ell, r])$ and $([p^{\prime}, q^{\prime}], [\ell^{\prime}, r^{\prime}])$ 
    because $K > \sum_{w = 1}^{h+1} \lfloor \mu(w) \rfloor$ follows from Lemma~\ref{lem:recover_basic_property}~\ref{enum:recover_basic_property:1}. 
    $K = K^{\prime}$ follows from Lemma~\ref{lem:dynamic_RR_subset}~\ref{enum:dynamic_RR_subset:lcp}. 
    Therefore, Lemma~\ref{lem:dynamic_f_recover} can be proved by the approach used to 
    prove Lemma~\ref{lem:recover_super_property} and Lemma~\ref{lem:m_recover_equality}. 
\end{proof}

\subsubsection{Proof of Lemma~\ref{lem:RB_IA_proceeding_formula}}\label{subsubsec:proof_RB_IA_proceeding_formula}
For proving Lemma~\ref{subsubsec:proof_RB_IA_proceeding_formula}, 
we leverage the set $\Delta(h, b)$ of intervals introduced in Section~\ref{subsec:RR_delta}. 
Here, $b$ is a position of sequence $S^{h}$. 
Similar to sequence $A(s, e)$ of intervals, 
the set $\Delta(h, b)$ is changed by inserting the given character $c$ into input string $T$. 
To distinguish between the set $\Delta(h, b)$ before and after changing $T$, 
let $\Delta^{\prime}(h, b)$ be the set $\Delta(h, b)$ after changing $T$ 
(i.e., the set $\Delta^{\prime}(h, b)$ is defined using RLSLP $\mathcal{G}^{R}_{\ins}$ instead of RLSLP $\mathcal{G}^{R}$).

\begin{proposition}\label{prop:dynamicP_I_last_syncro}
Consider two intervals $[s, e] \subseteq [1, n]$ ($s < e$) and $[s^{\prime}, e^{\prime}] \subseteq [1, n+1]$ ($s^{\prime} < e^{\prime}$) in 
two strings $T$ and $T^{\prime}$, respectively. 
Let $(h, b)$ be a pair of an integer $h \in [0, H]$ and a position $b \in [1, |S^{h}|]$ of sequence $S^{h}$ satisfying 
$[s, e] \in \Delta(h, b)$. 
If $T[s..e] = T^{\prime}[s^{\prime}..e^{\prime}]$ holds, 
then there exists a position $b^{\prime} \in [1, |S^{h}_{\ins}|]$ of sequence $S^{h}_{\ins}$ satisfying the following two conditions: 
\begin{enumerate}[label=\textbf{(\roman*)}]
    \item $[s^{\prime}, e^{\prime}] \in \Delta^{\prime}(h, b^{\prime})$; 
    \item let $T[x^{h}_{b}..y^{h}_{b}]$ be the substring of input string $T$ derived from the $b$-th nonterminal $S^{h}[b]$ of 
    sequence $S^{h}$. 
    Similarly, let $T^{\prime}[x^{\prime h}_{b}..y^{\prime h}_{b}]$ be the substring of string $T^{\prime}$ 
    derived from the $b^{\prime}$-th nonterminal $S^{h}_{\ins}[b^{\prime}]$ of sequence $S^{h}_{\ins}$.    
    Then, $|[s, x^{h}_{b}]| = |[s^{\prime}, x^{\prime h}_{b}]|$.    
\end{enumerate}
\end{proposition}
\begin{proof}
Let $[s^{0}, e^{0}]$, $[s^{1}, e^{1}]$, $\ldots$, $[s^{k}, e^{k}]$ and 
$[s^{\prime 0}, e^{\prime 0}]$, $[s^{\prime 1}, e^{\prime 1}]$, $\ldots$, $[s^{\prime k^{\prime}}, e^{\prime k^{\prime}}]$ 
be two sequences $A(s, e)$ and $A^{\prime}(s^{\prime}, e^{\prime})$ of intervals, respectively. 
Since $[s, e] \in \Delta(h, b)$, 
$k = h$ and $s^{h} = b$ follow from the definition of the set $\Delta(h, b)$. 
Lemma~\ref{lem:dynamicP_rec_function_syncro_property}~\ref{enum:dynamicP_rec_function_syncro_property:2} shows that 
$k = k^{\prime}$ holds. 
Let $T[x^{h}_{s}..y^{h}_{s}]$ be the substring derived from the $s^{h}$-th nonterminal of sequence $S^{h}$ in input string $T$. 
Similarly, 
let $T^{\prime}[x^{\prime h}_{s}..y^{\prime h}_{s}]$ be the substring  
derived from the $s^{\prime h}$-th nonterminal of sequence $S^{h}_{\ins}$ 
in input string $T^{\prime}$. 
Lemma~\ref{lem:dynamicP_rec_function_syncro_property}~\ref{enum:dynamicP_rec_function_syncro_property:1} shows that 
$|[s, x^{h}_{s}]| = |[s^{\prime}, x^{\prime h}_{s}]|$ holds. 

We prove Proposition~\ref{prop:dynamicP_I_last_syncro}.
$[s^{\prime}, e^{\prime}] \in \Delta^{\prime}(h, s^{\prime h})$ follows from 
the definition of the set $\Delta^{\prime}(h, s^{\prime h})$. 
$|[s, x^{h}_{b}]| = |[s^{\prime}, x^{\prime h}_{s}]|$ 
follows from $|[s, x^{h}_{s}]| = |[s^{\prime}, x^{\prime h}_{s}]|$ and 
$x^{h}_{s} = x^{h}_{b}$. 
Therefore, we obtain Proposition~\ref{prop:dynamicP_I_last_syncro}. 
\end{proof}

\begin{proposition}\label{prop:dynamicB_I_last_syncro}
Consider two intervals $[s, e] \subseteq [1, n]$ ($s < e$) and $[s^{\prime}, e^{\prime}] \subseteq [1, n+1]$ ($s^{\prime} < e^{\prime}$) in 
two strings $T$ and $T^{\prime}$, respectively. 
Let $(h, b^{\prime})$ be a pair of an integer $h \in [0, H^{\prime}]$ and a position $b^{\prime} \in [1, |S^{h}|]$ of sequence $S^{h}_{\ins}$ satisfying 
$[s, e] \in \Delta^{\prime}(h, b^{\prime})$. 
If $T[s..e] = T^{\prime}[s^{\prime}..e^{\prime}]$ holds, 
then there exists a position $b \in [1, |S^{h}|]$ of sequence $S^{h}$ satisfying the following two conditions: 
\begin{enumerate}[label=\textbf{(\roman*)}]
    \item $[s, e] \in \Delta(h, b)$; 
    \item let $T[x^{h}_{b}..y^{h}_{b}]$ be the substring of input string $T$ derived from the $b$-th nonterminal $S^{h}[b]$ of 
    sequence $S^{h}$. 
    Similarly, let $T^{\prime}[x^{\prime h}_{b}..y^{\prime h}_{b}]$ be the substring of string $T^{\prime}$ 
    derived from the $b^{\prime}$-th nonterminal $S^{h}_{\ins}[b^{\prime}]$ of sequence $S^{h}_{\ins}$.    
    Then, $|[s, x^{h}_{b}]| = |[s^{\prime}, x^{\prime h}_{b}]|$.    
\end{enumerate}
\end{proposition}
\begin{proof}
    Proposition~\ref{prop:dynamicB_I_last_syncro} is symmetric to Proposition~\ref{prop:dynamicP_I_last_syncro}. 
    We can prove Proposition~\ref{prop:dynamicB_I_last_syncro} using the same approach as for Proposition~\ref{prop:dynamicP_I_last_syncro}.    
\end{proof}

\begin{proof}[Proof of Lemma~\ref{lem:RB_IA_proceeding_formula}]
From Definition~\ref{def:RR_Delta}, 
there exists a position $b \in [1, |S^{h}|]$ in sequence $S^{h}$ 
satisfying the following four conditions: 
\begin{itemize}
\item $p = \min \{ i \mid [i, j] \in \Delta(h, b) \}$; 
\item $q = \max \{ i \mid [i, j] \in \Delta(h, b) \}$; 
\item $\ell = \min \{ j \mid [i, j] \in \Delta(h, b) \}$; 
\item $r = \max \{ j \mid [i, j] \in \Delta(h, b) \}$. 
\end{itemize}
Lemma~\ref{lem:IA_maximal_lemma} shows that 
the set $\Delta(h, b)$ contains interval $[q, r]$. 
Let $T[x^{h}_{b}..y^{h}_{b}]$ be the substring of input string $T$ derived from the $b$-th nonterminal $S^{h}[b]$ of 
sequence $S^{h}$. 
Because of $T[q..r] = T^{\prime}[i^{\prime} + |[p, q]|-1..j^{\prime}]$, 
Proposition~\ref{prop:dynamicP_I_last_syncro} shows that 
there exists a position $b_{1}$ of sequence $S^{h}_{\ins}$ satisfying the following two conditions: 
\begin{itemize}
    \item $[i^{\prime} + |[p, q]|-1, j^{\prime}] \in \Delta^{\prime}(h, b_{1})$;
    \item let $T^{\prime}[x^{\prime h}_{b_{1}}..y^{\prime h}_{b_{1}}]$ be the substring of string $T^{\prime}$ 
    derived from the $b_{1}$-th nonterminal $S^{h}_{\ins}[b_{1}]$ of sequence $S^{h}_{\ins}$. 
    Then, $|[q, x^{h}_{b}]| = |[i^{\prime} + |[p, q]|-1, x^{\prime h}_{b_{1}}]|$ holds. 
\end{itemize}
\end{proof}

In contrast, 
Lemma~\ref{lem:IA_maximal_lemma} shows that 
the set $\Delta(h, b)$ contains interval $[p, \ell]$. 
Because of $T[p..\ell] = T^{\prime}[i^{\prime}..j^{\prime} - |[\ell, r]| + 1]$, 
Proposition~\ref{prop:dynamicP_I_last_syncro} shows that 
there exists a position $b_{2}$ of sequence $S^{h}_{\ins}$ satisfying the following two conditions: 
\begin{itemize}
    \item $[i^{\prime}, j^{\prime} - |[\ell, r]| + 1] \in \Delta^{\prime}(h, b_{2})$;
    \item let $T^{\prime}[x^{\prime h}_{b_{2}}..y^{\prime h}_{b_{2}}]$ be the substring of string $T^{\prime}$ 
    derived from the $b_{2}$-th nonterminal $S^{h}_{\ins}[b_{2}]$ of sequence $S^{h}_{\ins}$. 
    Then, $|[p, x^{h}_{b}]| = |[i^{\prime}, x^{\prime h}_{b_{2}}]|$ holds. 
\end{itemize}

We prove $b_{1} = b_{2}$. 
$x^{\prime h}_{b_{1}} = i^{\prime} + |[p, x^{h}_{b}]| - 1$ follows from 
$|[q, x^{h}_{b}]| = |[i^{\prime} + |[p, q]|-1, x^{\prime h}_{b_{1}}]|$. 
$x^{\prime h}_{b_{2}} = i^{\prime} + |[p, x^{h}_{b}]| - 1$ follows from $|[p, x^{h}_{b}]| = |[i^{\prime}, x^{\prime h}_{b_{2}}]|$. 
Because of $x^{\prime h}_{b_{1}} = x^{\prime h}_{b_{2}}$, 
the two positions $b_{1}$ and $b_{2}$ are the same. 

We define four positions $p^{\prime}$, $q^{\prime}$, $\ell^{\prime}$, and $r^{\prime}$ as follows: 
\begin{itemize}
\item $p^{\prime} = \min \{ i \mid [i, j] \in \Delta^{\prime}(h, b_{1}) \}$; 
\item $q^{\prime} = \max \{ i \mid [i, j] \in \Delta^{\prime}(h, b_{1}) \}$; 
\item $\ell^{\prime} = \min \{ j \mid [i, j] \in \Delta^{\prime}(h, b_{1}) \}$; 
\item $r^{\prime} = \max \{ j \mid [i, j] \in \Delta^{\prime}(h, b_{1}) \}$. 
\end{itemize}

From Definition~\ref{def:RR_Delta}, 
set $\Psi^{\prime}_{\RR}$ contains interval attractor $([p^{\prime}, q^{\prime}], [\ell^{\prime}, r^{\prime}])$. 
Therefore, Lemma~\ref{lem:RB_IA_proceeding_formula} holds if 
$p^{\prime} = i^{\prime}$, 
$q^{\prime} = i^{\prime} + |[p, q]| - 1$, 
$\ell^{\prime} = j^{\prime} - |[\ell, r]| + 1$, 
and $r^{\prime} = j^{\prime}$ hold.

\textbf{Proof of $p^{\prime} = i^{\prime}$.}
We prove $p^{\prime} = i^{\prime}$ by contradiction. 
$p^{\prime} \leq i^{\prime}$ follows from $[i^{\prime}, j^{\prime} - |[\ell, r]| + 1] \in \Delta^{\prime}(h, b_{1})$. 
We assume that $p^{\prime} \neq i^{\prime}$ holds. 
Then, $p^{\prime} < i^{\prime}$ follows from $p^{\prime} \neq i^{\prime}$ and $p^{\prime} \leq i^{\prime}$. 
Lemma~\ref{lem:IA_maximal_lemma} shows that 
set $\Delta^{\prime}(h, b_{1})$ contains interval $[i^{\prime}-1, j^{\prime}]$ 
because $p^{\prime} < i^{\prime}$ and $[i^{\prime} + |[p, q]|-1, j^{\prime}] \in \Delta^{\prime}(h, b_{1})$ hold. 
Proposition~\ref{prop:dynamicB_I_last_syncro} shows that 
$[p-1, r] \in \Delta(h, b)$ holds 
because $T[p - 1..r] = T^{\prime}[i^{\prime}-1..j^{\prime}]$ 
and $[i^{\prime}-1, j^{\prime}] \in \Delta^{\prime}(h, b_{1})$ hold. 

On the other hand, 
$[p-1, r] \not \in \Delta(h, b)$ follows from $p = \min \{ i \mid [i, j] \in \Delta(h, b) \}$. 
The two facts $[p-1, r] \in \Delta(h, b)$ and $[p-1, r] \not \in \Delta(h, b)$ yield a contradiction. 
Therefore, $p^{\prime} = p + \epsilon$ must hold. 

\textbf{Proof of $r^{\prime} = j^{\prime}$.}
We can prove $r^{\prime} = j^{\prime}$ using the same approach as for $p^{\prime} = i^{\prime}$.

\textbf{Proof of $q^{\prime} = i^{\prime} + |[p, q]| - 1$.}
We prove $q^{\prime} = i^{\prime} + |[p, q]| - 1$ by contradiction. 
$q^{\prime} \geq i^{\prime} + |[p, q]| - 1$ follows from $[i^{\prime} + |[p, q]|-1, j^{\prime}] \in \Delta^{\prime}(h, b_{1})$.
We assume that $q^{\prime} \neq i^{\prime} + |[p, q]| - 1$ holds. 
Then, $q^{\prime} > i^{\prime} + |[p, q]| - 1$ follows from $q^{\prime} \neq i^{\prime} + |[p, q]| - 1$ and $q^{\prime} \geq i^{\prime} + |[p, q]| - 1$. 

Lemma~\ref{lem:IA_maximal_lemma} shows that 
set $\Delta^{\prime}(h, b_{1})$ contains interval $[i^{\prime} + |[p, q]|, r^{\prime}]$. 
Proposition~\ref{prop:dynamicB_I_last_syncro} shows that 
$[q+1, r] \in \Delta(h, b)$ holds 
because $T[q+1..r] = T^{\prime}[i^{\prime} + |[p, q]|..r^{\prime}]$ and 
$[i^{\prime} + |[p, q]|, r^{\prime}] \in \Delta^{\prime}(h, b_{1})$. 

On the other hand, 
$[q+1, r] \not \in \Delta(h, b)$ follows from $q = \max \{ i \mid [i, j] \in \Delta(h, b) \}$. 
The two facts $[q+1, r] \in \Delta(h, b)$ and $[q+1, r] \not \in \Delta(h, b)$ yield a contradiction. 
Therefore, $q^{\prime} = i^{\prime} + |[p, q]| - 1$ must hold. 

\textbf{Proof of $\ell^{\prime} = j^{\prime} - |[\ell, r]| + 1$.}
We can prove $\ell^{\prime} = j^{\prime} - |[\ell, r]| + 1$ using the same approach as for $q^{\prime} = i^{\prime} + |[p, q]| - 1$.
Finally, $([i^{\prime}, i^{\prime} + |[p, q]| - 1], [j^{\prime} - |[\ell, r]| + 1, j^{\prime}]) \in \Psi^{\prime}_{\RR}$ holds. 

\subsubsection{Proof of Lemma~\ref{lem:dynamic_IA_super_correspondence}}\label{subsubsec:proof_dynamic_IA_super_correspondence}
From Definition~\ref{def:RR_Delta}, 
there exists a position $b \in [1, |S^{h}|]$ in sequence $S^{h}$ 
satisfying the following four conditions: 
\begin{itemize}
\item $p = \min \{ i \mid [i, j] \in \Delta(h, b) \}$; 
\item $q = \max \{ i \mid [i, j] \in \Delta(h, b) \}$; 
\item $\ell = \min \{ j \mid [i, j] \in \Delta(h, b) \}$; 
\item $r = \max \{ j \mid [i, j] \in \Delta(h, b) \}$. 
\end{itemize}

Similarly, 
there exists a position $b^{\prime} \in [1, |S^{h}_{\ins}|]$ in sequence $S^{h}_{\ins}$ 
satisfying the following four conditions: 
\begin{itemize}
\item $p^{\prime} = \min \{ i \mid [i, j] \in \Delta^{\prime}(h^{\prime}, b^{\prime}) \}$; 
\item $q^{\prime} = \max \{ i \mid [i, j] \in \Delta^{\prime}(h^{\prime}, b^{\prime}) \}$; 
\item $\ell^{\prime} = \min \{ j \mid [i, j] \in \Delta^{\prime}(h^{\prime}, b^{\prime}) \}$; 
\item $r^{\prime} = \max \{ j \mid [i, j] \in \Delta^{\prime}(h^{\prime}, b^{\prime}) \}$. 
\end{itemize}

We prove $h = h^{\prime}$ by contradiction. 
We assume that $h \neq h^{\prime}$ holds. 
$[p, q] \in \Delta(h, b)$ and $[p^{\prime}, q^{\prime}] \in \Delta^{\prime}(h^{\prime}, b^{\prime})$ 
follow from Lemma~\ref{lem:IA_maximal_lemma}. 
Because of $T[p..q] = T^{\prime}[p^{\prime}..q^{\prime}]$, 
Proposition~\ref{prop:dynamicP_I_last_syncro} shows that 
there exists a position $\hat{b}$ of sequence $S^{h}_{\ins}$ satisfying $[p^{\prime}, q^{\prime}] \in \Delta^{\prime}(h, \hat{b})$. 
$\Delta^{\prime}(h^{\prime}, b^{\prime}) \cap \Delta^{\prime}(h, \hat{b}) \neq \emptyset$ follows from 
$[p^{\prime}, q^{\prime}] \in \Delta^{\prime}(h^{\prime}, b^{\prime})$ and $[p^{\prime}, q^{\prime}] \in \Delta^{\prime}(h, \hat{b})$. 
On the other hand, $\Delta^{\prime}(h^{\prime}, b^{\prime}) \cap \Delta^{\prime}(h, \hat{b}) = \emptyset$ holds follows from the definition of set $\Delta^{\prime}(h^{\prime}, b^{\prime})$. 
The two facts $\Delta^{\prime}(h^{\prime}, b^{\prime}) \cap \Delta^{\prime}(h, \hat{b}) \neq \emptyset$ 
and $\Delta^{\prime}(h^{\prime}, b^{\prime}) \cap \Delta^{\prime}(h, \hat{b})$ yield a contradiction. 
Therefore, $h = h^{\prime}$ must hold. 
Similarly, $b^{\prime} = \hat{b}$ can be proved using the same approach. 

\textbf{Proof of Lemma~\ref{lem:dynamic_IA_super_correspondence}(i).}
From Lemma~\ref{lem:RB_IA_proceeding_formula}, 
set $\Psi^{\prime}_{\RR}$ contains interval attractor $([p^{\prime}, p^{\prime} + |[p, q]| - 1], [r^{\prime} - |[\ell, r]| + 1, r^{\prime}])$. 
We prove $([p^{\prime}, p^{\prime} + |[p, q]| - 1], [r^{\prime} - |[\ell, r]| + 1, r^{\prime}]) = ([p^{\prime}, q^{\prime}], [\ell^{\prime}, r^{\prime}])$ by contradiction. 
We assume that $([p^{\prime}, p^{\prime} + |[p, q]| - 1], [r^{\prime} - |[\ell, r]| + 1, r^{\prime}]) \neq ([p^{\prime}, q^{\prime}], [\ell^{\prime}, r^{\prime}])$ holds. 
Then, Lemma~\ref{lem:IA_super_basic_property}~\ref{enum:IA_super_basic_property:3} shows that $[p^{\prime}, r^{\prime}] \neq [p^{\prime}, r^{\prime}]$ holds. 
The two facts $[p^{\prime}, r^{\prime}] \neq [p^{\prime}, r^{\prime}]$ and $[p^{\prime}, r^{\prime}] = [p^{\prime}, r^{\prime}]$ yield a contradiction. 
Therefore, $([p^{\prime}, p^{\prime} + |[p, q]| - 1], [r^{\prime} - |[\ell, r]| + 1, r^{\prime}]) = ([p^{\prime}, q^{\prime}], [\ell^{\prime}, r^{\prime}])$ must hold. 

We prove Lemma~\ref{lem:dynamic_IA_super_correspondence}(i). 
$|[p, q]| = |[p^{\prime}, q^{\prime}]|$ follows from $p^{\prime} + |[p, q]| - 1 = q^{\prime}$. 
$|[\ell, r]| = |[\ell^{\prime}, r^{\prime}]|$ follows from $r^{\prime} - |[\ell, r]| + 1 = \ell^{\prime}$. 
Therefore, Lemma~\ref{lem:dynamic_IA_super_correspondence}(i) holds. 

\textbf{Proof of Lemma~\ref{lem:dynamic_IA_super_correspondence}(ii).}
We already proved $h = h^{\prime}$. 

\textbf{Proof of Lemma~\ref{lem:dynamic_IA_super_correspondence}(iii).}
We prove $|[p, \gamma]| = |[p^{\prime}, \gamma^{\prime}]|$. 
Let $T[x^{h}_{b}..y^{h}_{b}]$ be the substring of input string $T$ derived from the $b$-th nonterminal $S^{h}[b]$ of sequence $S^{h}$. 
Similarly, let $T^{\prime}[x^{\prime h}_{b}..y^{\prime h}_{b}]$ be the substring of string $T^{\prime}$ 
derived from the $b^{\prime}$-th nonterminal $S^{h}_{\ins}[b^{\prime}]$ of sequence $S^{h}_{\ins}$.    
Then, $|[p, x^{h}_{b}]| = |[p^{\prime}, x^{\prime h}_{b}]|$ follows from Proposition~\ref{prop:dynamicP_I_last_syncro}. 
$\gamma = p + |[p, x^{h}_{b}]| - 1$ and $\gamma^{\prime} = p + |[p^{\prime}, x^{\prime h}_{b}]| - 1$ follow from 
the definition of attractor position. 
Therefore, $|[p, \gamma]| = |[p^{\prime}, \gamma^{\prime}]|$ follows from 
$\gamma = p + |[p, x^{h}_{b}]| - 1$, 
$\gamma^{\prime} = p + |[p^{\prime}, x^{\prime h}_{b}]| - 1$, and $|[p, x^{h}_{b}]| = |[p^{\prime}, x^{\prime h}_{b}]|$. 

We prove Lemma~\ref{lem:dynamic_IA_super_correspondence}(iii). 
$T[p-1..\gamma-1] = T^{\prime}[p^{\prime}-1..\gamma^{\prime}-1]$ follows form 
$T[p-1..r+1] = T^{\prime}[p^{\prime}-1..r^{\prime}+1]$ and $|[p, \gamma]| = |[p^{\prime}, \gamma^{\prime}]|$. 
$T[\gamma..r+1] = T^{\prime}[\gamma^{\prime}..r^{\prime}+1]$ follows from 
$T[p-1..r+1] = T^{\prime}[p^{\prime}-1..r^{\prime}+1]$ and $T[p-1..\gamma-1] = T^{\prime}[p^{\prime}-1..\gamma^{\prime}-1]$. 
Therefore, Lemma~\ref{lem:dynamic_IA_super_correspondence}(iii) holds. 

\textbf{Proof of Lemma~\ref{lem:dynamic_IA_super_correspondence}(iv).}
%\textbf{Proof of Lemma~\ref{lem:dynamic_IA_super_correspondence}(iv) for $|[\gamma, r]| > \sum_{w = 1}^{h+1} \lfloor \mu(w) \rfloor$.}
$|[\gamma, r]|, |[\gamma^{\prime}, r^{\prime}]| > \sum_{w = 1}^{h+1} \lfloor \mu(w) \rfloor$ 
follows from $T[\gamma..r+1] = T^{\prime}[\gamma^{\prime}..r^{\prime}+1]$ 
and $|[\gamma, r]| > \sum_{w = 1}^{h+1} \lfloor \mu(w) \rfloor$. 
$|\lcp(T[\gamma..r], T[\gamma^{\prime}..r^{\prime}])| > \sum_{w = 1}^{h+1} \lfloor \mu(w) \rfloor$ follows from 
$T[\gamma..r+1] = T^{\prime}[\gamma^{\prime}..r^{\prime}+1]$ and $|[\gamma, r]|, |[\gamma^{\prime}, r^{\prime}]| > \sum_{w = 1}^{h+1} \lfloor \mu(w) \rfloor$. 

Because of $|[\gamma, r]| > \sum_{w = 1}^{h+1} \lfloor \mu(w) \rfloor$, 
the associated string $C$ of the interval attractor $([p, q], [\ell, r])$ is defined as $C = \val(S^{h}[b])$. 
$|C| \leq \lfloor \mu(h+1) \rfloor$ follows from Lemma~\ref{lem:associated_string_C}~\ref{enum:associated_string_C:1}. 
Similarly, 
the associated string $C^{\prime}$ of the interval attractor $([p^{\prime}, q^{\prime}], [\ell^{\prime}, r^{\prime}])$ is defined as 
$C^{\prime} = \val(S^{h}_{\ins}[b^{\prime}])$, 
and $|C^{\prime}| \leq \lfloor \mu(h+1) \rfloor$ holds.

Let $\alpha = \min \{ |C|, |C^{\prime}| \}$. 
Then, $\alpha \leq \lfloor \mu(h+1) \rfloor$ holds. 
We can apply Lemma~\ref{lem:dynamic_rr_property}~\ref{enum:dynamic_rr_property:right} to the two nonterminals 
$S^{h}[b]$ and $S^{h}_{\ins}[b^{\prime}]$ 
if the following condition holds: 
$T[\gamma..\gamma + \alpha + \sum_{w = 1}^{h} \lfloor \mu(w) \rfloor] = T^{\prime}[\gamma^{\prime}..\gamma^{\prime} + \alpha + \sum_{w = 1}^{h} \lfloor \mu(w) \rfloor]$. 
Since $\alpha \leq \lfloor \mu(h+1) \rfloor$ and $|\lcp(T[\gamma..r], T[\gamma^{\prime}..r^{\prime}])| > \sum_{w = 1}^{h+1} \lfloor \mu(w) \rfloor$, 
this conditions is satisfied. 
Lemma~\ref{lem:dynamic_rr_property}~\ref{enum:dynamic_rr_property:right} shows that $C = C^{\prime}$. 
Therefore, we obtain Lemma~\ref{lem:dynamic_IA_super_correspondence}(iv). 

\subsection{Update of Sampling Subset}\label{subsec:update_sampling_subset}
This subsection shows that we can obtain a sampling subset for RLSLP $\mathcal{G}^{R}_{\ins}$ 
by modifying the interval attractors of the sampling subset $\Psi_{\samp}$ for RLSLP $\mathcal{G}^{R}$, 
which is introduced in Section~\ref{subsec:sampling_subset}. 

\paragraph{Set $\mathcal{Z}^{\symA}$ of strings.}
For this subsection, 
we introduce a set $\mathcal{Z}^{\symA} \subseteq \Sigma^{+}$ of $k$ strings $Z^{\symA}_{1}, Z^{\symA}_{2}, \ldots, Z^{\symA}_{k}$ ($Z^{\symA}_{1} \prec Z^{\symA}_{2} \prec \cdots \prec Z^{\symA}_{k}$). 
This set consists of strings such that 
each string $Z^{\symA}_{s} \in \mathcal{Z}^{\symA}$ satisfies 
$\Psi_{\samp} \cap \Psi_{\OLD} \cap \Psi_{\str}(Z^{\symA}_{s}) \neq \emptyset$ 
and $(\Psi_{\RR} \setminus \Psi_{\OLD}) \cap \Psi_{\str}(Z^{\symA}_{s}) \neq \emptyset$ 
(i.e., $\mathcal{Z}^{\symA} = \{ Z \in \Sigma^{+} \mid (\Psi_{\samp} \cap \Psi_{\OLD} \cap \Psi_{\str}(Z) \neq \emptyset) \land ((\Psi_{\RR} \setminus \Psi_{\OLD}) \cap \Psi_{\str}(Z) \neq \emptyset)\}$). 

The following lemma states properties of set $\mathcal{Z}^{\symA}$. 
\begin{lemma}\label{lem:Z_samp_A_property}
Consider the $m$ interval attractors $([p_{1}, q_{1}]$, $[\ell_{1}, r_{1}])$, 
$([p_{2}, q_{2}]$, $[\ell_{2}, r_{2}])$, $\ldots$, $([p_{m}, q_{m}]$, $[\ell_{m}, r_{m}])$ 
of subset $\Psi_{\str}(Z^{\symA}_{s})$ for a string $Z^{\symA}_{s}$ in set $\mathcal{Z}^{\symA}$. 
Then, the following three statements hold: 

\begin{enumerate}[label=\textbf{(\roman*)}]
    \item \label{enum:Z_samp_A_property:1} $m \geq 2$;
    \item \label{enum:Z_samp_A_property:2} $r_{i} \leq n-1$ and $p_{i} \geq 2$ for all $i \in [1, m]$;    
    \item \label{enum:Z_samp_A_property:3} $\Psi_{\str}(Z^{\symA}_{s}) \cap \Psi_{\run} = \emptyset$.
\end{enumerate}
\end{lemma}
\begin{proof}
    The proof of Lemma~\ref{lem:Z_samp_A_property} is as follows. 

    \textbf{Proof of Lemma~\ref{lem:Z_samp_A_property}(i).}
    The definition of set $Z^{\symA}_{s}$ indicates that 
    two disjoint subsets $\Psi_{\samp} \cap \Psi_{\OLD}$ and 
    $\Psi_{\RR} \setminus \Psi_{\OLD}$ contain interval attractors such that 
    each interval attractor is contained in subset $\Psi_{\str}(Z^{\symA}_{s})$.
    Therefore, we obtain $m \geq 2$. 

    \textbf{Proof of Lemma~\ref{lem:Z_samp_A_property}(ii).}
    Because of $([p_{i}, q_{i}]$, $[\ell_{i}, r_{i}]) \in \Psi_{\str}(Z^{\symA}_{s})$, 
    $Z^{\symA}_{s} = T[p_{i}-1..r_{i}+1]$ follows from the definition of the subset $\Psi_{\str}(Z^{\symA}_{s})$. 
    We prove $r_{i} \leq n-1$ and $p_{i} \geq 2$ by contradiction. 
    We assume that $[p_{i}-1, r_{i}+1] \not \subseteq [1, n]$ holds. 
    Then, Lemma~\ref{lem:psi_str_occ_property} shows that $|\Psi_{\str}(Z^{\symA}_{s})| = 1$ holds. 
    On the other hand, $|\Psi_{\str}(Z^{\symA}_{s})| \geq 2$ follows from $m \geq 2$. 
    The two facts $|\Psi_{\str}(Z^{\symA}_{s})| = 1$ and $|\Psi_{\str}(Z^{\symA}_{s})| \geq 2$ yield a contradiction. 
    Therefore, $[p_{i}-1, r_{i}+1] \subseteq [1, n]$ must hold, 
    i.e., $r_{i} \leq n-1$ and $p_{i} \geq 2$ hold. 
    
    \textbf{Proof of Lemma~\ref{lem:Z_samp_A_property}(iii).}
    We prove $\Psi_{\str}(Z^{\symA}_{s}) \cap \Psi_{\run} = \emptyset$. 
    $\Psi_{\samp} \cap \Psi_{\OLD} \cap \Psi_{\str}(Z^{\symA}_{s}) \neq \emptyset$ 
    follows from the definition of set $\mathcal{Z}^{\symA}$. 
    Let $([p, q], [\ell, r])$ be an interval attractor in set $\Psi_{\samp} \cap \Psi_{\OLD} \cap \Psi_{\str}(Z^{\symA}_{s})$.     
    Because of $([p, q], [\ell, r]) \in \Psi_{\samp}$, 
    Lemma~\ref{lem:samp_basic_property}~\ref{enum:samp_basic_property:3} shows that 
    subset $\Psi_{\run}$ does not contain the interval attractor $([p, q], [\ell, r])$. 
    For each interval attractor $([p_{i}, q_{i}], [\ell_{i}, r_{i}]) \in \Psi_{\str}(Z^{\symA}_{s})$, 
    we can apply 
    Lemma~\ref{lem:psi_equality_basic_property}~\ref{enum:psi_equality_basic_property:4} 
    to the two interval attractors $([p, q], [\ell, r])$ and $([p_{i}, q_{i}], [\ell_{i}, r_{i}])$ 
    because $T[p-1..r+1] = T[p_{i}-1..r_{i}+1]$ holds. 
    Lemma~\ref{lem:psi_equality_basic_property}~\ref{enum:psi_equality_basic_property:4} shows that 
    the subset $\Psi_{\run}$ does not contain the interval attractor $([p_{i}, q_{i}], [\ell_{i}, r_{i}])$. 
    Therefore, $\Psi_{\str}(Z^{\symA}_{s}) \cap \Psi_{\run} = \emptyset$ holds. 

    Next, we prove $(\Psi_{\OLD} \setminus \Psi_{\run}) \cap \Psi_{\str}(Z^{\symA}_{s}) = \Psi_{\OLD} \cap \Psi_{\str}(Z^{\symA}_{s})$ by contradiction. 
    We assume that $(\Psi_{\OLD} \setminus \Psi_{\run}) \cap \Psi_{\str}(Z^{\symA}_{s}) \neq \Psi_{\OLD} \cap \Psi_{\str}(Z^{\symA}_{s})$ holds. 
    Then, $\Psi_{\str}(Z^{\symA}_{s}) \cap \Psi_{\run} \neq \emptyset$ holds. 
    The two facts $\Psi_{\str}(Z^{\symA}_{s}) \cap \Psi_{\run} = \emptyset$ and $\Psi_{\str}(Z^{\symA}_{s}) \cap \Psi_{\run} \neq \emptyset$ yield a contradiction. 
    Therefore, $(\Psi_{\OLD} \setminus \Psi_{\run}) \cap \Psi_{\str}(Z^{\symA}_{s}) = \Psi_{\OLD} \cap \Psi_{\str}(Z^{\symA}_{s})$ must hold. 

\end{proof}

For each string $Z^{\symA}_{s} \in \mathcal{Z}^{\symA}$, 
the following lemma shows that an interval attractor of set $(\Psi_{\RR} \setminus \Psi_{\OLD}) \cap \Psi_{\str}(Z^{\symA}_{s})$ can be obtained by capture query.

\begin{lemma}\label{lem:dynamic_samp_substitute}
    Consider the sa-interval $[b^{\symA}_{s}, b^{\symA}_{s} + |\Occ(T, Z^{\symA}_{s})| - 1]$ 
    of a string $Z^{\symA}_{s}$ in set $\mathcal{Z}^{\symA}$. 
    Let $x^{\symA}_{s} \in [1, n]$ be the smallest position in the suffix array $\SA$
    satisfying the following two conditions: 
    \begin{itemize}
    \item $x^{\symA}_{s} \in [b^{\symA}_{s}, b^{\symA}_{s} + |\Occ(T, Z^{\symA}_{s})| - 1]$;
    \item $\SA[x^{\symA}_{s}] + 1 \not \in \{ p \mid ([p, q], [\ell, r]) \in (\Psi_{\OLD} \setminus \Psi_{\run}) \cap \Psi_{\str}(Z^{\symA}_{s}) \}$.
    \end{itemize}
    Then, the following two statements hold:  
    \begin{enumerate}[label=\textbf{(\roman*)}]
    \item \label{enum:dynamic_samp_substitute:1} the position $x^{\symA}_{s}$ exists, and $x^{\symA}_{s} - b^{\symA}_{s} \leq |(\Psi_{\OLD} \setminus \Psi_{\run}) \cap \Psi_{\str}(Z^{\symA}_{s})|$;
    \item \label{enum:dynamic_samp_substitute:2} 
    $I_{\capture}(\SA[x^{\symA}_{s}]+1, \SA[x^{\symA}_{s}] + |Z^{\symA}_{s}| - 2) \in (\Psi_{\RR} \setminus \Psi_{\OLD}) \cap \Psi_{\str}(Z^{\symA}_{s})$ 
    for interval attractor $I_{\capture}(\SA[x^{\symA}_{s}]+1, \SA[x^{\symA}_{s}] + |Z^{\symA}_{s}| - 2)$. 
    \end{enumerate}
\end{lemma}
\begin{proof}
    Consider the $m$ interval attractors 
    Let $([p_{1}, q_{1}]$, $[\ell_{1}, r_{1}])$, 
    $([p_{2}, q_{2}]$, $[\ell_{2}, r_{2}])$, $\ldots$, $([p_{m}, q_{m}]$, $[\ell_{m}, r_{m}])$ introduced in 
    Lemma~\ref{lem:Z_samp_A_property}. 
    For simplicity, 
    we assume that 
    $T[p_{1}-1..n] \prec T[p_{2}-1..n] \prec \cdots \prec T[p_{m}-1..n]$ holds. 
        
    Let $\alpha$ be the smallest integer in set $[1, m]$ satisfying 
    $([p_{\alpha}, q_{\alpha}]$, $[\ell_{\alpha}, r_{\alpha}]) \in \Psi_{\RR} \setminus \Psi_{\OLD}$. 
    Such integer $\alpha$ exists 
    because $(\Psi_{\RR} \setminus \Psi_{\OLD}) \cap \Psi_{\str}(Z^{\symA}_{s}) \neq \emptyset$ 
    follows from the definition of the set $\mathcal{Z}^{\symA}$.

    The following three statements are used to prove Lemma~\ref{lem:dynamic_samp_substitute}. 

    \begin{enumerate}[label=\textbf{(\Alph*)}]
    \item $\SA[b^{\symA}_{s} + i - 1] + 1 = p_{i}$ for all $i \in [1, m]$; 
    \item the $(b^{\symA}_{s} + \alpha - 1)$-th position of the suffix array $\SA$ satisfies the two conditions of Lemma~\ref{lem:dynamic_samp_substitute};
    \item $x^{\symA}_{s} = b^{\symA}_{s} + \alpha - 1$.
    \end{enumerate}
        
    \textbf{Proof of statement (A).}    
    We apply Lemma~\ref{lem:psi_str_occ_property} to the interval attractor $([p_{1}, q_{1}]$, $[\ell_{1}, r_{1}])$. 
    Here, $Z^{\symA}_{s} = T[p_{1}-1..r_{1}+1]$ follows from the definition of the subset $\Psi_{\str}(Z^{\symA}_{s})$. 
    Because of $r_{1} \leq n-1$ and $p_{1} \geq 2$ (Lemma~\ref{lem:Z_samp_A_property}~\ref{enum:Z_samp_A_property:2}), 
    Lemma~\ref{lem:psi_str_occ_property} shows that 
    $\Occ(T, Z^{\symA}_{s}) = \{ p_{1} - 1, p_{2} - 1, \ldots, p_{m} - 1 \}$ holds. 
        
    $\{ \SA[b^{\symA}_{s} + i - 1] \mid i \in [1, m]  \} = \{ p_{i} - 1 \mid i \in [1, m]  \}$ 
    follows from $\{ p_{1} - 1, p_{2} - 1, \ldots, p_{m} - 1 \} = \Occ(T, Z^{\symA}_{s})$. 
    $T[\SA[b^{\symA}_{s}]..n] \prec T[\SA[b^{\symA}_{s} + 1]..n] \prec \cdots \prec T[\SA[b^{\symA}_{s} + m - 1]..n]$ 
    follows from the definition of suffix array. 
    On the other hand, $T[p_{1}-1..n] \prec T[p_{2}-1..n] \prec \cdots \prec T[p_{m}-1..n]$ holds. 
    Therefore, 
    $\SA[b^{\symA}_{s}] + 1 = p_{1}$, 
    $\SA[b^{\symA}_{s} + 1] + 1 = p_{2}$, $\ldots$, 
    $\SA[b^{\symA}_{s} + m - 1] + 1 = p_{m}$ follow from 
    (a) $\{ \SA[b^{\symA}_{s} + i - 1] \mid i \in [1, m]  \} = \{ p_{i} - 1 \mid i \in [1, m]  \}$, 
    (b) $T[\SA[b^{\symA}_{s}]..n] \prec T[\SA[b^{\symA}_{s} + 1]..n] \prec \cdots \prec T[\SA[b^{\symA}_{s} + m - 1]..n]$, 
    and (c) $T[p_{1}-1..n] \prec T[p_{2}-1..n] \prec \cdots \prec T[p_{m}-1..n]$. 

    \textbf{Proof of statement (B).}
    Since $\Occ(T, Z^{\symA}_{s}) = \{ p_{1} - 1, p_{2} - 1, \ldots, p_{m} - 1 \}$, 
    the position $b^{\symA}_{s} + \alpha - 1$ of the suffix array $\SA$ satisfies the first condition of Lemma~\ref{lem:dynamic_samp_substitute} 
    (i.e., $b^{\symA}_{s} + \alpha - 1 \in [b^{\symA}_{s}, b^{\symA}_{s} + |\Occ(T, Z^{\symA}_{s})| - 1]$). 
        
    We show that 
    the $(b^{\symA}_{s} + \alpha - 1)$-th position of the suffix array $\SA$ satisfies the second condition of Lemma~\ref{lem:dynamic_samp_substitute} 
    by contradiction. 
    We assume that the position $b^{\symA}_{s} + \alpha - 1$ of the suffix array $\SA$ does not satisfy the second condition of Lemma~\ref{lem:dynamic_samp_substitute}
    (i.e., $\SA[b^{\symA}_{s} + \alpha - 1] + 1 \in \{ p \mid ([p, q], [\ell, r]) \in (\Psi_{\OLD} \setminus \Psi_{\run}) \cap \Psi_{\str}(Z^{\symA}_{s}) \}$ holds). 
    Under the assumption, 
    set $\Psi_{\OLD} \setminus \Psi_{\run}$ contains an interval attractor $([p, q], [\ell, r])$ satisfying $\SA[b^{\symA}_{s} + \alpha - 1] + 1 = p$ and $T[p-1..r+1] = Z^{\symA}_{s}$. 
    Statement (D) shows that $\SA[b^{\symA}_{s} + \alpha - 1] + 1 = p_{\alpha}$ holds. 
    $p_{\alpha} = p$ follows from $\SA[b^{\symA}_{s} + \alpha - 1] + 1 = p$ and $\SA[b^{\symA}_{s} + \alpha - 1] + 1 = p_{\alpha}$.      
    $[p_{\alpha}, r_{\alpha}] = [p, r]$ follows from $p_{\alpha} = p$, $|[p-1, r+1]| = |Z^{\symA}_{s}|$, and $|[p_{\alpha}-1, r_{\alpha}+1]| = |Z^{\symA}_{s}|$.

    On the other hand, 
    $([p, q], [\ell, r]) \neq ([p_{\alpha}, q_{\alpha}]$, $[\ell_{\alpha}, r_{\alpha}])$ holds because 
    $([p, q], [\ell, r]) \in \Psi_{\OLD}$ and $([p_{\alpha}, q_{\alpha}]$, $[\ell_{\alpha}, r_{\alpha}]) \not \in \Psi_{\OLD}$. 
    Because of $([p, q], [\ell, r]) \neq ([p_{\alpha}, q_{\alpha}]$, $[\ell_{\alpha}, r_{\alpha}])$,         
    Lemma~\ref{lem:IA_super_basic_property}~\ref{enum:IA_super_basic_property:3} shows that 
    $[p, r] \neq [p_{\alpha}, r_{\alpha}]$ holds. 
    The two facts $[p_{\alpha}, r_{\alpha}] = [p, r]$ and $[p, r] \neq [p_{\alpha}, r_{\alpha}]$ yield a contradiction.     
    Therefore, 
    the position $b^{\symA}_{s} + \alpha - 1$ of the suffix array $\SA$ must satisfy the second condition of Lemma~\ref{lem:dynamic_samp_substitute}. 

    \textbf{Proof of statement (C).}    
    We already showed that  
    the $(b^{\symA}_{s} + \alpha - 1)$-th position of the suffix array $\SA$ 
    satisfies the two conditions of Lemma~\ref{lem:dynamic_samp_substitute}. 
    Therefore, $x^{\symA}_{s} \leq b^{\symA}_{s} + \alpha - 1$ holds. 
    
    We prove $x^{\symA}_{s} = b^{\symA}_{s} + \alpha - 1$ by contradiction. 
    We assume that $x^{\symA}_{s} \neq b^{\symA}_{s} + \alpha - 1$ holds. 
    Because of $x^{\symA}_{s} \leq b^{\symA}_{s} + \alpha - 1$, 
    there exists an integer $\alpha^{\prime} \in [1, \alpha-1]$ satisfying 
    the two conditions of Lemma~\ref{lem:dynamic_samp_substitute}. 
    $\SA[b^{\symA}_{s} + \alpha^{\prime} - 1] + 1 \not \in \{ p \mid ([p, q], [\ell, r]) \in (\Psi_{\OLD} \setminus \Psi_{\run}) \cap \Psi_{\str}(Z^{\symA}_{s})  \}$ 
    follows from the second condition of Lemma~\ref{lem:dynamic_samp_substitute}. 
    Statement (A) shows that $\SA[b^{\symA}_{s} + \alpha^{\prime} - 1] + 1 = p_{\alpha^{\prime}}$ holds. 
    Therefore, $([p_{\alpha^{\prime}}, q_{\alpha^{\prime}}]$, $[\ell_{\alpha^{\prime}}, r_{\alpha^{\prime}}]) \not \in \Psi_{\OLD} \setminus \Psi_{\run}$ 
    follows from 
    $\SA[b^{\symA}_{s} + \alpha^{\prime} - 1] + 1 \not \in \{ p \mid ([p, q], [\ell, r]) \in (\Psi_{\OLD} \setminus \Psi_{\run}) \cap \Psi_{\str}(Z^{\symA}_{s}) \}$ and 
    $\SA[b^{\symA}_{s} + \alpha^{\prime} - 1] + 1 = p_{\alpha^{\prime}}$. 
    
    On the other hand, $([p_{\alpha^{\prime}}, q_{\alpha^{\prime}}]$, $[\ell_{\alpha^{\prime}}, r_{\alpha^{\prime}}]) \in \Psi_{\OLD}$ 
    follows from the definition of the integer $\alpha$. 
    $([p_{\alpha^{\prime}}, q_{\alpha^{\prime}}]$, $[\ell_{\alpha^{\prime}}, r_{\alpha^{\prime}}]) \not \in \Psi_{\run}$ follows from 
    $([p_{\alpha^{\prime}}, q_{\alpha^{\prime}}]$, $[\ell_{\alpha^{\prime}}, r_{\alpha^{\prime}}]) \in \Psi_{\str}(Z^{\symA}_{s})$ and $\Psi_{\str}(Z^{\symA}_{s}) \cap \Psi_{\run} = \emptyset$. 
    Therefore, 
    $([p_{\alpha^{\prime}}, q_{\alpha^{\prime}}]$, $[\ell_{\alpha^{\prime}}, r_{\alpha^{\prime}}]) \in \Psi_{\OLD} \setminus \Psi_{\run}$ holds.
    
    The two facts $([p_{\alpha^{\prime}}, q_{\alpha^{\prime}}]$, $[\ell_{\alpha^{\prime}}, r_{\alpha^{\prime}}]) \not \in \Psi_{\OLD} \setminus \Psi_{\run}$ 
    and $([p_{\alpha^{\prime}}, q_{\alpha^{\prime}}]$, $[\ell_{\alpha^{\prime}}, r_{\alpha^{\prime}}]) \in \Psi_{\OLD} \setminus \Psi_{\run}$ yield a contradiction. 
    Therefore, $x^{\symA}_{s} = b^{\symA}_{s} + \alpha - 1$ must hold. 
    
    \textbf{Proof of Lemma~\ref{lem:dynamic_samp_substitute}(i).}
    The position $x^{\symA}_{s}$ exists because of $x^{\symA}_{s} = b^{\symA}_{s} + \alpha - 1$. 
    We prove 
    $x^{\symA}_{s} - b^{\symA}_{s} \leq |(\Psi_{\OLD} \setminus \Psi_{\run}) \cap \Psi_{\str}(Z^{\symA}_{s})|$ 
    (i.e., $\alpha - 1 \leq |(\Psi_{\OLD} \setminus \Psi_{\run}) \cap \Psi_{\str}(Z^{\symA}_{s})|$). 
    For each integer $i \in [1, \alpha-1]$, 
    $([p_{i}, q_{i}]$, $[\ell_{i}, r_{i}]) \in \Psi_{\OLD} \cap \Psi_{\str}(Z^{\symA}_{s})$ 
    follows from the definition of the integer $\alpha$. 
    $([p_{i}, q_{i}]$, $[\ell_{i}, r_{i}]) \in (\Psi_{\OLD} \setminus \Psi_{\run}) \cap \Psi_{\str}(Z^{\symA}_{s})$ 
    follows from 
    $([p_{i}, q_{i}]$, $[\ell_{i}, r_{i}]) \in \Psi_{\OLD} \cap \Psi_{\str}(Z^{\symA}_{s})$ 
    and $\Psi_{\str}(Z^{\symA}_{s}) \cap \Psi_{\run} = \emptyset$ (Lemma~\ref{lem:Z_samp_A_property}~\ref{enum:Z_samp_A_property:3}).
    Therefore, $\alpha - 1 \leq |(\Psi_{\OLD} \setminus \Psi_{\run}) \cap \Psi_{\str}(Z^{\symA}_{s})|$ holds.  

    \textbf{Proof of Lemma~\ref{lem:dynamic_samp_substitute}(ii).}
    %We prove $\CAPQ([\SA[x^{\symA}_{s}]+1, \SA[x^{\symA}_{s}] + |Z^{\symA}_{s}| - 2]) = ([p_{\alpha}, q_{\alpha}]$, $[\ell_{\alpha}, r_{\alpha}])$. 
    Consider interval attractor $I_{\capture}(p_{\alpha}, r_{\alpha})$. 
    Then, Lemma~\ref{lem:IA_maximal_lemma} shows that 
    $I_{\capture}(p_{\alpha}, r_{\alpha}) = ([p_{\alpha}, q_{\alpha}], [\ell_{\alpha}, r_{\alpha}])$. 
    $p_{\alpha} = \SA[x^{\symA}_{s}]+1$ holds 
    because $\SA[b^{\symA}_{s} + \alpha - 1] + 1 = p_{\alpha}$ (statement (A)) 
    and $x^{\symA}_{s} = b^{\symA}_{s} + \alpha - 1$ (statement (C)). 
    $r_{\alpha} = \SA[x^{\symA}_{s}] + |Z^{\symA}_{s}| - 2$ follows from 
    $p_{\alpha} = \SA[x^{\symA}_{s}]+1$ and $|[p_{\alpha}-1, r_{\alpha}]| = |Z^{\symA}_{s}|$. 
    Therefore, 
    $I_{\capture}(\SA[x^{\symA}_{s}]+1, \SA[x^{\symA}_{s}] + |Z^{\symA}_{s}| - 2) = ([p_{\alpha}, q_{\alpha}]$, $[\ell_{\alpha}, r_{\alpha}])$ holds. 
    
    $([p_{\alpha}, q_{\alpha}]$, $[\ell_{\alpha}, r_{\alpha}]) \in (\Psi_{\RR} \setminus \Psi_{\OLD}) \cap \Psi_{\str}(Z^{\symA}_{s})$ follows from the definition of the integer $\alpha$. 
    Therefore, Lemma~\ref{lem:dynamic_samp_substitute} holds. 
\end{proof}

The following lemma states 
the relationships between two sets $\mathcal{Z}^{\symA}$ and $\Psi_{\OLD}$.

\begin{lemma}\label{lem:finding_samp_A_str}
Let $\Psi^{\symA}_{\OLD}$ be a subset of set $\Psi_{\RR}$ such that 
each interval attractor $([p, q], [\ell, r]) \in \Psi^{\symA}_{\OLD}$ 
satisfies the following two conditions: 
\begin{enumerate}[label=\textbf{(\roman*)}]
    \item $([p, q], [\ell, r]) \in (\Psi_{\OLD} \setminus \Psi_{\run}) \cap \Psi_{\samp}$;
    \item $|\Psi_{\str}(T[p-1..r+1])| - |\Psi_{\str}(T[p-1..r+1]) \cap (\Psi_{\OLD} \setminus \Psi_{\run})| \geq 1$.
\end{enumerate}
Then, $|\Psi^{\symA}_{\OLD}| = |\mathcal{Z}^{\symA}|$ 
and $\mathcal{Z}^{\symA} = \{ T[p-1..r+1] \mid ([p, q], [\ell, r]) \in \Psi^{\symA}_{\OLD} \}$ hold. 
\end{lemma}
\begin{proof}
    Lemma~\ref{lem:finding_samp_A_str} follows from the following three statements: 
    \begin{enumerate}[label=\textbf{(\Alph*)}]
    \item $T[p-1..r+1] \neq T[p^{\prime}-1..r^{\prime}+1]$ for any pair of two interval attractors 
    $([p, q], [\ell, r]), ([p^{\prime}, q^{\prime}], [\ell^{\prime}, r^{\prime}]) \in \Psi^{\symA}_{\OLD}$;    
    \item $\mathcal{Z}^{\symA} \subseteq \{ T[p-1..r+1] \mid ([p, q], [\ell, r]) \in \Psi^{\symA}_{\OLD} \}$;
    \item $\mathcal{Z}^{\symA} \supseteq \{ T[p-1..r+1] \mid ([p, q], [\ell, r]) \in \Psi^{\symA}_{\OLD} \}$.
    \end{enumerate}

    \textbf{Proof of statement (A).}
    Because of $([p, q], [\ell, r]), ([p^{\prime}, q^{\prime}], [\ell^{\prime}, r^{\prime}]) \in \Psi_{\samp}$, 
    $T[p-1..r+1] \neq T[p^{\prime}-1..r^{\prime}+1]$ follows from the definition of sampling subset $\Psi_{\samp}$. 
    
    \textbf{Proof of statement (B).}
    Consider a string $Z^{\symA}_{s}$ in set $\mathcal{Z}^{\symA}$. 
    Then, 
    $\Psi_{\samp} \cap \Psi_{\OLD} \cap \Psi_{\str}(Z^{\symA}_{s}) \neq \emptyset$ 
    and $(\Psi_{\RR} \setminus \Psi_{\OLD}) \cap \Psi_{\str}(Z^{\symA}_{s}) \neq \emptyset$ 
    follow from the definition of set $\mathcal{Z}^{\symA}$. 
    Set $\Psi_{\samp} \cap \Psi_{\OLD} \cap \Psi_{\str}(Z^{\symA}_{s})$ contains an interval attractor $([p, q], [\ell, r])$. 
    Similarly, 
    set $(\Psi_{\RR} \setminus \Psi_{\OLD}) \cap \Psi_{\str}(Z^{\symA}_{s})$ contains 
    an interval attractor $([p^{\prime}, q^{\prime}], [\ell^{\prime}, r^{\prime}])$.     
    Here, $([p, q], [\ell, r]) \neq ([p^{\prime}, q^{\prime}], [\ell^{\prime}, r^{\prime}])$ 
    follows from $([p, q], [\ell, r]) \in \Psi_{\OLD}$ 
    and $([p^{\prime}, q^{\prime}], [\ell^{\prime}, r^{\prime}]) \not \in \Psi_{\OLD}$. 
    $T[p-1..r+1] = Z^{\symA}_{s}$ follows from the definition of the subset $\Psi_{\str}(Z^{\symA}_{s})$. 

    We prove $([p, q], [\ell, r]) \in (\Psi_{\OLD} \setminus \Psi_{\run}) \cap \Psi_{\samp}$ 
    and $|\Psi_{\str}(Z^{\symA}_{s})| - |\Psi_{\str}(Z^{\symA}_{s}) \cap (\Psi_{\OLD} \setminus \Psi_{\run})| \geq 1$. 
    $\Psi_{\str}(Z^{\symA}_{s}) \cap \Psi_{\run} = \emptyset$ follows from Lemma~\ref{lem:Z_samp_A_property}~\ref{enum:Z_samp_A_property:3}. 
    $([p, q], [\ell, r]) \not \in \Psi_{\run}$ follows from 
    $([p, q], [\ell, r]) \in \Psi_{\str}(Z^{\symA}_{s})$ and $\Psi_{\str}(Z^{\symA}_{s}) \cap \Psi_{\run} = \emptyset$. 
    $([p, q], [\ell, r]) \in (\Psi_{\OLD} \setminus \Psi_{\run}) \cap \Psi_{\samp}$ follows from 
    $([p, q], [\ell, r]) \in \Psi_{\samp} \cap \Psi_{\OLD}$ and $([p, q], [\ell, r]) \not \in \Psi_{\run}$.
    $\Psi_{\str}(Z^{\symA}_{s}) \supset \Psi_{\str}(Z^{\symA}_{s}) \cap \Psi_{\OLD}$ 
    follows from 
    $([p, q], [\ell, r])$, $([p^{\prime}, q^{\prime}]$, $[\ell^{\prime}, r^{\prime}]) \in \Psi_{\str}(Z^{\symA}_{s})$ 
    and $([p^{\prime}, q^{\prime}], [\ell^{\prime}, r^{\prime}]) \not \in \Psi_{\OLD}$.     
    $\Psi_{\str}(Z^{\symA}_{s}) \cap \Psi_{\OLD} = \Psi_{\str}(Z^{\symA}_{s}) \cap (\Psi_{\OLD} \setminus \Psi_{\run})$ 
    holds because of $\Psi_{\str}(Z^{\symA}_{s}) \cap \Psi_{\run} = \emptyset$. 
    Therefore, 
    $|\Psi_{\str}(Z^{\symA}_{s})| - |\Psi_{\str}(Z^{\symA}_{s}) \cap (\Psi_{\OLD} \setminus \Psi_{\run})| \geq 1$ 
    follows from $\Psi_{\str}(Z^{\symA}_{s}) \supset \Psi_{\str}(Z^{\symA}_{s}) \cap \Psi_{\OLD}$ 
    and $\Psi_{\str}(Z^{\symA}_{s}) \cap \Psi_{\OLD} = \Psi_{\str}(Z^{\symA}_{s}) \cap (\Psi_{\OLD} \setminus \Psi_{\run})$.     
    
    We prove statement (B). 
    $([p, q], [\ell, r]) \in \Psi^{\symA}_{\OLD}$ follows from 
    $([p, q], [\ell, r]) \in (\Psi_{\OLD} \setminus \Psi_{\run}) \cap \Psi_{\samp}$ 
    and $|\Psi_{\str}(Z^{\symA}_{s})| - |\Psi_{\str}(Z^{\symA}_{s}) \cap (\Psi_{\OLD} \setminus \Psi_{\run})| \geq 1$ 
    (i.e., $|\Psi_{\str}(T[p-1..r+1])| - |\Psi_{\str}(T[p-1..r+1] ) \cap (\Psi_{\OLD} \setminus \Psi_{\run})| \geq 1$).    
    $Z^{\symA}_{s} \in \{ T[\hat{p}-1..\hat{r}+1] \mid ([\hat{p}, \hat{q}], [\hat{\ell}, \hat{r}]) \in \Psi^{\symA}_{\OLD} \}$ follows from $([p, q], [\ell, r]) \in \Psi^{\symA}_{\OLD}$ and $T[p-1..r+1] = Z^{\symA}_{s}$. 
    Therefore, statement (B) holds. 

    \textbf{Proof of statement (C).}
    Consider an interval attractor $([p, q], [\ell, r])$ in set $\Psi^{\symA}_{\OLD}$. 
    Then, $([p, q], [\ell, r]) \in (\Psi_{\OLD} \setminus \Psi_{\run}) \cap \Psi_{\samp}$ 
    and $|\Psi_{\str}(T[p-1..r+1])| - |\Psi_{\str}(T[p-1..r+1]) \cap (\Psi_{\OLD} \setminus \Psi_{\run})| \geq 1$ hold. 
    $\Psi_{\samp} \cap \Psi_{\OLD} \cap \Psi_{\str}(T[p-1..r+1]) \neq \emptyset$ 
    follows from $([p, q], [\ell, r]) \in \Psi_{\OLD} \cap \Psi_{\samp}$ and 
    $([p, q], [\ell, r]) \in \Psi_{\str}(T[p-1..r+1])$. 

    We prove $\Psi_{\str}(T[p-1..r+1]) \cap \Psi_{\run} = \emptyset$. 
    %Because of $([p, q], [\ell, r]) \in \Psi_{\samp}$, 
    %Lemma~\ref{lem:samp_basic_property}~\ref{enum:samp_basic_property:3} shows that 
    %subset $\Psi_{\run}$ does not contain the interval attractor $([p, q], [\ell, r])$. 
    For each interval attractor $([\hat{p}, \hat{q}], [\hat{\ell}, \hat{r}]) \in \Psi_{\str}(T[p-1..r+1])$, 
    we can apply 
    Lemma~\ref{lem:psi_equality_basic_property}~\ref{enum:psi_equality_basic_property:4} 
    to the two interval attractors $([p, q], [\ell, r])$ and $([\hat{p}, \hat{q}], [\hat{\ell}, \hat{r}])$ 
    because $T[p-1..r+1] = T[\hat{p}-1..\hat{r}+1]$ holds. 
    Because of $([p, q], [\ell, r]) \not \in \Psi_{\run}$, 
    Lemma~\ref{lem:psi_equality_basic_property}~\ref{enum:psi_equality_basic_property:4} shows that 
    the subset $\Psi_{\run}$ does not contain the interval attractor $([\hat{p}, \hat{q}], [\hat{\ell}, \hat{r}])$.
    Therefore, $\Psi_{\str}(T[p-1..r+1]) \cap \Psi_{\run} = \emptyset$ holds. 

    We prove $(\Psi_{\RR} \setminus \Psi_{\OLD}) \cap \Psi_{\str}(T[p-1..r+1]) \neq \emptyset$.
    $\Psi_{\str}(T[p-1..r+1]) \cap (\Psi_{\OLD} \setminus \Psi_{\run}) = \Psi_{\str}(T[p-1..r+1]) \cap \Psi_{\OLD}$ holds because of $\Psi_{\str}(T[p-1..r+1]) \cap \Psi_{\run} = \emptyset$. 
    $|\Psi_{\str}(T[p-1..r+1])| - |\Psi_{\str}(T[p-1..r+1]) \cap \Psi_{\OLD}| \geq 1$ 
    follows from 
    $|\Psi_{\str}(T[p-1..r+1])| - |\Psi_{\str}(T[p-1..r+1]) \cap (\Psi_{\OLD} \setminus \Psi_{\run})| \geq 1$ 
    and $\Psi_{\str}(T[p-1..r+1]) \cap (\Psi_{\OLD} \setminus \Psi_{\run}) = \Psi_{\str}(T[p-1..r+1]) \cap \Psi_{\OLD}$. 
    Because of $|\Psi_{\str}(T[p-1..r+1])| - |\Psi_{\str}(T[p-1..r+1]) \cap \Psi_{\OLD}| \geq 1$, 
    set $(\Psi_{\RR} \setminus \Psi_{\OLD}) \cap \Psi_{\str}(T[p-1..r+1])$ contains an interval attractor. 
    Therefore, $(\Psi_{\RR} \setminus \Psi_{\OLD}) \cap \Psi_{\str}(T[p-1..r+1]) \neq \emptyset$ holds. 

    We proved $\Psi_{\samp} \cap \Psi_{\OLD} \cap \Psi_{\str}(T[p-1..r+1]) \neq \emptyset$ 
    and $(\Psi_{\RR} \setminus \Psi_{\OLD}) \cap \Psi_{\str}(T[p-1..r+1]) \neq \emptyset$. 
    $T[p-1..r+1] \in \mathcal{Z}^{\symA}$ follows from the definition of the subset $\mathcal{Z}^{\symA}$. 
    Therefore, statement (C) holds.

\end{proof}

\paragraph{Set $\mathcal{Z}^{\symB}$ of strings.}
We introduce a set $\mathcal{Z}^{\symB} \subseteq \Sigma^{+}$ of strings. 
This set consists of strings such that 
each string $Z \in \mathcal{Z}^{\symB}$ satisfies 
$(\Psi^{\prime}_{\NEW} \setminus \Psi^{\prime}_{\run}) \cap \Psi^{\prime}_{\str}(Z) \neq \emptyset$ 
and $(\Psi_{\RR} \setminus \Psi_{\OLD}) \cap \Psi_{\str}(Z) = \emptyset$ 
(i.e., $\mathcal{Z}^{\symB} = \{ Z \in \Sigma^{+} \mid ((\Psi^{\prime}_{\NEW} \setminus \Psi^{\prime}_{\run}) \cap \Psi^{\prime}_{\str}(Z) \neq \emptyset) \land ((\Psi_{\RR} \setminus \Psi_{\OLD}) \cap \Psi_{\str}(Z) = \emptyset) \}$). 
For this subsection, 
let $Z^{\symB}_{1}, Z^{\symB}_{2}, \ldots, Z^{\symB}_{k^{\prime}}$ ($Z^{\symB}_{1} \prec Z^{\symB}_{2} \prec \cdots \prec Z^{\symB}_{k^{\prime}}$) be the strings of set $\mathcal{Z}^{\symB}$. 

The following lemma states the relationship between two sets $\mathcal{Z}^{\symB}$ and $\Psi^{\prime}_{\NEW}$. 

\begin{lemma}\label{lem:finding_samp_B_str}
Let $\Psi^{\prime}_{\NEW, \symB}$ be a subset of set $\Psi^{\prime}_{\RR}$ such that 
each interval attractor $([p, q], [\ell, r]) \in \Psi^{\prime}_{\NEW, \symB}$ 
satisfies the following two conditions: 
\begin{enumerate}[label=\textbf{(\roman*)}]
    \item $([p, q], [\ell, r]) \in \Psi^{\prime}_{\NEW} \setminus \Psi^{\prime}_{\run}$;
    \item $|\Psi_{\str}(T^{\prime}[p-1..r+1])| = |\Psi_{\str}(T^{\prime}[p-1..r+1]) \cap (\Psi_{\OLD} \setminus \Psi_{\run})|$.
\end{enumerate}
Then, $\mathcal{Z}^{\symB} = \{ T^{\prime}[p-1..r+1] \mid ([p, q], [\ell, r]) \in \Psi^{\prime}_{\NEW, \symB} \}$ holds. 
\end{lemma}
\begin{proof}
    Lemma~\ref{lem:finding_samp_B_str} follows from the following two statements: 
    \begin{enumerate}[label=\textbf{(\Alph*)}] 
    \item $\mathcal{Z}^{\symB} \subseteq \{ T^{\prime}[p-1..r+1] \mid ([p, q], [\ell, r]) \in \Psi^{\prime}_{\NEW, \symB} \}$;
    \item $\mathcal{Z}^{\symB} \supseteq \{ T^{\prime}[p-1..r+1] \mid ([p, q], [\ell, r]) \in \Psi^{\prime}_{\NEW, \symB} \}$.
    \end{enumerate}

\textbf{Proof of statement (A).}
Consider a string $Z^{\symB}_{s} \in \mathcal{Z}^{\symB}$. 
Then, 
$(\Psi^{\prime}_{\NEW} \setminus \Psi^{\prime}_{\run}) \cap \Psi^{\prime}_{\str}(Z^{\symB}_{s}) \neq \emptyset$ 
and $(\Psi_{\RR} \setminus \Psi_{\OLD}) \cap \Psi_{\str}(Z^{\symB}_{s}) = \emptyset$ hold. 
Set $(\Psi^{\prime}_{\NEW} \setminus \Psi^{\prime}_{\run}) \cap \Psi^{\prime}_{\str}(Z^{\symB}_{s})$ contains an interval attractor $([p, q], [\ell, r])$. 
Here, $T^{\prime}[p-1..r+1] = Z^{\symB}_{s}$ follows from the definition of the subset $\Psi_{\str}(Z^{\symB}_{s})$. 

We prove $([p, q], [\ell, r]) \in \Psi^{\prime}_{\NEW, \symB}$.
$\Psi_{\str}(Z^{\symB}_{s}) = \Psi_{\str}(Z^{\symB}_{s}) \cap (\Psi_{\OLD} \setminus \Psi_{\run})$ 
follows from $(\Psi_{\RR} \setminus \Psi_{\OLD}) \cap \Psi_{\str}(Z^{\symB}_{s}) = \emptyset$. 
$|\Psi_{\str}(T^{\prime}[p-1..r+1])| = |\Psi_{\str}(T^{\prime}[p-1..r+1]) \cap (\Psi_{\OLD} \setminus \Psi_{\run})|$  
follows from 
$\Psi_{\str}(Z^{\symB}_{s}) = \Psi_{\str}(Z^{\symB}_{s}) \cap (\Psi_{\OLD} \setminus \Psi_{\run})$ 
and $T^{\prime}[p-1..r+1] = Z^{\symB}_{s}$. 
Therefore, $([p, q], [\ell, r]) \in \Psi^{\prime}_{\NEW, \symB}$ follows from 
$([p, q], [\ell, r]) \in \Psi^{\prime}_{\NEW} \setminus \Psi^{\prime}_{\run}$ 
and $|\Psi_{\str}(T^{\prime}[p-1..r+1])| = |\Psi_{\str}(T^{\prime}[p-1..r+1]) \cap (\Psi_{\OLD} \setminus \Psi_{\run})|$. 

$Z^{\symB}_{s} \in \{ T^{\prime}[\hat{p}-1..\hat{r}+1] \mid ([\hat{p}, \hat{q}], [\hat{\ell}, \hat{r}]) \in \Psi^{\prime}_{\NEW, \symB} \}$ follows from 
$([p, q], [\ell, r]) \in \Psi^{\prime}_{\NEW, \symB}$ and $T^{\prime}[p-1..r+1] = Z^{\symB}_{s}$. 
Therefore, statement (A) holds. 

\textbf{Proof of statement (B).}
Consider an interval attractor $([p, q], [\ell, r]) \in \Psi^{\prime}_{\NEW, \symB}$. 
Then, $([p, q]$, $[\ell, r]) \in \Psi^{\prime}_{\NEW} \setminus \Psi^{\prime}_{\run}$ 
and $|\Psi_{\str}(T^{\prime}[p-1..r+1])| = |\Psi_{\str}(T^{\prime}[p-1..r+1]) \cap (\Psi_{\OLD} \setminus \Psi_{\run})|$ hold. 
$(\Psi^{\prime}_{\NEW} \setminus \Psi^{\prime}_{\run}) \cap \Psi^{\prime}_{\str}(T^{\prime}[p-1..r+1]) \neq \emptyset$ follows from $([p, q], [\ell, r]) \in \Psi^{\prime}_{\NEW} \setminus \Psi^{\prime}_{\run}$ 
and $([p, q], [\ell, r]) \in \Psi^{\prime}_{\str}(T^{\prime}[p-1..r+1])$. 

We prove $(\Psi_{\RR} \setminus \Psi_{\OLD}) \cap \Psi_{\str}(T^{\prime}[p-1..r+1]) = \emptyset$ by contradiction. 
We assume that $(\Psi_{\RR} \setminus \Psi_{\OLD}) \cap \Psi_{\str}(T^{\prime}[p-1..r+1]) \neq \emptyset$ holds. 
Then, the set $(\Psi_{\RR} \setminus \Psi_{\OLD}) \cap \Psi_{\str}(T^{\prime}[p-1..r+1])$ contains an interval attractor $([p^{\prime}, q^{\prime}], [\ell^{\prime}, r^{\prime}])$. 
$([p^{\prime}, q^{\prime}], [\ell^{\prime}, r^{\prime}]) \not \in \Psi_{\str}(T^{\prime}[p-1..r+1]) \cap (\Psi_{\OLD} \setminus \Psi_{\run})$ holds because of $([p^{\prime}, q^{\prime}], [\ell^{\prime}, r^{\prime}]) \not \in \Psi_{\OLD}$. 
$\Psi_{\str}(T^{\prime}[p-1..r+1]) \supset \Psi_{\str}(T^{\prime}[p-1..r+1]) \cap (\Psi_{\OLD} \setminus \Psi_{\run})$ follows from $([p^{\prime}, q^{\prime}], [\ell^{\prime}, r^{\prime}]) \in \Psi_{\str}(T^{\prime}[p-1..r+1])$ and $([p^{\prime}, q^{\prime}], [\ell^{\prime}, r^{\prime}]) \not \in \Psi_{\str}(T^{\prime}[p-1..r+1]) \cap (\Psi_{\OLD} \setminus \Psi_{\run})$. 
$|\Psi_{\str}(T^{\prime}[p-1..r+1])| \neq |\Psi_{\str}(T^{\prime}[p-1..r+1]) \cap (\Psi_{\OLD} \setminus \Psi_{\run})|$ follows from $\Psi_{\str}(T^{\prime}[p-1..r+1]) \supset \Psi_{\str}(T^{\prime}[p-1..r+1]) \cap (\Psi_{\OLD} \setminus \Psi_{\run})$. 
The two facts $|\Psi_{\str}(T^{\prime}[p-1..r+1])| = |\Psi_{\str}(T^{\prime}[p-1..r+1]) \cap (\Psi_{\OLD} \setminus \Psi_{\run})|$ and $|\Psi_{\str}(T^{\prime}[p-1..r+1])| \neq |\Psi_{\str}(T^{\prime}[p-1..r+1]) \cap (\Psi_{\OLD} \setminus \Psi_{\run})|$ yield a contradiction. 
Therefore, $(\Psi_{\RR} \setminus \Psi_{\OLD}) \cap \Psi_{\str}(T^{\prime}[p-1..r+1]) = \emptyset$ must hold. 

We proved $(\Psi^{\prime}_{\NEW} \setminus \Psi^{\prime}_{\run}) \cap \Psi^{\prime}_{\str}(T^{\prime}[p-1..r+1]) \neq \emptyset$ and $(\Psi_{\RR} \setminus \Psi_{\OLD}) \cap \Psi_{\str}(T^{\prime}[p-1..r+1]) = \emptyset$. 
$T^{\prime}[p-1..r+1] \in \mathcal{Z}^{\symB}$ follows from the definition of the set $\mathcal{Z}^{\symB}$. 
Therefore, statement (B) holds. 

\end{proof}

For each string $Z^{\symB}_{s} \in \mathcal{Z}^{\symB}$, 
the following lemma shows that set $(\Psi^{\prime}_{\NEW} \setminus \Psi^{\prime}_{\run}) \cap \Psi^{\prime}_{\str}(Z^{\symB}_{s})$ is not empty. 

\begin{lemma}\label{lem:dynamic_samp_new}
    Consider a string $Z^{\symB}_{s}$ in set $\mathcal{Z}^{\symB}$. 
    Let $([p_{s, 1}, q_{s, 1}], [\ell_{s, 1}, r_{s, 1}])$, $([p_{s, 2}, q_{s, 2}], [\ell_{s, 2}, r_{s, 2}])$, $\ldots$, $([p_{s, m}, q_{s, m}], [\ell_{s, m}, r_{s, m}])$ 
    ($p_{s, 1} < p_{s, 2} < \cdots < p_{s, m}$)
    be the interval attractors of set $(\Psi^{\prime}_{\NEW} \setminus \Psi^{\prime}_{\run}) \cap \Psi^{\prime}_{\str}(Z^{\symB}_{s})$. 
    Then, $m \geq 1$ holds. 
\end{lemma}
\begin{proof}
    $m \geq 1$ follows from the definition of the set $\mathcal{Z}^{\symB}$. 
\end{proof}

\paragraph{Three sets $\Psi^{\symA}$, $\Psi^{\symA^\prime}$, and $\Psi^{\prime \symB}$ of interval attractors.}
We introduce two subsets $\Psi^{\symA}$ of set $\Psi_{\RR}$ using 
the set $\mathcal{Z}^{\symA}$ of strings. 
For each string $Z^{\symA}_{s} \in \mathcal{Z}^{\symA}$, 
set $\Psi_{\samp} \cap \Psi_{\OLD}$ contains an interval attractor $([p^{\symA}_{s}, q^{\symA}_{s}], [\ell^{\symA}_{s}, r^{\symA}_{s}])$ satisfying $T[p^{\symA}_{s}-1..r^{\symA}_{s}+1] = Z^{\symA}_{s}$ 
because of $\Psi_{\samp} \cap \Psi_{\OLD} \cap \Psi_{\str}(Z) \neq \emptyset$. 
The subset $\Psi^{\symA}$ consists of 
the $k$ interval attractors $([p^{\symA}_{1}, q^{\symA}_{1}], [\ell^{\symA}_{1}, r^{\symA}_{1}])$, 
$([p^{\symA}_{2}, q^{\symA}_{2}], [\ell^{\symA}_{2}, r^{\symA}_{2}])$, 
$\ldots$, $([p^{\symA}_{k}, q^{\symA}_{k}], [\ell^{\symA}_{k}, r^{\symA}_{k}])$. 

Next, we introduce a subset $\Psi^{\symA^\prime}$ of set $\Psi_{\RR}$ using Lemma~\ref{lem:dynamic_samp_substitute}. 
For each string $Z^{\symA}_{s} \in \mathcal{Z}^{\symA}$, 
Lemma~\ref{lem:dynamic_samp_substitute} shows that 
set $(\Psi_{\RR} \setminus \Psi_{\OLD}) \cap \Psi_{\str}(Z^{\symA}_{s})$ contains 
an interval attractor $([p^{\symA^{\prime}}_{s}, q^{\symA^{\prime}}_{s}], [\ell^{\symA^{\prime}}_{s}, r^{\symA^{\prime}}_{s}])$ 
satisfying $([p^{\symA^{\prime}}_{s}, q^{\symA^{\prime}}_{s}], [\ell^{\symA^{\prime}}_{s}, r^{\symA^{\prime}}_{s}]) = I_{\capture}(\SA[x^{\symA}_{s}]+1, \SA[x^{\symA}_{s}] + |Z^{\symA}_{s}| - 2)$.  
The subset $\Psi^{\symA^\prime}$ consists of 
the $k$ interval attractors $([p^{\symA^{\prime}}_{1}, q^{\symA^{\prime}}_{1}], [\ell^{\symA^{\prime}}_{1}, r^{\symA^{\prime}}_{1}])$, 
$([p^{\symA^{\prime}}_{2}, q^{\symA^{\prime}}_{2}], [\ell^{\symA^{\prime}}_{2}, r^{\symA^{\prime}}_{2}])$, 
$\ldots$, $([p^{\symA^{\prime}}_{k}, q^{\symA^{\prime}}_{k}], [\ell^{\symA^{\prime}}_{k}, r^{\symA^{\prime}}_{k}])$. 
In addition, 
let $\epsilon^{\symA^{\prime}}_{s} = 0$ if $([p^{\symA^{\prime}}_{s}, q^{\symA^{\prime}}_{s}], [\ell^{\symA^{\prime}}_{s}, r^{\symA^{\prime}}_{s}]) \in \Psi_{\LEFT}$. 
Otherwise, let $\epsilon^{\symA^{\prime}}_{s} = 1$. 
Then, Theorem~\ref{theo:RS_LEFT_RIGHT} shows that 
interval attractor $([p^{\symA^{\prime}}_{s} + \epsilon^{\symA^{\prime}}_{s}, q^{\symA^{\prime}}_{s} + \epsilon^{\symA^{\prime}}_{s}], [\ell^{\symA^{\prime}}_{s} + \epsilon^{\symA^{\prime}}_{s}, r^{\symA^{\prime}}_{s} + \epsilon^{\symA^{\prime}}_{s}])$ is contained in $\Psi^{\prime}_{\LEFT}$ or $\Psi^{\prime}_{\RIGHT}$. 

Next, we introduce a subset $\Psi^{\prime \symB}$ of set $\Psi^{\prime}_{\RR}$ 
using the set $\mathcal{Z}^{\symB}$ of strings. 
For each string $Z^{\symB}_{s} \in \mathcal{Z}^{\symB}$, 
let $([p^{\symB}_{s}, q^{\symB}_{s}], [\ell^{\symB}_{s}, r^{\symB}_{s}])$ 
be the interval attractor 
$([p_{s, 1}, q_{s, 1}], [\ell_{s, 1}, r_{s, 1}])$ introduced in Lemma~\ref{lem:dynamic_samp_new}. 
Then, 
the subset $\Psi^{\prime \symB}$ consists of $k^{\prime}$ interval attractors
$([p^{\symB}_{1}, q^{\symB}_{1}], [\ell^{\symB}_{1}, r^{\symB}_{1}])$, 
$([p^{\symB}_{2}, q^{\symB}_{2}], [\ell^{\symB}_{2}, r^{\symB}_{2}])$, 
$\ldots$, $([p^{\symB}_{k^{\prime}}, q^{\symB}_{k^{\prime}}], [\ell^{\symB}_{k^{\prime}}, r^{\symB}_{k^{\prime}}])$. 

\paragraph{Sets $\Psi^{\prime}_{\samp}$ of interval attractors.}
We define a set $\Psi^{\prime}_{\samp}$ of interval attractors as 
the union of the following four sets: 
\begin{enumerate}[label=\textbf{(\roman*)}]
    \item $\Psi_{\samp} \cap \Psi_{\LEFT}$;
    \item $\{ ([p + 1, q + 1], [\ell + 1, r + 1]) \mid ([p, q], [\ell, r]) \in \Psi_{\samp} \cap \Psi_{\RIGHT} \}$;
    \item $\{ ([p^{\symA^{\prime}}_{s} + \epsilon^{\symA^{\prime}}_{s}, q^{\symA^{\prime}}_{s} + \epsilon^{\symA^{\prime}}_{s}], [\ell^{\symA^{\prime}}_{s} + \epsilon^{\symA^{\prime}}_{s}, r^{\symA^{\prime}}_{s} + \epsilon^{\symA^{\prime}}_{s}]) \mid s \in [1, k] \}$;
    \item $\Psi^{\prime \symB}$. 
\end{enumerate}

The following lemma shows that the set $\Psi^{\prime}_{\samp}$ is a subset of set $\Psi^{\prime}_{\RR}$. 
\begin{lemma}\label{lem:dynamic_samp_formula_sub1}
Set $\Psi^{\prime}_{\samp}$ is a subset of set $\Psi^{\prime}_{\RR}$.
\end{lemma}
\begin{proof}
    Theorem~\ref{theo:RS_LEFT_RIGHT} indicates that 
    set $\Psi_{\samp} \cap \Psi_{\LEFT}$ is a subset of set $\Psi^{\prime}_{\RR}$. 
    Similarly, 
    set $\{ ([p + 1, q + 1], [\ell + 1, r + 1]) \mid ([p, q], [\ell, r]) \in \Psi_{\samp} \cap \Psi_{\RIGHT} \}$ 
    is a subset of set $\Psi^{\prime}_{\RR}$. 
    We already showed that each interval attractor $([p^{\symA^{\prime}}_{s} + \epsilon^{\symA^{\prime}}_{s}, q^{\symA^{\prime}}_{s} + \epsilon^{\symA^{\prime}}_{s}], [\ell^{\symA^{\prime}}_{s} + \epsilon^{\symA^{\prime}}_{s}, r^{\symA^{\prime}}_{s} + \epsilon^{\symA^{\prime}}_{s}])$ is contained in $\Psi^{\prime}_{\LEFT}$ or $\Psi^{\prime}_{\RIGHT}$. 
    Set $\Psi^{\prime \symB}$ is a subset of set $\Psi^{\prime}_{\RR}$. 
    Therefore, set $\Psi^{\prime}_{\samp}$ is a subset of set $\Psi^{\prime}_{\RR}$. 
\end{proof}

The following theorem shows that 
we can obtain a sampling subset for RLSLP $\mathcal{G}^{R}_{\ins}$ by modifying the sampling subset $\Psi_{\samp}$ for $\mathcal{G}^{R}$ using two subsets $\Psi^{\symA^\prime}$ and $\Psi^{\prime \symB}$. 
\begin{theorem}\label{theo:dynamic_samp_formula}
    Set $\Psi^{\prime}_{\samp}$ is a sampling subset for RLSLP $\mathcal{G}^{R}_{\ins}$, i.e., 
    the following three statements hold: 
\begin{enumerate}[label=\textbf{(\Alph*)}]
    \item $T^{\prime}[p_{1}-1..r_{1}+1] \neq T^{\prime}[p_{2}-1..r_{2}+1]$ 
    for any pair of two interval attractors $([p_{1}, q_{1}], [\ell_{1}, r_{1}]), ([p_{2}$, $q_{2}], [\ell_{2}, r_{2}])$ 
    in set $\Psi^{\prime}_{\samp}$; 
    \item for each interval attractor $([p_{1}, q_{1}], [\ell_{1}, r_{1}]) \in \Psi^{\prime}_{\samp}$, 
    set $\Psi_{\leftmost}^{\prime} \setminus \Psi^{\prime}_{\run}$ contains an interval attractor 
    $([p_{2}, q_{2}], [\ell_{2}, r_{2}])$ satisfying 
    $T^{\prime}[p_{1}-1..r_{1}+1] = T^{\prime}[p_{2}-1..r_{2}+1]$; 
    \item for each interval attractor $([p_{1}, q_{1}], [\ell_{1}, r_{1}]) \in \Psi_{\leftmost}^{\prime} \setminus \Psi_{\run}^{\prime}$, 
    set $\Psi^{\prime}_{\samp}$ contains an interval attractor 
    $([p_{2}, q_{2}], [\ell_{2}, r_{2}])$ satisfying 
    $T^{\prime}[p_{1}-1..r_{1}+1] = T^{\prime}[p_{2}-1..r_{2}+1]$. 
\end{enumerate}
    
\end{theorem}
\begin{proof}
    See Section~\ref{subsubsec:proof_dynamic_samp_formula}.     
\end{proof}

The following lemma states the relationship between 
two sampling subsets $\Psi_{\samp}$ and $\Psi^{\prime}_{\samp}$. 
\begin{lemma}\label{lem:dynamic_samp_back_formula}
Consider an interval attractor $([p^{\prime}, q^{\prime}], [\ell^{\prime}, r^{\prime}]) \in \Psi^{\prime}_{\samp}$ satisfying 
$([p^{\prime}, q^{\prime}], [\ell^{\prime}, r^{\prime}]) \not \in \Psi^{\prime \symB}$. 
Then, the sampling subset $\Psi_{\samp}$ contains an interval attractor 
$([p, q], [\ell, r])$ satisfying 
$T[p-1..r+1] = T^{\prime}[p^{\prime}-1..r^{\prime}+1]$ and 
$([p, q], [\ell, r]) \not \in \Psi_{\OLD} \setminus \Psi^{\symA}$. 
\end{lemma}
\begin{proof}
From the definition of the sampling subset $\Psi^{\prime \symB}$, 
the interval attractor $([p^{\prime}, q^{\prime}], [\ell^{\prime}, r^{\prime}])$ satisfies  
at least one of the following three conditions: 
\begin{enumerate}[label=\textbf{(\roman*)}]
    \item $([p^{\prime}, q^{\prime}], [\ell^{\prime}, r^{\prime}]) \in \Psi_{\samp} \cap \Psi_{\LEFT}$;
    \item $([p^{\prime}, q^{\prime}], [\ell^{\prime}, r^{\prime}]) \in \{ ([p + 1, q + 1], [\ell + 1, r + 1]) \mid ([p, q], [\ell, r]) \in \Psi_{\samp} \cap \Psi_{\RIGHT} \}$;
    \item $([p^{\prime}, q^{\prime}], [\ell^{\prime}, r^{\prime}]) \in \{ ([p^{\symA^{\prime}}_{s} + \epsilon^{\symA^{\prime}}_{s}, q^{\symA^{\prime}}_{s} + \epsilon^{\symA^{\prime}}_{s}], [\ell^{\symA^{\prime}}_{s} + \epsilon^{\symA^{\prime}}_{s}, r^{\symA^{\prime}}_{s} + \epsilon^{\symA^{\prime}}_{s}]) \mid s \in [1, k] \}$. 
\end{enumerate}

\paragraph{Proof of Lemma~\ref{lem:dynamic_samp_back_formula} for condition (i).}
In this case, set $\Psi_{\samp} \cap \Psi_{\LEFT}$ contains 
an interval attractor $([p, q], [\ell, r])$ satisfying 
$([p, q], [\ell, r]) = ([p^{\prime}, q^{\prime}], [\ell^{\prime}, r^{\prime}])$. 
This interval attractor satisfies 
$T[p-1..r+1] = T^{\prime}[p^{\prime}-1..r^{\prime}+1]$ 
and $([p, q], [\ell, r]) \not \in \Psi_{\OLD}$. 
Therefore, Lemma~\ref{lem:dynamic_samp_back_formula} holds. 

\paragraph{Proof of Lemma~\ref{lem:dynamic_samp_back_formula} for condition (ii).}
In this case, 
Lemma~\ref{lem:dynamic_samp_back_formula} can be proved using the same approach as for condition (i). 

\paragraph{Proof of Lemma~\ref{lem:dynamic_samp_back_formula} for condition (iii).}
In this case, 
subset $\Psi^{\symA^\prime}$ contains an interval attractor 
$([p^{\symA^{\prime}}_{s}, q^{\symA^{\prime}}_{s}], [\ell^{\symA^{\prime}}_{s}, r^{\symA^{\prime}}_{s}])$ 
satisfying $([p^{\symA^{\prime}}_{s} + \epsilon^{\symA^{\prime}}_{s}, q^{\symA^{\prime}}_{s} + \epsilon^{\symA^{\prime}}_{s}], [\ell^{\symA^{\prime}}_{s} + \epsilon^{\symA^{\prime}}_{s}, r^{\symA^{\prime}}_{s} + \epsilon^{\symA^{\prime}}_{s}]) = ([p^{\prime}, q^{\prime}], [\ell^{\prime}, r^{\prime}])$. 
Lemma~\ref{lem:dynamic_samp_substitute} shows that 
$([p^{\symA^{\prime}}_{s}, q^{\symA^{\prime}}_{s}], [\ell^{\symA^{\prime}}_{s}, r^{\symA^{\prime}}_{s}]) \in (\Psi_{\RR} \setminus \Psi_{\OLD}) \cap \Psi_{\str}(Z^{\symA}_{s})$ holds. 
Here, $T[p^{\symA^{\prime}}_{s}-1..r^{\symA^{\prime}}_{s}+1] = Z^{\symA}_{s}$ follows from 
$([p^{\symA^{\prime}}_{s}, q^{\symA^{\prime}}_{s}], [\ell^{\symA^{\prime}}_{s}, r^{\symA^{\prime}}_{s}]) \in \Psi_{\str}(Z^{\symA}_{s})$. 

The subset $\Psi^{\symA}$ contains interval attractor $([p^{\symA}_{s}, q^{\symA}_{s}], [\ell^{\symA}_{s}, r^{\symA}_{s}])$. 
$T[p^{\symA}_{s}-1..r^{\symA}_{s}+1] = Z^{\symA}_{s}$ 
and $([p^{\symA}_{s}, q^{\symA}_{s}], [\ell^{\symA}_{s}, r^{\symA}_{s}]) \in \Psi_{\samp} \cap \Psi_{\OLD}$ 
follow from the definition of the subset $\Psi^{\symA}$. 
Therefore, 
Lemma~\ref{lem:dynamic_samp_back_formula} holds if the following two statements hold: 
(A) $T[p^{\symA}_{s}-1..r^{\symA}_{s}+1] = T^{\prime}[p^{\prime}-1..r^{\prime}+1]$; 
(B) $([p^{\symA}_{s}, q^{\symA}_{s}]$, $[\ell^{\symA}_{s}, r^{\symA}_{s}]) \not \in \Psi_{\OLD} \setminus \Psi^{\symA}$.

%Consider subset $\Psi^{\symA} = \{ ([p_{\symA, g}, q_{\symA, g}], [\ell_{\symA, g}, r_{\symA, g}]) \mid g \in [1, k] \}$. 

We prove statement (A). 
$([p^{\symA^{\prime}}_{s}, q^{\symA^{\prime}}_{s}], [\ell^{\symA^{\prime}}_{s}, r^{\symA^{\prime}}_{s}]) \in \Psi_{\LEFT} \cup \Psi_{\RIGHT}$ follows from 
$([p^{\symA^{\prime}}_{s}, q^{\symA^{\prime}}_{s}], [\ell^{\symA^{\prime}}_{s}, r^{\symA^{\prime}}_{s}]) \not \in \Psi_{\OLD}$. 
In this case, 
$T[p^{\symA^{\prime}}_{s}-1..r^{\symA^{\prime}}_{s}+1] = T^{\prime}[p^{\symA^{\prime}}_{s}-1+\epsilon^{\symA^{\prime}}_{s}..r^{\symA^{\prime}}_{s}+1+\epsilon^{\symA^{\prime}}_{s}]$ hold. 
Therefore, $T[p^{\symA}_{s}-1..r^{\symA}_{s}+1] = T^{\prime}[p^{\prime}-1..r^{\prime}+1]$ follows from 
$T[p^{\symA}_{s}-1..r^{\symA}_{s}+1] = T^{\prime}[p^{\symA^{\prime}}_{s}-1+\epsilon^{\symA^{\prime}}_{s}..r^{\symA^{\prime}}_{s}+1+\epsilon^{\symA^{\prime}}_{s}]$ 
and $T^{\prime}[p^{\symA^{\prime}}_{s}-1+\epsilon^{\symA^{\prime}}_{s}..r^{\symA^{\prime}}_{s}+1+\epsilon^{\symA^{\prime}}_{s}] = T^{\prime}[p^{\prime}-1..r^{\prime}+1]$. 

Statement (B) follows from 
$([p^{\symA}_{s}, q^{\symA}_{s}], [\ell^{\symA}_{s}, r^{\symA}_{s}]) \in \Psi_{\samp} \cap \Psi_{\OLD}$ 
and $([p^{\symA}_{s}, q^{\symA}_{s}], [\ell^{\symA}_{s}, r^{\symA}_{s}]) \in \Psi^{\symA}$. 
Therefore, Lemma~\ref{lem:dynamic_samp_back_formula} holds. 
\end{proof}

\subsubsection{Proof of Theorem~\ref{theo:dynamic_samp_formula}}\label{subsubsec:proof_dynamic_samp_formula}
%The following proposition shows that 
%set $\Psi^{\prime}_{\samp}$ is a subset of set $\Psi^{\prime}_{\RR}$.

For this proof, 
let $\mathcal{Z}^{L} = \{ T[p-1..r+1] \mid ([p, q], [\ell, r]) \in \Psi_{\samp} \cap \Psi_{\LEFT} \}$ and $\mathcal{Z}^{R} = \{ T[p-1..r+1] \mid ([p, q], [\ell, r]) \in \Psi_{\samp} \cap \Psi_{\RIGHT} \}$. 
The following proposition states the relationship among 
four sets $\mathcal{Z}^{L}$, $\mathcal{Z}^{R}$, $\mathcal{Z}^{\symA}$, and $\mathcal{Z}^{\symB}$.

\begin{proposition}\label{prop:dynamic_samp_formula_sub2}
The following five statements hold: 
\begin{enumerate}[label=\textbf{(\roman*)}]
    \item \label{enum:dynamic_samp_formula_sub2:1} $(\Psi_{\samp} \setminus \Psi_{\OLD}) \cap \Psi_{\str}(Z) \neq \emptyset$ for each string $Z \in \mathcal{Z}^{L} \cup \mathcal{Z}^{R}$;
    \item \label{enum:dynamic_samp_formula_sub2:2} $\mathcal{Z}^{L} \cap \mathcal{Z}^{R} = \emptyset$;
    \item \label{enum:dynamic_samp_formula_sub2:3} $(\mathcal{Z}^{L} \cup \mathcal{Z}^{R}) \cap \mathcal{Z}^{\symA} = \emptyset$;
    \item \label{enum:dynamic_samp_formula_sub2:4} $(\mathcal{Z}^{L} \cup \mathcal{Z}^{R}) \cap \mathcal{Z}^{\symB} = \emptyset$;
    \item \label{enum:dynamic_samp_formula_sub2:5} $\mathcal{Z}^{\symA} \cap \mathcal{Z}^{\symB} = \emptyset$.
\end{enumerate}
\end{proposition}
\begin{proof}    
    The proof of Proposition~\ref{prop:dynamic_samp_formula_sub2} is as follows. 
    
    \textbf{Proof of Proposition~\ref{prop:dynamic_samp_formula_sub2}(i).}    
    Set $\Psi_{\samp} \cap \Psi_{\LEFT}$ or $\Psi_{\samp} \cap \Psi_{\RIGHT}$ 
    contains an interval attractor $([p, q], [\ell, r])$ satisfying 
    $T[p-1..r+1] = Z$. 
    $([p, q], [\ell, r]) \in \Psi_{\str}(Z)$ follows from $T[p-1..r+1] = Z$. 
    $([p, q], [\ell, r]) \not \in \Psi_{\OLD}$ follows from 
    $([p, q], [\ell, r]) \in \Psi_{\LEFT} \cup \Psi_{\RIGHT}$. 
    Therefore, $([p, q], [\ell, r]) \in (\Psi_{\samp} \setminus \Psi_{\OLD}) \cap \Psi_{\str}(Z)$ holds 
    (i.e., $(\Psi_{\samp} \setminus \Psi_{\OLD}) \cap \Psi_{\str}(Z) \neq \emptyset$). 

    \textbf{Proof of Proposition~\ref{prop:dynamic_samp_formula_sub2}(ii).}
    We prove $\mathcal{Z}^{L} \cap \mathcal{Z}^{R} = \emptyset$ by contradiction. 
    We assume that $\mathcal{Z}^{L} \cap \mathcal{Z}^{R} \neq \emptyset$ holds. 
    Then, set $\mathcal{Z}^{L} \cap \mathcal{Z}^{R}$ contains a string $Z$. 
    Because of $Z \in \mathcal{Z}^{L}$, 
    set $\Psi_{\samp} \cap \Psi_{\LEFT}$ contains an interval attractor $([p, q], [\ell, r])$ satisfying 
    $T[p-1..r+1] = Z$. 
    Similarly, 
    set $\Psi_{\samp} \cap \Psi_{\RIGHT}$ contains an interval attractor $([p^{\prime}, q^{\prime}], [\ell^{\prime}, r^{\prime}])$ satisfying 
    $T[p^{\prime}-1..r^{\prime}+1] = Z$. 
    Therefore, we obtain $T[p-1..r+1] = T[p^{\prime}-1..r^{\prime}+1]$. 

    On the other hand,     
    $([p, q], [\ell, r]) \neq ([p^{\prime}, q^{\prime}], [\ell^{\prime}, r^{\prime}])$ holds 
    because 
    the two sets $\Psi_{\LEFT}$ and $\Psi_{\RIGHT}$ are mutually disjoint.     
    $T[p-1..r+1] \neq T[p^{\prime}-1..r^{\prime}+1]$ follows from the definition of 
    sampling subset $\Psi_{\samp}$. 
    The two facts $T[p-1..r+1] = T[p^{\prime}-1..r^{\prime}+1]$ and $T[p-1..r+1] \neq T[p^{\prime}-1..r^{\prime}+1]$ yield a contradiction. 
    Therefore, $\mathcal{Z}^{L} \cap \mathcal{Z}^{R} = \emptyset$ must hold. 
        
    \textbf{Proof of Proposition~\ref{prop:dynamic_samp_formula_sub2}(iii).}
    We prove $(\mathcal{Z}^{L} \cup \mathcal{Z}^{R}) \cap \mathcal{Z}^{\symA} = \emptyset$ by contradiction. 
    We assume that $(\mathcal{Z}^{L} \cup \mathcal{Z}^{R}) \cap \mathcal{Z}^{\symA} \neq \emptyset$ holds. 
    Then, set $(\mathcal{Z}^{L} \cup \mathcal{Z}^{R}) \cap \mathcal{Z}^{\symA}$ contains a string $Z$. 
    $\Psi_{\samp} \cap \Psi_{\OLD} \cap \Psi_{\str}(Z) \neq \emptyset$ follows from the definition of set $\mathcal{Z}^{\symA}$. 
    Therefore, 
    set $\Psi_{\samp} \cap \Psi_{\OLD} \cap \Psi_{\str}(Z)$ contains 
    an interval attractor $([p, q], [\ell, r])$ be an interval attractor satisfying 
    $T[p-1..r+1] = Z$. 

    Because of $Z \in \mathcal{Z}^{L} \cup \mathcal{Z}^{R}$, 
    Proposition~\ref{prop:dynamic_samp_formula_sub2}~\ref{enum:dynamic_samp_formula_sub2:1} shows that 
    set $\Psi_{\samp} \setminus \Psi_{\OLD}$ contains 
    an interval attractor $([p^{\prime}, q^{\prime}], [\ell^{\prime}, r^{\prime}])$ 
    satisfying $T[p^{\prime}-1..r^{\prime}+1] = Z$.
    $T[p-1..r+1] = T[p^{\prime}-1..r^{\prime}+1]$ follows from 
    $T[p-1..r+1] = Z$ and $T[p^{\prime}-1..r^{\prime}+1] = Z$. 

    On the other hand,     
    $([p, q], [\ell, r]) \neq ([p^{\prime}, q^{\prime}], [\ell^{\prime}, r^{\prime}])$ follows from 
    $([p, q], [\ell, r]) \in \Psi_{\OLD}$ and $([p^{\prime}, q^{\prime}]$, $[\ell^{\prime}, r^{\prime}]) \not \in \Psi_{\OLD}$. 
    $T[p-1..r+1] \neq T[p^{\prime}-1..r^{\prime}+1]$ follows from sampling subset $\Psi_{\samp}$. 
    The two facts $T[p-1..r+1] = T[p^{\prime}-1..r^{\prime}+1]$ and $T[p-1..r+1] \neq T[p^{\prime}-1..r^{\prime}+1]$ yield a contradiction. 
    Therefore, $(\mathcal{Z}^{L} \cup \mathcal{Z}^{R}) \cap \mathcal{Z}^{\symA} = \emptyset$ must hold.

    \textbf{Proof of Proposition~\ref{prop:dynamic_samp_formula_sub2}(iv).}
    We prove $(\mathcal{Z}^{L} \cup \mathcal{Z}^{R}) \cap \mathcal{Z}^{\symB} = \emptyset$ by contradiction. 
    We assume that $(\mathcal{Z}^{L} \cup \mathcal{Z}^{R}) \cap \mathcal{Z}^{\symB} \neq \emptyset$ holds. 
    Then, set $(\mathcal{Z}^{L} \cup \mathcal{Z}^{R}) \cap \mathcal{Z}^{\symB}$ contains a string $Z$. 
    Proposition~\ref{prop:dynamic_samp_formula_sub2}~\ref{enum:dynamic_samp_formula_sub2:1} shows that $(\Psi_{\samp} \setminus \Psi_{\OLD}) \cap \Psi_{\str}(Z) \neq \emptyset$ holds. 
    On the other hand, 
    $(\Psi_{\samp} \setminus \Psi_{\OLD}) \cap \Psi_{\str}(Z) = \emptyset$ follows from the definition of the set $\mathcal{Z}^{\symB}$. 
    The two facts $(\Psi_{\samp} \setminus \Psi_{\OLD}) \cap \Psi_{\str}(Z) \neq \emptyset$ and $(\Psi_{\samp} \setminus \Psi_{\OLD}) \cap \Psi_{\str}(Z) = \emptyset$ yield a contradiction. 
    Therefore, $(\mathcal{Z}^{L} \cup \mathcal{Z}^{R}) \cap \mathcal{Z}^{\symB} = \emptyset$ must hold. 

    \textbf{Proof of Proposition~\ref{prop:dynamic_samp_formula_sub2}(v).}
    We prove $\mathcal{Z}^{\symA} \cap \mathcal{Z}^{\symB} = \emptyset$ by contradiction. 
    We assume that $\mathcal{Z}^{\symA} \cap \mathcal{Z}^{\symB} \neq \emptyset$ holds. 
    Then, set $\mathcal{Z}^{\symA} \cap \mathcal{Z}^{\symB}$ contains a string $Z$. 
    $(\Psi_{\samp} \setminus \Psi_{\OLD}) \cap \Psi_{\str}(Z) \neq \emptyset$ follows from the definition of the set $\mathcal{Z}^{\symA}$. 
    On the other hand, 
    $(\Psi_{\samp} \setminus \Psi_{\OLD}) \cap \Psi_{\str}(Z) = \emptyset$ follows from the definition of the set $\mathcal{Z}^{\symB}$. 
    The two facts $(\Psi_{\samp} \setminus \Psi_{\OLD}) \cap \Psi_{\str}(Z) \neq \emptyset$ and $(\Psi_{\samp} \setminus \Psi_{\OLD}) \cap \Psi_{\str}(Z) = \emptyset$ yield a contradiction. 
    Therefore, $\mathcal{Z}^{\symA} \cap \mathcal{Z}^{\symB} = \emptyset$ must hold.
\end{proof}

The following proposition states the relationship among interval attractors in set $\Psi^{\prime}_{\samp}$. 
\begin{proposition}\label{prop:dynamic_samp_formula_sub3}
Consider two interval attractors $([p_{1}, q_{1}], [\ell_{1}, r_{1}])$ and 
$([p_{2}, q_{2}], [\ell_{2}, r_{2}])$ in set $\Psi^{\prime}_{\samp}$. 
Then, $T^{\prime}[p_{1}-1..r_{1}+1] \neq T^{\prime}[p_{2}-1..r_{2}+1]$ holds if 
at least one of the following four conditions: 
\begin{enumerate}[label=\textbf{(\roman*)}]
    \item $([p_{1}, q_{1}], [\ell_{1}, r_{1}]), ([p_{2}, q_{2}], [\ell_{2}, r_{2}]) \in \Psi_{\samp} \cap \Psi_{\LEFT}$;
    \item $([p_{1}, q_{1}], [\ell_{1}, r_{1}]), ([p_{2}, q_{2}], [\ell_{2}, r_{2}]) \in \{ ([p + 1, q + 1], [\ell + 1, r + 1]) \mid ([p, q], [\ell, r]) \in \Psi_{\samp} \cap \Psi_{\RIGHT} \}$;
    \item 
    $([p_{1}, q_{1}], [\ell_{1}, r_{1}]), ([p_{2}, q_{2}], [\ell_{2}, r_{2}]) \in \{ ([p^{\symA^{\prime}}_{s} + \epsilon^{\symA^{\prime}}_{s}, q^{\symA^{\prime}}_{s} + \epsilon^{\symA^{\prime}}_{s}], [\ell^{\symA^{\prime}}_{s} + \epsilon^{\symA^{\prime}}_{s}, r^{\symA^{\prime}}_{s} + \epsilon^{\symA^{\prime}}_{s}]) \mid s \in [1, k] \}$ for set $\mathcal{Z}^{\symA} = \{ Z^{\symA}_{1}, Z^{\symA}_{2}, \ldots, Z^{\symA}_{k} \}$ of strings;
    \item 
    $([p_{1}, q_{1}], [\ell_{1}, r_{1}]), ([p_{2}, q_{2}], [\ell_{2}, r_{2}]) \in \{ ([p^{\symB}_{s}, q^{\symB}_{s}], [\ell^{\symB}_{s}, r^{\symB}_{s}]) \mid s \in [1, k^{\prime}] \}$ for set $\mathcal{Z}^{\symB} = \{ Z^{\symB}_{1}, Z^{\symB}_{2}$, $\ldots$, $Z^{\symB}_{k^{\prime}} \}$ of strings.
\end{enumerate}
\end{proposition}
\begin{proof}
    The proof of Proposition~\ref{prop:dynamic_samp_formula_sub3} is as follows.

    \paragraph{Proof of Proposition~\ref{prop:dynamic_samp_formula_sub3} for condition (i).}
    $T[p_{1}-1..r_{1}+1] \neq T[p_{2}-1..r_{2}+1]$ follows from the definition of the sampling subset $\Psi_{\samp}$.     
    Because of $([p_{1}, q_{1}], [\ell_{1}, r_{1}]) \in \Psi_{\LEFT}$, 
    $T^{\prime}[p_{1}-1..r_{1}+1] = T[p_{1}-1..r_{1}+1]$ holds. 
    Similarly, 
    $T^{\prime}[p_{2}-1..r_{2}+1] = T[p_{2}-1..r_{2}+1]$ holds. 
    Therefore, 
    we obtain $T^{\prime}[p_{1}-1..r_{1}+1] \neq T^{\prime}[p_{2}-1..r_{2}+1]$. 

    \paragraph{Proof of Proposition~\ref{prop:dynamic_samp_formula_sub3} for condition (ii).}
    We can prove $T^{\prime}[p_{1}-1..r_{1}+1] \neq T^{\prime}[p_{2}-1..r_{2}+1]$ 
    in a similar way as the proof for condition (i).

    \paragraph{Proof of Proposition~\ref{prop:dynamic_samp_formula_sub3} for condition (iii).}
    In this case, 
    there exists a pair of two integers $s, s^{\prime} \in [1, k]$ ($s \neq s^{\prime}$)
    satisfying 
    $([p_{1}, q_{1}], [\ell_{1}, r_{1}]) = ([p^{\symA^{\prime}}_{s} + \epsilon^{\symA^{\prime}}_{s}, q^{\symA^{\prime}}_{s} + \epsilon^{\symA^{\prime}}_{s}], [\ell^{\symA^{\prime}}_{s} + \epsilon^{\symA^{\prime}}_{s}, r^{\symA^{\prime}}_{s} + \epsilon^{\symA^{\prime}}_{s}])$ 
    and 
    $([p_{2}, q_{2}], [\ell_{2}, r_{2}]) = ([p^{\symA^{\prime}}_{s^{\prime}} + \epsilon^{\symA^{\prime}}_{s^{\prime}}, q^{\symA^{\prime}}_{s^{\prime}} + \epsilon^{\symA^{\prime}}_{s^{\prime}}], [\ell^{\symA^{\prime}}_{s^{\prime}} + \epsilon^{\symA^{\prime}}_{s^{\prime}}, r^{\symA^{\prime}}_{s^{\prime}} + \epsilon^{\symA^{\prime}}_{s^{\prime}}])$. 
    Here, $T^{\prime}[p^{\symA^{\prime}}_{s} - 1 + \epsilon^{\symA^{\prime}}_{s}..r^{\symA^{\prime}}_{s} + 1 + \epsilon^{\symA^{\prime}}_{s}] = Z^{\symA}_{s}$ 
    and $T^{\prime}[p^{\symA^{\prime}}_{s^{\prime}} - 1 + \epsilon^{\symA^{\prime}}_{s^{\prime}}..r^{\symA^{\prime}}_{s^{\prime}} + 1 + \epsilon^{\symA^{\prime}}_{s^{\prime}}] = Z^{\symA}_{s^{\prime}}$. 
    $T^{\prime}[p^{\symA^{\prime}}_{s} - 1 + \epsilon^{\symA^{\prime}}_{s}..r^{\symA^{\prime}}_{s} + 1 + \epsilon^{\symA^{\prime}}_{s}] \neq T^{\prime}[p^{\symA^{\prime}}_{s^{\prime}} - 1 + \epsilon^{\symA^{\prime}}_{s^{\prime}}..r^{\symA^{\prime}}_{s^{\prime}} + 1 + \epsilon^{\symA^{\prime}}_{s^{\prime}}]$ holds 
    because $Z^{\symA}_{s} \neq Z^{\symA}_{s^{\prime}}$.     
    Therefore, $T^{\prime}[p_{1}-1..r_{1}+1] \neq T^{\prime}[p_{2}-1..r_{2}+1]$ 
    follows from 
    (a) $T^{\prime}[p^{\symA^{\prime}}_{s} - 1 + \epsilon^{\symA^{\prime}}_{s}..r^{\symA^{\prime}}_{s} + 1 + \epsilon^{\symA^{\prime}}_{s}] \neq T^{\prime}[p^{\symA^{\prime}}_{s^{\prime}} - 1 + \epsilon^{\symA^{\prime}}_{s^{\prime}}..r^{\symA^{\prime}}_{s^{\prime}} + 1 + \epsilon^{\symA^{\prime}}_{s^{\prime}}]$, 
    (b) $[p_{1}, r_{1}] = [p^{\symA^{\prime}}_{s}, r^{\symA^{\prime}}_{s}]$, 
    and (c) $[p_{2}, r_{2}] = [p^{\symA^{\prime}}_{s^{\prime}}, r^{\symA^{\prime}}_{s^{\prime}}]$. 
    
    \paragraph{Proof of Proposition~\ref{prop:dynamic_samp_formula_sub3} for condition (iv).}
    In this case, 
    there exists a pair of two integers $s, s^{\prime} \in [1, k^{\prime}]$ ($s \neq s^{\prime}$)
    satisfying 
    $([p_{1}, q_{1}], [\ell_{1}, r_{1}]) = ([p^{\symB}_{s}, q^{\symB}_{s}], [\ell^{\symB}_{s}, r^{\symB}_{s}])$ 
    and 
    $([p_{2}, q_{2}], [\ell_{2}, r_{2}]) = ([p^{\symB}_{s^{\prime}}, q^{\symB}_{s^{\prime}}], [\ell^{\symB}_{s^{\prime}}, r^{\symB}_{s^{\prime}}])$. 
    Here, $([p^{\symB}_{s}, q^{\symB}_{s}], [\ell^{\symB}_{s}, r^{\symB}_{s}]) \in \Psi_{\str}(Z^{\symB}_{s})$,  
    $([p^{\symB}_{s^{\prime}}, q^{\symB}_{s^{\prime}}], [\ell^{\symB}_{s^{\prime}}, r^{\symB}_{s^{\prime}}]) \in \Psi_{\str}(Z^{\symB}_{s^{\prime}})$, and $Z^{\symB}_{s} \neq Z^{\symB}_{s^{\prime}}$ hold. 
    $T^{\prime}[p^{\symB}_{s}-1..r^{\symB}_{s}+1] = Z^{\symB}_{s}$ follows from the definition of the subset $\Psi_{\str}(Z^{\symB}_{s})$. 
    Similarly, 
    $T^{\prime}[p^{\symB}_{s}-1..r^{\symB}_{s}+1] = Z^{\symB}_{s}$ follows from the definition of the subset $\Psi_{\str}(Z^{\symB}_{s})$.

\end{proof}

We prove Theorem~\ref{theo:dynamic_samp_formula} using 
Proposition~\ref{prop:dynamic_samp_formula_sub2} and Proposition~\ref{prop:dynamic_samp_formula_sub3}. 

\begin{proof}[Proof of Theorem~\ref{theo:dynamic_samp_formula}(A)]
The interval attractor $([p_{1}, q_{1}], [\ell_{1}, r_{1}])$ satisfies one of the following four conditions: 
\begin{enumerate}[label=\textbf{(\alph*)}]
    \item $([p_{1}, q_{1}], [\ell_{1}, r_{1}]) \in \Psi_{\samp} \cap \Psi_{\LEFT}$;
    \item $([p_{1}, q_{1}], [\ell_{1}, r_{1}]) \in \{ ([p + 1, q + 1], [\ell + 1, r + 1]) \mid ([p, q], [\ell, r]) \in \Psi_{\samp} \cap \Psi_{\RIGHT} \}$;
    \item $([p_{1}, q_{1}], [\ell_{1}, r_{1}]) \in \{ ([p^{\symA^{\prime}}_{s} + \epsilon^{\symA^{\prime}}_{s}, q^{\symA^{\prime}}_{s} + \epsilon^{\symA^{\prime}}_{s}], [\ell^{\symA^{\prime}}_{s} + \epsilon^{\symA^{\prime}}_{s}, r^{\symA^{\prime}}_{s} + \epsilon^{\symA^{\prime}}_{s}]) \mid s \in [1, k] \}$;
    \item $([p_{1}, q_{1}], [\ell_{1}, r_{1}]) \in \{ ([p^{\symB}_{s}, q^{\symB}_{s}], [\ell^{\symB}_{s}, r^{\symB}_{s}]) \mid s \in [1, k^{\prime}] \}$.
\end{enumerate}

For condition (a), 
$T[p_{1}-1..r_{1}+1] \in \mathcal{Z}^{L}$ holds. 
If $([p_{2}, q_{2}], [\ell_{2}, r_{2}]) \in \Psi_{\samp} \cap \Psi_{\LEFT}$, 
then Proposition~\ref{prop:dynamic_samp_formula_sub3} shows that 
$T^{\prime}[p_{1}-1..r_{1}+1] \neq T^{\prime}[p_{2}-1..r_{2}+1]$ holds. 
Otherwise, 
$T^{\prime}[p_{2}-1..r_{2}+1] \in \mathcal{Z}^{R} \cup \mathcal{Z}^{\symA} \cup \mathcal{Z}^{\symB}$ holds. 
Proposition~\ref{prop:dynamic_samp_formula_sub2} indicates that 
$\mathcal{Z}^{L} \cap (\mathcal{Z}^{R} \cup \mathcal{Z}^{\symA} \cup \mathcal{Z}^{\symB}) = \emptyset$ holds. 
Therefore, 
$T^{\prime}[p_{1}-1..r_{1}+1] \neq T^{\prime}[p_{2}-1..r_{2}+1]$ holds. 

For condition (b), 
$T^{\prime}[p_{1}-1..r_{1}+1] \in \mathcal{Z}^{R}$ holds. 
If $([p_{2}, q_{2}], [\ell_{2}, r_{2}]) \in \{ ([p + 1, q + 1], [\ell + 1, r + 1]) \mid ([p, q], [\ell, r]) \in \Psi_{\samp} \cap \Psi_{\RIGHT} \}$ holds, 
then Proposition~\ref{prop:dynamic_samp_formula_sub3} shows that 
$T^{\prime}[p_{1}-1..r_{1}+1] \neq T^{\prime}[p_{2}-1..r_{2}+1]$ holds. 
Otherwise, 
$T^{\prime}[p_{1}-1..r_{1}+1] \neq T^{\prime}[p_{2}-1..r_{2}+1]$ can be proved using the same approach as for condition (a). 

For condition (c), 
$T^{\prime}[p_{1}-1..r_{1}+1] \in \mathcal{Z}^{\symA}$ holds. 
If $([p_{2}, q_{2}], [\ell_{2}, r_{2}]) \in \{ ([p^{\symA^{\prime}}_{s} + \epsilon^{\symA^{\prime}}_{s}, q^{\symA^{\prime}}_{s} + \epsilon^{\symA^{\prime}}_{s}], [\ell^{\symA^{\prime}}_{s} + \epsilon^{\symA^{\prime}}_{s}, r^{\symA^{\prime}}_{s} + \epsilon^{\symA^{\prime}}_{s}]) \mid s \in [1, k] \}$ holds, 
then Proposition~\ref{prop:dynamic_samp_formula_sub3} shows that 
$T^{\prime}[p_{1}-1..r_{1}+1] \neq T^{\prime}[p_{2}-1..r_{2}+1]$ holds. 
Otherwise, 
$T^{\prime}[p_{1}-1..r_{1}+1] \neq T^{\prime}[p_{2}-1..r_{2}+1]$ can be proved using the same approach as for condition (a). 

For condition (d), 
$T^{\prime}[p_{1}-1..r_{1}+1] \in \mathcal{Z}^{\symB}$ holds. 
If $([p_{2}, q_{2}], [\ell_{2}, r_{2}]) \in \{ ([p^{\symB}_{s}, q^{\symB}_{s}], [\ell^{\symB}_{s}, r^{\symB}_{s}]) \mid s \in [1, k^{\prime}] \}$ holds, 
then Proposition~\ref{prop:dynamic_samp_formula_sub3} shows that 
$T^{\prime}[p_{1}-1..r_{1}+1] \neq T^{\prime}[p_{2}-1..r_{2}+1]$ holds. 
Otherwise, 
$T^{\prime}[p_{1}-1..r_{1}+1] \neq T^{\prime}[p_{2}-1..r_{2}+1]$ can be proved using the same approach as for condition (a). 

Finally, $T^{\prime}[p_{1}-1..r_{1}+1] \neq T^{\prime}[p_{2}-1..r_{2}+1]$ always holds. 
\end{proof}

\begin{proof}[Proof of Theorem~\ref{theo:dynamic_samp_formula}(B)]
Similar to the proof of Theorem~\ref{theo:dynamic_samp_formula}(A), 
we consider four conditions of the interval attractor $([p_{1}, q_{1}], [\ell_{1}, r_{1}])$, i.e., 
\begin{enumerate}[label=\textbf{(\alph*)}]
    \item $([p_{1}, q_{1}], [\ell_{1}, r_{1}]) \in \Psi_{\samp} \cap \Psi_{\LEFT}$;
    \item $([p_{1}, q_{1}], [\ell_{1}, r_{1}]) \in \{ ([p + 1, q + 1], [\ell + 1, r + 1]) \mid ([p, q], [\ell, r]) \in \Psi_{\samp} \cap \Psi_{\RIGHT} \}$;
    \item $([p_{1}, q_{1}], [\ell_{1}, r_{1}]) \in \{ ([p^{\symA^{\prime}}_{s} + \epsilon^{\symA^{\prime}}_{s}, q^{\symA^{\prime}}_{s} + \epsilon^{\symA^{\prime}}_{s}], [\ell^{\symA^{\prime}}_{s} + \epsilon^{\symA^{\prime}}_{s}, r^{\symA^{\prime}}_{s} + \epsilon^{\symA^{\prime}}_{s}]) \mid s \in [1, k] \}$;
    \item $([p_{1}, q_{1}], [\ell_{1}, r_{1}]) \in \{ ([p^{\symB}_{s}, q^{\symB}_{s}], [\ell^{\symB}_{s}, r^{\symB}_{s}]) \mid s \in [1, k^{\prime}] \}$.
\end{enumerate}

For condition (a), 
set $\Psi_{\samp} \cap \Psi_{\LEFT}$ contains the interval attractor $([p_{1}, q_{1}], [\ell_{1}, r_{1}])$. 
$([p_{1}, q_{1}], [\ell_{1}$, $r_{1}]) \not \in \Psi_{\run}$ follows from 
Lemma~\ref{lem:samp_basic_property}~\ref{enum:samp_basic_property:3}. 
We apply Lemma~\ref{lem:dynamic_RR_subset}~\ref{enum:dynamic_RR_subset:run} 
to the interval attractor $([p_{1}, q_{1}], [\ell_{1}, r_{1}])$. 
Then, $([p_{1}, q_{1}], [\ell_{1}, r_{1}]) \not \in \Psi^{\prime}_{\run}$ holds. 
Because of $([p_{1}, q_{1}], [\ell_{1}, r_{1}]) \in \Psi^{\prime}_{\RR}$, 
Lemma~\ref{lem:lm_basic_property}~\ref{enum:lm_basic_property:2} shows that 
set $\Psi^{\prime}_{\leftmost}$ contains 
an interval attractor $([p_{2}, q_{2}], [\ell_{2}, r_{2}])$ satisfying 
$T^{\prime}[p_{1}-1..r_{1}+1] = T^{\prime}[p_{2}-1..r_{2}+1]$. 
We apply 
Lemma~\ref{lem:psi_equality_basic_property}~\ref{enum:psi_equality_basic_property:4} 
to the two interval attractors $([p_{1}, q_{1}], [\ell_{1}, r_{1}])$ and $([p_{2}, q_{2}], [\ell_{2}, r_{2}])$. 
Then, $([p_{2}, q_{2}], [\ell_{2}, r_{2}]) \not \in \Psi^{\prime}_{\run}$ holds 
because of $([p_{1}, q_{1}], [\ell_{1}, r_{1}]) \not \in \Psi^{\prime}_{\run}$. 
Therefore, Theorem~\ref{theo:dynamic_samp_formula}(B) holds for condition (a). 

For condition (b), 
we can prove Theorem~\ref{theo:dynamic_samp_formula}(B) using the same approach as for condition (a).
For condition (c), 
there exists an integer $s \in [1, k]$ satisfying 
$([p_{1}, q_{1}], [\ell_{1}, r_{1}]) = ([p^{\symA^{\prime}}_{s} + \epsilon^{\symA^{\prime}}_{s}, q^{\symA^{\prime}}_{s} + \epsilon^{\symA^{\prime}}_{s}], [\ell^{\symA^{\prime}}_{s} + \epsilon^{\symA^{\prime}}_{s}, r^{\symA^{\prime}}_{s} + \epsilon^{\symA^{\prime}}_{s}])$. 
Lemma~\ref{lem:dynamic_samp_substitute} shows that 
$([p^{\symA^{\prime}}_{s}, q^{\symA^{\prime}}_{s}], [\ell^{\symA^{\prime}}_{s}, r^{\symA^{\prime}}_{s}]) \in (\Psi_{\RR} \setminus \Psi_{\OLD}) \cap \Psi_{\str}(Z^{\symA}_{s})$. 
Here, $T[p^{\symA^{\prime}}_{s}-1..r^{\symA^{\prime}}_{s}+1] = Z^{\symA}_{s}$ follows from the definition of the subset $\Psi_{\str}(Z^{\symA}_{s})$. 

$\Psi_{\samp} \cap \Psi_{\OLD} \cap \Psi_{\str}(Z^{\symA}_{s}) \neq \emptyset$ 
follows from the definition of the set $\mathcal{Z}^{\symA}$. 
Because of 
$\Psi_{\samp} \cap \Psi_{\OLD} \cap \Psi_{\str}(Z^{\symA}_{s}) \neq \emptyset$, 
the sampling subset $\Psi_{\samp}$ contains an interval attractor 
$([p_{3}, q_{3}], [\ell_{3}, r_{3}])$ satisfying $T[p_{3}-1..r_{3}+1] = Z^{\symA}_{s}$. 
$([p_{3}, q_{3}], [\ell_{3}, r_{3}]) \not \in \Psi_{\run}$ follows from 
Lemma~\ref{lem:samp_basic_property}~\ref{enum:samp_basic_property:3}. 
We can apply Lemma~\ref{lem:psi_equality_basic_property}~\ref{enum:psi_equality_basic_property:4} 
to the two interval attractors $([p^{\symA^{\prime}}_{s}, q^{\symA^{\prime}}_{s}], [\ell^{\symA^{\prime}}_{s}, r^{\symA^{\prime}}_{s}])$ and $([p_{3}, q_{3}], [\ell_{3}, r_{3}])$ because $T[p^{\symA^{\prime}}_{s}-1..r^{\symA^{\prime}}_{s}+1] = T[p_{3}-1..r_{3}+1]$ holds. 
This lemma shows that $([p^{\symA^{\prime}}_{s}, q^{\symA^{\prime}}_{s}], [\ell^{\symA^{\prime}}_{s}, r^{\symA^{\prime}}_{s}]) \not \in \Psi_{\run}$ holds 
because of $([p_{3}, q_{3}], [\ell_{3}, r_{3}]) \not \in \Psi_{\run}$. 
We apply Lemma~\ref{lem:dynamic_RR_subset}~\ref{enum:dynamic_RR_subset:run} 
to the interval attractor $([p^{\symA^{\prime}}_{s}, q^{\symA^{\prime}}_{s}], [\ell^{\symA^{\prime}}_{s}, r^{\symA^{\prime}}_{s}])$. 
Then, $([p^{\symA^{\prime}}_{s} + \epsilon^{\symA^{\prime}}_{s}, q^{\symA^{\prime}}_{s} + \epsilon^{\symA^{\prime}}_{s}], [\ell^{\symA^{\prime}}_{s} + \epsilon^{\symA^{\prime}}_{s}, r^{\symA^{\prime}}_{s} + \epsilon^{\symA^{\prime}}_{s}]) \not \in \Psi^{\prime}_{\run}$ holds 
(i.e., $([p_{1}, q_{1}], [\ell_{1}, r_{1}]) \not \in \Psi^{\prime}_{\run}$). 

Lemma~\ref{lem:lm_basic_property}~\ref{enum:lm_basic_property:2} shows that 
set $\Psi^{\prime}_{\leftmost}$ contains interval attractor 
an interval attractor $([p_{2}, q_{2}]$, $[\ell_{2}, r_{2}])$ satisfying 
$T^{\prime}[p_{1}-1..r_{1}+1] = T^{\prime}[p_{2}-1..r_{2}+1]$. 
We apply 
Lemma~\ref{lem:psi_equality_basic_property}~\ref{enum:psi_equality_basic_property:4} 
to the two interval attractors $([p_{1}, q_{1}], [\ell_{1}, r_{1}])$ and $([p_{2}, q_{2}], [\ell_{2}, r_{2}])$. 
Then, $([p_{2}, q_{2}], [\ell_{2}, r_{2}]) \not \in \Psi^{\prime}_{\run}$ holds 
because of $([p_{1}, q_{1}], [\ell_{1}, r_{1}]) \not \in \Psi^{\prime}_{\run}$. 
Therefore, statement (B) holds for condition (c). 

For condition (d), 
there exists an integer $s \in [1, k]$ satisfying 
$([p_{1}, q_{1}], [\ell_{1}, r_{1}]) = ([p^{\symB}_{s}, q^{\symB}_{s}]$, $[\ell^{\symB}_{s}, r^{\symB}_{s}])$. 
Here, $([p^{\symB}_{s}, q^{\symB}_{s}], [\ell^{\symB}_{s}, r^{\symB}_{s}]) \in (\Psi^{\prime}_{\NEW} \setminus \Psi^{\prime}_{\run}) \cap \Psi^{\prime}_{\str}(Z^{\symB}_{s})$ holds (see Lemma~\ref{lem:dynamic_samp_new}). 
Because of $([p_{1}, q_{1}], [\ell_{1}, r_{1}]) \in \Psi^{\prime}_{\str}(Z^{\symB}_{s})$, 
$T^{\prime}[p_{1}-1..r_{1}+1] = Z^{\symB}_{s}$ follows from the definition of the subset $\Psi^{\prime}_{\str}(Z^{\symB}_{s})$. 
Lemma~\ref{lem:lm_basic_property}~\ref{enum:lm_basic_property:2} shows that 
set $\Psi^{\prime}_{\leftmost}$ contains interval attractor 
an interval attractor $([p_{2}, q_{2}], [\ell_{2}, r_{2}])$ satisfying 
$T^{\prime}[p_{1}-1..r_{1}+1] = T^{\prime}[p_{2}-1..r_{2}+1]$. 
$T^{\prime}[p_{2}-1..r_{2}+1] = Z^{\symB}_{s}$ follows from 
$T^{\prime}[p_{1}-1..r_{1}+1] = T^{\prime}[p_{2}-1..r_{2}+1]$ 
and $T^{\prime}[p_{1}-1..r_{1}+1] = Z^{\symB}_{s}$. 

We can apply Lemma~\ref{lem:psi_equality_basic_property}~\ref{enum:psi_equality_basic_property:4} 
to the two interval attractors $([p_{1}, q_{1}], [\ell_{1}, r_{1}])$ and $([p_{2}, q_{2}], [\ell_{2}, r_{2}])$ because $T^{\prime}[p_{1}-1..r_{1}+1] = T^{\prime}[p_{2}-1..r_{2}+1]$ holds. 
This lemma shows that $([p_{2}, q_{2}], [\ell_{2}, r_{2}]) \not \in \Psi^{\prime}_{\run}$ holds 
because of $([p_{1}, q_{1}], [\ell_{1}, r_{1}]) \not \in \Psi^{\prime}_{\run}$. 
Therefore, statement (B) holds for condition (d). 

%$(\Psi^{\prime}_{\NEW} \setminus \Psi^{\prime}_{\run}) \cap \Psi^{\prime}_{\str}(Z^{\symB}_{s}) \neq \emptyset$ 
%follows from the definition of the set $\mathcal{Z}^{\symB}$. 
%Because of $(\Psi^{\prime}_{\NEW} \setminus \Psi^{\prime}_{\run}) \cap \Psi^{\prime}_{\str}(Z^{\symB}_{s}) \neq \emptyset$, 
%set $\Psi^{\prime}_{\RR}$ contains an interval attractor $([p_{3}, q_{3}], [\ell_{3}, r_{3}])$ 
%satisfying $([p_{3}, q_{3}], [\ell_{3}$, $r_{3}]) \not \in \Psi^{\prime}_{\run}$ and 
%$T^{\prime}[p_{3}-1..r_{3}+1] = Z^{\symB}_{s}$. 

Finally, statement (B) always holds. 
\end{proof}

\begin{proof}[Proof of Theorem~\ref{theo:dynamic_samp_formula}(C)]
There exist two cases: 
$(\Psi_{\RR} \setminus \Psi_{\OLD}) \cap \Psi_{\str}(T^{\prime}[p_{1}-1..r_{1}+1]) \neq \emptyset$, or not.

\paragraph{Proof of Theorem~\ref{theo:dynamic_samp_formula}(C) for $(\Psi_{\RR} \setminus \Psi_{\OLD}) \cap \Psi_{\str}(T^{\prime}[p_{1}-1..r_{1}+1]) \neq \emptyset$.}
In this case, 
set $(\Psi_{\RR} \setminus \Psi_{\OLD})$ contains an interval attractor $([p_{3}, q_{3}], [\ell_{3}, r_{3}])$ 
satisfying $T[p_{3}-1..r_{3}+1] = T^{\prime}[p_{1}-1..r_{1}+1]$. 
Let $\epsilon_{3} = 0$ if $([p_{3}, q_{3}], [\ell_{3}, r_{3}]) \in \Psi_{\LEFT}$; 
otherwise let $\epsilon_{3} = 1$. 
Then, Theorem~\ref{theo:RS_LEFT_RIGHT} shows that 
$([p_{3}+\epsilon_{3}, q_{3}+\epsilon_{3}], [\ell_{3}+\epsilon_{3}, r_{3}+\epsilon_{3}]) \in \Psi^{\prime}_{\LEFT} \cup \Psi^{\prime}_{\RIGHT}$. 
Here, $T[p_{3}-1..r_{3}+1] = T^{\prime}[p_{3}+\epsilon_{3}-1..r_{3}+\epsilon_{3}+1]$ follows from Corollary~\ref{cor:RB_IA_before_after}~\ref{enum:RB_IA_before_after:1}. 

We apply Lemma~\ref{lem:psi_equality_basic_property}~\ref{enum:psi_equality_basic_property:4} 
to the two interval attractors $([p_{3}+\epsilon_{3}, q_{3}+\epsilon_{3}], [\ell_{3}+\epsilon_{3}, r_{3}+\epsilon_{3}])$ and $([p_{1}, q_{1}], [\ell_{1}, r_{1}])$.   
Lemma~\ref{lem:psi_equality_basic_property}~\ref{enum:psi_equality_basic_property:4} shows that 
$([p_{3}+\epsilon_{3}, q_{3}+\epsilon_{3}], [\ell_{3}+\epsilon_{3}, r_{3}+\epsilon_{3}]) \not \in \Psi^{\prime}_{\run}$ holds 
because of $([p_{1}, q_{1}], [\ell_{1}, r_{1}]) \not \in \Psi^{\prime}_{\run}$. 
We apply Lemma~\ref{lem:dynamic_RR_subset}~\ref{enum:dynamic_RR_subset:run} 
to the two interval attractor $([p_{3}, q_{3}], [\ell_{3}, r_{3}])$ and $([p_{3}+\epsilon_{3}, q_{3}+\epsilon_{3}], [\ell_{3}+\epsilon_{3}, r_{3}+\epsilon_{3}])$. 
Then, $([p_{3}, q_{3}], [\ell_{3}, r_{3}]) \not \in \Psi_{\run}$ holds. 
Therefore, the subset $\Psi_{\run}$ does not contain the interval attractor $([p_{3}, q_{3}], [\ell_{3}, r_{3}])$. 

Since $([p_{3}, q_{3}], [\ell_{3}, r_{3}]) \in \Psi_{\RR} \setminus \Psi_{\run}$, 
Lemma~\ref{lem:samp_basic_property}~\ref{enum:samp_basic_property:2} shows that 
the sampling subset $\Psi_{\samp} \setminus \Psi_{\run}$ contains an interval attractor 
$([p_{4}, q_{4}], [\ell_{4}, r_{4}])$ satisfying $T[p_{4}-1..r_{4}+1] = T[p_{3}-1..r_{3}+1]$.
Because of $\Psi_{\RR} = \Psi_{\LEFT} \cup \Psi_{\OLD} \cup \Psi_{\RIGHT}$, 
the interval attractor $([p_{4}, q_{4}], [\ell_{4}, r_{4}])$ satisfies at least one of the following three conditions: 
\begin{enumerate}[label=\textbf{(\alph*)}]
    \item $([p_{4}, q_{4}], [\ell_{4}, r_{4}]) \in \Psi_{\LEFT}$;
    \item $([p_{4}, q_{4}], [\ell_{4}, r_{4}]) \in \Psi_{\RIGHT}$;
    \item $([p_{4}, q_{4}], [\ell_{4}, r_{4}]) \in \Psi_{\OLD}$. 
\end{enumerate}

For condition (a), 
$([p_{4}, q_{4}], [\ell_{4}, r_{4}]) \in \Psi^{\prime}_{\samp}$ follows from the definition of the set $\Psi^{\prime}_{\samp}$. 
Here, $T[p_{4}-1..r_{4}+1] = T^{\prime}[p_{4}-1..r_{4}+1]$ 
follows from Corollary~\ref{cor:RB_IA_before_after}~\ref{enum:RB_IA_before_after:1}. 
$T^{\prime}[p_{4}-1..r_{4}+1] = T^{\prime}[p_{1}-1..r_{1}+1]$ follows from 
$T[p_{4}-1..r_{4}+1] = T^{\prime}[p_{4}-1..r_{4}+1]$, 
$T[p_{4}-1..r_{4}+1] = T[p_{3}-1..r_{3}+1]$, 
and $T[p_{3}-1..r_{3}+1] = T^{\prime}[p_{1}-1..r_{1}+1]$. 
Therefore, 
$([p_{4}, q_{4}], [\ell_{4}, r_{4}]) \in \Psi^{\prime}_{\samp}$ 
and $T^{\prime}[p_{4}-1..r_{4}+1] = T^{\prime}[p_{1}-1..r_{1}+1]$ hold. 

For condition (b), 
we can prove  
$([p_{4}+1, q_{4}+1], [\ell_{4}+1, r_{4}+1]) \in \Psi^{\prime}_{\samp}$ 
and $T^{\prime}[p_{4}..r_{4}+2] = T^{\prime}[p_{1}-1..r_{1}+1]$ 
using the same approach as for condition (a). 

For condition (c), 
$([p_{4}, q_{4}], [\ell_{4}, r_{4}]) \in \Psi_{\str}(T^{\prime}[p_{1}-1..r_{1}+1])$ holds 
because of $T[p_{4}-1..r_{4}+1] = T^{\prime}[p_{1}-1..r_{1}+1]$. 
$\Psi_{\samp} \cap \Psi_{\OLD} \cap \Psi_{\str}(T^{\prime}[p_{1}-1..r_{1}+1]) \neq \emptyset$ 
follows from $([p_{4}, q_{4}], [\ell_{4}, r_{4}]) \in \Psi_{\samp} \cap \Psi_{\OLD} \cap \Psi_{\str}(T^{\prime}[p_{1}-1..r_{1}+1])$. 
$T^{\prime}[p_{1}-1..r_{1}+1] \in \mathcal{Z}^{\symA}$ follows from 
the definition of the set $\mathcal{Z}^{\symA}$ 
because 
$\Psi_{\samp} \cap \Psi_{\OLD} \cap \Psi_{\str}(T^{\prime}[p_{1}-1..r_{1}+1]) \neq \emptyset$ 
and $(\Psi_{\RR} \setminus \Psi_{\OLD}) \cap \Psi_{\str}(T^{\prime}[p_{1}-1..r_{1}+1]) \neq \emptyset$.
In other words, set $\mathcal{Z}^{\symA}$ contains the string $T^{\prime}[p_{1}-1..r_{1}+1]$ as 
a string $Z^{\symA}_{s}$. 
Lemma~\ref{lem:dynamic_samp_substitute} shows that 
$([p^{\symA^{\prime}}_{s}, q^{\symA^{\prime}}_{s}], [\ell^{\symA^{\prime}}_{s}, r^{\symA^{\prime}}_{s}]) \in (\Psi_{\RR} \setminus \Psi_{\OLD}) \cap \Psi_{\str}(Z^{\symA}_{s})$ holds.
Because of $([p^{\symA^{\prime}}_{s}, q^{\symA^{\prime}}_{s}], [\ell^{\symA^{\prime}}_{s}, r^{\symA^{\prime}}_{s}]) \in \Psi_{\str}(Z^{\symA}_{s})$, 
$T[p^{\symA^{\prime}}_{s}-1..r^{\symA^{\prime}}_{s}+1] = Z^{\symA}_{s}$ holds. 

$([p^{\symA^{\prime}}_{s}, q^{\symA^{\prime}}_{s}], [\ell^{\symA^{\prime}}_{s}, r^{\symA^{\prime}}_{s}]) \in \Psi_{\LEFT} \cup \Psi_{\RIGHT}$ holds 
because $([p^{\symA^{\prime}}_{s}, q^{\symA^{\prime}}_{s}], [\ell^{\symA^{\prime}}_{s}, r^{\symA^{\prime}}_{s}]) \not \in \Psi_{\OLD}$. 
Corollary~\ref{cor:RB_IA_before_after}~\ref{enum:RB_IA_before_after:1} shows that 
$T[p^{\symA^{\prime}}_{s}-1..r^{\symA^{\prime}}_{s}+1] = T^{\prime}[p^{\symA^{\prime}}_{s} + \epsilon^{\symA^{\prime}}_{s} - 1..r^{\symA^{\prime}}_{s} + \epsilon^{\symA^{\prime}}_{s} + 1]$ holds. 
Set $\Psi^{\prime}_{\samp}$ contains interval attractor 
$([p^{\symA^{\prime}}_{s} + \epsilon^{\symA^{\prime}}_{s}, q^{\symA^{\prime}}_{s} + \epsilon^{\symA^{\prime}}_{s}], [\ell^{\symA^{\prime}}_{s} + \epsilon^{\symA^{\prime}}_{s}, r^{\symA^{\prime}}_{s} + \epsilon^{\symA^{\prime}}_{s}])$. 
$T^{\prime}[p^{\symA^{\prime}}_{s} + \epsilon^{\symA^{\prime}}_{s} - 1..r^{\symA^{\prime}}_{s} + \epsilon^{\symA^{\prime}}_{s} + 1] = T^{\prime}[p_{1}-1..r_{1}+1]$ 
follows from 
$T[p^{\symA^{\prime}}_{s}-1..r^{\symA^{\prime}}_{s}+1] = T^{\prime}[p^{\symA^{\prime}}_{s} + \epsilon^{\symA^{\prime}}_{s} - 1..r^{\symA^{\prime}}_{s} + \epsilon^{\symA^{\prime}}_{s} + 1]$, 
$T[p^{\symA^{\prime}}_{s}-1..r^{\symA^{\prime}}_{s}+1] = Z^{\symA}_{s}$, and $Z^{\symA}_{s} = T^{\prime}[p_{1}-1..r_{1}+1]$. 

Finally, 
set $\Psi^{\prime}_{\samp}$ always contains an interval attractor 
$([p_{2}, q_{2}], [\ell_{2}, r_{2}])$ satisfying $T^{\prime}[p_{1}-1..r_{1}+1] = T^{\prime}[p_{2}-1..r_{2}+1]$, 
i.e., Theorem~\ref{theo:dynamic_samp_formula}(C) holds. 

\paragraph{Proof of Theorem~\ref{theo:dynamic_samp_formula}(C) for $(\Psi_{\RR} \setminus \Psi_{\OLD}) \cap \Psi_{\str}(T^{\prime}[p_{1}-1..r_{1}+1]) = \emptyset$.}
The interval attractor $([p_{1}, q_{1}], [\ell_{1}, r_{1}])$ is contained in set $(\Psi^{\prime}_{\RR} \setminus \Psi_{\run}^{\prime}) \cap \Psi_{\str}(T^{\prime}[p_{1}-1..r_{1}+1])$. 
We prove $(\Psi^{\prime}_{\NEW} \setminus \Psi^{\prime}_{\run}) \cap \Psi^{\prime}_{\str}(T^{\prime}[p_{1}-1..r_{1}+1]) \neq \emptyset$ by contradiction. 
We assume that 
$(\Psi^{\prime}_{\NEW} \setminus \Psi^{\prime}_{\run}) \cap \Psi^{\prime}_{\str}(T^{\prime}[p_{1}-1..r_{1}+1]) = \emptyset$ holds. 
Then, $([p_{1}, q_{1}], [\ell_{1}, r_{1}]) \in (\Psi^{\prime}_{\LEFT} \cup \Psi^{\prime}_{\RIGHT}) \setminus \Psi^{\prime}_{\run}$ holds 
because set $(\Psi^{\prime}_{\RR} \setminus \Psi_{\run}^{\prime})$ can be divided to 
three subsets $(\Psi^{\prime}_{\LEFT} \setminus \Psi^{\prime}_{\run}) \cup (\Psi^{\prime}_{\NEW} \setminus \Psi^{\prime}_{\run}) \cup (\Psi^{\prime}_{\RIGHT} \setminus \Psi^{\prime}_{\run})$.

Let $\epsilon_{1} = 0$ if $([p_{1}, q_{1}], [\ell_{1}, r_{1}]) \in \Psi^{\prime}_{\LEFT}$; 
otherwise, let $\epsilon_{1} = 1$. 
Because of $([p_{1}, q_{1}], [\ell_{1}, r_{1}]) \in \Psi^{\prime}_{\LEFT} \cup \Psi^{\prime}_{\RIGHT}$, 
set $\Psi_{\LEFT} \cup \Psi_{\RIGHT}$ contains 
interval attractor $([p_{1}-\epsilon_{1}, q_{1}-\epsilon_{1}], [\ell_{1}-\epsilon_{1}, r_{1}- \epsilon_{1}])$. 
Here, $T[p_{1}-\epsilon_{1}-1..r_{1}-\epsilon_{1}+1] = T^{\prime}[p_{1}-1..r_{1}+1]$ follows from 
Corollary~\ref{cor:RB_IA_after_before}~\ref{enum:RB_IA_after_before:1}, 
and hence, $([p_{1}- \epsilon_{1}, q_{1}- \epsilon_{1}], [\ell_{1}- \epsilon_{1}, r_{1}- \epsilon_{1}]) \in \Psi_{\str}(T^{\prime}[p_{1}-1..r_{1}+1])$. 
Because of $([p_{1}- \epsilon_{1}, q_{1}- \epsilon_{1}], [\ell_{1}- \epsilon_{1}, r_{1}- \epsilon_{1}]) \in \Psi_{\LEFT} \cup \Psi_{\RIGHT}$, 
the interval attractor $([p_{1}- \epsilon_{1}, q_{1}- \epsilon_{1}], [\ell_{1}- \epsilon_{1}, r_{1}- \epsilon_{1}])$ is contained in $\Psi_{\RR} \setminus \Psi_{\OLD}$. 
Therefore, 
$(\Psi_{\RR} \setminus \Psi_{\OLD}) \cap \Psi_{\str}(T^{\prime}[p_{1}-1..r_{1}+1]) \neq \emptyset$ 
follows from 
$([p_{1}- \epsilon_{1}, q_{1}- \epsilon_{1}], [\ell_{1}- \epsilon_{1}, r_{1}- \epsilon_{1}]) \in \Psi_{\str}(T^{\prime}[p_{1}-1..r_{1}+1])$ and 
$([p_{1}- \epsilon_{1}, q_{1}- \epsilon_{1}], [\ell_{1}- \epsilon_{1}, r_{1}- \epsilon_{1}]) \in \Psi_{\RR} \setminus \Psi_{\OLD}$. 

The two facts $(\Psi_{\RR} \setminus \Psi_{\OLD}) \cap \Psi_{\str}(T^{\prime}[p_{1}-1..r_{1}+1]) \neq \emptyset$ and $(\Psi_{\RR} \setminus \Psi_{\OLD}) \cap \Psi_{\str}(T^{\prime}[p_{1}-1..r_{1}+1]) = \emptyset$ yield a contradiction. 
Therefore, $(\Psi^{\prime}_{\NEW} \setminus \Psi^{\prime}_{\run}) \cap \Psi^{\prime}_{\str}(T^{\prime}[p_{1}-1..r_{1}+1]) \neq \emptyset$ must hold. 

We prove Theorem~\ref{theo:dynamic_samp_formula}(C). 
$T^{\prime}[p_{1}-1..r_{1}+1] \in \mathcal{Z}^{\symB}$ follows from the definition of the set $\mathcal{Z}^{\symB}$ 
because $(\Psi^{\prime}_{\NEW} \setminus \Psi^{\prime}_{\run}) \cap \Psi^{\prime}_{\str}(T^{\prime}[p_{1}-1..r_{1}+1]) \neq \emptyset$ and $(\Psi_{\RR} \setminus \Psi_{\OLD}) \cap \Psi_{\str}(T^{\prime}[p_{1}-1..r_{1}+1]) = \emptyset$. 
In other words, the set $\mathcal{Z}^{\symB}$ contains the string $T^{\prime}[p_{1}-1..r_{1}+1]$ as a string $Z^{\symB}_{s}$. 
Two sets $\Psi^{\prime}_{\samp}$ and $\Psi^{\prime}_{\str}(Z^{\symB}_{s})$ 
contain interval attractor $([p^{\symB}_{s}, q^{\symB}_{s}], [\ell^{\symB}_{s}, r^{\symB}_{s}])$. 
$T^{\prime}[p^{\symB}_{s}-1..r^{\symB}_{s}+1] = Z^{\symB}_{s}$ (i.e., $T^{\prime}[p^{\symB}_{s}-1..r^{\symB}_{s}+1] = T^{\prime}[p_{1}-1..r_{1}+1]$) follows from the definition of the subset $\Psi^{\prime}_{\str}(Z^{\symB}_{s})$. 
Therefore, Theorem~\ref{theo:dynamic_samp_formula}(C) holds. 
\end{proof}

%We proved statement (A), statement (B), and statement (C). 
%Therefore, 
%set $\Psi^{\prime}_{\samp}$ is a sampling subset for RLSLP $\mathcal{G}^{R}_{\ins}$. 

%\subsubsection{Proof of Lemma~\ref{lem:dynamic_samp_back_formula}}\label{subsubsec:proof_dynamic_samp_back_formula}

\subsection{Computation of Interval Attractors}\label{subsec:comp_IA_for_update}
This subsection presents algorithms computing interval attractors used to update the dynamic data structures built on set $\Psi_{\RR}$, which are introduced in Section~\ref{sec:other_query} and Section~\ref{sec:RSC_query}. 
The following lemma is the summary of this subsection. 

\begin{lemma}\label{lem:dynamic_IA_summary}
Using the dynamic data structures introduced in 
Section~\ref{sec:recompression}, Section~\ref{sec:other_query}, and Section~\ref{sec:RSC_query}, 
we can compute the following sets of interval attractors in expected $O((\max \{H, H^{\prime}$, $\log$ $(nHH^{\prime}) \})^{8})$ time: 
\begin{enumerate}[label=\textbf{(\roman*)}]
    \item $\Psi_{\OLD} \setminus \Psi_{\run}$;
    \item $\Psi^{\prime}_{\NEW} \setminus \Psi^{\prime}_{\run}$;
    \item the three sets $\Psi^{\symA}$, $\Psi^{\symA^{\prime}}$ and $\Psi^{\prime \symB}$ introduced in Section~\ref{subsec:update_sampling_subset};
    \item $\Psi_{\samp} \cap \Psi_{\OLD}$ and $(\Psi_{\samp} \cap \Psi_{\OLD}) \setminus \Psi^{\symA}$.
\end{enumerate}
\end{lemma}
\begin{proof}
    See Section~\ref{subsubsec:computation_lambda_ov}, Section~\ref{subsubsec:computation_samp_AB}, Section~\ref{subsubsec:computation_samp_C}, and Section~\ref{subsubsec:computation_samp_cap_old}.
\end{proof}

The following lemma states properties of the interval attractors stated in Lemma~\ref{lem:dynamic_IA_summary}.

\begin{lemma}\label{lem:dynamic_psi_overlap}
    Let $\Psi$ and $\Psi^{\prime}$ be the results of two overlap queries 
    $\OVQ([\lambda-2, \lambda+1])$ executed on set $\Psi_{\RR}$ 
    and $\OVQ([\lambda-2, \lambda+2])$ executed on set $\Psi^{\prime}_{\RR}$, respectively, 
    for the insertion position $\lambda$.     
    The following three statements hold: 
\begin{enumerate}[label=\textbf{(\roman*)}]
    \item \label{enum:dynamic_psi_overlap:1}
    $\Psi_{\OLD} \setminus \Psi_{\run} = \Psi$, 
    $|\Psi_{\OLD} \setminus \Psi_{\run}| = O(n^{2})$, 
    and $\mathbb{E}[|\Psi_{\OLD} \setminus \Psi_{\run}|] = O(H + \log n)$.
    \item \label{enum:dynamic_psi_overlap:2}
    $\Psi^{\prime}_{\NEW} \setminus \Psi^{\prime}_{\run} = \Psi^{\prime}$, 
    $|\Psi^{\prime}_{\NEW} \setminus \Psi^{\prime}_{\run}| = O(n^{2})$, 
    and $\mathbb{E}[|\Psi^{\prime}_{\NEW} \setminus \Psi^{\prime}_{\run}|] = O(H^{\prime} + \log n)$;
    \item \label{enum:dynamic_psi_overlap:3} $\Psi_{\samp} \cap \Psi_{\OLD} = \Psi_{\samp} \cap (\Psi_{\OLD} \setminus \Psi_{\run})$, 
    $|\Psi_{\samp} \cap \Psi_{\OLD}| = O(n^{2})$, 
    and $\mathbb{E}[|\Psi_{\samp} \cap \Psi_{\OLD}|] = O(H + \log n)$.
\end{enumerate}
\end{lemma}
\begin{proof}
    The proof of Lemma~\ref{lem:dynamic_psi_overlap} is as follows. 

    \textbf{Proof of Lemma~\ref{lem:dynamic_psi_overlap}(i).}
    $\Psi_{\OLD} \setminus \Psi_{\run} = \Psi$ follows from 
    the definitions of set $\Psi_{\OLD}$ and overlap query. 
    $|\Psi_{\OLD} \setminus \Psi_{\run}| = O(n^{2})$ follows from 
    $\Psi_{\OLD} \setminus \Psi_{\run} \subseteq \Psi_{\RR}$ and $|\Psi_{\RR}| = O(n^{2})$ (Lemma~\ref{lem:non_comp_IA_size}) hold. 
    Lemma~\ref{lem:overlap_count} shows that the expected value of $|\Psi|$ is $O(H + \log n)$. 
    Therefore, we obtain $\mathbb{E}[|\Psi_{\OLD} \cap \Psi_{\run}|] = O(H + \log n)$. 

    \textbf{Proof of Lemma~\ref{lem:dynamic_psi_overlap}(ii).}
    Lemma~\ref{lem:dynamic_psi_overlap}~\ref{enum:dynamic_psi_overlap:2} can be proved using the same approach as for Lemma~\ref{lem:dynamic_psi_overlap}~\ref{enum:dynamic_psi_overlap:1}.

    \textbf{Proof of Lemma~\ref{lem:dynamic_psi_overlap}(iii).}
    $\Psi_{\samp} \cap \Psi_{\run} = \emptyset$ follows from Lemma~\ref{lem:samp_basic_property}~\ref{enum:samp_basic_property:3}. 
    $\Psi_{\samp} \cap \Psi_{\OLD} = \Psi_{\samp} \cap (\Psi_{\OLD} \setminus \Psi_{\run})$ 
    follows from $\Psi_{\samp} \cap \Psi_{\run} = \emptyset$. 
    $|\Psi_{\samp} \cap \Psi_{\OLD}| = O(n^{2})$ follows from 
    $\Psi_{\samp} \cap \Psi_{\OLD} \subseteq \Psi_{\OLD} \setminus \Psi_{\run}$ 
    and $|\Psi_{\OLD} \setminus \Psi_{\run}| = O(n^{2})$. 
    $\mathbb{E}[|\Psi_{\OLD} \setminus \Psi_{\run}|] = O(H + \log n)$ follows from 
    $\Psi_{\samp} \cap \Psi_{\OLD} \subseteq \Psi_{\OLD} \setminus \Psi_{\run}$ 
    and $\mathbb{E}[|\Psi_{\OLD} \setminus \Psi_{\run}|] = O(H + \log n)$.     
\end{proof}

\subsubsection{Computation of Two Sets \texorpdfstring{$\Psi_{\OLD} \setminus \Psi_{\run}$}{} and \texorpdfstring{$\Psi^{\prime}_{\NEW} \setminus \Psi^{\prime}_{\run}$}{}}\label{subsubsec:computation_lambda_ov}
We present the algorithm computing two sets $\Psi_{\OLD} \setminus \Psi_{\run}$ and $\Psi^{\prime}_{\NEW} \setminus \Psi^{\prime}_{\run}$ of interval attractors using the dynamic data structures for the RR-DAG of RLSLP $\mathcal{G}^{R}$. 
We leverage Lemma~\ref{lem:dynamic_psi_overlap} for computing the two sets $\Psi_{\OLD} \setminus \Psi_{\run}$ and $\Psi^{\prime}_{\NEW} \setminus \Psi^{\prime}_{\run}$. 
Lemma~\ref{lem:dynamic_psi_overlap}~\ref{enum:dynamic_psi_overlap:1} shows that 
the subset $\Psi_{\OLD} \setminus \Psi_{\run}$ can be obtained by overlap query $\OVQ([\lambda-2, \lambda+1])$ 
on set $\Psi_{\RR}$ for the insertion position $\lambda$. 
Similarly, Lemma~\ref{lem:dynamic_psi_overlap}~\ref{enum:dynamic_samp_substitute:2} shows that 
the subset $\Psi^{\prime}_{\NEW} \setminus \Psi^{\prime}_{\run}$ can be obtained by overlap query $\OVQ([\lambda-2, \lambda+2])$ on set $\Psi^{\prime}_{\RR}$. 
Therefore, the two sets $\Psi_{\OLD} \setminus \Psi_{\run}$ and $\Psi^{\prime}_{\NEW} \setminus \Psi^{\prime}_{\run}$ can be obtained by two overlap queries.

We compute two sets $\Psi_{\OLD} \setminus \Psi_{\run}$ and $\Psi^{\prime}_{\NEW} \setminus \Psi^{\prime}_{\run}$ 
in four phases. 

\paragraph{Phase (i).}
In the first phase, we compute set $\Psi_{\OLD} \setminus \Psi_{\run}$ by overlap query $\OVQ([\lambda-2, \lambda+1])$. 
This overlap query takes expected $O((H + \log n) H^{3} \log n)$ time. 

\paragraph{Phase (ii).}
In the second phase, 
we obtain the dynamic data structures for the RR-DAG of RLSLP $\mathcal{G}^{R}_{\ins}$ 
by modifying the dynamic data structures for the RR-DAG of RLSLP $\mathcal{G}^{R}$. 
This update takes $O((\max \{H, H^{\prime}, \log (nHH^{\prime}) \})^{3})$ time (Lemma~\ref{lem:dynamic_rrdag_summary}). 

\paragraph{Phase (iii).}
In the third phase, 
we compute set $\Psi^{\prime}_{\NEW} \setminus \Psi^{\prime}_{\run}$ of interval attractors by overlap query $\OVQ([\lambda-2, \lambda+2])$ on set $\Psi^{\prime}_{\RR}$. 
This overlap query takes expected $O((H^{\prime} + \log n) H^{\prime 3} \log n)$ time. 

\paragraph{Phase (iv).}
In the fourth phase, 
we obtain the dynamic data structures for the RR-DAG of RLSLP $\mathcal{G}^{R}$ 
by modifying the dynamic data structures for the RR-DAG of RLSLP $\mathcal{G}^{R}_{\ins}$. 
This update takes $O((\max \{H, H^{\prime}, \log (nHH^{\prime}) \})^{3})$ time (Lemma~\ref{lem:dynamic_rrdag_summary}). 
Therefore, 
the two sets $\Psi_{\OLD} \setminus \Psi_{\run}$ and $\Psi^{\prime}_{\NEW} \setminus \Psi^{\prime}_{\run}$ can be computed in 
expected $O((\max \{H, H^{\prime}, \log (nHH^{\prime}) \})^{5})$ time. 

%Therefore, set $\Psi_{\OLD} \setminus \Psi_{\run}$ can be computed in expected $O((H + \log n) H^{3} \log n)$ time by the first phase. 
%Set $\Psi^{\prime}_{\NEW} \setminus \Psi^{\prime}_{\run}$ can be computed in expected $O((H^{\prime} + \log n) H^{\prime 3} \log n + (H^{2} + H^{\prime 2}) \max \{ H$, $H^{\prime}, \log (nH) \})$ time by the second, third, and fourth phases. 

\subsubsection{Computation of Two Subsets \texorpdfstring{$\Psi^{\symA}$}{} and \texorpdfstring{$\Psi^{\symA^{\prime}}$}{}}\label{subsubsec:computation_samp_AB}
We present the algorithm computing 
the two subsets 
$\Psi^{\symA} = \{ ([p^{\symA}_{1}, q^{\symA}_{1}], [\ell^{\symA}_{1}, r^{\symA}_{1}])$, $([p^{\symA}_{2}, q^{\symA}_{2}]$, $[\ell^{\symA}_{2}, r^{\symA}_{2}])$, 
$\ldots$, $([p^{\symA}_{k}, q^{\symA}_{k}], [\ell^{\symA}_{k}, r^{\symA}_{k}]) \}$ 
and $\Psi^{\symA^{\prime}} = \{ ([p^{\symA^{\prime}}_{1}, q^{\symA^{\prime}}_{1}], [\ell^{\symA^{\prime}}_{1}, r^{\symA^{\prime}}_{1}])$, $([p^{\symA^{\prime}}_{2}, q^{\symA^{\prime}}_{2}]$, $[\ell^{\symA^{\prime}}_{2}$, $r^{\symA^{\prime}}_{2}])$, 
$\ldots$, $([p^{\symA^{\prime}}_{k}, q^{\symA^{\prime}}_{k}], [\ell^{\symA^{\prime}}_{k}, r^{\symA^{\prime}}_{k}]) \}$ 
introduced in Section~\ref{subsec:update_sampling_subset}. 
This algorithm requires the dynamic data structures introduced in Section~\ref{sec:recompression}, Section~\ref{sec:other_query}, and Section~\ref{sec:RSC_query} 
for RLSLP $\mathcal{G}^{R}$. 

For computing the set $\Psi^{\symA}$, 
we leverage the set $\mathcal{Z}^{\symA} = \{ Z^{\symA}_{1}, Z^{\symA}_{2}, \ldots, Z^{\symA}_{k} \}$ of string introduced in 
Section~\ref{subsec:update_sampling_subset}. 
Consider the interval attractors $([p_{1}, q_{1}], [\ell_{1}, r_{1}])$, $([p_{2}, q_{2}], [\ell_{2}, r_{2}])$, $\ldots$, $([p_{m}, q_{m}], [\ell_{m}, r_{m}])$ of set $\Psi_{\OLD} \setminus \Psi_{\run}$. 
Here, the $m$ interval attractors are sorted so that $T[p_{1}-1..r_{1}+1] \preceq T[p_{2}-1..r_{2}+1] \preceq \cdots \preceq T[p_{m}-1..r_{m}+1]$. 
Let $\{ \pi_{1}, \pi_{2}, \ldots, \pi_{w}\}$ be a subset of set $\{ 1, 2, \ldots, m \}$ such that 
each integer $\pi_{s}$ satisfies the following two conditions: 
\begin{itemize}
    \item $([p_{\pi_{s}}, q_{\pi_{s}}], [\ell_{\pi_{s}}, r_{\pi_{s}}]) \in \Psi_{\samp}$;
    \item $|\Psi_{\str}(T[p_{\pi_{s}}-1..r_{\pi_{s}}+1])| - |\Psi_{\str}(T[p_{\pi_{s}}-1..r_{\pi_{s}}+1]) \cap (\Psi_{\OLD} \setminus \Psi_{\run})| \geq 1$. 
\end{itemize} 
Then, Lemma~\ref{lem:finding_samp_A_str} indicates that 
$w = k$, 
$Z^{\symA}_{1} = T[p_{\pi_{1}}-1..r_{\pi_{1}}+1]$, 
$Z^{\symA}_{2} = T[p_{\pi_{1}}-1..r_{\pi_{1}}+1]$, 
$\ldots$, 
$Z^{\symA}_{k} = T[p_{\pi_{k}}-1..r_{\pi_{k}}+1]$ hold. 
In addition, for each integer $s \in [1, k]$, 
$([p^{\symA}_{s}, q^{\symA}_{s}], [\ell^{\symA}_{s}, r^{\symA}_{s}]) = ([p_{\pi_{s}}, q_{\pi_{s}}], [\ell_{\pi_{s}}, r_{\pi_{s}}])$ 
follows from the definition of the subset $\Psi^{\symA}$. 
Therefore, the subset $\Psi^{\symA}$ can be computed by 
set $\Psi_{\OLD} \setminus \Psi_{\run}$ and subset $\{ \pi_{1}, \pi_{2}, \ldots, \pi_{w}\}$. 

Next, we leverage capture query for computing the set $\Psi^{\symA^{\prime}}$. 
Let $[\eta_{s}, \eta^{\prime}_{s}]$ be the sa-interval of the string $T[p_{s}-1..r_{s}+1]$ in the suffix array $\SA$ of input string $T$ for each integer $s \in [1, m]$. 
Let $x^{\symA}_{s} \in [1, n]$ be the smallest position in the suffix array $\SA$ satisfying 
the following two conditions: 
\begin{itemize}
    \item $x^{\symA}_{s} \in [\eta_{s}, \eta^{\prime}_{s}]$;
    \item $\SA[x^{\symA}_{s}] + 1 \not \in \{ p_{s} \mid s \in [1, m] \}$. 
\end{itemize} 
Then, for each integer $s \in [1, k]$, 
Lemma~\ref{lem:dynamic_samp_substitute} shows that 
the position $x^{\symA}_{\pi_{s}}$ exists, 
and capture query $\CAPQ([\SA[x^{\symA}_{\pi_{s}}]+1, \SA[x^{\symA}_{\pi_{s}}] + |Z^{\symA}_{s}| - 2])$ returns 
interval attractor $([p^{\symA^{\prime}}_{s}, q^{\symA^{\prime}}_{s}]$, $[\ell^{\symA^{\prime}}_{s}, r^{\symA^{\prime}}_{s}])$. 
Therefore, 
we can compute the two subsets $\Psi^{\symA}$ and $\Psi^{\symA^{\prime}}$ by processing the interval attractors of the set $\Psi_{\OLD} \setminus \Psi_{\run}$.

The algorithm computing the two subsets $\Psi^{\symA}$ and $\Psi^{\symA^{\prime}}$ consists of the following eight phases. 

\paragraph{Phase (i).}
In the first phase, we compute set $\Psi_{\OLD} \setminus \Psi_{\run}$ of interval attractors 
in expected $O((\max \{H, H^{\prime}, \log (nHH^{\prime}) \})^{5})$ time by the algorithm presented in Section~\ref{subsubsec:computation_lambda_ov}. 
After computing the set $\Psi_{\OLD} \setminus \Psi_{\run}$, 
we sort the interval attractors $([p_{1}, q_{1}], [\ell_{1}, r_{1}])$, $([p_{2}, q_{2}], [\ell_{2}, r_{2}])$, 
$\ldots$, $([p_{m}, q_{m}], [\ell_{m}, r_{m}])$ of the set $\Psi_{\OLD} \setminus \Psi_{\run}$ so that $T[p_{1}-1..r_{1}+1] \preceq T[p_{2}-1..r_{2}+1] \preceq \cdots \preceq T[p_{m}-1..r_{m}+1]$. 
These interval attractors can be sorted in $O(m H \log m)$ time using $O(m \log m)$ LCE and random access queries on input string $T$. 
Lemma~\ref{lem:dynamic_psi_overlap}~\ref{enum:dynamic_psi_overlap:1} shows that 
$\log m = O(\log n)$ holds, 
and the expected value of $m$ is $O(H + \log n)$. 
Therefore, the first phase takes expected $O((\max \{H, H^{\prime}, \log (nHH^{\prime}) \})^{5})$ time in total. 

\paragraph{Phase (ii).}
Consider the interval attractors 
$([p^{\prime}_{1}, q^{\prime}_{1}], [\ell^{\prime}_{1}, r^{\prime}_{1}])$, 
$([p^{\prime}_{2}, q^{\prime}_{2}], [\ell^{\prime}_{2}, r^{\prime}_{2}])$, 
$\ldots$, $([p^{\prime}_{d}, q^{\prime}_{d}], [\ell^{\prime}_{d}, r^{\prime}_{d}])$ of the sampling subset $\Psi_{\samp}$. 
Here, 
the $d$ interval attractors are sorted in lexicographic order of the $d$ strings 
$T[p^{\prime}_{1}-1..r^{\prime}_{1}+1]$, $T[p^{\prime}_{2}-1..r^{\prime}_{2}+1]$, $\ldots$, $T[p^{\prime}_{d}-1..r^{\prime}_{d}+1]$ 
(i.e., $T[p^{\prime}_{1}-1..r^{\prime}_{1}+1] \prec T[p^{\prime}_{2}-1..r^{\prime}_{2}+1] \cdots \prec T[p^{\prime}_{d}-1..r^{\prime}_{d}+1]$). 
For each interval attractor $([p_{s}, q_{s}], [\ell_{s}, r_{s}]) \in \Psi_{\OLD} \setminus \Psi_{\run}$, 
Lemma~\ref{lem:samp_basic_property}~\ref{enum:samp_basic_property:2} shows that 
there exists an integer $\xi_{s} \in [1, d]$ satisfying 
$T[p^{\prime}_{\xi_{s}}-1..r^{\prime}_{\xi_{s}}+1] = T[p_{s}-1..r_{s}+1]$. 

In the second phase, 
we compute the $\xi_{s}$-th interval attractor $([p^{\prime}_{\xi_{s}}, q^{\prime}_{\xi_{s}}]$, 
$[\ell^{\prime}_{\xi_{s}}, r^{\prime}_{\xi_{s}}])$ 
for each interval attractor $([p_{s}, q_{s}], [\ell_{s}, r_{s}]) \in \Psi_{\OLD} \setminus \Psi_{\run}$. 
We find the interval attractor $([p^{\prime}_{\xi_{s}}, q^{\prime}_{\xi_{s}}], [\ell^{\prime}_{\xi_{s}}$, $r^{\prime}_{\xi_{s}}])$ by binary search on the interval attractors $([p^{\prime}_{1}, q^{\prime}_{1}], [\ell^{\prime}_{1}, r^{\prime}_{1}])$, 
$([p^{\prime}_{2}, q^{\prime}_{2}], [\ell^{\prime}_{2}, r^{\prime}_{2}])$, 
$\ldots$, $([p^{\prime}_{d}, q^{\prime}_{d}], [\ell^{\prime}_{d}, r^{\prime}_{d}])$. 
This binary search can be executed using $O(\log d)$ sample, LCE, and random access queries. 
Here, $\log d = O(\log n)$ holds (see Section~\ref{subsubsec:computation_delta_samp}). 
The binary search takes $O(\log^{2} n + H \log n)$ time. 
Therefore, 
the running time of the second phase is $O(m (\log^{2} n + H \log n))$, 
i.e., this phase takes expected $O((H + \log n) (H + \log n)\log n)$ time. 

\paragraph{Phase (iii).}
In the third phase, 
we verify whether each interval attractor $([p_{s}, q_{s}], [\ell_{s}, r_{s}]) \in \Psi_{\OLD} \setminus \Psi_{\run}$ is contained in the sampling subset $\Psi_{\samp}$ or not. 
For this verification, 
we leverage the result of the second phase. 
If $([p_{s}, q_{s}], [\ell_{s}, r_{s}]) = ([p^{\prime}_{\xi_{s}}, q^{\prime}_{\xi_{s}}]$, 
$[\ell^{\prime}_{\xi_{s}}, r^{\prime}_{\xi_{s}}])$, 
then, $([p_{s}, q_{s}], [\ell_{s}, r_{s}]) \in \Psi_{\samp}$ holds. 
Otherwise, $([p_{s}, q_{s}], [\ell_{s}, r_{s}]) \not \in \Psi_{\samp}$ holds. 
We can verify whether 
$([p_{s}, q_{s}], [\ell_{s}, r_{s}]) = ([p^{\prime}_{\xi_{s}}, q^{\prime}_{\xi_{s}}], [\ell^{\prime}_{\xi_{s}}, r^{\prime}_{\xi_{s}}])$ or not in $O(1)$ time. 
Therefore, 
the running time of the third phase is $O(m)$, 
i.e., this phase takes expected $O(H + \log n)$ time.

\paragraph{Phase (iv).}
In the fourth phase, 
we compute the size of subset $\Psi_{\str}(T[p_{s}-1..r_{s}+1])$ for each interval attractor $([p_{s}, q_{s}], [\ell_{s}, r_{s}]) \in \Psi_{\OLD} \setminus \Psi_{\run}$. 
If $[p_{s}-1, r_{s}+1] \not \subseteq [1, n]$, 
then Lemma~\ref{lem:psi_str_occ_property} shows that $|\Psi_{\str}(T[p_{s}-1..r_{s}+1])| = 1$ holds. 
Otherwise (i.e., $[p_{s}-1, r_{s}+1] \subseteq [1, n]$), $|\Psi_{\str}(T[p_{s}-1..r_{s}+1])| = |\Occ(T, T[p_{s}-1..r_{s}+1])|$ holds. 
In this case, $|\Occ(T, T[p_{s}-1..r_{s}+1])| = |[\eta_{s}, \eta^{\prime}_{s}]|$ holds for the sa-interval $[\eta_{s}, \eta^{\prime}_{s}]$ of string $T[p_{s}-1..r_{s}+1]$. 
Therefore, we can compute the size of subset $\Psi_{\str}(T[p_{s}-1..r_{s}+1])$ by computing the sa-interval of string $T[p_{s}-1..r_{s}+1]$. 

We compute the position $\eta_{s}$ by binary search on the suffix array. 
This binary search finds the smallest position $i$ of the suffix array satisfying 
$T[\SA[i]..\SA[i] + |[p_{s}-1, r_{s}+1]| - 1] = T[p_{s}-1..r_{s}+1]$. 
Then, $\eta_{s} = i$ holds. 
This binary search can be executed using $O(\log n)$ SA, LCE, and random access queries. 
The bottleneck of this binary search is SA query, which takes $O(H^{3} \log^{2} n + H \log^{6} n)$ time. 
Therefore, we can compute the position $\eta$ in $O(H^{3} \log^{3} n + H \log^{7} n)$ time. 
Similarly, the position $\eta^{\prime}_{s}$ can be computed in the same time using the same approach. 

Finally, 
the running time of the fourth phase is $O(m (H^{3} \log^{3} n + H \log^{7} n))$, 
i.e., this phase takes expected $O((H + \log n) (H^{3} \log^{3} n + H \log^{7} n))$ time.

\paragraph{Phase (v).}
In the fifth phase, 
we compute the size of set $\Psi_{\str}(T[p_{s}-1..r_{s}+1]) \cap (\Psi_{\OLD} \setminus \Psi_{\run})$ for each interval attractor $([p_{s}, q_{s}], [\ell_{s}, r_{s}]) \in \Psi_{\OLD} \setminus \Psi_{\run}$. 
This size can be computed by counting interval attractors in set $\Psi_{\OLD} \setminus \Psi_{\run}$ 
such that each interval attractor $([p_{s^{\prime}}, q_{s^{\prime}}], [\ell_{s^{\prime}}, r_{s^{\prime}}])$ 
satisfies $T[p_{s^{\prime}}-1..r_{s^{\prime}}+1] = T[p_{s}-1..r_{s}+1]$. 
Formally, let $\kappa$ be the smallest integer in set $[1, s]$ satisfying $T[p_{x}-1..r_{x}+1] = T[p_{s}-1..r_{s}+1]$. 
Similarly, let $\kappa^{\prime}$ be the largest integer in set $[s, m]$ satisfying $T[p_{y}-1..r_{y}+1] = T[p_{s}-1..r_{s}+1]$. 
Then, $|\Psi_{\str}(T[p_{s}-1..r_{s}+1]) \cap (\Psi_{\OLD} \setminus \Psi_{\run})| = \kappa^{\prime} - \kappa + 1$ 
holds because of $T[p_{1}-1..r_{1}+1] \preceq T[p_{2}-1..r_{2}+1] \preceq \cdots \preceq T[p_{m}-1..r_{m}+1]$. 

We compute the two integers $\kappa$ and $\kappa^{\prime}$ by binary search on the $m$ interval attractors $([p_{1}, q_{1}]$, $[\ell_{1}, r_{1}])$, $([p_{2}, q_{2}], [\ell_{2}, r_{2}])$, $\ldots$, $([p_{m}, q_{m}], [\ell_{m}, r_{m}])$. 
For each interval attractor $([p_{s^{\prime}}, q_{s^{\prime}}], [\ell_{s^{\prime}}, r_{s^{\prime}}]) \in \Psi_{\OLD} \setminus \Psi_{\run}$, 
we can verify whether $T[p_{s^{\prime}}-1..r_{s^{\prime}}+1] = T[p_{s}-1..r_{s}+1]$ or not by one LCE query. 
Similarly, 
we can verify whether $T[p_{s^{\prime}}-1..r_{s^{\prime}}+1] \prec T[p_{s}-1..r_{s}+1]$ or not by one LCE query and at most two random access queries. 
Therefore, this binary search can be executed using $O(\log m)$ LCE and random access queries. 

The running time of the fifth phase is $O(m H \log m)$, 
i.e., this phase takes expected $O((H + \log n) H \log n)$ time.

\paragraph{Phase (vi).}
In the sixth phase, 
we compute the $k$ integers $\pi_{1}, \pi_{2}, \ldots, \pi_{k}$. 
We can verify whether each integer $s \in [1, m]$ is contained in set $\{ \pi_{1}, \pi_{2}, \ldots, \pi_{k} \}$ 
in $O(1)$ time using the results of the third, fourth, and fifth phases. 
Therefore, the running time of the sixth phase is $O(m)$, 
i.e., this phase takes expected $O(H + \log n)$ time. 

After the sixth phase, 
we obtain subset $\Psi^{\symA}$ because 
$([p^{\symA}_{s}, q^{\symA}_{s}], [\ell^{\symA}_{s}, r^{\symA}_{s}]) = ([p_{\pi_{s}}, q_{\pi_{s}}], [\ell_{\pi_{s}}, r_{\pi_{s}}])$ holds for each integer $s \in [1, k]$. 

\paragraph{Phase (vii).}
In the seventh phase, 
we compute the position $x^{\symA}_{\pi_{s}}$ for each integer $s \in [1, k]$. 
Let $\alpha_{s} \geq 1$ be the smallest integer satisfying 
$\SA[\eta_{\pi_{s}} + \alpha_{s} - 1] + 1 \not \in \{ p_{s^{\prime}} \mid s^{\prime} \in [1, m] \}$. 
Then, $x^{\symA}_{\pi_{s}} = \eta_{\pi_{s}} + \alpha_{s} - 1$ holds. 
Therefore, the position $x^{\symA}_{\pi_{s}}$ can be computed by verifying whether 
$\SA[\eta_{\pi_{s}} + i - 1] + 1 \not \in \{ p_{s^{\prime}} \mid s^{\prime} \in [1, m] \}$ or not 
for each integer $i \in [1, \alpha_{s}]$. 

The $(\eta_{\pi_{s}} + i - 1)$-th value of the suffix array $\SA$ can be obtained by SA query, 
which takes $O(H^{3} \log^{2} n + H \log^{6} n)$ time. 
We verify whether $\SA[\eta_{\pi_{s}} + i - 1] + 1 \not \in \{ p_{s^{\prime}} \mid s^{\prime} \in [1, m] \}$ or not by binary search on the $m$ integers $p_{1}$, $p_{2}$, $\ldots$, $p_{m}$. 
This binary search takes $O(\log n)$ time because of $m = O(n^{2})$. 
Therefore, we can compute the position $x^{\symA}_{\pi_{s}}$ in $O(\alpha_{s} H^{3} \log^{2} n + H \log^{6} n)$ time. 

The running time of the seventh phase is $O((\Sigma_{s = 1}^{k} \alpha_{s}) (H^{3} \log^{2} n + H \log^{6} n))$ in total. 
We prove $\Sigma_{s = 1}^{k} \alpha_{s} \leq 2m$. 
$\alpha_{s} = 1 + x^{\symA}_{\pi_{s}} - \eta_{\pi_{s}}$ follows from $x^{\symA}_{\pi_{s}} = \eta_{\pi_{s}} + \alpha_{s} - 1$. 
Lemma~\ref{lem:dynamic_samp_substitute}~\ref{enum:dynamic_samp_substitute:1} shows that 
$x^{\symA}_{\pi_{s}} - \eta_{\pi_{s}} \leq |(\Psi_{\OLD} \setminus \Psi_{\run}) \cap \Psi_{\str}(Z^{\symA}_{s})|$. 
$\Sigma_{s = 1}^{k} x^{\symA}_{\pi_{s}} - \eta_{\pi_{s}} \leq m$ holds 
because (A) 
each set $(\Psi_{\OLD} \setminus \Psi_{\run}) \cap \Psi_{\str}(Z^{\symA}_{s})$ is a subset of set $\{ ([p_{s^{\prime}}, q_{s^{\prime}}], [\ell_{s^{\prime}}, r_{s^{\prime}}]) \in s^{\prime} \in [1, m]\}$, 
and (B) $\Psi_{\str}(Z^{\symA}_{s}) \cap \Psi_{\str}(Z_{s^{\prime}}) = \emptyset$ holds 
for any pair of two strings $Z^{\symA}_{s}, Z_{s^{\prime}} \in \Psi^{\symA^{\prime}}$. 
Therefore, $\Sigma_{s = 1}^{k} \alpha_{s} \leq 2m$ follows from the following equation: 
\begin{equation*}
    \begin{split}
        \Sigma_{s = 1}^{k} \alpha_{s} &= k + (\Sigma_{s = 1}^{k} x^{\symA}_{\pi_{s}} - \eta_{\pi_{s}}) \\
        &\leq m + m \\
        &= 2m.
    \end{split}    
\end{equation*}

Finally, the running time of the seventh phase can be bounded by $O(m (H^{3} \log^{2} n + H \log^{6} n))$, 
i.e., this phase takes expected $O((H + \log n) (H^{3} \log^{2} n + H \log^{6} n))$ time.

\paragraph{Phase (viii).}
In the eighth phase, 
we obtain interval attractor $([p^{\symA^{\prime}}_{s}, q^{\symA^{\prime}}_{s}], [\ell^{\symA^{\prime}}_{s}, r^{\symA^{\prime}}_{s}])$ by 
capture query $\CAPQ([\SA[x^{\symA}_{\pi_{s}}]+1, \SA[x^{\symA}_{\pi_{s}}] + |Z^{\symA}_{s}| - 2])$ 
for each integer $s \in [1, k]$. 
This capture query takes $O(H^{2} \log n)$ time. 
Because of $k \leq m$, 
the running time of the eighth phase is $O(m H^{2} \log n)$ time,  
i.e., this phase takes expected $O((H + \log n) H^{2} \log n)$ time.

Finally, 
the eight phases take expected $O((\max \{H, H^{\prime}, \log (nHH^{\prime}) \})^{8})$ time in total.

%%%%%%%%%%%%%%%%%%%%%%%%%%%%%%%%%%%%%%%%%%%%%%%%%%%%%%
\subsubsection{Computation of Subset \texorpdfstring{$\Psi^{\prime \symB}$}{}}\label{subsubsec:computation_samp_C}
We present the algorithm computing the subset $\Psi^{\prime \symB} = \{ ([p^{\symB}_{1}, q^{\symB}_{1}], [\ell^{\symB}_{1}, r^{\symB}_{1}])$, $([p^{\symB}_{2}, q^{\symB}_{2}]$, $[\ell^{\symB}_{2}, r^{\symB}_{2}])$, 
$\ldots$, $([p^{\symB}_{k^{\prime}}, q^{\symB}_{k^{\prime}}], [\ell^{\symB}_{k^{\prime}}, r^{\symB}_{k^{\prime}}]) \}$ introduced in Section~\ref{subsec:update_sampling_subset}. 
This algorithm requires the dynamic data structures introduced in Section~\ref{sec:recompression}, Section~\ref{sec:other_query}, and Section~\ref{sec:RSC_query} 
for RLSLP $\mathcal{G}^{R}$. 

For computing the subset $\Psi^{\prime \symB}$, 
we leverage Lemma~\ref{lem:finding_samp_B_str} and Lemma~\ref{lem:dynamic_samp_new}. 
Consider the interval attractors $([p^{\prime}_{1}, q^{\prime}_{1}], [\ell^{\prime}_{1}, r^{\prime}_{1}])$, $([p^{\prime}_{2}, q^{\prime}_{2}], [\ell^{\prime}_{2}, r^{\prime}_{2}])$, $\ldots$, $([p^{\prime}_{m^{\prime}}, q^{\prime}_{m^{\prime}}], [\ell^{\prime}_{m^{\prime}}, r^{\prime}_{m^{\prime}}])$ of set $\Psi^{\prime}_{\NEW} \setminus \Psi^{\prime}_{\run}$. 
Here, the $m^{\prime}$ interval attractors are sorted so that $T^{\prime}[p^{\prime}_{1}-1..r^{\prime}_{1}+1] \preceq T^{\prime}[p^{\prime}_{2}-1..r^{\prime}_{2}+1] \preceq \cdots \preceq T^{\prime}[p^{\prime}_{m^{\prime}}-1..r^{\prime}_{m^{\prime}}+1]$. 
If the set $\Psi^{\prime}_{\NEW} \setminus \Psi^{\prime}_{\run}$ contains two distinct interval attractors 
$([p^{\prime}_{s}, q^{\prime}_{s}], [\ell^{\prime}_{s}, r^{\prime}_{s}])$ and $([p^{\prime}_{s^{\prime}}, q^{\prime}_{s^{\prime}}], [\ell^{\prime}_{s^{\prime}}, r^{\prime}_{s^{\prime}}])$ 
satisfying $T^{\prime}[p^{\prime}_{s}-1..r^{\prime}_{s}+1] = T^{\prime}[p^{\prime}_{s^{\prime}}-1..r^{\prime}_{s^{\prime}}+1]$, 
then the interval attractor $([p^{\prime}_{s}, q^{\prime}_{s}], [\ell^{\prime}_{s}, r^{\prime}_{s}])$ precedes the interval attractor $([p^{\prime}_{s^{\prime}}, q^{\prime}_{s^{\prime}}], [\ell^{\prime}_{s^{\prime}}, r^{\prime}_{s^{\prime}}])$ if and only if $p^{\prime}_{s} < p^{\prime}_{s^{\prime}}$. 

Let $\{ \pi_{1}, \pi_{2}, \ldots, \pi_{w}\}$ be a subset of set $\{ 1, 2, \ldots, m \}$ such that 
each integer $\pi_{s}$ satisfies the following two conditions: 
\begin{itemize}
    \item $|\Psi_{\str}(T^{\prime}[p^{\prime}_{\pi_{s}}-1..r^{\prime}_{\pi_{s}}+1])| = |\Psi_{\str}(T^{\prime}[p^{\prime}_{\pi_{s}}-1..r^{\prime}_{\pi_{s}}+1]) \cap (\Psi_{\OLD} \setminus \Psi_{\run})|$;
    \item either (A) $\pi_{s} = 1$ or (B) $\pi_{s} \geq 2$ and $T^{\prime}[p^{\prime}_{\pi_{s}-1}-1..r^{\prime}_{\pi_{s}-1}+1] \prec T^{\prime}[p^{\prime}_{\pi_{s}}-1..r^{\prime}_{\pi_{s}}+1]$. 
\end{itemize}
Then, Lemma~\ref{lem:finding_samp_B_str} indicates that 
$w = k^{\prime}$, 
$Z^{\prime}_{1} = T^{\prime}[p^{\prime}_{\pi_{1}}-1..r^{\prime}_{\pi_{1}}+1]$, 
$Z^{\prime}_{2} = T^{\prime}[p^{\prime}_{\pi_{1}}-1..r^{\prime}_{\pi_{1}}+1]$, 
$\ldots$, 
$Z^{\prime}_{k^{\prime}} = T[p^{\prime}_{\pi_{k^{\prime}}}-1..r^{\prime}_{\pi_{k^{\prime}}}+1]$ hold 
for set $\mathcal{Z}^{\symB} = \{ Z^{\prime}_{1}, Z^{\prime}_{2}, \ldots, Z^{\prime}_{k^{\prime}} \}$ of strings. 
For each integer $s \in [1, k^{\prime}]$, 
Lemma~\ref{lem:dynamic_samp_new} shows that 
$([p^{\symB}_{s}, q^{\symB}_{s}], [\ell^{\symB}_{s}, r^{\symB}_{s}]) = ([p^{\prime}_{\pi_{s}}, q^{\prime}_{\pi_{s}}], [\ell^{\prime}_{\pi_{s}}, r^{\prime}_{\pi_{s}}])$ holds. 
Therefore, 
we can compute the subset $\Psi^{\prime \symB}$ by processing the interval attractors of the set $\Psi^{\prime}_{\NEW} \setminus \Psi^{\prime}_{\run}$. 

The algorithm computing the subset $\Psi^{\prime \symB}$ consists of the following eight phases. 
\paragraph{Phase (i).}
In the first phase, we compute set $\Psi^{\prime}_{\NEW} \setminus \Psi^{\prime}_{\run}$ of interval attractors 
in $O((\max \{H, H^{\prime}$, $\log (nHH^{\prime}) \})^{5})$ time by the algorithm presented in Section~\ref{subsubsec:computation_lambda_ov}. 
After computing the set $\Psi^{\prime}_{\NEW} \setminus \Psi^{\prime}_{\run}$, 
we obtain the dynamic data structures for the RR-DAG of RLSLP $\mathcal{G}^{R}_{\ins}$ 
by modifying the dynamic data structures for the RR-DAG of RLSLP $\mathcal{G}^{R}$. 
This update takes $O((\max \{H, H^{\prime}, \log (nHH^{\prime}) \})^{3})$ time (Lemma~\ref{lem:dynamic_rrdag_summary}). 

\paragraph{Phase (ii).}
In the second phase, 
we sort the interval attractors $([p^{\prime}_{1}, q^{\prime}_{1}], [\ell^{\prime}_{1}, r^{\prime}_{1}])$, $([p^{\prime}_{2}, q^{\prime}_{2}], [\ell^{\prime}_{2}, r^{\prime}_{2}])$, 
$\ldots$, $([p^{\prime}_{m^{\prime}}, q^{\prime}_{m^{\prime}}], [\ell^{\prime}_{m^{\prime}}, r^{\prime}_{m^{\prime}}])$ of the set $\Psi^{\prime}_{\NEW} \setminus \Psi^{\prime}_{\run}$ so that $T^{\prime}[p^{\prime}_{1}-1..r^{\prime}_{1}+1] \preceq T^{\prime}[p^{\prime}_{2}-1..r^{\prime}_{2}+1] \preceq \cdots \preceq T^{\prime}[p^{\prime}_{m^{\prime}}-1..r^{\prime}_{m^{\prime}}+1]$. 
These interval attractors can be sorted in $O(m^{\prime} H^{\prime} \log m^{\prime})$ time using $O(m^{\prime} \log m^{\prime})$ LCE and random access queries on input string $T^{\prime}$. 

Lemma~\ref{lem:dynamic_psi_overlap}~\ref{enum:dynamic_samp_substitute:2} shows that 
$\log m^{\prime} = O(\log n)$ holds, 
and the expected value of $m^{\prime}$ is $O(H^{\prime} + \log n)$. 
Therefore, the second phase takes expected $O((H^{\prime} + \log n) H^{\prime 2} \log n)$ in total. 

\paragraph{Phase (iii).}
In the third phase, 
we obtain the dynamic data structures for the RR-DAG of RLSLP $\mathcal{G}^{R}$ 
by modifying the dynamic data structures for the RR-DAG of RLSLP $\mathcal{G}^{R}_{\ins}$. 
This update takes $O((\max \{H, H^{\prime}, \log (nHH^{\prime}) \})^{3})$ time (Lemma~\ref{lem:dynamic_rrdag_summary}). 

\paragraph{Phase (iv).}
In the fourth phase, we compute 
the interval attractors $([p_{1}, q_{1}], [\ell_{1}, r_{1}])$, $([p_{2}, q_{2}]$, $[\ell_{2}, r_{2}])$, 
$\ldots$, $([p_{m}, q_{m}], [\ell_{m}, r_{m}])$ of set $\Psi_{\OLD} \setminus \Psi_{\run}$. 
We obtain these interval attractors using the first phase of the algorithm presented 
in Section~\ref{subsubsec:computation_samp_AB}. 
Therefore, 
the fourth phase takes expected $O((H + \log n) H^{3} \log n)$ time. 

\paragraph{Phase (v).}
For each interval attractor $([p^{\prime}_{s}, q^{\prime}_{s}], [\ell^{\prime}_{s}, r^{\prime}_{s}]) \in \Psi^{\prime}_{\NEW} \setminus \Psi^{\prime}_{\run}$, 
let $([\hat{p}_{\xi_{s}}, \hat{q}_{\xi_{s}}], [\hat{\ell}_{\xi_{s}}, \hat{r}_{\xi_{s}}])$ be an interval attractor in 
the sampling subset $\Psi_{\samp}$ satisfying $T[\hat{p}_{\xi_{s}}-1..\hat{r}_{\xi_{s}}+1] = T^{\prime}[p^{\prime}_{s}-1..r^{\prime}_{s}+1]$. 
Lemma~\ref{lem:samp_basic_property}~\ref{enum:samp_basic_property:2} indicates that 
the interval attractor $([\hat{p}_{\xi_{s}}, \hat{q}_{\xi_{s}}], [\hat{\ell}_{\xi_{s}}, \hat{r}_{\xi_{s}}])$ exists if 
$\Psi_{\str}(T^{\prime}[p^{\prime}_{s}-1..r^{\prime}_{s}+1]) \neq \emptyset$ holds; 
otherwise, 
the interval attractor $([\hat{p}_{\xi_{s}}, \hat{q}_{\xi_{s}}], [\hat{\ell}_{\xi_{s}}, \hat{r}_{\xi_{s}}])$ does not exist. 

In the fifth phase, we find the interval attractor $([\hat{p}_{\xi_{s}}, \hat{q}_{\xi_{s}}], [\hat{\ell}_{\xi_{s}}, \hat{r}_{\xi_{s}}])$ 
for each interval attractor $([p^{\prime}_{s}, q^{\prime}_{s}], [\ell^{\prime}_{s}, r^{\prime}_{s}]) \in \Psi^{\prime}_{\NEW} \setminus \Psi^{\prime}_{\run}$. 
The interval attractor $([\hat{p}_{\xi_{s}}, \hat{q}_{\xi_{s}}], [\hat{\ell}_{\xi_{s}}, \hat{r}_{\xi_{s}}])$ can be found using the same approach as for the second phase of the algorithm presented in Section~\ref{subsubsec:computation_samp_AB}. 
Therefore, 
the running time of the fifth phase is $O(m^{\prime} (\log^{2} n + H^{2} \log n))$, 
i.e., this phase takes expected $O((H^{\prime} + \log n) (H^{2} + \log n)\log n)$ time.

\paragraph{Phase (vi).}
In the sixth phase, 
we compute the size of subset $\Psi_{\str}(T^{\prime}[p^{\prime}_{s}-1..r^{\prime}_{s}+1])$ 
for each interval attractor $([p^{\prime}_{s}, q^{\prime}_{s}], [\ell^{\prime}_{s}, r^{\prime}_{s}]) \in \Psi^{\prime}_{\NEW} \setminus \Psi^{\prime}_{\run}$. 
The interval attractor $([p^{\prime}_{s}, q^{\prime}_{s}], [\ell^{\prime}_{s}, r^{\prime}_{s}])$ satisfies one of the following three conditions: 
\begin{enumerate}[label=\textbf{(\Alph*)}]
    \item interval attractor $([\hat{p}_{\xi_{s}}, \hat{q}_{\xi_{s}}], [\hat{\ell}_{\xi_{s}}, \hat{r}_{\xi_{s}}])$ does not exist;
    \item interval attractor $([\hat{p}_{\xi_{s}}, \hat{q}_{\xi_{s}}], [\hat{\ell}_{\xi_{s}}, \hat{r}_{\xi_{s}}])$ exists, 
    and $[\hat{p}_{\xi_{s}}-1, \hat{r}_{\xi_{s}}+1] \not \subseteq [1, n]$;
    \item interval attractor $([\hat{p}_{\xi_{s}}, \hat{q}_{\xi_{s}}], [\hat{\ell}_{\xi_{s}}, \hat{r}_{\xi_{s}}])$ exists, 
    and $[\hat{p}_{\xi_{s}}-1, \hat{r}_{\xi_{s}}+1] \subseteq [1, n]$.    
\end{enumerate}

For condition (A), 
$\Psi_{\str}(T^{\prime}[p^{\prime}_{s}-1..r^{\prime}_{s}+1]) = \emptyset$ holds (see Phase (v)).
For condition (B), 
$\Psi_{\str}(T^{\prime}[p^{\prime}_{s}-1..r^{\prime}_{s}+1]) = \Psi_{\str}(T[\hat{p}_{\xi_{s}}-1..\hat{r}_{\xi_{s}}+1])$ holds because of $T[\hat{p}_{\xi_{s}}-1..\hat{r}_{\xi_{s}}+1] = T^{\prime}[p^{\prime}_{s}-1..r^{\prime}_{s}+1]$. 
Lemma~\ref{lem:psi_str_occ_property} shows that $|\Psi_{\str}(T[\hat{p}_{\xi_{s}}-1..\hat{r}_{\xi_{s}}+1])| = 1$ holds. 
Therefore, $|\Psi_{\str}(T^{\prime}[p^{\prime}_{s}-1..r^{\prime}_{s}+1])| = 1$ holds. 

For condition (C), 
$\Psi_{\str}(T^{\prime}[p^{\prime}_{s}-1..r^{\prime}_{s}+1]) = \Psi_{\str}(T[\hat{p}_{\xi_{s}}-1..\hat{r}_{\xi_{s}}+1])$ holds. 
Lemma~\ref{lem:psi_str_occ_property} shows that $|\Psi_{\str}(T[\hat{p}_{\xi_{s}}-1..\hat{r}_{\xi_{s}}+1])| = |\Occ(T, T[\hat{p}_{\xi_{s}}-1..\hat{r}_{\xi_{s}}+1])|$ holds. 
In this case, $|\Occ(T, T[\hat{p}_{\xi_{s}}-1..\hat{r}_{\xi_{s}}+1])| = |[\eta_{s}, \eta^{\prime}_{s}]|$ holds for the sa-interval $[\eta_{s}, \eta^{\prime}_{s}]$ of string $T[\hat{p}_{\xi_{s}}-1..\hat{r}_{\xi_{s}}+1]$. 
This sa-interval can be computed using the same approach for the fourth phase of the algorithm 
presented in Section~\ref{subsubsec:computation_samp_AB}. 
Therefore, the sixth phase can be executed in $O(m^{\prime} (H^{3} \log^{3} n + H \log^{7} n))$ time, 
i.e., this phase takes expected $O((H^{\prime} + \log n) (H^{3} \log^{3} n + H \log^{7} n))$ time.

\paragraph{Phase (vii).}
In the seventh phase, 
we compute the size of set $\Psi_{\str}(T^{\prime}[p^{\prime}_{s}-1..r^{\prime}_{s}+1]) \cap (\Psi_{\OLD} \setminus \Psi_{\run})$ for each interval attractor $([p^{\prime}_{s}, q^{\prime}_{s}], [\ell^{\prime}_{s}, r^{\prime}_{s}]) \in \Psi^{\prime}_{\NEW} \setminus \Psi^{\prime}_{\run}$. 
This computation can be executed using the same approach as for 
the fifth phase of the algorithm presented in Section~\ref{subsubsec:computation_samp_AB}. 
Therefore, 
the running time of the seventh phase is $O(m^{\prime} H^{2} \log m^{\prime})$, 
i.e., this phase takes expected $O((H^{\prime} + \log n) H^{2} \log n)$ time.

\paragraph{Phase (viii).}
In the eighth phase, 
we compute the $k$ integers $\pi_{1}, \pi_{2}, \ldots, \pi_{k}$. 
We can verify whether each integer $s \in [1, m^{\prime}]$ is contained in set $\{ \pi_{1}, \pi_{2}, \ldots, \pi_{k} \}$ 
in $O(1)$ time using the results of the sixth and seventh phases. 
Therefore, the running time of the eighth phase is $O(m^{\prime})$, 
i.e., this phase takes expected $O(H^{\prime} + \log n)$ time. 

Finally, we return set $\{ ([p^{\prime}_{\pi_{s}}, q^{\prime}_{\pi_{s}}], [\ell^{\prime}_{\pi_{s}}, r^{\prime}_{\pi_{s}}]) \mid s \in [1, k^{\prime}] \}$ as the subset $\Psi^{\prime \symB}$. 
The eight phases take expected $O((\max \{H, H^{\prime}, \log (nHH^{\prime}) \})^{8})$ time in total.

%%%%%%%%%%%%%%%%%%%%%%%%%%%%%%%%%%%%%%%%%%%%%%%%%%%%%%

\subsubsection{Computation of Two Sets \texorpdfstring{$\Psi_{\samp} \cap \Psi_{\OLD}$}{} and \texorpdfstring{$(\Psi_{\samp} \cap \Psi_{\OLD}) \setminus \Psi^{\symA}$}{}}\label{subsubsec:computation_samp_cap_old}
We present the algorithm computing two set $\Psi_{\samp} \cap \Psi_{\OLD}$ 
and $(\Psi_{\samp} \cap \Psi_{\OLD}) \setminus \Psi^{\symA}$. 
This algorithm requires the dynamic data structures introduced in Section~\ref{sec:recompression}, Section~\ref{sec:other_query}, and Section~\ref{sec:RSC_query} for RLSLP $\mathcal{G}^{R}$. 

We leverage Lemma~\ref{lem:dynamic_psi_overlap} for computing the set $\Psi_{\samp} \cap \Psi_{\OLD}$. 
Lemma~\ref{lem:dynamic_psi_overlap}~\ref{enum:dynamic_psi_overlap:3} shows that 
$\Psi_{\samp} \cap \Psi_{\OLD} = \Psi_{\samp} \cap (\Psi_{\OLD} \setminus \Psi_{\run})$ holds. 
Therefore, we can obtain the set $\Psi_{\samp} \cap \Psi_{\OLD}$ by verifying whether 
each interval attractor of set $\Psi_{\OLD} \setminus \Psi_{\run}$ is contained in the sampling subset $\Psi_{\samp}$. 

We verify whether each interval attractor of set $\Psi_{\OLD} \setminus \Psi_{\run}$ is contained in the sampling subset $\Psi_{\samp}$ using the first, second, third phases of the algorithm presented in Section~\ref{subsubsec:computation_samp_AB}. 
The three phases can be executed in expected $O((H + \log n) (H^{3} + \log n)\log n)$ time using the dynamic data structures for the RR-DAG of RLSLP $\mathcal{G}^{R}$ and sample query. 
Therefore, set $\Psi_{\OLD} \setminus \Psi_{\run}$ can be computed in expected $O((H + \log n) (H^{3} + \log n)\log n)$ time. 

Next, we compute $(\Psi_{\samp} \cap \Psi_{\OLD}) \setminus \Psi^{\symA}$ by removing 
the interval attractors of set $\Psi^{\symA}$ from set $\Psi_{\samp} \cap \Psi_{\OLD}$. 
The set $\Psi^{\symA}$ can be obtained in expected $O((\max \{H, H^{\prime}, \log (nHH^{\prime}) \})^{8})$ time by the algorithm presented in Section~\ref{subsubsec:computation_samp_AB}. 
Therefore, the set $(\Psi_{\samp} \cap \Psi_{\OLD}) \setminus \Psi^{\symA}$ can be computed in expected $O((\max \{H, H^{\prime}, \log (nHH^{\prime}) \})^{8})$ time. 

\subsection{Update of Data Structures for Sample Query}\label{subsec:update_sample_query}
This subsection explains how to update the dynamic data structures for sample query introduced in Section~\ref{subsubsec:sample_ds}. 
The following  lemma is the summary of this subsection.

\begin{lemma}\label{lem:dynamic_samp_summary}
Consider the two RLSLPs $\mathcal{G}^{R}$ and $\mathcal{G}^{R}_{\ins}$ of Theorem~\ref{theo:update1}, which derive input string $T$ and string $T^{\prime}$, respectively. 
The dynamic data structures of Section~\ref{subsubsec:sample_ds} can be updated 
in expected $O((\max \{H, H^{\prime}, \log (nHH^{\prime}) \})^{4})$ time 
after changing RLSLP $\mathcal{G}^{R}$ to $\mathcal{G}^{R}_{\ins}$. 
This update requires the dynamic data structures for 
the RR-DAG of RLSLP $\mathcal{G}^{R}$ (Section~\ref{subsubsec:rrdag_ds}) 
and the interval attractors obtained from Lemma~\ref{lem:dynamic_IA_summary}. 
%Section~\ref{sec:recompression}, Section~\ref{sec:other_query}, and Section~\ref{sec:RSC_query}. 
\end{lemma}
\begin{proof}
    See Section~\ref{subsubsec:dynamic_sample_query_algo}. 
\end{proof}

%\subsubsection{Computation of Two Sets \texorpdfstring{$\Psi_{\OLD} \setminus \Psi_{\run}$}{} and \texorpdfstring{$(\Psi_{\OLD} \setminus \Psi_{\run}) \setminus \Psi_{\samp}$}{}}\label{subsubsec:computation_old_new}

\subsubsection{Algorithm Updating Dynamic Data Structures for Sample Query}\label{subsubsec:dynamic_sample_query_algo}
We prove Lemma~\ref{lem:dynamic_samp_summary}, i.e., 
we show that the data structures for sample query (Section~\ref{subsubsec:sample_ds}) 
can be updated in expected $O((\max \{H, H^{\prime}, \log (nHH^{\prime}) \})^{4})$ time using the RR-DAG of RLSLP $\mathcal{G}^{R}$ (Section~\ref{subsubsec:rrdag_ds}) 
and the interval attractors obtained from Lemma~\ref{lem:dynamic_IA_summary}. 

%the dynamic data structures introduced in 
%Section~\ref{sec:recompression}, Section~\ref{sec:other_query}, and Section~\ref{sec:RSC_query}. 

For quickly updating the data structures for sample query, 
we leverage the relationship between two sampling subset $\Psi_{\samp}$ and $\Psi^{\prime}_{\samp}$ (See the definition of the sampling subset $\Psi^{\prime}_{\samp}$). 
Consider the following interval attractors:
\begin{itemize}
    \item the interval attractors 
    $([p_{1}, q_{1}], [\ell_{1}, r_{1}])$, 
    $([p_{2}, q_{2}], [\ell_{2}, r_{2}])$, 
    $\ldots$, $([p_{d}, q_{d}], [\ell_{d}, r_{d}])$ of the sampling subset $\Psi_{\samp}$. 
    Here, 
    the $d$ interval attractors are sorted in lexicographic order of the $d$ strings 
    $T[p_{1}-1..r_{1}+1]$, $T[p_{2}-1..r_{2}+1]$, $\ldots$, $T[p_{d}-1..r_{d}+1]$ 
    (i.e., $T[p_{1}-1..r_{1}+1] \prec T[p_{2}-1..r_{2}+1] \cdots \prec T[p_{d}-1..r_{d}+1]$); 
    \item the interval attractors 
    $([p^{\symA}_{1}, q^{\symA}_{1}], [\ell^{\symA}_{1}, r^{\symA}_{1}])$, $([p^{\symA}_{2}, q^{\symA}_{2}]$, $[\ell^{\symA}_{2}, r^{\symA}_{2}])$, $\ldots$, $([p^{\symA}_{k}, q^{\symA}_{k}], [\ell^{\symA}_{k}, r^{\symA}_{k}])$ of the subset $\Psi^{\symA}$ introduced in Section~\ref{subsec:update_sampling_subset}; 
    \item the interval attractors 
    $([p^{\symA^{\prime}}_{1}, q^{\symA^{\prime}}_{1}], [\ell^{\symA^{\prime}}_{1}, r^{\symA^{\prime}}_{1}])$, $([p^{\symA^{\prime}}_{2}, q^{\symA^{\prime}}_{2}]$, $[\ell^{\symA^{\prime}}_{2}, r^{\symA^{\prime}}_{2}])$, $\ldots$, $([p^{\symA^{\prime}}_{k}, q^{\symA^{\prime}}_{k}], [\ell^{\symA^{\prime}}_{k}, r^{\symA^{\prime}}_{k}])$ of the subset $\Psi^{\symA^{\prime}}$ introduced in Section~\ref{subsec:update_sampling_subset};
    \item the interval attractors 
    $([p^{\symB}_{1}, q^{\symB}_{1}], [\ell^{\symB}_{1}, r^{\symB}_{1}])$, $([p^{\symB}_{2}, q^{\symB}_{2}]$, $[\ell^{\symB}_{2}, r^{\symB}_{2}])$, 
    $\ldots$, $([p^{\symB}_{k^{\prime}}, q^{\symB}_{k^{\prime}}], [\ell^{\symB}_{k^{\prime}}, r^{\symB}_{k^{\prime}}])$ of the subset $\Psi^{\prime \symB}$ introduced in Section~\ref{subsec:update_sampling_subset}.     
\end{itemize}

From the definition of the sampling subset $\Psi^{\prime}_{\samp}$, 
this sampling subset can be obtained by modifying the sampling subset $\Psi_{\samp}$ as follows: 
\begin{description}
    \item [Modification 1:] each interval attractor $([p_{s}, q_{s}], [\ell_{s}, r_{s}])$ of set $\Psi_{\samp} \cap (\Psi_{\LEFT} \cup \Psi_{\RIGHT})$ is 
    replaced with interval attractor $([p_{s}+\epsilon_{s}, q_{s}+\epsilon_{s}], [\ell_{s}+\epsilon_{s}, r_{s}+\epsilon_{s}])$. 
    Here, let $\epsilon_{s} = 1$ if $([p_{s}, q_{s}], [\ell_{s}, r_{s}]) \in \Psi_{\RIGHT}$. 
    Otherwise, let $\epsilon_{s} = 0$;
    \item [Modification 2:]  
    subset $\Psi_{\samp}$ contains each interval attractor $([p^{\symA}_{s}, q^{\symA}_{s}], [\ell^{\symA}_{s}, r^{\symA}_{s}])$ of subset $\Psi^{\symA}$, 
    and this interval attractor is replaced with $([p^{\symA^{\prime}}_{s} + \epsilon^{\symA^{\prime}}_{s}, q^{\symA^{\prime}}_{s} + \epsilon^{\symA^{\prime}}_{s}], [\ell^{\symA^{\prime}}_{s} + \epsilon^{\symA^{\prime}}_{s}, r^{\symA^{\prime}}_{s} + \epsilon^{\symA^{\prime}}_{s}])$. 
    Here, $\epsilon^{\symA^{\prime}}_{s}$ is the integer introduced in Section~\ref{subsec:update_sampling_subset}; 
    \item [Modification 3:] each interval attractor of set $(\Psi_{\samp} \cap \Psi_{\OLD}) \setminus \Psi^{\symA}$ is removed from subset $\Psi_{\samp}$; 
    \item [Modification 4:] each interval attractor of subset $\Psi^{\prime \symB}$ is added to subset $\Psi_{\samp}$.
\end{description}

The dynamic data structures for sample query consists of the three doubly linked lists 
$\mathbf{L}_{\samp, 1}$, $\mathbf{L}_{\samp, 2}$, and $\mathbf{L}_{\samp, 3}$ introduced in Section~\ref{subsubsec:sample_ds}. 
Each element of these three doubly linked lists corresponds to an interval attractor of the sampling subset $\Psi_{\samp}$, 
and we need to appropriately update the three doubly linked lists based on the modification of the sampling subset $\Psi_{\samp}$. 
In the three doubly linked lists, 
the number of updated elements can be bounded by $O(|\Psi_{\OLD} \setminus \Psi_{\run}| + |\Psi^{\prime}_{\NEW} \setminus \Psi^{\prime}_{\run}|)$. 
Here, the expected value of $|\Psi_{\OLD} \setminus \Psi_{\run}|$ is $O(H + \log n)$ (see Lemma~\ref{lem:dynamic_psi_overlap}~\ref{enum:dynamic_psi_overlap:1}). 
Similarly, the expected value of $|\Psi^{\prime}_{\NEW} \setminus \Psi^{\prime}_{\run}|$ is $O(H^{\prime} + \log n)$ (see Lemma~\ref{lem:dynamic_psi_overlap}~\ref{enum:dynamic_psi_overlap:2}). 
Therefore, we can update the dynamic data structures for sample query in expected $O(\polylog(nHH^{\prime}))$ time. 

The algorithm updating the dynamic data structures for sample query consists of the following four phases.
%\paragraph{Phase (i).}
%In the first phase, 
%we compute set $(\Psi_{\samp} \cap \Psi_{\OLD}) \setminus \Psi^{\symA}$ by removing 
%the interval attractors of the subset $\Psi^{\symA}$ from the set $\Psi_{\samp} \cap \Psi_{\OLD}$. 
%Here, the two sets $\Psi_{\samp} \cap \Psi_{\OLD}$ and $\Psi^{\symA}$ 
%are obtained from Lemma~\ref{lem:dynamic_IA_summary}. 
%Let $\kappa = |\Psi_{\samp} \cap \Psi_{\OLD}|$. 
%Then, the first phase can be executed in $O((\kappa + k) \log (\kappa + k))$ time. 
%Lemma~\ref{lem:dynamic_psi_overlap} shows that $\log \kappa = O(\log n)$ holds, 
%and the expected value of $\kappa$ is $O(H + \log n)$.

%The subset $\Psi^{\symA^{\prime}}$ can be obtained in 
%expected $O((H + \log n) (H^{3} \log^{3} n + H \log^{7} n))$ time by the algorithm presented in 
%Section~\ref{subsubsec:computation_samp_AB}. 
%The set $((\Psi_{\OLD} \setminus \Psi_{\run}) \cap \Psi_{\samp}) \setminus \Psi^{\symA^{\prime}}$ can be obtained in 
%the same time by modifying the algorithm presented in Section~\ref{subsubsec:computation_samp_AB}. 
%The subset $\Psi^{\prime \symB}$ can be obtained in 
%expected $O((H^{\prime} + \log n) (H^{\prime 3} \log n + H^{3} \log^{3} n + H \log^{7} n))$ time 
%by the algorithm presented in Section~\ref{subsubsec:computation_samp_C}. 
%Therefore, 
%the first phase takes expected $O((H + H^{\prime} + \log n)(H^{\prime 3} \log n + H^{3} \log^{3} n + H \log^{7} n))$ time. 

\paragraph{Phase (i).}
In the first phase, we update three doubly linked lists $\mathbf{L}_{\samp, 1}$, $\mathbf{L}_{\samp, 2}$, $\mathbf{L}_{\samp, 3}$ based on Modification 1.  
First, we update the doubly linked list $\mathbf{L}_{\samp, 1}$ representing 
the sequence $\mathbf{Q}_{\samp} = v_{1}, v_{2}, \ldots, v_{d}$ of nodes introduced in Section~\ref{subsec:sample_query}. 
Here, each node $v_{s}$ corresponds to the $\phi_{s}$-th interval attractor $([p_{\phi_{s}}, q_{\phi_{s}}], [\ell_{\phi_{s}}, r_{\phi_{s}}])$ for the permutation $\phi_{1}, \phi_{2}, \ldots, \phi_{d}$ introduced in Section~\ref{subsec:sample_query}; 
the $s$-th element of the doubly linked list $\mathbf{L}_{\samp, 1}$ stores triplet $(q_{\phi_{s}} - p_{\phi_{s}}, \ell_{\phi_{s}} - p_{\phi_{s}}, r_{\phi_{s}} - p_{\phi_{s}})$. 

For each interval attractor $([p_{s}, q_{s}], [\ell_{s}, r_{s}]) \in \Psi_{\samp} \cap (\Psi_{\LEFT} \cup \Psi_{\RIGHT})$, 
sequence $\mathbf{Q}_{\samp}$ contains a node $v_{s^{\prime}}$ corresponding to the interval attractor 
(i.e., $\phi_{s^{\prime}} = s$).
After replacing the interval attractor $([p_{s}, q_{s}], [\ell_{s}, r_{s}])$ with interval attractor $([p_{s}+\epsilon_{s}, q_{s}+\epsilon_{s}], [\ell_{s}+\epsilon_{s}, r_{s}+\epsilon_{s}])$, 
the $s^{\prime}$-th node $v_{s^{\prime}}$ corresponds to the interval attractor $([p_{s}+\epsilon_{s}, q_{s}+\epsilon_{s}], [\ell_{s}+\epsilon_{s}, r_{s}+\epsilon_{s}])$, 
and the $s^{\prime}$-th element of the doubly linked list $\mathbf{L}_{\samp, 1}$ stores triplet $((q_{s} +\epsilon_{s}) - (p_{s} +\epsilon_{s}), (\ell_{s} +\epsilon_{s}) - (p_{s} +\epsilon_{s}), (r_{s} +\epsilon_{s}) - (p_{s} +\epsilon_{s}))$. 
This replacement does not change the doubly linked list $\mathbf{L}_{\samp, 1}$ 
because $q_{s} - p_{s} = (q_{s} +\epsilon_{s}) - (p_{s} +\epsilon_{s})$, 
$\ell_{s} - p_{s} = (\ell_{s} +\epsilon_{s}) - (p_{s} +\epsilon_{s})$, and 
$r_{s} - p_{s} = (r_{s} +\epsilon_{s}) - (p_{s} +\epsilon_{s})$.
Therefore, the doubly linked list $\mathbf{L}_{\samp, 1}$ can be updated in $O(1)$ time. 

Second, we update the doubly linked list $\mathbf{L}_{\samp, 2}$ representing 
the order of the $d$ interval attractors $([p_{1}, q_{1}], [\ell_{1}, r_{1}])$, 
$([p_{2}, q_{2}], [\ell_{2}, r_{2}])$, 
$\ldots$, $([p_{d}, q_{d}], [\ell_{d}, r_{d}])$. 
This doubly linked list $\mathbf{L}_{\samp, 2}$ consists of $d$ elements, 
and each $\phi_{s}$-th element stores a pointer to the $s$-th element of doubly linked list $\mathbf{L}_{\samp, 1}$. 

After replacing each interval attractor $([p_{s}, q_{s}], [\ell_{s}, r_{s}]) \in \Psi_{\samp} \cap (\Psi_{\LEFT} \cup \Psi_{\RIGHT})$ with interval attractor $([p_{s}+\epsilon_{s}, q_{s}+\epsilon_{s}], [\ell_{s}+\epsilon_{s}, r_{s}+\epsilon_{s}])$, the doubly linked list $\mathbf{L}_{\samp, 2}$ is not changed 
because Corollary~\ref{cor:RB_IA_before_after} shows that 
$T[p_{s}-1..r_{s}+1] = T^{\prime}[p_{s} +\epsilon_{s} - 1..r_{s} +\epsilon_{s} + 1]$ holds. 
Therefore, the doubly linked list $\mathbf{L}_{\samp, 2}$ can be updated in $O(1)$ time. 

Third, we update the doubly linked list $\mathbf{L}_{\samp, 3}$ representing 
the sequence $\mathbf{Q}_{\ofs} = g_{1}, g_{2}, \ldots, g_{d}$ of non-negative integers introduced in Section~\ref{subsec:sample_query}. 
Here, each integer $g_{s}$ is defined as $(p_{\phi_{s}} - p_{\phi_{s-1}})$, 
and let $p_{\phi_{0}} = 0$ for simplicity.
The doubly linked list $\mathbf{L}_{\samp, 3}$ consists of $d$ elements, 
and each $s$-th element stores the $s$-th integer $g_{s}$ of sequence $\mathbf{Q}_{\ofs}$. 

For appropriately updating the doubly linked list $\mathbf{L}_{\samp, 3}$, 
we leverage the smallest integer $w$ in $[1, d]$ satisfying $([p_{\phi_{w}}, q_{\phi_{w}}], [\ell_{\phi_{w}}, r_{\phi_{w}}]) \in \Psi_{\RIGHT}$. 
The first $w-1$ interval attractors $([p_{\phi_{1}}, q_{\phi_{1}}], [\ell_{\phi_{1}}$, $r_{\phi_{1}}])$, 
$([p_{\phi_{2}}, q_{\phi_{2}}], [\ell_{\phi_{2}}, r_{\phi_{2}}])$, 
$\ldots$, $([p_{\phi_{w-1}}, q_{\phi_{w-1}}], [\ell_{\phi_{w-1}}, r_{\phi_{w-1}}])$ are not contained 
in subset $\Psi_{\RIGHT}$. 
In contrast, 
the other interval attractors 
$([p_{\phi_{w}}, q_{\phi_{w}}], [\ell_{\phi_{w}}, r_{\phi_{w}}])$, 
$([p_{\phi_{w+1}}, q_{\phi_{w+1}}], [\ell_{\phi_{w+1}}, r_{\phi_{w+1}}])$, 
$\ldots$, $([p_{\phi_{d}}, q_{\phi_{d}}], [\ell_{\phi_{d}}, r_{\phi_{d}}])$ 
are contained in the subset $\Psi_{\RIGHT}$. 
This is because $p_{\phi_{1}} \leq p_{\phi_{2}} \leq \cdots \leq p_{\phi_{d}}$ holds, 
and $p_{\phi_{w}}-1 > \lambda$ follows from the definition of the subset $\Psi_{\RIGHT}$ 
for the insertion position $\lambda$. 

Consider the change of an integer $g_{i}$ 
after replacing each interval attractor $([p_{s}, q_{s}], [\ell_{s}, r_{s}]) \in \Psi_{\samp} \cap (\Psi_{\LEFT} \cup \Psi_{\RIGHT})$ with interval attractor $([p_{s}+\epsilon_{s}, q_{s}+\epsilon_{s}], [\ell_{s}+\epsilon_{s}, r_{s}+\epsilon_{s}])$. 
From the definition of the sequence $\mathbf{Q}_{\ofs}$, 
the integer $g_{i}$ is changed into $((p_{\phi_{i}} + \epsilon_{\phi_{i}}) - (p_{\phi_{i-1}} + \epsilon_{\phi_{i-1}}))$. 
Here, let $\epsilon_{\phi_{0}} = 0$ for simplicity. 
The integer $g_{i}$ satisfies one of the following three conditions: 
\begin{enumerate}[label=\textbf{(\alph*)}]
    \item $i \leq w-1$;
    \item $i \geq w+1$;
    \item $i = w$.
\end{enumerate}

For condition (a), 
the integer $g_{i}$ is not changed because 
$\epsilon_{\phi_{i}} = 0$ and $\epsilon_{\phi_{i-1}} = 0$.
For condition (b), 
the integer $g_{i}$ is not changed because 
$\epsilon_{\phi_{i}} = 1$ and $\epsilon_{\phi_{i-1}} = 1$. 
For condition (c), 
the integer $g_{i}$ is changed into $g_{i} + 1$ because 
$\epsilon_{\phi_{i}} = 1$ and $\epsilon_{\phi_{i-1}} = 0$. 
Therefore, the integer $g_{i}$ is changed if and only if $i = w$. 

The doubly linked list $\mathbf{L}_{\samp, 3}$ is updated by changing the integer $g_{w}$ stored in the $w$-th element of $\mathbf{L}_{\samp, 3}$ into $g_{w} + 1$. 
The integer $w$ can be computed in $O(\log d \log n)$ time using $O(\log d)$ sample queries. 
Here, $\log d = O(\log n)$ (see Section~\ref{subsec:sample_query}). 
The partial sum data structure built on doubly linked list $\mathbf{L}_{\samp, 3}$ is updated in $O(\log n)$ time 
based on the change of the integer $g_{w}$. 
Therefore, the update of doubly linked list $\mathbf{L}_{\samp, 3}$ takes $O(\log^{2} n)$ time. 

Finally, the first phase takes $O(\log^{2} n)$ time. 

\paragraph{Phase (ii).}
The sampling subset $\Psi_{\samp}$ contains  
each interval attractor $([p^{\symA}_{s}, q^{\symA}_{s}], [\ell^{\symA}_{s}, r^{\symA}_{s}])$ of 
the subset $\Psi^{\symA}$ as an interval attractor $([p_{i}, q_{i}], [\ell_{i}, r_{i}])$. 
By Modification 2, 
this interval attractor is replaced with 
interval attractor $([p^{\symA^{\prime}}_{s} + \epsilon^{\symA^{\prime}}_{s}, q^{\symA^{\prime}}_{s} + \epsilon^{\symA^{\prime}}_{s}], [\ell^{\symA^{\prime}}_{s} + \epsilon^{\symA^{\prime}}_{s}, r^{\symA^{\prime}}_{s} + \epsilon^{\symA^{\prime}}_{s}])$. 
Here, the two interval attractors $([p^{\symA}_{s}, q^{\symA}_{s}], [\ell^{\symA}_{s}, r^{\symA}_{s}])$ and 
$([p^{\symA^{\prime}}_{s}, q^{\symA^{\prime}}_{s}], [\ell^{\symA^{\prime}}_{s}, r^{\symA^{\prime}}_{s}])$ are obtained from the result of Lemma~\ref{lem:dynamic_IA_summary}.
In the second phase, we update three doubly linked lists $\mathbf{L}_{\samp, 1}$, $\mathbf{L}_{\samp, 2}$, $\mathbf{L}_{\samp, 3}$ based on this replacement. 

The second phase consists of four steps. 
In the first step, 
we find the interval attractor $([p_{i}, q_{i}], [\ell_{i}, r_{i}])$ by binary search on the $d$ interval attractors of the sampling subset $\Psi_{\samp}$. 
This binary search takes $O(\log^{2} n + H^{2} \log n)$ time (see the second phase of the algorithm presented in Section~\ref{subsubsec:computation_samp_AB}). 

In the second step, we update doubly linked list $\mathbf{L}_{\samp, 1}$. 
Sequence $\mathbf{Q}_{\samp}$ contains a node $v_{i^{\prime}}$ corresponding to 
the interval attractor $([p_{i}, q_{i}], [\ell_{i}, r_{i}])$ (i.e., $\phi_{i^{\prime}} = i$). 
Let $j$ be the smallest integer in set $[1, d]$ satisfying either of the following two conditions: 
\begin{enumerate}[label=\textbf{(\alph*)}]
    \item $p^{\symA^{\prime}}_{s} + \epsilon^{\symA^{\prime}}_{s} < p_{\phi_{j}} + \epsilon_{\phi_{j}}$; 
    \item $p^{\symA^{\prime}}_{s} + \epsilon^{\symA^{\prime}}_{s} = p_{\phi_{j}} + \epsilon_{\phi_{j}}$ and 
    $r^{\symA^{\prime}}_{s} + \epsilon^{\symA^{\prime}}_{s} \leq r_{\phi_{j}} + \epsilon_{\phi_{j}}$. 
\end{enumerate}
By replacing the interval attractor $([p_{i}, q_{i}], [\ell_{i}, r_{i}])$ with 
interval attractor $([p^{\symA^{\prime}}_{s} + \epsilon^{\symA^{\prime}}_{s}, q^{\symA^{\prime}}_{s} + \epsilon^{\symA^{\prime}}_{s}], [\ell^{\symA^{\prime}}_{s} + \epsilon^{\symA^{\prime}}_{s}, r^{\symA^{\prime}}_{s} + \epsilon^{\symA^{\prime}}_{s}])$, 
the $i^{\prime}$-th node $v_{i^{\prime}}$ is moved to the $j$-th position in sequence $\mathbf{Q}_{\samp}$, 
and this node corresponds to the interval attractor $([p^{\symA^{\prime}}_{s} + \epsilon^{\symA^{\prime}}_{s}, q^{\symA^{\prime}}_{s} + \epsilon^{\symA^{\prime}}_{s}], [\ell^{\symA^{\prime}}_{s} + \epsilon^{\symA^{\prime}}_{s}, r^{\symA^{\prime}}_{s} + \epsilon^{\symA^{\prime}}_{s}])$. 
The position $j$ can be found in $O(\log d \log n)$ by binary search on sequence $\mathbf{Q}_{\samp}$ 
because the interval attractor corresponding to a node of sequence $\mathbf{Q}_{\samp}$ can be computed in 
$O(\log n)$ time using the algorithm presented in Section~\ref{subsubsec:computation_delta_samp}. 
Therefore, the second step takes $O(\log^{2} n)$ time. 

In the third step, we update doubly linked list $\mathbf{L}_{\samp, 2}$. 
The doubly linked list $\mathbf{L}_{\samp, 2}$ is not changed by 
replacing the interval attractor $([p_{i}, q_{i}], [\ell_{i}, r_{i}])$ with 
interval attractor $([p^{\symA^{\prime}}_{s} + \epsilon^{\symA^{\prime}}_{s}, q^{\symA^{\prime}}_{s} + \epsilon^{\symA^{\prime}}_{s}], [\ell^{\symA^{\prime}}_{s} + \epsilon^{\symA^{\prime}}_{s}, r^{\symA^{\prime}}_{s} + \epsilon^{\symA^{\prime}}_{s}])$. 
This is because the two interval attractors represent the same string (i.e., $T[p_{i}-1..r_{i}+1] = T^{\prime}[p^{\symA^{\prime}}_{s} + \epsilon^{\symA^{\prime}}_{s} - 1..r^{\symA^{\prime}}_{s} + \epsilon^{\symA^{\prime}}_{s} + 1]$). 
Therefore, the third step takes $O(1)$ time. 

In the fourth step, we update doubly linked list $\mathbf{L}_{\samp, 3}$. 
By replacing the interval attractor $([p_{i}, q_{i}], [\ell_{i}, r_{i}])$ with 
interval attractor $([p^{\symA^{\prime}}_{s} + \epsilon^{\symA^{\prime}}_{s}, q^{\symA^{\prime}}_{s} + \epsilon^{\symA^{\prime}}_{s}], [\ell^{\symA^{\prime}}_{s} + \epsilon^{\symA^{\prime}}_{s}, r^{\symA^{\prime}}_{s} + \epsilon^{\symA^{\prime}}_{s}])$, 
sequence $\mathbf{Q}_{\ofs}$ is changed as follows: 
\begin{enumerate}[label=\textbf{(\arabic*)}]
    \item the $i^{\prime}$-th integer $g_{i^{\prime}}$ is removed from sequence $\mathbf{Q}_{\ofs}$;
    \item the $(i^{\prime}+1)$-th integer $g_{i^{\prime}+1}$ is changed to $(p_{\phi_{i^{\prime}+1}} + \epsilon_{\phi_{i^{\prime}+1}}) - (p_{\phi_{i^{\prime}-1}} + \epsilon_{\phi_{i^{\prime}-1}})$;
    \item the $j$-th integer $g_{j}$ is changed to $(p_{\phi_{j}} + \epsilon_{\phi_{j}}) - (p^{\symA^{\prime}}_{s} + \epsilon^{\symA^{\prime}}_{s})$;
    \item integer $(p^{\symA^{\prime}}_{s} + \epsilon^{\symA^{\prime}}_{s}) - (p_{\phi_{j-1}} + \epsilon_{\phi_{j-1}})$ 
    is inserted into sequence $\mathbf{Q}_{\ofs}$ as the $j$-th integer.     
\end{enumerate}
The doubly linked list $\mathbf{L}_{\samp, 3}$ is updated based on the change of sequence $\mathbf{Q}_{\ofs}$. 
This update of this doubly linked list can be executed in $O(\log^{2} n)$ time. 

Finally, 
the running time of the second phase is $O(k(\log^{2} n + H^{2} \log n))$ time in total. 
Here, $k \leq |\Psi_{\OLD} \setminus \Psi_{\run}|$, 
and the expected value of $|\Psi_{\OLD} \setminus \Psi_{\run}|$ is $O(H + \log n)$. 
Therefore, the second phase takes expected $O((H + \log n)(\log^{2} n + H^{2} \log n))$ time. 

\paragraph{Phase (iii).}
By Modification 3, 
each interval attractor $([p_{s}, q_{s}], [\ell_{s}, r_{s}])$ of set $(\Psi_{\samp} \cap \Psi_{\OLD}) \setminus \Psi^{\symA}$ is removed from subset $\Psi_{\samp}$. 
Here, the interval attractor $([p_{s}, q_{s}], [\ell_{s}, r_{s}])$ is obtained from 
the result of Lemma~\ref{lem:dynamic_IA_summary}. 
In the third phase, we update three doubly linked lists $\mathbf{L}_{\samp, 1}$, $\mathbf{L}_{\samp, 2}$, $\mathbf{L}_{\samp, 3}$ based on this removal. 

The third phase consists of three steps. 
In the first step, 
we update doubly linked list $\mathbf{L}_{\samp, 1}$. 
Sequence $\mathbf{Q}_{\samp}$ contains a node $v_{s^{\prime}}$ corresponding to 
the interval attractor $([p_{s}, q_{s}], [\ell_{s}, r_{s}])$. 
This node $v_{s^{\prime}}$ is removed from $\mathbf{Q}_{\samp}$. 
Similarly, the $s^{\prime}$-th element is removed from doubly linked list $\mathbf{L}_{\samp, 1}$. 
Similar to the third phase, 
the integer $s^{\prime}$ can be computed in $O(\log^{2} n)$ time. 
Therefore, the doubly linked list $\mathbf{L}_{\samp, 1}$ can be updated in $O(\log^{2} n)$ time. 

In the second step, 
we update doubly linked list $\mathbf{L}_{\samp, 2}$. 
Similar to the first step, 
the $s$-th element is removed from doubly linked list $\mathbf{L}_{\samp, 2}$. 
This removal can be executed in $O(\log^{2} n)$ time. 

In the third step, 
we update doubly linked list $\mathbf{L}_{\samp, 3}$. 
By the removal of node $v_{s^{\prime}}$ from $\mathbf{Q}_{\samp}$, 
the $s^{\prime}$-th integer $g_{s^{\prime}}$ is removed from sequence $\mathbf{Q}_{\ofs}$, 
and the $(s^{\prime}+1)$-th integer $g_{s^{\prime}+1}$ is changed into 
$(p_{\phi_{s^{\prime}+1}} + \epsilon_{\phi_{s^{\prime}+1}}) - (p_{\phi_{s^{\prime}-1}} + \epsilon_{\phi_{s^{\prime}-1}})$. 
The doubly linked list $\mathbf{L}_{\samp, 3}$ is updated based on the change of the sequence $\mathbf{Q}_{\ofs}$, and this update can be executed in $O(\log^{2} n)$ time. 

Finally, 
the running time of the third phase is $O(|(\Psi_{\samp} \cap \Psi_{\OLD}) \setminus \Psi^{\symA}|(\log^{2} n + H^{2} \log n))$ time in total. 
Here, $(\Psi_{\samp} \cap \Psi_{\OLD}) \setminus \Psi^{\symA^{\prime}} \subseteq \Psi_{\samp} \cap \Psi_{\OLD}$ holds, and $\mathbb{E}[|\Psi_{\samp} \cap \Psi_{\OLD}|] = O(H + \log n)$ (Lemma~\ref{lem:dynamic_psi_overlap}).
Therefore, the third phase takes expected $O((H + \log n) (\log^{2} n + H^{2} \log n))$ time. 

%$(\Psi_{\samp} \cap \Psi_{\OLD}) \setminus \Psi^{\symA^{\prime}} = ((\Psi_{\OLD} \setminus \Psi_{\run}) \cap \Psi_{\samp}) \setminus \Psi^{\symA^{\prime}}$. 

\paragraph{Phase (iv).}
By Modification 4, 
each interval attractor $([p^{\symB}_{s}, q^{\symB}_{s}], [\ell^{\symB}_{s}, r^{\symB}_{s}])$ of 
subset $\Psi^{\prime \symB}$ is added into subset $\Psi_{\samp}$. 
In the fourth phase, we update three doubly linked lists $\mathbf{L}_{\samp, 1}$, $\mathbf{L}_{\samp, 2}$, $\mathbf{L}_{\samp, 3}$ based on this addition. 

The fourth phase consists of three steps. 
In the first step, 
we obtain the dynamic data structures for the RR-DAG of RLSLP $\mathcal{G}^{R}_{\ins}$ 
by modifying the dynamic data structures for the RR-DAG of RLSLP $\mathcal{G}^{R}$. 
This update takes $O((\max \{H, H^{\prime}, \log (nHH^{\prime}) \})^{3})$ (Lemma~\ref{lem:dynamic_rrdag_summary}). 

In the second step, 
we update doubly linked list $\mathbf{L}_{\samp, 1}$. 
Let $j$ be the smallest integer in set $[1, d]$ satisfying either of the following two conditions: 
\begin{enumerate}[label=\textbf{(\alph*)}]
    \item $p^{\symB}_{s} < p_{\phi_{j}} + \epsilon_{\phi_{j}}$; 
    \item $p^{\symB}_{s} = p_{\phi_{j}} + \epsilon_{\phi_{j}}$ and 
    $r^{\symB}_{s} \leq r_{\phi_{j}} + \epsilon_{\phi_{j}}$. 
\end{enumerate}
Then, 
a new node $v$ is inserted into sequence $\mathbf{Q}_{\samp}$ as the $j$-th node, 
and this node corresponds to the interval attractor $([p^{\symB}_{s}, q^{\symB}_{s}], [\ell^{\symB}_{s}, r^{\symB}_{s}])$. 
Based on the insertion of the node $v$, doubly linked list $\mathbf{L}_{\samp, 1}$ is updated.
Similar to the third phase, 
the position $j$ can be found in $O(\log^{2} n)$ time by binary search on sequence $\mathbf{Q}_{\samp}$. 
Therefore, the second step takes $O(\log^{2} n)$ time. 

In the third step, 
we update doubly linked list $\mathbf{L}_{\samp, 2}$. 
Let $j^{\prime}$ be the smallest integer in set $[1, d]$ satisfying 
$T^{\prime}[p^{\symB}_{s} - 1, r^{\symB}_{s}+1] \prec T^{\prime}[p_{j^{\prime}} + \epsilon_{j^{\prime}} - 1, r_{j^{\prime}} + \epsilon_{j^{\prime}}+1]$. 
Then, a new element is inserted into doubly linked list $\mathbf{L}_{\samp, 2}$ as the $j^{\prime}$-th element, 
and it stores a pointer to the $j$-th element of doubly linked list $\mathbf{L}_{\samp, 1}$, which corresponds to the node $v$. 
We compute the position $j^{\prime}$ by binary search on the interval attractors 
$([p_{1} + \epsilon_{1}, q_{1} + \epsilon_{1}], [\ell_{1} + \epsilon_{1}, r_{1} + \epsilon_{1}])$, 
$([p_{2} + \epsilon_{2}, q_{2} + \epsilon_{2}], [\ell_{2} + \epsilon_{2}, r_{2} + \epsilon_{2}])$, $\ldots$, $([p_{d} + \epsilon_{d}, q_{d} + \epsilon_{d}], [\ell_{d} + \epsilon_{d}, r_{d} + \epsilon_{d}])$. 
Similar to the second phase of the algorithm presented in Section~\ref{subsubsec:computation_samp_AB}, 
this binary search can be executed using $O(\log n)$ sample, LCE, and random access queries. 
Therefore, the third step takes $O(\log^{2} n + H \log n)$ time. 

In the fourth step, 
we update doubly linked list $\mathbf{L}_{\samp, 3}$. 
By inserting node $v$ into $\mathbf{Q}_{\samp}$, 
the $j$-th integer $g_{j}$ is changed into 
$(p_{\phi_{j}} + \epsilon_{\phi_{j}}) - p^{\symB}_{s}$, 
and integer $p^{\symB}_{s} - (p_{\phi_{j-1}} + \epsilon_{\phi_{j-1}})$ is inserted into 
sequence $\mathbf{Q}_{\ofs}$ as the $j$-th integer. 
The doubly linked list $\mathbf{L}_{\samp, 3}$ is updated based on the change of the sequence $\mathbf{Q}_{\ofs}$, and this update can be executed in $O(\log^{2} n)$ time. 
Therefore, 
the running time of the fourth phase is expected $O((\max \{H, H^{\prime}, \log (nHH^{\prime}) \})^{3})$ time in total. 

Finally, 
the four phases takes expected $O((\max \{H, H^{\prime}, \log (nHH^{\prime}) \})^{4})$ time in total. 
Therefore, we obtain Lemma~\ref{lem:dynamic_samp_summary}. 

\paragraph{The changes of sequence $\mathbf{Q}_{\samp}$.}
Before updating the dynamic data structures for sample query, 
each node $v$ of sequence $\mathbf{Q}_{\samp}$ 
represents an interval attractor in sampling subset $\Psi_{\samp}$. 
If the node $v$ is not removed from the sequence $\mathbf{Q}_{\samp}$ by the update of sequence, 
then the interval attractor represented as the node $v$ is changed to an interval attractor in sampling subset $\Psi^{\prime}_{\samp}$. 
Formally, the following lemma holds. 

\begin{lemma}\label{lem:dynamic_samp_proceeding_formula}
Consider an interval attractor $([p, q], [\ell, r]) \in \Psi_{\samp}$ satisfying 
$([p, q], [\ell, r]) \not \in \Psi_{\OLD} \setminus \Psi^{\symA}$. 
Here, the $([p, q], [\ell, r])$ is represented as a node $u$ of sequence $\mathbf{Q}_{\samp}$.
After updating the dynamic data structures for sample query,  
the node $u$ represents an interval attractor 
$([p^{\prime}, q^{\prime}], [\ell^{\prime}, r^{\prime}])$ in set $\Psi^{\prime}_{\samp}$ 
satisfying (i) $T[p-1..r+1] = T^{\prime}[p^{\prime}-1..r^{\prime}+1]$, 
(ii) $([p^{\prime}, q^{\prime}], [\ell^{\prime}, r^{\prime}]) \not \in \Psi^{\prime \symB}$, 
and (iii) $([p^{\prime}, q^{\prime}], [\ell^{\prime}, r^{\prime}]) \not \in \Psi^{\prime}_{\NEW}$. 
\end{lemma}
\begin{proof}
    The interval attractor $([p, q], [\ell, r])$ satisfies one of the following three conditions: 
    \begin{enumerate}[label=\textbf{(\Alph*)}]
    \item $([p, q], [\ell, r]) \in \Psi_{\LEFT}$;
    \item $([p, q], [\ell, r]) \in \Psi_{\RIGHT}$;
    \item $([p, q], [\ell, r]) \in \Psi^{\symA}$.    
    \end{enumerate}

    \textbf{Proof of Lemma~\ref{lem:dynamic_samp_proceeding_formula} for condition (A).}
    After updating the dynamic data structures for sample query, 
    node $u$ represents interval attractor $([p, q], [\ell, r])$ in set $\Psi^{\prime}_{\RR}$. 
    $([p, q], [\ell, r]) \in \Psi^{\prime}_{\samp}$ follows from 
    the definition of the subset $\Psi^{\prime}_{\samp}$. 
    Because of $([p, q], [\ell, r]) \in \Psi_{\LEFT}$, 
    $T[p-1..r+1] = T^{\prime}[p-1..r+1]$ and 
    $([p, q], [\ell, r]) \in \Psi^{\prime}_{\LEFT}$.    
    $([p, q], [\ell, r]) \not \in \Psi^{\prime}_{\NEW}$ follows from 
    $([p, q], [\ell, r]) \in \Psi^{\prime}_{\LEFT}$.    
    $([p, q], [\ell, r]) \not \in \Psi^{\prime \symB}$ 
    follows from 
    $([p, q], [\ell, r]) \not \in \Psi^{\prime}_{\NEW}$ 
    and $\Psi^{\prime \symB} \subseteq \Psi^{\prime}_{\NEW}$. 
    
    \textbf{Proof of Lemma~\ref{lem:dynamic_samp_proceeding_formula} for condition (B).}
    In this case, 
    Lemma~\ref{lem:dynamic_samp_proceeding_formula} can be proved using the same approach as for condition (A).

    \textbf{Proof of Lemma~\ref{lem:dynamic_samp_proceeding_formula} for condition (C).}
    The subset $\Psi^{\symA}$ contains the interval attractor $([p, q], [\ell, r])$ 
    as $([p^{\symA}_{s}, q^{\symA}_{s}], [\ell^{\symA}_{s}, r^{\symA}_{s}])$.     
    From the definition of subset $\Psi^{\symA^{\prime}}$, 
    $T[p^{\symA}_{s}-1..r^{\symA}_{s}+1] = T[p^{\symA^{\prime}}_{s}-1..r^{\symA^{\prime}}_{s}+1]$ holds 
    for the interval attractor $([p^{\symA^{\prime}}_{s}, q^{\symA^{\prime}}_{s}], [\ell^{\symA^{\prime}}_{s}, r^{\symA^{\prime}}_{s}])$ 
    in the subset $\Psi^{\symA^{\prime}}$.     
    After updating the dynamic data structures for sample query, 
    node $u$ represents interval attractor $([p^{\symA^{\prime}}_{s} + \epsilon^{\symA^{\prime}}_{s}, q^{\symA^{\prime}}_{s}+ \epsilon^{\symA^{\prime}}_{s}], [\ell^{\symA^{\prime}}_{s}+ \epsilon^{\symA^{\prime}}_{s}, r^{\symA^{\prime}}_{s}+ \epsilon^{\symA^{\prime}}_{s}])$ in set $\Psi^{\prime}_{\RR}$, 
    and 
    $T[p^{\symA^{\prime}}_{s}-1..r^{\symA^{\prime}}_{s}+1] = T^{\prime}[p^{\symA^{\prime}}_{s} + \epsilon^{\symA^{\prime}}_{s} -1..r^{\symA^{\prime}}_{s} + \epsilon^{\symA^{\prime}}_{s} +1]$ holds. 
    $([p^{\symA^{\prime}}_{s} + \epsilon^{\symA^{\prime}}_{s}, q^{\symA^{\prime}}_{s}+ \epsilon^{\symA^{\prime}}_{s}], [\ell^{\symA^{\prime}}_{s}+ \epsilon^{\symA^{\prime}}_{s}, r^{\symA^{\prime}}_{s}+ \epsilon^{\symA^{\prime}}_{s}])\in \Psi^{\prime}_{\samp}$ follows from 
    the definition of the sampling subset $\Psi^{\prime}_{\samp}$. 

    We prove $([p^{\symA^{\prime}}_{s} + \epsilon^{\symA^{\prime}}_{s}, q^{\symA^{\prime}}_{s}+ \epsilon^{\symA^{\prime}}_{s}], [\ell^{\symA^{\prime}}_{s}+ \epsilon^{\symA^{\prime}}_{s}, r^{\symA^{\prime}}_{s}+ \epsilon^{\symA^{\prime}}_{s}]) \not \in \Psi^{\prime}_{\NEW}$. 
    $([p^{\symA^{\prime}}_{s}, q^{\symA^{\prime}}_{s}], [\ell^{\symA^{\prime}}_{s}, r^{\symA^{\prime}}_{s}]) \not \in \Psi_{\OLD}$ 
    follows from Lemma~\ref{lem:dynamic_samp_substitute}~\ref{enum:dynamic_samp_substitute:2}. 
    $([p^{\symA^{\prime}}_{s}, q^{\symA^{\prime}}_{s}], [\ell^{\symA^{\prime}}_{s}, r^{\symA^{\prime}}_{s}]) \in \Psi_{\LEFT} \cup \Psi_{\RIGHT}$  
    follows from $([p^{\symA^{\prime}}_{s}, q^{\symA^{\prime}}_{s}], [\ell^{\symA^{\prime}}_{s}, r^{\symA^{\prime}}_{s}]) \not \in \Psi_{\OLD}$. 
    Theorem~\ref{theo:RS_LEFT_RIGHT} shows that 
    $([p^{\symA^{\prime}}_{s} + \epsilon^{\symA^{\prime}}_{s}, q^{\symA^{\prime}}_{s} + \epsilon^{\symA^{\prime}}_{s}], [\ell^{\symA^{\prime}}_{s} + \epsilon^{\symA^{\prime}}_{s}, r^{\symA^{\prime}}_{s} + \epsilon^{\symA^{\prime}}_{s}]) \in \Psi^{\prime}_{\LEFT} \cup \Psi^{\prime}_{\RIGHT}$. 
    Therefore, $([p^{\symA^{\prime}}_{s} + \epsilon^{\symA^{\prime}}_{s}, q^{\symA^{\prime}}_{s}+ \epsilon^{\symA^{\prime}}_{s}], [\ell^{\symA^{\prime}}_{s}+ \epsilon^{\symA^{\prime}}_{s}, r^{\symA^{\prime}}_{s}+ \epsilon^{\symA^{\prime}}_{s}]) \not \in \Psi^{\prime}_{\NEW}$ follows from $([p^{\symA^{\prime}}_{s} + \epsilon^{\symA^{\prime}}_{s}, q^{\symA^{\prime}}_{s} + \epsilon^{\symA^{\prime}}_{s}], [\ell^{\symA^{\prime}}_{s} + \epsilon^{\symA^{\prime}}_{s}, r^{\symA^{\prime}}_{s} + \epsilon^{\symA^{\prime}}_{s}]) \in \Psi^{\prime}_{\LEFT} \cup \Psi^{\prime}_{\RIGHT}$. 

    $([p^{\symA^{\prime}}_{s} + \epsilon^{\symA^{\prime}}_{s}, q^{\symA^{\prime}}_{s}+ \epsilon^{\symA^{\prime}}_{s}], [\ell^{\symA^{\prime}}_{s}+ \epsilon^{\symA^{\prime}}_{s}, r^{\symA^{\prime}}_{s}+ \epsilon^{\symA^{\prime}}_{s}]) \not \in \Psi^{\prime \symB}$ 
    follows from 
    $([p^{\symA^{\prime}}_{s} + \epsilon^{\symA^{\prime}}_{s}, q^{\symA^{\prime}}_{s}+ \epsilon^{\symA^{\prime}}_{s}], [\ell^{\symA^{\prime}}_{s}+ \epsilon^{\symA^{\prime}}_{s}, r^{\symA^{\prime}}_{s}+ \epsilon^{\symA^{\prime}}_{s}]) \not \in \Psi^{\prime}_{\NEW}$ 
    and $\Psi^{\prime \symB} \subseteq \Psi^{\prime}_{\NEW}$. 
    Therefore, Lemma~\ref{lem:dynamic_samp_proceeding_formula} holds. 

\end{proof}

\subsection{Update of Data Structures for Bigram Search Query}\label{subsec:bis_update}
This subsection explains how to update the dynamic data structures for bigram search query introduced in Section~\ref{subsubsec:bis_ds}. 
The following lemma is the summary of this subsection.

\begin{lemma}\label{lem:dynamic_bis_summary}
Consider the two RLSLPs $\mathcal{G}^{R}$ and $\mathcal{G}^{R}_{\ins}$ of Theorem~\ref{theo:update1}, which derive input string $T$ and string $T^{\prime}$, respectively. 
The dynamic data structures of Section~\ref{subsubsec:sample_ds} can be updated 
in expected $O((\max \{H, H^{\prime}, \log (nHH^{\prime}) \})^{4})$ time 
after changing RLSLP $\mathcal{G}^{R}$ to $\mathcal{G}^{R}_{\ins}$. 
This update requires 
(i) the dynamic data structures for the RR-DAG of RLSLP $\mathcal{G}^{R}$ (Section~\ref{subsubsec:rrdag_ds}) 
and sample query (Section~\ref{subsec:sample_query}), 
and (ii) the interval attractors obtained from Lemma~\ref{lem:dynamic_IA_summary}. 
\end{lemma}
\begin{proof}
    See Section~\ref{subsubsec:dynamic_bis_query_algo}. 
\end{proof}

Consider the sequence $\mathbf{Q}_{\BiSQ} = u_{1}, u_{2}, \ldots, u_{k}$ 
and label function $\mathcal{L}_{\BiSQ}$ introduced in Section~\ref{subsec:bigram_search_query}. 
For this subsection, 
let $\mathbf{Q}^{\prime}_{\BiSQ} = u^{\prime}_{1}, u^{\prime}_{2}, \ldots, u^{\prime}_{k^{\prime}}$ 
and $\mathcal{L}^{\prime}_{\BiSQ}$ be the sequence $\mathbf{Q}_{\BiSQ}$ and 
label function $\mathcal{L}_{\BiSQ}$ after RLSLP $\mathcal{G}^{R}$ is changed to $\mathcal{G}^{R}_{\ins}$. 

From the definition of the sequence $\mathbf{Q}_{\BiSQ}$, 
each node $u_{s}$ corresponds to interval attractor $([p_{\varrho_{s}}, q_{\varrho_{s}}], [\ell_{\varrho_{s}}, r_{\varrho_{s}}])$ in set 
$\Psi_{0} \cap \Psi_{\samp} = \{ ([p_{1}, q_{1}], [\ell_{1}, r_{1}]), ([p_{2}, q_{2}], [\ell_{2}, r_{2}]), \ldots, ([p_{k}, q_{k}], [\ell_{k}, r_{k}]) \}$ for the permutation $\varrho_{1}, \varrho_{2}, \ldots, \varrho_{k}$ 
introduced in Section~\ref{subsec:bigram_search_query}. 

For this subsection, 
let $\varrho^{\prime}_{1}, \varrho^{\prime}_{2}, \ldots, \varrho^{\prime}_{k^{\prime}}$ be 
the permutation $\varrho_{1}, \varrho_{2}, \ldots, \varrho_{k}$ after RLSLP $\mathcal{G}^{R}$ is changed to $\mathcal{G}^{R}_{\ins}$. 
Then, 
each node $u^{\prime}_{s}$ corresponds to 
interval attractor $([p^{\prime}_{\varrho^{\prime}_{s}}, q^{\prime}_{\varrho^{\prime}_{s}}], [\ell^{\prime}_{\varrho^{\prime}_{s}}, r^{\prime}_{\varrho^{\prime}_{s}}])$ 
in set $\Psi^{\prime}_{0} \cap \Psi^{\prime}_{\samp} = \{ ([p^{\prime}_{1}, q^{\prime}_{1}], [\ell^{\prime}_{1}, r^{\prime}_{1}]), ([p^{\prime}_{2}, q^{\prime}_{2}], [\ell^{\prime}_{2}, r^{\prime}_{2}]), \ldots, ([p^{\prime}_{k^{\prime}}, q^{\prime}_{k^{\prime}}], [\ell^{\prime}_{k^{\prime}}, r^{\prime}_{k^{\prime}}]) \}$. 

We explain properties of set $\Psi_{0} \cap \Psi_{\samp}$, sequence $\mathbf{Q}_{\BiSQ}$, and label function $\mathcal{L}_{\BiSQ}$. 
The following lemma states the relationship between two sets $\Psi_{0} \cap \Psi_{\samp}$ and $\Psi^{\prime}_{0} \cap \Psi^{\prime}_{\samp}$. 

\begin{lemma}\label{lem:dynamic_bis_back_correspondence}
    Consider an interval attractor 
    $([p^{\prime}, q^{\prime}], [\ell^{\prime}, r^{\prime}])$ in set $\Psi^{\prime}_{0} \cap \Psi^{\prime}_{\samp}$ satisfying $([p^{\prime}, q^{\prime}]$, $[\ell^{\prime}, r^{\prime}]) \not \in \Psi^{\prime \symB}$. 
    Then, set $\Psi_{0} \cap \Psi_{\samp}$ contains an interval attractor 
    $([p, q], [\ell, r])$ satisfying 
    $T[p-1..r+1] = T^{\prime}[p^{\prime}-1..r^{\prime}+1]$ and 
    $([p, q], [\ell, r]) \not \in \Psi_{\OLD} \setminus \Psi^{\symA}$. 
\end{lemma}
\begin{proof}
    Lemma~\ref{lem:dynamic_samp_back_formula} shows that 
    sampling subset $\Psi_{\samp}$ contains an interval attractor 
    $([p, q], [\ell, r])$ satisfying 
    $T[p-1..r+1] = T^{\prime}[p^{\prime}-1..r^{\prime}+1]$ and 
    $([p, q], [\ell, r]) \not \in \Psi_{\OLD} \setminus \Psi^{\symA}$.     
    We apply Lemma~\ref{lem:dynamic_IA_super_correspondence} 
    to the two interval attractors $([p, q], [\ell, r])$ and $([p^{\prime}, q^{\prime}], [\ell^{\prime}, r^{\prime}])$. 
    Then, 
    Lemma~\ref{lem:dynamic_IA_super_correspondence} shows that 
    $([p, q], [\ell, r]) \in \Psi_{0}$ holds. 
    Therefore, Lemma~\ref{lem:dynamic_bis_back_correspondence} holds.     
\end{proof}

The following lemma states the change of 
the interval attractor represented as each node of sequence $\mathbf{Q}_{\BiSQ}$ after 
RLSLP $\mathcal{G}^{R}$ is changed to $\mathcal{G}^{R}_{\ins}$. 

\begin{lemma}\label{lem:dynamic_bis_proceeding_correspondence}
Consider a node $u_{s}$ of sequence $\mathbf{Q}_{\BiSQ} = u_{1}, u_{2}, \ldots, u_{k}$ 
satisfying $([p_{\varrho_{s}}, q_{\varrho_{s}}], [\ell_{\varrho_{s}}, r_{\varrho_{s}}]) \not \in \Psi_{\OLD} \setminus \Psi^{\symA}$. 
Here, the $s$-th element of the doubly linked list $\mathbf{L}_{\BiSQ}$ stores 
a pointer to the node $v$ of the sequence $\mathbf{Q}_{\samp}$ representing 
interval attractor $([p_{\varrho_{s}}, q_{\varrho_{s}}], [\ell_{\varrho_{s}}, r_{\varrho_{s}}])$. 
After updating the dynamic data structures for sample query,  
the node $v$ represents an interval attractor 
$([p, q], [\ell, r])$ in set $\Psi^{\prime}_{0} \cap \Psi^{\prime}_{\samp}$ 
satisfying $T[p_{\varrho_{s}}-1..r_{\varrho_{s}}+1] = T^{\prime}[p-1..r+1]$ 
and $([p, q], [\ell, r]) \not \in \Psi^{\prime \symB}$. 
\end{lemma}
\begin{proof}
After updating the dynamic data structures for sample query,  
Lemma~\ref{lem:dynamic_samp_proceeding_formula} shows that 
the node $v$ represents an interval attractor 
$([p, q], [\ell, r])$ in set $\Psi^{\prime}_{\samp}$ 
satisfying $T[p_{\varrho_{s}}-1..r_{\varrho_{s}}+1] = T^{\prime}[p-1..r+1]$ 
and $([p, q], [\ell, r]) \not \in \Psi^{\prime \symB}$. 
We apply Lemma~\ref{lem:dynamic_IA_super_correspondence} to the two interval attractors 
$([p_{\varrho_{s}}, q_{\varrho_{s}}], [\ell_{\varrho_{s}}, r_{\varrho_{s}}])$ and $([p, q], [\ell, r])$. 
Then, the lemma shows that $([p, q], [\ell, r]) \in \Psi^{\prime}_{0}$ holds. 
Therefore, Lemma~\ref{lem:dynamic_bis_proceeding_correspondence} holds.     
\end{proof}

The following two lemma states the relationship between the order of sequence $u_{1}$, $u_{2}$, $\ldots$, $u_{k}$ 
and the order of sequence $u^{\prime}_{1}$, $u^{\prime}_{2}$, $\ldots$, $u^{\prime}_{k^{\prime}}$. 

\begin{lemma}\label{lem:dynamic_bis_order}
    Consider two interval attractors $([p_{\varrho_{x}}, q_{\varrho_{x}}], [\ell_{\varrho_{x}}, r_{\varrho_{x}}])$ and $([p_{\varrho_{y}}, q_{\varrho_{y}}], [\ell_{\varrho_{y}}, r_{\varrho_{y}}])$ in set $\Psi_{0} \cap \Psi_{\samp}$ satisfying 
    $([p_{\varrho_{x}}, q_{\varrho_{x}}], [\ell_{\varrho_{x}}, r_{\varrho_{x}}]), ([p_{\varrho_{y}}, q_{\varrho_{y}}], [\ell_{\varrho_{y}}, r_{\varrho_{y}}]) \not \in \Psi_{\OLD} \setminus \Psi^{\symA}$. Here, Lemma~\ref{lem:dynamic_bis_proceeding_correspondence} shows that 
    set $\Psi^{\prime}_{0} \cap \Psi^{\prime}_{\samp}$ contains 
    an interval attractor $([p^{\prime}_{\varrho^{\prime}_{\alpha}}, q^{\prime}_{\varrho^{\prime}_{\alpha}}], [\ell^{\prime}_{\varrho^{\prime}_{\alpha}}, r^{\prime}_{\varrho^{\prime}_{\alpha}}])$ satisfying $T[p_{\varrho_{x}}-1..r_{\varrho_{x}}+1] = T^{\prime}[p^{\prime}_{\varrho^{\prime}_{\alpha}}-1..r^{\prime}_{\varrho^{\prime}_{\alpha}}+1]$. 
    Similarly, 
    set $\Psi^{\prime}_{0} \cap \Psi^{\prime}_{\samp}$ contains 
    an interval attractor $([p^{\prime}_{\varrho^{\prime}_{\beta}}, q^{\prime}_{\varrho^{\prime}_{\beta}}]$, 
    $[\ell^{\prime}_{\varrho^{\prime}_{\beta}}, r^{\prime}_{\varrho^{\prime}_{\beta}}])$ satisfying $T[p_{\varrho_{y}}-1..r_{\varrho_{y}}+1] = T^{\prime}[p^{\prime}_{\varrho^{\prime}_{\beta}}-1..r^{\prime}_{\varrho^{\prime}_{\beta}}+1]$. 
    Then, $x < y \Leftrightarrow \alpha < \beta$ holds.         
\end{lemma}
\begin{proof}
    From the definition of permutation $\varrho_{1}, \varrho_{2}, \ldots, \varrho_{k}$, 
    Lemma~\ref{lem:dynamic_bis_order} holds if 
    $T[q_{\varrho_{x}}..q_{\varrho_{x}}+1] = T^{\prime}[q^{\prime}_{\varrho^{\prime}_{\alpha}}..q^{\prime}_{\varrho^{\prime}_{\alpha}}+1]$ and 
    $T[q_{\varrho_{y}}..q_{\varrho_{y}}+1] = T^{\prime}[q^{\prime}_{\varrho^{\prime}_{\beta}}..q^{\prime}_{\varrho^{\prime}_{\beta}}+1]$ hold. 
    We apply Lemma~\ref{lem:dynamic_IA_super_correspondence} to the two interval attractors 
    $([p_{\varrho_{x}}, q_{\varrho_{x}}], [\ell_{\varrho_{x}}, r_{\varrho_{x}}])$ and $([p^{\prime}_{\varrho^{\prime}_{\alpha}}, q^{\prime}_{\varrho^{\prime}_{\alpha}}], [\ell^{\prime}_{\varrho^{\prime}_{\alpha}}, r^{\prime}_{\varrho^{\prime}_{\alpha}}])$. 
    Then, the lemma shows that $T[q_{\varrho_{x}}..q_{\varrho_{x}}+1] = T^{\prime}[q^{\prime}_{\varrho^{\prime}_{\alpha}}..q^{\prime}_{\varrho^{\prime}_{\alpha}}+1]$ holds. 
    Similarly, $T[q_{\varrho_{y}}..q_{\varrho_{y}}+1] = T^{\prime}[q^{\prime}_{\varrho^{\prime}_{\beta}}..q^{\prime}_{\varrho^{\prime}_{\beta}}+1]$ can be proved by applying Lemma~\ref{lem:dynamic_IA_super_correspondence} to 
    $([p_{\varrho_{y}}, q_{\varrho_{y}}], [\ell_{\varrho_{y}}, r_{\varrho_{y}}])$ and $([p^{\prime}_{\varrho^{\prime}_{\beta}}, q^{\prime}_{\varrho^{\prime}_{\beta}}], [\ell^{\prime}_{\varrho^{\prime}_{\beta}}, r^{\prime}_{\varrho^{\prime}_{\beta}}])$.
    Therefore, Lemma~\ref{lem:dynamic_bis_order} holds. 
\end{proof}

The following lemma states the relationship between two label functions 
$\mathcal{L}_{\BiSQ}$ and $\mathcal{L}^{\prime}_{\BiSQ}$.

\begin{lemma}\label{lem:dynamic_bis_label}    
    Let $u^{\prime}_{s^{\prime}}$ 
    be a node of sequence $\mathbf{Q}^{\prime}_{\BiSQ} = u^{\prime}_{1}, u^{\prime}_{2}, \ldots, u^{\prime}_{k^{\prime}}$. 
    Here, 
    Lemma~\ref{lem:dynamic_bis_back_correspondence} shows that 
    sequence $\mathbf{Q}_{\BiSQ} = u_{1}, u_{2}, \ldots, u_{k}$ contains a node $u_{s}$  
    satisfying $T[p_{\varrho_{s}}-1..r_{\varrho_{s}}+1] = T^{\prime}[p^{\prime}_{\varrho^{\prime}_{s^{\prime}}}-1..r^{\prime}_{\varrho^{\prime}_{s^{\prime}}}+1]$ 
    if $([p^{\prime}_{\varrho^{\prime}_{s^{\prime}}}, q^{\prime}_{\varrho^{\prime}_{s^{\prime}}}], [\ell^{\prime}_{\varrho^{\prime}_{s^{\prime}}}, r^{\prime}_{\varrho^{\prime}_{s^{\prime}}}]) \not \in \Psi^{\prime \symB}$. 
    Let $m^{\prime} = |f^{\prime}_{\recover}(([p^{\prime}_{\varrho^{\prime}_{s^{\prime}}}, q^{\prime}_{\varrho^{\prime}_{s^{\prime}}}], [\ell^{\prime}_{\varrho^{\prime}_{s^{\prime}}}, r^{\prime}_{\varrho^{\prime}_{s^{\prime}}}]))|$ if $([p^{\prime}_{\varrho^{\prime}_{s^{\prime}}}, q^{\prime}_{\varrho^{\prime}_{s^{\prime}}}]$, 
    $[\ell^{\prime}_{\varrho^{\prime}_{s^{\prime}}}, r^{\prime}_{\varrho^{\prime}_{s^{\prime}}}]) \in \Psi^{\prime}_{\source}$;
    otherwise $m^{\prime} = 1$.         
    If $([p^{\prime}_{\varrho^{\prime}_{s^{\prime}}}, q^{\prime}_{\varrho^{\prime}_{s^{\prime}}}], [\ell^{\prime}_{\varrho^{\prime}_{s^{\prime}}}, r^{\prime}_{\varrho^{\prime}_{s^{\prime}}}]) \not \in \Psi^{\prime \symB}$, 
    then the following equation holds: 
    \begin{equation*}
        \begin{split}
        \mathcal{L}^{\prime}_{\BiSQ}(u^{\prime}_{s^{\prime}}) &= \mathcal{L}_{\BiSQ}(u_{s}) - m^{\prime}|\Psi_{0} \cap (\Psi_{\OLD} \setminus \Psi_{\run}) \cap \Psi_{\str}(T[p_{\varrho_{s}}-1..r_{\varrho_{s}}+1])| \\
        &+ m^{\prime}|\Psi^{\prime}_{0} \cap (\Psi^{\prime}_{\NEW} \setminus \Psi^{\prime}_{\run}) \cap \Psi^{\prime}_{\str}(T[p_{\varrho_{s}}-1..r_{\varrho_{s}}+1])|.  
        \end{split}          
    \end{equation*}
    Otherwise (i.e., $([p^{\prime}_{\varrho^{\prime}_{s^{\prime}}}, q^{\prime}_{\varrho^{\prime}_{s^{\prime}}}], [\ell^{\prime}_{\varrho^{\prime}_{s^{\prime}}}, r^{\prime}_{\varrho^{\prime}_{s^{\prime}}}]) \in \Psi^{\prime \symB}$), 
    then the following equation holds: 
    \begin{equation*}
        \begin{split}
        \mathcal{L}^{\prime}_{\BiSQ}(u^{\prime}_{s^{\prime}}) &= m^{\prime}|\Psi^{\prime}_{0} \cap (\Psi^{\prime}_{\NEW} \setminus \Psi^{\prime}_{\run}) \cap \Psi^{\prime}_{\str}(T[p_{\varrho_{s}}-1..r_{\varrho_{s}}+1])|. 
        \end{split}          
    \end{equation*}
\end{lemma}
\begin{proof}
    Let $Z = T^{\prime}[p^{\prime}_{\varrho^{\prime}_{s^{\prime}}}-1..r^{\prime}_{\varrho^{\prime}_{s^{\prime}}}+1]$ for simplicity. 
    Because of $([p^{\prime}_{\varrho^{\prime}_{s^{\prime}}}, q^{\prime}_{\varrho^{\prime}_{s^{\prime}}}], [\ell^{\prime}_{\varrho^{\prime}_{s^{\prime}}}, r^{\prime}_{\varrho^{\prime}_{s^{\prime}}}]) \in \Psi^{\prime}_{\samp}$, 
    $([p^{\prime}_{\varrho^{\prime}_{s^{\prime}}}, q^{\prime}_{\varrho^{\prime}_{s^{\prime}}}], [\ell^{\prime}_{\varrho^{\prime}_{s^{\prime}}}, r^{\prime}_{\varrho^{\prime}_{s^{\prime}}}]) \not \in \Psi^{\prime}_{\run}$ 
    follows from Lemma~\ref{lem:samp_basic_property}~\ref{enum:samp_basic_property:3}. 
        
    \textbf{Proof of Lemma~\ref{lem:dynamic_bis_label} for $([p^{\prime}_{\varrho^{\prime}_{s^{\prime}}}, q^{\prime}_{\varrho^{\prime}_{s^{\prime}}}], [\ell^{\prime}_{\varrho^{\prime}_{s^{\prime}}}, r^{\prime}_{\varrho^{\prime}_{s^{\prime}}}]) \not \in \Psi^{\prime \symB}$.}
    Let $\alpha = |f_{\recover}(([p_{\varrho_{s}}, q_{\varrho_{s}}], [\ell_{\varrho_{s}}$, $r_{\varrho_{s}}]))|$ 
    and $\alpha^{\prime} = |f^{\prime}_{\recover}(([p^{\prime}_{\varrho^{\prime}_{s^{\prime}}}, q^{\prime}_{\varrho^{\prime}_{s^{\prime}}}], [\ell^{\prime}_{\varrho^{\prime}_{s^{\prime}}}, r^{\prime}_{\varrho^{\prime}_{s^{\prime}}}]))|$ for simplicity.     
    Let $m = \alpha$ if $([p_{\varrho_{s}}, q_{\varrho_{s}}]$, 
    $[\ell_{\varrho_{s}}, r_{\varrho_{s}}]) \in \Psi_{\source}$;
    otherwise $m = 1$. 
    
    We prove $m = m^{\prime}$. 
    If $([p_{\varrho_{s}}, q_{\varrho_{s}}], [\ell_{\varrho_{s}}, r_{\varrho_{s}}]) \in \Psi_{\source}$, 
    then $([p^{\prime}_{\varrho^{\prime}_{s^{\prime}}}, q^{\prime}_{\varrho^{\prime}_{s^{\prime}}}]$, 
    $[\ell^{\prime}_{\varrho^{\prime}_{s^{\prime}}}, r^{\prime}_{\varrho^{\prime}_{s^{\prime}}}]) \in \Psi^{\prime}_{\source}$ follows from Lemma~\ref{lem:dynamic_RR_subset}~\ref{enum:dynamic_RR_subset:source}. 
    Lemma~\ref{lem:dynamic_f_recover}~\ref{enum:dynamic_f_recover:X} shows that 
    $\alpha = \alpha^{\prime}$ holds. 
    Therefore, $m = m^{\prime}$ follows from $m = \alpha$, 
    $m^{\prime} = \alpha^{\prime}$, 
    and $\alpha = \alpha^{\prime}$. 
    
    Otherwise (i.e., $([p_{\varrho_{s}}, q_{\varrho_{s}}], [\ell_{\varrho_{s}}, r_{\varrho_{s}}]) \not \in \Psi_{\source}$), $m = 1$. 
    $m^{\prime} = 1$ holds because 
    $([p^{\prime}_{\varrho^{\prime}_{s^{\prime}}}, q^{\prime}_{\varrho^{\prime}_{s^{\prime}}}]$, 
    $[\ell^{\prime}_{\varrho^{\prime}_{s^{\prime}}}, r^{\prime}_{\varrho^{\prime}_{s^{\prime}}}]) \not \in \Psi^{\prime}_{\source}$ follows from Lemma~\ref{lem:dynamic_RR_subset}~\ref{enum:dynamic_RR_subset:source}. 
    Therefore, $m = m^{\prime}$ holds. 

    We prove Lemma~\ref{lem:dynamic_bis_label}. 
    $\mathcal{L}_{\BiSQ}(u_{s}) = m |\Psi_{\str}(Z)|$ follows from the definition of the label function $\mathcal{L}_{\BiSQ}$. 
    Similarly, 
    $\mathcal{L}^{\prime}_{\BiSQ}(u^{\prime}_{s^{\prime}}) = m^{\prime} |\Psi^{\prime}_{\str}(Z)|$ follows from the definition of the label function $\mathcal{L}^{\prime}_{\BiSQ}$. 
    $|\Psi^{\prime}_{\str}(Z)| = |\Psi_{\str}(Z)| - |\Psi_{0} \cap (\Psi_{\OLD} \setminus \Psi_{\run}) \cap \Psi_{\str}(Z)| + |\Psi^{\prime}_{0} \cap (\Psi^{\prime}_{\NEW} \setminus \Psi^{\prime}_{\run}) \cap \Psi^{\prime}_{\str}(Z)|$ follows from Corollary~\ref{cor:dynamic_str_formula}~\ref{enum:dynamic_str_formula:1}. 
    Therefore, the following equation holds: 
    \begin{equation*}
        \begin{split}
        \mathcal{L}^{\prime}_{\BiSQ}(u^{\prime}_{s^{\prime}}) &= m^{\prime} |\Psi^{\prime}_{\str}(Z)| \\
        &= m^{\prime} |\Psi_{\str}(Z)| - m^{\prime} |\Psi_{0} \cap (\Psi_{\OLD} \setminus \Psi_{\run}) \cap \Psi_{\str}(Z)| + m^{\prime} |\Psi^{\prime}_{0} \cap (\Psi^{\prime}_{\NEW} \setminus \Psi^{\prime}_{\run}) \cap \Psi^{\prime}_{\str}(Z)| \\
        &= m |\Psi_{\str}(Z)| - m^{\prime} |\Psi_{0} \cap (\Psi_{\OLD} \setminus \Psi_{\run}) \cap \Psi_{\str}(Z)| + m^{\prime} |\Psi^{\prime}_{0} \cap (\Psi^{\prime}_{\NEW} \setminus \Psi^{\prime}_{\run}) \cap \Psi^{\prime}_{\str}(Z)| \\
        &= \mathcal{L}_{\BiSQ}(u_{s}) - m^{\prime} |\Psi_{0} \cap (\Psi_{\OLD} \setminus \Psi_{\run}) \cap \Psi_{\str}(Z)| + m^{\prime} |\Psi^{\prime}_{0} \cap (\Psi^{\prime}_{\NEW} \setminus \Psi^{\prime}_{\run}) \cap \Psi^{\prime}_{\str}(Z)|.
        \end{split}          
    \end{equation*}

    \textbf{Proof of Lemma~\ref{lem:dynamic_bis_label} for $([p^{\prime}_{\varrho^{\prime}_{s^{\prime}}}, q^{\prime}_{\varrho^{\prime}_{s^{\prime}}}], [\ell^{\prime}_{\varrho^{\prime}_{s^{\prime}}}, r^{\prime}_{\varrho^{\prime}_{s^{\prime}}}]) \in \Psi^{\prime \symB}$.}
    Because of $([p^{\prime}_{\varrho^{\prime}_{s^{\prime}}}, q^{\prime}_{\varrho^{\prime}_{s^{\prime}}}], [\ell^{\prime}_{\varrho^{\prime}_{s^{\prime}}}, r^{\prime}_{\varrho^{\prime}_{s^{\prime}}}]) \in \Psi^{\prime \symB}$, 
    $(\Psi_{\RR} \setminus \Psi_{\OLD}) \cap \Psi_{\str}(Z) = \emptyset$ follows from the definition of set $\mathcal{Z}^{\symB}$. 
    Corollary~\ref{cor:dynamic_str_formula}~\ref{enum:dynamic_str_formula:2} shows that 
    $|\Psi^{\prime}_{\str}(Z)| = |\Psi^{\prime}_{0} \cap (\Psi^{\prime}_{\NEW} \setminus \Psi^{\prime}_{\run}) \cap \Psi^{\prime}_{\str}(Z)|$ holds. 
    Therefore, the following equation holds: 
    \begin{equation*}
        \begin{split}
        \mathcal{L}^{\prime}_{\BiSQ}(u^{\prime}_{s^{\prime}}) &= m^{\prime} |\Psi^{\prime}_{\str}(Z)| \\
        &= m^{\prime} |\Psi^{\prime}_{0} \cap (\Psi^{\prime}_{\NEW} \setminus \Psi^{\prime}_{\run}) \cap \Psi^{\prime}_{\str}(Z)|.
        \end{split}          
    \end{equation*}
\end{proof}

The following lemma states the relationship between two sets 
$\Psi_{0} \cap (\Psi_{\RR} \setminus \Psi_{\run})$ and $\Psi_{0} \cap \Psi_{\samp}$. 
\begin{lemma}\label{lem:varrho_search}
Consider an interval attractor $([p, q], [\ell, r]) \in \Psi_{0} \cap (\Psi_{\RR} \setminus \Psi_{\run})$. 
Let $x \in [1, k]$ be the smallest integer 
satisfying 
$T[q_{\varrho_{x}}..q_{\varrho_{x}}+1] = T[q..q+1]$ 
and $T[p_{\varrho_{x}}-1..r_{\varrho_{x}}+1] = T[p-1..r+1]$.
Then, the integer $x$ exists. 
\end{lemma}
\begin{proof}
    Because of $([p, q], [\ell, r]) \in \Psi_{\RR} \setminus \Psi_{\run}$,     
    Lemma~\ref{lem:samp_basic_property}~\ref{enum:samp_basic_property:2} shows that 
    sampling subset $\Psi_{\samp}$ contains an interval attractor $([p^{\prime}, q^{\prime}], [\ell^{\prime}, r^{\prime}])$ satisfying $T[p-1..r+1] = T[p^{\prime}-1..r^{\prime}+1]$. 
    Because of $([p, q], [\ell, r]) \in \Psi_{0}$, 
    Lemma~\ref{lem:psi_str_property}~\ref{enum:psi_str_property:2} shows that 
    $([p^{\prime}, q^{\prime}], [\ell^{\prime}, r^{\prime}]) \in \Psi_{0}$ holds. 
    Because of $([p^{\prime}, q^{\prime}], [\ell^{\prime}, r^{\prime}]) \in \Psi_{0} \cap \Psi_{\samp}$, 
    there exists an integer $x \in [1, k]$ satisfying 
    $([p_{\varrho_{x}}, q_{\varrho_{x}}], [\ell_{\varrho_{x}}, r_{\varrho_{x}}]) = ([p^{\prime}, q^{\prime}], [\ell^{\prime}, r^{\prime}])$. 

    Lemma~\ref{lem:psi_str_property}~\ref{enum:psi_str_property:3} indicates that 
    $T[q_{\varrho_{x}}..q_{\varrho_{x}}+1] = T[q..q+1]$ holds. 
    Therefore, Lemma~\ref{lem:varrho_search} follows from 
    $T[q_{\varrho_{x}}..q_{\varrho_{x}}+1] = T[q..q+1]$ and $T[p_{\varrho_{x}}-1..r_{\varrho_{x}}+1] = T[p-1..r+1]$. 
\end{proof}

\subsubsection{Algorithm}\label{subsubsec:dynamic_bis_query_algo}
We prove Lemma~\ref{lem:dynamic_bis_summary}, i.e., 
we show that the data structures for bigram search query (Section~\ref{subsubsec:bis_ds}) 
can be updated in expected $O((\max \{H, H^{\prime}, \log (nHH^{\prime}) \})^{4})$ time using 
(A) the dynamic data structures for the RR-DAG of RLSLP $\mathcal{G}^{R}$ (Section~\ref{subsubsec:rrdag_ds}) 
and sample query (Section~\ref{subsec:sample_query}), 
and (B) the interval attractors obtained from Lemma~\ref{lem:dynamic_IA_summary}. 

The data structures for bigram search query consists of the doubly linked list $\mathbf{L}_{\BiSQ}$ representing sequence $\mathbf{Q}_{\BiSQ} = u_{1}, u_{2}, \ldots, u_{k}$ of nodes. 
Lemma~\ref{lem:dynamic_bis_back_correspondence}, 
Lemma~\ref{lem:dynamic_bis_proceeding_correspondence}, 
and Lemma~\ref{lem:dynamic_bis_order} indicate that 
doubly linked list $\mathbf{L}_{\BiSQ}$ represents $\mathbf{Q}^{\prime}_{\BiSQ} = u^{\prime}_{1}, u^{\prime}_{2}, \ldots, u^{\prime}_{k^{\prime}}$ by modifying the doubly linked list as follows:
\begin{description}
    \item [Modification 1:] 
    For each interval attractor $([p_{s}, q_{s}], [\ell_{s}, r_{s}])$ of set $\Psi_{0} \cap ((\Psi_{\samp} \cap \Psi_{\OLD}) \setminus \Psi^{\symA})$, 
    let $u_{s^{\prime}}$ be the node corresponding the interval attractor $([p_{s}, q_{s}], [\ell_{s}, r_{s}])$ 
    (i.e., $\varrho_{s^{\prime}} = s$). 
    Then, node $u_{s^{\prime}}$ is removed from sequence $\mathbf{Q}_{\BiSQ}$, 
    and hence the $s^{\prime}$-th element is removed from doubly linked list $\mathbf{L}_{\BiSQ}$.
    \item [Modification 2:] 
    For each interval attractor of set $\Psi^{\prime}_{0} \cap \Psi^{\prime \symB}$, 
    a new node corresponding to the interval attractor is created and inserted into sequence $\mathbf{Q}_{\BiSQ}$ 
    at an appropriate position. 
    Therefore, a new element is inserted into sequence $\mathbf{Q}_{\BiSQ}$ at the same position.
\end{description}

The doubly linked list $\mathbf{L}_{\BiSQ}$ stores 
the integers obtained from label function $\mathcal{L}_{\BiSQ}$. 
Lemma~\ref{lem:dynamic_bis_label} indicates that 
these integers can be updated by modifying the doubly linked list $\mathbf{L}_{\BiSQ}$ as follows: 
\begin{description}
    \item [Modification 3:] 
    For each interval attractor $([p, q], [\ell, r]) \in \Psi_{0} \cap (\Psi_{\OLD} \setminus \Psi_{\run})$, 
    let $u_{s}$ be a node of sequence $\mathbf{Q}_{\BiSQ}$ satisfying 
    $T[p_{\varrho_{s}}-1..r_{\varrho_{s}}+1] = T[p-1..r+1]$. 
    Then, $|f_{\recover}(([p, q], [\ell, r]))|$ is subtracted from the integer 
    stored in the $s$-th element of doubly linked list $\mathbf{L}_{\BiSQ}$ 
    if $([p, q], [\ell, r]) \in \Psi_{\source}$. 
    Otherwise, $1$ is subtracted from the integer 
    stored in the $s$-th element;
    \item [Modification 4:] 
    After Modification 1 and Modification 2, 
    each $s$-th element of $\mathbf{L}_{\BiSQ}$ corresponds to the $s$-th node $u^{\prime}_{s}$ 
    of sequence $\mathbf{Q}^{\prime}_{\BiSQ}$. 
    For each interval attractor $([p, q], [\ell, r])$ of 
    set $\Psi^{\prime}_{0} \cap (\Psi^{\prime}_{\NEW} \setminus \Psi^{\prime}_{\run})$, 
    let $u^{\prime}_{s}$ be a node of sequence $\mathbf{Q}^{\prime}_{\BiSQ}$ satisfying 
    $T^{\prime}[p^{\prime}_{\varrho^{\prime}_{s}}-1..r^{\prime}_{\varrho^{\prime}_{s}}+1] = T^{\prime}[p-1..r+1]$. 
    Then, $|f^{\prime}_{\recover}(([p, q], [\ell, r]))|$ is subtracted from the integer 
    stored in the $s$-th element of doubly linked list $\mathbf{L}_{\BiSQ}$ 
    if $([p, q], [\ell, r]) \in \Psi^{\prime}_{\source}$. 
    Otherwise, $1$ is subtracted from the integer 
    stored in the $s$-th element.
\end{description}

Therefore, we can update the data structures for bigram search query by applying the above 
four modifications to doubly linked list $\mathbf{L}_{\BiSQ}$. 

The algorithm updating the dynamic data structures for bigram search query consists of the following six phases.

\paragraph{Phase (i).}
In the first phase, we compute two sets $\Psi_{0} \cap (\Psi_{\OLD} \setminus \Psi_{\run})$ 
and $\Psi_{0} \cap ((\Psi_{\samp} \cap \Psi_{\OLD}) \setminus \Psi^{\symA})$ of interval attractors. 
Here, two sets $\Psi_{\OLD} \setminus \Psi_{\run}$ and $(\Psi_{\samp} \cap \Psi_{\OLD}) \setminus \Psi^{\symA}$ are obtained from Lemma~\ref{lem:dynamic_IA_summary}. 

We obtain set $\Psi_{0} \cap (\Psi_{\OLD} \setminus \Psi_{\run})$ by 
verifying whether 
each interval attractor of set $\Psi_{\OLD} \setminus \Psi_{\run}$ is contained in 
set $\Psi_{0}$. 
This verification can be executed by level query. 
Similarly, we obtain set $\Psi_{0} \cap ((\Psi_{\samp} \cap \Psi_{\OLD}) \setminus \Psi^{\symA})$ by 
verifying whether 
each interval attractor of set $((\Psi_{\samp} \cap \Psi_{\OLD}) \setminus \Psi^{\symA})$ is contained in 
set $\Psi_{0}$. 
Therefore, we can obtain the two sets $\Psi_{0} \cap (\Psi_{\OLD} \setminus \Psi_{\run})$ 
and $\Psi_{0} \cap ((\Psi_{\samp} \cap \Psi_{\OLD}) \setminus \Psi^{\symA})$ by 
executing $O(|\Psi_{\OLD} \setminus \Psi_{\run}| + |\Psi_{\samp} \cap \Psi_{\OLD}|)$ level queries. 

$\mathbb{E}[|\Psi_{\samp} \cap \Psi_{\OLD}|] = O(H + \log n)$ 
and $\mathbb{E}[|\Psi_{\OLD} \setminus \Psi_{\run}|] = O(H + \log n)$ follow from Lemma~\ref{lem:dynamic_psi_overlap}. 
Each level query takes $O(H^{2})$ time. 
Therefore, the first phase takes expected $O((H + \log n) H^{2})$ time.

\paragraph{Phase (ii).}
In the second phase, 
we update doubly linked list $\mathbf{L}_{\BiSQ}$ by Modification 3. 
We execute two steps 
for each interval attractor $([p, q], [\ell, r]) \in \Psi_{0} \cap (\Psi_{\OLD} \setminus \Psi_{\run})$. 
In the first step, 
we find the node $u_{s}$ of sequence $\mathbf{Q}_{\BiSQ}$ satisfying 
$T[p_{\varrho_{s}}-1..r_{\varrho_{s}}+1] = T[p-1..r+1]$. 
We leverage Lemma~\ref{lem:varrho_search} for computing the integer $s$. 

Let $x \in [1, k]$ be the smallest integer satisfying 
$T[q_{\varrho_{x}}..q_{\varrho_{x}}+1] = T[q..q+1]$ 
and $T[p_{\varrho_{x}}-1..r_{\varrho_{x}}+1] = T[p-1..r+1]$. 
Then, $s = x$ holds, 
and Lemma~\ref{lem:varrho_search} shows that the integer $x$ exists. 
From the definition of the permutation $\varrho_{1}, \varrho_{2}, \ldots, \varrho_{k}$, 
the integer $x$ can be found by binary search on 
the $k$ interval attractors 
$([p_{\varrho_{1}}, q_{\varrho_{1}}], [\ell_{\varrho_{1}}, r_{\varrho_{1}}])$, 
$([p_{\varrho_{2}}, q_{\varrho_{2}}], [\ell_{\varrho_{2}}, r_{\varrho_{2}}])$, 
$\ldots$, 
$([p_{\varrho_{k}}, q_{\varrho_{k}}], [\ell_{\varrho_{k}}, r_{\varrho_{k}}])$. 
Here, each interval attractor $([p_{\varrho_{j}}, q_{\varrho_{j}}], [\ell_{\varrho_{j}}, r_{\varrho_{j}}])$ 
can be obtained in $O(\log n)$ time because 
(A) the $j$-th element of doubly linked list $\mathbf{L}_{\BiSQ}$ stores a pointer to 
sequence a node $v$ of $\mathbf{Q}_{\samp}$ corresponding to the interval attractor $([p_{\varrho_{j}}, q_{\varrho_{j}}], [\ell_{\varrho_{j}}, r_{\varrho_{j}}])$, 
and (B) we can recover the interval attractor $([p_{\varrho_{j}}, q_{\varrho_{j}}], [\ell_{\varrho_{j}}, r_{\varrho_{j}}])$ from the node $v$ in $O(\log n)$ time using the algorithm presented in Section~\ref{subsubsec:computation_delta_samp}. 

For finding the integer $x$ (i.e., $s$), 
we can access node $u_{s}$ in $O(\log n)$ time using 
the list indexing data structure built on the doubly linked list $\mathbf{L}_{\BiSQ}$. 
Therefore, the first step takes $O(H^{2} \log n + \log^{2} n)$ time. 

In the second step, 
$|f_{\recover}(([p, q], [\ell, r]))|$ is subtracted from the integer 
stored in the $s$-th element of doubly linked list $\mathbf{L}_{\BiSQ}$ 
if $([p, q], [\ell, r]) \in \Psi_{\source}$. 
Otherwise, $1$ is subtracted from the integer stored in the $s$-th element. 
We verify whether $([p, q], [\ell, r]) \in \Psi_{\source}$ or not by verify-source query $\sourceQ(([p, q], [\ell, r]))$, which takes $O(H^{2} \log n)$ time. 
The integer $|f_{\recover}(([p, q], [\ell, r]))|$ can be computed by 
r-size query $\rsizeQ(([p, q], [\ell, r]))$, which takes $O(H^{2})$ time. 
Therefore, the second step takes $O(H^{2} \log n)$ time. 

Finally, 
the running time of the second phase is $O(|\Psi_{\OLD} \setminus \Psi_{\run}| (H^{2} \log n + \log^{2} n))$, 
i.e., this phase takes expected $O((H + \log n) (H^{2} \log n + \log^{2} n))$ time. 

\paragraph{Phase (iii).}
In the third phase, 
we update doubly linked list $\mathbf{L}_{\BiSQ}$ by Modification 2. 
We execute two steps 
for each interval attractor $([p_{s}, q_{s}], [\ell_{s}, r_{s}])$ of set $\Psi_{0} \cap ((\Psi_{\samp} \cap \Psi_{\OLD}) \setminus \Psi^{\symA})$. 
In the first step, 
we find the node $u_{s^{\prime}}$ of sequence $\mathbf{Q}_{\BiSQ}$ satisfying 
$T[p_{\varrho_{s^{\prime}}}-1..r_{\varrho_{s^{\prime}}}+1] = T[p_{s}-1..r_{s}+1]$. 
Similar to the second phase, 
the node $u_{s^{\prime}}$ can be found in $O(H^{2} \log n + \log^{2} n)$ time. 
In the second step, 
the $s^{\prime}$-th element is removed from doubly linked list $\mathbf{L}_{\BiSQ}$. 
This removal takes $O(\log n)$ time. 

The running time of the third phase is $O(|\Psi_{\samp} \cap \Psi_{\OLD}| (H^{2} \log n + \log^{2} n))$ time in total, i.e., 
this phase takes expected $O((H + \log n) (H^{2} \log n + \log^{2} n))$ time. 

\paragraph{Phase (iv).}
In the fourth phase, 
we compute two sets $\Psi^{\prime}_{0} \cap (\Psi^{\prime}_{\NEW} \setminus \Psi^{\prime}_{\run})$ 
and $\Psi^{\prime}_{0} \cap \Psi^{\prime \symB}$ of interval attractors. 
Here, two sets $\Psi^{\prime}_{\NEW} \setminus \Psi^{\prime}_{\run}$ and $\Psi^{\prime \symB}$ are obtained from Lemma~\ref{lem:dynamic_IA_summary}. 
This phase is executed after updating the dynamic data structures for RR-DAG and sample query. 

We obtain two set $\Psi^{\prime}_{0} \cap (\Psi^{\prime}_{\NEW} \setminus \Psi^{\prime}_{\run})$ 
and $\Psi^{\prime}_{0} \cap \Psi^{\prime \symB}$ by verifying whether 
each interval attractor of two sets 
$\Psi^{\prime}_{\NEW} \setminus \Psi^{\prime}_{\run}$ and $\Psi^{\prime \symB}$ is contained 
in set $\Psi^{\prime}_{0}$. 
Similar to the first phase, 
this verification can be executed by $|\Psi^{\prime}_{\NEW} \setminus \Psi^{\prime}_{\run}| + |\Psi^{\prime \symB}|$ level queries on set $\Psi^{\prime}_{\RR}$. 
$\Psi^{\prime \symB} \subseteq \Psi^{\prime}_{\NEW} \setminus \Psi^{\prime}_{\run}$ holds, 
and $\mathbb{E}[|\Psi^{\prime}_{\NEW} \setminus \Psi^{\prime}_{\run}|] = O(H^{\prime} + \log n)$ follows from Lemma~\ref{lem:dynamic_psi_overlap}. 
Therefore, the fourth phase takes expected $O((H^{\prime} + \log n) H^{\prime 2})$ time.

\paragraph{Phase (v).}
In the fifth phase, 
we update doubly linked list $\mathbf{L}_{\BiSQ}$ by Modification 3. 

Let $\hat{u}_{1}$, $\hat{u}_{2}$, $\ldots$, $\hat{u}_{\hat{k}}$ be the nodes of sequence $\mathbf{Q}_{\BiSQ}$ after executing the fourth phase. 
Then, each node $\hat{u}_{s}$ corresponds to an interval attractor $([\hat{p}_{s}, \hat{q}_{s}], [\hat{\ell}_{s}, \hat{r}_{s}])$ in set $\Psi_{0} \cap \Psi_{\samp}$. 
The $\hat{k}$ nodes are sorted in the order of the permutation $\varrho^{\prime}_{1}$, $\varrho^{\prime}_{2}$, 
$\ldots$, $\varrho^{\prime}_{k^{\prime}}$. 
That is, for each integer $s \in [1, \hat{k}-1]$, 
node $\hat{u}_{s}$ satisfies either of the following two conditions: 
\begin{itemize}
    \item $T^{\prime}[\hat{q}_{s}..\hat{q}_{s}+1] \prec T^{\prime}[\hat{q}_{s+1}..\hat{q}_{s+1}+1]$; 
    \item $T^{\prime}[\hat{q}_{s}..\hat{q}_{s}+1] = T^{\prime}[\hat{q}_{s+1}..\hat{q}_{s+1}+1]$ and 
    $T^{\prime}[\hat{p}_{s}-1..\hat{r}_{s}+1] \prec T^{\prime}[\hat{p}_{s+1}-1..\hat{r}_{s+1}+1]$.     
\end{itemize}

We execute three steps 
for each interval attractor $([p, q], [\ell, r]) \in \Psi^{\prime}_{0} \cap \Psi^{\prime \symB}$. 
Let $s^{\prime}$ be the smallest integer in set $[1, \hat{k}]$ satisfying either of the following two conditions: 
\begin{itemize}
    \item $T^{\prime}[q..q+1] \prec T^{\prime}[\hat{q}_{s^{\prime}}..\hat{q}_{s^{\prime}}+1]$; 
    \item $T^{\prime}[q..q+1] = T^{\prime}[\hat{q}_{s^{\prime}}..\hat{q}_{s^{\prime}}+1]$ and 
    $T^{\prime}[p-1..r+1] \prec T^{\prime}[\hat{q}_{s^{\prime}}-1..\hat{r}_{s^{\prime}}+1]$.     
\end{itemize}
In the first step, 
we find the node $\hat{u}_{s^{\prime}}$ in sequence $\mathbf{Q}_{\BiSQ}$. 
Similar to the second phase, 
the integer $s^{\prime}$ can be found in $O(H^{2} \log n + \log^{2} n)$ time 
by binary search on the $\hat{k}$ interval attractors represented as 
the nodes of sequence $\mathbf{Q}_{\BiSQ}$. 

For the sequence $\mathbf{Q}_{\samp}$ introduced in Section~\ref{subsec:sample_query}, 
in the second step, 
we find the node $v$ of sequence $\mathbf{Q}_{\samp}$ 
corresponding to the interval attractor $([p, q], [\ell, r])$. 
The node $v$ can be found in $O(\log^{2} n)$ time by binary search on the nodes of sequence $\mathbf{Q}_{\samp}$ 
using the dynamic data structures for sample query. 

In the third step, 
we insert a new element into doubly linked list $\mathbf{L}_{\BiSQ}$ as the $s^{\prime}$-th element in 
$O(\log n)$ time, 
which corresponds to the interval attractor $([p, q], [\ell, r])$. 
This element stores zero and a pointer to the node $v$. 
The integer stored in this element is appropriately updated in the next phase. 

The running time of the fifth phase is $O(|\Psi^{\prime \symB}| (H^{2} \log n + \log^{2} n))$ time in total, i.e., 
this phase takes expected $O((H^{\prime} + \log n) (H^{2} \log n + \log^{2} n))$ time. 

After executing the fifth phase, 
sequence $\mathbf{Q}_{\BiSQ}$ is changed to $\mathbf{Q}^{\prime}_{\BiSQ}$, 
and each node $v^{\prime}_{s}$ of the sequence $\mathbf{Q}^{\prime}_{\BiSQ}$ 
is represented as the $s$-th element of doubly linked list $\mathbf{L}_{\BiSQ}$. 

\paragraph{Phase (vi).}
In the sixth phase, 
we update doubly linked list $\mathbf{L}_{\BiSQ}$ by Modification 4. 
We execute two steps 
for each interval attractor $([p, q], [\ell, r]) \in \Psi^{\prime}_{0} \cap (\Psi^{\prime}_{\NEW} \setminus \Psi^{\prime}_{\run})$. 
In the first step, 
we find the node $u^{\prime}_{s}$ of sequence $\mathbf{Q}_{\BiSQ}$ satisfying 
$T[p^{\prime}_{\varrho^{\prime}_{s}}-1..r^{\prime}_{\varrho^{\prime}_{s}}+1] = T[p-1..r+1]$. 
Similar to the second phase, 
this node $u^{\prime}_{s}$ can be found in $O(H^{\prime 2} \log n + \log^{2} n)$ time.

In the second step, 
$|f^{\prime}_{\recover}(([p, q], [\ell, r]))|$ is added to the integer 
stored in the $s$-th element of doubly linked list $\mathbf{L}_{\BiSQ}$ 
if $([p, q], [\ell, r]) \in \Psi^{\prime}_{\source}$. 
Otherwise, $1$ is added to from the integer stored in the $s$-th element. 
Similar to the second phase, 
the second step can be executed using verify-source and r-size queries on set $\Psi^{\prime}_{\RR}$. 
Therefore, the second step takes $O(H^{\prime 2} \log n)$ time. 

After executing the sixth phase, 
each $s$-th element of doubly linked list $\mathbf{L}_{\BiSQ}$ stores the integer obtained from 
label function $\mathcal{L}^{\prime}_{\BiSQ}(u^{\prime}_{s})$. 
The running time of the sixth phase is $O(|\Psi^{\prime}_{\NEW} \setminus \Psi^{\prime}_{\run}| (H^{\prime 2} \log n + \log^{2} n))$, 
i.e., this phase takes expected $O((H^{\prime} + \log n) (H^{\prime 2} \log n + \log^{2} n))$ time. 

Finally, we can update the dynamic data structures for bigram search query in expected $O((H + H^{\prime} + \log n) (H^{2} \log n + H^{\prime 2} \log n + \log^{2} n))$ time. 
Therefore, Lemma~\ref{lem:dynamic_bis_summary} holds. 

\subsection{Update of Weighted Points of Non-periodic Interval Attractors}
%\subsection{Key Observation for Updating Weighted Points for Non-periodic Interval Attractors}
This section explains the key observation for updating weighted points representing non-periodic interval attractors. 
Consider the set $\mathcal{J}^h \subseteq \mathcal{X} \times \mathcal{Y} \times \mathbb{Z} \times \mathbb{Z}$ of weighted points 
introduced in Section~\ref{subsec:weighted_point} for $h \in [0, H]$. 
Here, each element of $\mathcal{J}^h$ corresponds to a non-periodic interval attractor in the $h$-th level interval attractors $\Psi_{h}$. 
Similarly, let $\mathcal{J}^{\prime h} \subseteq \mathcal{X} \times \mathcal{Y} \times \mathbb{Z} \times \mathbb{Z}$ 
be a set weighted points such that 
each element of $\mathcal{J}^h$ corresponds to a non-periodic interval attractor in the $h$-th level interval attractors $\Psi^{\prime}_{h}$ for $h \in [0, H^{\prime}]$. 
Here, $H^{\prime}$ is the height of the derivation tree of RLSLP $\mathcal{G}^{R}_{\ins}$. 
For simplicity, 
let $\mathcal{J}^{h} = \emptyset$ for $h \geq H+1$ and $\mathcal{J}^{\prime h} = \emptyset$ for $h \geq H^{\prime}+1$. 

After updating RLSLP $\mathcal{G}^{R}$, 
the weights of each weighted point $(x, y, \alpha, \beta)$ in set $\mathcal{J}^{h}$ is changed to 
that of the weighted point with the same coordinates in $\mathcal{J}^{\prime h}$. 
We explain the relationship between the two sets $\mathcal{J}^{h}$ and $\mathcal{J}^{\prime h}$ 
using two functions $\kappa_{\ins}(x, y)$ and $\kappa^\prime_{\ins}(x, y)$ for any $(x, y) \in \mathcal{X} \times \mathcal{Y}$. 

The first function $\kappa_{\ins}(x, y)$ is defined as 
the number of non-periodic interval attractors in $\Psi_{\RR}$ that satisfies two specific conditions. 
For each $([p, q], [\ell, r]) \in \Psi_{\RR}$, the conditions are as follows:
(A) $([p, q], [\ell, r]) \in \Psi_{\OLD}$ for the subset $\Psi_{\OLD}$ introduced in Section~\ref{subsec:relationship_old_and_new};
(B) the weighted point representing interval attractor $([p, q], [\ell, r])$ is located at the same $(x, y)$ in the $\mathcal{X} \times \mathcal{Y}$ plane. 
Similarly, the second function $\kappa^\prime_{\ins}(x, y)$ is defined as the number of non-periodic interval attractors in $\Psi^{\prime}_{\RR}$ that satisfies two specific conditions. 
For each $([p^{\prime}, q^{\prime}], [\ell^{\prime}, r^{\prime}]) \in \Psi^{\prime}_{\RR}$, the conditions are as follows:
(a) $([p^{\prime}, q^{\prime}], [\ell^{\prime}, r^{\prime}]) \in \Psi^{\prime}_{\NEW}$ for the subset $\Psi^{\prime}_{\NEW}$ introduced in Section~\ref{subsec:relationship_old_and_new};
(b) the weighted point representing interval attractor $([p^{\prime}, q^{\prime}], [\ell^{\prime}, r^{\prime}])$ is located at the same $(x, y)$ in the $\mathcal{X} \times \mathcal{Y}$ plane. 

Set $\mathcal{J}^{\prime h}$ can be obtained by updating set $\mathcal{J}^{h}$ as follows. 
\begin{theorem}\label{theo:intro_weighted_point_formula_ins}
    Consider two sets $\mathcal{J}^{h}$ and $\mathcal{J}^{\prime h}$ of weighted points for an integer $h \in [0, \max \{ H, H^{\prime} \}]$. 
    Then, the following two statements hold. 
    \begin{enumerate}[label=\textbf{(\roman*)}]
    \item For each weighted point $(x, y, \alpha, \beta)$ in set $\mathcal{J}^{h}$, 
    set $\mathcal{J}^{\prime h}$ contains a weighted point $(x^{\prime}, y^{\prime}, \alpha^{\prime}, \beta^{\prime})$ 
    as $x = x^{\prime}$, $y = y^{\prime}$,  $\alpha^{\prime} = \alpha$ and $\beta^{\prime} = \beta - \kappa_{\ins}(x, y) + \kappa^{\prime}_{\ins}(x, y)$ if and only if $\beta - \kappa_{\ins}(x, y) + \kappa^{\prime}_{\ins}(x, y) \geq 1$. 
    \item Consider a weighted point $(x^{\prime}, y^{\prime}, \alpha^{\prime}, \beta^{\prime})$ in set $\mathcal{J}^{\prime h}$ such that the same point $(x^{\prime}, y^{\prime})$ with any weights is not contained in $\mathcal{J}^h$.      
    For any non-periodic interval attractor $([p^{\prime}, q^{\prime}], [\ell^{\prime}, r^{\prime}]) \in \Psi^{\prime}_{\RR}$ located at the same point $(x^\prime, y^\prime)$, condition $([p^{\prime}, q^{\prime}], [\ell^{\prime}, r^{\prime}]) \in \Psi^{\prime}_{\NEW}$ holds.
    \end{enumerate}
\end{theorem}
\begin{proof}
The proof of Theorem~\ref{theo:intro_weighted_point_formula_ins} is as follows. 

\textbf{Proof of Theorem~\ref{theo:intro_weighted_point_formula_ins}(i).}
The point of a non-periodic interval attractor $([p, q], [\ell, r]) \in \Psi_{\RR}$ is aggregated to the weighted point $(x, y, \alpha, \beta)$. 
Here, $x = T[p-1..\gamma-1]$ and $y = T[\gamma..r+1]$ hold for the attractor position $\gamma$ of the interval attractor $([p, q], [\ell, r])$. Let $d = |f_{\recover}(([p, q], [\ell, r]))|$ if $([p, q], [\ell, r]) \in \Psi_{\source}$; 
otherwise $d = 0$. 
From Lemma~\ref{lem:weight}~\ref{enum:weight:1}, Lemma~\ref{lem:psi_equality_basic_property}~\ref{enum:psi_equality_basic_property:5}, and 
Lemma~\ref{lem:samp_basic_property}~\ref{enum:samp_basic_property:2}, 
the first integer weight $\alpha$ of the weighted point $(x, y, \alpha, \beta)$ is equal to $1 + d$. 
Similarly, Lemma~\ref{lem:weight}~\ref{enum:weight:2} indicates that 
the second integer weight $\beta$ of the weighted point $(x, y, \alpha, \beta)$ is equal to $|\Psi_{\str}(T[p-1..r+1])|$. 

We prove $\kappa_{\ins}(x, y) = |\Psi_{\OLD} \cap \Psi_{\str}(Z)|$. 
$\kappa_{\ins}(x, y) = |(\Psi_{\OLD} \setminus \Psi_{\run}) \cap \Psi_{\str}(T[p-1..r+1])|$ can be proved using the fact that $T[p-1..r+1] = T[p^{\prime}-1..r^{\prime}+1] \Leftrightarrow T[p-1..\gamma-1] = T[p^{\prime}-1..\gamma^{\prime}-1]$ and $T[\gamma..r+1] = T[\gamma^{\prime}..r^{\prime}+1]$ holds for an interval attractor $([p^{\prime}, q^{\prime}], [\ell^{\prime}, r^{\prime}]) \in \Psi_{\RR}$ with attractor position $\gamma^{\prime}$. 
Lemma~\ref{lem:psi_equality_basic_property}~\ref{enum:psi_equality_basic_property:4} shows that 
$T[p-1..r+1] = T[p^{\prime}-1..r^{\prime}+1] \Rightarrow ([p^{\prime}, q^{\prime}], [\ell^{\prime}, r^{\prime}]) \not \in \Psi_{\run}$ holds. 
Therefore, we obtain the following equation: 
\begin{equation*}
        \begin{split}
            \kappa_{\ins}(x, y) &= |(\Psi_{\OLD} \setminus \Psi_{\run}) \cap \Psi_{\str}(T[p-1..r+1])| \\ 
            &= |\Psi_{\OLD} \cap \Psi_{\str}(T[p-1..r+1])|.            
        \end{split}
\end{equation*}
Similarly, we can prove $\kappa^{\prime}_{\ins}(x, y) = |\Psi^{\prime}_{\NEW} \cap \Psi^{\prime}_{\str}(Z)|$ using the same approach as for the proof of $\kappa_{\ins}(x, y) = |\Psi_{\OLD} \cap \Psi_{\str}(Z)|$. 

Let $m$ be the number of non-periodic interval attractors in set $\Psi^{\prime}_{h}$ such that 
each interval attractor $([p^{\prime}, q^{\prime}], [\ell^{\prime}, r^{\prime}])$ satisfies 
$T^{\prime}[p^{\prime}-1..r^{\prime}+1] = T[p-1..r+1]$ (i.e., $m = |(\Psi^{\prime}_{\str}(T[p-1..r+1]) \cap \Psi^{\prime}_{h}) \setminus \Psi^{\prime}_{\run}|$).
Here, 
$m = |(\Psi^{\prime}_{\str}(T[p-1..r+1])|$ holds 
because $|(\Psi^{\prime}_{\str}(T[p-1..r+1]) \cap \Psi^{\prime}_{h}) \setminus \Psi^{\prime}_{\run}| = |\Psi^{\prime}_{\str}(T[p-1..r+1]|$ follows from 
Lemma~\ref{lem:psi_equality_basic_property}~\ref{enum:psi_equality_basic_property:4} 
and 
Lemma~\ref{lem:psi_str_property}~\ref{enum:psi_str_property:2}. 
$|\Psi^{\prime}_{\str}(T[p-1..r+1])| = |\Psi_{\str}(Z)| - |\Psi_{\OLD} \cap \Psi_{\str}(Z)| + |\Psi^{\prime}_{\NEW} \cap \Psi^{\prime}_{\str}(Z)|$ follows from Lemma~\ref{lem:dynamic_and_basic_str_formula}~\ref{enum:dynamic_and_basic_str_formula:1}. 
Therefore, we obtain the following equation: 
\begin{equation*}
        \begin{split}
            m &= |\Psi^{\prime}_{\str}(T[p-1..r+1])| \\ 
            &= |\Psi_{\str}(T[p-1..r+1])| - |\Psi_{\OLD} \cap \Psi_{\str}(T[p-1..r+1])| + |\Psi^{\prime}_{\NEW} \cap \Psi^{\prime}_{\str}(T[p-1..r+1])| \\
            &= \beta - \kappa_{\ins}(x, y) + \kappa^{\prime}_{\ins}(x, y).            
        \end{split}
\end{equation*}

If $\beta - \kappa_{\ins}(x, y) + \kappa^{\prime}_{\ins}(x, y) \geq 1$ (i.e., $m \geq 1$), 
then 
set $\Psi^{\prime}_{\RR}$ contains a non-periodic interval attractor $([p^{\prime}, q^{\prime}], [\ell^{\prime}, r^{\prime}])$ such that 
it is aggregated into a point $(x, y)$ in $\mathcal{J}^{\prime h}$. 
This fact indicates that set $\mathcal{J}^{\prime h}$ contains 
a weighted point $(x^{\prime}, y^{\prime}, \alpha^{\prime}, \beta^{\prime})$ satisfying 
$x = x^{\prime}$ and $y = y^{\prime}$. 
Let $d^{\prime} = |f^{\prime}_{\recover}(([p^{\prime}, q^{\prime}], [\ell^{\prime}, r^{\prime}]))|$ if $([p^{\prime}, q^{\prime}], [\ell^{\prime}, r^{\prime}]) \in \Psi^{\prime}_{\source}$; 
otherwise $d^{\prime} = 0$. 
Similar to the weighted point $(x, y, \alpha, \beta)$, 
the weights $\alpha$ and $\beta$ of the weighted point $(x^{\prime}, y^{\prime}, \alpha^{\prime}, \beta^{\prime})$ are 
defined as $1+d^{\prime}$ and $|\Psi^{\prime}_{\str}(T[p-1..r+1])|$ (i.e., $m$), respectively. 
$\alpha = \alpha^{\prime}$ holds because $d = d^{\prime}$ follows from Lemma~\ref{lem:dynamic_f_recover}~\ref{enum:dynamic_f_recover:X}. 
$\beta^{\prime} = \beta - \kappa_{\ins}(x, y) + \kappa^{\prime}_{\ins}(x, y)$ follows from 
$\beta^{\prime} = m$ and $m = \beta - \kappa_{\ins}(x, y) + \kappa^{\prime}_{\ins}(x, y)$. 

Otherwise (i.e., $\beta - \kappa_{\ins}(x, y) + \kappa^{\prime}_{\ins}(x, y) = 0$), 
any non-periodic interval attractor of set $\Psi^{\prime}_{\RR}$ is not aggregated into a point $(x, y)$ in $\mathcal{J}^{\prime h}$. 
This fact indicates that 
set $\mathcal{J}^{\prime h}$ does not contain a weighted point $(x^{\prime}, y^{\prime}, \alpha^{\prime}, \beta^{\prime})$ satisfying $x = x^{\prime}$ and $y = y^{\prime}$. 
Therefore, we obtain Theorem~\ref{theo:intro_weighted_point_formula_ins}(i). 

\textbf{Proof of Theorem~\ref{theo:intro_weighted_point_formula_ins}(ii).}
The point of a non-periodic interval attractor $([p^{\prime}, q^{\prime}], [\ell^{\prime}, r^{\prime}]) \in \Psi^{\prime}_{\RR}$ is aggregated to the weighted point $(x^{\prime}, y^{\prime}, \alpha^{\prime}, \beta^{\prime})$. 
We prove $([p^{\prime}, q^{\prime}], [\ell^{\prime}, r^{\prime}]) \in \Psi^{\prime}_{\NEW}$ by contradiction. 
We assume that $([p^{\prime}, q^{\prime}], [\ell^{\prime}, r^{\prime}]) \not \in \Psi^{\prime}_{\NEW}$ holds (i.e., $([p^{\prime}, q^{\prime}], [\ell^{\prime}, r^{\prime}]) \in \Psi_{\LEFT} \cup \Psi_{\RIGHT}$). 
Then, Theorem~\ref{theo:RS_LEFT_RIGHT} shows that 
set $\Psi_{\RR}$ contains interval attractor $([p^{\prime}-\epsilon^{\prime}, q^{\prime}-\epsilon^{\prime}], [\ell^{\prime}-\epsilon^{\prime}, r^{\prime}-\epsilon^{\prime}])$. 
Here, $\epsilon^{\prime} = 1$ if $([p^{\prime}, q^{\prime}], [\ell^{\prime}, r^{\prime}]) \in \Psi_{\RIGHT}$; otherwise $\epsilon^{\prime} = 0$. 
Corollary~\ref{cor:RB_IA_after_before} shows that $T^{\prime}[p^{\prime}-1..r^{\prime}+1] = T[p^{\prime}-\epsilon^{\prime}-1..r^{\prime}-\epsilon^{\prime}+1]$ holds. 
Lemma~\ref{lem:dynamic_RR_subset}~\ref{enum:dynamic_RR_subset:run} shows that the interval attractor $([p^{\prime}-\epsilon^{\prime}, q^{\prime}-\epsilon^{\prime}], [\ell^{\prime}-\epsilon^{\prime}, r^{\prime}-\epsilon^{\prime}])$ is non-periodic, 
and Lemma~\ref{lem:dynamic_RR_subset}~\ref{enum:dynamic_RR_subset:H} shows that 
the interval attractor is contained in the $h$-th level interval attractors $\Psi_{h}$. 
In this case, the point of a non-periodic interval attractor $([p^{\prime}-\epsilon^{\prime}, q^{\prime}-\epsilon^{\prime}], [\ell^{\prime}-\epsilon^{\prime}, r^{\prime}-\epsilon^{\prime}])$ is located at the same $(x^{\prime}, y^{\prime})$ in the $\mathcal{X} \times \mathcal{Y}$ plane. 
Therefore, set $\mathcal{J}^h$ contains a weighted point $(x, y, \alpha, \beta)$ satisfying $x = x^{\prime}$ and $y = y^{\prime}$. 

The existence of the weighted point $(x, y, \alpha, \beta)$ contradicts the fact that 
the same point $(x^{\prime}, y^{\prime})$ with any weights is not contained in $\mathcal{J}^h$. 
Therefore, Theorem~\ref{theo:intro_weighted_point_formula_ins}(ii) must hold. 
\end{proof}

We give an intuitive explanation of Theorem~\ref{theo:intro_weighted_point_formula_ins}. 
The weights of each weighted point $(x, y, \alpha, \beta)$ in set $\mathcal{J}^{h}$ are updated using the non-periodic interval attractors in the two subsets $\Psi_{\OLD}$ and $\Psi^{\prime}_{\NEW}$. 
Consider a non-periodic interval attractor $([p, q], [\ell, r]) \in \Psi_{\OLD}$ aggregated to the weighted point $(x, y, \alpha, \beta)$. Here, $\beta$ is the weight representing the number of the interval attractors aggregated to the weighted point. 
The non-periodic interval attractor $([p, q], [\ell, r])$ is removed from set $\Psi_{\RR}$ by updating RLSLP $\mathcal{G}^{R}$, 
and hence, the weight $\beta$ decreases by one. If this weight $\beta$ is changed to zero, then 
the weighted point $(x, y, \alpha, \beta)$ is removed from the set $\mathcal{J}^{h}$. 
On the other hand, each non-periodic interval attractor $([p^{\prime}, q^{\prime}], [\ell^{\prime}, r^{\prime}])$ of the subset 
$\Psi^{\prime}_{\NEW}$ is added to set $\Psi^{\prime}_{\RR}$ by updating RLSLP $\mathcal{G}^{R}$. 
If the non-periodic interval attractor is aggregated into a weighted point $(x, y, \alpha, \beta)$ in set $\mathcal{J}^{h}$, 
then the weight $\beta$ increases by one. 
Otherwise (i.e., such weighted point $(x, y, \alpha, \beta)$ does not exist), 
we add a new weighted point for the non-periodic interval attractor into the set $\mathcal{J}^{h}$. 
Finally, we obtain set $\mathcal{J}^{\prime h}$ by processing the non-periodic interval attractors in the two subsets 
$\Psi_{\OLD}$ and $\Psi^{\prime}_{\NEW}$. 

The update time for set $\mathcal{J}^{h}$ depends on the number of interval attractors in the two subsets 
$\Psi_{\OLD}$ and $\Psi^{\prime}_{\NEW}$. 
The following theorem gives an upper bound on the number of such interval attractors. 
\begin{theorem}\label{theo:bound_overlapping_IA_ins}
Let $W$ be the number of non-periodic interval attractors $([p, q], [\ell, r])$ in set $\Psi_{\RR}$ satisfying $([p, q], [\ell, r]) \in \Psi_{\OLD}$. 
Similarly, 
let $W^{\prime}$ be the number of non-periodic interval attractors $([p, q], [\ell, r])$ in set $\Psi^{\prime}_{\RR}$ satisfying $([p, q], [\ell, r]) \in \Psi^{\prime}_{\NEW}$. 
Then, $\mathbb{E}[W] = O(H + \log n)$ and $\mathbb{E}[W^{\prime}] = O(H^{\prime} + \log n)$. 
\end{theorem}
\begin{proof}
Theorem~\ref{theo:bound_overlapping_IA_ins} follows from Lemma~\ref{lem:dynamic_psi_overlap}. 
\end{proof}

For an insertion operation, 
Theorem~\ref{theo:intro_weighted_point_formula} and Theorem~\ref{theo:bound_overlapping_IA} follow from 
Theorem~\ref{theo:intro_weighted_point_formula_ins} and Theorem~\ref{theo:bound_overlapping_IA_ins}, respectively. 
Theorem~\ref{theo:bound_overlapping_IA} indicates that the range-sum data structure built on set $\mathcal{J}^{h}$ 
can be updated in polylogarithmic expected time.
In Section~\ref{sec:RSC_query}, we showed that 
we can build dynamic data structures supporting RSC query using the weighted points obtained by transforming 
weighted points in each set $\mathcal{J}^{h}$. 
By leveraging the key observation explained in this section, 
we give an expected $O(\polylog(n))$-time algorithm for the dynamic data structures of Section~\ref{sec:RSC_query} 
in subsequent subsections.

\subsection{Update of Data Structures for RSC-A Subquery}\label{subsec:ra_update}
This subsection explains how to update the dynamic data structures for subquery $\RSCQA$ 
introduced in Section~\ref{subsubsec:JA_ds}. 
The following lemma is the summary of this subsection.

\begin{lemma}\label{lem:dynamic_JA_summary}
Consider the two RLSLPs $\mathcal{G}^{R}$ and $\mathcal{G}^{R}_{\ins}$ of Theorem~\ref{theo:update1}, which derive input string $T$ and string $T^{\prime}$, respectively. 
The dynamic data structures of Section~\ref{subsubsec:JA_ds} can be updated 
in expected $O((\max \{H, H^{\prime}, \log (nHH^{\prime}) \})^{5})$ time 
after changing RLSLP $\mathcal{G}^{R}$ to $\mathcal{G}^{R}_{\ins}$. 
This update requires 
(i) the dynamic data structures for the RR-DAG of RLSLP $\mathcal{G}^{R}$ (Section~\ref{subsubsec:rrdag_ds}) 
and sample query (Section~\ref{subsec:sample_query}), 
and (ii) the interval attractors obtained from Lemma~\ref{lem:dynamic_IA_summary}. 
\end{lemma}
\begin{proof}
    See Section~\ref{subsubsec:JA_update_algorithm}. 
\end{proof}

The dynamic data structures for subquery $\RSCQA$ store weighted points of the $(H+1)$ sets 
$\mathcal{J}_{A}(0)$, $\mathcal{J}_{A}(1)$, $\ldots$, $\mathcal{J}_{A}(H)$ introduced in Section~\ref{subsec:RSC_comp_A}. 
Here, each set $\mathcal{J}_{A}(h)$ consists of weighted points on 
grid $(\mathcal{X}_{A}(h), \mathcal{Y}_{A}(h))$ for the two ordered sets $\mathcal{X}_{A}(h) = \{ L_{1}, L_{2}, \ldots, L_{d} \}$ and $\mathcal{Y}_{A}(h) = \{ R_{1}, R_{2}, \ldots, R_{d^{\prime}} \}$ of strings introduced in Section~\ref{subsec:RSC_comp_A}. 
For each integer $h \in [0, H^{\prime}]$, 
let $\mathcal{J}^{\prime}_{A}(h)$ be the set $\mathcal{J}_{A}(h)$ of weighted points defined using set $\Psi^{\prime}_{\RR}$ instead of set $\Psi_{\RR}$. 
For simplicity, let $\mathcal{J}_{A}(H+1)$, $\mathcal{J}_{A}(H+2)$, $\ldots$, $\mathcal{J}_{A}(H^{\prime})$ be empty sets of weighted points if $H^{\prime} > H$. 
Similarly, let $\mathcal{J}^{\prime}_{A}(H^{\prime}+1)$, $\mathcal{J}^{\prime}_{A}(H^{\prime}+2)$, $\ldots$, $\mathcal{J}^{\prime}_{A}(H)$ be empty sets of weighted points if $H^{\prime} < H$. 

The following lemma shows that each set $\mathcal{J}^{\prime}_{A}(h)$ can be obtained by modifying set $\mathcal{J}_{A}(h)$. 

\begin{lemma}\label{lem:dynamic_JA_change}
Consider an integer $h$ in set $[0, H^{\prime}]$. 
For each weighted point $(x_{s}, y_{s}, w_{s}, e_{s}) \in \mathcal{J}_{A}(h)$, 
let $([p_{s}, q_{s}], [\ell_{s}, r_{s}]) \in \Psi_{h} \cap \Psi_{\samp}$ be 
the interval attractor corresponding to the weighted point $(x_{s}, y_{s}, w_{s}, e_{s})$, 
$\kappa_{s} = |\Psi_{h} \cap (\Psi_{\OLD} \setminus \Psi_{\run}) \cap \Psi_{\str}(e_{s})|$ 
and $\kappa^{\prime}_{s} = |\Psi^{\prime}_{h} \cap (\Psi^{\prime}_{\NEW} \setminus \Psi^{\prime}_{\run}) \cap \Psi^{\prime}_{\str}(e_{s})|$. 
Let $\gamma^{\prime}_{s}$ be the attractor position of 
each interval attractor $([p^{\prime}_{s}, q^{\prime}_{s}], [\ell^{\prime}_{s}, r^{\prime}_{s}]) \in \Psi^{\prime}_{h} \cap \Psi^{\prime \symB}$. 
Then, set $\mathcal{J}^{\prime}_{A}(h)$ of weighted points 
is equal to the union of the following two sets of weighted points: 
\begin{enumerate}[label=\textbf{(\roman*)}]
    \item $\{ (x_{s}, y_{s}, w_{s} - \kappa_{s} + \kappa^{\prime}_{s}, e_{s}) \mid (x_{s}, y_{s}, w_{s}, e_{s}) \in \mathcal{J}_{A}(h) \text{ s.t. } ([p_{s}, q_{s}], [\ell_{s}, r_{s}]) \not \in \Psi_{\OLD} \setminus \Psi^{\symA} \}$; 
    \item $\{ (\reverse(T^{\prime}[p^{\prime}_{s}-1..\gamma^{\prime}_{s}-1]), T^{\prime}[\gamma^{\prime}_{s}..r^{\prime}_{s}+1], |\Psi^{\prime}_{\str}(T^{\prime}[p^{\prime}_{s}-1..r^{\prime}_{s}+1])|, T^{\prime}[p^{\prime}_{s}-1..r^{\prime}_{s}+1]) \mid ([p^{\prime}_{s}, q^{\prime}_{s}], [\ell^{\prime}_{s}, r^{\prime}_{s}]) \in \Psi^{\prime}_{h} \cap \Psi^{\prime \symB} \}$. 
\end{enumerate}
\end{lemma}
\begin{proof}
Lemma~\ref{lem:dynamic_JA_change} follows from the following four statements: 
\begin{enumerate}[label=\textbf{(\Alph*)}]
    \item consider a weighted point $(x_{s}, y_{s}, w_{s}, e_{s}) \in \mathcal{J}_{A}(h)$ satisfying $([p_{s}, q_{s}], [\ell_{s}, r_{s}]) \not \in \Psi_{\OLD} \setminus \Psi^{\symA}$. 
    Then, $(x_{s}, y_{s}, w_{s} - \kappa_{s} + \kappa^{\prime}_{s}, e_{s}) \in \mathcal{J}^{\prime}_{A}(h)$;
    \item consider an interval attractor $([p^{\prime}_{s}, q^{\prime}_{s}], [\ell^{\prime}_{s}, r^{\prime}_{s}]) \in \Psi^{\prime}_{h} \cap \Psi^{\prime \symB}$. 
    Then, $(\reverse(T^{\prime}[p^{\prime}_{s}-1..\gamma^{\prime}_{s}-1]), T^{\prime}[\gamma^{\prime}_{s}..r^{\prime}_{s}+1], |\Psi^{\prime}_{\str}(T^{\prime}[p^{\prime}_{s}-1..r^{\prime}_{s}+1])|, T^{\prime}[p^{\prime}_{s}-1..r^{\prime}_{s}+1]) \in \mathcal{J}^{\prime}_{A}(h)$;
    \item consider a weighted point $(x^{\prime}, y^{\prime}, w^{\prime}, e^{\prime}) \in \mathcal{J}^{\prime}_{A}(h)$ satisfying $e^{\prime} \not \in \mathcal{Z}^{\symB}$ for 
    the set $\mathcal{Z}^{\symB}$ of strings introduced in Section~\ref{subsec:update_sampling_subset}. 
    Then, there exists a weighted point $(x_{s}, y_{s}, w_{s}, e_{s}) \in \mathcal{J}_{A}(h)$ satisfying 
    $([p_{s}, q_{s}], [\ell_{s}, r_{s}]) \not \in \Psi_{\OLD} \setminus \Psi^{\symA}$ 
    and $(x_{s}, y_{s}, w_{s} - \kappa_{s} + \kappa^{\prime}_{s}, e_{s}) = (x^{\prime}, y^{\prime}, w^{\prime}, e^{\prime})$;
    \item \label{enu:dynamic_JA_change:D} consider a weighted point $(x^{\prime}, y^{\prime}, w^{\prime}, e^{\prime}) \in \mathcal{J}^{\prime}_{A}(h)$ satisfying $e^{\prime} \in \mathcal{Z}^{\symB}$. 
    Then, there exists an interval attractor $([p^{\prime}_{s}, q^{\prime}_{s}], [\ell^{\prime}_{s}, r^{\prime}_{s}]) \in \Psi^{\prime}_{h} \cap \Psi^{\prime \symB}$ 
    satisfying $(\reverse(T^{\prime}[p^{\prime}_{s}-1..\gamma^{\prime}_{s}-1]), T^{\prime}[\gamma^{\prime}_{s}..r^{\prime}_{s}+1], |\Psi^{\prime}_{\str}(T^{\prime}[p^{\prime}_{s}-1..r^{\prime}_{s}+1])|, T^{\prime}[p^{\prime}_{s}-1..r^{\prime}_{s}+1]) = (x^{\prime}, y^{\prime}, w^{\prime}, e^{\prime})$. 
\end{enumerate}

\textbf{Proof of statement (A).}
Lemma~\ref{lem:dynamic_samp_proceeding_formula} shows that 
sampling subset $\Psi^{\prime}_{\samp}$ contains an interval attractor $([p^{\prime}, q^{\prime}], [\ell^{\prime}, r^{\prime}])$ satisfying $T[p_{s}-1..r_{s}+1] = T^{\prime}[p^{\prime}-1..r^{\prime}+1]$. 
We apply Lemma~\ref{lem:dynamic_IA_super_correspondence} to the two interval attractors 
$([p_{s}, q_{s}], [\ell_{s}, r_{s}])$ and $([p^{\prime}, q^{\prime}], [\ell^{\prime}, r^{\prime}])$. 
Then, the lemma shows that $([p^{\prime}, q^{\prime}], [\ell^{\prime}, r^{\prime}]) \in \Psi^{\prime}_{h}$, 
$T[p_{s}-1..\gamma_{s}-1] = T^{\prime}[p^{\prime}-1..\gamma^{\prime}-1]$, 
and $T[\gamma_{s}..r_{s}+1] = T^{\prime}[\gamma^{\prime}..r^{\prime}+1]$ 
for the attractor positions $\gamma_{s}$ and $\gamma^{\prime}$ of 
the two interval attractors $([p_{s}, q_{s}], [\ell_{s}, r_{s}])$ and $([p^{\prime}, q^{\prime}], [\ell^{\prime}, r^{\prime}])$, respectively. 
Because of $([p^{\prime}, q^{\prime}], [\ell^{\prime}, r^{\prime}]) \in \Psi^{\prime}_{h} \cap \Psi^{\prime}_{\samp}$, 
set $\mathcal{J}^{\prime}_{A}(h)$ contains the weighted point $(\reverse(T^{\prime}[p^{\prime}-1..\gamma^{\prime}-1]), T^{\prime}[\gamma^{\prime}..r^{\prime}+1], |\Psi^{\prime}_{\str}(T^{\prime}[p^{\prime}-1..r^{\prime}+1])|, T^{\prime}[p^{\prime}-1..r^{\prime}+1])$ corresponding to the interval attractor $([p^{\prime}, q^{\prime}], [\ell^{\prime}, r^{\prime}])$. 

$x_{s} = \reverse(T^{\prime}[p^{\prime}-1..\gamma^{\prime}-1])$ follows from 
$x_{s} = \reverse(T[p_{s}-1..\gamma_{s}-1])$ and $T[p_{s}-1..\gamma_{s}-1] = T^{\prime}[p^{\prime}-1..\gamma^{\prime}-1]$. 
$y_{s} = T^{\prime}[\gamma^{\prime}..r^{\prime}+1]$ follows from 
$y_{s} = T[\gamma_{s}..r_{s}+1]$ and 
$T[\gamma_{s}..r_{s}+1] = T^{\prime}[\gamma^{\prime}..r^{\prime}+1]$. 
$e_{s} = T^{\prime}[p^{\prime}-1..r^{\prime}+1]$ 
follows from $e_{s} = T[p_{s}-1..r_{s}+1]$ and $T[p_{s}-1..r_{s}+1] = T^{\prime}[p^{\prime}-1..r^{\prime}+1]$. 

We prove $w_{s} - \kappa_{s} + \kappa^{\prime}_{s} = |\Psi^{\prime}_{\str}(T^{\prime}[p^{\prime}-1..r^{\prime}+1])|$. 
Let $Z = T^{\prime}[p^{\prime}-1..r^{\prime}+1]$ for simplicity. 
Because of $([p^{\prime}, q^{\prime}], [\ell^{\prime}, r^{\prime}]) \in \Psi^{\prime}_{\samp}$, 
$([p^{\prime}, q^{\prime}], [\ell^{\prime}, r^{\prime}]) \not \in \Psi^{\prime}_{\run}$ follows from 
Lemma~\ref{lem:samp_basic_property}~\ref{enum:samp_basic_property:3}. 
$|\Psi^{\prime}_{\str}(Z)| = |\Psi_{\str}(Z)| - |\Psi_{h} \cap (\Psi_{\OLD} \setminus \Psi_{\run}) \cap \Psi_{\str}(Z)| + |\Psi^{\prime}_{h} \cap (\Psi^{\prime}_{\NEW} \setminus \Psi^{\prime}_{\run}) \cap \Psi^{\prime}_{\str}(Z)|$ follows from 
Corollary~\ref{cor:dynamic_str_formula}~\ref{enum:dynamic_str_formula:1}. 
Therefore, the following equation holds: 
\begin{equation*}
    \begin{split}
        |\Psi^{\prime}_{\str}(Z)| &= |\Psi_{\str}(Z)| - |\Psi_{h} \cap (\Psi_{\OLD} \setminus \Psi_{\run})  \cap \Psi_{\str}(Z)| + |\Psi^{\prime}_{h} \cap (\Psi^{\prime}_{\NEW} \setminus \Psi^{\prime}_{\run}) \cap \Psi^{\prime}_{\str}(Z)| \\
        &= w_{s} - \kappa_{s} + \kappa^{\prime}_{s}.
    \end{split}
\end{equation*}

We proved $(x_{s}, y_{s}, w_{s} - \kappa_{s} + \kappa^{\prime}_{s}, e_{s}) = (\reverse(T^{\prime}[p^{\prime}-1..\gamma^{\prime}-1]), T^{\prime}[\gamma^{\prime}..r^{\prime}+1], |\Psi^{\prime}_{\str}(T^{\prime}[p^{\prime}-1..r^{\prime}+1])|, T^{\prime}[p^{\prime}-1..r^{\prime}+1])$. 
Therefore, $(x_{s}, y_{s}, w_{s} - \kappa_{s} + \kappa^{\prime}_{s}, e_{s}) \in \mathcal{J}^{\prime}_{A}(h)$ holds. 

\textbf{Proof of statement (B).}
Set $\Psi^{\prime}_{h} \cap \Psi^{\prime}_{\samp}$ contains 
interval attractor $([p^{\prime}_{s}, q^{\prime}_{s}], [\ell^{\prime}_{s}, r^{\prime}_{s}])$ 
because Theorem~\ref{theo:dynamic_samp_formula} shows that 
$\Psi^{\prime \symB} \subseteq \Psi^{\prime}_{\samp}$ holds. 
Because of $([p^{\prime}_{s}, q^{\prime}_{s}], [\ell^{\prime}_{s}, r^{\prime}_{s}]) \in \Psi^{\prime}_{h} \cap \Psi^{\prime}_{\samp}$, 
$(\reverse(T^{\prime}[p^{\prime}_{s}-1..\gamma^{\prime}_{s}-1]), T^{\prime}[\gamma^{\prime}_{s}..r^{\prime}_{s}+1], |\Psi^{\prime}_{\str}(T^{\prime}[p^{\prime}_{s}-1..r^{\prime}_{s}+1])|, T^{\prime}[p^{\prime}_{s}-1..r^{\prime}_{s}+1]) \in \mathcal{J}^{\prime}_{A}(h)$ follows from the definition of the set $\mathcal{J}^{\prime}_{A}(h)$. 

\textbf{Proof of statement (C).}
The weighted point $(x^{\prime}, y^{\prime}, w^{\prime}, e^{\prime})$ corresponds to 
an interval attractor $([p^{\prime}, q^{\prime}], [\ell^{\prime}, r^{\prime}]) \in \Psi^{\prime}_{h} \cap \Psi^{\prime}_{\samp}$ satisfying $T^{\prime}[p^{\prime}-1..r^{\prime}+1] = e^{\prime}$. 
$([p^{\prime}, q^{\prime}], [\ell^{\prime}, r^{\prime}]) \not \in \Psi^{\prime \symB}$ follows from 
$T^{\prime}[p^{\prime}-1..r^{\prime}+1] \not \in \mathcal{Z}^{\symB}$. 
Because of $([p^{\prime}, q^{\prime}], [\ell^{\prime}, r^{\prime}]) \not \in \Psi^{\prime \symB}$, 
Lemma~\ref{lem:dynamic_samp_back_formula} shows that 
sampling subset $\Psi_{\samp}$ contains an interval attractor 
$([p, q], [\ell, r])$ satisfying 
$T[p-1..r+1] = T^{\prime}[p^{\prime}-1..r^{\prime}+1]$ and 
$([p, q], [\ell, r]) \not \in \Psi_{\OLD} \setminus \Psi^{\symA}$. 

We apply Lemma~\ref{lem:dynamic_IA_super_correspondence} to the two interval attractors 
$([p, q], [\ell, r])$ and $([p^{\prime}, q^{\prime}], [\ell^{\prime}, r^{\prime}])$. 
Then, the lemma shows that $([p, q], [\ell, r]) \in \Psi_{h}$ holds.  
Because of $([p, q], [\ell, r]) \in \Psi_{h} \cap \Psi_{\samp}$, 
set $\mathcal{J}_{A}(h)$ contains a weighted point $(x_{s}, y_{s}, w_{s}, e_{s})$ corresponding to  
the interval attractor $([p, q], [\ell, r])$ (i.e., $([p_{s}, q_{s}], [\ell_{s}, r_{s}]) = ([p, q], [\ell, r])$). 
Similar to the proof of statement (A), 
$(x_{s}, y_{s}, w_{s} - \kappa_{s} + \kappa^{\prime}_{s}, e_{s}) = (x^{\prime}, y^{\prime}, w^{\prime}, e^{\prime})$ can be proved. 
Therefore, statement (C) holds. 

\textbf{Proof of statement (D).}
The weighted point $(x^{\prime}, y^{\prime}, w^{\prime}, e^{\prime})$ corresponds to 
an interval attractor $([p^{\prime}, q^{\prime}], [\ell^{\prime}, r^{\prime}]) \in \Psi^{\prime}_{h} \cap \Psi^{\prime}_{\samp}$ satisfying $T^{\prime}[p^{\prime}-1..r^{\prime}+1] = e^{\prime}$. 
Here, $(\reverse(T^{\prime}[p^{\prime}-1..\gamma^{\prime}-1]), T^{\prime}[\gamma^{\prime}..r^{\prime}+1], |\Psi^{\prime}_{\str}(T^{\prime}[p^{\prime}-1..r^{\prime}+1])|, T^{\prime}[p^{\prime}-1..r^{\prime}+1]) = (x^{\prime}, y^{\prime}, w^{\prime}, e^{\prime})$ follows from the definition of set $\mathcal{J}^{\prime}_{A}(h)$. 

We prove $([p^{\prime}, q^{\prime}], [\ell^{\prime}, r^{\prime}]) \in \Psi^{\prime \symB}$ by contradiction. 
We assume that $([p^{\prime}, q^{\prime}], [\ell^{\prime}, r^{\prime}]) \not \in \Psi^{\prime \symB}$ holds. 
Because of $T^{\prime}[p^{\prime}-1..r^{\prime}+1] \in \mathcal{Z}^{\symB}$, 
set $\Psi^{\prime \symB}$ contains an interval attractor $([p, q], [\ell, r])$ satisfying 
$T^{\prime}[p-1..r+1] = T^{\prime}[p^{\prime}-1..r^{\prime}+1]$. 
Here, $([p, q], [\ell, r]) \in \Psi^{\prime}_{\samp}$ follows from $\Psi^{\prime \symB} \subseteq \Psi^{\prime}_{\samp}$. 
$T^{\prime}[p-1..r+1] \neq T^{\prime}[p^{\prime}-1..r^{\prime}+1]$ follows from 
from the definition of sampling subset (Definition~\ref{def:sampling_subset}) 
because $([p, q], [\ell, r]), ([p^{\prime}, q^{\prime}], [\ell^{\prime}, r^{\prime}]) \in \Psi^{\prime}_{\samp}$ 
and $([p, q], [\ell, r]) \neq ([p^{\prime}, q^{\prime}], [\ell^{\prime}, r^{\prime}])$. 
The two facts $T^{\prime}[p-1..r+1] = T^{\prime}[p^{\prime}-1..r^{\prime}+1]$ and $T^{\prime}[p-1..r+1] \neq T^{\prime}[p^{\prime}-1..r^{\prime}+1]$ yield a contradiction. 
Therefore, $([p^{\prime}, q^{\prime}], [\ell^{\prime}, r^{\prime}]) \in \Psi^{\prime \symB}$ must hold. 

We proved 
$([p^{\prime}, q^{\prime}], [\ell^{\prime}, r^{\prime}]) \in \Psi^{\prime}_{h} \cap \Psi^{\prime \symB}$ 
satisfying $(\reverse(T^{\prime}[p^{\prime}-1..\gamma^{\prime}-1]), T^{\prime}[\gamma^{\prime}..r^{\prime}+1], |\Psi^{\prime}_{\str}(T^{\prime}[p^{\prime}-1..r^{\prime}+1])|, T^{\prime}[p^{\prime}-1..r^{\prime}+1]) = (x^{\prime}, y^{\prime}, w^{\prime}, e^{\prime})$. 
Therefore statement (D) holds. 
\end{proof}

Consider the $d$ sequences $\mathbf{Q}^{X}_{A}(h, L_{1})$, $\mathbf{Q}^{X}_{A}(h, L_{2})$, $\ldots$, 
$\mathbf{Q}^{X}_{A}(h, L_{d})$ and $d^{\prime}$ sequences $\mathbf{Q}^{Y}_{A}(h, R_{1})$, 
$\mathbf{Q}^{Y}_{A}(h, R_{2})$, $\ldots$, $\mathbf{Q}^{Y}_{A}(h, R_{d^{\prime}})$ of weighted points 
introduced in Section~\ref{subsubsec:JA_X_ds} and Section~\ref{subsubsec:JA_Y_ds}. 
Here, each sequence $\mathbf{Q}^{X}_{A}(h, L_{s})$ is represented as the doubly linked list $\mathbf{X}_{A}(h, L_{s})$ introduced in Section~\ref{subsubsec:JA_X_ds}. 
Similarly, each sequence $\mathbf{Q}^{Y}_{A}(h, R_{s})$ is represented as the doubly linked list $\mathbf{Y}_{A}(h, R_{s})$ introduced in Section~\ref{subsubsec:JA_X_ds}. 
The following lemma states properties of these sequences of weighted points. 

\begin{lemma}\label{lem:dynamic_JA_move}
Consider an interval attractor $([p, q], [\ell, r]) \in \Psi_{h} \cap \Psi_{\samp}$ for an integer $h \in [0, H]$ satisfying $([p, q], [\ell, r]) \not \in \Psi_{\OLD} \setminus \Psi^{\symA}$. 
Here, set $\mathcal{J}_{A}(h)$ contains a weighted point $(x, y, w, e)$ corresponding to 
the interval attractor $([p, q], [\ell, r])$; 
the interval attractor $([p, q], [\ell, r])$ is represented as a node $u$ of the sequence $\mathbf{Q}_{\samp}$ introduced in Section~\ref{subsec:sample_query}. 
From the definition of sequence $\mathbf{Q}^{X}_{A}(h, x)$, 
there exists an integer $\alpha$ such that 
the sequence $\mathbf{Q}^{X}_{A}(h, x)$ contains weighted point $(x, y, w, e)$ as the $\alpha$-th element, 
and the $\alpha$-th element of doubly linked list $\mathbf{X}_{A}(h, x)$ stores a pointer to node $u$. 
Similarly, 
there exists an integer $\beta$ such that 
the sequence $\mathbf{Q}^{Y}_{A}(h, y)$ contains weighted point $(x, y, w, e)$ as the $\beta$-th element, 
and the $\beta$-th element of doubly linked list $\mathbf{Y}_{A}(h, y)$ stores a pointer to node $u$. 

After updating the dynamic data structures for sample query, 
node $u$ represents an interval attractor $([p^{\prime}, q^{\prime}], [\ell^{\prime}, r^{\prime}])$ in set $\Psi^{\prime}_{h} \cap \Psi^{\prime}_{\samp}$ satisfying the following three conditions: 
\begin{enumerate}[label=\textbf{(\roman*)}]
    \item $T[p-1..r+1] = T^{\prime}[p^{\prime}-1..r^{\prime}+1]$;
    \item $([p^{\prime}, q^{\prime}], [\ell^{\prime}, r^{\prime}]) \not \in \Psi^{\prime \symB}$;    
    \item $T[p-1..\gamma-1] = T^{\prime}[p^{\prime}-1..\gamma^{\prime}+1]$ 
    and $T[\gamma..r+1] = T^{\prime}[\gamma^{\prime}..r^{\prime}+1]$ for 
    the attractor positions $\gamma$ and $\gamma^{\prime}$ of two interval attractors 
    $([p, q], [\ell, r])$ and $([p^{\prime}, q^{\prime}], [\ell^{\prime}, r^{\prime}])$, respectively.
\end{enumerate}
\end{lemma}
\begin{proof}
Lemma~\ref{lem:dynamic_samp_proceeding_formula} shows that 
node $u$ represents an interval attractor $([p^{\prime}, q^{\prime}], [\ell^{\prime}, r^{\prime}])$ in set 
$\Psi^{\prime}_{\samp}$ 
satisfying $T[p-1..r+1] = T^{\prime}[p^{\prime}-1..r^{\prime}+1]$ 
and $([p^{\prime}, q^{\prime}], [\ell^{\prime}, r^{\prime}]) \not \in \Psi^{\prime \symB}$. 
We apply Lemma~\ref{lem:dynamic_IA_super_correspondence} to the two interval attractors 
$([p, q], [\ell, r])$ and $([p^{\prime}, q^{\prime}], [\ell^{\prime}, r^{\prime}])$. 
Then, the lemma shows that $([p^{\prime}, q^{\prime}], [\ell^{\prime}, r^{\prime}]) \in \Psi^{\prime}_{h}$, 
$T[p-1..\gamma-1] = T^{\prime}[p^{\prime}-1..\gamma^{\prime}-1]$, 
and $T[\gamma..r+1] = T^{\prime}[\gamma^{\prime}..r^{\prime}+1]$. 
Therefore, Lemma~\ref{lem:dynamic_JA_move} holds. 

\end{proof}

The following lemma shows that 
we can obtain the interval attractor corresponding to each weighted point of $\mathcal{J}_{A}(h)$ 
using the data structures for subquery $\RSCQA$. 
\begin{lemma}\label{lem:dynamic_JA_find}
Consider the weighted points 
$(x_{1}, y_{1}, w_{1}, e_{1})$, $(x_{2}, y_{2}, w_{2}, e_{2})$, $\ldots$, $(x_{m}, y_{m}, w_{m}, e_{m})$ 
($e_{1} \prec e_{2} \prec \cdots \prec e_{m}$) of set $\mathcal{J}_{A}(h)$ for an integer $h \in [0, H]$. 
Here, each weighted point $(x_{s}, y_{s}, w_{s}, e_{s})$ corresponds to an interval attractor in set $\Psi_{h} \cap \Psi_{\samp}$. 
For a given integer $s \in [1, m]$, 
we can compute the interval attractor corresponding to the $s$-th weighted point $(x_{s}, y_{s}, w_{s}, e_{s})$ in $O(H + \log n)$ time using the data structures for subquery $\RSCQA$.
\end{lemma}
\begin{proof}
The interval attractor $([p_{s}, q_{s}], [\ell_{s}, r_{s}])$ 
is represented as a node $u$ of the sequence $\mathbf{Q}_{\samp}$ introduced in Section~\ref{subsec:sample_query}. 
The interval attractor $([p_{s}, q_{s}], [\ell_{s}, r_{s}])$ can be obtained by 
the node $u$ and algorithm presented in Section~\ref{subsubsec:computation_delta_samp}. 

Consider the sequence $\mathbf{Q}^{X}_{A}(h, x_{s})$ of weighted points introduced in Section~\ref{subsubsec:JA_X_ds}. 
From the definition of the doubly linked list $\mathbf{X}_{A}(h, x_{s})$, 
the doubly linked list has an element $v$ storing a pointer to node $u$. 
Consider the doubly linked list representing set $\mathcal{J}_{A}(h)$, which is introduced in Section~\ref{subsubsec:JA_ds}. 
Then, the $s$-th element of the doubly linked list stores a pointer to the element $v$. 
Therefore, we can compute the interval attractor $([p_{s}, q_{s}], [\ell_{s}, r_{s}])$ 
using the pointer stored in the $s$-th element of the doubly linked list representing set $\mathcal{J}_{A}(h)$. 

We compute the interval attractor $([p_{s}, q_{s}], [\ell_{s}, r_{s}])$ in the following five steps: 
\begin{enumerate}[label=\textbf{(\roman*)}]
    \item access the doubly linked list representing set $\mathcal{J}_{A}(h)$ in $O(H)$ time;
    \item access the $s$-th element of the doubly linked list representing set $\mathcal{J}_{A}(h)$ in $O(\log n)$ time;    
    \item access the element $v$ of doubly linked list $\mathbf{X}_{A}(h, x_{s})$ 
    using the pointer stored in the $s$-th element of the doubly linked list representing set $\mathcal{J}_{A}(h)$;
    \item obtain the node $u$ of sequence $\mathbf{Q}_{\samp}$ using 
    the pointer stored in the element $v$;
    \item compute interval attractor $([p_{s}, q_{s}], [\ell_{s}, r_{s}])$ in $O(\log n)$ time 
    by the algorithm presented in Section~\ref{subsubsec:computation_delta_samp}. 
\end{enumerate}
Therefore, 
the interval attractor $([p_{s}, q_{s}], [\ell_{s}, r_{s}])$ can be computed in $O(H + \log n)$ time. 
\end{proof}

\subsubsection{Algorithm}\label{subsubsec:JA_update_algorithm}
We prove Lemma~\ref{lem:dynamic_JA_summary}, i.e., 
we show that the data structures for subquery $\RSCQA$ (Section~\ref{subsubsec:JA_ds}) 
can be updated in expected $O((\max \{H, H^{\prime}, \log (nHH^{\prime}) \})^{5})$ time using 
(A) the dynamic data structures for the RR-DAG of RLSLP $\mathcal{G}^{R}$ (Section~\ref{subsubsec:rrdag_ds}) 
and sample query (Section~\ref{subsec:sample_query}), 
and (B) the interval attractors obtained from Lemma~\ref{lem:dynamic_IA_summary}. 

Lemma~\ref{lem:dynamic_JA_change} and Lemma~\ref{lem:dynamic_JA_move} indicate that 
we can obtain $(H^{\prime}+1)$ sets $\mathcal{J}^{\prime}_{A}(0)$, $\mathcal{J}^{\prime}_{A}(1)$, 
$\ldots$, $\mathcal{J}^{\prime}_{A}(H^{\prime})$ of weighted points by 
modifying $(\max \{ H, H^{\prime} \}+1)$ sets $\mathcal{J}_{A}(0)$, $\mathcal{J}_{A}(1)$, $\ldots$, $\mathcal{J}_{A}(\max \{ H, H^{\prime} \} + 1)$ of weighted points as follows: 
\begin{description}
    \item [Modification 1:] 
    For each interval attractor $([p, q], [\ell, r]) \in \Psi_{\OLD} \setminus \Psi_{\run}$ 
    with level $h$, 
    set $\mathcal{J}_{A}(h)$ contains a weighted point $(x, y, w, e)$ satisfying $e = T[p-1..r+1]$, 
    and $1$ is subtracted from the weight $w$ of the weighted point. 
    \item [Modification 2:] 
    For each interval attractor $([p, q], [\ell, r]) \in \Psi_{\samp} \cap (\Psi_{\OLD} \setminus \Psi^{\symA})$ with level $h$,     
    set $\mathcal{J}_{A}(h)$ contains a weighted point $(x, y, w, e)$ corresponding to 
    the interval attractor $([p, q], [\ell, r])$, 
    and this weighted point is removed from the set $\mathcal{J}_{A}(h)$.
    \item [Modification 3:] 
    For each interval attractor $([p, q], [\ell, r]) \in \Psi^{\prime \symB}$ 
    with attractor position $\gamma$, 
    we create a weighted point $(\reverse(T^{\prime}[p-1..\gamma-1]), T^{\prime}[\gamma..r+1], 0, T^{\prime}[p-1..r+1])$ corresponding to the interval attractor $([p, q], [\ell, r])$. 
    The weighted point is added to set $\mathcal{J}_{A}(h)$ for the level $h$ of the interval attractor $([p, q], [\ell, r])$.
    \item [Modification 4:]
    This process is executed after Modification 2 and Modification 3. 
    For each interval attractor $([p, q], [\ell, r])$ of set $\Psi^{\prime}_{\NEW} \setminus \Psi^{\prime}_{\run}$ with level $h$, 
    set $\mathcal{J}_{A}(h)$ contains a weighted point $(x, y, w, e)$ satisfying $e = T^{\prime}[p-1..r+1]$, 
    and $1$ is added to the weight $w$ of the weighted point. 
\end{description}

We update the data structures for subquery $\RSCQA$ based on these modifications. 
Here, the four sets $\Psi_{\OLD} \setminus \Psi_{\run}$, $\Psi_{\samp} \cap (\Psi_{\OLD} \setminus \Psi^{\symA})$, $\Psi^{\prime \symB}$, and $\Psi^{\prime}_{\NEW} \setminus \Psi^{\prime}_{\run}$ are obtained from Lemma~\ref{lem:dynamic_IA_summary}. 
The algorithm updating the data structures for subquery $\RSCQA$ consists of the following four phases. 

\paragraph{Phase (i).}
In the first phase, 
we update the data structures for subquery $\RSCQA$ based on Modification 1. 
For each interval attractor $([p, q], [\ell, r]) \in \Psi_{\OLD} \setminus \Psi_{\run}$ with level $h$, 
let $(x_{1}, y_{1}, w_{1}, e_{1})$, $(x_{2}, y_{2}, w_{2}, e_{2})$, $\ldots$, $(x_{m}, y_{m}, w_{m}, e_{m})$ 
($e_{1} \prec e_{2} \prec \cdots \prec e_{m}$) be the weighted points of set $\mathcal{J}_{A}(h)$. 
Here, each weighted point $(x_{s}, y_{s}, w_{s}, e_{s})$ corresponds to an interval attractor in set $\Psi_{h} \cap \Psi_{\samp}$, 
and the weighted point $(x_{s}, y_{s}, w_{s}, e_{s})$ is represented as the $s$-th element of the doubly linked list representing the set $\mathcal{J}_{A}(h)$. 
Set $\mathcal{J}_{A}(h)$ contains a weighted point $(x_{j}, y_{j}, w_{j}, e_{j})$ satisfying $e_{j} = T[p-1..r+1]$, and $1$ is subtracted from the weight $w_{j}$. 

The first phase consists of three steps. 
In the first step, 
we compute the level $h$ of interval attractor $([p, q], [\ell, r])$ by the level query introduced in 
Section~\ref{subsec:level_query}. 
The first step takes $O(H^{2})$ time. 

In the second step, we compute the integer $j$ by binary search on 
the $m$ identifiers $e_{1}, e_{2}, \ldots, e_{m}$. 
This binary search needs $O(\log m)$ LCE and random access queries on string $T$. 
Here, $\log m = O(\log n)$ follows from Lemma~\ref{lem:JA_size}~\ref{enum:JA_size:3}. 
Each weighted point $(x_{j}, y_{j}, w_{j}, e_{j})$ corresponds to an interval attractor $([p_{j}, q_{j}], [\ell_{j}, r_{j}])$ satisfying $e_{j} = T[p_{j}-1..r_{j}+1]$, 
and this interval attractor can be computed in $O(H + \log n)$ time by Lemma~\ref{lem:dynamic_JA_find}. 
Therefore, the integer $j$ can be computed in $O(H \log n + \log^{2} n)$ time. 

In the third step, 
$1$ is subtracted from the weight $w_{j}$ stored in the $j$-th element of the doubly linked list representing the set $\mathcal{J}_{A}(h)$. 
This subtraction takes $O(\log^{4} m)$ time because 
we need to update the range-sum data structure built on set $\mathcal{J}_{A}(h)$. 

The first phase takes $O(|\Psi_{\OLD} \setminus \Psi_{\run}| (H^{2} \log n + \log^{4} n))$ time in total. 
Here, $\mathbb{E}[|\Psi_{\OLD} \setminus \Psi_{\run}|] = O(H + \log n)$ follows from Lemma~\ref{lem:dynamic_psi_overlap}. 
Therefore, this phase takes expected $O((H + \log n)(H^{2} \log n + \log^{4} n))$ time.

\paragraph{Phase (ii).}
In the second phase, 
we update the data structures for subquery $\RSCQA$ based on Modification 2. 
For each interval attractor $([p, q], [\ell, r]) \in \Psi_{\samp} \cap (\Psi_{\OLD} \setminus \Psi^{\symA})$ with level $h$, 
set $\mathcal{J}_{A}(h)$ contains a weighted point $(x_{j}, y_{j}, w_{j}, e_{j})$ corresponding to the interval attractor $([p, q], [\ell, r])$, 
and this weighted point is removed from the set $\mathcal{J}_{A}(h)$. 
By this removal, 
the data structures for subquery $\RSCQA$ are changed as follows: 
\begin{itemize}
    \item Consider the two doubly linked lists $\mathbf{X}_{A}(h, x_{j})$ and $\mathbf{L}^{X}_{A}(h)$ introduced in Section~\ref{subsubsec:JA_X_ds}. 
    The weighted point $(x_{j}, y_{j}, w_{j}, e_{j})$ corresponds to an element $u$ of doubly linked list $\mathbf{X}_{A}(h, x_{j})$, and the doubly linked list $\mathbf{X}_{A}(h, x_{j})$ corresponds to an element $u^{\prime}$ of $\mathbf{L}^{X}_{A}(h)$. 
    The element $u$ is removed from the doubly linked list $\mathbf{X}_{A}(h, x_{j})$. 
    If doubly linked list $\mathbf{X}_{A}(h, x_{j})$ is changed to an empty list by the removal of the element $u$, 
    then element $u^{\prime}$ is removed from doubly linked list $\mathbf{L}^{X}_{A}(h)$. 
    \item 
    Similarly, 
    consider the two doubly linked lists $\mathbf{Y}_{A}(h, y_{j})$ and $\mathbf{L}^{Y}_{A}(h)$ introduced in Section~\ref{subsubsec:JA_Y_ds}. 
    The weighted point $(x_{j}, y_{j}, w_{j}, e_{j})$ corresponds to an element $v$ of doubly linked list $\mathbf{Y}_{A}(h, y_{j})$, and the doubly linked list $\mathbf{Y}_{A}(h, y_{j})$ corresponds to an element $v^{\prime}$ of $\mathbf{L}^{Y}_{A}(h)$. 
    The element $v$ is removed from the doubly linked list $\mathbf{Y}_{A}(h, y_{j})$. 
    If doubly linked list $\mathbf{Y}_{A}(h, y_{j})$ is changed to an empty list by the removal of the element $v$, 
    then element $v^{\prime}$ is removed from doubly linked list $\mathbf{L}^{Y}_{A}(h)$. 
    \item The $j$-th element is removed from the doubly linked list representing the set $\mathcal{J}_{A}(h)$.    
\end{itemize}

The second phase consists of six steps. 
In the first step, 
we compute the level $h$ of interval attractor $([p, q], [\ell, r])$ by level query, 
which takes $O(H^{2})$ time. 

In the second step, we compute the integer $j$ by binary search on 
the $m$ identifiers $e_{1}, e_{2}, \ldots, e_{m}$. 
For an integer $s \in [1, m]$, 
$s = j$ holds only if $e_{s} = T[p-1..r+1]$ holds. 
Therefore, the integer $j$ can be computed in $O(H^{2} \log n + \log^{2} n)$ time 
using the same approach as for the second step of the first phase. 

In the third step, 
we remove the two elements $u^{\prime}$ and $v^{\prime}$ from two doubly linked lists 
$\mathbf{L}^{X}_{A}(h)$ and $\mathbf{Y}_{A}(h, y_{j})$, respectively. 
These removals can be executed in $O(\log n)$ time 
because the $j$-th element of the doubly linked list representing the set $\mathcal{J}_{A}(h)$ stores two pointers to the two elements $u$ and $v$. 

The fourth step is executed if  
doubly linked list $\mathbf{X}_{A}(h, x_{j})$ is changed to an empty list by the removal of the element $u$. 
In the fourth step, 
we remove element $u^{\prime}$ from doubly linked list $\mathbf{L}^{X}_{A}(h)$ in $O(\log n)$ time. 

The fifth step is executed if 
doubly linked list $\mathbf{Y}_{A}(h, y_{j})$ is changed to an empty list by the removal of the element $v$. 
In the fifth step, 
we remove element $v^{\prime}$ from doubly linked list $\mathbf{L}^{Y}_{A}(h)$ in $O(\log n)$ time. 

In the sixth step, we remove the $j$-th element from the doubly linked list representing the set $\mathcal{J}_{A}(h)$. This removal takes $O(\log^{4} m)$ time 
because we need to update the range-sum data structure built on set $\mathcal{J}_{A}(h)$. 

The second phase takes $O(|\Psi_{\samp} \cap \Psi_{\OLD}| (H^{2} \log n + \log^{4} n))$ time in total. 
Here, $\mathbb{E}[|\Psi_{\samp} \cap \Psi_{\OLD}|] = O(H + \log n)$ follows from Lemma~\ref{lem:dynamic_psi_overlap}. 
Therefore, this phase takes expected $O((H + \log n)(H^{2} \log n + \log^{4} n))$ time.

\paragraph{Phase (iii).}
In the third phase, 
we update the data structures for subquery $\RSCQA$ based on Modification 3. 
This phase is executed after updating the dynamic data structures for RR-DAG and sample query. 

Consider the sequence $\mathbf{Q}_{\samp}$ of nodes introduced in Section~\ref{subsec:sample_query}.
For each interval attractor $([p, q], [\ell, r]) \in \Psi^{\prime \symB}$, 
let $u$ be the node of the sequence $\mathbf{Q}_{\samp}$ representing the interval attractor $([p, q], [\ell, r])$. 
By Modification 3, 
a weighted point $(\reverse(T^{\prime}[p-1..\gamma-1]), T^{\prime}[\gamma..r+1], 0, T^{\prime}[p-1..r+1])$ 
is created, and it is added to set $\mathcal{J}_{A}(h)$ for the level $h$ and attractor position $\gamma$ of the interval attractor $([p, q], [\ell, r])$.
By this addition, 
the data structures for subquery $\RSCQA$ are changed as follows: 
\begin{itemize}
    \item Consider the strings $L_{1}, L_{2}, \ldots, L_{d}$ ($L_{1} \prec L_{2} \prec \cdots L_{d}$) 
    of ordered set $\mathcal{X}_{A}(h)$. 
    Let $\alpha$ be the smallest integer in set $[1, d]$ satisfying 
    $\reverse(T^{\prime}[p-1..\gamma-1]) \preceq L_{\alpha}$. 
    If $\reverse(T^{\prime}[p-1..\gamma-1]) \neq L_{\alpha}$, then 
    a new doubly linked list $\mathbf{X}_{A}(h, \reverse(T^{\prime}[p-1..\gamma-1]))$ is created,     
    and a new element is inserted into doubly linked list $\mathbf{L}^{X}_{A}(h)$ as the $\alpha$-th element. 
    Here, the new element of doubly linked list $\mathbf{L}^{X}_{A}(h)$ stores a pointer to the new doubly linked list $\mathbf{X}_{A}(h, \reverse(T^{\prime}[p-1..\gamma-1]))$. 
    \item 
    Consider the weighted points $(x^{\prime}_{1}, y^{\prime}_{1}, w^{\prime}_{1}, e^{\prime}_{1})$, 
    $(x^{\prime}_{2}, y^{\prime}_{2}, w^{\prime}_{2}, e^{\prime}_{2})$, 
    $\ldots$, $(x^{\prime}_{m^{\prime}}, y^{\prime}_{m^{\prime}}, w^{\prime}_{m^{\prime}}, e^{\prime}_{m^{\prime}})$ 
    of sequence $\mathbf{Q}^{X}_{A}(h, \reverse(T^{\prime}[p-1..\gamma-1]))$. 
    Let $\alpha^{\prime}$ be the largest integer in set $[1, m^{\prime}]$ satisfying 
    either (A) $y^{\prime}_{\alpha^{\prime}} \prec T^{\prime}[\gamma..r+1]$ 
    or (B) $y^{\prime}_{\alpha^{\prime}} = T^{\prime}[\gamma..r+1]$ and $e^{\prime}_{\alpha^{\prime}} \prec T^{\prime}[p-1..r+1]$.     
    Then, a new element $v$ is created, 
    and it is inserted into to doubly linked list $\mathbf{X}_{A}(h, \reverse(T^{\prime}[p-1..\gamma-1]))$ as the $\alpha^{\prime}+1$-th element.     
    Here, this element $v$ corresponds to the weighted point $(\reverse(T^{\prime}[p-1..\gamma-1]), T^{\prime}[\gamma..r+1], 0, T^{\prime}[p-1..r+1])$ 
    and stores a pointer to a node $u$ in the sequence $\mathbf{Q}_{\samp}$. 
    \item 
    Consider the strings $R_{1}, R_{2}, \ldots, R_{d^{\prime}}$ ($R_{1} \prec R_{2} \prec \cdots R_{d^{\prime}}$) 
    of ordered set $\mathcal{Y}_{A}(h)$. 
    Let $\beta$ be the smallest integer in set $[1, d^{\prime}]$ satisfying 
    $T^{\prime}[\gamma..r+1] \preceq R_{\beta}$. 
    If $T^{\prime}[\gamma..r+1] \neq R_{\beta}$, then 
    a new doubly linked list $\mathbf{Y}_{A}(h, T^{\prime}[\gamma..r+1])$ is created, 
    and a new element is inserted into doubly linked list $\mathbf{L}^{Y}_{A}(h)$ as the $\beta$-th element. 
    Here, the new element of doubly linked list $\mathbf{L}^{Y}_{A}(h)$ stores a pointer to the new doubly linked list $\mathbf{Y}_{A}(h, T^{\prime}[\gamma..r+1])$. 
    \item 
    Consider the weighted points $(\hat{x}_{1}, \hat{y}_{1}, \hat{w}_{1}, \hat{e}_{1})$, 
    $(\hat{x}_{2}, \hat{y}_{2}, \hat{w}_{2}, \hat{e}_{2})$, 
    $\ldots$, $(\hat{x}_{\hat{m}}, \hat{y}_{\hat{m}}, \hat{w}_{\hat{m}}, \hat{e}_{\hat{m}})$ 
    of sequence $\mathbf{Q}^{Y}_{A}(h, T^{\prime}[\gamma..r+1])$. 
    Let $\beta^{\prime}$ be the largest integer in set $[1, \hat{m}]$ satisfying 
    either (a) $\hat{x}_{\beta^{\prime}} \prec \reverse(T^{\prime}[p-1..\gamma-1])$ 
    or (b) $\hat{x}_{\beta^{\prime}} = \reverse(T^{\prime}[p-1..\gamma-1])$ and $\hat{e}_{\beta^{\prime}} \prec T^{\prime}[p-1..r+1]$. 
    Then, a new element $v^{\prime}$ is created, 
    and it is inserted into doubly linked list $\mathbf{Y}_{A}(h, T^{\prime}[\gamma..r+1])$ as 
    the $\beta^{\prime}+1$-th element.
    Here, this element $v^{\prime}$ corresponds to the weighted point $(\reverse(T^{\prime}[p-1..\gamma-1]), T^{\prime}[\gamma..r+1], 0, T^{\prime}[p-1..r+1])$ 
    and stores a pointer to the node $u$ in the sequence $\mathbf{Q}_{\samp}$. 
    \item 
    Let $j$ be the smallest integer in set $[1, m]$ satisfying 
    $e_{j} \prec T^{\prime}[p-1..r+1]$ for the $m$ weighted points 
    $(x_{1}, y_{1}, w_{1}, e_{1})$, $(x_{2}, y_{2}, w_{2}, e_{2})$, $\ldots$, $(x_{m}, y_{m}, w_{m}, e_{m})$ 
    of set $\mathcal{J}_{A}(h)$. 
    Then, a new element is inserted into the doubly linked list representing the set $\mathcal{J}_{A}(h)$ 
    as the $(j+1)$-th element. 
    Here, this element corresponds to weighted point $(\reverse(T^{\prime}[p-1..\gamma-1]), T^{\prime}[\gamma..r+1], 0, T^{\prime}[p-1..r+1])$ 
    and stores two pointers to the element $v$ of doubly linked list $\mathbf{X}_{A}(h, \reverse(T^{\prime}[p-1..\gamma-1]))$ and the element $v^{\prime}$ of doubly linked list $\mathbf{Y}_{A}(h, T^{\prime}[\gamma..r+1])$. 
\end{itemize}

We execute nine steps for each interval attractor $([p, q], [\ell, r]) \in \Psi^{\prime \symB}$. 
In the first step, 
we compute the level $h$ of interval attractor $([p, q], [\ell, r])$ by level query, 
which takes $O(H^{\prime 2})$ time. 

In the second step, we find the node $u$ in the sequence $\mathbf{Q}_{\samp}$. 
The node $u$ can be found in $O(\log^{2} n)$ time by binary search on the nodes of sequence $\mathbf{Q}_{\samp}$ 
using the dynamic data structures for sample query. 

In the third step, we find the smallest integer $\alpha$ by binary search on the strings of the ordered set $\mathcal{X}_{A}(h)$. 
This binary search needs $O(\log d)$ reversed LCE and random access queries. 
Here, $\log d = O(\log n)$ follows from Lemma~\ref{lem:JA_size}~\ref{enum:JA_size:3}. 
For each string $L_{s}$, 
we can obtain an interval $[g, g + |L_{s}| - 1]$ in string $T^{\prime}$ satisfying $\reverse(T^{\prime}[g..g + |L_{s}| - 1]) = L_{s}$ in $O(H^{\prime 2} + \log n)$ time 
by Lemma~\ref{lem:JA_X_queries}~\ref{enum:JA_X_queries:2}. 
Therefore, the smallest integer $\alpha$ can be found in $O(H^{\prime} \log n + \log^{2} n)$ time. 

The fourth step is executed if $\reverse(T^{\prime}[p-1..\gamma-1]) \neq L_{\alpha}$ holds. 
In the fourth step, we create a new doubly linked list $\mathbf{X}_{A}(h, \reverse(T^{\prime}[p-1..\gamma-1]))$ 
and insert a new element into doubly linked list $\mathbf{L}^{X}_{A}(h)$ as the $\alpha$-th element. 
This step can be executed in $O(\log n)$ time. 

In the fifth step, 
we create element $v$ and insert it into doubly linked list $\mathbf{X}_{A}(h, \reverse(T^{\prime}[p-1..\gamma-1]))$ as the $\alpha^{\prime}+1$-th element. 
The integer $\alpha^{\prime}$ is computed by binary search on the weighted points of sequence $\mathbf{Q}^{X}_{A}(h, \reverse(T^{\prime}[p-1..\gamma-1]))$. 
This binary search needs $O(\log m^{\prime})$ LCE and random access queries. 
Here, $\log m^{\prime} = O(\log n)$ holds because $m^{\prime} \leq m$ and $\log m = O(\log n)$. 
Therefore, the fifth step takes $O(H^{\prime} \log n + \log^{2} n)$ time. 

In the sixth step, we find the smallest integer $\beta$ by binary search on the strings of the ordered set $\mathcal{Y}_{A}(h)$. 
Similar to the third step, 
this binary search can be executed in $O(H^{\prime} \log n + \log^{2} n)$ time. 

The seventh step is executed if $T^{\prime}[\gamma..r+1] \neq R_{\beta}$ holds. 
In the seventh step, we create a new doubly linked list $\mathbf{Y}_{A}(h, T^{\prime}[\gamma..r+1])$ 
and insert a new element into doubly linked list $\mathbf{L}^{Y}_{A}(h)$ as the $\beta$-th element. 
Similar to the fourth step, 
the seventh step can be executed in $O(\log n)$ time. 

In the eighth step, 
we create element $v^{\prime}$ and insert it into doubly linked list $\mathbf{Y}_{A}(h, T^{\prime}[\gamma..r+1])$ as the $\beta^{\prime}+1$-th element. 
Similar to the fifth step, 
the eighth step can be executed in $O(H^{\prime} \log n + \log^{2} n)$ time. 

In the ninth step, 
we insert a new element into the doubly linked list representing the set $\mathcal{J}_{A}(h)$ 
as the $(j+1)$-th element. 
This insertion takes $O(\log^{4} m)$ time 
because we need to update the range-sum data structure built on set $\mathcal{J}_{A}(h)$. 
The integer $j$ is computed by binary search on the $m$ strings $e_{1}, e_{2}, \ldots, e_{m}$. 
Similar to the first phase,  
the integer $j$ can be computed in $O(H^{\prime} \log n + \log^{2} n)$ time. 
Therefore, the ninth step takes $O(H^{\prime} \log n + \log^{4} n)$ time. 

Finally, the third phase takes $O(|\Psi^{\prime \symB}| (H^{\prime} \log n + \log^{4} n))$ time. 
Here, $\Psi^{\prime \symB} \subseteq \Psi^{\prime}_{\NEW} \setminus \Psi^{\prime}_{\run}$ holds, 
and $\mathbb{E}[|\Psi^{\prime}_{\NEW} \setminus \Psi^{\prime}_{\run}|] = O(H^{\prime} + \log n)$ follows from 
Lemma~\ref{lem:dynamic_psi_overlap}~\ref{enum:dynamic_psi_overlap:2}. 
Therefore, this phase takes expected $O((H^{\prime} + \log n)(H^{\prime} \log n + \log^{4} n))$ time. 

\paragraph{Phase (iv).}
In the fourth phase, 
we update the data structures for subquery $\RSCQA$ based on Modification 4. 
For each interval attractor $([p, q], [\ell, r]) \in \Psi^{\prime}_{\NEW} \setminus \Psi^{\prime}_{\run}$ with level $h$, 
set $\mathcal{J}_{A}(h)$ contains a weighted point $(x_{j}, y_{j}, w_{j}, e_{j})$ satisfying $e_{j} = T^{\prime}[p-1..r+1]$, and $1$ is added to the weight $w_{j}$. 

The fourth phase consists of three steps. 
In the first step, 
we compute the level $h$ of interval attractor $([p, q], [\ell, r])$ by level query, 
which takes $O(H^{\prime 2})$ time. 

In the second step, we compute the integer $j$ by binary search on 
the $m$ identifiers $e_{1}, e_{2}, \ldots, e_{m}$. 
Similar to the second step of the first phase, 
the integer $j$ can be computed in $O(H^{\prime} \log n + \log^{2} n)$ time. 

In the third step, 
$1$ is added to the weight $w_{j}$ stored in the $j$-th element of the doubly linked list representing the set $\mathcal{J}_{A}(h)$. 
Similar to the third step of the first phase, 
this addition takes $O(\log^{4} m)$ time. 

The fourth phase takes $O(|\Psi^{\prime}_{\NEW} \setminus \Psi^{\prime}_{\run}| (H^{\prime 2} \log n + \log^{4} n))$ time in total. 
Here, $\mathbb{E}[|\Psi^{\prime}_{\NEW} \setminus \Psi^{\prime}_{\run}|] = O(H^{\prime} + \log n)$ follows from Lemma~\ref{lem:dynamic_psi_overlap}. 
Therefore, this phase takes expected $O((H^{\prime} + \log n)(H^{\prime 2} \log n + \log^{4} n))$ time. 

Finally, the four phases take expected $O((H + H^{\prime} + \log n)(H^{2} \log n + H^{\prime 2} \log n + \log^{4} n))$ time in total. 
Therefore, Lemma~\ref{lem:dynamic_JA_summary} holds.

\subsection{Update of Data Structures for RSC-B1 and RSC-B2 Subqueries}\label{subsec:rb_update}
This subsection explains how to update the dynamic data structures for subqueries $\RSCQBX$ and $\RSCQBY$ 
introduced in Section~\ref{subsubsec:JB_ds}. 
The following lemma is the summary of this subsection.

\begin{lemma}\label{lem:dynamic_JB_summary}
Consider the two RLSLPs $\mathcal{G}^{R}$ and $\mathcal{G}^{R}_{\ins}$ of Theorem~\ref{theo:update1}, which derive input string $T$ and string $T^{\prime}$, respectively. 
The dynamic data structures of Section~\ref{subsubsec:JB_ds} can be updated 
in expected $O((\max \{H, H^{\prime}, \log (nHH^{\prime}) \})^{5})$ time 
after changing RLSLP $\mathcal{G}^{R}$ to $\mathcal{G}^{R}_{\ins}$. 
This update requires 
(i) the dynamic data structures for the RR-DAG of RLSLP $\mathcal{G}^{R}$ (Section~\ref{subsubsec:rrdag_ds}) 
and sample query (Section~\ref{subsec:sample_query}), 
and (ii) the interval attractors obtained from Lemma~\ref{lem:dynamic_IA_summary}. 
\end{lemma}
\begin{proof}
    See Section~\ref{subsubsec:JB_update_algorithm}. 
\end{proof}

The dynamic data structures for subqueries $\RSCQBX$ and $\RSCQBY$ store weighted points of the $(H+1)$ sets 
$\mathcal{J}_{B}(0)$, $\mathcal{J}_{B}(1)$, $\ldots$, $\mathcal{J}_{B}(H)$ introduced in Section~\ref{subsec:RSC_comp_B}. 
Here, each set $\mathcal{J}_{B}(h)$ consists of weighted points on 
grid $(\mathcal{X}_{B}(h), \mathcal{Y}_{B}(h))$ for the two ordered sets $\mathcal{X}_{B}(h) = \{ L_{1}, L_{2}, \ldots, L_{d} \}$ and $\mathcal{Y}_{B}(h) = \{ R_{1}, R_{2}, \ldots, R_{d^{\prime}} \}$ of strings introduced in Section~\ref{subsec:RSC_comp_B}. 
For each integer $h \in [0, H^{\prime}]$, 
let $\mathcal{J}^{\prime}_{B}(h)$ be the set $\mathcal{J}_{B}(h)$ of weighted points defined using set $\Psi^{\prime}_{\RR}$ instead of set $\Psi_{\RR}$. 
For simplicity, let $\mathcal{J}_{B}(H+1)$, $\mathcal{J}_{B}(H+2)$, $\ldots$, $\mathcal{J}_{B}(H^{\prime})$ be empty sets of weighted points if $H^{\prime} > H$. 
Similarly, let $\mathcal{J}^{\prime}_{B}(H^{\prime}+1)$, $\mathcal{J}^{\prime}_{B}(H^{\prime}+2)$, $\ldots$, $\mathcal{J}^{\prime}_{B}(H)$ be empty sets of weighted points if $H^{\prime} < H$. 

The following lemma shows that each set $\mathcal{J}^{\prime}_{B}(h)$ can be obtained by modifying set $\mathcal{J}_{B}(h)$. 

\begin{lemma}\label{lem:dynamic_JB_change}
Consider an integer $h$ in set $[0, H^{\prime}]$. 
For each weighted point $(x_{s}, y_{s}, w_{s}, e_{s}) \in \mathcal{J}_{B}(h)$, 
let $([p_{s}, q_{s}], [\ell_{s}, r_{s}]) \in \Psi_{h} \cap \Psi_{\source} \cap \Psi_{\samp}$ be 
the interval attractor corresponding to the weighted point $(x_{s}, y_{s}, w_{s}, e_{s})$, 
$\kappa_{s} = |\Psi_{h} \cap \Psi_{\source} \cap (\Psi_{\OLD} \setminus \Psi_{\run}) \cap \Psi_{\str}(e_{s})|$,  
$\kappa^{\prime}_{s} = |\Psi^{\prime}_{h} \cap \Psi^{\prime}_{\source} \cap (\Psi^{\prime}_{\NEW} \setminus \Psi^{\prime}_{\run}) \cap \Psi^{\prime}_{\str}(e_{s})|$, 
and $\nu_{s} = |f_{\recover}(([p_{s}, q_{s}], [\ell_{s}, r_{s}]))|$. 
For each interval attractor $([p^{\prime}_{s}, q^{\prime}_{s}], [\ell^{\prime}_{s}, r^{\prime}_{s}]) \in \Psi^{\prime}_{h} \cap \Psi^{\prime}_{\source} \cap \Psi^{\prime \symB}$, 
let $\nu^{\prime}_{s} = |f^{\prime}_{\recover}(([p^{\prime}_{s}, q^{\prime}_{s}], [\ell^{\prime}_{s}, r^{\prime}_{s}]))|$. 
Here, Lemma~\ref{lem:mRecover_basic_property} shows that 
there exists an interval attractor $([\hat{p}_{s}, \hat{q}_{s}], [\hat{\ell}_{s}, \hat{r}_{s}]) \in \Psi^{\prime}_{\RR}$ satisfying $f^{\prime}_{\recover}(([p^{\prime}_{s}, q^{\prime}_{s}], [\ell^{\prime}_{s}, r^{\prime}_{s}])) \cap \Psi^{\prime}_{\mRecover} = \{ ([\hat{p}_{s}, \hat{q}_{s}], [\hat{\ell}_{s}, \hat{r}_{s}]) \}$. 
Then, set $\mathcal{J}^{\prime}_{B}(h)$ of weighted points 
is equal to the union of the following two sets of weighted points: 
\begin{enumerate}[label=\textbf{(\roman*)}]
    \item $\{ (x_{s}, y_{s}, w_{s} - \nu_{s} \kappa_{s} + \nu_{s} \kappa^{\prime}_{s}, e_{s}) \mid (x_{s}, y_{s}, w_{s}, e_{s}) \in \mathcal{J}_{B}(h) \text{ s.t. } ([p_{s}, q_{s}], [\ell_{s}, r_{s}]) \not \in \Psi_{\OLD} \setminus \Psi^{\symA} \}$;
    \item $\{ (\reverse(T^{\prime}[\hat{p}_{s}-1..\hat{\gamma}_{s}-1]), T^{\prime}[\hat{\gamma}_{s}..\hat{r}_{s}+1], \nu^{\prime}_{s} |\Psi^{\prime}_{\str}(T[p^{\prime}_{s}-1..r^{\prime}_{s}+1])|, T^{\prime}[p^{\prime}_{s}-1..r^{\prime}_{s}+1]) \mid ([p^{\prime}_{s}, q^{\prime}_{s}], [\ell^{\prime}_{s}$, $r^{\prime}_{s}]) \in \Psi^{\prime}_{h} \cap \Psi^{\prime}_{\source} \cap \Psi^{\prime \symB} \}$. 
    Here, $\hat{\gamma}_{s}$ is the attractor position of interval attractor $([\hat{p}_{s}, \hat{q}_{s}]$, 
    $[\hat{\ell}_{s}, \hat{r}_{s}])$. 
\end{enumerate}
\end{lemma}
\begin{proof}
For each weighted point $(x_{s}, y_{s}, w_{s}, e_{s}) \in \mathcal{J}_{B}(h)$, 
Lemma~\ref{lem:mRecover_basic_property} shows that 
there exists an interval attractor $([p_{A, s}, q_{A, s}], [\ell_{A, s}, r_{A, s}]) \in \Psi$ satisfying $f_{\recover}(([p_{s}, q_{s}]$, $[\ell_{s}, r_{s}])) \cap \Psi^{\prime}_{\mRecover} = \{ ([p_{A, s}, q_{A, s}], [\ell_{A, s}, r_{A, s}]) \}$. 
Let $\gamma_{A, s}$ be the attractor position of the interval attractor $([p_{A, s}, q_{A, s}]$, $[\ell_{A, s}, r_{A, s}])$. 

Lemma~\ref{lem:dynamic_JB_change} follows from the following four statements: 
\begin{enumerate}[label=\textbf{(\Alph*)}]
    \item consider a weighted point $(x_{s}, y_{s}, w_{s}, e_{s}) \in \mathcal{J}_{B}(h)$ satisfying $([p_{s}, q_{s}], [\ell_{s}, r_{s}]) \not \in \Psi_{\OLD} \setminus \Psi^{\symA}$. 
    Then, $(x_{s}, y_{s}, w_{s} - \nu_{s} \kappa_{s} + \nu_{s} \kappa^{\prime}_{s}, e_{s}) \in \mathcal{J}^{\prime}_{B}(h)$;
    \item consider an interval attractor $([p^{\prime}_{s}, q^{\prime}_{s}], [\ell^{\prime}_{s}, r^{\prime}_{s}]) \in \Psi^{\prime}_{h} \cap \Psi^{\prime}_{\source} \cap \Psi^{\prime \symB}$. 
    Then, $(\reverse(T^{\prime}[\hat{p}_{s}-1..\hat{\gamma}_{s}-1]), T^{\prime}[\hat{\gamma}_{s}..\hat{r}_{s}+1], \nu^{\prime}_{s} |\Psi^{\prime}_{\str}(T[p^{\prime}_{s}-1..r^{\prime}_{s}+1])|, T^{\prime}[p^{\prime}_{s}-1..r^{\prime}_{s}+1]) \in \mathcal{J}^{\prime}_{B}(h)$;
    \item consider a weighted point $(x^{\prime}, y^{\prime}, w^{\prime}, e^{\prime}) \in \mathcal{J}^{\prime}_{B}(h)$ satisfying $e^{\prime} \not \in \mathcal{Z}^{\symB}$ for 
    the set $\mathcal{Z}^{\symB}$ of strings introduced in Section~\ref{subsec:update_sampling_subset}. 
    Then, there exists a weighted point $(x_{s}, y_{s}, w_{s}, e_{s}) \in \mathcal{J}_{B}(h)$ satisfying 
    $([p_{s}, q_{s}], [\ell_{s}, r_{s}]) \not \in \Psi_{\OLD} \setminus \Psi^{\symA}$ 
    and $(x_{s}, y_{s}, w_{s} - \nu_{s} \kappa_{s} + \nu_{s} \kappa^{\prime}_{s}, e_{s}) = (x^{\prime}, y^{\prime}, w^{\prime}, e^{\prime})$;
    \item consider a weighted point $(x^{\prime}, y^{\prime}, w^{\prime}, e^{\prime}) \in \mathcal{J}^{\prime}_{B}(h)$ satisfying $e^{\prime} \in \mathcal{Z}^{\symB}$. 
    Then, there exists an interval attractor $([p^{\prime}_{s}, q^{\prime}_{s}], [\ell^{\prime}_{s}, r^{\prime}_{s}]) \in \Psi^{\prime}_{h} \cap \Psi^{\prime}_{\source} \cap \Psi^{\prime \symB}$ 
    satisfying $(\reverse(T^{\prime}[\hat{p}_{s}-1..\hat{\gamma}_{s}-1]), T^{\prime}[\hat{\gamma}_{s}..\hat{r}_{s}+1], \nu^{\prime}_{s} |\Psi^{\prime}_{\str}(T[p^{\prime}_{s}-1..r^{\prime}_{s}+1])|, T^{\prime}[p^{\prime}_{s}-1..r^{\prime}_{s}+1]) = (x^{\prime}, y^{\prime}, w^{\prime}, e^{\prime})$. 
\end{enumerate}

\textbf{Proof of statement (A).}
Lemma~\ref{lem:dynamic_samp_proceeding_formula} shows that 
sampling subset $\Psi^{\prime}_{\samp}$ contains an interval attractor $([p^{\prime}, q^{\prime}], [\ell^{\prime}, r^{\prime}])$ satisfying $T[p_{s}-1..r_{s}+1] = T^{\prime}[p^{\prime}-1..r^{\prime}+1]$. 
We apply Lemma~\ref{lem:dynamic_IA_super_correspondence} to the two interval attractors 
$([p_{s}, q_{s}], [\ell_{s}, r_{s}])$ and $([p^{\prime}, q^{\prime}], [\ell^{\prime}, r^{\prime}])$. 
Then, the lemma shows that $([p^{\prime}, q^{\prime}], [\ell^{\prime}, r^{\prime}]) \in \Psi^{\prime}_{h}$ holds. 
Similarly, we apply Lemma~\ref{lem:dynamic_RR_subset}~\ref{enum:dynamic_RR_subset:source} to the two interval attractors $([p_{s}, q_{s}], [\ell_{s}, r_{s}])$ and $([p^{\prime}, q^{\prime}], [\ell^{\prime}, r^{\prime}])$. 
Then, the lemma shows that $([p^{\prime}, q^{\prime}], [\ell^{\prime}, r^{\prime}]) \in \Psi^{\prime}_{\source}$ holds. 
Lemma~\ref{lem:mRecover_basic_property} shows that 
there exists an interval attractor $([\hat{p}, \hat{q}], [\hat{\ell}, \hat{r}]) \in \Psi^{\prime}_{\RR}$ satisfying $f^{\prime}_{\recover}(([p^{\prime}, q^{\prime}]$, $[\ell^{\prime}, r^{\prime}])) \cap \Psi^{\prime}_{\mRecover} = \{ ([\hat{p}, \hat{q}], [\hat{\ell}, \hat{r}]) \}$. 
Let $\nu^{\prime} = |f^{\prime}_{\recover}(([p^{\prime}, q^{\prime}]$, $[\ell^{\prime}, r^{\prime}]))|$ for simplicity. 
We apply Lemma~\ref{lem:dynamic_f_recover} to the two interval attractors 
$([p_{s}, q_{s}], [\ell_{s}, r_{s}])$ and $([p^{\prime}, q^{\prime}], [\ell^{\prime}, r^{\prime}])$. 
Then, the following two statements hold: 
\begin{itemize}
    \item $\nu_{s} = \nu^{\prime}$;
    \item 
    $T[p_{A, s}-1..r_{A, s}-1] = T^{\prime}[\hat{p}-1..\hat{r}+1]$. 
    In addition,     
    $T[p_{A, s}-1..\gamma_{A, s}-1] = T^{\prime}[\hat{p}-1..\hat{\gamma}+1]$ 
    and $T[\gamma_{A, s}..r_{A, s}+1] = T^{\prime}[\hat{\gamma}..\hat{r}+1]$ hold for 
    the attractor position $\hat{\gamma}$ of interval attractor $([\hat{p}, \hat{q}], [\hat{\ell}, \hat{r}])$. 
    This is because we can apply Lemma~\ref{lem:dynamic_IA_super_correspondence} to 
    the two interval attractors $([p_{A, s}, q_{A, s}], [\ell_{A, s}, r_{A, s}])$ and $([\hat{p}, \hat{q}], [\hat{\ell}, \hat{r}])$.
\end{itemize}

Because of $([p^{\prime}, q^{\prime}], [\ell^{\prime}, r^{\prime}]) \in \Psi^{\prime}_{h} \cap \Psi^{\prime}_{\source} \cap \Psi^{\prime}_{\samp}$, 
set $\mathcal{J}^{\prime}_{B}(h)$ contains weighted point $(\reverse(T^{\prime}[\hat{p}-1..\hat{\gamma}-1]), T^{\prime}[\hat{\gamma}..\hat{r}+1], \nu^{\prime} |\Psi^{\prime}_{\str}(T[p^{\prime}-1..r^{\prime}+1])|, T^{\prime}[p^{\prime}-1..r^{\prime}+1])$, which corresponds to 
the interval attractor $([p^{\prime}, q^{\prime}], [\ell^{\prime}, r^{\prime}])$. 

$x_{s} = \reverse(T^{\prime}[\hat{p}-1..\hat{\gamma}-1])$ follows from 
$x_{s} = \reverse(T[p_{A, s}-1..\gamma_{A, s}-1])$ and $T[p_{A, s}-1..\gamma_{A, s}-1] = T^{\prime}[\hat{p}-1..\hat{\gamma}-1]$. 
$y_{s} = T^{\prime}[\gamma^{\prime}..r^{\prime}+1]$ follows from 
$y_{s} = T^{\prime}[\hat{\gamma}..\hat{r}+1]$ and 
$T[\gamma_{A, s}..r_{A, s}+1] = T^{\prime}[\hat{\gamma}..\hat{r}+1]$. 
$e_{s} = T^{\prime}[p^{\prime}-1..r^{\prime}+1]$ 
follows from $e_{s} = T[p_{s}-1..r_{s}+1]$ and $T[p_{s}-1..r_{s}+1] = T^{\prime}[p^{\prime}-1..r^{\prime}+1]$. 

Let $Z = T^{\prime}[p^{\prime}-1..r^{\prime}+1]$ for simplicity. 
We prove $\Psi_{h} \cap (\Psi_{\OLD} \setminus \Psi_{\run}) \cap \Psi_{\str}(Z) = \Psi_{h} \cap \Psi_{\source} \cap (\Psi_{\OLD} \setminus \Psi_{\run}) \cap \Psi_{\str}(Z)$. 
Consider an interval attractor $([p, q], [\ell, r])$ in set $\Psi_{h} \cap (\Psi_{\OLD} \setminus \Psi_{\run}) \cap \Psi_{\str}(Z)$. 
Then, $T[p-1..r+1] = Z$ follows from the definition of subset $\Psi_{\str}(Z)$. 
Because of $T[p-1..r+1] = T[p_{s}-1..r_{s}+1]$ and $([p_{s}, q_{s}], [\ell_{s}, r_{s}]) \in \Psi_{\source}$, 
Lemma~\ref{lem:psi_equality_basic_property}~\ref{enum:psi_equality_basic_property:5} shows that 
$([p, q], [\ell, r]) \in \Psi_{\source}$ holds. 
Therefore, $\Psi_{h} \cap (\Psi_{\OLD} \setminus \Psi_{\run}) \cap \Psi_{\str}(Z) = \Psi_{h} \cap \Psi_{\source} \cap (\Psi_{\OLD} \setminus \Psi_{\run}) \cap \Psi_{\str}(Z)$ holds. 
Similarly, $\Psi^{\prime}_{h} \cap (\Psi^{\prime}_{\NEW} \setminus \Psi^{\prime}_{\run}) \cap \Psi^{\prime}_{\str}(Z) = \Psi^{\prime}_{h} \cap \Psi^{\prime}_{\source}  \cap (\Psi^{\prime}_{\NEW} \setminus \Psi^{\prime}_{\run}) \cap \Psi^{\prime}_{\str}(Z)$ holds.

We prove $w_{s} - \nu_{s} \kappa_{s} + \nu_{s} \kappa^{\prime}_{s} = \nu^{\prime} |\Psi^{\prime}_{\str}(T^{\prime}[p^{\prime}-1..r^{\prime}+1])|$. 
Because of $([p^{\prime}, q^{\prime}], [\ell^{\prime}, r^{\prime}]) \in \Psi^{\prime}_{\samp}$, 
$([p^{\prime}, q^{\prime}], [\ell^{\prime}, r^{\prime}]) \not \in \Psi^{\prime}_{\run}$ follows from 
Lemma~\ref{lem:samp_basic_property}~\ref{enum:samp_basic_property:3}. 
$|\Psi^{\prime}_{\str}(Z)| = |\Psi_{\str}(Z)| - |\Psi_{h} \cap (\Psi_{\OLD} \setminus \Psi_{\run}) \cap \Psi_{\str}(Z)| + |\Psi^{\prime}_{h} \cap (\Psi^{\prime}_{\NEW} \setminus \Psi^{\prime}_{\run}) \cap \Psi^{\prime}_{\str}(Z)|$ follows from 
Corollary~\ref{cor:dynamic_str_formula}~\ref{enum:dynamic_str_formula:1}. 
Therefore, the following equation holds: 
\begin{equation*}
    \begin{split}
        \nu^{\prime} |\Psi^{\prime}_{\str}(Z)| &= \nu_{s} |\Psi^{\prime}_{\str}(Z)| \\ 
        &= \nu_{s} (|\Psi_{\str}(Z)| - |\Psi_{h} \cap (\Psi_{\OLD} \setminus \Psi_{\run}) \cap \Psi_{\str}(Z)| + |\Psi^{\prime}_{h} \cap (\Psi^{\prime}_{\NEW} \setminus \Psi^{\prime}_{\run}) \cap \Psi^{\prime}_{\str}(Z)|) \\
        &= \nu_{s} (|\Psi_{\str}(Z)| - |\Psi_{h} \cap \Psi_{\source} \cap (\Psi_{\OLD} \setminus \Psi_{\run}) \cap \Psi_{\str}(Z)| \\
        &+ |\Psi^{\prime}_{h} \cap \Psi^{\prime}_{\source} \cap (\Psi^{\prime}_{\NEW} \setminus \Psi^{\prime}_{\run}) \cap \Psi^{\prime}_{\str}(Z)|) \\
        &= w_{s} - \nu_{s} \kappa_{s} + \nu_{s} \kappa^{\prime}_{s}.
    \end{split}
\end{equation*}

We proved $(x_{s}, y_{s}, w_{s} - \nu_{s} \kappa_{s} + \nu_{s} \kappa^{\prime}_{s}, e_{s}) = (\reverse(T^{\prime}[\hat{p}-1..\hat{\gamma}-1]), T^{\prime}[\hat{\gamma}..\hat{r}+1], \nu^{\prime} |\Psi^{\prime}_{\str}(T[p^{\prime}-1..r^{\prime}+1])|, T^{\prime}[p^{\prime}-1..r^{\prime}+1])$. 
Therefore, $(x_{s}, y_{s}, w_{s} - \nu_{s} \kappa_{s} + \nu_{s} \kappa^{\prime}_{s}, e_{s}) \in \mathcal{J}^{\prime}_{B}(h)$ holds. 

\textbf{Proof of statement (B).}
Set $\Psi^{\prime}_{h} \cap \Psi^{\prime}_{\source} \cap \Psi^{\prime}_{\samp}$ contains 
interval attractor $([p^{\prime}_{s}, q^{\prime}_{s}], [\ell^{\prime}_{s}, r^{\prime}_{s}])$ 
because Theorem~\ref{theo:dynamic_samp_formula} shows that 
$\Psi^{\prime \symB} \subseteq \Psi^{\prime}_{\samp}$ holds. 
Because of $([p^{\prime}_{s}, q^{\prime}_{s}], [\ell^{\prime}_{s}, r^{\prime}_{s}]) \in \Psi^{\prime}_{h} \cap \Psi^{\prime}_{\source} \cap \Psi^{\prime}_{\samp}$, 
$(\reverse(T^{\prime}[\hat{p}_{s}-1..\hat{\gamma}_{s}-1]), T^{\prime}[\hat{\gamma}_{s}..\hat{r}_{s}+1], \nu^{\prime}_{s} |\Psi^{\prime}_{\str}(T[p^{\prime}_{s}-1..r^{\prime}_{s}+1])|, T^{\prime}[p^{\prime}_{s}-1..r^{\prime}_{s}+1]) \in \mathcal{J}^{\prime}_{B}(h)$ follows from the definition of the set $\mathcal{J}^{\prime}_{B}(h)$.

\textbf{Proof of statement (C).}
The weighted point $(x^{\prime}, y^{\prime}, w^{\prime}, e^{\prime})$ corresponds to 
an interval attractor $([p^{\prime}, q^{\prime}], [\ell^{\prime}, r^{\prime}]) \in \Psi^{\prime}_{h} \cap \Psi^{\prime}_{\source} \cap \Psi^{\prime}_{\samp}$ satisfying $T^{\prime}[p^{\prime}-1..r^{\prime}+1] = e^{\prime}$. 
$([p^{\prime}, q^{\prime}], [\ell^{\prime}, r^{\prime}]) \not \in \Psi^{\prime \symB}$ follows from 
$T^{\prime}[p^{\prime}-1..r^{\prime}+1] \not \in \mathcal{Z}^{\symB}$. 
Because of $([p^{\prime}, q^{\prime}], [\ell^{\prime}, r^{\prime}]) \not \in \Psi^{\prime \symB}$, 
Lemma~\ref{lem:dynamic_samp_back_formula} shows that 
sampling subset $\Psi_{\samp}$ contains an interval attractor 
$([p, q], [\ell, r])$ satisfying 
$T[p-1..r+1] = T^{\prime}[p^{\prime}-1..r^{\prime}+1]$ and 
$([p, q], [\ell, r]) \not \in \Psi_{\OLD} \setminus \Psi^{\symA}$. 

We apply Lemma~\ref{lem:dynamic_IA_super_correspondence} to the two interval attractors 
$([p, q], [\ell, r])$ and $([p^{\prime}, q^{\prime}], [\ell^{\prime}, r^{\prime}])$. 
Then, the lemma shows that $([p, q], [\ell, r]) \in \Psi_{h}$ holds. 
Similarly, we apply Lemma~\ref{lem:dynamic_RR_subset}~\ref{enum:dynamic_RR_subset:source} to the two interval attractors $([p, q], [\ell, r])$ and $([p^{\prime}, q^{\prime}], [\ell^{\prime}, r^{\prime}])$. 
Then, the lemma shows that $([p, q], [\ell, r]) \in \Psi_{\source}$ holds. 
Because of $([p, q], [\ell, r]) \in \Psi_{h} \cap \Psi_{\source} \cap \Psi_{\samp}$, 
set $\mathcal{J}_{B}(h)$ contains a weighted point $(x_{s}, y_{s}, w_{s}, e_{s})$ corresponding to  
the interval attractor $([p, q], [\ell, r])$ (i.e., $([p_{s}, q_{s}], [\ell_{s}, r_{s}]) = ([p, q], [\ell, r])$). 
Similar to the proof of statement (A), 
$(x_{s}, y_{s}, w_{s} - \nu_{s} \kappa_{s} + \nu_{s} \kappa^{\prime}_{s}, e_{s}) = (x^{\prime}, y^{\prime}, w^{\prime}, e^{\prime})$ can be proved. 
Therefore, statement (C) holds.

\textbf{Proof of statement (D).}
The weighted point $(x^{\prime}, y^{\prime}, w^{\prime}, e^{\prime})$ corresponds to 
an interval attractor $([p^{\prime}, q^{\prime}], [\ell^{\prime}, r^{\prime}]) \in \Psi^{\prime}_{h} \cap \Psi^{\prime}_{\source} \cap \Psi^{\prime}_{\samp}$ satisfying $T^{\prime}[p^{\prime}-1..r^{\prime}+1] = e^{\prime}$. 
$([p^{\prime}, q^{\prime}], [\ell^{\prime}, r^{\prime}]) \in \Psi^{\prime \symB}$ can be proved using the same approach as for statement \ref{enu:dynamic_JA_change:D} in the proof of Lemma~\ref{lem:dynamic_JA_change}. 
Because of $([p^{\prime}, q^{\prime}], [\ell^{\prime}, r^{\prime}]) \in \Psi^{\prime}_{h} \cap \Psi^{\prime}_{\source} \cap \Psi^{\prime \symB}$, 
set $\Psi^{\prime}_{h} \cap \Psi^{\prime}_{\source} \cap \Psi^{\prime \symB}$ contains the interval attractor $([p^{\prime}, q^{\prime}], [\ell^{\prime}, r^{\prime}])$ as an interval attractor $([p^{\prime}_{s}, q^{\prime}_{s}], [\ell^{\prime}_{s}, r^{\prime}_{s}])$. 
$(\reverse(T^{\prime}[\hat{p}_{s}-1..\hat{\gamma}_{s}-1]), T^{\prime}[\hat{\gamma}_{s}..\hat{r}_{s}+1], \nu^{\prime}_{s} |\Psi^{\prime}_{\str}(T[p^{\prime}_{s}-1..r^{\prime}_{s}+1])|, T^{\prime}[p^{\prime}_{s}-1..r^{\prime}_{s}+1]) = (x^{\prime}, y^{\prime}, w^{\prime}, e^{\prime})$ follows from the definition of set $\mathcal{J}^{\prime}_{B}(h)$. 
Therefore, statement (D) holds. 
\end{proof}

Consider the $d$ sequences $\mathbf{Q}^{X}_{B}(h, L_{1})$, $\mathbf{Q}^{X}_{B}(h, L_{2})$, $\ldots$, 
$\mathbf{Q}^{X}_{B}(h, L_{d})$ and $d^{\prime}$ sequences $\mathbf{Q}^{Y}_{B}(h, R_{1})$, 
$\mathbf{Q}^{Y}_{B}(h, R_{2})$, $\ldots$, $\mathbf{Q}^{Y}_{B}(h, R_{d^{\prime}})$ of weighted points 
introduced in Section~\ref{subsubsec:JB_X_ds} and Section~\ref{subsubsec:JB_Y_ds}. 
Here, each sequence $\mathbf{Q}^{X}_{B}(h, L_{s})$ is represented as the doubly linked list $\mathbf{X}_{B}(h, L_{s})$ introduced in Section~\ref{subsubsec:JB_X_ds}. 
Similarly, each sequence $\mathbf{Q}^{Y}_{B}(h, R_{s})$ is represented as the doubly linked list $\mathbf{Y}_{B}(h, R_{s})$ introduced in Section~\ref{subsubsec:JB_X_ds}. 
The following lemma states properties of these sequences of weighted points. 

\begin{lemma}\label{lem:dynamic_JB_move}
Consider an interval attractor $([p_{1}, q_{1}], [\ell_{1}, r_{1}]) \in \Psi_{h} \cap \Psi_{\source} \cap \Psi_{\samp}$ for an integer $h \in [0, H]$ satisfying $([p_{1}, q_{1}], [\ell_{1}, r_{1}]) \not \in \Psi_{\OLD} \setminus \Psi^{\symA}$. 
Here, set $\mathcal{J}_{B}(h)$ contains a weighted point $(x, y, w, e)$ corresponding to 
the interval attractor $([p_{1}, q_{1}], [\ell_{1}, r_{1}])$; 
the interval attractor $([p_{1}, q_{1}]$, $[\ell_{1}, r_{1}])$ is represented as a node $u$ of the sequence $\mathbf{Q}_{\samp}$ introduced in Section~\ref{subsec:sample_query}; 
Lemma~\ref{lem:mRecover_basic_property} shows that 
there exists an interval attractor $([\hat{p}_{1}, \hat{q}_{1}], [\hat{\ell}_{1}, \hat{r}_{1}]) \in \Psi_{\RR}$ satisfying $f_{\recover}(([p_{1}, q_{1}]$, $[\ell_{1}, r_{1}])) \cap \Psi_{\mRecover} = \{ ([\hat{p}_{1}, \hat{q}_{1}], [\hat{\ell}_{1}, \hat{r}_{1}]) \}$. 

From the definition of sequence $\mathbf{Q}^{X}_{B}(h, x)$, 
there exists an integer $\alpha$ such that 
the sequence $\mathbf{Q}^{X}_{B}(h, x)$ contains weighted point $(x, y, w, e)$ as the $\alpha$-th element, 
and the $\alpha$-th element of doubly linked list $\mathbf{X}_{B}(h, x)$ stores a pointer to node $u$. 
Similarly, 
there exists an integer $\beta$ such that 
the sequence $\mathbf{Q}^{Y}_{B}(h, y)$ contains weighted point $(x, y, w, e)$ as the $\beta$-th element, 
and the $\beta$-th element of doubly linked list $\mathbf{Y}_{B}(h, y)$ stores a pointer to node $u$. 

After updating the dynamic data structures for sample query, 
node $u$ represents an interval attractor $([p_{2}, q_{2}], [\ell_{2}, r_{2}])$ in set $\Psi^{\prime}_{h} \cap \Psi^{\prime}_{\source} \cap \Psi^{\prime}_{\samp}$ satisfying the following three conditions: 
\begin{enumerate}[label=\textbf{(\roman*)}]
    \item $T[p_{1}-1..r_{1}+1] = T^{\prime}[p_{2}-1..r_{2}+1]$;
    \item $([p_{2}, q_{2}], [\ell_{2}, r_{2}]) \not \in \Psi^{\prime \symB}$;    
    \item let $([\hat{p}_{2}, \hat{q}_{2}], [\hat{\ell}_{2}, \hat{r}_{2}]) \in \Psi^{\prime}_{\RR}$ be an interval attractor satisfying $f^{\prime}_{\recover}(([p_{2}, q_{2}], [\ell_{2}, r_{2}])) \cap \Psi^{\prime}_{\mRecover} = \{ ([\hat{p}_{2}, \hat{q}_{2}], [\hat{\ell}_{2}, \hat{r}_{2}]) \}$. 
    Then, 
    $T[\hat{p}_{1}-1..\hat{\gamma}_{1}-1] = T^{\prime}[\hat{p}_{2}-1..\hat{\gamma}_{2}+1]$ 
    and $T[\hat{\gamma}_{1}..\hat{r}_{1}+1] = T^{\prime}[\hat{\gamma}_{2}..\hat{r}_{1}+1]$ for 
    the attractor positions $\hat{\gamma}_{1}$ and $\hat{\gamma}_{2}$ of two interval attractors 
    $([\hat{p}_{1}, \hat{q}_{1}], [\hat{\ell}_{1}, \hat{r}_{1}])$ and $([\hat{p}_{2}, \hat{q}_{2}], [\hat{\ell}_{2}, \hat{r}_{2}])$, respectively.
\end{enumerate}
\end{lemma}
\begin{proof}
Lemma~\ref{lem:dynamic_samp_proceeding_formula} shows that 
node $u$ represents an interval attractor $([p_{2}, q_{2}], [\ell_{2}, r_{2}])$ in set 
$\Psi^{\prime}_{\samp}$ 
satisfying $T[p_{1}-1..r_{1}+1] = T^{\prime}[p_{2}-1..r_{2}+1]$ 
and $([p_{2}, q_{2}], [\ell_{2}, r_{2}]) \not \in \Psi^{\prime \symB}$. 
We apply Lemma~\ref{lem:dynamic_IA_super_correspondence} to the two interval attractors 
$([p_{1}, q_{1}], [\ell_{1}, r_{1}])$ and $([p_{2}, q_{2}], [\ell_{2}, r_{2}])$. 
Then, the lemma shows that $([p_{2}, q_{2}], [\ell_{2}, r_{2}]) \in \Psi^{\prime}_{h}$ holds. 
Similarly, we apply Lemma~\ref{lem:dynamic_RR_subset}~\ref{enum:dynamic_RR_subset:source} to the two interval attractors $([p_{1}, q_{1}], [\ell_{1}, r_{1}])$ and $([p_{2}, q_{2}], [\ell_{2}, r_{2}])$. 
Then, the lemma shows that $([p_{2}, q_{2}], [\ell_{2}, r_{2}]) \in \Psi^{\prime}_{\source}$ holds. 

Because of $([p_{2}, q_{2}], [\ell_{2}, r_{2}]) \in \Psi^{\prime}_{\source}$, 
Lemma~\ref{lem:mRecover_basic_property} shows that 
there exists an interval attractor $([\hat{p}_{2}, \hat{q}_{2}], [\hat{\ell}_{2}, \hat{r}_{2}]) \in \Psi^{\prime}_{\RR}$ satisfying $f^{\prime}_{\recover}(([p_{2}, q_{2}]$, $[\ell_{2}, r_{2}])) \cap \Psi_{\mRecover} = \{ ([\hat{p}_{2}, \hat{q}_{2}], [\hat{\ell}_{2}, \hat{r}_{2}]) \}$. 
We apply Lemma~\ref{lem:dynamic_f_recover} to the two interval attractors 
$([p_{1}, q_{1}], [\ell_{1}, r_{1}])$ and $([p_{2}, q_{2}], [\ell_{2}, r_{2}])$. 
Then, the lemma shows that 
$T[\hat{p}_{1}-1..\hat{\gamma}_{1}-1] = T^{\prime}[\hat{p}_{2}-1..\hat{\gamma}_{2}+1]$ 
and $T[\hat{\gamma}_{1}..\hat{r}_{1}+1] = T^{\prime}[\hat{\gamma}_{2}..\hat{r}_{1}+1]$ hold for 
the attractor positions $\hat{\gamma}_{1}$ and $\hat{\gamma}_{2}$ of two interval attractors 
$([\hat{p}_{1}, \hat{q}_{1}], [\hat{\ell}_{1}, \hat{r}_{1}])$ and $([\hat{p}_{2}, \hat{q}_{2}], [\hat{\ell}_{2}, \hat{r}_{2}])$, respectively. 
Therefore, Lemma~\ref{lem:dynamic_JB_move} holds. 
\end{proof}

\subsubsection{Algorithm}\label{subsubsec:JB_update_algorithm}
We prove Lemma~\ref{lem:dynamic_JB_summary}, i.e., 
we show that the data structures for subqueries $\RSCQBX$ and $\RSCQBY$ (Section~\ref{subsubsec:JB_ds}) 
can be updated in expected $O((\max \{H, H^{\prime}, \log (nHH^{\prime}) \})^{5})$ time using 
(A) the dynamic data structures for the RR-DAG of RLSLP $\mathcal{G}^{R}$ (Section~\ref{subsubsec:rrdag_ds}) 
and sample query (Section~\ref{subsec:sample_query}), 
and (B) the interval attractors obtained from Lemma~\ref{lem:dynamic_IA_summary}. 

Lemma~\ref{lem:dynamic_JB_change} and Lemma~\ref{lem:dynamic_JB_move} indicate that 
we can obtain $(H^{\prime}+1)$ sets $\mathcal{J}^{\prime}_{B}(0)$, $\mathcal{J}^{\prime}_{B}(1)$, 
$\ldots$, $\mathcal{J}^{\prime}_{B}(H^{\prime})$ of weighted points by 
modifying $(\max \{ H, H^{\prime} \}+1)$ sets $\mathcal{J}_{B}(0)$, $\mathcal{J}_{B}(1)$, $\ldots$, $\mathcal{J}_{B}(\max \{ H, H^{\prime} \} + 1)$ of weighted points as follows: 
\begin{description}
    \item [Modification 1:] 
    For each interval attractor $([p, q], [\ell, r]) \in \Psi_{\source} \cap (\Psi_{\OLD} \setminus \Psi_{\run})$ 
    with level $h$, 
    set $\mathcal{J}_{B}(h)$ contains a weighted point $(x, y, w, e)$ satisfying $e = T[p-1..r+1]$, 
    and $|f_{\recover}(([p, q], [\ell, r]))|$ is subtracted from the weight $w$ of the weighted point. 
    \item [Modification 2:] 
    For each interval attractor $([p, q], [\ell, r]) \in \Psi_{\source} \cap \Psi_{\samp} \cap (\Psi_{\OLD} \setminus \Psi^{\symA})$ with level $h$,     
    set $\mathcal{J}_{B}(h)$ contains a weighted point $(x, y, w, e)$ corresponding to 
    the interval attractor $([p, q], [\ell, r])$, 
    and this weighted point is removed from the set $\mathcal{J}_{B}(h)$.
    \item [Modification 3:] 
    For each interval attractor $([p, q], [\ell, r]) \in \Psi^{\prime}_{\source} \cap \Psi^{\prime \symB}$, 
    Lemma~\ref{lem:mRecover_basic_property} shows that 
    there exists an interval attractor $([\hat{p}, \hat{q}], [\hat{\ell}, \hat{r}]) \in \Psi^{\prime}_{\RR}$ satisfying $f^{\prime}_{\recover}(([p, q], [\ell, r])) \cap \Psi^{\prime}_{\mRecover} = \{ ([\hat{p}, \hat{q}], [\hat{\ell}, \hat{r}]) \}$. 
    We create a weighted point $(\reverse(T^{\prime}[\hat{p}-1..\hat{\gamma}-1]), T^{\prime}[\hat{\gamma}..\hat{r}+1], 0, T^{\prime}[p-1..r+1])$ corresponding to the interval attractor $([p, q], [\ell, r])$ 
    for the attractor position $\hat{\gamma}$ of interval attractor $([\hat{p}, \hat{q}], [\hat{\ell}, \hat{r}])$. 
    The weighted point is added to set $\mathcal{J}_{B}(h)$ for the level $h$ of the interval attractor $([p, q], [\ell, r])$.
    \item [Modification 4:]
    This process is executed after Modification 2 and Modification 3. 
    For each interval attractor $([p, q], [\ell, r])$ of set $\Psi^{\prime}_{\source} \cap (\Psi^{\prime}_{\NEW} \setminus \Psi^{\prime}_{\run})$ with level $h$, 
    set $\mathcal{J}_{B}(h)$ contains a weighted point $(x, y, w, e)$ satisfying $e = T^{\prime}[p-1..r+1]$, 
    and $|f^{\prime}_{\recover}(([p, q], [\ell, r]))|$ is added to the weight $w$ of the weighted point. 
\end{description}

Similar to the data structures for subquery $\RSCQA$, 
we update the data structures for subqueries $\RSCQBX$ and $\RSCQBY$ based on these modifications. 
The algorithm updating the data structures for subqueries $\RSCQBX$ and $\RSCQBY$ consists of the following six phases. 

\paragraph{Phase (i).}
In the first phase, 
we compute two sets $\Psi_{\source} \cap (\Psi_{\OLD} \setminus \Psi_{\run})$ 
and $\Psi_{\source} \cap \Psi_{\samp} \cap (\Psi_{\OLD} \setminus \Psi^{\symA})$ of interval attractors. 
Here, two sets $\Psi_{\OLD} \setminus \Psi_{\run}$ and $\Psi_{\samp} \cap (\Psi_{\OLD} \setminus \Psi^{\symA})$ are obtained from Lemma~\ref{lem:dynamic_IA_summary}. 

We obtain set $\Psi_{\source} \cap (\Psi_{\OLD} \setminus \Psi_{\run})$  by 
verifying whether 
each interval attractor of set $\Psi_{\OLD} \setminus \Psi_{\run}$ is contained in subset $\Psi_{\source}$. 
This verification can be executed by the verify-source query introduced in Section~\ref{subsec:verify_source_query}. 
Similarly, we obtain set $\Psi_{\source} \cap \Psi_{\samp} \cap (\Psi_{\OLD} \setminus \Psi^{\symA})$ by 
verifying whether 
each interval attractor of set $\Psi_{\samp} \cap (\Psi_{\OLD} \setminus \Psi^{\symA})$ is contained in 
set $\Psi_{\source}$. 
Therefore, we can obtain the two sets $\Psi_{\source} \cap (\Psi_{\OLD} \setminus \Psi_{\run})$ 
and $\Psi_{\source} \cap \Psi_{\samp} \cap (\Psi_{\OLD} \setminus \Psi^{\symA})$ by 
executing $O(|\Psi_{\OLD} \setminus \Psi_{\run}| + |\Psi_{\samp} \cap \Psi_{\OLD}|)$ verify-source queries. 

$\mathbb{E}[|\Psi_{\samp} \cap \Psi_{\OLD}|] = O(H + \log n)$ 
and $\mathbb{E}[|\Psi_{\OLD} \setminus \Psi_{\run}|] = O(H + \log n)$ follow from Lemma~\ref{lem:dynamic_psi_overlap}. 
Each verify-source query takes $O(H^{2} \log n)$ time. 
Therefore, the first phase takes expected $O((H + \log n) H^{2} \log n)$ time. 

\paragraph{Phase (ii).}
We update the data structures for subqueries $\RSCQBX$ and $\RSCQBY$ based on Modification 1. 
For each interval attractor $([p, q], [\ell, r]) \in \Psi_{\source} \cap \Psi_{\samp} \cap (\Psi_{\OLD} \setminus \Psi^{\symA})$ with level $h$, 
let $(x_{1}, y_{1}, w_{1}, e_{1})$, $(x_{2}, y_{2}, w_{2}, e_{2})$, $\ldots$, $(x_{m}, y_{m}, w_{m}, e_{m})$ 
($e_{1} \prec e_{2} \prec \cdots \prec e_{m}$) be the weighted points of set $\mathcal{J}_{B}(h)$. 
Here, each weighted point $(x_{s}, y_{s}, w_{s}, e_{s})$ corresponds to an interval attractor in set $\Psi_{h} \cap \Psi_{\source} \cap \Psi_{\samp}$, 
and the weighted point $(x_{s}, y_{s}, w_{s}, e_{s})$ is represented as the $s$-th element of the doubly linked list representing the set $\mathcal{J}_{B}(h)$. 
Set $\mathcal{J}_{B}(h)$ contains a weighted point $(x_{j}, y_{j}, w_{j}, e_{j})$ satisfying $e_{j} = T[p-1..r+1]$, and $|f_{\recover}(([p, q], [\ell, r]))|$ is subtracted from the weight $w_{j}$ by Modification 1. 

The first phase consists of three steps. 
In the first step, 
we compute the level $h$ of interval attractor $([p, q], [\ell, r])$ by the level query introduced in 
Section~\ref{subsec:level_query}. 
The first step takes $O(H^{2})$ time. 

In the second step, we compute the integer $j$ by binary search on 
the $m$ identifiers $e_{1}, e_{2}, \ldots, e_{m}$. 
Here, $\log m = O(\log n)$ follows from Lemma~\ref{lem:JB_size}~\ref{enum:JB_size:3}. 
Similar to the first phase of the algorithm presented in Section~\ref{subsubsec:JA_update_algorithm}, 
the integer $j$ can be computed in $O(H^{2} \log n + \log^{2} n)$ time. 

In the third step, 
$|f_{\recover}(([p, q], [\ell, r]))|$ is subtracted from the weight $w_{j}$ stored in the $j$-th element of the doubly linked list representing the set $\mathcal{J}_{B}(h)$. 
This subtraction takes $O(\log^{4} m)$ time because 
we need to update the range-sum data structure built on set $\mathcal{J}_{B}(h)$. 
The integer $|f_{\recover}(([p, q], [\ell, r]))|$ can be computed by the r-size query $\rsizeQ(([p, q], [\ell, r]))$ introduced in Section~\ref{subsec:rsize_query}. 
The r-size query takes $O(H^{2})$ time. 
Therefore, the third step takes $O(H^{2} + \log^{4} n)$ time. 

The second phase takes $O(|\Psi_{\OLD} \setminus \Psi_{\run}| (H^{2} \log n + \log^{4} n))$ time in total, 
i.e., this phase takes expected $O((H + \log n)(H^{2} \log n + \log^{4} n))$ time.

\paragraph{Phase (iii).}
In the third phase, 
we update the data structures for subqueries $\RSCQBX$ and $\RSCQBY$ based on Modification 2. 
For each interval attractor $([p, q], [\ell, r]) \in \Psi_{\source} \cap \Psi_{\samp} \cap (\Psi_{\OLD} \setminus \Psi^{\symA})$ with level $h$, 
set $\mathcal{J}_{B}(h)$ contains a weighted point $(x_{j}, y_{j}, w_{j}, e_{j})$ corresponding to the interval attractor $([p, q], [\ell, r])$, 
and this weighted point is removed from the set $\mathcal{J}_{B}(h)$. 
By this removal, 
the data structures for subqueries $\RSCQBX$ and $\RSCQBY$ are changed as follows: 
\begin{itemize}
    \item Consider the two doubly linked lists $\mathbf{X}_{B}(h, x_{j})$ and $\mathbf{L}^{X}_{B}(h)$ introduced in Section~\ref{subsubsec:JB_X_ds}. 
    The weighted point $(x_{j}, y_{j}, w_{j}, e_{j})$ corresponds to an element $u$ of doubly linked list $\mathbf{X}_{B}(h, x_{j})$, and the doubly linked list $\mathbf{X}_{B}(h, x_{j})$ corresponds to an element $u^{\prime}$ of $\mathbf{L}^{X}_{B}(h)$. 
    The element $u$ is removed from the doubly linked list $\mathbf{X}_{B}(h, x_{j})$. 
    If doubly linked list $\mathbf{X}_{B}(h, x_{j})$ is changed to an empty list by the removal of the element $u$, 
    then element $u^{\prime}$ is removed from doubly linked list $\mathbf{L}^{X}_{B}(h)$. 
    \item 
    Similarly, 
    consider the two doubly linked lists $\mathbf{Y}_{B}(h, y_{j})$ and $\mathbf{L}^{Y}_{B}(h)$ introduced in Section~\ref{subsubsec:JB_Y_ds}. 
    The weighted point $(x_{j}, y_{j}, w_{j}, e_{j})$ corresponds to an element $v$ of doubly linked list $\mathbf{Y}_{B}(h, y_{j})$, and the doubly linked list $\mathbf{Y}_{B}(h, y_{j})$ corresponds to an element $v^{\prime}$ of $\mathbf{L}^{Y}_{B}(h)$. 
    The element $v$ is removed from the doubly linked list $\mathbf{Y}_{B}(h, y_{j})$. 
    If doubly linked list $\mathbf{Y}_{B}(h, y_{j})$ is changed to an empty list by the removal of the element $v$, 
    then element $v^{\prime}$ is removed from doubly linked list $\mathbf{L}^{Y}_{B}(h)$. 
    \item The $j$-th element is removed from the doubly linked list representing the set $\mathcal{J}_{B}(h)$.    
\end{itemize}

We execute the third phase using the same approach as for 
the second phase of the algorithm presented in Section~\ref{subsubsec:JA_update_algorithm}. 
The third phase takes $O(|\Psi_{\samp} \cap \Psi_{\OLD}| (H^{2} \log n + \log^{4} n))$ time in total, 
i.e., this phase takes expected $O((H + \log n)(H^{2} \log n + \log^{4} n))$ time.

\paragraph{Phase (iv).}
In the fourth phase, 
we compute two sets $\Psi^{\prime}_{\source} \cap (\Psi^{\prime}_{\NEW} \setminus \Psi^{\prime}_{\run})$ 
and $\Psi^{\prime}_{\source} \cap \Psi^{\prime \symB}$ of interval attractors. 
Here, two sets $\Psi^{\prime}_{\NEW} \setminus \Psi^{\prime}_{\run}$ and $\Psi^{\prime \symB}$ are obtained from Lemma~\ref{lem:dynamic_IA_summary}. 
This phase is executed after updating the dynamic data structures for RR-DAG and sample query. 

Similar to the first phase, 
we obtain two sets 
$\Psi^{\prime}_{\source} \cap (\Psi^{\prime}_{\NEW} \setminus \Psi^{\prime}_{\run})$ and $\Psi^{\prime}_{\source} \cap \Psi^{\prime \symB}$ by verifying whether 
each interval attractor of two sets $\Psi^{\prime}_{\NEW} \setminus \Psi^{\prime}_{\run}$ and $\Psi^{\prime \symB}$ is contained in subset $\Psi^{\prime}_{\source}$. 
This verification can be executed by $|\Psi^{\prime}_{\NEW} \setminus \Psi^{\prime}_{\run}| + |\Psi^{\prime \symB}|$ verify-source queries on set $\Psi^{\prime}_{\RR}$. 
$\Psi^{\prime \symB} \subseteq \Psi^{\prime}_{\NEW} \setminus \Psi^{\prime}_{\run}$ holds, 
and $\mathbb{E}[|\Psi^{\prime}_{\NEW} \setminus \Psi^{\prime}_{\run}|] = O(H^{\prime} + \log n)$ follows from Lemma~\ref{lem:dynamic_psi_overlap}. 
Therefore, the fourth phase takes expected $O((H^{\prime} + \log n) H^{\prime 2} \log n)$ time.

\paragraph{Phase (v).}
In the fifth phase, 
we update the data structures for subqueries $\RSCQBX$ and $\RSCQBY$ based on Modification 3. 
Consider the sequence $\mathbf{Q}_{\samp}$ of nodes introduced in Section~\ref{subsec:sample_query}.
For each interval attractor $([p, q], [\ell, r]) \in \Psi^{\prime}_{\source} \cap \Psi^{\prime \symB}$ with level $h$, 
let $u$ be the node of the sequence $\mathbf{Q}_{\samp}$ representing 
the interval attractor $([p, q], [\ell, r])$. 
Lemma~\ref{lem:mRecover_basic_property} shows that 
there exists an interval attractor $([\hat{p}, \hat{q}], [\hat{\ell}, \hat{r}]) \in \Psi^{\prime}_{\RR}$ satisfying $f^{\prime}_{\recover}(([p, q], [\ell, r])) \cap \Psi^{\prime}_{\mRecover} = \{ ([\hat{p}, \hat{q}], [\hat{\ell}, \hat{r}]) \}$. 
By Modification 3, 
a weighted point $(\reverse(T^{\prime}[\hat{p}-1..\hat{\gamma}-1]), T^{\prime}[\hat{\gamma}..\hat{r}+1], 0, T^{\prime}[p-1..r+1])$  
is created, and it is added to set $\mathcal{J}_{B}(h)$ for the attractor position $\hat{\gamma}$ of the interval attractor $([p, q], [\ell, r])$.
By this addition, 
the data structures for subqueries $\RSCQBX$ and $\RSCQBY$ are changed, 
which are similar to the change of the data structures for subquery $\RSCQA$ at 
the third phase of the algorithm presented in Section~\ref{subsubsec:JA_update_algorithm}. 

We update the data structures for subqueries $\RSCQBX$ and $\RSCQBY$ using a similar approach as for 
the third phase of the algorithm presented in Section~\ref{subsubsec:JA_update_algorithm}. 
Then, the third phase can be executed in $O(|\Psi^{\prime \symB}| (H^{\prime 2} \log n + \log^{4} n))$ time, i.e., 
this phase takes expected $O((H^{\prime} + \log n)(H^{\prime 2} \log n + \log^{4} n))$ time. 

\paragraph{Phase (vi).}
In the sixth phase, 
we update the data structures for subqueries $\RSCQBX$ and $\RSCQBY$ based on Modification 4. 
For each interval attractor $([p, q], [\ell, r])$ of set $\Psi^{\prime}_{\source} \cap (\Psi^{\prime}_{\NEW} \setminus \Psi^{\prime}_{\run})$ with level $h$, 
set $\mathcal{J}_{B}(h)$ contains a weighted point $(x_{j}, y_{j}, w_{j}, e_{j})$ satisfying $e_{j} = T^{\prime}[p-1..r+1]$. 
By Modification 4, 
$|f^{\prime}_{\recover}(([p, q], [\ell, r]))|$ is added to the weight $w_{j}$ of the weighted point. 

The sixth phase consists of three steps. 
In the first step, 
we compute the level $h$ of interval attractor $([p, q], [\ell, r])$ by level query, 
which takes $O(H^{\prime 2})$ time. 

In the second step, we compute the integer $j$ by binary search on 
the $m$ identifiers $e_{1}, e_{2}, \ldots, e_{m}$. 
Similar to the first step of the second phase, 
the integer $j$ can be computed in $O(H^{\prime 2} \log n + \log^{2} n)$ time. 

In the third step, 
$|f^{\prime}_{\recover}(([p, q], [\ell, r]))|$ is added to the weight $w_{j}$ stored in the $j$-th element of the doubly linked list representing the set $\mathcal{J}_{B}(h)$. 
Similar to the third step of the second phase, 
this addition takes $O(\log^{4} m)$ time, 
and the integer $|f_{\recover}(([p, q], [\ell, r]))|$ can be computed in $O(H^{\prime 2})$ time.
Therefore, the third step takes $O(H^{\prime 2} + \log^{4} n)$ time. 

The sixth phase takes $O(|\Psi^{\prime}_{\NEW} \setminus \Psi^{\prime}_{\run}| (H^{\prime 2} \log n + \log^{4} n))$ time in total, i.e.,  
this phase takes expected $O((H^{\prime} + \log n)(H^{\prime 2} \log n + \log^{4} n))$ time. 

Finally, the six phases take expected $O((H + H^{\prime} + \log n)(H^{2} \log n + H^{\prime 2} \log n + \log^{4} n))$ time in total. 
Therefore, Lemma~\ref{lem:dynamic_JB_summary} holds.

\subsection{Update of Data Structures for RSC-C1 Subquery}\label{subsec:rc1_update}
This subsection explains how to update the dynamic data structures for subquery $\RSCQCX$ 
introduced in Section~\ref{subsubsec:JC1_ds}. 
The following lemma is the summary of this subsection.

\begin{lemma}\label{lem:dynamic_JC1_summary}
Consider the two RLSLPs $\mathcal{G}^{R}$ and $\mathcal{G}^{R}_{\ins}$ of Theorem~\ref{theo:update1}, which derive input string $T$ and string $T^{\prime}$, respectively. 
The dynamic data structures of Section~\ref{subsubsec:JC1_ds} can be updated 
in expected $O((\max \{H, H^{\prime}, \log (nHH^{\prime}) \})^{5})$ time 
after changing RLSLP $\mathcal{G}^{R}$ to $\mathcal{G}^{R}_{\ins}$. 
This update requires 
(i) the dynamic data structures for the RR-DAG of RLSLP $\mathcal{G}^{R}$ (Section~\ref{subsubsec:rrdag_ds}) 
and sample query (Section~\ref{subsec:sample_query}), 
and (ii) the interval attractors obtained from Lemma~\ref{lem:dynamic_IA_summary}. 
\end{lemma}
\begin{proof}
    See Section~\ref{subsubsec:JC1_update_algorithm}. 
\end{proof}

Let $(h_{1}, C_{1}, M_{1})$, $(h_{2}, C_{2}, M_{2})$, $\ldots$, $(h_{z}, C_{z}, M_{z})$ be the triplets in the ordered set $\mathcal{T}_{C}$ introduced in Section~\ref{subsubsec:TC1_ds}. 
Then, 
the dynamic data structures for subquery $\RSCQCX$ store weighted points of 
the $m$ sets $\mathcal{J}_{C}(h_{1}, C_{1}, M_{1})$, $\mathcal{J}_{C}(h_{2}, C_{2}, M_{2})$, 
$\ldots$, $\mathcal{J}_{C}(h_{z}, C_{z}, M_{z})$ introduced in Section~\ref{subsec:RSC_comp_C1}. 

For a triplet of an integer $h \in [0, H^{\prime}]$, a string $C \in \Sigma^{+}$, and an integer $M \geq 0$, 
let $\mathcal{J}^{\prime}_{C}(h, C, M)$ be the set $\mathcal{J}_{C}(h, C, M)$ of weighted points defined using set $\Psi^{\prime}_{\RR}$ instead of set $\Psi_{\RR}$. 
For simplicity, 
let $\mathcal{J}_{C}(h, C, M)$ be an empty set of weighted points if $h > H$. 
Let $\mathcal{T}^{\prime}_{C} = \{ (h^{\prime}_{1}, C^{\prime}_{1}, M^{\prime}_{1})$, $(h^{\prime}_{2}, C^{\prime}_{2}, M^{\prime}_{2})$, $\ldots$, $(h^{\prime}_{z^{\prime}}, C_{z^{\prime}}, M_{z^{\prime}}) \}$ 
be the ordered set $\mathcal{T}_{C}$ defined using set $\Psi^{\prime}_{\RR}$ instead of set $\Psi_{\RR}$. 

The following lemma shows that each set $\mathcal{J}^{\prime}_{C}(h, C, M)$ can be obtained by modifying set $\mathcal{J}_{C}(h, C, M)$. 

\begin{lemma}\label{lem:dynamic_JC1_change}
Consider a triplet of an integer $h \in [0, H^{\prime}]$, a string $C \in \Sigma^{+}$, and an integer $M \geq 0$. 
For each weighted point $(x_{s}, y_{s}, w_{s}, e_{s}) \in \mathcal{J}_{C}(h, C, M)$, 
let $([p_{s}, q_{s}], [\ell_{s}, r_{s}]) \in \Psi_{h} \cap \Psi_{\source} \cap \Psi_{\centerset}(C) \cap \Psi_{\modulo}(M) \cap \Psi_{\preceding} \cap \Psi_{\samp}$ be 
the interval attractor corresponding to the weighted point $(x_{s}, y_{s}, w_{s}, e_{s})$, 
$\kappa_{s} = |\Psi_{h} \cap \Psi_{\source} \cap \Psi_{\centerset}(C) \cap \Psi_{\modulo}(M) \cap \Psi_{\preceding} \cap (\Psi_{\OLD} \setminus \Psi_{\run}) \cap \Psi_{\str}(e_{s})|$, 
and 
$\kappa^{\prime}_{s} = |\Psi^{\prime}_{h} \cap \Psi^{\prime}_{\source} \cap \Psi^{\prime}_{\centerset}(C) \cap \Psi^{\prime}_{\modulo}(M) \cap \Psi^{\prime}_{\preceding} \cap (\Psi^{\prime}_{\NEW} \setminus \Psi^{\prime}_{\run}) \cap \Psi^{\prime}_{\str}(e_{s})|$. 
For each interval attractor $([p^{\prime}_{s}, q^{\prime}_{s}], [\ell^{\prime}_{s}, r^{\prime}_{s}]) \in \Psi^{\prime}_{h} \cap \Psi^{\prime}_{\source} \cap \Psi^{\prime}_{\centerset}(C) \cap \Psi^{\prime}_{\modulo}(M) \cap \Psi^{\prime}_{\preceding} \cap \Psi^{\prime \symB}$ with interval attractor $\gamma^{\prime}_{s}$, 
let $\nu^{\prime}_{s} = |f^{\prime}_{\recover}(([p^{\prime}_{s}, q^{\prime}_{s}], [\ell^{\prime}_{s}, r^{\prime}_{s}]))|$ 
and $K^{\prime}_{s}$ be the length of the longest common prefix between two strings $T^{\prime}[\gamma^{\prime}_{s}..r^{\prime}_{s}]$ and $C^{n+1}$. 
Then, set $\mathcal{J}^{\prime}_{C}(h, C, M)$ of weighted points 
is equal to the union of the following two sets of weighted points: 
\begin{enumerate}[label=\textbf{(\roman*)}]
    \item $\{ (x_{s}, y_{s}, w_{s} - \kappa_{s} + \kappa^{\prime}_{s}, e_{s}) \mid (x_{s}, y_{s}, w_{s}, e_{s}) \in \mathcal{J}_{C}(h, C, M) \text{ s.t. } ([p_{s}, q_{s}], [\ell_{s}, r_{s}]) \not \in \Psi_{\OLD} \setminus \Psi^{\symA} \}$;
    \item $\{ (\nu^{\prime}_{s}, T^{\prime}[\gamma^{\prime}_{s} + K^{\prime}_{s}..r^{\prime}_{s} + 1]$, $|\Psi^{\prime}_{\str}(T^{\prime}[p^{\prime}_{s}-1..r^{\prime}_{s}+1])|, T^{\prime}[p^{\prime}_{s}-1..r^{\prime}_{s}+1]) \mid ([p^{\prime}_{s}, q^{\prime}_{s}], [\ell^{\prime}_{s}$, $r^{\prime}_{s}]) \in \Psi^{\prime}_{h} \cap \Psi^{\prime}_{\source} \cap \Psi^{\prime}_{\centerset}(C) \cap \Psi^{\prime}_{\modulo}(M) \cap \Psi^{\prime}_{\preceding} \cap \Psi^{\prime \symB} \}$. 
\end{enumerate}
\end{lemma}
\begin{proof}
Lemma~\ref{lem:dynamic_JC1_change} follows from the following four statements: 
\begin{enumerate}[label=\textbf{(\Alph*)}]
    \item consider a weighted point $(x_{s}, y_{s}, w_{s}, e_{s}) \in \mathcal{J}_{C}(h, C, M)$ satisfying $([p_{s}, q_{s}], [\ell_{s}, r_{s}]) \not \in \Psi_{\OLD} \setminus \Psi^{\symA}$. 
    Then, $(x_{s}, y_{s}, w_{s} - \kappa_{s} + \kappa^{\prime}_{s}, e_{s}) \in \mathcal{J}^{\prime}_{C}(h, C, M)$;
    \item consider an interval attractor $([p^{\prime}_{s}, q^{\prime}_{s}], [\ell^{\prime}_{s}, r^{\prime}_{s}]) \in \Psi^{\prime}_{h} \cap \Psi^{\prime}_{\source} \cap \Psi^{\prime}_{\centerset}(C) \cap \Psi^{\prime}_{\modulo}(M) \cap \Psi^{\prime}_{\preceding} \cap \Psi^{\prime \symB}$. 
    Then, $(\nu^{\prime}_{s}, T^{\prime}[\gamma^{\prime}_{s} + K^{\prime}_{s}..r^{\prime}_{s} + 1]$, $|\Psi^{\prime}_{\str}(T^{\prime}[p^{\prime}_{s}-1..r^{\prime}_{s}+1])|, T^{\prime}[p^{\prime}_{s}-1..r^{\prime}_{s}+1]) \in \mathcal{J}^{\prime}_{C}(h, C, M)$;
    \item consider a weighted point $(x^{\prime}, y^{\prime}, w^{\prime}, e^{\prime}) \in \mathcal{J}^{\prime}_{C}(h, C, M)$ satisfying $e^{\prime} \not \in \mathcal{Z}^{\symB}$ for 
    the set $\mathcal{Z}^{\symB}$ of strings introduced in Section~\ref{subsec:update_sampling_subset}. 
    Then, there exists a weighted point $(x_{s}, y_{s}, w_{s}$, $e_{s}) \in \mathcal{J}_{C}(h, C, M)$ satisfying 
    $([p_{s}, q_{s}], [\ell_{s}, r_{s}]) \not \in \Psi_{\OLD} \setminus \Psi^{\symA}$ 
    and $(x_{s}, y_{s}, w_{s} - \kappa_{s} + \kappa^{\prime}_{s}, e_{s}) = (x^{\prime}, y^{\prime}, w^{\prime}, e^{\prime})$;
    \item consider a weighted point $(x^{\prime}, y^{\prime}, w^{\prime}, e^{\prime}) \in \mathcal{J}^{\prime}_{C}(h, C, M)$ satisfying $e^{\prime} \in \mathcal{Z}^{\symB}$. 
    Then, there exists an interval attractor $([p^{\prime}_{s}, q^{\prime}_{s}], [\ell^{\prime}_{s}, r^{\prime}_{s}]) \in \Psi^{\prime}_{h} \cap \Psi^{\prime}_{\source} \cap \Psi^{\prime}_{\centerset}(C) \cap \Psi^{\prime}_{\modulo}(M) \cap \Psi^{\prime}_{\preceding} \cap \Psi^{\prime \symB}$ 
    satisfying $(\nu^{\prime}_{s}, T^{\prime}[\gamma^{\prime}_{s} + K^{\prime}_{s}..r^{\prime}_{s} + 1]$, $|\Psi^{\prime}_{\str}(T^{\prime}[p^{\prime}_{s}-1..r^{\prime}_{s}+1])|, T^{\prime}[p^{\prime}_{s}-1..r^{\prime}_{s}+1]) = (x^{\prime}, y^{\prime}, w^{\prime}, e^{\prime})$. 
\end{enumerate}

\textbf{Proof of statement (A).}
Lemma~\ref{lem:dynamic_samp_proceeding_formula} shows that 
sampling subset $\Psi^{\prime}_{\samp}$ contains an interval attractor $([p^{\prime}, q^{\prime}], [\ell^{\prime}, r^{\prime}])$ satisfying $T[p_{s}-1..r_{s}+1] = T^{\prime}[p^{\prime}-1..r^{\prime}+1]$. 
We apply Lemma~\ref{lem:dynamic_IA_super_correspondence} to the two interval attractors 
$([p_{s}, q_{s}], [\ell_{s}, r_{s}])$ and $([p^{\prime}, q^{\prime}], [\ell^{\prime}, r^{\prime}])$. 
Then, the lemma shows that $([p^{\prime}, q^{\prime}], [\ell^{\prime}, r^{\prime}]) \in \Psi^{\prime}_{h} \cap \Psi^{\prime}_{\centerset}(C)$ holds.  
Similarly, we apply Lemma~\ref{lem:dynamic_RR_subset} to the two interval attractors $([p_{s}, q_{s}], [\ell_{s}, r_{s}])$ and $([p^{\prime}, q^{\prime}], [\ell^{\prime}, r^{\prime}])$. 
Then, the lemma shows that $([p^{\prime}, q^{\prime}], [\ell^{\prime}, r^{\prime}]) \in \Psi^{\prime}_{\source} \cap \Psi^{\prime}_{\modulo}(M) \cap \Psi^{\prime}_{\preceding}$ holds. 

Let $\nu^{\prime} = |f^{\prime}_{\recover}(([p^{\prime}, q^{\prime}], [\ell^{\prime}, r^{\prime}]))|$ 
and $K^{\prime}$ be the length of the longest common prefix between two strings $T^{\prime}[\gamma^{\prime}..r^{\prime}]$ and $C^{n+1}$ 
for the attractor position $\gamma^{\prime}$ of the interval attractor $([p^{\prime}, q^{\prime}], [\ell^{\prime}, r^{\prime}])$. 
Because of $([p^{\prime}, q^{\prime}], [\ell^{\prime}, r^{\prime}]) \in \Psi^{\prime}_{h} \cap \Psi^{\prime}_{\source} \cap \Psi^{\prime}_{\centerset}(C) \cap \Psi^{\prime}_{\modulo}(M) \cap \Psi^{\prime}_{\preceding} \cap \Psi^{\prime}_{\samp}$, 
set $\mathcal{J}^{\prime}_{C}(h, C, M)$ contains weighted point $(\nu^{\prime}, T^{\prime}[\gamma^{\prime} + K^{\prime}..r^{\prime} + 1]$, $|\Psi^{\prime}_{\str}(T^{\prime}[p^{\prime}-1..r^{\prime}+1])|, T^{\prime}[p^{\prime}-1..r^{\prime}+1])$, which corresponds to 
the interval attractor $([p^{\prime}, q^{\prime}], [\ell^{\prime}, r^{\prime}])$. 

$x_{s} = \nu^{\prime}$ follows from 
$x_{s} = |f_{\recover}(([p_{s}, q_{s}], [\ell_{s}, r_{s}]))|$ holds, 
and $|f_{\recover}(([p_{s}, q_{s}], [\ell_{s}, r_{s}]))| = \nu^{\prime}$ follows from 
Lemma~\ref{lem:dynamic_f_recover}~\ref{enum:dynamic_f_recover:X}. 
Let $K_{s}$ be the length of the longest common prefix between two strings $T[\gamma_{s}..r_{s}]$ and $C^{n+1}$ 
for the attractor position $\gamma_{s}$ of the interval attractor $([p_{s}, q_{s}], [\ell_{s}, r_{s}])$.
Then, 
$y_{s} = T^{\prime}[\gamma^{\prime} + K^{\prime}..r^{\prime} + 1]$ holds 
because (a) $y_{s} = T[\gamma_{s} + K_{s}..r_{s} + 1]$, 
(b) $T[\gamma_{s}..r_{s} + 1] = T^{\prime}[\gamma^{\prime}..r^{\prime} + 1]$ follows from Lemma~\ref{lem:dynamic_IA_super_correspondence}, 
and (c) $K = K^{\prime}$ follows from Lemma~\ref{lem:dynamic_RR_subset}~\ref{enum:dynamic_RR_subset:lcp}. 
$e_{s} = T[p_{s}-1..r_{s}+1] = T^{\prime}[p^{\prime}-1..r^{\prime}+1]$ 
follows from $e_{s} = T[p_{s}-1..r_{s}+1]$ and $T[p_{s}-1..r_{s}+1] = T^{\prime}[p^{\prime}-1..r^{\prime}+1]$. 

Let $Z = T^{\prime}[p^{\prime}-1..r^{\prime}+1]$ for simplicity. 
We prove $\Psi_{h} \cap (\Psi_{\OLD} \setminus \Psi_{\run}) \cap \Psi_{\str}(Z) = \Psi_{h} \cap \Psi_{\source} \cap \Psi_{\centerset}(C) \cap \Psi_{\modulo}(M) \cap \Psi_{\preceding} \cap (\Psi_{\OLD} \setminus \Psi_{\run}) \cap \Psi_{\str}(Z)$. 
Consider an interval attractor $([p, q], [\ell, r])$ in set $\Psi_{h} \cap (\Psi_{\OLD} \setminus \Psi_{\run}) \cap \Psi_{\str}(Z)$. 
Then, $T[p-1..r+1] = Z$ follows from the definition of subset $\Psi_{\str}(Z)$. 
Because of $T[p-1..r+1] = T[p_{s}-1..r_{s}+1]$, 
we can apply Lemma~\ref{lem:psi_equality_basic_property} to the two interval attractors 
$([p_{s}, q_{s}], [\ell_{s}, r_{s}])$ and $([p, q], [\ell, r])$. 
Because of $([p_{s}, q_{s}], [\ell_{s}, r_{s}]) \in \Psi_{\source} \cap \Psi_{\centerset}(C) \cap \Psi_{\modulo}(M) \cap \Psi_{\preceding}$, 
Lemma~\ref{lem:psi_equality_basic_property} shows that $([p, q], [\ell, r]) \in \Psi_{\source} \cap \Psi_{\centerset}(C) \cap \Psi_{\modulo}(M) \cap \Psi_{\preceding}$ holds. 
Therefore, $\Psi_{h} \cap (\Psi_{\OLD} \setminus \Psi_{\run}) \cap \Psi_{\str}(Z) = \Psi_{h} \cap \Psi_{\source} \cap \Psi_{\centerset}(C) \cap \Psi_{\modulo}(M) \cap \Psi_{\preceding} \cap (\Psi_{\OLD} \setminus \Psi_{\run}) \cap \Psi_{\str}(Z)$ holds. 
Similarly, $\Psi^{\prime}_{h} \cap (\Psi^{\prime}_{\NEW} \setminus \Psi^{\prime}_{\run}) \cap \Psi^{\prime}_{\str}(Z) = \Psi^{\prime}_{h} \cap \Psi^{\prime}_{\source} \cap \Psi^{\prime}_{\centerset}(C) \cap \Psi^{\prime}_{\modulo}(M) \cap \Psi^{\prime}_{\preceding} \cap (\Psi^{\prime}_{\NEW} \setminus \Psi^{\prime}_{\run}) \cap \Psi^{\prime}_{\str}(Z)$ holds.

We prove $w_{s} - \kappa_{s} + \kappa^{\prime}_{s} = |\Psi^{\prime}_{\str}(Z)|$. 
Because of $([p^{\prime}, q^{\prime}], [\ell^{\prime}, r^{\prime}]) \in \Psi^{\prime}_{\samp}$, 
$([p^{\prime}, q^{\prime}], [\ell^{\prime}, r^{\prime}]) \not \in \Psi^{\prime}_{\run}$ follows from 
Lemma~\ref{lem:samp_basic_property}~\ref{enum:samp_basic_property:3}. 
$|\Psi^{\prime}_{\str}(Z)| = |\Psi_{\str}(Z)| - |\Psi_{h} \cap (\Psi_{\OLD} \setminus \Psi_{\run}) \cap \Psi_{\str}(Z)| + |\Psi^{\prime}_{h} \cap (\Psi^{\prime}_{\NEW} \setminus \Psi^{\prime}_{\run}) \cap \Psi^{\prime}_{\str}(Z)|$ follows from 
Corollary~\ref{cor:dynamic_str_formula}~\ref{enum:dynamic_str_formula:1}. 
Therefore, the following equation holds: 
\begin{equation*}
    \begin{split}
        |\Psi^{\prime}_{\str}(Z)| &=  |\Psi_{\str}(Z)| - |\Psi_{h} \cap (\Psi_{\OLD} \setminus \Psi_{\run}) \cap \Psi_{\str}(Z)| + |\Psi^{\prime}_{h} \cap (\Psi^{\prime}_{\NEW} \setminus \Psi^{\prime}_{\run}) \cap \Psi^{\prime}_{\str}(Z)| \\
        &= |\Psi_{\str}(Z)| - |\Psi_{h} \cap \Psi_{\source} \cap \Psi_{\centerset}(C) \cap \Psi_{\modulo}(M) \cap \Psi_{\preceding} \cap (\Psi_{\OLD} \setminus \Psi_{\run}) \cap \Psi_{\str}(Z)| \\
        &+ |\Psi^{\prime}_{h} \cap \Psi^{\prime}_{\source} \cap \Psi^{\prime}_{\centerset}(C) \cap \Psi^{\prime}_{\modulo}(M) \cap \Psi^{\prime}_{\preceding} \cap (\Psi^{\prime}_{\NEW} \setminus \Psi^{\prime}_{\run}) \cap \Psi^{\prime}_{\str}(Z)| \\
        &= w_{s} - \kappa_{s} + \kappa^{\prime}_{s}.
    \end{split}
\end{equation*}

We proved $(x_{s}, y_{s}, w_{s} - \kappa_{s} + \kappa^{\prime}_{s}, e_{s}) = (\nu^{\prime}, T^{\prime}[\gamma^{\prime} + K^{\prime}..r^{\prime} + 1]$, $|\Psi^{\prime}_{\str}(T^{\prime}[p^{\prime}-1..r^{\prime}+1])|, T^{\prime}[p^{\prime}-1..r^{\prime}+1])$. 
Therefore, $(x_{s}, y_{s}, w_{s} - \kappa_{s} + \kappa^{\prime}_{s}, e_{s}) \in \mathcal{J}^{\prime}_{C}(h, C, M)$ holds. 

\textbf{Proof of statement (B).}
Set $\Psi^{\prime}_{h} \cap \Psi^{\prime}_{\source} \cap \Psi^{\prime}_{\centerset}(C) \cap \Psi^{\prime}_{\modulo}(M) \cap \Psi^{\prime}_{\preceding} \cap \Psi^{\prime}_{\samp}$ contains 
interval attractor $([p^{\prime}_{s}, q^{\prime}_{s}], [\ell^{\prime}_{s}, r^{\prime}_{s}])$ 
because Theorem~\ref{theo:dynamic_samp_formula} shows that 
$\Psi^{\prime \symB} \subseteq \Psi^{\prime}_{\samp}$ holds. 
Because of $([p^{\prime}_{s}, q^{\prime}_{s}], [\ell^{\prime}_{s}, r^{\prime}_{s}]) \in \Psi^{\prime}_{h} \cap \Psi^{\prime}_{\source} \cap \Psi^{\prime}_{\centerset}(C) \cap \Psi^{\prime}_{\modulo}(M) \cap \Psi^{\prime}_{\preceding} \cap \Psi^{\prime}_{\samp}$, 
$(\nu^{\prime}_{s}, T^{\prime}[\gamma^{\prime}_{s} + K^{\prime}_{s}..r^{\prime}_{s} + 1]$, $|\Psi^{\prime}_{\str}(T^{\prime}[p^{\prime}_{s}-1..r^{\prime}_{s}+1])|, T^{\prime}[p^{\prime}_{s}-1..r^{\prime}_{s}+1]) \in \mathcal{J}^{\prime}_{C}(h, C, M)$ follows from the definition of the set $\mathcal{J}^{\prime}_{C}(h, C, M)$. 

\textbf{Proof of statement (C).}
The weighted point $(x^{\prime}, y^{\prime}, w^{\prime}, e^{\prime})$ corresponds to 
an interval attractor $([p^{\prime}, q^{\prime}], [\ell^{\prime}, r^{\prime}]) \in \Psi^{\prime}_{h} \cap \Psi^{\prime}_{\source} \cap \Psi^{\prime}_{\centerset}(C) \cap \Psi^{\prime}_{\modulo}(M) \cap \Psi^{\prime}_{\preceding} \cap \Psi^{\prime}_{\samp}$ satisfying $T^{\prime}[p^{\prime}-1..r^{\prime}+1] = e^{\prime}$. 
$([p^{\prime}, q^{\prime}], [\ell^{\prime}, r^{\prime}]) \not \in \Psi^{\prime \symB}$ follows from 
$T^{\prime}[p^{\prime}-1..r^{\prime}+1] \not \in \mathcal{Z}^{\symB}$. 
Because of $([p^{\prime}, q^{\prime}], [\ell^{\prime}, r^{\prime}]) \not \in \Psi^{\prime \symB}$, 
Lemma~\ref{lem:dynamic_samp_back_formula} shows that 
sampling subset $\Psi_{\samp}$ contains an interval attractor 
$([p, q], [\ell, r])$ satisfying 
$T[p-1..r+1] = T^{\prime}[p^{\prime}-1..r^{\prime}+1]$ and 
$([p, q], [\ell, r]) \not \in \Psi_{\OLD} \setminus \Psi^{\symA}$. 

We apply Lemma~\ref{lem:dynamic_IA_super_correspondence} to the two interval attractors 
$([p, q], [\ell, r])$ and $([p^{\prime}, q^{\prime}], [\ell^{\prime}, r^{\prime}])$. 
Then, the lemma shows that $([p, q], [\ell, r]) \in \Psi_{h}$ holds. 
Similarly, we apply Lemma~\ref{lem:dynamic_RR_subset} to the two interval attractors $([p, q], [\ell, r])$ and $([p^{\prime}, q^{\prime}], [\ell^{\prime}, r^{\prime}])$. 
Then, the lemma shows that $([p, q], [\ell, r]) \in \Psi_{\source} \cap \Psi_{\centerset}(C) \cap \Psi_{\modulo}(M) \cap \Psi_{\preceding}$ holds. 
Because of $([p, q], [\ell, r]) \in \Psi_{h} \cap \Psi_{\source} \cap \Psi_{\centerset}(C) \cap \Psi_{\modulo}(M) \cap \Psi_{\preceding} \cap \Psi_{\samp}$, 
set $\mathcal{J}_{C}(h, C, M)$ contains a weighted point $(x_{s}, y_{s}, w_{s}, e_{s})$ corresponding to  
the interval attractor $([p, q], [\ell, r])$ (i.e., $([p_{s}, q_{s}], [\ell_{s}, r_{s}]) = ([p, q], [\ell, r])$). 
Similar to the proof of statement (A), 
$(x_{s}, y_{s}, w_{s} - \kappa_{s} + \kappa^{\prime}_{s}, e_{s}) = (x^{\prime}, y^{\prime}, w^{\prime}, e^{\prime})$ can be proved. 
Therefore, statement (C) holds. 

\textbf{Proof of statement (D).}
The weighted point $(x^{\prime}, y^{\prime}, w^{\prime}, e^{\prime})$ corresponds to 
an interval attractor $([p^{\prime}, q^{\prime}], [\ell^{\prime}, r^{\prime}]) \in \Psi^{\prime}_{h} \cap \Psi^{\prime}_{\source} \cap \Psi^{\prime}_{\centerset}(C) \cap \Psi^{\prime}_{\modulo}(M) \cap \Psi^{\prime}_{\preceding} \cap \Psi^{\prime}_{\samp}$ satisfying $T^{\prime}[p^{\prime}-1..r^{\prime}+1] = e^{\prime}$. 
$([p^{\prime}, q^{\prime}], [\ell^{\prime}, r^{\prime}]) \in \Psi^{\prime \symB}$ can be proved using the same approach as for statement \ref{enu:dynamic_JA_change:D} in the proof of Lemma~\ref{lem:dynamic_JA_change}. 
Because of $([p^{\prime}, q^{\prime}], [\ell^{\prime}, r^{\prime}]) \in \Psi^{\prime}_{h} \cap \Psi^{\prime}_{\source} \cap \Psi^{\prime}_{\centerset}(C) \cap \Psi^{\prime}_{\modulo}(M) \cap \Psi^{\prime}_{\preceding} \cap \Psi^{\prime \symB}$, 
set $\Psi^{\prime}_{h} \cap \Psi^{\prime}_{\source} \cap \Psi^{\prime}_{\centerset}(C) \cap \Psi^{\prime}_{\modulo}(M) \cap \Psi^{\prime}_{\preceding} \cap \Psi^{\prime \symB}$ contains the interval attractor $([p^{\prime}, q^{\prime}], [\ell^{\prime}, r^{\prime}])$ as an interval attractor 
$([p^{\prime}_{s}, q^{\prime}_{s}], [\ell^{\prime}_{s}, r^{\prime}_{s}])$. 
$(\nu^{\prime}_{s}, T^{\prime}[\gamma^{\prime}_{s} + K^{\prime}_{s}..r^{\prime}_{s} + 1]$, $|\Psi^{\prime}_{\str}(T^{\prime}[p^{\prime}_{s}-1..r^{\prime}_{s}+1])|, T^{\prime}[p^{\prime}_{s}-1..r^{\prime}_{s}+1]) = (x^{\prime}, y^{\prime}, w^{\prime}, e^{\prime})$ follows from the definition of set $\mathcal{J}^{\prime}_{C}(h, C, M)$. 
Therefore, statement (D) holds. 
\end{proof}

Consider the $d$ sequences $\mathbf{Q}^{Y}_{C}(h, C, M, R_{1})$, $\mathbf{Q}^{Y}_{C}(h, C, M, R_{2})$, 
$\ldots$, $\mathbf{Q}^{Y}_{C}(h, C, M, R_{d})$ of weighted points 
introduced in Section~\ref{subsubsec:JC1_Y_ds} 
for a triplet $(h, C, M) \in \mathcal{T}_{C}$ 
and the ordered set $\mathcal{Y}_{C}(h, C, M) = \{ R_{1}, R_{2}, \ldots, R_{d} \}$ of strings introduced in Section~\ref{subsec:RSC_comp_C1}. 
Here, each sequence $\mathbf{Q}^{Y}_{C}(h, C, M, R_{s})$ is represented as the doubly linked list $\mathbf{Y}_{C}(h, C, M, y)$ introduced in Section~\ref{subsubsec:JC1_Y_ds}. 
The following lemma states properties of these sequences of weighted points. 

\begin{lemma}\label{lem:dynamic_JC1_move}
For a triplet $(h, C, M) \in \mathcal{T}_{C}$, 
consider an interval attractor $([p, q], [\ell, r]) \in \Psi_{h} \cap \Psi_{\source} \cap \Psi_{\centerset}(C) \cap \Psi_{\modulo}(M) \cap \Psi_{\preceding} \cap \Psi_{\samp}$ satisfying $([p, q], [\ell, r]) \not \in \Psi_{\OLD} \setminus \Psi^{\symA}$. 
Here, set $\mathcal{J}_{C}(h, C, M)$ contains a weighted point $(x, y, w, e)$ corresponding to 
the interval attractor $([p, q], [\ell, r])$; 
the interval attractor $([p, q], [\ell, r])$ is represented as a node $u$ of the sequence $\mathbf{Q}_{\samp}$ introduced in Section~\ref{subsec:sample_query}. 
From the definition of sequence $\mathbf{Q}^{Y}_{C}(h, C, M, y)$, 
there exists an integer $\alpha$ such that 
the sequence $\mathbf{Q}^{Y}_{C}(h, C, M, y)$ contains weighted point $(x, y, w, e)$ as the $\alpha$-th element, 
and the $\alpha$-th element of doubly linked list $\mathbf{Y}_{C}(h, C, M, y)$ stores a pointer to node $u$. 

After updating the dynamic data structures for sample query, 
node $u$ represents an interval attractor $([p^{\prime}, q^{\prime}], [\ell^{\prime}, r^{\prime}])$ in set $\Psi^{\prime}_{h} \cap \Psi^{\prime}_{\source} \cap \Psi^{\prime}_{\centerset}(C) \cap \Psi^{\prime}_{\modulo}(M) \cap \Psi^{\prime}_{\preceding} \cap \Psi^{\prime}_{\samp}$ satisfying the following four conditions: 
\begin{enumerate}[label=\textbf{(\roman*)}]
    \item $T[p-1..r+1] = T^{\prime}[p^{\prime}-1..r^{\prime}+1]$;
    \item $([p^{\prime}, q^{\prime}], [\ell^{\prime}, r^{\prime}]) \not \in \Psi^{\prime \symB}$; 
    \item $|f_{\recover}(([p, q], [\ell, r]))| = |f^{\prime}_{\recover}(([p^{\prime}, q^{\prime}], [\ell^{\prime}, r^{\prime}]))|$;     
    \item 
    let $K = |\lcp(T[\gamma..r], C^{n+1})|$ and $K^{\prime} = |\lcp(T^{\prime}[\gamma^{\prime}..r^{\prime}], C^{n+1})|$ for the attractor positions $\gamma$ and $\gamma^{\prime}$ 
    of the two interval attractors 
    $([p, q], [\ell, r])$ and $([p^{\prime}, q^{\prime}], [\ell^{\prime}, r^{\prime}])$, respectively. 
    Then, $T[\gamma + K..r + 1] = T^{\prime}[\gamma^{\prime} + K^{\prime}..r^{\prime} + 1]$.     
\end{enumerate}
\end{lemma}
\begin{proof}
Lemma~\ref{lem:dynamic_samp_proceeding_formula} shows that 
node $u$ represents an interval attractor $([p^{\prime}, q^{\prime}], [\ell^{\prime}, r^{\prime}])$ in set 
$\Psi^{\prime}_{\samp}$ 
satisfying $T[p-1..r+1] = T^{\prime}[p^{\prime}-1..r^{\prime}+1]$ 
and $([p^{\prime}, q^{\prime}], [\ell^{\prime}, r^{\prime}]) \not \in \Psi^{\prime \symB}$. 
We apply Lemma~\ref{lem:dynamic_IA_super_correspondence} to the two interval attractors 
$([p, q], [\ell, r])$ and $([p^{\prime}, q^{\prime}], [\ell^{\prime}, r^{\prime}])$. 
Then, the lemma shows that $([p^{\prime}, q^{\prime}], [\ell^{\prime}, r^{\prime}]) \in \Psi^{\prime}_{h}$ 
and $T[\gamma..r+1] = T^{\prime}[\gamma^{\prime}..r^{\prime}+1]$ hold. 
Similarly, we apply Lemma~\ref{lem:dynamic_RR_subset} 
to the two interval attractors $([p, q], [\ell, r])$ and $([p^{\prime}, q^{\prime}], [\ell^{\prime}, r^{\prime}])$. 
Then, the lemma shows that $([p^{\prime}, q^{\prime}], [\ell^{\prime}, r^{\prime}]) \in \Psi^{\prime}_{\source} \cap \Psi^{\prime}_{\centerset}(C) \cap \Psi^{\prime}_{\modulo}(M) \cap \Psi^{\prime}_{\preceding}$ 
and $K = K^{\prime}$ hold. 
$T[\gamma + K..r + 1] = T^{\prime}[\gamma^{\prime} + K^{\prime}..r^{\prime} + 1]$ follows from 
$T[\gamma..r+1] = T^{\prime}[\gamma^{\prime}..r^{\prime}+1]$ and $K = K^{\prime}$. 
$|f_{\recover}(([p, q], [\ell, r]))| = |f^{\prime}_{\recover}(([p^{\prime}, q^{\prime}], [\ell^{\prime}, r^{\prime}]))|$ follows from 
Lemma~\ref{lem:dynamic_f_recover}~\ref{enum:dynamic_f_recover:X}. 
Therefore, Lemma~\ref{lem:dynamic_JC1_move} holds. 
\end{proof}

\subsubsection{Algorithm}\label{subsubsec:JC1_update_algorithm}
We prove Lemma~\ref{lem:dynamic_JC1_summary}, i.e., 
we show that the data structures for subquery $\RSCQCX$ (Section~\ref{subsubsec:JB_ds}) 
can be updated in expected $O((\max \{H, H^{\prime}, \log (nHH^{\prime}) \})^{5})$ time using 
(A) the dynamic data structures for the RR-DAG of RLSLP $\mathcal{G}^{R}$ (Section~\ref{subsubsec:rrdag_ds}) 
and sample query (Section~\ref{subsec:sample_query}), 
and (B) the interval attractors obtained from Lemma~\ref{lem:dynamic_IA_summary}. 

Consider two ordered sets $\mathcal{T}_{C} = \{ (h_{1}, C_{1}, M_{1})$, $(h_{2}, C_{2}, M_{2})$, $\ldots$, $(h_{z}, C_{z}, M_{z}) \}$ and $\mathcal{T}^{\prime}_{C} = \{ (h^{\prime}_{1}, C^{\prime}_{1}, M^{\prime}_{1})$, $(h^{\prime}_{2}, C^{\prime}_{2}, M^{\prime}_{2})$, $\ldots$, $(h^{\prime}_{z^{\prime}}, C_{z^{\prime}}, M_{z^{\prime}}) \}$. 
Lemma~\ref{lem:dynamic_JC1_change} and Lemma~\ref{lem:dynamic_JC1_move} indicate that 
we can obtain $m^{\prime}$ sets $\mathcal{J}^{\prime}_{C}(h^{\prime}_{1}, C^{\prime}_{1}, M^{\prime}_{1})$, $\mathcal{J}^{\prime}_{C}(h^{\prime}_{2}, C^{\prime}_{2}, M^{\prime}_{2})$, $\ldots$, $\mathcal{J}^{\prime}_{C}(h^{\prime}_{z^{\prime}}, C^{\prime}_{z^{\prime}}, M^{\prime}_{z^{\prime}})$ of weighted points by 
modifying $m$ sets 
$\mathcal{J}_{C}(h_{1}, C_{1}, M_{1})$, $\mathcal{J}_{C}(h_{2}, C_{2}, M_{2})$, 
$\ldots$, $\mathcal{J}_{C}(h_{z}, C_{z}, M_{z})$ as follows: 

\begin{description}
    \item [Modification 1:] 
    For each interval attractor $([p, q], [\ell, r]) \in \Psi_{\source} \cap \Psi_{\preceding} \cap (\Psi_{\OLD} \setminus \Psi_{\run})$ 
    with level $h$ and associated string $C$, 
    let $M \geq 0$ be an integer satisfying $([p, q], [\ell, r]) \in \Psi_{\modulo}(M)$. 
    Then, 
    set $\mathcal{J}_{C}(h, C, M)$ contains a weighted point $(x, y, w, e)$ satisfying $e = T[p-1..r+1]$, 
    and $1$ is subtracted from the weight $w$ of the weighted point.  
    \item [Modification 2:] 
    For each interval attractor $([p, q], [\ell, r]) \in \Psi_{\source} \cap \Psi_{\preceding} \cap \Psi_{\samp} \cap (\Psi_{\OLD} \setminus \Psi^{\symA})$ with level $h$ and associated string $C$,     
    let $M \geq 0$ be an integer satisfying $([p, q], [\ell, r]) \in \Psi_{\modulo}(M)$. 
    Then, 
    set $\mathcal{J}_{C}(h, C, M)$ contains a weighted point $(x, y, w, e)$ corresponding to 
    the interval attractor $([p, q], [\ell, r])$, 
    and this weighted point is removed from the set $\mathcal{J}_{C}(h, C, M)$.
    \item [Modification 3:] 
    For each interval attractor $([p, q], [\ell, r]) \in \Psi^{\prime}_{\source} \cap \Psi^{\prime}_{\preceding} \cap \Psi^{\prime \symB}$, 
    let $K \geq 0$ and $M \geq 0$ be two integers 
    satisfying $([p, q], [\ell, r]) \in \Psi^{\prime}_{\lcp}(K)$ and $([p, q], [\ell, r]) \in \Psi^{\prime}_{\modulo}(M)$, 
    respectively.     
    We create a weighted point 
    $(|f^{\prime}_{\recover}(([p, q], [\ell, r]))|$, $T^{\prime}[\gamma + K..r + 1], 0, T^{\prime}[p-1..r+1])$, 
    and it is added to set $\mathcal{J}_{C}(h, C, M)$ for 
    the level $h$, attractor position $\gamma$, and associated string $C$ of the interval attractor $([p, q], [\ell, r])$.
    \item [Modification 4:]
    This process is executed after Modification 2 and Modification 3. 
    For each interval attractor $([p, q], [\ell, r])$ of set $\Psi^{\prime}_{\source} \cap \Psi^{\prime}_{\preceding} \cap (\Psi^{\prime}_{\NEW} \setminus \Psi^{\prime}_{\run})$ with level $h$ and associated string $C$, 
    let $M \geq 0$ be an integer satisfying $([p, q], [\ell, r]) \in \Psi^{\prime}_{\modulo}(M)$. 
    Then, 
    set $\mathcal{J}_{C}(h, C, M)$ contains a weighted point $(x, y, w, e)$ satisfying $e = T^{\prime}[p-1..r+1]$, 
    and $1$ is added to the weight $w$ of the weighted point. 
\end{description}

Similar to the data structures for subquery $\RSCQA$, 
we update the data structures for subquery $\RSCQCX$ based on these modifications. 
The algorithm updating the data structures for subquery $\RSCQCX$ consists of the following six phases. 

\paragraph{Phase (i).}
In the first phase, 
we compute two sets $\Psi_{\source} \cap \Psi_{\preceding} \cap (\Psi_{\OLD} \setminus \Psi_{\run})$ 
and $\Psi_{\source} \cap \Psi_{\samp} \cap \Psi_{\preceding} \cap (\Psi_{\OLD} \setminus \Psi^{\symA})$ of interval attractors. 
Here, two sets $\Psi_{\OLD} \setminus \Psi_{\run}$ and $\Psi_{\samp} \cap (\Psi_{\OLD} \setminus \Psi^{\symA})$ are obtained from Lemma~\ref{lem:dynamic_IA_summary}. 

We obtain set $\Psi_{\source} \cap \Psi_{\preceding} \cap (\Psi_{\OLD} \setminus \Psi_{\run})$  by 
verifying whether 
each interval attractor of set $\Psi_{\OLD} \setminus \Psi_{\run}$ is contained in set $\Psi_{\source} \cap \Psi_{\preceding}$. 
This verification can be executed by verify-source (Section~\ref{subsec:verify_source_query}) and 
verify-prec (Section~\ref{subsec:verify_prec_query}) queries. 
Similarly, we obtain set $\Psi_{\source} \cap \Psi_{\preceding} \cap \Psi_{\samp} \cap (\Psi_{\OLD} \setminus \Psi^{\symA})$ by 
verifying whether 
each interval attractor of set $\Psi_{\samp} \cap (\Psi_{\OLD} \setminus \Psi^{\symA})$ is contained in 
set $\Psi_{\source} \cap \Psi_{\preceding}$. 
Therefore, we can obtain the two sets $\Psi_{\source} \cap \Psi_{\preceding} \cap (\Psi_{\OLD} \setminus \Psi_{\run})$ 
and $\Psi_{\source} \cap \Psi_{\preceding} \cap \Psi_{\samp} \cap (\Psi_{\OLD} \setminus \Psi^{\symA})$ by 
executing $O(|\Psi_{\OLD} \setminus \Psi_{\run}| + |\Psi_{\samp} \cap \Psi_{\OLD}|)$ verify-source and verify-prec queries. 
$\mathbb{E}[|\Psi_{\samp} \cap \Psi_{\OLD}|] = O(H + \log n)$ 
and $\mathbb{E}[|\Psi_{\OLD} \setminus \Psi_{\run}|] = O(H + \log n)$ follow from Lemma~\ref{lem:dynamic_psi_overlap}. 
Verify-source and verify-prec queries take $O(H^{2} \log n)$ time. 
Therefore, the first phase takes expected $O((H + \log n) H^{2} \log n)$ time. 

\paragraph{Phase (ii).}
We update the data structures for subquery $\RSCQCX$ based on Modification 1. 
Consider an interval attractor $([p, q], [\ell, r]) \in \Psi_{\source} \cap \Psi_{\preceding} \cap (\Psi_{\OLD} \setminus \Psi_{\run})$ with level $h$ and associated string $C$. 
Let $M \geq 0$ be an integer satisfying $([p, q], [\ell, r]) \in \Psi_{\modulo}(M)$. 
Then, set $\mathcal{T}_{C}$ contains triplet $(h, C, M)$ as a triplet $(h_{t}, C_{t}, M_{t})$. 
Let $(x_{1}, y_{1}, w_{1}, e_{1})$, $(x_{2}, y_{2}, w_{2}, e_{2})$, $\ldots$, $(x_{m}, y_{m}, w_{m}, e_{m})$ 
($e_{1} \prec e_{2} \prec \cdots \prec e_{m}$) be the weighted points of set $\mathcal{J}_{C}(h, C, M)$. 
Here, each weighted point $(x_{s}, y_{s}, w_{s}, e_{s})$ corresponds to an interval attractor in set $\Psi_{h} \cap \Psi_{\source} \cap \Psi_{\centerset}(C) \cap \Psi_{\modulo}(M) \cap \Psi_{\preceding} \cap \Psi_{\samp}$, 
and the weighted point $(x_{s}, y_{s}, w_{s}, e_{s})$ is represented as the $s$-th element of the doubly linked list $\mathbf{L}_{C}(h, C, M)$ representing the set $\mathcal{J}_{C}(h, C, M)$. 
Set $\mathcal{J}_{C}(h, C, M)$ contains a weighted point $(x_{j}, y_{j}, w_{j}, e_{j})$ satisfying $e_{j} = T[p-1..r+1]$, and $1$ is subtracted from the weight $w_{j}$ by Modification 1. 

The second phase consists of four steps for each interval attractor $([p, q], [\ell, r]) \in \Psi_{\source} \cap \Psi_{\preceding} \cap \Psi_{\samp} \cap (\Psi_{\OLD} \setminus \Psi^{\symA})$. 
Let $K = |\lcp(T[\gamma..r], C^{n+1})|$ for the attractor position $\gamma$ of 
the interval attractor $([p, q], [\ell, r])$. 
In the first step, 
we compute the level $h$, attractor position $\gamma$, $|C|$, $K$, $M$, and an interval $[\alpha, \alpha^{\prime}]$ satisfying 
$T[\alpha..\alpha^{\prime}] = C$. 
The level $h$ can be computed in $O(H^{2})$ time by 
the level query $\levelQ(([p, q], [\ell, r]))$ introduced in Section~\ref{subsec:level_query}. 
The attractor position $\gamma$ can be computed in $O(H^{2})$ time by 
the attractor position query $\attrQ(([p, q], [\ell, r]))$ introduced in Section~\ref{subsec:attr_pos_query}. 
The length $|C|$ of the associated string $C$ can be computed in $O(H^{2})$ time by 
the C-length query $\clenQ(([p, q], [\ell, r]))$ introduced in Section~\ref{subsec:C_length_query}. 
The integer $K$ can be computed in $O(H^{2})$ time by 
the C-LCP query $\clcpQ(([p, q]$, $[\ell, r]))$ introduced in Section~\ref{subsec:C_lcp_query}. 
The integer $M$ can be computed in $O(H)$ time using the two integers $K$ and $|C|$ 
because $M = (K - (2 + \sum_{w = 1}^{h+3} \lfloor \mu(w) \rfloor)) \mod |C|$ follows from 
the definition of subset $\Psi_{\modulo}(M)$. 
The interval $[\alpha, \alpha^{\prime}]$ can be computed in $O(1)$ time 
because $T[\gamma..\gamma + |C| - 1] = C$ holds. 
Therefore, the first step takes $O(H^{2})$ time. 

In the second step, 
we obtain a pointer to doubly linked list $\mathbf{L}_{C}(h, C, M)$. 
The pointer to doubly linked list $\mathbf{L}_{C}(h, C, M)$ is stored in 
the $t$-th element of the doubly linked list representing set $\mathcal{T}_{C}$. 
The $t$-th element can be obtained in $O(H^{2} \log n + \log^{2} n)$ time by Lemma~\ref{lem:TC1_queries}~\ref{enum:TC1_queries:3}. 
Therefore, the second step takes $O(H^{2} \log n + \log^{2} n)$ time. 

In the third step, 
we compute the integer $j$ by binary search on the $m$ identifiers $e_{1}, e_{2}, \ldots, e_{m}$. 
Here, $\log m = O(\log n)$ follows from Lemma~\ref{lem:JC1_size}~\ref{enum:JC1_size:2}. 
Similar to the first phase of the algorithm presented in Section~\ref{subsubsec:JA_update_algorithm}, 
the integer $j$ can be computed in $O(H^{2} \log n + \log^{2} n)$ time. 

In the fourth step, 
$1$ is subtracted from the weight $w_{j}$ stored in the $j$-th element of doubly linked list $\mathbf{L}_{C}(h, C, M)$. 
This subtraction takes $O(\log^{4} m)$ time because 
we need to update the range-sum data structure built on set $\mathcal{J}_{C}(h, C, M)$. 

The second phase takes $O(|\Psi_{\OLD} \setminus \Psi_{\run}| (H^{2} \log n + \log^{4} n))$ time, 
i.e., this phase takes expected $O((H + \log n)(H^{2} \log n + \log^{4} n))$ time.

\paragraph{Phase (iii).}
In the third phase, 
we update the data structures for subquery $\RSCQCX$ based on Modification 2. 
Consider an interval attractor $([p, q], [\ell, r]) \in \Psi_{\source} \cap \Psi_{\preceding} \cap \Psi_{\samp} \cap (\Psi_{\OLD} \setminus \Psi^{\symA})$ with level $h$ and associated string $C$. 
Let $M \geq 0$ be an integer satisfying $([p, q], [\ell, r]) \in \Psi_{\modulo}(M)$. 
Then, set $\mathcal{T}_{C}$ contains triplet $(h, C, M)$ as a triplet $(h_{t}, C_{t}, M_{t})$, 
and 
set $\mathcal{J}_{C}(h_{t}, C_{t}, M_{t})$ contains a weighted point $(x_{j}, y_{j}, w_{j}, e_{j})$ corresponding to the interval attractor $([p, q], [\ell, r])$. 
This weighted point $(x_{j}, y_{j}, w_{j}, e_{j})$ is removed from the set $\mathcal{J}_{C}(h_{t}, C_{t}, M_{t})$ by Modification 2. 
By this removal, 
the data structures for subquery $\RSCQCX$ are changed as follows: 
\begin{itemize}
    \item Consider the two doubly linked lists 
    $\mathbf{Y}_{C}(h, C, M, y_{j})$ and $\mathbf{L}^{Y}_{C}(h, C, M)$ introduced in Section~\ref{subsubsec:JC1_Y_ds}. 
    The weighted point $(x_{j}, y_{j}, w_{j}, e_{j})$ corresponds to an element $u$ of doubly linked list $\mathbf{Y}_{C}(h, C, M, y_{j})$, and the doubly linked list $\mathbf{Y}_{C}(h, C, M, y_{j})$ corresponds to an element $u^{\prime}$ of $\mathbf{L}^{Y}_{C}(h, C, M)$. 
    The element $u$ is removed from the doubly linked list $\mathbf{Y}_{C}(h, C, M, y_{j})$. 
    If doubly linked list $\mathbf{Y}_{C}(h, C, M, y_{j})$ is changed to an empty list by the removal of the element $u$, 
    then element $u^{\prime}$ is removed from doubly linked list $\mathbf{L}^{Y}_{C}(h, C, M)$. 
    \item The $j$-th element is removed from doubly linked list $\mathbf{L}_{C}(h, C, M)$. 
    If doubly linked list $\mathbf{L}_{C}(h, C, M)$ is changed to an empty list by the removal of the $j$-th element, 
    then the $t$-th element is removed from the doubly linked list representing set $\mathcal{T}_{C}$.     
\end{itemize}

The third phase consists of seven steps. 
In the first step, 
we compute the level $h$, integer $M$, and an interval $[\alpha, \alpha^{\prime}]$ satisfying 
$T[\alpha..\alpha^{\prime}] = C$. 
Similar to the first step of the second phase, 
this step can be executed in $O(H^{2})$ time. 

In the second step, 
we obtain a pointer to doubly linked list $\mathbf{L}_{C}(h, C, M)$. 
The pointer to doubly linked list $\mathbf{L}_{C}(h, C, M)$ is stored in 
the $t$-th element of the doubly linked list representing set $\mathcal{T}_{C}$. 
Similar to the second step of the second phase, 
this step can be executed in $O(H^{2} \log n + \log^{2} n)$ time. 

In the third step, 
we compute the integer $j$ by binary search on the $m$ identifiers $e_{1}, e_{2}, \ldots, e_{m}$. 
Similar to the third step of the second phase, 
this step can be executed in $O(H^{2} \log n + \log^{2} n)$ time. 

In the fourth step, 
we remove element $u$ from the doubly linked list $\mathbf{Y}_{C}(h, C, M, y_{j})$. 
The $j$-th element of $\mathbf{L}^{Y}_{C}(h, C, M)$ has a pointer to the element $u$. 
Therefore, the fourth step can be executed in $O(\log n)$ time. 

The fifth step is executed if 
doubly linked list $\mathbf{Y}_{C}(h, C, M, y_{j})$ is changed to an empty list by the removal of the element $u$. 
In the fifth step, 
we remove element $u^{\prime}$ from doubly linked list $\mathbf{L}^{Y}_{C}(h, C, M)$. 
This step can be executed in $O(\log n)$ time. 

In the sixth step, 
we remove the $j$-th element from doubly linked list $\mathbf{L}_{C}(h, C, M)$. 
This removal takes $O(\log^{4} m)$ time because 
we need to update the range-sum data structure built on set $\mathcal{J}_{C}(h, C, M)$. 

The seventh step is executed if 
doubly linked list $\mathbf{L}_{C}(h, C, M)$ is changed to an empty list by the removal of the $j$-th element. 
In the seventh step, 
we remove the $t$-th element from the doubly linked list representing set $\mathcal{T}_{C}$. 
This step takes $O(\log n)$ time. 

Finally, the third step takes $O(|\Psi_{\samp} \cap \Psi_{\OLD}| (H^{2} \log n + \log^{4} n))$ time, 
i.e., this phase takes expected $O((H + \log n)(H^{2} \log n + \log^{4} n))$ time.

\paragraph{Phase (iv).}
In the fourth phase, 
we compute two sets $\Psi^{\prime}_{\source} \cap \Psi^{\prime}_{\preceding} \cap (\Psi^{\prime}_{\NEW} \setminus \Psi^{\prime}_{\run})$ 
and $\Psi^{\prime}_{\source} \cap \Psi^{\prime}_{\preceding} \cap \Psi^{\prime \symB}$ of interval attractors. 
Here, two sets $\Psi^{\prime}_{\NEW} \setminus \Psi^{\prime}_{\run}$ and $\Psi^{\prime \symB}$ are obtained from Lemma~\ref{lem:dynamic_IA_summary}. 
This phase is executed after updating the dynamic data structures for RR-DAG and sample query. 

Similar to the first phase, 
we obtain two sets 
$\Psi^{\prime}_{\source} \cap \Psi^{\prime}_{\preceding} \cap (\Psi^{\prime}_{\NEW} \setminus \Psi^{\prime}_{\run})$ and $\Psi^{\prime}_{\source} \cap \Psi^{\prime}_{\preceding} \cap \Psi^{\prime \symB}$ by verifying whether 
each interval attractor of two sets $\Psi^{\prime}_{\NEW} \setminus \Psi^{\prime}_{\run}$ and $\Psi^{\prime \symB}$ is contained in set $\Psi^{\prime}_{\source} \cap \Psi^{\prime}_{\preceding}$. 
This verification can be executed by $|\Psi^{\prime}_{\NEW} \setminus \Psi^{\prime}_{\run}| + |\Psi^{\prime \symB}|$ verify-source and verify-prec queries on set $\Psi^{\prime}_{\RR}$. 
$\Psi^{\prime \symB} \subseteq \Psi^{\prime}_{\NEW} \setminus \Psi^{\prime}_{\run}$ holds, 
and $\mathbb{E}[|\Psi^{\prime}_{\NEW} \setminus \Psi^{\prime}_{\run}|] = O(H^{\prime} + \log n)$ follows from Lemma~\ref{lem:dynamic_psi_overlap}. 
Therefore, the fourth phase takes expected $O((H^{\prime} + \log n) H^{\prime 2} \log n)$ time.

\paragraph{Phase (v).}
In the fifth phase, 
we update the data structures for subquery $\RSCQCX$ based on Modification 3. 
Consider the sequence $\mathbf{Q}_{\samp}$ of nodes introduced in Section~\ref{subsec:sample_query}.
For each interval attractor $([p, q], [\ell, r]) \in \Psi^{\prime}_{\source} \cap \Psi^{\prime}_{\preceding} \cap \Psi^{\prime \symB}$, 
let $u$ be the node of the sequence $\mathbf{Q}_{\samp}$ representing 
the interval attractor $([p, q], [\ell, r])$. 
Let $K = |\lcp(T[\gamma..r], C^{n+1})|$ 
and $M \geq 0$ be an integer satisfying $([p, q], [\ell, r]) \in \Psi^{\prime}_{\modulo}(M)$ 
for the attractor position $\gamma$ and associated string $C$ of the interval attractor $([p, q], [\ell, r])$. 
By Modification 3, 
a weighted point $(|f^{\prime}_{\recover}(([p, q], [\ell, r]))|$, $T^{\prime}[\gamma + K..r + 1], 0, T^{\prime}[p-1..r+1])$ is created, 
and it is added to set $\mathcal{J}_{C}(h, C, M)$ 
for the level $h$ of the interval attractor $([p, q], [\ell, r])$. 
By this addition, 
the data structures for subquery $\RSCQCX$ are changed as follows: 
\begin{itemize}
    \item 
    Consider the strings $R_{1}, R_{2}, \ldots, R_{d}$ ($R_{1} \prec R_{2} \prec \cdots R_{d}$) 
    of ordered set $\mathcal{Y}_{C}(h, C, M)$. 
    Let $\beta$ be the smallest integer in set $[1, d]$ satisfying 
    $T^{\prime}[\gamma + K..r + 1] \preceq R_{\beta}$. 
    If $T^{\prime}[\gamma + K..r + 1] \neq R_{\beta}$, then 
    a new doubly linked list $\mathbf{Y}_{C}(h, C, M, T^{\prime}[\gamma + K..r+1])$ is created, 
    and a new element is inserted into doubly linked list $\mathbf{L}^{Y}_{C}(h, C, M)$ 
    as the $\beta$-th element. 
    Here, the new element of doubly linked list $\mathbf{L}^{Y}_{C}(h, C, M)$ stores a pointer to the new doubly linked list $\mathbf{Y}_{C}(h, C, M, T^{\prime}[\gamma + K..r+1])$. 
    \item 
    Consider the weighted points $(x^{\prime}_{1}, y^{\prime}_{1}, w^{\prime}_{1}, e^{\prime}_{1})$, 
    $(x^{\prime}_{2}, y^{\prime}_{2}, w^{\prime}_{2}, e^{\prime}_{2})$, 
    $\ldots$, $(x^{\prime}_{m^{\prime}}, y^{\prime}_{m^{\prime}}, w^{\prime}_{m^{\prime}}, e^{\prime}_{m^{\prime}})$ 
    of sequence $\mathbf{Q}^{Y}_{C}(h, C, M, T^{\prime}[\gamma + K..r+1])$. 
    Let $\beta^{\prime}$ be the largest integer in set $[1, m^{\prime}]$ satisfying 
    either (A) $\hat{x}_{\beta^{\prime}} < |f^{\prime}_{\recover}(([p, q], [\ell, r]))|$ 
    or (B) $\hat{x}_{\beta^{\prime}} = |f^{\prime}_{\recover}(([p, q], [\ell, r]))|$ and $\hat{e}_{\beta^{\prime}} \prec T^{\prime}[p-1..r+1]$. 
    Then, a new element $v$ is created, 
    and it is inserted into doubly linked list $\mathbf{Y}_{C}(h, C, M, T^{\prime}[\gamma + K..r+1])$ as 
    the $\beta^{\prime}+1$-th element.
    Here, this element $v$ corresponds to the weighted point $(|f^{\prime}_{\recover}(([p, q], [\ell, r]))|$, $T^{\prime}[\gamma + K..r + 1], 0, T^{\prime}[p-1..r+1])$
    and stores a pointer to the node $u$ in the sequence $\mathbf{Q}_{\samp}$. 
    \item 
    Let $t$ be the smallest integer in set $[1, z]$ satisfying 
    one of the following three conditions is satisfied: 
    (a) $h < h_{t}$; 
    (b) $h = h_{t}$ and $C \prec C_{t}$; 
    (c) $h = h_{t}$, $C = C_{t}$, and $M \leq M_{t}$. 
    If $(h, C, M) \neq (h_{t}, C_{t}, M_{t})$, 
    then 
    two new doubly linked lists $\mathbf{L}_{C}(h, C, M)$ and $\mathbf{L}^{Y}_{C}(h, C, M)$ are created, 
    and a new element is inserted into the doubly linked list representing $\mathcal{T}_{C}$ 
    as the $t$-th element. 
    Here, the new element stores two pointers to 
    the two doubly linked lists $\mathbf{L}_{C}(h, C, M)$ and $\mathbf{L}^{Y}_{C}(h, C, M)$.     
    \item 
    Let $j$ be the smallest integer in set $[1, m]$ satisfying 
    $e_{j} \prec T^{\prime}[p-1..r+1]$ for the $m$ weighted points 
    $(x_{1}, y_{1}, w_{1}, e_{1})$, $(x_{2}, y_{2}, w_{2}, e_{2})$, $\ldots$, $(x_{m}, y_{m}, w_{m}, e_{m})$ 
    of set $\mathcal{J}_{C}(h, C, M)$. 
    Then, a new element is inserted into doubly linked list $\mathbf{L}_{C}(h, C, M)$ as the $(j+1)$-th element. 
    Here, this element corresponds to weighted point $(|f^{\prime}_{\recover}(([p, q], [\ell, r]))|$, $T^{\prime}[\gamma + K..r + 1], 0, T^{\prime}[p-1..r+1])$ 
    and stores a pointer to the element $v$. 
\end{itemize}

We execute eight steps for each interval attractor $([p, q], [\ell, r]) \in \Psi^{\prime}_{\source} \cap \Psi^{\prime}_{\preceding} \cap \Psi^{\prime \symB}$. 
In the first step, 
we compute the level $h$, attractor position $\gamma$, $K$, $M$, and an interval $[\alpha, \alpha^{\prime}]$ satisfying 
$T^{\prime}[\alpha..\alpha^{\prime}] = C$. 
Similar to the first step of the second phase, 
this step can be executed in $O(H^{\prime 2})$ time. 

In the second step, we find the node $u$ in the sequence $\mathbf{Q}_{\samp}$. 
The node $u$ can be found in $O(\log^{2} n)$ time by binary search on the nodes of sequence $\mathbf{Q}_{\samp}$ 
using the dynamic data structures for sample query. 

In the third step, 
we compute the integer $t$ by binary search on 
the $m$ triplets of the ordered set $\mathcal{T}_{C}$. 
This binary search can be executed in $O(\log z (H^{\prime 2} + \log n))$ time using Lemma~\ref{lem:TC1_queries}~\ref{enum:TC1_queries:2}. Here, $\log z = O(\log n)$ follows from 
Lemma~\ref{lem:TC1_size}~\ref{enum:TC1_size:3}. 
Therefore, the third step can be executed in $O(H^{\prime 2} \log n + \log^{2} n)$ time. 

The fourth step is executed if $(h, C, M) \neq (h_{t}, C_{t}, M_{t})$ holds. 
In the fourth step, 
we create two new doubly linked lists $\mathbf{L}_{C}(h, C, M)$ and $\mathbf{L}^{Y}_{C}(h, C, M)$.  
A new element is inserted into the doubly linked list representing $\mathcal{T}_{C}$ as the $t$-th element. 
The fourth step can be executed in $O(\log n)$ time. 

In the fifth step, 
we find the smallest integer $\beta$ by binary search on the strings of the ordered set $\mathcal{Y}_{C}(h, C, M)$. 
Similar to the third phase of the algorithm presented in Section~\ref{subsubsec:JA_update_algorithm}, 
this binary search can be executed in $O(H^{\prime 2} \log n + \log^{2} n)$ time. 

The sixth step is executed if $T^{\prime}[\gamma + K..r + 1] \neq R_{\beta}$ holds. 
In the sixth step, we create a new doubly linked list $\mathbf{Y}_{C}(h, T^{\prime}[\gamma + K..r + 1])$ 
and insert a new element into doubly linked list $\mathbf{L}^{Y}_{C}(h, C, M)$ as the $\beta$-th element. 
This step can be executed in $O(\log n)$ time. 

In the seventh step, 
we create element $v$ and insert it into doubly linked list $\mathbf{Y}_{C}(h, C, M, T^{\prime}[\gamma + K..r + 1])$ as the $\beta^{\prime}+1$-th element. 
The integer $\beta^{\prime}$ is computed by binary search on 
the weighted points of sequence $\mathbf{Q}^{Y}_{C}(h, C, M, T^{\prime}[\gamma + K..r+1])$. 
Similar to the third phase of the algorithm presented in Section~\ref{subsubsec:JA_update_algorithm}, 
this binary search can be executed in $O(H^{\prime 2} \log n + \log^{2} n)$ time.
Therefore, the seventh step takes $O(H^{\prime 2} \log n + \log^{2} n)$ time. 

In the eighth step, 
we insert a new element into doubly linked list $\mathbf{L}_{C}(h, C, M)$ as the $(j+1)$-th element. 
This insertion takes $O(\log^{4} m)$ time 
because we need to update the range-sum data structure built on set $\mathcal{J}_{C}(h, C, M)$. 
The integer $j$ is computed by binary search on the $m$ strings $e_{1}, e_{2}, \ldots, e_{m}$. 
Similar to the second phase, 
the integer $j$ can be computed in $O(H^{\prime 2} \log n + \log^{2} n)$ time. 
Therefore, the eighth step takes $O(H^{\prime 2} \log n + \log^{4} n)$ time. 

Finally, the fifth phase takes $O(|\Psi^{\prime \symB}| (H^{\prime 2} \log n + \log^{4} n))$ time, 
i.e., this phase takes expected $O((H^{\prime} + \log n)(H^{\prime 2} \log n + \log^{4} n))$ time. 

\paragraph{Phase (vi).}
In the sixth phase, 
we update the data structures for subquery $\RSCQCX$ based on Modification 4. 
For each interval attractor $([p, q], [\ell, r])$ of set $\Psi^{\prime}_{\source} \cap \Psi^{\prime}_{\preceding} \cap (\Psi^{\prime}_{\NEW} \setminus \Psi^{\prime}_{\run})$, 
let $M \geq 0$ be an integer satisfying $([p, q], [\ell, r]) \in \Psi^{\prime}_{\modulo}(M)$. 
Then, set $\mathcal{T}_{C}$ contains triplet $(h, C, M)$ as a triplet $(h_{t}, C_{t}, M_{t})$ 
for the level $h$ and associated string $C$ of interval attractor $([p, q], [\ell, r])$.
Set $\mathcal{J}_{C}(h_{t}, C_{t}, M_{t})$ contains a weighted point $(x_{j}, y_{j}, w_{j}, e_{j})$ satisfying $e_{j} = T^{\prime}[p-1..r+1]$, and $1$ is added to the weight $w_{j}$ by Modification 4. 

The sixth phase consists of four steps for each interval attractor $([p, q], [\ell, r]) \in \Psi^{\prime}_{\source} \cap \Psi^{\prime}_{\preceding} \cap (\Psi^{\prime}_{\NEW} \setminus \Psi^{\prime}_{\run})$. 
In the first step, 
we compute the level $h$, integer $M$, and an interval $[\alpha, \alpha^{\prime}]$ satisfying 
$T^{\prime}[\alpha..\alpha^{\prime}] = C$. 
Similar to the first step of the second phase, 
this step can be executed in $O(H^{\prime 2})$ time. 

In the second step, 
we obtain a pointer to doubly linked list $\mathbf{L}_{C}(h, C, M)$. 
The pointer to doubly linked list $\mathbf{L}_{C}(h, C, M)$ is stored in 
the $t$-th element of the doubly linked list representing set $\mathcal{T}_{C}$. 
Similar to the second step of the second phase, 
this step can be executed in $O(H^{\prime 2} \log n + \log^{2} n)$ time. 

In the third step, 
we compute the integer $j$ by binary search on the $m$ identifiers $e_{1}, e_{2}, \ldots, e_{m}$. 
Similar to the third step of the second phase, 
this step can be executed in $O(H^{\prime 2} \log n + \log^{2} n)$ time. 

In the fourth step, 
$1$ is added to the weight $w_{j}$ stored in the $j$-th element of doubly linked list $\mathbf{L}_{C}(h, C, M)$. 
This addition takes $O(\log^{4} m)$ time because 
we need to update the range-sum data structure built on set $\mathcal{J}_{C}(h, C, M)$. 

The sixth phase takes $O(|\Psi^{\prime}_{\NEW} \setminus \Psi^{\prime}_{\run}| (H^{\prime 2} \log n + \log^{4} n))$ time, i.e., this phase takes expected $O((H^{\prime} + \log n)(H^{\prime 2} \log n + \log^{4} n))$ time. 

Finally, the six phases take expected $O((H + H^{\prime} + \log n)(H^{2} \log n + H^{\prime 2} \log n + \log^{4} n))$ time in total. Therefore, Lemma~\ref{lem:dynamic_JC1_summary} holds.

%%%%%%%%%%%%%%%%%%%%%%%%%%%%%

\subsection{Update of Data Structures for RSC-C2 Subquery}\label{subsec:rc2_update}
This subsection explains how to update the dynamic data structures for subquery $\RSCQCY$ introduced in Section~\ref{subsubsec:JC2_ds}. 
The dynamic data structures for subquery $\RSCQCY$ are symmetric to 
the dynamic data structures for subquery $\RSCQCX$. 
Therefore, we can update the dynamic data structures for subquery $\RSCQCY$ using the same approach as for 
the dynamic data structures for subquery $\RSCQCX$, which is explained in Section~\ref{subsec:rc1_update}. 
Formally, we obtain the following lemma. 

\begin{lemma}\label{lem:dynamic_JC2_summary}
Consider the two RLSLPs $\mathcal{G}^{R}$ and $\mathcal{G}^{R}_{\ins}$ of Theorem~\ref{theo:update1}, which derive input string $T$ and string $T^{\prime}$, respectively. 
The dynamic data structures of Section~\ref{subsubsec:JC2_ds} can be updated 
in expected $O((\max \{H, H^{\prime}, \log (nHH^{\prime}) \})^{5})$ time 
after changing RLSLP $\mathcal{G}^{R}$ to $\mathcal{G}^{R}_{\ins}$. 
This update requires 
(i) the dynamic data structures for the RR-DAG of RLSLP $\mathcal{G}^{R}$ (Section~\ref{subsubsec:rrdag_ds}) 
and sample query (Section~\ref{subsec:sample_query}), 
and (ii) the interval attractors obtained from Lemma~\ref{lem:dynamic_IA_summary}. 
\end{lemma}
\begin{proof}
    This lemma can be proved using the same approach as for Lemma~\ref{lem:dynamic_JC1_summary}.
\end{proof}

\subsection{Update of Data Structures for RSC-D1 Subquery}\label{subsec:rd1_update}
This subsection explains how to update the dynamic data structures for subquery $\RSCQDX$ 
introduced in Section~\ref{subsubsec:JD1_ds}. 
The following lemma is the summary of this subsection.

\begin{lemma}\label{lem:dynamic_JD1_summary}
Consider the two RLSLPs $\mathcal{G}^{R}$ and $\mathcal{G}^{R}_{\ins}$ of Theorem~\ref{theo:update1}, which derive input string $T$ and string $T^{\prime}$, respectively. 
The dynamic data structures of Section~\ref{subsubsec:JD1_ds} can be updated 
in expected $O((\max \{H, H^{\prime}, \log (nHH^{\prime}) \})^{5})$ time 
after changing RLSLP $\mathcal{G}^{R}$ to $\mathcal{G}^{R}_{\ins}$. 
This update requires 
(i) the dynamic data structures for the RR-DAG of RLSLP $\mathcal{G}^{R}$ (Section~\ref{subsubsec:rrdag_ds}) 
and sample query (Section~\ref{subsec:sample_query}), 
and (ii) the interval attractors obtained from Lemma~\ref{lem:dynamic_IA_summary}. 
\end{lemma}
\begin{proof}
    See Section~\ref{subsubsec:JD1_update_algorithm}. 
\end{proof}

Let $(h_{1}, C_{1})$, $(h_{2}, C_{2})$, $\ldots$, $(h_{z}, C_{z})$ 
be the pairs in the ordered set $\mathcal{T}_{D}$ introduced in Section~\ref{subsubsec:TC1_ds}. 
Then, 
the dynamic data structures for subquery $\RSCQDX$ store weighted points of 
the $2m$ sets $\mathcal{J}^{\prime}_{D}(h^{\prime}_{1}, C^{\prime}_{1})$, 
$\mathcal{J}^{\prime}_{E}(h^{\prime}_{1}, C^{\prime}_{1})$
$\mathcal{J}^{\prime}_{D}(h^{\prime}_{2}, C^{\prime}_{2})$, 
$\mathcal{J}^{\prime}_{E}(h^{\prime}_{2}, C^{\prime}_{2})$, 
$\ldots$, 
$\mathcal{J}^{\prime}_{D}(h^{\prime}_{z^{\prime}}, C^{\prime}_{z^{\prime}})$, 
$\mathcal{J}^{\prime}_{E}(h^{\prime}_{z^{\prime}}, C^{\prime}_{z^{\prime}})$ 
introduced in Section~\ref{subsec:RSC_comp_D1}. 

For a pair of an integer $h \in [0, H^{\prime}]$ and a string $C \in \Sigma^{+}$, 
let $\mathcal{J}^{\prime}_{D}(h, C)$ (respectively, $\mathcal{J}^{\prime}_{E}(h, C)$) be the set $\mathcal{J}_{D}(h, C)$ (respectively, $\mathcal{J}_{E}(h, C)$) of weighted points 
defined using set $\Psi^{\prime}_{\RR}$ instead of set $\Psi_{\RR}$. 
For simplicity, 
let $\mathcal{J}_{D}(h, C)$ and $\mathcal{J}_{E}(h, C)$ be empty sets of weighted points if $h > H$. 
Let $\mathcal{T}^{\prime}_{D} = \{ (h^{\prime}_{1}, C^{\prime}_{1})$, $(h^{\prime}_{2}, C^{\prime}_{2})$, $\ldots$, $(h^{\prime}_{z^{\prime}}, C_{z^{\prime}}) \}$ 
be the ordered set $\mathcal{T}_{D}$ defined using set $\Psi^{\prime}_{\RR}$ instead of set $\Psi_{\RR}$. 

The following lemma shows that each set $\mathcal{J}^{\prime}_{D}(h, C)$ can be obtained by modifying set $\mathcal{J}_{D}(h, C)$. 

\begin{lemma}\label{lem:dynamic_JD1D_change}
Consider a pair of an integer $h \in [0, H^{\prime}]$ and a string $C \in \Sigma^{+}$. 
For each weighted point $(x_{s}, y_{s}, w_{s}, e_{s}) \in \mathcal{J}_{D}(h, C)$, 
let $([p_{s}, q_{s}], [\ell_{s}, r_{s}]) \in \Psi_{h} \cap \Psi_{\source} \cap \Psi_{\centerset}(C) \cap \Psi_{\preceding} \cap \Psi_{\samp}$ be 
the interval attractor corresponding to the weighted point $(x_{s}, y_{s}, w_{s}, e_{s})$, 
$\kappa_{s} = |\Psi_{h} \cap \Psi_{\source} \cap \Psi_{\centerset}(C) \cap \Psi_{\preceding} \cap (\Psi_{\OLD} \setminus \Psi_{\run}) \cap \Psi_{\str}(e_{s})|$, 
and 
$\kappa^{\prime}_{s} = |\Psi^{\prime}_{h} \cap \Psi^{\prime}_{\source} \cap \Psi^{\prime}_{\centerset}(C) \cap \Psi^{\prime}_{\preceding} \cap (\Psi^{\prime}_{\NEW} \setminus \Psi^{\prime}_{\run}) \cap \Psi^{\prime}_{\str}(e_{s})|$. 
For each interval attractor $([p^{\prime}_{s}, q^{\prime}_{s}], [\ell^{\prime}_{s}, r^{\prime}_{s}]) \in \Psi^{\prime}_{h} \cap \Psi^{\prime}_{\source} \cap \Psi^{\prime}_{\centerset}(C) \cap \Psi^{\prime}_{\preceding} \cap \Psi^{\prime \symB}$ with interval attractor $\gamma^{\prime}_{s}$, 
let $\nu^{\prime}_{s} = |f^{\prime}_{\recover}(([p^{\prime}_{s}, q^{\prime}_{s}], [\ell^{\prime}_{s}, r^{\prime}_{s}]))|$ 
and $M^{\prime}_{s} \geq 0$ be an integer satisfying $([p^{\prime}_{s}, q^{\prime}_{s}], [\ell^{\prime}_{s}, r^{\prime}_{s}]) \in \Psi^{\prime}_{\modulo}(M^{\prime}_{s})$. 
Then, set $\mathcal{J}^{\prime}_{D}(h, C)$ of weighted points 
is equal to the union of the following two sets of weighted points: 
\begin{enumerate}[label=\textbf{(\roman*)}]
    \item $\{ (x_{s}, y_{s}, w_{s} - \kappa_{s} + \kappa^{\prime}_{s}, e_{s}) \mid (x_{s}, y_{s}, w_{s}, e_{s}) \in \mathcal{J}_{D}(h, C) \text{ s.t. } ([p_{s}, q_{s}], [\ell_{s}, r_{s}]) \not \in \Psi_{\OLD} \setminus \Psi^{\symA} \}$;
    \item $\{ (\nu^{\prime}_{s}, M^{\prime}_{s}$, $|\Psi^{\prime}_{\str}(T^{\prime}[p^{\prime}_{s}-1..r^{\prime}_{s}+1])|, T^{\prime}[p^{\prime}_{s}-1..r^{\prime}_{s}+1]) \mid ([p^{\prime}_{s}, q^{\prime}_{s}], [\ell^{\prime}_{s}$, $r^{\prime}_{s}]) \in \Psi^{\prime}_{h} \cap \Psi^{\prime}_{\source} \cap \Psi^{\prime}_{\centerset}(C) \cap \Psi^{\prime}_{\preceding} \cap \Psi^{\prime \symB} \}$. 
\end{enumerate}
\end{lemma}
\begin{proof}
Lemma~\ref{lem:dynamic_JD1D_change} can be proved using the same approach as for Lemma~\ref{lem:dynamic_JC1_change}.
\end{proof}

Similarly, 
the following lemma shows that each set $\mathcal{J}^{\prime}_{E}(h, C)$ can be obtained by modifying set $\mathcal{J}_{E}(h, C)$. 

\begin{lemma}\label{lem:dynamic_JD1E_change}
Consider a pair of an integer $h \in [0, H^{\prime}]$ and a string $C \in \Sigma^{+}$. 
For each weighted point $(x_{s}, y_{s}, w_{s}, e_{s}) \in \mathcal{J}_{E}(h, C)$, 
let $([p_{s}, q_{s}], [\ell_{s}, r_{s}]) \in \Psi_{h} \cap \Psi_{\source} \cap \Psi_{\centerset}(C) \cap \Psi_{\preceding} \cap \Psi_{\samp}$ be 
the interval attractor corresponding to the weighted point $(x_{s}, y_{s}, w_{s}, e_{s})$, 
$\nu_{s} = |f_{\recover}(([p_{s}, q_{s}], [\ell_{s}, r_{s}]))|$, 
$\kappa_{s} = |\Psi_{h} \cap \Psi_{\source} \cap \Psi_{\centerset}(C) \cap \Psi_{\preceding} \cap (\Psi_{\OLD} \setminus \Psi_{\run}) \cap \Psi_{\str}(e_{s})|$, 
and 
$\kappa^{\prime}_{s} = |\Psi^{\prime}_{h} \cap \Psi^{\prime}_{\source} \cap \Psi^{\prime}_{\centerset}(C) \cap \Psi^{\prime}_{\preceding} \cap (\Psi^{\prime}_{\NEW} \setminus \Psi^{\prime}_{\run}) \cap \Psi^{\prime}_{\str}(e_{s})|$. 
For each interval attractor $([p^{\prime}_{s}, q^{\prime}_{s}], [\ell^{\prime}_{s}, r^{\prime}_{s}]) \in \Psi^{\prime}_{h} \cap \Psi^{\prime}_{\source} \cap \Psi^{\prime}_{\centerset}(C) \cap \Psi^{\prime}_{\preceding} \cap \Psi^{\prime \symB}$ with interval attractor $\gamma^{\prime}_{s}$, 
let $\nu^{\prime}_{s} = |f^{\prime}_{\recover}(([p^{\prime}_{s}, q^{\prime}_{s}], [\ell^{\prime}_{s}, r^{\prime}_{s}]))|$ 
and $M^{\prime}_{s} \geq 0$ be an integer satisfying $([p^{\prime}_{s}, q^{\prime}_{s}], [\ell^{\prime}_{s}, r^{\prime}_{s}]) \in \Psi^{\prime}_{\modulo}(M^{\prime}_{s})$. 
Then, set $\mathcal{J}^{\prime}_{E}(h, C)$ of weighted points 
is equal to the union of the following two sets of weighted points: 
\begin{enumerate}[label=\textbf{(\roman*)}]
    \item $\{ (x_{s}, y_{s}, w_{s} - \nu_{s} \kappa_{s} + \nu_{s} \kappa^{\prime}_{s}, e_{s}) \mid (x_{s}, y_{s}, w_{s}, e_{s}) \in \mathcal{J}_{E}(h, C) \text{ s.t. } ([p_{s}, q_{s}], [\ell_{s}, r_{s}]) \not \in \Psi_{\OLD} \setminus \Psi^{\symA} \}$;
    \item $\{ (\nu^{\prime}_{s}, M^{\prime}_{s}$, $\nu^{\prime}_{s} |\Psi^{\prime}_{\str}(T^{\prime}[p^{\prime}_{s}-1..r^{\prime}_{s}+1])|, T^{\prime}[p^{\prime}_{s}-1..r^{\prime}_{s}+1]) \mid ([p^{\prime}_{s}, q^{\prime}_{s}], [\ell^{\prime}_{s}, r^{\prime}_{s}]) \in \Psi^{\prime}_{h} \cap \Psi^{\prime}_{\source} \cap \Psi^{\prime}_{\centerset}(C) \cap \Psi^{\prime}_{\preceding} \cap \Psi^{\prime \symB} \}$. 
\end{enumerate}
\end{lemma}
\begin{proof}
Lemma~\ref{lem:dynamic_JD1E_change} can be proved using the same approach as for Lemma~\ref{lem:dynamic_JC1_change}.
\end{proof}

For each pair $(h, C) \in \mathcal{T}_{D}$, 
two sets $\mathcal{J}_{D}(h, C)$ and $\mathcal{J}_{E}(h, C)$ are represented as 
the two doubly linked lists $\mathbf{L}_{D}(h, C)$ and $\mathbf{L}_{E}(h, C)$ introduced in Section~\ref{subsubsec:JD1_ds}. 
The following lemma states properties of these sets of weighted points. 

\begin{lemma}\label{lem:dynamic_JD1_move}
For a pair $(h, C) \in \mathcal{T}_{D}$, 
consider an interval attractor $([p, q], [\ell, r]) \in \Psi_{h} \cap \Psi_{\source} \cap \Psi_{\centerset}(C) \cap \Psi_{\preceding} \cap \Psi_{\samp}$ satisfying $([p, q], [\ell, r]) \not \in \Psi_{\OLD} \setminus \Psi^{\symA}$. 
Here, set $\mathcal{J}_{D}(h, C)$ (respectively, $\mathcal{J}_{E}(h, C)$) contains 
a weighted point $(x, y, w, e)$ (respectively, $(x^{\prime}, y^{\prime}, w^{\prime}, e^{\prime})$) corresponding to the interval attractor $([p, q], [\ell, r])$; 
the interval attractor $([p, q], [\ell, r])$ is represented as a node $u$ of the sequence $\mathbf{Q}_{\samp}$ introduced in Section~\ref{subsec:sample_query}; 
doubly linked list $\mathbf{L}_{D}(h, C)$ (respectively, $\mathbf{L}_{E}(h, C)$) stores 
the weighted point $(x, y, w, e)$ (respectively, $(x^{\prime}, y^{\prime}, w^{\prime}, e^{\prime})$) as 
an element, and the element stores a pointer to node $u$. 

After updating the dynamic data structures for sample query, 
node $u$ represents an interval attractor $([p^{\prime}, q^{\prime}], [\ell^{\prime}, r^{\prime}])$ in set $\Psi^{\prime}_{h} \cap \Psi^{\prime}_{\source} \cap \Psi^{\prime}_{\centerset}(C) \cap \Psi^{\prime}_{\preceding} \cap \Psi^{\prime}_{\samp}$ satisfying the following four conditions: 
\begin{enumerate}[label=\textbf{(\roman*)}]
    \item $T[p-1..r+1] = T^{\prime}[p^{\prime}-1..r^{\prime}+1]$;
    \item $([p^{\prime}, q^{\prime}], [\ell^{\prime}, r^{\prime}]) \not \in \Psi^{\prime \symB}$; 
    \item $|f_{\recover}(([p, q], [\ell, r]))| = |f^{\prime}_{\recover}(([p^{\prime}, q^{\prime}], [\ell^{\prime}, r^{\prime}]))|$;     
    \item 
    let $M \geq 0$ and $M^{\prime} \geq 0$ be two integers satisfying 
    $([p, q], [\ell, r]) \in \Psi_{\modulo}(M)$ and $([p^{\prime}, q^{\prime}], [\ell^{\prime}, r^{\prime}]) \in \Psi^{\prime}_{\modulo}(M^{\prime})$, respectively. 
    Then, $M = M^{\prime}$.
\end{enumerate}
\end{lemma}
\begin{proof}
Lemma~\ref{lem:dynamic_JD1_move} can be proved using the same approach as for Lemma~\ref{lem:dynamic_JC1_move}.
\end{proof}

\subsubsection{Algorithm}\label{subsubsec:JD1_update_algorithm}
We prove Lemma~\ref{lem:dynamic_JD1_summary}, i.e., 
we show that the data structures for subquery $\RSCQDX$ (Section~\ref{subsubsec:JB_ds}) 
can be updated in expected $O((\max \{H, H^{\prime}, \log (nHH^{\prime}) \})^{5})$ time using 
(A) the dynamic data structures for the RR-DAG of RLSLP $\mathcal{G}^{R}$ (Section~\ref{subsubsec:rrdag_ds}) 
and sample query (Section~\ref{subsec:sample_query}), 
and (B) the interval attractors obtained from Lemma~\ref{lem:dynamic_IA_summary}. 

Consider two ordered sets $\mathcal{T}_{D} = \{ (h_{1}, C_{1})$, $(h_{2}, C_{2})$, $\ldots$, $(h_{z}, C_{z}) \}$ and $\mathcal{T}^{\prime}_{D} = \{ (h^{\prime}_{1}, C^{\prime}_{1})$, $(h^{\prime}_{2}, C^{\prime}_{2})$, $\ldots$, $(h^{\prime}_{z^{\prime}}, C_{z^{\prime}}) \}$. 
Lemma~\ref{lem:dynamic_JD1D_change}, Lemma~\ref{lem:dynamic_JD1E_change}, 
Lemma~\ref{lem:dynamic_JC1_move} indicate that 
we can obtain $2m^{\prime}$ sets 
$\mathcal{J}^{\prime}_{D}(h^{\prime}_{1}, C^{\prime}_{1})$, 
$\mathcal{J}^{\prime}_{E}(h^{\prime}_{1}, C^{\prime}_{1})$
$\mathcal{J}^{\prime}_{D}(h^{\prime}_{2}, C^{\prime}_{2})$, 
$\mathcal{J}^{\prime}_{E}(h^{\prime}_{2}, C^{\prime}_{2})$, 
$\ldots$, 
$\mathcal{J}^{\prime}_{D}(h^{\prime}_{z^{\prime}}, C^{\prime}_{z^{\prime}})$, 
$\mathcal{J}^{\prime}_{E}(h^{\prime}_{z^{\prime}}, C^{\prime}_{z^{\prime}})$ 
of weighted points by 
modifying $2m$ sets 
$\mathcal{J}_{D}(h_{1}, C_{1})$, 
$\mathcal{J}_{E}(h_{1}, C_{1})$, 
$\mathcal{J}_{D}(h_{2}, C_{2})$, 
$\mathcal{J}_{E}(h_{2}, C_{2})$, 
$\ldots$, 
$\mathcal{J}_{D}(h_{z}, C_{z})$, 
$\mathcal{J}_{E}(h_{z}, C_{z})$ 
as follows: 
\begin{description}
    \item [Modification 1:] 
    For each interval attractor $([p, q], [\ell, r]) \in \Psi_{\source} \cap (\Psi_{\OLD} \setminus \Psi_{\run})$ 
    with level $h$ and associated string $C$, 
    set $\mathcal{J}_{D}(h, C)$ contains a weighted point $(x, y, w, e)$ 
    satisfying $e = T[p-1..r+1]$, 
    and $1$ is subtracted from the weight $w$ of the weighted point. 
    Similarly, 
    set $\mathcal{J}_{E}(h, C)$ contains a weighted point $(x^{\prime}, y^{\prime}, w^{\prime}, e^{\prime})$ 
    satisfying $e^{\prime} = T[p-1..r+1]$, 
    and $f_{\recover}([p^{\prime}, q^{\prime}], [\ell^{\prime}, r^{\prime}])$ is subtracted from the weight $w^{\prime}$ of the weighted point.     
    \item [Modification 2:] 
    For each interval attractor $([p, q], [\ell, r]) \in \Psi_{\source} \cap \Psi_{\samp} \cap (\Psi_{\OLD} \setminus \Psi^{\symA})$ with level $h$ and associated string $C$,     
    set $\mathcal{J}_{D}(h, C)$ (respectively, $\mathcal{J}_{E}(h, C)$) contains 
    a weighted point corresponding to 
    the interval attractor $([p, q], [\ell, r])$, 
    and this weighted point is removed from the set $\mathcal{J}_{D}(h, C)$ (respectively, $\mathcal{J}_{E}(h, C)$). 
    \item [Modification 3:] 
    For each interval attractor $([p, q], [\ell, r]) \in \Psi^{\prime}_{\source} \cap \Psi^{\prime \symB}$, 
    let $M \geq 0$ be an integer 
    satisfying $([p, q], [\ell, r]) \in \Psi^{\prime}_{\modulo}(M)$.     
    We create a weighted point 
    $(|f^{\prime}_{\recover}(([p, q], [\ell, r]))|$, $M, 0, T^{\prime}[p-1..r+1])$, 
    and it is added to two set $\mathcal{J}_{D}(h, C)$ and $\mathcal{J}_{E}(h, C)$ 
    for the level $h$ and associated string $C$ of the interval attractor $([p, q], [\ell, r])$.
    \item [Modification 4:]
    This process is executed after Modification 2 and Modification 3. 
    For each interval attractor $([p, q], [\ell, r])$ of set $\Psi^{\prime}_{\source} \cap (\Psi^{\prime}_{\NEW} \setminus \Psi^{\prime}_{\run})$ with level $h$ and associated string $C$, 
    set $\mathcal{J}_{D}(h, C)$ contains a weighted point $(x, y, w, e)$ 
    satisfying $e = T^{\prime}[p-1..r+1]$, 
    and $1$ is added to the weight $w$ of the weighted point. 
    Similarly, 
    set $\mathcal{J}_{E}(h, C)$ contains a weighted point $(x^{\prime}, y^{\prime}, w^{\prime}, e^{\prime})$ 
    satisfying $e^{\prime} = T^{\prime}[p-1..r+1]$, 
    and $|f^{\prime}_{\recover}(([p, q], [\ell, r]))|$ is added to the weight $w^{\prime}$ of the weighted point. 
\end{description}

Similar to the data structures for subquery $\RSCQA$, 
we update the data structures for subquery $\RSCQDX$ based on these modifications. 
The algorithm updating the data structures for subquery $\RSCQDX$ consists of the following six phases. 

\paragraph{Phase (i).}
In the first phase, 
we compute two sets $\Psi_{\source} \cap \Psi_{\preceding} \cap (\Psi_{\OLD} \setminus \Psi_{\run})$ 
and $\Psi_{\source} \cap \Psi_{\samp} \cap \Psi_{\preceding} \cap (\Psi_{\OLD} \setminus \Psi^{\symA})$ of interval attractors by the first phase of the algorithm presented in Section~\ref{subsubsec:JC1_update_algorithm}. 
This phase takes expected $O((H + \log n) H^{2} \log n)$ time. 

\paragraph{Phase (ii).}
We update the data structures for subquery $\RSCQDX$ based on Modification 1. 
Consider an interval attractor $([p, q], [\ell, r]) \in \Psi_{\source} \cap \Psi_{\preceding} \cap (\Psi_{\OLD} \setminus \Psi_{\run})$ with level $h$ and associated string $C$. 
Then, set $\mathcal{T}_{D}$ contains pair $(h, C)$ as a pair $(h_{t}, C_{t})$. 
Let $(x_{1}, y_{1}, w_{1}, e_{1})$, $(x_{2}, y_{2}, w_{2}, e_{2})$, $\ldots$, $(x_{m}, y_{m}$, $w_{m}, e_{m})$ 
($e_{1} \prec e_{2} \prec \cdots \prec e_{m}$) be the weighted points of set $\mathcal{J}_{D}(h, C)$. 
Here, each weighted point $(x_{s}, y_{s}, w_{s}, e_{s})$ corresponds to an interval attractor in set $\Psi_{h} \cap \Psi_{\source} \cap \Psi_{\centerset}(C) \cap \Psi_{\preceding} \cap \Psi_{\samp}$, 
and the weighted point $(x_{s}, y_{s}, w_{s}, e_{s})$ is represented as the $s$-th element of the doubly linked list $\mathbf{L}_{D}(h, C)$ representing the set $\mathcal{J}_{D}(h, C)$. 
Set $\mathcal{J}_{D}(h, C)$ contains a weighted point $(x_{j}, y_{j}, w_{j}, e_{j})$ satisfying $e_{j} = T[p-1..r+1]$, and $1$ is subtracted from the weight $w_{j}$ by Modification 1. 

Similarly, 
set $\mathcal{J}_{E}(h, C)$ contains $m$ weighted points 
$(x^{\prime}_{1}, y^{\prime}_{1}, w^{\prime}_{1}, e^{\prime}_{1})$, $(x^{\prime}_{2}, y^{\prime}_{2}, w^{\prime}_{2}, e^{\prime}_{2})$, $\ldots$, $(x^{\prime}_{m}, y^{\prime}_{m}$, $w^{\prime}_{m}, e^{\prime}_{m})$ 
($e^{\prime}_{1} \prec e^{\prime}_{2} \prec \cdots \prec e^{\prime}_{m}$). 
Each weighted point $(x^{\prime}_{s}, y^{\prime}_{s}, w^{\prime}_{s}, e^{\prime}_{s})$ is represented as the $s$-th element of the doubly linked list $\mathbf{L}_{E}(h, C)$ representing the set $\mathcal{J}_{E}(h, C)$. 
$|f_{\recover}(([p, q], [\ell, r]))|$ is subtracted from the weight $w^{\prime}_{j}$ by Modification 1 
for the $j$-th weighted point $(x^{\prime}_{j}, y^{\prime}_{j}, w^{\prime}_{j}, e^{\prime}_{j})$. 

The second phase consists of four steps for each interval attractor $([p, q], [\ell, r]) \in \Psi_{\source} \cap \Psi_{\preceding} \cap \Psi_{\samp} \cap (\Psi_{\OLD} \setminus \Psi^{\symA})$. 
%Let $M \geq 0$ be an integer satisfying $([p, q], [\ell, r]) \in \Psi_{\modulo}(M)$. 
In the first step, 
we compute the level $h$ and an interval $[\alpha, \alpha^{\prime}]$ satisfying 
$T[\alpha..\alpha^{\prime}] = C$. 
Similar to the second phase of the algorithm presented in Section~\ref{subsubsec:JC1_update_algorithm}, 
the level $h$ and interval $[\alpha, \alpha^{\prime}]$ can be computed in $O(H^{2})$ time. 

In the second step, 
we obtain two pointers to two doubly linked lists $\mathbf{L}_{D}(h, C)$ and $\mathbf{L}_{E}(h, C)$. 
These pointers are stored in 
the $t$-th element of the doubly linked list representing set $\mathcal{T}_{D}$. 
The $t$-th element can be obtained in $O(H^{2} \log n + \log^{2} n)$ time by Lemma~\ref{lem:TD1_queries}~\ref{enum:TD1_queries:3}. 
Therefore, the second step takes $O(H^{2} \log n + \log^{2} n)$ time. 

In the third step, 
we compute the integer $j$ by binary search on the $m$ identifiers $e_{1}, e_{2}, \ldots, e_{m}$. 
Here, $\log m = O(\log n)$ follows from Lemma~\ref{lem:TD1_size}~\ref{enum:TD1_size:4}. 
Similar to the first phase of the algorithm presented in Section~\ref{subsubsec:JA_update_algorithm}, 
the integer $j$ can be computed in $O(H^{2} \log n + \log^{2} n)$ time. 

In the fourth step, 
$1$ (respectively, $|f_{\recover}(([p, q], [\ell, r]))|$) 
is subtracted from the weight $w_{j}$ (respectively, $w^{\prime}_{j}$) stored 
in the $j$-th element of doubly linked list $\mathbf{L}_{D}(h, C)$ (respectively, $\mathbf{L}_{E}(h, C)$). 
These subtractions take $O(\log^{4} m)$ time because 
we need to update the range-sum data structures built on two sets 
$\mathcal{J}_{D}(h, C)$ and $\mathcal{J}_{E}(h, C)$. 
The integer $|f_{\recover}(([p, q], [\ell, r]))|$ is computed in $O(H^{2})$ time 
by the r-size query $\rsizeQ(([p, q], [\ell, r]))$ introduced in Section~\ref{subsec:rsize_query}. 

The second phase takes $O(|\Psi_{\OLD} \setminus \Psi_{\run}| (H^{2} \log n + \log^{4} n))$ time. 
Here, 
$\mathbb{E}[|\Psi_{\OLD} \setminus \Psi_{\run}|] = O(H + \log n)$ follows 
from Lemma~\ref{lem:dynamic_psi_overlap}. 
Therefore, this phase takes expected $O((H + \log n)(H^{2} \log n + \log^{4} n))$ time.

\paragraph{Phase (iii).}
In the third phase, 
we update the data structures for subquery $\RSCQDX$ based on Modification 2. 
Consider an interval attractor $([p, q], [\ell, r]) \in \Psi_{\source} \cap \Psi_{\preceding} \cap \Psi_{\samp} \cap (\Psi_{\OLD} \setminus \Psi^{\symA})$ with level $h$ and associated string $C$. 
Then, set $\mathcal{T}_{D}$ contains pair $(h, C, M)$ as a pair $(h_{t}, C_{t})$, 
and 
set $\mathcal{J}_{D}(h, C)$ contains a weighted point $(x_{j}, y_{j}, w_{j}, e_{j})$ corresponding to the interval attractor $([p, q], [\ell, r])$, 
and the weighted point $(x^{\prime}_{j}, y^{\prime}_{j}, w^{\prime}_{j}, e^{\prime}_{j})$ of 
set $\mathcal{J}_{E}(h, C)$ corresponds to the interval attractor $([p, q], [\ell, r])$. 
The two weighted points $(x_{j}, y_{j}, w_{j}, e_{j})$ and $(x^{\prime}_{j}, y^{\prime}_{j}, w^{\prime}_{j}, e^{\prime}_{j})$ are removed from 
the two sets $\mathcal{J}_{D}(h, C)$ and $\mathcal{J}_{E}(h, C)$, respectively, by Modification 2. 
If set $\mathcal{J}_{D}(h, C)$ is changed to an empty set by the removal of 
the weighted point $(x_{j}, y_{j}, w_{j}, e_{j})$, 
then pair $(h_{t}, C_{t})$ is removed from set $\mathcal{T}_{D}$. 

The third phase consists of five steps for each interval attractor $([p, q], [\ell, r]) \in \Psi_{\source} \cap \Psi_{\preceding} \cap \Psi_{\samp} \cap (\Psi_{\OLD} \setminus \Psi^{\symA})$. 
In the first step, 
we compute the level $h$ and an interval $[\alpha, \alpha^{\prime}]$ satisfying $T[\alpha..\alpha^{\prime}] = C$. 
Similar to the second phase, this step can be executed in $O(H^{2})$ time. 

In the second step, 
we obtain two pointers to two doubly linked lists $\mathbf{L}_{D}(h, C)$ and $\mathbf{L}_{E}(h, C)$. 
These pointers are stored in 
the $t$-th element of the doubly linked list representing set $\mathcal{T}_{D}$. 
Similar to the second phase, 
the $t$-th element can obtained in $O(H^{2} \log n + \log^{2} n)$ time. 

In the third step, 
we compute the integer $j$ by binary search on the $m$ identifiers $e_{1}, e_{2}, \ldots, e_{m}$. 
Similar to the second phase, 
the integer $j$ can be computed in $O(H^{2} \log n + \log^{2} n)$ time. 

In the fourth step, 
we remove the $j$-th element from doubly linked list $\mathbf{L}_{D}(h, C)$. 
Similarly, we remove the $j$-th element from doubly linked list $\mathbf{L}_{E}(h, C)$. 
These removals take $O(\log^{4} m)$ time because 
we need to update the range-sum data structures built on two sets 
$\mathcal{J}_{D}(h, C)$ and $\mathcal{J}_{E}(h, C)$. 

The fifth step is executed if doubly linked list $\mathbf{L}_{D}(h, C)$ has no element. 
In the fifth step, 
we remove the $t$-th element from the doubly linked list representing set $\mathcal{T}_{D}$. 
The fifth step takes $O(\log z)$ time. 
Here, $\log z = O(\log n)$ follows from Lemma~\ref{lem:TD1_size}~\ref{enum:TD1_size:3}.

Finally, the third step takes $O(|\Psi_{\samp} \cap \Psi_{\OLD}| (H^{2} \log n + \log^{4} n))$ time. 
Here, $\mathbb{E}[|\Psi_{\samp} \cap \Psi_{\OLD}|] = O(H + \log n)$ follows 
from Lemma~\ref{lem:dynamic_psi_overlap}. 
Therefore, this phase takes expected $O((H + \log n)(H^{2} \log n + \log^{4} n))$ time.

\paragraph{Phase (iv).}
The fourth phase is executed after updating the dynamic data structures for RR-DAG and sample query. 
In the fourth phase, 
we compute two sets $\Psi^{\prime}_{\source} \cap \Psi^{\prime}_{\preceding} \cap (\Psi^{\prime}_{\NEW} \setminus \Psi^{\prime}_{\run})$ 
and $\Psi^{\prime}_{\source} \cap \Psi^{\prime}_{\preceding} \cap \Psi^{\prime \symB}$ of interval attractors by the fourth phase of the algorithm presented in Section~\ref{subsubsec:JC1_update_algorithm}. 
This phase takes expected $O((H^{\prime} + \log n) H^{\prime 2} \log n)$ time. 

\paragraph{Phase (v).}
In the fifth phase, 
we update the data structures for subquery $\RSCQDX$ based on Modification 3. 
Consider the sequence $\mathbf{Q}_{\samp}$ of nodes introduced in Section~\ref{subsec:sample_query}.
For each interval attractor $([p, q], [\ell, r]) \in \Psi^{\prime}_{\source} \cap \Psi^{\prime}_{\preceding} \cap \Psi^{\prime \symB}$, 
let $M \geq 0$ be an integer satisfying $([p, q], [\ell, r]) \in \Psi^{\prime}_{\modulo}(M)$. 
By Modification 3, 
a weighted point $(|f^{\prime}_{\recover}(([p, q], [\ell, r]))|$, $M, 0, T^{\prime}[p-1..r+1])$ is created, 
and it is added to two set $\mathcal{J}_{D}(h, C)$ and $\mathcal{J}_{E}(h, C)$ 
for the level $h$ and associated string $C$ of the interval attractor $([p, q], [\ell, r])$.
By this addition, 
the data structures for subquery $\RSCQDX$ are changed as follows: 
\begin{itemize}
    \item 
    Let $t$ be the smallest integer in set $[1, z]$ satisfying 
    either 
    (a) $h < h_{t}$ or (b) $h = h_{t}$ and $C \preceq C_{t}$. 
    If $(h, C) \neq (h_{t}, C_{t})$, 
    then two new doubly linked lists $\mathbf{L}_{D}(h, C)$ and $\mathbf{L}_{E}(h, C)$ are created, 
    and a new element is inserted into the doubly linked list representing $\mathcal{T}_{D}$ 
    as the $t$-th element. 
    Here, the new element stores two pointers to 
    the two doubly linked lists $\mathbf{L}_{D}(h, C)$ and $\mathbf{L}_{E}(h, C)$.     
    \item 
    Let $u$ be the node of the sequence $\mathbf{Q}_{\samp}$ representing 
    the interval attractor $([p, q], [\ell, r])$, 
    and let $j$ be the smallest integer in set $[1, m]$ satisfying 
    $e_{j} \prec T^{\prime}[p-1..r+1]$ for the $m$ weighted points 
    $(x_{1}, y_{1}, w_{1}, e_{1})$, $(x_{2}, y_{2}, w_{2}, e_{2})$, $\ldots$, $(x_{m}, y_{m}, w_{m}, e_{m})$ 
    of set $\mathcal{J}_{D}(h, C)$. 
    Then, a new element $v$ is inserted into doubly linked list $\mathbf{L}_{D}(h, C)$ as the $(j+1)$-th element. 
    Here, this element $v$ corresponds to weighted point $(|f^{\prime}_{\recover}(([p, q], [\ell, r]))|$, $M, 0, T^{\prime}[p-1..r+1])$ and stores a pointer to the element $u$. 
    The x-coordinate, y-coordinate, and weight of the $(|f^{\prime}_{\recover}(([p, q], [\ell, r]))|$, $M, 0, T^{\prime}[p-1..r+1])$ are stored in the element $v$.     
    Similarly, 
    a new element $v^{\prime}$ is inserted into doubly linked list $\mathbf{L}_{E}(h, C)$ as the $(j+1)$-th element. 
\end{itemize}

We execute five steps for each interval attractor $([p, q], [\ell, r]) \in \Psi^{\prime}_{\source} \cap \Psi^{\prime}_{\preceding} \cap \Psi^{\prime \symB}$. 
In the first step, 
we compute the level $h$, $|f^{\prime}_{\recover}(([p, q], [\ell, r]))|$, and an interval $[\alpha, \alpha^{\prime}]$ satisfying $T^{\prime}[\alpha..\alpha^{\prime}] = C$. 
Similar to the second phase, 
the level $h$ and interval $[\alpha, \alpha^{\prime}]$ can be computed in $O(H^{\prime 2})$ time. 
The integer $|f^{\prime}_{\recover}(([p, q], [\ell, r]))|$ is computed in $O(H^{\prime 2})$ time 
by the r-size query $\rsizeQ(([p, q], [\ell, r]))$. 

In the second step, we find the node $u$ in the sequence $\mathbf{Q}_{\samp}$. 
The node $u$ can be found in $O(\log^{2} n)$ time by binary search on the nodes of sequence $\mathbf{Q}_{\samp}$ 
using the dynamic data structures for sample query. 

In the third step, 
we compute the integer $t$ by binary search on 
the $m$ pairs of the ordered set $\mathcal{T}_{D}$. 
This binary search can be executed in $O(\log z (H^{\prime 2} + \log n))$ time using Lemma~\ref{lem:TD1_queries}~\ref{enum:TD1_queries:2}. 
Therefore, the third step can be executed in $O(H^{\prime 2} \log n + \log^{2} n)$ time. 

The fourth step is executed if $(h, C) \neq (h_{t}, C_{t})$ holds. 
In the fourth step, 
we create two new doubly linked lists $\mathbf{L}_{D}(h, C)$ and $\mathbf{L}_{E}(h, C)$.  
A new element is inserted into the doubly linked list representing $\mathcal{T}_{D}$ as the $t$-th element. 
The fourth step can be executed in $O(\log n)$ time. 

In the fifth step, 
we insert a new element into doubly linked list $\mathbf{L}_{D}(h, C)$ as the $(j+1)$-th element. 
Similarly, 
we insert a new element into doubly linked list $\mathbf{L}_{E}(h, C)$ as the $(j+1)$-th element. 
These insertions takes $O(\log^{4} m)$ time 
because we need to update the range-sum data structure built on two sets $\mathcal{J}_{D}(h, C)$ 
and $\mathcal{J}_{E}(h, C)$. 
The integer $j$ is computed by binary search on the $m$ strings $e_{1}, e_{2}, \ldots, e_{m}$. 
Similar to the second phase, 
the integer $j$ can be computed in $O(H^{\prime 2} \log n + \log^{2} n)$ time. 
Therefore, the fifth step takes $O(H^{\prime 2} \log n + \log^{4} n)$ time. 

Finally, the fifth phase takes $O(|\Psi^{\prime \symB}| (H^{\prime 2} \log n + \log^{4} n))$ time. 
Here, $\Psi^{\prime \symB} \subseteq \Psi^{\prime}_{\NEW} \setminus \Psi^{\prime}_{\run}$ holds, 
and $\mathbb{E}[|\Psi^{\prime}_{\NEW} \setminus \Psi^{\prime}_{\run}|] = O(H^{\prime} + \log n)$ follows from Lemma~\ref{lem:dynamic_psi_overlap}. 
Therefore, this phase takes expected $O((H^{\prime} + \log n)(H^{\prime 2} \log n + \log^{4} n))$ time. 

\paragraph{Phase (vi).}
In the sixth phase, 
we update the data structures for subquery $\RSCQDX$ based on Modification 4. 
For each interval attractor $([p, q], [\ell, r])$ of set $\Psi^{\prime}_{\source} \cap \Psi^{\prime}_{\preceding} \cap (\Psi^{\prime}_{\NEW} \setminus \Psi^{\prime}_{\run})$ with level $h$ and associated string $C$,  
set $\mathcal{T}_{D}$ contains pair $(h, C)$ as a pair $(h_{t}, C_{t})$. 
Set $\mathcal{J}_{D}(h, C)$ contains a weighted point $(x_{j}, y_{j}, w_{j}, e_{j})$ satisfying $e_{j} = T^{\prime}[p-1..r+1]$, and $1$ is added to the weight $w_{j}$ by Modification 3. 
Similarly, 
the $j$-th weighted point $(x^{\prime}_{j}, y^{\prime}_{j}, w^{\prime}_{j}, e^{\prime}_{j})$ of set $\mathcal{J}_{E}(h, C)$ satisfies $e_{j} = T^{\prime}[p-1..r+1]$, 
and 
$|f^{\prime}_{\recover}(([p, q], [\ell, r]))|$ is added to the weight $w^{\prime}_{j}$ by Modification 4. 

The sixth phase consists of four steps for each interval attractor $([p, q], [\ell, r]) \in \Psi^{\prime}_{\source} \cap \Psi^{\prime}_{\preceding} \cap (\Psi^{\prime}_{\NEW} \setminus \Psi^{\prime}_{\run})$. 
In the first step, 
we compute the level $h$, $|f^{\prime}_{\recover}(([p, q], [\ell, r]))|$, and an interval $[\alpha, \alpha^{\prime}]$ satisfying $T^{\prime}[\alpha..\alpha^{\prime}] = C$. 
Similar to the fifth phase, 
this step can be executed in $O(H^{\prime 2})$ time. 

In the second step, 
we obtain two pointers to two doubly linked lists $\mathbf{L}_{D}(h, C)$ and $\mathbf{L}_{E}(h, C)$. 
These pointers are stored in 
the $t$-th element of the doubly linked list representing set $\mathcal{T}_{D}$. 
Similar to the second step of the second phase, 
this step can be executed in $O(H^{\prime 2} \log n + \log^{2} n)$ time. 

In the third step, 
we compute the integer $j$ by binary search on the $m$ identifiers $e_{1}, e_{2}, \ldots, e_{m}$. 
Similar to the third step of the second phase, 
this step can be executed in $O(H^{\prime 2} \log n + \log^{2} n)$ time. 

In the fourth step, 
$1$ is added to the weight $w_{j}$ stored in the $j$-th element of doubly linked list $\mathbf{L}_{D}(h, C)$. 
Similarly, 
$|f^{\prime}_{\recover}(([p, q], [\ell, r]))|$ is added to the weight $w^{\prime}_{j}$ stored in the $j$-th element of doubly linked list $\mathbf{L}_{E}(h, C)$. 
These additions take $O(\log^{4} m)$ time because 
we need to update the range-sum data structure built on two sets $\mathcal{J}_{D}(h, C)$ 
and $\mathcal{J}_{E}(h, C)$. 

The sixth phase takes $O(|\Psi^{\prime}_{\NEW} \setminus \Psi^{\prime}_{\run}| (H^{\prime 2} \log n + \log^{4} n))$ time, i.e., this phase takes expected $O((H^{\prime} + \log n)(H^{\prime 2} \log n + \log^{4} n))$ time. 

Finally, the six phases take expected $O((H + H^{\prime} + \log n)(H^{2} \log n + H^{\prime 2} \log n + \log^{4} n))$ time in total. Therefore, Lemma~\ref{lem:dynamic_JD1_summary} holds. 

%%%%%%%%%%%%%%%%%%%%%%%%%%%%%%%%%%%%%%%%%%%%%%%%%%%%%%%%%%%%%%%%%%%%%%

\subsection{Update of Data Structures for RSC-D2 Subquery}\label{subsec:rd2_update}
This subsection explains how to update the dynamic data structures for subquery $\RSCQDY$ 
introduced in Section~\ref{subsubsec:JD2_ds}. 
The dynamic data structures for subquery $\RSCQDY$ are symmetric to 
the dynamic data structures for subquery $\RSCQDX$. 
Therefore, we can update the dynamic data structures for subquery $\RSCQDX$ using the same approach as for 
the dynamic data structures for subquery $\RSCQDX$, which is explained in Section~\ref{subsec:rd1_update}. 
Formally, we obtain the following lemma. 

\begin{lemma}\label{lem:dynamic_JD2_summary}
Consider the two RLSLPs $\mathcal{G}^{R}$ and $\mathcal{G}^{R}_{\ins}$ of Theorem~\ref{theo:update1}, which derive input string $T$ and string $T^{\prime}$, respectively. 
The dynamic data structures of Section~\ref{subsubsec:JD2_ds} can be updated 
in expected $O((\max \{H, H^{\prime}, \log (nHH^{\prime}) \})^{5})$ time 
after changing RLSLP $\mathcal{G}^{R}$ to $\mathcal{G}^{R}_{\ins}$. 
This update requires 
(i) the dynamic data structures for the RR-DAG of RLSLP $\mathcal{G}^{R}$ (Section~\ref{subsubsec:rrdag_ds}) 
and sample query (Section~\ref{subsec:sample_query}), 
and (ii) the interval attractors obtained from Lemma~\ref{lem:dynamic_IA_summary}. 
\end{lemma}
\begin{proof}
    This lemma can be proved using the same approach as for Lemma~\ref{lem:dynamic_JD1_summary}.
\end{proof}

\subsection{Update of Data Structures for SA, ISA, RA, and LCE Queries}\label{subsec:update_dyn_ins}
We show that 
we can update the data structures of Section~\ref{subsubsec:rrdag_ds}, Section~\ref{sec:other_query}, and Section~\ref{sec:RSC_query} 
in expected $O((\max \{H, H^{\prime}, \log (nHH^{\prime}) \})^{8})$ time. 
We presented algorithms updating each of our data structures in Section~\ref{subsubsec:update_rr_dag}, 
Section~\ref{subsec:update_sample_query}
Section~\ref{subsec:bis_update}, 
Section~\ref{subsec:ra_update}, 
Section~\ref{subsec:rb_update}, 
Section~\ref{subsec:rc1_update}, 
Section~\ref{subsec:rc2_update}, 
Section~\ref{subsec:rd1_update}, 
and Section~\ref{subsec:rc2_update}.
Therefore, 
we can update the data structures of Section~\ref{subsubsec:rrdag_ds}, Section~\ref{sec:other_query}, and Section~\ref{sec:RSC_query} by combining these algorithms. 

We update the data structures of Section~\ref{subsubsec:rrdag_ds}, Section~\ref{sec:other_query}, and Section~\ref{sec:RSC_query} in the following two steps: 
\begin{enumerate}[label=\textbf{(\roman*)}]
    \item compute seven sets (A) $\Psi_{\OLD} \setminus \Psi_{\run}$, (B) $\Psi^{\prime}_{\NEW} \setminus \Psi^{\prime}_{\run}$, (C) $\Psi^{\symA}$, (D) $\Psi^{\symA^{\prime}}$, (E) $\Psi^{\prime \symB}$, (F) $\Psi_{\samp} \cap \Psi_{\OLD}$, and (G) $(\Psi_{\samp} \cap \Psi_{\OLD}) \setminus \Psi^{\symA}$ of interval attractors in expected $O((\max \{H$, $H^{\prime}, \log (nHH^{\prime}) \})^{8})$ time by Lemma~\ref{lem:dynamic_IA_summary};
    \item update the data structures of Section~\ref{subsubsec:rrdag_ds}, Section~\ref{sec:other_query}, and Section~\ref{sec:RSC_query} 
    in expected $O((\max \{H$, $H^{\prime}, \log (nHH^{\prime}) \})^{5})$ time 
    by Lemma~\ref{lem:dynamic_rrdag_summary}, Lemma~\ref{lem:dynamic_samp_summary}, Lemma~\ref{subsec:bis_update}, Lemma~\ref{lem:dynamic_JA_summary}, Lemma~\ref{lem:dynamic_JB_summary}, 
    Lemma~\ref{lem:dynamic_JC1_summary}, Lemma~\ref{lem:dynamic_JC2_summary}, Lemma~\ref{lem:dynamic_JD1_summary}, and Lemma~\ref{lem:dynamic_JD2_summary}.
\end{enumerate}

The two steps take expected $O((\max \{H, H^{\prime}, \log (nHH^{\prime}) \})^{8})$ time in total. 
Therefore, we obtain the following theorem.  

\begin{theorem}\label{theo:update_SA_ISA_insertion}
Consider the two RLSLPs $\mathcal{G}^{R}$ and $\mathcal{G}^{R}_{\ins}$ of Theorem~\ref{theo:update1}, which derive input string $T$ and string $T^{\prime}$, respectively. 
After modifying the RLSLP $\mathcal{G}^{R}$ to $\mathcal{G}^{R}_{\ins}$, 
we can update the dynamic data structures introduced in Section~\ref{subsubsec:rrdag_ds}, Section~\ref{sec:other_query}, and Section~\ref{sec:RSC_query}, 
and this update takes expected $O((\max \{H, H^{\prime}, \log (nHH^{\prime}) \})^{8})$ time. 
\end{theorem}
%\begin{proof}
%    See Section~\ref{subsec:update_dyn_ins}. 
%\end{proof}

\section{Queries and Update with Polylogarithmic Time}\label{sec:summary}
This section shows our main results stated in Section~\ref{sec:introduction}, 
i.e., 
for input string $T$ of length $n$, 
we present dynamic data structure of expected $\delta$-optimal space supporting the following queries in $O(\polylog (n))$ time: 
\begin{itemize}
    \item SA, ISA, RA, and LCE queries; 
    \item update this data structure for insertion and deletion of a single character in the input string. 
    This update fails with probability $O(n^{-w})$, where a user-defined constant parameter $w \geq 2$.    
\end{itemize}

For supporting SA, ISA, RA, LCE queries, 
we store the following data structures: 
\begin{enumerate}[label=\textbf{(\roman*)}]
    \item the dynamic data structures for the RR-DAG of RLSLP $\mathcal{G}^{R}$ deriving input string $T$, which are introduced in Section~\ref{subsubsec:rrdag_ds}. 
    These data structures require $O(|\mathcal{U}_{\RR}|B)$ bits of space 
    for machine word size $B$ and the number $|\mathcal{U}_{\RR}|$ of nodes in the RR-DAG; 
    \item the dynamic data structures for sample query, which are introduced in Section~\ref{subsubsec:sample_ds}. 
    These data structures require $O(|\Psi_{\samp}|B)$ bits of space 
    for the sampling subset $\Psi_{\samp}$ introduced in Section~\ref{subsec:sampling_subset};
    \item the dynamic data structures for bigram search query, which are introduced in Section~\ref{subsubsec:bis_ds}. 
    These data structures require $O(|\Psi_{\samp}|B)$ bits of space;
    \item the dynamic data structures for RSC query, which are introduced in Section~\ref{subsec:answer_rsc_query}.     
    These data structures require $O((|\Psi_{\samp}| + |\mathcal{U}_{\RR}|) B)$ bits of space. 
\end{enumerate}

These dynamic data structures depend on RLSLP $\mathcal{G}^{R}$. 
For the height $H$ of the derivation tree of RLSLP $\mathcal{G}^{R}$, 
we assume that $H \leq 2(w+1) \log_{8/7} (4n) + 2$ holds to execute queries in $O(\polylog(n))$ time.  
If the RLSLP $\mathcal{G}^{R}$ does not satisfy this assumption, 
then we reconstruct the RLSLP by the restricted recompression until the obtained RLSLP satisfies the assumption. 

\paragraph{The size of our data structure.}
Our data structure requires $O((|\Psi_{\samp}| + |\mathcal{U}_{\RR}|) B)$ bits of space in total. 
Lemma~\ref{lem:samp_basic_property}~\ref{enum:samp_basic_property:4} shows that  
the expected value of $|\Psi_{\samp}|$ is $O(\delta \log \frac{n \log \sigma}{\delta \log n})$ 
for the alphabet size $\sigma$ and substring complexity $\delta$ of string $T$. 
Lemma~\ref{lem:rrdag_node_size} ensures that the expected value of $|\mathcal{U}_{\RR}|$ is $O(\delta \log \frac{n \log \sigma}{\delta \log n})$. 
Therefore, our data structure requires expected $O(\delta \log \frac{n \log \sigma}{\delta \log n} B)$ bits of space. 
If $B = O(\log n)$ holds, 
then our data structure achieves expected $\delta$-optimal space. 

\paragraph{Supporting SA, ISA, RA, and LCE queries.}
RA and LCE queries can be answered in $O(H)$ time and $O(H)$ time, respectively, by Theorem~\ref{theo:rr_dag_summary}.
SA and ISA queries can be answered in $O(H^{3} \log^{2} n + H \log^{6} n)$ time and $O(H^{3} \log n + H \log^{4} n)$ time, respectively, 
by Theorem~\ref{theo:sa_isa_queries}. 
Because of $H = O(\log n)$, 
our data structure can support SA, ISA, RA, and LCE queries in $O(\log^{7} n)$ time, $O(\log^{5} n)$ time, $O(\log n)$ time, and $O(\log n)$ time, respectively. 

\paragraph{Update of our data structure for insertion.}
Consider the string $T^{\prime}$ obtained by inserting a single character into input string $T$. 
In Section~\ref{sec:insertion}, 
we showed that RLSLP $\mathcal{G}^{R}$ can be changed into RLSLP $\mathcal{G}^{R}_{\ins}$ deriving string $T^{\prime}$ 
by the dynamic restricted recompression presented in Section~\ref{subsec:rlslp_update} (Theorem~\ref{theo:update1}). 
After that, 
we showed that the data structures for SA, ISA, RA, and LCE queries can be updated in $O((\max \{H, H^{\prime}, \log (nHH^{\prime}) \})^{8})$ time 
(Theorem~\ref{theo:update_SA_ISA_insertion}) 
for the height $H^{\prime}$ of the derivation tree of RLSLP $\mathcal{G}^{R}_{\ins}$. 
We leverage these results for updating our data structure in $O(\polylog(n))$ time. 

The algorithm updating our data structure consists of the following three steps: 
\begin{enumerate}[label=\textbf{(\Alph*)}]
    \item verify whether $H^{\prime} \leq 2(w+1) \log_{8/7} (4|T^{\prime}|) + 2$ or not 
    in $O(H^{2} w + w^{3} \log (nH))$ time by Theorem~\ref{theo:update1}~\ref{enum:update1:4}. 
    If $H^{\prime} \leq 2(w+1) \log_{8/7} (4|T^{\prime}|) + 2$ holds, 
    then proceed to the next step. 
    Otherwise, this update fails;
    \item change RLSLP $\mathcal{G}^{R}$ to $\mathcal{G}^{R}_{\ins}$ in $O(H^{2} H^{\prime} + H^{\prime 3} \log (nH))$ time by Theorem~\ref{theo:update1}~\ref{enum:update1:2};
    \item update our data structure in $O((\max \{H, H^{\prime}, \log (nHH^{\prime}) \})^{8})$ time by Theorem~\ref{theo:update_SA_ISA_insertion}.
\end{enumerate}
Because of $H, H^{\prime} = O(\log n)$, 
the three steps take $O(\log^{8} n)$ time in total. 
This update fails with probability $O(n^{-w})$  
because Lemma~\ref{lem:tree_height} shows that $H^{\prime} \leq 2(w+1) \log_{8/7} (4|T^{\prime}|) + 2$ holds with probability at least $1 - (1/n^{w})$. 

\paragraph{Update of our data structure for deletion.}
Consider the string $T^{\prime \prime}$ obtained by deleting a single character from input string $T$. 
It is easy to modify the dynamic restricted recompression so that 
RLSLP $\mathcal{G}^{R}$ is changed to an RLSLP $\mathcal{G}^{R}_{\del}$ deriving string $T^{\prime \prime}$. 
After that, 
the data structures for SA, ISA, RA, and LCE queries can be updated using the same approach as for 
an insertion of a single character to the input string. 
Therefore, 
we can update our data structure in $O(\log^{8} n)$ time using the same approach as for insertion. 
Similarly, 
this update fails with probability $O(n^{-w})$. 

\paragraph{Construction.}
We gradually construct our data structure built on input string $T$ 
while inserting $n$ characters $T[2], T[3], T[4], \ldots, T[n]$ to string $T[1..1]$. 
This construction algorithm consists of $n-1$ updates for insertion. 
If an update fails, 
then we execute either of the following two steps: 
\begin{enumerate}[label=\textbf{(\alph*)}]
    \item If the failure occurs while string $T[1..1]$ is changed to $T[1.. \lceil \sqrt{n} \rceil]$, 
    then we restart the construction algorithm of our data structure from the beginning. 
    \item Otherwise (i.e., the failure occurs while $T[1.. \lceil \sqrt{n} \rceil]$ is changed to $T$), 
    the construction algorithm fails. 
\end{enumerate}

The construction time of our data structure depends on the number of updates and restarts. 
The following lemma that 
this construction algorithm takes expected $O(n \log^{8} n)$ time with a high probabilistic guarantee. 

%The following lemma states the construction time. 

\begin{lemma}\label{lem:construction_time}
For input string $T$ of length $n$, 
the construction of our data structure takes expected $O(n \log^{8} n)$ time with probability at least $1 - (1 / n^{\omega(w)})$.
\end{lemma}
\begin{proof}    
    We obtain Lemma~\ref{lem:construction_time} by modifying the proof of Lemma A.1 in \cite{DBLP:journals/corr/GawrychowskiKKL15}. 
    The following three statements are used to prove Lemma~\ref{lem:construction_time}.
    \begin{enumerate}[label=\textbf{(\roman*)}]
    \item the probability that the construction of our data structure restarts at most $\log n^{w}$ times is at least $1 - (1 / n^{w})$; 
    \item the construction of our data structure successes with probability at least $1 - (2 / n^{(w-2) / 2})$.
    \end{enumerate}

    \textbf{Proof of statement (i).}
    %For each $k \geq 0$, 
    Let $a_{0}$ be the probability that the construction algorithm of our data structure does not restart. 
    Then, $a_{0}$ is at least $1 - \sum_{k = 1}^{\infty} \frac{1}{2k^{2}}$ 
    because for each $i \in [1, n-1]$, 
    the insertion of the $(i+1)$-th character $T[i+1]$ into string $T[1..i]$ 
    successes with probability at least $1 - (1/(i+1)^{2w})$. 
    Here, $\sum_{k = 1}^{\infty} \frac{1}{2k^{2}} = \frac{\pi^{2}}{12}$. 
    The probability that the construction algorithm of our data structure restarts at most $\log n^{w}$ times 
    is at least $1 - (1 - a_{0})^{\log n^{w}}$. 
    $(1 - a_{0})^{\log n^{w}} \leq 1 / n^{w}$ follows from the following equation: 
    \begin{equation*}
    \begin{split}
        (1 - a_{0})^{\log n^{w}} &= (\frac{\pi^{2}}{12})^{\log n^{w}} \\
        & \leq (\frac{4}{5})^{\log n^{w}} \\
        & \leq (\frac{4}{5})^{\log_{5/4} n^{w}} \\
        &= \frac{1}{n^{w}}.        
    \end{split}
    \end{equation*}
    Therefore, statement (i) holds. 

    \textbf{Proof of statement (ii).}
    %Consider the last $n-1$ updates used in the construction algorithm of our data structure. 
    If the construction algorithm fails, 
    then an update fails while $T[1.. \lceil \sqrt{n} \rceil]$ is changed to $T$, 
    and it occurs with probability at most $\sum_{i = \lceil \sqrt{n} \rceil + 1}^{n} 1/(i^{2w})$. 
    Because of $\lceil \sqrt{n} \rceil + 1 \geq 2$ and $w \geq 2$, 
    $\sum_{i = \lceil \sqrt{n} \rceil + 1}^{n} 1/(i^{2w}) \leq 1 / (2 n^{(w-2) / 2})$ follows from the following equation:    
    \begin{equation*}
    \begin{split}
        \sum_{i = \lceil \sqrt{n} \rceil + 1}^{n} \frac{1}{i^{2w}} &\leq \sum_{i = \lceil \sqrt{n} \rceil + 1}^{n} \frac{1}{2i^{w}} \\
        & \leq \frac{n}{2 (\sqrt{n})^{w}} \\
        &= \frac{1}{2 n^{(w-2) / 2}}.
    \end{split}
    \end{equation*}

    \textbf{Proof of Lemma~\ref{lem:construction_time}.}
    Consider the case that the construction algorithm of our data structure successes 
    after the algorithm restarts at most $\log n^{w}$ times. 
    In this case, the construction algorithm runs in expected $O(n \log^{8} n)$ time 
    because the construction algorithm is updated at most $n + \sqrt{n} \log n^{w}$ times, 
    and each update takes expected $O(\log^{8} n)$ time.     
    From statement (i) and statement (ii), 
    the probability that the case occurs is at least $(1 - (1 / n^{w}))(1 - (1 / (2 n^{(w-2) / 2})))$. 
    Here, $(1 - (1 / n^{w}))(1 - (1 / (2 n^{(w-2) / 2}))) \geq 1 - (2 / n^{(w-2) / 2})$ follows from the following equation:     
    \begin{equation*}
    \begin{split}
        (1 - \frac{1}{n^{w}})(1 - \frac{1}{2 n^{(w-2) / 2}}) &\geq 1 - \frac{1}{n^{w}} - \frac{1}{2 n^{(w-2) / 2}} \\
        &\geq 1 - \frac{2}{2 n^{(w-2) / 2}}.
    \end{split}
    \end{equation*}
    Therefore, we obtain Lemma~\ref{lem:construction_time}. 
\end{proof}

Finally, we obtain the following theorem. 
\begin{theorem}
    For an input string $T$ of length $n$ with an alphabet size $\sigma$ and a user-defined constant parameter $w \geq 2$, 
    we can build a dynamic data structure satisfying the following four statements: 
\begin{enumerate}[label=\textbf{(\roman*)}]
    \item it requires expected $O(\delta \log \frac{n \log \sigma}{\delta \log n} B)$ bits of space 
    for the substring complexity $\delta$ of $T$ and machine word size $B \geq \log n$. 
    If $B = \Theta(\log n)$, then this data structure can be stored in expected $\delta$-optimal space;
    \item it supports 
    SA, ISA, RA, and LCE queries in $O(\log^{7} n)$ time, $O(\log^{5} n)$ time, $O(\log n)$ time, and $O(\log n)$ time, respectively; 
    \item it can be updated in expected $O(\log^{8} n)$ time for insertion and deletion of a single character in string $T$. 
    This update fails with probability $O(n^{-w})$; 
    \item it can be constructed in expected $O(n \log^{8} n)$ time by processing the input string $T$. 
    This construction fails with probability $O(n^{-(w-2)/2})$.
\end{enumerate}    
\end{theorem}

\subsection{ISA query in static setting.}
In the static setting (i.e., input string $T$ is not changed), 
we can support ISA query in $O(\log^{4} n)$ time by modifying our data structure. 
This result improves upon the previous result of $O(\log^{4+\epsilon} n)$ time achieved by $\delta$-SA~\cite{doi:10.1137/0217026}
for any given constant $\epsilon > 0$. 
Note that our data structure requires expected $\delta$-optimal space, 
whereas their data structure requires strict $\delta$-optimal space. 

We modify our data structure as follows. 
If dynamic updates are not required, 
we can replace the range-sum data structure in our data structure with its static version.
The static range-sum data structure requires $O(m B)$ bits of space and 
supports range-sum query on $m$ weighted points in a two-dimensional grid in $O(\log^{2+\epsilon} n)$ time. 
This range-sum data structure assumes that no two weighted points have identical coordinates, an assumption satisfied by our data structure. 
Replacing the range-sum data structure with its static version can be done in $O(n \log^{8} n)$ time. 

We show that the modified data structure supports ISA queries in $O(\log^{4} n)$ time. 
The main bottleneck of the ISA query is $O(\log n)$ RSC queries, 
and the bottleneck of each RSC query is $O(1)$ range-sum queries, 
which require $O(H^{2} \log n + \log^{4} n)$ time. Here, $O(\log^{4} n)$ represents the running time of range-sum query, 
which can be reduced to $O(\log^{2+\epsilon} n)$. 
The height $H$ is bounded by $O(\log n)$. 
Consequently, we can answer ISA queries in $O(\log^{4} n)$ time in the static setting, yielding the following theorem.

\begin{theorem}
    For an input string $T$ of length $n$ with an alphabet size $\sigma$ and a user-defined constant parameter $w \geq 2$, 
    we can build a static data structure of expected $\delta$-optimal space 
    supporting ISA query in $O(\log^{4} n)$ time. 
    The construction of this data structure takes expected $O(n \log^{8} n)$ time 
    and fails with probability $O(n^{-(w-2)/2})$.
\end{theorem}

\medskip

%Sets the bibliography style to UNSRT and imports the 
%bibliography file "sample.bib".
\bibliographystyle{alphaurl}
\bibliography{ref}

\end{document}